\documentclass[useAMS,usenatbib]{mn2e}

\usepackage[pdftex]{graphicx}
\usepackage{txfonts}
\usepackage{url}

\usepackage{epstopdf}
\usepackage{subfig}
\usepackage{longtable,lscape}

\usepackage[amssymb,mediumspace,thickqspace]{SIunits}
\addunit{\mol}{mol}
\addunit{\erg}{erg}
\addunit{\cm}{cm}
\addunit{\gramm}{g}

\def\arcmin{\hbox{$^\prime$}}
\def\arcsec{\hbox{$^{\prime\prime}$}}

\newcommand{\um}{\,\micro\metre}
\newcommand{\umm}{\,\micro\metre\,}
\newcommand{\pc}{\,{\rm pc}}

\newcommand{\nh}{N_\mathrm{H}}
\newcommand{\sigbg}{\sigma_\mathrm{BG}}

\newcommand{\Fsas}{F_\nu^\mathrm{SAS}}
\newcommand{\ssas}{\sigma_\nu^\mathrm{SAS}}

\newcommand{\Fgau}{F_\nu^\mathrm{Gau}}
\newcommand{\Fpsf}{F_\nu^\mathrm{PSF}}
\newcommand{\sgau}{\sigma_\nu^\mathrm{Gau}}
\newcommand{\spsf}{\sigma_\nu^\mathrm{PSF}}

\newcommand{\Fas}{F_\nu^\mathrm{AS}}

\newcommand{\dunr}{d_\mathrm{unr}}

\newcommand{\Flar}{F_\nu^\mathrm{lar}}

\newcommand{\FNnuc}{F_\nu^\mathrm{nuc}(12\,\mu\mathrm{m})}
\newcommand{\FNsas}{F_\nu^\mathrm{SAS}(12\,\mu\mathrm{m})}

\newcommand{\FNtot}{F_\nu^\mathrm{tot}(12\,\mu\mathrm{m})}

\newcommand{\LN}{\nu L_\nu(12\,\mu\mathrm{m})}

\newcommand{\FQnuc}{F_\nu^\mathrm{nuc}(18\,\mu\mathrm{m})}
\newcommand{\FQsas}{F_\nu^\mathrm{SAS}(18\,\mu\mathrm{m})}

\newcommand{\Rmed}{R^\mathrm{SAS}_\mathrm{AS}}
\newcommand{\Rlar}{R^\mathrm{SAS}_\mathrm{tot}}

\newcommand{\aRmed}{\langle R^\mathrm{SAS}_\mathrm{AS}\rangle}

\newcommand{\spitzerr}{{\it Spitzer}\ }                 
\newcommand{\spitzer}{{\it Spitzer}}                    

\newcommand{\irass}{{\it IRAS}\ }
\newcommand{\iras}{{\it IRAS}}
\newcommand{\isoo}{{\it ISO}\ }
\newcommand{\iso}{{\it ISO}}

\newcommand{\swift}{\textit{Swift}}
\newcommand{\wise}{\textit{WISE}}
\newcommand{\wisee}{\textit{WISE}\ }

\newcommand{\neii}{[Ne\,II]\ }
\newcommand{\nev}{[Ne\,V]\ }
\newcommand{\oiv}{[O\,IV]\ }
\newcommand{\oiii}{[O\,III]\ }

 \title[The subarcsecond MIR view of local AGN]{The subarcsecond mid-infrared view of local active galactic nuclei\\ -- I. The \textit{N}- and \textit{Q}-band imaging atlas 
\thanks{Based on European Southern Observatory (ESO) observing programmes 
60.A-9242, 
074.A-9016,
075.B-0182,
075.B-0621,
 075.B-0631,
 075.B-0727, 
 075.B-0791, 
 075.B-0844, 
 076.B-0194, 
 076.B-0468, 
 076.B-0599, 
 076.B-0621, 
 076.B-0656, 
 076.B-0696, 
 076.B-0743, 
 077.B-0060, 
 077.B-0135, 
 077.B-0137, 
 077.B-0728, 
 078.B-0020, 
 078.B-0173, 
 078.B-0255, 
 078.B-0303, 
 080.B-0240, 
 080.B-0860,
 081.B-0182, 
 082.B-0299, 
 083.B-0239, 
 083.B-0452, 
 083.B-0536, 
 083.B-0592, 
 084.B-0366,
 084.B-0606, 
 084.B-0974, 
 085.B-0251, 
 085.B-0639, 
 086.B-0242, 
 086.B-0257, 
 086.B-0349, 
 086.B-0479, 
 086.B-0919, 
 087.B-0746, 
 382.A-0604, 
 382.B-0732, 
 384.B-0143, 
 384.B-0887, 
 384.B-0943, 
 385.B-0051, 
 385.B-0896, 
 385.B-0981 and 
 386.B-0026.
}}

   \author[D. Asmus et al.]{D.~Asmus,$^{1,2}$\thanks{E-mail: asmus@mpifr-bonn.mpg.de}
          S.~F.~H\"onig,$^{3,2,4}$
          P.~Gandhi,$^{5,6}$        
          A.~Smette$^7$
          and
          W.~J.~Duschl$^{2,8}$\\
         $^1$Max-Planck-Institut f\"ur Radioastronomie, 
           Auf dem H\"ugel 69, D-53121 Bonn, Germany              
         \\
         $^2$Institut f\"ur Theoretische Physik und Astrophysik,
             Christian- Albrechts-Universit\"at zu Kiel, Leibnizstr. 15, D-24098 Kiel, Germany
         \\
             $^3$Dark Cosmology Centre, Niels Bohr Institute, University of Copenhagen, Juliane Maries Vej 30, DK-2100 Copenhagen, Denmark
         \\
             $^4$UCSB Department of Physics, Broida Hall, Santa Barbara, CA 93106-9530, USA
         \\
             $^5$Department of Physics, Durham University, South Road, Durham, DH1 3LE, UK 
         \\
             $^6$Institute of Space and Astronautical Science (ISAS), Japan, Aerospace Exploration Agency,  3-1-1 Yoshinodai, chuo-ku,\\ Sagamihara, Kanagawa 252-5210, Japan
         \\
             $^7$European  Southern Observatory, Casilla 19001, Santiago 19, Chile
         \\
             $^8$Steward Observatory, The University of Arizona, 933 N. Cherry Ave, Tucson, AZ 85721, USA
              }

   \date{Accepted 2014 January 7. Received 2013 December 19; in original form 2013 September 18}

\pagerange{\pageref{firstpage}--\pageref{lastpage}} \pubyear{2013}
\begin{document}

\label{firstpage}

\maketitle

\begin{abstract}
We present the first subarcsecond-resolution mid-infrared (MIR) atlas of local active galactic nuclei (AGN).
Our atlas contains 253 AGN with a median redshift of $z=$ 0.016, and includes all publicly available MIR imaging performed to date with ground-based 8-m class telescopes, a total of 895 independent measurements.
Of these, more than $60\%$ are published here for the first time.
We detect extended nuclear emission in at least 21$\%$ of the objects, while another 19$\%$ appear clearly point-like, and the remaining objects cannot be constrained.
Where present, elongated nuclear emission aligns with the ionization cones in Seyferts.
Subarcsecond resolution allows us to isolate the AGN emission on scales of a few tens of parsecs and to obtain nuclear photometry in multiple filters for the objects. 
Median spectral energy distributions (SEDs) for the different optical AGN types are constructed and individual MIR 12 and 18\umm continuum luminosities are computed.
These range over more than six orders of magnitude. 
In comparison to the arcsecond-scale MIR emission as probed by {\it Spitzer}, the continuum emission is much lower on subarcsecond scales in many cases.
The silicate feature strength is similar on both scales and generally appears in emission (absorption) in type~I (II) AGN.
However, the polycyclic aromatic hydrocarbon emission appears weaker or absent on subarcsecond scales.
The differences of the MIR SEDs on both scales are particularly large for AGN/starburst composites and close-by (and weak) AGN. 
The nucleus dominates over the total emission of the galaxy only at luminosities $\gtrsim 10^{44}\,$erg\,s$^{-1}$.
The AGN MIR atlas is well suited not only for detailed investigation of individual sources but also for statistical studies of AGN unification.  
\end{abstract}


%

\section{Introduction}
Dust plays a key role for our understanding of active galactic nuclei (AGN) and the associated processes in the vicinity of the supermassive black holes. 
It is typically assumed that a large fraction of these objects are surrounded by a dusty obscuring structure, commonly dubbed the ``torus'' \citep{antonucci_spectropolarimetry_1985,antonucci_extended_1985,laing_sidedness_1988,barthel_is_1989,antonucci_unified_1993, urry_unified_1995}.
Depending on the inclination of this torus with respect to the line of sight, in this simple scheme, an observer can see the central parts, in particular the accretion disc, (type~I AGN) or not (type~II AGN). 
The dust in the torus can be directly observed at mid-infrared (MIR) wavelengths.
The majority of past AGN MIR studies used the {\it Infrared Space Observatory} (\textit{ISO};  \citealt{kessler_infrared_1996}) and {\it Spitzer Space Telescope} \citep{werner_spitzer_2004}, which provide superb sensitivity but only limited spatial resolution at arcsecond scales. 
Therefore, the \isoo and \spitzerr data can be significantly contaminated by non-AGN emission, i.e., from circum nuclear star formation.
Ground-based MIR instruments on 8-m class telescopes, on the other hand, provide subarcsecond resolution (here $\lesssim 0.4\arcsec$).  
In spite of their lower sensitivities, an increasing number of recent studies demonstrate the feasibility and power of ground-based MIR AGN observations \citep[e.g., ][]{gorjian_10_2004, grossan_high_2004, siebenmorgen_mid-infrared_2004, galliano_mid-infrared_2005, alonso-herrero_high_2006, horst_small_2006, haas_visir_2007,horst_mid_2008, siebenmorgen_nuclear_2008, horst_mid-infrared_2009, gandhi_resolving_2009, levenson_isotropic_2009, ramos_almeida_infrared_2009, honig_dusty_2010-1, reunanen_vlt_2010, van_der_wolk_dust_2010, alonso-herrero_torus_2011, asmus_mid-infrared_2011, imanishi_subaru_2011, ramos_almeida_testing_2011, mason_nuclear_2012}.
While  star formation can mostly be separated from the AGN with the high angular resolution of the largest telescopes, the AGN itself remains basically unresolved. 
Apart from the dusty torus, several other AGN components can emit significant radiation at MIR wavelengths: the accretion disc (outer: thermal and inner: non-thermal), the jet (non-thermal), and the narrow emission line clouds (thermal).    
Only the latter component has been resolved in a few cases (e.g., NGC\,1068; see \citealt{mason_spatially_2006}).
Therefore, decomposition of the spectral energy distributions (SEDs) and analysis of correlations of the emission at different wavelengths have to be used to disentangle and constrain the MIR emission properties of the various AGN components.
The big advantage of high angular resolution MIR data for this purpose in particular is that no non-AGN components have to be taken into account for the modelling. 
However, because of the large amount of time required for ground-based MIR observations, only small AGN samples have been investigated so far. 
The main goal of this paper is to provide a large and representative AGN sample using all available subarcsecond MIR imaging data.
These are used to build broad-band SEDs and make AGN continuum emission estimates, which will be used  in future works for multiwavelength analysis.

This paper is structured as follows: 
in Section~\ref{sec:sample}, we present the sample selection and properties, while Sections~\ref{sec:obs} and \ref{sec:met} describe the data acquisition and the analysis methods, respectively. 
In Section~\ref{sec:res}, the results are presented starting with the characterization of the subarcsecond MIR morphology and the extraction of the nuclear photometry.
Next, we compare the subarcsecond photometry to the arcsecond-scale emission as seen by \spitzerr in Section~\ref{sec:N/I} as a measure for the circum nuclear, intermediate scales. 
Combining the subarcsecond and arcsecond data, we construct median MIR SEDs and estimate the nuclear 12 and 18\umm continuum in Sections~\ref{sec:SED} and \ref{sec:nuc_emi}.
The last results section (\ref{sec:glob}) presents a comparison of the nuclear to the large-scale (total) MIR emission of the galaxies.
We summarize and conclude this work in Section~\ref{sec:concl}.
Finally, App.~\ref{app:tab}, and \ref{app:indi} contain the data tables and descriptions and results for individual objects, respectively.
All data are available also in electronic form through the Virtual Observatory (VO) hosted at the German Virtual Observatory (GAVO): \url{http://dc.g-vo.org/sasmirala}, and through the Strasbourg astronomical Data Center (CDS).
The object descriptions from App.~\ref{app:indi} are as well available on \url{http://dc.g-vo.org/sasmirala}.
For better legibility, we abridge the longest object identifiers by skipping the declination designator, e.g., 2MASX\,J03565655--4041453 becomes 2MASX\,J03565655.

\section{Sample selection and properties}\label{sec:sample}
We began this work with the uniform sample  of 102 local AGN from \cite{winter_x-ray_2009} as a systematic extension of our previous work \citep{horst_small_2006,horst_mid_2008,gandhi_resolving_2009}.
This sample is flux limited in the hardest X-ray band of 14--195\,keV and has been selected from nine months of observations with the \swift/Burst Alert Telescope (BAT; \citealt{tueller_swift_2008}).
Of these 102 AGN, we observed a subsample of 81 sources with ground-based high angular resolution MIR instruments.
These are the Very Large Telescope (VLT) mounted  Spectrometer and Imager for the Mid-infrared (VISIR; \citealt{lagage_successful_2004}), Gemini/Michelle \citep{glasse_michelle_1997} and the \textit{Subaru} mounted Cooled Mid-Infrared Camera and Spectrometer (COMICS; \citealt{kataza_comics:_2000}). 
In the following, we refer to this sample simply as the ``BAT AGN sample''.
Despite its selection method, the BAT AGN sample underrepresents the Compton-thick obscured objects. 
Furthermore, the rather high lower limit in the 14--195\,keV band of $\gtrsim 2 \times 10^{-11}\mathrm{erg}\,\mathrm{s}^{-1}\mathrm{cm}^{-2}$  excludes most low-luminosity AGN, which represent the majority of the nearby population.
For these reasons, we complemented the BAT AGN sample with all local AGN that have imaging observations available from any high angular resolution MIR instrument with a public archive.
Apart from VLT/VISIR, Gemini/Michelle and \textit{Subaru}/COMICS, this is the Thermal-Region Camera Spectrograph mounted on Gemini-South (T-ReCS;  \citealt{telesco_gatircam:_1998}). 
We searched their archives for observations of galactic nuclei based on the target name in the observation specification. 
Therefore, we cannot rule out that a few objects have been missed despite careful manual checking. 
Non-successful observations due to, e.g., technical failures, have been discarded from this candidate list.

In order to identify the AGN among the observed galactic nuclei we retrieved their optical classifications from the NASA Extragalactic Database (NED) and the AGN catalogue from \cite{veron-cetty_catalogue_2010}.
We discarded all objects without an optical AGN classification. 
Furthermore, this work focuses on the local population of AGN. 
We define local with a redshift $z < 0.4$, because in our selection, there are no Seyferts beyond that redshift but only  QSOs or blazars.
Thus, all objects with $z > 0.4$ were discarded.
These selection criteria provide a total sample of 253 objects, including three double AGN, i.e., merger systems containing two active nuclei (Mrk\,266, NGC\,3690 and NGC\,6240). 
Note that 3C\,321 possibly also contains a double AGN as discussed in App.~\ref{app:3C321}.
The confirmed double AGN are treated separately throughout this work, e.g., Mrk\,266SW and Mrk\,266NE.
Note that we treat NGC\,3393 as a single AGN because the close nuclei are not clearly resolved in the subarcsecond-resolution images (separation$\sim0.5\arcsec$;  \citealt{fabbiano_close_2011}).
We refer to the sample of 253 objects as  the ``total sample''.
It has a median redshift of $z=$ 0.016.
The individual objects are listed in Table~\ref{tab:basic} along with their basic properties, while all references and descriptions  for individual objects are collected in App.~\ref{app:indi}.

For the remainder of this paper, the various AGN types are summarized into the following optical groups:
\begin{description}
 \item -- type~I: Seyferts with broad emission lines (contains Sy\,1, Sy\,1.2, Sy\,1.5, Sy\,1/1.5, Sy\,1.5/L and Sy\,1n);
 \item -- intermediate type Seyferts (type~Ii; contains Sy\,1.8, Sy\,1.9 and Sy\,1.5/2);
 \item -- type~II: Seyferts without broad emission lines (in unpolarized light; contains Sy\,1.8/2, Sy\,1.9/2 and Sy\,2:); 
 \item -- LINERs (low-ionization nuclear emission line regions;  contains L, L:);
 \item -- AGN/starburst composites (AGN/SB comp; contains Cp, Cp:, and L/H; see \citealt{yuan_role_2010}).
\end{description}
The different classifications given in parentheses above are based on the presence and ratios of several optical emission lines (see e.g., \citealt{veron-cetty_catalogue_2010}).
For example, type~I objects have broad Balmer lines.
They are further separated into several Seyfert sub classes depending on the flux ratio of the total H$\beta$ to [O\,III]$\lambda$5007 lines, $R^{\mathrm{H}\beta}_\mathrm{[O\,III]}$.
The intermediate type AGN have $R^{\mathrm{H}\beta}_\mathrm{[O\,III]}<0.33$ and a weak broad component in  at least one Balmer line visible, while in type~II, no broad components are visible for any Balmer line. 
Furthermore, Seyferts are distinguished from AGN/starburst composites and LINERs through their ionization power and hardness as measured by the [O\,III]$\lambda$5007/H$\beta$ and [N\,II] $\lambda$6583/H$\alpha$ line ratios in the BPT \citep{baldwin_classification_1981} diagram \citep{kewley_host_2006,yuan_role_2010}.
In our group definitions, multiple classifications such as Sy\,1/1.5 mean that this object has been classified both as Sy\,1 and Sy\,1.5 by different works in the literature, while the suffix ``:'', e.g., L:, means that the classification is uncertain.
The Cp group includes objects with powerful nuclear starbursts comparable to the AGN, which can possibly affect or even dominate the MIR emission at subarcsecond scales.
Note that many objects have various and partly contradicting classifications in the literature, often caused by differences in either the used slit widths, data quality, or the data modelling. 
Objects with both AGN and circum nuclear or nuclear star formation are especially affected by this.
In particular, there are three objects (NGC\,3660, NGC\,4992 and NGC\,6251) with contradicting optical classifications that prevent us from placing them into any of the five optical groups above.
Note that for only a few objects, intrinsic changes of the optical type have been verified (e.g., Mrk\,1018, \citealt{cohen_variability_1986}; NGC\,7582, \citealt{aretxaga_seyfert_1999}).
In these cases, the most recent type is adopted.

\begin{figure}
   \centering
   \includegraphics[angle=0,width=6cm]{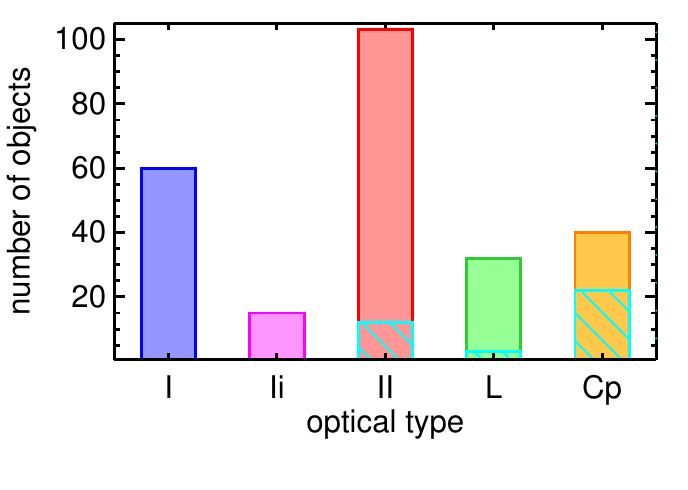}
    \caption{
             Distribution of optical classes for the total sample.
             The corresponding fractions of uncertain AGN  are displayed in as a cyan hatched histogram. 
            }
   \label{fig:type_hist}
\end{figure}

Furthermore, for a number of objects in the total sample, the optical evidence for the presence of an AGN is ambiguous, resulting in both active and non-active nuclear classifications in different published works.
Many AGN/starburst composites (e.g., NGC\,253 and NGC\,1614) and some type~II and LINER objects are affected.
Such objects are treated as ``uncertain AGN'' if they also lack evidence for AGN activity at other wavelengths, e.g., hard X-ray or radio point source.
This is decided on an individual object basis and discussed in App.~\ref{app:indi} for affected objects.
The total sample contains 38 uncertain AGN.
The distribution of the former into the different optical groups is given in Fig.~\ref{fig:type_hist}.

\begin{figure}
   \centering
   \includegraphics[angle=0,width=8cm]{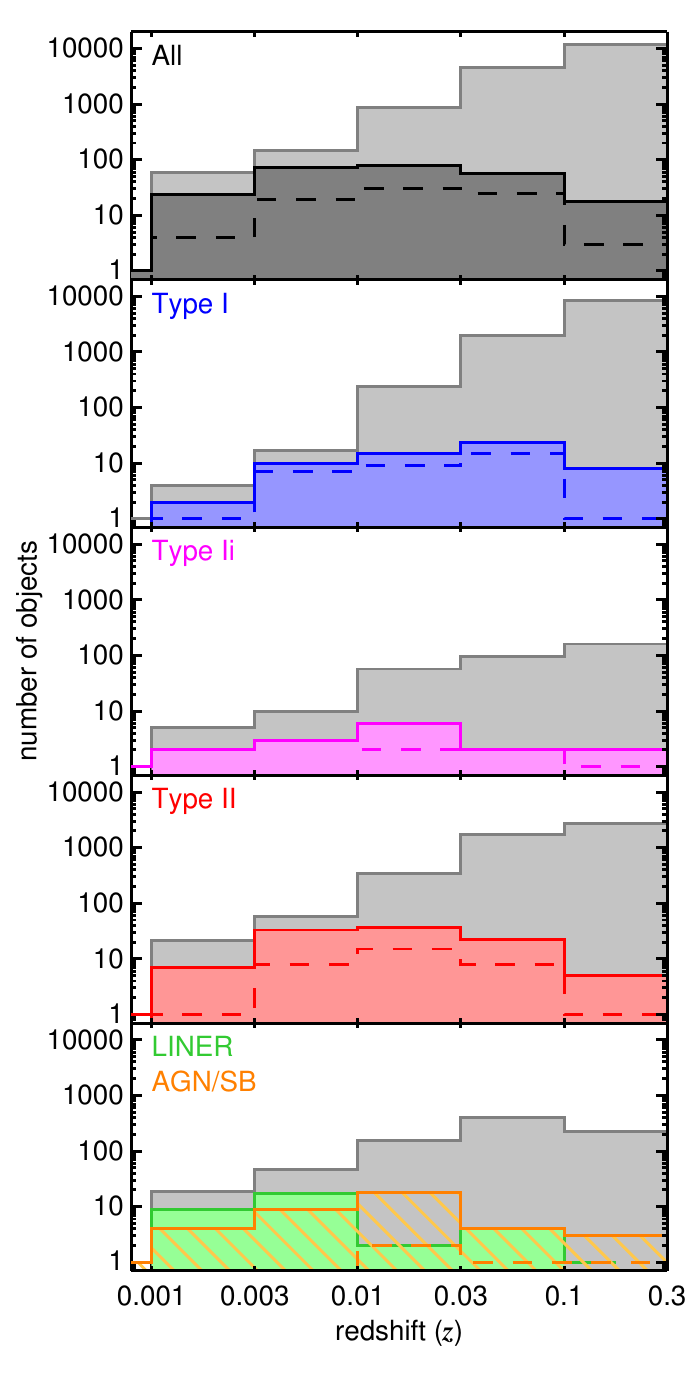}
    \caption{
             Redshift distribution for the total sample (top) and separated by optical classes (below).
             The light grey histograms are the corresponding distributions from the complete \citeauthor{veron-cetty_catalogue_2010} catalogue (2010; except AGN/starburst composites). 
             The dashed-lined empty histograms represent the BAT AGN sample only.         
             }
   \label{fig:opt_comp}
\end{figure}

We display the redshift distribution of the total sample in Fig.~\ref{fig:opt_comp} along with that of the AGN catalogue by \cite{veron-cetty_catalogue_2010}.
This catalogue is a compilation of all optical AGN from the literature and contains almost 170,000 objects. 
Its distribution increases approximately as a power-law with redshift for type~I, Ii and II AGN.
Only for LINERs, the increase with redshift saturates in the examined range, a selection effect caused by the intrinsically lower luminosities of this class.
The majority of AGN at low redshift are type~II AGN, while at higher redshift the majority of AGN are type~I. 
This trend is also present in the total AGN sample. 
Furthermore, up to a redshift of $z=$ 0.01, our sample comprises more than 1/3 of the known total optical AGN population as approximated by the 13th \citeauthor {veron-cetty_catalogue_2010} catalogue: i.e., 79 of 226 AGN in the catalogue with $z=$ 0.01 were observed with subarcsecond MIR imaging.     

Finally, object distances are collected from NED.
For the majority of AGN, the redshift-based luminosity distances are adopted. 
These are  corrected for the Earth's motion relative to the cosmic microwave background reference frame with $H_0 = 67.3, \Omega_\mathrm{m} = 0.315$ and $\Omega_\mathrm{vac} = 0.685$ \citep{planck_collaboration_planck_2013}.
However, for 74 of the closest AGN, redshift-independent measurements  are adopted.
The corresponding references are given in App.~\ref{app:indi} for individual objects.
Table~\ref{tab:sam} provides an overview of object numbers and other properties for both the original BAT sample and the total sample.
\begin{table}
\caption{Overview of sample properties}
\label{tab:sam}      
\begin{center}
\begin{tabular}{l c c c }        
\hline\hline                 
Property/class  & Quantity & BAT &  All \\    
\hline\smallskip  
All objects & number & 81 & 253 \\
Uncertain AGN & number & 0 & 38\\\hline

Type~I & number & 33 & 60 \\\smallskip
Type~Ii & number & 10 & 15 \\\smallskip
Type~II & number & 33 & 103 \\\smallskip
LINER & number &  0 & 32 \\\smallskip
AGN/SB comp.& number & 4 &  40 \\
Unclassified & number & 1 & 3 \\
\hline 
\smallskip Redshift & med$^\mathrm{max}_\mathrm{min}$ & 0.022$^{0.156}_{0.001}$ & 0.016$^{0.357}_{0}$ \\
Distance (Mpc) & med$^\mathrm{max}_\mathrm{min}$ & 96.2$^{781}_{3.8}$ & 71.7$^{1967}_{3.2}$ \\
\hline 
\smallskip Nuc. non-detections & number & 2  & 49 \\
\smallskip Nuc. point-like & number & 23 & 38 \\ 
\smallskip Nuc. extended & number & 15 & 43 \\
\smallskip Nuc. possibly. ext & number & 19 & 53 \\
\smallskip Nuc. ext. unknown & number & 22 & 70 \\
\hline
\smallskip $\FNsas$ (mJy) & med$^\mathrm{max}_\mathrm{min}$ & 129.0$^{1397}_{8.7}$ & 68.7$^{8923}_{1.6}$\\ 
\smallskip $\aRmed$ & med$\pm\sigma$ & 0.86$\pm$0.25 & 0.75$\pm$0.33 \\
\smallskip $\Rlar$ & med$\pm\sigma$ & 0.51$\pm$0.35& 0.32$\pm$0.43\\  
$\log \LN$ (erg\,s$^{-1}$) & med$^\mathrm{max}_\mathrm{min}$ & 43.5$^{44.7}_{39.7}$ & 43.2$^{45.7}_{39.7}$\\
\hline                                   
\end{tabular}
\end{center}

{\it -- Notes:} 
For the nuclear subarcsecond-scale classifications, see Section~\ref{sec:morph}; $\FNsas$: nuclear subarcsecond-scale 12\umm flux density (see Sections~\ref{sec:cont} and \ref{sec:nuc_emi}); $\aRmed$: average ratio of the nuclear subarcsecond to arcsecond flux density (see Section~\ref{sec:N/I}); $\Rlar$: ratio of the nuclear 12\umm subarcsecond to total flux density (see Section~\ref{sec:glob}); $\LN$: nuclear subarcsecond-scale 12\umm luminosity (see Section~\ref{sec:nuc_emi}).    
\end{table}

\section{Data acquisition}\label{sec:obs}
For all selected AGN described in the previous section, we collected all publicly available $N$- and $Q$-band filter images of the COMICS, Michelle, T-ReCS and VISIR instruments that were obtained in standard chopping and nodding mode. 
This included a few acquisition images as well.
Some images were discarded because of observational failures, data reduction failures, or redundancy (in case of observations in better conditions with similar filters). 
The observations were carried out in a large number of different programmes between 2003 December and 2011 June and employed various narrow and intermediate band filters.
They provided a total of 895 independent MIR imaging observations (795 in $N$ band and 99 in $Q$ band), of which 587 are, to our knowledge, unpublished (552 have no AGN flux measurements published).
Six images obtained contain double AGN.
All observations are listed along with their relevant properties in Table~\ref{tab:obs}.

Usually, a photometric standard star was observed within the two hours prior to or following each science target for calibration purposes.
These standard stars were selected from the MIR standard star catalogue from \cite{cohen_spectral_1999}.
The individual observatories provide corresponding absolute fluxes in their various filters. 
The given uncertainties on these values are $\le 10\,\%$ and often even much lower.

In general, the science targets were observed at a median airmass of 1.17 (standard deviation $\sigma = 0.24$) and a median MIR image quality of 0.36\arcsec\, ($\sigma = 0.12$) in the $N$-band and 0.53\arcsec\, ($\sigma = 0.15$) in the $Q$-band as measured from the full width at half maxima (FWHMs) of the standard stars. 
The following sections describe specific information for each instrument.

\subsection{VISIR observations}\label{sec:vobs}
The majority of subarcsecond-resolution MIR images were obtained with VISIR (622, of which 48 are in $Q$-band), which correspond to 171 AGN from the total sample and most objects from the BAT AGN sample (73).
The VISIR programmes included in this work are listed in Table~\ref{tab:vobs} with the principal investigator, number of images, number of objects and the corresponding wavelength bands of the filters used. 

VISIR was mounted at the Cassegrain focus of the UT3 telescope of the ESO-VLT on Cerro Paranal, Chile.
Observations were carried out either in parallel or perpendicular chopping-and-nodding mode  with different chopping throws, but Vmostly of 8\arcsec.
Such small chopping throws are usually not problematic because the diffuse host galaxy contribution in the MIR is often less than 10$\%$ at a distance of 8\arcsec\ from the nucleus and is therefore negligible for our investigation of the nuclear point sources.
In addition, the density of compact bright star-forming knots around the nuclei is mostly very low in the MIR. 
Therefore, chopping-induced over subtraction is very unlikely.

Most images were obtained using the small field of view (0.075\arcsec/pixel; $19.2\arcsec \times 19.2\arcsec$).
Only a few used the large one (0.127\arcsec/pixel; $32.5\arcsec \times 32.5\arcsec$).
Both fields of view used the same DRS $256 \times 256$ detector. 
Depending on the length of the chop throw, the maximum size of the sub images without overlap was restricted, e.g.,  for a throw of 8\arcsec\, to the inner 4\arcsec\, around the centre of the images. 

The sensitivity of a VISIR observation was measured as part of the standard calibration plan during instrument operation from the standard star before or after the science target. 
For most observations, the achieved sensitivities were better than, or equal to, the median values given in the VISIR manual for the corresponding filters.
The sensitivity was $>30\%$ worse than median only in 27 cases ($3.8\%$), i.e.,  the ambient conditions were poor for those observations. 
In addition, a count-to-flux conversion factor was computed for each standard star observation, which in general stayed constant within a $\sigma$ of $\sim 10\%$. 
However, the Q1 and Q2 filters showed a larger scatter ($\sim 25\%$), which was caused by the stronger dependency on the column density of precipitable water vapour in the atmosphere.
There were no strong outliers with respect to the conversion factors. 

The data reduction was performed with the ESO delivered VISIR pipeline using default parameters. 

\begin{table}
\caption{VISIR programmes used} 
\label{tab:vobs} 
\centering 
\begin{tabular}{l l c c c} 
\hline\hline 

Prog. ID & PI & \#im & \#obj & Bands \\ 
\hline 
60.A-9242	&	Lagage	&	16	&	13	&	N,Q	\\
074.A-9016	&	Lagage	&	4	&	2	&	N	\\
075.B-0182	&	Schaerer	&	10	&	2	&	N	\\
075.B-0621	&	Barthel	&	3	&	3	&	N	\\
075.B-0631	&	Wold	&	1	&	1	&	N	\\
075.B-0727	&	Snijders	&	3	&	2	&	N,Q	\\
075.B-0791	&	Van der Werf	&	2	&	1	&	N	\\
 075.B-0844	&	Horst	&	25	&	8	&	N	\\
 076.B-0194	&	Barthel	&	2	&	2	&	N	\\
 076.B-0468	&	Duc	&	4	&	1	&	N	\\
 076.B-0599	&	Reunanen	&	14	&	6	&	N,Q	\\
 076.B-0621	&	Jaffe	&	2	&	1	&	N	\\
 076.B-0656	&	Wold	&	4	&	3	&	N,Q	\\
 076.B-0696	&	Wold	&	8	&	4	&	N	\\
 076.B-0743	&	Jaffe	&	7	&	1	&	N	\\
 077.B-0060	&	Boeker	&	5	&	2	&	N,Q	\\
 077.B-0135	&	Barthel	&	6	&	6	&	N	\\
 077.B-0137	&	Horst	&	50	&	21	&	N	\\
 077.B-0728	&	Reunanen	&	19	&	9	&	N,Q	\\
 078.B-0020	&	Barthel	&	11	&	11	&	N	\\
 078.B-0173	&	Snijders	&	1	&	1	&	Q	\\
 078.B-0255	&	Tristram	&	6	&	2	&	N	\\
 078.B-0303	&	H\"onig	&	12	&	3	&	N,Q	\\
 080.B-0240	&	H\"onig	&	10	&	3	&	N,Q	\\
 080.B-0860	&	Horst	&	71	&	16	&	N	\\
 081.B-0182	&	D\'iaz-Santos	&	4	&	1	&	N	\\
 082.B-0299	&	H\"onig	&	7	&	2	&	N	\\
 382.A-0604	&	Treister	&	40	&	8	&	N	\\
 382.B-0732	&	H\"onig	&	11	&	3	&	N,Q	\\
 083.B-0239	&	H\"onig	&	7	&	3	&	N	\\
 083.B-0452	&	Kishimoto	&	12	&	5	&	N	\\
 083.B-0536	&	Asmus	&	30	&	17	&	N	\\
 083.B-0592	&	D\'iaz-Santos	&	4	&	1	&	N	\\
 084.B-0366	&	Kishimoto	&	11	&	6	&	N	\\
 084.B-0606	&	Asmus	&	118	&	41	&	N	\\
 084.B-0974	&	Gonz\'alez-Mart\'in	&	11	&	9	&	N	\\
  384.B-0143	&	H\"onig	&	2	&	1	&	N	\\
 384.B-0887	&	Asmus	&	2	&	1	&	N	\\
 384.B-0943	&	Reunanen	&	16	&	14	&	N	\\
 085.B-0251	&	H\"onig	&	12	&	5	&	N	\\
 085.B-0639	&	Asmus	&	20	&	13	&	N,Q	\\
  385.B-0051	&	H\"onig	&	5	&	1	&	N	\\
 385.B-0896	&	Tristram	&	4	&	1	&	N	\\
 385.B-0981	&	Van der Wolk	&	3	&	3	&	N	\\
 086.B-0242	&	H\"onig	&	20	&	2	&	N	\\
 086.B-0257	&	Tristram	&	4	&	1	&	N	\\
 086.B-0349	&	Asmus	&	15	&	7	&	N,Q	\\
 086.B-0479	&	Asmus	&	2	&	1	&	N	\\
 086.B-0919	&	Tristram	&	4	&	2	&	N	\\
  386.B-0026 	&	Tristram	&	6	&	1	&	N	\\
 087.B-0746	&	Tristram	&	4	&	1	&	N	\\

\hline 
\end{tabular}
\end{table}

\subsection{Michelle observations}\label{sec:mobs}
Michelle was used to observe 33 AGN from the total sample, of which 10 are BAT AGN, with 62 $N$- and $Q$-band images. 
The corresponding programmes  are listed in Table~\ref{tab:mobs}. 

Michelle was mounted at the Cassegrain focus of the Gemini-North Telescope on Mauna Kea in Hawaii.
It possessed a Si:As IBC detector with a format of $320\times240$ pixels and one field of view for imaging mode (0.1005\arcsec/pixel; $32\arcsec \times 24\arcsec$).
All observations were performed in parallel chopping and nodding mode while only the central (double) beam was guided. 
Therefore, the negative sub images were not used, apart from NGC\,3690W, which was only visible as a negative image in the same frame as NGC\,3690E (see App.~\ref{app:indi} for further details on this system). 
The chop was typically between 10 and 15\arcsec, and the conversion factors computed from the standard stars were constant within $10\%$, with the largest variations again in the $Q$-band (Qa filter).

Data reduction was performed using the \textsc{iraf}-based \textsc{midir}
pipeline provided by the Gemini Observatory.

\begin{table}
\caption{Michelle programmes used} 
\label{tab:mobs} 
\centering 
\begin{tabular}{l l c c c} 
\hline\hline 

Prog. ID & PI & \#im & \#obj & Bands \\ 
\hline 
  GN-2003B-Q-72      	&	Beck	&	2	&	2	&	N	\\
 GN-2006A-Q-11      	&	Packham	&	8	&	4	&	N,Q	\\
 GN-2006A-Q-30      	&	Mason	&	4	&	2	&	N	\\
 GN-2006B-Q-18      	&	Packham	&	1	&	1	&	N	\\
 GN-2006B-Q-19      	&	Perlman	&	2	&	1	&	N	\\
 GN-2007A-Q-49      	&	Mason	&	1	&	1	&	N	\\
 GN-2007A-Q-93      	&	Mason	&	2	&	2	&	N	\\
 GN-2008A-Q-43      	&	Mason	&	8	&	8	&	N	\\
 GN-2009B-Q-61      	&	Pastoriza	&	1	&	1	&	N	\\
 GN-2010A-C-7       	&	Gandhi	&	23	&	13	&	N	\\
 GN-2011A-Q-55      	&	Mason	&	10	&	3	&	N,Q	\\

\hline 
\end{tabular}
\end{table}

\subsection{T-ReCS observations}\label{sec:tobs}
The second largest fraction of AGN observations were performed with T-ReCS, namely 70 AGN from the total sample (17 BAT AGN) with 124 images.
Of these images, 39 are in $Q$-band (Table~\ref{tab:tobs}).

T-ReCS, formerly mounted on the Gemini-South telescope on Cerro Pachon in Chile, is very similar to Michelle and uses a Raytheon $320\times 240$ pixel Si:As IBC detector.
T-ReCS also observed by default  in parallel chop/node mode with guiding only in the centre position. 
The typical chop throws were between 10\arcsec\, and 15\arcsec, while the imaging field of view was a bit smaller than that of Michelle (0.09\arcsec/pixel; $28.8\arcsec \times 21.6\arcsec$).
The flux conversion factors of the T-ReCS standard star observations showed standard deviations of the order of $20\%$ for the N, Qa and Si-6 filters and $<10\%$ for the Si-2 and Si-5 filters.

Similar to Michelle, the \textsc{midir}
pipeline was used for the T-ReCS data reduction.

\begin{table}
\caption{T-ReCS programmes used} 
\label{tab:tobs} 
\centering 
\begin{tabular}{l l c c c} 
\hline\hline 

Prog. ID & PI & \#im & \#obj & Bands \\ 
\hline 
 GS-2003B-DD-4      	&	Packham	&	8	&	4	&	N,Q	\\
 GS-2003B-Q-44      	&	Lira	&	1	&	1	&	N	\\
 GS-2004A-C-2       	&	Roche	&	6	&	4	&	N	\\
 GS-2004A-Q-14      	&	Lira	&	14	&	14	&	N	\\
 GS-2004A-Q-41      	&	Matthews	&	3	&	1	&	N,Q	\\
 GS-2005A-Q-50      	&	Lumsden	&	2	&	2	&	N	\\
 GS-2005A-Q-6       	&	Radomski	&	7	&	3	&	N,Q	\\
 GS-2005B-DD-6      	&	Radomski	&	3	&	2	&	N,Q	\\
 GS-2005B-Q-10      	&	Packham	&	2	&	2	&	N	\\
 GS-2005B-Q-29      	&	Mason	&	2	&	1	&	N,Q	\\
 GS-2006A-Q-62      	&	Radomski	&	2	&	1	&	N,Q	\\
 GS-2006B-Q-9       	&	Packham	&	2	&	1	&	N	\\
 GS-2007A-DD-7      	&	Radomski	&	10	&	8	&	Q	\\
 GS-2007A-Q-40      	&	Perlman	&	5	&	2	&	N,Q	\\
 GS-2007B-Q-203     	&	Mason	&	16	&	8	&	N,Q	\\
GS-2007B-Q-215  	&	Perlman	&	2	&	1	&	N	\\
 GS-2008A-Q-55      	&	Roche	&	5	&	2	&	N	\\
 GS-2008A-Q-58      	&	Perlman	&	1	&	1	&	N	\\
 GS-2008B-C-2       	&	Imanishi	&	4	&	4	&	Q	\\
 GS-2008B-Q-36      	&	Radomski	&	3	&	3	&	N	\\
 GS-2008B-Q-46      	&	Binette	&	2	&	1	&	N	\\
 GS-2008B-Q-63      	&	Melbourne	&	1	&	1	&	N	\\
 GS-2009A-Q-59      	&	Radomski	&	2	&	1	&	N,Q	\\
 GS-2009B-C-3       	&	Imanishi	&	4	&	3	&	Q	\\
 GS-2009B-Q-43      	&	Levenson	&	6	&	3	&	N,Q	\\
 GS-2010A-Q-10      	&	Mason	&	1	&	1	&	N	\\
 GS-2010B-C-4       	&	Imanishi	&	1	&	1	&	Q	\\
 GS-2010B-Q-3       	&	Bauer	&	1	&	1	&	N	\\
 GS-2010B-Q-71      	&	Levenson	&	8	&	4	&	N,Q	\\

\hline 
\end{tabular}
\end{table}

\subsection{COMICS observations}\label{sec:cobs}
The fourth instrument used, COMICS, provides 39 $N$- and $Q$-band images for 33 different AGN in total (3 BAT AGN).
Table~\ref{tab:cobs} lists the corresponding observing programmes. 

COMICS is mounted on the \textit{Subaru} Telescope, Mauna Kea in Hawaii, also at the Cassegrain focus.
It has  the largest field of view and pixel scale in imaging mode (0.13\arcsec/pixel; $42\arcsec \times 32\arcsec$). 
It possesses a $320 \times 240$ pixel Si:As BIB detector for imaging.
As opposed to the other instruments, it does not have a fixed chop/nod pattern.
The nod parameters are manually chosen by the observer. 
In addition, only parts of the detector can be optionally read out. 
Most observations used small chopping throws of 10\arcsec\, and the N11.7 and Q17.7 filters.
For those, the conversion factors were constant to within $10$ and $15\%$, respectively.

We reduced the COMICS raw data with our own \textsc{idl} pipeline, which simply combines a given set of chopped images into one final image. 
Before combining, compromised images (high background, strong residuals, etc.) were excluded.

\begin{table}
\caption{COMICS programmes used} 
\label{tab:cobs} 
\centering 
\begin{tabular}{l l c c c} 
\hline\hline 

Prog. ID & PI & \#im & \#obj & Bands \\ 
\hline 
 o00001             	&	Yamashita	&	2	&	1	&	N	\\
 o05107             	&	Packham	&	1	&	1	&	N	\\
 o05405             	&	Tomono	&	2	&	1	&	N	\\
 o06139             	&	Imanishi	&	7	&	7	&	N	\\
 o07146             	&	Imanishi	&	3	&	3	&	Q	\\
 o09102             	&	Imanishi	&	7	&	5	&	N,Q	\\
 o09108             	&	Gandhi	&	17	&	15	&	N	\\

\hline 
\end{tabular}
\end{table}

\subsection{Complementary \textit{Spitzer}
data}\label{sec:spi}
We complement our ground-based MIR data with \textit{Spitzer Space Telescope} data from the archive.
Specifically,  $5.8$ and $8.0\,\mu$m images from the Infrared Array Camera  (IRAC; \citealt{fazio_infrared_2004}) and $24\,\mu$m images from the Multiband Imaging Photometer (MIPS; \citealt{rieke_multiband_2004} and spectra from the  Infrared Spectrograph (IRS; \citealt{houck_infrared_2004}) are used as probes for the arcsecond-scale structures with $\sim 4\arcsec$ resolution.
For the purpose of this work, the post-basic calibrated data (PBCD) products with flux uncertainties of the order of $10\%$ are sufficient.
The IRAC flux uncertainty for most objects is actually much lower and dominated by the systematics of the absolute flux calibration of $\sim3\%$ \citep{reach_absolute_2005}. 
All together, 232 out of 253 objects ($92\%$) from the total sample have \spitzerr coverage with at least one of its instruments.

\subsubsection{IRAC and MIPS photometry}
IRAC 5.8 and $8.0\,\mu$m images are available for 192 of the 253 AGN ($76\%$).
However, the nuclear source in the IRAC $5.8\,\mu$m PBCD images is saturated in 10 cases and 29 cases in the IRAC 8.0\umm band. 
MIPS 24\umm PBCD images  are available for 192 objects ($76\%$) and only in two cases are the nuclei saturated (Circinus and NGC\,253). 
Owing to the space-borne nature of these observations with a superb sensitivity, many more emission sources are usually visible in the images compared to ground-based MIR images. 
This is particularly the case for the nearby galaxies, for which the intrinsic resolution is sufficient to resolve the galactic structures  (94 out of 210 objects, $45\%$).
Following mainly the host galaxy morphology in the optical, the objects show either spiral-like MIR clumpy structures or elliptical diffuse emission around the nuclei. 
The IRAC and MIPS images are shown and discussed for individual objects in App.~\ref{app:indi}.

\subsubsection{IRS spectra}\label{app:irs}
IRS PBCD spectra could be retrieved for 226 of 253 objects ($89\%$) and are mainly in low-resolution staring mode with automatic background subtraction (159).
A further 59 objects were observed in low-resolution mapping mode.
We performed the background subtraction manually for these by using the spectra extracted from the off-nuclear positions.
Finally, seven objects only have high-resolution staring mode spectra available, for which no further background subtraction was performed (ESO\,253-3, NGC\,613, NGC\,1068, NGC\,3166, NGC\,3169, NGC\,3368 and NGC\,7626).
All PBCD spectra were computed using an automatic point-source extraction, which can be problematic with extended and complex sources. 
Therefore, we compared the IRS spectra with the better-isolated nuclear photometry of IRAC and MIPS and scaled the spectra with this photometry  whenever significant offset between both occurred.
However, 30 objects with IRS spectra ($14\%$) had no IRAC or MIPS photometry available.
The IRS spectra are shown and discussed for  individual objects in App.~\ref{app:indi}.

\section{Nuclear size and flux determinations}\label{sec:met}
\subsection{Image analysis}\label{sec:ana}
The following analysis of the reduced images was performed uniformly for all instruments using a custom developed \textsc{idl} software package, \textsc{mirphot}.
In this package, both science target and standard star data are processed in exactly the same way in order to ensure consistency. 
First, additional image cleaning is done: a sigma-clipping and, optionally, a row/column median cleaning for readout-induced artefacts.
\textsc{mirphot} carries out two different flux measurements, which are based on fitting of the nuclear compact sources in the negative and positive sub images with a 2D Gaussian plus a constant term.
For the fitting, the \textsc{mpfit} \textsc{idl} package is utilized \citep{markwardt_non-linear_2009}.  
The first resulting flux is then simply the integral of the fitted 2D Gaussian without the constant term, which is then  normalized with the conversion factor that has been obtained from the standard star in the same way.
Although the point spread function (PSF) of the telescopes is not exactly a Gaussian, this does not significantly affect the measurement because standard star and science target are treated equally.
In the following, this flux is called the ``Gaussian flux'', $\Fgau$, and is comparable to fluxes obtained with optimal aperture photometry.
The advantage of this method is that it isolates the nuclear component well from the local background.
However, we constrain the FWHM of the fit to be  $\le 1\arcsec$  in cases of significant extended emission, because the latter could otherwise dominate the fit.
Still, the Gaussian flux might not represent the AGN emission well in such cases.
Therefore, we adopt a second flux measurement, called the ``PSF flux'', $\Fpsf$, which is based on the same Gaussian fit but uses the FWHM from the standard star instead. 
For this reason, $\Fpsf$, can be a better measurement for the unresolved component in an extended nucleus (and also for some faint detections).
In addition, the peak value can be scaled manually, such that the residual of the remaining extended emission becomes ``flat''. 
This technique is similar to those applied in, e.g., \cite{ramos_almeida_infrared_2009} and \cite{reunanen_vlt_2010}.
For compact sources with no significant extended emission, both measurements, $\Fgau$ and $\Fpsf$, provide similar results. 
Note, however, that the PSF in ground-based MIR imaging can be quite variable mainly due to rapidly changing ambient conditions (e.g., \citealt{radomski_gemini_2008}).
This has led in several cases to broader FWHMs for the standard star than for the corresponding science objects (see Section~\ref{sec:res}). 
Therefore, comparisons between the beam shapes of science target and standard star can be misleading and the PSF fluxes are not always reliable. 
We use $\Fpsf$ only for clearly extended objects as determined in Section~\ref{sec:morph}.

For the VISIR and COMICS images, usually all sub images (positive and negative) can be used and a statistical error can be derived by comparing the fluxes measured from the single beams for each target. 
In case of low signal-to-noise ratio (S/N) detections in VISIR images, the sub images can be precombined before the flux measurements are performed.
In practice, this often increases the S/N, in particular for perpendicularly chopped images.
All VISIR and COMICS sub images were postcombined using the fitted beam positions. 
In general, the FWHM for the nuclear sources in VISIR and COMICS images are either taken from the pre- or postcombined images --whichever provides the higher S/N. 
Note that the usage of the negative sub images for VISIR and COMICS does not provide an advantage in terms of S/N compared to the positive double sub images of T-ReCS and Michelle owing to the background-limited nature of ground-based MIR observations (e.g., CanariCam manual, Sections~II.III). 

For non-detected objects, \textsc{mirphot} calculates upper limits using a simulated Gaussian-like source with the FWHM of the standard star and a peak height equal to three times the standard deviation of the local background, $\sigbg$.
These upper limits usually agree with expectations based on the exposure time and the sensitivity measured from the standard star.

The conversion factor was not corrected for large airmass differences between the standard star and science, which occurred only for few observations in the sample.
On the other hand, the FWHM increases with increasing airmass, which is particularly significant for those objects observed at an airmass of $\sim 2$, e.g., NGC\,5548, where the FWHM is increased by approximately $15\%$ for both $N$- and $Q$-band observations. 
The individual values of $\Fgau$ and $\Fpsf$ are listed in Table~\ref{tab:obs}.

We used \textsc{mirphot} as well to measure the nuclear sources in the non-saturated IRAC and MIPS PBCD images, if present.
For these flux measurements, $\Fas$, corresponding correction factors for losses due  to the Gauss fitting were applied, $\Fas = f_\mathrm{cor} \Fgau$ ($f_\mathrm{cor}(\mathrm{I}\,5.8\,\um)=1.271 \pm 0.032$; $f_\mathrm{cor}(\mathrm{I}\,8.0\,\um)=1.415 \pm 0.023$; $f_\mathrm{cor}(\mathrm{M}\,24\,\um)=1.516 \pm 0.019$).
These factors were computed from IRAC and MIPS standard star observations and allow us to reconstruct the source flux with an accuracy of $\lesssim 3\%$.
For 38 mostly nearby objects, the nuclear source is dominated by extended emission without a clear unresolved component. 
In those cases, we simply measured the flux with an aperture of 4\arcsec\, in diameter for IRAC and 7\arcsec\, for MIPS images and applied corresponding correcting factors for aperture losses ($f_\mathrm{cor}(\mathrm{I}\,5.8\,\um)=1.604 \pm 0.011$; $f_\mathrm{cor}(\mathrm{I}\,8.0\,\um)=1.783 \pm 0.017$; $f_\mathrm{cor}(\mathrm{M}\,24\,\um)=3.070 \pm 0.052$).

\subsection{Extension determination}\label{sec:ext}
The extension of the detected nuclei can be quantified by investigating the shape of the corresponding Gaussian fits, e.g., the ratio of the FWHMs of the science to calibration objects.
For this purpose, we distinguish between four different classifications:
\begin{description}
 \item -- point-like nucleus: without extended emission;
 \item -- unknown nuclear extension: insufficient data;
 \item -- possibly extended nucleus: marginal extension, or significant extension but only one epoch available;
 \item -- extended nucleus: significant and consistent extension in multiple epochs.
\end{description}
The distinction between extended and possibly extended is necessary because of the possibly  unstable PSFs during ground-based MIR observations as already mentioned in Section~\ref{sec:ana}.
A number of instrumental factors can generally affect the measured FWHM of the sources. 
The most important are:
technical uncertainties, e.g., the secondary mirror not returning to exactly the same position after chopping or nodding,
changing weather conditions, in particular for long observations or long time differences between standard star and science target,
and finally the S/N of the detection.
In order to make the estimates robust against these effects, we chose a significance threshold of $10\%$, i.e., if the FWHM of the science target is $>10\%$ larger than that of the standard star, the detected nucleus is significantly extended in that individual observation. 
Furthermore, the detected nucleus is regarded as marginally extended if the FWHM(science target) $- 2\sigma$(FWHM) $>$ FWHM(standard star) with $\sigma$(FWHM) being the measurement uncertainty (which is in many cases $\sim1\%$).
Otherwise, the nucleus remained unresolved.
However, an object is only classified as point-like if there is at least one unresolved detection with an FWHM equal to or smaller than the median FWHM of all standard stars taken in this filter.
All the remaining objects are classified as ``unknown extension''.
The nuclear extension determination as described here is applied in Section~\ref{sec:nuc_ext}.

\subsection{Continuum emission estimation}\label{sec:cont}
For, e.g., multiwavelength analysis and comparisons, we need to unify the very heterogeneous photometry of different objects into one or two characteristic flux values.
The nuclear subarcsecond-scale continuum flux densities at 12 and 18\umm restframe wavelengths, $\FNsas$ and $\FQsas$ are well suited as uniform brightness measures for the sample. 
Here, the following recipe is used, consisting of the following methods with decreasing priority: 
\begin{enumerate}
 \item using the measurement-uncertainty weighted mean of nuclear fluxes, $\Fgau$ (or $\Fpsf$), of filters with matching central wavelengths directly (for $\FNsas$ and $\FQsas$); 
 \item  using synthetic fluxes for objects that have MIR spectra available and exhibit similar flux levels in the spectra and the subarcsecond-scale photometry (for $\FNsas$ and $\FQsas$); 
 \item interpolation between the weighted means of $\Fgau$ (or $\Fpsf$) of filters with  central wavelengths shortwards and longwards (for $\FNsas$ and $\FQsas$);  
\item  or extrapolation of those measurements spectrally closest to the desired restframe wavelength (only for $\FNsas$).
\end{enumerate}
For method (iv), the local spectral slopes from the corresponding median MIR SEDs from the previous section were used, and their dispersion contributed to the final flux uncertainty.
The extrapolation method has to be used in particular for those cases, where, no subarcsecond-resolution $N$-band photometry is available but only $Q$-band photometry and method (ii) can not be used.   
Note that for the computation of $\FQsas$, we only use direct filter measurements or synthetic photometry from spectra, i.e., methods (i) and (ii), because of the large dispersion seen in the MIR SEDs in this range. 
For objects that remained undetected in the subarcsecond-resolution observations, the \spitzer/IRS spectra can possibly provide more constraining limits.
In these cases, IRS spectra are used for the calculation of the corresponding upper limit on $\FNsas$ and $\FQsas$ according to method (ii).

Furthermore, strong silicate $\sim10$  and $\sim18\um$ features due to  Si--O stretching and O--Si--O bending, respectively (e.g., \citealt{draine_interstellar_2003}) can significantly affect the measured $\FNsas$ and $\FQsas$. 
The nuclear continuum estimates can only be corrected for this effect with the help of spectra covering the silicate features and surrounding wavelength regions.
In those cases, we estimate the underlying continuum emission by interpolating the surrounding continuum regions at 5.5 and 13.5\umm for $\FNsas$ and 13.5 and 21\umm for $\FQsas$.
Because $\FNsas$ and $\FQsas$ are directly estimated in the object rest frames, no $K$-correction is necessary.
The individual values of $\FNsas$ and $\FQsas$ are listed in Table~\ref{tab:basic} and are used in Sections~\ref{sec:nuc_emi} and \ref{sec:glob}.

\section{Results}\label{sec:res}
\begin{figure*}
   \centering
   \includegraphics[angle=0,width=18cm]{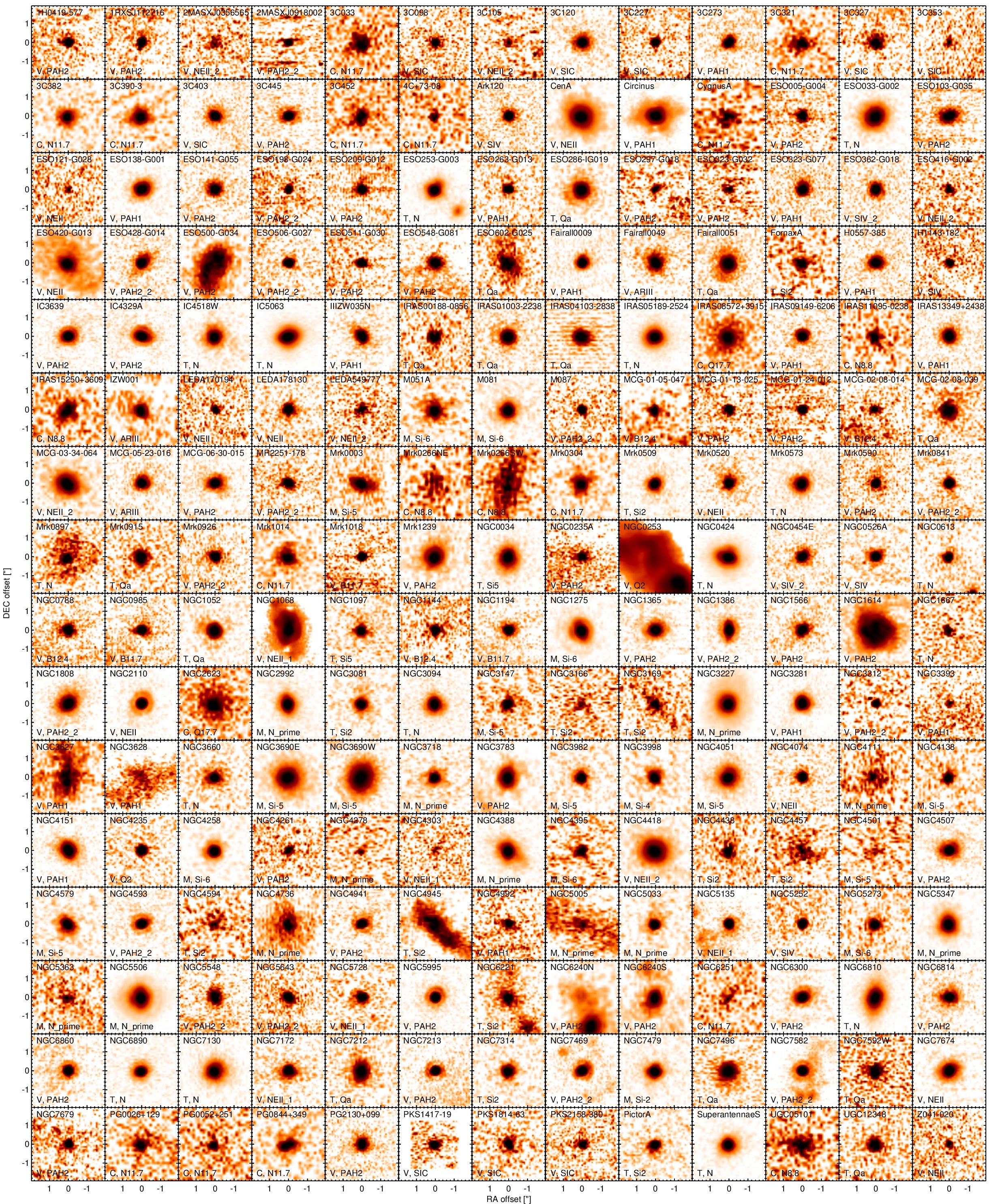}
    \caption{
             Subarcsecond-resolution MIR images of all objects with detected emission. 
             For each object only the image with the  highest-S/N is shown.
             Displayed are the inner $4\arcsec$ with North up and East to the left. The colour scaling is logarithmic with white corresponding to median background and black to the $1\%$ of pixels with the highest flux values.
             The labels in the bottom left state instrument and filter names (C: COMICS, M: Michelle, T: T-ReCS, V: VISIR).            
            }
   \label{fig:ima_col}
\end{figure*}

The highest S/N images of all objects with detected emission are displayed in Fig.~\ref{fig:ima_col}.
In 204 ($81\%$) of the 253 observed galactic nuclei, compact emission sources were detected ($> 3\sigma$). 
In four cases ($2\%$), MIR emission structures were detected, but  a clear compact nuclear component was not found (ESO\,500-34, NGC\,253, NGC\,1614 and NGC\,3628).
No MIR emission at all was detected in the rest of the cases (45 objects, $17\%$). 
From the BAT AGN, only two objects were not detected (NGC\,612 and LEDA\,13946). 
Out of the 49 objects without clearly detected nuclei, 21 ($43\%$) are uncertain AGN. 
The reduced images are available in electronic form through the VO and  \url{http://dc.g-vo.org/sasmirala}.

\subsection{Morphology at subarcsecond resolution}\label{sec:morph}
The MIR subarcsecond-resolution images are shown in App.~\ref{app:indi} for individual objects.
The circum nuclear morphology  was first assessed by eye for the detected objects. 
In the majority of cases, no non-nuclear emission was detected in the central 4\,arcsecond region.
However, a number of objects show extended emission, which we separated into four groups according to the apparent shape of the extended emission (see Fig.~\ref{fig:ima_ext}): 
\begin{figure*}
   \centering
   \includegraphics[angle=0,width=15cm]{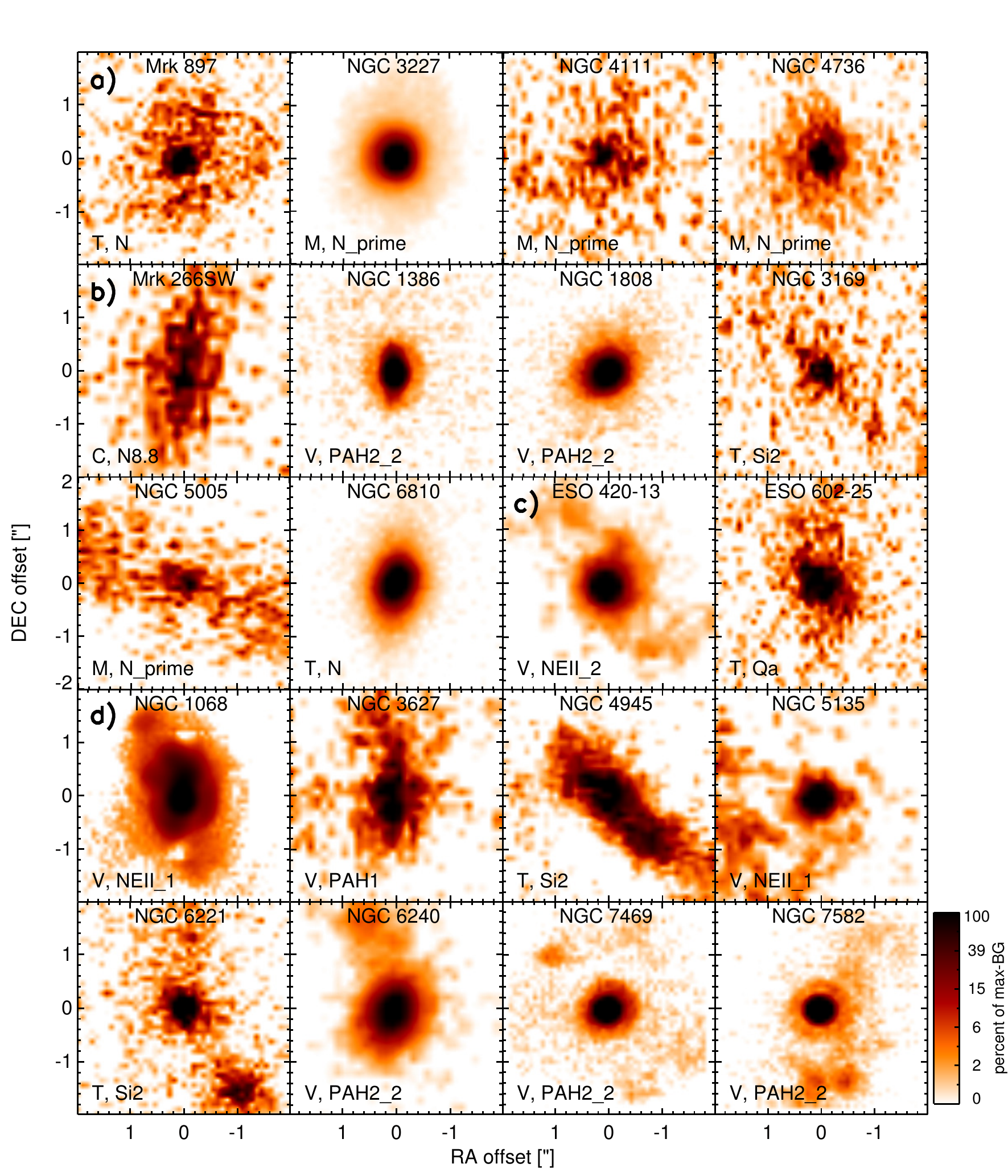}
    \caption{
             Examples of nuclei with significant extended/non-nuclear emission. 
             The images are group by apparent morphology: 
             first row: (a) circular extended emission;
             second row and third row first half: (b) elliptical or bar-like extended emission;
             third row second half: (c) spiral-like extended emission;
             fourth and fifth row: (d) complex extended emission.
             Displayed are the inner $4\arcsec$ with North up and East to the left. The colour scaling is logarithmic with white corresponding to median background and black to the $1\%$ of pixels with the highest flux values.         
            }
   \label{fig:ima_ext}
\end{figure*}
\begin{description}
 \item (a) circular: compact source embedded within diffuse circular emission (e.g., first row in Fig.~\ref{fig:ima_ext});
 \item (b) elliptical: compact source embedded within elliptical or bar-like emission structure ( e.g., second row and third row first half in Fig.~\ref{fig:ima_ext});
 \item (c) spiral: compact source within spiral structure (spir., e.g., third row second half in Fig.~\ref{fig:ima_ext}) and
 \item (d) complex: complex morphology possibly with multiple compact sources (e.g., bottom two rows in Fig.~\ref{fig:ima_ext}).
\end{description}
The MIR morphological classifications are listed for individual objects in Table~\ref{tab:basic}.
Note that these classifications are only distinguishing the apparent geometry: e.g., bar-like extended objects from group (b) might intrinsically be edge-on discs (group a) or spirals (group c).
The S/N of the individual subarcsecond-resolution images is mostly low  so that extended emission significantly fainter than the nuclei is not detected, but could certainly be present.
This is illustrated by the cases of NGC\,2992 and NGC\,3227, which have very deep Michelle images, showing faint extended emission, which remained undetected in the less deep VISIR images of the same sources (App.~\ref{app:NGC2992} and \ref{app:NGC3227}).
In five ($2\%$) of the extended objects from group (d), no clear nuclear MIR counterpart could be identified, although extended MIR emission was detected (ESO\,500-34, NGC\,253, NGC\,1614, NGC\,3628 and NGC\,7552). 
Four of them are uncertain AGN with signs of intense star formation.
Therefore, the presence of a real AGN in these objects is doubtful, and we derive upper limits for any point-like AGN emission using the extended emission as an upper bound.

\subsubsection{Nuclear extension at subarcsecond scales}\label{sec:nuc_ext}
The nuclear extension was further quantified by applying the method described in Section~\ref{sec:ext}.
Of the 204 detected objects, 38 ($19\%$) show point-like nuclei, 54 ($27\%$) are possibly extended and 43 ($21\%$) show clearly extended nuclei, while the extension remains unknown for the remaining sources.  
Among the latter are the 49 non-detected objects, 12 weakly detected objects, for which the FWHM had to be fixed during the fitting (Section~\ref{sec:ana}), and 58 objects, which could neither be classified as point-like or extended, or show contradictory nuclear extensions in different observations.
Up to $48\%$ of the total sample is at least possibly extended at $\sim0.4\arcsec\,$ resolution, which, however, needs further confirmation with deeper MIR imaging.
Note that several standard star observations have larger measured FWHMs than the FWHMs of the corresponding science targets. 
The most severe cases occur for 3C\,98, 3C\,353, 3C\,382, 3C\,403, ESO\,121-28, ESO\,198-24, IRAS\,09149--6206, IRAS\,11095--0238, NGC\,526A, NGC\,985, NGC\,4235 and PG\,0052+251.
This is an indication of the variability of the MIR sky and illustrates the limits of this kind of analysis.

Out of the 18 detected uncertain AGN, six have clearly extended nuclei, and a further three are possibly extended, while only one object appears point-like (NGC\,3094).
Thus, the fraction of extended cases is particularly high among the uncertain AGN, which indicates that the MIR emission of these objects is heavily affected or dominated by stellar processes. 
However, the MIR images are not deep enough to exclude the presence of an AGN if no unresolved core component was detected.
The extension classifications of individual objects are further discussed in  App.~\ref{app:indi}.

\begin{figure}
   \centering
   \includegraphics[angle=0,width=7cm]{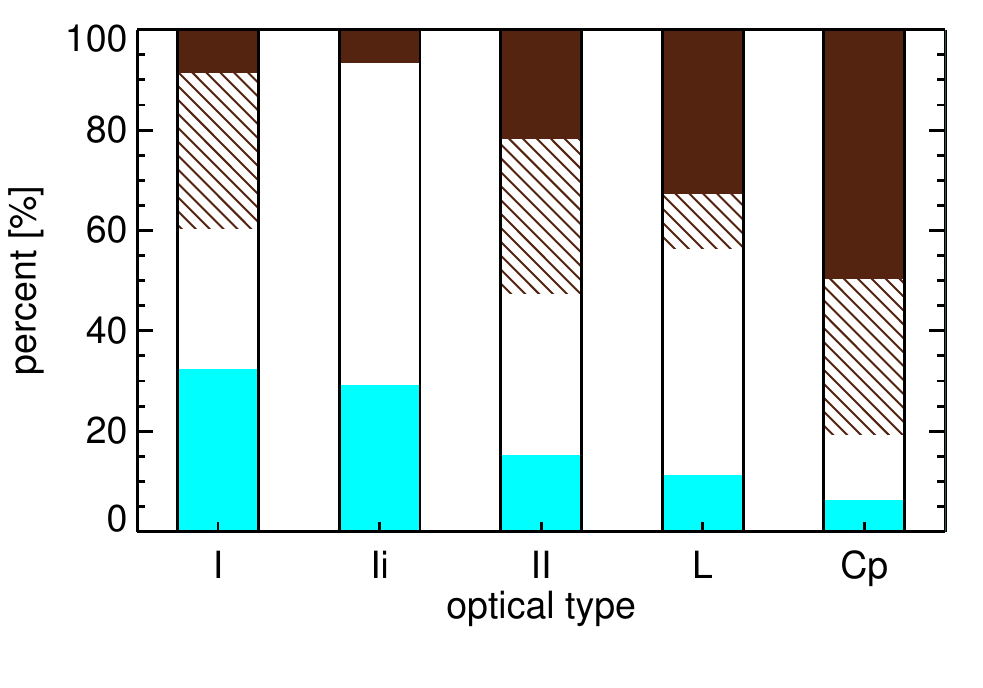}
    \caption{
             Distribution of the fractions of point-like and extended confirmed AGN for the different optical groups. 
             Solid cyan coloured bars are point-like objects, 
             brown hatched bars are possibly extended objects, solid brown bars are clearly extended objects, and the white bars are objects with unknown extension.
             See Section~\ref{sec:morph} for explanation.  
            }
   \label{fig:ext_type}
\end{figure}

The distribution of nuclear extension of confirmed AGN  separated by optical classification is displayed in Fig.~\ref{fig:ext_type}. 
The fraction of point-like sources is highest in type~I objects and decreases towards type~II, LINER and AGN/starburst composite objects.
This trend is exactly opposite for the fraction of extended objects, which is highest for AGN/starburst composites and decreases towards type~I objects.
In particular, the majority of the composites are extended or possibly extended at subarcsecond resolution.
However, the higher average object distance for the type~I sources could be responsible for the higher fraction of unresolved cases, at least partially. 
In addition, the type~II objects tend to live in more actively star-forming galaxies (Section~\ref{sec:spitzer}), leading to more extended emission also at subarcsecond scales. 

\subsubsection{Alignment with ionization cones}\label{sec:NLR}
\begin{figure}
   \centering
   \includegraphics[angle=0,width=6cm]{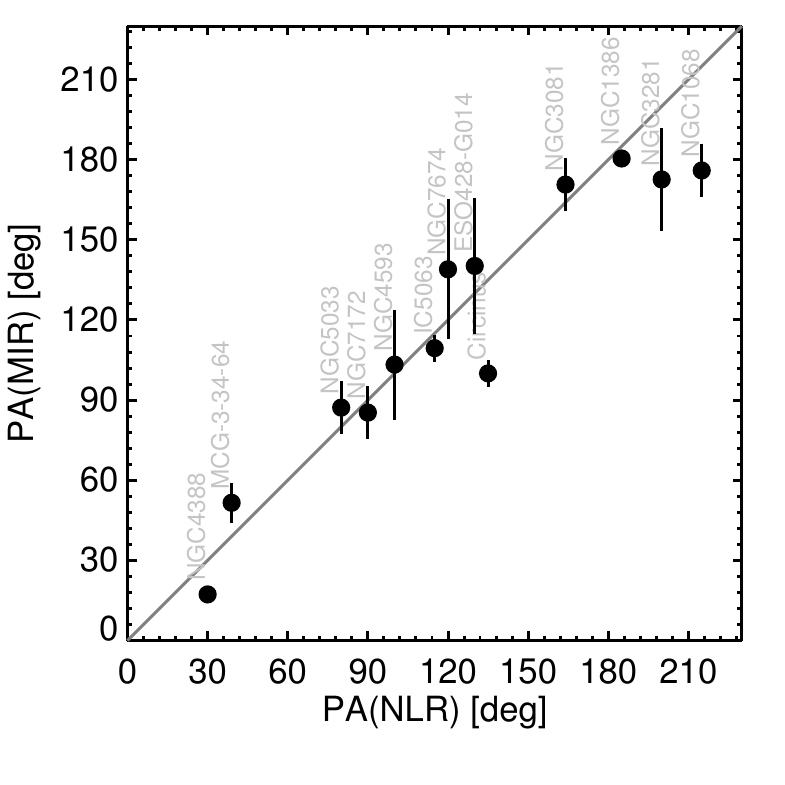}
    \caption{
             Comparison of the MIR and NLR PAs for the clearly elongated Seyferts with published PA(NLR) measurements. 
             The grey solid line marks the one-to-one correspondence.
            }
   \label{fig:PA}
\end{figure}

Among the confirmed Seyferts, 14 show elongated nuclei, most of which were observed on multiple nights with consistent position angles (PAs). 
The majority of the elongated nuclei belong to the type~II group, i.e., the system axis is expected to be approximately perpendicular to the line of sight.
Such an elongation can be caused by significant dust in the ionization cones along the system axis, i.e., the narrow-line region (NLR; e.g., \citealt{radomski_resolved_2003,schweitzer_extended_2008}).
Therefore, we compare the PAs of the MIR elongation with those of the NLR cones for all objects with available NLR PAs in the literature in  Fig.~\ref{fig:PA} (e.g., \citealt{schmitt_hubble_2003} and references in App.~\ref{app:indi}).
The PAs match for most of the objects: the Spearman rank is $\rho=0.92$ (null-hypothesis probability: $\log p = -5.2$), and the median difference between both PAs is $10\degree$.
We conclude that the scenario of significant dust in the ionization cones is thus confirmed.
Note however that the subarcsecond-extended emission generally does not dominate the MIR emission of the objects.

Elongated nuclei are also found in the three LINERs NGC\,3169, NGC\,5005 and NGC\,5363, but in these cases the elongation is aligned with the host galaxy axis and major dust lanes, instead of the NLR cones.
Therefore, the MIR observations were probably not sufficiently deep to detect the faint NLR emission in these low-luminosity systems. 

\subsection{Nuclear subarcsecond-scale photometry}\label{sec:phot}
\begin{figure}
   \centering
   \includegraphics[angle=0,width=7.5cm]{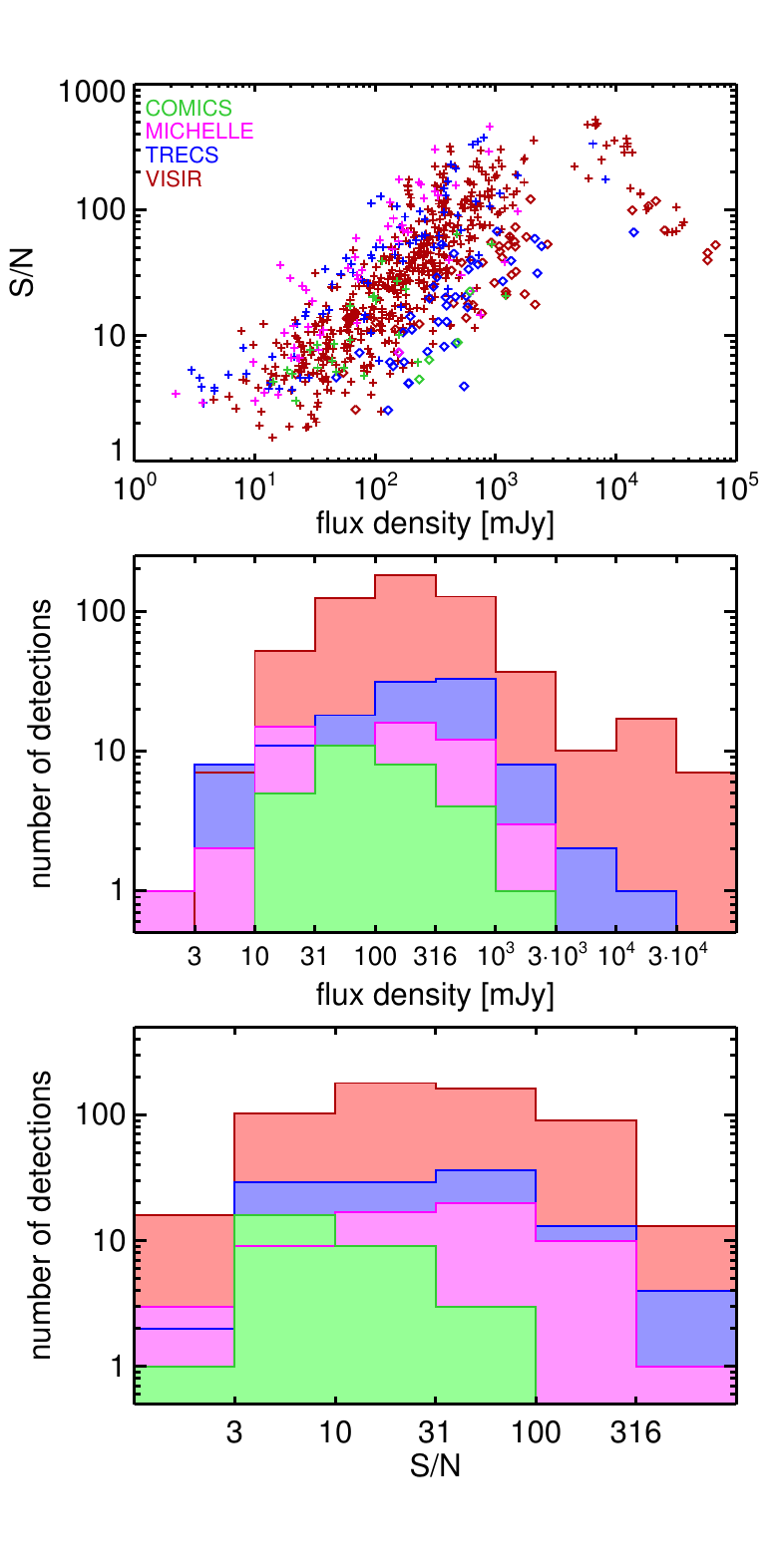}
    \caption{
             Relation of obtained S/N with measured Gaussian fluxes, $\Fgau$ for all 764 images with a detected nucleus (top).
             Crosses represent $N$-band filters and diamonds $Q$-band filters.
             The different instruments are colour-coded.
             The distributions of the fluxes and S/N separated by instrument are displayed below.
             Many object have been observed in multiple epochs and thus appear multiple times in these graphs.
            }
   \label{fig:sn_flux}
\end{figure}

We measured $\Fgau$ and $\Fpsf$ for all objects with \textsc{mirphot} as described in Section~\ref{sec:ana}.
The individual flux values are listed in Table~\ref{tab:obs}.
Fig.~\ref{fig:sn_flux} shows the distributions of the S/N of the detections, the measured $\Fgau$, and their relation (for each instrument). 
The measured fluxes are distributed similarly for all the instruments. 
Michelle observed slightly more AGN at lower fluxes than the other instruments. 
In addition, the S/N values are distributed equally for all objects, while the distribution for COMICS data is more concentrated towards lower S/N values.
For similar fluxes, the S/N of $Q$-band measurements is systematically lower than for the $N$-band measurements, as expected from the worse sensitivity in the $Q$-band wavelength region. 
Furthermore, Michelle and T-ReCS achieved higher S/N than VISIR and COMICS for similar flux levels. 

We also collected flux measurements that are published in the literature based on the observations used.
In general, these values agree  with our values within the measurement uncertainties. 
There are a couple of outliers, in particular in $Q$-band filters, which are mainly caused by the different measurement methods. 
Individual cases are discussed in App.~\ref{app:indi}.

For further analysis, we define the
nuclear subarcsecond flux, $\Fsas$, based on $\Fgau$ and $\Fpsf$ and the nuclear extension as determined in Section~\ref{sec:nuc_ext} . 
We prefer $\Fpsf$ over $\Fgau$ only for clearly extended objects, while we use the average of both for possibly extended objects.
For the second case, the corresponding flux uncertainty is set to $\ssas := 0.5 ((\Fgau+\sgau) - (\Fpsf - \spsf))$ with $\sgau$ and $\spsf$ being the $1\sigma$ uncertainties of $\Fgau$ and $\Fpsf$. 
Note that by definition, possibly extended objects have $\Fpsf$ smaller than $\Fgau$.

\subsection{Arcsecond-scale MIR spectral energy distributions}\label{sec:spitzer}
\begin{figure}
   \centering
   \includegraphics[angle=0,width=8.5cm]{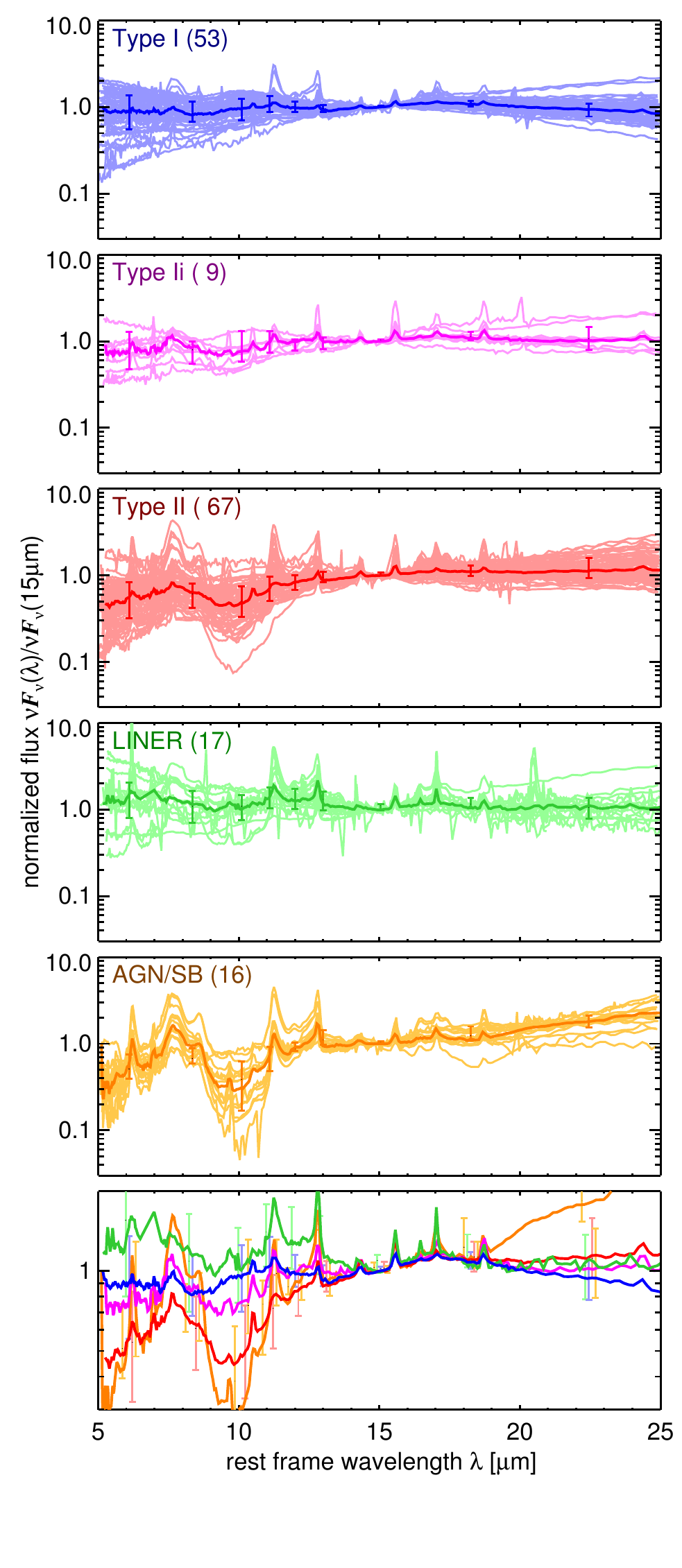}
    \caption{
             Arcsecond-scale median MIR spectra of the optical AGN types for the confirmed AGN from the total sample. 
             Each of the top five panels shows all individual IRS spectra with a decent S/N normalized at $15\,\mu$m in light colours, while the median spectrum is shown in normal colour with vertical error bars indicating the 2/3 fraction below and above the median.
             For this purpose, the same binning as in Fig.~\ref{fig:MIRSED12} is used.
             The bottom panel shows the comparison of the median spectra.            
            }
   \label{fig:IRSSED}
\end{figure}

In order to further characterize the nuclear MIR emission of our sources, we complement the nuclear subarcsecond-scale photometry with data from the \textit{Spitzer} as described in Section~\ref{sec:spi}.
The IRS spectra show a wide range  both in spectral continuum slope and spectral features (Fig.~\ref{fig:IRSSED}).
Of the 224 objects with IRS spectra, 56 ($25\%$) have significant silicate $\sim9.7\um$ emission, while 95 ($43\%$) have significant silicate absorption.
Significant polycyclic aromatic hydrocarbon (PAH) $11.3\um$ emission is found in 133 objects ($59\%$; equivalent width $>50\,$nm), while it appears weakly in an additional 45 ($20\%$; equivalent width $>20\,$nm).
Other PAH features occur at, e.g., 6.2, 7.62, 8.6, 11.3 and $12.7\um$ and are presumably caused by C--C and C--H stretching and bending vibrations in complex carbon molecules \citep{draine_optical_1984,puget_new_1989}.
The spectra of typical star formation regions show the silicate features in absorption in addition to strong PAH emission \citep[e.g.,][]{brandl_mid-infrared_2006}.
A significant fraction of the  IRS spectra fall into this category ($\sim 60$ objects, $27\%$). 
In particular 18 of the uncertain AGN ($47\%$) have star formation-like IRS spectra.
Other objects show an extremely strong silicate absorption feature with, at most, weak (or absorbed) PAH emission (14 objects, $6\%$; e.g., Circinus, ESO\,103-35, ESO\,286-19, ESO\,506-27, IRAS\,08572+3915,  IRAS\,11095--0238, IRAS\,15250+3609, NGC\,1194, NGC\,3281, NGC\,4418, NGC\,4992 and NGC\,5506). 
Finally, the last small group of objects exhibits no sign of nuclear activity in the MIR but shows blue spectra in $\nu F_\nu$ space with weak silicate emission, consistent with emission from old stellar populations (eight objects, e.g., NGC\,3379, NGC\,4472, NGC\,4594, NGC\,4636, NGC\,4698 and NGC\,5813; cf. \citealt{bressan_spitzer_2006}).
Detailed investigations of the arcsecond-scale spectra are beyond the scope of this work, and we refer to such studies in the literature (e.g., \citealt{gallimore_infrared_2010,sargsyan_infrared_2011,stierwalt_mid-infrared_2013} and references in App.~\ref{app:indi}).

Instead, we compute the arcsecond-scale median MIR spectra for the confirmed AGN with good-S/N IRS spectra separated by optical types. 
The wavelength regime around $15\,\mu$m is chosen for the normalization because it appears to be the least affected by the various spectral features.
Note, however, that the results change only slightly when 12 or 18\umm is chosen for the normalization. 
It becomes obvious that, of all AGN types, type~I AGN show the least dispersion in their MIR SEDs (average relative standard deviation = 23.5$\%$; Fig.~\ref{fig:IRSSED}), while this dispersion is much larger for both type~II (33.3$\%$) and LINER AGN (35.7$\%$). 
Type~Is have rather flat spectral slopes in $\nu F_\nu$ space, while the slope for type~IIs is red and intermediate for LINERs. 
Furthermore, the silicate feature is in strongest absorption for type~IIs and is again intermediate in LINERs. 
As expected, the star formation contribution is increasing from type\,II and LINER AGN to AGN/starburst composites, while it appears to be negligible in most type~I AGN.
However, all classes contain a number of outliers, e.g., three type~I objects (H0557-385, NGC\,1566 and NGC\,3227) have the silicate feature in absorption, while it is in emission in at least three type~II AGN (NGC\,424, NGC\,2110 and NGC\,3147).

\subsection{Comparison of subarcsecond to arcsecond scales}\label{sec:N/I}
The nuclear subarcsecond-scale photometry can now be compared with  the \spitzerr  arcsecond-scale data.
We also include $N$-band spectra from T-ReCS and VISIR for 36 AGN
\citep{honig_dusty_2010-1,gonzalez-martin_dust_2013}.
They generally agree with the subarcsecond-scale photometry well, apart from a few cases where the spectra have higher flux levels (e.g., VISIR: ESO\,323-77, ESO\,428-14, IC\,5063, MCG-3-34-64; T-ReCS: NGC\,5506).
In the case of the VISIR spectra, this is because all of these objects have extended emission, while the spectral extraction of \cite{honig_dusty_2010-1} used a fixed width of 0.75\arcsec\, with no correction for extended emission.
On the other hand, \cite{gonzalez-martin_dust_2013} used a PSF extraction for their T-ReCS spectra. 
The resulting difference  can also be seen in two of the three objects with both VISIR and T-ReCS spectra, where the flux levels are higher in the VISIR spectra (IC\,5063 and NGC\,7582, App.~\ref{app:IC5063} and \ref{app:NGC7582}).

All the MIR data together are displayed for individual objects in App.~\ref{app:indi}.
We find that the silicate feature strengths are similar in both the subarcsecond and arcsecond-scale SEDs.
The only possible exceptions are  ESO\,297-18 and NGC\,1365, which show no silicate features at subarcsecond resolution, while they are in absorption in the IRS spectra.
Contrary to the silicate features, the PAH features are significantly weaker or even absent at subarcsecond scales compared to the arcsecond scales in all 34 cases with strong PAH emission and sufficient spectral coverage at subarcsecond resolution to constrain the PAH feature strength.
NGC\,1808 is the only exception with strong PAH emission also on subarcsecond scales.
The PAH emission is completely absent at subarcsecond resolution in, e.g.,  the type~I AGN: ESO\,323-77, ESO\,362-18, Mrk\,520, NGC\,5033, NGC\,7469, NGC\,7213  and in the type~IIs: Cen\,A, ESO\,5-4, ESO\,297-18, NGC\,1144,  NGC\,1365, NGC\,2110, NGC\,5135, NGC\,5728, NGC\,7582 and the LINER NGC\,4579.
Weak PAH emission is still visible at subarcsecond resolution in seven objects (MCG-6-30-15, Mrk\,509, NGC\,1566, NGC\,3227, NGC\,5273, NGC\,6240S and NGC\,7130).
This result confirms that, in general, PAH emission is weak or absent in the subarcsecond-scale environment of AGN, contrary to its common appearance on arcsecond scales (see also, e.g., \citealt{honig_dusty_2010-1,gonzalez-martin_dust_2013}).

Furthermore, there are four objects, for which the continuum emission slope is significantly different at subarcsecond scales compared to the arcsecond scales: ESO\,297-18 has a blue slope at subarcsecond scales in $\nu F_\nu$ space (red at arcsecond scales), while the LINERs Fornax\,A, M87 and NGC\,4261 have  red slopes (blue at arcsecond scales).
Note that there might be more cases but the subarcsecond-scale wavelength coverage is not sufficient for a slope determination in many cases.

For a quantitative comparison of the general flux levels on subarcsecond and arcsecond scales, we perform synthetic photometry on the IRS spectra, obtaining the arcsecond-scale fluxes: $\Fas$. 
The transmission curves of the corresponding subarcsecond-scale filters are used for this purpose, and a flux uncertainty of $10\%$ is adopted.

This value is typically much larger than the given uncertainties in the spectra.
We minimize uncertainties due to possible slit losses and positioning issues through the scaling  the IRS spectra with the nuclear IRAC and MIPS photometry whenever significant differences between the flux levels of IRS and IRAC/MIPS occur.

The corresponding flux ratios between the subarcsecond and arcsecond scales can be computed: $\Rmed = \Fsas / \Fas$.
We aim for simple comparisons and, thus, use only the average of the flux ratios of different filter measurements for each object: $\aRmed$. 

\begin{figure}
   \centering
   \includegraphics[angle=0,width=8.5cm]{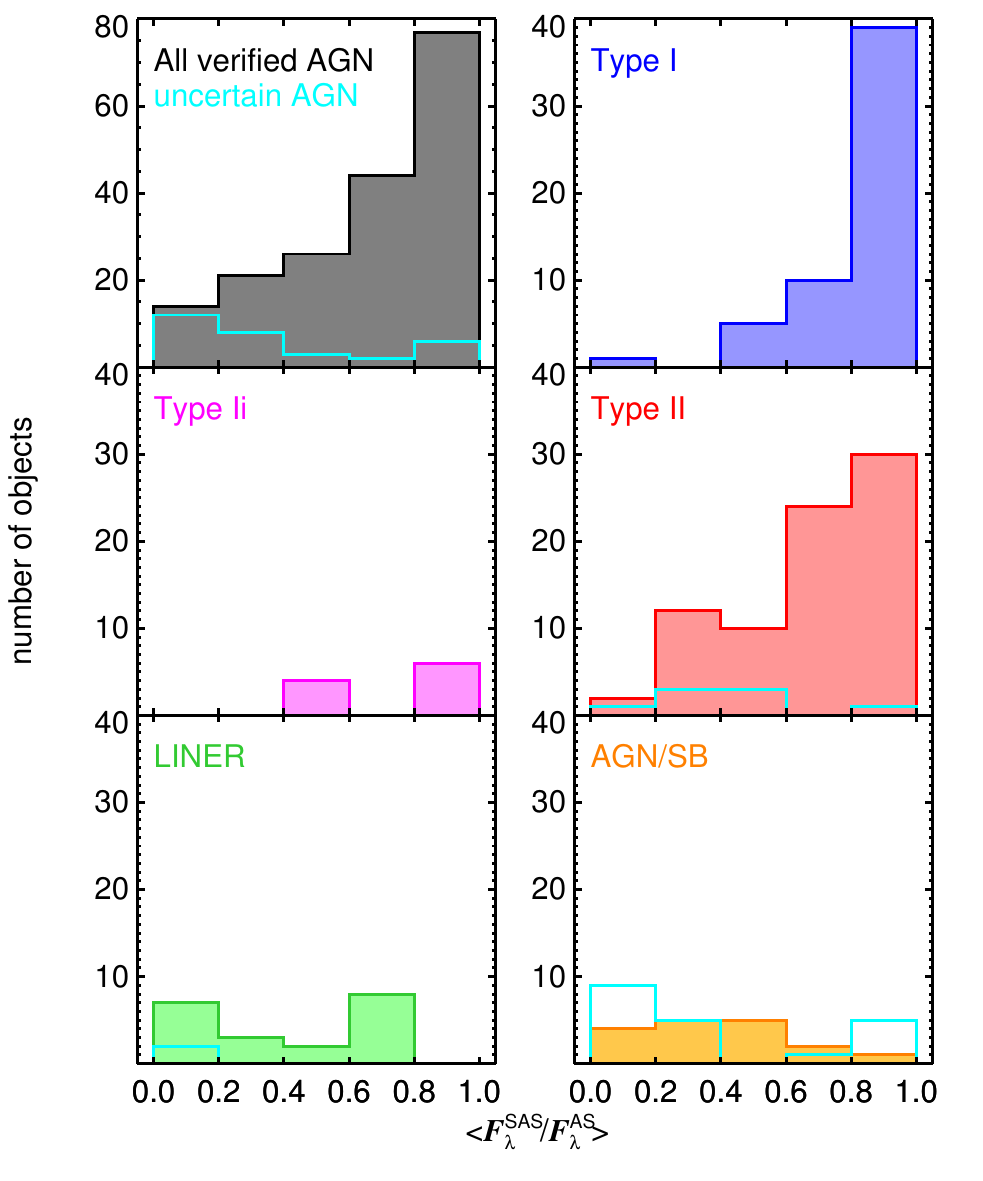}
    \caption{
             Distributions of the average subarcsecond to arcsecond-scale flux ratios for the AGN MIR atlas and the different optical classes.
             The uncertain AGN are displayed separately as cyan empty histograms.
            }
   \label{fig:flux_ratio}
\end{figure}

\begin{figure}
   \centering
   \includegraphics[angle=0,width=8.5cm]{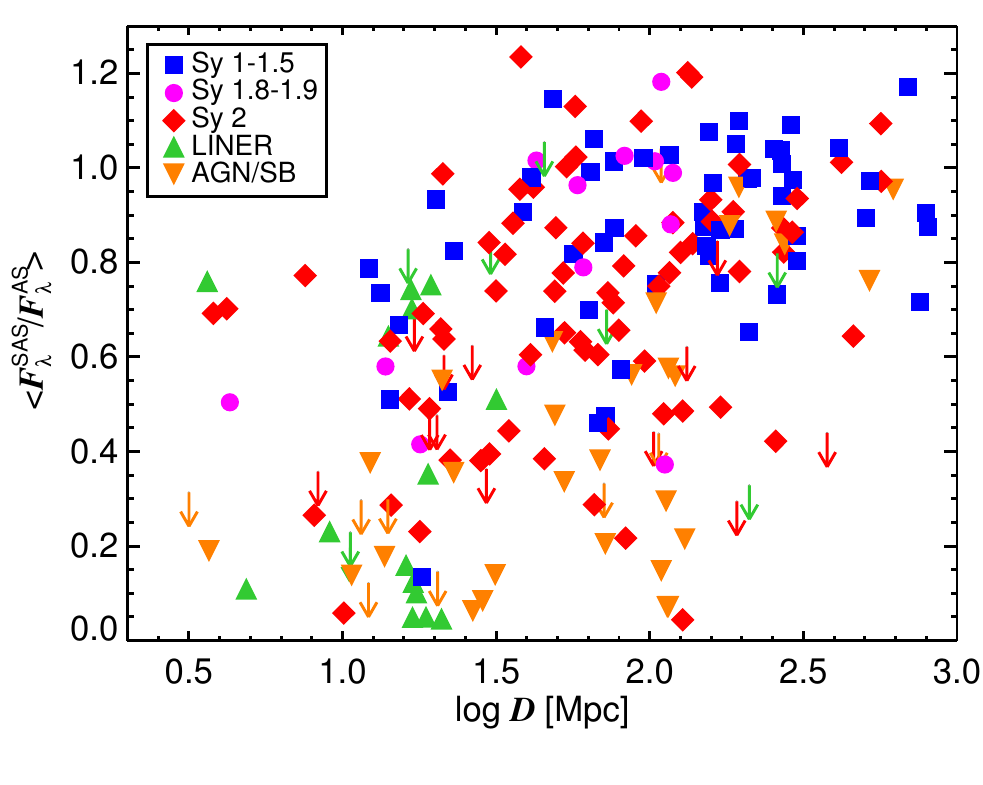}
    \caption{
             Relation of the average subarcsecond to arcsecond-scale flux ratios with the object distance for the AGN MIR atlas.
             Blue squares are type~I AGN, magenta circles type~Ii, red diamonds type~II, green bottom-heavy triangles LINERs, and orange top-heavy triangles AGN/starburst composites.
             The arrows mark upper limits due to non-detections at subarcsecond scales.
            }
   \label{fig:ratio_dist}
\end{figure}

Fig.~\ref{fig:flux_ratio} shows the distribution of $\aRmed$ for the total sample separated into the different optical classes. 
The majority of AGN have flux ratios $\aRmed > 0.5$ ($75\%$; the median flux ratio is $\sim 0.74$).
However, a significant fraction of objects have very small flux ratios, e.g., $10\%$ with $\aRmed < 0.25$ and $4\%$ with $\aRmed < 0.1$.
The most extreme cases with detected nuclei are the LINERs NGC\,3169, NGC\,5005 and NGC\,5363 and the Sy\,2 Mrk\,266SW with $\aRmed < 0.05$, i.e., their MIR emission is completely host dominated on arcsecond scales.  
In general, the LINERs and AGN/starburst composites have the largest fraction of low flux ratios, followed by the type~II AGN.
Once more, type~Ii appear intermediate between type~I and II objects, although their distribution suffers from small-number statistics.
It has been demonstrated in Fig.~\ref{fig:opt_comp} that LINERs and type~II objects are at smaller distances than type~I objects and, thus, such an effect is at least partially expected.
Therefore, we examine the relation of the flux ratio with the object distance, $D$, in Fig.~\ref{fig:ratio_dist}.
Indeed, an increasing trend of $\aRmed$ with $D$ is found despite a large scatter (Spearman rank $\rho = 0.52$; null-hypothesis probability $\log p = -13.4$). 
In addition, we can also compare the $\aRmed$ distributions for type~I  and type~II AGN in different distance bins.
At intermediate distances ($\sim30$ to 100\,Mpc) for example, the median flux ratio of type~I objects is 0.87, while it is 0.74 albeit with a large dispersion of $\sim0.23$ each.
At large distances ($> 150\,$Mpc), the difference of both types becomes smaller but still persists ($0.97 \pm 0.13$ versus $0.87 \pm 0.27$).
This indicates  that the different subarcsecond- to arcsecond-scale flux ratios of type~I and type~II AGN is not caused by distance effects alone.

In summary, the 0.4\arcsec-scale MIR SEDs are significantly different than those on 4\arcsec-scales ($\aRmed < 0.5$) in a significant fraction of the  AGN ($31\%$). 
This difference is strongest in AGN/starburst composites and LINER and type~II AGN and probably arises from a combination of two effects: (1) type~I objects are on average more distant and (2) the MIR contribution of circum nuclear star formation on arcsecond scales is not only higher in composites but also in LINERs and type~II AGN (as indicated by the PAHs in their MIR spectra; see also, e.g., \citealt{maiolino_new_1995,melendez_constraining_2008}).

\subsection{Subarcsecond-scale MIR spectral energy distributions}\label{sec:SED}
For those AGN that were observed and detected in multiple filters, their subarcsecond-scale MIR SEDs can roughly be constructed using the nuclear fluxes, $\Fsas$.
The individual MIR SEDs are displayed for each object in App.~\ref{app:indi}.
In general, the silicate feature appears at least weakly in emission in type~I objects, while it is strong in absorption in type~II and AGN/starburst composites.
The latter show the strongest silicate 10\umm absorption on average.
From the type~I AGN, H\,0557--385 and possibly NGC\,1566, are the only objects with significant silicate 10\umm absorption at subarcsecond scales.
On the other hand, there are at least two type~II AGN with significant silicate emission at nuclear scales (NGC\,2110 and NGC\,4258).
Three additional objects exhibit silicate emission on arcsecond scales and have insufficient but silicate-emission compatible subarcsecond-scale data (NGC\,424, NGC\,3147 and NGC\,4472).
LINER and intermediate-type AGN lie between these extremes, with some objects showing silicate emission and others showing silicate absorption.
The deep silicate $10\,\mu$m absorption features have recently been associated with foreground dust in the host galaxy \citep{goulding_deep_2012}.
Indeed, we find deep silicate $10\,\mu$m features in the MIR SEDs either in highly inclined spiral galaxies ($50\%$ of the cases), or heavily disturbed/interacting systems ($40\%$).
Interestingly, all low-inclination undisturbed galaxies with deep silicate absorption in their nuclear MIR SEDs are barred spirals (NGC\,3094, NGC\,5135, NGC\,6300, NGC\,7479 and NGC\,7552).
However, the converse is not true, i.e., not all highly inclined or barred spirals show deep silicate absorption features in their nuclear MIR SEDs. 
None of the objects with deep silicate features belongs to the type~I AGN group, as expected.
The available nuclear subarcsecond-scale photometry is mostly not sufficient for a more  quantitative analysis of the silicate features and, thus, is omitted here.

\begin{figure}
   \centering
   \includegraphics[angle=0,width=8.5cm]{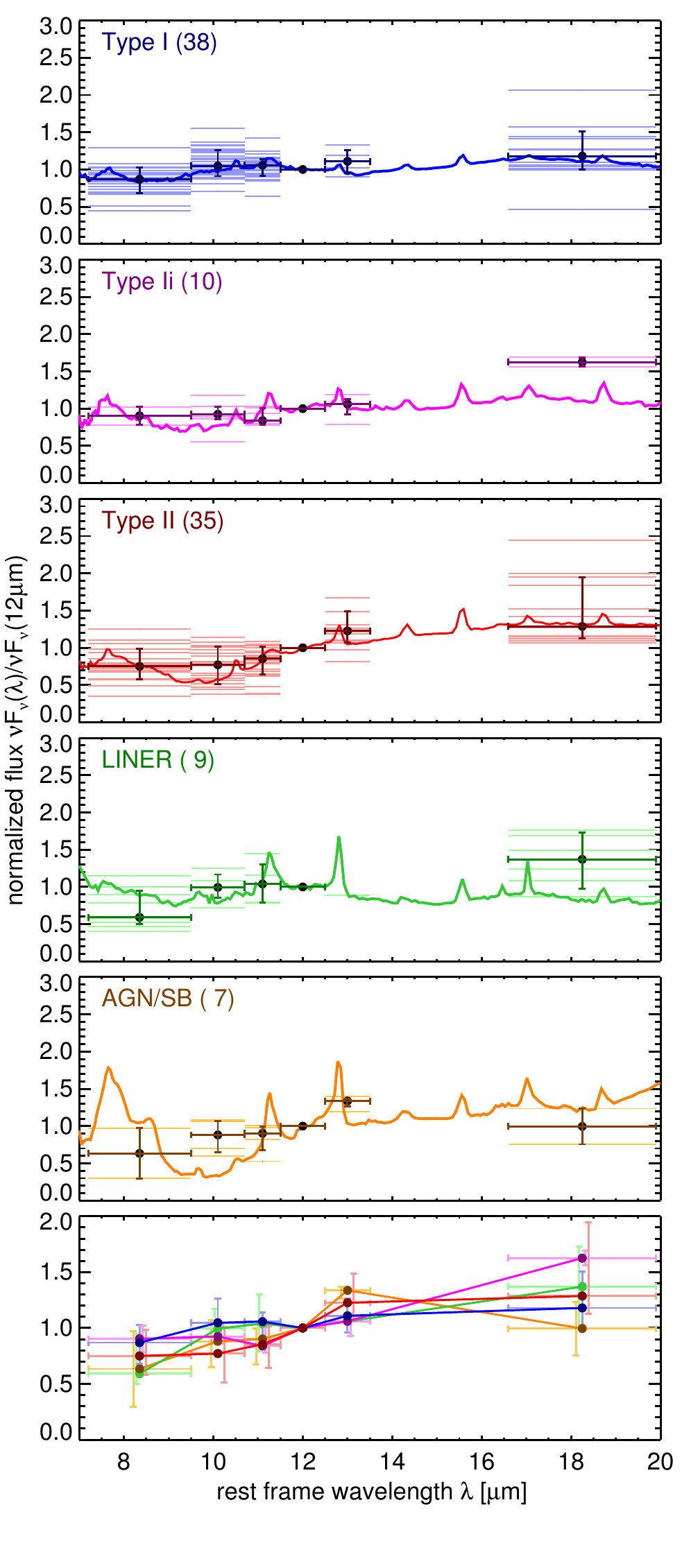}
    \caption{
             Subarcsecond-scale MIR SEDs of the different optical AGN types with normalization at 12\um.
             Solid lines indicate the median IRS spectrum for each AGN type, using the normalized spectra shown in
             Fig.~\ref{fig:IRSSED}.
             The individual SEDs are displayed as horizontal bars with a width corresponding to the chosen binning for the filters.
             The median SEDs are displayed as solid circles with vertical error bars indicating the 2/3 fraction below and above the median. 
             The bottom plot shows the comparison of the median spectra for different optical types.  
            }
   \label{fig:MIRSED12}
\end{figure}

We combine the individual MIR SEDs in order to create median MIR SEDs for the different AGN types.
For this purpose, a wavelength binning is chosen such that the silicate 10\umm and PAH 11.3\umm features can be probed with respect to the continuum emission short- and longwards of these features.
Furthermore, the MIR SEDs are flux normalized at the filter that is centred in the restframe range of  12\umm (11.5 to 12.5\um).  
The reason for choosing this range for the normalization is that the 12\umm flux has been shown to be a good indicator of the intrinsic AGN continuum emission \citep[e.g.,][]{gandhi_resolving_2009}, and that most objects have subarcsecond-scale measurements available in this range. 
Normalization at $15\,\mu$m would probably be superior, but this range is not accessible from the ground.
A normalization at 18\umm provides qualitatively similar results. 
The combination of all the individual MIR SEDs is displayed in Fig.~\ref{fig:MIRSED12}.
In general, the dispersion of the individual MIR SEDs is very large consistent with what is seen in the arcsecond-scale spectra (Fig.~\ref{fig:IRSSED}).
In addition, the subarcsecond-scale SEDs of  type~Ii AGN, LINER and AGN/starburst composites suffer from small numbers of available measurements.  
Owing to the large dispersion in the SEDs of all optical types, no strong statements about differences in the general (median) SEDs can be made, but we note the following trends.
The median subarcsecond-scale MIR SEDs reflect well  the silicate feature properties already noted from the individual spectra above, i.e., the feature is absent or weakly in emission in the median SEDs of type~I and LINER AGN and in absorption in type~II AGN and AGN/starburst composites.  
In addition, the median SEDs of type~Ii and II AGN are slightly redder than those of type~I objects. 
When comparing the median subarcsecond-scale MIR SEDs to the arcsecond-scale MIR SEDs (Fig.~\ref{fig:MIRSED12}), there is good agreement between the  SED shapes for type~I and type~II AGN.
However, differences are found for the AGN/starburst MIR SEDs, where either the silicate 10\umm absorption feature is shallower or the surrounding PAH features fainter at subarcsecond-scales. 
This indicates that at this resolution, the AGN is significantly better isolated from the surrounding star formation, which dominates the arcsecond-scale SEDs in many composites.

\subsection{Nuclear subarcsecond-scale continuum emission and size constraints}\label{sec:nuc_emi}
As described in Section~\ref{sec:cont}, we have calculated the nuclear subarcsecond-scale continuum flux densities at 12\umm and 18\umm restframe wavelengths, $\FNsas$ and $\FQsas$.
$\FNsas$ has been computed for all 253 objects, while $\FQsas$ could be computed for 67 objects.
For 11 non-detected objects, the  more constraining IRS fluxes are used (method (ii); 3C\,29, 3C\,317, 3C\, 424, Hydra\,A, IC\,883, LEDA\,13946, NGC\,4636, NGC\,4698, NGC\,7626, PKS\,1932--46 and PKS\,2354--35).
In 32 objects with strong silicate features, the arcsecond and subarcsecond-scale MIR SEDs are similar, and thus, the IRS spectra can be used for to correct for these features (see Section~\ref{sec:cont}).
For another eight objects with strong silicate features, the subarcsecond-scale $N$-band spectra from \cite{gonzalez-martin_dust_2013} and \cite{honig_dusty_2010-1} are used for computing the silicate correction (IC\,4518W, NGC\,1386, NGC\,3094, NGC\,4418, NGC\,7172, NGC\,7213, NGC\,7479 and NGC\,7582). 
Finally, for three objects, no $N$-band photometry is available but only $Q$-band photometry that is accordingly used to extrapolate the $\FNsas$ values (method d; ESO\,602-25,  MCG-2-8-39 and NGC\,7592W).
The nuclear fluxes $\FNsas$ and $\FQsas$ correspond to our best estimates of the AGN MIR continuum emission, excluding the outer NLR as detected in some objects (Section~\ref{sec:NLR}). 
We constrain the projected size of this emission region at the target distance with the minimum PSF FWHM(major axis) measured for each object, $\dunr$.
These should be regarded as upper limits on the intrinsic AGN MIR size only.
The AGN torus is expected to remain unresolved in the MIR with 8-m class telescopes \citep{honig_dusty_2010}. 
Furthermore, the upper limits on the nuclear fluxes of non-detected nuclei correspond to an AGN MIR size of $\dunr$.  
The individual values of $\FNsas$, $\FQsas$ and $\dunr$ are given in Table~\ref{tab:basic}.

\begin{figure}
   \centering
   \includegraphics[angle=0,width=8.5cm]{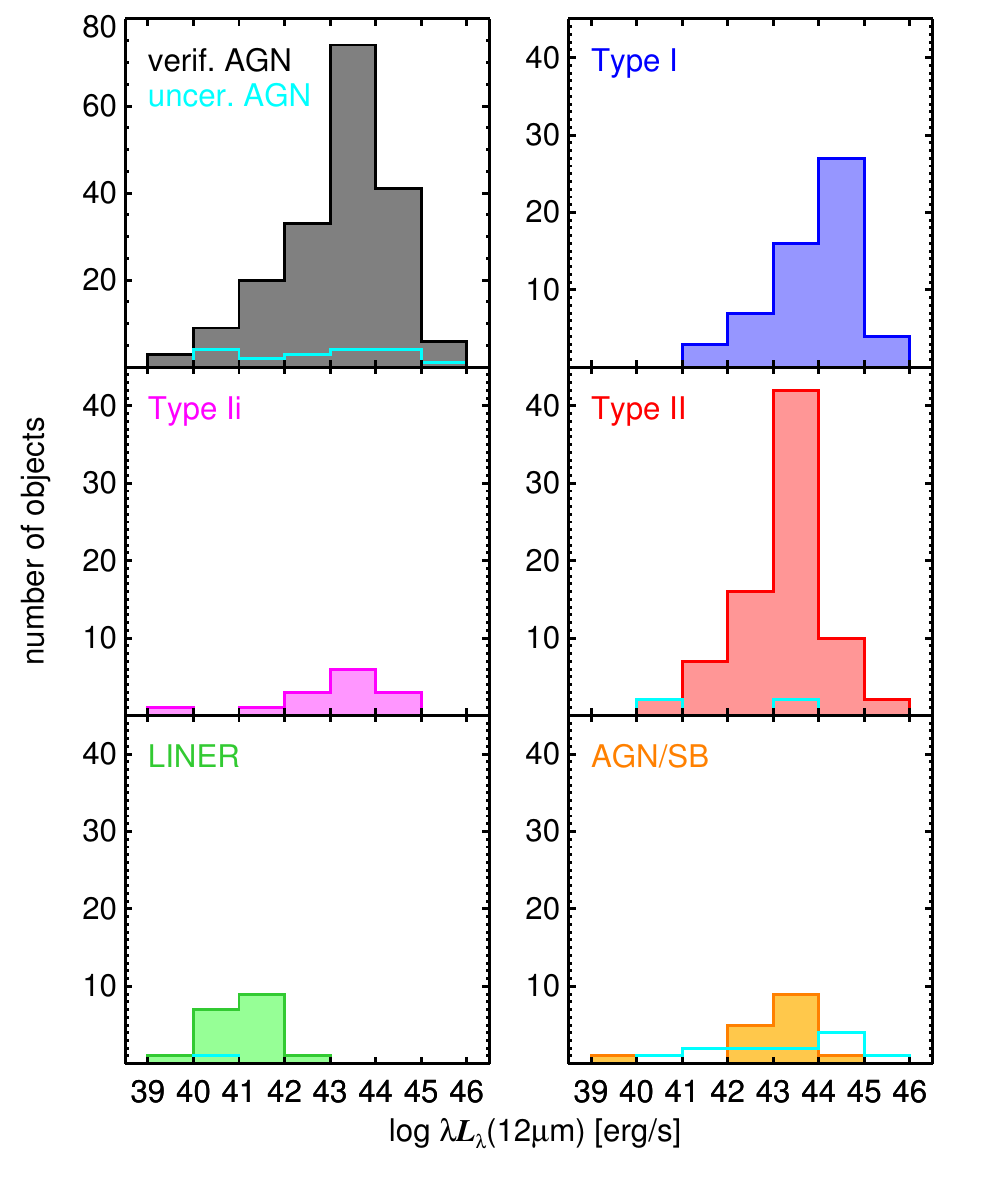}
    \caption{
             Distribution of the nuclear 12\umm luminosity for the detected AGN from the MIR atlas.
             Symbols and colours are as in Fig.~\ref{fig:flux_ratio}.
            }
   \label{fig:L12_hist}
\end{figure}

Using $\Fsas$ and the collected object distances $D$, we compute the nuclear 12\umm luminosities, $\LN$ for the whole sample.
The distribution of $\LN$ for the detected objects separated by optical class is displayed in Fig.~\ref{fig:L12_hist}.
Similar to the finding in other works \citep[e.g.,][]{winter_x-ray_2009}, type~I AGN are distributed at higher luminosities (median $\log \LN/\mathrm{erg}/\mathrm{s} = 44.2$) with respect to the type~II AGN (median $\log \LN/\mathrm{erg}/\mathrm{s} = 43.4$).
The distribution of type~Ii AGN is again intermediate between those of type~I and type~II AGN (median $\log \LN/\mathrm{erg}/\mathrm{s} = 43.5$).
LINERs populate only the lowest luminosity bins (median $\log \LN/\mathrm{erg}/\mathrm{s} = 41.2$), while AGN/starburst composites can be found mainly at higher luminosities (median $\log \LN/\mathrm{erg}/\mathrm{s} = 43.3$).
They are dominated by the ultra luminous infrared galaxies (ULIRGs).
The uncertain AGN are more or less equally spread over the whole luminosity range because they consist mainly of nearby little-active galaxies and the more distant ULIRGs.
Note that these distribution trends might not apply to the AGN classes in general because of the statistically incomplete nature of the total sample.

\begin{figure}
   \centering
   \includegraphics[angle=0,width=8cm]{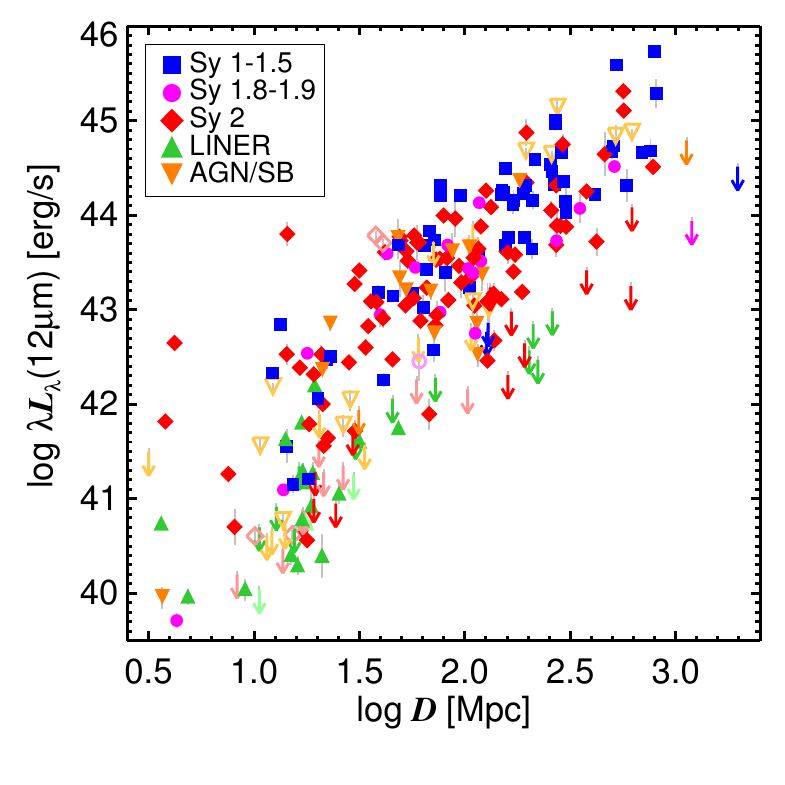}
    \caption{
             Relation of the nuclear subarcsecond-scale 12\umm luminosity to object distance. 
             Symbols and colours are as in Fig.~\ref{fig:ratio_dist}. 
             In addition, light-coloured symbols indicate uncertain AGN.
            }
   \label{fig:lum_dist}
\end{figure}

To further characterize the total sample, we investigate the relation of $\LN$ versus $D$ in Fig.~\ref{fig:lum_dist}.
The relatively low sensitivity of the ground-based MIR data results in a relatively narrow luminosity spread of detected objects of roughly one order of magnitude for given distances over most of the distance range.
This is because there are no known highly luminous AGN at small distances, while low-luminosity AGN at large distances have expected  MIR fluxes less than a few millijansky.
These could probably not be detected from the ground and were thus not observed.   
The luminosity range increases to approximately two orders of magnitude only at the shortest distances, albeit with decreasing object density. 
This result indicates well the coupling of the object luminosity with the distance in the AGN MIR atlas, which will affect any analysis based on this sample. 

%
%
%
%
%
%

\subsection{Comparison of AGN to total galaxy emission}\label{sec:glob}
The computed nuclear MIR emission of the previous section allows for a  comparison of the AGN  to the total emission MIR emission of the galaxies, i.e., AGN plus host.
For this purpose, we collected 12\umm fluxes, $\FNtot$, obtained with the \irass satellite \citep{neugebauer_infrared_1984} from the literature as a proxy for the total MIR emission of each galaxy.
These are mainly from NED, \cite{sanders_iras_2003,golombek_iras_1988,rush_extended_1993,rice_catalog_1988} and \cite{sanders_luminous_1996}.
For 208 ($83\%$) of the AGN from the total sample, \irass measurements are available.
Of those, 143 ($69\%$) were detected  at 12\um. 
We compare the nuclear and total fluxes through their ratio, $\Rlar$.
Note that the band pass of the \irass 12\umm filter comprises approximately the whole $N$-band. 
Therefore, the silicate feature and the spectral continuum slope strongly affect $\FNtot$ and lead to significantly lower $\Flar$ than $\Fsas$ for 17 objects.
Two additional objects have $\Rlar > 1$ but silicate strongly in emission and relatively flat $N$-band continuum (Mrk\,304 and PG\,0026+129; from \citealt{sanders_continuum_1989}). 
The reason for the high ratios of the quasar-like objects might be long-term MIR variability.

\begin{figure*}
   \includegraphics[angle=0,width=14cm]{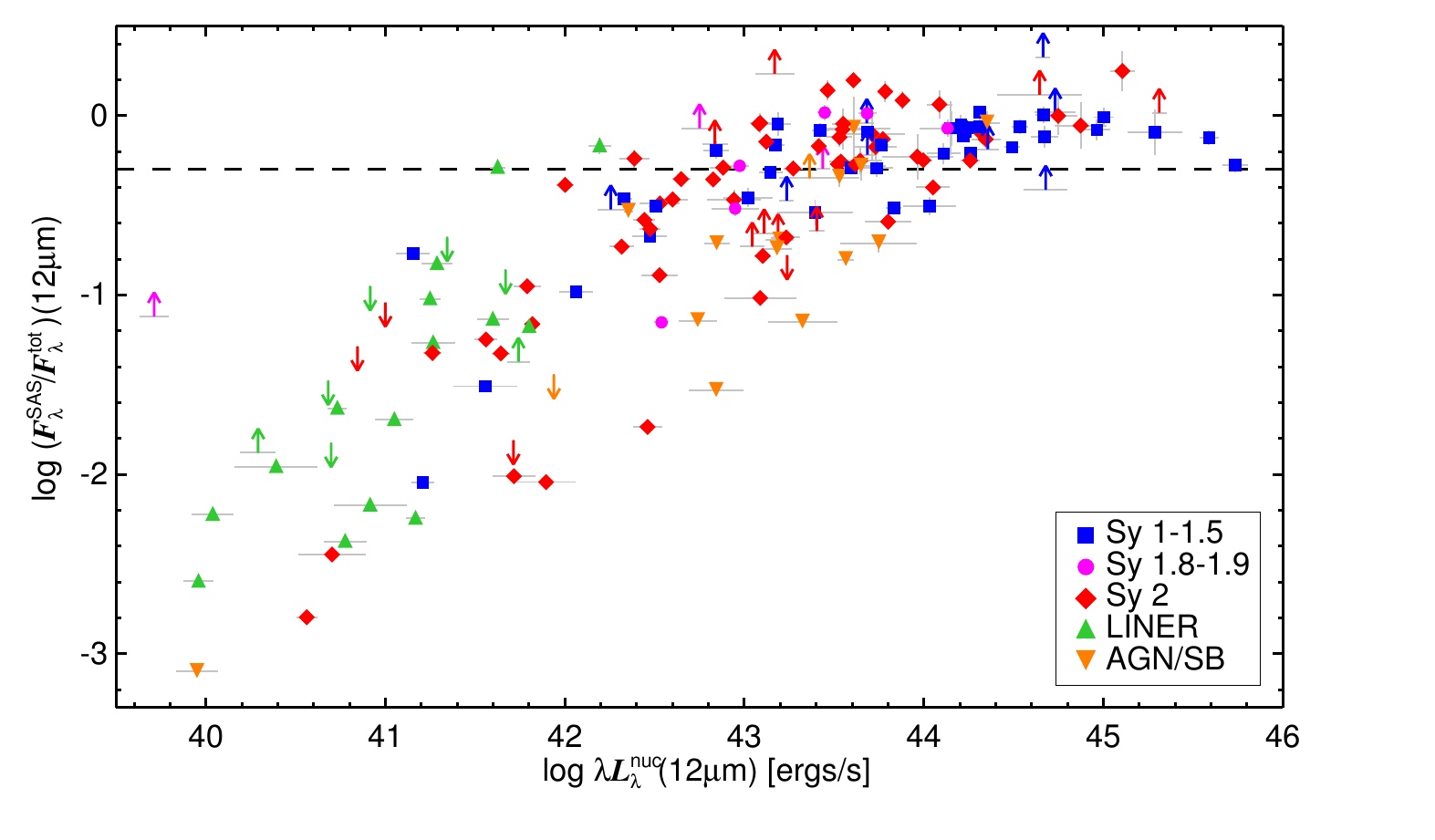}
    \caption{
             Relation of the ratio of the nuclear to the large-scale  12\umm fluxes with the nuclear 12\umm  luminosity for the confirmed AGN in the total sample.
             Symbols and colours are as in Fig.~\ref{fig:ratio_dist}.
             The dashed horizontal line marks $\Rlar = 0.5$.
            }
   \label{fig:tot}
\end{figure*}

Looking at differences for the optical AGN classes, we find that $\Rlar > 0.5$ for $> 66\%$ of all type~I AGN.
In other words, the AGN dominates the total 12\umm emission in these cases.
This occurs in $< 50\%$ of the type~II AGN and $< 20 \%$ of the LINERs.
In general, nearby spiral galaxies hosting low-luminosity AGN exhibit the lowest $\Rlar$ (e.g., M51a, NGC\,4501 and NGC\,4736), while the highest ratios are exhibited by the most powerful and distant AGN in the sample (e.g., 3C\,273).
As already mentioned before, this behaviour can be partly caused by distance effects.
Indeed, $\Rlar$ correlates well with the object distance $D$ (Spearman rank $\rho = 0.62, \log p = -14.4$). 
On the other hand, there is also a strong correlation of $\Rlar$  with the nuclear luminosity, $\LN$ with a dispersion of 0.45\,dex and a slope of $\sim 0.6$ (Spearman rank $\rho = 0.78, \log p = -27.4$; Fig.~\ref{fig:tot}).  
A correlation is,  of course, expected from involving the same quantity on both axes.
However, this result agrees with earlier works using the nuclear X-ray luminosity instead of $\LN$ \citep[e.g.,][]{vasudevan_power_2010,asmus_mid-infrared_2011}.
Therefore, $\Rlar$ is probably governed by the nuclear luminosity rather than by the distance.
On average, the nuclear subarcsecond-scale 12\umm emission dominates the large-scale emission for $\LN \gtrsim 2 \times 10^{43}\,$erg\,s$^{-1}$, although there are objects with $\Rlar < 0.5$ up to $10^{44}\,$erg\,s$^{-1}$.
This result is in agreement with \cite{prieto_spectral_2010} and \cite{rosario_mid-infrared_2013} and once more emphasizes the importance of high angular resolution data for AGN investigations.

\section{Conclusions}\label{sec:concl}
We presented the first subarcsecond MIR atlas of local AGN comprising 253 objects with a median redshift of $z=$ 0.016.
It consists of all 895 publicly available images of optical AGN with $z<0.4$ from the four instruments VLT/VISIR, Gemini/Michelle, Gemini/T-ReCS and \textit{Subaru}/COMICS.
This atlas includes $80\%$ of the sources from the uniform nine-month BAT AGN catalogue and features 62 type~I, 16 type~Ii, 103 type~II, 32 LINER AGN and 37 AGN/starburst composites, covering an MIR luminosity range of $\sim 10^{39}$ to $10^{46}$\,erg\,s$^{-1}$.
Among these are 38 uncertain AGN.
We determined the source extension and measured the unresolved nuclear fluxes. 
Out of the 253 AGN, 204 were detected and appear mostly compact at subarcsecond scales.
However, for $18\%$ of the objects, extended and non-nuclear emission on arcsecond scales was detected.
All Seyferts with elongated nuclear emission possess PAs comparable to those of the ionization (NLR) cones, which verifies the presence of warm dust in these regions.
The uncertain AGN have a particularly high fraction  of both  non-detections and extended emission, indicating that compact MIR nuclei are absent or at least very weak in many cases.
We used \spitzerr to complement the subarcsecond-scale photometry with lower-angular resolution spectra and photometry, which allow us to investigate arcsecond-scale morphology and SEDs.
The arcsecond-scale emission is significantly affected by non-AGN emission, as evident from the large fraction of sources with PAH emission.
On the other hand, the PAH emission is absent or at least much weaker on nuclear scales, while the silicate features are mostly similar.
The silicate features commonly appear in emission in optical type~I AGN and in absorption in type~II AGN on both scales despite a few outliers. 
The deepest silicate absorption features are found in AGN/starburst composites.
The ubiquitous presence of this feature demonstrates the presence of large amounts of dust in the majority of the observed galactic nuclei (or along the line of sight towards it).
However, this does not constrain the dominant MIR emission source in AGN.

Furthermore, the general subarcsecond-scale flux levels are less than half compared to arcsecond scales in a significant fraction of the objects ($31\%$).
Apart from the AGN/starburst composites, type~II AGN and LINERs are  mostly affected, which also have a higher star formation contribution on intermediate scales than that measured on type~I objects. 

Finally, we  computed the nuclear subarcsecond-scale continuum luminosities at 12 and 18\umm  for individual AGN by combining all available MIR data and correcting for the silicate features whenever possible. 
We compared the nuclear MIR luminosities with the global MIR emission of the host galaxies and AGN represented through the \irass photometry and verified that only at the highest luminosities do AGN dominate the total MIR output of the systems.
Our calculated nuclear continuum luminosities can easily be used for future multiwavelength analysis as best estimates for the AGN-related MIR emission.
 
The AGN MIR atlas represents a significant part of the local AGN population with all optically active subclasses, making it a suitable sample for AGN unification studies, despite its heterogeneous selection.
These will be subject of the forthcoming papers of this series.
\section*{Acknowledgements}
We thank the anonymous referee for their constructive comments. 
We are also very thankful to M.~Demleitner and the GAVO Team for the superb support with the online publication of the data.
DA thanks K.~Zelenevskiy and K.~Tristram for helping to improve the manuscript.
DA acknowledges the JAXA ITYF scheme for visitor support to ISAS in 2012 August.
PG acknowledges support from STFC (grant reference ST/J003697/1).
Based in part on observations obtained at the Gemini Observatory, which is operated by the 
Association of Universities for Research in Astronomy, Inc., under a cooperative agreement 
with the NSF on behalf of the Gemini partnership: the National Science Foundation 
(United States), the National Research Council (Canada), CONICYT (Chile), the Australian 
Research Council (Australia), Minist\'{e}rio da Ci\^{e}ncia, Tecnologia e Inova\c{c}\~{a}o 
(Brazil) and Ministerio de Ciencia, Tecnolog\'{i}a e Innovaci\'{o}n Productiva (Argentina).
Based in part on data collected at \textit{Subaru} Telescope, which is operated by the National Astronomical Observatory of Japan.
This research made use of the NASA/IPAC Extragalactic Database
(NED), which is operated by the Jet Propulsion Laboratory, California Institute
of Technology, under contract with the National Aeronautics and Space
Administration. 
This work is based in part on observations made with the Spitzer Space Telescope, which is operated by the Jet Propulsion Laboratory, California Institute of Technology under a contract with NASA.
This publication makes use of data products from the Two Micron All Sky Survey, which is a joint project of the University of Massachusetts and the Infrared Processing and Analysis Center/California Institute of Technology, funded by the National Aeronautics and Space Administration and the National Science Foundation.
The Digitized Sky Surveys were produced at the Space Telescope Science Institute under U.S. Government grant NAG W-2166. The images of these surveys are based on photographic data obtained using the Oschin Schmidt Telescope on Palomar Mountain and the UK Schmidt Telescope. The plates were processed into the present compressed digital form with the permission of these institutions.

%
\bibliographystyle{mn2e} 
\bibliography{my_lib_ref.bib} 

\appendix

\section{Tables}\label{app:tab}
{\onecolumn
\footnotesize
\begin{longtable}{c		c		l		c		c		c			c			c					c				c		c c}        
\caption{\label{tab:basic} Basic properties for all AGN of this work.}\\
\hline\hline	 
Section	&	BAT	&		&		&		&	Opt.		&	Nuc.	&					&					&		&		&		\\
	&	ID	&	Object	&	Redshift	&	$D$	&	class		&	morph.	&		$ \FNnuc$			&		$ \FQnuc$			&	$\dunr$ 	&	$\aRmed$	&	$\Rlar$	\\
	&		&		&		&	(Mpc)	&			&		&		 (mJy)			&		 (mJy)			&	(pc)	&		&		\\
(1)	&	(2)	&	(3)	&	(4)	&	(5)	&	(6)		&	(7)	&		(8)			&		(9)			&	(10)	&	(11)	&	(12)	\\

\hline																									
\endfirsthead
 \caption{continued.}\\
\hline\hline		
Section	&	BAT	&		&		&		&	Opt.		&	Nuc.	&					&					&		&		&		\\
	&	ID	&	Object	&	Redshift	&	$D$	&	class		&	morph.	&		$ \FNnuc$			&		$ \FQnuc$			&	$\dunr$ 	&	$\aRmed$	&	$\Rlar$	\\
	&		&		&		&	(Mpc)	&			&		&		 (mJy)			&		 (mJy)			&	(pc)	&		&		\\
(1)	&	(2)	&	(3)	&	(4)	&	(5)	&	(6)		&	(7)	&		(8)			&		(9)			&	(10)	&	(11)	&	(12)	\\

\hline																									
\endhead
\hline
\endfoot
 \ref{app:1H0419-577}	&	31	&	1H 0419-577	&	0.1040	&	499.0	&	1.5		&	-	&		62	$\pm$	12	&					&	712	&		&	0.762	\\
\ref{app:1RXSJ112716-6+190914}	&	78	&	1RXS J112716.6	&	0.1055	&	512.0	&	1.8		&	?	&		41	$\pm$	5	&					&	802	&		&		\\
\ref{app:2MASXJ03565655-4041453}	&	28	&	2MASX J03565655	&	0.0748	&	351.0	&	1.9		&	?	&		31	$\pm$	13	&					&	541	&		&		\\
\ref{app:2MASXJ09180027+0425066}	&	64	&	2MASX J09180027	&	0.1564	&	781.0	&	2		&	?	&		18	$\pm$	4	&					&	965	&		&		\\
\ref{app:3C029}	&		&	3C 29	&	0.0450	&	202.0	&	L		&	?	&	$\le$	3			&					&	371	&		&		\\
\ref{app:3C033}	&		&	3C 33	&	0.0597	&	273.0	&	2		&	o	&		35	$\pm$	18	&					&	839	&	0.82	&		\\
\ref{app:3C078}	&		&	3C 78	&	0.0287	&	127.0	&	1		&	?	&	$\le$	12			&					&	241	&		&	0.122	\\
\ref{app:3C093}	&		&	3C 93	&	0.3571	&	1967.0	&	1		&	?	&	$\le$	3			&					&	2135	&		&		\\
\ref{app:3C098}	&		&	3C 98	&	0.0305	&	137.0	&	2		&	?	&		25	$\pm$	7	&					&	304	&	1.19	&	1.258	\\
\ref{app:3C105}	&	29	&	3C 105	&	0.0890	&	421.0	&	2		&	?	&		10	$\pm$	4	&					&	593	&	1.01	&		\\
\ref{app:3C120}	&	32	&	3C 120	&	0.0330	&	150.0	&	1.5		&	?	&		266	$\pm$	7	&					&	200	&	0.88	&	0.619	\\
\ref{app:3C135}	&		&	3C 135	&	0.1274	&	620.0	&	2		&	?	&	$\le$	11			&					&	881	&		&		\\
\ref{app:3C227}	&		&	3C 227	&	0.0858	&	414.0	&	1.5		&	-	&		32	$\pm$	3	&					&	595	&	1.04	&		\\
\ref{app:3C264}	&		&	3C 264	&	0.0217	&	103.0	&	2:	*	&	?	&	$\le$	4			&					&	191	&	0.34	&	0.039	\\
\ref{app:3C273}	&		&	3C 273	&	0.1583	&	792.0	&	1		&	-	&		289	$\pm$	51	&					&	1095	&	0.90	&	0.528	\\
\ref{app:3C285}	&		&	3C 285	&	0.0794	&	378.0	&	2		&	?	&	$\le$	6			&					&	626	&	0.34	&	0.062	\\
\ref{app:3C293}	&		&	3C 293	&	0.0450	&	211.0	&	L		&	?	&	$\le$	5			&					&	373	&	0.23	&	0.201	\\
\ref{app:3C305}	&		&	3C 305	&	0.0416	&	192.0	&	2		&	?	&	$\le$	4			&					&	329	&	0.19	&	0.200	\\
\ref{app:3C317}	&		&	3C 317	&	0.0345	&	160.0	&	2/L		&	?	&	$\le$	3			&					&	310	&		&	0.033	\\
\ref{app:3C321}	&		&	3C 321	&	0.0961	&	460.0	&	2		&	o	&		70	$\pm$	50	&					&	782	&	0.64	&	0.822	\\
\ref{app:3C327}	&		&	3C 327	&	0.1048	&	505.0	&	1		&	o	&		71	$\pm$	21	&					&	1114	&	0.89	&	0.832	\\
\ref{app:3C353}	&		&	3C 353	&	0.0304	&	138.0	&	2/L		&	?	&		8	$\pm$	2	&					&	292	&	0.84	&		\\
\ref{app:3C382}	&	125	&	3C 382	&	0.0579	&	267.0	&	1		&	-	&		100	$\pm$	10	&					&	485	&	1.04	&		\\
\ref{app:3C390-3}	&	127	&	3C 390.3	&	0.0561	&	259.0	&	1.5		&	o	&		144	$\pm$	58	&					&	433	&	0.73	&		\\
\ref{app:3C403}	&		&	3C 403	&	0.0590	&	271.0	&	2		&	?	&		94	$\pm$	13	&					&	485	&	0.87	&	0.805	\\
\ref{app:3C424}	&		&	3C 424	&	0.1270	&	613.0	&	2		&	?	&	$\le$	2			&					&	982	&		&		\\
\ref{app:3C445}	&		&	3C 445	&	0.0559	&	254.0	&	1.5		&	?	&		180	$\pm$	19	&					&	357	&	1.04	&	0.866	\\
\ref{app:3C449}	&		&	3C 449	&	0.0171	&	72.4	&	L		&	?	&	$\le$	12			&					&	242	&	0.60	&	0.334	\\
\ref{app:3C452}	&	149	&	3C 452	&	0.0811	&	378.0	&	2		&	?	&		42	$\pm$	9	&					&	1039	&	1.40	&		\\
\ref{app:3C459}	&		&	3C 459	&	0.2201	&	1125.0	&	Cp		&	?	&	$\le$	16			&					&	1999	&		&	0.147	\\
\ref{app:4C+73-08}	&		&	4C +73.08	&	0.0581	&	271.0	&	2		&	?	&		22	$\pm$	8	&					&	421	&		&		\\
\ref{app:Ark120}	&	38	&	Ark 120	&	0.0327	&	149.0	&	1		&	-	&		247	$\pm$	14	&					&	200	&	0.91	&	0.773	\\
\ref{app:CenA}	&	104	&	Cen A	&	0.0018	&	3.8	&	2		&	circ.	&		1524	$\pm$	152	&		2298	$\pm$	294	&	5.8	&	0.69	&	0.069	\\
\ref{app:Circinus}	&		&	Circinus	&	0.0014	&	4.2	&	2		&	ellip.	&		8327	$\pm$	1049	&		14182	$\pm$	1467	&	5.34	&	0.70	&	0.443	\\
\ref{app:CygnusA}	&		&	Cygnus A	&	0.0561	&	257.0	&	2		&	circ.	&		57	$\pm$	14	&					&	626	&	0.42	&	0.398	\\
\ref{app:ESO005-G004}	&	49	&	ESO 5-4	&	0.0062	&	22.4	&	2		&	?	&		29	$\pm$	3	&		69	$\pm$	25	&	34.9	&	0.38	&	0.047	\\
\ref{app:ESO033-G002}	&		&	ESO 33-2	&	0.0181	&	82.3	&	2		&	o	&		174	$\pm$	62	&					&	276	&	0.79	&	0.838	\\
\ref{app:ESO103-G035}	&	126	&	ESO 103-35	&	0.0133	&	59.5	&	2		&	o	&		484	$\pm$	259	&					&	81	&	0.63	&	0.791	\\
\ref{app:ESO121-G028}	&	51	&	ESO 121-28	&	0.0405	&	187.0	&	2		&	-	&		15	$\pm$	1	&					&	285	&	0.91	&	0.203	\\
\ref{app:ESO138-G001}	&		&	ESO 138-1	&	0.0091	&	41.8	&	2		&	?	&		760	$\pm$	36	&					&	61.8	&	0.96	&	1.573	\\
\ref{app:ESO141-G055}	&		&	ESO 141-55	&	0.0371	&	169.0	&	1.2		&	o	&		149	$\pm$	36	&					&	289	&	0.76	&	0.615	\\
\ref{app:ESO198-G024}	&	18	&	ESO 198-24	&	0.0455	&	208.0	&	1		&	-	&		33	$\pm$	10	&					&	305	&	0.98	&		\\
\ref{app:ESO209-G012}	&		&	ESO 209-12	&	0.0405	&	189.0	&	1.5		&	?	&		164	$\pm$	23	&					&	250	&	0.87	&	0.812	\\
\ref{app:ESO253-G003}	&		&	ESO 253-3	&	0.0425	&	196.0	&	2		&	?	&		193	$\pm$	40	&					&	364	&	0.78	&	0.738	\\
\ref{app:ESO263-G013}	&		&	ESO 263-13	&	0.0335	&	158.0	&	2		&	-	&		56	$\pm$	4	&					&	223	&	0.93	&		\\
\ref{app:ESO286-IG019}	&		&	ESO 286-19	&	0.0430	&	195.0	&	Cp:	*	&	?	&		412	$\pm$	41	&		808	$\pm$	81	&	497	&	0.96	&	1.471	\\
\ref{app:ESO297-G018}	&	8	&	ESO 297-18	&	0.0252	&	111.0	&	2		&	?	&		31	$\pm$	5	&					&	171	&	0.48	&	0.181	\\
\ref{app:ESO323-G032}	&		&	ESO 323-32	&	0.0160	&	76.4	&	1.9		&	?	&		57	$\pm$	7	&					&	121	&		&	0.525	\\
\ref{app:ESO323-G077}	&		&	ESO 323-77	&	0.0150	&	71.8	&	1.2		&	circ.	&		347	$\pm$	79	&					&	89.3	&	0.48	&	0.510	\\
\ref{app:ESO362-G018}	&	39	&	ESO 362-18	&	0.0124	&	56.5	&	1.5		&	-	&		157	$\pm$	19	&					&	88.1	&	0.82	&	0.685	\\
\ref{app:ESO416-G002}	&	17	&	ESO 416-2	&	0.0591	&	272.0	&	1.9		&	?	&		25	$\pm$	3	&					&	504	&		&		\\
\ref{app:ESO420-G013}	&		&	ESO 420-13	&	0.0119	&	52.7	&	Cp		&	spir.	&		192	$\pm$	39	&					&	95	&	0.33	&	0.202	\\
\ref{app:ESO428-G014}	&		&	ESO 428-14	&	0.0057	&	28.2	&	2		&	ellip.	&		115	$\pm$	17	&		308	$\pm$	12	&	42.8	&	0.38	&	0.262	\\
\ref{app:ESO500-G034}	&		&	ESO 500-34	&	0.0122	&	60.2	&	Cp	*	&	comp.	&	$\le$	47			&					&	121	&		&	0.122	\\
\ref{app:ESO506-G027}	&	95	&	ESO 506-27	&	0.0250	&	119.0	&	2		&	-	&		180	$\pm$	18	&					&	183	&	0.88	&	1.215	\\
\ref{app:ESO511-G030}	&	113	&	ESO 511-30	&	0.0224	&	105.0	&	1		&	o	&		52	$\pm$	5	&					&	149	&	0.75	&	0.325	\\
\ref{app:ESO548-G081}	&	25	&	ESO 548-81	&	0.0145	&	63.5	&	1		&	o	&		87	$\pm$	33	&					&	103	&	0.70	&	0.349	\\
\ref{app:ESO602-G025}	&		&	ESO 602-25	&	0.0250	&	109.0	&	Cp:	*	&	spir.	&		34	$\pm$	13	&		43	$\pm$	7	&	297	&	0.14	&	0.124	\\
\ref{app:Fairall0009}	&	5	&	Fairall 9	&	0.0470	&	215.0	&	1.2		&	-	&		280	$\pm$	28	&					&	246	&	0.98	&		\\
\ref{app:Fairall0049}	&		&	Fairall 49	&	0.0200	&	90.1	&	2		&	o	&		368	$\pm$	213	&					&	193	&	0.86	&	0.590	\\
\ref{app:Fairall0051}	&		&	Fairall 51	&	0.0142	&	64.1	&	1.5		&	?	&		391	$\pm$	39	&		712	$\pm$	22	&	188	&	0.99	&	0.838	\\
\ref{app:FornaxA}	&		&	Fornax A	&	0.0059	&	19.0	&	L		&	o	&		17	$\pm$	6	&					&	36.2	&	0.35	&	0.054	\\
\ref{app:H0557-385}	&	47	&	H 0557-385	&	0.0339	&	156.0	&	1.2		&	?	&		426	$\pm$	43	&					&	251	&	1.08	&	0.666	\\
\ref{app:H1143-182}	&	82	&	H1143-182	&	0.0329	&	156.0	&	1.5		&	-	&		67	$\pm$	7	&					&	229	&	0.81	&	0.573	\\
\ref{app:HydraA}	&		&	Hydra A	&	0.0549	&	260.0	&	L		&	?	&	$\le$	5			&					&	379	&	0.72	&	0.159	\\
\ref{app:IC0883}	&		&	IC 883	&	0.0233	&	109.0	&	Cp	*	&	?	&	$\le$	222			&	$\le$	496			&	318	&	0.94	&	0.888	\\
\ref{app:IC1459}	&		&	IC 1459	&	0.0060	&	30.3	&	L		&	?	&	$\le$	17			&					&	44	&	0.75	&	0.138	\\
\ref{app:IC3639}	&		&	IC 3639	&	0.0109	&	53.6	&	2		&	-	&		386	$\pm$	39	&					&	84.6	&	1.00	&	0.536	\\
\ref{app:IC4329A}	&	108	&	IC 4329A	&	0.0161	&	76.5	&	1.2		&	o	&		1158	$\pm$	99	&		2220	$\pm$	142	&	114	&	1.01	&	1.043	\\
\ref{app:IC4518W}	&		&	IC 4518W	&	0.0163	&	76.1	&	2		&	o	&		199	$\pm$	32	&					&	196	&	0.71	&	0.554	\\
\ref{app:IC5063}	&	138	&	IC 5063	&	0.0113	&	49.1	&	2		&	ellip.	&		821	$\pm$	58	&		1850	$\pm$	37	&	82	&	0.74	&	0.739	\\
\ref{app:IIIZW035N}	&		&	III Zw 35N	&	0.0277	&	121.0	&	Cp		&	o	&		53	$\pm$	25	&					&	257	&	0.56	&	0.379	\\
\ref{app:IRAS00188-0856}	&		&	IRAS 00188-0856	&	0.1284	&	620.0	&	Cp	*	&	?	&		64	$\pm$	6	&		136	$\pm$	14	&	1485	&	0.95	&	0.551	\\
\ref{app:IRAS01003-2238}	&		&	IRAS 01003-2238	&	0.1178	&	565.0	&	2		&	?	&		215	$\pm$	21	&		402	$\pm$	40	&	1287	&	1.09	&	0.947	\\
\ref{app:IRAS04103-2838}	&		&	IRAS 04103-2838	&	0.1175	&	567.0	&	2		&	o	&		133	$\pm$	22	&		357	$\pm$	36	&	1398	&	0.97	&	1.773	\\
\ref{app:IRAS05189-2524}	&		&	IRAS 05189-2524	&	0.0426	&	196.0	&	2		&	o	&		651	$\pm$	259	&					&	359	&	1.01	&	0.880	\\
\ref{app:IRAS08572+3915}	&		&	IRAS 08572+3915	&	0.0584	&	275.0	&	Cp	*	&	o	&		602	$\pm$	132	&		1007	$\pm$	101	&	744	&	0.83	&	1.825	\\
\ref{app:IRAS09149-6206}	&		&	IRAS 09149-6206	&	0.0573	&	269.0	&	1		&	-	&		465	$\pm$	65	&					&	331	&	0.94	&	0.979	\\
\ref{app:IRAS11095-0238}	&		&	IRAS 11095-0238	&	0.1066	&	519.0	&	Cp:	*	&	?	&		84	$\pm$	29	&					&	1081	&	0.76	&	0.619	\\
\ref{app:IRAS13349+2438}	&		&	IRAS 13349+2438	&	0.1076	&	522.0	&	1n		&	?	&		476	$\pm$	63	&					&	893	&	0.97	&	0.755	\\
\ref{app:IRAS15250+3609}	&		&	IRAS 15250+3609	&	0.0552	&	258.0	&	Cp:	*	&	?	&		217	$\pm$	22	&		555	$\pm$	55	&	620	&	0.88	&	1.359	\\
\ref{app:IZW001}	&		&	I Zw 1	&	0.0589	&	269.0	&	1n		&	?	&		426	$\pm$	77	&					&	368	&	1.01	&	0.832	\\
\ref{app:LEDA013946}	&	27	&	LEDA 13946	&	0.0365	&	166.0	&	2		&	?	&	$\le$	12			&					&	266	&	0.74	&		\\
\ref{app:LEDA170194}	&	96	&	LEDA 170194	&	0.0367	&	173.0	&	2		&	-	&		42	$\pm$	8	&					&	288	&		&		\\
\ref{app:LEDA178130}	&	36	&	LEDA 178130	&	0.0350	&	160.0	&	2		&	o	&		52	$\pm$	17	&					&	284	&	0.89	&		\\
\ref{app:LEDA549777}	&	53	&	LEDA 549777	&	0.0610	&	284.1	&	2		&	?	&		33	$\pm$	5	&					&	887	&		&		\\
\ref{app:M051A}	&		&	M51a	&	0.0015	&	8.1	&	2		&	o	&		26	$\pm$	14	&					&	22.8	&	0.26	&	0.004	\\
\ref{app:M081}	&		&	M81	&	-1.0E-4	&	3.6	&	L		&	?	&		137	$\pm$	18	&		193	$\pm$	25	&	7.69	&	0.76	&	0.023	\\
\ref{app:M087}	&		&	M87	&	0.0044	&	16.7	&	L		&	-	&		21	$\pm$	3	&	$\le$	46			&	25.9	&	0.74	&	0.095	\\
\ref{app:MCG-01-05-047}	&		&	MCG-1-5-47	&	0.0172	&	73.4	&	2		&	?	&		53	$\pm$	22	&					&	92.2	&	0.45	&	0.340	\\
\ref{app:MCG-01-13-025}	&	34	&	MCG-1-13-25	&	0.0159	&	71.2	&	1.2		&	o	&		25	$\pm$	8	&					&	100	&	0.84	&		\\
\ref{app:MCG-01-24-012}	&	65	&	MCG-1-24-12	&	0.0196	&	93.8	&	2		&	-	&		111	$\pm$	11	&					&	129	&	1.10	&	1.383	\\
\ref{app:MCG-02-08-014}	&		&	MCG-2-8-14	&	0.0168	&	72.5	&	2		&	?	&		44	$\pm$	8	&					&	107	&		&	0.623	\\
\ref{app:MCG-02-08-039}	&		&	MCG-2-8-39	&	0.0299	&	133.0	&	2		&	-	&		231	$\pm$	41	&		369	$\pm$	50	&	356	&	1.20	&	1.154	\\
\ref{app:MCG-03-34-064}	&	103	&	MCG-3-34-64	&	0.0165	&	79.3	&	1.8/2		&	ellip.	&		531	$\pm$	65	&					&	116	&	0.66	&	0.564	\\
\ref{app:MCG-05-23-016}	&	69	&	MCG-5-23-16	&	0.0085	&	42.8	&	1.9		&	-	&		714	$\pm$	71	&		1457	$\pm$	52	&	60.3	&	1.02	&		\\
\ref{app:MCG-06-30-015}	&	105	&	MCG-6-30-15	&	0.0077	&	38.8	&	1.5		&	o	&		341	$\pm$	63	&					&	48.9	&	0.91	&	0.897	\\
\ref{app:MR2251-178}	&	151	&	MR 2251-178	&	0.0640	&	293.0	&	1.5		&	?	&		90	$\pm$	14	&					&	388	&	0.97	&	0.595	\\
\ref{app:Mrk0003}	&	50	&	Mrk 3	&	0.0135	&	60.6	&	2		&	o	&		448	$\pm$	120	&					&	137	&	0.84	&		\\
\ref{app:Mrk0266NE}	&		&	Mrk 266NE	&	0.0280	&	130.0	&	Cp	*	&	o	&		20	$\pm$	14	&					&	428	&	0.21	&	0.062	\\
\ref{app:Mrk0266SW}	&		&	Mrk 266SW	&	0.0276	&	128.0	&	2		&	ellip.	&		6	$\pm$	1	&					&	421	&	0.04	&	0.018	\\
\ref{app:Mrk0304}	&		&	Mrk 304	&	0.0658	&	301.0	&	1		&	o	&		52	$\pm$	25	&					&	511	&	0.86	&	0.851	\\
\ref{app:Mrk0509}	&	137	&	Mrk 509	&	0.0344	&	153.0	&	1.5		&	-	&		256	$\pm$	29	&		463	$\pm$	25	&	269	&	0.84	&	0.855	\\
\ref{app:Mrk0520}	&	144	&	Mrk 520	&	0.0266	&	115.0	&	Cp		&	o	&		86	$\pm$	24	&					&	196	&	0.57	&	0.452	\\
\ref{app:Mrk0573}	&		&	Mrk 573	&	0.0172	&	73.1	&	2		&	?	&		212	$\pm$	38	&		271	$\pm$	50	&	193	&	0.74	&	0.758	\\
\ref{app:Mrk0590}	&	12	&	Mrk 590	&	0.0264	&	116.0	&	1		&	-	&		98	$\pm$	10	&					&	164	&	1.03	&	0.511	\\
\ref{app:Mrk0841}	&	116	&	Mrk 841	&	0.0364	&	170.0	&	1.5		&	o	&		163	$\pm$	48	&					&	236	&	0.87	&	0.851	\\
\ref{app:Mrk0897}	&		&	Mrk 897	&	0.0263	&	115.0	&	Cp		&	circ.	&		8	$\pm$	3	&					&	214	&	0.07	&		\\
\ref{app:Mrk0915}	&		&	Mrk 915	&	0.0241	&	104.0	&	1.9		&	?	&		85	$\pm$	8	&		201	$\pm$	16	&	276	&	1.01	&	0.483	\\
\ref{app:Mrk0926}	&	153	&	Mrk 926	&	0.0469	&	210.0	&	1.5		&	o	&		107	$\pm$	27	&					&	287	&	0.65	&		\\
\ref{app:Mrk0937}	&		&	Mrk 937	&	0.0295	&	129.0	&	1		&	?	&	$\le$	15			&					&	175	&		&	0.103	\\
\ref{app:Mrk1014}	&		&	Mrk 1014	&	0.1631	&	807.0	&	1.5		&	o	&		99	$\pm$	42	&					&	1187	&	0.87	&	0.809	\\
\ref{app:Mrk1018}	&	10	&	Mrk 1018	&	0.0424	&	191.0	&	1		&	-	&		53	$\pm$	5	&					&	267	&	1.05	&		\\
\ref{app:Mrk1239}	&		&	Mrk 1239	&	0.0199	&	95.4	&	1n		&	o	&		574	$\pm$	95	&		916	$\pm$	41	&	142	&	1.02	&	0.884	\\
\ref{app:NGC0034}	&		&	NGC 34	&	0.0196	&	83.5	&	2		&	ellip.	&		58	$\pm$	6	&					&	158	&	0.22	&	0.165	\\
\ref{app:NGC0235A}	&	1	&	NGC 235A	&	0.0222	&	96.2	&	2		&	o	&		73	$\pm$	34	&					&	158	&	0.59	&		\\
\ref{app:NGC0253}	&		&	NGC 253	&	8.0E-4	&	3.2	&	Cp:	*	&	comp.	&	$\le$	1039			&	$\le$	2093			&	5.57	&	0.21	&	0.025	\\
\ref{app:NGC0424}	&		&	NGC 424	&	0.0118	&	49.5	&	2		&	o	&		736	$\pm$	223	&					&	80.4	&	0.87	&	0.669	\\
\ref{app:NGC0454E}	&	4	&	NGC 454E	&	0.0121	&	52.3	&	2		&	o	&		137	$\pm$	32	&					&	91.2	&	0.78	&		\\
\ref{app:NGC0526A}	&	6	&	NGC 526A	&	0.0191	&	82.8	&	1.9		&	-	&		236	$\pm$	29	&					&	130	&	1.03	&	1.028	\\
\ref{app:NGC0612}	&	7	&	NGC 612	&	0.0298	&	132.0	&	2		&	?	&	$\le$	33			&					&	270	&	0.52	&	0.166	\\
\ref{app:NGC0613}	&		&	NGC 613	&	0.0049	&	26.6	&	Cp	*	&	?	&		28	$\pm$	8	&					&	47.6	&	0.06	&	0.012	\\
\ref{app:NGC0676}	&		&	NGC 676	&	0.0050	&	19.5	&	2		&	?	&	$\le$	17			&					&	28.8	&		&	0.471	\\
\ref{app:NGC0788}	&	9	&	NGC 788	&	0.0136	&	57.2	&	2		&	-	&		134	$\pm$	16	&					&	88.2	&	1.13	&	0.715	\\
\ref{app:NGC0985}	&	16	&	NGC 985	&	0.0431	&	195.0	&	1.5		&	-	&		180	$\pm$	28	&					&	322	&	1.10	&	0.866	\\
\ref{app:NGC1052}	&		&	NGC 1052	&	0.0050	&	19.4	&	L		&	?	&		138	$\pm$	22	&		355	$\pm$	30	&	31.7	&	0.75	&	0.678	\\
\ref{app:NGC1068}	&		&	NGC 1068	&	0.0038	&	14.4	&	1.8/2		&	comp.	&		10206	$\pm$	3567	&		19950	$\pm$	537	&	24.1	&	0.29	&	0.256	\\
\ref{app:NGC1097}	&		&	NGC 1097	&	0.0042	&	17.0	&	L		&	circ.	&		17	$\pm$	2	&		46	$\pm$	11	&	27.1	&	0.12	&	0.006	\\
\ref{app:NGC1144}	&	20	&	NGC 1144	&	0.0288	&	128.0	&	2		&	?	&		25	$\pm$	15	&					&	185	&	0.49	&	0.096	\\
\ref{app:NGC1194}	&		&	NGC 1194	&	0.0136	&	58.2	&	1.9		&	-	&		277	$\pm$	28	&					&	87	&	0.96	&	1.040	\\
\ref{app:NGC1275}	&		&	NGC 1275	&	0.0176	&	76.8	&	1.5/L		&	?	&		917	$\pm$	92	&					&	129	&	0.87	&	0.865	\\
\ref{app:NGC1365}	&	24	&	NGC 1365	&	0.0055	&	17.9	&	1.8		&	?	&		361	$\pm$	36	&					&	29.8	&	0.42	&	0.070	\\
\ref{app:NGC1386}	&		&	NGC 1386	&	0.0029	&	16.5	&	2		&	ellip.	&		299	$\pm$	62	&		427	$\pm$	48	&	31.4	&	0.51	&	0.576	\\
\ref{app:NGC1433}	&		&	NGC 1433	&	0.0036	&	8.3	&	2:	*	&	?	&	$\le$	8			&					&	15.9	&	0.26	&	0.032	\\
\ref{app:NGC1553}	&		&	NGC 1553	&	0.0036	&	16.4	&	L		&	?	&	$\le$	27			&					&	52.6	&	0.73	&	0.209	\\
\ref{app:NGC1566}	&		&	NGC 1566	&	0.0050	&	14.3	&	1.5		&	?	&		59	$\pm$	30	&		146	$\pm$	29	&	18.4	&	0.51	&	0.031	\\
\ref{app:NGC1614}	&		&	NGC 1614	&	0.0159	&	71.0	&	Cp:	*	&	spir.	&	$\le$	346			&	$\le$	654			&	130	&	0.23	&	0.250	\\
\ref{app:NGC1667}	&		&	NGC 1667	&	0.0152	&	67.8	&	2		&	o	&		6	$\pm$	3	&					&	127	&	0.61	&	0.009	\\
\ref{app:NGC1808}	&		&	NGC 1808	&	0.0033	&	12.3	&	Cp:	*	&	ellip.	&		329	$\pm$	34	&					&	38.2	&	0.37	&	0.061	\\
\ref{app:NGC2110}	&	45	&	NGC 2110	&	0.0078	&	35.9	&	2		&	o	&		318	$\pm$	47	&		519	$\pm$	28	&	53.7	&	0.88	&	0.910	\\
\ref{app:NGC2623}	&		&	NGC 2623	&	0.0185	&	87.3	&	Cp		&	o	&		178	$\pm$	106	&		334	$\pm$	178	&	257	&	0.56	&	0.850	\\
\ref{app:NGC2992}	&	68	&	NGC 2992	&	0.0077	&	39.7	&	1.5/2		&	spir.	&		191	$\pm$	68	&		542	$\pm$	21	&	63.2	&	0.58	&	0.304	\\
\ref{app:NGC3081}	&	70	&	NGC 3081	&	0.0080	&	40.9	&	2		&	o	&		162	$\pm$	41	&		353	$\pm$	132	&	62.3	&	0.60	&		\\
\ref{app:NGC3094}	&		&	NGC 3094	&	0.0080	&	40.8	&	2:	*	&	-	&		1023	$\pm$	102	&					&	79.1	&		&	1.247	\\
\ref{app:NGC3147}	&		&	NGC 3147	&	0.0093	&	30.1	&	2		&	o	&		19	$\pm$	6	&					&	78.4	&	0.39	&	0.010	\\
\ref{app:NGC3166}	&		&	NGC 3166	&	0.0045	&	25.3	&	L		&	?	&		6	$\pm$	2	&					&	60.8	&		&	0.020	\\
\ref{app:NGC3169}	&		&	NGC 3169	&	0.0041	&	18.7	&	L		&	ellip.	&		8	$\pm$	5	&					&	31	&	0.05	&	0.007	\\
\ref{app:NGC3185}	&		&	NGC 3185	&	0.0041	&	20.3	&	2:	*	&	?	&	$\le$	31			&					&	50.5	&	0.38	&	0.202	\\
\ref{app:NGC3227}	&	71	&	NGC 3227	&	0.0039	&	22.1	&	1.5		&	circ.	&		201	$\pm$	50	&		772	$\pm$	47	&	27.4	&	0.53	&	0.213	\\
\ref{app:NGC3281}	&	72	&	NGC 3281	&	0.0107	&	52.8	&	2		&	ellip.	&		486	$\pm$	50	&		870	$\pm$	33	&	78.2	&	0.65	&	0.535	\\
\ref{app:NGC3312}	&		&	NGC 3312	&	0.0096	&	48.4	&	L		&	-	&		8	$\pm$	1	&					&	71.2	&		&	0.042	\\
\ref{app:NGC3368}	&		&	NGC 3368	&	0.0030	&	10.6	&	L		&	?	&	$\le$	15			&					&	21.1	&	0.10	&	0.015	\\
\ref{app:NGC3379}	&		&	NGC 3379	&	0.0030	&	10.6	&	L	*	&	?	&	$\le$	3			&					&	15.6	&	0.13	&	0.012	\\
\ref{app:NGC3393}	&		&	NGC 3393	&	0.0125	&	61.6	&	2		&	?	&		67	$\pm$	13	&					&	77	&	0.61	&	0.513	\\
\ref{app:NGC3486}	&		&	NGC 3486	&	0.0023	&	13.7	&	2/L	*	&	?	&	$\le$	5			&					&	28.1	&		&	0.008	\\
\ref{app:NGC3521}	&		&	NGC 3521	&	0.0027	&	11.5	&	L/H	*	&	?	&	$\le$	10			&					&	19.1	&	0.20	&	0.002	\\
\ref{app:NGC3607}	&		&	NGC 3607	&	0.0032	&	21.4	&	2/L	*	&	?	&	$\le$	14			&					&	49.3	&	0.50	&		\\
\ref{app:NGC3623}	&		&	NGC 3623	&	0.0027	&	12.8	&	L		&	?	&	$\le$	17			&					&	24.5	&		&	0.111	\\
\ref{app:NGC3627}	&		&	NGC 3627	&	0.0024	&	10.1	&	2/L	*	&	comp.	&		13	$\pm$	4	&					&	25.1	&	0.06	&	0.003	\\
\ref{app:NGC3628}	&		&	NGC 3628	&	0.0028	&	12.2	&	L/H	*	&	comp.	&	$\le$	10			&					&	22.9	&	0.02	&	0.003	\\
\ref{app:NGC3660}	&		&	NGC 3660	&	0.0123	&	60.8	&	1.8/2/L/H	*	&	o	&		26	$\pm$	15	&					&	127	&	0.79	&	0.132	\\
\ref{app:NGC3690E}	&		&	NGC 3690E	&	0.0104	&	49.1	&	Cp		&	o	&		280	$\pm$	157	&					&	126	&	0.47	&	0.070	\\
\ref{app:NGC3690W}	&		&	NGC 3690W	&	0.0104	&	48.2	&	Cp		&	o	&		773	$\pm$	492	&					&	124	&	0.63	&	0.195	\\
\ref{app:NGC3718}	&		&	NGC 3718	&	0.0033	&	17.0	&	L		&	?	&		22	$\pm$	5	&					&	35	&		&	0.150	\\
\ref{app:NGC3783}	&	79	&	NGC 3783	&	0.0097	&	48.4	&	1.5		&	?	&		682	$\pm$	46	&		1449	$\pm$	106	&	59.5	&	1.15	&	0.812	\\
\ref{app:NGC3982}	&		&	NGC 3982	&	0.0037	&	21.4	&	2		&	?	&		27	$\pm$	4	&					&	46	&	0.64	&	0.056	\\
\ref{app:NGC3998}	&		&	NGC 3998	&	0.0035	&	14.1	&	L		&	?	&		71	$\pm$	8	&		153	$\pm$	15	&	24	&	0.64	&	0.514	\\
\ref{app:NGC4051}	&	85	&	NGC 4051	&	0.0023	&	12.2	&	1n		&	?	&		464	$\pm$	48	&					&	27	&	0.79	&	0.344	\\
\ref{app:NGC4074}	&	86	&	NGC 4074	&	0.0224	&	107.0	&	2		&	o	&		68	$\pm$	19	&					&	170	&	0.75	&		\\
\ref{app:NGC4111}	&		&	NGC 4111	&	0.0027	&	15.0	&	L		&	circ.	&		4	$\pm$	1	&					&	37.9	&		&		\\
\ref{app:NGC4138}	&	88	&	NGC 4138	&	0.0030	&	13.8	&	1.9		&	?	&		21	$\pm$	3	&					&	30.5	&	0.58	&		\\
\ref{app:NGC4151}	&	89	&	NGC 4151	&	0.0033	&	13.3	&	1.5		&	o	&		1287	$\pm$	234	&					&	29.8	&	0.74	&	0.641	\\
\ref{app:NGC4235}	&		&	NGC 4235	&	0.0080	&	41.2	&	1.2		&	-	&		36	$\pm$	7	&		58	$\pm$	7	&	65.3	&	0.98	&	0.285	\\
\ref{app:NGC4258}	&		&	NGC 4258	&	0.0015	&	7.6	&	2		&	?	&		107	$\pm$	14	&					&	14.5	&	0.77	&	0.048	\\
\ref{app:NGC4261}	&		&	NGC 4261	&	0.0075	&	31.7	&	L		&	?	&		13	$\pm$	3	&		22	$\pm$	1	&	48.1	&	0.51	&	0.073	\\
\ref{app:NGC4278}	&		&	NGC 4278	&	0.0022	&	16.1	&	L		&	?	&		3	$\pm$	1	&					&	30.7	&	0.16	&	0.013	\\
\ref{app:NGC4303}	&		&	NGC 4303	&	0.0052	&	15.2	&	2	*	&	?	&		6	$\pm$	1	&					&	25.9	&		&	0.002	\\
\ref{app:NGC4374}	&		&	NGC 4374	&	0.0035	&	17.1	&	2/L		&	?	&	$\le$	8			&					&	29.5	&		&	0.052	\\
\ref{app:NGC4388}	&	91	&	NGC 4388	&	0.0084	&	19.2	&	2		&	ellip.	&		188	$\pm$	33	&		763	$\pm$	26	&	33.1	&	0.49	&	0.186	\\
\ref{app:NGC4395}	&	92	&	NGC 4395	&	0.0011	&	4.3	&	1.8		&	?	&		10	$\pm$	2	&					&	8.73	&	0.50	&	0.075	\\
\ref{app:NGC4418}	&		&	NGC 4418	&	0.0073	&	37.8	&	2	*	&	?	&		1427	$\pm$	168	&		3552	$\pm$	355	&	58.7	&	0.95	&	1.441	\\
\ref{app:NGC4438}	&		&	NGC 4438	&	2.0E-4	&	13.7	&	L/H	*	&	?	&		10	$\pm$	3	&					&	30.5	&	0.17	&	0.049	\\
\ref{app:NGC4457}	&		&	NGC 4457	&	0.0029	&	17.4	&	L	*	&	circ.	&		6	$\pm$	2	&					&	33.9	&	0.10	&	0.019	\\
\ref{app:NGC4472}	&		&	NGC 4472	&	0.0033	&	17.1	&	2/L	*	&	?	&	$\le$	9			&					&	29.2	&	0.58	&	0.066	\\
\ref{app:NGC4501}	&		&	NGC 4501	&	0.0076	&	17.9	&	2		&	?	&		4	$\pm$	1	&					&	36.4	&	0.23	&	0.002	\\
\ref{app:NGC4507}	&	94	&	NGC 4507	&	0.0118	&	57.5	&	2		&	-	&		623	$\pm$	64	&					&	90.7	&	1.02	&	1.363	\\
\ref{app:NGC4579}	&		&	NGC 4579	&	0.0051	&	16.8	&	L		&	?	&		75	$\pm$	5	&		101	$\pm$	11	&	25.6	&	0.70	&	0.067	\\
\ref{app:NGC4593}	&	97	&	NGC 4593	&	0.0090	&	45.6	&	1		&	ellip.	&		227	$\pm$	38	&		352	$\pm$	8	&	61.1	&	0.66	&	0.484	\\
\ref{app:NGC4594}	&		&	NGC 4594	&	0.0034	&	9.1	&	L		&	o	&		4	$\pm$	1	&					&	14.4	&	0.23	&	0.006	\\
\ref{app:NGC4636}	&		&	NGC 4636	&	0.0031	&	15.6	&	L		&	?	&	$\le$	7			&					&	23.9	&		&	0.033	\\
\ref{app:NGC4698}	&		&	NGC 4698	&	0.0034	&	24.4	&	2		&	?	&	$\le$	5			&					&	41.5	&		&	0.032	\\
\ref{app:NGC4736}	&		&	NGC 4736	&	0.0010	&	4.9	&	L		&	circ.	&		13	$\pm$	3	&		57	$\pm$	19	&	13.1	&	0.11	&	0.003	\\
\ref{app:NGC4746}	&		&	NGC 4746	&	0.0059	&	33.5	&	L/H	*	&	?	&	$\le$	11			&					&	59.5	&		&	0.029	\\
\ref{app:NGC4785}	&		&	NGC 4785	&	0.0123	&	59.0	&	2	*	&	?	&	$\le$	17			&					&	105	&		&	0.056	\\
\ref{app:NGC4941}	&		&	NGC 4941	&	0.0037	&	21.2	&	2		&	-	&		76	$\pm$	9	&		232	$\pm$	17	&	28.8	&	0.99	&	0.411	\\
\ref{app:NGC4945}	&		&	NGC 4945	&	0.0019	&	3.7	&	Cp		&	comp.	&		22	$\pm$	7	&	$\le$	95			&	5.68	&	0.19	&	0.001	\\
\ref{app:NGC4992}	&	102	&	NGC 4992	&	0.0251	&	119.0	&	1/2/L/N		&	?	&		80	$\pm$	18	&					&	182	&	0.99	&		\\
\ref{app:NGC5005}	&		&	NGC 5005	&	0.0032	&	16.9	&	L		&	ellip.	&		7	$\pm$	2	&					&	57.5	&	0.05	&	0.004	\\
\ref{app:NGC5033}	&		&	NGC 5033	&	0.0029	&	18.1	&	1.2		&	ellip.	&		16	$\pm$	3	&					&	32.8	&	0.13	&	0.009	\\
\ref{app:NGC5135}	&		&	NGC 5135	&	0.0137	&	66.0	&	2		&	-	&		132	$\pm$	26	&					&	97.8	&	0.29	&	0.209	\\
\ref{app:NGC5252}	&	106	&	NGC 5252	&	0.0230	&	109.0	&	1.9		&	?	&		69	$\pm$	7	&					&	157	&	1.18	&		\\
\ref{app:NGC5258}	&		&	NGC 5258	&	0.0225	&	107.0	&	L/H	*	&	?	&	$\le$	19			&					&	181	&	0.34	&	0.034	\\
\ref{app:NGC5273}	&		&	NGC 5273	&	0.0035	&	15.3	&	1.5		&	o	&		20	$\pm$	5	&					&	30.3	&	0.67	&	0.170	\\
\ref{app:NGC5347}	&		&	NGC 5347	&	0.0078	&	38.1	&	2		&	?	&		278	$\pm$	28	&		567	$\pm$	57	&	94.1	&	1.23	&	0.900	\\
\ref{app:NGC5363}	&		&	NGC 5363	&	0.0038	&	21.0	&	L		&	ellip.	&		2	$\pm$	1	&					&	35	&	0.04	&	0.011	\\
\ref{app:NGC5427}	&		&	NGC 5427	&	0.0087	&	29.4	&	2		&	?	&	$\le$	20			&					&	58.4	&	0.26	&	0.015	\\
\ref{app:NGC5506}	&	110	&	NGC 5506	&	0.0062	&	31.6	&	2		&	circ.	&		871	$\pm$	66	&		1697	$\pm$	143	&	49.9	&	0.74	&	0.675	\\
\ref{app:NGC5548}	&	112	&	NGC 5548	&	0.0172	&	80.7	&	1.5		&	o	&		124	$\pm$	77	&					&	132	&	0.57	&	0.288	\\
\ref{app:NGC5643}	&		&	NGC 5643	&	0.0040	&	20.9	&	2		&	o	&		254	$\pm$	69	&					&	29.9	&	0.66	&	0.129	\\
\ref{app:NGC5728}	&	115	&	NGC 5728	&	0.0094	&	45.4	&	1.9/2		&	circ.	&		49	$\pm$	7	&		123	$\pm$	34	&	76.7	&	0.38	&	0.234	\\
\ref{app:NGC5813}	&		&	NGC 5813	&	0.0066	&	29.7	&	L:	*	&	?	&	$\le$	7			&					&	49.1	&		&	0.223	\\
\ref{app:NGC5866}	&		&	NGC 5866	&	0.0022	&	14.1	&	L/H	*	&	?	&	$\le$	9			&					&	41.4	&	0.20	&	0.025	\\
\ref{app:NGC5953}	&		&	NGC 5953	&	0.0066	&	31.4	&	Cp		&	?	&	$\le$	29			&					&	59.6	&	0.14	&	0.036	\\
\ref{app:NGC5995}	&		&	NGC 5995	&	0.0252	&	117.0	&	1.9		&	-	&		332	$\pm$	47	&					&	175	&	0.88	&	0.850	\\
\ref{app:NGC6221}	&		&	NGC 6221	&	0.0050	&	10.7	&	Cp	*	&	comp.	&		104	$\pm$	21	&		190	$\pm$	44	&	21	&	0.13	&	0.033	\\
\ref{app:NGC6240N}	&	120	&	NGC 6240N	&	0.0251	&	114.0	&	Cp		&	comp.	&		17	$\pm$	7	&	$\le$	75			&	274	&	0.07	&	0.029	\\
\ref{app:NGC6240S}	&	120	&	NGC 6240S	&	0.0249	&	113.0	&	Cp		&	comp.	&		93	$\pm$	10	&		370	$\pm$	73	&	294	&	0.29	&	0.157	\\
\ref{app:NGC6251}	&		&	NGC 6251	&	0.0247	&	112.0	&	1/2/L		&	?	&		14	$\pm$	4	&					&	197	&	0.37	&	0.716	\\
\ref{app:NGC6300}	&		&	NGC 6300	&	0.0037	&	14.3	&	2		&	o	&		554	$\pm$	162	&					&	20	&	0.63	&	0.324	\\
\ref{app:NGC6810}	&		&	NGC 6810	&	0.0068	&	28.6	&	Cp	*	&	ellip.	&		44	$\pm$	13	&					&	48.9	&	0.08	&	0.035	\\
\ref{app:NGC6814}	&	129	&	NGC 6814	&	0.0052	&	20.1	&	1.5		&	-	&		96	$\pm$	24	&		150	$\pm$	15	&	33.5	&	0.93	&	0.104	\\
\ref{app:NGC6860}	&	133	&	NGC 6860	&	0.0149	&	65.8	&	1.5		&	-	&		206	$\pm$	24	&		304	$\pm$	14	&	105	&	1.06	&	0.824	\\
\ref{app:NGC6890}	&		&	NGC 6890	&	0.0081	&	33.8	&	1.9/2		&	o	&		117	$\pm$	26	&					&	56.9	&	0.82	&	0.341	\\
\ref{app:NGC7130}	&		&	NGC 7130	&	0.0162	&	68.9	&	Cp		&	circ.	&		105	$\pm$	21	&					&	100	&	0.38	&	0.180	\\
\ref{app:NGC7172}	&	145	&	NGC 7172	&	0.0087	&	34.8	&	2		&	ellip.	&		185	$\pm$	19	&					&	52.2	&	0.44	&	0.441	\\
\ref{app:NGC7212}	&		&	NGC 7212	&	0.0267	&	116.0	&	2		&	o	&		110	$\pm$	37	&		317	$\pm$	82	&	311	&	0.78	&	0.560	\\
\ref{app:NGC7213}	&	146	&	NGC 7213	&	0.0058	&	23.0	&	1.5/L		&	-	&		203	$\pm$	20	&		403	$\pm$	58	&	32.9	&	0.82	&	0.313	\\
\ref{app:NGC7314}	&	147	&	NGC 7314	&	0.0048	&	18.3	&	1.9/2		&	circ.	&		62	$\pm$	12	&		176	$\pm$	12	&	34.5	&	0.69	&	0.112	\\
\ref{app:NGC7469}	&	152	&	NGC 7469	&	0.0163	&	67.9	&	1/1.5		&	comp.	&		485	$\pm$	53	&		1174	$\pm$	155	&	109	&	0.46	&	0.305	\\
\ref{app:NGC7479}	&		&	NGC 7479	&	0.0079	&	30.0	&	2		&	-	&		695	$\pm$	95	&					&	54.5	&	0.84	&	0.507	\\
\ref{app:NGC7496}	&		&	NGC 7496	&	0.0055	&	21.1	&	Cp		&	circ.	&		170	$\pm$	11	&		389	$\pm$	21	&	53.3	&	0.55	&	0.292	\\
\ref{app:NGC7552}	&		&	NGC 7552	&	0.0054	&	20.4	&	L/H	*	&	?	&	$\le$	63			&	$\le$	100			&	39.3	&	0.05	&	0.017	\\
\ref{app:NGC7582}	&	154	&	NGC 7582	&	0.0053	&	23.0	&	Cp		&	comp.	&		443	$\pm$	79	&		568	$\pm$	48	&	33.9	&	0.35	&	0.193	\\
\ref{app:NGC7590}	&		&	NGC 7590	&	0.0053	&	26.5	&	2:	*	&	?	&	$\le$	11			&					&	43.5	&	0.52	&	0.015	\\
\ref{app:NGC7592W}	&		&	NGC 7592W	&	0.0246	&	105.0	&	Cp		&	?	&		135	$\pm$	50	&		173	$\pm$	22	&	276	&	0.71	&	0.521	\\
\ref{app:NGC7626}	&		&	NGC 7626	&	0.0114	&	45.4	&	L:		&	?	&	$\le$	18			&					&	73.2	&	0.95	&	0.376	\\
\ref{app:NGC7674}	&		&	NGC 7674	&	0.0289	&	126.0	&	2		&	ellip.	&		382	$\pm$	52	&		841	$\pm$	17	&	175	&	0.82	&	0.562	\\
\ref{app:NGC7679}	&		&	NGC 7679	&	0.0171	&	71.7	&	Cp		&	?	&		36	$\pm$	10	&					&	105	&	0.20	&	0.072	\\
\ref{app:NGC7743}	&		&	NGC 7743	&	0.0057	&	19.2	&	2/L		&	?	&	$\le$	9			&					&	28.8	&	0.38	&	0.090	\\
\ref{app:PG0026+129}	&		&	PG 0026+129	&	0.1420	&	691.0	&	1.2		&	?	&		32	$\pm$	3	&					&	1421	&	1.17	&	1.796	\\
\ref{app:PG0052+251}	&		&	PG 0052+251	&	0.1545	&	759.0	&	1.2		&	?	&		28	$\pm$	9	&					&	1328	&	0.72	&	0.346	\\
\ref{app:PG0844+349}	&		&	PG 0844+349	&	0.0640	&	302.0	&	1		&	o	&		39	$\pm$	16	&					&	643	&	0.80	&	0.313	\\
\ref{app:PG2130+099}	&		&	PG 2130+099	&	0.0630	&	288.0	&	1.5		&	?	&		188	$\pm$	21	&					&	451	&	1.09	&	1.010	\\
\ref{app:PictorA}	&	40	&	Pictor A	&	0.0351	&	161.0	&	1.5/L		&	?	&		75	$\pm$	7	&		128	$\pm$	20	&	249	&	0.97	&	0.684	\\
\ref{app:PKS1417-19}	&		&	PKS 1417-19	&	0.1200	&	586.0	&	1.5		&	o	&		20	$\pm$	10	&					&	844	&		&		\\
\ref{app:PKS1814-63}	&		&	PKS 1814-63	&	0.0647	&	302.0	&	2		&	?	&		28	$\pm$	5	&					&	595	&	0.94	&		\\
\ref{app:PKS1932-46}	&		&	PKS 1932-46	&	0.2307	&	1191.0	&	1.9		&	?	&	$\le$	2			&					&	2034	&		&		\\
\ref{app:PKS2158-380}	&		&	PKS 2158-380	&	0.0334	&	149.0	&	2		&	?	&		19	$\pm$	3	&					&	284	&		&	0.213	\\
\ref{app:PKS2354-35}	&		&	PKS 2354-35	&	0.0491	&	222.0	&	L		&	?	&	$\le$	2			&					&	346	&		&		\\
\ref{app:SuperantennaeS}	&		&	Superantennae S	&	0.0624	&	291.0	&	2		&	o	&		221	$\pm$	63	&		533	$\pm$	151	&	440	&	0.86	&	0.998	\\
\ref{app:UGC05101}	&		&	UGC 5101	&	0.0394	&	182.0	&	Cp		&	ellip.	&		227	$\pm$	41	&		230	$\pm$	42	&	597	&	0.87	&	0.908	\\
\ref{app:UGC12348}	&		&	UGC 12348	&	0.0255	&	110.0	&	2		&	o	&		98	$\pm$	16	&		158	$\pm$	39	&	282	&		&	0.901	\\
\ref{app:Z041-020}	&	83	&	Z 41-20	&	0.0360	&	170.0	&	2		&	?	&		29	$\pm$	3	&					&	238	&	0.49	&	0.222	\\
 
\end{longtable} 
\normalsize
 {\it -- Notes:}
 (1) Section in App.~\ref{app:indi}
 (2) BAT ID number from the nine-month AGN catalogue according to \cite{winter_x-ray_2009};
 (3) commonly used object name;
 (4) redshift from NED;
 (5) distance, mostly redshift-based luminosity distance, corrected for the Earth's motion relative to the cosmic microwave background reference frame with $H_0 = 67.3, \Omega_\mathrm{m} = 0.315,$ and $\Omega_\mathrm{vac} = 0.685$ \citep{planck_collaboration_planck_2013} or redshift-independent distance (see App.~\ref{app:indi} for individual references);
 (6) optical AGN classification, objects marked with a * mark uncertain AGN (see App.~\ref{app:indi} for individual references);
 (7) nuclear morphology at subarcsecond resolution: ``-'' point-like, ``o'' possibly extended, ``?'' unknown extension,  ``circ.'' spherical extended, ``ellip.'' elliptical and bar-like extended, ``spir.'' spiral-like extended, ``comp.'' complex extended emission (see Section~\ref{sec:morph});
 (8) nuclear subarcsecond-scale monochromatic flux density at restframe 12\umm  in mJy (see Section~\ref{sec:nuc_emi}); 
 (9) nuclear subarcsecond-scale monochromatic flux density at restframe 18\umm in mJy (see Section~\ref{sec:nuc_emi});
 (10) size constraint on the nuclear unresolved emission in pc, which is set equal to the minimum PSF FWHM(major axis) measured for the object.  
 (11) average nuclear subarcsecond to arcsecond flux ratio (see Section~\ref{sec:N/I});
 (12) nuclear subarcsecond to total flux ratio (see Section~\ref{sec:glob}).
 This table is also be available through the VO, CDS and \url{http://dc.g-vo.org/sasmirala}.
 
}
\twocolumn

{\onecolumn
\begin{landscape}
\scriptsize
\begin{longtable}{l c c c c c c c c c c c c c c c c }        
\caption{\label{tab:obs} Observational parameters and results for all AGN.}\\
\hline\hline	 
        &	 &         &      &        &                          &      &  Exp. & STD & STD & STD & SCI & SCI & SCI &  & \\
Object  & Instr. & Ob. ID. & Obs. date & Filter & $\lambda_\mathrm{rest}$ & HWHM & time & Maj. & Min. & PA & Maj. & Min. & PA & $\Fgau$  & $\Fpsf$  \\    
        &     &   & {\scriptsize(YY-MM-DD)} &  & \um & \um &(s) & (\arcsec) & (\arcsec) & ($\degree$) & (\arcsec) & (\arcsec) & ($\degree$) & (mJy) & (mJy) \\
        (1)	&	(2)	&	(3)	&	(4)	&	(5)	&	(6)	&	(7)	&	(8)	&		(9)			&		(10)			&	(11)	&	(12)	& (13) & (14) & (15) & (16) \\
\hline	
\endfirsthead
 \caption{continued.}\\
\hline\hline	
        &	 &         &      &        &                          &      &  Exp. & STD & STD & STD & SCI & SCI & SCI &  & \\
Object  & Instr. & Ob. ID. & Obs. date & Filter & $\lambda_\mathrm{rest}$ & HWHM & time & Maj. & Min. & PA & Maj. & Min. & PA & $\Fgau$  & $\Fpsf$  \\    
        &     &   & {\scriptsize(YY-MM-DD)} &  & \um & \um &(s) & (\arcsec) & (\arcsec) & ($\degree$) & (\arcsec) & (\arcsec) & ($\degree$) &  (mJy) & (mJy) \\
                (1)	&	(2)	&	(3)	&	(4)	&	(5)	&	(6)	&	(7)	&	(8)	&		(9)			&		(10)			&	(11)	&	(12)	& (13) & (14) & (15) & (16) \\
\hline	
\hline											\endhead
\hline
\endfoot
1H0419-577 & VISIR &  084.B-0606(A)  & 09-11-29 & NEII\_1 & 12.27 &  0.18 &      360 &  0.38 &  0.35 &      126 &  0.36 &  0.35 &      171 &     63.2 $\pm$     5.3 &     64.9 $\pm$     3.6\\
1H0419-577 & VISIR &  084.B-0606(A)  & 09-11-29 & NEII\_2 & 13.04 &  0.22 &      543 &  0.36 &  0.36 &      128 &  0.40 &  0.36 &       23 &     62.1 $\pm$    10.2 &     53.5 $\pm$     4.3\\
1H0419-577 & VISIR &  084.B-0606(A)  & 09-11-29 & PAH2 & 11.25 &  0.59 &      181 &  0.37 &  0.33 &      108 &  0.36 &  0.33 &       80 &     67.2 $\pm$    13.0 &     70.7 $\pm$     8.2\\
1RXSJ112716 & VISIR &  084.B-0606(A)  & 10-01-26 & NEII\_1 & 12.27 &  0.18 &     1442 &  0.56 &  0.48 &       95 &  0.51 &  0.41 &        2 &     39.0 $\pm$    15.1 &     53.8 $\pm$     7.2\\
1RXSJ112716 & VISIR &  084.B-0606(A)  & 10-03-23 & NEII\_2 & 13.04 &  0.22 &     1991 &  0.40 &  0.37 &       66 &  0.58 &  0.51 &      166 &     41.1 $\pm$     2.6 &     20.6 $\pm$     2.0\\
1RXSJ112716 & VISIR &  084.B-0606(A)  & 10-01-26 & PAH2 & 11.25 &  0.59 &      362 &  0.42 &  0.39 &       12 &  0.47 &  0.34 &      157 &     36.4 $\pm$     8.9 &     35.6 $\pm$     8.8\\
2MASXJ0356565 & VISIR &  084.B-0606(A)  & 09-09-30 & NEII\_1 & 12.27 &  0.18 &      540 &  0.43 &  0.40 &       10 &  0.77 &  0.58 &      119 &     26.5 $\pm$    11.2 &     10.0 $\pm$    11.2\\
2MASXJ0356565 & VISIR &  084.B-0606(A)  & 09-09-30 & NEII\_2 & 13.04 &  0.22 &      901 &  0.41 &  0.40 &       16 &  0.71 &  0.60 &       85 &     33.6 $\pm$     7.6 &     13.0 $\pm$     7.6\\
2MASXJ0356565 & VISIR &  084.B-0606(A)  & 09-09-30 & SIV\_2 & 10.77 &  0.19 &      362 &  0.37 &  0.34 &      171 &  0.69 &  0.33 &      120 &     23.8 $\pm$     7.7 &     13.0 $\pm$     7.7\\
2MASXJ0918002 & VISIR &  084.B-0606(A)  & 10-01-05 & PAH2\_2 & 11.88 &  0.37 &     1986 &  0.34 &  0.31 &       94 &  0.42 &  0.31 &       97 &     16.6 $\pm$     0.6 &     13.4 $\pm$     0.7\\
3C029 & VISIR &  078.B-0020(A)  & 06-12-27 & SIC & 11.85 &  2.34 &     1803 &  0.41 &  0.35 &       86 &  $\dots$  &  $\dots$  &  $\dots$  &  $\le$    11.1   &  $\le$    11.1  \\
3C033 & COMICS &  o09108  & 09-07-14 & N11.7 & 11.74 &  0.53 &     3208 &  0.71 &  0.52 &       72 &  1.00 &  0.85 &      121 &     47.4 $\pm$     5.0 &     20.8 $\pm$     3.0\\
\end{longtable} 
\normalsize
 {\it -- Notes:}
 (1) Object name;
 (2) instrument;
 (3) observational programme ID (see Section~\ref{sec:obs});
 (4) date of observations in year-month-day format;
 (5) filter name;
 (6) filter central wavelength in \um;
 (7) filter half width at half maximum in \um;
 (8) total exposure time on target in seconds;
 (9) standard star major axis FWHM in arcsecond;
 (10) standard star minor axis FWHM in arcsecond;
 (11) standard star PA of the major axis from north to east;
 (12) science target major axis FWHM in arcsecond (constrained to $\le1\arcsec$);
 (13) science target minor axis FWHM in arcsecond;
 (14) science target PA of the major axis from north to east; 
 (15) nuclear flux density with Gaussian measurement method (Section~\ref{sec:met});
 (16) nuclear flux density with PSF-scaling measurement method (Section~\ref{sec:met}). 
 The full table is available only in the online version.
 This table is also be available through the VO, CDS and \url{http://dc.g-vo.org/sasmirala}.
 
\end{landscape}
}
\twocolumn 

\twocolumn[\begin{@twocolumnfalse}
\section{The individual objects}\label{app:indi}
This Appendix section contains descriptions of the individual objects of the total sample with particular emphasis on their MIR properties. 
Along with the new subarcsecond-resolution MIR images, also the available $\sim 5$ to $25\,\mu$m \spitzerr data, as well as the resulting MIR SED are presented and mostly qualitatively described.
Here, we concentrate mostly on the MIR continuum and the PAH and silicate features.
Discussion and references of the MIR emission line properties are beyond the scope of this work. 
In order to allow for a better understanding of the  discussed MIR properties, additional general information like optical, radio and emission-line morphologies are provided, as well as a large-scale optical image from the Digitized Sky Survey (DSS).
However, the discussion and references of the non-MIR properties are not intended to be complete in any sense but rather meant as a summarizing overview for each object.
The sections are ordered alphabetically using the most common object name, while other commonly used names are also given in the section titles.
The object descriptions are also available on \url{http://dc.g-vo.org/sasmirala} including SED plots in $f_\nu$-space.

\subsection{1H\,0419-577 -- IRAS\,F04250-5718}\label{app:1H0419-577}
1H\,0419-577 is a radio-quiet low-luminosity quasar at a redshift of $z=$ 0.1004 ($D \sim 499$\,Mpc) with an optical Sy\,1.5 classification \citep{veron-cetty_catalogue_2010}.
It was first discovered with the \textit{HEAO-1} satellite in X-rays \citep{brissenden_multiwaveband_1987}.
Since then, it has mainly been studied in X-rays (e.g., \citealt{fabian_x-ray_2005,turner_suzaku_2009}), and is also included in the nine-month BAT sample.
No \spitzer/IRAC or IRS data are available for 1H\,0419-577 but MIPS shows an unresolved source. 
We observed this object with VISIR in three narrow $N$-band filters in 2009 and detected an unresolved nucleus without any further host emission.
The corresponding nuclear photometry indicates a blue $N$-band SED possibly due to silicate 10$\,\mu$m emission.
Combined with the MIPS 24$\,\mu$m photometry, the resulting broad-band MIR spectral slope appears to be flat or blue in $\nu F_\nu$-space.
\newline\end{@twocolumnfalse}]

\begin{figure}
    \centering
    \includegraphics[angle=0,width=8.500cm]{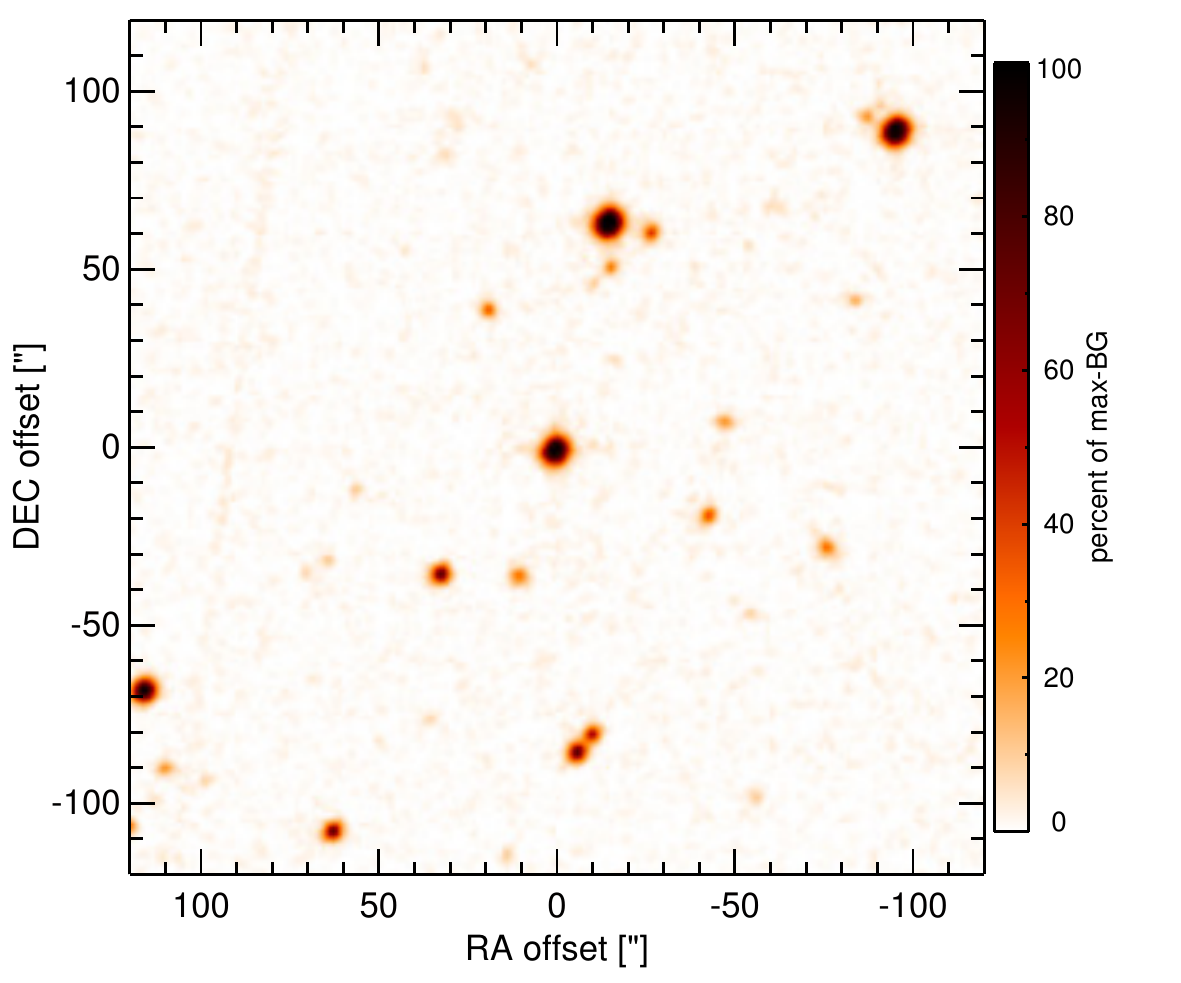}
     \caption{\label{fig:OPTim_first}
              Optical image (DSS, red filter) of 1H\,0419-577. 
              Displayed are the central $4\arcmin$ with North up and East to the left. 
              The colour scaling is linear with white corresponding to the median background and black to the $0.01\%$ pixels with the highest intensity. 
            }
 \end{figure}
\begin{figure}
   \centering
   \includegraphics[angle=0,height=3.11cm]{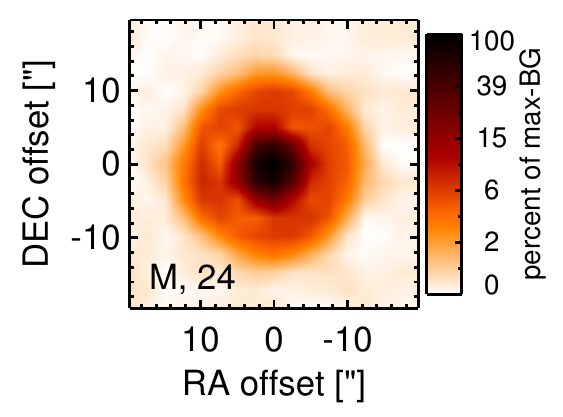} 
    \caption{\label{fig:INTim_first}
             \spitzerr MIR images of 1H\,0419-577. 
             Displayed are the inner $40\arcsec$ with North up and East to the left. The colour scaling is logarithmic with white corresponding to median background and black to the $0.1\%$ pixels with the highest intensity.
             The label in the bottom left states instrument and central wavelength of the filter in $\mu$m (I: IRAC, M: MIPS).
           }
\end{figure}
\begin{figure}
   \centering
   \includegraphics[angle=0,height=3.11cm]{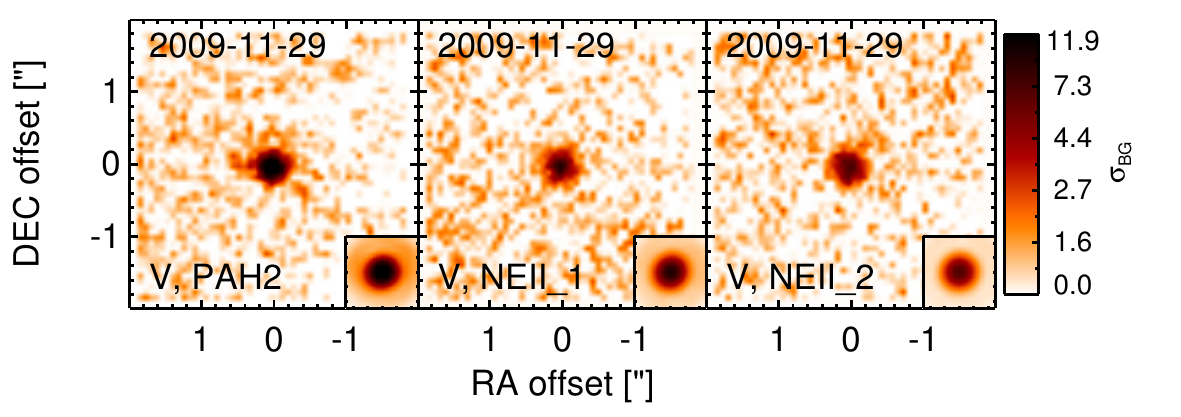}
    \caption{\label{fig:HARim_first}
             Subarcsecond-resolution MIR images of 1H\,0419-577 sorted by increasing filter wavelength. 
             Displayed are the inner $4\arcsec$ with North up and East to the left. 
             The colour scaling is logarithmic with white corresponding to median background and black to the $75\%$ of the highest intensity of all images in units of $\sigbg$.
             The inset image shows the central arcsecond of the PSF from the calibrator star, scaled to match the science target.
             The labels in the bottom left state instrument and filter names (C: COMICS, M: Michelle, T: T-ReCS, V: VISIR).
           }
\end{figure}
\begin{figure}
   \centering
  \includegraphics[angle=0,width=8.50cm]{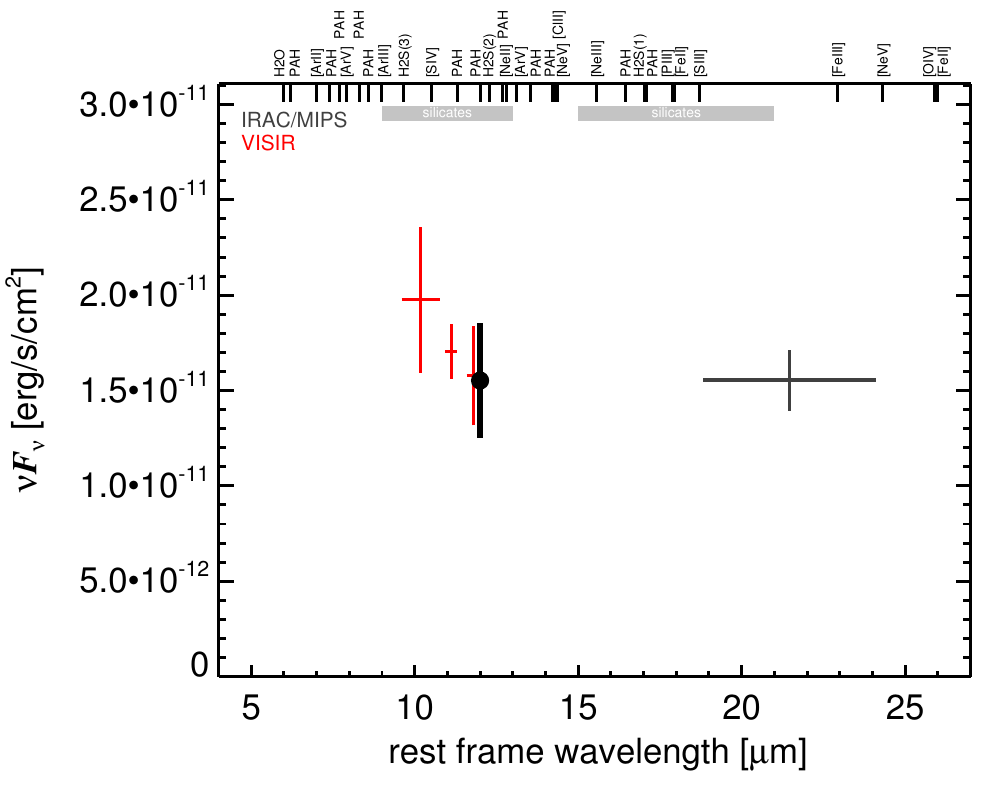}
   \caption{\label{fig:MIRSED_first}
      MIR SED of 1H\,0419-577. 
      The description  of the symbols (if present) is the following.
      Grey crosses and  solid lines mark the \spitzer/IRAC, MIPS and IRS data. 
      The colour coding of the other symbols is: 
      green for COMICS, magenta for Michelle, blue for T-ReCS and red for VISIR data.
      Darker-coloured solid lines mark spectra of the corresponding instrument.
      The black filled circles mark the nuclear 12 and $18\,\mu$m  continuum emission estimate from the data.
      The ticks on the top axis mark positions of common MIR emission lines, while the light grey horizontal bars mark wavelength ranges affected by the silicate 10 and 18$\mu$m features.     
   }
\end{figure}

 \clearpage

\twocolumn[\begin{@twocolumnfalse}
\subsection{1RXS\,J112716.6+190914 -- 2MASX\,J11271632+1909198}\label{app:1RXSJ112716-6+190914}
1RXS\,J112716.6 is a radio-quiet low-luminosity quasar at a redshift of $z=$ 0.1055 ($D \sim 512$\,Mpc) with an optical Sy\,1.8 classification \citep{veron-cetty_catalogue_2010}, discovered in the \textit{ROSAT} all sky survey \citep{voges_rosat_1999}.
Since its discovery, it has been studied little but it is a member of the nine-month BAT AGN sample.
No \spitzerr data are available for this object.
We observed 1RXS\,J112716.6 with VISIR in three narrow $N$-band filters in 2010 and weakly detected a compact nucleus.
The VISIR observations in the NEII\_1 filter are affected by poor ambient conditions as indicated from the poor sensitivity obtained from the corresponding standard star during that night. 
1RXS\,J112716.6 appears rather point-like in the other two filters and features a flat $N$-band spectral slope in $\nu F_\nu$-space.
\newline\end{@twocolumnfalse}]

\begin{figure}
   \centering
   \includegraphics[angle=0,width=8.500cm]{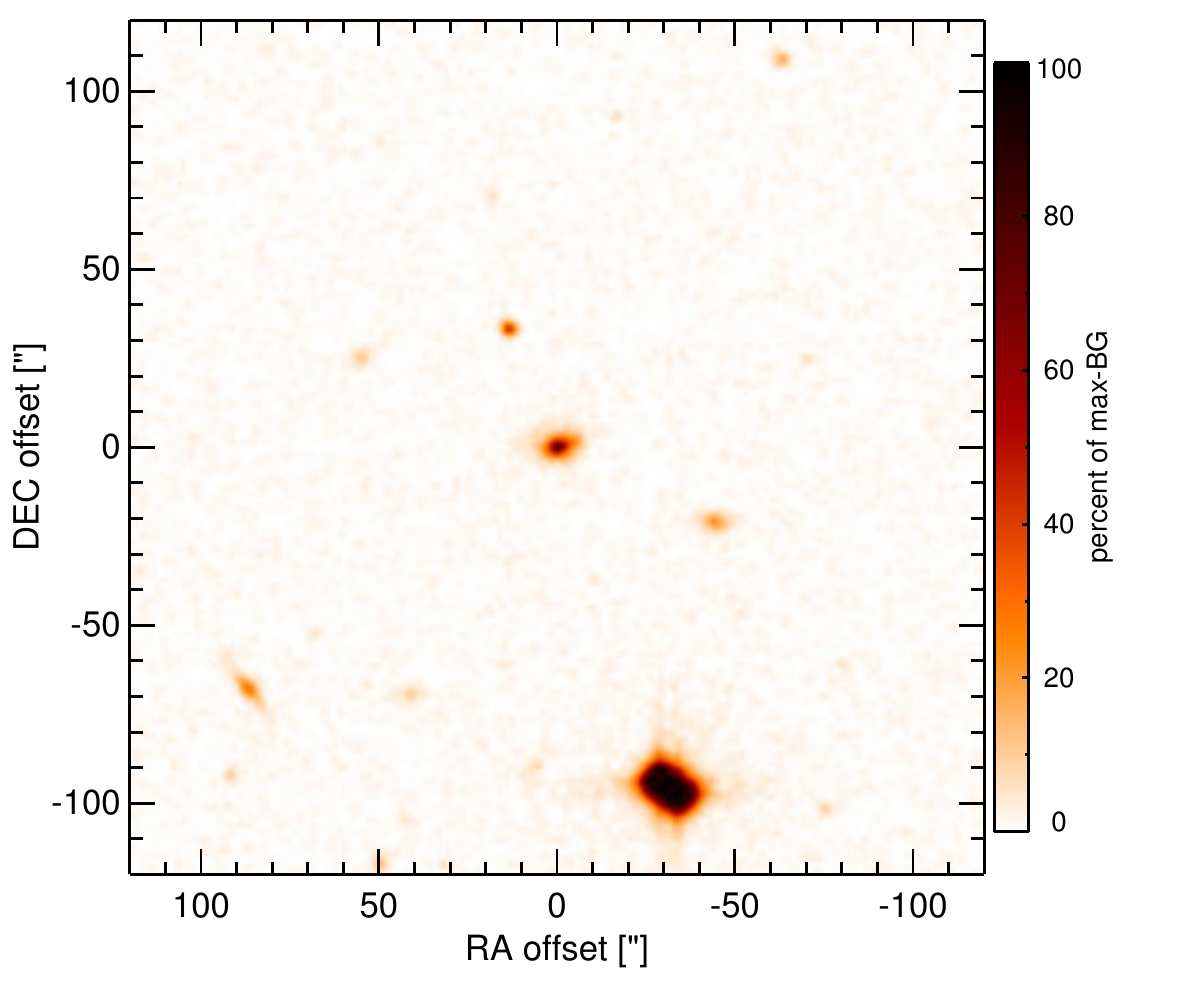}
    \caption{\label{fig:OPTim_1RXSJ112716-6+190914}
             Optical image (DSS, red filter) of 1RXS\,J112716.6+190914. Displayed are the central $4\arcmin$ with North up and East to the left. 
              The colour scaling is linear with white corresponding to the median background and black to the $0.01\%$ pixels with the highest intensity.  
           }
\end{figure}
\begin{figure}
   \centering
   \includegraphics[angle=0,height=3.11cm]{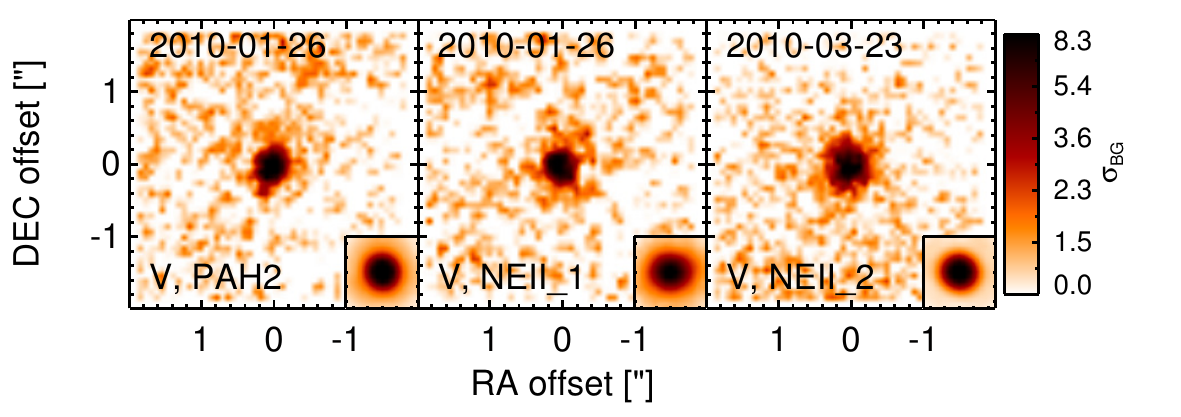}
    \caption{\label{fig:HARim_1RXSJ112716-6+190914}
             Subarcsecond-resolution MIR images of 1RXS\,J112716.6+190914 sorted by increasing filter wavelength. 
             Displayed are the inner $4\arcsec$ with North up and East to the left. 
             The colour scaling is logarithmic with white corresponding to median background and black to the $75\%$ of the highest intensity of all images in units of $\sigbg$.
             The inset image shows the central arcsecond of the PSF from the calibrator star, scaled to match the science target.
             The labels in the bottom left state instrument and filter names (C: COMICS, M: Michelle, T: T-ReCS, V: VISIR).
           }
\end{figure}
\begin{figure}
   \centering
   \includegraphics[angle=0,width=8.50cm]{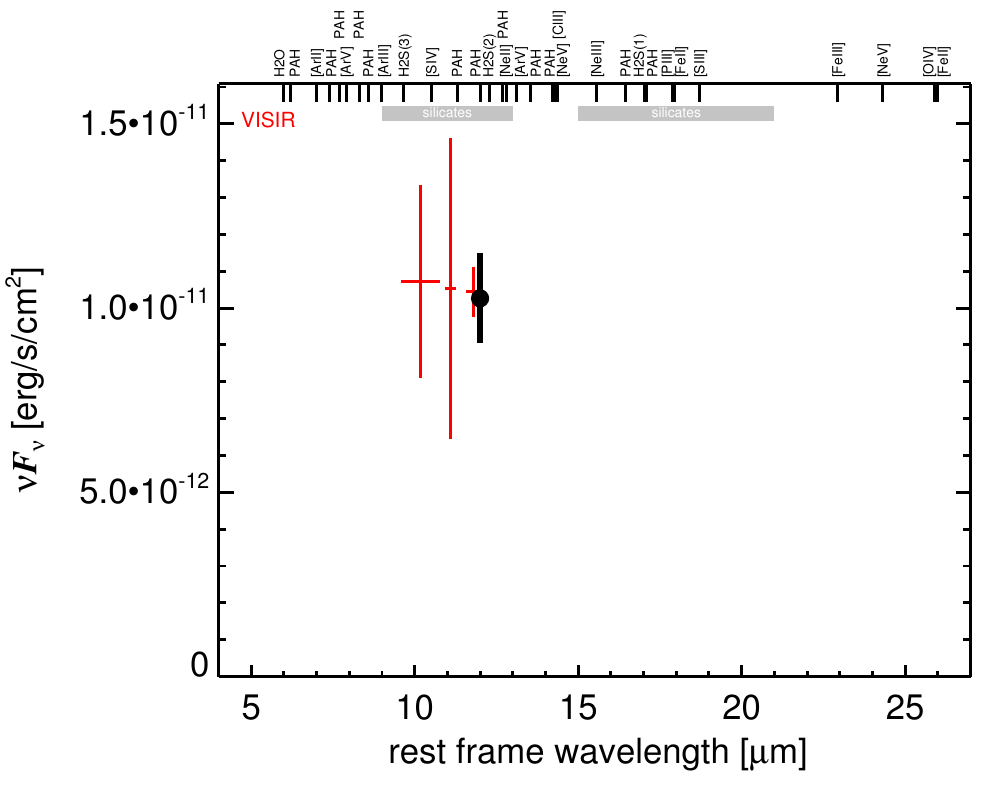}
   \caption{\label{fig:MISED_1RXSJ112716-6+190914}
      MIR SED of 1RXS\,J112716.6+190914. The description  of the symbols (if present) is the following.
      Grey crosses and  solid lines mark the \spitzer/IRAC, MIPS and IRS data. 
      The colour coding of the other symbols is: 
      green for COMICS, magenta for Michelle, blue for T-ReCS and red for VISIR data.
      Darker-coloured solid lines mark spectra of the corresponding instrument.
      The black filled circles mark the nuclear 12 and $18\,\mu$m  continuum emission estimate from the data.
      The ticks on the top axis mark positions of common MIR emission lines, while the light grey horizontal bars mark wavelength ranges affected by the silicate 10 and 18$\mu$m features.     
   }
\end{figure}
\clearpage

\twocolumn[\begin{@twocolumnfalse}  
\subsection{2MASX\,J03565655-4041453}\label{app:2MASXJ03565655-4041453}
2MASX\,J03565655 is a galaxy, possibly a spiral, at a redshift of $z=$ 0.0748 ($D \sim 324$\,Mpc) with an active nucleus optically classified as a Sy\,1.9.
The AGN was discovered only recently through \textit{Swift} observations \citep{tueller_swift_2008} and thus belongs to the nine-month BAT AGN sample. 
No \spitzerr data are available for 2MASX\,J03565655, and it appears very compact in the \textit{WISE} images.
We observed this object with VISIR in three narrow $N$-band filters in 2010 and detected a very faint nucleus.
The low S/N prohibits any meaningful analysis of the nuclear extension.
The corresponding nuclear photometry indicates a red $N$-band SED possibly due to silicate absorption.
\newline\end{@twocolumnfalse}]

\begin{figure}
   \centering
   \includegraphics[angle=0,width=8.500cm]{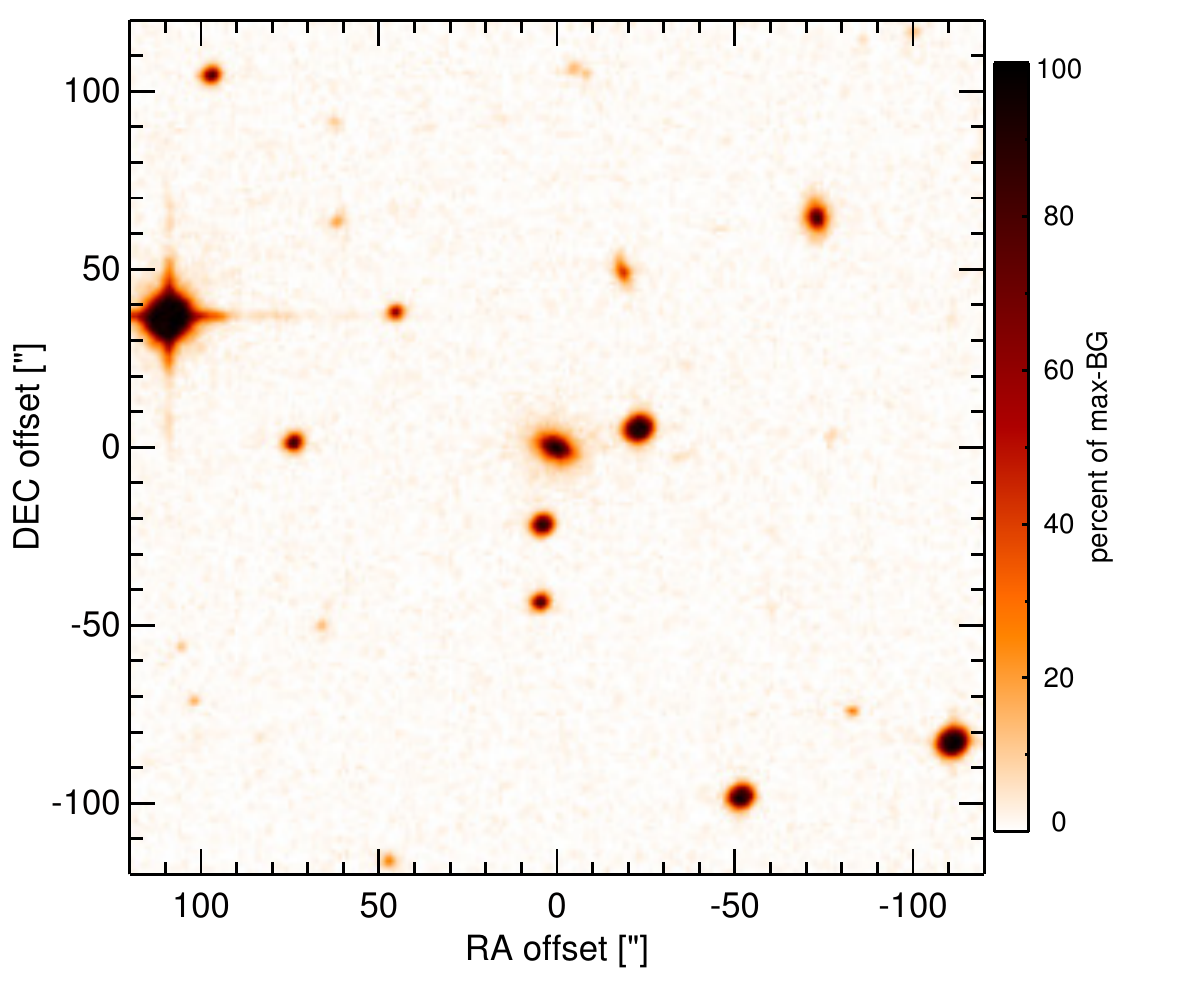}
    \caption{\label{fig:OPTim_2MASXJ03565655-4041453}
             Optical image (DSS, red filter) of 2MASX\,J03565655-4041453. Displayed are the central $4\arcmin$ with North up and East to the left. 
              The colour scaling is linear with white corresponding to the median background and black to the $0.01\%$ pixels with the highest intensity.  
           }
\end{figure}
\begin{figure}
   \centering
   \includegraphics[angle=0,height=3.11cm]{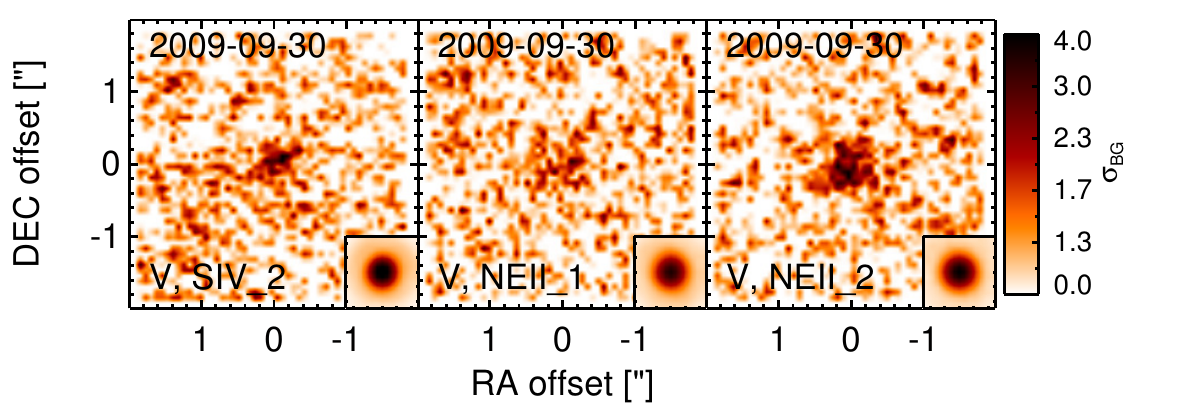}
    \caption{\label{fig:HARim_2MASXJ03565655-4041453}
             Subarcsecond-resolution MIR images of 2MASX\,J03565655-4041453 sorted by increasing filter wavelength. 
             Displayed are the inner $4\arcsec$ with North up and East to the left. 
             The colour scaling is logarithmic with white corresponding to median background and black to the $75\%$ of the highest intensity of all images in units of $\sigbg$.
             The inset image shows the central arcsecond of the PSF from the calibrator star, scaled to match the science target.
             The labels in the bottom left state instrument and filter names (C: COMICS, M: Michelle, T: T-ReCS, V: VISIR).
           }
\end{figure}
\begin{figure}
   \centering
   \includegraphics[angle=0,width=8.50cm]{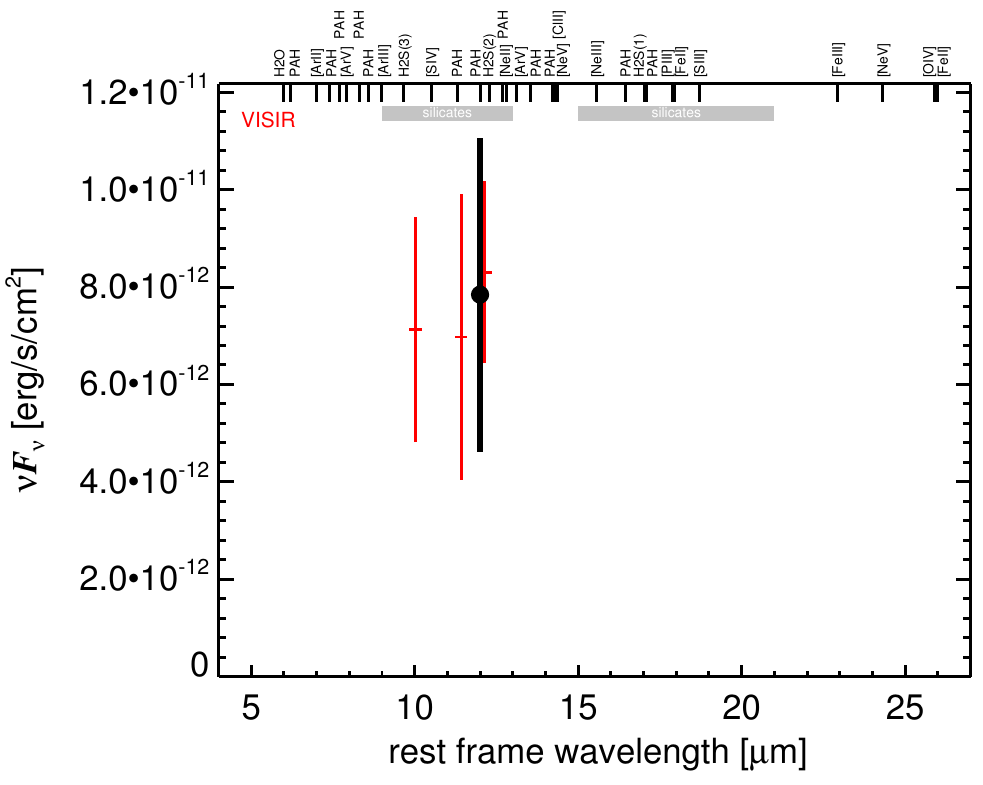}
   \caption{\label{fig:MISED_2MASXJ03565655-4041453}
      MIR SED of 2MASX\,J03565655-4041453. The description  of the symbols (if present) is the following.
      Grey crosses and  solid lines mark the \spitzer/IRAC, MIPS and IRS data. 
      The colour coding of the other symbols is: 
      green for COMICS, magenta for Michelle, blue for T-ReCS and red for VISIR data.
      Darker-coloured solid lines mark spectra of the corresponding instrument.
      The black filled circles mark the nuclear 12 and $18\,\mu$m  continuum emission estimate from the data.
      The ticks on the top axis mark positions of common MIR emission lines, while the light grey horizontal bars mark wavelength ranges affected by the silicate 10 and 18$\mu$m features.     
   }
\end{figure}
\clearpage

\twocolumn[\begin{@twocolumnfalse}  
\subsection{2MASX\,J09180027+0425066}\label{app:2MASXJ09180027+0425066}
2MASX\,J09180027 is a quasar at a redshift of $z=$ 0.1564 ($D \sim 781$\,Mpc) with an optical Sy\,2 classification \citep{veron-cetty_catalogue_2010}.
It was only recently identified in the SDSS data \citep{reyes_space_2008}. 
2MASX\,J09180027 is also a member of the nine-month BAT AGN sample, but does not appear in later BAT catalogues.
No \spitzerr data are available for 2MASX\,J09180027 and it appears point-like in the \wisee images.
We observed this object with VISIR in the NEII\_2 filter in 2010 and weakly detected a compact MIR source (the horizontal stripes in the image are instrumental artefacts).
\newline\end{@twocolumnfalse}]

\begin{figure}
   \centering
   \includegraphics[angle=0,width=8.500cm]{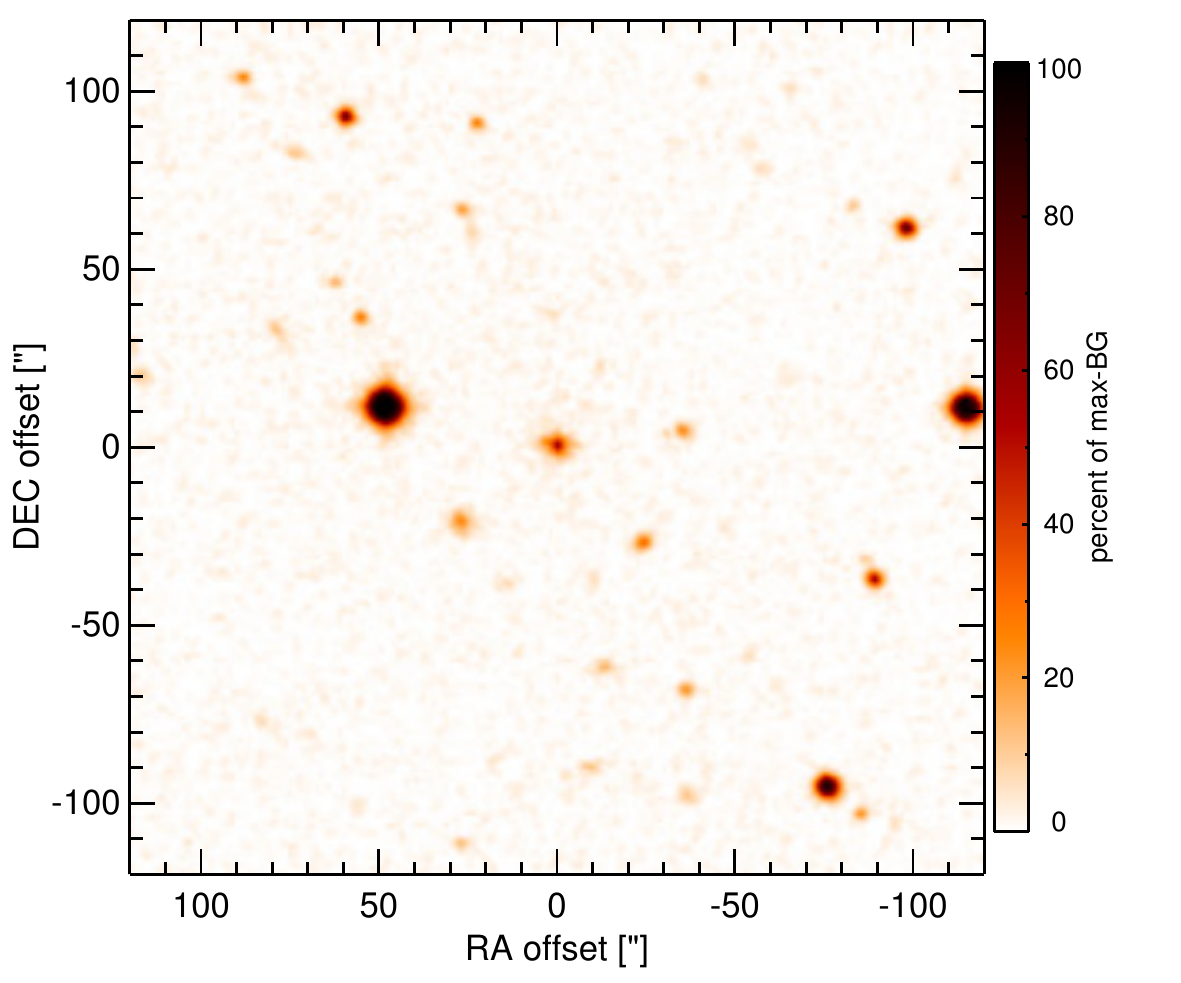}
    \caption{\label{fig:OPTim_2MASXJ09180027+0425066}
             Optical image (DSS, red filter) of  2MASX\,J09180027+0425066. Displayed are the central $4\arcmin$ with North up and East to the left. 
              The colour scaling is linear with white corresponding to the median background and black to the $0.01\%$ pixels with the highest intensity.  
           }
\end{figure}
\begin{figure}
   \centering
   \includegraphics[angle=0,height=3.11cm]{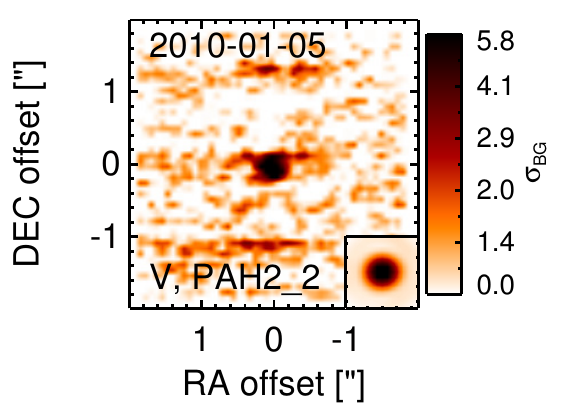}
    \caption{\label{fig:HARim_2MASXJ09180027+0425066}
             Subarcsecond-resolution MIR images of  2MASX\,J09180027+0425066 sorted by increasing filter wavelength. 
             Displayed are the inner $4\arcsec$ with North up and East to the left. 
             The colour scaling is logarithmic with white corresponding to median background and black to the $75\%$ of the highest intensity of all images in units of $\sigbg$.
             The inset image shows the central arcsecond of the PSF from the calibrator star, scaled to match the science target.
             The labels in the bottom left state instrument and filter names (C: COMICS, M: Michelle, T: T-ReCS, V: VISIR).
           }
\end{figure}
\begin{figure}
   \centering
   \includegraphics[angle=0,width=8.50cm]{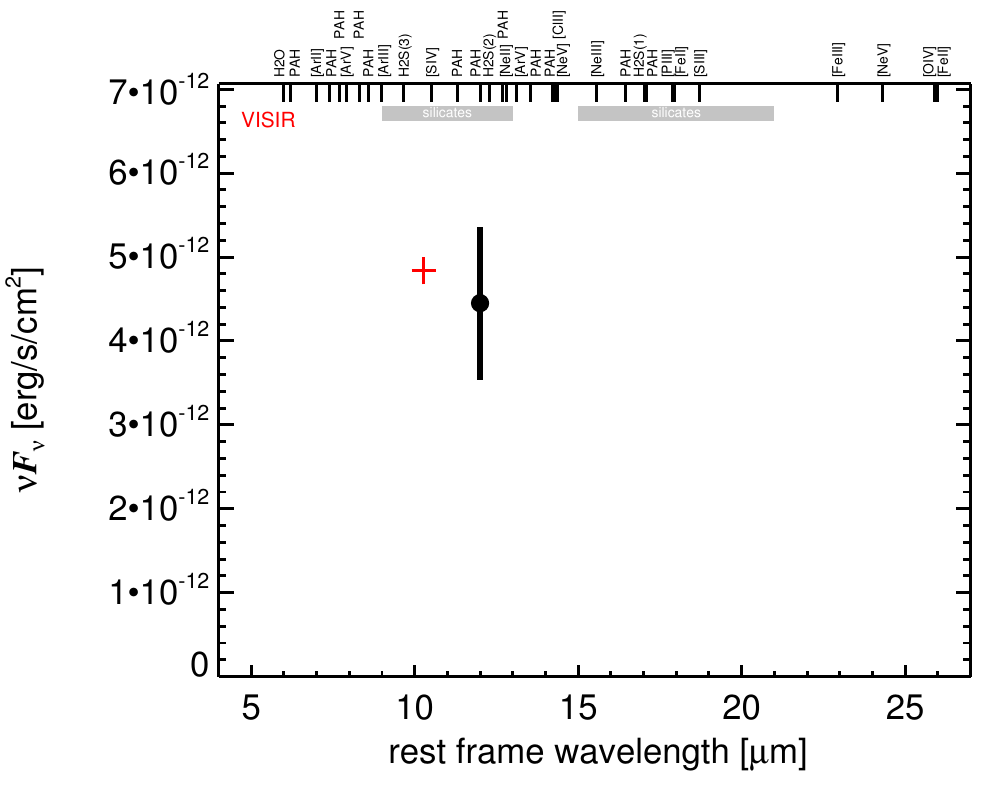}
   \caption{\label{fig:MISED_2MASXJ09180027+0425066}
      MIR SED of  2MASX\,J09180027+0425066. The description  of the symbols (if present) is the following.
      Grey crosses and  solid lines mark the \spitzer/IRAC, MIPS and IRS data. 
      The colour coding of the other symbols is: 
      green for COMICS, magenta for Michelle, blue for T-ReCS and red for VISIR data.
      Darker-coloured solid lines mark spectra of the corresponding instrument.
      The black filled circles mark the nuclear 12 and $18\,\mu$m  continuum emission estimate from the data.
      The ticks on the top axis mark positions of common MIR emission lines, while the light grey horizontal bars mark wavelength ranges affected by the silicate 10 and 18$\mu$m features.     
   }
\end{figure}
\clearpage

\twocolumn[\begin{@twocolumnfalse}  
\subsection{3C\,29 -- UGC\,595}\label{app:3C029}
3C\,29 is a radio-loud AGN in an elliptical host galaxy at a redshift of $z=$ 0.045 ($D \sim 202$\,Mpc) with a FR\,I radio morphology and an optical LINER classification \citep{veron-cetty_catalogue_2010}. 
It features supergalactic-scale biconical radio lobes in  the north-south directions (PA$\sim165\degree$; \citealt{morganti_radio_1993}). 
The \spitzer/IRAC and MIPS images are dominated by compact nuclear emission with only weak extended components in the IRAC 5.8 and 8$\,\mu$m images.
The \spitzer/IRS LR staring mode spectrum suffers from low S/N but indicates silicate emission and a shallow blue spectral slope in $\nu F_\nu$-space (see also \citealt{leipski_spitzer_2009}).
These authors attribute most of the MIR emission to non-thermal processes.
3C\,29 was observed with VISIR in the SIC filter in 2006 but remained undetected \citep{van_der_wolk_dust_2010}.
Our corresponding derived upper limit for the nuclear flux is almost four times higher than that of \cite{van_der_wolk_dust_2010} and also higher than the \spitzerr spectrophotometry.
Therefore, we use the latter to derive the upper limit of the nuclear $12\,\mu$m continuum emission.  
\newline\end{@twocolumnfalse}]

\begin{figure}
   \centering
   \includegraphics[angle=0,width=8.500cm]{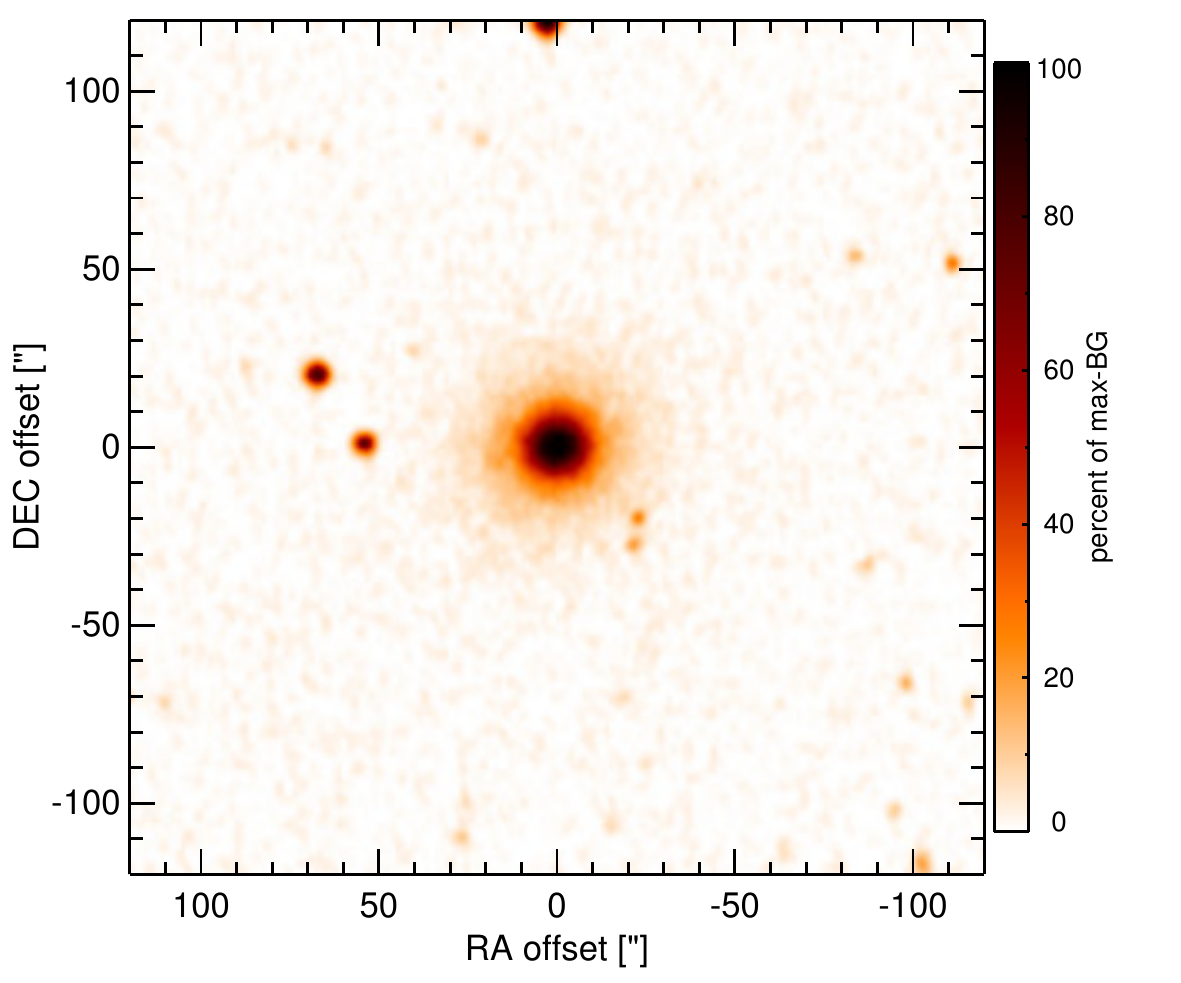}
    \caption{\label{fig:OPTim_3C029}
             Optical image (DSS, red filter) of 3C\,29. Displayed are the central $4\arcmin$ with North up and East to the left. 
              The colour scaling is linear with white corresponding to the median background and black to the $0.01\%$ pixels with the highest intensity.  
           }
\end{figure}
\begin{figure}
   \centering
   \includegraphics[angle=0,height=3.11cm]{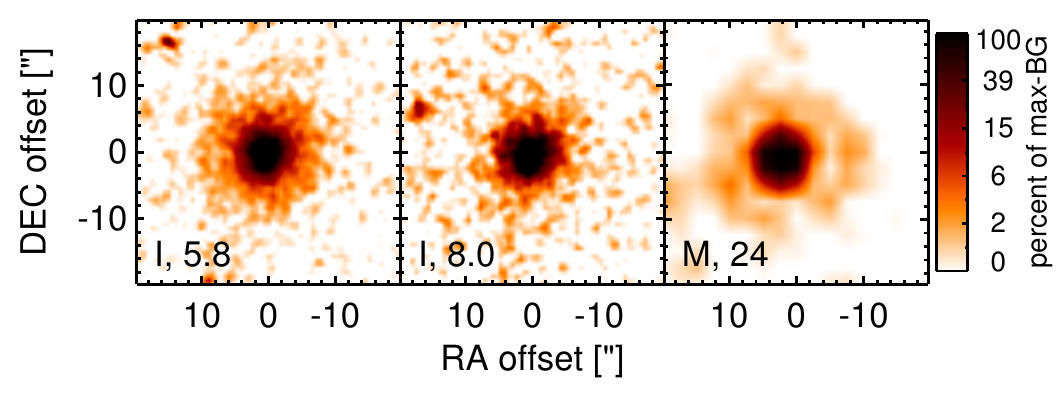}
    \caption{\label{fig:INTim_3C029}
             \spitzerr MIR images of 3C\,29. Displayed are the inner $40\arcsec$ with North up and East to the left. The colour scaling is logarithmic with white corresponding to median background and black to the $0.1\%$ pixels with the highest intensity.
             The label in the bottom left states instrument and central wavelength of the filter in $\mu$m (I: IRAC, M: MIPS). 
           }
\end{figure}
\begin{figure}
   \centering
   \includegraphics[angle=0,width=8.50cm]{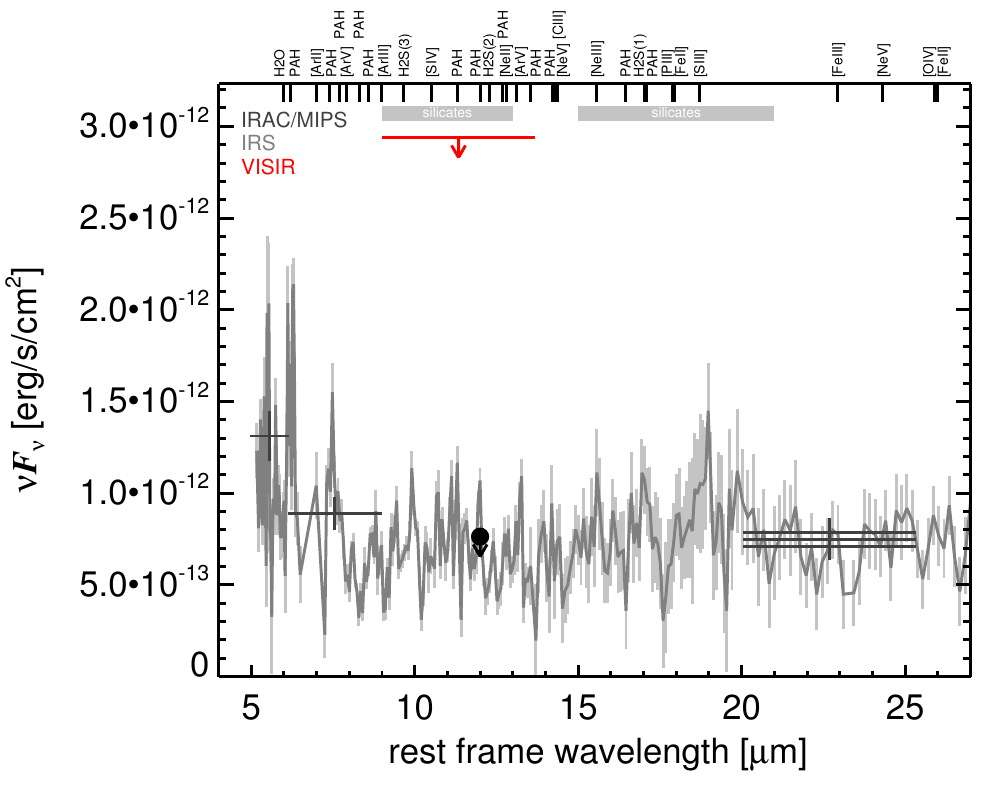}
   \caption{\label{fig:MISED_3C029}
      MIR SED of 3C\,29. The description  of the symbols (if present) is the following.
      Grey crosses and  solid lines mark the \spitzer/IRAC, MIPS and IRS data. 
      The colour coding of the other symbols is: 
      green for COMICS, magenta for Michelle, blue for T-ReCS and red for VISIR data.
      Darker-coloured solid lines mark spectra of the corresponding instrument.
      The black filled circles mark the nuclear 12 and $18\,\mu$m  continuum emission estimate from the data.
      The ticks on the top axis mark positions of common MIR emission lines, while the light grey horizontal bars mark wavelength ranges affected by the silicate 10 and 18$\mu$m features.     
   }
\end{figure}
\clearpage

\twocolumn[\begin{@twocolumnfalse}  
\subsection{3C\,33 -- LEDA\,4088}\label{app:3C033}
3C\,33 is a FR\,II radio object identified with the galaxy LEDA\,4088 at a redshift of $z=$ 0.0597 ($D \sim 273$\,Mpc).
It hosts a Sy\,2 AGN \citep{koski_spectrophotometry_1978} with detected polarized broad lines \citep{cohen_polarimetry_1999}.
It is also an X-ray ``buried" AGN candidate \citep{noguchi_new_2009}.
3C\,33 features classical FR\,II supergalactic-scale radio lobes in  the north-south directions with two symmetrical jets being visible also at parsec scales (PA$\sim20\degree$; e.g., \citealt{leahy_vla_1991,giovannini_bologna_2005}).
\spitzer/IRAC, IRS and MIPS data are available for this source.
It appears as a MIR point source in the IRAC $5.8$ and $8.0\,\mu$m and MIPS $24\,\mu$m images. 
Our nuclear MIPS $24\,\mu$m photometry provides a flux consistent with \cite{dicken_origin_2010}.
The IRS LR staring-mode spectrum exhibits weak silicate $10\,\mu$m absorption, weak PAH emission and strong forbidden emission lines (see also \citealt{ogle_spitzer_2006}).
The red MIR continuum peaks at around $20\,\mu$m and is attributed to thermal dust emission by \cite{ogle_spitzer_2006}. 
We observed 3C\,33 with COMICS in the N11.7 filter in 2009 and weakly detected a compact nucleus, which appears to be marginally resolved (FWHM$\sim0.92\arcsec\sim1\,$kpc).
However, at least a second epoch of deep subarcsecond MIR imaging is required to confirm this extension.
The nuclear COMICS photometric flux agrees with the \spitzerr spectrophotometry, but it would be significantly lower if the presence of subarcsecond-extended emission can be verified. 
\newline\end{@twocolumnfalse}]

\begin{figure}
   \centering
   \includegraphics[angle=0,width=8.500cm]{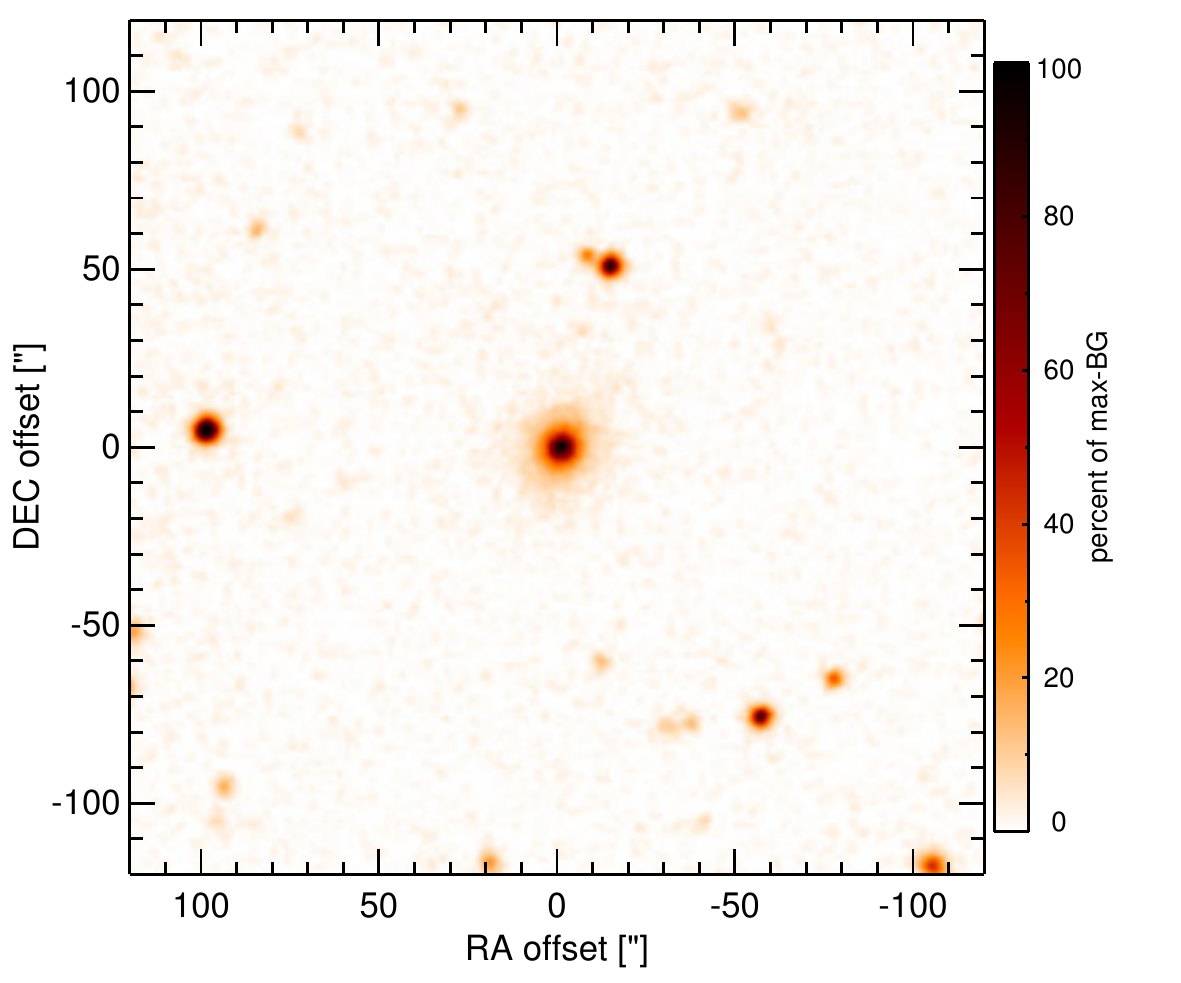}
    \caption{\label{fig:OPTim_3C033}
             Optical image (DSS, red filter) of 3C\,33. Displayed are the central $4\arcmin$ with North up and East to the left. 
              The colour scaling is linear with white corresponding to the median background and black to the $0.01\%$ pixels with the highest intensity.  
           }
\end{figure}
\begin{figure}
   \centering
   \includegraphics[angle=0,width=8.500cm]{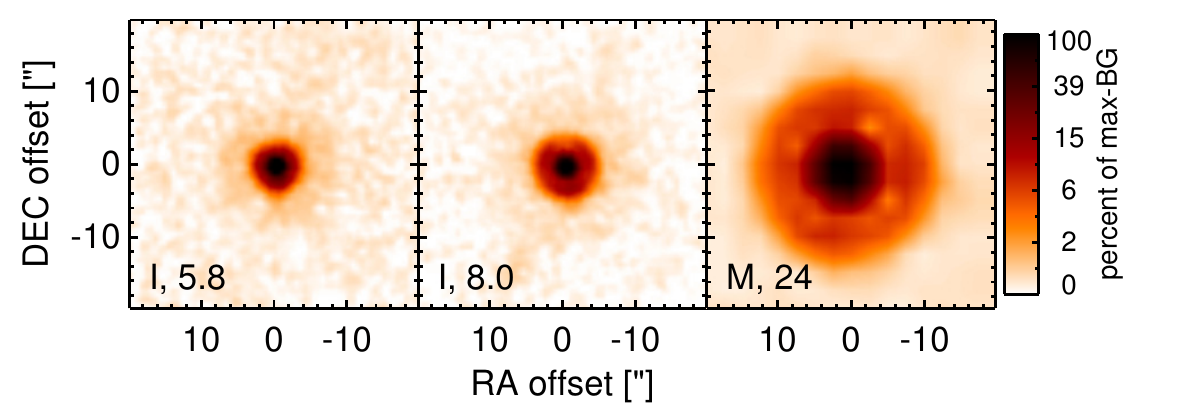}
    \caption{\label{fig:INTim_3C033}
             \spitzerr MIR images of 3C\,33. Displayed are the inner $40\arcsec$ with North up and East to the left. The colour scaling is logarithmic with white corresponding to median background and black to the $0.1\%$ pixels with the highest intensity.
             The label in the bottom left states instrument and central wavelength of the filter in $\mu$m (I: IRAC, M: MIPS). 
           }
\end{figure}
\begin{figure}
   \centering
   \includegraphics[angle=0,height=3.11cm]{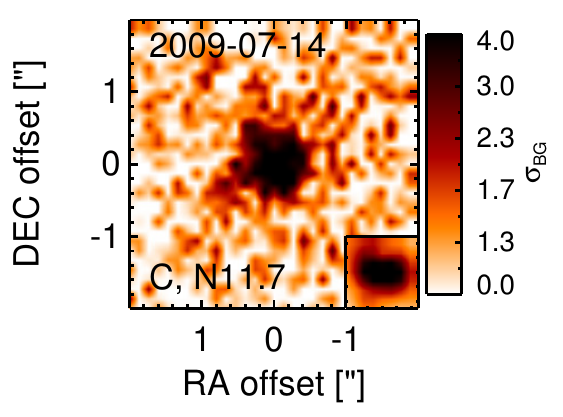}
    \caption{\label{fig:HARim_3C033}
             Subarcsecond-resolution MIR images of 3C\,33 sorted by increasing filter wavelength. 
             Displayed are the inner $4\arcsec$ with North up and East to the left. 
             The colour scaling is logarithmic with white corresponding to median background and black to the $75\%$ of the highest intensity of all images in units of $\sigbg$.
             The inset image shows the central arcsecond of the PSF from the calibrator star, scaled to match the science target.
             The labels in the bottom left state instrument and filter names (C: COMICS, M: Michelle, T: T-ReCS, V: VISIR).
           }
\end{figure}
\begin{figure}
   \centering
   \includegraphics[angle=0,width=8.50cm]{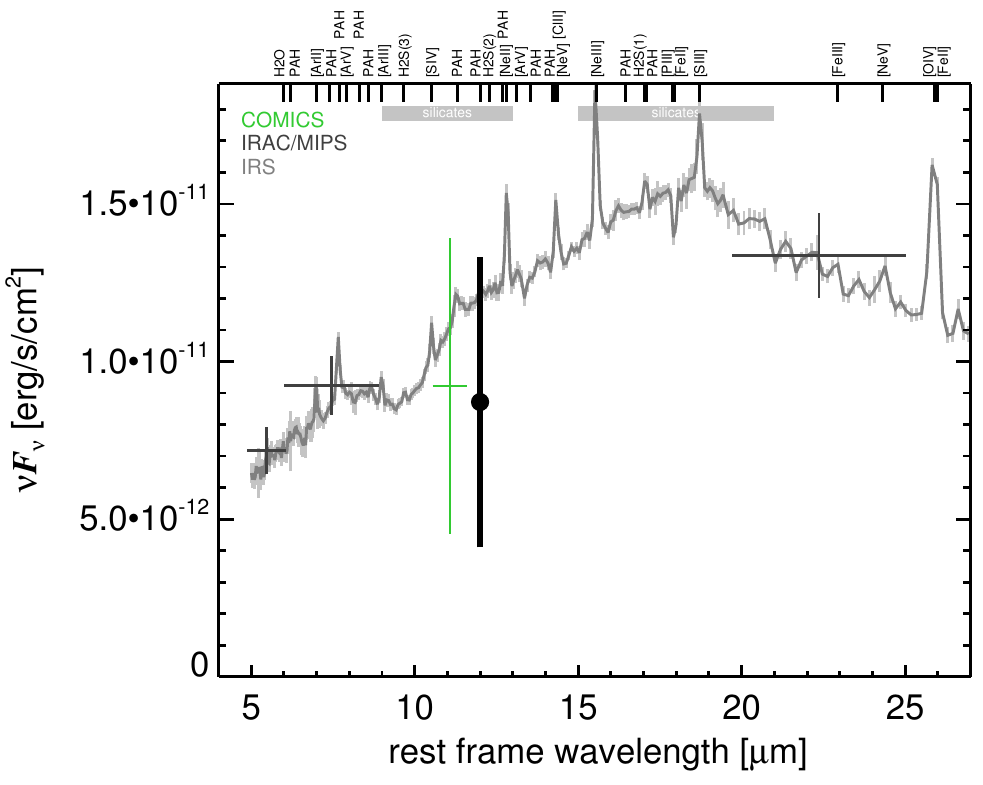}
   \caption{\label{fig:MISED_3C033}
      MIR SED of 3C\,33. The description  of the symbols (if present) is the following.
      Grey crosses and  solid lines mark the \spitzer/IRAC, MIPS and IRS data. 
      The colour coding of the other symbols is: 
      green for COMICS, magenta for Michelle, blue for T-ReCS and red for VISIR data.
      Darker-coloured solid lines mark spectra of the corresponding instrument.
      The black filled circles mark the nuclear 12 and $18\,\mu$m  continuum emission estimate from the data.
      The ticks on the top axis mark positions of common MIR emission lines, while the light grey horizontal bars mark wavelength ranges affected by the silicate 10 and 18$\mu$m features.     
   }
\end{figure}
\clearpage

\twocolumn[\begin{@twocolumnfalse}  
\subsection{3C\,78 -- NGC\,1218}\label{app:3C078}
3C\,78 is a FR\,I radio source coinciding with the early-type spiral galaxy NGC\,1218 at a redshift of $z=$ 0.0287 ($D\sim 127$\,Mpc), which is a radio-loud Sy\,1 AGN. 
A prominent synchrotron jet, similar to M87, was detected at radio and optical wavelengths (PA$\sim 50\degree$; \citealt{unger_kiloparsec_1984,sparks_discovery_1995}), which coincides with the inner optical emission line structure \citep{baum_extended_1988}.
The first MIR observations of 3C\,78 were performed by \cite{heckman_infrared_1983}. 
In addition to \iras, \spitzer/IRAC and MIPS images are also available, which show a compact MIR nucleus embedded within diffuse lenticular host emission.
No IRS spectrum is available but the nuclear IRAC  $5.8$ and $8.0\,\mu$m and MIPS $24\,\mu$m fluxes indicate a blue spectral slope in $\nu F_\nu$-space.
3C\,78 was observed with VISIR in the SIC filter in 2006 but remained undetected \citep{van_der_wolk_dust_2010}.
Our derived flux upper limit from the same data is significantly higher that of \cite{van_der_wolk_dust_2010}.
The fact that the upper limit is lower than the \spitzerr photometry indicates in any case that the total MIR emission of 3C\,78 is dominated by non-nuclear emission.
\newline\end{@twocolumnfalse}]

\begin{figure}
   \centering
   \includegraphics[angle=0,width=8.500cm]{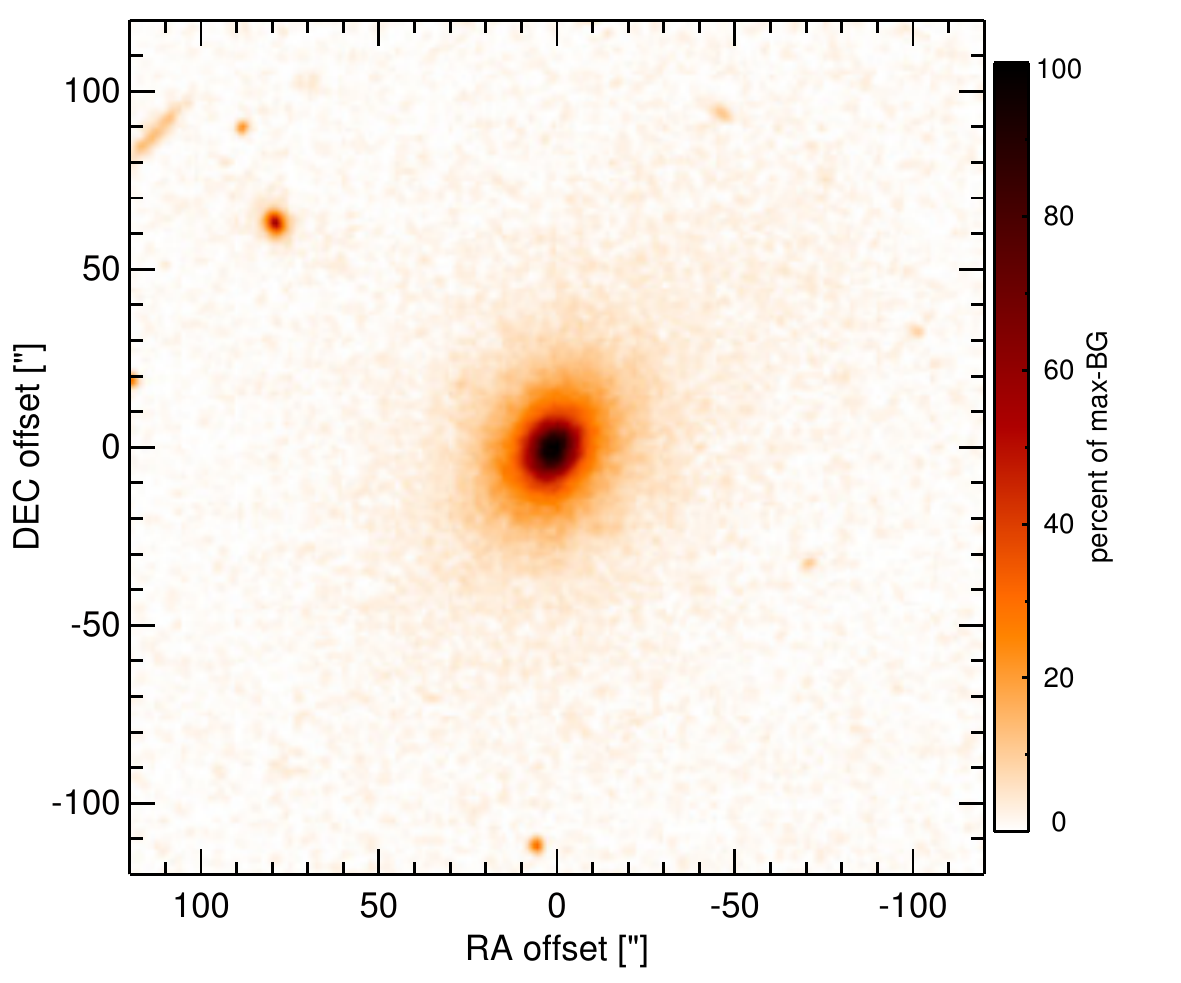}
    \caption{\label{fig:OPTim_3C078}
             Optical image (DSS, red filter) of 3C\,78. Displayed are the central $4\arcmin$ with North up and East to the left. 
              The colour scaling is linear with white corresponding to the median background and black to the $0.01\%$ pixels with the highest intensity.  
           }
\end{figure}
\begin{figure}
   \centering
   \includegraphics[angle=0,height=3.11cm]{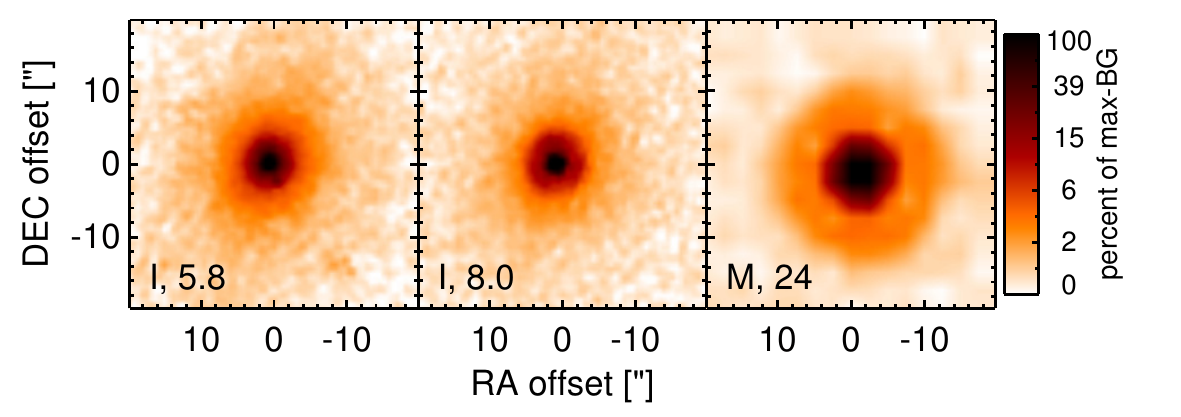}
    \caption{\label{fig:INTim_3C078}
             \spitzerr MIR images of 3C\,78. Displayed are the inner $40\arcsec$ with North up and East to the left. The colour scaling is logarithmic with white corresponding to median background and black to the $0.1\%$ pixels with the highest intensity.
             The label in the bottom left states instrument and central wavelength of the filter in $\mu$m (I: IRAC, M: MIPS). 
           }
\end{figure}
\begin{figure}
   \centering
   \includegraphics[angle=0,width=8.50cm]{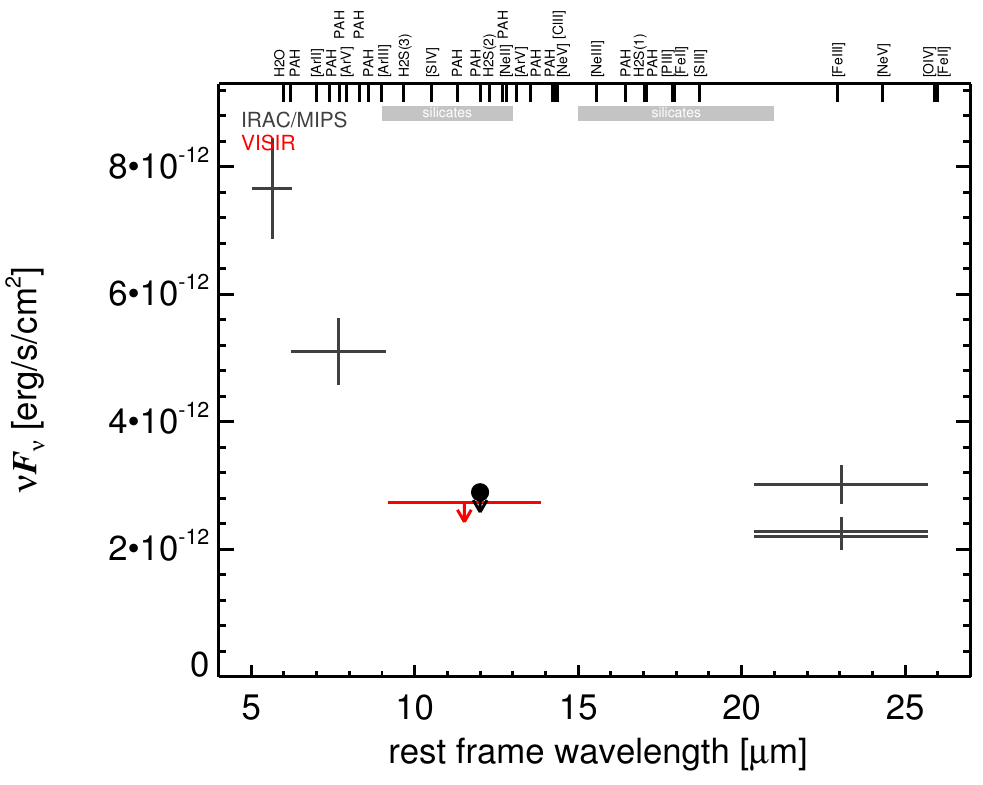}
   \caption{\label{fig:MISED_3C078}
      MIR SED of 3C\,78. The description  of the symbols (if present) is the following.
      Grey crosses and  solid lines mark the \spitzer/IRAC, MIPS and IRS data. 
      The colour coding of the other symbols is: 
      green for COMICS, magenta for Michelle, blue for T-ReCS and red for VISIR data.
      Darker-coloured solid lines mark spectra of the corresponding instrument.
      The black filled circles mark the nuclear 12 and $18\,\mu$m  continuum emission estimate from the data.
      The ticks on the top axis mark positions of common MIR emission lines, while the light grey horizontal bars mark wavelength ranges affected by the silicate 10 and 18$\mu$m features.     
}

\end{figure}
\clearpage

\twocolumn[\begin{@twocolumnfalse}  
\subsection{3C\,93 -- PKS\,0340+04}\label{app:3C093}
3C\,93 is a radio-loud quasar with an optical Sy\,1.0 classification \citep{veron-cetty_catalogue_2010} at a redshift of $z=$ 0.3571 ($D \sim 1967$\,Mpc) and thus the most distant object in the AGN MIR atlas.
It features symmetric biconical radio lobes on supergalactic scales in the north-east and south-west directions but no clear detections of collimated jets (PA$\sim45\degree$;  FR\,II;  e.g., \citealt{price_vla_1993,bogers_high_1994}).
No \spitzerr data are available for 3C\,93, and it appears as a faint point source in the \wisee images.  
This object was observed with VISIR in the broad SIC filter in 2006 but remained undetected by \cite{van_der_wolk_dust_2010}.
Our derived upper limit on the MIR flux is two times higher than theirs. 
However, the \wisee band 3 ($\sim12\,\mu$m) flux is lower than both, and, thus, is used as an upper limit for the estimate for the nuclear $12\,\mu$m continuum emission of 3C\,93.
\newline\end{@twocolumnfalse}]

\begin{figure}
   \centering
   \includegraphics[angle=0,width=8.500cm]{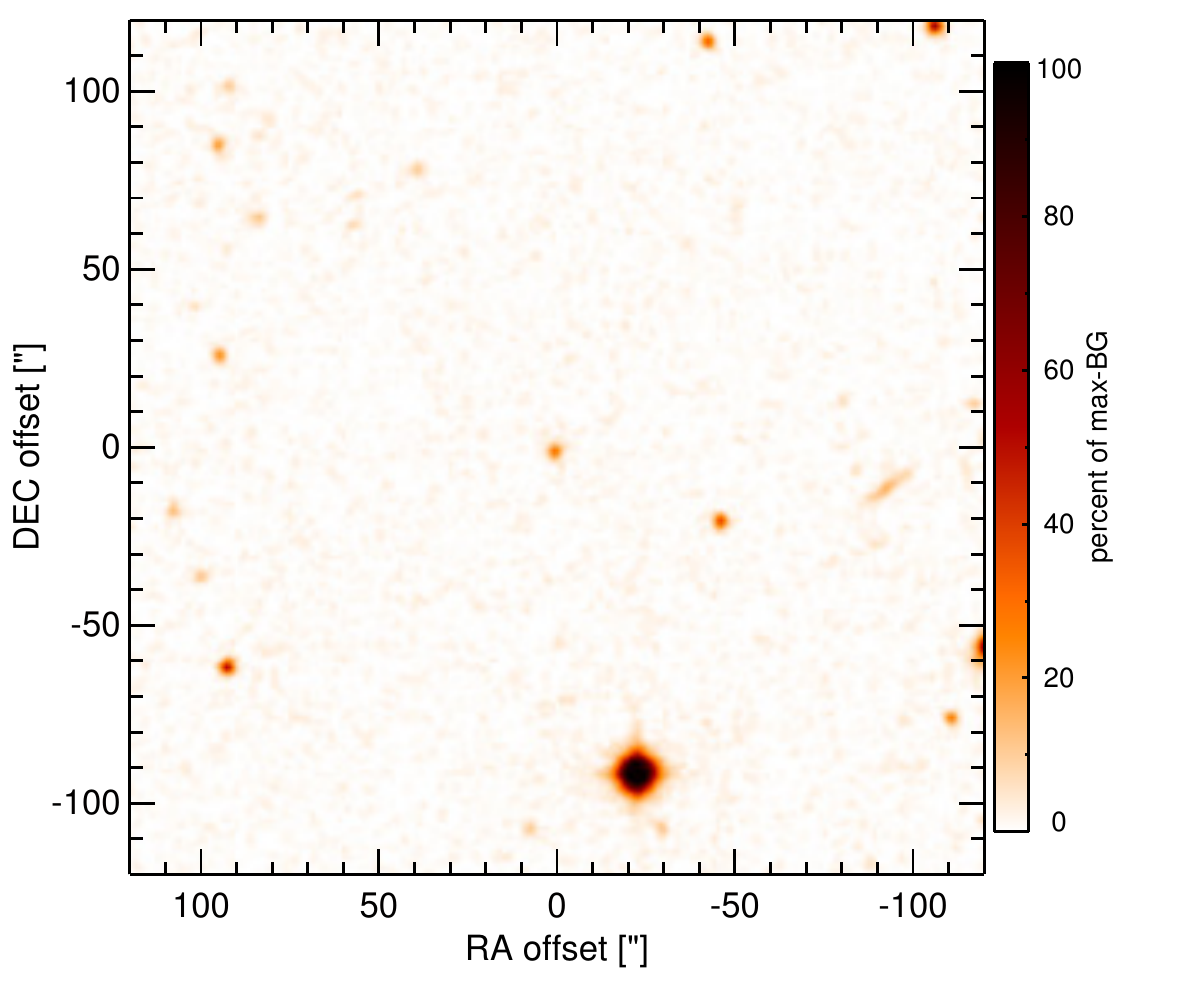}
    \caption{\label{fig:OPTim_3C093}
             Optical image (DSS, red filter) of 3C\,93. Displayed are the central $4\arcmin$ with North up and East to the left. 
              The colour scaling is linear with white corresponding to the median background and black to the $0.01\%$ pixels with the highest intensity.  
           }
\end{figure}
\begin{figure}
   \centering
   \includegraphics[angle=0,width=8.50cm]{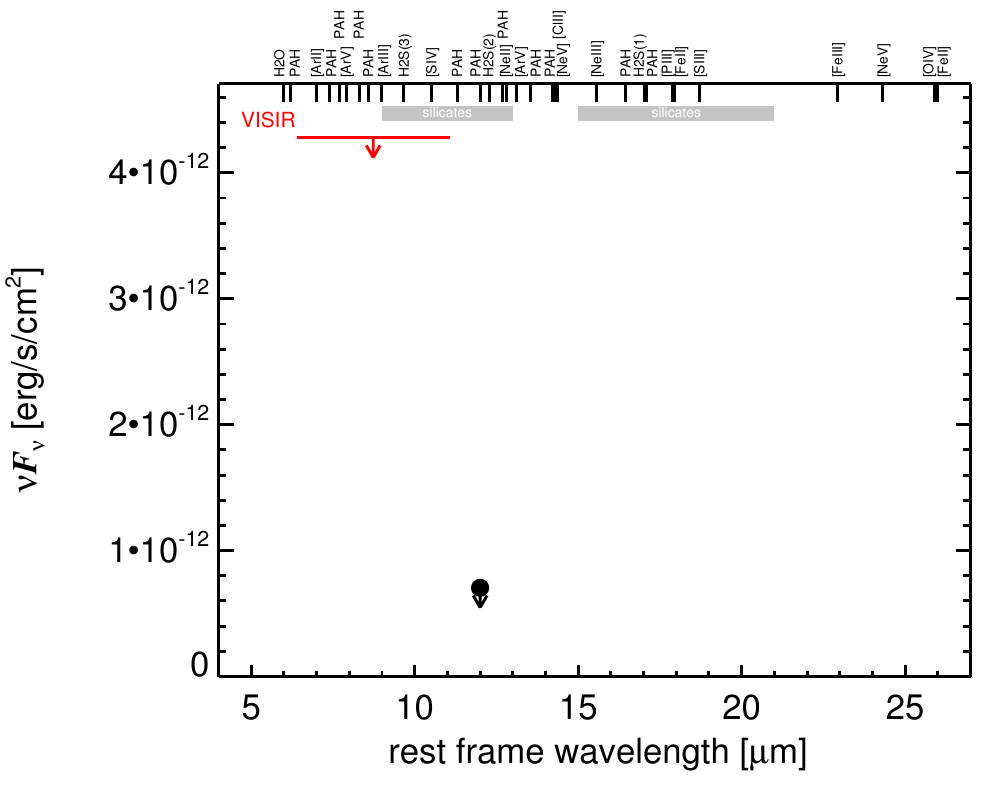}
   \caption{\label{fig:MISED_3C093}
      MIR SED of 3C\,93. The description  of the symbols (if present) is the following.
      Grey crosses and  solid lines mark the \spitzer/IRAC, MIPS and IRS data. 
      The colour coding of the other symbols is: 
      green for COMICS, magenta for Michelle, blue for T-ReCS and red for VISIR data.
      Darker-coloured solid lines mark spectra of the corresponding instrument.
      The black filled circles mark the nuclear 12 and $18\,\mu$m  continuum emission estimate from the data.
      The ticks on the top axis mark positions of common MIR emission lines, while the light grey horizontal bars mark wavelength ranges affected by the silicate 10 and 18$\mu$m features.     
   }
\end{figure}
\clearpage

\twocolumn[\begin{@twocolumnfalse}  
\subsection{3C\,98 -- LEDA\,14213}\label{app:3C098}
3C\,98 is a FR\,II radio source identified with the elliptical galaxy LEDA\,14213 at a redshift of $z=$ 0.0305 ($D \sim 137$\,Mpc).
It contains a radio-loud Sy\,2 AGN \citep{veron-cetty_catalogue_2010}.
It features the classical FR\,II super-galactic radio lobes in  the north-south directions with a collimated jet extending into the northern lobe (PA$\sim20\degree$; e.g., \citealt{baum_extended_1988,leahy_study_1997}).
The first MIR observations of 3C\,98 were carried out with IRTF and only tentatively detected nuclear emission \citep{elvis_1-20_1984}.
3C\,98 was also observed with \textit{ISO}/ISOCAM \citep{siebenmorgen_isocam_2004} and \spitzerr/IRAC, IRS and MIPS.
A compact nucleus was detected in the corresponding images with possible weak diffuse emission in the MIPS $24\,\mu$m image.
Our nuclear MIPS 24$\,\mu$m photometry is consistent with \cite{dicken_origin_2010}.
The IRS LR mapping-mode spectrum suffers from low S/N and does not display any obvious spectral features apart from a shallow emission peak at $\sim18\,\mu$m in $\nu F_\nu$-space (see also \citealt{dicken_spitzer_2012}). 
3C\,98 was imaged with VISIR in the broad SIC filter in 2005, and a compact nucleus was weakly detected  \citep{van_der_wolk_dust_2010}.
The low S/N does not allow for any robust conclusion about the nuclear extension at subarcsecond scales in the MIR.
Our nuclear SIC flux is consistent with \cite{van_der_wolk_dust_2010} and the \spitzerr spectrophotometry.
Note that the these authors attribute the nuclear MIR emission to thermal processes. 
\newline\end{@twocolumnfalse}]
\begin{figure}
   \centering
   \includegraphics[angle=0,width=8.500cm]{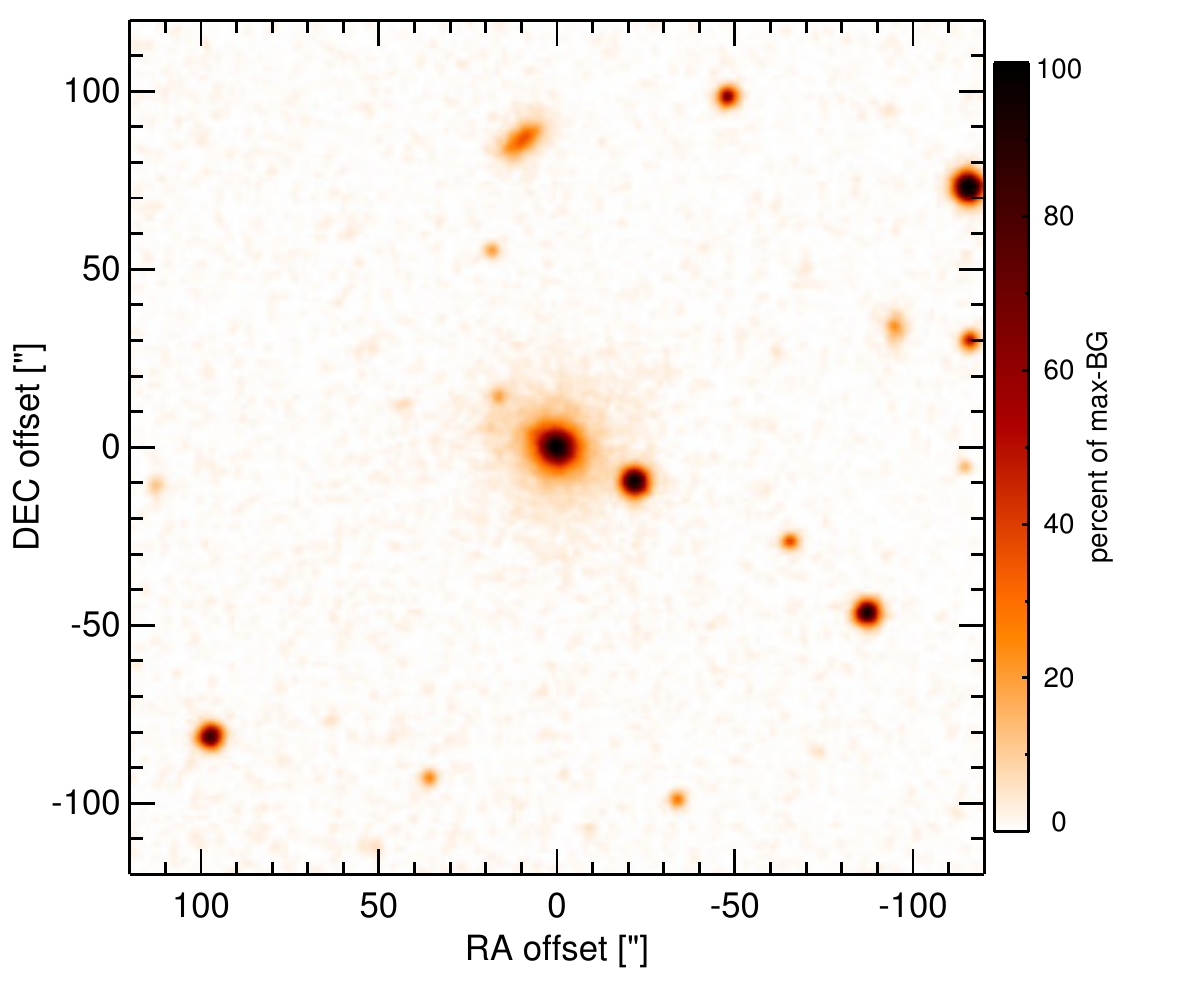}
    \caption{\label{fig:OPTim_3C098}
             Optical image (DSS, red filter) of 3C\,98. Displayed are the central $4\arcmin$ with North up and East to the left. 
              The colour scaling is linear with white corresponding to the median background and black to the $0.01\%$ pixels with the highest intensity.  
           }
\end{figure}
\begin{figure}
   \centering
   \includegraphics[angle=0,height=3.11cm]{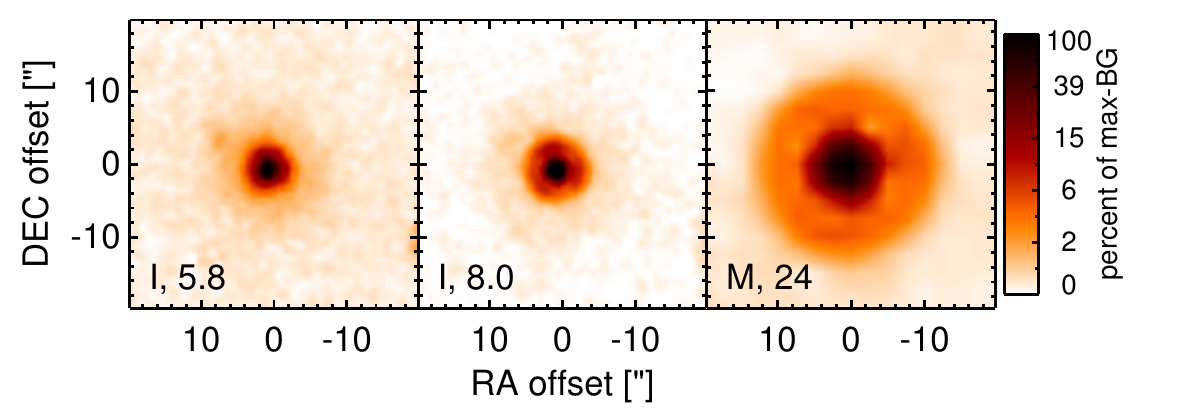}
    \caption{\label{fig:INTim_3C098}
             \spitzerr MIR images of 3C\,98. Displayed are the inner $40\arcsec$ with North up and East to the left. The colour scaling is logarithmic with white corresponding to median background and black to the $0.1\%$ pixels with the highest intensity.
             The label in the bottom left states instrument and central wavelength of the filter in $\mu$m (I: IRAC, M: MIPS). 
           }
\end{figure}
\begin{figure}
   \centering
   \includegraphics[angle=0,height=3.11cm]{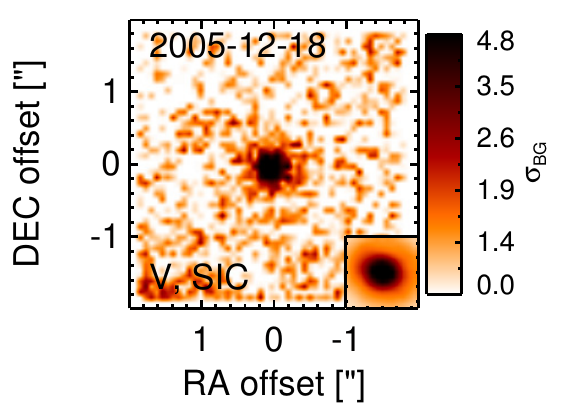}
    \caption{\label{fig:HARim_3C098}
             Subarcsecond-resolution MIR images of 3C\,98 sorted by increasing filter wavelength. 
             Displayed are the inner $4\arcsec$ with North up and East to the left. 
             The colour scaling is logarithmic with white corresponding to median background and black to the $75\%$ of the highest intensity of all images in units of $\sigbg$.
             The inset image shows the central arcsecond of the PSF from the calibrator star, scaled to match the science target.
             The labels in the bottom left state instrument and filter names (C: COMICS, M: Michelle, T: T-ReCS, V: VISIR).
           }
\end{figure}
\begin{figure}
   \centering
   \includegraphics[angle=0,width=8.50cm]{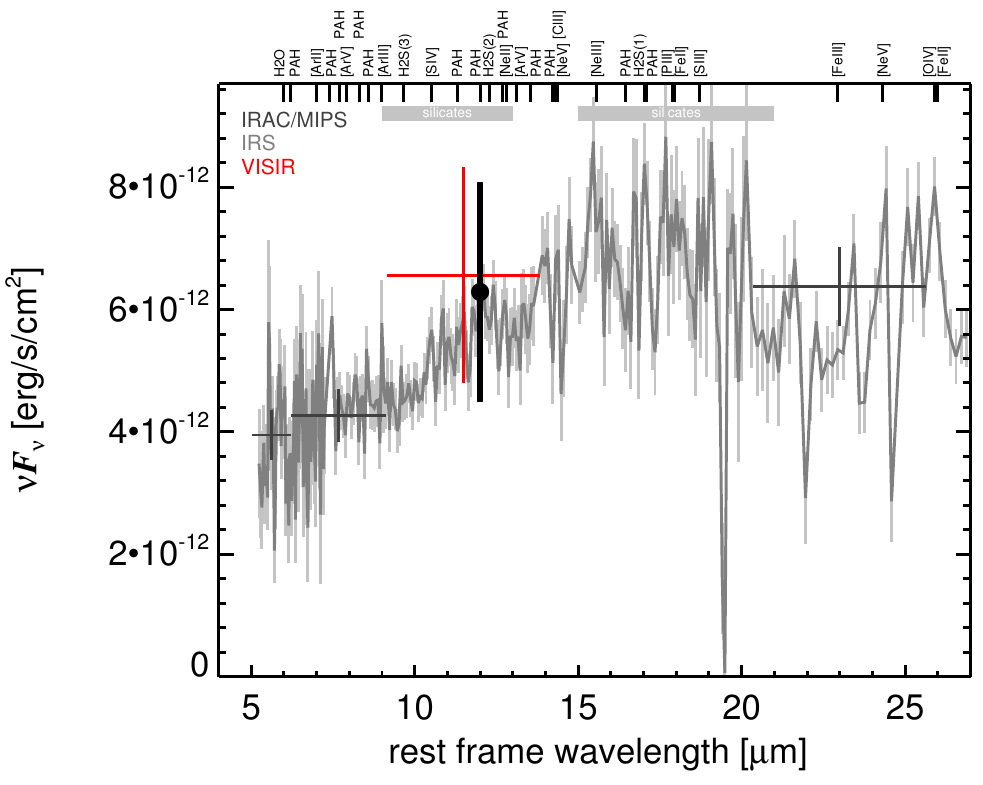}
   \caption{\label{fig:MISED_3C098}
      MIR SED of 3C\,98. The description  of the symbols (if present) is the following.
      Grey crosses and  solid lines mark the \spitzer/IRAC, MIPS and IRS data. 
      The colour coding of the other symbols is: 
      green for COMICS, magenta for Michelle, blue for T-ReCS and red for VISIR data.
      Darker-coloured solid lines mark spectra of the corresponding instrument.
      The black filled circles mark the nuclear 12 and $18\,\mu$m  continuum emission estimate from the data.
      The ticks on the top axis mark positions of common MIR emission lines, while the light grey horizontal bars mark wavelength ranges affected by the silicate 10 and 18$\mu$m features.     
   }
\end{figure}
\clearpage

\twocolumn[\begin{@twocolumnfalse}  
\subsection{3C\,105 -- IRAS\,04047+0332 -- LEDA\,14492}\label{app:3C105}
3C\,105 is a FR\,II radio source identified with the elliptical galaxy LEDA\,14492 at a redshift of $z=$ 0.089 ($D \sim 421$\,Mpc).
It contains a radio-loud Sy\,2 AGN \citep{veron-cetty_catalogue_2010} that belongs to the nine-month BAT AGN sample.
It features the classical FR\,II super-galactic radio lobes in north-west and south-east directions without a collimated jet evident (PA$\sim20\degree$; e.g., \citealt{baum_extended_1988,leahy_study_1997}).
The first successful MIR observations of 3C\,105 were performed with \spitzer/IRS and MIPS, and the corresponding images show a point source.
Our nuclear photometry agrees with \cite{dicken_origin_2008}.
Despite the very low S/N of the IRS LR mapping-mode spectrum, silicate $10\,\mu$m absorption is visible together with a red spectral slope in $\nu F_\nu$-space (see also \citealt{dicken_spitzer_2012}).
3C\,105 was observed with VISIR in the broad SIC filter in 2005 \citep{van_der_wolk_dust_2010} and in three narrow $N$-band filters in 2009 (this work).
While the nucleus remained undetected in the SIC image, a compact nucleus is weakly detected in both NEII\_2 images.
The low S/N, however, does not allow for any robust conclusion about the nuclear extension at subarcsecond scales in the MIR.
Our derived flux upper limit for the SIC image is several times higher than the value of \cite{van_der_wolk_dust_2010} for unknown reasons.
Our nuclear VISIR fluxes are in general consistent with the  \spitzerr spectrophotometry.
\newline\end{@twocolumnfalse}]

\begin{figure}
   \centering
   \includegraphics[angle=0,width=8.500cm]{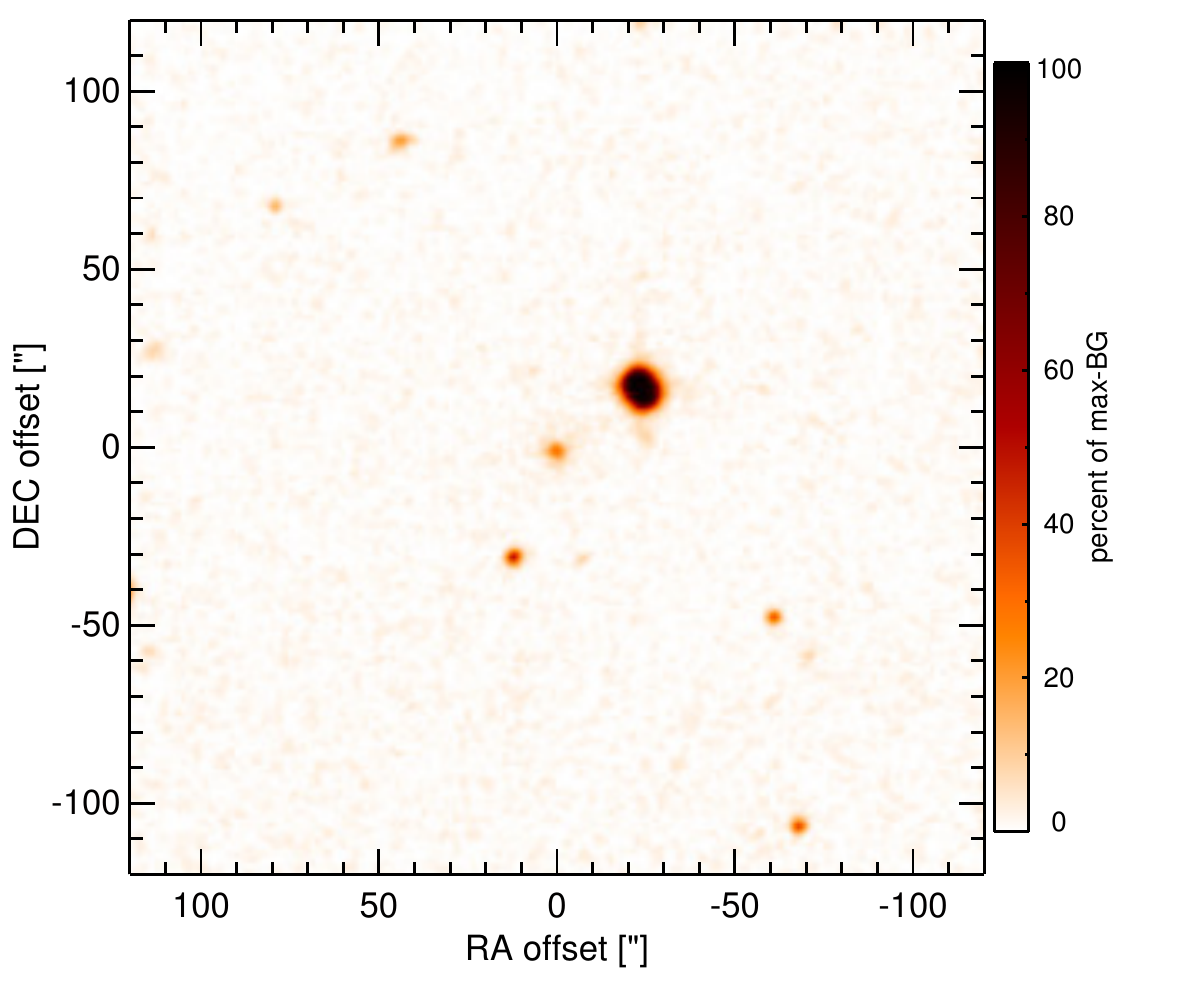}
    \caption{\label{fig:OPTim_3C105}
             Optical image (DSS, red filter) of 3C\,105. Displayed are the central $4\arcmin$ with North up and East to the left. 
              The colour scaling is linear with white corresponding to the median background and black to the $0.01\%$ pixels with the highest intensity.  
           }
\end{figure}
\begin{figure}
   \centering
   \includegraphics[angle=0,height=3.11cm]{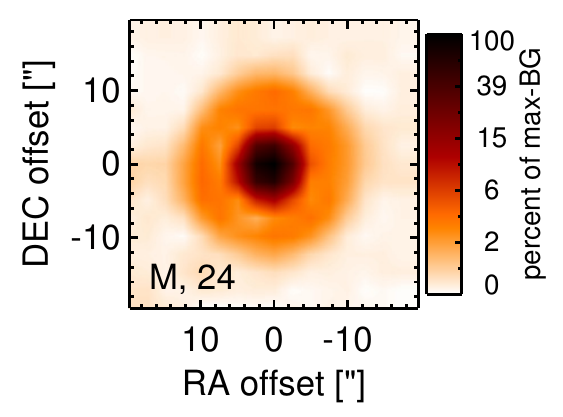}
    \caption{\label{fig:INTim_3C105}
             \spitzerr MIR images of 3C\,105. Displayed are the inner $40\arcsec$ with North up and East to the left. The colour scaling is logarithmic with white corresponding to median background and black to the $0.1\%$ pixels with the highest intensity.
             The label in the bottom left states instrument and central wavelength of the filter in $\mu$m (I: IRAC, M: MIPS). 
           }
\end{figure}
\begin{figure}
   \centering
   \includegraphics[angle=0,height=3.11cm]{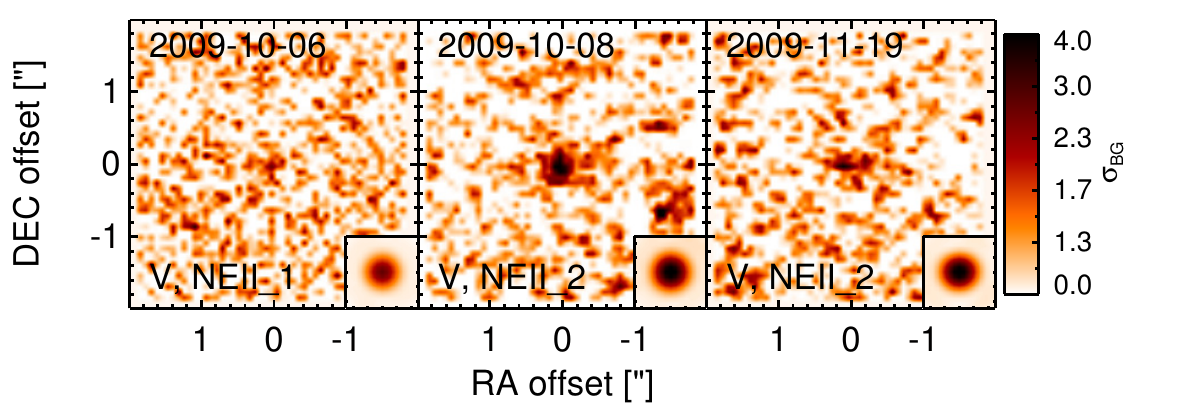}
    \caption{\label{fig:HARim_3C105}
             Subarcsecond-resolution MIR images of 3C\,105 sorted by increasing filter wavelength. 
             Displayed are the inner $4\arcsec$ with North up and East to the left. 
             The colour scaling is logarithmic with white corresponding to median background and black to the $75\%$ of the highest intensity of all images in units of $\sigbg$.
             The inset image shows the central arcsecond of the PSF from the calibrator star, scaled to match the science target.
             The labels in the bottom left state instrument and filter names (C: COMICS, M: Michelle, T: T-ReCS, V: VISIR).
           }
\end{figure}
\begin{figure}
   \centering
   \includegraphics[angle=0,width=8.50cm]{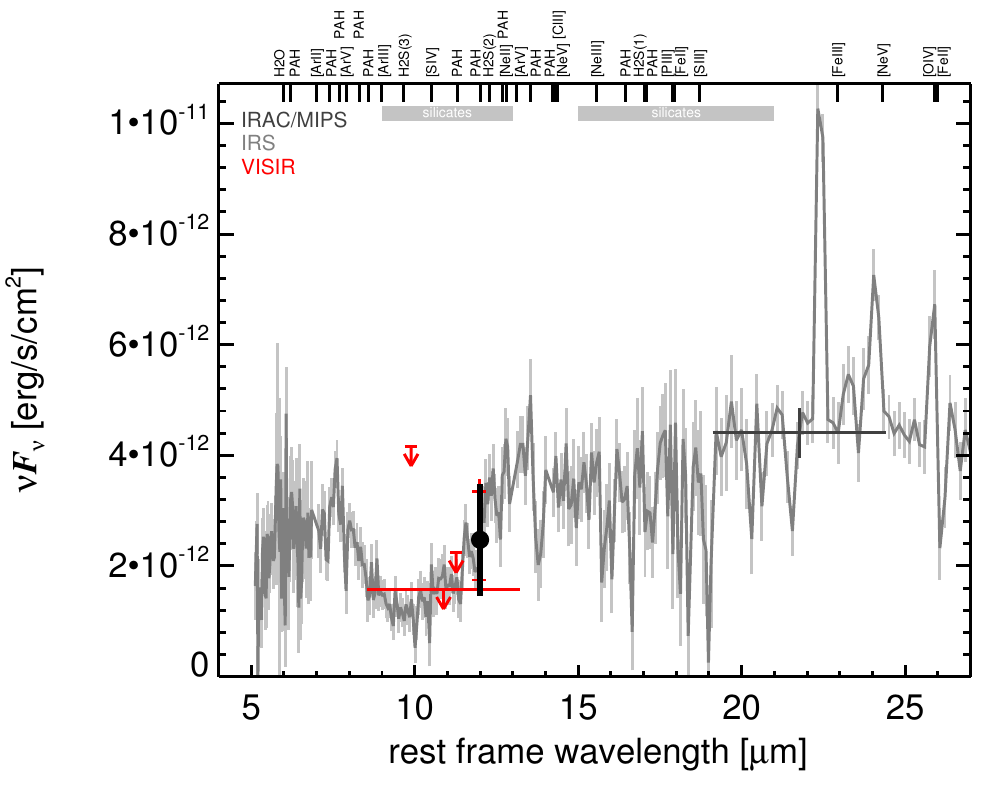}
   \caption{\label{fig:MISED_3C105}
      MIR SED of 3C\,105. The description  of the symbols (if present) is the following.
      Grey crosses and  solid lines mark the \spitzer/IRAC, MIPS and IRS data. 
      The colour coding of the other symbols is: 
      green for COMICS, magenta for Michelle, blue for T-ReCS and red for VISIR data.
      Darker-coloured solid lines mark spectra of the corresponding instrument.
      The black filled circles mark the nuclear 12 and $18\,\mu$m  continuum emission estimate from the data.
      The ticks on the top axis mark positions of common MIR emission lines, while the light grey horizontal bars mark wavelength ranges affected by the silicate 10 and 18$\mu$m features.     
   }
\end{figure}

\clearpage

\twocolumn[\begin{@twocolumnfalse}  
\subsection{3C\,120 -- Mrk\,1506}\label{app:3C120}
3C\,120 is a flat-spectrum FR\,I radio source coinciding with the lenticular galaxy Mrk\,1506 at a redshift of $z=$ 0.033 ($D \sim 150$\,Mpc).
It hosts a well-studied radio-loud Sy\,1.5 AGN \citep{veron-cetty_catalogue_2010}, which also belongs to the nine-month BAT AGN sample.
It is strongly variable at most wavelengths and possesses a prominent one-sided and bended jet extending from $\sim1$\,pc to $\sim100\,$kpc scales (PA$\sim -95\degree$; \citealt{walker_radio_1987}).
In addition, an extended NLR with complex morphology and extension parallel to the radio jet is present \citep{hua_forbidden_1988}.
3C\,120 was first observed at MIR wavelengths by \cite{kleinmann_observations_1970} and \cite{rieke_infrared_1972}, followed by \cite{rieke_infrared_1978,roche_8-13_1984,sparks_infrared_1986}.
The first subarcsecond resolution observations were performed with Palomar 5\,m/MIRLIN in 1999, where a point-like nucleus was detected \citep{gorjian_10_2004}.
3C\,120 was also observed with \spitzer/IRAC, IRS and MIPS and appears as compact source without any significant extended emission in the corresponding images.
The IRS LR staring-mode spectrum is AGN-dominated and resembles a typical type~I AGN SED with silicate emission, strong forbidden high-ionization lines, only weak PAH emission and a rather flat spectral slope in $\nu F_\nu$-space (see also \citealt{thompson_dust_2009,leipski_spitzer_2009, wu_spitzer/irs_2009,tommasin_spitzer-irs_2010, mullaney_defining_2011}). 
3C\,120 was observed with VISIR in the SIC filter in 2006 \citep{van_der_wolk_dust_2010}, in three narrow $N$-band filters in 2009 (this work) and in the PAH1 filter in 2010 (unpublished, to our knowledge).
A compact nucleus without further host or jet emission was detected in all images.
The nuclear source seems slightly extended in all cases (FWHM $\sim0.43\arcsec \sim 0.3$\,kpc), except in the sharpest image (SIC). 
Furthermore, the position angles do not agree for the different observations.
Therefore, the general nuclear extension at subarcsecond scales in the MIR remains uncertain.
The measured nuclear VISIR fluxes are on average $\sim 12\%$ lower than those exhibited by \spitzerr and show significant dispersion.
In particular, the oldest measurement, which is in the SIC filter, has the highest flux (consistent with the value in \citealt{van_der_wolk_dust_2010}), while the most recent measurement (PAH1) exhibits the lowest flux.
The comparison with the historic measurements indicates  possible MIR flux variations of the order of $\sim25\%$ during the last four decades, which however is further complicated by the uncertain nuclear extension.
\newline\end{@twocolumnfalse}]

\begin{figure}
   \centering
   \includegraphics[angle=0,width=8.500cm]{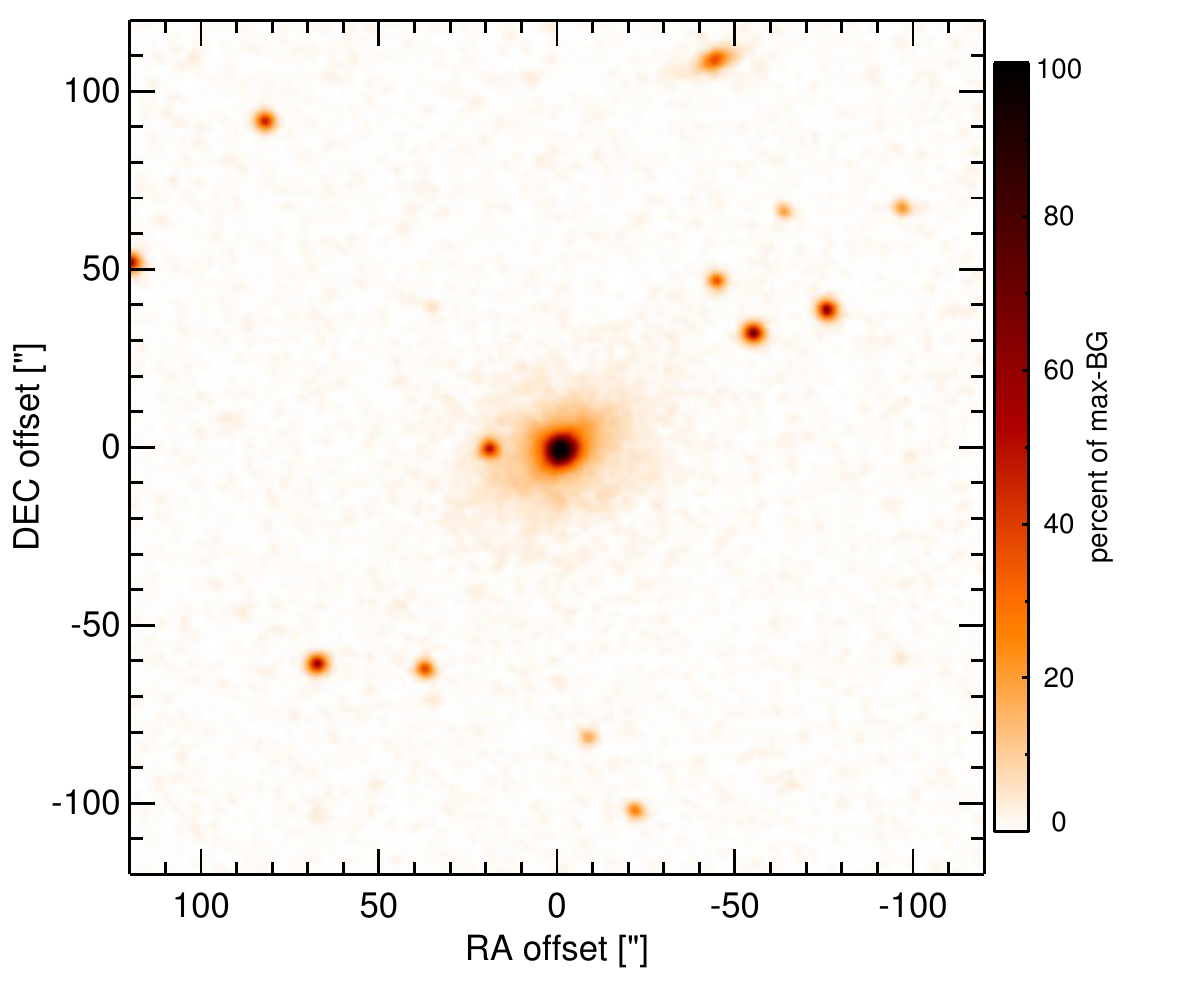}
    \caption{\label{fig:OPTim_3C120}
             Optical image (DSS, red filter) of 3C\,120. Displayed are the central $4\arcmin$ with North up and East to the left. 
              The colour scaling is linear with white corresponding to the median background and black to the $0.01\%$ pixels with the highest intensity.  
           }
\end{figure}
\begin{figure}
   \centering
   \includegraphics[angle=0,height=3.11cm]{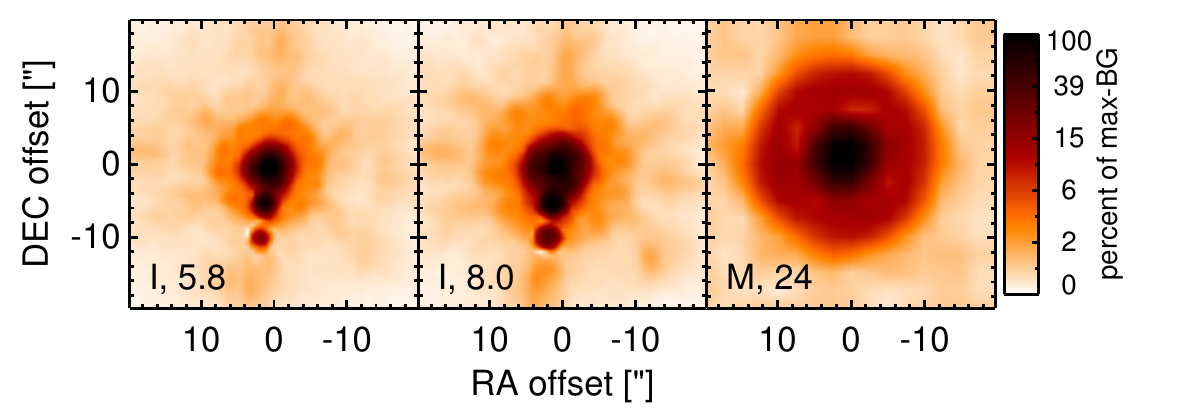}
    \caption{\label{fig:INTim_3C120}
             \spitzerr MIR images of 3C\,120. Displayed are the inner $40\arcsec$ with North up and East to the left. The colour scaling is logarithmic with white corresponding to median background and black to the $0.1\%$ pixels with the highest intensity.
             The label in the bottom left states instrument and central wavelength of the filter in $\mu$m (I: IRAC, M: MIPS).
             Note that the apparent off-nuclear compact sources in the IRAC $5.8$ and $8.0\,\mu$m images are instrumental artefacts. 
           }
\end{figure}
\begin{figure}
   \centering
   \includegraphics[angle=0,width=8.500cm]{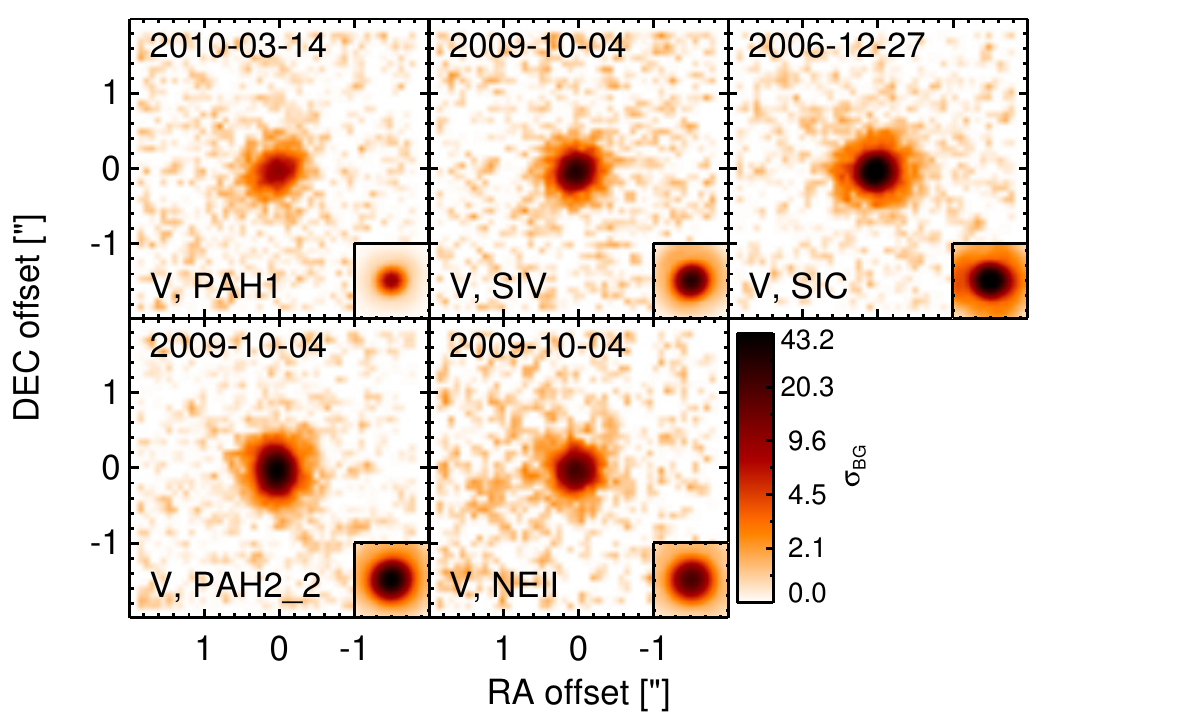}
    \caption{\label{fig:HARim_3C120}
             Subarcsecond-resolution MIR images of 3C\,120 sorted by increasing filter wavelength. 
             Displayed are the inner $4\arcsec$ with North up and East to the left. 
             The colour scaling is logarithmic with white corresponding to median background and black to the $75\%$ of the highest intensity of all images in units of $\sigbg$.
             The inset image shows the central arcsecond of the PSF from the calibrator star, scaled to match the science target.
             The labels in the bottom left state instrument and filter names (C: COMICS, M: Michelle, T: T-ReCS, V: VISIR).
           }
\end{figure}
\begin{figure}
   \centering
   \includegraphics[angle=0,width=8.50cm]{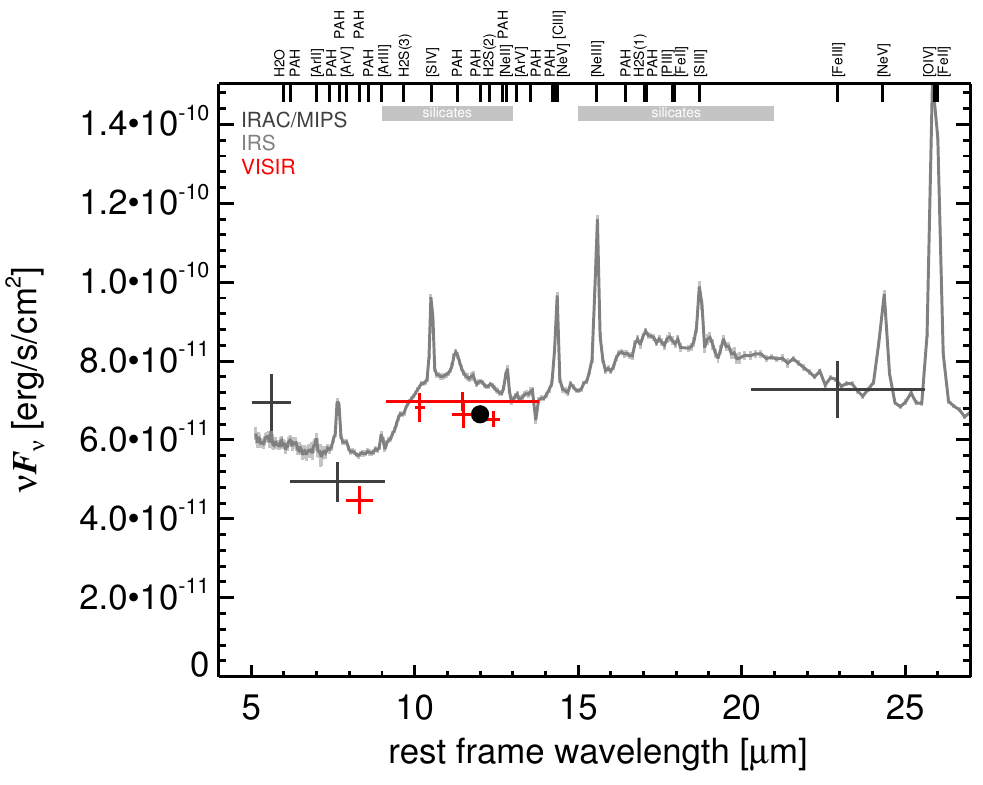}
   \caption{\label{fig:MISED_3C120}
      MIR SED of 3C\,120. The description  of the symbols (if present) is the following.
      Grey crosses and  solid lines mark the \spitzer/IRAC, MIPS and IRS data. 
      The colour coding of the other symbols is: 
      green for COMICS, magenta for Michelle, blue for T-ReCS and red for VISIR data.
      Darker-coloured solid lines mark spectra of the corresponding instrument.
      The black filled circles mark the nuclear 12 and $18\,\mu$m  continuum emission estimate from the data.
      The ticks on the top axis mark positions of common MIR emission lines, while the light grey horizontal bars mark wavelength ranges affected by the silicate 10 and 18$\mu$m features.     
   }
\end{figure}
\clearpage

\twocolumn[\begin{@twocolumnfalse}  
\subsection{3C\,135 -- LEDA\,16952}\label{app:3C135}
3C\,135 is a FR\,II radio source identified with the elliptical galaxy LEDA\,16952 at a redshift of $z=$ 0.1274 ($D \sim 620$\,Mpc).
It contains a radio-loud Sy\,2 nucleus \citep{veron-cetty_catalogue_1996} with polarized broad emission lines \citep{cohen_polarimetry_1999}.
3C\,135 features the classical FR\,II supergalactic-scale biconical radio lobes in the east-west directions (PA$\sim75\degree$; e.g., \citealt{leahy_study_1997}).
No \spitzerr data are available for 3C\,135, and it appears as a point source in the \wisee images.
3C\,135 was imaged with VISIR in the broad SIC filter in 2006 but remained undetected \citep{van_der_wolk_dust_2010}.
Our derived upper limit on the nuclear MIR flux is two times higher than the value given by \cite{van_der_wolk_dust_2010}.
\newline\end{@twocolumnfalse}]

\begin{figure}
   \centering
   \includegraphics[angle=0,width=8.500cm]{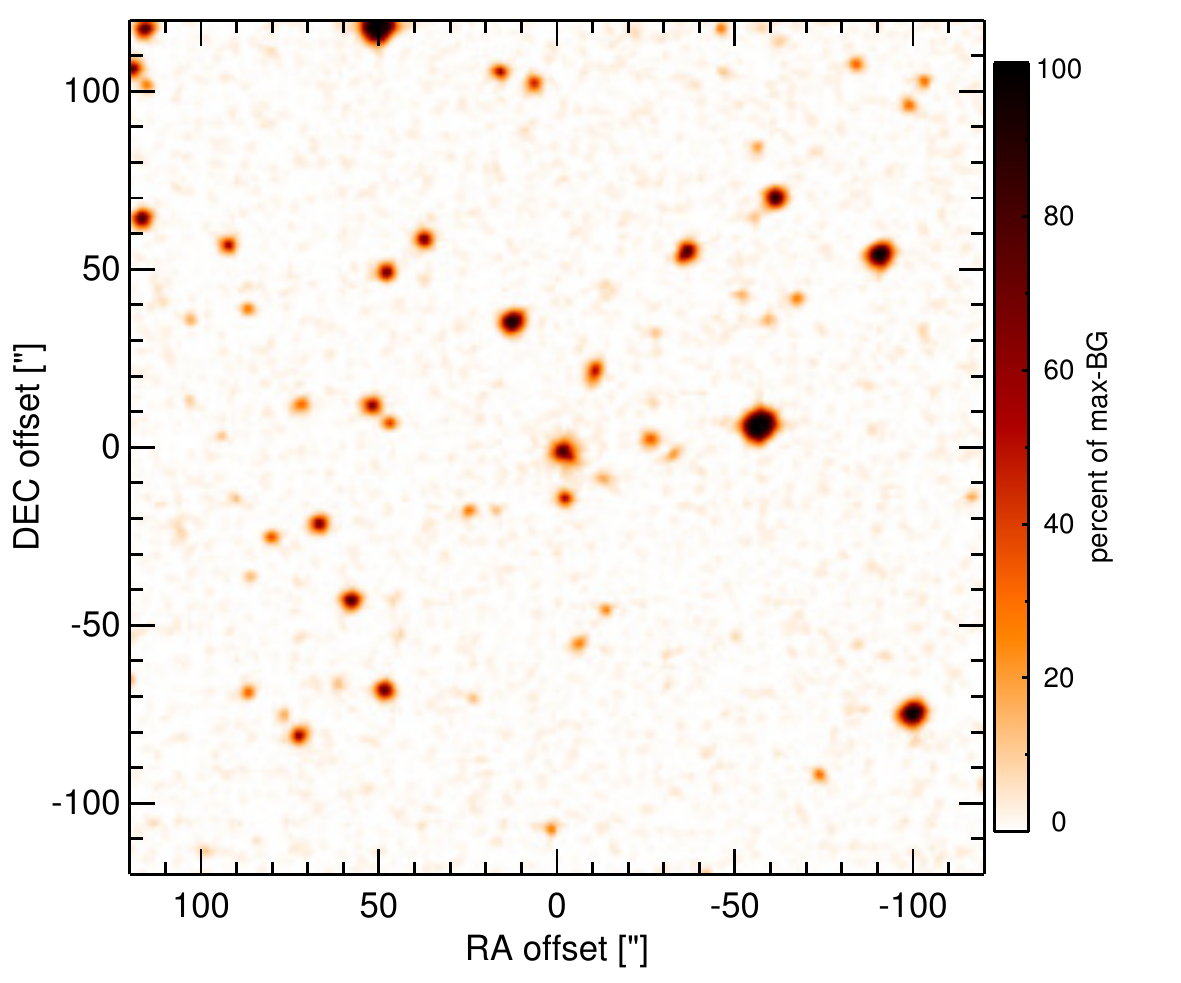}
    \caption{\label{fig:OPTim_3C135}
             Optical image (DSS, red filter) of 3C\,135. Displayed are the central $4\arcmin$ with North up and East to the left. 
              The colour scaling is linear with white corresponding to the median background and black to the $0.01\%$ pixels with the highest intensity.  
           }
\end{figure}
\begin{figure}
   \centering
   \includegraphics[angle=0,width=8.50cm]{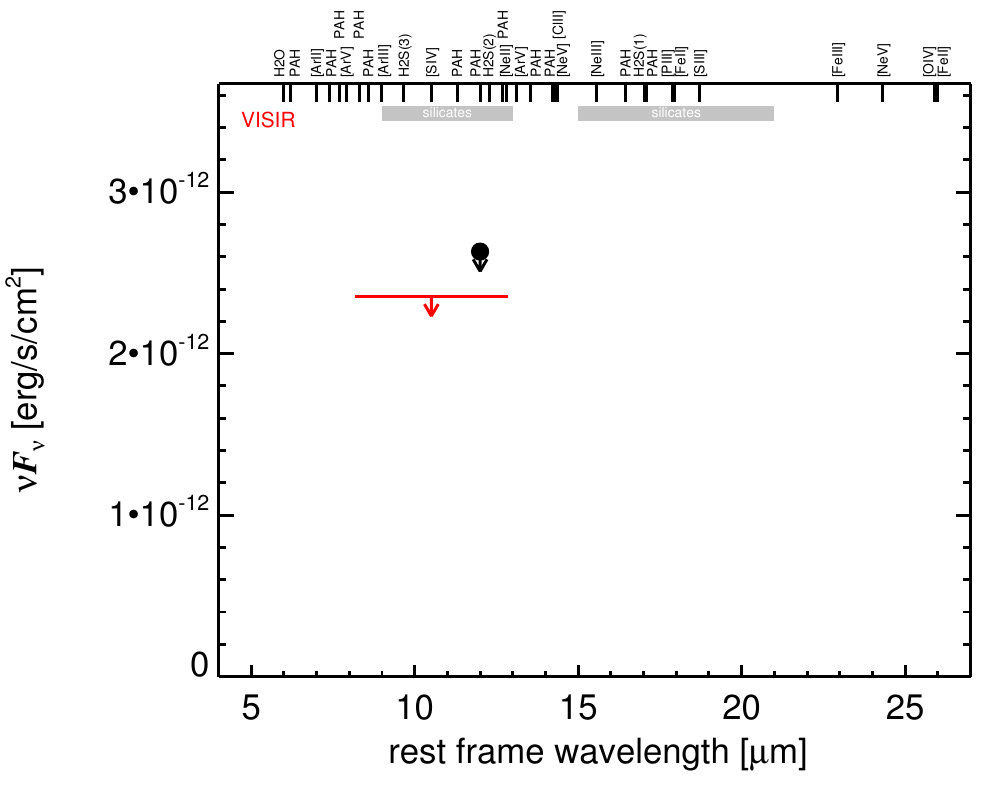}
   \caption{\label{fig:MISED_3C135}
      MIR SED of 3C\,135. The description  of the symbols (if present) is the following.
      Grey crosses and  solid lines mark the \spitzer/IRAC, MIPS and IRS data. 
      The colour coding of the other symbols is: 
      green for COMICS, magenta for Michelle, blue for T-ReCS and red for VISIR data.
      Darker-coloured solid lines mark spectra of the corresponding instrument.
      The black filled circles mark the nuclear 12 and $18\,\mu$m  continuum emission estimate from the data.
      The ticks on the top axis mark positions of common MIR emission lines, while the light grey horizontal bars mark wavelength ranges affected by the silicate 10 and 18$\mu$m features.     
   }
\end{figure}
\clearpage

\twocolumn[\begin{@twocolumnfalse}  
\subsection{3C\,227 -- PKS\,0945+07 -- SDSS\,J094745.14+072520.5}\label{app:3C227}
3C\,227 is a FR\,II radio source identified with the elliptical galaxy SDSS\,J094745.14+072520.5 at a redshift of $z=$ 0.0858 ($D \sim 414$\,Mpc), containing a 
radio-loud Sy\,1.5 nucleus \citep{veron-cetty_catalogue_2010}.
In addition to the supergalactic-scale biconical radio lobes in the east-west directions (PA$\sim90\degree$; e.g., \citealt{black_study_1992}), 3C\,227 possesses one of the most extended emission line regions with our line of sight grazing the edge of the AGN ionization cone ($> 200$ kpc; \citealt{baum_extended_1988,prieto_extended_1993}).
\spitzer/IRS and MIPS data are available for this object, in which it appears point-like.
Our nuclear MIPS photometry matches the value given in \cite{dicken_origin_2008}.
The IRS LR mapping mode spectrum suffers from low S/N but clearly exhibits silicate emission and a shallow blue spectral slope in $\nu F_\nu$-space (see also \citealt{dicken_spitzer_2012}).
3C\,227 was imaged with VISIR in the SIC filter in 2006, and a compact nucleus was detected  \citep{van_der_wolk_dust_2010}.
Our nuclear VISIR flux is consistent with the value by \cite{van_der_wolk_dust_2010} and the \spitzerr spectrophotometry.
The silicate emission feature demonstrates the presence of warm dust in the central $\sim 0.6$\,kpc of 3C\,227.
\newline\end{@twocolumnfalse}]

\begin{figure}
   \centering
   \includegraphics[angle=0,width=8.500cm]{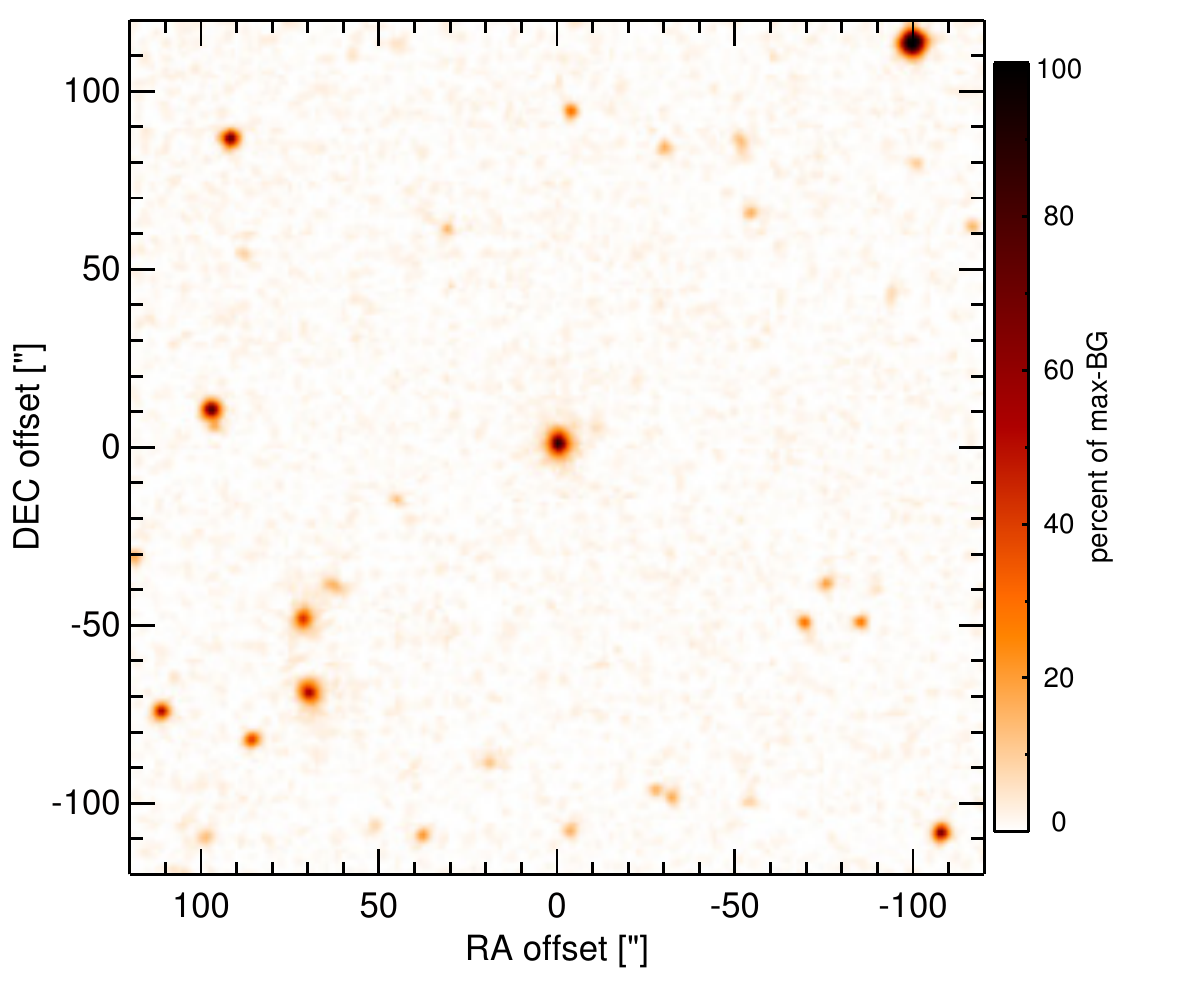}
    \caption{\label{fig:OPTim_3C227}
             Optical image (DSS, red filter) of 3C\,227. Displayed are the central $4\arcmin$ with North up and East to the left. 
              The colour scaling is linear with white corresponding to the median background and black to the $0.01\%$ pixels with the highest intensity.  
           }
\end{figure}
\begin{figure}
   \centering
   \includegraphics[angle=0,height=3.11cm]{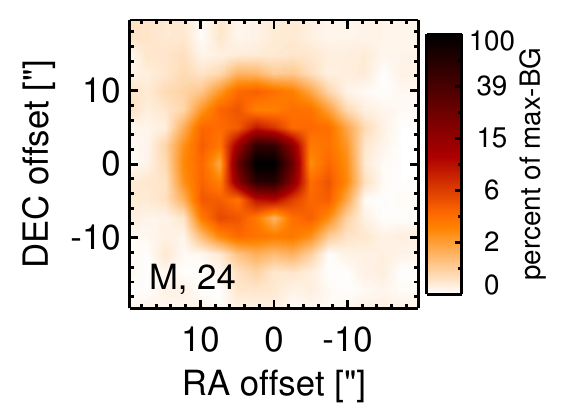}
    \caption{\label{fig:INTim_3C227}
             \spitzerr MIR images of 3C\,227. Displayed are the inner $40\arcsec$ with North up and East to the left. The colour scaling is logarithmic with white corresponding to median background and black to the $0.1\%$ pixels with the highest intensity.
             The label in the bottom left states instrument and central wavelength of the filter in $\mu$m (I: IRAC, M: MIPS). 
           }
\end{figure}
\begin{figure}
   \centering
   \includegraphics[angle=0,height=3.11cm]{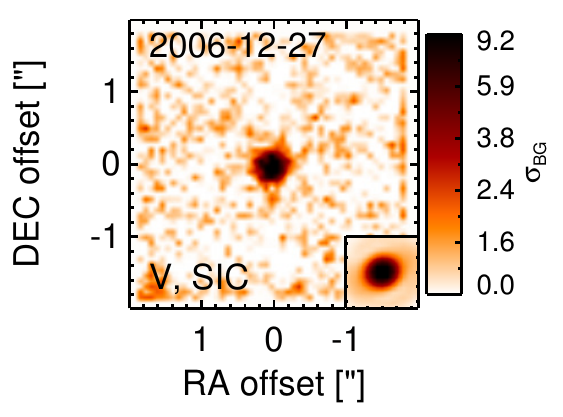}
    \caption{\label{fig:HARim_3C227}
             Subarcsecond-resolution MIR images of 3C\,227 sorted by increasing filter wavelength. 
             Displayed are the inner $4\arcsec$ with North up and East to the left. 
             The colour scaling is logarithmic with white corresponding to median background and black to the $75\%$ of the highest intensity of all images in units of $\sigbg$.
             The inset image shows the central arcsecond of the PSF from the calibrator star, scaled to match the science target.
             The labels in the bottom left state instrument and filter names (C: COMICS, M: Michelle, T: T-ReCS, V: VISIR).
           }
\end{figure}
\begin{figure}
   \centering
   \includegraphics[angle=0,width=8.50cm]{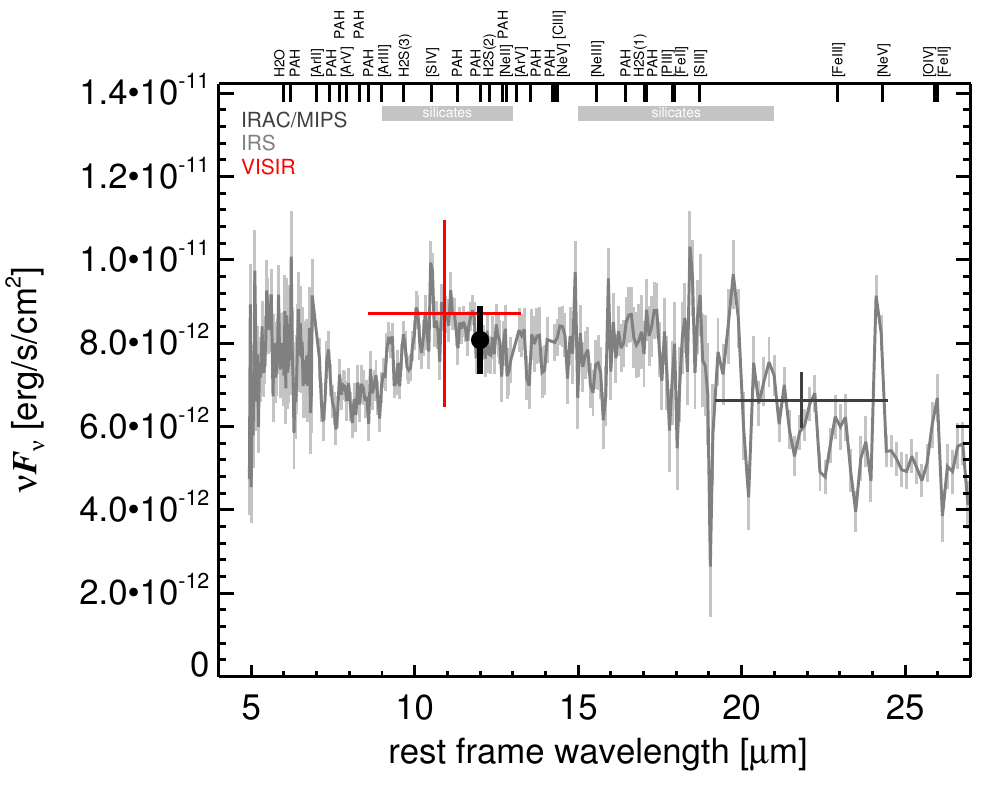}
   \caption{\label{fig:MISED_3C227}
      MIR SED of 3C\,227. The description  of the symbols (if present) is the following.
      Grey crosses and  solid lines mark the \spitzer/IRAC, MIPS and IRS data. 
      The colour coding of the other symbols is: 
      green for COMICS, magenta for Michelle, blue for T-ReCS and red for VISIR data.
      Darker-coloured solid lines mark spectra of the corresponding instrument.
      The black filled circles mark the nuclear 12 and $18\,\mu$m  continuum emission estimate from the data.
      The ticks on the top axis mark positions of common MIR emission lines, while the light grey horizontal bars mark wavelength ranges affected by the silicate 10 and 18$\mu$m features.     
   }
\end{figure}
\clearpage

\twocolumn[\begin{@twocolumnfalse}  
\subsection{3C\,264 -- NGC\,3862}\label{app:3C264}
3C\,264 is a FR\,I radio source coinciding with the elliptical galaxy NGC\,3982 at a redshift of $z=$ 0.0217 ($D \sim 103\,$Mpc), containing a little-studied  radio-loud AGN.
\cite{veron-cetty_catalogue_2010} do not give any optical classification for this object.
Based on the absence of broad emission lines \citep{buttiglione_optical_2009}, we assume a tentative Sy\,2/LINER classification for 3C\,264. 
The first successful $N$-band photometry was performed with the MMT by \cite{maiolino_low-luminosity_1995}.
\spitzer/IRAC, IRS and MIPS data are available for 3C\,264, which appears compact but resolved in the corresponding images. 
The IRS LR staring-mode spectrum exhibits PAH emission and a blue spectral slope in $\nu F_\nu$-space (see also \citealt{leipski_spitzer_2009}).
The PAH emission indicates significant star-formation in this object, which, however, has to be still rather moderate, considering the absence of large amounts of cool dust.  
We made two attempts to detect the nucleus of 3C\,264 with COMICS N11.7 imaging in 2009 but without success. 
The derived flux upper limit is $68\%$ lower than the \spitzerr fluxes, indicating that  extended host emission indeed dominates the MIR at arcsecond scales, while the AGN itself is much fainter.
\newline\end{@twocolumnfalse}]

\begin{figure}
   \centering
   \includegraphics[angle=0,width=8.500cm]{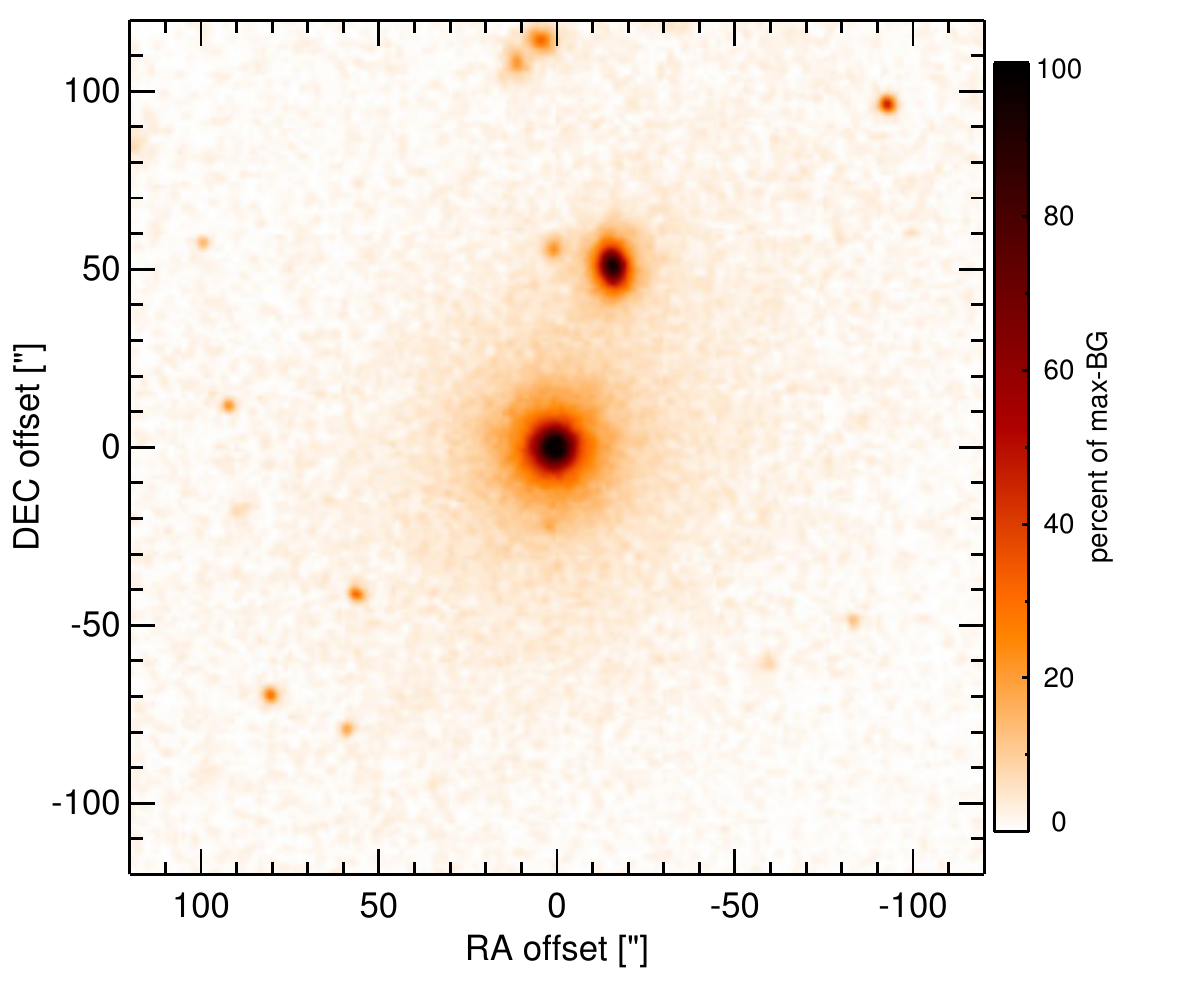}
    \caption{\label{fig:OPTim_3C264}
             Optical image (DSS, red filter) of 3C\,264. Displayed are the central $4\arcmin$ with North up and East to the left. 
              The colour scaling is linear with white corresponding to the median background and black to the $0.01\%$ pixels with the highest intensity.  
           }
\end{figure}
\begin{figure}
   \centering
   \includegraphics[angle=0,height=3.11cm]{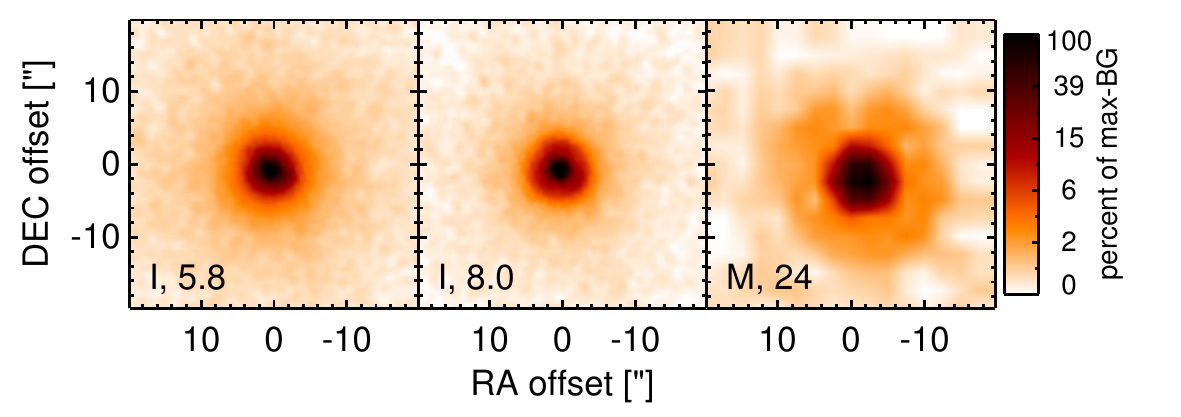}
    \caption{\label{fig:INTim_3C264}
             \spitzerr MIR images of 3C\,264. Displayed are the inner $40\arcsec$ with North up and East to the left. The colour scaling is logarithmic with white corresponding to median background and black to the $0.1\%$ pixels with the highest intensity.
             The label in the bottom left states instrument and central wavelength of the filter in $\mu$m (I: IRAC, M: MIPS). 
           }
\end{figure}
\begin{figure}
   \centering
   \includegraphics[angle=0,width=8.50cm]{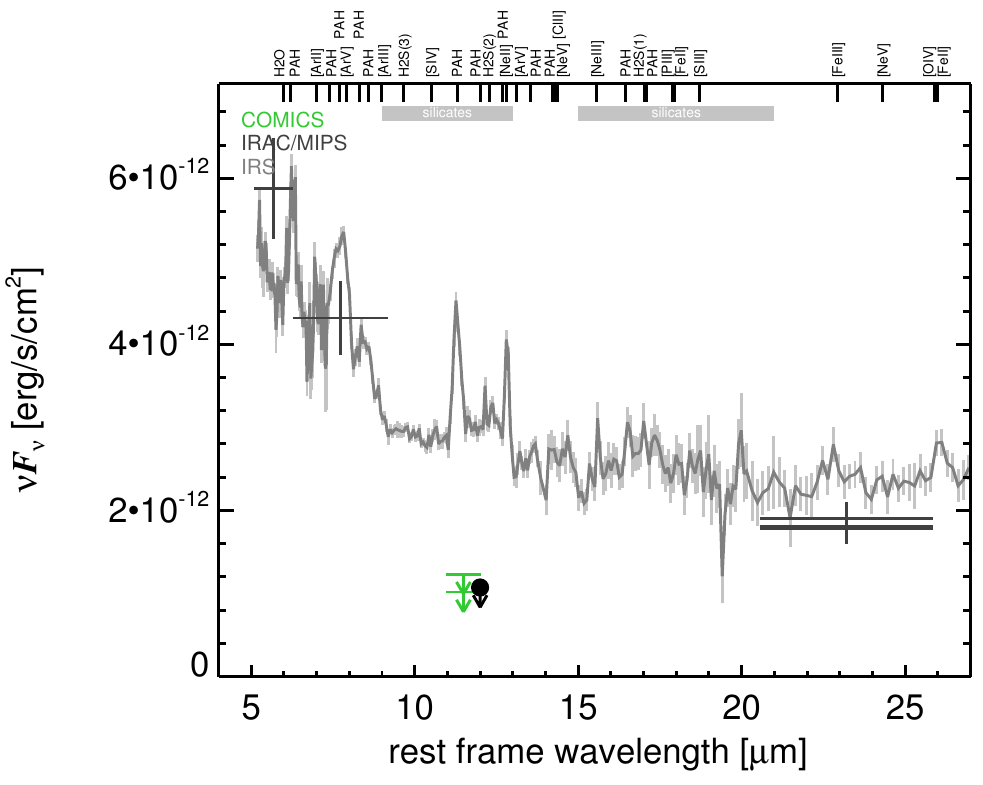}
   \caption{\label{fig:MISED_3C264}
      MIR SED of 3C\,264. The description  of the symbols (if present) is the following.
      Grey crosses and  solid lines mark the \spitzer/IRAC, MIPS and IRS data. 
      The colour coding of the other symbols is: 
      green for COMICS, magenta for Michelle, blue for T-ReCS and red for VISIR data.
      Darker-coloured solid lines mark spectra of the corresponding instrument.
      The black filled circles mark the nuclear 12 and $18\,\mu$m  continuum emission estimate from the data.
      The ticks on the top axis mark positions of common MIR emission lines, while the light grey horizontal bars mark wavelength ranges affected by the silicate 10 and 18$\mu$m features.     
   }
\end{figure}
\clearpage

\twocolumn[\begin{@twocolumnfalse}  
\subsection{3C\,273 -- PKS\,1226+02}\label{app:3C273}
3C\,273 is the first discovered  radio-loud quasar \citep{schmidt_3c_1963} with a redshift of $z=$ 0.1583 ($D \sim 792$\,Mpc), a flat radio spectrum, and a Sy\,1 optical classification \citep{veron-cetty_catalogue_2010}.
It is one of the best studied AGN with a one-sided jet oriented to the south-west (see \citealt{courvoisier_bright_1998} for a review). 
The source emission appears beamed, and thus 3C\,273 is also classified as a blazar.
This object is strongly variable at all frequencies (e.g., \citealt{soldi_multiwavelength_2008}), including the MIR where the variability is up to $60\%$ \citep{neugebauer_infrared_1984}.
It is also among the first AGN at MIR wavelengths with a few  decades of $N$-band coverage \citep{low_spectrum_1965, kleinmann_infrared_1970,  rieke_infrared_1972, sitko_0.35-3.5_1982,roche_8-13_1984,neugebauer_continuum_1987}.
The first subarcsecond-resolution $N$-band images were obtained with MIRLIN \citep{gorjian_10_2004} and TIMMI2 \citep{raban_core_2008}.
3C\,273 also has multiple-epoch coverage in all \spitzerr instruments. 
The first IRS spectrum was published by \cite{hao_detection_2005} and shows weak silicate $10\,\mu$m  emission and no PAH features, which indicates the 
presence of dust, although the SED is generally expected to be synchrotron-dominated.
IRAC images were taken in 2007 and 2009, while MIPS was used to observe 3C\,273 in 2004, 2006, 2007, and 2009.
We find a maximum MIR variability of $17\%$ around the median in the MIPS $24\,\mu$m band.
Our measurement from the 2004 data agrees with  \cite{dicken_origin_2008}.
The jet is visible in the IRAC and MIPS images, while the nucleus is  otherwise compact (see also \citealt{uchiyama_shedding_2006}).
VISIR images were take in multiple $N$-band filters in 2007 and 2011. 
The point-like nucleus was detected but no jet emission. 
Our flux measurements from the 2007 data agree with the published values from these data in \cite{tristram_parsec-scale_2009}.
Note the different beam shape and elongation direction of the central source in the three different nights of VISIR observations. 
However, the nucleus is unresolved in the sharpest and highest S/N images, and thus is classified as unresolved at subarcsecond scales in the MIR.
3C\,273 was also observed with MIR interferometry with MIDI, where it appears only marginally resolved with a maximum size of the emitter of $\lesssim 16$\,pc ($\sim 12$\,mas;  \citealt{tristram_parsec-scale_2009,burtscher_diversity_2013}). 
Thus, the elongations seen in the VISIR images can not be real.
The flux differences between the \spitzerr and VISIR measurements are most likely caused by variability.
To compute the nuclear $12\,\mu$m continuum emission we extrapolate from the PAH2\_2 measurements assuming the median type~I MIR SED.
\newline\end{@twocolumnfalse}]

\begin{figure}
   \centering
   \includegraphics[angle=0,width=8.500cm]{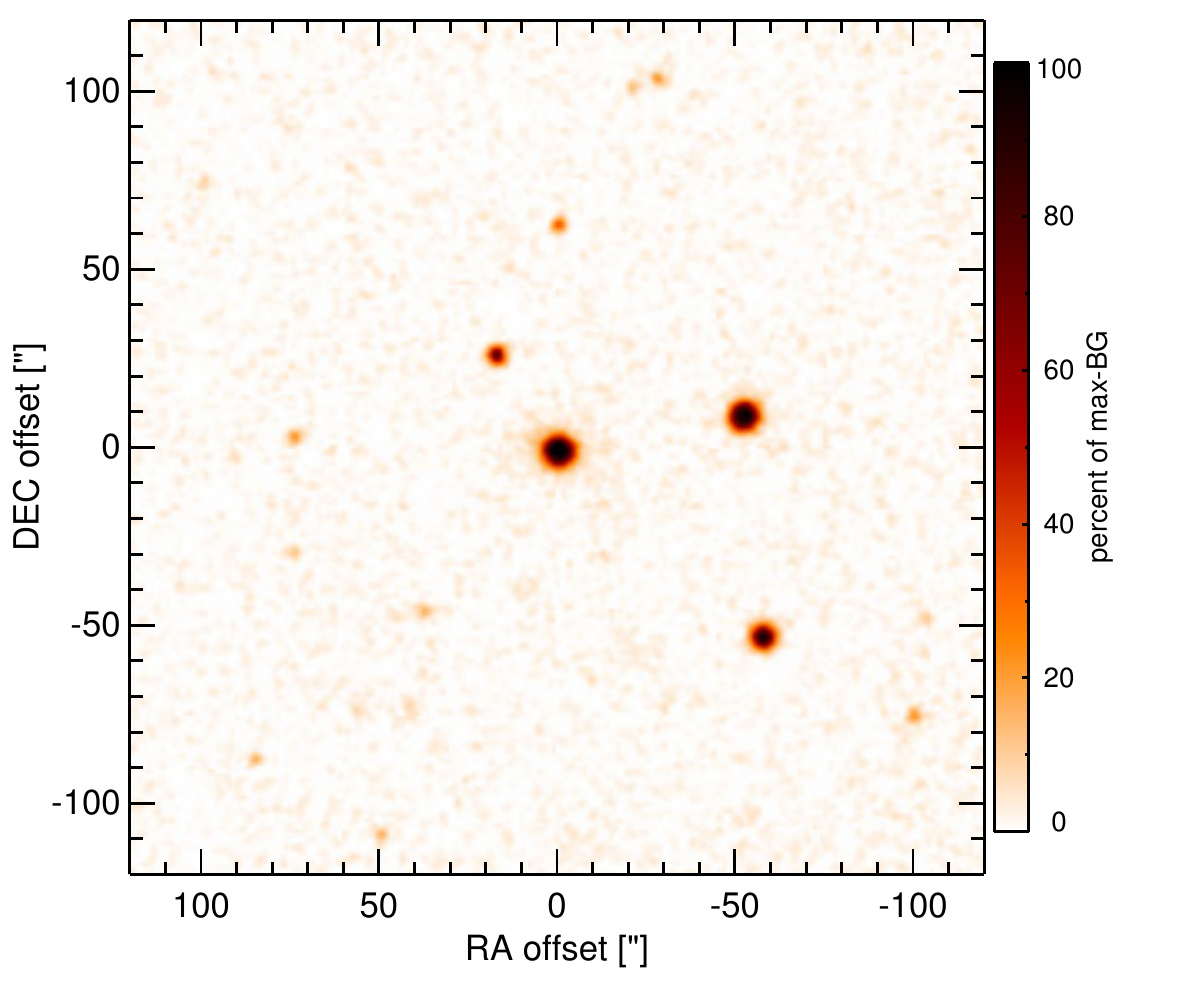}
    \caption{\label{fig:OPTim_3C273}
             Optical image (DSS, red filter) of 3C\,273. Displayed are the central $4\arcmin$ with North up and East to the left. 
              The colour scaling is linear with white corresponding to the median background and black to the $0.01\%$ pixels with the highest intensity.  
           }
\end{figure}
\begin{figure}
   \centering
   \includegraphics[angle=0,height=3.11cm]{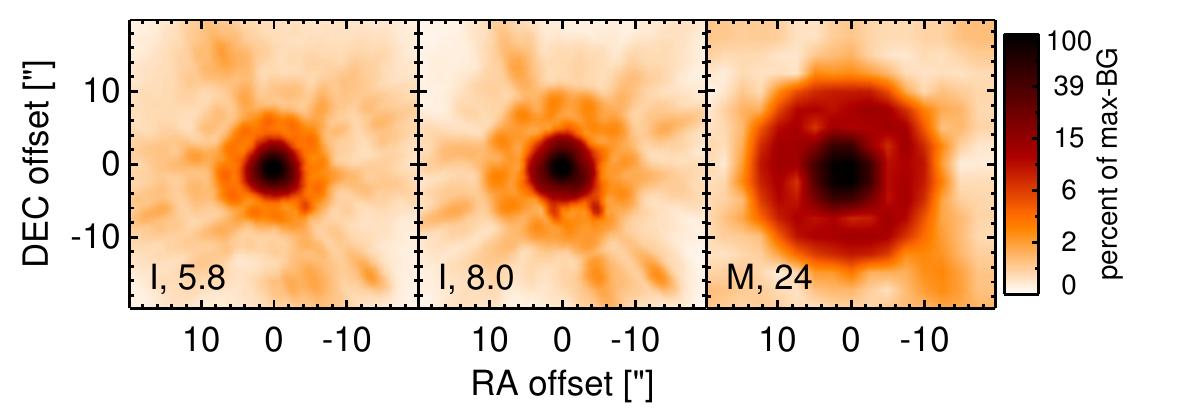}
    \caption{\label{fig:INTim_3C273}
             \spitzerr MIR images of 3C\,273. Displayed are the inner $40\arcsec$ with North up and East to the left. The colour scaling is logarithmic with white corresponding to median background and black to the $0.1\%$ pixels with the highest intensity.
             The label in the bottom left states instrument and central wavelength of the filter in $\mu$m (I: IRAC, M: MIPS). 
           }
\end{figure}
\begin{figure}
   \centering
   \includegraphics[angle=0,width=8.500cm]{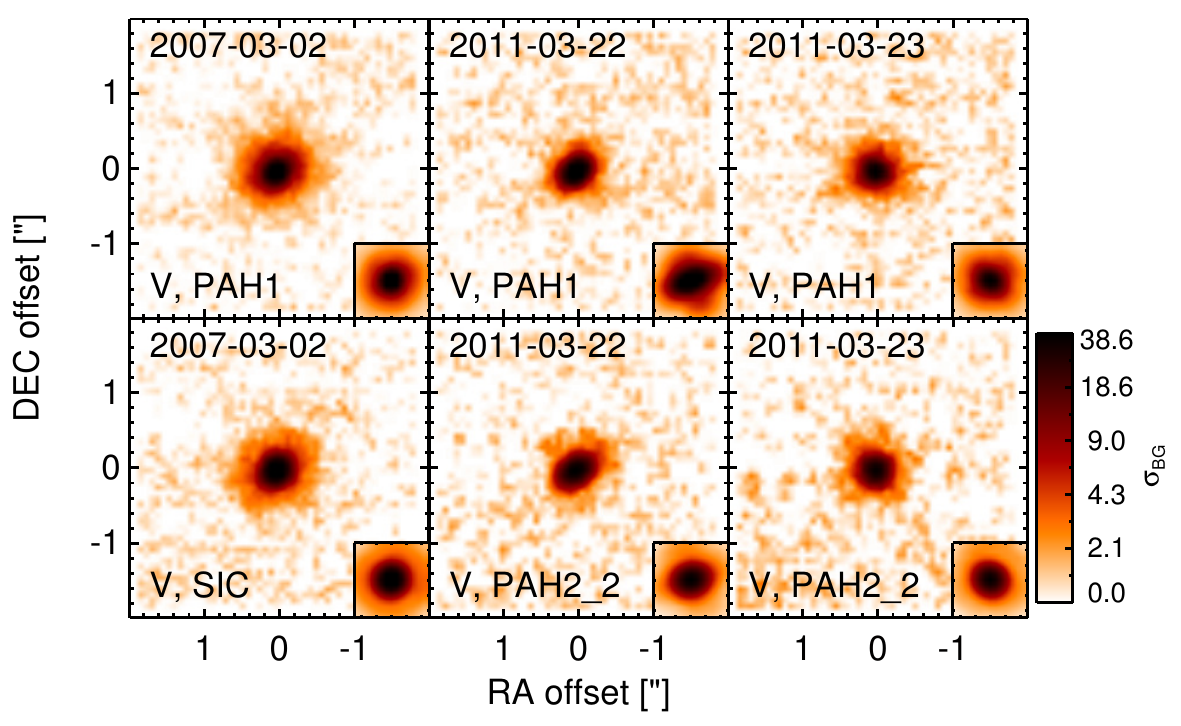}
    \caption{\label{fig:HARim_3C273}
             Subarcsecond-resolution MIR images of 3C\,273 sorted by increasing filter wavelength. 
             Displayed are the inner $4\arcsec$ with North up and East to the left. 
             The colour scaling is logarithmic with white corresponding to median background and black to the $75\%$ of the highest intensity of all images in units of $\sigbg$.
             The inset image shows the central arcsecond of the PSF from the calibrator star, scaled to match the science target.
             The labels in the bottom left state instrument and filter names (C: COMICS, M: Michelle, T: T-ReCS, V: VISIR).
           }
\end{figure}
\begin{figure}
   \centering
   \includegraphics[angle=0,width=8.50cm]{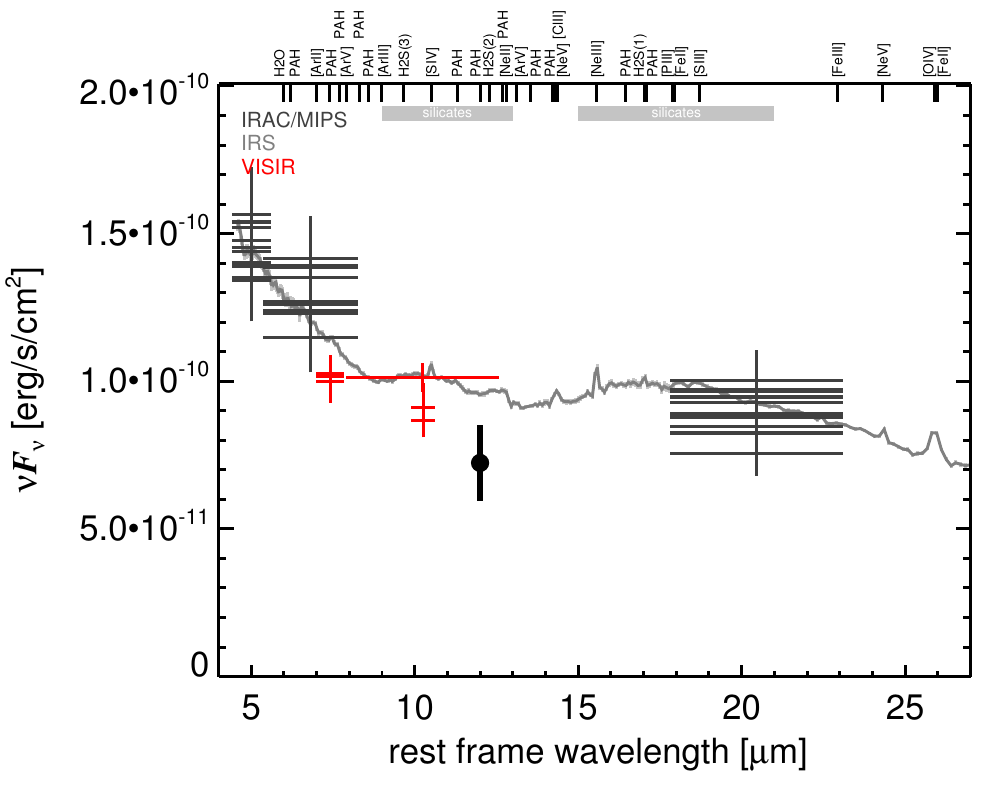}
   \caption{\label{fig:MISED_3C273}
      MIR SED of 3C\,273. The description  of the symbols (if present) is the following.
      Grey crosses and  solid lines mark the \spitzer/IRAC, MIPS and IRS data. 
      The colour coding of the other symbols is: 
      green for COMICS, magenta for Michelle, blue for T-ReCS and red for VISIR data.
      Darker-coloured solid lines mark spectra of the corresponding instrument.
      The black filled circles mark the nuclear 12 and $18\,\mu$m  continuum emission estimate from the data.
      The ticks on the top axis mark positions of common MIR emission lines, while the light grey horizontal bars mark wavelength ranges affected by the silicate 10 and 18$\mu$m features.     
   }
\end{figure}
\clearpage

\twocolumn[\begin{@twocolumnfalse}  
\subsection{3C\,285 -- IRAS\,F13190+425 -- LEDA\,46625}\label{app:3C285}
3C\,285 is a FR\,II radio source identified with the irregular galaxy LEDA\,46625 (e.g., \citealt{allen_ultraviolet_2002}) at a redshift of $z=$ 0.0794 ($D \sim 378$\,Mpc), containing a radio-loud Sy\,2 nucleus \citep{veron-cetty_catalogue_2010}.
Its supergalactic-scale biconical radio lobes extend an in the east-west directions (PA$\sim75\degree$; e.g., \citealt{leahy_bridges_1984}).
The first successful MIR detection of 3C\,285 was achieved with \spitzer/IRAC, IRS and MIPS, where it appears nearly unresolved. 
Our nuclear MIPS $24\,\mu$m flux agrees with \cite{dicken_origin_2010}.
The IRS LR mapping-mode spectrum suffers from low S/N but shows silicate $10\,\mu$m absorption, PAH emission and a red spectral slope in $\nu F_\nu$-space, indicating significant star formation (see also \citealt{dicken_spitzer_2012}).
We attempted to detect 3C\,285 with COMICS in the N11.7 filter in 2009 but without success.
The corresponding upper limit  of the nuclear N11.7 flux is $\sim 64\%$ lower than the \spitzerr spectrophotometry.
Therefore, star formation dominates the arcsecond-scale MIR SED of 3C\,285, while it is resolved-out at subarcsecond resolution. 
\newline\end{@twocolumnfalse}]

\begin{figure}
   \centering
   \includegraphics[angle=0,width=8.500cm]{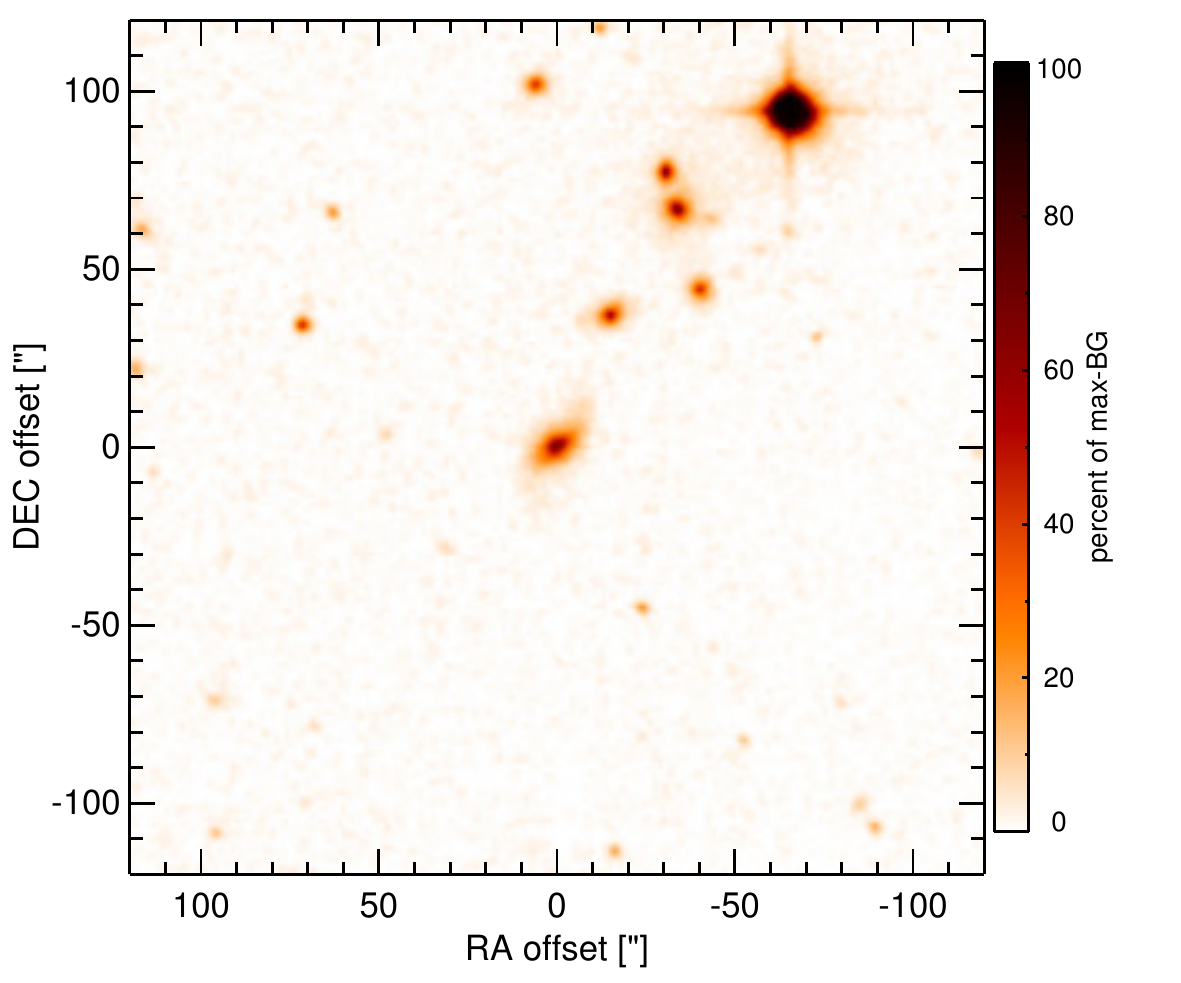}
    \caption{\label{fig:OPTim_3C285}
             Optical image (DSS, red filter) of 3C\,285. Displayed are the central $4\arcmin$ with North up and East to the left. 
              The colour scaling is linear with white corresponding to the median background and black to the $0.01\%$ pixels with the highest intensity.  
           }
\end{figure}
\begin{figure}
   \centering
   \includegraphics[angle=0,height=3.11cm]{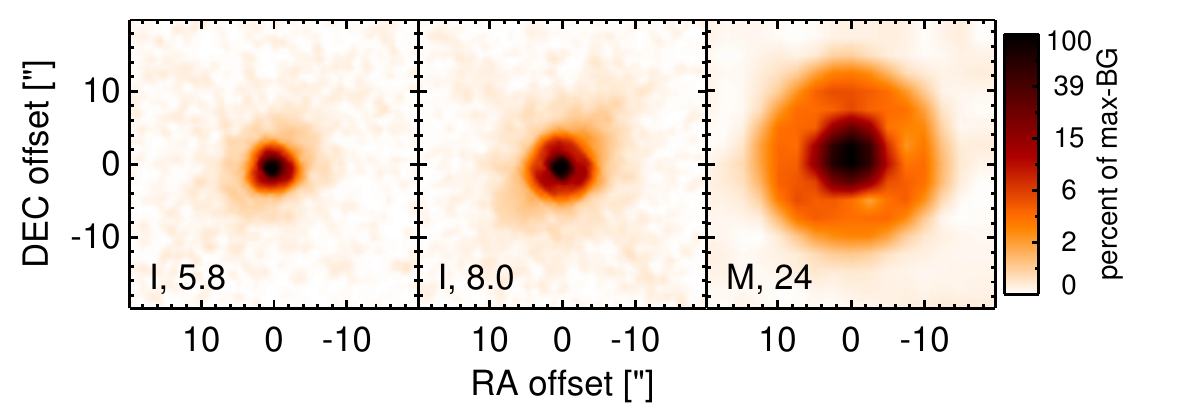}
    \caption{\label{fig:INTim_3C285}
             \spitzerr MIR images of 3C\,285. Displayed are the inner $40\arcsec$ with North up and East to the left. The colour scaling is logarithmic with white corresponding to median background and black to the $0.1\%$ pixels with the highest intensity.
             The label in the bottom left states instrument and central wavelength of the filter in $\mu$m (I: IRAC, M: MIPS). 
           }
\end{figure}
\begin{figure}
   \centering
   \includegraphics[angle=0,width=8.50cm]{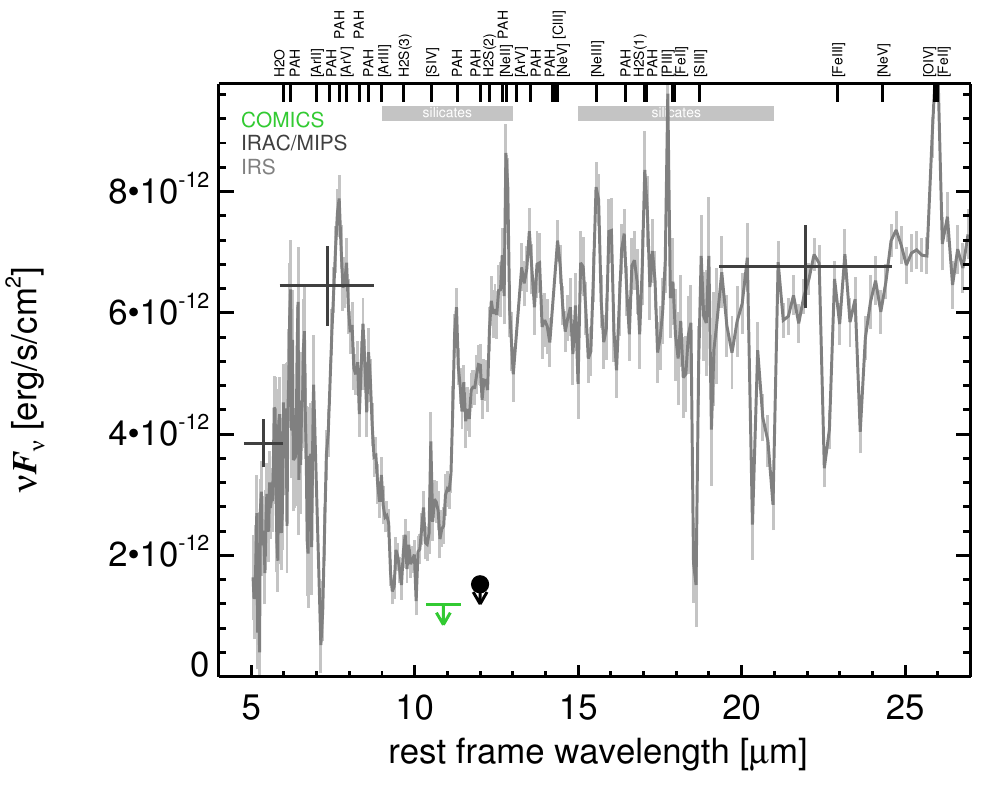}
   \caption{\label{fig:MISED_3C285}
      MIR SED of 3C\,285. The description  of the symbols (if present) is the following.
      Grey crosses and  solid lines mark the \spitzer/IRAC, MIPS and IRS data. 
      The colour coding of the other symbols is: 
      green for COMICS, magenta for Michelle, blue for T-ReCS and red for VISIR data.
      Darker-coloured solid lines mark spectra of the corresponding instrument.
      The black filled circles mark the nuclear 12 and $18\,\mu$m  continuum emission estimate from the data.
      The ticks on the top axis mark positions of common MIR emission lines, while the light grey horizontal bars mark wavelength ranges affected by the silicate 10 and 18$\mu$m features.     
   }
\end{figure}
\clearpage

\twocolumn[\begin{@twocolumnfalse}  
\subsection{3C\,293 -- UGC\,8782}\label{app:3C293}
3C\,293 is a FR\,I radio source coinciding with the irregular galaxy UGC\,8782 at a redshift of $z=$ 0.045 ($D \sim 311$\,Mpc) and an AGN optically classified as a LINER \citep{veron-cetty_catalogue_2010}.
It features a complex radio morphology with asymmetrical supergalactic-scale radio lobes and a misaligned inner two-sided jet (PA$\sim-55\degree$ and PA$\sim-85\degree$ respectively; e.g., \citealt{akujor_two-sided_1996,beswick_high-resolution_2004,giovannini_bologna_2005}).
The jet is even visible in the UV, optical and near-infrared \citep{floyd_jet_2006}.
Furthermore, the nuclear region of the heavily disturbed galaxy is obscured by several dust lanes roughly in  the north-south directions \citep{martel_hubble_1999}.
The first ground-based MIR observations of 3C\,293 were performed with IRTF, resulting in a tentative detection of nuclear emission \citep{elvis_1-20_1984,impey_infrared_1990}. 
Later, the object was followed up with \iso/ISOCAM \citep{siebenmorgen_isocam_2004} and \spitzer/IRAC, IRS and MIPS observations.
In addition to a compact nucleus, flattened host galaxy emission  with a diameter of $\sim 14$\,kpc is visible in the IRAC and MIPS images.
Our nuclear MIPS $24\,\mu$m flux agrees with \cite{dicken_origin_2010}.
The IRS LR staring-mode spectrum is dominated by strong PAH emission and silicate $10\,\mu$m absorption and has a rather flat spectral slope in $\nu F_\nu$-space (see also \citealt{shi_aromatic_2007,leipski_spitzer_2009,ogle_jet-powered_2010,dicken_spitzer_2012}).
Thus, the arcsecond-scale MIR SED is star-formation dominated. 
3C\,293 remained undetected in our COMICS  N11.7 imaging observations in 2009.
The derived upper limit on the nuclear N11.7 flux is $\sim87\%$ lower than the \spitzerr spectrophotometry.
Therefore, the dominating star formation must be extended and is mostly resolved out at subarcsecond resolution, while any AGN contribution to the MIR emission of 3C\,293 has to be small.
\newline\end{@twocolumnfalse}]

\begin{figure}
   \centering
   \includegraphics[angle=0,width=8.500cm]{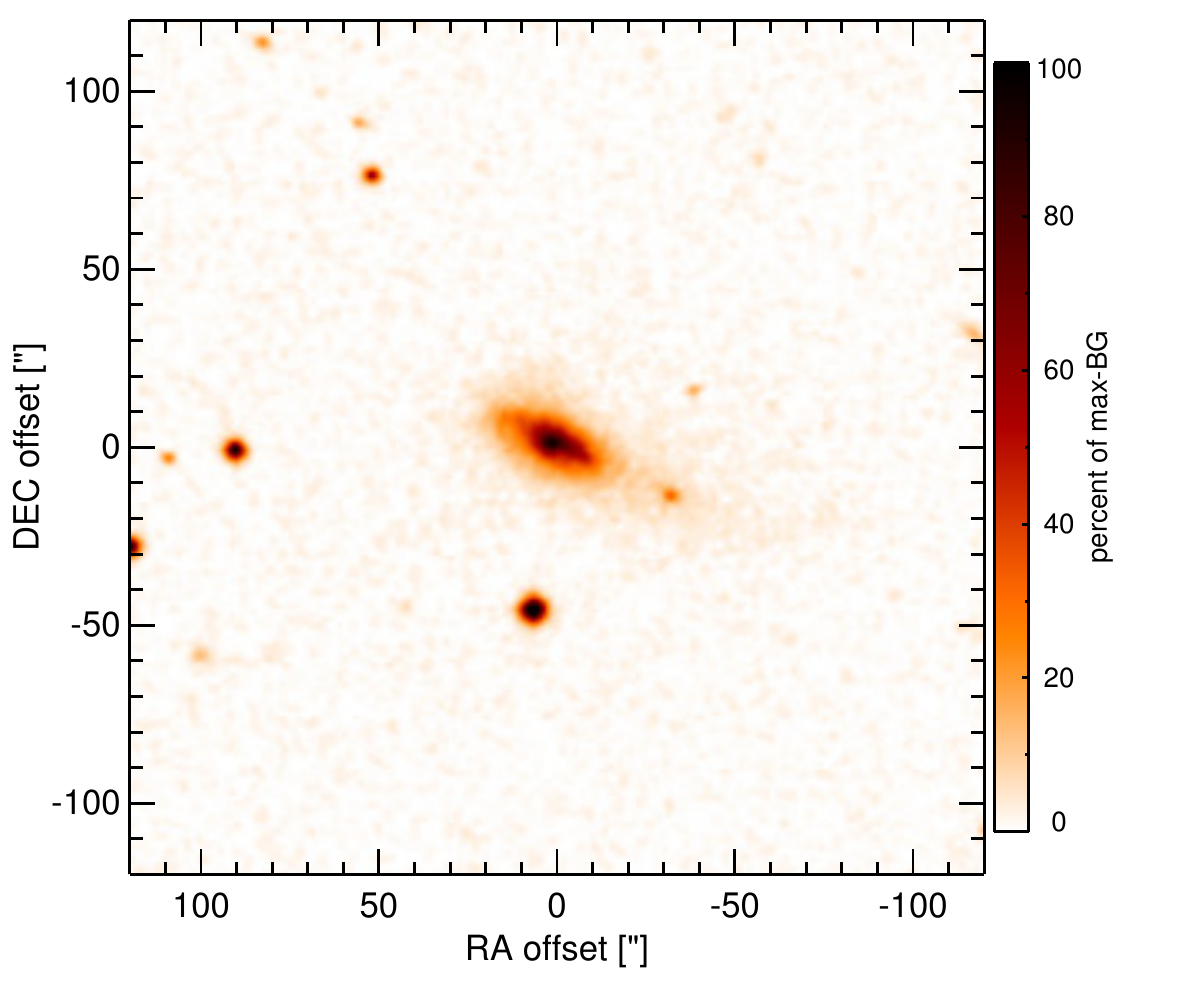}
    \caption{\label{fig:OPTim_3C293}
             Optical image (DSS, red filter) of 3C\,293. Displayed are the central $4\arcmin$ with North up and East to the left. 
              The colour scaling is linear with white corresponding to the median background and black to the $0.01\%$ pixels with the highest intensity.  
           }
\end{figure}
\begin{figure}
   \centering
   \includegraphics[angle=0,height=3.11cm]{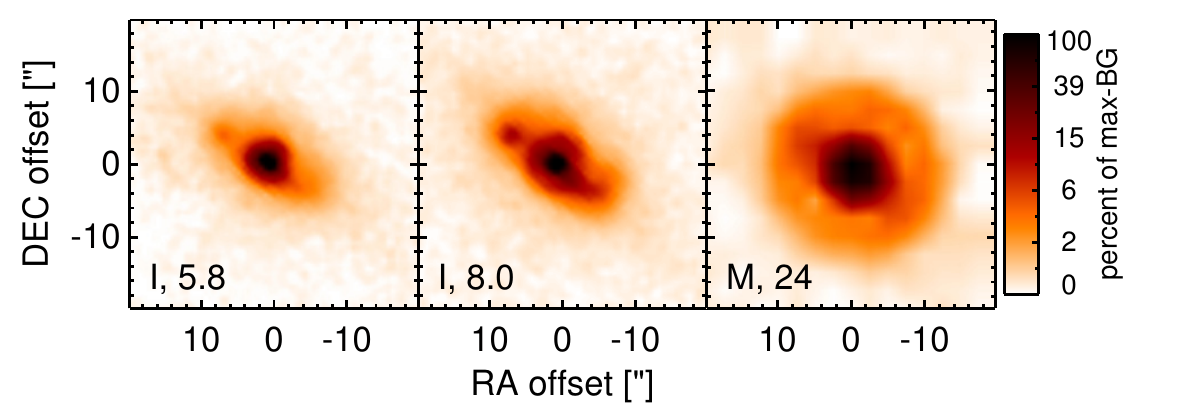}
    \caption{\label{fig:INTim_3C293}
             \spitzerr MIR images of 3C\,293. Displayed are the inner $40\arcsec$ with North up and East to the left. The colour scaling is logarithmic with white corresponding to median background and black to the $0.1\%$ pixels with the highest intensity.
             The label in the bottom left states instrument and central wavelength of the filter in $\mu$m (I: IRAC, M: MIPS). 
           }
\end{figure}
\begin{figure}
   \centering
   \includegraphics[angle=0,width=8.50cm]{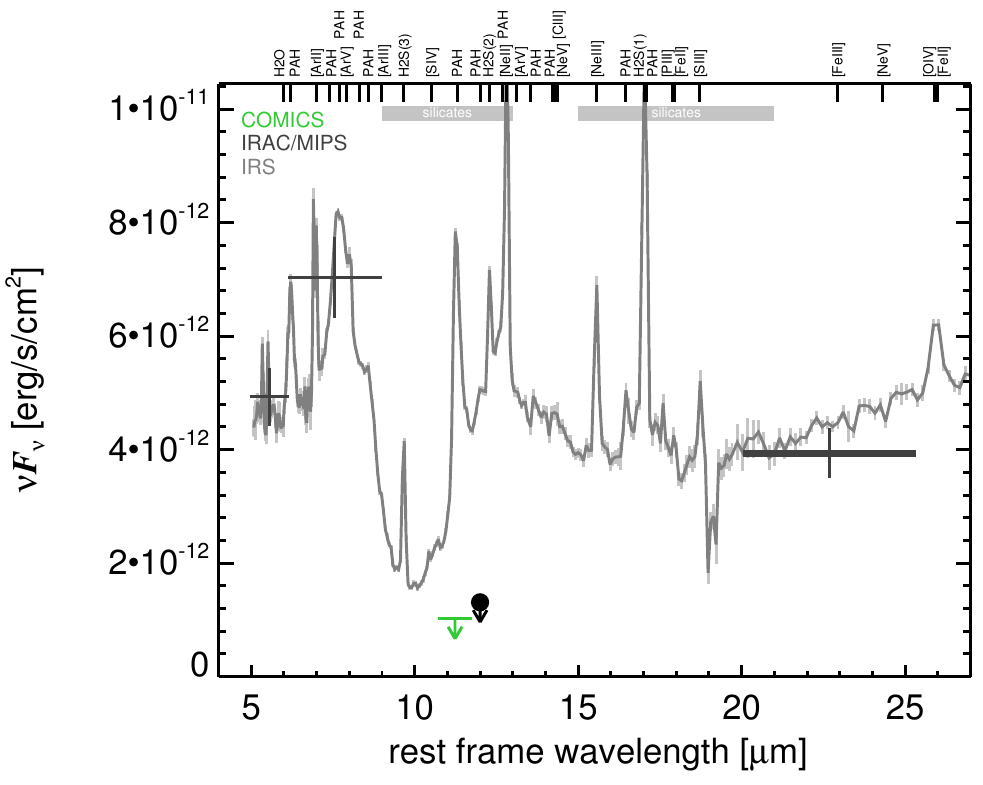}
   \caption{\label{fig:MISED_3C293}
      MIR SED of 3C\,293. The description  of the symbols (if present) is the following.
      Grey crosses and  solid lines mark the \spitzer/IRAC, MIPS and IRS data. 
      The colour coding of the other symbols is: 
      green for COMICS, magenta for Michelle, blue for T-ReCS and red for VISIR data.
      Darker-coloured solid lines mark spectra of the corresponding instrument.
      The black filled circles mark the nuclear 12 and $18\,\mu$m  continuum emission estimate from the data.
      The ticks on the top axis mark positions of common MIR emission lines, while the light grey horizontal bars mark wavelength ranges affected by the silicate 10 and 18$\mu$m features.     
   }
\end{figure}
\clearpage

\twocolumn[\begin{@twocolumnfalse}  
\subsection{3C\,305 -- IC\,1065}\label{app:3C305}
3C\,305 is a steep-spectrum FR\,I radio source coinciding with the peculiar galaxy IC\,1065 at a redshift of $z=$ 0.0416 ($D \sim 192$\,Mpc) and 
a radio-loud Sy\,2 nucleus \citep{veron-cetty_catalogue_2010}.
It possesses a compact radio morphology with both FR\,I and FR\,II characteristics, featuring a two-sided jet extending along a north-east axis to the biconical radio lobes, which are separated by $\sim4\arcsec\sim3.4\,$kpc (PA$\sim54\degree$; \citealt{heckman_optical_1982,jackson_observations_2003}).
The first successful MIR $N$-band observation was performed with \iso/ISOCAM \citep{siebenmorgen_isocam_2004} and followed up with \spitzer/IRAC, IRS and MIPS observations.
The corresponding images show a compact nucleus embedded within elliptically shaped host emission.
In addition, a compact source of unknown nature us visible $\sim10\arcsec\sim8.6\,$kpc to the west (PA$\sim243\degree$; 2MASS\,J14492029+6316107).
Our nuclear MIPS $24\,\mu$m flux agrees with \cite{dicken_origin_2010}.
The IRS LR mapping-mode spectrum suffers from low S/N but indicates silicate $10\,\mu$m absorption and possible PAH emission. 
A better S/N IRS spectrum is shown in \cite{dicken_spitzer_2012} with the same spectral features indicating star formation.
3C\,305 remained undetected in our COMICS imaging observations of 2009 in the N11.7 filter.
The resulting upper limit on the nuclear N11.7 flux is $\sim80\%$ lower than the \spitzerr spectrophotometry.
Therefore, the arcsecond-scale MIR SED is completely dominated by star formation, which is resolved out at subarcsecond resolution, leaving little room for significant MIR emission from the AGN. 
\newline\end{@twocolumnfalse}]

\begin{figure}
   \centering
   \includegraphics[angle=0,width=8.500cm]{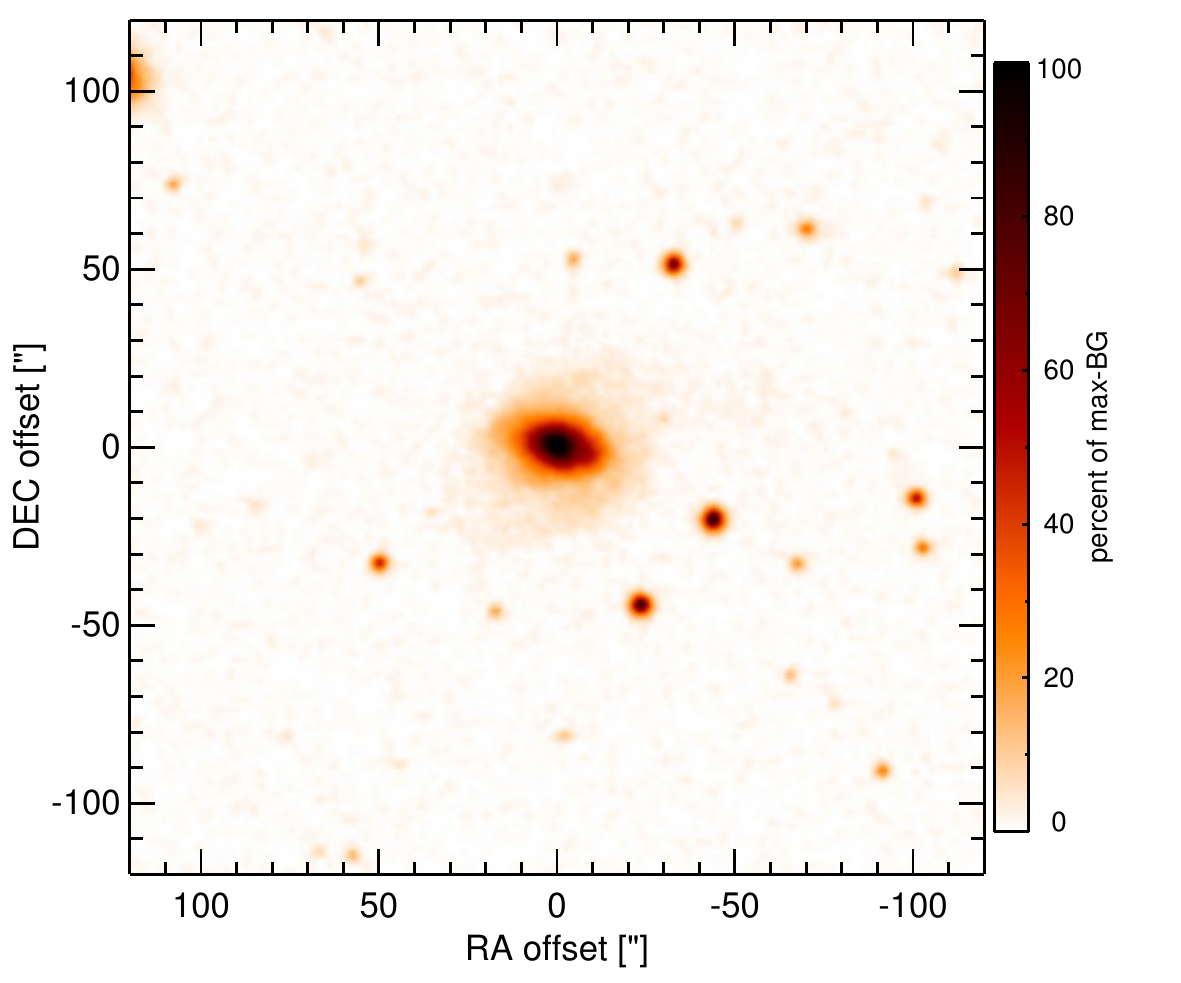}
    \caption{\label{fig:OPTim_3C305}
             Optical image (DSS, red filter) of 3C\,305. Displayed are the central $4\arcmin$ with North up and East to the left. 
              The colour scaling is linear with white corresponding to the median background and black to the $0.01\%$ pixels with the highest intensity.  
           }
\end{figure}
\begin{figure}
   \centering
   \includegraphics[angle=0,height=3.11cm]{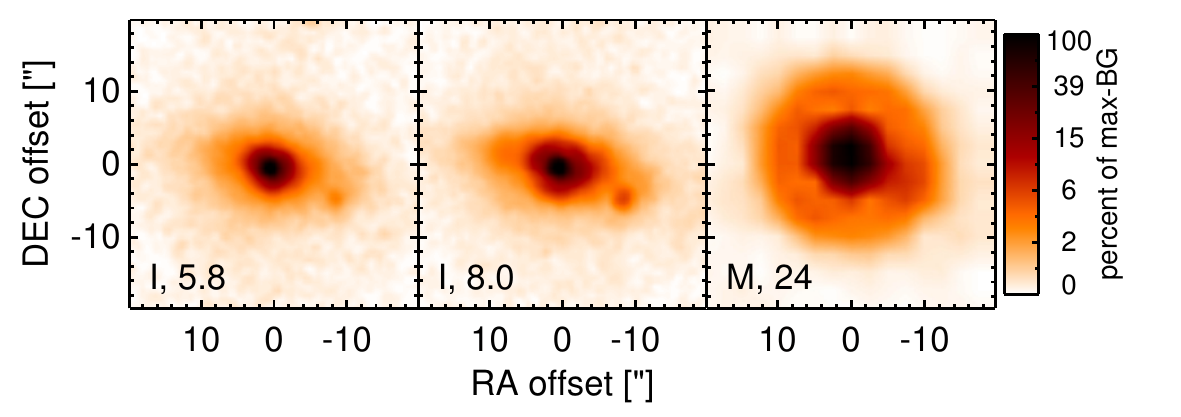}
    \caption{\label{fig:INTim_3C305}
             \spitzerr MIR images of 3C\,305. Displayed are the inner $40\arcsec$ with North up and East to the left. The colour scaling is logarithmic with white corresponding to median background and black to the $0.1\%$ pixels with the highest intensity.
             The label in the bottom left states instrument and central wavelength of the filter in $\mu$m (I: IRAC, M: MIPS). 
           }
\end{figure}
\begin{figure}
   \centering
   \includegraphics[angle=0,width=8.50cm]{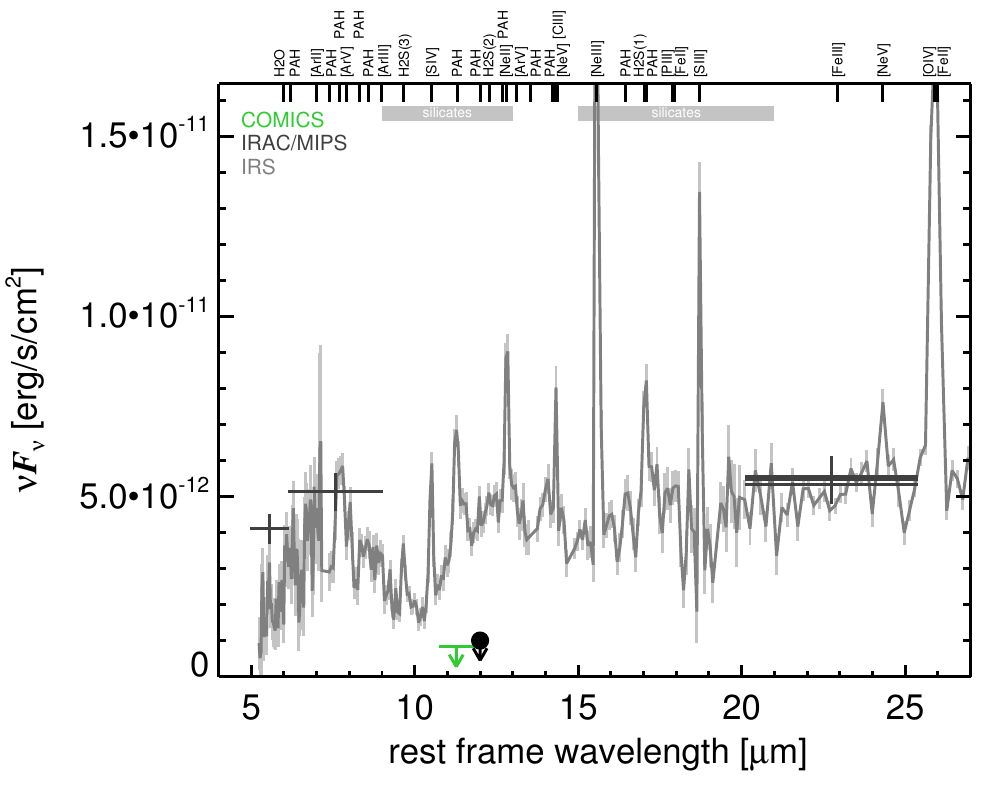}
   \caption{\label{fig:MISED_3C305}
      MIR SED of 3C\,305. The description  of the symbols (if present) is the following.
      Grey crosses and  solid lines mark the \spitzer/IRAC, MIPS and IRS data. 
      The colour coding of the other symbols is: 
      green for COMICS, magenta for Michelle, blue for T-ReCS and red for VISIR data.
      Darker-coloured solid lines mark spectra of the corresponding instrument.
      The black filled circles mark the nuclear 12 and $18\,\mu$m  continuum emission estimate from the data.
      The ticks on the top axis mark positions of common MIR emission lines, while the light grey horizontal bars mark wavelength ranges affected by the silicate 10 and 18$\mu$m features.     
   }
\end{figure}
\clearpage

\twocolumn[\begin{@twocolumnfalse}  
\subsection{3C\,317 -- UGC\,9799 -- Abell\,2052}\label{app:3C317}
3C\,317 is an ultra-steep spectrum FR\,I radio source in the early-type galaxy UGC\,9799 at a redshift of $z=$ 0.0345 ($D \sim 160$\,Mpc), the central galaxy of the cluster Abell\,2052.
It contains a radio-loud AGN, optically classified either as Sy\,2 \citep{veron-cetty_catalogue_2010} or LINER \citep{simpson_emission-line_1996}. 
Instead of twin jets or lobes, 3C\,317 displays an amorphous radio structure with a bright core \citep{zhao_3c_1993}.
At parsec-scales, a young radio source is found in the nucleus, suggestive of recently-restarted activity \citep{venturi_radio_2004}.  
The first successful MIR observations of 3C\,285 were performed with \spitzer/IRAC, IRS and MIPS, and the corresponding images show a compact nucleus embedded within diffuse elliptical host emission (see also  \citealt{quillen_infrared_2008}).
In addition, a compact source was detected at a distance of $\sim7\arcsec$ north-east of the nucleus.
It corresponds to the galaxy SDSS\,J151644.80+070123.3 and X-ray source EXSS\,1514.2+0712.
Our nuclear IRAC and MIPS fluxes are lower than those published in \cite{quillen_infrared_2008} because they used an aperture of 12.2\arcsec. 
The IRS LR staring-mode spectrum does not exhibit any significant silicate or PAH features and has a shallow blue spectral slope in $\nu F_\nu$-space (see also  \citealt{leipski_spitzer_2009,ogle_jet-powered_2010}).
VISIR imaging of 3C\,317 in the SIC filter was performed in 2006 but the nucleus remained undetected \citep{van_der_wolk_dust_2010}.
Our corresponding upper limit on the nuclear SIC flux is several times higher than that given by \cite{van_der_wolk_dust_2010}.
However, the most constraining upper limit is given by the \spitzerr spectrophotometry.
We, thus, use this for the upper limit of the nuclear 12$\,\mu$m continuum emission of 3C\,317.
\newline\end{@twocolumnfalse}]

\begin{figure}
   \centering
   \includegraphics[angle=0,width=8.500cm]{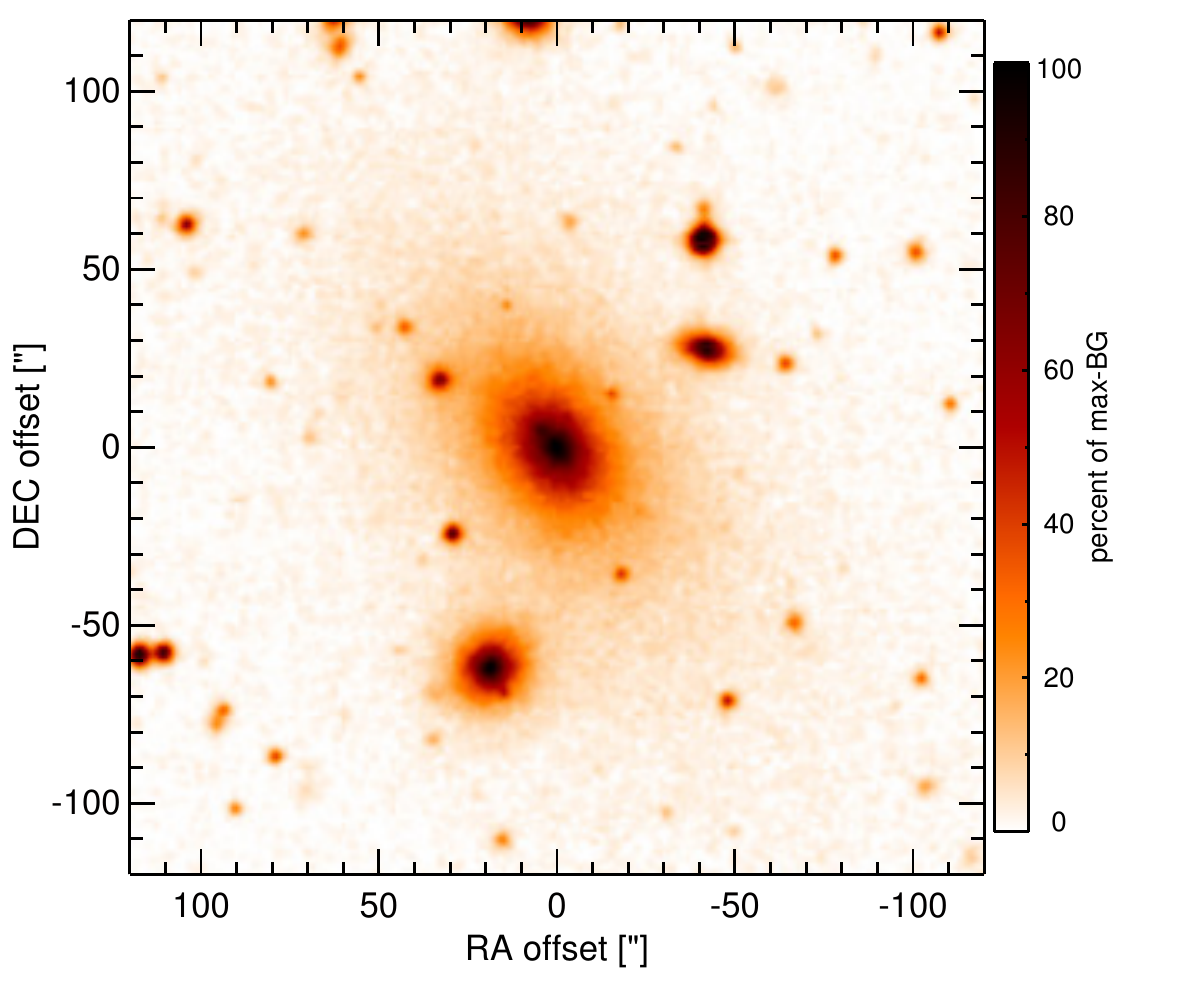}
    \caption{\label{fig:OPTim_3C317}
             Optical image (DSS, red filter) of 3C\,317. Displayed are the central $4\arcmin$ with North up and East to the left. 
              The colour scaling is linear with white corresponding to the median background and black to the $0.01\%$ pixels with the highest intensity.  
           }
\end{figure}
\begin{figure}
   \centering
   \includegraphics[angle=0,height=3.11cm]{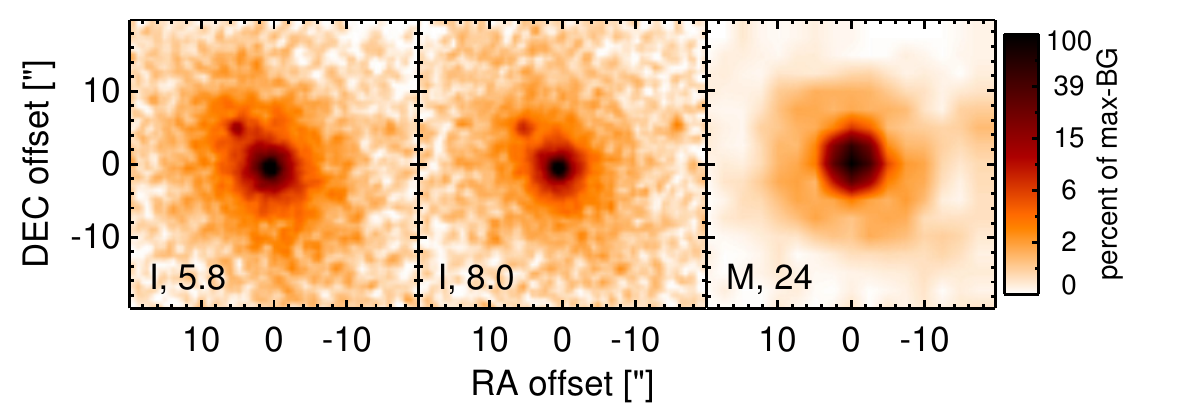}
    \caption{\label{fig:INTim_3C317}
             \spitzerr MIR images of 3C\,317. Displayed are the inner $40\arcsec$ with North up and East to the left. The colour scaling is logarithmic with white corresponding to median background and black to the $0.1\%$ pixels with the highest intensity.
             The label in the bottom left states instrument and central wavelength of the filter in $\mu$m (I: IRAC, M: MIPS). 
           }
\end{figure}
\begin{figure}
   \centering
   \includegraphics[angle=0,width=8.50cm]{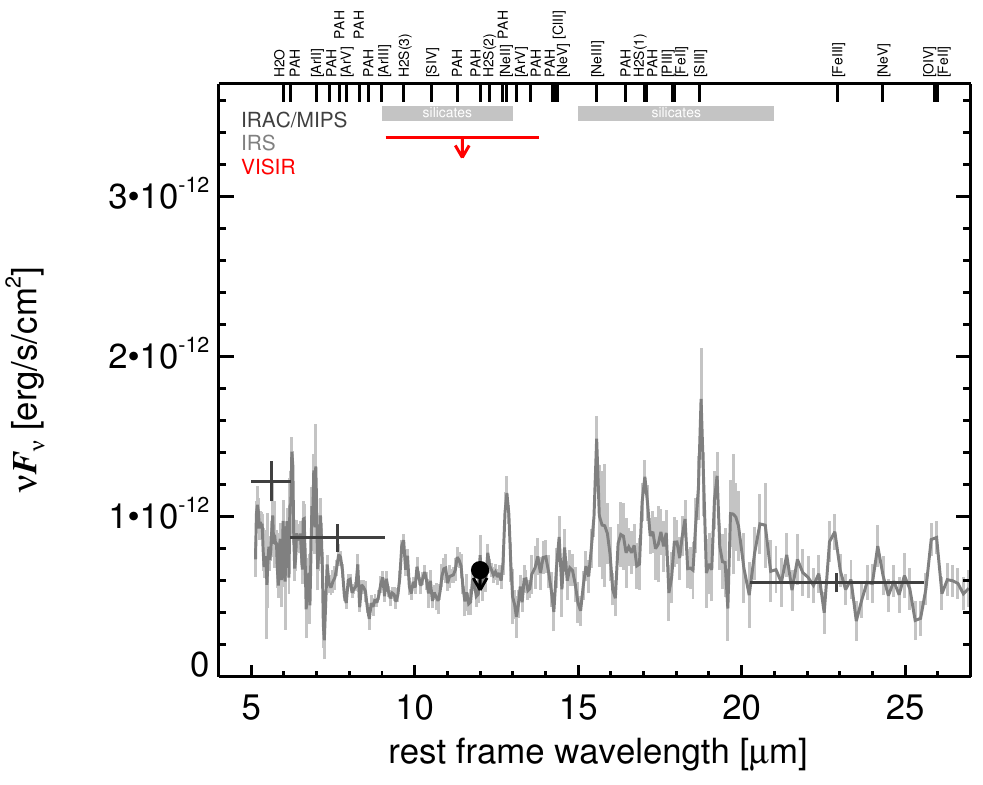}
   \caption{\label{fig:MISED_3C317}
      MIR SED of 3C\,317. The description  of the symbols (if present) is the following.
      Grey crosses and  solid lines mark the \spitzer/IRAC, MIPS and IRS data. 
      The colour coding of the other symbols is: 
      green for COMICS, magenta for Michelle, blue for T-ReCS and red for VISIR data.
      Darker-coloured solid lines mark spectra of the corresponding instrument.
      The black filled circles mark the nuclear 12 and $18\,\mu$m  continuum emission estimate from the data.
      The ticks on the top axis mark positions of common MIR emission lines, while the light grey horizontal bars mark wavelength ranges affected by the silicate 10 and 18$\mu$m features.     
   }
\end{figure}
\clearpage

\twocolumn[\begin{@twocolumnfalse}  
\subsection{3C\,321 -- IRAS\,15295+2414 -- LEDA\,55317}\label{app:3C321}
3C\,321 is a  FR\,II radio source coinciding with the peculiar galaxy LEDA\,55317 at a redshift of $z=$ 0.0961 ($D \sim 460$\,Mpc).
It contains two apparent nuclei at optical wavelengths with a projected nuclear separation of $\sim 3.5\arcsec$ ($\sim 6.5\,$kpc) in the south-east direction (PA$\sim-54\degree$; \citealt{roche_optical/ultraviolet_2000}).
Note that, in addition, there is a foreground star $\sim12\arcsec$ to the north of the system.
The compact radio nucleus can be associated with the south-eastern component \citep{baum_extended_1988}, which is obscured by a dust lane along a north-eastern axis in the optical (PA$\sim60\degree$; \citep{hurt_ultraviolet_1999,martel_hubble_1999}.
It is optically classified as a Sy\,2 \citep{filippenko_gravitational_1987} and exhibits polarized broad emission lines \citep{young_scattered_1996}.
It is unclear, whether the north-western component is another galaxy with its own AGN, as argued by \cite{filippenko_gravitational_1987}, \cite{roche_optical/ultraviolet_2000} and \cite{evans_radio_2008}, or if the two apparent nuclei are in fact scattered light originating from a true nucleus in between them, as argued by \cite{draper_optical_1993},  \cite{young_scattered_1996} and \cite{hurt_ultraviolet_1999}.
3C\,321 features supergalactic scale radio lobes along the north-west axis separated by $\sim500\,$kpc with a jet emanating from the south-eastern nucleus to the north-western radio hot spot  (PA$\sim-45\degree$; e.g., \citealt{leahy_bridges_1984,baum_extended_1988}).

The first successful MIR $N$-band observation was with ISOCAM \citep{siebenmorgen_isocam_2004}.
\spitzer/IRAC IRS and MIPS observations followed. 
3C\,321 appears only marginally resolved in the IRAC $5.8$ and $8.0\,\mu$m and MIPS $24\,\mu$m images with the compact south-eastern nucleus dominating and a slight extension towards the north-western nucleus, parallel to the jet axis.
Our nuclear MIPS 24\,$\mu$m photometry agrees with the value published in \cite{dicken_origin_2010}.
The IRS LR staring spectrum shows strong silicate $10\,\mu$m absorption, very weak PAH emission, strong forbidden high-ionization lines, and a red spectral slope in $\nu F_\nu$-space  (see also \citealt{haas_spitzer_2005,shi_aromatic_2007}). 
Therefore, star formation appears to be weak in the central few kiloparsecs of 3C\,321, which would be atypical for a galaxy merger.
We observed 3C\,321 with COMICS in the N11.7 filter in 2009 and weakly detected one compact nucleus, which appears slightly elongated (PA$\sim90\degree$; FWHM $\sim 1.3$\,kpc).
However, without at least a second epoch of deep subarcsecond MIR imaging this extension remains unverified.
We identify the detected nucleus with the south-eastern component, which is also dominating the \spitzerr data. 
The corresponding nuclear N11.7 flux matches the \spitzerr spectrophotometry, but it would be significantly lower if the presence of subarcsecond-extended emission can be verified.
For now, the IRS spectrum is used to compute the nuclear 12$\,\mu$m continuum emission estimate corrected for the silicate absorption feature.
The resulting flux is then scaled by half of the ratio between $\Fpsf$ and $\Fgau$ to account for the possible nuclear extension.
No sign of the north-western nucleus was detected in the COMICS image, which, therefore, has to be at least two times fainter.
\newline\end{@twocolumnfalse}]

\begin{figure}
   \centering
   \includegraphics[angle=0,width=8.500cm]{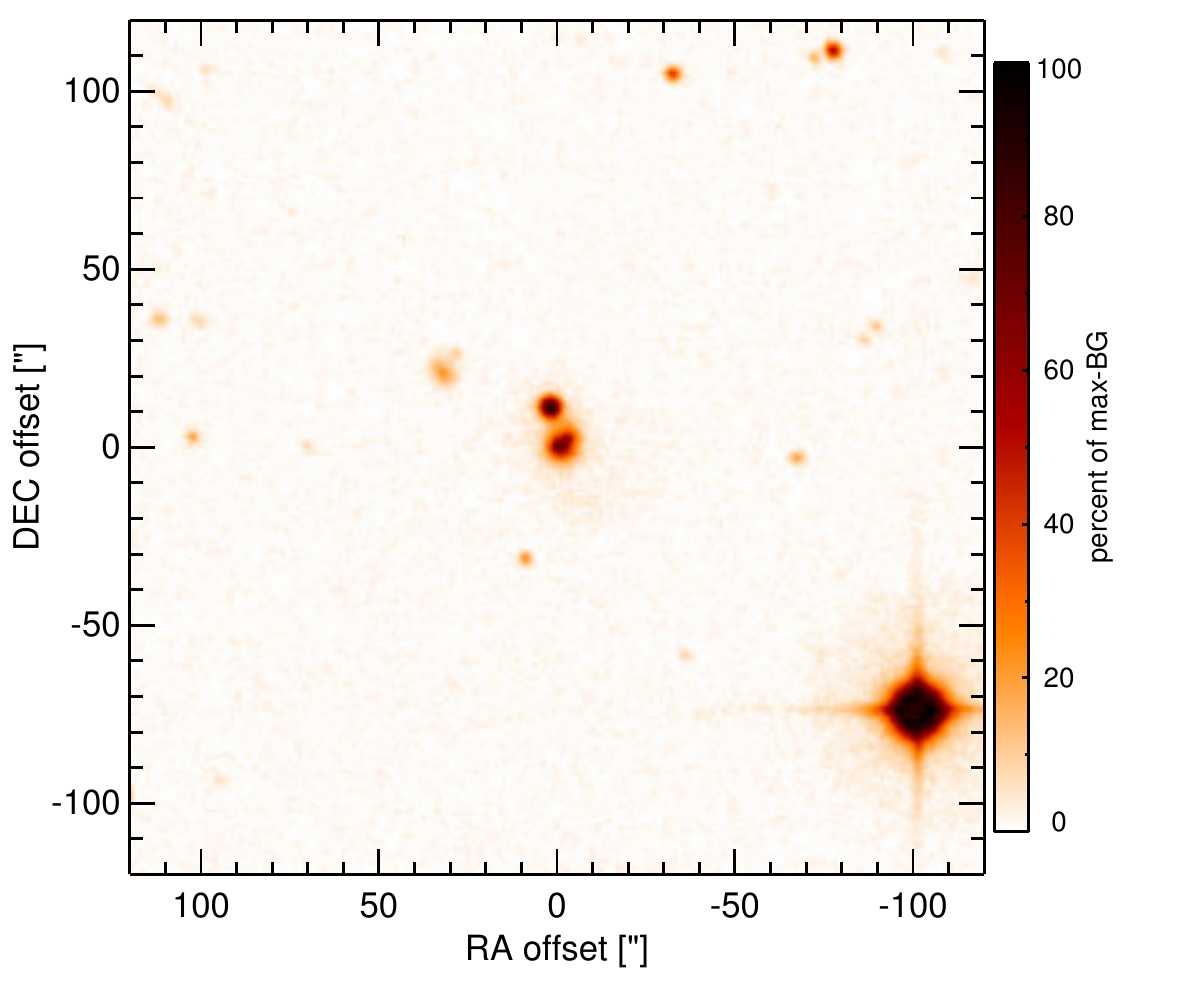}
    \caption{\label{fig:OPTim_3C321}
             Optical image (DSS, red filter) of 3C\,321. Displayed are the central $4\arcmin$ with North up and East to the left. 
              The colour scaling is linear with white corresponding to the median background and black to the $0.01\%$ pixels with the highest intensity.  
           }
\end{figure}
\begin{figure}
   \centering
   \includegraphics[angle=0,height=3.11cm]{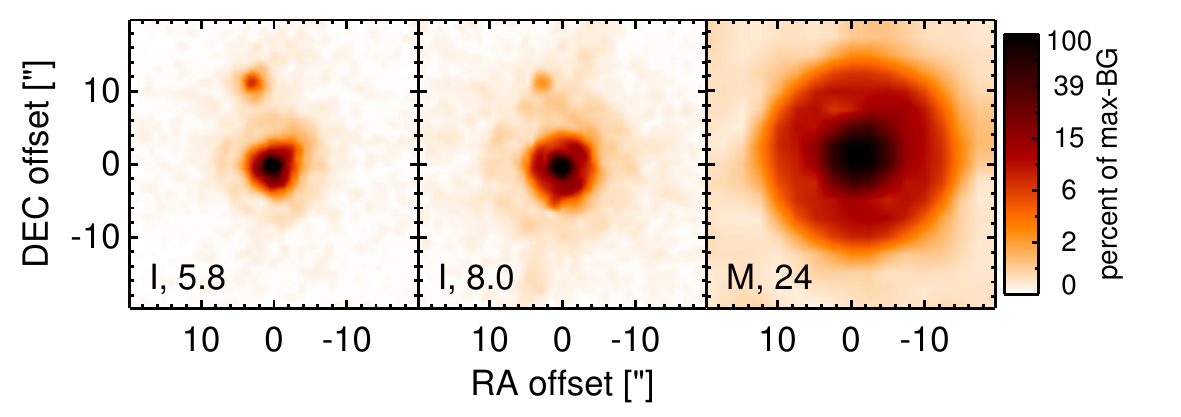}
    \caption{\label{fig:INTim_3C321}
             \spitzerr MIR images of 3C\,321. Displayed are the inner $40\arcsec$ with North up and East to the left. The colour scaling is logarithmic with white corresponding to median background and black to the $0.1\%$ pixels with the highest intensity.
             The label in the bottom left states instrument and central wavelength of the filter in $\mu$m (I: IRAC, M: MIPS). 
           }
\end{figure}
\begin{figure}
   \centering
   \includegraphics[angle=0,height=3.11cm]{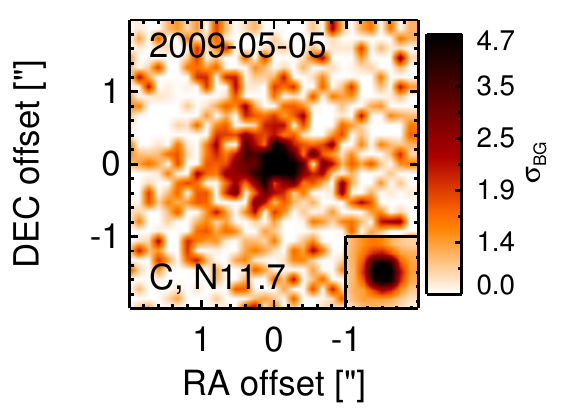}
    \caption{\label{fig:HARim_3C321}
             Subarcsecond-resolution MIR images of 3C\,321 sorted by increasing filter wavelength. 
             Displayed are the inner $4\arcsec$ with North up and East to the left. 
             The colour scaling is logarithmic with white corresponding to median background and black to the $75\%$ of the highest intensity of all images in units of $\sigbg$.
             The inset image shows the central arcsecond of the PSF from the calibrator star, scaled to match the science target.
             The labels in the bottom left state instrument and filter names (C: COMICS, M: Michelle, T: T-ReCS, V: VISIR).
           }
\end{figure}
\begin{figure}
   \centering
   \includegraphics[angle=0,width=8.50cm]{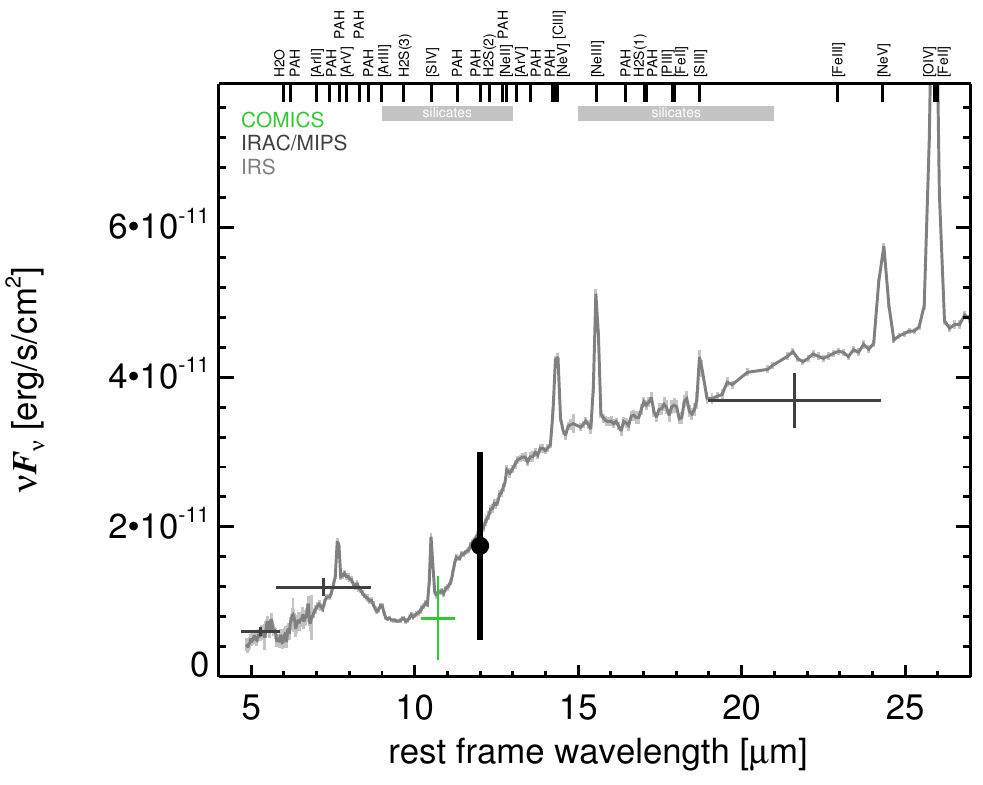}
   \caption{\label{fig:MISED_3C321}
      MIR SED of 3C\,321. The description  of the symbols (if present) is the following.
      Grey crosses and  solid lines mark the \spitzer/IRAC, MIPS and IRS data. 
      The colour coding of the other symbols is: 
      green for COMICS, magenta for Michelle, blue for T-ReCS and red for VISIR data.
      Darker-coloured solid lines mark spectra of the corresponding instrument.
      The black filled circles mark the nuclear 12 and $18\,\mu$m  continuum emission estimate from the data.
      The ticks on the top axis mark positions of common MIR emission lines, while the light grey horizontal bars mark wavelength ranges affected by the silicate 10 and 18$\mu$m features.     
   }
\end{figure}
\clearpage

\twocolumn[\begin{@twocolumnfalse}  
\subsection{3C\,327 -- IRAS\,15599+0206 -- LEDA\,56797}\label{app:3C327}
3C\,327 is a FR\,II radio source identified with the flattened elliptical galaxy LEDA\,56797 at a redshift of $z=$ 0.1048 ($D \sim 505$\,Mpc) with a Sy\,1 nucleus \citep{veron-cetty_catalogue_2010}. 
3C\,327 features supergalactic-scale double radio lobes in the east-west directions without prominent jets (PA$\sim100\degree$; e.g., \citealt{baum_extended_1988,leahy_study_1997}).
3C\,327 was first detected in the MIR with \iras, and was followed up with \spitzer/IRS and MIPS observations.
A point-like nucleus was detected in the MIPS $24\,\mu$m image.
Our corresponding nuclear MIPS $24\,\mu$m flux agrees with the literature values by  \cite{shi_far-infrared_2005} and \cite{dicken_origin_2008}.
The IRS LR staring-mode spectrum possibly shows weak silicate $10\,\mu$m  absorption, strong forbidden high ionization lines and a red spectral slope in $\nu F_\nu$-space but no PAH emission (see also \citealt{dicken_spitzer_2012}). 
Silicate absorption would be atypical for a type~I AGN.
3C\,327 was imaged with VISIR in the broad SIC filter in 2005, and a compact nucleus was weakly detected \citep{van_der_wolk_dust_2010}.
It appears extended in the image (FWHM $\sim 1.2$\,kpc) but a second epoch of deep subarcsecond MIR imaging is required to verify this extension.
\cite{van_der_wolk_dust_2010} conclude that the observed MIR emission must be dominated by thermal dust emission because of the high excess above the predicted synchrotron emission from extrapolating the radio wavelengths.  
Our nuclear SIC flux is $20\%$ lower than theirs but matches the \spitzerr spectrophotometry.
\newline\end{@twocolumnfalse}]

\begin{figure}
   \centering
   \includegraphics[angle=0,width=8.500cm]{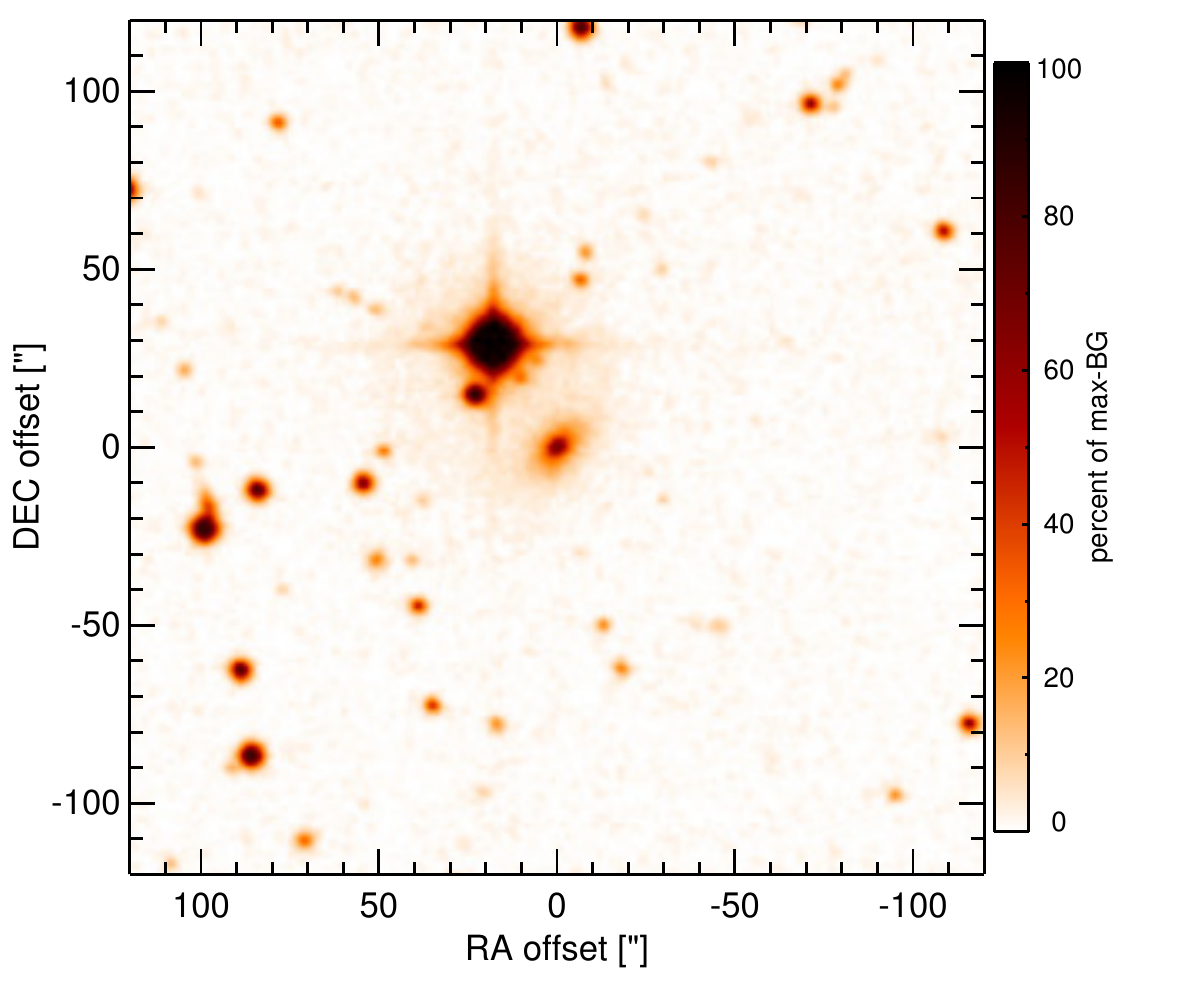}
    \caption{\label{fig:OPTim_3C327}
             Optical image (DSS, red filter) of 3C\,327. Displayed are the central $4\arcmin$ with North up and East to the left. 
              The colour scaling is linear with white corresponding to the median background and black to the $0.01\%$ pixels with the highest intensity.  
           }
\end{figure}
\begin{figure}
   \centering
   \includegraphics[angle=0,height=3.11cm]{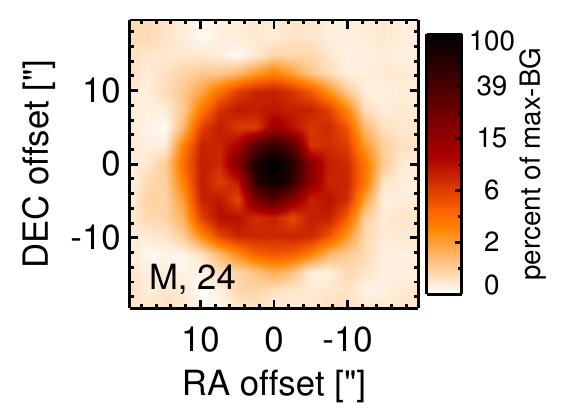}
    \caption{\label{fig:INTim_3C327}
             \spitzerr MIR images of 3C\,327. Displayed are the inner $40\arcsec$ with North up and East to the left. The colour scaling is logarithmic with white corresponding to median background and black to the $0.1\%$ pixels with the highest intensity.
             The label in the bottom left states instrument and central wavelength of the filter in $\mu$m (I: IRAC, M: MIPS). 
           }
\end{figure}
\begin{figure}
   \centering
   \includegraphics[angle=0,height=3.11cm]{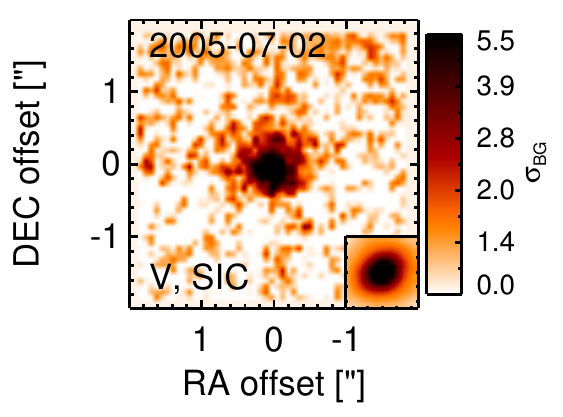}
    \caption{\label{fig:HARim_3C327}
             Subarcsecond-resolution MIR images of 3C\,327 sorted by increasing filter wavelength. 
             Displayed are the inner $4\arcsec$ with North up and East to the left. 
             The colour scaling is logarithmic with white corresponding to median background and black to the $75\%$ of the highest intensity of all images in units of $\sigbg$.
             The inset image shows the central arcsecond of the PSF from the calibrator star, scaled to match the science target.
             The labels in the bottom left state instrument and filter names (C: COMICS, M: Michelle, T: T-ReCS, V: VISIR).
           }
\end{figure}
\begin{figure}
   \centering
   \includegraphics[angle=0,width=8.50cm]{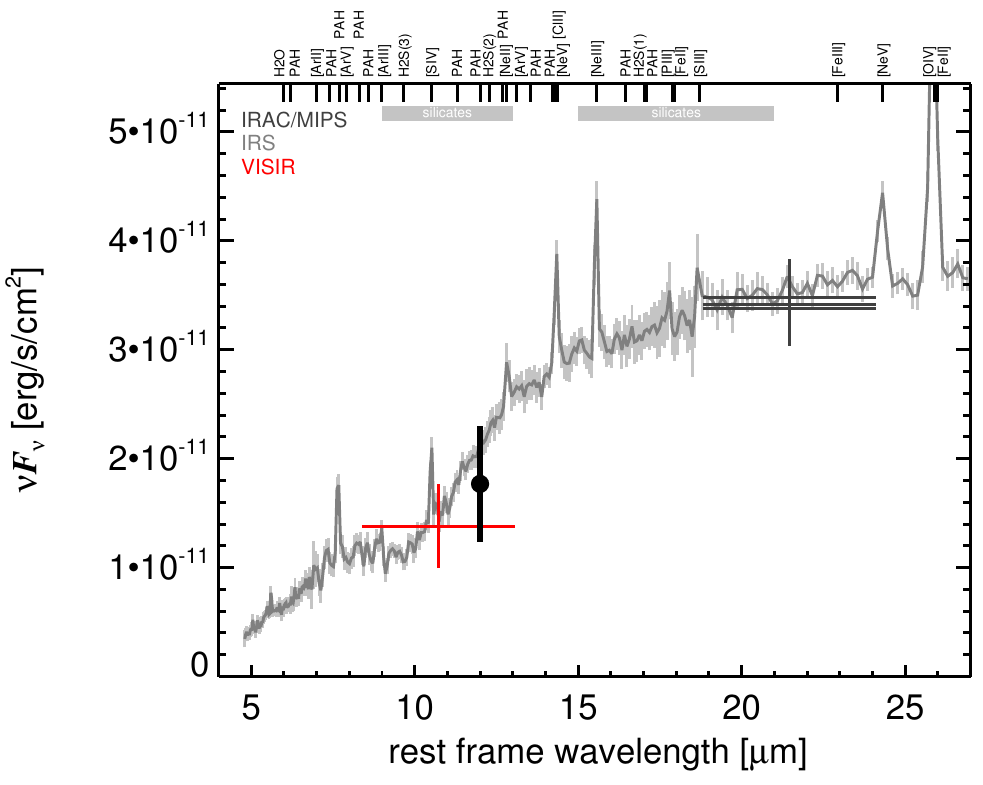}
   \caption{\label{fig:MISED_3C327}
      MIR SED of 3C\,327. The description  of the symbols (if present) is the following.
      Grey crosses and  solid lines mark the \spitzer/IRAC, MIPS and IRS data. 
      The colour coding of the other symbols is: 
      green for COMICS, magenta for Michelle, blue for T-ReCS and red for VISIR data.
      Darker-coloured solid lines mark spectra of the corresponding instrument.
      The black filled circles mark the nuclear 12 and $18\,\mu$m  continuum emission estimate from the data.
      The ticks on the top axis mark positions of common MIR emission lines, while the light grey horizontal bars mark wavelength ranges affected by the silicate 10 and 18$\mu$m features.     
   }
\end{figure}
\clearpage

\twocolumn[\begin{@twocolumnfalse}  
\subsection{3C\,353 -- LEDA\,60102}\label{app:3C353}
3C\,353 is a FR\,II radio source identified with the giant elliptical galaxy LEDA\,60102 \citep{madrid_hubble_2006} at a redshift of $z=$ 0.0304 ($D \sim 138$\,Mpc) with a Sy\,2 nucleus \citep{veron-cetty_catalogue_2010}.
It features supergalactic-scale biconical radio lobes in the east-west directions with both jets being visible (PA$\sim85\degree$; e.g., \citealt{morganti_radio_1993,swain_internal_1998}).
3C\,353 was not detected with \irass and appears as a weak compact source in the \wisee images.
In addition, an, to our knowledge, unpublished \spitzer/IRS LR mapping-mode spectrum is available but suffers from a very low S/N.
No spectral features can be discerned clearly apart from a shallow emission peak around $\sim 18\,\mu$m in $\nu F_\nu$-space (and possibly silicate emission). 
\cite{van_der_wolk_dust_2010} performed VISIR imaging in the SIC filter and claimed a non-detection with a flux upper limit of 4\,mJy.
However, in our reanalysis of the image, we weakly detect a compact emission source at the expected beam positions. 
No conclusion about the nuclear morphology can be drawn from this low S/N data, however.
Our corresponding nuclear SIC flux measurement of 8\,mJy matches the IRS flux levels as well as the \wisee band~3 flux.\newline\end{@twocolumnfalse}]

\begin{figure}
   \centering
   \includegraphics[angle=0,width=8.500cm]{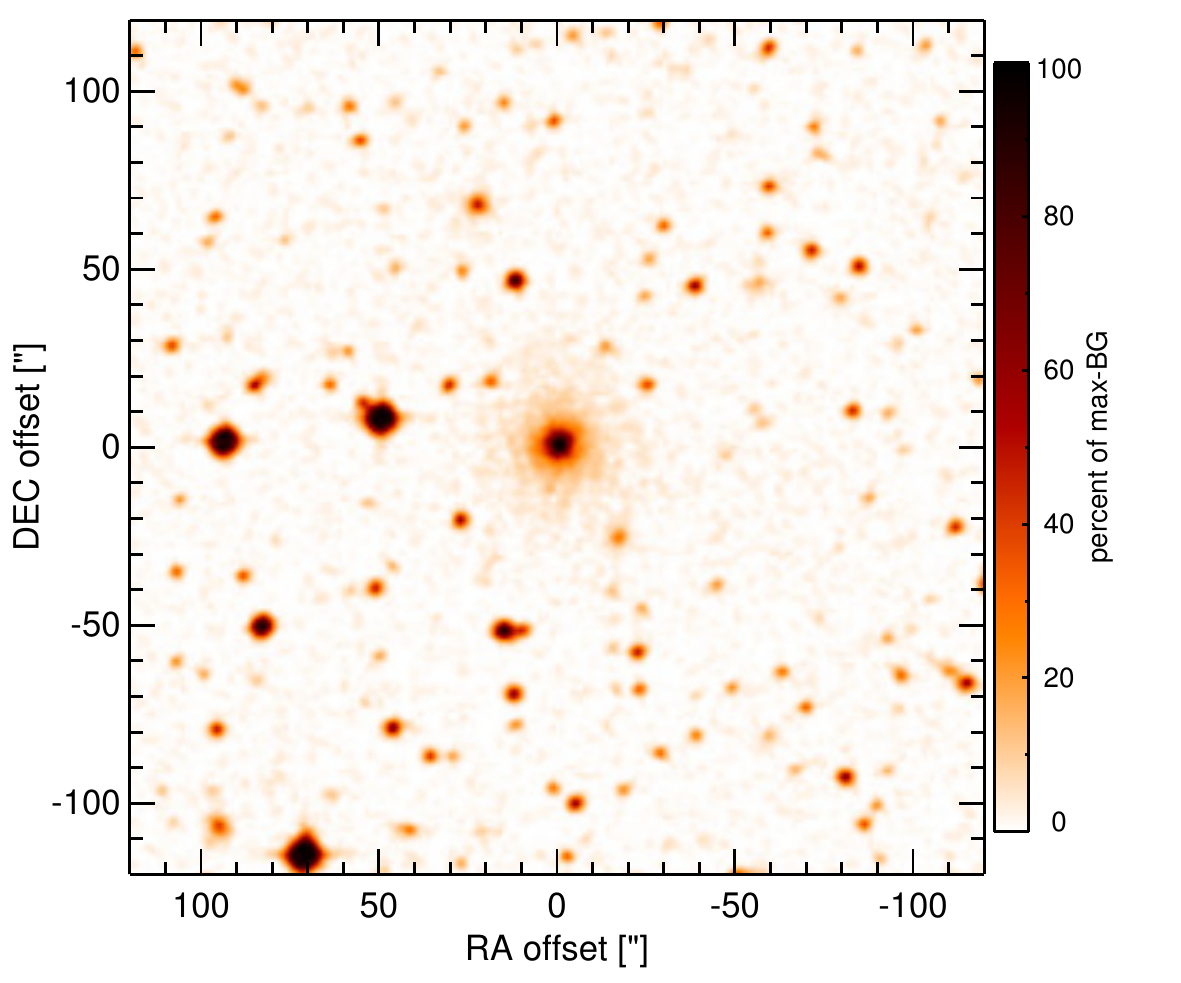}
    \caption{\label{fig:OPTim_3C353}
             Optical image (DSS, red filter) of 3C\,353. Displayed are the central $4\arcmin$ with North up and East to the left. 
              The colour scaling is linear with white corresponding to the median background and black to the $0.01\%$ pixels with the highest intensity.  
           }
\end{figure}
\begin{figure}
   \centering
   \includegraphics[angle=0,height=3.11cm]{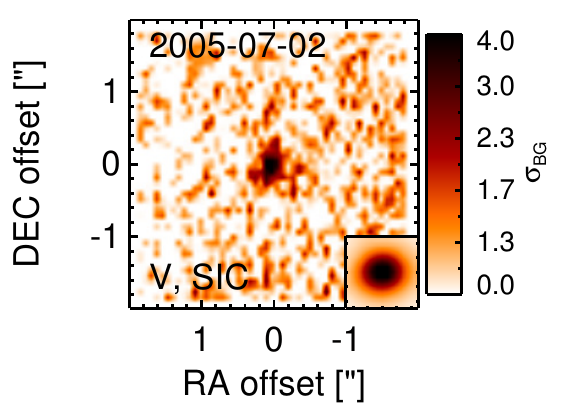}
    \caption{\label{fig:HARim_3C353}
             Subarcsecond-resolution MIR images of 3C\,353 sorted by increasing filter wavelength. 
             Displayed are the inner $4\arcsec$ with North up and East to the left. 
             The colour scaling is logarithmic with white corresponding to median background and black to the $75\%$ of the highest intensity of all images in units of $\sigbg$.
             The inset image shows the central arcsecond of the PSF from the calibrator star, scaled to match the science target.
             The labels in the bottom left state instrument and filter names (C: COMICS, M: Michelle, T: T-ReCS, V: VISIR).
           }
\end{figure}
\begin{figure}
   \centering
   \includegraphics[angle=0,width=8.50cm]{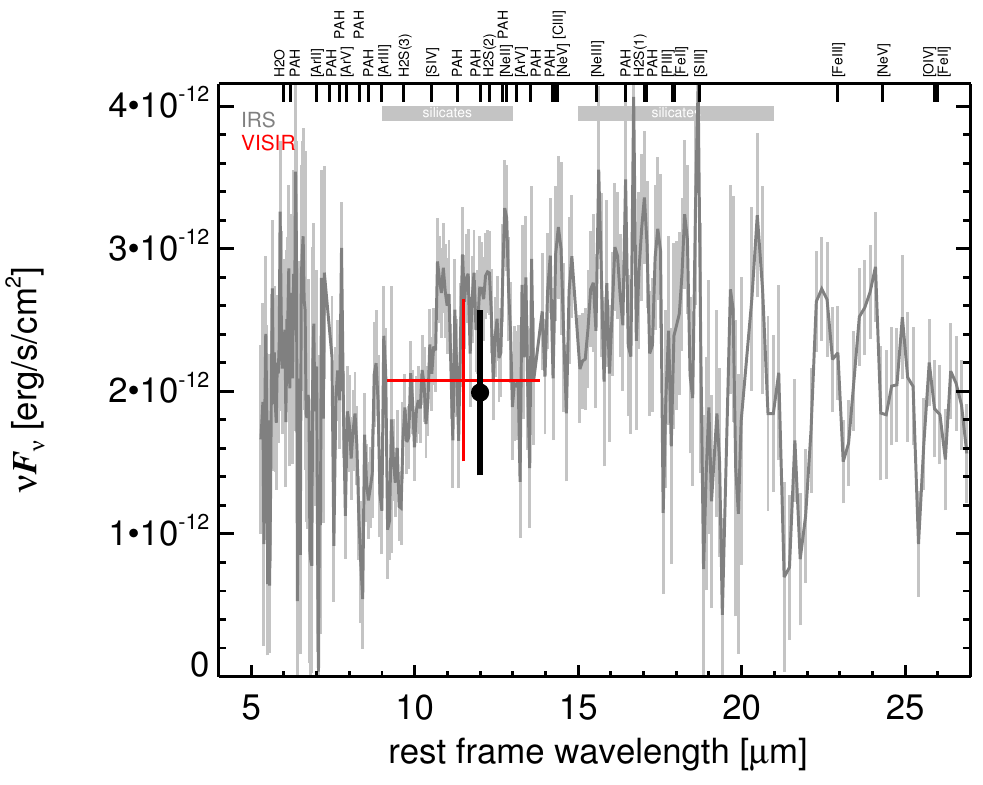}
   \caption{\label{fig:MISED_3C353}
      MIR SED of 3C\,353. The description  of the symbols (if present) is the following.
      Grey crosses and  solid lines mark the \spitzer/IRAC, MIPS and IRS data. 
      The colour coding of the other symbols is: 
      green for COMICS, magenta for Michelle, blue for T-ReCS and red for VISIR data.
      Darker-coloured solid lines mark spectra of the corresponding instrument.
      The black filled circles mark the nuclear 12 and $18\,\mu$m  continuum emission estimate from the data.
      The ticks on the top axis mark positions of common MIR emission lines, while the light grey horizontal bars mark wavelength ranges affected by the silicate 10 and 18$\mu$m features.     
   }
\end{figure}
\clearpage

\twocolumn[\begin{@twocolumnfalse}  
\subsection{3C\,382 -- Z\,173-14}\label{app:3C382}
3C\,382 is a flat-spectrum FR\,II radio source identified with the elliptical galaxy Z\,173-14 \citep{martel_hubble_1999} at a redshift of $z=$ 0.0579 ($D \sim 267$\,Mpc).
It contains an AGN with a Sy\,1 classification  \citep{veron-cetty_catalogue_2010} that is a member of the nine-month BAT AGN sample.
3C\,382 features supergalactic-scale biconical radio lobes with a prominent jet extending to the north-eastern lobe (PA$\sim50\degree$; e.g., \citealt{black_study_1992}).
The first $N$-band photometry of 3C\,382 was performed in 1980 
\citep{puschell_visual-infrared_1981}, followed by \cite{heckman_infrared_1983} and \cite{impey_infrared_1990}.
The MIR emission level has monotonically increased over the last few decades.
The \iso/ISOCAM photometry from 1996 provides a $\sim 50\%$ higher flux than that from 1980 \citep{clavel_2.5-11_2000,siebenmorgen_isocam_2004,ramos_almeida_mid-infrared_2007}.
The \spitzer/IRS, IRAC and MIPS data taken in 2005 again exhibits $\sim 50\%$ higher fluxes (see  \citealt{dicken_origin_2010} for MIPS $24\,\mu$m). 
3C\,382 is nearly unresolved in all \spitzerr images.
The IRS LR staring-mode spectrum shows strong silicate emission features and a blue spectral slope in $\nu F_\nu$-space but no PAH or other line emission  (see also \citealt{dicken_spitzer_2012}). 
Our COMICS N11.7 observation from 2009 shows an unresolved nuclear source, and the corresponding nuclear photometry is $\sim 4\%$ higher and, thus, is still consistent with the \spitzerr spectrophotometry.
Therefore, the silicate emission originates in the projected subarcsecond-scale nucleus, and thermal dust emission contributes significantly to the nuclear MIR emission.
We correct our $12\,\mu$m continuum emission estimate for the strong silicate emission using the IRS spectrum. 
\newline\end{@twocolumnfalse}]

\begin{figure}
   \centering
   \includegraphics[angle=0,width=8.500cm]{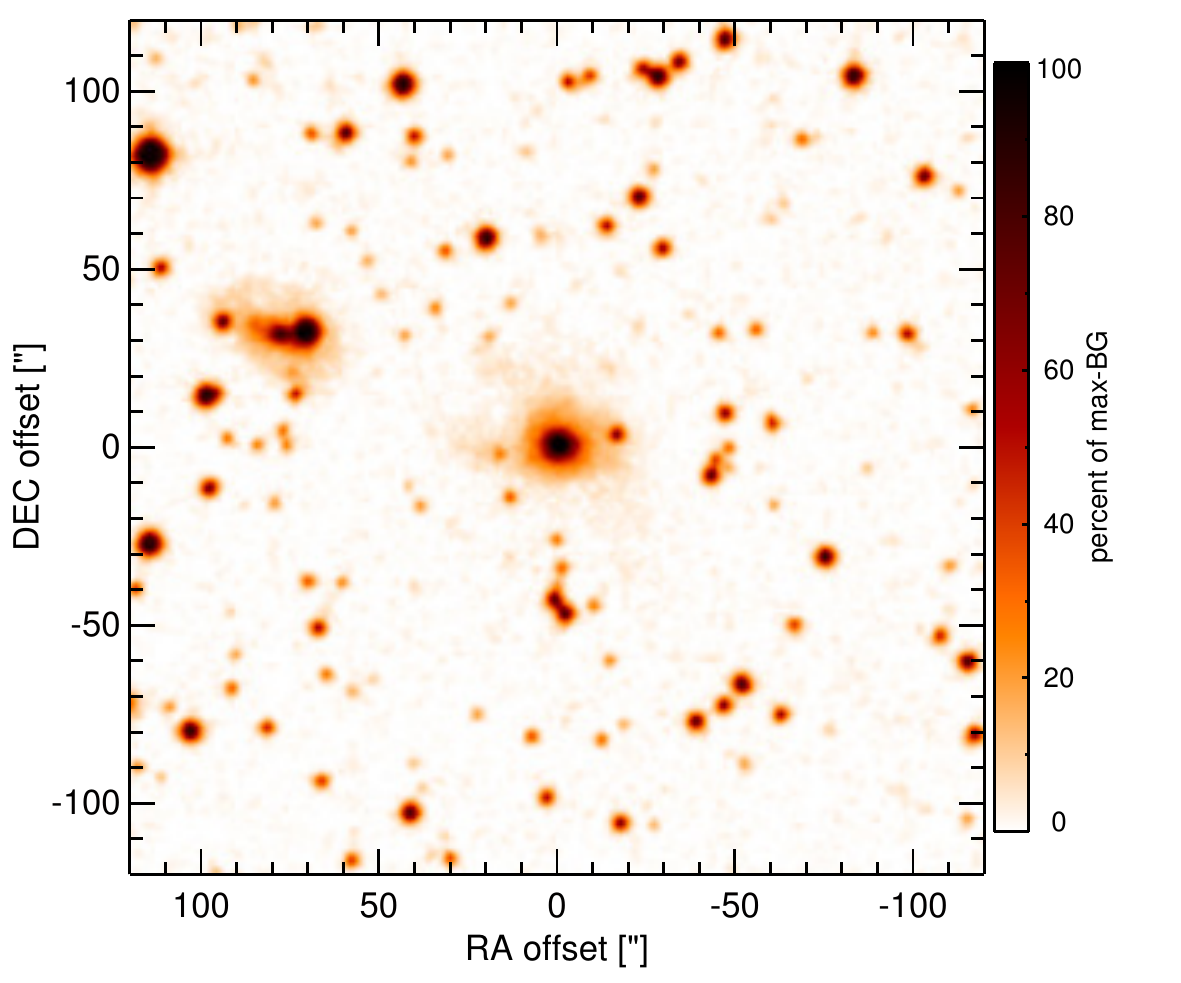}
    \caption{\label{fig:OPTim_3C382}
             Optical image (DSS, red filter) of 3C\,382. Displayed are the central $4\arcmin$ with North up and East to the left. 
              The colour scaling is linear with white corresponding to the median background and black to the $0.01\%$ pixels with the highest intensity.  
           }
\end{figure}
\begin{figure}
   \centering
   \includegraphics[angle=0,height=3.11cm]{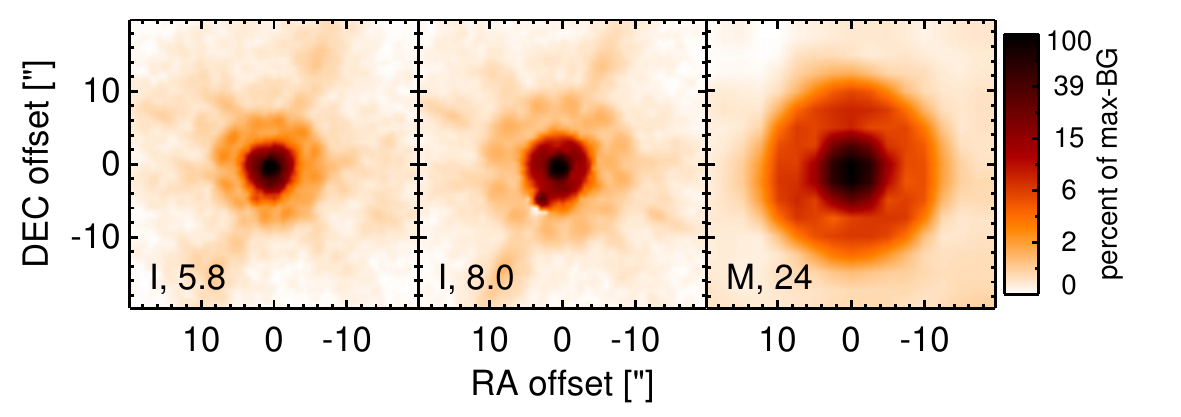}
    \caption{\label{fig:INTim_3C382}
             \spitzerr MIR images of 3C\,382. Displayed are the inner $40\arcsec$ with North up and East to the left. The colour scaling is logarithmic with white corresponding to median background and black to the $0.1\%$ pixels with the highest intensity.
             The label in the bottom left states instrument and central wavelength of the filter in $\mu$m (I: IRAC, M: MIPS). 
             Note that the apparent off-nuclear compact sources in the IRAC $5.8$ and $8.0\,\mu$m images are instrumental artefacts. 
           }
\end{figure}
\begin{figure}
   \centering
   \includegraphics[angle=0,height=3.11cm]{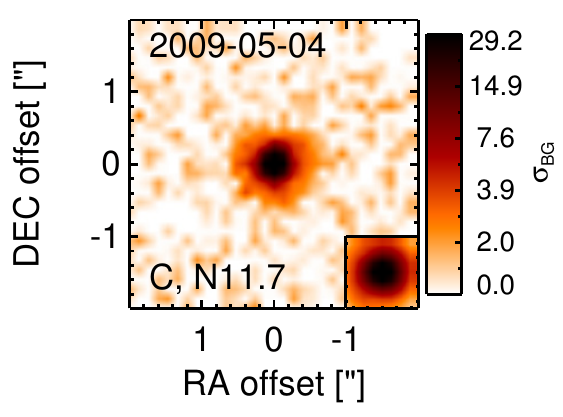}
    \caption{\label{fig:HARim_3C382}
             Subarcsecond-resolution MIR images of 3C\,382 sorted by increasing filter wavelength. 
             Displayed are the inner $4\arcsec$ with North up and East to the left. 
             The colour scaling is logarithmic with white corresponding to median background and black to the $75\%$ of the highest intensity of all images in units of $\sigbg$.
             The inset image shows the central arcsecond of the PSF from the calibrator star, scaled to match the science target.
             The labels in the bottom left state instrument and filter names (C: COMICS, M: Michelle, T: T-ReCS, V: VISIR).
           }
\end{figure}
\begin{figure}
   \centering
   \includegraphics[angle=0,width=8.50cm]{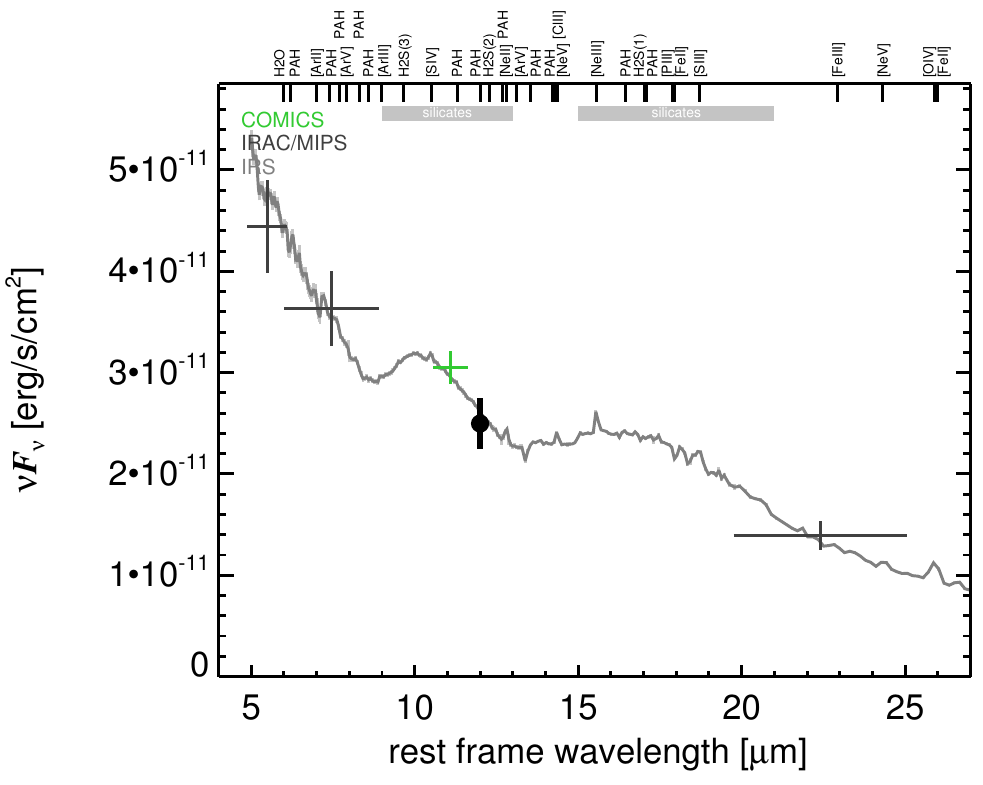}
   \caption{\label{fig:MISED_3C382}
      MIR SED of 3C\,382. The description  of the symbols (if present) is the following.
      Grey crosses and  solid lines mark the \spitzer/IRAC, MIPS and IRS data. 
      The colour coding of the other symbols is: 
      green for COMICS, magenta for Michelle, blue for T-ReCS and red for VISIR data.
      Darker-coloured solid lines mark spectra of the corresponding instrument.
      The black filled circles mark the nuclear 12 and $18\,\mu$m  continuum emission estimate from the data.
      The ticks on the top axis mark positions of common MIR emission lines, while the light grey horizontal bars mark wavelength ranges affected by the silicate 10 and 18$\mu$m features.     
   }
\end{figure}
\clearpage

\twocolumn[\begin{@twocolumnfalse}  
\subsection{3C\,390.3 -- VII\,Zw\,838}\label{app:3C390-3}
3C\,390.3 is a flat-spectrum FR\,II radio source identified with the elliptical galaxy VII\,Zw\,838 \citep{martel_hubble_1999} at a redshift of $z=$ 0.0561 ($D \sim 259$\,Mpc).
It contains a  variable AGN with optical Sy\,1.5 classification \citep{veron-cetty_catalogue_2010} and blazar characteristics that9 is also  a member of the nine-month BAT AGN sample.
3C\,390.3 features supergalactic-scale biconical radio lobes with a faint one-sided jet extending to the north-eastern lobe (PA$\sim145\degree$; e.g., \citealt{leahy_jets_1995}).
Early $N$-band imaging was performed with \isoo \citep{siebenmorgen_isocam_2004}.
3C\,390.3 was also observed with \spitzer/IRAC, IRS and MIPS and remains nearly unresolved in all images.
Our nuclear MIPS 24\,$\mu$m photometry agrees with the value given in  \cite{dicken_origin_2010}.
The IRS LR staring-mode spectrum shows silicate emission, only weak forbidden emission lines and an emission peak at $\sim19\,\mu$m in $\nu F_\nu$-space but no PAH features (see also \citealt{shi_9.7_2006,mullaney_defining_2011}). 
We observed 3C\,390.3 with COMICS in the N11.7 filter in 2009 and detected a compact source that appears to be resolved (FWHM $\sim0.5\arcsec\sim 0.5$\,kpc).
However, a second epoch of deep subarcsecond MIR imaging is required to verify this extension.
Notably, the \spitzerr flux levels are mismatched.
The MIPS ($24\,\mu$m) photometry and IRS spectrum from 2004 are lower than the IRAC ($5.8$ and $8.0\,\mu$m) photometry from 2008, providing one assumes a constant MIR SED shape.
Therefore, 3C\,390.3 appears to have increased its MIR emission in the last few years (similar to 3C\,382; see also \citealt{edelson_far-infrared_1987}).
For a better comparison with our COMICS N11.7 photometry, we scale the IRS spectrum according to the IRAC fluxes and find consistent flux levels between COMICS and \spitzer. 
This also indicates that the silicate emission arises in the subarcsecond-scale nucleus, and thus thermal dust must contribute significantly to the nuclear MIR emission.
Therefore, we use the IRS spectrum to compute the nuclear 12$\,\mu$m continuum emission estimate corrected for the silicate feature.
Note however, that the nuclear flux would be $\sim50\%$ lower than the \spitzerr spectrophotometry if the presence of subarcsecond-extended emission can be verified.
For now, the resulting synthetic flux is then scaled by half of the ratio between $\Fpsf$ and $\Fgau$ to account for the possibly nuclear extension.
\newline\end{@twocolumnfalse}]

\begin{figure}
   \centering
   \includegraphics[angle=0,width=8.500cm]{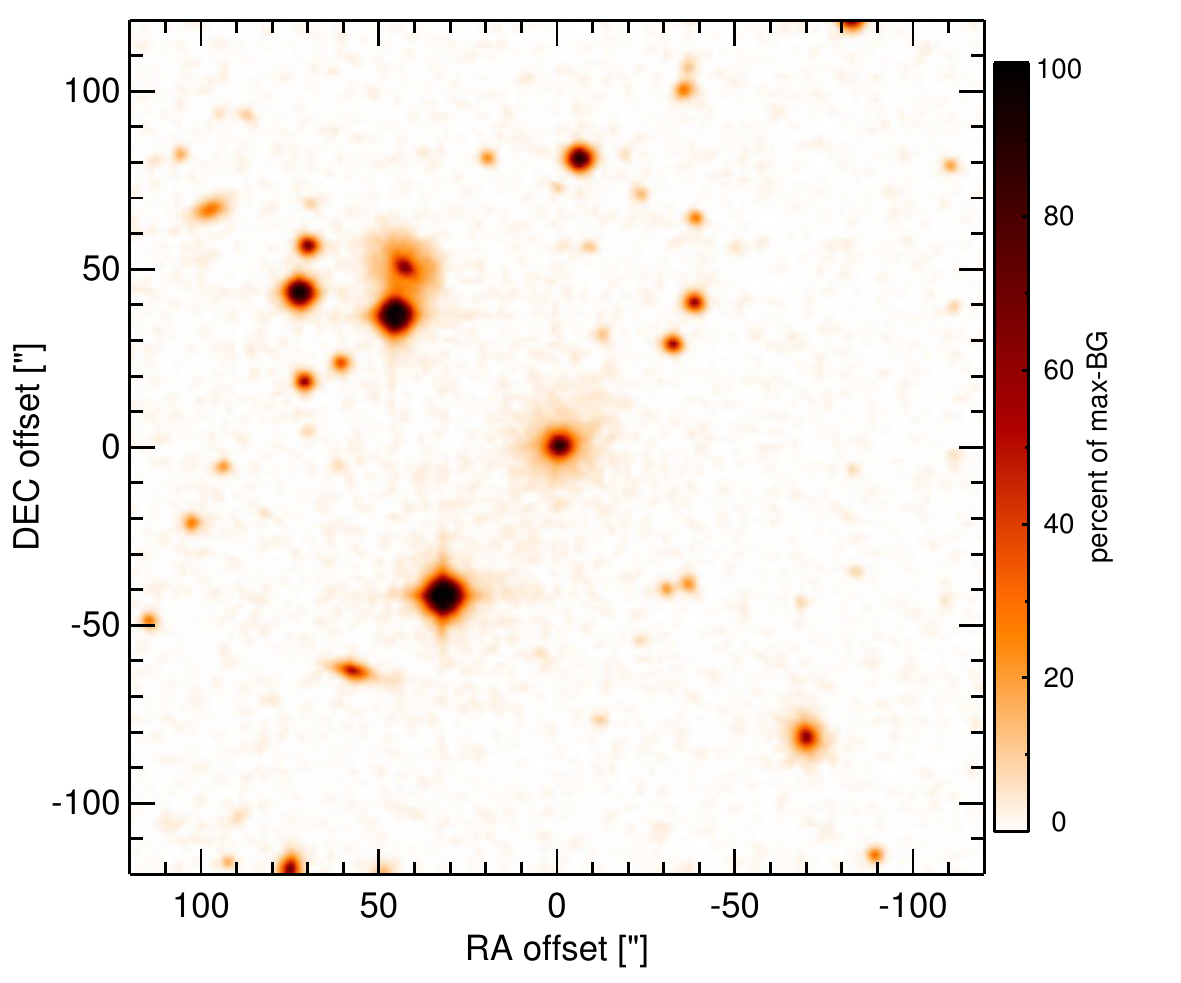}
    \caption{\label{fig:OPTim_3C390-3}
             Optical image (DSS, red filter) of 3C\,390.3. Displayed are the central $4\arcmin$ with North up and East to the left. 
              The colour scaling is linear with white corresponding to the median background and black to the $0.01\%$ pixels with the highest intensity.  
           }
\end{figure}
\begin{figure}
   \centering
   \includegraphics[angle=0,height=3.11cm]{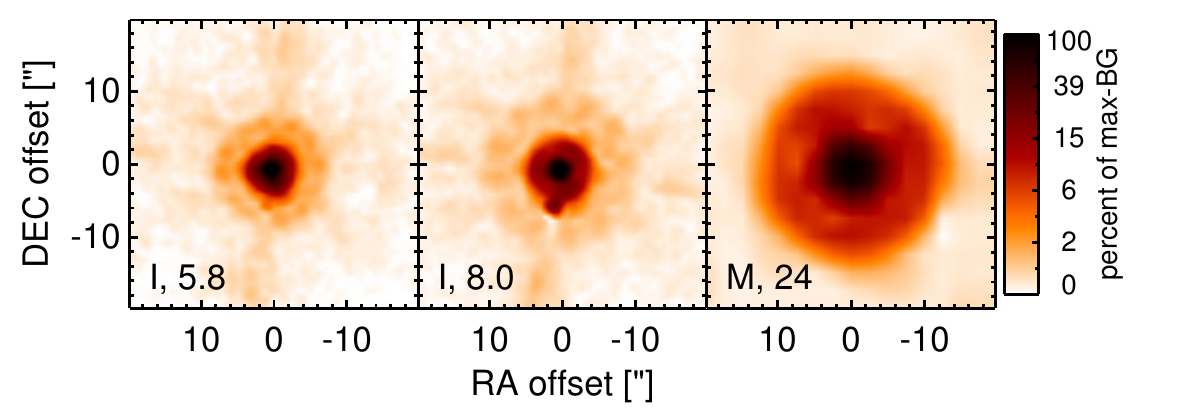}
    \caption{\label{fig:INTim_3C390-3}
             \spitzerr MIR images of 3C\,390.3. Displayed are the inner $40\arcsec$ with North up and East to the left. The colour scaling is logarithmic with white corresponding to median background and black to the $0.1\%$ pixels with the highest intensity.
             The label in the bottom left states instrument and central wavelength of the filter in $\mu$m (I: IRAC, M: MIPS).
             Note that the apparent off-nuclear compact sources in the IRAC $5.8$ and $8.0\,\mu$m images are instrumental artefacts. 
           }
\end{figure}
\begin{figure}
   \centering
   \includegraphics[angle=0,height=3.11cm]{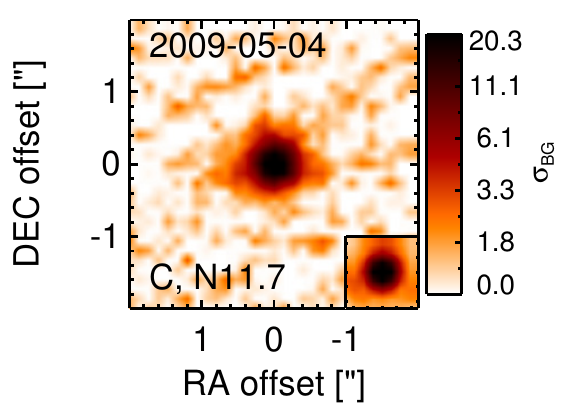}
    \caption{\label{fig:HARim_3C390-3}
             Subarcsecond-resolution MIR images of 3C\,390.3 sorted by increasing filter wavelength. 
             Displayed are the inner $4\arcsec$ with North up and East to the left. 
             The colour scaling is logarithmic with white corresponding to median background and black to the $75\%$ of the highest intensity of all images in units of $\sigbg$.
             The inset image shows the central arcsecond of the PSF from the calibrator star, scaled to match the science target.
             The labels in the bottom left state instrument and filter names (C: COMICS, M: Michelle, T: T-ReCS, V: VISIR).
           }
\end{figure}
\begin{figure}
   \centering
   \includegraphics[angle=0,width=8.50cm]{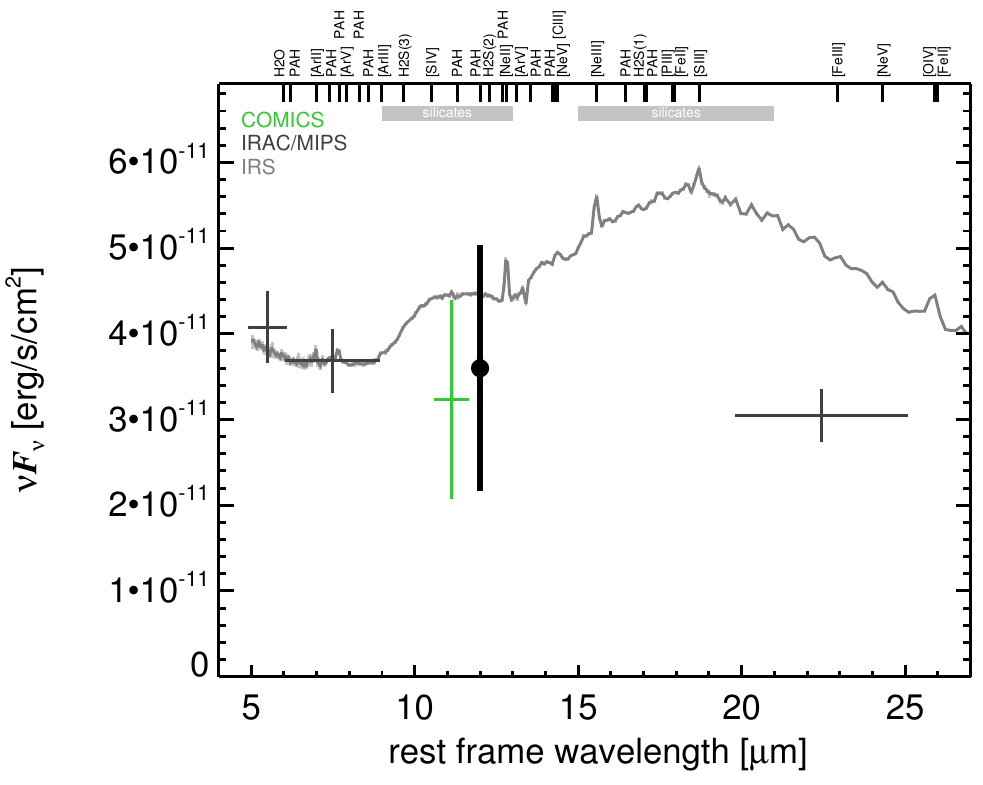}
   \caption{\label{fig:MISED_3C390-3}
      MIR SED of 3C\,390.3. The description  of the symbols (if present) is the following.
      Grey crosses and  solid lines mark the \spitzer/IRAC, MIPS and IRS data. 
      The colour coding of the other symbols is: 
      green for COMICS, magenta for Michelle, blue for T-ReCS and red for VISIR data.
      Darker-coloured solid lines mark spectra of the corresponding instrument.
      The black filled circles mark the nuclear 12 and $18\,\mu$m  continuum emission estimate from the data.
      The ticks on the top axis mark positions of common MIR emission lines, while the light grey horizontal bars mark wavelength ranges affected by the silicate 10 and 18$\mu$m features.     
   }
\end{figure}
\clearpage

\twocolumn[\begin{@twocolumnfalse}  
\subsection{3C\,403 -- IRAS\,F19497+0222 -- LEDA\,63758}\label{app:3C403}
3C\,403 is a FR\,II radio source identified with the early-type galaxy LEDA\,63758 \citep{martel_hubble_1999} at a redshift of $z=$ 0.059 ($D \sim 271$\,Mpc) with a Sy\,2 nucleus \citep{veron-cetty_catalogue_2010}.
It features a complex super-galactic-scale X-shaped radio morphology with a twin jet feeding biconical radio lobes (PA$\sim70\degree$; e.g., \citealt{black_study_1992}), and a nuclear water mega-maser \citep{tarchi_discovery_2003,tarchi_innermost_2007}.
After first being detected in the MIR with \iras, 3C\,403 was followed up with \spitzer/IRAC, IRS and MIPS and is only marginally resolved in all images.
Our nuclear MIPS (24\,$\mu$m) photometry agrees with the value given in  \cite{dicken_origin_2008}.
The IRS LR mapping-mode spectrum appears relatively featureless without significant silicate or PAH features but a curved red slope in $\nu F_\nu$-space.
A more carefully reduced version is published by \cite{dicken_spitzer_2012} and shows weak silicate emission, unexpected for a type~II AGN. 
\cite{van_der_wolk_dust_2010} performed VISIR imaging of 3C\,403 in the SIC filter in 2006 and detected a compact nucleus.
The observations were not diffraction-limited.
Thus, it remains unclear, whether the nucleus is resolved at subarcsecond resolution in the MIR.
Our nuclear SIC flux is consistent with \cite{van_der_wolk_dust_2010} and is $\sim 10\%$ lower than the \spitzerr spectrophotometry.
\newline\end{@twocolumnfalse}]

\begin{figure}
   \centering
   \includegraphics[angle=0,width=8.500cm]{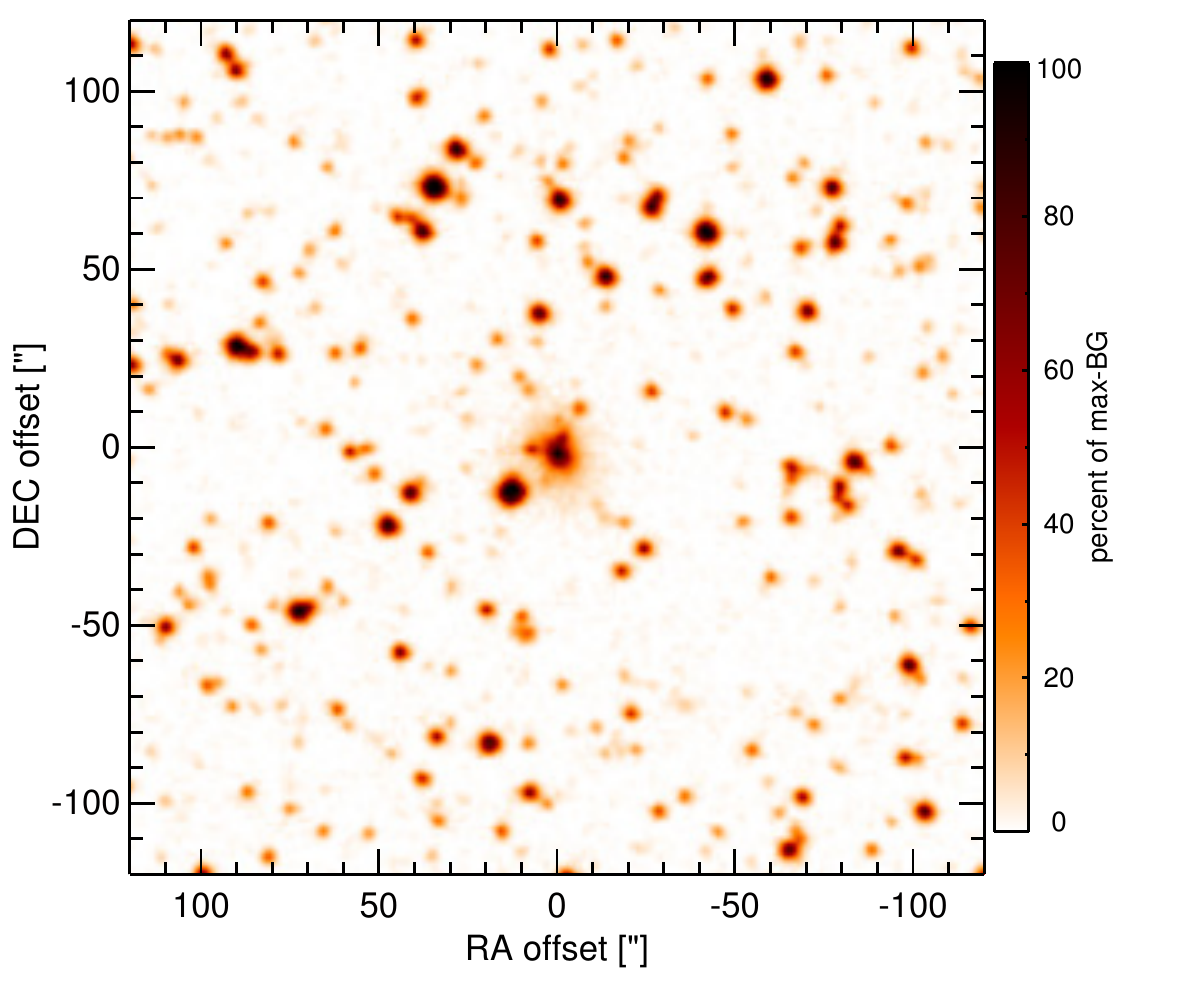}
    \caption{\label{fig:OPTim_3C403}
             Optical image (DSS, red filter) of 3C\,403. Displayed are the central $4\arcmin$ with North up and East to the left. 
              The colour scaling is linear with white corresponding to the median background and black to the $0.01\%$ pixels with the highest intensity.  
           }
\end{figure}
\begin{figure}
   \centering
   \includegraphics[angle=0,height=3.11cm]{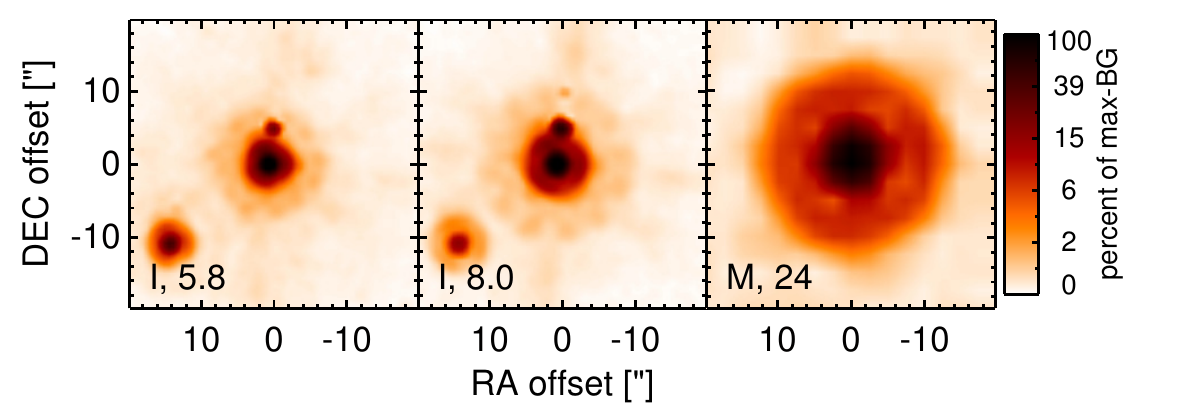}
    \caption{\label{fig:INTim_3C403}
             \spitzerr MIR images of 3C\,403. Displayed are the inner $40\arcsec$ with North up and East to the left. The colour scaling is logarithmic with white corresponding to median background and black to the $0.1\%$ pixels with the highest intensity.
             The label in the bottom left states instrument and central wavelength of the filter in $\mu$m (I: IRAC, M: MIPS). 
             Note that the apparent off-nuclear compact sources to the north in the IRAC $5.8$ and $8.0\,\mu$m images are instrumental artefacts. 
           }
\end{figure}
\begin{figure}
   \centering
   \includegraphics[angle=0,height=3.11cm]{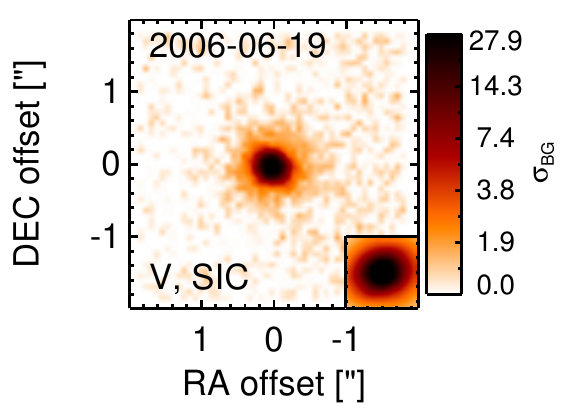}
    \caption{\label{fig:HARim_3C403}
             Subarcsecond-resolution MIR images of 3C\,403 sorted by increasing filter wavelength. 
             Displayed are the inner $4\arcsec$ with North up and East to the left. 
             The colour scaling is logarithmic with white corresponding to median background and black to the $75\%$ of the highest intensity of all images in units of $\sigbg$.
             The inset image shows the central arcsecond of the PSF from the calibrator star, scaled to match the science target.
             The labels in the bottom left state instrument and filter names (C: COMICS, M: Michelle, T: T-ReCS, V: VISIR).
           }
\end{figure}
\begin{figure}
   \centering
   \includegraphics[angle=0,width=8.50cm]{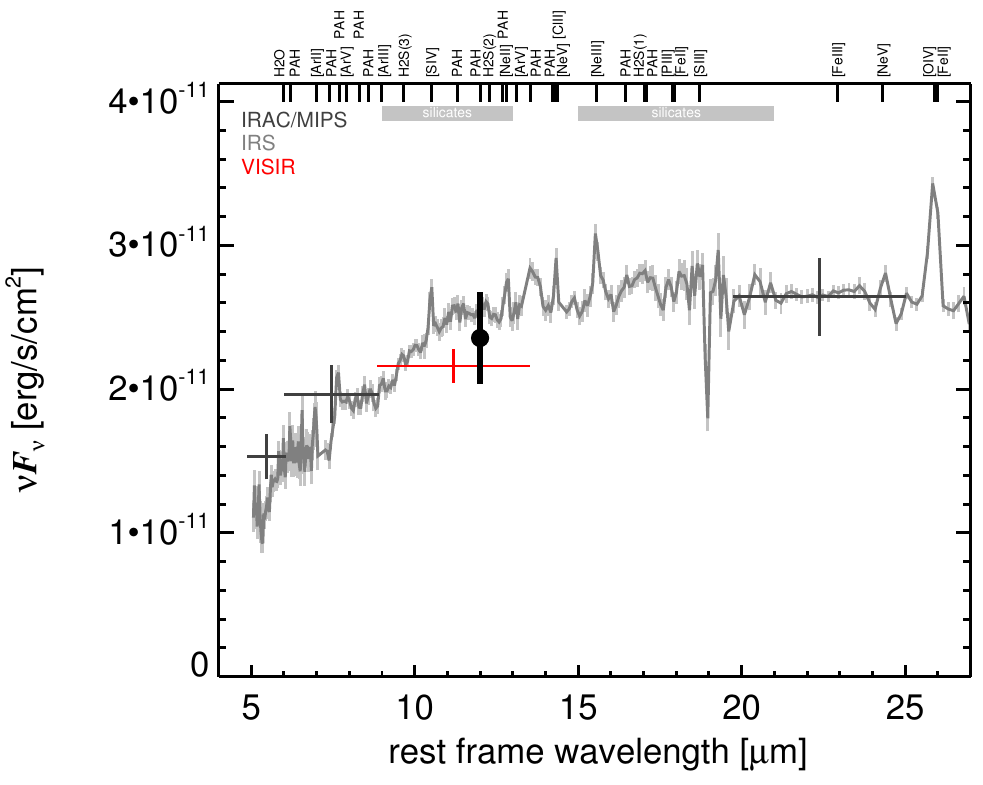}
   \caption{\label{fig:MISED_3C403}
      MIR SED of 3C\,403. The description  of the symbols (if present) is the following.
      Grey crosses and  solid lines mark the \spitzer/IRAC, MIPS and IRS data. 
      The colour coding of the other symbols is: 
      green for COMICS, magenta for Michelle, blue for T-ReCS and red for VISIR data.
      Darker-coloured solid lines mark spectra of the corresponding instrument.
      The black filled circles mark the nuclear 12 and $18\,\mu$m  continuum emission estimate from the data.
      The ticks on the top axis mark positions of common MIR emission lines, while the light grey horizontal bars mark wavelength ranges affected by the silicate 10 and 18$\mu$m features.     
   }
\end{figure}
\clearpage

\twocolumn[\begin{@twocolumnfalse}  
\subsection{3C\,424 -- PKS\,2045+06 -- LEDA\,65411}\label{app:3C424}
3C\,424 is a borderline FR\,I/II radio source identified with the early-type  galaxy LEDA\,65411 at a redshift of $z=$ 0.127 ($D \sim 613$\,Mpc) with a Sy\,2 nucleus \citep{veron-cetty_catalogue_2010}.
It features an unusual super-galactic scale radio morphology with faint smooth lobes and faint jets along the north-south axis (PA$\sim -25\degree$; e.g., \citealt{black_study_1992}). 
3C\,424 remained undetected in \irass and appears as a weak point source in the first three \wisee bands.
It becomes increasingly fainter towards longer wavelengths and is not detected in the \wisee band~4 ($24\,\mu$m).
The object was observed with \spitzer/IRS in LR staring-mode.
The corresponding spectrum has a low S/N and appears featureless with a flat slope in $\nu F_\nu$-space, except for  H$_2$ emission (see also \citealt{leipski_spitzer_2009,ogle_jet-powered_2010}).
3C\,424 was observed with VISIR in the SIC filter in 2006 but no nucleus was detected \citep{van_der_wolk_dust_2010}. 
Our corresponding upper limit on the nuclear flux is a factor of two higher than that given by \cite{van_der_wolk_dust_2010}.
However, the IRS spectrum consistent with the \wisee band~3 flux provides the tightest constraint on the nuclear MIR emission and is thus also used to derive our 12$\,\mu$m continuum emission estimate. 
\newline\end{@twocolumnfalse}]

\begin{figure}
   \centering
   \includegraphics[angle=0,width=8.500cm]{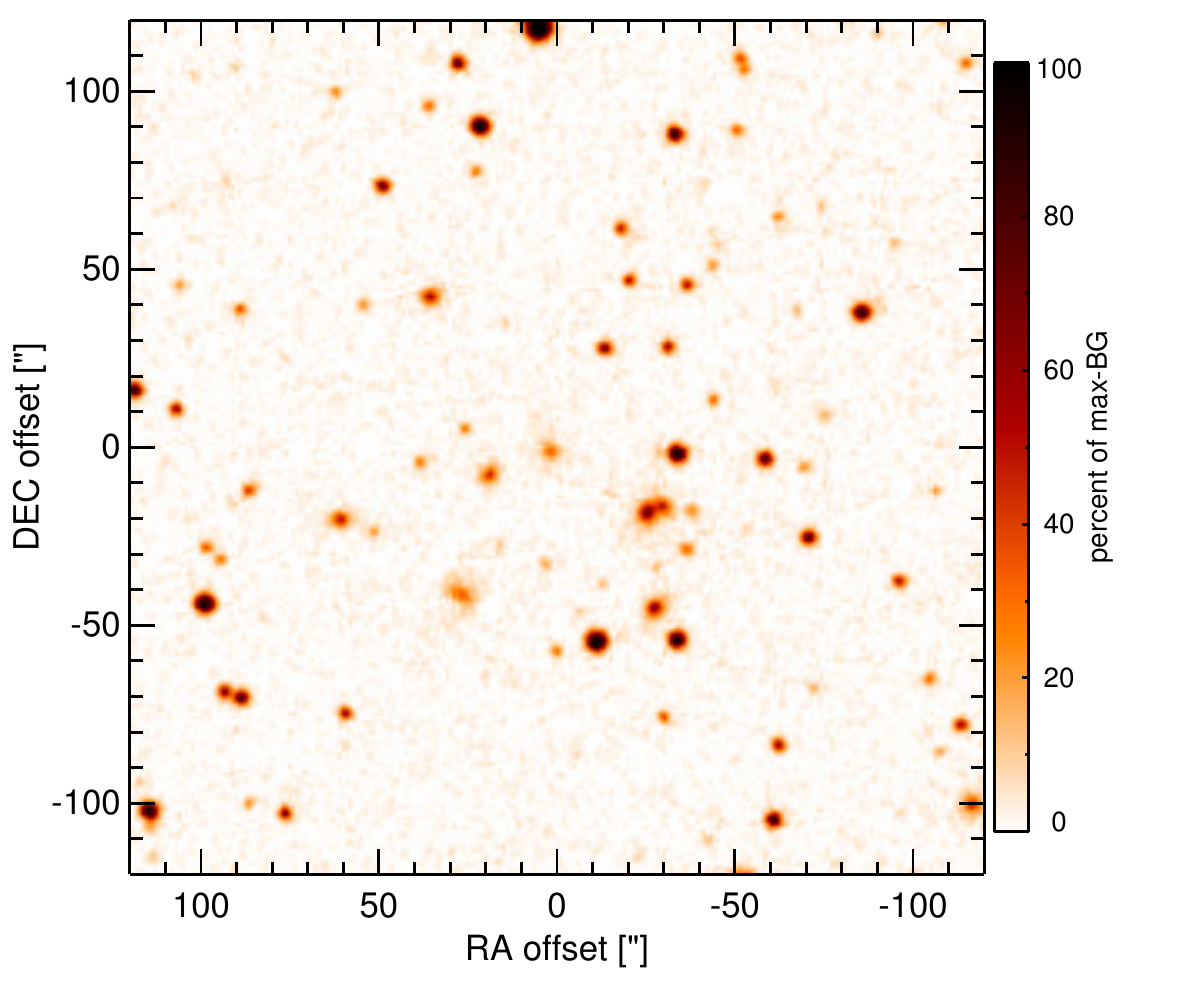}
    \caption{\label{fig:OPTim_3C424}
             Optical image (DSS, red filter) of 3C\,424. Displayed are the central $4\arcmin$ with North up and East to the left. 
              The colour scaling is linear with white corresponding to the median background and black to the $0.01\%$ pixels with the highest intensity.  
           }
\end{figure}
\begin{figure}
   \centering
   \includegraphics[angle=0,width=8.50cm]{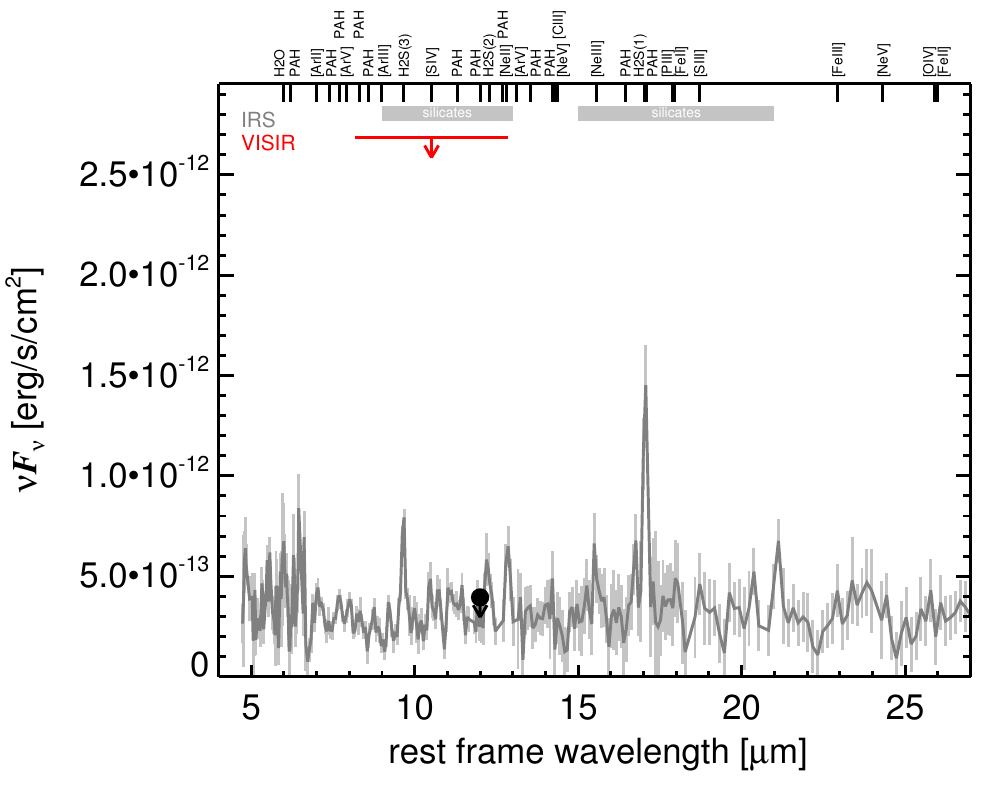}
   \caption{\label{fig:MISED_3C424}
      MIR SED of 3C\,424. The description  of the symbols (if present) is the following.
      Grey crosses and  solid lines mark the \spitzer/IRAC, MIPS and IRS data. 
      The colour coding of the other symbols is: 
      green for COMICS, magenta for Michelle, blue for T-ReCS and red for VISIR data.
      Darker-coloured solid lines mark spectra of the corresponding instrument.
      The black filled circles mark the nuclear 12 and $18\,\mu$m  continuum emission estimate from the data.
      The ticks on the top axis mark positions of common MIR emission lines, while the light grey horizontal bars mark wavelength ranges affected by the silicate 10 and 18$\mu$m features.     
   }
\end{figure}
\clearpage

\twocolumn[\begin{@twocolumnfalse}  
\subsection{3C\,445 -- LEDA\,68751 -- IRAS\,F22212-0221}\label{app:3C445}
3C\,445 is a FR\,II radio source identified with the elliptical galaxy LEDA\,68751 \citep{madrid_hubble_2006} at a redshift of $z=$ 0.0559 ($D \sim 254$\,Mpc) with a Sy\,1.5 nucleus \citep{veron-cetty_catalogue_2010}.
It features the typical supergalactic-scale biconical radio lobes with only a partial jet in the direction to the southern lobe (PA$\sim -10\degree$; e.g., \citealt{leahy_study_1997}).
After first being detected in the MIR with \iras, 3C\,445 was followed up with \iso/ISOCAM \citep{siebenmorgen_isocam_2004} and Palomar 5\,m/MIRLIN  \citep{gorjian_10_2004}, where it appeared point-like.
3C\,445 was also observed with \spitzer/IRS and MIPS in 2004 and IRAC in 2008. 
The MIR emission remains nearly unresolved in all \spitzerr images, and our nuclear MIPS 24$\,\mu$m flux agrees with that given in \cite{dicken_origin_2008}.
Interestingly, there is a discrepancy between the IRS and MIPS fluxes while IRS and IRAC fluxes agree. 
The IRS LR staring-mode spectrum shows strong silicate emission and a blue spectral slope in $\nu F_\nu$-space but no PAH emission (see also \citealt{haas_spitzer_2005,shi_9.7_2006,tommasin_spitzer-irs_2010,dicken_spitzer_2012}).
VISIR $N$-band imaging of 3C\,445 was performed in 2005 \citep{van_der_wolk_dust_2010} and 2006 \citep{horst_mid_2008,horst_mid-infrared_2009}.
T-ReCS Si2 imaging and $N$-band spectroscopy followed in  2010 \citep{gonzalez-martin_dust_2013}. 
The nuclear emission of 3C\,445 appears possibly marginally resolved in the VISIR and T-ReCS images (FWHM $\sim 410$\,pc) but with inconsistent position angles.
Thus, it remains uncertain, whether 3C\,445 is indeed resolved at subarcsecond scales in the MIR.
Our nuclear photometry is consistent with the values published by \cite{horst_mid_2008} and \cite{van_der_wolk_dust_2010}.
The T-ReCS spectrum verifies that the strong silicate emission originates on the projected nuclear scales ($< 360\,$pc) and indicates strong thermal dust emission in 3C\,445.
\newline\end{@twocolumnfalse}]

\begin{figure}
   \centering
   \includegraphics[angle=0,width=8.500cm]{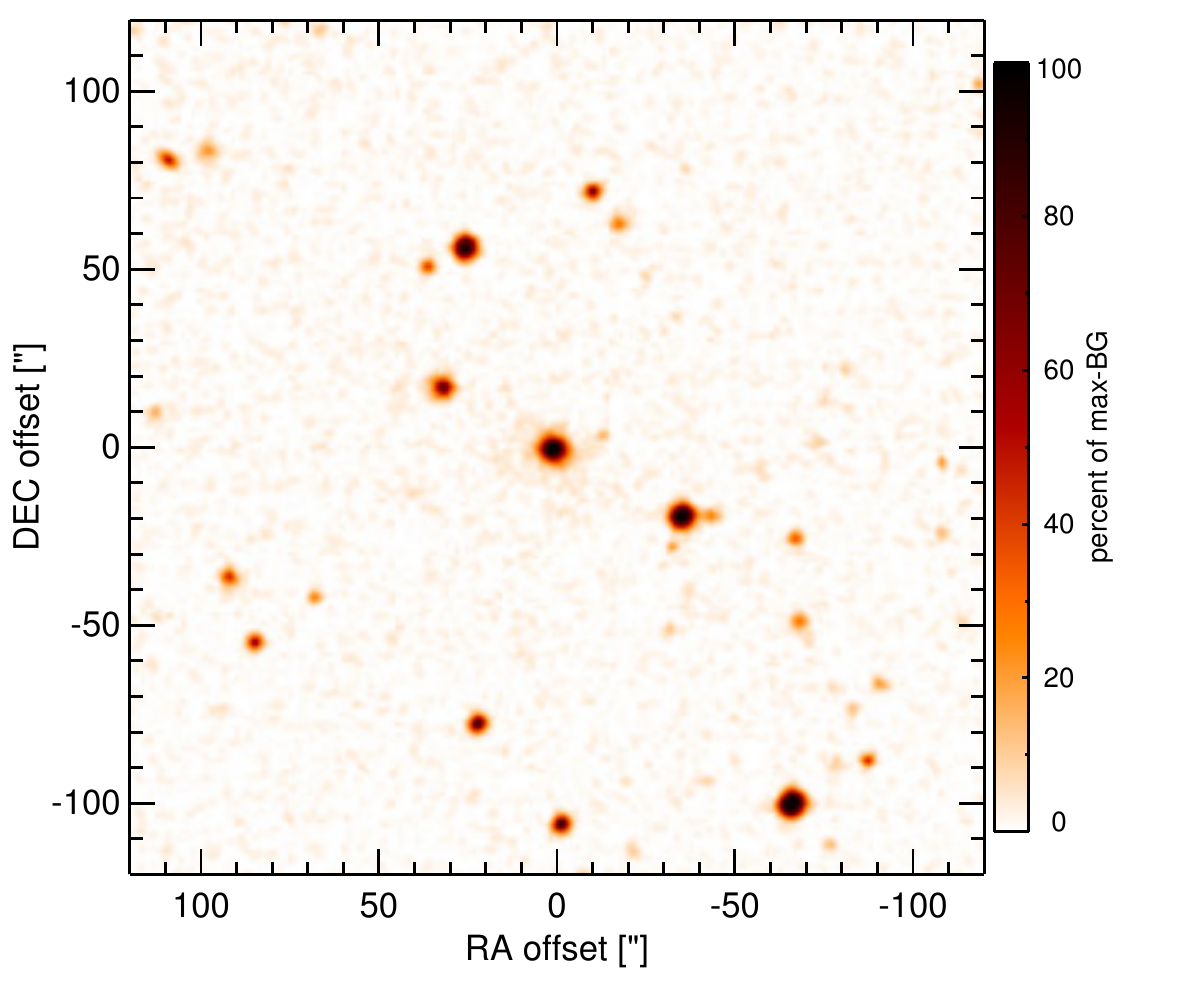}
    \caption{\label{fig:OPTim_3C445}
             Optical image (DSS, red filter) of 3C\,445. Displayed are the central $4\arcmin$ with North up and East to the left. 
              The colour scaling is linear with white corresponding to the median background and black to the $0.01\%$ pixels with the highest intensity.  
           }
\end{figure}
\begin{figure}
   \centering
   \includegraphics[angle=0,height=3.11cm]{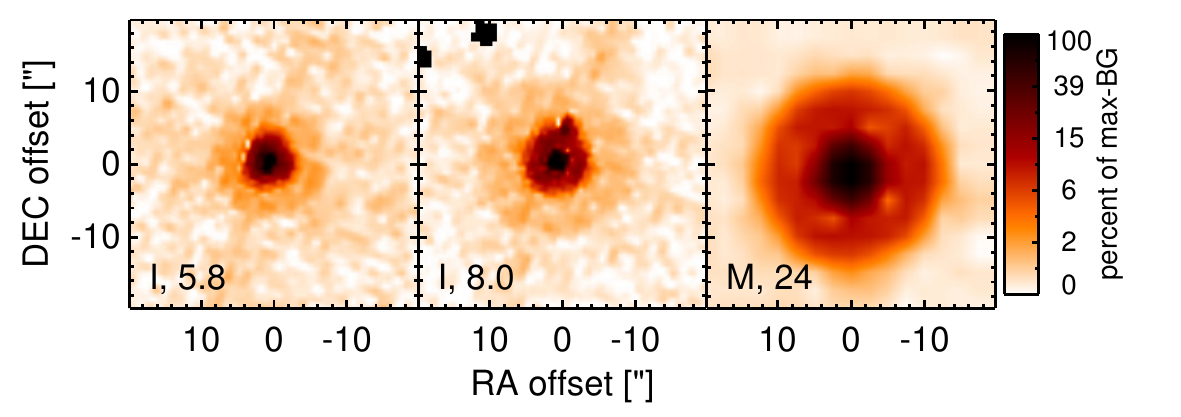}
    \caption{\label{fig:INTim_3C445}
             \spitzerr MIR images of 3C\,445. Displayed are the inner $40\arcsec$ with North up and East to the left. The colour scaling is logarithmic with white corresponding to median background and black to the $0.1\%$ pixels with the highest intensity.
             The label in the bottom left states instrument and central wavelength of the filter in $\mu$m (I: IRAC, M: MIPS). 
           }
\end{figure}
\begin{figure}
   \centering
   \includegraphics[angle=0,width=8.500cm]{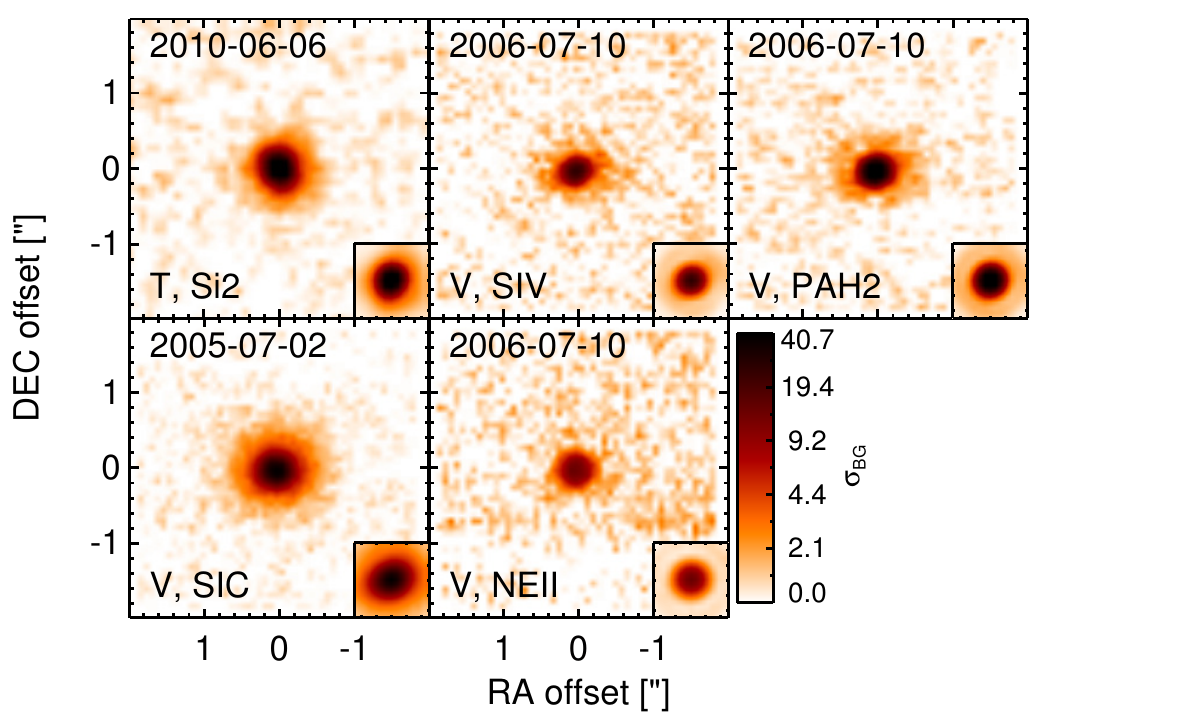}
    \caption{\label{fig:HARim_3C445}
             Subarcsecond-resolution MIR images of 3C\,445 sorted by increasing filter wavelength. 
             Displayed are the inner $4\arcsec$ with North up and East to the left. 
             The colour scaling is logarithmic with white corresponding to median background and black to the $75\%$ of the highest intensity of all images in units of $\sigbg$.
             The inset image shows the central arcsecond of the PSF from the calibrator star, scaled to match the science target.
             The labels in the bottom left state instrument and filter names (C: COMICS, M: Michelle, T: T-ReCS, V: VISIR).
           }
\end{figure}
\begin{figure}
   \centering
   \includegraphics[angle=0,width=8.50cm]{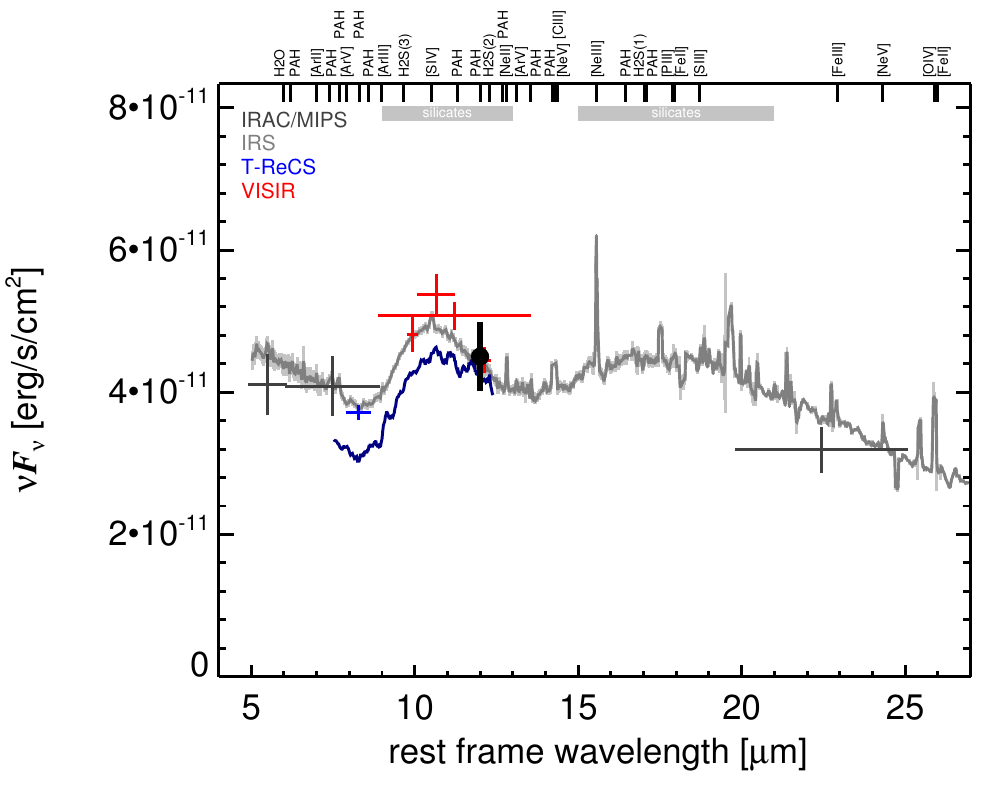}
   \caption{\label{fig:MISED_3C445}
      MIR SED of 3C\,445. The description  of the symbols (if present) is the following.
      Grey crosses and  solid lines mark the \spitzer/IRAC, MIPS and IRS data. 
      The colour coding of the other symbols is: 
      green for COMICS, magenta for Michelle, blue for T-ReCS and red for VISIR data.
      Darker-coloured solid lines mark spectra of the corresponding instrument.
      The black filled circles mark the nuclear 12 and $18\,\mu$m  continuum emission estimate from the data.
      The ticks on the top axis mark positions of common MIR emission lines, while the light grey horizontal bars mark wavelength ranges affected by the silicate 10 and 18$\mu$m features.     
   }
\end{figure}
\clearpage

\twocolumn[\begin{@twocolumnfalse}  
\subsection{3C\,449  -- Z\,514-050 -- UGC\,12064}\label{app:3C449}
3C\,449 is a FR\,I radio source identified with the elliptical galaxy UGC\,12064  at a redshift of $z=$ 0.0171 ($D \sim 72.4$\,Mpc), the southern component of a close pair of galaxies. 
It contains an active nucleus with optical LINER classification \citep{buttiglione_optical_2009,buttiglione_optical_2010}.
3C\,449 features supergalactic-scale twisted twin jets along the north-south axis (PA$\sim9\degree$; \citealt{perley_structure_1979,feretti_vla_1999}) and a nuclear dust disc $\sim1.5\,$kpc in diameter (PA$\sim155\degree$; \citealt{martel_hubble_1999,tremblay_warped_2006}).
Note that there is a smaller companion galaxy in projection $\sim38\arcsec\sim13$\,kpc to the north (PA$\sim 13\degree$) of UGC\,12064.
The first detection in the $N$-band was achieved with IRTF \citep{sparks_infrared_1986,impey_infrared_1990} and then followed up with \iso/ISOCAM \citep{siebenmorgen_isocam_2004}.
3C\,449 was also observed with \spitzer/IRAC, IRS and MIPS, and a compact nucleus embedded in diffuse host emission was detected in the corresponding images.
The high-resolution mode IRS spectrum is extremely noisy and does not allow us to discern spectral features apart from a shallow blue spectral continuum slope in $\nu F_\nu$-space.
Note, however, that no background subtraction was performed for this spectrum.  
Silicate 10$\,\mu$m emission is possibly present.
We observed 3C\,449 with COMICS in the N11.7 filter in 2009 but did not detect the object.
The corresponding upper limit on the nuclear N11.7 flux is roughly consistent with the \spitzerr spectrophotometry.
\newline\end{@twocolumnfalse}]

\begin{figure}
   \centering
   \includegraphics[angle=0,width=8.500cm]{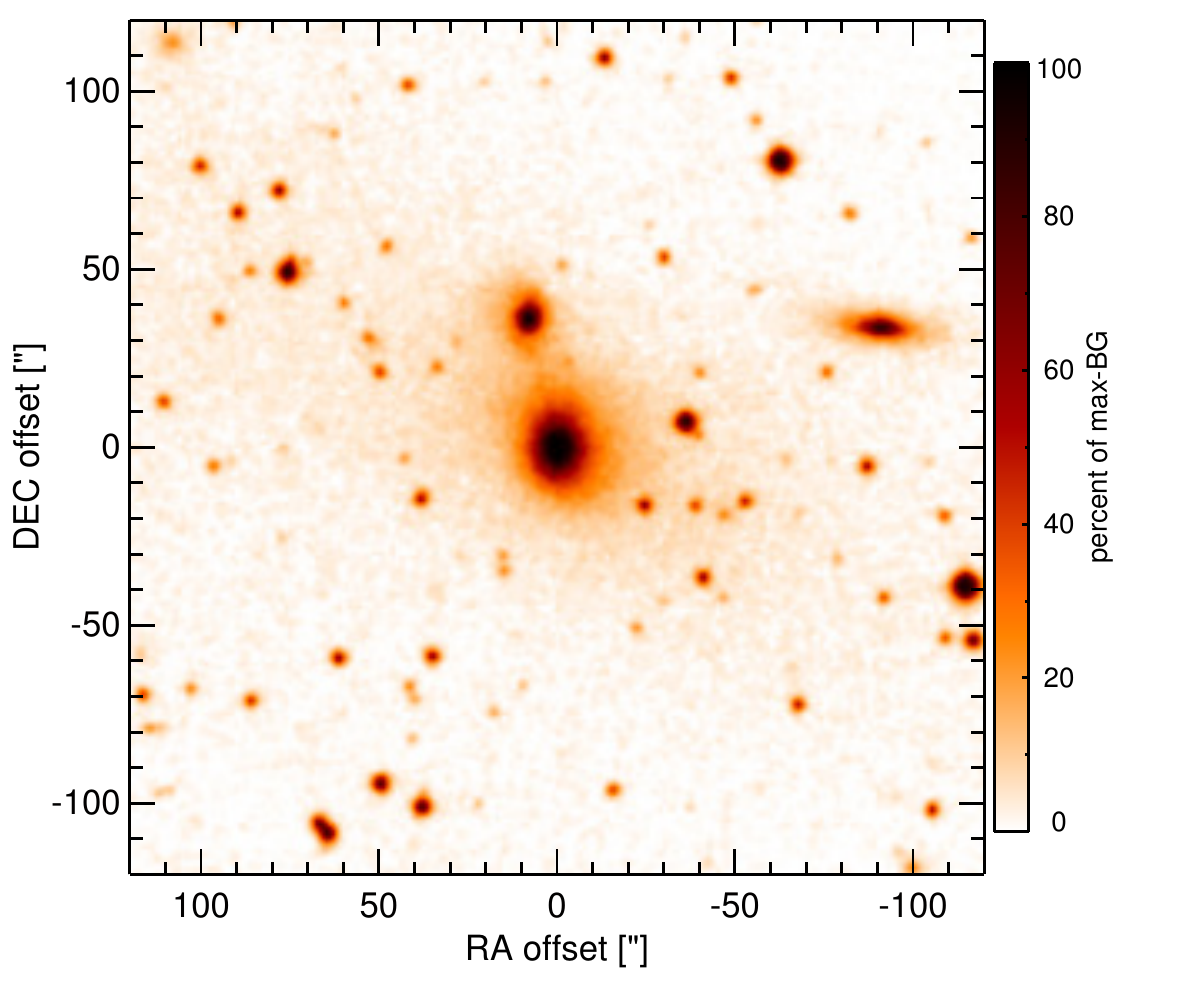}
    \caption{\label{fig:OPTim_3C449}
             Optical image (DSS, red filter) of 3C\,449. Displayed are the central $4\arcmin$ with North up and East to the left. 
              The colour scaling is linear with white corresponding to the median background and black to the $0.01\%$ pixels with the highest intensity.  
           }
\end{figure}
\begin{figure}
   \centering
   \includegraphics[angle=0,height=3.11cm]{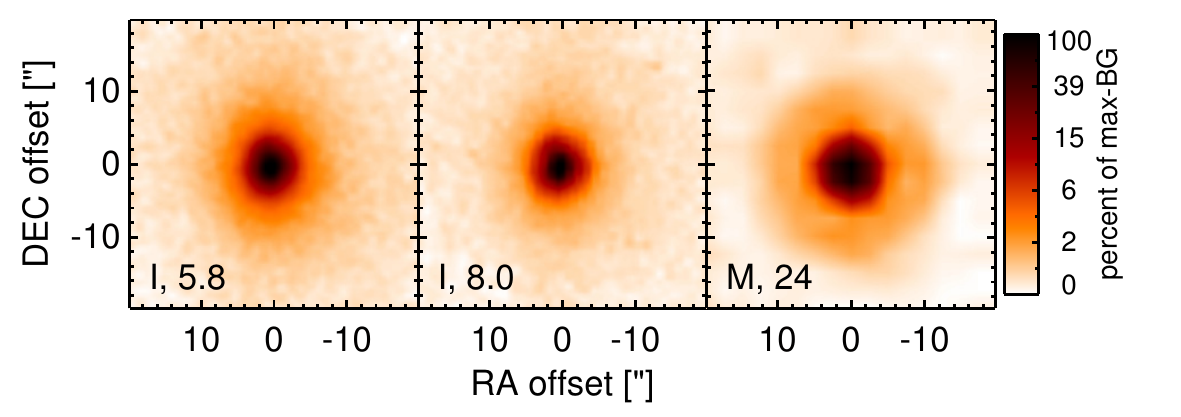}
    \caption{\label{fig:INTim_3C449}
             \spitzerr MIR images of 3C\,449. Displayed are the inner $40\arcsec$ with North up and East to the left. The colour scaling is logarithmic with white corresponding to median background and black to the $0.1\%$ pixels with the highest intensity.
             The label in the bottom left states instrument and central wavelength of the filter in $\mu$m (I: IRAC, M: MIPS). 
           }
\end{figure}
\begin{figure}
   \centering
   \includegraphics[angle=0,width=8.50cm]{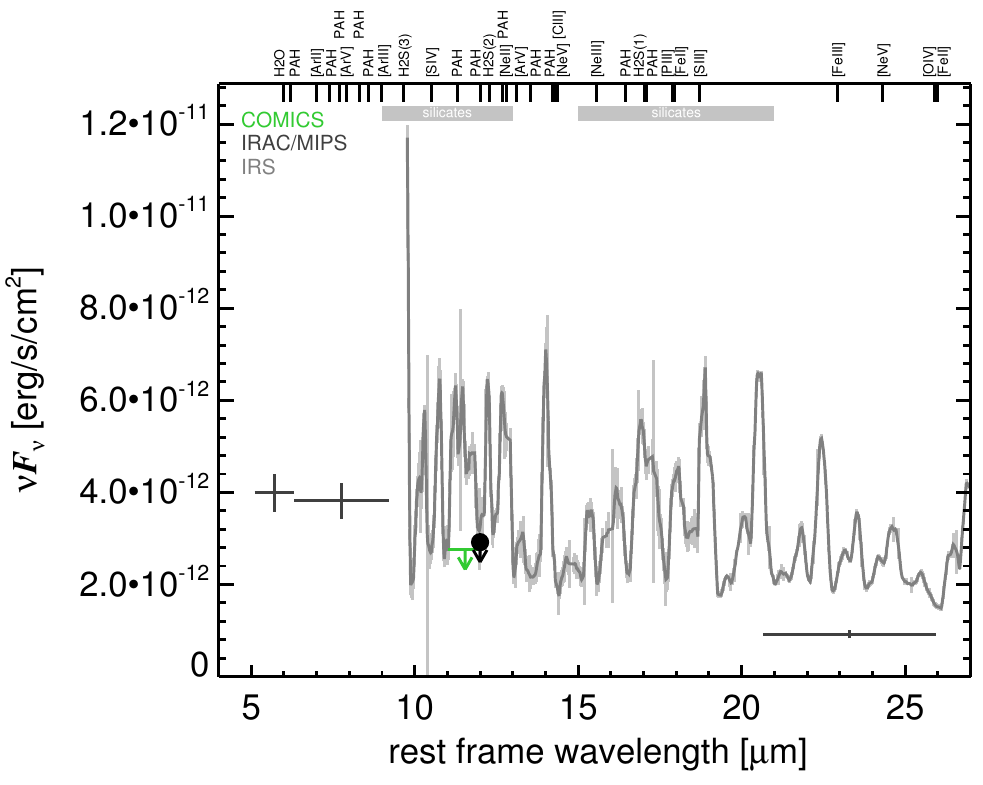}
   \caption{\label{fig:MISED_3C449}
      MIR SED of 3C\,449. The description  of the symbols (if present) is the following.
      Grey crosses and  solid lines mark the \spitzer/IRAC, MIPS and IRS data. 
      The colour coding of the other symbols is: 
      green for COMICS, magenta for Michelle, blue for T-ReCS and red for VISIR data.
      Darker-coloured solid lines mark spectra of the corresponding instrument.
      The black filled circles mark the nuclear 12 and $18\,\mu$m  continuum emission estimate from the data.
      The ticks on the top axis mark positions of common MIR emission lines, while the light grey horizontal bars mark wavelength ranges affected by the silicate 10 and 18$\mu$m features.     
   }
\end{figure}
\clearpage

\twocolumn[\begin{@twocolumnfalse}  
\subsection{3C\,452 -- LEDA\,69671}\label{app:3C452}
3C\,452 is a FR\,II radio source identified with the elliptical galaxy LEDA\,69671 \citep{martel_hubble_1999} at a redshift of $z=$ 0.0811 ($D \sim 378$\,Mpc).
It contains an AGN with optical Sy\,2 classification  \citep{veron-cetty_catalogue_2010}  that belongs to the nine-month BAT AGN sample.
The radio morphology is typical for a FR\,II source, with the supergalactic-scale biconical radio lobes connected to the nucleus by a twin jet along the east-west axis (PA$\sim75\degree$; e.g., \citep{black_study_1992}).
Recently, even larger, megaparsec-scale fossil radio lobes have been discovered along the same projected axis, indicating recurrent AGN activity in 3C\,452 \citep{sirothia_discovery_2013}. 
Note that there is a foreground star $\sim 12\arcsec$ to the west of the galaxy.
The first successful MIR observations of 3C\,452 were performed with \iso/ISOCAM \citep{siebenmorgen_isocam_2004}.
These observations were later followed up with \spitzer/IRAC, IRS and MIPS.
3C\,452 is nearly unresolved in all \spitzerr images.
Our nuclear MIPS 24\,$\mu$m photometry agrees with the value published in \cite{dicken_origin_2010}.
The IRS LR staring-mode spectrum is relatively featureless with weak silicate $10\,\mu$m absorption and an emission peak at at $\sim19\,\mu$m in $\nu F_\nu$-space but no PAH features \citep{ogle_spitzer_2006}.
Our COMICS N11.7 image taken in 2009 shows a weakly detected source with uncertain extension at subarcsecond scales.
The photometric flux is $\sim 30\%$ higher than expected from the \spitzerr data, albeit with a $20\%$ uncertainty. 
\newline\end{@twocolumnfalse}]

\begin{figure}
   \centering
   \includegraphics[angle=0,width=8.500cm]{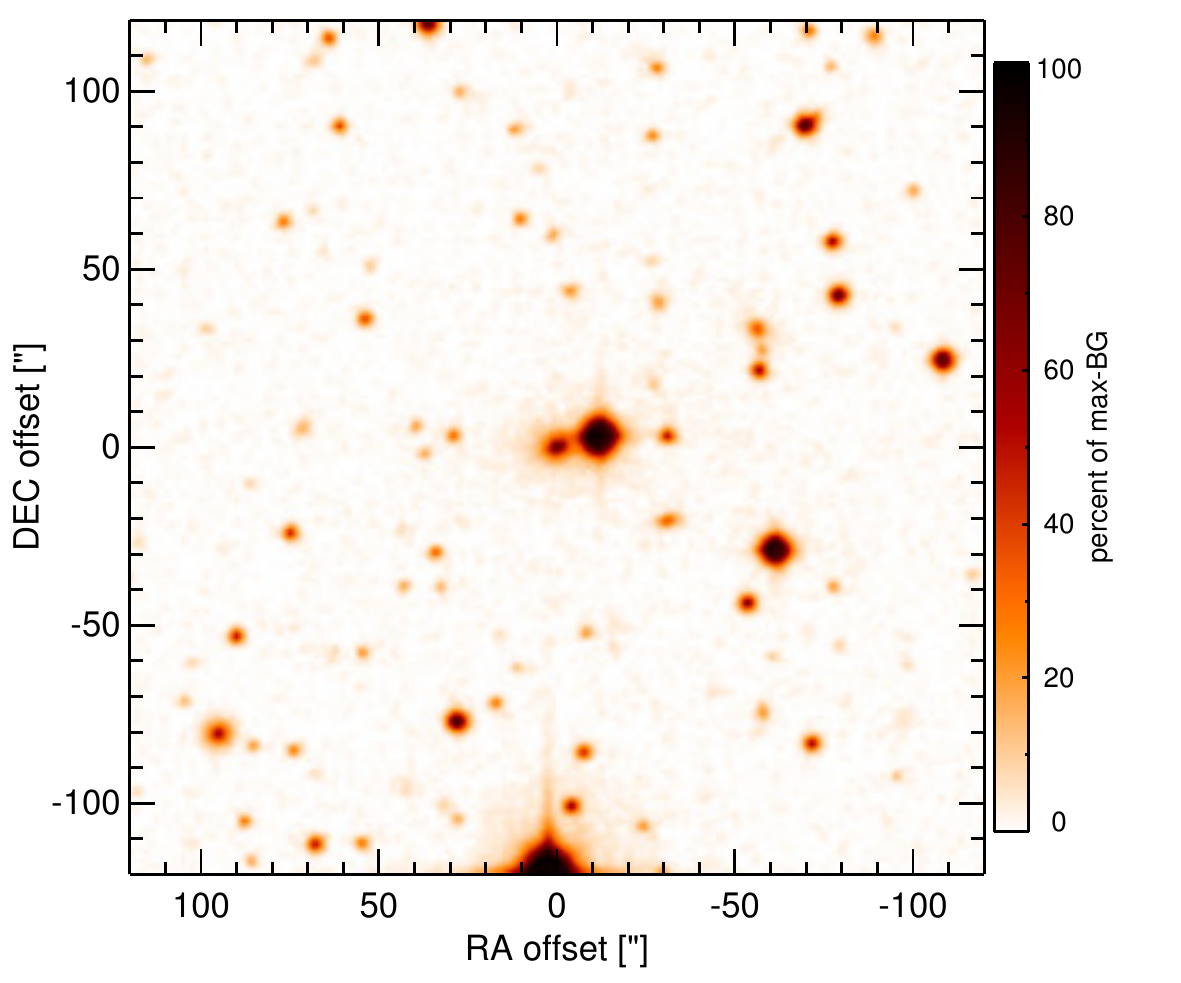}
    \caption{\label{fig:OPTim_3C452}
             Optical image (DSS, red filter) of 3C\,452. Displayed are the central $4\arcmin$ with North up and East to the left. 
              The colour scaling is linear with white corresponding to the median background and black to the $0.01\%$ pixels with the highest intensity.  
           }
\end{figure}
\begin{figure}
   \centering
   \includegraphics[angle=0,height=3.11cm]{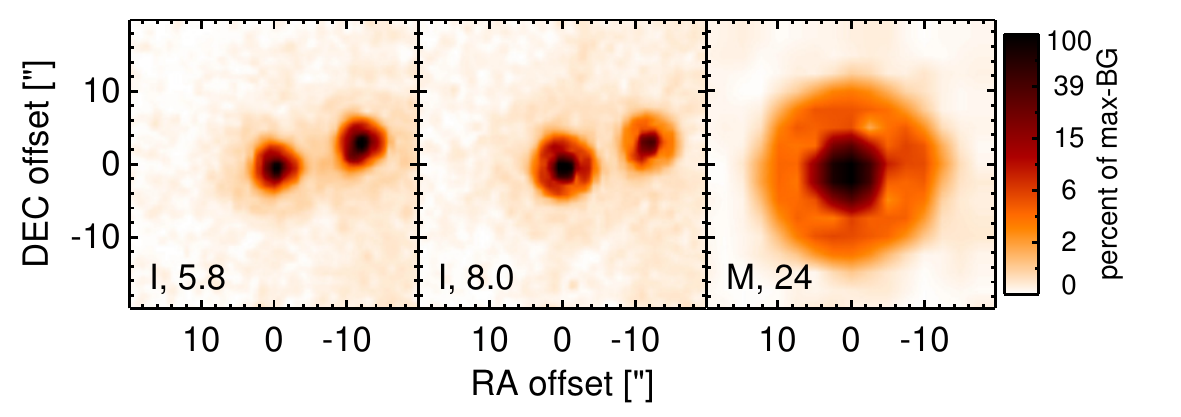}
    \caption{\label{fig:INTim_3C452}
             \spitzerr MIR images of 3C\,452. Displayed are the inner $40\arcsec$ with North up and East to the left. The colour scaling is logarithmic with white corresponding to median background and black to the $0.1\%$ pixels with the highest intensity.
             The label in the bottom left states instrument and central wavelength of the filter in $\mu$m (I: IRAC, M: MIPS). 
           }
\end{figure}
\begin{figure}
   \centering
   \includegraphics[angle=0,height=3.11cm]{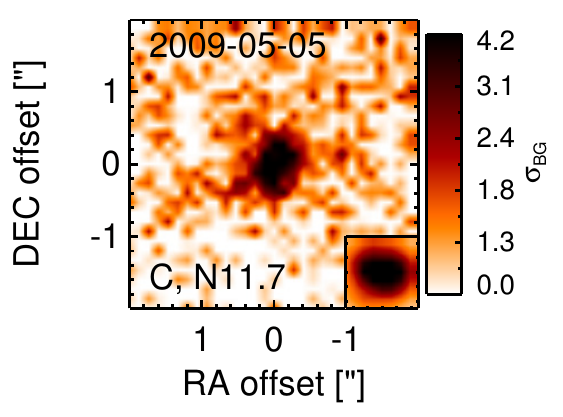}
    \caption{\label{fig:HARim_3C452}
             Subarcsecond-resolution MIR images of 3C\,452 sorted by increasing filter wavelength. 
             Displayed are the inner $4\arcsec$ with North up and East to the left. 
             The colour scaling is logarithmic with white corresponding to median background and black to the $75\%$ of the highest intensity of all images in units of $\sigbg$.
             The inset image shows the central arcsecond of the PSF from the calibrator star, scaled to match the science target.
             The labels in the bottom left state instrument and filter names (C: COMICS, M: Michelle, T: T-ReCS, V: VISIR).
           }
\end{figure}
\begin{figure}
   \centering
   \includegraphics[angle=0,width=8.50cm]{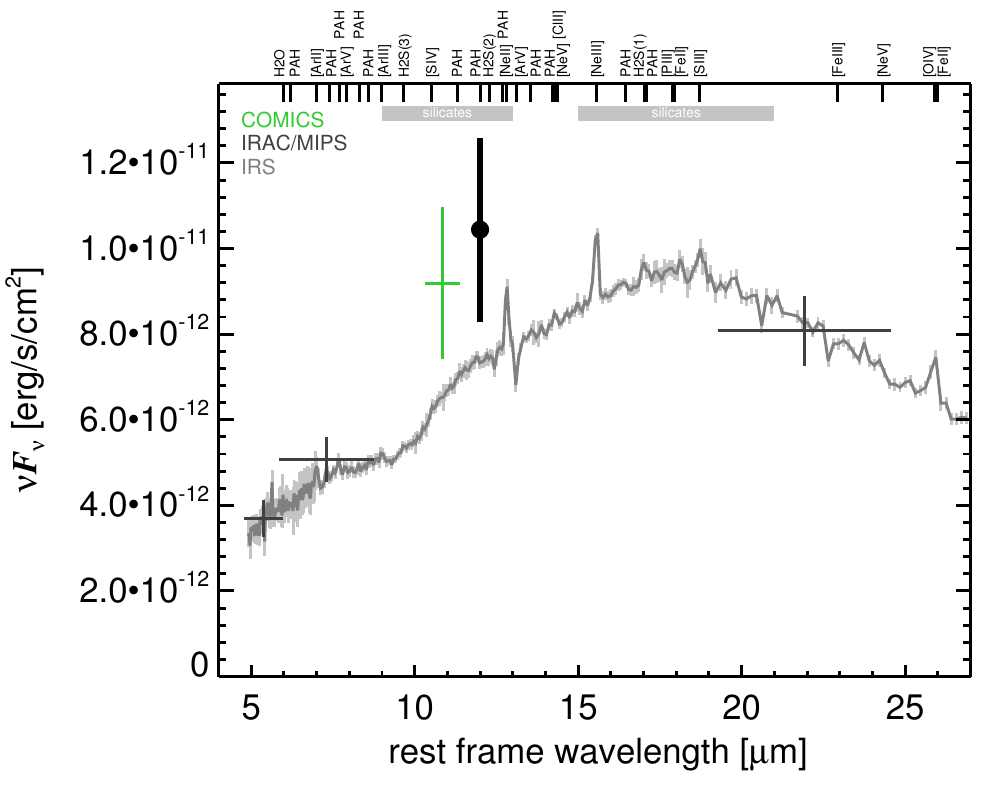}
   \caption{\label{fig:MISED_3C452}
      MIR SED of 3C\,452. The description  of the symbols (if present) is the following.
      Grey crosses and  solid lines mark the \spitzer/IRAC, MIPS and IRS data. 
      The colour coding of the other symbols is: 
      green for COMICS, magenta for Michelle, blue for T-ReCS and red for VISIR data.
      Darker-coloured solid lines mark spectra of the corresponding instrument.
      The black filled circles mark the nuclear 12 and $18\,\mu$m  continuum emission estimate from the data.
      The ticks on the top axis mark positions of common MIR emission lines, while the light grey horizontal bars mark wavelength ranges affected by the silicate 10 and 18$\mu$m features.     
   }
\end{figure}
\clearpage

\twocolumn[\begin{@twocolumnfalse}  
\subsection{3C\,459 -- IRAS\,23140+0348 -- LEDA\,70899}\label{app:3C459}
3C\,459 is a FR\,II radio source identified with the infrared ultra-luminous, disturbed early-type galaxy LEDA\,70899 \citep{zheng_hst_1999} at a redshift of $z=$ 0.2201 ($D \sim 1125$\,Mpc).
Its optical classification is controversial in the literature: it is classified as a type~I AGN by \cite{buttiglione_optical_2009,buttiglione_optical_2010}, based on which \cite{veron-cetty_catalogue_2010} presumably classified it as a Sy\,1, while \cite{tadhunter_optical_1993}, \cite{eracleous_doubled-peaked_1994}, and \cite{buchanan_radio-excess_2006} do not find evidence for broad lines and, thus, classify it as a type~II object.
Finally, \cite{zheng_hst_1999} classify it as a LINER.
Based on the presence of an intense starburst in 3C\,459 \citep{thomasson_3c459:_2003}, we treat this object as an AGN/starburst composite.
3C\,459 features an asymmetric supergalactic-scale biconical radio morphology along the east-west axis (PA$\sim95\degree$; \citealt{ulvestad_radio_1985,morganti_radio_1993}).
The first successful MIR observations were performed with \isoo/ISOCAM \citep{siebenmorgen_isocam_2004}.
This was then followed up with \spitzer/IRS and MIPS, where it appears unresolved.
Our MIPS 24\,$\mu$m photometry agrees with the value given in  \cite{dicken_origin_2008}.
The IRS LR staring-mode spectrum was first published in \cite{haas_spitzer_2005} and shows silicate absorption, strong PAH emission, and a red spectral slope in $\nu F_\nu$-space, as expected for intense star formation.
The VISIR SIC observation from 2010 has, to our knowledge, not yet been published.
Unfortunately, 3C\,459 remains undetected in the corresponding image. 
The derived upper limit on the nuclear SIC flux  is consistent with the \spitzerr spectrophotometry.
\newline\end{@twocolumnfalse}]

\begin{figure}
   \centering
   \includegraphics[angle=0,width=8.500cm]{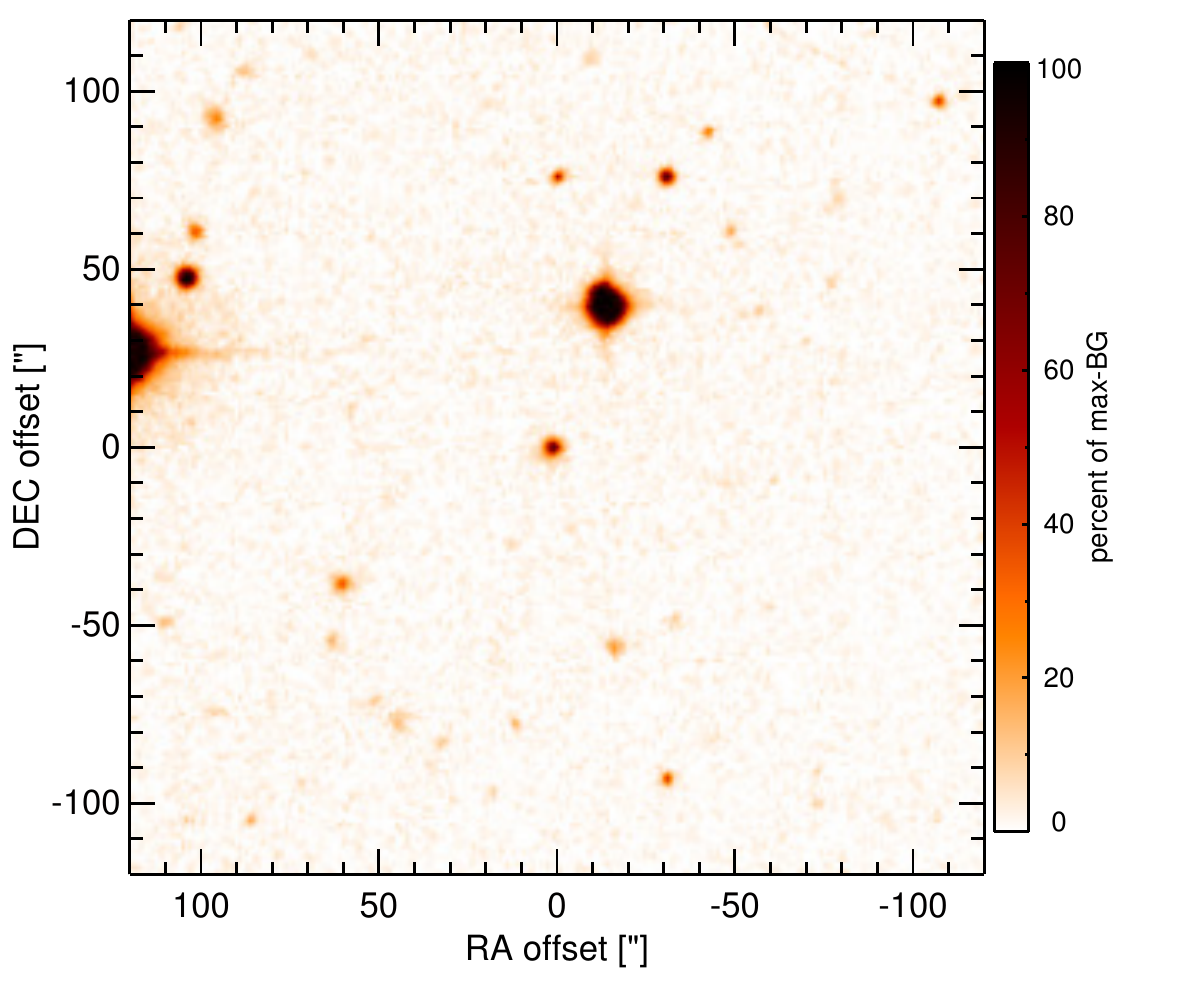}
    \caption{\label{fig:OPTim_3C459}
             Optical image (DSS, red filter) of 3C\,459. Displayed are the central $4\arcmin$ with North up and East to the left. 
              The colour scaling is linear with white corresponding to the median background and black to the $0.01\%$ pixels with the highest intensity.  
           }
\end{figure}
\begin{figure}
   \centering
   \includegraphics[angle=0,height=3.11cm]{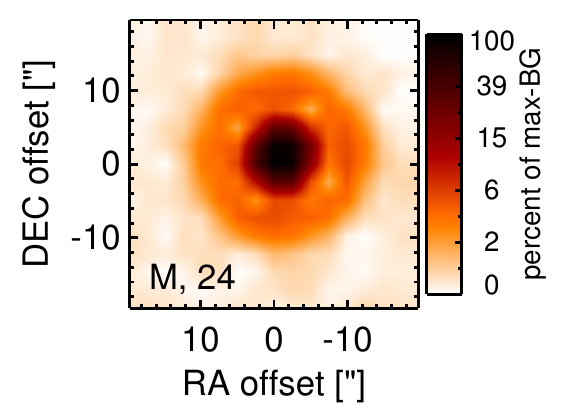}
    \caption{\label{fig:INTim_3C459}
             \spitzerr MIR images of 3C\,459. Displayed are the inner $40\arcsec$ with North up and East to the left. The colour scaling is logarithmic with white corresponding to median background and black to the $0.1\%$ pixels with the highest intensity.
             The label in the bottom left states instrument and central wavelength of the filter in $\mu$m (I: IRAC, M: MIPS). 
           }
\end{figure}
\begin{figure}
   \centering
   \includegraphics[angle=0,width=8.50cm]{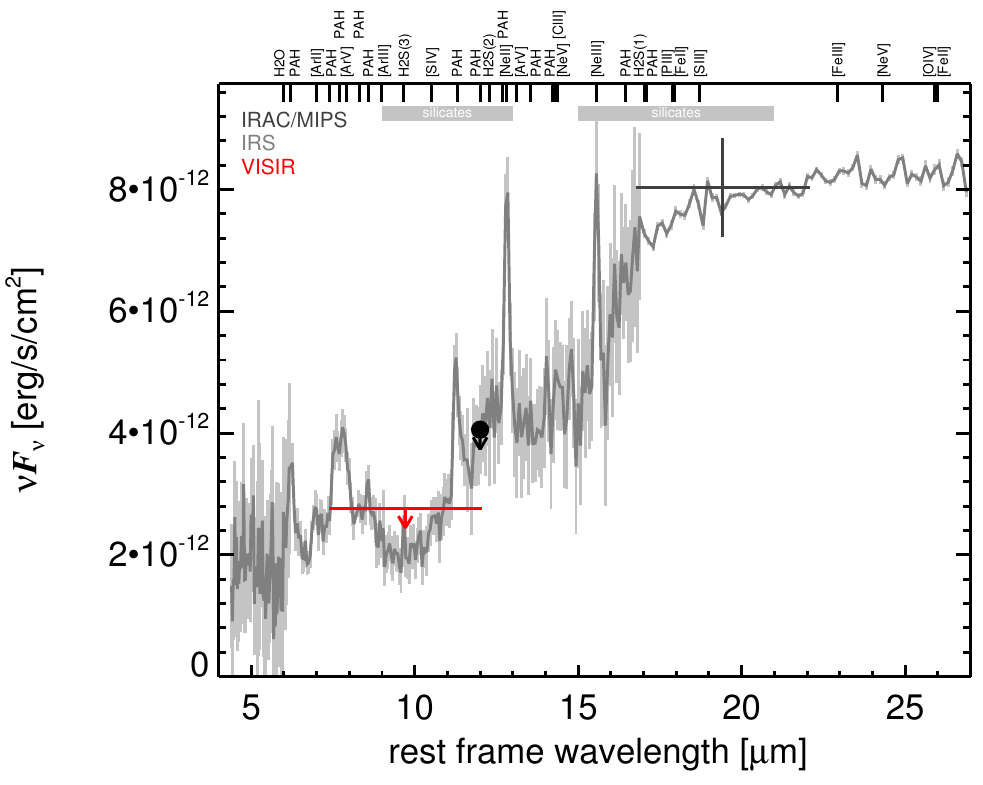}
   \caption{\label{fig:MISED_3C459}
      MIR SED of 3C\,459. The description  of the symbols (if present) is the following.
      Grey crosses and  solid lines mark the \spitzer/IRAC, MIPS and IRS data. 
      The colour coding of the other symbols is: 
      green for COMICS, magenta for Michelle, blue for T-ReCS and red for VISIR data.
      Darker-coloured solid lines mark spectra of the corresponding instrument.
      The black filled circles mark the nuclear 12 and $18\,\mu$m  continuum emission estimate from the data.
      The ticks on the top axis mark positions of common MIR emission lines, while the light grey horizontal bars mark wavelength ranges affected by the silicate 10 and 18$\mu$m features.     
   }
\end{figure}
\clearpage

\twocolumn[\begin{@twocolumnfalse}  
\subsection{4C\,+73.08  -- VII\,Zw\,292}\label{app:4C+73-08}
4C\,+73.08 is a FR\,II radio source identified with the compact galaxy VII\,Zw\,292 \citep{mayer_multifrequency_1979} at a redshift of $z=$ 0.0581 ($D \sim 271$\,Mpc) with a little-studied AGN.
We treat the AGN as Sy\,2, based on the existence of only narrow emission lines \citep{saunders_spectrophotometry_1989}.
It features megaparsec-scale radio lobes (PA$\sim65\degree$; \citealt{mayer_multifrequency_1979}).
4C\,+73.08 was first detected in the MIR with \spitzer/IRAC and MIPS, where it appears point-like.
Our MIPS 24\,$\mu$m photometry agrees with the value given in  \cite{dicken_origin_2010}.
No MIR spectrum is available for this object.
The \spitzerr photometry indicates a red spectral slope in the MIR in $\nu F_\nu$-space.
We observed 4C\,+73.08 with COMICS in the N11.7 filter in 2009 and weakly detected a compact source.
The low S/N of the detection does not allow for any conclusion about   the nuclear extension on subarcsecond scales in the MIR.
The nuclear N11.7 flux is consistent with the \spitzerr photometry. 
\newline\end{@twocolumnfalse}]

\begin{figure}
   \centering
   \includegraphics[angle=0,width=8.500cm]{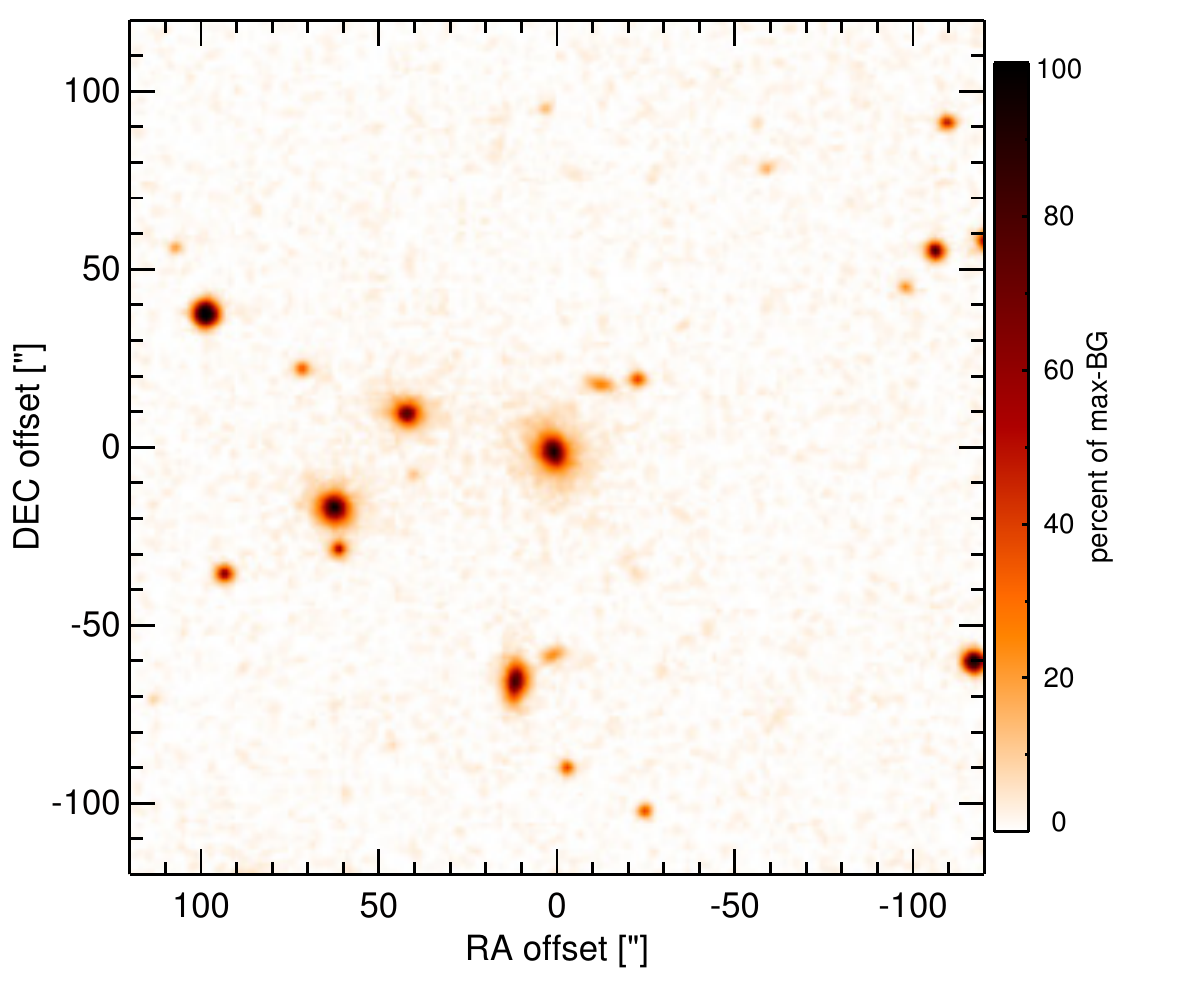}
    \caption{\label{fig:OPTim_4C+73-08}
             Optical image (DSS, red filter) of 4C\,+73.08. Displayed are the central $4\arcmin$ with North up and East to the left. 
              The colour scaling is linear with white corresponding to the median background and black to the $0.01\%$ pixels with the highest intensity.  
           }
\end{figure}
\begin{figure}
   \centering
   \includegraphics[angle=0,height=3.11cm]{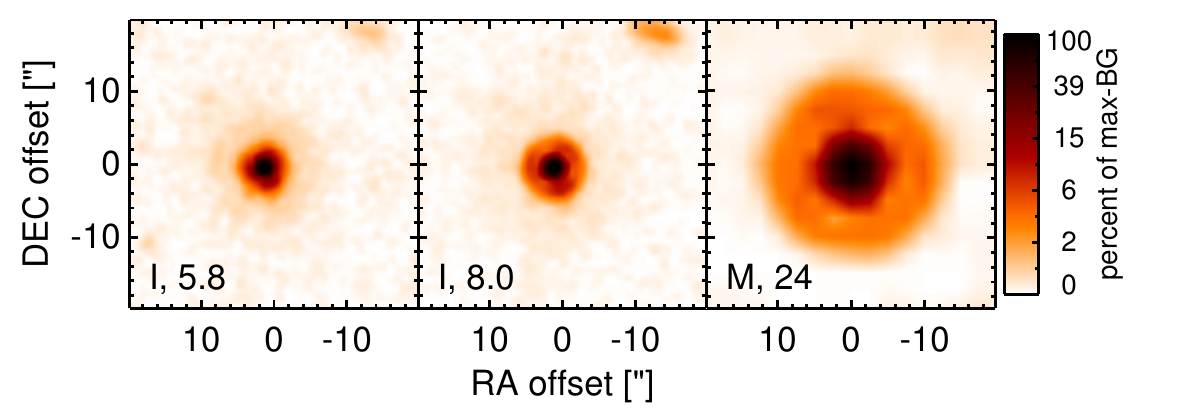}
    \caption{\label{fig:INTim_4C+73-08}
             \spitzerr MIR images of 4C\,+73.08. Displayed are the inner $40\arcsec$ with North up and East to the left. The colour scaling is logarithmic with white corresponding to median background and black to the $0.1\%$ pixels with the highest intensity.
             The label in the bottom left states instrument and central wavelength of the filter in $\mu$m (I: IRAC, M: MIPS). 
           }
\end{figure}
\begin{figure}
   \centering
   \includegraphics[angle=0,height=3.11cm]{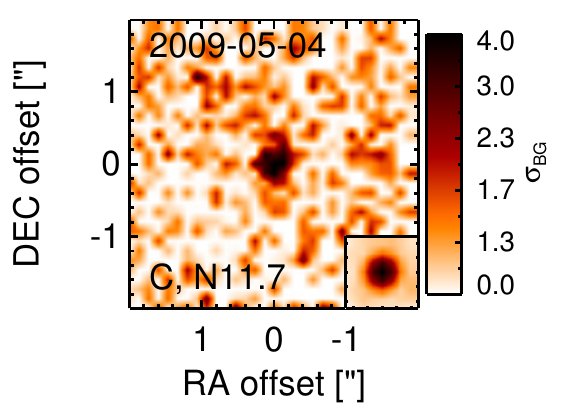}
    \caption{\label{fig:HARim_4C+73-08}
             Subarcsecond-resolution MIR images of 4C\,+73.08 sorted by increasing filter wavelength. 
             Displayed are the inner $4\arcsec$ with North up and East to the left. 
             The colour scaling is logarithmic with white corresponding to median background and black to the $75\%$ of the highest intensity of all images in units of $\sigbg$.
             The inset image shows the central arcsecond of the PSF from the calibrator star, scaled to match the science target.
             The labels in the bottom left state instrument and filter names (C: COMICS, M: Michelle, T: T-ReCS, V: VISIR).
           }
\end{figure}
\begin{figure}
   \centering
   \includegraphics[angle=0,width=8.50cm]{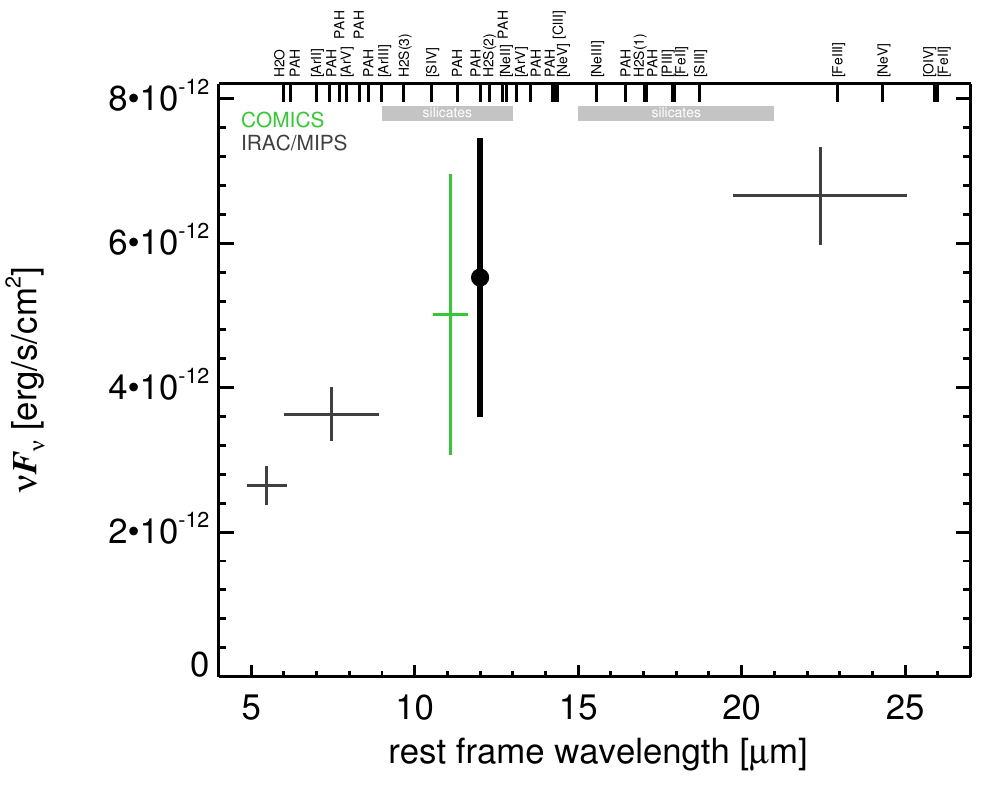}
   \caption{\label{fig:MISED_4C+73-08}
      MIR SED of 4C\,+73.08. The description  of the symbols (if present) is the following.
      Grey crosses and  solid lines mark the \spitzer/IRAC, MIPS and IRS data. 
      The colour coding of the other symbols is: 
      green for COMICS, magenta for Michelle, blue for T-ReCS and red for VISIR data.
      Darker-coloured solid lines mark spectra of the corresponding instrument.
      The black filled circles mark the nuclear 12 and $18\,\mu$m  continuum emission estimate from the data.
      The ticks on the top axis mark positions of common MIR emission lines, while the light grey horizontal bars mark wavelength ranges affected by the silicate 10 and 18$\mu$m features.     
   }
\end{figure}
\clearpage

\twocolumn[\begin{@twocolumnfalse}  
\subsection{Ark\,120 -- Mrk\,1095 -- UGC\,3271}\label{app:Ark120}
Ark\,120 is a spiral galaxy at a redshift of $z=$ 0.0327 ($D \sim 149$\,Mpc) hosting a radio-quiet Sy\,1 nucleus \citep{veron-cetty_catalogue_2010} that belongs to the nine-month BAT AGN sample.
The NLR morphology appears halo-like \citep{mulchaey_emission-line_1996}.
The first ground-based $N-$ and $Q$-band photometry of the nucleus were obtained with IRTF in 1983 and UKIRT in 1984 \citep{ward_continuum_1987}.
After \iras, MIR observations were carried out with \iso/ISOCAM \citep{ramos_almeida_mid-infrared_2007} and \spitzer/IRAC, IRS and MIPS, where Ark\,120 appears as an unresolved nucleus without any further host emission being detected.
The IRS LR staring-mode spectrum shows strong silicate emission,  PAH emission and a blue spectral slope in $\nu F_\nu$-space (see also \citealt{sargsyan_infrared_2011}). 
The nucleus of Ark\,120 was observed with VISIR in 2009 and 2010  in a total of five $N$-band filter images.
An unresolved nucleus without further host emission was detected in all cases.
The corresponding nuclear fluxes of the three 2009 measurements match the \spitzerr spectrophotometry, while the 2010 measurements are $\sim 20\%$ lower for unknown reasons.
\newline\end{@twocolumnfalse}]

\begin{figure}
   \centering
   \includegraphics[angle=0,width=8.500cm]{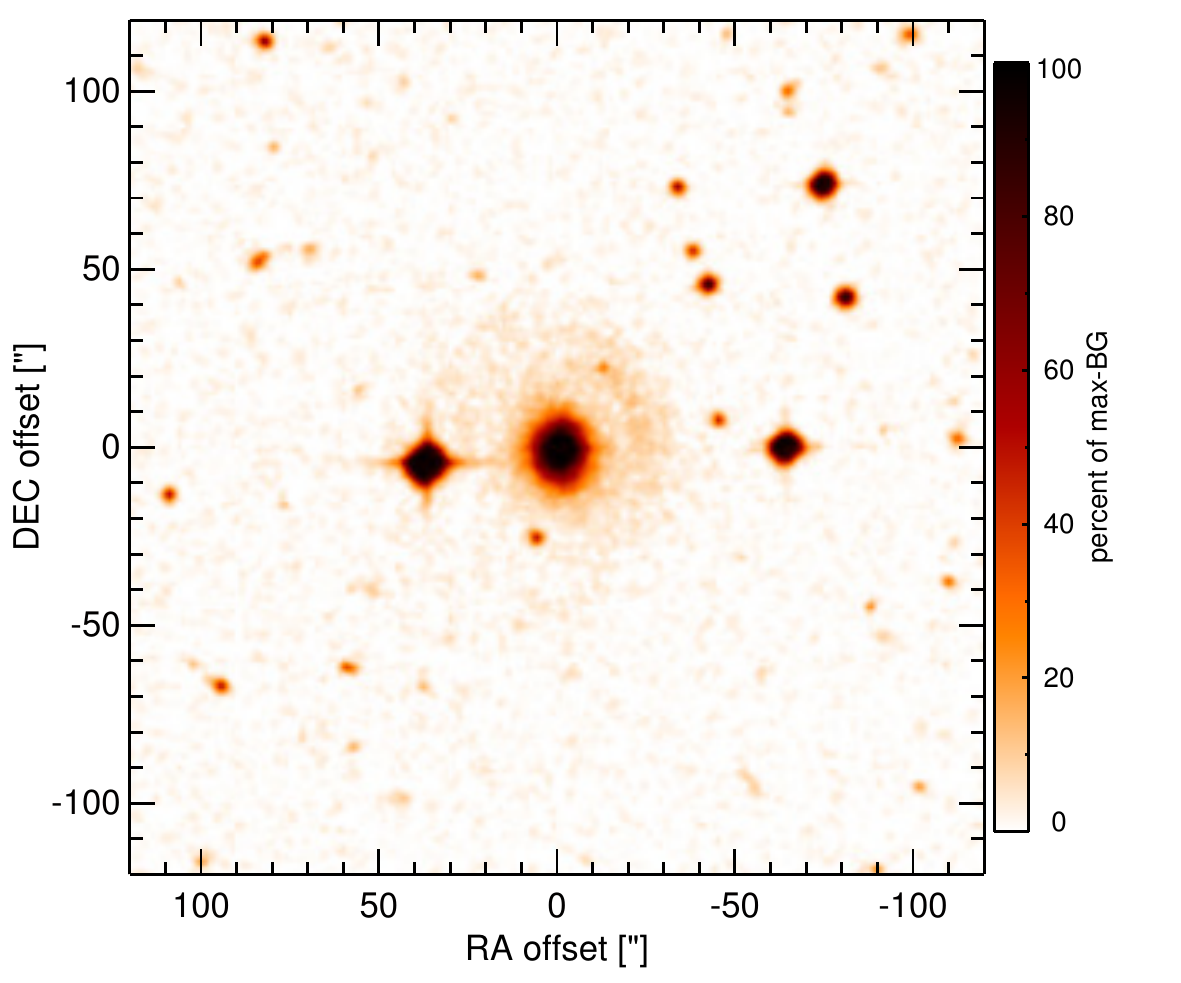}
    \caption{\label{fig:OPTim_Ark120}
             Optical image (DSS, red filter) of Ark\,120. Displayed are the central $4\arcmin$ with North up and East to the left. 
              The colour scaling is linear with white corresponding to the median background and black to the $0.01\%$ pixels with the highest intensity.  
           }
\end{figure}
\begin{figure}
   \centering
   \includegraphics[angle=0,height=3.11cm]{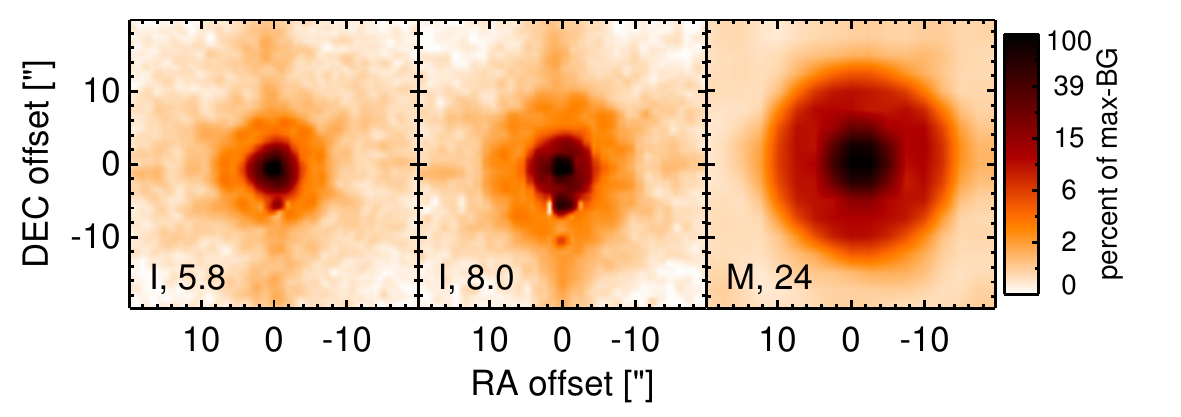}
    \caption{\label{fig:INTim_Ark120}
             \spitzerr MIR images of Ark\,120. Displayed are the inner $40\arcsec$ with North up and East to the left. The colour scaling is logarithmic with white corresponding to median background and black to the $0.1\%$ pixels with the highest intensity.
             The label in the bottom left states instrument and central wavelength of the filter in $\mu$m (I: IRAC, M: MIPS). 
             Note that the apparent off-nuclear compact sources in the IRAC $5.8$ and $8.0\,\mu$m images are instrumental artefacts. 
           }
\end{figure}
\begin{figure}
   \centering
   \includegraphics[angle=0,width=8.500cm]{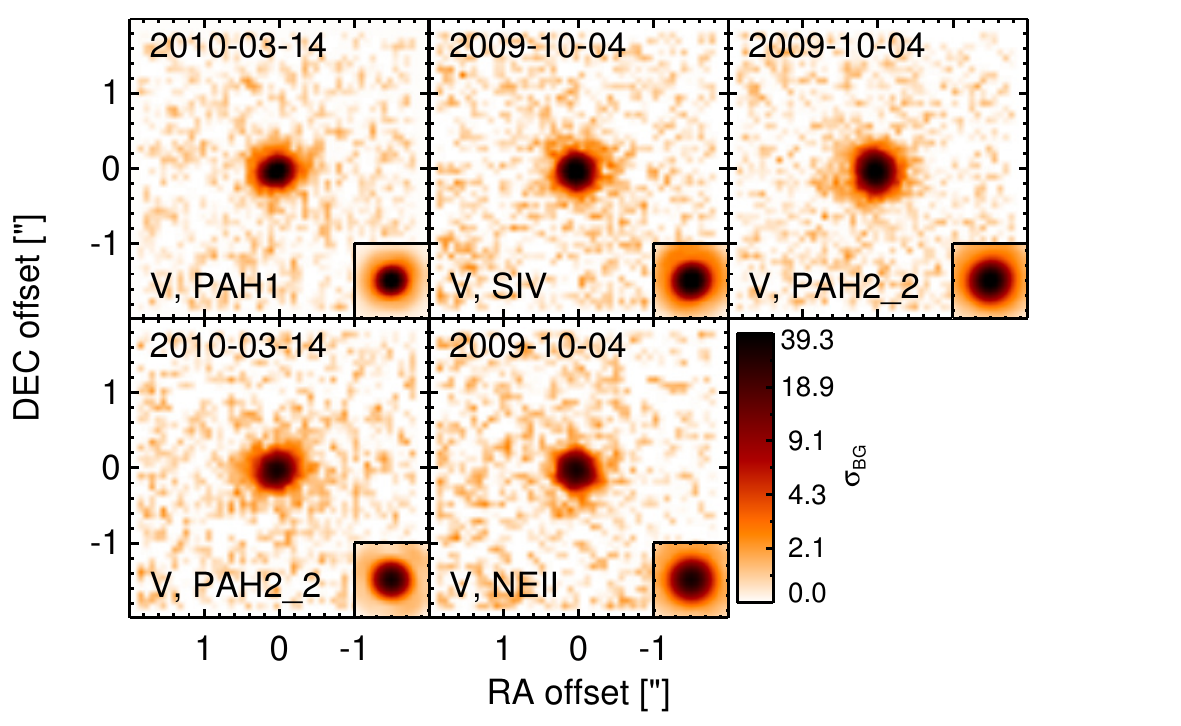}
    \caption{\label{fig:HARim_Ark120}
             Subarcsecond-resolution MIR images of Ark\,120 sorted by increasing filter wavelength. 
             Displayed are the inner $4\arcsec$ with North up and East to the left. 
             The colour scaling is logarithmic with white corresponding to median background and black to the $75\%$ of the highest intensity of all images in units of $\sigbg$.
             The inset image shows the central arcsecond of the PSF from the calibrator star, scaled to match the science target.
             The labels in the bottom left state instrument and filter names (C: COMICS, M: Michelle, T: T-ReCS, V: VISIR).
           }
\end{figure}
\begin{figure}
   \centering
   \includegraphics[angle=0,width=8.50cm]{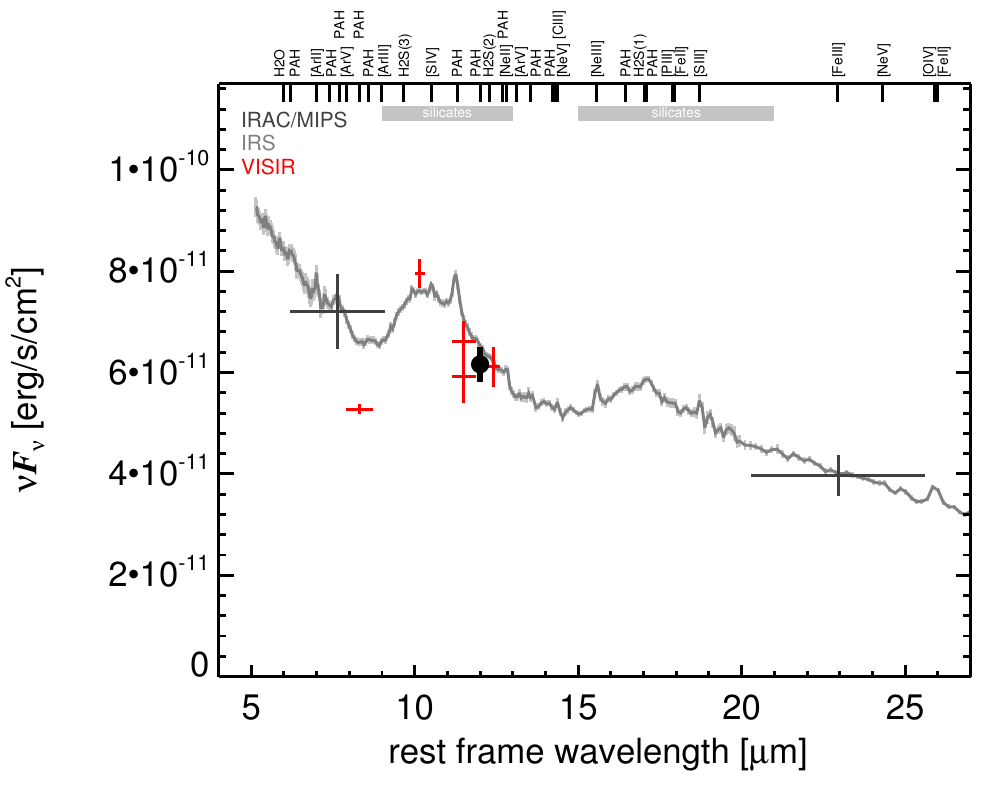}
   \caption{\label{fig:MISED_Ark120}
      MIR SED of Ark\,120. The description  of the symbols (if present) is the following.
      Grey crosses and  solid lines mark the \spitzer/IRAC, MIPS and IRS data. 
      The colour coding of the other symbols is: 
      green for COMICS, magenta for Michelle, blue for T-ReCS and red for VISIR data.
      Darker-coloured solid lines mark spectra of the corresponding instrument.
      The black filled circles mark the nuclear 12 and $18\,\mu$m  continuum emission estimate from the data.
      The ticks on the top axis mark positions of common MIR emission lines, while the light grey horizontal bars mark wavelength ranges affected by the silicate 10 and 18$\mu$m features.     
   }
\end{figure}
\clearpage

\twocolumn[\begin{@twocolumnfalse}  
\subsection{Cen\,A -- NGC\,5128}\label{app:CenA}
Cen\,A is the closest and one of the best-studied AGN, hosted in a S0 peculiar galaxy ($D = 3.8 \pm 0.1$\,Mpc; \citealt{harris_distance_2010}; see \citealt{israel_centaurus_1998,morganti_many_2010} for reviews).
It is usually treated as a radio-loud Sy\,2 (e.g., NED), albeit being completely obscured in the optical. 
Cen\,A has a supergalactic-scale FR\,I radio morphology with biconical lobes and a twin jet (inner PA$\sim55\degree$; e.g., \citealt{schreier_detection_1981}). 
The source may be an ongoing major merger between a massive elliptical and a spiral galaxy, as indicated by the peculiar morphology such as the warped lane of dust and gas (PA$\sim122\degree$; \citealt{dufour_picture_1979}).
Cen\,A was among the first AGN to be observed at MIR wavelengths, with repeated $N$-band photometry available over several decades \citep{becklin_infrared_1971, kleinmann_10-micron_1974, grasdalen_infrared_1976, telesco_extended_1978, frogel_8-13_1982, krabbe_n-band_2001, karovska_spatially_2003, siebenmorgen_mid-infrared_2004}.
It was also observed with \spitzer/IRAC, IRS and MIPS multiple times.
The host galaxy, in particular the warped dust disc, and the large-scale jet are clearly visible in the MIPS image \citep{quillen_spitzer_2006, hardcastle_infrared_2006}.
The PBCD IRAC images are saturated in the nuclei and, thus, are not used for the nuclear photometry.
The IRS LR staring-mode  spectra show deep silicate (10\,$\mu$m) absorption and weak PAH emission (see also \citealt{weedman_mid-infrared_2005}). 
Cen\,A was also among the first AGN to be observed with 8-meter class telescopes in the MIR.
\cite{hardcastle_infrared_2006} and \cite{radomski_gemini_2008} presented T-ReCS $N-$ and $Q$-band imaging taken in 2004 in various filters. they found an unresolved nucleus embedded in diffuse extended emission.
In addition, VISIR $N-$ and $Q$-band imaging was performed by \cite{horst_mid_2008}, \cite{reunanen_vlt_2010} and \cite{van_der_wolk_dust_2010} in 2006.
We also analysed VISIR $N$-band images taken in 2010 that are unpublished, to our knowledge.
The nuclear morphology agrees with the one found in the pioneering T-ReCS observations in all cases, although the extended component is not always clearly detected.
The resulting nuclear fluxes of our reanalysis of the T-ReCS and VISIR images generally agree with the literature values when using similar measurement methods. 
\cite{gonzalez-martin_dust_2013} published a T-ReCS $N$-band spectrum, which is similar to the IRS spectrum but without PAH emission features and with marginally lower continuum flux. 
Combining all the high-angular resolution measurements, there is  dispersion on the order of $\sim 20\%$ that possibly indicates  variability of the nuclear MIR emission of Cen\,A.
This indicates that the MIR emission may be affected by synchrotron emission from the jet base. 
Recent MIR interferometric observations of Cen\,A with MIDI show an unresolved ($<1$\,pc) central source that is attributed to jet emission, and an extended possibly disc-like component, perpendicular to the jet axis \citep{meisenheimer_resolving_2007,burtscher_resolving_2010,burtscher_diversity_2013}.
\newline\end{@twocolumnfalse}]

\begin{figure}
   \centering
   \includegraphics[angle=0,width=8.500cm]{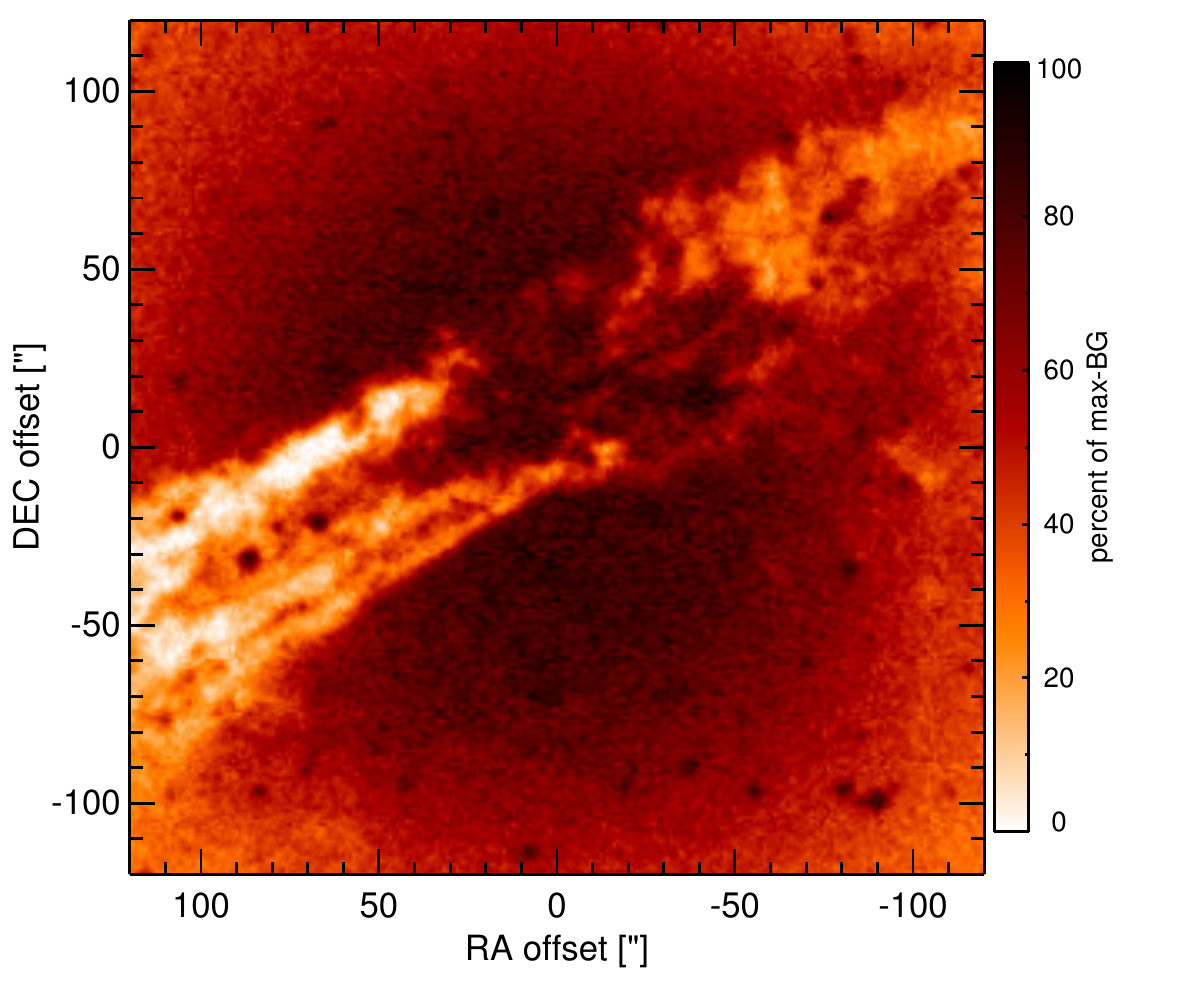}
    \caption{\label{fig:OPTim_CenA}
             Optical image (DSS, red filter) of Cen\,A. Displayed are the central $4\arcmin$ with North up and East to the left. 
              The colour scaling is linear with white corresponding to the median background and black to the $0.01\%$ pixels with the highest intensity.  
           }
\end{figure}
\begin{figure}
   \centering
   \includegraphics[angle=0,height=3.11cm]{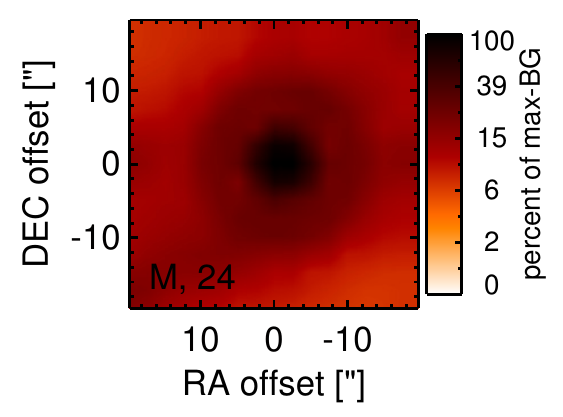}
    \caption{\label{fig:INTim_CenA}
             \spitzerr MIR images of Cen\,A. Displayed are the inner $40\arcsec$ with North up and East to the left. The colour scaling is logarithmic with white corresponding to median background and black to the $0.1\%$ pixels with the highest intensity.
             The label in the bottom left states instrument and central wavelength of the filter in $\mu$m (I: IRAC, M: MIPS). 
           }
\end{figure}
\begin{figure}
   \centering
   \includegraphics[angle=0,width=8.500cm]{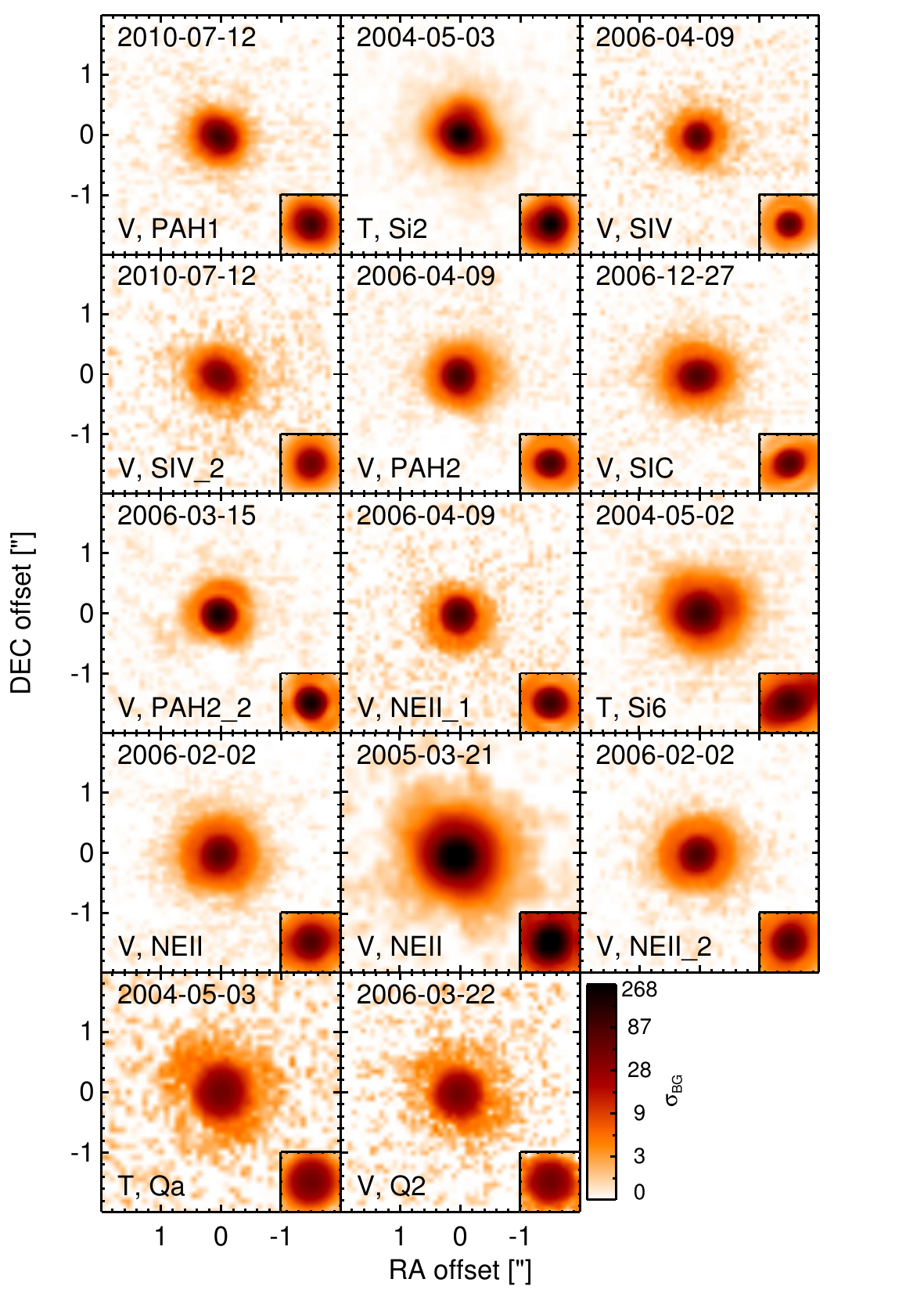}
    \caption{\label{fig:HARim_CenA}
             Subarcsecond-resolution MIR images of Cen\,A sorted by increasing filter wavelength. 
             Displayed are the inner $4\arcsec$ with North up and East to the left. 
             The colour scaling is logarithmic with white corresponding to median background and black to the $75\%$ of the highest intensity of all images in units of $\sigbg$.
             The inset image shows the central arcsecond of the PSF from the calibrator star, scaled to match the science target.
             The labels in the bottom left state instrument and filter names (C: COMICS, M: Michelle, T: T-ReCS, V: VISIR).
           }
\end{figure}
\begin{figure}
   \centering
   \includegraphics[angle=0,width=8.50cm]{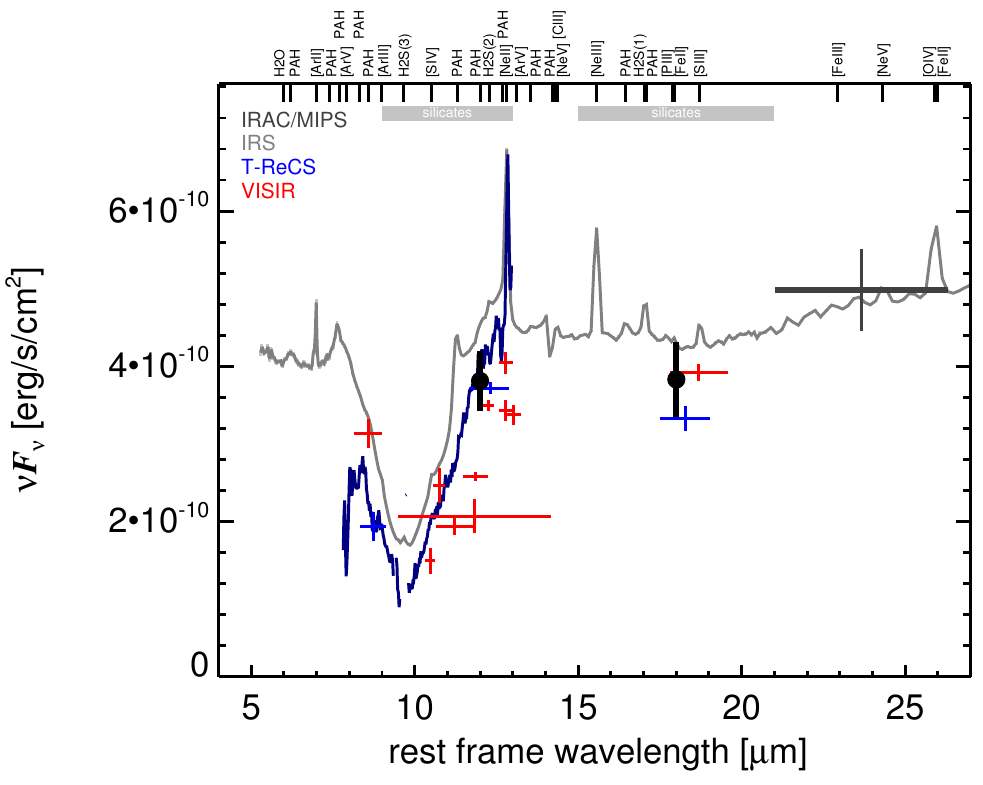}
   \caption{\label{fig:MISED_CenA}
      MIR SED of Cen\,A. The description  of the symbols (if present) is the following.
      Grey crosses and  solid lines mark the \spitzer/IRAC, MIPS and IRS data. 
      The colour coding of the other symbols is: 
      green for COMICS, magenta for Michelle, blue for T-ReCS and red for VISIR data.
      Darker-coloured solid lines mark spectra of the corresponding instrument.
      The black filled circles mark the nuclear 12 and $18\,\mu$m  continuum emission estimate from the data.
      The ticks on the top axis mark positions of common MIR emission lines, while the light grey horizontal bars mark wavelength ranges affected by the silicate 10 and 18$\mu$m features.     
   }
\end{figure}
\clearpage

\twocolumn[\begin{@twocolumnfalse}  
\subsection{Circinus -- ESO\,97-13}\label{app:Circinus}
The highly inclined Circinus spiral galaxy, which lies within $10\degree\,$ of the Galactic plane, harbours one of the nearest AGN at $\sim 4.2 \pm 0.8$\,Mpc \citep{tully_extragalactic_2009}.
It is optically classified as a Sy\,2 with polarized broad lines \citep{oliva_spectropolarimetry_1998}.
A nuclear water-maser disc (PA$\sim56\degree$; \citealt{greenhill_warped_2003}) and a recent starburst \citep{maiolino_seyfert_1998,sanchez_sinfoni_2006} have been found on parsec-scales, while a one-sided kiloparsec-scale ionization cone is also present in this object (PA$\sim-44\degree$; \citealt{marconi_prominent_1994,wilson_hubble_2000}).
The first ground-based MIR observations were preformed by \cite{moorwood_infrared_1984}, and first $N$-band images were taken by \cite{siebenmorgen_origin_1997,siebenmorgen_mid-infrared_2004} and \cite{krabbe_n-band_2001} with ESO 3.6m/TIMMI, TIMMI2 and ESO MPI 2.2m/MANIAC respectively.
The images show a nuclear point source with weak diffuse emission and marginally-extended emission on 20\,pc (1\arcsec) scale in  the north-south directions.
However, this morphology could not be verified by \cite{galliano_mid-infrared_2005} based on their $N$- and $Q$-band TIMMI2 images.
Circinus was also observed with \spitzer/IRAC, IRS and MIPS but its nucleus is saturated in all IRAC and MIPS PBCD images and thus, they are not analysed.
The IRS LR staring-mode PBCD spectrum is also partly saturated but shows a steeply rising spectral slope towards longer wavelengths and a deep silicate $10\,\mu$m absorption feature without strong PAH emission.  
These spectral characteristics are verified to be unchanged for the nucleus at subarcsecond scales by the T-ReCS spectra published in \cite{roche_mid-infrared_2006}.
The T-ReCS and VISIR $N$- and $Q$-band images obtained by \cite{packham_extended_2005} and \cite{reunanen_vlt_2010} show an east-west elongation with $\sim4\arcsec$ diameter roughly coincident with the ionization cones (PA$\sim100\degree$). 
Our reanalysis of these T-ReCS and VISIR images provides consistent results when using similar measurement methods.
In addition, we analyse 20 VISIR $N$- and $Q$-band images (unpublished, as far as we are aware). 
Owing to the extended emission evident in the images, we measure only the unresolved nuclear fluxes.
Note that the T-ReCS N and VISIR Q2 measurements are treated as upper limit on the unresolved flux because they lack a suitable standard star observations for the PSF determination.
The average photometric flux level is $\sim 30\%$ lower than that of the IRS and T-ReCS spectra from 2004, the latter taken from \cite{gonzalez-martin_dust_2013}. 
Therefore, we do not attempt to correct our 12 and $18\,\mu$m continuum flux estimates for the silicate features.
The variation in the individual photometric measurements of the different epochs indicates MIR variability in Circinus but requires a more detailed analysis.
Note that MIR interferometric measurements of Circinus further resolve its nuclear MIR structure to consist of a polar elongated extended structure, and a compact flattened disc-like structure perpendicular to the ionization cones (\citealt{tristram_resolving_2007,tristram_dusty_2013}).
\newline\end{@twocolumnfalse}]

\begin{figure}
   \centering
   \includegraphics[angle=0,width=8.500cm]{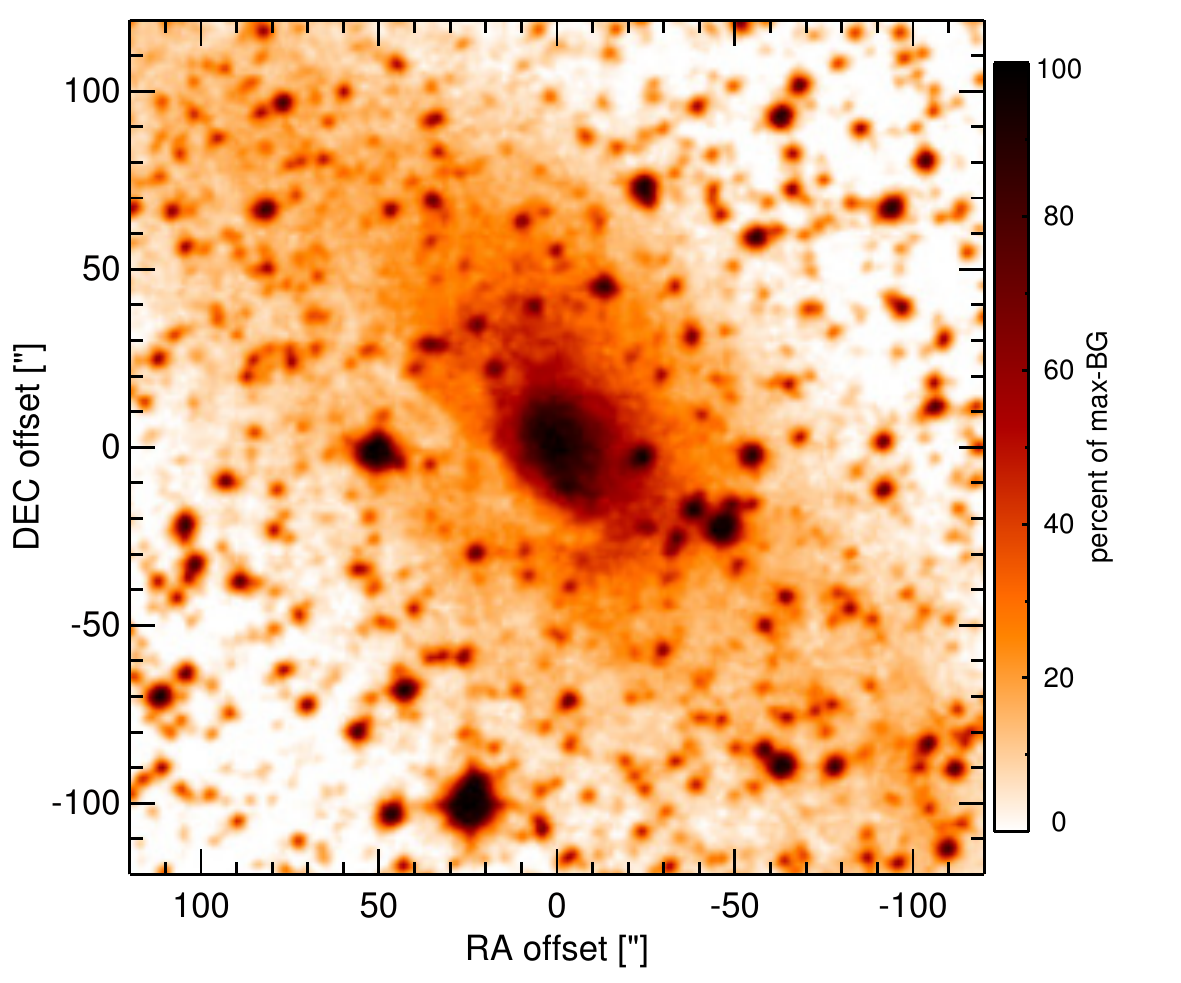}
    \caption{\label{fig:OPTim_Circinus}
             Optical image (DSS, red filter) of Circinus. Displayed are the central $4\arcmin$ with North up and East to the left. 
              The colour scaling is linear with white corresponding to the median background and black to the $0.01\%$ pixels with the highest intensity.  
           }
\end{figure}
\begin{figure}
   \centering
   \includegraphics[angle=0,height=3.11cm]{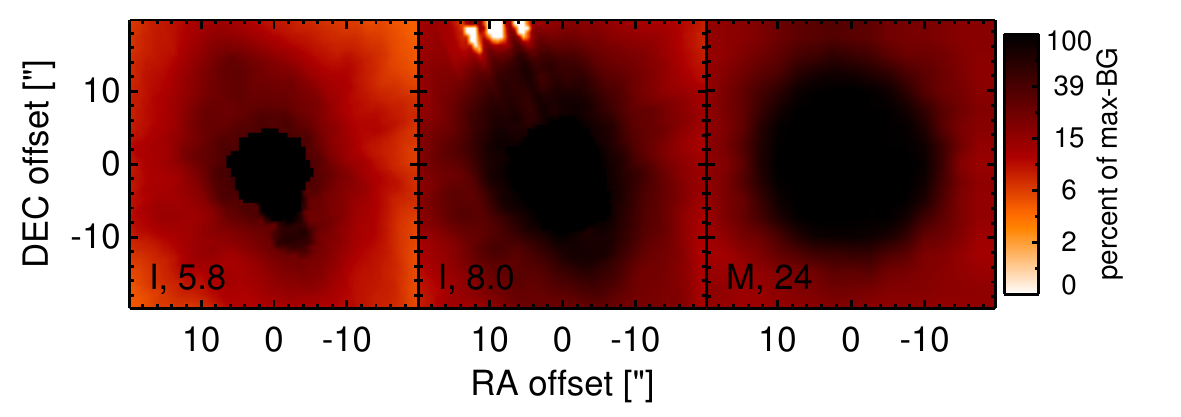}
    \caption{\label{fig:INTim_Circinus}
             \spitzerr MIR images of Circinus. Displayed are the inner $40\arcsec$ with North up and East to the left. The colour scaling is logarithmic with white corresponding to median background and black to the $0.1\%$ pixels with the highest intensity.
             The label in the bottom left states instrument and central wavelength of the filter in $\mu$m (I: IRAC, M: MIPS).
             Note that the images are completely saturated in the centres.
           }
\end{figure}
\begin{figure}
   \centering
   \includegraphics[angle=0,width=8.500cm]{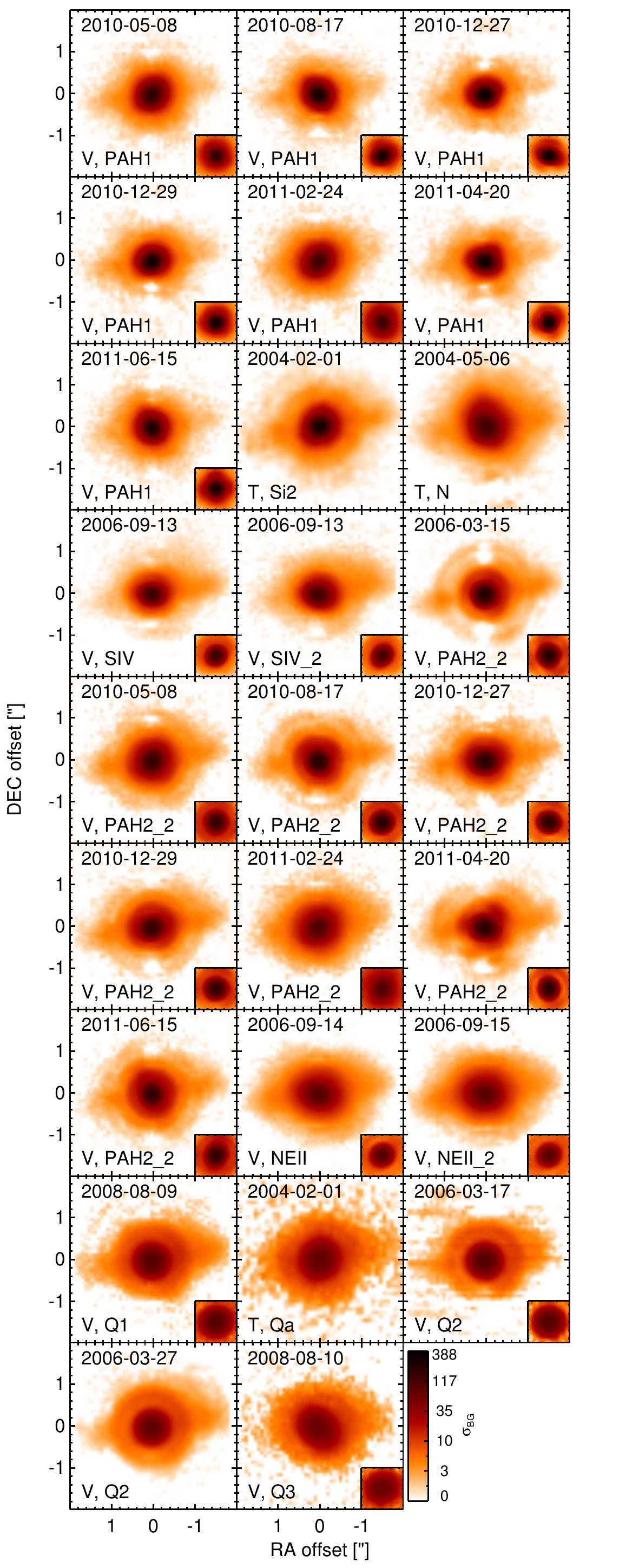}
    \caption{\label{fig:HARim_Circinus}
             Subarcsecond-resolution MIR images of Circinus sorted by increasing filter wavelength. 
             Displayed are the inner $4\arcsec$ with North up and East to the left. 
             The colour scaling is logarithmic with white corresponding to median background and black to the $75\%$ of the highest intensity of all images in units of $\sigbg$.
             The inset image shows the central arcsecond of the PSF from the calibrator star, scaled to match the science target.
             The labels in the bottom left state instrument and filter names (C: COMICS, M: Michelle, T: T-ReCS, V: VISIR).
           }
\end{figure}
\begin{figure}
   \centering
   \includegraphics[angle=0,width=8.50cm]{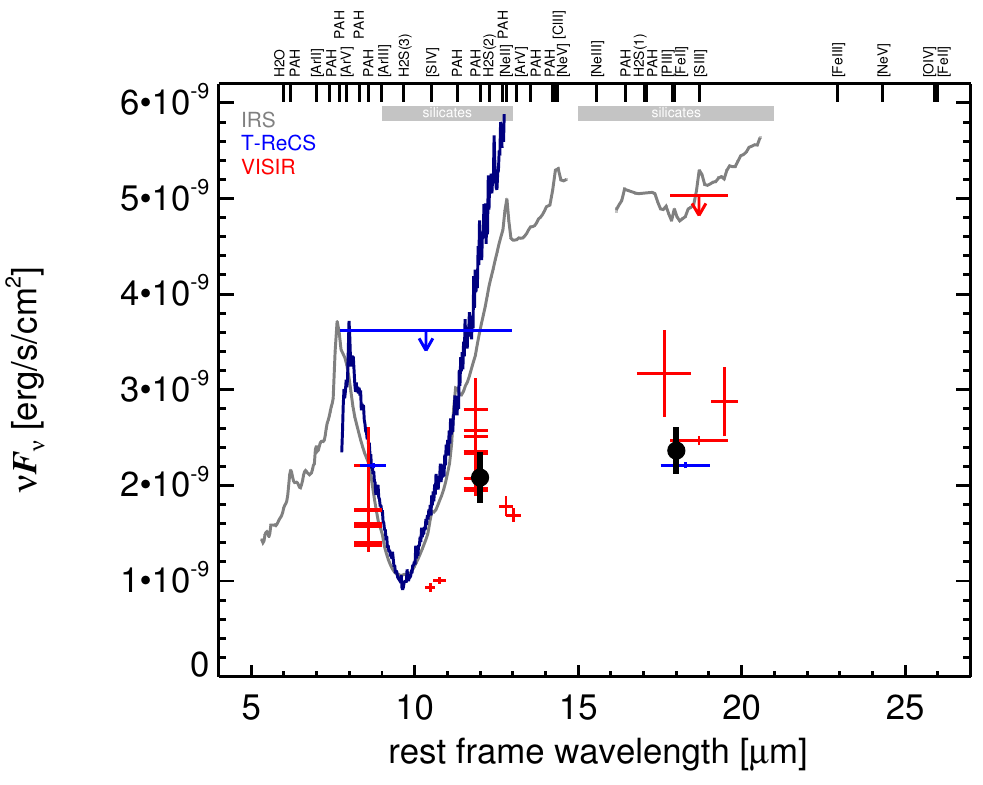}
   \caption{\label{fig:MISED_Circinus}
      MIR SED of Circinus. The description  of the symbols (if present) is the following.
      Grey crosses and  solid lines mark the \spitzer/IRAC, MIPS and IRS data. 
      The colour coding of the other symbols is: 
      green for COMICS, magenta for Michelle, blue for T-ReCS and red for VISIR data.
      Darker-coloured solid lines mark spectra of the corresponding instrument.
      The black filled circles mark the nuclear 12 and $18\,\mu$m  continuum emission estimate from the data.
      The ticks on the top axis mark positions of common MIR emission lines, while the light grey horizontal bars mark wavelength ranges affected by the silicate 10 and 18$\mu$m features.     
   }
\end{figure}
\clearpage

\begin{figure}
   \centering
   \includegraphics[angle=0,width=8.50cm]{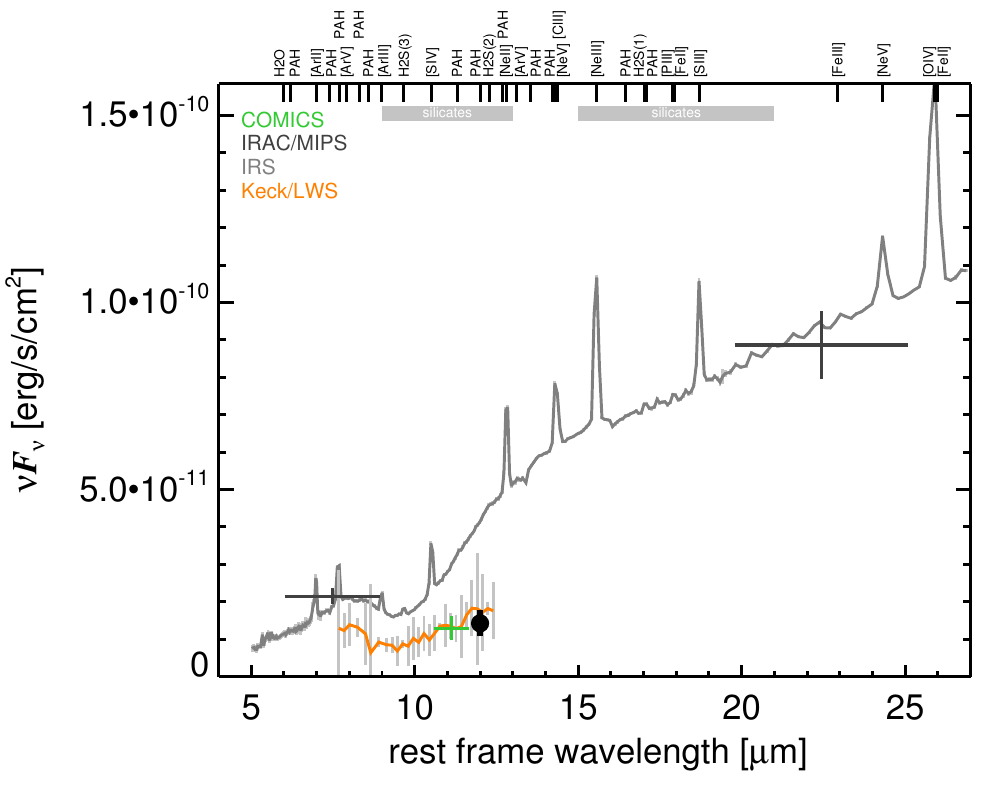}
   \caption{\label{fig:MISED_CygnusA}
      MIR SED of Cygnus\,A. The description  of the symbols (if present) is the following.
      Grey crosses and  solid lines mark the \spitzer/IRAC, MIPS and IRS data. 
      The colour coding of the other symbols is: 
      green for COMICS, magenta for Michelle, blue for T-ReCS and red for VISIR data.
      Darker-coloured solid lines mark spectra of the corresponding instrument.
      The black filled circles mark the nuclear 12 and $18\,\mu$m  continuum emission estimate from the data.
      The ticks on the top axis mark positions of common MIR emission lines, while the light grey horizontal bars mark wavelength ranges affected by the silicate 10 and 18$\mu$m features.     
   }
\end{figure}

\twocolumn[\begin{@twocolumnfalse}  
\subsection{Cygnus\,A  --   3C\,405 -- MCG+7-41-3}\label{app:CygnusA}
Cygnus\,A is a FR\,II radio source identified with the dusty elliptical galaxy MCG+7-41-3 at a redshift of $z=$ 0.0561 ($D \sim 257$\,Mpc; see \citealt{carilli_cygnus_1996} for a review).
It contains a Sy\,2 nucleus \citep{tueller_swift_2008} with broad emission lines in polarized light \citep{ogle_scattered_1997}.
It features the classical supergalactic-scale biconical radio lobes fed by highly collimated jets (PA$\sim105\degree$; e.g., \citealt{perley_jet_1984}), and kiloparsec-scale ionization cones parallel to the jet axis \citep{jackson_cygnus_1998,canalizo_adaptive_2003}.
Interestingly, recently-discovered evidence indicates previous phases of AGN activity in Cygnus\,A \citep{steenbrugge_multiwavelength_2008,steenbrugge_multiwavelength_2010,chon_discovery_2012}.
Early ground-based $N$-band observations were performed by \cite{rieke_infrared_1972}, followed by \irass \citep{edelson_far-infrared_1987} and \isoo observations \citep{haas_far-infrared_1998}.
The first subarcsecond-scale MIR images were obtained with Keck/OSCIR in 1998 \citep{radomski_high-resolution_2002} and Keck/LWS in 1999 \citep{whysong_hidden_2001,whysong_thermal_2004}.
The images show a compact nucleus with V-shaped emission aligned with the ionization cone and extending $\sim 0.5\arcsec\sim0.6\,$kpc to the east. 
In addition, \cite{imanishi_9.7_2000} performed Keck/LWS LR $N$-band spectroscopy in 1999 and detected prominent silicate 10$\,\mu$m absorption without any PAH emission.
Cygnus\,A was also observed with \spitzer/IRAC, IRS and MIPS and appears compact in the corresponding images.  
Note that no IRAC $5.8\,\mu$m image is available.
Our nuclear IRAC 8.0\,$\mu$m flux is consistent with the value published in \cite{privon_modeling_2012}.
The IRS LR staring-mode spectrum shows silicate $10\,\mu$m absorption, strong forbidden emission lines and a red spectral slope in $\nu F_\nu$-space but no PAH features (see also \citealt{shi_9.7_2006,privon_modeling_2012}).
Cygnus\,A was observed with COMICS in the N11.7 filter in 2005, and a weak possibly extended nuclear MIR source was detected (FWHM $\sim 0.7$\,kpc). 
The low S/N of the detection renders this extension uncertain, but it is consistent with the Keck results.
Therefore, we classify the nucleus of Cygnus\,A as extended on subarcsecond scales in the MIR.
The corresponding unresolved nuclear N11.7 flux is consistent with the Keck results and  $\sim 58\%$ lower than the \spitzerr spectrophotometry.
In Fig.~\ref{fig:MISED_CygnusA}, we also show for comparison the Keck spectrum from \cite{imanishi_9.7_2000}, which has been scaled with the COMICS photometry. 
\newline\end{@twocolumnfalse}]

\begin{figure}
   \centering
   \includegraphics[angle=0,width=8.500cm]{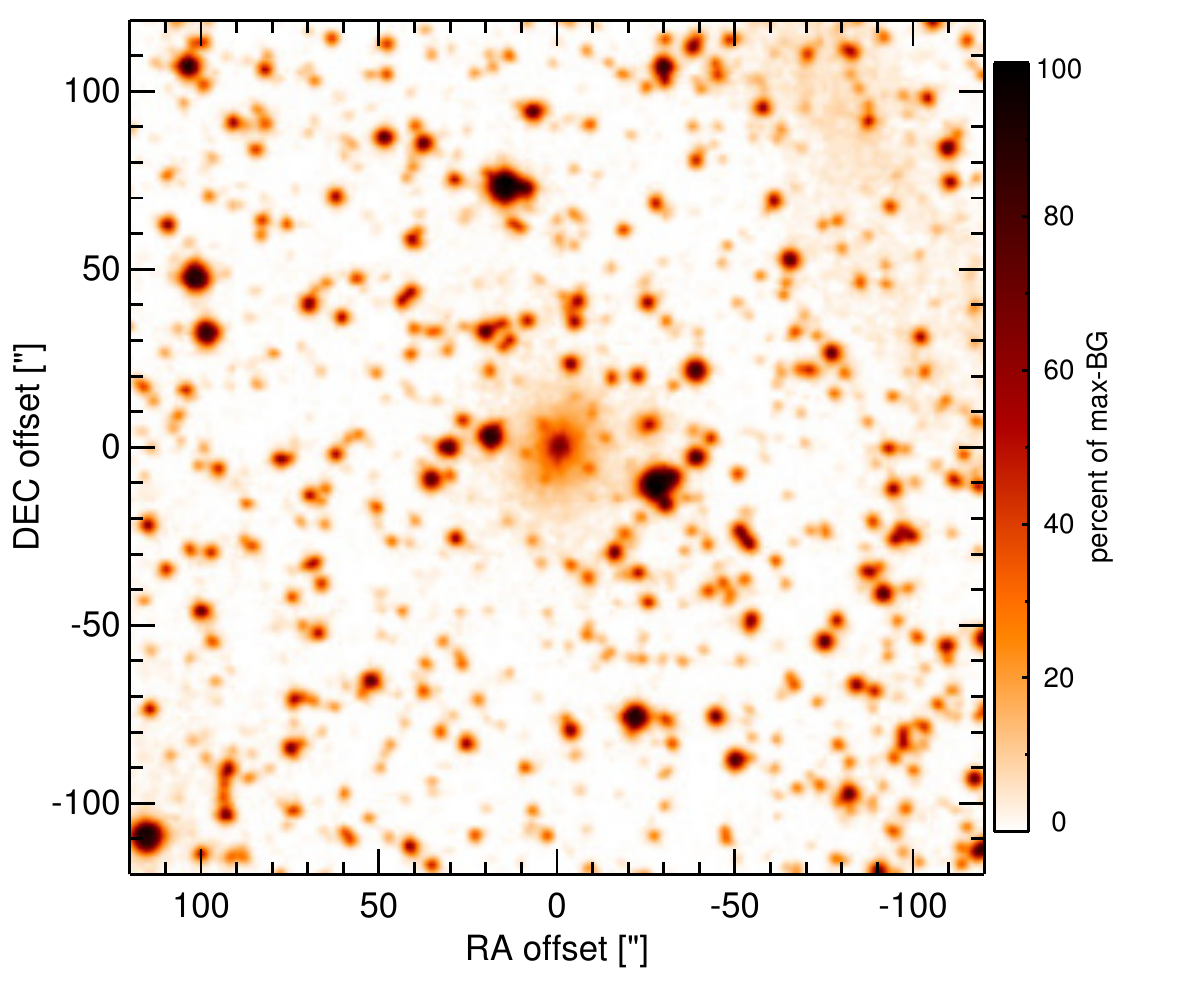}
    \caption{\label{fig:OPTim_CygnusA}
             Optical image (DSS, red filter) of Cygnus\,A. Displayed are the central $4\arcmin$ with North up and East to the left. 
              The colour scaling is linear with white corresponding to the median background and black to the $0.01\%$ pixels with the highest intensity.  
           }
\end{figure}
\begin{figure}
   \centering
   \includegraphics[angle=0,height=3.11cm]{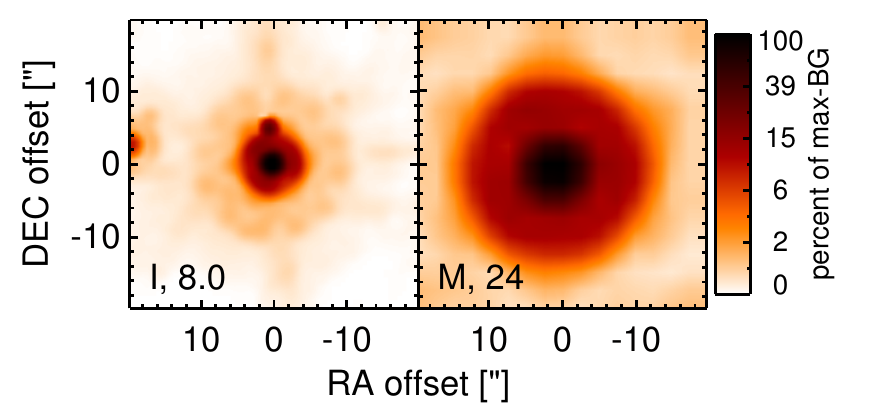}
    \caption{\label{fig:INTim_CygnusA}
             \spitzerr MIR images of Cygnus\,A. Displayed are the inner $40\arcsec$ with North up and East to the left. The colour scaling is logarithmic with white corresponding to median background and black to the $0.1\%$ pixels with the highest intensity.
             The label in the bottom left states instrument and central wavelength of the filter in $\mu$m (I: IRAC, M: MIPS). 
             Note that the apparent off-nuclear compact source in the IRAC $8.0\,\mu$m image is an instrumental artefact.
           }
\end{figure}
\begin{figure}
   \centering
   \includegraphics[angle=0,height=3.11cm]{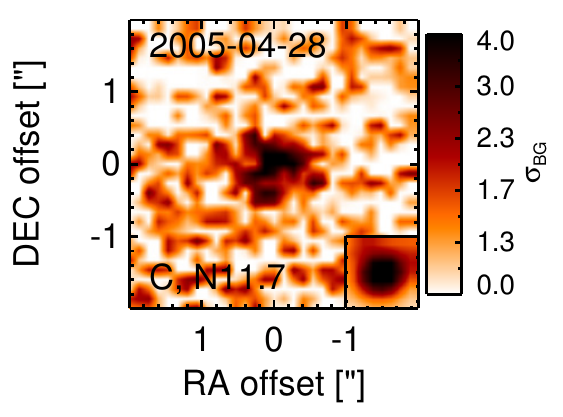}
    \caption{\label{fig:HARim_CygnusA}
             Subarcsecond-resolution MIR images of Cygnus\,A sorted by increasing filter wavelength. 
             Displayed are the inner $4\arcsec$ with North up and East to the left. 
             The colour scaling is logarithmic with white corresponding to median background and black to the $75\%$ of the highest intensity of all images in units of $\sigbg$.
             The inset image shows the central arcsecond of the PSF from the calibrator star, scaled to match the science target.
             The labels in the bottom left state instrument and filter names (C: COMICS, M: Michelle, T: T-ReCS, V: VISIR).
           }
\end{figure}
\clearpage

\twocolumn[\begin{@twocolumnfalse}  
\subsection{ESO\,5-4 -- Swift\,J0601.9-8636}\label{app:ESO005-G004}
ESO\,5-4 is an edge-on spiral galaxy at a distance of $D=$ 22.4\,Mpc \citep{tully_nearby_1988} hosting a Sy\,2 nucleus \citep{veron-cetty_catalogue_2010}.
It is a member of the nine-month BAT AGN sample and has only recently been discovered as an AGN \citep{landi_agn_2007}, which is presumably related to the high nuclear obscuration in this X-ray ``buried" AGN \citep{ueda_suzaku_2007}.
After first being detected in the MIR with \iras, ESO\,5-4 was also observed with \spitzer/IRAC and IRS.
Edge-on host galaxy emission and a compact nucleus are visible in the IRAC ($5.8$ and $8.0\,\mu$m) images.
Their nuclear photometry agrees with the IRS LR staring-mode spectrum well, which shows deep silicate 10$\,\mu$m absorption and strong PAH emission lines. 
The IRS spectrum resembles a typical star formation MIR SED. 
ESO\,5-4 was observed by us with VLT/VISIR in 2007 and 2011 in narrow $N$- and $Q$-band filters.
The first $N$-band photometry (NEII\_2) has already been published in \cite{gandhi_resolving_2009} with a lower flux than in our low-S/N optimized analyses presented.
Owing to the deep silicate absorption feature, the MIR nucleus has only been detected shortwards of 9$\,\mu$m and longwards of 11$\,\mu$m, where the average flux level is $\sim 62\%$ lower than the \spitzerr spectrophotometry.
The extension of this nuclear source remains uncertain because of the low detection significance, in particular at short wavelengths.
The high flux in the NEII\_2 filter indicates strong \neii $\lambda 12.8\,\mu$m emission from the nuclear region. 
However, note that the photometric sensitivity of this (and the SIV\_2) observation was $20\%$ worse than expected from the median sensitivity as measured from the corresponding standard star. 
Because of the uncertain nuclear MIR SED, we do not attempt to correct our 12$\,\mu$m continuum flux estimate for the silicate feature.
\newline\end{@twocolumnfalse}]

\begin{figure}
   \centering
   \includegraphics[angle=0,width=8.500cm]{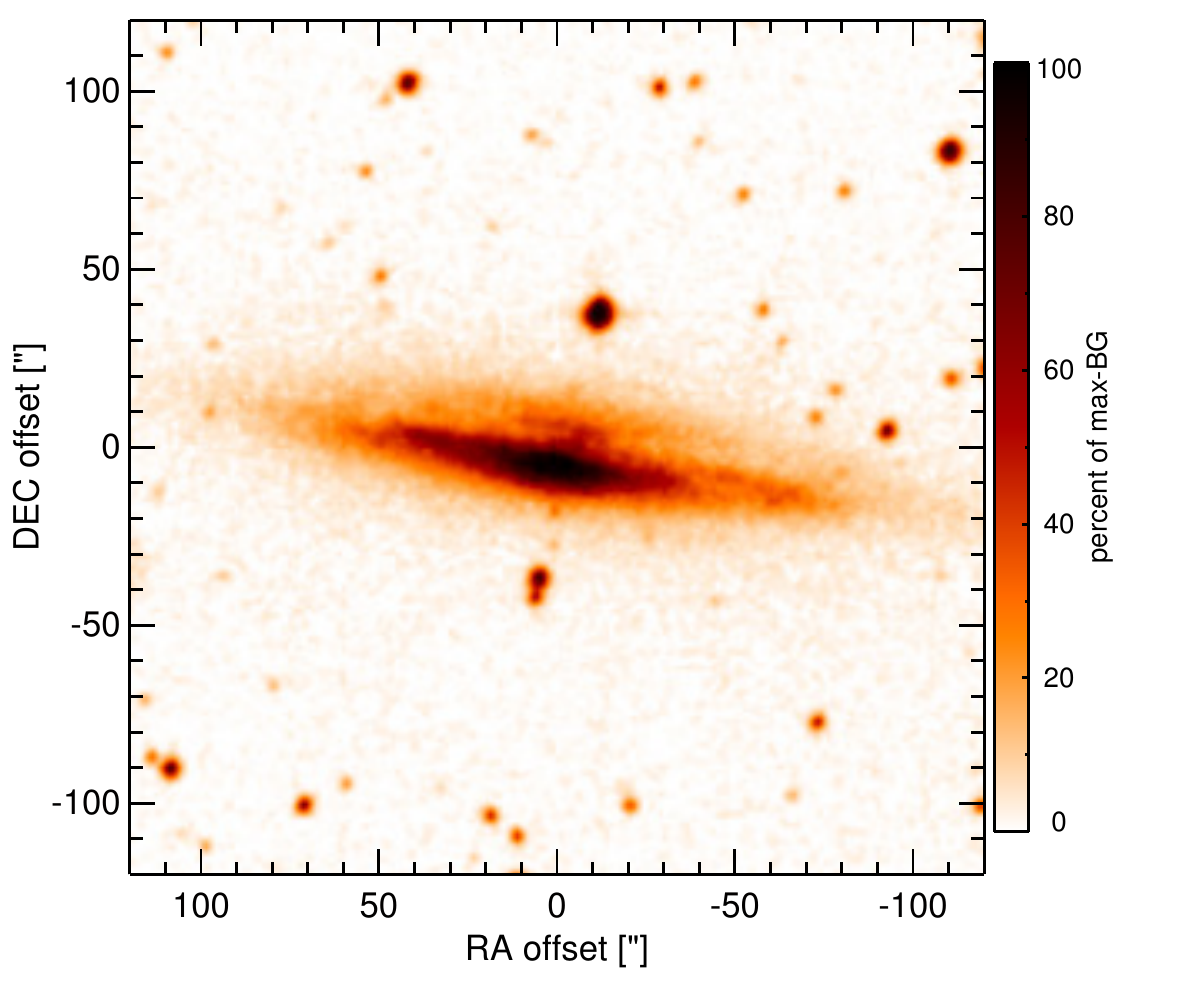}
    \caption{\label{fig:OPTim_ESO005-G004}
             Optical image (DSS, red filter) of ESO\,5-4. Displayed are the central $4\arcmin$ with North up and East to the left. 
              The colour scaling is linear with white corresponding to the median background and black to the $0.01\%$ pixels with the highest intensity.  
           }
\end{figure}
\begin{figure}
   \centering
   \includegraphics[angle=0,height=3.11cm]{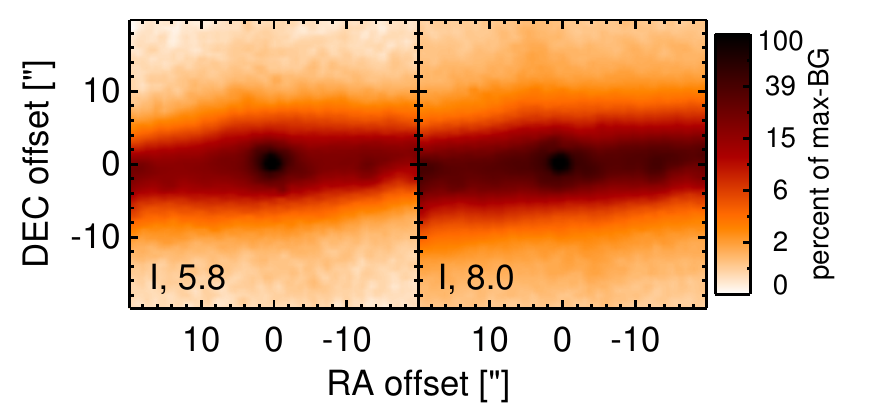}
    \caption{\label{fig:INTim_ESO005-G004}
             \spitzerr MIR images of ESO\,5-4. Displayed are the inner $40\arcsec$ with North up and East to the left. The colour scaling is logarithmic with white corresponding to median background and black to the $0.1\%$ pixels with the highest intensity.
             The label in the bottom left states instrument and central wavelength of the filter in $\mu$m (I: IRAC, M: MIPS). 
           }
\end{figure}
\begin{figure}
   \centering
   \includegraphics[angle=0,width=8.500cm]{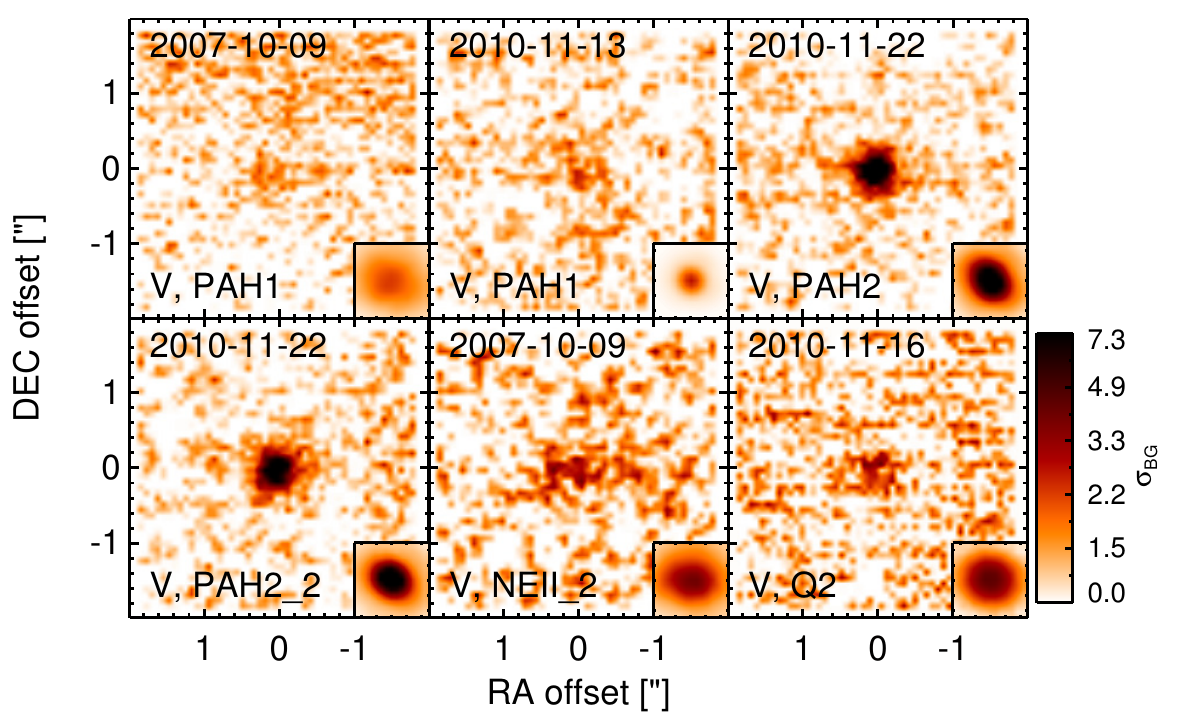}
    \caption{\label{fig:HARim_ESO005-G004}
             Subarcsecond-resolution MIR images of ESO\,5-4 sorted by increasing filter wavelength. 
             Displayed are the inner $4\arcsec$ with North up and East to the left. 
             The colour scaling is logarithmic with white corresponding to median background and black to the $75\%$ of the highest intensity of all images in units of $\sigbg$.
             The inset image shows the central arcsecond of the PSF from the calibrator star, scaled to match the science target.
             The labels in the bottom left state instrument and filter names (C: COMICS, M: Michelle, T: T-ReCS, V: VISIR).
           }
\end{figure}
\begin{figure}
   \centering
   \includegraphics[angle=0,width=8.50cm]{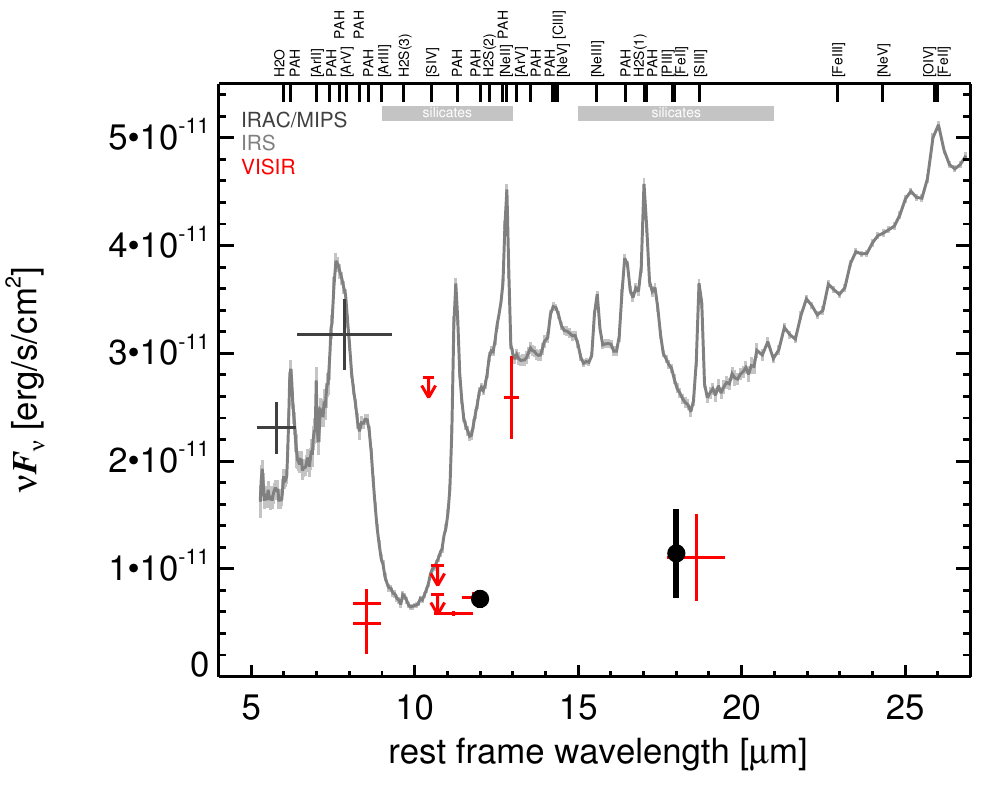}
   \caption{\label{fig:MISED_ESO005-G004}
      MIR SED of ESO\,5-4. The description  of the symbols (if present) is the following.
      Grey crosses and  solid lines mark the \spitzer/IRAC, MIPS and IRS data. 
      The colour coding of the other symbols is: 
      green for COMICS, magenta for Michelle, blue for T-ReCS and red for VISIR data.
      Darker-coloured solid lines mark spectra of the corresponding instrument.
      The black filled circles mark the nuclear 12 and $18\,\mu$m  continuum emission estimate from the data.
      The ticks on the top axis mark positions of common MIR emission lines, while the light grey horizontal bars mark wavelength ranges affected by the silicate 10 and 18$\mu$m features.     
   }
\end{figure}
\clearpage

\twocolumn[\begin{@twocolumnfalse}  
\subsection{ESO\,33-2 -- IRAS\,04575-7537}\label{app:ESO033-G002}
ESO\,33-2 is a spiral galaxy at a redshift of $z=$ 0.0181 ($D\sim 82.3\,$Mpc) hosting a radio-quiet Sy\,2 nucleus \citep{veron-cetty_catalogue_2010}.
The \oiii emission is extended by $\sim1.5\arcsec\sim0.6\,$kpc in  the north-south directions (PA$\sim5\degree$; \citealt{schmitt_hubble_2003}).
After first being detected in the MIR with \iras, ESO\,33-2 was observed with \spitzer/IRAC, IRS and MIPS and appears almost point-like without non-nuclear emission in all corresponding images.
IRAC photometry was published by \cite{gallimore_infrared_2010} and agrees with our IRAC $5.8$ and $8.0\,\mu$m measurements well.
The IRS LR mapping-mode spectrum displays weak silicate 10$\,\mu$m absorption, no significant PAH emission features, and a shallow emission peak at $\sim12\,\mu$m in $\nu F_\nu$-space (see also \citealt{buchanan_spitzer_2006,wu_spitzer/irs_2009,tommasin_spitzer_2008,tommasin_spitzer-irs_2010,gallimore_infrared_2010}). 
ESO\,33-2 was observed in the broad $N$-band filter with T-ReCS in 2004.
A compact nuclear source was detected. 
The emission is possibly extended (FWHM $\sim 0.85\arcsec \sim 300\,$pc), but this has to be verified with at least another epoch of subarcsecond MIR imaging.
The measured nuclear flux agrees with the \spitzerr spectrophotometry well, but it would be significantly lower if the presence of subarcsecond-extended emission can be verified.
\newline\end{@twocolumnfalse}]

\begin{figure}
   \centering
   \includegraphics[angle=0,width=8.500cm]{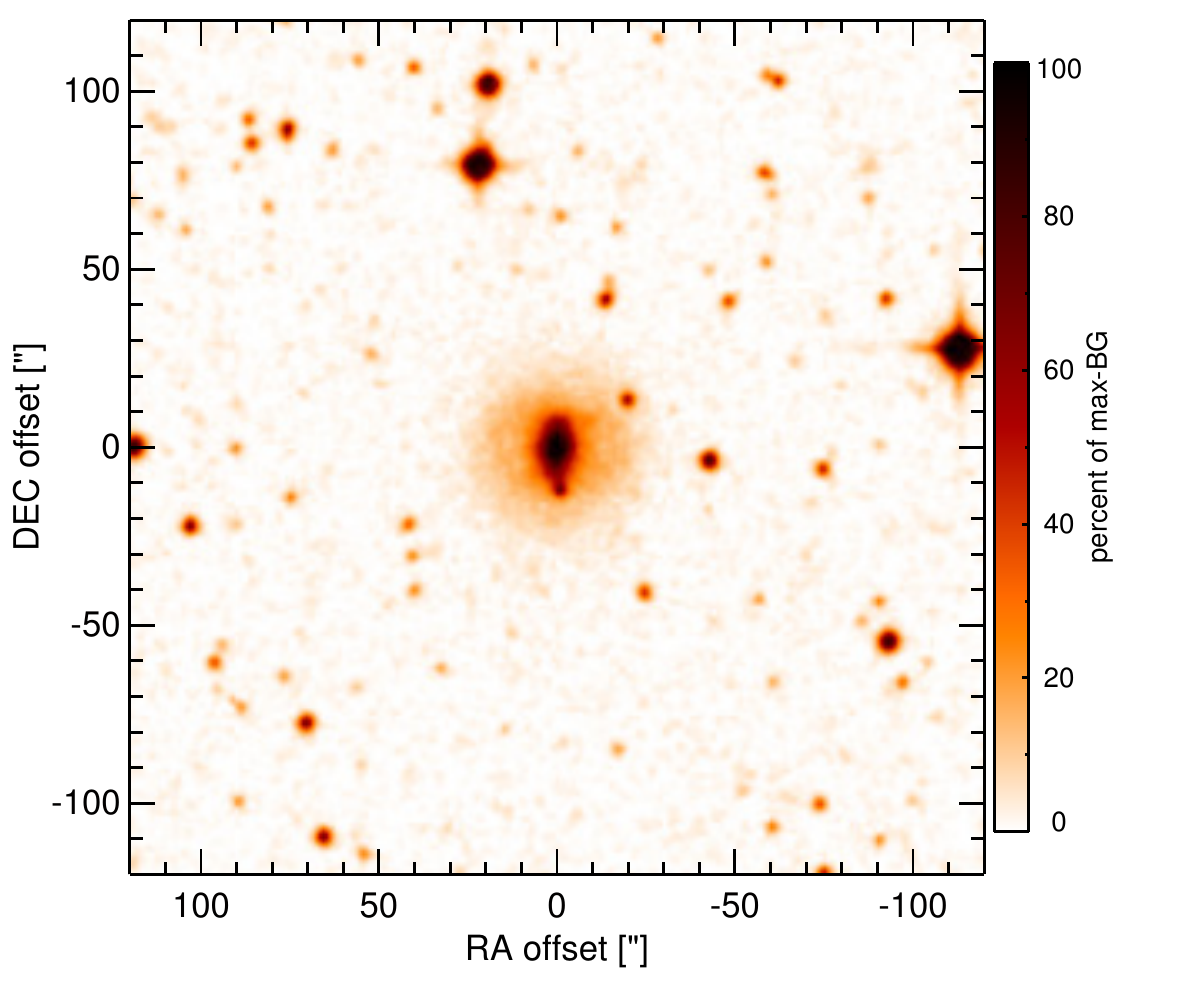}
    \caption{\label{fig:OPTim_ESO033-G002}
             Optical image (DSS, red filter) of ESO\,33-2. Displayed are the central $4\arcmin$ with North up and East to the left. 
              The colour scaling is linear with white corresponding to the median background and black to the $0.01\%$ pixels with the highest intensity.  
           }
\end{figure}
\begin{figure}
   \centering
   \includegraphics[angle=0,height=3.11cm]{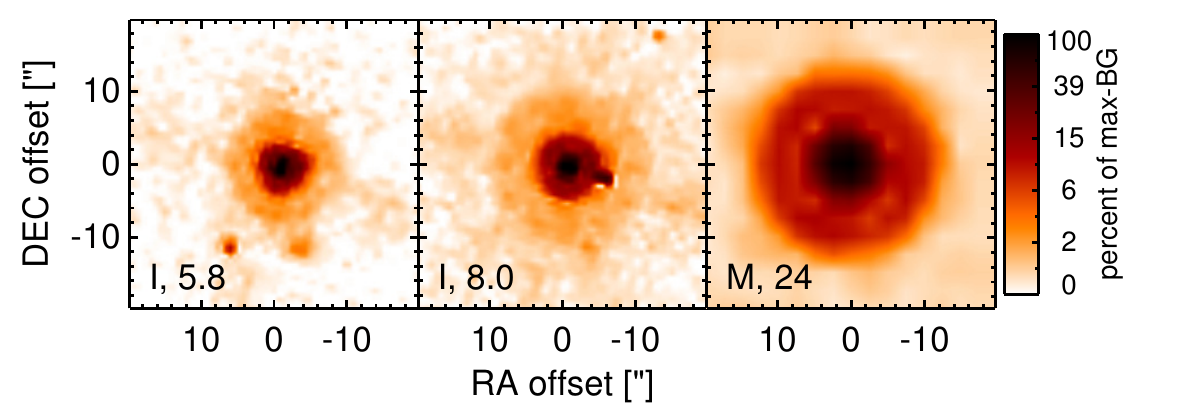}
    \caption{\label{fig:INTim_ESO033-G002}
             \spitzerr MIR images of ESO\,33-2. Displayed are the inner $40\arcsec$ with North up and East to the left. The colour scaling is logarithmic with white corresponding to median background and black to the $0.1\%$ pixels with the highest intensity.
             The label in the bottom left states instrument and central wavelength of the filter in $\mu$m (I: IRAC, M: MIPS). 
             Note that the apparent off-nuclear compact source in the IRAC $8.0\,\mu$m image is an instrumental artefact.
           }
\end{figure}
\begin{figure}
   \centering
   \includegraphics[angle=0,height=3.11cm]{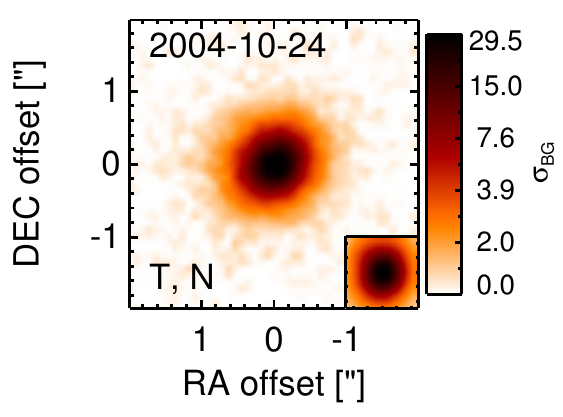}
    \caption{\label{fig:HARim_ESO033-G002}
             Subarcsecond-resolution MIR images of ESO\,33-2 sorted by increasing filter wavelength. 
             Displayed are the inner $4\arcsec$ with North up and East to the left. 
             The colour scaling is logarithmic with white corresponding to median background and black to the $75\%$ of the highest intensity of all images in units of $\sigbg$.
             The inset image shows the central arcsecond of the PSF from the calibrator star, scaled to match the science target.
             The labels in the bottom left state instrument and filter names (C: COMICS, M: Michelle, T: T-ReCS, V: VISIR).
           }
\end{figure}
\begin{figure}
   \centering
   \includegraphics[angle=0,width=8.50cm]{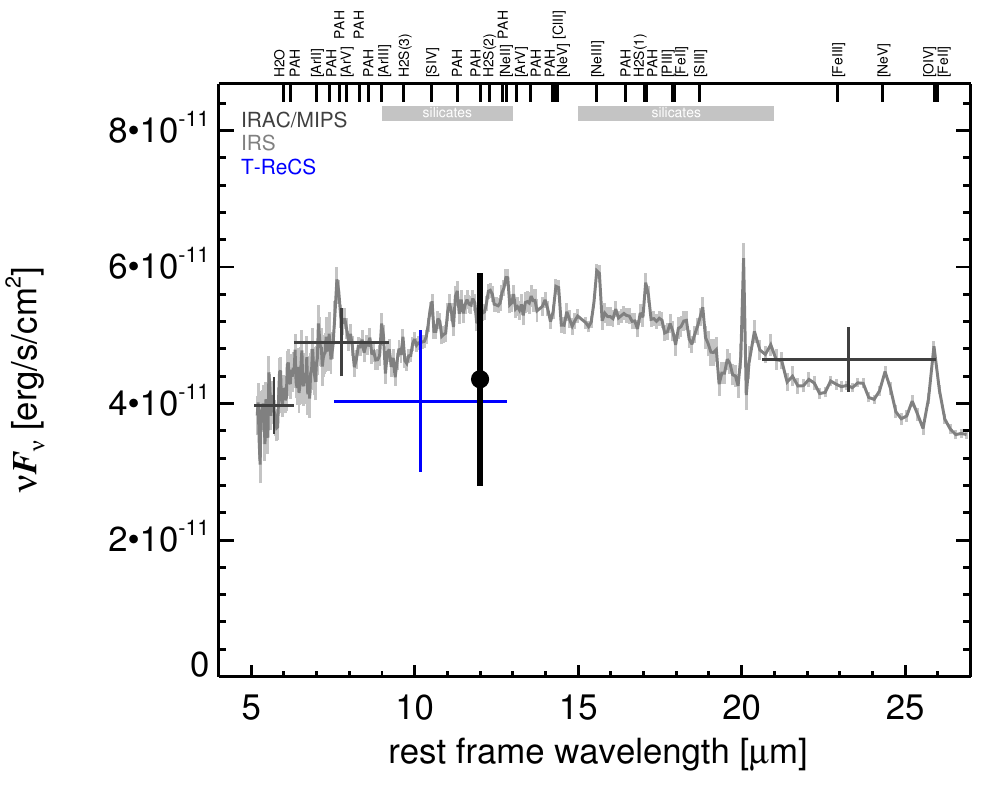}
   \caption{\label{fig:MISED_ESO033-G002}
      MIR SED of ESO\,33-2. The description  of the symbols (if present) is the following.
      Grey crosses and  solid lines mark the \spitzer/IRAC, MIPS and IRS data. 
      The colour coding of the other symbols is: 
      green for COMICS, magenta for Michelle, blue for T-ReCS and red for VISIR data.
      Darker-coloured solid lines mark spectra of the corresponding instrument.
      The black filled circles mark the nuclear 12 and $18\,\mu$m  continuum emission estimate from the data.
      The ticks on the top axis mark positions of common MIR emission lines, while the light grey horizontal bars mark wavelength ranges affected by the silicate 10 and 18$\mu$m features.     
   }
\end{figure}
\clearpage

\twocolumn[\begin{@twocolumnfalse}  
\subsection{ESO\,103-35 -- IRAS\,18333-6528}\label{app:ESO103-G035}
ESO\,103-35 is an inclined early-type  galaxy at a redshift of $z=$ 0.0133 ($D\sim 59.5\,$Mpc) hosting a radio-quiet Sy\,2 nucleus \citep{veron-cetty_catalogue_2010}.
It belongs to the BAT nine-month AGN sample and is an X-ray ``buried" AGN candidate \citep{noguchi_new_2009}.
The nucleus is obscured in the UV possibly by crossing dust lanes \citep{munoz_marin_atlas_2007}.
Furthermore, a H$_2$O mega-maser was discovered in the nucleus of ESO\,103-35 \citep{braatz_survey_1996}.
\cite{glass_long-term_2004} finds no significant variability in the infrared.
After first being detected in the MIR with \iras, a ground-based LR $N$-band spectrum was obtained with UKIRT \citep{roche_atlas_1991} before ESO\,103-35 was followed up with \spitzer/IRAC, IRS and MIPS.
It appears very compact in the \spitzerr images with no extended host galaxy emission detected.
The IRAC $8\,\mu$m PBCD image  is partly saturated and no nuclear flux measurement was performed. 
Our MIPS 24$\,\mu$m flux is significantly higher than that of \cite{temi_spitzer_2009} but in agreement with the other \spitzerr data.
The IRS LR staring-mode spectrum shows a deep silicate 10$\,\mu$m absorption feature and a red spectral slope in $\nu F_\nu$-space but  no PAH emission  (see also \citealt{shi_9.7_2006, mullaney_defining_2011}).
VISIR images in four $N$-band filters in a single epoch were taken and analysed by \cite{gandhi_resolving_2009}.
ESO\,103-35 is possibly elongated in the east-west directions (FWHM$\sim0.6\arcsec \sim 160\,$pc).
A T-ReCS spectrum from \cite{gonzalez-martin_dust_2013} agrees  with our VISIR photometry well. 
The subarcsecond flux level is $\sim 37\%$ lower than the arcsecond resolution level probed by \spitzer, while the SED shape is  similar at both scales.
We corrected our nuclear $12\,\mu$m continuum flux estimate for the silicate absorption by using the T-ReCS spectrum. 
Note however, that the nuclear flux level would be even lower if the presence of subarcsecond-extended emission can be verified.
\newline\end{@twocolumnfalse}]

\begin{figure}
   \centering
   \includegraphics[angle=0,width=8.500cm]{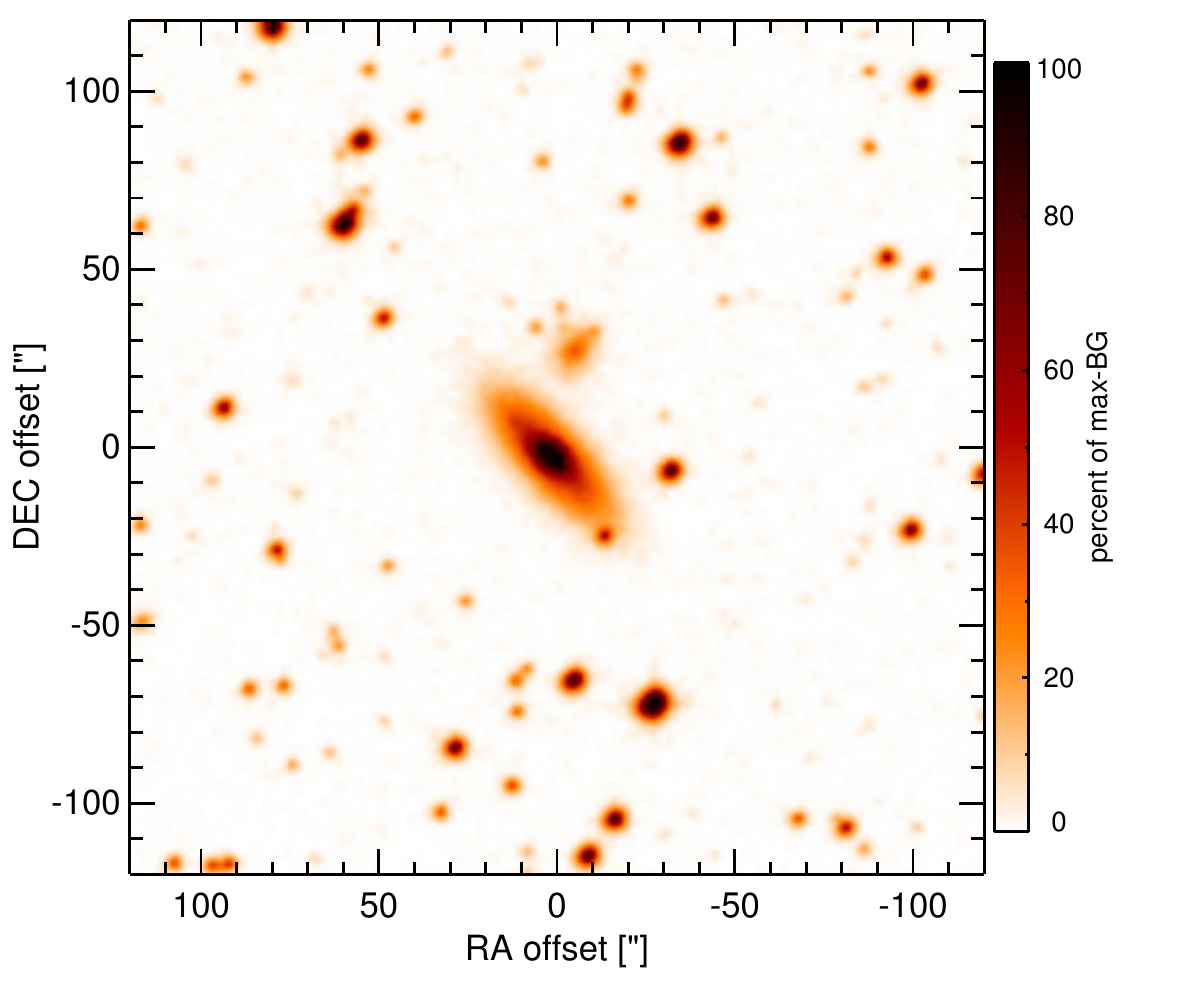}
    \caption{\label{fig:OPTim_ESO103-G035}
             Optical image (DSS, red filter) of ESO\,103-35. Displayed are the central $4\arcmin$ with North up and East to the left. 
              The colour scaling is linear with white corresponding to the median background and black to the $0.01\%$ pixels with the highest intensity.  
           }
\end{figure}
\begin{figure}
   \centering
   \includegraphics[angle=0,height=3.11cm]{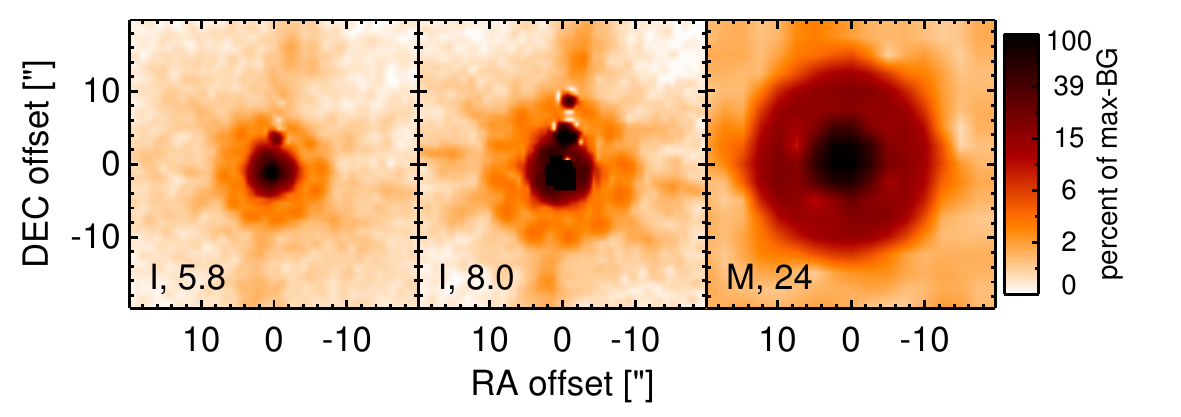}
    \caption{\label{fig:INTim_ESO103-G035}
             \spitzerr MIR images of ESO\,103-35. Displayed are the inner $40\arcsec$ with North up and East to the left. The colour scaling is logarithmic with white corresponding to median background and black to the $0.1\%$ pixels with the highest intensity.
             The label in the bottom left states instrument and central wavelength of the filter in $\mu$m (I: IRAC, M: MIPS).
             Note that the apparent off-nuclear compact sources in the IRAC $5.8$ and $8.0\,\mu$m images are instrumental artefacts.
           }
\end{figure}
\begin{figure}
   \centering
   \includegraphics[angle=0,width=8.500cm]{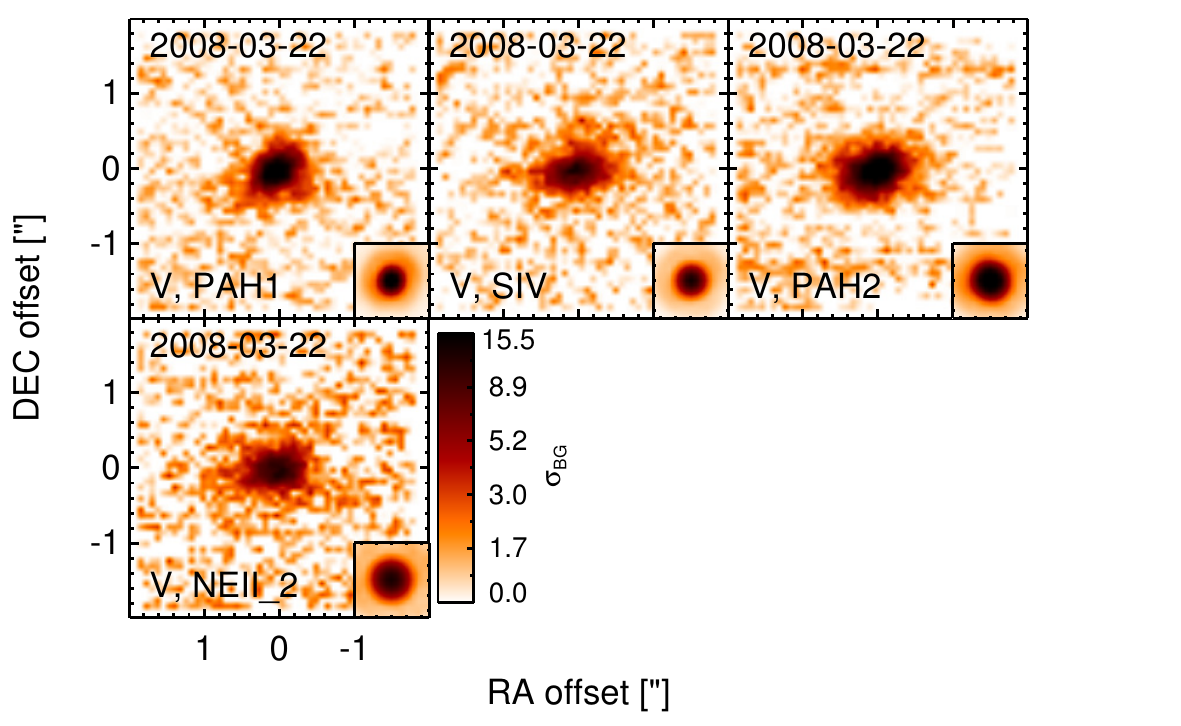}
    \caption{\label{fig:HARim_ESO103-G035}
             Subarcsecond-resolution MIR images of ESO\,103-35 sorted by increasing filter wavelength. 
             Displayed are the inner $4\arcsec$ with North up and East to the left. 
             The colour scaling is logarithmic with white corresponding to median background and black to the $75\%$ of the highest intensity of all images in units of $\sigbg$.
             The inset image shows the central arcsecond of the PSF from the calibrator star, scaled to match the science target.
             The labels in the bottom left state instrument and filter names (C: COMICS, M: Michelle, T: T-ReCS, V: VISIR).
           }
\end{figure}
\begin{figure}
   \centering
   \includegraphics[angle=0,width=8.50cm]{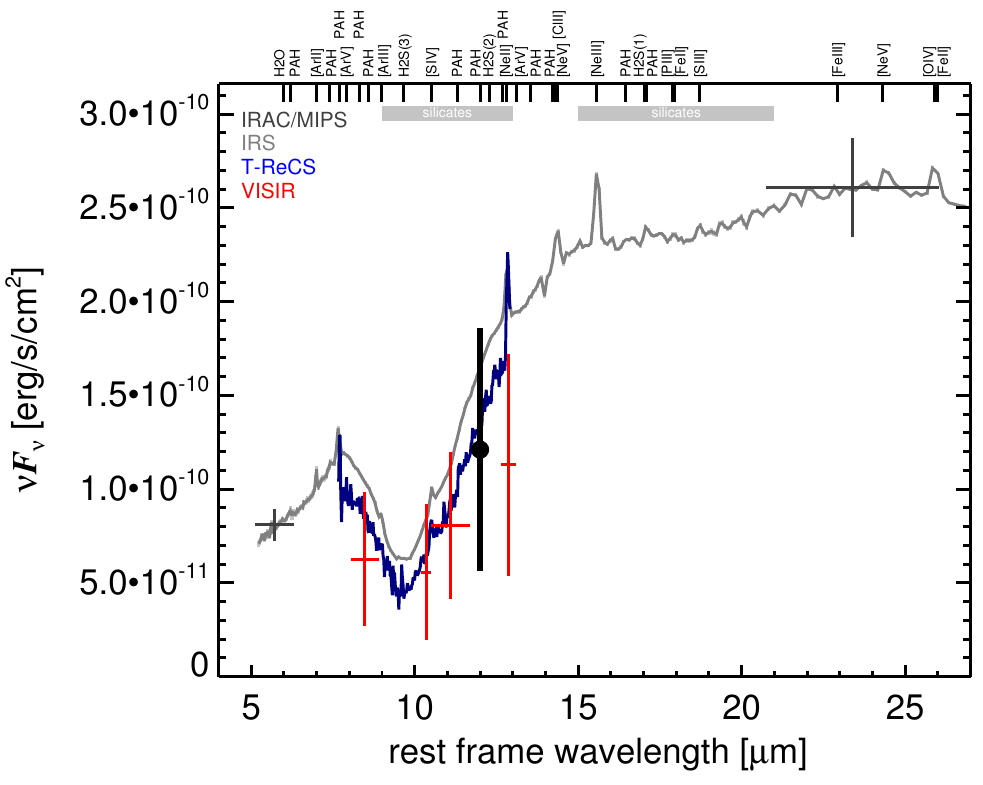}
   \caption{\label{fig:MISED_ESO103-G035}
      MIR SED of ESO\,103-35. The description  of the symbols (if present) is the following.
      Grey crosses and  solid lines mark the \spitzer/IRAC, MIPS and IRS data. 
      The colour coding of the other symbols is: 
      green for COMICS, magenta for Michelle, blue for T-ReCS and red for VISIR data.
      Darker-coloured solid lines mark spectra of the corresponding instrument.
      The black filled circles mark the nuclear 12 and $18\,\mu$m  continuum emission estimate from the data.
      The ticks on the top axis mark positions of common MIR emission lines, while the light grey horizontal bars mark wavelength ranges affected by the silicate 10 and 18$\mu$m features.     
   }
\end{figure}
\clearpage

\twocolumn[\begin{@twocolumnfalse}  
\subsection{ESO\,121-28 -- IRAS\,F06230-6057}\label{app:ESO121-G028}
ESO\,121-28 is a spiral galaxy at a redshift of $z=$ 0.0405 ($D\sim187\,$Mpc), interacting with a compact galaxy to the south.
It was only recently discovered to host a Sy\,2 nucleus \citep{donzelli_spectroscopic_2000}, which remained unstudied until it was detected in the \swift/BAT survey \citep{tueller_swift_2008} and became a member of the nine-month BAT AGN sample.
After its first but faint MIR detection with \iras, ESO\,121-28 was also observed with \spitzer/IRS.
No IRAC or MIPS images are available, and the object appears compact without any structure in the \wisee images.
The rather noisy IRS LR staring-mode spectrum has a red spectral slope in $\nu F_\nu$-space with prominent silicate 10$\,\mu$m absorption but no PAH emission (see also \citealt{sargsyan_infrared_2011}).
We observed ESO\,121-28 with VISIR in 2009 in three narrow $N$-band filters during one night and weakly detected an unresolved nucleus. 
The VISIR photometry agrees with the IRS spectrum well.
Therefore, we use this spectrum to correct our $12\,\mu$m continuum flux estimate for the silicate absorption.
\newline\end{@twocolumnfalse}]

\begin{figure}
   \centering
   \includegraphics[angle=0,width=8.500cm]{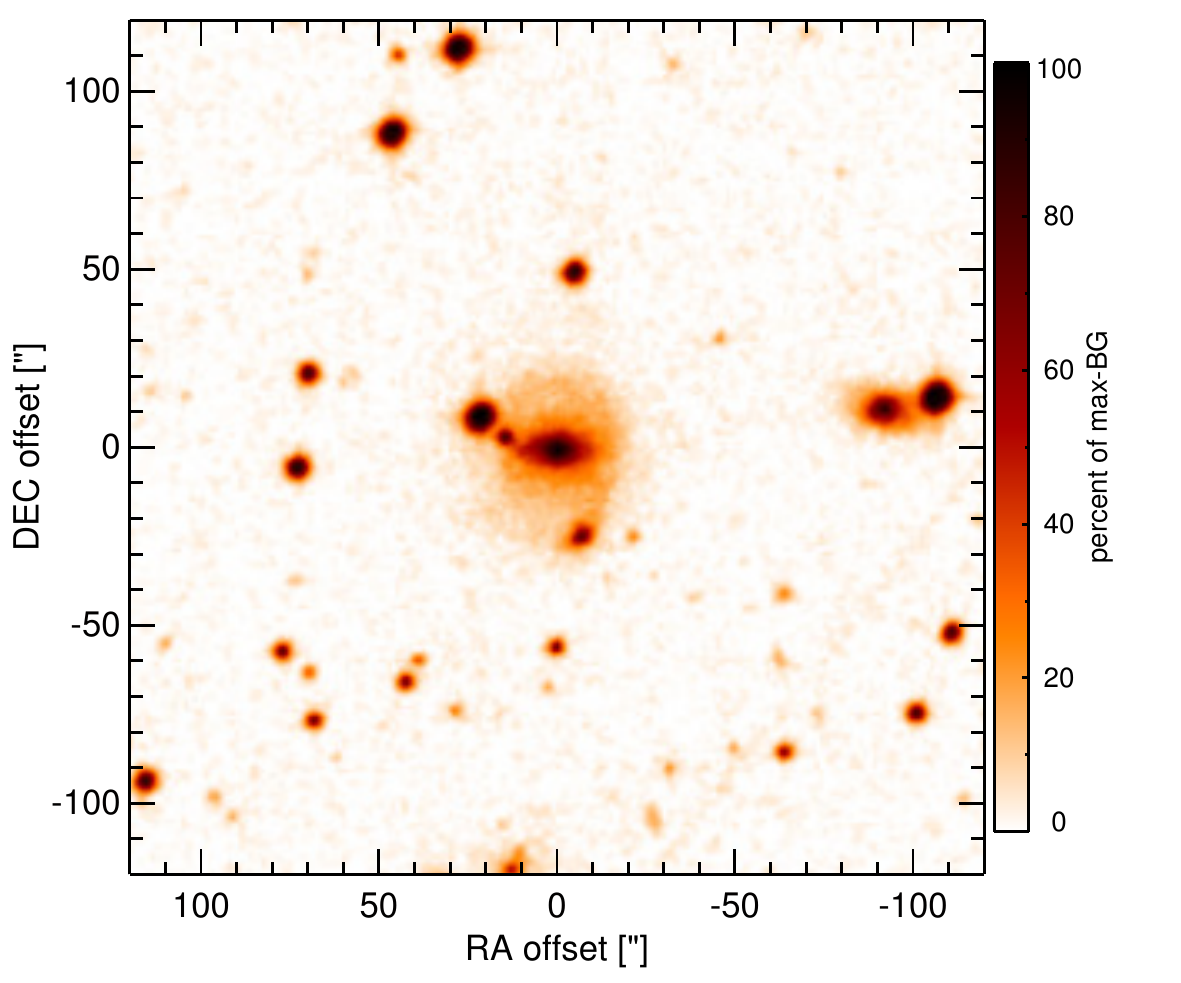}
    \caption{\label{fig:OPTim_ESO121-G028}
             Optical image (DSS, red filter) of ESO\,121-28. Displayed are the central $4\arcmin$ with North up and East to the left. 
              The colour scaling is linear with white corresponding to the median background and black to the $0.01\%$ pixels with the highest intensity.  
           }
\end{figure}
\begin{figure}
   \centering
   \includegraphics[angle=0,height=3.11cm]{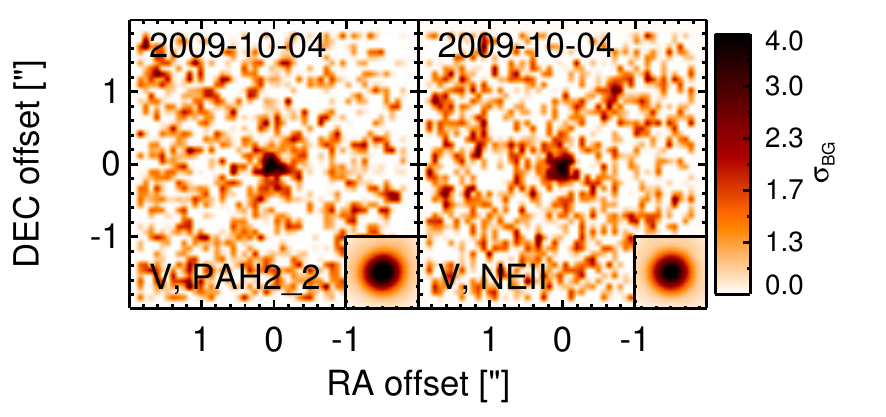}
    \caption{\label{fig:HARim_ESO121-G028}
             Subarcsecond-resolution MIR images of ESO\,121-28 sorted by increasing filter wavelength. 
             Displayed are the inner $4\arcsec$ with North up and East to the left. 
             The colour scaling is logarithmic with white corresponding to median background and black to the $75\%$ of the highest intensity of all images in units of $\sigbg$.
             The inset image shows the central arcsecond of the PSF from the calibrator star, scaled to match the science target.
             The labels in the bottom left state instrument and filter names (C: COMICS, M: Michelle, T: T-ReCS, V: VISIR).
           }
\end{figure}
\begin{figure}
   \centering
   \includegraphics[angle=0,width=8.50cm]{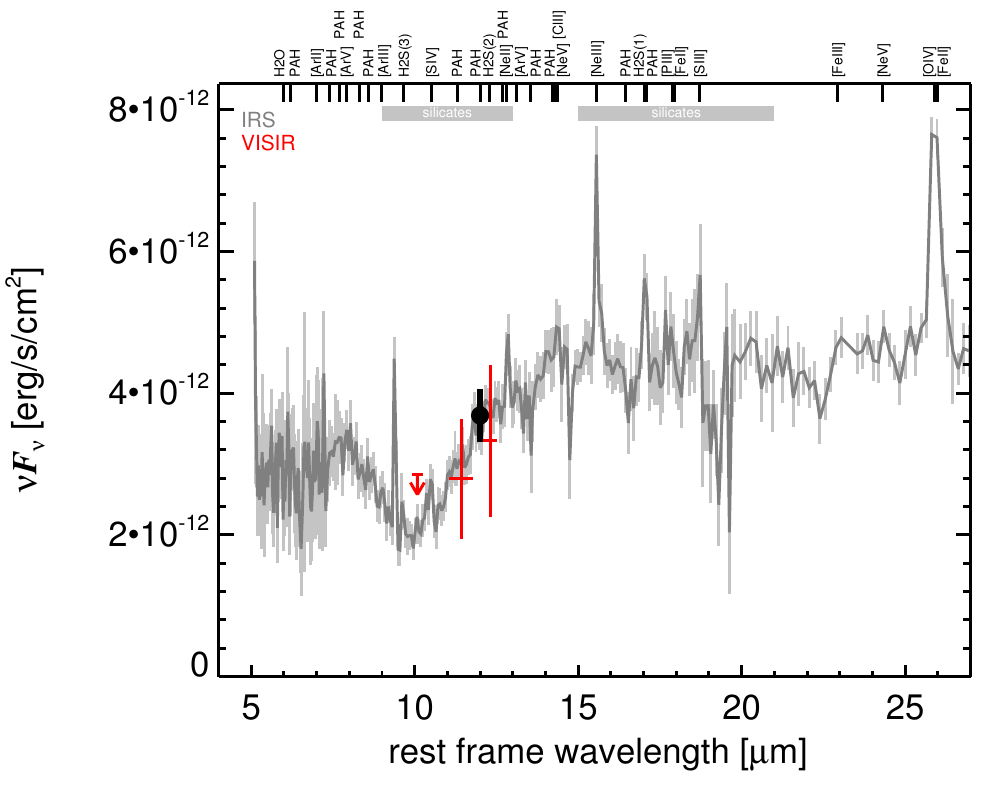}
   \caption{\label{fig:MISED_ESO121-G028}
      MIR SED of ESO\,121-28. The description  of the symbols (if present) is the following.
      Grey crosses and  solid lines mark the \spitzer/IRAC, MIPS and IRS data. 
      The colour coding of the other symbols is: 
      green for COMICS, magenta for Michelle, blue for T-ReCS and red for VISIR data.
      Darker-coloured solid lines mark spectra of the corresponding instrument.
      The black filled circles mark the nuclear 12 and $18\,\mu$m  continuum emission estimate from the data.
      The ticks on the top axis mark positions of common MIR emission lines, while the light grey horizontal bars mark wavelength ranges affected by the silicate 10 and 18$\mu$m features.     
   }
\end{figure}
\clearpage

\twocolumn[\begin{@twocolumnfalse}  
\subsection{ESO\,138-1 -- IRAS\,16470-5909}\label{app:ESO138-G001}
ESO\,138-1 is an early-type galaxy at a redshift of $z=$ 0.0091 ($D\sim 38.6$\,Mpc) hosting a radio-quiet Sy\,2 nucleus \citep{veron-cetty_catalogue_2010}.
It features possibly elongated radio emission on scales of $\sim0.5$\,kpc along the north-east axis (PA$\sim35\degree$; \citealt{morganti_radio_1999}), while the NLR has a complex morphology with jet-like extended emission into the north-western direction on kiloparsec-scales (PA$\sim-95\degree$;  \citealt{schmitt_anisotropic_1995}) and another extended emission feature $\sim1\arcsec\sim0.2$\,kpc towards the south-east (PA$\sim135\degree$;  \citealt{ferruit_hubble_2000}).
After first being detected in the MIR with \iras, ESO\,138-1 was followed up with \spitzer/IRAC, IRS and MIPS observations.
The nucleus appears nearly unresolved in the \spitzerr images without any detected non-nuclear emission. 
The IRS LR staring-mode spectrum shows weak PAH emission and the silicate features in emission, which is surprising for a Sy\,2, in particular because it is obscured by a Compton-thick column density of gas in X-rays (e.g., \citealt{piconcelli_x-ray_2011}; see also \citealt{goulding_deep_2012}) 
The MIR spectrum peaks at $\sim 18\,\mu$m in $\nu F_\nu$-space.
We observed ESO 138-1 with VISIR in several narrow $N$-band filters in 2008 and 2010.
A compact MIR nucleus was detected in all the images with no signs of non-nuclear emission.
The nucleus appears to be marginally resolved in the images  of both epochs but with inconsistent position angles ($\sim110\degree$ and $\sim 160\degree$ respectively). 
Thus, it remains uncertain, whether the nuclear MIR emission in ESO\,138-1 is really resolved or not.
Of those, two measurements were published already in \cite{gandhi_resolving_2009}. 
Our analysis of all images yields flux values consistent with the previous work and the \spitzerr spectrophotometry, albeit systematically lower by $\sim 5\%$.
Therefore, we do not apply any silicate correction to our nuclear $12\,\mu$m continuum flux estimate.
\newline\end{@twocolumnfalse}]

\begin{figure}
   \centering
   \includegraphics[angle=0,width=8.500cm]{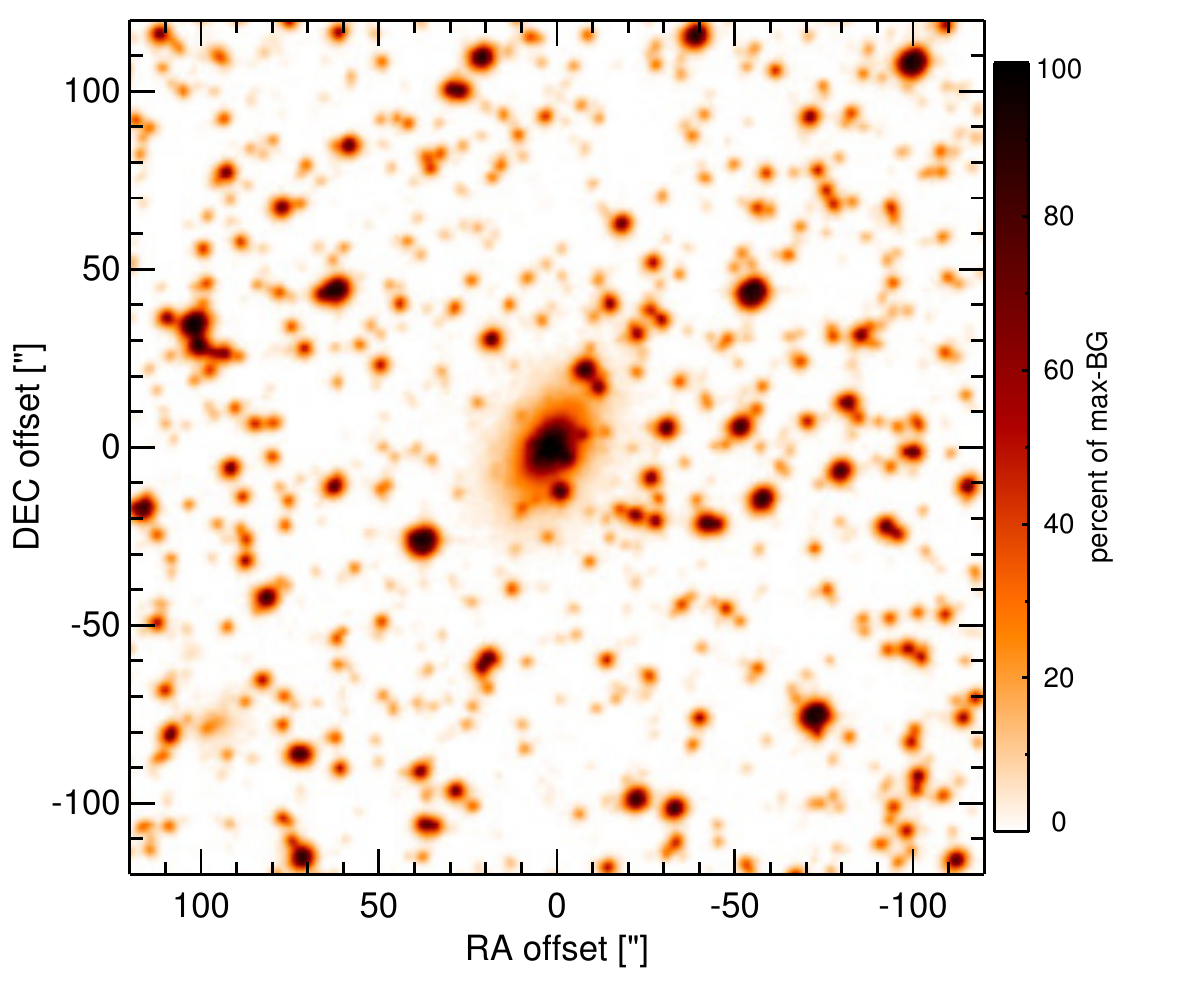}
    \caption{\label{fig:OPTim_ESO138-G001}
             Optical image (DSS, red filter) of ESO\,138-1. Displayed are the central $4\arcmin$ with North up and East to the left. 
              The colour scaling is linear with white corresponding to the median background and black to the $0.01\%$ pixels with the highest intensity.  
           }
\end{figure}
\begin{figure}
   \centering
   \includegraphics[angle=0,height=3.11cm]{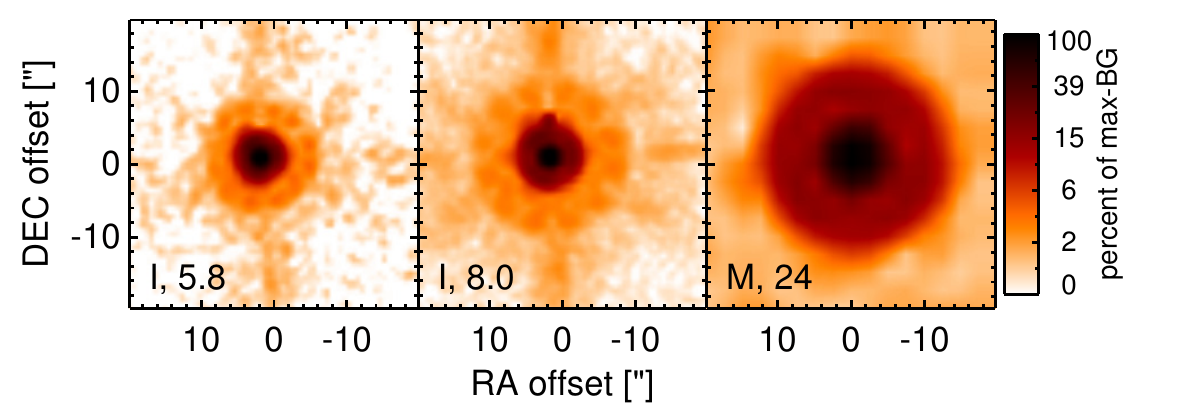}
    \caption{\label{fig:INTim_ESO138-G001}
             \spitzerr MIR images of ESO\,138-1. Displayed are the inner $40\arcsec$ with North up and East to the left. The colour scaling is logarithmic with white corresponding to median background and black to the $0.1\%$ pixels with the highest intensity.
             The label in the bottom left states instrument and central wavelength of the filter in $\mu$m (I: IRAC, M: MIPS). 
           }
\end{figure}
\begin{figure}
   \centering
   \includegraphics[angle=0,width=8.500cm]{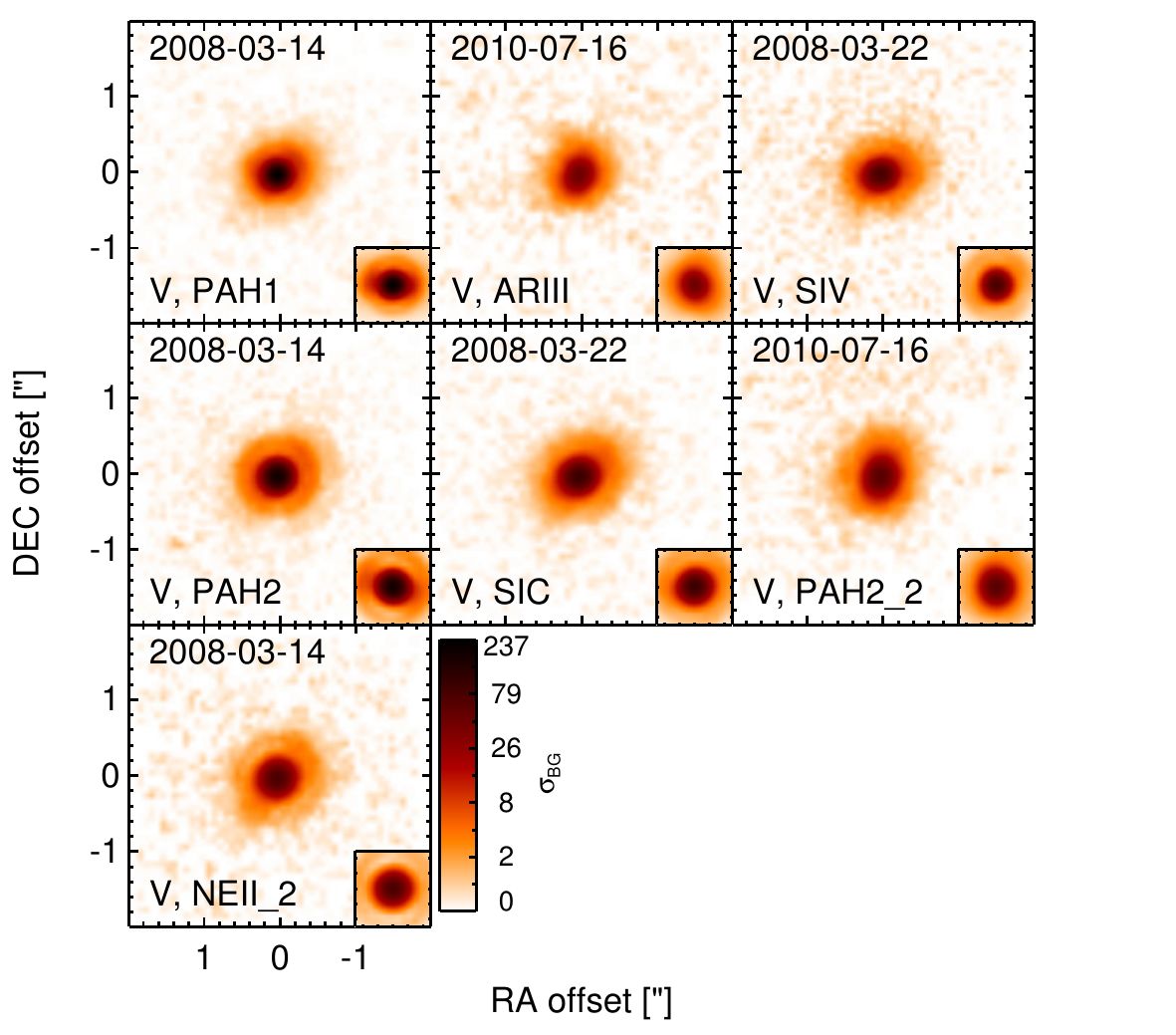}
    \caption{\label{fig:HARim_ESO138-G001}
             Subarcsecond-resolution MIR images of ESO\,138-1 sorted by increasing filter wavelength. 
             Displayed are the inner $4\arcsec$ with North up and East to the left. 
             The colour scaling is logarithmic with white corresponding to median background and black to the $75\%$ of the highest intensity of all images in units of $\sigbg$.
             The inset image shows the central arcsecond of the PSF from the calibrator star, scaled to match the science target.
             The labels in the bottom left state instrument and filter names (C: COMICS, M: Michelle, T: T-ReCS, V: VISIR).
           }
\end{figure}
\begin{figure}
   \centering
   \includegraphics[angle=0,width=8.50cm]{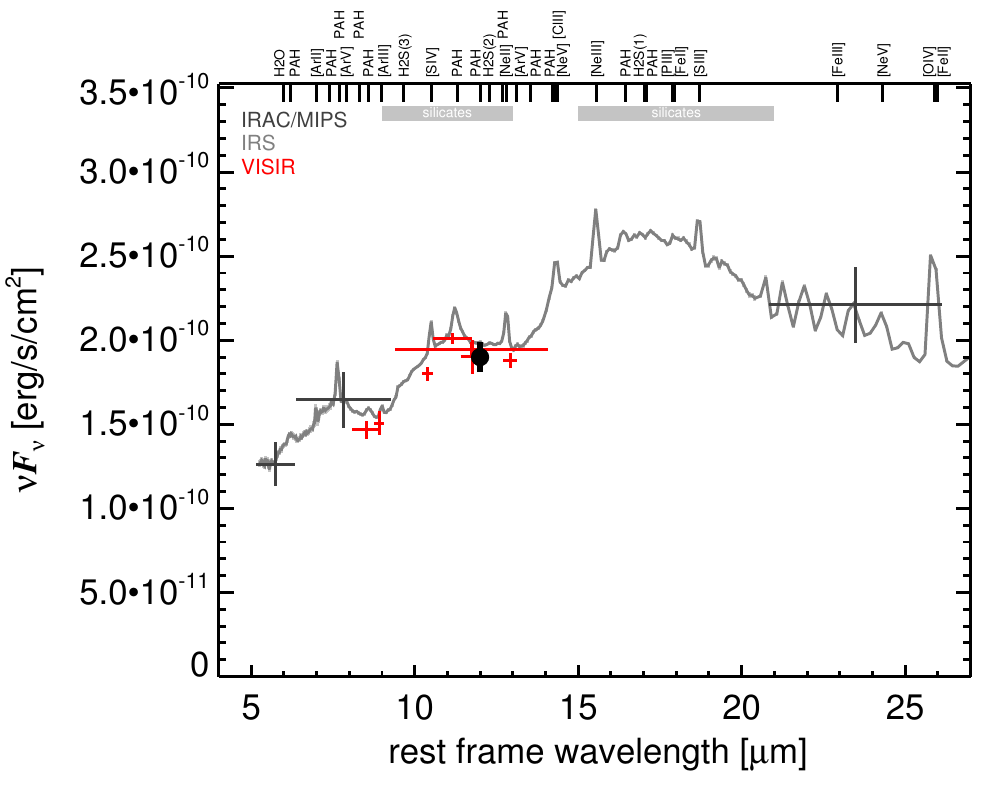}
   \caption{\label{fig:MISED_ESO138-G001}
      MIR SED of ESO\,138-1. The description  of the symbols (if present) is the following.
      Grey crosses and  solid lines mark the \spitzer/IRAC, MIPS and IRS data. 
      The colour coding of the other symbols is: 
      green for COMICS, magenta for Michelle, blue for T-ReCS and red for VISIR data.
      Darker-coloured solid lines mark spectra of the corresponding instrument.
      The black filled circles mark the nuclear 12 and $18\,\mu$m  continuum emission estimate from the data.
      The ticks on the top axis mark positions of common MIR emission lines, while the light grey horizontal bars mark wavelength ranges affected by the silicate 10 and 18$\mu$m features.     
   }
\end{figure}
\clearpage


\twocolumn[\begin{@twocolumnfalse}  
\subsection{ESO\,141-55 -- IRAS\,19169-5845}\label{app:ESO141-G055}
ESO\,141-55 is a rather face-on spiral galaxy at a redshift of $z=$ 0.0371 ($D\sim 156$\,Mpc) with a Sy\,1.2  nucleus \citep{veron-cetty_catalogue_2010}.
The first photometric $N$-band measurements of ESO\,141-55 were performed by \cite{glass_mid-infrared_1982} with the ESO infrared photometer system at the 3\,m telescope.
Later, it was observed with \iso/ISOCAM and \spitzer/IRAC and IRS.
The ISOCAM total fluxes published by \citep{ramos_almeida_mid-infrared_2007} agree with the \spitzerr data well.
The IRAC $5.8$ and $8.0\,\mu$m images  show a nearly unresolved nucleus, and the spiral arms are only weakly detected in the $8.0\,\mu$m image.
The IRS LR staring mode spectrum shows strong silicate 10 and $18\,\mu$m emission and weak PAH features (see \citealt{tommasin_spitzer_2008} for an IRS HR spectrum). 
The MIR slope is blue in $\nu F_\nu$-space.
We observed ESO\,141-55 with VISIR in 2006 (one epoch) in three narrow $N$-band filters and detected a compact MIR nucleus without any sign of significant extended or non-nuclear emission (see also \citealt{horst_mid-infrared_2009}).
However, the PSF comparison indicates that the nucleus is possibly marginally resolved (FWHM $\sim 0.42\arcsec \sim 300\,$pc).
Our new measurement of the nuclear fluxes is consistent with the previously published values \citep{horst_mid_2008}.
The subarcsecond fluxes are on average $\sim 24\%$ lower than the flux levels of the \spitzerr spectrophotometry and would be even lower if the presence of subarcsecond-extended emission can be verified.
Thus, no silicate correction is attempted in our $12\,\mu$m continuum flux estimate.
\newline\end{@twocolumnfalse}]

\begin{figure}
   \centering
   \includegraphics[angle=0,width=8.500cm]{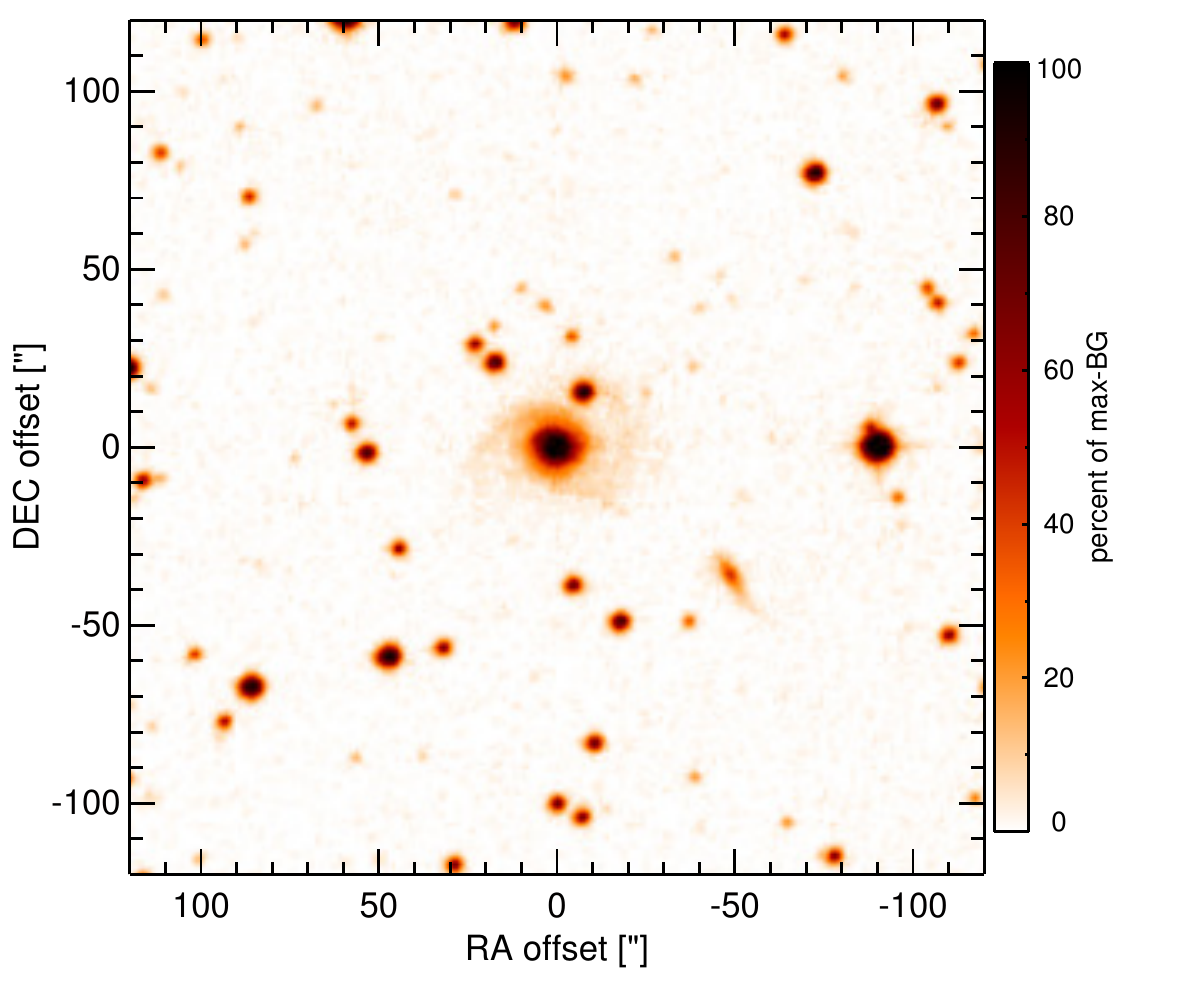}
    \caption{\label{fig:OPTim_ESO141-G055}
             Optical image (DSS, red filter) of ESO\,141-55. Displayed are the central $4\arcmin$ with North up and East to the left. 
              The colour scaling is linear with white corresponding to the median background and black to the $0.01\%$ pixels with the highest intensity.  
           }
\end{figure}
\begin{figure}
   \centering
   \includegraphics[angle=0,height=3.11cm]{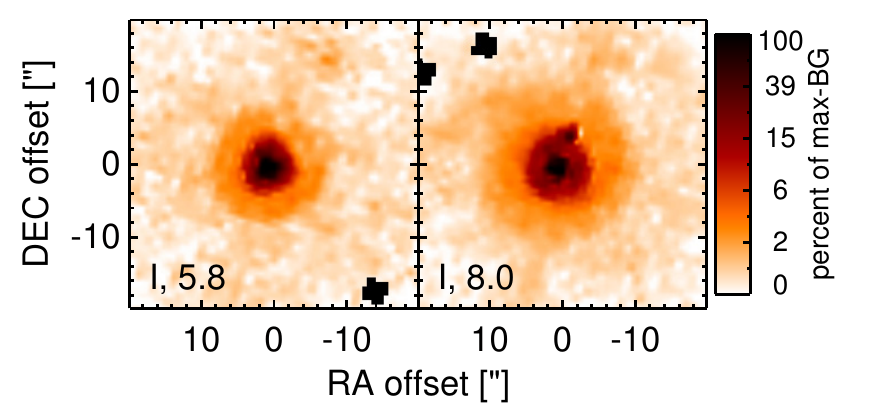}
    \caption{\label{fig:INTim_ESO141-G055}
             \spitzerr MIR images of ESO\,141-55. Displayed are the inner $40\arcsec$ with North up and East to the left. The colour scaling is logarithmic with white corresponding to median background and black to the $0.1\%$ pixels with the highest intensity.
             The label in the bottom left states instrument and central wavelength of the filter in $\mu$m (I: IRAC, M: MIPS). 
             Note that the apparent off-nuclear compact sources in the IRAC $5.8$ and $8.0\,\mu$m images are instrumental artefacts.
           }
\end{figure}
\begin{figure}
   \centering
   \includegraphics[angle=0,height=3.11cm]{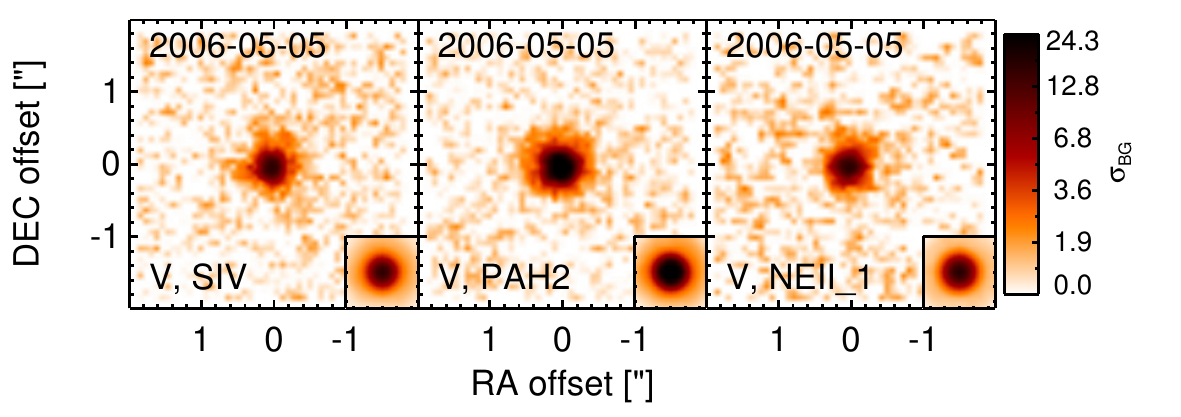}
    \caption{\label{fig:HARim_ESO141-G055}
             Subarcsecond-resolution MIR images of ESO\,141-55 sorted by increasing filter wavelength. 
             Displayed are the inner $4\arcsec$ with North up and East to the left. 
             The colour scaling is logarithmic with white corresponding to median background and black to the $75\%$ of the highest intensity of all images in units of $\sigbg$.
             The inset image shows the central arcsecond of the PSF from the calibrator star, scaled to match the science target.
             The labels in the bottom left state instrument and filter names (C: COMICS, M: Michelle, T: T-ReCS, V: VISIR).
           }
\end{figure}
\begin{figure}
   \centering
   \includegraphics[angle=0,width=8.50cm]{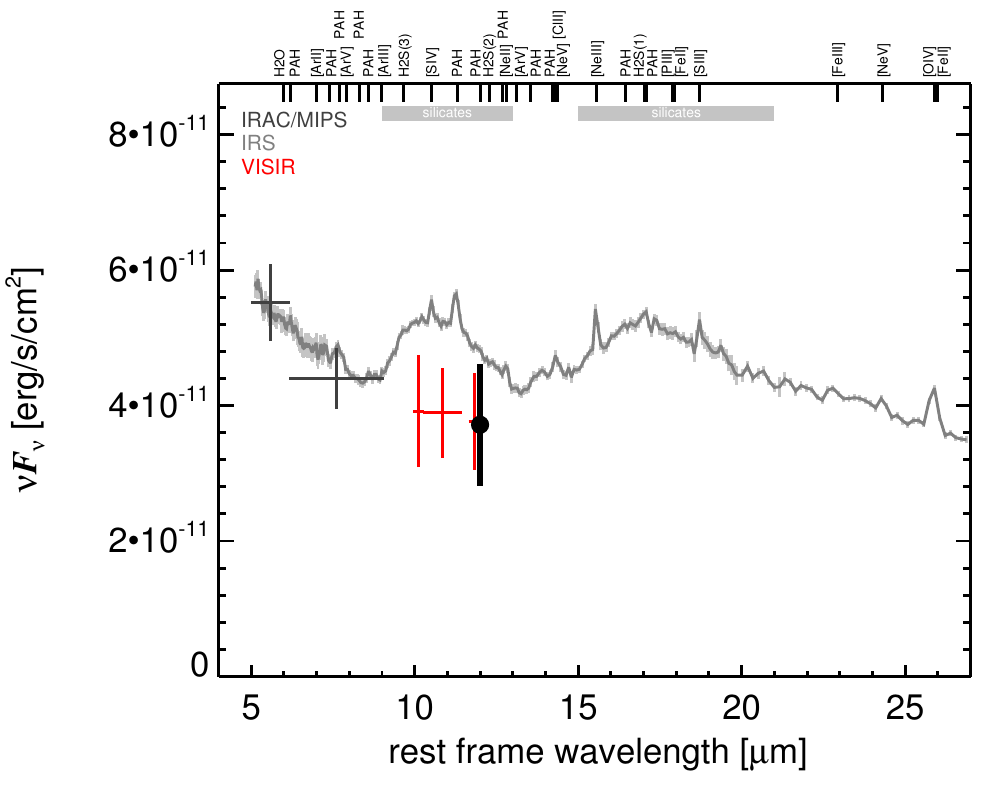}
   \caption{\label{fig:MISED_ESO141-G055}
      MIR SED of ESO\,141-55. The description  of the symbols (if present) is the following.
      Grey crosses and  solid lines mark the \spitzer/IRAC, MIPS and IRS data. 
      The colour coding of the other symbols is: 
      green for COMICS, magenta for Michelle, blue for T-ReCS and red for VISIR data.
      Darker-coloured solid lines mark spectra of the corresponding instrument.
      The black filled circles mark the nuclear 12 and $18\,\mu$m  continuum emission estimate from the data.
      The ticks on the top axis mark positions of common MIR emission lines, while the light grey horizontal bars mark wavelength ranges affected by the silicate 10 and 18$\mu$m features.     
   }
\end{figure}
\clearpage

\twocolumn[\begin{@twocolumnfalse}  
\subsection{ESO\,198-24 -- LEDA\,9998}\label{app:ESO198-G024}
ESO\,198-24 is an elliptical galaxy at a redshift of $z=$ 0.0455 ($D\sim 192\,$Mpc) hosting a Sy\,1.0 nucleus \citep{veron-cetty_catalogue_2010}, which is a member of the nine-month BAT AGN sample.
It was observed with \spitzer/IRAC, IRS and MIPS and appears nearly unresolved in the corresponding images. 
The IRS LR staring-mode spectrum agrees with the IRAC and MIPS photometry well without further scaling and shows strong silicate 10 and $18\,\mu$m emission with weak PAH features. 
The MIR spectral slope is rather flat with a maximum at $\sim 18\,\mu$m in $\nu F_\nu$-space.
We observed ESO\,198-24 with VISIR in 2009 in three narrow $N$-band filters spread over two nights.
A compact MIR nucleus without any sign of extended or non-nuclear emission was weakly detected. 
The measured subarcsecond fluxes agree with the \spitzerr spectrophotometry and the $12\,\mu$m continuum flux estimate is correspondingly corrected for the silicate emission feature.
\newline\end{@twocolumnfalse}]

\begin{figure}
   \centering
   \includegraphics[angle=0,width=8.500cm]{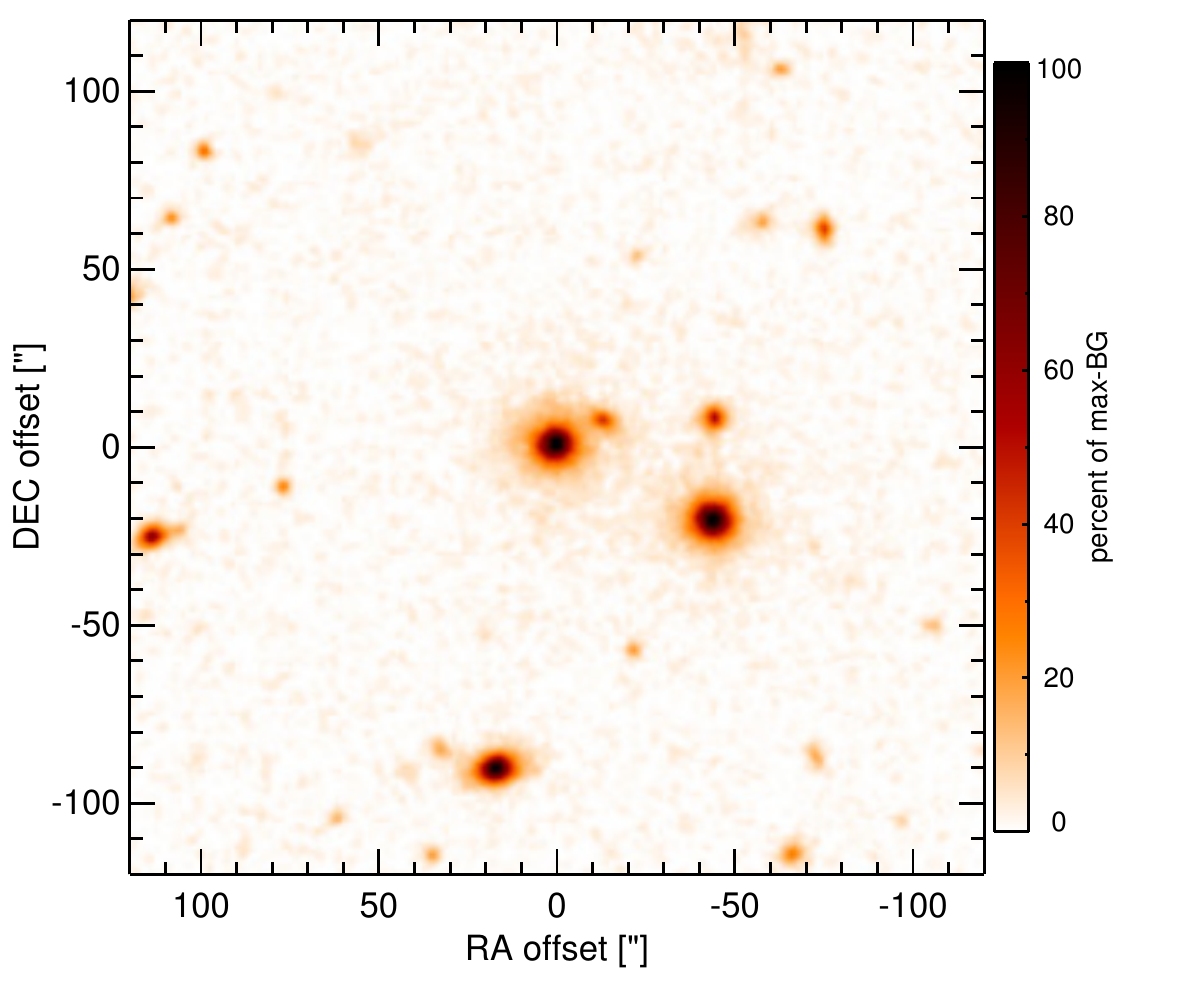}
    \caption{\label{fig:OPTim_ESO198-G024}
             Optical image (DSS, red filter) of ESO\,198-24. Displayed are the central $4\arcmin$ with North up and East to the left. 
              The colour scaling is linear with white corresponding to the median background and black to the $0.01\%$ pixels with the highest intensity.  
           }
\end{figure}
\begin{figure}
   \centering
   \includegraphics[angle=0,height=3.11cm]{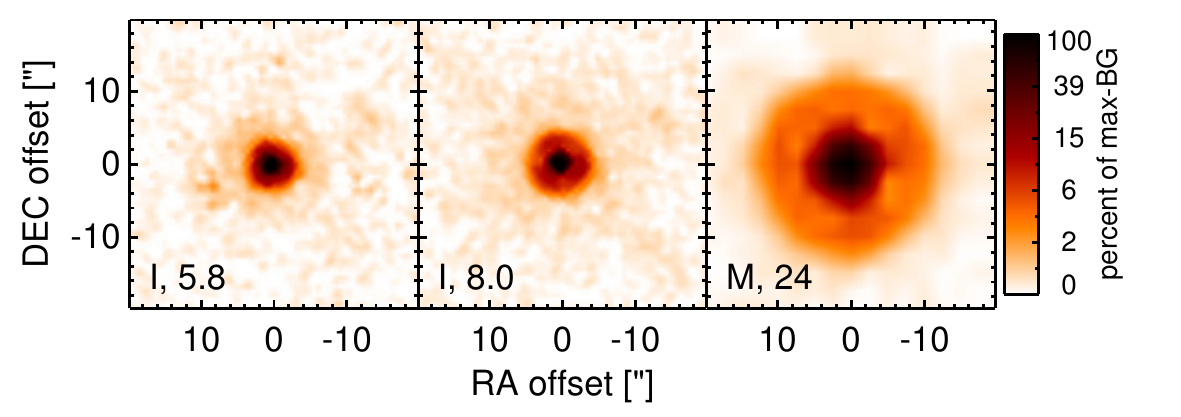}
    \caption{\label{fig:INTim_ESO198-G024}
             \spitzerr MIR images of ESO\,198-24. Displayed are the inner $40\arcsec$ with North up and East to the left. The colour scaling is logarithmic with white corresponding to median background and black to the $0.1\%$ pixels with the highest intensity.
             The label in the bottom left states instrument and central wavelength of the filter in $\mu$m (I: IRAC, M: MIPS). 
           }
\end{figure}
\begin{figure}
   \centering
   \includegraphics[angle=0,height=3.11cm]{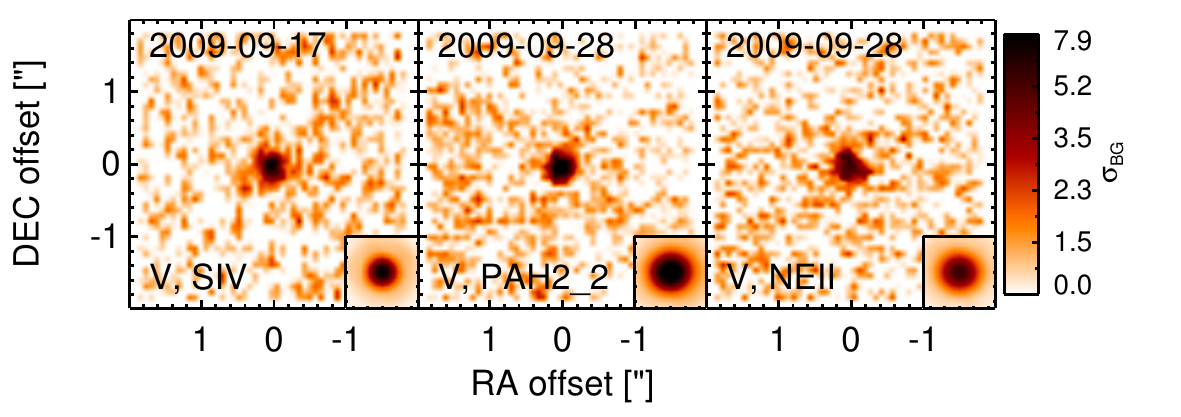}
    \caption{\label{fig:HARim_ESO198-G024}
             Subarcsecond-resolution MIR images of ESO\,198-24 sorted by increasing filter wavelength. 
             Displayed are the inner $4\arcsec$ with North up and East to the left. 
             The colour scaling is logarithmic with white corresponding to median background and black to the $75\%$ of the highest intensity of all images in units of $\sigbg$.
             The inset image shows the central arcsecond of the PSF from the calibrator star, scaled to match the science target.
             The labels in the bottom left state instrument and filter names (C: COMICS, M: Michelle, T: T-ReCS, V: VISIR).
           }
\end{figure}
\begin{figure}
   \centering
   \includegraphics[angle=0,width=8.50cm]{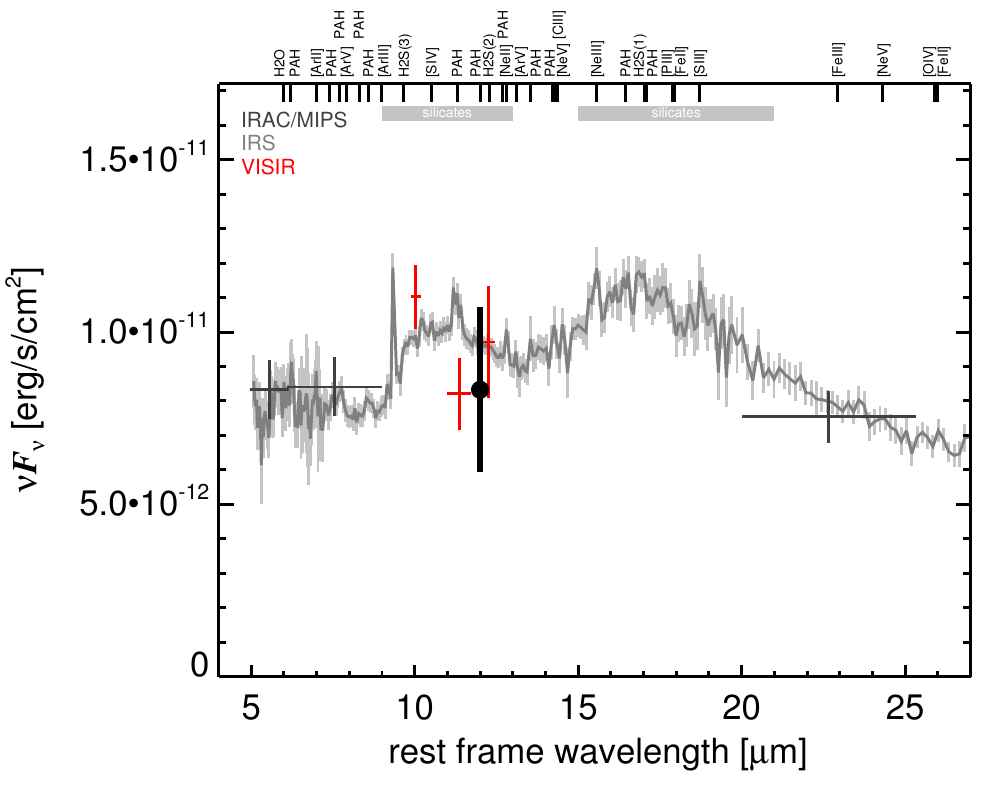}
   \caption{\label{fig:MISED_ESO198-G024}
      MIR SED of ESO\,198-24. The description  of the symbols (if present) is the following.
      Grey crosses and  solid lines mark the \spitzer/IRAC, MIPS and IRS data. 
      The colour coding of the other symbols is: 
      green for COMICS, magenta for Michelle, blue for T-ReCS and red for VISIR data.
      Darker-coloured solid lines mark spectra of the corresponding instrument.
      The black filled circles mark the nuclear 12 and $18\,\mu$m  continuum emission estimate from the data.
      The ticks on the top axis mark positions of common MIR emission lines, while the light grey horizontal bars mark wavelength ranges affected by the silicate 10 and 18$\mu$m features.     
   }
\end{figure}
\clearpage

\twocolumn[\begin{@twocolumnfalse}  
\subsection{ESO\,209-12 -- IRAS\,08005-4938}\label{app:ESO209-G012}
ESO\,209-12 is a highly inclined spiral galaxy at a redshift of $z=$ 0.0405 ($D\sim175\,$Mpc) hosting an AGN with optical Sy\,1.5 classification \citep{veron-cetty_catalogue_2010}.
It was observed with \spitzer/IRAC and IRS, with which a compact MIR nucleus with extended host galaxy emission was detected.
The shortest wavelength part of the IRS LR mapping-mode spectrum is missing, but the other part of the spectrum matches with the IRAC $5.8$ and $8.0\,\mu$m photometry.
Weak silicate and PAH emission is evident, while the spectral slope is rather flat in $\nu F_\nu$-space with a maximum at $\sim 18\,\mu$m.
We observed ESO\,209-12 with VISIR in 2007 in four narrow $N$-band filters (one epoch) and detected the nucleus but no host galaxy emission. 
The nucleus is unresolved in the sharpest images (PAH1 and SIV) and is thus classified as unresolved at subarcsecond scales in the MIR.
Nuclear photometry of the PAH2 and NEII images was already published  in \cite{gandhi_resolving_2009}.
Our new measurements are consistent with the previous values. 
The average subarcsecond flux level is $\sim 13\%$ lower than the \spitzerr spectrophotometry.
\newline\end{@twocolumnfalse}]

\begin{figure}
   \centering
   \includegraphics[angle=0,width=8.500cm]{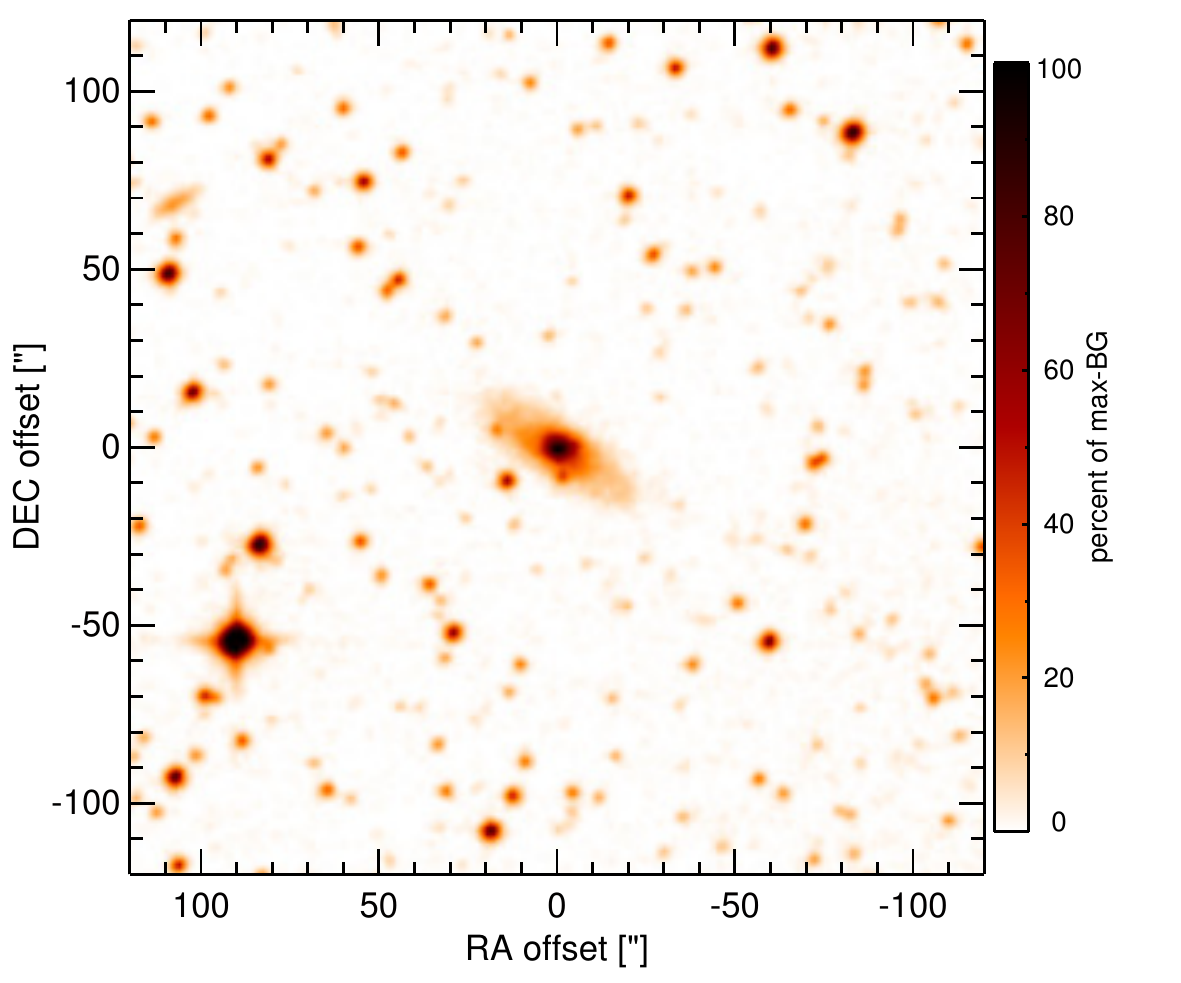}
    \caption{\label{fig:OPTim_ESO209-G012}
             Optical image (DSS, red filter) of ESO\,209-12. Displayed are the central $4\arcmin$ with North up and East to the left. 
              The colour scaling is linear with white corresponding to the median background and black to the $0.01\%$ pixels with the highest intensity.  
           }
\end{figure}
\begin{figure}
   \centering
   \includegraphics[angle=0,height=3.11cm]{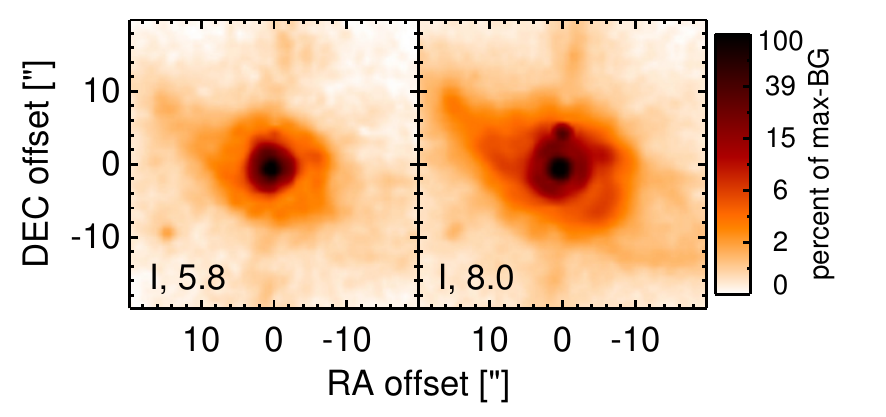}
    \caption{\label{fig:INTim_ESO209-G012}
             \spitzerr MIR images of ESO\,209-12. Displayed are the inner $40\arcsec$ with North up and East to the left. The colour scaling is logarithmic with white corresponding to median background and black to the $0.1\%$ pixels with the highest intensity.
             The label in the bottom left states instrument and central wavelength of the filter in $\mu$m (I: IRAC, M: MIPS). 
             Note that the apparent off-nuclear compact source in the IRAC $8.0\,\mu$m image is an instrumental artefact.
           }
\end{figure}
\begin{figure}
   \centering
   \includegraphics[angle=0,width=8.500cm]{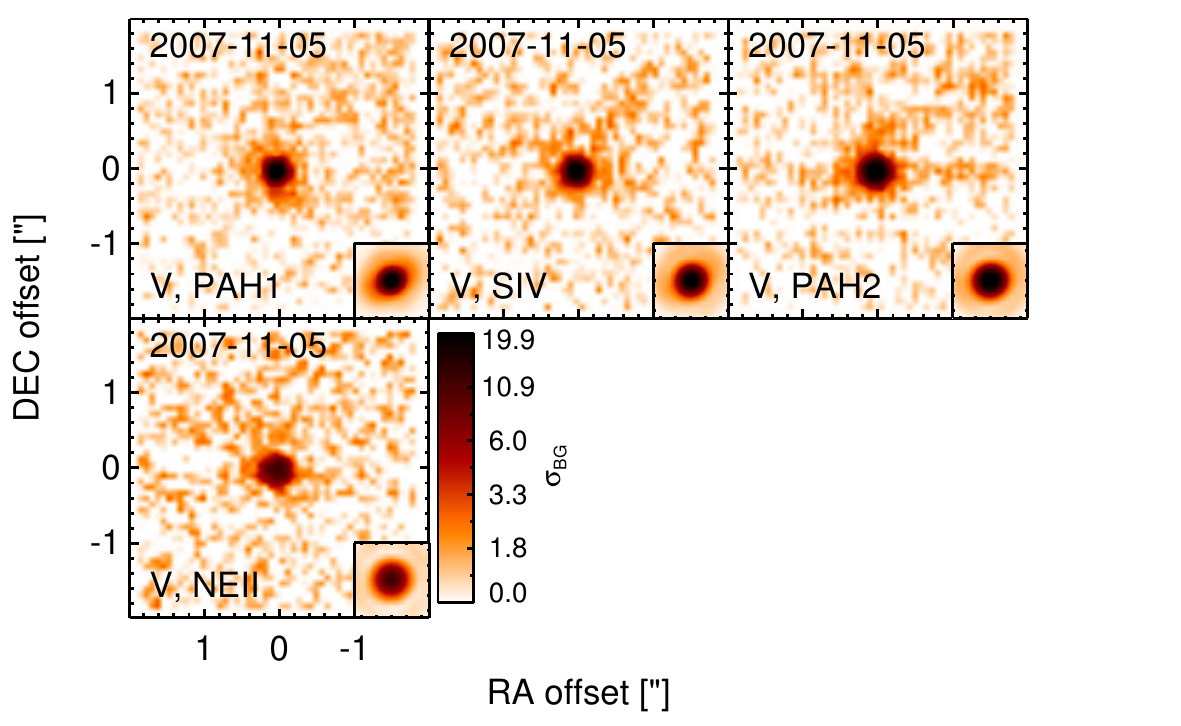}
    \caption{\label{fig:HARim_ESO209-G012}
             Subarcsecond-resolution MIR images of ESO\,209-12 sorted by increasing filter wavelength. 
             Displayed are the inner $4\arcsec$ with North up and East to the left. 
             The colour scaling is logarithmic with white corresponding to median background and black to the $75\%$ of the highest intensity of all images in units of $\sigbg$.
             The inset image shows the central arcsecond of the PSF from the calibrator star, scaled to match the science target.
             The labels in the bottom left state instrument and filter names (C: COMICS, M: Michelle, T: T-ReCS, V: VISIR).
           }
\end{figure}
\begin{figure}
   \centering
   \includegraphics[angle=0,width=8.50cm]{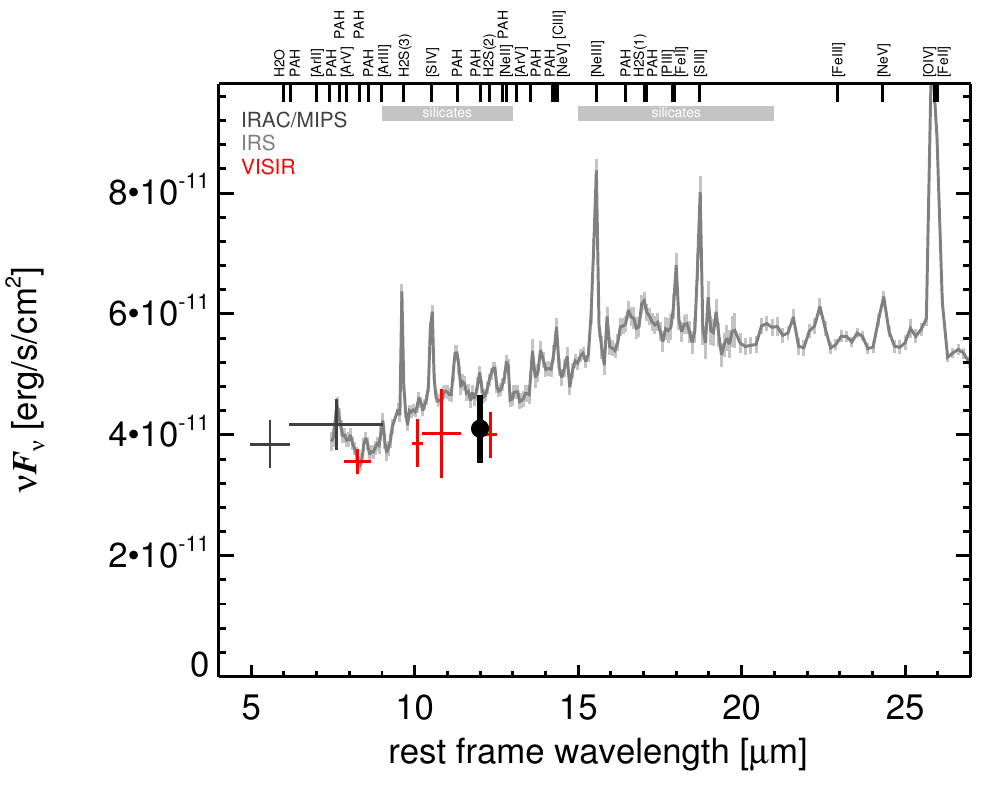}
   \caption{\label{fig:MISED_ESO209-G012}
      MIR SED of ESO\,209-12. The description  of the symbols (if present) is the following.
      Grey crosses and  solid lines mark the \spitzer/IRAC, MIPS and IRS data. 
      The colour coding of the other symbols is: 
      green for COMICS, magenta for Michelle, blue for T-ReCS and red for VISIR data.
      Darker-coloured solid lines mark spectra of the corresponding instrument.
      The black filled circles mark the nuclear 12 and $18\,\mu$m  continuum emission estimate from the data.
      The ticks on the top axis mark positions of common MIR emission lines, while the light grey horizontal bars mark wavelength ranges affected by the silicate 10 and 18$\mu$m features.     
   }
\end{figure}
\clearpage

\twocolumn[\begin{@twocolumnfalse}  
\subsection{ESO\,253-3 -- IRAS\,05238-4602}\label{app:ESO253-G003}
ESO\,253-3 is a spiral galaxy at a redshift of $z=$ 0.0425 ($D\sim181\,$Mpc) hosting an AGN with optical Sy\,2 classification \citep{veron-cetty_catalogue_2010}.
It was observed with \spitzer/IRAC and IRS, where a nucleus elongated to the north-east with  faint extended host galaxy emission being detected.
The IRS HR staring-mode spectrum does not overlap with the IRAC photometry but displays possible lower flux levels.
The MIR spectral slope is red in $\nu F_\nu$-space without any identifiable silicate or PAH features. 
Note, however, that no background subtraction was performed for this spectrum.
ESO\,253-3 was observed with T-ReCS in the broad $N$-band filter in 2004 (unpublished, to our knowledge).
In the T-ReCS image, the nucleus (elongated in IRAC) is resolved into two compact sources with a separation of $\sim 1.7\arcsec \sim  1.4\,$kpc in the north-east direction. 
We assume the north-eastern source, ESO\,253-3NE, which is significantly brighter, to be the AGN.
ESO\,253-3NE is possibly extended (FWHM $\sim 0.44\arcsec \sim 360\,$pc; PA$\sim125\degree$) but at least a second epoch of subarcsecond MIR imaging is required to verify this extension. 
Its flux agrees with the IRS spectrum and therefore we use the latter to estimate the $12\,\mu$m continuum flux.
The nature of the double MIR nucleus of ESO\,253-3 remains unknown. 
It could possibly be a post galaxy merger similar to NGC\,6240 but this scenario is not supported by the IRS spectrum.
\newline\end{@twocolumnfalse}]

\begin{figure}
   \centering
   \includegraphics[angle=0,width=8.500cm]{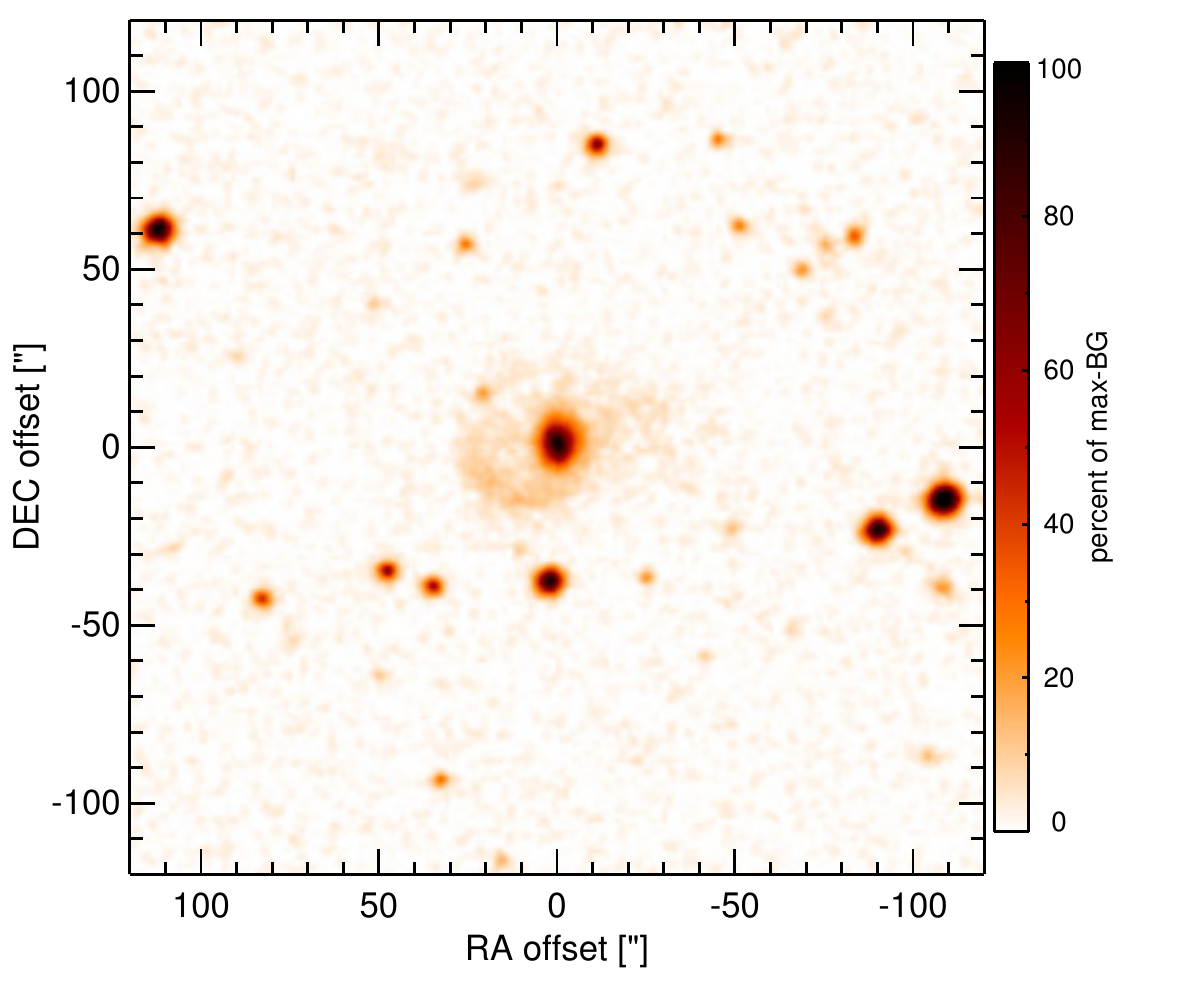}
    \caption{\label{fig:OPTim_ESO253-G003}
             Optical image (DSS, red filter) of ESO\,253-3. Displayed are the central $4\arcmin$ with North up and East to the left. 
              The colour scaling is linear with white corresponding to the median background and black to the $0.01\%$ pixels with the highest intensity.  
           }
\end{figure}
\begin{figure}
   \centering
   \includegraphics[angle=0,height=3.11cm]{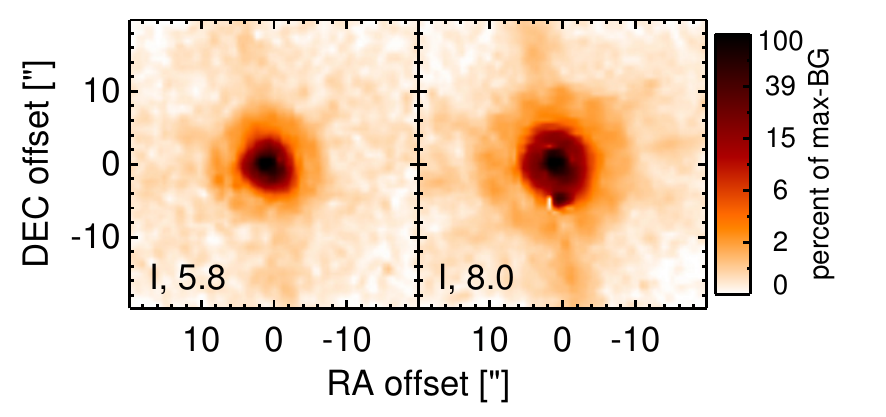}
    \caption{\label{fig:INTim_ESO253-G003}
             \spitzerr MIR images of ESO\,253-3. Displayed are the inner $40\arcsec$ with North up and East to the left. The colour scaling is logarithmic with white corresponding to median background and black to the $0.1\%$ pixels with the highest intensity.
             The label in the bottom left states instrument and central wavelength of the filter in $\mu$m (I: IRAC, M: MIPS).
             Note that the apparent off-nuclear compact source in the IRAC $8.0\,\mu$m image is an instrumental artefact.
           }
\end{figure}
\begin{figure}
   \centering
   \includegraphics[angle=0,height=3.11cm]{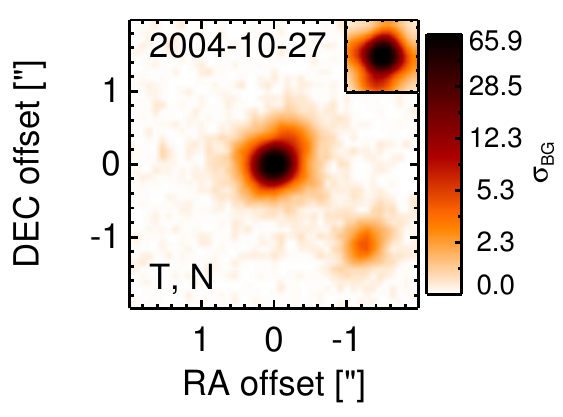}
    \caption{\label{fig:HARim_ESO253-G003}
             Subarcsecond-resolution MIR images of ESO\,253-3 sorted by increasing filter wavelength. 
             Displayed are the inner $4\arcsec$ with North up and East to the left. 
             The colour scaling is logarithmic with white corresponding to median background and black to the $75\%$ of the highest intensity of all images in units of $\sigbg$.
             The inset image shows the central arcsecond of the PSF from the calibrator star, scaled to match the science target.
             The labels in the bottom left state instrument and filter names (C: COMICS, M: Michelle, T: T-ReCS, V: VISIR).
           }
\end{figure}
\begin{figure}
   \centering
   \includegraphics[angle=0,width=8.50cm]{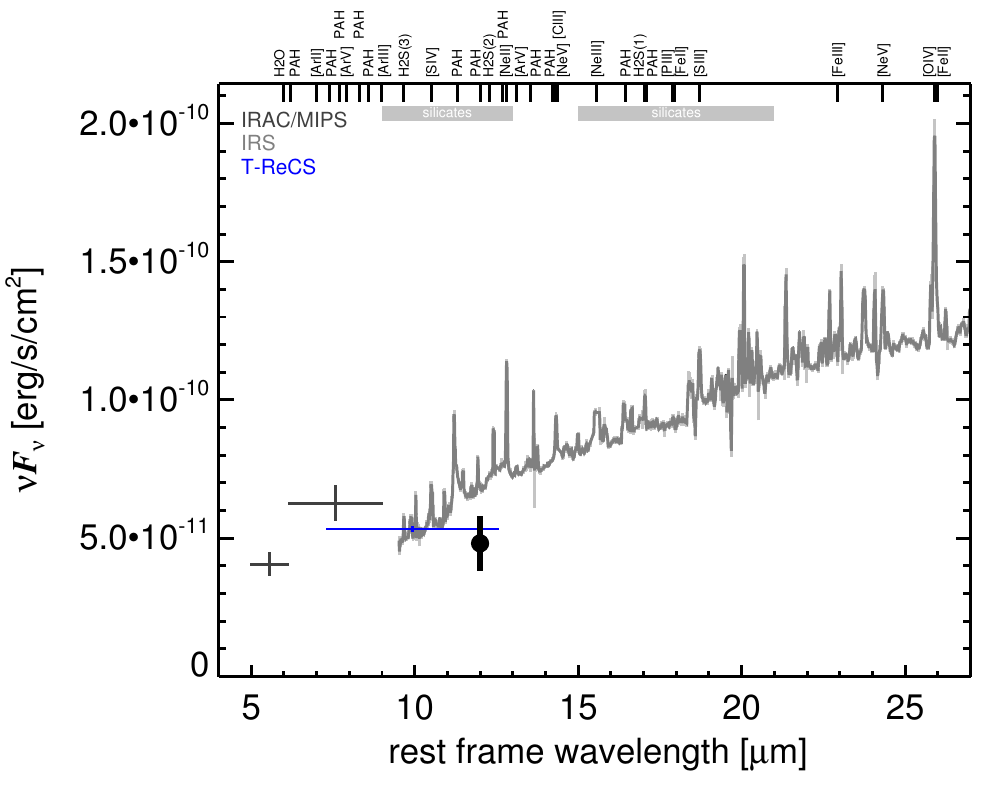}
   \caption{\label{fig:MISED_ESO253-G003}
      MIR SED of ESO\,253-3. The description  of the symbols (if present) is the following.
      Grey crosses and  solid lines mark the \spitzer/IRAC, MIPS and IRS data. 
      The colour coding of the other symbols is: 
      green for COMICS, magenta for Michelle, blue for T-ReCS and red for VISIR data.
      Darker-coloured solid lines mark spectra of the corresponding instrument.
      The black filled circles mark the nuclear 12 and $18\,\mu$m  continuum emission estimate from the data.
      The ticks on the top axis mark positions of common MIR emission lines, while the light grey horizontal bars mark wavelength ranges affected by the silicate 10 and 18$\mu$m features.     
   }
\end{figure}
\clearpage

\twocolumn[\begin{@twocolumnfalse}  
\subsection{ESO\,263-13 -- Fairall\,427}\label{app:ESO263-G013}
ESO\,263-13 is a face-on spiral galaxy at a redshift of $z=$ 0.0335 ($D \sim 146\,$Mpc) and hosts a Sy\,2 nucleus \citep{veron-cetty_catalogue_2010}.
It was observed with \spitzer/IRAC and IRS.
In the IRAC images, a nearly unresolved MIR nucleus with weak spiral-like host galaxy emission was detected.
The IRS LR mapping-mode spectrum lacks the lowest wavelength setting and has only a low S/N.
Therefore, the presence of silicate and/or PAH features remains uncertain.
The MIR spectrum peaks at $\sim 18\,\mu$m in $\nu F_\nu$-space.
We observed ESO\,263-13 with VISIR in 2007 and 2008 in six different narrow $N$-band filters (9 observations spread over three epochs).
A compact MIR nucleus was detected in all images without any sign of extended host emission. 
The nucleus is unresolved in the sharpest SIV, PAH1 and PAH2 images.
Therefore, it is classified as unresolved in the MIR at subarcsecond scales.
The SIC and NEII filter photometry has already been published in \cite{gandhi_resolving_2009}, who found $\sim 10\%$ lower fluxes than we found in our reanalysis of the same data. 
Our new subarcsecond flux values are consistent with the \spitzerr spectrophotometry but are systematically  $\sim 7\%$ lower .
\newline\end{@twocolumnfalse}]

\begin{figure}
   \centering
   \includegraphics[angle=0,width=8.500cm]{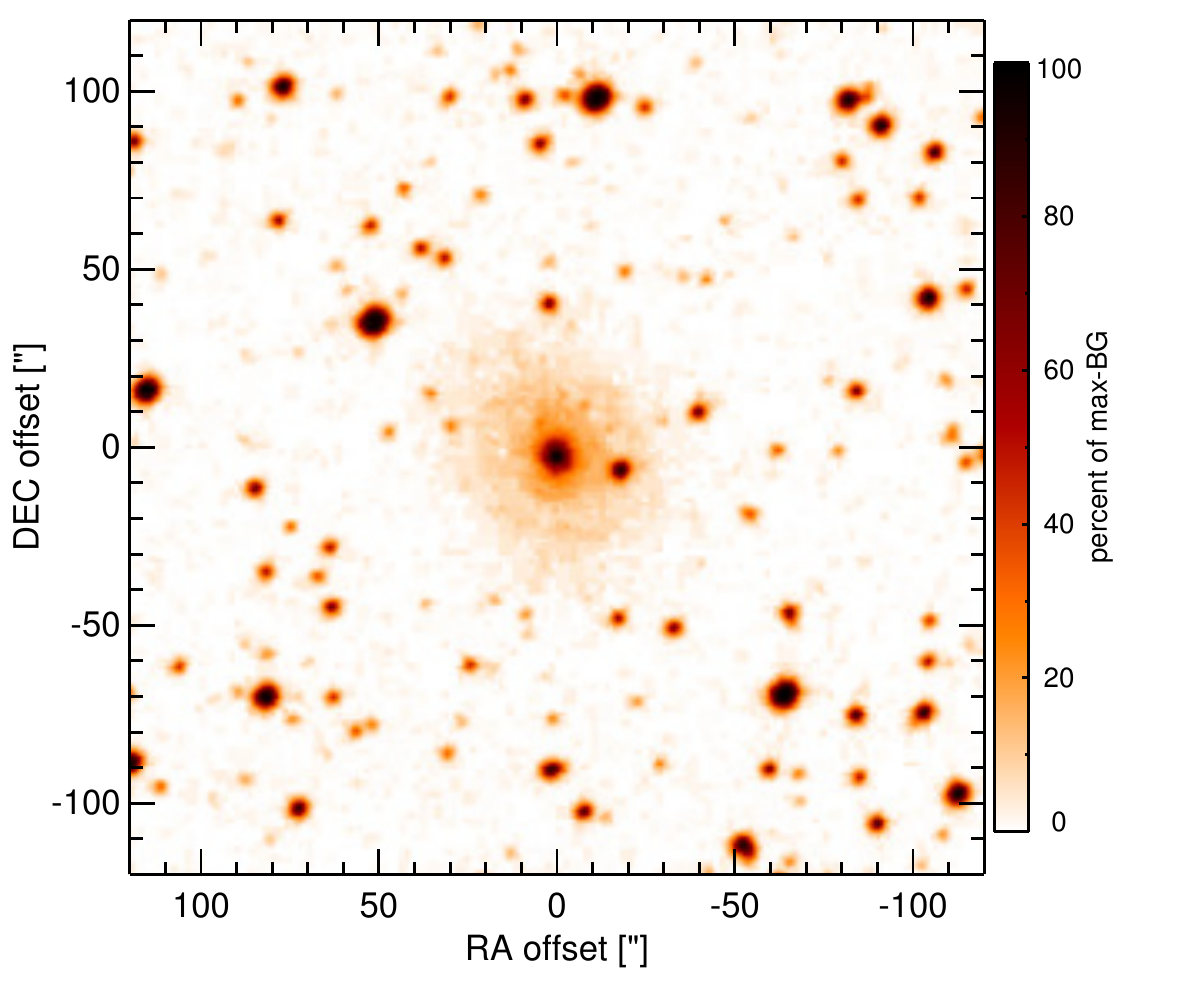}
    \caption{\label{fig:OPTim_ESO263-G013}
             Optical image (DSS, red filter) of ESO\,263-13. Displayed are the central $4\arcmin$ with North up and East to the left. 
              The colour scaling is linear with white corresponding to the median background and black to the $0.01\%$ pixels with the highest intensity.  
           }
\end{figure}
\begin{figure}
   \centering
   \includegraphics[angle=0,height=3.11cm]{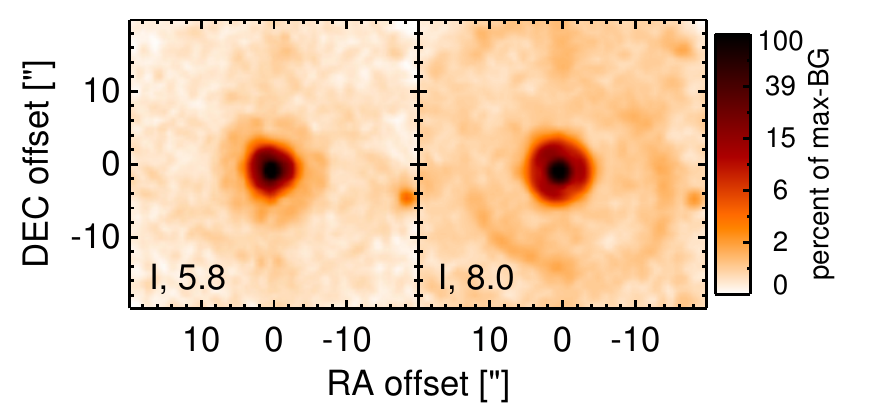}
    \caption{\label{fig:INTim_ESO263-G013}
             \spitzerr MIR images of ESO\,263-13. Displayed are the inner $40\arcsec$ with North up and East to the left. The colour scaling is logarithmic with white corresponding to median background and black to the $0.1\%$ pixels with the highest intensity.
             The label in the bottom left states instrument and central wavelength of the filter in $\mu$m (I: IRAC, M: MIPS). 
           }
\end{figure}
\begin{figure}
   \centering
   \includegraphics[angle=0,width=8.500cm]{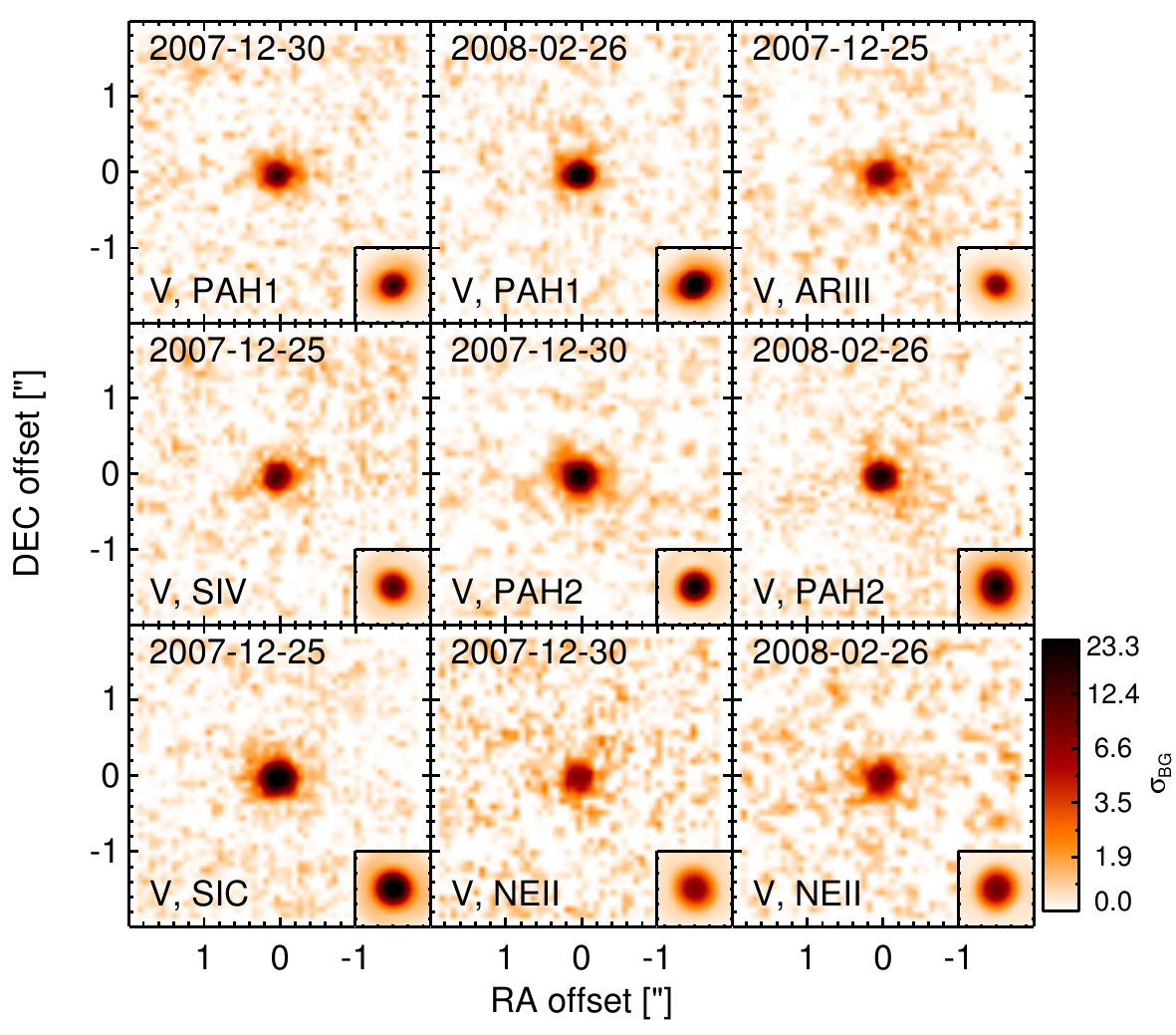}
    \caption{\label{fig:HARim_ESO263-G013}
             Subarcsecond-resolution MIR images of ESO\,263-13 sorted by increasing filter wavelength. 
             Displayed are the inner $4\arcsec$ with North up and East to the left. 
             The colour scaling is logarithmic with white corresponding to median background and black to the $75\%$ of the highest intensity of all images in units of $\sigbg$.
             The inset image shows the central arcsecond of the PSF from the calibrator star, scaled to match the science target.
             The labels in the bottom left state instrument and filter names (C: COMICS, M: Michelle, T: T-ReCS, V: VISIR).
           }
\end{figure}
\begin{figure}
   \centering
   \includegraphics[angle=0,width=8.50cm]{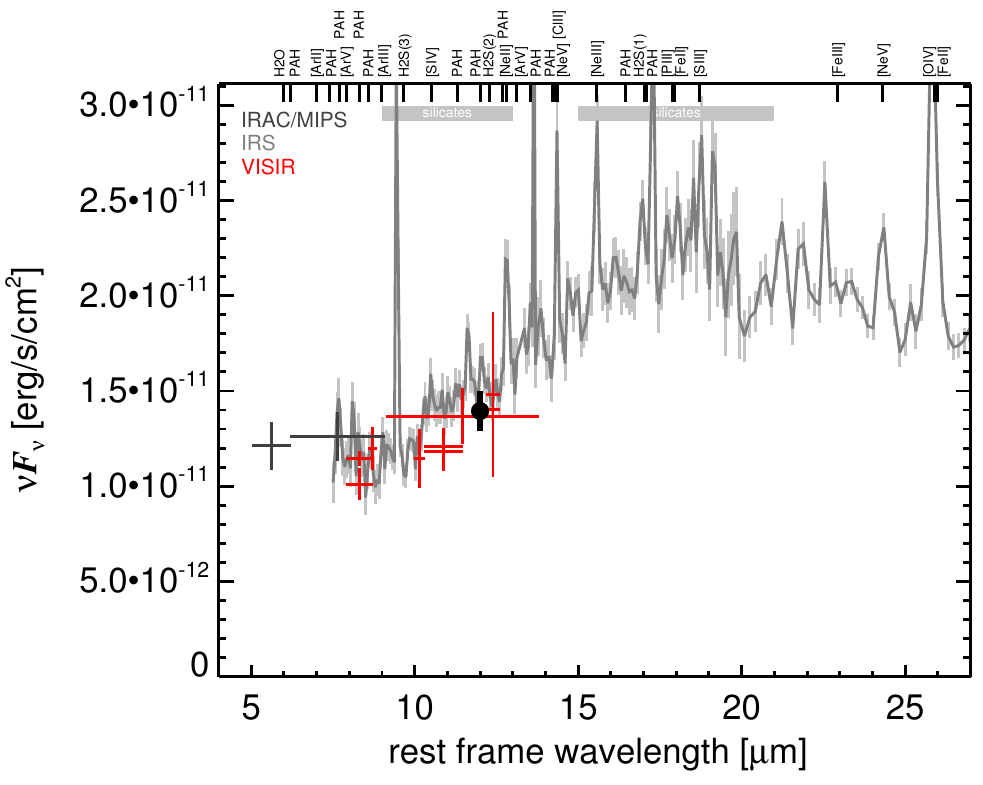}
   \caption{\label{fig:MISED_ESO263-G013}
      MIR SED of ESO\,263-13. The description  of the symbols (if present) is the following.
      Grey crosses and  solid lines mark the \spitzer/IRAC, MIPS and IRS data. 
      The colour coding of the other symbols is: 
      green for COMICS, magenta for Michelle, blue for T-ReCS and red for VISIR data.
      Darker-coloured solid lines mark spectra of the corresponding instrument.
      The black filled circles mark the nuclear 12 and $18\,\mu$m  continuum emission estimate from the data.
      The ticks on the top axis mark positions of common MIR emission lines, while the light grey horizontal bars mark wavelength ranges affected by the silicate 10 and 18$\mu$m features.     
   }
\end{figure}
\clearpage

\twocolumn[\begin{@twocolumnfalse}  
\subsection{ESO\,286-19 -- IRAS\,20551-4250}\label{app:ESO286-IG019}
ESO\,286-19 is a spiral-like merger system at a redshift of $z=$ 0.043 ($D \sim 180$\,Mpc) classified as an ULIRG with an H\,II optical nucleus \citep{johansson_extremely_1991,kim_optical_1995,veilleux_optical_1995}. 
However, multiple lines of evidence suggest that ESO\,286-19 harbours a highly obscured AGN \citep{pernechele_spectropolarimetric_2003,risaliti_unveiling_2006,brightman_xmm-newton_2011,imanishi_subaru_2011}.
In the MIR, it was observed with \spitzer/IRAC, IRS and MIPS, where it appears as a compact source in all images.
The IRAC $5.8$ and $8.0\,\mu$m and MIPS $24\,\mu$m photometry matches well the IRS LR staring-mode spectrum, which was published in \cite{spoon_detection_2006}.
The arcsecond-scale MIR SED is dominated by the extremely deep silicate 10 and $18\,\mu$m silicate absorption features, and possesses a red slope in  $\nu F_\nu$-space.
PAH emission is present as well and indicates the presence of star formation in ESO\,286-19 (see also \citealt{farrah_high-resolution_2007} for an IRS HR staring-mode spectrum).
The absence of any strong emission features indicates that putative MIR emission-line producing regions  are heavily extincted (similar to, e.g., NGC\,4945; \citealt{perez-beaupuits_deeply_2011}).
One subarcsecond-resolution $Q$-band image with T-ReCS was taken in 2008 \citep{imanishi_subaru_2011}.
A compact MIR nucleus without any extended host emission was detected. 
The source is possibly marginally resolved (FWHM $\sim 0.6\arcsec \sim 480\,$pc), however, in absence of other subarcsecond MIR images, the nuclear extension remains uncertain.
Our reanalysis of the $Q$-band image provides a consistent nuclear flux with the published value and also matches the \spitzerr spectrophotometry.
Therefore, we correct our 12 and $18\,\mu$m continuum flux estimates for the silicate absorption features.
\newline\end{@twocolumnfalse}]

\begin{figure}
   \centering
   \includegraphics[angle=0,width=8.500cm]{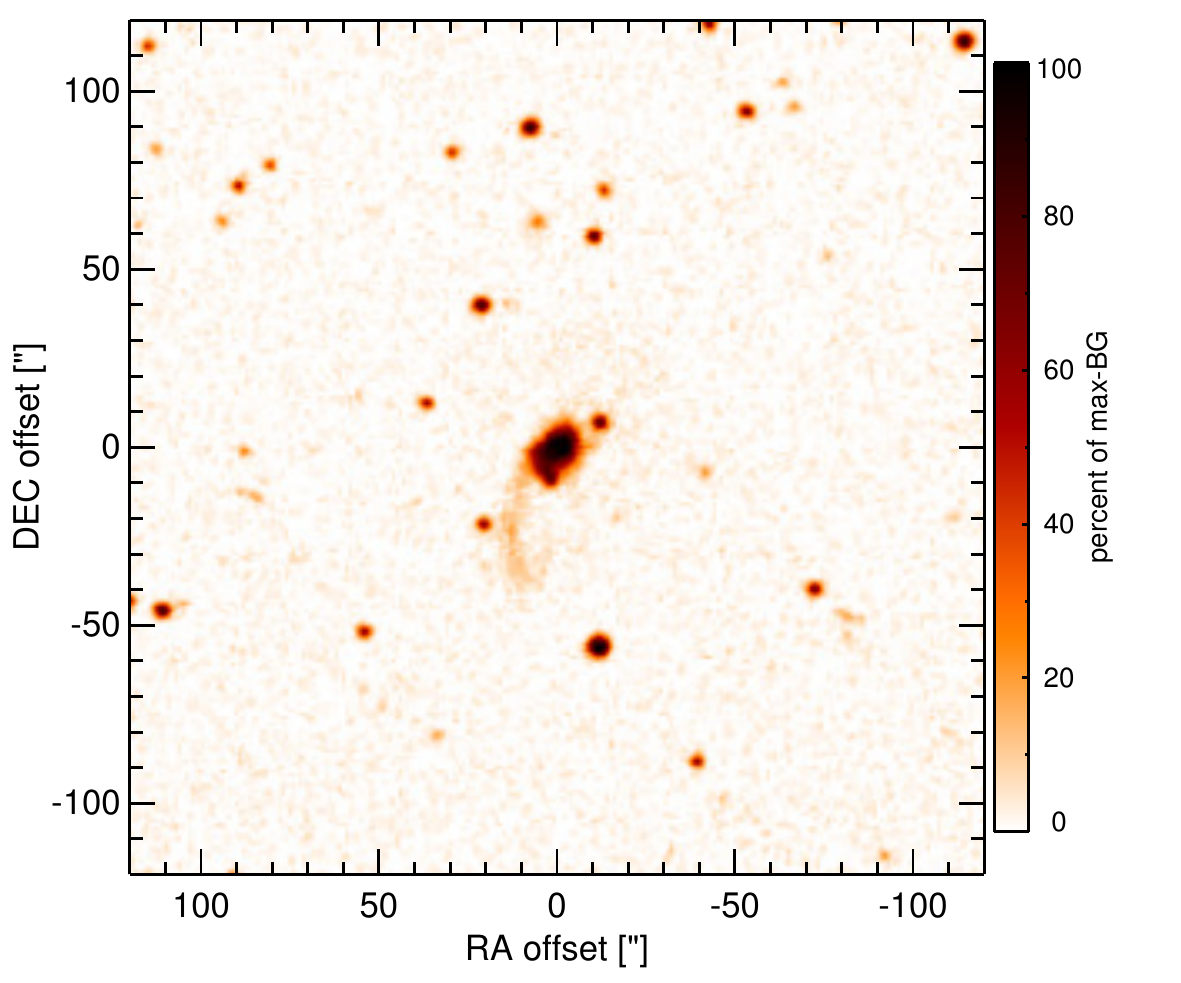}
    \caption{\label{fig:OPTim_ESO286-IG019}
             Optical image (DSS, red filter) of ESO\,286-19. Displayed are the central $4\arcmin$ with North up and East to the left. 
              The colour scaling is linear with white corresponding to the median background and black to the $0.01\%$ pixels with the highest intensity.  
           }
\end{figure}
\begin{figure}
   \centering
   \includegraphics[angle=0,height=3.11cm]{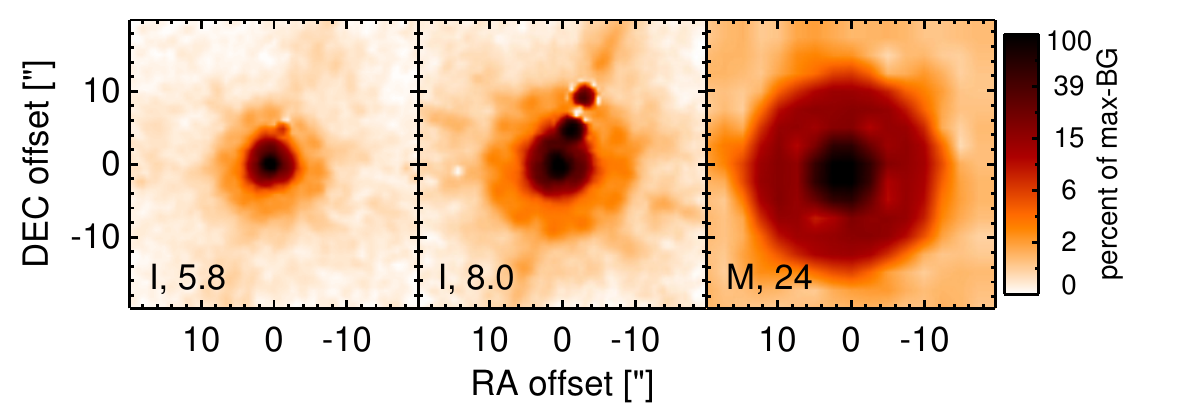}
    \caption{\label{fig:INTim_ESO286-IG019}
             \spitzerr MIR images of ESO\,286-19. Displayed are the inner $40\arcsec$ with North up and East to the left. The colour scaling is logarithmic with white corresponding to median background and black to the $0.1\%$ pixels with the highest intensity.
             The label in the bottom left states instrument and central wavelength of the filter in $\mu$m (I: IRAC, M: MIPS).
             Note that the apparent off-nuclear compact sources in the IRAC 5.8 and $8.0\,\mu$m images are instrumental artefacts.
           }
\end{figure}
\begin{figure}
   \centering
   \includegraphics[angle=0,height=3.11cm]{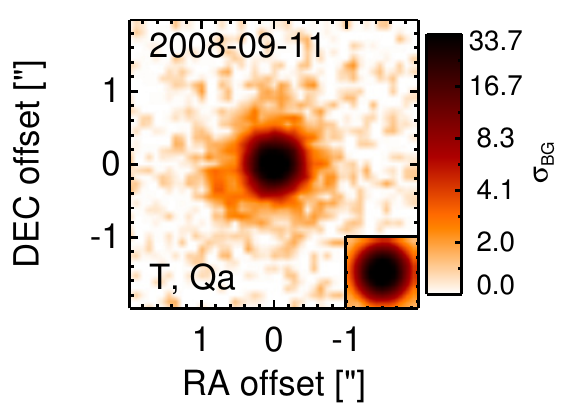}
    \caption{\label{fig:HARim_ESO286-IG019}
             Subarcsecond-resolution MIR images of ESO\,286-19 sorted by increasing filter wavelength. 
             Displayed are the inner $4\arcsec$ with North up and East to the left. 
             The colour scaling is logarithmic with white corresponding to median background and black to the $75\%$ of the highest intensity of all images in units of $\sigbg$.
             The inset image shows the central arcsecond of the PSF from the calibrator star, scaled to match the science target.
             The labels in the bottom left state instrument and filter names (C: COMICS, M: Michelle, T: T-ReCS, V: VISIR).
           }
\end{figure}
\begin{figure}
   \centering
   \includegraphics[angle=0,width=8.50cm]{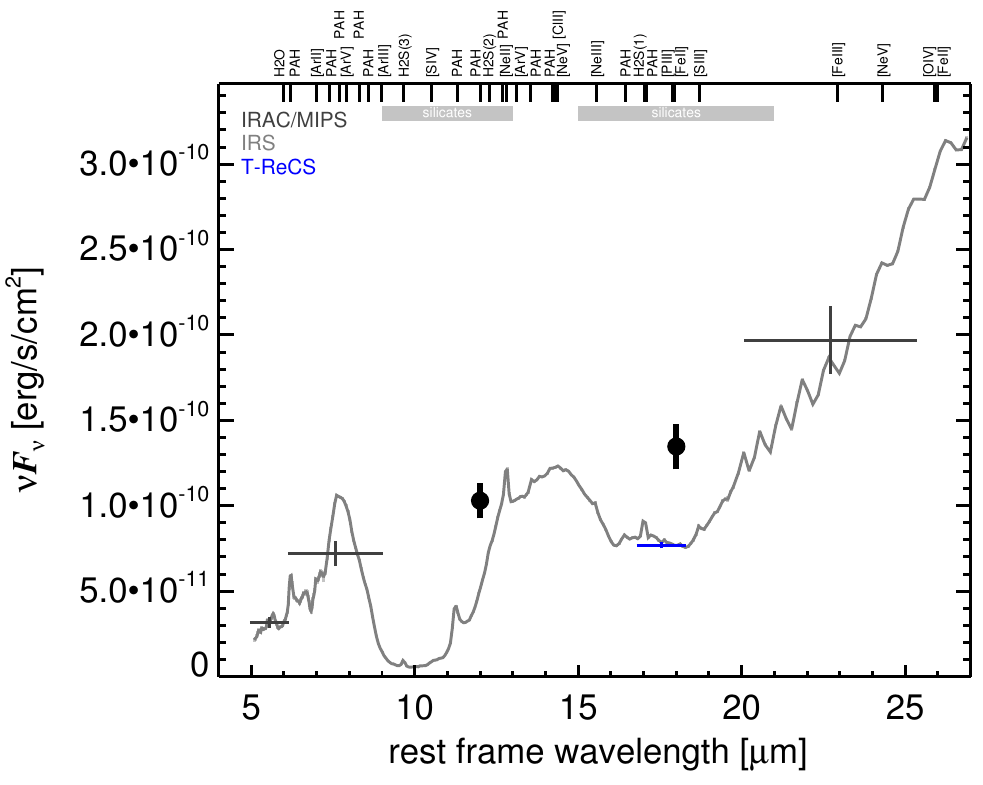}
   \caption{\label{fig:MISED_ESO286-IG019}
      MIR SED of ESO\,286-19. The description  of the symbols (if present) is the following.
      Grey crosses and  solid lines mark the \spitzer/IRAC, MIPS and IRS data. 
      The colour coding of the other symbols is: 
      green for COMICS, magenta for Michelle, blue for T-ReCS and red for VISIR data.
      Darker-coloured solid lines mark spectra of the corresponding instrument.
      The black filled circles mark the nuclear 12 and $18\,\mu$m  continuum emission estimate from the data.
      The ticks on the top axis mark positions of common MIR emission lines, while the light grey horizontal bars mark wavelength ranges affected by the silicate 10 and 18$\mu$m features.     
   }
\end{figure}
\clearpage

\twocolumn[\begin{@twocolumnfalse}  
\subsection{ESO\,297-18 -- Swift\,J0138.6-4001 }\label{app:ESO297-G018}
ESO\,297-18 is a highly inclined spiral galaxy at a redshift of $z=$ 0.0252 ($D \sim103$\,Mpc) that harbours an AGN with optical Sy\,2 classification \citep{veron-cetty_catalogue_2010}.
This buried AGN \citep{ueda_suzaku_2007} belongs to the nine-month BAT AGN sample.
It was observed with \spitzer/IRAC and IRS and appears as a compact nuclear source with extended host emission tracing the inclined spiral in the IRAC $5.8$ and $8.0\,\mu$m images.
The IRS LR staring-mode spectrum shows a prominent silicate $10\,\mu$m absorption feature with PAH and  \neii emission, indicating significant star formation. 
The MIR spectrum peaks around $\sim 15\,\mu$m in $\nu F_\nu$-space.
We observed ESO\,297-18 in 2007 with VISIR in four narrow $N$-band filters during one night. 
The MIR nucleus is only weakly detected in all images without any sign of host emission. 
It appears very compact, but the low S/N prevents any extension analysis.
Our new measurement of the nuclear fluxes is consistent with the values for the SIV and PAH2 filters, which have already been published in \cite{gandhi_resolving_2009}.
Interestingly, these fluxes are on average $\sim 52\%$ lower than those of the \spitzerr spectrophotometry and indicate a different MIR SED slope without the silicate $10\,\mu$m absorption feature.
\newline\end{@twocolumnfalse}]

\begin{figure}
   \centering
   \includegraphics[angle=0,width=8.500cm]{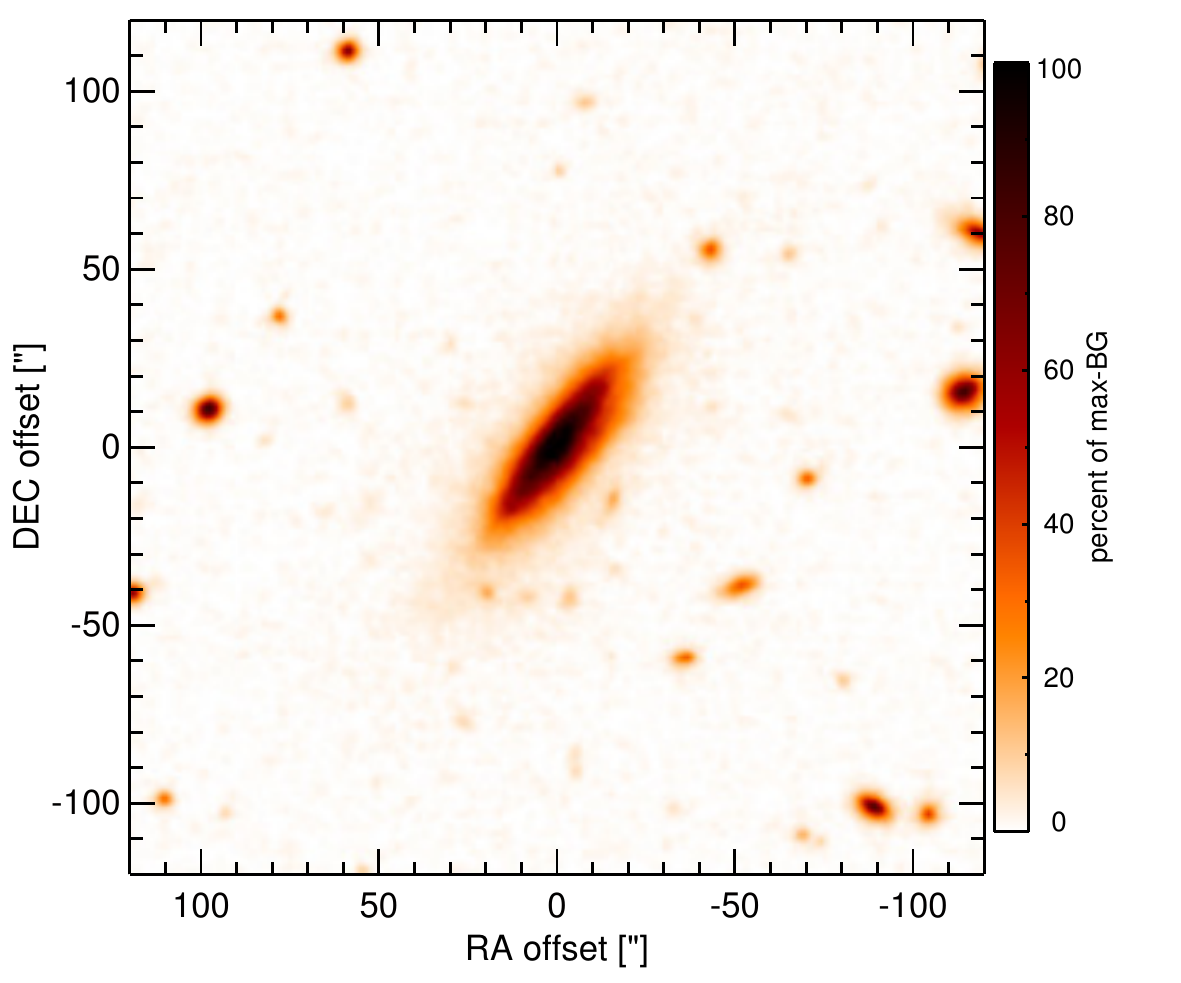}
    \caption{\label{fig:OPTim_ESO297-G018}
             Optical image (DSS, red filter) of ESO\,297-18. Displayed are the central $4\arcmin$ with North up and East to the left. 
              The colour scaling is linear with white corresponding to the median background and black to the $0.01\%$ pixels with the highest intensity.  
           }
\end{figure}
\begin{figure}
   \centering
   \includegraphics[angle=0,height=3.11cm]{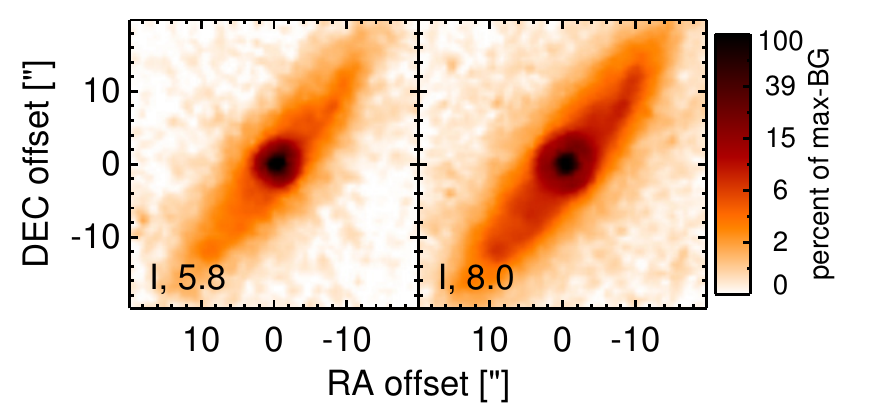}
    \caption{\label{fig:INTim_ESO297-G018}
             \spitzerr MIR images of ESO\,297-18. Displayed are the inner $40\arcsec$ with North up and East to the left. The colour scaling is logarithmic with white corresponding to median background and black to the $0.1\%$ pixels with the highest intensity.
             The label in the bottom left states instrument and central wavelength of the filter in $\mu$m (I: IRAC, M: MIPS). 
           }
\end{figure}
\begin{figure}
   \centering
   \includegraphics[angle=0,width=8.500cm]{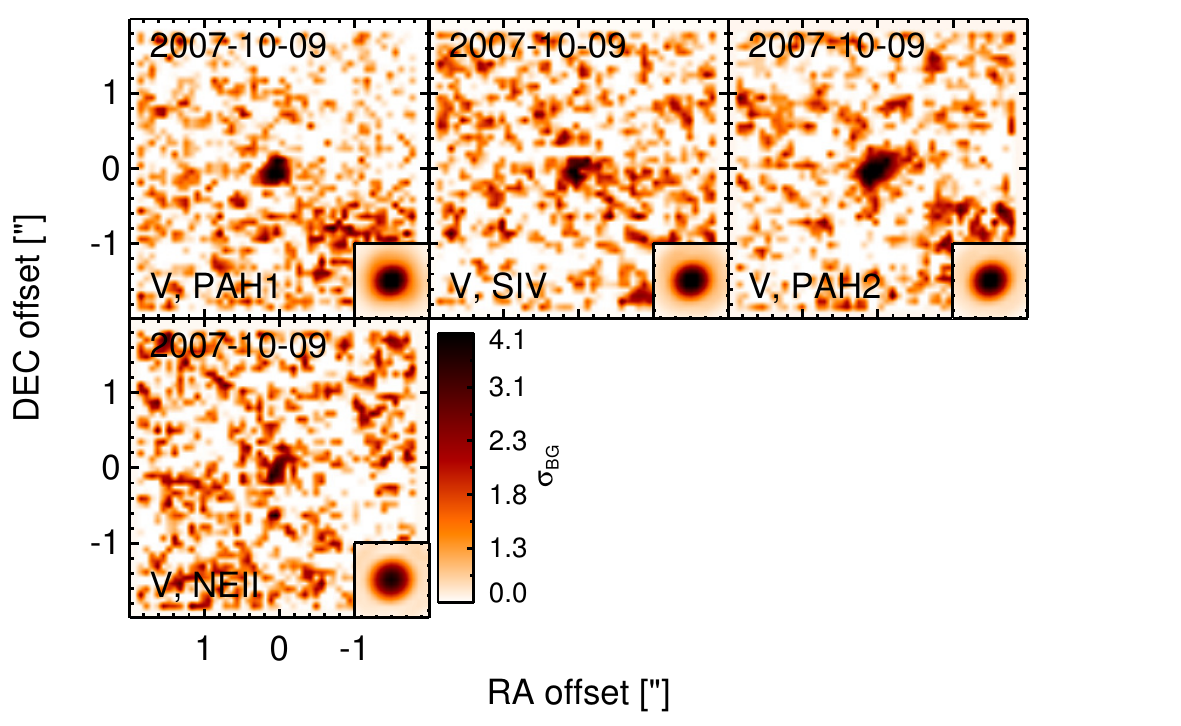}
    \caption{\label{fig:HARim_ESO297-G018}
             Subarcsecond-resolution MIR images of ESO\,297-18 sorted by increasing filter wavelength. 
             Displayed are the inner $4\arcsec$ with North up and East to the left. 
             The colour scaling is logarithmic with white corresponding to median background and black to the $75\%$ of the highest intensity of all images in units of $\sigbg$.
             The inset image shows the central arcsecond of the PSF from the calibrator star, scaled to match the science target.
             The labels in the bottom left state instrument and filter names (C: COMICS, M: Michelle, T: T-ReCS, V: VISIR).
           }
\end{figure}
\begin{figure}
   \centering
   \includegraphics[angle=0,width=8.50cm]{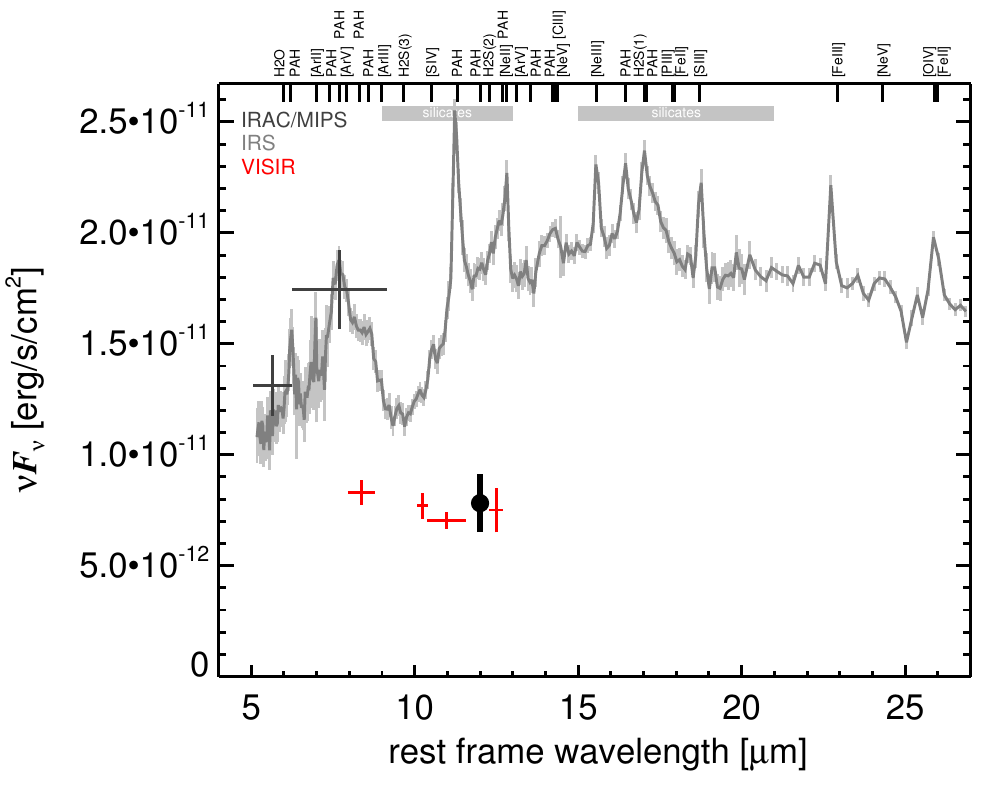}
   \caption{\label{fig:MISED_ESO297-G018}
      MIR SED of ESO\,297-18. The description  of the symbols (if present) is the following.
      Grey crosses and  solid lines mark the \spitzer/IRAC, MIPS and IRS data. 
      The colour coding of the other symbols is: 
      green for COMICS, magenta for Michelle, blue for T-ReCS and red for VISIR data.
      Darker-coloured solid lines mark spectra of the corresponding instrument.
      The black filled circles mark the nuclear 12 and $18\,\mu$m  continuum emission estimate from the data.
      The ticks on the top axis mark positions of common MIR emission lines, while the light grey horizontal bars mark wavelength ranges affected by the silicate 10 and 18$\mu$m features.     
   }
\end{figure}
\clearpage

\twocolumn[\begin{@twocolumnfalse}  
\subsection{ESO\,323-32 -- Fairall\,315}\label{app:ESO323-G032}
ESO\,323-32 is a spiral galaxy at a redshift of $z=$ 0.0160 ($D\sim70.5$\,Mpc) hosting an AGN with optical Sy\,1.9 classification \citep{veron-cetty_catalogue_2010}.
No \spitzerr data are available for this object.
We observed ESO\,323-32 with VISIR in 2008 in four narrow $N$-band filters during one night.
A compact MIR nucleus without any host emission was detected in the PAH1, PAH2 and NEII filter images but not in SIV, which indicates the presence of a deep silicate $10\,\mu$m absorption feature.
The low S/N prevents us to perform an extension analyses of the source. 
Our new measurement of the nuclear flux in the PAH2 filter is a factor two lower than the previously published value in \cite{gandhi_resolving_2009}, which is due to the new measurement method optimized for low S/N. 
\newline\end{@twocolumnfalse}]

\begin{figure}
   \centering
   \includegraphics[angle=0,width=8.500cm]{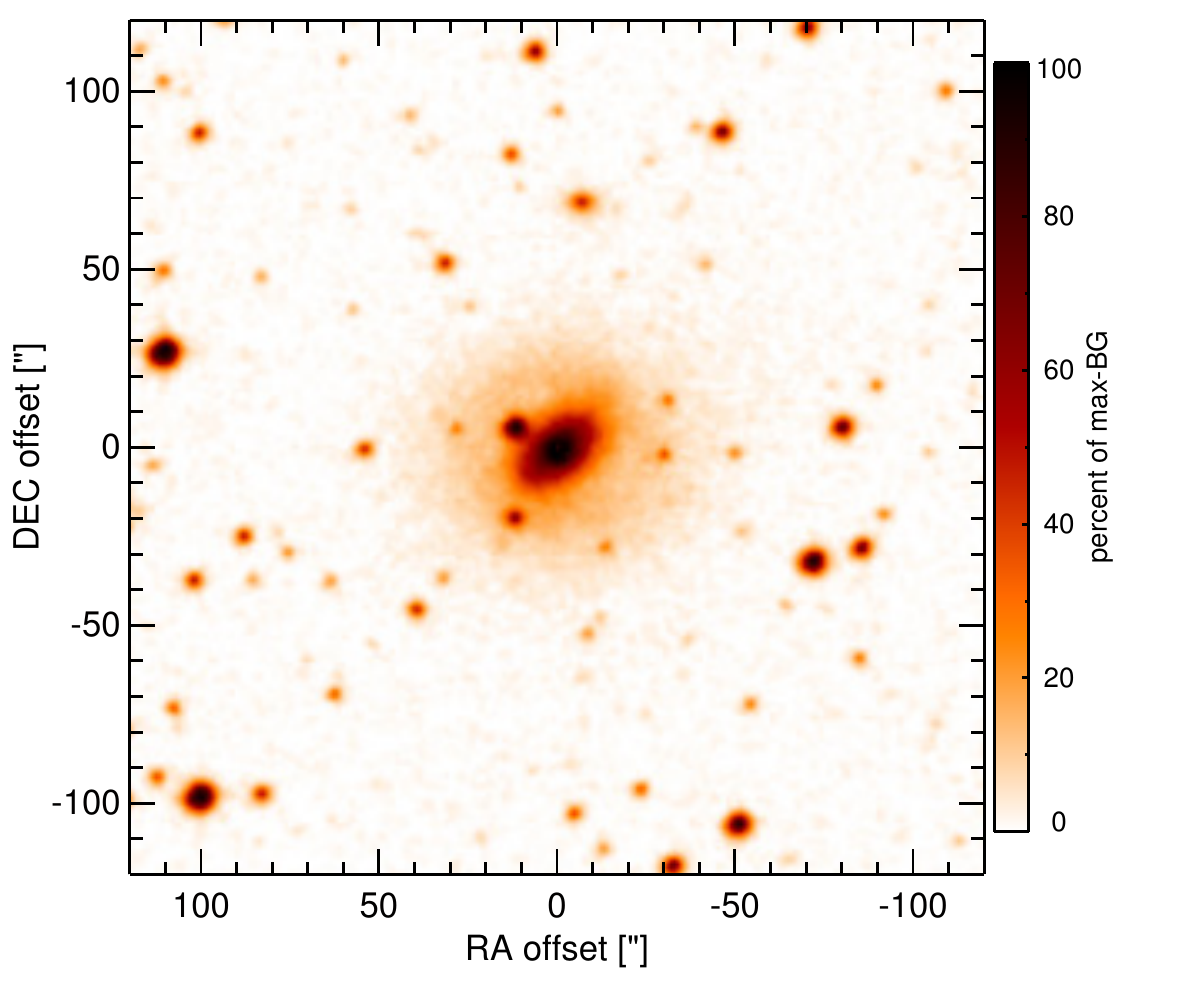}
    \caption{\label{fig:OPTim_ESO323-G032}
             Optical image (DSS, red filter) of ESO\,323-32. Displayed are the central $4\arcmin$ with North up and East to the left. 
              The colour scaling is linear with white corresponding to the median background and black to the $0.01\%$ pixels with the highest intensity.  
           }
\end{figure}
\begin{figure}
   \centering
   \includegraphics[angle=0,height=3.11cm]{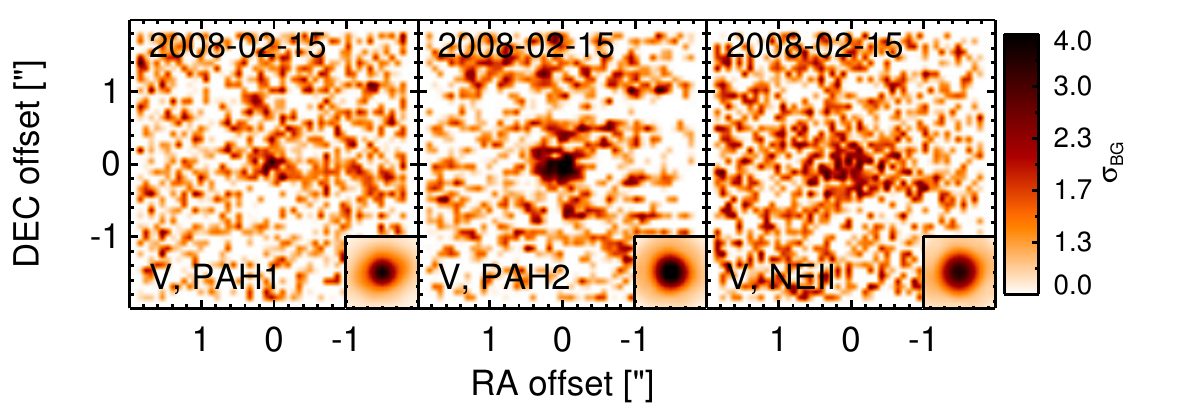}
    \caption{\label{fig:HARim_ESO323-G032}
             Subarcsecond-resolution MIR images of ESO\,323-32 sorted by increasing filter wavelength. 
             Displayed are the inner $4\arcsec$ with North up and East to the left. 
             The colour scaling is logarithmic with white corresponding to median background and black to the $75\%$ of the highest intensity of all images in units of $\sigbg$.
             The inset image shows the central arcsecond of the PSF from the calibrator star, scaled to match the science target.
             The labels in the bottom left state instrument and filter names (C: COMICS, M: Michelle, T: T-ReCS, V: VISIR).
           }
\end{figure}
\begin{figure}
   \centering
   \includegraphics[angle=0,width=8.50cm]{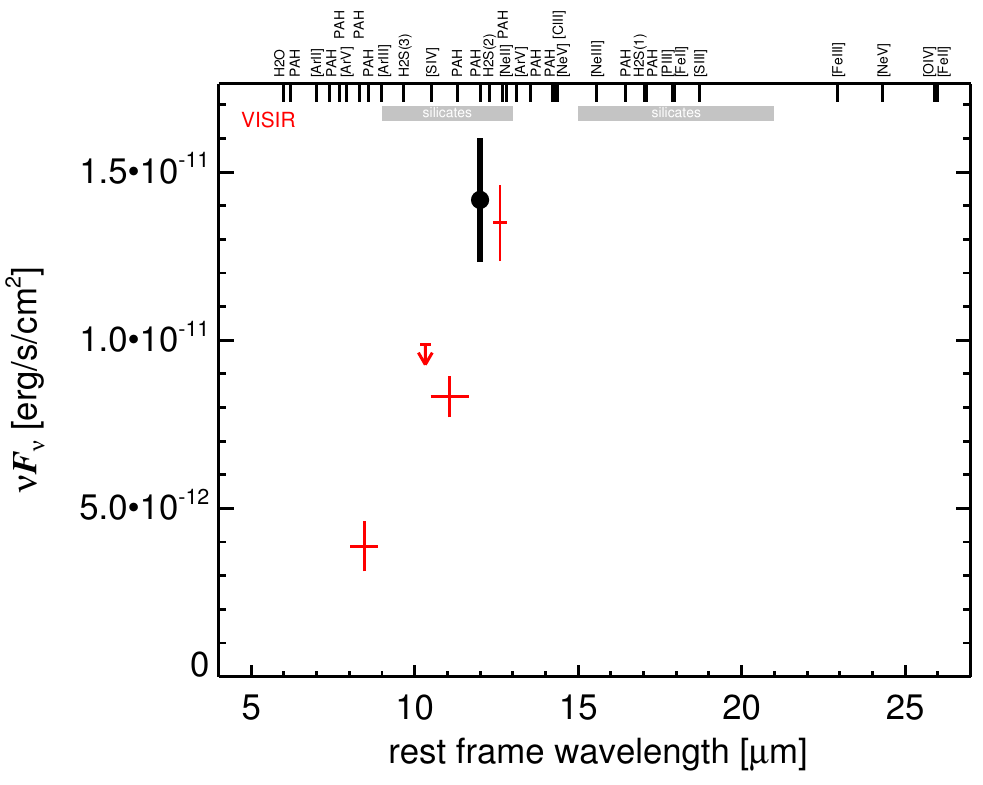}
   \caption{\label{fig:MISED_ESO323-G032}
      MIR SED of ESO\,323-32. The description  of the symbols (if present) is the following.
      Grey crosses and  solid lines mark the \spitzer/IRAC, MIPS and IRS data. 
      The colour coding of the other symbols is: 
      green for COMICS, magenta for Michelle, blue for T-ReCS and red for VISIR data.
      Darker-coloured solid lines mark spectra of the corresponding instrument.
      The black filled circles mark the nuclear 12 and $18\,\mu$m  continuum emission estimate from the data.
      The ticks on the top axis mark positions of common MIR emission lines, while the light grey horizontal bars mark wavelength ranges affected by the silicate 10 and 18$\mu$m features.     
   }
\end{figure}
\clearpage

\twocolumn[\begin{@twocolumnfalse}  
\subsection{ESO\,323-77 -- MCG-7-27-32}\label{app:ESO323-G077}
ESO\,323-77 is a spiral galaxy at a redshift of $z=$ 0.0150 ($D\sim66.3$\,Mpc) with an AGN with optical Sy\,1.2 classification \citep{veron-cetty_catalogue_2010} that appears partly obscured \citep{schmid_spectropolarimetry_2003}.
It was observed with \spitzer/IRAC, IRS and MIPS and appears compact in all images.
The IRS LR staring-mode spectrum exhibits weak silicate $10\,\mu$m absorption and strong PAH emission, which indicates significant star formation.
The MIR spectral slope is rather flat in $\nu F_\nu$-space.
We observed ESO\,323-77 with VISIR in three different narrow-$N$ filters in 2009 (two images) and 2010 (two images). 
A compact MIR nucleus was detected in all images without any sign of extended host galaxy emission.
Quantitative extension analysis shows that the source is marginally resolved in all images (FWHM $\sim 0.39\arcsec \sim 120\,$pc; PA$\sim 100\degree$).
The VISIR photometry from 2009 together and a VISIR $N$-band spectrum were published by \cite{honig_dusty_2010-1}.
The VISIR spectrum differs significantly from the IRS spectrum in having lower flux levels but no PAH emission and no silicate absorption, which indicates that most of the star-formation related emission is resolved out at subarcsecond resolution. 
Our new Gaussian flux measurements agree  with the previous values well, while the nuclear fluxes using the PSF measurement are significantly lower, in particular at the shortest wavelengths. 
Compared to the \spitzerr spectrophotometry, the nuclear fluxes are on average $\sim 52\%$ lower.
MIDI interferometric observations of ESO\,323-77, however, found a compact source with a size of $\sim2$\,pc to dominate the nuclear MIR emission \citep{burtscher_diversity_2013}.
\newline\end{@twocolumnfalse}]

\begin{figure}
   \centering
   \includegraphics[angle=0,width=8.500cm]{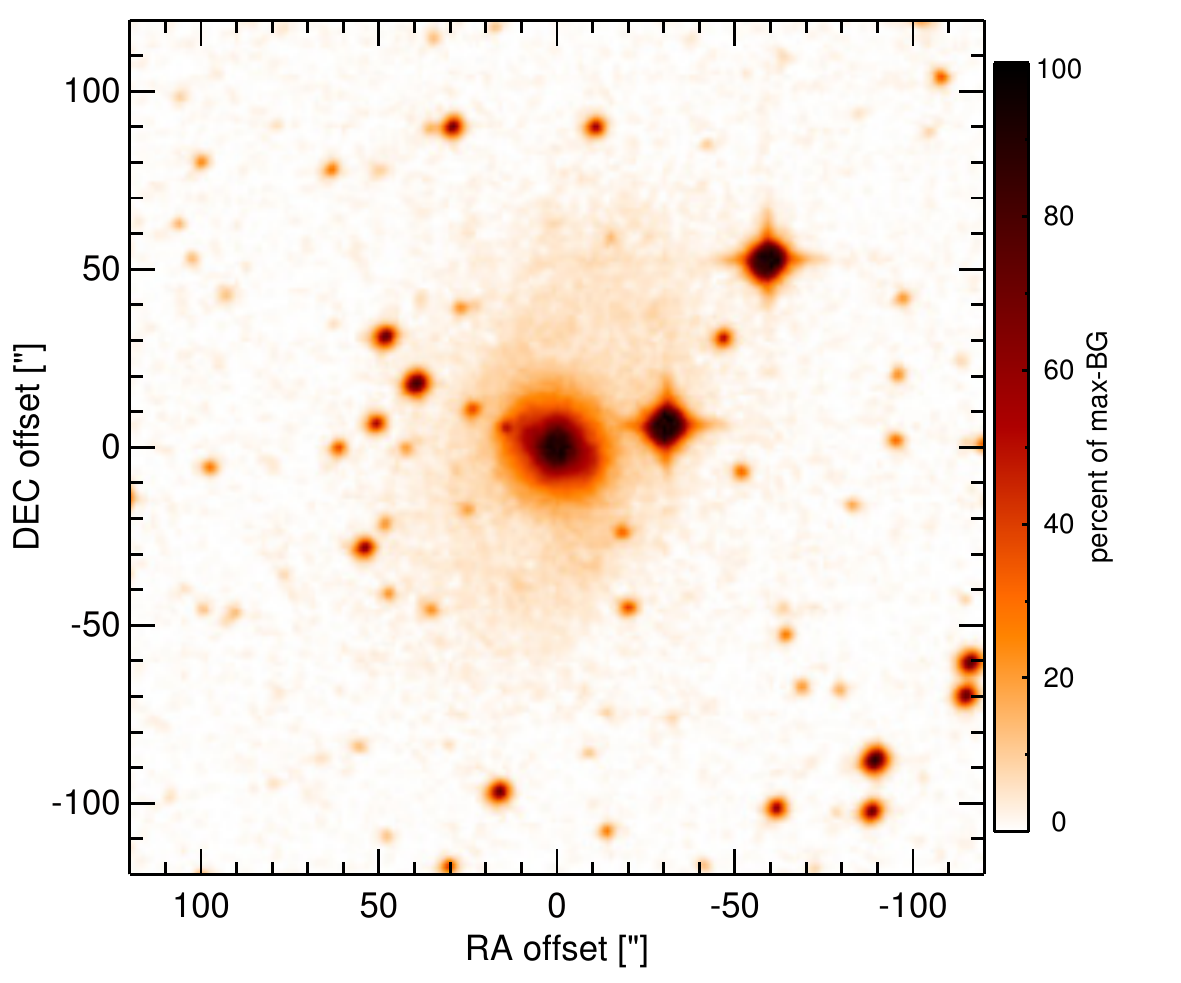}
    \caption{\label{fig:OPTim_ESO323-G077}
             Optical image (DSS, red filter) of ESO\,323-77. Displayed are the central $4\arcmin$ with North up and East to the left. 
              The colour scaling is linear with white corresponding to the median background and black to the $0.01\%$ pixels with the highest intensity.  
           }
\end{figure}
\begin{figure}
   \centering
   \includegraphics[angle=0,height=3.11cm]{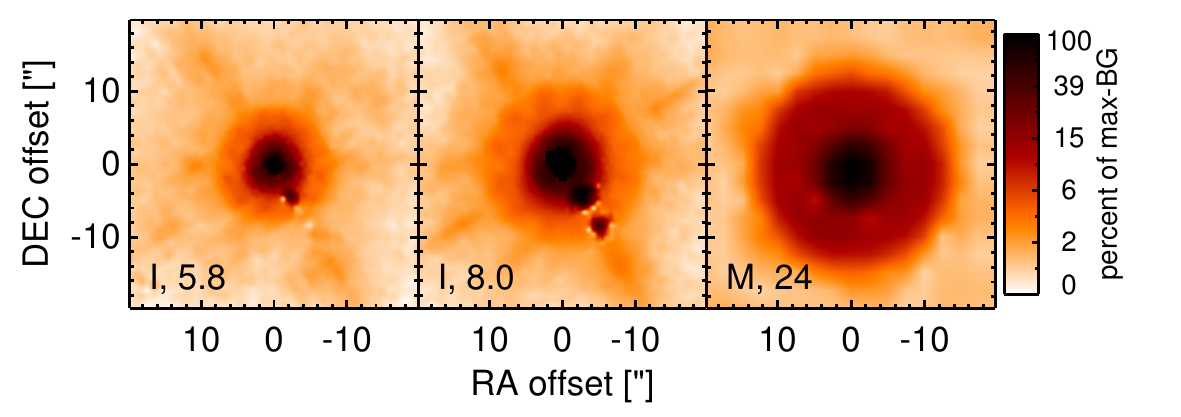}
    \caption{\label{fig:INTim_ESO323-G077}
             \spitzerr MIR images of ESO\,323-77. Displayed are the inner $40\arcsec$ with North up and East to the left. The colour scaling is logarithmic with white corresponding to median background and black to the $0.1\%$ pixels with the highest intensity.
             The label in the bottom left states instrument and central wavelength of the filter in $\mu$m (I: IRAC, M: MIPS).
             Note that the apparent off-nuclear compact sources in the IRAC 5.8 and $8.0\,\mu$m images are instrumental artefacts.
           }
\end{figure}
\begin{figure}
   \centering
   \includegraphics[angle=0,width=8.500cm]{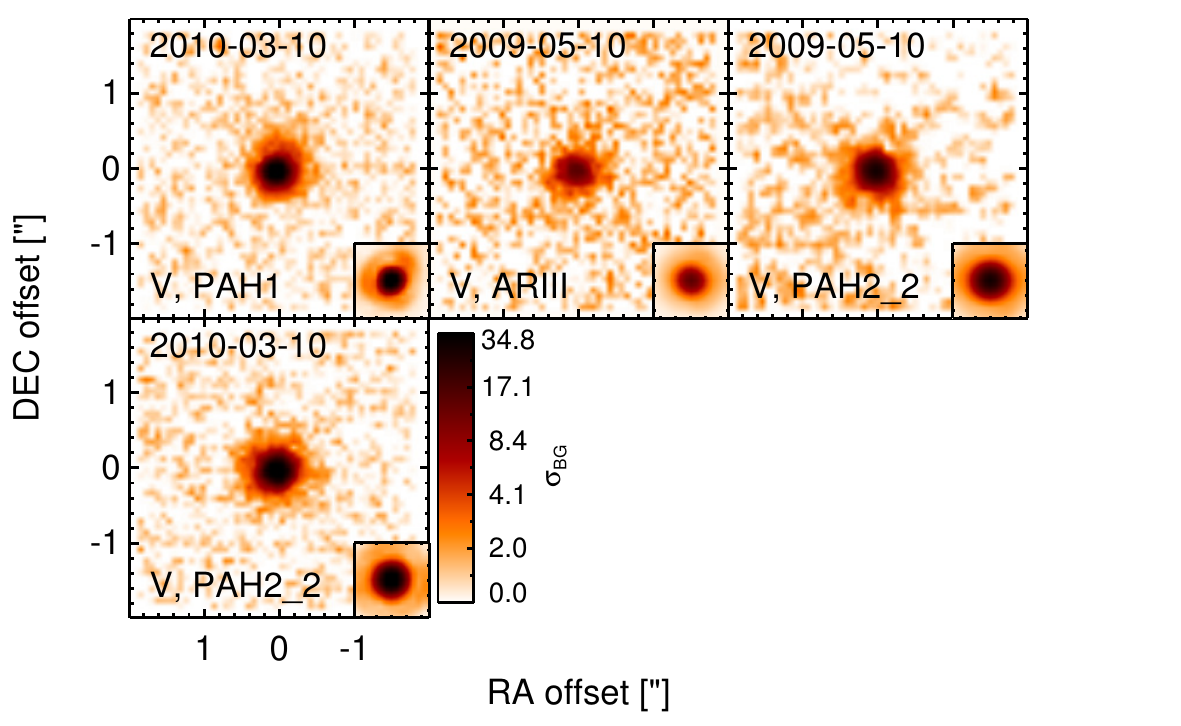}
    \caption{\label{fig:HARim_ESO323-G077}
             Subarcsecond-resolution MIR images of ESO\,323-77 sorted by increasing filter wavelength. 
             Displayed are the inner $4\arcsec$ with North up and East to the left. 
             The colour scaling is logarithmic with white corresponding to median background and black to the $75\%$ of the highest intensity of all images in units of $\sigbg$.
             The inset image shows the central arcsecond of the PSF from the calibrator star, scaled to match the science target.
             The labels in the bottom left state instrument and filter names (C: COMICS, M: Michelle, T: T-ReCS, V: VISIR).
           }
\end{figure}
\begin{figure}
   \centering
   \includegraphics[angle=0,width=8.50cm]{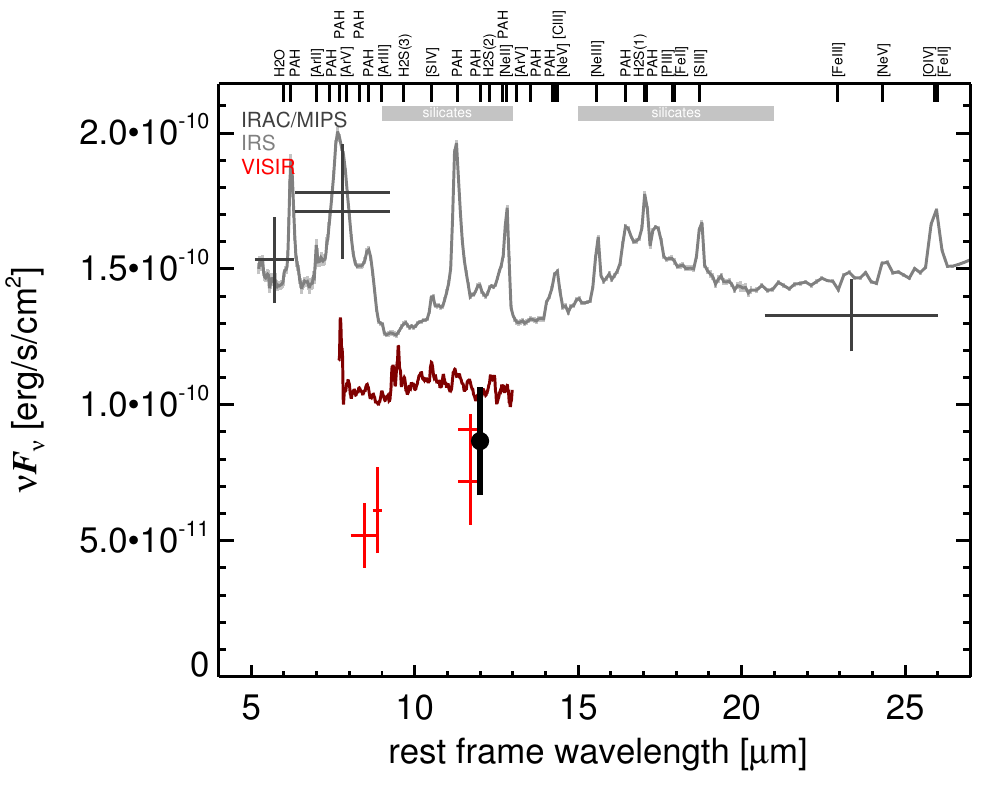}
   \caption{\label{fig:MISED_ESO323-G077}
      MIR SED of ESO\,323-77. The description  of the symbols (if present) is the following.
      Grey crosses and  solid lines mark the \spitzer/IRAC, MIPS and IRS data. 
      The colour coding of the other symbols is: 
      green for COMICS, magenta for Michelle, blue for T-ReCS and red for VISIR data.
      Darker-coloured solid lines mark spectra of the corresponding instrument.
      The black filled circles mark the nuclear 12 and $18\,\mu$m  continuum emission estimate from the data.
      The ticks on the top axis mark positions of common MIR emission lines, while the light grey horizontal bars mark wavelength ranges affected by the silicate 10 and 18$\mu$m features.     
   }
\end{figure}
\clearpage

\twocolumn[\begin{@twocolumnfalse}  
\subsection{ESO\,362-18 -- MCG-5-13-17}\label{app:ESO362-G018}
ESO\,362-18 is a perturbed spiral galaxy at a redshift of $z=$ 0.0124 ($D\sim52.1$\,Mpc) hosting a Sy\,1.5 nucleus \citep{veron-cetty_catalogue_2010} with an extended narrow-line region \citep{fraquelli_extended_2000}, which is a member of the nine-month BAT AGN sample.
It was observed with Palomar 5\,m/MIRLIN in 1999 in a broad $N$-band filter with a marginally resolved MIR nucleus being detected \citep{gorjian_10_2004}.
ESO\,362-18 was also observed with \spitzer/IRAC, IRS and MIPS.
The IRAC $5.8$ and $8.0\,\mu$m images show weak host galaxy emission despite  streak-like artefacts apart from the compact nucleus, which is also detected in the MIPS $24\,\mu$m image.
Our nuclear IRAC $5.8$ and $8.0\,\mu$m photometry yields significantly lower fluxes than those published in \cite{gallimore_infrared_2010}, which is probably caused by the different measurement method and the surrounding host emission.
The IRS LR mapping spectrum shows strong PAH emission features and peaks at $\sim 18\,\mu$m in  $\nu F_\nu$-space (see also \citealt{wu_spitzer/irs_2009,gallimore_infrared_2010}).
ESO\,362-18 was observed once with T-ReCS in one narrow $N$-band filter in 2008.
Note that no standard star could be identified for this observation, and we  use the median conversion factor for the absolute flux calibration of the T-ReCS measurement.
In addition, we observed it with VISIR in in three narrow $N$-band filters in 2009. 
The nucleus is clearly detected in all images, while  surrounding host emission might be marginally detected in particular in the Si2 and NEII\_1 images.
The nucleus is unresolved in the sharpest image (PAH2) and thus classified as unresolved in the MIR at subarcsecond scales.
The measured nuclear fluxes of the VISIR images agree with the \spitzerr spectrophotometry except for the PAH2 measurement, which indicates that the PAH 11.3\,$\mu$m feature is absent in the subarcsecond MIR spectrum.
This also agrees with the T-ReCS Si2 measurement, which indicates the PAH  8.6\,$\mu$m is absent as well.
We conclude that the star formation affecting the \spitzerr spectrophotometry is probably not originating from the inner $\sim85$\,pc of ESO\,362-18.
\newline\end{@twocolumnfalse}]

\begin{figure}
   \centering
   \includegraphics[angle=0,width=8.500cm]{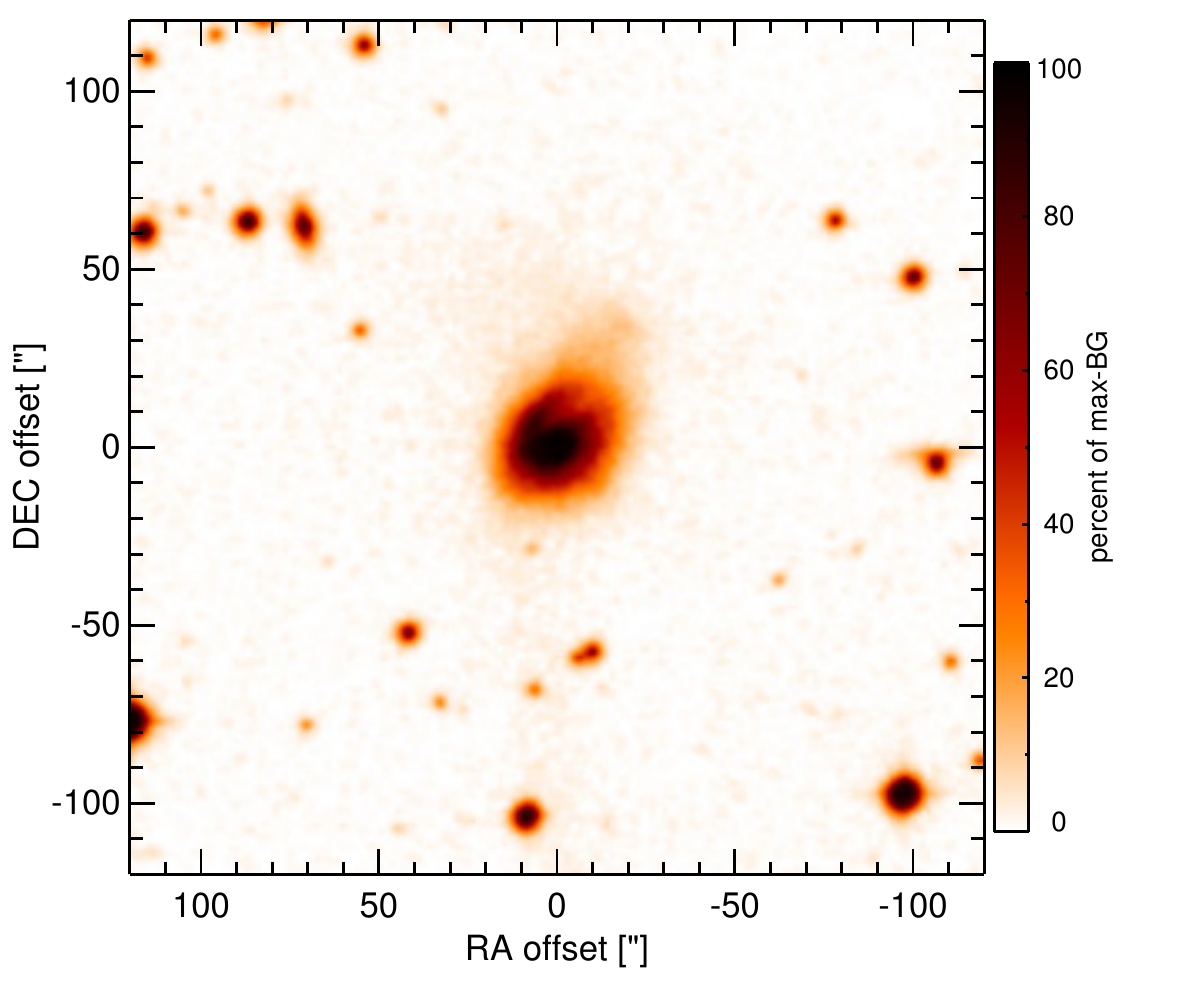}
    \caption{\label{fig:OPTim_ESO362-G018}
             Optical image (DSS, red filter) of ESO\,362-18. Displayed are the central $4\arcmin$ with North up and East to the left. 
              The colour scaling is linear with white corresponding to the median background and black to the $0.01\%$ pixels with the highest intensity.  
           }
\end{figure}
\begin{figure}
   \centering
   \includegraphics[angle=0,height=3.11cm]{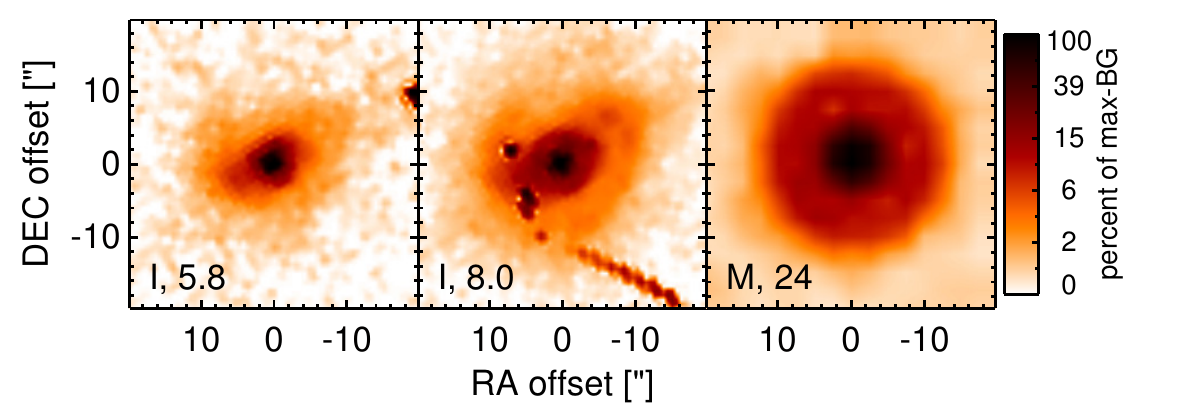}
    \caption{\label{fig:INTim_ESO362-G018}
             \spitzerr MIR images of ESO\,362-18. Displayed are the inner $40\arcsec$ with North up and East to the left. The colour scaling is logarithmic with white corresponding to median background and black to the $0.1\%$ pixels with the highest intensity.
             The label in the bottom left states instrument and central wavelength of the filter in $\mu$m (I: IRAC, M: MIPS).
             Note that the IRAC $8.0\,\mu$m image suffers from several instrumental artefacts.
           }
\end{figure}
\begin{figure}
   \centering
   \includegraphics[angle=0,width=8.500cm]{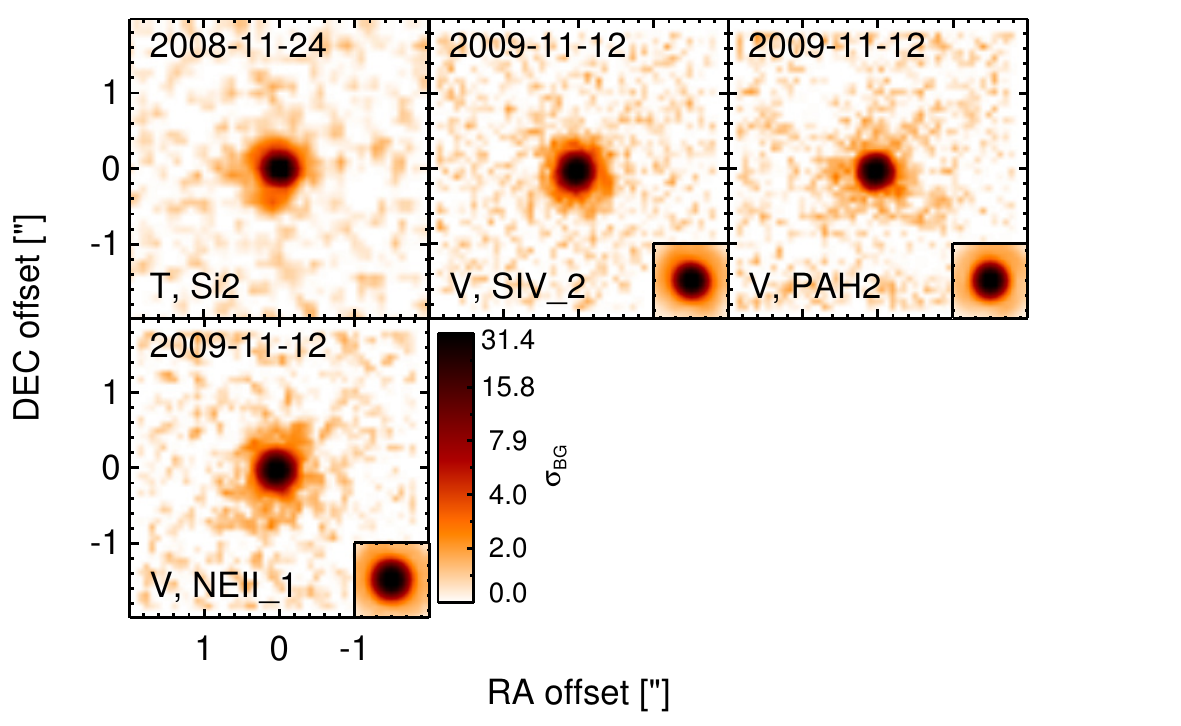}
    \caption{\label{fig:HARim_ESO362-G018}
             Subarcsecond-resolution MIR images of ESO\,362-18 sorted by increasing filter wavelength. 
             Displayed are the inner $4\arcsec$ with North up and East to the left. 
             The colour scaling is logarithmic with white corresponding to median background and black to the $75\%$ of the highest intensity of all images in units of $\sigbg$.
             The inset image shows the central arcsecond of the PSF from the calibrator star, scaled to match the science target.
             The labels in the bottom left state instrument and filter names (C: COMICS, M: Michelle, T: T-ReCS, V: VISIR).
           }
\end{figure}
\begin{figure}
   \centering
   \includegraphics[angle=0,width=8.50cm]{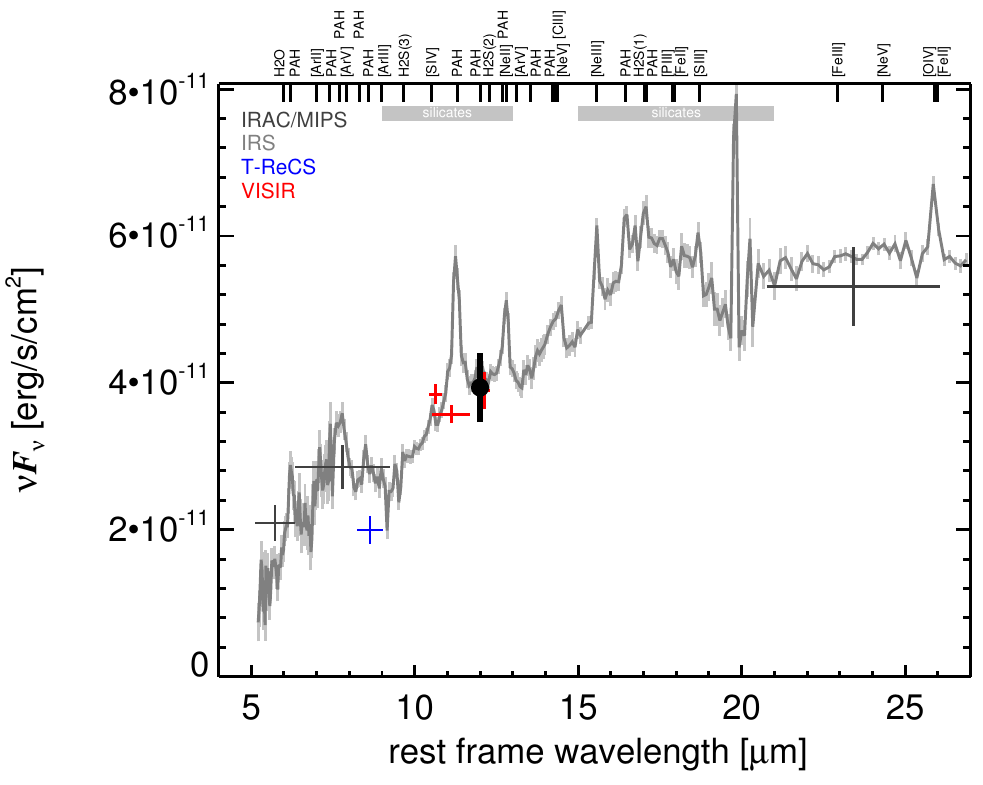}
   \caption{\label{fig:MISED_ESO362-G018}
      MIR SED of ESO\,362-18. The description  of the symbols (if present) is the following.
      Grey crosses and  solid lines mark the \spitzer/IRAC, MIPS and IRS data. 
      The colour coding of the other symbols is: 
      green for COMICS, magenta for Michelle, blue for T-ReCS and red for VISIR data.
      Darker-coloured solid lines mark spectra of the corresponding instrument.
      The black filled circles mark the nuclear 12 and $18\,\mu$m  continuum emission estimate from the data.
      The ticks on the top axis mark positions of common MIR emission lines, while the light grey horizontal bars mark wavelength ranges affected by the silicate 10 and 18$\mu$m features.     
   }
\end{figure}
\clearpage

\twocolumn[\begin{@twocolumnfalse}  
\subsection{ESO\,416-2 -- MCG-5-7-7}\label{app:ESO416-G002}
ESO\,416-2 is a radio-loud spiral galaxy at a redshift of $z=$ 0.0591 ($D \sim 251$\,Mpc) hosting a Sy\,1.9 nucleus \citep{veron-cetty_catalogue_2010} that belongs to the nine-month BAT AGN sample.
Our VISIR observations in three narrow $N$-band filters during one night in 2009 are the first MIR observations of ESO\,416-2.
A compact nucleus was weakly detected in all images.
Due to the low S/N of the detection no extension analysis can be performed, and no conclusion about the properties of MIR SED can be drawn. 
\newline\end{@twocolumnfalse}]

\begin{figure}
   \centering
   \includegraphics[angle=0,width=8.500cm]{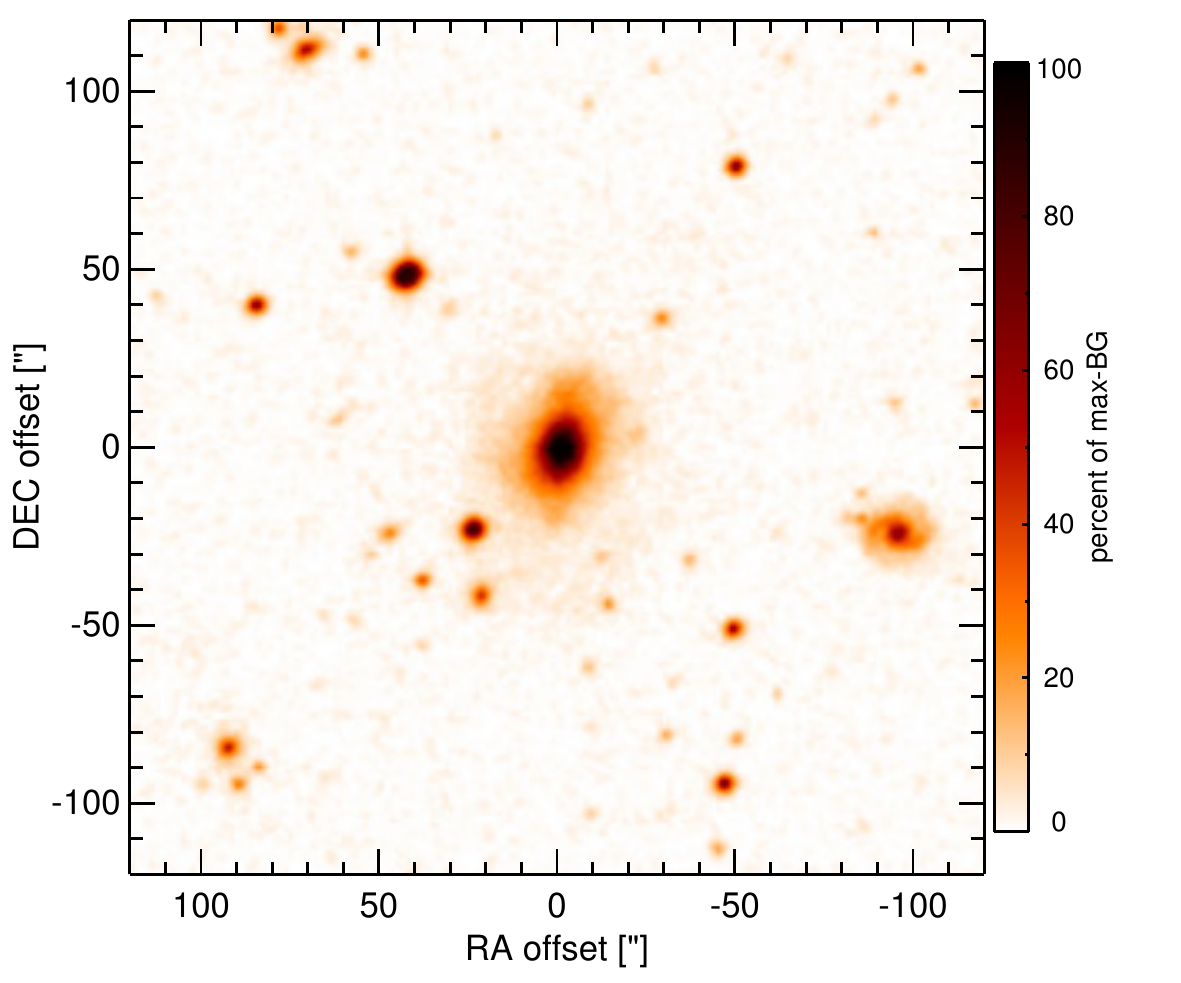}
    \caption{\label{fig:OPTim_ESO416-G002}
             Optical image (DSS, red filter) of ESO\,416-2. Displayed are the central $4\arcmin$ with North up and East to the left. 
              The colour scaling is linear with white corresponding to the median background and black to the $0.01\%$ pixels with the highest intensity.  
           }
\end{figure}
\begin{figure}
   \centering
   \includegraphics[angle=0,height=3.11cm]{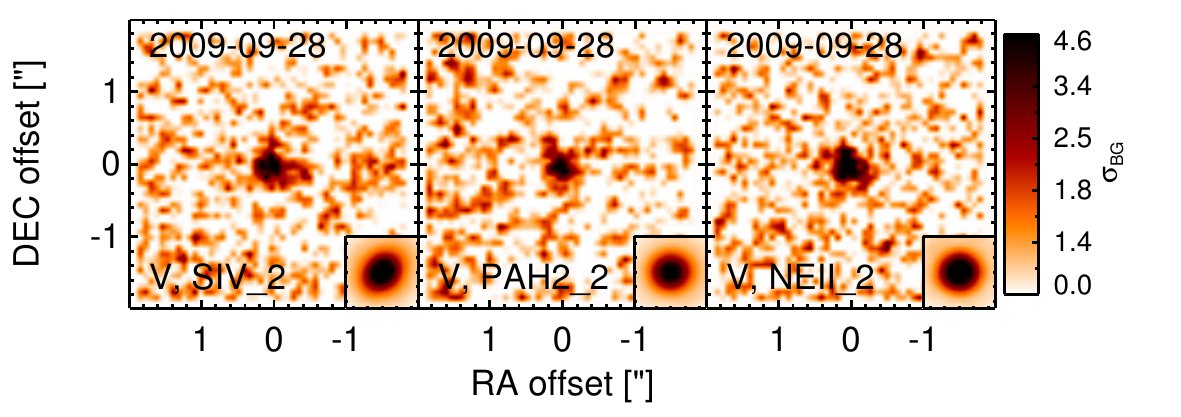}
    \caption{\label{fig:HARim_ESO416-G002}
             Subarcsecond-resolution MIR images of ESO\,416-2 sorted by increasing filter wavelength. 
             Displayed are the inner $4\arcsec$ with North up and East to the left. 
             The colour scaling is logarithmic with white corresponding to median background and black to the $75\%$ of the highest intensity of all images in units of $\sigbg$.
             The inset image shows the central arcsecond of the PSF from the calibrator star, scaled to match the science target.
             The labels in the bottom left state instrument and filter names (C: COMICS, M: Michelle, T: T-ReCS, V: VISIR).
           }
\end{figure}
\begin{figure}
   \centering
   \includegraphics[angle=0,width=8.50cm]{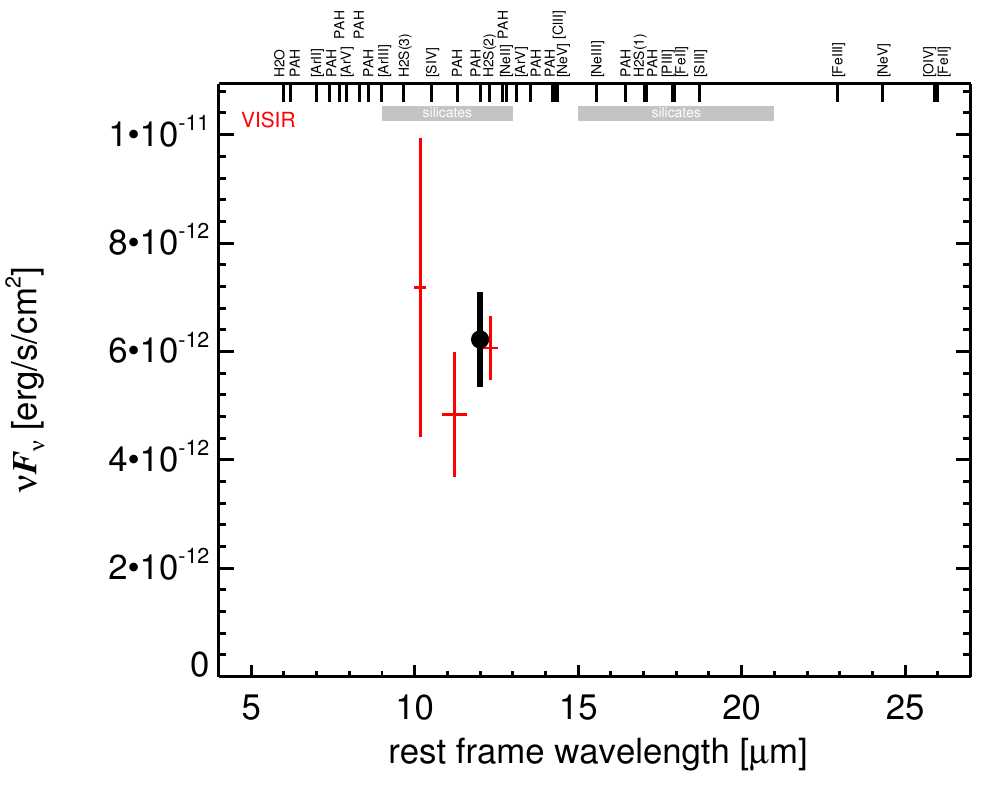}
   \caption{\label{fig:MISED_ESO416-G002}
      MIR SED of ESO\,416-2. The description  of the symbols (if present) is the following.
      Grey crosses and  solid lines mark the \spitzer/IRAC, MIPS and IRS data. 
      The colour coding of the other symbols is: 
      green for COMICS, magenta for Michelle, blue for T-ReCS and red for VISIR data.
      Darker-coloured solid lines mark spectra of the corresponding instrument.
      The black filled circles mark the nuclear 12 and $18\,\mu$m  continuum emission estimate from the data.
      The ticks on the top axis mark positions of common MIR emission lines, while the light grey horizontal bars mark wavelength ranges affected by the silicate 10 and 18$\mu$m features.     
   }
\end{figure}
\clearpage

\twocolumn[\begin{@twocolumnfalse}  
\subsection{ESO\,420-13 -- MCG-5-11-6}\label{app:ESO420-G013}
ESO\,420-13 is an infrared-luminous early-type galaxy at a redshift of $z=$ 0.0119 ($D\sim48.6$\,Mpc) hosting an AGN and a strong starburst.
The AGN is optically classified either as Sy\,2 \citep{veron-cetty_catalogue_2010} of an H\,II nucleus \citep{yuan_role_2010}.
Therefore, we treat this object as AGN/starburst composite. 
It was observed with \spitzer/IRAC, IRS and MIPS and appears compact without significantly extended host emission in the corresponding images.
The nucleus is partly saturated in the IRAC $8.0\,\mu$m PBCD image.
The IRS LR staring-mode spectrum shows very strong PAH emission, silicate absorption and a red slope in $\nu F_\nu$-space.
Thus, the \spitzerr MIR spectrum seems to be star formation dominated.
ESO\,420-13 was observed with VISIR in three $N$ band filters during one night in 2006 (unpublished, to our knowledge).
A compact MIR nucleus embedded in  extended host emission is visible in the images.
The extended emission has a spiral shape of $\sim 3.8\arcsec \sim 880$\,pc in diameter with one arm reaching out to the north-east and the other to the south-west. 
These structures become more prominent at longer wavelengths.
The PSF photometry of the nucleus yields fluxes that are on average $\sim 67\%$ lower than the \spitzerr spectrophotometry.
The resulting shape of the subarcsecond MIR SED suggests a much lower star formation contribution at scales less than 100\,pc in ESO\,420-13.
\newline\end{@twocolumnfalse}]

\begin{figure}
   \centering
   \includegraphics[angle=0,width=8.500cm]{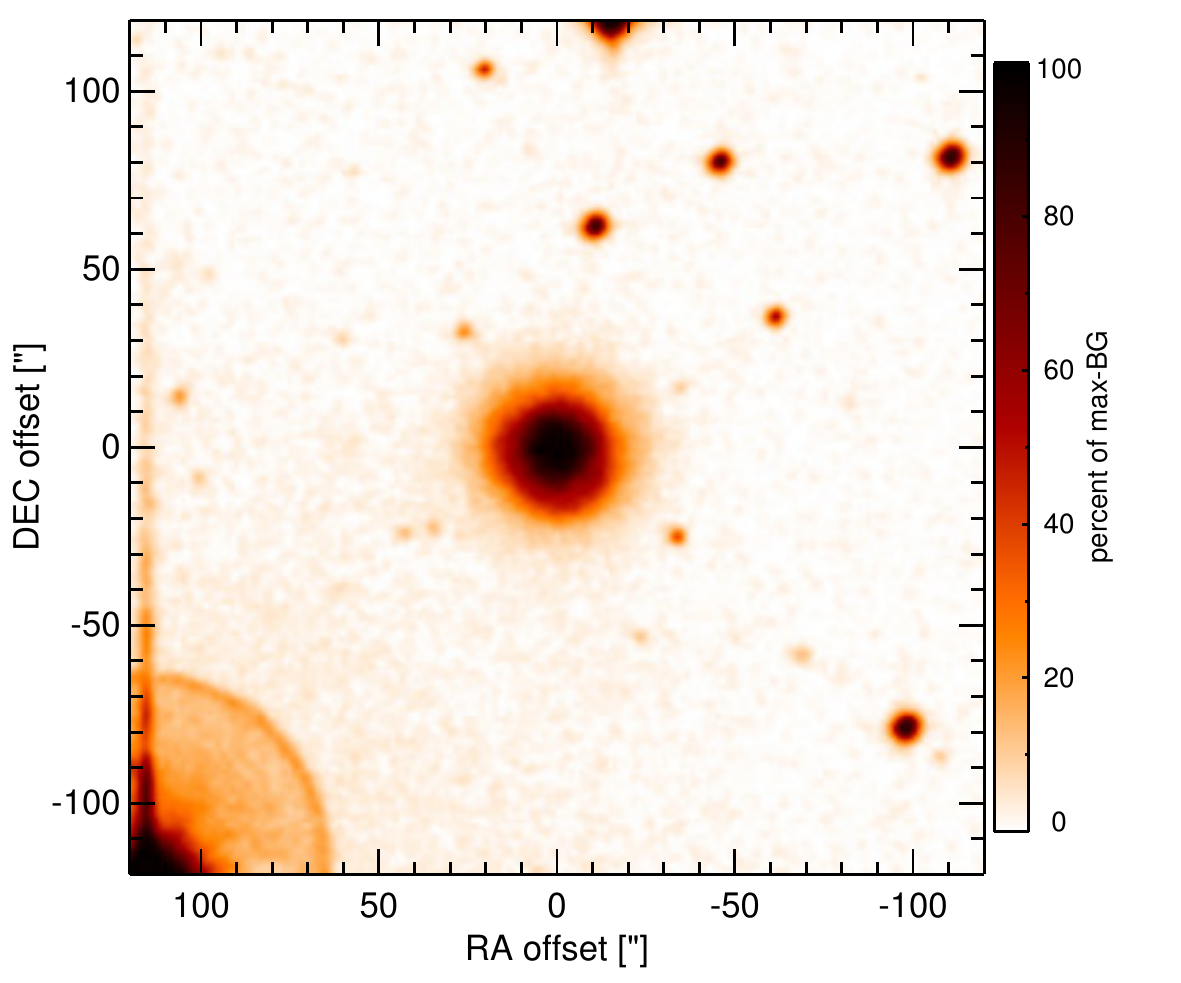}
    \caption{\label{fig:OPTim_ESO420-G013}
             Optical image (DSS, red filter) of ESO\,420-13. Displayed are the central $4\arcmin$ with North up and East to the left. 
              The colour scaling is linear with white corresponding to the median background and black to the $0.01\%$ pixels with the highest intensity.  
           }
\end{figure}
\begin{figure}
   \centering
   \includegraphics[angle=0,height=3.11cm]{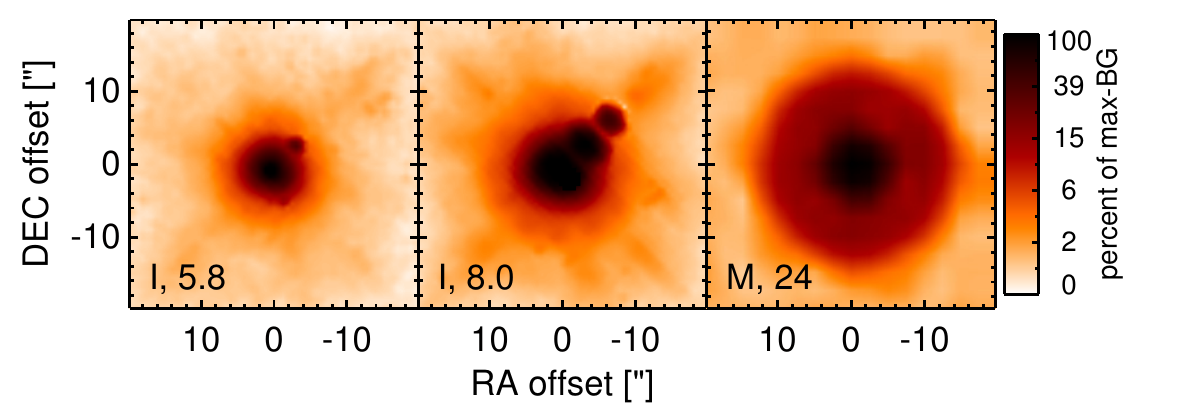}
    \caption{\label{fig:INTim_ESO420-G013}
             \spitzerr MIR images of ESO\,420-13. Displayed are the inner $40\arcsec$ with North up and East to the left. The colour scaling is logarithmic with white corresponding to median background and black to the $0.1\%$ pixels with the highest intensity.
             The label in the bottom left states instrument and central wavelength of the filter in $\mu$m (I: IRAC, M: MIPS).
             Note that the apparent off-nuclear compact sources in the IRAC 5.8 and $8.0\,\mu$m images are instrumental artefacts.
           }
\end{figure}
\begin{figure}
   \centering
   \includegraphics[angle=0,height=3.11cm]{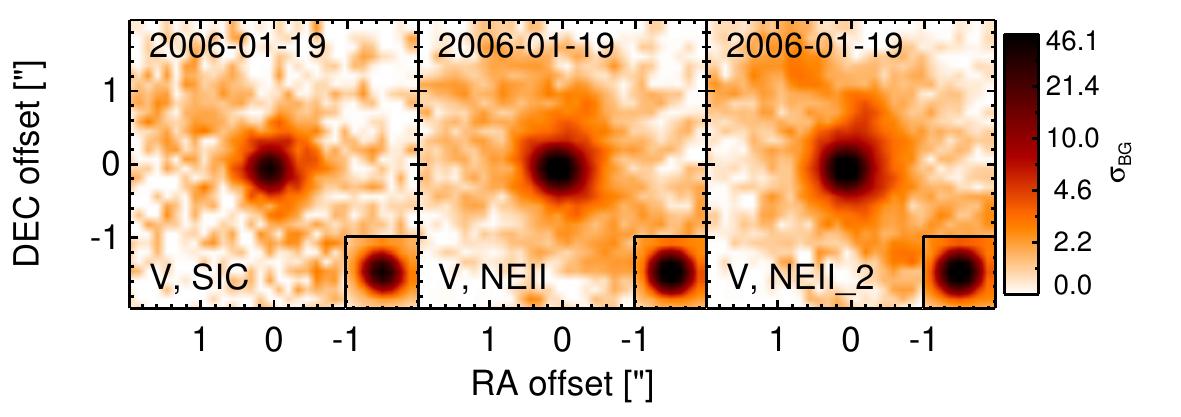}
    \caption{\label{fig:HARim_ESO420-G013}
             Subarcsecond-resolution MIR images of ESO\,420-13 sorted by increasing filter wavelength. 
             Displayed are the inner $4\arcsec$ with North up and East to the left. 
             The colour scaling is logarithmic with white corresponding to median background and black to the $75\%$ of the highest intensity of all images in units of $\sigbg$.
             The inset image shows the central arcsecond of the PSF from the calibrator star, scaled to match the science target.
             The labels in the bottom left state instrument and filter names (C: COMICS, M: Michelle, T: T-ReCS, V: VISIR).
           }
\end{figure}
\begin{figure}
   \centering
   \includegraphics[angle=0,width=8.50cm]{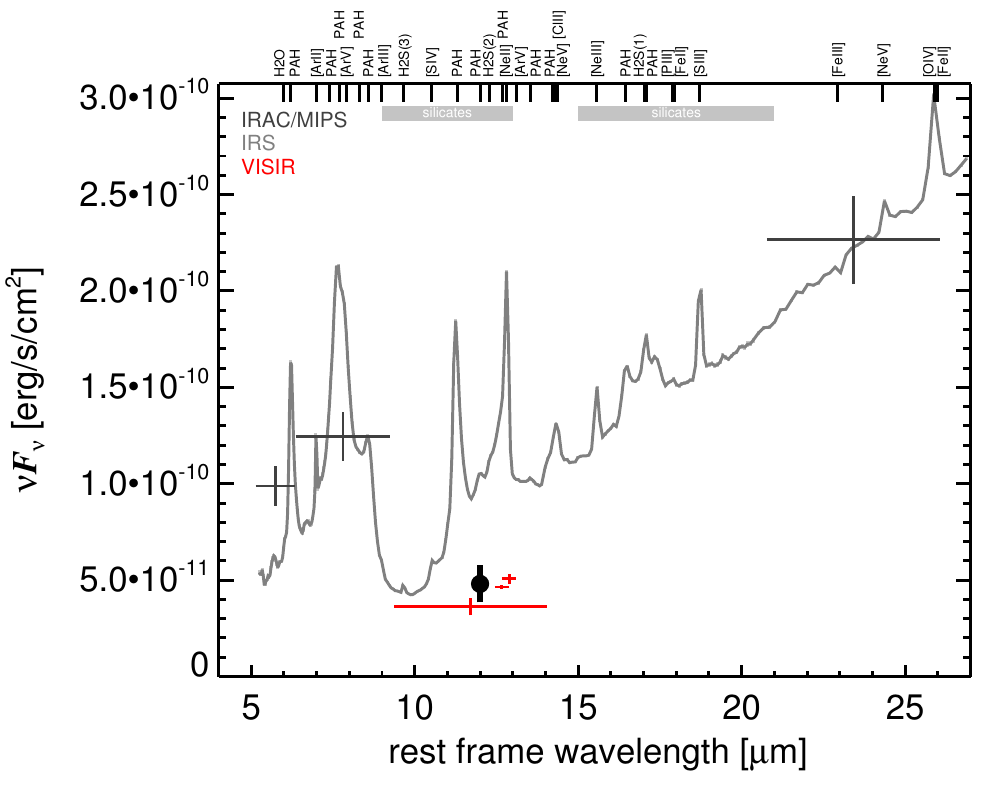}
   \caption{\label{fig:MISED_ESO420-G013}
      MIR SED of ESO\,420-13. The description  of the symbols (if present) is the following.
      Grey crosses and  solid lines mark the \spitzer/IRAC, MIPS and IRS data. 
      The colour coding of the other symbols is: 
      green for COMICS, magenta for Michelle, blue for T-ReCS and red for VISIR data.
      Darker-coloured solid lines mark spectra of the corresponding instrument.
      The black filled circles mark the nuclear 12 and $18\,\mu$m  continuum emission estimate from the data.
      The ticks on the top axis mark positions of common MIR emission lines, while the light grey horizontal bars mark wavelength ranges affected by the silicate 10 and 18$\mu$m features.     
   }
\end{figure}
\clearpage

\twocolumn[\begin{@twocolumnfalse}  
\subsection{ESO\,428-G014 -- MCG-5-18-2}\label{app:ESO428-G014}
ESO\,428-14 is a spiral galaxy at a redshift of $z=$ 0.0057 ($D\sim26\,$Mpc) close to the Galactic plane.
It is hosting a Sy\,2 nucleus \citep{veron-cetty_catalogue_2010} and possesses a bended jet-like radio structure \citep{ulvestad_radio_1989} aligned with a cone-like extended narrow-line region (PA$\sim-50degre$; \citealt{wilson_0714_1989,falcke_helical_1996}).
ESO\,428-14 was observed with \spitzer/IRAC, IRS and MIPS.
The compact MIR nucleus detected in all images is embedded in weak extended host emission along the north-south direction visible in the IRAC $5.8$ and $8.0\,\mu$m images.
Two epochs of MIPS $24\,\mu$m images (2004 and 2006) are available.
Our flux measurement of the nuclear source agrees with the value published in \cite{temi_spitzer_2009}.
The IRS LR staring-mode spectrum shows strong forbidden line emission, PAH features and silicate $10\,\mu$m absorption, indicating a star formation contribution (see also \citealt{shi_9.7_2006}). 
The MIR spectrum rises steeply and peaks around $\sim 18\,\mu$m in $\nu F_\nu$-space.
We observed ESO\,428-14 with VISIR in three narrow $N$-band and one $Q$-band filter during one night in 2007 and also obtained a VISIR $N$-band spectrum \citep{honig_dusty_2010-1}.
All images show a compact nuclear source embedded in increasingly prominent extended bar-like emission towards the longer wavelength filters. 
In the $Q$-band image (unpublished, to our knowledge), a knot $\sim1\arcsec$ south-east of the nucleus becomes visible.
The extended emission is oriented along the north-west direction (PA$\sim140\degree$), aligned with the NLR emission and host galaxy major axis. 
The VISIR spectrum possesses the same continuum shape, silicate absorption and forbidden emission lines as the IRS spectrum. 
However, the continuum flux level is lower and the PAH emission is also absent, indicating that the subarcsecond scales do not have significant star formation contribution.
While \cite{honig_dusty_2010-1} have presented the VISIR spectrophotometry integrated over the inner 0.75\arcsec, we now isolate the unresolved fluxes, which are significantly lower.
In particular, the average subarcsecond nuclear fluxes are $\sim60\%$ lower than the values of the \spitzerr spectrophotometry.
\newline\end{@twocolumnfalse}]

\begin{figure}
   \centering
   \includegraphics[angle=0,width=8.500cm]{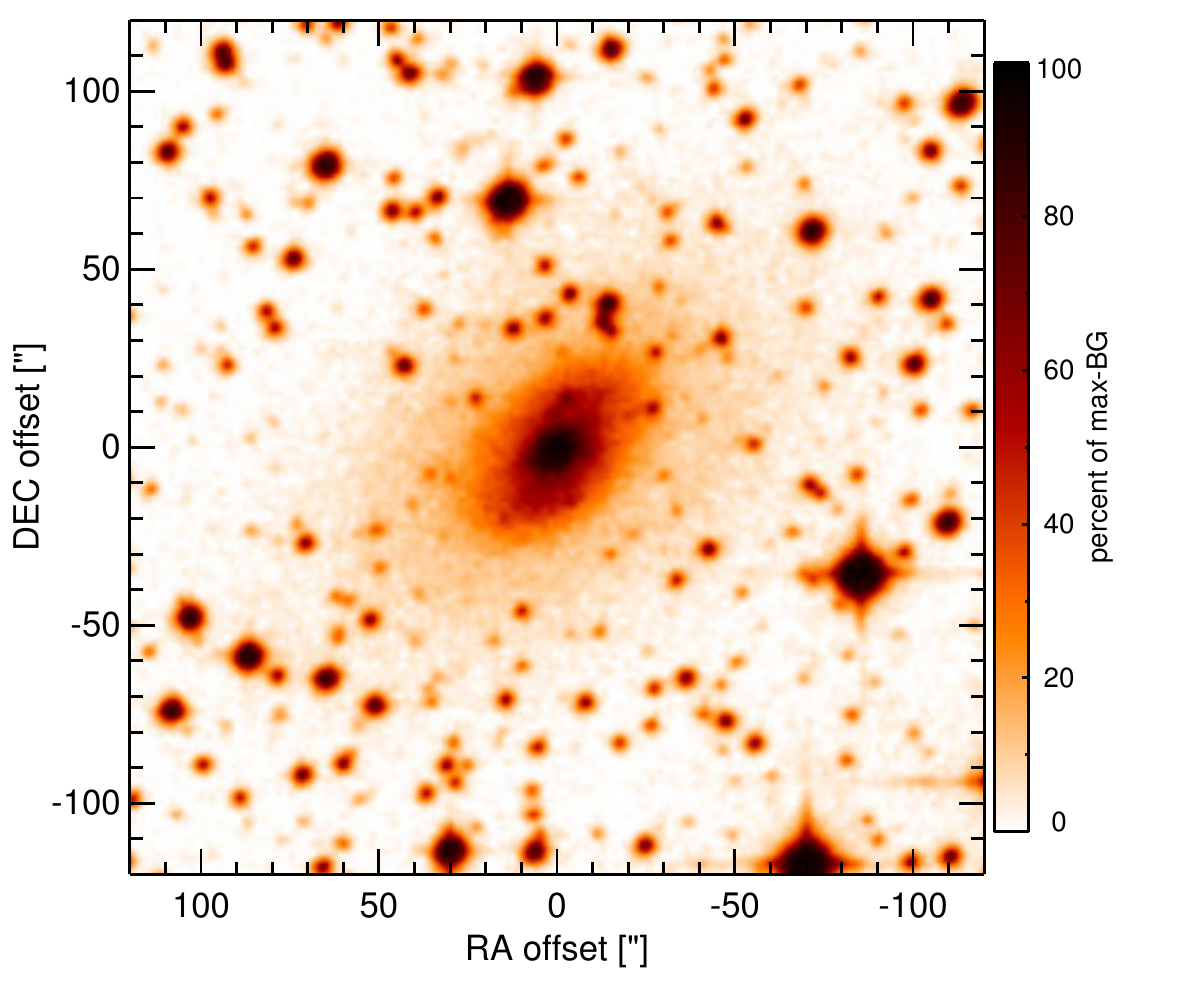}
    \caption{\label{fig:OPTim_ESO428-G014}
             Optical image (DSS, red filter) of ESO\,428-14. Displayed are the central $4\arcmin$ with North up and East to the left. 
              The colour scaling is linear with white corresponding to the median background and black to the $0.01\%$ pixels with the highest intensity.  
           }
\end{figure}
\begin{figure}
   \centering
   \includegraphics[angle=0,height=3.11cm]{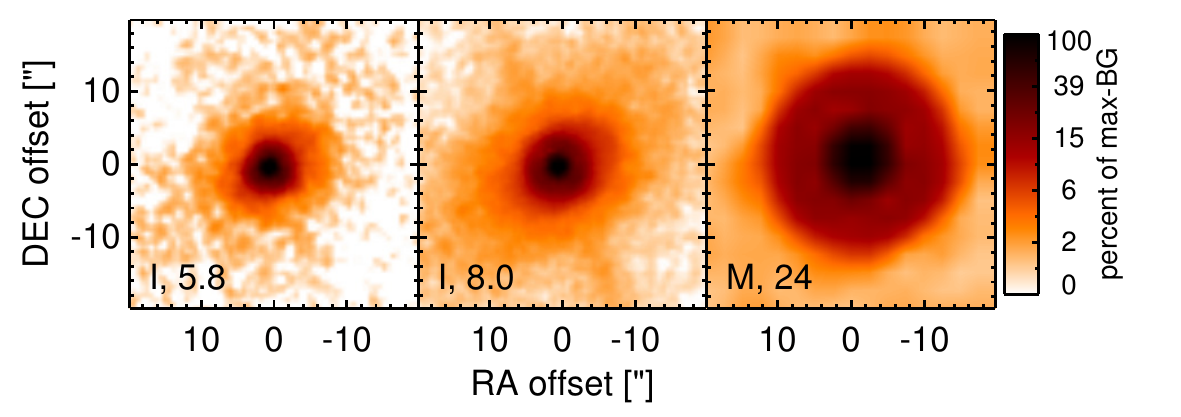}
    \caption{\label{fig:INTim_ESO428-G014}
             \spitzerr MIR images of ESO\,428-14. Displayed are the inner $40\arcsec$ with North up and East to the left. The colour scaling is logarithmic with white corresponding to median background and black to the $0.1\%$ pixels with the highest intensity.
             The label in the bottom left states instrument and central wavelength of the filter in $\mu$m (I: IRAC, M: MIPS). 
           }
\end{figure}
\begin{figure}
   \centering
   \includegraphics[angle=0,width=8.500cm]{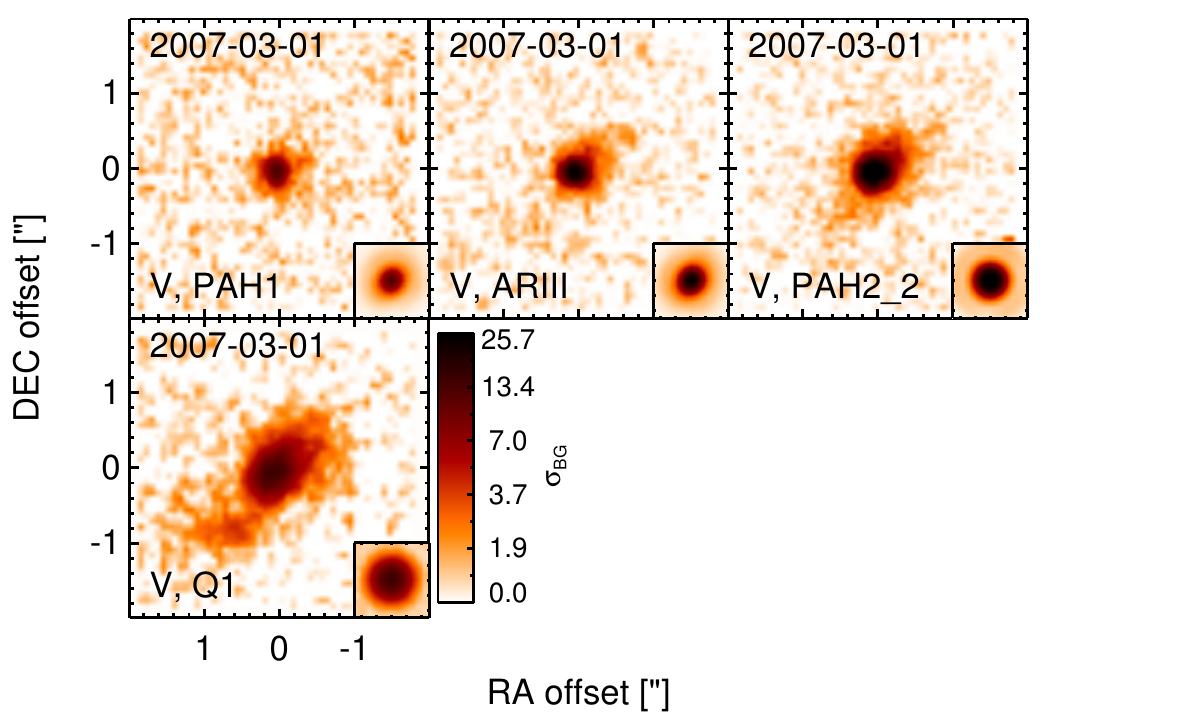}
    \caption{\label{fig:HARim_ESO428-G014}
             Subarcsecond-resolution MIR images of ESO\,428-14 sorted by increasing filter wavelength. 
             Displayed are the inner $4\arcsec$ with North up and East to the left. 
             The colour scaling is logarithmic with white corresponding to median background and black to the $75\%$ of the highest intensity of all images in units of $\sigbg$.
             The inset image shows the central arcsecond of the PSF from the calibrator star, scaled to match the science target.
             The labels in the bottom left state instrument and filter names (C: COMICS, M: Michelle, T: T-ReCS, V: VISIR).
           }
\end{figure}
\begin{figure}
   \centering
   \includegraphics[angle=0,width=8.50cm]{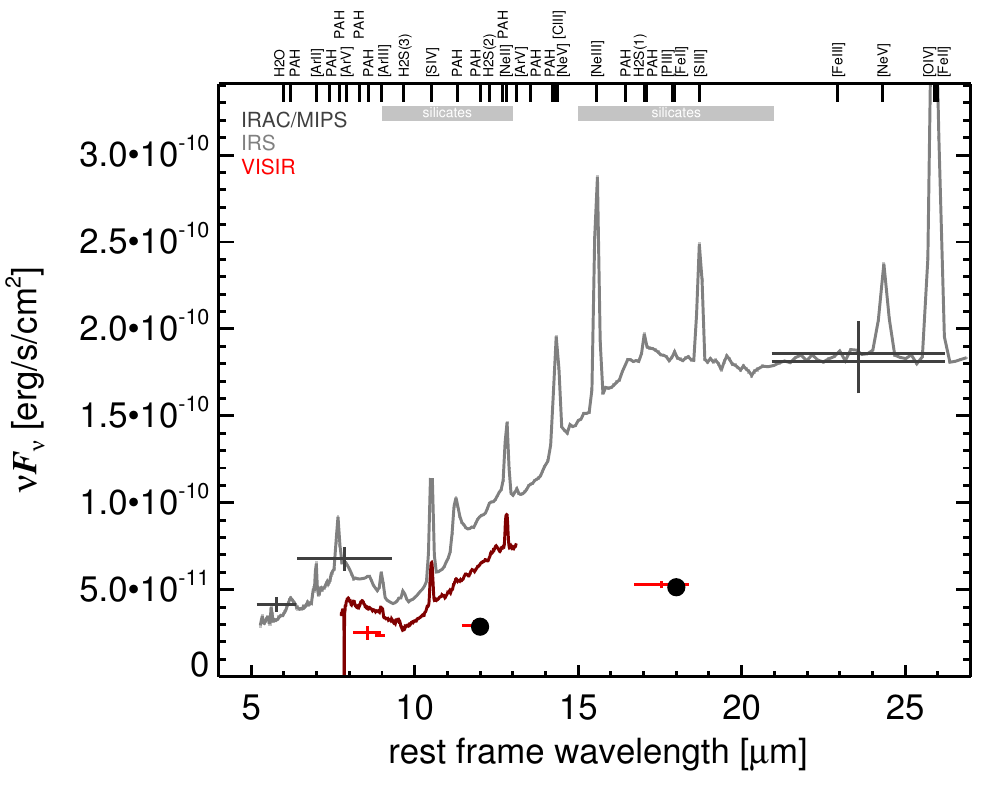}
   \caption{\label{fig:MISED_ESO428-G014}
      MIR SED of ESO\,428-14. The description  of the symbols (if present) is the following.
      Grey crosses and  solid lines mark the \spitzer/IRAC, MIPS and IRS data. 
      The colour coding of the other symbols is: 
      green for COMICS, magenta for Michelle, blue for T-ReCS and red for VISIR data.
      Darker-coloured solid lines mark spectra of the corresponding instrument.
      The black filled circles mark the nuclear 12 and $18\,\mu$m  continuum emission estimate from the data.
      The ticks on the top axis mark positions of common MIR emission lines, while the light grey horizontal bars mark wavelength ranges affected by the silicate 10 and 18$\mu$m features.     
   }
\end{figure}
\clearpage

\twocolumn[\begin{@twocolumnfalse}  
\subsection{ESO\,500-34 -- MCG-4-25-6}\label{app:ESO500-G034}
ESO\,500-34 is a spiral galaxy at a redshift of $z=$ 0.0122 ($D\sim55.6\,$Mpc) with an active nucleus containing a starburst and possibly an AGN \citep{hill_starburst_1999}.
It is optically  classified either as a Sy\,2  or an H\,II \citep{yuan_role_2010}, and we treat it as uncertain AGN/starburst composite.
There is no evidence for the presence of an AGN available at X-ray or radio wavelengths. 
No \spitzerr observations are available, and it appears as an elongated source along the host galaxy major axis, in particular in the short-wavelength \wisee bands.
ESO\,500-34 has only been imaged once with VISIR in 2005 in one narrow $N$-band filter \citep{siebenmorgen_nuclear_2008}.
An elongated extended emission source without any clear nuclear component was detected. 
The major axis is roughly in  the north-south directions ($\sim 3\arcsec \sim 800\,$pc) with the emission peak on the northern side. 
Because no AGN component can be identified, \cite{siebenmorgen_nuclear_2008} conclude that the primary MIR emission source is star formation. 
We use the unresolved flux in the emission peak position to estimate an upper limit on any AGN in ESO\,500-34. 
MIR spectroscopy and further imaging will be necessary to decide on the nature of the MIR emission in this object.
\newline\end{@twocolumnfalse}]

\begin{figure}
   \centering
   \includegraphics[angle=0,width=8.500cm]{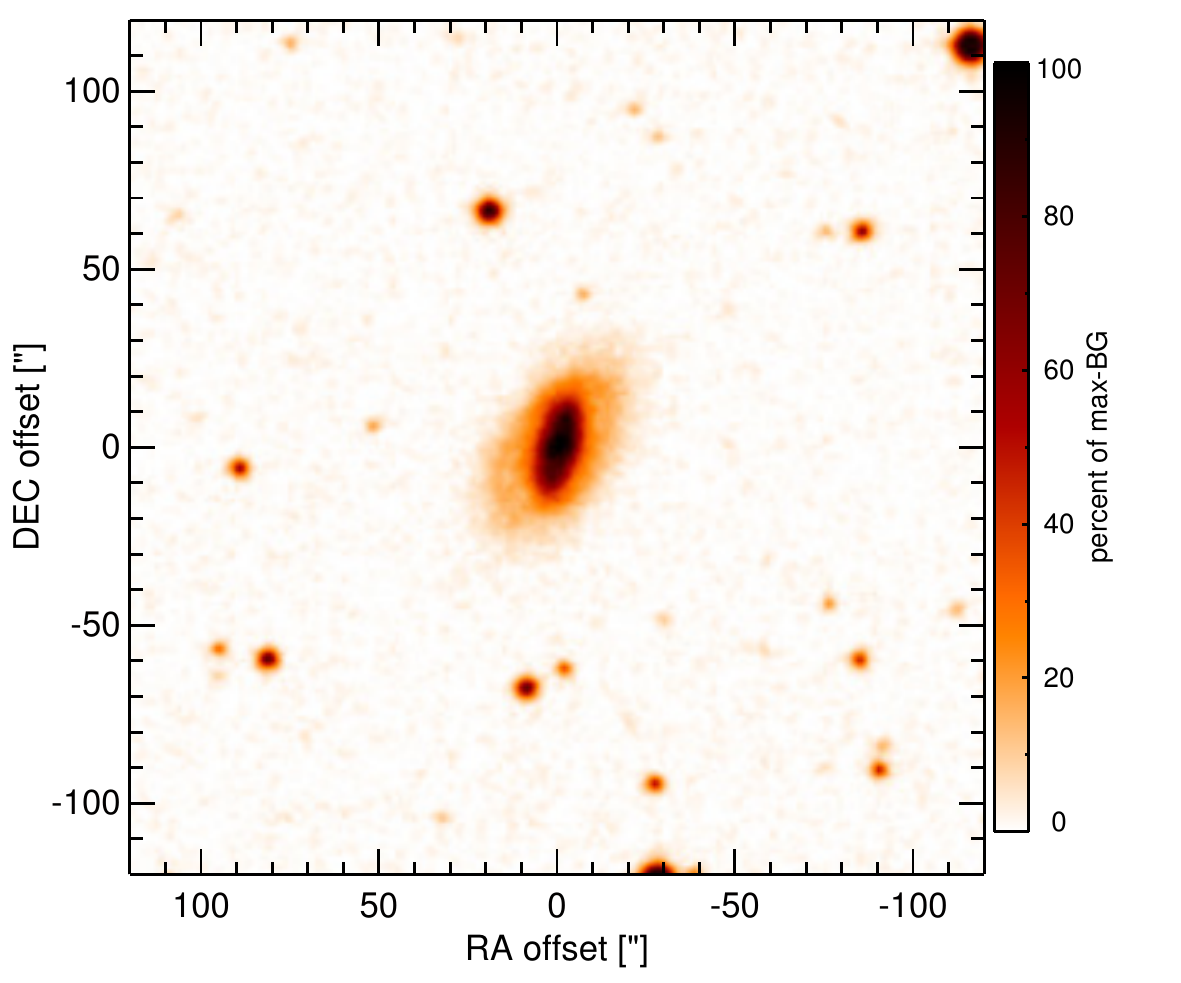}
    \caption{\label{fig:OPTim_ESO500-G034}
             Optical image (DSS, red filter) of ESO\,500-34. Displayed are the central $4\arcmin$ with North up and East to the left. 
              The colour scaling is linear with white corresponding to the median background and black to the $0.01\%$ pixels with the highest intensity.  
           }
\end{figure}
\begin{figure}
   \centering
   \includegraphics[angle=0,height=3.11cm]{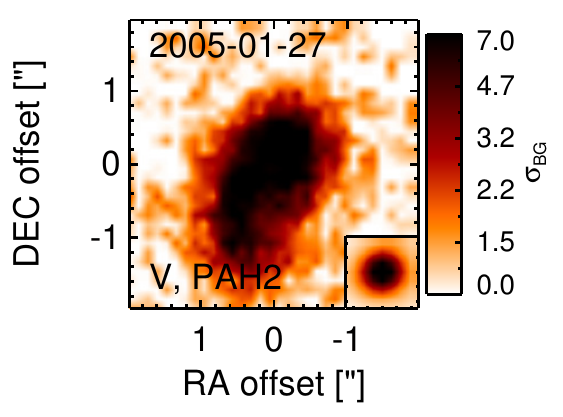}
    \caption{\label{fig:HARim_ESO500-G034}
             Subarcsecond-resolution MIR images of ESO\,500-34 sorted by increasing filter wavelength. 
             Displayed are the inner $4\arcsecond$ with North up and East to the left. 
             The colour scaling is logarithmic with white corresponding to median background and black to the $75\%$ of the highest intensity of all images in units of $\sigbg$.
             The inset image shows the central arcsecond of the PSF from the calibrator star, scaled to match the science target.
             The labels in the bottom left state instrument and filter names (C: COMICS, M: Michelle, T: T-ReCS, V: VISIR).
           }
\end{figure}
\begin{figure}
   \centering
   \includegraphics[angle=0,width=8.50cm]{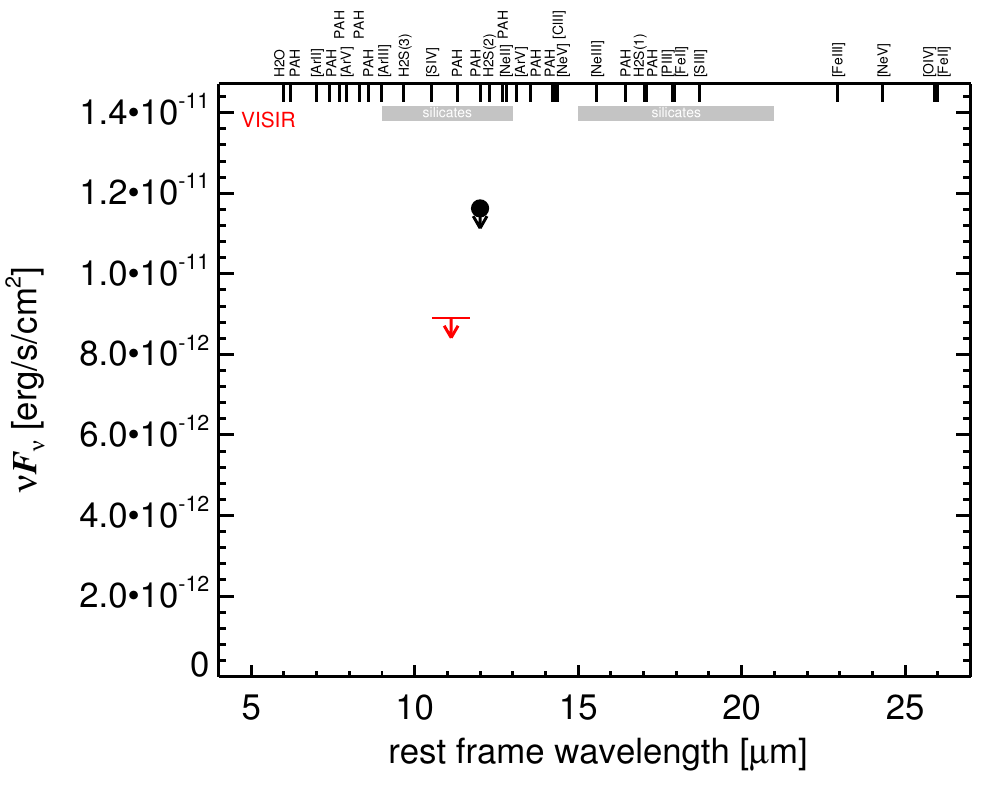}
   \caption{\label{fig:MISED_ESO500-G034}
      MIR SED of ESO\,500-34. The description  of the symbols (if present) is the following.
      Grey crosses and  solid lines mark the \spitzer/IRAC, MIPS and IRS data. 
      The colour coding of the other symbols is: 
      green for COMICS, magenta for Michelle, blue for T-ReCS and red for VISIR data.
      Darker-coloured solid lines mark spectra of the corresponding instrument.
      The black filled circles mark the nuclear 12 and $18\,\mu$m  continuum emission estimate from the data.
      The ticks on the top axis mark positions of common MIR emission lines, while the light grey horizontal bars mark wavelength ranges affected by the silicate 10 and 18$\mu$m features.     
   }
\end{figure}
\clearpage

\twocolumn[\begin{@twocolumnfalse}  
\subsection{ESO\,506-27 -- MCG-4-30-7}\label{app:ESO506-G027}
ESO\,506-27 is an edge-on spiral galaxy at a redshift of $z=$ 0.0250 ($D\sim 110$\,Mpc) hosting a Sy\,2 nucleus \citep{veron-cetty_catalogue_2010}.
It belongs to the nine-month BAT AGN sample and is an X-ray ``buried" AGN candidate \citep{noguchi_new_2009}.
ESO\,506-27 was observed with \spitzer/IRS, and the LR staring-mode spectrum shows an extremely deep silicate $10\,\mu$m absorption feature (see also \citealt{mullaney_defining_2011}).
Silicate $18\,\mu$m absorption and PAH emission seems present as well but less prominent.
The general MIR spectral slope is blue in $\nu F_\nu$-space.
We observed ESO\,506-27 with VISIR in three narrow $N$-band filters in one night of 2010 and detected a compact MIR nucleus without any sign of extended host emission.
The nucleus is unresolved in the sharper SIV and PAH2\_2 images and thus classified as unresolved in the MIR at subarcsecond scales.
The corresponding photometric measurements agree with the IRS spectrum and indicate the same deep silicate feature. 
Therefore, we use the IRS spectrum to correct our 12$\,\mu$m continuum emission estimate for the silicate feature.
\newline\end{@twocolumnfalse}]

\begin{figure}
   \centering
   \includegraphics[angle=0,width=8.500cm]{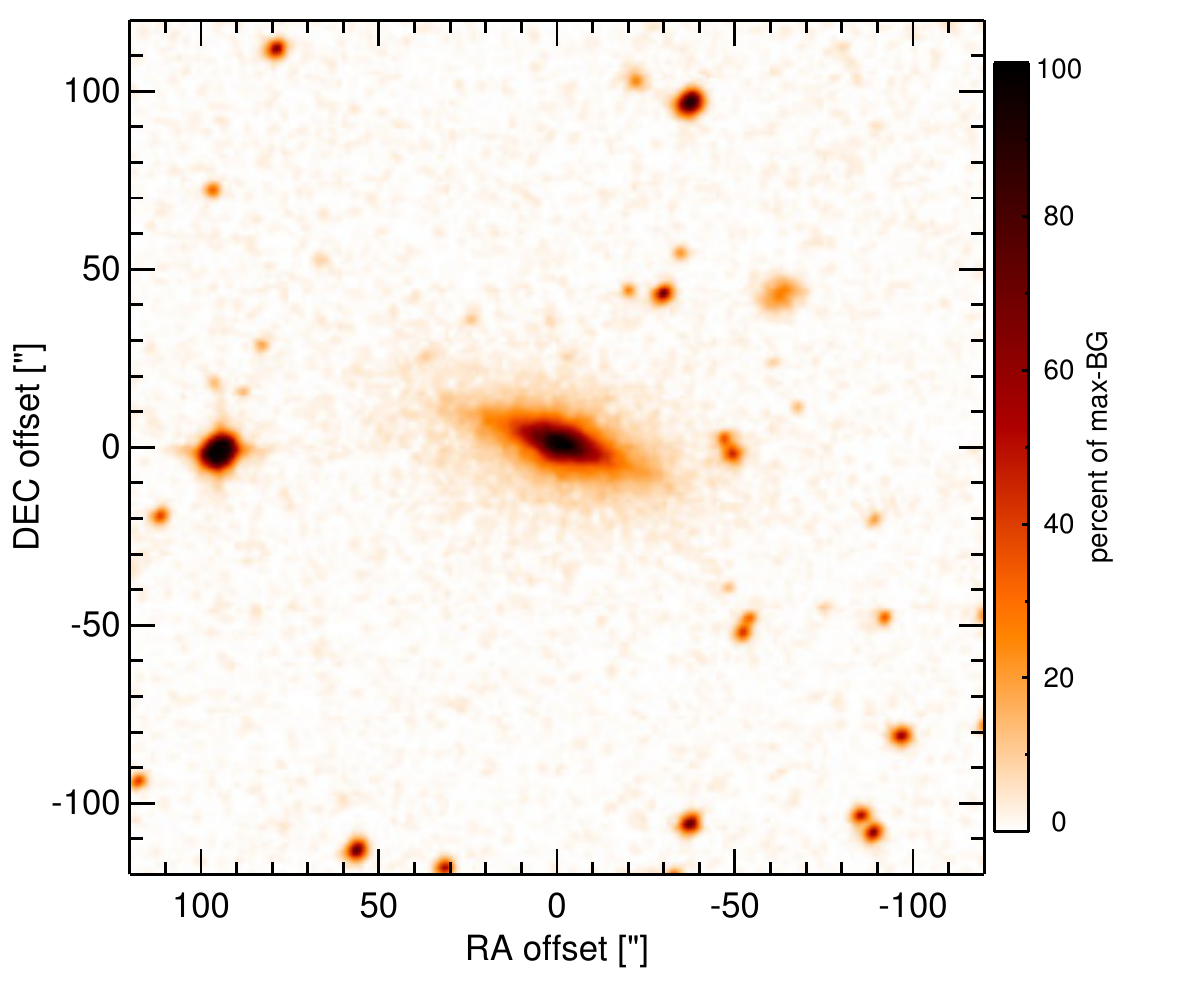}
    \caption{\label{fig:OPTim_ESO506-G027}
             Optical image (DSS, red filter) of ESO\,506-27. Displayed are the central $4\arcmin$ with North up and East to the left. 
              The colour scaling is linear with white corresponding to the median background and black to the $0.01\%$ pixels with the highest intensity.  
           }
\end{figure}
\begin{figure}
   \centering
   \includegraphics[angle=0,height=3.11cm]{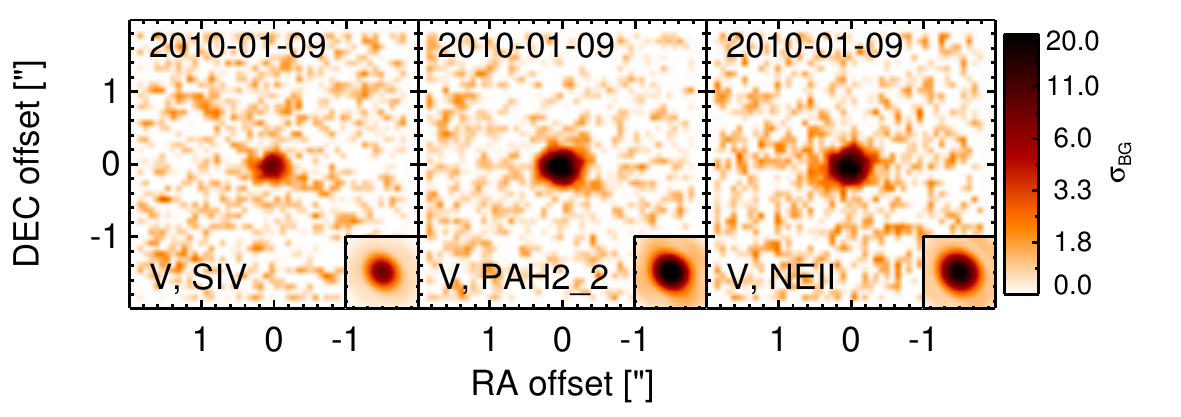}
    \caption{\label{fig:HARim_ESO506-G027}
             Subarcsecond-resolution MIR images of ESO\,506-27 sorted by increasing filter wavelength. 
             Displayed are the inner $4\arcsec$ with North up and East to the left. 
             The colour scaling is logarithmic with white corresponding to median background and black to the $75\%$ of the highest intensity of all images in units of $\sigbg$.
             The inset image shows the central arcsecond of the PSF from the calibrator star, scaled to match the science target.
             The labels in the bottom left state instrument and filter names (C: COMICS, M: Michelle, T: T-ReCS, V: VISIR).
           }
\end{figure}
\begin{figure}
   \centering
   \includegraphics[angle=0,width=8.50cm]{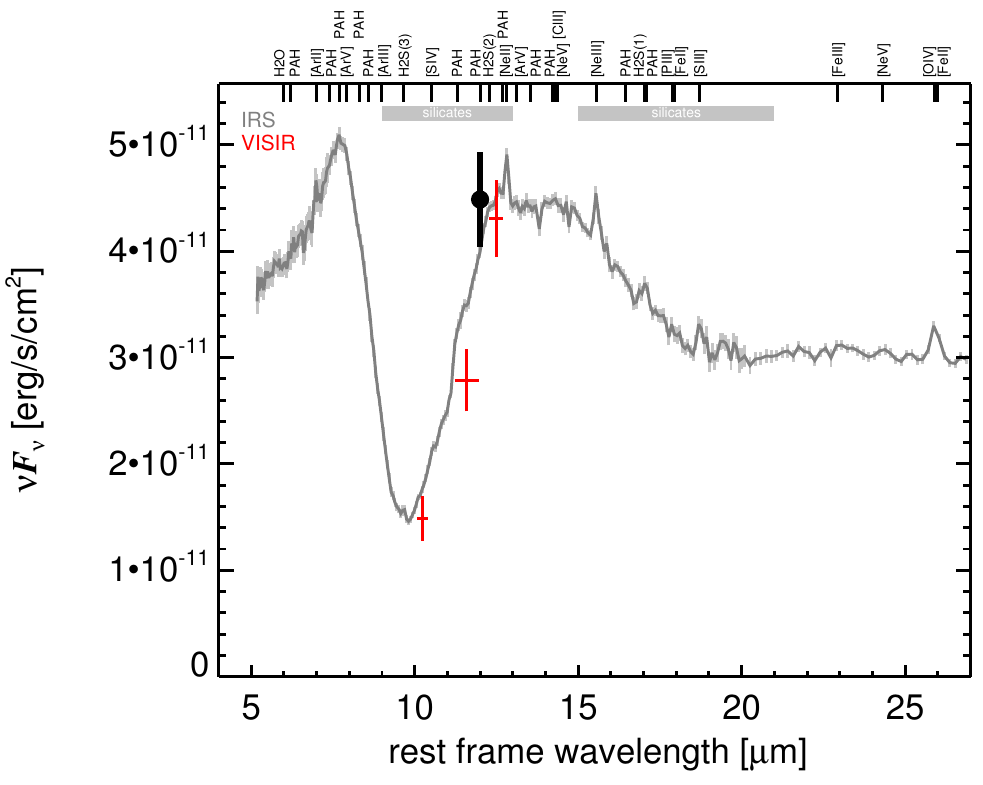}
   \caption{\label{fig:MISED_ESO506-G027}
      MIR SED of ESO\,506-27. The description  of the symbols (if present) is the following.
      Grey crosses and  solid lines mark the \spitzer/IRAC, MIPS and IRS data. 
      The colour coding of the other symbols is: 
      green for COMICS, magenta for Michelle, blue for T-ReCS and red for VISIR data.
      Darker-coloured solid lines mark spectra of the corresponding instrument.
      The black filled circles mark the nuclear 12 and $18\,\mu$m  continuum emission estimate from the data.
      The ticks on the top axis mark positions of common MIR emission lines, while the light grey horizontal bars mark wavelength ranges affected by the silicate 10 and 18$\mu$m features.     
   }
\end{figure}
\clearpage

\twocolumn[\begin{@twocolumnfalse}  
\subsection{ESO\,511-30 -- MCG-4-34-10}\label{app:ESO511-G030}
ESO\,511-30 is a face-on spiral galaxy at a redshift of $z=$ 0.0224 ($D\sim97.1\,$Mpc) hosting a Sy\,1 nucleus \citep{veron-cetty_catalogue_2010}, belonging to the nine-month BAT AGN sample.
It was observed with \spitzer/IRS and MIPS and appears compact in the MIPS $24\,\mu$m image.
The IRS LR staring-mode spectrum exhibits prominent silicate 10 and 18$\,\mu$m emission and a blue spectral slope in $\nu F_\nu$-space (see also \citealt{sargsyan_infrared_2011}).
We observed ESO\,511-30 with VISIR in three narrow $N$-band filters in one night of 2010 and detected a compact MIR nucleus without any sign of extended host emission.
The nuclear emission appears elongated in the SIV and PAH2 filter images (FWHM(major axis)$\sim 0.38\arcsec \sim 170\,$pc, PA$\sim80\degree$), while the S/N in the NEII filter image is too low for an extension analysis.
A second epoch of subarcsecond MIR observations is necessary to confirm this elongation to be source intrinsic.
The photometry in the SIV and PAH2 filters is consistent with the \spitzerr  spectrophotometry, while the NEII flux is significantly lower. 
We employ the IRS spectrum to correct our 12\,$\mu$m continuum emission estimate for the silicate emission.
Note that, in particular, the nuclear SIV and PAH2 fluxes would be significantly lower if the presence of subarcsecond-extended emission can be verified.
\newline\end{@twocolumnfalse}]

\begin{figure}
   \centering
   \includegraphics[angle=0,width=8.500cm]{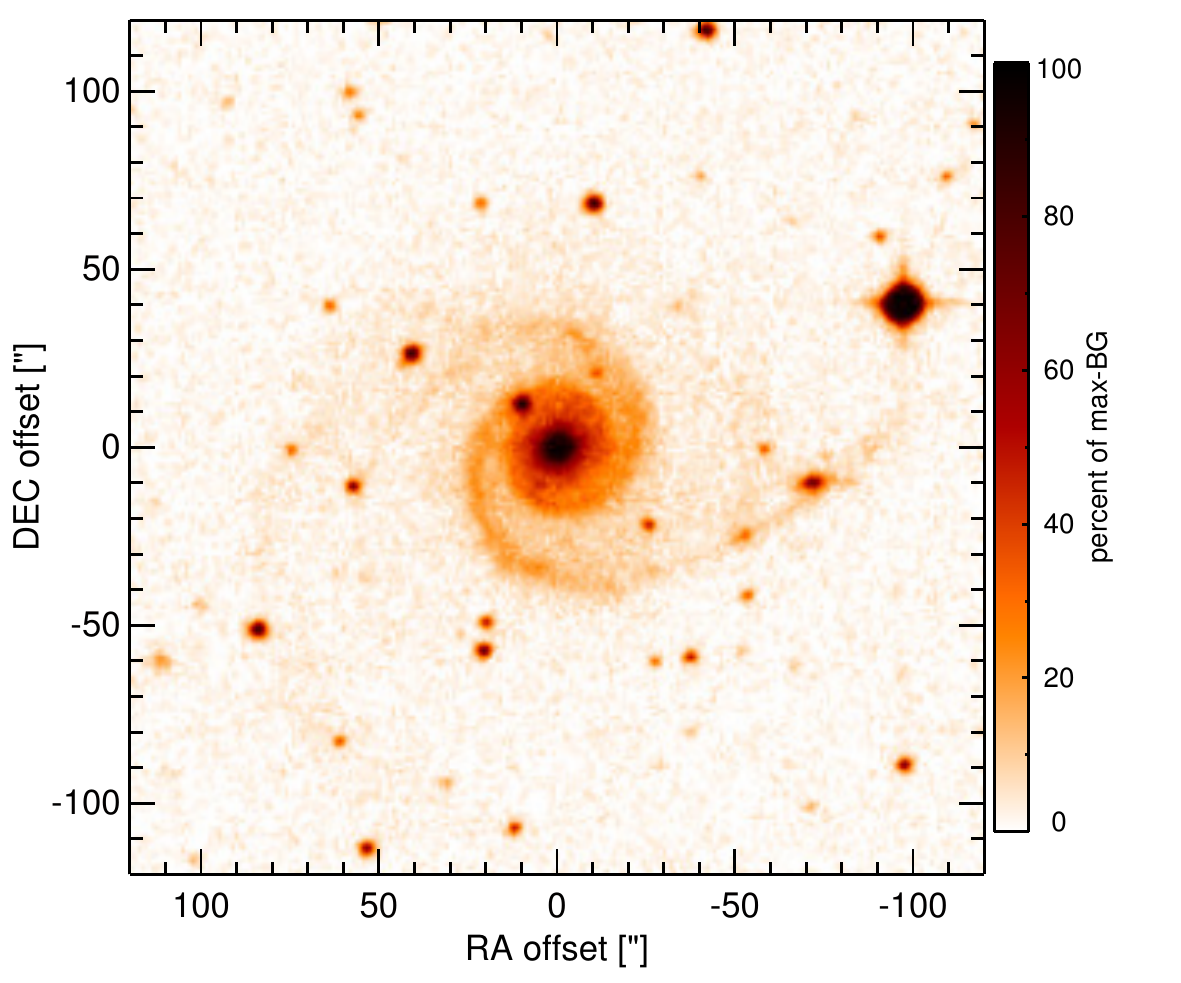}
    \caption{\label{fig:OPTim_ESO511-G030}
             Optical image (DSS, red filter) of ESO\,511-30. Displayed are the central $4\arcmin$ with North up and East to the left. 
              The colour scaling is linear with white corresponding to the median background and black to the $0.01\%$ pixels with the highest intensity.  
           }
\end{figure}
\begin{figure}
   \centering
   \includegraphics[angle=0,height=3.11cm]{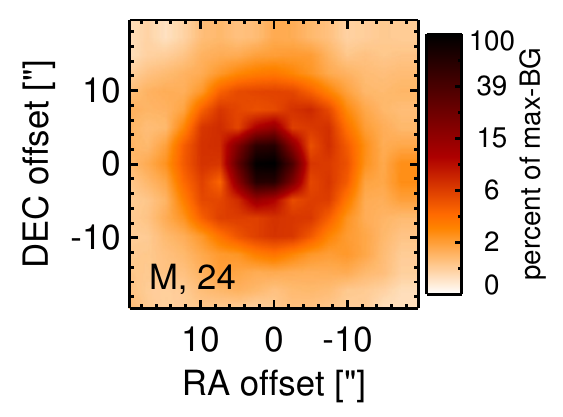}
    \caption{\label{fig:INTim_ESO511-G030}
             \spitzerr MIR images of ESO\,511-30. Displayed are the inner $40\arcsec$ with North up and East to the left. The colour scaling is logarithmic with white corresponding to median background and black to the $0.1\%$ pixels with the highest intensity.
             The label in the bottom left states instrument and central wavelength of the filter in $\mu$m (I: IRAC, M: MIPS). 
           }
\end{figure}
\begin{figure}
   \centering
   \includegraphics[angle=0,height=3.11cm]{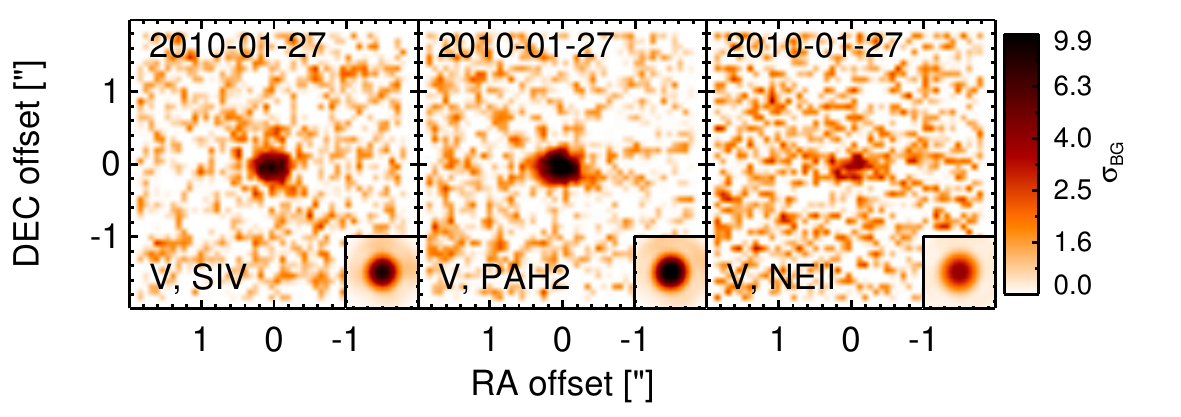}
    \caption{\label{fig:HARim_ESO511-G030}
             Subarcsecond-resolution MIR images of ESO\,511-30 sorted by increasing filter wavelength. 
             Displayed are the inner $4\arcsec$ with North up and East to the left. 
             The colour scaling is logarithmic with white corresponding to median background and black to the $75\%$ of the highest intensity of all images in units of $\sigbg$.
             The inset image shows the central arcsecond of the PSF from the calibrator star, scaled to match the science target.
             The labels in the bottom left state instrument and filter names (C: COMICS, M: Michelle, T: T-ReCS, V: VISIR).
           }
\end{figure}
\begin{figure}
   \centering
   \includegraphics[angle=0,width=8.50cm]{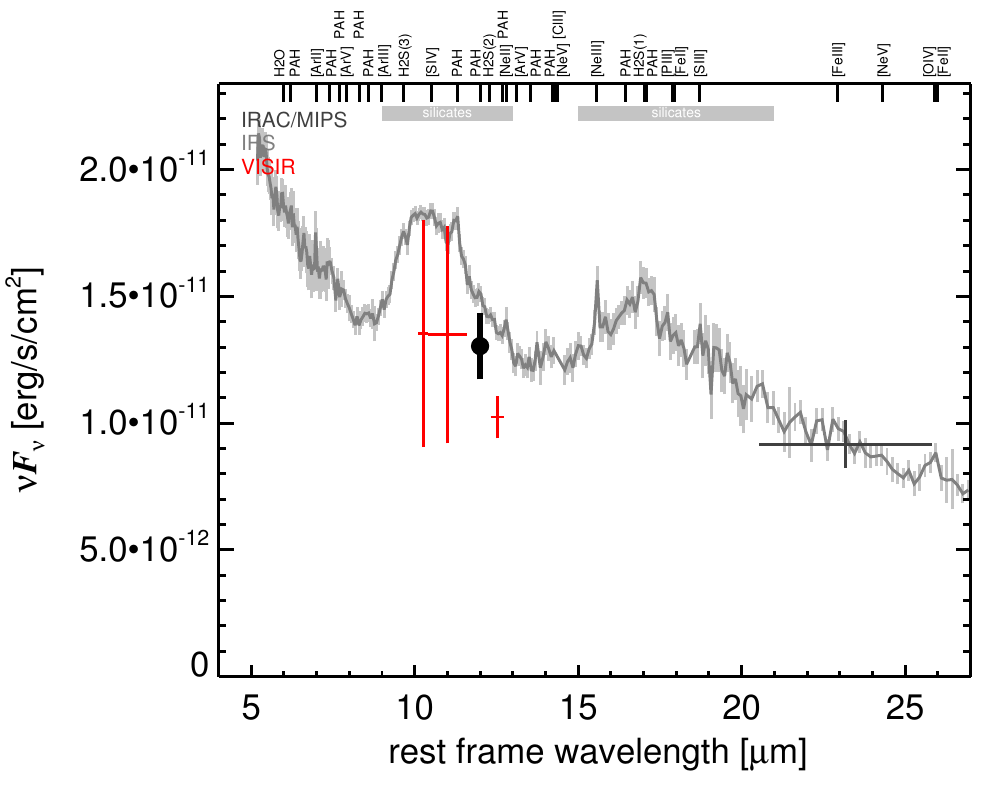}
   \caption{\label{fig:MISED_ESO511-G030}
      MIR SED of ESO\,511-30. The description  of the symbols (if present) is the following.
      Grey crosses and  solid lines mark the \spitzer/IRAC, MIPS and IRS data. 
      The colour coding of the other symbols is: 
      green for COMICS, magenta for Michelle, blue for T-ReCS and red for VISIR data.
      Darker-coloured solid lines mark spectra of the corresponding instrument.
      The black filled circles mark the nuclear 12 and $18\,\mu$m  continuum emission estimate from the data.
      The ticks on the top axis mark positions of common MIR emission lines, while the light grey horizontal bars mark wavelength ranges affected by the silicate 10 and 18$\mu$m features.     
   }
\end{figure}
\clearpage

\twocolumn[\begin{@twocolumnfalse}  
\subsection{ESO\,548-81 -- MCG-4-9-52}\label{app:ESO548-G081}
ESO\,548-81 is a face-on spiral galaxy at a redshift of $z=$ 0.0145 ($D\sim58.6$\,Mpc) hosting a Sy\,1 nucleus \citep{veron-cetty_catalogue_2010} that belongs to the nine-month BAT AGN sample.
It was observed with \spitzer/IRS, and the LR staring-mode spectrum shows prominent silicate 10 and $18\,\mu$m emission (see also \citealt{mullaney_defining_2011}).
The general MIR spectral slope is blue in $\nu F_\nu$-space.
We observed ESO\,548-81 with VISIR in three narrow $N$-band filters in one night of 2009 and detected a compact MIR nucleus without any sign of extended host emission.
The nucleus appears marginally resolved in all images (FWHM $\sim 0.37\arcsec \sim 100\,$pc) but at least a second epoch is needed to confirm this extension. 
The VISIR photometric measurements are consistent with the IRS spectrum, but they would be significantly lower if the presence of subarcsecond-extended emission can be verified.
\newline\end{@twocolumnfalse}]

\begin{figure}
   \centering
   \includegraphics[angle=0,width=8.500cm]{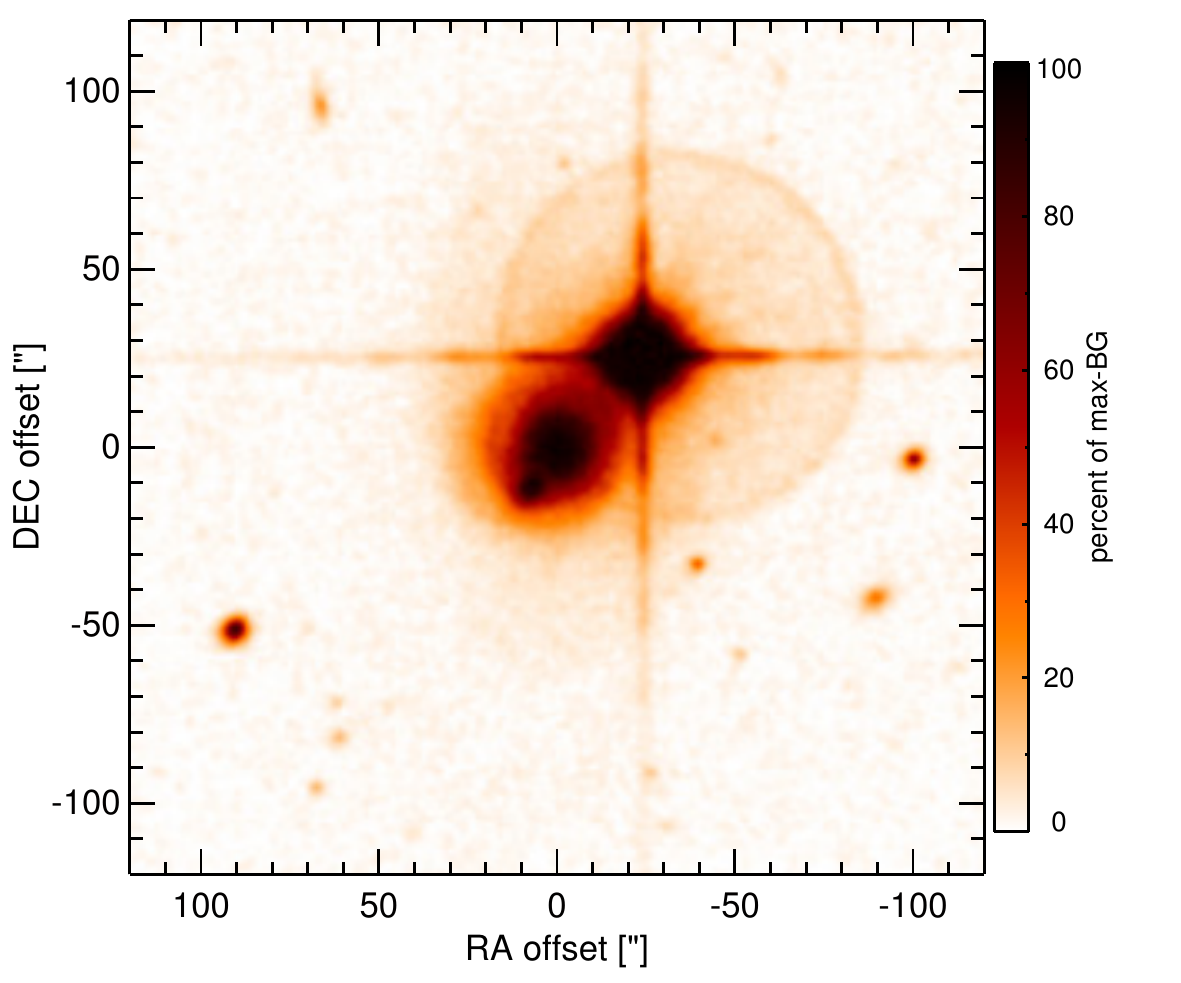}
    \caption{\label{fig:OPTim_ESO548-G081}
             Optical image (DSS, red filter) of ESO\,548-81. Displayed are the central $4\arcmin$ with North up and East to the left. 
              The colour scaling is linear with white corresponding to the median background and black to the $0.01\%$ pixels with the highest intensity.  
           }
\end{figure}
\begin{figure}
   \centering
   \includegraphics[angle=0,height=3.11cm]{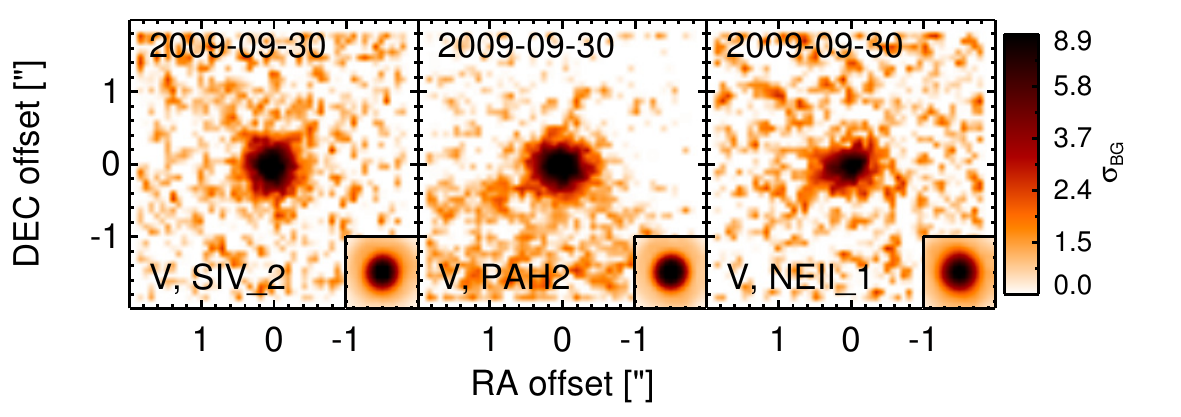}
    \caption{\label{fig:HARim_ESO548-G081}
             Subarcsecond-resolution MIR images of ESO\,548-81 sorted by increasing filter wavelength. 
             Displayed are the inner $4\arcsec$ with North up and East to the left. 
             The colour scaling is logarithmic with white corresponding to median background and black to the $75\%$ of the highest intensity of all images in units of $\sigbg$.
             The inset image shows the central arcsecond of the PSF from the calibrator star, scaled to match the science target.
             The labels in the bottom left state instrument and filter names (C: COMICS, M: Michelle, T: T-ReCS, V: VISIR).
           }
\end{figure}
\begin{figure}
   \centering
   \includegraphics[angle=0,width=8.50cm]{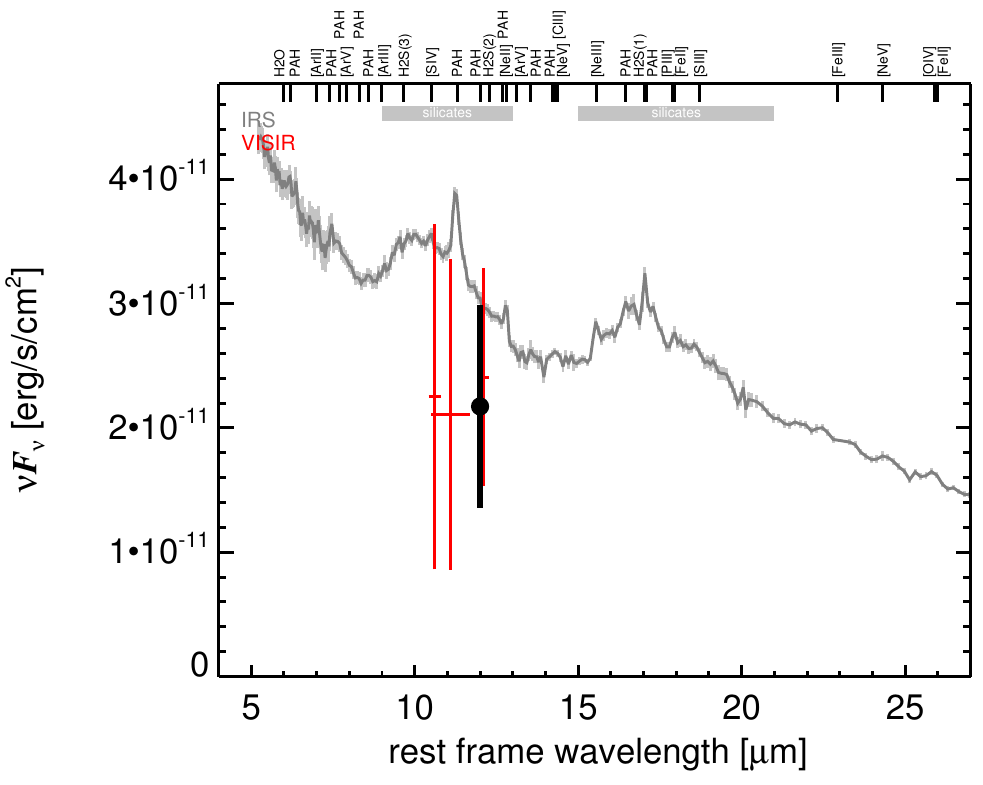}
   \caption{\label{fig:MISED_ESO548-G081}
      MIR SED of ESO\,548-81. The description  of the symbols (if present) is the following.
      Grey crosses and  solid lines mark the \spitzer/IRAC, MIPS and IRS data. 
      The colour coding of the other symbols is: 
      green for COMICS, magenta for Michelle, blue for T-ReCS and red for VISIR data.
      Darker-coloured solid lines mark spectra of the corresponding instrument.
      The black filled circles mark the nuclear 12 and $18\,\mu$m  continuum emission estimate from the data.
      The ticks on the top axis mark positions of common MIR emission lines, while the light grey horizontal bars mark wavelength ranges affected by the silicate 10 and 18$\mu$m features.     
   }
\end{figure}
\clearpage

\twocolumn[\begin{@twocolumnfalse}  
\subsection{ESO\,602-25 -- IRAS\,22287-1917}\label{app:ESO602-G025}
ESO\,602-25 is an infrared-luminous spiral galaxy at a redshift of $z=$ 0.025 ($D\sim100\,$Mpc) with an active nucleus optically classified either as a H\,II or Sy\,2 \citep{veron_agns_1997}, LINER \citep{kim_optical_1995,veilleux_optical_1995} or a AGN/starburst composite \citep{yuan_role_2010}.
Because we are not aware of any AGN evidence from other than optical wavelengths we treat it as uncertain AGN/starburst composite. 
\spitzer/IRAC, IRS and MIPS observations are available, and a compact nucleus embedded within extended host emission is visible in the IRAC $5.8$ and $8.0\,\mu$m and MIPS $24\,\mu$m images.
Thus, our photometry of the nuclear component alone provides lower values than published in \cite{u_spectral_2012}.
The IRS LR staring-mode spectrum is dominated by star formation with strong PAH emission, silicate absorption and a red spectral slope in $\nu F_\nu$-space (see also \citealt{sargsyan_infrared_2011}).
ESO\,602-25 was imaged with T-ReCS in the Qa filter in 2009 \citep{imanishi_subaru_2011}.
The MIR nucleus was weakly detected and consists of an unresolved component embedded within emission with a north-south $\sim 2\arcsec$ ($\sim 0.9$\,kpc) extent coinciding with the host major axis.
Thus, our measured flux of the unresolved nuclear component is significantly lower than the total flux value of \cite{imanishi_subaru_2011}, and $86\%$ lower than \spitzerr spectrophotometry.
These results indicate that an AGN embedded within intense star formation might indeed be present in ESO\,602-25.
We extrapolate from the Qa measurement towards shorter wavelengths in order to compute the nuclear $12\,\mu$m continuum emission estimate as described in Sect.~\ref{sec:cont}.
\newline\end{@twocolumnfalse}]

\begin{figure}
   \centering
   \includegraphics[angle=0,width=8.500cm]{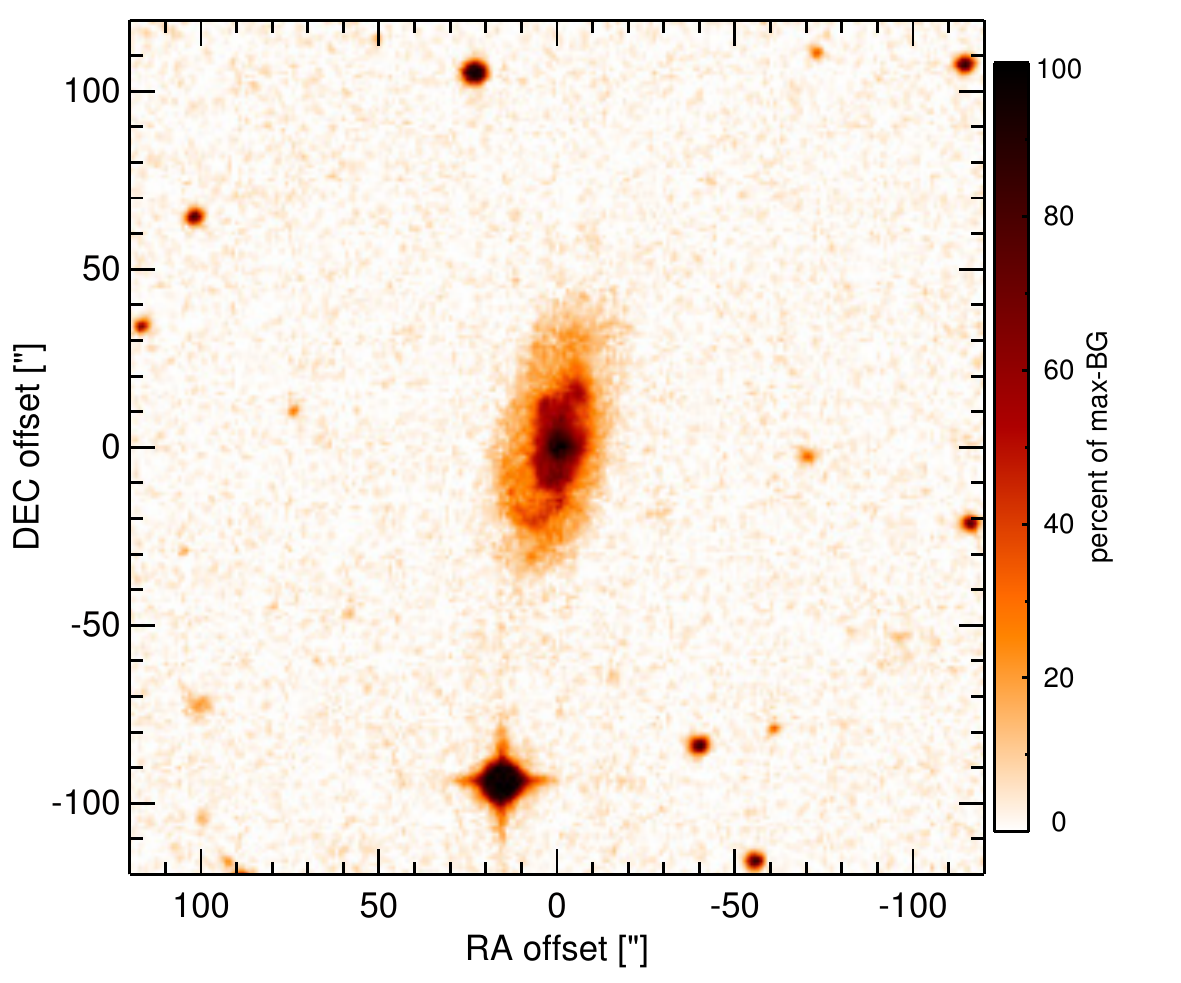}
    \caption{\label{fig:OPTim_ESO602-G025}
             Optical image (DSS, red filter) of ESO\,602-25. Displayed are the central $4\arcmin$ with North up and East to the left. 
              The colour scaling is linear with white corresponding to the median background and black to the $0.01\%$ pixels with the highest intensity.  
           }
\end{figure}
\begin{figure}
   \centering
   \includegraphics[angle=0,height=3.11cm]{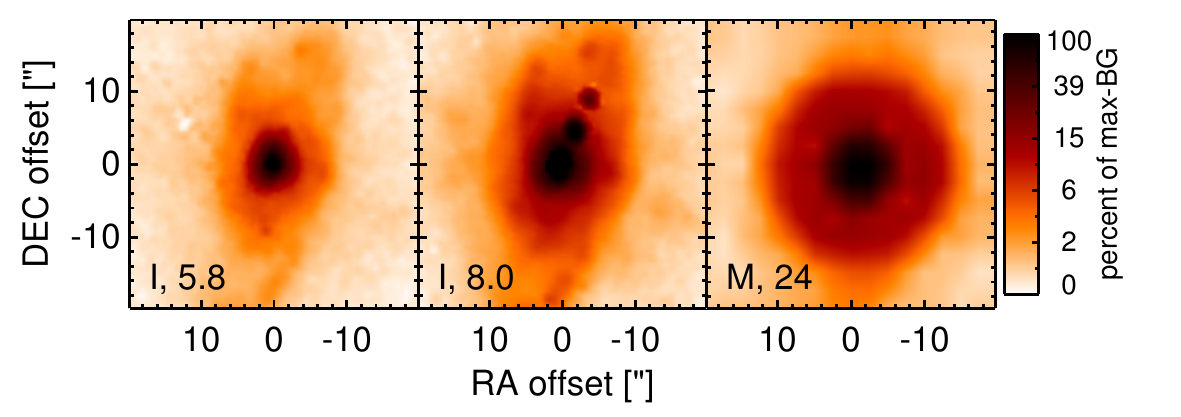}
    \caption{\label{fig:INTim_ESO602-G025}
             \spitzerr MIR images of ESO\,602-25. Displayed are the inner $40\arcsec$ with North up and East to the left. The colour scaling is logarithmic with white corresponding to median background and black to the $0.1\%$ pixels with the highest intensity.
             The label in the bottom left states instrument and central wavelength of the filter in $\mu$m (I: IRAC, M: MIPS). 
             Note that the apparent off-nuclear compact sources in the IRAC $8.0\,\mu$m image are instrumental artefacts.
           }
\end{figure}
\begin{figure}
   \centering
   \includegraphics[angle=0,height=3.11cm]{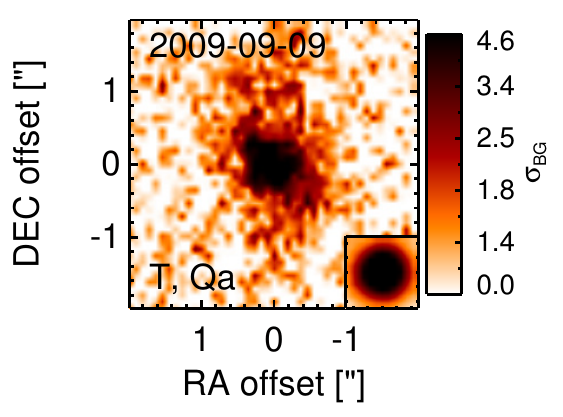}
    \caption{\label{fig:HARim_ESO602-G025}
             Subarcsecond-resolution MIR images of ESO\,602-25 sorted by increasing filter wavelength. 
             Displayed are the inner $4\arcsec$ with North up and East to the left. 
             The colour scaling is logarithmic with white corresponding to median background and black to the $75\%$ of the highest intensity of all images in units of $\sigbg$.
             The inset image shows the central arcsecond of the PSF from the calibrator star, scaled to match the science target.
             The labels in the bottom left state instrument and filter names (C: COMICS, M: Michelle, T: T-ReCS, V: VISIR).
           }
\end{figure}
\begin{figure}
   \centering
   \includegraphics[angle=0,width=8.50cm]{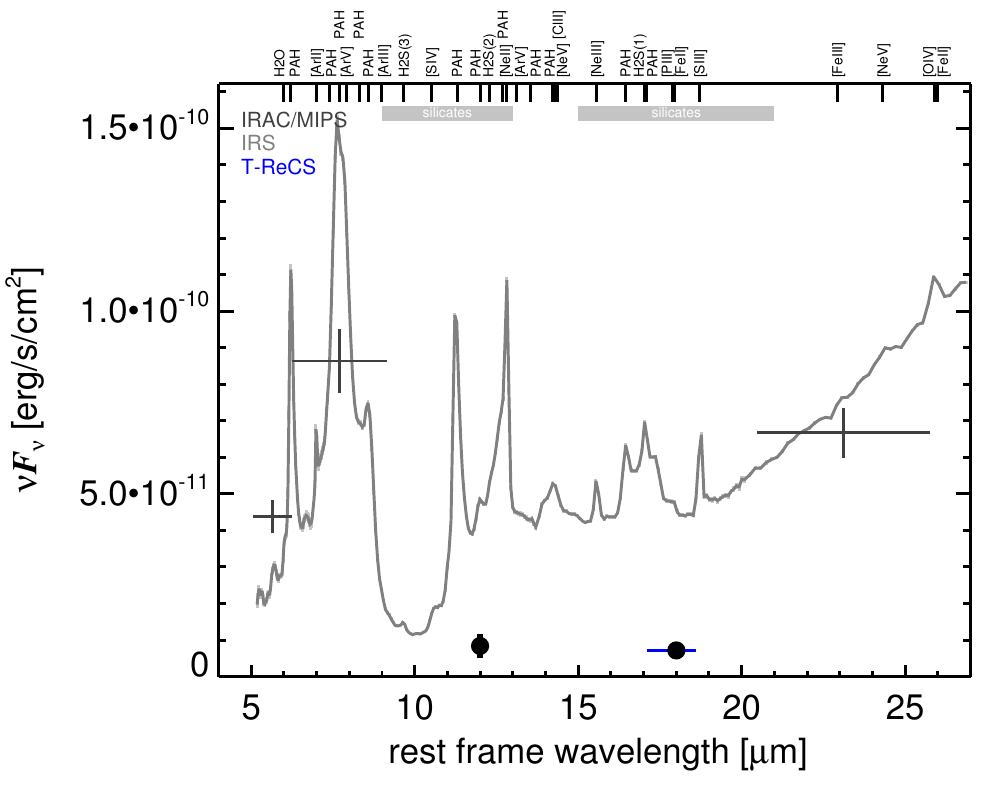}
   \caption{\label{fig:MISED_ESO602-G025}
      MIR SED of ESO\,602-25. The description  of the symbols (if present) is the following.
      Grey crosses and  solid lines mark the \spitzer/IRAC, MIPS and IRS data. 
      The colour coding of the other symbols is: 
      green for COMICS, magenta for Michelle, blue for T-ReCS and red for VISIR data.
      Darker-coloured solid lines mark spectra of the corresponding instrument.
      The black filled circles mark the nuclear 12 and $18\,\mu$m  continuum emission estimate from the data.
      The ticks on the top axis mark positions of common MIR emission lines, while the light grey horizontal bars mark wavelength ranges affected by the silicate 10 and 18$\mu$m features.     
   }
\end{figure}
\clearpage

\twocolumn[\begin{@twocolumnfalse}  
\subsection{Fairall\,9 -- ESO 113-45}\label{app:Fairall0009}
Fairall\,9 is a spiral galaxy behind the Magellanic Stream at a redshift of $z=$ 0.0470 ($D\sim199\,$Mpc) with a Sy\,1.2 nucleus \citep{veron-cetty_catalogue_2010} that belongs to the nine-month BAT AGN sample. 
Apart from \iras, first $N$-band observations of Fairall\,9 were performed with \isoo \citep{ramos_almeida_mid-infrared_2007}, where an unresolved MIR nucleus was detected.
It was also observed with \spitzer/IRAC, IRS and MIPS and appears also as a nearly-unresolved nuclear source without any extended host emission being detected.
Three epochs of MIPS $24\,\mu$m images are available (2004, 2006, and 2010), which all provide the same source flux. 
The IRS LR staring-mode spectrum shows silicate 10 and 18$\,\mu$m emission, weak PAH emission and a blue spectral slope in $\nu F_\nu$-space (see also \citealt{shi_9.7_2006}).
We observed Fairall\,9 with VISIR in seven different narrow $N$-band filters in 17 observations spread over 2005 and 2010.
The data from 2005 were analysed in \cite{horst_small_2006,horst_mid-infrared_2009}.
A compact nuclear MIR source without any extended host emission is visible in all images. 
The nucleus is unresolved in the sharpest PAH1 images and thus classified as unresolved in the MIR at subarcsecond resolution.
The VISIR photometry agrees in general well with the \spitzerr spectrophotometry and the \isoo measurements.
Only the PAH2\_2 flux from 2010 is significantly lower for unknown reasons.
\newline\end{@twocolumnfalse}]

\begin{figure}
   \centering
   \includegraphics[angle=0,width=8.500cm]{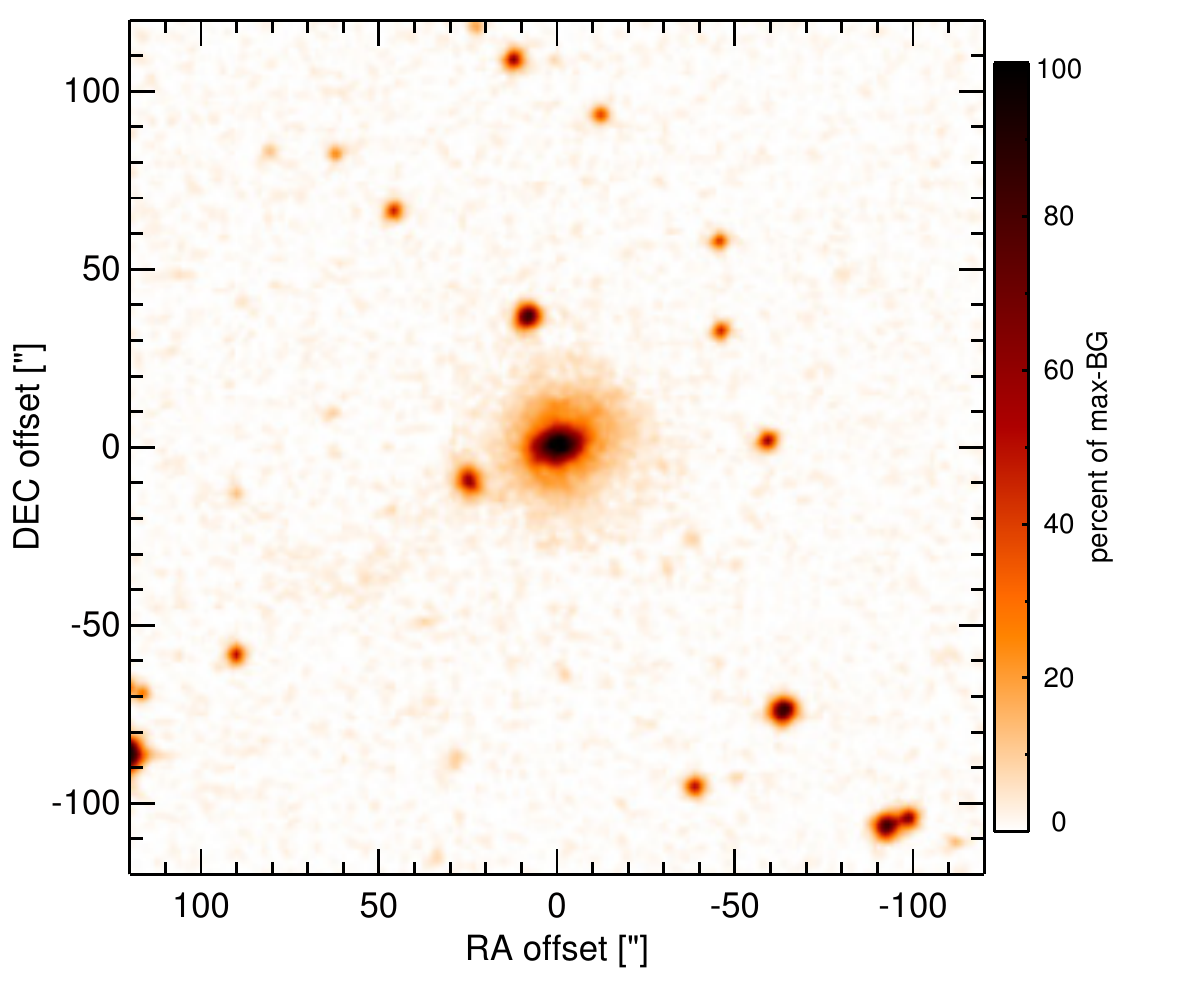}
    \caption{\label{fig:OPTim_Fairall0009}
             Optical image (DSS, red filter) of Fairall\,9. Displayed are the central $4\arcmin$ with North up and East to the left. 
              The colour scaling is linear with white corresponding to the median background and black to the $0.01\%$ pixels with the highest intensity.  
           }
\end{figure}
\begin{figure}
   \centering
   \includegraphics[angle=0,height=3.11cm]{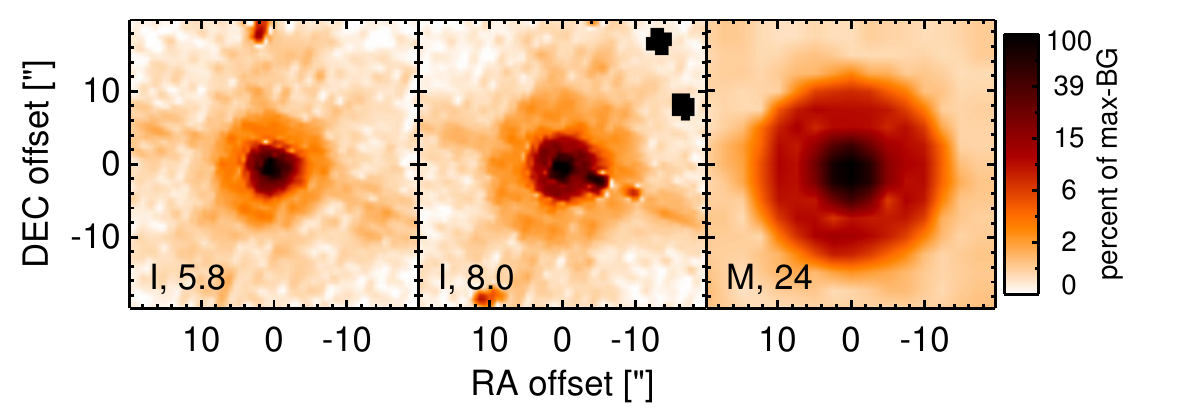}
    \caption{\label{fig:INTim_Fairall0009}
             \spitzerr MIR images of Fairall\,9. Displayed are the inner $40\arcsec$ with North up and East to the left. The colour scaling is logarithmic with white corresponding to median background and black to the $0.1\%$ pixels with the highest intensity.
             The label in the bottom left states instrument and central wavelength of the filter in $\mu$m (I: IRAC, M: MIPS).
             Note that the apparent off-nuclear compact sources in the IRAC $8.0\,\mu$m image are instrumental artefacts.
           }
\end{figure}
\begin{figure}
   \centering
   \includegraphics[angle=0,width=8.500cm]{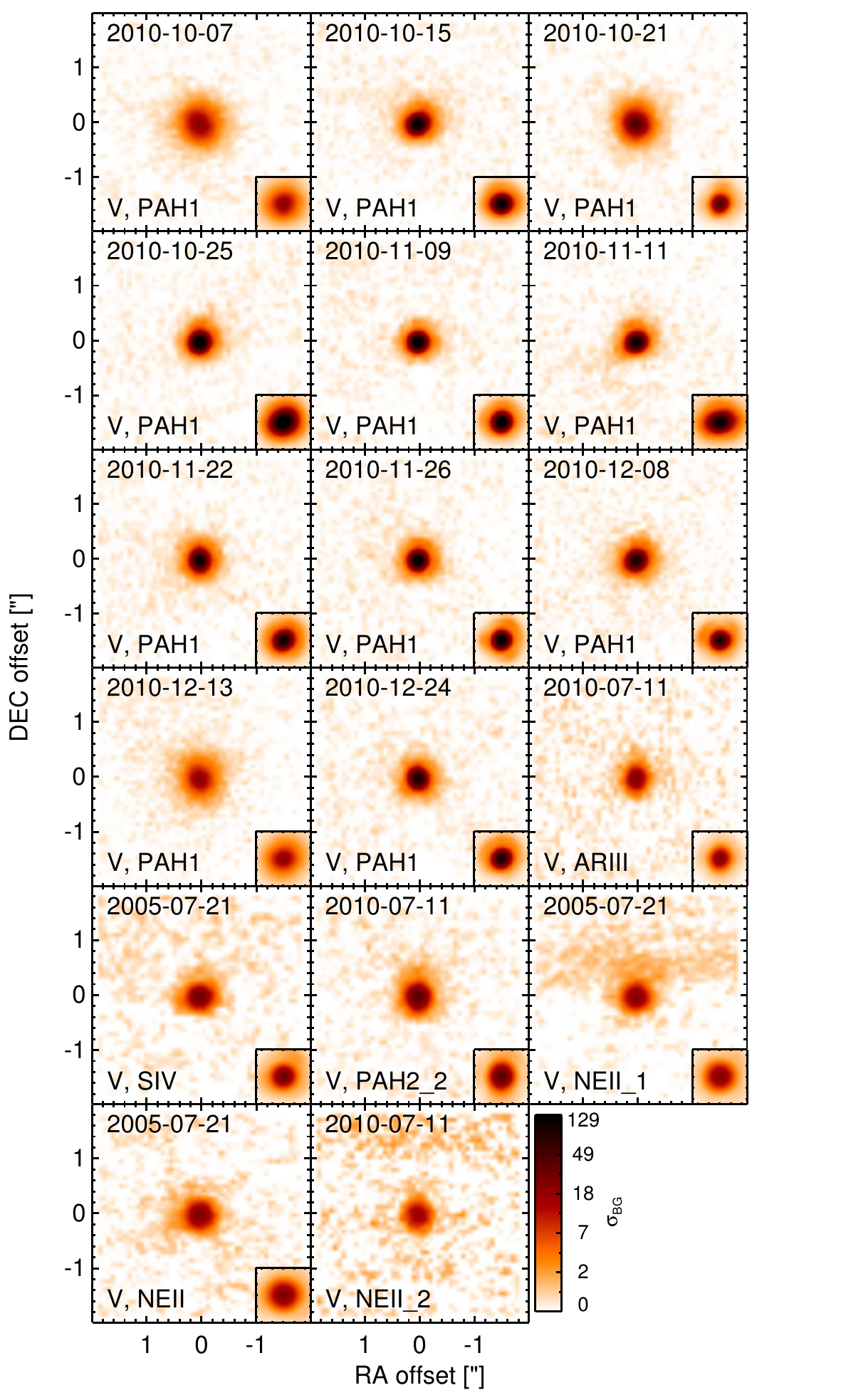}
    \caption{\label{fig:HARim_Fairall0009}
             Subarcsecond-resolution MIR images of Fairall\,9 sorted by increasing filter wavelength. 
             Displayed are the inner $4\arcsec$ with North up and East to the left. 
             The colour scaling is logarithmic with white corresponding to median background and black to the $75\%$ of the highest intensity of all images in units of $\sigbg$.
             The inset image shows the central arcsecond of the PSF from the calibrator star, scaled to match the science target.
             The labels in the bottom left state instrument and filter names (C: COMICS, M: Michelle, T: T-ReCS, V: VISIR).
           }
\end{figure}
\begin{figure}
   \centering
   \includegraphics[angle=0,width=8.50cm]{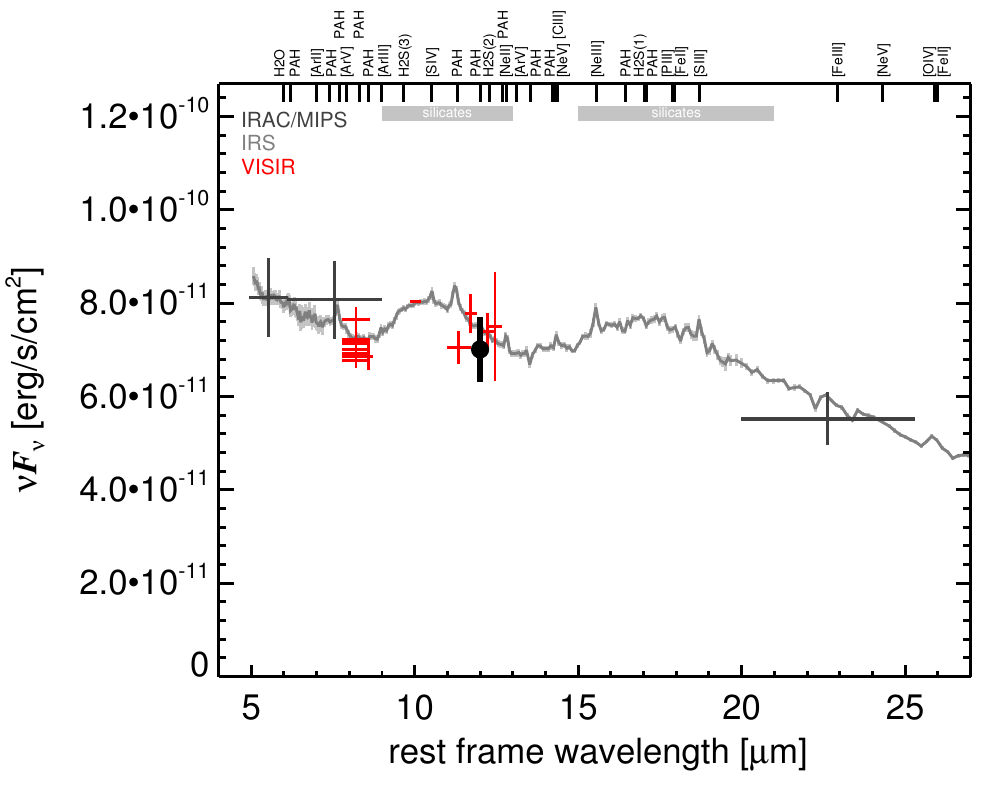}
   \caption{\label{fig:MISED_Fairall0009}
      MIR SED of Fairall\,9. The description  of the symbols (if present) is the following.
      Grey crosses and  solid lines mark the \spitzer/IRAC, MIPS and IRS data. 
      The colour coding of the other symbols is: 
      green for COMICS, magenta for Michelle, blue for T-ReCS and red for VISIR data.
      Darker-coloured solid lines mark spectra of the corresponding instrument.
      The black filled circles mark the nuclear 12 and $18\,\mu$m  continuum emission estimate from the data.
      The ticks on the top axis mark positions of common MIR emission lines, while the light grey horizontal bars mark wavelength ranges affected by the silicate 10 and 18$\mu$m features.     
   }
\end{figure}
\clearpage

\twocolumn[\begin{@twocolumnfalse}  
\subsection{Fairall\,49 -- IRAS\,18325-5926}\label{app:Fairall0049}
Fairall\,49 is an early-type galaxy at a redshift of $z=$ 0.0200 ($D\sim83.2$\,Mpc) with a Sy\,2 nucleus \citep{trippe_multi-wavelength_2010} with polarized broad lines \citep{lumsden_spectropolarimetry_2004}.
It was observed with \spitzer/IRS and MIPS and appears as an unresolved nuclear source without extended host emission in the MIPS $24\,\mu$m images from 2004 and 2006.
The measured fluxes agree  with the IRS LR staring-mode spectrum well, which shows silicate $10\,\mu$m absorption and PAH emission.
Thus,  star formation is probably affecting the \spitzerr data.
The MIR spectrum is rather flat with a peak at $\sim 17\,\mu$m in $\nu F_\nu$-space.
We observed Fairall\,49 with VISIR in three narrow $N$-band filters on one night in 2010.
The detected MIR nucleus appears elongated in particular in the ARIII and PAH2\_2 images (FWHM(major axis)$\sim 0.7\arcsec \sim 270\,$pc; PA$\sim 40\degree$).
At least another epoch of subarcsecond MIR imaging will be necessary to confirm this extension. 
The nuclear VISIR photometry is consistent with the \spitzerr spectrophotometry, but it would be significantly lower if the presence of subarcsecond-extended emission can be verified.
\newline\end{@twocolumnfalse}]

\begin{figure}
   \centering
   \includegraphics[angle=0,width=8.500cm]{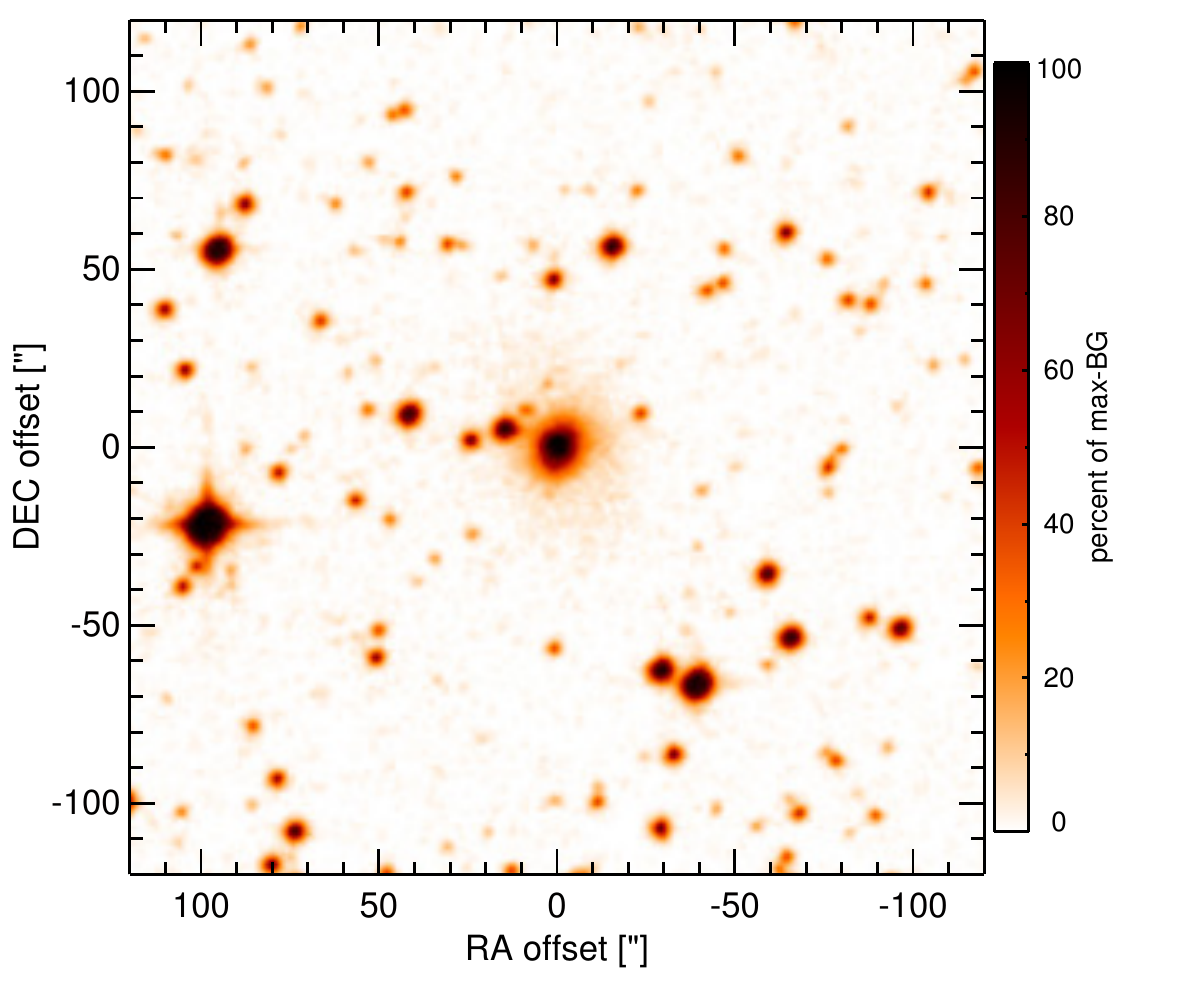}
    \caption{\label{fig:OPTim_Fairall0049}
             Optical image (DSS, red filter) of Fairall\,49. Displayed are the central $4\arcmin$ with North up and East to the left. 
              The colour scaling is linear with white corresponding to the median background and black to the $0.01\%$ pixels with the highest intensity.  
           }
\end{figure}
\begin{figure}
   \centering
   \includegraphics[angle=0,height=3.11cm]{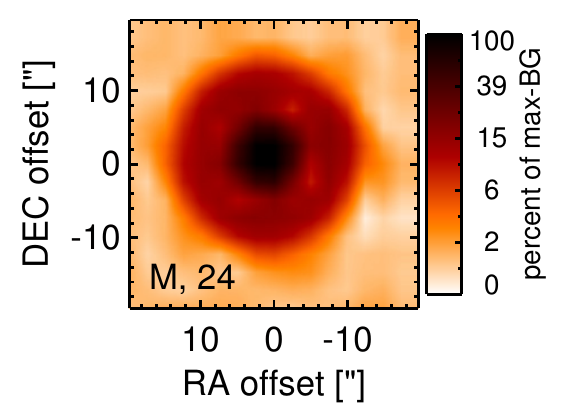}
    \caption{\label{fig:INTim_Fairall0049}
             \spitzerr MIR images of Fairall\,49. Displayed are the inner $40\arcsec$ with North up and East to the left. The colour scaling is logarithmic with white corresponding to median background and black to the $0.1\%$ pixels with the highest intensity.
             The label in the bottom left states instrument and central wavelength of the filter in $\mu$m (I: IRAC, M: MIPS). 
           }
\end{figure}
\begin{figure}
   \centering
   \includegraphics[angle=0,height=3.11cm]{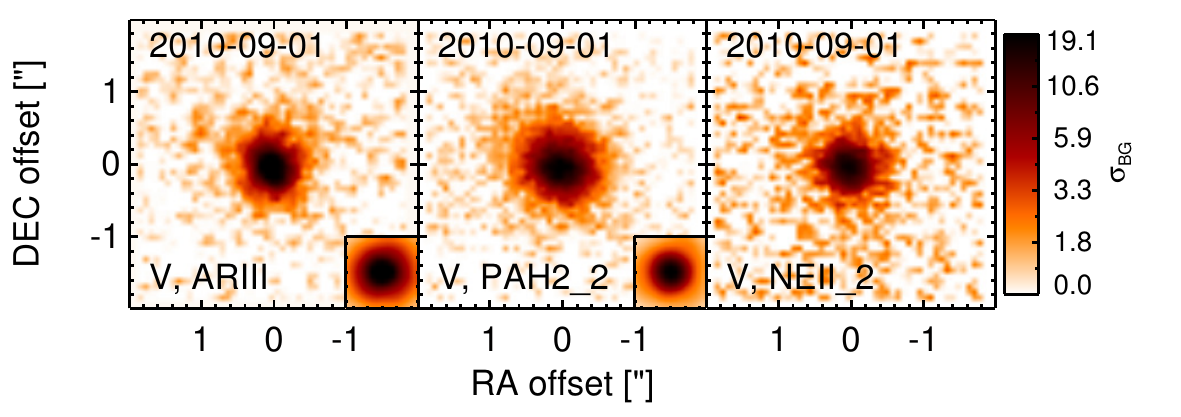}
    \caption{\label{fig:HARim_Fairall0049}
             Subarcsecond-resolution MIR images of Fairall\,49 sorted by increasing filter wavelength. 
             Displayed are the inner $4\arcsec$ with North up and East to the left. 
             The colour scaling is logarithmic with white corresponding to median background and black to the $75\%$ of the highest intensity of all images in units of $\sigbg$.
             The inset image shows the central arcsecond of the PSF from the calibrator star, scaled to match the science target.
             The labels in the bottom left state instrument and filter names (C: COMICS, M: Michelle, T: T-ReCS, V: VISIR).
           }
\end{figure}
\begin{figure}
   \centering
   \includegraphics[angle=0,width=8.50cm]{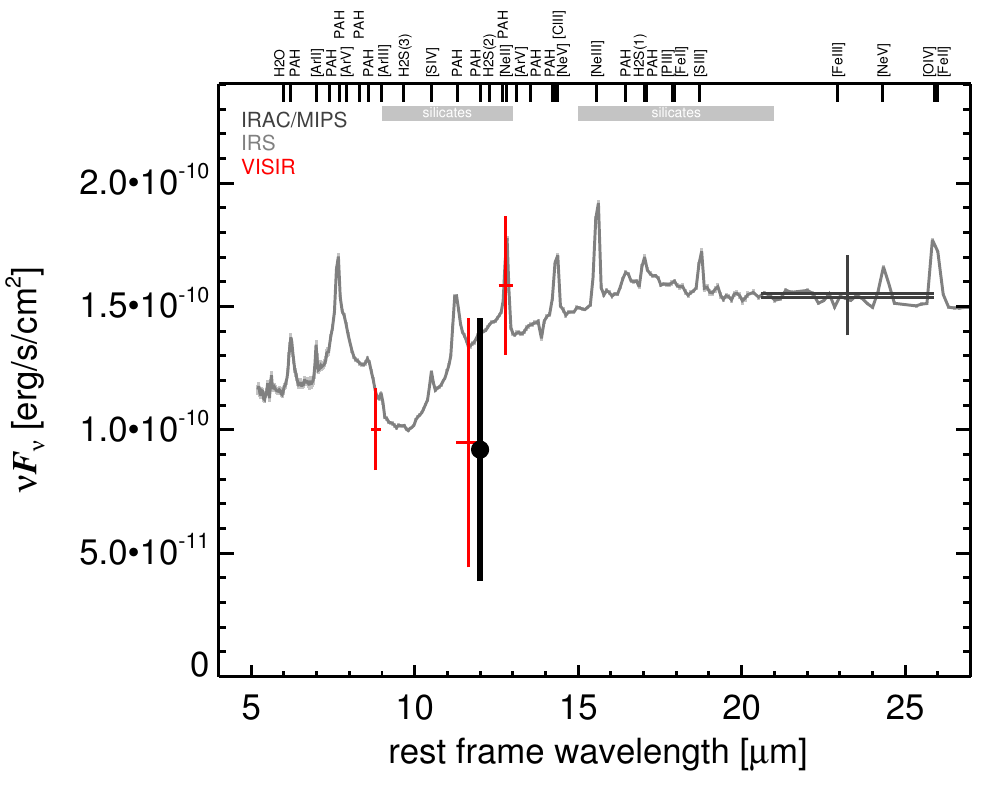}
   \caption{\label{fig:MISED_Fairall0049}
      MIR SED of Fairall\,49. The description  of the symbols (if present) is the following.
      Grey crosses and  solid lines mark the \spitzer/IRAC, MIPS and IRS data. 
      The colour coding of the other symbols is: 
      green for COMICS, magenta for Michelle, blue for T-ReCS and red for VISIR data.
      Darker-coloured solid lines mark spectra of the corresponding instrument.
      The black filled circles mark the nuclear 12 and $18\,\mu$m  continuum emission estimate from the data.
      The ticks on the top axis mark positions of common MIR emission lines, while the light grey horizontal bars mark wavelength ranges affected by the silicate 10 and 18$\mu$m features. 
      }
\end{figure}
\clearpage

\twocolumn[\begin{@twocolumnfalse}  
\subsection{Fairall\,51 -- ESO\,140-43}\label{app:Fairall0051}
Fairall\,51 is a barred spiral galaxy at a distance of $D=$ 64.1\,Mpc (\citealt{springob_erratum:_2009}; $z=0.0142$) hosting an AGN with optical Sy\,1.5 classification \citep{veron-cetty_catalogue_2010}.
Its extended narrow line region is aligned with the major axis of the galaxy \citep{schmitt_hubble_2003} and shows the same extinction as the broad-line region, indicating dust in the foreground \citep{bennert_size_2006}.
The first $N$-band photometry of Fairall\,51 was carried out by \cite{glass_mid-infrared_1982} with the ESO infrared photometer system at the 3\,m telescope.
Since then, it was observed with \spitzer/IRS in LR staring mode.
The MIR spectrum displays silicate 10 and 18\,$\mu$m emission and  PAH emission with a rather flat spectral slope in $\nu F_\nu$-space (see also \citealt{sargsyan_infrared_2011}).
The only subarcsecond MIR observation of Fairall\,51 was carried out with T-ReCS in a $Q$-band filter in 2007 (unpublished, to our knowledge).
A compact MIR nucleus was detected without any extended host emission. 
Our photometry measurement agrees with the flux level of IRS spectrum and thus we use the latter to compute the nuclear 12\,$\mu$m continuum emission estimate.
\newline\end{@twocolumnfalse}]

\begin{figure}
   \centering
   \includegraphics[angle=0,width=8.500cm]{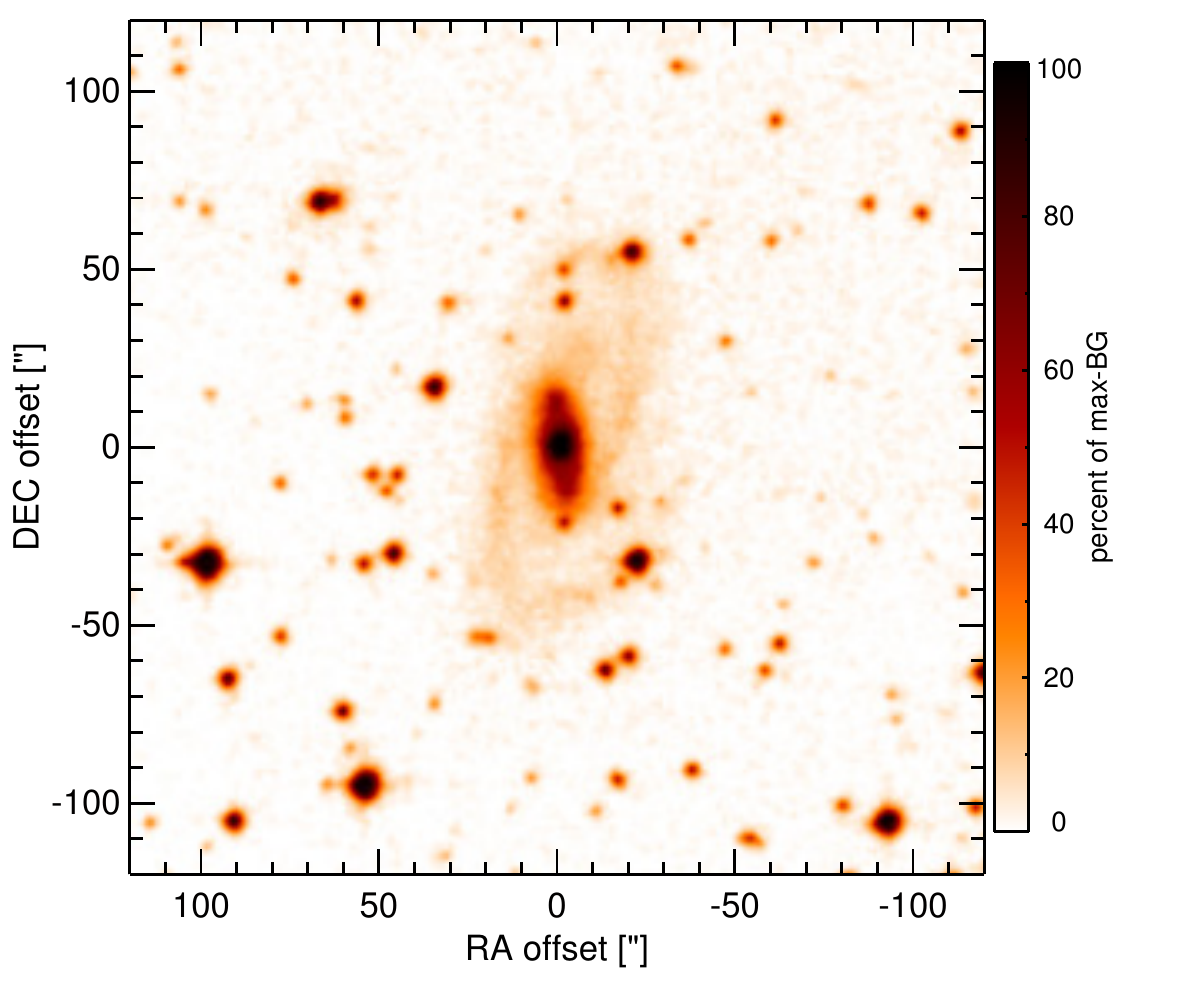}
    \caption{\label{fig:OPTim_Fairall0051}
             Optical image (DSS, red filter) of Fairall\,51. Displayed are the central $4\arcmin$ with North up and East to the left. 
              The colour scaling is linear with white corresponding to the median background and black to the $0.01\%$ pixels with the highest intensity.  
           }
\end{figure}
\begin{figure}
   \centering
   \includegraphics[angle=0,height=3.11cm]{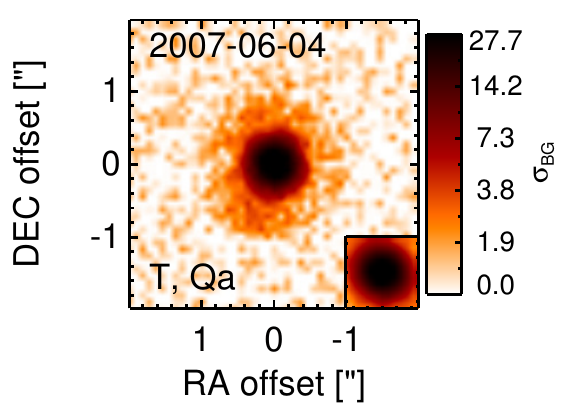}
    \caption{\label{fig:HARim_Fairall0051}
             Subarcsecond-resolution MIR images of Fairall\,51 sorted by increasing filter wavelength. 
             Displayed are the inner $4\arcsec$ with North up and East to the left. 
             The colour scaling is logarithmic with white corresponding to median background and black to the $75\%$ of the highest intensity of all images in units of $\sigbg$.
             The inset image shows the central arcsecond of the PSF from the calibrator star, scaled to match the science target.
             The labels in the bottom left state instrument and filter names (C: COMICS, M: Michelle, T: T-ReCS, V: VISIR).
           }
\end{figure}
\begin{figure}
   \centering
   \includegraphics[angle=0,width=8.50cm]{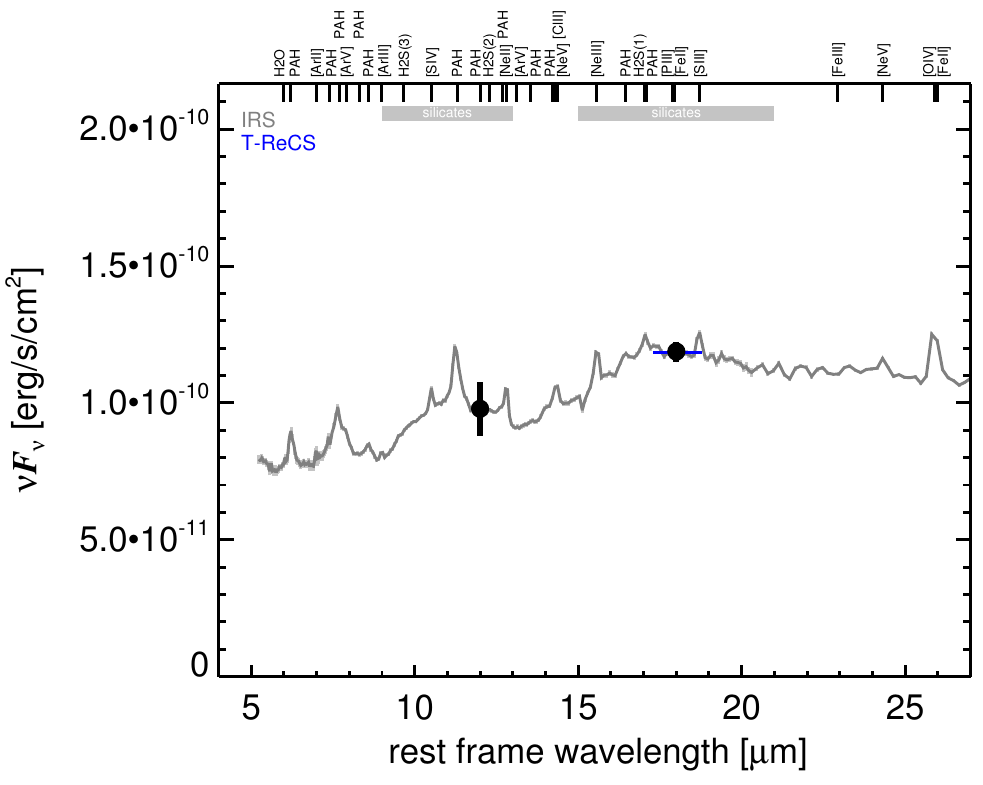}
   \caption{\label{fig:MISED_Fairall0051}
      MIR SED of Fairall\,51. The description  of the symbols (if present) is the following.
      Grey crosses and  solid lines mark the \spitzer/IRAC, MIPS and IRS data. 
      The colour coding of the other symbols is: 
      green for COMICS, magenta for Michelle, blue for T-ReCS and red for VISIR data.
      Darker-coloured solid lines mark spectra of the corresponding instrument.
      The black filled circles mark the nuclear 12 and $18\,\mu$m  continuum emission estimate from the data.
      The ticks on the top axis mark positions of common MIR emission lines, while the light grey horizontal bars mark wavelength ranges affected by the silicate 10 and 18$\mu$m features.}
\end{figure}
\clearpage

\twocolumn[\begin{@twocolumnfalse}  
\subsection{Fornax\,A -- NGC\,1316}\label{app:FornaxA}
Fornax\,A is a FR\,I-like radio source coinciding with the peculiar galaxy NGC\,1316 at a distance of $D=$ $\sim 19$\,Mpc (NED redshift-independent median).
It contains a LINER-like nucleus \citep{bregman_o_2005} and features large diffuse radio lobes.
There are various indications for Fornax\,A being a merger remnant \citep{schweizer_optical_1980,goudfrooij_star_2001} with powerful nuclear activity until relatively recently \citep{iyomoto_declined_1998}.
The first $N$-band observations of Fornax\,A were carried out by \cite{rieke_infrared_1978} and \cite{sparks_infrared_1986}, and later by \cite{xilouris_dust_2004} with \iso/ISOCAM.
Since then, it was also observed with \spitzer/IRAC, IRS and MIPS.
In the IRAC $5.8$ and $8.0\,\mu$m and MIPS 24\,$\mu$m images, an extended elliptical structure was detected in the nucleus (major axis in the north-east direction) with  perpendicularly extended weak emission (see also \citealt{temi_ages_2005,lanz_constraining_2010}). 
Our nuclear IRAC $5.8$ and $8.0\,\mu$m and MIPS 24\,$\mu$m photometry yields significantly lower fluxes than the global flux measurements of \cite{dale_infrared_2005} and \cite{munoz-mateos_radial_2009}, and significantly higher values than those from the nuclear decomposition performed by \cite{lanz_constraining_2010}.
Owing to this complex extended MIR structure, the PBCD IRS LR mapping-mode spectrum is mostly unusable (see \citealt{smith_mid-infrared_2007,dale_spitzer_2009} for a proper spectrum).
We show only the short-wavelength part, which indicates silicate 10\,$\mu$m and PAH emission.
Fornax\,A was observed at subarcsecond resolution with VISIR in the broad $N$-band filter SIC in 2006 and PAH2\_2 in 2009, and the nucleus remained undetected on both occasions.
The SIC measurement was analysed by \cite{van_der_wolk_dust_2010} obtaining a much lower upper limit than we derive from the same data.
In addition, Fornax\,A was observed with T-ReCS in two $N$-band filters in 2007 where a compact MIR nucleus was weakly detected (unpublished, to our knowledge). 
The nucleus appears possibly resolved (FWHM(major axis) $\sim 0.6\arcsec \sim 55\,$pc; PA$\sim90\degree$).
However, the S/N is very low and a second epoch is necessary to verify this extension.
The measured nuclear fluxes are on average $\sim63\%$ lower than the \spitzerr spectrophotometry and consistent with our VISIR SIC upper limit but not with the PAH2\_2 upper limit, for unknown reasons.
\newline\end{@twocolumnfalse}]

\begin{figure}
   \centering
   \includegraphics[angle=0,width=8.500cm]{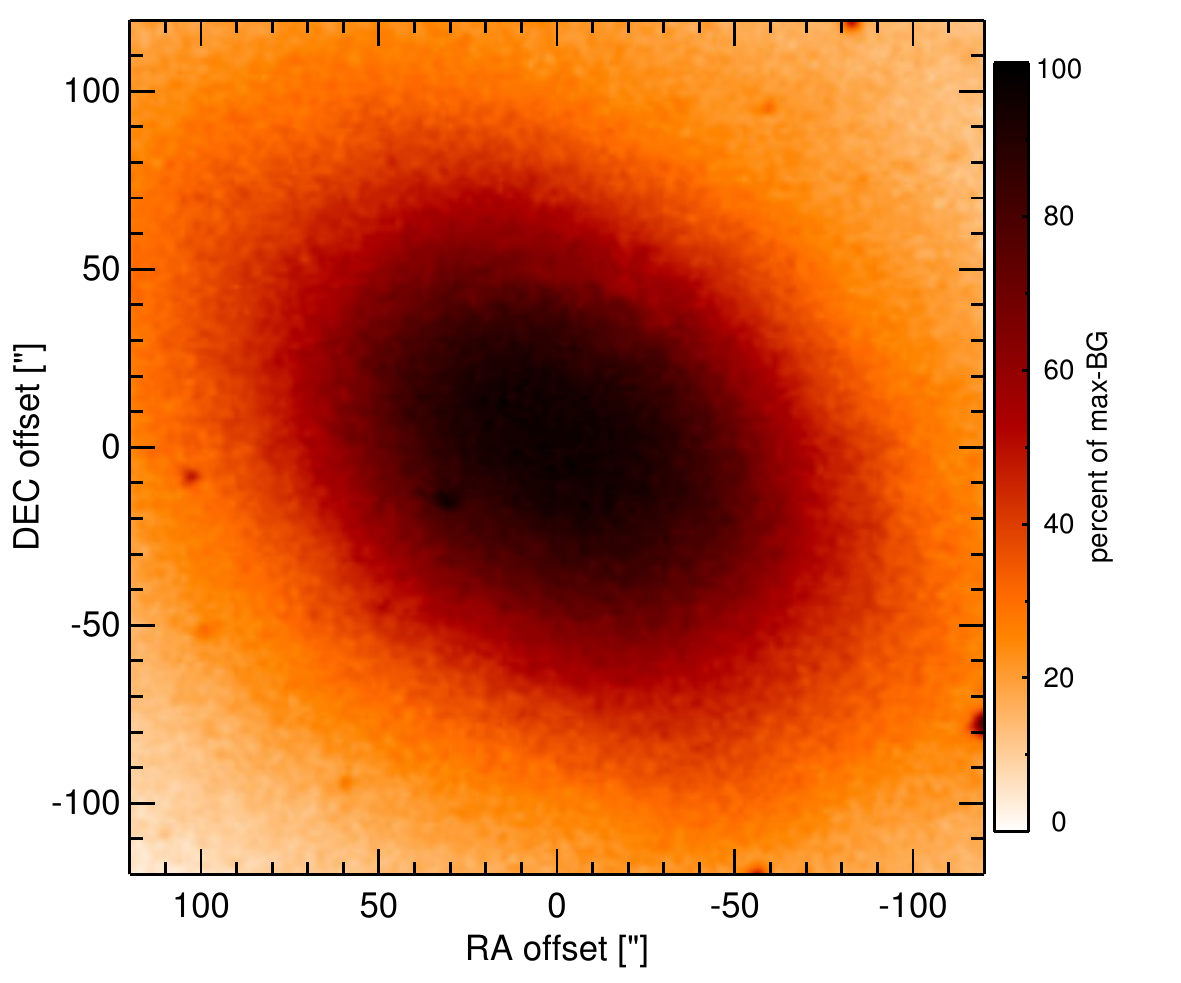}
    \caption{\label{fig:OPTim_FornaxA}
             Optical image (DSS, red filter) of Fornax\,A. Displayed are the central $4\arcmin$ with North up and East to the left. 
              The colour scaling is linear with white corresponding to the median background and black to the $0.01\%$ pixels with the highest intensity.  
           }
\end{figure}
\begin{figure}
   \centering
   \includegraphics[angle=0,height=3.11cm]{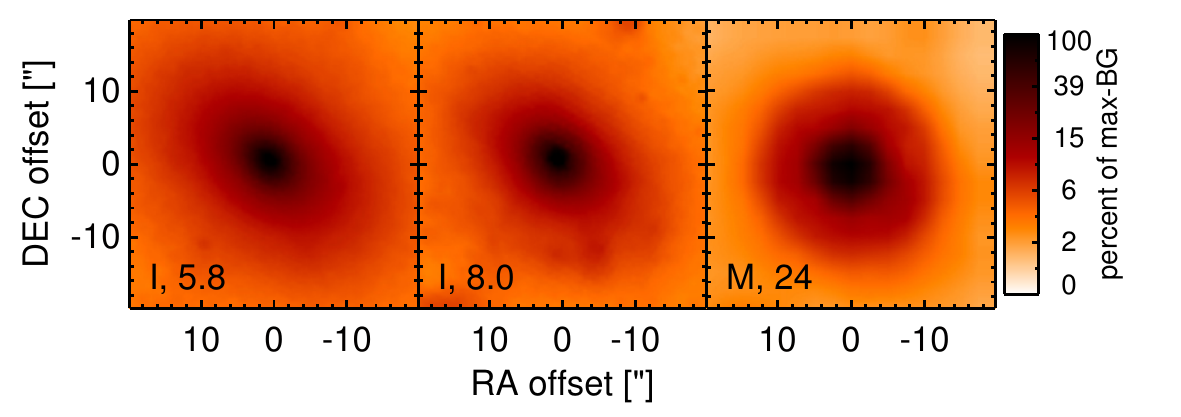}
    \caption{\label{fig:INTim_FornaxA}
             \spitzerr MIR images of Fornax\,A. Displayed are the inner $40\arcsec$ with North up and East to the left. The colour scaling is logarithmic with white corresponding to median background and black to the $0.1\%$ pixels with the highest intensity.
             The label in the bottom left states instrument and central wavelength of the filter in $\mu$m (I: IRAC, M: MIPS). 
           }
\end{figure}
\begin{figure}
   \centering
   \includegraphics[angle=0,height=3.11cm]{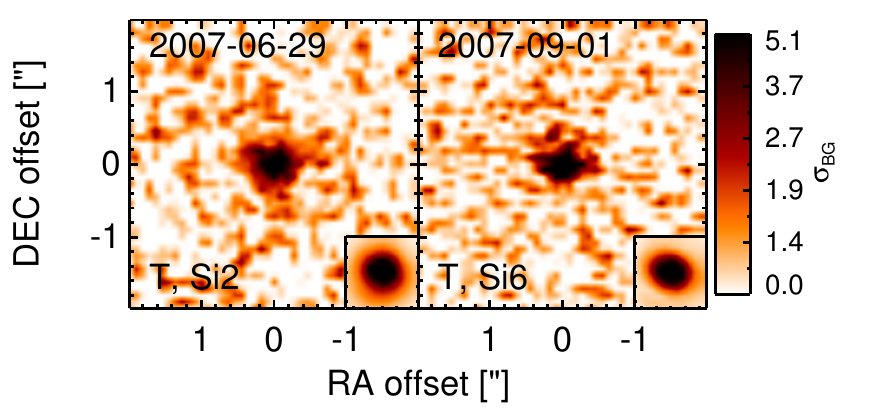}
    \caption{\label{fig:HARim_FornaxA}
             Subarcsecond-resolution MIR images of Fornax\,A sorted by increasing filter wavelength. 
             Displayed are the inner $4\arcsec$ with North up and East to the left. 
             The colour scaling is logarithmic with white corresponding to median background and black to the $75\%$ of the highest intensity of all images in units of $\sigbg$.
             The inset image shows the central arcsecond of the PSF from the calibrator star, scaled to match the science target.
             The labels in the bottom left state instrument and filter names (C: COMICS, M: Michelle, T: T-ReCS, V: VISIR).
           }
\end{figure}
\begin{figure}
   \centering
   \includegraphics[angle=0,width=8.50cm]{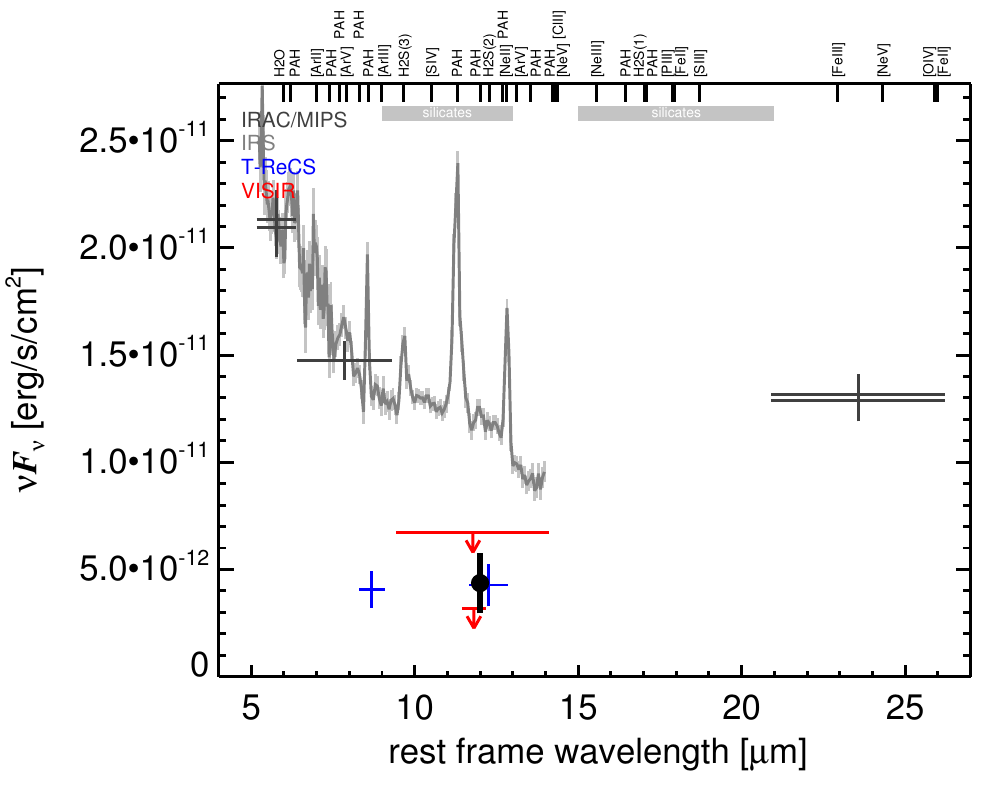}
   \caption{\label{fig:MISED_FornaxA}
      MIR SED of Fornax\,A. The description  of the symbols (if present) is the following.
      Grey crosses and  solid lines mark the \spitzer/IRAC, MIPS and IRS data. 
      The colour coding of the other symbols is: 
      green for COMICS, magenta for Michelle, blue for T-ReCS and red for VISIR data.
      Darker-coloured solid lines mark spectra of the corresponding instrument.
      The black filled circles mark the nuclear 12 and $18\,\mu$m  continuum emission estimate from the data.
      The ticks on the top axis mark positions of common MIR emission lines, while the light grey horizontal bars mark wavelength ranges affected by the silicate 10 and 18$\mu$m features.}
\end{figure}
\clearpage

\twocolumn[\begin{@twocolumnfalse}  
\subsection{H\,0557-385 -- LEDA 75476 -- EXO 055620-3820.2 -- 2MASX\,J05580206-3820043}\label{app:H0557-385}
H\,0557-385 is a Sy\,1.2 object \citep{veron-cetty_catalogue_2010} in the highly-inclined early-type galaxy LEDA\,75476 at a redshift of $z=$ 0.0339 ($D\sim144$\,Mpc) that belongs to the nine-month BAT AGN sample.
This object was observed with \spitzer/IRAC, IRS and MIPS where it appears as a nearly unresolved nuclear MIR source without any extended host emission. 
The PBCD IRAC $8.0\,\mu$m image is partly saturated and is not used.
Our IRAC $5.8$ photometry matches the values published in \cite{kishimoto_mapping_2011}.
The IRS LR staring-mode spectrum shows  silicate  $10\,\mu$m absorption, which is unusual for a type\,I AGN but matches the partial obscuration scenario proposed by \cite{longinotti_obscuring_2009} for H\,0557-385.
The spectral slope is blue in $\nu F_\nu$-space (see also \citealt{mullaney_defining_2011}).
H\,0557-385 was observed with VISIR in 2009 by us in three narrow $N$-band filters and by \cite{kishimoto_mapping_2011} in another two.
A compact nuclear MIR source was detected in all images, which appears marginally resolved in the SIV filter image (FWHM $\sim 0.46\arcsec \sim 300\,$pc) but not in the PAH1 and NEII images.
Therefore, it remains uncertain, whether the MIR nucleus of H\,0557-385 is resolved at subarcsecond resolution. 
Our photometry is consistent with the values of \cite{kishimoto_mapping_2011} and the \spitzerr spectrophotometry.
Note that \cite{kishimoto_mapping_2011} marginally resolve the nuclear MIR emission of H\,0557-385 with MIDI interferometric observations (see also \citealt{burtscher_diversity_2013}).
\newline\end{@twocolumnfalse}]

\begin{figure}
   \centering
   \includegraphics[angle=0,width=8.500cm]{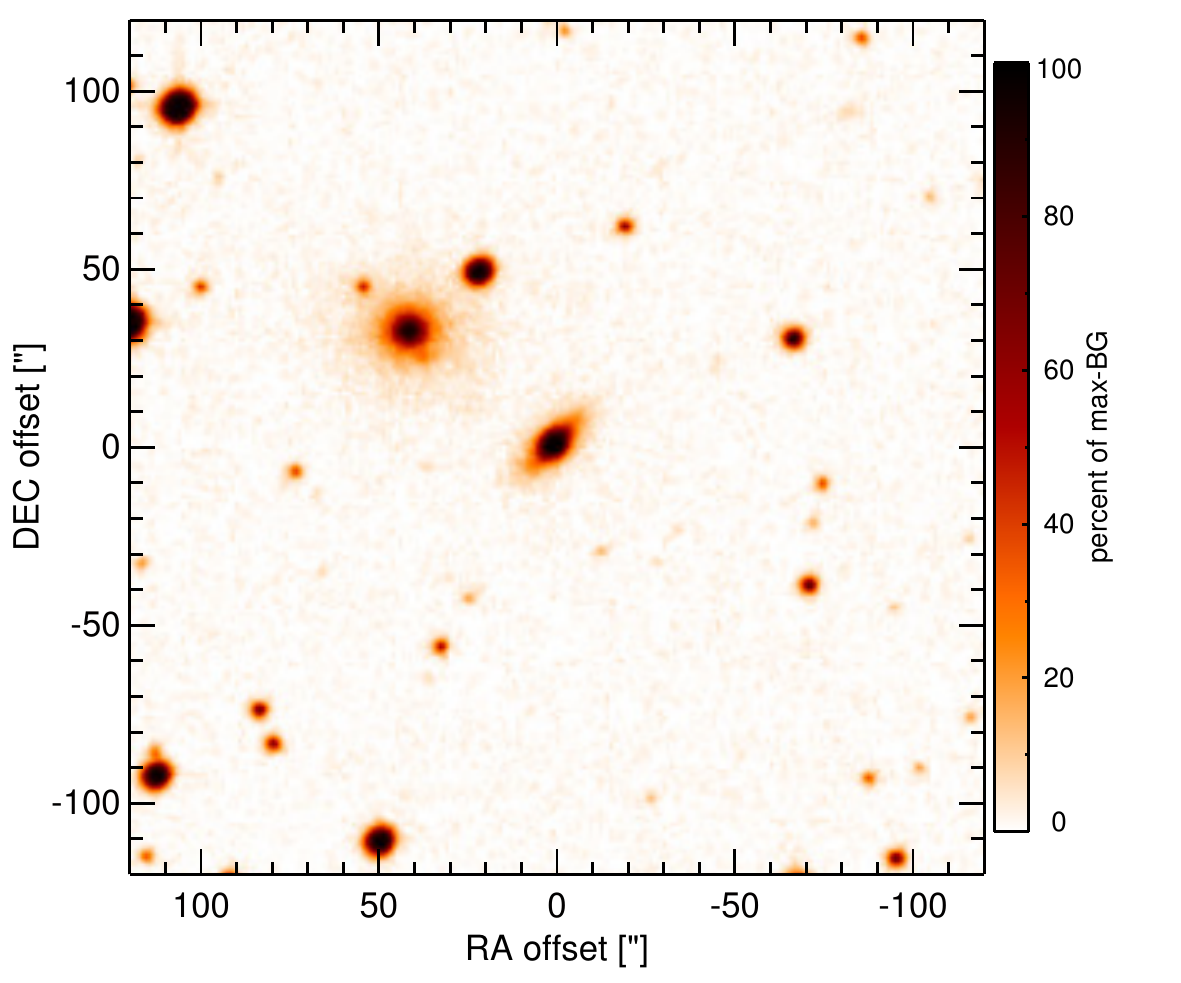}
    \caption{\label{fig:OPTim_H0557-385}
             Optical image (DSS, red filter) of H\,0557-385. Displayed are the central $4\arcmin$ with North up and East to the left. 
              The colour scaling is linear with white corresponding to the median background and black to the $0.01\%$ pixels with the highest intensity.  
           }
\end{figure}
\begin{figure}
   \centering
   \includegraphics[angle=0,height=3.11cm]{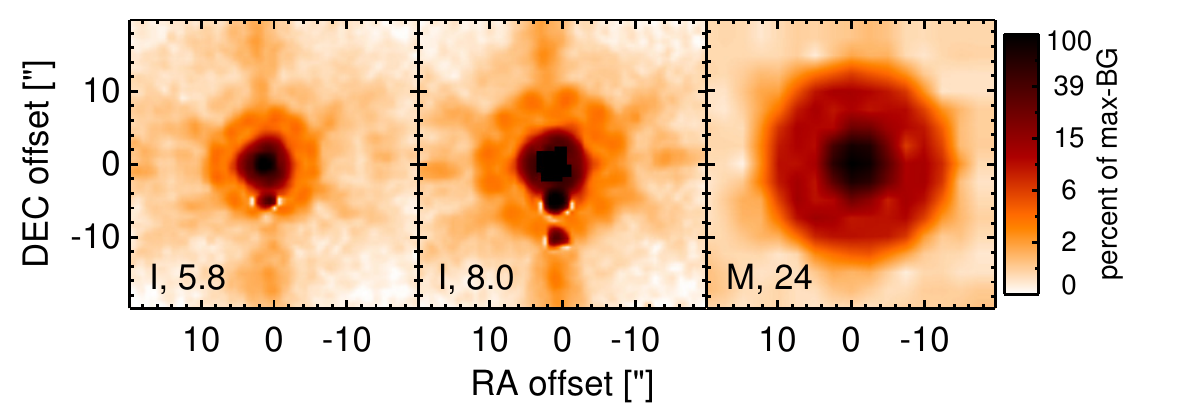}
    \caption{\label{fig:INTim_H0557-385}
             \spitzerr MIR images of H\,0557-385. Displayed are the inner $40\arcsec$ with North up and East to the left. The colour scaling is logarithmic with white corresponding to median background and black to the $0.1\%$ pixels with the highest intensity.
             The label in the bottom left states instrument and central wavelength of the filter in $\mu$m (I: IRAC, M: MIPS). 
             Note that the apparent off-nuclear compact sources in the IRAC 5.8 and $8.0\,\mu$m images are instrumental artefacts.
           }
\end{figure}
\begin{figure}
   \centering
   \includegraphics[angle=0,width=8.500cm]{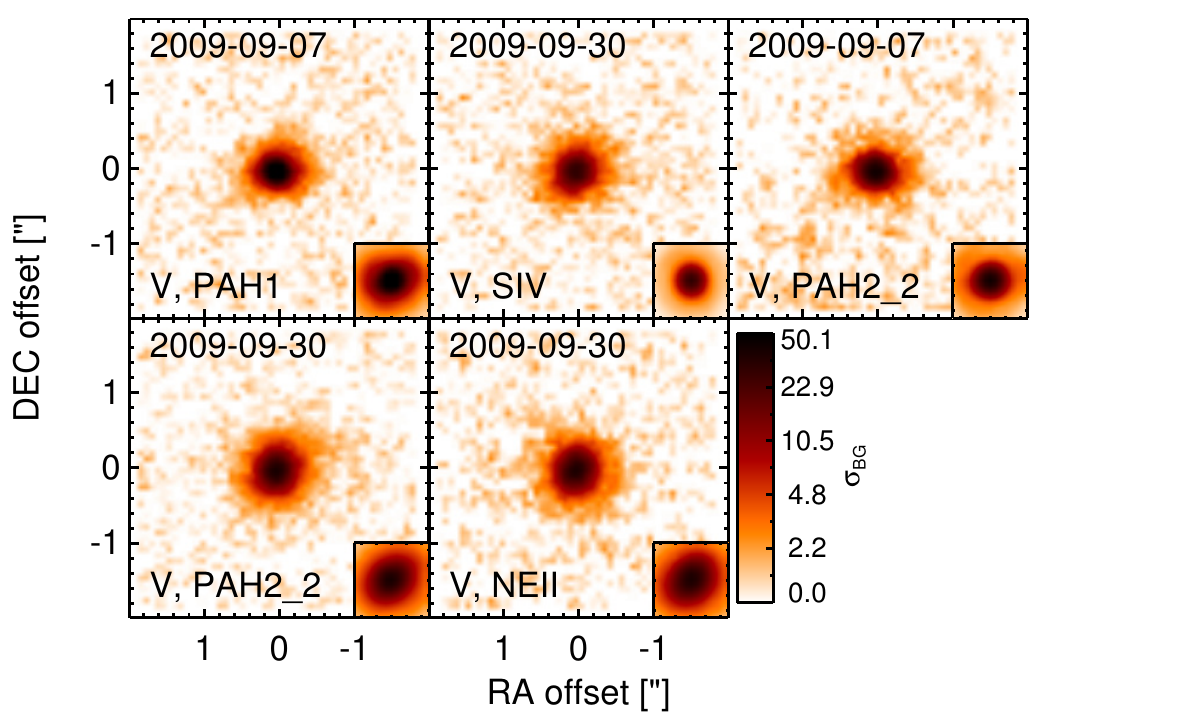}
    \caption{\label{fig:HARim_H0557-385}
             Subarcsecond-resolution MIR images of H\,0557-385 sorted by increasing filter wavelength. 
             Displayed are the inner $4\arcsec$ with North up and East to the left. 
             The colour scaling is logarithmic with white corresponding to median background and black to the $75\%$ of the highest intensity of all images in units of $\sigbg$.
             The inset image shows the central arcsecond of the PSF from the calibrator star, scaled to match the science target.
             The labels in the bottom left state instrument and filter names (C: COMICS, M: Michelle, T: T-ReCS, V: VISIR).
           }
\end{figure}
\begin{figure}
   \centering
   \includegraphics[angle=0,width=8.50cm]{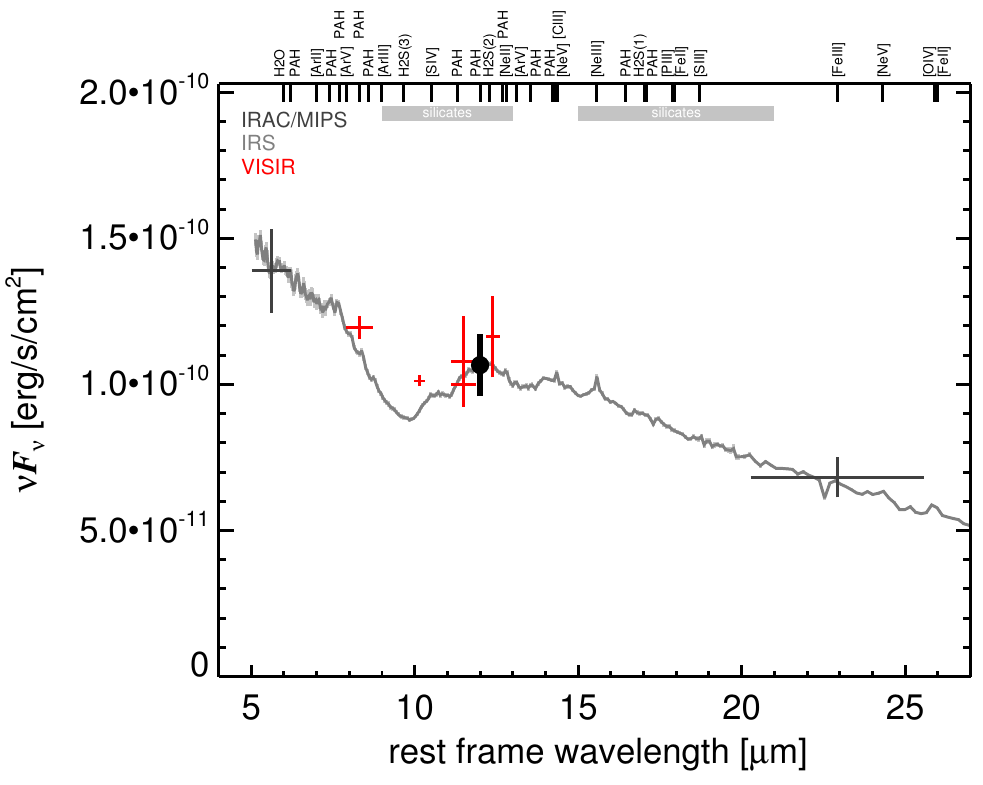}
   \caption{\label{fig:MISED_H0557-385}
      MIR SED of H\,0557-385. The description  of the symbols (if present) is the following.
      Grey crosses and  solid lines mark the \spitzer/IRAC, MIPS and IRS data. 
      The colour coding of the other symbols is: 
      green for COMICS, magenta for Michelle, blue for T-ReCS and red for VISIR data.
      Darker-coloured solid lines mark spectra of the corresponding instrument.
      The black filled circles mark the nuclear 12 and $18\,\mu$m  continuum emission estimate from the data.
      The ticks on the top axis mark positions of common MIR emission lines, while the light grey horizontal bars mark wavelength ranges affected by the silicate 10 and 18$\mu$m features.}
\end{figure}
\clearpage

\twocolumn[\begin{@twocolumnfalse}  
\subsection{H\,1143-182 -- LEDA\,88639 -- 2MASX\,J11454045-1827149}\label{app:H1143-182}
H\,1143-182 ($z = 0.0329$: $D \sim 144\,$Mpc) was discovered as an AGN in X-rays by \cite{remillard_discovery_1986}.
It is classified as a Sy\,1.5 \citep{veron-cetty_catalogue_2010}, and belongs to the 9 month BAT AGN sample.
The host galaxy is the low-inclination spiral galaxy LEDA\,88639.
A \spitzer/IRS LR staring-mode spectrum is available and was analysed in \cite{sargsyan_infrared_2011}.
It shows silicate emission and only very weak PAH emission. 
In our three VISIR images, the nucleus of H\,1143-182 appears very compact.
The SIV filter image possesses the best S/N and also exhibits higher fluxes than those of the other filters, which suffer from low S/N.
Therefore, the SIV filter flux is more reliable and also agrees with the IRS spectrum. 
The latter is used to compute the 12$\,\mu$m nuclear continuum emission estimate corrected for the silicate $10\,\mu$m emission feature.
\newline\end{@twocolumnfalse}]

\begin{figure}
   \centering
   \includegraphics[angle=0,width=8.500cm]{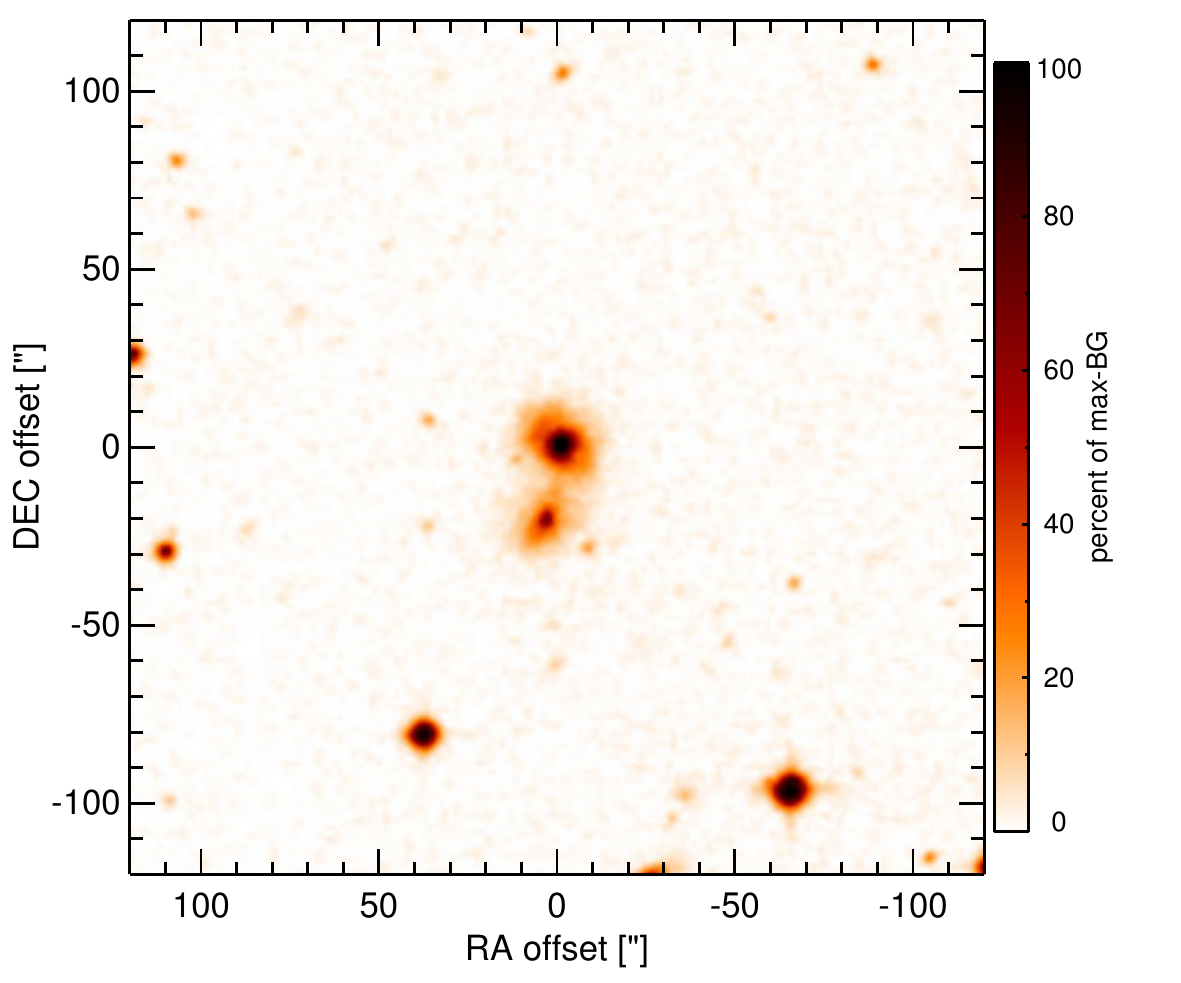}
    \caption{\label{fig:OPTim_H1143-182}
             Optical image (DSS, red filter) of H\,1143-182. Displayed are the central $4\arcmin$ with North up and East to the left. 
              The colour scaling is linear with white corresponding to the median background and black to the $0.01\%$ pixels with the highest intensity.  
           }
\end{figure}
\begin{figure}
   \centering
   \includegraphics[angle=0,height=3.11cm]{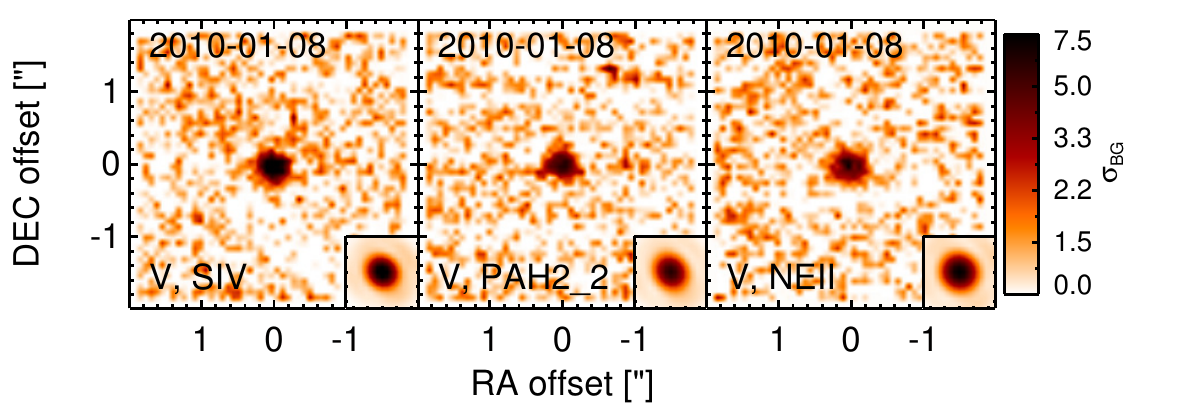}
    \caption{\label{fig:HARim_H1143-182}
             Subarcsecond-resolution MIR images of H\,1143-182 sorted by increasing filter wavelength. 
             Displayed are the inner $4\arcsec$ with North up and East to the left. 
             The colour scaling is logarithmic with white corresponding to median background and black to the $75\%$ of the highest intensity of all images in units of $\sigbg$.
             The inset image shows the central arcsecond of the PSF from the calibrator star, scaled to match the science target.
             The labels in the bottom left state instrument and filter names (C: COMICS, M: Michelle, T: T-ReCS, V: VISIR).
           }
\end{figure}
\begin{figure}
   \centering
   \includegraphics[angle=0,width=8.50cm]{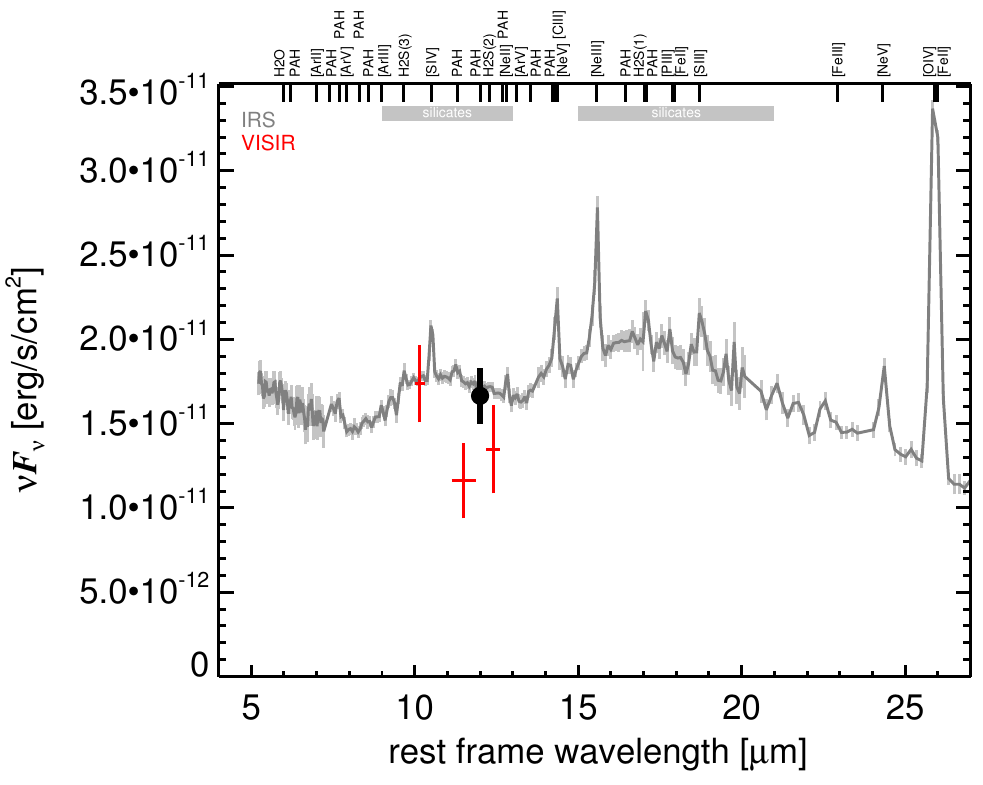}
   \caption{\label{fig:MISED_H1143-182}
      MIR SED of H\,1143-182. The description  of the symbols (if present) is the following.
      Grey crosses and  solid lines mark the \spitzer/IRAC, MIPS and IRS data. 
      The colour coding of the other symbols is: 
      green for COMICS, magenta for Michelle, blue for T-ReCS and red for VISIR data.
      Darker-coloured solid lines mark spectra of the corresponding instrument.
      The black filled circles mark the nuclear 12 and $18\,\mu$m  continuum emission estimate from the data.
      The ticks on the top axis mark positions of common MIR emission lines, while the light grey horizontal bars mark wavelength ranges affected by the silicate 10 and 18$\mu$m features.}
\end{figure}
\clearpage

\twocolumn[\begin{@twocolumnfalse}  
\subsection{Hydra\,A -- 3C 218 -- MCG-2-24-7}\label{app:HydraA}
Hydra\,A is a FR\,I radio source coinciding with the early-type  galaxy MCG-2-24-7 at a redshift of $z=$ 0.0549 ($D \sim240$\,Mpc) with a LINER nucleus \citep{veron-cetty_catalogue_2010}.
The first MIR detection of Hydra\,A was with \spitzer/MIPS \citep{shi_far-infrared_2005}.
Our photometric measurement of the MIPS $24\,\mu$m matches their published fluxes. 
It was also observed with \spitzer/IRAC and IRS.
In the IRAC $8\,\mu$m image, the elliptical host galaxy emission is weakly detected. 
In addition, the nuclear source appears elongated in east-south-east direction.
The low S/N IRS LR staring-mode spectrum shows silicate $10\,\mu$m absorption and PAH emission, i.e., star formation (see also \citealt{shi_9.7_2006}). 
In fact, \cite{leipski_spitzer_2009} argue that the MIR SED of this object can be explained with pure star formation and no AGN contribution. 
Hydra\,A remained undetected in the VISIR PAH1 observation and the derived flux upper limit is lower than, but still consistent with, the \spitzerr data. 
\newline\end{@twocolumnfalse}]

\begin{figure}
   \centering
   \includegraphics[angle=0,width=8.500cm]{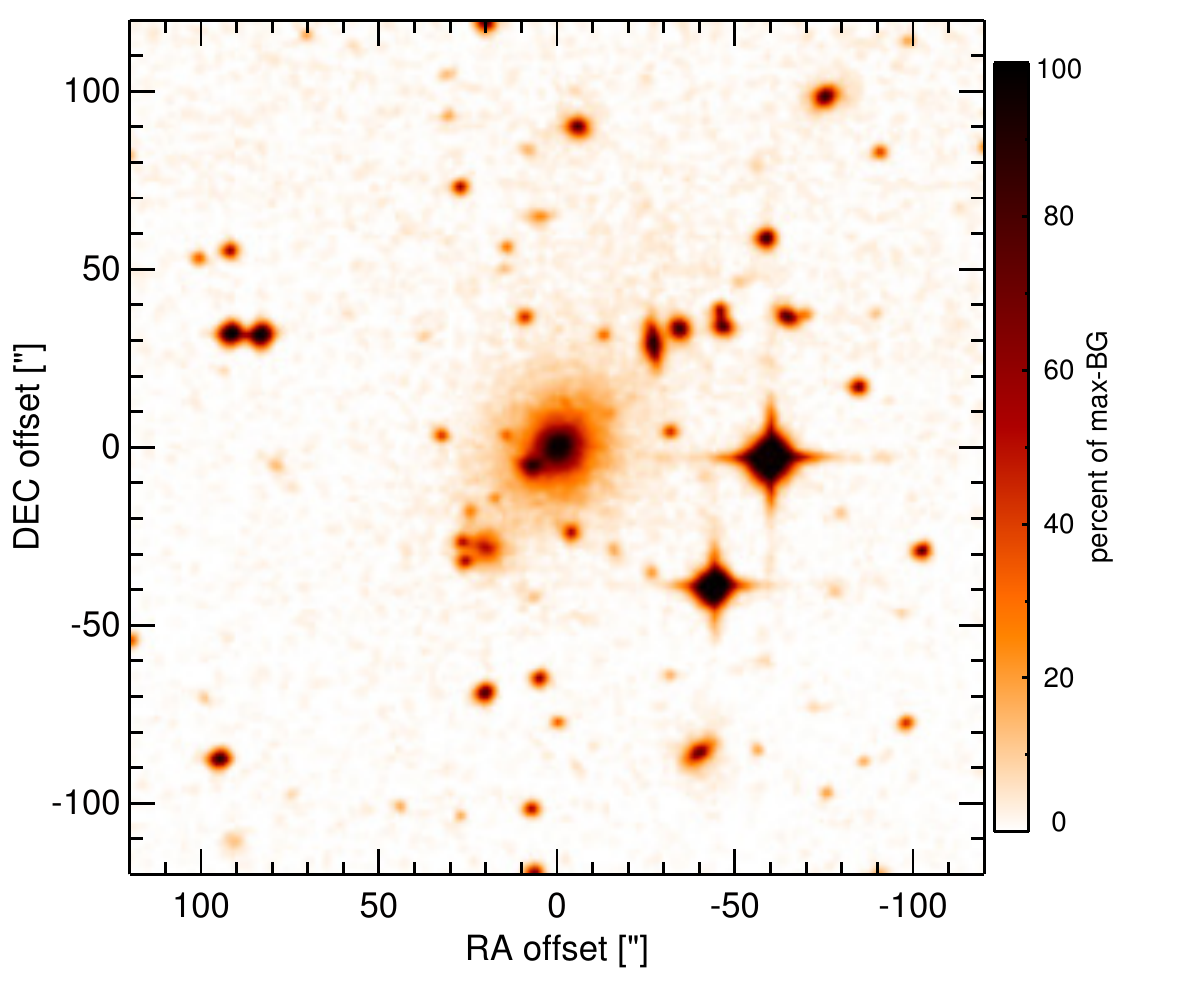}
    \caption{\label{fig:OPTim_HydraA}
             Optical image (DSS, red filter) of Hydra\,A. Displayed are the central $4\arcmin$ with North up and East to the left. 
              The colour scaling is linear with white corresponding to the median background and black to the $0.01\%$ pixels with the highest intensity.  
           }
\end{figure}
\begin{figure}
   \centering
   \includegraphics[angle=0,height=3.11cm]{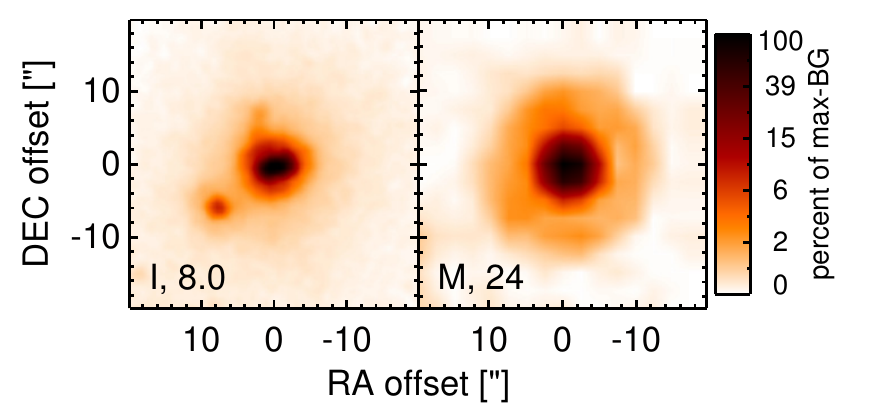}
    \caption{\label{fig:INTim_HydraA}
             \spitzerr MIR images of Hydra\,A. Displayed are the inner $40\arcsec$ with North up and East to the left. The colour scaling is logarithmic with white corresponding to median background and black to the $0.1\%$ pixels with the highest intensity.
             The label in the bottom left states instrument and central wavelength of the filter in $\mu$m (I: IRAC, M: MIPS). 
           }
\end{figure}
\begin{figure}
   \centering
   \includegraphics[angle=0,width=8.50cm]{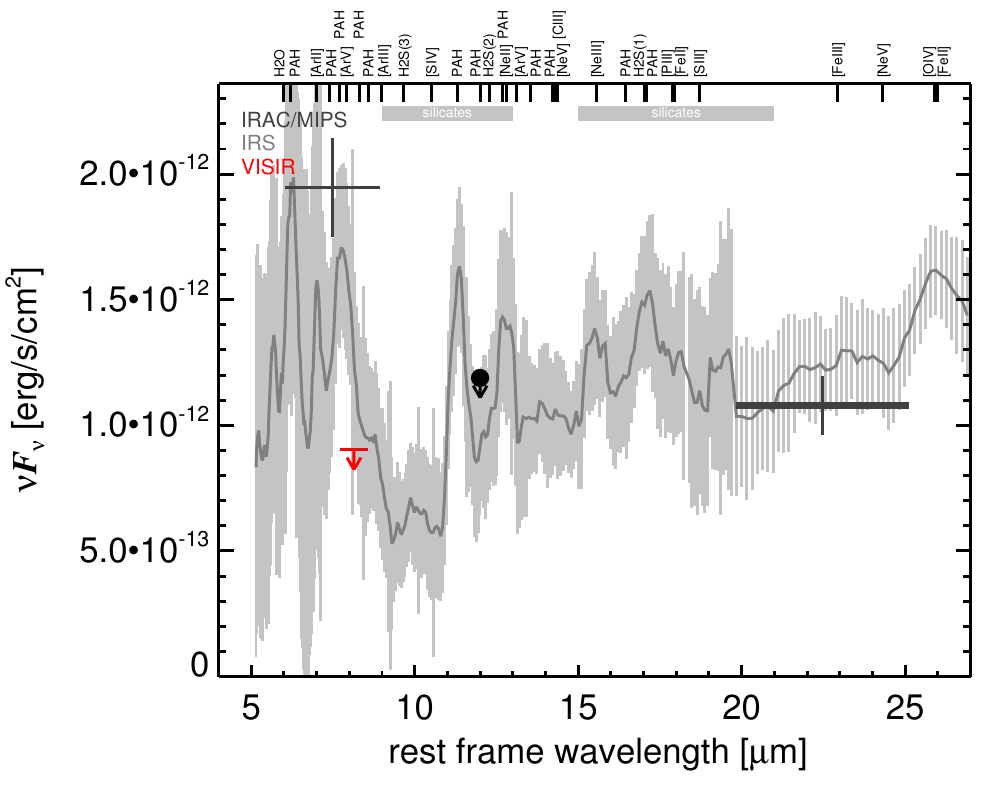}
   \caption{\label{fig:MISED_HydraA}
      MIR SED of Hydra\,A. The description  of the symbols (if present) is the following.
      Grey crosses and  solid lines mark the \spitzer/IRAC, MIPS and IRS data. 
      The colour coding of the other symbols is: 
      green for COMICS, magenta for Michelle, blue for T-ReCS and red for VISIR data.
      Darker-coloured solid lines mark spectra of the corresponding instrument.
      The black filled circles mark the nuclear 12 and $18\,\mu$m  continuum emission estimate from the data.
      The ticks on the top axis mark positions of common MIR emission lines, while the light grey horizontal bars mark wavelength ranges affected by the silicate 10 and 18$\mu$m features.}
\end{figure}
\clearpage

\twocolumn[\begin{@twocolumnfalse}  
\subsection{IC\,883 -- APG\,193 -- UGC\,8387}\label{app:IC0883}
IC\,883 is an infrared-luminous  galaxy with a peculiar morphology at a redshift of $z=$ 0.0233 ($D \sim 101\,$Mpc), which has an active nucleus optically classified either as LINER \citep{kim_optical_1995,veilleux_optical_1995} or as AGN/starburst composite \citep{yuan_role_2010}.
We treat this object as AGN/starburst composite, and because there is no strong evidence for an AGN from other wavelengths, we treat it as an uncertain AGN (but see below). 
IC\,883 was already studied at high angular resolution with Keck/LWS \citep{soifer_high-resolution_2001} in four $N$-band filters.
An elongated, complex MIR structure was detected (FWHM $1.8\arcsec \times 0.7\arcsec$, PA $135\degree$).
Only one of the two nuclei (with $\sim 1\arcsec$) separation is visible in the $N$-band.
IC\,883 was also observed with \spitzer/IRAC, IRS and MIPS. 
The same MIR morphology is indicated in the IRAC $5.8$ and $8\,\mu$m images, while the object appears rather compact in MIPS $24\,\mu$m as expected from the lower spatial resolution.
The 4\arcsec\, aperture fluxes from \cite{soifer_high-resolution_2001} match our \spitzerr spectrophotometry. 
The IRS LR staring-mode spectrum resembles a typical star-formation SED with strong PAH emission, silicate absorption and a steep rise in flux for wavelengths $> 20\,\mu$m (see also \citealt{vega_modelling_2008}).
\cite{imanishi_subaru_2011} observed IC\,883 with COMICS Q17.7 and detected an extended nucleus.
Because of the low S/N of this detection in our reanalysis of the COMICS data (2.3), we provide only an upper limit for the nuclear MIR emission. 
The absence of a clear MIR nucleus and the star-formation-like MIR SED cast  doubt on the AGN nature of IC\,883, although \nev  (see also \citealt{dudik_spitzer_2009}) and a compact radio source \citep{romero-canizales_e-merlin_2012} are detected.
\newline\end{@twocolumnfalse}]

\begin{figure}
   \centering
   \includegraphics[angle=0,width=8.500cm]{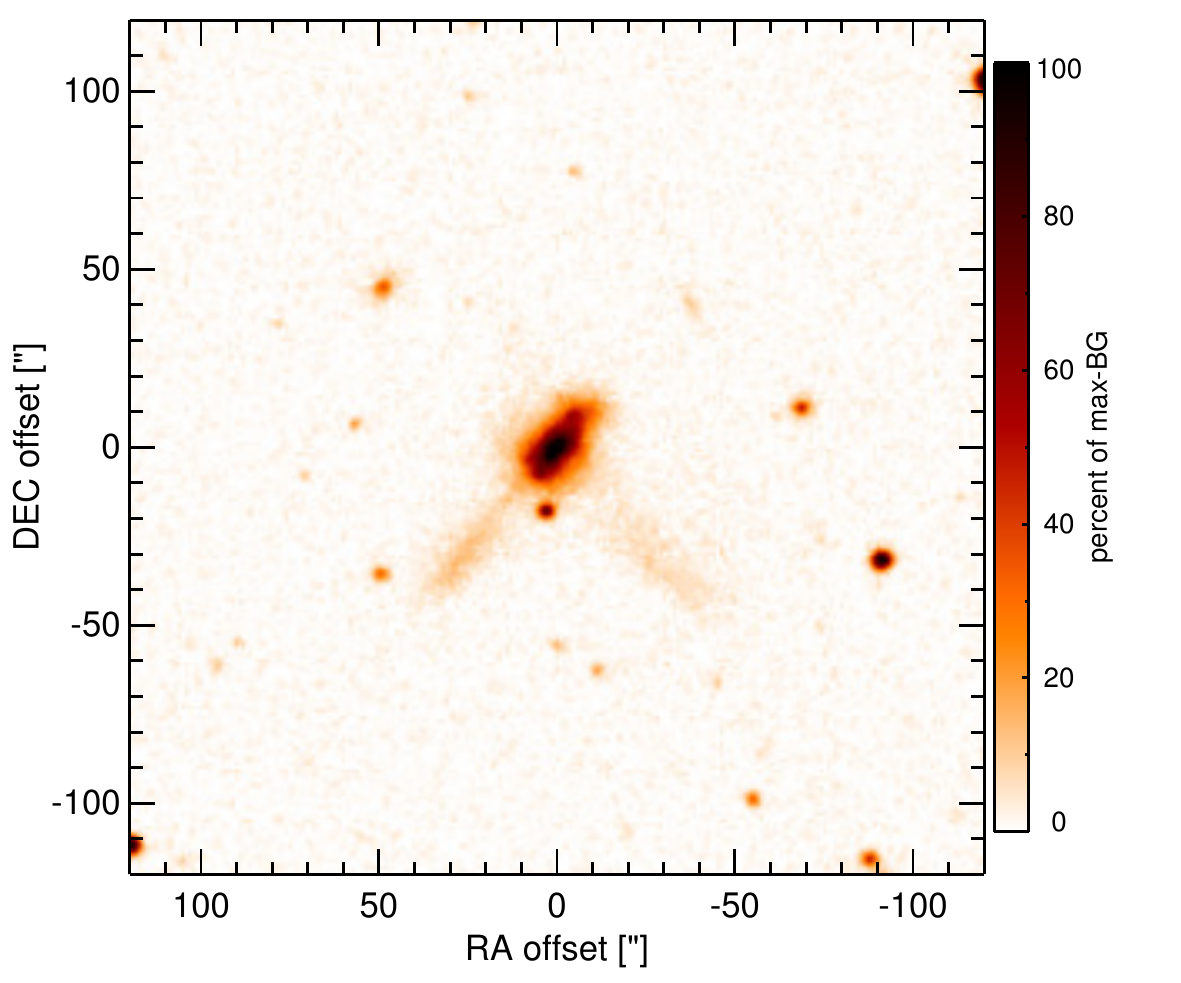}
    \caption{\label{fig:OPTim_IC0883}
             Optical image (DSS, red filter) of IC\,883. Displayed are the central $4\arcmin$ with North up and East to the left. 
              The colour scaling is linear with white corresponding to the median background and black to the $0.01\%$ pixels with the highest intensity.  
           }
\end{figure}
\begin{figure}
   \centering
   \includegraphics[angle=0,height=3.11cm]{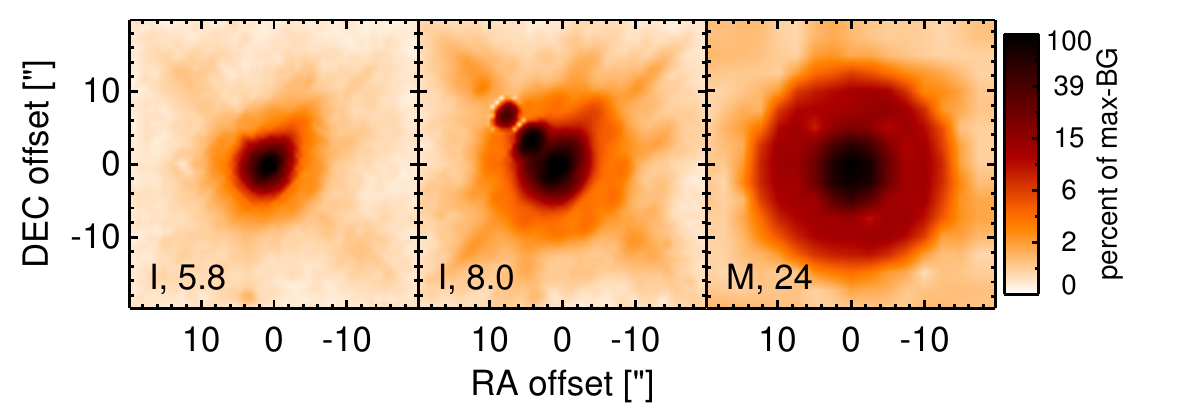}
    \caption{\label{fig:INTim_IC0883}
             \spitzerr MIR images of IC\,883. Displayed are the inner $40\arcsec$ with North up and East to the left. The colour scaling is logarithmic with white corresponding to median background and black to the $0.1\%$ pixels with the highest intensity.
             The label in the bottom left states instrument and central wavelength of the filter in $\mu$m (I: IRAC, M: MIPS). 
             Note that the apparent off-nuclear compact sources in the IRAC $8.0\,\mu$m image are instrumental artefacts.
           }
\end{figure}
\begin{figure}
   \centering
   \includegraphics[angle=0,width=8.50cm]{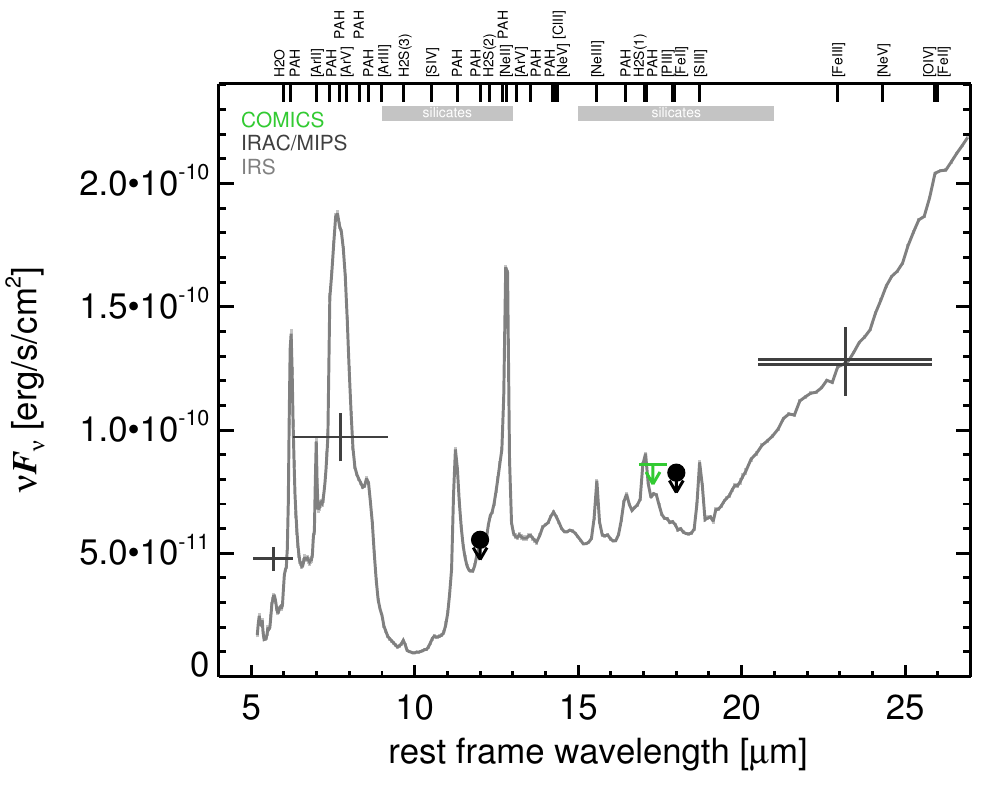}
   \caption{\label{fig:MISED_IC0883}
      MIR SED of IC\,883. The description  of the symbols (if present) is the following.
      Grey crosses and  solid lines mark the \spitzer/IRAC, MIPS and IRS data. 
      The colour coding of the other symbols is: 
      green for COMICS, magenta for Michelle, blue for T-ReCS and red for VISIR data.
      Darker-coloured solid lines mark spectra of the corresponding instrument.
      The black filled circles mark the nuclear 12 and $18\,\mu$m  continuum emission estimate from the data.
      The ticks on the top axis mark positions of common MIR emission lines, while the light grey horizontal bars mark wavelength ranges affected by the silicate 10 and 18$\mu$m features.}
\end{figure}
\clearpage

\twocolumn[\begin{@twocolumnfalse}  
\subsection{IC\,1459}\label{app:IC1459}
IC\,1459 is a giant elliptical galaxy at a distance of $D=$ 30.3\,Mpc ($z=0.006$; \citealt{blakeslee_synthesis_2001}) with a moderately radio-loud LINER nucleus \citep{veron-cetty_catalogue_2010}.
The detection of a compact flat-spectrum radio core verifies the AGN nature of IC\,1459 \citep{slee_parsecscale_1994}. 
Its stellar core is counter-rotating, which indicates a past merger \citep{franx_counterrotating_1988}, and hosts one of the most massive black holes of the nearby galaxy population ($\log M_\mathrm{BH}/M_\odot = 9 \pm 0.4$; \citealt{cappellari_counterrotating_2002}).
IC\,1459 was first observed in the $N$-band with IRTF in 1983 \citep{sparks_infrared_1986} and thereafter with \iras, \isoo \citep{temi_ages_2005} and \spitzer/IRAC, IRS and MIPS.
In the  IRAC $5.8$ and $8.0\,\mu$m and MIPS $24\,\mu$m images, an extended nuclear structure was detected with a PA of $\sim0\degree$.
Because we isolate the nuclear component in our photometric measurements, the resulting fluxes are significantly lower than the values published in \cite{temi_spitzer_2009} for MIPS $24\,\mu$m.
The IRS LR mapping-mode spectrum has a relatively low S/N but indicates silicate $10\,\mu$m and PAH $11.3\,\mu$m emission and an extremely blue spectral slope at short wavelengths in $\nu F_\nu$-space.
Thus, no prominent AGN contribution is evident in the MIR spectrum of IC\,1459, which instead resembles more the SED of an old stellar population (compare e.g., \citealt{panuzzo_nearby_2011}).
We observed IC\,1459 with VISIR in two narrow $N$-band filters in 2009 \citep{asmus_mid-infrared_2011} but did not detect any nuclear MIR emission.
This is also the case for an additional, to our knowledge, unpublished VISIR observation from the same year.
The derived upper limits are significantly below the \spitzerr spectrophotometric flux levels in agreement with the latter arising from extended host emission.
\newline\end{@twocolumnfalse}]

\begin{figure}
   \centering
   \includegraphics[angle=0,width=8.500cm]{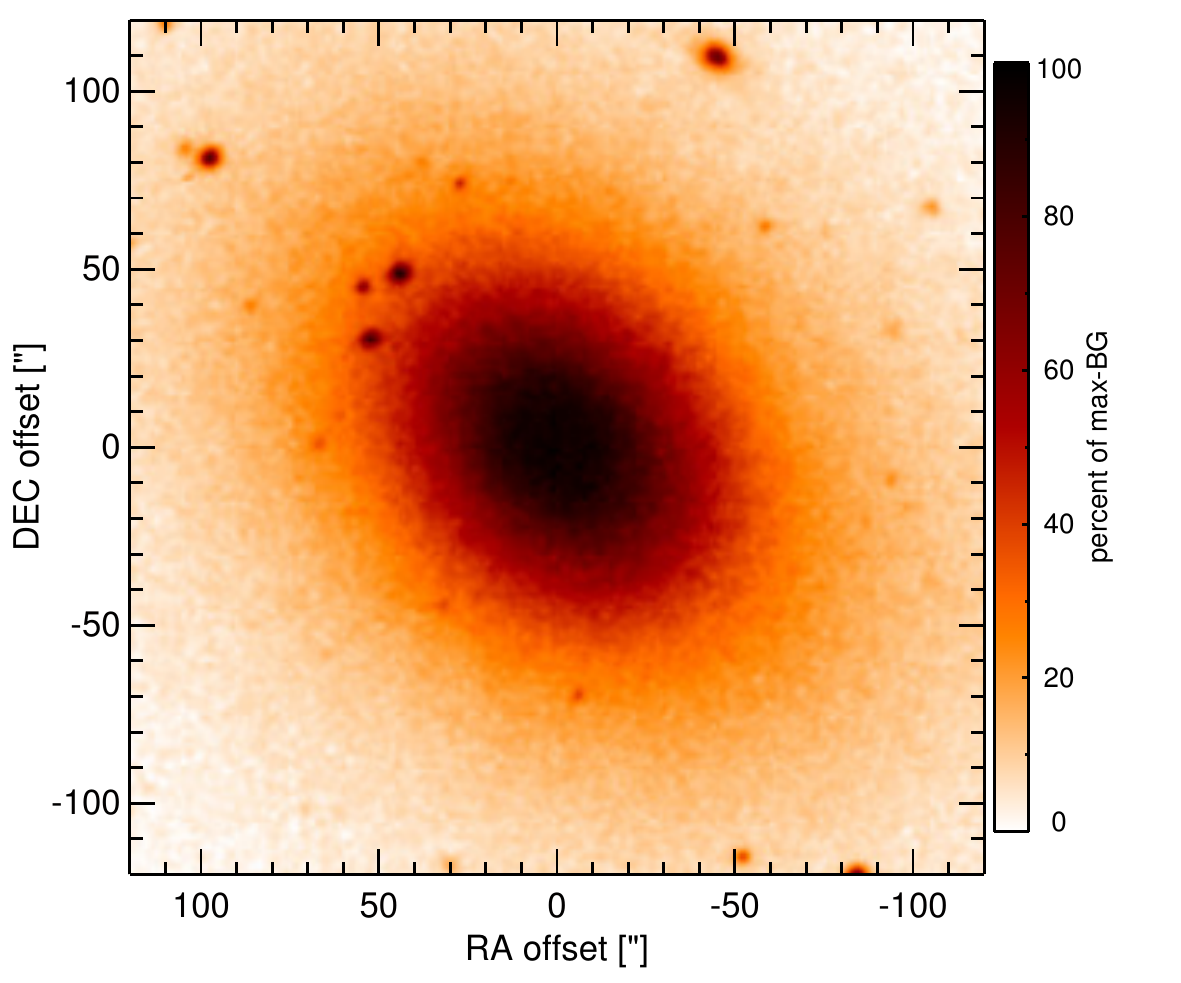}
    \caption{\label{fig:OPTim_IC1459}
             Optical image (DSS, red filter) of IC\,1459. Displayed are the central $4\arcmin$ with North up and East to the left. 
              The colour scaling is linear with white corresponding to the median background and black to the $0.01\%$ pixels with the highest intensity.  
           }
\end{figure}
\begin{figure}
   \centering
   \includegraphics[angle=0,height=3.11cm]{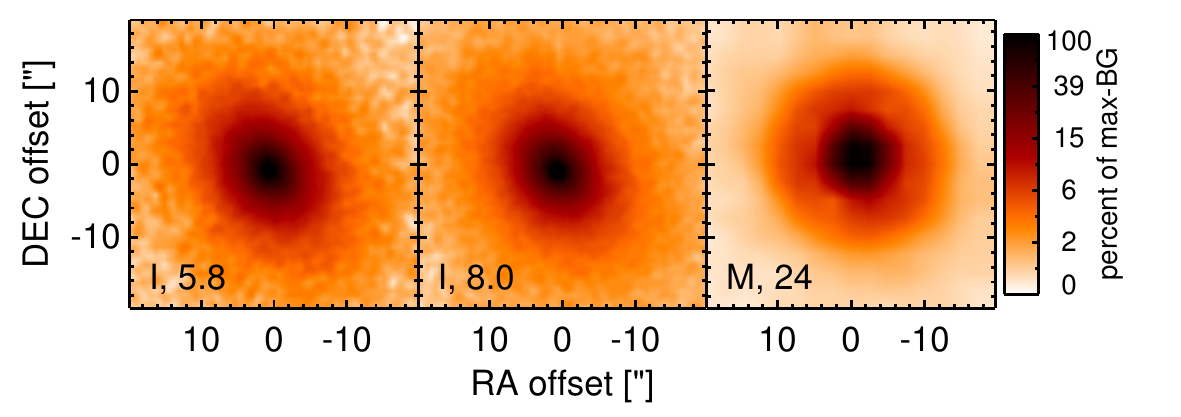}
    \caption{\label{fig:INTim_IC1459}
             \spitzerr MIR images of IC\,1459. Displayed are the inner $40\arcsec$ with North up and East to the left. The colour scaling is logarithmic with white corresponding to median background and black to the $0.1\%$ pixels with the highest intensity.
             The label in the bottom left states instrument and central wavelength of the filter in $\mu$m (I: IRAC, M: MIPS). 
           }
\end{figure}
\begin{figure}
   \centering
   \includegraphics[angle=0,width=8.50cm]{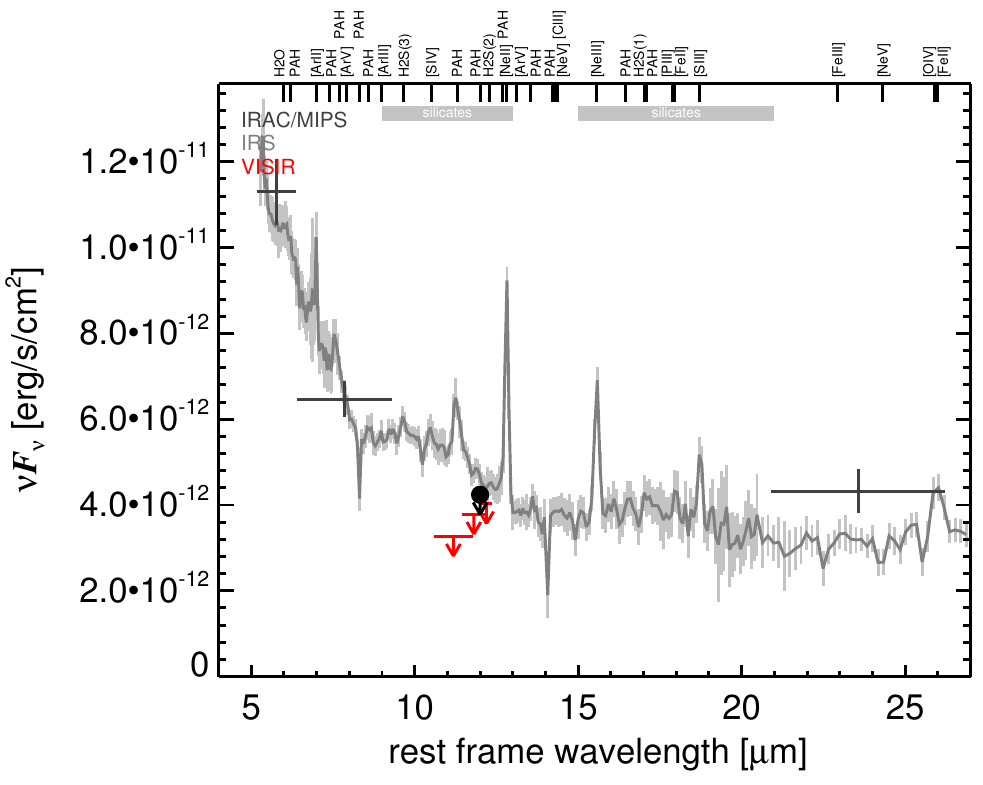}
   \caption{\label{fig:MISED_IC1459}
      MIR SED of IC\,1459. The description  of the symbols (if present) is the following.
      Grey crosses and  solid lines mark the \spitzer/IRAC, MIPS and IRS data. 
      The colour coding of the other symbols is: 
      green for COMICS, magenta for Michelle, blue for T-ReCS and red for VISIR data.
      Darker-coloured solid lines mark spectra of the corresponding instrument.
      The black filled circles mark the nuclear 12 and $18\,\mu$m  continuum emission estimate from the data.
      The ticks on the top axis mark positions of common MIR emission lines, while the light grey horizontal bars mark wavelength ranges affected by the silicate 10 and 18$\mu$m features.}
\end{figure}
\clearpage

\twocolumn[\begin{@twocolumnfalse}  
\subsection{IC\,3639 -- TOL\,1238-364 -- Fairall\,312 -- ESO\,381-8}\label{app:IC3639}
IC\,3639 is a face-on spiral galaxy at a redshift of $z=$ 0.0109 ($D\sim49.5$\,Mpc) with an AGN and a recent nuclear starburst \citep{gonzalez_delgado_ultraviolet-optical_1998,gonzalez_delgado_nuclear_2001}.
The AGN is optically classified as a Sy\,2 \citep{kewley_optical_2001} with polarized broad lines being detected \citep{heisler_visibility_1997}.
According to \cite{yuan_role_2010}, the AGN is dominating the emission, and we do not treat IC\,3639 as AGN/starburst composite.
It was observed with \spitzer/IRAC, IRS and MIPS and shows a clear spiral-like emission structure in the IRAC $5.8$ and $8.0\,\mu$m and MIPS $24\,\mu$m images with decreasing prominence, which is dominated by a central marginally resolved source. 
While our IRAC $5.8\,\mu$m flux matches the value of \cite{gallimore_infrared_2010}, our IRAC $8.0\,\mu$m flux is significantly lower but matches the flux levels of the IRS LR mapping-mode spectrum. 
The latter displays a MIR SED with silicate $10\,\mu$m absorption and PAH emission, indicating a star-formation contribution.
The spectral slope is rather red and peaks at $\sim25\,\mu$m in $\nu F_\nu$-space (see also \citealt{buchanan_spitzer_2006,tommasin_spitzer_2008,deo_mid-infrared_2009,wu_spitzer/irs_2009}).
IC\,3639 was observed with T-ReCS in a broad $N$-band filter in 2004 \citep{videla_nuclear_2013} and with VISIR in three narrow $N$-band filters in 2008 \citep{gandhi_resolving_2009}.
A compact MIR nucleus without any detection of extended host emission is visible in all images. 
The nucleus is possibly marginally resolved in the NEII\_2 filter image (FWHM $\sim 0.39\arcsec \sim 91\,$pc) but not in the others.
In particular in the PAH2 filter image, the nucleus is unresolved with a FWHM$<0.34\arcsec$. 
Thus, we classify IC\,3639 as unresolved in the MIR at subarcsecond resolution in general.
We obtain a flux that is ten times higher than \cite{videla_nuclear_2013} for the T-ReCS--detected nucleus, which is presumably caused by a typo in that paper. 
Our reanalysis of the PAH2 and NEII\_2 images yields flux values consistent with \cite{gandhi_resolving_2009}.
Our subarcsecond photometry agrees also with the \spitzerr spectrophotometry, indicating that at least part of the star formation must  occur in the projected central $\sim80$\,pc as already pointed out by \cite{gonzalez_delgado_ultraviolet-optical_1998}.
Interestingly, the nucleus of IC\,3639 was not detected during MIR interferometric observations with MIDI on a short baseline.
This indicates that the nuclear MIR emission is dominated by very extended structures at milliarcsecond scales \citep{burtscher_diversity_2013}.
\newline\end{@twocolumnfalse}]

\begin{figure}
   \centering
   \includegraphics[angle=0,width=8.500cm]{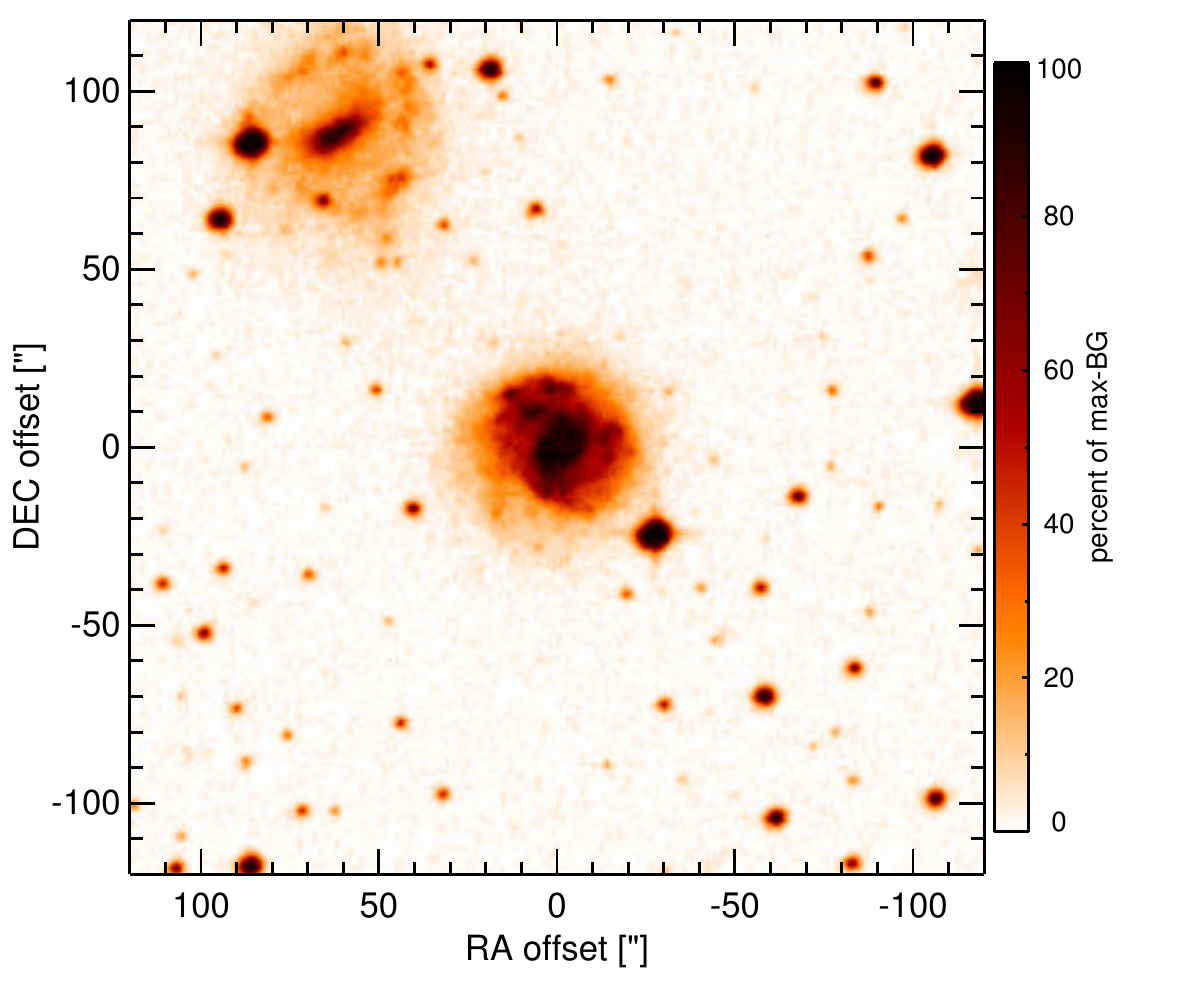}
    \caption{\label{fig:OPTim_IC3639}
             Optical image (DSS, red filter) of IC\,3639. Displayed are the central $4\arcmin$ with North up and East to the left. 
              The colour scaling is linear with white corresponding to the median background and black to the $0.01\%$ pixels with the highest intensity.  
           }
\end{figure}
\begin{figure}
   \centering
   \includegraphics[angle=0,height=3.11cm]{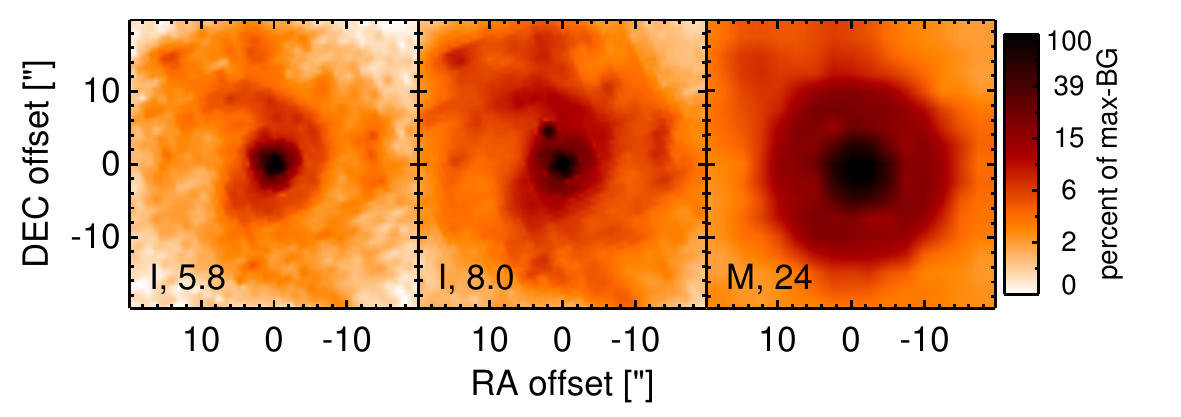}
    \caption{\label{fig:INTim_IC3639}
             \spitzerr MIR images of IC\,3639. Displayed are the inner $40\arcsec$ with North up and East to the left. The colour scaling is logarithmic with white corresponding to median background and black to the $0.1\%$ pixels with the highest intensity.
             The label in the bottom left states instrument and central wavelength of the filter in $\mu$m (I: IRAC, M: MIPS).\
             Note that the apparent off-nuclear compact source in the IRAC $8.0\,\mu$m image is an instrumental artefact.
           }
\end{figure}
\begin{figure}
   \centering
   \includegraphics[angle=0,width=8.500cm]{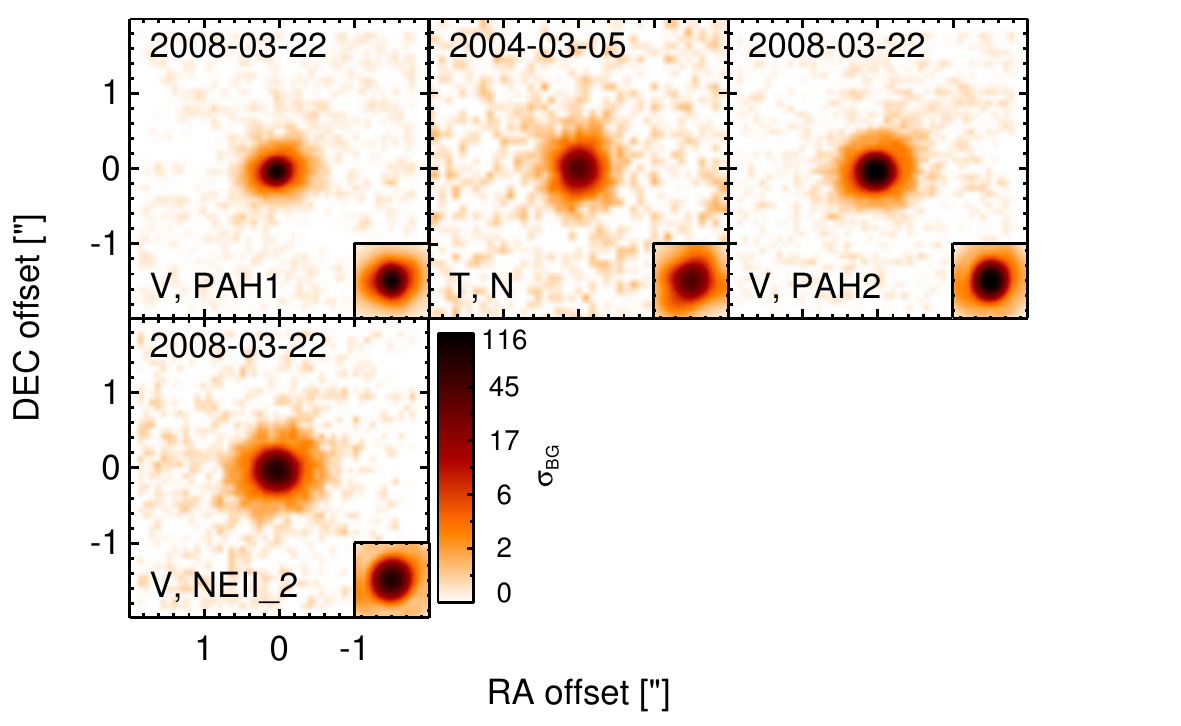}
    \caption{\label{fig:HARim_IC3639}
             Subarcsecond-resolution MIR images of IC\,3639 sorted by increasing filter wavelength. 
             Displayed are the inner $4\arcsec$ with North up and East to the left. 
             The colour scaling is logarithmic with white corresponding to median background and black to the $75\%$ of the highest intensity of all images in units of $\sigbg$.
             The inset image shows the central arcsecond of the PSF from the calibrator star, scaled to match the science target.
             The labels in the bottom left state instrument and filter names (C: COMICS, M: Michelle, T: T-ReCS, V: VISIR).
           }
\end{figure}
\begin{figure}
   \centering
   \includegraphics[angle=0,width=8.50cm]{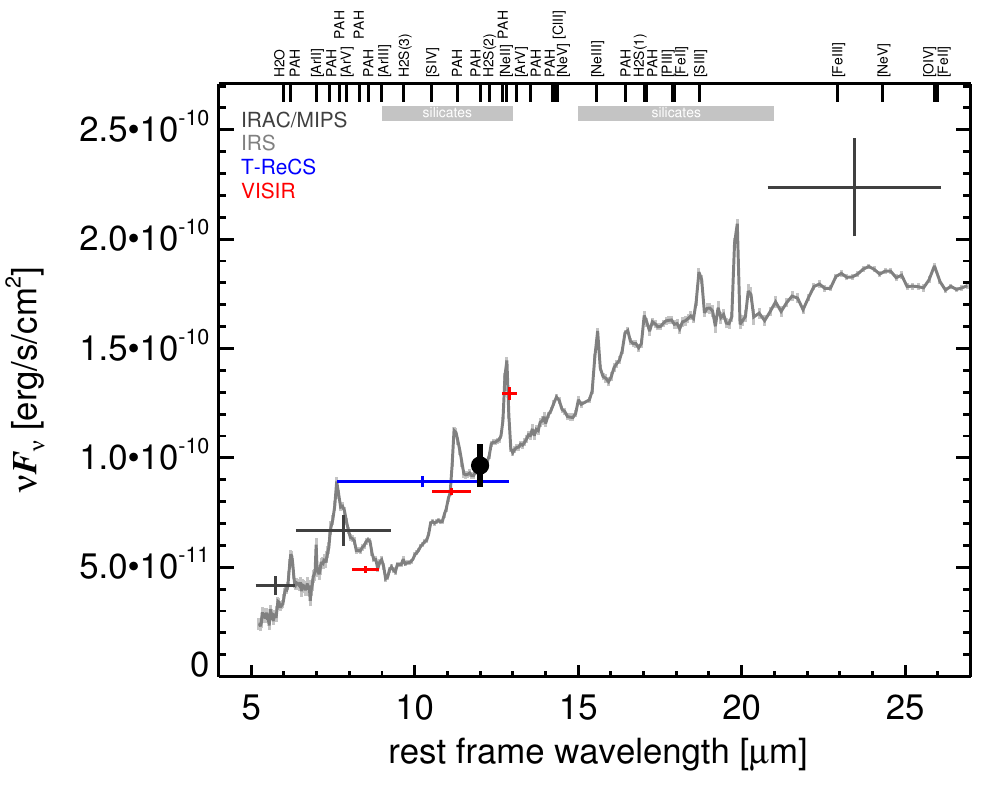}
   \caption{\label{fig:MISED_IC3639}
      MIR SED of IC\,3639. The description  of the symbols (if present) is the following.
      Grey crosses and  solid lines mark the \spitzer/IRAC, MIPS and IRS data. 
      The colour coding of the other symbols is: 
      green for COMICS, magenta for Michelle, blue for T-ReCS and red for VISIR data.
      Darker-coloured solid lines mark spectra of the corresponding instrument.
      The black filled circles mark the nuclear 12 and $18\,\mu$m  continuum emission estimate from the data.
      The ticks on the top axis mark positions of common MIR emission lines, while the light grey horizontal bars mark wavelength ranges affected by the silicate 10 and 18$\mu$m features.}
\end{figure}
\clearpage

\twocolumn[\begin{@twocolumnfalse}  
\subsection{IC\,4329A}\label{app:IC4329A}
IC\,4329A is an edge-on spiral galaxy at a redshift of $z=$ 0.0161 ($D\sim 70.6$\,Mpc) with a Sy\,1.2 nucleus \citep{veron-cetty_catalogue_2010} that belongs to the nine-month BAT AGN sample.
According to its high luminosity, IC\,4329A is the nearest quasar-like object \citep{wilson_ic_1979}.
The nucleus is partly obscured by a prominent dust lane, but the X-ray properties indicate additional absorption closer to the AGN \citep{steenbrugge_xmm-newton_2005}.
Pioneering MIR observations of IC\,4329A were performed by \cite{kleinmann_10-micron_1974}, \cite{rieke_infrared_1978}, \cite{roche_8-13_1984} and \cite{ward_continuum_1987}.
Apart from \iras, it was also observed with \iso/ISOCAM but appeared basically unresolved \citep{ramos_almeida_mid-infrared_2007}.
The first sub-arcsecond resolution images were made with Palomar 5\,m/MIRLIN in 2000 \citep{gorjian_10_2004} and with the ESO 3.6\,m/TIMMI2 in 2002 \citep{siebenmorgen_mid-infrared_2004}.
An unresolved MIR nucleus was detected.
IC\,4329A was also observed with \spitzer/IRAC, IRS and MIPS.
The unresolved MIR nucleus is saturated in the IRAC $5.8$ and $8.0\,\mu$m PBCD images, which therefore are not analysed (but see \citealt{gallimore_infrared_2010}).  
The IRS LR staring-mode spectrum displays weak silicate 10 and $18\,\mu$m emission and peaks at $\sim17\,\mu$m in $\nu F_\nu$-space (see also \citealt{buchanan_spitzer_2006,shi_9.7_2006,wu_spitzer/irs_2009,tommasin_spitzer-irs_2010,gallimore_infrared_2010,mullaney_defining_2011}).
We observed IC\,4329A with VISIR in two narrow $N$-band filters and LR $N$-band spectroscopy in 2009 \citep{honig_dusty_2010-1} and in three additional filters in 2010. 
It was also observed with T-ReCS in the $N-$ and $Q$-bands in 2009 (unpublished, to our knowledge).
A compact but elongated MIR nucleus was consistently detected  in all images (FWHM(major axis) $\sim 0.4\arcsec \sim 130\,$pc; PA$\sim60\degree$).
The VISIR and T-ReCS spectrophotometry is, in general, consistent with the \spitzerr spectrophotometry. 
However, at the shortest wavelengths, the VISIR spectrum increasingly deviates from the other data towards lower flux levels.
In addition, the T-ReCS $Q$-band flux is significantly above the \spitzerr flux level.
Combining all the $N$-band measurements from 1974 until 2010, the $10\,\,\mu$m flux of IC\,4329A  has possibly varied by $\sim 20\%$.
Its nuclear MIR emission could be further resolved with MIDI interferometric observations and was modelled as two approximately equally bright components, one extending by $\sim20\,$pc, and one remaining unresolved ($\lesssim 2$\,pc; \citealt{burtscher_diversity_2013}).
\newline\end{@twocolumnfalse}]

\begin{figure}
   \centering
   \includegraphics[angle=0,width=8.500cm]{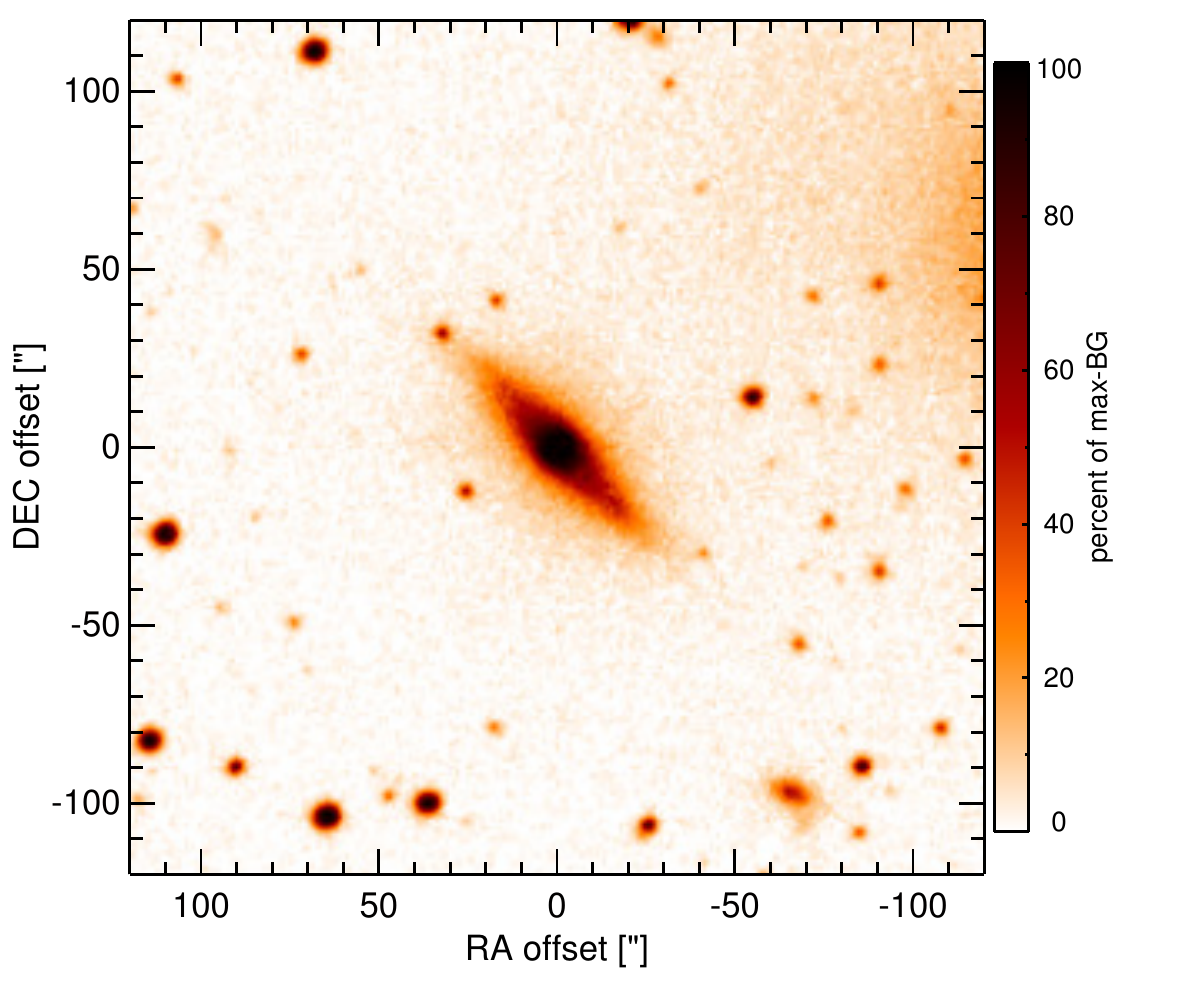}
    \caption{\label{fig:OPTim_IC4329A}
             Optical image (DSS, red filter) of IC\,4329A. Displayed are the central $4\arcmin$ with North up and East to the left. 
              The colour scaling is linear with white corresponding to the median background and black to the $0.01\%$ pixels with the highest intensity.  
           }
\end{figure}
\begin{figure}
   \centering
   \includegraphics[angle=0,height=3.11cm]{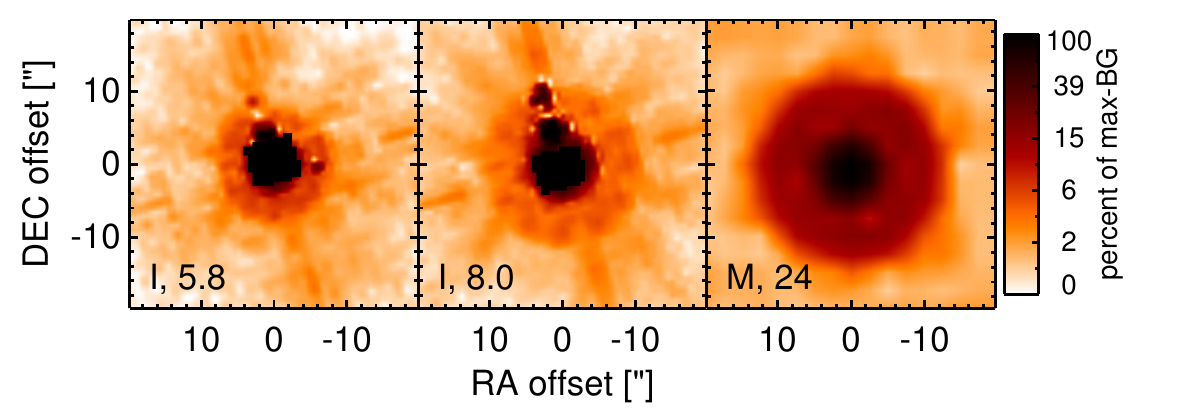}
    \caption{\label{fig:INTim_IC4329A}
             \spitzerr MIR images of IC\,4329A. Displayed are the inner $40\arcsec$ with North up and East to the left. The colour scaling is logarithmic with white corresponding to median background and black to the $0.1\%$ pixels with the highest intensity.
             The label in the bottom left states instrument and central wavelength of the filter in $\mu$m (I: IRAC, M: MIPS).
             Note that the apparent off-nuclear compact sources in the IRAC 5.8 and $8.0\,\mu$m images are instrumental artefacts.
           }
\end{figure}
\begin{figure}
   \centering
   \includegraphics[angle=0,width=8.500cm]{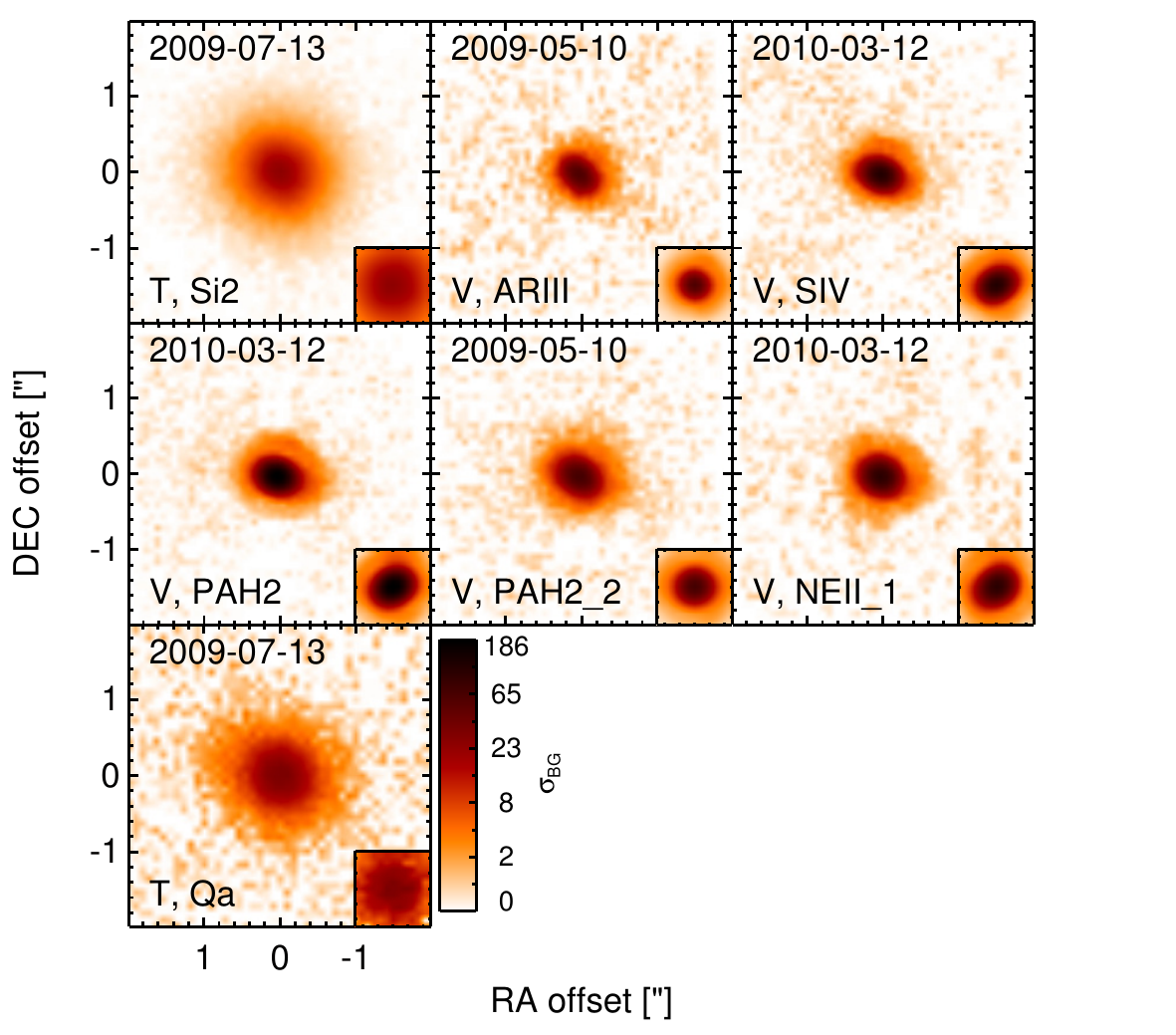}
    \caption{\label{fig:HARim_IC4329A}
             Subarcsecond-resolution MIR images of IC\,4329A sorted by increasing filter wavelength. 
             Displayed are the inner $4\arcsec$ with North up and East to the left. 
             The colour scaling is logarithmic with white corresponding to median background and black to the $75\%$ of the highest intensity of all images in units of $\sigbg$.
             The inset image shows the central arcsecond of the PSF from the calibrator star, scaled to match the science target.
             The labels in the bottom left state instrument and filter names (C: COMICS, M: Michelle, T: T-ReCS, V: VISIR).
           }
\end{figure}
\begin{figure}
   \centering
   \includegraphics[angle=0,width=8.50cm]{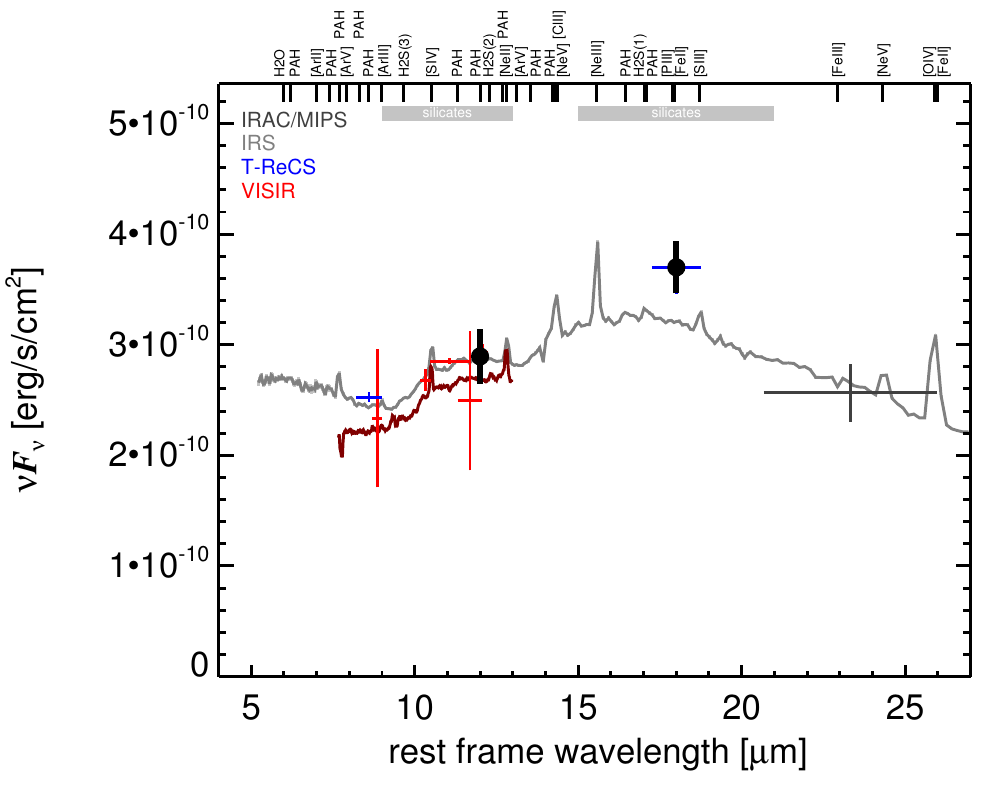}
   \caption{\label{fig:MISED_IC4329A}
      MIR SED of IC\,4329A. The description  of the symbols (if present) is the following.
      Grey crosses and  solid lines mark the \spitzer/IRAC, MIPS and IRS data. 
      The colour coding of the other symbols is: 
      green for COMICS, magenta for Michelle, blue for T-ReCS and red for VISIR data.
      Darker-coloured solid lines mark spectra of the corresponding instrument.
      The black filled circles mark the nuclear 12 and $18\,\mu$m  continuum emission estimate from the data.
      The ticks on the top axis mark positions of common MIR emission lines, while the light grey horizontal bars mark wavelength ranges affected by the silicate 10 and 18$\mu$m features.}
\end{figure}
\clearpage

\twocolumn[\begin{@twocolumnfalse}  
\subsection{IC\,4518W -- IC\,4518A -- MCG-7-31-1}\label{app:IC4518W}
IC\,4518W is the western galaxy of an interacting pair with a nuclear separation of $\sim 38\arcsec$ at a redshift of $z=$ 0.0163 ($D\sim70.2\,$Mpc) hosting a Sy\,2 nucleus \citep{veron-cetty_catalogue_2010}.
It was observed with \spitzer/IRAC, IRS and MIPS, where a compact MIR nucleus without any host emission was detected.
Note that IC\,4518E is also detected and extended in the IRAC and MIPS images.
The IRS LR staring-mode PBCD spectrum matches well the IRAC $5.8$ and $8.0\,\mu$m and MIPS $24\,\mu$m photometry and shows deep silicate $10\,\mu$m absorption (possibly also at $18\,\mu$m) and PAH emission with a steep red slope in $\nu F_\nu$-space.
The \spitzerr data thus contain   significant star-formation contribution.
IC\,4518W was observed with T-ReCS in the N filter in 2006 where a compact MIR nucleus possibly embedded in weak extended emission was detected \citep{alonso-herrero_high_2006}. 
Our reanalysis of the image yields a nuclear flux consistent with the value of \cite{alonso-herrero_high_2006}.
In addition, a LR T-ReCS spectrum was taken of IC\,4518W in 2006 \citep{diaz-santos_high_2010}, which shows a lower continuum flux level than the IRS spectrum without any PAH features but at least the same silicate $10\,\mu$m depth.
Therefore, the star formation contribution at subarcsecond resolution is probably minor in the nucleus of IC\,4518W (see also \citealt{alonso-herrero_local_2012}).
We use the T-ReCS spectrum to compute the nuclear 12$\,\mu$m continuum emission estimate corrected for the silicate absorption.
\newline\end{@twocolumnfalse}]

\begin{figure}
   \centering
   \includegraphics[angle=0,width=8.500cm]{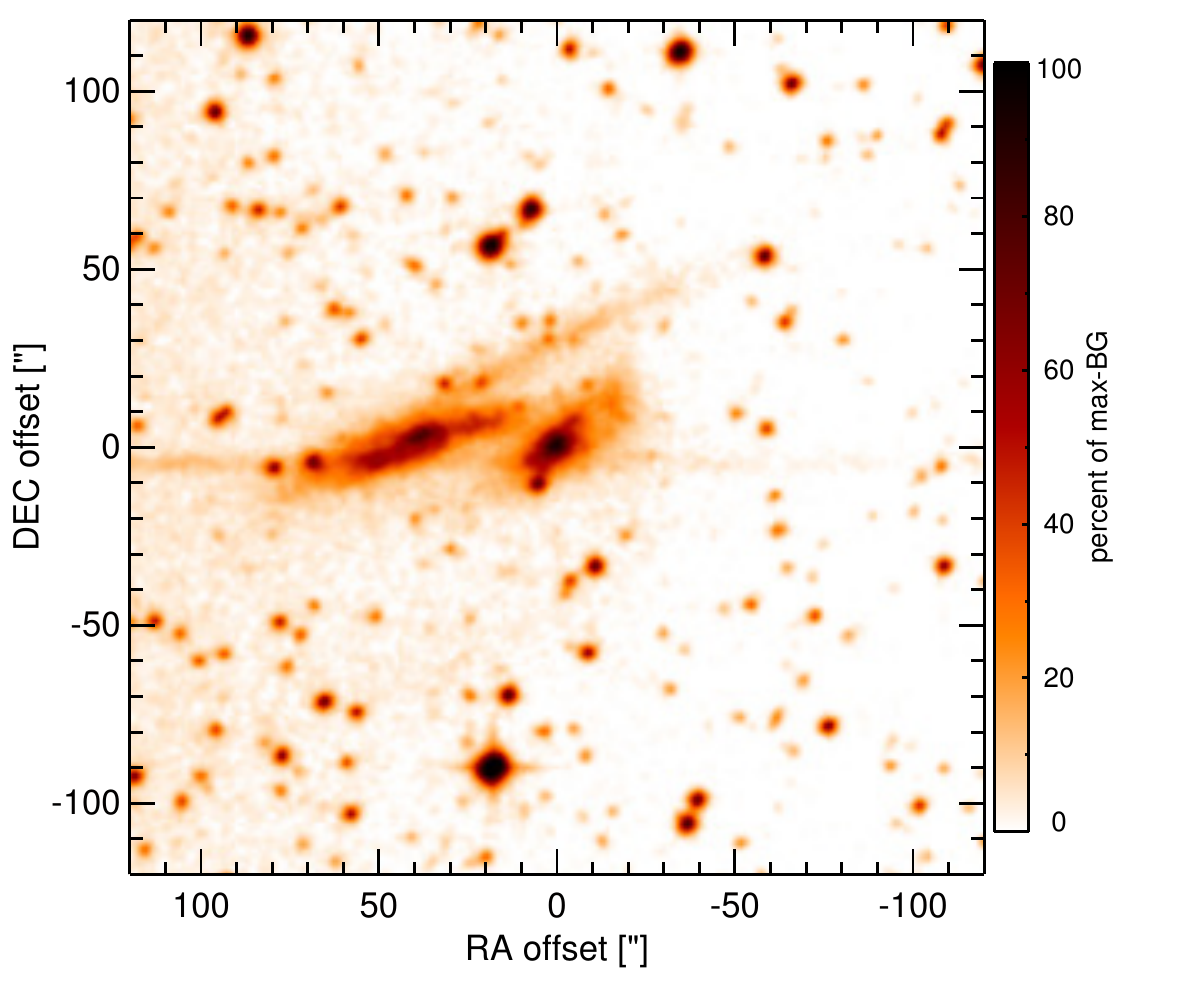}
    \caption{\label{fig:OPTim_IC4518W}
             Optical image (DSS, red filter) of IC\,4518W. Displayed are the central $4\arcmin$ with North up and East to the left. 
              The colour scaling is linear with white corresponding to the median background and black to the $0.01\%$ pixels with the highest intensity.  
           }
\end{figure}
\begin{figure}
   \centering
   \includegraphics[angle=0,height=3.11cm]{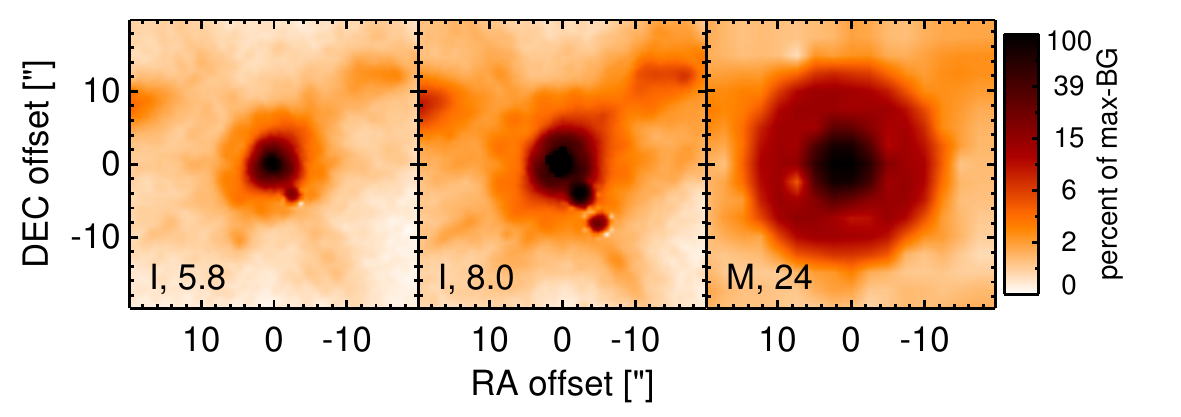}
    \caption{\label{fig:INTim_IC4518W}
             \spitzerr MIR images of IC\,4518W. Displayed are the inner $40\arcsec$ with North up and East to the left. The colour scaling is logarithmic with white corresponding to median background and black to the $0.1\%$ pixels with the highest intensity.
             The label in the bottom left states instrument and central wavelength of the filter in $\mu$m (I: IRAC, M: MIPS).
             Note that the apparent off-nuclear compact sources in the IRAC 5.8 and $8.0\,\mu$m images are instrumental artefacts.
           }
\end{figure}
\begin{figure}
   \centering
   \includegraphics[angle=0,height=3.11cm]{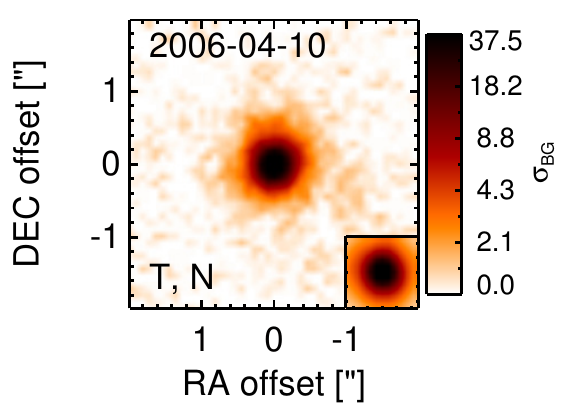}
    \caption{\label{fig:HARim_IC4518W}
             Subarcsecond-resolution MIR images of IC\,4518W sorted by increasing filter wavelength. 
             Displayed are the inner $4\arcsec$ with North up and East to the left. 
             The colour scaling is logarithmic with white corresponding to median background and black to the $75\%$ of the highest intensity of all images in units of $\sigbg$.
             The inset image shows the central arcsecond of the PSF from the calibrator star, scaled to match the science target.
             The labels in the bottom left state instrument and filter names (C: COMICS, M: Michelle, T: T-ReCS, V: VISIR).
           }
\end{figure}
\begin{figure}
   \centering
   \includegraphics[angle=0,width=8.50cm]{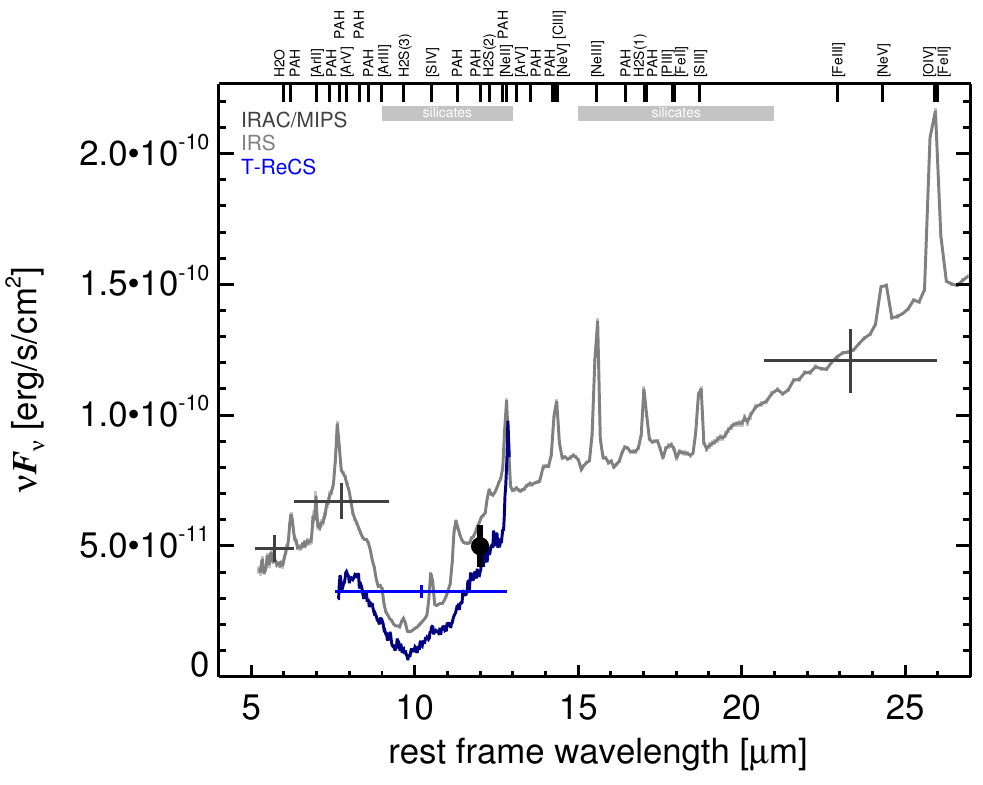}
   \caption{\label{fig:MISED_IC4518W}
      MIR SED of IC\,4518W. The description  of the symbols (if present) is the following.
      Grey crosses and  solid lines mark the \spitzer/IRAC, MIPS and IRS data. 
      The colour coding of the other symbols is: 
      green for COMICS, magenta for Michelle, blue for T-ReCS and red for VISIR data.
      Darker-coloured solid lines mark spectra of the corresponding instrument.
      The black filled circles mark the nuclear 12 and $18\,\mu$m  continuum emission estimate from the data.
      The ticks on the top axis mark positions of common MIR emission lines, while the light grey horizontal bars mark wavelength ranges affected by the silicate 10 and 18$\mu$m features.}
\end{figure}
\clearpage

\twocolumn[\begin{@twocolumnfalse}  
\subsection{IC\,5063 -- PKS\,2048-57 -- IRAS\,20481-5715}\label{app:IC5063}
IC\,5063 is a peculiar galaxy with both spiral and elliptical properties at a redshift of $z=$ 0.0113 ($D\sim45.3\,$Mpc) with a Sy\,2 nucleus \citep{kewley_optical_2001} with polarized broad lines being detected \citep{lumsden_spectropolarimetry_2004}.
It belongs to the nine-month BAT AGN sample, is luminous at radio wavelengths and possesses an extended NLR (PA$\sim-65\degree$; \citealt{morganti_radio_1998,morganti_radio_1999,schmitt_hubble_2003}).
\cite{colina_ic_1991} propose IC\,5063 to be the remnant of a recent merger, while \cite{martini_circumnuclear_2003} speculate that the nuclear obscuration might be caused by foreground dust lanes. 
After \iras, $N$-band observations of IC\,5063 were performed with the ESO 3.6\,m/TIMMI2 in 2002, where an unresolved MIR nucleus was detected \citep{siebenmorgen_mid-infrared_2004}.
It was also observed with \spitzer/IRAC, IRS and MIPS, where the MIR nucleus is nearly unresolved, and no further host emission was detected.
Note that the IRAC $8.0\,\mu$m PBCD image is partly saturated and thus is not used (but see \citealt{gallimore_infrared_2010}).
Our nuclear photometry of the IRAC $5.8\,\mu$m data agrees with \cite{gallimore_infrared_2010}, while our MIPS $24\,\mu$m flux is significantly higher than the value in \cite{temi_spitzer_2009} but agrees with the IRAC and IRS data.
The IRS LR staring-mode spectrum displays silicate $10\,\mu$m absorption and weak PAH emission.
The spectral slope is red but flattening towards longer wavelengths in $\nu F_\nu$-space (see also \citealt{wu_spitzer/irs_2009,tommasin_spitzer-irs_2010,gallimore_infrared_2010,mullaney_defining_2011}). 
IC\,5063 was observed with T-ReCS in imaging mode in three $N-$ and $Q$-band filters in 2004 and 2005 \citep{videla_nuclear_2013,ramos_almeida_infrared_2009}, and in LR spectroscopy mode in the $N$-band in 2005 \citep{young_spatially_2007,gonzalez-martin_dust_2013}.
We imaged IC\,5063 with VISIR in four narrow $N$-band filters in 2005 and 2006, and also took a LR $N$-band spectrum in 2006 \citep{horst_small_2006,horst_mid_2008,honig_dusty_2010-1}.
In all T-ReCS and VISIR images, a compact  but consistently elongated MIR nucleus (FWHM(major axis)$\sim 0.52\arcsec \sim 110\,$pc; PA$\sim 107\degree$) without any host further emission was detected.
The direction of this elongation coincides with the extended  \oiii emission (\citealt{schmitt_hubble_2003}; see also \citealt{honig_dusty_2010-1}).
Our Gaussian flux values of the subarcsecond photometry are consistent with \cite{horst_small_2006,horst_mid_2008} and \cite{videla_nuclear_2013}, while for the Si2 filter our PSF flux agrees with the one from \cite{ramos_almeida_infrared_2009}, only our Qa flux is significantly lower. 
The VISIR spectrum from \cite{honig_dusty_2010-1} extracted over 0.75\arcsec\, agrees with the \spitzerr spectrophotometry well, while the  nuclear subarcsecond PSF photometry is on average $\sim 25\%$ lower.
However, the T-ReCS spectrum from \cite{gonzalez-martin_dust_2013} has flux levels even lower than our photometry for unknown reasons. 
Taking into account all the subarcsecond MIR data of IC\,5063, the nuclear brightness might have varied up to $\sim25\%$ in the $N$-band from 2005 to 2006. 
\newline\end{@twocolumnfalse}]

\begin{figure}
   \centering
   \includegraphics[angle=0,width=8.500cm]{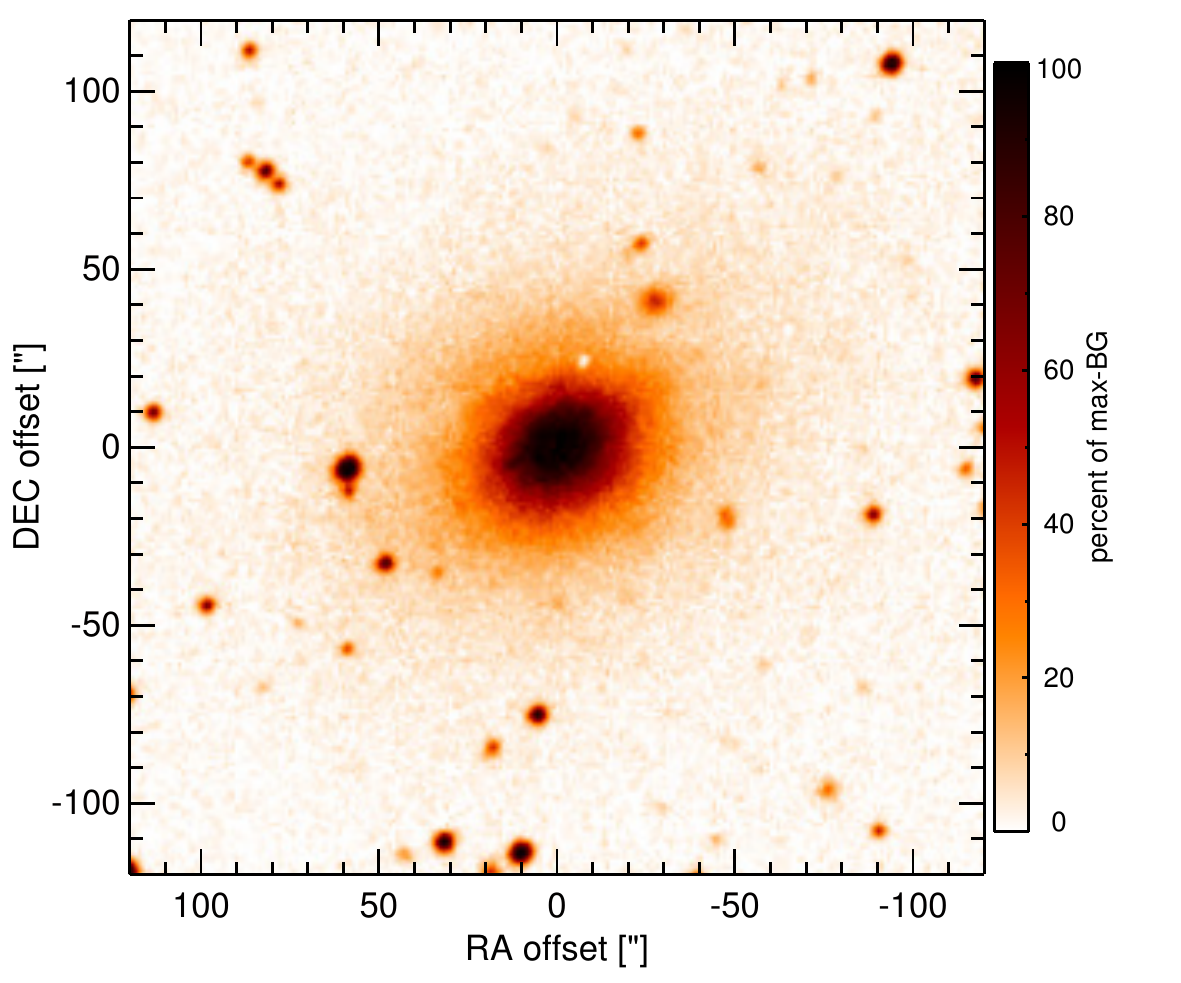}
    \caption{\label{fig:OPTim_IC5063}
             Optical image (DSS, red filter) of IC\,5063. Displayed are the central $4\arcmin$ with North up and East to the left. 
              The colour scaling is linear with white corresponding to the median background and black to the $0.01\%$ pixels with the highest intensity.  
           }
\end{figure}
\begin{figure}
   \centering
   \includegraphics[angle=0,height=3.11cm]{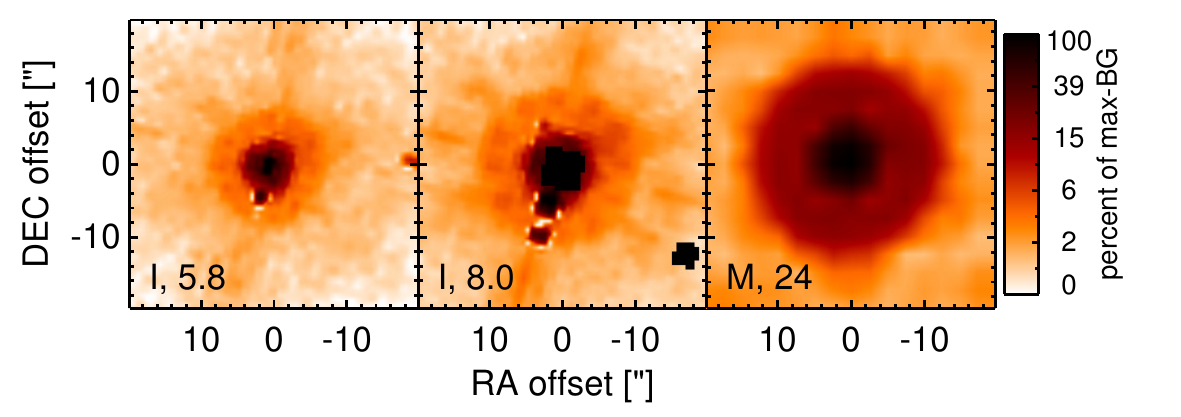}
    \caption{\label{fig:INTim_IC5063}
             \spitzerr MIR images of IC\,5063. Displayed are the inner $40\arcsec$ with North up and East to the left. The colour scaling is logarithmic with white corresponding to median background and black to the $0.1\%$ pixels with the highest intensity.
             The label in the bottom left states instrument and central wavelength of the filter in $\mu$m (I: IRAC, M: MIPS). 
             Note that the apparent off-nuclear compact sources in the IRAC 5.8 and $8.0\,\mu$m images are instrumental artefacts.
           }
\end{figure}
\begin{figure}
   \centering
   \includegraphics[angle=0,width=8.500cm]{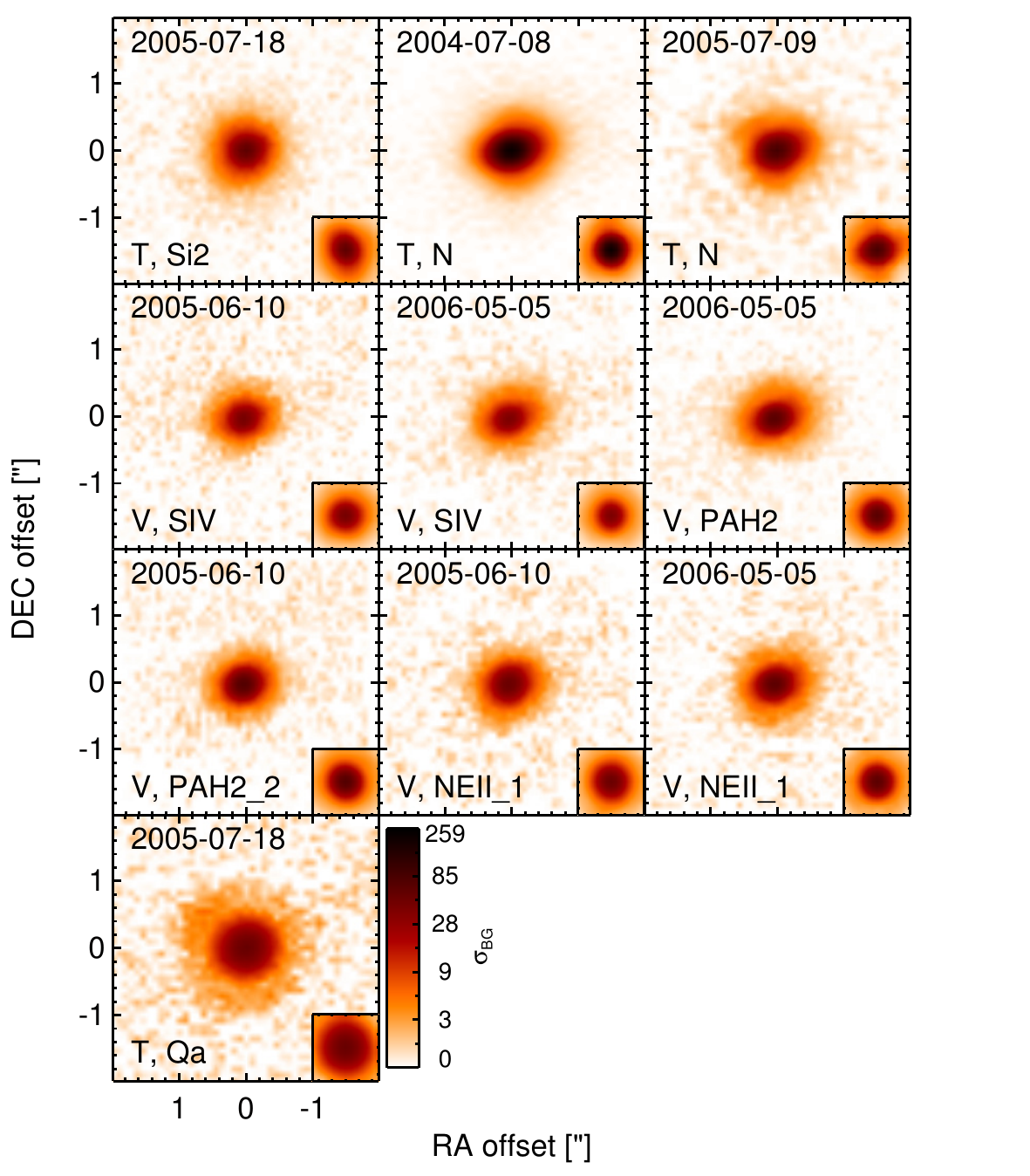}
    \caption{\label{fig:HARim_IC5063}
             Subarcsecond-resolution MIR images of IC\,5063 sorted by increasing filter wavelength. 
             Displayed are the inner $4\arcsec$ with North up and East to the left. 
             The colour scaling is logarithmic with white corresponding to median background and black to the $75\%$ of the highest intensity of all images in units of $\sigbg$.
             The inset image shows the central arcsecond of the PSF from the calibrator star, scaled to match the science target.
             The labels in the bottom left state instrument and filter names (C: COMICS, M: Michelle, T: T-ReCS, V: VISIR).
           }
\end{figure}
\begin{figure}
   \centering
   \includegraphics[angle=0,width=8.50cm]{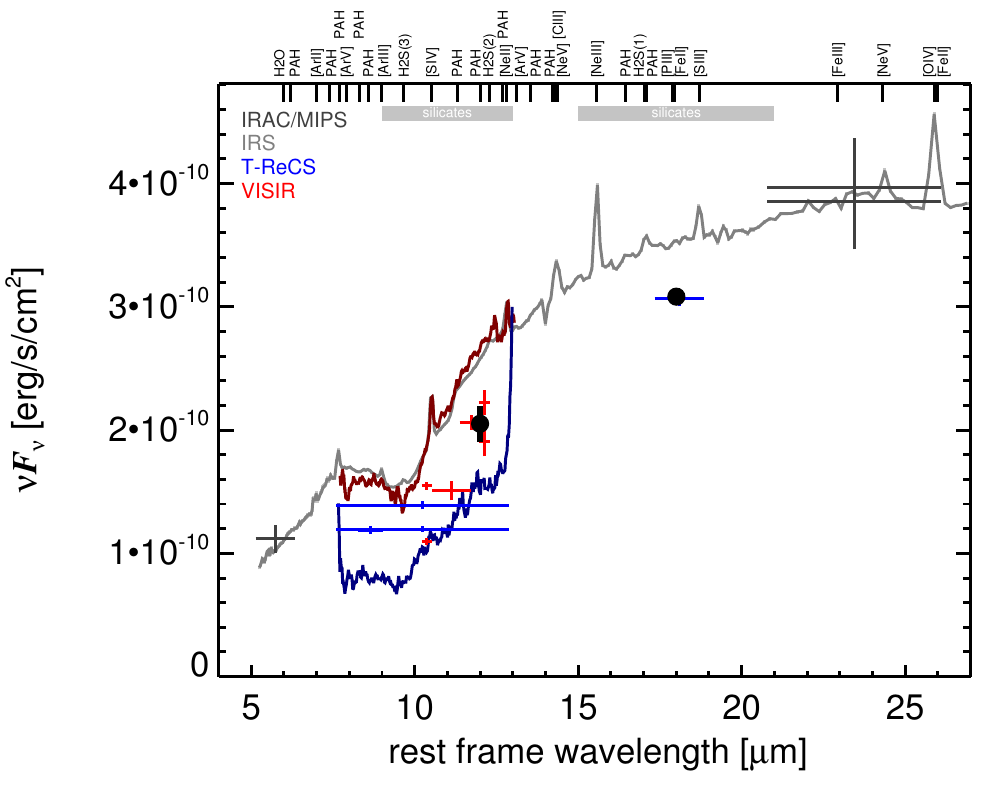}
   \caption{\label{fig:MISED_IC5063}
      MIR SED of IC\,5063. The description  of the symbols (if present) is the following.
      Grey crosses and  solid lines mark the \spitzer/IRAC, MIPS and IRS data. 
      The colour coding of the other symbols is: 
      green for COMICS, magenta for Michelle, blue for T-ReCS and red for VISIR data.
      Darker-coloured solid lines mark spectra of the corresponding instrument.
      The black filled circles mark the nuclear 12 and $18\,\mu$m  continuum emission estimate from the data.
      The ticks on the top axis mark positions of common MIR emission lines, while the light grey horizontal bars mark wavelength ranges affected by the silicate 10 and 18$\mu$m features.}
\end{figure}
\clearpage

\twocolumn[\begin{@twocolumnfalse}  
\subsection{III\,Zw\,35N -- IRAS\,01418+1651}\label{app:IIIZW035N}
III\,Zw\,35 is a close pair of galaxies (10\arcsec\ separation, PA$\sim20\degree$). 
The northern galaxy, III\,Zw\,35N, is an infrared-luminous  galaxy at a redshift of $z=$ 0.0277 ($D\sim121$\,Mpc) hosting an AGN, and a OH mega-maser \citep{chapman_oh_1986,chapman_combined_1990}.
The AGN is optically classified either as Sy\,2 \citep{veron-cetty_catalogue_2010} or  LINER \citep{veilleux_optical_1995} or as borderline Sy\,2/H\,II \citep{yuan_role_2010}.
Intense star formation appears to be present in III\,Zw\,35N \citep{pihlstrom_evn_2001,hattori_study_2004}, and thus we treat this object as AGN/starburst composite.
III\,Zw\,35 was observed with \spitzer/IRAC, IRS and MIPS and a marginally resolved nucleus was detected in all images.
The southern component is weakly visible as elongated emission in the IRAC $5.8$ and $8.0\,\mu$m images (nuclear separation$\sim8.5\arcsec\sim4.8\,$kpc, PA$\sim200\degree$).
Our IRAC and MIPS photometry for the northern nucleus matches the values published in \cite{u_spectral_2012}.
The IRS LR staring-mode spectrum shows deep silicate $10\,\mu$m and weak silicate $18\,\mu$m absorption, strong PAH emission and a red spectral slope in $\nu F_\nu$-space.
The MIR SED indicates a highly obscured nucleus with significant star formation on arcsecond scales (see also \citealt{willett_mid-infrared_2011}) as already suggested by \cite{chapman_combined_1990} based on different data.
III\,Zw\,35 was observed with VISIR in the PAH1 filter in 2009 (unpublished, to our knowledge).
A compact MIR nucleus was detected which appears elongated with an orientation roughly coinciding with the optical major axis (FWHM(major axis) $\sim 0.61\arcsec \sim 340\,$pc; PA$\sim30\degree$).
However, at least a second epoch of subarcsecond observations are needed to verify this extension. 
The southern component remains undetected in the image. 
The measured nuclear PAH1 flux is $44\%$ lower than the \spitzerr spectrophotometry and indicates that the PAH emission might be significantly lower in the inner $\sim 300$\,pc.
\newline\end{@twocolumnfalse}]

\begin{figure}
   \centering
   \includegraphics[angle=0,width=8.500cm]{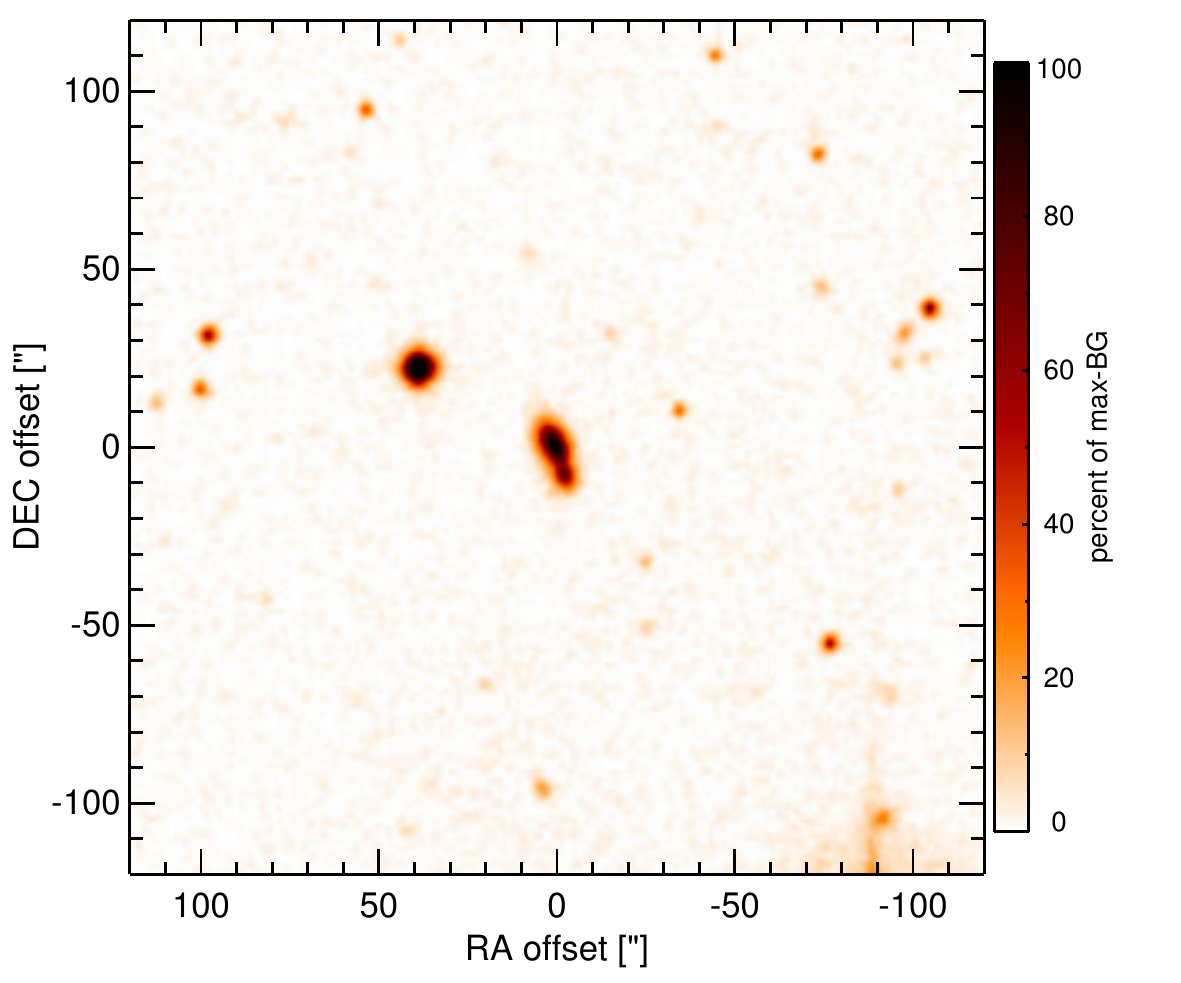}
    \caption{\label{fig:OPTim_IIIZW035N}
             Optical image (DSS, red filter) of III\,Zw\,35N. Displayed are the central $4\arcmin$ with North up and East to the left. 
              The colour scaling is linear with white corresponding to the median background and black to the $0.01\%$ pixels with the highest intensity.  
           }
\end{figure}
\begin{figure}
   \centering
   \includegraphics[angle=0,height=3.11cm]{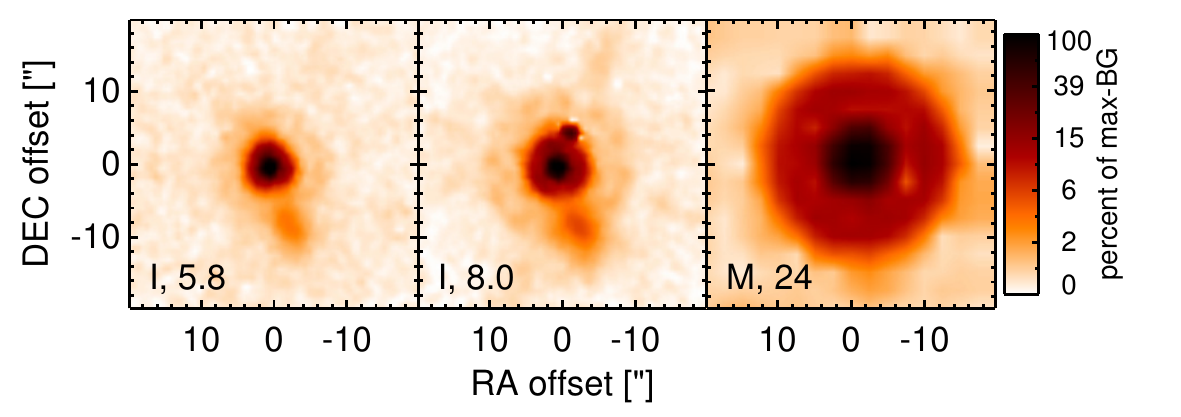}
    \caption{\label{fig:INTim_IIIZW035N}
             \spitzerr MIR images of III\,Zw\,35N. Displayed are the inner $40\arcsec$ with North up and East to the left. The colour scaling is logarithmic with white corresponding to median background and black to the $0.1\%$ pixels with the highest intensity.
             The label in the bottom left states instrument and central wavelength of the filter in $\mu$m (I: IRAC, M: MIPS).
             Note that the apparent off-nuclear compact source in the IRAC $8.0\,\mu$m image is an instrumental artefact.
           }
\end{figure}
\begin{figure}
   \centering
   \includegraphics[angle=0,height=3.11cm]{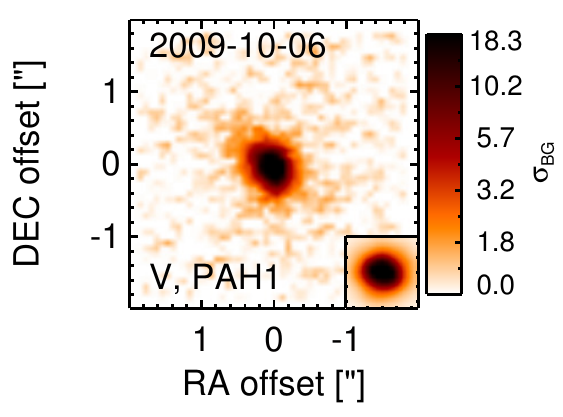}
    \caption{\label{fig:HARim_IIIZW035N}
             Subarcsecond-resolution MIR images of III\,Zw\,35N sorted by increasing filter wavelength. 
             Displayed are the inner $4\arcsec$ with North up and East to the left. 
             The colour scaling is logarithmic with white corresponding to median background and black to the $75\%$ of the highest intensity of all images in units of $\sigbg$.
             The inset image shows the central arcsecond of the PSF from the calibrator star, scaled to match the science target.
             The labels in the bottom left state instrument and filter names (C: COMICS, M: Michelle, T: T-ReCS, V: VISIR).
           }
\end{figure}
\begin{figure}
   \centering
   \includegraphics[angle=0,width=8.50cm]{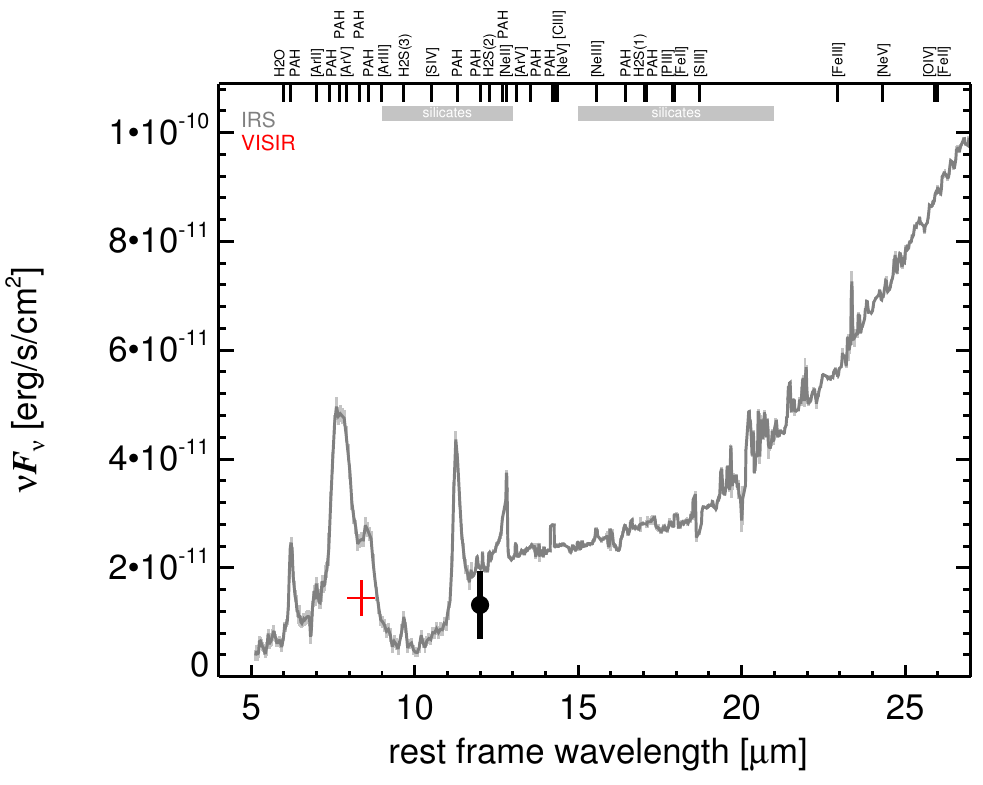}
   \caption{\label{fig:MISED_IIIZW035N}
      MIR SED of III\,Zw\,35N. The description  of the symbols (if present) is the following.
      Grey crosses and  solid lines mark the \spitzer/IRAC, MIPS and IRS data. 
      The colour coding of the other symbols is: 
      green for COMICS, magenta for Michelle, blue for T-ReCS and red for VISIR data.
      Darker-coloured solid lines mark spectra of the corresponding instrument.
      The black filled circles mark the nuclear 12 and $18\,\mu$m  continuum emission estimate from the data.
      The ticks on the top axis mark positions of common MIR emission lines, while the light grey horizontal bars mark wavelength ranges affected by the silicate 10 and 18$\mu$m features.}
\end{figure}
\clearpage

\twocolumn[\begin{@twocolumnfalse}  
\subsection{IRAS\,00188-0856 -- 2MASX\,J00212652-0839261}\label{app:IRAS00188-0856}
IRAS\,00188-0856 is an ultra-luminous infrared galaxy at a redshift of $z=$ 0.1284 ($D\sim574$\,Mpc) with an optical nuclear classification as a LINER \citep{kim_optical_1995,veilleux_optical_1995,veilleux_new_1999} or as borderline Sy\,2/starburst \citep{yuan_role_2010}.
Here, we treat it as AGN/starburst composite.
It was observed with \spitzer/IRAC and IRS and appears nearly unresolved in all corresponding images.
The IRS LR staring-mode spectrum shows extremely deep silicate  $10\,\mu$m absorption and  silicate $18\,\mu$m absorption with PAH emission and a steep red spectral slope in $\nu F_\nu$-space (see also \citealt{imanishi_spitzer_2007}).
The MIR SED thus indicates strong star formation but also a highly obscured active nucleus in agreement with the results of \cite{imanishi_spitzer_2007}.
IRAS\,00188-0856 was observed with T-ReCS in the Qa filter in 2009 and a compact MIR nucleus was weakly detected \citep{imanishi_subaru_2011}.
The low S/N prevents us from the extension analysis.
Our measured nuclear Qa flux is consistent with the value from \cite{imanishi_subaru_2011} and matches well the \spitzerr spectrophotometry.
Therefore, we use the IRS spectrum to compute 12 and 18$\,\mu$m continuum emission estimates, which are corrected for the silicate features.
\newline\end{@twocolumnfalse}]

\begin{figure}
   \centering
   \includegraphics[angle=0,width=8.500cm]{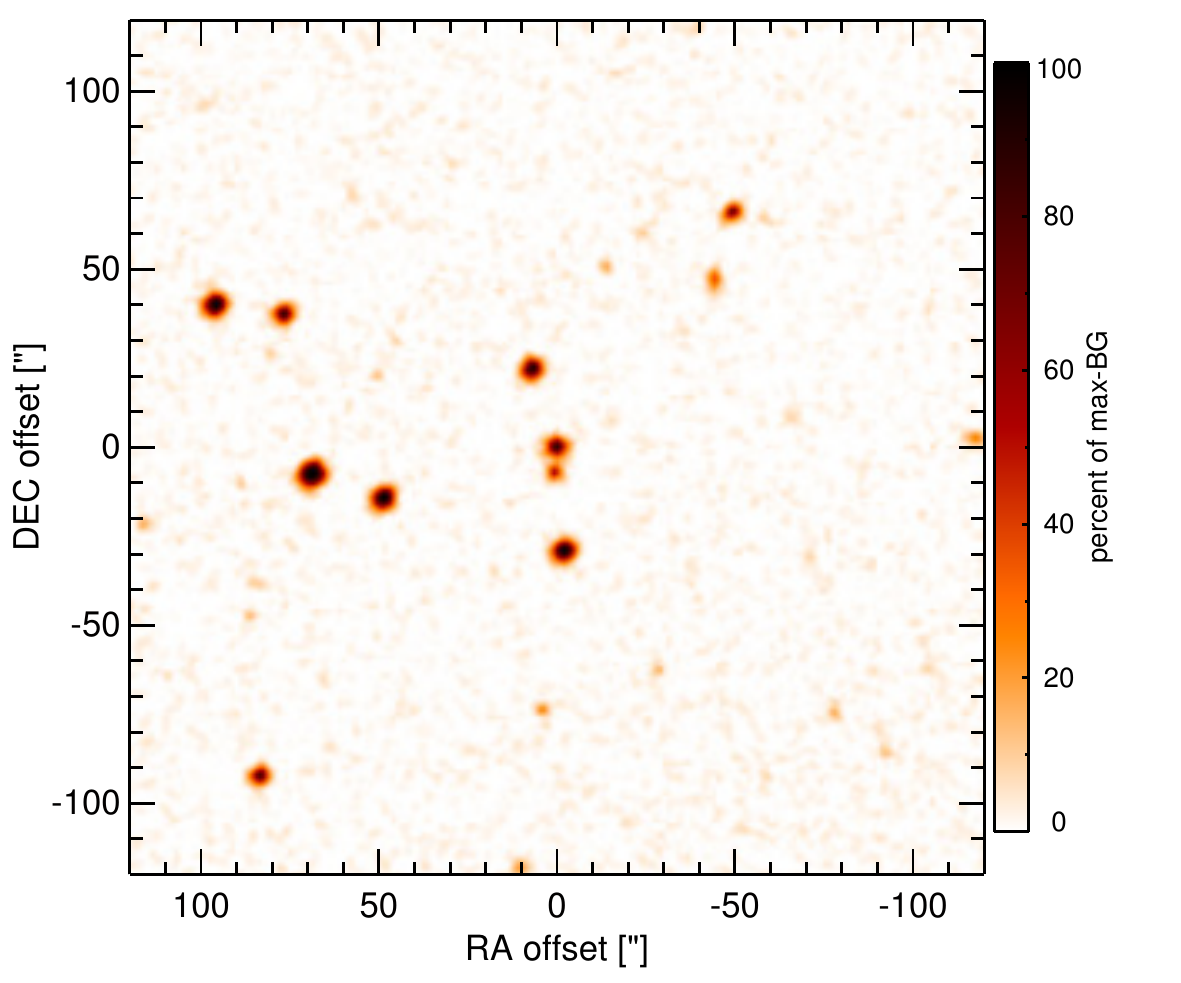}
    \caption{\label{fig:OPTim_IRAS00188-0856}
             Optical image (DSS, red filter) of IRAS\,00188-0856. Displayed are the central $4\arcmin$ with North up and East to the left. 
              The colour scaling is linear with white corresponding to the median background and black to the $0.01\%$ pixels with the highest intensity.  
           }
\end{figure}
\begin{figure}
   \centering
   \includegraphics[angle=0,height=3.11cm]{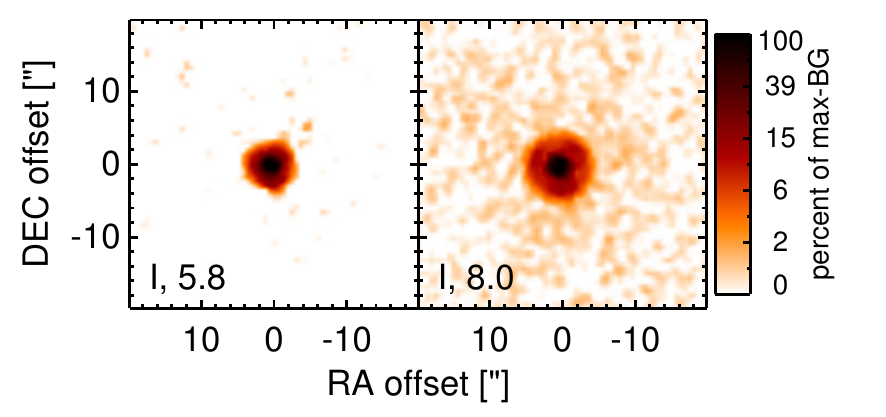}
    \caption{\label{fig:INTim_IRAS00188-0856}
             \spitzerr MIR images of IRAS\,00188-0856. Displayed are the inner $40\arcsec$ with North up and East to the left. The colour scaling is logarithmic with white corresponding to median background and black to the $0.1\%$ pixels with the highest intensity.
             The label in the bottom left states instrument and central wavelength of the filter in $\mu$m (I: IRAC, M: MIPS). 
           }
\end{figure}
\begin{figure}
   \centering
   \includegraphics[angle=0,height=3.11cm]{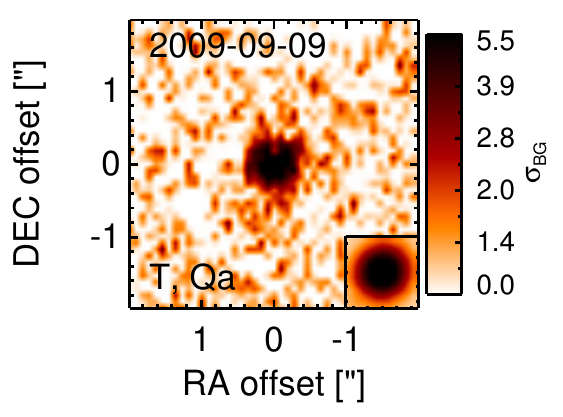}
    \caption{\label{fig:HARim_IRAS00188-0856}
             Subarcsecond-resolution MIR images of IRAS\,00188-0856 sorted by increasing filter wavelength. 
             Displayed are the inner $4\arcsec$ with North up and East to the left. 
             The colour scaling is logarithmic with white corresponding to median background and black to the $75\%$ of the highest intensity of all images in units of $\sigbg$.
             The inset image shows the central arcsecond of the PSF from the calibrator star, scaled to match the science target.
             The labels in the bottom left state instrument and filter names (C: COMICS, M: Michelle, T: T-ReCS, V: VISIR).
           }
\end{figure}
\begin{figure}
   \centering
   \includegraphics[angle=0,width=8.50cm]{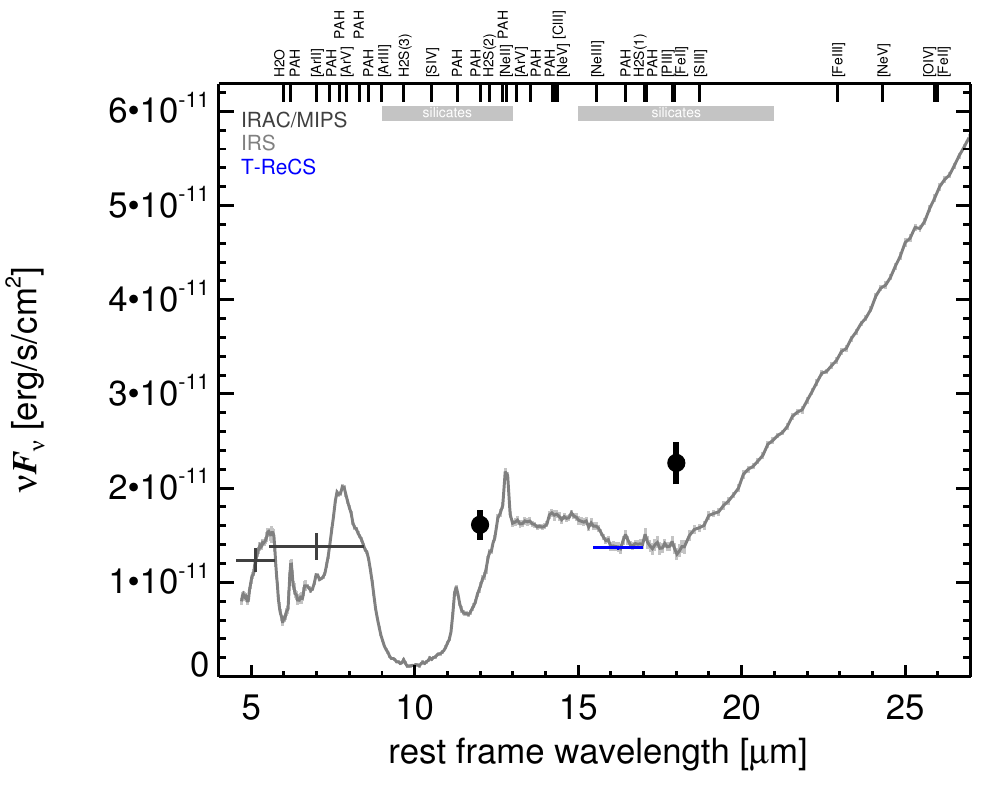}
   \caption{\label{fig:MISED_IRAS00188-0856}
      MIR SED of IRAS\,00188-0856. The description  of the symbols (if present) is the following.
      Grey crosses and  solid lines mark the \spitzer/IRAC, MIPS and IRS data. 
      The colour coding of the other symbols is: 
      green for COMICS, magenta for Michelle, blue for T-ReCS and red for VISIR data.
      Darker-coloured solid lines mark spectra of the corresponding instrument.
      The black filled circles mark the nuclear 12 and $18\,\mu$m  continuum emission estimate from the data.
      The ticks on the top axis mark positions of common MIR emission lines, while the light grey horizontal bars mark wavelength ranges affected by the silicate 10 and 18$\mu$m features.}
\end{figure}
\clearpage

\twocolumn[\begin{@twocolumnfalse}  
\subsection{IRAS\,01003-2238 -- LEDA\,3730}\label{app:IRAS01003-2238}
IRAS\,01003-2238 is classified as an ultra-luminous infrared galaxy and a Wolf-Rayet galaxy indicating a recent starburst \citep{armus_detection_1988}.
It has a redshift of $z=$ 0.1178 ($D\sim523$\,Mpc) and hosts a Sy\,2 nucleus, which appears to dominate over the starburst \citep{yuan_role_2010} .
IRAS\,01003-2238 was observed with \spitzer/IRAC and IRS and appears nearly unresolved in the IRAC $5.8\,\mu$m image.
The IRS LR staring-mode spectrum shows deep silicate  10 and $18\,\mu$m absorption and only weak PAH emission (see also \citealt{imanishi_spitzer_2007,farrah_high-resolution_2007}). 
The spectral slope is red in $\nu F_\nu$-space.
IRAS\,0103-2238 was observed with T-ReCS in the Qa filter in 2008 and a compact MIR nucleus was detected \citep{imanishi_subaru_2011}.
Our measured nuclear Qa flux is consistent with the value from \cite{imanishi_subaru_2011} but is interestingly $10\%$ higher than the \spitzerr spectrophotometry, i.e., basically at the level of the continuum without the silicate 18$\,\mu$m absorption.
This indicates that this absorption is possibly not caused in the inner kiloparsec of IRAS\,01003-2238.
For the 12$\,\mu$m continuum emission estimate, we use the IRS spectrum to correct for the silicate 10$\,\mu$m feature.
\newline\end{@twocolumnfalse}]

\begin{figure}
   \centering
   \includegraphics[angle=0,width=8.500cm]{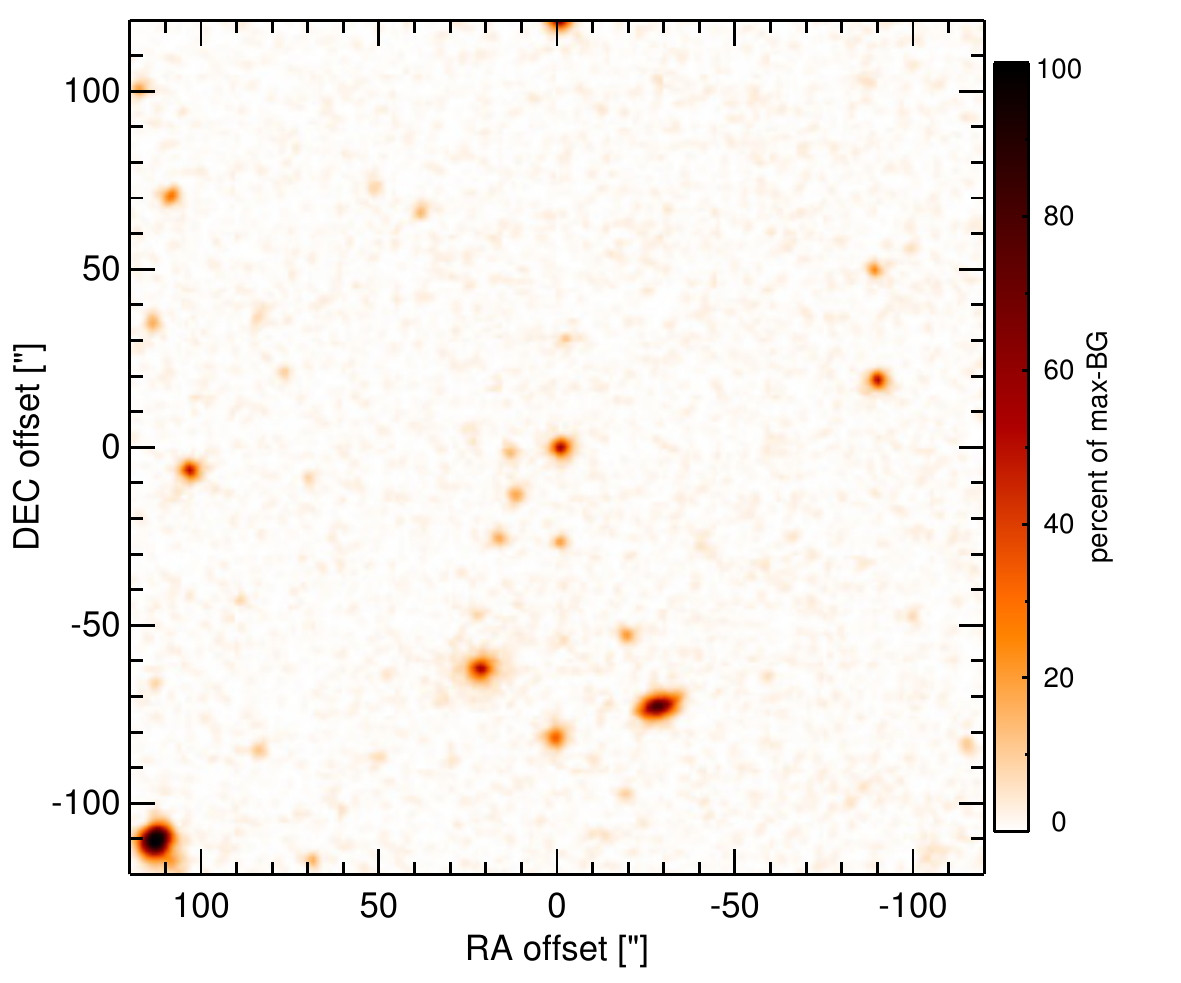}
    \caption{\label{fig:OPTim_IRAS01003-2238}
             Optical image (DSS, red filter) of IRAS\,01003-2238. Displayed are the central $4\arcmin$ with North up and East to the left. 
              The colour scaling is linear with white corresponding to the median background and black to the $0.01\%$ pixels with the highest intensity.  
           }
\end{figure}
\begin{figure}
   \centering
   \includegraphics[angle=0,height=3.11cm]{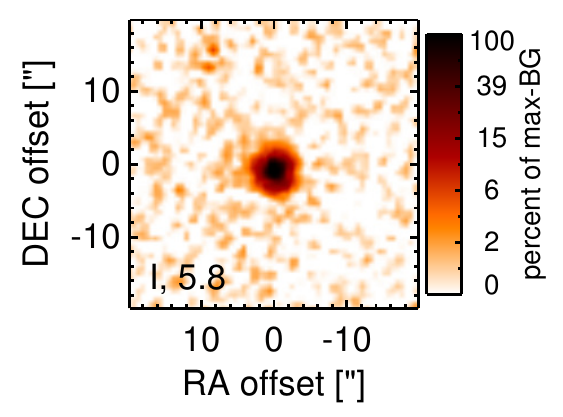}
    \caption{\label{fig:INTim_IRAS01003-2238}
             \spitzerr MIR images of IRAS\,01003-2238. Displayed are the inner $40\arcsec$ with North up and East to the left. The colour scaling is logarithmic with white corresponding to median background and black to the $0.1\%$ pixels with the highest intensity.
             The label in the bottom left states instrument and central wavelength of the filter in $\mu$m (I: IRAC, M: MIPS). 
           }
\end{figure}
\begin{figure}
   \centering
   \includegraphics[angle=0,height=3.11cm]{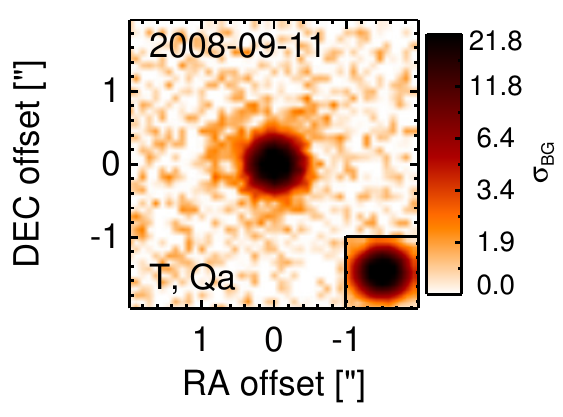}
    \caption{\label{fig:HARim_IRAS01003-2238}
             Subarcsecond-resolution MIR images of IRAS\,01003-2238 sorted by increasing filter wavelength. 
             Displayed are the inner $4\arcsec$ with North up and East to the left. 
             The colour scaling is logarithmic with white corresponding to median background and black to the $75\%$ of the highest intensity of all images in units of $\sigbg$.
             The inset image shows the central arcsecond of the PSF from the calibrator star, scaled to match the science target.
             The labels in the bottom left state instrument and filter names (C: COMICS, M: Michelle, T: T-ReCS, V: VISIR).
           }
\end{figure}
\begin{figure}
   \centering
   \includegraphics[angle=0,width=8.50cm]{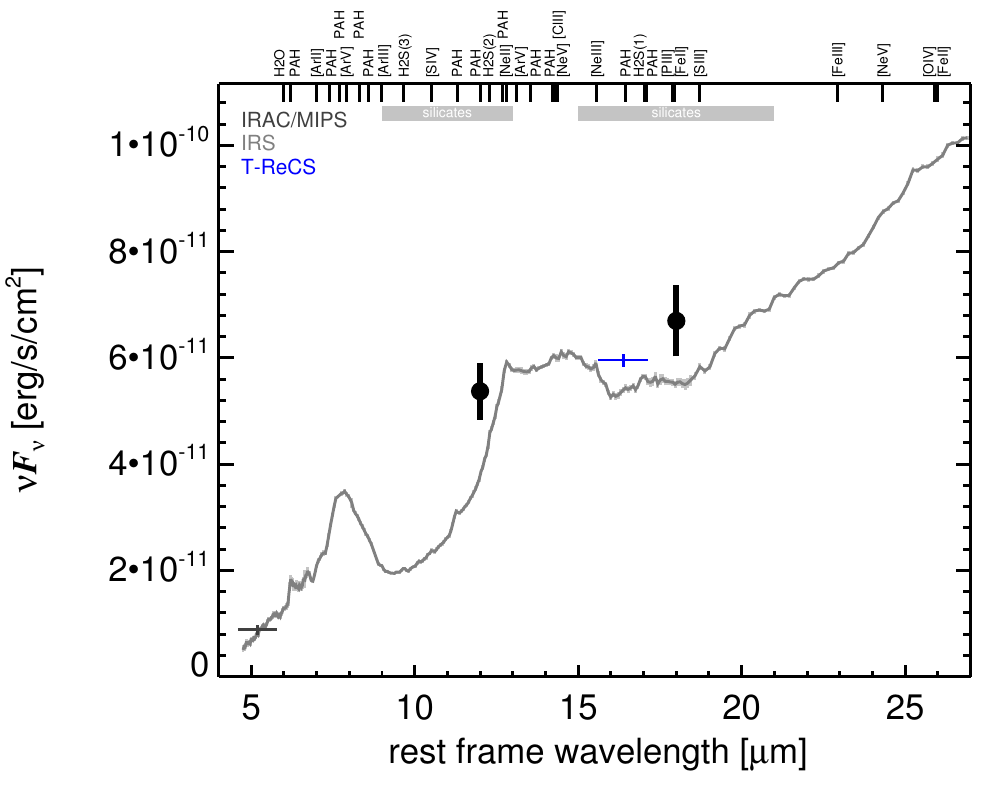}
   \caption{\label{fig:MISED_IRAS01003-2238}
      MIR SED of IRAS\,01003-2238. The description  of the symbols (if present) is the following.
      Grey crosses and  solid lines mark the \spitzer/IRAC, MIPS and IRS data. 
      The colour coding of the other symbols is: 
      green for COMICS, magenta for Michelle, blue for T-ReCS and red for VISIR data.
      Darker-coloured solid lines mark spectra of the corresponding instrument.
      The black filled circles mark the nuclear 12 and $18\,\mu$m  continuum emission estimate from the data.
      The ticks on the top axis mark positions of common MIR emission lines, while the light grey horizontal bars mark wavelength ranges affected by the silicate 10 and 18$\mu$m features.}
\end{figure}
\clearpage

\twocolumn[\begin{@twocolumnfalse}  
\subsection{IRAS\,04103-2838 -- 2MASX\,J04121945-2830252}\label{app:IRAS04103-2838}
IRAS\,04103-2838 is a disturbed ultra-luminous infrared galaxy at a redshift of $z=$ 0.1175 with an active nucleus classified optically either as H\,II or LINER \citep{veilleux_optical_1999,veilleux_new_1999} but also as Sy\,2 \citep{yuan_role_2010}.
The presence of a Compton-thick AGN was demonstrated through deep X-ray observations \citep{teng_xmm-newton_2008}, confirming the type~II nature of this source.
IRAS\,04103-2838 was observed with \spitzer/IRS in LR staring mode.
The MIR spectrum indicates star formation with silicate  $10\,\mu$m absorption, PAH emission and a red spectral slope in $\nu F_\nu$-space (see also \citealt{imanishi_spitzer_2007}).
IRAS\,04103-2838 was observed with T-ReCS in the Qa filter in 2008 and a compact MIR nucleus was detected \citep{imanishi_subaru_2011}.
The nucleus is possibly marginally resolved (FWHM $\sim 0.63\arcsec \sim 1.3\,$kpc), but this needs to be verified with at least a second epoch of subarcsecond MIR imaging.
Our measured nuclear Qa flux is consistent with the value from \cite{imanishi_subaru_2011} and also the IRS spectrum.
Therefore, we use the latter to compute the nuclear $12\,\mu$m continuum emission estimates corrected for the silicate 10\,$\mu$m feature.
\newline\end{@twocolumnfalse}]

\begin{figure}
   \centering
   \includegraphics[angle=0,width=8.500cm]{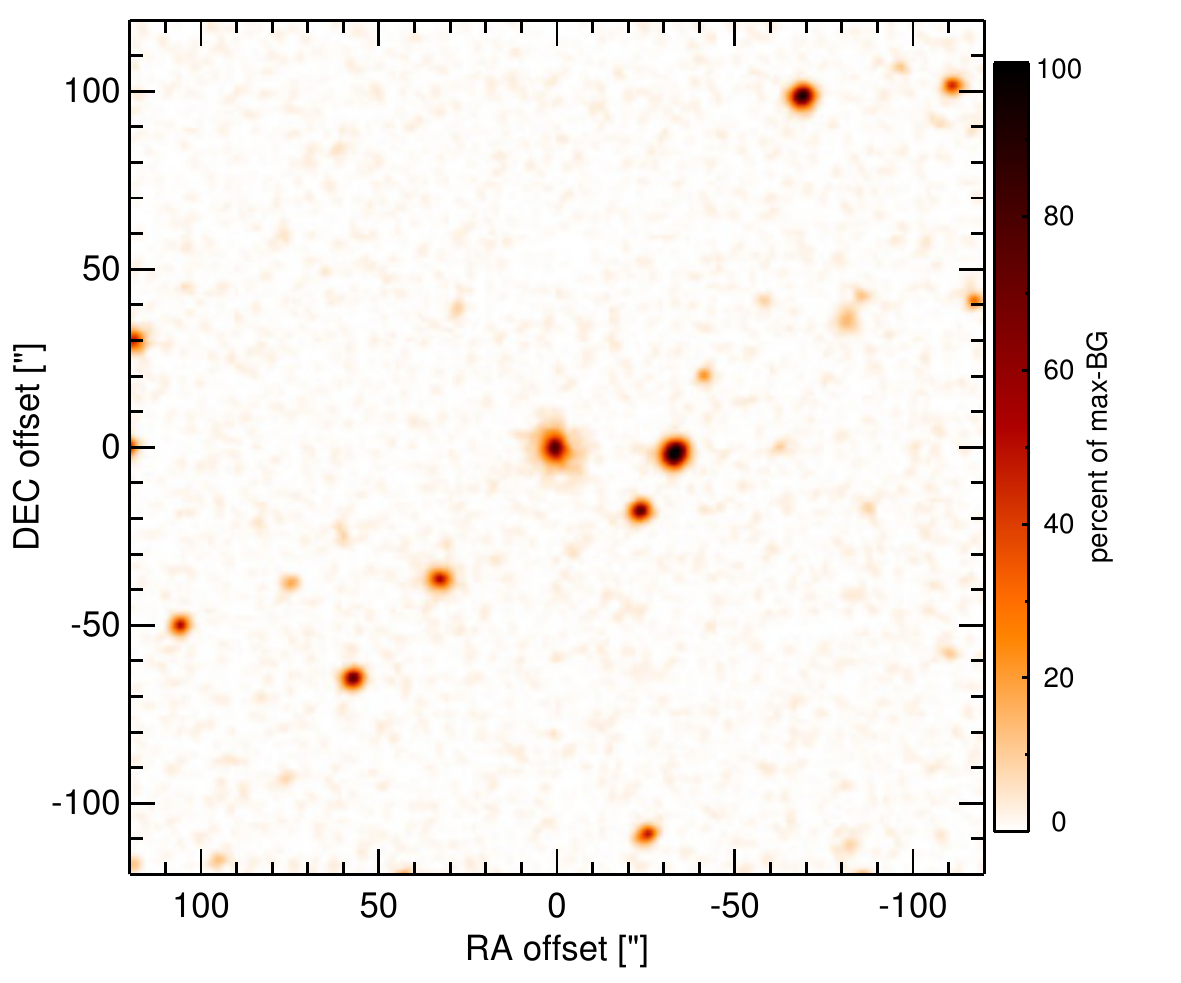}
    \caption{\label{fig:OPTim_IRAS04103-2838}
             Optical image (DSS, red filter) of IRAS\,04103-2838. Displayed are the central $4\arcmin$ with North up and East to the left. 
              The colour scaling is linear with white corresponding to the median background and black to the $0.01\%$ pixels with the highest intensity.  
           }
\end{figure}
\begin{figure}
   \centering
   \includegraphics[angle=0,height=3.11cm]{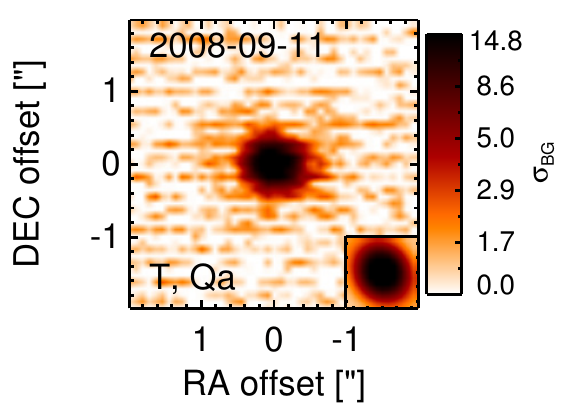}
    \caption{\label{fig:HARim_IRAS04103-2838}
             Subarcsecond-resolution MIR images of IRAS\,04103-2838 sorted by increasing filter wavelength. 
             Displayed are the inner $4\arcsec$ with North up and East to the left. 
             The colour scaling is logarithmic with white corresponding to median background and black to the $75\%$ of the highest intensity of all images in units of $\sigbg$.
             The inset image shows the central arcsecond of the PSF from the calibrator star, scaled to match the science target.
             The labels in the bottom left state instrument and filter names (C: COMICS, M: Michelle, T: T-ReCS, V: VISIR).
           }
\end{figure}
\begin{figure}
   \centering
   \includegraphics[angle=0,width=8.50cm]{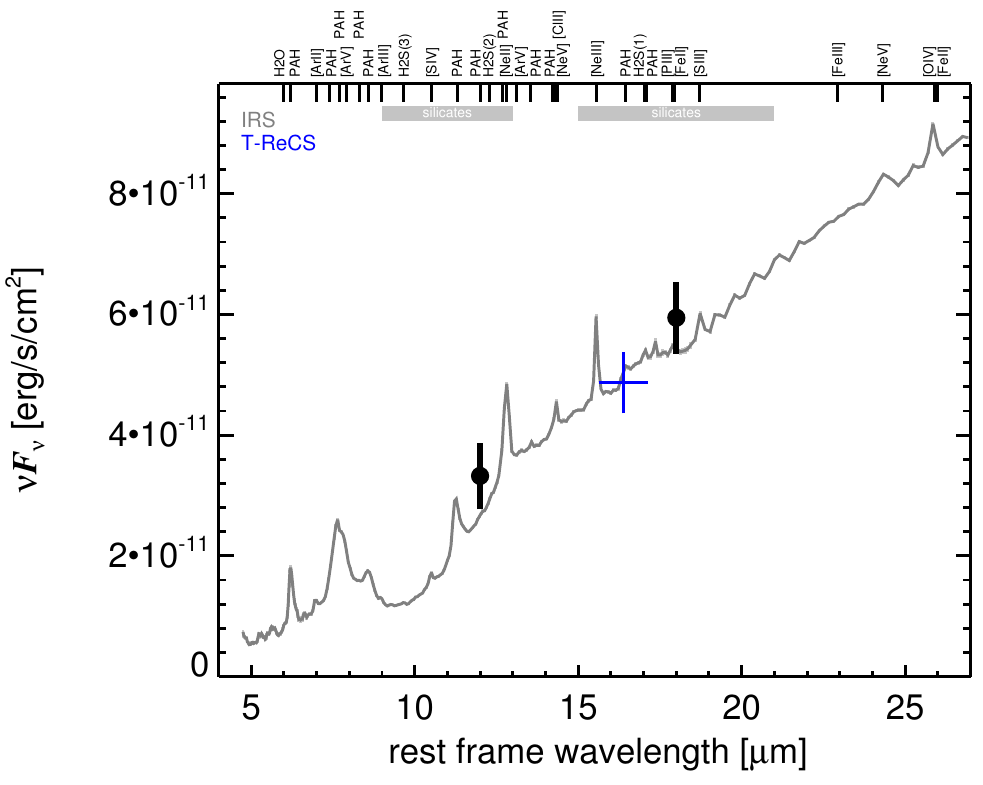}
   \caption{\label{fig:MISED_IRAS04103-2838}
      MIR SED of IRAS\,04103-2838. The description  of the symbols (if present) is the following.
      Grey crosses and  solid lines mark the \spitzer/IRAC, MIPS and IRS data. 
      The colour coding of the other symbols is: 
      green for COMICS, magenta for Michelle, blue for T-ReCS and red for VISIR data.
      Darker-coloured solid lines mark spectra of the corresponding instrument.
      The black filled circles mark the nuclear 12 and $18\,\mu$m  continuum emission estimate from the data.
      The ticks on the top axis mark positions of common MIR emission lines, while the light grey horizontal bars mark wavelength ranges affected by the silicate 10 and 18$\mu$m features.}
\end{figure}
\clearpage

\twocolumn[\begin{@twocolumnfalse}  
\subsection{IRAS\,05189-2524 -- LEDA\,17155 -- 2MASX\,J05210136-2521450}\label{app:IRAS05189-2524}
IRAS\,05189-2524 is possibly a late stage merger system of two spirals \citep{sanders_ultraluminous_1988} and is classified as an ultra-luminous infrared galaxy  at a redshift of $z=$ 0.0426 ($D\sim181$\,Mpc) hosting a Sy\,2 nucleus \citep{veilleux_optical_1995} with polarized broad emission lines \citep{young_polarimetry_1996}.
After its discovery with \iras, this object was observed with
Palomar 5\,m \citep{carico_iras_1988}, MMT \citep{maiolino_new_1995}, \isoo \citep{klaas_infrared_2001,ramos_almeida_mid-infrared_2007}.
The first high-angular resolution $N$-band images were obtained with Keck/MIRLIN in 1998 \citep{soifer_high_2000}, Palomar 5\,m/MIRLIN in 1999 \citep{gorjian_10_2004} and ESO 3.6\,m/TIMMI2 in 2002 \citep{siebenmorgen_mid-infrared_2004}.
An unresolved MIR nucleus was detected in the Keck/MIRLIN and TIMMI2 images, while the source remained undetected in the Palomar 5\,m/MIRLIN image.
The \spitzer/IRAC $5.8$ and $8.0\,\mu$m and MIPS $24\,\mu$m images also show a marginally resolved nucleus without any host emission.
Note that the IRAC $8.0\,\mu$m PBCD image is saturated and not used analysed (but see \citealt{u_spectral_2012}).
Our IRAC $5.8$ and MIPS $24\,\mu$m photometry agrees with the values in \cite{u_spectral_2012}.
The IRS LR staring-mode spectrum shows silicate 10 and $18\,\mu$m absorption, PAH emission and a red spectral slope in $\nu F_\nu$-space, and thus indicates significant star-formation contribution (see also \citealt{farrah_high-resolution_2007,wu_spitzer/irs_2009}).
This was already pointed out by \cite{dudley_new_1999} and \cite{farrah_starburst_2003}. 
IRAS\,05189-2524 was imaged with T-ReCS in the N filter in 2004 and with VISIR in two narrow $N$-band filters in 2010 (unpublished, to our knowledge).
In all images a compact MIR nucleus without further host emission was detected. 
The nucleus is possibly marginally resolved in the N filter image (FWHM(major axis) $\sim 0.47\arcsec \sim 380\,$pc; PA$\sim 95\degree$), while no extension analysis is possible for the VISIR image because no appropriate standard star was observed. 
However, the PA in the VISIR images agrees with the one in the T-ReCS image, albeit with a smaller axis ratio.
At least another epoch of subarcsecond-resolution MIR images is necessary to verify this extension.
Our N filter photometry agrees with the value published in \cite{videla_nuclear_2013}, while both the T-ReCS and VISIR fluxes are consistent with the \spitzerr spectrophotometry.
We  use the IRS spectrum to compute the nuclear $12\,\mu$m continuum emission estimate corrected for the silicate feature.
Note that the nuclear MIR emission of IRAS\,05189-2524  was  resolved in interferometric observations with MIDI, which show that approximately half of the MIR emission originates from an extended region of $\sim30\,$pc size \citep{tristram_parsec-scale_2009,burtscher_diversity_2013}.
\newline\end{@twocolumnfalse}]

\begin{figure}
   \centering
   \includegraphics[angle=0,width=8.500cm]{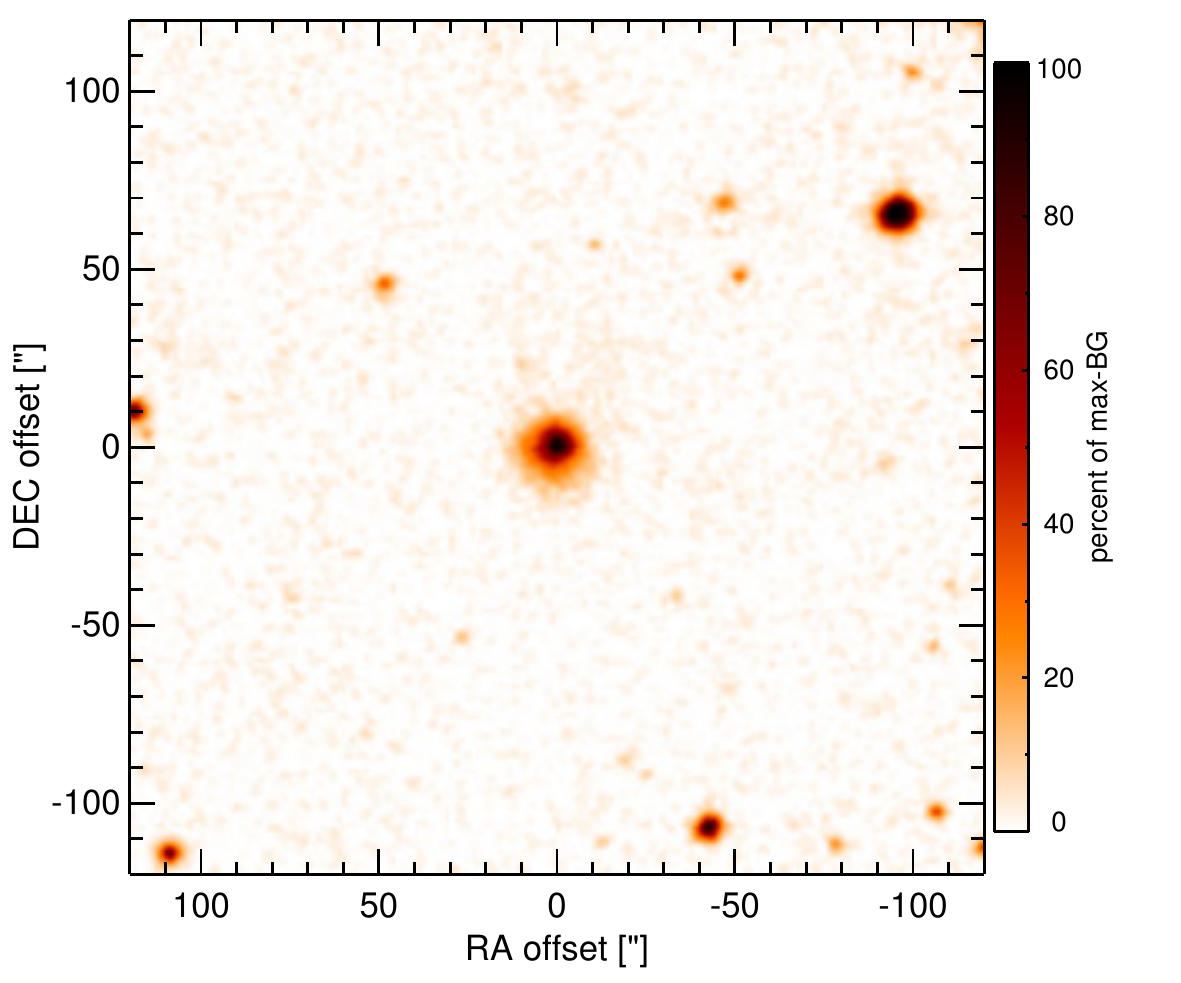}
    \caption{\label{fig:OPTim_IRAS05189-2524}
             Optical image (DSS, red filter) of IRAS\,05189-2524. Displayed are the central $4\arcmin$ with North up and East to the left. 
              The colour scaling is linear with white corresponding to the median background and black to the $0.01\%$ pixels with the highest intensity.  
           }
\end{figure}
\begin{figure}
   \centering
   \includegraphics[angle=0,height=3.11cm]{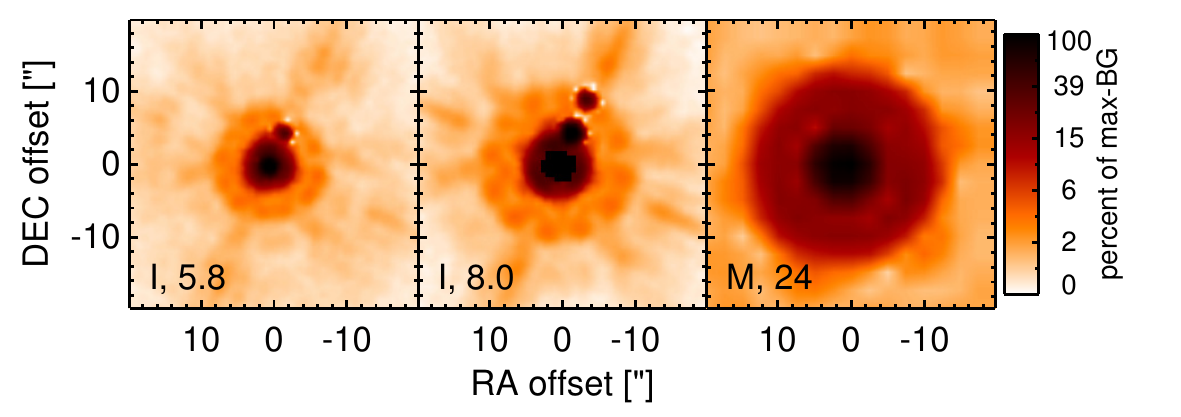}
    \caption{\label{fig:INTim_IRAS05189-2524}
             \spitzerr MIR images of IRAS\,05189-2524. Displayed are the inner $40\arcsec$ with North up and East to the left. The colour scaling is logarithmic with white corresponding to median background and black to the $0.1\%$ pixels with the highest intensity.
             The label in the bottom left states instrument and central wavelength of the filter in $\mu$m (I: IRAC, M: MIPS).
             Note that the apparent off-nuclear compact sources in the IRAC 5.8 and $8.0\,\mu$m images are instrumental artefacts.
           }
\end{figure}
\begin{figure}
   \centering
   \includegraphics[angle=0,height=3.11cm]{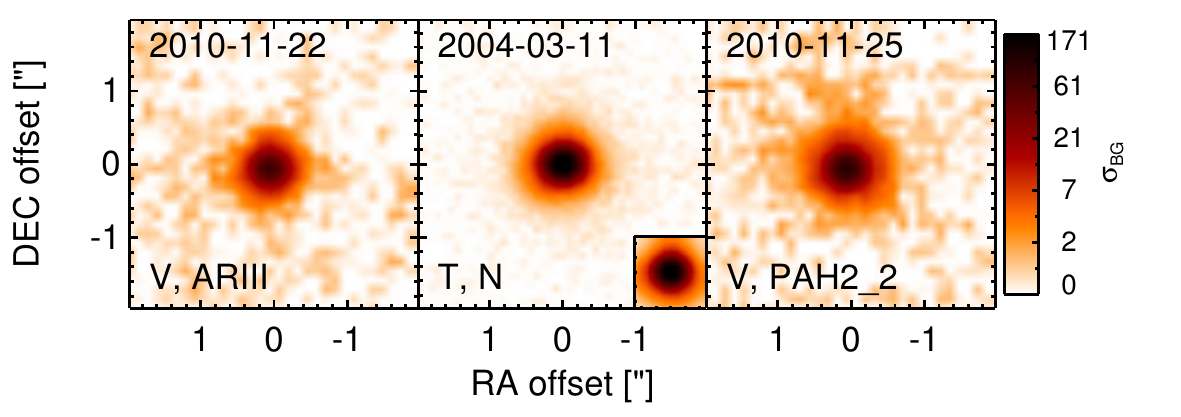}
    \caption{\label{fig:HARim_IRAS05189-2524}
             Subarcsecond-resolution MIR images of IRAS\,05189-2524 sorted by increasing filter wavelength. 
             Displayed are the inner $4\arcsec$ with North up and East to the left. 
             The colour scaling is logarithmic with white corresponding to median background and black to the $75\%$ of the highest intensity of all images in units of $\sigbg$.
             The inset image shows the central arcsecond of the PSF from the calibrator star, scaled to match the science target.
             The labels in the bottom left state instrument and filter names (C: COMICS, M: Michelle, T: T-ReCS, V: VISIR).
           }
\end{figure}
\begin{figure}
   \centering
   \includegraphics[angle=0,width=8.50cm]{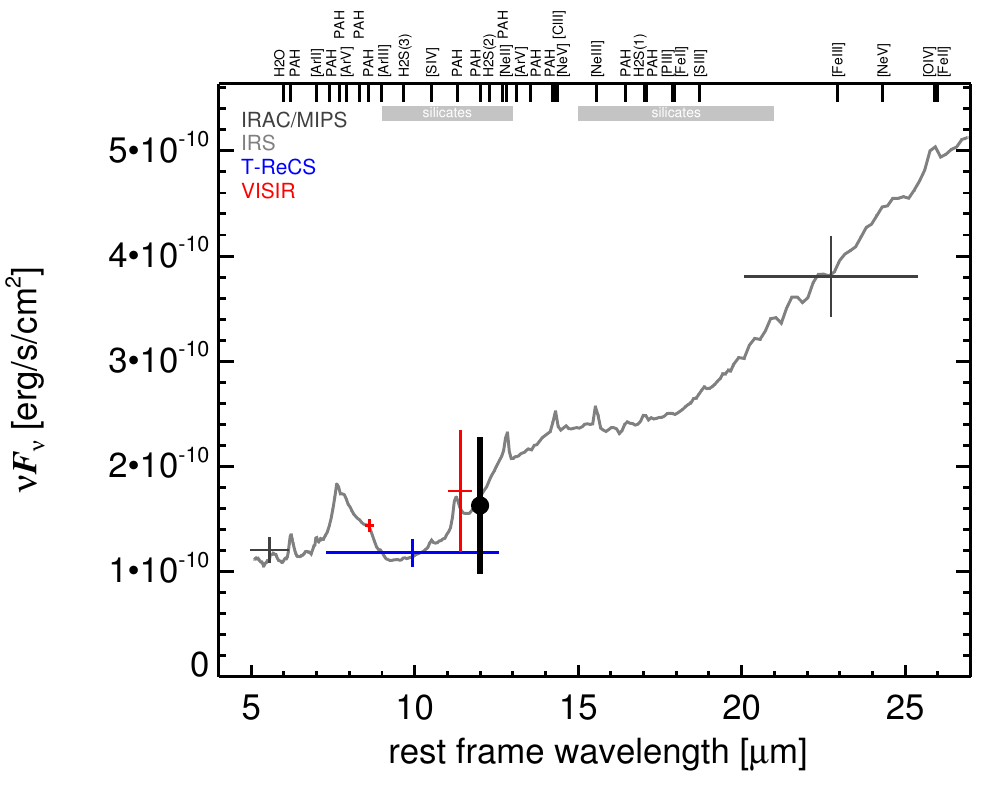}
   \caption{\label{fig:MISED_IRAS05189-2524}
      MIR SED of IRAS\,05189-2524. The description  of the symbols (if present) is the following.
      Grey crosses and  solid lines mark the \spitzer/IRAC, MIPS and IRS data. 
      The colour coding of the other symbols is: 
      green for COMICS, magenta for Michelle, blue for T-ReCS and red for VISIR data.
      Darker-coloured solid lines mark spectra of the corresponding instrument.
      The black filled circles mark the nuclear 12 and $18\,\mu$m  continuum emission estimate from the data.
      The ticks on the top axis mark positions of common MIR emission lines, while the light grey horizontal bars mark wavelength ranges affected by the silicate 10 and 18$\mu$m features.}
\end{figure}
\clearpage

\twocolumn[\begin{@twocolumnfalse}  
\subsection{IRAS\,08572+3915}\label{app:IRAS08572+3915}
IRAS\,08572+3915 is a galaxy pair with two overlapping spirals with a separation of $\sim6\arcsec$ ($\sim 6\,$kpc) \citep{sanders_ultraluminous_1988}.
The north-western galaxy is an ultra-luminous infrared galaxy at a redshift of $z=$ 0.0584 ($D\sim254\,$Mpc) with an active nucleus  classified optically either as H\,II \citep{veron-cetty_catalogue_2010}, LINER \citep{kim_optical_1995,veilleux_optical_1995,veilleux_new_1999}, or Sy\,2/starburst composite \citep{yuan_role_2010}.
We adopt the latter classification.
After its discovery with \iras, IRAS\,08572+3915 was observed in the $N$-band by \cite{carico_iras_1988} with the Palomar 5\,m, \cite{wynn-williams_luminous_1993} with IRTF/BOLO1 in 1986/87, \cite{miles_high-resolution_1996} with Palomar 5\,m/SpectroCam-10 in 1993, by \cite{soifer_high_2000} with Keck/LWS and MIRLIN in 1998,  and \cite{gorjian_10_2004} with Palomar 5\,m/MIRLIN in 2000.
IRAS\,08572+3915 remained unresolved in all these observations. 
It was also observed with \spitzer/IRAC, IRS and MIPS, where a  marginally resolved MIR nucleus without any other host emission was detected. 
In particular, the south-eastern galaxy remains undetected.
Note that the IRAC $8.0\,\mu$m PBCD image is partly saturated and thus not used.
Our IRAC $5.8\,\mu$m and MIPS $24\,\mu$m photometry matches the values published in \cite{u_spectral_2012}.
The IRS LR staring mode spectrum exhibits extremely deep silicate  10 and $18\,\mu$m absorption, no prominent PAH emission and a red spectral slope in $\nu F_\nu$-space (see also \citealt{spoon_detection_2006,levenson_deep_2007,imanishi_spitzer_2007}).
The properties of the MIR SED  favour the existence of a highly obscured AGN in IRAS\,08572+3915 as proposed by \cite{imanishi_energy_2000,imanishi_spitzer_2007}.
IRAS\,08572+3915 was observed with COMICS in the Q17.7 filter in 2008 \citep{imanishi_subaru_2011}.
We find the detected MIR nucleus to be possibly marginally resolved (FWHM $\sim 0.63\arcsec \sim 690\,$pc), which however needs to be verified with at least another epoch of subarcsecond MIR imaging.
Our nuclear Q17.7 flux is significantly higher than the value published in \cite{imanishi_subaru_2011} but agrees with the \spitzerr spectrophotometry.
Therefore, we use the latter to compute our 12 and 18$\,\mu$m continuum emission estimates corrected for the silicate absorption. 
The absence of any strong emission features indicates that putative MIR emission-line producing regions  are heavily extincted (similar to, e.g., NGC\,4945; \citealt{perez-beaupuits_deeply_2011}).
\newline\end{@twocolumnfalse}]

\begin{figure}
   \centering
   \includegraphics[angle=0,width=8.500cm]{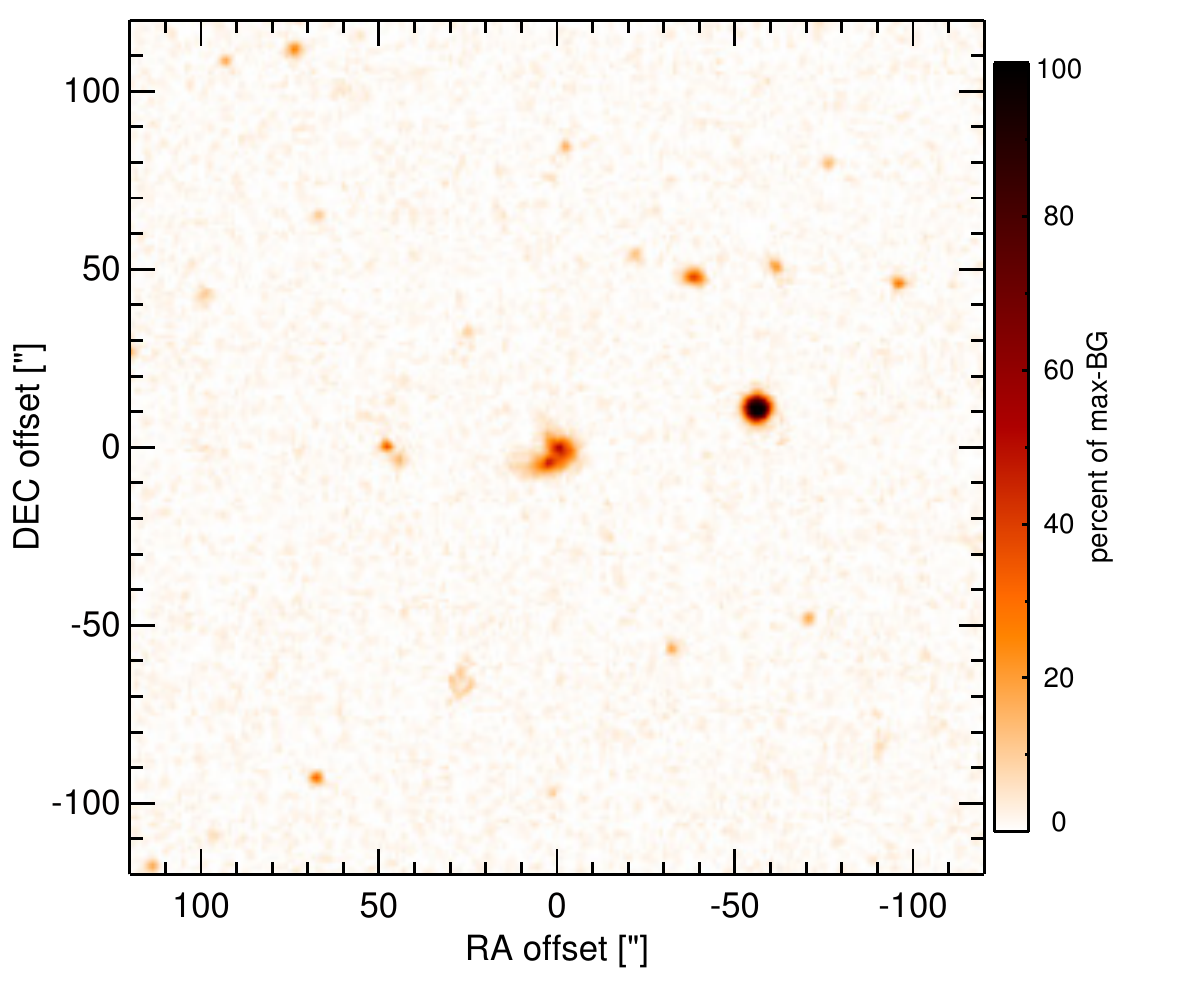}
    \caption{\label{fig:OPTim_IRAS08572+3915}
             Optical image (DSS, red filter) of IRAS\,08572+3915. Displayed are the central $4\arcmin$ with North up and East to the left. 
              The colour scaling is linear with white corresponding to the median background and black to the $0.01\%$ pixels with the highest intensity.  
           }
\end{figure}
\begin{figure}
   \centering
   \includegraphics[angle=0,height=3.11cm]{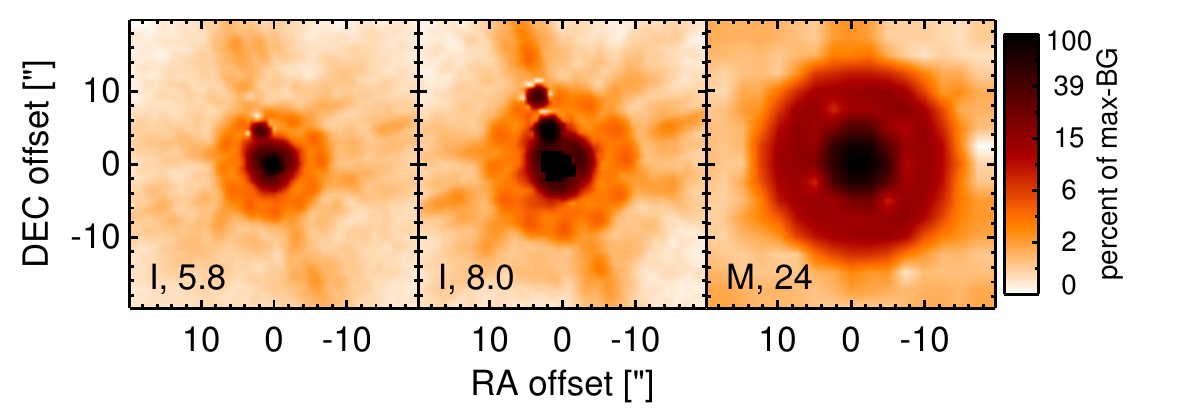}
    \caption{\label{fig:INTim_IRAS08572+3915}
             \spitzerr MIR images of IRAS\,08572+3915. Displayed are the inner $40\arcsec$ with North up and East to the left. The colour scaling is logarithmic with white corresponding to median background and black to the $0.1\%$ pixels with the highest intensity.
             The label in the bottom left states instrument and central wavelength of the filter in $\mu$m (I: IRAC, M: MIPS).
             Note that the apparent off-nuclear compact sources in the IRAC 5.8 and $8.0\,\mu$m images are instrumental artefacts.
           }
\end{figure}
\begin{figure}
   \centering
   \includegraphics[angle=0,height=3.11cm]{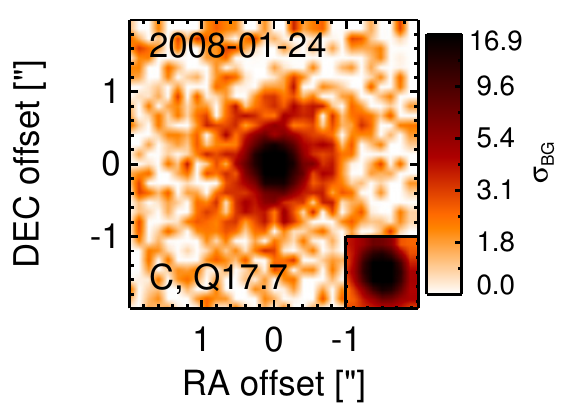}
    \caption{\label{fig:HARim_IRAS08572+3915}
             Subarcsecond-resolution MIR images of IRAS\,08572+3915 sorted by increasing filter wavelength. 
             Displayed are the inner $4\arcsec$ with North up and East to the left. 
             The colour scaling is logarithmic with white corresponding to median background and black to the $75\%$ of the highest intensity of all images in units of $\sigbg$.
             The inset image shows the central arcsecond of the PSF from the calibrator star, scaled to match the science target.
             The labels in the bottom left state instrument and filter names (C: COMICS, M: Michelle, T: T-ReCS, V: VISIR).
           }
\end{figure}
\begin{figure}
   \centering
   \includegraphics[angle=0,width=8.50cm]{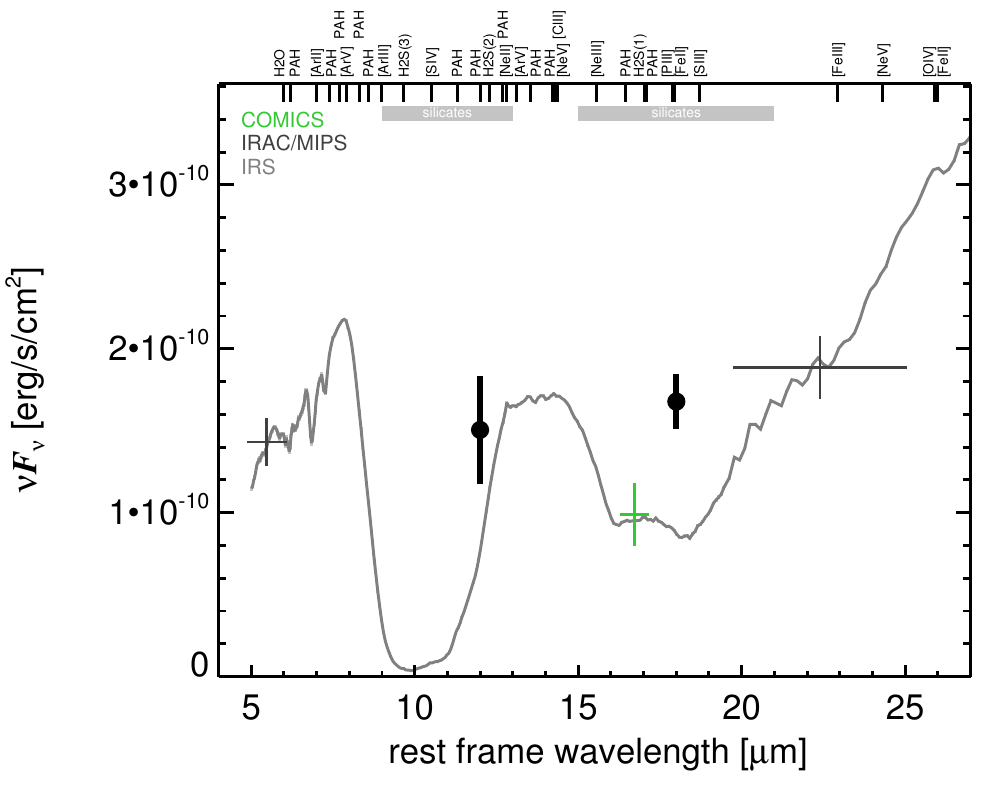}
   \caption{\label{fig:MISED_IRAS08572+3915}
      MIR SED of IRAS\,08572+3915. The description  of the symbols (if present) is the following.
      Grey crosses and  solid lines mark the \spitzer/IRAC, MIPS and IRS data. 
      The colour coding of the other symbols is: 
      green for COMICS, magenta for Michelle, blue for T-ReCS and red for VISIR data.
      Darker-coloured solid lines mark spectra of the corresponding instrument.
      The black filled circles mark the nuclear 12 and $18\,\mu$m  continuum emission estimate from the data.
      The ticks on the top axis mark positions of common MIR emission lines, while the light grey horizontal bars mark wavelength ranges affected by the silicate 10 and 18$\mu$m features.}
\end{figure}
\clearpage

\twocolumn[\begin{@twocolumnfalse}  
\subsection{IRAS\,09149-6206 -- LEDA 90443}\label{app:IRAS09149-6206}
IRAS\,09149-6206 is a Seyfert galaxy at a low Galactic latitude and a redshift of $z=$ 0.0573 ($D\sim249$\,Mpc) hosting a Sy\,1 nucleus \citep{perez_iras_1989}.
It was observed with \spitzer/IRAC, IRS and MIPS and appears slightly resolved in all corresponding images.
Our IRAC $5.8$ and $8.0\,\mu$m photometry matches the values published in \cite{kishimoto_mapping_2011}.
The IRS LR mapping-mode spectrum exhibits silicate  10 and $18\,\mu$m emission, no prominent PAH emission and a flat spectral slope in $\nu F_\nu$-space.
IRAS\,09149-6206 was observed with VISIR in two narrow $N$-band filters in 2009 where the MIR nucleus remained unresolved.
Our measurement of the nuclear fluxes also agrees with the values published in \cite{kishimoto_mapping_2011} and the \spitzerr spectrophotometry.
The latter authors also observed the object interferometrically with MIDI using several baselines and could marginally resolve the nuclear MIR structure of IRAS\,09149-6206 (see also \citealt{burtscher_diversity_2013}). 
\newline\end{@twocolumnfalse}]

\begin{figure}
   \centering
   \includegraphics[angle=0,width=8.500cm]{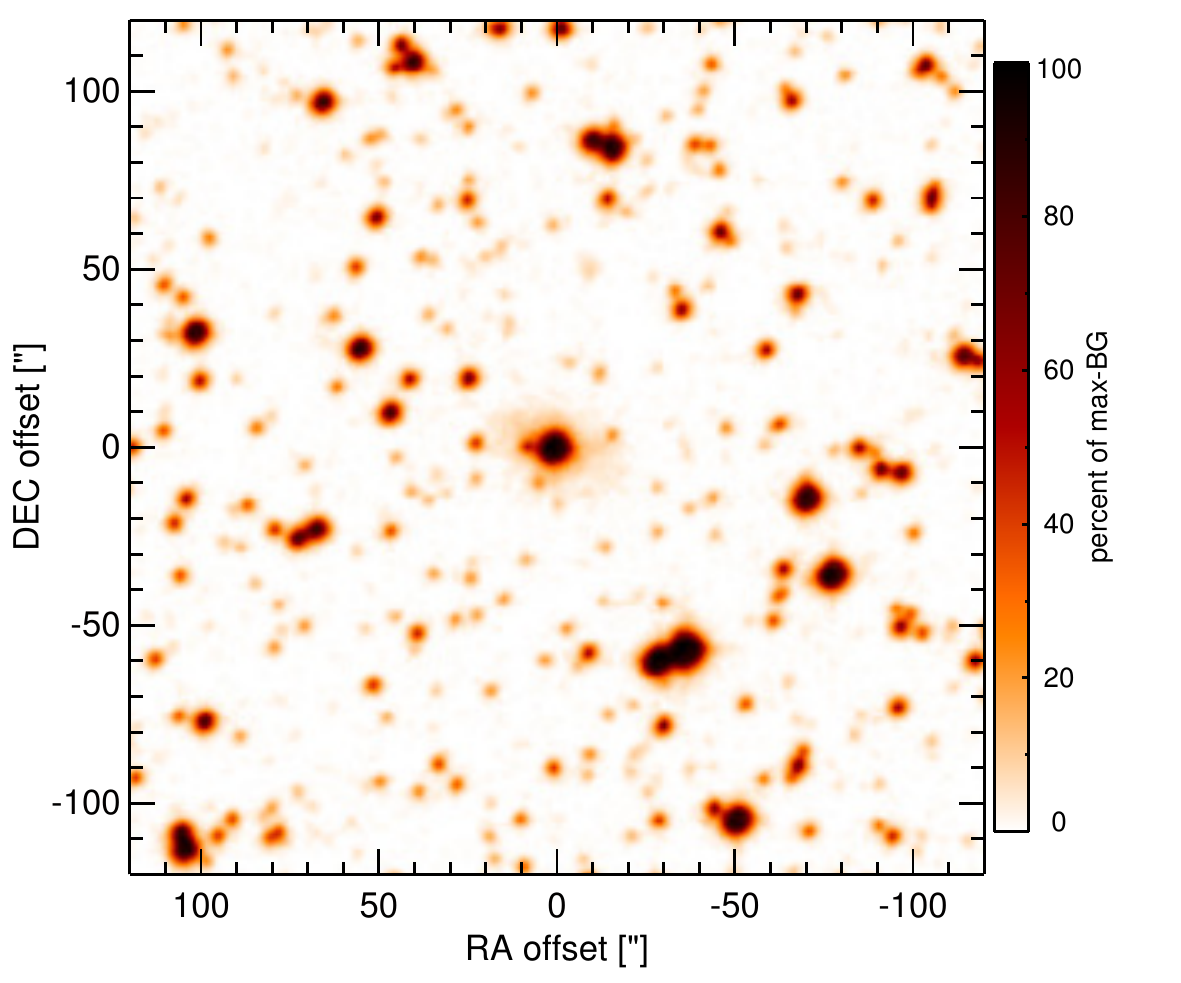}
    \caption{\label{fig:OPTim_IRAS09149-6206}
             Optical image (DSS, red filter) of IRAS\,09149-6206. Displayed are the central $4\arcmin$ with North up and East to the left. 
              The colour scaling is linear with white corresponding to the median background and black to the $0.01\%$ pixels with the highest intensity.  
           }
\end{figure}
\begin{figure}
   \centering
   \includegraphics[angle=0,height=3.11cm]{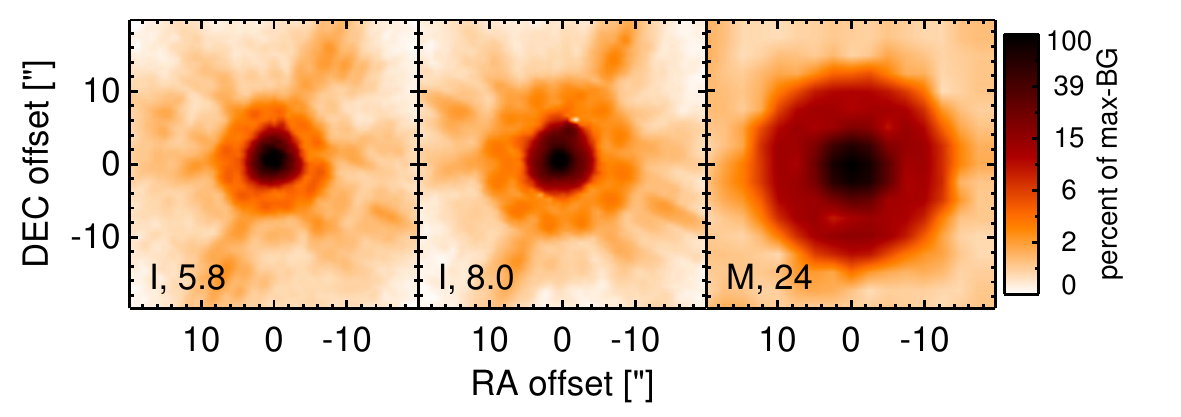}
    \caption{\label{fig:INTim_IRAS09149-6206}
             \spitzerr MIR images of IRAS\,09149-6206. Displayed are the inner $40\arcsec$ with North up and East to the left. The colour scaling is logarithmic with white corresponding to median background and black to the $0.1\%$ pixels with the highest intensity.
             The label in the bottom left states instrument and central wavelength of the filter in $\mu$m (I: IRAC, M: MIPS). 
           }
\end{figure}
\begin{figure}
   \centering
   \includegraphics[angle=0,height=3.11cm]{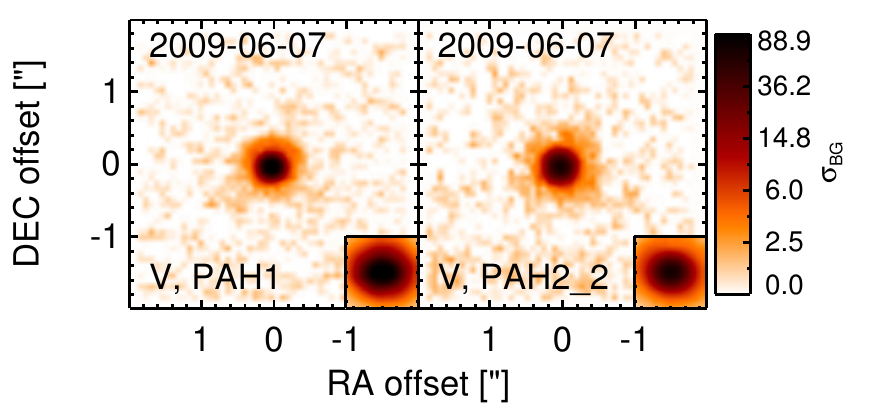}
    \caption{\label{fig:HARim_IRAS09149-6206}
             Subarcsecond-resolution MIR images of IRAS\,09149-6206 sorted by increasing filter wavelength. 
             Displayed are the inner $4\arcsec$ with North up and East to the left. 
             The colour scaling is logarithmic with white corresponding to median background and black to the $75\%$ of the highest intensity of all images in units of $\sigbg$.
             The inset image shows the central arcsecond of the PSF from the calibrator star, scaled to match the science target.
             The labels in the bottom left state instrument and filter names (C: COMICS, M: Michelle, T: T-ReCS, V: VISIR).
           }
\end{figure}
\begin{figure}
   \centering
   \includegraphics[angle=0,width=8.50cm]{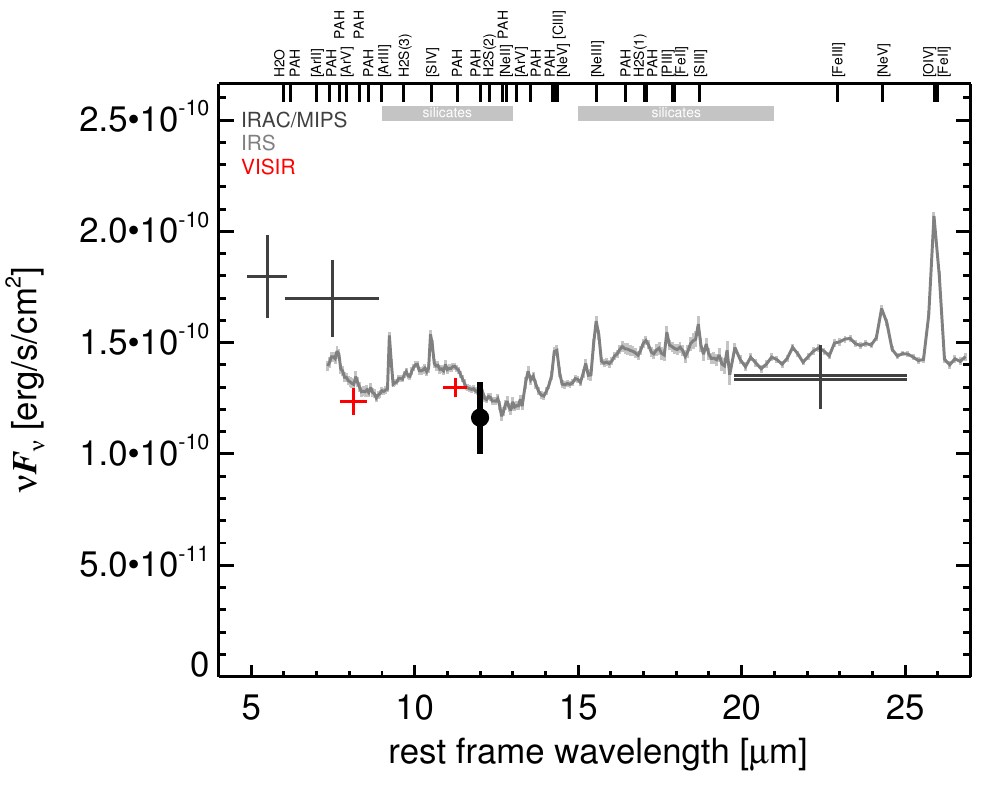}
   \caption{\label{fig:MISED_IRAS09149-6206}
      MIR SED of IRAS\,09149-6206. The description  of the symbols (if present) is the following.
      Grey crosses and  solid lines mark the \spitzer/IRAC, MIPS and IRS data. 
      The colour coding of the other symbols is: 
      green for COMICS, magenta for Michelle, blue for T-ReCS and red for VISIR data.
      Darker-coloured solid lines mark spectra of the corresponding instrument.
      The black filled circles mark the nuclear 12 and $18\,\mu$m  continuum emission estimate from the data.
      The ticks on the top axis mark positions of common MIR emission lines, while the light grey horizontal bars mark wavelength ranges affected by the silicate 10 and 18$\mu$m features.}
\end{figure}
\clearpage

\twocolumn[\begin{@twocolumnfalse}  
\subsection{IRAS\,11095-0238 -- LCRS\,B110930.3-023804}\label{app:IRAS11095-0238}
IRAS\,11095-0238 is an ultra-luminous infrared galaxy at a redshift of $z=$ 0.1066 ($D\sim480$\,Mpc) which might represent the late stage of a merger with two nuclei separated by $\sim 0.5\arcsec$ ($\sim 0.9$\,kpc; PA$\sim45\degree$ \citealt{duc_southern_1997,bushouse_ultraluminous_2002}).
Optically the object is classified either as LINER \citep{veron-cetty_catalogue_2010} or as AGN/starburst composite \citep{yuan_role_2010}. 
We conservatively treat this object as uncertain AGN/starburst composite  in the absence of multiwavelength evidence for the presence of an AGN.
IRAS\,11095-0238 was observed with \spitzer/IRAC and IRS and appears rather point-like in the corresponding images.
The IRS LR staring-mode spectrum shows extremely deep silicate  10 and  $18\,\mu$m absorption, PAH emission and an extremely red spectral slope in $\nu F_\nu$-space (see also \citealt{imanishi_spitzer_2007,farrah_high-resolution_2007}).
Thus, the MIR SED appears very similar to IRAS\,08572+3915 and suggests the presence of a deeply buried AGN as proposed by \cite{imanishi_spitzer_2007}.
The absence of any strong emission features indicates that putative MIR emission-line producing regions  are heavily extincted (similar to, e.g., NGC\,4945; \citealt{perez-beaupuits_deeply_2011}).
COMICS observations of this object were performed in 2009 in the N8.8 filter (unpublished, to our knowledge), where a compact MIR nucleus was weakly detected. 
The second nucleus is possibly detected to the south-west but with less than $3\sigma_\mathrm{BG}$ significance.
The N8.8 flux of the detected nucleus is $24\%$ lower than the \spitzerr spectrophotometry and possibly indicates a lower PAH emission from this nucleus.
\newline\end{@twocolumnfalse}]

\begin{figure}
   \centering
   \includegraphics[angle=0,width=8.500cm]{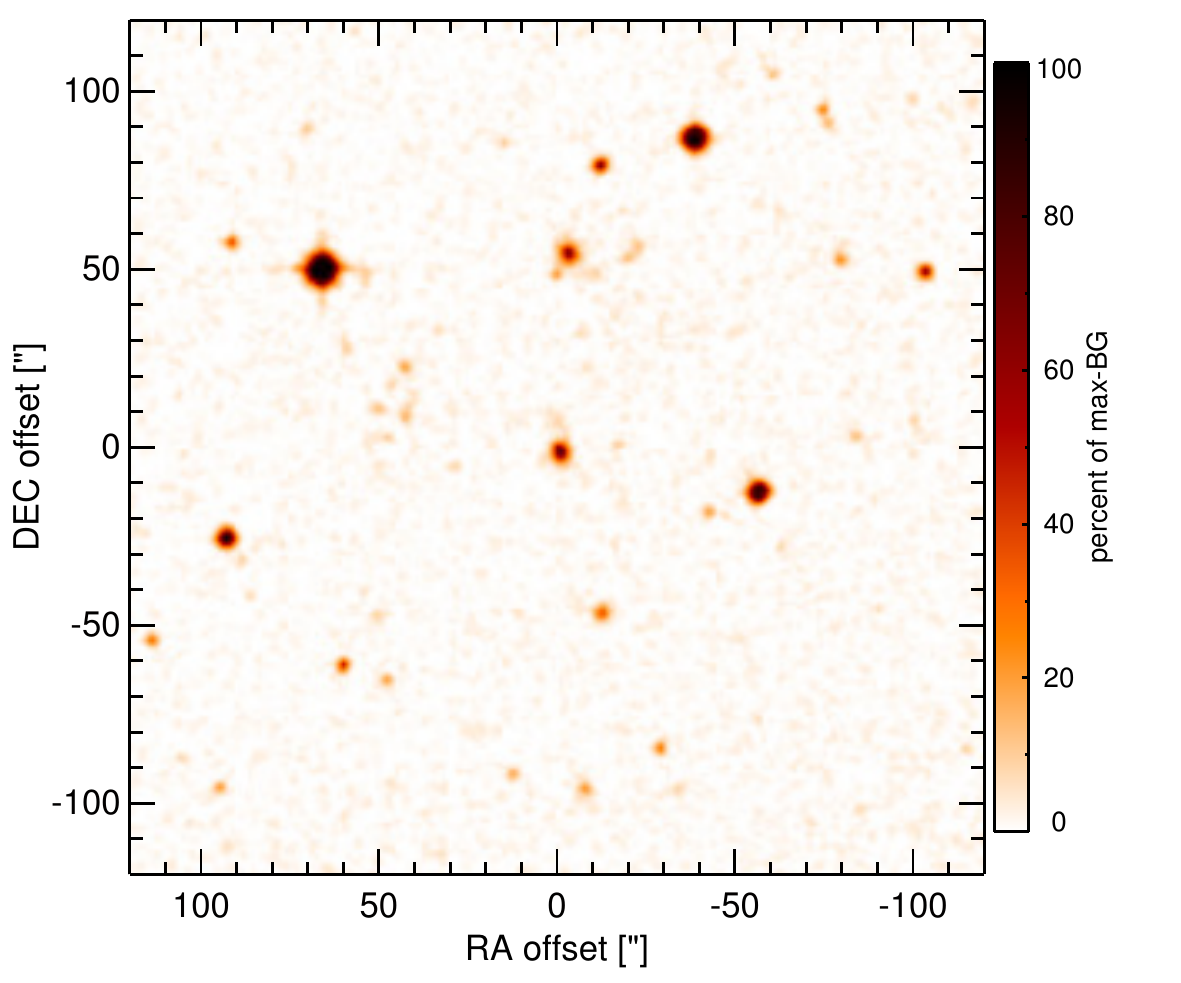}
    \caption{\label{fig:OPTim_IRAS11095-0238}
             Optical image (DSS, red filter) of IRAS\,11095-0238. Displayed are the central $4\arcmin$ with North up and East to the left. 
              The colour scaling is linear with white corresponding to the median background and black to the $0.01\%$ pixels with the highest intensity.  
           }
\end{figure}
\begin{figure}
   \centering
   \includegraphics[angle=0,height=3.11cm]{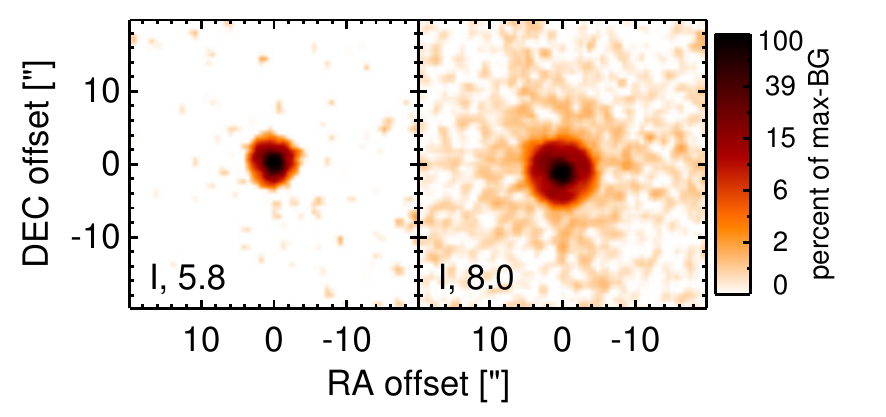}
    \caption{\label{fig:INTim_IRAS11095-0238}
             \spitzerr MIR images of IRAS\,11095-0238. Displayed are the inner $40\arcsec$ with North up and East to the left. The colour scaling is logarithmic with white corresponding to median background and black to the $0.1\%$ pixels with the highest intensity.
             The label in the bottom left states instrument and central wavelength of the filter in $\mu$m (I: IRAC, M: MIPS). 
           }
\end{figure}
\begin{figure}
   \centering
   \includegraphics[angle=0,height=3.11cm]{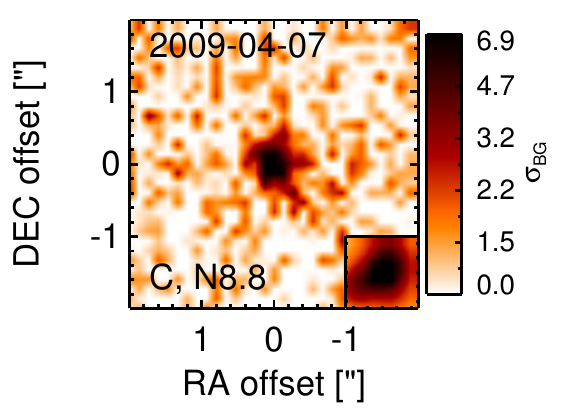}
    \caption{\label{fig:HARim_IRAS11095-0238}
             Subarcsecond-resolution MIR images of IRAS\,11095-0238 sorted by increasing filter wavelength. 
             Displayed are the inner $4\arcsec$ with North up and East to the left. 
             The colour scaling is logarithmic with white corresponding to median background and black to the $75\%$ of the highest intensity of all images in units of $\sigbg$.
             The inset image shows the central arcsecond of the PSF from the calibrator star, scaled to match the science target.
             The labels in the bottom left state instrument and filter names (C: COMICS, M: Michelle, T: T-ReCS, V: VISIR).
           }
\end{figure}
\begin{figure}
   \centering
   \includegraphics[angle=0,width=8.50cm]{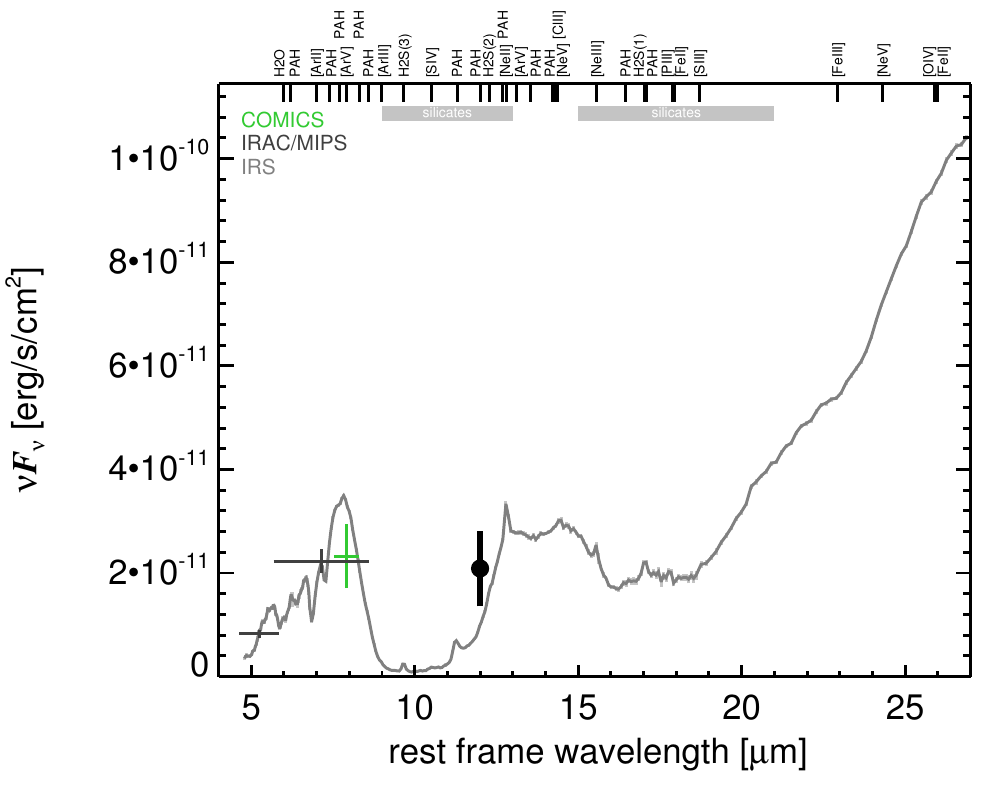}
   \caption{\label{fig:MISED_IRAS11095-0238}
      MIR SED of IRAS\,11095-0238. The description  of the symbols (if present) is the following.
      Grey crosses and  solid lines mark the \spitzer/IRAC, MIPS and IRS data. 
      The colour coding of the other symbols is: 
      green for COMICS, magenta for Michelle, blue for T-ReCS and red for VISIR data.
      Darker-coloured solid lines mark spectra of the corresponding instrument.
      The black filled circles mark the nuclear 12 and $18\,\mu$m  continuum emission estimate from the data.
      The ticks on the top axis mark positions of common MIR emission lines, while the light grey horizontal bars mark wavelength ranges affected by the silicate 10 and 18$\mu$m features.}
\end{figure}
\clearpage

\twocolumn[\begin{@twocolumnfalse}  
\subsection{IRAS\,13349+2438 -- [HB89]\,1334+246}\label{app:IRAS13349+2438}
IRAS\,13349+2438 is a quasar and narrow-line Sy\,1 galaxy at a redshift of $z=$ 0.1076 ($D\sim483$\,Mpc; see \citealt{lee_ionized_2013} for a dedicated comprehensive multiwavelength study).
It was the first radio-quiet quasar discovered in the infrared by \cite{beichman_discovery_1986}, who observed this object with Palomar 5\,m/BOLO1 in 1985 in the $N$-band.
In 2000, IRAS\,13349+2438 was observed with Palomar 5\,m/MIRLIN \citep{gorjian_10_2004}, followed by \spitzer/IRC in 2005 and \spitzer/IRAC in 2009.
A slightly resolved MIR nucleus was detected in all cases.
Our IRAC $5.8$ and $8.0\,\mu$m photometry is consistent with the values published in \cite{kishimoto_mapping_2011} but significantly higher than the IRS LR staring mode spectrum.
The latter is relatively featureless, showing weak silicate $10\,\mu$m emission, no prominent PAH emission, and a blue spectral slope in $\nu F_\nu$-space (see also \citealt{wu_spitzer/irs_2009,lee_ionized_2013}).
VISIR observations of IRAS\,13349+2438 in two narrow $N$-band filters were performed in 2009.
The MIR nucleus has a smaller FWHM than the corresponding standard star but still larger than the resolution limit. 
Therefore, it remains uncertain, whether the nucleus is marginally resolved in the MIR at subarcsecond resolution.
The nuclear photometry matches the IRS spectrum and the values in \cite{kishimoto_mapping_2011}.
These authors have also observed IRAS\,13349+2438 interferometrically with MIDI using several baselines and marginally resolved the MIR nucleus (but see \citealt{burtscher_diversity_2013}).
\newline\end{@twocolumnfalse}]

\begin{figure}
   \centering
   \includegraphics[angle=0,width=8.500cm]{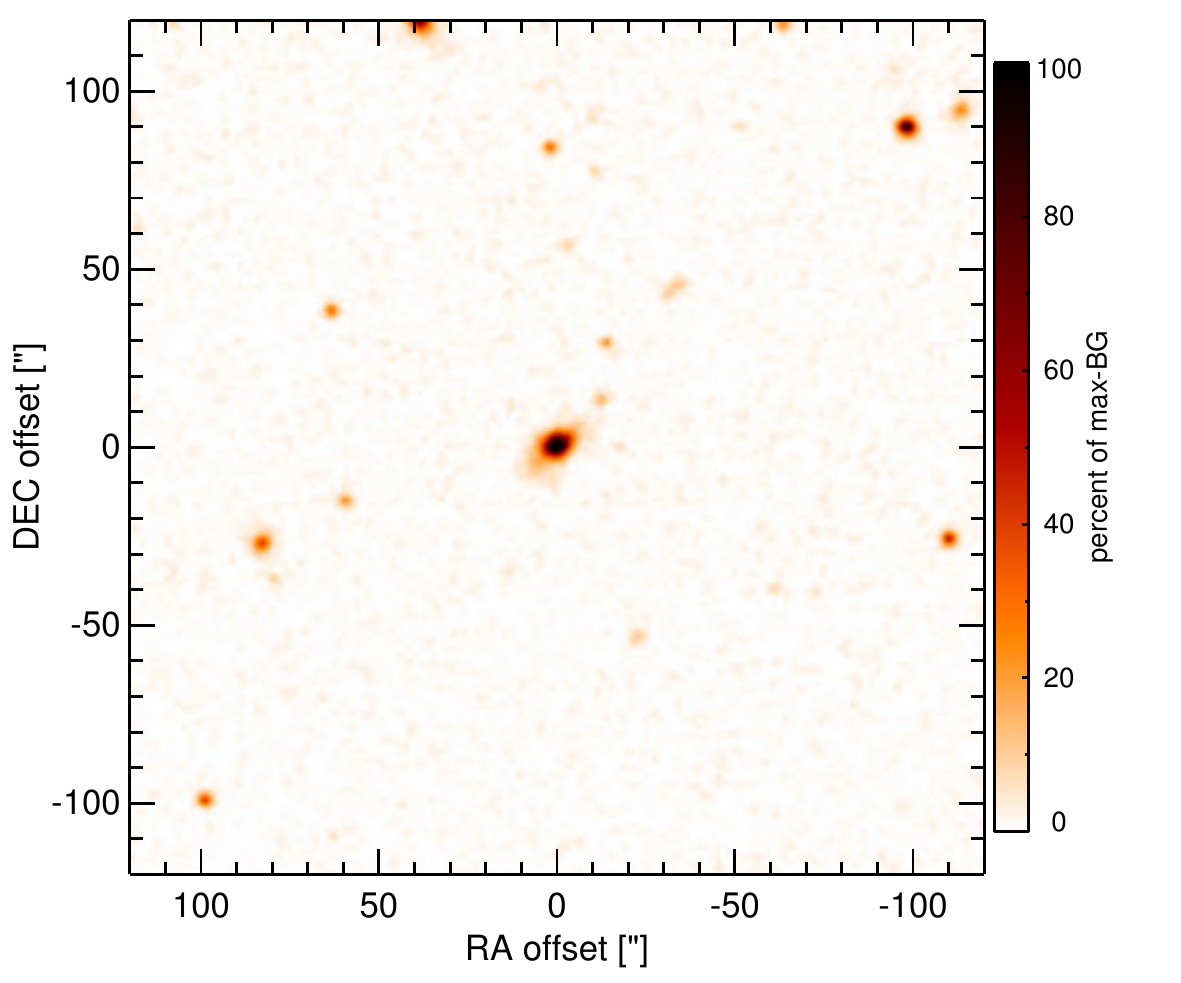}
    \caption{\label{fig:OPTim_IRAS13349+2438}
             Optical image (DSS, red filter) of IRAS\,13349+2438. Displayed are the central $4\arcmin$ with North up and East to the left. 
              The colour scaling is linear with white corresponding to the median background and black to the $0.01\%$ pixels with the highest intensity.  
           }
\end{figure}
\begin{figure}
   \centering
   \includegraphics[angle=0,height=3.11cm]{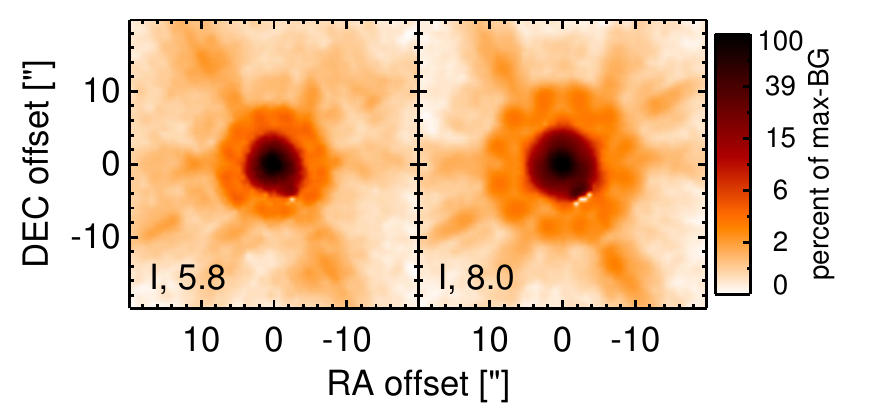}
    \caption{\label{fig:INTim_IRAS13349+2438}
             \spitzerr MIR images of IRAS\,13349+2438. Displayed are the inner $40\arcsec$ with North up and East to the left. The colour scaling is logarithmic with white corresponding to median background and black to the $0.1\%$ pixels with the highest intensity.
             The label in the bottom left states instrument and central wavelength of the filter in $\mu$m (I: IRAC, M: MIPS). 
           }
\end{figure}
\begin{figure}
   \centering
   \includegraphics[angle=0,height=3.11cm]{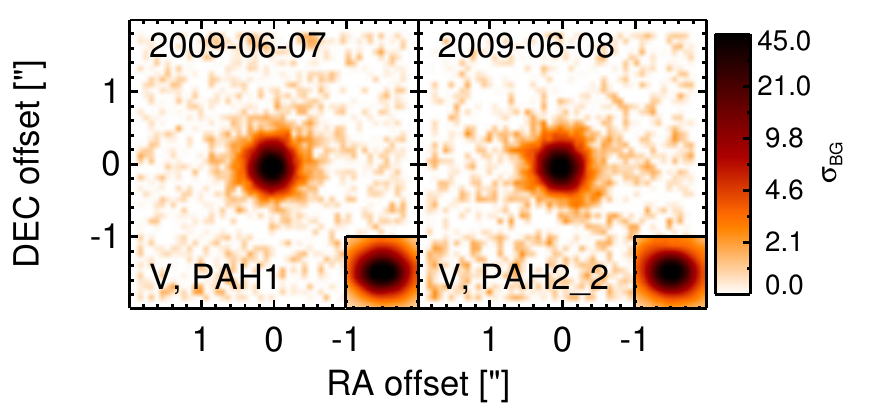}
    \caption{\label{fig:HARim_IRAS13349+2438}
             Subarcsecond-resolution MIR images of IRAS\,13349+2438 sorted by increasing filter wavelength. 
             Displayed are the inner $4\arcsec$ with North up and East to the left. 
             The colour scaling is logarithmic with white corresponding to median background and black to the $75\%$ of the highest intensity of all images in units of $\sigbg$.
             The inset image shows the central arcsecond of the PSF from the calibrator star, scaled to match the science target.
             The labels in the bottom left state instrument and filter names (C: COMICS, M: Michelle, T: T-ReCS, V: VISIR).
           }
\end{figure}
\begin{figure}
   \centering
   \includegraphics[angle=0,width=8.50cm]{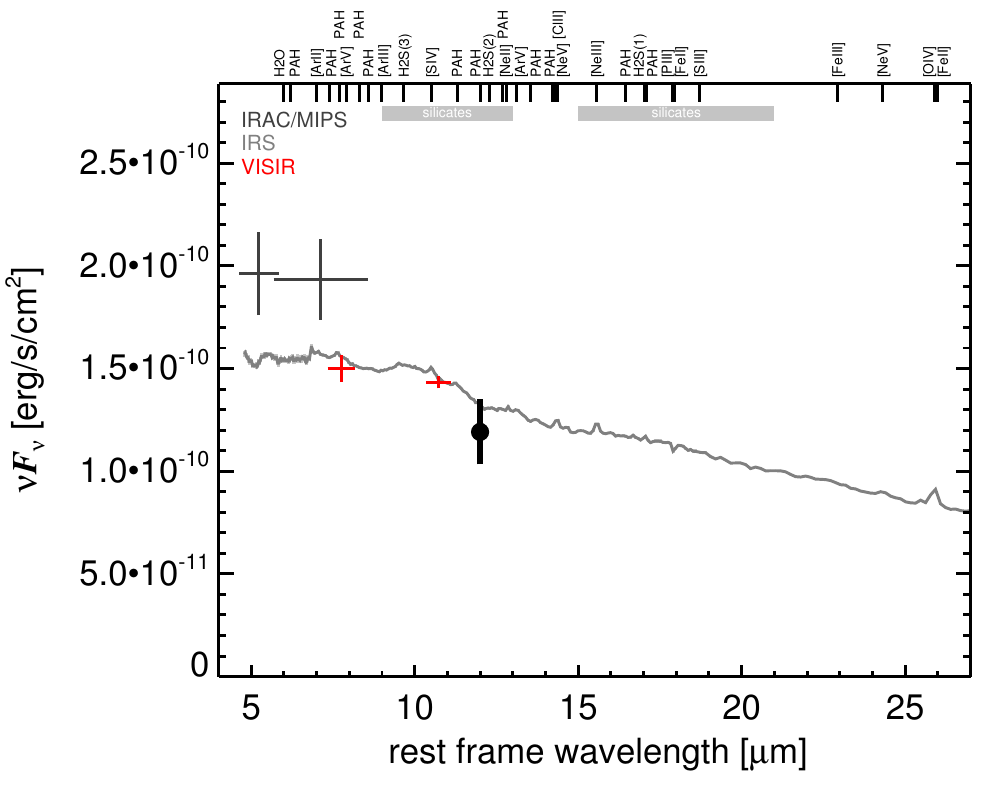}
   \caption{\label{fig:MISED_IRAS13349+2438}
      MIR SED of IRAS\,13349+2438. The description  of the symbols (if present) is the following.
      Grey crosses and  solid lines mark the \spitzer/IRAC, MIPS and IRS data. 
      The colour coding of the other symbols is: 
      green for COMICS, magenta for Michelle, blue for T-ReCS and red for VISIR data.
      Darker-coloured solid lines mark spectra of the corresponding instrument.
      The black filled circles mark the nuclear 12 and $18\,\mu$m  continuum emission estimate from the data.
      The ticks on the top axis mark positions of common MIR emission lines, while the light grey horizontal bars mark wavelength ranges affected by the silicate 10 and 18$\mu$m features.}
\end{figure}
\clearpage

\twocolumn[\begin{@twocolumnfalse}  
\subsection{IRAS\,15250+3609 -- LEDA 55114}\label{app:IRAS15250+3609}
IRAS\,15250+3609 is a peculiar ultra-luminous infrared galaxy at a redshift of $z=$ 0.0552 ($D\sim$238\,Mpc) with an active nucleus optically classified as a composite LINER and H\,II \citep{kim_optical_1995,veilleux_optical_1995,veilleux_new_1999} and AGN/starburst composite \citep{farrah_starburst_2003,yuan_role_2010}.
We conservatively treat this object as uncertain AGN/starburst composite  in the absence of multiwavelength evidence for the presence of an AGN.
A putative second, fainter nucleus 0.7\arcsec\ ($\sim 0.8$\,kpc) south-east is visible in the NICMOS images \citep{scoville_nicmos_2000}.
After \irass, the object was observed in the MIR with Palomar 5\,m \citep{carico_iras_1988}, with \iso/ISOPHOT and ISOCAM \citep{klaas_infrared_2001,siebenmorgen_mid_2001}, and then with    \spitzer/IRAC, IRS and MIPS.
The nucleus of IRAS\,15250+3609 remains nearly unresolved in all images.
The IRS LR staring-mode spectrum shows extremely deep silicate  10 and  $18\,\mu$m absorption,  PAH emission and an extremely red spectral slope in $\nu F_\nu$-space (see also \citealt{spoon_detection_2006,farrah_high-resolution_2007,imanishi_subaru_2011}).
Thus, the MIR SED appears very similar to IRAS\,11095-0238 and suggests the presence of a deeply buried AGN with significant star formation as proposed by \cite{farrah_starburst_2003}.
The absence of any strong emission features indicates that putative MIR emission-line producing regions  are heavily extincted (similar to, e.g., NGC\,4945; \citealt{perez-beaupuits_deeply_2011}).
COMICS observations of this source were performed in 2009 in the N8.8 and Q17.7 filters and a compact MIR nucleus is weakly detected in both images.
The PSF seems to have been very unstable during these observations, and together with the low S/N, no statement about the nuclear extension can be made.
The measured Q17.7 flux is significantly higher than the value published in \cite{imanishi_subaru_2011} but the N8.8 and Q17.7 fluxes agrees with the \spitzerr spectrophotometry well.
Therefore, we use the IRS spectrum to compute the 12 and $18\,\mu$m continuum emission estimates corrected for the silicate features.
\newline\end{@twocolumnfalse}]

\begin{figure}
   \centering
   \includegraphics[angle=0,width=8.500cm]{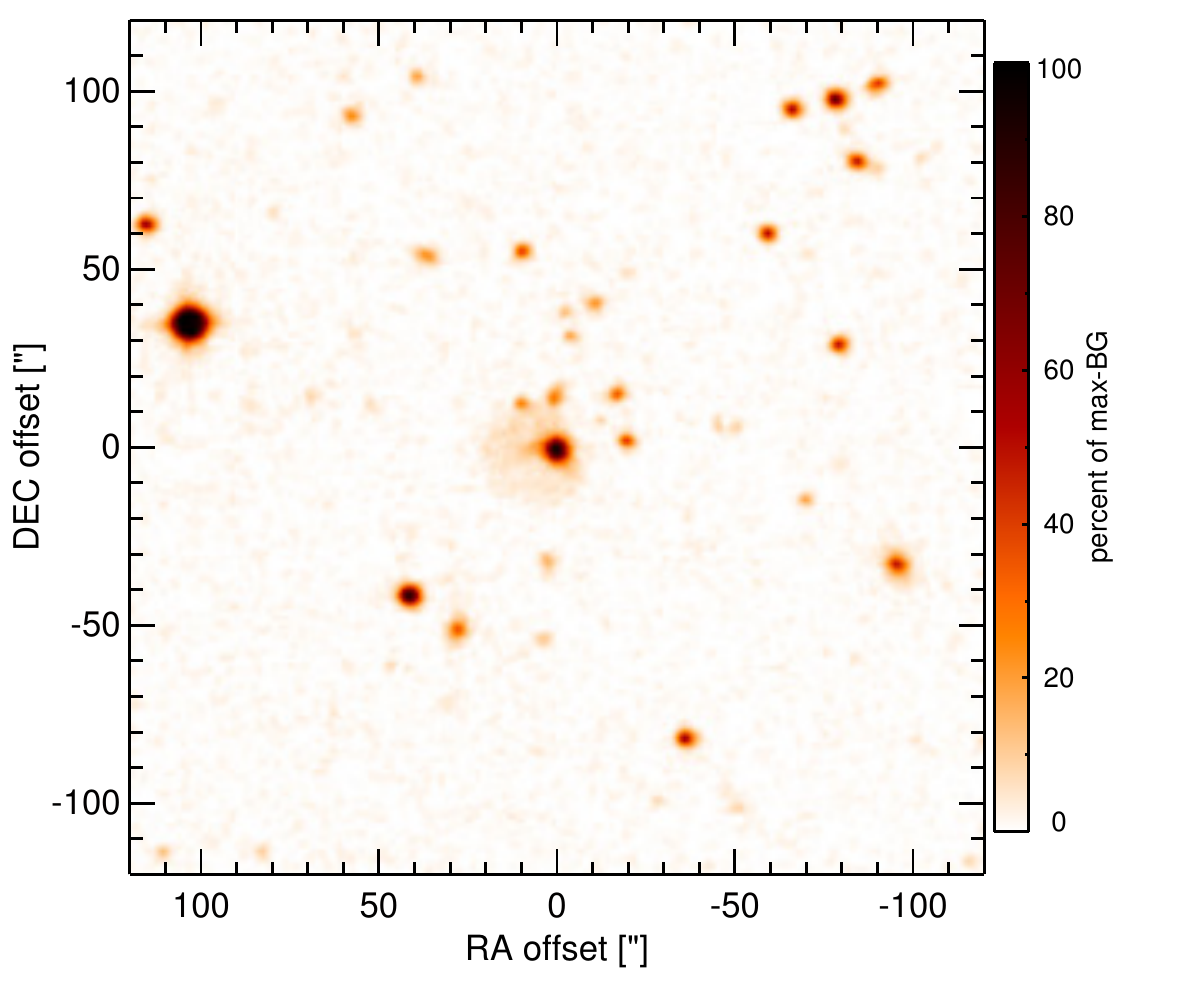}
    \caption{\label{fig:OPTim_IRAS15250+3609}
             Optical image (DSS, red filter) of IRAS\,15250+3609. Displayed are the central $4\arcmin$ with North up and East to the left. 
              The colour scaling is linear with white corresponding to the median background and black to the $0.01\%$ pixels with the highest intensity.  
           }
\end{figure}
\begin{figure}
   \centering
   \includegraphics[angle=0,height=3.11cm]{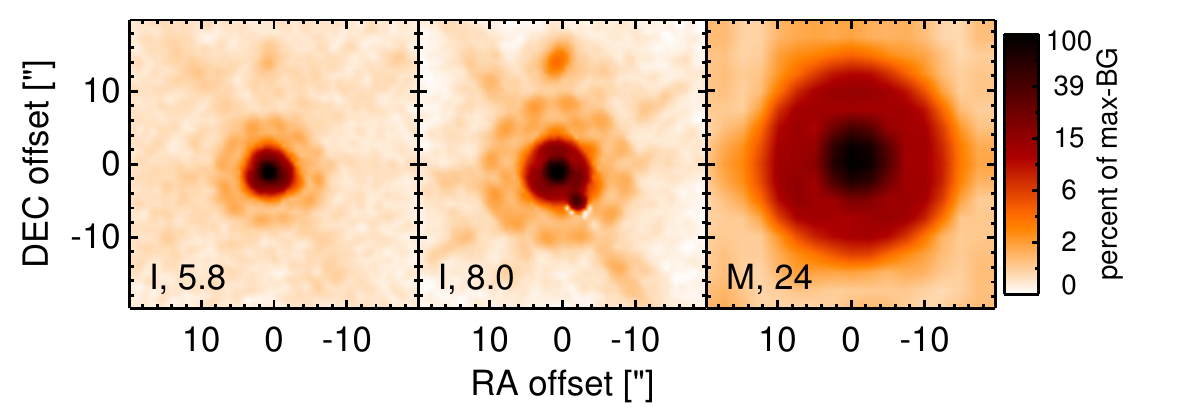}
    \caption{\label{fig:INTim_IRAS15250+3609}
             \spitzerr MIR images of IRAS\,15250+3609. Displayed are the inner $40\arcsec$ with North up and East to the left. The colour scaling is logarithmic with white corresponding to median background and black to the $0.1\%$ pixels with the highest intensity.
             The label in the bottom left states instrument and central wavelength of the filter in $\mu$m (I: IRAC, M: MIPS). 
             Note that the apparent off-nuclear compact source in the IRAC $8.0\,\mu$m image is an instrumental artefact.
           }
\end{figure}
\begin{figure}
   \centering
   \includegraphics[angle=0,height=3.11cm]{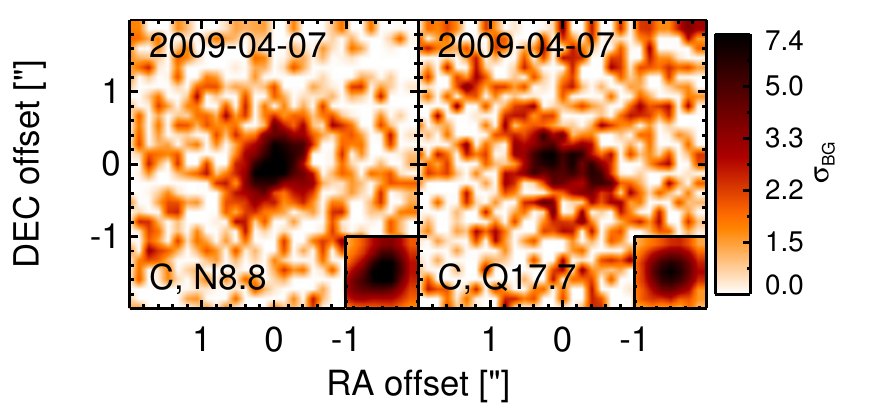}
    \caption{\label{fig:HARim_IRAS15250+3609}
             Subarcsecond-resolution MIR images of IRAS\,15250+3609 sorted by increasing filter wavelength. 
             Displayed are the inner $4\arcsec$ with North up and East to the left. 
             The colour scaling is logarithmic with white corresponding to median background and black to the $75\%$ of the highest intensity of all images in units of $\sigbg$.
             The inset image shows the central arcsecond of the PSF from the calibrator star, scaled to match the science target.
             The labels in the bottom left state instrument and filter names (C: COMICS, M: Michelle, T: T-ReCS, V: VISIR).
           }
\end{figure}
\begin{figure}
   \centering
   \includegraphics[angle=0,width=8.50cm]{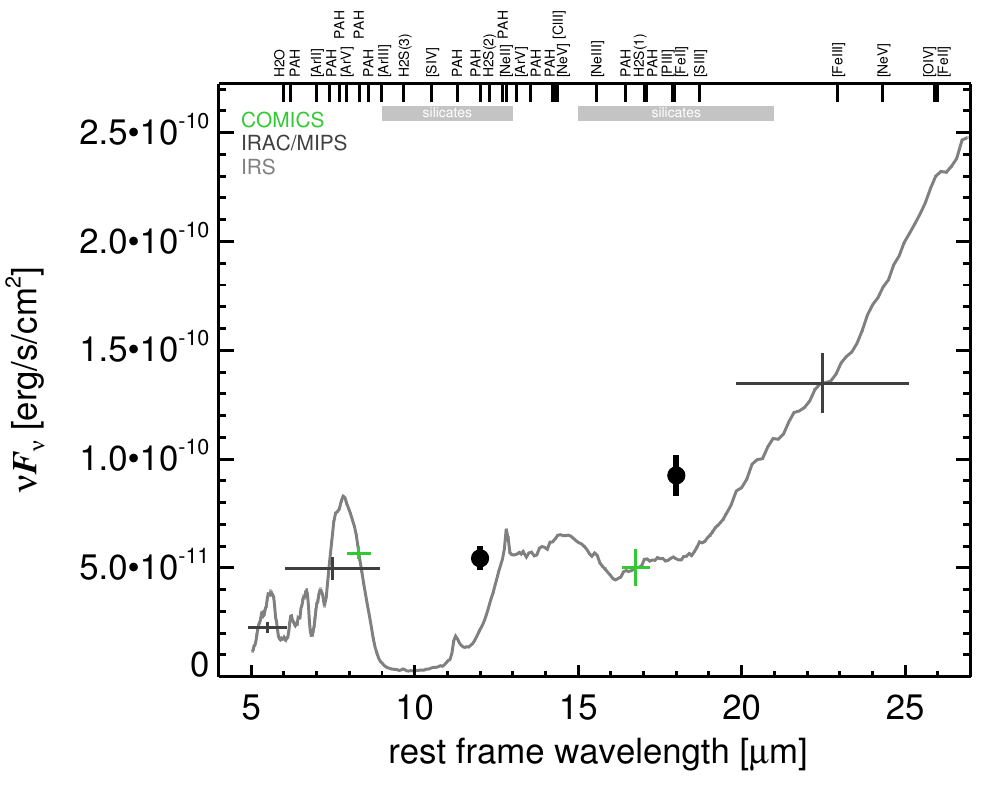}
   \caption{\label{fig:MISED_IRAS15250+3609}
      MIR SED of IRAS\,15250+3609. The description  of the symbols (if present) is the following.
      Grey crosses and  solid lines mark the \spitzer/IRAC, MIPS and IRS data. 
      The colour coding of the other symbols is: 
      green for COMICS, magenta for Michelle, blue for T-ReCS and red for VISIR data.
      Darker-coloured solid lines mark spectra of the corresponding instrument.
      The black filled circles mark the nuclear 12 and $18\,\mu$m  continuum emission estimate from the data.
      The ticks on the top axis mark positions of common MIR emission lines, while the light grey horizontal bars mark wavelength ranges affected by the silicate 10 and 18$\mu$m features.}
\end{figure}
\clearpage

\twocolumn[\begin{@twocolumnfalse}  
\subsection{I\,Zw\,1 -- UGC\,545 -- Mrk\,1502}\label{app:IZW001}
I\,Zw\,1 is a compact spiral galaxy at a redshift of $z=$ 0.0589 ($D\sim248$\,Mpc) with a Sy\,1n nucleus \citep{veron-cetty_catalogue_2010}, and a nuclear starburst ring \citep{schinnerer_molecular_1998,sosa-brito_integral_2001}.
Early MIR observation of this object were carried out by \cite{rieke_infrared_1972},\cite{rieke_infrared_1978}, \cite{lebofsky_extinction_1979}, \cite{roche_8-13_1984},\cite{sanders_continuum_1989},  \cite{elvis_atlas_1994}, and \cite{haas_dust_2000}.
In addition, the source was monitored over the $J$ to $N$ bands with approximately yearly coverage from 1987 to 1998 and no significant variability was found \citep{neugebauer_variability_1999}.
The first sub-arcsecond resolution $N$-band imaging of I\,Zw\,1 was carried out with Palomar 5\,m/MIRLIN in 1999 \citep{gorjian_10_2004}, Keck/LWS in 2000 \citep{soifer_high_2004}, and ESO 3.6\,m/TIMMI2 in 2001 \citep{galliano_mid-infrared_2005}.
In all cases an unresolved MIR nucleus without any host emission was detected, which is also the case for the \spitzer/IRAC 5.8$\,\mu$m and MIPS 24$\,\mu$m images.
The \spitzer/IRS LR staring-mode spectrum exhibits prominent silicate  10 and $18\,\mu$m emission, no significant PAH emission and a flat spectral slope in $\nu F_\nu$-space (see also \citealt{hao_detection_2005,weedman_mid-infrared_2005,schweitzer_extended_2008}).
The absence of the PAH emission is surprising with respect to the scenario of a nuclear starburst ring in I\,Zw\,1.
Subarcsecond-resolution $N$-band observation were performed with COMICS in the N11.7 filter in 2006 \citep{imanishi_subaru_2011} and with VISIR in two narrow $N$-band filters in 2010 (unpublished, to our knowledge).
None of observations was carried out under diffraction-limited conditions, and in particular the N11.7 image suffers from PSF instability.
However, the nucleus appears very compact in the ARIII acquisition image for which, unfortunately, no appropriate standard star observation is available. 
We adopt the nucleus to be unresolved as also found in the previous sub-arcsecond diffraction-limited observations mentioned above, and use the total Gaussian measurements for the nuclear flux. 
Note that our N11.7 flux is significantly higher  than the value in \cite{imanishi_subaru_2011} (probably owing to the issue above) but consistent with the VISIR measurements and the \spitzerr spectrophotometry. 
Note that I\,Zw\,1 was partially resolved in MIR interferometric observations with MIDI but with a dominating unresolved component smaller than 2.7\,pc \citep{burtscher_diversity_2013}.
\newline\end{@twocolumnfalse}]

\begin{figure}
   \centering
   \includegraphics[angle=0,width=8.500cm]{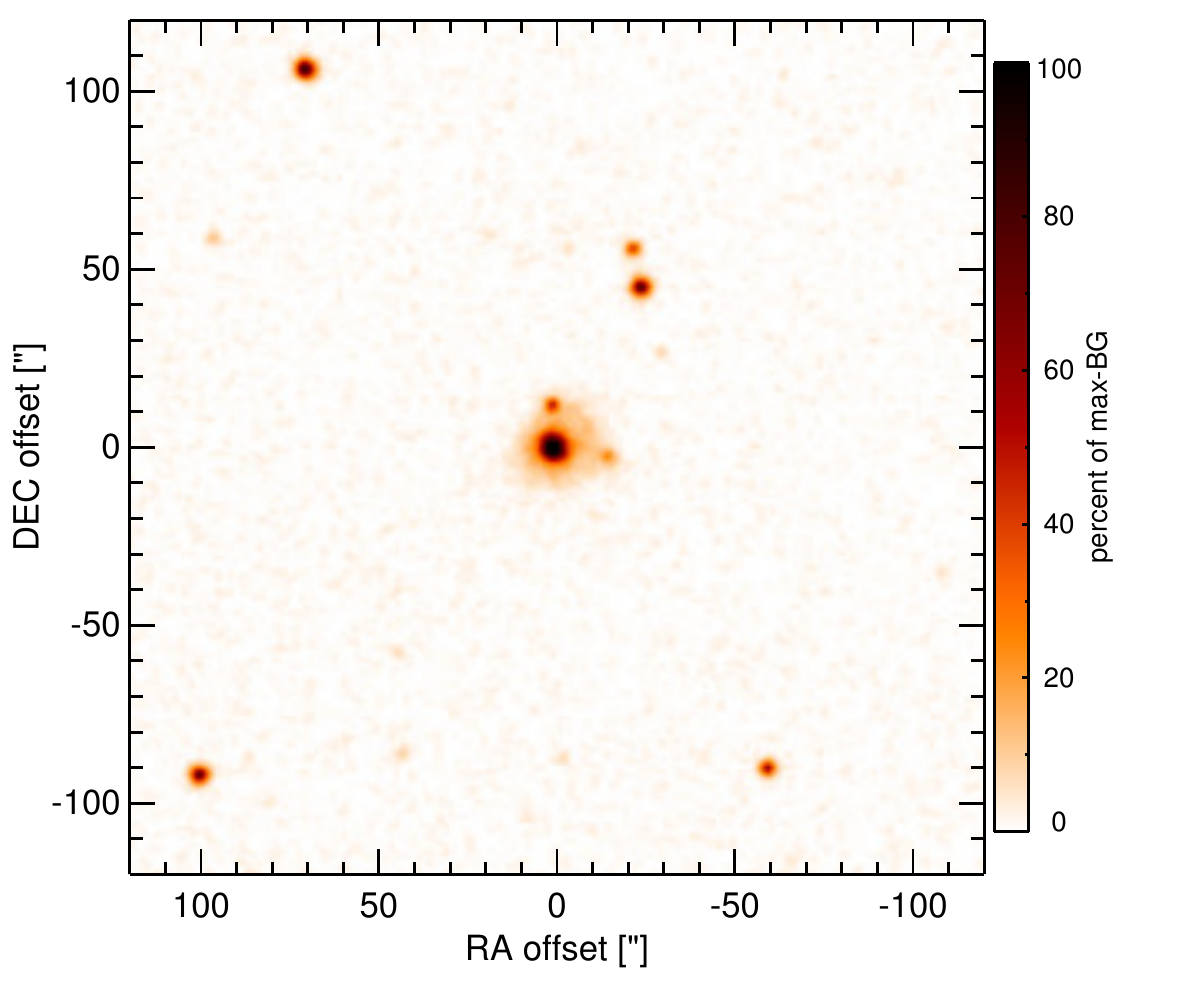}
    \caption{\label{fig:OPTim_IZW001}
             Optical image (DSS, red filter) of I\,Zw\,1. Displayed are the central $4\arcmin$ with North up and East to the left. 
              The colour scaling is linear with white corresponding to the median background and black to the $0.01\%$ pixels with the highest intensity.  
           }
\end{figure}
\begin{figure}
   \centering
   \includegraphics[angle=0,height=3.11cm]{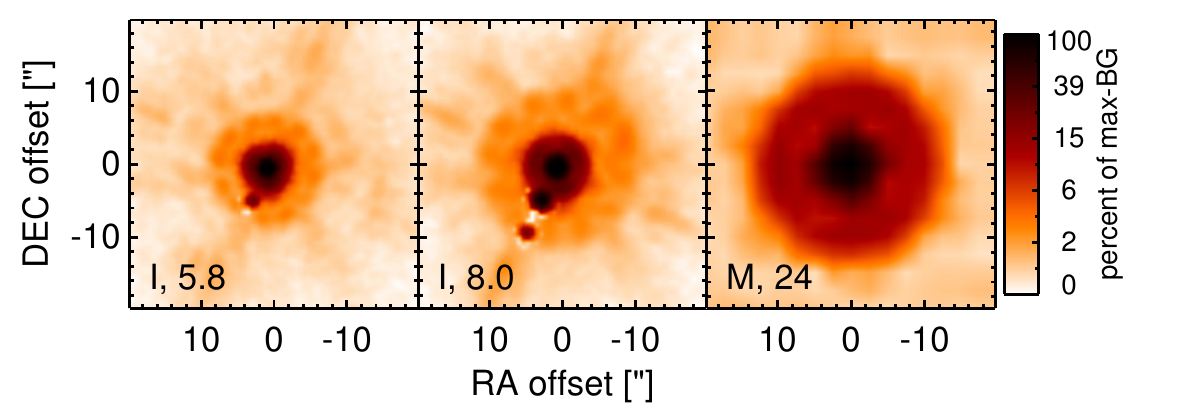}
    \caption{\label{fig:INTim_IZW001}
             \spitzerr MIR images of I\,Zw\,1. Displayed are the inner $40\arcsec$ with North up and East to the left. The colour scaling is logarithmic with white corresponding to median background and black to the $0.1\%$ pixels with the highest intensity.
             The label in the bottom left states instrument and central wavelength of the filter in $\mu$m (I: IRAC, M: MIPS).
             Note that the apparent off-nuclear compact sources in the IRAC 5.8 and $8.0\,\mu$m images are instrumental artefacts.
           }
\end{figure}
\begin{figure}
   \centering
   \includegraphics[angle=0,height=3.11cm]{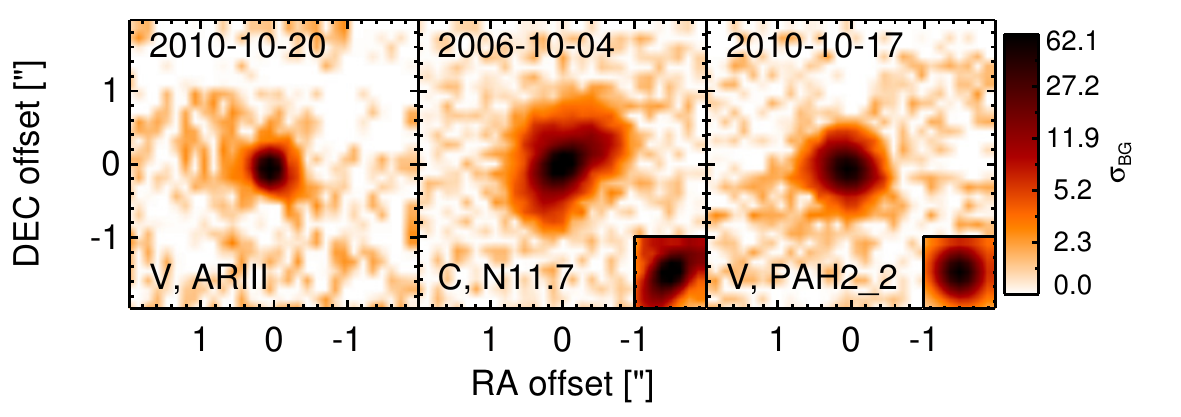}
    \caption{\label{fig:HARim_IZW001}
             Subarcsecond-resolution MIR images of I\,Zw\,1 sorted by increasing filter wavelength. 
             Displayed are the inner $4\arcsec$ with North up and East to the left. 
             The colour scaling is logarithmic with white corresponding to median background and black to the $75\%$ of the highest intensity of all images in units of $\sigbg$.
             The inset image shows the central arcsecond of the PSF from the calibrator star, scaled to match the science target.
             The labels in the bottom left state instrument and filter names (C: COMICS, M: Michelle, T: T-ReCS, V: VISIR).
           }
\end{figure}
\begin{figure}
   \centering
   \includegraphics[angle=0,width=8.50cm]{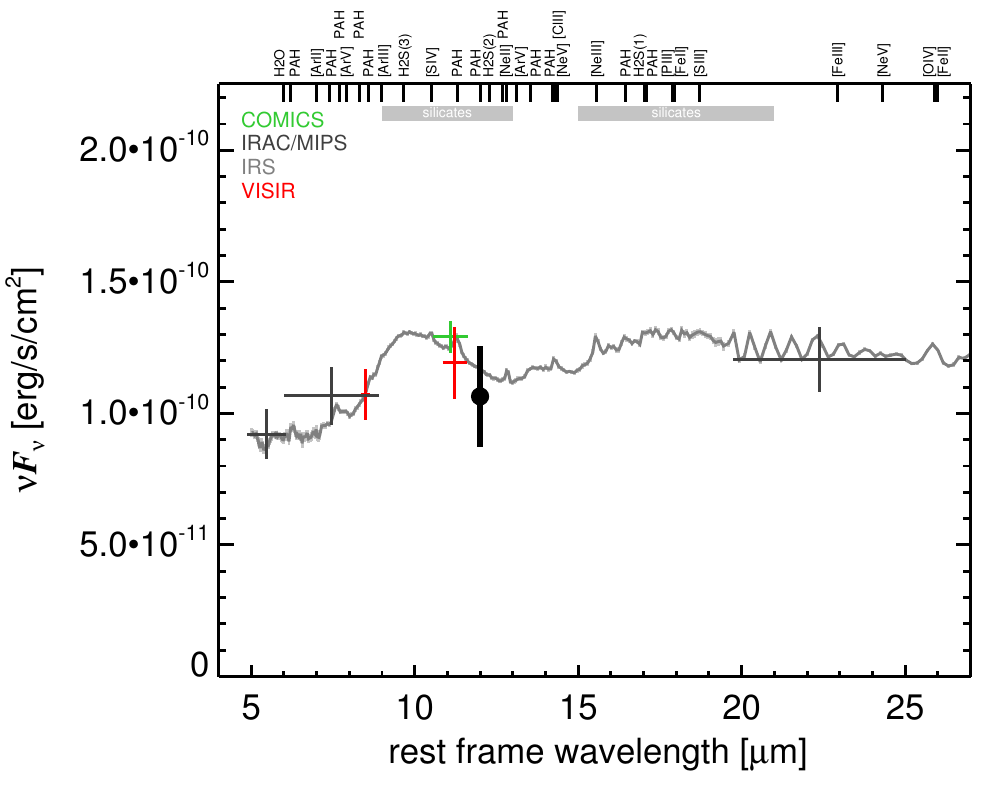}
   \caption{\label{fig:MISED_IZW001}
      MIR SED of I\,Zw\,1. The description  of the symbols (if present) is the following.
      Grey crosses and  solid lines mark the \spitzer/IRAC, MIPS and IRS data. 
      The colour coding of the other symbols is: 
      green for COMICS, magenta for Michelle, blue for T-ReCS and red for VISIR data.
      Darker-coloured solid lines mark spectra of the corresponding instrument.
      The black filled circles mark the nuclear 12 and $18\,\mu$m  continuum emission estimate from the data.
      The ticks on the top axis mark positions of common MIR emission lines, while the light grey horizontal bars mark wavelength ranges affected by the silicate 10 and 18$\mu$m features.}
\end{figure}
\clearpage

\twocolumn[\begin{@twocolumnfalse}  
\subsection{LEDA\,13946 -- 2MASX\,J03502377-5018354}\label{app:LEDA013946}
LEDA\,13946 is the smaller of a pair of two interacting spiral galaxies at a redshift of $z=$ 0.0365 ($D\sim166\,$Mpc) with ESO\,201-4 being the northern component (nuclear separation $\sim28\arcsec$).
It harbours a little-studied Sy\,2 nucleus \citep{parisi_accurate_2009} that belongs to the nine-month BAT AGN sample.
LEDA\,13946 has no reported MIR detection in the literature and appears as a compact source blended with ESO\,201-4 in the \wisee images.
A \spitzer/IRS LR staring-mode spectrum is available but suffers from low S/N.
It indicates strong PAH emission, at best weak silicate absorption, and a red spectral slope in $\nu F_\nu$-space, i.e., dominating star formation.
We observed LEDA\,13946 with VISIR in three narrow $N$-band filters in 2009 but failed to detect any emission.
The corresponding flux upper limits are less stringent than the IRS spectrum.
Therefore, we use the latter to compute an upper limit on the 12$\,\mu$m continuum emission estimate on the AGN in LEDA\,13946.
\newline\end{@twocolumnfalse}]

\begin{figure}
   \centering
   \includegraphics[angle=0,width=8.500cm]{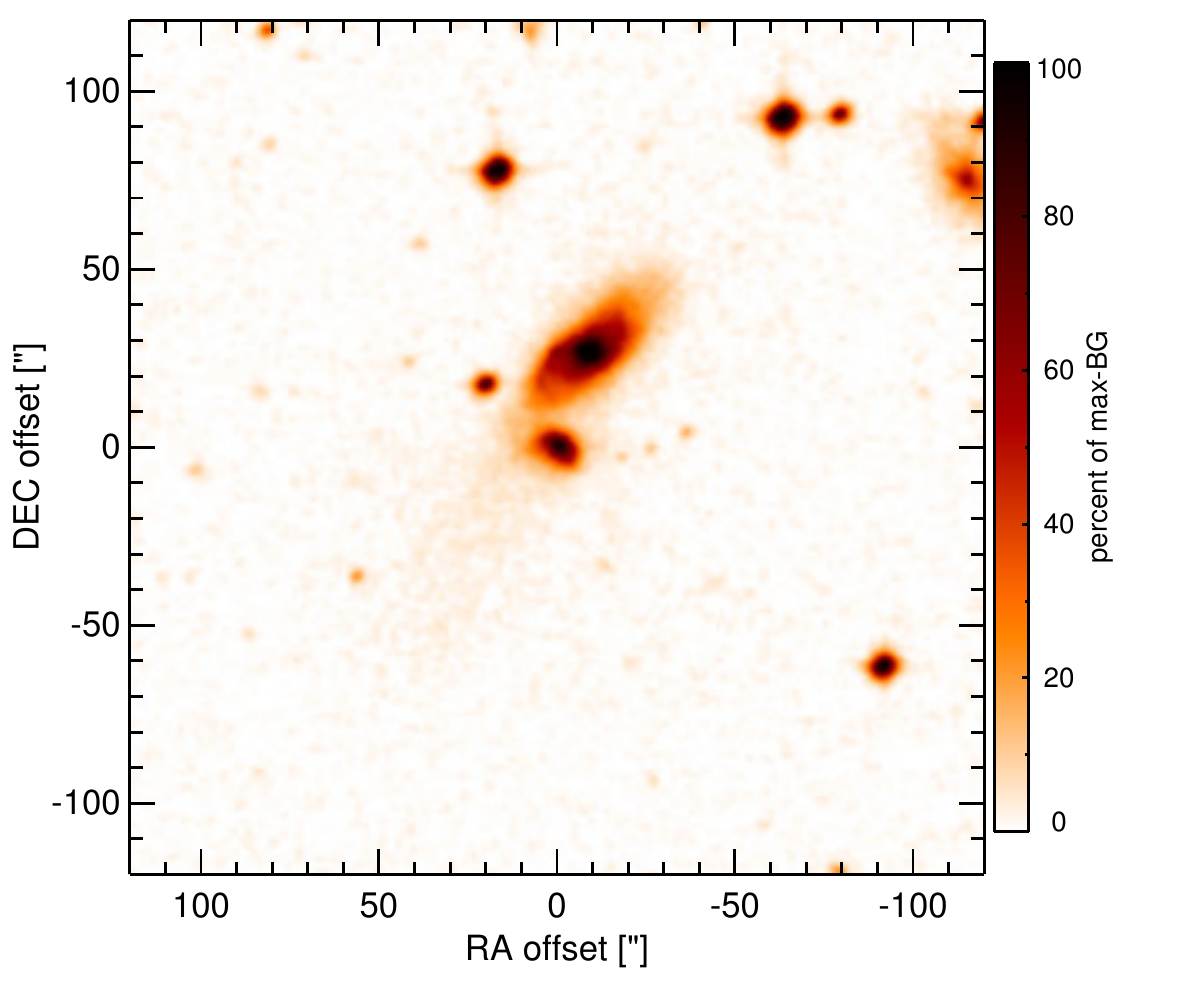}
    \caption{\label{fig:OPTim_LEDA013946}
             Optical image (DSS, red filter) of LEDA\,13946. Displayed are the central $4\arcmin$ with North up and East to the left. 
              The colour scaling is linear with white corresponding to the median background and black to the $0.01\%$ pixels with the highest intensity.  
           }
\end{figure}
\begin{figure}
   \centering
   \includegraphics[angle=0,width=8.50cm]{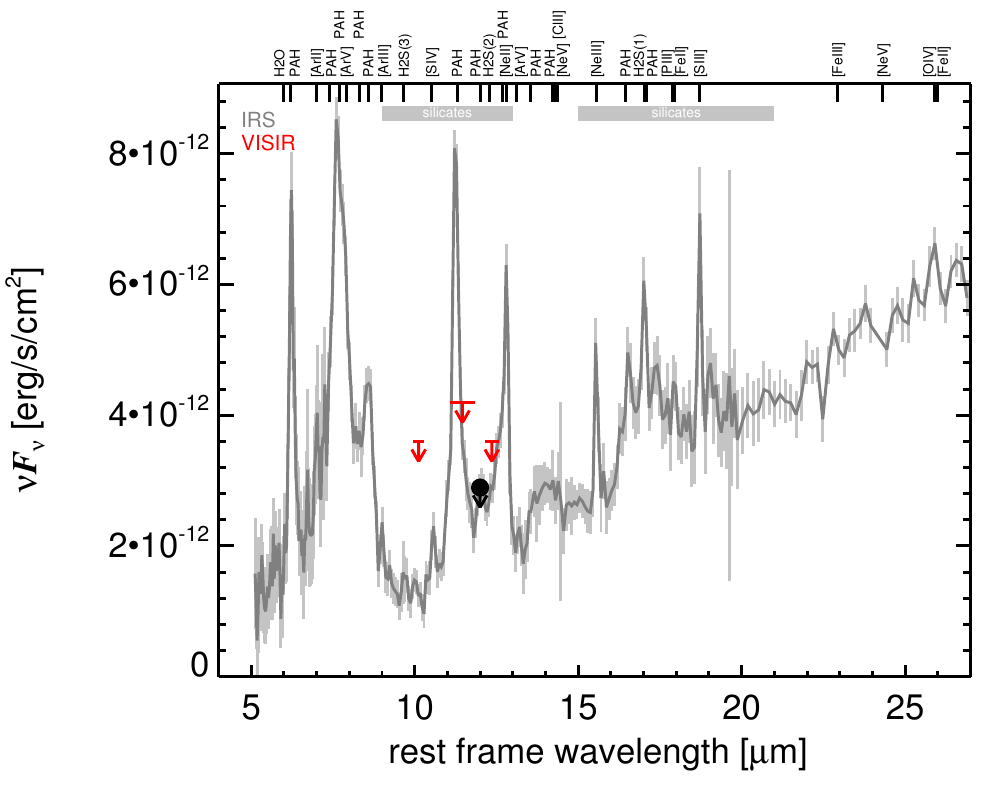}
   \caption{\label{fig:MISED_LEDA013946}
      MIR SED of LEDA\,13946. The description  of the symbols (if present) is the following.
      Grey crosses and  solid lines mark the \spitzer/IRAC, MIPS and IRS data. 
      The colour coding of the other symbols is: 
      green for COMICS, magenta for Michelle, blue for T-ReCS and red for VISIR data.
      Darker-coloured solid lines mark spectra of the corresponding instrument.
      The black filled circles mark the nuclear 12 and $18\,\mu$m  continuum emission estimate from the data.
      The ticks on the top axis mark positions of common MIR emission lines, while the light grey horizontal bars mark wavelength ranges affected by the silicate 10 and 18$\mu$m features.}
\end{figure}
\clearpage

\twocolumn[\begin{@twocolumnfalse}  
\subsection{LEDA\,170194 -- XSS\,J12389-1614 -- 2MASX\,J12390630-1610472}\label{app:LEDA170194}
LEDA\,170194 is a low-inclination spiral galaxy at a redshift of $z=$ 0.0367 ($D\sim173\,$Mpc) with a little-studied Sy\,2 nucleus \citep{veron-cetty_catalogue_2010}, which was discovered with  \textit{RXTE} and \textit{INTEGRAL} in X-rays \citep{sazonov_identification_2005,masetti_unveiling_2006-1}.
It also belongs to the nine-month BAT AGN sample.
No MIR observations are reported in the literature and no archival \spitzerr data are available for LEDA\,170194.
It appears as a compact source in the \wisee images.
We observed this object with VISIR in three narrow $N$-band filters in 2009 and detected an unresolved nucleus without further host emission.
The flux uncertainties are too large to constrain the nuclear MIR SED.
\newline\end{@twocolumnfalse}]

\begin{figure}
   \centering
   \includegraphics[angle=0,width=8.500cm]{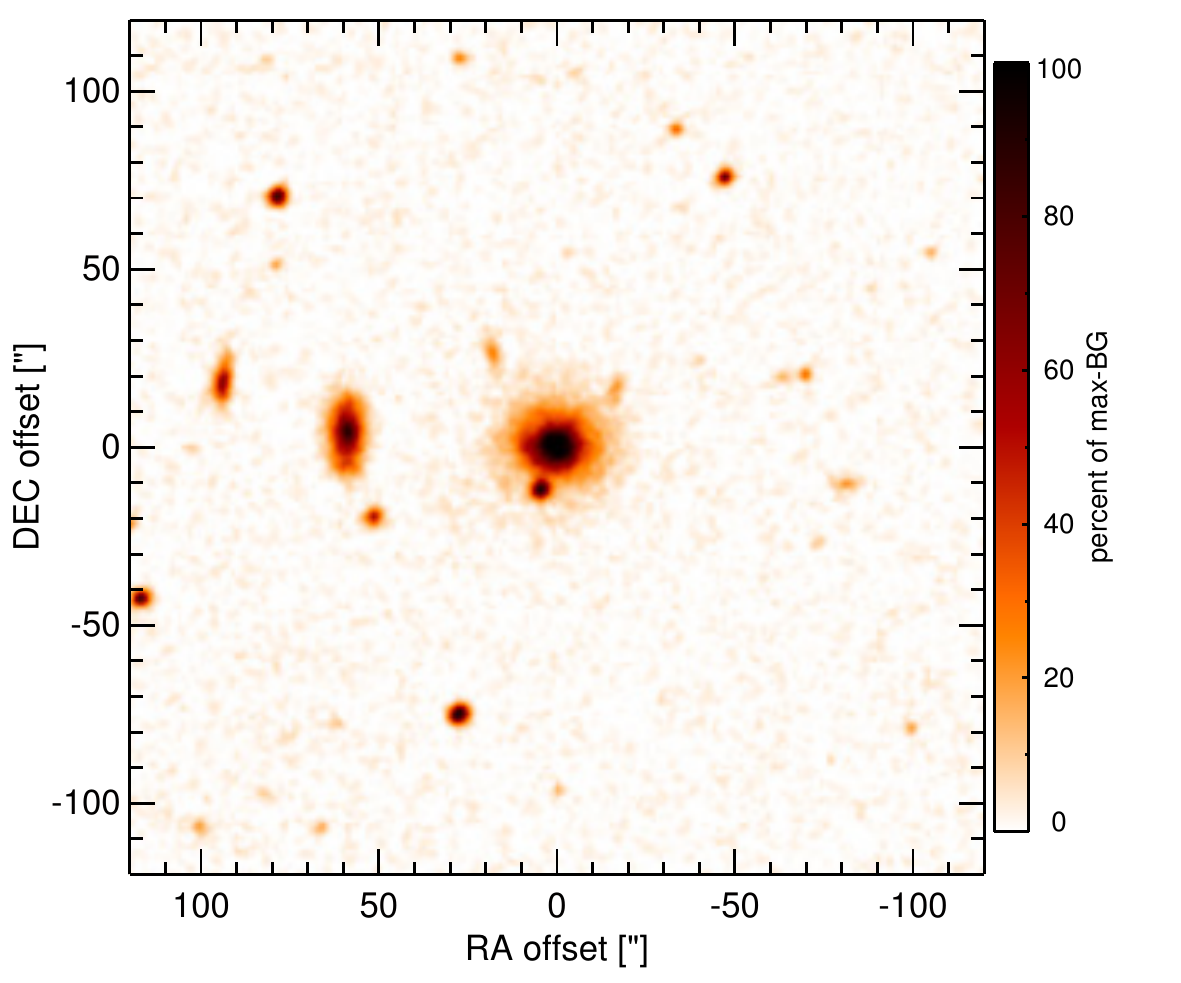}
    \caption{\label{fig:OPTim_LEDA170194}
             Optical image (DSS, red filter) of LEDA\,170194. Displayed are the central $4\arcmin$ with North up and East to the left. 
              The colour scaling is linear with white corresponding to the median background and black to the $0.01\%$ pixels with the highest intensity.  
           }
\end{figure}
\begin{figure}
   \centering
   \includegraphics[angle=0,height=3.11cm]{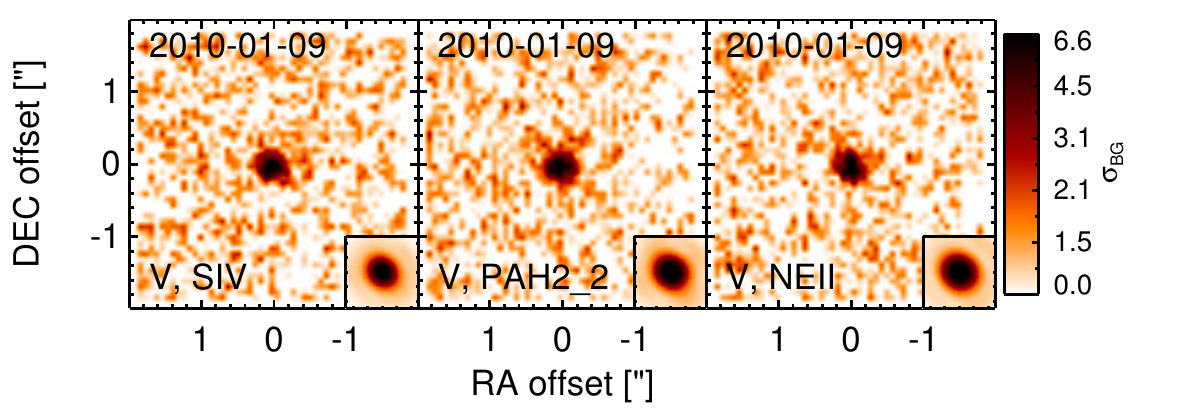}
    \caption{\label{fig:HARim_LEDA170194}
             Subarcsecond-resolution MIR images of LEDA\,170194 sorted by increasing filter wavelength. 
             Displayed are the inner $4\arcsec$ with North up and East to the left. 
             The colour scaling is logarithmic with white corresponding to median background and black to the $75\%$ of the highest intensity of all images in units of $\sigbg$.
             The inset image shows the central arcsecond of the PSF from the calibrator star, scaled to match the science target.
             The labels in the bottom left state instrument and filter names (C: COMICS, M: Michelle, T: T-ReCS, V: VISIR).
           }
\end{figure}
\begin{figure}
   \centering
   \includegraphics[angle=0,width=8.50cm]{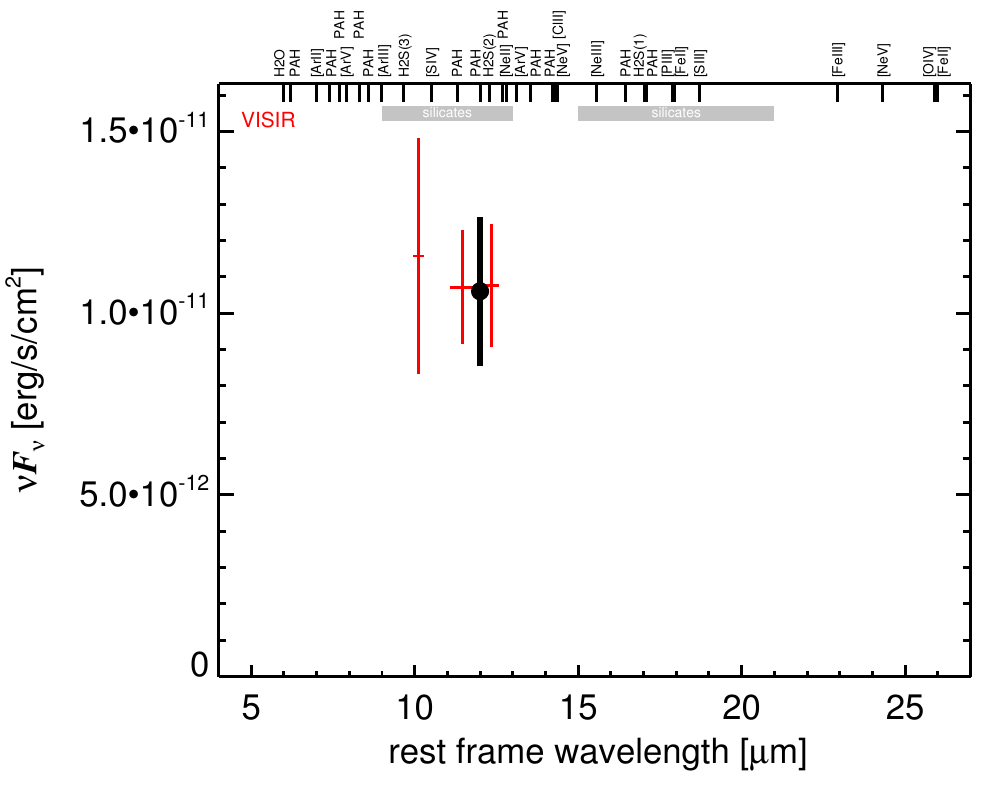}
   \caption{\label{fig:MISED_LEDA170194}
      MIR SED of LEDA\,170194. The description  of the symbols (if present) is the following.
      Grey crosses and  solid lines mark the \spitzer/IRAC, MIPS and IRS data. 
      The colour coding of the other symbols is: 
      green for COMICS, magenta for Michelle, blue for T-ReCS and red for VISIR data.
      Darker-coloured solid lines mark spectra of the corresponding instrument.
      The black filled circles mark the nuclear 12 and $18\,\mu$m  continuum emission estimate from the data.
      The ticks on the top axis mark positions of common MIR emission lines, while the light grey horizontal bars mark wavelength ranges affected by the silicate 10 and 18$\mu$m features.}
\end{figure}
\clearpage

\twocolumn[\begin{@twocolumnfalse}  
\subsection{LEDA\,178130 -- XSS\,J05054-2348 -- 2MASX\,J05054575-2351139}\label{app:LEDA178130}
LEDA\,178130 is a morphologically-undetermined galaxy at a redshift of $z=$ 0.0350 ($D\sim160\,$Mpc) with a little-studied Sy\,2 nucleus \citep{veron-cetty_catalogue_2010} discovered with \textit{RXTE} in X-rays \citep{revnivtsev_identification_2006,parisi_accurate_2009}.
It belongs as well to the nine-month BAT AGN sample.
No MIR observations are reported but an archival \spitzer/IRS LR staring-mode spectrum is available.
LEDA\,178130 appears as a compact source in \wise.
The IRS spectrum exhibits, unexpectedly to a type~II AGN, silicate emission, very weak PAH features and an emission peak at $\sim18\,\mu$m in $\nu F_\nu$-space.
The arcsecond-scale MIR SED thus rather indicates an unobscured AGN without significant star formation. 
We imaged this object with VISIR in three narrow $N$-band filters in 2009 and detected a possibly marginally resolved nucleus without any further host emission (FWHM$\sim0.5\arcsec\sim0.4\,$kpc).
However, at least another epoch of subarcsecond MIR imaging is required to confirm this extension.
The nuclear VISIR photometry is consistent with the IRS spectrum, but it would be significantly lower if the presence of subarcsecond-extended emission can be verified.
\newline\end{@twocolumnfalse}]

\begin{figure}
   \centering
   \includegraphics[angle=0,width=8.500cm]{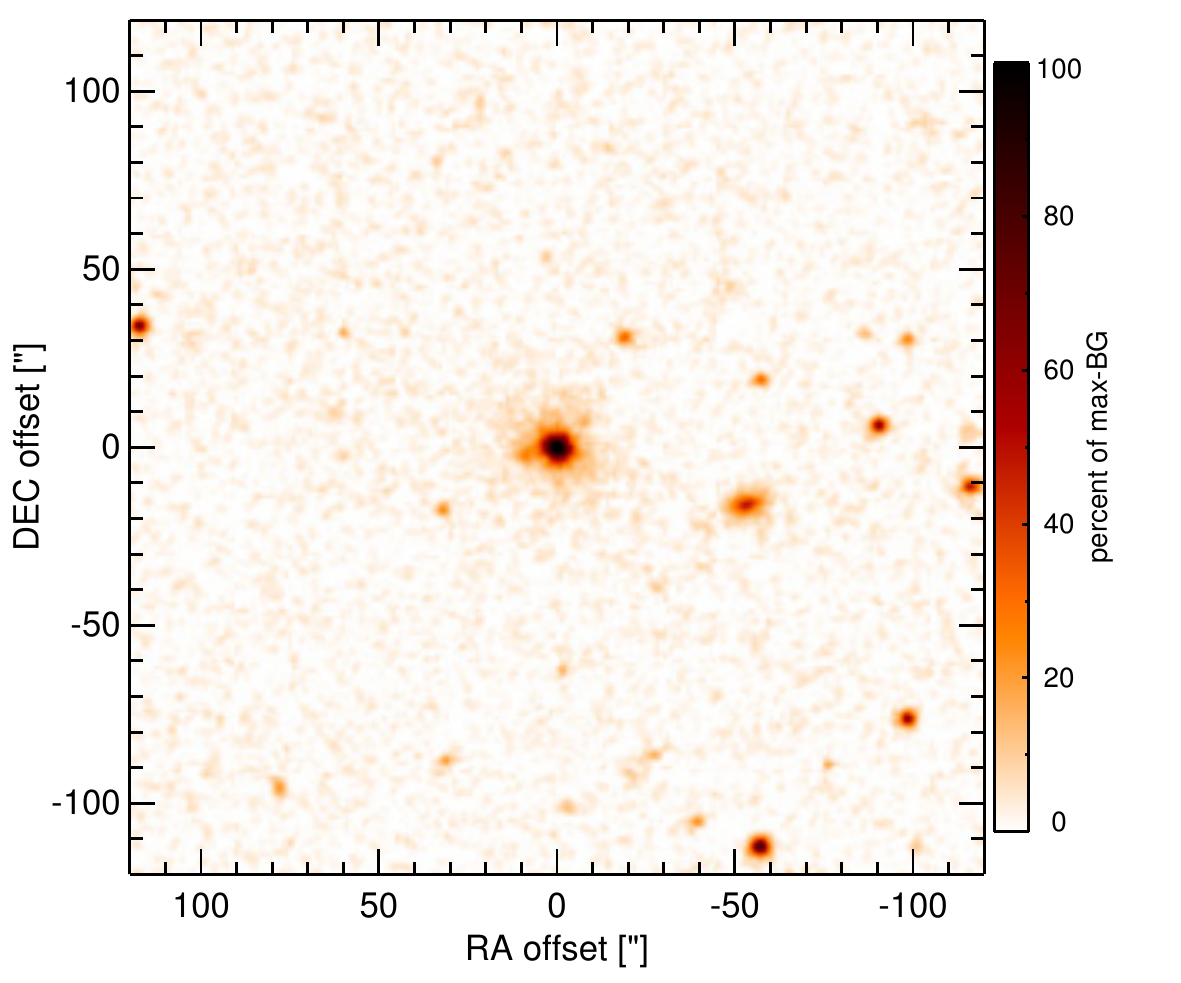}
    \caption{\label{fig:OPTim_LEDA178130}
             Optical image (DSS, red filter) of LEDA\,178130. Displayed are the central $4\arcmin$ with North up and East to the left. 
              The colour scaling is linear with white corresponding to the median background and black to the $0.01\%$ pixels with the highest intensity.  
           }
\end{figure}
\begin{figure}
   \centering
   \includegraphics[angle=0,height=3.11cm]{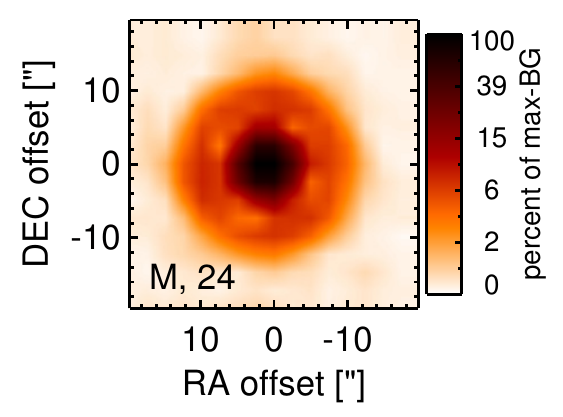}
    \caption{\label{fig:INTim_LEDA178130}
             \spitzerr MIR images of LEDA\,178130. Displayed are the inner $40\arcsec$ with North up and East to the left. The colour scaling is logarithmic with white corresponding to median background and black to the $0.1\%$ pixels with the highest intensity.
             The label in the bottom left states instrument and central wavelength of the filter in $\mu$m (I: IRAC, M: MIPS). 
           }
\end{figure}
\begin{figure}
   \centering
   \includegraphics[angle=0,height=3.11cm]{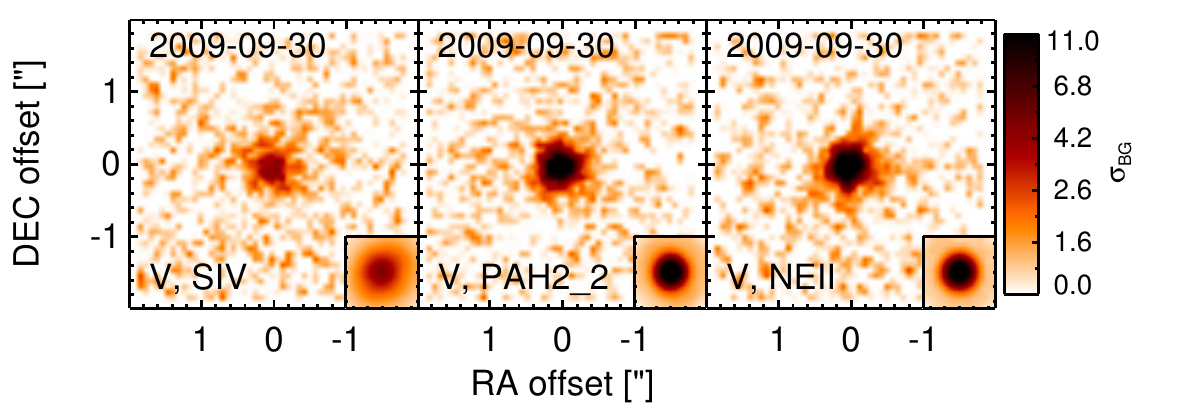}
    \caption{\label{fig:HARim_LEDA178130}
             Subarcsecond-resolution MIR images of LEDA\,178130 sorted by increasing filter wavelength. 
             Displayed are the inner $4\arcsec$ with North up and East to the left. 
             The colour scaling is logarithmic with white corresponding to median background and black to the $75\%$ of the highest intensity of all images in units of $\sigbg$.
             The inset image shows the central arcsecond of the PSF from the calibrator star, scaled to match the science target.
             The labels in the bottom left state instrument and filter names (C: COMICS, M: Michelle, T: T-ReCS, V: VISIR).
           }
\end{figure}
\begin{figure}
   \centering
   \includegraphics[angle=0,width=8.50cm]{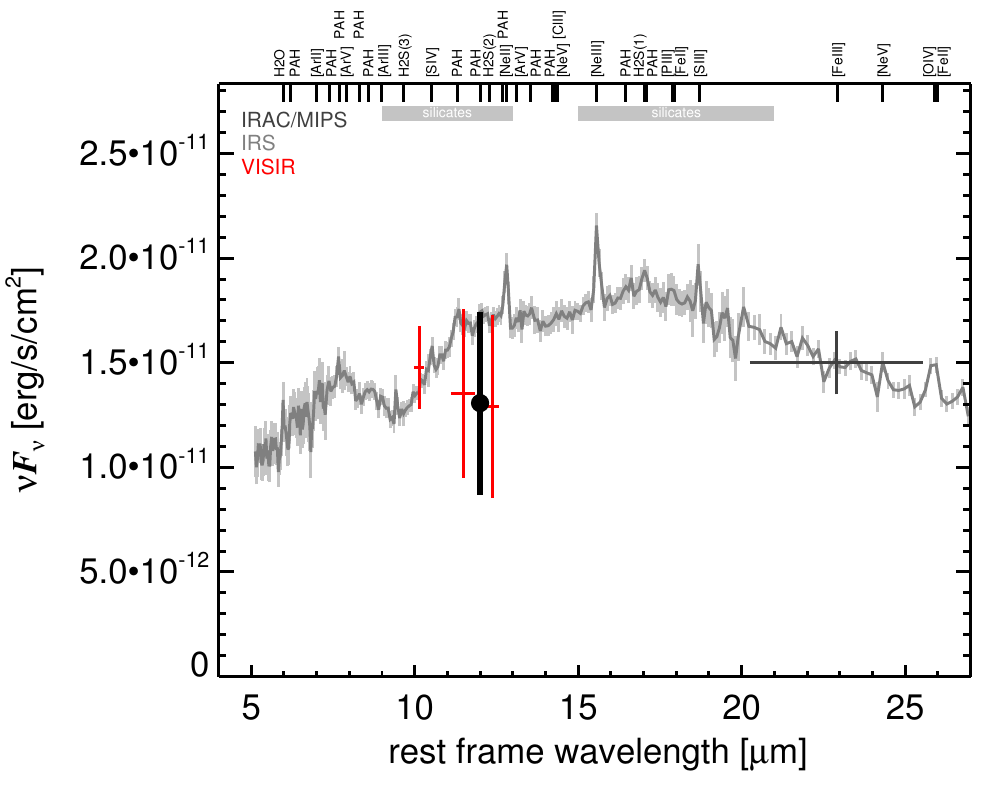}
   \caption{\label{fig:MISED_LEDA178130}
      MIR SED of LEDA\,178130. The description  of the symbols (if present) is the following.
      Grey crosses and  solid lines mark the \spitzer/IRAC, MIPS and IRS data. 
      The colour coding of the other symbols is: 
      green for COMICS, magenta for Michelle, blue for T-ReCS and red for VISIR data.
      Darker-coloured solid lines mark spectra of the corresponding instrument.
      The black filled circles mark the nuclear 12 and $18\,\mu$m  continuum emission estimate from the data.
      The ticks on the top axis mark positions of common MIR emission lines, while the light grey horizontal bars mark wavelength ranges affected by the silicate 10 and 18$\mu$m features.}
\end{figure}
\clearpage

\twocolumn[\begin{@twocolumnfalse}  
\subsection{LEDA\,549777 -- 2MASX\,J06403799-4321211}\label{app:LEDA549777}
The galaxy LEDA\,549777 at a redshift of $z=$ 0.061 ($D \sim 258$\,Mpc) was discovered to contain an AGN with \swift/BAT observations \citep{tueller_swift_2008}, which thus belongs to the 9 month BAT AGN sample.
It was classified as a Sy\,2.0 by \cite{parisi_accurate_2009}.
No \spitzerr observations have targeted LEDA\,549777.
It appears as a compact source in the \wisee images except the band~3 image, which shows a faint ring-like structure with $\sim1\arcmin\sim80\,$kpc diameter surrounding the nucleus. 
We observed LEDA\,549777 with VISIR in the NEII\_2 filter in 2009 and weakly detected a compact MIR source.
The low S/N of the detection and lack of data does not allow  us to determine the nuclear extension at subarcsecond scales in the MIR.
The nuclear NEII\_2 flux is consistent with the \wisee band~3 ($\sim12\,\mu$m) flux. 
\newline\end{@twocolumnfalse}]

\begin{figure}
   \centering
   \includegraphics[angle=0,width=8.500cm]{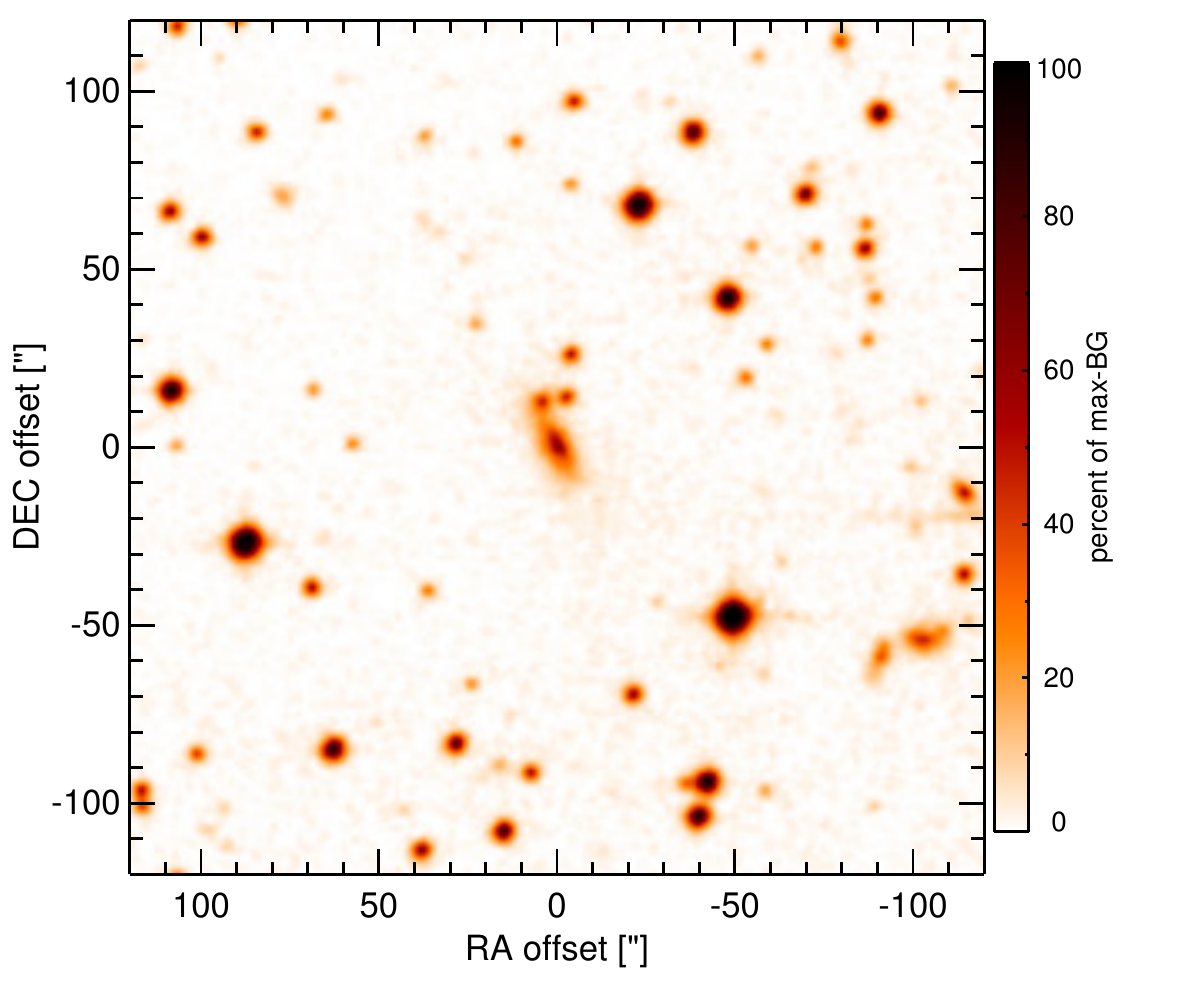}
    \caption{\label{fig:OPTim_LEDA549777}
             Optical image (DSS, red filter) of LEDA\,549777. Displayed are the central $4\arcmin$ with North up and East to the left. 
              The colour scaling is linear with white corresponding to the median background and black to the $0.01\%$ pixels with the highest intensity.  
           }
\end{figure}
\begin{figure}
   \centering
   \includegraphics[angle=0,height=3.11cm]{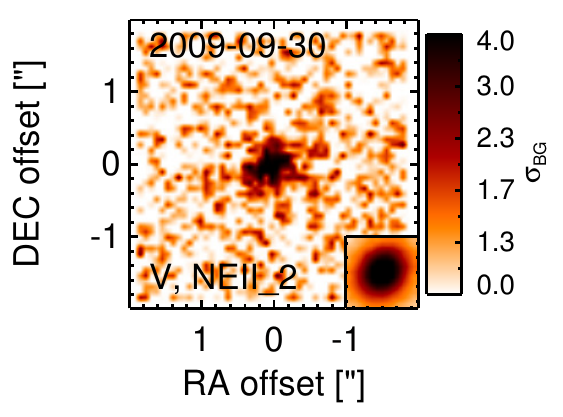}
    \caption{\label{fig:HARim_LEDA549777}
             Subarcsecond-resolution MIR images of LEDA\,549777 sorted by increasing filter wavelength. 
             Displayed are the inner $4\arcsec$ with North up and East to the left. 
             The colour scaling is logarithmic with white corresponding to median background and black to the $75\%$ of the highest intensity of all images in units of $\sigbg$.
             The inset image shows the central arcsecond of the PSF from the calibrator star, scaled to match the science target.
             The labels in the bottom left state instrument and filter names (C: COMICS, M: Michelle, T: T-ReCS, V: VISIR).
           }
\end{figure}
\begin{figure}
   \centering
   \includegraphics[angle=0,width=8.50cm]{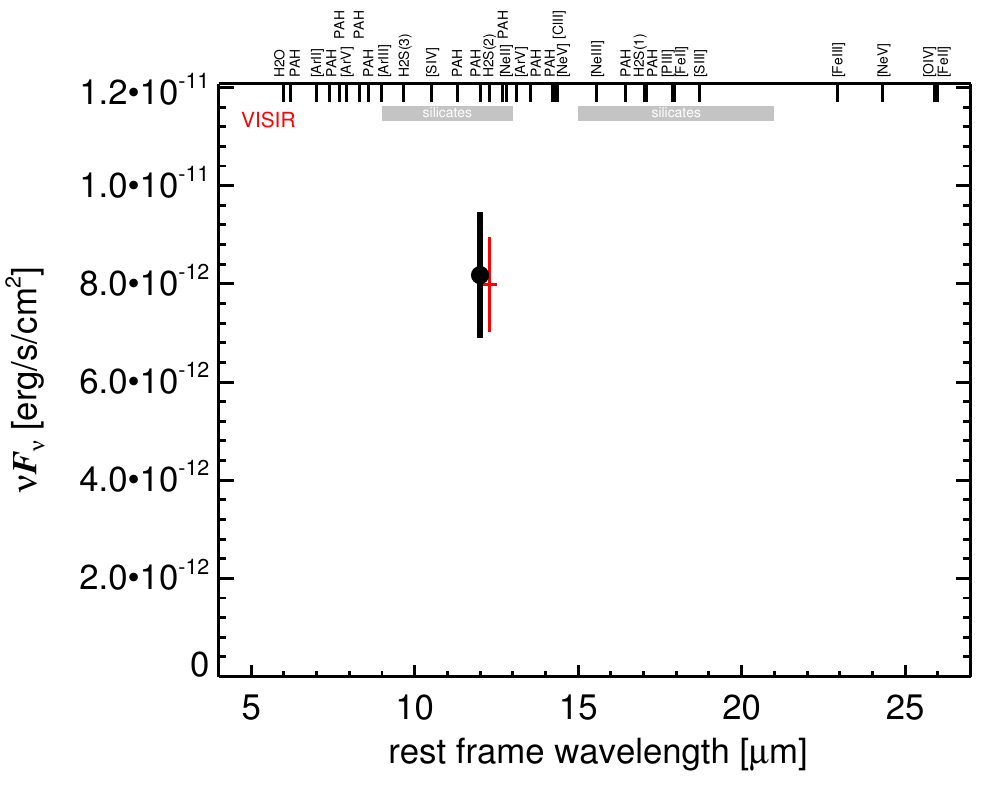}
   \caption{\label{fig:MISED_LEDA549777}
      MIR SED of LEDA\,549777. The description  of the symbols (if present) is the following.
      Grey crosses and  solid lines mark the \spitzer/IRAC, MIPS and IRS data. 
      The colour coding of the other symbols is: 
      green for COMICS, magenta for Michelle, blue for T-ReCS and red for VISIR data.
      Darker-coloured solid lines mark spectra of the corresponding instrument.
      The black filled circles mark the nuclear 12 and $18\,\mu$m  continuum emission estimate from the data.
      The ticks on the top axis mark positions of common MIR emission lines, while the light grey horizontal bars mark wavelength ranges affected by the silicate 10 and 18$\mu$m features.}
\end{figure}
\clearpage

\twocolumn[\begin{@twocolumnfalse}  
\subsection{M51a -- NGC\,5194 -- Whirlpool Galaxy}\label{app:M051A}
M51a is a face-on grand-design spiral galaxy at a distance of $D=$ $8.09 \pm 0.87$\,Mpc  (NED redshift-independent median) interacting with the smaller M51b galaxy.
It hosts a Sy\,2 nucleus \citep{veron-cetty_catalogue_2010} sometimes also classified as a LINER \citep{heckman_optical_1980} and possesses a biconical NLR, a weak radio jet \citep{ford_bubbles_1985,ho_radio_2001} and a nuclear water maser \citep{hagiwara_water_2001,hagiwara_low-luminosity_2007}.
The first MIR observations of M51a were performed by \cite{kleinmann_infrared_1970} and \cite{rieke_10_1978}, whereas the nucleus remained undetected. 
The \irass and \isoo observations on the other hand were dominated by the circum-nuclear star formation (e.g., \citealt{sauvage_isocam_1996,roussel_atlas_2001,forster_schreiber_warm_2004}).
The nucleus of M51a remained also undetected in the first subarcsecond MIR observations with Palomar 5\,m/MIRLIN \citep{gorjian_10_2004}.
Finally, the \spitzer/IRAC and MIPS images show a compact nucleus in the centre of the extended spiral emission of M51a.
Because we measure the photometry of the nuclear component only, our IRAC $5.8$ and $8.0\,\mu$m and MIPS $24\,\mu$m fluxes are significantly lower than the values in the literature (e.g., \citealt{dale_infrared_2005,munoz-mateos_radial_2009}).
Owing to the complex emission morphology of M51a in the MIR, the IRS LR mapping-mode PBCD spectrum is not reliable.
However, it roughly matches the IRAC and MIPS photometry and shows strong PAH emission with possibly weak silicate absorption and a red spectral slope in $\nu F_\nu$-space, indicating star formation in the central 4\arcsec $\sim 160$\,pc region (see e.g., \citealt{smith_mid-infrared_2007,wu_spitzer/irs_2009,goulding_towards_2009} for accurate versions).
We observed M51a with Michelle in 2010 in two $N$-band filters and detected a possibly marginally resolved nucleus but no further host emission (FWHM $\sim 0.8\arcsec \sim 31\,$pc).
However, the observations were made under bad conditions and at least a second epoch is required to verify this extension.
The corresponding nuclear photometry is $\sim73\%$ lower than the IRS spectrum but still consistent with pure star formation, which also matches the detection of PAH 3.3\,$\mu$m emission in the inner $\sim 0.8\arcsec \sim 31\,$pc of M51a \citep{oi_comparison_2010}.
\newline\end{@twocolumnfalse}]

\begin{figure}
   \centering
   \includegraphics[angle=0,width=8.500cm]{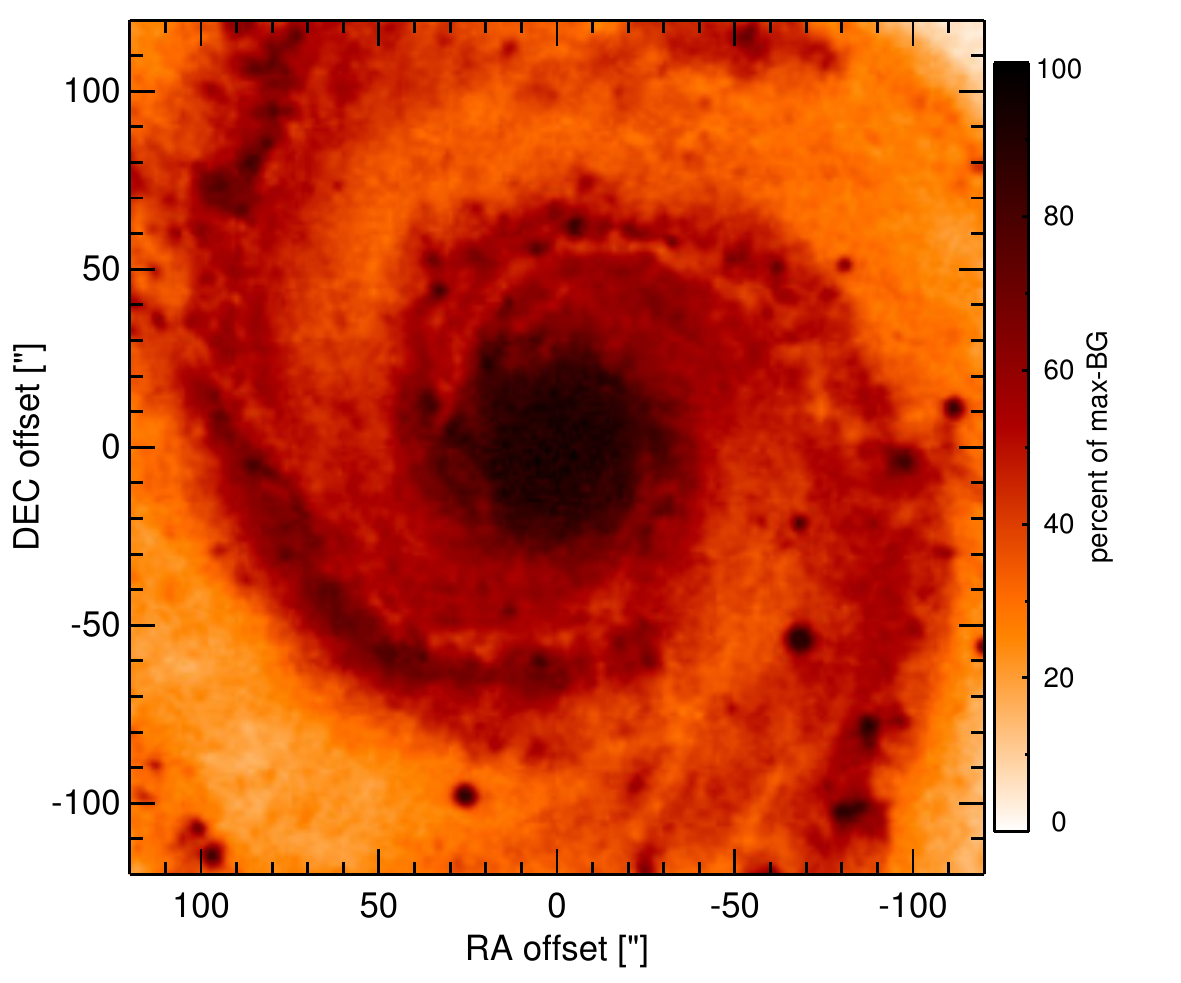}
    \caption{\label{fig:OPTim_M051A}
             Optical image (DSS, red filter) of M51a. Displayed are the central $4\arcmin$ with North up and East to the left. 
              The colour scaling is linear with white corresponding to the median background and black to the $0.01\%$ pixels with the highest intensity.  
           }
\end{figure}
\begin{figure}
   \centering
   \includegraphics[angle=0,height=3.11cm]{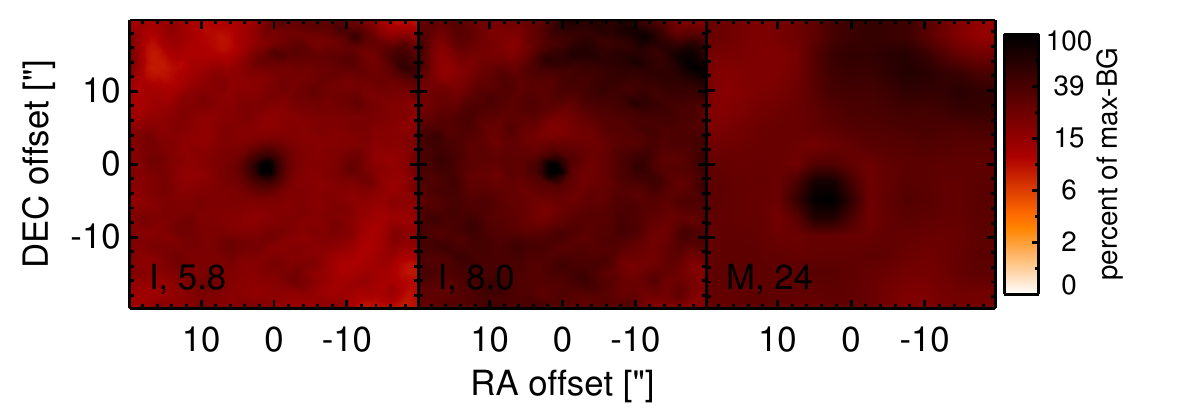}
    \caption{\label{fig:INTim_M051A}
             \spitzerr MIR images of M51a. Displayed are the inner $40\arcsec$ with North up and East to the left. The colour scaling is logarithmic with white corresponding to median background and black to the $0.1\%$ pixels with the highest intensity.
             The label in the bottom left states instrument and central wavelength of the filter in $\mu$m (I: IRAC, M: MIPS). 
           }
\end{figure}
\begin{figure}
   \centering
   \includegraphics[angle=0,height=3.11cm]{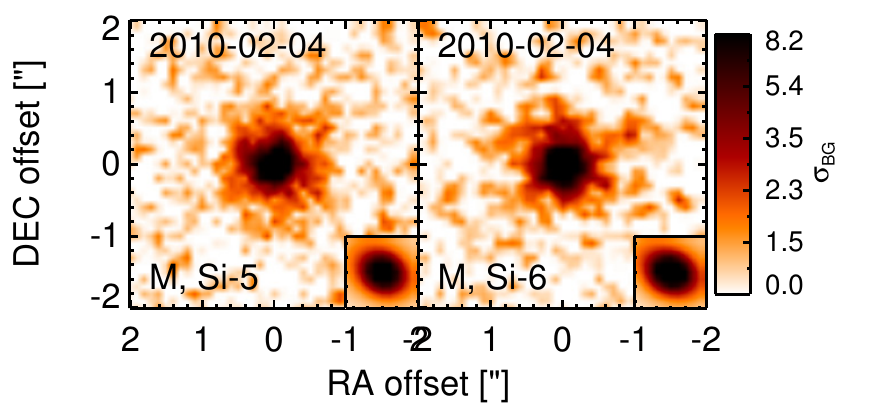}
    \caption{\label{fig:HARim_M051A}
             Subarcsecond-resolution MIR images of M51a sorted by increasing filter wavelength. 
             Displayed are the inner $4\arcsec$ with North up and East to the left. 
             The colour scaling is logarithmic with white corresponding to median background and black to the $75\%$ of the highest intensity of all images in units of $\sigbg$.
             The inset image shows the central arcsecond of the PSF from the calibrator star, scaled to match the science target.
             The labels in the bottom left state instrument and filter names (C: COMICS, M: Michelle, T: T-ReCS, V: VISIR).
           }
\end{figure}
\begin{figure}
   \centering
   \includegraphics[angle=0,width=8.50cm]{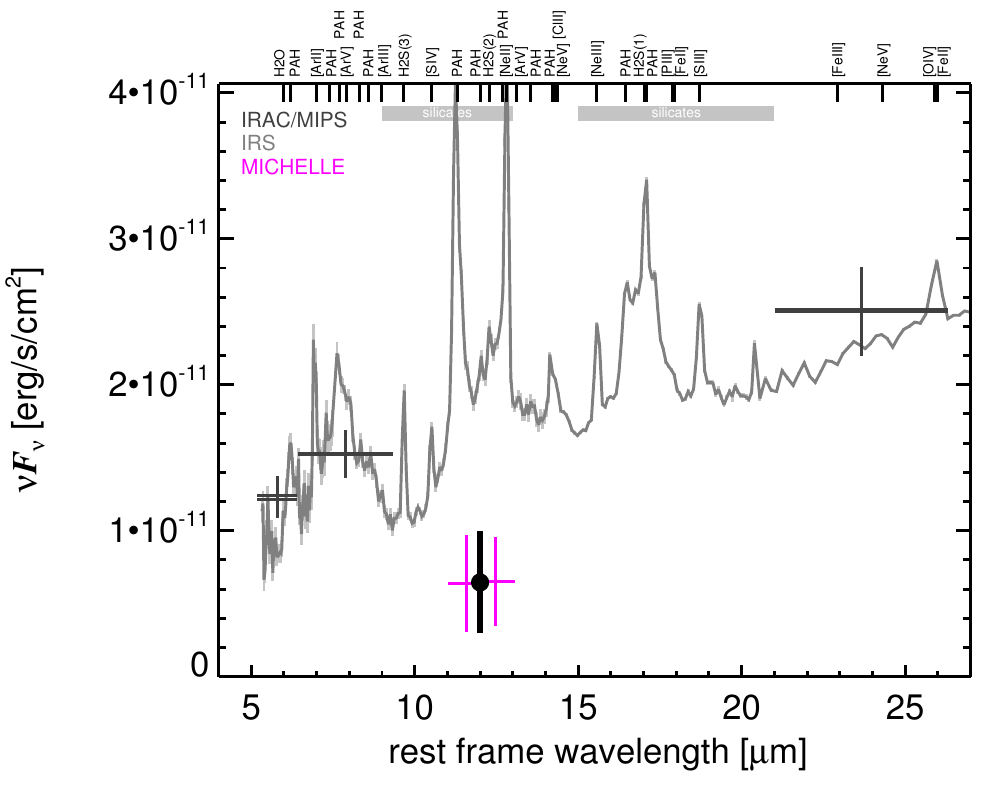}
   \caption{\label{fig:MISED_M051A}
      MIR SED of M51a. The description  of the symbols (if present) is the following.
      Grey crosses and  solid lines mark the \spitzer/IRAC, MIPS and IRS data. 
      The colour coding of the other symbols is: 
      green for COMICS, magenta for Michelle, blue for T-ReCS and red for VISIR data.
      Darker-coloured solid lines mark spectra of the corresponding instrument.
      The black filled circles mark the nuclear 12 and $18\,\mu$m  continuum emission estimate from the data.
      The ticks on the top axis mark positions of common MIR emission lines, while the light grey horizontal bars mark wavelength ranges affected by the silicate 10 and 18$\mu$m features.}
\end{figure}
\clearpage

\twocolumn[\begin{@twocolumnfalse}  
\subsection{M81 -- NGC\,3031}\label{app:M081}
M81 is a nearby grand-design spiral galaxy at a distance of $D=$ $3,6 \pm 0.5$\,Mpc (NED redshift-independent median) with one of the intrinsically weakest known AGN, which is also a flat-spectrum radio source (e.g., \citealt{ho_radio_2001}).
It exhibits both LINER and Sy\,1.8 characteristics \cite{filippenko_detailed_1988,ho_new_1996} and extraordinary low obscuration ($\nh \sim 5\cdot 10^{20}\,\cm^{-2}$; \citealt{page_x-ray_2003}). 
Motivated by its similarity to Sgr A$^*$, an extensive multiwavelength campaign was performed on M\,81 by \cite{markoff_results_2008} in order to compare and characterize SED and the short-term (daily) variability that occurs in this object (in addition to the yearly variations). 
They found striking similarities to the hard state of X-ray binaries.
The first MIR observations of M81 were performed by \cite{kleinmann_infrared_1970}, \cite{rieke_10_1978} and \cite{dyck_photometry_1978}, followed by first subarcsecond-resolution $N$-band imaging with Palomar 5\,m/MIRLIN in 1999 \citep{grossan_high-resolution_2001,gorjian_10_2004} with a possibly resolved nucleus and a significantly higher than the previous measurements.
M81 was followed up by \cite{soifer_high_2004} with Keck/LWS in the SIC filter in 2000 (and 2002 but under non-photometric conditions by \citealt{grossan_high_2004}), who find the nucleus marginally resolved (FWHM$=0.36\arcsec \sim 7$\,pc).
\spitzer/IRAC has observed M81 seven times between 2003 and 2005 and \spitzer/MIPS was used for three observations between 2003 and 2004. 
In all cases, a compact nucleus embedded in extended host emission was detected (see also \citealt{willner_infrared_2004,smith_anomalous_2010}).
Because we measure the photometry of the nuclear component only, our IRAC $5.8$ and $8.0\,\mu$m and MIPS $24\,\mu$m fluxes are significantly lower than the values in the literature (e.g., \citealt{dale_infrared_2005,munoz-mateos_radial_2009}).
However, our nuclear IRAC fluxes are a factor of two higher than those of \cite{willner_infrared_2004} for unknown reasons, but are on the other hand roughly consistent with the \spitzer/IRS LR mapping mode spectrum. 
The latter shows very strong silicate emission and also a prominent PAH 11.3$\,\mu$m feature but no other significant PAH features (see e.g., \citealt{smith_anomalous_2010,wu_spitzer/irs_2009,dale_spitzer_2009,mason_nuclear_2012}). 
The steep spectral slope at the short wavelength side of the silicate $10\,\mu$m is most likely the cause for the apparent flux variations reported earlier as already noted by \cite{smith_anomalous_2010}.

We observed the nucleus of M81 with Michelle in two $N$-band filters in 2010.
Additional Michelle imaging in three $N$ and one $Q$-band filters was performed in 2011 \citep{mason_nuclear_2012}. 
A compact MIR nucleus was detected in all Michelle images, which appears marginally resolved in our Si-5 and Si-6 images but not in the Si-2 image from 2011 which shows the highest angular resolution.
Therefore, it remains uncertain, whether the nucleus of M81 is in general resolved at subarcsecond resolution in the MIR but the results of \cite{mason_nuclear_2012} suggest it to be rather unresolved.
Our nuclear flux measurements are on average $\sim24\%$ lower than the \spitzerr spectrophotometry and consistent with previous measurements but for unknown reasons systematically $\sim 20\%$ higher than the values of \cite{mason_nuclear_2012} for the same data.
The silicate $10\,\mu$m feature is reproduced with the subarcsecond data and thus originates in the projected central $\sim7$\,pc of the nucleus of M81.
Therefore, we use the IRS spectrum to correct our 12 and $18\,\mu$m continuum emission estimate for the silicate features.
\newline\end{@twocolumnfalse}]

\begin{figure}
   \centering
   \includegraphics[angle=0,width=8.500cm]{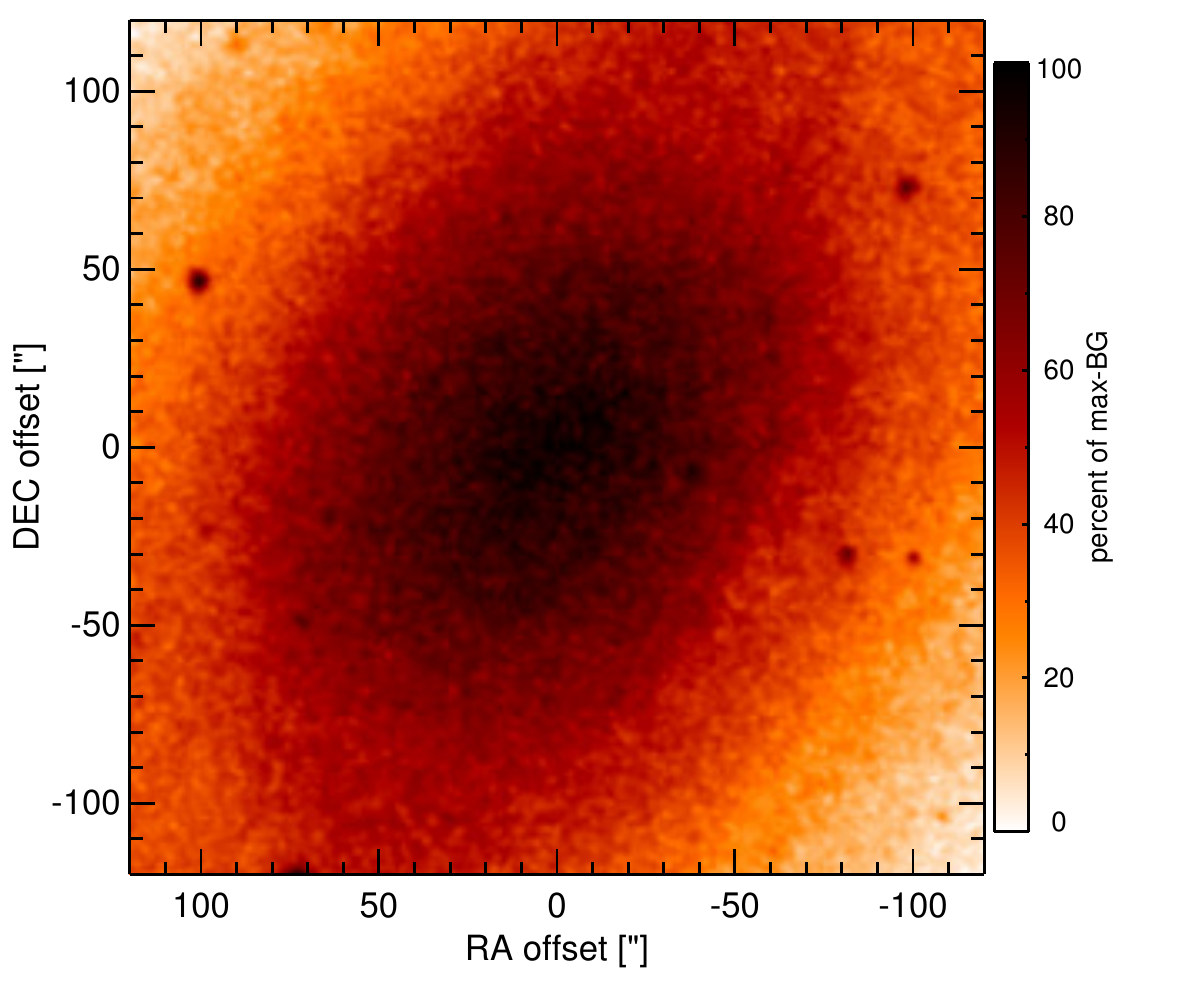}
    \caption{\label{fig:OPTim_M081}
             Optical image (DSS, red filter) of M81. Displayed are the central $4\arcmin$ with North up and East to the left. 
              The colour scaling is linear with white corresponding to the median background and black to the $0.01\%$ pixels with the highest intensity.  
           }
\end{figure}
\begin{figure}
   \centering
   \includegraphics[angle=0,height=3.11cm]{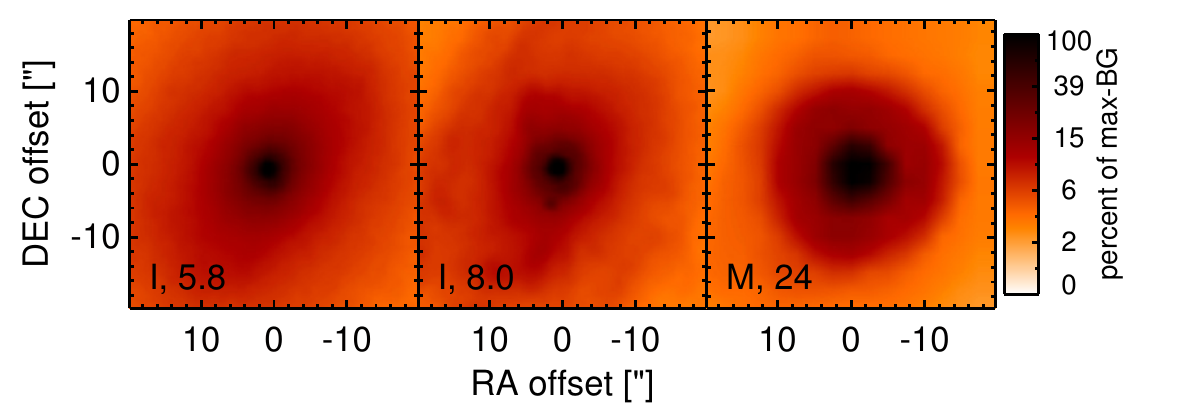}
    \caption{\label{fig:INTim_M081}
             \spitzerr MIR images of M81. Displayed are the inner $40\arcsec$ with North up and East to the left. The colour scaling is logarithmic with white corresponding to median background and black to the $0.1\%$ pixels with the highest intensity.
             The label in the bottom left states instrument and central wavelength of the filter in $\mu$m (I: IRAC, M: MIPS). 
             Note that the apparent off-nuclear compact source in the IRAC $8.0\,\mu$m image is an instrumental artefact.
           }
\end{figure}
\begin{figure}
   \centering
   \includegraphics[angle=0,width=8.500cm]{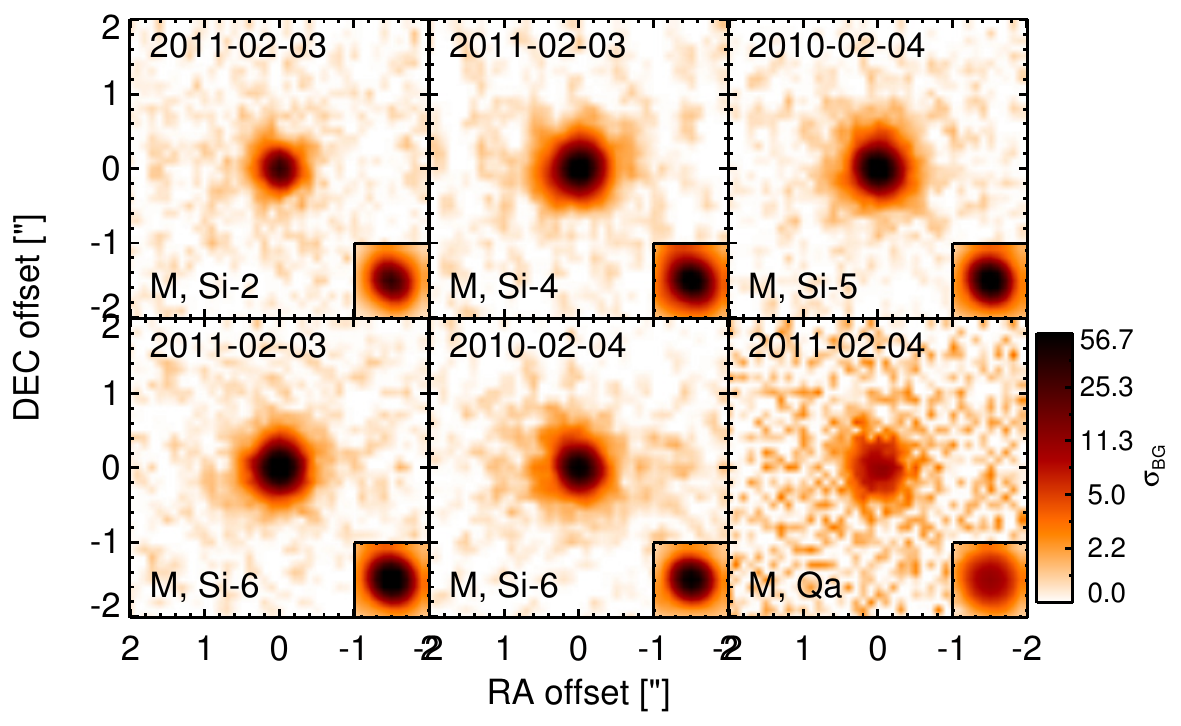}
    \caption{\label{fig:HARim_M081}
             Subarcsecond-resolution MIR images of M81 sorted by increasing filter wavelength. 
             Displayed are the inner $4\arcsec$ with North up and East to the left. 
             The colour scaling is logarithmic with white corresponding to median background and black to the $75\%$ of the highest intensity of all images in units of $\sigbg$.
             The inset image shows the central arcsecond of the PSF from the calibrator star, scaled to match the science target.
             The labels in the bottom left state instrument and filter names (C: COMICS, M: Michelle, T: T-ReCS, V: VISIR).
           }
\end{figure}
\begin{figure}
   \centering
   \includegraphics[angle=0,width=8.50cm]{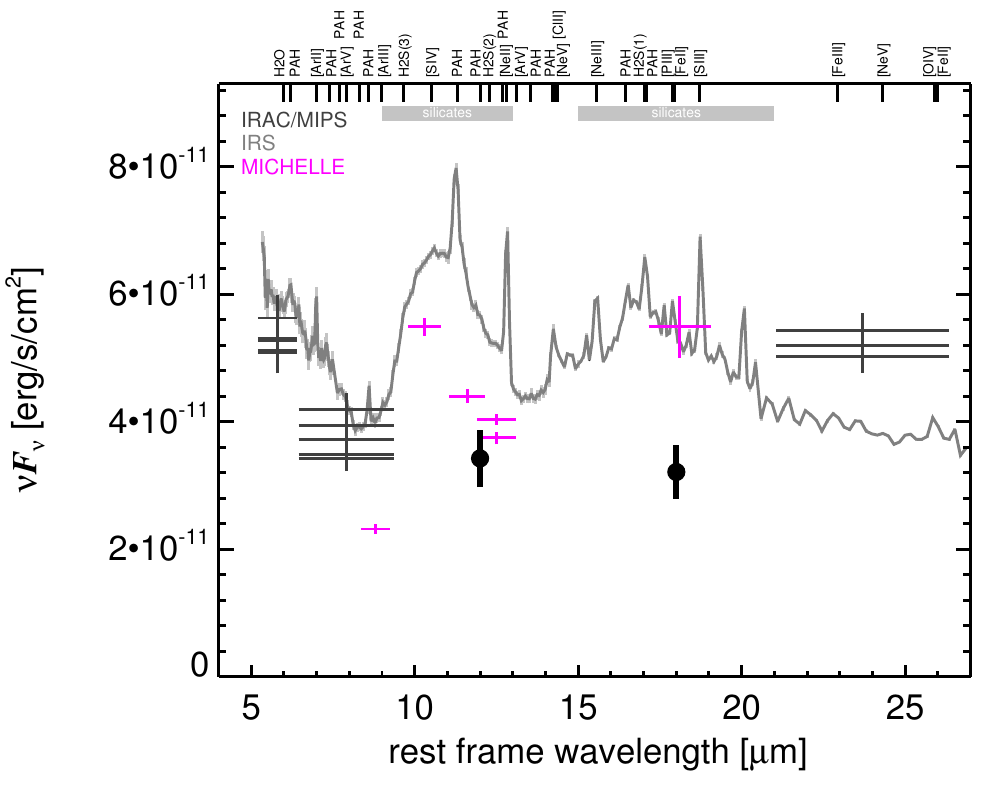}
   \caption{\label{fig:MISED_M081}
      MIR SED of M81. The description  of the symbols (if present) is the following.
      Grey crosses and  solid lines mark the \spitzer/IRAC, MIPS and IRS data. 
      The colour coding of the other symbols is: 
      green for COMICS, magenta for Michelle, blue for T-ReCS and red for VISIR data.
      Darker-coloured solid lines mark spectra of the corresponding instrument.
      The black filled circles mark the nuclear 12 and $18\,\mu$m  continuum emission estimate from the data.
      The ticks on the top axis mark positions of common MIR emission lines, while the light grey horizontal bars mark wavelength ranges affected by the silicate 10 and 18$\mu$m features.}
\end{figure}
\clearpage

\twocolumn[\begin{@twocolumnfalse}  
\subsection{M87 -- NGC\,4486 -- Vir\,A -- 3C\,274 -- VCC\,1316}\label{app:M087}
M87 is the central giant elliptical galaxy in the Virgo cluster at a distance of $D=$ $16.7 \pm 2.6$\,Mpc with a FR\,I radio morphology, a radio-loud LINER nucleus \citep{veron-cetty_catalogue_2010} and a prominent jet visible at all wavelengths (see \citealt{meisenheimer_synchrotron_1996} for a summary on the jet properties).
It contains one of the most massive black hole of the local galaxy population ($\log M_\mathrm{BH}/M_\odot = 9.82 \pm 0.03$; \citealt{gebhardt_black_2011}).
The first MIR photometry of M87 was attempted by \cite{kleinmann_observations_1970} but the nucleus remained undetected in $N-$ and $Q$-band.
The first successful detection was achieved by \cite{rieke_infrared_1972} in $N$-band while the jet still remained undetected \citep{kinman_optical_1974}.
Additional $N$-band photometry of the nucleus was obtained over then next years by \cite{rieke_infrared_1978}, \cite{puschell_nonstellar_1981}, \cite{heckman_infrared_1983} and \cite{impey_infrared_1986}. 
Over the course of this decade, the four different measurements with small apertures show an apparent variation of $\sim30\%$ around the median flux. 
Note that the intrinsic variation probably is much smaller because of the calibration uncertainties, the different measurement methods and filter band-passes.
After \iras, also \iso/ISOCAM imaged M\,87 in various MIR filters detecting both the nucleus and the jet \citep{siebenmorgen_isocam_2004,xilouris_dust_2004}. 
The first subarcsecond MIR imaging of the nuclear region was performed with Keck/LWS in 2000 \citep{whysong_hidden_2001,whysong_thermal_2004} and Gemini/OSCIR in 2001 \citep{perlman_deep_2001}, in which a compact nucleus and several jet knots was detected.
The same morphology is also seen in the \spitzer/IRAC and MIPS images in addition to diffuse elliptical host emission embedding nucleus and jet (see also \citealt{perlman_mid-infrared_2007,shi_thermal_2007}).
Two epoch of IRAC $8.0\,\mu$m images (2005 and 2008) and four epochs of MIPS $24\,\mu$m images (2004, 2005, 2008 and 2009) are available and interestingly show different fluxes for the nuclear component. 
Because we measure the nuclear component only, our  IRAC $5.8$ and $8.0\,\mu$m  and MIPS $24\,\mu$m fluxes are in general much lower than in \cite{shi_thermal_2007} and \cite{temi_spitzer_2009}, while  the nuclear MIPS $24\,\mu$m flux of the former work is consistent with ours.
Note that the \spitzer/IRS LR staring-mode spectrum from 2005 matches the corresponding IRAC photometry.
It shows weak silicate emission and a red spectral slope  in $\nu F_\nu$-space but no PAH features (see also \citealt{bressan_spitzer_2006,perlman_mid-infrared_2007,leipski_spitzer_2009}).
The silicate feature  gives evidence for the presence of dust in the central $\sim 320\,$pc of M87. 
\cite{buson_role_2009} however show that this silicate emission can be totally accounted for by the old stellar population.

M87 was imaged with VISIR in the PAH2\_2 and Q2 filters in 2006 by \cite{reunanen_vlt_2010} and in two additional narrow $N$-band filters in 2011 (this work). 
We also performed COMICS imaging in the N11.7 filter in 2009.
Because all these subarcsecond-resolution images are shallower than the OSCIR ones, only the MIR nucleus was detected as a point source in the $N$-band images, while it remained undetected in Q2.
Our nuclear photometry is significantly higher than the values in \cite{reunanen_vlt_2010} but consistent with the previous subarcsecond measurements (see also \citealt{asmus_mid-infrared_2011}) and on average $\sim 29\%$ lower than the \spitzerr spectrophotometry.
Unfortunately, the spectral coverage and the precision of the subarcsecond photometry is not sufficient to check the existence of the silicate feature at this resolution.
On the other hand, we note that our values fall onto the residual  MIR SED from \cite{buson_role_2009} after subtraction of the host emission template. 
This power-law MIR SED can be well explained only with non-thermal (synchrotron) emission and thus, the existence of a dusty AGN torus in M87 becomes increasingly unlikely.
\newline\end{@twocolumnfalse}]

\begin{figure}
   \centering
   \includegraphics[angle=0,width=8.500cm]{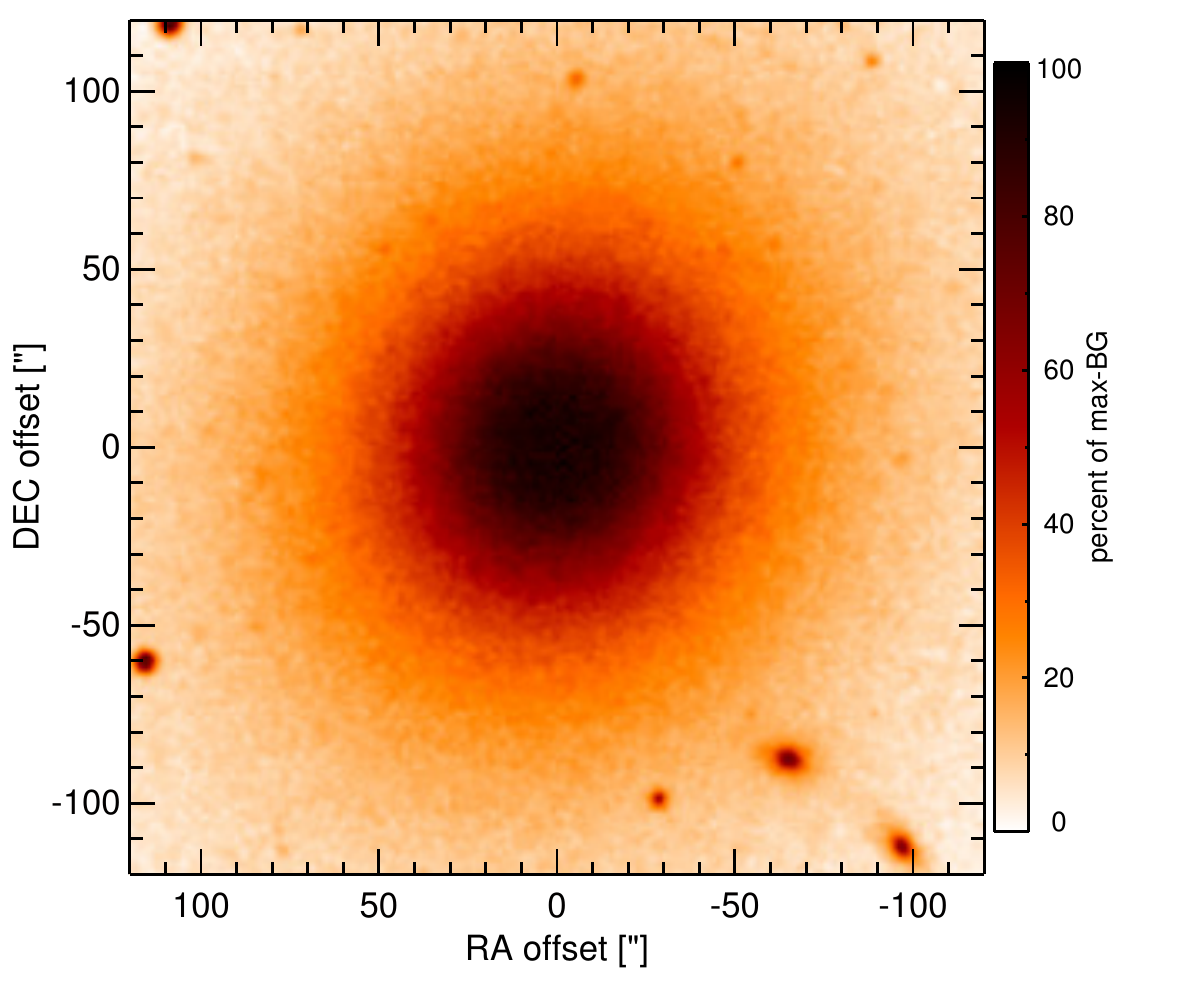}
    \caption{\label{fig:OPTim_M087}
             Optical image (DSS, red filter) of M87. Displayed are the central $4\arcmin$ with North up and East to the left. 
              The colour scaling is linear with white corresponding to the median background and black to the $0.01\%$ pixels with the highest intensity.  
           }
\end{figure}
\begin{figure}
   \centering
   \includegraphics[angle=0,height=3.11cm]{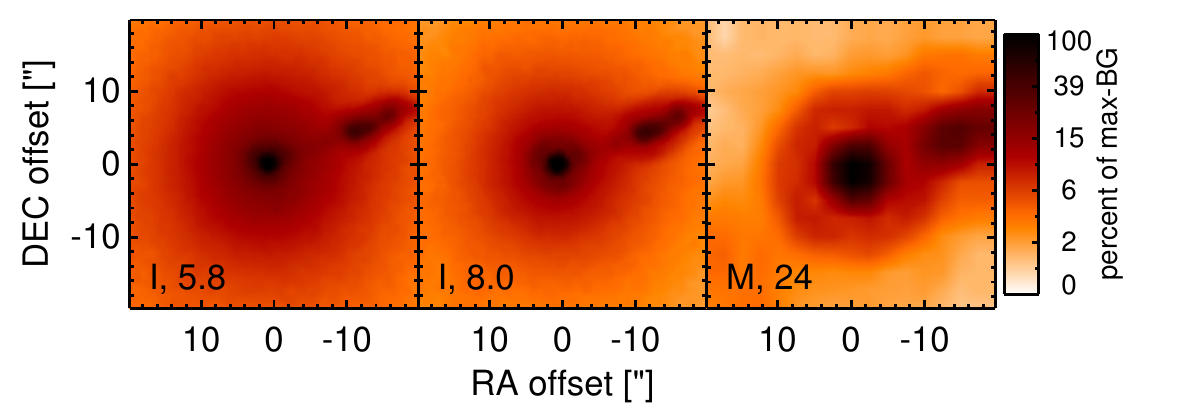}
    \caption{\label{fig:INTim_M087}
             \spitzerr MIR images of M87. Displayed are the inner $40\arcsec$ with North up and East to the left. The colour scaling is logarithmic with white corresponding to median background and black to the $0.1\%$ pixels with the highest intensity.
             The label in the bottom left states instrument and central wavelength of the filter in $\mu$m (I: IRAC, M: MIPS). 
           }
\end{figure}
\begin{figure}
   \centering
   \includegraphics[angle=0,width=8.500cm]{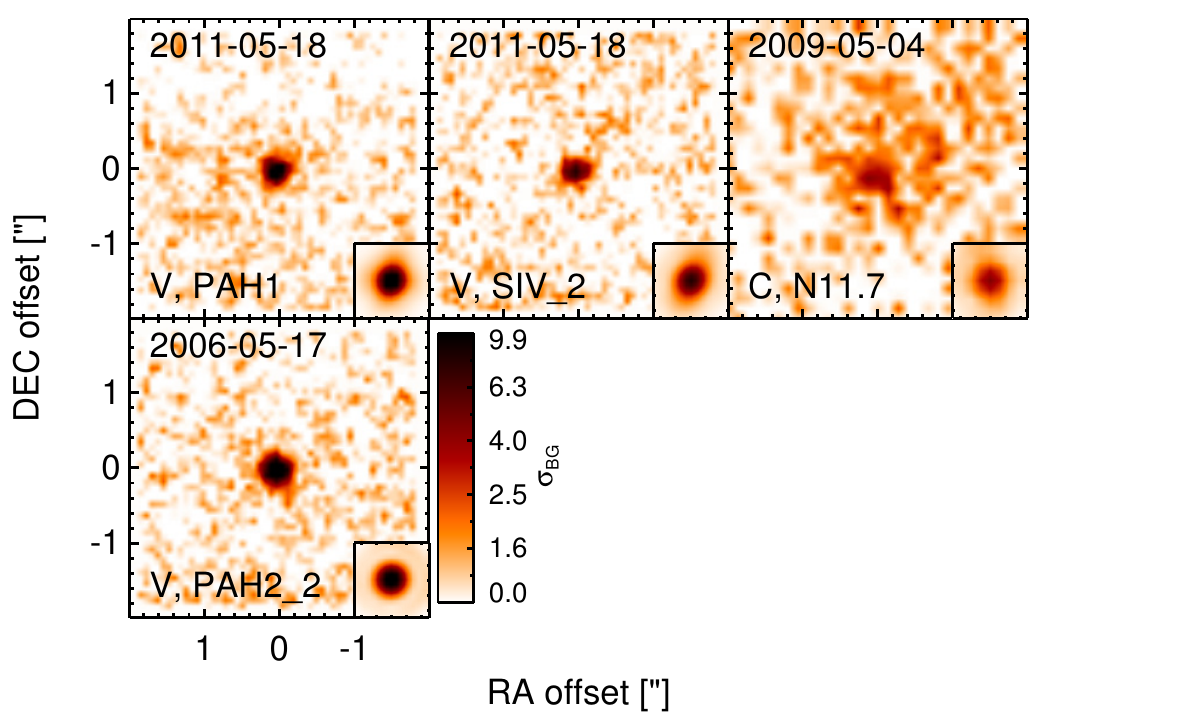}
    \caption{\label{fig:HARim_M087}
             Subarcsecond-resolution MIR images of M87 sorted by increasing filter wavelength. 
             Displayed are the inner $4\arcsec$ with North up and East to the left. 
             The colour scaling is logarithmic with white corresponding to median background and black to the $75\%$ of the highest intensity of all images in units of $\sigbg$.
             The inset image shows the central arcsecond of the PSF from the calibrator star, scaled to match the science target.
             The labels in the bottom left state instrument and filter names (C: COMICS, M: Michelle, T: T-ReCS, V: VISIR).
           }
\end{figure}
\begin{figure}
   \centering
   \includegraphics[angle=0,width=8.50cm]{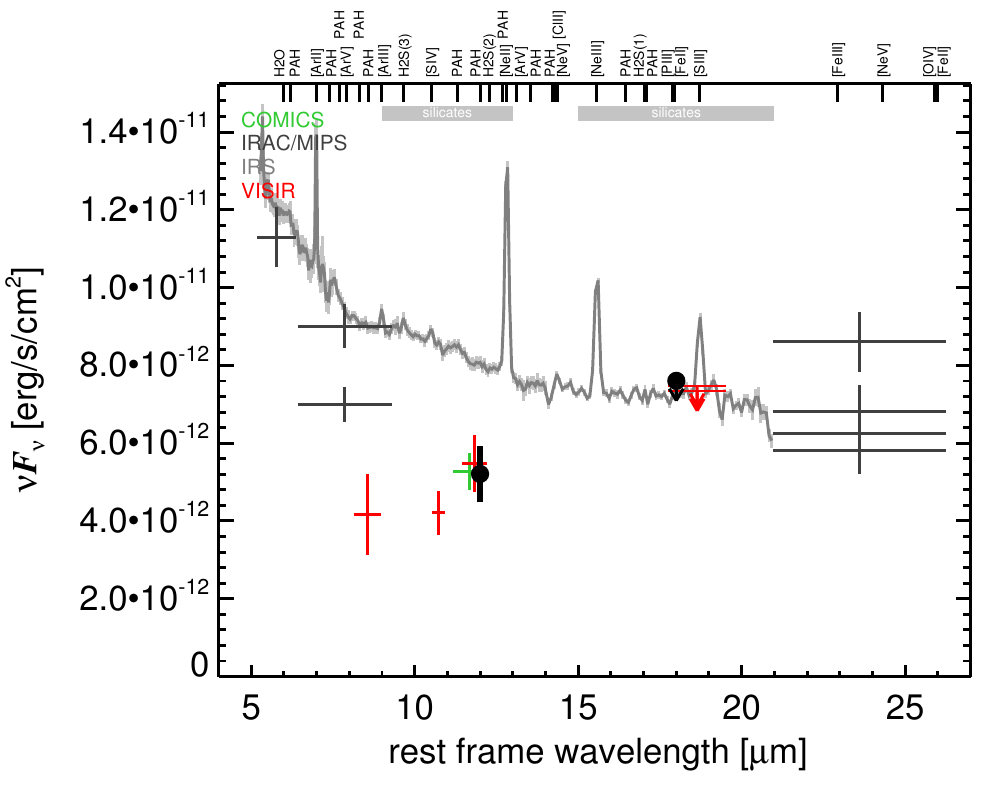}
   \caption{\label{fig:MISED_M087}
      MIR SED of M87. The description  of the symbols (if present) is the following.
      Grey crosses and  solid lines mark the \spitzer/IRAC, MIPS and IRS data. 
      The colour coding of the other symbols is: 
      green for COMICS, magenta for Michelle, blue for T-ReCS and red for VISIR data.
      Darker-coloured solid lines mark spectra of the corresponding instrument.
      The black filled circles mark the nuclear 12 and $18\,\mu$m  continuum emission estimate from the data.
      The ticks on the top axis mark positions of common MIR emission lines, while the light grey horizontal bars mark wavelength ranges affected by the silicate 10 and 18$\mu$m features.}
\end{figure}
\clearpage

\twocolumn[\begin{@twocolumnfalse}  
\subsection{MCG-1-5-47 -- IGR\,J01528-0326}\label{app:MCG-01-05-047}
MCG-1-5-47 is an edge-on spiral galaxy at a redshift of $z=$ 0.0172 ($D\sim67.7\,$Mpc) with a Sy\,2 nucleus \citep{veron-cetty_catalogue_2010}, which was first discovered in hardest X-rays with \textit{INTEGRAL} \citep{bird_third_2007,bodaghee_description_2007,masetti_unveiling_2008}.
\spitzer/IRAC and IRS observations of this object are available, and a compact nuclear source embedded in the narrow host emission is visible in the IRAC images. 
The IRS LR/HR staring-mode PBCD spectrum is very noisy and partly flagged in the $N$-band. 
No spectral features apart from the continuum slope, which peaks at $\sim 15\,\mu$m, can be identified.
MCG-1-5-47 was observed with VISIR in five narrow-band filters covering the whole $N$-band in 2008 (unpublished, to our knowledge).
A compact MIR nucleus is only detected in the two longest-wavelength filters, which appears possibly marginally resolved in the B11.7 filter (FWHM $\sim 0.4\arcsec \sim 130\,$pc) but not in the B12.4 filter image. 
Therefore, it remains uncertain, whether the nucleus of MCG-1-5-47 is truly resolved at subarcsecond resolution in the MIR.
The nuclear photometry is on average $\sim 57\%$ lower than the \spitzerr spectrophotometry.
The fact that the nucleus has not been detected at shorter wavelengths might indicate the presence of a silicate $10\,\mu$m absorption feature in the projected central $\sim100$\,pc of MCG-1-5-4.
\newline\end{@twocolumnfalse}]

\begin{figure}
   \centering
   \includegraphics[angle=0,width=8.500cm]{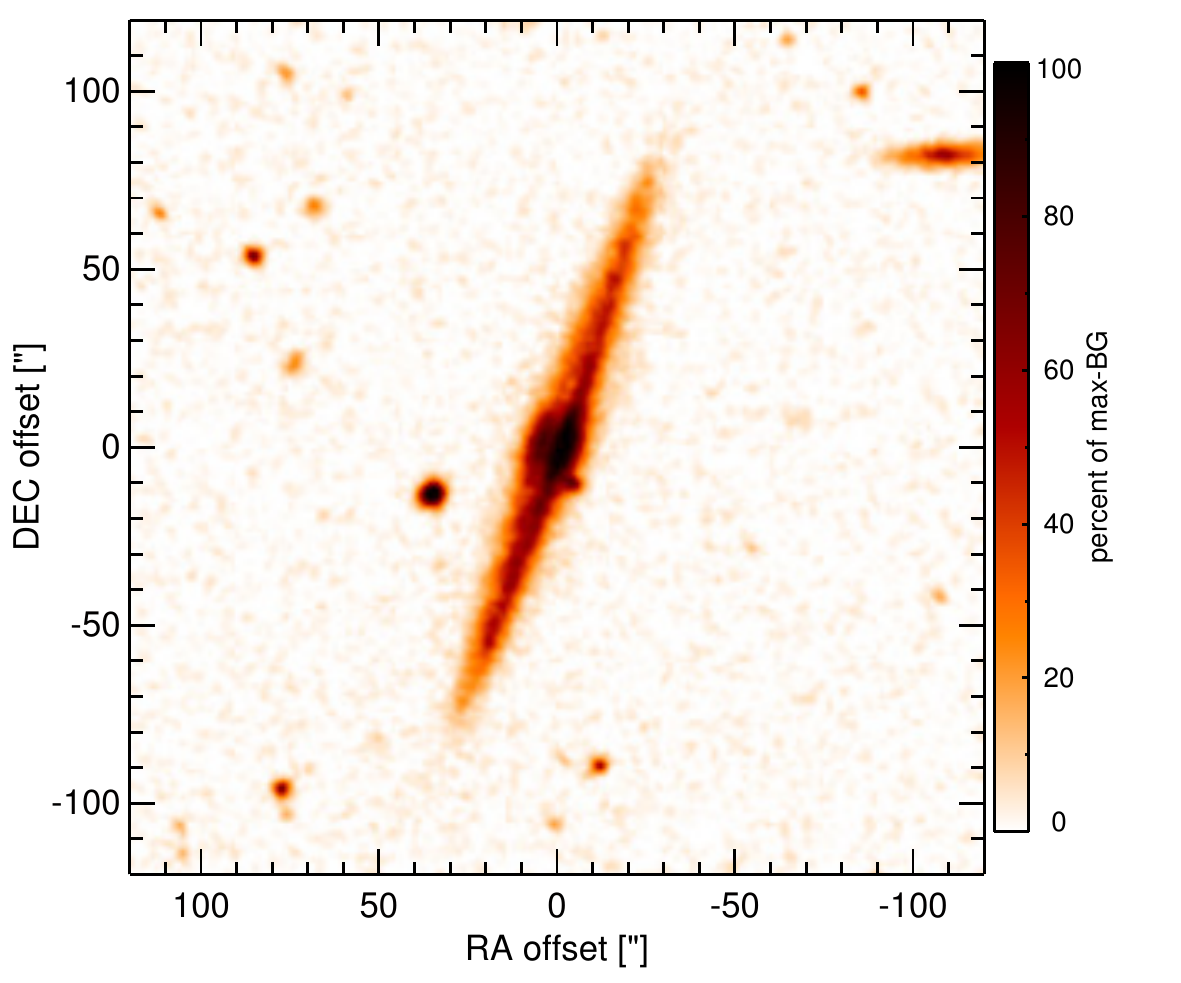}
    \caption{\label{fig:OPTim_MCG-01-05-047}
             Optical image (DSS, red filter) of MCG-1-5-47. Displayed are the central $4\arcmin$ with North up and East to the left. 
              The colour scaling is linear with white corresponding to the median background and black to the $0.01\%$ pixels with the highest intensity.  
           }
\end{figure}
\begin{figure}
   \centering
   \includegraphics[angle=0,height=3.11cm]{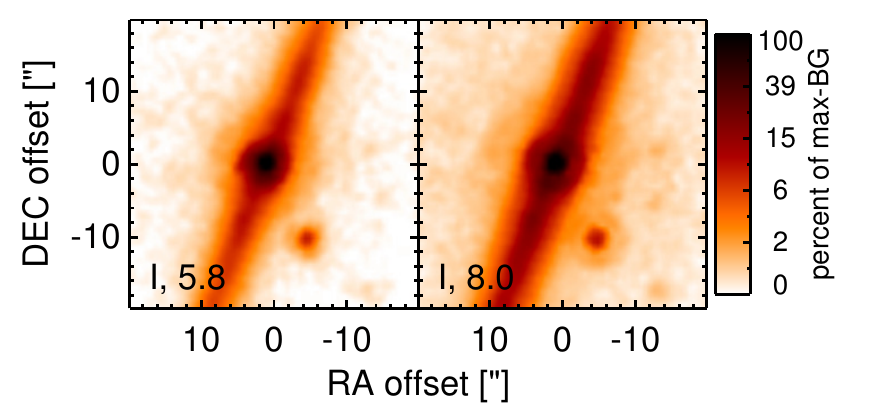}
    \caption{\label{fig:INTim_MCG-01-05-047}
             \spitzerr MIR images of MCG-1-5-47. Displayed are the inner $40\arcsec$ with North up and East to the left. The colour scaling is logarithmic with white corresponding to median background and black to the $0.1\%$ pixels with the highest intensity.
             The label in the bottom left states instrument and central wavelength of the filter in $\mu$m (I: IRAC, M: MIPS). 
           }
\end{figure}
\begin{figure}
   \centering
   \includegraphics[angle=0,height=3.11cm]{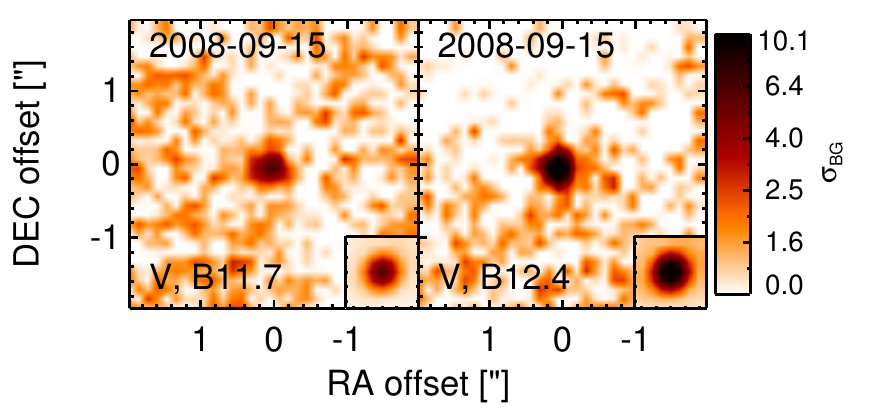}
    \caption{\label{fig:HARim_MCG-01-05-047}
             Subarcsecond-resolution MIR images of MCG-1-5-47 sorted by increasing filter wavelength. 
             Displayed are the inner $4\arcsec$ with North up and East to the left. 
             The colour scaling is logarithmic with white corresponding to median background and black to the $75\%$ of the highest intensity of all images in units of $\sigbg$.
             The inset image shows the central arcsecond of the PSF from the calibrator star, scaled to match the science target.
             The labels in the bottom left state instrument and filter names (C: COMICS, M: Michelle, T: T-ReCS, V: VISIR).
           }
\end{figure}
\begin{figure}
   \centering
   \includegraphics[angle=0,width=8.50cm]{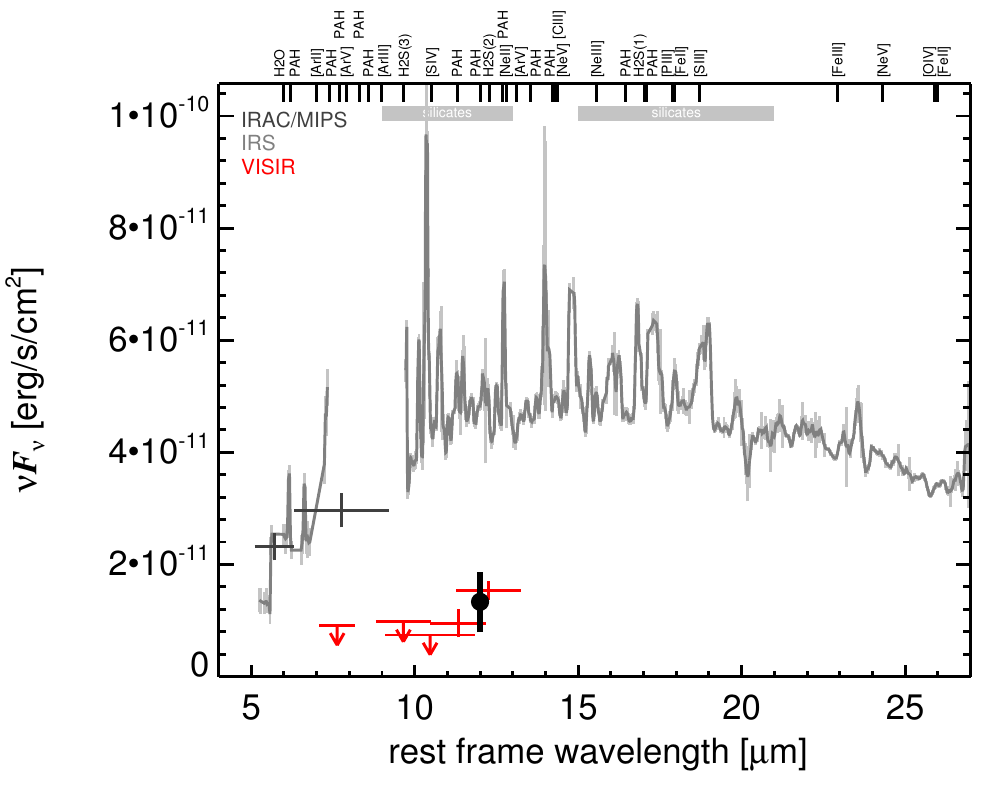}
   \caption{\label{fig:MISED_MCG-01-05-047}
      MIR SED of MCG-1-5-47. The description  of the symbols (if present) is the following.
      Grey crosses and  solid lines mark the \spitzer/IRAC, MIPS and IRS data. 
      The colour coding of the other symbols is: 
      green for COMICS, magenta for Michelle, blue for T-ReCS and red for VISIR data.
      Darker-coloured solid lines mark spectra of the corresponding instrument.
      The black filled circles mark the nuclear 12 and $18\,\mu$m  continuum emission estimate from the data.
      The ticks on the top axis mark positions of common MIR emission lines, while the light grey horizontal bars mark wavelength ranges affected by the silicate 10 and 18$\mu$m features.}
\end{figure}
\clearpage

\twocolumn[\begin{@twocolumnfalse}  
\subsection{MCG-1-13-25 -- H\,0448-041}\label{app:MCG-01-13-025}
MCG-1-13-25 is a spiral galaxy at a redshift of $z=$ 0.0159 ($D\sim$65.7\,Mpc) with a Sy\,1.2 nucleus \citep{pietsch_new_1998}, which was first discovered with \textit{HEAO} \citep{brissenden_multiwaveband_1987}.
It is a member of the nine-month BAT AGN sample. 
\spitzer/IRAC, IRS and MIPS observations of MCG-1-13-25 are available and show a nearly unresolved  source in the corresponding images.
The IRS LR staring-mode spectrum exhibits strong silicate 10 and $18\,\mu$m emission with a weak PAH $11.3\,\mu$m feature and a slightly blue spectral slope in $\nu F_\nu$-space (see also \citealt{weaver_mid-infrared_2010}).
We observed MCG-1-13-25 with VISIR in three narrow $N$-band filters in 2009 and weakly detected a possibly marginally resolved MIR nucleus (FWHM $\sim 0.41\arcsec \sim 130\,$pc).
However, at least a second epoch with deeper subarcsecond-resolution MIR images is required to verify this extension.
Our nuclear photometry provides fluxes consistent with the \spitzerr spectrophotometry.
Therefore, we use the IRS spectrum to estimate the nuclear $12\,\mu$m continuum emission corrected for the silicate feature.
Note however, that the nuclear fluxes would be significantly lower if the presence of subarcsecond-extended emission can be verified
For now, the resulting synthetic flux is then scaled by half of the ratio between $\Fpsf$ and $\Fgau$ to account for the possibly nuclear extension.
\newline\end{@twocolumnfalse}]

\begin{figure}
   \centering
   \includegraphics[angle=0,width=8.500cm]{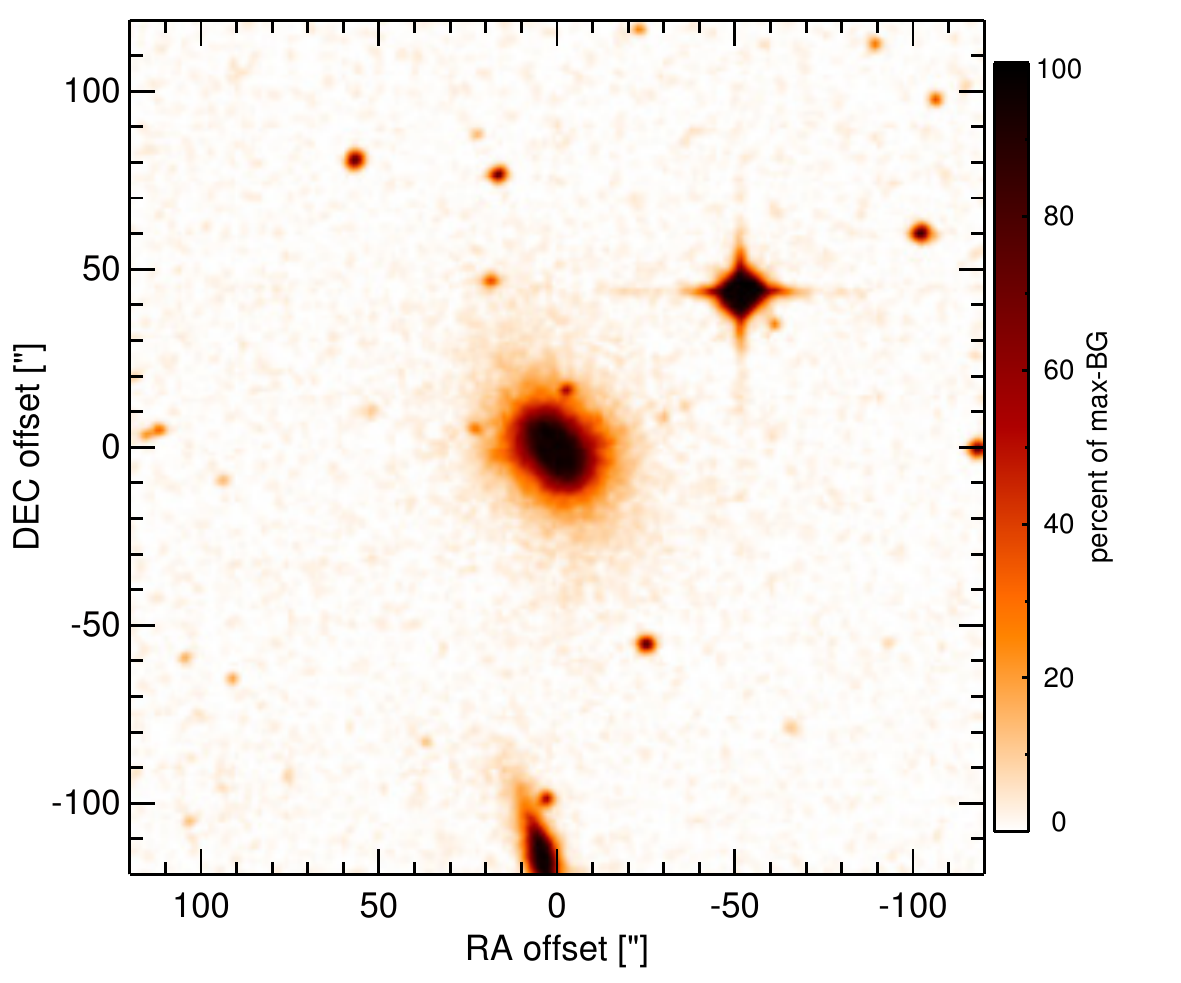}
    \caption{\label{fig:OPTim_MCG-01-13-025}
             Optical image (DSS, red filter) of MCG-1-13-25. Displayed are the central $4\arcmin$ with North up and East to the left. 
              The colour scaling is linear with white corresponding to the median background and black to the $0.01\%$ pixels with the highest intensity.  
           }
\end{figure}
\begin{figure}
   \centering
   \includegraphics[angle=0,height=3.11cm]{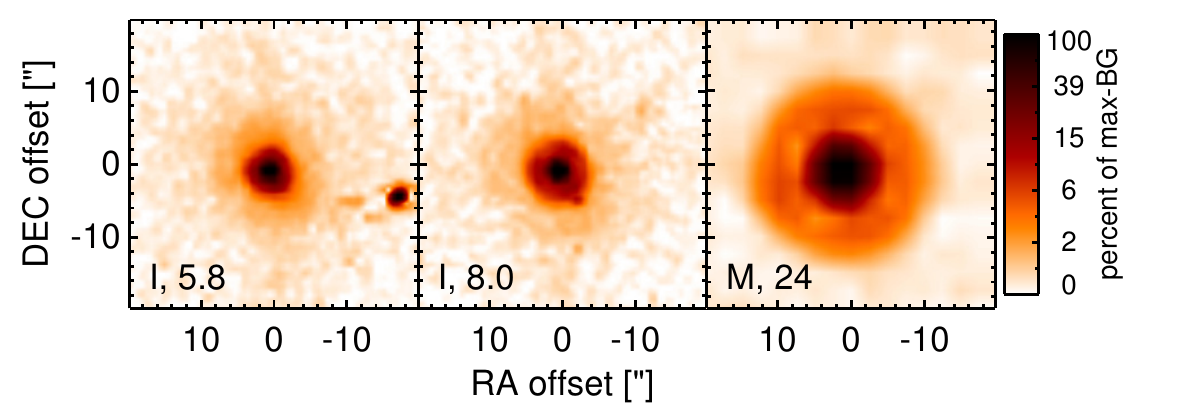}
    \caption{\label{fig:INTim_MCG-01-13-025}
             \spitzerr MIR images of MCG-1-13-25. Displayed are the inner $40\arcsec$ with North up and East to the left. The colour scaling is logarithmic with white corresponding to median background and black to the $0.1\%$ pixels with the highest intensity.
             The label in the bottom left states instrument and central wavelength of the filter in $\mu$m (I: IRAC, M: MIPS). 
           }
\end{figure}
\begin{figure}
   \centering
   \includegraphics[angle=0,height=3.11cm]{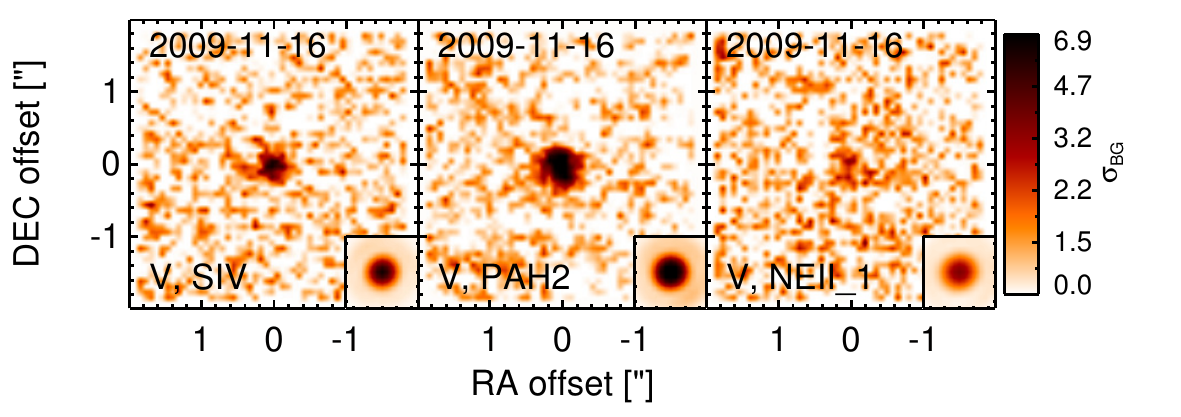}
    \caption{\label{fig:HARim_MCG-01-13-025}
             Subarcsecond-resolution MIR images of MCG-1-13-25 sorted by increasing filter wavelength. 
             Displayed are the inner $4\arcsec$ with North up and East to the left. 
             The colour scaling is logarithmic with white corresponding to median background and black to the $75\%$ of the highest intensity of all images in units of $\sigbg$.
             The inset image shows the central arcsecond of the PSF from the calibrator star, scaled to match the science target.
             The labels in the bottom left state instrument and filter names (C: COMICS, M: Michelle, T: T-ReCS, V: VISIR).
           }
\end{figure}
\begin{figure}
   \centering
   \includegraphics[angle=0,width=8.50cm]{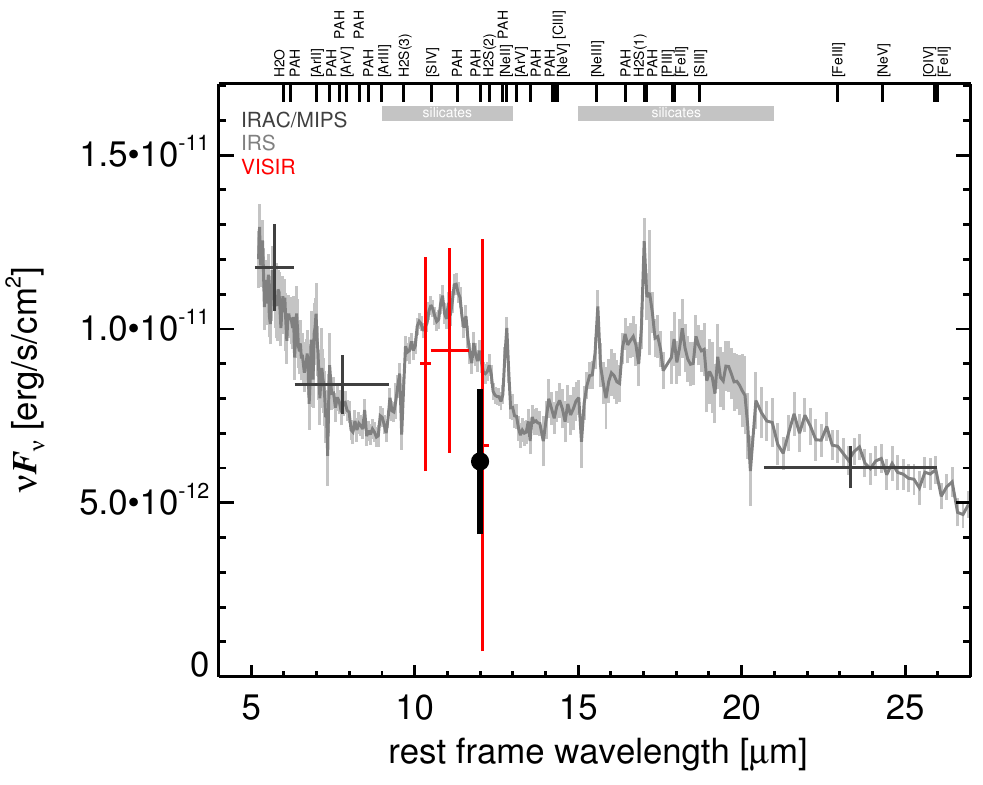}
   \caption{\label{fig:MISED_MCG-01-13-025}
      MIR SED of MCG-1-13-25. The description  of the symbols (if present) is the following.
      Grey crosses and  solid lines mark the \spitzer/IRAC, MIPS and IRS data. 
      The colour coding of the other symbols is: 
      green for COMICS, magenta for Michelle, blue for T-ReCS and red for VISIR data.
      Darker-coloured solid lines mark spectra of the corresponding instrument.
      The black filled circles mark the nuclear 12 and $18\,\mu$m  continuum emission estimate from the data.
      The ticks on the top axis mark positions of common MIR emission lines, while the light grey horizontal bars mark wavelength ranges affected by the silicate 10 and 18$\mu$m features.}
\end{figure}
\clearpage

\twocolumn[\begin{@twocolumnfalse}  
\subsection{MCG-1-24-12 -- H\,0917-074}\label{app:MCG-01-24-012}
MCG-1-24-12 is a spiral galaxy at a redshift of $z=$ 0.0196 ($D\sim86.6\,$Mpc) with a Sy\,2 nucleus \citep{de_grijp_warm_1992}, which was first discovered with \textit{HEAO} \citep{piccinotti_complete_1982,malizia_bepposax/pds_2002}.
It belongs to the nine-month BAT AGN sample and has an extended NLR region with a PA$=75\degree$ \citep{schmitt_hubble_2003}.
Only \spitzer/IRS observations of MCG-1-24-12 are available.
The corresponding LR  staring-mode spectrum exhibits deep silicate $10\,\mu$m absorption and weak PAH emission and a red spectral slope in $\nu F_\nu$-space (see also \citealt{mullaney_defining_2011}).
In the \wisee images, the object appears elliptically extended along the galaxy major axis. 
We observed MCG-1-24-12 with VISIR in three narrow $N$-band filters in 2009 and detected an unresolved MIR nucleus.
Our nuclear photometry provides fluxes consistent with the \spitzerr spectrophotometry.
Therefore, we use the IRS spectrum to estimate the nuclear $12\,\mu$m continuum emission.
\newline\end{@twocolumnfalse}]

\begin{figure}
   \centering
   \includegraphics[angle=0,width=8.500cm]{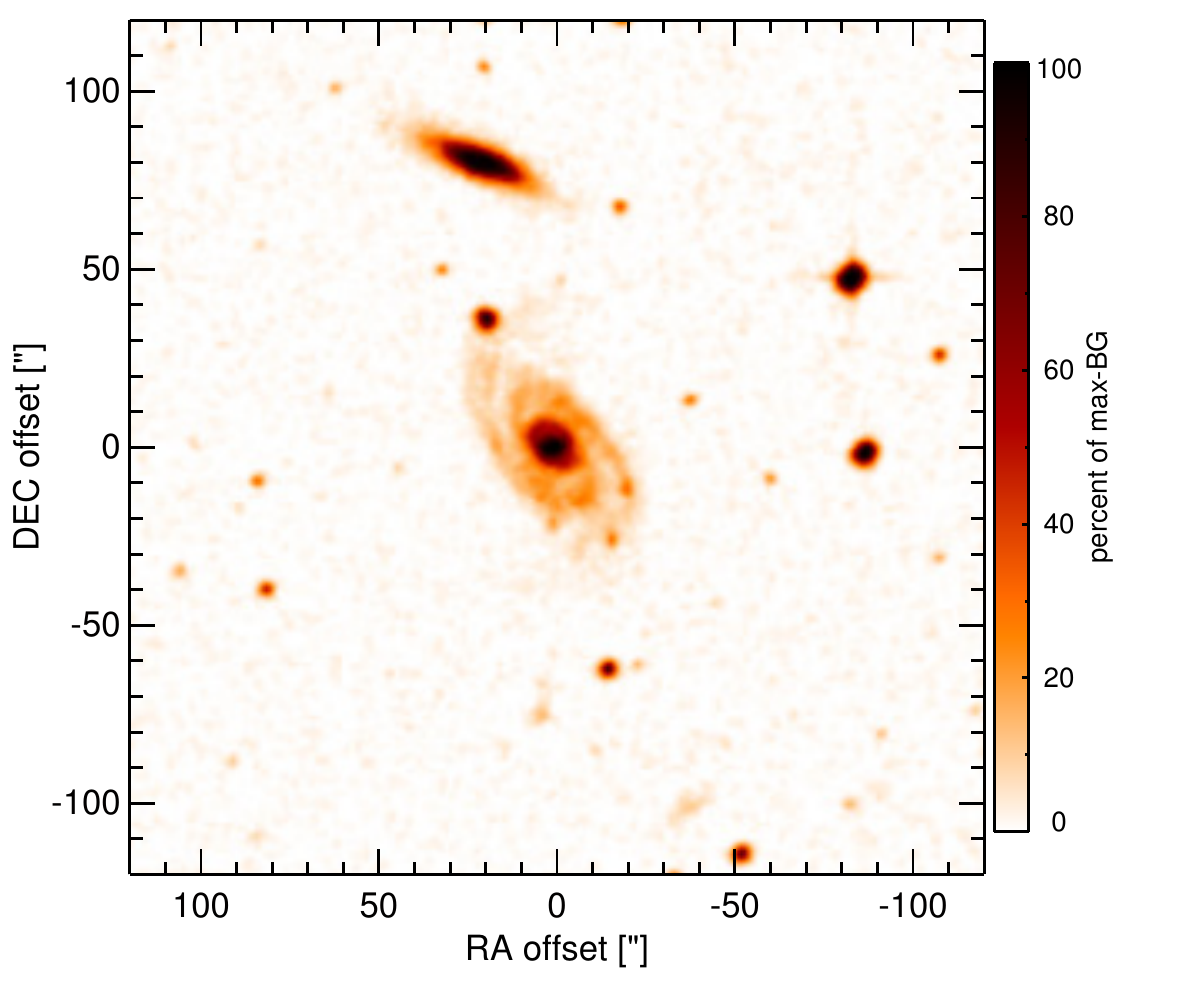}
    \caption{\label{fig:OPTim_MCG-01-24-012}
             Optical image (DSS, red filter) of MCG-1-24-12. Displayed are the central $4\arcmin$ with North up and East to the left. 
              The colour scaling is linear with white corresponding to the median background and black to the $0.01\%$ pixels with the highest intensity.  
           }
\end{figure}
\begin{figure}
   \centering
   \includegraphics[angle=0,height=3.11cm]{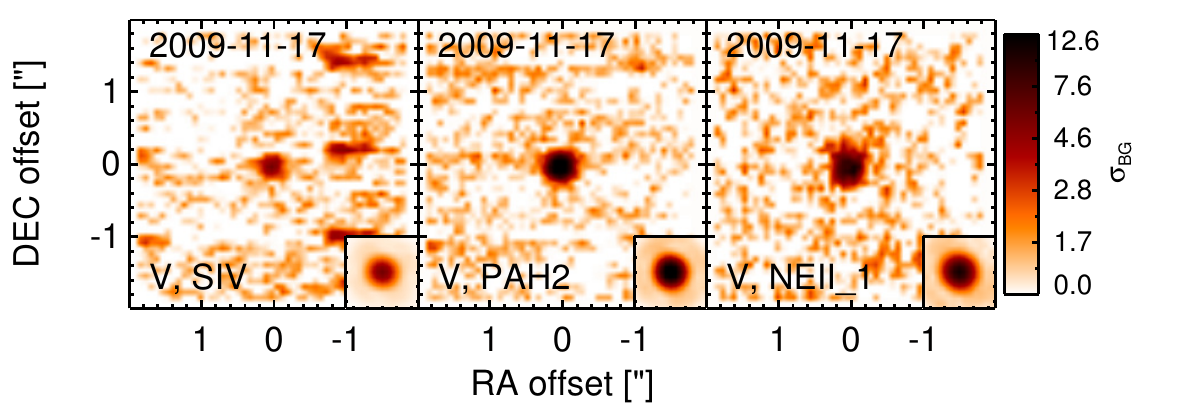}
    \caption{\label{fig:HARim_MCG-01-24-012}
             Subarcsecond-resolution MIR images of MCG-1-24-12 sorted by increasing filter wavelength. 
             Displayed are the inner $4\arcsec$ with North up and East to the left. 
             The colour scaling is logarithmic with white corresponding to median background and black to the $75\%$ of the highest intensity of all images in units of $\sigbg$.
             The inset image shows the central arcsecond of the PSF from the calibrator star, scaled to match the science target.
             The labels in the bottom left state instrument and filter names (C: COMICS, M: Michelle, T: T-ReCS, V: VISIR).
           }
\end{figure}
\begin{figure}
   \centering
   \includegraphics[angle=0,width=8.50cm]{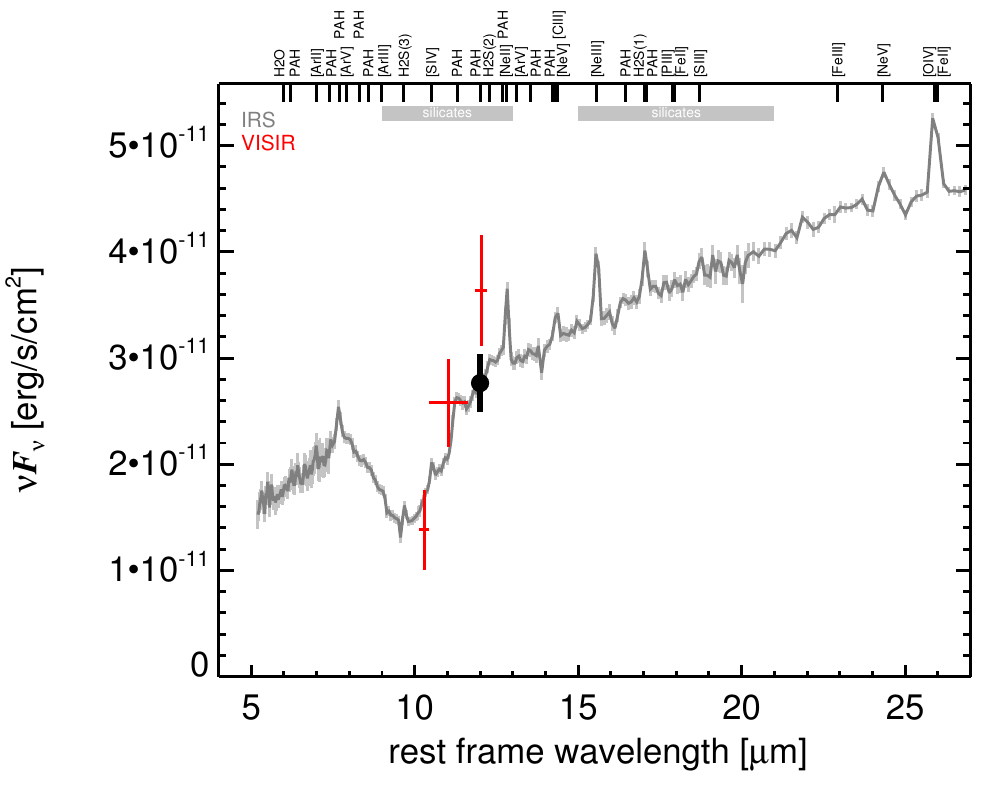}
   \caption{\label{fig:MISED_MCG-01-24-012}
      MIR SED of MCG-1-24-12. The description  of the symbols (if present) is the following.
      Grey crosses and  solid lines mark the \spitzer/IRAC, MIPS and IRS data. 
      The colour coding of the other symbols is: 
      green for COMICS, magenta for Michelle, blue for T-ReCS and red for VISIR data.
      Darker-coloured solid lines mark spectra of the corresponding instrument.
      The black filled circles mark the nuclear 12 and $18\,\mu$m  continuum emission estimate from the data.
      The ticks on the top axis mark positions of common MIR emission lines, while the light grey horizontal bars mark wavelength ranges affected by the silicate 10 and 18$\mu$m features.}
\end{figure}
\clearpage

\twocolumn[\begin{@twocolumnfalse}  
\subsection{MCG-2-8-14 -- IRAS\,F02499-0842}\label{app:MCG-02-08-014}
MCG-2-8-14 is an edge-on spiral galaxy at a redshift of $z=$ 0.0168 ($D\sim$66.9\,Mpc) with a Sy\,2 nucleus \citep{veron-cetty_catalogue_2010}, which was discovered in the SDSS \citep{hao_active_2005}. 
No \spitzerr observations are available for MCG-2-8-14, which appears extended in the \wisee images with similar morphology as in the optical images.
The object was observed with VISIR in five narrow $N$-band filters in 2009 (unpublished, to our knowledge).
A compact MIR nucleus is detect at shortest and longest wavelengths, while the source remained undetected in the central two filters.
The low S/N of the detection prohibits any conclusion about the MIR subarcsecond morphology.
The nuclear photometry indicates the presence of a deep silicate $10\,\mu$m absorption feature in the projected central $\sim 100\,$pc of MCG-2-8-14.
\newline\end{@twocolumnfalse}]

\begin{figure}
   \centering
   \includegraphics[angle=0,width=8.500cm]{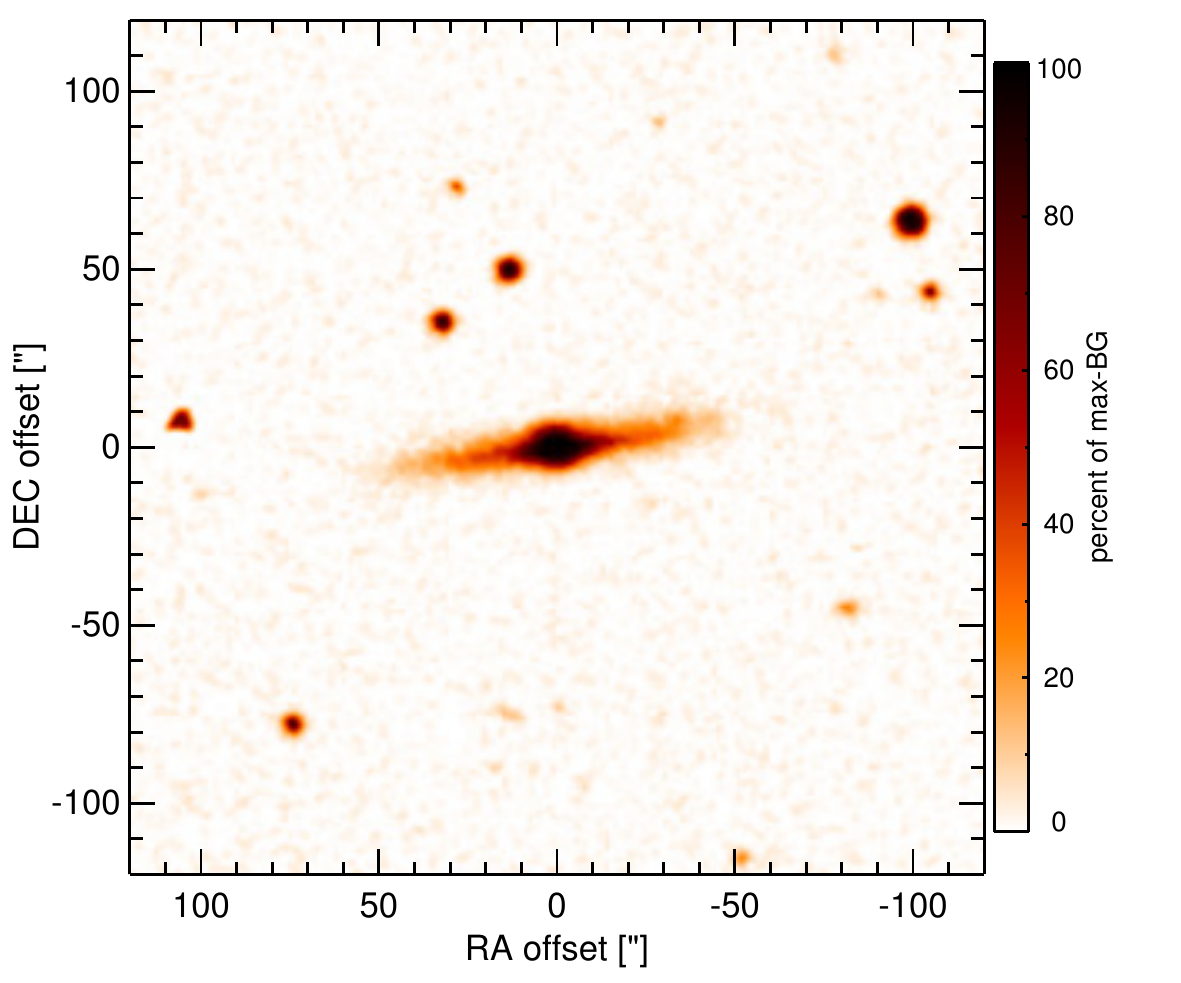}
    \caption{\label{fig:OPTim_MCG-02-08-014}
             Optical image (DSS, red filter) of MCG-2-8-14. Displayed are the central $4\arcmin$ with North up and East to the left. 
              The colour scaling is linear with white corresponding to the median background and black to the $0.01\%$ pixels with the highest intensity.  
           }
\end{figure}
\begin{figure}
   \centering
   \includegraphics[angle=0,height=3.11cm]{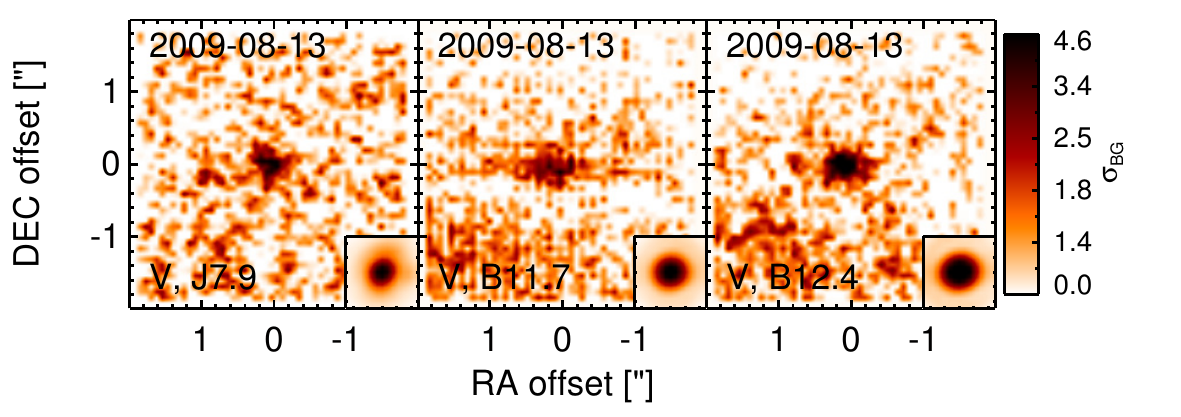}
    \caption{\label{fig:HARim_MCG-02-08-014}
             Subarcsecond-resolution MIR images of MCG-2-8-14 sorted by increasing filter wavelength. 
             Displayed are the inner $4\arcsec$ with North up and East to the left. 
             The colour scaling is logarithmic with white corresponding to median background and black to the $75\%$ of the highest intensity of all images in units of $\sigbg$.
             The inset image shows the central arcsecond of the PSF from the calibrator star, scaled to match the science target.
             The labels in the bottom left state instrument and filter names (C: COMICS, M: Michelle, T: T-ReCS, V: VISIR).
           }
\end{figure}
\begin{figure}
   \centering
   \includegraphics[angle=0,width=8.50cm]{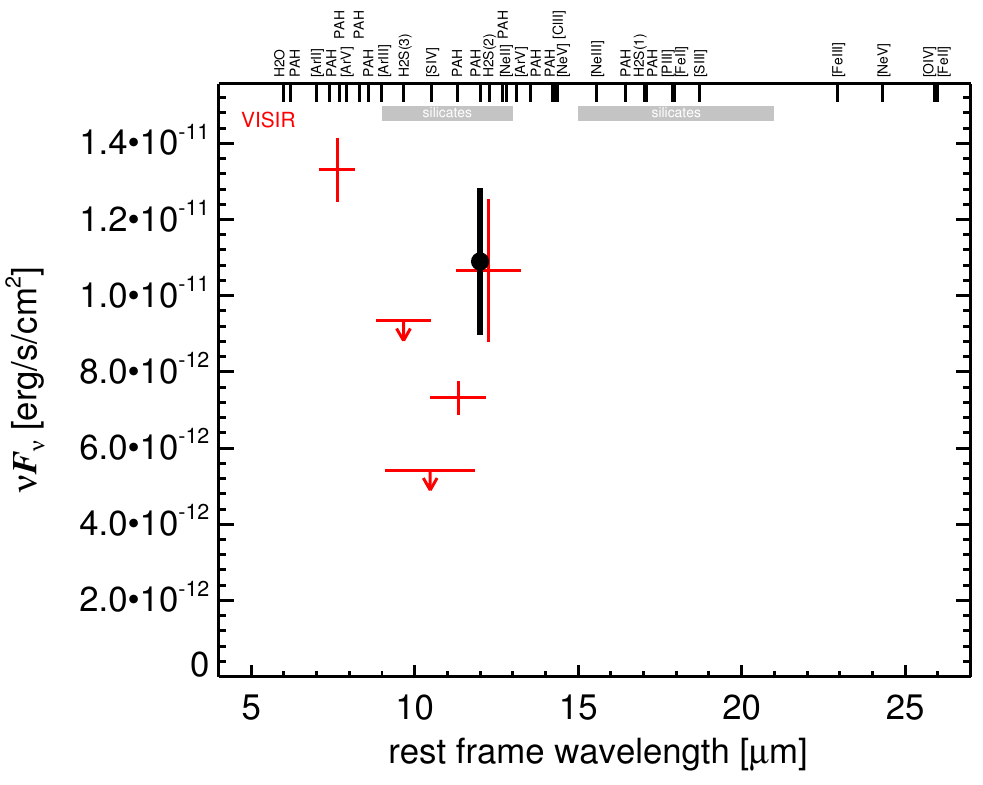}
   \caption{\label{fig:MISED_MCG-02-08-014}
      MIR SED of MCG-2-8-14. The description  of the symbols (if present) is the following.
      Grey crosses and  solid lines mark the \spitzer/IRAC, MIPS and IRS data. 
      The colour coding of the other symbols is: 
      green for COMICS, magenta for Michelle, blue for T-ReCS and red for VISIR data.
      Darker-coloured solid lines mark spectra of the corresponding instrument.
      The black filled circles mark the nuclear 12 and $18\,\mu$m  continuum emission estimate from the data.
      The ticks on the top axis mark positions of common MIR emission lines, while the light grey horizontal bars mark wavelength ranges affected by the silicate 10 and 18$\mu$m features.}
\end{figure}
\clearpage

\twocolumn[\begin{@twocolumnfalse}  
\subsection{MCG-2-8-39 -- IRAS\,02580-1136}\label{app:MCG-02-08-039}
MCG-2-8-39 is a possible merger system with spiral-like morphology and two nuclei (separation: $\sim 13\arcsec \sim 7.3\,$kpc; PA$\sim 10\degree$; \citealt{heisler_double-nucleus_1989}) at a redshift of $z=$ 0.0299 ($D\sim$123\,Mpc).
The brighter northern nucleus\,A is identified with a Sy\,2 nucleus with polarized broad emission lines \citep{tran_hidden_2001}, while nucleus\,B has the optical spectrum of a H\,II region \citep{heisler_double-nucleus_1989}.
After its detection with \iras, MCG-2-8-39 was observed with the MMT bolometer in $N$-band \citep{maiolino_new_1995} and with \spitzer/IRAC and IRS.
In the IRAC images, compact MIR nuclei are detected in positions A and B with weak extended host emission. 
Nucleus\,B is significantly weaker than nucleus\,A, for which our IRAC $5.8$ and $8.0\,\mu$m is consistent with \cite{gallimore_infrared_2010}. 
The IRS LR mapping-mode spectrum of nucleus\,A is relatively featureless apart from a possible silicate $18\,\mu$m emission feature which coincides with the peak of the MIR SED (see also \citealt{buchanan_spitzer_2006,wu_spitzer/irs_2009,tommasin_spitzer-irs_2010,gallimore_infrared_2010}).
MCG-2-8-39 was observed with T-ReCS in the Qa filter during two nights in 2007.
In the images, only nucleus\,A was detected as a point source with no signs of any extended host emission.
Our nuclear photometry is interestingly $\sim23\%$ higher than the \spitzerr spectrophotometry, while the latter is also lower than the MMT photometry by \cite{maiolino_new_1995}.
This points at a problem in the flux calibration of the \spitzerr PBCD data rather than any variability.
Therefore, we extrapolate from the Qa measurement towards shorter wavelengths in order to compute the nuclear $12\,\mu$m continuum emission estimate as described in Sect.~\ref{sec:cont}.
\newline\end{@twocolumnfalse}]

\begin{figure}
   \centering
   \includegraphics[angle=0,width=8.500cm]{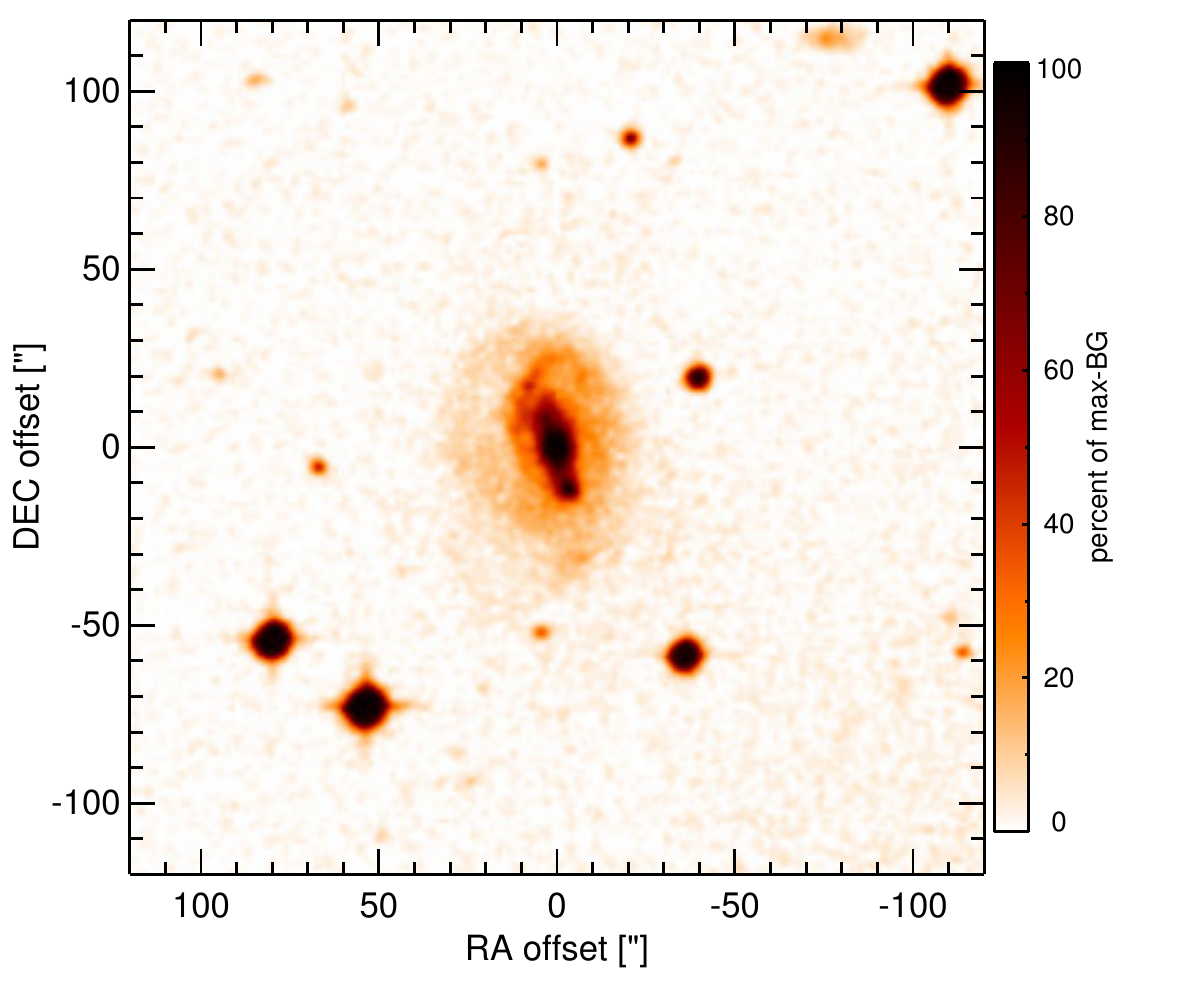}
    \caption{\label{fig:OPTim_MCG-02-08-039}
             Optical image (DSS, red filter) of MCG-2-8-39. Displayed are the central $4\arcmin$ with North up and East to the left. 
              The colour scaling is linear with white corresponding to the median background and black to the $0.01\%$ pixels with the highest intensity.  
           }
\end{figure}
\begin{figure}
   \centering
   \includegraphics[angle=0,height=3.11cm]{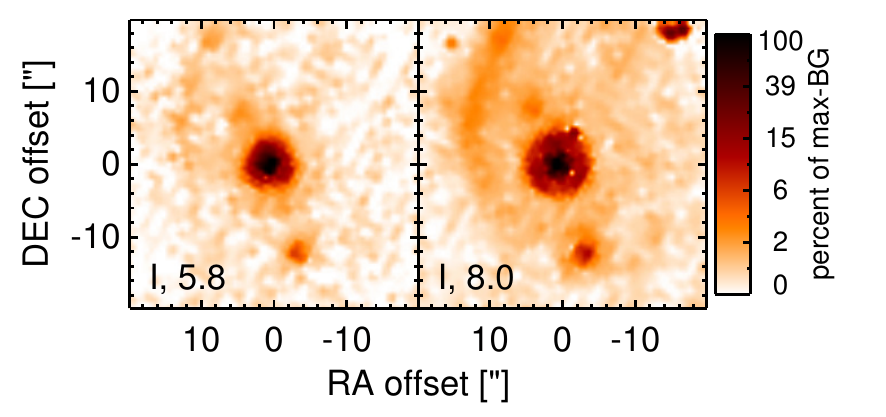}
    \caption{\label{fig:INTim_MCG-02-08-039}
             \spitzerr MIR images of MCG-2-8-39. Displayed are the inner $40\arcsec$ with North up and East to the left. The colour scaling is logarithmic with white corresponding to median background and black to the $0.1\%$ pixels with the highest intensity.
             The label in the bottom left states instrument and central wavelength of the filter in $\mu$m (I: IRAC, M: MIPS). 
           }
\end{figure}
\begin{figure}
   \centering
   \includegraphics[angle=0,height=3.11cm]{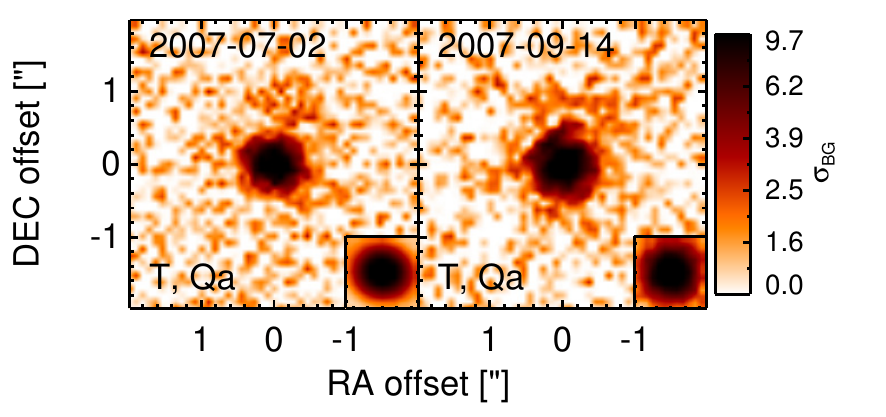}
    \caption{\label{fig:HARim_MCG-02-08-039}
             Subarcsecond-resolution MIR images of MCG-2-8-39 sorted by increasing filter wavelength. 
             Displayed are the inner $4\arcsec$ with North up and East to the left. 
             The colour scaling is logarithmic with white corresponding to median background and black to the $75\%$ of the highest intensity of all images in units of $\sigbg$.
             The inset image shows the central arcsecond of the PSF from the calibrator star, scaled to match the science target.
             The labels in the bottom left state instrument and filter names (C: COMICS, M: Michelle, T: T-ReCS, V: VISIR).
           }
\end{figure}
\begin{figure}
   \centering
   \includegraphics[angle=0,width=8.50cm]{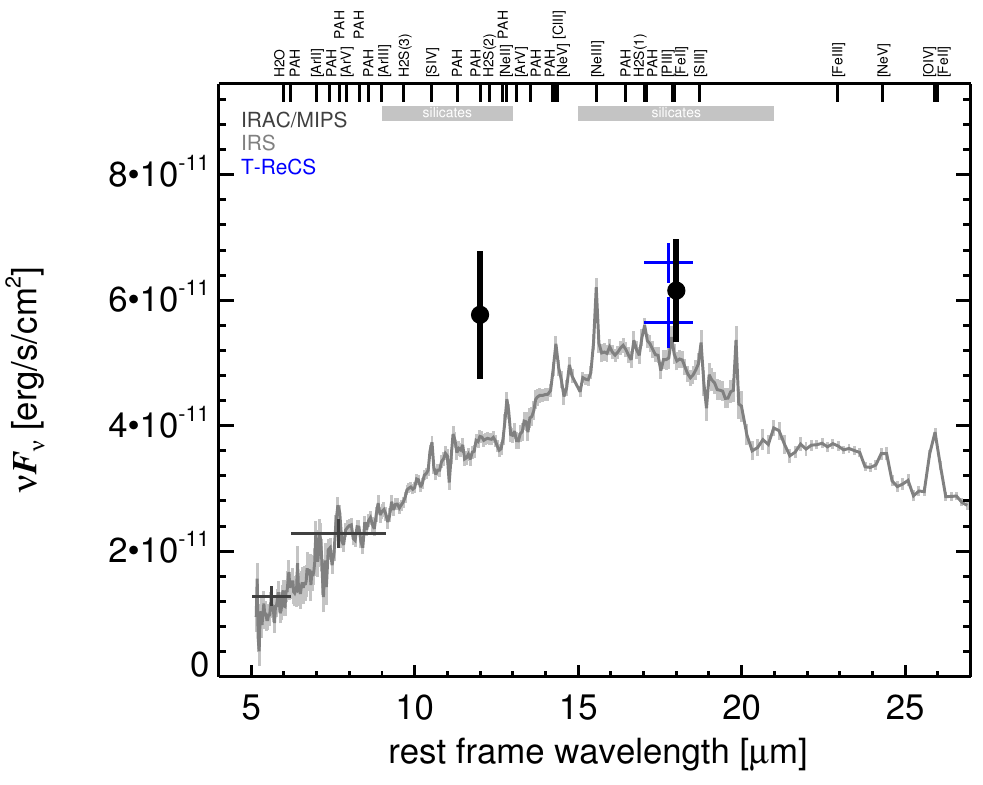}
   \caption{\label{fig:MISED_MCG-02-08-039}
      MIR SED of MCG-2-8-39. The description  of the symbols (if present) is the following.
      Grey crosses and  solid lines mark the \spitzer/IRAC, MIPS and IRS data. 
      The colour coding of the other symbols is: 
      green for COMICS, magenta for Michelle, blue for T-ReCS and red for VISIR data.
      Darker-coloured solid lines mark spectra of the corresponding instrument.
      The black filled circles mark the nuclear 12 and $18\,\mu$m  continuum emission estimate from the data.
      The ticks on the top axis mark positions of common MIR emission lines, while the light grey horizontal bars mark wavelength ranges affected by the silicate 10 and 18$\mu$m features.}
\end{figure}
\clearpage

\twocolumn[\begin{@twocolumnfalse}  
\subsection{MCG-3-34-64 -- IRAS\,13197-1627}\label{app:MCG-03-34-064}
MCG-3-34-64 is an early-type infrared-luminous spiral galaxy at a redshift of $z=$ 0.0165 ($D\sim$\,Mpc) with a Sy\,1.8-2 nucleus \citep{aguero_spectral_1994,de_grijp_warm_1992} with polarized broad emission lines \citep{young_polarimetry_1996}.
It is a member of the nine-month BAT AGN sample and possesses a jet visible in radio with an inner PA of $\sim 39\degree$ \citep{schmitt_jet_2001}.
Not that MCG-3-34-64 is often confused with or erroneously called MCG-3-34-63.
After its detection through \irass \citep{osterbrock_optical_1985,de_robertis_optical_1988}, MCG-3-34-64 was observed in the $N$-band with IRTF \cite{hill_infrared_1988} and with Palomar 5\,m/MIRLIN \citep{gorjian_10_2004}.
In the \spitzer/IRAC and MIPS images, the nucleus appears marginally resolved but without significantly extended host emission.
Our nuclear IRAC $5.8\,\mu$m and MIPS $24\,\mu$m photometry is consistent with \cite{u_spectral_2012}, while the PBCD IRAC $8.0\,\mu$m image is saturated and not analysed.
The IRS LR staring-mode spectrum exhibits silicate $10\,\mu$m absorption, weak PAH emission, and a red spectral slope with a peak at $\sim 17\,\mu$m (see also \citealt{tommasin_spitzer-irs_2010,mullaney_defining_2011}).
We observed MCG-3-34-64 with VISIR in three narrow $N$-band filters in 2006 \citep{horst_mid_2008,horst_mid-infrared_2009}, followed by a VISIR LR $N$-band spectrum in 2008 \citep{honig_dusty_2010-1}.
In addition, a VISIR NEII\_2 image from 2006 is available (unpublished, to our knowledge).
As already discussed in \cite{honig_dusty_2010-1}, the MIR nucleus detected in all images appears elongated (FWHM(major axis)$\sim0.52\arcsec \sim 180$\,pc; PA$\sim51\degree$), which roughly coincides with the radio jet.
The VISIR spectrophotometry of the total nuclear flux presented in \cite{honig_dusty_2010-1} agrees with the \spitzerr spectrophotometry, while our measurement of the unresolved nuclear fluxes alone is on average $\sim 34\%$ lower. 
The silicate absorption feature seems to be reproduced by the latter, which indicates an origin in the projected central $\sim 120$\,pc.
Note that the NEII\_2 flux is much closer to the total nuclear flux because it was measured during bad MIR seeing conditions and contains the \neii emission line.
\newline\end{@twocolumnfalse}]

\begin{figure}
   \centering
   \includegraphics[angle=0,width=8.500cm]{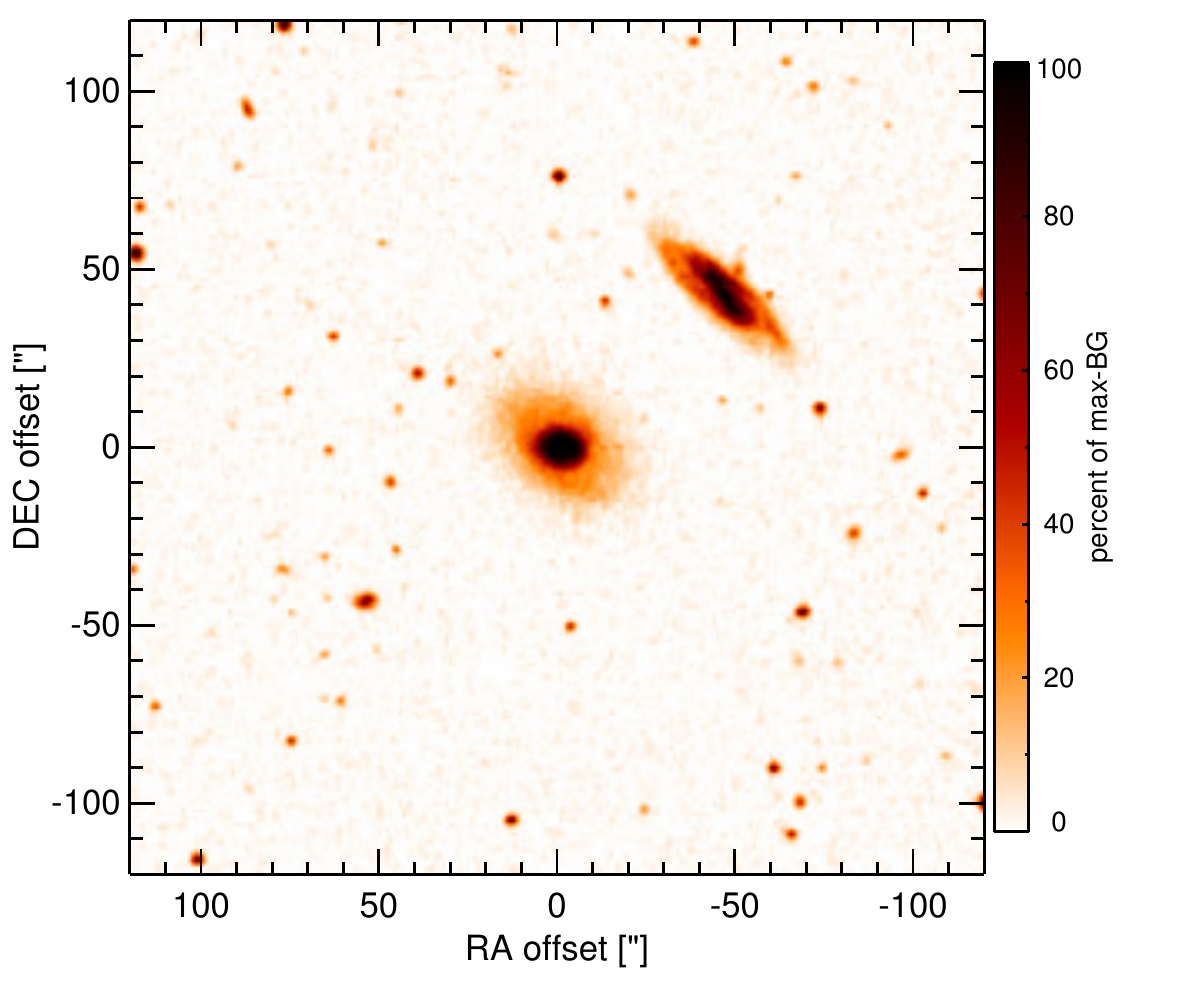}
    \caption{\label{fig:OPTim_MCG-03-34-064}
             Optical image (DSS, red filter) of MCG-3-34-64. Displayed are the central $4\arcmin$ with North up and East to the left. 
              The colour scaling is linear with white corresponding to the median background and black to the $0.01\%$ pixels with the highest intensity.  
           }
\end{figure}
\begin{figure}
   \centering
   \includegraphics[angle=0,height=3.11cm]{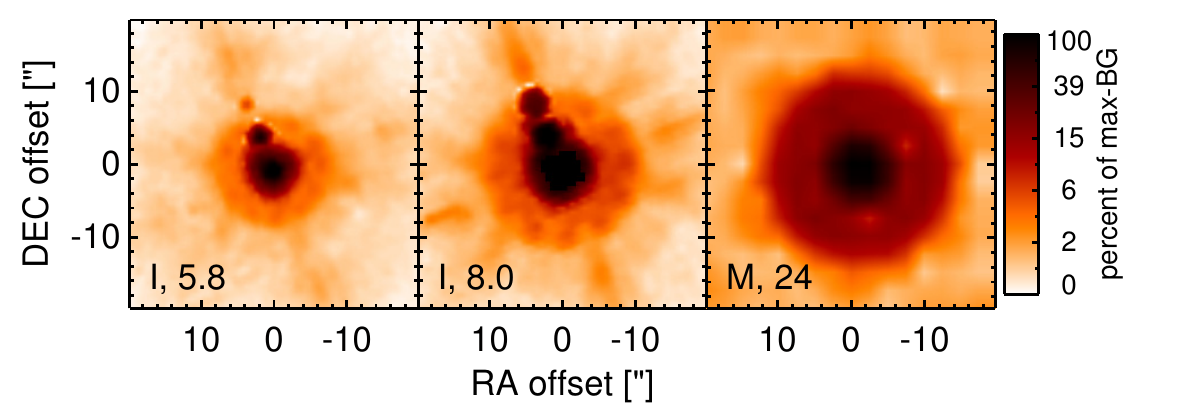}
    \caption{\label{fig:INTim_MCG-03-34-064}
             \spitzerr MIR images of MCG-3-34-64. Displayed are the inner $40\arcsec$ with North up and East to the left. The colour scaling is logarithmic with white corresponding to median background and black to the $0.1\%$ pixels with the highest intensity.
             The label in the bottom left states instrument and central wavelength of the filter in $\mu$m (I: IRAC, M: MIPS).
             Note that the apparent off-nuclear compact sources in the IRAC 5.8 and $8.0\,\mu$m images are instrumental artefacts.
           }
\end{figure}
\begin{figure}
   \centering
   \includegraphics[angle=0,width=8.500cm]{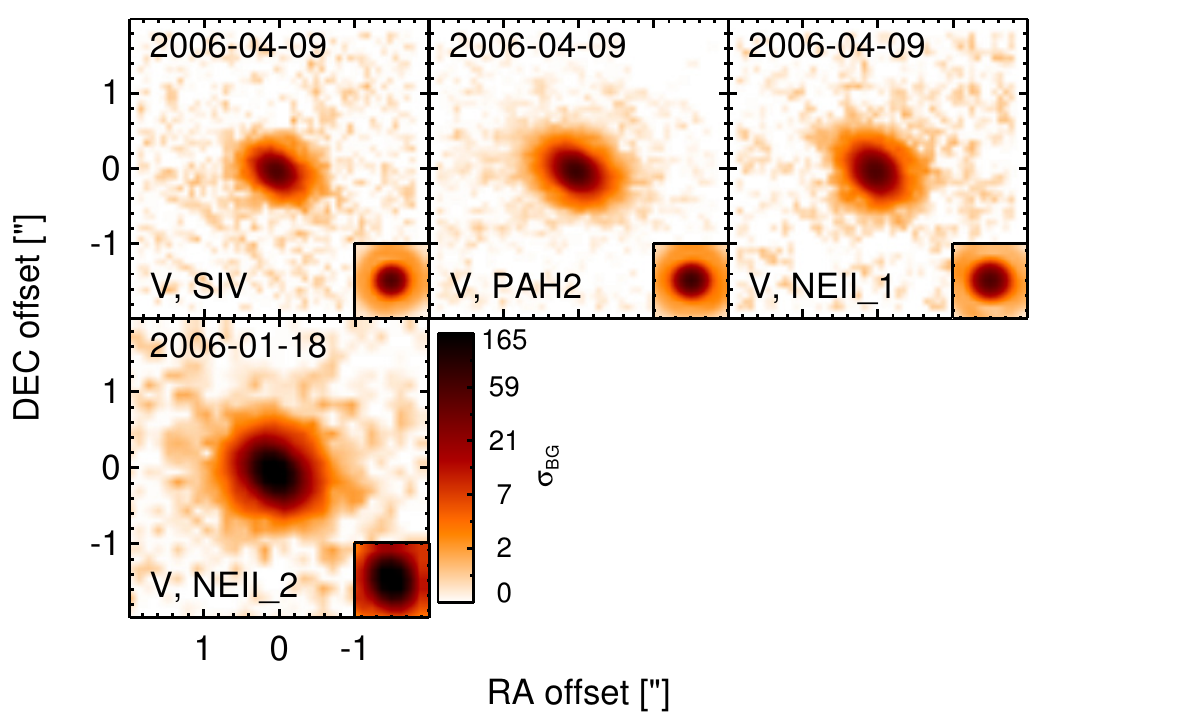}
    \caption{\label{fig:HARim_MCG-03-34-064}
             Subarcsecond-resolution MIR images of MCG-3-34-64 sorted by increasing filter wavelength. 
             Displayed are the inner $4\arcsec$ with North up and East to the left. 
             The colour scaling is logarithmic with white corresponding to median background and black to the $75\%$ of the highest intensity of all images in units of $\sigbg$.
             The inset image shows the central arcsecond of the PSF from the calibrator star, scaled to match the science target.
             The labels in the bottom left state instrument and filter names (C: COMICS, M: Michelle, T: T-ReCS, V: VISIR).
           }
\end{figure}
\begin{figure}
   \centering
   \includegraphics[angle=0,width=8.50cm]{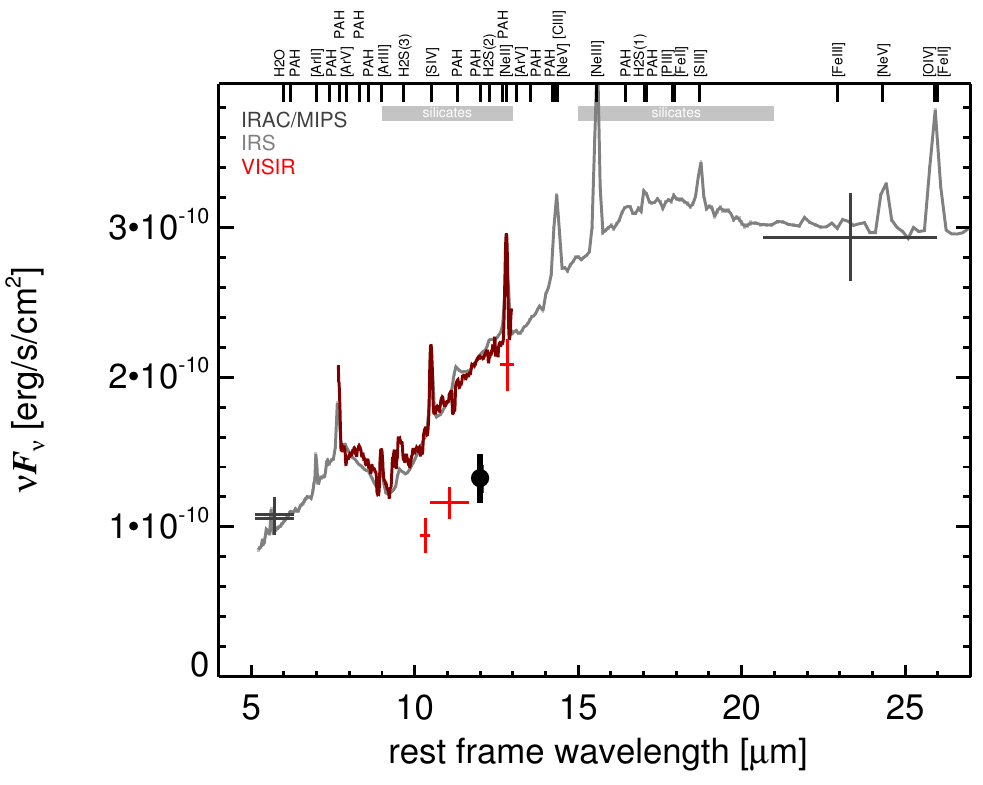}
   \caption{\label{fig:MISED_MCG-03-34-064}
      MIR SED of MCG-3-34-64. The description  of the symbols (if present) is the following.
      Grey crosses and  solid lines mark the \spitzer/IRAC, MIPS and IRS data. 
      The colour coding of the other symbols is: 
      green for COMICS, magenta for Michelle, blue for T-ReCS and red for VISIR data.
      Darker-coloured solid lines mark spectra of the corresponding instrument.
      The black filled circles mark the nuclear 12 and $18\,\mu$m  continuum emission estimate from the data.
      The ticks on the top axis mark positions of common MIR emission lines, while the light grey horizontal bars mark wavelength ranges affected by the silicate 10 and 18$\mu$m features.}
\end{figure}
\clearpage

\twocolumn[\begin{@twocolumnfalse}  
\subsection{MCG-5-23-16 -- ESO\,434-40 -- A\,0945-307}\label{app:MCG-05-23-016}
MCG-5-23-16 is an early-type galaxy at a redshift of $z=$ 0.0085 ($D\sim 39.5\,$Mpc) with an AGN classified as a Sy\,1.9 \citep{wilson_kinematics_1985} with broad emission lines in the near-infrared \citep{goodrich_infrared_1994}.
It possesses an extended NLR  on each side of the nucleus ($\sim 1\arcsec \sim 190$\,pc; PA$\sim40\degree$; \citealt{ferruit_hubble_2000}).
MCG-5-23-16 was extensively studied in the X-rays and belongs to the nine-month BAT AGN sample.
The first MIR observations were performed by \cite{frogel_8-13_1982} and \cite{glass_mid-infrared_1982}, followed by first subarcsecond-resolution $N$-band imaging with ESO 3.6\,m/TIMMI2 in 2002 \citep{raban_core_2008}.
An unresolved MIR nucleus was detected in the TIMMI2 images, which is also the case for the \spitzer/IRAC images where no host emission is evident.
The IRS LR staring-mode spectrum shows silicate $10\,\mu$m absorption, very weak PAH emission, and a red spectral slope with an emission peak at $\sim 18\,\mu$m in $\nu F_\nu$-space (see also \citealt{weaver_mid-infrared_2010}).
MCG-05-23-16 was observed with VISIR in three different narrow $N$-band filters in 2006 and 2007 \citep{honig_dusty_2010-1,reunanen_vlt_2010} and in two $Q$-band filters in 2005 and 2007 (\citealt{reunanen_vlt_2010}; and unpublished, to our knowledge).
In addition, a VISIR LR $N$-band spectrum was obtained in 2007 \citep{honig_dusty_2010-1}.
An unresolved MIR nucleus was detected in all images taken under good conditions, while the PAH1 and Q2 images suffer from non-diffraction-limited conditions.
The VISIR spectrophotometry is consistent with the \spitzerr spectrophotometry and also the older MIR photometry mentioned above.
This indicates that there have been no significant flux changes over the last $\sim 25$\,years. 
Note that the nucleus of MCG-5-23-16 was partly resolved with MIDI interferometric observations, which provides size constraints of 1.5 to 4\,pc for the MIR emitter \citep{tristram_parsec-scale_2009,burtscher_diversity_2013}.
\newline\end{@twocolumnfalse}]

\begin{figure}
   \centering
   \includegraphics[angle=0,width=8.500cm]{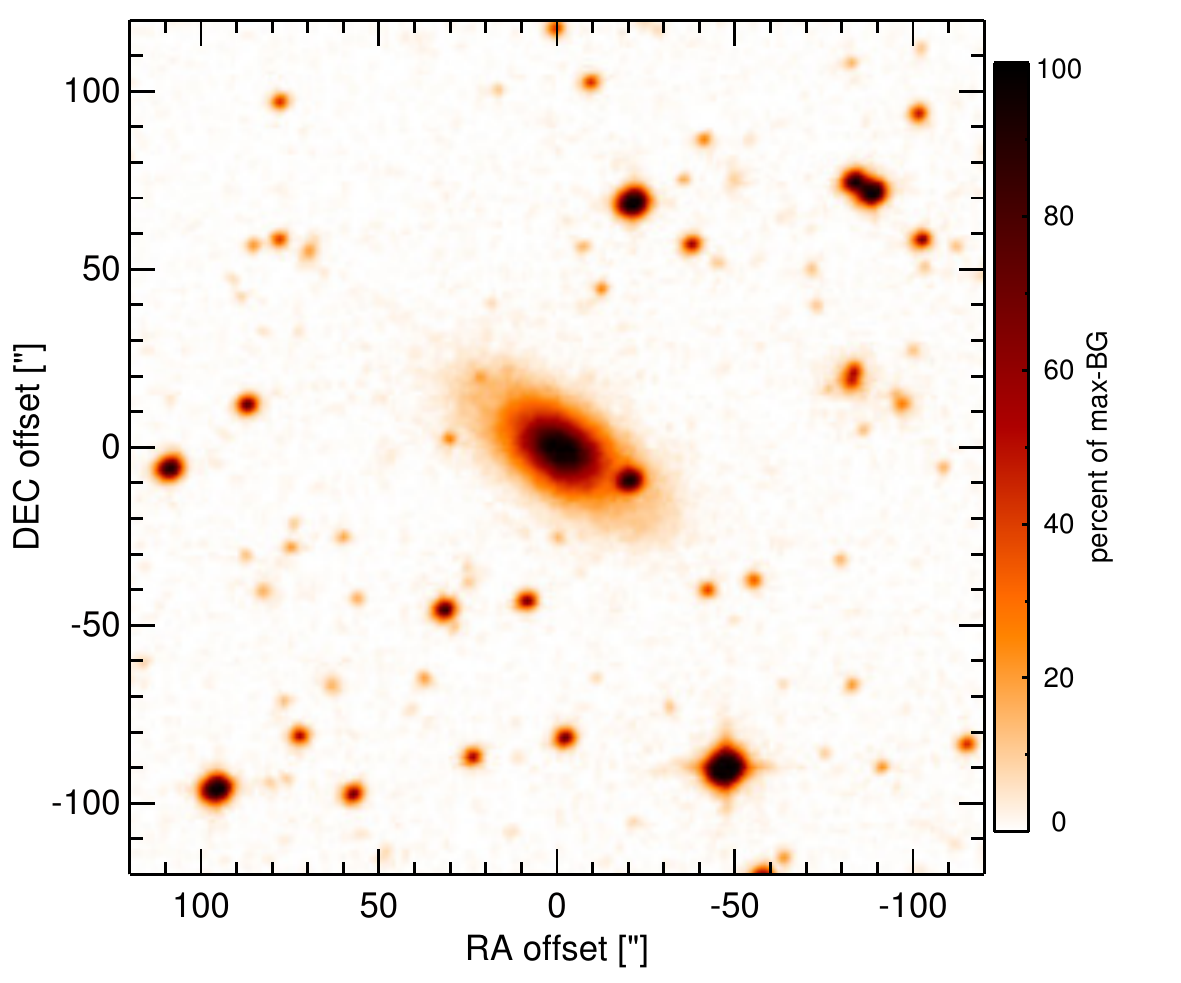}
    \caption{\label{fig:OPTim_MCG-05-23-016}
             Optical image (DSS, red filter) of MCG-5-23-16. Displayed are the central $4\arcmin$ with North up and East to the left. 
              The colour scaling is linear with white corresponding to the median background and black to the $0.01\%$ pixels with the highest intensity.  
           }
\end{figure}
\begin{figure}
   \centering
   \includegraphics[angle=0,height=3.11cm]{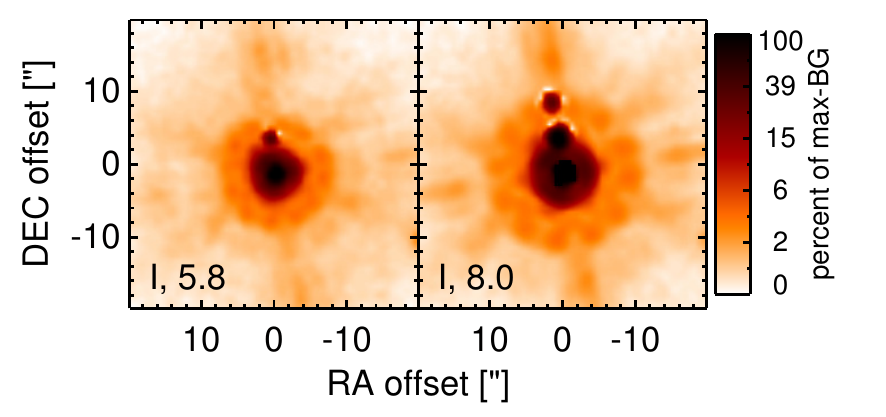}
    \caption{\label{fig:INTim_MCG-05-23-016}
             \spitzerr MIR images of MCG-5-23-16. Displayed are the inner $40\arcsec$ with North up and East to the left. The colour scaling is logarithmic with white corresponding to median background and black to the $0.1\%$ pixels with the highest intensity.
             The label in the bottom left states instrument and central wavelength of the filter in $\mu$m (I: IRAC, M: MIPS). 
             Note that the apparent off-nuclear compact sources in the IRAC 5.8 and $8.0\,\mu$m images are instrumental artefacts.
           }
\end{figure}
\begin{figure}
   \centering
   \includegraphics[angle=0,width=8.500cm]{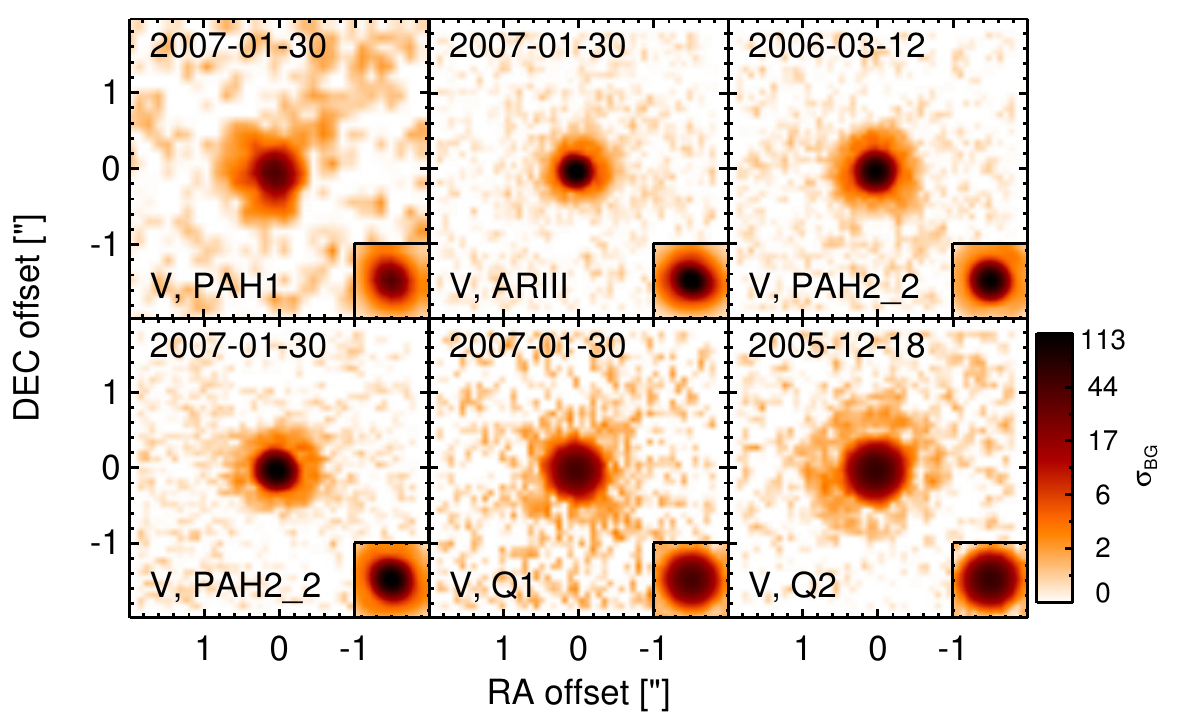}
    \caption{\label{fig:HARim_MCG-05-23-016}
             Subarcsecond-resolution MIR images of MCG-5-23-16 sorted by increasing filter wavelength. 
             Displayed are the inner $4\arcsec$ with North up and East to the left. 
             The colour scaling is logarithmic with white corresponding to median background and black to the $75\%$ of the highest intensity of all images in units of $\sigbg$.
             The inset image shows the central arcsecond of the PSF from the calibrator star, scaled to match the science target.
             The labels in the bottom left state instrument and filter names (C: COMICS, M: Michelle, T: T-ReCS, V: VISIR).
           }
\end{figure}
\begin{figure}
   \centering
   \includegraphics[angle=0,width=8.50cm]{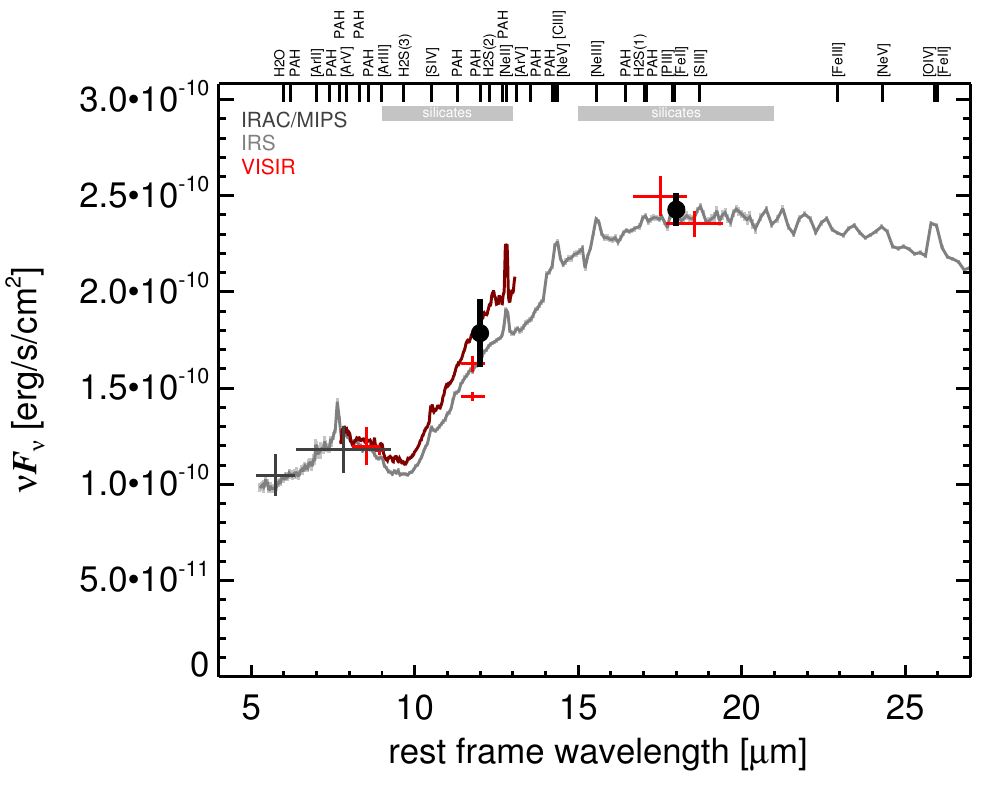}
   \caption{\label{fig:MISED_MCG-05-23-016}
      MIR SED of MCG-5-23-16. The description  of the symbols (if present) is the following.
      Grey crosses and  solid lines mark the \spitzer/IRAC, MIPS and IRS data. 
      The colour coding of the other symbols is: 
      green for COMICS, magenta for Michelle, blue for T-ReCS and red for VISIR data.
      Darker-coloured solid lines mark spectra of the corresponding instrument.
      The black filled circles mark the nuclear 12 and $18\,\mu$m  continuum emission estimate from the data.
      The ticks on the top axis mark positions of common MIR emission lines, while the light grey horizontal bars mark wavelength ranges affected by the silicate 10 and 18$\mu$m features.}
\end{figure}
\clearpage

\twocolumn[\begin{@twocolumnfalse}  
\subsection{MCG-6-30-15 -- ESO\,383-35}\label{app:MCG-06-30-015}
MCG-6-30-15 is an elongated lenticular galaxy at a redshift of $z=$ 0.0077 ($D\sim35.8$\,Mpc) with a Sy\,1.5 nucleus \citep{veron-cetty_catalogue_2010}, which was discovered in X-rays \citep{pineda_mcg_1978,pineda_x-ray_1980} and since then has been extensively studied in X-rays.
MCG-6-30-15 also belongs to the nine-month BAT AGN sample.
It possesses a NLR extending several hundreds of parsec with a PA$\sim115\degree$ \citep{ferruit_hubble_2000,schmitt_hubble_2003}, while the radio emission is unresolved \citep{nagar_radio_1999}.
The first MIR observations of MCG-6-30-15 were performed by \cite{glass_mid-infrared_1982} and \cite{ward_continuum_1987}.
After \iras, \iso/ISOCAM produced MIR images in 1996 \citep{ramos_almeida_mid-infrared_2007}, while the first subarcsecond-resolution $N$-band imaging was obtained with Palomar 5\,m/MIRLIN in 2000 by \cite{gorjian_10_2004}. 
MCG-6-30-15 appears rather point-like in the \spitzer/IRAC and MIPS images and our IRAC $5.8$ and $8.0\,\mu$m photometry agrees with the values by \cite{gallimore_infrared_2010}.
The \spitzer/IRS LR staring mode spectrum shows silicate 10 and $18\,\mu$m emission, weak PAH features and a rather flat spectral slope with an emission peak at $\sim 18\,\mu$m  in $\nu F_\nu$-space (see also \citealt{buchanan_spitzer_2006,wu_spitzer/irs_2009,tommasin_spitzer-irs_2010,gallimore_infrared_2010,mullaney_defining_2011}).
We observed MCG-6-30-15 with VISIR in three narrow $N$-band filters in 2006 \citep{horst_mid_2008,horst_mid-infrared_2009} and obtained a VISIR LR $N$-band spectrum in 2009, while additional two narrow $N$-band images from 2010 are available (unpublished, to our knowledge).
A compact MIR nucleus was detected in all images, which is possibly marginally resolved (FWHM $\sim 0.37\arcsec \sim 63\,$pc; PA$\sim100\degree$).
This would roughly coincide with the NLR extension.
However, the uncertainties are too high to definitely classify the nucleus of MCG-6-30-15 as extended at subarcsecond resolution in the MIR.
Our nuclear VISIR photometry is consistent with the \spitzerr spectrophotometry and the VISIR spectrum from \cite{honig_dusty_2010-1}.
Apart from the significantly higher MIRLIN measurement, the MIR photometry obtained during the last $\sim 25$\,years provides consistent results, indicating that MCG-6-30-15 is not significantly variable in the MIR.
 \newline\end{@twocolumnfalse}]

\begin{figure}
   \centering
   \includegraphics[angle=0,width=8.500cm]{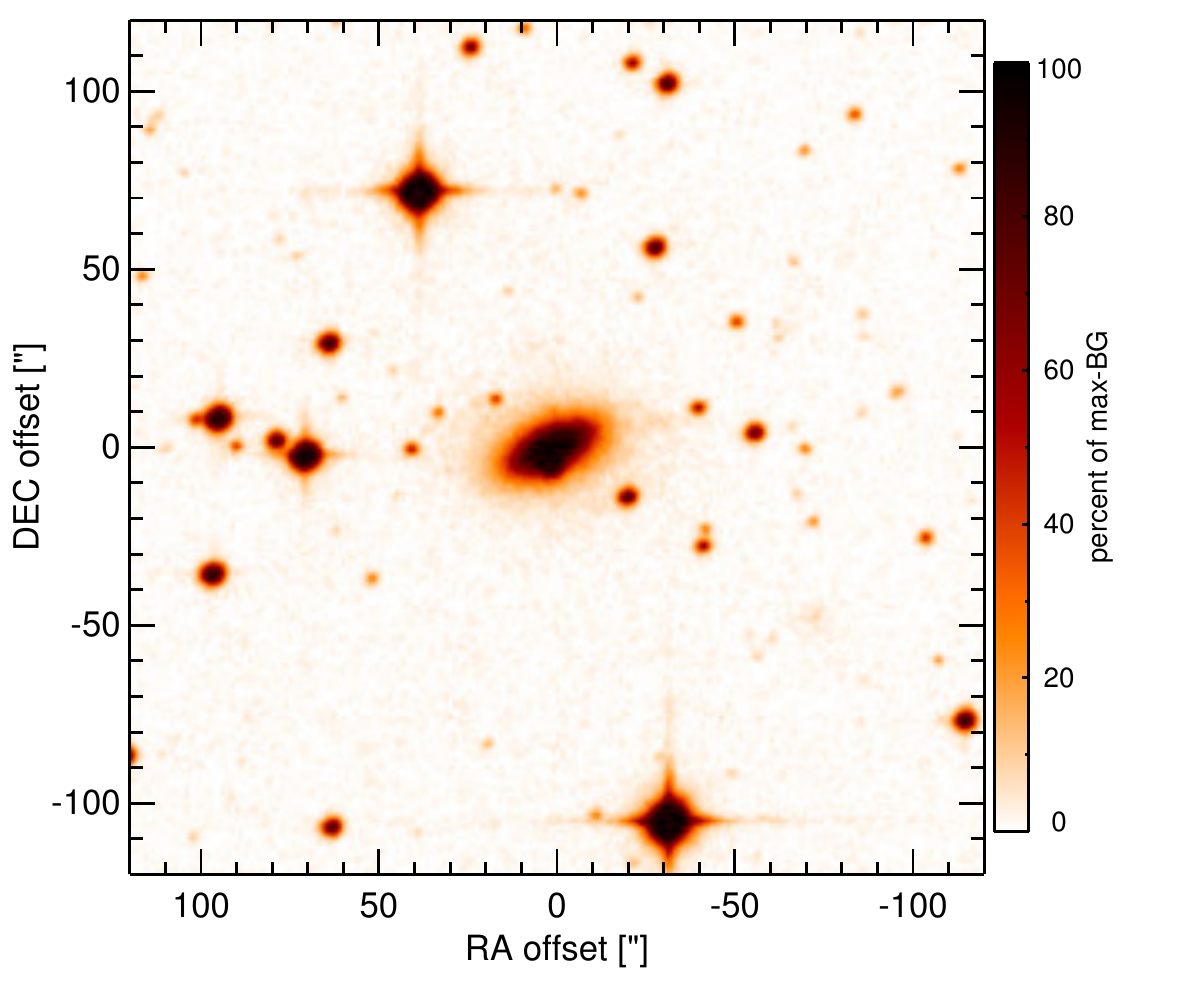}
    \caption{\label{fig:OPTim_MCG-06-30-015}
             Optical image (DSS, red filter) of MCG-6-30-15. Displayed are the central $4\arcmin$ with North up and East to the left. 
              The colour scaling is linear with white corresponding to the median background and black to the $0.01\%$ pixels with the highest intensity.  
           }
\end{figure}
\begin{figure}
   \centering
   \includegraphics[angle=0,height=3.11cm]{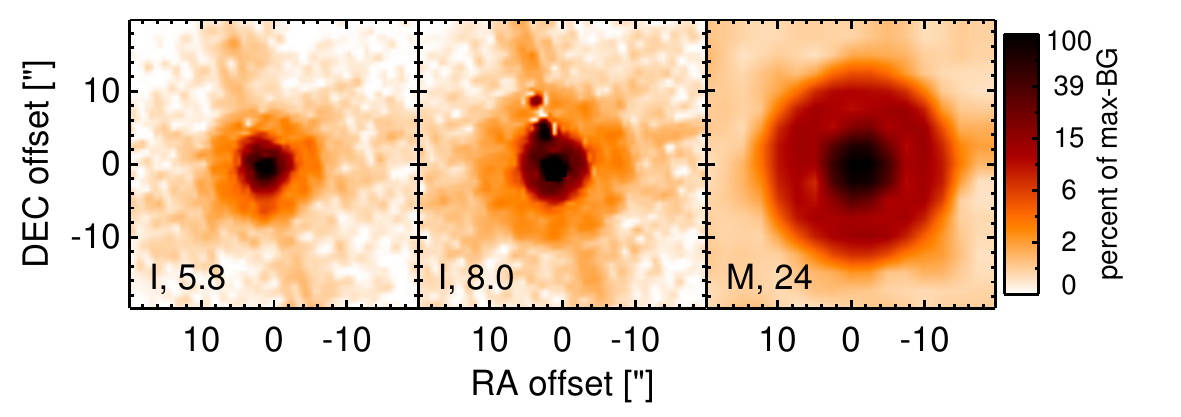}
    \caption{\label{fig:INTim_MCG-06-30-015}
             \spitzerr MIR images of MCG-6-30-15. Displayed are the inner $40\arcsec$ with North up and East to the left. The colour scaling is logarithmic with white corresponding to median background and black to the $0.1\%$ pixels with the highest intensity.
             The label in the bottom left states instrument and central wavelength of the filter in $\mu$m (I: IRAC, M: MIPS). 
             Note that the apparent off-nuclear compact sources in the IRAC $8.0\,\mu$m image are instrumental artefacts.
           }
\end{figure}
\begin{figure}
   \centering
   \includegraphics[angle=0,width=8.500cm]{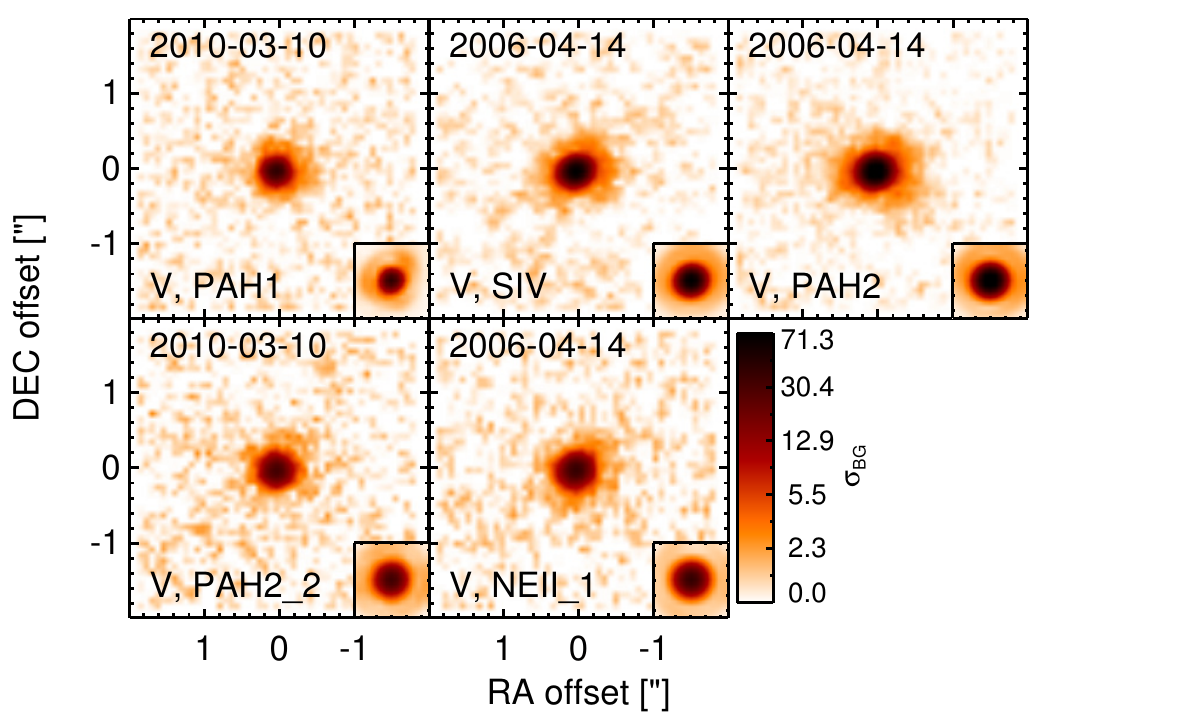}
    \caption{\label{fig:HARim_MCG-06-30-015}
             Subarcsecond-resolution MIR images of MCG-6-30-15 sorted by increasing filter wavelength. 
             Displayed are the inner $4\arcsec$ with North up and East to the left. 
             The colour scaling is logarithmic with white corresponding to median background and black to the $75\%$ of the highest intensity of all images in units of $\sigbg$.
             The inset image shows the central arcsecond of the PSF from the calibrator star, scaled to match the science target.
             The labels in the bottom left state instrument and filter names (C: COMICS, M: Michelle, T: T-ReCS, V: VISIR).
           }
\end{figure}
\begin{figure}
   \centering
   \includegraphics[angle=0,width=8.50cm]{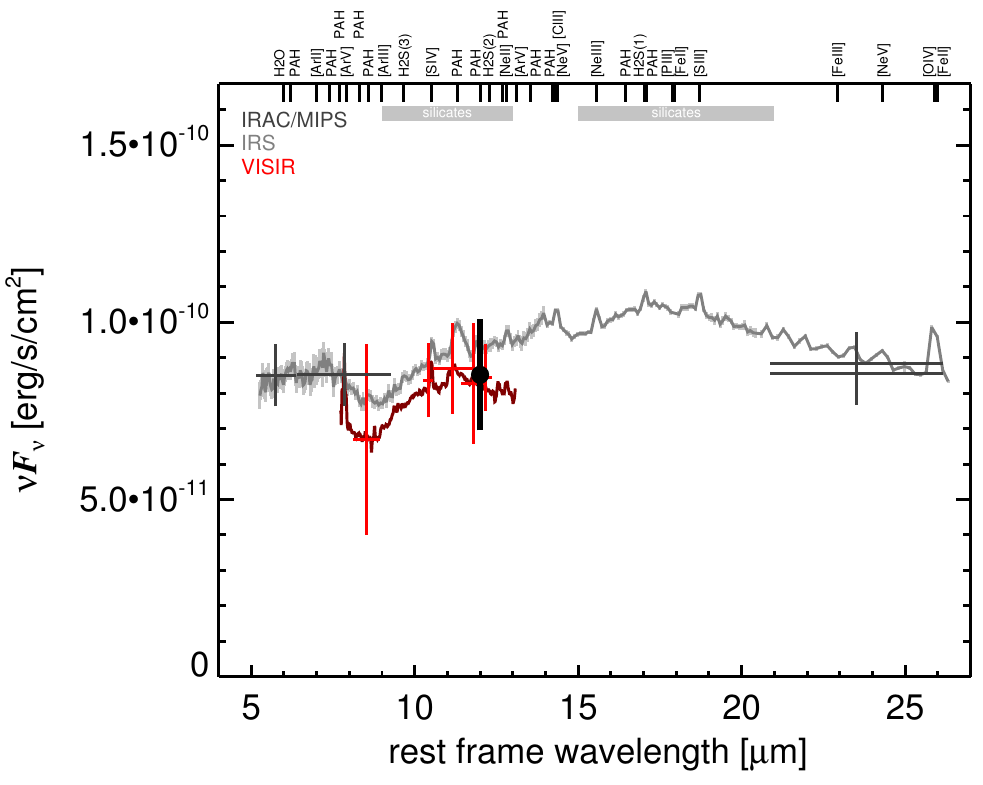}
   \caption{\label{fig:MISED_MCG-06-30-015}
      MIR SED of MCG-6-30-15. The description  of the symbols (if present) is the following.
      Grey crosses and  solid lines mark the \spitzer/IRAC, MIPS and IRS data. 
      The colour coding of the other symbols is: 
      green for COMICS, magenta for Michelle, blue for T-ReCS and red for VISIR data.
      Darker-coloured solid lines mark spectra of the corresponding instrument.
      The black filled circles mark the nuclear 12 and $18\,\mu$m  continuum emission estimate from the data.
      The ticks on the top axis mark positions of common MIR emission lines, while the light grey horizontal bars mark wavelength ranges affected by the silicate 10 and 18$\mu$m features.}
\end{figure}
\clearpage

\twocolumn[\begin{@twocolumnfalse}  
\subsection{MR\,2251-178}\label{app:MR2251-178}
MR\,2251-178 is a quasar-regime AGN at a redshift of $z=$ 0.0640 ($D\sim271$\,Mpc) with an optical Sy\,1.5 classification \citep{veron-cetty_catalogue_2010}.
It was the first quasar discovered in X-rays and the second quasar with X-ray detection after 3C\,273 \citep{ricker_discovery_1978}.
Since then it has been extensively studied in X-rays and also belongs to the nine-month BAT AGN sample.
Initial MIR observations were performed by \cite{soifer_infrared_1979}, followed by \cite{elvis_atlas_1994}.
MR\,2251-178 appears point-like in the \spitzer/IRAC and MIPS images. 
In the IRS LR mapping-mode spectrum, significant silicate $10\,\mu$m emission can be identified, while the continuum has a flat spectral slope in $\nu F_\nu$-space.
We observed MR\,2251-178 with VISIR in three narrow $N$-band filters in 2009.
The weakly detected MIR nucleus appears extended and elongated in all images, in particular the SIV\_2 filter, but with inconsistent position angles between $40$ and $90\degree$. 
Therefore, it remains uncertain whether MR\,2251-178 is truly extended in the MIR at subarcsecond resolution, and at least another epoch of deeper MIR imaging is needed.
The VISIR photometry agrees with the \spitzerr spectrophotometry and indicates the same strong silicate $10\,\mu$m emission feature, which is also consistent with the previous MIR photometry described above. 
\newline\end{@twocolumnfalse}]

\begin{figure}
   \centering
   \includegraphics[angle=0,width=8.500cm]{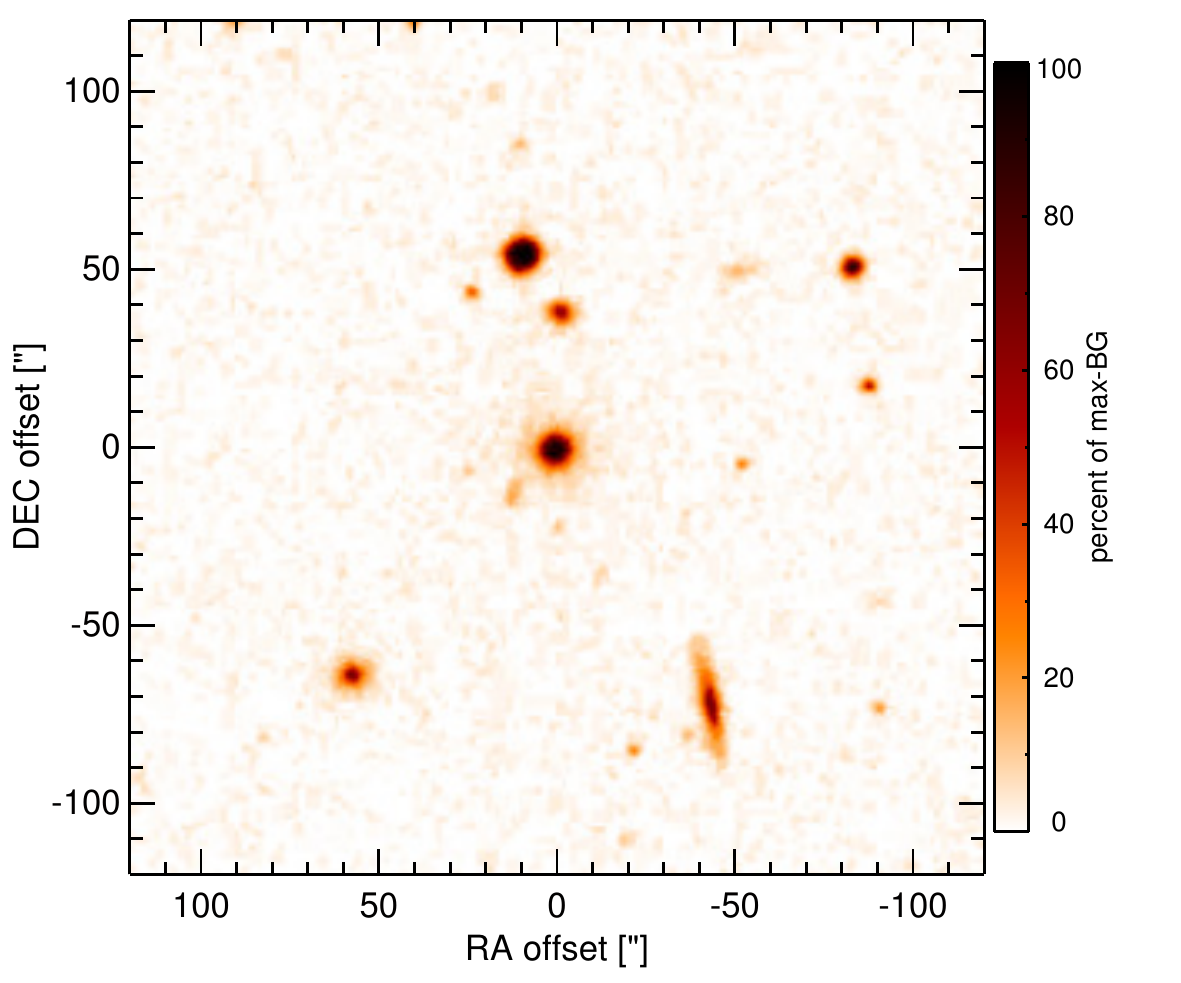}
    \caption{\label{fig:OPTim_MR2251-178}
             Optical image (DSS, red filter) of MR\,2251-178. Displayed are the central $4\arcmin$ with North up and East to the left. 
              The colour scaling is linear with white corresponding to the median background and black to the $0.01\%$ pixels with the highest intensity.  
           }
\end{figure}
\begin{figure}
   \centering
   \includegraphics[angle=0,height=3.11cm]{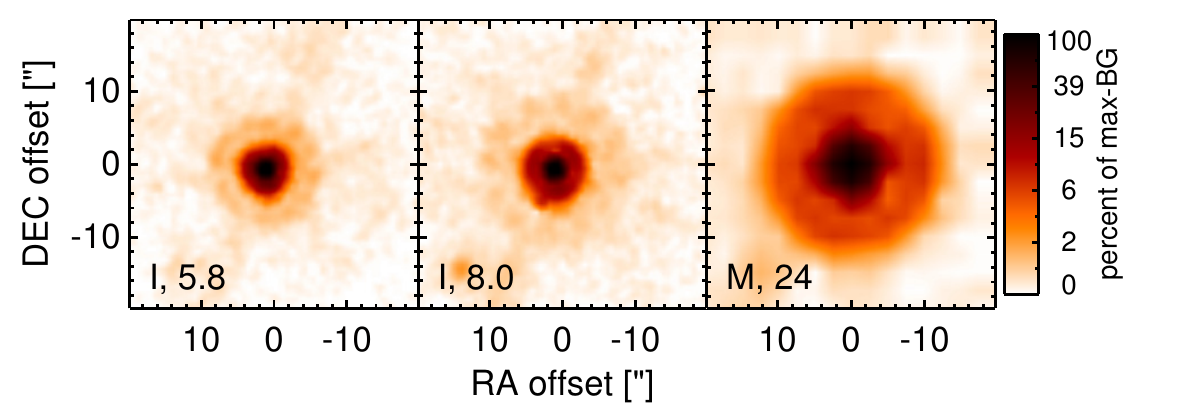}
    \caption{\label{fig:INTim_MR2251-178}
             \spitzerr MIR images of MR\,2251-178. Displayed are the inner $40\arcsec$ with North up and East to the left. The colour scaling is logarithmic with white corresponding to median background and black to the $0.1\%$ pixels with the highest intensity.
             The label in the bottom left states instrument and central wavelength of the filter in $\mu$m (I: IRAC, M: MIPS). 
           }
\end{figure}
\begin{figure}
   \centering
   \includegraphics[angle=0,height=3.11cm]{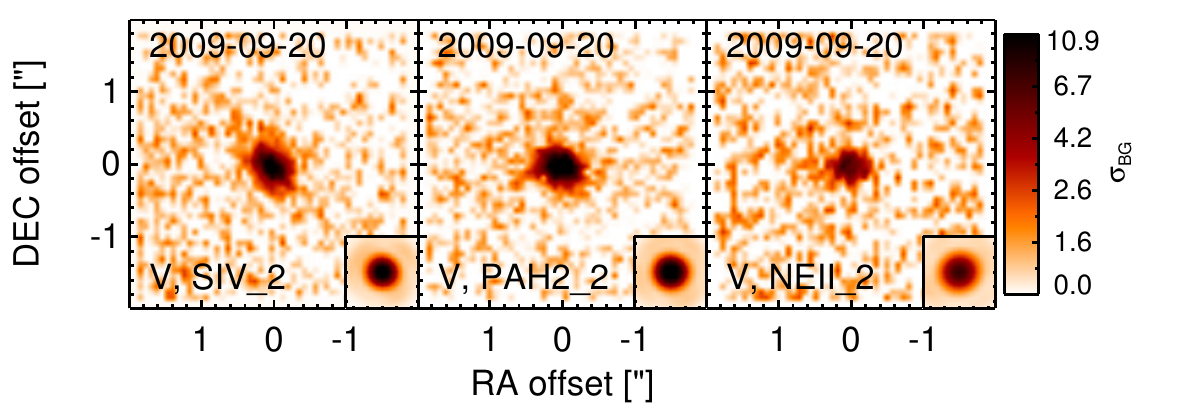}
    \caption{\label{fig:HARim_MR2251-178}
             Subarcsecond-resolution MIR images of MR\,2251-178 sorted by increasing filter wavelength. 
             Displayed are the inner $4\arcsec$ with North up and East to the left. 
             The colour scaling is logarithmic with white corresponding to median background and black to the $75\%$ of the highest intensity of all images in units of $\sigbg$.
             The inset image shows the central arcsecond of the PSF from the calibrator star, scaled to match the science target.
             The labels in the bottom left state instrument and filter names (C: COMICS, M: Michelle, T: T-ReCS, V: VISIR).
           }
\end{figure}
\begin{figure}
   \centering
   \includegraphics[angle=0,width=8.50cm]{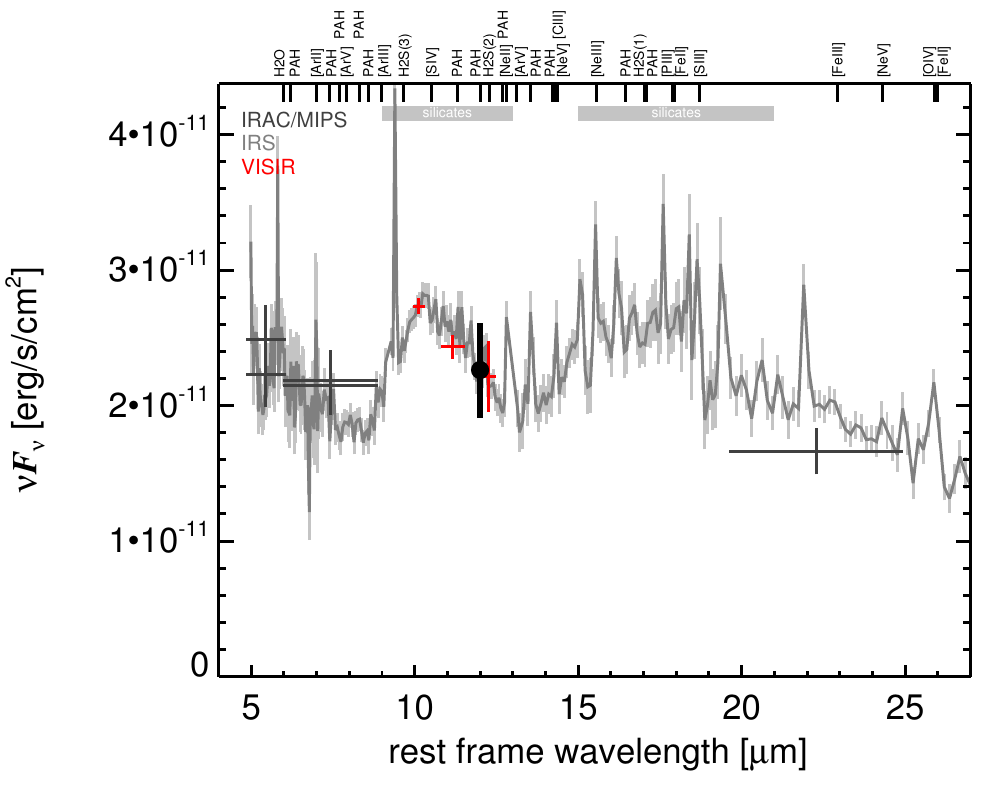}
   \caption{\label{fig:MISED_MR2251-178}
      MIR SED of MR\,2251-178. The description  of the symbols (if present) is the following.
      Grey crosses and  solid lines mark the \spitzer/IRAC, MIPS and IRS data. 
      The colour coding of the other symbols is: 
      green for COMICS, magenta for Michelle, blue for T-ReCS and red for VISIR data.
      Darker-coloured solid lines mark spectra of the corresponding instrument.
      The black filled circles mark the nuclear 12 and $18\,\mu$m  continuum emission estimate from the data.
      The ticks on the top axis mark positions of common MIR emission lines, while the light grey horizontal bars mark wavelength ranges affected by the silicate 10 and 18$\mu$m features.}
\end{figure}
\clearpage

\twocolumn[\begin{@twocolumnfalse}  
\subsection{Mrk\,3 -- UGC\,3426}\label{app:Mrk0003}
Mrk\,3 is an early-type galaxy at a redshift of $z=$ 0.0135 ($D\sim56\,$Mpc) with an AGN classified as a Sy\,2 \citep{khachikian_atlas_1974} with polarized broad emission lines \citep{miller_spectropolarimetry_1990}.
It is also a member of the nine-month BAT AGN sample.
Mrk\,3 possesses a slightly bended, collimated radio jet with a total extend of $\sim 2\arcsec$ ($\sim530$\pc; PA$\sim84\degree$; \citealt{kukula_high-resolution_1993}), and an extended NLR with $\sim2.1\arcsec$ diameter, which coincides with the radio emission ($\sim555$\pc; PA$\sim70\degree$; \citealt{capetti_morphology_1995,schmitt_comparison_1996}).
In addition, water maser emission was detected in Mrk\,3 \citep{braatz_green_2004}.
The first MIR observations of Mrk\,3 were performed by \cite{neugebauer_optical_1976} and \cite{rieke_infrared_1978}, followed by \iso/ISOCAM observations after \irass \citep{ramos_almeida_mid-infrared_2007}.
In the \spitzer/IRAC and MIPS images, Mrk\,3 appears rather point-like, and the \spitzer/IRS LR staring-mode spectrum shows weak silicate $10\,\mu$m absorption, strong forbidden emission lines, and a steep spectral slope with an emission peak at $\sim 18\,\mu$m in $\nu F_\nu$-space (see also \citealt{weedman_mid-infrared_2005,deo_mid-infrared_2009,mullaney_defining_2011}).
Mrk\,3 was observed with Michelle in the Si-5 filter in 2010 (unpublished, to our knowledge).
An elongated nucleus was detected (FWHM(major axis)$\sim 0.65\arcsec \sim 170\,$pc; PA$\sim70\degree$), which matches the NLR and radio elongation.
However, at least a second epoch of subarcsecond MIR imaging is required to verify this extension.
Our total nuclear photometry is consistent with the \spitzerr spectrophotometry.
\newline\end{@twocolumnfalse}]

\begin{figure}
   \centering
   \includegraphics[angle=0,width=8.500cm]{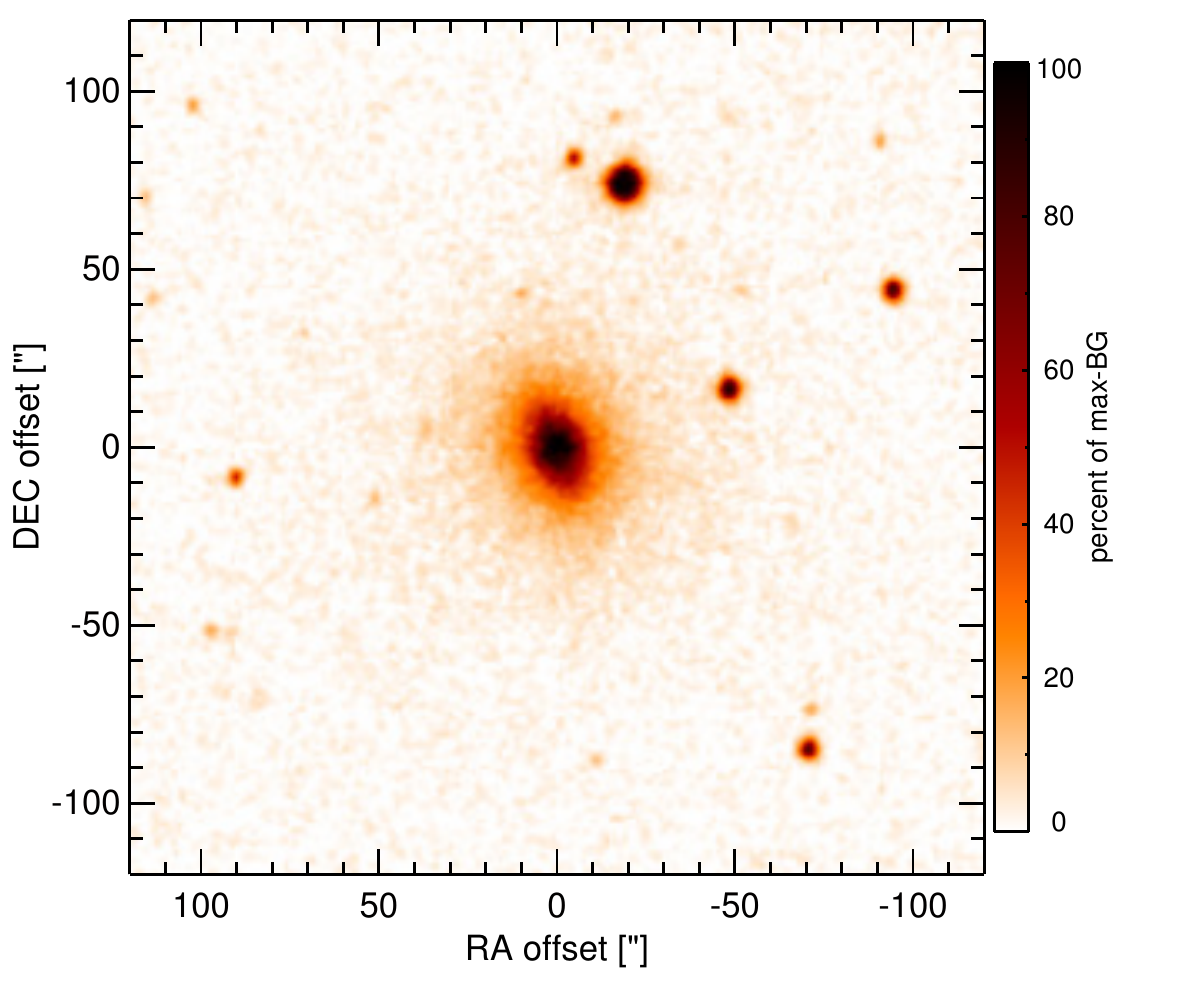}
    \caption{\label{fig:OPTim_Mrk0003}
             Optical image (DSS, red filter) of Mrk\,3. Displayed are the central $4\arcmin$ with North up and East to the left. 
              The colour scaling is linear with white corresponding to the median background and black to the $0.01\%$ pixels with the highest intensity.  
           }
\end{figure}
\begin{figure}
   \centering
   \includegraphics[angle=0,height=3.11cm]{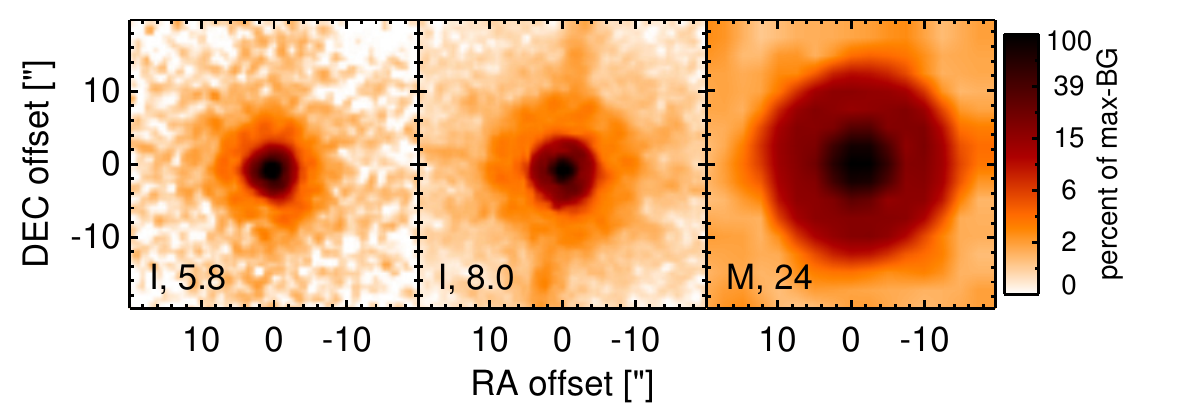}
    \caption{\label{fig:INTim_Mrk0003}
             \spitzerr MIR images of Mrk\,3. Displayed are the inner $40\arcsec$ with North up and East to the left. The colour scaling is logarithmic with white corresponding to median background and black to the $0.1\%$ pixels with the highest intensity.
             The label in the bottom left states instrument and central wavelength of the filter in $\mu$m (I: IRAC, M: MIPS). 
           }
\end{figure}
\begin{figure}
   \centering
   \includegraphics[angle=0,height=3.11cm]{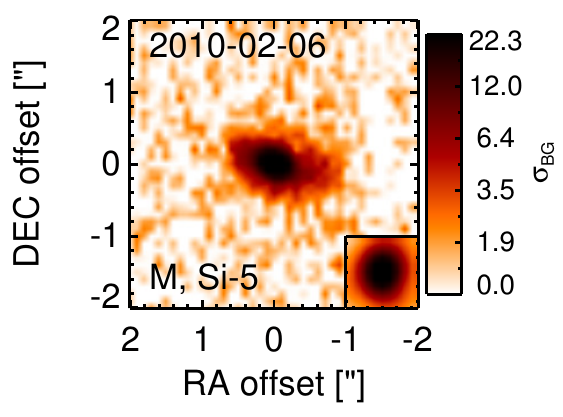}
    \caption{\label{fig:HARim_Mrk0003}
             Subarcsecond-resolution MIR images of Mrk\,3 sorted by increasing filter wavelength. 
             Displayed are the inner $4\arcsec$ with North up and East to the left. 
             The colour scaling is logarithmic with white corresponding to median background and black to the $75\%$ of the highest intensity of all images in units of $\sigbg$.
             The inset image shows the central arcsecond of the PSF from the calibrator star, scaled to match the science target.
             The labels in the bottom left state instrument and filter names (C: COMICS, M: Michelle, T: T-ReCS, V: VISIR).
           }
\end{figure}
\begin{figure}
   \centering
   \includegraphics[angle=0,width=8.50cm]{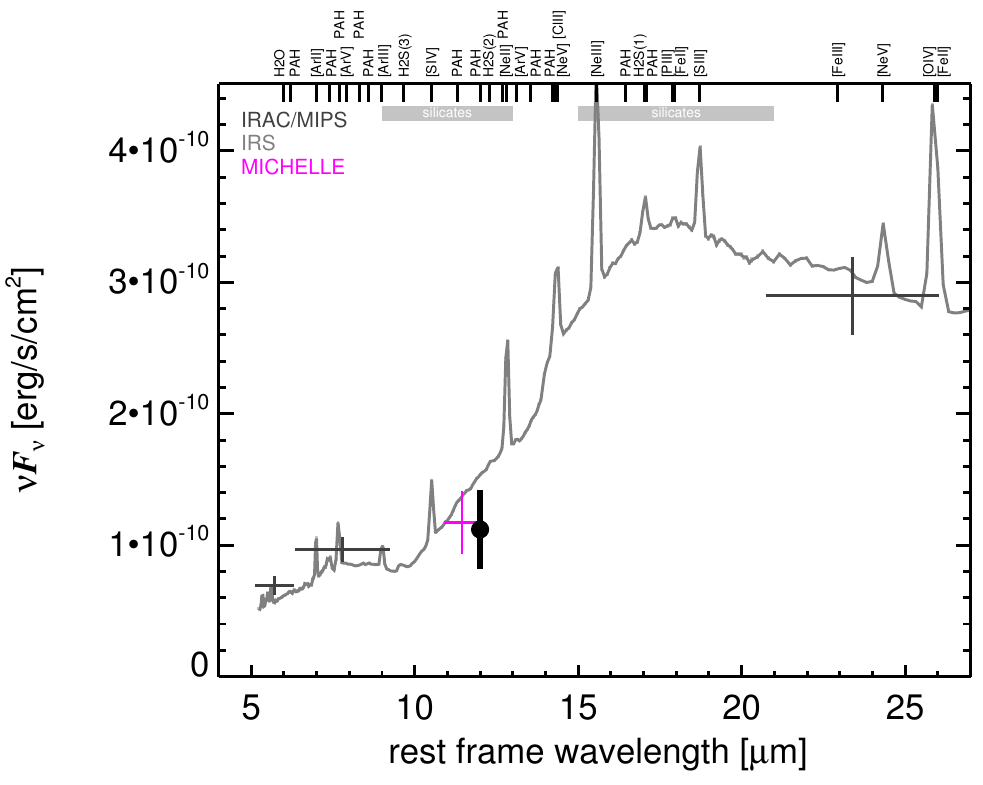}
   \caption{\label{fig:MISED_Mrk0003}
      MIR SED of Mrk\,3. The description  of the symbols (if present) is the following.
      Grey crosses and  solid lines mark the \spitzer/IRAC, MIPS and IRS data. 
      The colour coding of the other symbols is: 
      green for COMICS, magenta for Michelle, blue for T-ReCS and red for VISIR data.
      Darker-coloured solid lines mark spectra of the corresponding instrument.
      The black filled circles mark the nuclear 12 and $18\,\mu$m  continuum emission estimate from the data.
      The ticks on the top axis mark positions of common MIR emission lines, while the light grey horizontal bars mark wavelength ranges affected by the silicate 10 and 18$\mu$m features.}
\end{figure}
\clearpage

\twocolumn[\begin{@twocolumnfalse}  
\subsection{Mrk\,266 -- NGC\,5256}\label{app:Mrk0266NE}\label{app:Mrk0266SW}
Mrk\,266 is an infrared-luminous merger system consisting of two spirals at a redshift 0.0279 ($D\sim119$\,Mpc) with their nuclei separated by  $\sim10\arcsec$\, ($\sim5.5$\,kpc; PA$\sim 30\degree$; \citealt{kollatschny_double_1984}; see \citealt{mazzarella_investigation_2012} for a recent detailed study).
The nucleus, Mrk\,266SW is optically classified as a Sy\,2 \citep{veron-cetty_catalogue_2010}, while the north-eastern, Mrk\,266NE has LINER and AGN/starburst composite classifications \citep{yuan_role_2010}.
Water maser emission was detected in Mrk\,266 \citep{braatz_green_2004}.
The first MIR observations of Mrk\,266NE were performed with the Palomar 5\,m bolometer \citep{carico_iras_1988}.
It is unclear which nucleus was targeted by the following MMT bolometer MIR observations published in \cite{maiolino_new_1995}.
\iso/ISOCAM observations of Mrk\,266 are presented in \cite{ramos_almeida_mid-infrared_2007}, while the first subarcsecond-resolution $N$-band imaging performed with Palomar 5\,m/MIRLIN is reported in \cite{gorjian_10_2004} but without any description of the morphology or which nucleus was measured.
\spitzer/IRAC, IRS and MIPS data for both nuclei are presented and analysed in \cite{mazzarella_investigation_2012}.
The \spitzer/IRAC and MIPS images show both nuclei as compact sources.
The brighter southern nucleus is however slightly elongated (major axis$\sim 7\arcsec \sim 3.8\,$kpc; PA$\sim170\degree$).
The IRS LR mapping-mode spectra of both nuclei resemble typical star formation MIR SEDs with weak silicate $10\,\mu$m absorption and PAH emission features, while the features are weaker for the northern nucleus.
Mrk\,266 was observed with COMICS in the N8.8 filter in 2009 \citep{imanishi_subaru_2011}. 
Both nuclei are weakly detected in the image, the Mrk\,266NE appears compact but fainter than Mrk\,266SW, which is clearly extended (major axis $\sim 3.5\arcsec \sim 1.9\,$kpc; PA$\sim170\degree$)).
The low S/N of the detection prevents the extension analysis of Mrk\,266NE.
Our reanalysis of the image provides nuclear fluxes consistent with \cite{imanishi_subaru_2011} and significantly lower than the \spitzerr spectrophotometry. 
\newline\end{@twocolumnfalse}]

\begin{figure}
   \centering
   \includegraphics[angle=0,width=8.500cm]{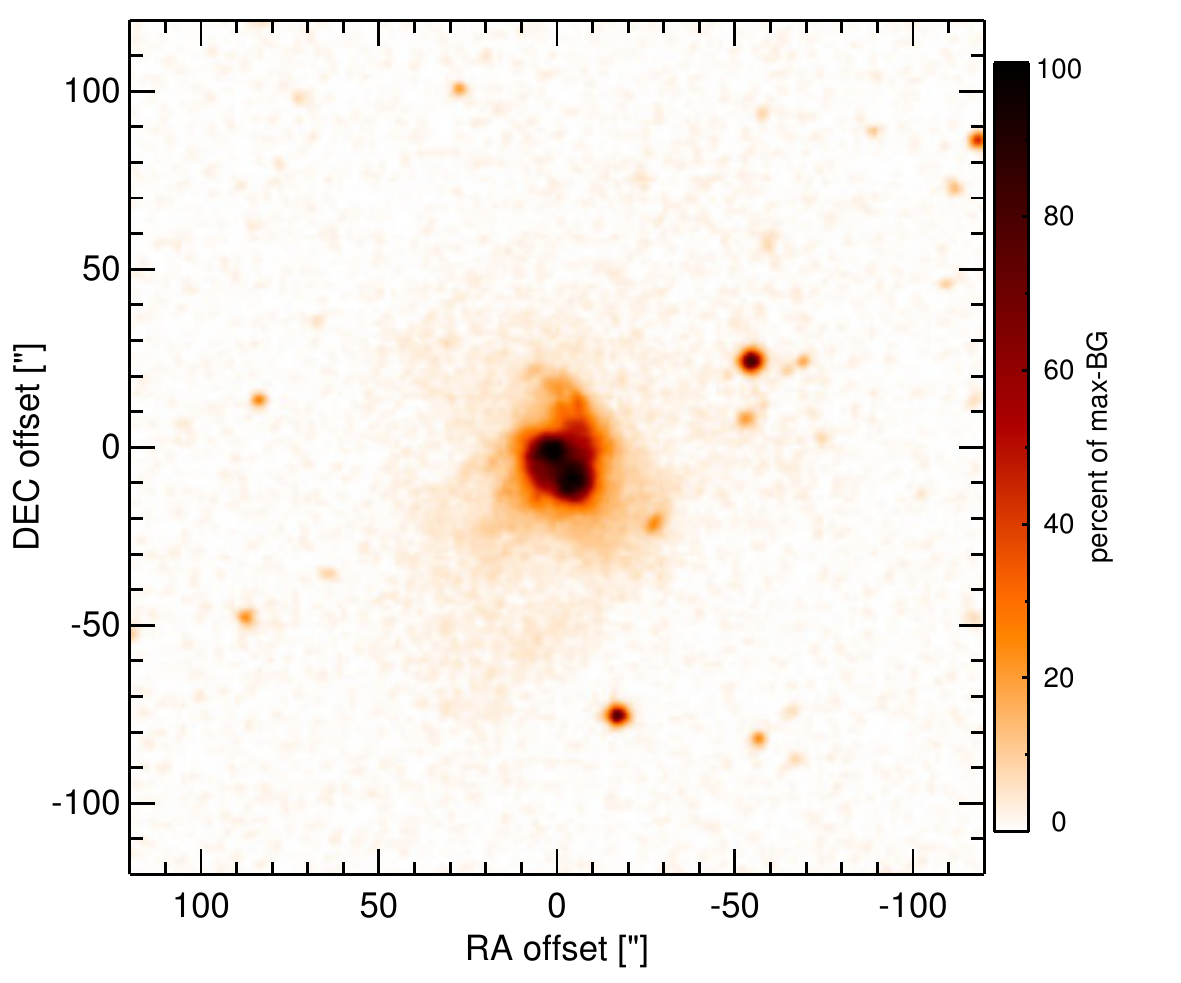}
    \caption{\label{fig:OPTim_Mrk0266NE}
             Optical image (DSS, red filter) of Mrk\,266NE. Displayed are the central $4\arcmin$ with North up and East to the left. 
              The colour scaling is linear with white corresponding to the median background and black to the $0.01\%$ pixels with the highest intensity.  
           }
\end{figure}
\begin{figure}
   \centering
   \includegraphics[angle=0,height=3.11cm]{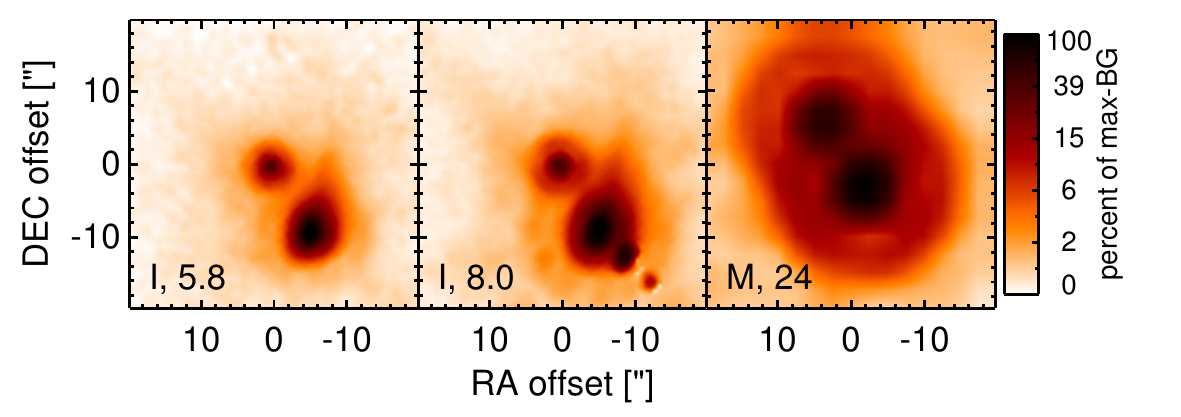}
    \caption{\label{fig:INTim_Mrk0266NE}
             \spitzerr MIR images of Mrk\,266NE. Displayed are the inner $40\arcsec$ with North up and East to the left. The colour scaling is logarithmic with white corresponding to median background and black to the $0.1\%$ pixels with the highest intensity.
             The label in the bottom left states instrument and central wavelength of the filter in $\mu$m (I: IRAC, M: MIPS).
             Note that the two apparent off-nuclear compact sources to the south-west in the IRAC $8.0\,\mu$m image are instrumental artefacts.
           }
\end{figure}
\begin{figure}
   \centering
   \includegraphics[angle=0,height=3.11cm]{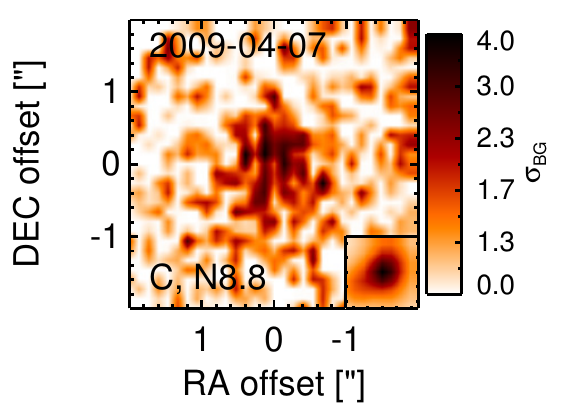}
    \caption{\label{fig:HARim_Mrk0266NE}
             Subarcsecond-resolution MIR images of Mrk\,266NE sorted by increasing filter wavelength. 
             Displayed are the inner $4\arcsec$ with North up and East to the left. 
             The colour scaling is logarithmic with white corresponding to median background and black to the $75\%$ of the highest intensity of all images in units of $\sigbg$.
             The inset image shows the central arcsecond of the PSF from the calibrator star, scaled to match the science target.
             The labels in the bottom left state instrument and filter names (C: COMICS, M: Michelle, T: T-ReCS, V: VISIR).
           }
\end{figure}
\begin{figure}
   \centering
   \includegraphics[angle=0,width=8.50cm]{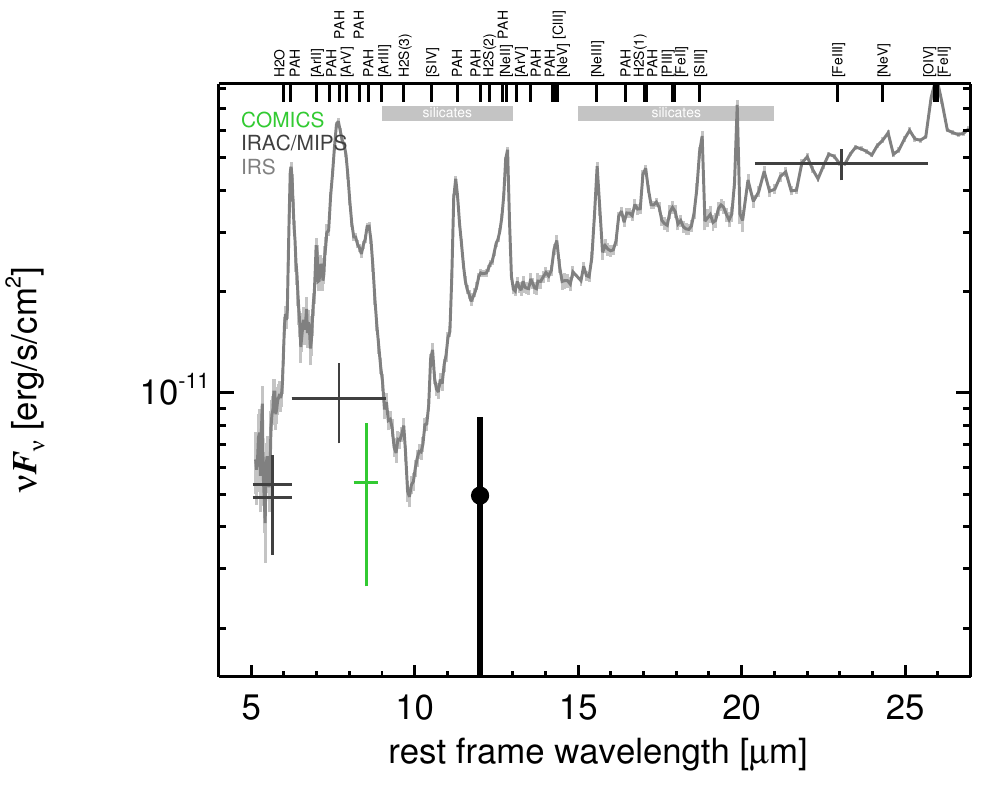}
   \caption{\label{fig:MISED_Mrk0266NE}
      MIR SED of Mrk\,266NE. The description  of the symbols (if present) is the following.
      Grey crosses and  solid lines mark the \spitzer/IRAC, MIPS and IRS data. 
      The colour coding of the other symbols is: 
      green for COMICS, magenta for Michelle, blue for T-ReCS and red for VISIR data.
      Darker-coloured solid lines mark spectra of the corresponding instrument.
      The black filled circles mark the nuclear 12 and $18\,\mu$m  continuum emission estimate from the data.
      The ticks on the top axis mark positions of common MIR emission lines, while the light grey horizontal bars mark wavelength ranges affected by the silicate 10 and 18$\mu$m features.}
\end{figure}

\begin{figure}
   \centering
   \includegraphics[angle=0,height=3.11cm]{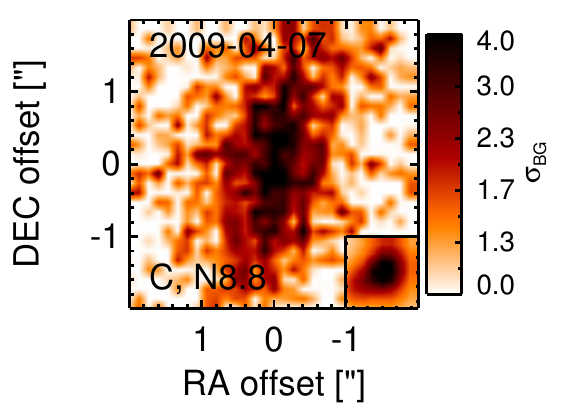}
    \caption{\label{fig:HARim_Mrk0266SW}
             Subarcsecond-resolution MIR images of Mrk\,266SW sorted by increasing filter wavelength. 
             Displayed are the inner $4\arcsec$ with North up and East to the left. 
             The colour scaling is logarithmic with white corresponding to median background and black to the $75\%$ of the highest intensity of all images in units of $\sigbg$.
             The inset image shows the central arcsecond of the PSF from the calibrator star, scaled to match the science target.
             The labels in the bottom left state instrument and filter names (C: COMICS, M: Michelle, T: T-ReCS, V: VISIR).
           }
\end{figure}
\begin{figure}
   \centering
   \includegraphics[angle=0,width=8.50cm]{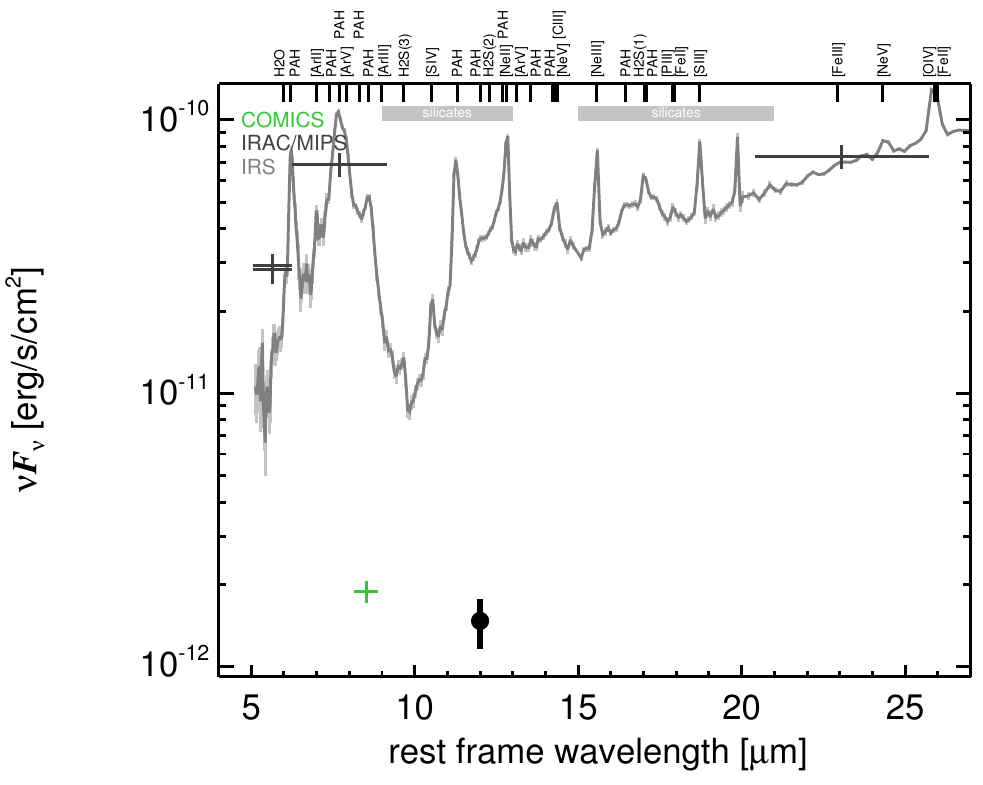}
   \caption{\label{fig:MISED_Mrk0266SW}
      MIR SED of Mrk\,266SW. The description  of the symbols (if present) is the following.
      Grey crosses and  solid lines mark the \spitzer/IRAC, MIPS and IRS data. 
      The colour coding of the other symbols is: 
      green for COMICS, magenta for Michelle, blue for T-ReCS and red for VISIR data.
      Darker-coloured solid lines mark spectra of the corresponding instrument.
      The black filled circles mark the nuclear 12 and $18\,\mu$m  continuum emission estimate from the data.
      The ticks on the top axis mark positions of common MIR emission lines, while the light grey horizontal bars mark wavelength ranges affected by the silicate 10 and 18$\mu$m features.}
\end{figure}
\clearpage

\twocolumn[\begin{@twocolumnfalse}  
\subsection{Mrk\,304 -- PG\,2214+139}\label{app:Mrk0304}
Mrk\,304 is an ultra-luminous infrared, extremely compact galaxy at a redshift of $z=$ 0.066 ($D\sim301\,$Mpc), hosting a radio-quiet borderline quasar with Sy\,1 classification \citep{veron-cetty_catalogue_2010}.
No subarcsecond-resolution radio observations are reported in the literature, only a low-resolution point source was detected \citep{kellermann_radio_1994}.
Pioneering MIR observations were performed by \cite{rieke_infrared_1978}, followed by \spitzer/IRS and MIPS and appears as a compact source in the corresponding MIPS 24\,$\mu$m image.
The IRS LR staring-mode spectrum exhibits prominent silicate 10 and 18$\,\mu$m emission and a blue spectral slope in $\nu F_\nu$-space but no PAH features (see also \citealt{schweitzer_spitzer_2006,schweitzer_extended_2008,shi_aromatic_2007}).
Mrk\,304 was imaged with COMICS in the N11.7 filter in 2006 \citep{imanishi_subaru_2011}, and a compact nucleus without further host emission was detected.
The nucleus appears to be marginally resolved (FWHM$\sim0.56\arcsec\sim0.7\,$kpc) but at least another epoch of subarcsecond MIR imaging is required to verify this extension.
Our nuclear N11.7 photometry is $\sim25\%$ higher than the value by \cite{imanishi_subaru_2011} for unknown reasons, and also higher than but still consistent with the \spitzerr spectrophotometry.
Therefore, we use the latter to compute the nuclear 12\,$\mu$m continuum emission estimate corrected for the silicate feature.
Note however, that the nuclear flux would be significantly lower if the presence of subarcsecond-extended emission can be verified.
For now, the resulting synthetic flux is  scaled by half of the ratio between $\Fpsf$ and $\Fgau$ to account for the possibly nuclear extension.
\newline\end{@twocolumnfalse}]

\begin{figure}
   \centering
   \includegraphics[angle=0,width=8.500cm]{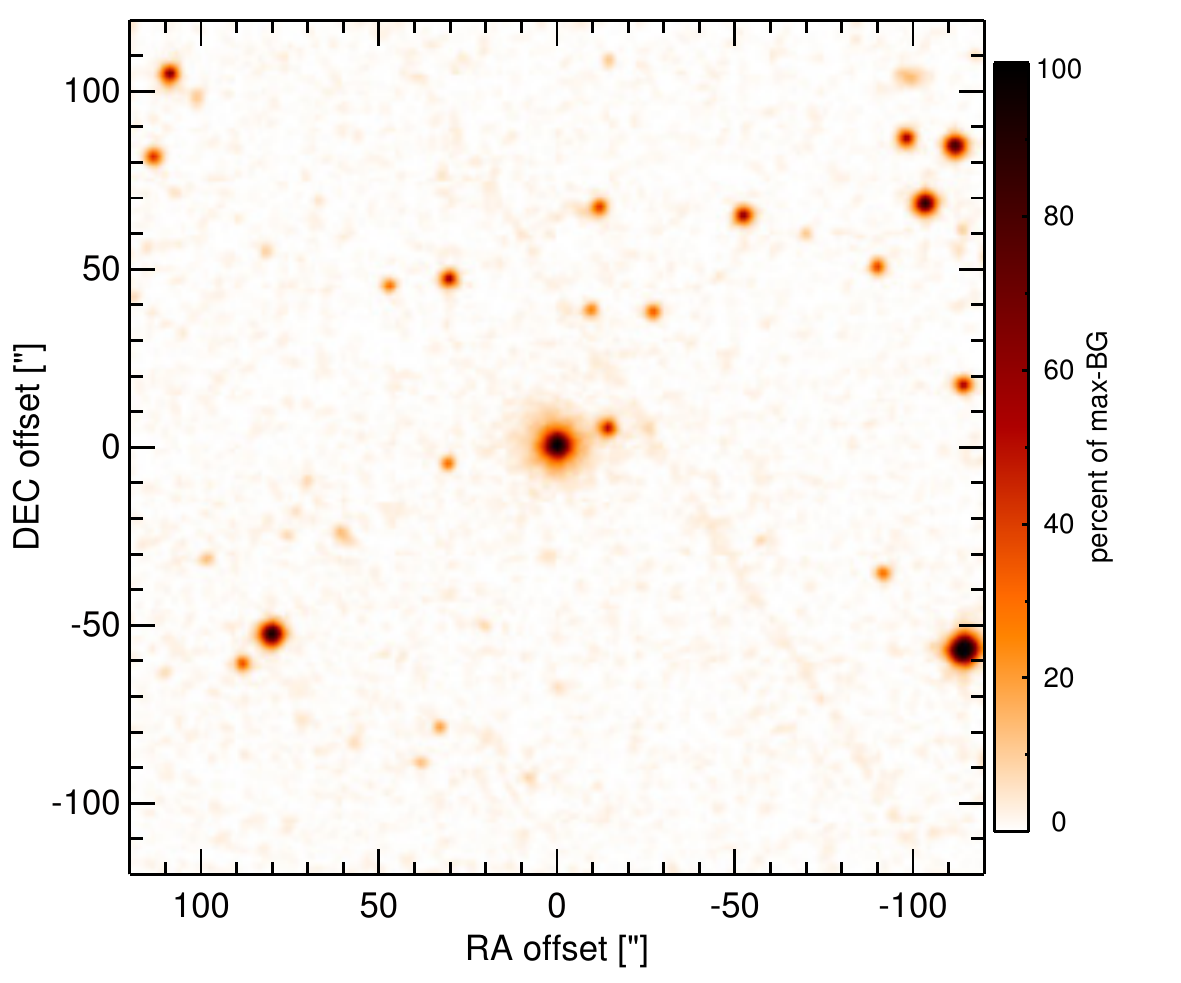}
    \caption{\label{fig:OPTim_Mrk0304}
             Optical image (DSS, red filter) of Mrk\,304. Displayed are the central $4\arcmin$ with North up and East to the left. 
              The colour scaling is linear with white corresponding to the median background and black to the $0.01\%$ pixels with the highest intensity.  
           }
\end{figure}
\begin{figure}
   \centering
   \includegraphics[angle=0,height=3.11cm]{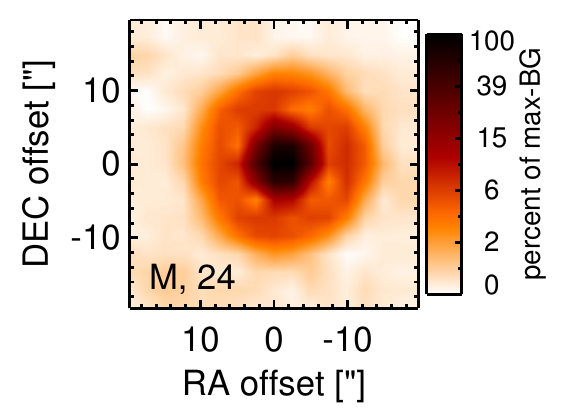}
    \caption{\label{fig:INTim_Mrk0304}
             \spitzerr MIR images of Mrk\,304. Displayed are the inner $40\arcsec$ with North up and East to the left. The colour scaling is logarithmic with white corresponding to median background and black to the $0.1\%$ pixels with the highest intensity.
             The label in the bottom left states instrument and central wavelength of the filter in $\mu$m (I: IRAC, M: MIPS). 
           }
\end{figure}
\begin{figure}
   \centering
   \includegraphics[angle=0,height=3.11cm]{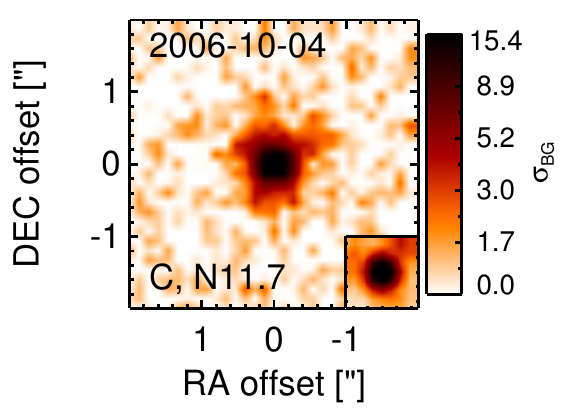}
    \caption{\label{fig:HARim_Mrk0304}
             Subarcsecond-resolution MIR images of Mrk\,304 sorted by increasing filter wavelength. 
             Displayed are the inner $4\arcsec$ with North up and East to the left. 
             The colour scaling is logarithmic with white corresponding to median background and black to the $75\%$ of the highest intensity of all images in units of $\sigbg$.
             The inset image shows the central arcsecond of the PSF from the calibrator star, scaled to match the science target.
             The labels in the bottom left state instrument and filter names (C: COMICS, M: Michelle, T: T-ReCS, V: VISIR).
           }
\end{figure}
\begin{figure}
   \centering
   \includegraphics[angle=0,width=8.50cm]{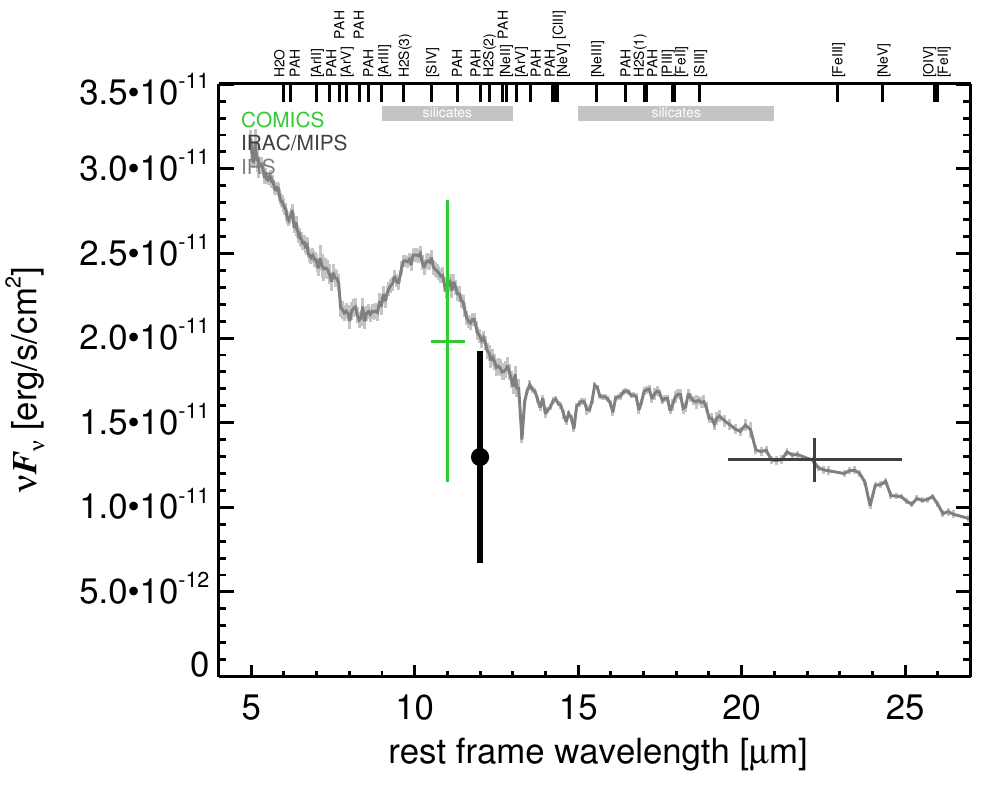}
   \caption{\label{fig:MISED_Mrk0304}
      MIR SED of Mrk\,304. The description  of the symbols (if present) is the following.
      Grey crosses and  solid lines mark the \spitzer/IRAC, MIPS and IRS data. 
      The colour coding of the other symbols is: 
      green for COMICS, magenta for Michelle, blue for T-ReCS and red for VISIR data.
      Darker-coloured solid lines mark spectra of the corresponding instrument.
      The black filled circles mark the nuclear 12 and $18\,\mu$m  continuum emission estimate from the data.
      The ticks on the top axis mark positions of common MIR emission lines, while the light grey horizontal bars mark wavelength ranges affected by the silicate 10 and 18$\mu$m features.}
\end{figure}
\clearpage

\twocolumn[\begin{@twocolumnfalse}  
\subsection{Mrk\,509}\label{app:Mrk0509}
Mrk\,509 is an early-type galaxy at a redshift of $z=$ 0.0344 ($D\sim141$\,Mpc) with a bright Sy\,1.5 nucleus \citep{veron-cetty_catalogue_2010} that has been extensively studied at all wavelengths (see \citealt{kaastra_multiwavelength_2011} for a recent detailed study).
The nucleus is known to be variable at almost all wavelengths, also in the infrared \citep{glass_long-term_2004}.
Mrk\,509 possesses a very extended NLR ($\sim10\arcsec \sim 6.4\,$kpc; \citealt{phillips_outflow_1983}) and also belongs to the nine-month BAT AGN sample.
The first $N$ band photometry was attempted in 1975 by \cite{allen_near-infrared_1976} but the nucleus remained undetected due to bad ambient conditions.
\cite{rieke_infrared_1978} report the first $N$-band detection, followed by $N$-band photometry with the ESO 3.6\,m  in 1981 \citep{glass_mid-infrared_1982}, UKIRT and IRTF spectrophotometry also in 1981 \citep{roche_8-13_1984},  $Q$ band photometry with UKIRT in 1982 \citep{ward_continuum_1987}, and MMT $N$-band photometry \citep{maiolino_new_1995}.
After \iras, the first MIR images of Mrk\,509 were obtained with \iso/ISOCAM in 1996 \citep{ramos_almeida_mid-infrared_2007}, and with subarcsecond resolution with Palomar 5\,m/MIRLIN \citep{gorjian_10_2004} and ESO 3.6\,m/TIMMI2 in 2002 \citep{siebenmorgen_mid-infrared_2004}.
Mrk\,509 appears essentially as point source in the $N$-band images, while \cite{siebenmorgen_mid-infrared_2004} claim also very weak extended emission to the south-east. 
No significant extended emission is visible in the \spitzer/IRAC and MIPS images. 
The \spitzer/IRS LR staring-mode spectrum exhibits silicate 10 and $18\,\mu$m emission, PAH features and a rather flat spectral slope in $\nu F_\nu$-space with the emission peaking at $\sim18\,\mu$m (see also \citealt{shi_9.7_2006,wu_spitzer/irs_2009,tommasin_spitzer-irs_2010,mullaney_defining_2011}).
We observed Mrk\,509 with VISIR in three narrow $N$-band filters in 2006 \citep{horst_mid_2008,horst_mid-infrared_2009} and obtained a VISIR LR $N$-band spectrum in 2008 \citep{honig_dusty_2010-1}.
In addition, T-ReCS imaging in two $N$-band filters and the Qa filter from 2008 and 2010 are available (unpublished, to our knowledge).
In all images taken under good seeing conditions, an unresolved nucleus without any other host emission was detected. 
Solely in the Si2 image from 2010 the nucleus has a wider FWHM than the standard star but seeing was worse than in the Si-2 image from 2008.
Our reanalysis of the VISIR images provides fluxes consistent with \cite{horst_mid_2008}. 
In general all subarcsecond photometry is consistent with the VISIR spectrum and on average $\sim 16\%$ lower than the \spitzerr spectrophotometry.
This together with the absence of the PAH features in the subarcsecond data indicates a minor contamination of the \spitzerr data by star formation, which occurs outside the central $\sim230$\,pc of Mrk\,509.
Considering these results, all MIR flux measurements during the last $\sim 30$ years yield consistent results within the uncertainties introduced by different filters, apertures and instruments. 
Therefore, we conclude that Mrk\,509 is not significantly variable in the $N$-band.
\newline\end{@twocolumnfalse}]

\begin{figure}
   \centering
   \includegraphics[angle=0,width=8.500cm]{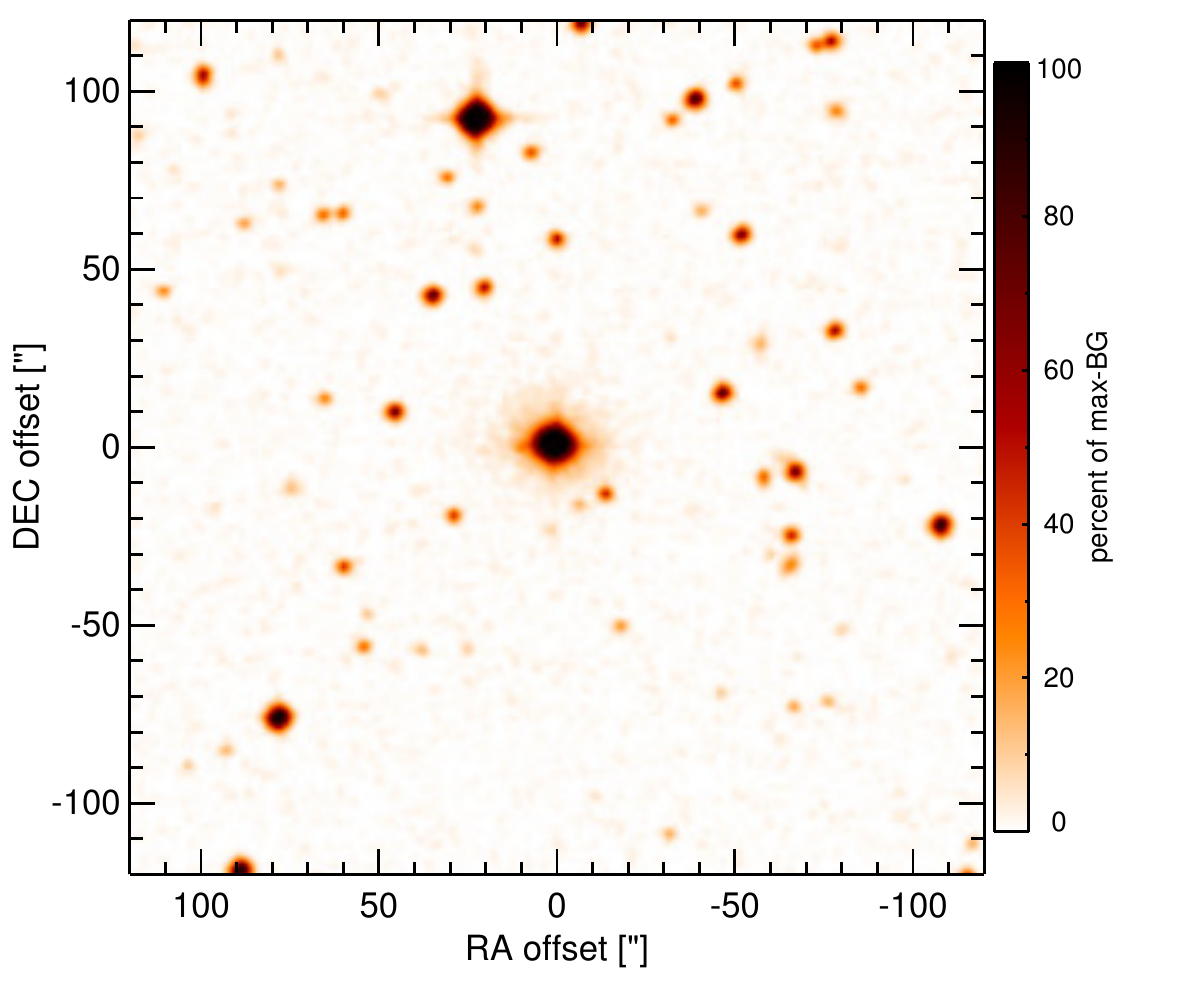}
    \caption{\label{fig:OPTim_Mrk0509}
             Optical image (DSS, red filter) of Mrk\,509. Displayed are the central $4\arcmin$ with North up and East to the left. 
              The colour scaling is linear with white corresponding to the median background and black to the $0.01\%$ pixels with the highest intensity.  
           }
\end{figure}
\begin{figure}
   \centering
   \includegraphics[angle=0,height=3.11cm]{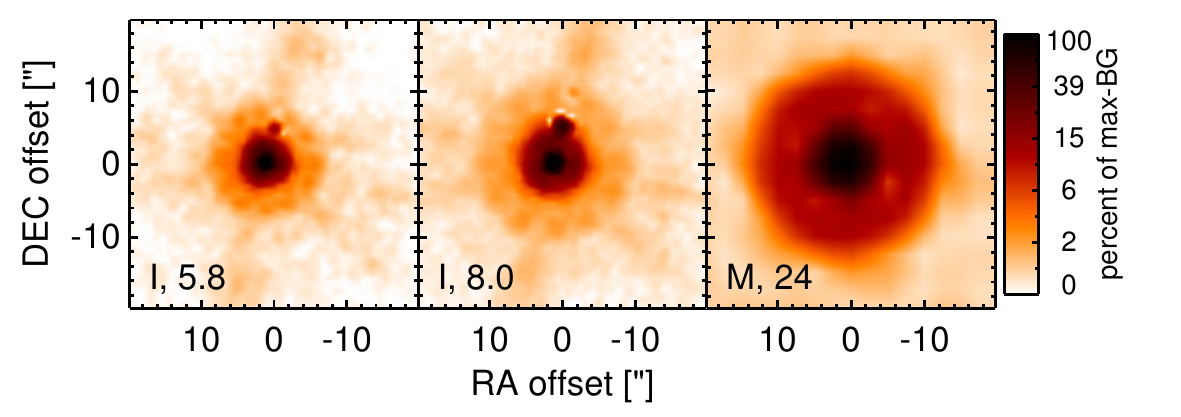}
    \caption{\label{fig:INTim_Mrk0509}
             \spitzerr MIR images of Mrk\,509. Displayed are the inner $40\arcsec$ with North up and East to the left. The colour scaling is logarithmic with white corresponding to median background and black to the $0.1\%$ pixels with the highest intensity.
             The label in the bottom left states instrument and central wavelength of the filter in $\mu$m (I: IRAC, M: MIPS).
             Note that the apparent off-nuclear compact source in the IRAC 5.8 and $8.0\,\mu$m images is an instrumental artefact.
           }
\end{figure}
\begin{figure}
   \centering
   \includegraphics[angle=0,width=8.500cm]{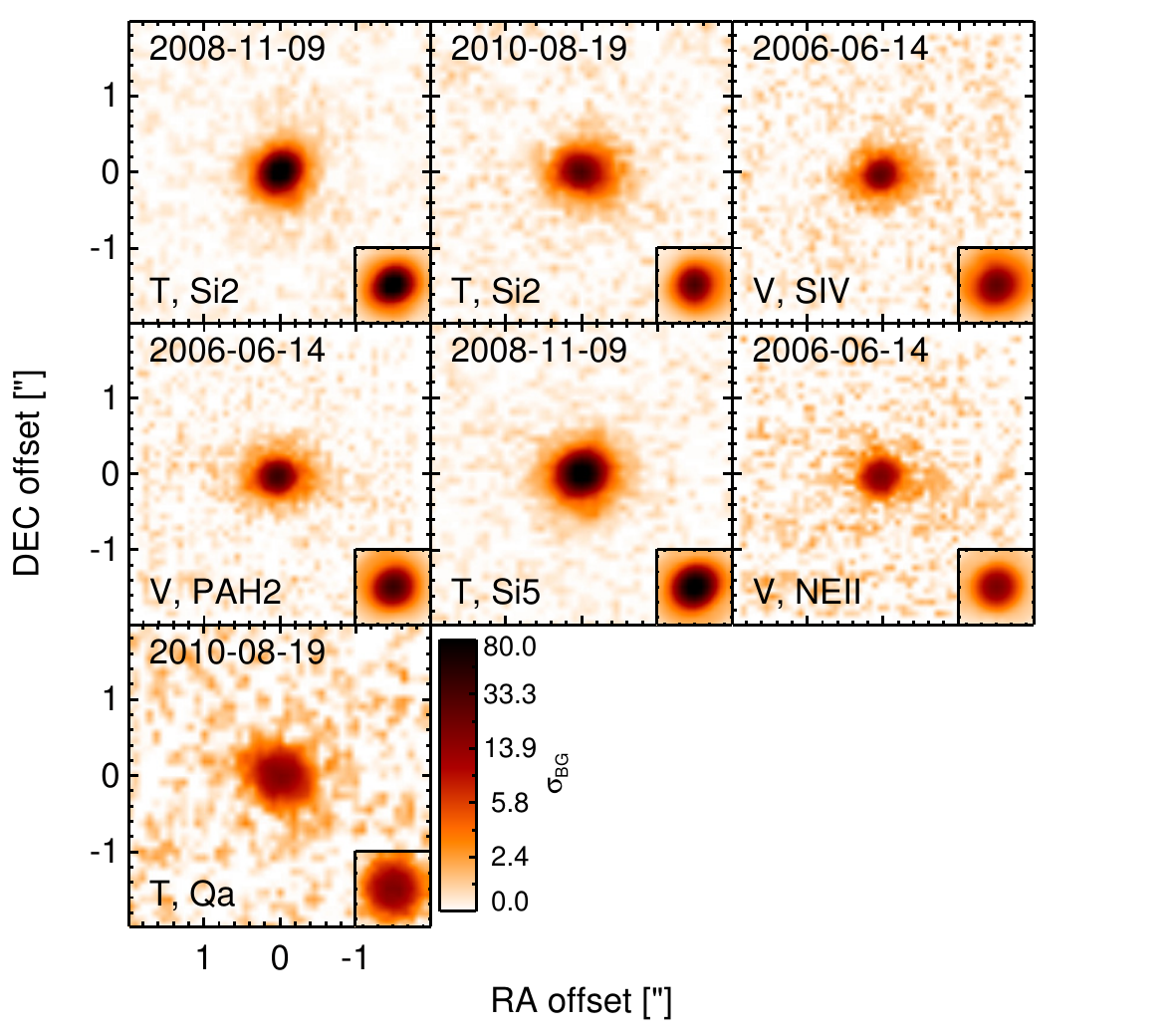}
    \caption{\label{fig:HARim_Mrk0509}
             Subarcsecond-resolution MIR images of Mrk\,509 sorted by increasing filter wavelength. 
             Displayed are the inner $4\arcsec$ with North up and East to the left. 
             The colour scaling is logarithmic with white corresponding to median background and black to the $75\%$ of the highest intensity of all images in units of $\sigbg$.
             The inset image shows the central arcsecond of the PSF from the calibrator star, scaled to match the science target.
             The labels in the bottom left state instrument and filter names (C: COMICS, M: Michelle, T: T-ReCS, V: VISIR).
           }
\end{figure}
\begin{figure}
   \centering
   \includegraphics[angle=0,width=8.50cm]{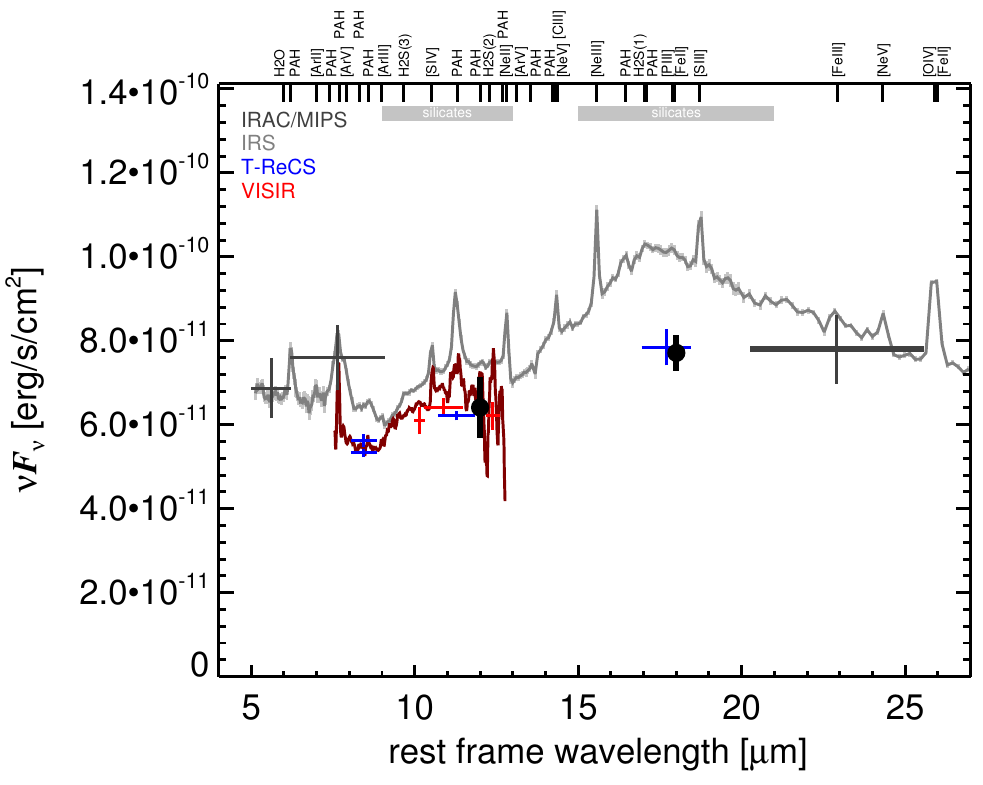}
   \caption{\label{fig:MISED_Mrk0509}
      MIR SED of Mrk\,509. The description  of the symbols (if present) is the following.
      Grey crosses and  solid lines mark the \spitzer/IRAC, MIPS and IRS data. 
      The colour coding of the other symbols is: 
      green for COMICS, magenta for Michelle, blue for T-ReCS and red for VISIR data.
      Darker-coloured solid lines mark spectra of the corresponding instrument.
      The black filled circles mark the nuclear 12 and $18\,\mu$m  continuum emission estimate from the data.
      The ticks on the top axis mark positions of common MIR emission lines, while the light grey horizontal bars mark wavelength ranges affected by the silicate 10 and 18$\mu$m features.}
\end{figure}
\clearpage

\twocolumn[\begin{@twocolumnfalse}  
\subsection{Mrk\,520 -- UGC\,11871}\label{app:Mrk0520}
Mrk\,520 is a peculiar galaxy at a redshift of $z=$ 0.0266 ($D\sim115\,$Mpc) with wide tidal arms and a compact object (LEDA\,200376) superimposed $\sim21\arcsec$ to the south-east, which is most likely a star (see 2MASS images).
Mrk\,520 contains a little-studied Sy\,1.9 nucleus \citep{lonsdale_vlbi_1992} that belongs to nine-month BAT AGN sample.
Furthermore, the nucleus is classified as an AGN/starburst composite by \cite{winter_optical_2010}, and we adopt this classification.
It appears extended at radio wavelengths on kiloparsec scales, mainly in  the north-south directions \citep{lonsdale_vlbi_1992}. 
Mrk\,520 was first detected in the MIR with \irass and followed up with \isoo \citep{dennefeld_iso_2003} and \spitzer/IRS.
The corresponding IRS LR staring-mode spectrum is dominated by PAH emission with silicate 10\,$\mu$m absorption and a red spectral slope in $\nu F_\nu$-space (see also \citealt{sargsyan_infrared_2011}).
Thus, the arcsecond-scale MIR SED  appears to be star-formation dominated but the detection of strong \nev emission points at the existence of a powerful AGN in Mrk\,520 as well.
We observed Mrk\,520 with VISIR in three narrow $N$-band filters in 2009 and detected a possibly marginally resolved nucleus (FWHM $\sim 0.43\arcsec \sim 230\,$pc).
However, at least another epoch of subarcsecond MIR imaging is required to verify this extension.
The nuclear VISIR photometry is on average $\sim 43\%$ lower than the \spitzerr spectrophotometry and indicates that at least part of the star-formation emission is excluded at subarcsecond resolution.
\newline\end{@twocolumnfalse}]

\begin{figure}
   \centering
   \includegraphics[angle=0,width=8.500cm]{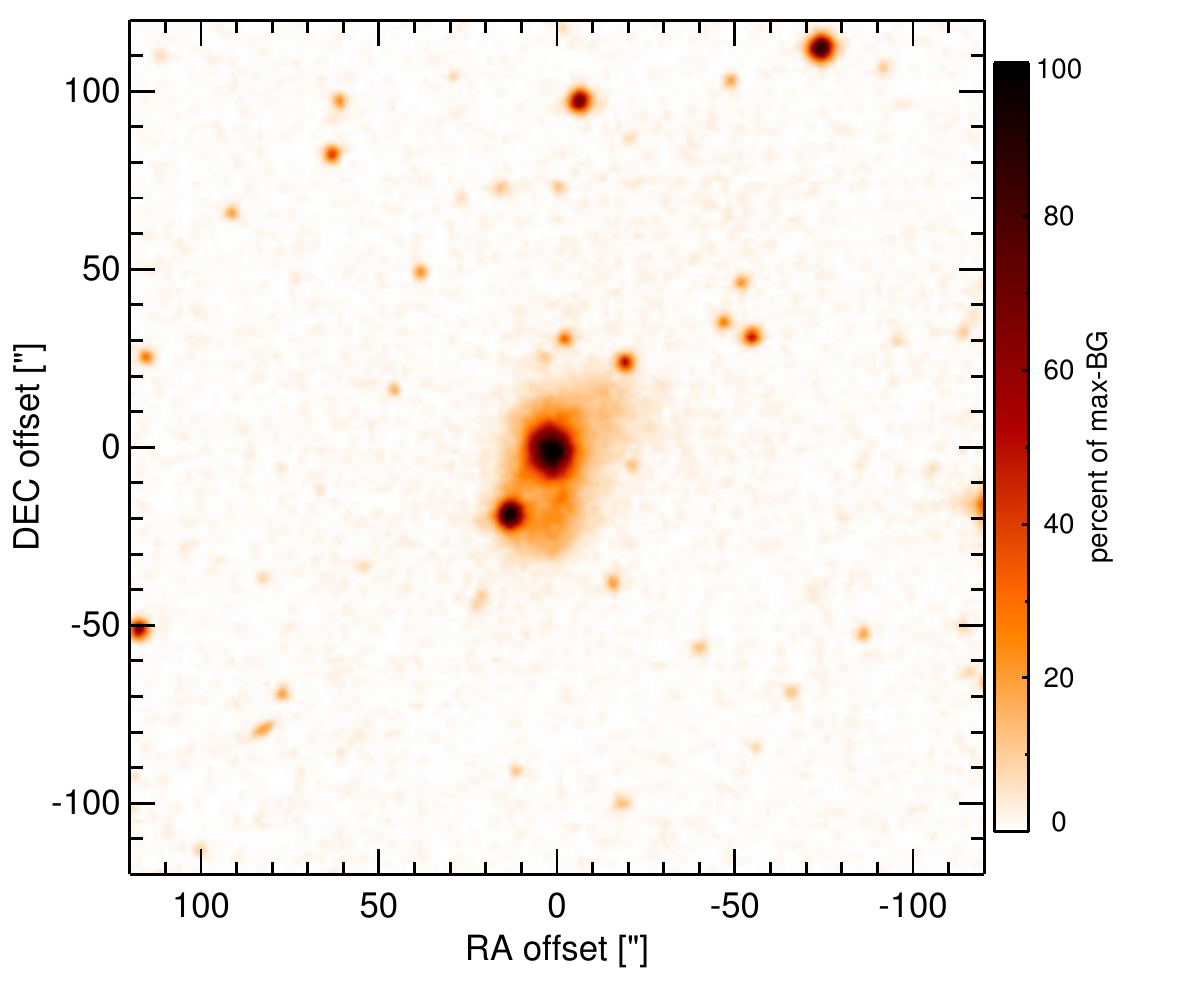}
    \caption{\label{fig:OPTim_Mrk0520}
             Optical image (DSS, red filter) of Mrk\,520. Displayed are the central $4\arcmin$ with North up and East to the left. 
              The colour scaling is linear with white corresponding to the median background and black to the $0.01\%$ pixels with the highest intensity.  
           }
\end{figure}
\begin{figure}
   \centering
   \includegraphics[angle=0,height=3.11cm]{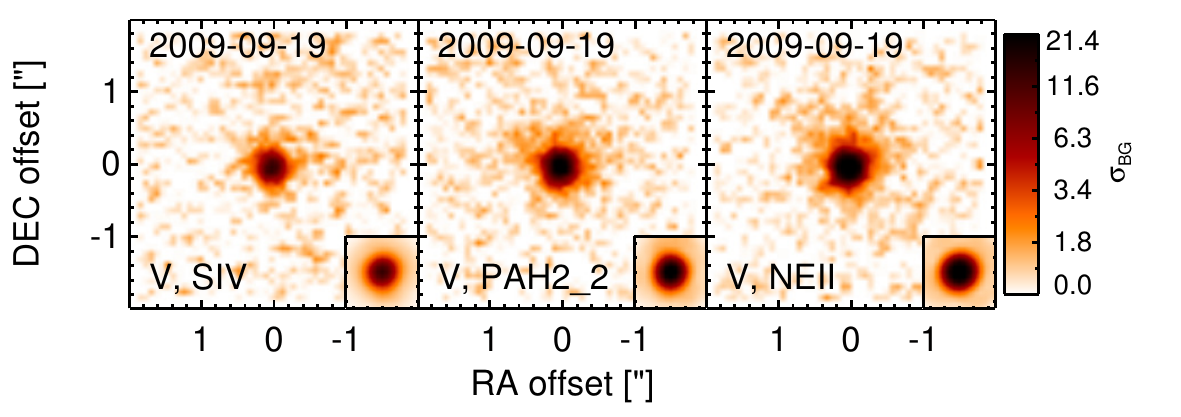}
    \caption{\label{fig:HARim_Mrk0520}
             Subarcsecond-resolution MIR images of Mrk\,520 sorted by increasing filter wavelength. 
             Displayed are the inner $4\arcsec$ with North up and East to the left. 
             The colour scaling is logarithmic with white corresponding to median background and black to the $75\%$ of the highest intensity of all images in units of $\sigbg$.
             The inset image shows the central arcsecond of the PSF from the calibrator star, scaled to match the science target.
             The labels in the bottom left state instrument and filter names (C: COMICS, M: Michelle, T: T-ReCS, V: VISIR).
           }
\end{figure}
\begin{figure}
   \centering
   \includegraphics[angle=0,width=8.50cm]{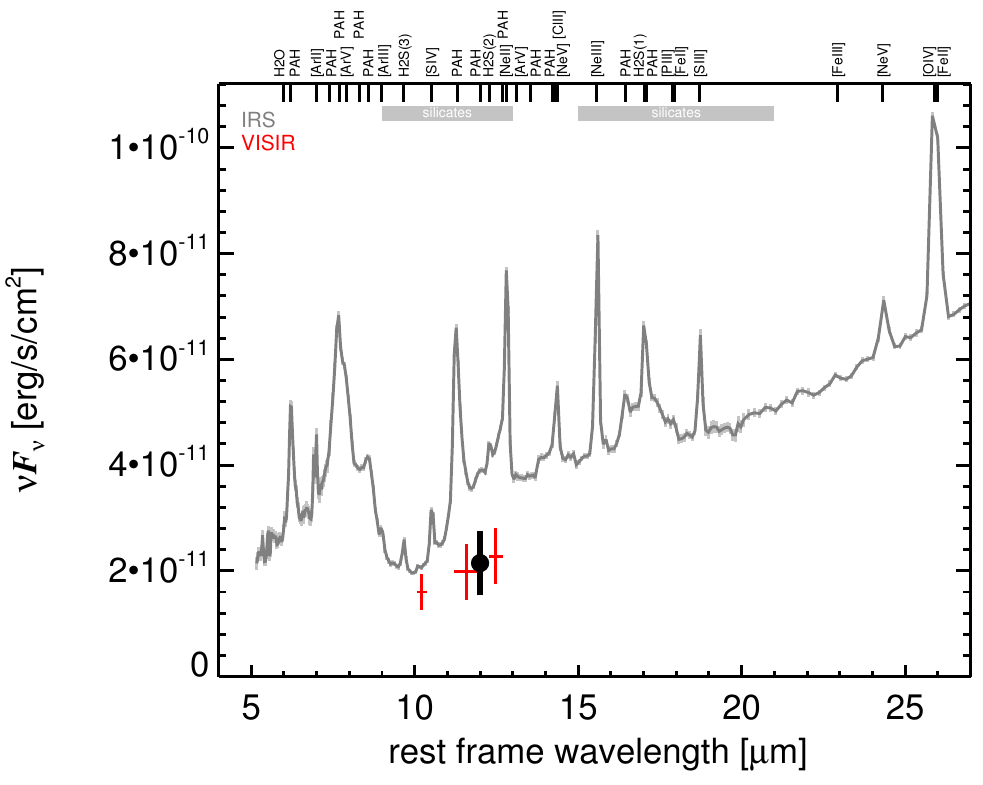}
   \caption{\label{fig:MISED_Mrk0520}
      MIR SED of Mrk\,520. The description  of the symbols (if present) is the following.
      Grey crosses and  solid lines mark the \spitzer/IRAC, MIPS and IRS data. 
      The colour coding of the other symbols is: 
      green for COMICS, magenta for Michelle, blue for T-ReCS and red for VISIR data.
      Darker-coloured solid lines mark spectra of the corresponding instrument.
      The black filled circles mark the nuclear 12 and $18\,\mu$m  continuum emission estimate from the data.
      The ticks on the top axis mark positions of common MIR emission lines, while the light grey horizontal bars mark wavelength ranges affected by the silicate 10 and 18$\mu$m features.}
\end{figure}
\clearpage

\twocolumn[\begin{@twocolumnfalse}  
\subsection{Mrk\,573 -- UGC\,1214}\label{app:Mrk0573}
Mrk\,573 is a spiral galaxy at a redshift of $z=$ 0.0172 ($D\sim$67.4\,Mpc) with a Sy\,2 nucleus \citep{osterbrock_spectroscopic_1993} with broad emission lines in polarized light \citep{nagao_detection_2004}.
It features an extended biconical NLR coinciding with extended radio emission (diameter$\sim8.9\arcsec\sim2.8\,$kpc; PA$\sim 124\degree$; e.g., \citealt{ulvestad_radio_1984-1,pogge_imaging_1995,schmitt_comparison_1996}).
The AGN has recently also been classified as an obscured narrow-line Seyfert~1 (Sy\,1n) by \cite{ramos_almeida_unveiling_2008}.
After \iras, $N$-band photometry was performed with IRTF in 1985 \citep{edelson_broad-band_1987}.
No \spitzer/IRAC or MIPS images are available for Mrk\,573 but a \spitzer/IRS LR staring mode spectrum, which exhibits strong forbidden emission lines, very weak silicate $10\,\mu$m absorption and PAH emission, and peaks at $\sim 18\,\mu$m in $\nu F_\nu$-space, i.e., the MIR SED seems AGN dominated.
Mrk\,573 was observed with T-ReCS in the N and Qa filters in 2003 \citep{ramos_almeida_infrared_2009}.
A compact MIR nucleus was detected in both images but no extension analysis could be performed because we failed to identify matching standard star observations.
Therefore, we use corresponding median conversion factors for the two filters to measure the nuclear fluxes, which are consistent with \cite{ramos_almeida_infrared_2009}.
Note that our Qa flux is $37\%$ lower but the nucleus is barely detected leading to large flux uncertainties.
The N filter flux is consistent with the IRTF photometry and the \spitzerr spectrophotometry, but the Qa filter flux is $53\%$ lower.
This, indicates significant cold dust contribution to the \spitzerr data outside the central $\sim 150$\,pc, even for the higher Qa value of \cite{ramos_almeida_infrared_2009}.
\newline\end{@twocolumnfalse}]

\begin{figure}
   \centering
   \includegraphics[angle=0,width=8.500cm]{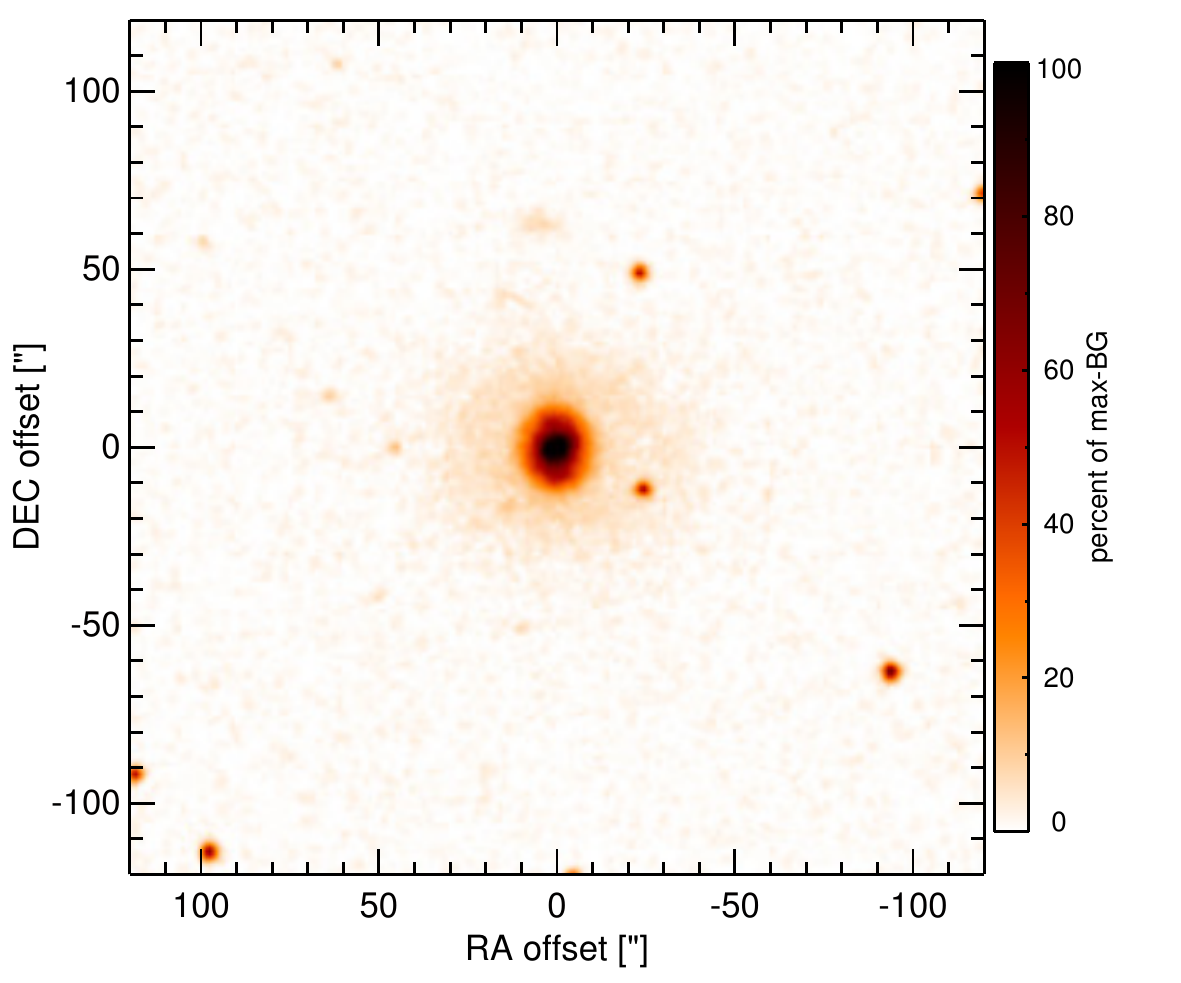}
    \caption{\label{fig:OPTim_Mrk0573}
             Optical image (DSS, red filter) of Mrk\,573. Displayed are the central $4\arcmin$ with North up and East to the left. 
              The colour scaling is linear with white corresponding to the median background and black to the $0.01\%$ pixels with the highest intensity.  
           }
\end{figure}
\begin{figure}
   \centering
   \includegraphics[angle=0,height=3.11cm]{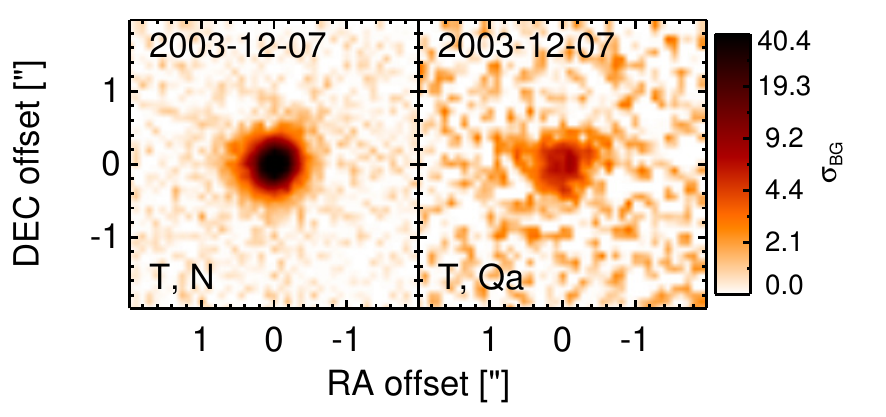}
    \caption{\label{fig:HARim_Mrk0573}
             Subarcsecond-resolution MIR images of Mrk\,573 sorted by increasing filter wavelength. 
             Displayed are the inner $4\arcsec$ with North up and East to the left. 
             The colour scaling is logarithmic with white corresponding to median background and black to the $75\%$ of the highest intensity of all images in units of $\sigbg$.
             The labels in the bottom left state instrument and filter names (C: COMICS, M: Michelle, T: T-ReCS, V: VISIR).
           }
\end{figure}
\begin{figure}
   \centering
   \includegraphics[angle=0,width=8.50cm]{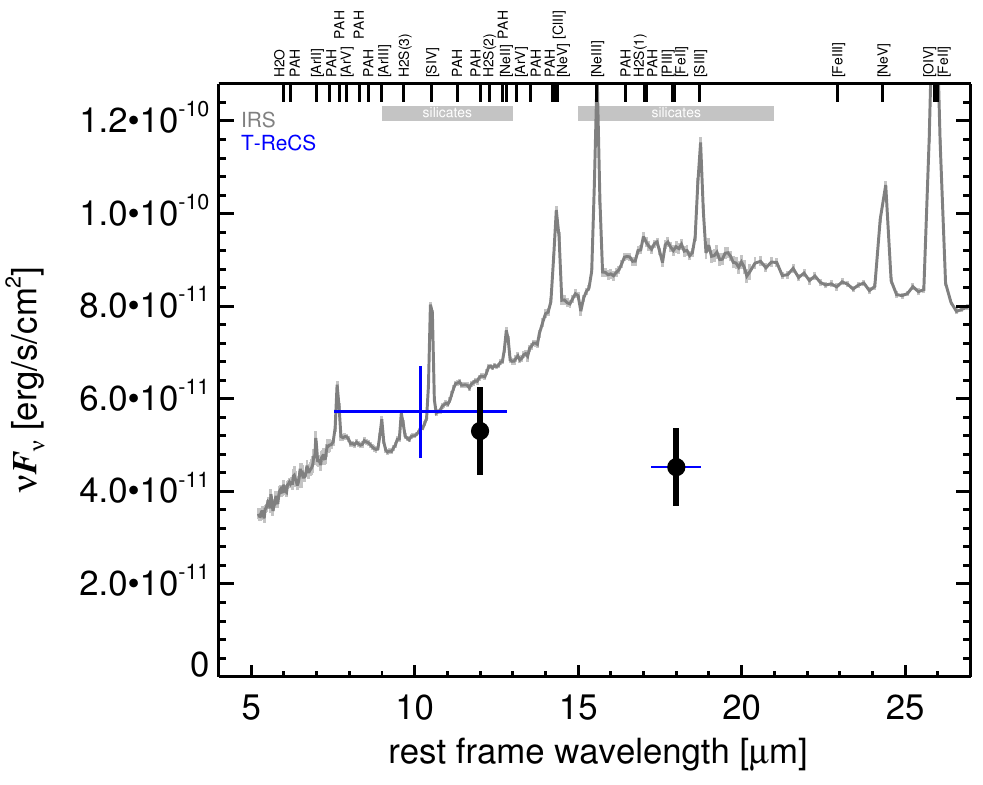}
   \caption{\label{fig:MISED_Mrk0573}
      MIR SED of Mrk\,573. The description  of the symbols (if present) is the following.
      Grey crosses and  solid lines mark the \spitzer/IRAC, MIPS and IRS data. 
      The colour coding of the other symbols is: 
      green for COMICS, magenta for Michelle, blue for T-ReCS and red for VISIR data.
      Darker-coloured solid lines mark spectra of the corresponding instrument.
      The black filled circles mark the nuclear 12 and $18\,\mu$m  continuum emission estimate from the data.
      The ticks on the top axis mark positions of common MIR emission lines, while the light grey horizontal bars mark wavelength ranges affected by the silicate 10 and 18$\mu$m features.}
\end{figure}
\clearpage

\twocolumn[\begin{@twocolumnfalse}  
\subsection{Mrk\,590 -- NGC\,863}\label{app:Mrk0590}
Mrk\,590 is a spiral galaxy at a redshift of $z=$ 0.0264 ($D\sim107\,$Mpc) with a Sy\,1.0 nucleus \citep{veron-cetty_catalogue_2010} and an extended NLR (diameter$\sim1.5\arcsec\sim750\,$pc; PA$\sim-5\degree$; \citealt{schmitt_hubble_2003}) and an unresolved radio core \citep{schmitt_jet_2001}.
It belongs to the nine-month BAT AGN sample.
Since \iras, the source was observed in the MIR with the MMT \citep{maiolino_new_1995}, with \iso/ISOCAM \citep{ramos_almeida_mid-infrared_2007} and \spitzer/IRS (e.g., \citealt{dasyra_high-ionization_2008,weaver_mid-infrared_2010}) and MIPS.
Mrk\,590 appears compact in the MIPS $24\,\mu$m image, while the 
IRS LR staring mode spectrum shows silicate 10 and $108\,\mu$m emission, a weak PAH 11.3$\,\mu$m feature and an emission peak at $\sim 17\,\mu$m in $\nu f_\nu$-space, i.e., the MIR SED seems AGN dominated.
We observed Mrk\,590 with VISIR in three narrow $N$-band filters in 2006 \citep{horst_mid_2008,horst_mid-infrared_2009}.
An unresolved MIR nucleus without any sign of host emission was detected in all images.
Our reanalysis of the images provides nuclear fluxes consistent with \cite{horst_mid_2008} and the \spitzerr spectrophotometry.
Note that the MMT and \isoo photometric fluxes are $\sim 50\%$ higher than the more recent measurements, which is either caused by host emission outside of the central $\sim 4\arcsec \sim 2\,$kpc or a decrease of the nuclear flux over the last decade.
\newline\end{@twocolumnfalse}]

\begin{figure}
   \centering
   \includegraphics[angle=0,width=8.500cm]{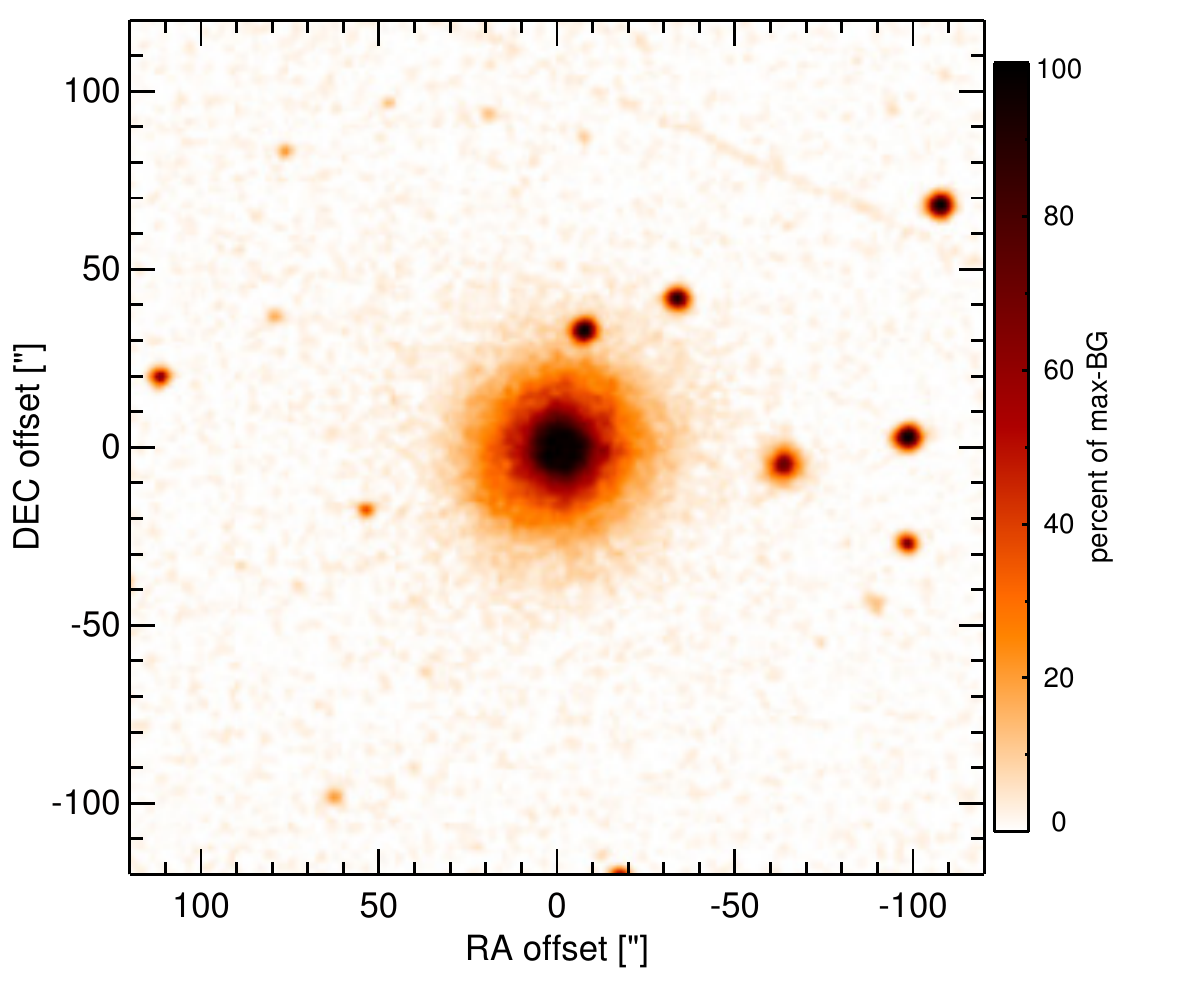}
    \caption{\label{fig:OPTim_Mrk0590}
             Optical image (DSS, red filter) of Mrk\,590. Displayed are the central $4\arcmin$ with North up and East to the left. 
              The colour scaling is linear with white corresponding to the median background and black to the $0.01\%$ pixels with the highest intensity.  
           }
\end{figure}
\begin{figure}
   \centering
   \includegraphics[angle=0,height=3.11cm]{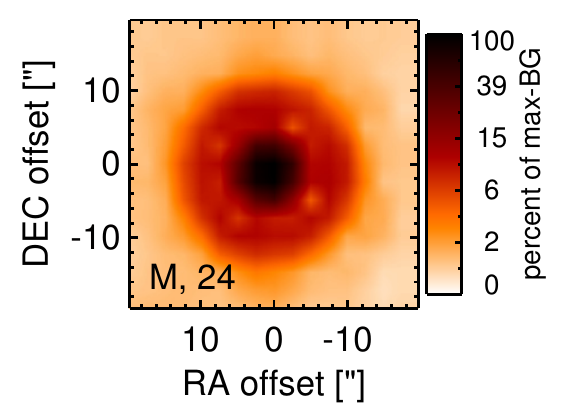}
    \caption{\label{fig:INTim_Mrk0590}
             \spitzerr MIR images of Mrk\,590. Displayed are the inner $40\arcsec$ with North up and East to the left. The colour scaling is logarithmic with white corresponding to median background and black to the $0.1\%$ pixels with the highest intensity.
             The label in the bottom left states instrument and central wavelength of the filter in $\mu$m (I: IRAC, M: MIPS). 
           }
\end{figure}
\begin{figure}
   \centering
   \includegraphics[angle=0,height=3.11cm]{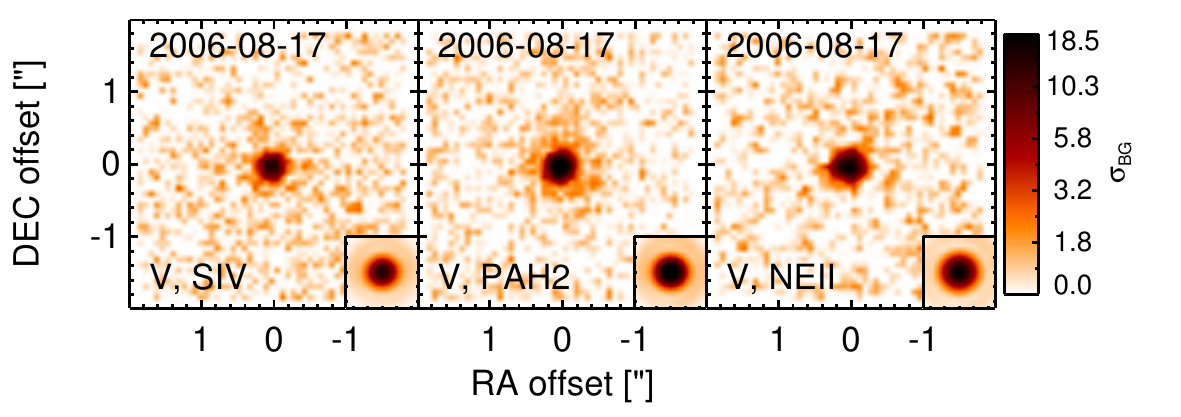}
    \caption{\label{fig:HARim_Mrk0590}
             Subarcsecond-resolution MIR images of Mrk\,590 sorted by increasing filter wavelength. 
             Displayed are the inner $4\arcsec$ with North up and East to the left. 
             The colour scaling is logarithmic with white corresponding to median background and black to the $75\%$ of the highest intensity of all images in units of $\sigbg$.
             The inset image shows the central arcsecond of the PSF from the calibrator star, scaled to match the science target.
             The labels in the bottom left state instrument and filter names (C: COMICS, M: Michelle, T: T-ReCS, V: VISIR).
           }
\end{figure}
\begin{figure}
   \centering
   \includegraphics[angle=0,width=8.50cm]{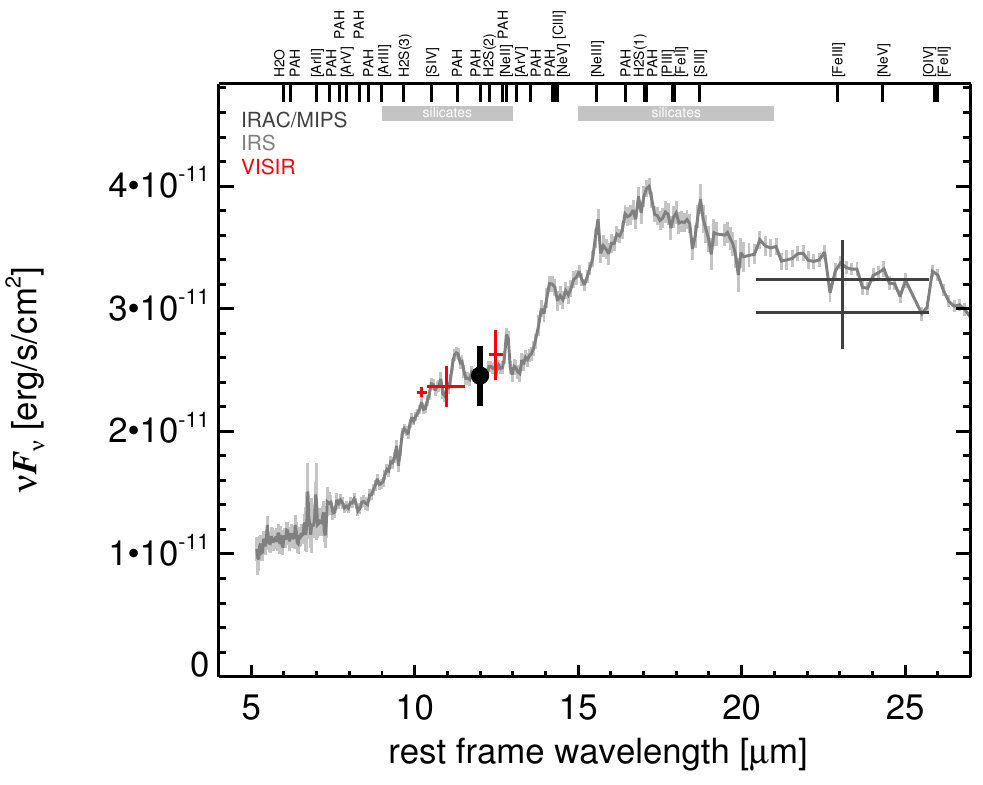}
   \caption{\label{fig:MISED_Mrk0590}
      MIR SED of Mrk\,590. The description  of the symbols (if present) is the following.
      Grey crosses and  solid lines mark the \spitzer/IRAC, MIPS and IRS data. 
      The colour coding of the other symbols is: 
      green for COMICS, magenta for Michelle, blue for T-ReCS and red for VISIR data.
      Darker-coloured solid lines mark spectra of the corresponding instrument.
      The black filled circles mark the nuclear 12 and $18\,\mu$m  continuum emission estimate from the data.
      The ticks on the top axis mark positions of common MIR emission lines, while the light grey horizontal bars mark wavelength ranges affected by the silicate 10 and 18$\mu$m features.}
\end{figure}
\clearpage

\twocolumn[\begin{@twocolumnfalse}  
\subsection{Mrk\,841 -- PG\,1501+106}\label{app:Mrk0841}
Mrk\,841 is an early-type galaxy at a redshift of $z=$ 0.0364 ($D\sim157\,$Mpc) with a radio-quiet quasar-regime AGN optically classified as  a Sy\,1.5 \citep{veron-cetty_catalogue_2010}.
It has mainly been studies at X-rays and also belongs to the nine-month BAT AGN sample.
Apart from \iras, Mrk\,841 was observed at MIR wavelengths by \cite{neugebauer_continuum_1987}, \cite{ward_continuum_1987}, \cite{elvis_atlas_1994}, \cite{maiolino_new_1995} and \cite{neugebauer_variability_1999} with photometers, while \isoo imaging is reported in \cite{haas_iso_2003} and \cite{ramos_almeida_mid-infrared_2007}. 
The $N$-band photometry indicates a flux decrease by $\sim17\%$ from 1981 to 1996 (see also \citealt{neugebauer_variability_1999}).
A decreasing trend is also found in the \irass $12\,\mu$m flux monitoring during 1983 reported in \cite{edelson_far-infrared_1987}. 
Mrk\,841 was also observed with \spitzer/IRAC, IRS and MIPS, and appears very compact in the corresponding images.
The IRS LR staring-mode spectrum exhibits very weak silicate 10 and $18\,\mu$m emission, almost no PAH emission, and an emission peak at $\sim 18\,\mu$m in $\nu F_\nu$-space, i.e., the MIR SED seems totally AGN dominated. 
We observed Mrk\,841 with VISIR in three narrow $N$-band filter in 2010 and detected a compact MIR nucleus without any host emission (see also \citealt{dudik_mid-infrared_2007,cao_mid-infrared_2008}). 
The nucleus appears possibly marginally resolved in all images (FWHM$\sim0.36\arcsec\sim250$\,pc) but at least a second epoch of subarcsecond MIR imaging is required to confirm this finding.
The nuclear photometry marginally agrees with the \spitzerr spectrophotometry and the previous $N$-band photometry obtained in the 1990s.
Note however, that the nuclear fluxes would be significantly lower if the presence of subarcsecond-extended emission can be verified.
\newline\end{@twocolumnfalse}]

\begin{figure}
   \centering
   \includegraphics[angle=0,width=8.500cm]{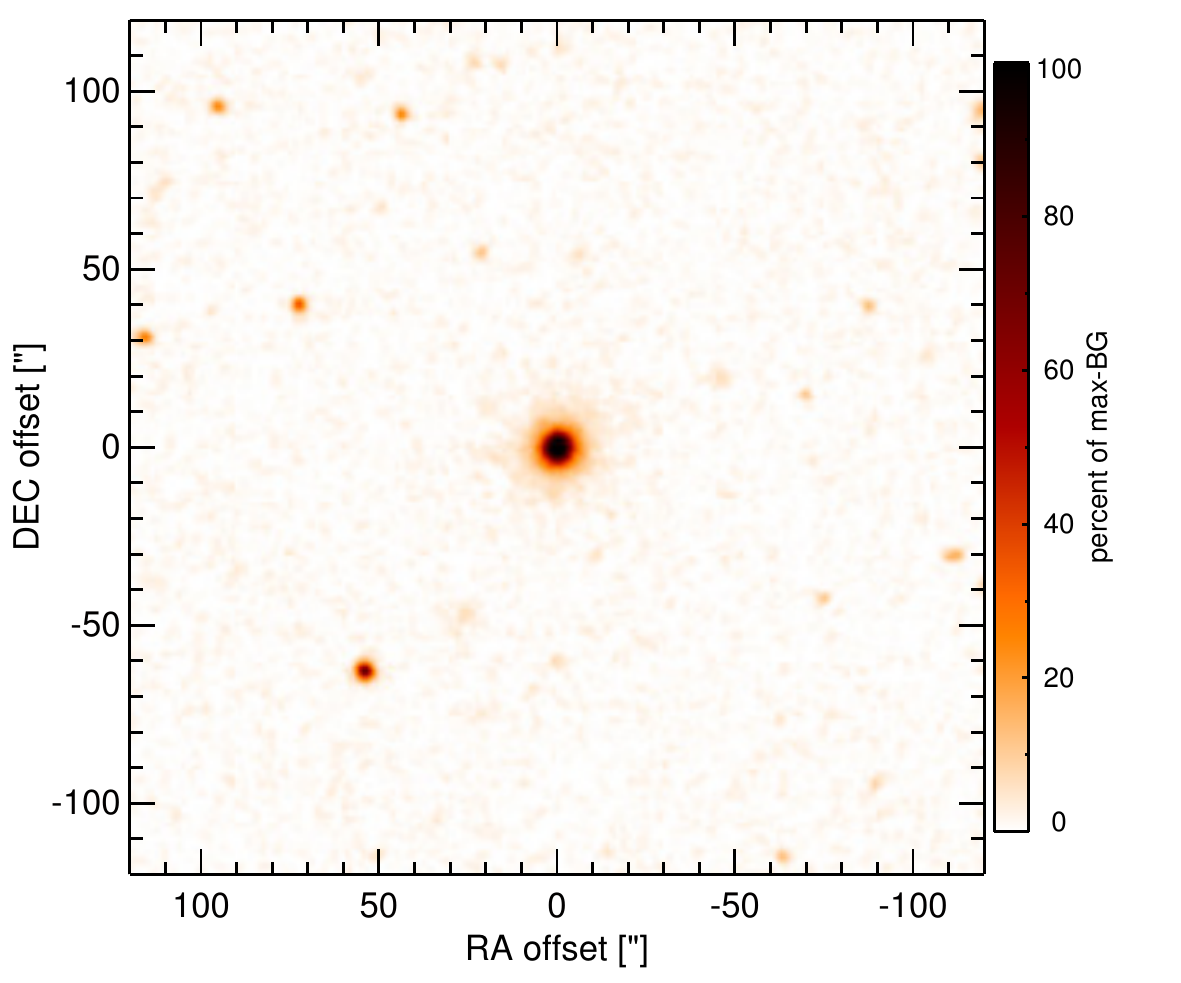}
    \caption{\label{fig:OPTim_Mrk0841}
             Optical image (DSS, red filter) of Mrk\,841. Displayed are the central $4\arcmin$ with North up and East to the left. 
              The colour scaling is linear with white corresponding to the median background and black to the $0.01\%$ pixels with the highest intensity.  
           }
\end{figure}
\begin{figure}
   \centering
   \includegraphics[angle=0,height=3.11cm]{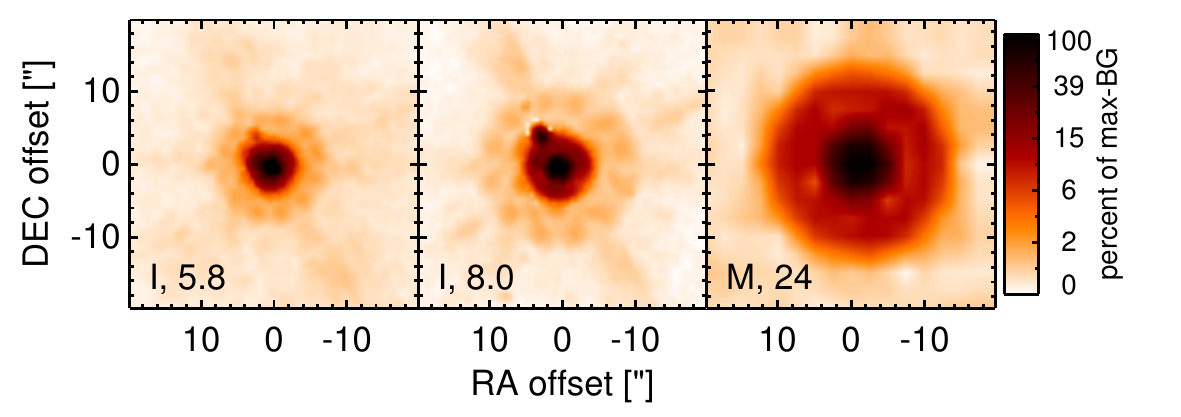}
    \caption{\label{fig:INTim_Mrk0841}
             \spitzerr MIR images of Mrk\,841. Displayed are the inner $40\arcsec$ with North up and East to the left. The colour scaling is logarithmic with white corresponding to median background and black to the $0.1\%$ pixels with the highest intensity.
             The label in the bottom left states instrument and central wavelength of the filter in $\mu$m (I: IRAC, M: MIPS).
             Note that the apparent off-nuclear compact source in the IRAC $8.0\,\mu$m image is an instrumental artefact.
           }
\end{figure}
\begin{figure}
   \centering
   \includegraphics[angle=0,height=3.11cm]{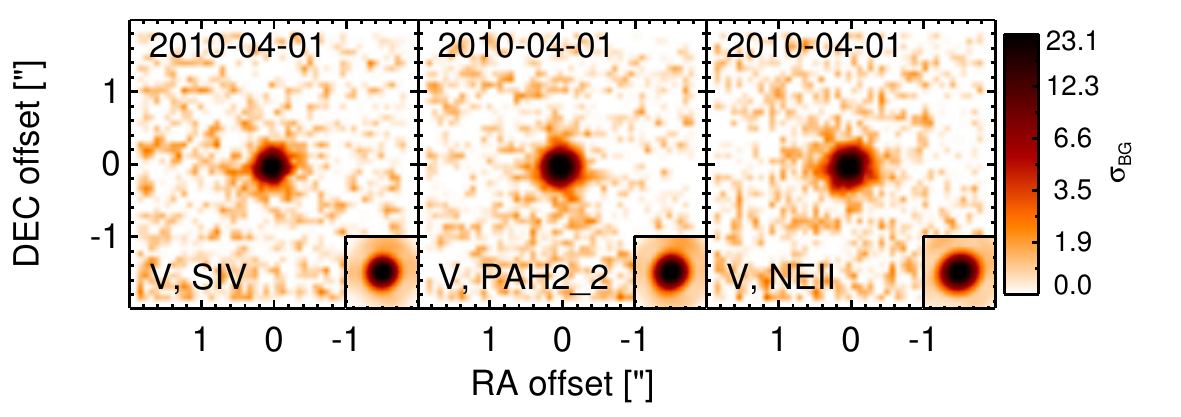}
    \caption{\label{fig:HARim_Mrk0841}
             Subarcsecond-resolution MIR images of Mrk\,841 sorted by increasing filter wavelength. 
             Displayed are the inner $4\arcsec$ with North up and East to the left. 
             The colour scaling is logarithmic with white corresponding to median background and black to the $75\%$ of the highest intensity of all images in units of $\sigbg$.
             The inset image shows the central arcsecond of the PSF from the calibrator star, scaled to match the science target.
             The labels in the bottom left state instrument and filter names (C: COMICS, M: Michelle, T: T-ReCS, V: VISIR).
           }
\end{figure}
\begin{figure}
   \centering
   \includegraphics[angle=0,width=8.50cm]{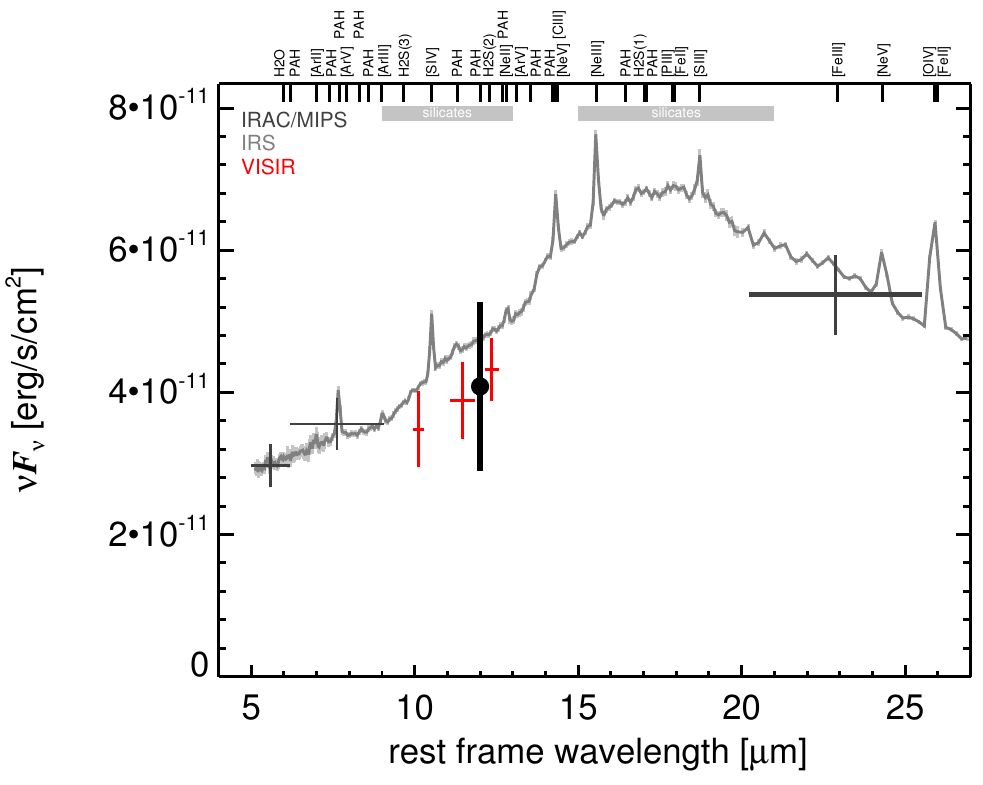}
   \caption{\label{fig:MISED_Mrk0841}
      MIR SED of Mrk\,841. The description  of the symbols (if present) is the following.
      Grey crosses and  solid lines mark the \spitzer/IRAC, MIPS and IRS data. 
      The colour coding of the other symbols is: 
      green for COMICS, magenta for Michelle, blue for T-ReCS and red for VISIR data.
      Darker-coloured solid lines mark spectra of the corresponding instrument.
      The black filled circles mark the nuclear 12 and $18\,\mu$m  continuum emission estimate from the data.
      The ticks on the top axis mark positions of common MIR emission lines, while the light grey horizontal bars mark wavelength ranges affected by the silicate 10 and 18$\mu$m features.}
\end{figure}
\clearpage

\twocolumn[\begin{@twocolumnfalse}  
\subsection{Mrk\,897 -- UGC\,11680 NED02}\label{app:Mrk0897}
Mrk\,897 is a compact early-type galaxy at a redshift of $z=$ 0.0263 ($D\sim 107\,$Mpc) with an active nucleus classified optically either as Sy\,2 \citep{veron-cetty_catalogue_2010} or AGN/starburst composite (\citealt{yuan_role_2010}; adopted).
The galaxy is interacting with UGC\,11680 NED01, which is 1.1\arcmin\,($\sim32$\,kpc) towards the west, forming the galaxy pair CPG\,552 (UGC\,11680).
A compact radio core was detected in Mrk\,897 (e.g., \citealt{thean_high-resolution_2000}).
The first MIR detection  was reported by \cite{maiolino_new_1995} with MMT $N$-band photometry, while it remained undetected in the Palomar 5\,m/MIRLIN subarcsecond imaging \citep{gorjian_10_2004}.
Mrk\,897 was also observed with \iso/ISOCAM \citep{domingue_multiwavelength_2003}, and \spitzer/IRS and MIPS.
It appears compact in the corresponding images.
The IRS LR mapping-mode spectrum is very noisy but indicates PAH emission and weak silicate  $10\,\mu$m absorption, indicating significant star formation (see also \citealt{tommasin_spitzer_2008}).
The MIR SED peaks at $\sim 18\,\mu$m in $\nu F_\nu$-space.
Mrk\,897 was observed with T-ReCS in the N filter in 2003 \citep{videla_nuclear_2013}, and a compact MIR nucleus embedded within weak diffuse emission is weakly detected.
The extended emission is roughly circular with a diameter of $\sim 2\arcsec \sim 1\,$kpc.
We measure the flux of the unresolved nuclear emission alone, which is consistent with \cite{videla_nuclear_2013} and only $7\%$ of the \spitzerr spectrophotometry.
These results suggest indeed the existence of an AGN deeply embedded within a bright starburst region dominating the MIR emission of Mrk\,897.
\newline\end{@twocolumnfalse}]

\begin{figure}
   \centering
   \includegraphics[angle=0,width=8.500cm]{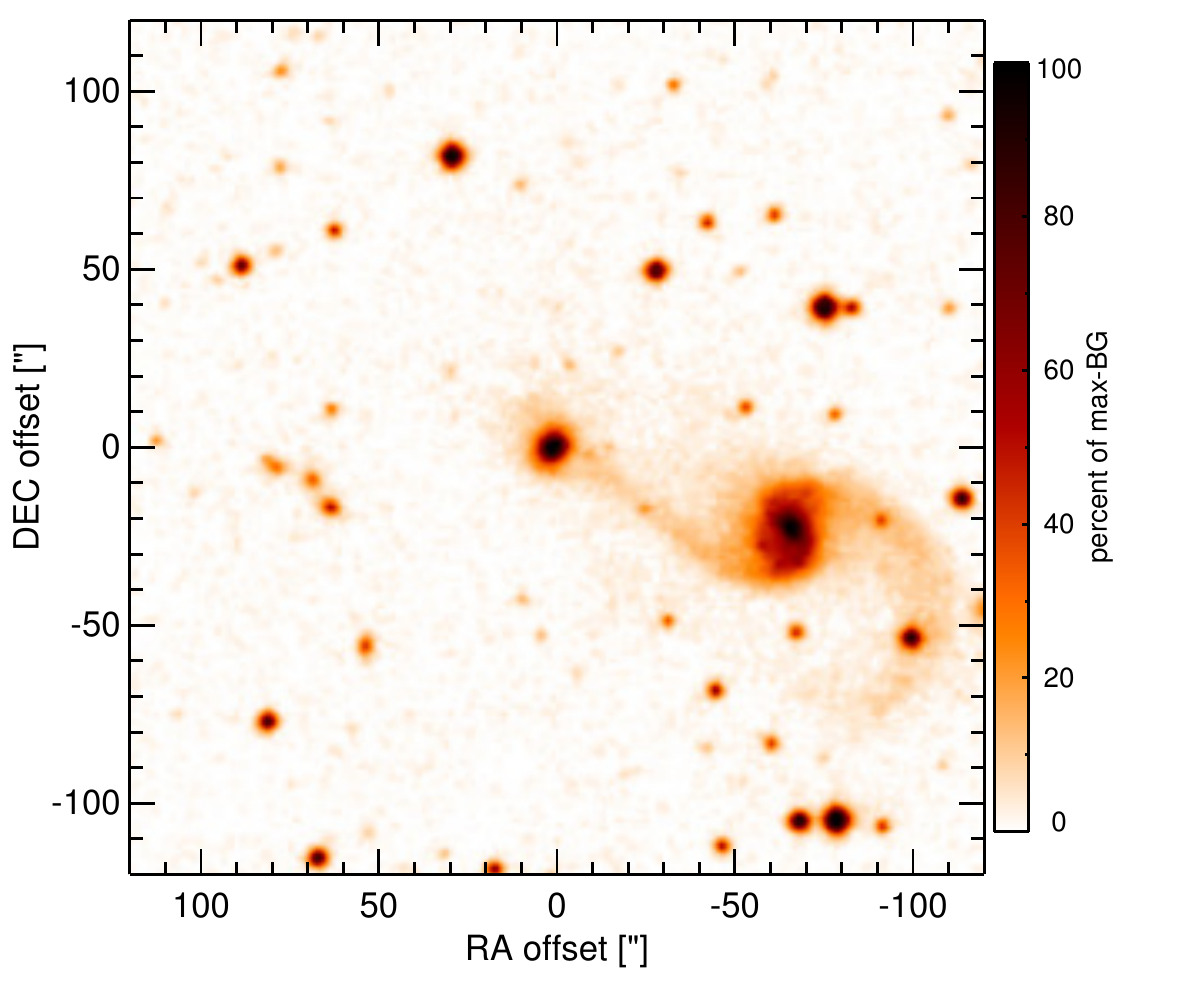}
    \caption{\label{fig:OPTim_Mrk0897}
             Optical image (DSS, red filter) of Mrk\,897. Displayed are the central $4\arcmin$ with North up and East to the left. 
              The colour scaling is linear with white corresponding to the median background and black to the $0.01\%$ pixels with the highest intensity.  
           }
\end{figure}
\begin{figure}
   \centering
   \includegraphics[angle=0,height=3.11cm]{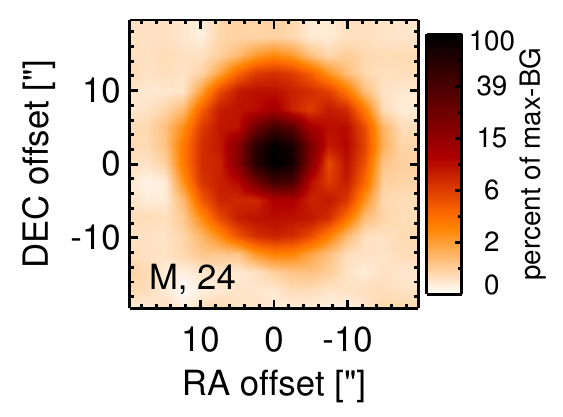}
    \caption{\label{fig:INTim_Mrk0897}
             \spitzerr MIR images of Mrk\,897. Displayed are the inner $40\arcsec$ with North up and East to the left. The colour scaling is logarithmic with white corresponding to median background and black to the $0.1\%$ pixels with the highest intensity.
             The label in the bottom left states instrument and central wavelength of the filter in $\mu$m (I: IRAC, M: MIPS). 
           }
\end{figure}
\begin{figure}
   \centering
   \includegraphics[angle=0,height=3.11cm]{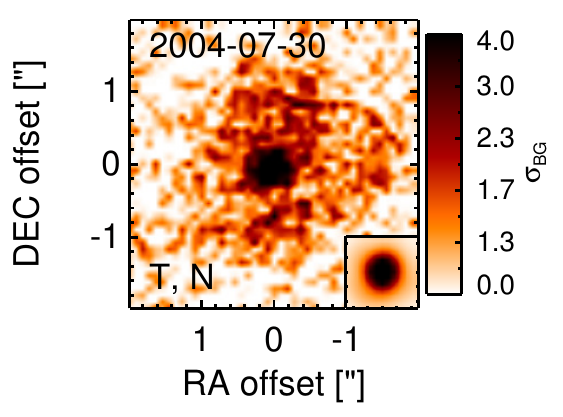}
    \caption{\label{fig:HARim_Mrk0897}
             Subarcsecond-resolution MIR images of Mrk\,897 sorted by increasing filter wavelength. 
             Displayed are the inner $4\arcsec$ with North up and East to the left. 
             The colour scaling is logarithmic with white corresponding to median background and black to the $75\%$ of the highest intensity of all images in units of $\sigbg$.
             The inset image shows the central arcsecond of the PSF from the calibrator star, scaled to match the science target.
             The labels in the bottom left state instrument and filter names (C: COMICS, M: Michelle, T: T-ReCS, V: VISIR).
           }
\end{figure}
\begin{figure}
   \centering
   \includegraphics[angle=0,width=8.50cm]{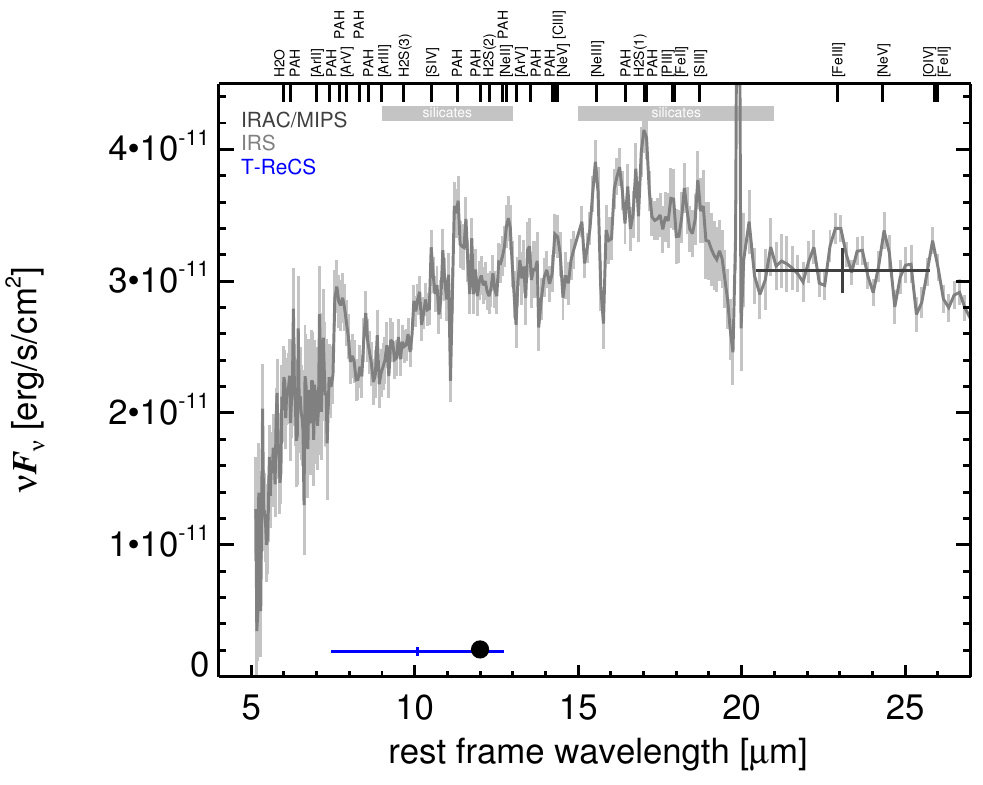}
   \caption{\label{fig:MISED_Mrk0897}
      MIR SED of Mrk\,897. The description  of the symbols (if present) is the following.
      Grey crosses and  solid lines mark the \spitzer/IRAC, MIPS and IRS data. 
      The colour coding of the other symbols is: 
      green for COMICS, magenta for Michelle, blue for T-ReCS and red for VISIR data.
      Darker-coloured solid lines mark spectra of the corresponding instrument.
      The black filled circles mark the nuclear 12 and $18\,\mu$m  continuum emission estimate from the data.
      The ticks on the top axis mark positions of common MIR emission lines, while the light grey horizontal bars mark wavelength ranges affected by the silicate 10 and 18$\mu$m features.}
\end{figure}
\clearpage

\twocolumn[\begin{@twocolumnfalse}  
\subsection{Mrk\,915 -- MCG-2-57-23}\label{app:Mrk0915}
Mrk\,915 is an inclined spiral galaxy at a redshift of $z=$ 0.0241 ($D\sim96.1\,$Mpc) with an AGN either classified as a Sy\,1.5  \citep{bennert_size_2006}, Sy\,1.8 or Sy\,1.9 \citep{trippe_multi-wavelength_2010}.
We treat it as Sy\,1.9 following the argumentation of \citeauthor{trippe_multi-wavelength_2010}.
It possesses an extended NLR (diameter$\sim4\arcsec\sim1.8\,$kpc; PA$\sim5\degree$; \citealt{schmitt_hubble_2003}), and appears essentially unresolved at radio wavelengths with only a weak emission blob to the south-west \citep{schmitt_jet_2001}.
Apart from \iras, Mrk\,915 was observed with \spitzer/IRS and the LR staring-mode spectrum shows strong forbidden emission lines, weak silicate 10 and  $18\,\mu$m emission, a weak PAH 11.3$\,\mu$m feature and an emission peak at $\sim 18\,\mu$m  in $\nu F_\nu$-space (see also \citealt{sargsyan_infrared_2011}).
The MIR SED thus appears to be AGN-dominated.
Mrk\,915 was observed with T-ReCS in the Qa filter in 2007 (unpublished, to our knowledge), and a compact nucleus was detected in the image.
The nucleus is possibly marginally extended but without at least a second epoch of MIR imaging the MIR subarcsecond morphology remains uncertain.
The measured nuclear Qa flux is consistent with the IRS spectrum and therefore, we use the latter to compute the $12\,\mu$m continuum emission estimate for the nucleus of Mrk\,915.
\newline\end{@twocolumnfalse}]

\begin{figure}
   \centering
   \includegraphics[angle=0,width=8.500cm]{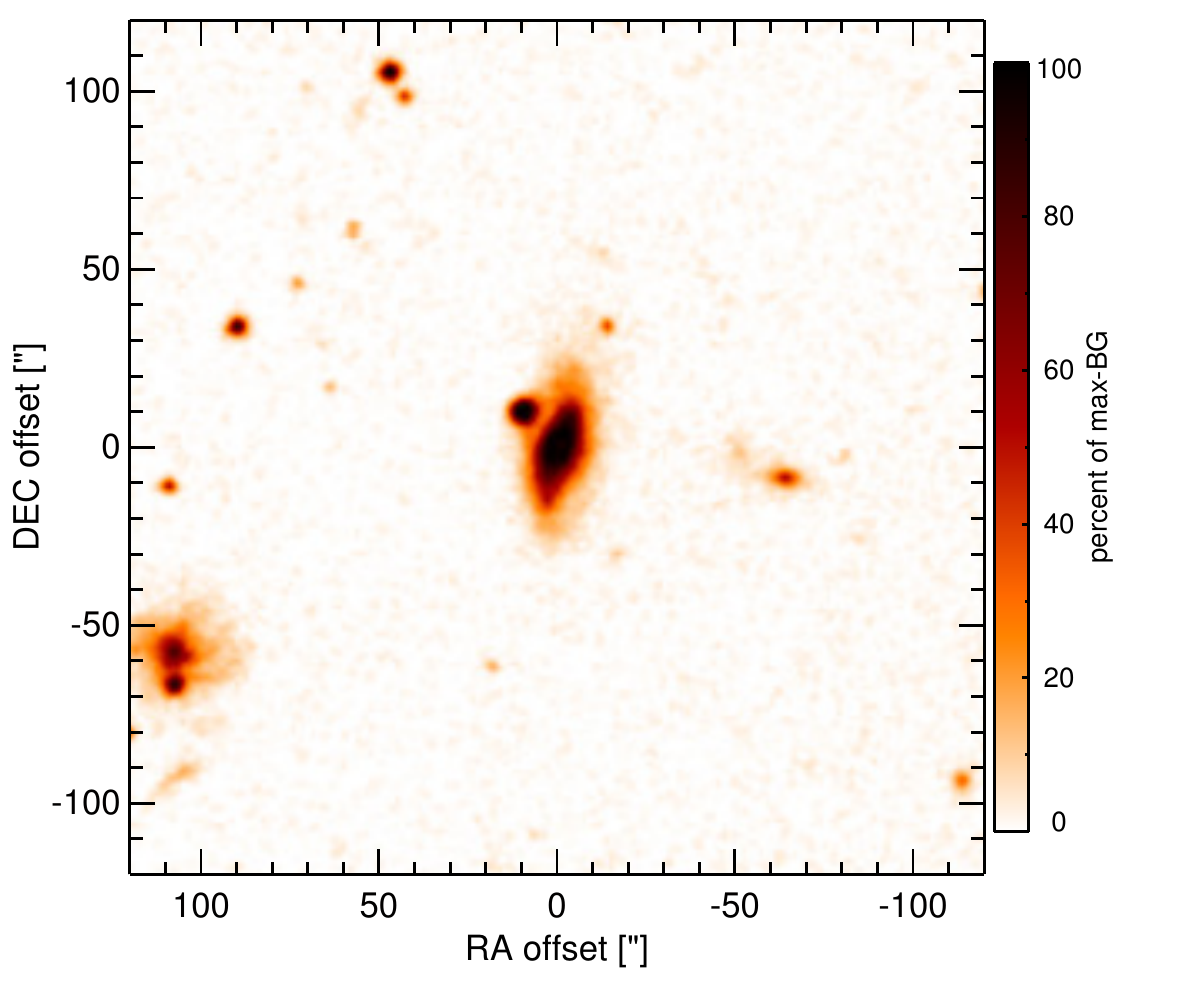}
    \caption{\label{fig:OPTim_Mrk0915}
             Optical image (DSS, red filter) of Mrk\,915. Displayed are the central $4\arcmin$ with North up and East to the left. 
              The colour scaling is linear with white corresponding to the median background and black to the $0.01\%$ pixels with the highest intensity.  
           }
\end{figure}
\begin{figure}
   \centering
   \includegraphics[angle=0,height=3.11cm]{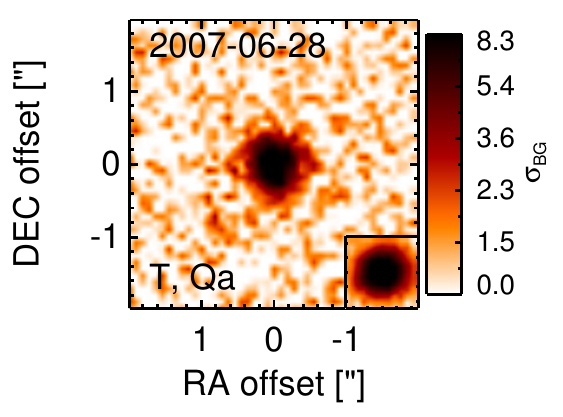}
    \caption{\label{fig:HARim_Mrk0915}
             Subarcsecond-resolution MIR images of Mrk\,915 sorted by increasing filter wavelength. 
             Displayed are the inner $4\arcsec$ with North up and East to the left. 
             The colour scaling is logarithmic with white corresponding to median background and black to the $75\%$ of the highest intensity of all images in units of $\sigbg$.
             The inset image shows the central arcsecond of the PSF from the calibrator star, scaled to match the science target.
             The labels in the bottom left state instrument and filter names (C: COMICS, M: Michelle, T: T-ReCS, V: VISIR).
           }
\end{figure}
\begin{figure}
   \centering
   \includegraphics[angle=0,width=8.50cm]{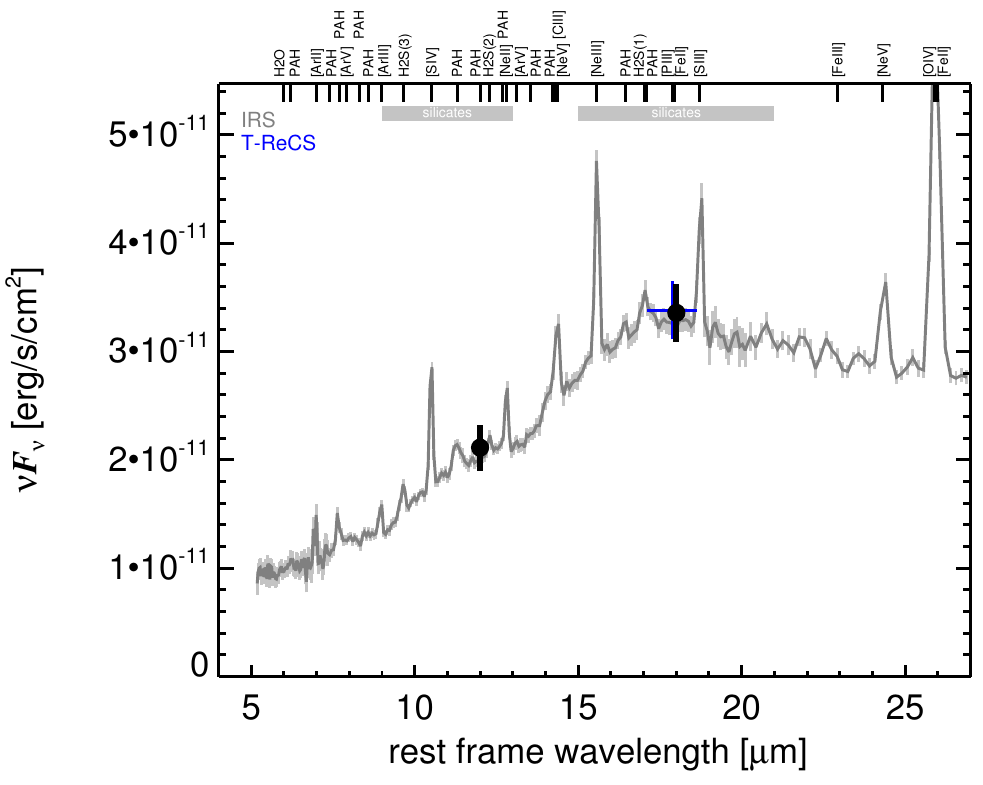}
   \caption{\label{fig:MISED_Mrk0915}
      MIR SED of Mrk\,915. The description  of the symbols (if present) is the following.
      Grey crosses and  solid lines mark the \spitzer/IRAC, MIPS and IRS data. 
      The colour coding of the other symbols is: 
      green for COMICS, magenta for Michelle, blue for T-ReCS and red for VISIR data.
      Darker-coloured solid lines mark spectra of the corresponding instrument.
      The black filled circles mark the nuclear 12 and $18\,\mu$m  continuum emission estimate from the data.
      The ticks on the top axis mark positions of common MIR emission lines, while the light grey horizontal bars mark wavelength ranges affected by the silicate 10 and 18$\mu$m features.}
\end{figure}
\clearpage

\twocolumn[\begin{@twocolumnfalse}  
\subsection{Mrk\,926 -- MCG-2-58-22}\label{app:Mrk0926}
Mrk\,926 is a low-inclination early-type spiral galaxy at a redshift of $z=$ 0.0469 ($D\sim 194$\,Mpc) with a Sy\,1.5 nucleus \citep{veron-cetty_catalogue_2010}, belonging to the nine-month BAT AGN sample.
It possesses a compact radio core with very weak extension in east-west direction of one arcsecond ($\sim 0.85$\,kpc) scale. \citep{mundell_parsec-scale_2000}.
Mrk\,926 was observed with \spitzer/IRAC, IRS and MIPS and appears point-like in the corresponding images.
The IRS LR staring-mode spectrum exhibits silicate 10 and  $18\,\mu$m emission, a weak PAH 11.3$\,\mu$m feature, and an emission peak at $\sim 18\,\mu$m in $\nu F_\nu$-space (see also \citealt{shi_9.7_2006}). 
We observed Mrk\,926 with VISIR in three narrow $N$-band filters in 2009 and detected a compact MIR nucleus without any host emission in all images.
The nucleus appears marginally extended in all images (FWHM $\sim 0.43\arcsec \sim 370\,$pc) with significant but inconsistent elongations.
Therefore, it is uncertain, whether this extension is object-intrinsic or caused by an unstable PSF and at least a second epoch of subarcsecond MIR imaging is required to check.
The nuclear photometry is on average $\sim 35\%$ lower than the \spitzerr spectrophotometry.
Note that the MIPS $24\,\mu$m flux is significantly below the IRS spectrum and thus might indicate contamination of cold dust in the  host affecting the latter.
\newline\end{@twocolumnfalse}]

\begin{figure}
   \centering
   \includegraphics[angle=0,height=3.11cm]{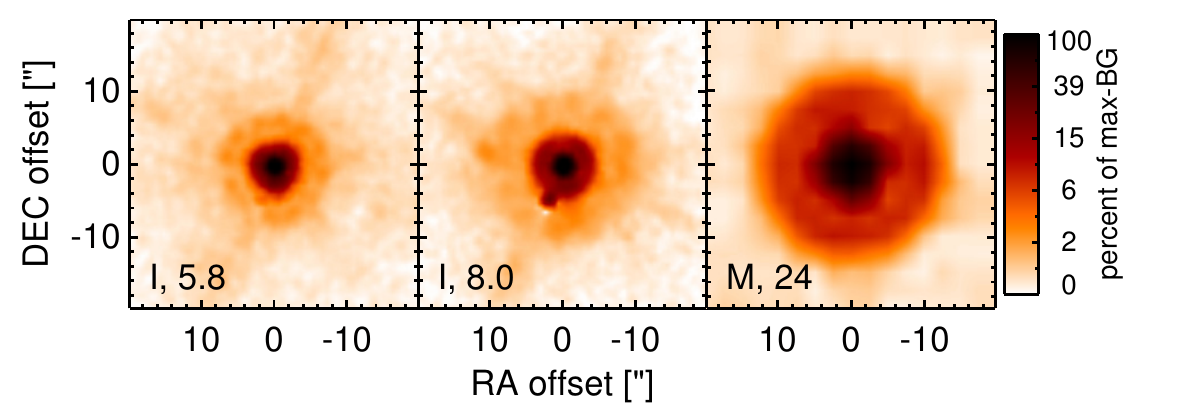}
    \caption{\label{fig:INTim_Mrk0926}
             \spitzerr MIR images of Mrk\,926. Displayed are the inner $40\arcsec$ with North up and East to the left. The colour scaling is logarithmic with white corresponding to median background and black to the $0.1\%$ pixels with the highest intensity.
             The label in the bottom left states instrument and central wavelength of the filter in $\mu$m (I: IRAC, M: MIPS).
             Note that the apparent off-nuclear compact source in the IRAC $8.0\,\mu$m image is an instrumental artefact.
           }
\end{figure}
\begin{figure}
   \centering
   \includegraphics[angle=0,height=3.11cm]{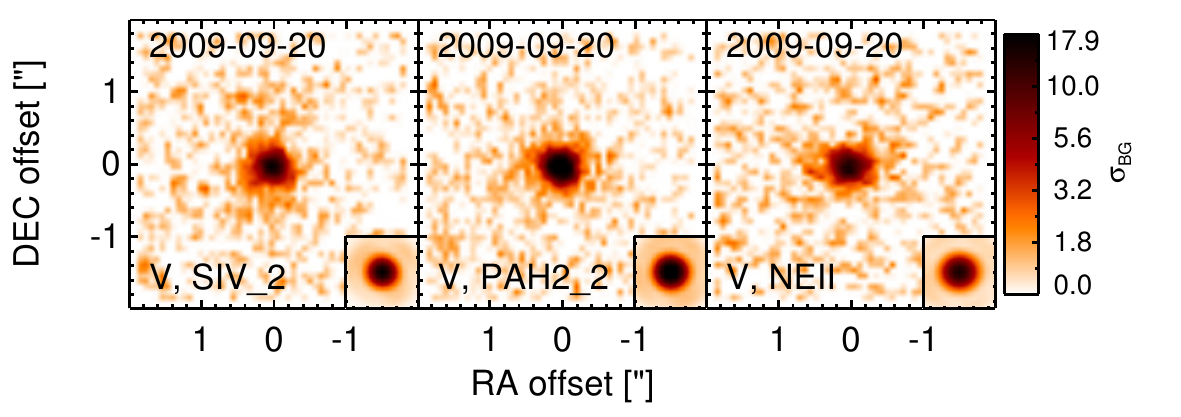}
    \caption{\label{fig:HARim_Mrk0926}
             Subarcsecond-resolution MIR images of Mrk\,926 sorted by increasing filter wavelength. 
             Displayed are the inner $4\arcsec$ with North up and East to the left. 
             The colour scaling is logarithmic with white corresponding to median background and black to the $75\%$ of the highest intensity of all images in units of $\sigbg$.
             The inset image shows the central arcsecond of the PSF from the calibrator star, scaled to match the science target.
             The labels in the bottom left state instrument and filter names (C: COMICS, M: Michelle, T: T-ReCS, V: VISIR).
           }
\end{figure}
\begin{figure}
   \centering
   \includegraphics[angle=0,width=8.50cm]{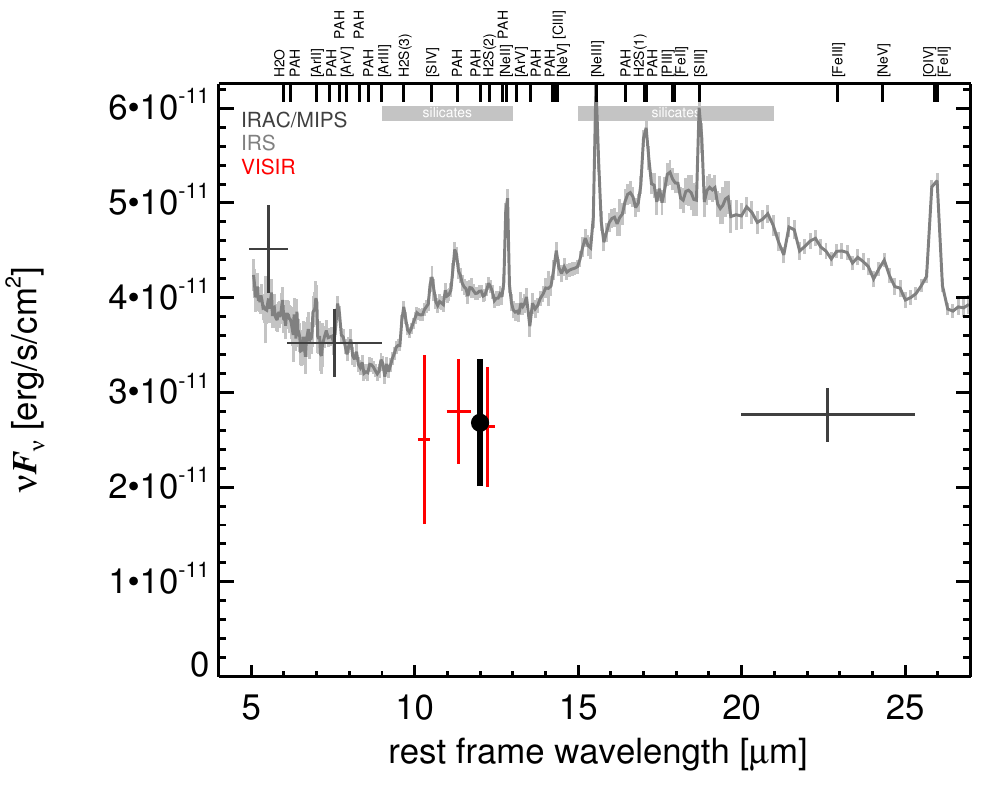}
   \caption{\label{fig:MISED_Mrk0926}
      MIR SED of Mrk\,926. The description  of the symbols (if present) is the following.
      Grey crosses and  solid lines mark the \spitzer/IRAC, MIPS and IRS data. 
      The colour coding of the other symbols is: 
      green for COMICS, magenta for Michelle, blue for T-ReCS and red for VISIR data.
      Darker-coloured solid lines mark spectra of the corresponding instrument.
      The black filled circles mark the nuclear 12 and $18\,\mu$m  continuum emission estimate from the data.
      The ticks on the top axis mark positions of common MIR emission lines, while the light grey horizontal bars mark wavelength ranges affected by the silicate 10 and 18$\mu$m features.}
\end{figure}
\clearpage

\twocolumn[\begin{@twocolumnfalse}  
\subsection{Mrk\,937 -- MCG-1-1-43}\label{app:Mrk0937}
Mrk\,937 is a face-on spiral galaxy at a redshift of $z=$ 0.0295 ($D\sim 119$\,Mpc) with an AGN with optical Sy\,1 classification \citep{veron-cetty_catalogue_2010}.
This object remained undetected with \irass at $12\,\mu$m, and no \spitzerr observations are available.
We observed Mrk\,937 in two narrow $N$-band filters in 2006 but did not detect any MIR emission either. 
Mrk\,937 appears as a compact source with \textit{WISE} with a $12\,\mu$m flux of 22\,mJy, which is higher than our VISIR upper limits.
This indicates that the total MIR emission of Mrk\,937 might be dominated by non-AGN-related radiation.
\newline\end{@twocolumnfalse}]

\begin{figure}
   \centering
   \includegraphics[angle=0,width=8.500cm]{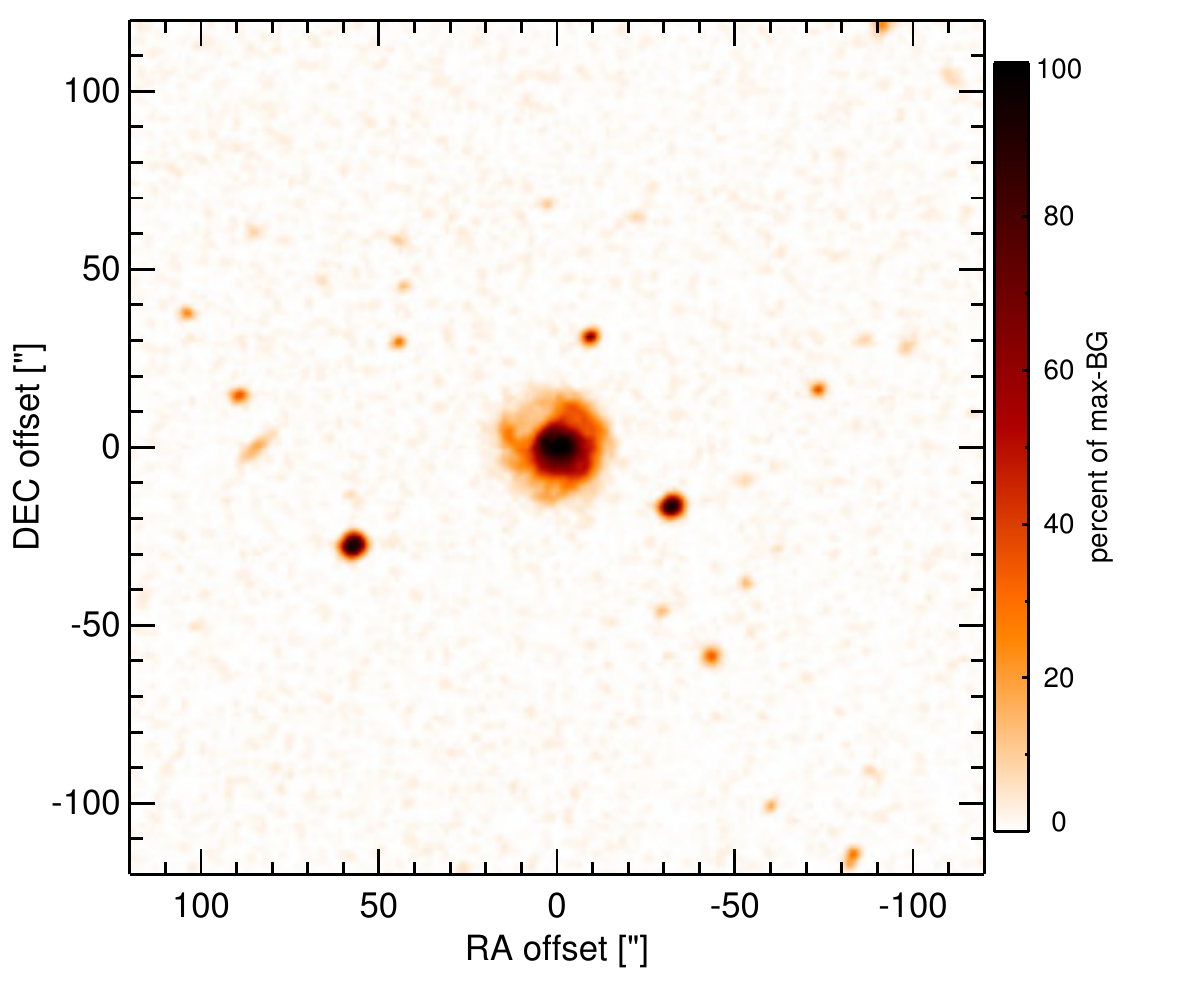}
    \caption{\label{fig:OPTim_Mrk0937}
             Optical image (DSS, red filter) of Mrk\,937. Displayed are the central $4\arcmin$ with North up and East to the left. 
              The colour scaling is linear with white corresponding to the median background and black to the $0.01\%$ pixels with the highest intensity.  
           }
\end{figure}
\begin{figure}
   \centering
   \includegraphics[angle=0,width=8.50cm]{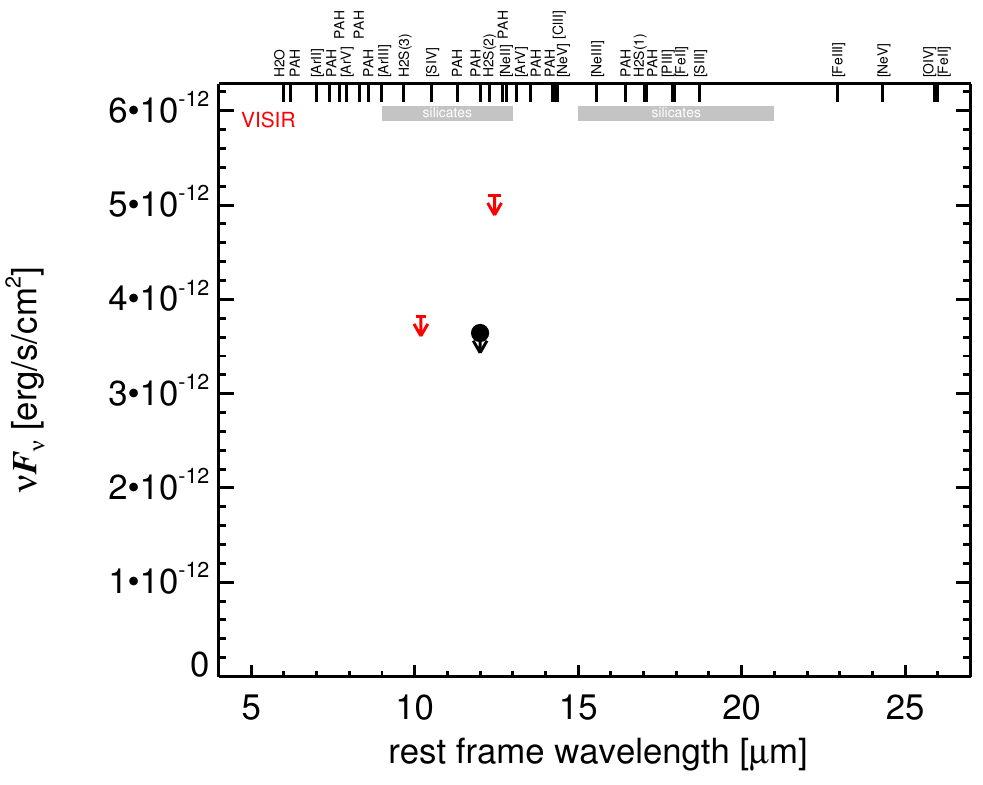}
   \caption{\label{fig:MISED_Mrk0937}
      MIR SED of Mrk\,937. The description  of the symbols (if present) is the following.
      Grey crosses and  solid lines mark the \spitzer/IRAC, MIPS and IRS data. 
      The colour coding of the other symbols is: 
      green for COMICS, magenta for Michelle, blue for T-ReCS and red for VISIR data.
      Darker-coloured solid lines mark spectra of the corresponding instrument.
      The black filled circles mark the nuclear 12 and $18\,\mu$m  continuum emission estimate from the data.
      The ticks on the top axis mark positions of common MIR emission lines, while the light grey horizontal bars mark wavelength ranges affected by the silicate 10 and 18$\mu$m features.}
\end{figure}
\clearpage

\twocolumn[\begin{@twocolumnfalse}  
\subsection{Mrk\,1014 -- PG 0157+001 -- IRAS\,01572+0009}\label{app:Mrk1014}
Mrk\,1014 is a disturbed face-on ultra-luminous infrared spiral galaxy at a redshift of $z=$ 0.1631 ($D\sim$748\,Mpc) with a radio-quiet quasar-type AGN optically classified as a Sy\,1.5 \citep{veron-cetty_catalogue_2010}.
It features a very extended and complex NLR with two arcsecond ($\sim 5.4\,$kpc; PA$\sim90\degree$; e.g., \citealt{bennert_size_2002}) coinciding with the radio emission structure (e.g., \citealt{leipski_radio_2006}).
After the discovery of its MIR brightness through \iras, Mrk\,1014 was monitored with ground-based $N$-band photometry \citep{neugebauer_continuum_1987, neugebauer_variability_1999}.
It was also observed with \iso in different modes \citep{rigopoulou_large_1999,sturm_mid-infrared_2002,haas_iso_2003} and \spitzer/IRAC and IRS.
Mark\,1014 appears point-like in all corresponding images, while the IRS LR staring-mode spectrum shows extremely weak silicate 10$\,\mu$m emission, weak PAH features and a red slope in $\nu F_\nu$-space (see also \citealt{armus_observations_2004,armus_observations_2007,farrah_high-resolution_2007,cao_mid-infrared_2008}).
Thus, the MIR SED appears AGN-dominated.
Mrk\,1014 was observed with COMICS in the N11.7 filter in 2006 \citep{imanishi_subaru_2011}.
In the image, a marginally resolved and elongated MIR nucleus was detected (FWHM $\sim 0.58\arcsec \sim 1.9\,$pc; PA$\sim63\degree$).
However, at least a second epoch of subarcsecond MIR imaging is required to verify this extension.
Our measured nuclear N11.7 flux is $25\%$ higher than that of \cite{imanishi_subaru_2011} but consistent with the previous $N$-band photometry and the \spitzerr spectrophotometry.
Therefore, we use the IRS spectrum to compute the nuclear 12$\,\mu$m continuum emission estimate. 
\newline\end{@twocolumnfalse}]

\begin{figure}
   \centering
   \includegraphics[angle=0,width=8.500cm]{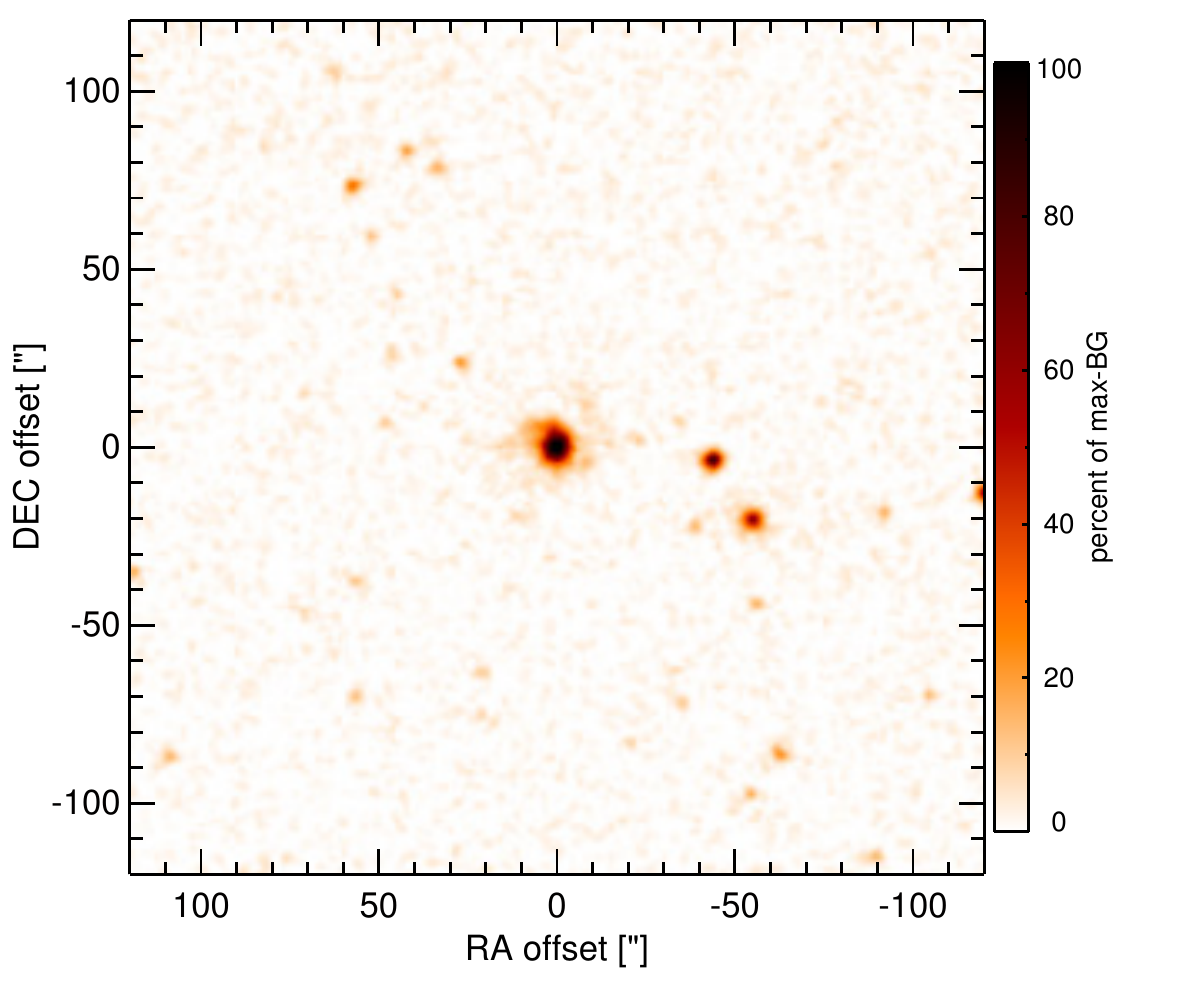}
    \caption{\label{fig:OPTim_Mrk1014}
             Optical image (DSS, red filter) of Mrk\,1014. Displayed are the central $4\arcmin$ with North up and East to the left. 
              The colour scaling is linear with white corresponding to the median background and black to the $0.01\%$ pixels with the highest intensity.  
           }
\end{figure}
\begin{figure}
   \centering
   \includegraphics[angle=0,height=3.11cm]{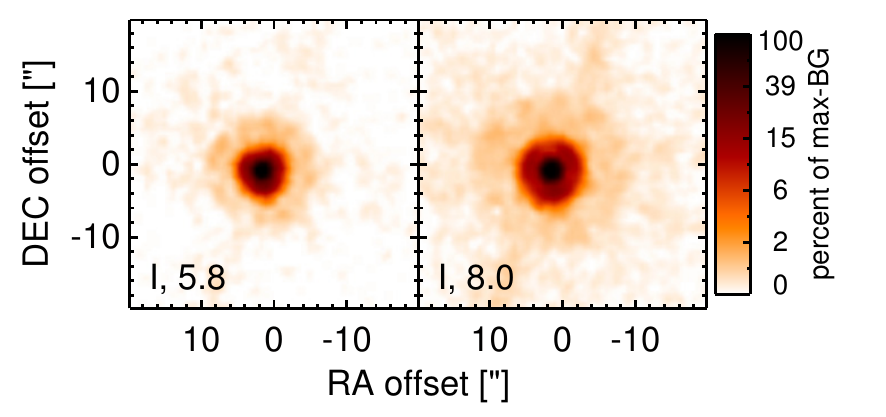}
    \caption{\label{fig:INTim_Mrk1014}
             \spitzerr MIR images of Mrk\,1014. Displayed are the inner $40\arcsec$ with North up and East to the left. The colour scaling is logarithmic with white corresponding to median background and black to the $0.1\%$ pixels with the highest intensity.
             The label in the bottom left states instrument and central wavelength of the filter in $\mu$m (I: IRAC, M: MIPS). 
           }
\end{figure}
\begin{figure}
   \centering
   \includegraphics[angle=0,height=3.11cm]{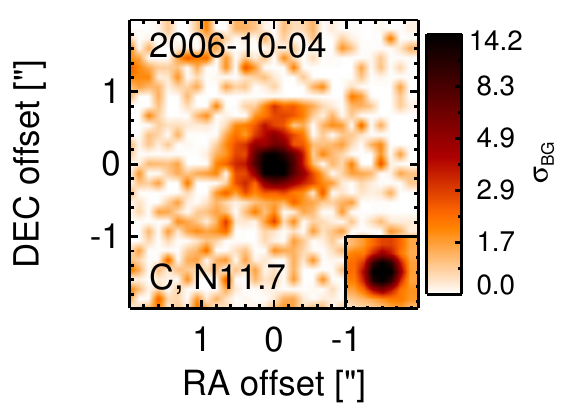}
    \caption{\label{fig:HARim_Mrk1014}
             Subarcsecond-resolution MIR images of Mrk\,1014 sorted by increasing filter wavelength. 
             Displayed are the inner $4\arcsec$ with North up and East to the left. 
             The colour scaling is logarithmic with white corresponding to median background and black to the $75\%$ of the highest intensity of all images in units of $\sigbg$.
             The inset image shows the central arcsecond of the PSF from the calibrator star, scaled to match the science target.
             The labels in the bottom left state instrument and filter names (C: COMICS, M: Michelle, T: T-ReCS, V: VISIR).
           }
\end{figure}
\begin{figure}
   \centering
   \includegraphics[angle=0,width=8.50cm]{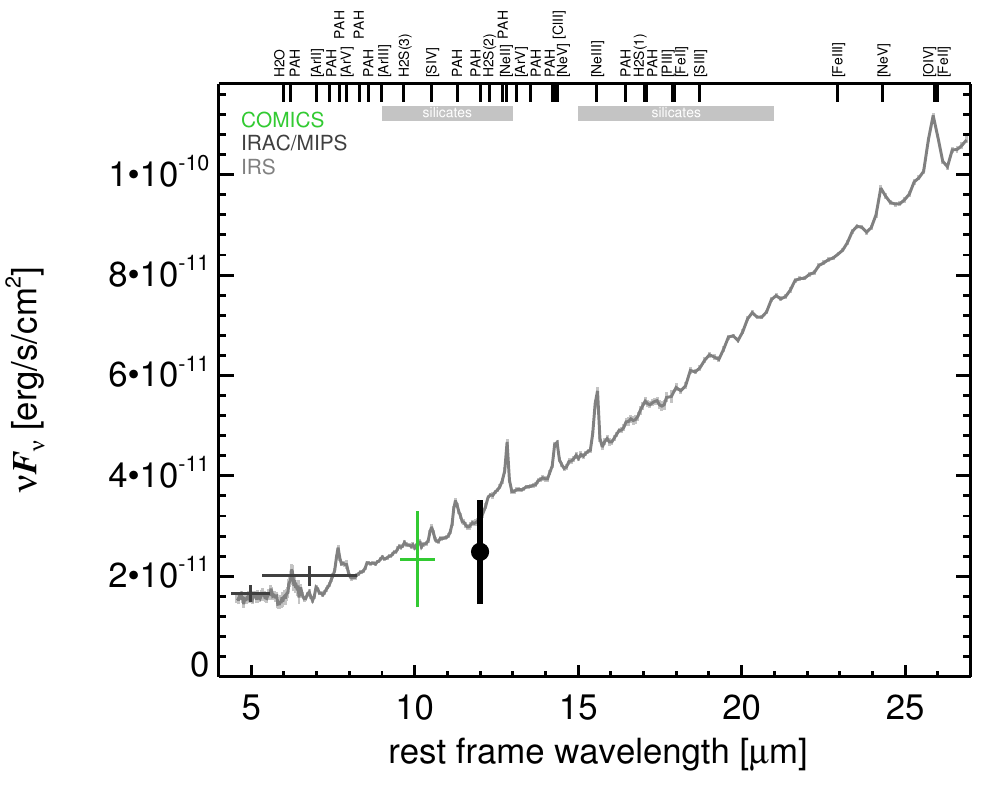}
   \caption{\label{fig:MISED_Mrk1014}
      MIR SED of Mrk\,1014. The description  of the symbols (if present) is the following.
      Grey crosses and  solid lines mark the \spitzer/IRAC, MIPS and IRS data. 
      The colour coding of the other symbols is: 
      green for COMICS, magenta for Michelle, blue for T-ReCS and red for VISIR data.
      Darker-coloured solid lines mark spectra of the corresponding instrument.
      The black filled circles mark the nuclear 12 and $18\,\mu$m  continuum emission estimate from the data.
      The ticks on the top axis mark positions of common MIR emission lines, while the light grey horizontal bars mark wavelength ranges affected by the silicate 10 and 18$\mu$m features.}
\end{figure}
\clearpage

\twocolumn[\begin{@twocolumnfalse}  
\subsection{Mrk\,1018 -- UGC\,1597}\label{app:Mrk1018}
Mrk\,1018 is a lenticular galaxy at a redshift of $z=$ 0.0424 ($D\sim191\,$Mpc)
with an active nucleus that belongs to the nine-month BAT AGN sample.
It was initially optically classified as  a Sy\,1.9 but intrinsically changed to Sy\,1.0 type in 1984 \citep{cohen_variability_1986,trippe_multi-wavelength_2010}.
At radio wavelengths, the nucleus appears unresolved at subarcsecond resolution \citep{ulvestad_radio_1986}.
The first successful MIR observation of Mrk\,1018 was performed with IRTF \citep{rudy_infrared_1985}, followed by \spitzer/IRS spectroscopy.
The corresponding IRS LR staring-mode spectrum exhibits silicate 10 and 18\,$\mu$m emission, no significant PAH features and a blue spectral slope in $\nu F_\nu$-space.
Thus, the arcsecond-scale MIR SED  appears to be dominated by an unobscured AGN.
The object appears compact in the \wisee images.
Mrk\,1018 was imaged with VISIR in five filters covering the whole $N$-band in 2009 (unpublished, to our knowledge) and an unresolved nucleus was detected in all cases.
The corresponding nuclear photometry agrees with the IRS spectrum, and therefore we use the latter to compute the nuclear $12\,\mu$m continuum emission estimate corrected for the silicate feature.
\newline\end{@twocolumnfalse}]

\begin{figure}
   \centering
   \includegraphics[angle=0,width=8.500cm]{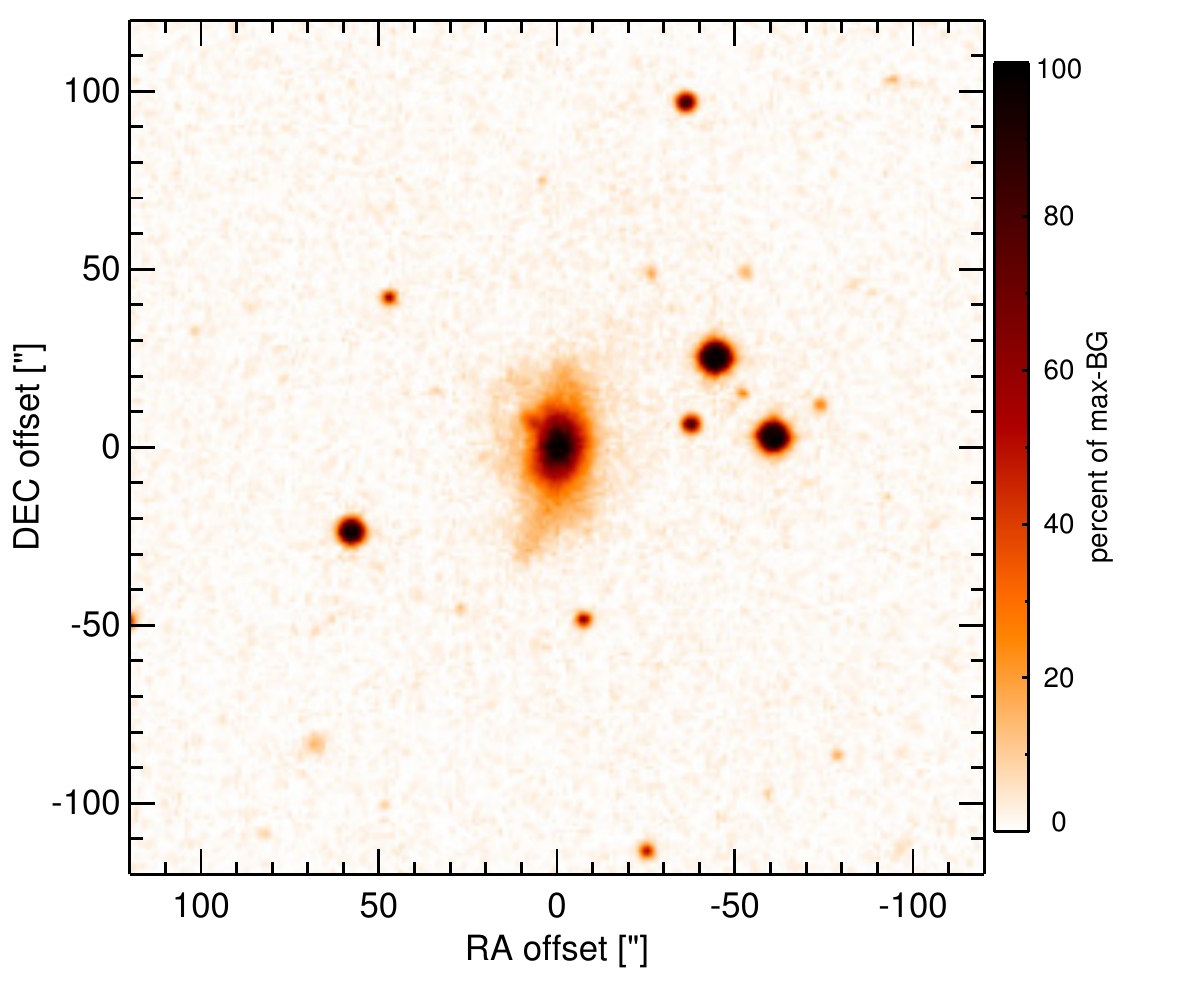}
    \caption{\label{fig:OPTim_Mrk1018}
             Optical image (DSS, red filter) of Mrk\,1018. Displayed are the central $4\arcmin$ with North up and East to the left. 
              The colour scaling is linear with white corresponding to the median background and black to the $0.01\%$ pixels with the highest intensity.  
           }
\end{figure}
\begin{figure}
   \centering
   \includegraphics[angle=0,width=8.500cm]{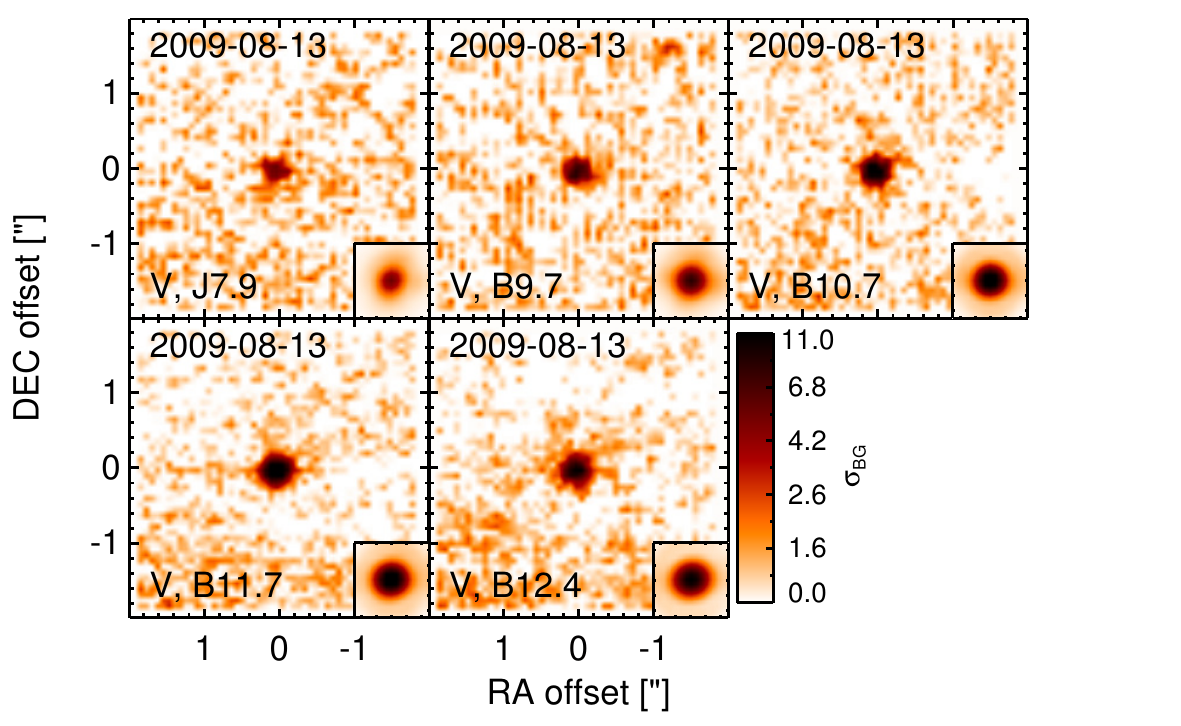}
    \caption{\label{fig:HARim_Mrk1018}
             Subarcsecond-resolution MIR images of Mrk\,1018 sorted by increasing filter wavelength. 
             Displayed are the inner $4\arcsec$ with North up and East to the left. 
             The colour scaling is logarithmic with white corresponding to median background and black to the $75\%$ of the highest intensity of all images in units of $\sigbg$.
             The inset image shows the central arcsecond of the PSF from the calibrator star, scaled to match the science target.
             The labels in the bottom left state instrument and filter names (C: COMICS, M: Michelle, T: T-ReCS, V: VISIR).
           }
\end{figure}
\begin{figure}
   \centering
   \includegraphics[angle=0,width=8.50cm]{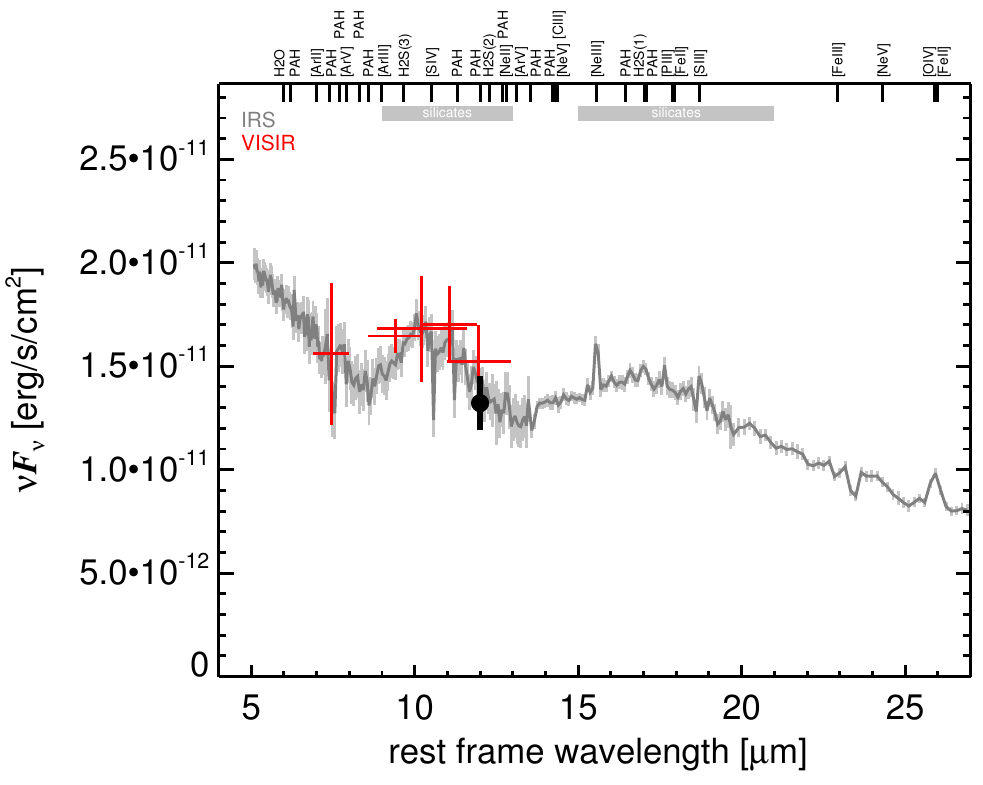}
   \caption{\label{fig:MISED_Mrk1018}
      MIR SED of Mrk\,1018. The description  of the symbols (if present) is the following.
      Grey crosses and  solid lines mark the \spitzer/IRAC, MIPS and IRS data. 
      The colour coding of the other symbols is: 
      green for COMICS, magenta for Michelle, blue for T-ReCS and red for VISIR data.
      Darker-coloured solid lines mark spectra of the corresponding instrument.
      The black filled circles mark the nuclear 12 and $18\,\mu$m  continuum emission estimate from the data.
      The ticks on the top axis mark positions of common MIR emission lines, while the light grey horizontal bars mark wavelength ranges affected by the silicate 10 and 18$\mu$m features.}
\end{figure}
\clearpage

\twocolumn[\begin{@twocolumnfalse}  
\subsection{Mrk\,1239 -- IRAS\,09497-0122}\label{app:Mrk1239}
Mrk\,1239 is an elliptical galaxy at a redshift of $z=$ 0.0199 ($D\sim$88\,Mpc) with an AGN classified as a Sy\,1.5 and Sy\,1n \citep{veron-cetty_catalogue_2010}.
This object has a steep radio spectrum with an unresolved core \citep{ulvestad_radio_1995} and an extended NLR (diameter$\sim 10\arcsec\sim4.1$\,kpc; PA$\sim 11\degree$; \citealt{mulchaey_emission-line_1996}).
After \iras, ground-based MIR observations of Mrk\,1239 were performed by \cite{maiolino_new_1995}.
The first subarcsecond-resolution $N$-band imaging was done with Palomar 5\,m/MIRLIN \citep{gorjian_10_2004}, and with ESO 3.6\,m/TIMMI2 \citep{raban_core_2008}.
In the MIRLIN image, a compact MIR nucleus with extended emission with $\sim 1\arcsec$ to the north-west was detected, while in the TIMMI2 image   elongated emission rather coinciding with the NLR was detected.
Then, the object appears rather point-like in the \spitzer/IRAC images.
The nuclear source is saturated in the PBCD version of the latter and thus not analysed (but see \citealt{gallimore_infrared_2010}).
In addition, a \spitzer/IRS LR staring mode spectrum is of Mrk\,1239 is available and shows silicate 10 and $18\,\mu$m emission, very weak PAH features and a blue spectral slope in $\nu F_\nu$-space (see also \citealt{wu_spitzer/irs_2009,tommasin_spitzer-irs_2010,gallimore_infrared_2010}).
Mrk\,1239 was observed with VISIR in PAH2 in 2005 \citep{haas_visir_2007}, and with PAH2\_2 and Q2 in 2006 \citep{reunanen_vlt_2010}.
The MIR nucleus appears compact in all images but possibly marginally resolved in the sharpest image (PAH2\_2; FWHM$\sim0.33\arcsec$; PA$\sim85\degree$).
Our nuclear MIR photometry is on average $\sim 10\%$ higher than the literature values but mostly consistent with the IRS spectrum. 
Neither the structure of the MIRLIN nor of the TIMMI2 image are visible in the VISIR images. 
Therefore, it remains unclear, whether weak non-nuclear emission truly exists in Mrk\,1239. 
Interestingly, the MIRLIN flux is at least $50\%$ lower than all the other measurements. 
Note that the nucleus of Mrk\,1239 was also been observed interferometrically with MIDI, where it remained essentially unresolved, implying a physical size of less than 7\,pc for the MIR emitter \citep{tristram_parsec-scale_2009,burtscher_diversity_2013}.
\newline\end{@twocolumnfalse}]

\begin{figure}
   \centering
   \includegraphics[angle=0,width=8.500cm]{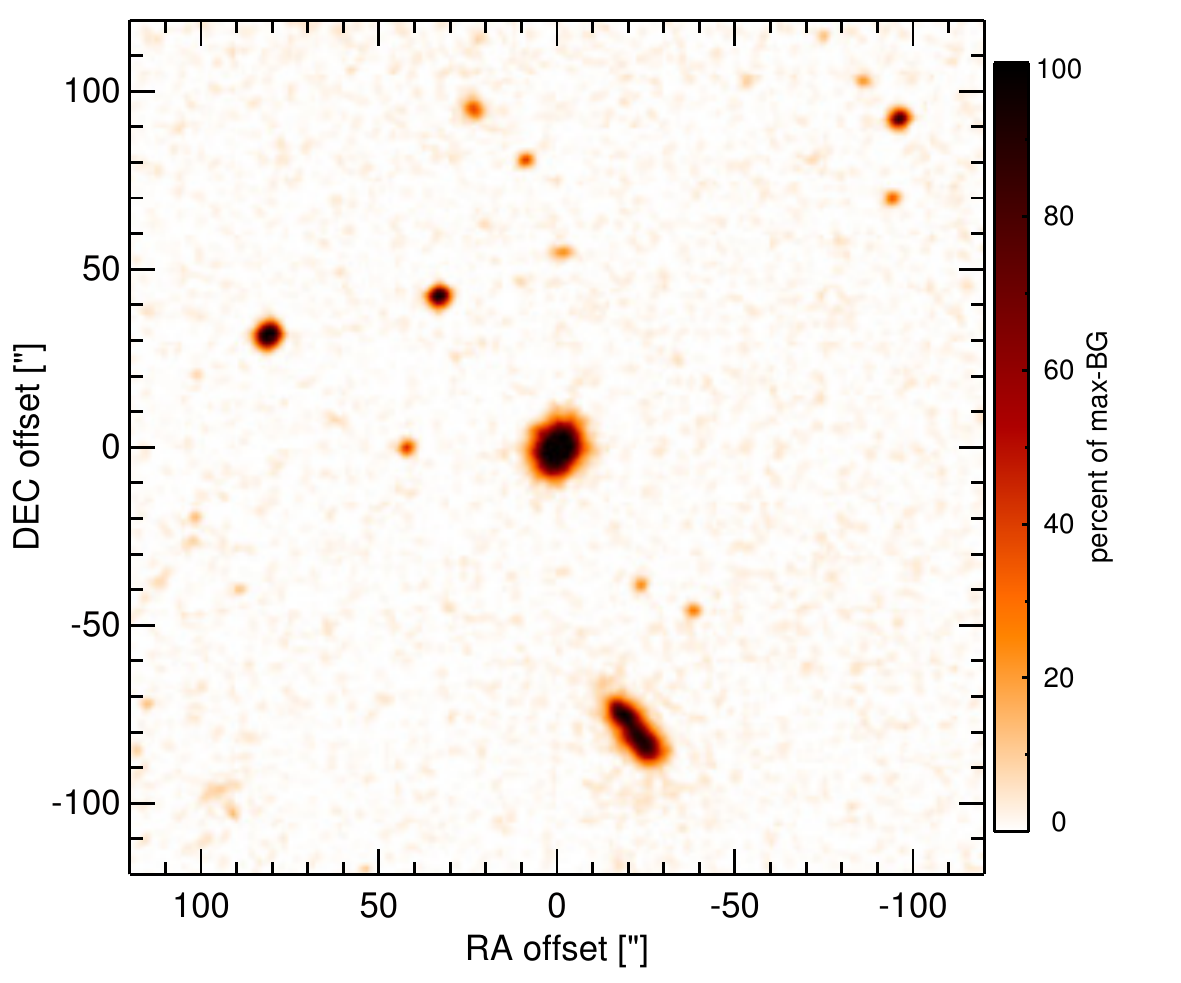}
    \caption{\label{fig:OPTim_Mrk1239}
             Optical image (DSS, red filter) of Mrk\,1239. Displayed are the central $4\arcmin$ with North up and East to the left. 
              The colour scaling is linear with white corresponding to the median background and black to the $0.01\%$ pixels with the highest intensity.  
           }
\end{figure}
\begin{figure}
   \centering
   \includegraphics[angle=0,height=3.11cm]{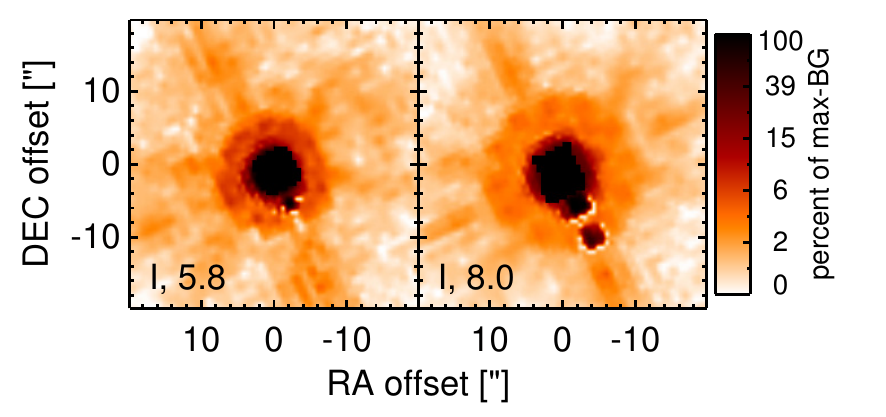}
    \caption{\label{fig:INTim_Mrk1239}
             \spitzerr MIR images of Mrk\,1239. Displayed are the inner $40\arcsec$ with North up and East to the left. The colour scaling is logarithmic with white corresponding to median background and black to the $0.1\%$ pixels with the highest intensity.
             The label in the bottom left states instrument and central wavelength of the filter in $\mu$m (I: IRAC, M: MIPS).
             Note that the apparent off-nuclear compact sources in the IRAC 5.8 and $8.0\,\mu$m images are instrumental artefacts.
           }
\end{figure}
\begin{figure}
   \centering
   \includegraphics[angle=0,height=3.11cm]{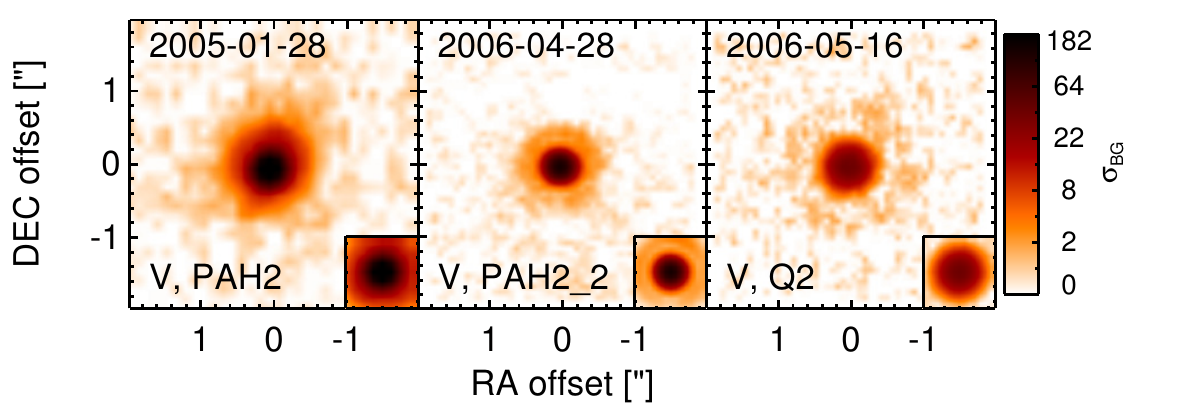}
    \caption{\label{fig:HARim_Mrk1239}
             Subarcsecond-resolution MIR images of Mrk\,1239 sorted by increasing filter wavelength. 
             Displayed are the inner $4\arcsec$ with North up and East to the left. 
             The colour scaling is logarithmic with white corresponding to median background and black to the $75\%$ of the highest intensity of all images in units of $\sigbg$.
             The inset image shows the central arcsecond of the PSF from the calibrator star, scaled to match the science target.
             The labels in the bottom left state instrument and filter names (C: COMICS, M: Michelle, T: T-ReCS, V: VISIR).
           }
\end{figure}
\begin{figure}
   \centering
   \includegraphics[angle=0,width=8.50cm]{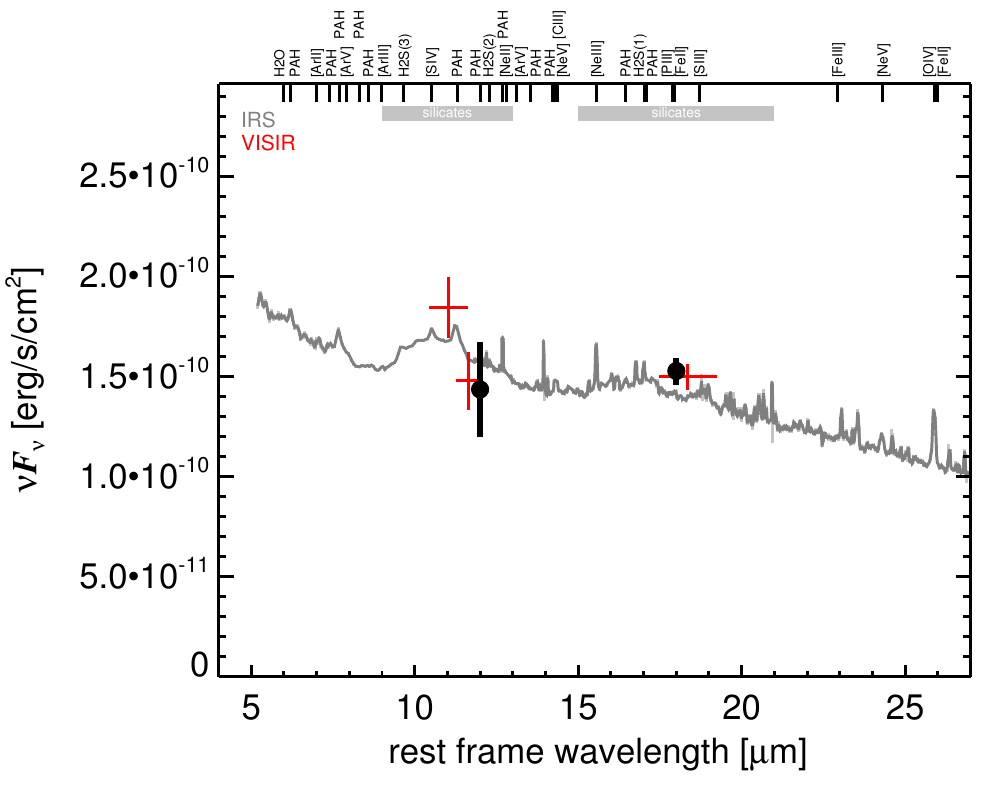}
   \caption{\label{fig:MISED_Mrk1239}
      MIR SED of Mrk\,1239. The description  of the symbols (if present) is the following.
      Grey crosses and  solid lines mark the \spitzer/IRAC, MIPS and IRS data. 
      The colour coding of the other symbols is: 
      green for COMICS, magenta for Michelle, blue for T-ReCS and red for VISIR data.
      Darker-coloured solid lines mark spectra of the corresponding instrument.
      The black filled circles mark the nuclear 12 and $18\,\mu$m  continuum emission estimate from the data.
      The ticks on the top axis mark positions of common MIR emission lines, while the light grey horizontal bars mark wavelength ranges affected by the silicate 10 and 18$\mu$m features.}
\end{figure}
\clearpage

\twocolumn[\begin{@twocolumnfalse}  
\subsection{NGC\,34 -- NGC\,17 -- Mrk\,938}\label{app:NGC0034}
NGC\,34 is an infrared-luminous late-stage merger system (nuclei separation $\sim17\arcsec; \sim 6\,$kpc; PA$\sim-30 \degree$; see \citealt{schweizer_remnant_2007} for a detailed study) at a redshift of $z=$ 0.0196 ($D\sim77\,$Mpc) with an active nucleus controversially classified either as a Sy\,2 or H\,II \citep{mazzarella_optical_1993}.
The nucleus is resolved and bright at radio wavelengths (PA$\sim 125\degree$; \citealt{nagar_radio_1999}), while the \oiii emission is weak and rather decentralized favouring a non-AGN nucleus (\citealt{mulchaey_emission-line_1996}; see also \citealt{riffel_0.8-2.4_2006}).
However, the most recent classification by \cite{yuan_role_2010} is Sy\,2 (see also \citealt{goncalves_agns_1999}), while X-ray observations revealed the presence of a Compton-thick obscured AGN \citep{shu_investigating_2007}, which might explain the previous contradicting results.
In addition, a nuclear disc-like water maser was detected \citep{greenhill_discovery_2009}.
After the discovery of its infrared brightness, NGC\,34 was observed with ground-based MIR instruments \citep{carico_iras_1988,keto_high_1991,maiolino_new_1995}, and first subarcsecond-resolution $N$-band images were obtained with Palomar 5\,m/SpectroCam-10 \citep{miles_high-resolution_1996}, followed by Palomar 5\,m/MIRLIN \citep{gorjian_10_2004}.
In these images, a compact MIR nucleus with  emission extending $\sim 1.2\arcsec \sim430$\,pc to the south was detected, while the nuclear emission is also extended in the  \spitzer/IRAC and MIPS images (PA$\sim0\degree$).
Because we measure the nuclear component only, our IRAC $5.8$ and $8.0\,\mu$m  fluxes are significantly lower compared to the values in \cite{gallimore_infrared_2010}.
The IRS LR mapping-mode spectrum appears to be dominated by star formation with strong PAH features, silicate  $10\,\mu$m absorption and a red steep spectral slope in $\nu F_\nu$-space (see also \citealt{buchanan_spitzer_2006,wu_spitzer/irs_2009,gallimore_infrared_2010,tommasin_spitzer-irs_2010}).
NGC\,34 was observed with T-ReCS in the Si5 filter in 2008 (unpublished, to our knowledge). 
An extended elongated MIR nucleus consistent with the previous MIR observations was detected in the image (FWHM(major axis) $\sim 0.71\arcsec \sim 260\,$pc; PA$\sim 173\degree$). 
The unresolved nuclear Si5 flux is $\sim 25\%$ of the \spitzerr spectrophotometry, while the total Si5 flux agrees with the latter.
Therefore, we conclude that the MIR emission of NGC\,34 is dominated by a nuclear starburst on $300$\,pc scales, while the AGN contributes only up to $25\%$.
\newline\end{@twocolumnfalse}]

\begin{figure}
   \centering
   \includegraphics[angle=0,width=8.500cm]{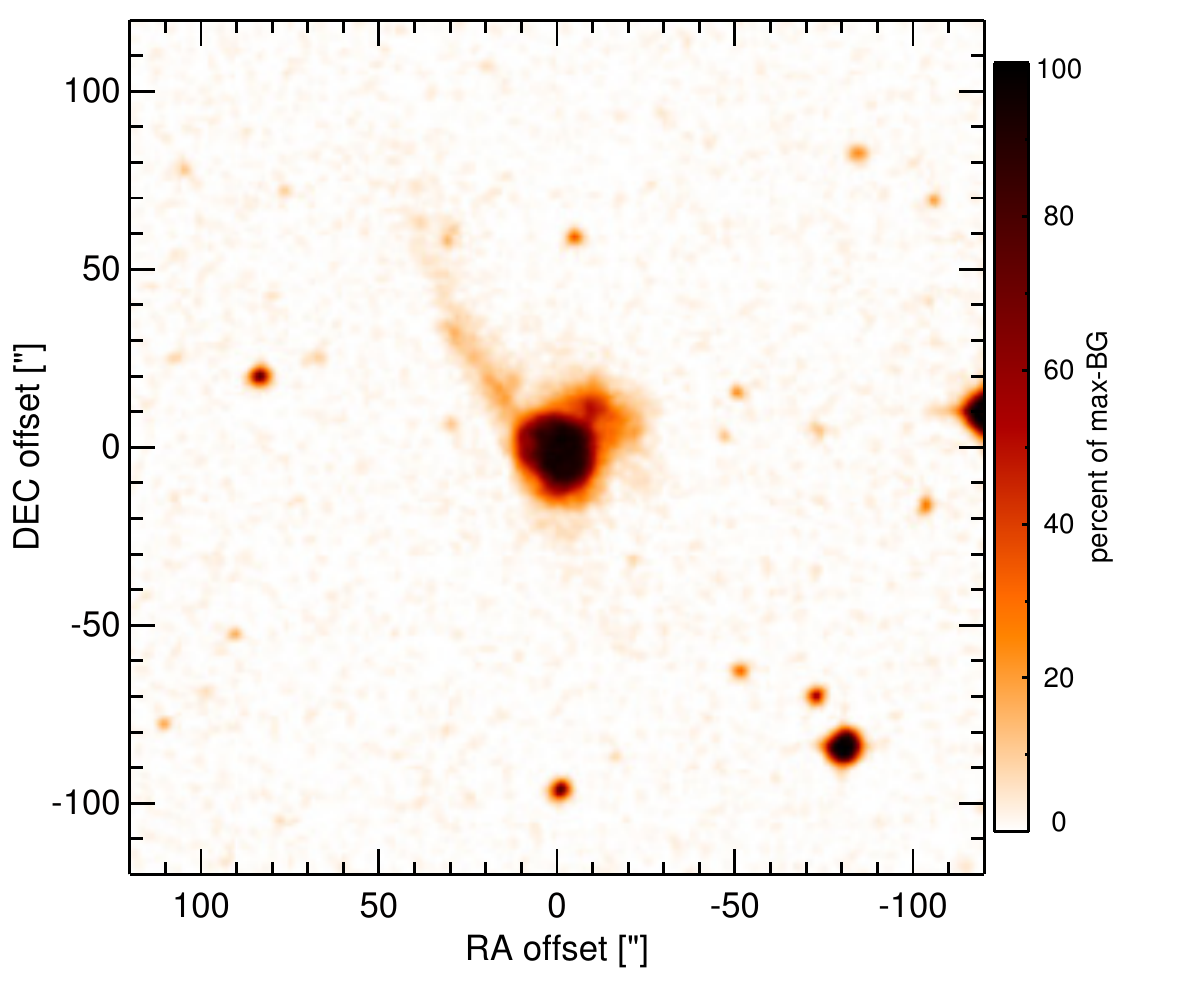}
    \caption{\label{fig:OPTim_NGC0034}
             Optical image (DSS, red filter) of NGC\,34. Displayed are the central $4\arcmin$ with North up and East to the left. 
              The colour scaling is linear with white corresponding to the median background and black to the $0.01\%$ pixels with the highest intensity.  
           }
\end{figure}
\begin{figure}
   \centering
   \includegraphics[angle=0,height=3.11cm]{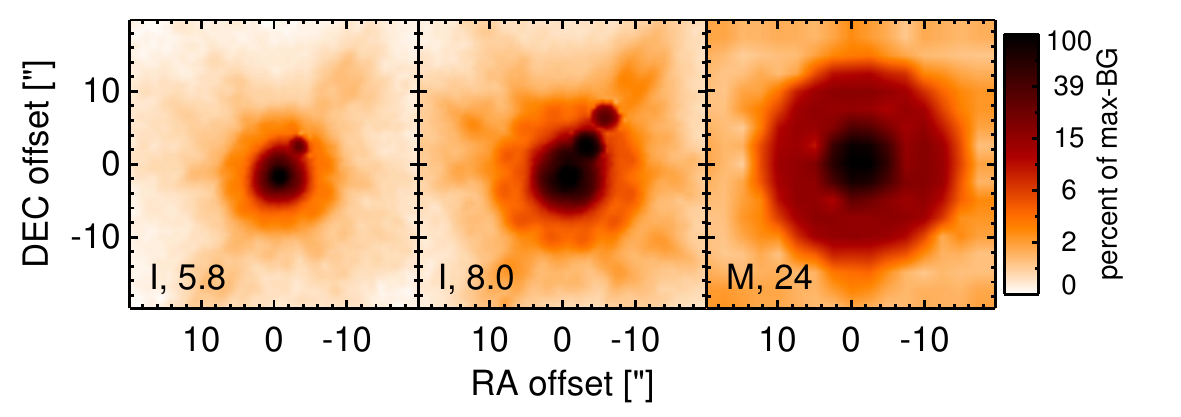}
    \caption{\label{fig:INTim_NGC0034}
             \spitzerr MIR images of NGC\,34. Displayed are the inner $40\arcsec$ with North up and East to the left. The colour scaling is logarithmic with white corresponding to median background and black to the $0.1\%$ pixels with the highest intensity.
             The label in the bottom left states instrument and central wavelength of the filter in $\mu$m (I: IRAC, M: MIPS).
            Note that the apparent off-nuclear compact sources in the IRAC 5.8 and $8.0\,\mu$m images are instrumental artefacts. 
           }
\end{figure}
\begin{figure}
   \centering
   \includegraphics[angle=0,height=3.11cm]{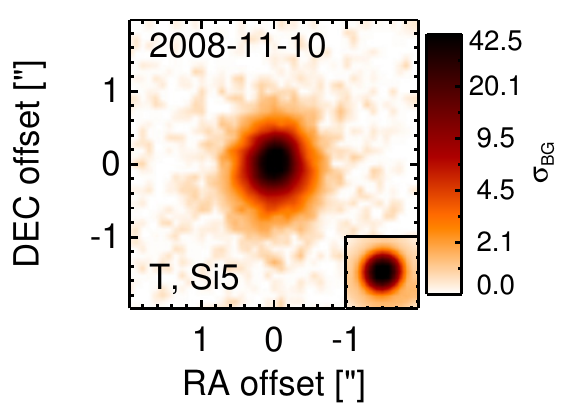}
    \caption{\label{fig:HARim_NGC0034}
             Subarcsecond-resolution MIR images of NGC\,34 sorted by increasing filter wavelength. 
             Displayed are the inner $4\arcsec$ with North up and East to the left. 
             The colour scaling is logarithmic with white corresponding to median background and black to the $75\%$ of the highest intensity of all images in units of $\sigbg$.
             The inset image shows the central arcsecond of the PSF from the calibrator star, scaled to match the science target.
             The labels in the bottom left state instrument and filter names (C: COMICS, M: Michelle, T: T-ReCS, V: VISIR).
           }
\end{figure}
\begin{figure}
   \centering
   \includegraphics[angle=0,width=8.50cm]{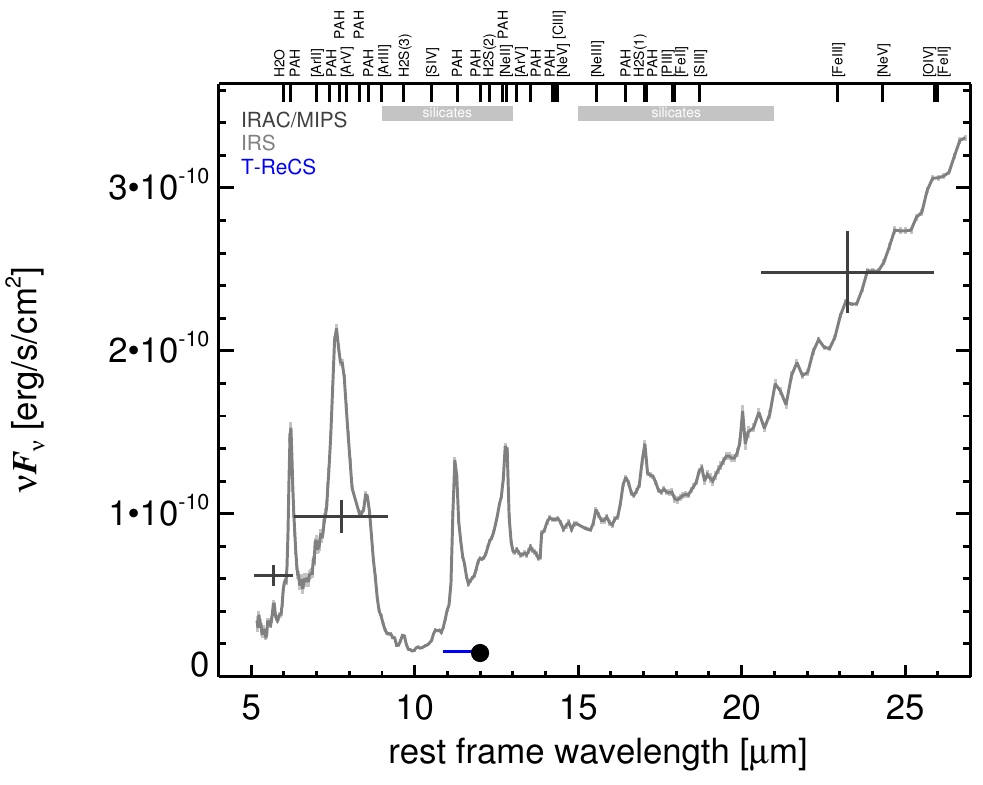}
   \caption{\label{fig:MISED_NGC0034}
      MIR SED of NGC\,34. The description  of the symbols (if present) is the following.
      Grey crosses and  solid lines mark the \spitzer/IRAC, MIPS and IRS data. 
      The colour coding of the other symbols is: 
      green for COMICS, magenta for Michelle, blue for T-ReCS and red for VISIR data.
      Darker-coloured solid lines mark spectra of the corresponding instrument.
      The black filled circles mark the nuclear 12 and $18\,\mu$m  continuum emission estimate from the data.
      The ticks on the top axis mark positions of common MIR emission lines, while the light grey horizontal bars mark wavelength ranges affected by the silicate 10 and 18$\mu$m features.}
\end{figure}
\clearpage

\twocolumn[\begin{@twocolumnfalse}  
\subsection{NGC\,235A}\label{app:NGC0235A}
NGC\,235A is an early-type spiral galaxy in an interacting pair at a redshift of $z=$ 0.0222 ($D\sim 88.7\,$Mpc) with an AGN controversially classified either as Sy\,1.0 \citep{maia_new_1987,veron-cetty_catalogue_2010} or Sy\,2 (e.g., \citealt{keel_seyfert_1996, tueller_swift_2008}), see the discussion in \cite{kondratko_discovery_2006}.
The fact that the nucleus is highly obscured in X-rays favours the Sy\,2 classification  also adopted \citep{winter_x-ray_2009}.
It features extended \oiii emission in east-west direction ($\sim 9\arcsec \sim 3.7$\,kpc; PA$\sim80\degree$; \citealt{mulchaey_emission-line_1996}) with a slightly different radio extension (PA$\sim43\degree$; \citealt{nagar_radio_1999}).
In addition, a nuclear water maser was detected \citep{kondratko_discovery_2006}.
NGC\,235A belongs to the nine-month BAT AGN sample.
It was observed with \spitzer/IRS in LR staring mode, and the corresponding spectrum shows strong PAH emission, weak silicate  $10\,\mu$m absorption and a red spectral slope in $\nu F_\nu$-space, i.e., is star formation dominated.
The interacting system is resolved only in the short wavelength bands of \wisee with NGC\,235A dominating the MIR emission in all bands.
We observed NGC\,235A with VISIR in three narrow $N$-band filters in 2009.
In all images, a compact MIR nucleus is weakly detected, which appears marginally resolved in the PAH2 and NEII filters consistent with the \oiii emission orientation (FWHM $\sim 0.5\arcsec \sim 210\,$pc; PA $\sim 80\degree$). 
However, at least a second epoch of deeper subarcsecond-resolution MIR images is required to confirm this extension.
The VISIR photometry is on average $\sim 41\%$ lower than the IRS spectrum flux levels.
Therefore, we estimate that star formation contributes significantly to the projected central $\sim 150$\,pc MIR emission in NGC\,235A. 
\newline\end{@twocolumnfalse}]

\begin{figure}
   \centering
   \includegraphics[angle=0,width=8.500cm]{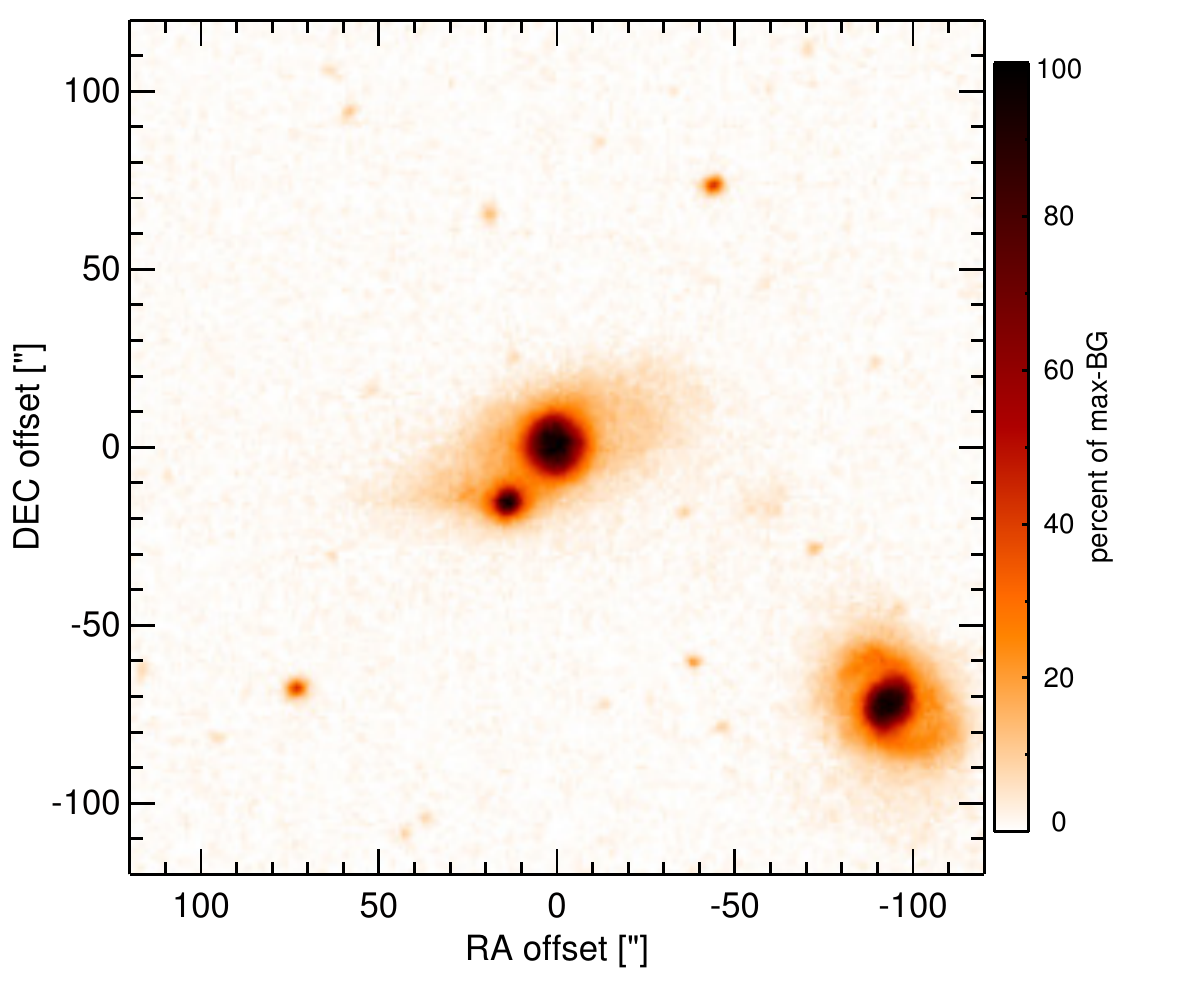}
    \caption{\label{fig:OPTim_NGC0235A}
             Optical image (DSS, red filter) of NGC\,235A. Displayed are the central $4\arcmin$ with North up and East to the left. 
              The colour scaling is linear with white corresponding to the median background and black to the $0.01\%$ pixels with the highest intensity.  
           }
\end{figure}
\begin{figure}
   \centering
   \includegraphics[angle=0,height=3.11cm]{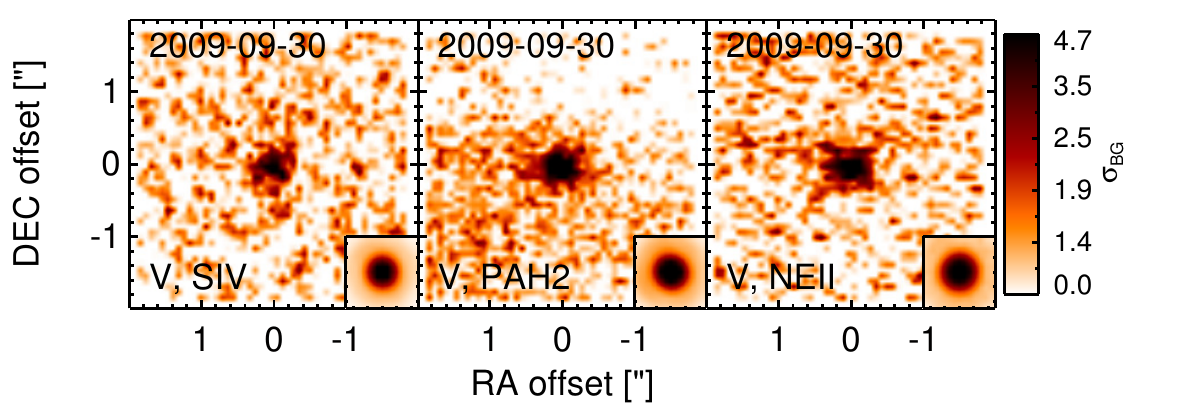}
    \caption{\label{fig:HARim_NGC0235A}
             Subarcsecond-resolution MIR images of NGC\,235A sorted by increasing filter wavelength. 
             Displayed are the inner $4\arcsec$ with North up and East to the left. 
             The colour scaling is logarithmic with white corresponding to median background and black to the $75\%$ of the highest intensity of all images in units of $\sigbg$.
             The inset image shows the central arcsecond of the PSF from the calibrator star, scaled to match the science target.
             The labels in the bottom left state instrument and filter names (C: COMICS, M: Michelle, T: T-ReCS, V: VISIR).
           }
\end{figure}
\begin{figure}
   \centering
   \includegraphics[angle=0,width=8.50cm]{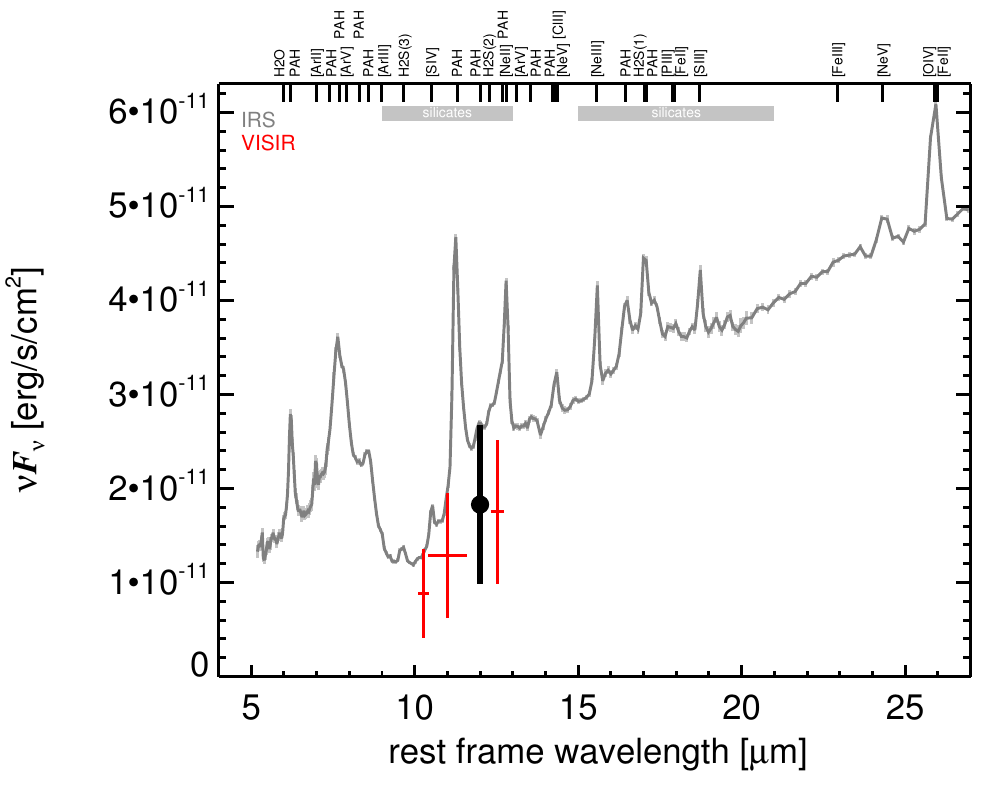}
   \caption{\label{fig:MISED_NGC0235A}
      MIR SED of NGC\,235A. The description  of the symbols (if present) is the following.
      Grey crosses and  solid lines mark the \spitzer/IRAC, MIPS and IRS data. 
      The colour coding of the other symbols is: 
      green for COMICS, magenta for Michelle, blue for T-ReCS and red for VISIR data.
      Darker-coloured solid lines mark spectra of the corresponding instrument.
      The black filled circles mark the nuclear 12 and $18\,\mu$m  continuum emission estimate from the data.
      The ticks on the top axis mark positions of common MIR emission lines, while the light grey horizontal bars mark wavelength ranges affected by the silicate 10 and 18$\mu$m features.}
\end{figure}
\clearpage

\twocolumn[\begin{@twocolumnfalse}  
\subsection{NGC\,253}\label{app:NGC0253}
NGC\,253 is an edge-on spiral galaxy at a distance of $D=$ $3.2 \pm 0.6$\,Mpc (NED redshift-independent median) with an active nucleus dominated by a powerful starburst \citep{engelbracht_nuclear_1998}, which according to the optical classification, is a starburst/LINER transition nucleus.
Narrow-line \oiii emission \citep{tadhunter_optical_1993}, a compact non-thermal radio source, TH2 \citep{turner_1_1985}, and an obscured compact X-ray source, X-1 \citep{weaver_chandra_2002}, are detected in the nucleus, suggesting the presence of an obscured AGN (see also \citealt{mohan_very_2002}).
On the other hand, no compact nuclear source could be detected with VLBI in radio \citep{brunthaler_evidence_2009} or with subarcsecond MIR imaging \citep{fernandez-ontiveros_nucleus_2009}.
In addition, \cite{muller-sanchez_stellar_2010} argue that the positions of TH2 and X-1 are not consistent with each other, i.e., instead of a buried AGN, the nucleus could  also be Sgr A$^*$-like when associated with TH2 only.
NGC\,253 was discovered as a bright MIR source by \cite{becklin_infrared_1973} and since then target of a large number of $N-$ and $Q$-band photometric and spectroscopic studies \citep{rieke_nucleus_1975,gillett_observations_1975,rieke_10_1978,beck_ne_1979,roche_8-13-micron_1985,ho_excess_1989,pina_12_1992,keto_mid-infrared_1993,telesco_genesis_1993,boeker_mid-infrared_1998,engelbracht_nuclear_1998,dudley_8-13_1999,keto_super-star_1999,rigopoulou_large_1999,sturm_iso-sws_2000,forster_schreiber_isocam_2003}.
The first subarcsecond-resolution $N$-band image was obtained by \cite{galliano_mid-infrared_2005} with ESO 3.6\,m/TIMMI2, where the nuclear region of NGC\,253 is resolved in six sources embedded within diffuse emission with $\sim 15\arcsec \sim 225$\,pc north-east extent (see also \citealt{raban_core_2008}).
Notably, the brighter MIR source, M1 does not coincide with the nucleus, instead the latter might correspond to M3.
Such an extended structure is also indicated in the IRAC $5.8$ and $8.0\,\mu$m and MIPS $24\,\mu$m PBCD images, which are completely saturated and thus not analysed (but see \citealt{dale_spitzer_2009}).
Owing to the complex extended MIR morphology of NGC\,253, the IRS LR PBCD spectrum is not reliable and only qualitatively examined.
The MIR SED is prototypical for star formation with strong PAH features, silicate $10\,\mu$m absorption an a red spectral slope in $\nu F_\nu$-space (see also \citealt{devost_spitzer_2004,goulding_towards_2009}).
The nuclear region of NGC\,253 was also observed with VISIR in PAH2 and PAH2\_2 in 2004 and in SIV and Q2 in 2005. 
The Q2 image was published by \cite{fernandez-ontiveros_nucleus_2009} along with the Q2 and PAH2\_2 fluxes of the emission knots.
The VISIR images show the same MIR morphology as the previous TIMMI2 images. 
We perform manually scaled PSF photometry at the approximated expected nuclear position to derive upper limits on any AGN emission component.
These fluxes are on average $\sim 21\%$ of the \spitzerr spectrophotometry, which demonstrates that the star formation completely dominates the nuclear MIR emission of NGC\,253. 
Therefore, it is also not surprising that no compact MIR source could be detected with MIDI \citep{tristram_parsec-scale_2009}.
\newline\end{@twocolumnfalse}]

\begin{figure}
   \centering
   \includegraphics[angle=0,width=8.500cm]{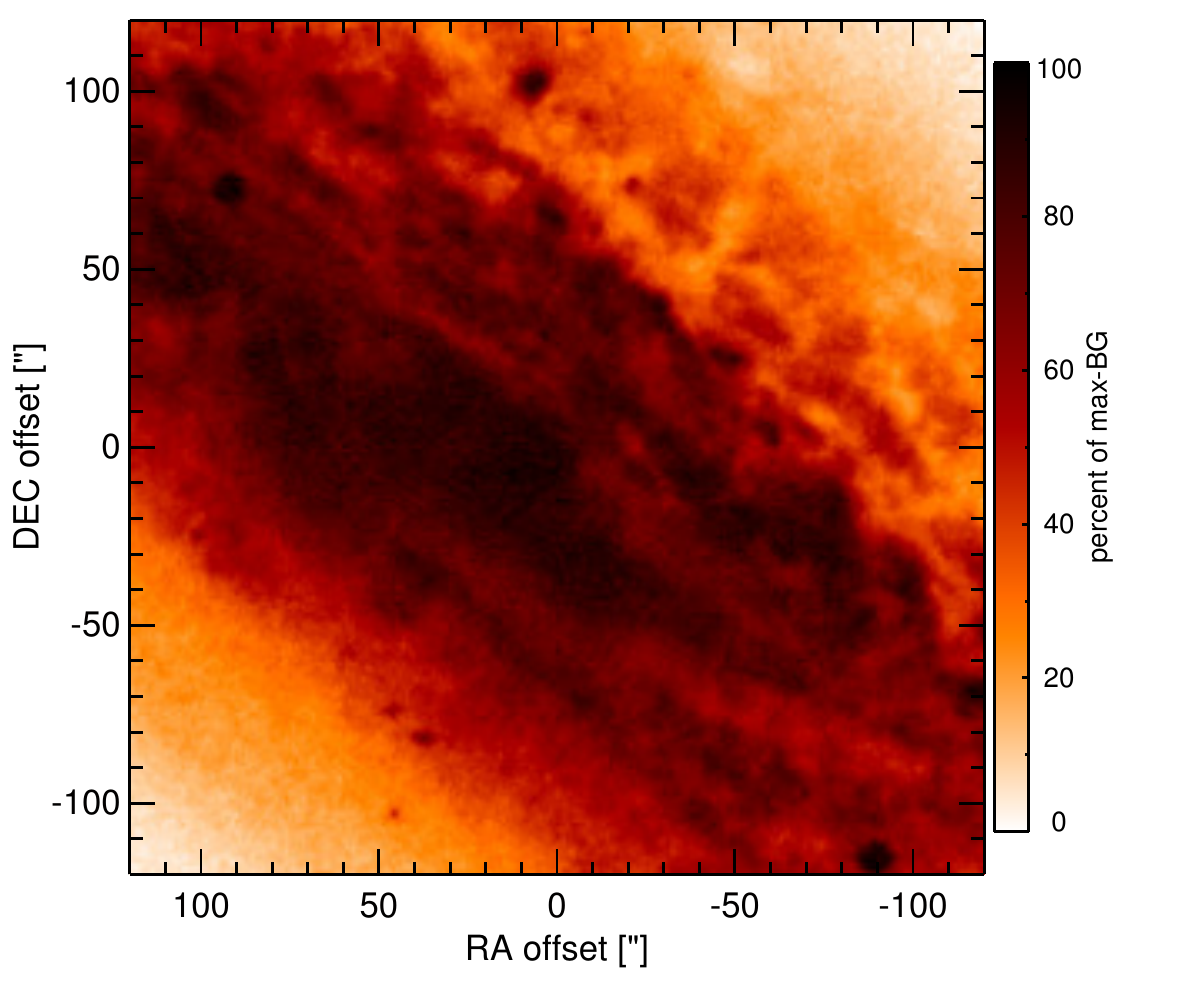}
    \caption{\label{fig:OPTim_NGC0253}
             Optical image (DSS, red filter) of NGC\,253. Displayed are the central $4\arcmin$ with North up and East to the left. 
              The colour scaling is linear with white corresponding to the median background and black to the $0.01\%$ pixels with the highest intensity.  
           }
\end{figure}
\begin{figure}
   \centering
   \includegraphics[angle=0,height=3.11cm]{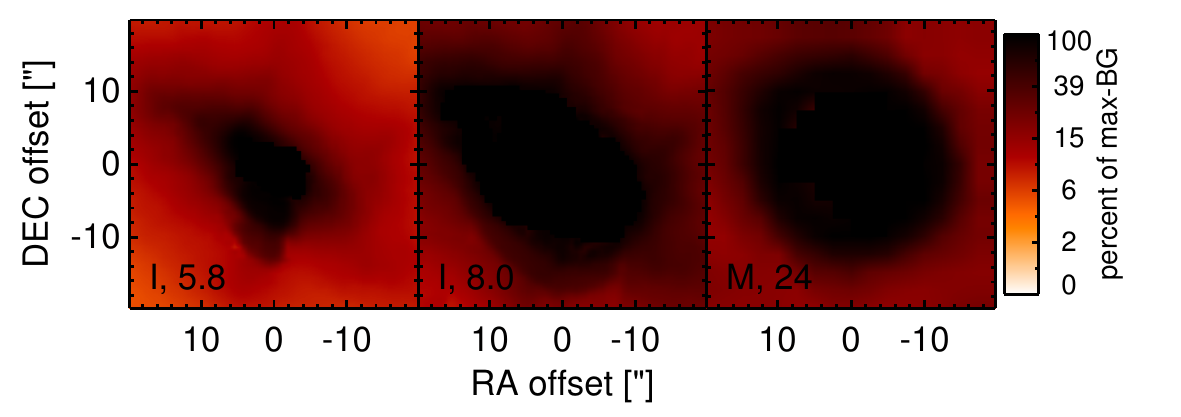}
    \caption{\label{fig:INTim_NGC0253}
             \spitzerr MIR images of NGC\,253. Displayed are the inner $40\arcsec$ with North up and East to the left. The colour scaling is logarithmic with white corresponding to median background and black to the $0.1\%$ pixels with the highest intensity.
             The label in the bottom left states instrument and central wavelength of the filter in $\mu$m (I: IRAC, M: MIPS). 
             Note that the central region in the images is completely saturated.
           }
\end{figure}
\begin{figure}
   \centering
   \includegraphics[angle=0,width=8.500cm]{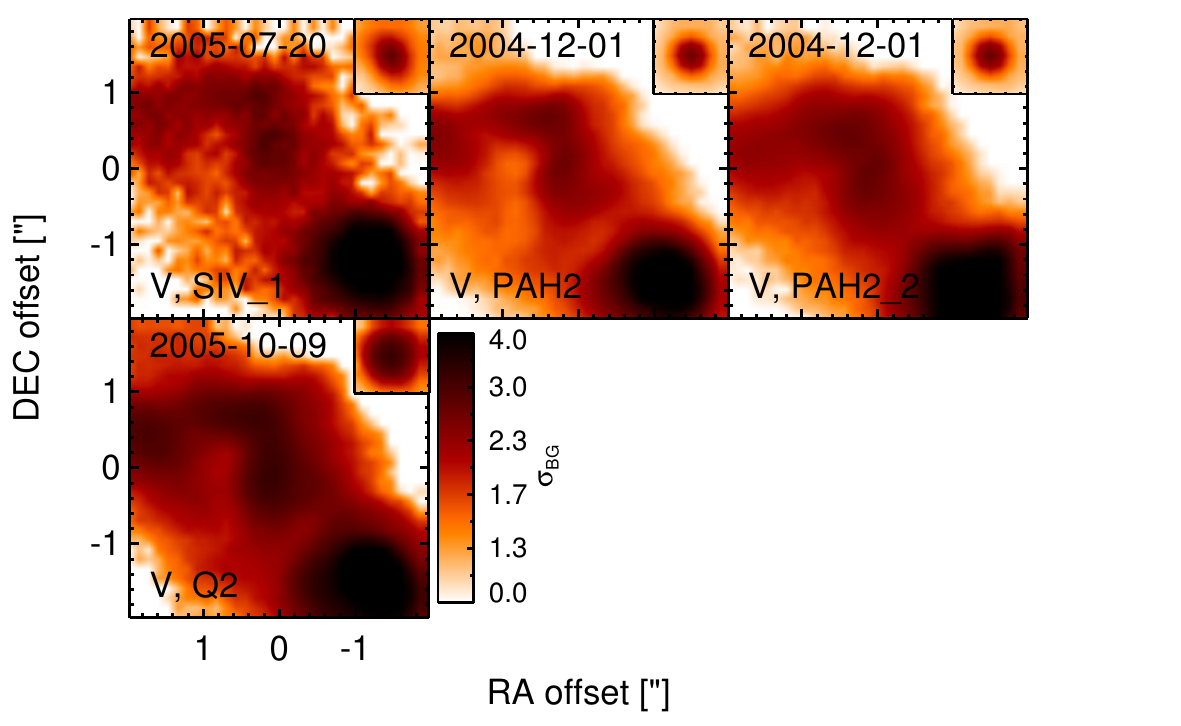}
    \caption{\label{fig:HARim_NGC0253}
             Subarcsecond-resolution MIR images of NGC\,253 sorted by increasing filter wavelength. 
             Displayed are the inner $4\arcsec$ with North up and East to the left. 
             The colour scaling is logarithmic with white corresponding to median background and black to the $75\%$ of the highest intensity of all images in units of $\sigbg$.
             The inset image shows the central arcsecond of the PSF from the calibrator star, scaled to match the science target.
             The labels in the bottom left state instrument and filter names (C: COMICS, M: Michelle, T: T-ReCS, V: VISIR).
           }
\end{figure}
\begin{figure}
   \centering
   \includegraphics[angle=0,width=8.50cm]{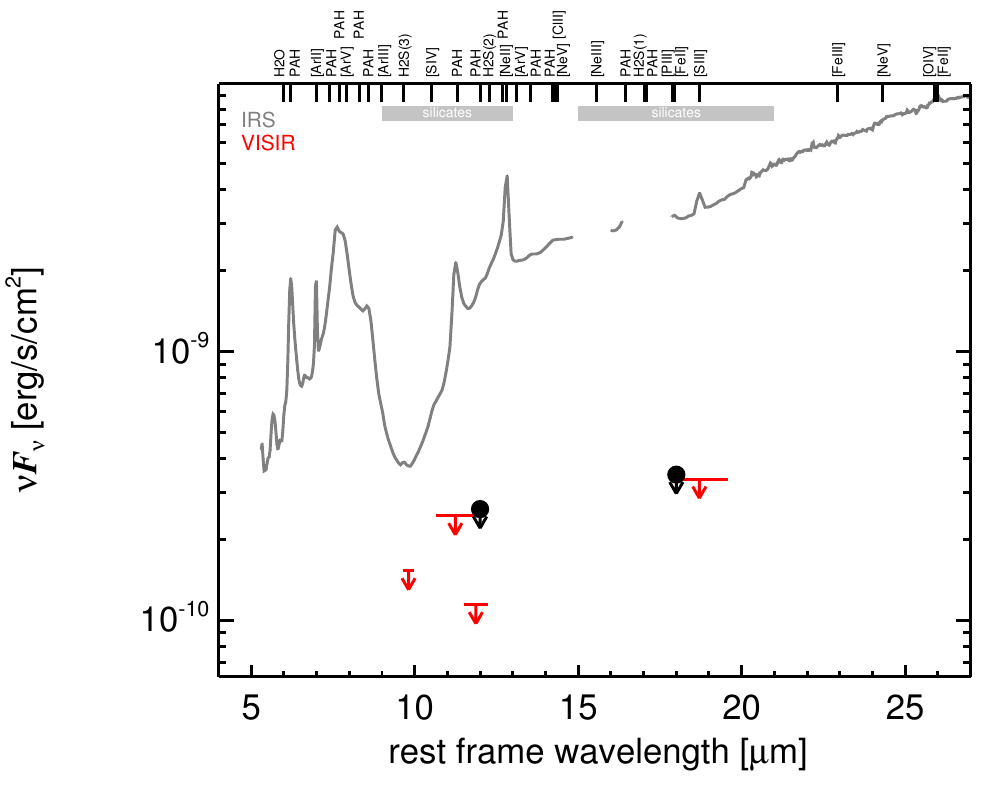}
   \caption{\label{fig:MISED_NGC0253}
      MIR SED of NGC\,253. The description  of the symbols (if present) is the following.
      Grey crosses and  solid lines mark the \spitzer/IRAC, MIPS and IRS data. 
      The colour coding of the other symbols is: 
      green for COMICS, magenta for Michelle, blue for T-ReCS and red for VISIR data.
      Darker-coloured solid lines mark spectra of the corresponding instrument.
      The black filled circles mark the nuclear 12 and $18\,\mu$m  continuum emission estimate from the data.
      The ticks on the top axis mark positions of common MIR emission lines, while the light grey horizontal bars mark wavelength ranges affected by the silicate 10 and 18$\mu$m features.}
\end{figure}
\clearpage


\twocolumn[\begin{@twocolumnfalse}  
\subsection{NGC\,424 -- Tololo\,0109-383}\label{app:NGC0424}
NGC\,424 is an inclined spiral galaxy at a redshift of $z=$ 0.0118 ($D\sim 49.5$\,Mpc) with an AGN controversially classified as a Sy\,1 \citep{murayama_high-ionization_1998} or as Sy\,2 \citep{smith_emission-line_1975}.
Broad emission lines are detected in polarized light \citep{moran_frequency_2000}. 
The fact that the nucleus is highly obscured and probably even Compton-thick in X-rays favours the Sy\,2 classification \citep{collinge_compton-thick_2000}.
The nucleus of NGC\,424 is radio-quiet and unresolved \citep{mundell_parsec-scale_2000}.
After \iras, NGC\,424 was observed with \spitzer/IRAC, IRS and MIPS.
It appears very compact in the IRAC and MIPS PBCD images but is saturated in the former.
Therefore, the  IRAC $5.8$ and $8.0\,\mu$m data are not analysed (but see \citealt{gallimore_infrared_2010}).
The IRS LR mapping mode spectrum exhibits  silicate 10 and $18\,\mu$m emission and a blue spectral slope in $\nu F_\nu$-space but no PAH 11.3$\,\mu$m feature (see also \citealt{buchanan_spitzer_2006,tommasin_spitzer_2008, wu_spitzer/irs_2009,gallimore_infrared_2010,tommasin_spitzer-irs_2010}.
Interestingly, the MIR SED thus resembles more that of an unobscured AGN.
NGC\,424 was observed with T-ReCS in the N filter in 2003 (unpublished, to our knowledge).
An elongated nucleus was detected in the image (FWHM $\sim 0.45\arcsec \sim 106\,$pc; PA$\sim80\degree$).
However, this finding needs to be verified by at least another epoch of subarcsecond MIR imaging.
The measured N filter flux is consistent with the \spitzerr spectrophotometry, but it would be significantly lower if the presence of subarcsecond-extended emission can be verified.
Note that the MIR nucleus of NGC\,424 could be resolved with MIDI interferometric observations, where it shows an elongation along the polar direction of the system (diameter$\sim 4$\,pc; PA$\sim 153\degree$; \citealt{honig_parsec-scale_2012}).
\newline\end{@twocolumnfalse}]

\begin{figure}
   \centering
   \includegraphics[angle=0,width=8.500cm]{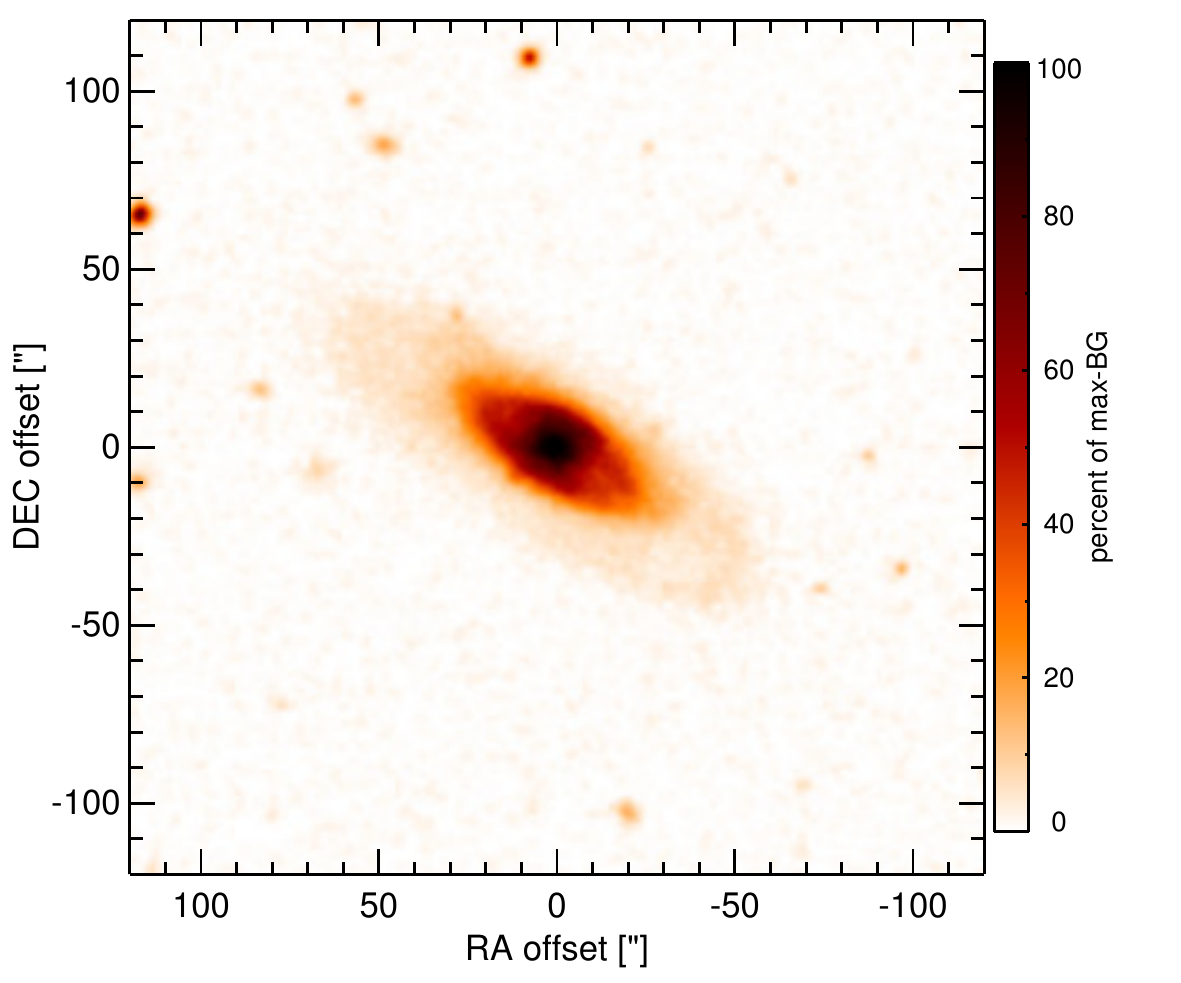}
    \caption{\label{fig:OPTim_NGC0424}
             Optical image (DSS, red filter) of NGC\,424. Displayed are the central $4\arcmin$ with North up and East to the left. 
              The colour scaling is linear with white corresponding to the median background and black to the $0.01\%$ pixels with the highest intensity.  
           }
\end{figure}
\begin{figure}
   \centering
   \includegraphics[angle=0,height=3.11cm]{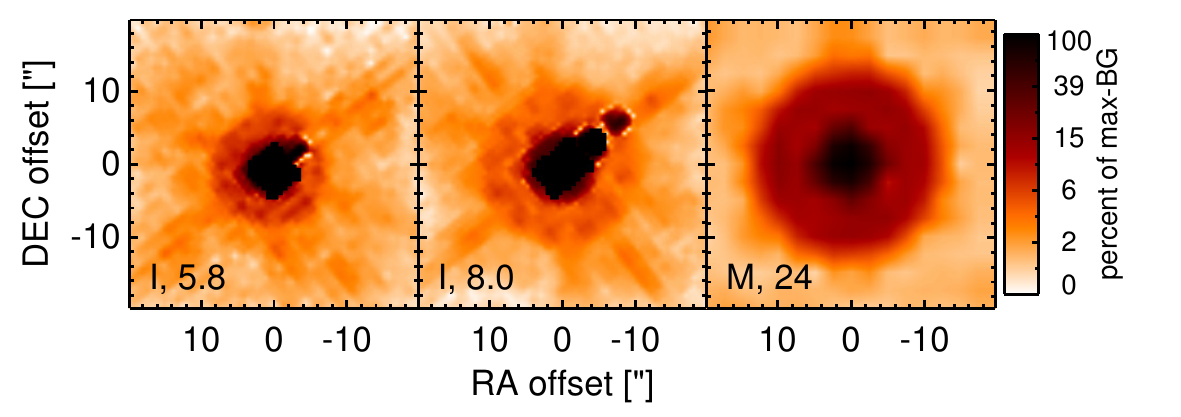}
    \caption{\label{fig:INTim_NGC0424}
             \spitzerr MIR images of NGC\,424. Displayed are the inner $40\arcsec$ with North up and East to the left. The colour scaling is logarithmic with white corresponding to median background and black to the $0.1\%$ pixels with the highest intensity.
             The label in the bottom left states instrument and central wavelength of the filter in $\mu$m (I: IRAC, M: MIPS). 
             Note that the apparent off-nuclear compact sources in the IRAC 5.8 and $8.0\,\mu$m images are instrumental artefacts.
           }
\end{figure}
\begin{figure}
   \centering
   \includegraphics[angle=0,height=3.11cm]{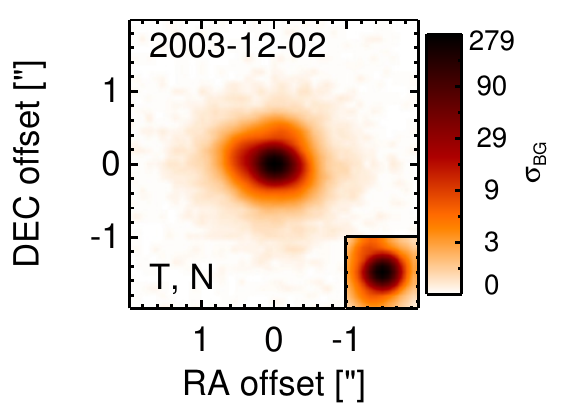}
    \caption{\label{fig:HARim_NGC0424}
             Subarcsecond-resolution MIR images of NGC\,424 sorted by increasing filter wavelength. 
             Displayed are the inner $4\arcsec$ with North up and East to the left. 
             The colour scaling is logarithmic with white corresponding to median background and black to the $75\%$ of the highest intensity of all images in units of $\sigbg$.
             The inset image shows the central arcsecond of the PSF from the calibrator star, scaled to match the science target.
             The labels in the bottom left state instrument and filter names (C: COMICS, M: Michelle, T: T-ReCS, V: VISIR).
           }
\end{figure}
\begin{figure}
   \centering
   \includegraphics[angle=0,width=8.50cm]{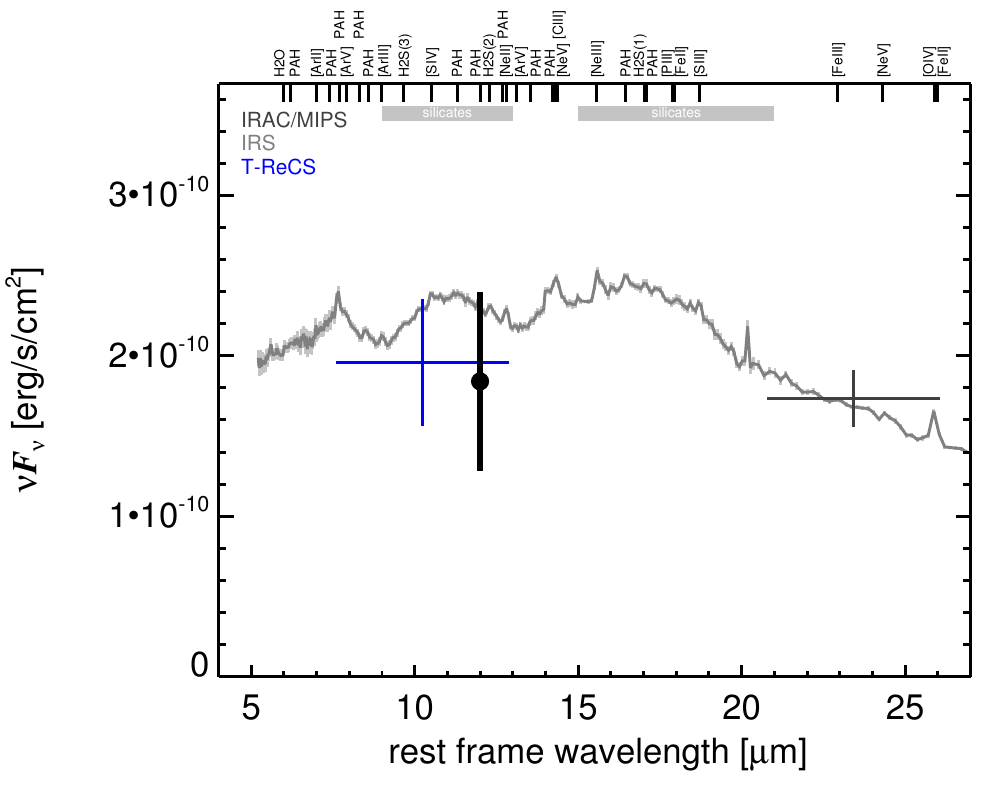}
   \caption{\label{fig:MISED_NGC0424}
      MIR SED of NGC\,424. The description  of the symbols (if present) is the following.
      Grey crosses and  solid lines mark the \spitzer/IRAC, MIPS and IRS data. 
      The colour coding of the other symbols is: 
      green for COMICS, magenta for Michelle, blue for T-ReCS and red for VISIR data.
      Darker-coloured solid lines mark spectra of the corresponding instrument.
      The black filled circles mark the nuclear 12 and $18\,\mu$m  continuum emission estimate from the data.
      The ticks on the top axis mark positions of common MIR emission lines, while the light grey horizontal bars mark wavelength ranges affected by the silicate 10 and 18$\mu$m features.}
\end{figure}
\clearpage

\twocolumn[\begin{@twocolumnfalse}  
\subsection{NGC\,454E}\label{app:NGC0454E}
NGC\,454 is a strongly interacting system of two galaxies (nuclei separation$\sim 28\arcsec\sim 7$\,kpc; PA$\sim70\degree$; \citep{johansson_activity_1988}) at a redshift of $z=$ 0.0122 ($D\sim 52.3\,$Mpc).
The eastern component, NGC\,454E, is elliptical, more regular and brighter than the western starburst component.
NGC\,454E hosts a Sy\,2 nucleus \citep{veron-cetty_catalogue_2010} that belongs to the nine-month BAT AGN sample.
Its X-ray obscuration was  observed to change significantly  \citep{marchese_ngc_2012}.
After \iras, NGC\,454E was observed with \spitzer/IRS in LR staring mode.
The spectrum exhibits a flat spectral slope in $\nu F_\nu$-space, silicate  $10\,\mu$m absorption and PAH 11.3$\,\mu$m emission, i.e., significant star formation (see also \citealt{weaver_mid-infrared_2010,sargsyan_infrared_2011}).
The interacting system is resolved in the \wisee images with NGC\,454E dominating the MIR emission in all bands. 
NGC\,454W appears bow-like with a bright compact source at the closest point to NGC\,454E (PA$\sim245\degree$; distance $\sim0.5\arcmin\sim7\,$kpc).
We observed NGC\,454E with VISIR in three narrow $N$-band filters in 2009 and detected an elongated MIR nucleus  in all cases (FWHM $\sim 0.43 \arcsec \sim 107\,$pc; PA$\sim 120\degree$).
However, at least a second epoch of subarcsecond MIR imaging is required to verify this extension.
The nuclear VISIR photometry is in general consistent with the IRS spectrum but with less PAH $11\,\mu$m emission indicating a lower star formation contribution towards the MIR emission of the projected central $\sim 100$\,pc of NGC\,454E. 
Note that the nuclear fluxes would be significantly lower if the presence of subarcsecond-extended emission can be verified.
\newline\end{@twocolumnfalse}]

\begin{figure}
   \centering
   \includegraphics[angle=0,width=8.500cm]{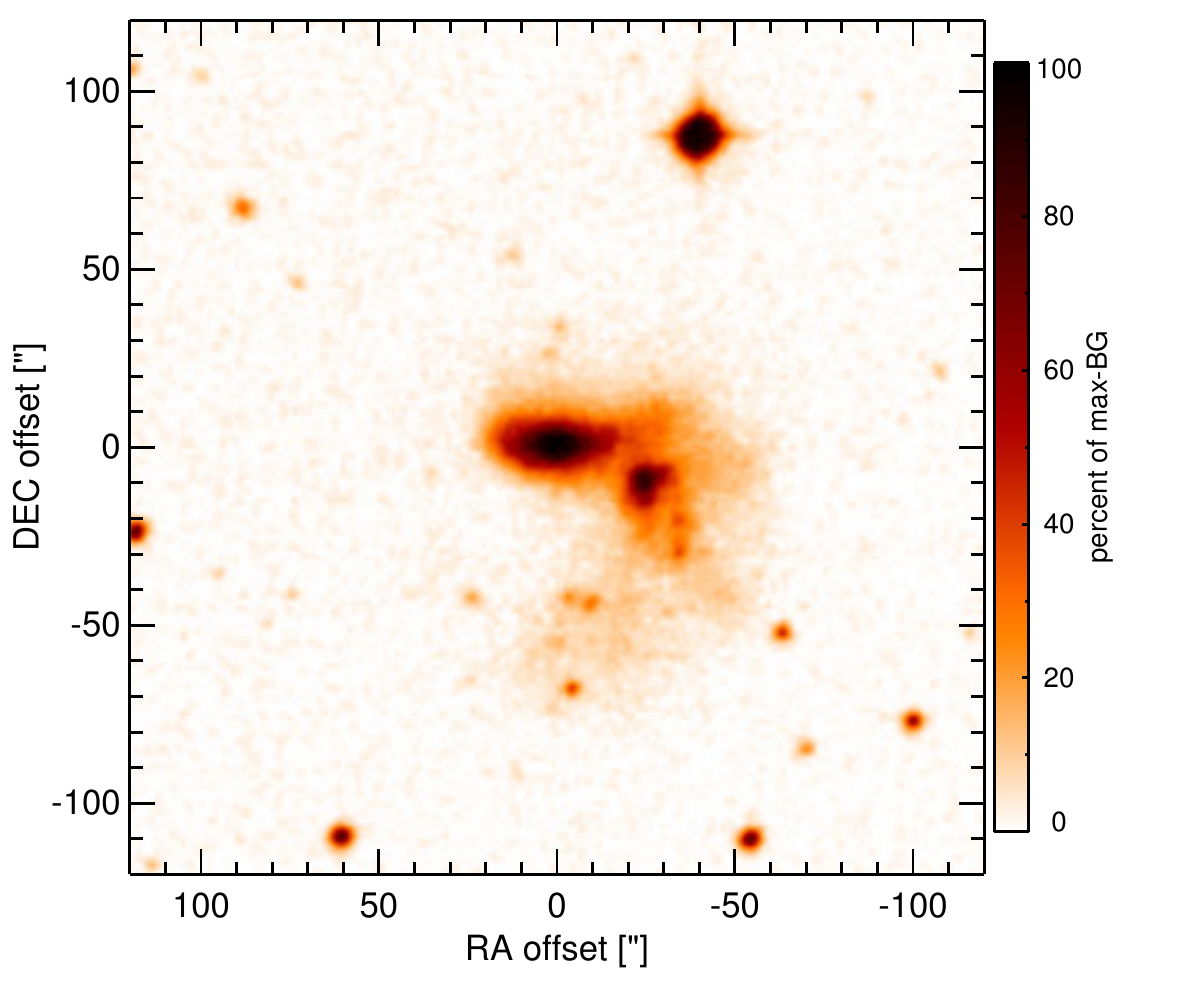}
    \caption{\label{fig:OPTim_NGC0454E}
             Optical image (DSS, red filter) of NGC\,454E. Displayed are the central $4\arcmin$ with North up and East to the left. 
              The colour scaling is linear with white corresponding to the median background and black to the $0.01\%$ pixels with the highest intensity.  
           }
\end{figure}
\begin{figure}
   \centering
   \includegraphics[angle=0,height=3.11cm]{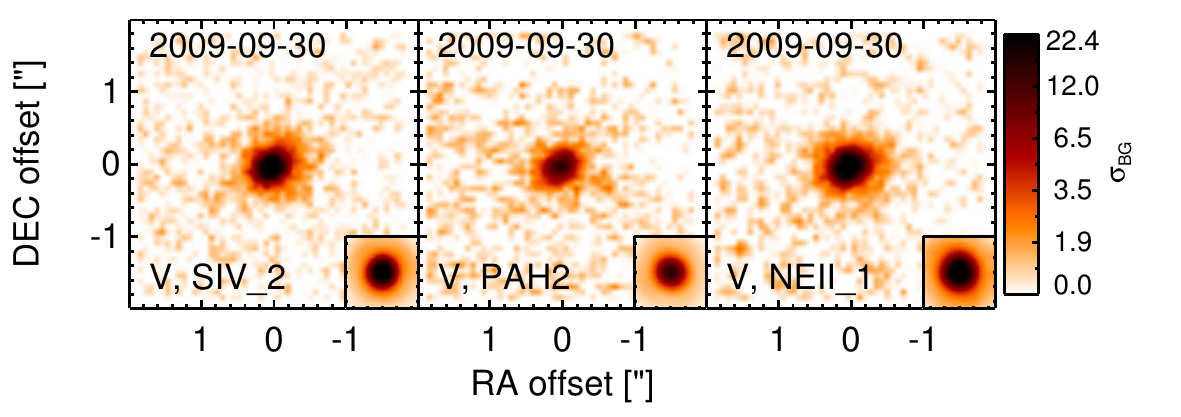}
    \caption{\label{fig:HARim_NGC0454E}
             Subarcsecond-resolution MIR images of NGC\,454E sorted by increasing filter wavelength. 
             Displayed are the inner $4\arcsec$ with North up and East to the left. 
             The colour scaling is logarithmic with white corresponding to median background and black to the $75\%$ of the highest intensity of all images in units of $\sigbg$.
             The inset image shows the central arcsecond of the PSF from the calibrator star, scaled to match the science target.
             The labels in the bottom left state instrument and filter names (C: COMICS, M: Michelle, T: T-ReCS, V: VISIR).
           }
\end{figure}
\begin{figure}
   \centering
   \includegraphics[angle=0,width=8.50cm]{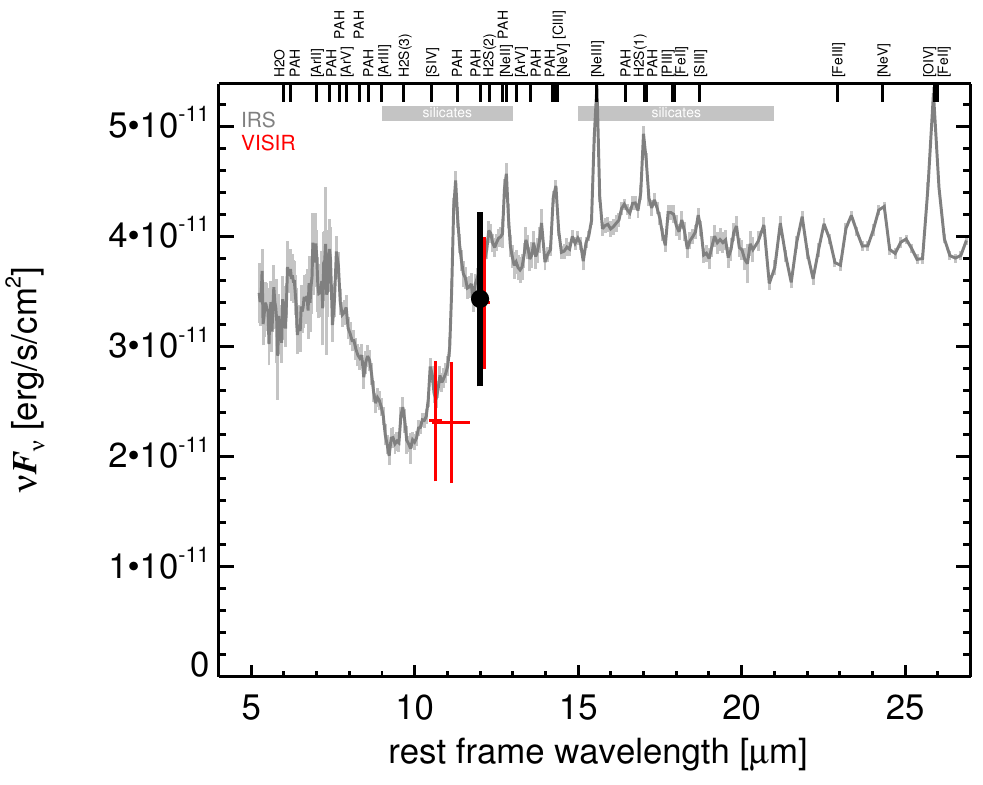}
   \caption{\label{fig:MISED_NGC0454E}
      MIR SED of NGC\,454E. The description  of the symbols (if present) is the following.
      Grey crosses and  solid lines mark the \spitzer/IRAC, MIPS and IRS data. 
      The colour coding of the other symbols is: 
      green for COMICS, magenta for Michelle, blue for T-ReCS and red for VISIR data.
      Darker-coloured solid lines mark spectra of the corresponding instrument.
      The black filled circles mark the nuclear 12 and $18\,\mu$m  continuum emission estimate from the data.
      The ticks on the top axis mark positions of common MIR emission lines, while the light grey horizontal bars mark wavelength ranges affected by the silicate 10 and 18$\mu$m features.}
\end{figure}
\clearpage

\twocolumn[\begin{@twocolumnfalse}  
\subsection{NGC\,526A}\label{app:NGC0526A}
NGC\,526A is the western elliptical component of an strongly interacting pair of galaxies at a redshift of $z=$ 0.0191 ($D\sim82.8$\,Mpc) with a Sy\,1.9 nucleus \citep{veron-cetty_catalogue_2010}, belonging to the nine-month BAT AGN sample.
Note that the nucleus has formally be classified as a Sy\,1.5 \citep{veron-cetty_catalogue_1996}.
It possesses a very extended NLR (several kpc diameter; PA$\sim 130\degree$;  \citealt{mulchaey_emission-line_1996,bennert_size_2006}), coinciding with the extended radio morphology \citep{nagar_radio_1999}.
Apart from \iras, NGC\,526A was also observed with \isoo \citep{clavel_2.5-11_2000,ramos_almeida_mid-infrared_2007} and \spitzer/IRAC, IRS and MIPS.
It appears point-like in the  IRAC $5.8$ and $8.0\,\mu$m and MIPS $24\,\mu$m images, and our corresponding IRAC fluxes are consistent with the values in \cite{gallimore_infrared_2010}.
The IRS LR staring-mode spectrum shows silicate 10 and $18\,\mu$m emission and a rather blue spectral slope in $\nu F_\nu$-space but no PAH 11.3\,$\mu$m feature (see also \citealt{shi_9.7_2006,wu_spitzer/irs_2009,tommasin_spitzer-irs_2010,gallimore_infrared_2010}).
Interestingly, the MIR SED thus resembles more on of an unobscured AGN.
We observed NGC\,526A with VISIR in two narrow $N$-band filters in 2005 \citep{horst_small_2006,horst_mid-infrared_2009}.
The MIR nucleus appears unresolved without any other host emission in both images and our reanalysis provides fluxes consistent with the previous values and the \spitzerr spectrophotometry. 
 \newline\end{@twocolumnfalse}]

\begin{figure}
   \centering
   \includegraphics[angle=0,width=8.500cm]{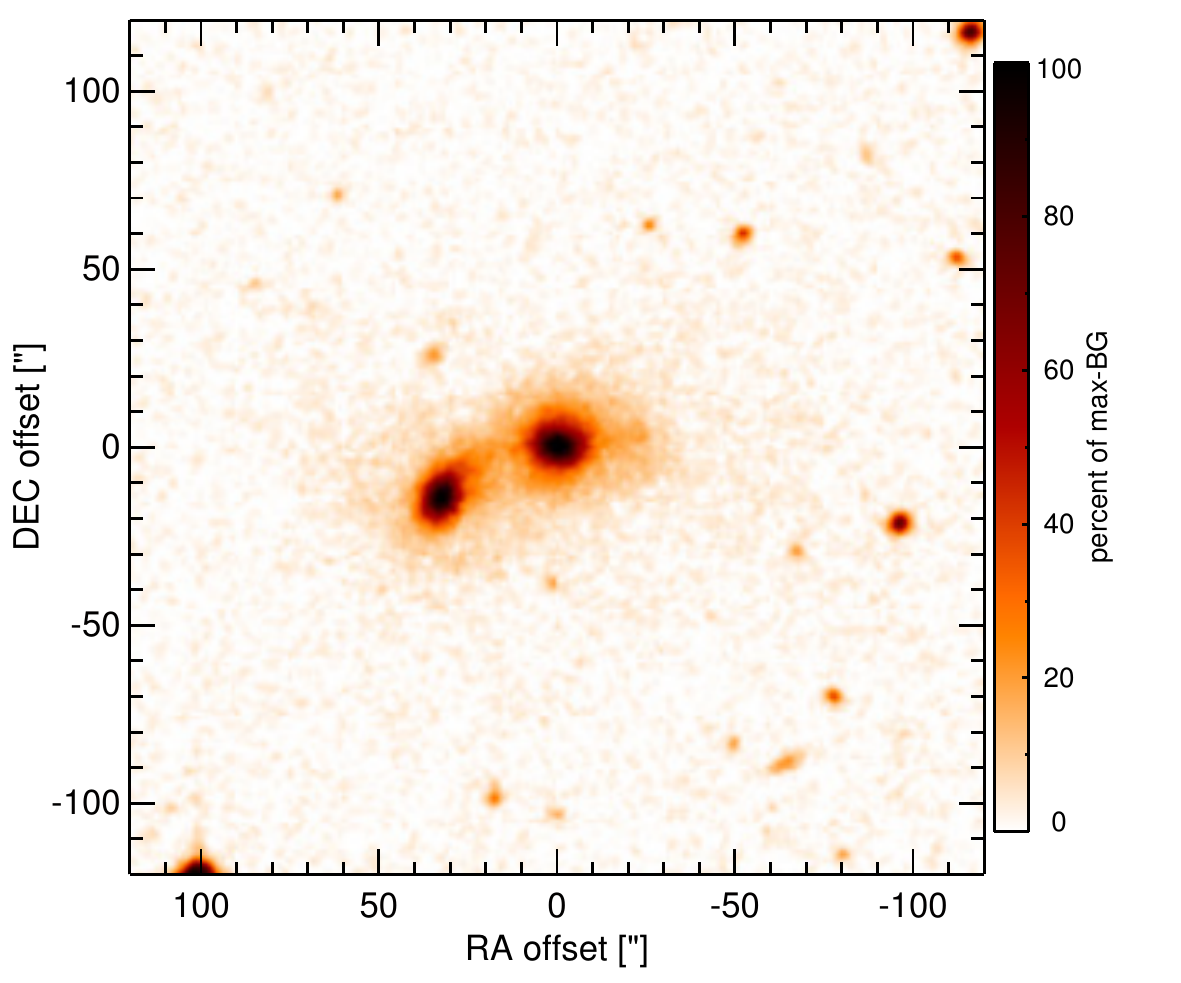}
    \caption{\label{fig:OPTim_NGC0526A}
             Optical image (DSS, red filter) of NGC\,526A. Displayed are the central $4\arcmin$ with North up and East to the left. 
              The colour scaling is linear with white corresponding to the median background and black to the $0.01\%$ pixels with the highest intensity.  
           }
\end{figure}
\begin{figure}
   \centering
   \includegraphics[angle=0,height=3.11cm]{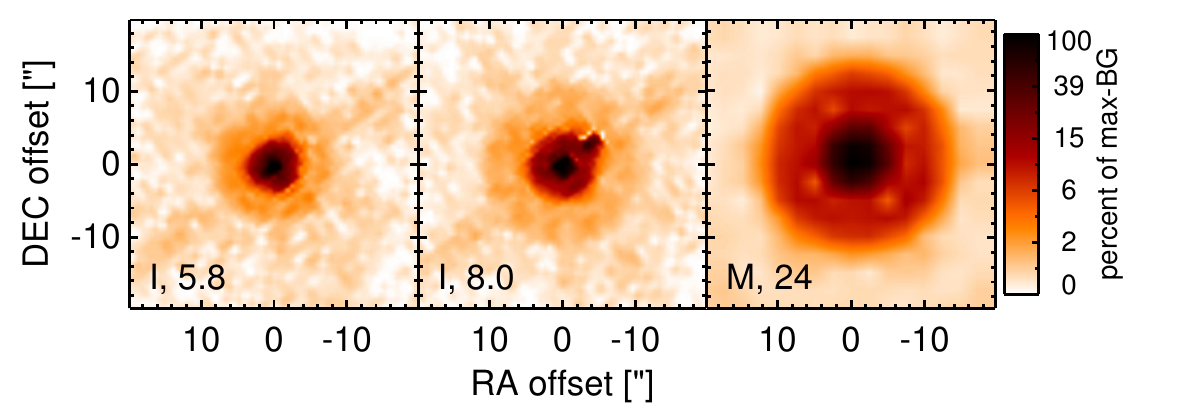}
    \caption{\label{fig:INTim_NGC0526A}
             \spitzerr MIR images of NGC\,526A. Displayed are the inner $40\arcsec$ with North up and East to the left. The colour scaling is logarithmic with white corresponding to median background and black to the $0.1\%$ pixels with the highest intensity.
             The label in the bottom left states instrument and central wavelength of the filter in $\mu$m (I: IRAC, M: MIPS). 
             Note that the apparent off-nuclear compact source in the IRAC $8.0\,\mu$m image is an instrumental artefact.
           }
\end{figure}
\begin{figure}
   \centering
   \includegraphics[angle=0,height=3.11cm]{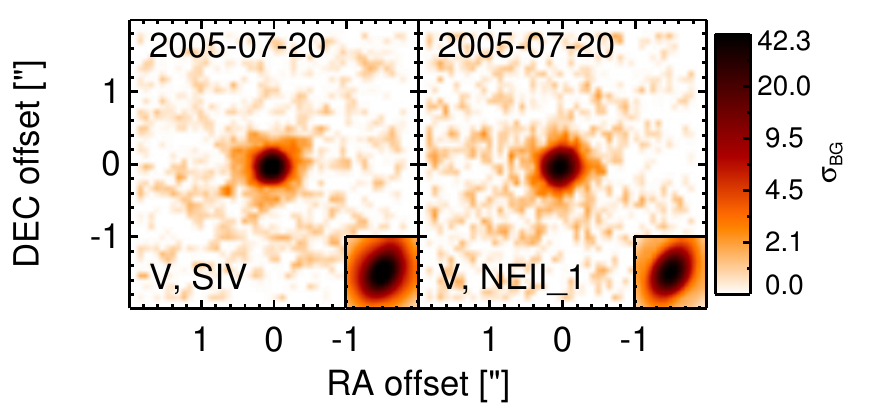}
    \caption{\label{fig:HARim_NGC0526A}
             Subarcsecond-resolution MIR images of NGC\,526A sorted by increasing filter wavelength. 
             Displayed are the inner $4\arcsec$ with North up and East to the left. 
             The colour scaling is logarithmic with white corresponding to median background and black to the $75\%$ of the highest intensity of all images in units of $\sigbg$.
             The inset image shows the central arcsecond of the PSF from the calibrator star, scaled to match the science target.
             The labels in the bottom left state instrument and filter names (C: COMICS, M: Michelle, T: T-ReCS, V: VISIR).
           }
\end{figure}
\begin{figure}
   \centering
   \includegraphics[angle=0,width=8.50cm]{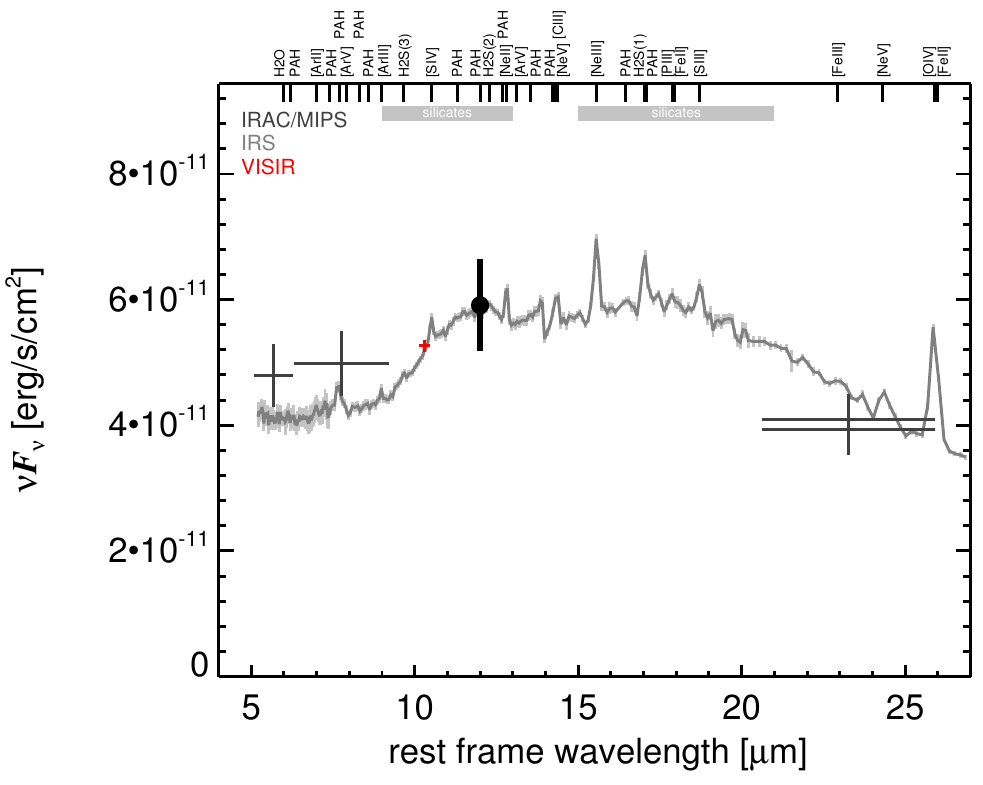}
   \caption{\label{fig:MISED_NGC0526A}
      MIR SED of NGC\,526A. The description  of the symbols (if present) is the following.
      Grey crosses and  solid lines mark the \spitzer/IRAC, MIPS and IRS data. 
      The colour coding of the other symbols is: 
      green for COMICS, magenta for Michelle, blue for T-ReCS and red for VISIR data.
      Darker-coloured solid lines mark spectra of the corresponding instrument.
      The black filled circles mark the nuclear 12 and $18\,\mu$m  continuum emission estimate from the data.
      The ticks on the top axis mark positions of common MIR emission lines, while the light grey horizontal bars mark wavelength ranges affected by the silicate 10 and 18$\mu$m features.}
\end{figure}
\clearpage

\twocolumn[\begin{@twocolumnfalse}  
\subsection{NGC\,612 -- PKS\,0131-36}\label{app:NGC0612}
NGC\,612 is a highly inclined, radio-loud  early-type galaxy at a redshift of $z=$ 0.0298 ($D\sim$132\,Mpc) with hybrid radio morphology \citep{gopal-krishna_extragalactic_2000} and a Sy\,2 nucleus \citep{veron-cetty_catalogue_2010} that belongs to the nine-month BAT AGN sample.
After \iras, NGC\,612 was observed with \spitzer/IRS in LR staring mode.
The spectrum indicates dominant and intense star formation with strong PAH features, silicate 10$\,\mu$m absorption and a slightly red spectral slope in $\nu F_\nu$-space.
In the \wisee images, NGC\,612 appears  elliptically elongated along the galaxy major axis.
We observed the object with VISIR in three narrow $N$-band filters in 2009 but did not detect any nuclear MIR emission.
Our derived upper limits are on average $\sim 49\%$ below the IRS spectrum, which indicates that the MIR emission of NGC\,612 is indeed dominated by star formation with only a weak AGN.
\newline\end{@twocolumnfalse}]

\begin{figure}
   \centering
   \includegraphics[angle=0,width=8.500cm]{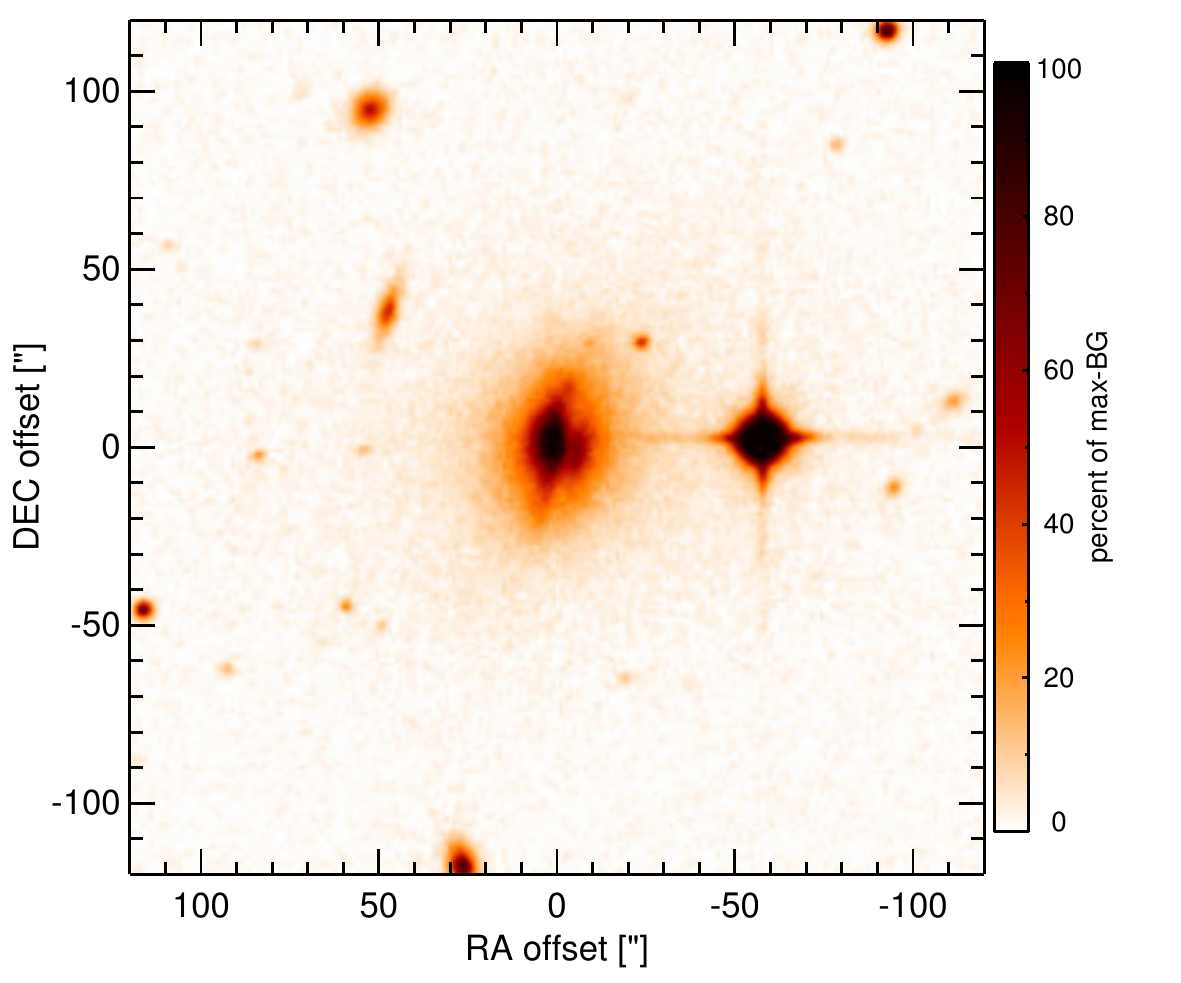}
    \caption{\label{fig:OPTim_NGC0612}
             Optical image (DSS, red filter) of NGC\,612. Displayed are the central $4\arcmin$ with North up and East to the left. 
              The colour scaling is linear with white corresponding to the median background and black to the $0.01\%$ pixels with the highest intensity.  
           }
\end{figure}
\begin{figure}
   \centering
   \includegraphics[angle=0,width=8.50cm]{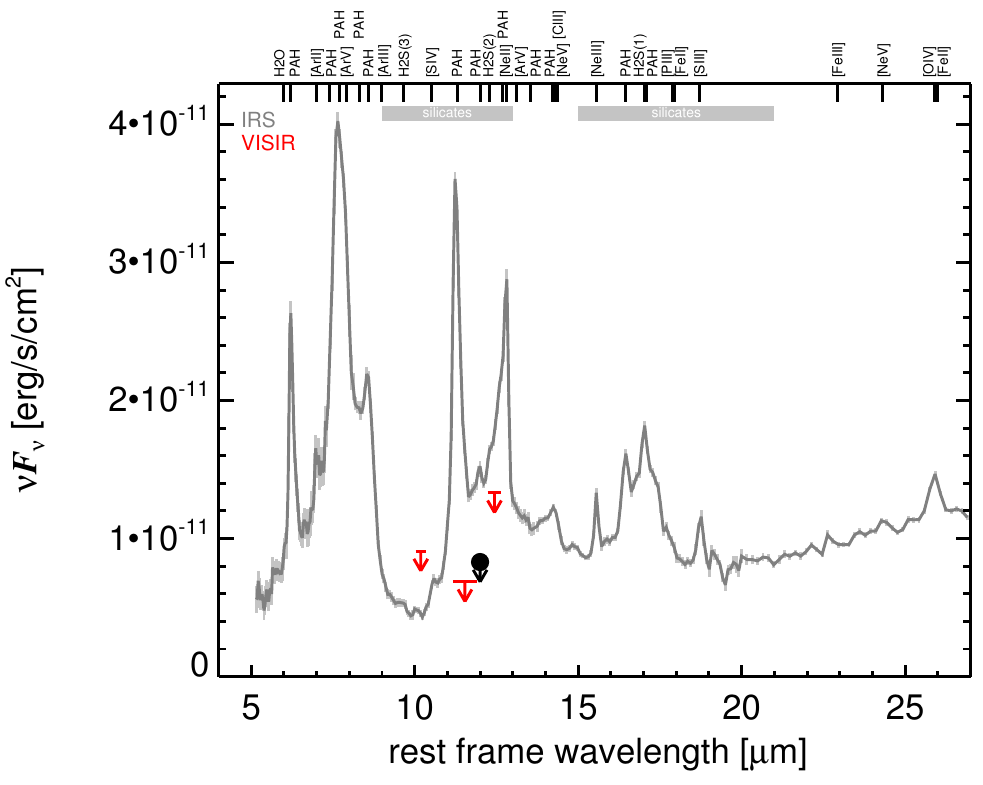}
   \caption{\label{fig:MISED_NGC0612}
      MIR SED of NGC\,612. The description  of the symbols (if present) is the following.
      Grey crosses and  solid lines mark the \spitzer/IRAC, MIPS and IRS data. 
      The colour coding of the other symbols is: 
      green for COMICS, magenta for Michelle, blue for T-ReCS and red for VISIR data.
      Darker-coloured solid lines mark spectra of the corresponding instrument.
      The black filled circles mark the nuclear 12 and $18\,\mu$m  continuum emission estimate from the data.
      The ticks on the top axis mark positions of common MIR emission lines, while the light grey horizontal bars mark wavelength ranges affected by the silicate 10 and 18$\mu$m features.}
\end{figure}
\clearpage

\twocolumn[\begin{@twocolumnfalse}  
\subsection{NGC\,613}\label{app:NGC0613}
NGC\,613 is a late-type spiral galaxy at a distance of $D=$ $ 26.6\pm 5.3$\,Mpc (NED redshift-independent median) with an active nucleus classified as an AGN/starburst composite \citep{veron_agns_1997} and a surrounding elliptical starburst ring (major axis$\sim 7\arcsec \sim900$\,pc; PA$\sim110\degree$; \citealt{hummel_central_1992}).
It shows a complex \oiii and radio morphology, possibly caused by an AGN (\citealt{hummel_central_1987}; see also \citealt{boker_sinfoni_2008}).
In addition, nuclear water maser emission was detected \citep{kondratko_discovery_2006}. 
After \iras, NGC\,613 was observed with \isoo \citep{roussel_atlas_2001} and \spitzer/IRS and MIPS.
In the  MIPS $24\,\mu$m image the nucleus is blended with the elliptical starburst ring. 
The IRS HR staring-mode spectrum is as well affected by the starburst ring and exhibits strong PAH emission and a red spectral slope in $\nu F_\nu$-space (see also \citealt{goulding_towards_2009}). 
NGC\,613 was observed with T-ReCS in the N filter in 2010 (unpublished, to our knowledge).
In this image, the starburst ring and a compact MIR nucleus are clearly resolved. 
The nucleus appears possibly marginally resolved (FWHM $\sim 0.39\arcsec \sim 50\,$pc) but this is not sufficient to robustly determined the nuclear subarcsecond MIR morphology.
The nuclear flux determined by T-ReCS photometry  is only $6\%$ of the \spitzerr spectrophotometry demonstrating the dominance of the starburst ring in the MIR. 
The detection of the compact MIR nucleus is another indication  for an AGN in NGC\,613 in agreement with the results of \cite{goulding_towards_2009}.
\newline\end{@twocolumnfalse}]

\begin{figure}
   \centering
   \includegraphics[angle=0,width=8.500cm]{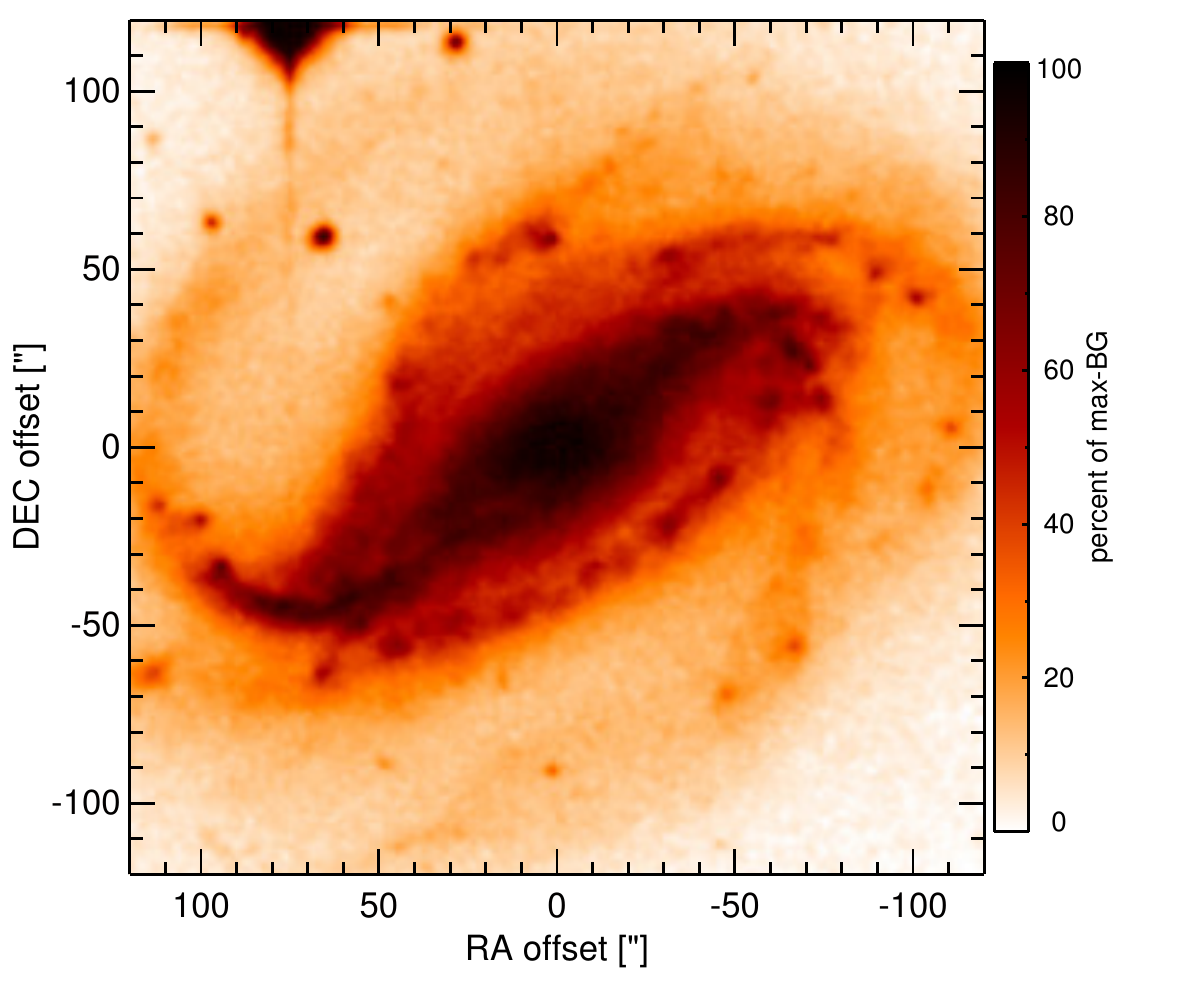}
    \caption{\label{fig:OPTim_NGC0613}
             Optical image (DSS, red filter) of NGC\,613. Displayed are the central $4\arcmin$ with North up and East to the left. 
              The colour scaling is linear with white corresponding to the median background and black to the $0.01\%$ pixels with the highest intensity.  
           }
\end{figure}
\begin{figure}
   \centering
   \includegraphics[angle=0,height=3.11cm]{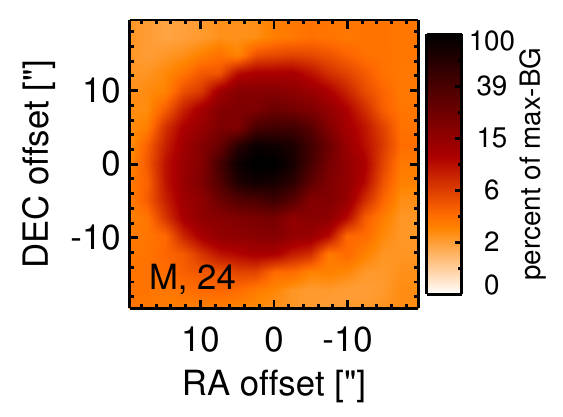}
    \caption{\label{fig:INTim_NGC0613}
             \spitzerr MIR images of NGC\,613. Displayed are the inner $40\arcsec$ with North up and East to the left. The colour scaling is logarithmic with white corresponding to median background and black to the $0.1\%$ pixels with the highest intensity.
             The label in the bottom left states instrument and central wavelength of the filter in $\mu$m (I: IRAC, M: MIPS). 
           }
\end{figure}
\begin{figure}
   \centering
   \includegraphics[angle=0,height=3.11cm]{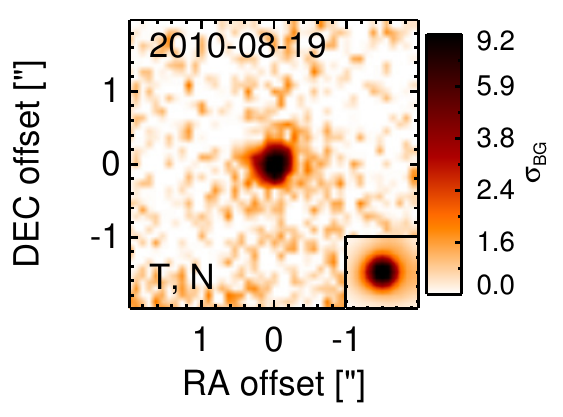}
    \caption{\label{fig:HARim_NGC0613}
             Subarcsecond-resolution MIR images of NGC\,613 sorted by increasing filter wavelength. 
             Displayed are the inner $4\arcsec$ with North up and East to the left. 
             The colour scaling is logarithmic with white corresponding to median background and black to the $75\%$ of the highest intensity of all images in units of $\sigbg$.
             The inset image shows the central arcsecond of the PSF from the calibrator star, scaled to match the science target.
             The labels in the bottom left state instrument and filter names (C: COMICS, M: Michelle, T: T-ReCS, V: VISIR).
           }
\end{figure}
\begin{figure}
   \centering
   \includegraphics[angle=0,width=8.50cm]{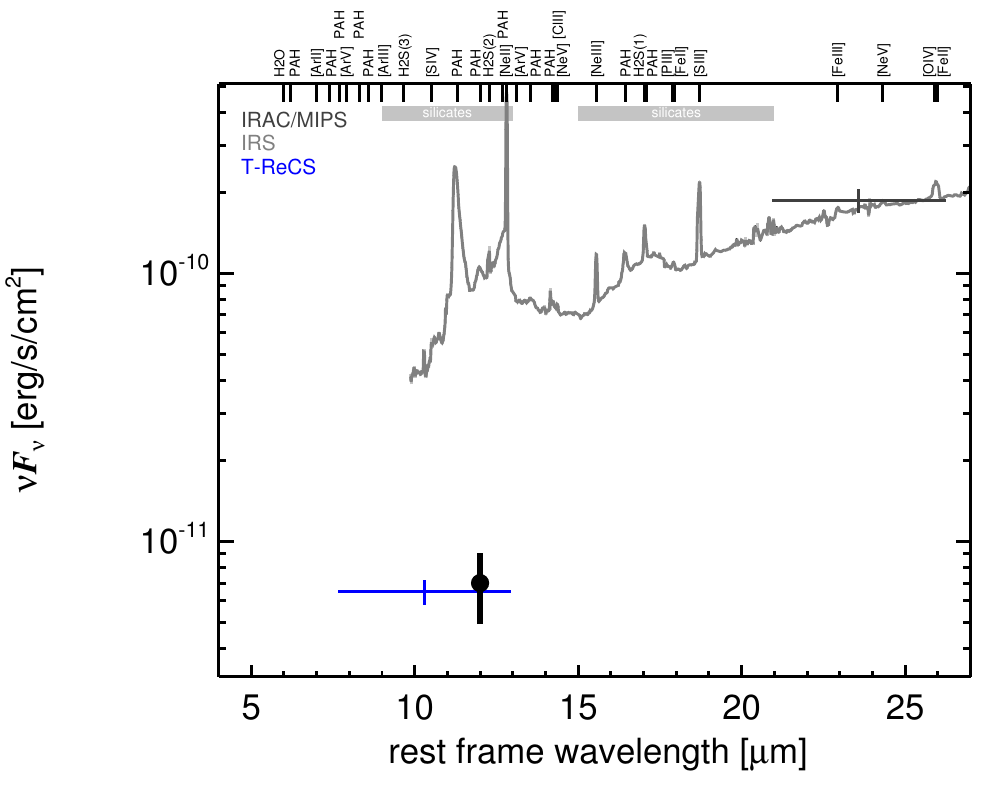}
   \caption{\label{fig:MISED_NGC0613}
      MIR SED of NGC\,613. The description  of the symbols (if present) is the following.
      Grey crosses and  solid lines mark the \spitzer/IRAC, MIPS and IRS data. 
      The colour coding of the other symbols is: 
      green for COMICS, magenta for Michelle, blue for T-ReCS and red for VISIR data.
      Darker-coloured solid lines mark spectra of the corresponding instrument.
      The black filled circles mark the nuclear 12 and $18\,\mu$m  continuum emission estimate from the data.
      The ticks on the top axis mark positions of common MIR emission lines, while the light grey horizontal bars mark wavelength ranges affected by the silicate 10 and 18$\mu$m features.}
\end{figure}
\clearpage

\twocolumn[\begin{@twocolumnfalse}  
\subsection{NGC\,676}\label{app:NGC0676}
NGC\,676 is an edge-on spiral galaxy at a distance of $D=$ $19.5 \pm 8.7$\,Mpc \citep{tully_nearby_1988} with a Sy\,2 nucleus \citep{veron-cetty_catalogue_2010}, which is undetected at radio wavelengths \citep{ho_detection_2001}. 
No \spitzerr observations are available for NGC\,676.
In the \wisee images, it appears elliptically elongated along the galaxy major axis.
We observed NGC\,676 with VISIR in PAH2 and NEII\_1 in 2009 and in PAH2 and PAH2\_2 in 2010 \citep{asmus_mid-infrared_2011} but the nucleus remained undetected in all cases.
The foreground star 5\arcsec south of the galaxy centres allows us to derive an accurate upper limit for the nuclear emission in NGC\,676. 
This non-detection does not exclude the presence of an AGN because the upper limit on the flux is consistent with the one expected from the MIR--X-ray luminosity correlation (see \citealt{asmus_mid-infrared_2011}).
\newline\end{@twocolumnfalse}]

\begin{figure}
   \centering
   \includegraphics[angle=0,width=8.500cm]{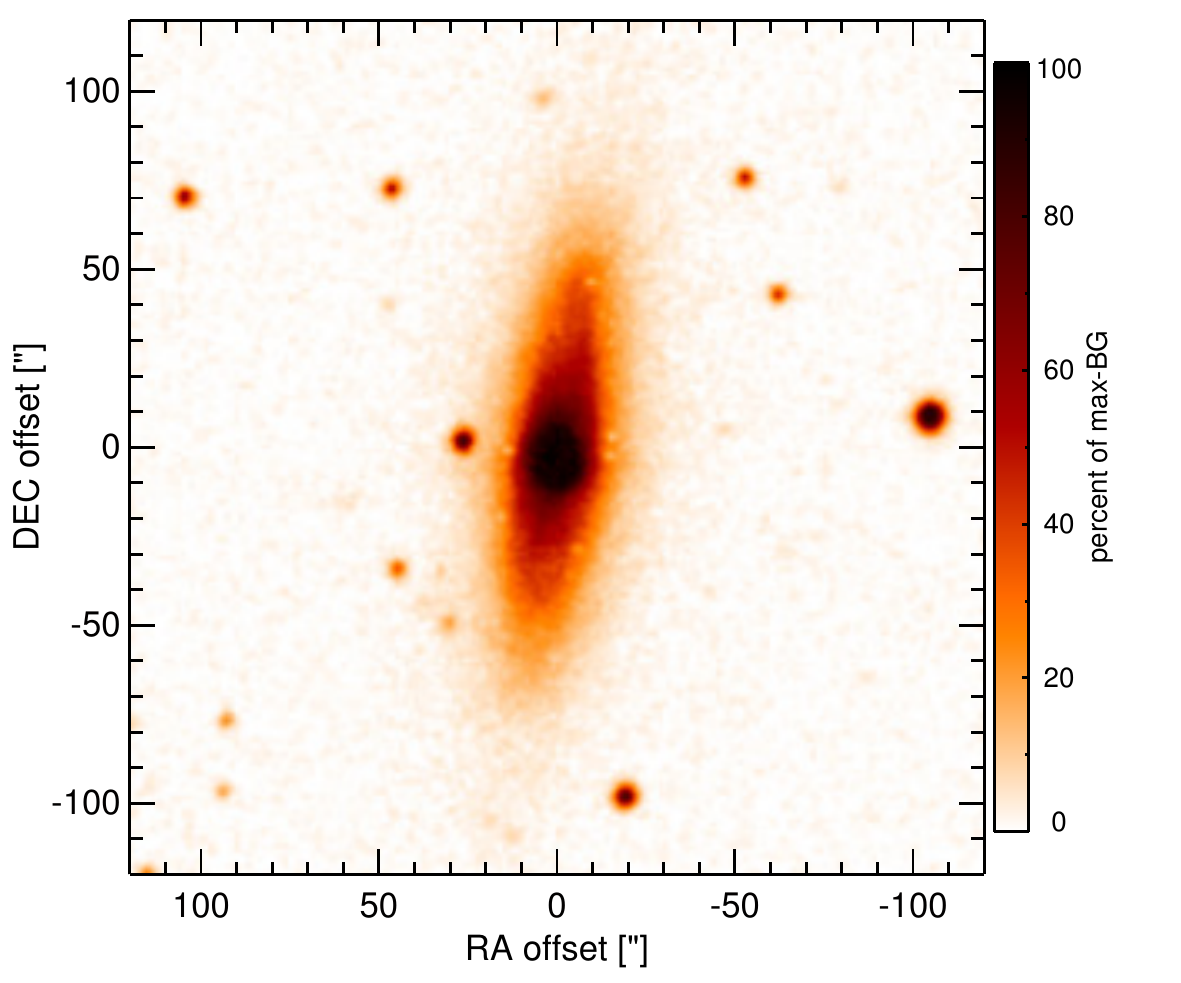}
    \caption{\label{fig:OPTim_NGC0676}
             Optical image (DSS, red filter) of NGC\,676. Displayed are the central $4\arcmin$ with North up and East to the left. 
              The colour scaling is linear with white corresponding to the median background and black to the $0.01\%$ pixels with the highest intensity.  
           }
\end{figure}
\begin{figure}
   \centering
   \includegraphics[angle=0,width=8.50cm]{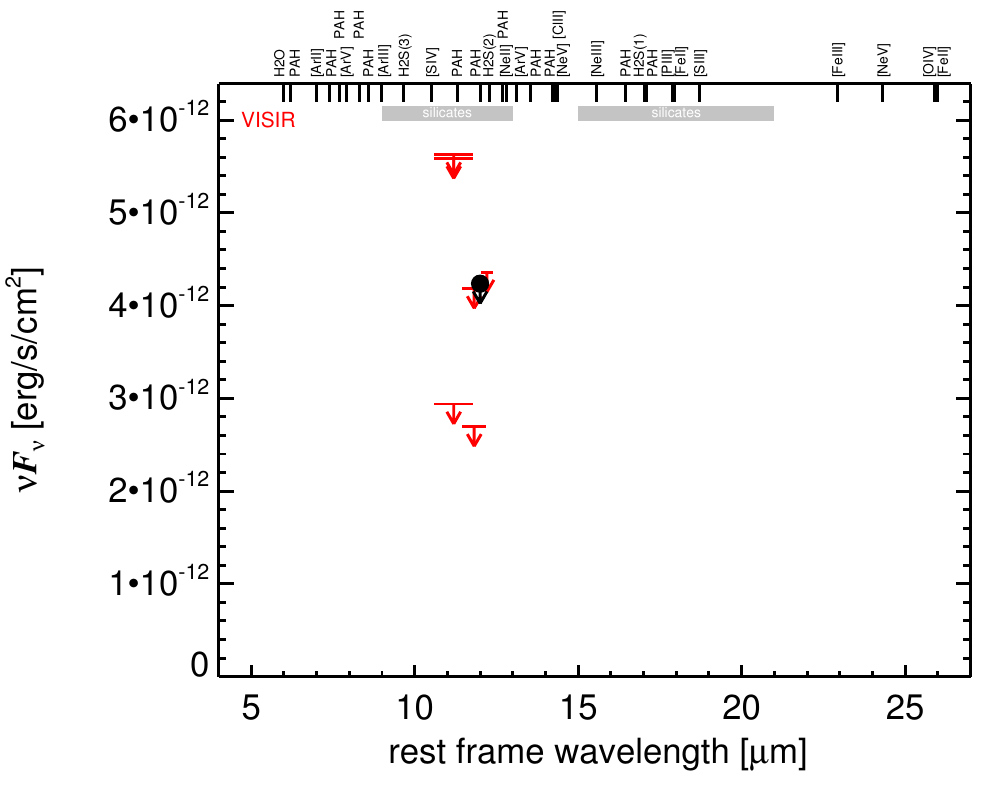}
   \caption{\label{fig:MISED_NGC0676}
      MIR SED of NGC\,676. The description  of the symbols (if present) is the following.
      Grey crosses and  solid lines mark the \spitzer/IRAC, MIPS and IRS data. 
      The colour coding of the other symbols is: 
      green for COMICS, magenta for Michelle, blue for T-ReCS and red for VISIR data.
      Darker-coloured solid lines mark spectra of the corresponding instrument.
      The black filled circles mark the nuclear 12 and $18\,\mu$m  continuum emission estimate from the data.
      The ticks on the top axis mark positions of common MIR emission lines, while the light grey horizontal bars mark wavelength ranges affected by the silicate 10 and 18$\mu$m features.}
\end{figure}
\clearpage

\twocolumn[\begin{@twocolumnfalse}  
\subsection{NGC\,788}\label{app:NGC0788}
NGC\,788 is a low-inclination spiral galaxy at a redshift of $z=$ 0.0136 ($D\sim$57.2\,Mpc) with a Sy\,2 nucleus with polarized broad emission lines \citep{kay_hidden_1998}.
The AGN was discovered by \cite{huchra_new_1982} and belongs to the nine-month BAT AGN sample..
At radio wavelengths the nucleus is slightly elongated (PA$\sim62\degree$; \citealt{nagar_radio_1999}).
It also possesses visible ionization cones \citep{martini_circumnuclear_2003}.
NGC\,788 has observed with \spitzer/IRAC, IRS and MIPS and appears as a compact MIR nucleus.
In the IRAC $8.0\,\mu$m image the host galaxy is weakly detected as well.
The IRS LR staring-mode spectrum exhibits silicate  $10\,\mu$m absorption, weak PAH emission and an emission peak at $\sim 18\,\mu$m (see also \citealt{deo_mid-infrared_2009,weaver_mid-infrared_2010}).
NGC\,788 was observed with VISIR in five $N$-band filters in 2008,  and an unresolved MIR nucleus was detected in all cases (unpublished, to our knowledge).
The nuclear VISIR photometry provides fluxes consistent with the \spitzerr spectrophotometry.
\newline\end{@twocolumnfalse}]

\begin{figure}
   \centering
   \includegraphics[angle=0,width=8.500cm]{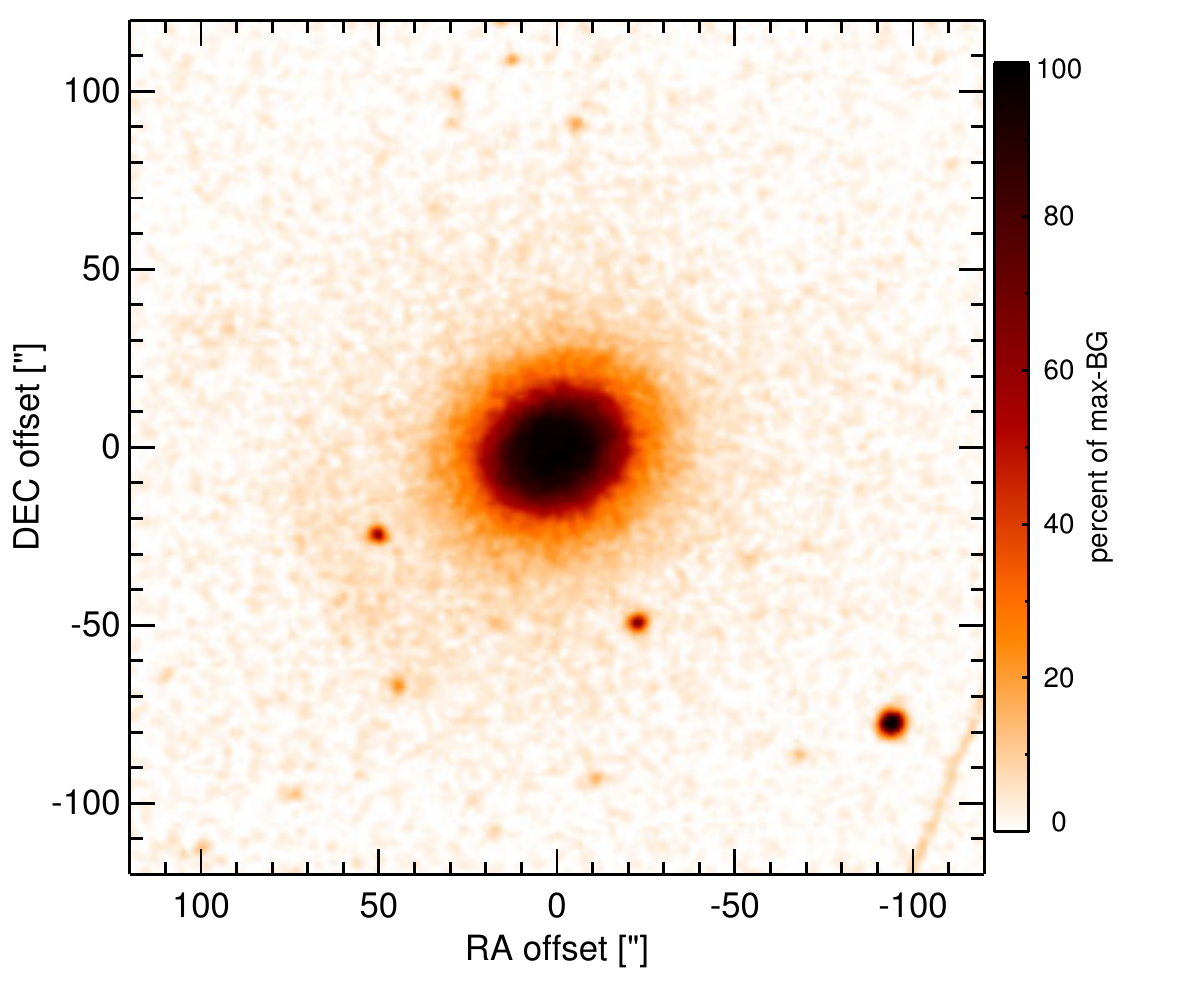}
    \caption{\label{fig:OPTim_NGC0788}
             Optical image (DSS, red filter) of NGC\,788. Displayed are the central $4\arcmin$ with North up and East to the left. 
              The colour scaling is linear with white corresponding to the median background and black to the $0.01\%$ pixels with the highest intensity.  
           }
\end{figure}
\begin{figure}
   \centering
   \includegraphics[angle=0,height=3.11cm]{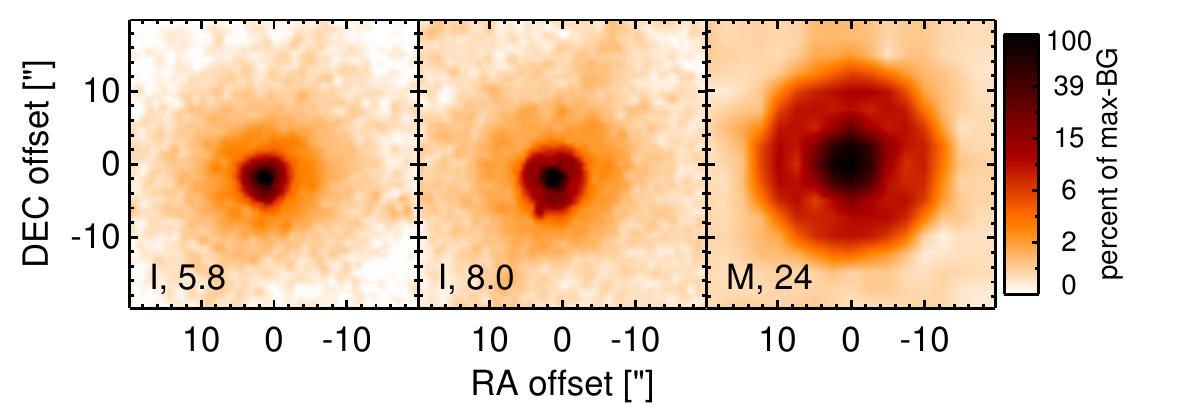}
    \caption{\label{fig:INTim_NGC0788}
             \spitzerr MIR images of NGC\,788. Displayed are the inner $40\arcsec$ with North up and East to the left. The colour scaling is logarithmic with white corresponding to median background and black to the $0.1\%$ pixels with the highest intensity.
             The label in the bottom left states instrument and central wavelength of the filter in $\mu$m (I: IRAC, M: MIPS). 
             Note that the apparent off-nuclear compact source in the IRAC $8.0\,\mu$m image is an instrumental artefact.
           }
\end{figure}
\begin{figure}
   \centering
   \includegraphics[angle=0,width=8.500cm]{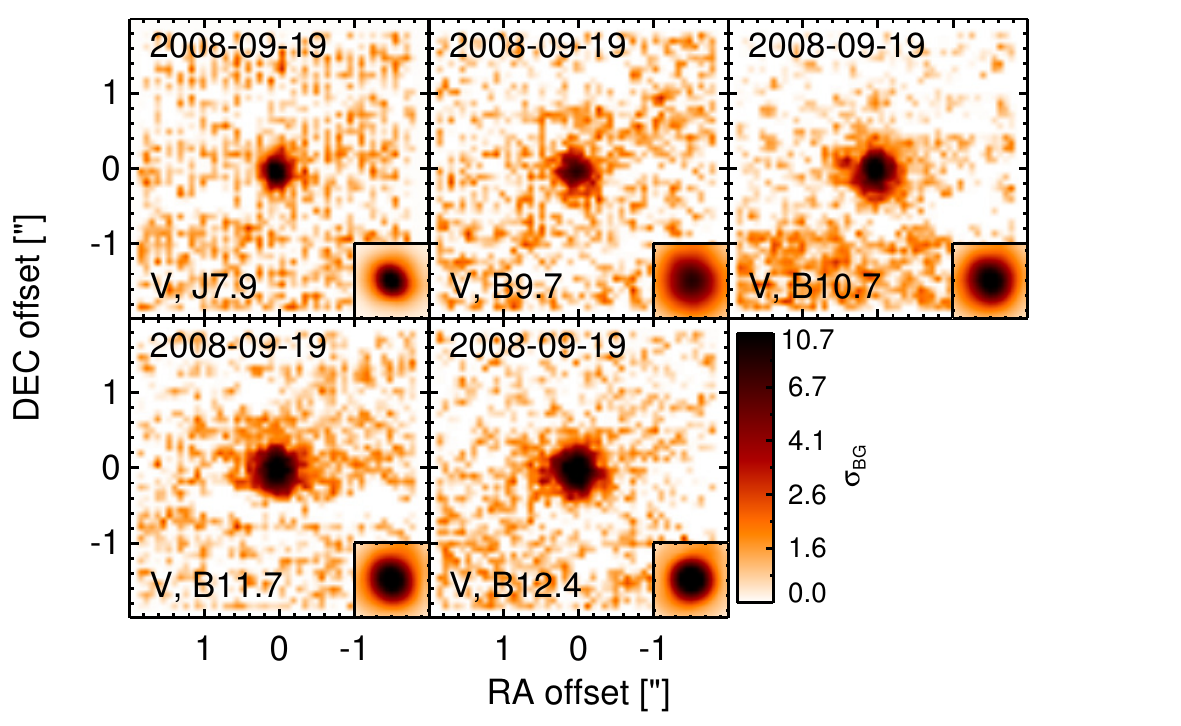}
    \caption{\label{fig:HARim_NGC0788}
             Subarcsecond-resolution MIR images of NGC\,788 sorted by increasing filter wavelength. 
             Displayed are the inner $4\arcsec$ with North up and East to the left. 
             The colour scaling is logarithmic with white corresponding to median background and black to the $75\%$ of the highest intensity of all images in units of $\sigbg$.
             The inset image shows the central arcsecond of the PSF from the calibrator star, scaled to match the science target.
             The labels in the bottom left state instrument and filter names (C: COMICS, M: Michelle, T: T-ReCS, V: VISIR).
           }
\end{figure}
\begin{figure}
   \centering
   \includegraphics[angle=0,width=8.50cm]{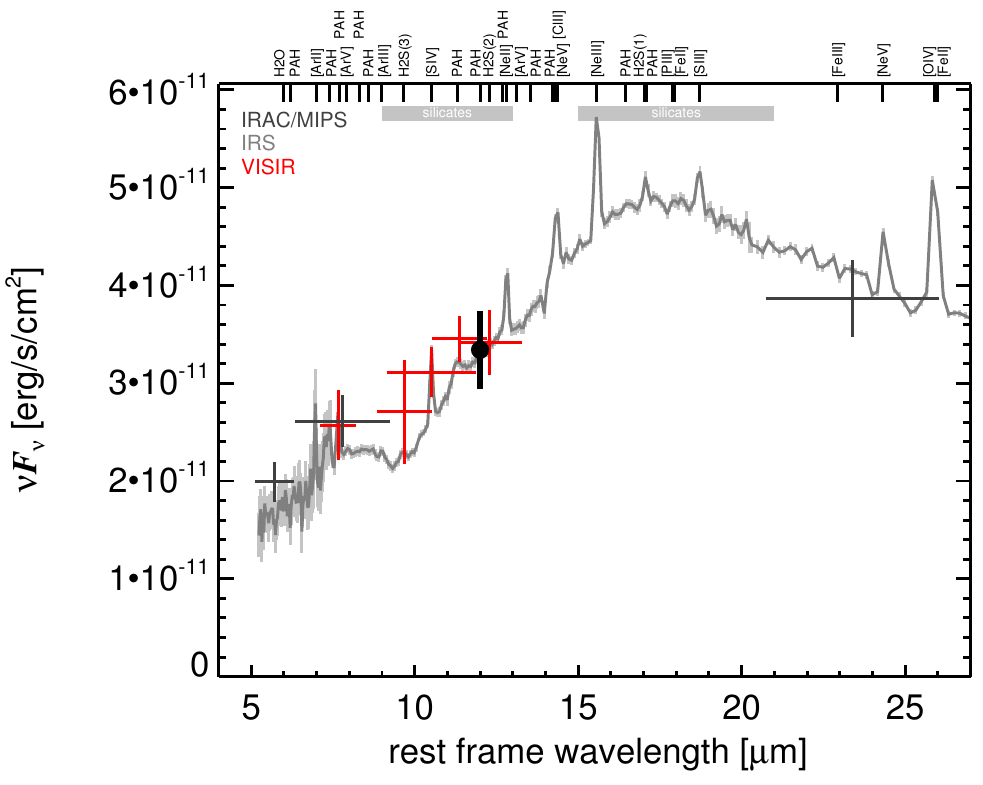}
   \caption{\label{fig:MISED_NGC0788}
      MIR SED of NGC\,788. The description  of the symbols (if present) is the following.
      Grey crosses and  solid lines mark the \spitzer/IRAC, MIPS and IRS data. 
      The colour coding of the other symbols is: 
      green for COMICS, magenta for Michelle, blue for T-ReCS and red for VISIR data.
      Darker-coloured solid lines mark spectra of the corresponding instrument.
      The black filled circles mark the nuclear 12 and $18\,\mu$m  continuum emission estimate from the data.
      The ticks on the top axis mark positions of common MIR emission lines, while the light grey horizontal bars mark wavelength ranges affected by the silicate 10 and 18$\mu$m features.}
\end{figure}
\clearpage

\twocolumn[\begin{@twocolumnfalse}  
\subsection{NGC\,985 -- Mrk\,1048}\label{app:NGC0985}
NGC\,985 is a peculiar ring-like late-stage merger system, at a redshift of $z=$ 0.0431 ($D\sim195\,$Mpc).
The two nuclei have a separation of $3.5\arcsec$ ($\sim 3$\,kpc; PA$\sim 110\degree$; \citealt{appleton_infrared_1993}).
The eastern nucleus is classified a Sy\,1.5 \citep{veron-cetty_catalogue_2010}, belonging to the nine-month BAT AGN sample.
After \iras, $N$-band photometry with MMT \citep{maiolino_new_1995} and MIR imaging with \isoo \citep{appleton_mid-infrared_2002} was obtained.
NGC\,985 was also observed with \spitzer/IRAC and IRS, and both nuclei are blended in the IRAC images, resulting in a slightly elongated compact MIR source with part of the ring-like host structure also visible.
The IRS LR staring-mode spectrum exhibits possibly weak silicate  10 and $18\,\mu$m emission,  PAH emission and an emission peak at $\sim 17\,\mu$m (see also \citealt{mullaney_defining_2011}).
NGC\,985 was observed with VISIR in five $N$-band filters in 2008 (unpublished, to our knowledge), and only one unresolved nucleus was detected in all images but B9.7, where the nucleus appears slightly extended. 
The nuclear VISIR photometry is consistent with the \spitzerr spectrophotometry, which indicates that the AGN dominates the MIR emission in the central 3\,kpc of NGC\,985.
\newline\end{@twocolumnfalse}]

\begin{figure}
   \centering
   \includegraphics[angle=0,width=8.500cm]{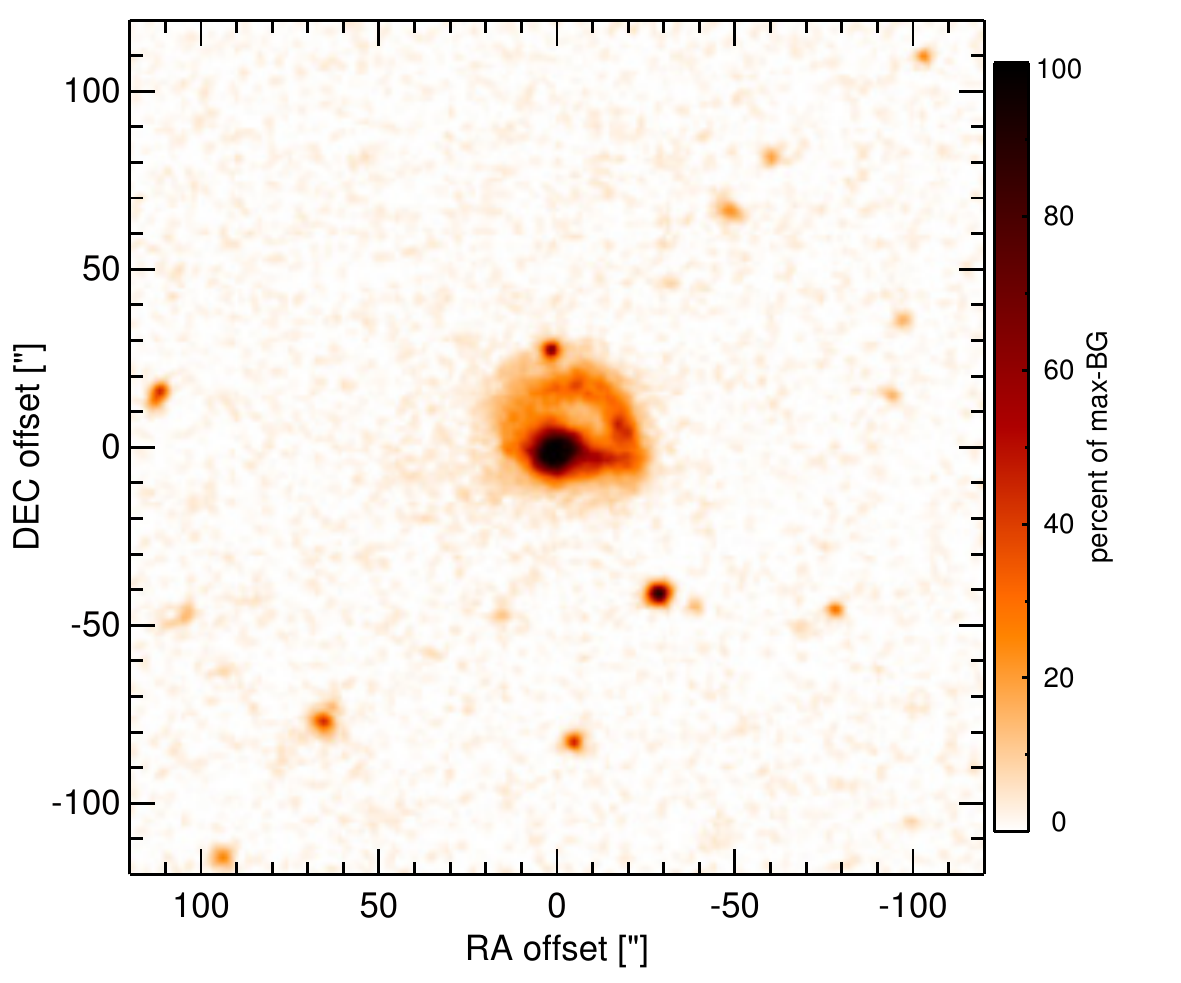}
    \caption{\label{fig:OPTim_NGC0985}
             Optical image (DSS, red filter) of NGC\,985. Displayed are the central $4\arcmin$ with North up and East to the left. 
              The colour scaling is linear with white corresponding to the median background and black to the $0.01\%$ pixels with the highest intensity.  
           }
\end{figure}
\begin{figure}
   \centering
   \includegraphics[angle=0,height=3.11cm]{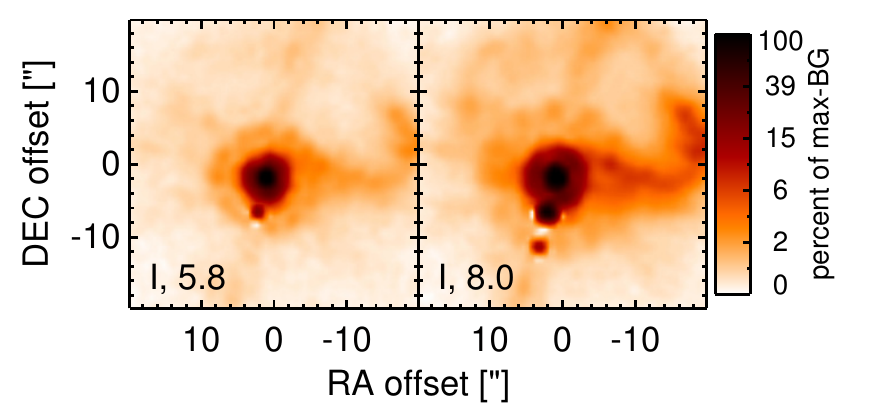}
    \caption{\label{fig:INTim_NGC0985}
             \spitzerr MIR images of NGC\,985. Displayed are the inner $40\arcsec$ with North up and East to the left. The colour scaling is logarithmic with white corresponding to median background and black to the $0.1\%$ pixels with the highest intensity.
             The label in the bottom left states instrument and central wavelength of the filter in $\mu$m (I: IRAC, M: MIPS). 
             Note that the apparent off-nuclear compact sources in the IRAC 5.8 and $8.0\,\mu$m images are instrumental artefacts.
           }
\end{figure}
\begin{figure}
   \centering
   \includegraphics[angle=0,width=8.500cm]{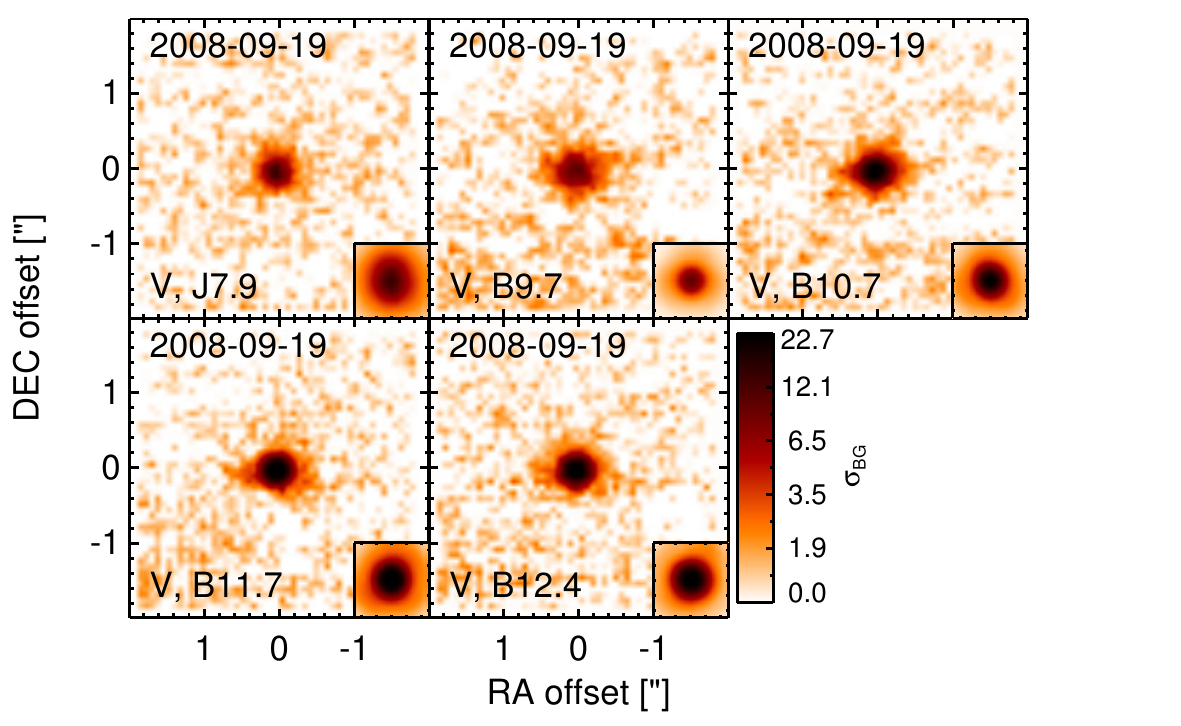}
    \caption{\label{fig:HARim_NGC0985}
             Subarcsecond-resolution MIR images of NGC\,985 sorted by increasing filter wavelength. 
             Displayed are the inner $4\arcsec$ with North up and East to the left. 
             The colour scaling is logarithmic with white corresponding to median background and black to the $75\%$ of the highest intensity of all images in units of $\sigbg$.
             The inset image shows the central arcsecond of the PSF from the calibrator star, scaled to match the science target.
             The labels in the bottom left state instrument and filter names (C: COMICS, M: Michelle, T: T-ReCS, V: VISIR).
           }
\end{figure}
\begin{figure}
   \centering
   \includegraphics[angle=0,width=8.50cm]{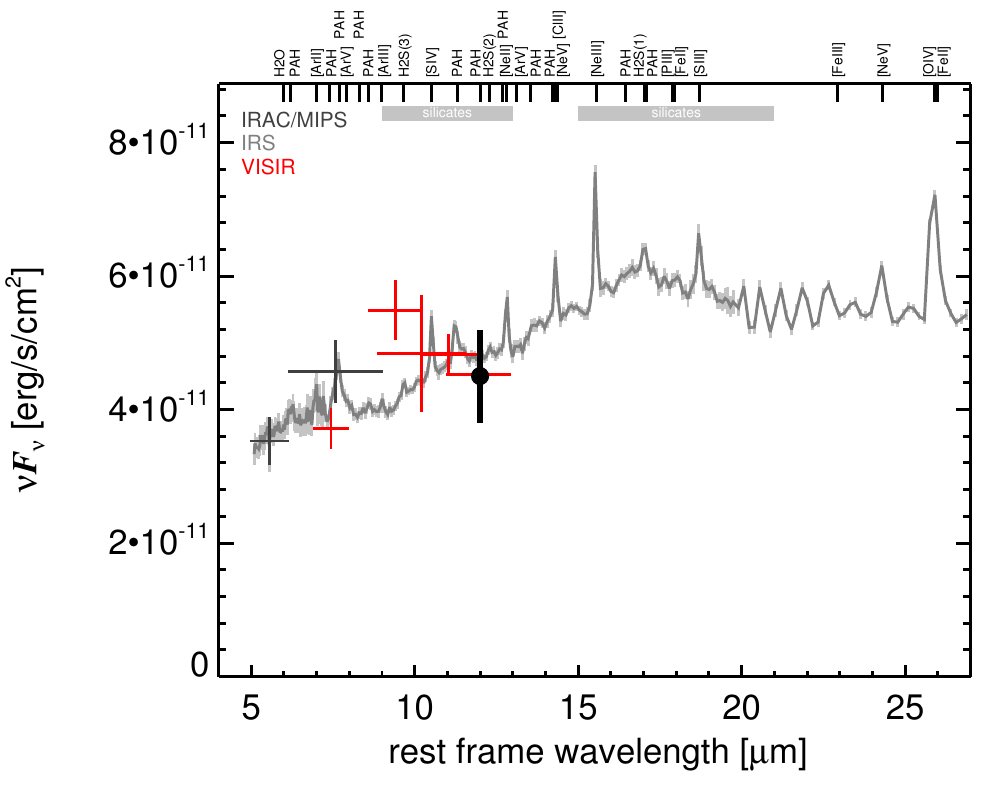}
   \caption{\label{fig:MISED_NGC0985}
      MIR SED of NGC\,985. The description  of the symbols (if present) is the following.
      Grey crosses and  solid lines mark the \spitzer/IRAC, MIPS and IRS data. 
      The colour coding of the other symbols is: 
      green for COMICS, magenta for Michelle, blue for T-ReCS and red for VISIR data.
      Darker-coloured solid lines mark spectra of the corresponding instrument.
      The black filled circles mark the nuclear 12 and $18\,\mu$m  continuum emission estimate from the data.
      The ticks on the top axis mark positions of common MIR emission lines, while the light grey horizontal bars mark wavelength ranges affected by the silicate 10 and 18$\mu$m features.}
\end{figure}
\clearpage

\twocolumn[\begin{@twocolumnfalse}  
\subsection{NGC\,1052 -- PKS\,0238-084}\label{app:NGC1052}
NGC\,1052 is a dusty elliptical galaxy at a distance of $D=$ $19.4\pm1.5$\,Mpc (NED redshift-independent median) with a radio-loud LINER nucleus with broad polarized lines, a biconical jet (diameter$\sim3$\,pc; PA$\sim66\degree$; \citealt{kellermann_sub-milliarcsecond_1998, vermeulen_shroud_2003}) and a flat radio spectrum showing variability by a factor of two in flux  \citep{tornikoski_high_2000}.
A subparsec-scale dense plasma torus and a water maser around the nucleus have been detected with VLBI radio observations \citep{braatz_discovery_1994,kameno_dense_2001,kadler_twin-jet_2004,sawada-satoh_positional_2008}.
NGC\,1052 is commonly regarded as the prototypical LINER and has been studied extensively, in particular at radio wavelengths.
In the MIR, it was first observed by \cite{kleinmann_observations_1970}, \cite{rieke_infrared_1972}, \cite{rieke_nonthermal_1982}, \cite{becklin_infrared_1982}, \cite{willner_infrared_1985}, \cite{impey_infrared_1986}, and \cite{roche_atlas_1991}.
In addition, NGC\,1052 was also observed with \iso \citep{sugai_mid-infrared_2000,malhotra_probing_2000,xilouris_dust_2004} and \spitzer/IRAC and IRS \citep{tang_infrared-red_2009,dudik_spitzer_2009}.
In the  IRAC $5.8$ and $8.0\,\mu$m images, a compact MIR nucleus embedded within diffuse host emission was detected.
Because we measure the nuclear component only, our IRAC fluxes are significantly lower than the total fluxes in \cite{tang_infrared-red_2009}.
The IRS LR staring-mode spectrum exhibits silicate  10 and $18\,\mu$m emission, a weak PAH 11.3\,$\mu$m feature and an emission peak at $
\sim 18\,\mu$m (see also \citealt{mason_nuclear_2012}). 
NGC\,1052 was observed with T-ReCS in the Si2 and Qa filters in 2007 \citep{mason_nuclear_2012}, and with VISIR in six $N$-band filters in 2009 (unpublished, to our knowledge).
In addition, we observed it also with VISIR in PAH1 and Q2 in 2010.
A compact MIR nucleus without any other emission was detected in all images.
Compared to the corresponding standard star, the nucleus appears resolved in the B10.7, B12.4, PAH1 and Si2 images but not in the others. 
Apart from the Qa one, none of the observations was performed under diffraction-limited conditions and the obtained FWHMs and PAs are inconsistent. 
Therefore, it remains uncertain, whether the nucleus of NGC\,1052 is in general resolved at subarcsecond resolution in the MIR. 
The nuclear photometry is on average $\sim 25\%$ lower than the \spitzerr spectrophotometry, which is indicates that the latter is still affected by diffuse host emission.
This extended emission has affected the historical MIR photometry on various levels and thus complicates any comparison of the individual measurements
However, we note that in general larger aperture measurements provide higher flux, while comparable apertures yield comparable fluxes.
It seems therefore that NGC\,1052 has not significantly changed it MIR brightness over the last 40\,years. 
The general MIR SED shape suggests that the silicate emission is originating from the projected subarcsecond-scale nucleus.
Thus, warm significant dust is present in the projected central $\sim 40$\,pc of NGC\,1052. 
\newline\end{@twocolumnfalse}]

\begin{figure}
   \centering
   \includegraphics[angle=0,width=8.500cm]{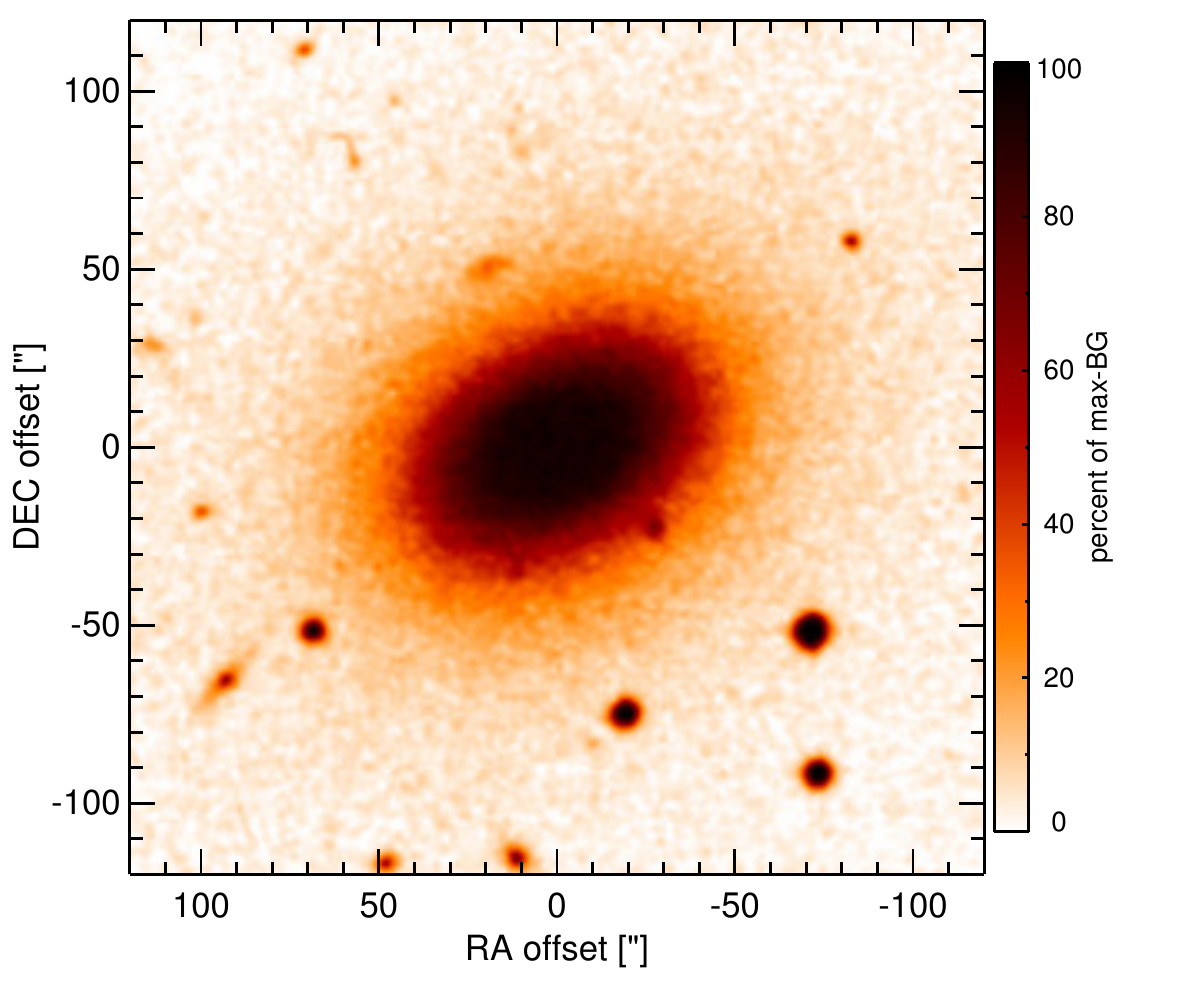}
    \caption{\label{fig:OPTim_NGC1052}
             Optical image (DSS, red filter) of NGC\,1052. Displayed are the central $4\arcmin$ with North up and East to the left. 
              The colour scaling is linear with white corresponding to the median background and black to the $0.01\%$ pixels with the highest intensity.  
           }
\end{figure}
\begin{figure}
   \centering
   \includegraphics[angle=0,height=3.11cm]{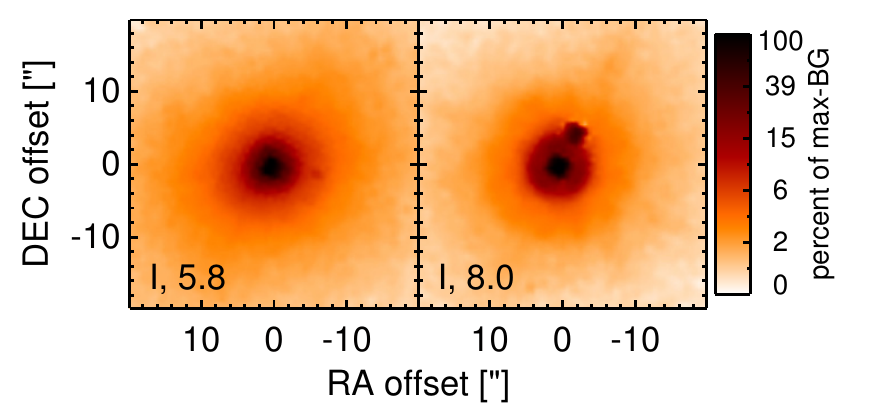}
    \caption{\label{fig:INTim_NGC1052}
             \spitzerr MIR images of NGC\,1052. Displayed are the inner $40\arcsec$ with North up and East to the left. The colour scaling is logarithmic with white corresponding to median background and black to the $0.1\%$ pixels with the highest intensity.
             The label in the bottom left states instrument and central wavelength of the filter in $\mu$m (I: IRAC, M: MIPS).
             Note that the apparent off-nuclear compact source in the IRAC $8.0\,\mu$m image is an instrumental artefact.
           }
\end{figure}
\begin{figure}
   \centering
   \includegraphics[angle=0,width=8.500cm]{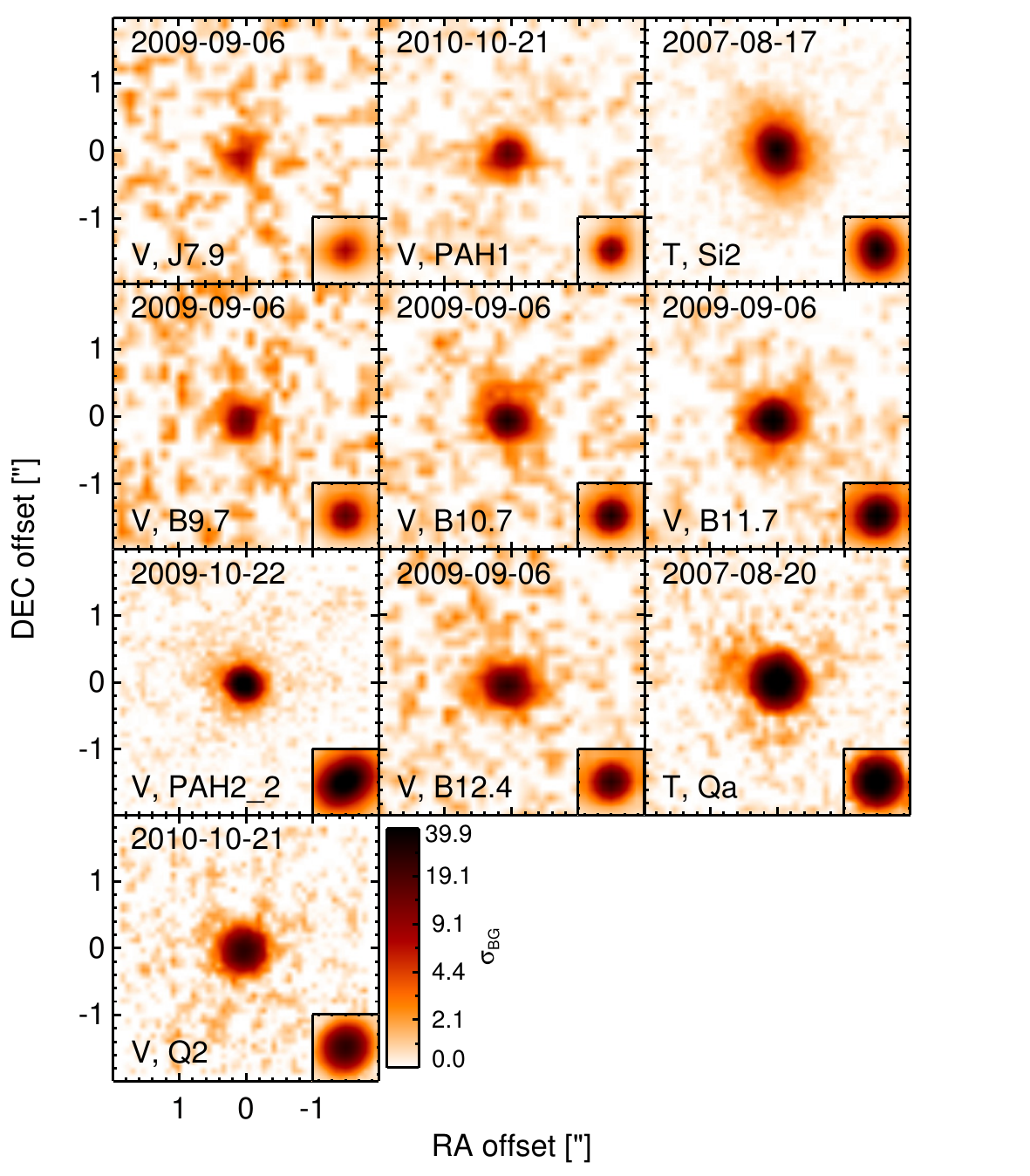}
    \caption{\label{fig:HARim_NGC1052}
             Subarcsecond-resolution MIR images of NGC\,1052 sorted by increasing filter wavelength. 
             Displayed are the inner $4\arcsec$ with North up and East to the left. 
             The colour scaling is logarithmic with white corresponding to median background and black to the $75\%$ of the highest intensity of all images in units of $\sigbg$.
             The inset image shows the central arcsecond of the PSF from the calibrator star, scaled to match the science target.
             The labels in the bottom left state instrument and filter names (C: COMICS, M: Michelle, T: T-ReCS, V: VISIR).
           }
\end{figure}
\begin{figure}
   \centering
   \includegraphics[angle=0,width=8.50cm]{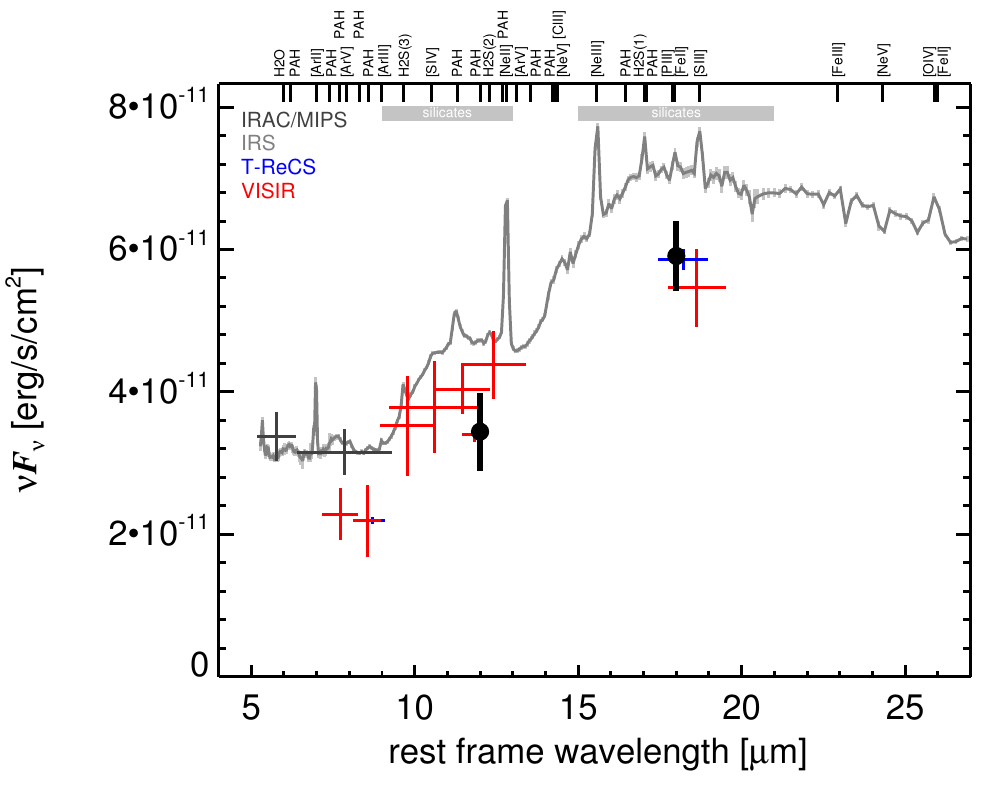}
   \caption{\label{fig:MISED_NGC1052}
      MIR SED of NGC\,1052. The description  of the symbols (if present) is the following.
      Grey crosses and  solid lines mark the \spitzer/IRAC, MIPS and IRS data. 
      The colour coding of the other symbols is: 
      green for COMICS, magenta for Michelle, blue for T-ReCS and red for VISIR data.
      Darker-coloured solid lines mark spectra of the corresponding instrument.
      The black filled circles mark the nuclear 12 and $18\,\mu$m  continuum emission estimate from the data.
      The ticks on the top axis mark positions of common MIR emission lines, while the light grey horizontal bars mark wavelength ranges affected by the silicate 10 and 18$\mu$m features.}
\end{figure}
\clearpage

\twocolumn[\begin{@twocolumnfalse}  
\subsection{NGC\,1068 -- M77 -- 3C\,71}\label{app:NGC1068}
NGC\,1068 is a face-on spiral-like luminous infrared galaxy at a distance $14.4 \pm 3.0 $\,Mpc (NED redshift-independent median) with an AGN optically classified either as Sy\,1.8-2   \citep{ho_search_1997-1,ho_search_1997,osterbrock_spectroscopic_1993} and has polarized broad emission lines \citep{antonucci_spectropolarimetry_1985}.
NGC\,1068 is one of the original six Seyfert galaxies \citep{seyfert_nuclear_1943} and thus its AGN  is one of the most-studied at all wavelengths (see \citealt{honig_high-spatial_2008} for a recent detailed multiwavelength study), and the discovery of polarized broad emission lines in NGC\,1068 led to the formulation of the unification scheme for AGN \citep{antonucci_spectropolarimetry_1985}.
It possesses a cone-like NLR in northern direction (length$\sim 10\arcsec \sim 700$\,pc; PA$\sim 35\degree$; \citealt{pogge_extended_1988,evans_hst_1991}) roughly coinciding with a bended radio jet \citep{wilson_radio_1982,gallimore_subarcsecond_1996-1}, and nuclear H$_2$O and OH maser emission perpendicularly orientated to the jet (PA$\sim94\degree$; \citealt{claussen_water-vapor_1984,greenhill_vlbi_1996,gallimore_h_1996}).
Intense circum-nuclear star formation out to radii of $\sim 1$\,kpc was detected as well \citep{telesco_luminous_1984,balick_footprints_1985}.

NGC\,1068 possesses the MIR-brightest active nucleus in terms of flux.
The first MIR photometry of NGC\,1068 was obtained by \cite{kleinmann_infrared_1970}.
Since then a wealth of studies employing MIR photometry and spectroscopy with bolometers, including mapping observations, were performed \citep{kleinmann_infrared_1970,neugebauer_infrared_1971,rieke_variability_1972,rieke_infrared_1972,becklin_size_1973,jameson_infrared_1974,jameson_infrared_1974-1,stein_observations_1974,rieke_infrared_1975,kleinmann_8-13_1976,lebofsky_infrared_1978,rieke_10_1978,lebofsky_extinction_1979,telesco_extended_1980,houck_medium-resolution_1980,malkan_stellar_1983,telesco_luminous_1984,roche_8-13_1984}.
NGC 1068 was also observed by all space-based MIR facilities: e.g., \isoo (e.g., \citealt{rigopoulou_large_1999,lutz_iso-sws_2000,alexander_infrared_2000,le_floch_mid-infrared_2001,siebenmorgen_isocam_2004}) and \spitzerr (e.g., \citealt{goulding_towards_2009}).
The first ground-based $N$-band images were obtained by \cite{tresch-fienberg_structure_1987} with the Goddard IR array camera mounted on IRTF, revealing an extended complex structure, which was later resolved further with the first subarcsecond resolution image obtained with an upgraded Goddard Infrared Array Camera (\citealt{braatz_high-resolution_1993}; see also \citealt{cameron_subarcsecond_1993}).
Further improved MIR images with better instruments and larger telescopes followed \citep{bock_high-resolution_1998,bock_high_2000,alloin_0.6_2000, siebenmorgen_mid-infrared_2004,gorjian_10_2004,galliano_mid-infrared_2005,poncelet_original_2007}.
In addition, spectroscopic \citep{rhee_first_2006,mason_spatially_2006,poncelet_dynamics_2008} and polarimetric 
\citep{knacke_infrared_1974,aitken_infrared_1984,lumsden_near-_1999,packham_gemini_2007} studies were dedicated to the nuclear region of NGC\,1068.
Finally, MIR interferometric observations with MIDI were performed as well \citep{jaffe_central_2004,poncelet_new_2006,raban_resolving_2009}.

We reanalysed all the eleven conventional VISIR $N-$ and $Q$-band filter images, which were obtained in 2004 \citep{galliano_mid-infrared_2005-1}, in 2005 \citep{poncelet_dynamics_2008}, and  in 2008 (this work).
Some of the major results of all the previous MIR studies and our own can be summarized as follows. 
A compact north-south elongated MIR nucleus embedded within extended emission was detected in all subarcsecond-resolution images.
From the core, a thin structure extends $\sim 0.5\arcsec$ (35\,pc) to the north with a PA$\sim-7\degree$ before it bends to the east with a PA$\sim20\degree$ extending another $\sim1.3\arcsec$ (90\,pc). 
This structure follows the radio jet along the western side and coincides with the western part of the bright ionization cone emission.
In addition an emission clump is located east of this structure at a distance of $\sim 0.6\arcsec$ (42\,pc; PA$\sim30\degree$) coinciding with an eastern bright \oiii emission region.
To the south, another thin structure extends with a PA$\sim190\degree$ for $\sim1.3\arcsec$ (90\,pc).
The southern structure coincides as well with \oiii emission while there is no radio counterpart.
In particular in the deep VISIR $12\,\mu$m images, additional but weak emission knots can be seen at larger distances ($\sim 3.3\arcsec \sim 230\,$pc) towards the north-east and south-west along the same position angles and also aligned with \oiii emission.
Therefore, it seems that the non-nuclear MIR emission in general coincides with bright \oiii clouds while bracketing the radio emission.
The presence of the silicate feature and MIR polarization due to dichroic emission indicates that the MIR emission is caused by dust \citep{packham_gemini_2007}.
The MIR emission line kinematics indicate that the material is outflowing along the edges of the ionization cone \citep{poncelet_dynamics_2008}.
It remains uncertain which parts of the MIR emission structure are actually associated with the obscuring torus, but most works favour that it is unresolved in the subarcsecond-resolution images as in other local AGN.
Indeed, the extensive MIR interferometric observations with MIDI have found a parsec-sized structure elongated along the system axis \citep{raban_resolving_2009}.
We isolate only the unresolved nuclear component in NGC\,1068 by manual PSF scaling in order to leave a roughly smooth residual.
This provides fluxes much lower than the total MIR emission on arcsecond scales, e.g., on average $\sim80$ lower than the \spitzerr/IRS HR staring-mode spectrum.
As expected our fluxes are in between the nuclear Michelle $N$-band spectrum of \citeauthor{mason_spatially_2006} (2006; dark-magenta in Fig.~\ref{fig:MISED_NGC1068}) and the MIDI correlated flux spectrum of \cite{raban_resolving_2009}.
The nuclear photometry also indicates a deep silicate $10\,\mu$m absorption feature and a declining flux at wavelengths $>18\,\mu$m in $\nu F_\nu$-space.
This agrees with the spectroscopic studies, which found that the silicate absorption becomes stronger towards the nucleus. 
The complex source morphology complicates any analysis of flux variations. which is therefore omitted.  
In summary, we note that the nuclear MIR emission of NGC\,1068 is both the brightest and most complex compared to the other local AGN on subarcsecond scales, and it is unclear, whether the results from NGC\,1068 can be applied to other AGN in general.
 \newline\end{@twocolumnfalse}]

\begin{figure}
   \centering
   \includegraphics[angle=0,width=8.500cm]{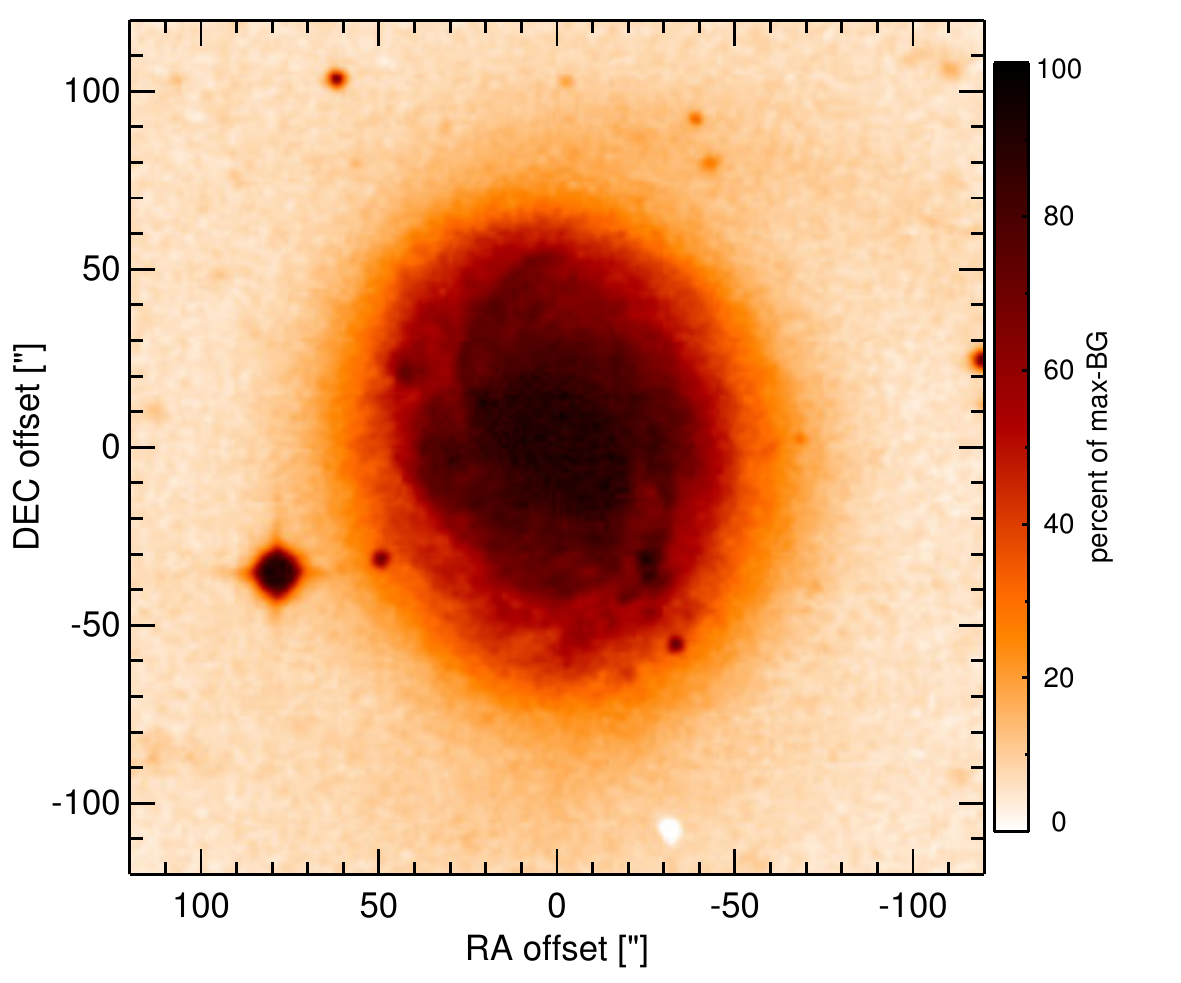}
    \caption{\label{fig:OPTim_NGC1068}
             Optical image (DSS, red filter) of NGC\,1068. Displayed are the central $4\arcmin$ with North up and East to the left. 
              The colour scaling is linear with white corresponding to the median background and black to the $0.01\%$ pixels with the highest intensity.  
           }
\end{figure}
\begin{figure}
   \centering
   \includegraphics[angle=0,height=3.11cm]{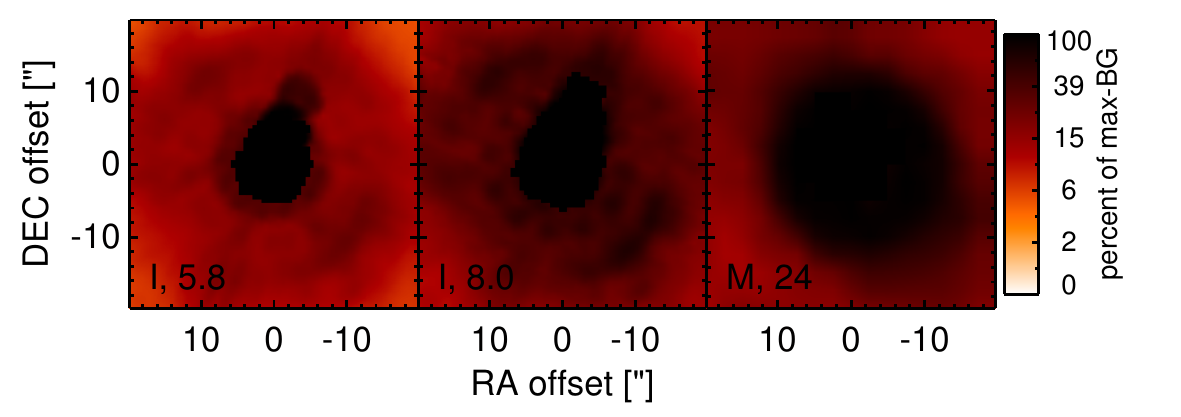}
    \caption{\label{fig:INTim_NGC1068}
             \spitzerr MIR images of NGC\,1068. Displayed are the inner $40\arcsec$ with North up and East to the left. The colour scaling is logarithmic with white corresponding to median background and black to the $0.1\%$ pixels with the highest intensity.
             The label in the bottom left states instrument and central wavelength of the filter in $\mu$m (I: IRAC, M: MIPS).
             Note that the central region in the images is completely saturated.
           }
\end{figure}
\begin{figure}
   \centering
   \includegraphics[angle=0,width=8.500cm]{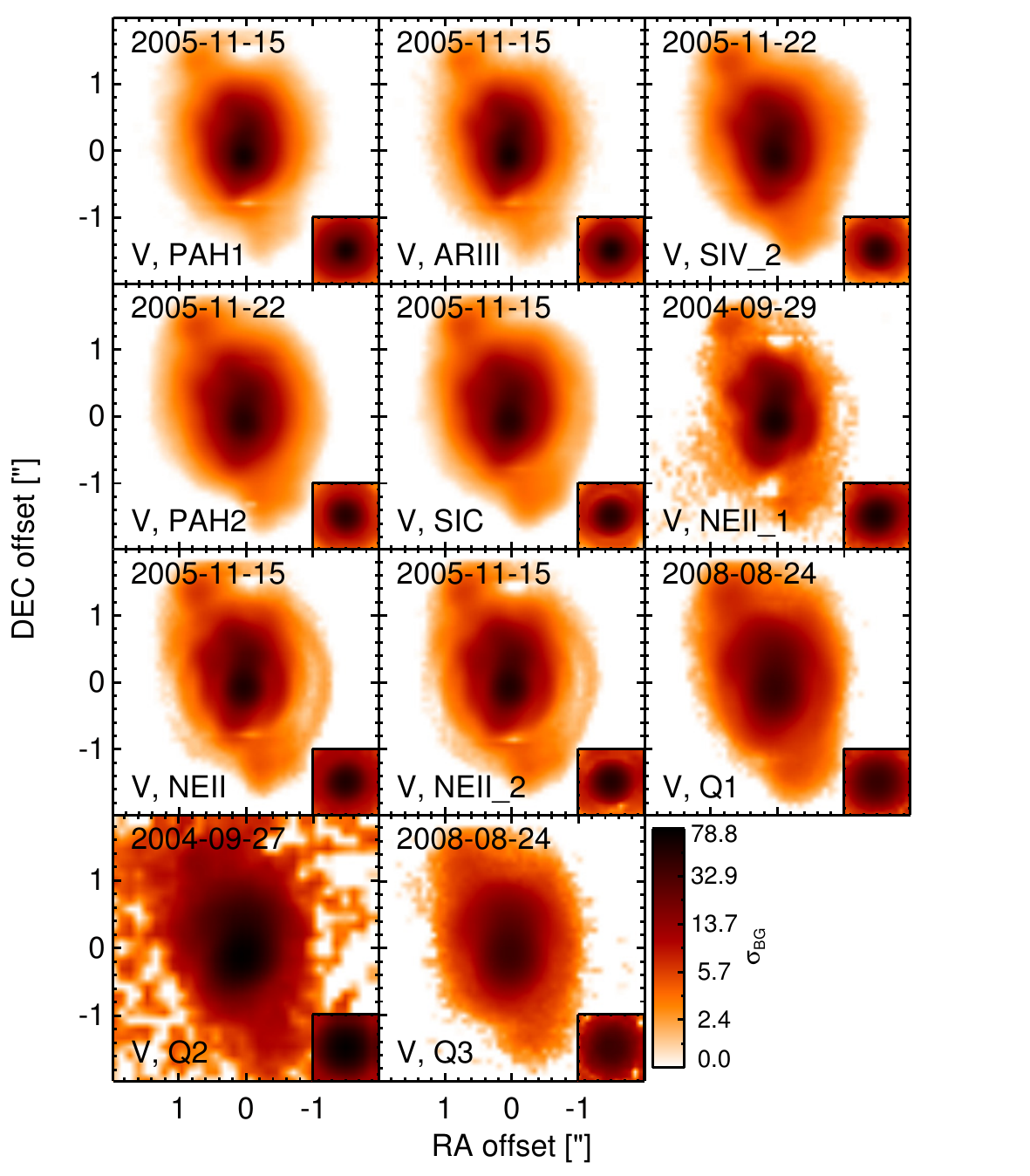}
    \caption{\label{fig:HARim_NGC1068}
             Subarcsecond-resolution MIR images of NGC\,1068 sorted by increasing filter wavelength. 
             Displayed are the inner $4\arcsec$ with North up and East to the left. 
             The colour scaling is logarithmic with white corresponding to median background and black to the $75\%$ of the highest intensity of all images in units of $\sigbg$.
             The inset image shows the central arcsecond of the PSF from the calibrator star, scaled to match the science target.
             The labels in the bottom left state instrument and filter names (C: COMICS, M: Michelle, T: T-ReCS, V: VISIR).
           }
\end{figure}
\begin{figure}
   \centering
   \includegraphics[angle=0,width=8.50cm]{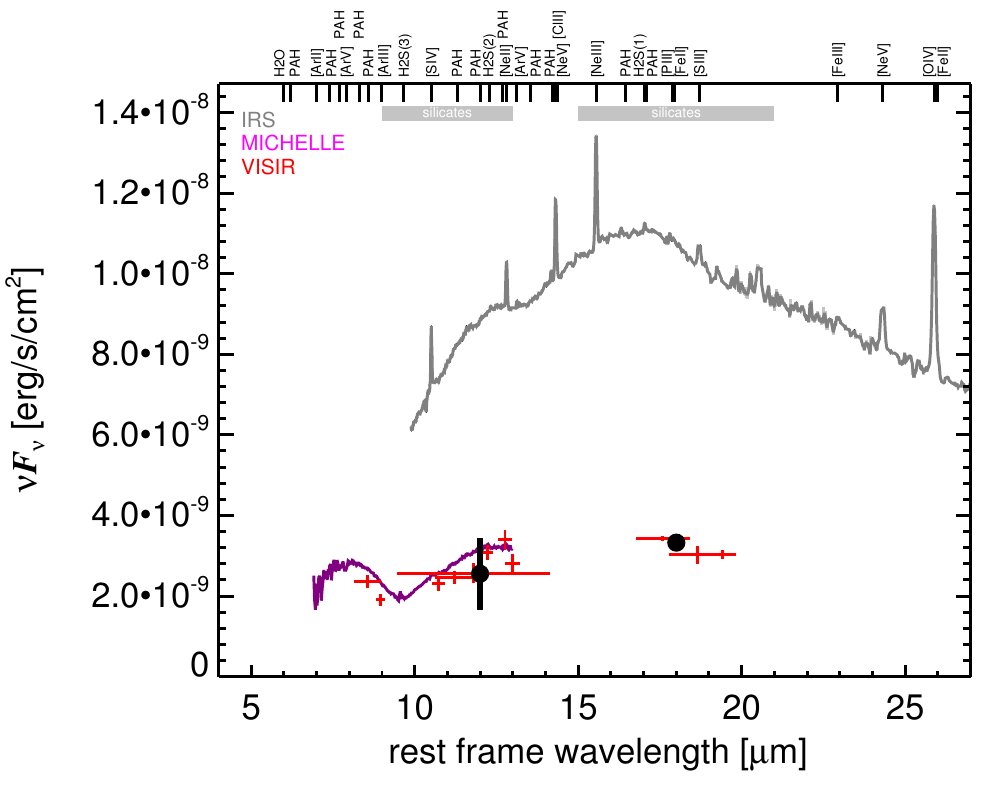}
   \caption{\label{fig:MISED_NGC1068}
      MIR SED of NGC\,1068. The description  of the symbols (if present) is the following.
      Grey crosses and  solid lines mark the \spitzer/IRAC, MIPS and IRS data. 
      The colour coding of the other symbols is: 
      green for COMICS, magenta for Michelle, blue for T-ReCS and red for VISIR data.
      Darker-coloured solid lines mark spectra of the corresponding instrument.
      The black filled circles mark the nuclear 12 and $18\,\mu$m  continuum emission estimate from the data.
      The ticks on the top axis mark positions of common MIR emission lines, while the light grey horizontal bars mark wavelength ranges affected by the silicate 10 and 18$\mu$m features.}
\end{figure}
\clearpage

\twocolumn[\begin{@twocolumnfalse}  
\subsection{NGC\,1097}\label{app:NGC1097}
NGC\,1097 is a low-inclination grand-design spiral galaxy at a distance of $D=$ $ 17.0\pm 5.3$\,Mpc (NED redshift-independent median) with a prominent nuclear starburst ring (diameter$\sim 18\arcsec \sim 1.5$\,kpc; \citealt{hummel_circumnuclear_1987}) and an active nucleus both containing a LINER/Seyfert nucleus with broad double-peaked Balmer lines \citep{storchi-bergmann_double-peaked_1993}, and a young massive nuclear star cluster (\citealt{storchi-bergmann_evidence_2005}; see \cite{nemmen_radiatively_2006} and \cite{mason_dust_2007} for detailed studies of the AGN).
The nucleus is also a flat-spectrum radio source \citep{wolstencroft_vla_1984}.
A system of two symmetrical jet-like systems was found around the nucleus (e.g., \citealt{lorre_enhancement_1978}).
However, these seem to be merger or interaction remnants of stellar origin \citep{carter_jets_1984,wehrle_nature_1997,higdon_minor-merger_2003}. 
NGC\,1097 was first observed at MIR wavelengths by \cite{rieke_10_1978},  \cite{telesco_ngc_1981} and \cite{telesco_genesis_1993}.
After \iras, \isoo observations \citep{roussel_atlas_2001,ramos_almeida_mid-infrared_2007} and first subarcsecond-resolution $N$-band imaging with Palomar 5\,m/MIRLIN followed \citep{gorjian_10_2004}.
The nucleus remained undetected in the latter.
In the \spitzer/IRAC and MIPS images, the compact but resolved nucleus and the bright starburst ring are detected.
The latter outshines the nucleus with increasing wavelength.
Because me measure the nuclear component only, our  IRAC $5.8$ and $8.0\,\mu$m and MIPS $24\,\mu$m fluxes are significantly lower than in the literature (e.g., \citealt{dale_infrared_2005,gallimore_infrared_2010}).
The IRS LR staring-mode spectrum appears completely star-formation dominated with strong PAH features, silicate  $10\,\mu$m absorption and a red spectral slope in $\nu F_\nu$-space (see also \citealt{brandl_mid-infrared_2006,mason_dust_2007,smith_mid-infrared_2007,wu_spitzer/irs_2009,bernard-salas_spitzer_2009,tommasin_spitzer-irs_2010,gallimore_infrared_2010}).
At subarcsecond-resolution, the nuclear region of NGC\,1097 was observed with VISIR in five narrow $N$-band filters and one $Q$-band filter in 2005 (unpublished, to our knowledge), 2006 \citep{horst_mid_2008,horst_mid-infrared_2009,reunanen_vlt_2010}, 2009 \citep{asmus_mid-infrared_2011} and 2010 (unpublished, to our knowledge).
In addition, \cite{mason_dust_2007,mason_nuclear_2012} used T-ReCS to acquire both a Si5 and a Qa filter image in 2005.
The compact MIR nucleus and at least part of the starburst ring are detected in all images, albeit partly with very low S/N (e.g., PAH1).
The ring is not discussed (but see \citealt{mason_dust_2007}).
In all images with sufficient S/N (but Qa), the nucleus is marginally resolved (FWHM $\sim 0.45\arcsec \sim 37\,$pc). 
Therefore, we measure only the unresolved nuclear component which provides fluxes lower than total nuclear fluxes in \cite{horst_mid_2008} and \cite{asmus_mid-infrared_2011}, but consistent with the PSF fluxes in \cite{reunanen_vlt_2010} and \cite{mason_nuclear_2012}.
The subarcsecond fluxes are on average $\sim90\%$ lower than the \spitzerr spectrophotometry, which indicates that even the central region inside the starburst ring is totally dominated by diffuse star formation.
The nuclear MIR SED indicates that the PAH 11.3$\,\mu$m feature is also present in the projected central $\sim 30$\,pc. 
Any AGN contribution in the MIR must thus be very low as already indicated by the absence of \oiv and \nev emission \citep{bernard-salas_spitzer_2009}.
\newline\end{@twocolumnfalse}]

\begin{figure}
   \centering
   \includegraphics[angle=0,width=8.500cm]{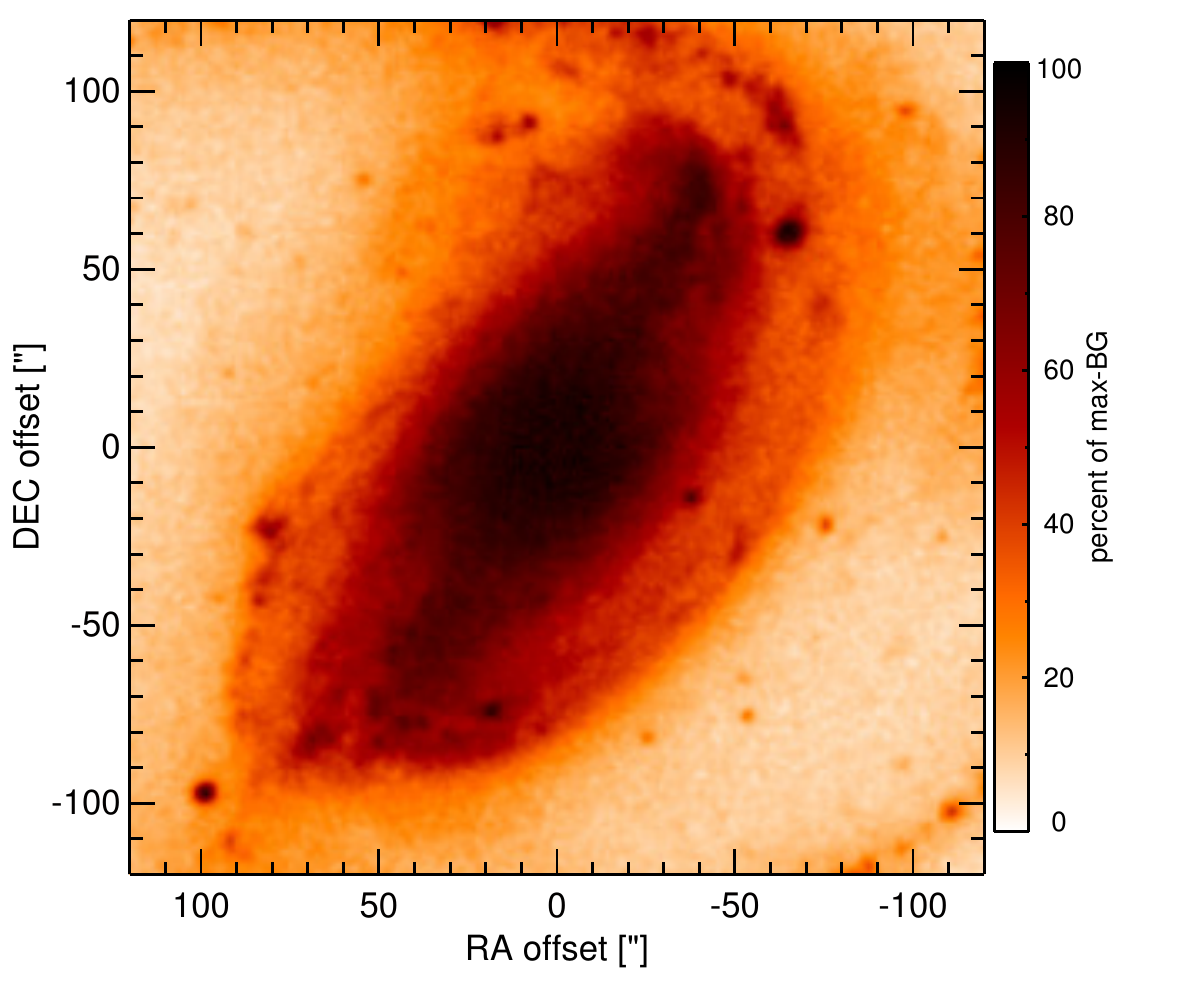}
    \caption{\label{fig:OPTim_NGC1097}
             Optical image (DSS, red filter) of NGC\,1097. Displayed are the central $4\arcmin$ with North up and East to the left. 
              The colour scaling is linear with white corresponding to the median background and black to the $0.01\%$ pixels with the highest intensity.  
           }
\end{figure}
\begin{figure}
   \centering
   \includegraphics[angle=0,height=3.11cm]{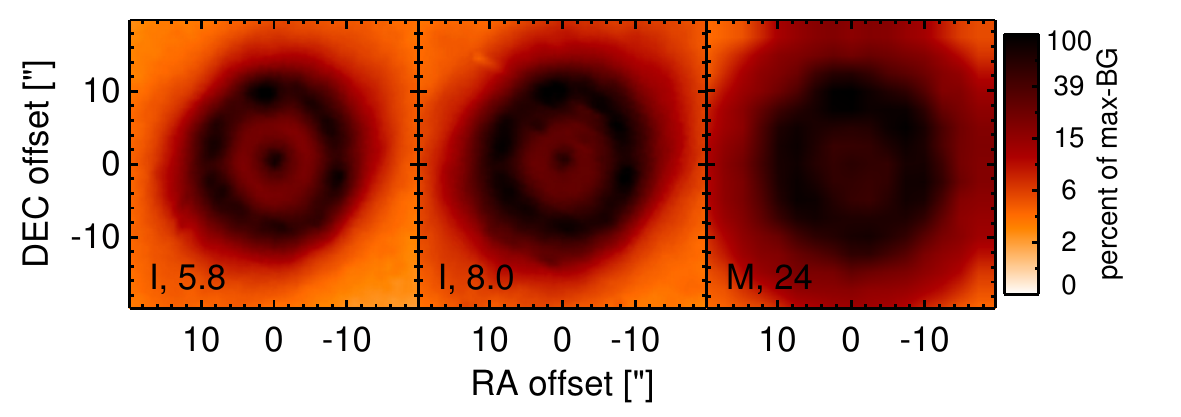}
    \caption{\label{fig:INTim_NGC1097}
             \spitzerr MIR images of NGC\,1097. Displayed are the inner $40\arcsec$ with North up and East to the left. The colour scaling is logarithmic with white corresponding to median background and black to the $0.1\%$ pixels with the highest intensity.
             The label in the bottom left states instrument and central wavelength of the filter in $\mu$m (I: IRAC, M: MIPS). 
           }
\end{figure}
\begin{figure}
   \centering
   \includegraphics[angle=0,width=8.500cm]{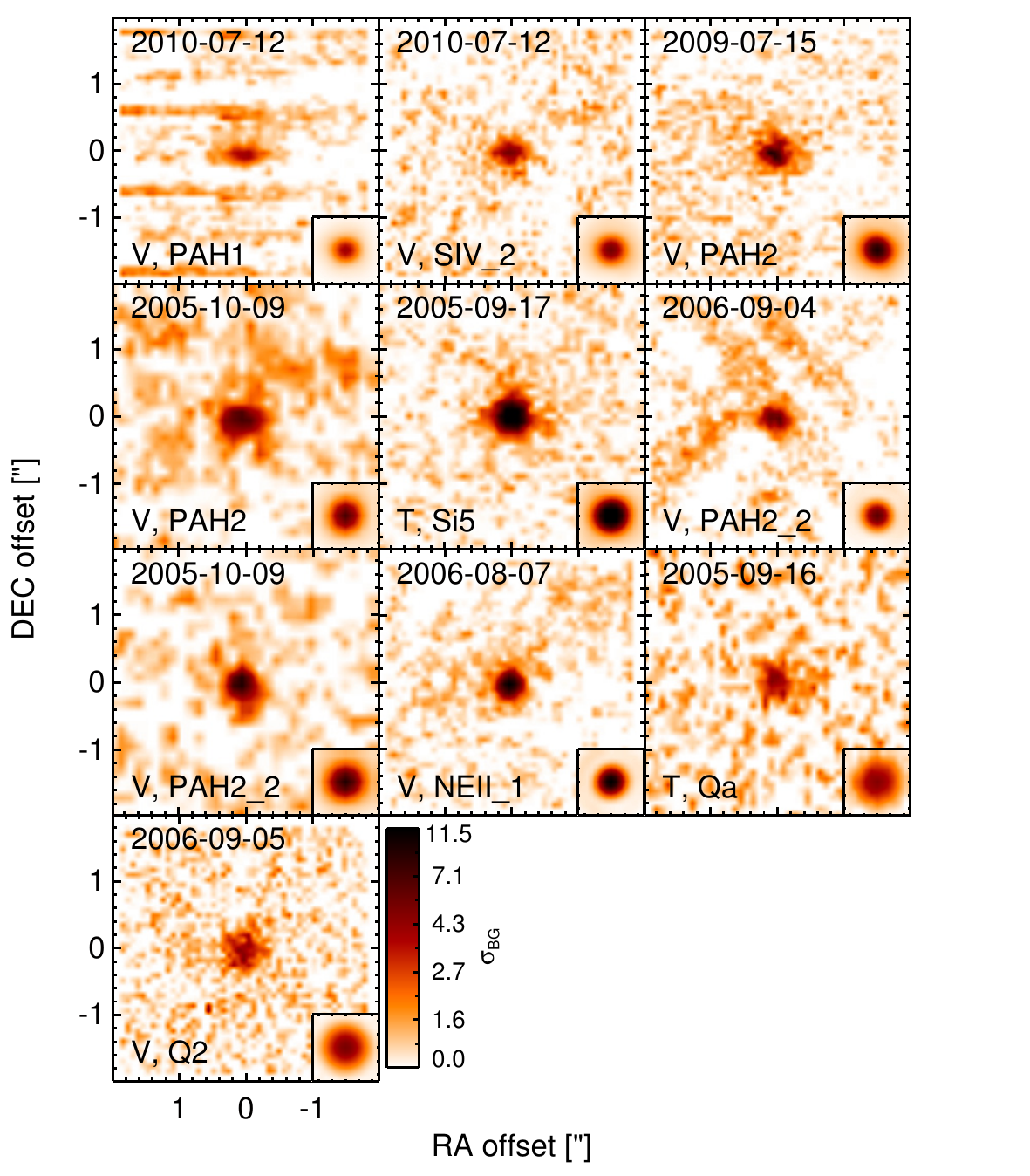}
    \caption{\label{fig:HARim_NGC1097}
             Subarcsecond-resolution MIR images of NGC\,1097 sorted by increasing filter wavelength. 
             Displayed are the inner $4\arcsec$ with North up and East to the left. 
             The colour scaling is logarithmic with white corresponding to median background and black to the $75\%$ of the highest intensity of all images in units of $\sigbg$.
             The inset image shows the central arcsecond of the PSF from the calibrator star, scaled to match the science target.
             The labels in the bottom left state instrument and filter names (C: COMICS, M: Michelle, T: T-ReCS, V: VISIR).
           }
\end{figure}
\begin{figure}
   \centering
   \includegraphics[angle=0,width=8.50cm]{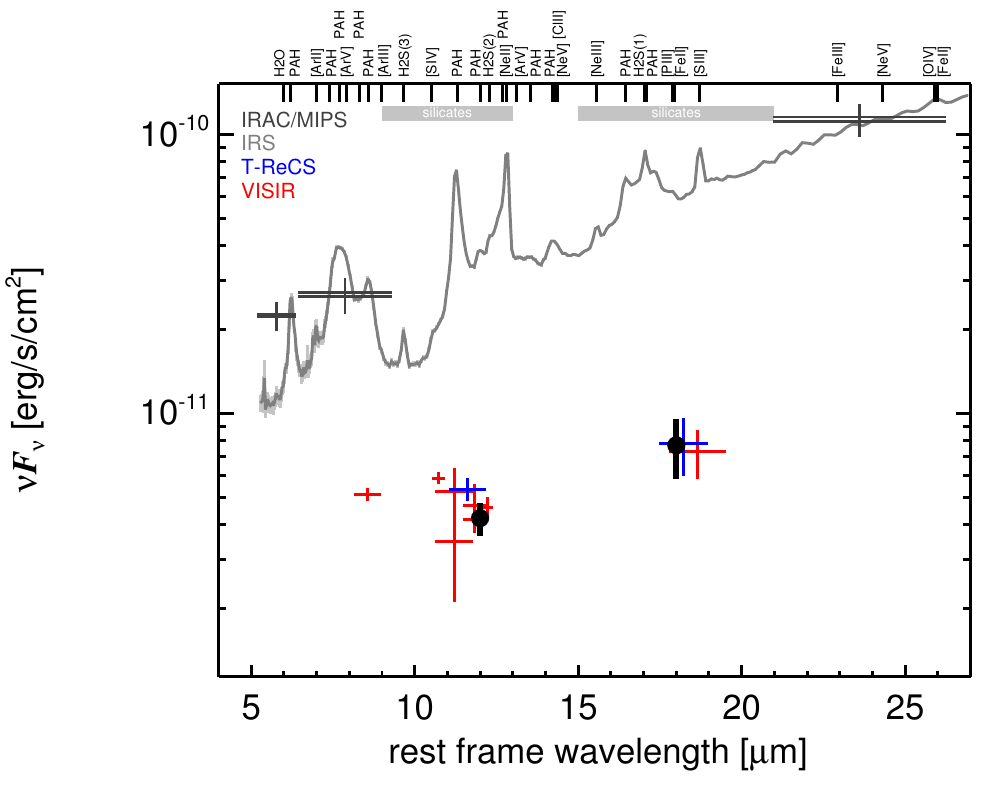}
   \caption{\label{fig:MISED_NGC1097}
      MIR SED of NGC\,1097. The description  of the symbols (if present) is the following.
      Grey crosses and  solid lines mark the \spitzer/IRAC, MIPS and IRS data. 
      The colour coding of the other symbols is: 
      green for COMICS, magenta for Michelle, blue for T-ReCS and red for VISIR data.
      Darker-coloured solid lines mark spectra of the corresponding instrument.
      The black filled circles mark the nuclear 12 and $18\,\mu$m  continuum emission estimate from the data.
      The ticks on the top axis mark positions of common MIR emission lines, while the light grey horizontal bars mark wavelength ranges affected by the silicate 10 and 18$\mu$m features.}
\end{figure}
\clearpage

\twocolumn[\begin{@twocolumnfalse}  
\subsection{NGC\,1144 -- NGC\,1142 -- Arp\,118}\label{app:NGC1144}
NGC\,1144 is a disturbed spiral galaxy at a redshift of $z=$ 0.0288 ($D\sim128$\,Mpc) interacting with the elliptical NGC\,1143 (also named NGC\,1141) with a nuclear separation of $\sim 44\arcsec \sim 26\,$kpc towards the north-west \citep{hippelein_arp_1989}.
It possesses a Sy\,2 nucleus \citep{veron-cetty_catalogue_2010} that belongs to the nine-month BAT AGN sample.
Apart from \iras, NGC\,1142 was observed in the MIR with IRTF \citep{edelson_broad-band_1987}, with Palomar 5\,m \citep{carico_iras_1988}, with \isoo \citep{charmandaris_mid-infrared_2001,ramos_almeida_mid-infrared_2007} and with \spitzer/IRAC, IRS and MIPS.
In the IRAC images, a compact MIR nucleus embedded within extended complex host emission was detected.
In addition, a bright compact off-nuclear source is visible $\sim 9\arcsec \sim 5.3\,$kpc away from the nucleus to the west (PA$\sim -104\degree$).
This region is identified as a giant H\,II region, labelled D in \cite{hippelein_arp_1989}.
Its brightness increases with wavelength and dominates over the nucleus in the MIPS $24\,\mu$m image.
The IRS LR mapping-mode spectrum is completely star-formation dominated with strong PAH features, silicate  $10\,\mu$m absorption and a red spectral slope in $\nu F_\nu$-space (see also \citealt{wu_spitzer/irs_2009,tommasin_spitzer-irs_2010}).
However, \nev is also detected verifying the presence of an AGN in NGC\,1144.
The nucleus was observed with VISIR in eight $N$-band filters in 2009 (unpublished, to our knowledge) and remained undetected in four filters, in particular at short wavelengths.
The other images show a weakly detected compact MIR nucleus with uncertain extension. 
The core of the H\,II region is as well weakly detected in the VISIR images but not analysed.
The nuclear MIR photometry is on average $\sim 47\%$ lower than the \spitzerr spectrophotometry, and in particular the PAH emission seems to be absent at subarcsecond resolution.
We conclude that the AGN contributes less than half of the MIR luminosity from the central $\sim 2\,$kpc of NGC\,1144.
\newline\end{@twocolumnfalse}]

\begin{figure}
   \centering
   \includegraphics[angle=0,width=8.500cm]{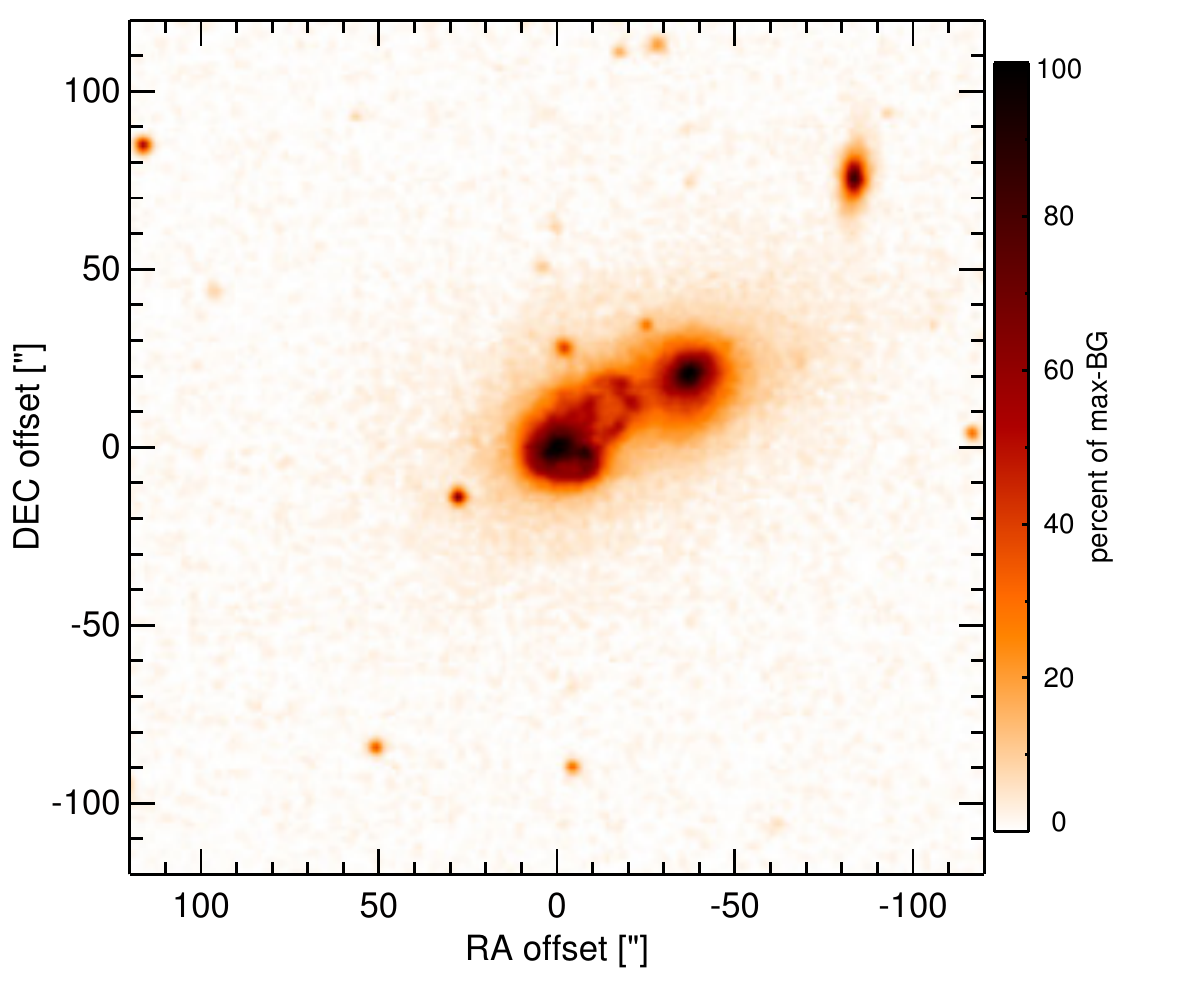}
    \caption{\label{fig:OPTim_NGC1144}
             Optical image (DSS, red filter) of NGC\,1144. Displayed are the central $4\arcmin$ with North up and East to the left. 
              The colour scaling is linear with white corresponding to the median background and black to the $0.01\%$ pixels with the highest intensity.  
           }
\end{figure}
\begin{figure}
   \centering
   \includegraphics[angle=0,height=3.11cm]{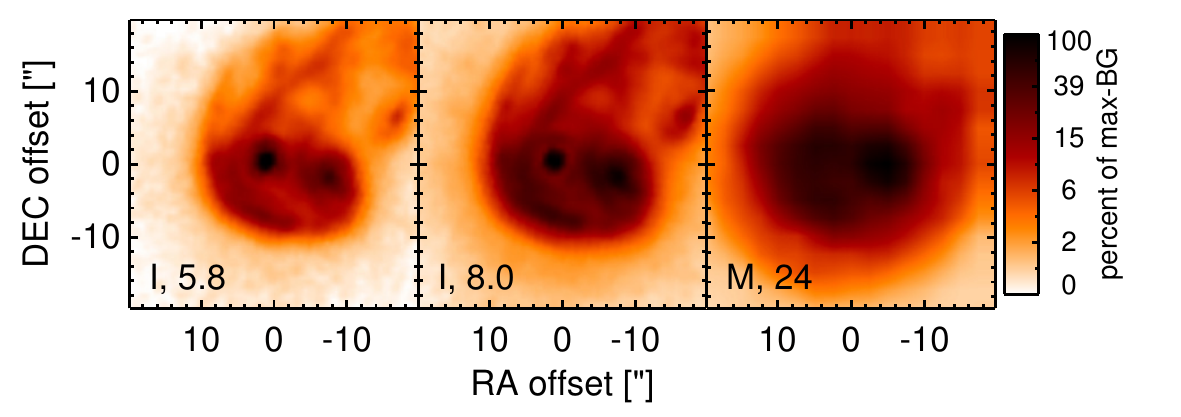}
    \caption{\label{fig:INTim_NGC1144}
             \spitzerr MIR images of NGC\,1144. Displayed are the inner $40\arcsec$ with North up and East to the left. The colour scaling is logarithmic with white corresponding to median background and black to the $0.1\%$ pixels with the highest intensity.
             The label in the bottom left states instrument and central wavelength of the filter in $\mu$m (I: IRAC, M: MIPS).
             Note that the apparent off-nuclear compact sources to the north-west in the IRAC $8.0\,\mu$m image are instrumental artefacts.
           }
\end{figure}
\begin{figure}
   \centering
   \includegraphics[angle=0,width=8.500cm]{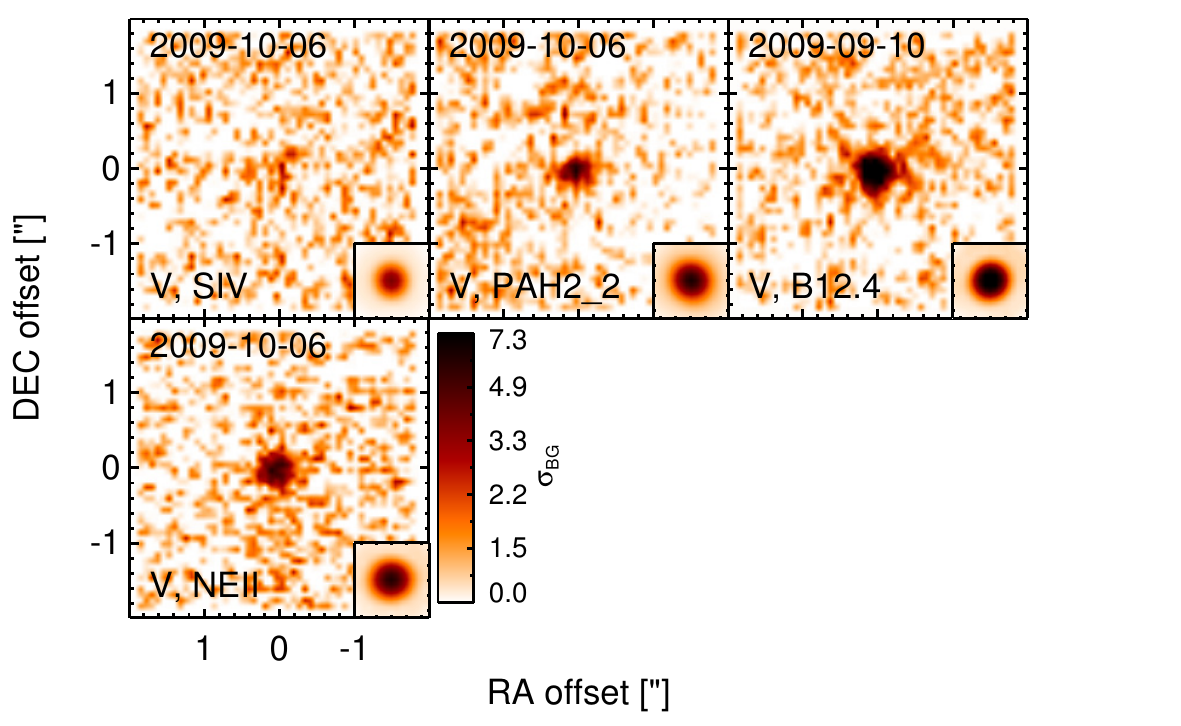}
    \caption{\label{fig:HARim_NGC1144}
             Subarcsecond-resolution MIR images of NGC\,1144 sorted by increasing filter wavelength. 
             Displayed are the inner $4\arcsec$ with North up and East to the left. 
             The colour scaling is logarithmic with white corresponding to median background and black to the $75\%$ of the highest intensity of all images in units of $\sigbg$.
             The inset image shows the central arcsecond of the PSF from the calibrator star, scaled to match the science target.
             The labels in the bottom left state instrument and filter names (C: COMICS, M: Michelle, T: T-ReCS, V: VISIR).
           }
\end{figure}
\begin{figure}
   \centering
   \includegraphics[angle=0,width=8.50cm]{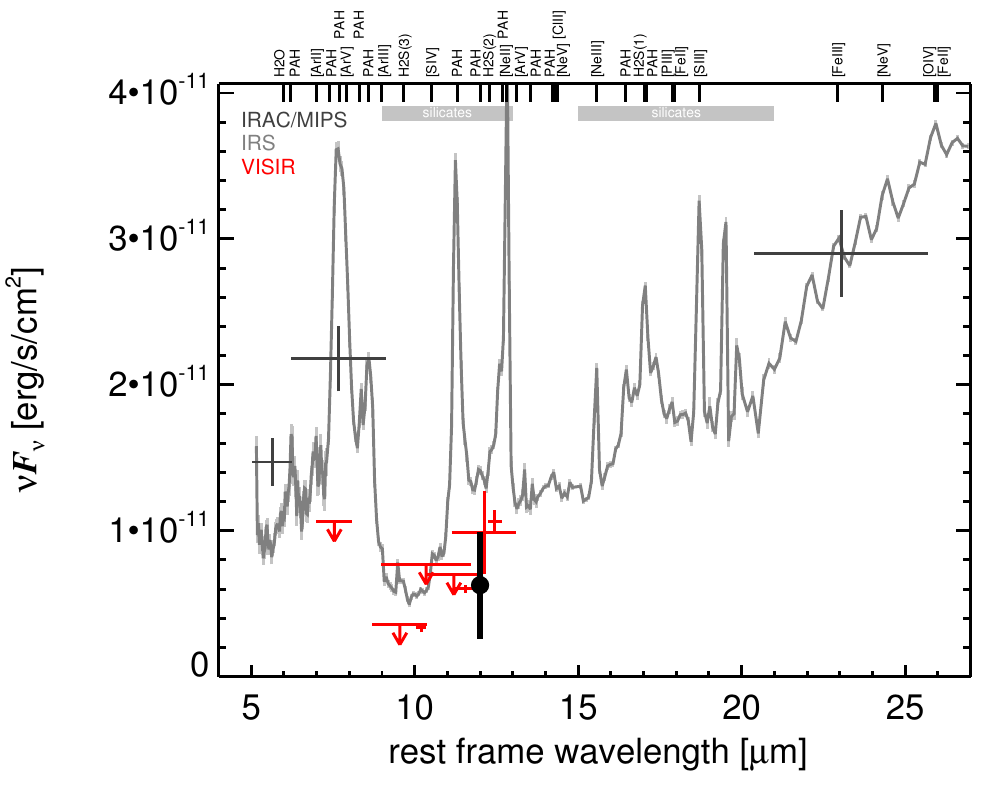}
   \caption{\label{fig:MISED_NGC1144}
      MIR SED of NGC\,1144. The description  of the symbols (if present) is the following.
      Grey crosses and  solid lines mark the \spitzer/IRAC, MIPS and IRS data. 
      The colour coding of the other symbols is: 
      green for COMICS, magenta for Michelle, blue for T-ReCS and red for VISIR data.
      Darker-coloured solid lines mark spectra of the corresponding instrument.
      The black filled circles mark the nuclear 12 and $18\,\mu$m  continuum emission estimate from the data.
      The ticks on the top axis mark positions of common MIR emission lines, while the light grey horizontal bars mark wavelength ranges affected by the silicate 10 and 18$\mu$m features.}
\end{figure}
\clearpage

\twocolumn[\begin{@twocolumnfalse}  
\subsection{NGC\,1194 -- UGC\,2514}\label{app:NGC1194}
NGC\,1194 is a highly-inclined early-type spiral galaxy at a redshift of $z=$ 0.0136 ($D\sim58.2$\,Mpc) with an AGN, optically classified as  a Sy\,1.9  \citep{veron-cetty_catalogue_2010}. 
It features one-sided weak extended radio emission cospatial with \oiii emission (extension$\sim1\arcsec\sim$0.3\,kpc; PA$\sim-124\degree$;  \citealt{schmitt_jet_2001,schmitt_hubble_2003}).
After \iras, NGC\,1194 was observed in the MIR with Palomar 5\,m/MIRLIN \citep{gorjian_10_2004} and \spitzer/IRAC and IRS.
The IRAC images, it appears as a compact but slightly resolved source. 
Our IRAC $5.8$ and $8.0\,\mu$m are consistent with the values presented in \cite{gallimore_infrared_2010}.
The IRS LR mapping-mode spectrum exhibits a deep silicate  $10\,\mu$m absorption feature, possibly  short-wavelength PAH emission, and a  shallow blue spectral slope in $\nu F_\nu$-space (see also \citealt{buchanan_spitzer_2006,deo_spitzer_2007,deo_mid-infrared_2009,wu_spitzer/irs_2009,tommasin_spitzer-irs_2010,gallimore_infrared_2010}).
NGC\,1194 was observed with VISIR in five $N$-band filters in 2009 (unpublished, to our knowledge) and an unresolved nucleus without any host emission was detected in all images.
The nuclear photometry in general agrees with the \spitzerr spectrophotometry except at $\sim 8\,\mu$m.
Therefore, the silicate feature originates in the projected central $\sim 100\,$pc of NGC\,1194, and we correct our 12\,$\mu$m continuum emission estimate for it by using the IRS spectrum.
The apparent absence of nuclear PAH emission  agrees with the findings of \cite{imanishi_compact_2003} and together with the MIR compactness disfavours significant nuclear star formation in NGC\,1194 as suggested by \cite{davies_molecular_2005}.
\newline\end{@twocolumnfalse}]

\begin{figure}
   \centering
   \includegraphics[angle=0,width=8.500cm]{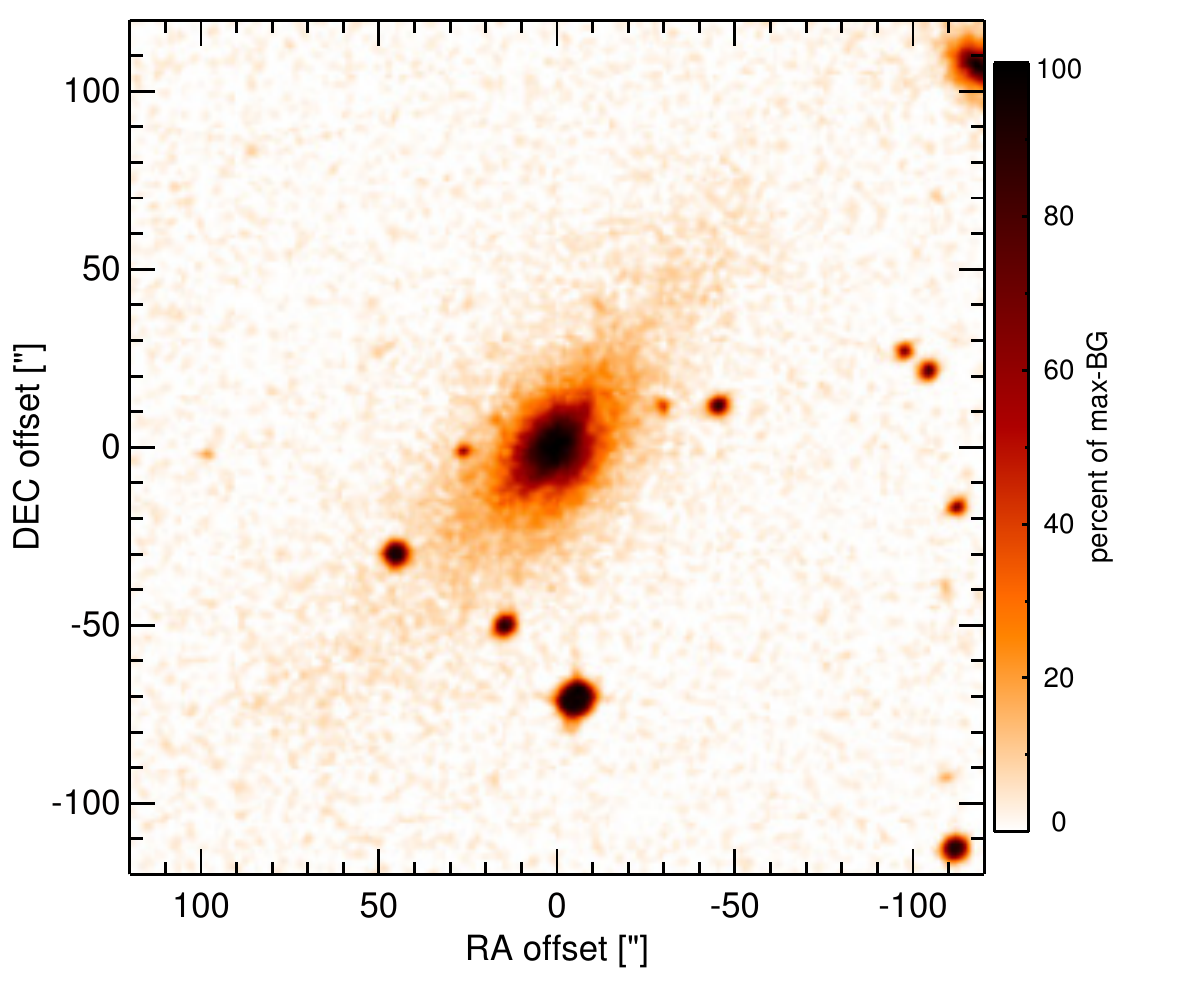}
    \caption{\label{fig:OPTim_NGC1194}
             Optical image (DSS, red filter) of NGC\,1194. Displayed are the central $4\arcmin$ with North up and East to the left. 
              The colour scaling is linear with white corresponding to the median background and black to the $0.01\%$ pixels with the highest intensity.  
           }
\end{figure}
\begin{figure}
   \centering
   \includegraphics[angle=0,height=3.11cm]{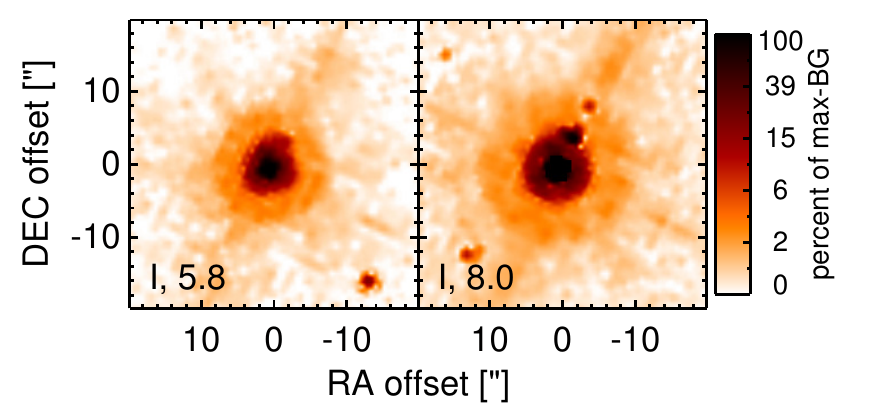}
    \caption{\label{fig:INTim_NGC1194}
             \spitzerr MIR images of NGC\,1194. Displayed are the inner $40\arcsec$ with North up and East to the left. The colour scaling is logarithmic with white corresponding to median background and black to the $0.1\%$ pixels with the highest intensity.
             The label in the bottom left states instrument and central wavelength of the filter in $\mu$m (I: IRAC, M: MIPS). 
           }
\end{figure}
\begin{figure}
   \centering
   \includegraphics[angle=0,width=8.500cm]{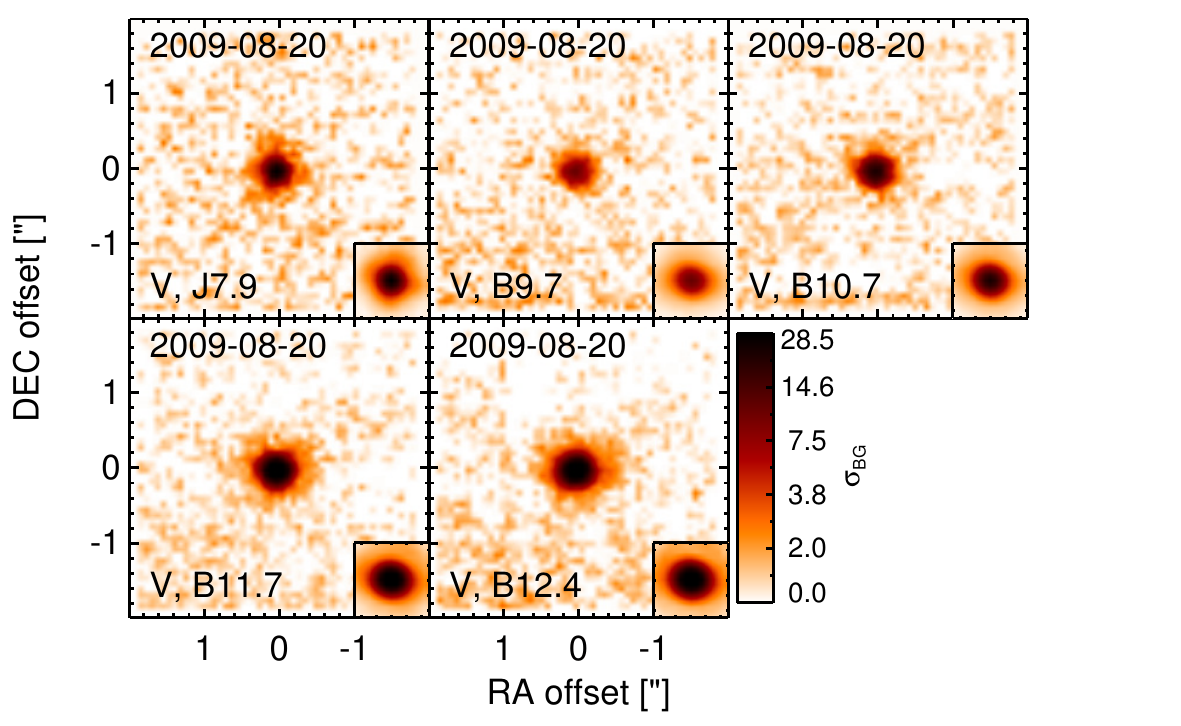}
    \caption{\label{fig:HARim_NGC1194}
             Subarcsecond-resolution MIR images of NGC\,1194 sorted by increasing filter wavelength. 
             Displayed are the inner $4\arcsec$ with North up and East to the left. 
             The colour scaling is logarithmic with white corresponding to median background and black to the $75\%$ of the highest intensity of all images in units of $\sigbg$.
             The inset image shows the central arcsecond of the PSF from the calibrator star, scaled to match the science target.
             The labels in the bottom left state instrument and filter names (C: COMICS, M: Michelle, T: T-ReCS, V: VISIR).
           }
\end{figure}
\begin{figure}
   \centering
   \includegraphics[angle=0,width=8.50cm]{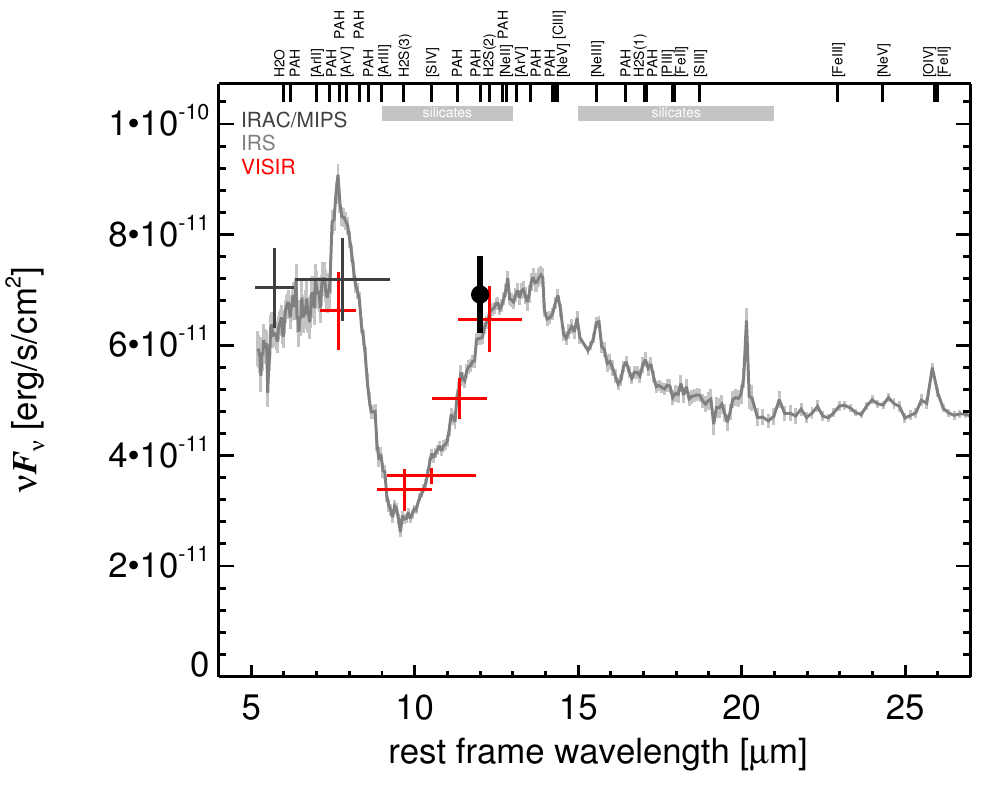}
   \caption{\label{fig:MISED_NGC1194}
      MIR SED of NGC\,1194. The description  of the symbols (if present) is the following.
      Grey crosses and  solid lines mark the \spitzer/IRAC, MIPS and IRS data. 
      The colour coding of the other symbols is: 
      green for COMICS, magenta for Michelle, blue for T-ReCS and red for VISIR data.
      Darker-coloured solid lines mark spectra of the corresponding instrument.
      The black filled circles mark the nuclear 12 and $18\,\mu$m  continuum emission estimate from the data.
      The ticks on the top axis mark positions of common MIR emission lines, while the light grey horizontal bars mark wavelength ranges affected by the silicate 10 and 18$\mu$m features.}
\end{figure}
\clearpage

\twocolumn[\begin{@twocolumnfalse}  
\subsection{NGC\,1275 -- 3C\,84 -- Perseus\,A}\label{app:NGC1275}
NGC\,1275 is an infrared-luminous, radio-loud peculiar giant elliptical  galaxy at a redshift of $z=$ 0.0175 ($D\sim72.7\,$Mpc) with a FR\,I radio morphology and prominent galactic-scale gas filaments. 
This well-studied object is the central galaxy in the Perseus cluster and one of the original six Seyfert galaxies (\citealt{seyfert_nuclear_1943}; see \citealt{scharwachter_kinematics_2013} for a recent detailed study).
It is presumably in a post-merger state \citep{holtzman_planetary_1992,conselice_nature_2001} and features complex optical spectral properties, leading to many different classifications like Sy\,2, borderline Sy\,1.5/LINER, BL Lac and even Sy/H\,II composite \citep{khachikian_atlas_1974,ho_search_1997,veron_ngc1275_1978,veron_agns_1997}.
The nucleus is a compact source with a flat spectrum at radio wavelengths, and a pair of galactic-scale jets emerge from it with a PA$\sim160\degree$ in the central kiloparsec region \citep{pedlar_radio_1990,walker_detection_1994,vermeulen_discovery_1994}.
Pioneering MIR observations of NGC\,1275 were performed by \cite{low_proceedings_1968} and \cite{kleinmann_observations_1970}, followed by \cite{rieke_variability_1972,rieke_infrared_1978}, \cite{lebofsky_extinction_1979}, \cite{aitken_question_1981}, \cite{malkan_stellar_1983}, \cite{gear_thermal_1985},  and \cite{roche_atlas_1991}.
The first subarcsecond-resolution $N$-band image obtained with Keck/LWS show an unresolved nucleus without any extended emission \citep{soifer_high_2003}.
NGC\,1275 was also observed with \isoo \citep{rigopoulou_large_1999,tran_isocam-cvf_2001,siebenmorgen_isocam_2004,temi_cold_2004} and \spitzer/IRAC, IRS and MIPS in the MIR.
The corresponding IRAC and MIPS images are dominated by a bright compact nucleus.
The IRS LR staring-mode spectrum exhibits silicate 10 and 18\,$\mu$m emission, a very weak PAH 11.3\,$\mu$m features and a steep red spectral slope in $\nu F_\nu$-space (see also \citealt{weedman_mid-infrared_2005,shi_aromatic_2007,leipski_spitzer_2009,mullaney_defining_2011}).
Thus, the arcsecond-scale MIR SED  verifies the existence of large amounts of AGN-heated dust in NGC\,1275, while any star-formation contribution is at least minor \citep{leipski_spitzer_2009}.
NGC\,1275 was observed with Michelle in the Si-2 and Si-6 filters in 2006 and with COMICS in the N11.7 and NEII filters in 2009 (unpublished, to our knowledge).
A bright compact nucleus without further host emission was detected in all images.
It appears to be marginally resolved in the sharpest image (Si-2) but displays inconsistent nuclear shapes between the Michelle and COMICS images.
Therefore, its MIR extension at subarcsecond resolution remains uncertain.
Our nuclear photometry is generally consistent with the \spitzerr spectrophotometry with the Michelle fluxes being systematically lower and the COMICS flux systematically higher.
We use the IRS spectrum to compute the nuclear 12$\,\mu$m continuum emission estimate corrected for the silicate feature.
Comparison with the historical $N$-band photometry shows apparent flux variations on the order of $\sim16\%$ during the last $\sim40$ years.
This indicates  intrinsic variability in the $N$-band emission of NGC\,1275 despite uncertainties and systematics of the different instruments, filters and measurement methods.
\newline\end{@twocolumnfalse}]

\begin{figure}
   \centering
   \includegraphics[angle=0,width=8.500cm]{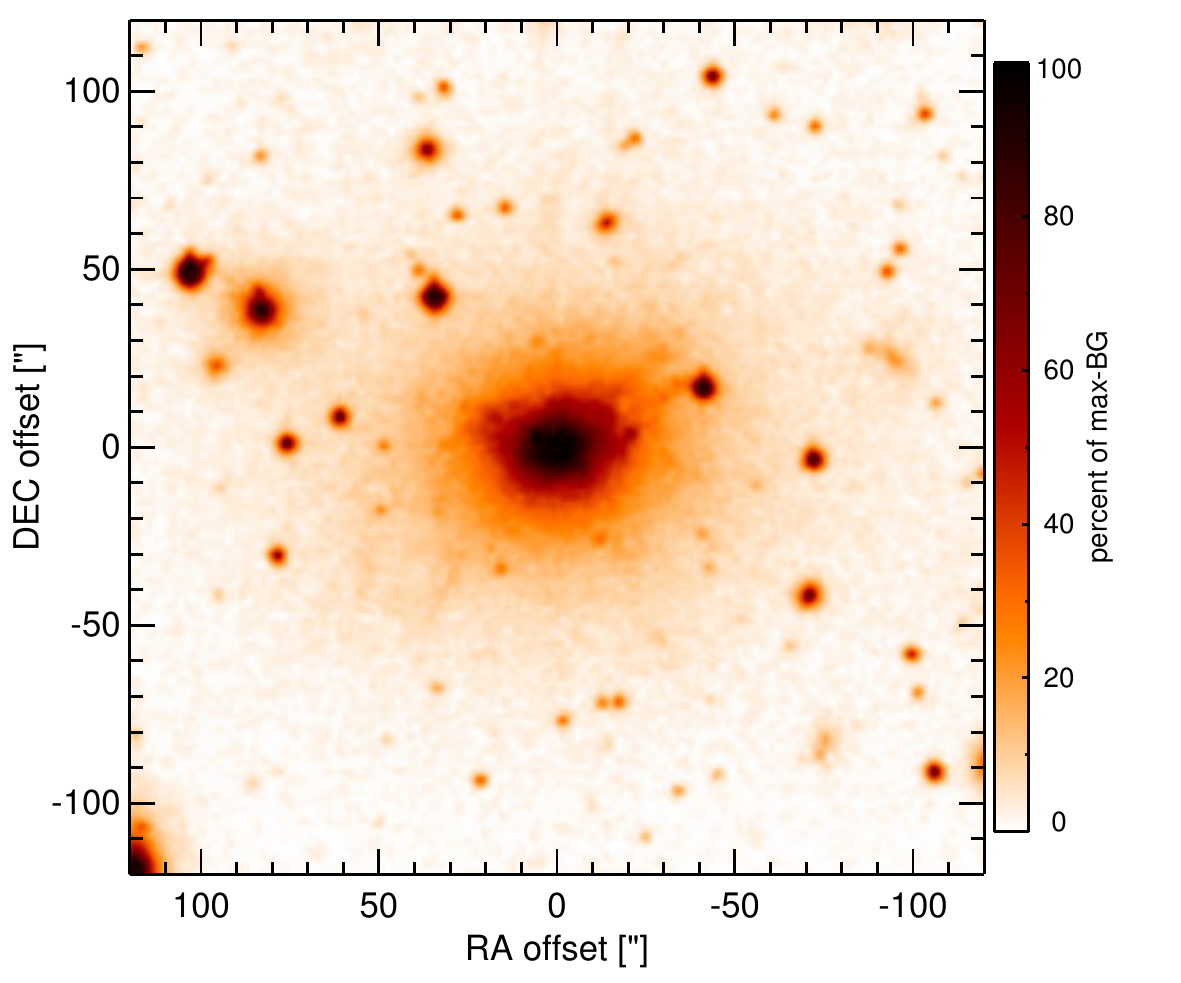}
    \caption{\label{fig:OPTim_NGC1275}
             Optical image (DSS, red filter) of NGC\,1275. Displayed are the central $4\arcmin$ with North up and East to the left. 
              The colour scaling is linear with white corresponding to the median background and black to the $0.01\%$ pixels with the highest intensity.  
           }
\end{figure}
\begin{figure}
   \centering
   \includegraphics[angle=0,height=3.11cm]{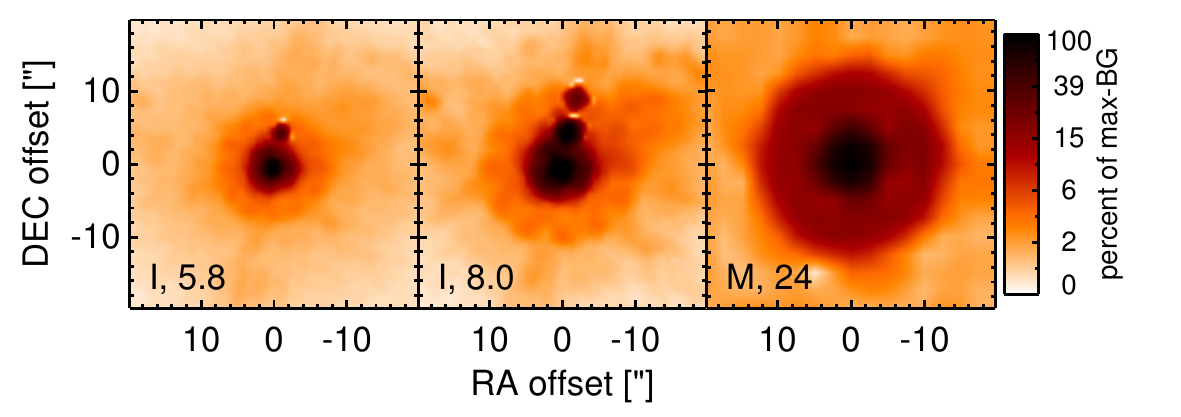}
    \caption{\label{fig:INTim_NGC1275}
             \spitzerr MIR images of NGC\,1275. Displayed are the inner $40\arcsec$ with North up and East to the left. The colour scaling is logarithmic with white corresponding to median background and black to the $0.1\%$ pixels with the highest intensity.
             The label in the bottom left states instrument and central wavelength of the filter in $\mu$m (I: IRAC, M: MIPS).
             Note that the apparent off-nuclear compact sources in the IRAC 5.8 and $8.0\,\mu$m images are instrumental artefacts.
           }
\end{figure}
\begin{figure}
   \centering
   \includegraphics[angle=0,width=8.500cm]{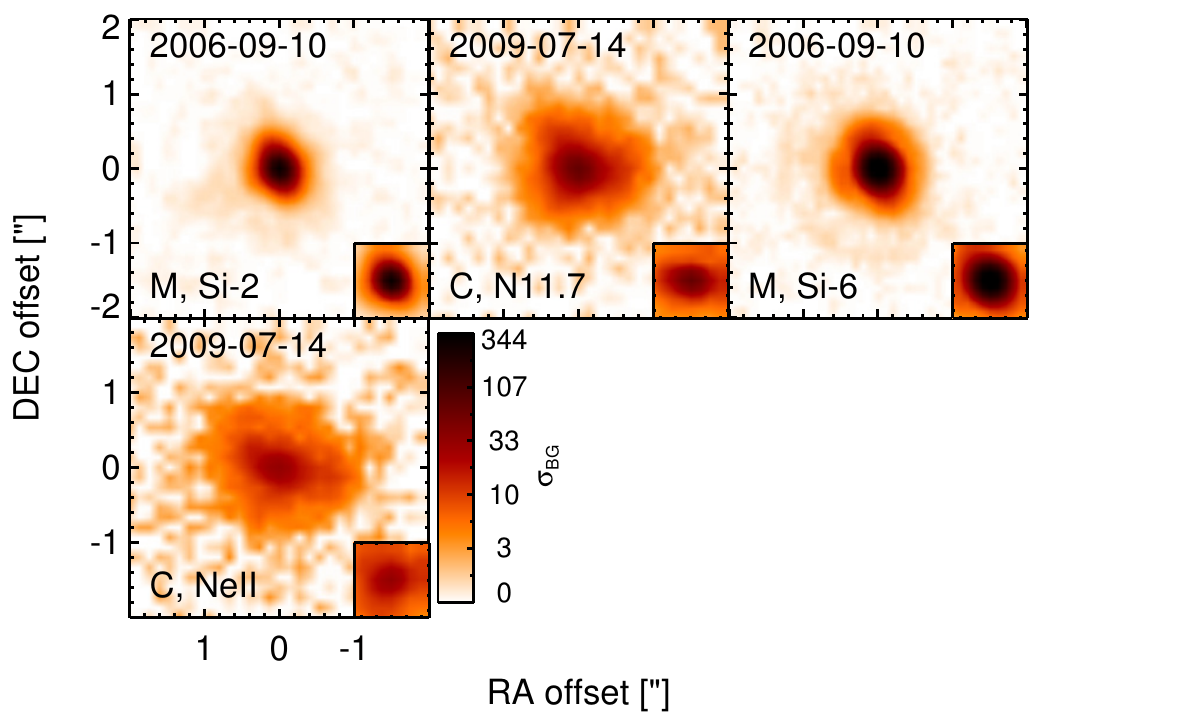}
    \caption{\label{fig:HARim_NGC1275}
             Subarcsecond-resolution MIR images of NGC\,1275 sorted by increasing filter wavelength. 
             Displayed are the inner $4\arcsec$ with North up and East to the left. 
             The colour scaling is logarithmic with white corresponding to median background and black to the $75\%$ of the highest intensity of all images in units of $\sigbg$.
             The inset image shows the central arcsecond of the PSF from the calibrator star, scaled to match the science target.
             The labels in the bottom left state instrument and filter names (C: COMICS, M: Michelle, T: T-ReCS, V: VISIR).
           }
\end{figure}
\begin{figure}
   \centering
   \includegraphics[angle=0,width=8.50cm]{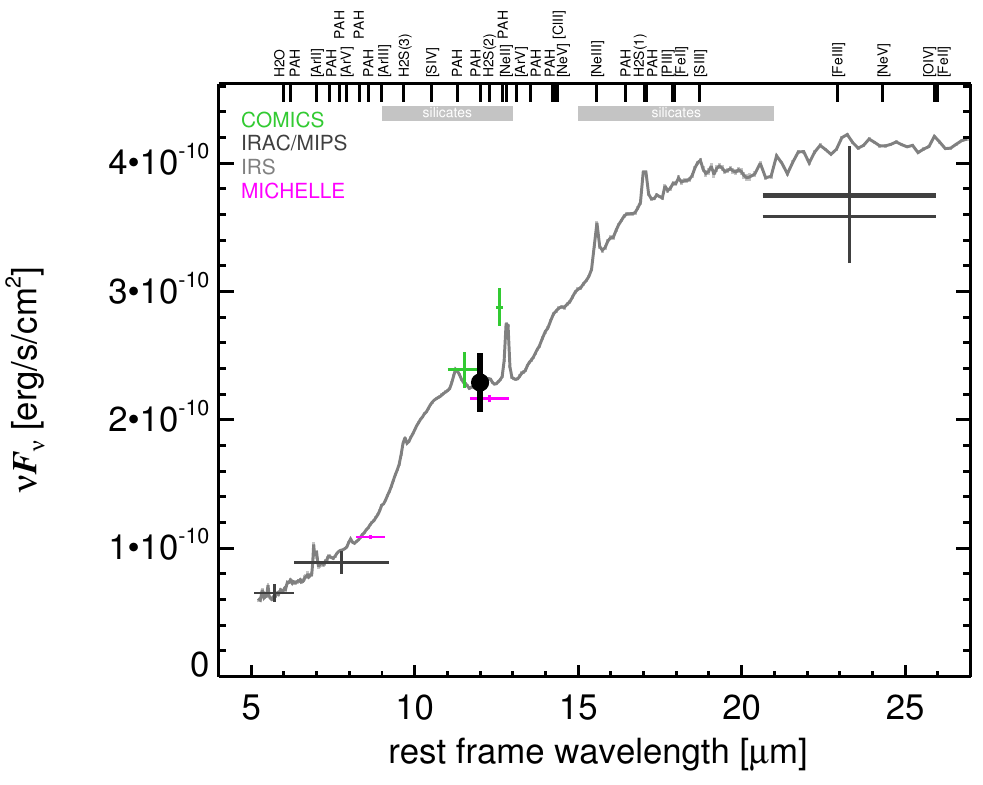}
   \caption{\label{fig:MISED_NGC1275}
      MIR SED of NGC\,1275. The description  of the symbols (if present) is the following.
      Grey crosses and  solid lines mark the \spitzer/IRAC, MIPS and IRS data. 
      The colour coding of the other symbols is: 
      green for COMICS, magenta for Michelle, blue for T-ReCS and red for VISIR data.
      Darker-coloured solid lines mark spectra of the corresponding instrument.
      The black filled circles mark the nuclear 12 and $18\,\mu$m  continuum emission estimate from the data.
      The ticks on the top axis mark positions of common MIR emission lines, while the light grey horizontal bars mark wavelength ranges affected by the silicate 10 and 18$\mu$m features.}
\end{figure}
\clearpage

\twocolumn[\begin{@twocolumnfalse}  
\subsection{NGC\,1365}\label{app:NGC1365}
NGC\,1365 is an infrared-luminous low-inclination barred spiral galaxy at a distance of $D=$ $17.9\pm2.7$\,Mpc (NED redshift-independent median) with an AGN and intense circum-nuclear star formation (see \citealt{lindblad_ngc_1999} for a global review, and \citealt{alonso-herrero_probing_2012} for a dedicated infrared study of the nuclear region).
The AGN has been classified optically either as Sy\,1.5 \citep{schulz_clues_1999}, Sy\,1.8 \citep{edmunds_nuclear_1982} or as Sy\,1.9/H\,II \citep{trippe_multi-wavelength_2010}. 
At radio wavelengths, the most prominent feature is the kiloparsec-scale star formation ring ($8\arcsec \times 20\arcsec$; PA$\sim30\degree$, \citealt{sandqvist_radio_1982}), which contains numerous bright super star clusters \citep{kristen_imaging_1997,galliano_mid-infrared_2005}. 
The nucleus is only a faint radio source  with a possible jet emanating $\sim5\arcsec\sim440$\,pc towards the south-east with a PA$\sim125\degree$   \citep{sandqvist_central_1995}.
This jet coincides with the one-side cone-like outflowing \oiii emission \citep{phillips_remarkable_1983,jorsater_kinematics_1984,storchi-bergmann_detection_1991,kristen_imaging_1997}.
At X-ray wavelengths, the nucleus is highly variable with dramatic changes in obscuration, which suggests that X-ray absorption originates from the BLR clouds  \citep{risaliti_hard_2000-1,risaliti_rapid_2005,risaliti_xmm-newton_2009}.
NGC\,1365 also belongs to the nine-month BAT AGN sample.
The first $N$-band photometric and spectrophotometric observations of NGC\,1365 were carried by \cite{frogel_8-13_1982} \cite{roche_8-13_1984} and \cite{devereux_spatial_1987}.
After the \irass images, \cite{telesco_genesis_1993} obtained an $N$-band map with the NASA MSFC bolometer array at the IRTF, which showed the extended complex MIR morphology with the bright off-nuclear super clusters as one blended source.
NGC\,1365 was also observed with \isoo \citep{roussel_atlas_2001,forster_schreiber_warm_2004} but the nuclear region is not sufficiently resolved to isolate the AGN.
The first subarcsecond resolution $N$-band images of the nuclear region were obtained with CTIO 4\,m/OSCIR in 1998 \citep{ramos_almeida_infrared_2009}, and in 2001 and 2002 with ESO 3.6\,m/TIMMI2 \citep{siebenmorgen_mid-infrared_2004,galliano_mid-infrared_2005,raban_core_2008}.
Three particularly bright compact MIR sources inside the starburst ring $\sim 9\arcsec\sim780\,$pc north-west of the nucleus were detected.
The nucleus itself appeared slightly elongated along the east-west direction.
Despite lower angular resolution, the same morphology is visible in the \spitzer/IRAC and MIPS images.
The nucleus becomes fainter compared to the star formation with increasing wavelengths (see also \citealt{alonso-herrero_probing_2012}).
The PBCD IRAC $8.0\,\mu$m image is partly saturated and not used (see \citealt{gallimore_infrared_2010}), while our IRAC $5.8\,\mu$m flux measurements of the nuclear component is significantly lower than the value published in \cite{gallimore_infrared_2010}.
The IRS LR staring-mode spectrum is star-formation dominated with prominent PAH features, possibly silicate $10\,\mu$m absorption and a red spectral slope in $\nu F_\nu$-space, although also the AGN-indicative \nev was detected (see also \citealt{buchanan_spitzer_2006,brandl_mid-infrared_2006,dudik_mid-infrared_2007,wu_spitzer/irs_2009,bernard-salas_spitzer_2009,tommasin_spitzer-irs_2010,gallimore_infrared_2010,alonso-herrero_probing_2012}).
The nuclear region of NGC\,1365 was observed with VISIR in the NEII filter in 2004 \citep{galliano_extremely_2008} and in three additional narrow $N$-band filters in 2009 (this work). 
We focus only on the nucleus and refer to \cite{galliano_extremely_2008} for a detailed analysis of the non-nuclear emission as seen by VISIR.
In addition, a T-ReCS LR $N$-band spectrum was obtained \citep{gonzalez-martin_dust_2013}, as well as Si2 and Qa images, which were still not publically available at the time of writing \citep{alonso-herrero_probing_2012}.
In all the VISIR images, a slightly elongated MIR nucleus  but with inconsistent position angles was detected. 
Therefore, we classify its MIR extension as uncertain.
Our nuclear VISIR photometry is on average $\sim 58\%$ lower than the \spitzerr spectrophotometry, lower than the TIMMI2 photometry, consistent with the OSCIR photometry, and $\sim 25\%$ higher than the PSF-extracted T-ReCS spectrum. 
This indicates, the nucleus is indeed slightly extended in the VISIR images.
Interestingly, the T-ReCS spectrum is basically free of any spectral features (similar to the TIMMI2 spectrum by \citealt{siebenmorgen_mid-infrared_2004}).
Thus, it most likely traces uncontaminated AGN emission, and we use the T-ReCS spectrum for the $12\,\mu$m continuum emission estimate.
We conclude that star formation dominates the MIR emission in the central 350\,pc of NGC\,1365.
Note that the MIR nucleus appears slightly resolved and elongated with a size $\lesssim 2$\,pc in the MIR interferometric MIDI observations described by \cite{tristram_parsec-scale_2009} and \cite{burtscher_diversity_2013}. 
\newline\end{@twocolumnfalse}]

\begin{figure}
   \centering
   \includegraphics[angle=0,width=8.500cm]{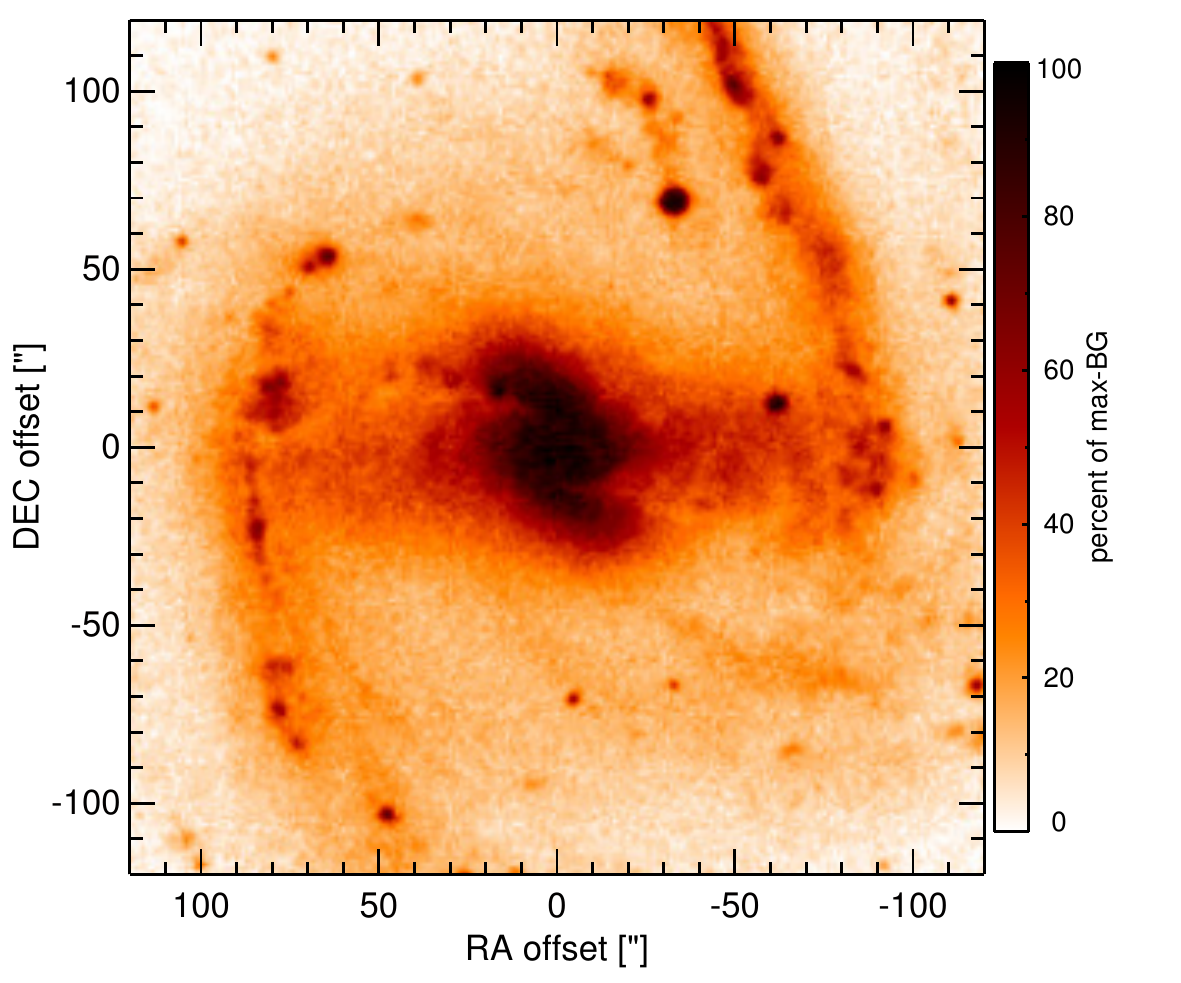}
    \caption{\label{fig:OPTim_NGC1365}
             Optical image (DSS, red filter) of NGC\,1365. Displayed are the central $4\arcmin$ with North up and East to the left. 
              The colour scaling is linear with white corresponding to the median background and black to the $0.01\%$ pixels with the highest intensity.  
           }
\end{figure}
\begin{figure}
   \centering
   \includegraphics[angle=0,height=3.11cm]{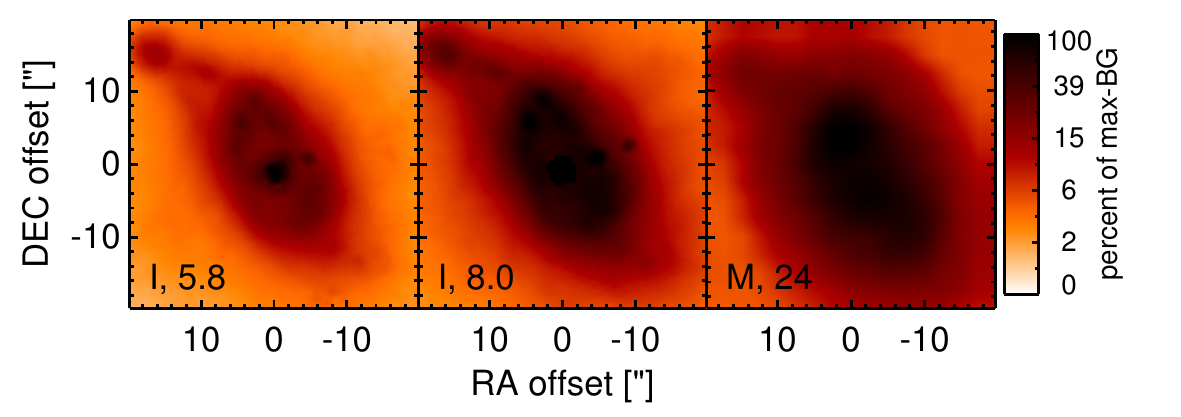}
    \caption{\label{fig:INTim_NGC1365}
             \spitzerr MIR images of NGC\,1365. Displayed are the inner $40\arcsec$ with North up and East to the left. The colour scaling is logarithmic with white corresponding to median background and black to the $0.1\%$ pixels with the highest intensity.
             The label in the bottom left states instrument and central wavelength of the filter in $\mu$m (I: IRAC, M: MIPS).
             Note that the apparent off-nuclear compact sources in the IRAC 5.8 and $8.0\,\mu$m images are instrumental artefacts.
           }
\end{figure}
\begin{figure}
   \centering
   \includegraphics[angle=0,width=8.500cm]{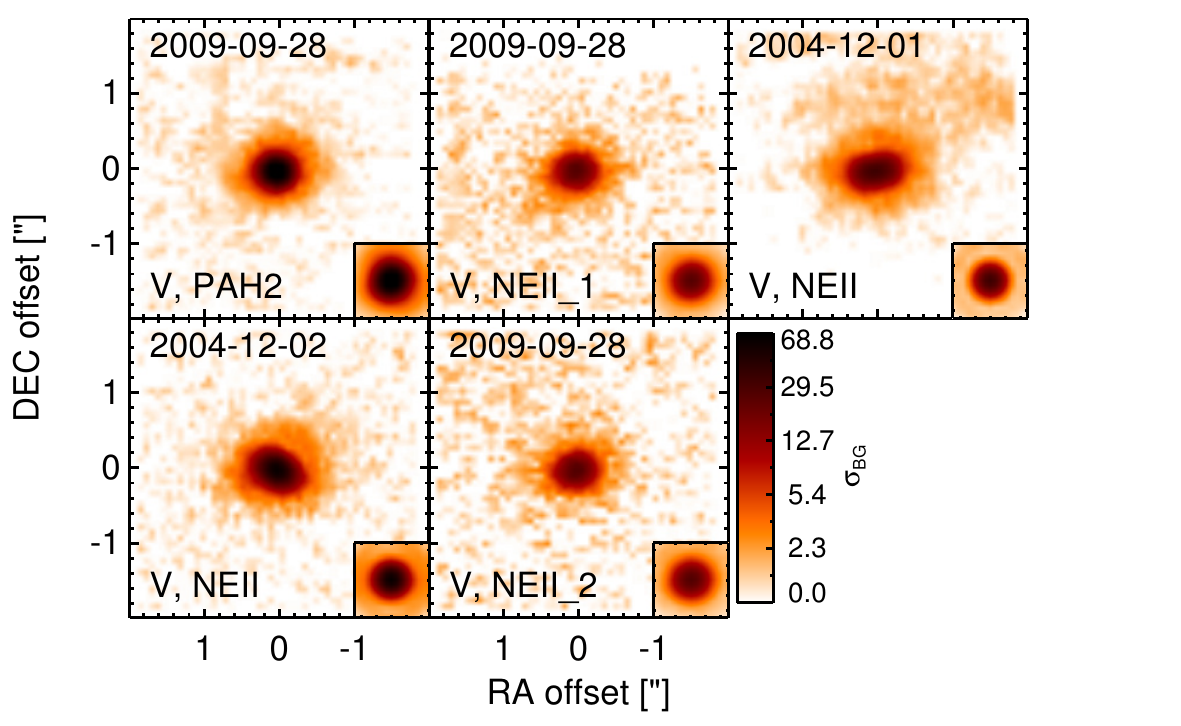}
    \caption{\label{fig:HARim_NGC1365}
             Subarcsecond-resolution MIR images of NGC\,1365 sorted by increasing filter wavelength. 
             Displayed are the inner $4\arcsec$ with North up and East to the left. 
             The colour scaling is logarithmic with white corresponding to median background and black to the $75\%$ of the highest intensity of all images in units of $\sigbg$.
             The inset image shows the central arcsecond of the PSF from the calibrator star, scaled to match the science target.
             The labels in the bottom left state instrument and filter names (C: COMICS, M: Michelle, T: T-ReCS, V: VISIR).
           }
\end{figure}
\begin{figure}
   \centering
   \includegraphics[angle=0,width=8.50cm]{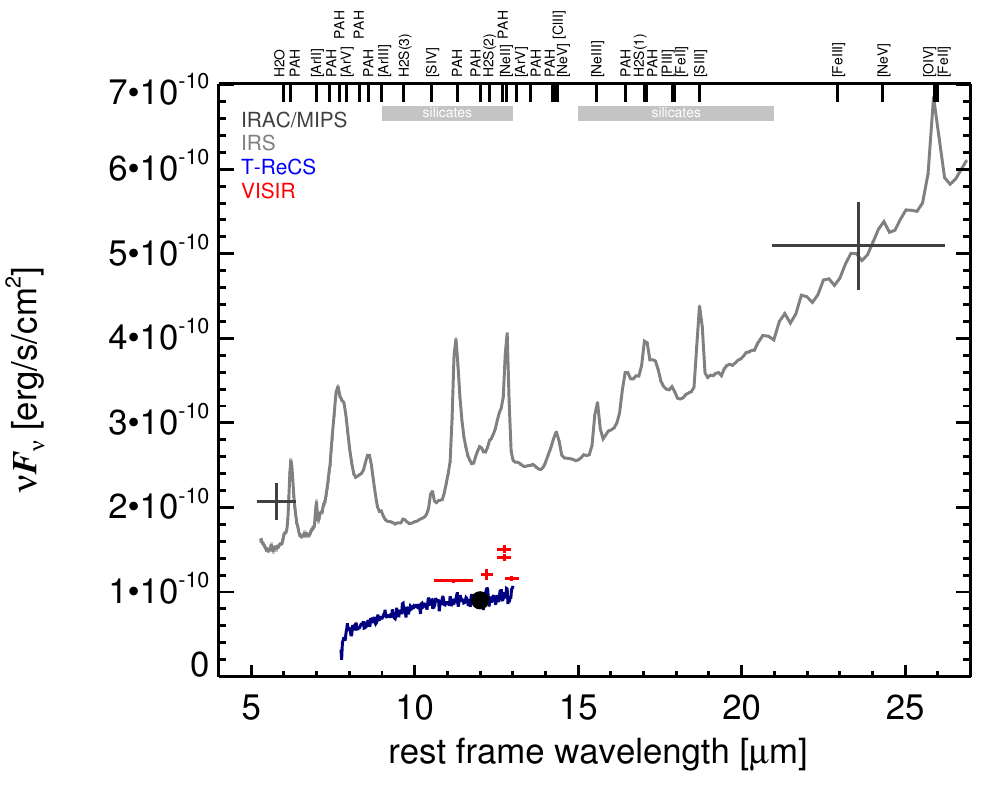}
   \caption{\label{fig:MISED_NGC1365}
      MIR SED of NGC\,1365. The description  of the symbols (if present) is the following.
      Grey crosses and  solid lines mark the \spitzer/IRAC, MIPS and IRS data. 
      The colour coding of the other symbols is: 
      green for COMICS, magenta for Michelle, blue for T-ReCS and red for VISIR data.
      Darker-coloured solid lines mark spectra of the corresponding instrument.
      The black filled circles mark the nuclear 12 and $18\,\mu$m  continuum emission estimate from the data.
      The ticks on the top axis mark positions of common MIR emission lines, while the light grey horizontal bars mark wavelength ranges affected by the silicate 10 and 18$\mu$m features.}
\end{figure}
\clearpage

\twocolumn[\begin{@twocolumnfalse}  
\subsection{NGC\,1386}\label{app:NGC1386}
NGC\,1386  is an edge-on spiral galaxy in the Fornax cluster at a distance of $D=$ $16.5 \pm 0.8$\,Mpc (NED redshift independent median) with a Sy\,2 nucleus with broad lines in the near-infrared \citep{reunanen_near-infrared_2002} that belongs to the nine-month BAT AGN sample.
The nuclear radio source is slightly extended to the south (PA$\sim170\degree$; \citealt{nagar_radio_1999}), and is surrounded by a biconical north-south extended NLR, which has been extensively studied (major axis$\sim6\arcsec\sim480\,$pc; PA$\sim 5\degree$; e.g.,  \citealt{weaver_kinematics_1991,ferruit_hubble_2000,schmitt_hubble_2003,bennert_size_2006-1}).
In addition, a nuclear H$_2$O mega-maser was detected \citep{braatz_survey_1996,braatz_survey_1997}.
The first MIR observations of NGC\,1386 were performed by \cite{phillips_infrared_1980}, followed by \cite{sparks_infrared_1986} and \cite{devereux_spatial_1987}.
Later, NGC\,1386 was also observed with \isoo \citep{clavel_2.5-11_2000,ramos_almeida_mid-infrared_2007} and ESO 3.6\,m/TIMMI2 \citep{siebenmorgen_mid-infrared_2004}.
The MIR nucleus appeared slightly resolved in the TIMMI2 image.
The \spitzer/IRAC, IRS and MIPS images, the MIR nucleus appears nearly unresolved except in the IRAC $8.0\,\mu$m image, where the nucleus is slightly elongated along the north-south direction. 
In addition, faint spiral-like host emission is visible in the IRAC images.
Our IRAC $5.8$ and $8.0\,\mu$m nuclear photometry is $\sim 25\%$ lower than the values by \cite{gallimore_infrared_2010} but in rough agreement with the \spitzer/IRS LR staring mode spectrum, while the MIPS $24\,\mu$m agrees with the value by \cite{temi_spitzer_2009}.
The IRS spectrum is dominated by the deep silicate  $10\,\mu$m and a moderate silicate $18\,\mu$m absorption features in addition to  PAH emission, prominent forbidden emission lines and a shallow spectral slope in $\nu F_\nu$-space (see also \citealt{shi_9.7_2006,wu_spitzer/irs_2009,deo_mid-infrared_2009,tommasin_spitzer-irs_2010,gallimore_infrared_2010}).
NGC\,1386 was imaged with T-ReCS in the broad N and Q filters in 2003 \citep{ramos_almeida_infrared_2009}, and an LR $N$-band spectrum  was presented in \cite{gonzalez-martin_dust_2013}.
and with VISIR in PAH2\_2 and Q2 in 2006 \citep{reunanen_vlt_2010} as well as in PAH1 and SIV\_2 in 2010 (this work).
In all but the latest two VISIR images, a $\sim1\arcsec$ (80\,pc) north-south extended nucleus is visible (PA$\sim0\degree$).
The reason why this extended feature is not visible in the PAH1 and SIV\_2 images is presumably the bad MIR seeing during these observations. 
On the other hand, it is possible that it is intrinsically weaker at shorter wavelengths. 
The extended emission is aligned with the NLR and thus presumably originates from hot optically thin dust in the NLR as already suggested by \cite{bennert_size_2006-1}.
We perform manual PSF-scaling to determine only the unresolved nuclear flux in the VISIR images and obtain values consistent with \cite{reunanen_vlt_2010}, while no matching standard star could be retrieved for the T-ReCS images.
Therefore, we treat the T-ReCS measurements as upper limits.
The resulting VISIR nuclear photometry is on average $\sim 50\%$ lower than the \spitzerr spectrophotometry and agrees with the T-ReCS spectrum, which exhibits the same silicate $10\mu$m absorption but no PAH emission compared to the IRS spectrum. 
Therefore, we use the T-ReCS spectrum to estimate the nuclear 12$\,\mu$m continuum emission corrected for the silicate feature. 
The subarcsecond-resolution $Q$-band in relation to the $N$-band values indicate that the silicate $18\,\mu$m absorption has at least the same strength as in the IRS spectrum at subarcsecond resolution. 
We conclude that the nuclear obscuring structure contributes less than half of the MIR emission in the central 300\,pc of NGC\,1386, while the other part is caused by a dusty NLR and star formation.
\newline\end{@twocolumnfalse}]

\begin{figure}
   \centering
   \includegraphics[angle=0,width=8.500cm]{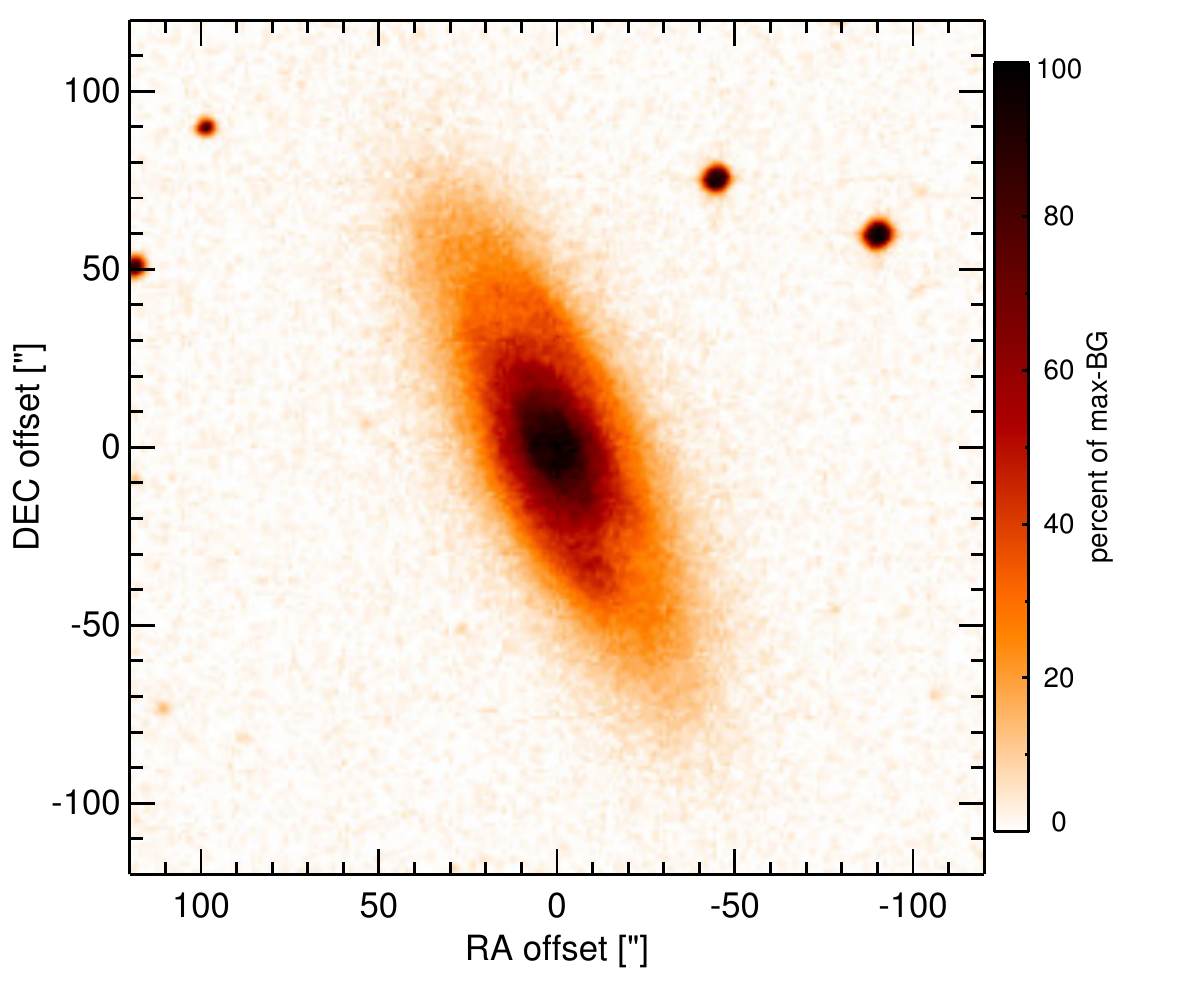}
    \caption{\label{fig:OPTim_NGC1386}
             Optical image (DSS, red filter) of NGC\,1386. Displayed are the central $4\arcmin$ with North up and East to the left. 
              The colour scaling is linear with white corresponding to the median background and black to the $0.01\%$ pixels with the highest intensity.  
           }
\end{figure}
\begin{figure}
   \centering
   \includegraphics[angle=0,height=3.11cm]{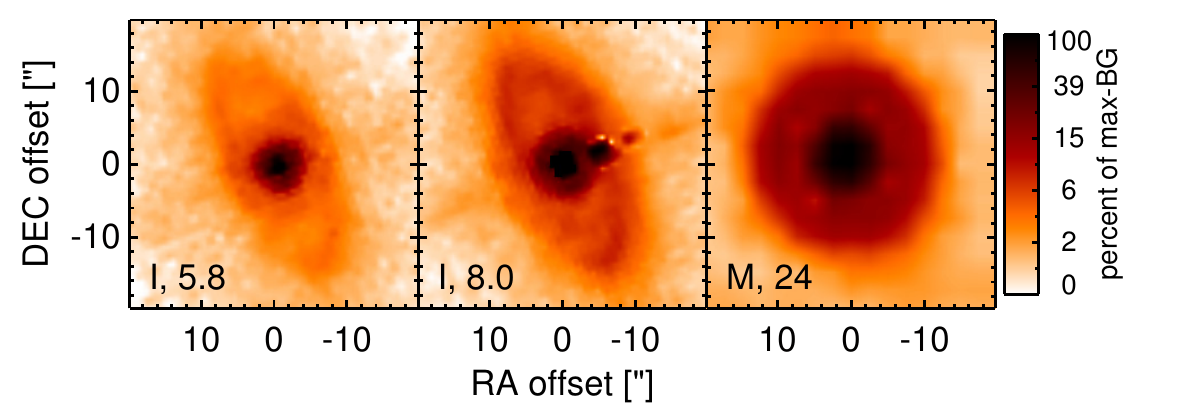}
    \caption{\label{fig:INTim_NGC1386}
             \spitzerr MIR images of NGC\,1386. Displayed are the inner $40\arcsec$ with North up and East to the left. The colour scaling is logarithmic with white corresponding to median background and black to the $0.1\%$ pixels with the highest intensity.
             The label in the bottom left states instrument and central wavelength of the filter in $\mu$m (I: IRAC, M: MIPS). 
             Note that the apparent off-nuclear compact sources in the IRAC $8.0\,\mu$m image are instrumental artefacts.
           }
\end{figure}
\begin{figure}
   \centering
   \includegraphics[angle=0,width=8.500cm]{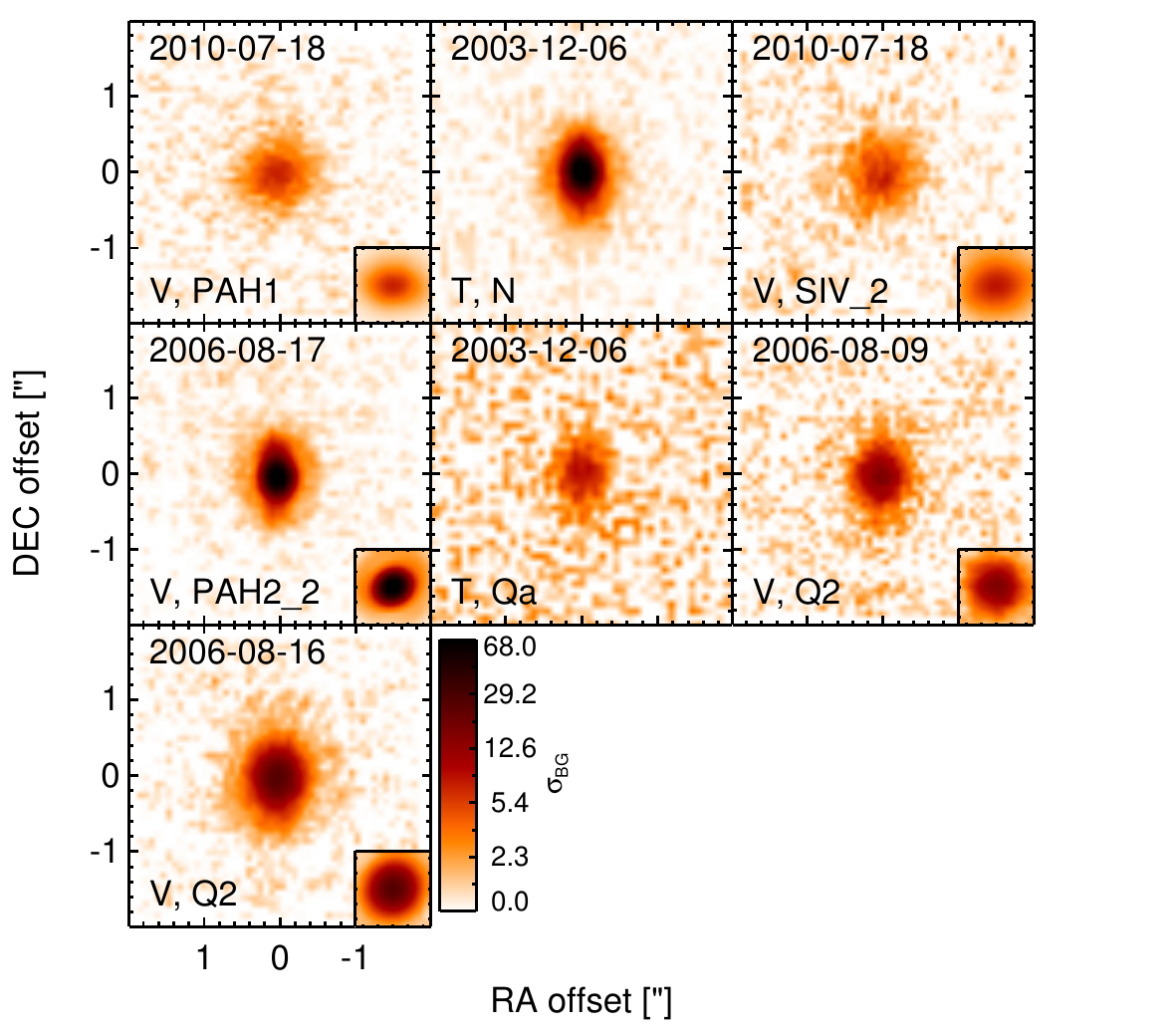}
    \caption{\label{fig:HARim_NGC1386}
             Subarcsecond-resolution MIR images of NGC\,1386 sorted by increasing filter wavelength. 
             Displayed are the inner $4\arcsec$ with North up and East to the left. 
             The colour scaling is logarithmic with white corresponding to median background and black to the $75\%$ of the highest intensity of all images in units of $\sigbg$.
             The inset image shows the central arcsecond of the PSF from the calibrator star, scaled to match the science target.
             The labels in the bottom left state instrument and filter names (C: COMICS, M: Michelle, T: T-ReCS, V: VISIR).
           }
\end{figure}
\begin{figure}
   \centering
   \includegraphics[angle=0,width=8.50cm]{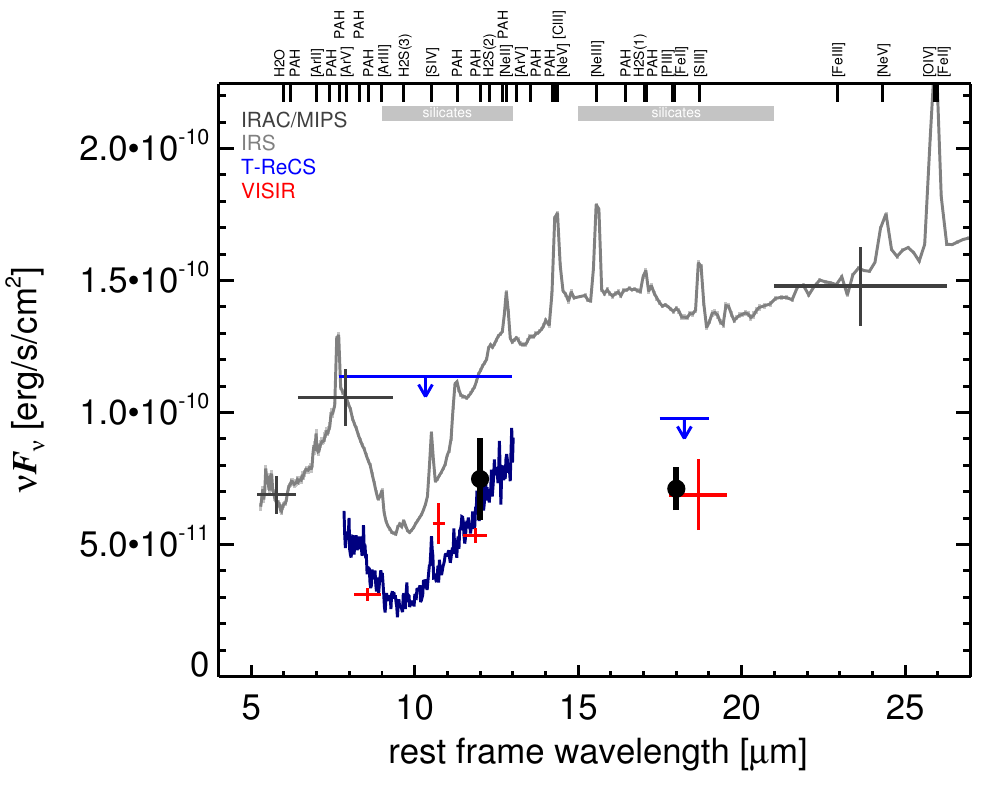}
   \caption{\label{fig:MISED_NGC1386}
      MIR SED of NGC\,1386. The description  of the symbols (if present) is the following.
      Grey crosses and  solid lines mark the \spitzer/IRAC, MIPS and IRS data. 
      The colour coding of the other symbols is: 
      green for COMICS, magenta for Michelle, blue for T-ReCS and red for VISIR data.
      Darker-coloured solid lines mark spectra of the corresponding instrument.
      The black filled circles mark the nuclear 12 and $18\,\mu$m  continuum emission estimate from the data.
      The ticks on the top axis mark positions of common MIR emission lines, while the light grey horizontal bars mark wavelength ranges affected by the silicate 10 and 18$\mu$m features.}
\end{figure}
\clearpage

\twocolumn[\begin{@twocolumnfalse}  
\subsection{NGC\,1433}\label{app:NGC1433}
NGC\,1433 is a low-inclination barred spiral galaxy at a distance of $D=$ $8.3 \pm 2.3$\,Mpc \citep{tully_extragalactic_2009} with a possibly active nucleus, which has been classified as a Sy\,2.0 \citep{veron-cetty_study_1986} or a LINER \citep{storchi-bergmann_ultraviolet_1995}.
This AGN has barely been studied and in particular no X-ray works are available.
At radio wavelengths only a low-angular resolution measurement was performed, which detected a weak non-thermal nuclear compact source (diameter$\sim4\arcmin$; PA$\sim90\degree$; \citealt{harnett_radio_1987}).
We  conservatively treat NGC\,1433 as uncertain AGN.
In the MIR, NGC\,1433 was first detected with \irass and was followed up with \isoo \citep{roussel_atlas_2001} and \spitzer/IRAC, IRS and MIPS.
Bright host galaxy emission with an elliptical core but no distinct unresolved component was detected in the corresponding IRAC and MIPS images.
The IRS LR staring-mode spectrum is completely dominated by strong PAH emission features with a rather flat underlying continuum.
None of the AGN-indicative high-ionization lines (e.g., \nev) are clearly detected \citep{pereira-santaella_mid-infrared_2010}.
The nuclear region of NGC\,1433 was imaged with T-ReCS in the Si2 filter in 2005 and with VISIR in the PAH2\_2 filter in 2009 (both unpublished, to our knowledge).
The derived flux upper limits are $\sim 76\%$ lower than the \spitzerr spectrophotometry.
Therefore, the presence of an AGN in NGC\,1433 is highly uncertain and in any case much weaker than the host emission of the central $\sim150$\,pc.
\newline\end{@twocolumnfalse}]

\begin{figure}
   \centering
   \includegraphics[angle=0,width=8.500cm]{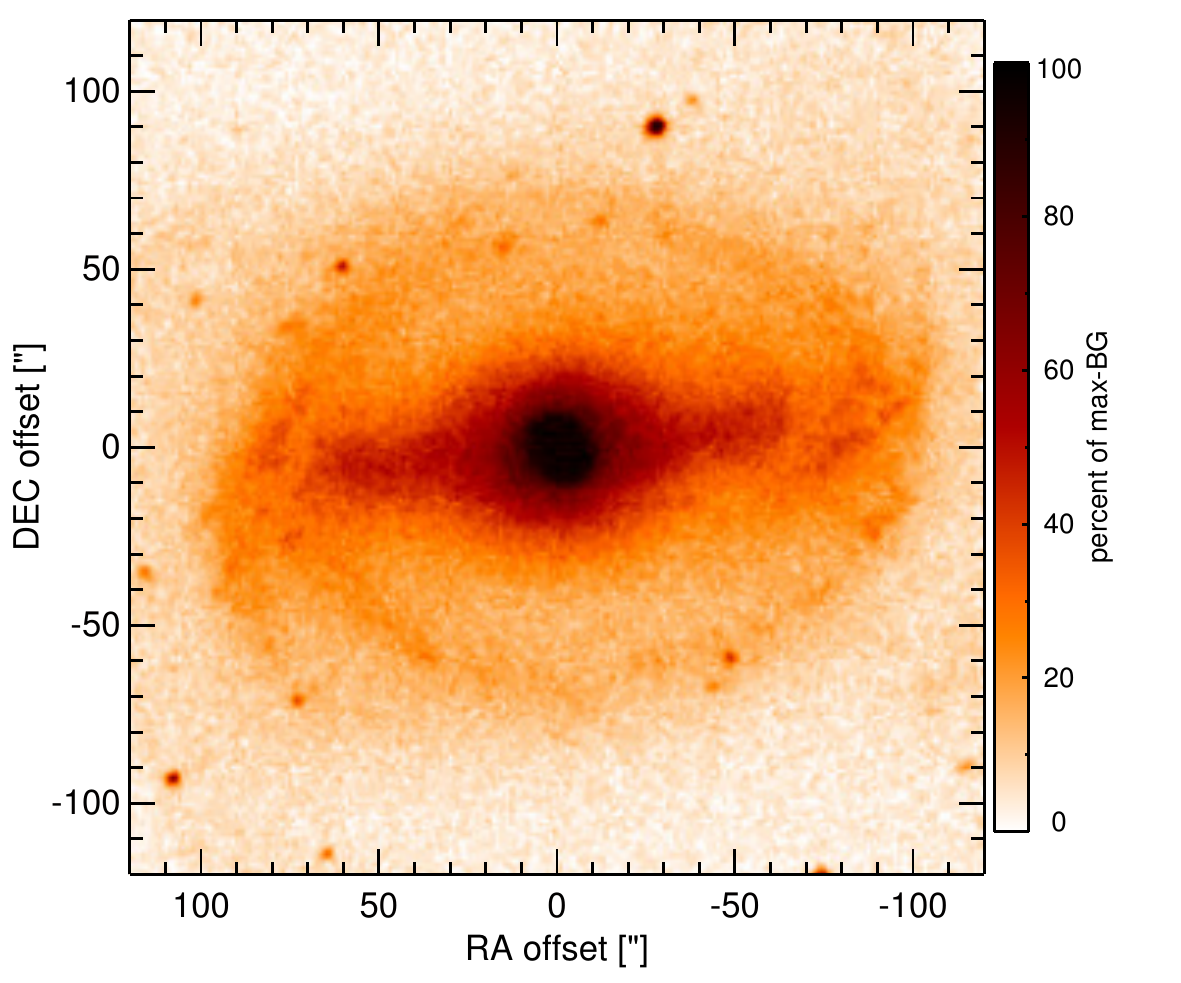}
    \caption{\label{fig:OPTim_NGC1433}
             Optical image (DSS, red filter) of NGC\,1433. Displayed are the central $4\arcmin$ with North up and East to the left. 
              The colour scaling is linear with white corresponding to the median background and black to the $0.01\%$ pixels with the highest intensity.  
           }
\end{figure}
\begin{figure}
   \centering
   \includegraphics[angle=0,height=3.11cm]{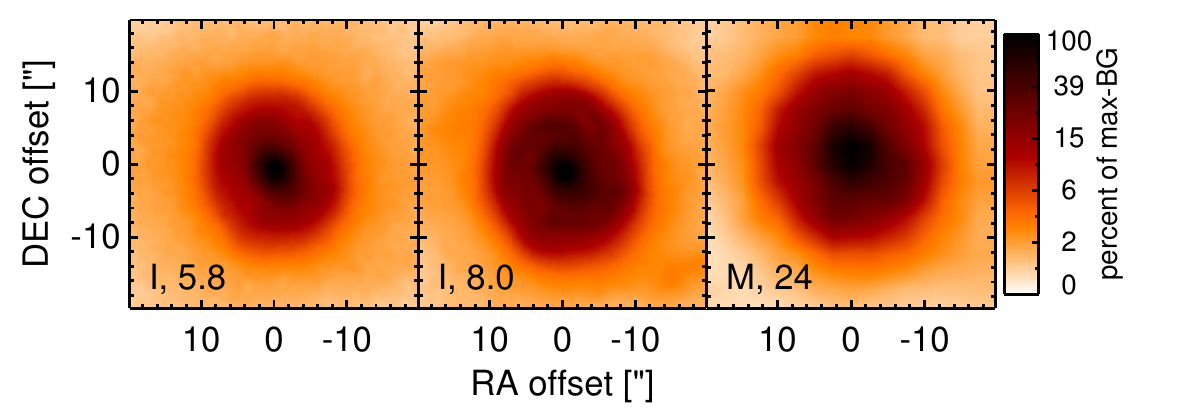}
    \caption{\label{fig:INTim_NGC1433}
             \spitzerr MIR images of NGC\,1433. Displayed are the inner $40\arcsec$ with North up and East to the left. The colour scaling is logarithmic with white corresponding to median background and black to the $0.1\%$ pixels with the highest intensity.
             The label in the bottom left states instrument and central wavelength of the filter in $\mu$m (I: IRAC, M: MIPS). 
           }
\end{figure}
\begin{figure}
   \centering
   \includegraphics[angle=0,width=8.50cm]{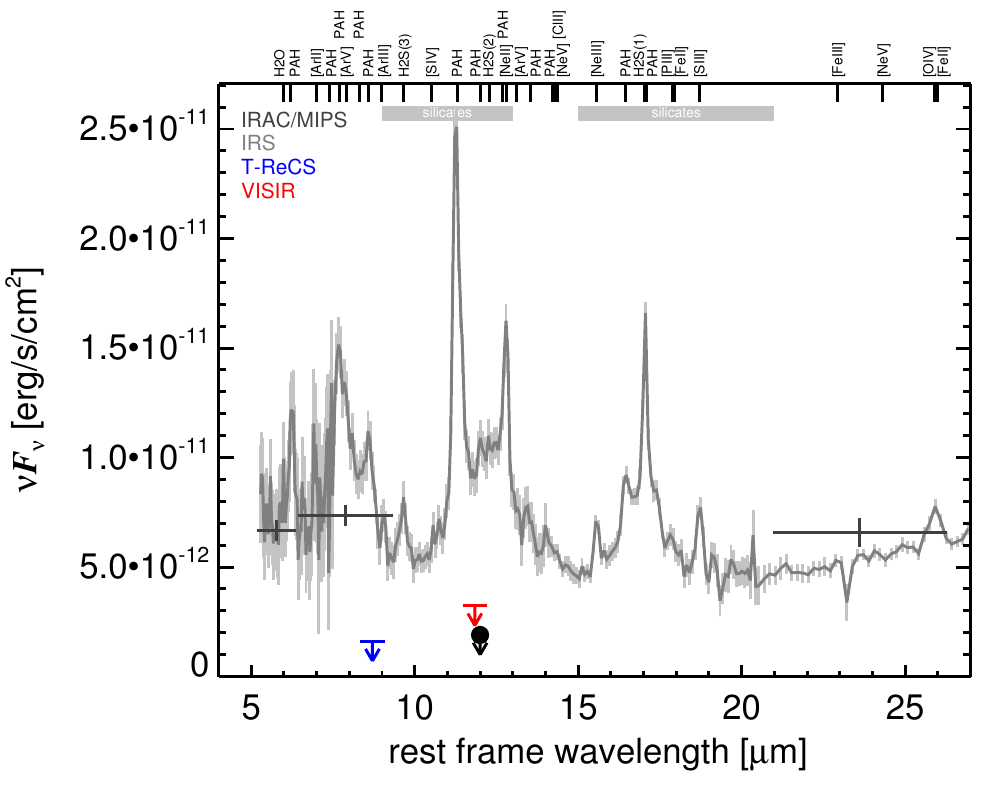}
   \caption{\label{fig:MISED_NGC1433}
      MIR SED of NGC\,1433. The description  of the symbols (if present) is the following.
      Grey crosses and  solid lines mark the \spitzer/IRAC, MIPS and IRS data. 
      The colour coding of the other symbols is: 
      green for COMICS, magenta for Michelle, blue for T-ReCS and red for VISIR data.
      Darker-coloured solid lines mark spectra of the corresponding instrument.
      The black filled circles mark the nuclear 12 and $18\,\mu$m  continuum emission estimate from the data.
      The ticks on the top axis mark positions of common MIR emission lines, while the light grey horizontal bars mark wavelength ranges affected by the silicate 10 and 18$\mu$m features.}
\end{figure}
\clearpage

\twocolumn[\begin{@twocolumnfalse}  
\subsection{NGC\,1553}\label{app:NGC1553}
NGC\,1553 is an inclined early-type spiral galaxy interacting with NGC\,1549 (11.5\arcmin\, away) at a distance of $D=$ $16.4 \pm 5.1$\,Mpc (NED redshift-independent median) with an active nucleus, classified as a LINER or transition object \citep{flohic_central_2006}.
The presence of an AGN in NGC\,1553 is supported by the detection of a compact source in radio \citep{harnett_radio_1987} and in X-rays \citep{blanton_diffuse_2001,flohic_central_2006}.
After its detection in \iras, NGC\,1553 was observed with \isoo \citep{temi_cold_2004} and with \spitzer/IRS and MIPS.
In the MIPS $24\,\mu$m image, a compact north-west elongated MIR nucleus (PA$\sim-30\degree$) embedded within prominent diffuse host emission was detected. 
Because we measure only the unresolved nuclear component, our MIPS $24\,\mu$m flux value is significantly lower than in \cite{temi_spitzer_2009}.
The IRS LR staring-mode spectrum exhibits silicate $10\,\mu$m  and possibly silicate 18$\,\mu$m emission, a prominent PAH 11.3$\,\mu$m feature and a blue spectral slope in $\nu F_\nu$-space, indicating mainly passive host emission (see also \citealt{panuzzo_nearby_2011}).
The nuclear region of NGC\,1553 was observed with VISIR in the PAH2\_2 filter in 2009 (unpublished, to our knowledge) but no nuclear source was detected.
Our derived flux upper limit is significantly lower than the \spitzerr spectrophotometry.
Therefore, from the subarcsecond MIR point of view no definite conclusion can be drawn about the presence of an AGN in NGC\,1553 but note that \cite{panuzzo_nearby_2011} have detected  weak \nev emission in the IRS spectrum.
\newline\end{@twocolumnfalse}]

\begin{figure}
   \centering
   \includegraphics[angle=0,width=8.500cm]{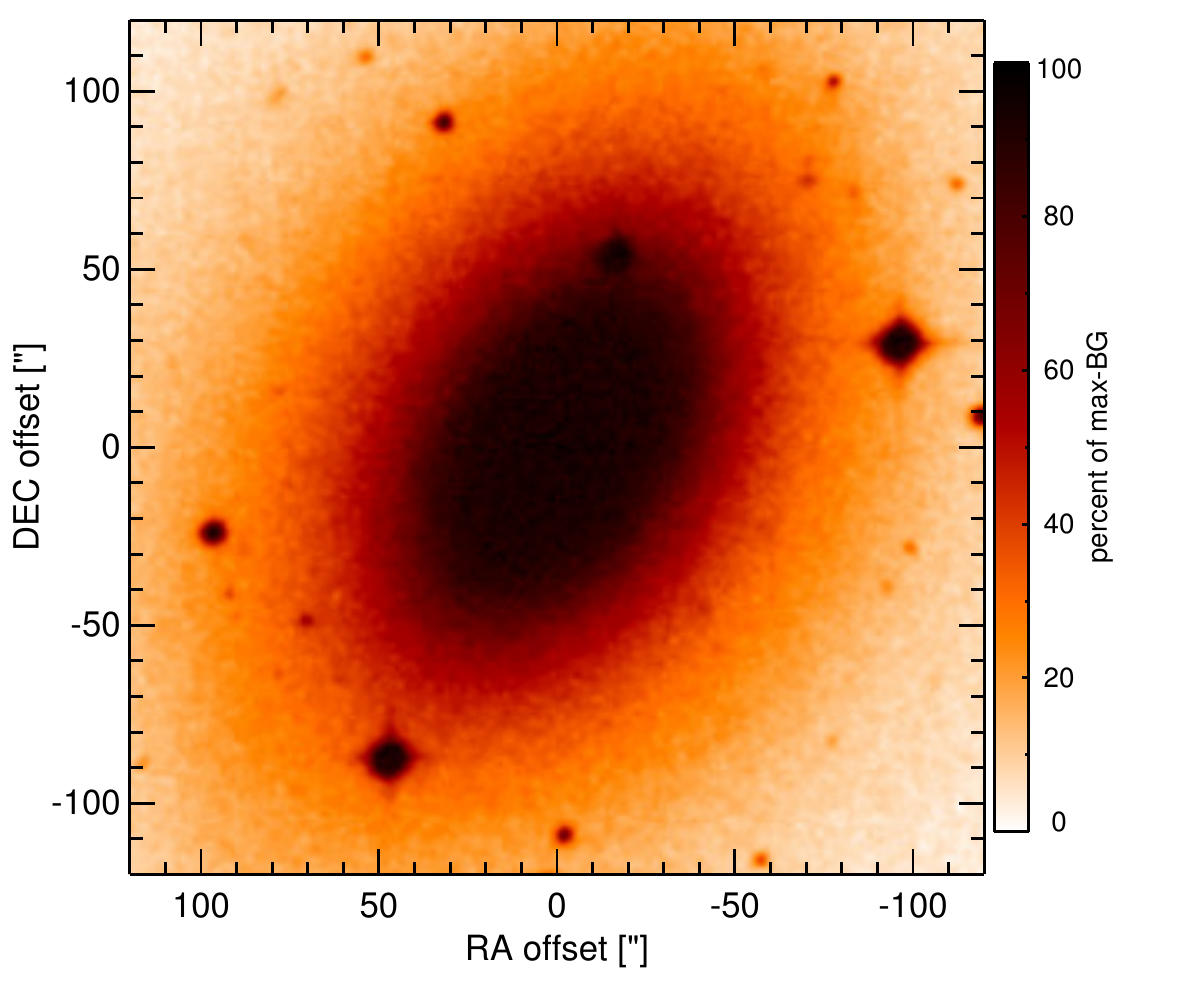}
    \caption{\label{fig:OPTim_NGC1553}
             Optical image (DSS, red filter) of NGC\,1553. Displayed are the central $4\arcmin$ with North up and East to the left. 
              The colour scaling is linear with white corresponding to the median background and black to the $0.01\%$ pixels with the highest intensity.  
           }
\end{figure}
\begin{figure}
   \centering
   \includegraphics[angle=0,height=3.11cm]{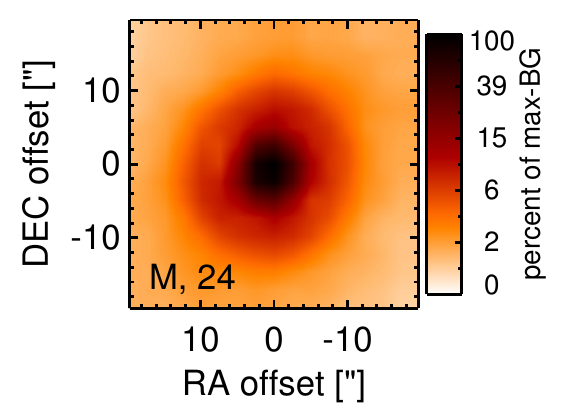}
    \caption{\label{fig:INTim_NGC1553}
             \spitzerr MIR images of NGC\,1553. Displayed are the inner $40\arcsec$ with North up and East to the left. The colour scaling is logarithmic with white corresponding to median background and black to the $0.1\%$ pixels with the highest intensity.
             The label in the bottom left states instrument and central wavelength of the filter in $\mu$m (I: IRAC, M: MIPS). 
           }
\end{figure}
\begin{figure}
   \centering
   \includegraphics[angle=0,width=8.50cm]{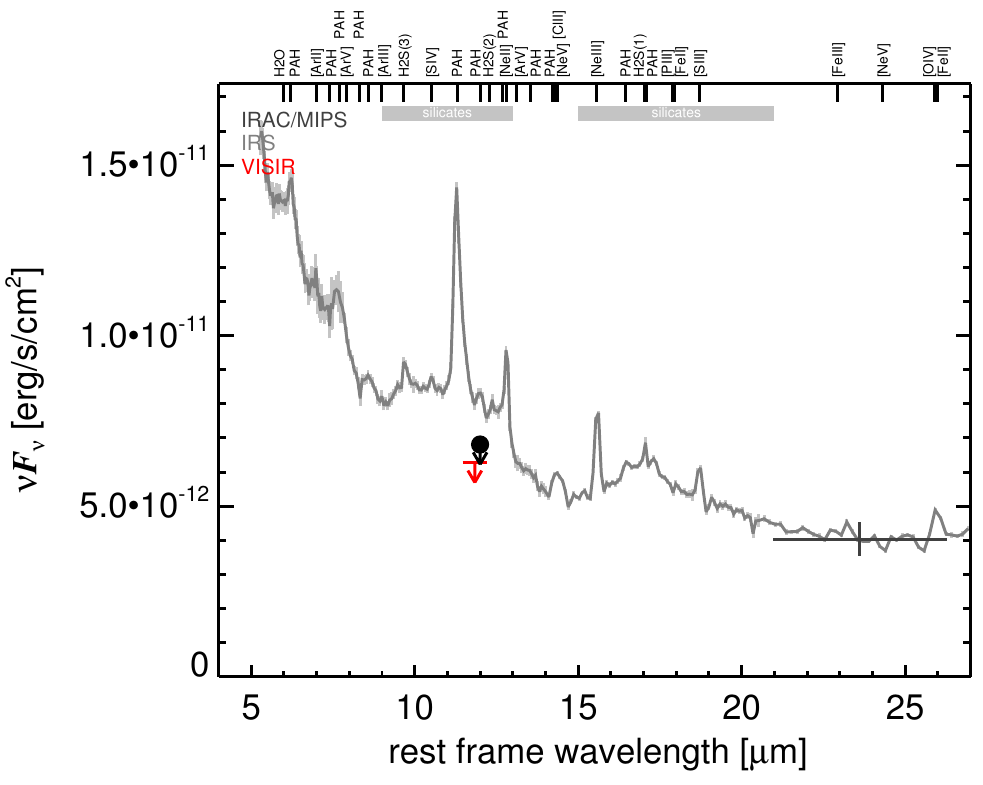}
   \caption{\label{fig:MISED_NGC1553}
      MIR SED of NGC\,1553. The description  of the symbols (if present) is the following.
      Grey crosses and  solid lines mark the \spitzer/IRAC, MIPS and IRS data. 
      The colour coding of the other symbols is: 
      green for COMICS, magenta for Michelle, blue for T-ReCS and red for VISIR data.
      Darker-coloured solid lines mark spectra of the corresponding instrument.
      The black filled circles mark the nuclear 12 and $18\,\mu$m  continuum emission estimate from the data.
      The ticks on the top axis mark positions of common MIR emission lines, while the light grey horizontal bars mark wavelength ranges affected by the silicate 10 and 18$\mu$m features.}
\end{figure}
\clearpage

\twocolumn[\begin{@twocolumnfalse}  
\subsection{NGC\,1566}\label{app:NGC1566}
NGC\,1566 is a face-on grand-design spiral galaxy at a distance of $D=$ $14.3 \pm 5.9$\,Mpc (average of \citealt{tully_nearby_1988}, \cite{willick_homogeneous_1997} and $D_L$ from NED) with a Sy\,1.5 nucleus,
which in the past showed spectral variation ranging between Sy\,1.2 and Sy\,1.9 \citep{alloin_recent_1985,alloin_recurrent_1986}.
It appears point-like at radio wavelengths (e.g., \citealt{morganti_radio_1999}).
The NLR forms in projection a short but wide cone extending $\sim0.5\arcsec\sim35\,$pc to the south-east (PA$\sim135\degree$; e.g., \citealt{schmitt_comparison_1996}). 
After its detection in \iras, NGC\,1566 was observed in the MIR with 
\isoo \citep{clavel_2.5-11_2000,ramos_almeida_mid-infrared_2007} and ESO MPI 2.2\,m/MANIAC \citep{krabbe_n-band_2001}, where the nucleus appeared unresolved.
In the IRAC and MIPS images, a compact MIR nucleus embedded within diffuse extended emission and on larger scale surrounded by spiral-like emission was detected. 
Me measure the nuclear component only, which provides IRAC $5.8$ and $8.0\,\mu$m and MIPS $24\,\mu$m fluxes significantly lower than the values by e.g., \cite{dale_infrared_2005,munoz-mateos_radial_2009,gallimore_infrared_2010}.
The IRS LR mapping-mode spectrum shows strong PAH emission, possible weak silicate $10\,\mu$m absorption and a rather flat spectral slope in $\nu F_\nu$-space (see \citealt{smith_mid-infrared_2007,wu_spitzer/irs_2009,gallimore_infrared_2010} for more accurate versions).
Thus, the MIR SED at arcsecond scales appears star-formation dominated.
The nuclear region of NGC\,1566 was observed with T-ReCS in the Si2 and Qa filters in 2005 \citep{ramos_almeida_infrared_2009}, and with VISIR in total in five narrow $N$-band and one $Q$-band filters spread over 2006 \citep{reunanen_vlt_2010}, 2009 \citep{asmus_mid-infrared_2011} and 2010 (this work).
In all cases, a compact MIR nucleus without any other host emission was detected. 
Note that for the T-ReCS images no corresponding standard star could be retrieved, and thus median flux conversion factors are used. 
The nucleus appears marginally resolved in all VISIR images except in the diffraction-limited PAH2\_2 image from 2006 but with inconsistent position angles. 
Therefore, it remains uncertain, whether the MIR nucleus of NGC\,1566 is in general extended at subarcsecond resolution.
On the other hand, the $N$-band images with the largest FWHM values (PAH2 and NEII\_1; $\sim 0.5\arcsec$) also exhibit two times higher fluxes than in the SIV\_2 and PAH2\_2 filters with small FWHM values. 
It is thus possible, that the PAH2 and NEII\_1 filters are significantly affected by extended emission, e.g., PAH. 
Deep subarcsecond-resolution MIR images covering the $N$-band obtained under good conditions are required to verify the nuclear MIR structure. 
The nuclear photometry in general is consistent with the previously published values and on average $\sim 49\%$ lower than the \spitzerr spectrophotometry.
We conclude that star formation dominates the MIR emission form the projected central $\sim250\,$pc in NGC\,1566.
\newline\end{@twocolumnfalse}]

\begin{figure}
   \centering
   \includegraphics[angle=0,width=8.500cm]{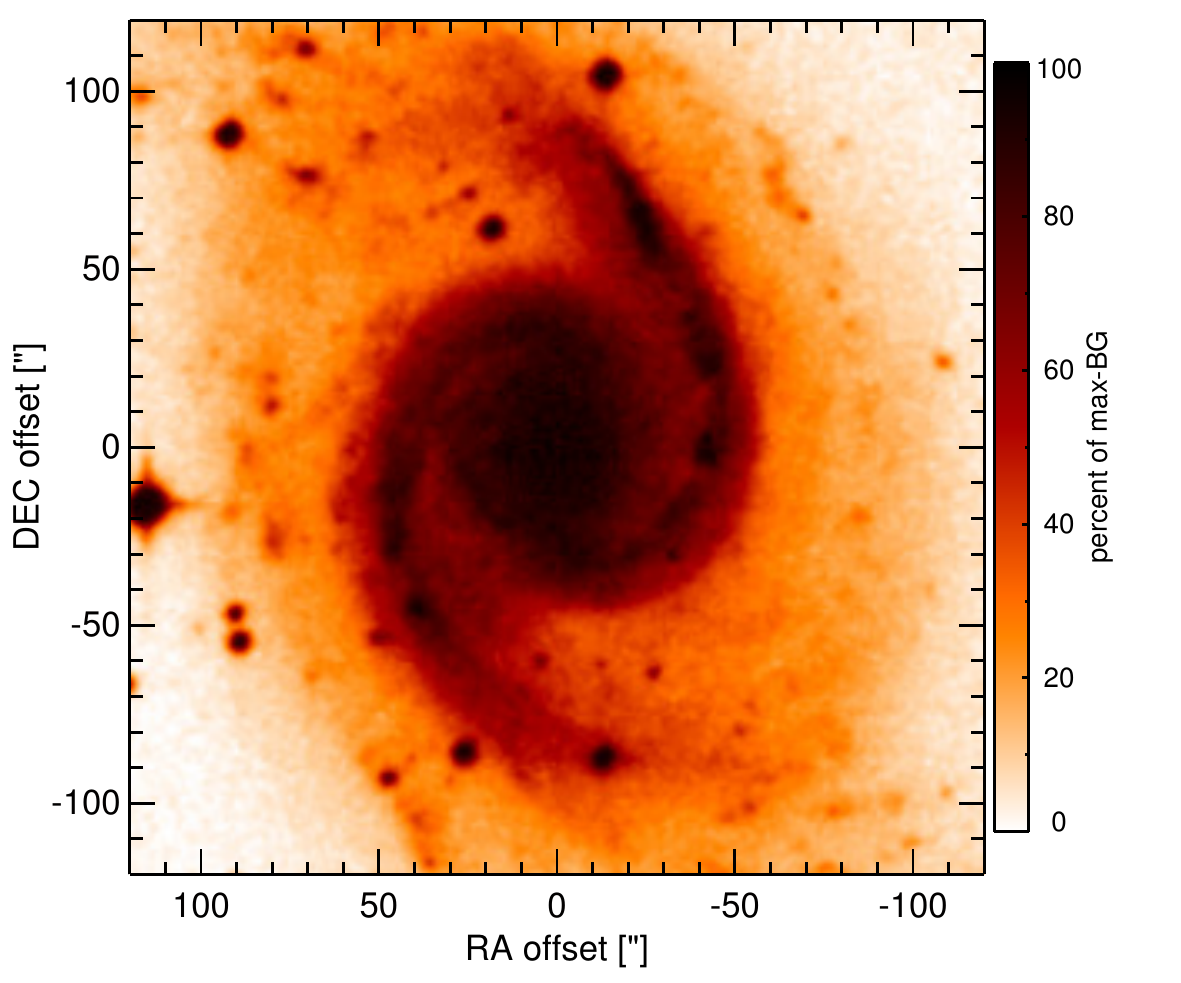}
    \caption{\label{fig:OPTim_NGC1566}
             Optical image (DSS, red filter) of NGC\,1566. Displayed are the central $4\arcmin$ with North up and East to the left. 
              The colour scaling is linear with white corresponding to the median background and black to the $0.01\%$ pixels with the highest intensity.  
           }
\end{figure}
\begin{figure}
   \centering
   \includegraphics[angle=0,height=3.11cm]{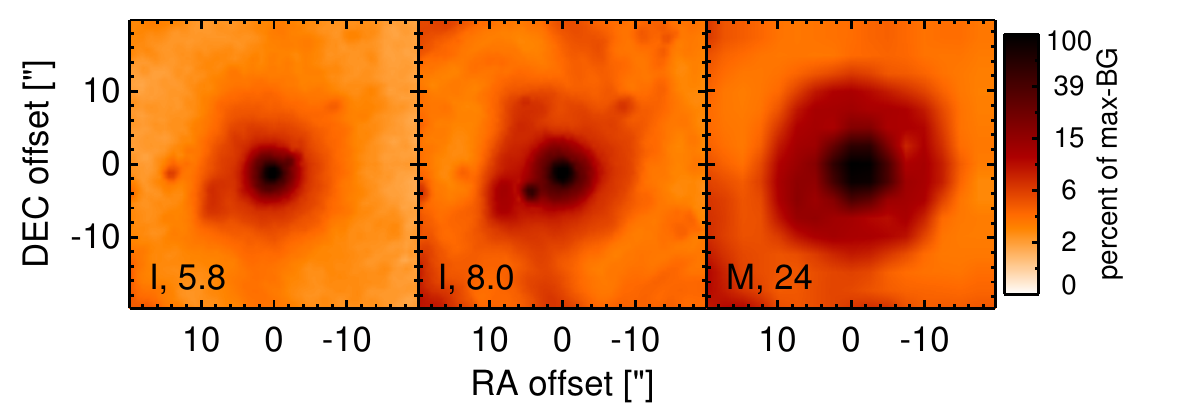}
    \caption{\label{fig:INTim_NGC1566}
             \spitzerr MIR images of NGC\,1566. Displayed are the inner $40\arcsec$ with North up and East to the left. The colour scaling is logarithmic with white corresponding to median background and black to the $0.1\%$ pixels with the highest intensity.
             The label in the bottom left states instrument and central wavelength of the filter in $\mu$m (I: IRAC, M: MIPS).
             Note that the apparent off-nuclear compact source in the IRAC $8.0\,\mu$m image is an instrumental artefact.
           }
\end{figure}
\begin{figure}
   \centering
   \includegraphics[angle=0,width=8.500cm]{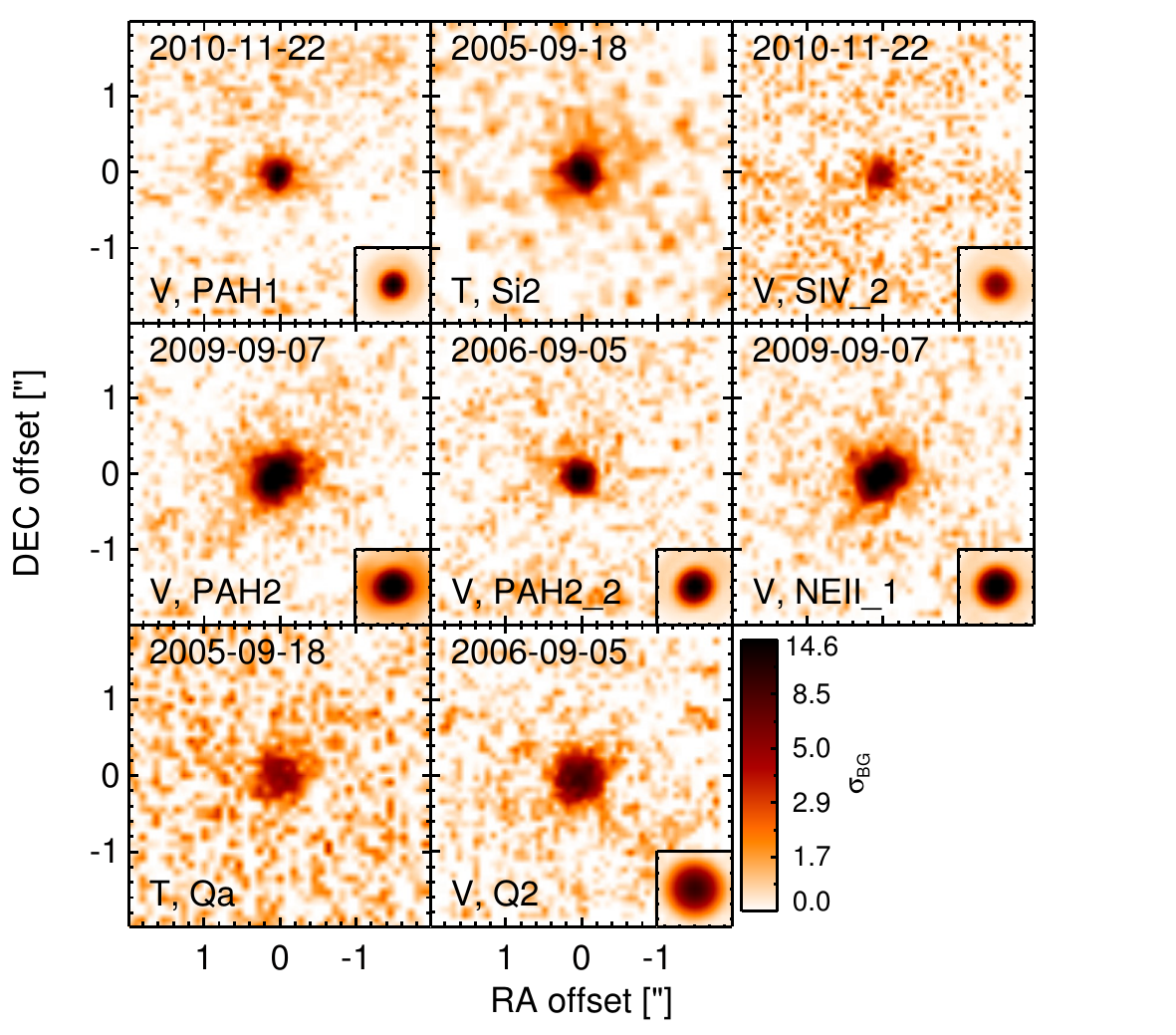}
    \caption{\label{fig:HARim_NGC1566}
             Subarcsecond-resolution MIR images of NGC\,1566 sorted by increasing filter wavelength. 
             Displayed are the inner $4\arcsec$ with North up and East to the left. 
             The colour scaling is logarithmic with white corresponding to median background and black to the $75\%$ of the highest intensity of all images in units of $\sigbg$.
             The inset image shows the central arcsecond of the PSF from the calibrator star, scaled to match the science target.
             The labels in the bottom left state instrument and filter names (C: COMICS, M: Michelle, T: T-ReCS, V: VISIR).
           }
\end{figure}
\begin{figure}
   \centering
   \includegraphics[angle=0,width=8.50cm]{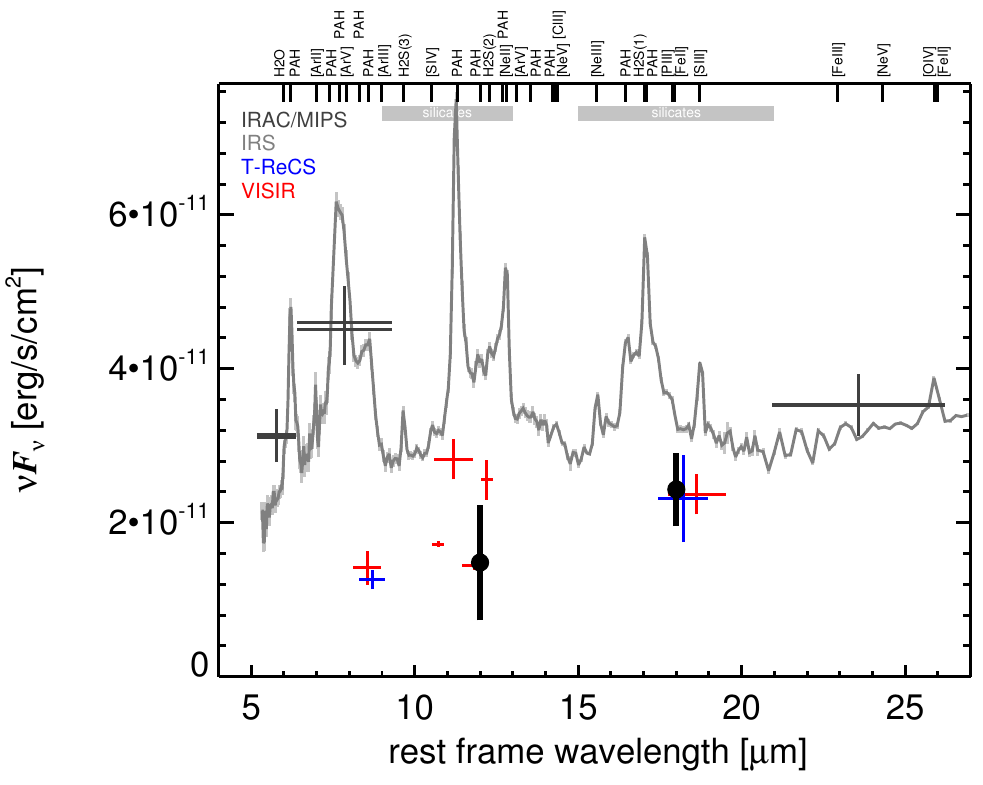}
   \caption{\label{fig:MISED_NGC1566}
      MIR SED of NGC\,1566. The description  of the symbols (if present) is the following.
      Grey crosses and  solid lines mark the \spitzer/IRAC, MIPS and IRS data. 
      The colour coding of the other symbols is: 
      green for COMICS, magenta for Michelle, blue for T-ReCS and red for VISIR data.
      Darker-coloured solid lines mark spectra of the corresponding instrument.
      The black filled circles mark the nuclear 12 and $18\,\mu$m  continuum emission estimate from the data.
      The ticks on the top axis mark positions of common MIR emission lines, while the light grey horizontal bars mark wavelength ranges affected by the silicate 10 and 18$\mu$m features.}
\end{figure}
\clearpage

\twocolumn[\begin{@twocolumnfalse}  
\subsection{NGC\,1614 -- Mrk\,617 -- Arp\,186}\label{app:NGC1614}
NGC\,1614 is a infrared-luminous heavily interacting spiral galaxy at a redshift of $z=$ 0.0159 ($D\sim71$\,Mpc) with an active starburst nucleus, which possibly contains an AGN (see \citealt{vaisanen_nuclear_2012} for a recent detailed study).
Optically, the nucleus has been classified as either an H\,II region (e.g., \citealt{veron-cetty_catalogue_2010}), a possible LINER \citep{kim_optical_1995,veilleux_optical_1995} or an AGN/starburst composite \citep{yuan_role_2010}.
Indications for an obscured AGN come from X-ray spectroscopy \citep{risaliti_hard_2000}, while at radio wavelengths no compact nuclear source was detected (e.g., \citealt{hill_starburst_2001}).
However, \cite{vaisanen_nuclear_2012} rule out the presence of an obscured AGN based on the infrared spectral properties like absorption features and forbidden emission line ratios.
The starburst nucleus is surrounded by a bright starburst ring with $\sim2\arcsec\,$ ($\sim650$\,pc) diameter \citep{alonso-herrero_ngc_2001}.
Note that \cite{vaisanen_nuclear_2012} argue that instead of a late-stage merger with a double nucleus, the companion galaxy survived the collision and is now more than 20\,kpc away to the south-west.
The first MIR observations of NGC\,1614 were performed by \cite{rieke_infrared_1972} followed by \cite{lebofsky_extinction_1979}, \cite{aitken_question_1981} and \cite{lonsdale_infrared_1984}, \cite{phillips_8-13_1984}, \cite{lawrence_observations_1985}, \cite{carico_iras_1988}, \cite{wright_recent_1988}, and \cite{wynn-williams_luminous_1993} after \iras.
\cite{keto_subarcsecond_1992} obtained the first subarcsecond $N$-band image of NGC\,1614 finding a complex resolved structure, which was further resolved into a roughly ring-like structure coinciding with the starburst ring by \cite{miles_high-resolution_1996} using Palomar 5\,m/Spectral-10 and \cite{soifer_high-resolution_2001} using Keck/MIRLIN.
\spitzerr lacks the angular resolution to resolve the ring, which appears embedded within additional host emission in the corresponding IRAC and MIPS images. 
Furthermore, the IRAC 8.0$\,\mu$m PBCD image is saturated in the centre and thus not used.
We measure the nuclear flux in the IRAC $5.8\,\mu$m and MIPS $24\,\mu$m images and scale the IRS LR staring-mode spectrum accordingly.
The resulting \spitzerr MIR SED is to be regarded as an upper limit on the arcsecond-scale MIR emission, and the fluxes are still significantly lower than published in e.g., \cite{engelbracht_metallicity_2008, u_spectral_2012}.
It displays typical star formation features with strong PAH emission, weak silicate $10\,\mu$m absorption, and a red spectral slope in $\nu F_\nu$-space (see also \citealt{brandl_mid-infrared_2006,bernard-salas_spitzer_2009,vaisanen_nuclear_2012}).
In addition, NGC\,1614 was observed with VISIR in the PAH2 filter in 2004 \citep{siebenmorgen_nuclear_2008} and in the Q2 filter in 2006 (unpublished, to our knowledge), as well as with T-ReCS in the Si2 filter in 2006 \citep{diaz-santos_understanding_2008}, and with COMICS in the Q17.7 filter in 2008 \citep{imanishi_subaru_2011}.
The bright star burst ring was detected in all image despite varying S/N and resolution.
The ring has not been fully resolved with a distinguishable nucleus in any case except possibly the Si2 image.
Therefore, we just derive upper limits on the nuclear flux through PSF-scaling at the central position.
These are on average $\sim 80\%$ lower than the \spitzerr spectrophotometry. 
We conclude that from the subarcsecond MIR point of view there is no evidence for an AGN, which, if present, is the cause of less than $20\%$ of the total MIR emission in the central $\sim1\,$kpc.
\newline\end{@twocolumnfalse}]

\begin{figure}
   \centering
   \includegraphics[angle=0,width=8.500cm]{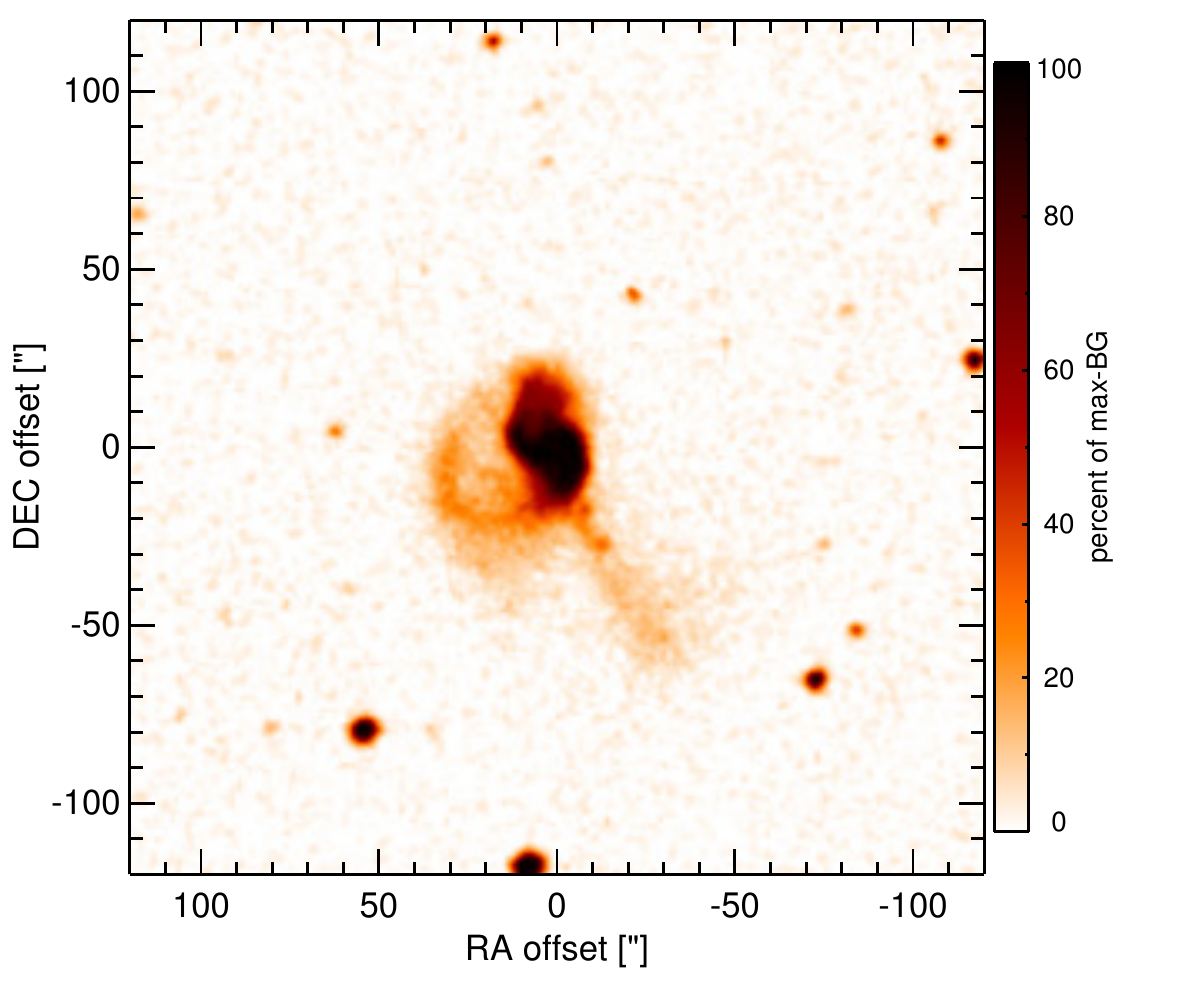}
    \caption{\label{fig:OPTim_NGC1614}
             Optical image (DSS, red filter) of NGC\,1614. Displayed are the central $4\arcmin$ with North up and East to the left. 
              The colour scaling is linear with white corresponding to the median background and black to the $0.01\%$ pixels with the highest intensity.  
           }
\end{figure}
\begin{figure}
   \centering
   \includegraphics[angle=0,height=3.11cm]{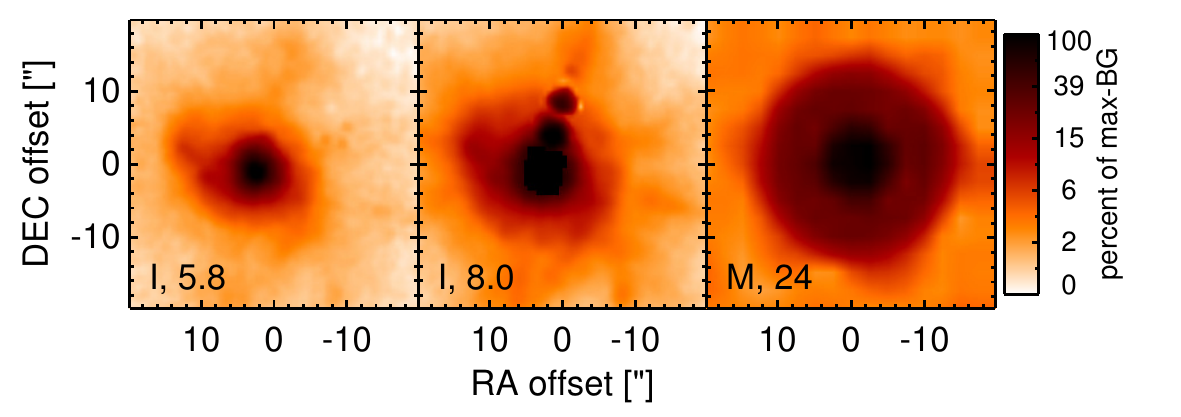}
    \caption{\label{fig:INTim_NGC1614}
             \spitzerr MIR images of NGC\,1614. Displayed are the inner $40\arcsec$ with North up and East to the left. The colour scaling is logarithmic with white corresponding to median background and black to the $0.1\%$ pixels with the highest intensity.
             The label in the bottom left states instrument and central wavelength of the filter in $\mu$m (I: IRAC, M: MIPS).
             Note that the apparent off-nuclear compact sources in the IRAC $8.0\,\mu$m image are instrumental artefacts.
           }
\end{figure}
\begin{figure}
   \centering
   \includegraphics[angle=0,width=8.500cm]{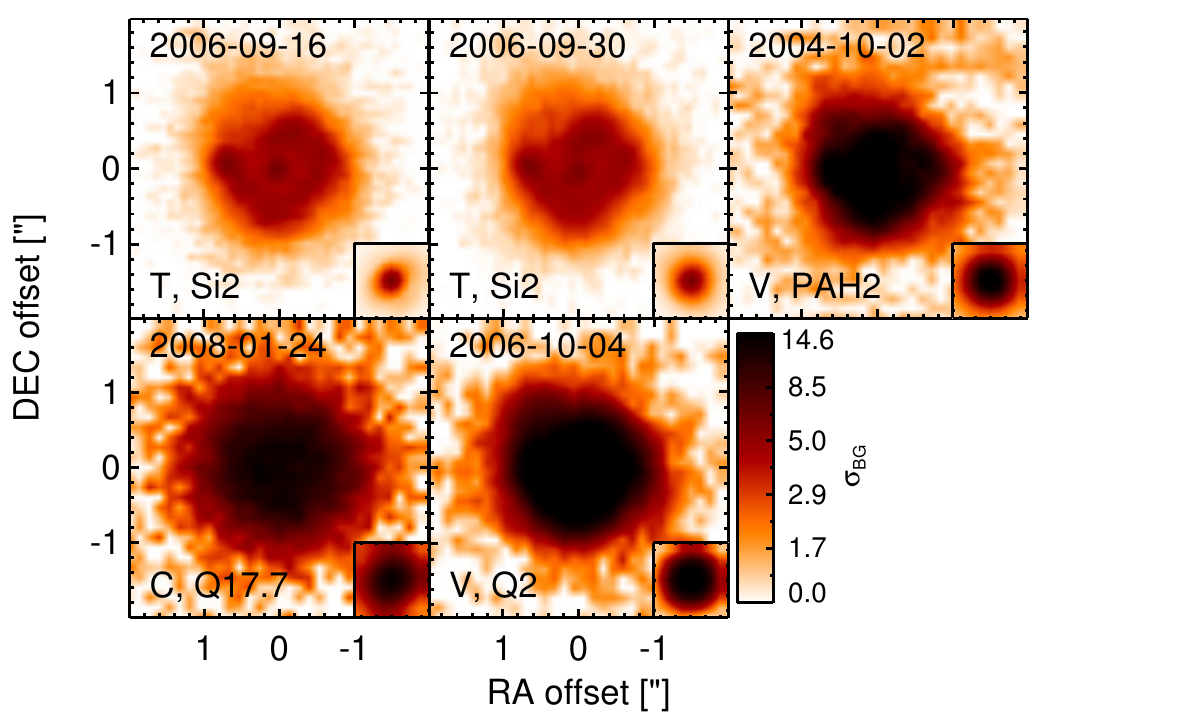}
    \caption{\label{fig:HARim_NGC1614}
             Subarcsecond-resolution MIR images of NGC1614 sorted by increasing filter wavelength. 
             Displayed are the inner $4\arcsecond$ with North up and East to the left. 
             The colour scaling is logarithmic with white corresponding to median background and black to the $75\%$ of the highest intensity of all images in units of $\sigbg$.
             The inset image shows the central arcsecond of the PSF from the calibrator star, scaled to match the science target.
             The labels in the bottom left state instrument and filter names (C: COMICS, M: Michelle, T: T-ReCS, V: VISIR).
           }
\end{figure}
\begin{figure}
   \centering
   \includegraphics[angle=0,width=8.50cm]{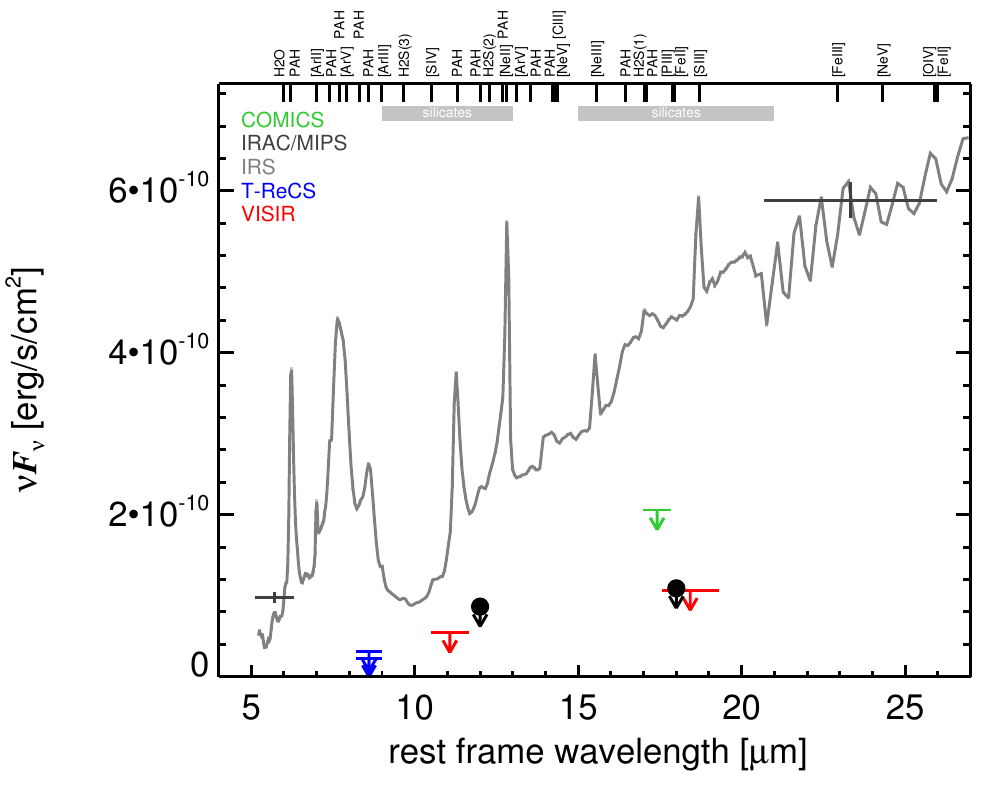}
   \caption{\label{fig:MISED_NGC1614}
      MIR SED of NGC\,1614. The description  of the symbols (if present) is the following.
      Grey crosses and  solid lines mark the \spitzer/IRAC, MIPS and IRS data. 
      The colour coding of the other symbols is: 
      green for COMICS, magenta for Michelle, blue for T-ReCS and red for VISIR data.
      Darker-coloured solid lines mark spectra of the corresponding instrument.
      The black filled circles mark the nuclear 12 and $18\,\mu$m  continuum emission estimate from the data.
      The ticks on the top axis mark positions of common MIR emission lines, while the light grey horizontal bars mark wavelength ranges affected by the silicate 10 and 18$\mu$m features.}
\end{figure}
\clearpage

\twocolumn[\begin{@twocolumnfalse}  
\subsection{NGC\,1667}\label{app:NGC1667}
NGC\,1667 is a low-inclination infrared-luminous spiral galaxy at a redshift of $z=$ 0.0152 ($D\sim 67.8$\,Mpc) with a Sy\,2 nucleus \citep{veron-cetty_catalogue_2010}, which is also detected as compact core in radio (e.g., \citealt{ho_radio_2001}).
After its detection in \irass, NGC\,1667 was measured in the $N$-band but uncertainly detected with the MMT \citep{maiolino_new_1995}.
The nucleus remained undetected in the subarcsecond-resolution $N$-band imaging with Palomar 5\,m/MIRLIN \citep{gorjian_10_2004}. 
NGC\,1667 was also observed with \isoo \citep{clavel_2.5-11_2000,ramos_almeida_mid-infrared_2007} and \spitzer/IRAC, IRS and MIPS.
In the corresponding IRAC and MIPS images, a compact nucleus embedded within diffuse host emission and surrounded by further spiral-like host emission was detected.
The latter is brighter than the nucleus in the IRAC 8\,$\mu$m and MIPS 24\,$\mu$m images.
We measure the nuclear component only providing IRAC $5.8$ and $8.0\,\mu$m fluxes significantly lower than the values in \cite{gallimore_infrared_2010} but in rough agreement with the IRS LR mapping-mode spectrum.
The IRS spectrum suffers from low S/N but indicates prominent PAH emission and a red spectral slope in $\nu F_\nu$-space, reminiscent of star formation (see also \citealt{wu_spitzer/irs_2009,tommasin_spitzer-irs_2010,gallimore_infrared_2010}).
NGC\,1667 was observed with T-ReCS in the broad N filter in 2004 (unpublished, to our knowledge), and with VISIR in two narrow $N$-band filters in 2009 \citep{asmus_mid-infrared_2011}.
In the T-ReCS image, a compact MIR nucleus is weakly detected, while it remained undetected in the VISIR images.
The MIR nucleus is possibly marginally resolved with a FWHM(major axis)$\sim 0.52\arcsec \sim 170\,$pc and a PA$\sim 54\degree$.
However, at least a second epoch with a detection in the MIR at subarcsecond resolution is required to confirm this extension.
The nuclear N flux is $\sim43\%$ lower than the \spitzerr spectrophotometry, which indicates that star formation dominates the total MIR emission in the central kiloparsec of NGC\,1667.
\newline\end{@twocolumnfalse}]

\begin{figure}
   \centering
   \includegraphics[angle=0,width=8.500cm]{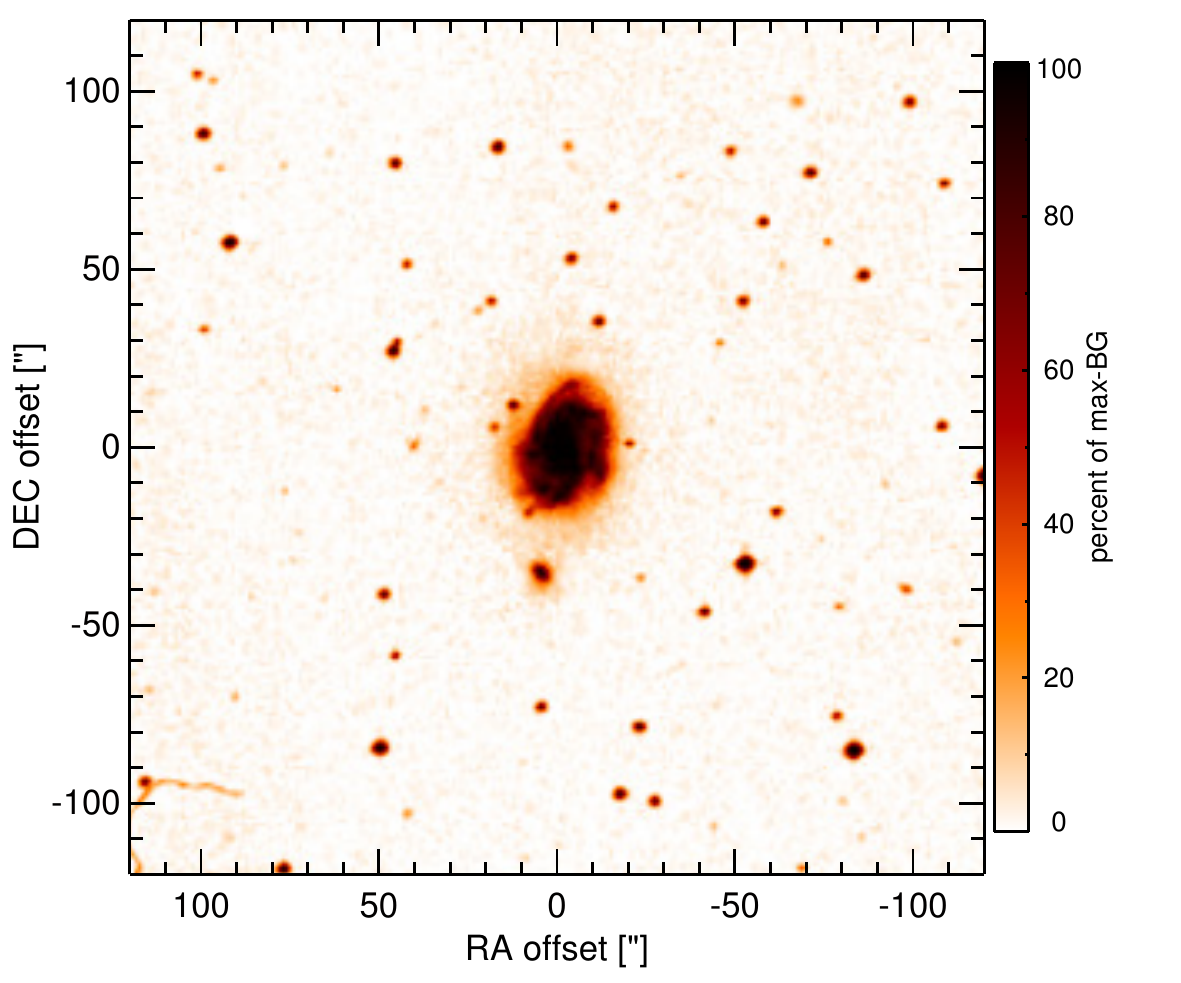}
    \caption{\label{fig:OPTim_NGC1667}
             Optical image (DSS, red filter) of NGC\,1667. Displayed are the central $4\arcmin$ with North up and East to the left. 
              The colour scaling is linear with white corresponding to the median background and black to the $0.01\%$ pixels with the highest intensity.  
           }
\end{figure}
\begin{figure}
   \centering
   \includegraphics[angle=0,height=3.11cm]{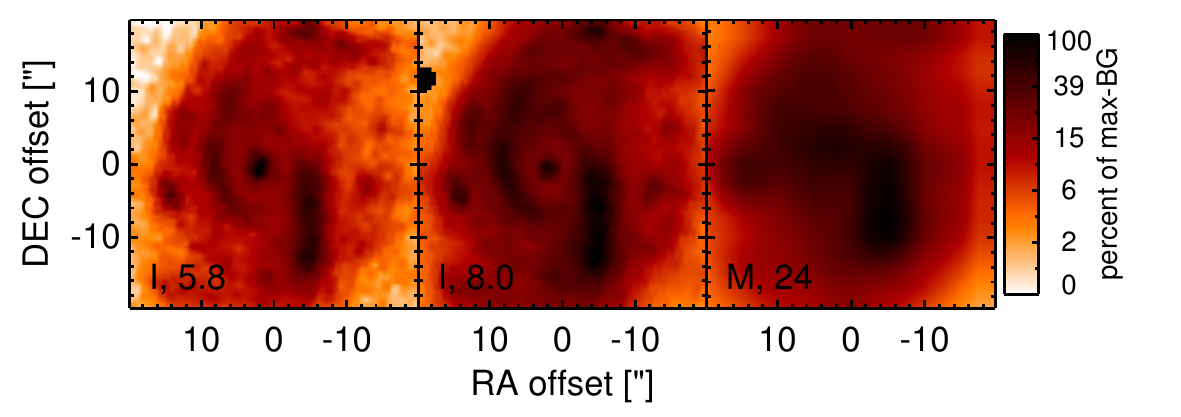}
    \caption{\label{fig:INTim_NGC1667}
             \spitzerr MIR images of NGC\,1667. Displayed are the inner $40\arcsec$ with North up and East to the left. The colour scaling is logarithmic with white corresponding to median background and black to the $0.1\%$ pixels with the highest intensity.
             The label in the bottom left states instrument and central wavelength of the filter in $\mu$m (I: IRAC, M: MIPS). 
           }
\end{figure}
\begin{figure}
   \centering
   \includegraphics[angle=0,width=8.50cm]{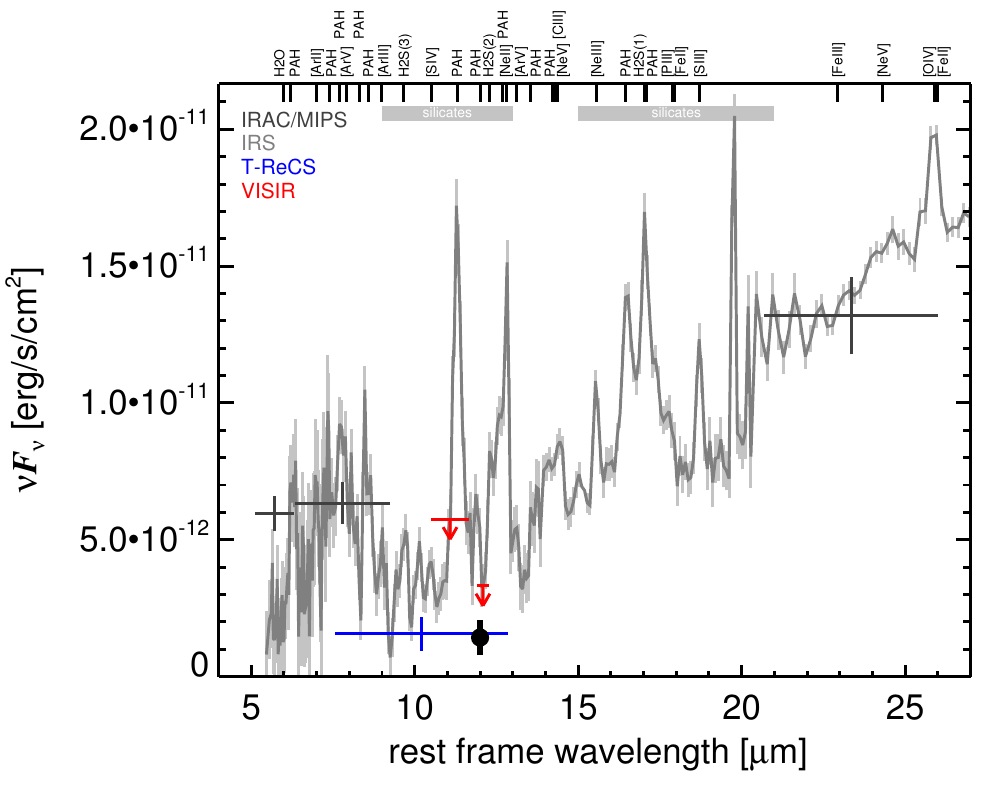}
   \caption{\label{fig:MISED_NGC1667}
      MIR SED of NGC\,1667. The description  of the symbols (if present) is the following.
      Grey crosses and  solid lines mark the \spitzer/IRAC, MIPS and IRS data. 
      The colour coding of the other symbols is: 
      green for COMICS, magenta for Michelle, blue for T-ReCS and red for VISIR data.
      Darker-coloured solid lines mark spectra of the corresponding instrument.
      The black filled circles mark the nuclear 12 and $18\,\mu$m  continuum emission estimate from the data.
      The ticks on the top axis mark positions of common MIR emission lines, while the light grey horizontal bars mark wavelength ranges affected by the silicate 10 and 18$\mu$m features.}
\end{figure}
\clearpage

\twocolumn[\begin{@twocolumnfalse}  
\subsection{NGC\,1808}\label{app:NGC1808}
NGC\,1808 is an inclined peculiar spiral galaxy at a distance of $D=$ $12.3\pm 2.2\,$Mpc \citep{tully_extragalactic_2009} with an active nucleus containing a prominent starburst an possibly an AGN (see e.g., \citealt{krabbe_near-infrared_1994}).
Optically, the nucleus has been classified as a Sy\,2 \citep{veron-cetty_ngc_1985} or as an H\,II region \citep{yuan_role_2010}.
The detection of an obscured long-term variable hard X-ray source supports the AGN scenario \citep{awaki_asca_1996,jimenez-bailon_x-ray_2005}.
At radio wavelengths a number of compact non-thermal radio sources was detected in the centre of NGC\,1808, most of which are associated with supernovae remnants (e.g., \citealt{collison_radio_1994}).
The brightest central radio source is marginally extended with uncertain nature.
\cite{forbes_ngc_1992} conclude that the optical and radio properties can be explained with supernovae embedded within an H\,II region (see also \citealt{kotilainen_near-infrared_1996}).
MIR forbidden emission line diagnostics neither support the presence of an obscured AGN \citep{goulding_towards_2009}.
The first MIR observations of NGC\,1808 were performed by \cite{frogel_8-13_1982}, \cite{phillips_8-13_1984}, and \cite{roche_atlas_1991}.
After \iras, the first ground-based $N$-band map of the nuclear region were made by \cite{telesco_genesis_1993}, showing a large extended MIR structure (PA$\sim-40\degree$).
\isoo observations followed \citep{laurent_mid-infrared_2000,siebenmorgen_infrared_2001,forster_schreiber_isocam_2003}.
The nuclear region of NGC\,1808 was later imaged in the MIR at arcsecond resolution with ESO 2.2\,m/MANIAC \citep{krabbe_n-band_2001}, with ESO 3.6\,m/TIMMI2 \citep{galliano_mid-infrared_2005} and with CTIO 4\,m/OSCIR \citep{ramos_almeida_infrared_2009}.
The deep TIMMI2 image further resolves the previously detected structure into several clumpy sources embedded within diffuse emission with a dominating elongated MIR nucleus (major axis$\sim 1.5\arcsec\sim\,90$pc; PA$\sim-45\degree$).
\spitzerr lacks the angular resolution to resolve the complex nuclear structure and only the compact nucleus embedded within the extended emission is visible in the corresponding IRAC and MIPS images. 
Furthermore, the IRAC 8.0$\,\mu$m PBCD image is saturated in the centre and thus not used.
We measure the nuclear component in the IRAC $5.8\,\mu$m and MIPS $24\,\mu$m images.
The IRS LR staring-mode spectrum displays typical star formation features with strong PAH emission, possible silicate $10\,\mu$m absorption, and a red spectral slope in $\nu F_\nu$-space (see also \citealt{goulding_towards_2009,gonzalez-martin_dust_2013}).
Here, we report VISIR imaging in the PAH2\_2 filter performed in 2009 (unpublished, to our knowledge), which shows a morphology very similar to the previous TIMMI2 image despite higher angular resolution.
In particular, the nucleus again appears elongated (FWHM(major axis)$\sim 0.64\arcsec\sim38$\,pc; PA$\sim-50\degree$).
The unresolved nuclear component flux is  $\sim 63\%$ lower than the \spitzerr spectrophotometry and consistent with the LR $N$-band T-ReCS spectrum by \citep{gonzalez-martin_dust_2013}.
The latter exhibits still PAH emission albeit weaker, which indicates that even the subarcsecond measurements are still star-formation contaminated. 
From the subarcsecond MIR point of view, we cannot exclude the existence of an AGN in NGC\,1808 but agree with previous works that the MIR emission in the central $\sim200$\,pc is star formation dominated.
\newline\end{@twocolumnfalse}]

\begin{figure}
   \centering
   \includegraphics[angle=0,width=8.500cm]{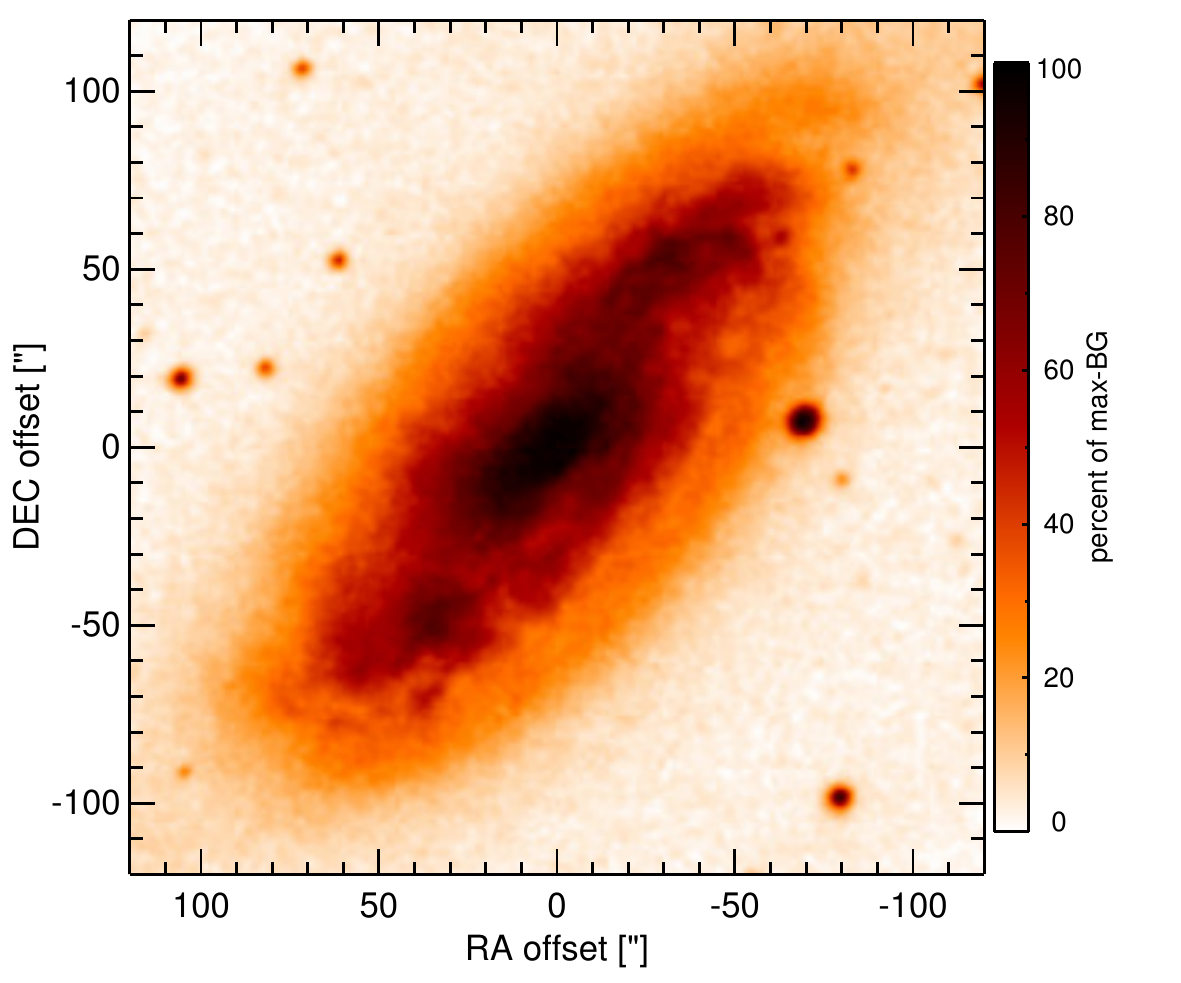}
    \caption{\label{fig:OPTim_NGC1808}
             Optical image (DSS, red filter) of NGC\,1808. Displayed are the central $4\arcmin$ with North up and East to the left. 
              The colour scaling is linear with white corresponding to the median background and black to the $0.01\%$ pixels with the highest intensity.  
           }
\end{figure}
\begin{figure}
   \centering
   \includegraphics[angle=0,height=3.11cm]{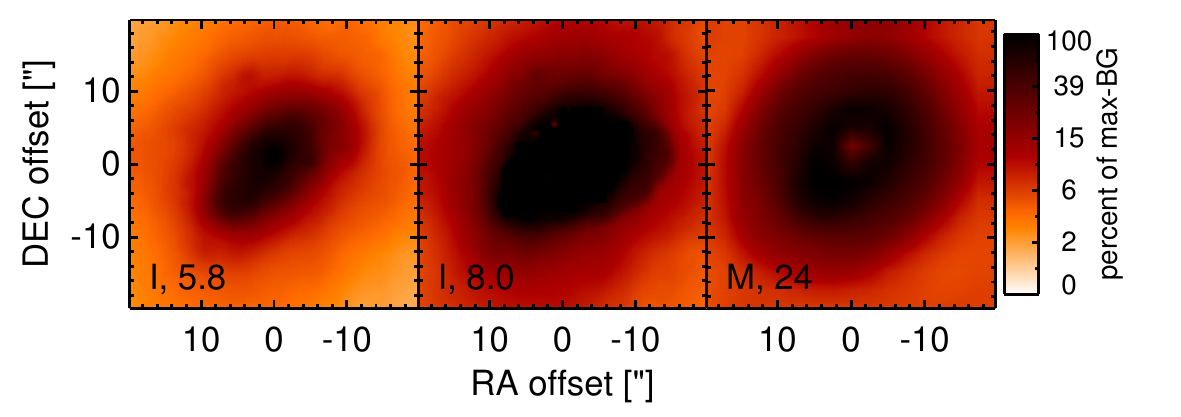}
    \caption{\label{fig:INTim_NGC1808}
             \spitzerr MIR images of NGC\,1808. Displayed are the inner $40\arcsec$ with North up and East to the left. The colour scaling is logarithmic with white corresponding to median background and black to the $0.1\%$ pixels with the highest intensity.
             The label in the bottom left states instrument and central wavelength of the filter in $\mu$m (I: IRAC, M: MIPS).
             Note that the central region in the IRAC $8.0\,\mu$m image is completely saturated.
           }
\end{figure}
\begin{figure}
   \centering
   \includegraphics[angle=0,height=3.11cm]{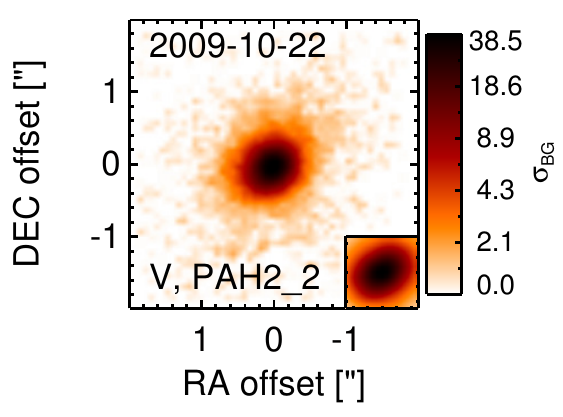}
    \caption{\label{fig:HARim_NGC1808}
             Subarcsecond-resolution MIR images of NGC\,1808 sorted by increasing filter wavelength. 
             Displayed are the inner $4\arcsec$ with North up and East to the left. 
             The colour scaling is logarithmic with white corresponding to median background and black to the $75\%$ of the highest intensity of all images in units of $\sigbg$.
             The inset image shows the central arcsecond of the PSF from the calibrator star, scaled to match the science target.
             The labels in the bottom left state instrument and filter names (C: COMICS, M: Michelle, T: T-ReCS, V: VISIR).
           }
\end{figure}
\begin{figure}
   \centering
   \includegraphics[angle=0,width=8.50cm]{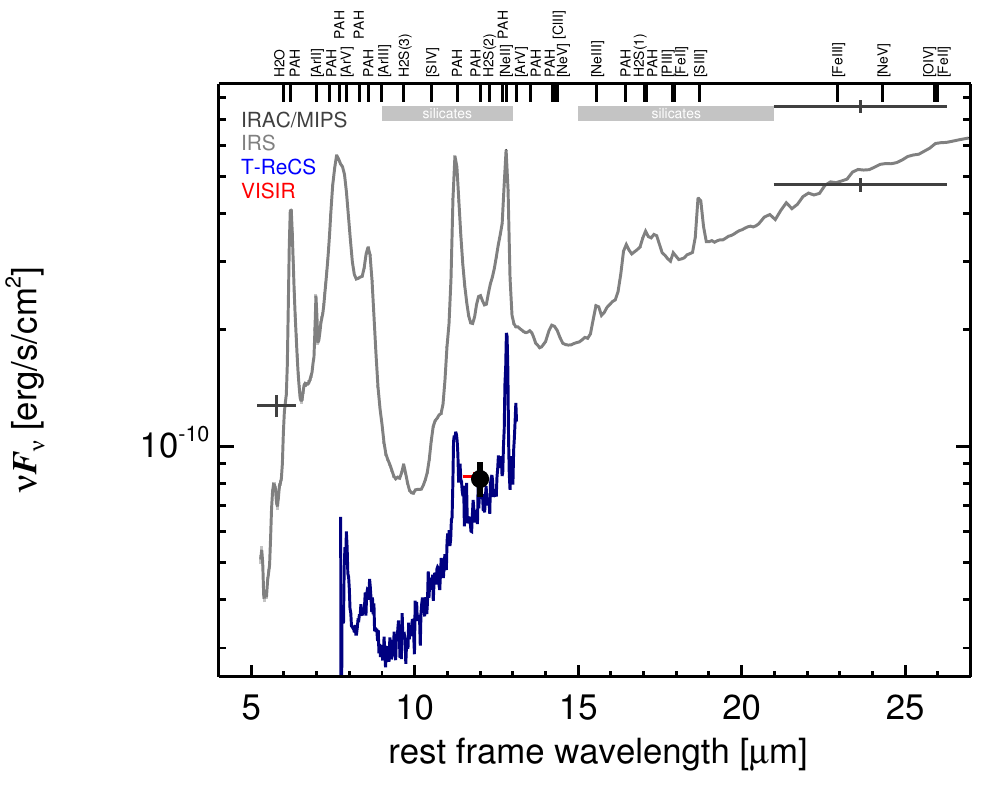}
   \caption{\label{fig:MISED_NGC1808}
      MIR SED of NGC\,1808. The description  of the symbols (if present) is the following.
      Grey crosses and  solid lines mark the \spitzer/IRAC, MIPS and IRS data. 
      The colour coding of the other symbols is: 
      green for COMICS, magenta for Michelle, blue for T-ReCS and red for VISIR data.
      Darker-coloured solid lines mark spectra of the corresponding instrument.
      The black filled circles mark the nuclear 12 and $18\,\mu$m  continuum emission estimate from the data.
      The ticks on the top axis mark positions of common MIR emission lines, while the light grey horizontal bars mark wavelength ranges affected by the silicate 10 and 18$\mu$m features.}
\end{figure}
\clearpage

\twocolumn[\begin{@twocolumnfalse}  
\subsection{NGC\,2110}\label{app:NGC2110}
NGC\,2110 is a spiral galaxy at a distance of $D=$ $\sim33\,$Mpc ($z\sim 0.0078$) with a low-luminosity Sy\,2 nucleus \citep{bradt_ngc_1978} with broad emission lines in polarized and near-infrared light \citep{moran_transient_2007,reunanen_near-infrared_2003}.
The AGN is at the border to the LINER regime \citep{kewley_optical_2001,yuan_role_2010} and belongs to the nine-month BAT AGN sample.
Radio observations with the VLA found a S-shaped jet structure that extends in total 4\arcsec\ symmetrically in  the north-south directions \citep{ulvestad_nuclear_1983,nagar_radio_1999}.
This implies that the AGN is seen edge-on, which is also supported by 
the NLR structure following roughly the jet structure \citep{mulchaey_hubble_1994} and the polarization orientation of the broad H$\alpha$ emission \citep{moran_transient_2007}.
Early MIR observations were performed by \cite{roche_8-13_1984} and \cite{lawrence_observations_1985}.
NGC\,2110 was observed with \spitzer/IRS and MIPS, while no IRAC images are available.
It appears compact in MIPS $24\,\mu$m, and our flux measurement agrees with the value given by \cite{temi_spitzer_2009}.
The IRS spectrum shows strong silicate emission (both at 10 and $18\,\mu$m), forbidden emission lines and also PAH $11.3\,\mu$m emission (see also \citealt{shi_9.7_2006, mason_origin_2009}).
The first high angular resolution MIR study of NGC\,2110 was done by \cite{mason_origin_2009}, comparing the \spitzerr data to Michelle $N$-band spectrophotometry. 
Our reanalysis of the broad $N$-band Michelle image yields  results consistent with theirs.
In addition, also VISIR $N$-band spectrophotometry was performed \citep{honig_dusty_2010}, and again our reanalysis is consistent with the original work.
Furthermore, a $Q$-band and two narrow $N$-band VISIR images were taken by our group. 
NGC\,2110 appears compact in all HR MIR images without any non-nuclear emission being detected neither from the jet nor the host galaxy. 
Our quantitative analysis indicates that the nucleus is possibly marginally resolved with a position angle $\sim 90\degree$ (FWHM at $12\,\mu$m $\sim 57$\,pc $\sim 0.36$). 
The general MIR flux level of the HR data is $\sim 10\%$ lower than in \spitzerr and the PAH $11.3\,\mu$m feature is absent (see also \citealt{esquej_nuclear_2014}).
NGC\,2110 is one of the most prominent cases of those few Seyfert 2s that show strong silicate emission (see also \citealt{mason_origin_2009}).
Interestingly, the MIR flux level of NGC\,2110 has been rising since 2006 (VISIR first epoch) over 2007 (Michelle) to 2010 (VISIR second epoch). 
\newline\end{@twocolumnfalse}]

\begin{figure}
   \centering
   \includegraphics[angle=0,width=8.500cm]{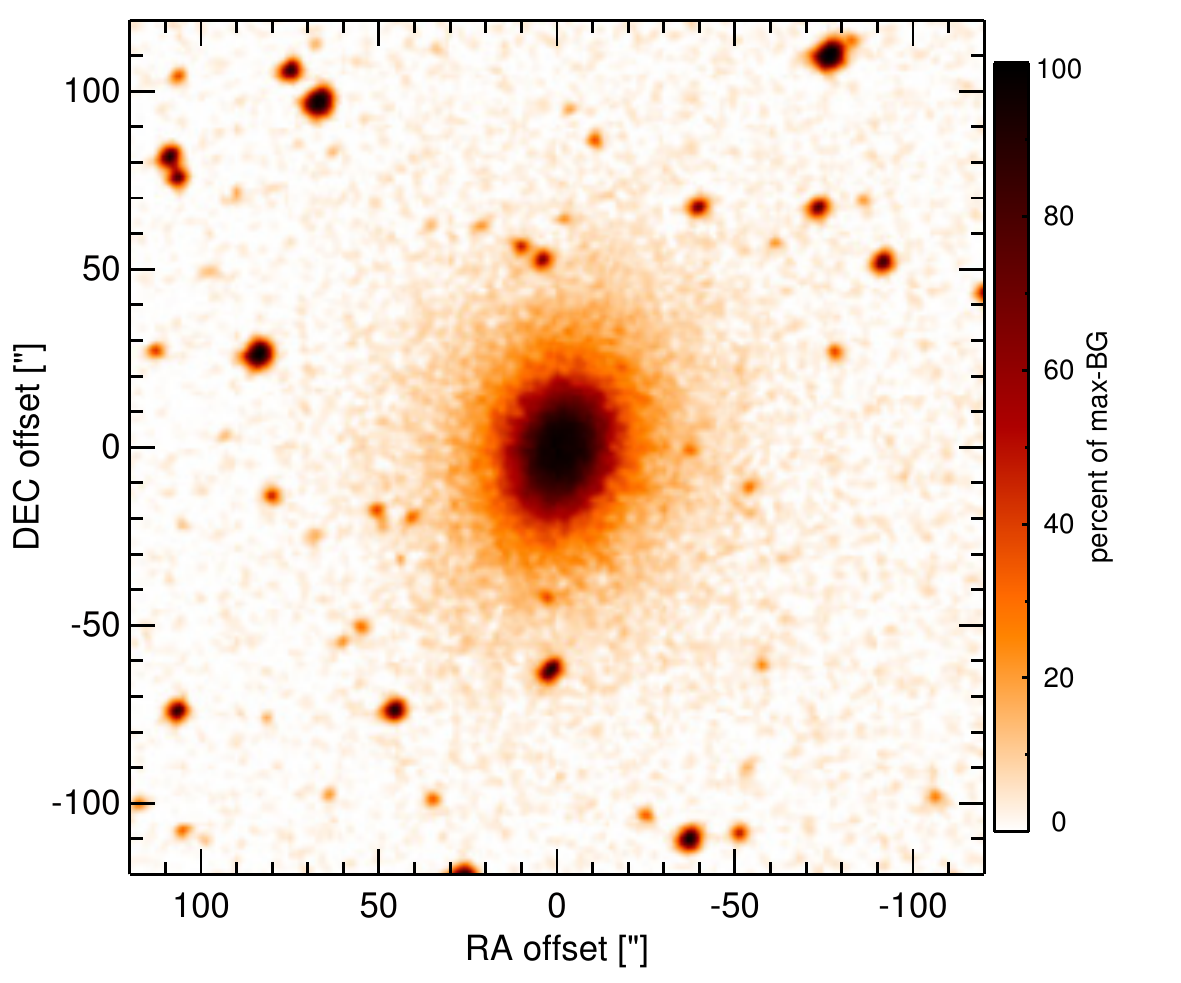}
    \caption{\label{fig:OPTim_NGC2110}
             Optical image (DSS, red filter) of NGC\,2110. Displayed are the central $4\arcmin$ with North up and East to the left. 
              The colour scaling is linear with white corresponding to the median background and black to the $0.01\%$ pixels with the highest intensity.  
           }
\end{figure}
\begin{figure}
   \centering
   \includegraphics[angle=0,height=3.11cm]{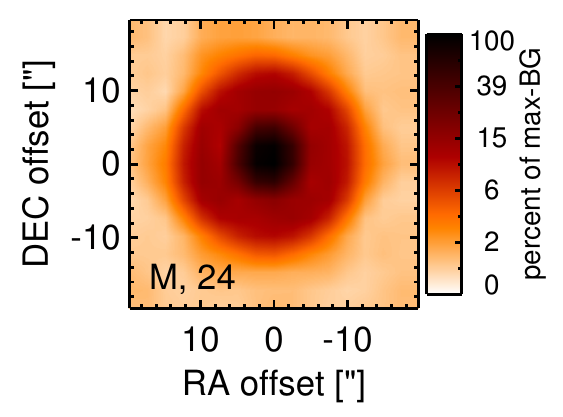}
    \caption{\label{fig:INTim_NGC2110}
             \spitzerr MIR images of NGC\,2110. Displayed are the inner $40\arcsec$ with North up and East to the left. The colour scaling is logarithmic with white corresponding to median background and black to the $0.1\%$ pixels with the highest intensity.
             The label in the bottom left states instrument and central wavelength of the filter in $\mu$m (I: IRAC, M: MIPS). 
           }
\end{figure}
\begin{figure}
   \centering
   \includegraphics[angle=0,width=8.500cm]{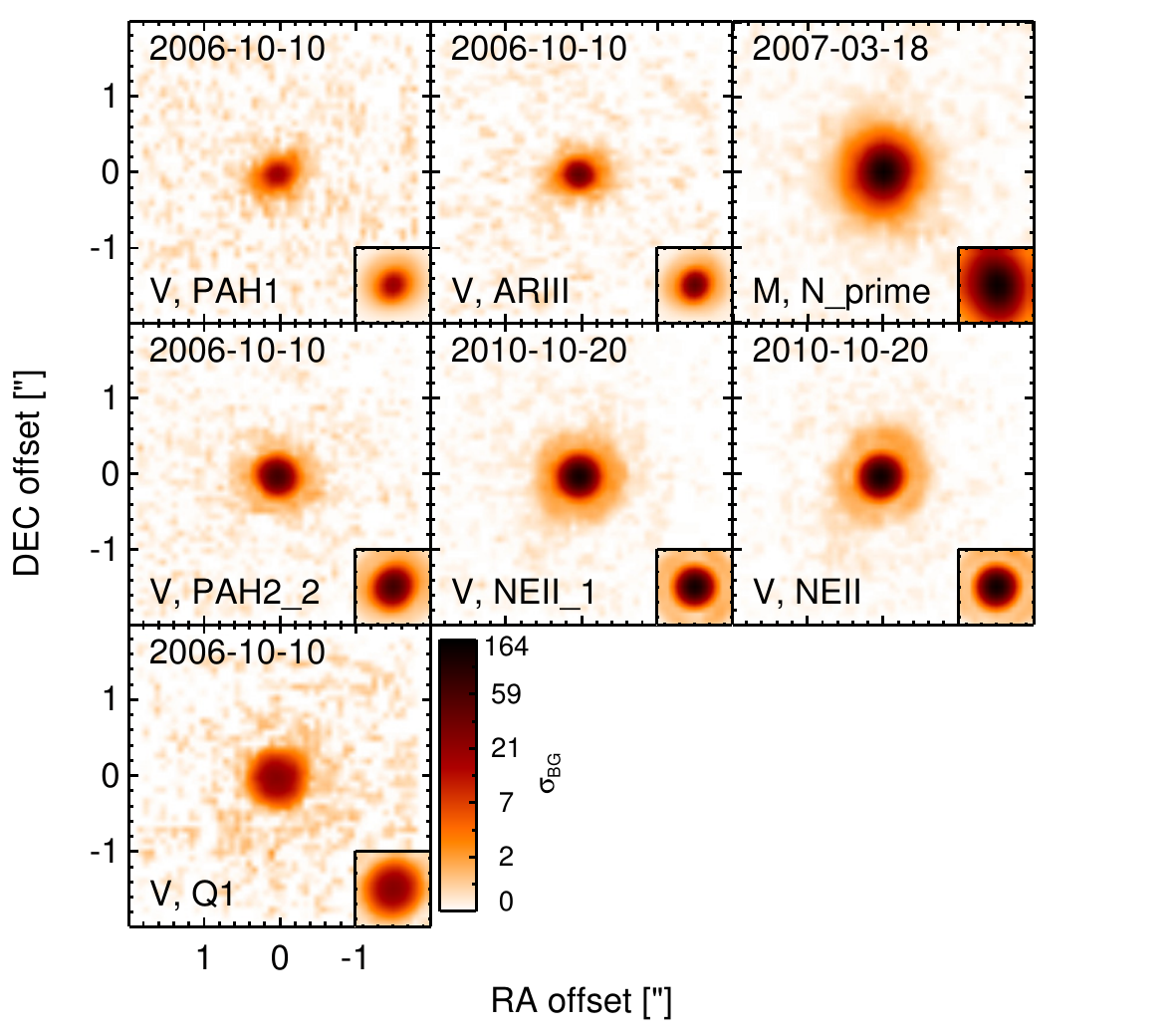}
    \caption{\label{fig:HARim_NGC2110}
             Subarcsecond-resolution MIR images of NGC\,2110 sorted by increasing filter wavelength. 
             Displayed are the inner $4\arcsec$ with North up and East to the left. 
             The colour scaling is logarithmic with white corresponding to median background and black to the $75\%$ of the highest intensity of all images in units of $\sigbg$.
             The inset image shows the central arcsecond of the PSF from the calibrator star, scaled to match the science target.
             The labels in the bottom left state instrument and filter names (C: COMICS, M: Michelle, T: T-ReCS, V: VISIR).
           }
\end{figure}
\begin{figure}
   \centering
   \includegraphics[angle=0,width=8.50cm]{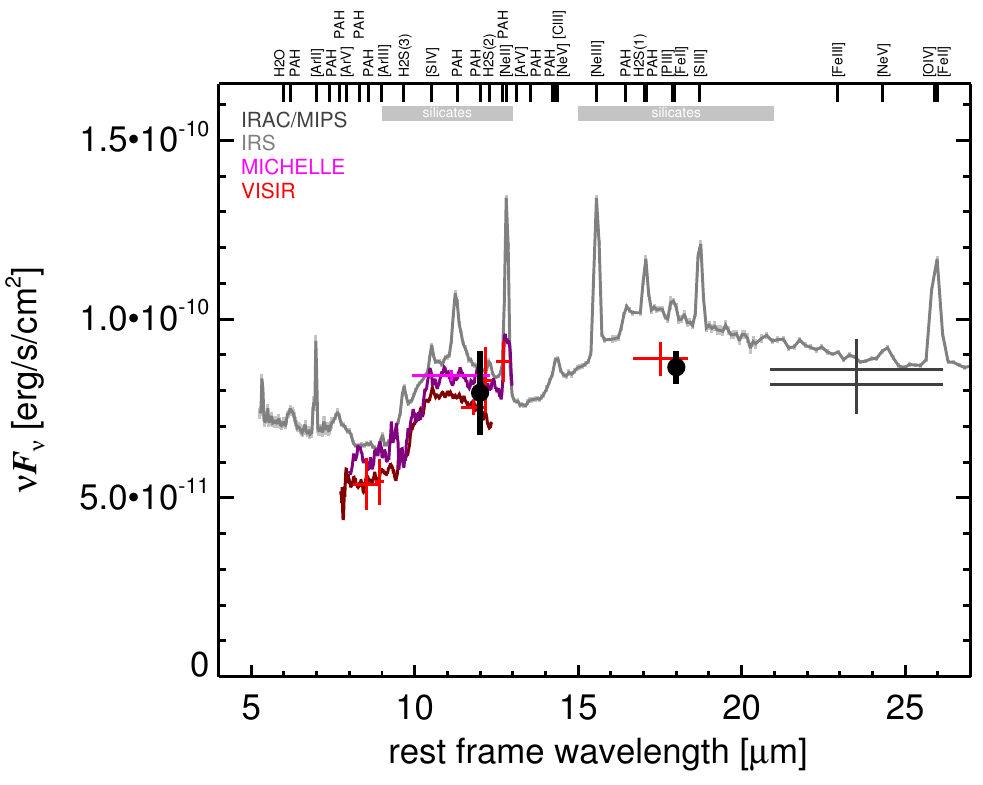}
   \caption{\label{fig:MISED_NGC2110}
      MIR SED of NGC\,2110. The description  of the symbols (if present) is the following.
      Grey crosses and  solid lines mark the \spitzer/IRAC, MIPS and IRS data. 
      The colour coding of the other symbols is: 
      green for COMICS, magenta for Michelle, blue for T-ReCS and red for VISIR data.
      Darker-coloured solid lines mark spectra of the corresponding instrument.
      The black filled circles mark the nuclear 12 and $18\,\mu$m  continuum emission estimate from the data.
      The ticks on the top axis mark positions of common MIR emission lines, while the light grey horizontal bars mark wavelength ranges affected by the silicate 10 and 18$\mu$m features.}
\end{figure}
\clearpage

\twocolumn[\begin{@twocolumnfalse}  
\subsection{NGC\,2623 -- Arp\,243 -- VV\,79 -- UGC\,4509}\label{app:NGC2623}
NGC\,2623 is an infrared-luminous advanced merger system at a redshift of $z=$ 0.0185 ($D\sim87.3\,$Mpc) with only one identified nucleus (see \citealt{evans_off-nuclear_2008} for a recent multiwavelength study of the system).
An AGN was detected at X-ray and radio wavelengths inside the compact central starburst \citep{maiolino_elusive_2003,lonsdale_starburst-agn_1993} but is optically ''elusive'' with a spectral properties similar to NGC\,4945 \citep{maiolino_elusive_2003}.
It has also been classified as a LINER or AGN/starburst composite (e.g., \citealt{lipari_infrared_2004}).
In addition, the detection of \nev provides further evidence for an AGN in NGC\,2623 \citep{dudik_mid-infrared_2007}.
After the discovery of its MIR brightness through \iras, NGC\,2623 was observed by \cite{carico_iras_1988}, \cite{wright_recent_1988}, \cite{wynn-williams_luminous_1993},  \cite{bushouse_distribution_1998}, and \cite{dudley_new_1999}.
The first subarcsecond-resolution $N$-band images were made with Keck/LWS \citep{soifer_high-resolution_2001} where a marginally resolved MIR nucleus with east-west elongation coinciding with the radio morphology was detected (major axis$\sim2\arcsec\sim0.8\,$kpc; PA$\sim  90\degree$).
In the \spitzer/IRAC and MIPS images, NGC\,2623 appears a compact MIR nucleus with  faint host emission, which becomes weaker towards longer wavelengths.
Our nuclear IRAC $5.8$ and $8.0\,\mu$m and MIPS $24\,\mu$m fluxes agree in general with the values by \cite{u_spectral_2012}.
The IRS LR staring-mode spectrum exhibits deep silicate  $10\,\mu$m and possible silicate $18\,\mu$m absorption features, prominent PAH emission and a steep red spectral slope in $\nu F_\nu$-space (see also \citealt{brandl_mid-infrared_2006,bernard-salas_spitzer_2009}).
NGC\,2623 was observed with COMICS in the Q17.7 filter in 2008 \citep{imanishi_subaru_2011} and a marginally resolved MIR nucleus was detected ($\sim 1\arcsec \sim 0.4\,$kpc) but with a position angle of $\sim 50\degree$.
Because this is inconsistent with the previous Keck/LWS images, we classify NGC\,2623 as possibly extended only. 
Our remeasured nuclear Q17.7 flux is significantly higher than the value by \cite{imanishi_subaru_2011} but consistent with the \spitzerr spectrophotometry and the Keck/LWS data.
Therefore, we use the IRS spectrum to calculate the 12\,$\mu$m continuum emission estimate corrected for the silicate feature.
Note however, that the nuclear fluxes would be significantly lower if the presence of subarcsecond-extended emission can be verified.
Owing to the presence of PAH emission in the IRS spectrum and the object distance, the subarcsecond MIR values of NGC\,2623 are presumably heavily star-formation contaminated.
\newline\end{@twocolumnfalse}]

\begin{figure}
   \centering
   \includegraphics[angle=0,width=8.500cm]{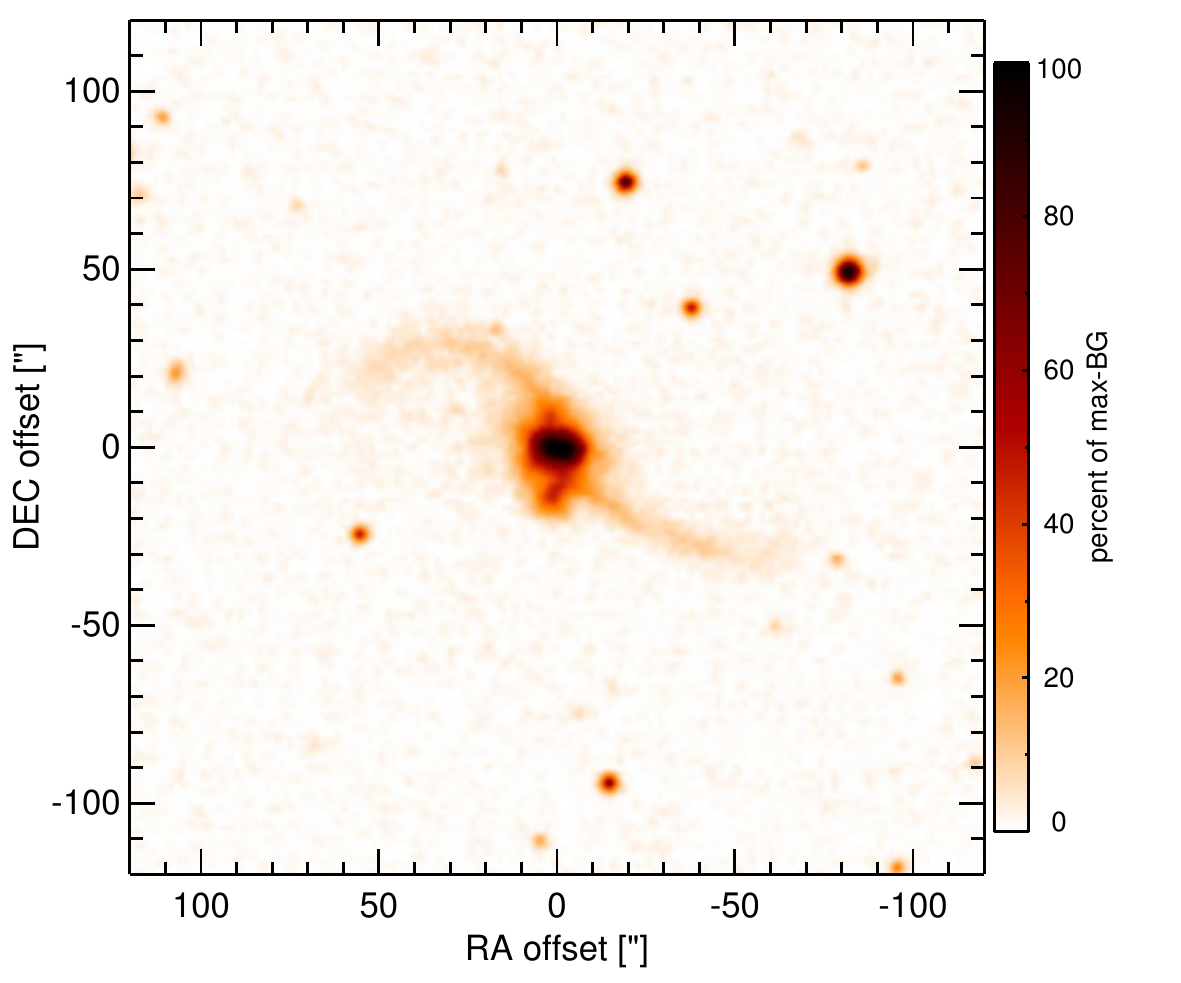}
    \caption{\label{fig:OPTim_NGC2623}
             Optical image (DSS, red filter) of NGC\,2623. Displayed are the central $4\arcmin$ with North up and East to the left. 
              The colour scaling is linear with white corresponding to the median background and black to the $0.01\%$ pixels with the highest intensity.  
           }
\end{figure}
\begin{figure}
   \centering
   \includegraphics[angle=0,height=3.11cm]{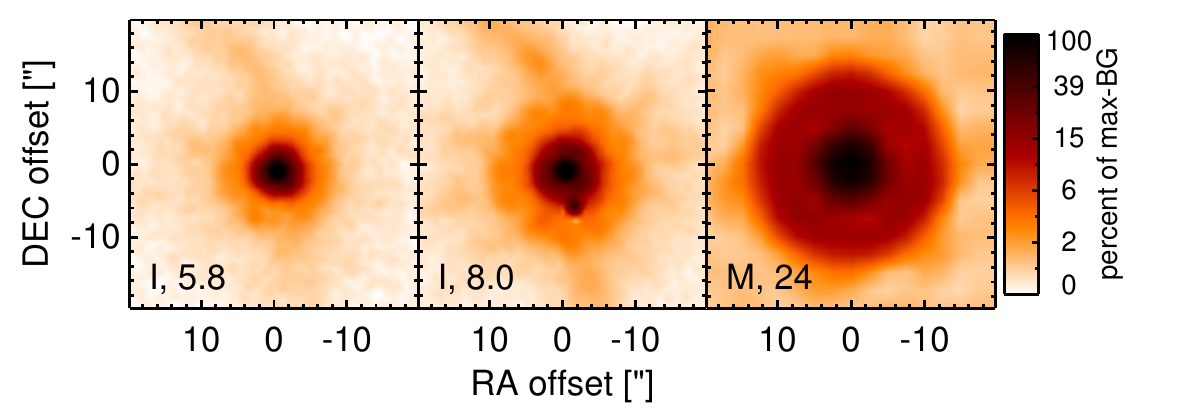}
    \caption{\label{fig:INTim_NGC2623}
             \spitzerr MIR images of NGC\,2623. Displayed are the inner $40\arcsec$ with North up and East to the left. The colour scaling is logarithmic with white corresponding to median background and black to the $0.1\%$ pixels with the highest intensity.
             The label in the bottom left states instrument and central wavelength of the filter in $\mu$m (I: IRAC, M: MIPS). 
             Note that the apparent off-nuclear compact source in the IRAC $8.0\,\mu$m image is an instrumental artefact.
           }
\end{figure}
\begin{figure}
   \centering
   \includegraphics[angle=0,height=3.11cm]{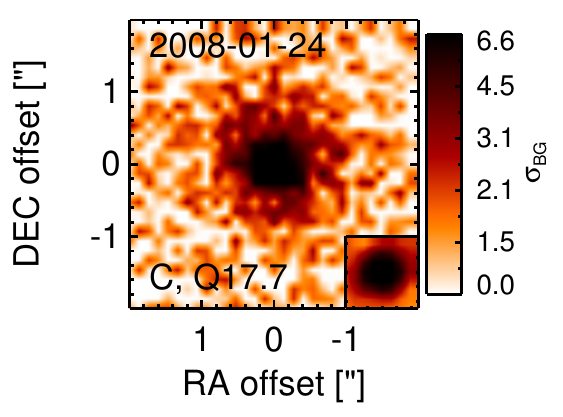}
    \caption{\label{fig:HARim_NGC2623}
             Subarcsecond-resolution MIR images of NGC\,2623 sorted by increasing filter wavelength. 
             Displayed are the inner $4\arcsec$ with North up and East to the left. 
             The colour scaling is logarithmic with white corresponding to median background and black to the $75\%$ of the highest intensity of all images in units of $\sigbg$.
             The inset image shows the central arcsecond of the PSF from the calibrator star, scaled to match the science target.
             The labels in the bottom left state instrument and filter names (C: COMICS, M: Michelle, T: T-ReCS, V: VISIR).
           }
\end{figure}
\begin{figure}
   \centering
   \includegraphics[angle=0,width=8.50cm]{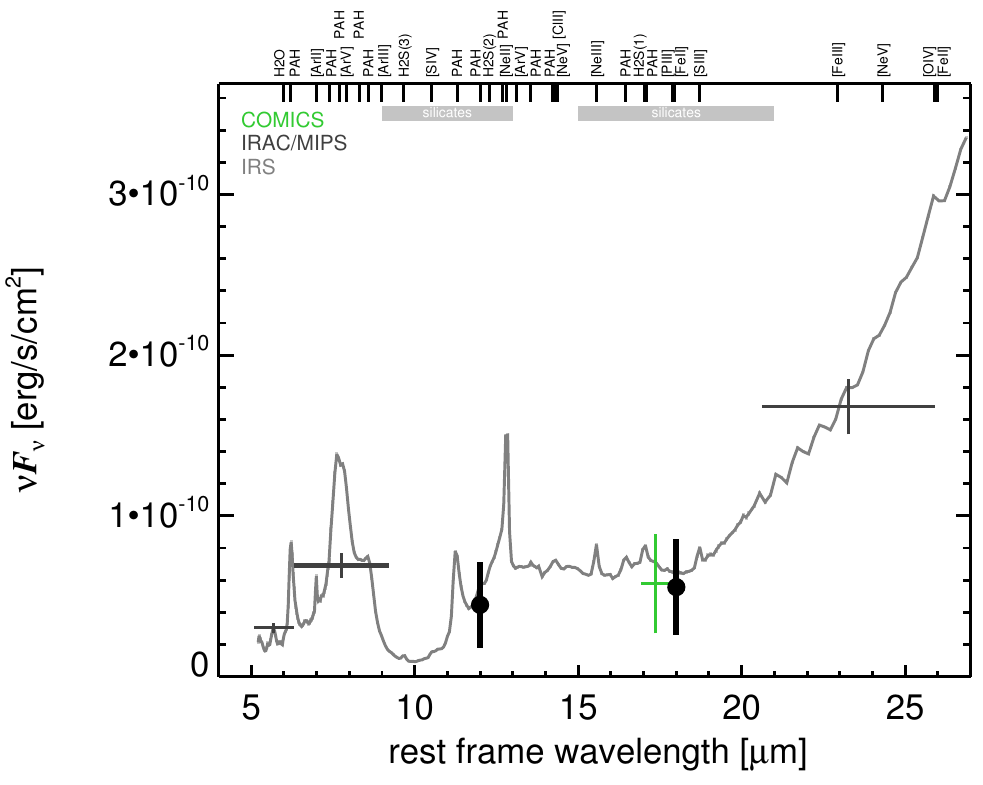}
   \caption{\label{fig:MISED_NGC2623}
      MIR SED of NGC\,2623. The description  of the symbols (if present) is the following.
      Grey crosses and  solid lines mark the \spitzer/IRAC, MIPS and IRS data. 
      The colour coding of the other symbols is: 
      green for COMICS, magenta for Michelle, blue for T-ReCS and red for VISIR data.
      Darker-coloured solid lines mark spectra of the corresponding instrument.
      The black filled circles mark the nuclear 12 and $18\,\mu$m  continuum emission estimate from the data.
      The ticks on the top axis mark positions of common MIR emission lines, while the light grey horizontal bars mark wavelength ranges affected by the silicate 10 and 18$\mu$m features.}
\end{figure}
\clearpage

\twocolumn[\begin{@twocolumnfalse}  
\subsection{NGC\,2992}\label{app:NGC2992}
NGC\,2992 is a highly-inclined spiral galaxy interacting with NGC\,2993 2.9\arcmin\, to the south at a redshift of $z=$ 0.0077 ($D\sim39.7\,$Mpc) with a well-studied AGN belonging to the nine-month BAT AGN sample.
This AGN exhibits huge X-ray variations (factor of $\sim20$; e.g., \citealt{murphy_monitoring_2007}) and has changed its optical type between Sy\,1.5 and Sy\,2.0, which is presumably caused by intrinsic changes in the ionizing continuum rather than varying obscuration \citep{gilli_variability_2000,trippe_long-term_2008}.
The nucleus is obscured by a crossing prominent dust lane instead \citep{ward_new_1980}.
The AGN is surrounded by an extended biconical NLR with a total extent of $\sim 5\,$kpc (PA$\sim125\degree$;e.g., \citealt{allen_physical_1999}).
At radio wavelengths, a compact radio core and a prominent double loop structure extending in total 8\arcsec\,($\sim1.5\,$kpc) to the north-west and south-east (PA$\sim154\degree$; \citealt{ulvestad_radio_1984,wehrle_radio_1988}), which are possibly generated by an AGN-driven jet \citep{colbert_large-scale_1998,veilleux_biconical_2001}.
The first MIR observations were performed by \cite{glass_mid-infrared_1982}, \cite{lonsdale_infrared_1984}, followed by \cite{aitken_8-13_1985}, \cite{ward_continuum_1987}, and \cite{roche_atlas_1991}.
Subarcsecond $N$-band imaging was then performed with Palomar 5\,m/MIRLIN \citep{gorjian_10_2004}, with Keck/LWS \citep{soifer_high_2004}, and with ESO 3.6\,m/TIMMI2 \citep{galliano_mid-infrared_2005}.
In all cases an unresolved MIR nucleus without extended host emission was detected.
In the \spitzer/IRAC images lenticular extended host emission was detected in addiction to the MIR nucleus, while NGC\,2992 appears  compact in the MIPS images.
We extract the nuclear component resulting in an IRAC $5.8$ flux consistent with \cite{gallimore_infrared_2010}, while the $8.0\,\mu$m value is significantly lower.
The IRS LR staring-mode spectrum exhibits silicate $10\,\mu$m absorption, prominent PAH features and forbidden emission lines, and a generally red spectral slope in $\nu F_\nu$-space (see also \citealt{wu_spitzer/irs_2009,deo_mid-infrared_2009,gallimore_infrared_2010}).
Thus, it appears that NGC\,2992 is significantly affected by star formation on arcsecond scales, although  on nuclear scales \cite{friedrich_adaptive_2010} constrain star formation to be minor.
NGC\,2992 was repeatedly observed with VISIR in four narrow $N$-band filters spread over 2004, 2005, 2006 and 2009.
The first two measurements were published in \cite{haas_visir_2007}. 
In addition, Michelle imaging in the N' and Qa filters was performed in 2006 \citep{ramos_almeida_infrared_2009}.
A compact MIR nucleus was detected in all cases, while the deepest image (N') shows very faint edge-on spiral-like emission extending from the nucleus similar to the IRAC $8\,\mu$m image (PA$\sim30\degree$; diameter$\sim13\arcsec\sim2.5\,$kpc).
The nucleus itself is extended in  the north-south directions in that image as well (PA$\sim5\degree$; FWHM(major axis)$\sim0.49\arcsec\sim90$\,pc).
It appears resolved also in the sharpest image (NEII\_1 from 2006-02-16) but with a slightly larger position angle (PA$\sim12\degree$).
Thus, we classify the nucleus of NGC\,2992 as resolved in the MIR at subarcsecond resolution.
On average, the nuclear PSF photometry is $\sim 42\%$ lower than the \spitzerr spectrophotometry. 
However, we notice that the VISIR measurements from 2009 are systematically $\sim 48\%$ lower than the VISIR measurements from 2006 in the same filters. 
The Michelle and early VISIR total fluxes are rather consistent with the \spitzerr data, and possibly also with the historical measurements.
Such a comparison is however complicated by the strong MIR features in the $N$-band at arcsecond scales.
Thus, it is possible that NGC\,2992 exhibits long-term MIR flux variations similar to shorter wavelengths (e.g., \citealt{glass_variability_1997}).
For these reasons, we are also unable to constrain the spectral features as silicate and PAH at subarcsecond resolution, despite the good $N$-band coverage.
\newline\end{@twocolumnfalse}]

\begin{figure}
   \centering
   \includegraphics[angle=0,width=8.500cm]{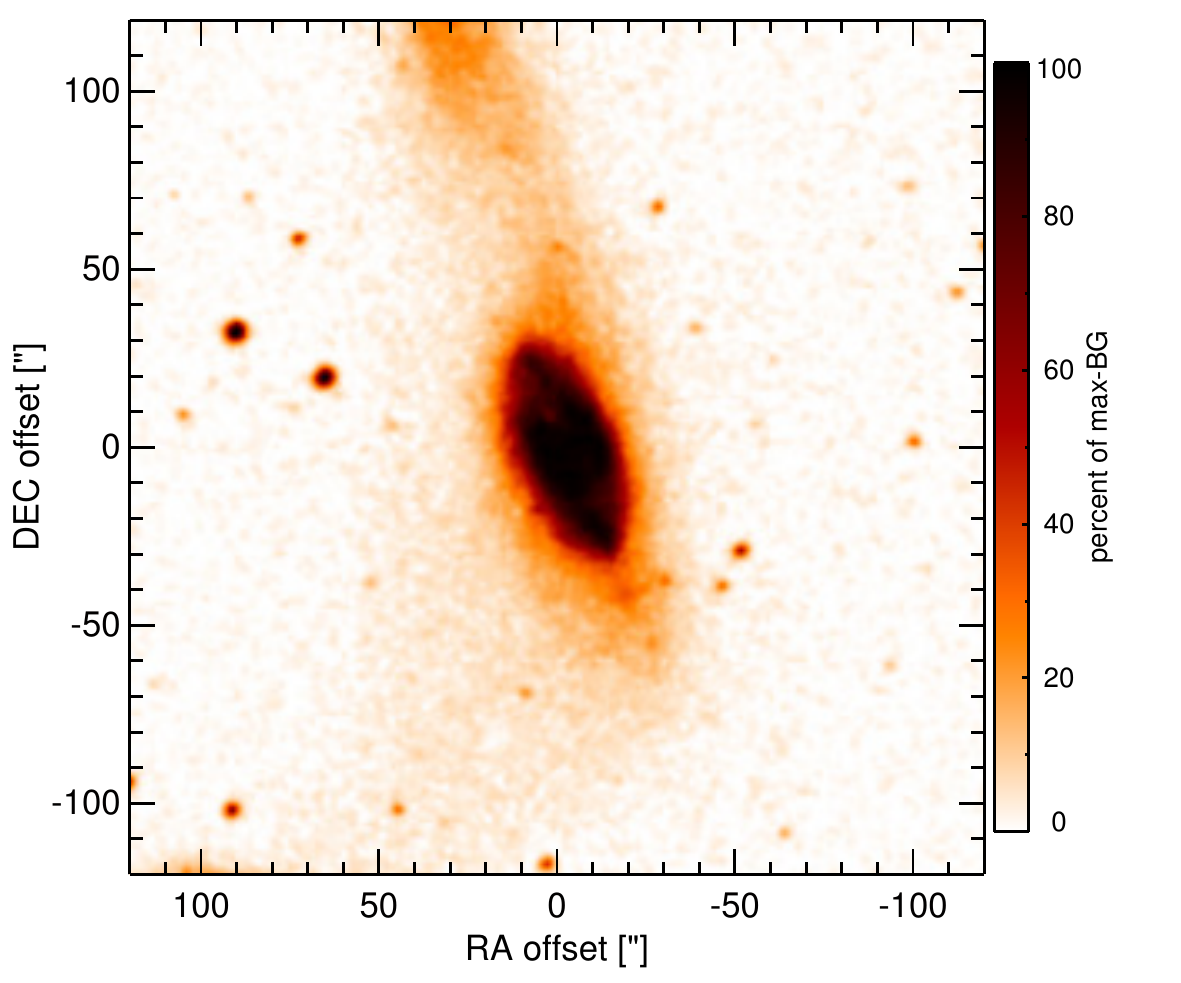}
    \caption{\label{fig:OPTim_NGC2992}
             Optical image (DSS, red filter) of NGC\,2992. Displayed are the central $4\arcmin$ with North up and East to the left. 
              The colour scaling is linear with white corresponding to the median background and black to the $0.01\%$ pixels with the highest intensity.  
           }
\end{figure}
\begin{figure}
   \centering
   \includegraphics[angle=0,height=3.11cm]{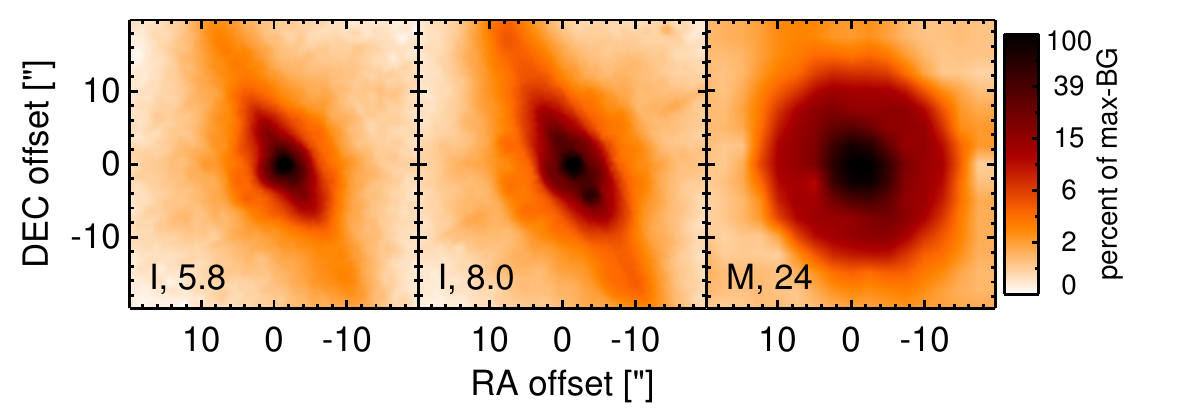}
    \caption{\label{fig:INTim_NGC2992}
             \spitzerr MIR images of NGC\,2992. Displayed are the inner $40\arcsec$ with North up and East to the left. The colour scaling is logarithmic with white corresponding to median background and black to the $0.1\%$ pixels with the highest intensity.
             The label in the bottom left states instrument and central wavelength of the filter in $\mu$m (I: IRAC, M: MIPS).
             Note that the apparent off-nuclear compact source in the IRAC $8.0\,\mu$m image is an instrumental artefact.
           }
\end{figure}
\begin{figure}
   \centering
   \includegraphics[angle=0,width=8.500cm]{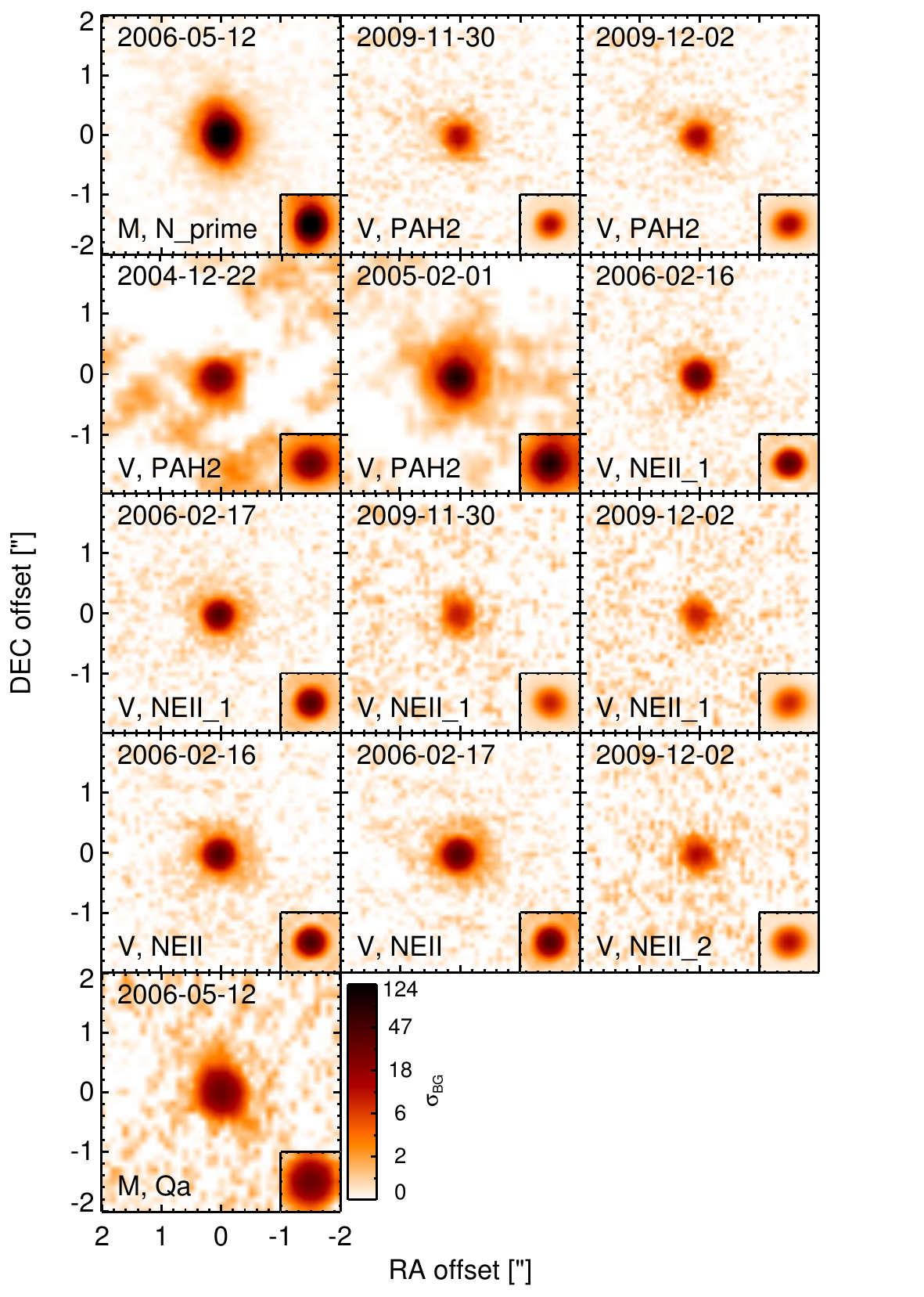}
    \caption{\label{fig:HARim_NGC2992}
             Subarcsecond-resolution MIR images of NGC\,2992 sorted by increasing filter wavelength. 
             Displayed are the inner $4\arcsec$ with North up and East to the left. 
             The colour scaling is logarithmic with white corresponding to median background and black to the $75\%$ of the highest intensity of all images in units of $\sigbg$.
             The inset image shows the central arcsecond of the PSF from the calibrator star, scaled to match the science target.
             The labels in the bottom left state instrument and filter names (C: COMICS, M: Michelle, T: T-ReCS, V: VISIR).
           }
\end{figure}
\begin{figure}
   \centering
   \includegraphics[angle=0,width=8.50cm]{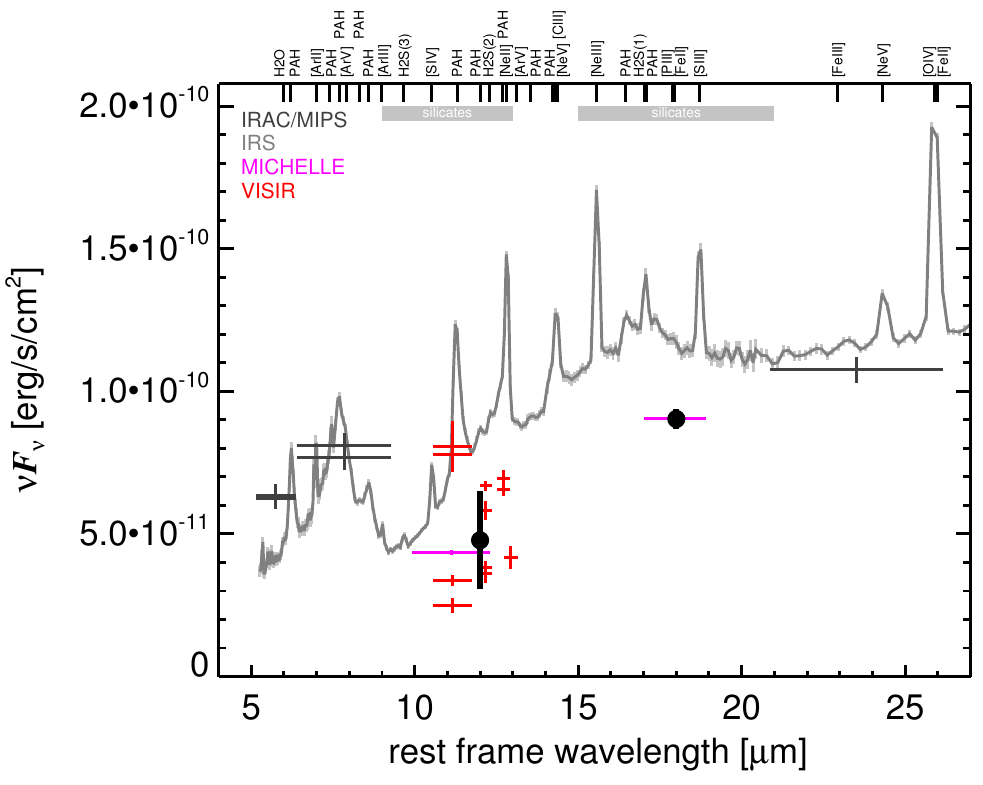}
   \caption{\label{fig:MISED_NGC2992}
      MIR SED of NGC\,2992. The description  of the symbols (if present) is the following.
      Grey crosses and  solid lines mark the \spitzer/IRAC, MIPS and IRS data. 
      The colour coding of the other symbols is: 
      green for COMICS, magenta for Michelle, blue for T-ReCS and red for VISIR data.
      Darker-coloured solid lines mark spectra of the corresponding instrument.
      The black filled circles mark the nuclear 12 and $18\,\mu$m  continuum emission estimate from the data.
      The ticks on the top axis mark positions of common MIR emission lines, while the light grey horizontal bars mark wavelength ranges affected by the silicate 10 and 18$\mu$m features.}
\end{figure}
\clearpage

\twocolumn[\begin{@twocolumnfalse}  
\subsection{NGC\,3081}\label{app:NGC3081}
NGC\,3081 is a low-inclination barred spiral galaxy at a redshift of $z=$ 0.008 ($D\sim40.9\,$Mpc) with a Sy\,2 nucleus \citep{phillips_nearby_1983,durret_narrow_1986} with polarized broad emission lines \citep{moran_frequency_2000}.
It is an X-ray ``buried" AGN candidate \citep{eguchi_suzaku_2011} and belongs to the nine-month BAT AGN sample.
The AGN very compact in radio possibly elongated with PA$\sim 164\degree$; \citealt{nagar_radio_1999} and possesses a biconical NLR extending $\sim4\arcsec\sim780\,$pc along a similar PA \citep{ferruit_hubble_2000}.
The first $N$-band observations were performed by \cite{krabbe_n-band_2001} who report a point-like MIR nucleus without any host emission being detected.
In the \spitzer/IRAC and MIPS images, the compact nucleus is embedded within weak lenticular host emission (north-south direction) and surrounded by a large-scale ring, which becomes increasingly prominent with increasing wavelength.
The \spitzer/IRS LR staring-mode spectrum exhibits possible weak silicate  $10\,\mu$m absorption, weak PAH emission, prominent forbidden emission lines and an emission peak at $\sim17\,\mu$m in $\nu F_\nu$-space (see also \citealt{shi_9.7_2006,deo_mid-infrared_2009,weaver_mid-infrared_2010}).
Thus, the arcsecond-scale MIR SED  seems to contain only minor star formation contribution.
NGC\,3081 was observed with T-ReCS in the Si2 and Qa filters in 2006 \citep{ramos_almeida_infrared_2009} and with VISIR in four narrow $N$-band filters in 2007, two of which a published in \cite{gandhi_resolving_2009}.
A compact MIR nucleus was detected in all images albeit with low S/N in most VISIR images, while extended biconical emission is visible in the Qa image (diameter$\sim2\arcsec\sim400\,pc$; PA$\sim170\degree$).
This extended emission is consistent with the radio and NLR structures.
Owing to the low S/N, it remains uncertain, whether this structure is present also at shorter wavelengths. 
Note that extended emission might also be present in the Si2 image but with a different PA.
Our T-ReCS nuclear photometry is consistent with \cite{ramos_almeida_infrared_2009}, while this is not the case for the VISIR fluxes compared to \cite{gandhi_resolving_2009}, which is caused by our improved measurement technique for faint detections. 
In addition, our photometry agrees with the T-ReCS LR $N$-band spectrum published by \cite{gonzalez-martin_dust_2013} well and is on average $\sim 39\%$ lower than the \spitzerr spectrophotometry.
Note that the T-ReCS spectrum does not exhibit any PAH emission, which indicates that at subarcsecond resolution, we are probing the pure AGN-related MIR emission of NGC\,3081.
\newline\end{@twocolumnfalse}]

\begin{figure}
   \centering
   \includegraphics[angle=0,width=8.500cm]{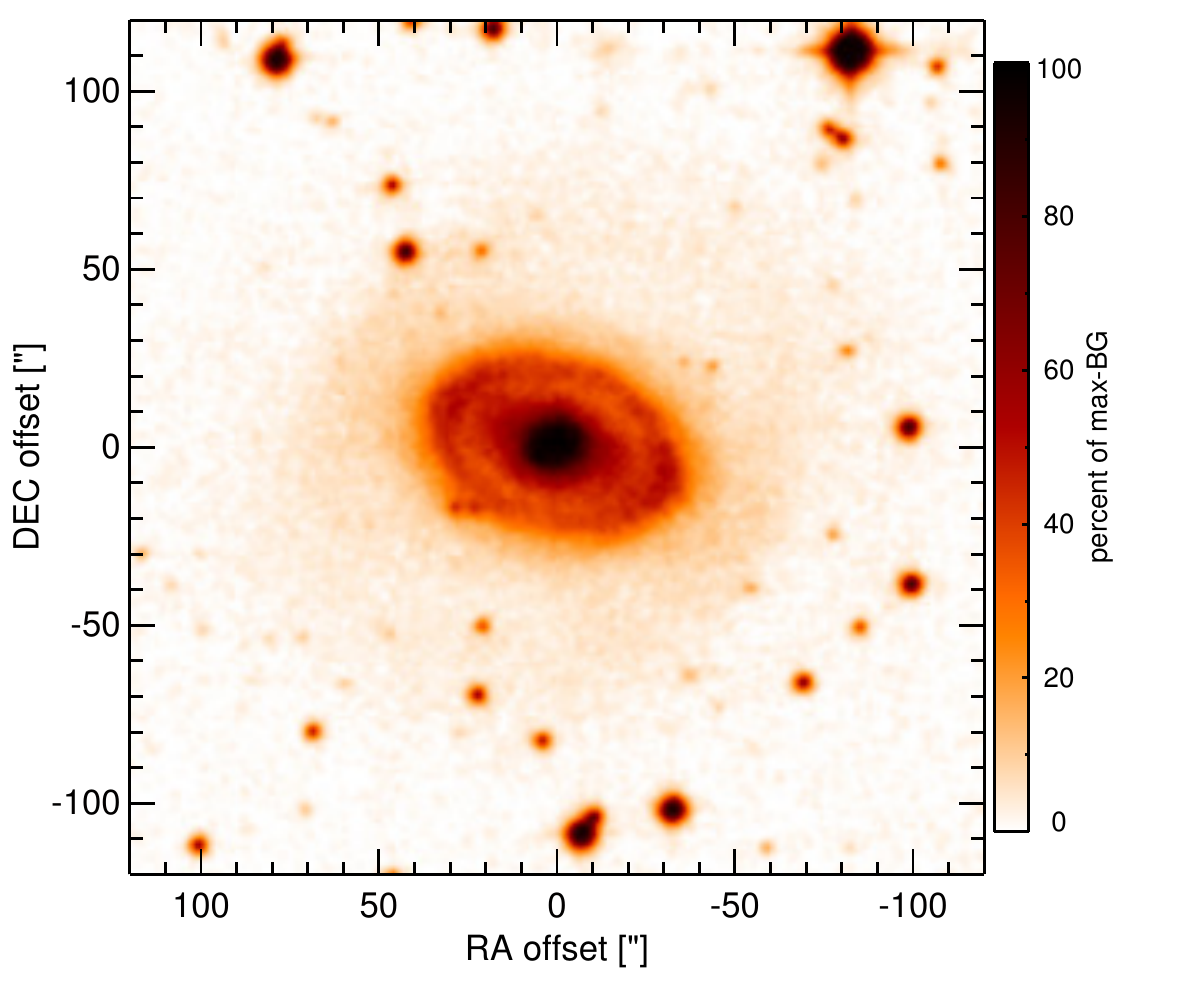}
    \caption{\label{fig:OPTim_NGC3081}
             Optical image (DSS, red filter) of NGC\,3081. Displayed are the central $4\arcmin$ with North up and East to the left. 
              The colour scaling is linear with white corresponding to the median background and black to the $0.01\%$ pixels with the highest intensity.  
           }
\end{figure}
\begin{figure}
   \centering
   \includegraphics[angle=0,height=3.11cm]{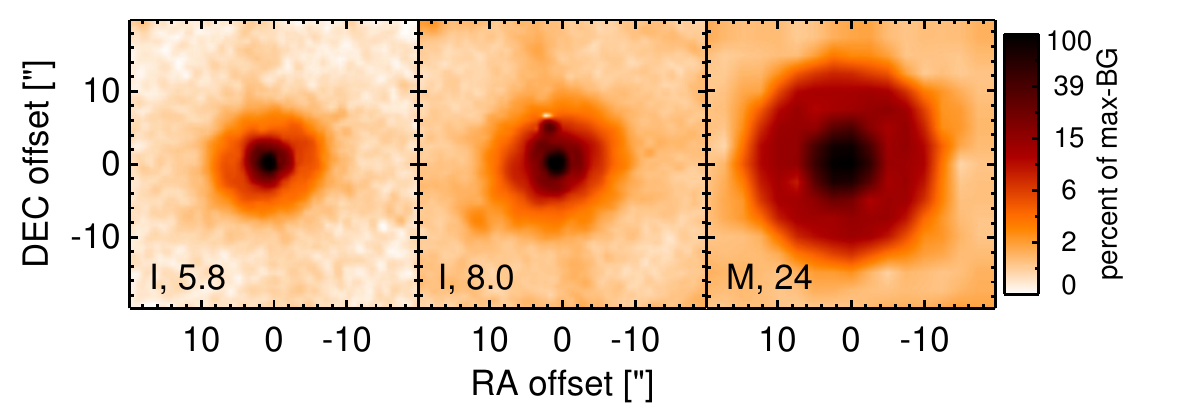}
    \caption{\label{fig:INTim_NGC3081}
             \spitzerr MIR images of NGC\,3081. Displayed are the inner $40\arcsec$ with North up and East to the left. The colour scaling is logarithmic with white corresponding to median background and black to the $0.1\%$ pixels with the highest intensity.
             The label in the bottom left states instrument and central wavelength of the filter in $\mu$m (I: IRAC, M: MIPS). 
             Note that the apparent off-nuclear compact source in the IRAC $8.0\,\mu$m image is an instrumental artefact.
           }
\end{figure}
\begin{figure}
   \centering
   \includegraphics[angle=0,width=8.500cm]{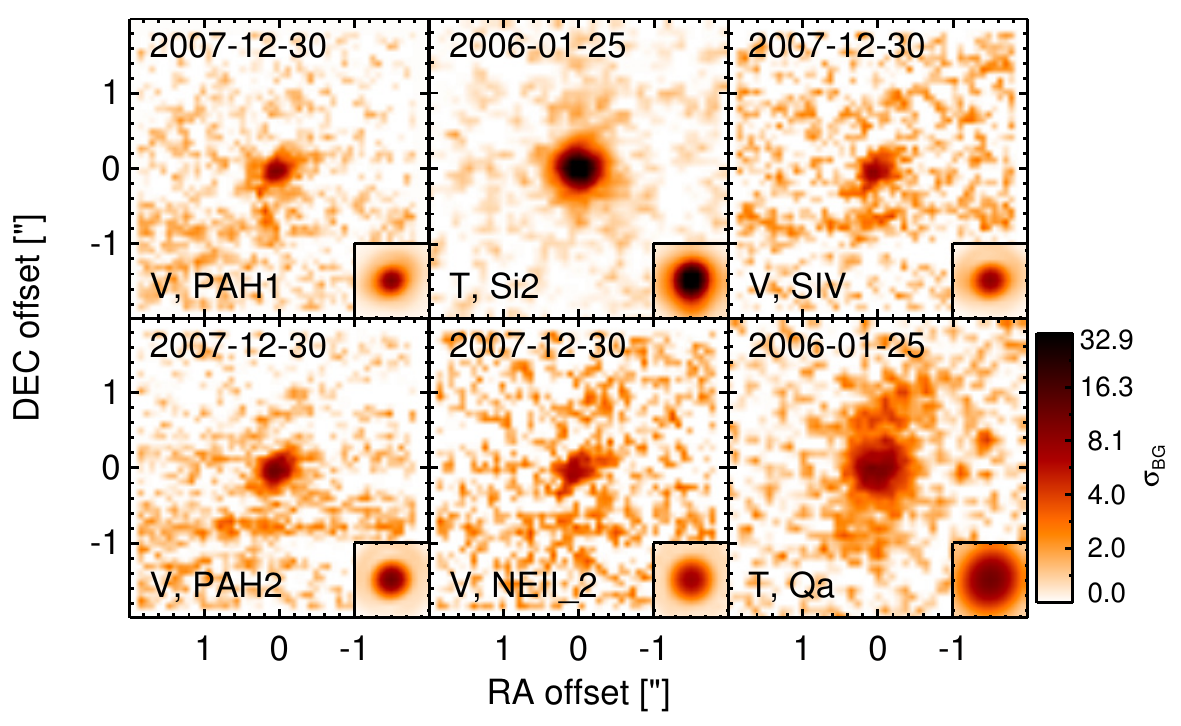}
    \caption{\label{fig:HARim_NGC3081}
             Subarcsecond-resolution MIR images of NGC\,3081 sorted by increasing filter wavelength. 
             Displayed are the inner $4\arcsec$ with North up and East to the left. 
             The colour scaling is logarithmic with white corresponding to median background and black to the $75\%$ of the highest intensity of all images in units of $\sigbg$.
             The inset image shows the central arcsecond of the PSF from the calibrator star, scaled to match the science target.
             The labels in the bottom left state instrument and filter names (C: COMICS, M: Michelle, T: T-ReCS, V: VISIR).
           }
\end{figure}
\begin{figure}
   \centering
   \includegraphics[angle=0,width=8.50cm]{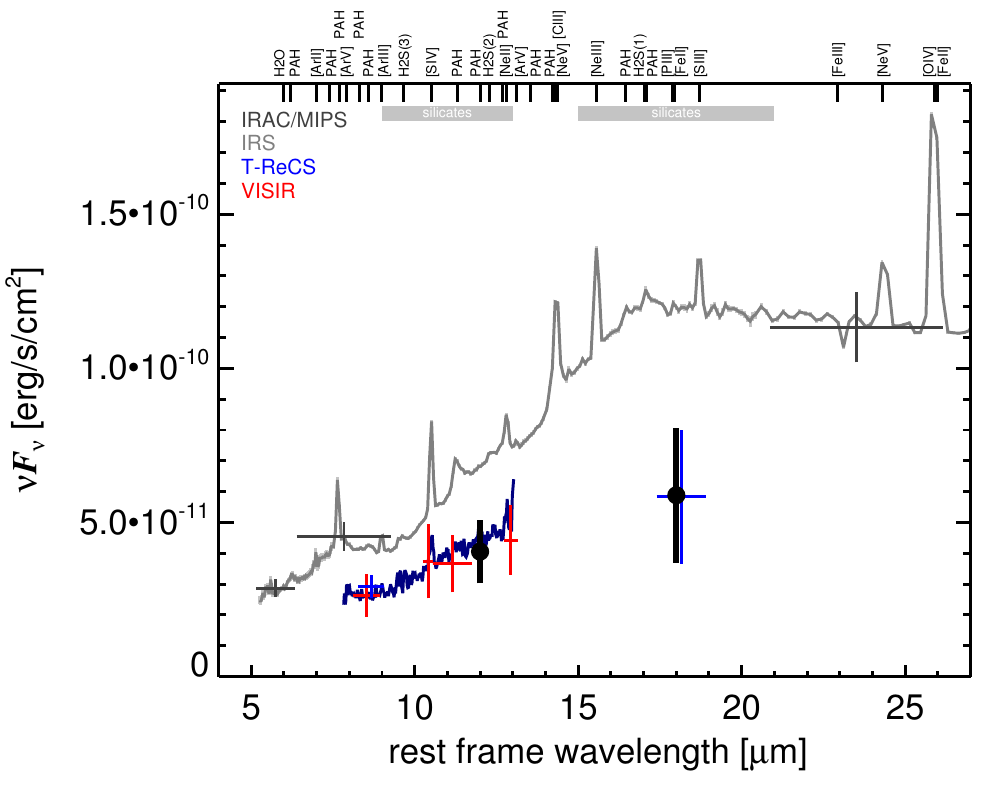}
   \caption{\label{fig:MISED_NGC3081}
      MIR SED of NGC\,3081. The description  of the symbols (if present) is the following.
      Grey crosses and  solid lines mark the \spitzer/IRAC, MIPS and IRS data. 
      The colour coding of the other symbols is: 
      green for COMICS, magenta for Michelle, blue for T-ReCS and red for VISIR data.
      Darker-coloured solid lines mark spectra of the corresponding instrument.
      The black filled circles mark the nuclear 12 and $18\,\mu$m  continuum emission estimate from the data.
      The ticks on the top axis mark positions of common MIR emission lines, while the light grey horizontal bars mark wavelength ranges affected by the silicate 10 and 18$\mu$m features.}
\end{figure}
\clearpage

\twocolumn[\begin{@twocolumnfalse}  
\subsection{NGC\,3094}\label{app:NGC3094}
NGC\,3094 is a low-inclination barred spiral galaxy at a redshift of $z=$ 0.008 ($D\sim40.8\,$Mpc) with an obscured AGN \citep{armus_long-slit_1989,imanishi_3.4-m_2000}.
Not much is known about this AGN, there are not X-ray measurements available, and only a low-resolution radio map showing extended emission \citep{condon_1.49_1990}.
We conservatively treat this object as uncertain type~II AGN.
After the discovery of its MIR brightness through \iras, NGC\,3094 was observed spectroscopically by \cite{roche_atlas_1991} who found a deep silicate feature dominating the whole $N$-band.
No \spitzerr data are available for this object but \cite{roche_silicate_2007} obtained a T-ReCS LR $N$-band spectrum verifying the deep silicate $10\,\mu$m absorption feature. 
In an acquisition image, the nucleus appear extended along the north-south direction with $\sim1\arcsec$ diameter.
Here, we report three additional T-ReCS images in the broad N filter obtained between 2008 and 2009 (unpublished, to our knowledge).
Again, the nucleus appears elongated north-south but the difference to the corresponding standard stars is not significant. 
Note that in the four \textit{WISE} bands, NGC\,3094 is also slightly elongated in the same direction.
Our nuclear photometry is consistent with the T-ReCS spectrum reanalysed by \cite{gonzalez-martin_dust_2013} and therefore we use the latter to correct our 12\,$\mu$m continuum emission estimate for the silicate absorption.
\newline\end{@twocolumnfalse}]

\begin{figure}
   \centering
   \includegraphics[angle=0,width=8.500cm]{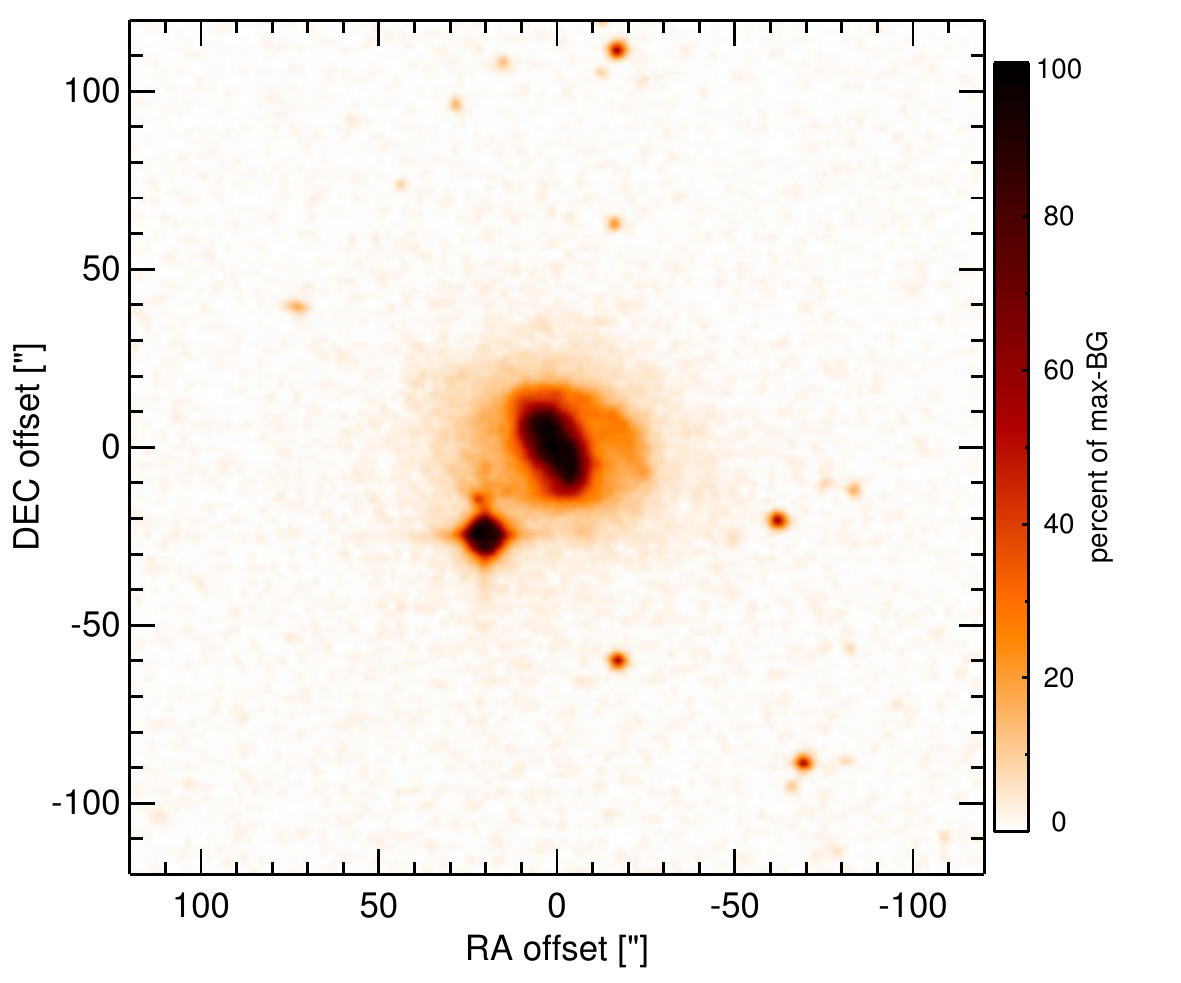}
    \caption{\label{fig:OPTim_NGC3094}
             Optical image (DSS, red filter) of NGC\,3094. Displayed are the central $4\arcmin$ with North up and East to the left. 
              The colour scaling is linear with white corresponding to the median background and black to the $0.01\%$ pixels with the highest intensity.  
           }
\end{figure}
\begin{figure}
   \centering
   \includegraphics[angle=0,height=3.11cm]{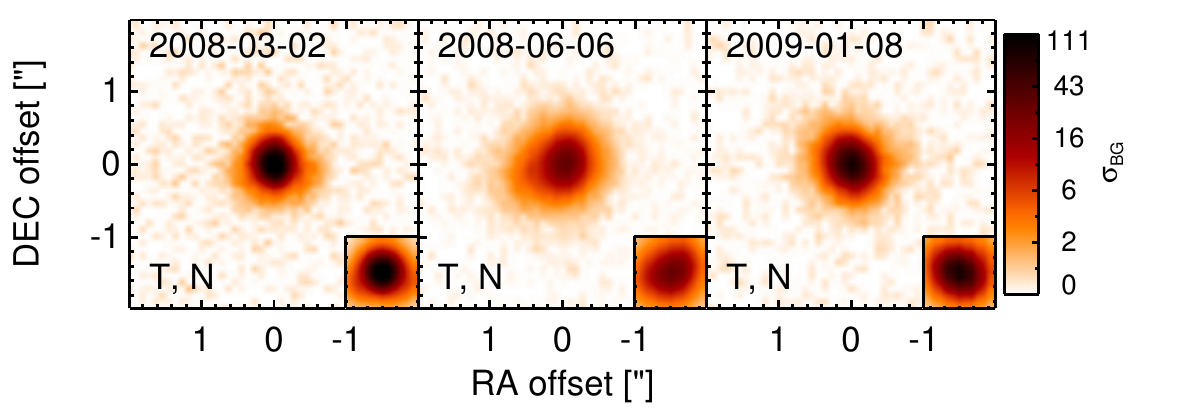}
    \caption{\label{fig:HARim_NGC3094}
             Subarcsecond-resolution MIR images of NGC\,3094 sorted by increasing filter wavelength. 
             Displayed are the inner $4\arcsec$ with North up and East to the left. 
             The colour scaling is logarithmic with white corresponding to median background and black to the $75\%$ of the highest intensity of all images in units of $\sigbg$.
             The inset image shows the central arcsecond of the PSF from the calibrator star, scaled to match the science target.
             The labels in the bottom left state instrument and filter names (C: COMICS, M: Michelle, T: T-ReCS, V: VISIR).
           }
\end{figure}
\begin{figure}
   \centering
   \includegraphics[angle=0,width=8.50cm]{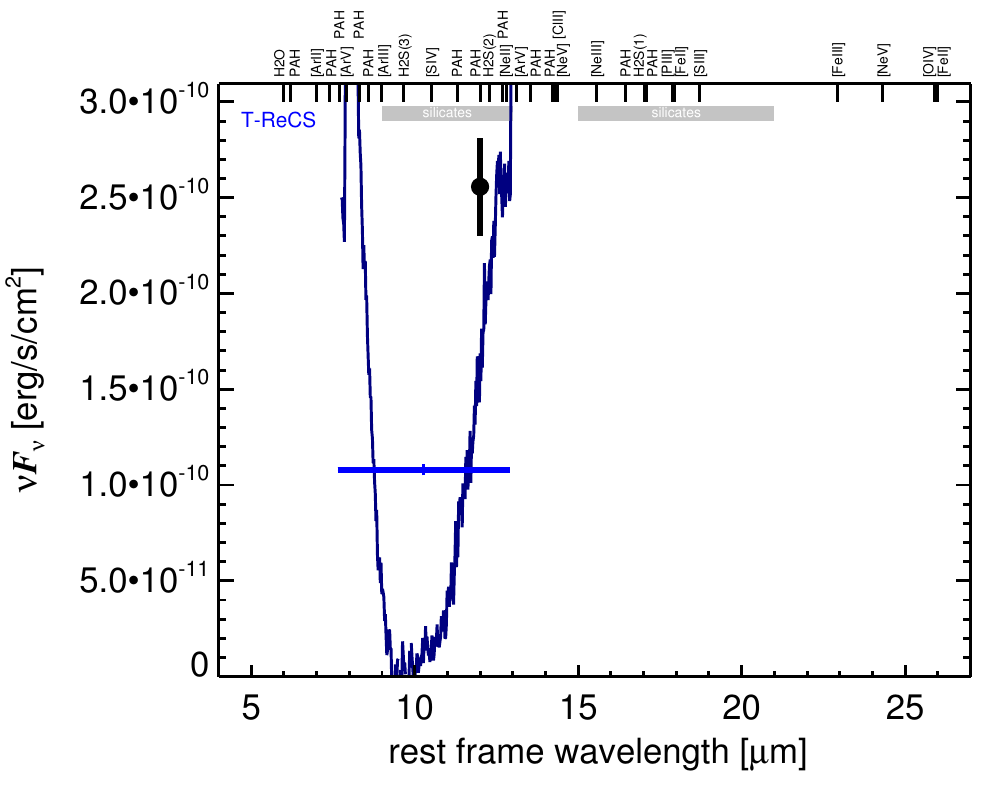}
   \caption{\label{fig:MISED_NGC3094}
      MIR SED of NGC\,3094. The description  of the symbols (if present) is the following.
      Grey crosses and  solid lines mark the \spitzer/IRAC, MIPS and IRS data. 
      The colour coding of the other symbols is: 
      green for COMICS, magenta for Michelle, blue for T-ReCS and red for VISIR data.
      Darker-coloured solid lines mark spectra of the corresponding instrument.
      The black filled circles mark the nuclear 12 and $18\,\mu$m  continuum emission estimate from the data.
      The ticks on the top axis mark positions of common MIR emission lines, while the light grey horizontal bars mark wavelength ranges affected by the silicate 10 and 18$\mu$m features.}
\end{figure}
\clearpage

\twocolumn[\begin{@twocolumnfalse}  
\subsection{NGC\,3147}\label{app:NGC3147}
NGC\,3147 is a face-on spiral galaxy at a distance of $D=$ $30.1 \pm 4.5$\,Mpc \citep{amanullah_spectra_2010} with a Sy\,2 nucleus \citep{veron-cetty_catalogue_2010}, which is the best known candidate of a ``true'' (unabsorbed) type~II Seyfert (e.g., \citealt{bianchi_ngc_2008,matt_suzaku_2012}).
Note that the detailed multiwavelength study by \cite{casasola_molecular_2008} found large amounts of inflowing gas in the central arcsecond ($\sim 150$\,pc) region of NGC\,3147.
The nucleus appears point-like at radio wavelengths even with VLBI \citep{ulvestad_origin_2001}.
After \iras, NGC\,3147 was observed with \spitzer/IRAC, IRS and MIPS.
A compact dominating nucleus embedded within the spiral-like host emission was detected in the corresponding IRAC and MIPS images (see also \citealt{pahre_mid-infrared_2004}).
The nuclear MIPS 24\,$\mu$m flux is consistent with the value published in \cite{shi_unobscured_2010}.
The IRS LR staring-mode spectrum suffers from low S/N but indicates silicate  $10\,\mu$m emission, PAH features and a shallow blue spectral slope in $\nu F_\nu$-space (see also \citealt{shi_unobscured_2010}).
Thus, the arcsecond-scale MIR SED is possibly host-dominated.
We observed the nuclear region of NGC\,3147 with Michelle in two $N$-band filters in 2010 and detected a compact MIR nucleus, which is possibly marginally resolved (FWHM(major axis)$\sim 0.69\arcsec \sim 100$\,pc; PA$\sim 150\degree$).
However, at least a second epoch with deep subarcsecond MIR imaging is required to confirm this extension.
Our nuclear photometry is on average $\sim 60\%$ lower than the \spitzerr spectrophotometry.
We conclude that the nucleus of NGC\,3147 is either jet-dominated as M87 or it possesses significant AGN-heated dust as expected from a torus.
\newline\end{@twocolumnfalse}]

\begin{figure}
   \centering
   \includegraphics[angle=0,width=8.500cm]{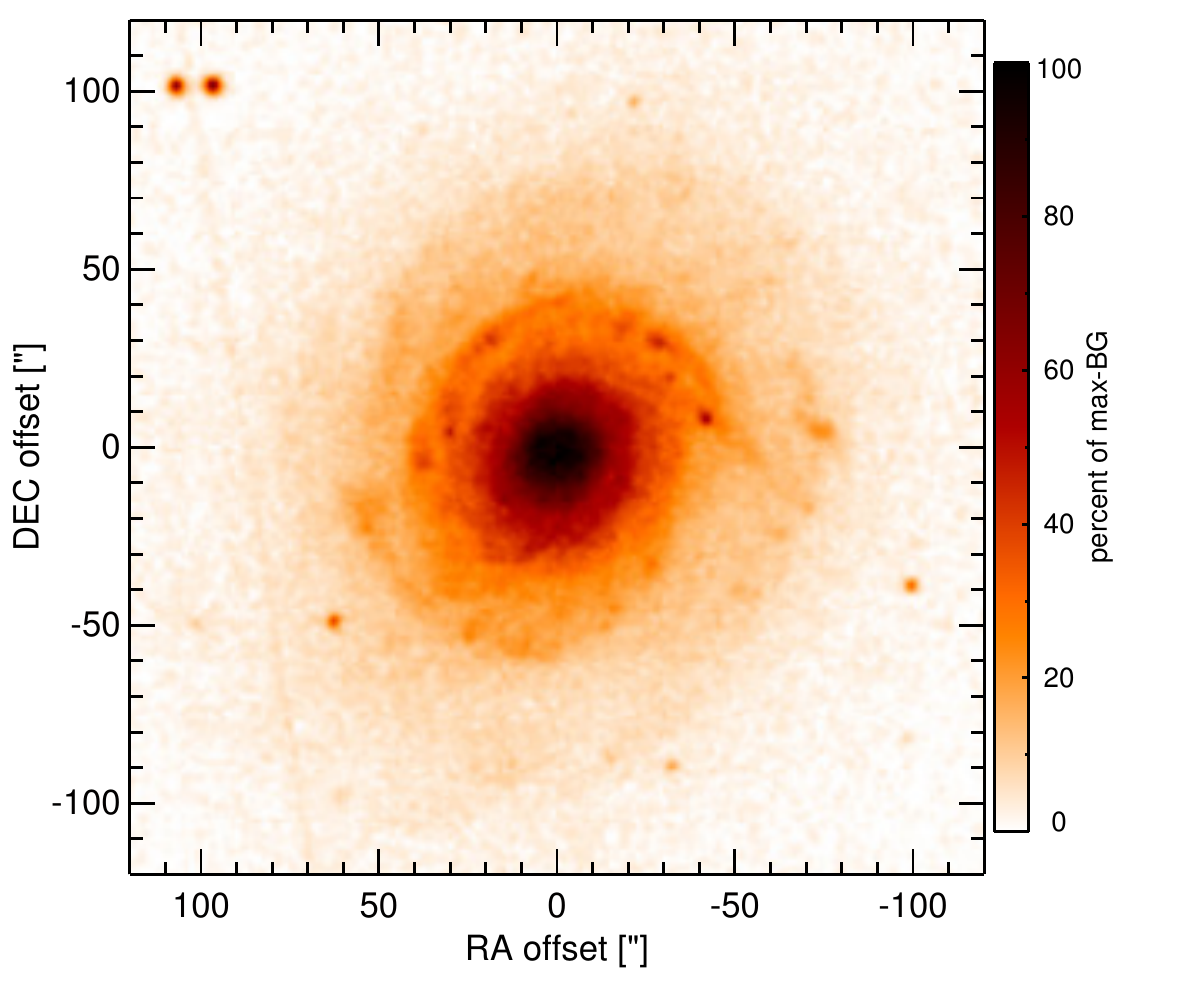}
    \caption{\label{fig:OPTim_NGC3147}
             Optical image (DSS, red filter) of NGC\,3147. Displayed are the central $4\arcmin$ with North up and East to the left. 
              The colour scaling is linear with white corresponding to the median background and black to the $0.01\%$ pixels with the highest intensity.  
           }
\end{figure}
\begin{figure}
   \centering
   \includegraphics[angle=0,height=3.11cm]{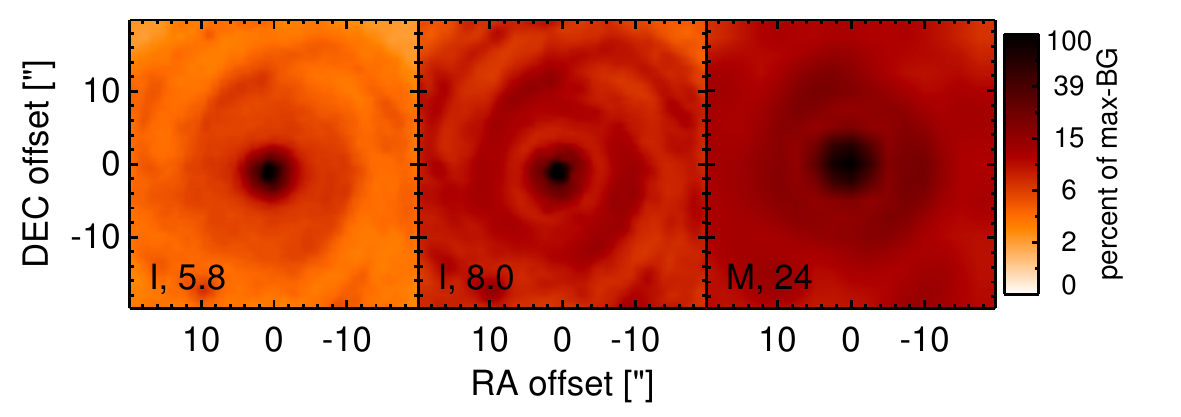}
    \caption{\label{fig:INTim_NGC3147}
             \spitzerr MIR images of NGC\,3147. Displayed are the inner $40\arcsec$ with North up and East to the left. The colour scaling is logarithmic with white corresponding to median background and black to the $0.1\%$ pixels with the highest intensity.
             The label in the bottom left states instrument and central wavelength of the filter in $\mu$m (I: IRAC, M: MIPS). 
           }
\end{figure}
\begin{figure}
   \centering
   \includegraphics[angle=0,height=3.11cm]{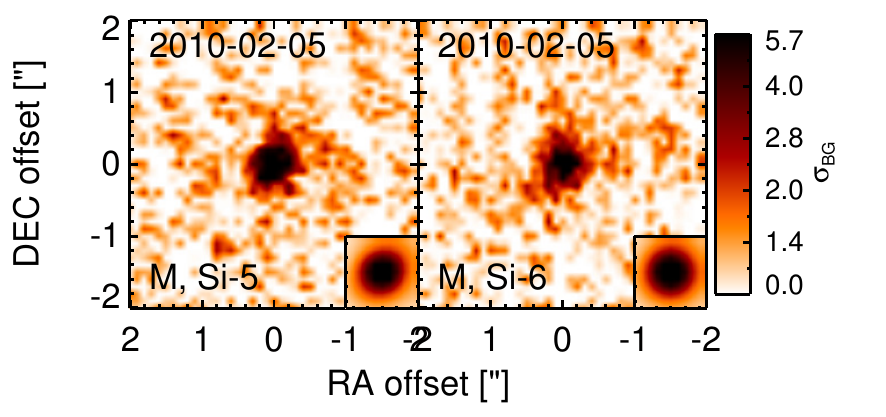}
    \caption{\label{fig:HARim_NGC3147}
             Subarcsecond-resolution MIR images of NGC\,3147 sorted by increasing filter wavelength. 
             Displayed are the inner $4\arcsec$ with North up and East to the left. 
             The colour scaling is logarithmic with white corresponding to median background and black to the $75\%$ of the highest intensity of all images in units of $\sigbg$.
             The inset image shows the central arcsecond of the PSF from the calibrator star, scaled to match the science target.
             The labels in the bottom left state instrument and filter names (C: COMICS, M: Michelle, T: T-ReCS, V: VISIR).
           }
\end{figure}
\begin{figure}
   \centering
   \includegraphics[angle=0,width=8.50cm]{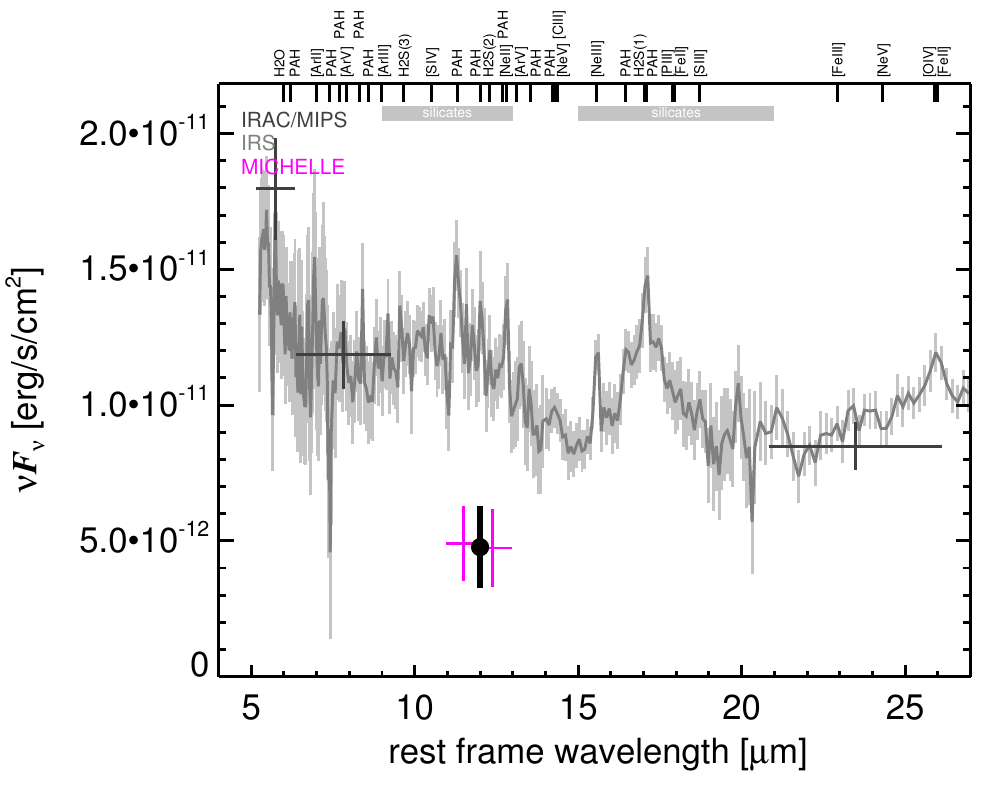}
   \caption{\label{fig:MISED_NGC3147}
      MIR SED of NGC\,3147. The description  of the symbols (if present) is the following.
      Grey crosses and  solid lines mark the \spitzer/IRAC, MIPS and IRS data. 
      The colour coding of the other symbols is: 
      green for COMICS, magenta for Michelle, blue for T-ReCS and red for VISIR data.
      Darker-coloured solid lines mark spectra of the corresponding instrument.
      The black filled circles mark the nuclear 12 and $18\,\mu$m  continuum emission estimate from the data.
      The ticks on the top axis mark positions of common MIR emission lines, while the light grey horizontal bars mark wavelength ranges affected by the silicate 10 and 18$\mu$m features.}
\end{figure}
\clearpage

\twocolumn[\begin{@twocolumnfalse}  
\subsection{NGC\,3166}\label{app:NGC3166}
NGC\,3166 is an inclined spiral galaxy at a redshift of $z=$ 
0.0045 ($D\sim25.3\,$Mpc), forming a possibly interacting group with NGC\,3165 and NGC\,3169.
It hosts a little-studied LINER nucleus \citep{ho_search_1997-1}, which remained undetected at radio wavelengths \citep{nagar_radio_2005} but is supported by the detection of \nev in the MIR \citep{dudik_spitzer_2009}.
After \iras, it was observed with IRTF \citep{devereux_infrared_1987}, and with \spitzer/IRAC, IRS and MIPS.
NGC\,3166 shows an elliptical extended nucleus embedded in spiral-like host emission in the corresponding IRAC and MIPS images.
The IRS HR staring-mode spectrum exhibits prominent a PAH 11.3$\,\mu$m feature on-top of silicate 10$\,\mu$m emission and a flat spectral slope in $\nu F_\nu$-space.
Note however that no background subtraction was performed for this spectrum.
The nuclear region of NGC\,3166 was observed with T-ReCS in the Si2 filter in 2007 \citep{mason_nuclear_2012} and a weak, compact MIR nucleus was detected.
The low S/N prevents us to perform a quantitative extension analysis.
Our re-measurement of the nuclear flux is consistent with the value by \citep{mason_nuclear_2012} and an order of magnitude lower than the \spitzerr spectrophotometry. 
We conclude that the AGN is very weak and completely dominated by host emission even on arcsecond scales ($\sim 0.4$\,kpc).
\newline\end{@twocolumnfalse}]

\begin{figure}
   \centering
   \includegraphics[angle=0,width=8.500cm]{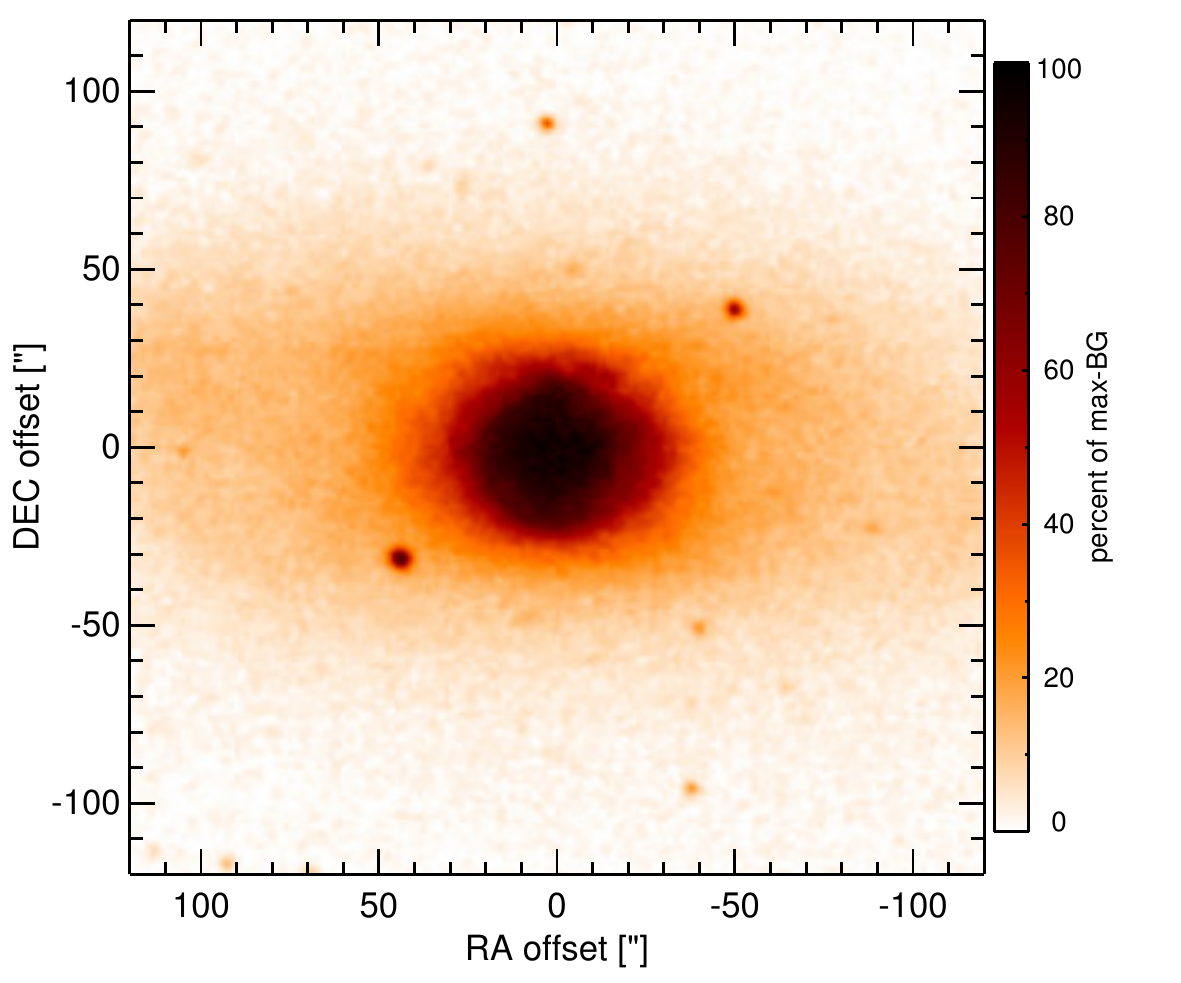}
    \caption{\label{fig:OPTim_NGC3166}
             Optical image (DSS, red filter) of NGC\,3166. Displayed are the central $4\arcmin$ with North up and East to the left. 
              The colour scaling is linear with white corresponding to the median background and black to the $0.01\%$ pixels with the highest intensity.  
           }
\end{figure}
\begin{figure}
   \centering
   \includegraphics[angle=0,height=3.11cm]{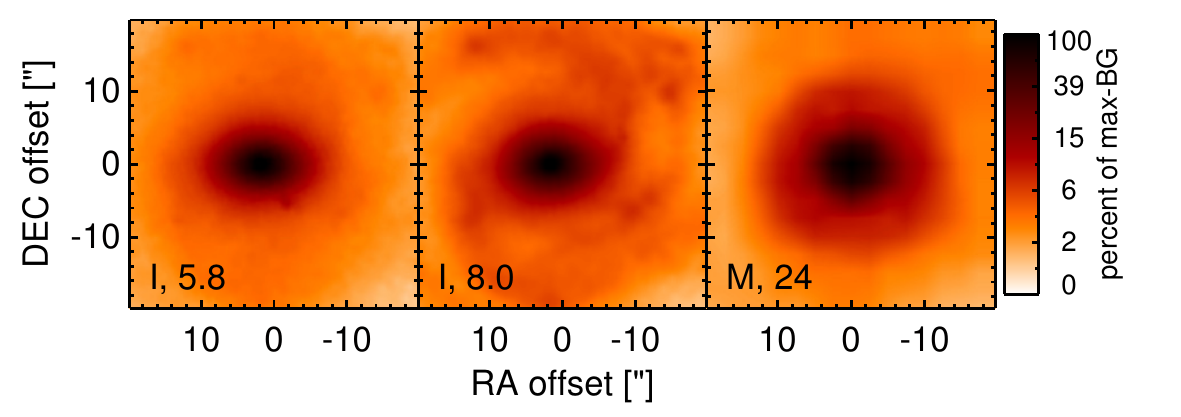}
    \caption{\label{fig:INTim_NGC3166}
             \spitzerr MIR images of NGC\,3166. Displayed are the inner $40\arcsec$ with North up and East to the left. The colour scaling is logarithmic with white corresponding to median background and black to the $0.1\%$ pixels with the highest intensity.
             The label in the bottom left states instrument and central wavelength of the filter in $\mu$m (I: IRAC, M: MIPS). 
           }
\end{figure}
\begin{figure}
   \centering
   \includegraphics[angle=0,height=3.11cm]{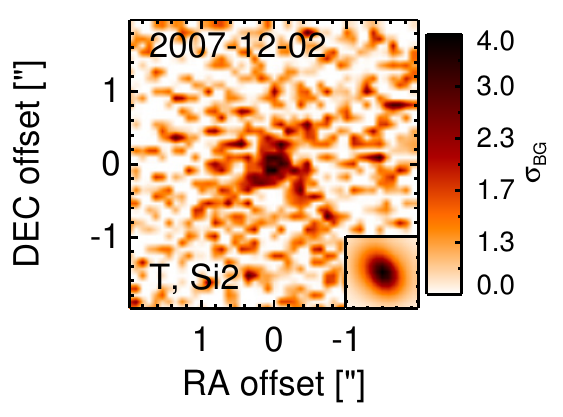}
    \caption{\label{fig:HARim_NGC3166}
             Subarcsecond-resolution MIR images of NGC\,3166 sorted by increasing filter wavelength. 
             Displayed are the inner $4\arcsec$ with North up and East to the left. 
             The colour scaling is logarithmic with white corresponding to median background and black to the $75\%$ of the highest intensity of all images in units of $\sigbg$.
             The inset image shows the central arcsecond of the PSF from the calibrator star, scaled to match the science target.
             The labels in the bottom left state instrument and filter names (C: COMICS, M: Michelle, T: T-ReCS, V: VISIR).
           }
\end{figure}
\begin{figure}
   \centering
   \includegraphics[angle=0,width=8.50cm]{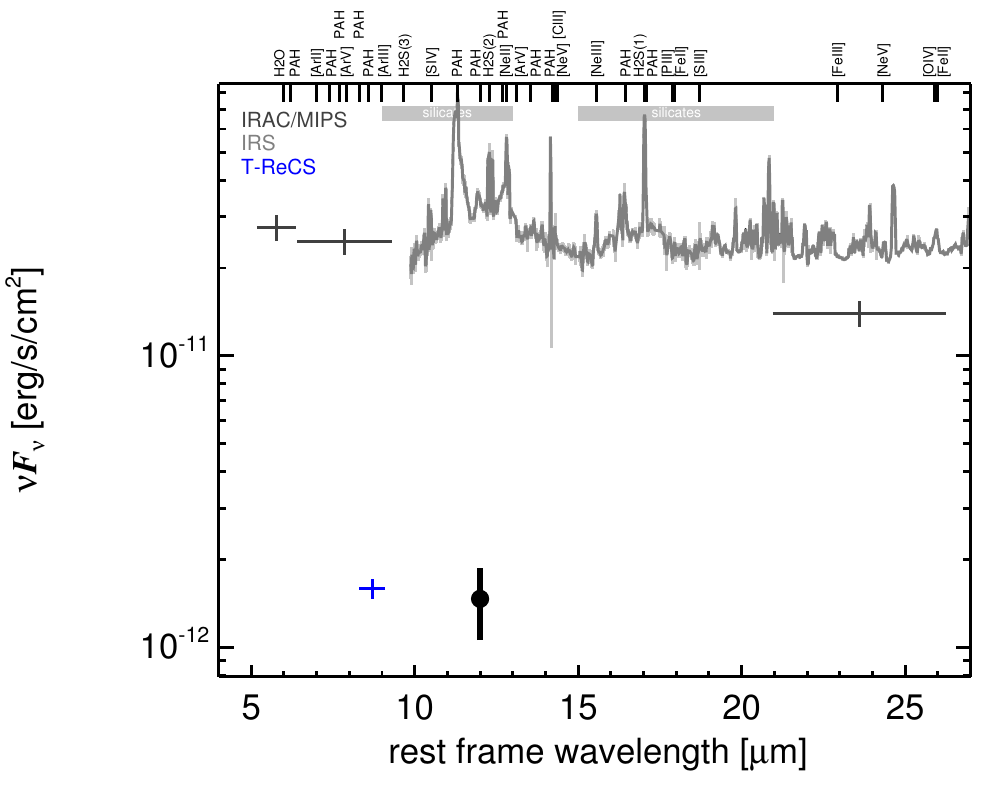}
   \caption{\label{fig:MISED_NGC3166}
      MIR SED of NGC\,3166. The description  of the symbols (if present) is the following.
      Grey crosses and  solid lines mark the \spitzer/IRAC, MIPS and IRS data. 
      The colour coding of the other symbols is: 
      green for COMICS, magenta for Michelle, blue for T-ReCS and red for VISIR data.
      Darker-coloured solid lines mark spectra of the corresponding instrument.
      The black filled circles mark the nuclear 12 and $18\,\mu$m  continuum emission estimate from the data.
      The ticks on the top axis mark positions of common MIR emission lines, while the light grey horizontal bars mark wavelength ranges affected by the silicate 10 and 18$\mu$m features.}
\end{figure}
\clearpage

\twocolumn[\begin{@twocolumnfalse}  
\subsection{NGC\,3169}\label{app:NGC3169}
NGC\,3169 is an inclined spiral galaxy at a distance of $D=$ $18.7 \pm 3.4$\,Mpc (NED redshift-independent median) with a LINER nucleus \citep{ho_search_1997-1}.
At radio wavelengths, the nucleus appears a compact source with biconical extended emission along a PA$\sim120\degree$ on arcsecond-scales \citep{hummel_effects_1987,nagar_radio_2000} and as unresolved on subarcsecond scales \citep{nagar_radio_2005}.
The extended NLR is cospatial with the arcsecond radio structure \citep{gonzalez_delgado_h_1997}.
After \iras, NGC\,3169 was observed in the MIR with IRTF \citep{devereux_infrared_1987} and \spitzer/IRAC, IRS and MIPS.
An elliptical elongated nucleus embedded within the spiral-like host emission was detected in the corresponding IRAC and MIPS images.
The IRS HR staring-mode spectrum exhibits prominent PAH 11.3\,$\mu$m emission, a flat spectral slope in $\nu F_\nu$-space, and silicate  emission.
Note however that no background subtraction was performed for this spectrum.
The AGN-indicative high-ionization emission line \nev is not clearly detected \citep{dudik_spitzer_2009}.
The nuclear region of NGC\,3169 was imaged with T-ReCS in the Si2 filter during two nights in 2007 \citep{mason_nuclear_2012} and with VISIR in the PAH2\_2 filter in 2010 (this work).
In all images, a compact MIR nucleus is weakly detected as well as biconical extended emission of $\sim 2\arcsec \sim 180$\,pc extent along a PA$\sim51\degree$. 
Although the S/N is very low, we assume that this structure is real because it appears consistent in three different images of two different instruments.
Surprisingly though, it is perpendicular to the radio and NLR emission.
Our nuclear photometry is generally consistent with \cite{mason_nuclear_2012} and on average $\sim 95\%$ lower than the \spitzerr spectrophotometry.
We conclude that the AGN in NGC\,3169 is totally dominated by the host emission in the MIR  on arcsecond scales ($\sim 350\,$pc). 
\newline\end{@twocolumnfalse}]

\begin{figure}
   \centering
   \includegraphics[angle=0,width=8.500cm]{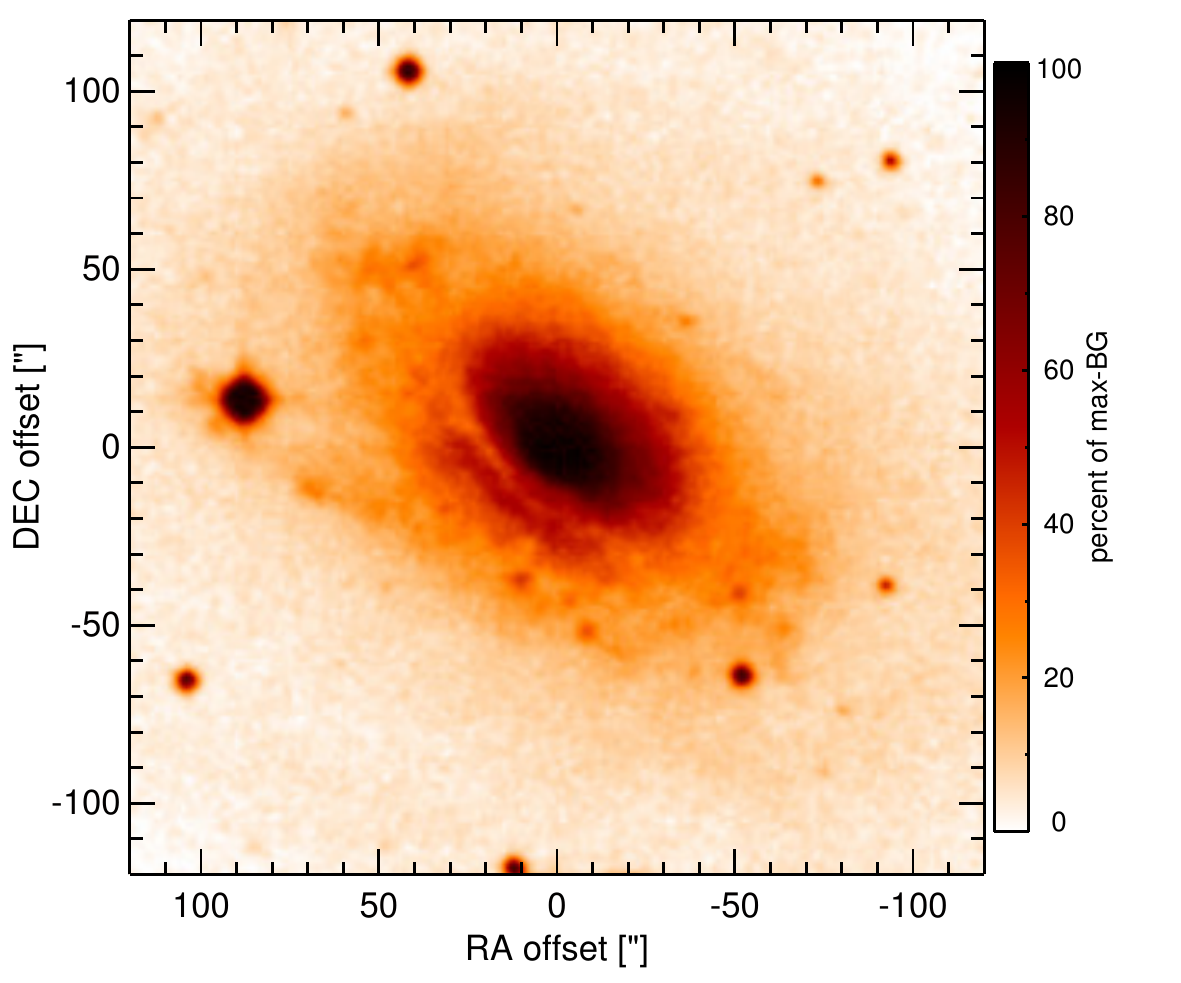}
    \caption{\label{fig:OPTim_NGC3169}
             Optical image (DSS, red filter) of NGC\,3169. Displayed are the central $4\arcmin$ with North up and East to the left. 
              The colour scaling is linear with white corresponding to the median background and black to the $0.01\%$ pixels with the highest intensity.  
           }
\end{figure}
\begin{figure}
   \centering
   \includegraphics[angle=0,height=3.11cm]{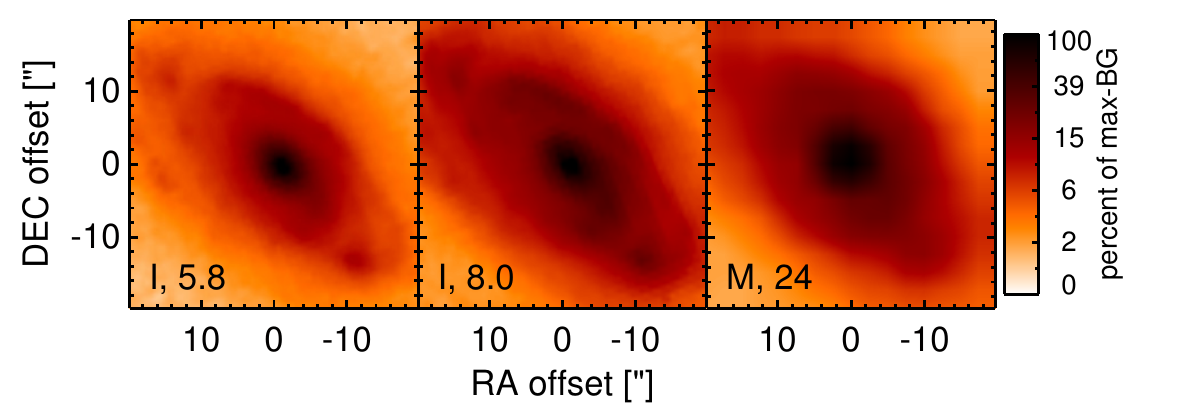}
    \caption{\label{fig:INTim_NGC3169}
             \spitzerr MIR images of NGC\,3169. Displayed are the inner $40\arcsec$ with North up and East to the left. The colour scaling is logarithmic with white corresponding to median background and black to the $0.1\%$ pixels with the highest intensity.
             The label in the bottom left states instrument and central wavelength of the filter in $\mu$m (I: IRAC, M: MIPS). 
           }
\end{figure}
\begin{figure}
   \centering
   \includegraphics[angle=0,height=3.11cm]{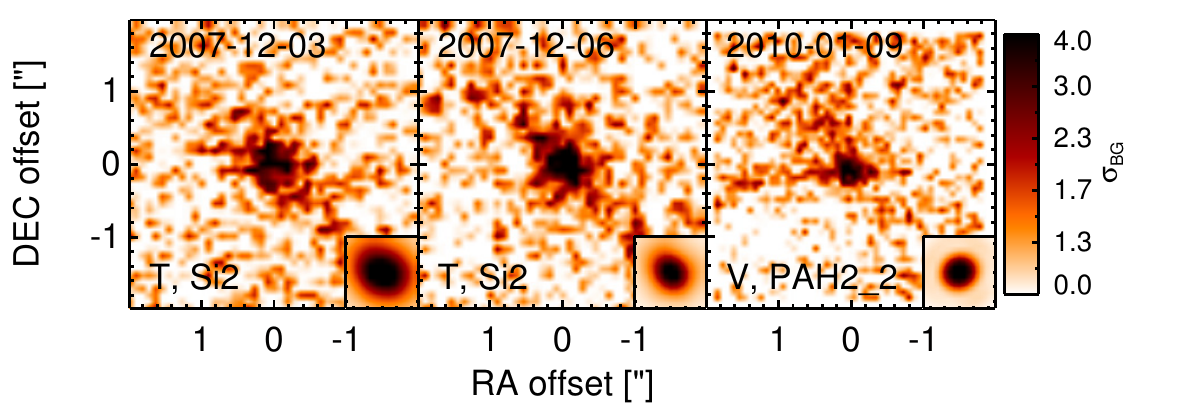}
    \caption{\label{fig:HARim_NGC3169}
             Subarcsecond-resolution MIR images of NGC\,3169 sorted by increasing filter wavelength. 
             Displayed are the inner $4\arcsec$ with North up and East to the left. 
             The colour scaling is logarithmic with white corresponding to median background and black to the $75\%$ of the highest intensity of all images in units of $\sigbg$.
             The inset image shows the central arcsecond of the PSF from the calibrator star, scaled to match the science target.
             The labels in the bottom left state instrument and filter names (C: COMICS, M: Michelle, T: T-ReCS, V: VISIR).
           }
\end{figure}
\begin{figure}
   \centering
   \includegraphics[angle=0,width=8.50cm]{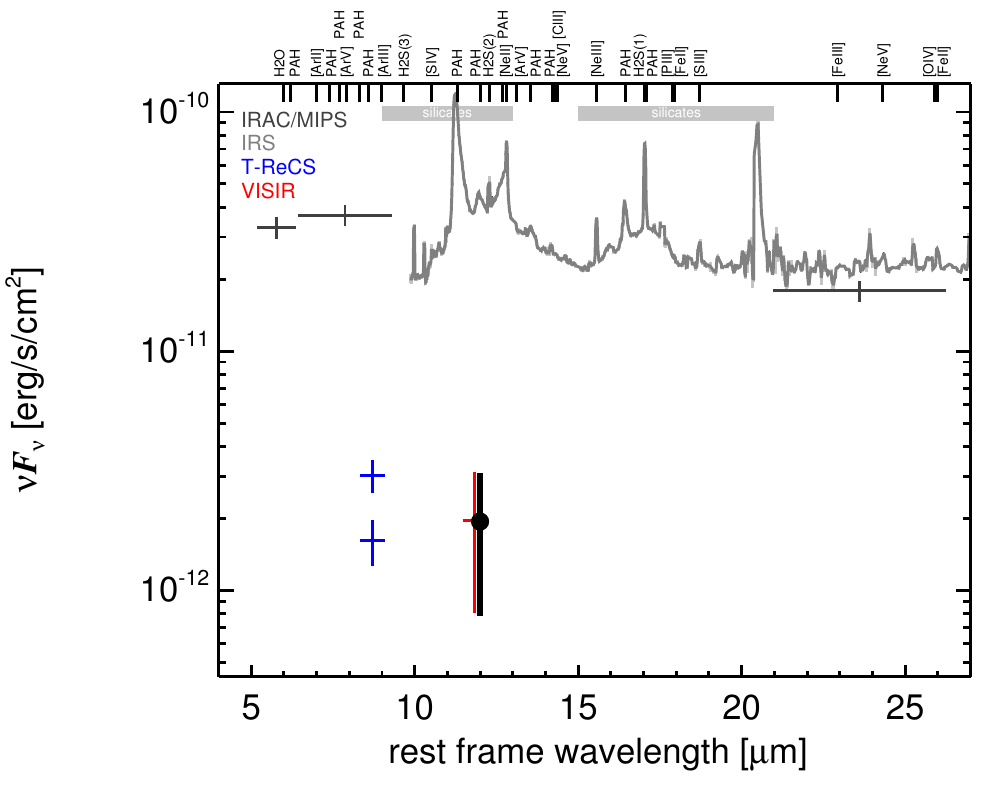}
   \caption{\label{fig:MISED_NGC3169}
      MIR SED of NGC\,3169. The description  of the symbols (if present) is the following.
      Grey crosses and  solid lines mark the \spitzer/IRAC, MIPS and IRS data. 
      The colour coding of the other symbols is: 
      green for COMICS, magenta for Michelle, blue for T-ReCS and red for VISIR data.
      Darker-coloured solid lines mark spectra of the corresponding instrument.
      The black filled circles mark the nuclear 12 and $18\,\mu$m  continuum emission estimate from the data.
      The ticks on the top axis mark positions of common MIR emission lines, while the light grey horizontal bars mark wavelength ranges affected by the silicate 10 and 18$\mu$m features.}
\end{figure}
\clearpage

\twocolumn[\begin{@twocolumnfalse}  
\subsection{NGC\,3185}\label{app:NGC3185}
NGC\,3185 is an inclined spiral galaxy at a distance of $D=$ $20.3 \pm 6.1$\,Mpc \citep{springob_erratum:_2009} with a possible AGN, optically classified as a borderline Sy\,2/H\,II nucleus \citep{ho_search_1997-1,goncalves_agns_1999}.
The nucleus is only marginally detected at radio wavelengths (e.g., \citealt{ho_radio_2001}).
The nucleus of NGC\,3185 was first observed with IRTF \citep{devereux_infrared_1987} and  MMT \citep{maiolino_new_1995}, in which it was weakly detected at $\sim 20\,$mJy.
In the \spitzer/IRAC and MIPS images, a compact nucleus embedded within the spiral-like host emission was detected.
The nucleus is extended in the IRAC images.
The \spitzer/IRS LR staring-mode spectrum is dominated by PAH emission linked to star formation, a possible weak silicate $10\,\mu$m absorption feature and a red spectral slope in $\nu F_\nu$-space.
At subarcsecond-resolution, the nuclear region of NGC\,3185 was observed with VISIR in the PAH1 filter in 2010 (unpublished, to our knowledge), but the nucleus remained undetected.
Our derived flux upper limit is $\sim62\%$ lower than the \spitzerr spectrophotometry, which is consistent with the historical measurements.
This result demonstrates that star formation is dominating the MIR emission of the central $\sim 0.4$\,kpc in NGC\,3185.
Note however that \cite{pereira-santaella_mid-infrared_2010} claims the detection of \nev, which supports the presence of an AGN in this object.
\newline\end{@twocolumnfalse}]

\begin{figure}
   \centering
   \includegraphics[angle=0,width=8.500cm]{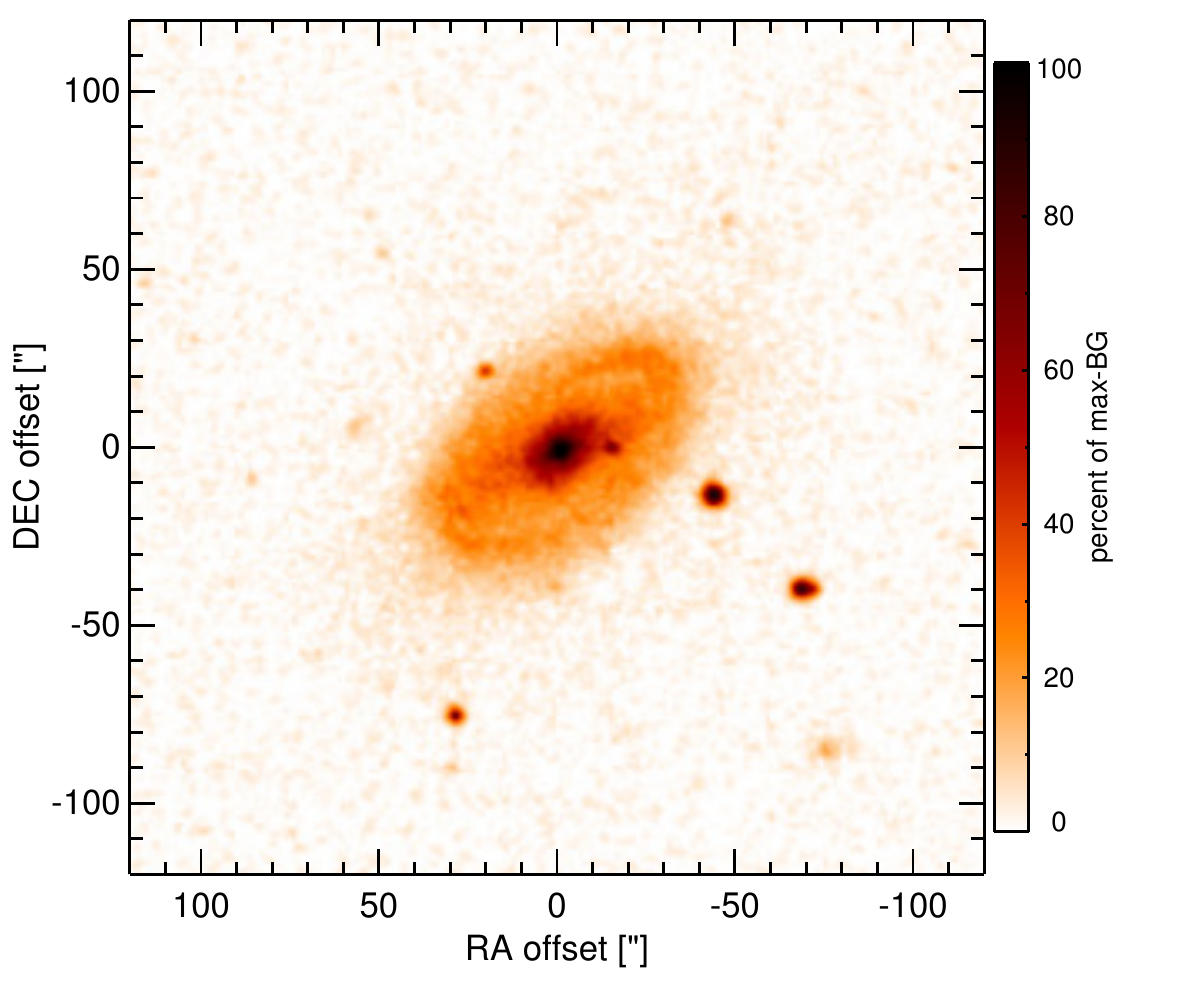}
    \caption{\label{fig:OPTim_NGC3185}
             Optical image (DSS, red filter) of NGC\,3185. Displayed are the central $4\arcmin$ with North up and East to the left. 
              The colour scaling is linear with white corresponding to the median background and black to the $0.01\%$ pixels with the highest intensity.  
           }
\end{figure}
\begin{figure}
   \centering
   \includegraphics[angle=0,height=3.11cm]{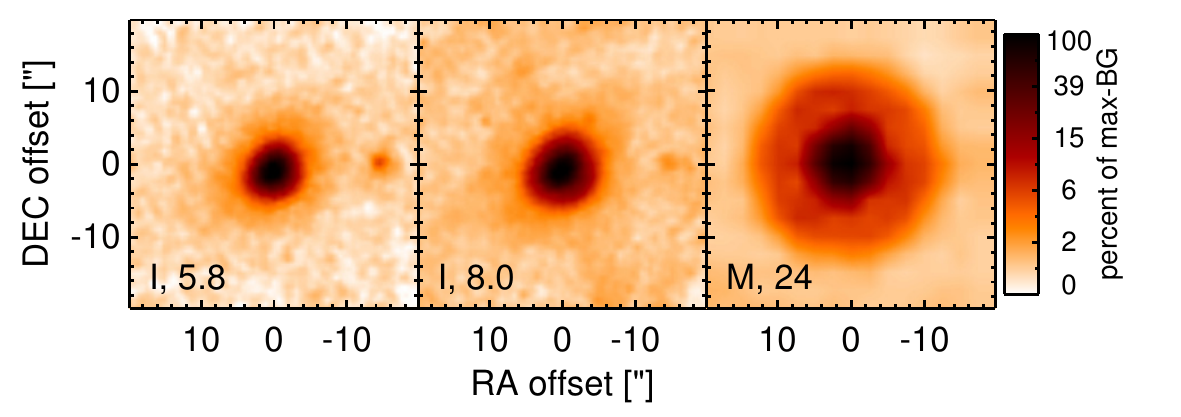}
    \caption{\label{fig:INTim_NGC3185}
             \spitzerr MIR images of NGC\,3185. Displayed are the inner $40\arcsec$ with North up and East to the left. The colour scaling is logarithmic with white corresponding to median background and black to the $0.1\%$ pixels with the highest intensity.
             The label in the bottom left states instrument and central wavelength of the filter in $\mu$m (I: IRAC, M: MIPS). 
           }
\end{figure}
\begin{figure}
   \centering
   \includegraphics[angle=0,width=8.50cm]{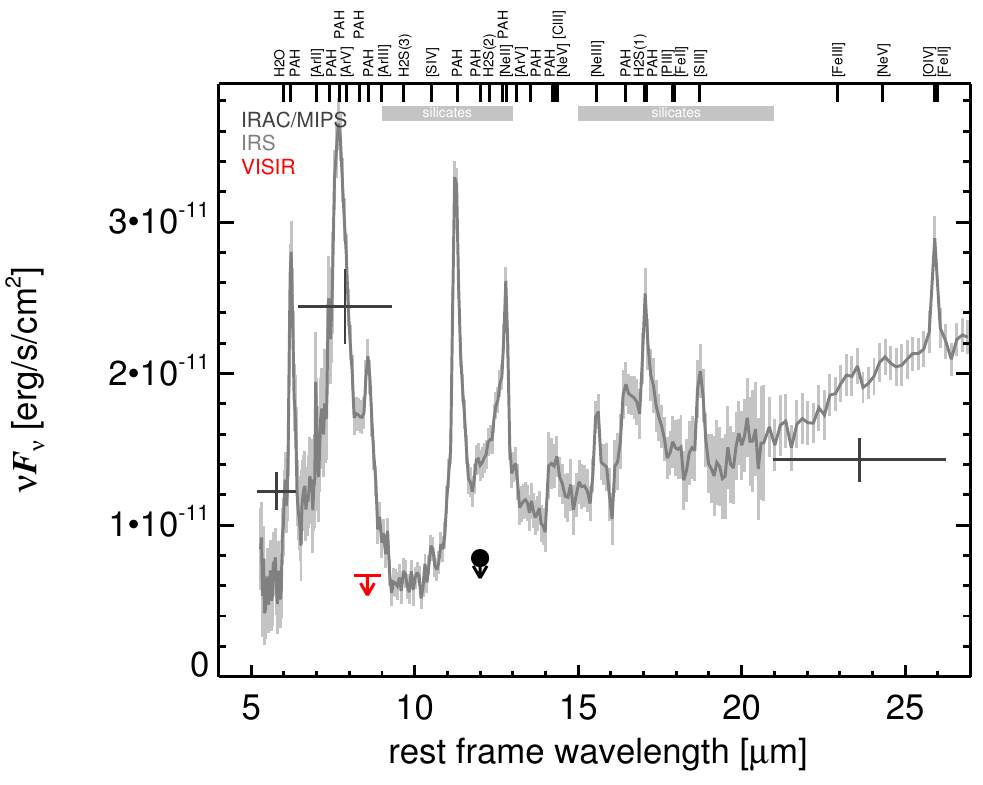}
   \caption{\label{fig:MISED_NGC3185}
      MIR SED of NGC\,3185. The description  of the symbols (if present) is the following.
      Grey crosses and  solid lines mark the \spitzer/IRAC, MIPS and IRS data. 
      The colour coding of the other symbols is: 
      green for COMICS, magenta for Michelle, blue for T-ReCS and red for VISIR data.
      Darker-coloured solid lines mark spectra of the corresponding instrument.
      The black filled circles mark the nuclear 12 and $18\,\mu$m  continuum emission estimate from the data.
      The ticks on the top axis mark positions of common MIR emission lines, while the light grey horizontal bars mark wavelength ranges affected by the silicate 10 and 18$\mu$m features.}
\end{figure}
\clearpage

\twocolumn[\begin{@twocolumnfalse}  
\subsection{NGC\,3227}\label{app:NGC3227}
NGC\,3227 is a low-inclination barred spiral galaxy at a redshift of $z=$ 0.0039 ($D\sim 22.1\,$Mpc) interacting with NGC\,3226.
It hosts a circum-nuclear starburst and a well-studied variable Sy\,1.5 nucleus \citep{veron-cetty_catalogue_2010} that belongs to the nine-month BAT AGN sample. 
In addition, NGC\,3227 possesses a complex extended NLR with \oiii emission extending to the north-east (PA$\sim30\degree$), which does not align with the double lobe radio emission along a PA$\sim-10\degree$ \citep{mundell_anisotropic_1995}. 
Furthermore, the nuclear H$\alpha$ emission extends along a PA$\sim65\degree$ \citep{gonzalez_delgado_circumnuclear_1997}. 
The first MIR observations of NGC\,3227 were carried out by \cite{kleinmann_infrared_1970} and followed by many others using bolometers \citep{rieke_infrared_1972,rieke_infrared_1978,lebofsky_extinction_1979,cutri_statistical_1985,devereux_infrared_1987, ward_continuum_1987, wright_recent_1988}.
Apart from \iras, also \isoo observations are available \citep{perez_garcia_mid-_1998,clavel_2.5-11_2000,domingue_multiwavelength_2003,ramos_almeida_mid-infrared_2007}.
The first subarcsecond $N$-band imaging of the nuclear region was performed with Palomar 5\,m/MIRLIN in 1999 by \cite{gorjian_10_2004}, where a compact MIR nucleus was detected.
The \spitzer/IRAC and MIPS images reveal S-shaped host emission extending from the dominating compact nucleus. 
Our nuclear IRAC $5.8\,\mu$m flux agrees with \cite{gallimore_infrared_2010}, while the IRAC $8.0\,\mu$m flux is significantly lower but in agreement with the \spitzer/IRS LR spectrum.
The latter exhibits prominent PAH features, forbidden emission lines and a red spectral slope in $\nu F_\nu$-space, while the presence of silicate features is uncertain (see also \citealt{wu_spitzer/irs_2009,gallimore_infrared_2010}).
The arcsecond-scale MIR emission of NGC\,3227 is thus severely affected or even dominated by star formation.  
The nuclear region was observed with Michelle in the N' filter in 2006 \citep{ramos_almeida_infrared_2009}, and with VISIR in four $N$ and one $Q$-band filter throughout 2006, 2008 and 2010 (partly published in \cite{honig_dusty_2010-1}).
A compact MIR nucleus was detected in all images, while the deepest image (N') shows faint elliptically extended emission along the host major axis surrounding the nucleus (PA$\sim-10\degree$; major axis diameter $\sim3.5\arcsec\sim0.4\,$kpc).
The nucleus itself appears marginally resolved but not significantly elongated (FWHM $\sim15\%$ larger than standard star) in all cases except the ARIII and N' images, the latter of which was taken in substandard MIR seeing conditions judged from the standard star FWHM.
Thus, we classify this object as extended and measure only the unresolved nuclear component, which is on average $\sim 50\%$ lower than the \spitzerr spectrophotometry and $\sim 30\%$ lower than the VISIR LR $N$-band spectrum presented in \cite{honig_dusty_2010-1}.
The latter is in turn consistent with our total nuclear fluxes and exhibits on average $\sim 30\%$ lower fluxes than the IRS spectrum.
The fact that the VISIR spectrum extracted from the central $\sim70$\,pc shows still  PAH emission indicates the presence of star formation on these scales, which is  consistent with the findings of \cite{rodriguez-ardila_hidden_2003} and \cite{davies_star-forming_2006}.  
In addition, the VISIR spectrum appears to show silicate $10\,\mu$m emission as expected for an unobscured AGN.
Finally, possible $N$-band flux variations of NGC\,3227 are $\lesssim 14\%$ in the last $\sim35\,$years from comparison of the historical and the recent data.
This is well within the expected dispersion only due to different filters, measurement methods and the feature-rich $N$-band SED.
Interestingly, the nucleus of NGC\,3227 was not detected during MIR interferometric observations with MIDI on a short baseline.
This indicates that the nuclear MIR emission is dominated by very extended structures at milliarcsecond scales \citep{burtscher_diversity_2013}.
\newline\end{@twocolumnfalse}]

\begin{figure}
   \centering
   \includegraphics[angle=0,width=8.500cm]{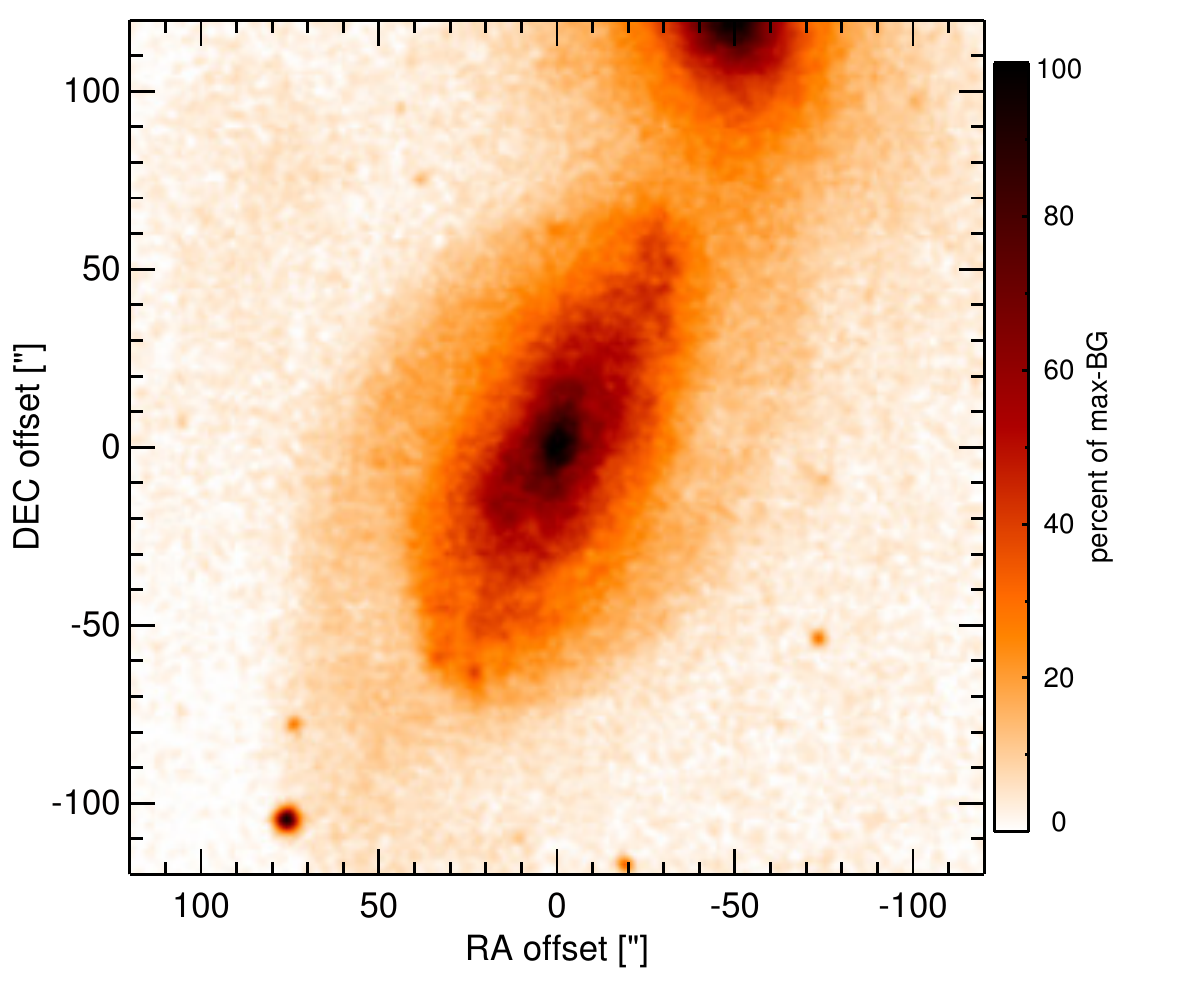}
    \caption{\label{fig:OPTim_NGC3227}
             Optical image (DSS, red filter) of NGC\,3227. Displayed are the central $4\arcmin$ with North up and East to the left. 
              The colour scaling is linear with white corresponding to the median background and black to the $0.01\%$ pixels with the highest intensity.  
           }
\end{figure}
\begin{figure}
   \centering
   \includegraphics[angle=0,height=3.11cm]{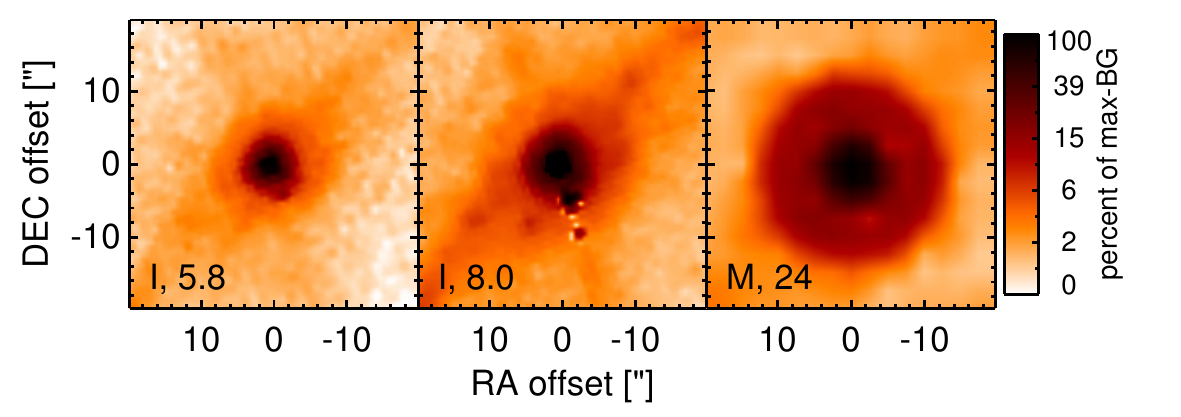}
    \caption{\label{fig:INTim_NGC3227}
             \spitzerr MIR images of NGC\,3227. Displayed are the inner $40\arcsec$ with North up and East to the left. The colour scaling is logarithmic with white corresponding to median background and black to the $0.1\%$ pixels with the highest intensity.
             The label in the bottom left states instrument and central wavelength of the filter in $\mu$m (I: IRAC, M: MIPS).
             Note that the apparent off-nuclear compact sources in the IRAC 5.8 and $8.0\,\mu$m images are instrumental artefacts.
           }
\end{figure}
\begin{figure}
   \centering
   \includegraphics[angle=0,width=8.500cm]{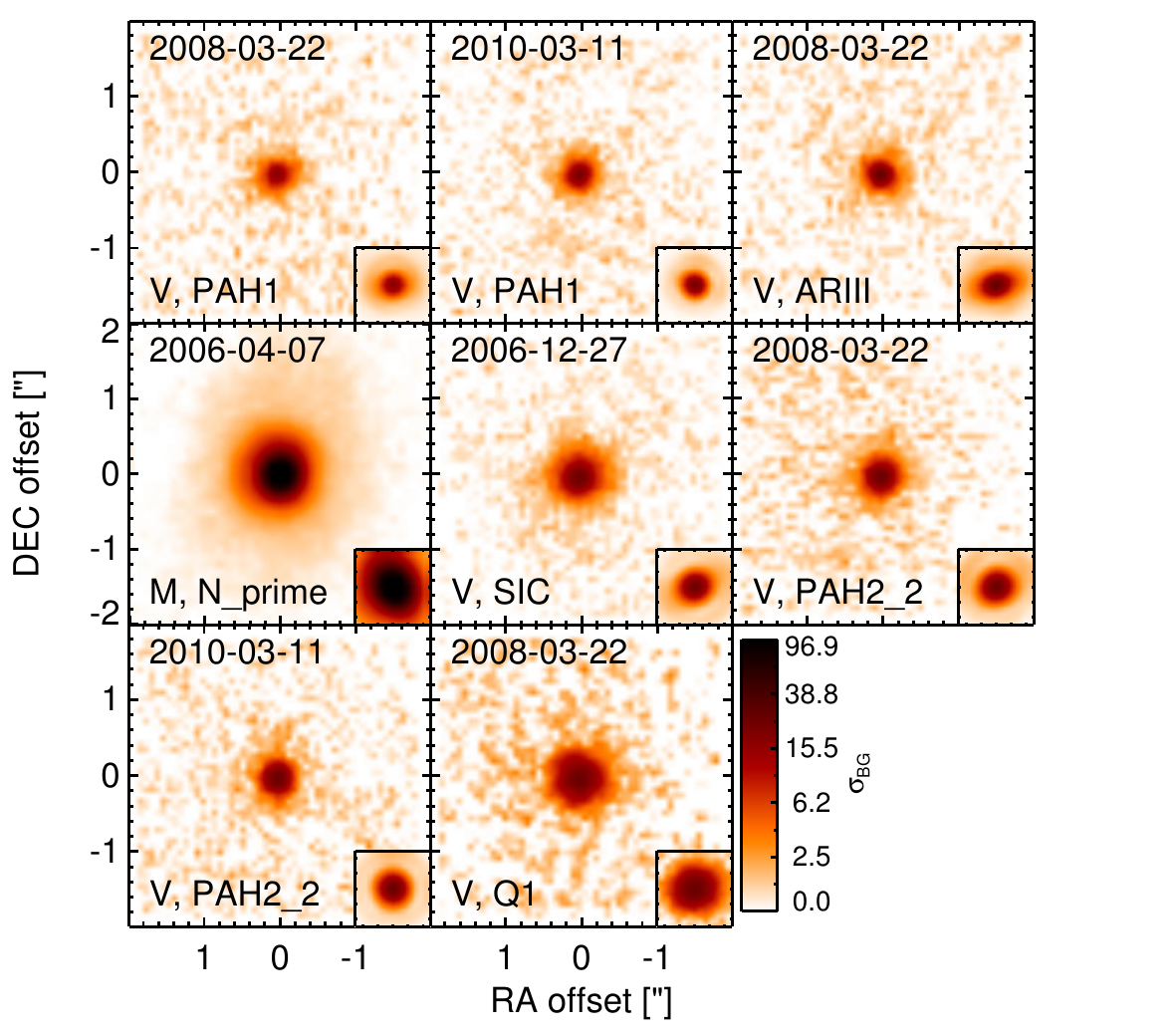}
    \caption{\label{fig:HARim_NGC3227}
             Subarcsecond-resolution MIR images of NGC\,3227 sorted by increasing filter wavelength. 
             Displayed are the inner $4\arcsec$ with North up and East to the left. 
             The colour scaling is logarithmic with white corresponding to median background and black to the $75\%$ of the highest intensity of all images in units of $\sigbg$.
             The inset image shows the central arcsecond of the PSF from the calibrator star, scaled to match the science target.
             The labels in the bottom left state instrument and filter names (C: COMICS, M: Michelle, T: T-ReCS, V: VISIR).
           }
\end{figure}
\begin{figure}
   \centering
   \includegraphics[angle=0,width=8.50cm]{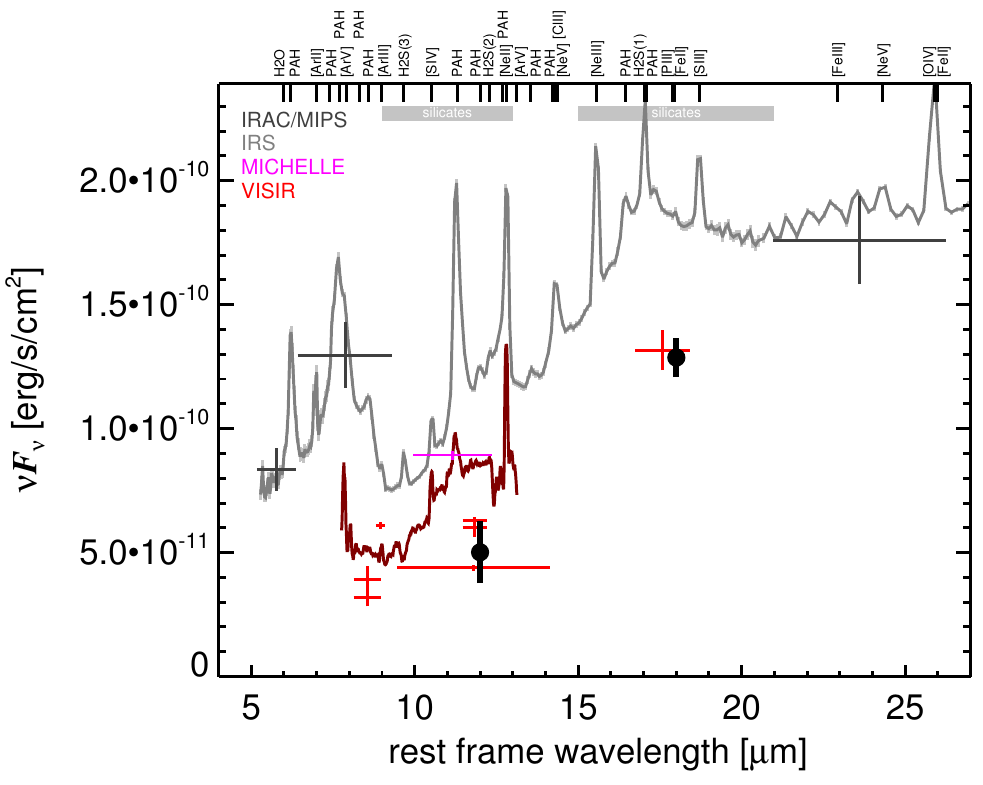}
   \caption{\label{fig:MISED_NGC3227}
      MIR SED of NGC\,3227. The description  of the symbols (if present) is the following.
      Grey crosses and  solid lines mark the \spitzer/IRAC, MIPS and IRS data. 
      The colour coding of the other symbols is: 
      green for COMICS, magenta for Michelle, blue for T-ReCS and red for VISIR data.
      Darker-coloured solid lines mark spectra of the corresponding instrument.
      The black filled circles mark the nuclear 12 and $18\,\mu$m  continuum emission estimate from the data.
      The ticks on the top axis mark positions of common MIR emission lines, while the light grey horizontal bars mark wavelength ranges affected by the silicate 10 and 18$\mu$m features.}
\end{figure}
\clearpage

\twocolumn[\begin{@twocolumnfalse}  
\subsection{NGC\,3281}\label{app:NGC3281}
NGC\,3281 is an highly-inclined spiral galaxy at a redshift of $z=$ 0.0107 ($D\sim52.8\,$Mpc) with a Sy\,2 nucleus \citep{veron-cetty_catalogue_2010} that belongs to the nine-month BAT AGN sample.
It features an extended cone-like NLR with a total length of $\sim6\arcsec\sim1.5\,$kpc along the north-south direction (PA$\sim20\degree$; \citealt{storchi-bergmann_ionization_1992,schmitt_hubble_2003}), while the nucleus is unresolved at radio wavelengths (e.g., \citealt{ulvestad_radio_1989,schmitt_jet_2001}). 
After the discovery of its MIR brightness by \iras, NGC\,3281 was imaged in $N$-band by \cite{krabbe_n-band_2001} and \cite{raban_core_2008} who detected a compact MIR nucleus.
A compact nucleus does also dominate the \spitzer/IRAC and MIPS images, while faint host emission is only visible in the IRAC $8.0\,\mu$m image.
The nucleus is partially saturated in the PBCD IRAC images and thus not analysed.
The IRS LR staring-mode spectrum is dominated by a deep silicate 10 and $18\,\mu$m absorption without significant PAH emission and a shallow red spectral slope in $\nu F_\nu$-space (see also \citealt{shi_9.7_2006,pereira-santaella_mid-infrared_2010,weaver_mid-infrared_2010}).
The nuclear region of NGC\,3281 was observed with T-ReCS in the broad N and Qa filters in 2004 \citep{ramos_almeida_infrared_2009}, and in the Si2 filter in 2008 (unpublished, to our knowledge).
In addition, we imaged it with VISIR in five $N$-band filters in 2008, two of which were analysed already in \cite{gandhi_resolving_2009}.
In all images, a marginally resolved MIR nucleus was detected (FWHM $\sim15\%$ larger than standard star; PA$\sim170\degree$).
The unresolved nuclear fluxes are on average $\sim 20\%$ lower than the values published in \cite{ramos_almeida_infrared_2009} and \cite{gandhi_resolving_2009}, $\sim 18\%$ lower than the \spitzerr spectrophotometry, and also systematically lower then the T-ReCS LR $N$-band spectrum from \cite{gonzalez-martin_dust_2013}.
The latter was first published by \cite{sales_compton-thick_2011} and is consistent with the IRS spectrum apart from increasingly lower fluxes towards short wavelengths. 
These results confirm that the silicate absorption originates in the projected central $<100$\,pc region, or somewhere along our line of side towards it.
Note that the nuclear MIR emission was further resolved with MIDI interferometric observations and was modelled as two components, one extending >5\,pc and one unresolved dominating the emission \citep{burtscher_diversity_2013}.
\newline\end{@twocolumnfalse}]

\begin{figure}
   \centering
   \includegraphics[angle=0,width=8.500cm]{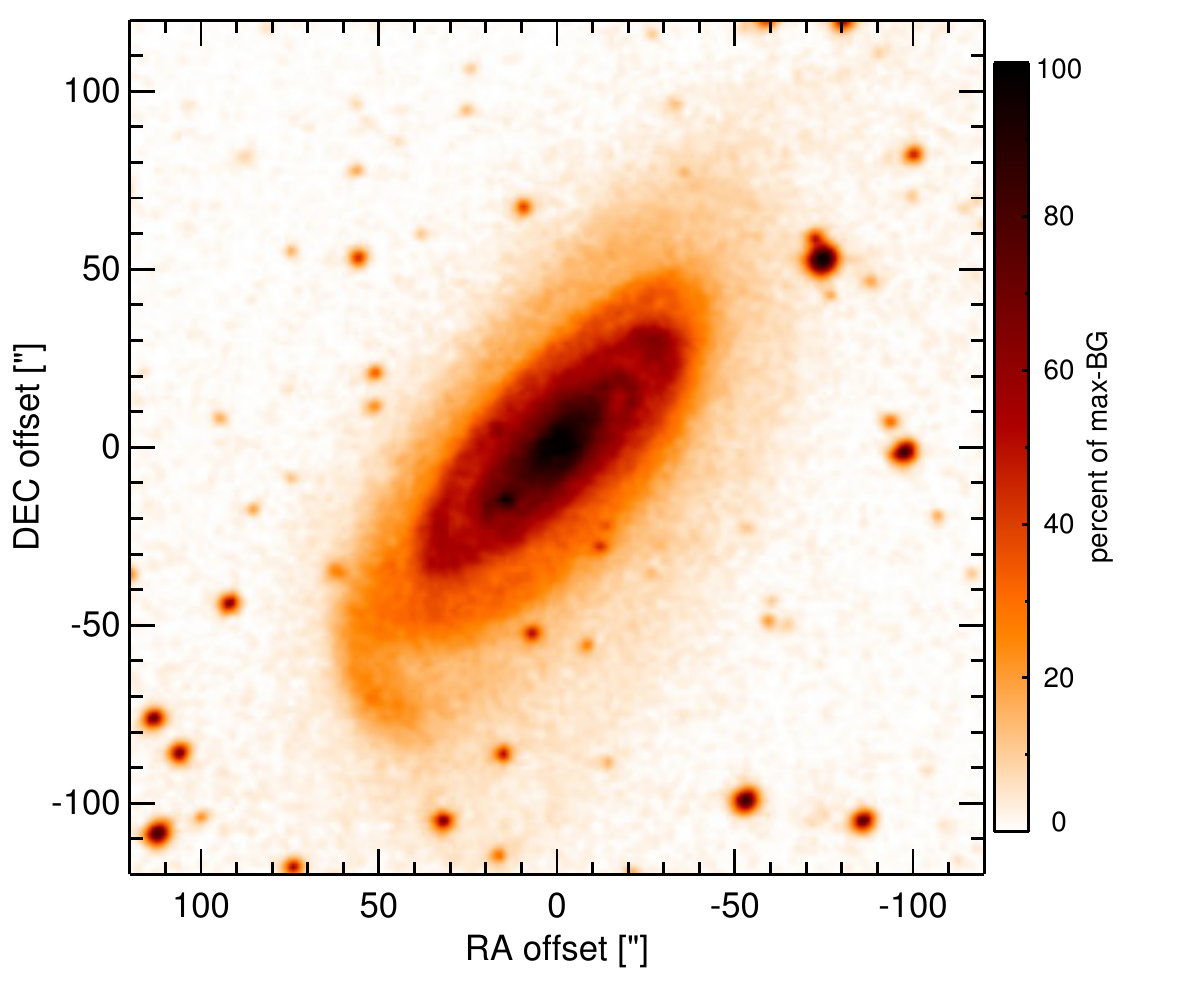}
    \caption{\label{fig:OPTim_NGC3281}
             Optical image (DSS, red filter) of NGC\,3281. Displayed are the central $4\arcmin$ with North up and East to the left. 
              The colour scaling is linear with white corresponding to the median background and black to the $0.01\%$ pixels with the highest intensity.  
           }
\end{figure}
\begin{figure}
   \centering
   \includegraphics[angle=0,height=3.11cm]{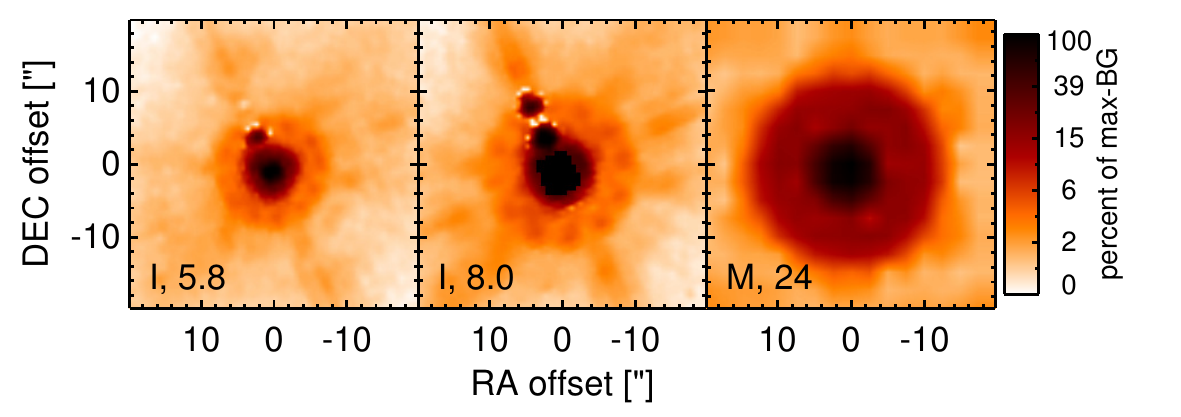}
    \caption{\label{fig:INTim_NGC3281}
             \spitzerr MIR images of NGC\,3281. Displayed are the inner $40\arcsec$ with North up and East to the left. The colour scaling is logarithmic with white corresponding to median background and black to the $0.1\%$ pixels with the highest intensity.
             The label in the bottom left states instrument and central wavelength of the filter in $\mu$m (I: IRAC, M: MIPS).
             Note that the apparent off-nuclear compact sources in the IRAC 5.8 and $8.0\,\mu$m images are instrumental artefacts.
           }
\end{figure}
\begin{figure}
   \centering
   \includegraphics[angle=0,width=8.500cm]{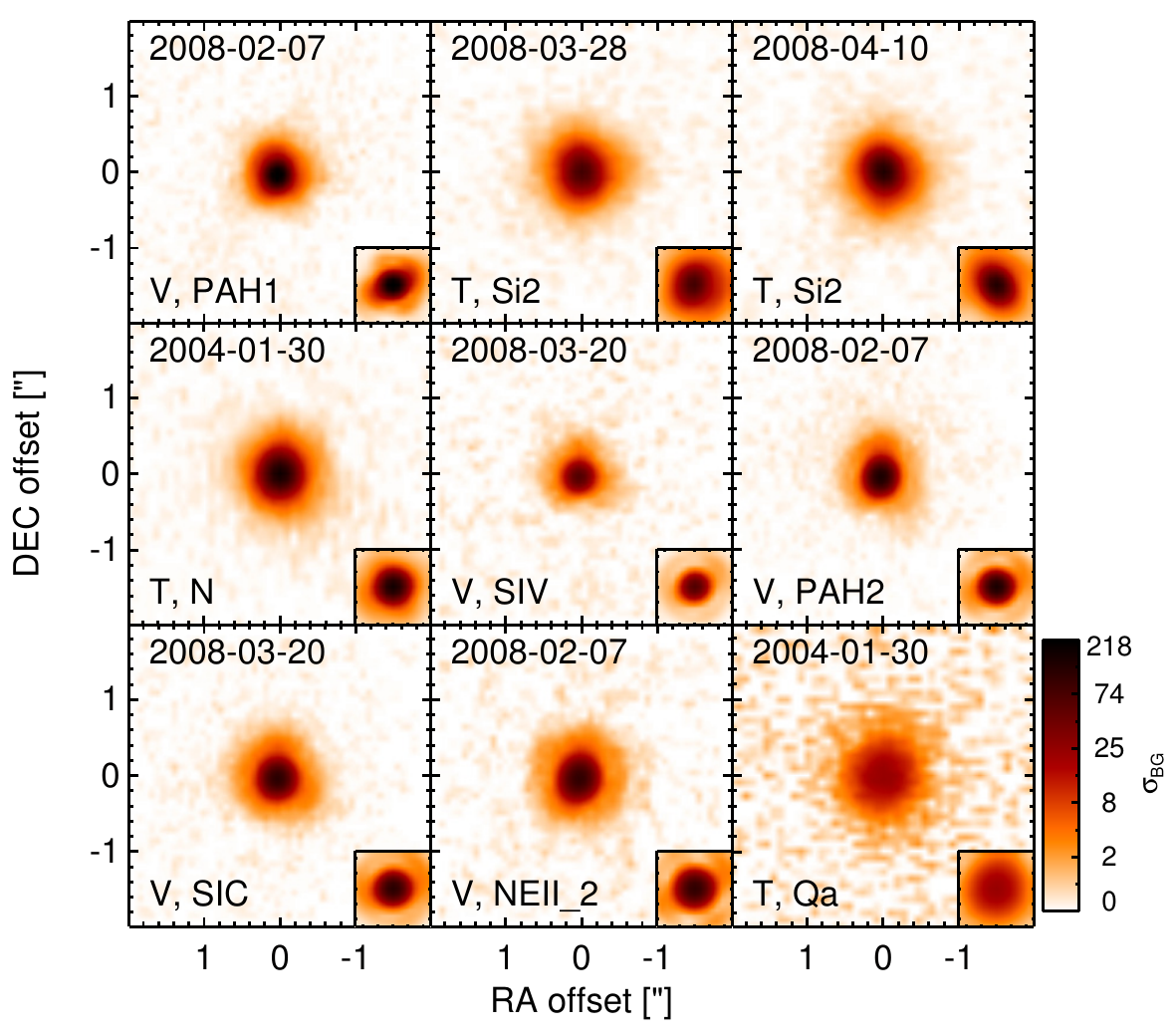}
    \caption{\label{fig:HARim_NGC3281}
             Subarcsecond-resolution MIR images of NGC\,3281 sorted by increasing filter wavelength. 
             Displayed are the inner $4\arcsec$ with North up and East to the left. 
             The colour scaling is logarithmic with white corresponding to median background and black to the $75\%$ of the highest intensity of all images in units of $\sigbg$.
             The inset image shows the central arcsecond of the PSF from the calibrator star, scaled to match the science target.
             The labels in the bottom left state instrument and filter names (C: COMICS, M: Michelle, T: T-ReCS, V: VISIR).
           }
\end{figure}
\begin{figure}
   \centering
   \includegraphics[angle=0,width=8.50cm]{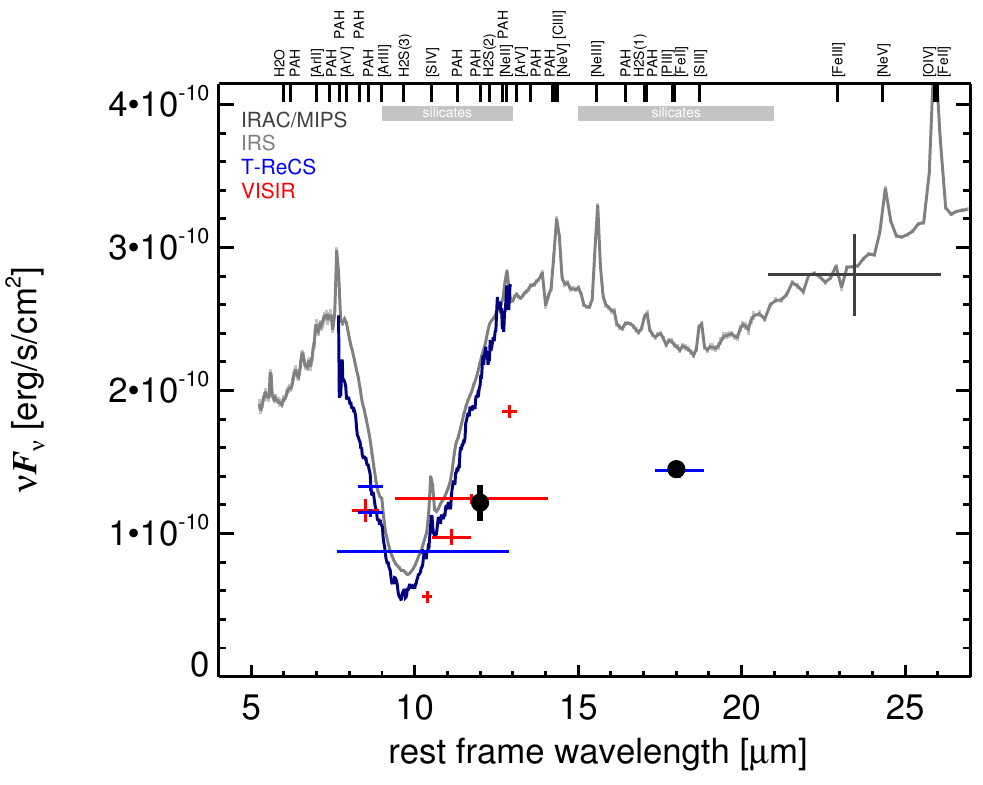}
   \caption{\label{fig:MISED_NGC3281}
      MIR SED of NGC\,3281. The description  of the symbols (if present) is the following.
      Grey crosses and  solid lines mark the \spitzer/IRAC, MIPS and IRS data. 
      The colour coding of the other symbols is: 
      green for COMICS, magenta for Michelle, blue for T-ReCS and red for VISIR data.
      Darker-coloured solid lines mark spectra of the corresponding instrument.
      The black filled circles mark the nuclear 12 and $18\,\mu$m  continuum emission estimate from the data.
      The ticks on the top axis mark positions of common MIR emission lines, while the light grey horizontal bars mark wavelength ranges affected by the silicate 10 and 18$\mu$m features.}
\end{figure}
\clearpage

\twocolumn[\begin{@twocolumnfalse}  
\subsection{NGC\,3312}\label{app:NGC3312}
NGC\,3312 is a highly-inclined disturbed spiral galaxy at a redshift of $z=$ 0.0096 ($D\sim48.4\,$Mpc) with a little-studied LINER nucleus \citep{veron-cetty_catalogue_2010}.
No \spitzerr observations are available. 
In the WISE images, a compact MIR nucleus embedded within spiral-like host emission is visible.
We observed the nuclear region of NGC\,3312 with VISIR in the PAH2 filter in 2009 but failed to clearly detect the nucleus \citep{asmus_mid-infrared_2011}. 
In 2010, we followed up with PAH2\_2 imaging and detected an unresolved nucleus without any further host emission. 
Owing to the lack of information, no conclusion about the nature of this object can be drawn.
\newline\end{@twocolumnfalse}]

\begin{figure}
   \centering
   \includegraphics[angle=0,width=8.500cm]{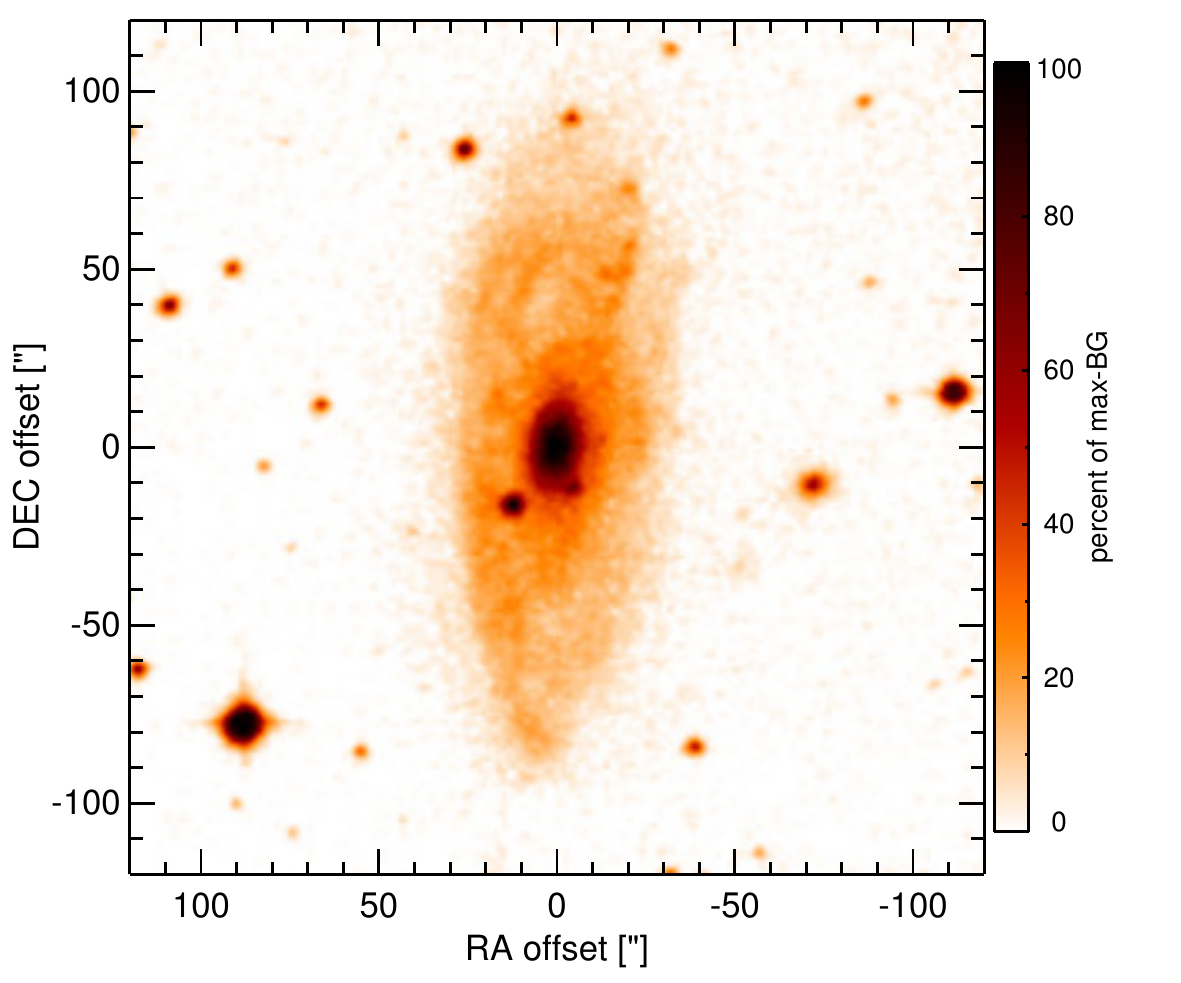}
    \caption{\label{fig:OPTim_NGC3312}
             Optical image (DSS, red filter) of NGC\,3312. Displayed are the central $4\arcmin$ with North up and East to the left. 
              The colour scaling is linear with white corresponding to the median background and black to the $0.01\%$ pixels with the highest intensity.  
           }
\end{figure}
\begin{figure}
   \centering
   \includegraphics[angle=0,height=3.11cm]{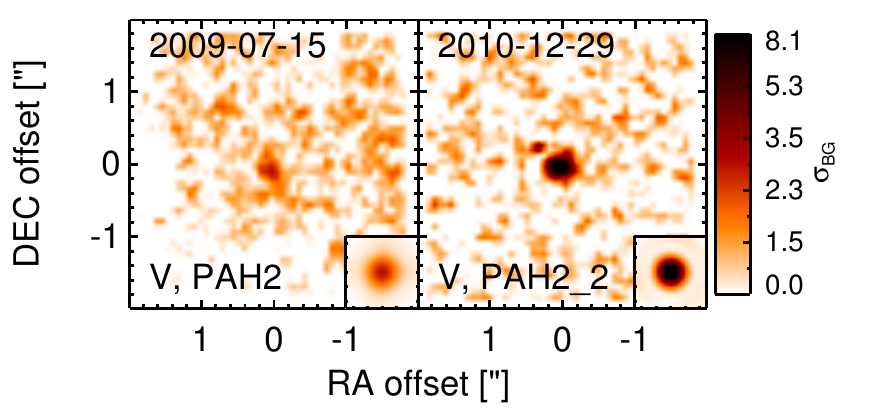}
    \caption{\label{fig:HARim_NGC3312}
             Subarcsecond-resolution MIR images of NGC\,3312 sorted by increasing filter wavelength. 
             Displayed are the inner $4\arcsec$ with North up and East to the left. 
             The colour scaling is logarithmic with white corresponding to median background and black to the $75\%$ of the highest intensity of all images in units of $\sigbg$.
             The inset image shows the central arcsecond of the PSF from the calibrator star, scaled to match the science target.
             The labels in the bottom left state instrument and filter names (C: COMICS, M: Michelle, T: T-ReCS, V: VISIR).
           }
\end{figure}
\begin{figure}
   \centering
   \includegraphics[angle=0,width=8.50cm]{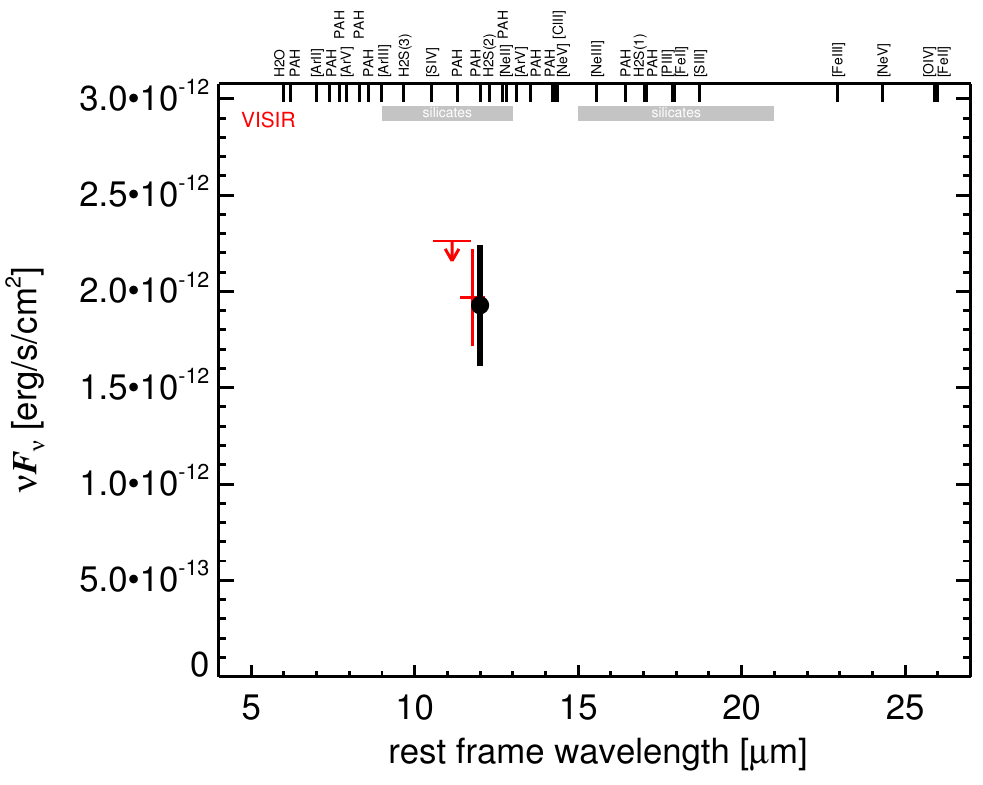}
   \caption{\label{fig:MISED_NGC3312}
      MIR SED of NGC\,3312. The description  of the symbols (if present) is the following.
      Grey crosses and  solid lines mark the \spitzer/IRAC, MIPS and IRS data. 
      The colour coding of the other symbols is: 
      green for COMICS, magenta for Michelle, blue for T-ReCS and red for VISIR data.
      Darker-coloured solid lines mark spectra of the corresponding instrument.
      The black filled circles mark the nuclear 12 and $18\,\mu$m  continuum emission estimate from the data.
      The ticks on the top axis mark positions of common MIR emission lines, while the light grey horizontal bars mark wavelength ranges affected by the silicate 10 and 18$\mu$m features.}
\end{figure}
\clearpage

\twocolumn[\begin{@twocolumnfalse}  
\subsection{NGC\,3368 -- M96}\label{app:NGC3368}

NGC\,3368 is a low-inclination spiral galaxy at a distance of $D=$ $10.6 \pm 1.6$\,Mpc (NED redshift-independent median) with an active nucleus optically classified as a LINER \citep{ho_search_1997-1}.
No compact radio core could be detect in the centre of this galaxy \citep{nagar_radio_2002}, although a compact UV and X-ray source was detected \citep{satyapal_joint_2004, chiaberge_hubble_2005,maoz_murmur_2005}.
The first $N$-band photometry of NGC\,3368 was obtained with IRTF \citep{cizdziel_multiaperture_1985}.
The \spitzer/IRAC and MIPS images show an extended nucleus embedded within the spiral-like host emission.
We measure the nuclear component in the  IRAC $5.8$ and $8.0\,\mu$m and MIPS $24\,\mu$m images, which  provides significantly lower values than the ones published in \cite{dudik_spitzer_2009}.
The IRS HR staring-mode PBCD spectrum is not very reliable owing to the complex MIR morphology of NGC\,3368 but matches the nuclear IRAC photometry.
Note also that no background subtraction was performed for this spectrum.
It exhibits strong PAH emission and weak silicate absorption typical for star formation (see also \citealt{goulding_towards_2009}). 
The nuclear region of NGC\,3368 was imaged with VISIR in the PAH2\_2 filter in 2010 (unpublished, to our knowledge).
The nucleus has not been detected in the image and our derived upper limit is $90\%$ lower than the \spitzerr spectrophotometry.
This indicates that star formation is dominating the nuclear MIR emission at $\sim200\,$pc scales.
\newline\end{@twocolumnfalse}]

\begin{figure}
   \centering
   \includegraphics[angle=0,width=8.500cm]{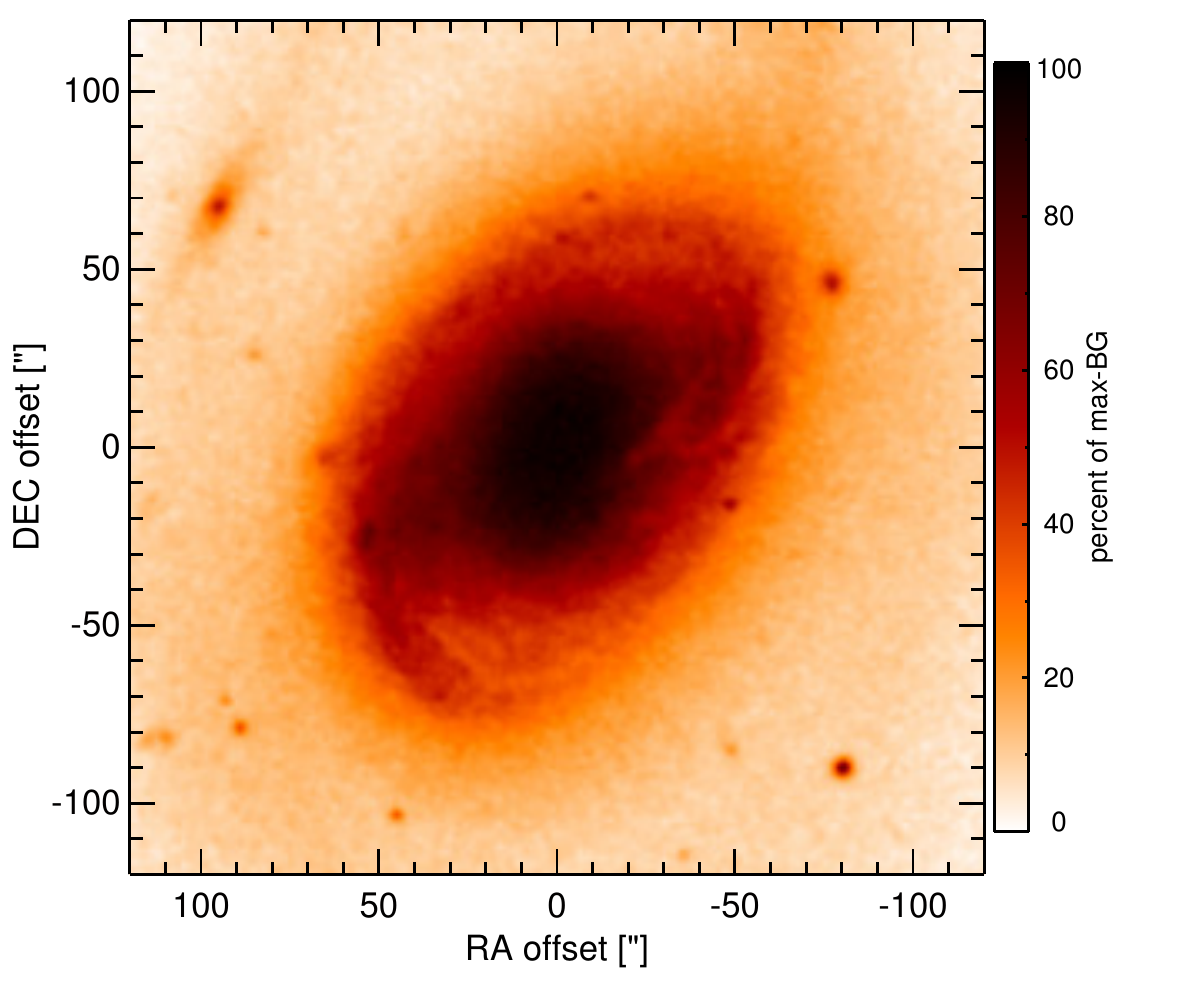}
    \caption{\label{fig:OPTim_NGC3368}
             Optical image (DSS, red filter) of NGC\,3368. Displayed are the central $4\arcmin$ with North up and East to the left. 
              The colour scaling is linear with white corresponding to the median background and black to the $0.01\%$ pixels with the highest intensity.  
           }
\end{figure}
\begin{figure}
   \centering
   \includegraphics[angle=0,height=3.11cm]{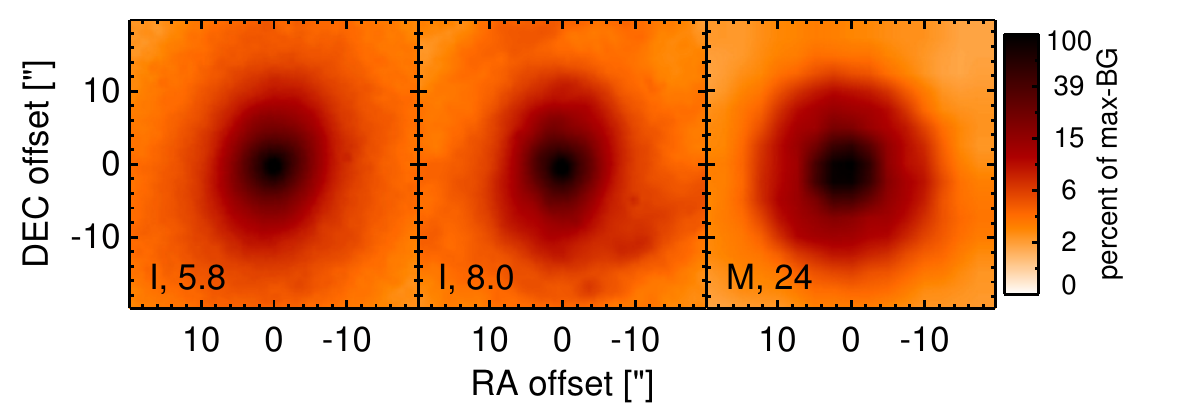}
    \caption{\label{fig:INTim_NGC3368}
             \spitzerr MIR images of NGC\,3368. Displayed are the inner $40\arcsec$ with North up and East to the left. The colour scaling is logarithmic with white corresponding to median background and black to the $0.1\%$ pixels with the highest intensity.
             The label in the bottom left states instrument and central wavelength of the filter in $\mu$m (I: IRAC, M: MIPS). 
           }
\end{figure}
\begin{figure}
   \centering
   \includegraphics[angle=0,width=8.50cm]{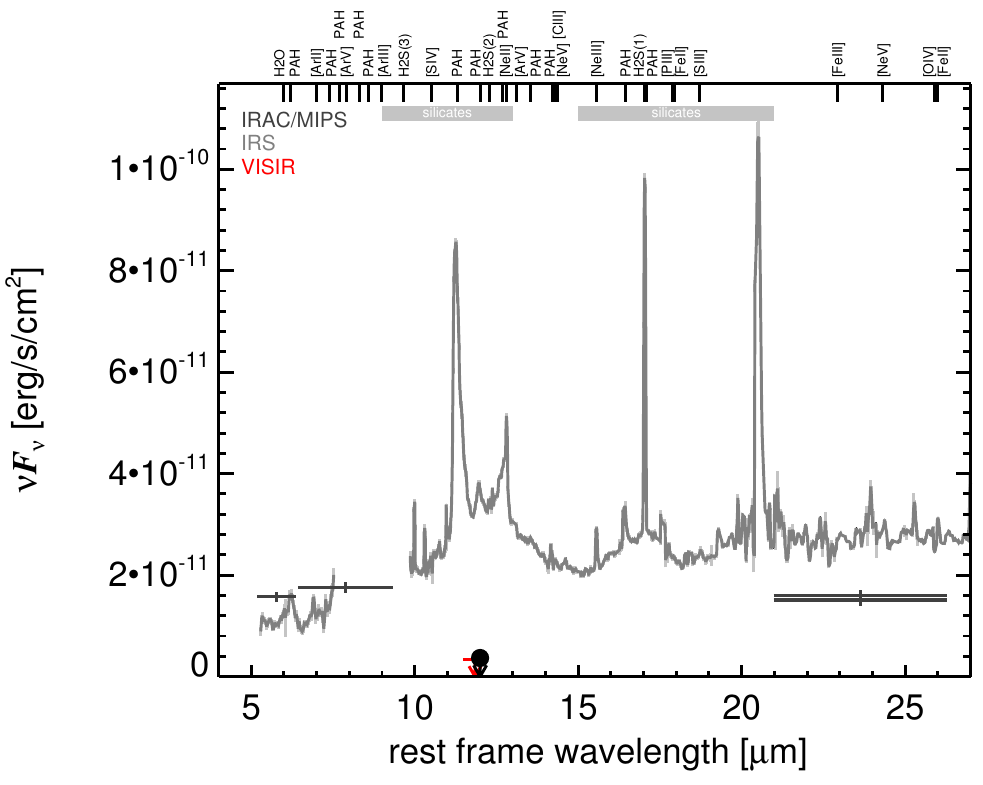}
   \caption{\label{fig:MISED_NGC3368}
      MIR SED of NGC\,3368. The description  of the symbols (if present) is the following.
      Grey crosses and  solid lines mark the \spitzer/IRAC, MIPS and IRS data. 
      The colour coding of the other symbols is: 
      green for COMICS, magenta for Michelle, blue for T-ReCS and red for VISIR data.
      Darker-coloured solid lines mark spectra of the corresponding instrument.
      The black filled circles mark the nuclear 12 and $18\,\mu$m  continuum emission estimate from the data.
      The ticks on the top axis mark positions of common MIR emission lines, while the light grey horizontal bars mark wavelength ranges affected by the silicate 10 and 18$\mu$m features.}
\end{figure}
\clearpage

\twocolumn[\begin{@twocolumnfalse}  
\subsection{NGC\,3379 -- M105}\label{app:NGC3379}
NGC\,3379 is an elliptical galaxy at a distance of $D=$ $10.6\pm1.3$\,Mpc (NED redshift-independent median) with an active nucleus classified as  a LINER/transition object \citep{ho_search_1997-1}.
An unresolved radio core was detected in the nucleus \citep{wrobel_radio-continuum_1991} along with several X-ray point sources of comparable brightness \citep{flohic_central_2006}.
The first MIR observations of NGC\,3379 were performed with IRTF in 1983 \citep{impey_infrared_1986}.
The IRAC and MIPS images show extended elliptical emission without a clearly separable nuclear component. 
However, we still measure the nuclear fluxes in the IRAC $5.8$ and $8.0\,\mu$m and MIPS $24\,\mu$m images.
Thus, our values are significantly lower than in the literature (e.g., \citealt{temi_spitzer_2009}).
The IRS LR spectrum possesses only a low S/N but indicates silicate emission and a blue spectral slope in $\nu F_\nu$-space but no PAH features, indicating the predominance of old stellar population emission (see also \citealt{bregman_ages_2006}).
The nuclear region of NGC\,3379 was observed with VISIR in the PAH1 filter in 2010 (unpublished, to our knowledge) but no compact nucleus could be detected.
The derived upper limit is $87\%$ lower than the \spitzerr spectrophotometry. 
Therefore, host emission completely dominates the MIR even at the central $\sim 200\,$pc of NGC\,3379 and it remains unclear, whether an AGN is present at all.
\newline\end{@twocolumnfalse}]

\begin{figure}
   \centering
   \includegraphics[angle=0,width=8.500cm]{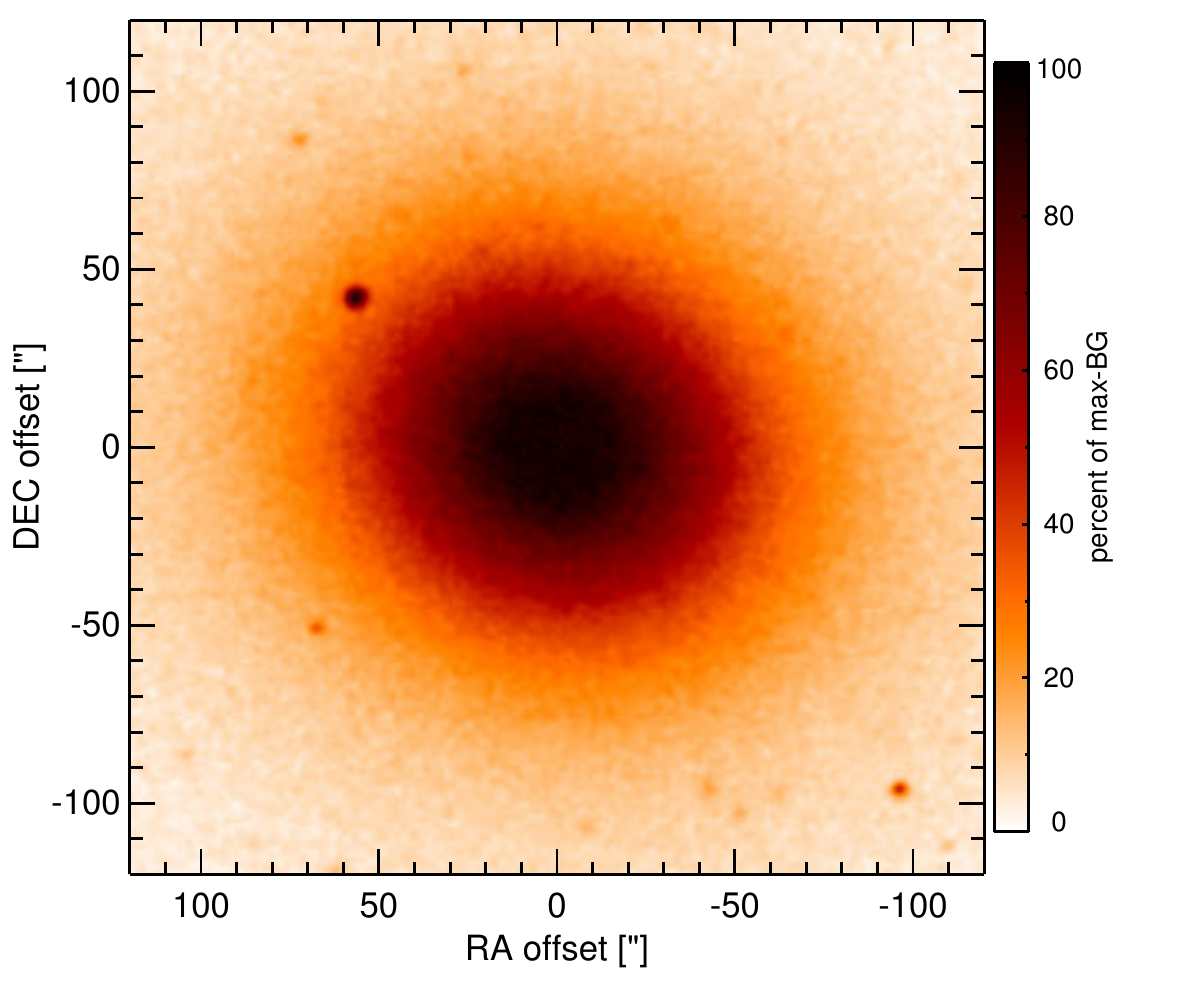}
    \caption{\label{fig:OPTim_NGC3379}
             Optical image (DSS, red filter) of NGC\,3379. Displayed are the central $4\arcmin$ with North up and East to the left. 
              The colour scaling is linear with white corresponding to the median background and black to the $0.01\%$ pixels with the highest intensity.  
           }
\end{figure}
\begin{figure}
   \centering
   \includegraphics[angle=0,height=3.11cm]{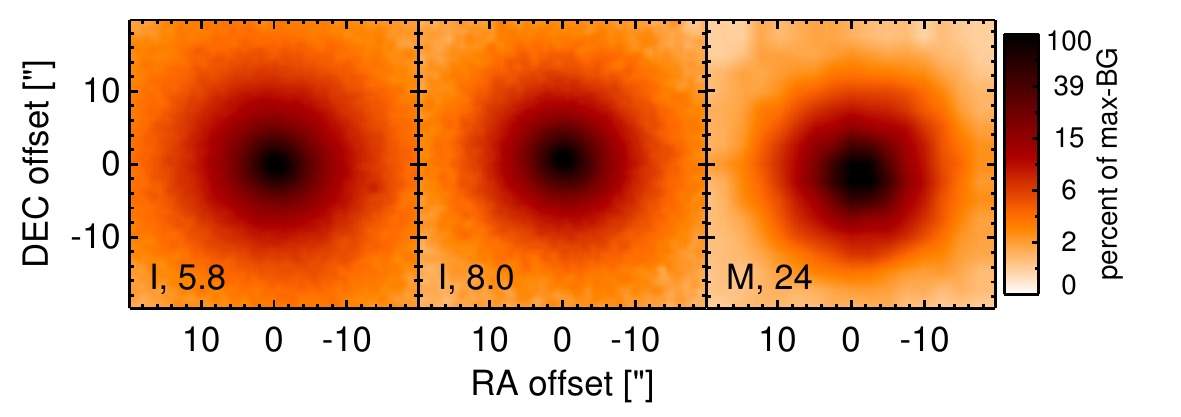}
    \caption{\label{fig:INTim_NGC3379}
             \spitzerr MIR images of NGC\,3379. Displayed are the inner $40\arcsec$ with North up and East to the left. The colour scaling is logarithmic with white corresponding to median background and black to the $0.1\%$ pixels with the highest intensity.
             The label in the bottom left states instrument and central wavelength of the filter in $\mu$m (I: IRAC, M: MIPS). 
           }
\end{figure}
\begin{figure}
   \centering
   \includegraphics[angle=0,width=8.50cm]{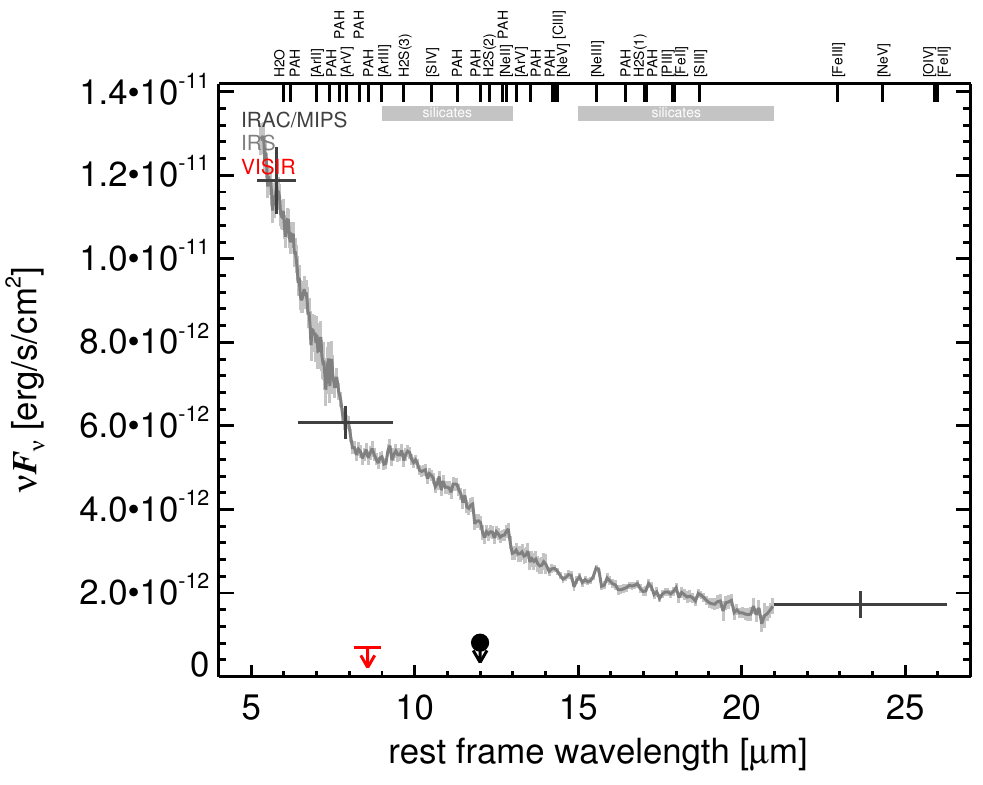}
   \caption{\label{fig:MISED_NGC3379}
      MIR SED of NGC\,3379. The description  of the symbols (if present) is the following.
      Grey crosses and  solid lines mark the \spitzer/IRAC, MIPS and IRS data. 
      The colour coding of the other symbols is: 
      green for COMICS, magenta for Michelle, blue for T-ReCS and red for VISIR data.
      Darker-coloured solid lines mark spectra of the corresponding instrument.
      The black filled circles mark the nuclear 12 and $18\,\mu$m  continuum emission estimate from the data.
      The ticks on the top axis mark positions of common MIR emission lines, while the light grey horizontal bars mark wavelength ranges affected by the silicate 10 and 18$\mu$m features.}
\end{figure}
\clearpage

\twocolumn[\begin{@twocolumnfalse}  
\subsection{NGC\,3393}\label{app:NGC3393}
NGC\,3393 is a face-on barred spiral galaxy at a redshift of $z=$ 0.0125 ($D\sim61.6\,$Mpc) with a close binary AGN (separation $\sim0.5\arcsec\sim150$\,pc; PA$\sim40\degree$; \citealt{fabbiano_close_2011}), which optically appears as a Sy\,2 \citep{veron-cetty_catalogue_2010}.
The binary AGN is presumably the result of a minor merger, which did not significantly affect the host morphology (see also \citealt{contini_evidence_2012}).
Both are surrounded by an S-shaped NLR extending $1.8\arcsec \sim 0.4\,$kpc on either side along a PA$\sim55\degree$ \citep{schmitt_comparison_1996,cooke_narrow-line_2000} coinciding with a double-sided jet along a PA$\sim50\degree$ \citep{morganti_radio_1999,schmitt_jet_2001}.
Water maser emission tracing the accretion disc  was detected oriented perpendicular to the jet and NLR (PA$\sim-34\degree$; \citealt{kondratko_discovery_2006,kondratko_parsec-scale_2008}).
Apart from \irass, NGC\,3393 was also observed with \spitzer/IRS and MIPS in the MIR.
A compact dominating nucleus embedded within the spiral-like host emission was detected in the MIPS $24\,\mu$m image.
The IRS LR staring-mode spectrum exhibits bright forbidden emission lines, possibly weak silicate $10\,\mu$m absorption, almost no PAH emission and a red spectral slope in $\nu F_\nu$-space (see also \citealt{shi_9.7_2006,sargsyan_infrared_2011}).
Therefore, star formation seems to be weak at arcsecond scales.
We observed the nuclear region of NGC\,3393 with VISIR in four different $N$-band filters in 2008 (two published in \citealt{gandhi_resolving_2009}), and weakly detected one compact source but no second. 
The low S/N of all observations prohibits any constraining results about the nuclear extension, which is however consistent with being unresolved.
We co-added all obtained 8 images centred on the detected nucleus, which results in an elongated source, which might comprise both nuclei (FWHM(major axis)$\sim0.4\arcsec$;PA$\sim53\degree$).
In this case, the emission peak would correspond to the south-western nucleus, which would also dominate the re-measured nuclear VISIR fluxes of the individual filters.
These are  consistent with the values in \cite{gandhi_resolving_2009} and on average $\sim39\%$ lower than the IRS spectrum. 
\newline\end{@twocolumnfalse}]

\begin{figure}
   \centering
   \includegraphics[angle=0,width=8.500cm]{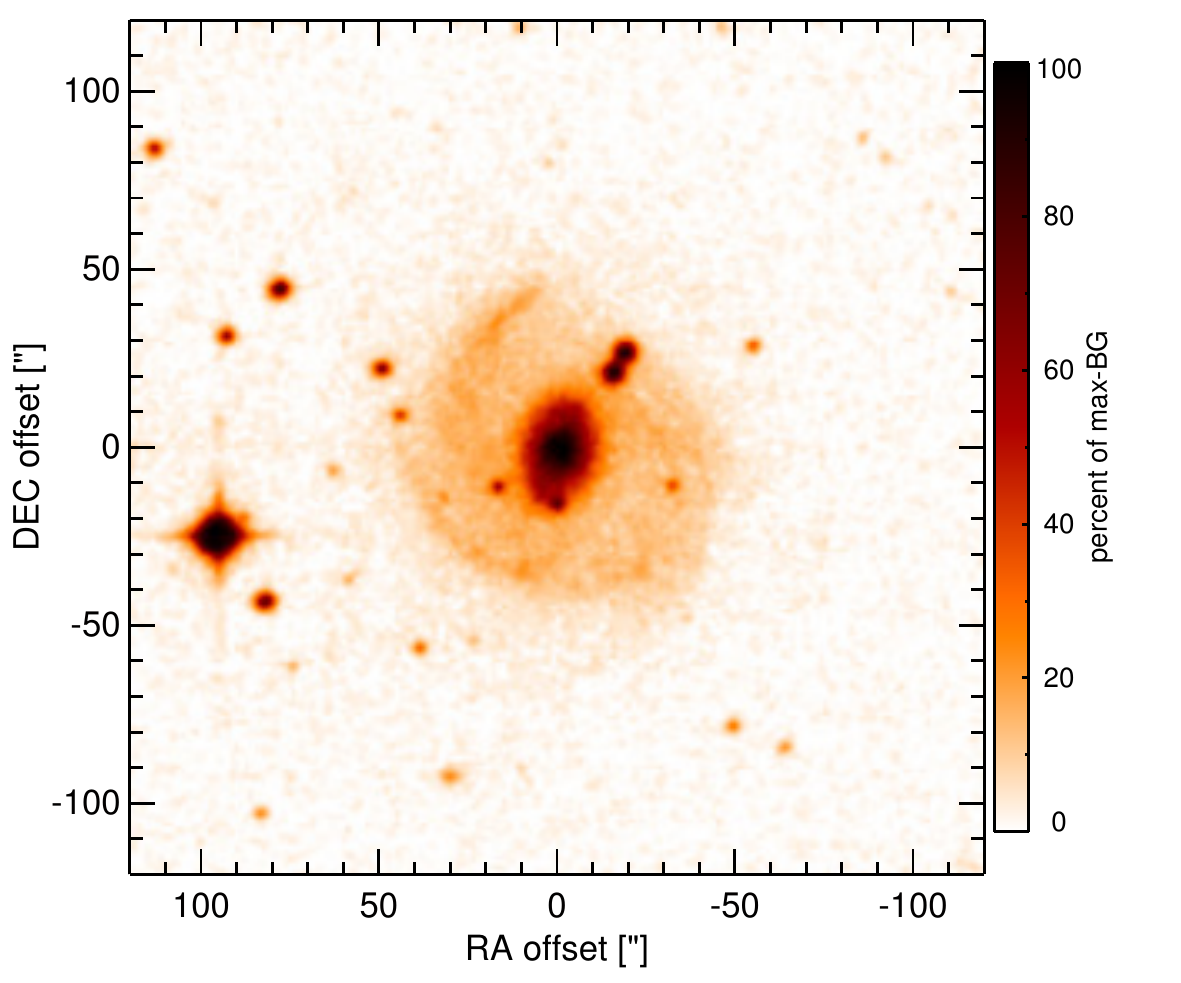}
    \caption{\label{fig:OPTim_NGC3393}
             Optical image (DSS, red filter) of NGC\,3393. Displayed are the central $4\arcmin$ with North up and East to the left. 
              The colour scaling is linear with white corresponding to the median background and black to the $0.01\%$ pixels with the highest intensity.  
           }
\end{figure}
\begin{figure}
   \centering
   \includegraphics[angle=0,height=3.11cm]{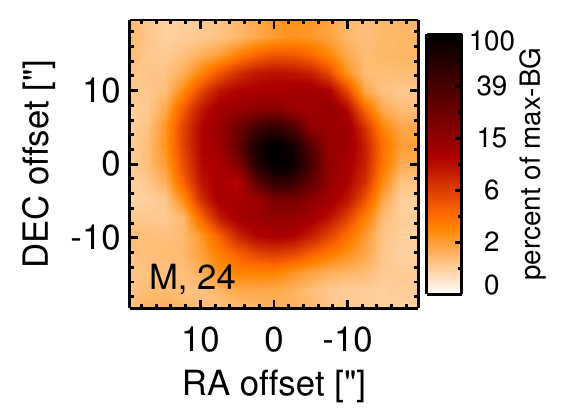}
    \caption{\label{fig:INTim_NGC3393}
             \spitzerr MIR images of NGC\,3393. Displayed are the inner $40\arcsec$ with North up and East to the left. The colour scaling is logarithmic with white corresponding to median background and black to the $0.1\%$ pixels with the highest intensity.
             The label in the bottom left states instrument and central wavelength of the filter in $\mu$m (I: IRAC, M: MIPS). 
           }
\end{figure}
\begin{figure}
   \centering
   \includegraphics[angle=0,width=8.500cm]{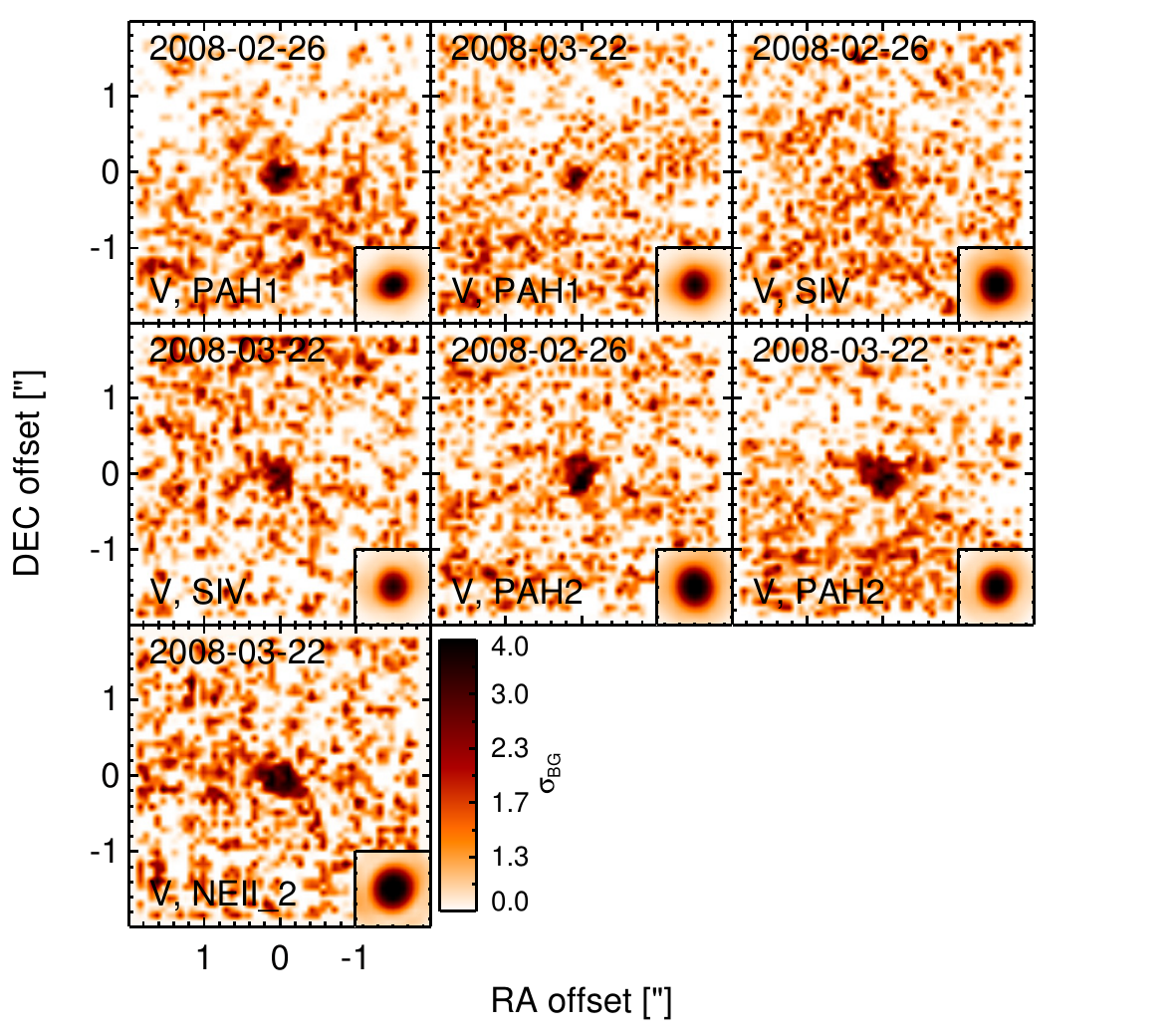}
    \caption{\label{fig:HARim_NGC3393}
             Subarcsecond-resolution MIR images of NGC\,3393 sorted by increasing filter wavelength. 
             Displayed are the inner $4\arcsec$ with North up and East to the left. 
             The colour scaling is logarithmic with white corresponding to median background and black to the $75\%$ of the highest intensity of all images in units of $\sigbg$.
             The inset image shows the central arcsecond of the PSF from the calibrator star, scaled to match the science target.
             The labels in the bottom left state instrument and filter names (C: COMICS, M: Michelle, T: T-ReCS, V: VISIR).
           }
\end{figure}
\begin{figure}
   \centering
   \includegraphics[angle=0,width=8.50cm]{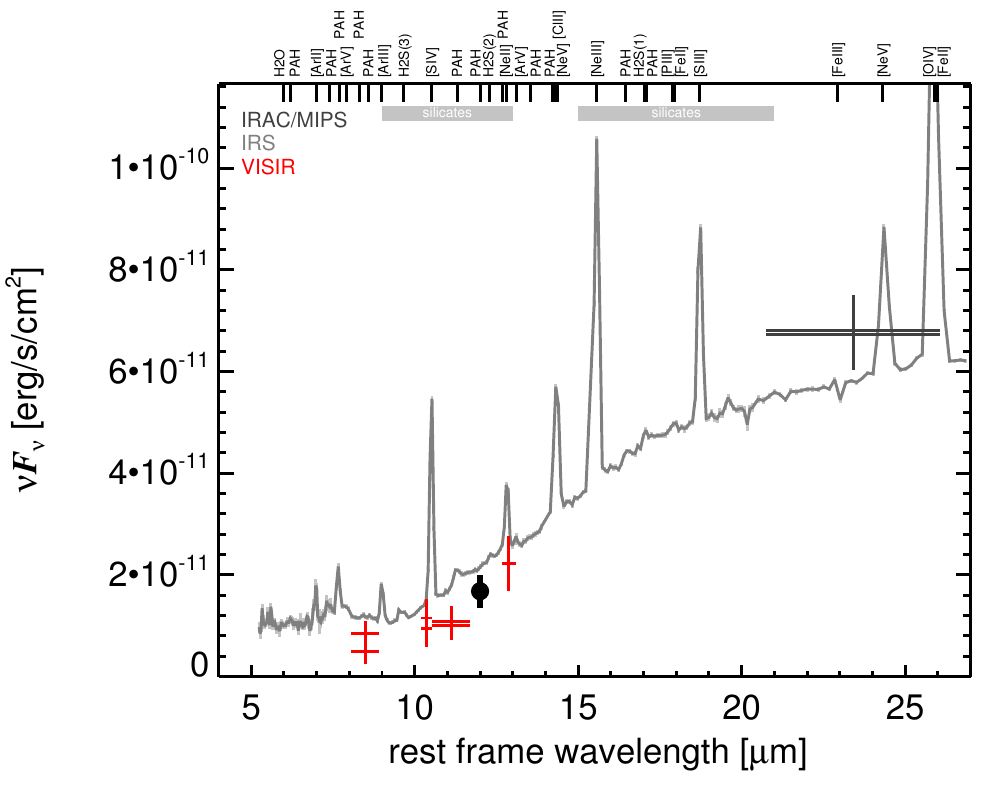}
   \caption{\label{fig:MISED_NGC3393}
      MIR SED of NGC\,3393. The description  of the symbols (if present) is the following.
      Grey crosses and  solid lines mark the \spitzer/IRAC, MIPS and IRS data. 
      The colour coding of the other symbols is: 
      green for COMICS, magenta for Michelle, blue for T-ReCS and red for VISIR data.
      Darker-coloured solid lines mark spectra of the corresponding instrument.
      The black filled circles mark the nuclear 12 and $18\,\mu$m  continuum emission estimate from the data.
      The ticks on the top axis mark positions of common MIR emission lines, while the light grey horizontal bars mark wavelength ranges affected by the silicate 10 and 18$\mu$m features.}
\end{figure}
\clearpage

\twocolumn[\begin{@twocolumnfalse}  
\subsection{NGC\,3486}\label{app:NGC3486}
NGC\,3486 is a face-on spiral galaxy at a distance of $D=$ $13.7 \pm 2.7\,$Mpc \citep{tully_extragalactic_2009} with a borderline Sy\,2.0/LINER nucleus, which might be powered by a compact central star cluster instead of an accreting SMBH \citep{maoz_low-luminosity_2007}.
It has not been detected at radio wavelengths \citep{ho_detection_2001} but in the X-rays, where it appears rather heavily absorbed \citep{brightman_nature_2008}.
The first attempt to detect NGC\,3486 in the MIR failed \citep{rieke_10_1978}.
After its detection in \iras, this object was also observed with \spitzer/IRAC, IRS and MIPS and appears as a compact nucleus embedded in bright ring-like host emission in the corresponding images.
This ring-like emission dominates in the MIPS $24\,\mu$m image.
The nucleus is extended in the IRAC images, and we measure only the central four-arcsecond region, which results in much lower fluxes than published for the total emission \citep{dale_spitzer_2009}.
The IRS LR staring-mode spectrum has a very low S/N and does not allow us to identify any spectral features. 
The general \spitzerr spectrophotometry indicates a blue SED dominated by old stellar emission.
We observed then nuclear region of NGC\,3486 with Michelle in two $N$-band filters in 2010 but nothing was detected in the images.
The derived flux upper limits are consistent with the \spitzerr spectrophotometry.
Thus, we can only conclude that the MIR emission of the central $\sim250\,$pc in NGC\,3486 is dominated by diffuse stellar emission.
\newline\end{@twocolumnfalse}]

\begin{figure}
   \centering
   \includegraphics[angle=0,width=8.500cm]{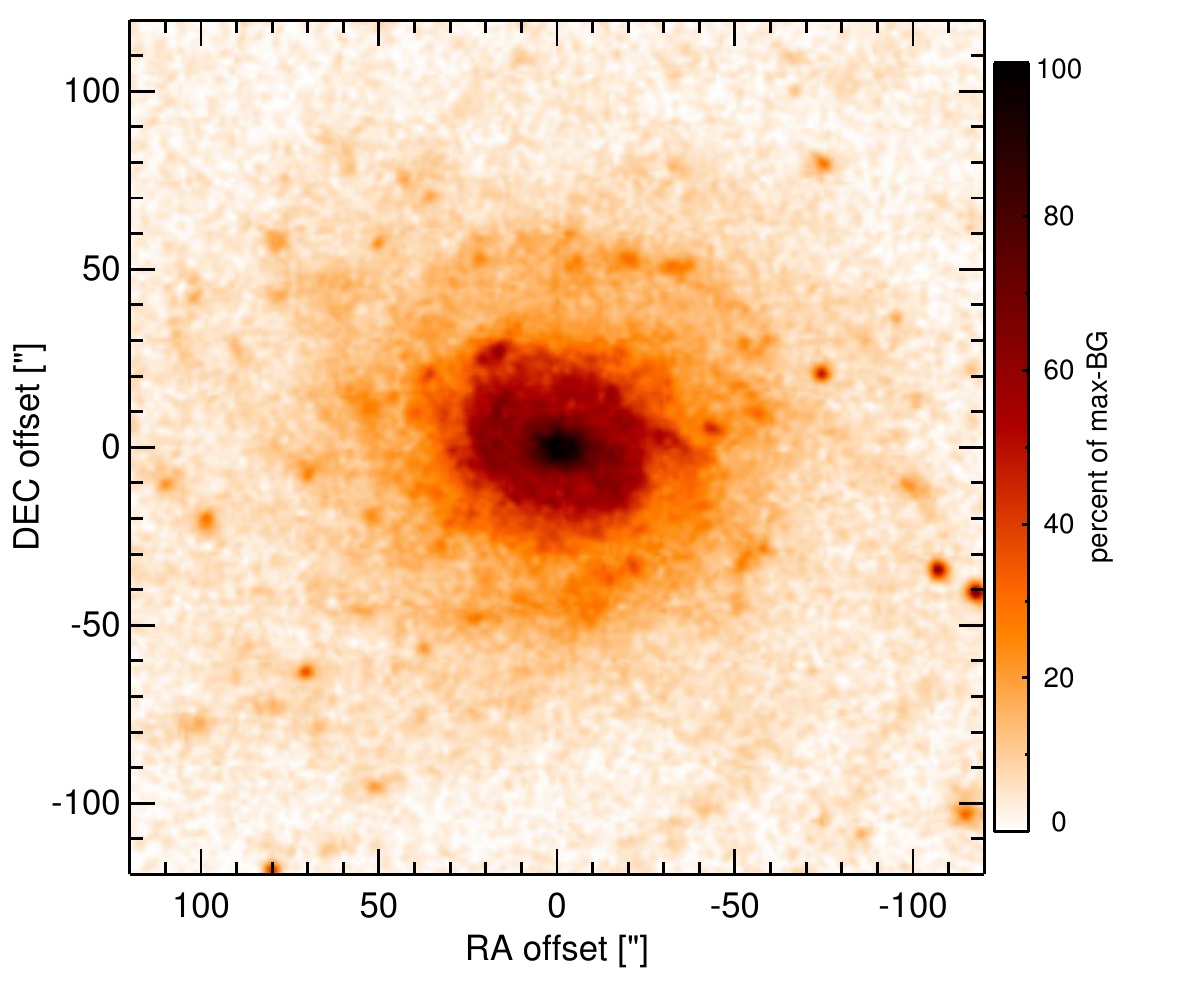}
    \caption{\label{fig:OPTim_NGC3486}
             Optical image (DSS, red filter) of NGC\,3486. Displayed are the central $4\arcmin$ with North up and East to the left. 
              The colour scaling is linear with white corresponding to the median background and black to the $0.01\%$ pixels with the highest intensity.  
           }
\end{figure}
\begin{figure}
   \centering
   \includegraphics[angle=0,height=3.11cm]{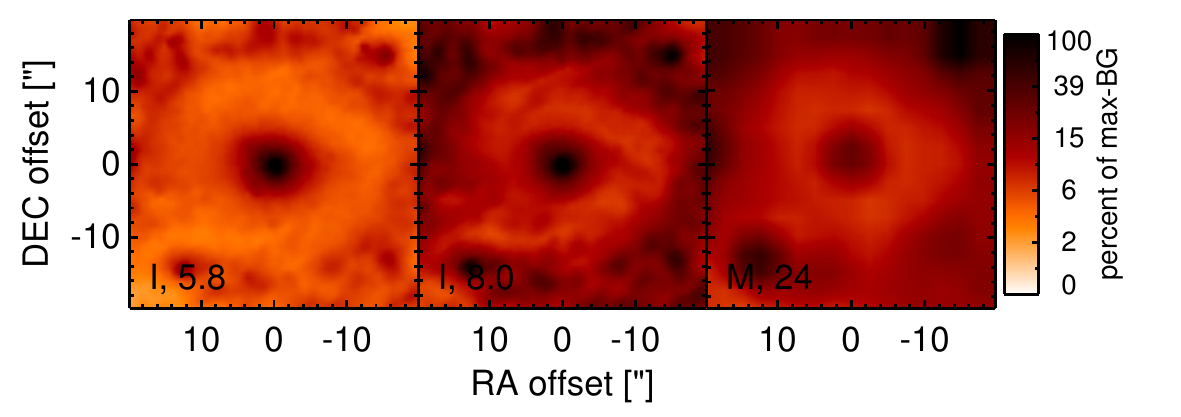}
    \caption{\label{fig:INTim_NGC3486}
             \spitzerr MIR images of NGC\,3486. Displayed are the inner $40\arcsec$ with North up and East to the left. The colour scaling is logarithmic with white corresponding to median background and black to the $0.1\%$ pixels with the highest intensity.
             The label in the bottom left states instrument and central wavelength of the filter in $\mu$m (I: IRAC, M: MIPS). 
           }
\end{figure}
\begin{figure}
   \centering
   \includegraphics[angle=0,width=8.50cm]{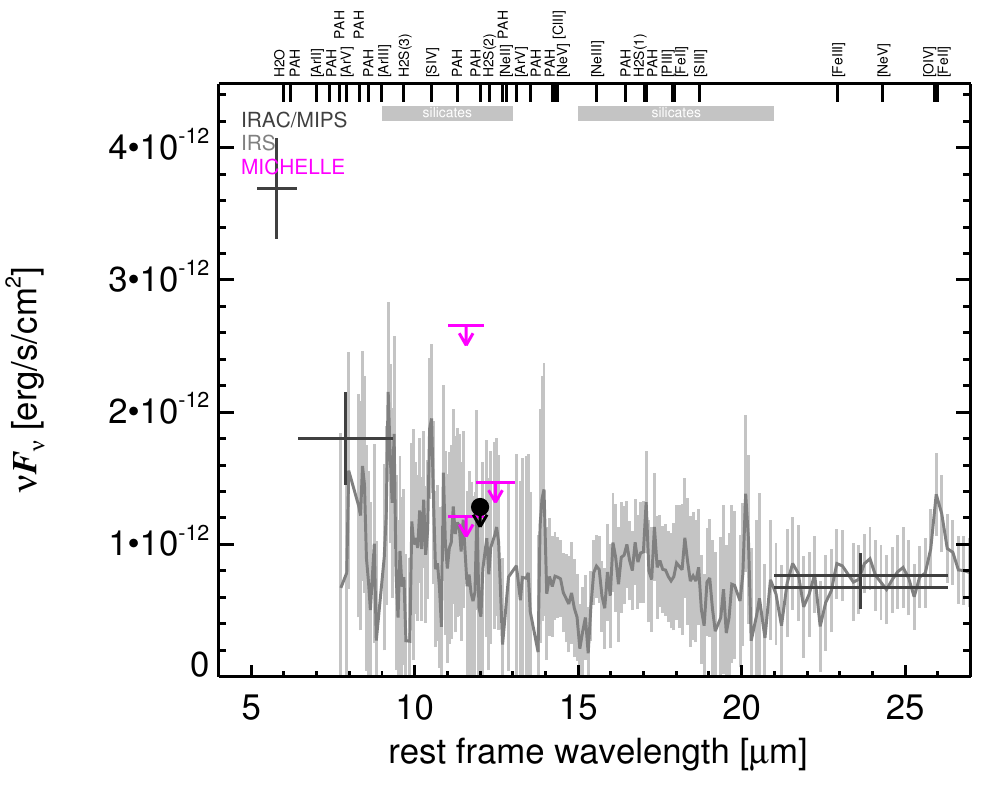}
   \caption{\label{fig:MISED_NGC3486}
      MIR SED of NGC\,3486. The description  of the symbols (if present) is the following.
      Grey crosses and  solid lines mark the \spitzer/IRAC, MIPS and IRS data. 
      The colour coding of the other symbols is: 
      green for COMICS, magenta for Michelle, blue for T-ReCS and red for VISIR data.
      Darker-coloured solid lines mark spectra of the corresponding instrument.
      The black filled circles mark the nuclear 12 and $18\,\mu$m  continuum emission estimate from the data.
      The ticks on the top axis mark positions of common MIR emission lines, while the light grey horizontal bars mark wavelength ranges affected by the silicate 10 and 18$\mu$m features.}
\end{figure}
\clearpage

\twocolumn[\begin{@twocolumnfalse}  
\subsection{NGC\,3521}\label{app:NGC3521}
NGC\,3521 is a nearby highly-inclined spiral galaxy at a distance of $D=$ $11.5 \pm 2.8\,$Mpc (NED redshift-independent median) with a little-studied active nucleus, classified optically as borderline H\,II/LINER \citep{ho_search_1997-1}.
A nuclear compact source was detected in X-rays \citep{zhang_census_2009}, while no subarcsecond-resolution radio observations have been published.
The first observed in the MIR by \cite{rieke_10_1978} but not detected, while \cite{cizdziel_multiaperture_1985} report a marginal detection.
The \spitzer/IRAC and MIPS images show an elongated nucleus embedded within diffuse host emission and surrounded by a large-scale ring. 
We measure the nuclear component resulting in flux much lower than the total fluxes in the literature (e.g., \citealt{dale_infrared_2005,marble_aromatic_2010}).
The \spitzer/IRS LR staring-mode PBCD spectrum for this object is unreliable due to the complex MIR morphology but indicates significant PAH emission, i.e., star formation (see also \citealt{goulding_towards_2009,marble_aromatic_2010}).
Note that \cite{dudik_spitzer_2009} claim the detection of the AGN-indicative \nev emission line, which however is inconsistent with the upper limit by \cite{goulding_towards_2009} for this line.
The nuclear region of NGC\,3521 was observed with VISIR in the PAH2\_2 filter in 2010 (unpublished, to our knowledge) but nothing was detected in the image.
Our derived flux upper limit is $\sim80\%$ lower than the flux levels of the \spitzerr data.
Therefore, there is no evidence for an AGN from the MIR point of view.
\newline\end{@twocolumnfalse}]

\begin{figure}
   \centering
   \includegraphics[angle=0,width=8.500cm]{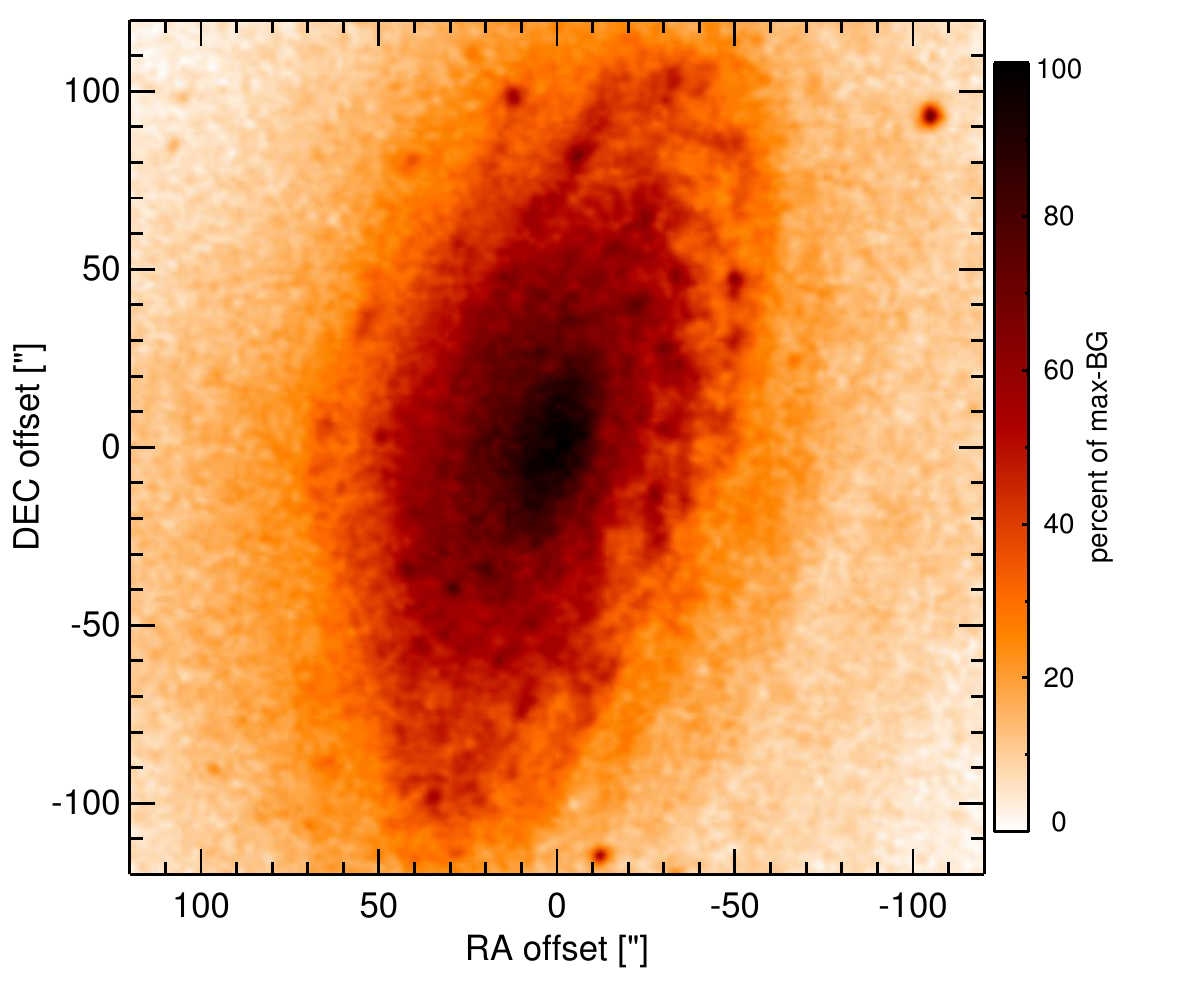}
    \caption{\label{fig:OPTim_NGC3521}
             Optical image (DSS, red filter) of NGC\,3521. Displayed are the central $4\arcmin$ with North up and East to the left. 
              The colour scaling is linear with white corresponding to the median background and black to the $0.01\%$ pixels with the highest intensity.  
           }
\end{figure}
\begin{figure}
   \centering
   \includegraphics[angle=0,height=3.11cm]{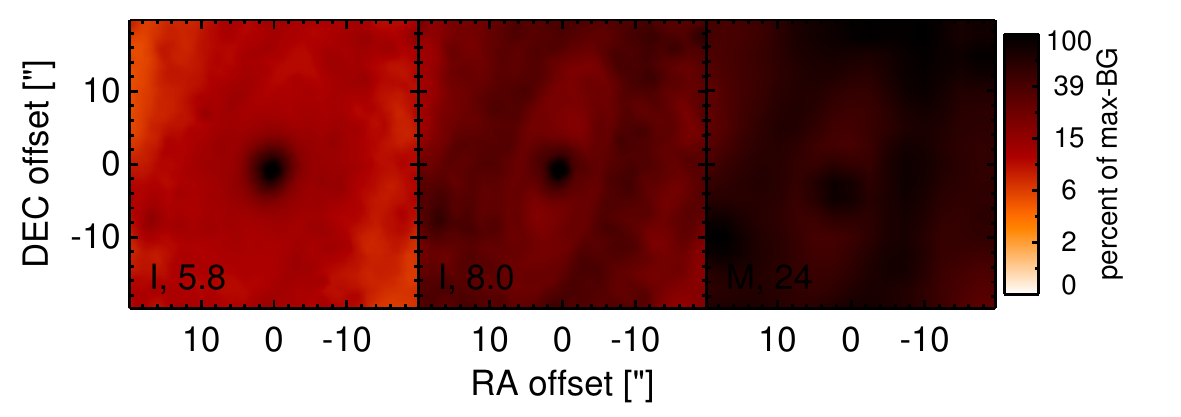}
    \caption{\label{fig:INTim_NGC3521}
             \spitzerr MIR images of NGC\,3521. Displayed are the inner $40\arcsec$ with North up and East to the left. The colour scaling is logarithmic with white corresponding to median background and black to the $0.1\%$ pixels with the highest intensity.
             The label in the bottom left states instrument and central wavelength of the filter in $\mu$m (I: IRAC, M: MIPS). 
           }
\end{figure}
\begin{figure}
   \centering
   \includegraphics[angle=0,width=8.50cm]{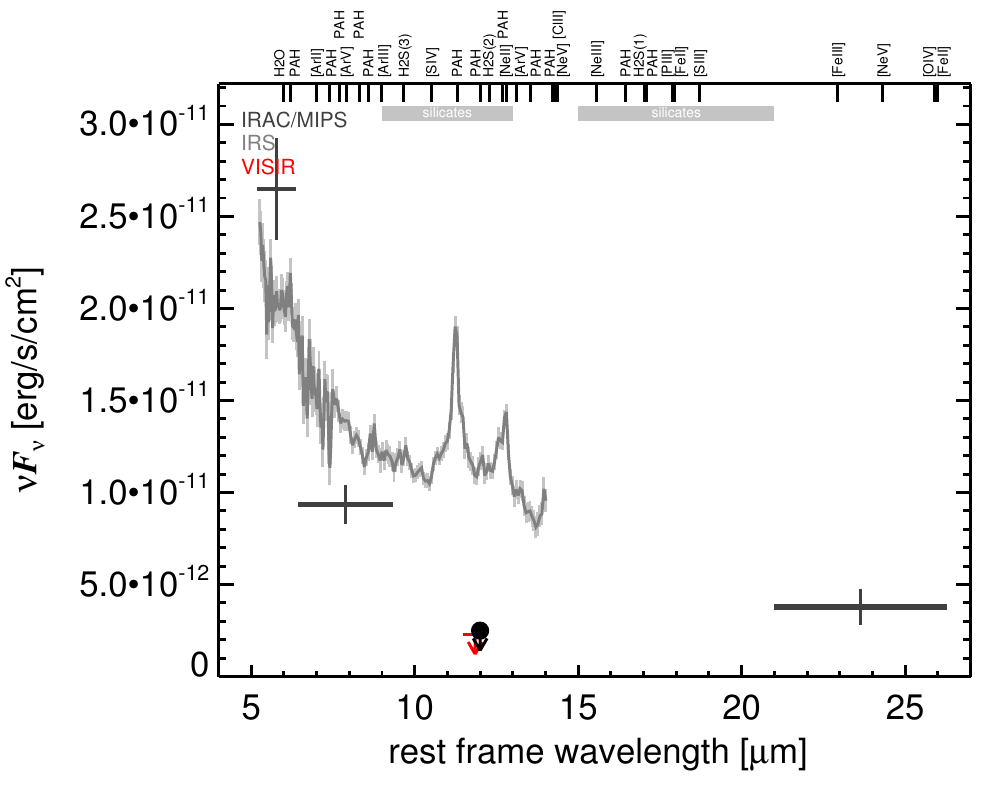}
   \caption{\label{fig:MISED_NGC3521}
      MIR SED of NGC\,3521. The description  of the symbols (if present) is the following.
      Grey crosses and  solid lines mark the \spitzer/IRAC, MIPS and IRS data. 
      The colour coding of the other symbols is: 
      green for COMICS, magenta for Michelle, blue for T-ReCS and red for VISIR data.
      Darker-coloured solid lines mark spectra of the corresponding instrument.
      The black filled circles mark the nuclear 12 and $18\,\mu$m  continuum emission estimate from the data.
      The ticks on the top axis mark positions of common MIR emission lines, while the light grey horizontal bars mark wavelength ranges affected by the silicate 10 and 18$\mu$m features.}
\end{figure}
\clearpage

\twocolumn[\begin{@twocolumnfalse}  
\subsection{NGC\,3607}\label{app:NGC3607}
NGC\,3607 is a low-inclination early-type spiral  galaxy at a distance of $D=$ $21.4 \pm 2.4\,$Mpc (NED redshift-independent median) with an AGN optically classified either as LINER \citep{ho_search_1997-1} or Sy\,2.0 \citep{veron-cetty_catalogue_2010}.
The nucleus is covered by a thick dust filament \citep{barth_search_1998}.
It was detected at radio wavelengths \citep{nagar_radio_2005} but not in X-rays \citep{terashima_x-ray_2002,flohic_central_2006}, although \cite{gonzalez-martin_fitting_2009} propose that the nucleus could be Compton-thick obscured based on the X-ray to \oiii flux ratio.
NGC\,3607 remained undetected in \irass and was only successfully observed in the MIR with \spitzer.
In the corresponding IRAC and MIPS images, it appears as an extended elliptical source without a distinct nuclear component.
The IRS LR mapping-mode PBCD spectrum contains only the shortest wavelength setting and possibly exhibits PAH emission.
The MIR SED from the IRAC and MIPS photometry rather indicates a blue spectral slope in $\nu F_\nu$-space consistent with old stellar emission.
The nuclear region of NGC\,3607 was imaged with VISIR in the PAH1 filter in 2010 (unpublished, to our knowledge) but nothing was detected.
Our derived flux upper limit is $\sim50\%$ lower than the IRAC 8.0$\,\mu$m flux.
Therefore, we conclude that host emission is presumably dominating the MIR in the central $\sim0.4\,$kpc of NGC\,3607, and the MIR data are not sufficient to either verify or disfavour the presence of an AGN.
\newline\end{@twocolumnfalse}]

\begin{figure}
   \centering
   \includegraphics[angle=0,width=8.500cm]{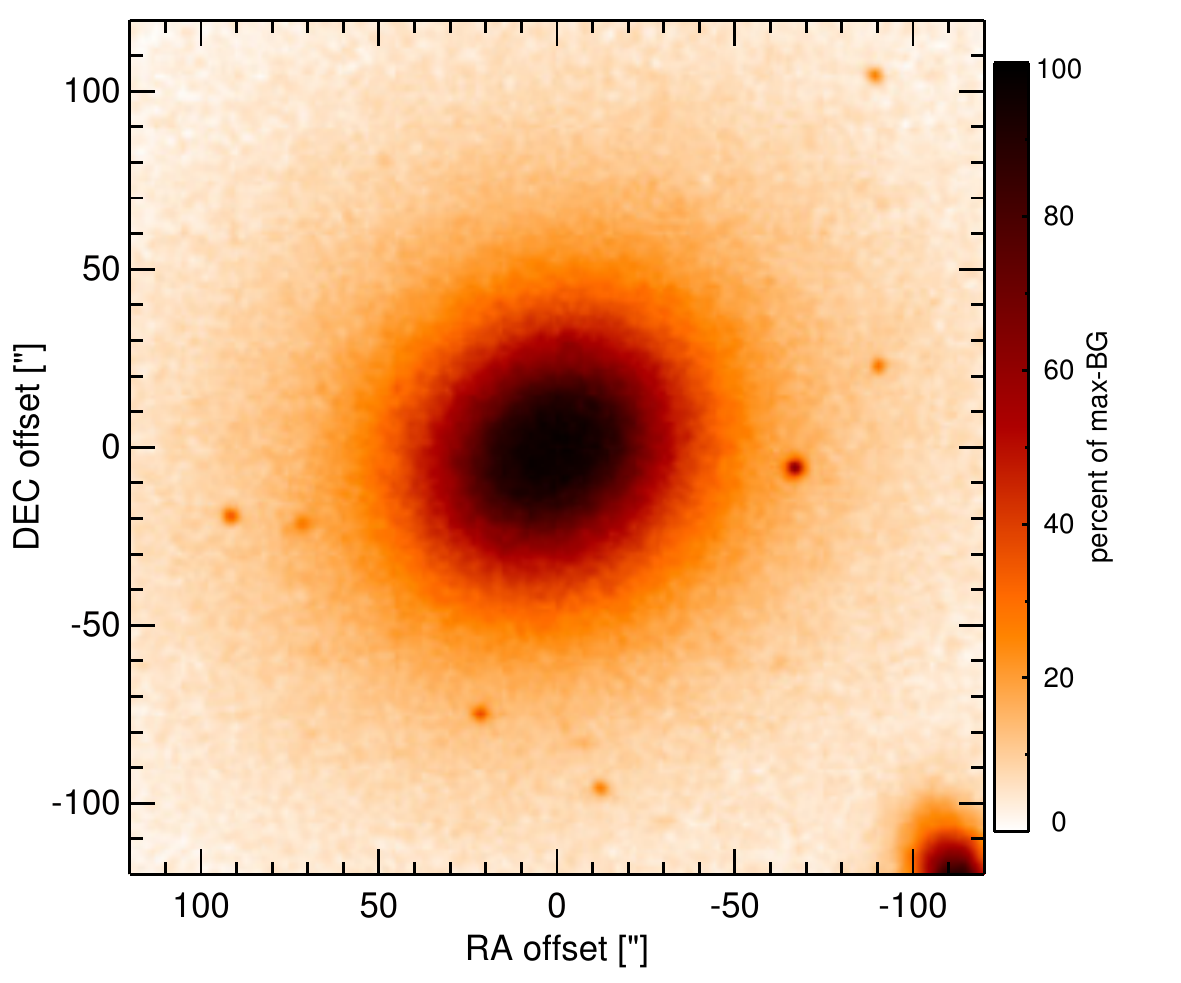}
    \caption{\label{fig:OPTim_NGC3607}
             Optical image (DSS, red filter) of NGC\,3607. Displayed are the central $4\arcmin$ with North up and East to the left. 
              The colour scaling is linear with white corresponding to the median background and black to the $0.01\%$ pixels with the highest intensity.  
           }
\end{figure}
\begin{figure}
   \centering
   \includegraphics[angle=0,height=3.11cm]{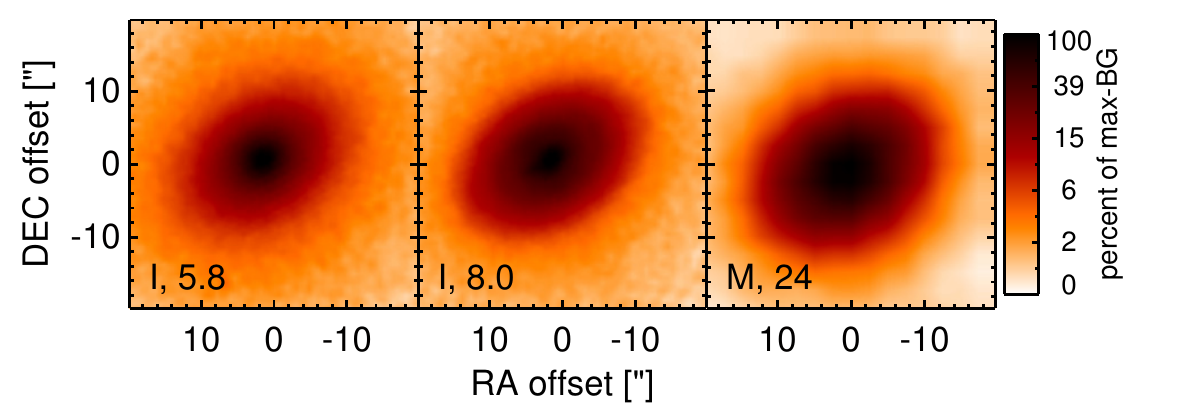}
    \caption{\label{fig:INTim_NGC3607}
             \spitzerr MIR images of NGC\,3607. Displayed are the inner $40\arcsec$ with North up and East to the left. The colour scaling is logarithmic with white corresponding to median background and black to the $0.1\%$ pixels with the highest intensity.
             The label in the bottom left states instrument and central wavelength of the filter in $\mu$m (I: IRAC, M: MIPS). 
           }
\end{figure}
\begin{figure}
   \centering
   \includegraphics[angle=0,width=8.50cm]{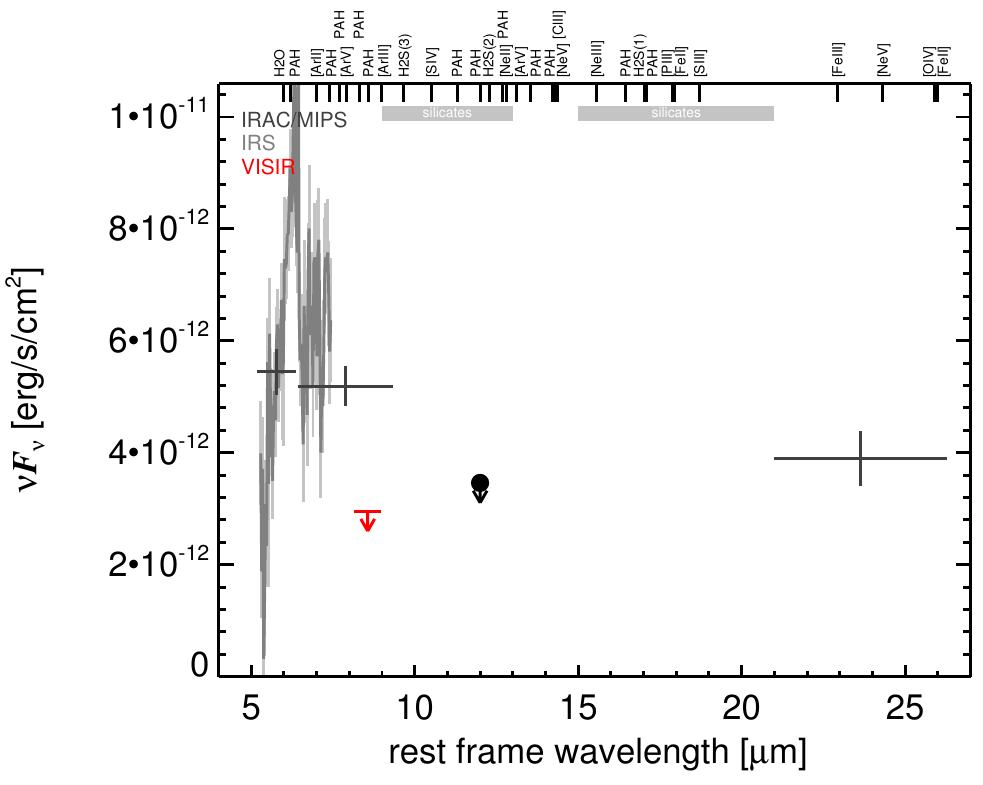}
   \caption{\label{fig:MISED_NGC3607}
      MIR SED of NGC\,3607. The description  of the symbols (if present) is the following.
      Grey crosses and  solid lines mark the \spitzer/IRAC, MIPS and IRS data. 
      The colour coding of the other symbols is: 
      green for COMICS, magenta for Michelle, blue for T-ReCS and red for VISIR data.
      Darker-coloured solid lines mark spectra of the corresponding instrument.
      The black filled circles mark the nuclear 12 and $18\,\mu$m  continuum emission estimate from the data.
      The ticks on the top axis mark positions of common MIR emission lines, while the light grey horizontal bars mark wavelength ranges affected by the silicate 10 and 18$\mu$m features.}
\end{figure}
\clearpage

\twocolumn[\begin{@twocolumnfalse}  
\subsection{NGC\,3623 -- M65}\label{app:NGC3623}
NGC\,3623 is an inclined spiral galaxy in the Leo Triplet at a distance of $D=$ $12.8 \pm 2.2$\,Mpc (NED redshift-independent median) with a LINER nucleus \citep{ho_search_1997-1}.
The first $N$-band observations of this object were performed by \cite{rieke_10_1978} but the nucleus remained undetected.
The \spitzer/IRAC and MIPS images show an extended nucleus embedded in large-scale host emission.
We attempted to isolate the nuclear emission component and accordingly our IRAC $5.8$ and $8.0\,\mu$m and MIPS $24\,\mu$m fluxes are significantly lower than the values in \cite{dale_spitzer_2009}.
Unfortunately, no \spitzer/IRS data are available for NGC\,3623.
Also the VISIR observation in the PAH2\_2 filter in 2010 failed to detect any nuclear emission (unpublished, to our knowledge).
Our derived upper limit is consistent with the \spitzerr photometry.
It remains unclear, whether NGC\,3623 harbours an AGN or not (see also \citealt{dudik_chandra_2005}).
\newline\end{@twocolumnfalse}]

\begin{figure}
   \centering
   \includegraphics[angle=0,width=8.500cm]{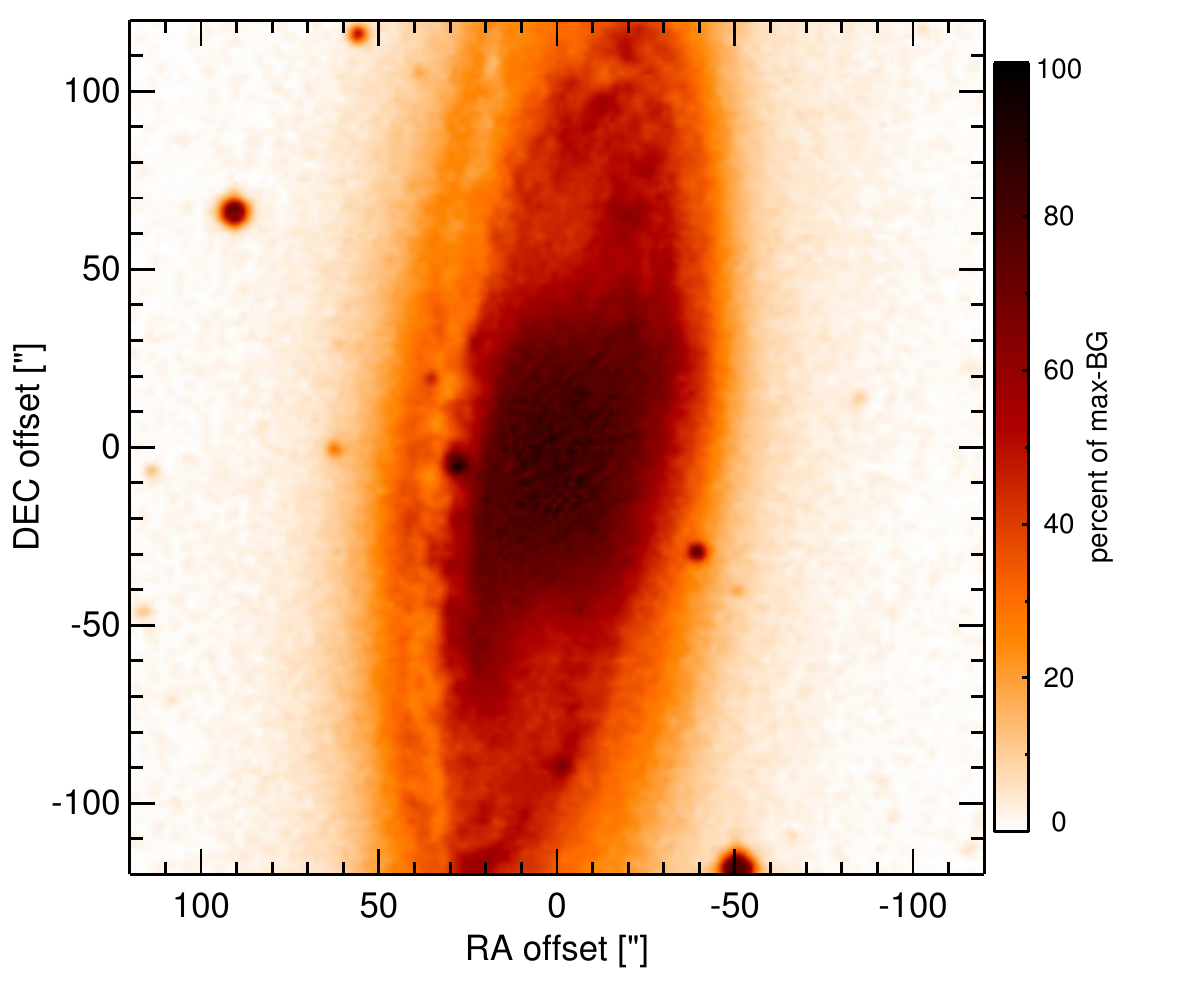}
    \caption{\label{fig:OPTim_NGC3623}
             Optical image (DSS, red filter) of NGC\,3623. Displayed are the central $4\arcmin$ with North up and East to the left. 
              The colour scaling is linear with white corresponding to the median background and black to the $0.01\%$ pixels with the highest intensity.  
           }
\end{figure}
\begin{figure}
   \centering
   \includegraphics[angle=0,height=3.11cm]{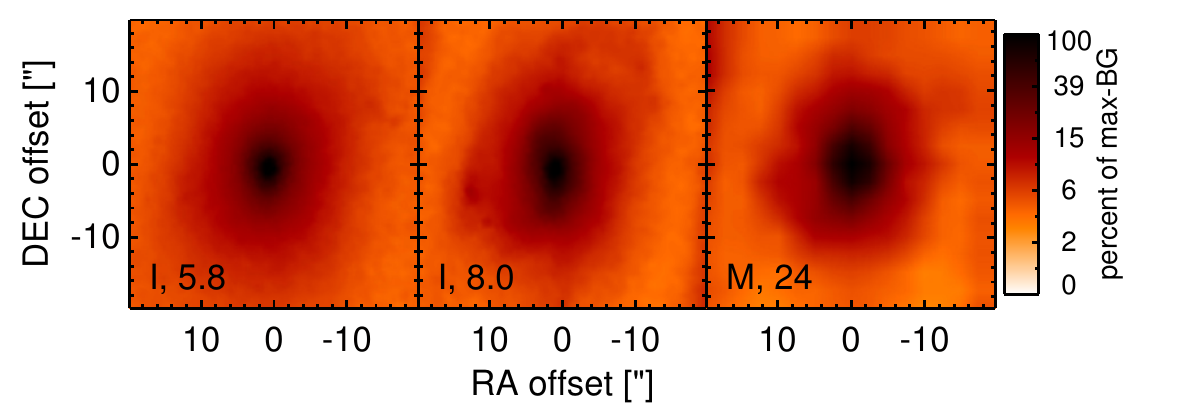}
    \caption{\label{fig:INTim_NGC3623}
             \spitzerr MIR images of NGC\,3623. Displayed are the inner $40\arcsec$ with North up and East to the left. The colour scaling is logarithmic with white corresponding to median background and black to the $0.1\%$ pixels with the highest intensity.
             The label in the bottom left states instrument and central wavelength of the filter in $\mu$m (I: IRAC, M: MIPS). 
           }
\end{figure}
\begin{figure}
   \centering
   \includegraphics[angle=0,width=8.50cm]{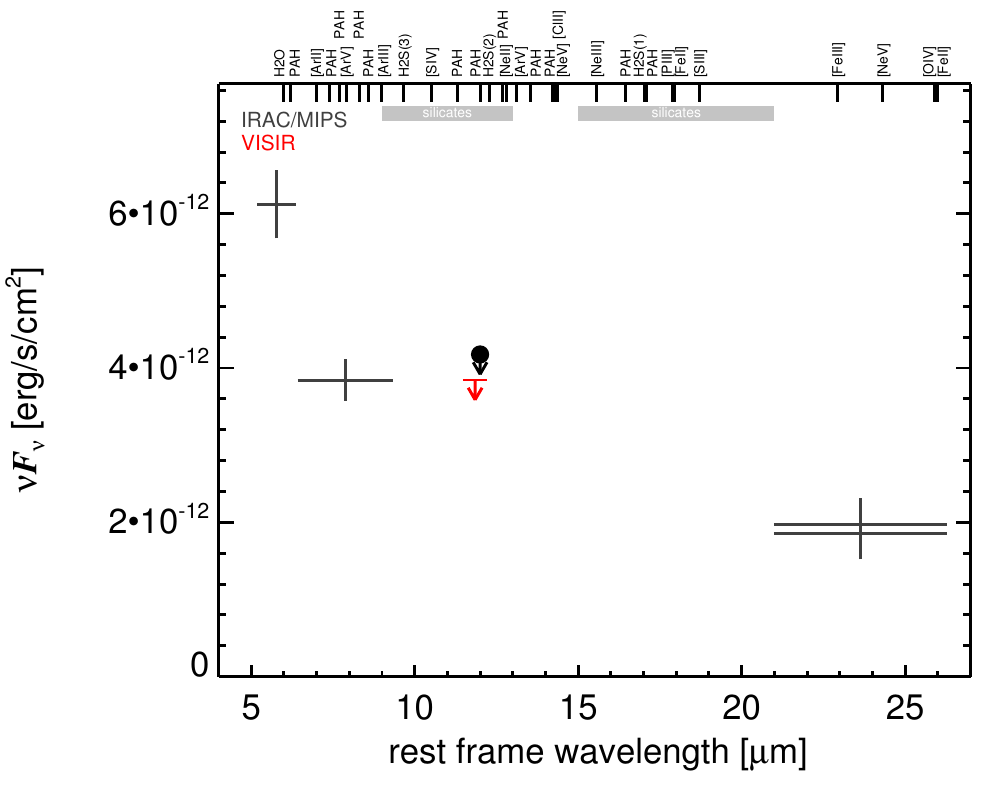}
   \caption{\label{fig:MISED_NGC3623}
      MIR SED of NGC\,3623. The description  of the symbols (if present) is the following.
      Grey crosses and  solid lines mark the \spitzer/IRAC, MIPS and IRS data. 
      The colour coding of the other symbols is: 
      green for COMICS, magenta for Michelle, blue for T-ReCS and red for VISIR data.
      Darker-coloured solid lines mark spectra of the corresponding instrument.
      The black filled circles mark the nuclear 12 and $18\,\mu$m  continuum emission estimate from the data.
      The ticks on the top axis mark positions of common MIR emission lines, while the light grey horizontal bars mark wavelength ranges affected by the silicate 10 and 18$\mu$m features.}
\end{figure}
\clearpage

\twocolumn[\begin{@twocolumnfalse}  
\subsection{NGC\,3627 -- M66}\label{app:NGC3627}
NGC\,3627 is a grand-design spiral galaxy in the Leo Triplet at a distance of $D=$ $10.1 \pm 1.7$\,Mpc (NED redshift-independent median) with an active nucleus either classified as a transition/Sy\,2 \citep{ho_search_1997-1} or a LINER \citep{veron-cetty_catalogue_2010}.
It is in tidal interaction with NGC\,3628 \citep{zhang_high-resolution_1993}.
The first $N$-band photometry was taken by \cite{rieke_infrared_1972} and \cite{rieke_10_1978}.
Since \iras, \spitzer/IRAC, IRS and MIPS observations followed.
The IRAC and MIPS images show a compact nucleus with surrounding spiral-like host emission.
Because we measure the photometry of  the nuclear component only, our IRAC $5.8$ and $8.0\,\mu$m and MIPS $24\,\mu$m fluxes are significantly lower than the values in the literature (e.g., \citealt{dale_infrared_2005,dale_spitzer_2009,munoz-mateos_radial_2009}).
Owing to the complex emission morphology of NGC\,3627 in the MIR, the IRS LR mapping-mode PBCD spectrum is not very reliable. 
However, it roughly matches the IRAC photometry and exhibits strong PAH emission with possibly weak silicate $10\,\mu$m absorption, i.e., star formation in the central 4\arcsec $\sim 200$\,pc region (see e.g., \citealt{dale_spitzer_2009,goulding_towards_2009}).
NGC\,3627 was observed with VISIR in the PAH1 filter in 2010 (unpublished, to our knowledge). 
An north-south elongated structure ($\sim 2\arcsec \sim 100\,$pc) matching the larger scale morphology seen in IRAC in orientation is weakly detected.
We measure the nuclear unresolved component with a flux only $6\%$ of that in \spitzer.
However, the unresolved component is not clearly separable from the extended emission.
Therefore, our nuclear $12\,\mu$m continuum emission estimate should be regarded as an upper limit for any AGN. 
The nucleus has not been detected in X-rays so far \citep{ho_detection_2001,filho_further_2004} but is clearly present at radio wavelengths \citep{filho_nature_2000}.
If an AGN is present in this object it is most likely heavily obscured and embedded in nuclear star formation (see also \citealt{gonzalez-martin_fitting_2009,masegosa_nature_2011}).
\newline\end{@twocolumnfalse}]

\begin{figure}
   \centering
   \includegraphics[angle=0,width=8.500cm]{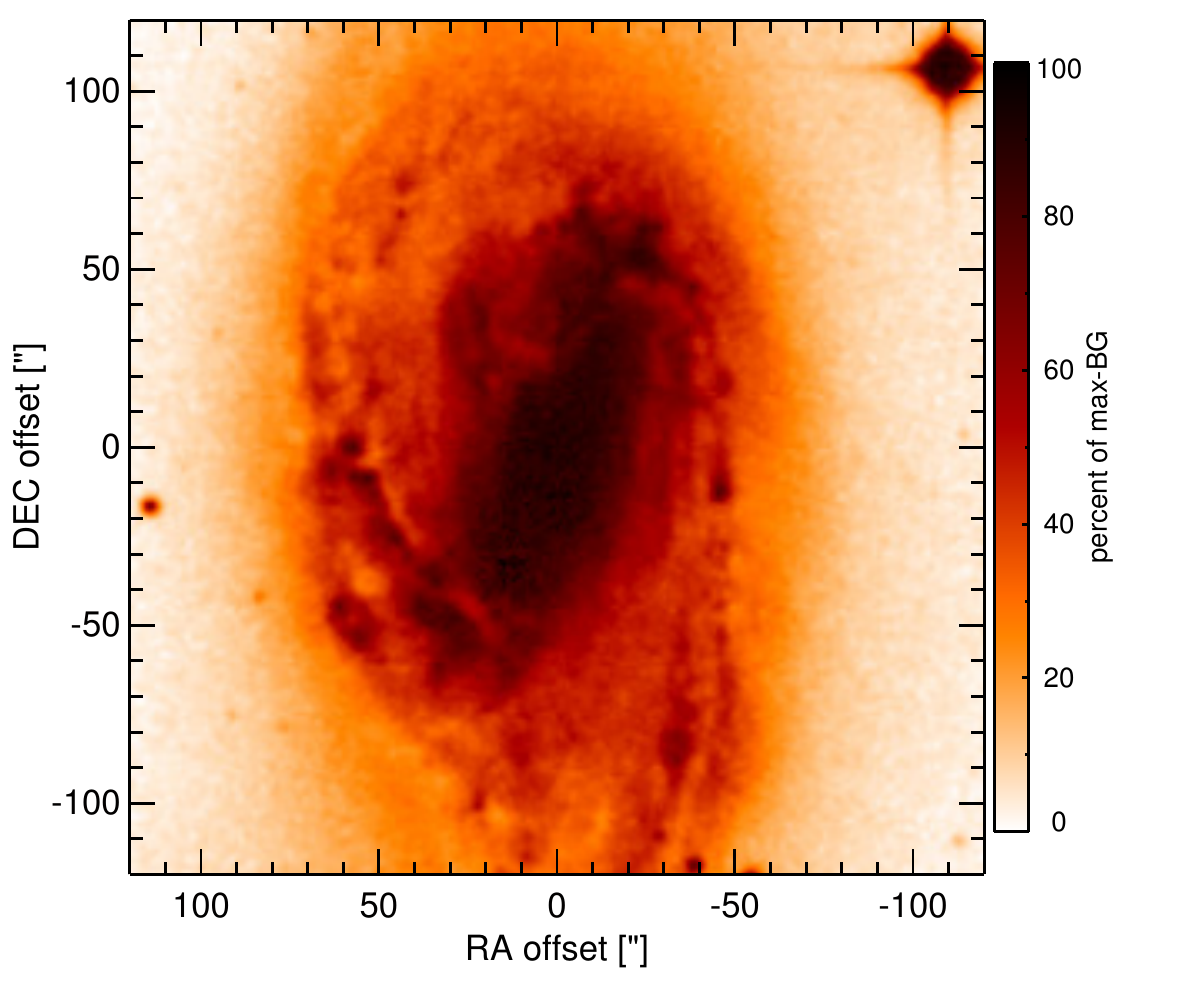}
    \caption{\label{fig:OPTim_NGC3627}
             Optical image (DSS, red filter) of NGC\,3627. Displayed are the central $4\arcmin$ with North up and East to the left. 
              The colour scaling is linear with white corresponding to the median background and black to the $0.01\%$ pixels with the highest intensity.  
           }
\end{figure}
\begin{figure}
   \centering
   \includegraphics[angle=0,height=3.11cm]{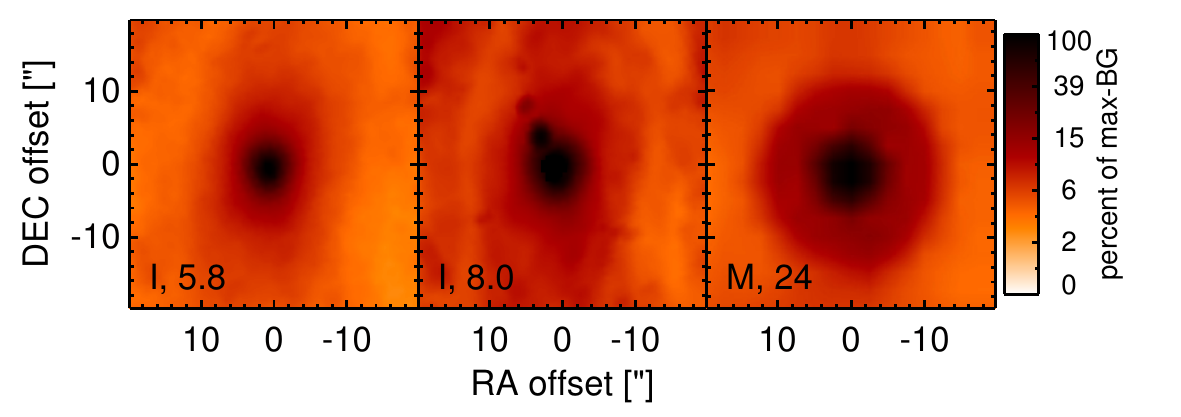}
    \caption{\label{fig:INTim_NGC3627}
             \spitzerr MIR images of NGC\,3627. Displayed are the inner $40\arcsec$ with North up and East to the left. The colour scaling is logarithmic with white corresponding to median background and black to the $0.1\%$ pixels with the highest intensity.
             The label in the bottom left states instrument and central wavelength of the filter in $\mu$m (I: IRAC, M: MIPS).
             Note that the apparent off-nuclear compact sources in the IRAC $8.0\,\mu$m image are instrumental artefacts.
           }
\end{figure}
\begin{figure}
   \centering
   \includegraphics[angle=0,height=3.11cm]{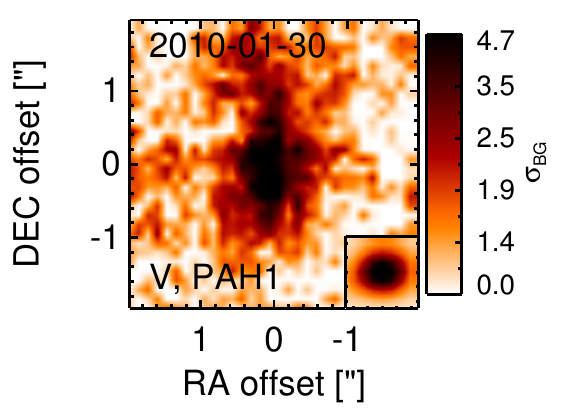}
    \caption{\label{fig:HARim_NGC3627}
             Subarcsecond-resolution MIR images of NGC\,3627 sorted by increasing filter wavelength. 
             Displayed are the inner $4\arcsec$ with North up and East to the left. 
             The colour scaling is logarithmic with white corresponding to median background and black to the $75\%$ of the highest intensity of all images in units of $\sigbg$.
             The inset image shows the central arcsecond of the PSF from the calibrator star, scaled to match the science target.
             The labels in the bottom left state instrument and filter names (C: COMICS, M: Michelle, T: T-ReCS, V: VISIR).
           }
\end{figure}
\begin{figure}
   \centering
   \includegraphics[angle=0,width=8.50cm]{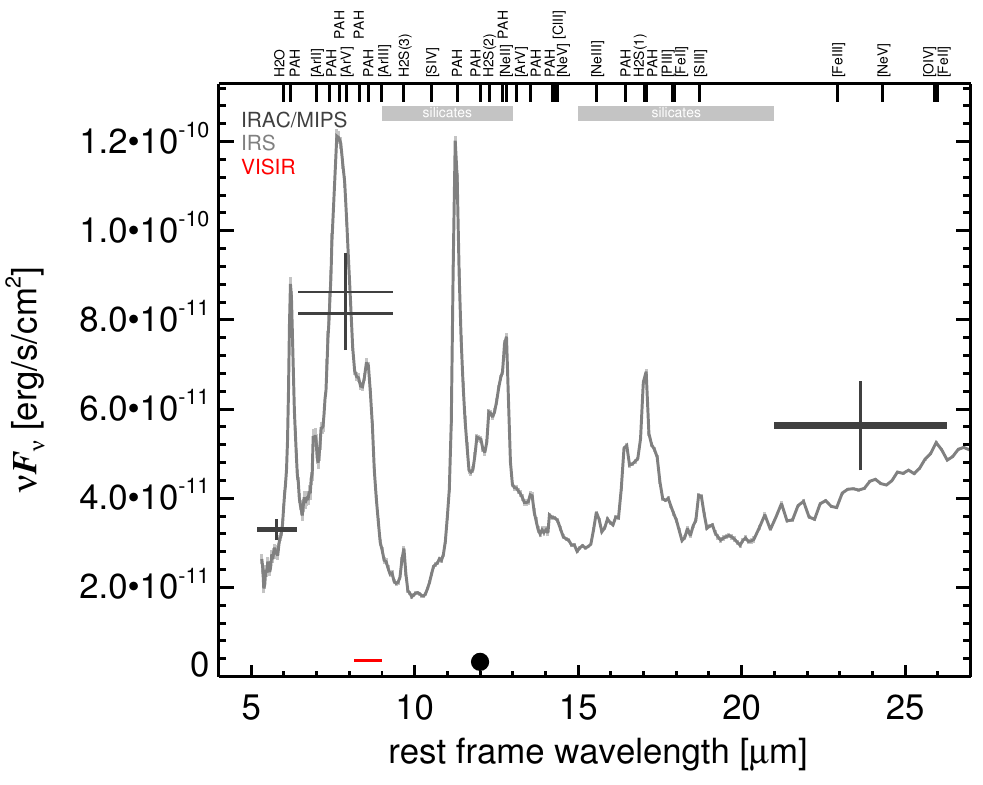}
   \caption{\label{fig:MISED_NGC3627}
      MIR SED of NGC\,3627. The description  of the symbols (if present) is the following.
      Grey crosses and  solid lines mark the \spitzer/IRAC, MIPS and IRS data. 
      The colour coding of the other symbols is: 
      green for COMICS, magenta for Michelle, blue for T-ReCS and red for VISIR data.
      Darker-coloured solid lines mark spectra of the corresponding instrument.
      The black filled circles mark the nuclear 12 and $18\,\mu$m  continuum emission estimate from the data.
      The ticks on the top axis mark positions of common MIR emission lines, while the light grey horizontal bars mark wavelength ranges affected by the silicate 10 and 18$\mu$m features.}
\end{figure}
\clearpage

\twocolumn[\begin{@twocolumnfalse}  
\subsection{NGC\,3628}\label{app:NGC3628}
NGC\,3628 is a nearby disturbed edge-on spiral galaxy in the Leo Triplet at distance of $D=$ $12.2 \pm 2.8\,$Mpc (NED redshift-independent median) with a prominent dust lane covering  an active starburst nucleus, which is optically classified as an AGN/H\,II transition object \citep{ho_search_1997}. 
Its association with a compact radio and X-ray source is controversial  \citep{filho_nature_2000,nagar_radio_2005,yaqoob_hard_1995,gonzalez-martin_x-ray_2006,flohic_central_2006}.
In addition, the detection of the AGN-indicative \nev is also controversial \citep{dudik_mid-infrared_2007,goulding_towards_2009}.
NGC\,3628 was observed with \spitzer/IRAC, IRS and MIPS, and the corresponding images show bright host galaxy emission with an elongated nucleus without distinguishable unresolved component. 
The IRS LR staring-mode spectrum exhibits deep silicate 10 and possibly also 18$\,\mu$m absorption, prominent PAH emission and a red spectral slope  in $\nu F_\nu$-space (see also \citealt{brandl_mid-infrared_2006,goulding_towards_2009,bernard-salas_spitzer_2009}).
Thus, the arcsecond-scale MIR SED both indicates star formation and heavy obscuration.
The nuclear region of NGC\,3628 was observed with VISIR in the PAH1 filter in 2010 (unpublished, to our knowledge).
The image shows an elongated MIR structure (major axis$\sim 7\arcsec\sim0.4$\,kpc; minimum axis$\sim1.7\arcsec\sim100\,$kpc; PA$\sim 102\degree$) with three emission blobs marking the eastern, western and northern borders. 
The uncertainty in the astrometry of the VISIR data together with the lack of multiwavelength morphological subarcsecond information prohibits the association of the putative AGN with any of the MIR sources.
Therefore, we measure the brightest emission blob (the northern) and use its unresolved flux portion as an upper limit for the putative AGN emission. 
This PAH1 flux makes up only $2\%$ of the corresponding flux from the \spitzerr spectrophotometry.
We conclude that any AGN contribution to the MIR emission of the central $\sim0.2$\,kpc is minor in NGC\,3628.
\newline\end{@twocolumnfalse}]

\begin{figure}
   \centering
   \includegraphics[angle=0,width=8.500cm]{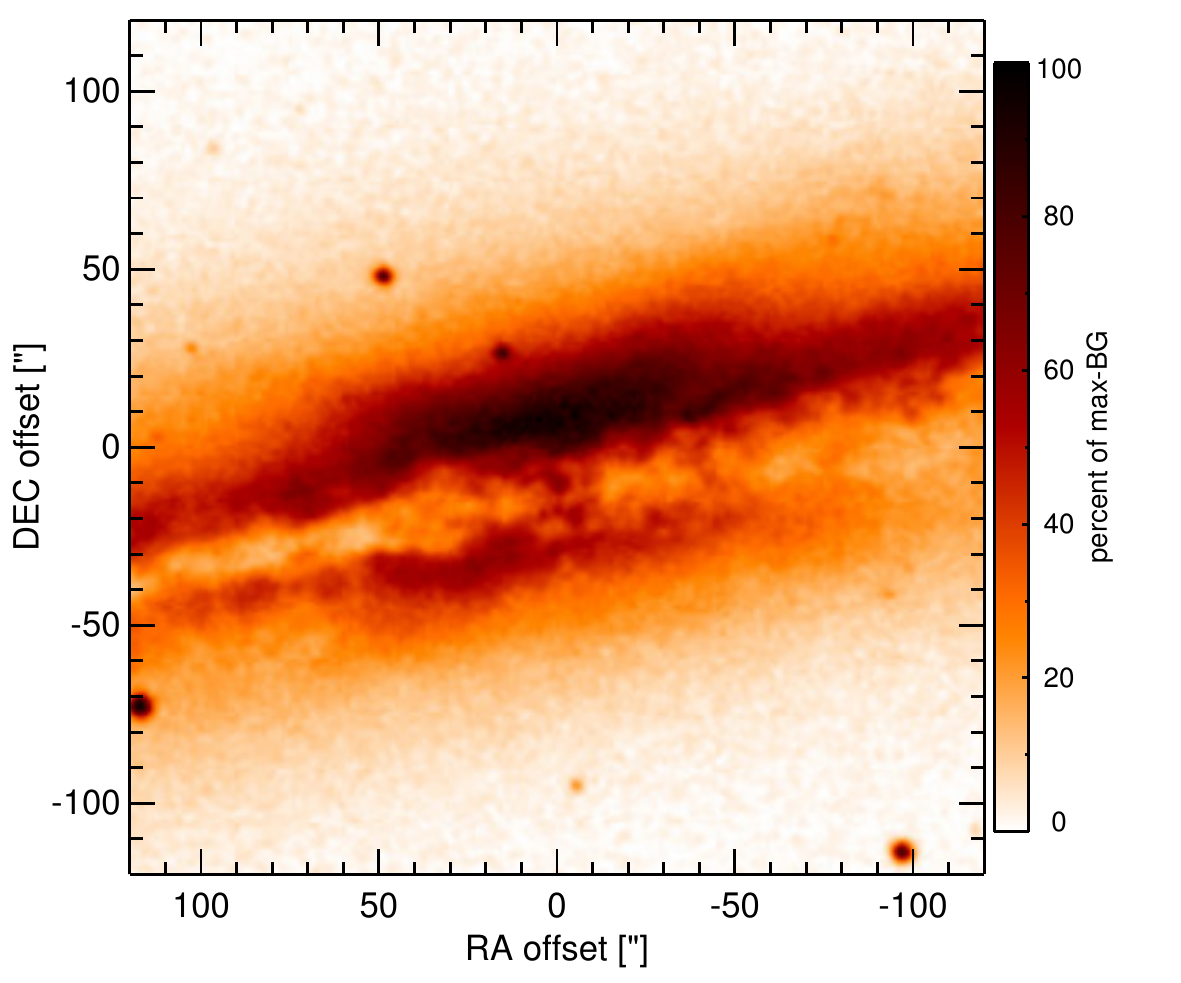}
    \caption{\label{fig:OPTim_NGC3628}
             Optical image (DSS, red filter) of NGC\,3628. Displayed are the central $4\arcmin$ with North up and East to the left. 
              The colour scaling is linear with white corresponding to the median background and black to the $0.01\%$ pixels with the highest intensity.  
           }
\end{figure}
\begin{figure}
   \centering
   \includegraphics[angle=0,height=3.11cm]{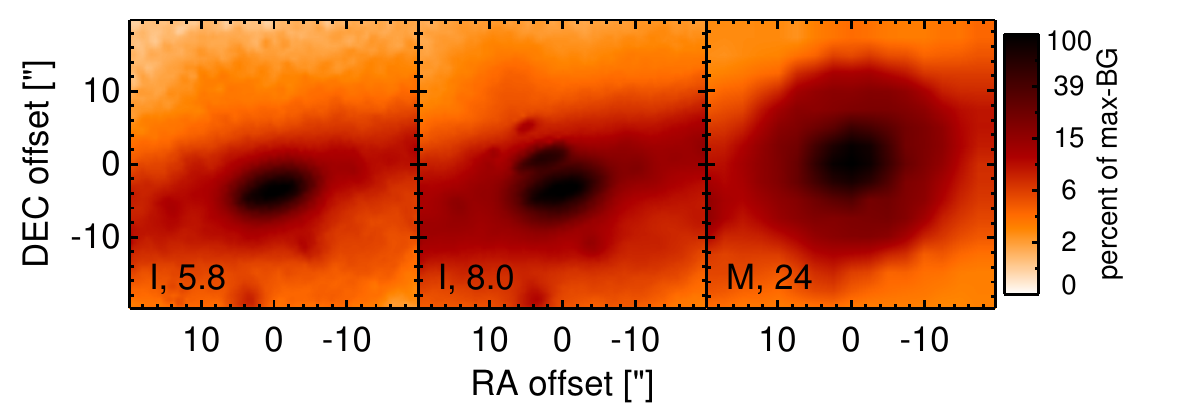}
    \caption{\label{fig:INTim_NGC3628}
             \spitzerr MIR images of NGC\,3628. Displayed are the inner $40\arcsec$ with North up and East to the left. The colour scaling is logarithmic with white corresponding to median background and black to the $0.1\%$ pixels with the highest intensity.
             The label in the bottom left states instrument and central wavelength of the filter in $\mu$m (I: IRAC, M: MIPS).
             Note that the apparent off-nuclear elongated source in the IRAC $8.0\,\mu$m image is an instrumental artefact.
           }
\end{figure}
\begin{figure}
   \centering
   \includegraphics[angle=0,height=3.11cm]{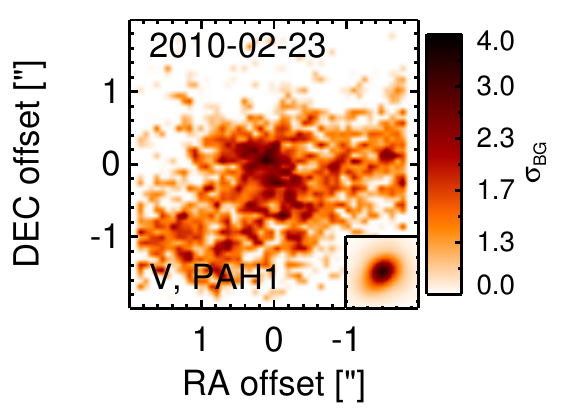}
    \caption{\label{fig:HARim_NGC3628}
             Subarcsecond-resolution MIR images of NGC\,3628 sorted by increasing filter wavelength. 
             Displayed are the inner $4\arcsec$ with North up and East to the left. 
             The colour scaling is logarithmic with white corresponding to median background and black to the $75\%$ of the highest intensity of all images in units of $\sigbg$.
             The inset image shows the central arcsecond of the PSF from the calibrator star, scaled to match the science target.
             The labels in the bottom left state instrument and filter names (C: COMICS, M: Michelle, T: T-ReCS, V: VISIR).
           }
\end{figure}
\begin{figure}
   \centering
   \includegraphics[angle=0,width=8.50cm]{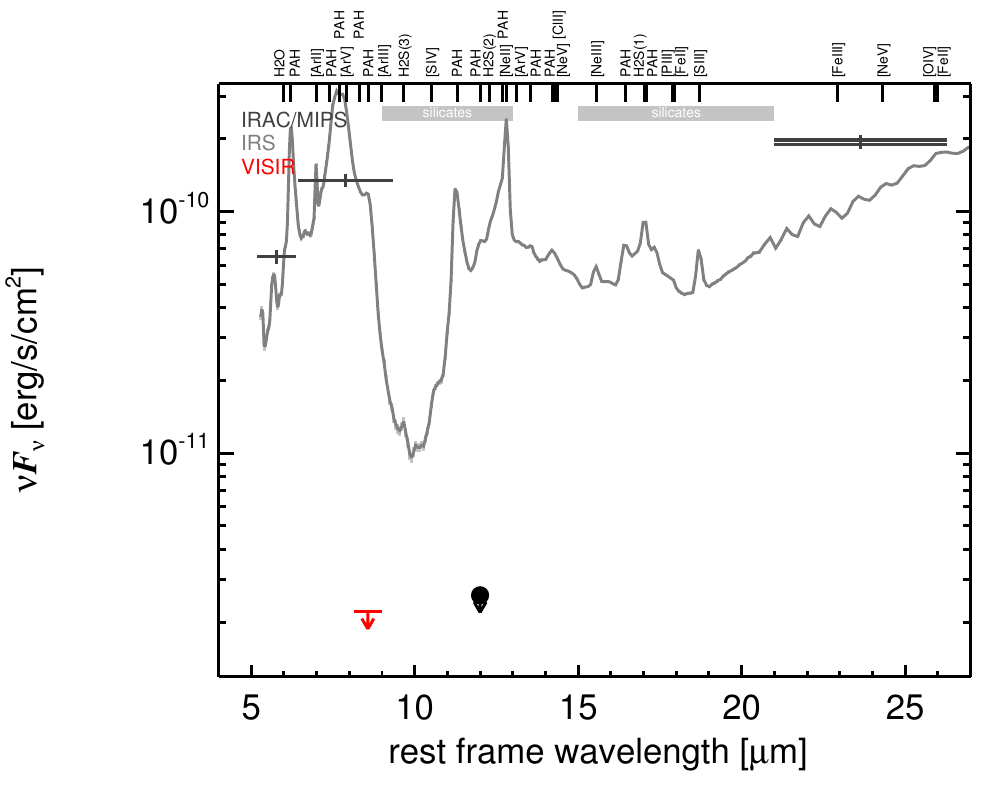}
   \caption{\label{fig:MISED_NGC3628}
      MIR SED of NGC\,3628. The description  of the symbols (if present) is the following.
      Grey crosses and  solid lines mark the \spitzer/IRAC, MIPS and IRS data. 
      The colour coding of the other symbols is: 
      green for COMICS, magenta for Michelle, blue for T-ReCS and red for VISIR data.
      Darker-coloured solid lines mark spectra of the corresponding instrument.
      The black filled circles mark the nuclear 12 and $18\,\mu$m  continuum emission estimate from the data.
      The ticks on the top axis mark positions of common MIR emission lines, while the light grey horizontal bars mark wavelength ranges affected by the silicate 10 and 18$\mu$m features.}
\end{figure}
\clearpage

\twocolumn[\begin{@twocolumnfalse}  
\subsection{NGC\,3660 -- Mrk\,1291}\label{app:NGC3660}
NGC\,3660 is a face-on spiral galaxy at a redshift of $z=$ 0.0123 ($D\sim60.8\,$Mpc) with an active nucleus, which is a ``true''-Seyfert~2 candidate (\citealt{brightman_nature_2008,bianchi_simultaneous_2012}; but see \citealt{shi_unobscured_2010}.
The optical classification of this AGN varies throughout the literature:
\cite{kollatschny_nuclear_1983} first classified it as a Seyfert/H\,II transition object, while later it was classified either as borderline LINER/H\,II \citep{contini_starbursts_1998},  Sy\,1.8 \citep{veron-cetty_catalogue_2010}, or Sy\,2.0 \citep{moran_classification_1996, bianchi_simultaneous_2012}.
This controversy might be related to the possible existence of a nuclear starburst  on the one hand \citep{imanishi_compact_2003}, and very weak broad emission lines on the other (see discussion in \citealt{shi_unobscured_2010} and \citealt{bianchi_simultaneous_2012}).  
The nucleus features a compact NLR \citep{gonzalez_delgado_h_1997} and radio core \citep{morganti_radio_1999}.
The first successful MIR observations of NGC\,3660 were reported by \cite{maiolino_new_1995}, while it remained undetected in the latter subarcsecond-resolution $N$-band imaging with Palomar 5\,m/MIRLIN \citep{gorjian_10_2004}.
The object was detected in the \spitzer/IRAC images where it appears as a compact nucleus surrounded by faint ring-like host emission on large scale. 
We measure the nuclear component, which provides an IRAC 5.8$\,\mu$m flux consistent with \cite{gallimore_infrared_2010}, while our IRAC $8.0\,\mu$m flux is significantly higher.
The \spitzer/IRS LR mapping-mode spectrum suffers from low S/N but indicates prominent PAH emission and a red spectral slope in $\nu F_\nu$-space (see also \citealt{tommasin_spitzer_2008,wu_spitzer/irs_2009,gallimore_infrared_2010, shi_unobscured_2010, tommasin_spitzer-irs_2010}), i.e., significant star formation on arcsecond scales.
The nuclear region of NGC\,3660 was imaged with T-ReCS in the broad N filter in 2004 (flux published in \citealt{videla_nuclear_2013}).
In the image, a marginally resolved nucleus is  detected (FWHM $\sim 0.64\arcsec \sim 180\,$pc; PA$\sim 96\degree$).
However, at least a second epoch of MIR subarcsecond imaging is required to verify this extension.
Our measurement of the nuclear flux is consistent with \cite{videla_nuclear_2013}, the \spitzerr spectrophotometry and the historical measurements.
Thus, presumably even the nuclear flux is significantly star formation contaminated. 
\newline\end{@twocolumnfalse}]

\begin{figure}
   \centering
   \includegraphics[angle=0,width=8.500cm]{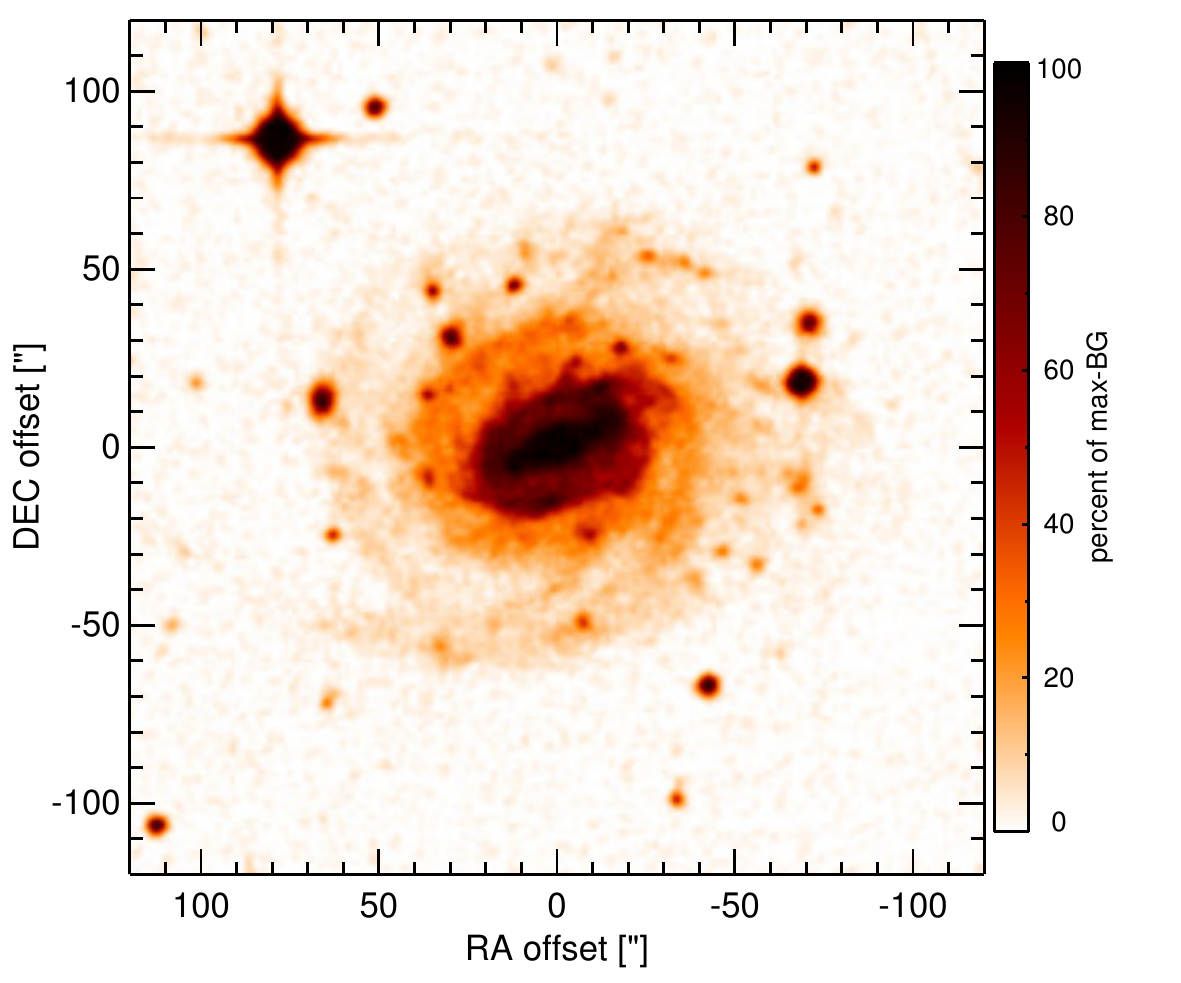}
    \caption{\label{fig:OPTim_NGC3660}
             Optical image (DSS, red filter) of NGC\,3660. Displayed are the central $4\arcmin$ with North up and East to the left. 
              The colour scaling is linear with white corresponding to the median background and black to the $0.01\%$ pixels with the highest intensity.  
           }
\end{figure}
\begin{figure}
   \centering
   \includegraphics[angle=0,height=3.11cm]{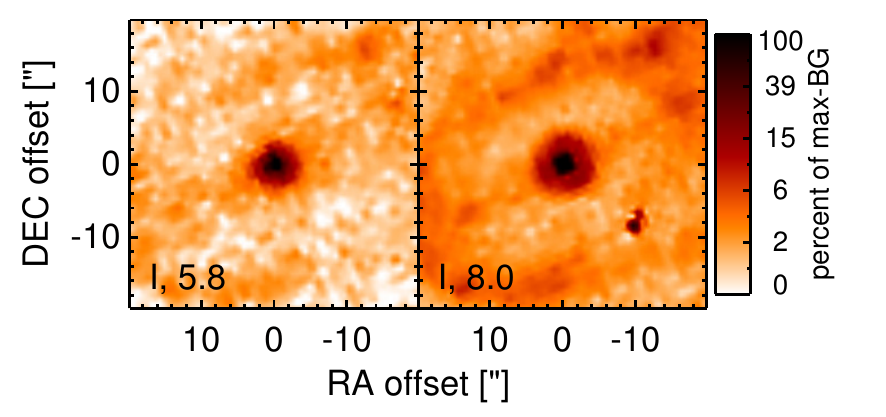}
    \caption{\label{fig:INTim_NGC3660}
             \spitzerr MIR images of NGC\,3660. Displayed are the inner $40\arcsec$ with North up and East to the left. The colour scaling is logarithmic with white corresponding to median background and black to the $0.1\%$ pixels with the highest intensity.
             The label in the bottom left states instrument and central wavelength of the filter in $\mu$m (I: IRAC, M: MIPS). 
           }
\end{figure}
\begin{figure}
   \centering
   \includegraphics[angle=0,height=3.11cm]{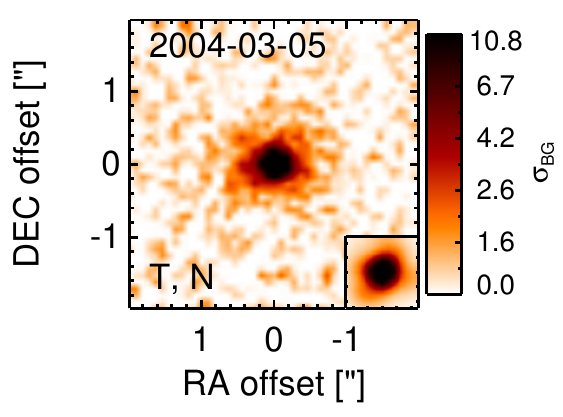}
    \caption{\label{fig:HARim_NGC3660}
             Subarcsecond-resolution MIR images of NGC\,3660 sorted by increasing filter wavelength. 
             Displayed are the inner $4\arcsec$ with North up and East to the left. 
             The colour scaling is logarithmic with white corresponding to median background and black to the $75\%$ of the highest intensity of all images in units of $\sigbg$.
             The inset image shows the central arcsecond of the PSF from the calibrator star, scaled to match the science target.
             The labels in the bottom left state instrument and filter names (C: COMICS, M: Michelle, T: T-ReCS, V: VISIR).
           }
\end{figure}
\begin{figure}
   \centering
   \includegraphics[angle=0,width=8.50cm]{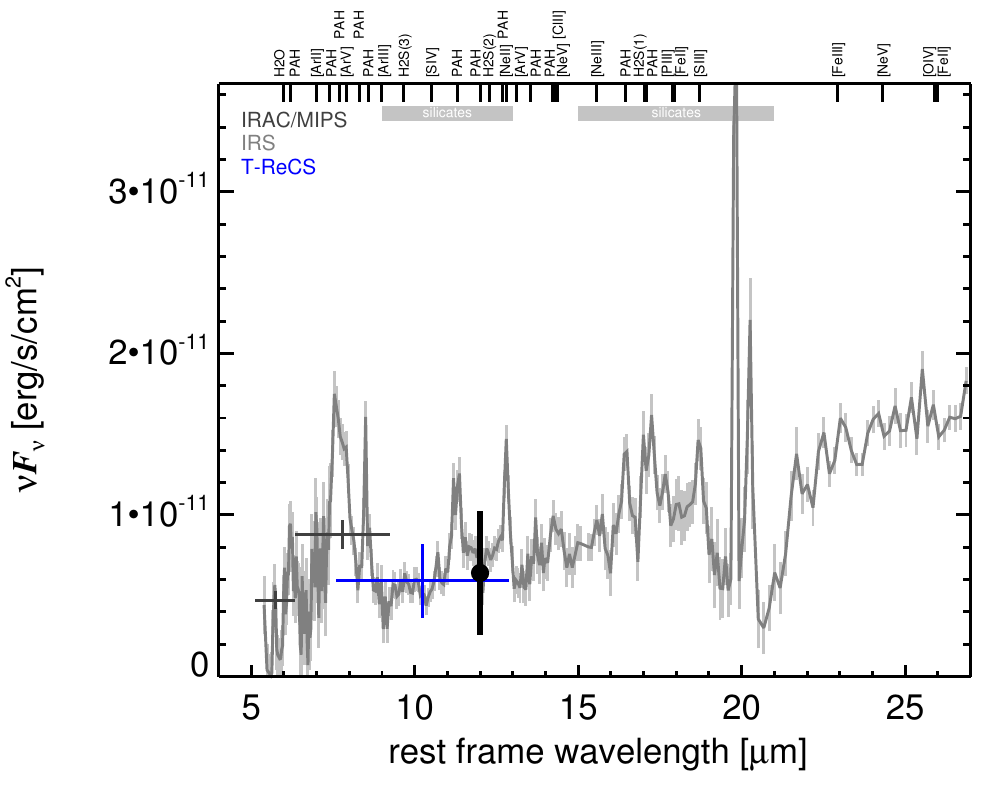}
   \caption{\label{fig:MISED_NGC3660}
      MIR SED of NGC\,3660. The description  of the symbols (if present) is the following.
      Grey crosses and  solid lines mark the \spitzer/IRAC, MIPS and IRS data. 
      The colour coding of the other symbols is: 
      green for COMICS, magenta for Michelle, blue for T-ReCS and red for VISIR data.
      Darker-coloured solid lines mark spectra of the corresponding instrument.
      The black filled circles mark the nuclear 12 and $18\,\mu$m  continuum emission estimate from the data.
      The ticks on the top axis mark positions of common MIR emission lines, while the light grey horizontal bars mark wavelength ranges affected by the silicate 10 and 18$\mu$m features.}
\end{figure}
\clearpage

\twocolumn[\begin{@twocolumnfalse}  
\subsection{NGC\,3690 -- Arp\,299 -- Mrk\,171}\label{app:NGC3690E}\label{app:NGC3690W}
NGC\,3690 is a pair of merging late-type galaxies at a distance of $D=$ $\sim 45\,$Mpc.  
The two nuclei are separated by 20\arcsec\, ($\sim 4.3\,$kpc) and have been designated with various names, leading to  confusion in the literature. 
Therefore, we refer to the western nucleus as NGC\,3690W, it is also called NGC\,3690A, Arp\,299B, UGC\,6472 and sometimes just NGC\,3690. 
We call the eastern nucleus correspondingly NGC\,3690E (NGC\,3690B, Arp\,299A, UGC\,6471), which is often wrongly called IC\,694 (see NED and \citealt{yamaoka_supernova_1998}).
NGC\,3690W is classified as a H\,II nucleus in \cite{veron-cetty_catalogue_2010}, while NGC\,3690E is not even included.
However, based on the evidence from deep X-ray, optical and radio observations, NGC\,3690W is possibly a Sy\,2 and NGC\,3690E possibly a LINER \citep{ballo_arp_2004,garcia-marin_integral_2006,perez-torres_serendipitous_2010}.
At the same time, the whole NGC\,3690 system is one of the most extreme cases of an extended burst of star formation known \citep{augarde_peculiar_1985} and has been studied extensively during the last 40 years.
Both NGC\,3690E and NGC\,3690W contain luminous water maser emission possibly associated with AGN activity \citep{henkel_new_2005,tarchi_new_2011}.
Early ground-based $N$-band photometry of NGC\,3690 was performed by \cite{rieke_infrared_1972,allen_near-infrared_1976,lebofsky_extinction_1979,gehrz_star_1983,carico_iras_1988,keto_infrared_1997} and \cite{miles_high-resolution_1996}.
Both nuclei are detected in the first $N$-band images and two additional compact sources (C and C') are found $\sim 8\arcsec$ ($\sim1.7$\,kpc) north of the NGC\,3690W nucleus (B1), while the whole system is embedded in diffuse MIR emission \citep{gehrz_star_1983}.
\cite{soifer_high-resolution_2001} present the first subarcsecond resolution $N-$ and $Q$-band images of NGC\,3690 obtained with Keck/LWS, which verify the morphology found earlier and further constrain the MIR emission of the two nuclei to be confined to $< 0.6\arcsec$ ($\sim130$\,pc).
The first MIR spectroscopic study using \textit{ISO} was \cite{gallais_dust_2004}.
They find strong PAH emission and silicate absorption in NGC3690E, indicating a deeply embedded starburst, and strong silicate but less PAH emission in  NGC3690W.
From this and the emission line ratios, they interpret that a deeply absorbed AGN exist in this nucleus.
The \spitzer/IRAC, IRS and MIPS data verifies the morphology and MIR SEDs of NGC\,3690. 
Note that the IRAC $8\,\mu$m image is saturated in both nuclei and thus we could not perform nuclear photometry in this waveband. 
A comprehensive study of the IRS spectra of NGC\,3690 can be found in \cite{alonso-herrero_extreme_2009}.
The system was observed with Gemini/Michelle in 2003 in the $N$-band filters Si-1 and Si-5, in which the pointing was centred on NGC\,3690E. 
Therefore, the nucleus of NGC\,3690W is only visible in one of the negative chopping positions. 
Because Michelle is unguided in these positions, we can only measure the total flux of the source but not constrain its morphology.
The nucleus of NGC\,3690E is possibly extended (FWHM$\sim 1\arcsec$, 215\,pc) but this remains to be verified with another epoch.
The Michelle fluxes of both nuclei marginally agree with the \spitzerr data, although the PAH emission seems to be weaker at the high angular resolution.
Note that the nuclear fluxes would be significantly lower if the presence of subarcsecond-extended emission can be verified.
\newline\end{@twocolumnfalse}]

\begin{figure}
   \centering
   \includegraphics[angle=0,width=8.500cm]{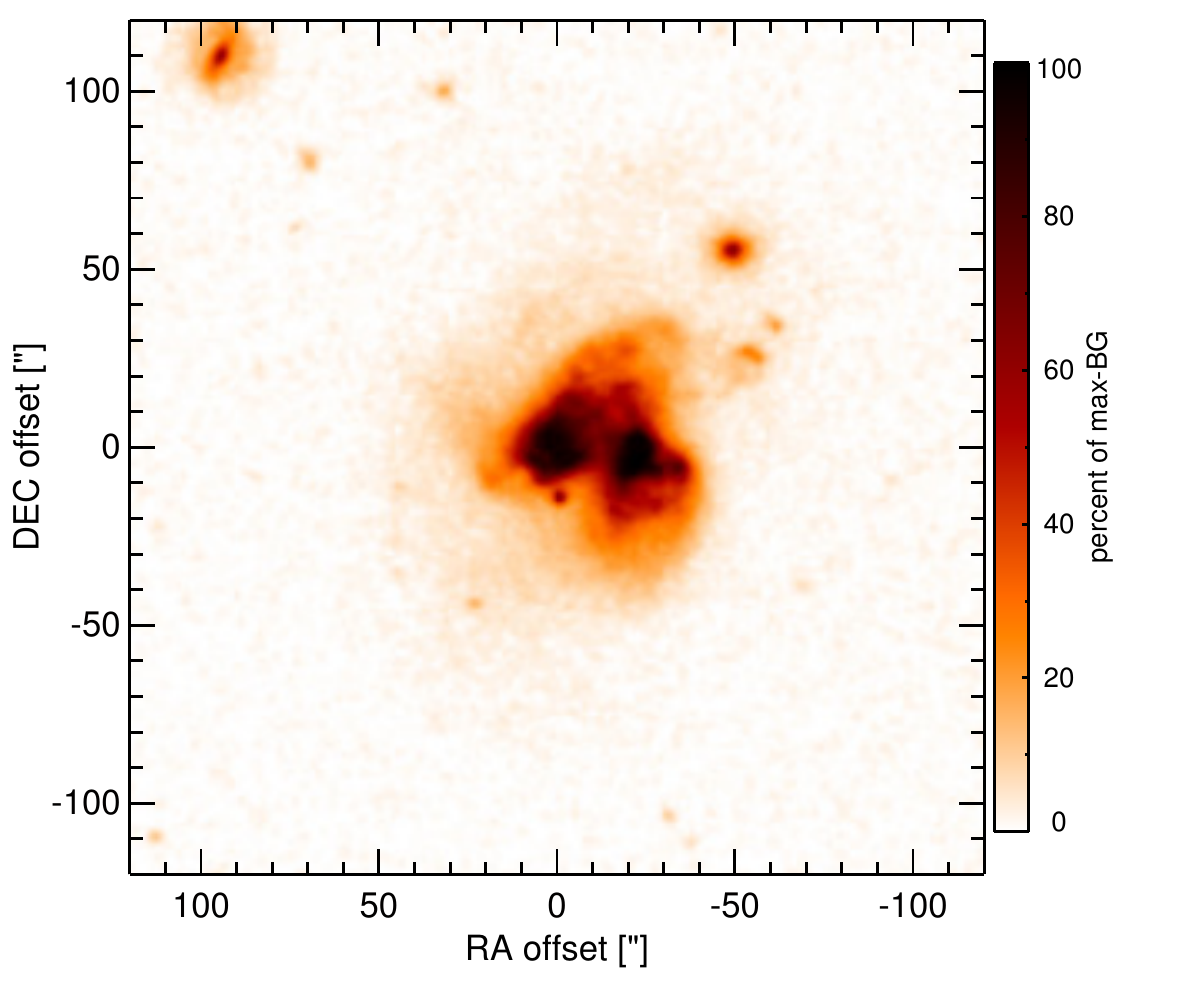}
    \caption{\label{fig:OPTim_NGC3690E}
             Optical image (DSS, red filter) of NGC\,3690E. Displayed are the central $4\arcmin$ with North up and East to the left. 
              The colour scaling is linear with white corresponding to the median background and black to the $0.01\%$ pixels with the highest intensity.  
           }
\end{figure}
\begin{figure}
   \centering
   \includegraphics[angle=0,height=3.11cm]{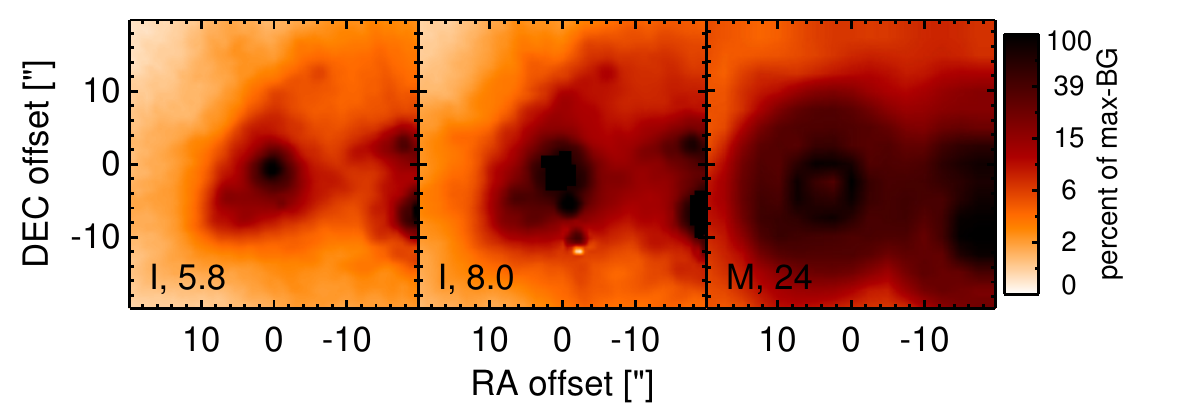}
    \caption{\label{fig:INTim_NGC3690E}
             \spitzerr MIR images of NGC\,3690E. Displayed are the inner $40\arcsec$ with North up and East to the left. The colour scaling is logarithmic with white corresponding to median background and black to the $0.1\%$ pixels with the highest intensity.
             The label in the bottom left states instrument and central wavelength of the filter in $\mu$m (I: IRAC, M: MIPS). 
             Note that the apparent off-nuclear compact sources in the IRAC $8.0\,\mu$m image are instrumental artefacts.
           }
\end{figure}
\begin{figure}
   \centering
   \includegraphics[angle=0,height=3.11cm]{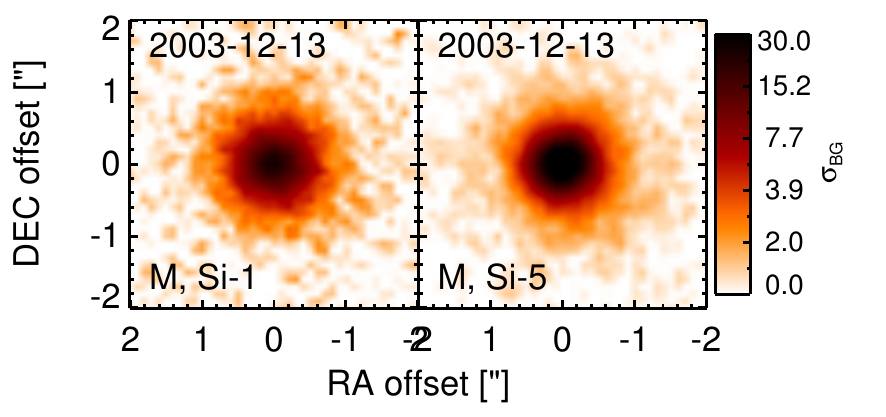}
    \caption{\label{fig:HARim_NGC3690E}
             Subarcsecond-resolution MIR images of NGC\,3690E sorted by increasing filter wavelength. 
             Displayed are the inner $4\arcsec$ with North up and East to the left. 
             The colour scaling is logarithmic with white corresponding to median background and black to the $75\%$ of the highest intensity of all images in units of $\sigbg$.
             The labels in the bottom left state instrument and filter names (C: COMICS, M: Michelle, T: T-ReCS, V: VISIR).
           }
\end{figure}
\begin{figure}
   \centering
   \includegraphics[angle=0,width=8.50cm]{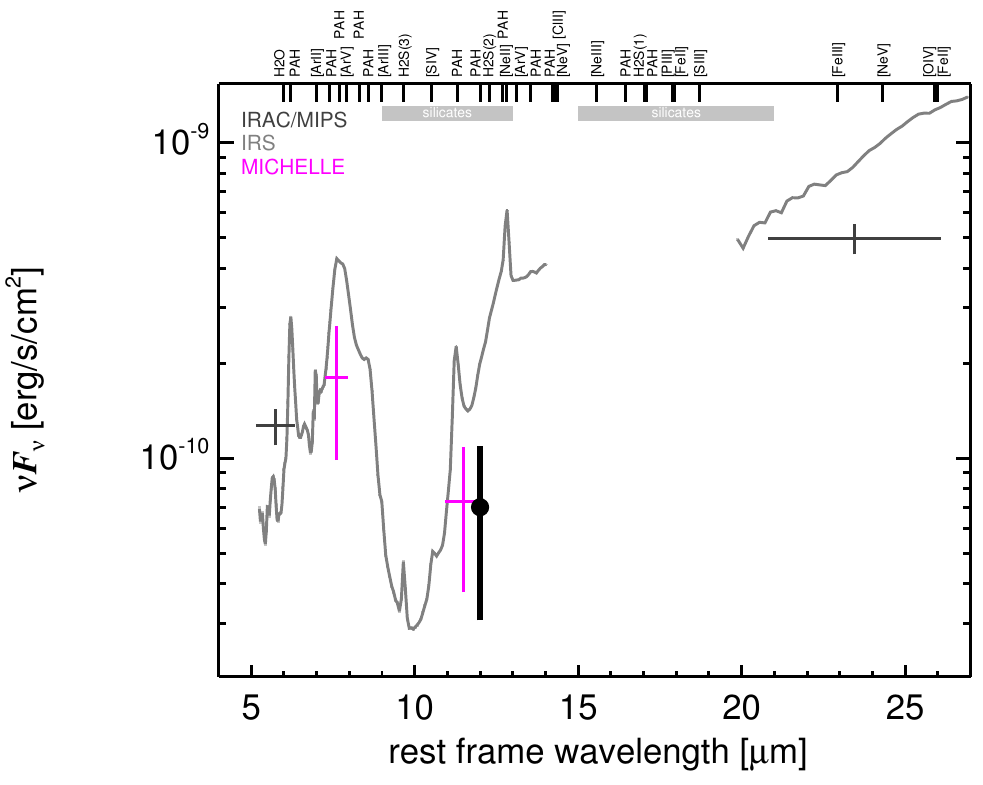}
   \caption{\label{fig:MISED_NGC3690E}
      MIR SED of NGC\,3690E. The description  of the symbols (if present) is the following.
      Grey crosses and  solid lines mark the \spitzer/IRAC, MIPS and IRS data. 
      The colour coding of the other symbols is: 
      green for COMICS, magenta for Michelle, blue for T-ReCS and red for VISIR data.
      Darker-coloured solid lines mark spectra of the corresponding instrument.
      The black filled circles mark the nuclear 12 and $18\,\mu$m  continuum emission estimate from the data.
      The ticks on the top axis mark positions of common MIR emission lines, while the light grey horizontal bars mark wavelength ranges affected by the silicate 10 and 18$\mu$m features.}
\end{figure}
\clearpage

\begin{figure}
   \centering
   \includegraphics[angle=0,height=3.11cm]{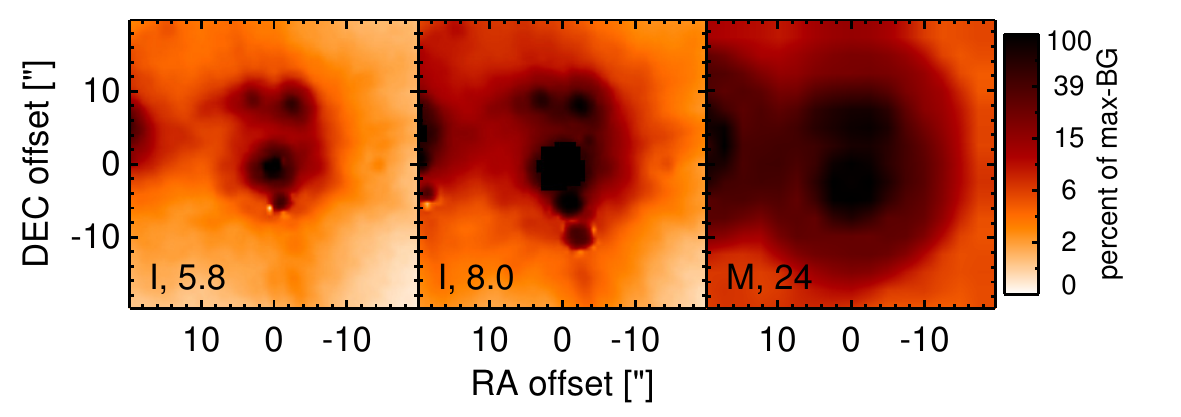}
    \caption{\label{fig:INTim_NGC3690W}
             \spitzerr MIR images of NGC\,3690W. Displayed are the inner $40\arcsec$ with North up and East to the left. The colour scaling is logarithmic with white corresponding to median background and black to the $0.1\%$ pixels with the highest intensity.
             The label in the bottom left states instrument and central wavelength of the filter in $\mu$m (I: IRAC, M: MIPS). 
             Note that the apparent off-nuclear compact sources in the IRAC $8.0\,\mu$m image are instrumental artefacts.
           }
\end{figure}
\begin{figure}
   \centering
   \includegraphics[angle=0,height=3.11cm]{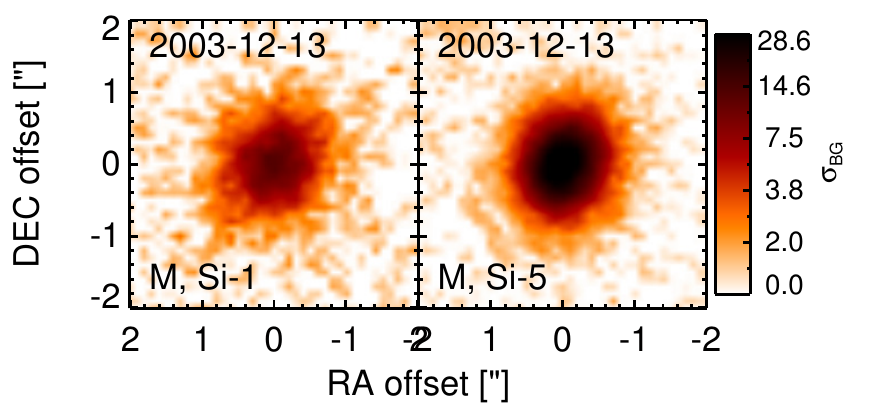}
    \caption{\label{fig:HARim_NGC3690W}
             Subarcsecond-resolution MIR images of NGC\,3690W sorted by increasing filter wavelength. 
             Displayed are the inner $4\arcsec$ with North up and East to the left. 
             The colour scaling is logarithmic with white corresponding to median background and black to the $75\%$ of the highest intensity of all images in units of $\sigbg$.
             The labels in the bottom left state instrument and filter names (C: COMICS, M: Michelle, T: T-ReCS, V: VISIR).
           }
\end{figure}
\begin{figure}
   \centering
   \includegraphics[angle=0,width=8.50cm]{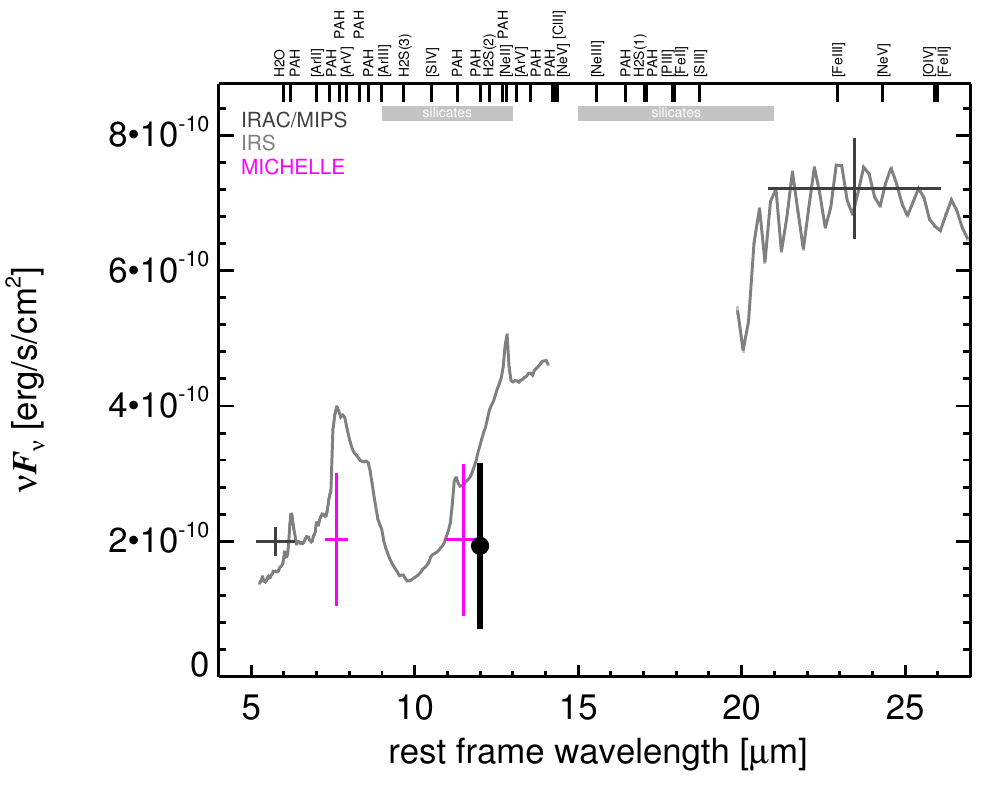}
   \caption{\label{fig:MISED_NGC3690W}
      MIR SED of NGC\,3690W. The description  of the symbols (if present) is the following.
      Grey crosses and  solid lines mark the \spitzer/IRAC, MIPS and IRS data. 
      The colour coding of the other symbols is: 
      green for COMICS, magenta for Michelle, blue for T-ReCS and red for VISIR data.
      Darker-coloured solid lines mark spectra of the corresponding instrument.
      The black filled circles mark the nuclear 12 and $18\,\mu$m  continuum emission estimate from the data.
      The ticks on the top axis mark positions of common MIR emission lines, while the light grey horizontal bars mark wavelength ranges affected by the silicate 10 and 18$\mu$m features.}
\end{figure}
\clearpage

\twocolumn[\begin{@twocolumnfalse}  
\subsection{NGC\,3718 -- Arp\,214}\label{app:NGC3718}
NGC\,3718 is a peculiar early-type spiral galaxy at a distance of $D=$ $17.0 \pm 3.4\,$Mpc \citep{tully_nearby_1988} with a broad-line LINER nucleus \citep{veron-cetty_catalogue_2010}, covered by a prominent dust lane, which completely absorbs the nuclear UV signature \citep{barth_search_1998}.
A compact radio source with a possible jet extending $0.5\arcsec\sim40\,$pc to the north-west was detected \citep{nagar_radio_2002,krips_nuclei_2007}.
Also only very weak extended NLR emission was detected \citep{pogge_narrow-line_2000}.
The first MIR observations of NGC\,3718 were performed with IRTF but the nucleus was not clearly detected \citep{lonsdale_infrared_1984,willner_infrared_1985,devereux_infrared_1987}.
The are no cryogenic \spitzerr observations available for NGC\,3718.
In the \textit{WISE} band~3 image, it appears a compact nuclear source embedded within elliptical host emission turning into an S-shaped bar at larger distances, resembling the dust lane. 
The nuclear region of NGC\,3718 was imaged with Michelle in the N' filter in 2008 \citep{mason_nuclear_2012}, and a compact MIR nucleus was detected.
The nucleus is possibly marginally resolved (FWHM $\sim 0.47\arcsec \sim 39\,$pc) but this is insufficient for a robust classification of the nuclear extension in the MIR at subarcsecond scales.
Our nuclear photometry is consistent with \cite{mason_nuclear_2012}.
\newline\end{@twocolumnfalse}]

\begin{figure}
   \centering
   \includegraphics[angle=0,width=8.500cm]{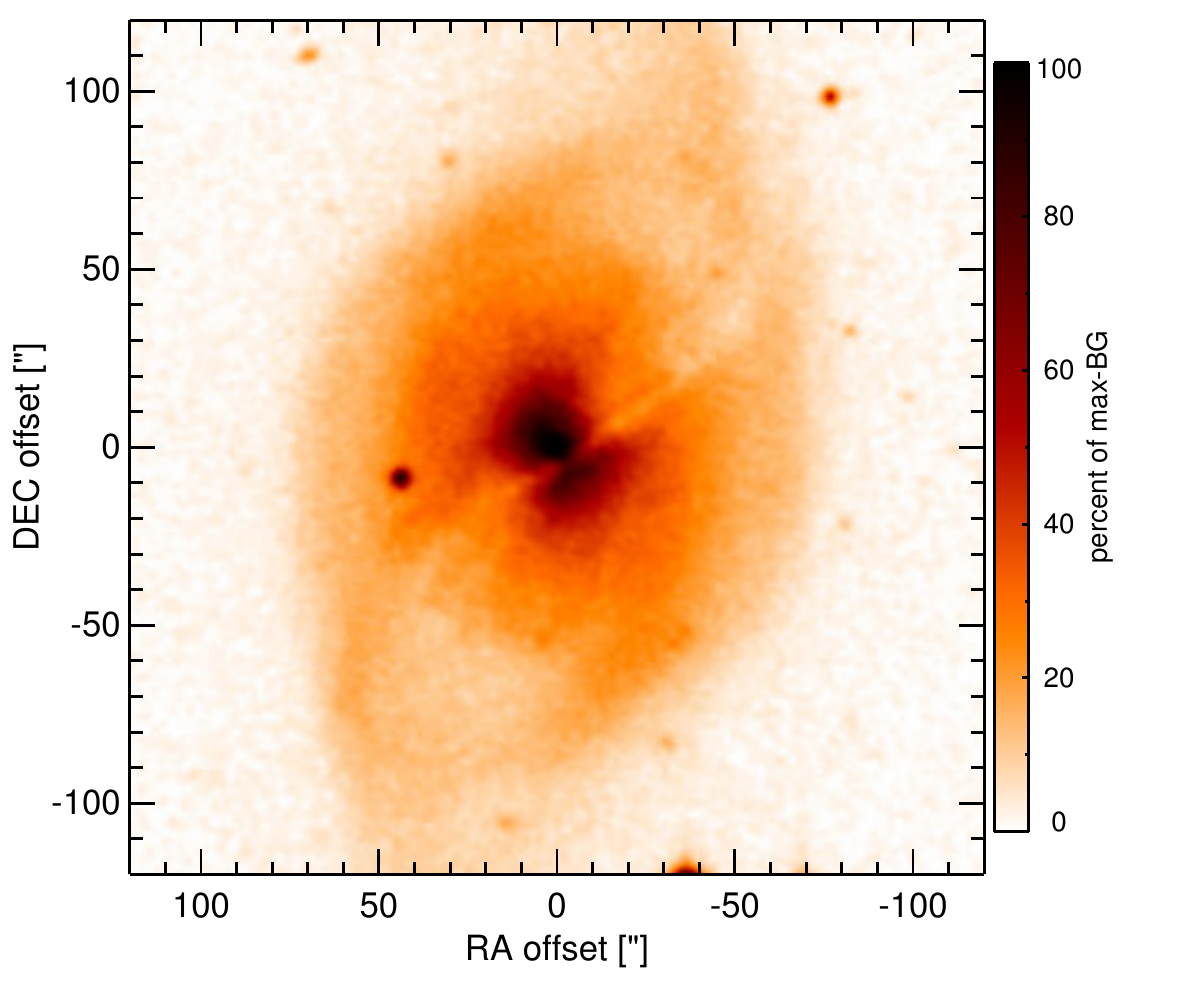}
    \caption{\label{fig:OPTim_NGC3718}
             Optical image (DSS, red filter) of NGC\,3718. Displayed are the central $4\arcmin$ with North up and East to the left. 
              The colour scaling is linear with white corresponding to the median background and black to the $0.01\%$ pixels with the highest intensity.  
           }
\end{figure}
\begin{figure}
   \centering
   \includegraphics[angle=0,height=3.11cm]{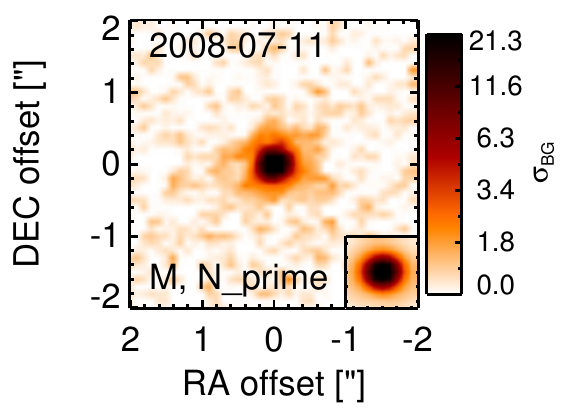}
    \caption{\label{fig:HARim_NGC3718}
             Subarcsecond-resolution MIR images of NGC\,3718 sorted by increasing filter wavelength. 
             Displayed are the inner $4\arcsec$ with North up and East to the left. 
             The colour scaling is logarithmic with white corresponding to median background and black to the $75\%$ of the highest intensity of all images in units of $\sigbg$.
             The inset image shows the central arcsecond of the PSF from the calibrator star, scaled to match the science target.
             The labels in the bottom left state instrument and filter names (C: COMICS, M: Michelle, T: T-ReCS, V: VISIR).
           }
\end{figure}
\begin{figure}
   \centering
   \includegraphics[angle=0,width=8.50cm]{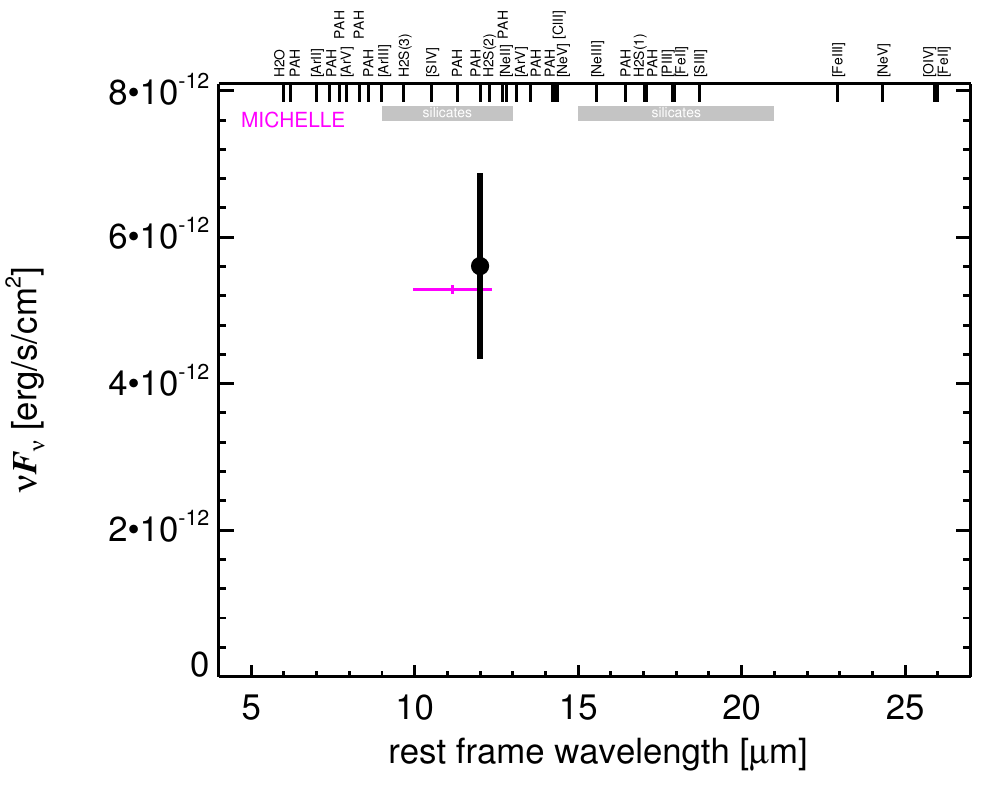}
   \caption{\label{fig:MISED_NGC3718}
      MIR SED of NGC\,3718. The description  of the symbols (if present) is the following.
      Grey crosses and  solid lines mark the \spitzer/IRAC, MIPS and IRS data. 
      The colour coding of the other symbols is: 
      green for COMICS, magenta for Michelle, blue for T-ReCS and red for VISIR data.
      Darker-coloured solid lines mark spectra of the corresponding instrument.
      The black filled circles mark the nuclear 12 and $18\,\mu$m  continuum emission estimate from the data.
      The ticks on the top axis mark positions of common MIR emission lines, while the light grey horizontal bars mark wavelength ranges affected by the silicate 10 and 18$\mu$m features.}
\end{figure}
\clearpage

\twocolumn[\begin{@twocolumnfalse}  
\subsection{NGC\,3783}\label{app:NGC3783}
NGC\,3783 is a barred face-on spiral galaxy at a redshift of $z=$ 0.0097 ($D\sim48.4$\,Mpc) with a well-studied variable Sy\,1.5 nucleus \citep{veron-cetty_catalogue_2010}, which also belongs to the nine-month BAT AGN sample.
The nucleus is unresolved at radio wavelengths \citep{morganti_radio_1999,kinney_jet_2000}, embedded within a halo-like extended NLR with 1.9\arcsec ($\sim 0.4\,$kpc) size \citep{schmitt_hubble_2003}.
The first MIR observations of NGC\,3873 were carried out by \cite{kleinmann_10-micron_1974}, followed by \cite{rieke_infrared_1978} \cite{frogel_8-13_1982}, \cite{glass_mid-infrared_1982}, \cite{devereux_spatial_1987}, and \cite{roche_atlas_1991}.
The first subarcsecond $N$-band image of NGC\,3783 was taken in 2001 with 3.6\,m/TIMMI2 \citep{raban_core_2008} where it appeared as point source.
The \spitzer/IRAC and MIPS images reveal  faint spiral-like host emission apart from the compact nucleus.
Note that the IRAC $8\,\mu$m PBCD image is saturated in the nucleus and thus not analysed.
The \spitzer/IRS LR staring-mode spectrum exhibits silicate  $10\,\mu$m and possibly silicate 18$\,\mu$m emission and an emission peak at $\sim20\,\mu$m in $\nu F_\nu$-space but no PAH features (see also \citealt{shi_9.7_2006,mullaney_defining_2011}).
The nuclear region of NGC\,3783 was extensively imaged with VISIR with imaging in six different narrow $N$-band filters in total and two $Q$ band filters between 2005 and 2011 (part of which published in \citealt{horst_small_2006,horst_mid-infrared_2009,haas_visir_2007,honig_dusty_2010-1,reunanen_vlt_2010,kishimoto_mapping_2011}).
In addition, a VISIR LR $N$-band spectrum was obtained \citep{honig_dusty_2010-1}.
A compact MIR nucleus without other host emission was detected in all cases.
In the sharpest image with the highest S/N taken in the PAH1 filter the nucleus is unresolved, while in the best PAH2\_2 image, it is possibly marginally resolved (FWHM $\sim 0.34\arcsec \sim 80\,$pc). 
This is also the case in the $Q$-band images. 
Therefore, NGC\,3783 might be marginally extended at longer MIR wavelengths, and we classify its extension as in general uncertain at subarcsecond resolution.
Our reanalysis of the already published images provides in general consistent value with the previous publications, except for \cite{reunanen_vlt_2010}, which have stated a $\sim 20\%$ lower PAH2\_2 flux value.
The multiple measurements in the PAH1 and PAH2\_2 filters indicate possible flux variations on the order of $20\%$ over six years, which will be analysed elsewhere (H\"onig et al., in prep.).
Similarly strong variations also appear in the historical data obtained in the 1970-1980s. 
In general, the VISIR spectrophotometry is consistent with the \spitzerr spectrophotometry, demonstrating that the AGN completely dominates the MIR emission of NGC\,3783 also on arcsecond scales ($\sim 1\,$kpc).
No significant star formation seems to be present in this object.
Note that the nucleus of NGC\,3783 has also been studied extensively with MIR interferometry using MIDI, in which the nuclear structure is resolved with an polar-elongated size of a few parsec   \citep{honig_dust_2013,beckert_probing_2008}.
\newline\end{@twocolumnfalse}]

\begin{figure}
   \centering
   \includegraphics[angle=0,width=8.500cm]{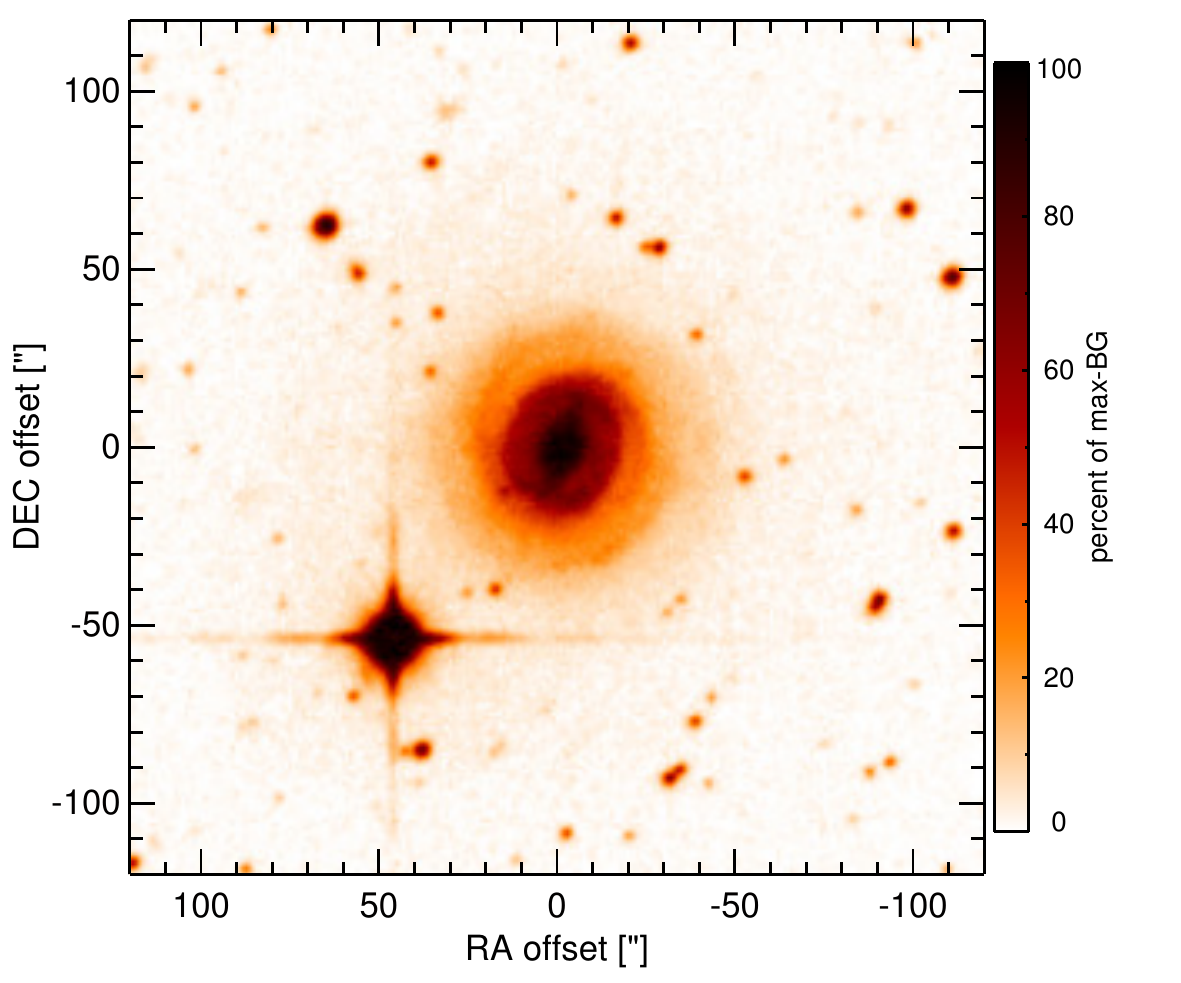}
    \caption{\label{fig:OPTim_NGC3783}
             Optical image (DSS, red filter) of NGC\,3783. Displayed are the central $4\arcmin$ with North up and East to the left. 
              The colour scaling is linear with white corresponding to the median background and black to the $0.01\%$ pixels with the highest intensity.  
           }
\end{figure}
\begin{figure}
   \centering
   \includegraphics[angle=0,height=3.11cm]{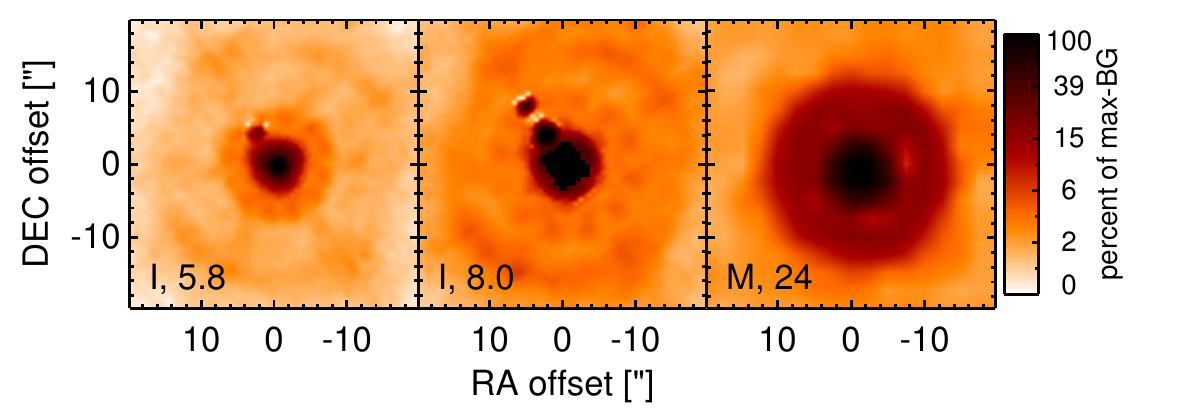}
    \caption{\label{fig:INTim_NGC3783}
             \spitzerr MIR images of NGC\,3783. Displayed are the inner $40\arcsec$ with North up and East to the left. The colour scaling is logarithmic with white corresponding to median background and black to the $0.1\%$ pixels with the highest intensity.
             The label in the bottom left states instrument and central wavelength of the filter in $\mu$m (I: IRAC, M: MIPS). 
             Note that the apparent off-nuclear compact sources in the IRAC 5.8 and $8.0\,\mu$m images are instrumental artefacts.
           }
\end{figure}
\begin{figure}
   \centering
   \includegraphics[angle=0,width=8.500cm]{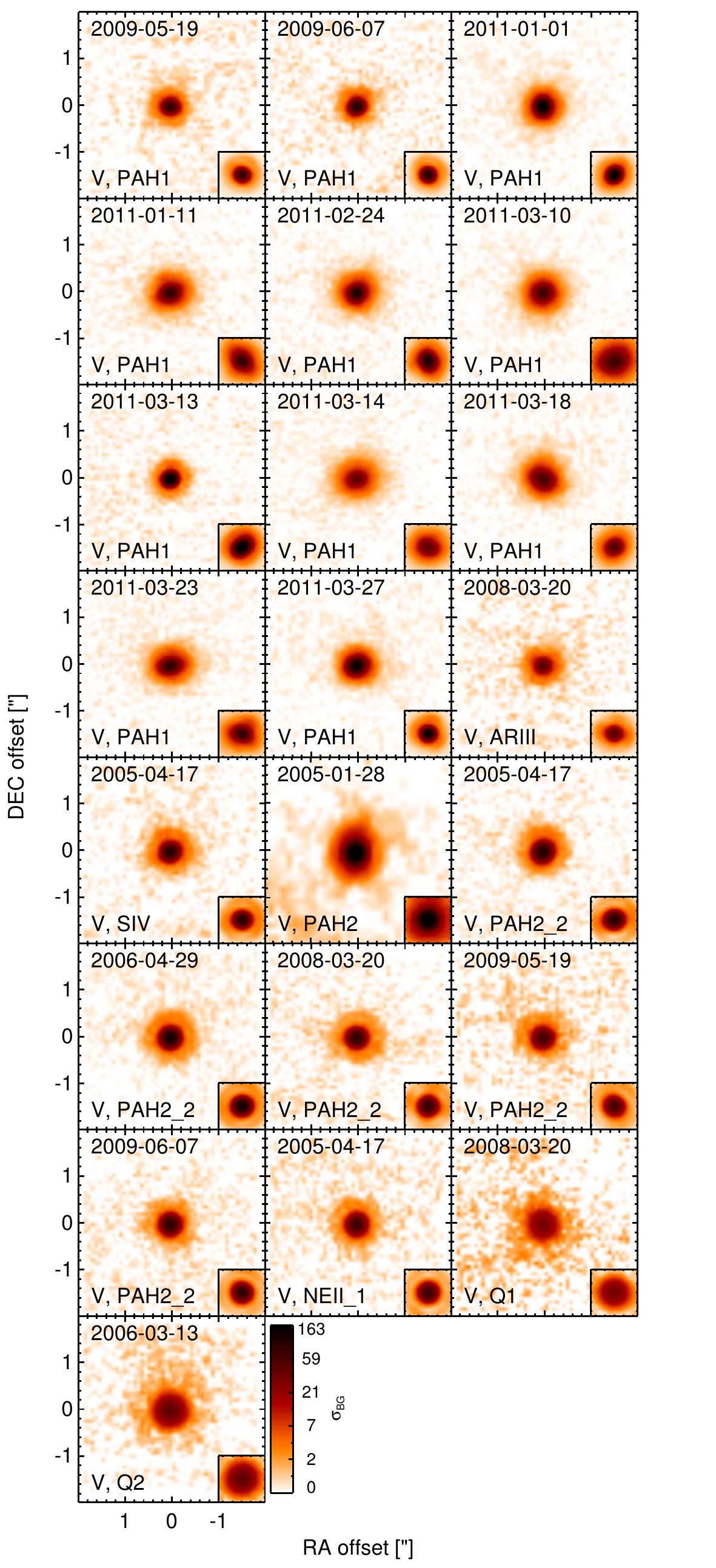}
    \caption{\label{fig:HARim_NGC3783}
             Subarcsecond-resolution MIR images of NGC\,3783 sorted by increasing filter wavelength. 
             Displayed are the inner $4\arcsec$ with North up and East to the left. 
             The colour scaling is logarithmic with white corresponding to median background and black to the $75\%$ of the highest intensity of all images in units of $\sigbg$.
             The inset image shows the central arcsecond of the PSF from the calibrator star, scaled to match the science target.
             The labels in the bottom left state instrument and filter names (C: COMICS, M: Michelle, T: T-ReCS, V: VISIR).
           }
\end{figure}
\begin{figure}
   \centering
   \includegraphics[angle=0,width=8.50cm]{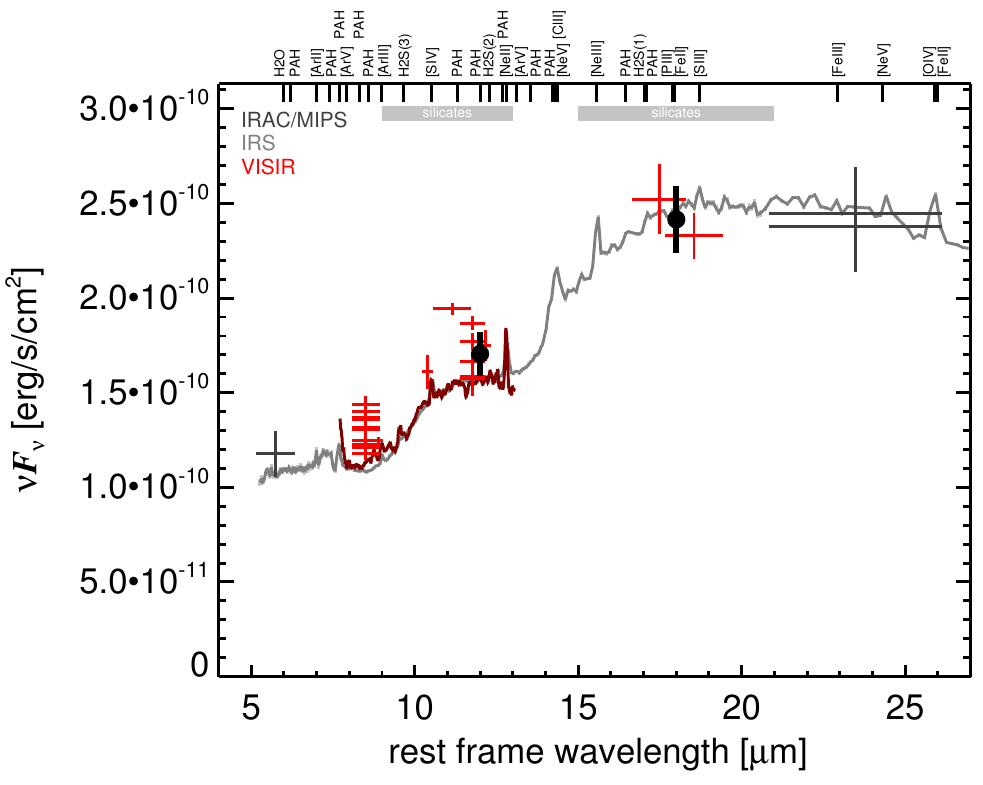}
   \caption{\label{fig:MISED_NGC3783}
      MIR SED of NGC\,3783. The description  of the symbols (if present) is the following.
      Grey crosses and  solid lines mark the \spitzer/IRAC, MIPS and IRS data. 
      The colour coding of the other symbols is: 
      green for COMICS, magenta for Michelle, blue for T-ReCS and red for VISIR data.
      Darker-coloured solid lines mark spectra of the corresponding instrument.
      The black filled circles mark the nuclear 12 and $18\,\mu$m  continuum emission estimate from the data.
      The ticks on the top axis mark positions of common MIR emission lines, while the light grey horizontal bars mark wavelength ranges affected by the silicate 10 and 18$\mu$m features.}
\end{figure}
\clearpage

\twocolumn[\begin{@twocolumnfalse}  
\subsection{NGC\,3982}\label{app:NGC3982}
NGC\,3982 is a face-on barred spiral galaxy at a distance of $D=$ $21.4 \pm 2.2\,$Mpc (NED redshift-independent median) with an AGN optically classified either as Sy\,1.9  or Sy\,2 \citep{trippe_multi-wavelength_2010}.
It features an unresolved NLR \citep{gonzalez_delgado_h_1997} and unresolved radio core \citep{ho_radio_2001}.
The first ground-based MIR observations of NGC\,3982 were carried out after \irass but the nucleus was not detected \citep{edelson_broad-band_1987,devereux_spatial_1987}. 
\cite{maiolino_new_1995} claim the first weak detection, while it again was undetected in the first subarcsecond MIR imaging with Palomar 5\,m/MIRLIN \citep{gorjian_10_2004}.
At low angular resolution, NGC\,3982 has been studied with \isoo (e.g., \citealt{ramos_almeida_mid-infrared_2007}) and \spitzer/IRAC and IRS.
The corresponding IRAC images show a compact nucleus embedded within relatively bright spiral-like host emission.
Therefore, our  IRAC $5.8$ and $8.0\,\mu$m fluxes of only the nuclear four-arcsecond region are much lower than the values of \cite{gallimore_infrared_2010}.
The nuclear IRS LR mapping-mode spectrum suffers from a low S/N but indicates  PAH emission and a red spectral slope in $\nu F_\nu$-space.
Note that at least part of the circum-nuclear star formation  is cancelled through our background subtraction (see also \citealt{buchanan_spitzer_2006,wu_spitzer/irs_2009,tommasin_spitzer-irs_2010,trippe_multi-wavelength_2010,gallimore_infrared_2010}).
We observed the nuclear region of NGC\,3982 with Michelle in two $N$-band filters in 2010 and detected a possibly marginally resolved nucleus without any further host emission (FWHM $\sim 0.48\arcsec \sim 50\,$pc).
However, the data are not sufficient for a robust classification of the nuclear extension in the MIR at subarcsecond resolution.
Our nuclear photometry is on average $\sim 36\%$ lower than the \spitzerr spectrophotometry verifying that the MIR emission of the arcsecond scale region is star formation contaminated.
Finally, these results are consistent with the previous MIR observations.
\newline\end{@twocolumnfalse}]

\begin{figure}
   \centering
   \includegraphics[angle=0,width=8.500cm]{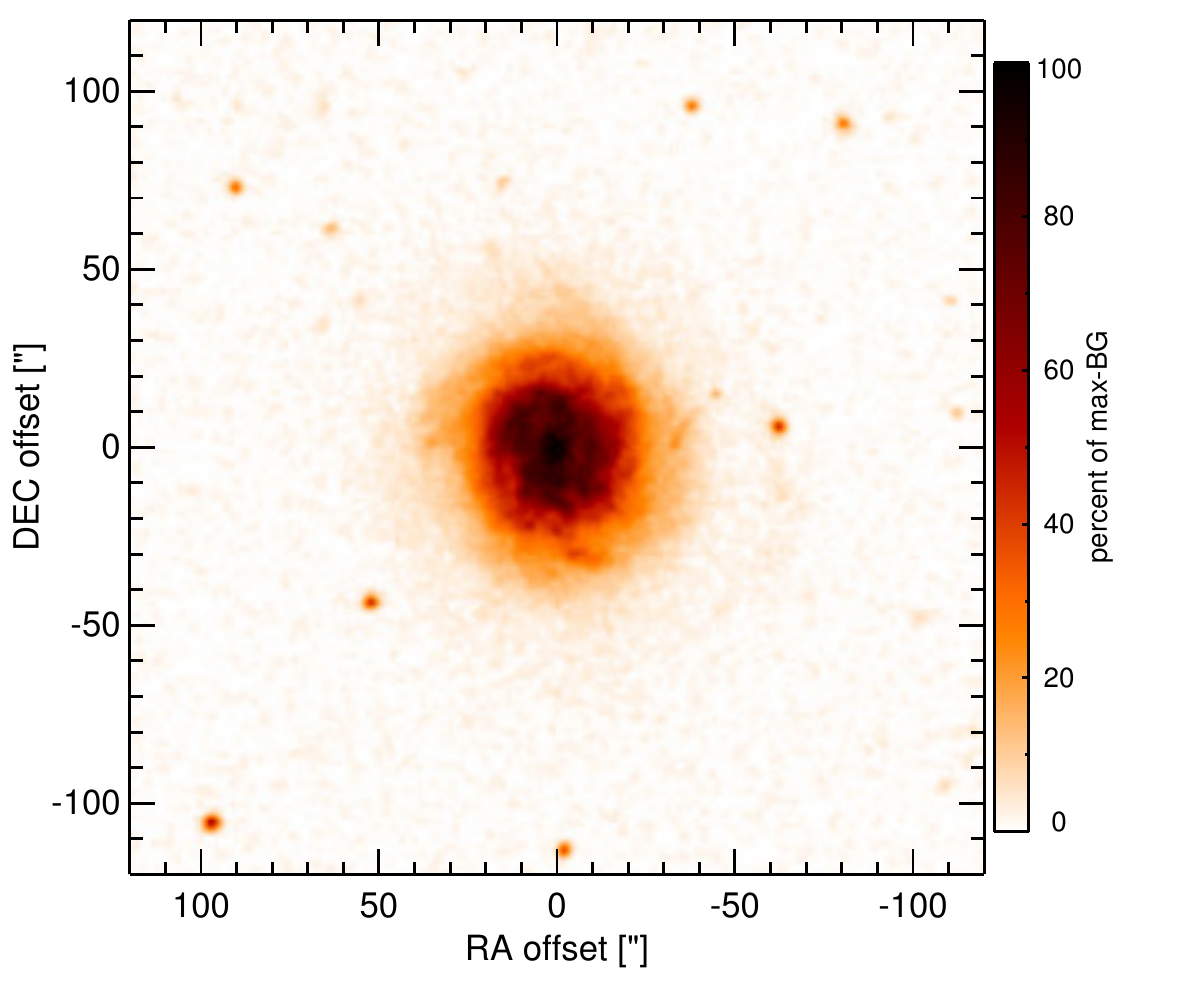}
    \caption{\label{fig:OPTim_NGC3982}
             Optical image (DSS, red filter) of NGC\,3982. Displayed are the central $4\arcmin$ with North up and East to the left. 
              The colour scaling is linear with white corresponding to the median background and black to the $0.01\%$ pixels with the highest intensity.  
           }
\end{figure}
\begin{figure}
   \centering
   \includegraphics[angle=0,height=3.11cm]{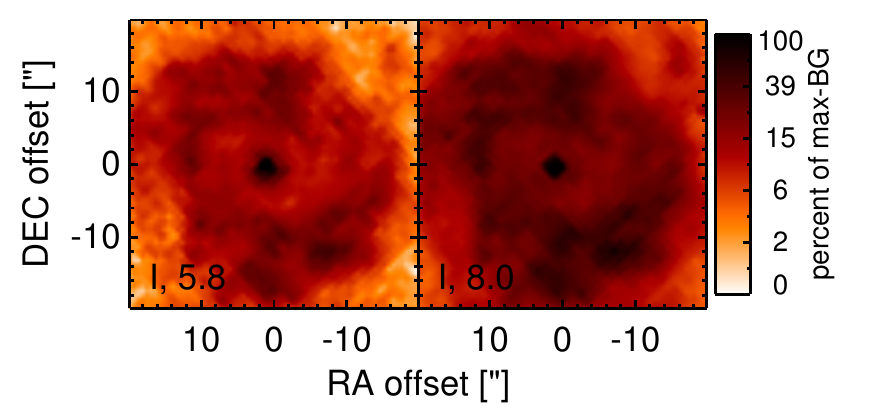}
    \caption{\label{fig:INTim_NGC3982}
             \spitzerr MIR images of NGC\,3982. Displayed are the inner $40\arcsec$ with North up and East to the left. The colour scaling is logarithmic with white corresponding to median background and black to the $0.1\%$ pixels with the highest intensity.
             The label in the bottom left states instrument and central wavelength of the filter in $\mu$m (I: IRAC, M: MIPS). 
           }
\end{figure}
\begin{figure}
   \centering
   \includegraphics[angle=0,height=3.11cm]{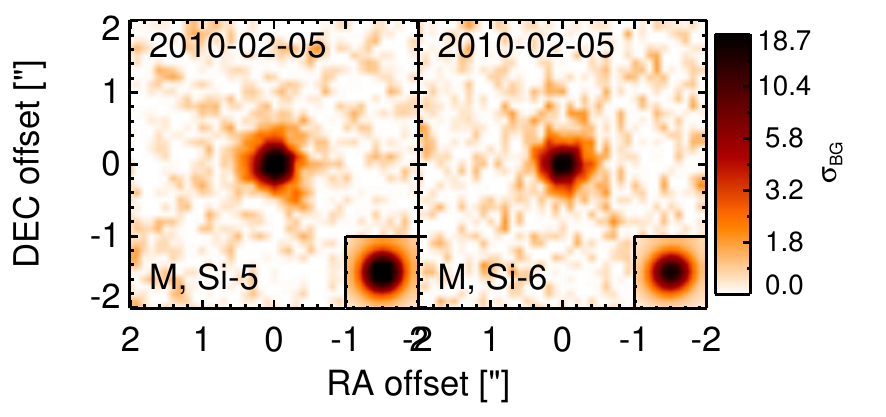}
    \caption{\label{fig:HARim_NGC3982}
             Subarcsecond-resolution MIR images of NGC\,3982 sorted by increasing filter wavelength. 
             Displayed are the inner $4\arcsec$ with North up and East to the left. 
             The colour scaling is logarithmic with white corresponding to median background and black to the $75\%$ of the highest intensity of all images in units of $\sigbg$.
             The inset image shows the central arcsecond of the PSF from the calibrator star, scaled to match the science target.
             The labels in the bottom left state instrument and filter names (C: COMICS, M: Michelle, T: T-ReCS, V: VISIR).
           }
\end{figure}
\begin{figure}
   \centering
   \includegraphics[angle=0,width=8.50cm]{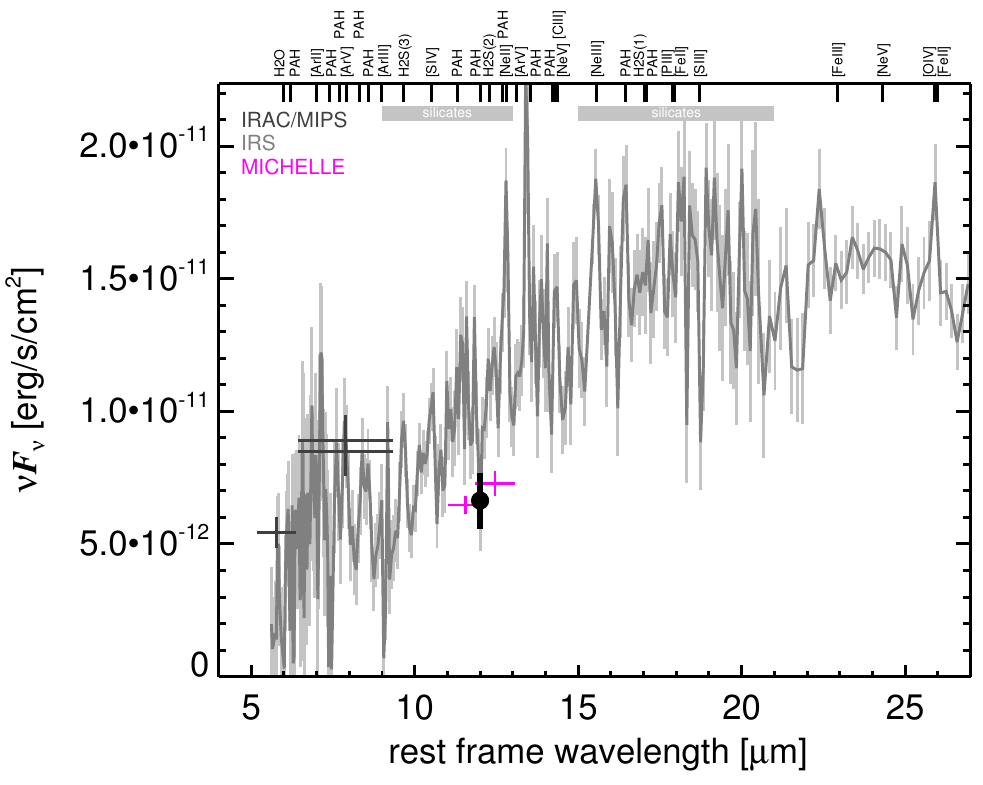}
   \caption{\label{fig:MISED_NGC3982}
      MIR SED of NGC\,3982. The description  of the symbols (if present) is the following.
      Grey crosses and  solid lines mark the \spitzer/IRAC, MIPS and IRS data. 
      The colour coding of the other symbols is: 
      green for COMICS, magenta for Michelle, blue for T-ReCS and red for VISIR data.
      Darker-coloured solid lines mark spectra of the corresponding instrument.
      The black filled circles mark the nuclear 12 and $18\,\mu$m  continuum emission estimate from the data.
      The ticks on the top axis mark positions of common MIR emission lines, while the light grey horizontal bars mark wavelength ranges affected by the silicate 10 and 18$\mu$m features.}
\end{figure}
\clearpage

\twocolumn[\begin{@twocolumnfalse}  
\subsection{NGC\,3998}\label{app:NGC3998}
NGC\,3998 is an early-type spiral galaxy at a distance of $D=$ $14.1 \pm 1.3\,$Mpc \citep{tonry_sbf_2001} with a variable radio-loud broad-line LINER nucleus \citep{veron-cetty_catalogue_2010}.
It features a compact flat-spectrum radio core with a jet-like northern extension (e.g., \citealt{filho_light-year_2002}), and a disc-like NLR with a major axis of $\sim 3\arcsec \sim 200\,$pc and a PA$\sim 90\degree$ \citep{pogge_narrow-line_2000}. 
The first ground-based MIR observations of NGC\,3998 were carried out with IRTF and UKIRT in 1983 and 1984 \citep{sparks_infrared_1986, willner_infrared_1985}.
In addition, it was observed with \isoo \citep{knapp_iso_1996} and \spitzer/IRAC, IRS and MIPS.
The corresponding images show a dominating nucleus embedded within elliptical emission.
The LR IRS spectrum exhibits strong silicate 10 and $18\,\mu$m emission features and a shallow blue spectral slope in $\nu F_\nu$-space but no PAH features (see also \citealt{sturm_silicate_2005,mason_nuclear_2012}).
The nuclear region of NGC\,3998 was observed with COMICS in two $N$-band filters in 2005 (unpublished, to our knowledge), and with Michelle in three $N$ and one $Q$-band filters in 2006 and 2011 \citep{mason_nuclear_2012}.
A compact nucleus without host emission was detected in all images but the shape and extent of the source is inconsistent in the different images.
Therefore, we classify its subarcsecond MIR extension as uncertain . 
Our nuclear  photometry provides values generally consistent with \cite{mason_nuclear_2012}, except for the Qa flux, which is stated $\sim 25\%$ lower in the latter work.
In addition, the subarcsecond photometry is on average $\sim 36\%$ lower than the \spitzerr spectrophotometry.
The strength of the silicate 10\,$\mu$m emission seems to be comparable though, which indicates that it is caused in the projected central $\sim25\,$pc of NGC\,3998.
Therefore, we use only the Si-6 flux for the 12\,$\mu$m continuum emission estimate.
 \newline\end{@twocolumnfalse}]

\begin{figure}
   \centering
   \includegraphics[angle=0,width=8.500cm]{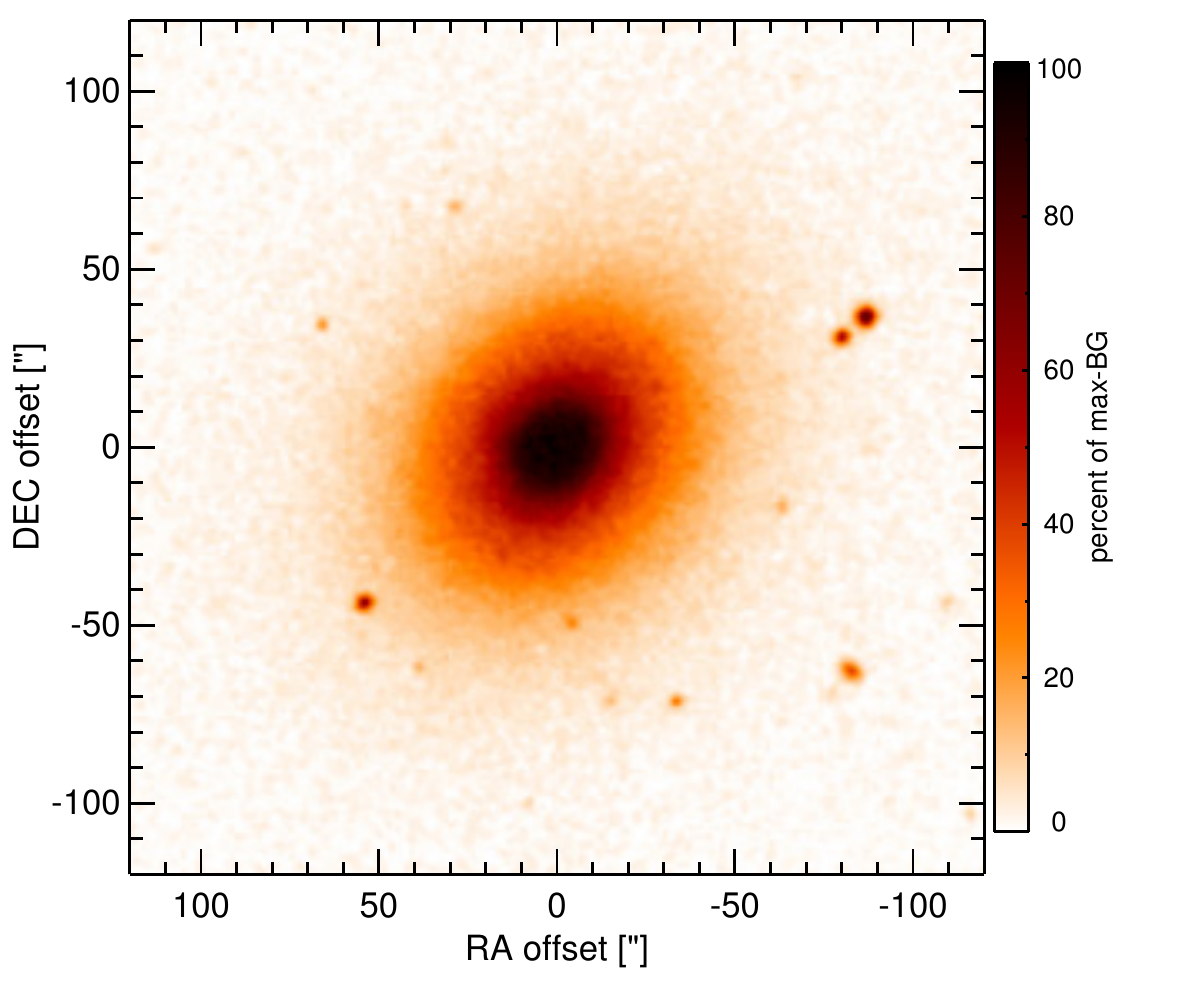}
    \caption{\label{fig:OPTim_NGC3998}
             Optical image (DSS, red filter) of NGC\,3998. Displayed are the central $4\arcmin$ with North up and East to the left. 
              The colour scaling is linear with white corresponding to the median background and black to the $0.01\%$ pixels with the highest intensity.  
           }
\end{figure}
\begin{figure}
   \centering
   \includegraphics[angle=0,height=3.11cm]{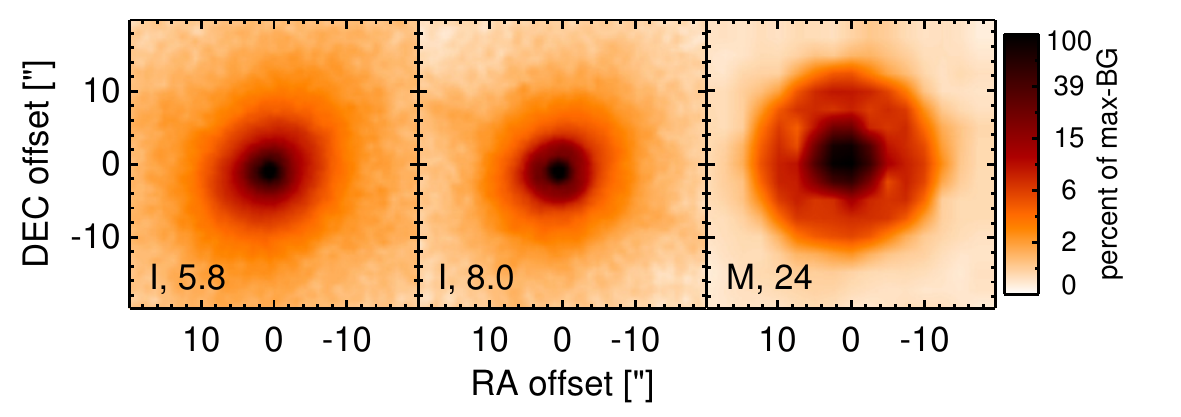}
    \caption{\label{fig:INTim_NGC3998}
             \spitzerr MIR images of NGC\,3998. Displayed are the inner $40\arcsec$ with North up and East to the left. The colour scaling is logarithmic with white corresponding to median background and black to the $0.1\%$ pixels with the highest intensity.
             The label in the bottom left states instrument and central wavelength of the filter in $\mu$m (I: IRAC, M: MIPS). 
           }
\end{figure}
\begin{figure}
   \centering
   \includegraphics[angle=0,width=8.500cm]{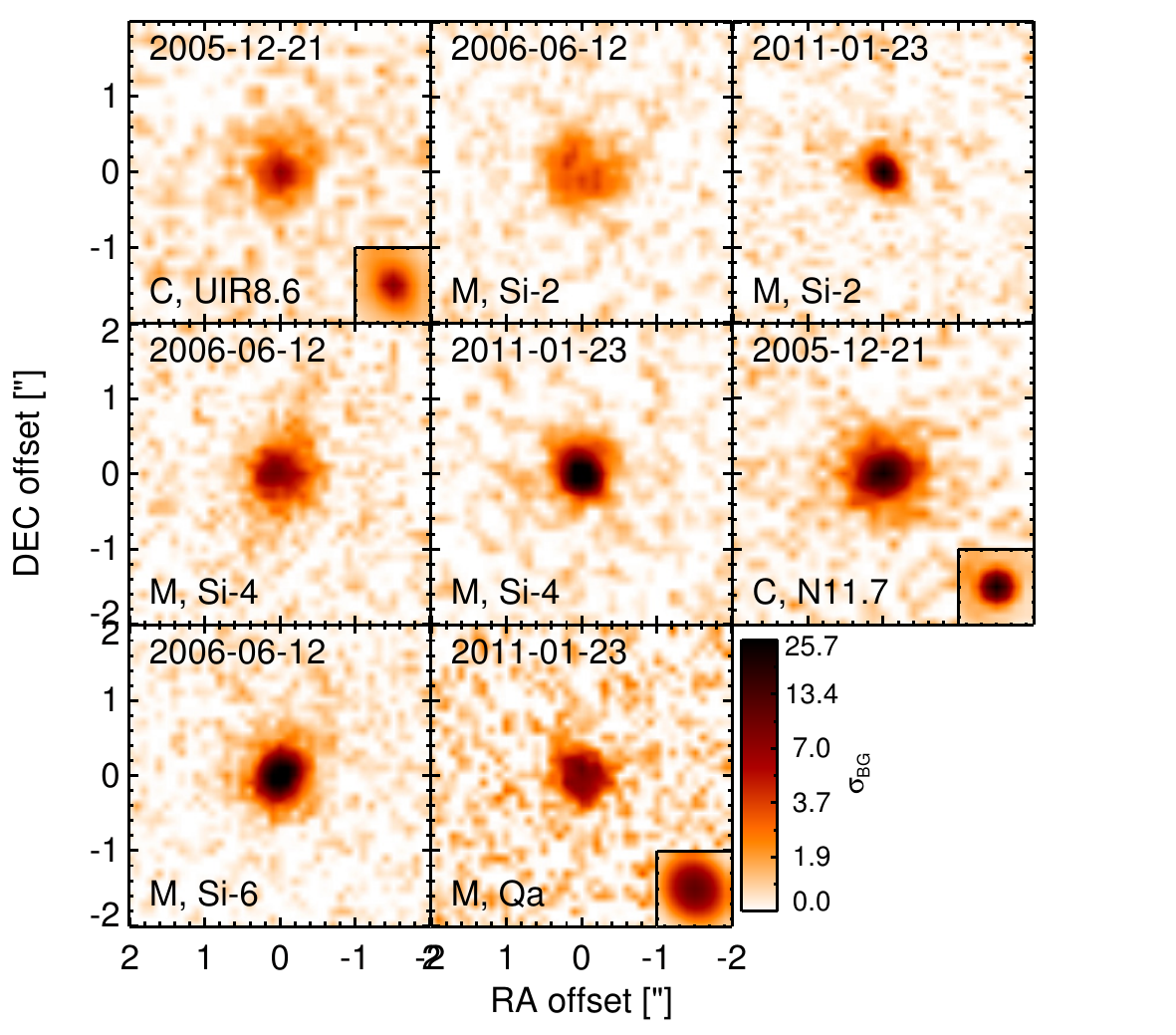}
    \caption{\label{fig:HARim_NGC3998}
             Subarcsecond-resolution MIR images of NGC\,3998 sorted by increasing filter wavelength. 
             Displayed are the inner $4\arcsec$ with North up and East to the left. 
             The colour scaling is logarithmic with white corresponding to median background and black to the $75\%$ of the highest intensity of all images in units of $\sigbg$.
             The inset image shows the central arcsecond of the PSF from the calibrator star, scaled to match the science target.
             The labels in the bottom left state instrument and filter names (C: COMICS, M: Michelle, T: T-ReCS, V: VISIR).
           }
\end{figure}
\begin{figure}
   \centering
   \includegraphics[angle=0,width=8.50cm]{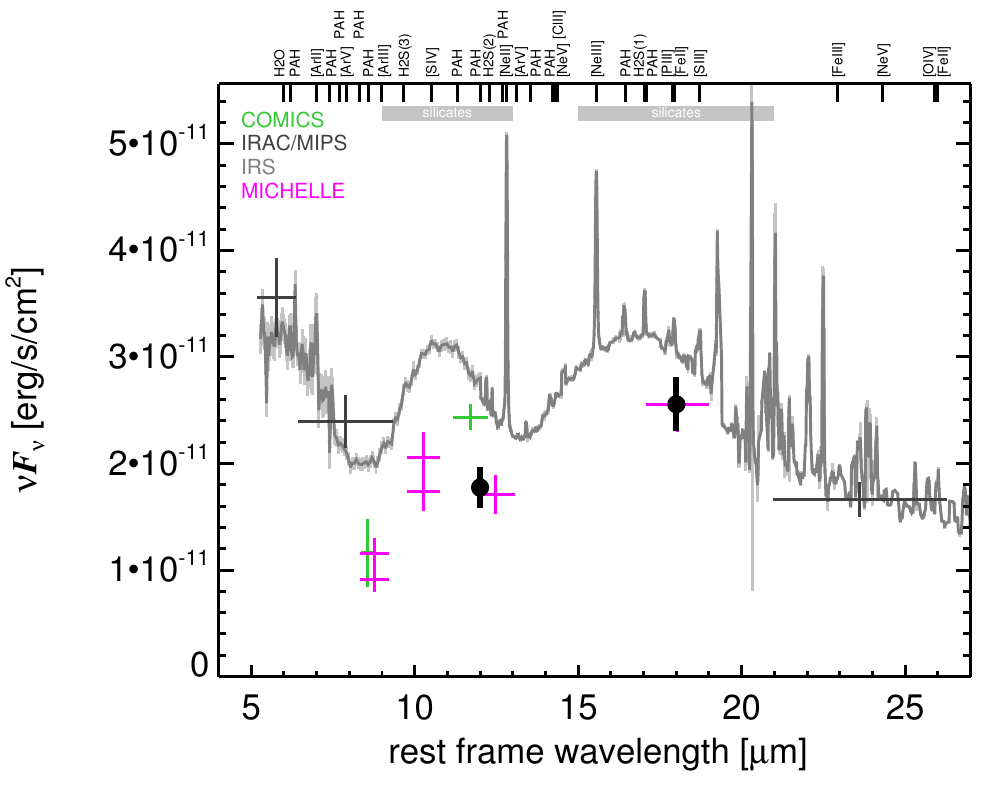}
   \caption{\label{fig:MISED_NGC3998}
      MIR SED of NGC\,3998. The description  of the symbols (if present) is the following.
      Grey crosses and  solid lines mark the \spitzer/IRAC, MIPS and IRS data. 
      The colour coding of the other symbols is: 
      green for COMICS, magenta for Michelle, blue for T-ReCS and red for VISIR data.
      Darker-coloured solid lines mark spectra of the corresponding instrument.
      The black filled circles mark the nuclear 12 and $18\,\mu$m  continuum emission estimate from the data.
      The ticks on the top axis mark positions of common MIR emission lines, while the light grey horizontal bars mark wavelength ranges affected by the silicate 10 and 18$\mu$m features.}
\end{figure}
\clearpage

\twocolumn[\begin{@twocolumnfalse}  
\subsection{NGC\,4051}\label{app:NGC4051}
NGC\,4051 is a low-inclination grand-design spiral galaxy at a distance of $D=$ $12.2\pm2.4\,$Mpc \citep{tully_extragalactic_2009} and belongs to the original six Seyfert galaxies \citep{seyfert_nuclear_1943}.
It contains a well-studied variable AGN (see \citealt{maitra_jet_2011} for a recent analysis if its SED), which is classified optically either as Sy\,1n \citep{veron-cetty_catalogue_2010} or Sy\,1.2 \citep{ho_search_1997-1}.
It is  a member of the nine-month BAT AGN sample.
The nucleus has a complex radio morphology with a double/triple source with $\sim0.4\arcsec$ separation in east-west direction \citep{ulvestad_radio_1984,kukula_high-resolution_1995} and a very weak central unresolved source \citep{giroletti_faintest_2009}.
A weak extended NLR component coincides with the radio structure (diameter$\sim1.2\arcsec\sim71\,$pc; PA$\sim 100\degree$; \citealt{schmitt_comparison_1996}).
In addition, very weak water maser emission was detected in the nucleus with uncertain relation to the AGN \citep{hagiwara_search_2003,hagiwara_low-luminosity_2007}.
The first successful MIR observations were performed by \cite{rieke_infrared_1972}, followed by \cite{rieke_infrared_1978} and \cite{lebofsky_extinction_1979}, \cite{aitken_8-13_1985},\cite{devereux_spatial_1987}, \cite{roche_atlas_1991}.
The first subarcsecond-resolution $N$-band image was obtained with Palomar 5\,m/MIRLIN in 2000 \citep{gorjian_10_2004}, where a point-like nucleus without host emission was detected.
In addition, \isoo (e.g., \citealt{ramos_almeida_mid-infrared_2007}) and \spitzer/IRAC, IRS and MIPS observations are available.
In the IRAC $5.8$ and $8.0\,\mu$m  and MIPS $24\,\mu$m images, the compact nucleus completely dominates the MIR emission, while the host is only weakly visible. 
The latter becomes brighter towards longer wavelengths.
Our nuclear IRAC $5.8$ and $8.0\,\mu$m photometry is consistent with \cite{gallimore_infrared_2010}.
The IRS LR staring-mode spectrum exhibits weak silicate 10 and $18\,\mu$m emission, prominent PAH features, and a shallow emission peak at $\sim18\mu$m in $\nu F_\nu$-space (see also \citealt{buchanan_spitzer_2006,wu_spitzer/irs_2009,deo_mid-infrared_2009,goulding_towards_2009,tommasin_spitzer-irs_2010}).
Note however that no background subtraction was performed for the long-wavelength part of the spectrum ($>14\,\mu$m).
The presence of the PAH emission indicates that the arcsecond-scale MIR SED is affected by star formation.
The \spitzerr spectrophotometry is consistent with the MIRLIN flux but significantly higher than the historical measurements.
We observed the nuclear region of NGC\,4051 with Michelle in the Si-5 filter in 2010 and detected a possibly marginally resolved nucleus without any other host emission (FWHM $\sim 0.44\arcsec \sim 26\,$pc).
However, the current data are not sufficient for a robust classification of the nuclear extension in the MIR at subarcsecond scales.
The nuclear photometry is $21\%$ lower than the \spitzerr spectrophotometry.
The current subarcsecond data are not sufficient to constrain the star formation contribution in the central $\sim25\,$pc.
\newline\end{@twocolumnfalse}]

\begin{figure}
   \centering
   \includegraphics[angle=0,width=8.500cm]{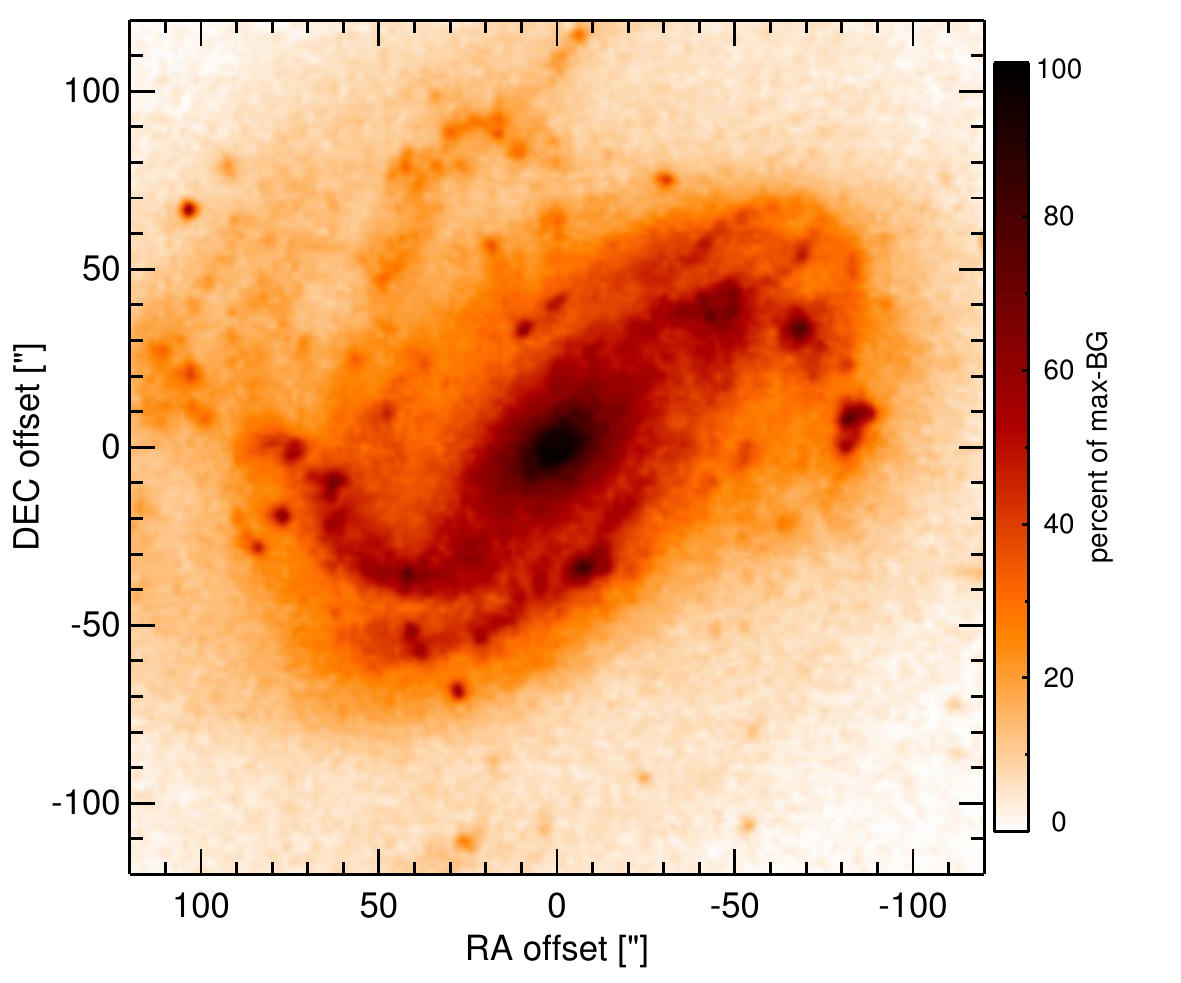}
    \caption{\label{fig:OPTim_NGC4051}
             Optical image (DSS, red filter) of NGC\,4051. Displayed are the central $4\arcmin$ with North up and East to the left. 
              The colour scaling is linear with white corresponding to the median background and black to the $0.01\%$ pixels with the highest intensity.  
           }
\end{figure}
\begin{figure}
   \centering
   \includegraphics[angle=0,height=3.11cm]{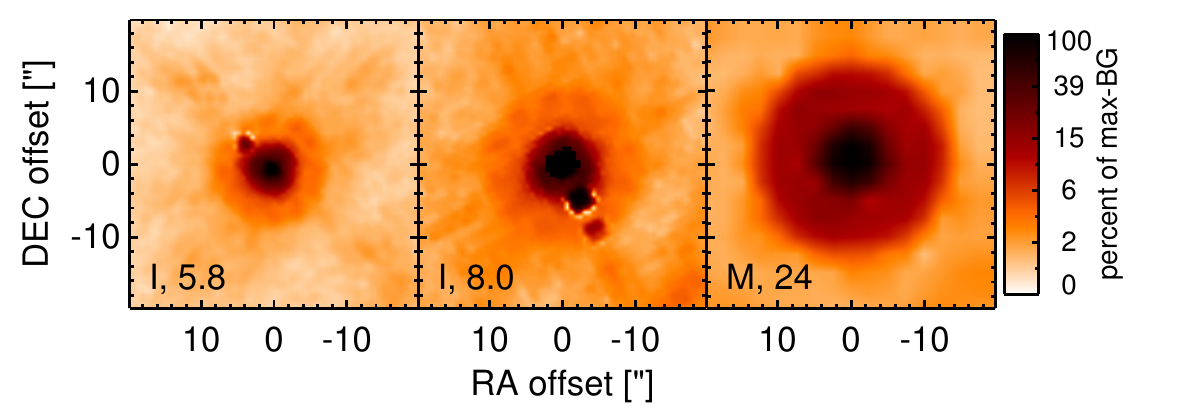}
    \caption{\label{fig:INTim_NGC4051}
             \spitzerr MIR images of NGC\,4051. Displayed are the inner $40\arcsec$ with North up and East to the left. The colour scaling is logarithmic with white corresponding to median background and black to the $0.1\%$ pixels with the highest intensity.
             The label in the bottom left states instrument and central wavelength of the filter in $\mu$m (I: IRAC, M: MIPS). 
             Note that the apparent off-nuclear compact sources in the IRAC 5.8 and $8.0\,\mu$m images are instrumental artefacts.
           }
\end{figure}
\begin{figure}
   \centering
   \includegraphics[angle=0,height=3.11cm]{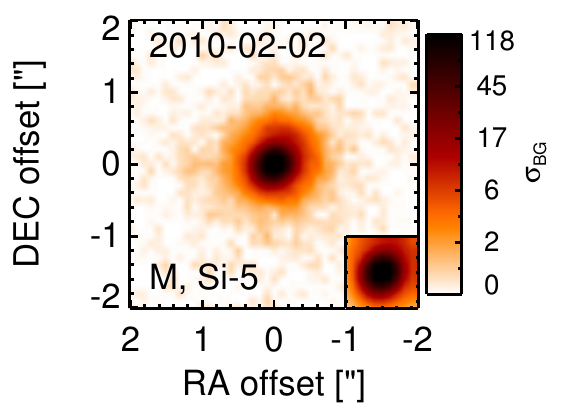}
    \caption{\label{fig:HARim_NGC4051}
             Subarcsecond-resolution MIR images of NGC\,4051 sorted by increasing filter wavelength. 
             Displayed are the inner $4\arcsec$ with North up and East to the left. 
             The colour scaling is logarithmic with white corresponding to median background and black to the $75\%$ of the highest intensity of all images in units of $\sigbg$.
             The inset image shows the central arcsecond of the PSF from the calibrator star, scaled to match the science target.
             The labels in the bottom left state instrument and filter names (C: COMICS, M: Michelle, T: T-ReCS, V: VISIR).
           }
\end{figure}
\begin{figure}
   \centering
   \includegraphics[angle=0,width=8.50cm]{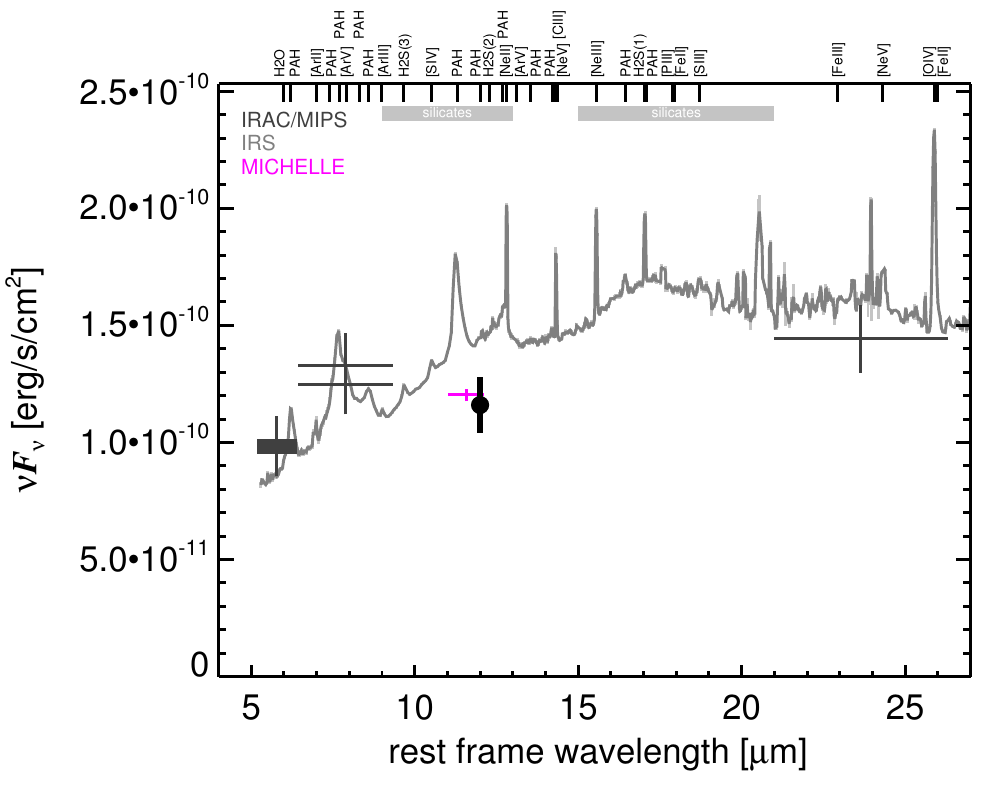}
   \caption{\label{fig:MISED_NGC4051}
      MIR SED of NGC\,4051. The description  of the symbols (if present) is the following.
      Grey crosses and  solid lines mark the \spitzer/IRAC, MIPS and IRS data. 
      The colour coding of the other symbols is: 
      green for COMICS, magenta for Michelle, blue for T-ReCS and red for VISIR data.
      Darker-coloured solid lines mark spectra of the corresponding instrument.
      The black filled circles mark the nuclear 12 and $18\,\mu$m  continuum emission estimate from the data.
      The ticks on the top axis mark positions of common MIR emission lines, while the light grey horizontal bars mark wavelength ranges affected by the silicate 10 and 18$\mu$m features.}
\end{figure}
\clearpage

\twocolumn[\begin{@twocolumnfalse}  
\subsection{NGC\,4074 -- Ark\,347}\label{app:NGC4074}
NGC\,4074 is a peculiar early-type galaxy with a high surface brightness at a redshift of $z=$ 0.0224 ($D\sim107$\,Mpc).
It contains a Sy\,2 nucleus \citep{veron-cetty_catalogue_2010} that belongs to the nine-month BAT AGN sample.
The NLR and radio emission appears to be mostly compact with weak extended radio emission along a PA$\sim131\degree$ \citep{mulchaey_emission-line_1996,nagar_radio_1999}.
NGC\,4074 was not detected by \irass and no other MIR observations of it have been published so far.
It appears unresolved in the \wisee band~3 and 4 images.
In addition, a \spitzer/IRS LR staring-mode spectrum is available, which shows silicate 10$\,\mu$m and possibly 18\,$\mu$m emission, weak PAH features, prominent forbidden ionization lines, and an emission peak at $\sim17\,\mu$m in $\nu F_\nu$-space.
Therefore, the arcsecond-scale MIR SED is clearly AGN dominated.
Interestingly, it indicates an unobscured AGN contrary to its optical classification.
We observed NGC\,4074 with VISIR in the NEII filter in 2010 and detected a marginally resolved nucleus without any further host emission (FWHM $\sim 0.4\arcsec \sim 200\,$pc).
However, the extension needs to be confirmed by at least a second epoch of subarcsecond MIR imaging.
The nuclear photometry is $25\%$ lower than the IRS flux levels.
Comparison with the \wisee band~3 flux indicates that the AGN dominates the total MIR emission of NGC\,4074.
\newline\end{@twocolumnfalse}]

\begin{figure}
   \centering
   \includegraphics[angle=0,width=8.500cm]{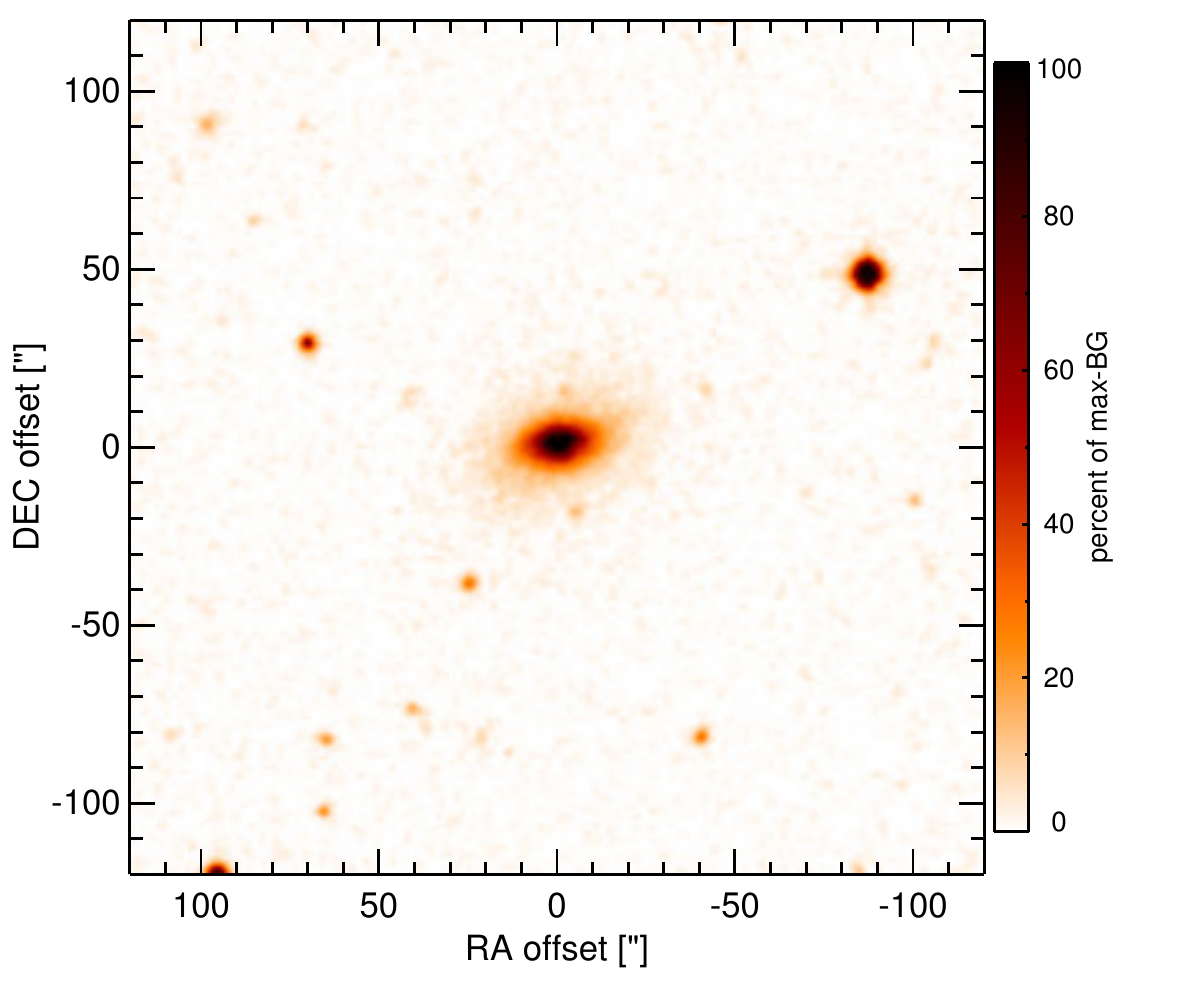}
    \caption{\label{fig:OPTim_NGC4074}
             Optical image (DSS, red filter) of NGC\,4074. Displayed are the central $4\arcmin$ with North up and East to the left. 
              The colour scaling is linear with white corresponding to the median background and black to the $0.01\%$ pixels with the highest intensity.  
           }
\end{figure}
\begin{figure}
   \centering
   \includegraphics[angle=0,height=3.11cm]{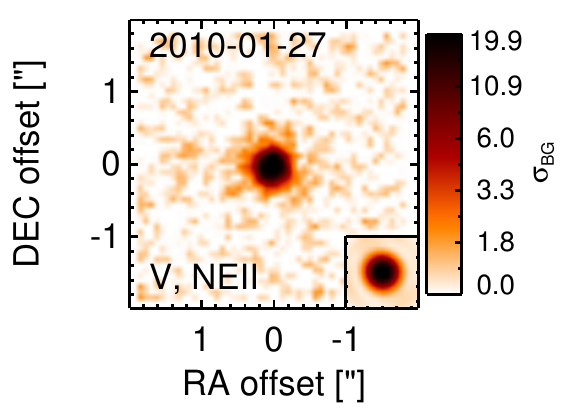}
    \caption{\label{fig:HARim_NGC4074}
             Subarcsecond-resolution MIR images of NGC\,4074 sorted by increasing filter wavelength. 
             Displayed are the inner $4\arcsec$ with North up and East to the left. 
             The colour scaling is logarithmic with white corresponding to median background and black to the $75\%$ of the highest intensity of all images in units of $\sigbg$.
             The inset image shows the central arcsecond of the PSF from the calibrator star, scaled to match the science target.
             The labels in the bottom left state instrument and filter names (C: COMICS, M: Michelle, T: T-ReCS, V: VISIR).
           }
\end{figure}
\begin{figure}
   \centering
   \includegraphics[angle=0,width=8.50cm]{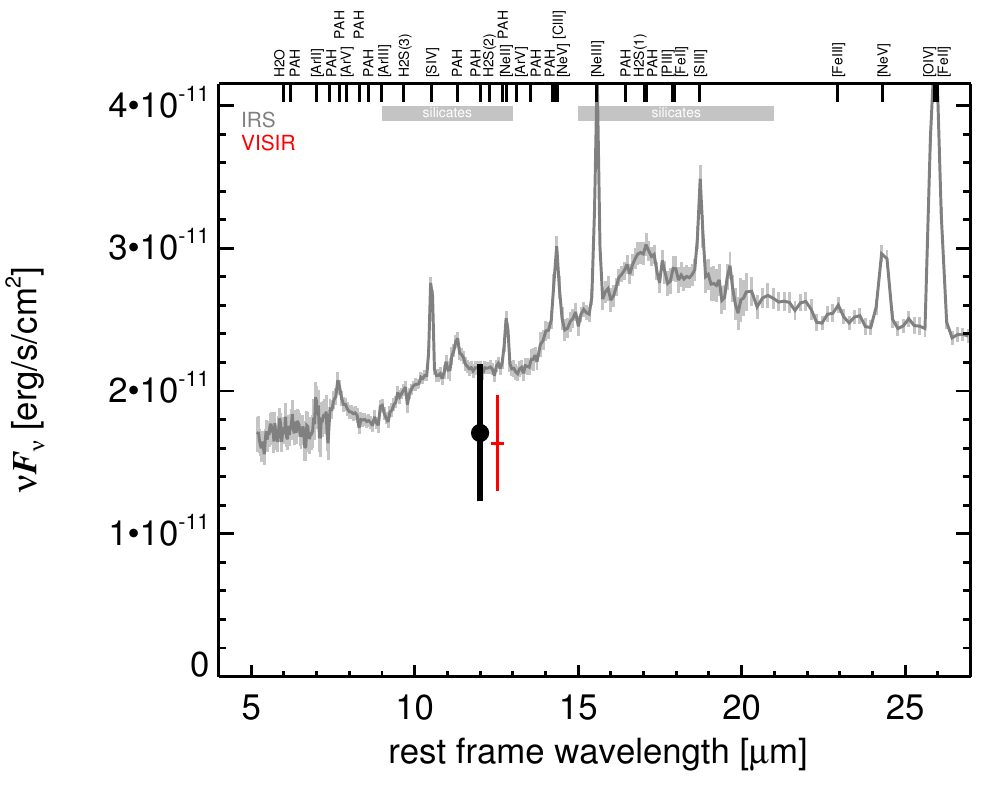}
   \caption{\label{fig:MISED_NGC4074}
      MIR SED of NGC\,4074. The description  of the symbols (if present) is the following.
      Grey crosses and  solid lines mark the \spitzer/IRAC, MIPS and IRS data. 
      The colour coding of the other symbols is: 
      green for COMICS, magenta for Michelle, blue for T-ReCS and red for VISIR data.
      Darker-coloured solid lines mark spectra of the corresponding instrument.
      The black filled circles mark the nuclear 12 and $18\,\mu$m  continuum emission estimate from the data.
      The ticks on the top axis mark positions of common MIR emission lines, while the light grey horizontal bars mark wavelength ranges affected by the silicate 10 and 18$\mu$m features.}
\end{figure}
\clearpage

\twocolumn[\begin{@twocolumnfalse}  
\subsection{NGC\,4111}\label{app:NGC4111}
NGC\,4111 is an edge-on early-type spiral galaxy at a distance of $D=$ $15 \pm 1.5\,$Mpc \citep{tonry_sbf_2001} with a LINER nucleus \citep{ho_search_1997-1}.
The nuclear remained undetected in subarcsecond-resolution radio observations \citep{nagar_radio_2000}, while it appears a compact hard source in X-rays \citep{gonzalez-martin_x-ray_2006,flohic_central_2006}.
The H$\alpha$ emission has a compact component and an extended halo with a major axis of $\sim6\arcsec\sim0.4\,$kpc length along a PA coincident with the galaxy major axis ($\sim150\degree$; \citealt{masegosa_nature_2011}).
The first successful MIR observation of NGC\,4111 was performed by \cite{willner_infrared_1985} who marginally detected the nucleus using the IRTF.
The object was not detected by \iras, while in the \wisee images an compact nuclear component embedded within the edge-on host emission is visible.
No \spitzerr data are available for NGC\,4111.
\cite{mason_nuclear_2012} have observed its nuclear region with Michelle in the N' filter in 2008 and weakly detected a compact nucleus embedded within clumpy extended emission of $\sim2\arcsec\sim 150\,$pc total diameter.
Our measurement of the unresolved nuclear component is consistent with the value given by \cite{mason_nuclear_2012}.
The lack of MIR data does not allow for any conclusions about the source nature.
\newline\end{@twocolumnfalse}]

\begin{figure}
   \centering
   \includegraphics[angle=0,width=8.500cm]{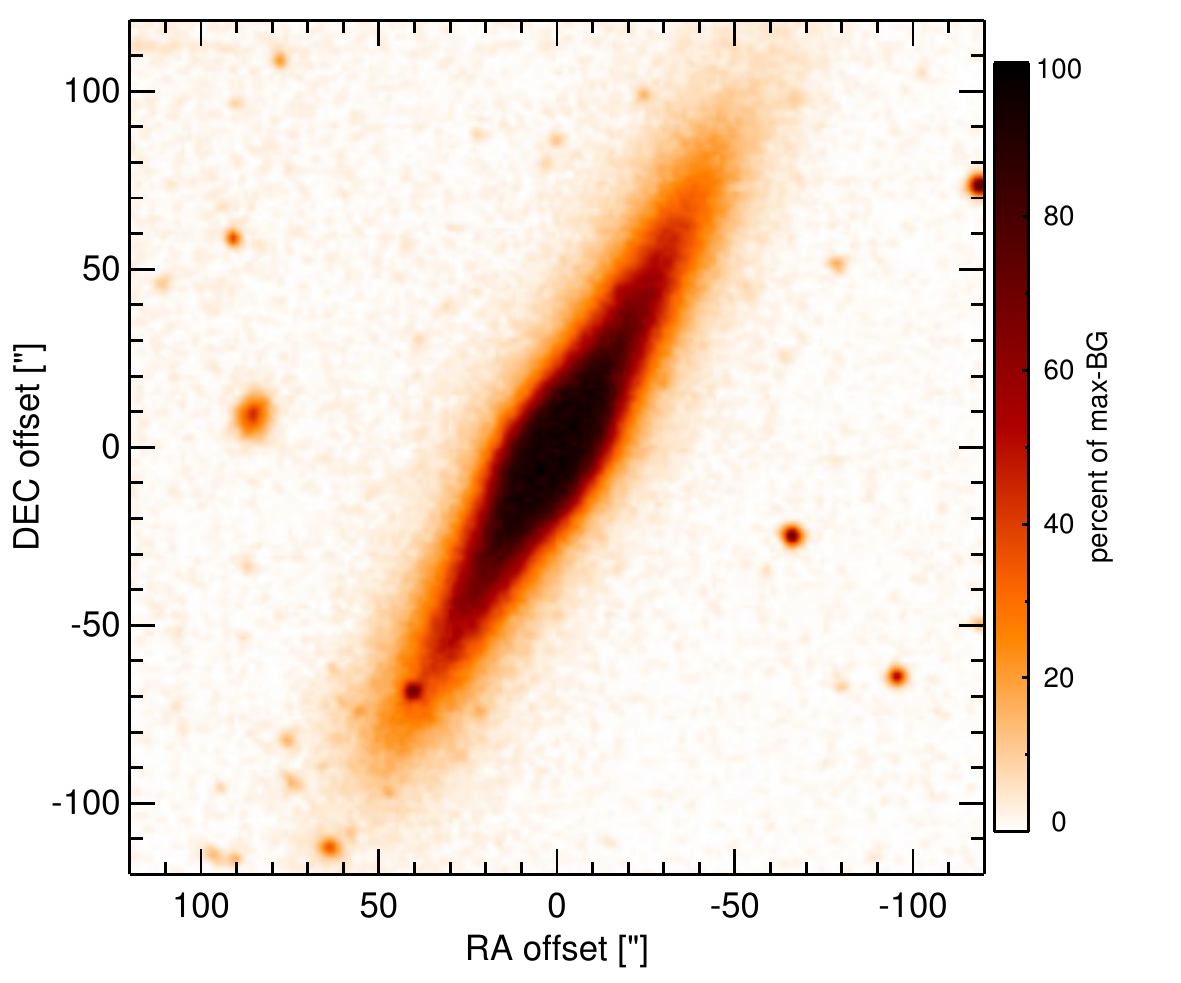}
    \caption{\label{fig:OPTim_NGC4111}
             Optical image (DSS, red filter) of NGC\,4111. Displayed are the central $4\arcmin$ with North up and East to the left. 
              The colour scaling is linear with white corresponding to the median background and black to the $0.01\%$ pixels with the highest intensity.  
           }
\end{figure}
\begin{figure}
   \centering
   \includegraphics[angle=0,height=3.11cm]{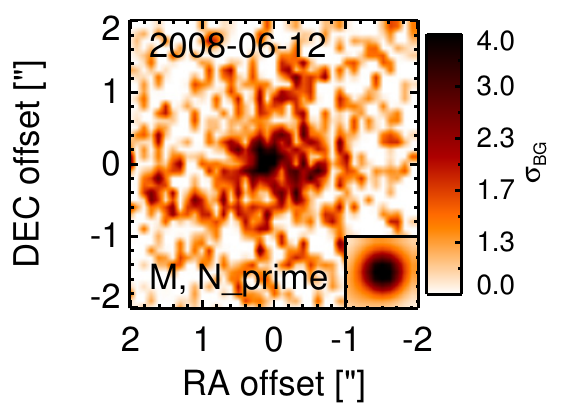}
    \caption{\label{fig:HARim_NGC4111}
             Subarcsecond-resolution MIR images of NGC\,4111 sorted by increasing filter wavelength. 
             Displayed are the inner $4\arcsec$ with North up and East to the left. 
             The colour scaling is logarithmic with white corresponding to median background and black to the $75\%$ of the highest intensity of all images in units of $\sigbg$.
             The inset image shows the central arcsecond of the PSF from the calibrator star, scaled to match the science target.
             The labels in the bottom left state instrument and filter names (C: COMICS, M: Michelle, T: T-ReCS, V: VISIR).
           }
\end{figure}
\begin{figure}
   \centering
   \includegraphics[angle=0,width=8.50cm]{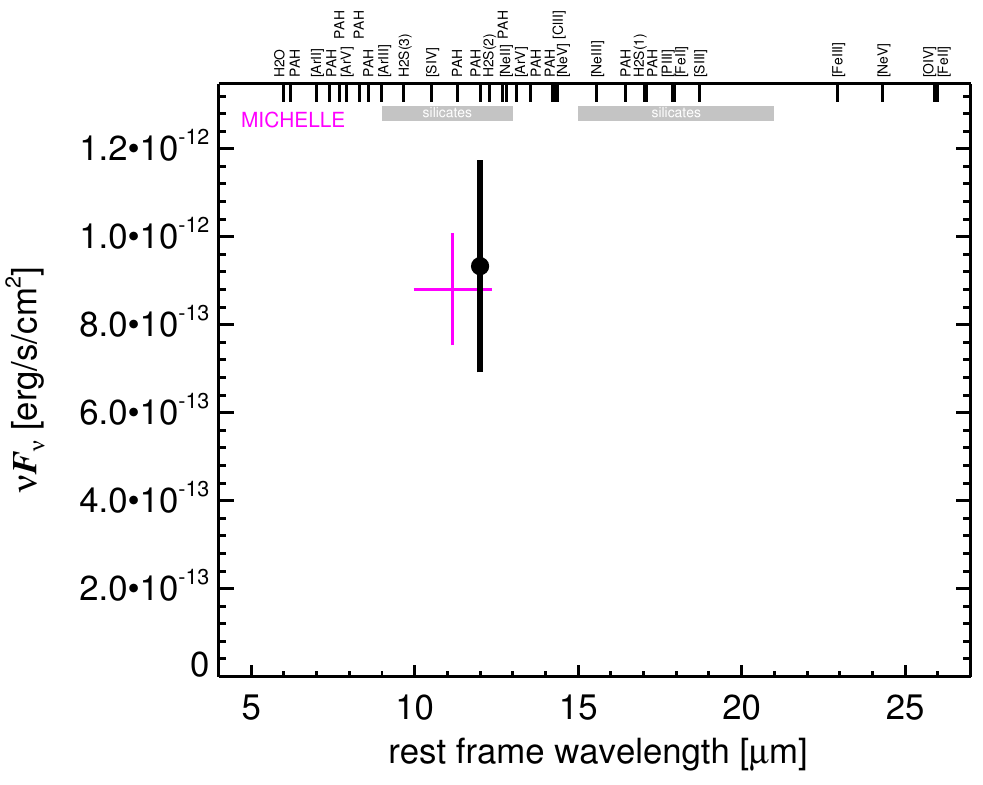}
   \caption{\label{fig:MISED_NGC4111}
      MIR SED of NGC\,4111. The description  of the symbols (if present) is the following.
      Grey crosses and  solid lines mark the \spitzer/IRAC, MIPS and IRS data. 
      The colour coding of the other symbols is: 
      green for COMICS, magenta for Michelle, blue for T-ReCS and red for VISIR data.
      Darker-coloured solid lines mark spectra of the corresponding instrument.
      The black filled circles mark the nuclear 12 and $18\,\mu$m  continuum emission estimate from the data.
      The ticks on the top axis mark positions of common MIR emission lines, while the light grey horizontal bars mark wavelength ranges affected by the silicate 10 and 18$\mu$m features.}
\end{figure}
\clearpage

\twocolumn[\begin{@twocolumnfalse}  
\subsection{NGC\,4138}\label{app:NGC4138}
NGC\,4138 is an early-type spiral galaxy at a distance of $D=$ $13.8 \pm 1.8\,$Mpc with a Sy\,1.9 nucleus \citep{veron-cetty_catalogue_2010} that belongs to the nine-month BAT AGN sample and is an X-ray ``buried" AGN candidate \citep{noguchi_new_2009}.
A weak compact radio nucleus was detected in NGC\,4138 \citep{ho_radio_2001}.
This object has not been detected with \irass but with \spitzer/IRAC, IRS and MIPS.
The corresponding images show a compact nucleus embedded within extended host emission and is surrounded by a starburst ring with a major axis diameter of $\sim39\arcsec \sim 2.6\,$kpc along north-west direction.
Therefore, our MIPS $24\,\mu$m flux of the unresolved nuclear component is significantly lower than the total flux by \cite{temi_spitzer_2009}, which includes the whole central region.
The IRS LR mapping-mode spectrum suffers from low S/N but indicates PAH emission, possibly weak silicate 10$\,\mu$m emission, and a shallow blue spectral slope in $\nu F_\nu$-space (
see \citealt{weaver_mid-infrared_2010} for measurements of  forbidden emission lines).
We observed the nuclear region of NGC\,4138 with Michelle in the Si-5 filter in 2010 and detected a possibly marginally extended MIR nucleus without further host emission (FWHM(major axis) $\sim 0.52\arcsec \sim 35\,$pc; PA$\sim 146\degree$).
However, the current data are not sufficient for a robust classification of the nuclear extension in the MIR at subarcsecond scales.
The nuclear photometry is $39\%$ lower than the IRS flux levels, providing additional evidence for significant star formation at least on arcsecond scale in NGC\,4138.
\newline\end{@twocolumnfalse}]

\begin{figure}
   \centering
   \includegraphics[angle=0,width=8.500cm]{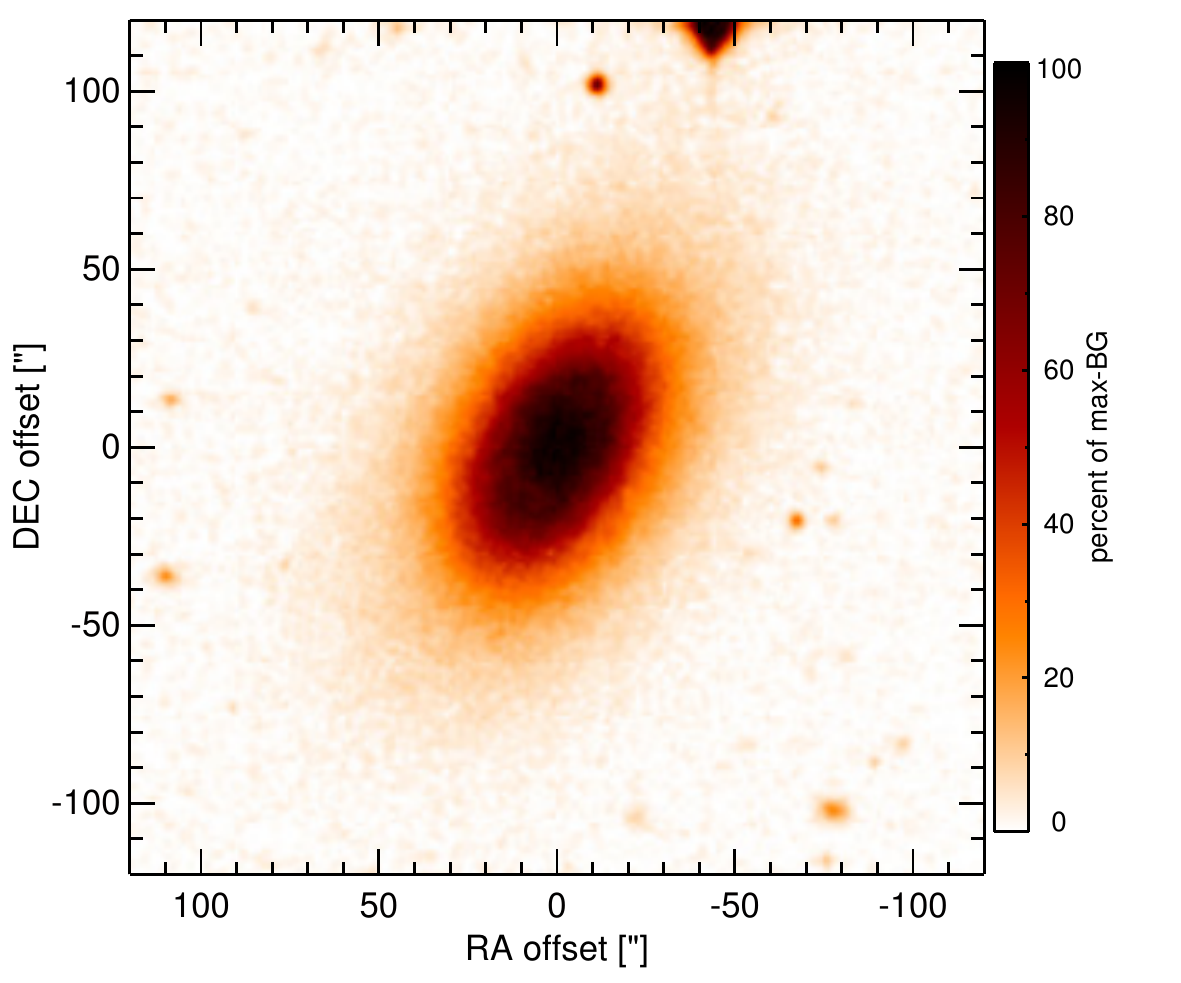}
    \caption{\label{fig:OPTim_NGC4138}
             Optical image (DSS, red filter) of NGC\,4138. Displayed are the central $4\arcmin$ with North up and East to the left. 
              The colour scaling is linear with white corresponding to the median background and black to the $0.01\%$ pixels with the highest intensity.  
           }
\end{figure}
\begin{figure}
   \centering
   \includegraphics[angle=0,height=3.11cm]{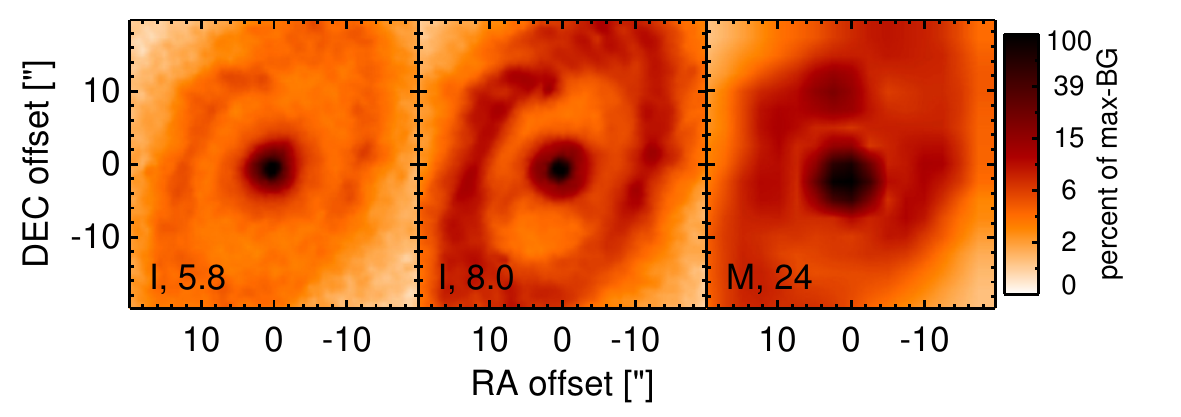}
    \caption{\label{fig:INTim_NGC4138}
             \spitzerr MIR images of NGC\,4138. Displayed are the inner $40\arcsec$ with North up and East to the left. The colour scaling is logarithmic with white corresponding to median background and black to the $0.1\%$ pixels with the highest intensity.
             The label in the bottom left states instrument and central wavelength of the filter in $\mu$m (I: IRAC, M: MIPS). 
           }
\end{figure}
\begin{figure}
   \centering
   \includegraphics[angle=0,height=3.11cm]{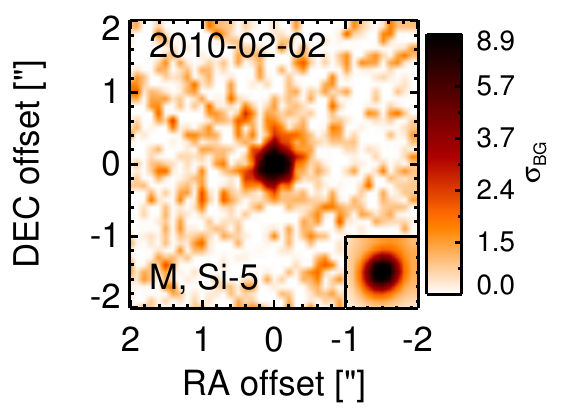}
    \caption{\label{fig:HARim_NGC4138}
             Subarcsecond-resolution MIR images of NGC\,4138 sorted by increasing filter wavelength. 
             Displayed are the inner $4\arcsec$ with North up and East to the left. 
             The colour scaling is logarithmic with white corresponding to median background and black to the $75\%$ of the highest intensity of all images in units of $\sigbg$.
             The inset image shows the central arcsecond of the PSF from the calibrator star, scaled to match the science target.
             The labels in the bottom left state instrument and filter names (C: COMICS, M: Michelle, T: T-ReCS, V: VISIR).
           }
\end{figure}
\begin{figure}
   \centering
   \includegraphics[angle=0,width=8.50cm]{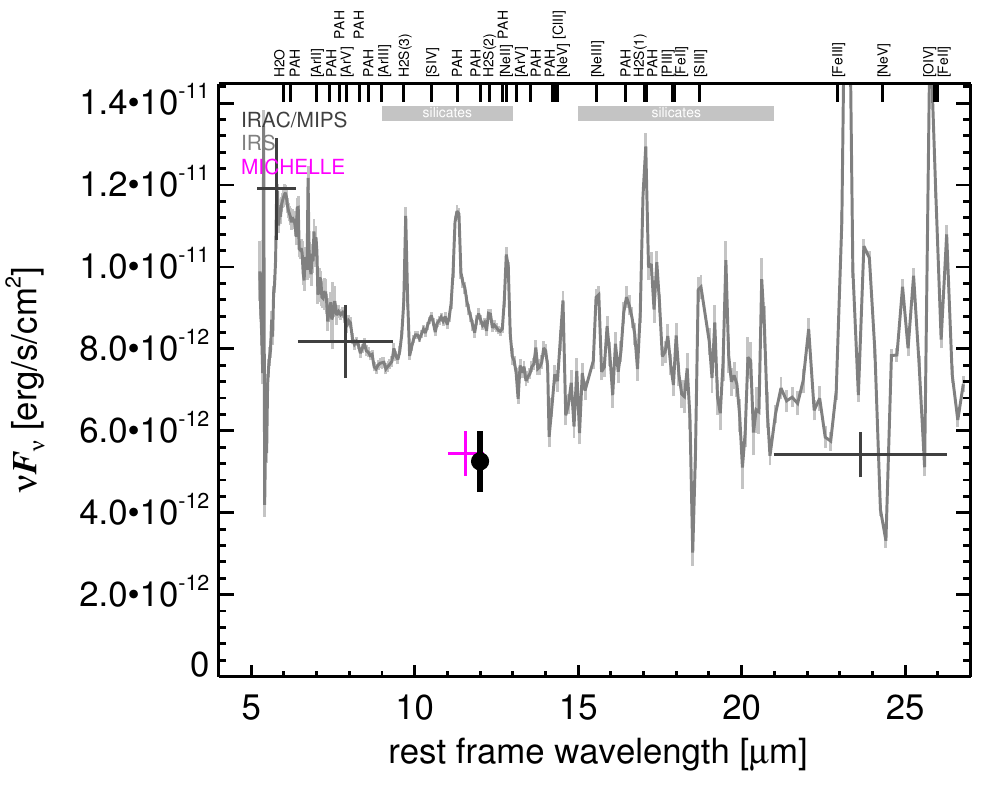}
   \caption{\label{fig:MISED_NGC4138}
      MIR SED of NGC\,4138. The description  of the symbols (if present) is the following.
      Grey crosses and  solid lines mark the \spitzer/IRAC, MIPS and IRS data. 
      The colour coding of the other symbols is: 
      green for COMICS, magenta for Michelle, blue for T-ReCS and red for VISIR data.
      Darker-coloured solid lines mark spectra of the corresponding instrument.
      The black filled circles mark the nuclear 12 and $18\,\mu$m  continuum emission estimate from the data.
      The ticks on the top axis mark positions of common MIR emission lines, while the light grey horizontal bars mark wavelength ranges affected by the silicate 10 and 18$\mu$m features.}
\end{figure}
\clearpage

\twocolumn[\begin{@twocolumnfalse}  
\subsection{NGC\,4151}\label{app:NGC4151}
NGC\,4151 is a nearly face-on barred spiral galaxy at a distance of $D=$ $13.3 \pm 1$\,Mpc \citep{mundell_nuclear_2003} with a Sy\,1.5 nucleus \citep{veron-cetty_catalogue_2010}.
This galaxy is one of the original six Seyfert galaxies \citep{seyfert_nuclear_1943} and host the flux-brightest, nearest and most variable type~I AGN.
Therefore, it belongs to the best-studied AGN at all wavelengths (see \citealt{ulrich_active_2000} for a review).
NGC\,4151 also belongs to the nine-month BAT AGN sample.
The AGN is surrounded by an extended biconical NLR with an inner extent of $\sim 5.5\arcsec\sim350\,$pc along a PA$\sim67\degree$ ( e.g., \citealt{pogge_circumnuclear_1989,perez_complex_1989,evans_hubble_1993}).
This roughly coincides with a two-sided radio outflow with total extent of $\sim600$\,pc along a PA$\sim77\degree$ \citep{wilson_radio_1982} and a collimated jet in the central $\sim100$\,pc \citep{mundell_nuclear_2003}.
In addition, nuclear water maser emission was detected in this object \citep{braatz_green_2004,tarchi_new_2011}.
Pioneering MIR observations of NGC\,4151 were performed by \cite{low_proceedings_1968}.
Early follow-up observations by \cite{stein_possible_1969}, \cite{kleinmann_observations_1970} and \cite{rieke_variability_1972} found apparent $N$-band flux variations of a factor of a few over the course of three years.
However, later measurements constrained possible $N$-band flux variations to $\lesssim 20\%$ 
\citep{stein_observations_1974, rieke_infrared_1978,lebofsky_extinction_1979,rieke_spectral_1981,ward_continuum_1987}.
The first $N$-band spectrum was presented by \cite{jones_dust_1984}, followed by \cite{aitken_8-13_1985} and \cite{roche_atlas_1991}, while \cite{neugebauer_size_1990} made a first attempt to measure the size of the MIR nucleus in NGC\,4151 with bolometer-based scanning method.
Its emission was also studied with \isoo \citep{rodriguez_espinosa_bimodal_1996,sturm_infrared_1999,alexander_infrared_1999,rigopoulou_large_1999}.
The first subarcsecond-resolution MIR images were obtained in 2000 with Palomar 5\,m/MIRLIN \citep{gorjian_10_2004} and Keck/LWS 
\citep{soifer_high_2003}, and in 2001 with Gemini/OSCIR 
\citep{radomski_resolved_2003}.
The images show a dominating unresolved MIR nucleus, while the OSCIR images also reveal extended emission ($\sim 3.5\arcsec \sim 220\,$pc) coinciding with the NLR (PA$\sim 60\degree$).
NGC\,4151 was also observed with \spitzer/IRAC, IRS and MIPS, and the corresponding images are completely dominated by the bright nucleus outshining the spiral-like host emission.
The IRAC $5.8$ and $8.0\,\mu$m PBCD images are saturated and thus not analysed (but see \citealt{gallimore_infrared_2010}).
The IRS LR staring-mode spectrum exhibits silicate 10 and $18\,\mu$m emission, very weak PAH features and an emission peak at $\sim18\,\mu$m  in $\nu F_\nu$-space (see also \citealt{weedman_mid-infrared_2005,buchanan_spitzer_2006,wu_spitzer/irs_2009,deo_mid-infrared_2009,gallimore_infrared_2010}).
Thus, the arcsecond-scale MIR SED appears AGN-dominated without significant star formation contribution.
The nuclear region of NGC\,4151 was imaged with Michelle in the Si-5 filter in 2007 (unpublished, to our knowledge), and with VISIR in the PAH1 and PAH2\_2 filter 2009 and 2010 \citep{kishimoto_mapping_2011}.
In addition, a Michelle LR $N$-band spectrum was obtained in 2007 \citealt{alonso-herrero_torus_2011, esquej_nuclear_2014}.
Owing to the high airmass, the nucleus appears elongated in the VISIR images similar to the corresponding standard star images. 
Therefore, the VISIR images cannot be used to investigate the source morphology.
In the Michelle image, the nucleus appears marginally resolved (FWHM $\sim 0.5\arcsec \sim 32\,$pc), while the extended NLR-associated emission seen in the OSCIR image is not visible.
At least a second epoch of deep low-airmass MIR subarcsecond imaging is required to verify the source morphology.
Our nuclear photometry is consistent with \cite{kishimoto_mapping_2011} and the Michelle spectrum, at least at longer wavelengths. 
The fluxes are on average $\sim 22\%$ lower than the \spitzerr spectrophotometry.
Comparison with the previous MIR photometry constrains any possible flux variation to $\lesssim 20\%$ in the $N$-band but a more detailed analysis has to be conducted.
Note the the nucleus of NGC\,4151 is resolved interferometrically with MIDI, and a multi-component structure with a few-parsec size is found \citep{burtscher_dust_2009,burtscher_diversity_2013,kishimoto_mapping_2011}.
\newline\end{@twocolumnfalse}]

\begin{figure}
   \centering
   \includegraphics[angle=0,width=8.500cm]{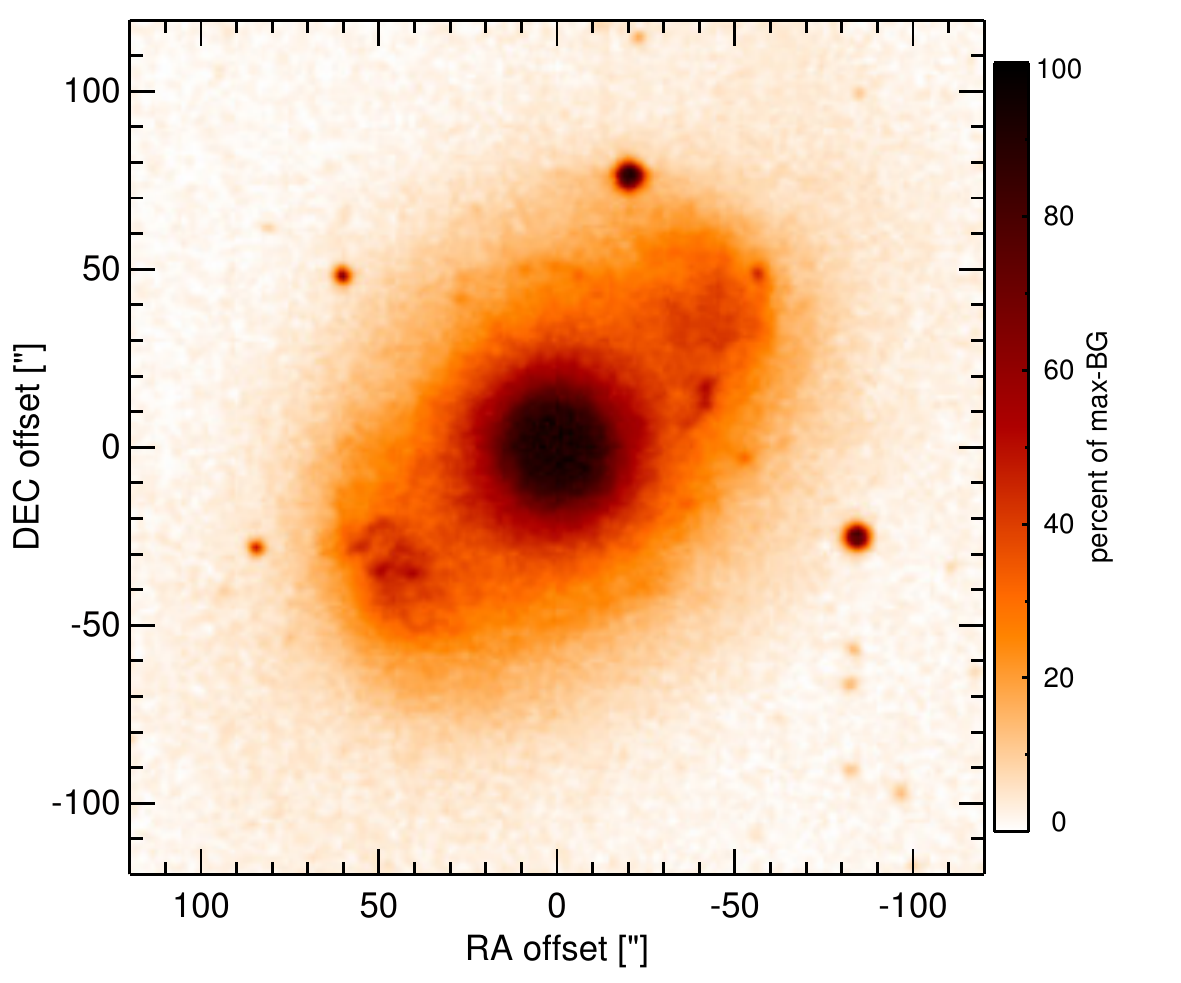}
    \caption{\label{fig:OPTim_NGC4151}
             Optical image (DSS, red filter) of NGC\,4151. Displayed are the central $4\arcmin$ with North up and East to the left. 
              The colour scaling is linear with white corresponding to the median background and black to the $0.01\%$ pixels with the highest intensity.  
           }
\end{figure}
\begin{figure}
   \centering
   \includegraphics[angle=0,height=3.11cm]{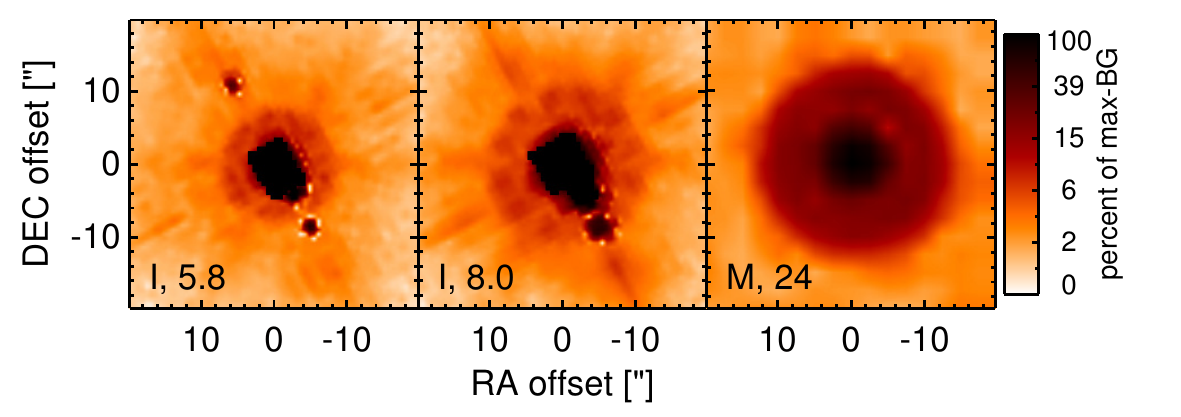}
    \caption{\label{fig:INTim_NGC4151}
             \spitzerr MIR images of NGC\,4151. Displayed are the inner $40\arcsec$ with North up and East to the left. The colour scaling is logarithmic with white corresponding to median background and black to the $0.1\%$ pixels with the highest intensity.
             The label in the bottom left states instrument and central wavelength of the filter in $\mu$m (I: IRAC, M: MIPS).
             Note that the apparent off-nuclear compact sources in the IRAC 5.8 and $8.0\,\mu$m images are instrumental artefacts.
           }
\end{figure}
\begin{figure}
   \centering
   \includegraphics[angle=0,width=8.500cm]{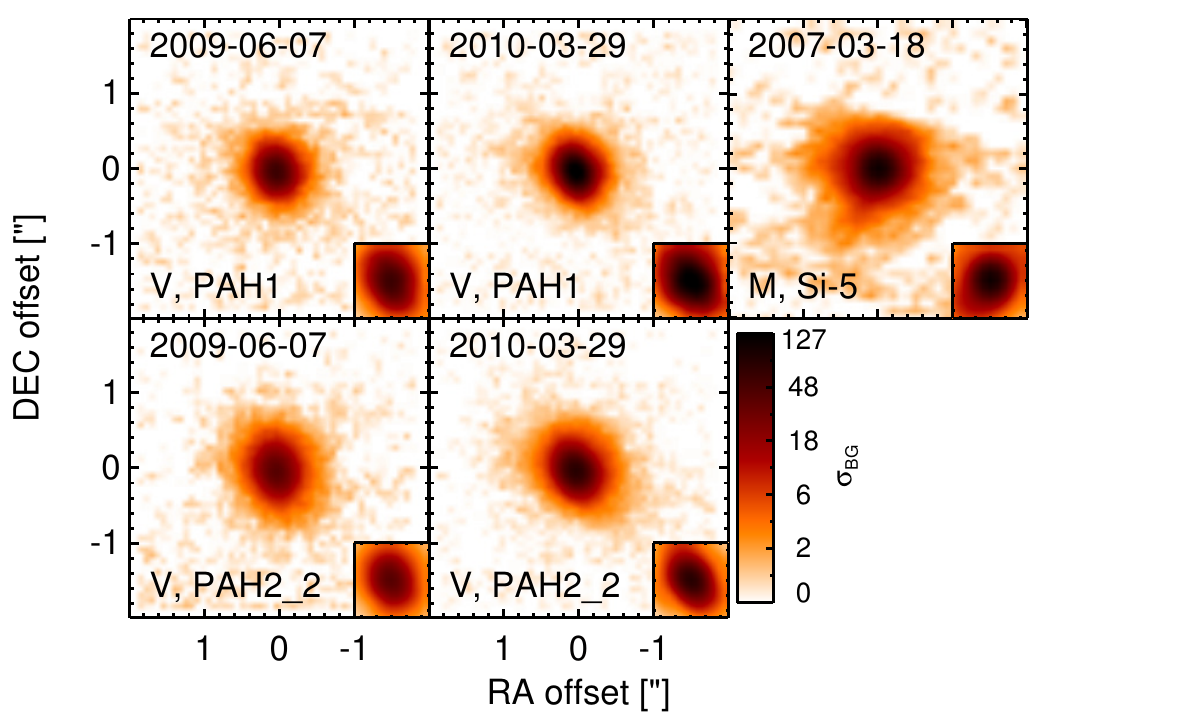}
    \caption{\label{fig:HARim_NGC4151}
             Subarcsecond-resolution MIR images of NGC\,4151 sorted by increasing filter wavelength. 
             Displayed are the inner $4\arcsec$ with North up and East to the left. 
             The colour scaling is logarithmic with white corresponding to median background and black to the $75\%$ of the highest intensity of all images in units of $\sigbg$.
             The inset image shows the central arcsecond of the PSF from the calibrator star, scaled to match the science target.
             The labels in the bottom left state instrument and filter names (C: COMICS, M: Michelle, T: T-ReCS, V: VISIR).
           }
\end{figure}
\begin{figure}
   \centering
   \includegraphics[angle=0,width=8.50cm]{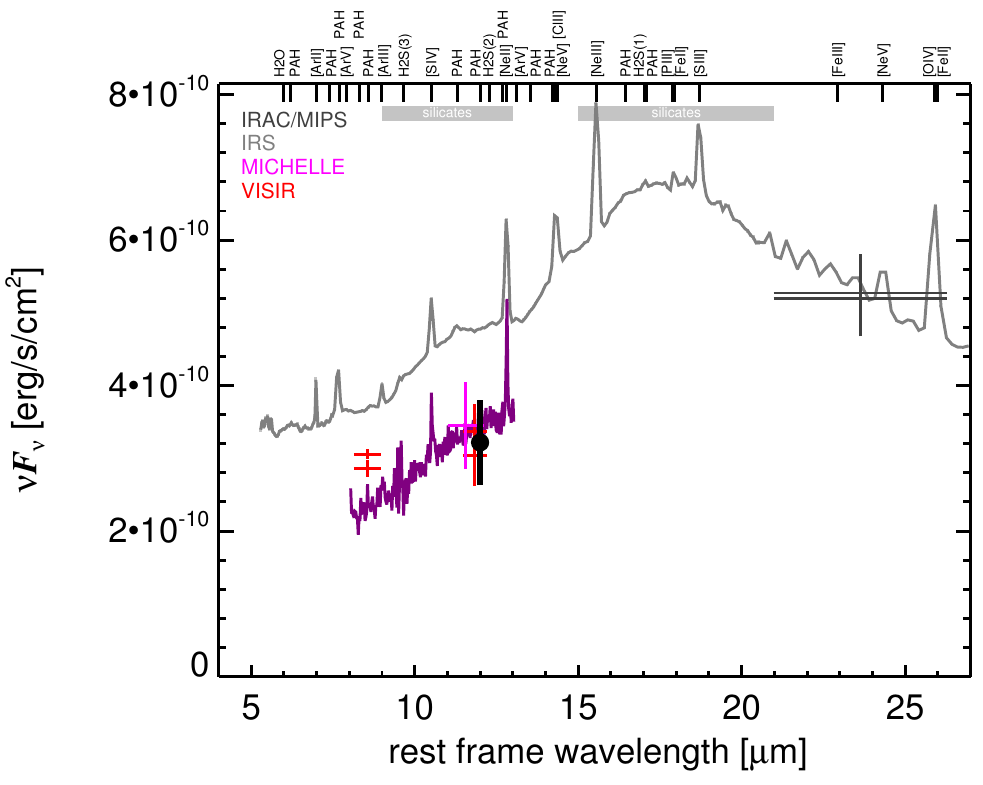}
   \caption{\label{fig:MISED_NGC4151}
      MIR SED of NGC\,4151. The description  of the symbols (if present) is the following.
      Grey crosses and  solid lines mark the \spitzer/IRAC, MIPS and IRS data. 
      The colour coding of the other symbols is: 
      green for COMICS, magenta for Michelle, blue for T-ReCS and red for VISIR data.
      Darker-coloured solid lines mark spectra of the corresponding instrument.
      The black filled circles mark the nuclear 12 and $18\,\mu$m  continuum emission estimate from the data.
      The ticks on the top axis mark positions of common MIR emission lines, while the light grey horizontal bars mark wavelength ranges affected by the silicate 10 and 18$\mu$m features.}
\end{figure}
\clearpage

\twocolumn[\begin{@twocolumnfalse}  
\subsection{NGC\,4235 -- VCC\,222}\label{app:NGC4235}
NGC\,4235 is a nearly edge-on spiral galaxy at a redshift of $z=$ 0.008 ($D\sim41.2\,$Mpc) with a Sy\,1.2 nucleus \citep{veron-cetty_catalogue_2010}, which is detected but appears unresolved at radio wavelengths \citep{ulvestad_radio_1984,kukula_high-resolution_1995}.
The NLR appears compact with a possible \oiii extension towards the north-east with a PA$\sim48\degree$ and a length of $\sim4.4\arcsec\sim0.9\,$kpc \citep{pogge_circumnuclear_1989}.
The first $N$-band detection of NGC\,4235 is reported in \cite{maiolino_new_1995}. 
In addition, \spitzer/IRAC, IRS and MIPS data are available and the corresponding images show a compact nucleus embedded within weak host emission. 
The IRS LR staring-mode spectrum suffers from a low S/N but indicates silicate 10 and $18\,\mu$m emission, very weak PAH emission and a blue spectral slope in  $\nu F_\nu$-space.
Therefore, star formation seems to be at best weak on arcsecond scales.
The MIR emission lines have been measured by \cite{weaver_mid-infrared_2010} and \cite{pereira-santaella_mid-infrared_2010}.
We observed the nuclear region of NGC\,4235 with VISIR in two narrow $N$-band filters in 2009 \citep{asmus_mid-infrared_2011} and followed up with two additional $N$ and one $Q$-band filter images in 2010.
In all cases, the nucleus appears as a point source without any further host emission being detected. 
Our reanalysis of the published images is consistent with previously reported flux values, while the VISIR photometry in general is consistent with the \spitzerr spectrophotometry.
\newline\end{@twocolumnfalse}]

\begin{figure}
   \centering
   \includegraphics[angle=0,width=8.500cm]{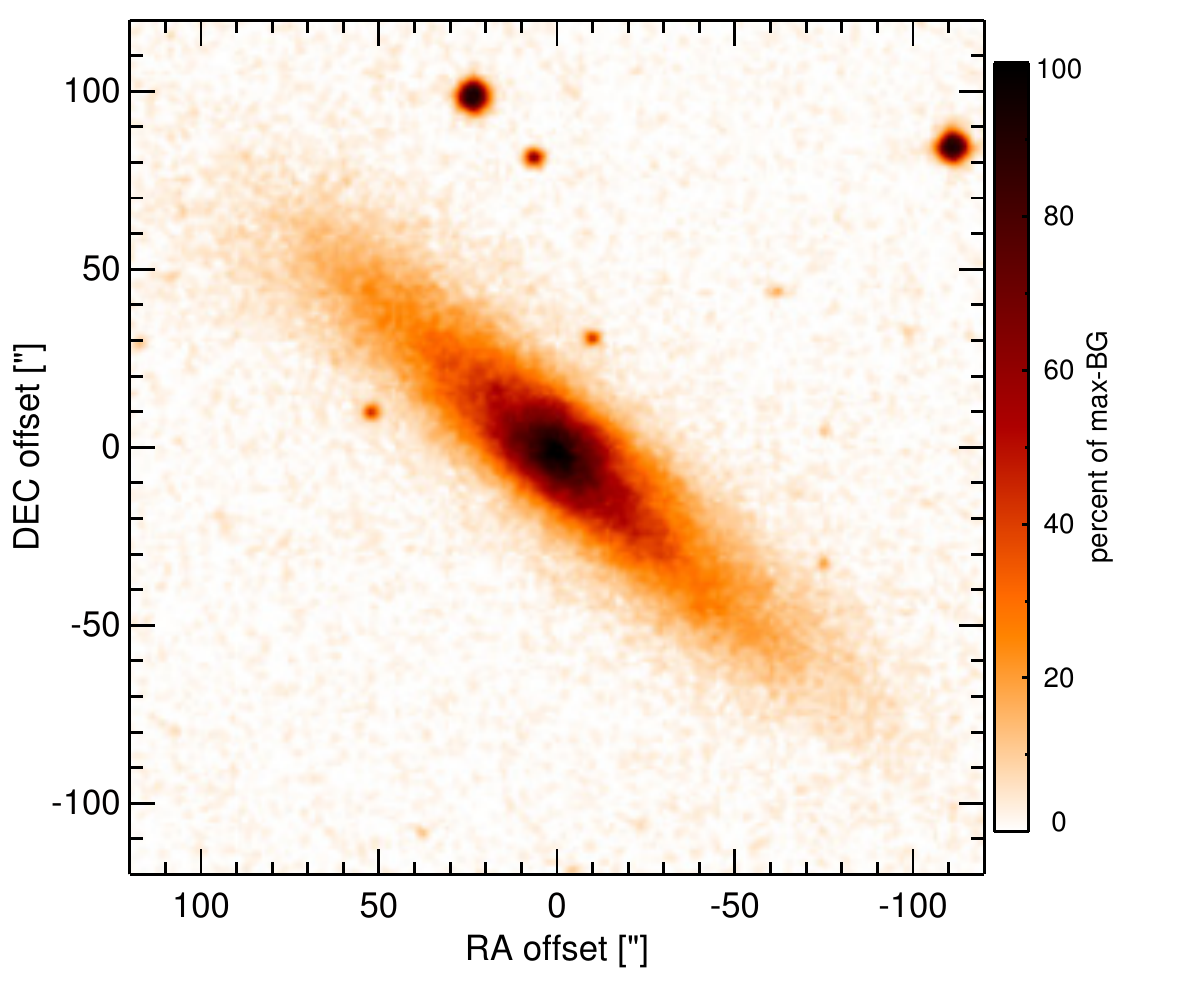}
    \caption{\label{fig:OPTim_NGC4235}
             Optical image (DSS, red filter) of NGC\,4235. Displayed are the central $4\arcmin$ with North up and East to the left. 
              The colour scaling is linear with white corresponding to the median background and black to the $0.01\%$ pixels with the highest intensity.  
           }
\end{figure}
\begin{figure}
   \centering
   \includegraphics[angle=0,height=3.11cm]{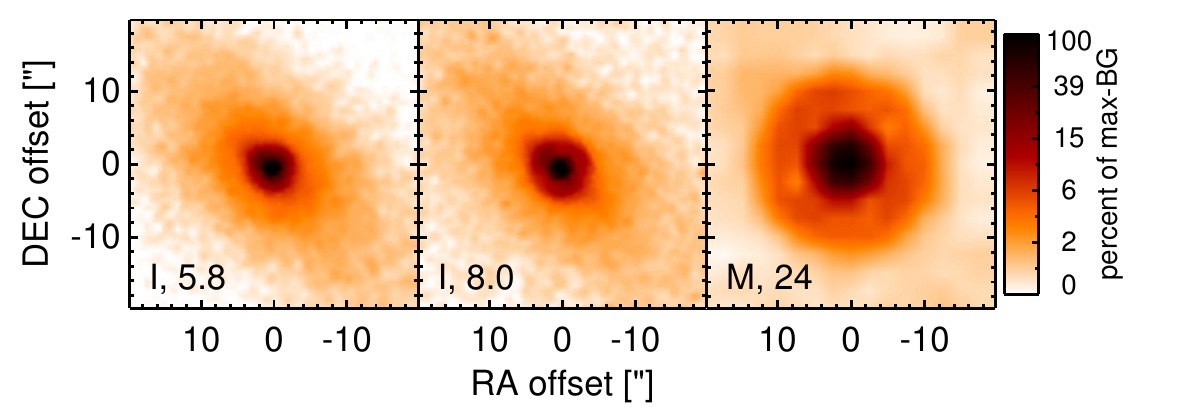}
    \caption{\label{fig:INTim_NGC4235}
             \spitzerr MIR images of NGC\,4235. Displayed are the inner $40\arcsec$ with North up and East to the left. The colour scaling is logarithmic with white corresponding to median background and black to the $0.1\%$ pixels with the highest intensity.
             The label in the bottom left states instrument and central wavelength of the filter in $\mu$m (I: IRAC, M: MIPS). 
           }
\end{figure}
\begin{figure}
   \centering
   \includegraphics[angle=0,width=8.500cm]{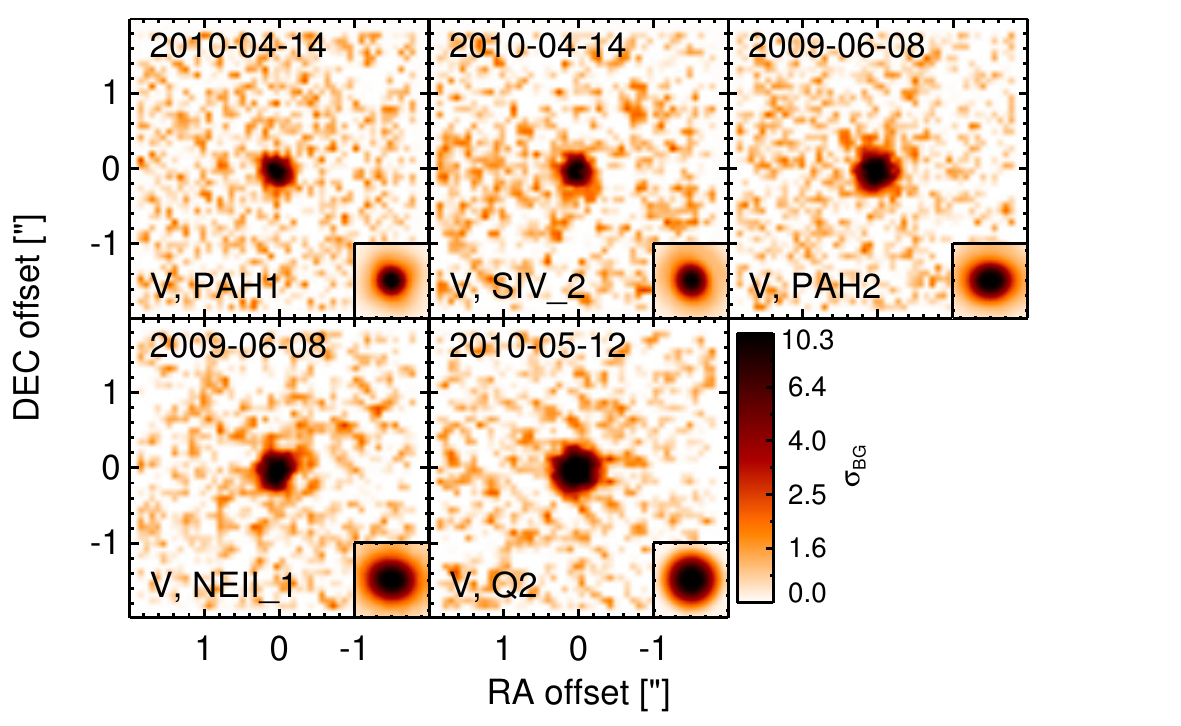}
    \caption{\label{fig:HARim_NGC4235}
             Subarcsecond-resolution MIR images of NGC\,4235 sorted by increasing filter wavelength. 
             Displayed are the inner $4\arcsec$ with North up and East to the left. 
             The colour scaling is logarithmic with white corresponding to median background and black to the $75\%$ of the highest intensity of all images in units of $\sigbg$.
             The inset image shows the central arcsecond of the PSF from the calibrator star, scaled to match the science target.
             The labels in the bottom left state instrument and filter names (C: COMICS, M: Michelle, T: T-ReCS, V: VISIR).
           }
\end{figure}
\begin{figure}
   \centering
   \includegraphics[angle=0,width=8.50cm]{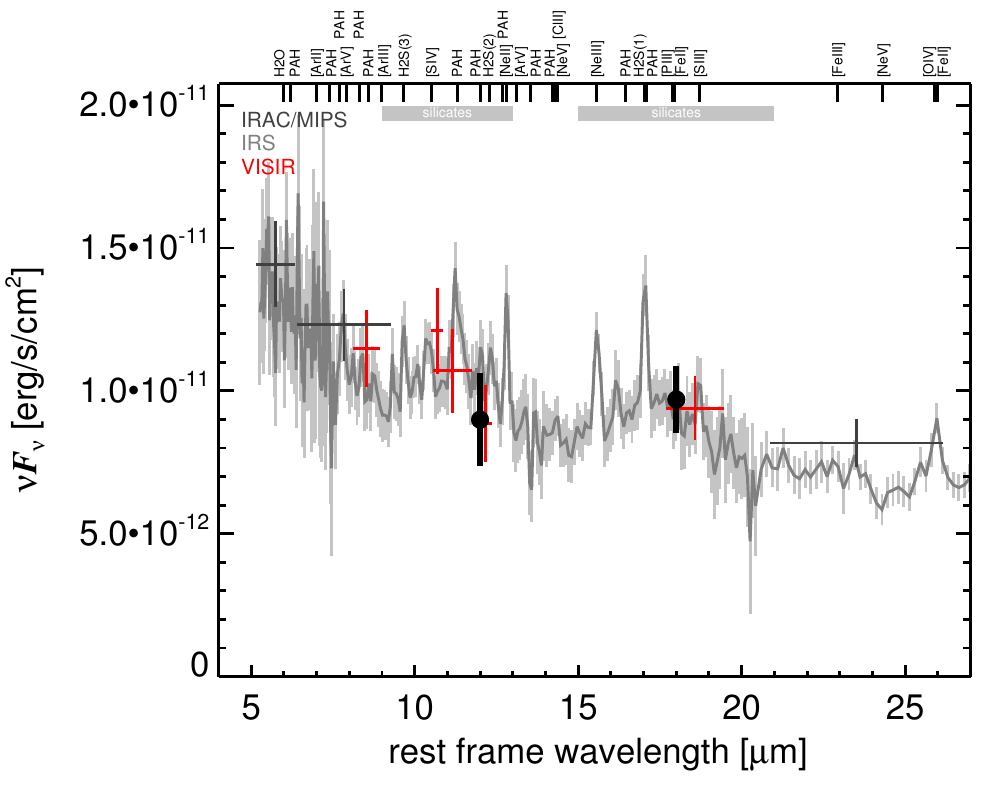}
   \caption{\label{fig:MISED_NGC4235}
      MIR SED of NGC\,4235. The description  of the symbols (if present) is the following.
      Grey crosses and  solid lines mark the \spitzer/IRAC, MIPS and IRS data. 
      The colour coding of the other symbols is: 
      green for COMICS, magenta for Michelle, blue for T-ReCS and red for VISIR data.
      Darker-coloured solid lines mark spectra of the corresponding instrument.
      The black filled circles mark the nuclear 12 and $18\,\mu$m  continuum emission estimate from the data.
      The ticks on the top axis mark positions of common MIR emission lines, while the light grey horizontal bars mark wavelength ranges affected by the silicate 10 and 18$\mu$m features.}
\end{figure}
\clearpage

\twocolumn[\begin{@twocolumnfalse}  
\subsection{NGC\,4258 -- M106}\label{app:NGC4258}
NGC\,4258 is an inclined grand-design spiral galaxy at a distance of $D=$ $7.6 \pm 0.8$\,Mpc (NED redshift-independent median) with a weak Sy\,2 nucleus \citep{veron-cetty_catalogue_2010}, a radio jet \citep{ford_bubbles_1986}, and a famous water maser (e.g., \citealt{nakai_extremely-high-velocity_1993,greenhill_detection_1995,herrnstein_vlba_1998}), allowing for accurate determination of the mass of the SMBH.
\cite{yuan_ngc_2002} present a detailed study of the AGN in NGC\,4258.
The first MIR observations of the nucleus were performed by \cite{rieke_10_1978} and \cite{dyck_photometry_1978} and first subarcsecond-resolution $N-$ and $Q$-band images were taken with Keck/MIRLIN in 1998 \cite{chary_high-resolution_2000}.
A point-like nucleus (FWHM$,0.5\arcsec$) without any other emission sources in the central 6\arcsec\, is visible in the images.
In the lower angular resolution \spitzer/IRAC and MIPS images on the other hand, spiral-like host emission surrounding the nucleus was detected. 
Because we measure the nuclear component only, our IRAC $5.8$ and $8.0\,\mu$m and MIPS $24\,\mu$m fluxes are much lower than in \cite{dale_spitzer_2009}.
The \spitzer/IRS LR staring-mode spectrum exhibits silicate 10 and $18\,\mu$m emission, a PAH 11.3$\,\mu$m feature and a flat spectral slope in $\nu F_\nu$-space.
We observed NGC\,4258 with Michelle in two $N$-band filters in 2010, and additional Michelle $N$-band imaging in two more $N$-band filters was carried out in 2011 (unpublished, to our knowledge).
Similar to the Keck imaging, only a compact MIR nucleus was detected in all the Michelle images.
The nucleus appears marginally resolved in the Si-4 filter image (FWHM $\sim 0.46\arcsec \sim 17\,$pc) but not in the Si-5 images. 
Therefore, it remains uncertain, whether the nucleus is in general resolved at subarcsecond resolution in the MIR.
Our nuclear photometry is on average $\sim 22\%$ lower than the \spitzerr spectrophotometry, while the silicate $10\,\mu$m emission feature remains roughly similar in strength at subarcsecond resolution. 
Therefore, we can correct our $12\,\mu$m emission estimate for the silicate feature.
\newline\end{@twocolumnfalse}]

\begin{figure}
   \centering
   \includegraphics[angle=0,width=8.500cm]{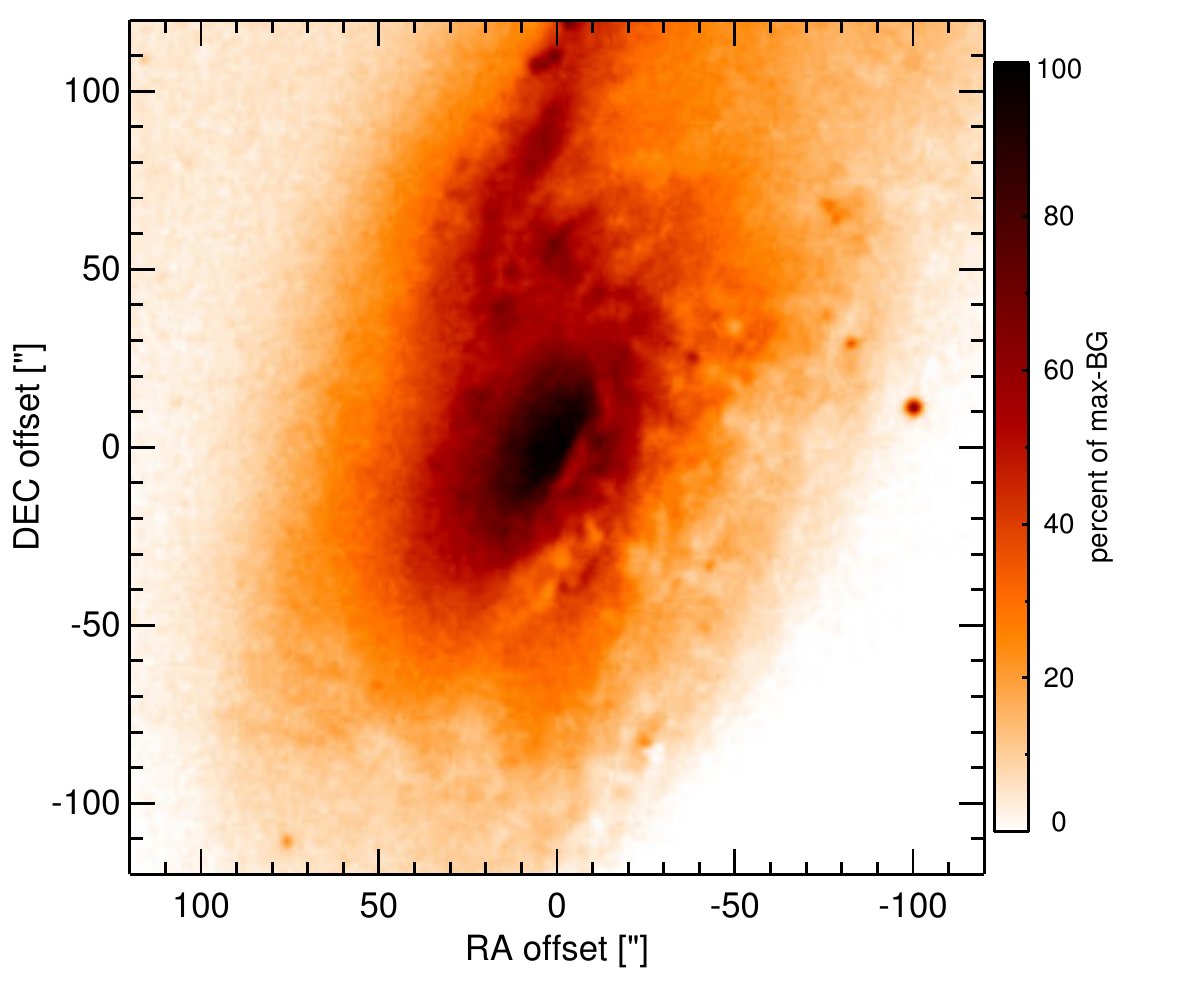}
    \caption{\label{fig:OPTim_NGC4258}
             Optical image (DSS, red filter) of NGC\,4258. Displayed are the central $4\arcmin$ with North up and East to the left. 
              The colour scaling is linear with white corresponding to the median background and black to the $0.01\%$ pixels with the highest intensity.  
           }
\end{figure}
\begin{figure}
   \centering
   \includegraphics[angle=0,height=3.11cm]{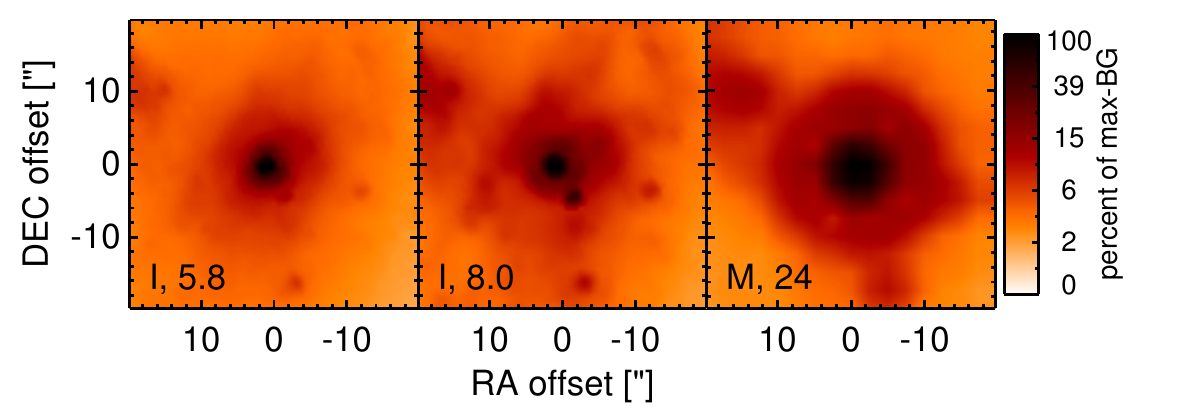}
    \caption{\label{fig:INTim_NGC4258}
             \spitzerr MIR images of NGC\,4258. Displayed are the inner $40\arcsec$ with North up and East to the left. The colour scaling is logarithmic with white corresponding to median background and black to the $0.1\%$ pixels with the highest intensity.
             The label in the bottom left states instrument and central wavelength of the filter in $\mu$m (I: IRAC, M: MIPS). 
             Note that the apparent off-nuclear compact source close to the nucleus in the IRAC $8.0\,\mu$m image is an instrumental artefact.
           }
\end{figure}
\begin{figure}
   \centering
   \includegraphics[angle=0,width=8.500cm]{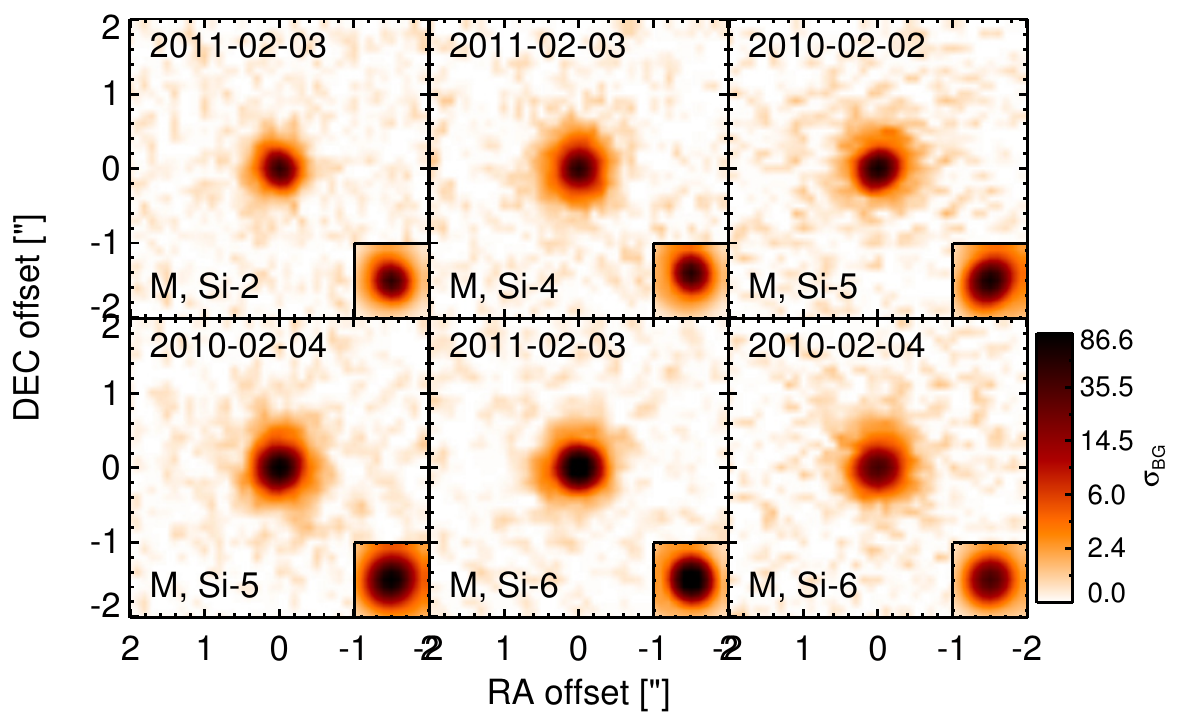}
    \caption{\label{fig:HARim_NGC4258}
             Subarcsecond-resolution MIR images of NGC\,4258 sorted by increasing filter wavelength. 
             Displayed are the inner $4\arcsec$ with North up and East to the left. 
             The colour scaling is logarithmic with white corresponding to median background and black to the $75\%$ of the highest intensity of all images in units of $\sigbg$.
             The inset image shows the central arcsecond of the PSF from the calibrator star, scaled to match the science target.
             The labels in the bottom left state instrument and filter names (C: COMICS, M: Michelle, T: T-ReCS, V: VISIR).
           }
\end{figure}
\begin{figure}
   \centering
   \includegraphics[angle=0,width=8.50cm]{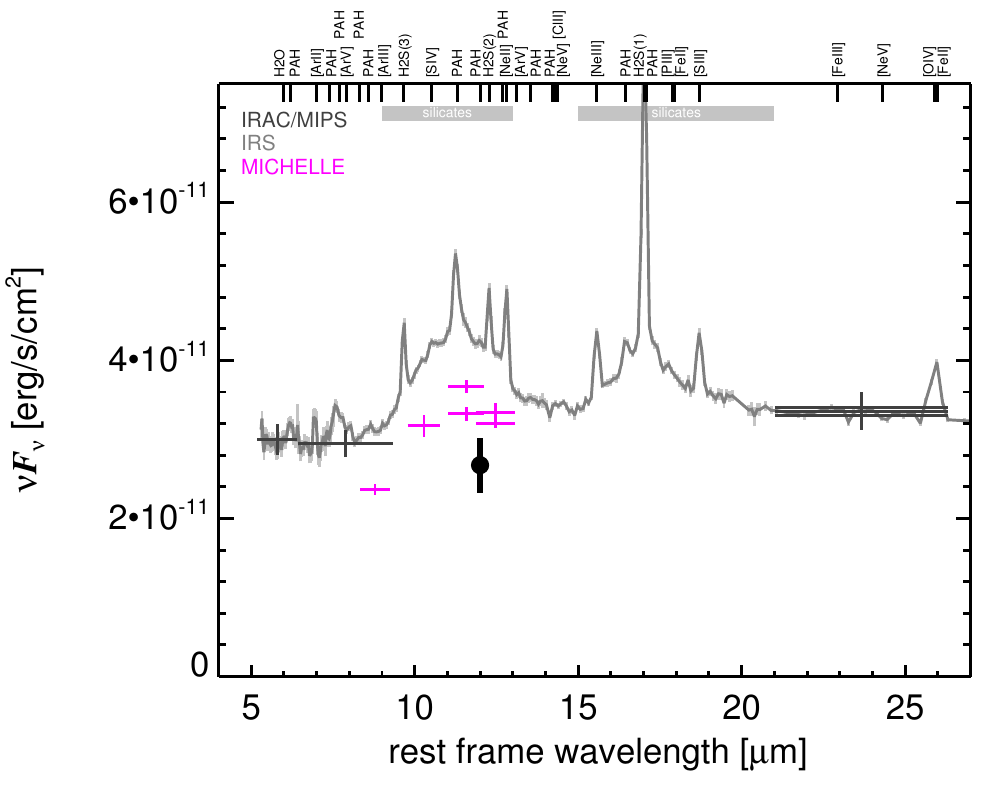}
   \caption{\label{fig:MISED_NGC4258}
      MIR SED of NGC\,4258. The description  of the symbols (if present) is the following.
      Grey crosses and  solid lines mark the \spitzer/IRAC, MIPS and IRS data. 
      The colour coding of the other symbols is: 
      green for COMICS, magenta for Michelle, blue for T-ReCS and red for VISIR data.
      Darker-coloured solid lines mark spectra of the corresponding instrument.
      The black filled circles mark the nuclear 12 and $18\,\mu$m  continuum emission estimate from the data.
      The ticks on the top axis mark positions of common MIR emission lines, while the light grey horizontal bars mark wavelength ranges affected by the silicate 10 and 18$\mu$m features.}
\end{figure}
\clearpage

\twocolumn[\begin{@twocolumnfalse}  
\subsection{NGC\,4261 --  3C\,270 -- VCC\,345}\label{app:NGC4261}
NGC\,4261 is a radio-loud elliptical galaxy in the Virgo cluster at a distance of $D=$ $31.7\pm3.2\,$Mpc (NED redshift-independent median) with a FR\,I radio morphology and a LINER nucleus with polarized broad emission lines \citep{barth_polarized_1999}.
It features a compact radio core and a pair of symmetric jets extending east-west (PA$\sim88\degree$; \citealt{birkinshaw_orientations_1985,jones_vlba_1997}), which are also visible in UV and X-ray images \citep{chiaberge_what_2003,zezas_chandra_2005}.
In addition to a compact NLR, a nearly edge-on nuclear dust disc roughly perpendicular to the jet axis was imaged (diameter$\sim1.7\arcsec\sim262\,$pc; PA$\sim-16\degree$; \citealt{jaffe_large_1993,jaffe_nuclear_1996}).
The first $N$-band detection of NGC\,4261 was reported by \cite{impey_infrared_1986}, followed up by \isoo \citep{siebenmorgen_isocam_2004} and \spitzer/IRAC, IRS and MIPS observations.
NGC\,4261 appears as a compact nucleus in the corresponding IRAC and MIPS images, while the nucleus is embedded within elliptical host emission in the IRAC images.
We measure the nuclear component which provides values in between the non-stellar and total flux values in \cite{tang_infrared-red_2009}.
The IRS LR staring-mode spectrum exhibits silicate 10$\,\mu$m emission, a weak PAH 11.3$\,\mu$m feature and a blue spectral slope in $\nu F_\nu$-space (see also \citealt{leipski_spitzer_2009}).
Therefore, the arcsecond-scale MIR SED is rather dominated by old-stellar and possibly AGN emission rather than star formation.
The nuclear region of NGC\,4261 was observed with COMICS in two $N$-band filters in 2005 (unpublished, to our knowledge), with T-ReCS in the Si2 filter in 2008 \citep{mason_nuclear_2012}, and with VISIR in total in six $N$ and one $Q$-band filter in 2006 \citep{van_der_wolk_dust_2010}, 2009 \citep{asmus_mid-infrared_2011} and 2010 (unpublished, to our knowledge).
A compact MIR nucleus is weakly detected in the T-ReCS and VISIR images (except for PAH2\_2), while it remained undetected in the COMICS images.
The low S/N of the images prohibits a quantitative extension analysis.
Our remeasured nuclear fluxes agree with the previous publications except for the SIC measurement as discussed already in \cite{asmus_mid-infrared_2011}. 
The nuclear photometry is on average $\sim44\%$ lower than the \spitzerr spectrophotometry and has a red spectral slope, which suggests that the subarcsecond MIR SED is AGN-dominated with most of the host emission being resolved out. 
We note that, on the one hand, the SIC flux is significantly higher and the PAH2\_2 and N11.7 upper limits are significantly lower than the rest of the nuclear photometry for unknown reasons. 
\newline\end{@twocolumnfalse}]

\begin{figure}
   \centering
   \includegraphics[angle=0,width=8.500cm]{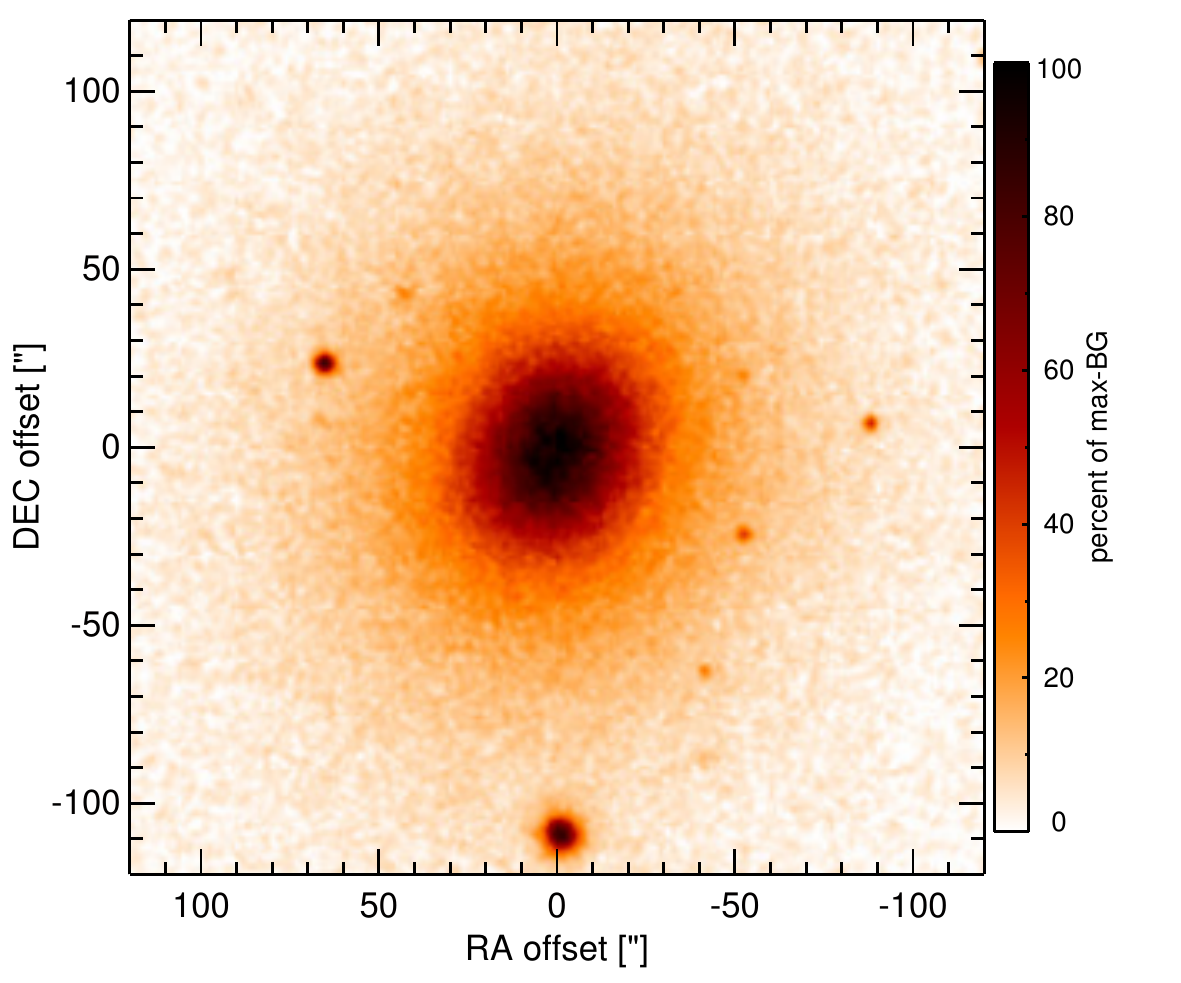}
    \caption{\label{fig:OPTim_NGC4261}
             Optical image (DSS, red filter) of NGC\,4261. Displayed are the central $4\arcmin$ with North up and East to the left. 
              The colour scaling is linear with white corresponding to the median background and black to the $0.01\%$ pixels with the highest intensity.  
           }
\end{figure}
\begin{figure}
   \centering
   \includegraphics[angle=0,height=3.11cm]{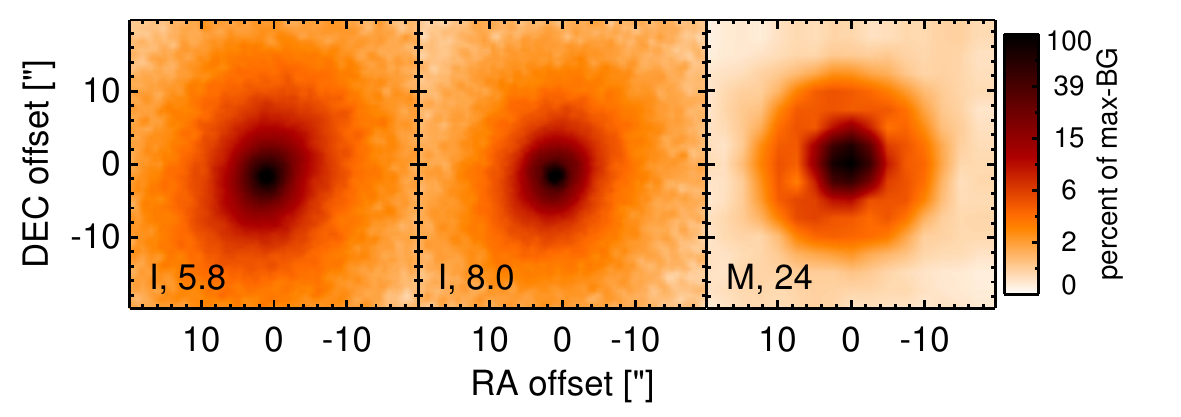}
    \caption{\label{fig:INTim_NGC4261}
             \spitzerr MIR images of NGC\,4261. Displayed are the inner $40\arcsec$ with North up and East to the left. The colour scaling is logarithmic with white corresponding to median background and black to the $0.1\%$ pixels with the highest intensity.
             The label in the bottom left states instrument and central wavelength of the filter in $\mu$m (I: IRAC, M: MIPS). 
           }
\end{figure}
\begin{figure}
   \centering
   \includegraphics[angle=0,width=8.500cm]{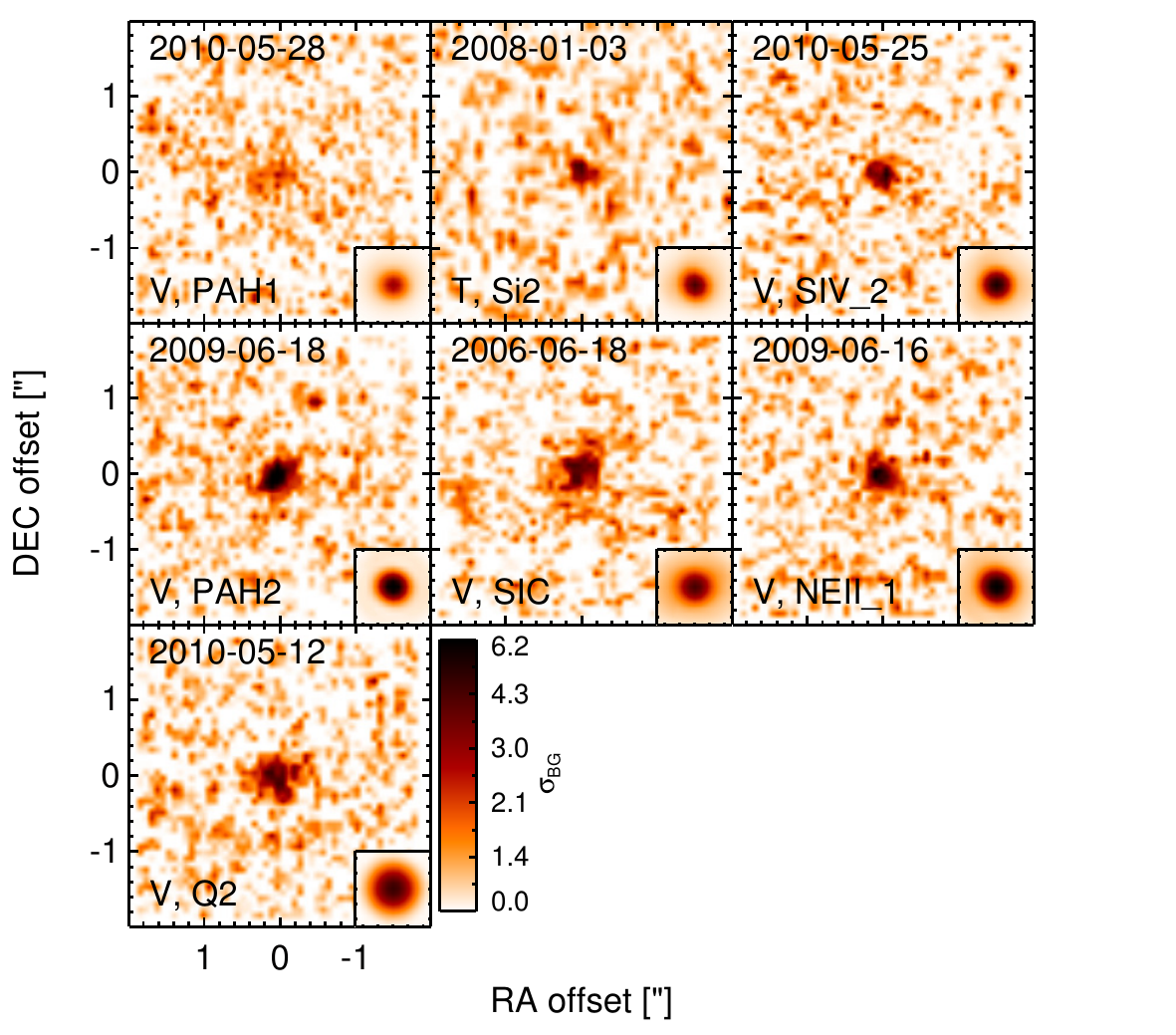}
    \caption{\label{fig:HARim_NGC4261}
             Subarcsecond-resolution MIR images of NGC\,4261 sorted by increasing filter wavelength. 
             Displayed are the inner $4\arcsec$ with North up and East to the left. 
             The colour scaling is logarithmic with white corresponding to median background and black to the $75\%$ of the highest intensity of all images in units of $\sigbg$.
             The inset image shows the central arcsecond of the PSF from the calibrator star, scaled to match the science target.
             The labels in the bottom left state instrument and filter names (C: COMICS, M: Michelle, T: T-ReCS, V: VISIR).
           }
\end{figure}
\begin{figure}
   \centering
   \includegraphics[angle=0,width=8.50cm]{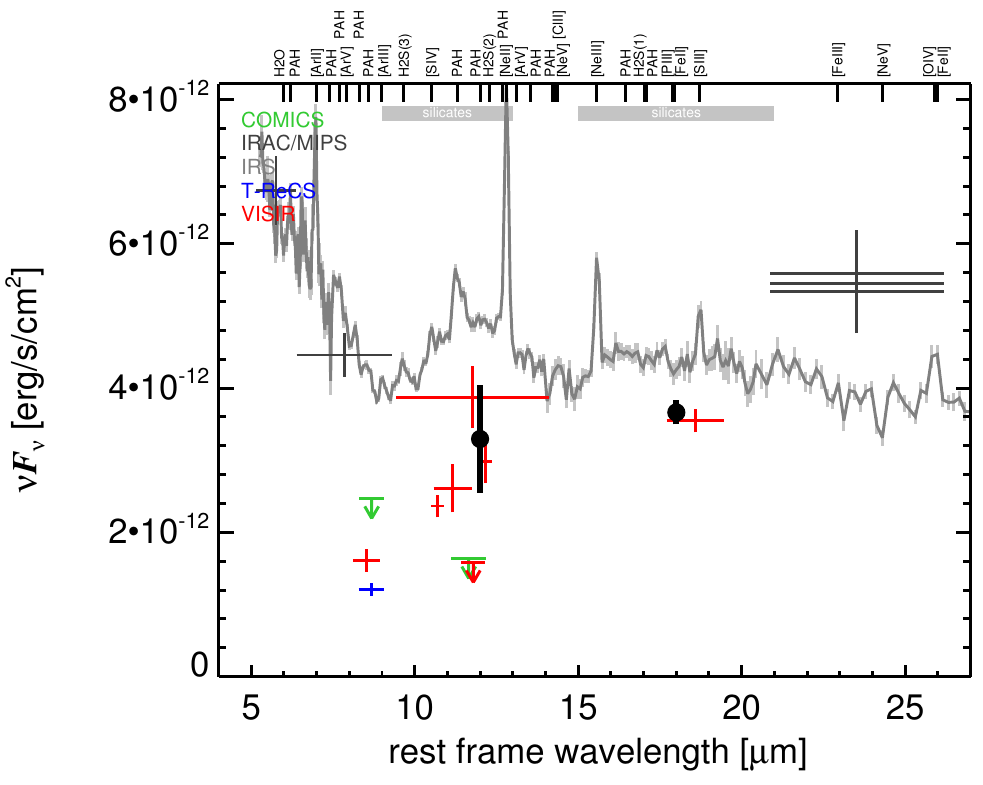}
   \caption{\label{fig:MISED_NGC4261}
      MIR SED of NGC\,4261. The description  of the symbols (if present) is the following.
      Grey crosses and  solid lines mark the \spitzer/IRAC, MIPS and IRS data. 
      The colour coding of the other symbols is: 
      green for COMICS, magenta for Michelle, blue for T-ReCS and red for VISIR data.
      Darker-coloured solid lines mark spectra of the corresponding instrument.
      The black filled circles mark the nuclear 12 and $18\,\mu$m  continuum emission estimate from the data.
      The ticks on the top axis mark positions of common MIR emission lines, while the light grey horizontal bars mark wavelength ranges affected by the silicate 10 and 18$\mu$m features.}
\end{figure}
\clearpage

\twocolumn[\begin{@twocolumnfalse}  
\subsection{NGC\,4278}\label{app:NGC4278}
NGC\,4278 is a radio-loud dusty elliptical galaxy at a distance of $D=$ $16.1\pm3.3\,$Mpc (NED redshift-independent median) with a broad-line LINER nucleus (see \citealt{pellegrini_agn_2012} for a recent detailed study). 
It features a compact flat-spectrum radio core with a parsec-scale two-sided jet (PA$\sim -40\degree$; \citealt{giroletti_two-sided_2005,nagar_radio_2005}). 
The \oiii emission extends to kiloparsec-scales (PA$\sim70\degree$; \citealt{sarzi_sauron_2006}).
The first unsuccessful attempt to detect the nucleus of NGC\,4278 in the MIR was made by \cite{puschell_nonstellar_1981}, while marginal detections are reported by \cite{heckman_infrared_1983} and \cite{impey_infrared_1986}.
NGC\,4278 was observed and detected by \iras, \isoo \citep{xilouris_dust_2004,temi_cold_2004}, and \spitzerr/IRAC, IRS and MIPS \citep{tang_infrared-red_2009,tang_multiphase_2011,dudik_spitzer_2009,pereira-santaella_mid-infrared_2010}.
The corresponding IRAC and MIPS images are dominated by extended host emission without a clearly separable unresolved nuclear component. 
Our nuclear  IRAC $5.8$ and $8.0\,\mu$m and MIPS $24\,\mu$m photometry provides values in between the total and nuclear excess fluxes reported in \cite{tang_infrared-red_2009}.
The nuclear-extracted IRS spectrum exhibits possible weak silicate 10 and $18\,\mu$m emission, a PAH 11.3\,$\mu$m feature and a blue curved spectral slope in $\nu F_\nu$-space (see also \citealt{tang_multiphase_2011,mason_nuclear_2012}).
Thus, the arcsecond-scale MIR SED is  dominated by old-stellar host emission without significant star formation.
The nuclear region of NGC\,4278 was observed with Michelle in the N' filter in 2008 \citep{mason_nuclear_2012}, and a compact nucleus is weakly detected. 
The low S/N of the detection prohibits any quantitative extension analysis.
Our reanalysis of the image provides a nuclear flux value consistent with \cite{mason_nuclear_2012} and $\sim80\%$ lower than the \spitzerr spectrophotometry.
\newline\end{@twocolumnfalse}]

\begin{figure}
   \centering
   \includegraphics[angle=0,width=8.500cm]{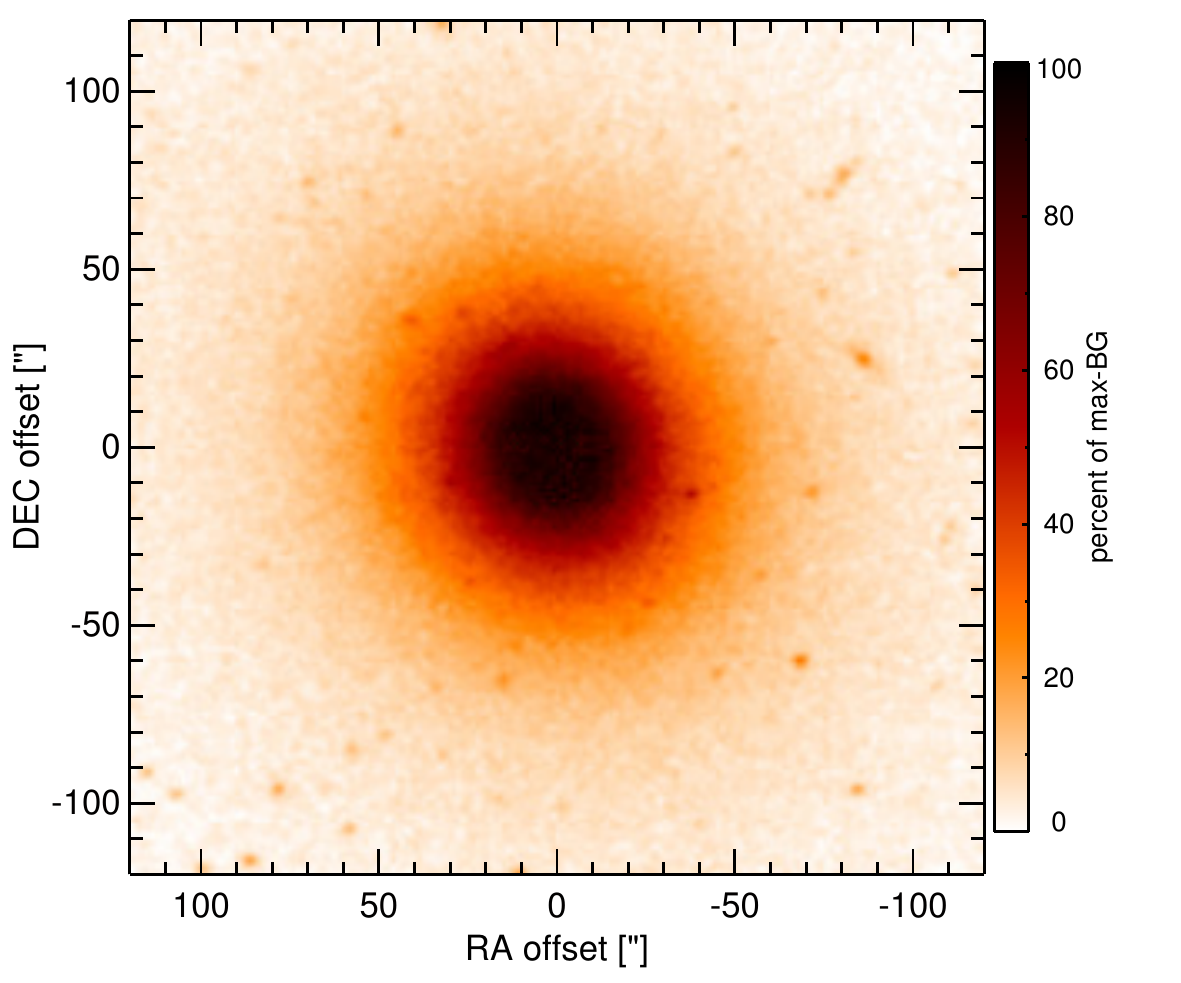}
    \caption{\label{fig:OPTim_NGC4278}
             Optical image (DSS, red filter) of NGC\,4278. Displayed are the central $4\arcmin$ with North up and East to the left. 
              The colour scaling is linear with white corresponding to the median background and black to the $0.01\%$ pixels with the highest intensity.  
           }
\end{figure}
\begin{figure}
   \centering
   \includegraphics[angle=0,height=3.11cm]{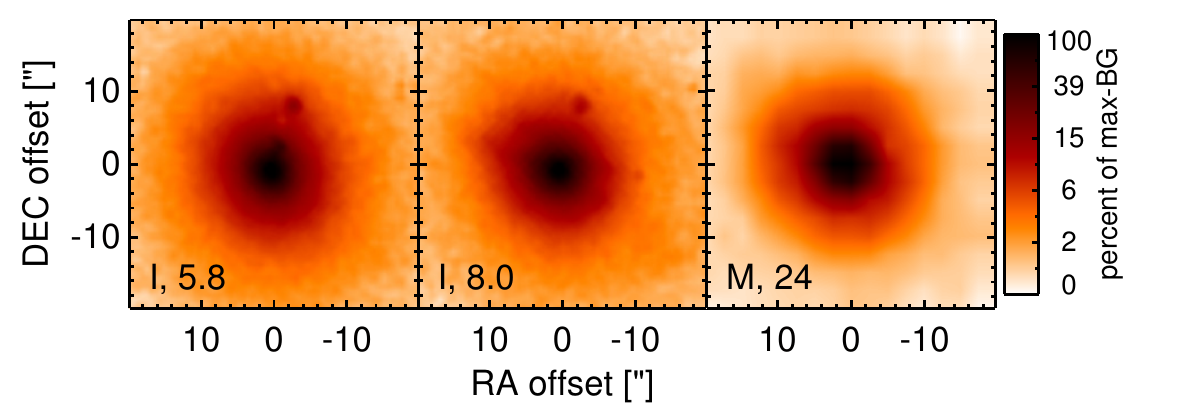}
    \caption{\label{fig:INTim_NGC4278}
             \spitzerr MIR images of NGC\,4278. Displayed are the inner $40\arcsec$ with North up and East to the left. The colour scaling is logarithmic with white corresponding to median background and black to the $0.1\%$ pixels with the highest intensity.
             The label in the bottom left states instrument and central wavelength of the filter in $\mu$m (I: IRAC, M: MIPS). 
           }
\end{figure}
\begin{figure}
   \centering
   \includegraphics[angle=0,height=3.11cm]{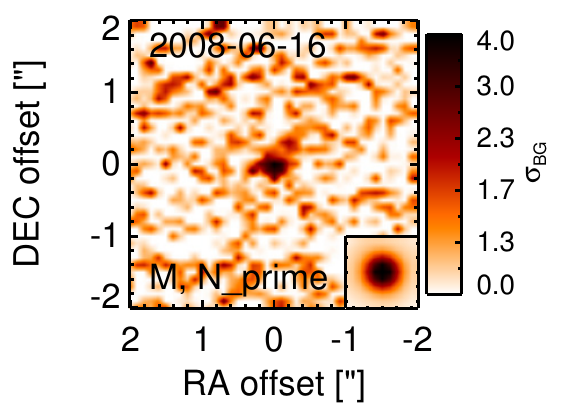}
    \caption{\label{fig:HARim_NGC4278}
             Subarcsecond-resolution MIR images of NGC\,4278 sorted by increasing filter wavelength. 
             Displayed are the inner $4\arcsec$ with North up and East to the left. 
             The colour scaling is logarithmic with white corresponding to median background and black to the $75\%$ of the highest intensity of all images in units of $\sigbg$.
             The inset image shows the central arcsecond of the PSF from the calibrator star, scaled to match the science target.
             The labels in the bottom left state instrument and filter names (C: COMICS, M: Michelle, T: T-ReCS, V: VISIR).
           }
\end{figure}
\begin{figure}
   \centering
   \includegraphics[angle=0,width=8.50cm]{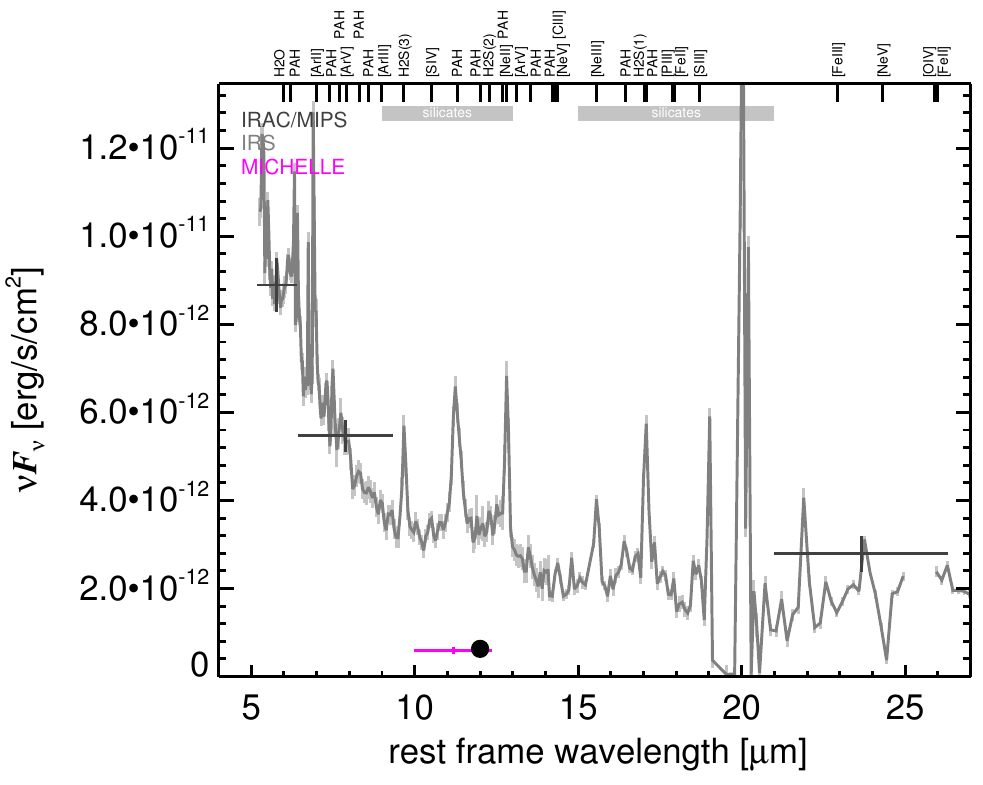}
   \caption{\label{fig:MISED_NGC4278}
      MIR SED of NGC\,4278. The description  of the symbols (if present) is the following.
      Grey crosses and  solid lines mark the \spitzer/IRAC, MIPS and IRS data. 
      The colour coding of the other symbols is: 
      green for COMICS, magenta for Michelle, blue for T-ReCS and red for VISIR data.
      Darker-coloured solid lines mark spectra of the corresponding instrument.
      The black filled circles mark the nuclear 12 and $18\,\mu$m  continuum emission estimate from the data.
      The ticks on the top axis mark positions of common MIR emission lines, while the light grey horizontal bars mark wavelength ranges affected by the silicate 10 and 18$\mu$m features.}
\end{figure}
\clearpage

\twocolumn[\begin{@twocolumnfalse}  
\subsection{NGC\,4303 -- M61 -- VCC\,508}\label{app:NGC4303}
NGC\,4303 is a face-on grand-design spiral galaxy in the Virgo cluster at a distance of $D=$ $15.2 \pm 3.1$\,Mpc \citep{tully_nearby_1988} with an active nucleus classified either as a H\,II \citep{ho_search_1997-1} or as a Sy\,2 \citep{veron-cetty_catalogue_2010}.
The nucleus contains a young super-star cluster, which appears to be the primary source of the ionizing radiation \citep{colina_detection_2002} with the presence of a weak AGN still possible \citep{jimenez-bailon_nuclear_2003}.
Therefore, we conservatively treat this object as an uncertain AGN.
The first $N$-band photometry of NGC\,4303 was performed by \cite{kleinmann_observations_1970}, \cite{rieke_infrared_1972} and \cite{rieke_10_1978}, followed by IRTF observations in 1980-82 \citep{scoville_10_1983,cizdziel_multiaperture_1985}.
All these measurements turn out to be dominated by circum-nuclear emission, which is visible in the \spitzer/IRAC and MIPS images in addition to the large scale spiral structure.
No unresolved component can be clearly distinguished from the nuclear MIR emission. 
Only the shortest wavelength setting of the IRS LR mapping-mode  spectrum is available for NGC\,4303. 
It roughly matches the IRAC photometry and indicates strong PAH emission, i.e., star formation in the central 4\arcsec $\sim 300$\,pc region.
We observed the nucleus of NGC\,4303 with VISIR in 2006 in the NEII\_1 filter \citep{horst_mid_2008}.
In our careful reanalysis of the data, we found the a compact MIR nucleus weakly detected with a flux ten times lower than all previous photometry including \spitzer.
The low S/N prevents any extension analysis.
Because this MIR emission from the central $\sim 30\,$pc is still more than an order of magnitude higher than expected from the MIR--X-ray correlation for AGN, it is probably dominated by the nuclear cluster (see also \citealt{asmus_mid-infrared_2011}).
\newline\end{@twocolumnfalse}]

\begin{figure}
   \centering
   \includegraphics[angle=0,width=8.500cm]{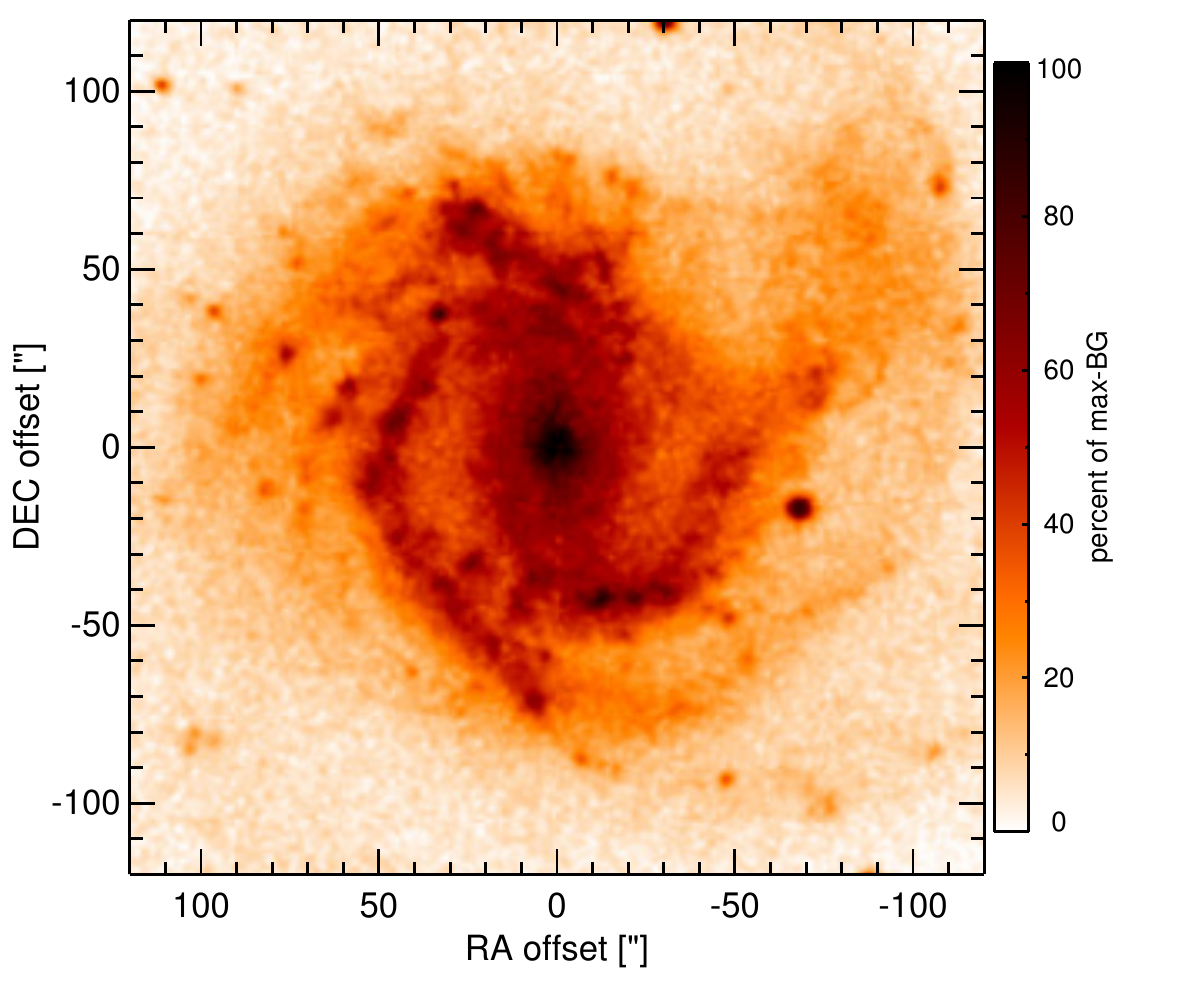}
    \caption{\label{fig:OPTim_NGC4303}
             Optical image (DSS, red filter) of NGC\,4303. Displayed are the central $4\arcmin$ with North up and East to the left. 
              The colour scaling is linear with white corresponding to the median background and black to the $0.01\%$ pixels with the highest intensity.  
           }
\end{figure}
\begin{figure}
   \centering
   \includegraphics[angle=0,height=3.11cm]{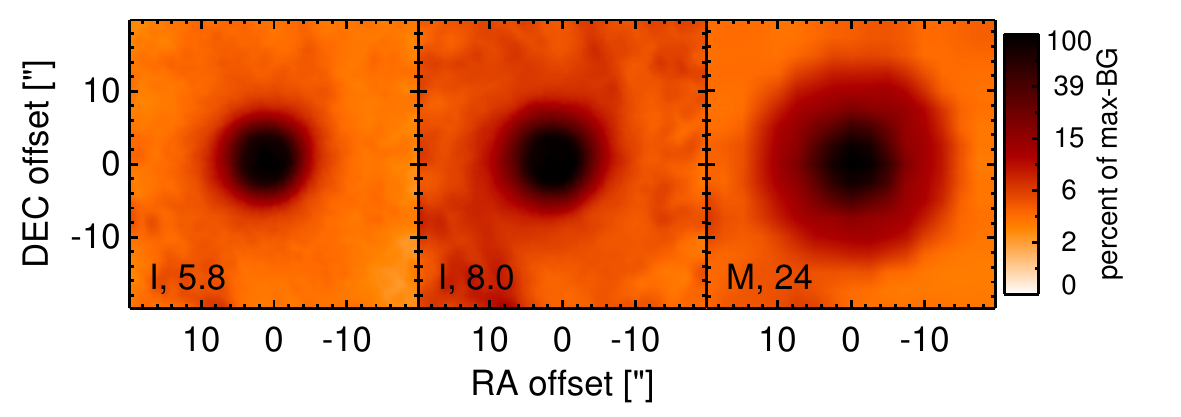}
    \caption{\label{fig:INTim_NGC4303}
             \spitzerr MIR images of NGC\,4303. Displayed are the inner $40\arcsec$ with North up and East to the left. The colour scaling is logarithmic with white corresponding to median background and black to the $0.1\%$ pixels with the highest intensity.
             The label in the bottom left states instrument and central wavelength of the filter in $\mu$m (I: IRAC, M: MIPS). 
           }
\end{figure}
\begin{figure}
   \centering
   \includegraphics[angle=0,height=3.11cm]{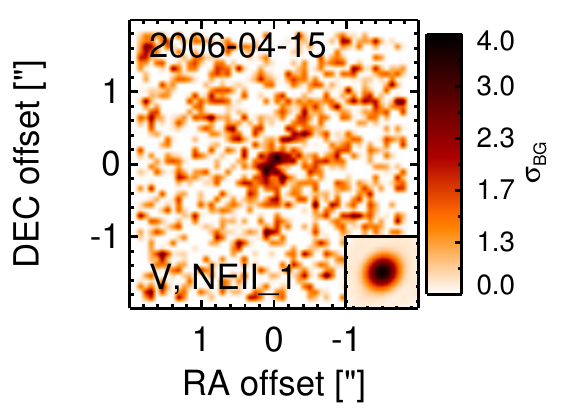}
    \caption{\label{fig:HARim_NGC4303}
             Subarcsecond-resolution MIR images of NGC\,4303 sorted by increasing filter wavelength. 
             Displayed are the inner $4\arcsec$ with North up and East to the left. 
             The colour scaling is logarithmic with white corresponding to median background and black to the $75\%$ of the highest intensity of all images in units of $\sigbg$.
             The inset image shows the central arcsecond of the PSF from the calibrator star, scaled to match the science target.
             The labels in the bottom left state instrument and filter names (C: COMICS, M: Michelle, T: T-ReCS, V: VISIR).
           }
\end{figure}
\begin{figure}
   \centering
   \includegraphics[angle=0,width=8.50cm]{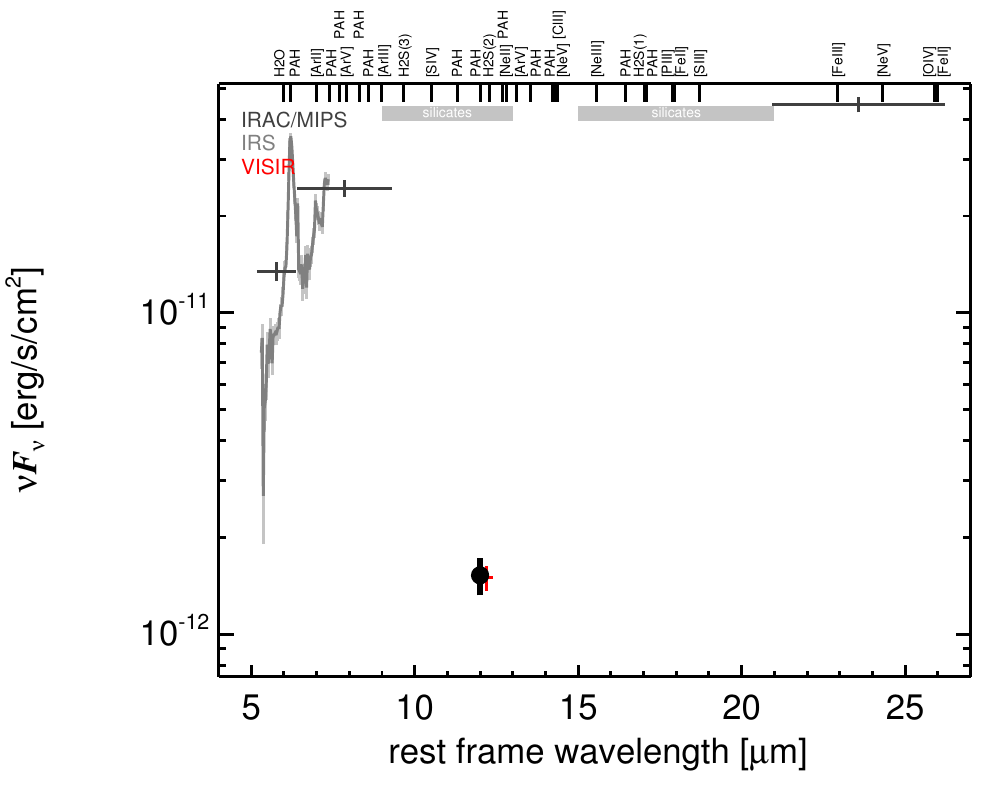}
   \caption{\label{fig:MISED_NGC4303}
      MIR SED of NGC\,4303. The description  of the symbols (if present) is the following.
      Grey crosses and  solid lines mark the \spitzer/IRAC, MIPS and IRS data. 
      The colour coding of the other symbols is: 
      green for COMICS, magenta for Michelle, blue for T-ReCS and red for VISIR data.
      Darker-coloured solid lines mark spectra of the corresponding instrument.
      The black filled circles mark the nuclear 12 and $18\,\mu$m  continuum emission estimate from the data.
      The ticks on the top axis mark positions of common MIR emission lines, while the light grey horizontal bars mark wavelength ranges affected by the silicate 10 and 18$\mu$m features.}
\end{figure}
\clearpage

\twocolumn[\begin{@twocolumnfalse}  
\subsection{NGC\,4374 -- M84 -- VCC\,763 -- 3C\,272.1}\label{app:NGC4374}
NGC\,4374 is a dusty giant elliptical in the Virgo cluster at a distance of $D=$ $17.1 \pm 2.4$\,Mpc (NED redshift-independent median) with FR\,I radio morphology and an radio-loud AGN classified either as LINER \citep{ho_search_1997-1} or Sy\,2 \citep{veron-cetty_catalogue_2010}.
Two parallel dust lanes are seen perpendicular to the radio jet axis with one crossing the nucleus (PA$\sim60\degree$; \citealt{hansen_properties_1985,jaffe_hubble_1994,bower_nuclear_1997}).
The first $N$-band photometry of NGC\,4374 was obtained with IRTF in 1983 \citep{impey_infrared_1986} where the nucleus was weakly detected at 21\,mJy (5.7\arcsec aperture).
After \iras, NGC\,4374 was also observed with \iso/ISOCAM where it appeared extended and diffuse \citep{ferrari_survey_2002,siebenmorgen_isocam_2004,xilouris_dust_2004}.
This is also the case in the IRAC $5.8$ and $8.0\,\mu$m: the nucleus appears as an extended emission peak without any clearly separable unresolved component.
Only in the MIPS $24\,\mu$m image, an unresolved component embedded in diffuse emission is dominating. 
We measured the nuclear component in the  IRAC $5.8$ and $8.0\,\mu$m and MIPS $24\,\mu$m images, which yields much lower values than the total fluxes published in the literature \citep{tang_infrared-red_2009,temi_spitzer_2009}.
For extended objects such as NGC\,4374, the PBCD IRS LR mapping mode spectrum is not very reliable but matches qualitatively the more accurate version by \cite{leipski_spitzer_2009}.
While the latter authors presented a complete MIR spectrum we could only retrieve the shortest wavelength setting from the \spitzerr archive.
The MIR SED from \citeauthor{leipski_spitzer_2009} exhibits a weak silicate $10\,\mu$m absorption feature and prominent PAH 11.3\,$\mu$m emission, i.e., star formation might be present. 
The nucleus of NGC\,4374 was observed with T-ReCS in the Si2 filter in 2008 during four nights (two of which are published in \citealt{mason_nuclear_2012}).
In addition, we observed it with COMICS in the N11.7 filter in 2009, and others performed VISIR PAH2\_2 imaging in 2010 (both unpublished, to our knowledge). 
In all these subarcsecond-resolution $N$-band images, the object remained undetected.
Together with the derived upper limit, which are significantly lower than the \spitzerr data, this indicates that the MIR emission of NGC\,4374 is totally host-dominated even at arcsecond scales.
\newline\end{@twocolumnfalse}]

\begin{figure}
   \centering
   \includegraphics[angle=0,width=8.500cm]{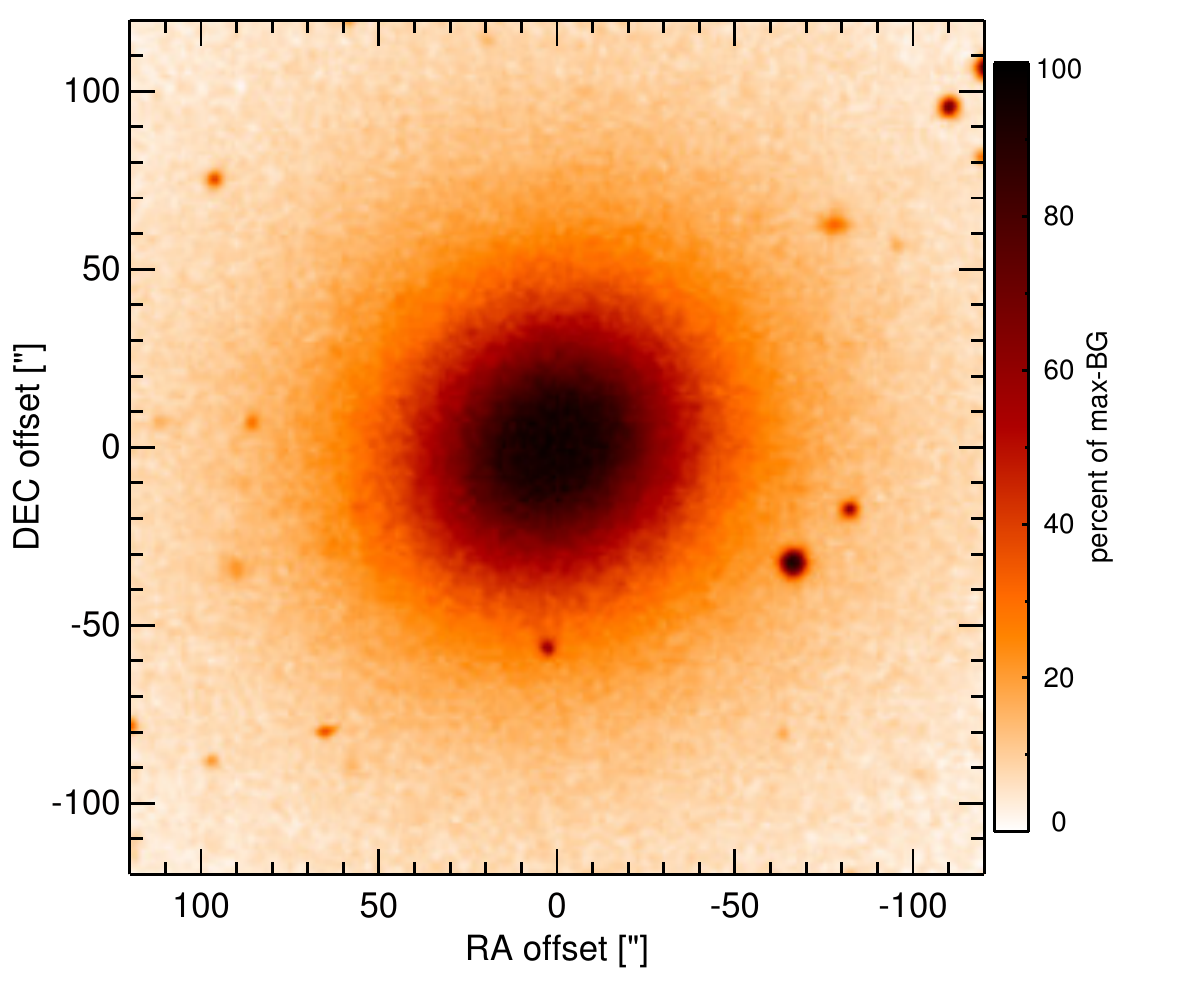}
    \caption{\label{fig:OPTim_NGC4374}
             Optical image (DSS, red filter) of NGC\,4374. Displayed are the central $4\arcmin$ with North up and East to the left. 
              The colour scaling is linear with white corresponding to the median background and black to the $0.01\%$ pixels with the highest intensity.  
           }
\end{figure}
\begin{figure}
   \centering
   \includegraphics[angle=0,height=3.11cm]{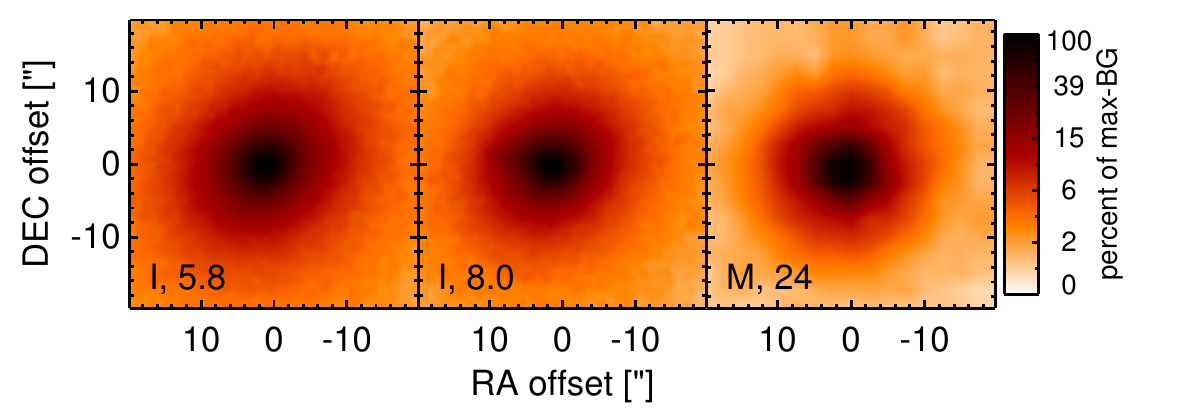}
    \caption{\label{fig:INTim_NGC4374}
             \spitzerr MIR images of NGC\,4374. Displayed are the inner $40\arcsec$ with North up and East to the left. The colour scaling is logarithmic with white corresponding to median background and black to the $0.1\%$ pixels with the highest intensity.
             The label in the bottom left states instrument and central wavelength of the filter in $\mu$m (I: IRAC, M: MIPS). 
           }
\end{figure}
\begin{figure}
   \centering
   \includegraphics[angle=0,width=8.50cm]{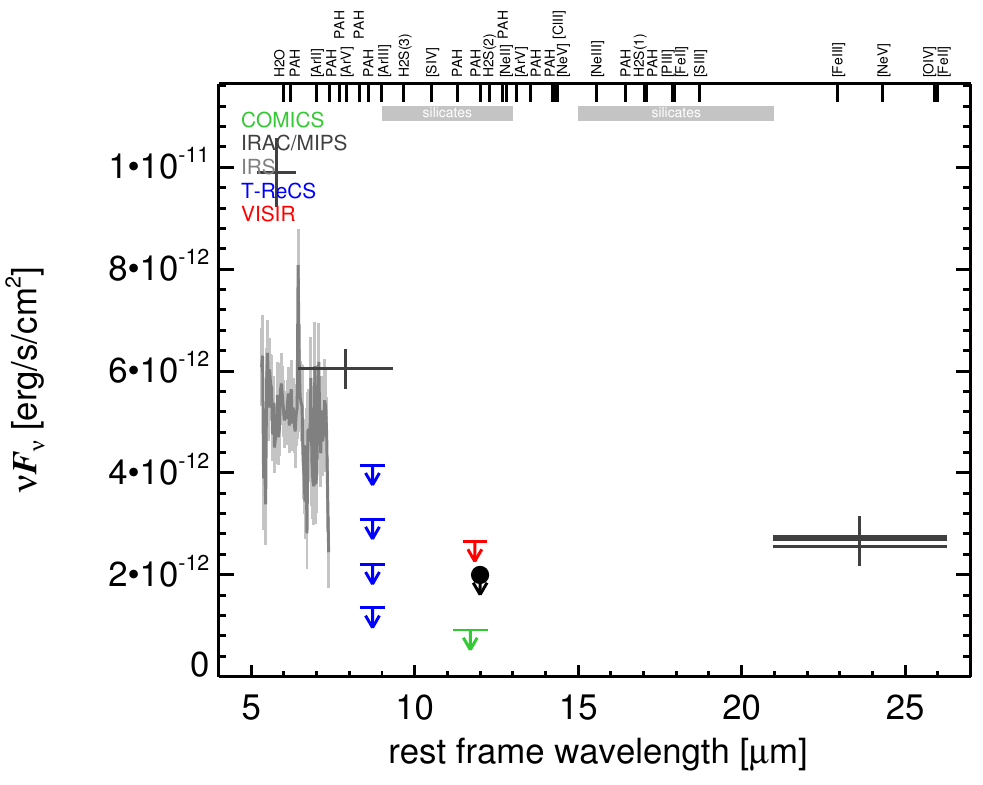}
   \caption{\label{fig:MISED_NGC4374}
      MIR SED of NGC\,4374. The description  of the symbols (if present) is the following.
      Grey crosses and  solid lines mark the \spitzer/IRAC, MIPS and IRS data. 
      The colour coding of the other symbols is: 
      green for COMICS, magenta for Michelle, blue for T-ReCS and red for VISIR data.
      Darker-coloured solid lines mark spectra of the corresponding instrument.
      The black filled circles mark the nuclear 12 and $18\,\mu$m  continuum emission estimate from the data.
      The ticks on the top axis mark positions of common MIR emission lines, while the light grey horizontal bars mark wavelength ranges affected by the silicate 10 and 18$\mu$m features.}
\end{figure}
\clearpage

\twocolumn[\begin{@twocolumnfalse}  
\subsection{NGC\,4388 -- VCC\,836}\label{app:NGC4388}
NGC\,4388 is an almost edge-on spiral galaxy in the Virgo cluster \citep{phillips_ngc_1982} at a distance of $D=$ $19.2\pm6.6\,$Mpc (NED redshift-independent median) with a Sy\,2 nucleus \citep{trippe_multi-wavelength_2010} with broad emission lines in polarized light \citep{shields_reality_1996}.
The AGN is one of the X-ray brightest objects of its class in terms of flux.
It is an X-ray ``buried" AGN candidate \citep{noguchi_new_2009} and belongs to the nine-month BAT AGN sample.
NGC\,4388 possesses a very extended NLR over several kiloparsecs towards the north-east and somewhat to the south-west  (PA$\sim30\degree$; e.g.,\citealt{colina_extended_1987,pogge_extended_1988,veilleux_galactic-scale_1999,yoshida_discovery_2002}), which roughly coincides with the radio morphology  (e.g., \citealt{stone_collimated_1988,hummel_anomalous_1991,falcke_hst_1998}).
It also features nuclear water maser disc \citep{braatz_green_2004}.
After first being detected in the MIR with \iras, NGC\,4388 was followed up with ground-based MIR instruments \citep{scoville_10_1983,devereux_infrared_1987,roche_atlas_1991,stone_collimated_1988}, and the space-based \isoo \citep{boselli_mid-ir_1998,boselli_mid-ir_2003,roussel_atlas_2001,ramos_almeida_mid-infrared_2007}.
The first subarcsecond $N$-band observations were performed with  Palomar 5\,m/MIRLIN in 2000 \citep{gorjian_10_2004} and ESO 3.6\,m/TIMMI2 in 2002 \citep{siebenmorgen_mid-infrared_2004}.
In both cases, an unresolved nucleus without any host emission was detected.
The host becomes visible as extended spiral-like emission in the \spitzer/IRAC and MIPS images, while a compact nucleus still dominates.
Our measured nuclear IRAC $5.8$ and $8.0\,\mu$m fluxes  are significantly lower than the values reported in \cite{gallimore_infrared_2010} but agree with the \spitzer/IRS LR staring mode spectrum.
The latter exhibits a deep silicate $10\,\mu$m absorption feature,  PAH emission and a steep red spectral slope in $\nu F_\nu$-space (see also \citealt{shi_9.7_2006,wu_spitzer/irs_2009,deo_mid-infrared_2009,tommasin_spitzer-irs_2010,gallimore_infrared_2010}).
The nuclear region of NGC\,4388 was observed with Michelle in the N' and Qa filters in 2006 \citep{ramos_almeida_infrared_2009}, and with VISIR in four narrow $N$-band filters in 2008 (partly published in \citealt{gandhi_resolving_2009}).
A marginally resolved nucleus was detected in all images but with uncertain morphology. 
In particular, a number of VISIR images suffer from low S/N $<10$.
The Michelle images have the highest S/N and show an emission ''finger`` extending $\sim 1.1\arcsec\sim 100\,$pc to the south-west, roughly consistent with the ionization cone (PA$\sim-135\degree$; see also \citealt{ramos_almeida_infrared_2009}).
We extract only the unresolved nuclear component, which provides fluxes consistent with \cite{ramos_almeida_infrared_2009} and \cite{gandhi_resolving_2009}.
The resulting nuclear MIR SED is on average $\sim51\%$ lower than the \spitzerr spectrophotometry and indicates a similar silicate absorption depth. 
This matches the TIMMI2 LR $N$-band spectrum from \cite{siebenmorgen_mid-infrared_2004}, which in addition does not possess any PAH feature emission.
Therefore, the arcsecond-scale MIR SED is significantly affected by star formation, which is located either around the nucleus or along our line of sight in the host disc.
\newline\end{@twocolumnfalse}]

\begin{figure}
   \centering
   \includegraphics[angle=0,width=8.500cm]{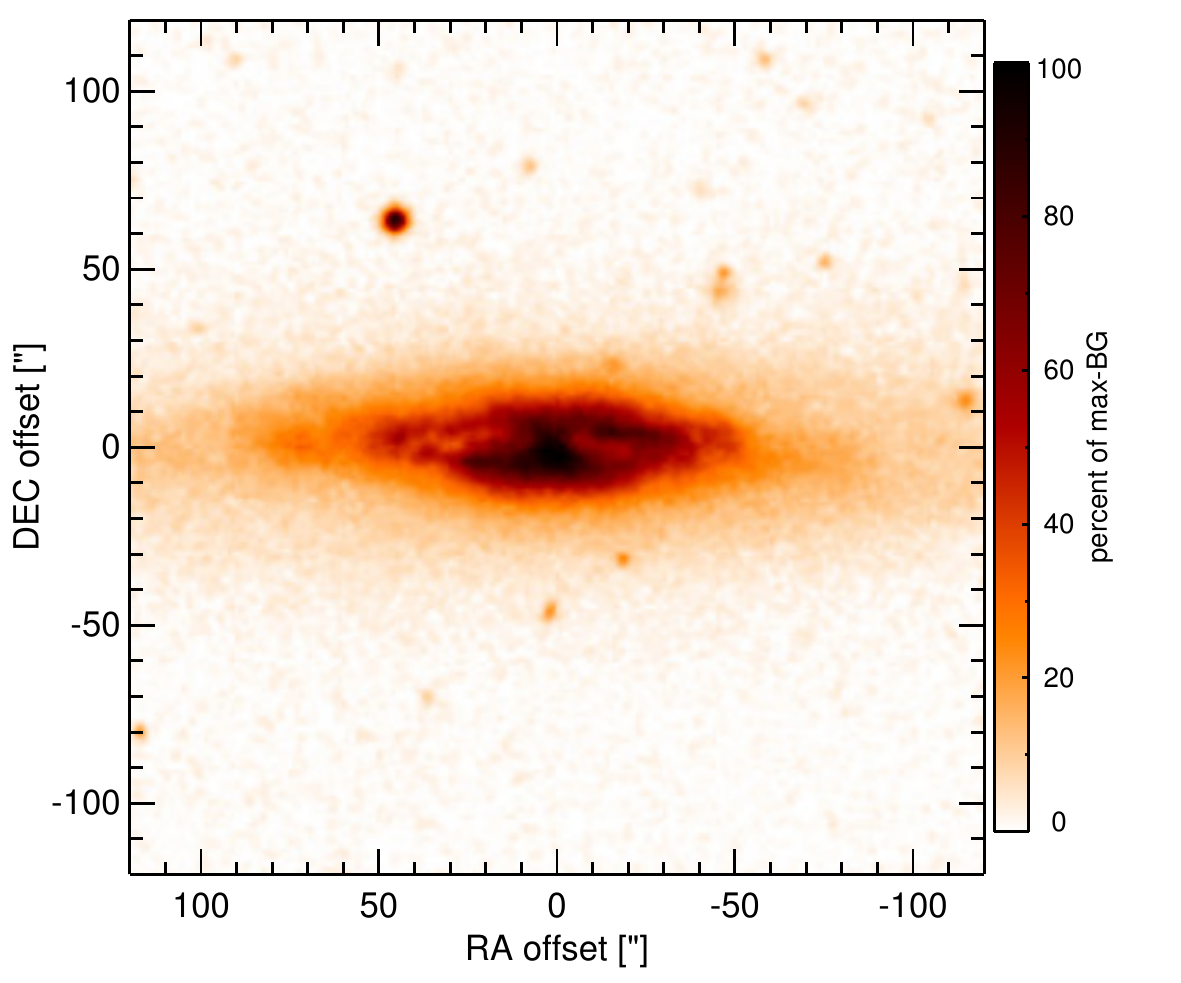}
    \caption{\label{fig:OPTim_NGC4388}
             Optical image (DSS, red filter) of NGC\,4388. Displayed are the central $4\arcmin$ with North up and East to the left. 
              The colour scaling is linear with white corresponding to the median background and black to the $0.01\%$ pixels with the highest intensity.  
           }
\end{figure}
\begin{figure}
   \centering
   \includegraphics[angle=0,height=3.11cm]{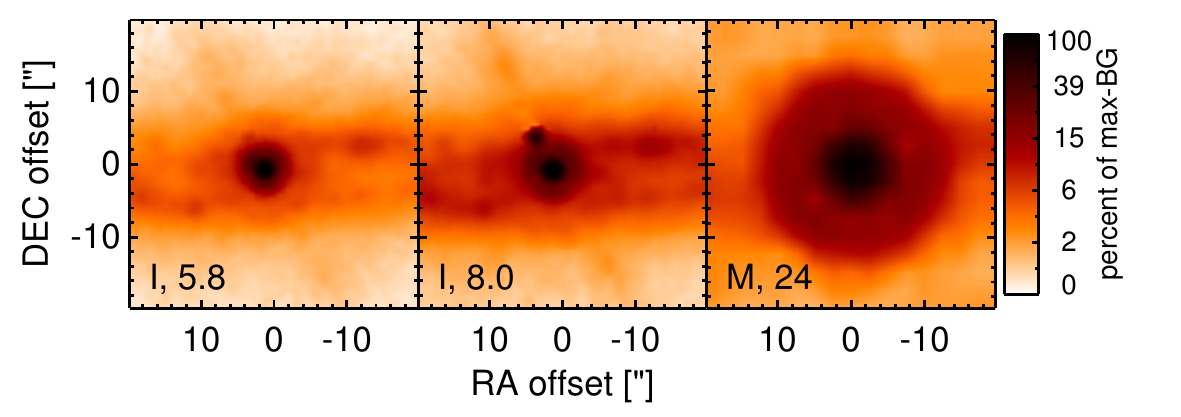}
    \caption{\label{fig:INTim_NGC4388}
             \spitzerr MIR images of NGC\,4388. Displayed are the inner $40\arcsec$ with North up and East to the left. The colour scaling is logarithmic with white corresponding to median background and black to the $0.1\%$ pixels with the highest intensity.
             The label in the bottom left states instrument and central wavelength of the filter in $\mu$m (I: IRAC, M: MIPS).
             Note that the apparent off-nuclear compact source in the IRAC $8.0\,\mu$m image is an instrumental artefact.
           }
\end{figure}
\begin{figure}
   \centering
   \includegraphics[angle=0,width=8.500cm]{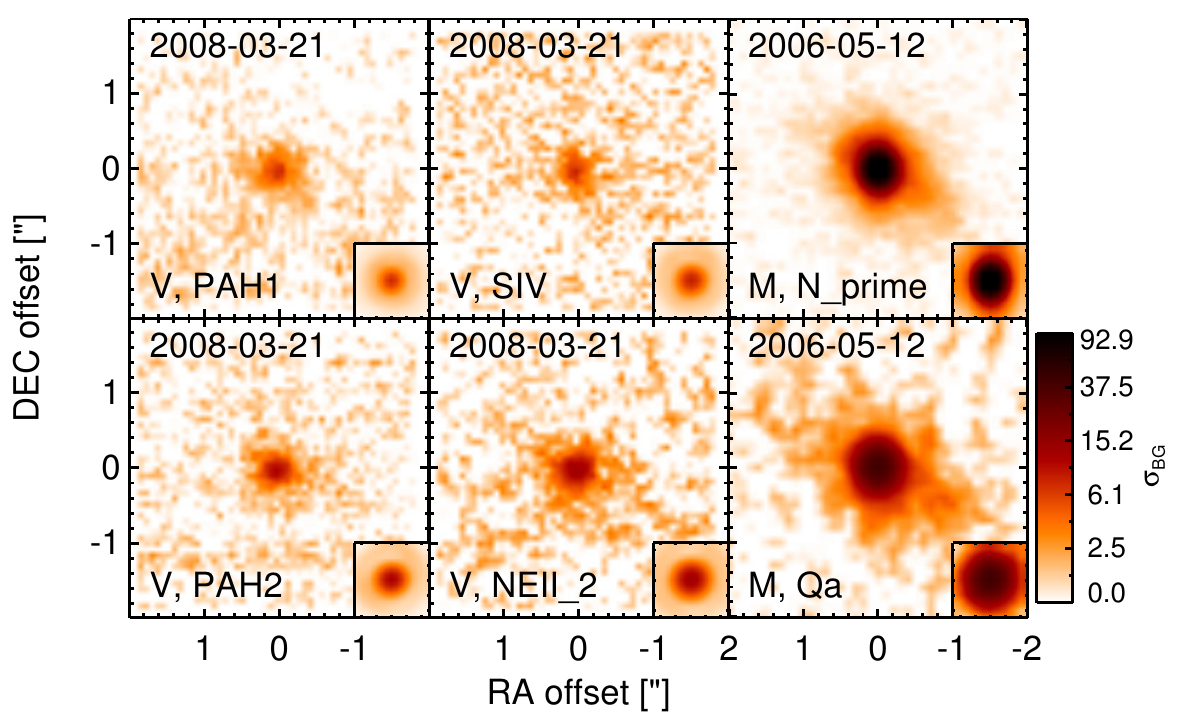}
    \caption{\label{fig:HARim_NGC4388}
             Subarcsecond-resolution MIR images of NGC\,4388 sorted by increasing filter wavelength. 
             Displayed are the inner $4\arcsec$ with North up and East to the left. 
             The colour scaling is logarithmic with white corresponding to median background and black to the $75\%$ of the highest intensity of all images in units of $\sigbg$.
             The inset image shows the central arcsecond of the PSF from the calibrator star, scaled to match the science target.
             The labels in the bottom left state instrument and filter names (C: COMICS, M: Michelle, T: T-ReCS, V: VISIR).
           }
\end{figure}
\begin{figure}
   \centering
   \includegraphics[angle=0,width=8.50cm]{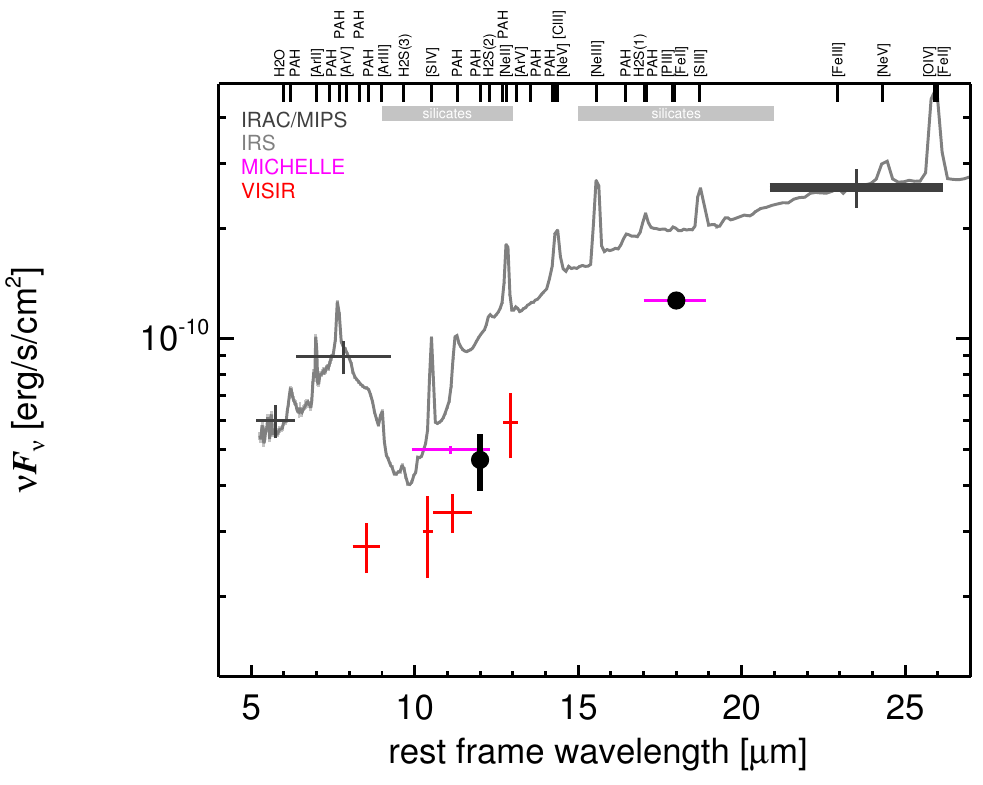}
   \caption{\label{fig:MISED_NGC4388}
      MIR SED of NGC\,4388. The description  of the symbols (if present) is the following.
      Grey crosses and  solid lines mark the \spitzer/IRAC, MIPS and IRS data. 
      The colour coding of the other symbols is: 
      green for COMICS, magenta for Michelle, blue for T-ReCS and red for VISIR data.
      Darker-coloured solid lines mark spectra of the corresponding instrument.
      The black filled circles mark the nuclear 12 and $18\,\mu$m  continuum emission estimate from the data.
      The ticks on the top axis mark positions of common MIR emission lines, while the light grey horizontal bars mark wavelength ranges affected by the silicate 10 and 18$\mu$m features.}
\end{figure}
\clearpage

\twocolumn[\begin{@twocolumnfalse}  
\subsection{NGC\,4395}\label{app:NGC4395}
NGC\,4395 is a close-by bulge-less dwarf spiral galaxy at a distance of $D=$ $4.3 \pm 1.1\,$Mpc (NED redshift-independent median) with a Sy\,1.8 nucleus \citep{veron-cetty_catalogue_2010} that belongs to nine-month BAT AGN sample.
This AGN is the least-luminous type~I AGN known \citep{filippenko_discovery_1989, filippenko_hst_1993} and possesses one of the lowest central black hole masses in an AGN, too \citep{filippenko_low-mass_2003,peterson_multiwavelength_2005}.
It is  highly variable at optical, UV and X-ray wavelengths (e.g., \citealt{cameron_correlated_2012}) and features a compact radio nucleus with a subparsec-scale outflow \citep{moran_nuclear_1999,ho_radio_2001,wrobel_inner_2001,wrobel_radio_2006}.
Furthermore, the AGN is embedded within a non-active elliptical nuclear star cluster with a diameter of $\sim0.4\arcsec\sim8\,$pc and a PA$\sim255\degree$, and complex bended \oiii emission with $\sim1\arcsec\sim20\,$pc along a PA$\sim270\degree$ \citep{matthews_wfpc2_1999,filippenko_low-mass_2003}.
The first $N$-band detection of NGC\,4395 is reported in \cite{maiolino_new_1995}, which was followed up with \isoo \citep{bendo_infrared_2002} and \spitzer/IRAC, IRS and MIPS observations.
The corresponding IRAC and MIPS images are dominated by a nuclear compact source, with the clumpy host emission being only weakly visible.
Still, our nuclear  IRAC $5.8$ and $8.0\,\mu$m and MIPS $24\,\mu$m fluxes are significantly lower than the total fluxes published in \cite{dale_spitzer_2009}.
The IRS LR staring-mode spectrum exhibits very weak PAH features, prominent forbidden emission lines, and a red spectral slope in $\nu F_\nu$-space (see also \citealt{marble_aromatic_2010,weaver_mid-infrared_2010,pereira-santaella_mid-infrared_2010}).
We observed the nuclear region of NGC\,4395 with Michelle in the Si-6 filter in 2010 and weakly detected a compact nucleus without further host emission.
The nucleus appears marginally resolved (FWHM $\sim 0.53\arcsec \sim 11\,$pc).
However, the current data are not sufficient for a robust classification of the nuclear extension in the MIR at subarcsecond scales.
The nuclear photometry is $\sim 50\%$ lower than the \spitzerr spectrophotometry.
The current MIR data are not sufficient to characterize the properties of the nuclear MIR emission. 
\newline\end{@twocolumnfalse}]

\begin{figure}
   \centering
   \includegraphics[angle=0,width=8.500cm]{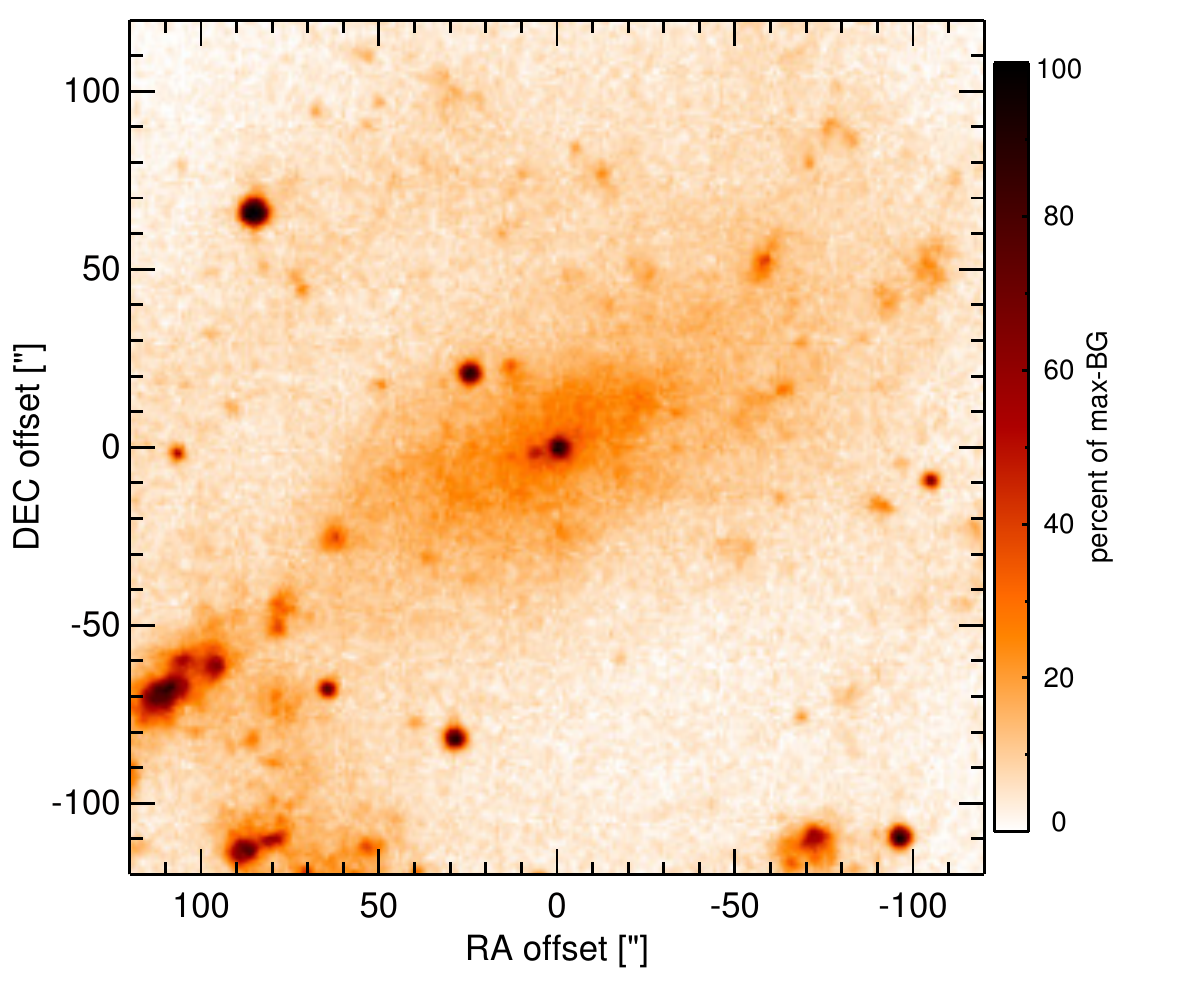}
    \caption{\label{fig:OPTim_NGC4395}
             Optical image (DSS, red filter) of NGC\,4395. Displayed are the central $4\arcmin$ with North up and East to the left. 
              The colour scaling is linear with white corresponding to the median background and black to the $0.01\%$ pixels with the highest intensity.  
           }
\end{figure}
\begin{figure}
   \centering
   \includegraphics[angle=0,height=3.11cm]{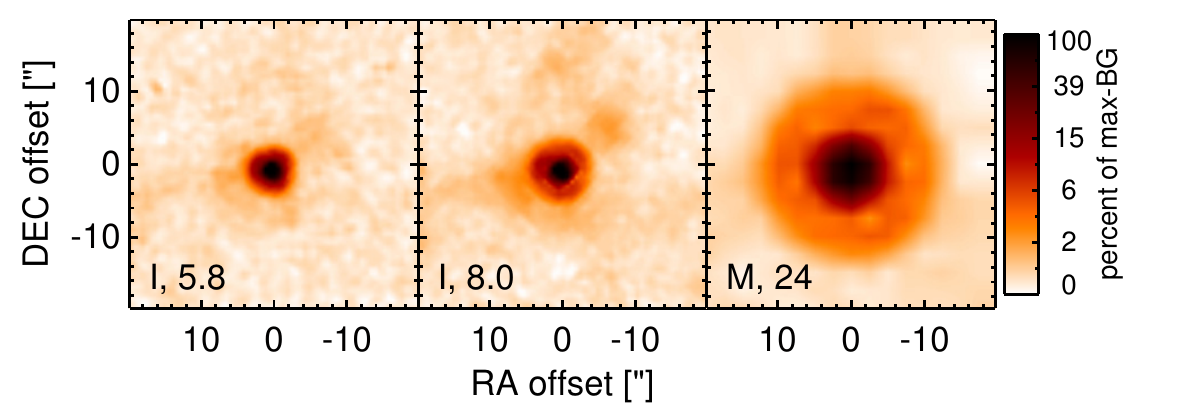}
    \caption{\label{fig:INTim_NGC4395}
             \spitzerr MIR images of NGC\,4395. Displayed are the inner $40\arcsec$ with North up and East to the left. The colour scaling is logarithmic with white corresponding to median background and black to the $0.1\%$ pixels with the highest intensity.
             The label in the bottom left states instrument and central wavelength of the filter in $\mu$m (I: IRAC, M: MIPS). 
           }
\end{figure}
\begin{figure}
   \centering
   \includegraphics[angle=0,height=3.11cm]{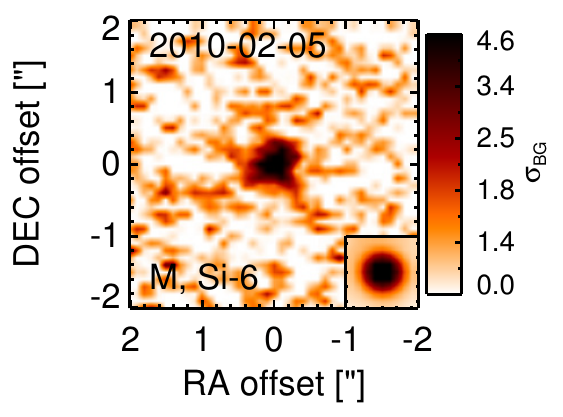}
    \caption{\label{fig:HARim_NGC4395}
             Subarcsecond-resolution MIR images of NGC\,4395 sorted by increasing filter wavelength. 
             Displayed are the inner $4\arcsec$ with North up and East to the left. 
             The colour scaling is logarithmic with white corresponding to median background and black to the $75\%$ of the highest intensity of all images in units of $\sigbg$.
             The inset image shows the central arcsecond of the PSF from the calibrator star, scaled to match the science target.
             The labels in the bottom left state instrument and filter names (C: COMICS, M: Michelle, T: T-ReCS, V: VISIR).
           }
\end{figure}
\begin{figure}
   \centering
   \includegraphics[angle=0,width=8.50cm]{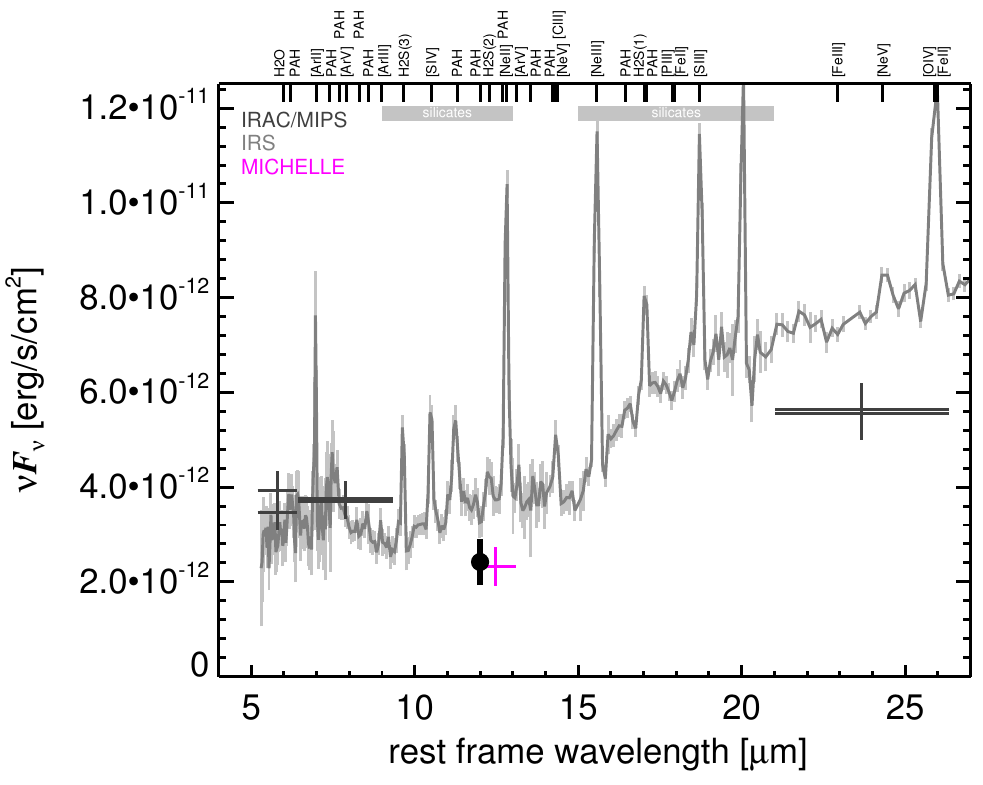}
   \caption{\label{fig:MISED_NGC4395}
      MIR SED of NGC\,4395. The description  of the symbols (if present) is the following.
      Grey crosses and  solid lines mark the \spitzer/IRAC, MIPS and IRS data. 
      The colour coding of the other symbols is: 
      green for COMICS, magenta for Michelle, blue for T-ReCS and red for VISIR data.
      Darker-coloured solid lines mark spectra of the corresponding instrument.
      The black filled circles mark the nuclear 12 and $18\,\mu$m  continuum emission estimate from the data.
      The ticks on the top axis mark positions of common MIR emission lines, while the light grey horizontal bars mark wavelength ranges affected by the silicate 10 and 18$\mu$m features.}
\end{figure}
\clearpage

\twocolumn[\begin{@twocolumnfalse}  
\subsection{NGC\,4418 -- NGC\,4355}\label{app:NGC4418}
NGC\,4418 is an infrared-luminous inclined spiral galaxy at a redshift of $z=$ 0.0073 ($D\sim37.8$\,Mpc) with a highly obscured active nucleus optically classified as  a Sy\,2.0 \citep{veron-cetty_catalogue_2010}.
The nucleus might also contain a compact massive starburst (e.g., \citealt{imanishi_near-infrared_2004,aalto_luminous_2007}); see \citealt{sakamoto_submillimeter_2013} for a recent detailed study).
Owing to the high obscuration, the AGN is difficult to verify at X-ray wavelengths \citep{maiolino_elusive_2003}.
Radio observations show a compact high-brightness temperature core \citep{condon_1.49_1990,kewley_compact_2000,baan_radio_2006}.
A kiloparsec-scale outflow cone along a PA$\sim15\degree$ was discovered \citep{sakamoto_submillimeter_2013}.
We conservatively treat NGC\,4418 as an uncertain AGN.
After the discovery of its infrared brightness with \iras, NGC\,4418 was followed up with ground-based MIR spectroscopy using UKIRT, which revealed an extremely deep silicate 10\,$\mu$m absorption feature and the absence of any MIR emission lines \citep{roche_ngc_1986}.
These authors, for the first time, suggested the presence of a deeply buried AGN in NGC\,4418.
Ground-based MIR photometry \citep{carico_iras_1988,wynn-williams_luminous_1993} and \isoo observations  \citep{dale_iso_2000,spoon_obscured_2001,lu_infrared_2003} were subsequently performed.
The first subarcsecond-resolution MIR images were obtained with Keck/MIRLIN in 1998 by \cite{evans_compact_2003}.
They show an unresolved nuclear source (except possibly at $\sim10\,\mu$m).
The corresponding nuclear photometry demonstrates that the silicate absorption is originating in the projected central  $\sim 0.3\arcsec \sim60\,$pc.
The nucleus is completely dominating the emission in the \spitzer/IRAC and MIPS PBCD images and saturates in the IRAC 8\,$\mu$m band.
The \spitzer/IRS LR staring-mode spectrum verifies the deep silicate 10$\,\mu$m absorption and absence of PAH and ionic emission lines, while in addition revealing  a strong silicate 18\,$\mu$m absorption feature and a very steep red spectral slope in $\nu F_\nu$-space (see also \citealt{dartois_carbonaceous_2007,stierwalt_mid-infrared_2013}).
Thus, this arcsecond-scale MIR SED does not provide any direct indication of star formation activity and rather favours the presence of a buried AGN.
However, the absence of any strong emission features indicates that putative MIR emission-line producing regions  are heavily extincted (similar to, e.g., NGC\,4945; \citealt{perez-beaupuits_deeply_2011}).
The nuclear region of NGC\,4418 was observed with T-ReCS in the N and Si6 filter in 2004 (unpublished, to our knowledge) and with VISIR in three $N$ and one $Q$-band filter between 2005 and 2006 (PAH2 flux published in \citealt{siebenmorgen_nuclear_2008}).
In all images, a compact MIR nucleus without further host emission was detected.
It appears unresolved in the PAH2 images and possibly resolved in the Q2 image.
However, in the other $N$-band images the apparent morphologies are contradicting and thus, we classify the general MIR extension as uncertain.
Our remeasured PAH2 flux is consistent with \cite{siebenmorgen_nuclear_2008}, while the nuclear  photometry generally is consistent with the \spitzerr spectrophotometry, the T-ReCS LR $N$-band spectrum from \cite{gonzalez-martin_dust_2013} and the previous MIR results.
Therefore, we can verify that the deep silicate absorption features are originating in the projected central $\sim 60\,$pc and correct our 12 and 18$\,\mu$m continuum emission estimates accordingly. 
\newline\end{@twocolumnfalse}]

\begin{figure}
   \centering
   \includegraphics[angle=0,width=8.500cm]{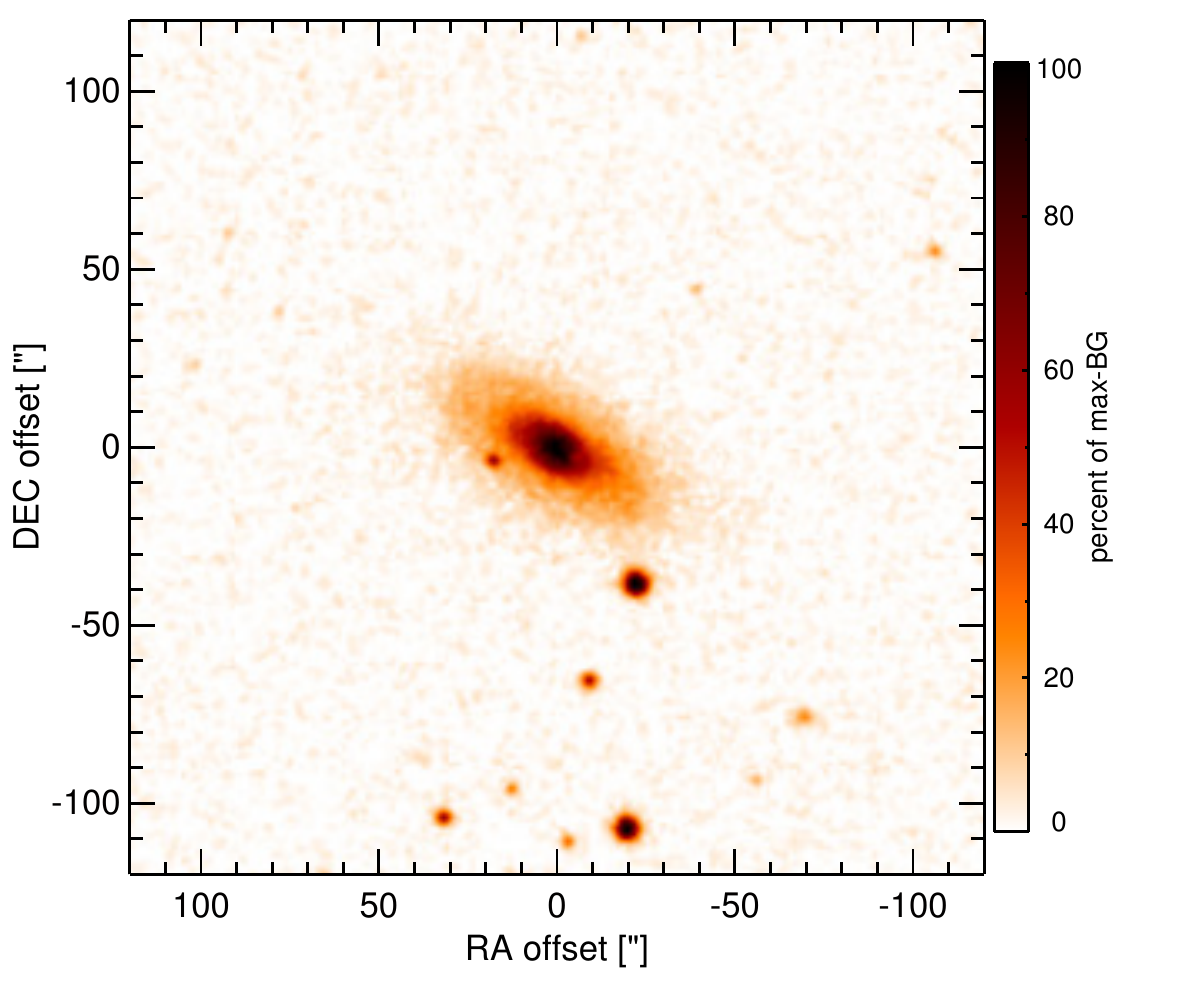}
    \caption{\label{fig:OPTim_NGC4418}
             Optical image (DSS, red filter) of NGC\,4418. Displayed are the central $4\arcmin$ with North up and East to the left. 
              The colour scaling is linear with white corresponding to the median background and black to the $0.01\%$ pixels with the highest intensity.  
           }
\end{figure}
\begin{figure}
   \centering
   \includegraphics[angle=0,height=3.11cm]{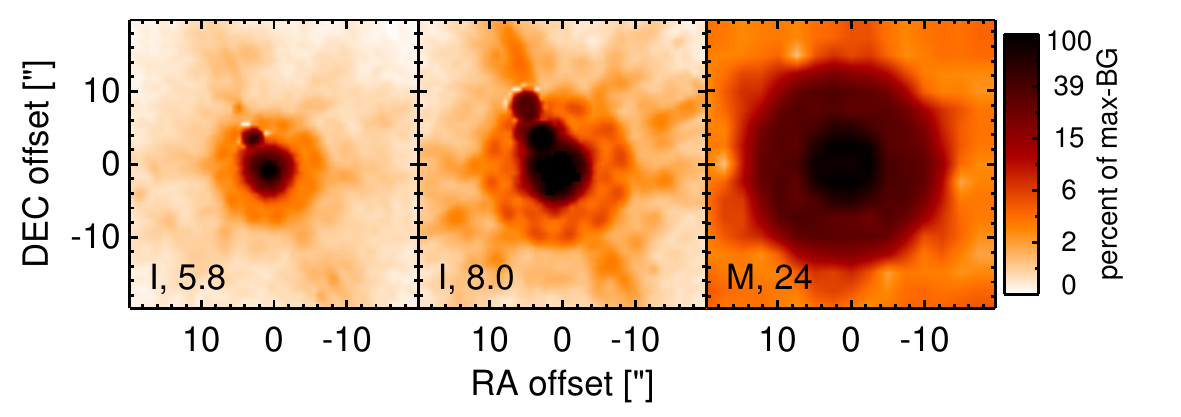}
    \caption{\label{fig:INTim_NGC4418}
             \spitzerr MIR images of NGC\,4418. Displayed are the inner $40\arcsec$ with North up and East to the left. The colour scaling is logarithmic with white corresponding to median background and black to the $0.1\%$ pixels with the highest intensity.
             The label in the bottom left states instrument and central wavelength of the filter in $\mu$m (I: IRAC, M: MIPS). 
             Note that the apparent off-nuclear compact sources in the IRAC 5.8 and $8.0\,\mu$m images are instrumental artefacts.
           }
\end{figure}
\begin{figure}
   \centering
   \includegraphics[angle=0,width=8.500cm]{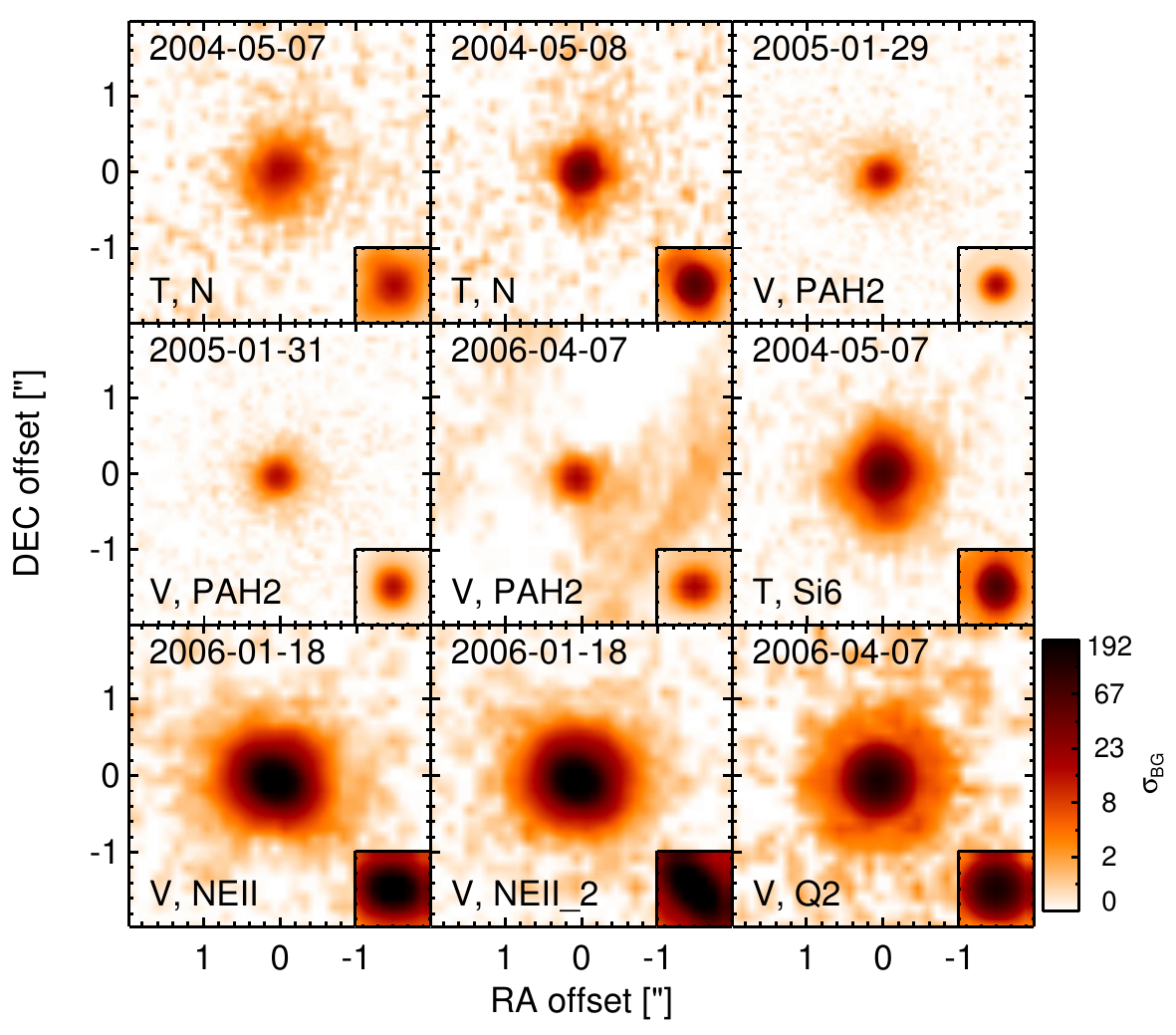}
    \caption{\label{fig:HARim_NGC4418}
             Subarcsecond-resolution MIR images of NGC\,4418 sorted by increasing filter wavelength. 
             Displayed are the inner $4\arcsec$ with North up and East to the left. 
             The colour scaling is logarithmic with white corresponding to median background and black to the $75\%$ of the highest intensity of all images in units of $\sigbg$.
             The inset image shows the central arcsecond of the PSF from the calibrator star, scaled to match the science target.
             The labels in the bottom left state instrument and filter names (C: COMICS, M: Michelle, T: T-ReCS, V: VISIR).
           }
\end{figure}
\begin{figure}
   \centering
   \includegraphics[angle=0,width=8.50cm]{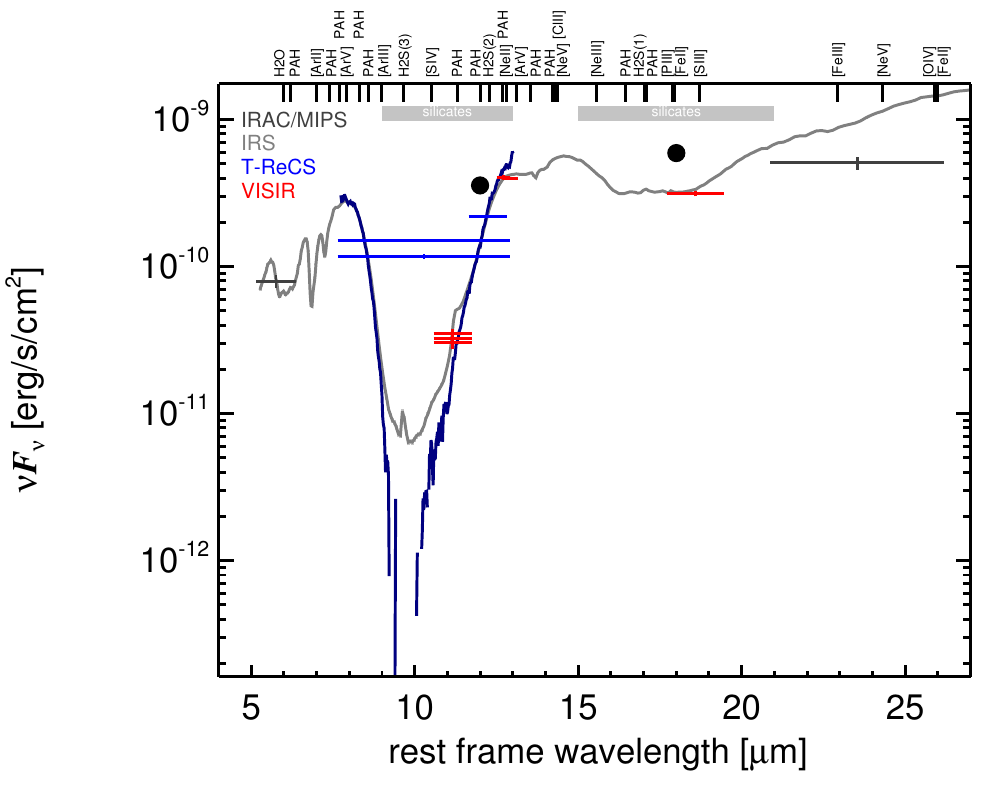}
   \caption{\label{fig:MISED_NGC4418}
      MIR SED of NGC\,4418. The description  of the symbols (if present) is the following.
      Grey crosses and  solid lines mark the \spitzer/IRAC, MIPS and IRS data. 
      The colour coding of the other symbols is: 
      green for COMICS, magenta for Michelle, blue for T-ReCS and red for VISIR data.
      Darker-coloured solid lines mark spectra of the corresponding instrument.
      The black filled circles mark the nuclear 12 and $18\,\mu$m  continuum emission estimate from the data.
      The ticks on the top axis mark positions of common MIR emission lines, while the light grey horizontal bars mark wavelength ranges affected by the silicate 10 and 18$\mu$m features.}
\end{figure}
\clearpage

\twocolumn[\begin{@twocolumnfalse}  
\subsection{NGC\,4438 -- VCC\,1043}\label{app:NGC4438}
NGC\,4438 is a highly-inclined disturbed spiral galaxy in the Virgo cluster at a distance of $D=$ $13.7\pm2.2$\,Mpc (NED redshift-independent median) with an active nucleus, which has initially be classified as a broad-line LINER \citep{ho_search_1997-1}.
However, \cite{kenney_hubble_2002} point out that the LINER-like emission is actually not coming from the nucleus, which rather shows optical emission line ratios like an H\,II nucleus.
Studies of the  high-resolution X-ray emission regarding the existence of an AGN are not conclusive \citep{machacek_chandra_2004,satyapal_link_2005,gonzalez-martin_x-ray_2006,gonzalez-martin_fitting_2009,flohic_central_2006}.
The presence of an AGN is supported by the presence of a compact radio source with a jet-like extension of $\sim0.3\arcsec\sim20\,$pc along a PA$\sim233\degree$ and two asymmetric radio lobes on kiloparsec scale \citep{hummel_anomalous_1991,hota_ngc_2007}. 
The western side of the bipolar structure is also seen in H$\alpha$  \citep{kenney_hubble_2002}.
We conservatively treat NGC\,4438 as  an uncertain AGN.
Early ground-based MIR photometry was performed by \cite{scoville_10_1983}, \cite{lonsdale_infrared_1984}, and \cite{lawrence_observations_1985}, followed by \isoo observations \citep{boselli_mid-ir_1998,roussel_atlas_2001}.
The first subarcsecond MIR image of NGC\,4438 was obtained with ESO 3.6\,m/TIMMI2 in 2002 \citep{perez_near-infrared_2009} and shows a compact nucleus embedded in $\sim3.5\arcsec\sim230\,$pc extended emission along the galaxy major axis (PA$\sim10\degree$).
The \spitzer/IRAC and MIPS images show a comparable morphology but are dominated by extended host emission.
The \spitzer/IRS LR staring-mode spectrum is dominated by strong PAH emission features with possibly weak silicate 10$\,\mu$m absorption and a flat spectral slope in $\nu F_\nu$-space (see also \citealt{mason_nuclear_2012}).
Thus, the arcsecond-scale MIR SED is star-formation dominated.
Note that \cite{dudik_spitzer_2009} report the detection of the AGN-indicative \nev emission line in the IRS spectrum.
The nuclear region of NGC\,4438 was observed with T-ReCS in the Si2 filter during two nights in 2008 \citep{mason_nuclear_2012}.
In both images, a compact nucleus is weakly detected.
Extended emission seems to be visible to the eye but the low S/N prohibits a quantitative analysis.
Our nuclear photometry is consistent with the PSF-flux reported in \cite{mason_nuclear_2012} and $\sim 83\%$ lower than the \spitzerr spectrophotometry.
Therefore, we conclude that the central $\sim250\,$pc are completely dominated by star formation emission in the MIR, while the subarcsecond-scale results are consistent with both the presence of an AGN or a nuclear starburst.
\newline\end{@twocolumnfalse}]

\begin{figure}
   \centering
   \includegraphics[angle=0,width=8.500cm]{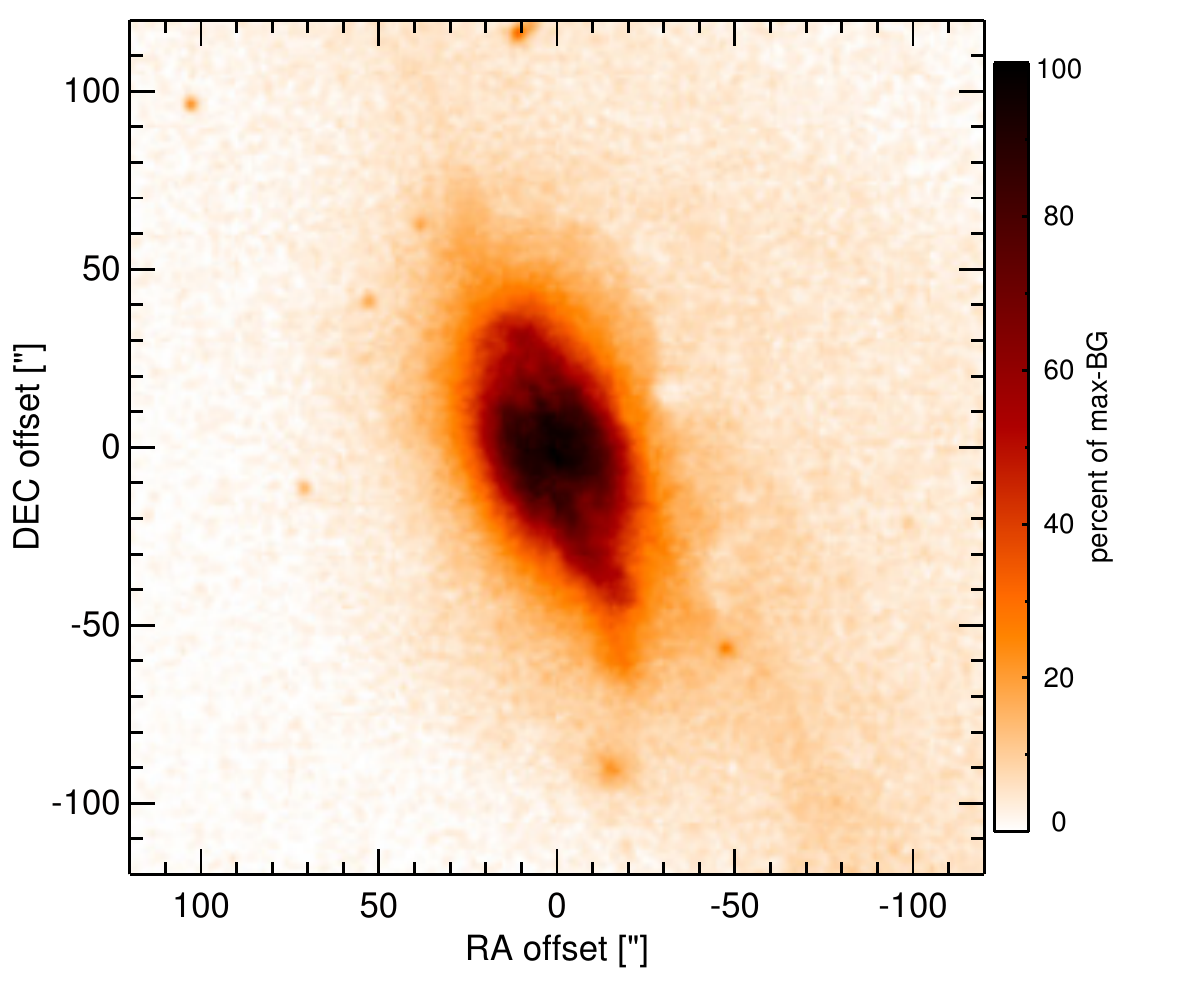}
    \caption{\label{fig:OPTim_NGC4438}
             Optical image (DSS, red filter) of NGC\,4438. Displayed are the central $4\arcmin$ with North up and East to the left. 
              The colour scaling is linear with white corresponding to the median background and black to the $0.01\%$ pixels with the highest intensity.  
           }
\end{figure}
\begin{figure}
   \centering
   \includegraphics[angle=0,height=3.11cm]{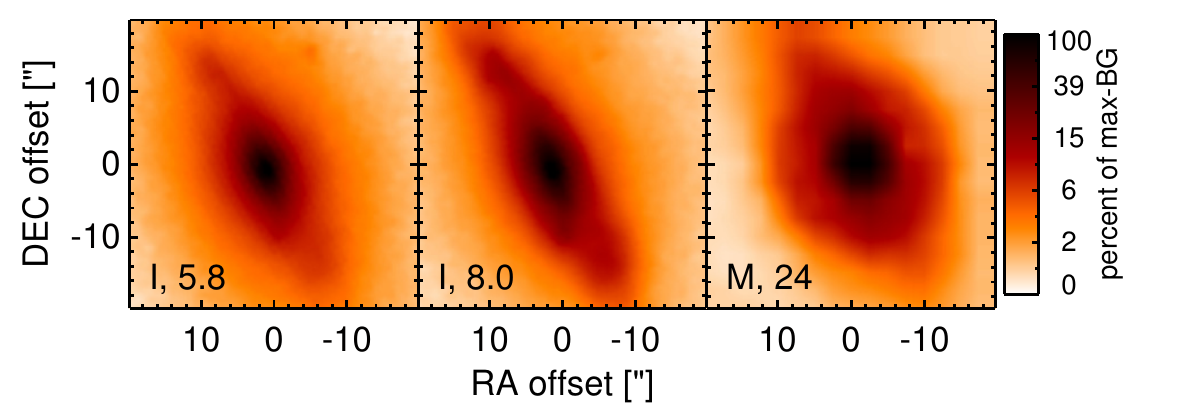}
    \caption{\label{fig:INTim_NGC4438}
             \spitzerr MIR images of NGC\,4438. Displayed are the inner $40\arcsec$ with North up and East to the left. The colour scaling is logarithmic with white corresponding to median background and black to the $0.1\%$ pixels with the highest intensity.
             The label in the bottom left states instrument and central wavelength of the filter in $\mu$m (I: IRAC, M: MIPS). 
           }
\end{figure}
\begin{figure}
   \centering
   \includegraphics[angle=0,height=3.11cm]{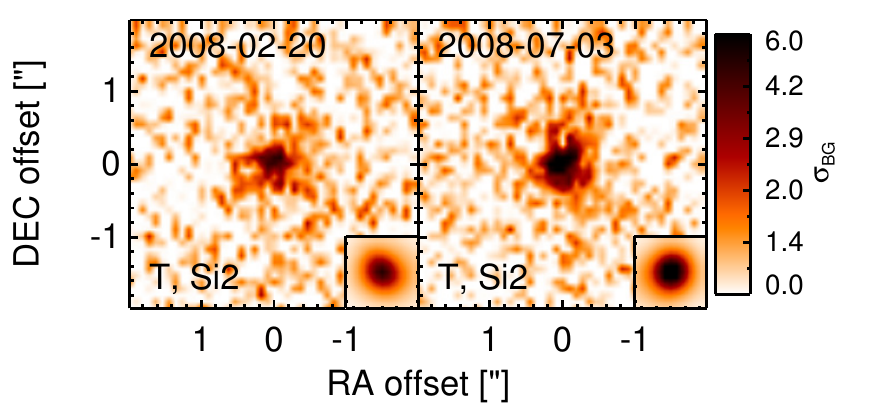}
    \caption{\label{fig:HARim_NGC4438}
             Subarcsecond-resolution MIR images of NGC\,4438 sorted by increasing filter wavelength. 
             Displayed are the inner $4\arcsec$ with North up and East to the left. 
             The colour scaling is logarithmic with white corresponding to median background and black to the $75\%$ of the highest intensity of all images in units of $\sigbg$.
             The inset image shows the central arcsecond of the PSF from the calibrator star, scaled to match the science target.
             The labels in the bottom left state instrument and filter names (C: COMICS, M: Michelle, T: T-ReCS, V: VISIR).
           }
\end{figure}
\begin{figure}
   \centering
   \includegraphics[angle=0,width=8.50cm]{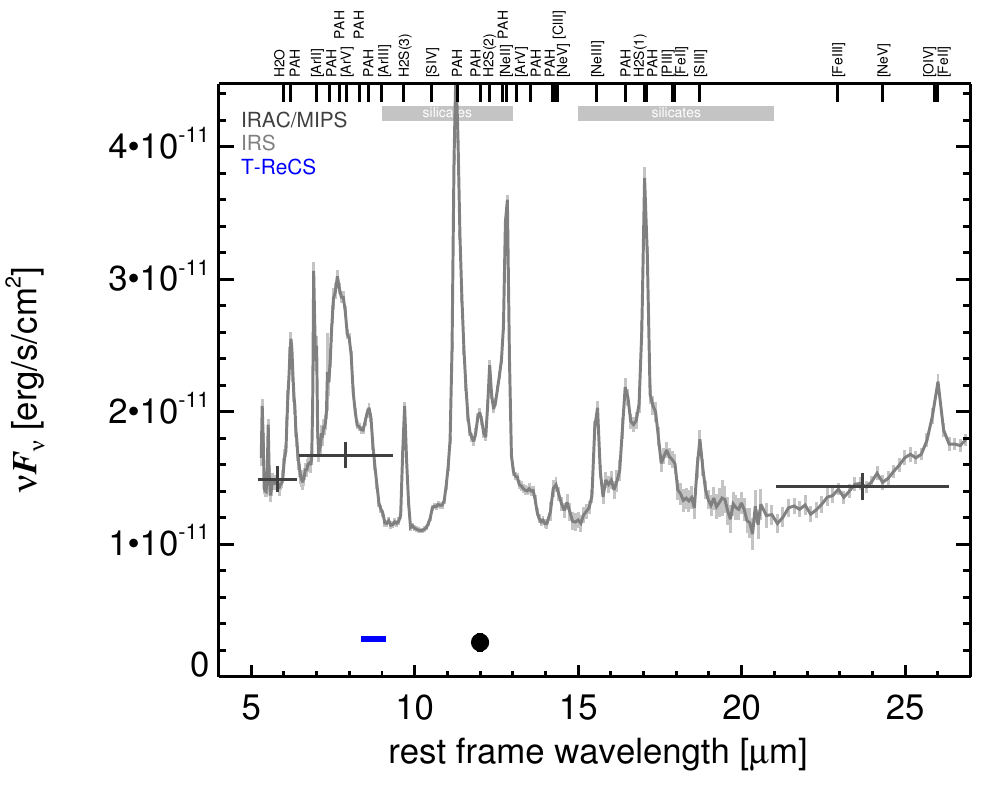}
   \caption{\label{fig:MISED_NGC4438}
      MIR SED of NGC\,4438. The description  of the symbols (if present) is the following.
      Grey crosses and  solid lines mark the \spitzer/IRAC, MIPS and IRS data. 
      The colour coding of the other symbols is: 
      green for COMICS, magenta for Michelle, blue for T-ReCS and red for VISIR data.
      Darker-coloured solid lines mark spectra of the corresponding instrument.
      The black filled circles mark the nuclear 12 and $18\,\mu$m  continuum emission estimate from the data.
      The ticks on the top axis mark positions of common MIR emission lines, while the light grey horizontal bars mark wavelength ranges affected by the silicate 10 and 18$\mu$m features.}
\end{figure}
\clearpage

\twocolumn[\begin{@twocolumnfalse}  
\subsection{NGC\,4457 -- VCC\,1145}\label{app:NGC4457}
NGC\,4457 is a low-inclination spiral galaxy in the Virgo cluster at a distance of $D=$ $17.4 \pm 3.5\,$Mpc \citep{tully_nearby_1988}, hosting a LINER nucleus \citep{ho_search_1997-1}.
The detection of a hard X-ray point source supports the presence of an AGN, while additional diffuse soft emission indicates an AGN/starburst composite nature \citep{satyapal_link_2005,gonzalez-martin_x-ray_2006,flohic_central_2006}.
A compact radio source at 6\,cm wavelength embedded within a halo with an extent of $\sim10\arcsec\sim840\,$pc was detected in the nucleus of NGC\,4457 (PA$\sim177\degree$; \citealt{van_der_hulst_structure_1981}) but remained undetected at 2\,cm \citep{nagar_radio_2002}.
We conservatively treat NGC\,4457 as an uncertain AGN.
The first ground-based MIR observations of NGC\,4457 were performed by \cite{scoville_10_1983}.
The \spitzer/IRAC and MIPS images reveal an extended nucleus embedded within a prominent spiral pattern.
The \spitzer/IRS staring-mode spectrum is dominated by strong PAH emission features with weak silicate 10\,$\mu$m absorption and a flat spectral slope in $\nu F_\nu$-space.
Thus, the arcsecond-scale MIR SED  is star-formation dominated.
Note that \cite{dudik_spitzer_2009} and \cite{pereira-santaella_mid-infrared_2010} report the absence of any significant AGN-indicative \nev emission line in the IRS spectrum.
The nuclear region of NGC\,4457 was observed with T-ReCS in the Si2 filter during two nights in 2008 \citep{mason_nuclear_2012}.
In both images, an $\sim1\arcsec\sim84\,$pc extended nucleus is weakly detected .
Our manually-scaled nuclear PSF photometry is consistent with the PSF-flux reported in \cite{mason_nuclear_2012} and $\sim 90\%$ lower than the \spitzerr spectrophotometry.
Therefore, we conclude that the central $\sim300\,$pc are completely dominated by star formation emission in the MIR.
The subarcsecond data indicates the presence of a nuclear starburst with a possible weak AGN contribution, for which the nuclear flux should be considered an upper limit.
\newline\end{@twocolumnfalse}]

\begin{figure}
   \centering
   \includegraphics[angle=0,width=8.500cm]{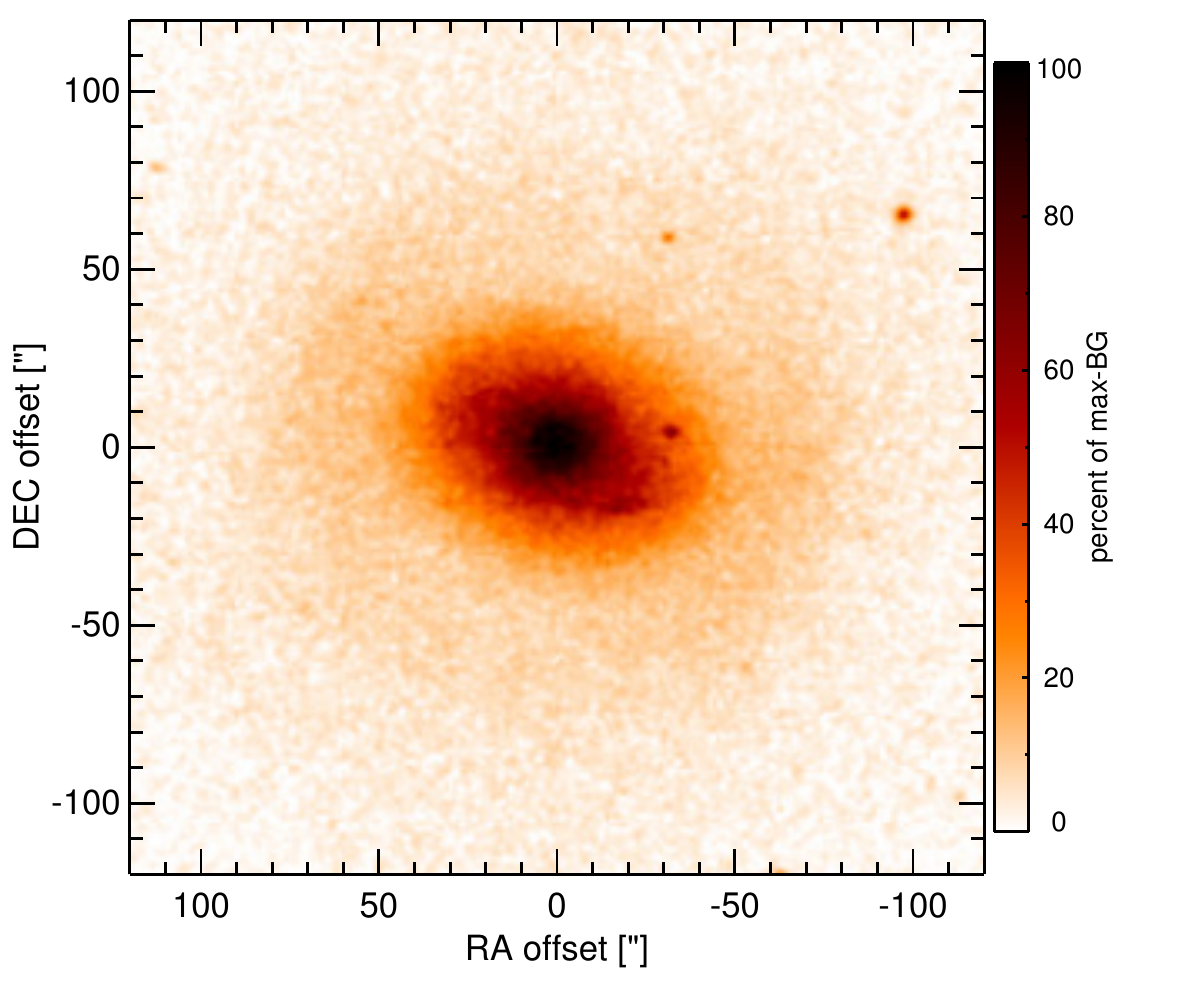}
    \caption{\label{fig:OPTim_NGC4457}
             Optical image (DSS, red filter) of NGC\,4457. Displayed are the central $4\arcmin$ with North up and East to the left. 
              The colour scaling is linear with white corresponding to the median background and black to the $0.01\%$ pixels with the highest intensity.  
           }
\end{figure}
\begin{figure}
   \centering
   \includegraphics[angle=0,height=3.11cm]{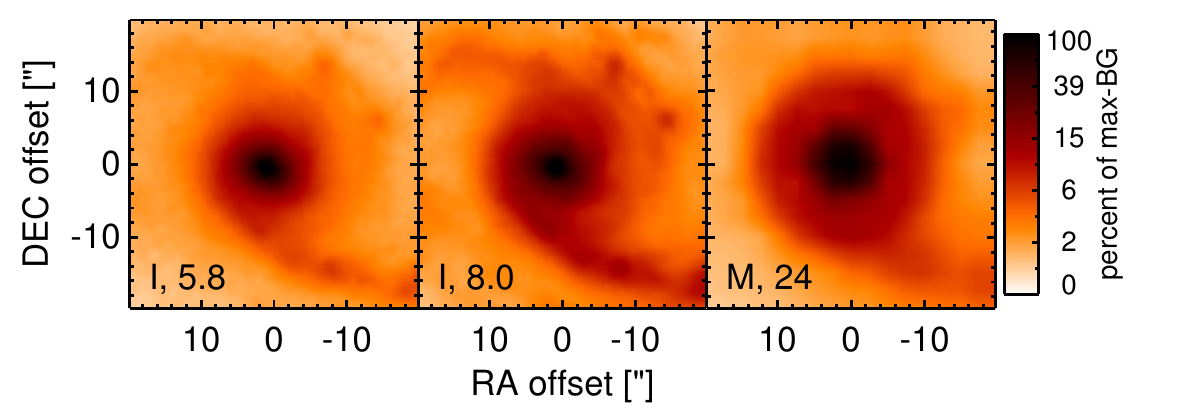}
    \caption{\label{fig:INTim_NGC4457}
             \spitzerr MIR images of NGC\,4457. Displayed are the inner $40\arcsec$ with North up and East to the left. The colour scaling is logarithmic with white corresponding to median background and black to the $0.1\%$ pixels with the highest intensity.
             The label in the bottom left states instrument and central wavelength of the filter in $\mu$m (I: IRAC, M: MIPS). 
           }
\end{figure}
\begin{figure}
   \centering
   \includegraphics[angle=0,height=3.11cm]{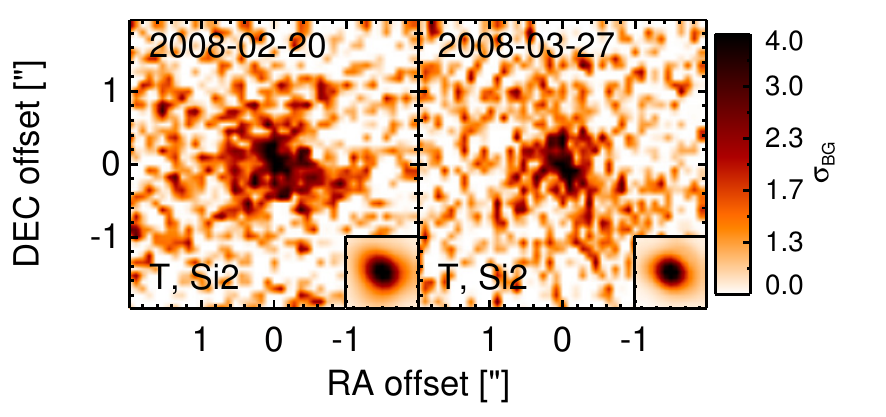}
    \caption{\label{fig:HARim_NGC4457}
             Subarcsecond-resolution MIR images of NGC\,4457 sorted by increasing filter wavelength. 
             Displayed are the inner $4\arcsec$ with North up and East to the left. 
             The colour scaling is logarithmic with white corresponding to median background and black to the $75\%$ of the highest intensity of all images in units of $\sigbg$.
             The inset image shows the central arcsecond of the PSF from the calibrator star, scaled to match the science target.
             The labels in the bottom left state instrument and filter names (C: COMICS, M: Michelle, T: T-ReCS, V: VISIR).
           }
\end{figure}
\begin{figure}
   \centering
   \includegraphics[angle=0,width=8.50cm]{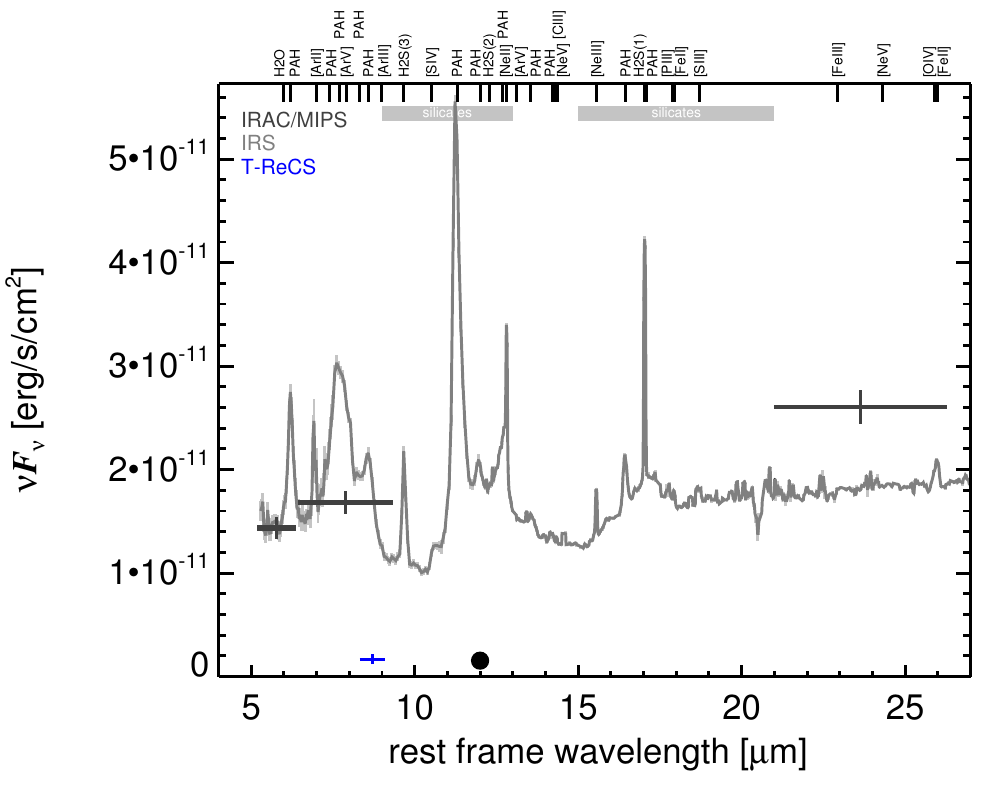}
   \caption{\label{fig:MISED_NGC4457}
      MIR SED of NGC\,4457. The description  of the symbols (if present) is the following.
      Grey crosses and  solid lines mark the \spitzer/IRAC, MIPS and IRS data. 
      The colour coding of the other symbols is: 
      green for COMICS, magenta for Michelle, blue for T-ReCS and red for VISIR data.
      Darker-coloured solid lines mark spectra of the corresponding instrument.
      The black filled circles mark the nuclear 12 and $18\,\mu$m  continuum emission estimate from the data.
      The ticks on the top axis mark positions of common MIR emission lines, while the light grey horizontal bars mark wavelength ranges affected by the silicate 10 and 18$\mu$m features.}
\end{figure}
\clearpage

\twocolumn[\begin{@twocolumnfalse}  
\subsection{NGC\,4472 -- M49 -- VCC\,1226}\label{app:NGC4472}

NGC\,4472 is an elliptical galaxy in the Virgo cluster at a distance of $D=$ 17.1\,Mpc \citep{mei_acs_2007} with possibly a Sy\,2 or LINER nucleus (\citealt{ho_search_1997-1}; NED).
A nuclear radio point source supports the existence of an AGN in this object \citep{anderson_size_2005}, while the X-ray data remains inconclusive so far \citep{maccarone_low-mass_2003}.
Apart from \iras, NGC\,4472 was observed with IRTF \citep{impey_infrared_1986} but was not unambiguously detected.
\iso/ISOCAM \citep{ferrari_survey_2002} and \spitzer/IRAC, IRS and MIPS observations were performed subsequently. 
In the IRAC and MIPS images, extended elliptical emission without any clear unresolved nucleus was detected.
Therefore, our IRAC $5.8$ and $8.0\,\mu$m and MIPS $24\,\mu$m photometry of the central four-arcsecond region (seven for MIPS) is significantly lower than the values published in \cite{temi_ages_2005}.
The IRS LR staring spectrum is dominated by passive host emission with a blue slope in $\nu F_\nu$-space, silicate $10\,\mu$m emission and no PAH features (see also \citealt{bregman_ages_2006}). 
We observed the nucleus of NGC\,4472 with VISIR in the NEII\_1 filter in 2006 but failed to detect any emission \citep{horst_mid_2008}.
Therefore, from the MIR point of view there is no clear evidence for the existence of an AGN in NGC\,4472.
However, the derived upper limit on the $12\,\mu$m continuum emission is not inconsistent with the existence of a faint AGN as expected from the MIR--X-ray correlation of AGN \citep{asmus_mid-infrared_2011}. 
\newline\end{@twocolumnfalse}]

\begin{figure}
   \centering
   \includegraphics[angle=0,width=8.500cm]{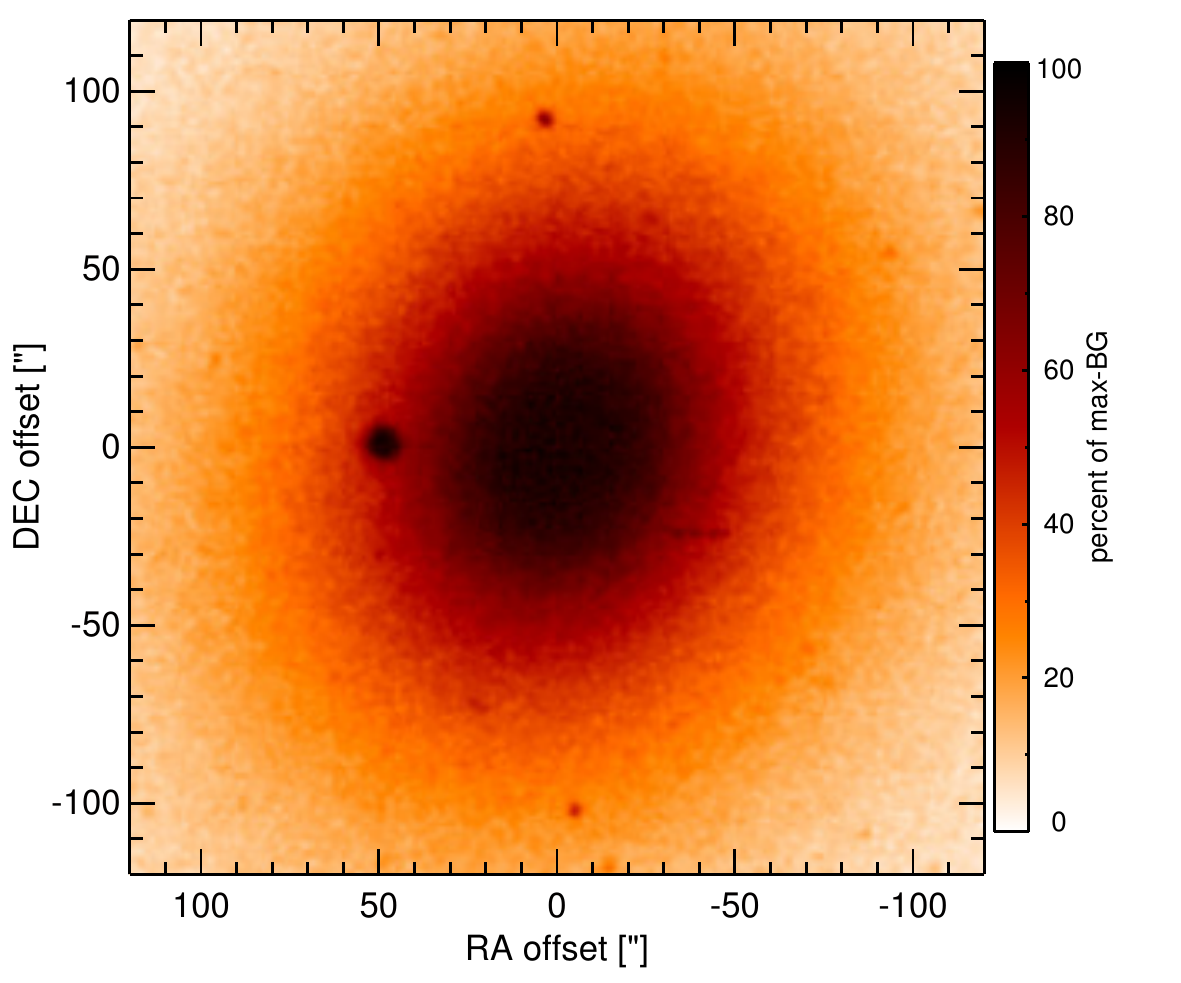}
    \caption{\label{fig:OPTim_NGC4472}
             Optical image (DSS, red filter) of NGC\,4472. Displayed are the central $4\arcmin$ with North up and East to the left. 
              The colour scaling is linear with white corresponding to the median background and black to the $0.01\%$ pixels with the highest intensity.  
           }
\end{figure}
\begin{figure}
   \centering
   \includegraphics[angle=0,height=3.11cm]{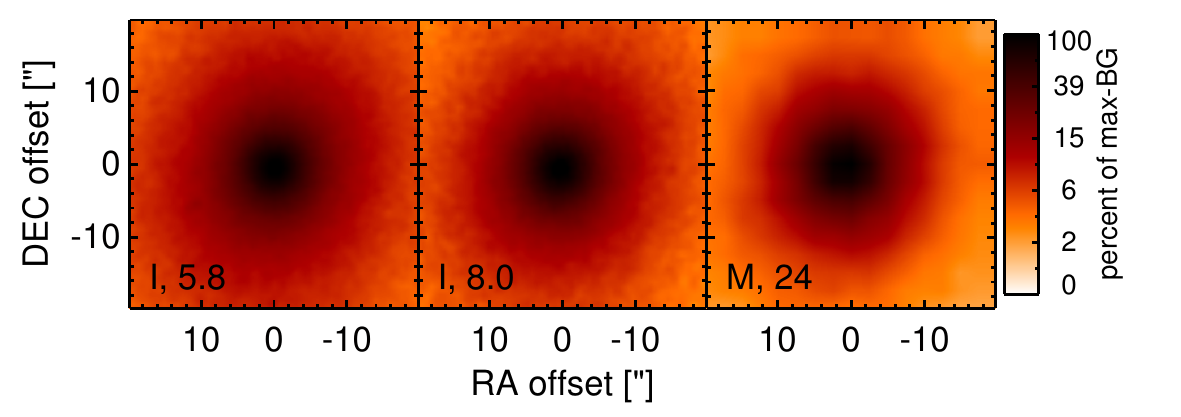}
    \caption{\label{fig:INTim_NGC4472}
             \spitzerr MIR images of NGC\,4472. Displayed are the inner $40\arcsec$ with North up and East to the left. The colour scaling is logarithmic with white corresponding to median background and black to the $0.1\%$ pixels with the highest intensity.
             The label in the bottom left states instrument and central wavelength of the filter in $\mu$m (I: IRAC, M: MIPS). 
           }
\end{figure}
\begin{figure}
   \centering
   \includegraphics[angle=0,width=8.50cm]{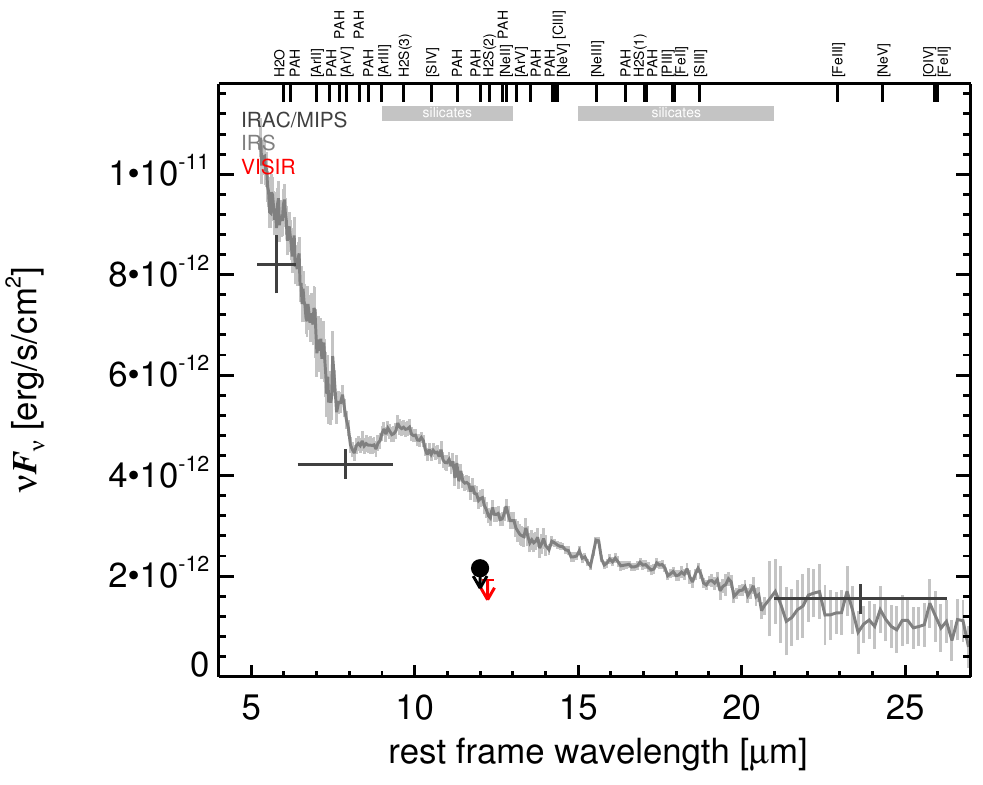}
   \caption{\label{fig:MISED_NGC4472}
      MIR SED of NGC\,4472. The description  of the symbols (if present) is the following.
      Grey crosses and  solid lines mark the \spitzer/IRAC, MIPS and IRS data. 
      The colour coding of the other symbols is: 
      green for COMICS, magenta for Michelle, blue for T-ReCS and red for VISIR data.
      Darker-coloured solid lines mark spectra of the corresponding instrument.
      The black filled circles mark the nuclear 12 and $18\,\mu$m  continuum emission estimate from the data.
      The ticks on the top axis mark positions of common MIR emission lines, while the light grey horizontal bars mark wavelength ranges affected by the silicate 10 and 18$\mu$m features.}
\end{figure}
\clearpage

\twocolumn[\begin{@twocolumnfalse}  
\subsection{NGC\,4501 -- M88 -- VCC\,1401}\label{app:NGC4501}
NGC\,4501 is an inclined spiral galaxy in the Virgo cluster at a distance of $D=$ $17.9 \pm 5.8$\,Mpc (NED redshift-independent median) that hosts a Sy\,2 nucleus \citep{veron-cetty_catalogue_2010}.
The first unsuccessful attempts to detect the nucleus in the MIR were made by \cite{rieke_10_1978} and \cite{scoville_10_1983}.
The first possible weak detection is reported by \cite{maiolino_new_1995} with the MMT bolometer, while the nucleus again remained undetected in the first subarcsecond-resolution $N$-band imaging observation with Palomar 5\,m/MIRLIN \citep{gorjian_10_2004}.
The \spitzer/IRAC images show an extended nucleus embedded in the spiral-like host emission, while the nucleus appears as a weak compact source in the MIPS $24\,\mu$m image.
Because we measure the flux of the central four-arcsecond region, our $5.8$ and $8.0\,\mu$m photometry is significantly lower than previously published values \citep{gallimore_infrared_2010}.
The IRS LR mapping-mode PBCD spectrum suffers from very low S/N and is probably not very reliable regarding the complex extended morphology of NGC\,4501 in the MIR.
However, it matches our IRAC and MIPS photometry and indicates weak silicate $10\,\mu$m absorption and PAH emission, i.e., star formation (see also \citealt{wu_spitzer/irs_2009,deo_mid-infrared_2009,gallimore_infrared_2010,tommasin_spitzer_2008}).
The nucleus of NGC\,4501 was observed with T-ReCS in the N filter in 2004 but has not been detected.
We observed it with Michelle in 2010 in the Si-5 filter and find a very weak compact detection with a flux of only $23\%$ of the \spitzerr spectrophotometry.
The low S/N prevents us to make any statement about the source extension.
The fact that our derived upper limit from the T-ReCS observation gives a lower flux indicates silicate absorption in the projected central $\sim30\,$pc of NGC\,4501, supporting the obscured scenario for its nucleus.
\newline\end{@twocolumnfalse}]

\begin{figure}
   \centering
   \includegraphics[angle=0,width=8.500cm]{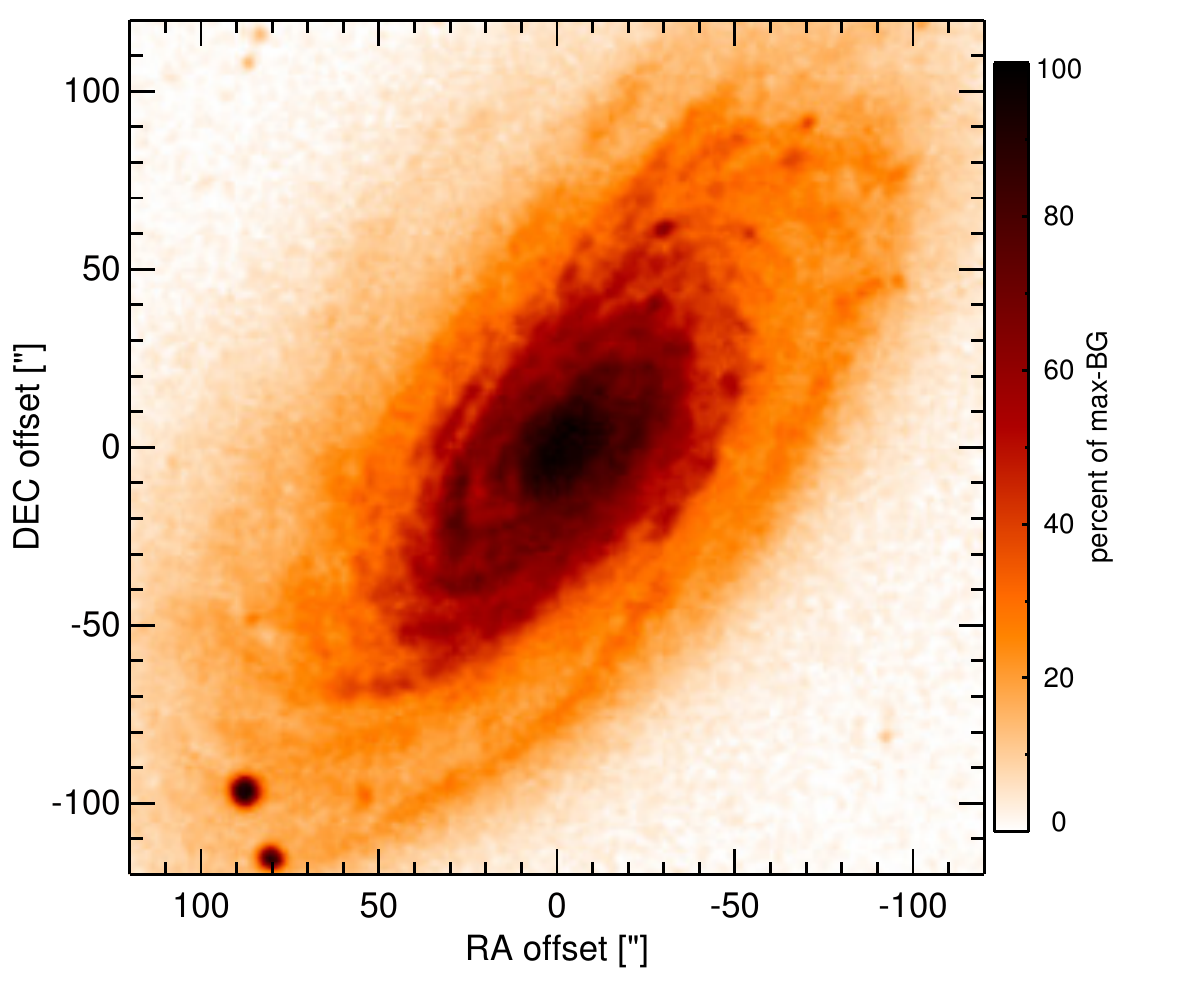}
    \caption{\label{fig:OPTim_NGC4501}
             Optical image (DSS, red filter) of NGC\,4501. Displayed are the central $4\arcmin$ with North up and East to the left. 
              The colour scaling is linear with white corresponding to the median background and black to the $0.01\%$ pixels with the highest intensity.  
           }
\end{figure}
\begin{figure}
   \centering
   \includegraphics[angle=0,height=3.11cm]{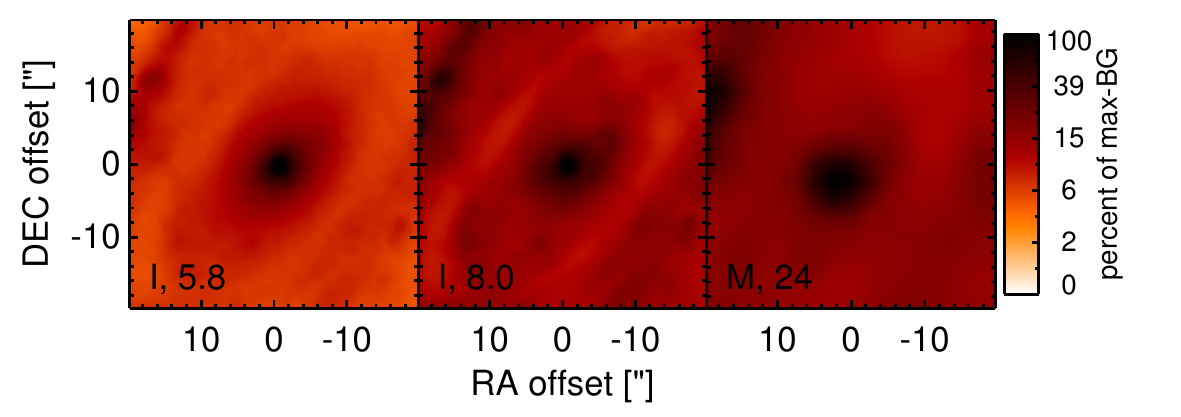}
    \caption{\label{fig:INTim_NGC4501}
             \spitzerr MIR images of NGC\,4501. Displayed are the inner $40\arcsec$ with North up and East to the left. The colour scaling is logarithmic with white corresponding to median background and black to the $0.1\%$ pixels with the highest intensity.
             The label in the bottom left states instrument and central wavelength of the filter in $\mu$m (I: IRAC, M: MIPS). 
           }
\end{figure}
\begin{figure}
   \centering
   \includegraphics[angle=0,height=3.11cm]{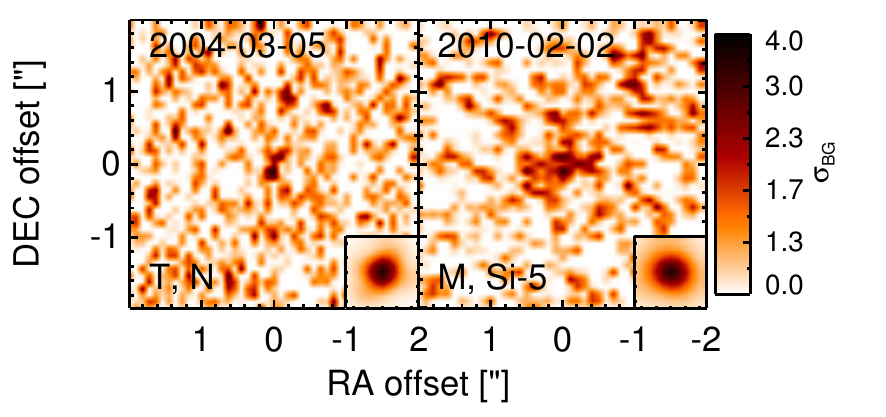}
    \caption{\label{fig:HARim_NGC4501}
             Subarcsecond-resolution MIR images of NGC\,4501 sorted by increasing filter wavelength. 
             Displayed are the inner $4\arcsec$ with North up and East to the left. 
             The colour scaling is logarithmic with white corresponding to median background and black to the $75\%$ of the highest intensity of all images in units of $\sigbg$.
             The inset image shows the central arcsecond of the PSF from the calibrator star, scaled to match the science target.
             The labels in the bottom left state instrument and filter names (C: COMICS, M: Michelle, T: T-ReCS, V: VISIR).
           }
\end{figure}
\begin{figure}
   \centering
   \includegraphics[angle=0,width=8.50cm]{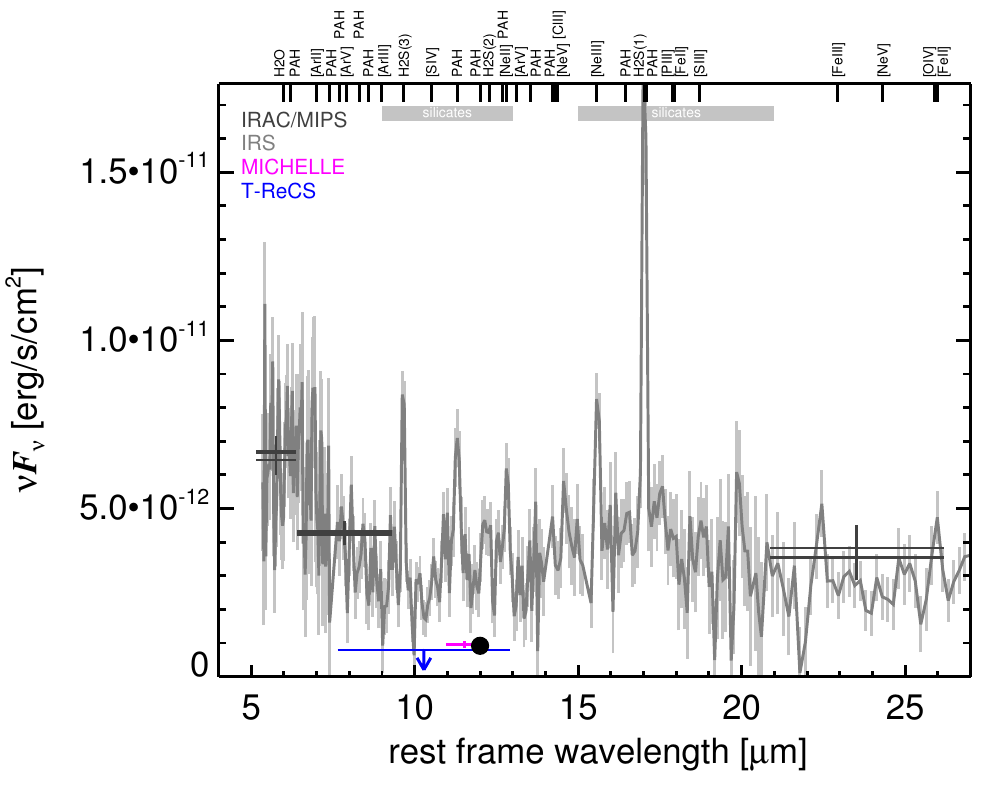}
   \caption{\label{fig:MISED_NGC4501}
      MIR SED of NGC\,4501. The description  of the symbols (if present) is the following.
      Grey crosses and  solid lines mark the \spitzer/IRAC, MIPS and IRS data. 
      The colour coding of the other symbols is: 
      green for COMICS, magenta for Michelle, blue for T-ReCS and red for VISIR data.
      Darker-coloured solid lines mark spectra of the corresponding instrument.
      The black filled circles mark the nuclear 12 and $18\,\mu$m  continuum emission estimate from the data.
      The ticks on the top axis mark positions of common MIR emission lines, while the light grey horizontal bars mark wavelength ranges affected by the silicate 10 and 18$\mu$m features.}
\end{figure}
\clearpage

\twocolumn[\begin{@twocolumnfalse}  
\subsection{NGC\,4507}\label{app:NGC4507}
NGC\,4507 is a face-on barred spiral galaxy at a redshift of $z=$ 0.0118 ($D\sim57.5\,$Mpc) that hosts a Sy\,2 nucleus \citep{kewley_optical_2001} with polarized broad emission lines \citep{moran_frequency_2000}.
The AGN is one of the X-ray flux-brightest type~II objects, belongs to the nine-month BAT AGN sample, and is also variable in X-rays.
Subarcsecond-resolution radio observations reveal a slightly elongated nucleus along a PA$\sim10\degree$ \citep{morganti_radio_1999}.
The NLR appears one-sided in \oiii along a PA$\sim-35\degree$ with an extent of $\sim2\arcsec\sim0.5\,$kpc \citep{schmitt_hubble_2003}.
The first ground-based MIR observations were performed by \cite{glass_mid-infrared_1982}, followed by \isoo observations \citep{clavel_2.5-11_2000,ramos_almeida_mid-infrared_2007}.
The \spitzer/IRAC and MIPS images are dominated by the bright nucleus embedded within the much weaker spiral-like host emission. 
Note that the IRAC 8.0\,$\mu$m PBCD image is saturated and, thus, not analysed.
Surprisingly, the \spitzer/IRS LR staring-mode spectrum exhibits weak silicate 10$\,\mu$m and possible 18\,$\mu$m emission more reminiscent for an unobscured AGN.
In addition, only weak PAH features are present and the continuum peaks at $\sim 18\,\mu$m in $\nu F_\nu$-space (see also \citealt{shi_9.7_2006,mullaney_defining_2011}).
Thus, the arcsecond-scale MIR emission of NGC\,4507 is dominated by the AGN without any significant contribution from star formation.
We observed the nuclear region of NGC\,4507 with VISIR in three narrow $N$-band filters in 2006 \citep{horst_mid_2008,horst_mid-infrared_2009} and obtained a LR $N$-band spectrum in 2008 \citep{honig_dusty_2010-1}.
In all images, an unresolved nucleus without any host emission was detected.
The remeasured nuclear fluxes agree with \cite{horst_mid_2008}, the VISIR spectrum and the \spitzerr spectrophotometry.
Note that the VISIR spectrum does not exhibit any PAH emission, which verifies that the nuclear MIR SED is not star-formation contaminated.
The nuclear MIR emission of NGC\,4507 has been resolved with MIDI interferometric observations and was modelled as two components with sizes of $\sim4$ and $\sim12$\,pc \citep{burtscher_diversity_2013}.
\newline\end{@twocolumnfalse}]

\begin{figure}
   \centering
   \includegraphics[angle=0,width=8.500cm]{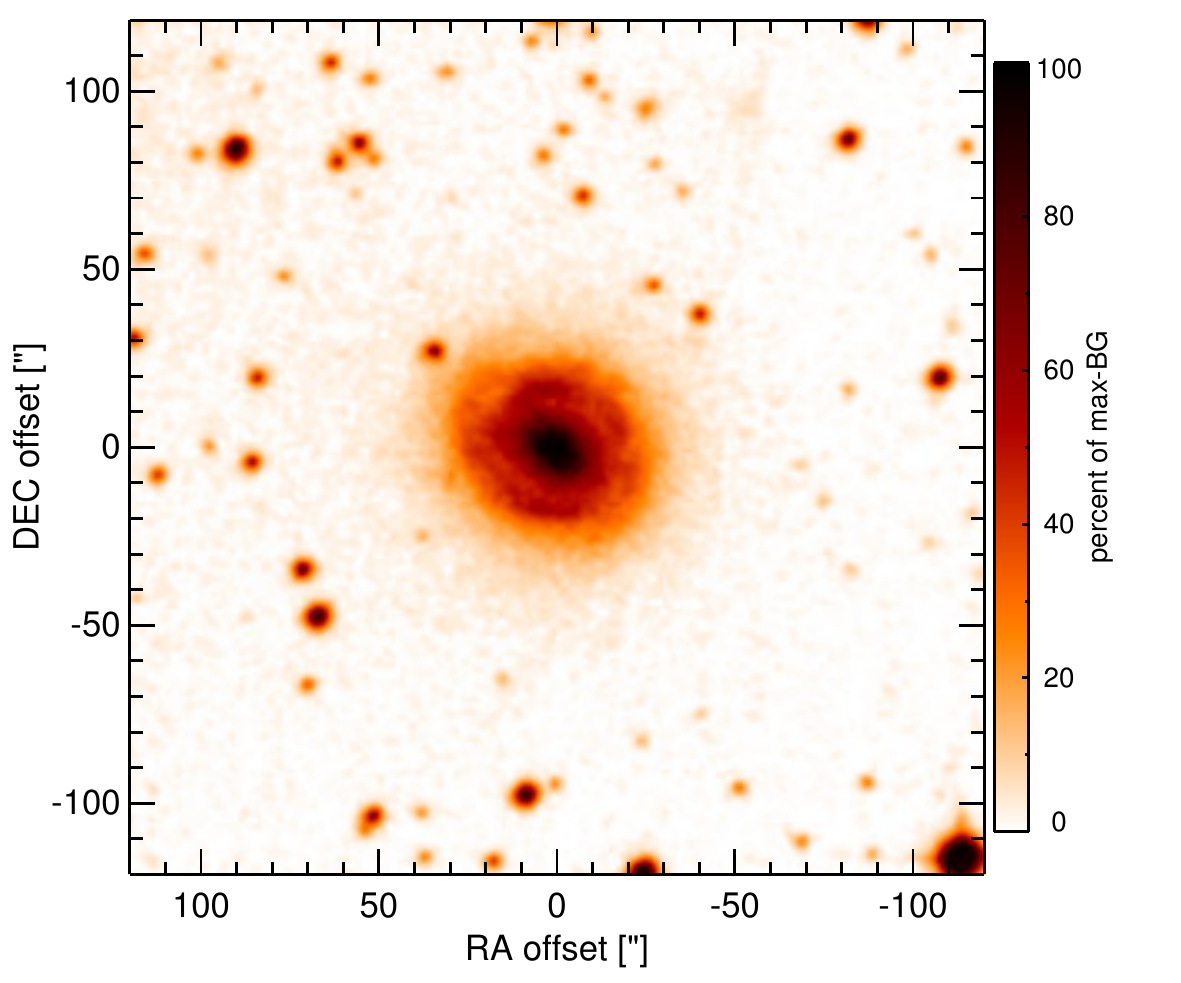}
    \caption{\label{fig:OPTim_NGC4507}
             Optical image (DSS, red filter) of NGC\,4507. Displayed are the central $4\arcmin$ with North up and East to the left. 
              The colour scaling is linear with white corresponding to the median background and black to the $0.01\%$ pixels with the highest intensity.  
           }
\end{figure}
\begin{figure}
   \centering
   \includegraphics[angle=0,height=3.11cm]{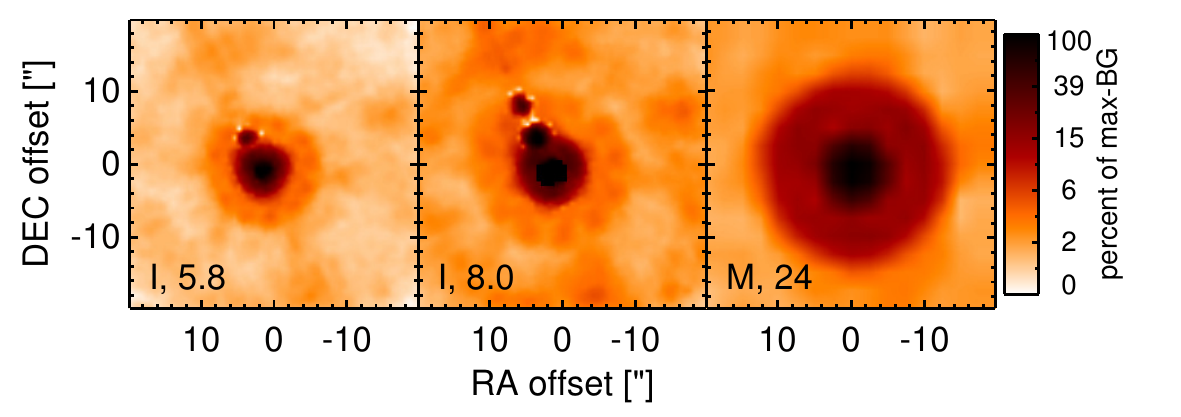}
    \caption{\label{fig:INTim_NGC4507}
             \spitzerr MIR images of NGC\,4507. Displayed are the inner $40\arcsec$ with North up and East to the left. The colour scaling is logarithmic with white corresponding to median background and black to the $0.1\%$ pixels with the highest intensity.
             The label in the bottom left states instrument and central wavelength of the filter in $\mu$m (I: IRAC, M: MIPS).
             Note that the apparent off-nuclear compact sources in the IRAC 5.8 and $8.0\,\mu$m images are instrumental artefacts.
           }
\end{figure}
\begin{figure}
   \centering
   \includegraphics[angle=0,height=3.11cm]{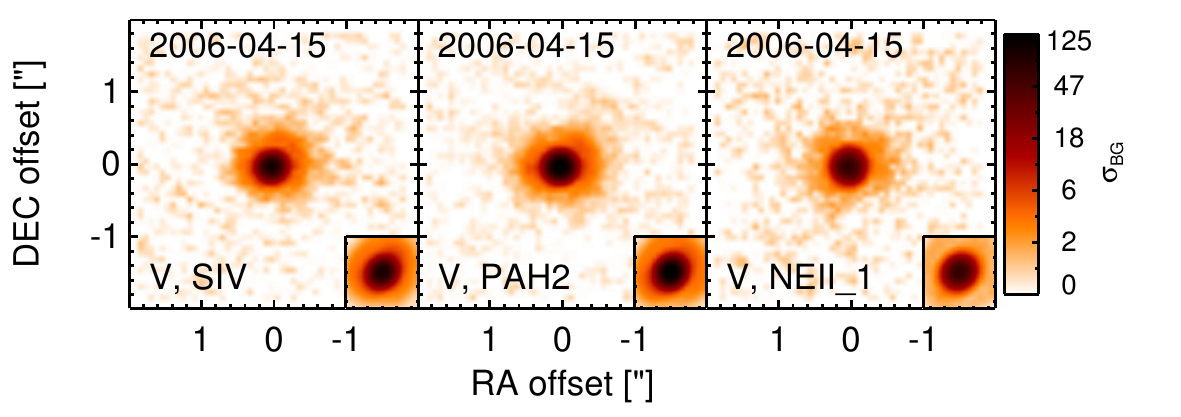}
    \caption{\label{fig:HARim_NGC4507}
             Subarcsecond-resolution MIR images of NGC\,4507 sorted by increasing filter wavelength. 
             Displayed are the inner $4\arcsec$ with North up and East to the left. 
             The colour scaling is logarithmic with white corresponding to median background and black to the $75\%$ of the highest intensity of all images in units of $\sigbg$.
             The inset image shows the central arcsecond of the PSF from the calibrator star, scaled to match the science target.
             The labels in the bottom left state instrument and filter names (C: COMICS, M: Michelle, T: T-ReCS, V: VISIR).
           }
\end{figure}
\begin{figure}
   \centering
   \includegraphics[angle=0,width=8.50cm]{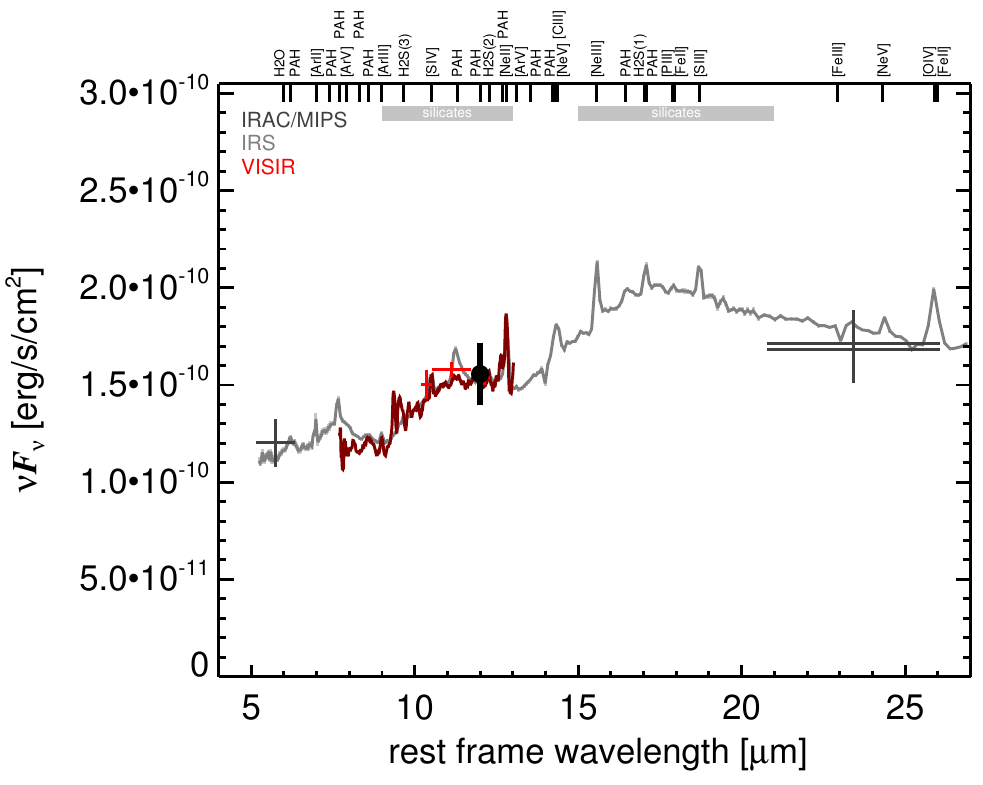}
   \caption{\label{fig:MISED_NGC4507}
      MIR SED of NGC\,4507. The description  of the symbols (if present) is the following.
      Grey crosses and  solid lines mark the \spitzer/IRAC, MIPS and IRS data. 
      The colour coding of the other symbols is: 
      green for COMICS, magenta for Michelle, blue for T-ReCS and red for VISIR data.
      Darker-coloured solid lines mark spectra of the corresponding instrument.
      The black filled circles mark the nuclear 12 and $18\,\mu$m  continuum emission estimate from the data.
      The ticks on the top axis mark positions of common MIR emission lines, while the light grey horizontal bars mark wavelength ranges affected by the silicate 10 and 18$\mu$m features.}
\end{figure}
\clearpage

\twocolumn[\begin{@twocolumnfalse}  
\subsection{NGC\,4579 -- M58}\label{app:NGC4579}
NGC\,4579 is a spiral galaxy in the Virgo cluster at a distance of $D=$ $16.8 \pm 3.4$\,Mpc \citep{tully_nearby_1988} with an active nucleus classified as a broad-line LINER/Sy\,1.9 transition nucleus \citep{ho_search_1997-1}.
A radio jet (PA$\sim135\degree$) and  star formation were detected in the nucleus \citep{ulvestad_origin_2001,contini_complex_2004}.
The first MIR observations of this object were performed with IRTF in 1980 \citep{cizdziel_multiaperture_1985}, in 1982 \citep{scoville_10_1983} and in 1983 \citep{lawrence_observations_1985}.
After \iras, NGC\,4579 was observed with \iso/ISOCAM in 1996, and an unresolved nucleus was detected at $15\,\mu$m with resolved surrounding host emission \citep{roussel_atlas_2001,ramos_almeida_mid-infrared_2007}.
The same MIR morphology is evident in the \spitzer/IRAC and MIPS images.
Because our IRAC $5.8$ and $8.0\,\mu$m and MIPS $24\,\mu$m photometry encompasses only the nuclear flux, the values are significantly lower than previously published fluxes of the same data (e.g., \citealt{dale_infrared_2005,smith_spitzer_2007,munoz-mateos_radial_2009,gallimore_infrared_2010}). 
Owing to the complex emission morphology of NGC\,4579 in the MIR, the IRS LR mapping-mode PBCD spectrum is not very reliable. 
However, it roughly matches the IRAC photometry and indicates silicate $10\,\mu$m, PAH emission and a flat spectral slope in $\nu F_\nu$-space, i.e., a composite AGN/star-formation SED (see e.g., \citealt{wu_spitzer/irs_2009,gallimore_infrared_2010,dale_spitzer_2009} for detailed versions).
We observed the core of NGC\,4579 with VISIR in several narrow $N$ and one $Q$-band filters in 2005, 2009 and 2010. 
The 2005 measurements were analysed and published in \cite{horst_small_2006,horst_mid-infrared_2009}.
In addition, we obtained Michelle imaging data in two $N$-band filters in 2010.
In all cases, a compact MIR nucleus without any extended host emission was detected. 
The nucleus appears marginally resolved in the Michelle images (FWHM(major axis) $\sim 0.57\arcsec \sim 47\,$pc; PA$\sim38\degree$), as well as possibly in the VISIR NEII\_1 and PAH2\_2 images, but not in any other image. 
Because the Michelle images were taken in bad ambient conditions, it remains uncertain, whether the nucleus is actually marginally resolved at subarcsecond resolution in the MIR. 
The nuclear fluxes are on average  $\sim29\%$ lower than the \spitzerr spectrophotometry.
The flux difference is particularly high at short wavelengths.
In addition, the PAH 11.3$\,\mu$m emission seems to be weaker or even absent at subarcsecond resolution, while the silicate $10\,\mu$m emission feature is still present and possibly even stronger.
These results indicate that part or all of the star formation affecting the \spitzerr spectrophotometry is not located in the inner $\sim 30$\,pc of the nucleus.
\newline\end{@twocolumnfalse}]

\begin{figure}
   \centering
   \includegraphics[angle=0,width=8.500cm]{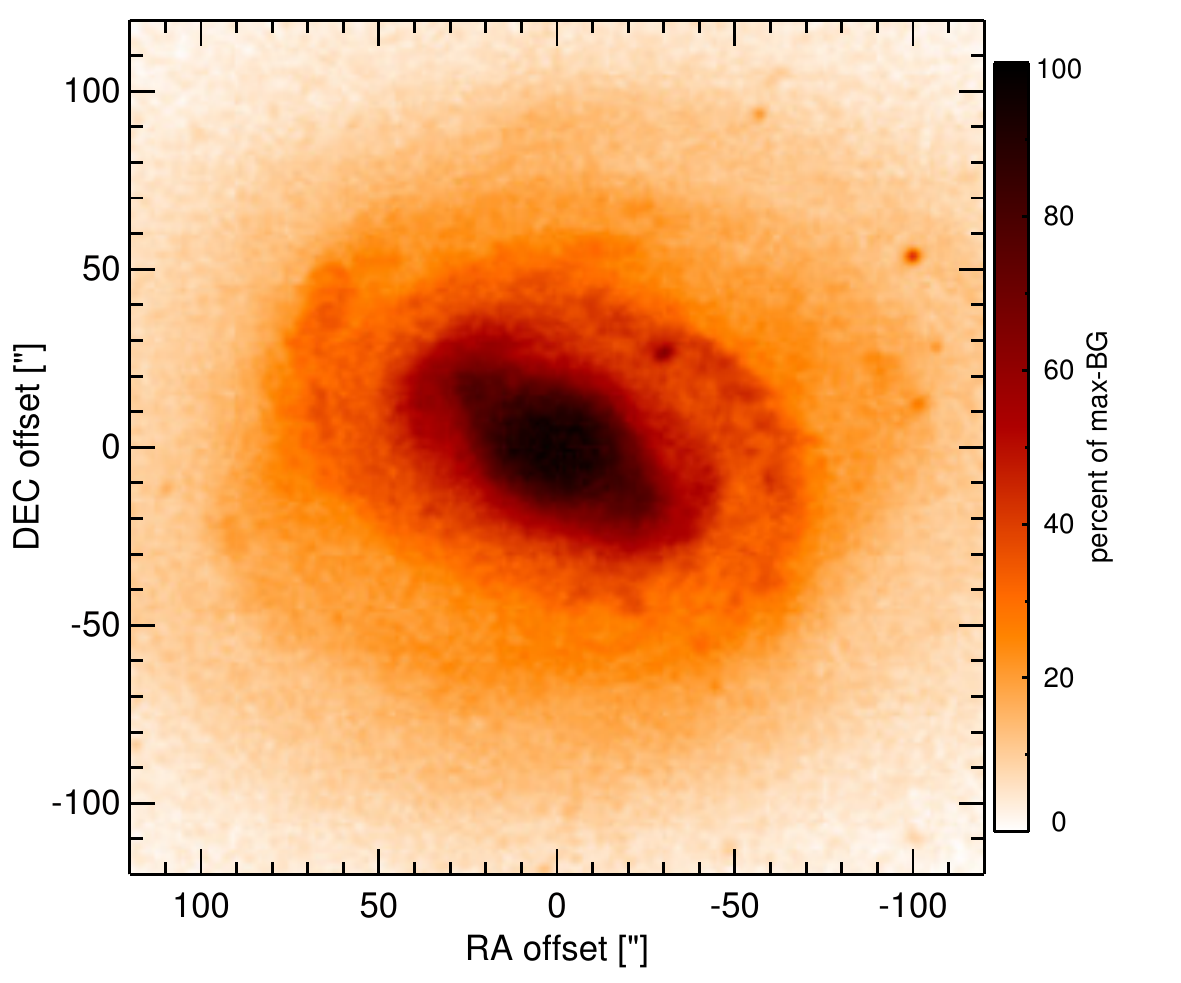}
    \caption{\label{fig:OPTim_NGC4579}
             Optical image (DSS, red filter) of NGC\,4579. Displayed are the central $4\arcmin$ with North up and East to the left. 
              The colour scaling is linear with white corresponding to the median background and black to the $0.01\%$ pixels with the highest intensity.  
           }
\end{figure}
\begin{figure}
   \centering
   \includegraphics[angle=0,height=3.11cm]{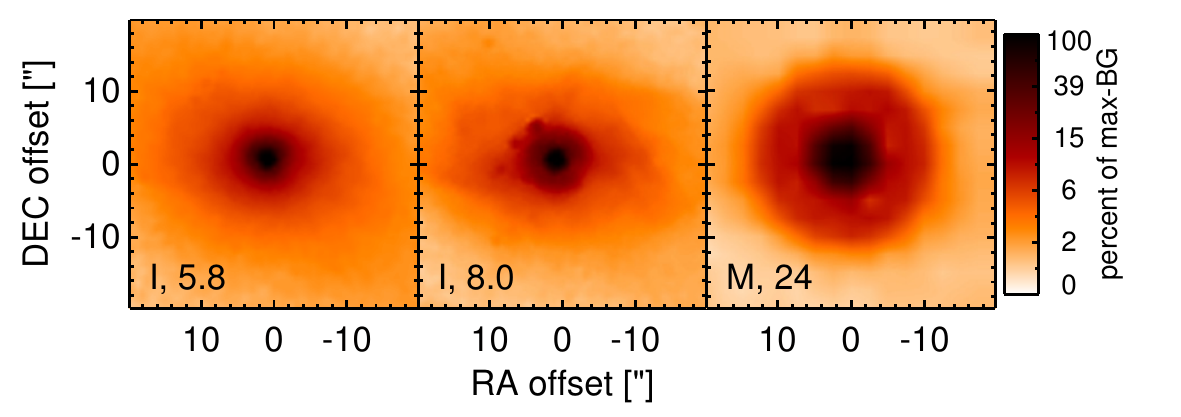}
    \caption{\label{fig:INTim_NGC4579}
             \spitzerr MIR images of NGC\,4579. Displayed are the inner $40\arcsec$ with North up and East to the left. The colour scaling is logarithmic with white corresponding to median background and black to the $0.1\%$ pixels with the highest intensity.
             The label in the bottom left states instrument and central wavelength of the filter in $\mu$m (I: IRAC, M: MIPS).
             Note that the apparent off-nuclear compact source in the IRAC $8.0\,\mu$m image is an instrumental artefact.
           }
\end{figure}
\begin{figure}
   \centering
   \includegraphics[angle=0,width=8.500cm]{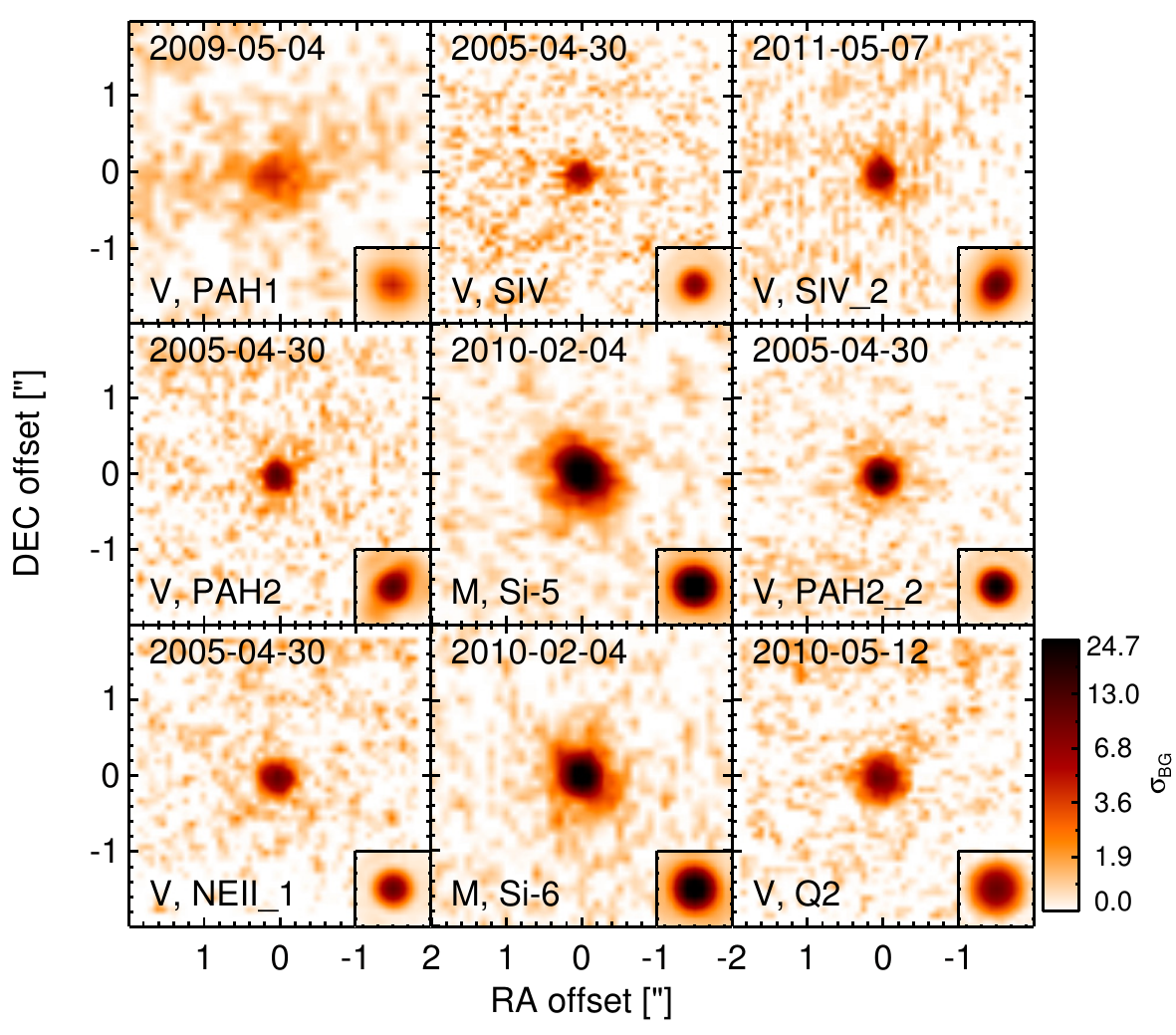}
    \caption{\label{fig:HARim_NGC4579}
             Subarcsecond-resolution MIR images of NGC\,4579 sorted by increasing filter wavelength. 
             Displayed are the inner $4\arcsec$ with North up and East to the left. 
             The colour scaling is logarithmic with white corresponding to median background and black to the $75\%$ of the highest intensity of all images in units of $\sigbg$.
             The inset image shows the central arcsecond of the PSF from the calibrator star, scaled to match the science target.
             The labels in the bottom left state instrument and filter names (C: COMICS, M: Michelle, T: T-ReCS, V: VISIR).
           }
\end{figure}
\begin{figure}
   \centering
   \includegraphics[angle=0,width=8.50cm]{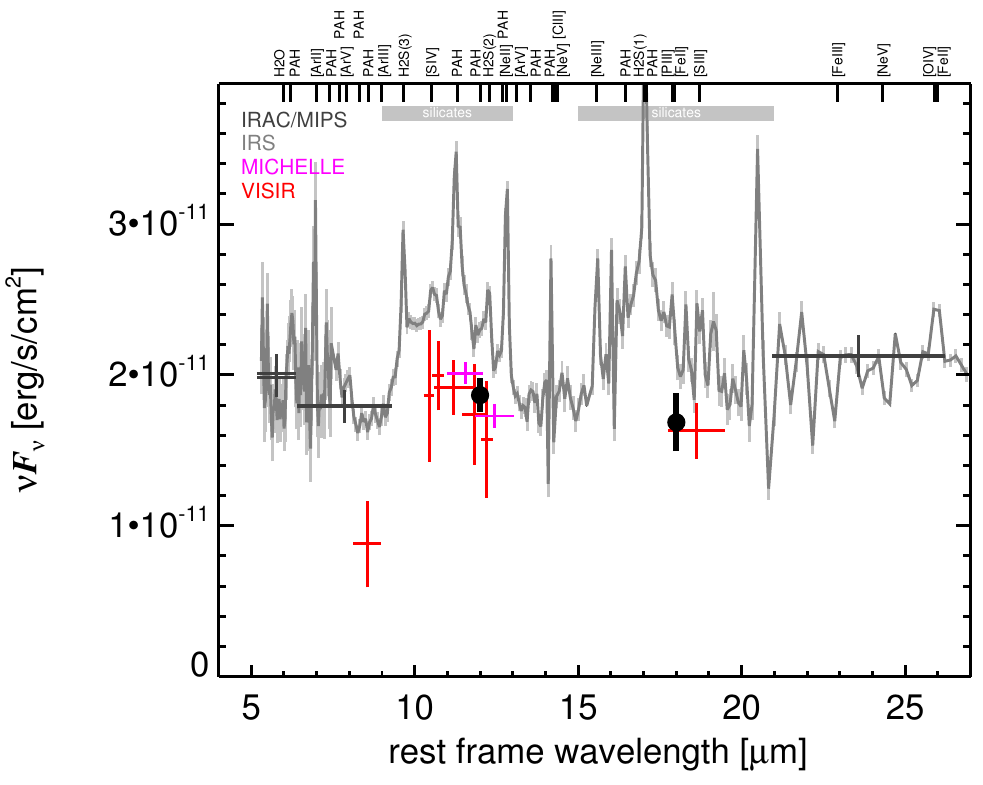}
   \caption{\label{fig:MISED_NGC4579}
      MIR SED of NGC\,4579. The description  of the symbols (if present) is the following.
      Grey crosses and  solid lines mark the \spitzer/IRAC, MIPS and IRS data. 
      The colour coding of the other symbols is: 
      green for COMICS, magenta for Michelle, blue for T-ReCS and red for VISIR data.
      Darker-coloured solid lines mark spectra of the corresponding instrument.
      The black filled circles mark the nuclear 12 and $18\,\mu$m  continuum emission estimate from the data.
      The ticks on the top axis mark positions of common MIR emission lines, while the light grey horizontal bars mark wavelength ranges affected by the silicate 10 and 18$\mu$m features.}
\end{figure}
\clearpage

\twocolumn[\begin{@twocolumnfalse}  
\subsection{NGC\,4593 -- Mrk\,1330}\label{app:NGC4593}
NGC\,4593 is a low-inclination barred spiral galaxy at a redshift of $z=$ 0.009 ($D\sim 45.6$\,Mpc) hosting a Sy\,1.0 nucleus \citep{veron-cetty_catalogue_2010} that belongs to the nine-month BAT AGN sample.
The AGN is highly variable at all wavelengths shorter than $L$-band (e.g., \citealt{santos-lleo_multifrequency_1994,santos-lleo_multifrequency_1995}) and possess a compact radio core without any evidence for a jet component
\citep{ulvestad_radio_1984,kinney_jet_2000}).
The NLR is  halo-like  with a slight elongation of $\sim1.7\arcsec \sim 370\,$pc (PA$\sim100\degree$; \citealt{schmitt_hubble_2003}).
The first ground-based $N-$ and $Q$-band photometry was obtained with UKIRT in 1982 and 1984 \citep{ward_continuum_1987}, and IRTF in 1986 \citep{devereux_spatial_1987}.
Additional MIR observations with the MMT \citep{maiolino_new_1995}, and \isoo \citep{clavel_2.5-11_2000,ramos_almeida_mid-infrared_2007} followed.
The first subarcsecond-resolution $N$-band images were taken with Palomar 5\,m/MIRLIN in 2000 \citep{gorjian_10_2004}.
The \spitzer/IRAC and MIPS images show a dominating compact nuclear source embedded within much weaker spiral-like host emission. 
Our nuclear IRAC $5.8$ and $8.0\,\mu$m fluxes are in reasonable agreement with \cite{gallimore_infrared_2010}.
The \spitzer/IRS spectrum exhibits weak silicate 10 and 18$\,\mu$m emission, a moderate PAH 11.3\,$\mu$m feature and a shallow blue spectral slope in $\nu F_\nu$-space (see also \citealt{shi_9.7_2006,wu_spitzer/irs_2009,tommasin_spitzer-irs_2010,gallimore_infrared_2010}).
Therefore, the arcsecond scales appear to be AGN-dominated.
We observed the nuclear region of NGC\,4593 with VISIR in six $N$ and one $Q$-band filter in 2005, 2008 and 2010 (partly published in \citealt{horst_mid_2008,horst_mid-infrared_2009,honig_dusty_2010-1}).
In addition, we obtained a VISIR LR $N$-band spectrum in 2008.
A compact nucleus was detected in all cases without further host emission.
It appears marginally resolved in many images including those with the best achieved angular resolution (FWHM $\sim10\%$ larger as standard star).
The PA is $\sim 110\degree$ in these cases, which would roughly match the NLR orientation.
Thus, we classify the nucleus as resolved at subarcsecond resolution in the MIR.
Our reanalysis of the published VISIR images provides  results consistent with the previous works, while the VISIR photometry is on average $\sim 22\%$ lower than the \spitzerr spectrophotometry.
Interestingly, the VISIR spectrum indicates a different spectral shape of the MIR SED at subarcsecond resolution, in particular a stronger silicate 10$\,\mu$m emission feature might be present. 
Furthermore, the nuclear photometry indicates possible flux variations on the order of $\sim20\%$ between 2005 and 2010, which is subject of a future a work (H\"onig et al., in prep.).
The nuclear MIR emission of NGC\,4593 has been further resolved with MIDI interferometric observations and was modelled as two equally bright components, one extending for tens of parsecs and one remaining unresolved \citep{burtscher_diversity_2013}.
\newline\end{@twocolumnfalse}]

\begin{figure}
   \centering
   \includegraphics[angle=0,width=8.500cm]{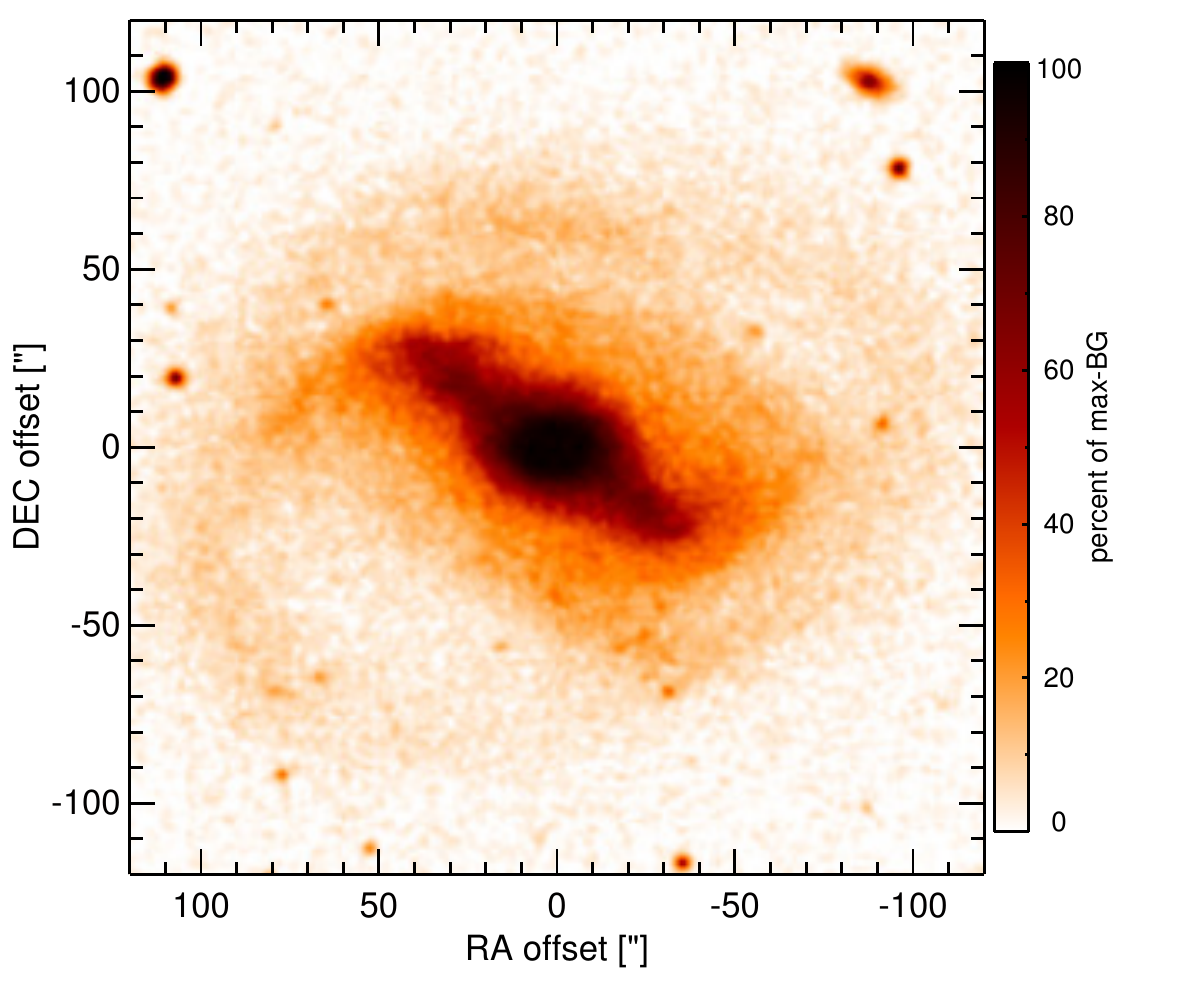}
    \caption{\label{fig:OPTim_NGC4593}
             Optical image (DSS, red filter) of NGC\,4593. Displayed are the central $4\arcmin$ with North up and East to the left. 
              The colour scaling is linear with white corresponding to the median background and black to the $0.01\%$ pixels with the highest intensity.  
           }
\end{figure}
\begin{figure}
   \centering
   \includegraphics[angle=0,height=3.11cm]{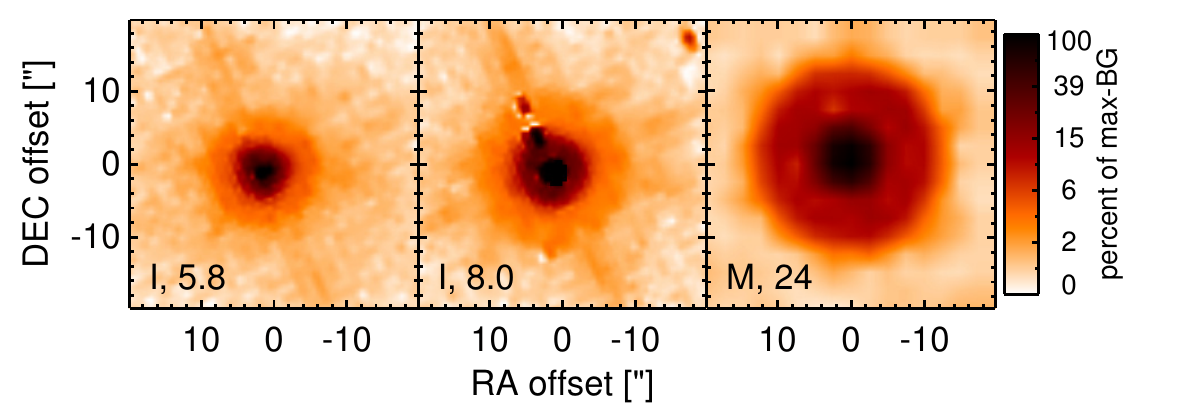}
    \caption{\label{fig:INTim_NGC4593}
             \spitzerr MIR images of NGC\,4593. Displayed are the inner $40\arcsec$ with North up and East to the left. The colour scaling is logarithmic with white corresponding to median background and black to the $0.1\%$ pixels with the highest intensity.
             The label in the bottom left states instrument and central wavelength of the filter in $\mu$m (I: IRAC, M: MIPS).
             Note that the apparent off-nuclear compact sources in the IRAC $8.0\,\mu$m image are instrumental artefacts.
           }
\end{figure}
\begin{figure}
   \centering
   \includegraphics[angle=0,width=8.500cm]{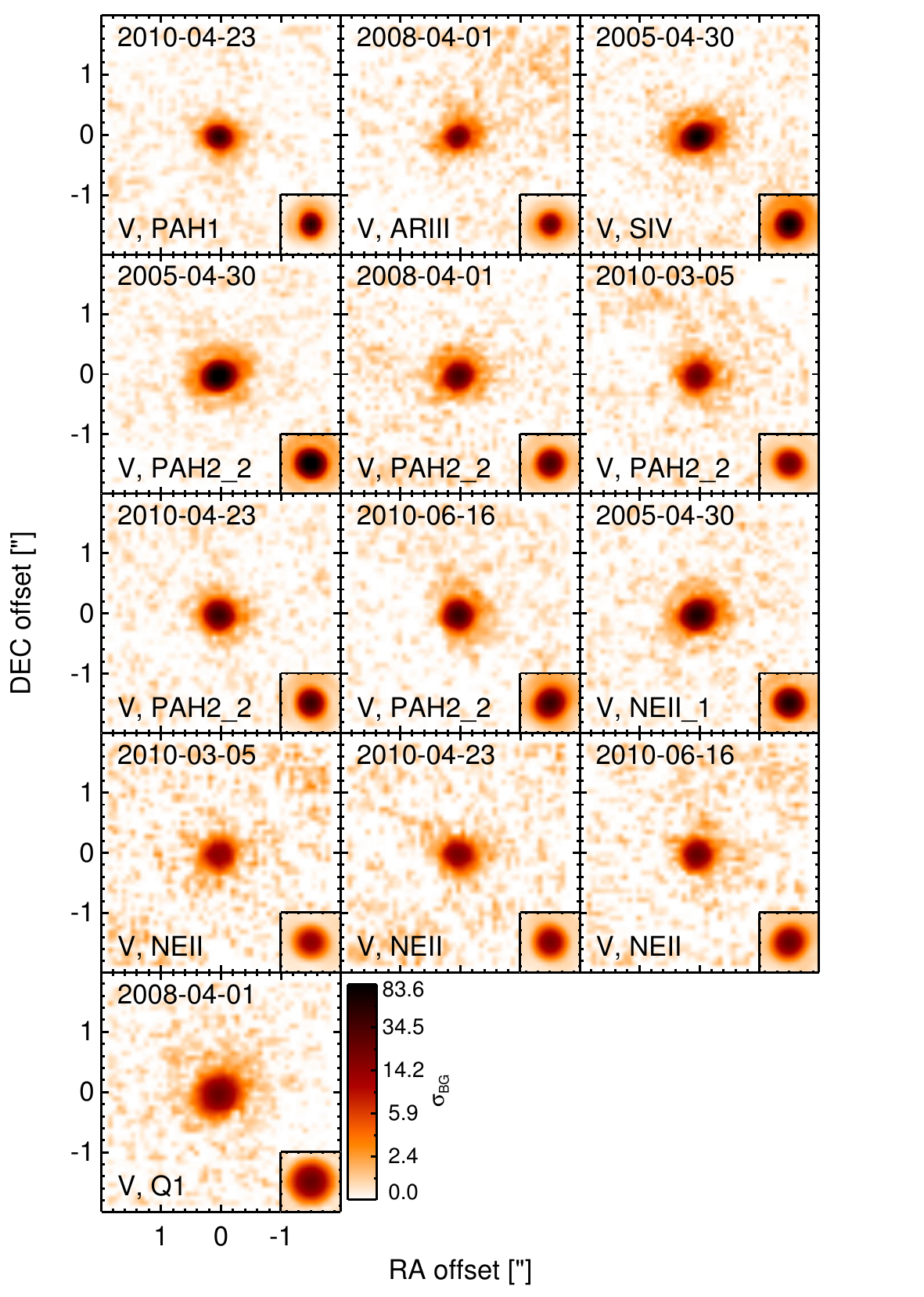}
    \caption{\label{fig:HARim_NGC4593}
             Subarcsecond-resolution MIR images of NGC\,4593 sorted by increasing filter wavelength. 
             Displayed are the inner $4\arcsec$ with North up and East to the left. 
             The colour scaling is logarithmic with white corresponding to median background and black to the $75\%$ of the highest intensity of all images in units of $\sigbg$.
             The inset image shows the central arcsecond of the PSF from the calibrator star, scaled to match the science target.
             The labels in the bottom left state instrument and filter names (C: COMICS, M: Michelle, T: T-ReCS, V: VISIR).
           }
\end{figure}
\begin{figure}
   \centering
   \includegraphics[angle=0,width=8.50cm]{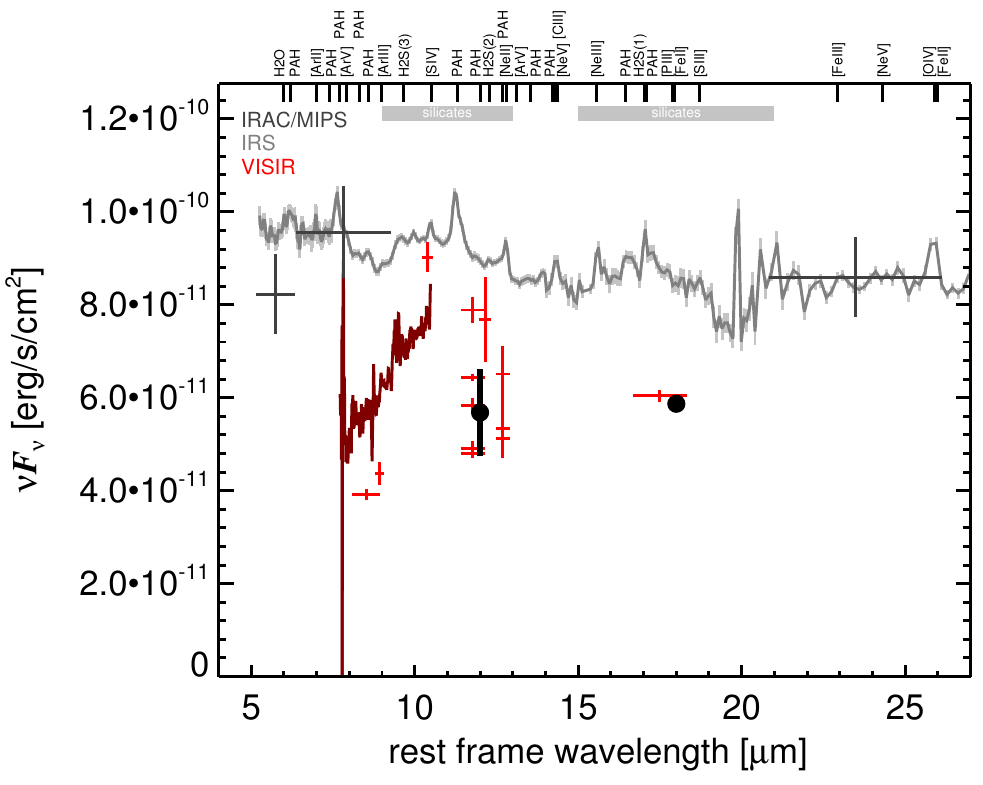}
   \caption{\label{fig:MISED_NGC4593}
      MIR SED of NGC\,4593. The description  of the symbols (if present) is the following.
      Grey crosses and  solid lines mark the \spitzer/IRAC, MIPS and IRS data. 
      The colour coding of the other symbols is: 
      green for COMICS, magenta for Michelle, blue for T-ReCS and red for VISIR data.
      Darker-coloured solid lines mark spectra of the corresponding instrument.
      The black filled circles mark the nuclear 12 and $18\,\mu$m  continuum emission estimate from the data.
      The ticks on the top axis mark positions of common MIR emission lines, while the light grey horizontal bars mark wavelength ranges affected by the silicate 10 and 18$\mu$m features.}
\end{figure}
\clearpage

\twocolumn[\begin{@twocolumnfalse}  
\subsection{NGC\,4594 -- Sombrero Galaxy -- M104}\label{app:NGC4594}
NGC\,4594 is an edge-on early-type spiral galaxy at a distance of $D=$ $9.1 \pm 3.4$\,Mpc (NED redshift-independent median) with an active nucleus classified as a LINER (previously classified as a Sy\,1.9; \citealt{veron-cetty_catalogue_2010}), which appears as a compact radio source with a flat spectrum \citep{shaffer_vlbi_1979,thean_high-resolution_2000}.
The AGN is also discussed as ``true''-Sy\,2 candidate \citep{shi_unobscured_2010}.
First unsuccessful attempts to observe the nucleus in the MIR were made by \cite{kleinmann_infrared_1970} and \cite{rieke_10_1978}.
\cite{maiolino_new_1995} first detected the nucleus with the MMT bolometer, while it remained undetected again in the first subarcsecond-resolution $N$-band imaging attempts with Palomar 5\,m/MIRLIN \citep{gorjian_10_2004} and ESO 3.6\,m/TIMMI2 \citep{raban_core_2008}.
Finally, the nucleus could be detected with Keck/LWS in $N$-band in 2003 where it appeared marginally resolved (FWHM$\sim 0,5\arcsec\sim22$\,pc; \citealt{grossan_high_2004}).
In the IRAC and MIPS images the nucleus appears extended and embedded within elliptical host emission.
The famous large-scale dust ring is also detected and becomes brighter towards longer wavelengths.
We measure the nuclear flux in the IRAC $5.8$ and $8.0\,\mu$m and MIPS $24\,\mu$m images, which results in values consistent with the nuclear fluxes in \cite{bendo_spitzer_2006}.
The IRS LR mapping-mode spectrum shows weak silicate and PAH emission and a blue spectral slope  in $\nu F_\nu$-space, i.e., it seems to be dominated by old stellar host emission with only minor star formation (see also \citealt{bendo_spitzer_2006, smith_mid-infrared_2007,wu_spitzer/irs_2009,shi_unobscured_2010,gallimore_infrared_2010,mason_nuclear_2012}).
NGC\,4594 was observed with VISIR in PAH2\_2 in 2006 \citep{reunanen_vlt_2010}, in PAH2 and NEII\_1 in 2009 \citep{asmus_mid-infrared_2011} and in PAH2\_2 again in 2010 (unpublished, to our knowledge).
The nucleus remained undetected in all VISIR images. 
However, it was also observed with T-ReCS in the Si2 filter in four different nights in 2007 where a compact MIR nucleus is always detected \citep{mason_nuclear_2012}.
The low S/N prevents us from performing a quantitative extension analysis, but owing to the fact that the nucleus appears extended in all subarcsecond-resolution images in which it was detected, we classify it as possibly extended.
Our measured nuclear photometry is consistent with the previously published values and on average $\sim 78\%$ lower than the \spitzerr spectrophotometry.
Therefore, we conclude that stellar emission completely dominates the MIR SED in the central 
$\sim 200$\,pc.
\newline\end{@twocolumnfalse}]

\begin{figure}
   \centering
   \includegraphics[angle=0,width=8.500cm]{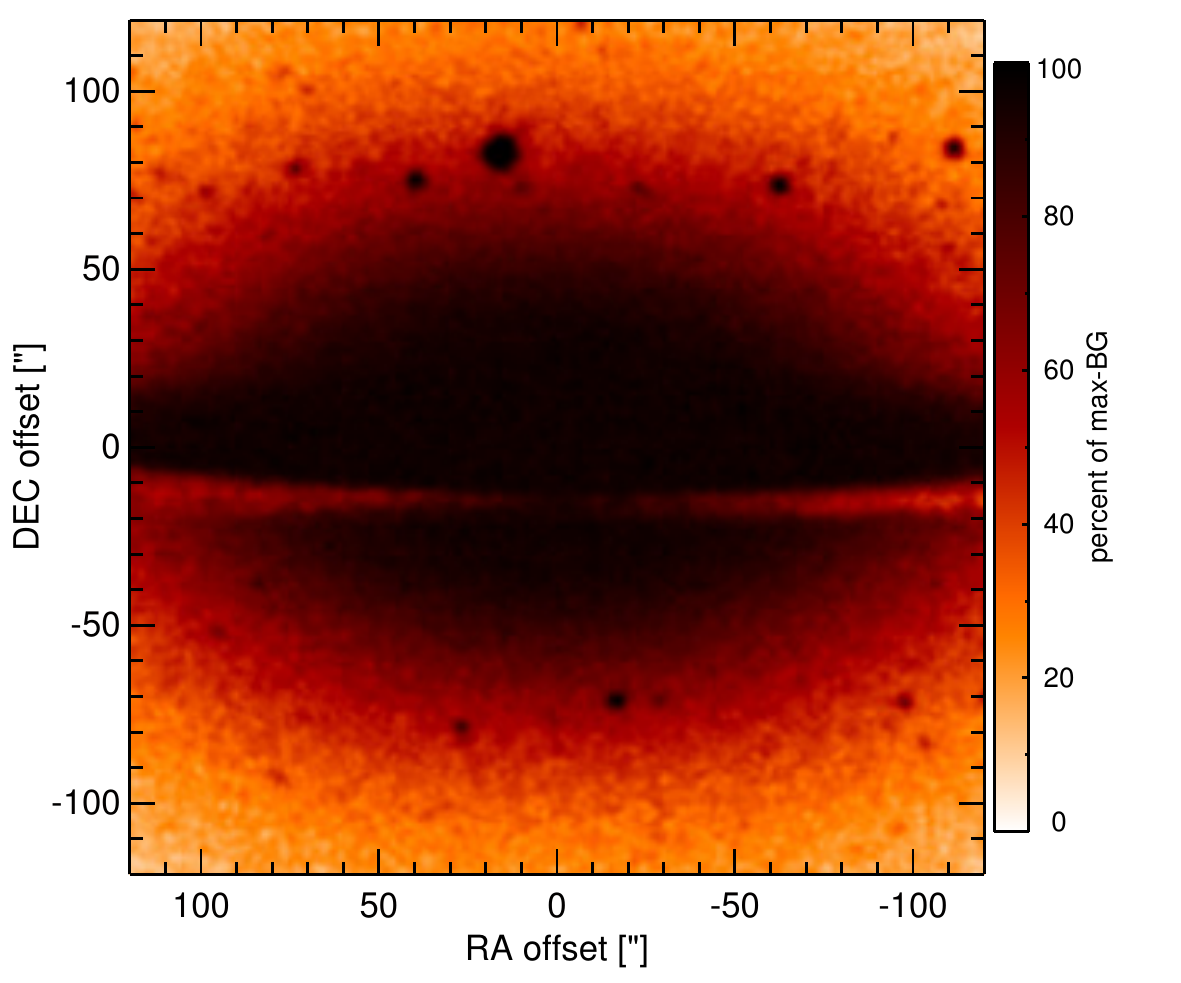}
    \caption{\label{fig:OPTim_NGC4594}
             Optical image (DSS, red filter) of NGC\,4594. Displayed are the central $4\arcmin$ with North up and East to the left. 
              The colour scaling is linear with white corresponding to the median background and black to the $0.01\%$ pixels with the highest intensity.  
           }
\end{figure}
\begin{figure}
   \centering
   \includegraphics[angle=0,height=3.11cm]{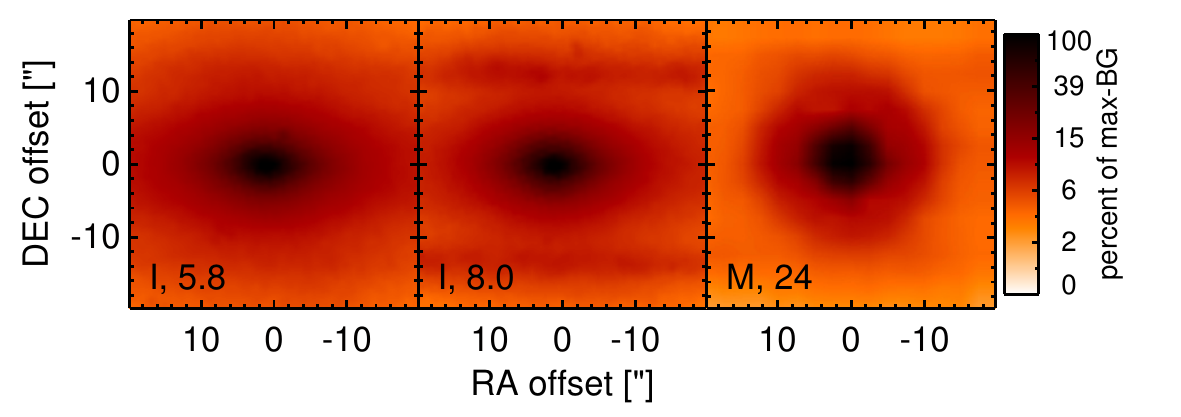}
    \caption{\label{fig:INTim_NGC4594}
             \spitzerr MIR images of NGC\,4594. Displayed are the inner $40\arcsec$ with North up and East to the left. The colour scaling is logarithmic with white corresponding to median background and black to the $0.1\%$ pixels with the highest intensity.
             The label in the bottom left states instrument and central wavelength of the filter in $\mu$m (I: IRAC, M: MIPS). 
           }
\end{figure}
\begin{figure}
   \centering
   \includegraphics[angle=0,width=8.500cm]{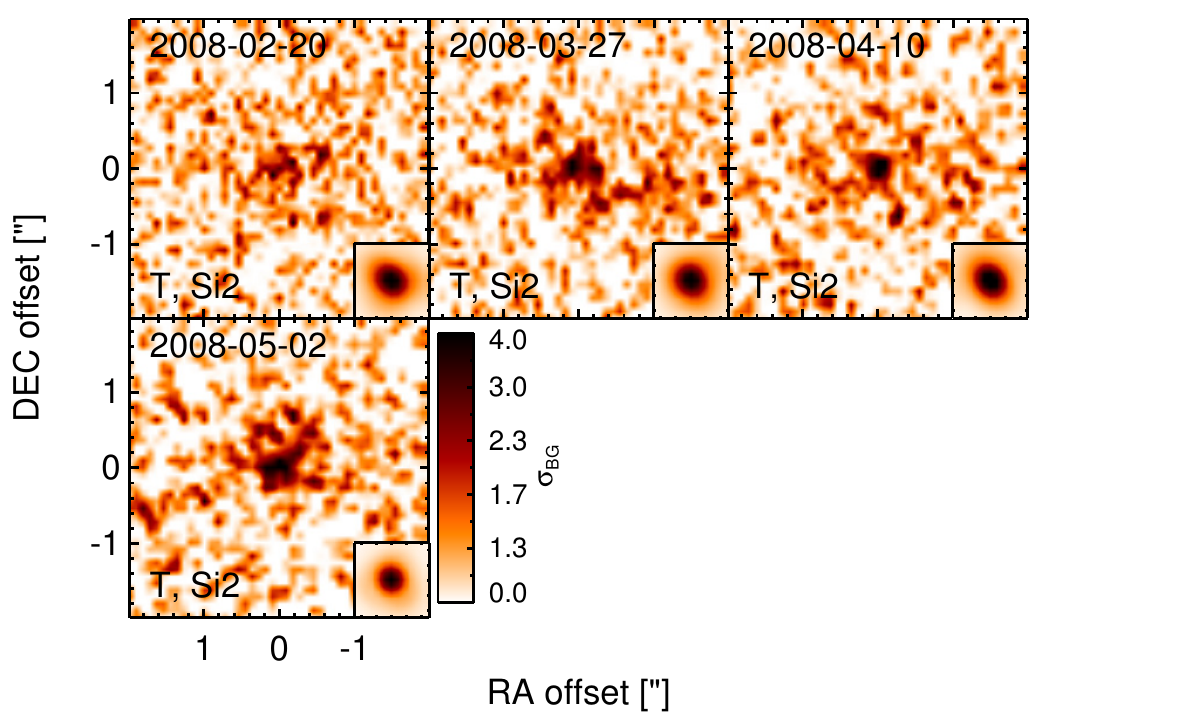}
    \caption{\label{fig:HARim_NGC4594}
             Subarcsecond-resolution MIR images of NGC\,4594 sorted by increasing filter wavelength. 
             Displayed are the inner $4\arcsec$ with North up and East to the left. 
             The colour scaling is logarithmic with white corresponding to median background and black to the $75\%$ of the highest intensity of all images in units of $\sigbg$.
             The inset image shows the central arcsecond of the PSF from the calibrator star, scaled to match the science target.
             The labels in the bottom left state instrument and filter names (C: COMICS, M: Michelle, T: T-ReCS, V: VISIR).
           }
\end{figure}
\begin{figure}
   \centering
   \includegraphics[angle=0,width=8.50cm]{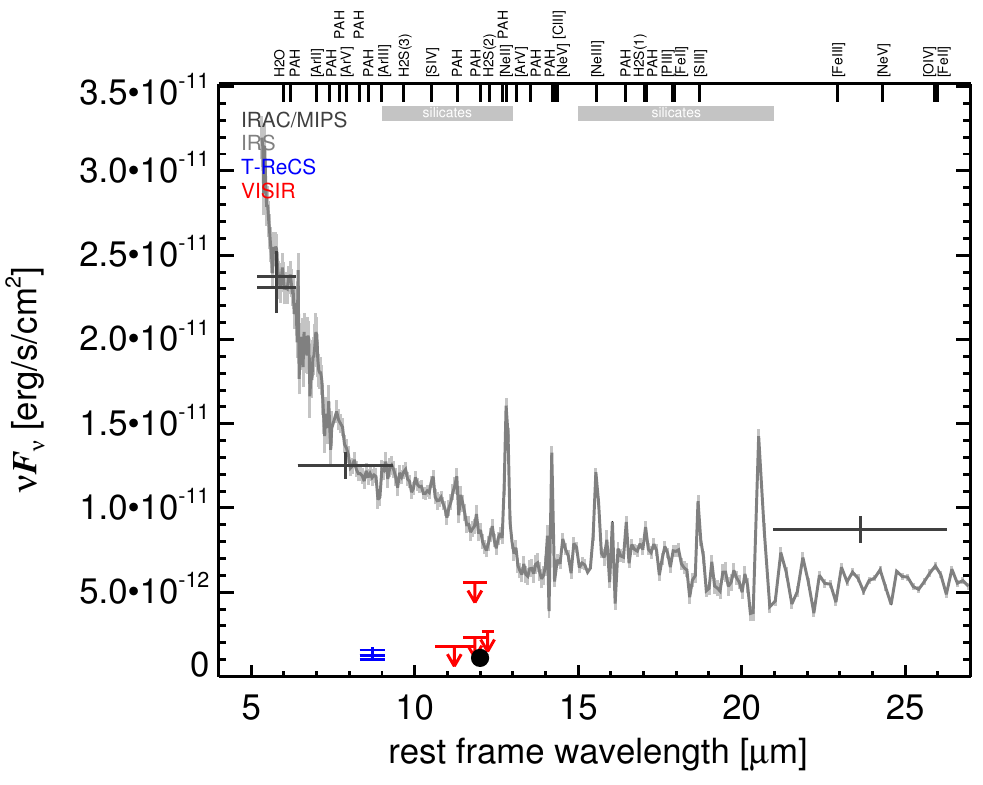}
   \caption{\label{fig:MISED_NGC4594}
      MIR SED of NGC\,4594. The description  of the symbols (if present) is the following.
      Grey crosses and  solid lines mark the \spitzer/IRAC, MIPS and IRS data. 
      The colour coding of the other symbols is: 
      green for COMICS, magenta for Michelle, blue for T-ReCS and red for VISIR data.
      Darker-coloured solid lines mark spectra of the corresponding instrument.
      The black filled circles mark the nuclear 12 and $18\,\mu$m  continuum emission estimate from the data.
      The ticks on the top axis mark positions of common MIR emission lines, while the light grey horizontal bars mark wavelength ranges affected by the silicate 10 and 18$\mu$m features.}
\end{figure}
\clearpage

\twocolumn[\begin{@twocolumnfalse}  
\subsection{NGC\,4636 -- VCC\,1939}\label{app:NGC4636}
NGC\,4636 is one of the X-ray brightest elliptical galaxies in the Virgo cluster at a distance of $D=$ $15.6 \pm 2.7\,$Mpc (NED redshift-independent median).
It hosts a broad-line LINER \citep{ho_search_1997-1}.
An X-ray point source consistent with the nuclear position is present, but appears very weakly and with a soft spectrum, which disfavours an AGN unless highly obscured  \citep{gonzalez-martin_x-ray_2006,gonzalez-martin_fitting_2009,flohic_central_2006,baldi_unusual_2009}.
In addition, NGC\,4636 features a weak compact radio source with kiloparsec-scale jet-like bicones (PA$\sim 40\degree$; \citealt{birkinshaw_orientations_1985,stanger_high-resolution_1986,nagar_radio_2000}).
The NLR has a compact and a wide outflow component, extending $\sim3\arcsec\sim220\,$pc towards the north (e.g., \citealt{masegosa_nature_2011}).
The first ground-based MIR observations of NGC\,4636 were performed by \cite{impey_infrared_1986} and \cite{devereux_infrared_1987}, but the nucleus remained undetected.
In addition, \isoo \citep{ferrari_survey_2002,athey_mid-infrared_2002,temi_cold_2004,temi_ages_2005} and \spitzer/IRAC, IRS and MIPS observed this object.
The corresponding IRAC and MIPS images show extended host emission without any separable nuclear component.
Our unresolved nuclear MIPS $24\,\mu$m flux measurement, therefore, provides  a significantly lower value than the total flux given in \cite{temi_spitzer_2009}.
The IRS LR staring-mode spectrum shows weak silicate 10 and $18\,\mu$m emission and a blue spectral slope in $\nu F_\nu$-space but no PAH features, typical for an inactive elliptical (see also \citealt{bressan_spitzer_2006}).
The nuclear region of NGC\,4636 was observed with VISIR in the PAH2\_2 during three nights in 2010,  but nothing was detected (unpublished, to our knowledge).
All derived flux upper limits are higher than the \spitzerr spectrophotometry, which we, therefore, use to derive an upper limit on the $12\,\mu$m continuum emission of any AGN in NGC\,4636.  
\newline\end{@twocolumnfalse}]

\begin{figure}
   \centering
   \includegraphics[angle=0,width=8.500cm]{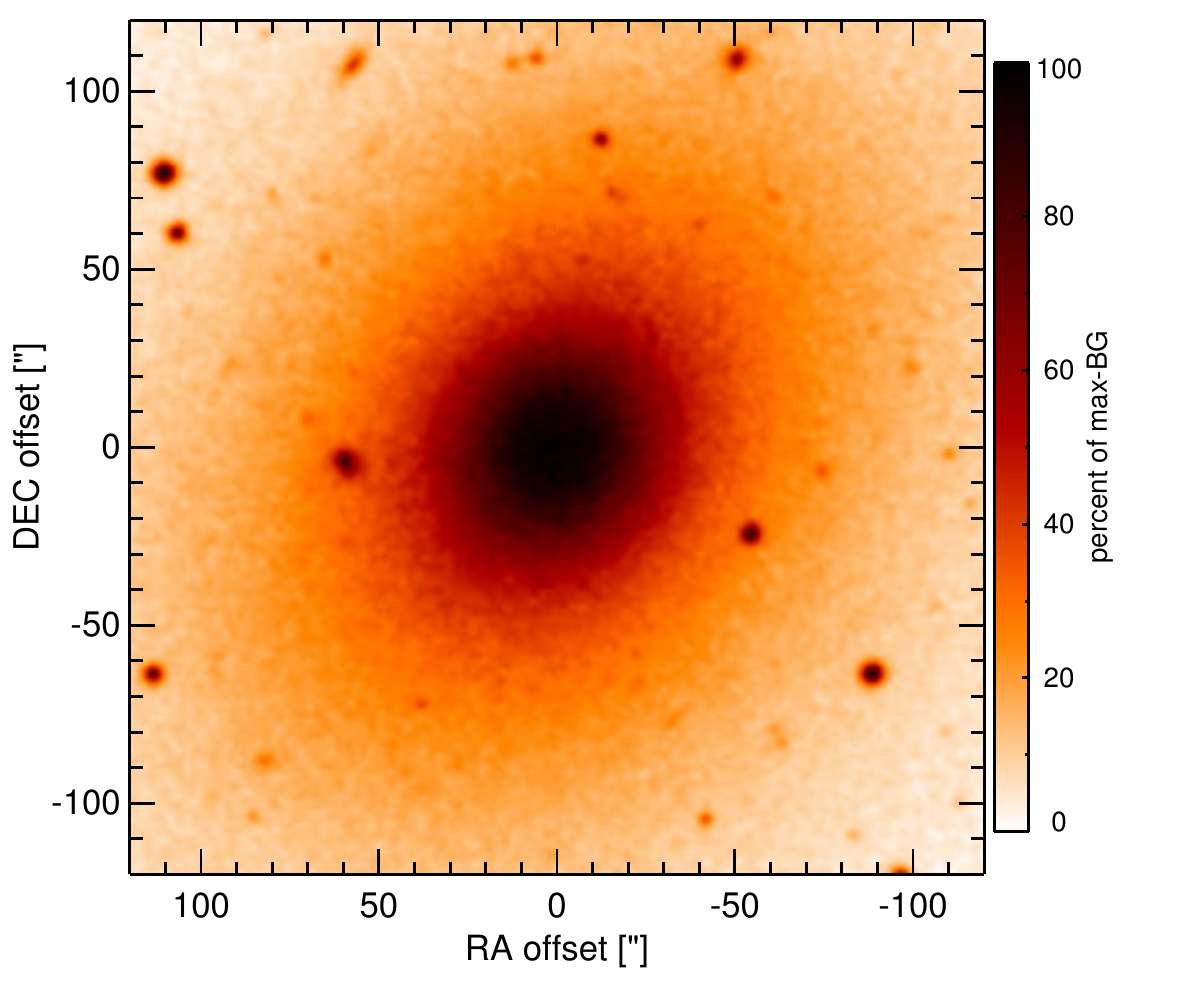}
    \caption{\label{fig:OPTim_NGC4636}
             Optical image (DSS, red filter) of NGC\,4636. Displayed are the central $4\arcmin$ with North up and East to the left. 
              The colour scaling is linear with white corresponding to the median background and black to the $0.01\%$ pixels with the highest intensity.  
           }
\end{figure}
\begin{figure}
   \centering
   \includegraphics[angle=0,height=3.11cm]{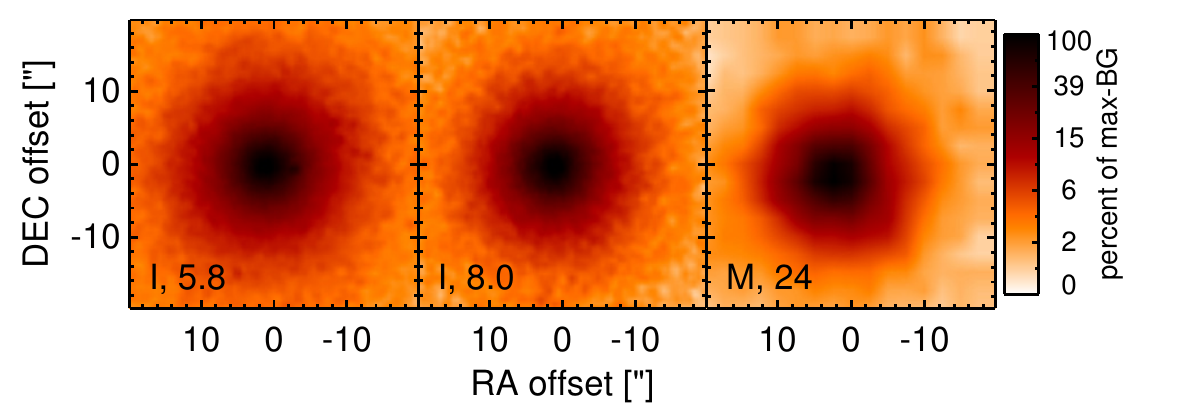}
    \caption{\label{fig:INTim_NGC4636}
             \spitzerr MIR images of NGC\,4636. Displayed are the inner $40\arcsec$ with North up and East to the left. The colour scaling is logarithmic with white corresponding to median background and black to the $0.1\%$ pixels with the highest intensity.
             The label in the bottom left states instrument and central wavelength of the filter in $\mu$m (I: IRAC, M: MIPS). 
           }
\end{figure}
\begin{figure}
   \centering
   \includegraphics[angle=0,width=8.50cm]{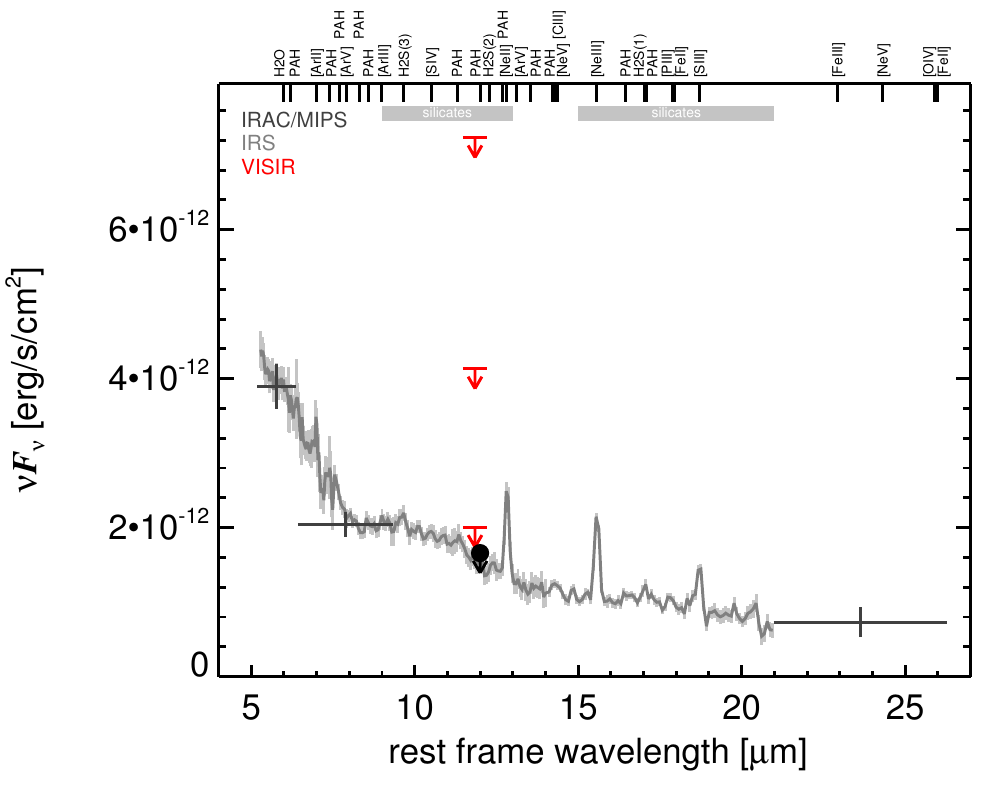}
   \caption{\label{fig:MISED_NGC4636}
      MIR SED of NGC\,4636. The description  of the symbols (if present) is the following.
      Grey crosses and  solid lines mark the \spitzer/IRAC, MIPS and IRS data. 
      The colour coding of the other symbols is: 
      green for COMICS, magenta for Michelle, blue for T-ReCS and red for VISIR data.
      Darker-coloured solid lines mark spectra of the corresponding instrument.
      The black filled circles mark the nuclear 12 and $18\,\mu$m  continuum emission estimate from the data.
      The ticks on the top axis mark positions of common MIR emission lines, while the light grey horizontal bars mark wavelength ranges affected by the silicate 10 and 18$\mu$m features.}
\end{figure}
\clearpage

\twocolumn[\begin{@twocolumnfalse}  
\subsection{NGC\,4698 -- VCC\,2070}\label{app:NGC4698}
NGC\,4698 is an inclined early-type spiral galaxy in the Virgo Cluster at a distance of $D=$ $24.4 \pm 6.7\,$Mpc (NED redshift-independent median) that hosts a Sy\,2.0 nucleus \citep{veron-cetty_catalogue_2010}.
The AGN is considered a ``true''-Seyfert~2 candidate (\citealt{pappa_x-ray_2001,georgantopoulos_chandra_2003}; but see \citealt{shi_unobscured_2010}).
However, the hard X-ray point source in its centre might in fact be Compton-thick \citep{gonzalez-martin_fitting_2009}.
A compact radio source was weakly detected \citep{ho_radio_2001}.
The first ground-based MIR detection attempts remained unsuccessful \citep{scoville_10_1983}.
NGC\,4698 was observed with the space-based \isoo \citep{boselli_mid-ir_1998} and \spitzer/IRAC, IRS and MIPS.
The corresponding IRAC and MIPS images show an extended nucleus embedded within large-scale ring-like host emission.
Our MIPS 24\,$\mu$m unresolved nuclear flux measurement agrees with \cite{shi_unobscured_2010}.
The IRS LR staring-mode spectrum suffers from low S/N but indicates weak silicate 10 and $18\,\mu$m emission, a very weak PAH 11.3$\mu$m feature and a blue spectral slope in $\nu F_\nu$-space (see also \citealt{shi_unobscured_2010}).
Therefore, the arcsecond-scale MIR SED does neither indicate intense star formation nor AGN activity.
We observed the nuclear region of NGC\,4698 with VISIR in the NEII\_1 filter in 2006 but did not detect any compact source \citep{horst_mid_2008}.
The corresponding flux upper limit is higher than the \spitzerr spectrophotometry, which thus is used  to derive an upper limit on the $12\,\mu$m continuum emission of the AGN in NGC\,4698.
As discussed already in \cite{asmus_mid-infrared_2011}, the nuclear MIR upper limit does not constrain the obscuration in this object.
\newline\end{@twocolumnfalse}]

\begin{figure}
   \centering
   \includegraphics[angle=0,width=8.500cm]{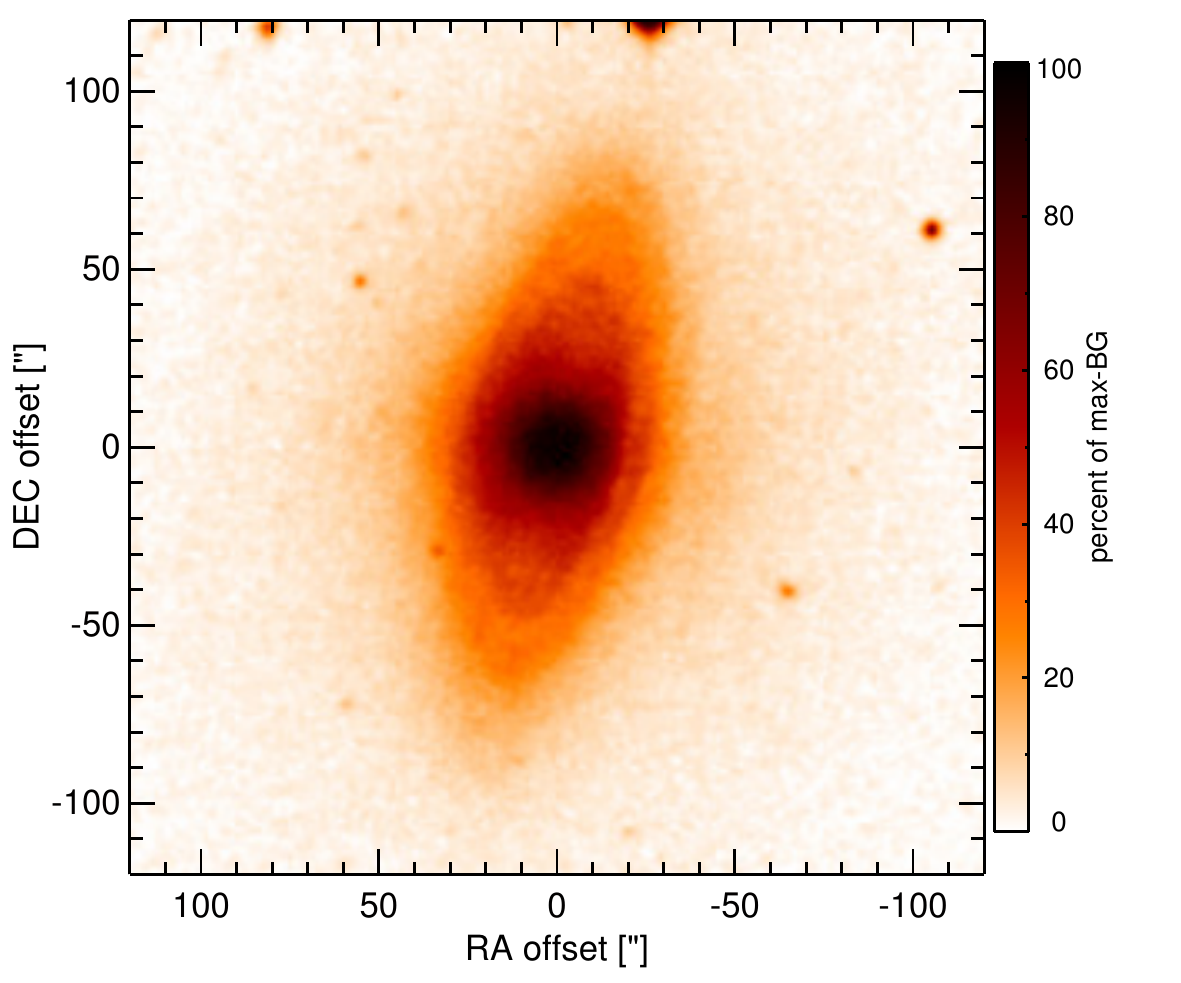}
    \caption{\label{fig:OPTim_NGC4698}
             Optical image (DSS, red filter) of NGC\,4698. Displayed are the central $4\arcmin$ with North up and East to the left. 
              The colour scaling is linear with white corresponding to the median background and black to the $0.01\%$ pixels with the highest intensity.  
           }
\end{figure}
\begin{figure}
   \centering
   \includegraphics[angle=0,height=3.11cm]{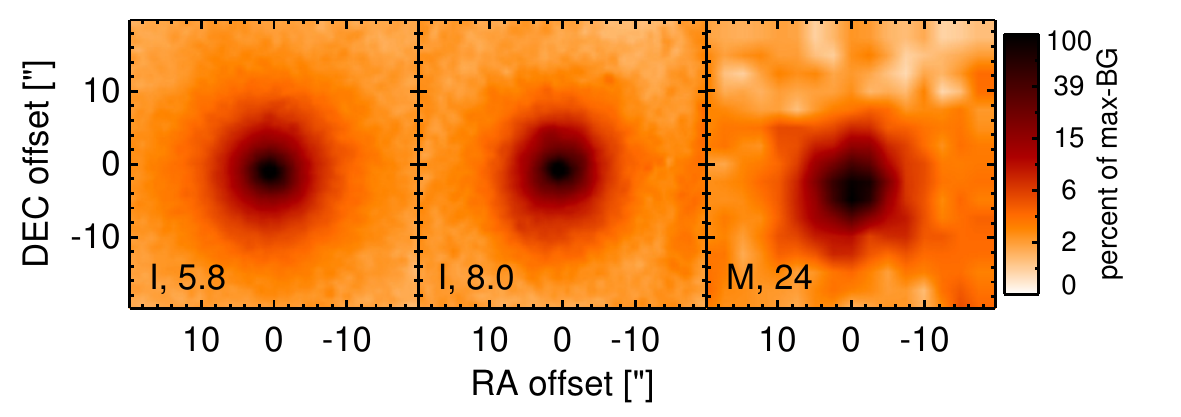}
    \caption{\label{fig:INTim_NGC4698}
             \spitzerr MIR images of NGC\,4698. Displayed are the inner $40\arcsec$ with North up and East to the left. The colour scaling is logarithmic with white corresponding to median background and black to the $0.1\%$ pixels with the highest intensity.
             The label in the bottom left states instrument and central wavelength of the filter in $\mu$m (I: IRAC, M: MIPS). 
           }
\end{figure}
\begin{figure}
   \centering
   \includegraphics[angle=0,width=8.50cm]{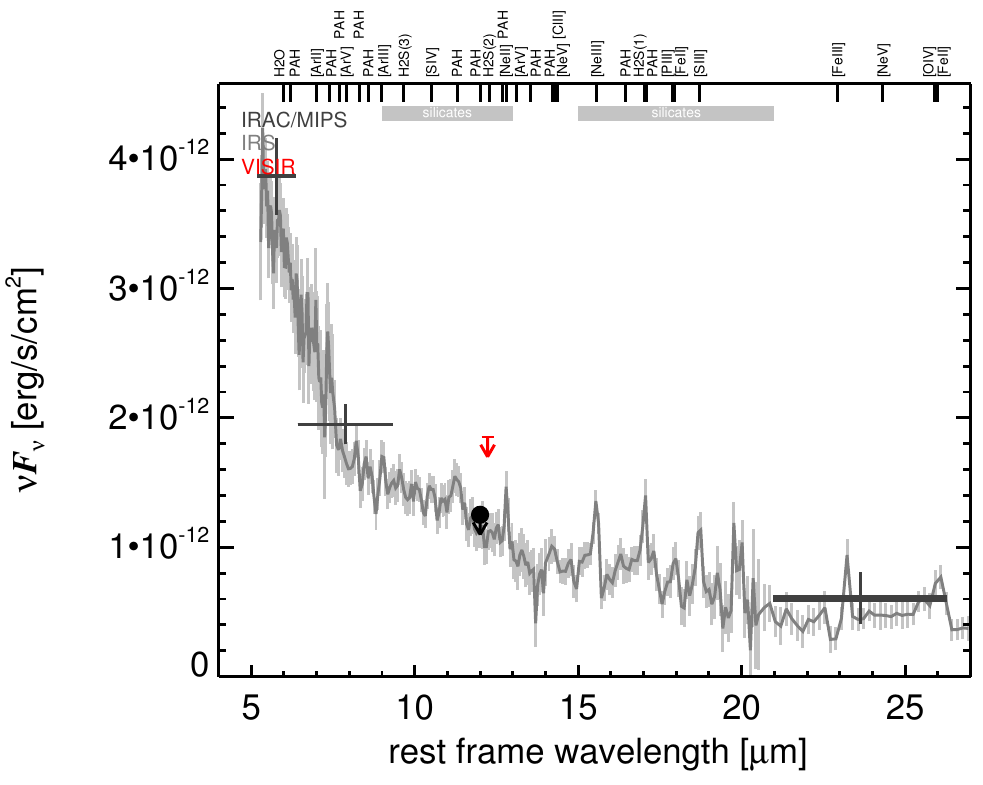}
   \caption{\label{fig:MISED_NGC4698}
      MIR SED of NGC\,4698. The description  of the symbols (if present) is the following.
      Grey crosses and  solid lines mark the \spitzer/IRAC, MIPS and IRS data. 
      The colour coding of the other symbols is: 
      green for COMICS, magenta for Michelle, blue for T-ReCS and red for VISIR data.
      Darker-coloured solid lines mark spectra of the corresponding instrument.
      The black filled circles mark the nuclear 12 and $18\,\mu$m  continuum emission estimate from the data.
      The ticks on the top axis mark positions of common MIR emission lines, while the light grey horizontal bars mark wavelength ranges affected by the silicate 10 and 18$\mu$m features.}
\end{figure}
\clearpage

\twocolumn[\begin{@twocolumnfalse}  
\subsection{NGC\,4736 -- M94}\label{app:NGC4736}
NGC\,4736 is a spiral galaxy at a distance of $D=$ $4.9 \pm 0.8$\,Mpc (NED redshift-independent median) with an active  nucleus classified as a LINER \citep{ho_search_1997-1}.
There are indications of a past powerful starburst in the nucleus \citep{walker_2_1988,taniguchi_poststarburst_1996} and a nuclear star cluster \citep{gonzalez_delgado_hst/wfpc2_2008}, which could also explain the observed optical emission lines.
However, the existence of a compact source in radio, UV and X-ray together with the  H$\alpha$ properties favour the presence of a low-luminosity AGN in NGC\,4736 \citep{nagar_radio_2005,kording_radio_2005,gonzalez-martin_x-ray_2009,masegosa_nature_2011}.
Therefore, we  conclude that an AGN is present in NGC\,4736.
In fact, it is possible that NGC\,4736 harbours a double AGN as suggested by the presence of two UV point sources (separation $2.5\arcsec \sim 60$\,pc, PA$\sim 170\degree$; \citealt{maoz_detection_1995,maoz_murmur_2005}).
The first MIR observations of NGC\,4736 are reported by \cite{kleinmann_infrared_1970}, \cite{rieke_infrared_1972}, \cite{rieke_infrared_1978}, and \cite{dyck_photometry_1978}.
After \iras, the galaxy was also observed with \iso/ISOCAM \citep{roussel_atlas_2001} and \spitzer/IRAC, IRS and MIPS.
The IRAC and MIPS images show an extended nucleus embedded within bright host emission. 
Because we measure the nuclear component only, our  IRAC $5.8$ and $8.0\,\mu$m  and MIPS $24\,\mu$m fluxes are much lower than the literature values (e.g., \citealt{dale_infrared_2005,smith_spitzer_2007,munoz-mateos_radial_2009}).
The IRS LR mapping-mode PBCD spectrum is not very reliable but matches qualitatively the more accurate versions in the literature \citep{goulding_towards_2009,mason_nuclear_2012}.
It exhibits strong PAH emission and weak silicate absorption, indicating star formation in the central $\sim 100$\,pc of NGC\,4736.
The nuclear region of NGC\,4736 was imaged with Michelle in the $N-$ and $Q$-band in 2007.
An extended nucleus (FWHM$>1\arcsec$) embedded within weak diffuse emission was detected \citep{mason_nuclear_2012}.
Our photometry of the unresolved component provides values consistent with \cite{mason_nuclear_2012}, while being on average $\sim89\%$ lower than the \spitzerr spectrophotometry.
Thus, the central $\sim 100$\,pc of NGC\,4736 are completely dominated by extended stellar emission.
Note that the second possible nucleus has not been detected yet in the MIR. 
\newline\end{@twocolumnfalse}]

\begin{figure}
   \centering
   \includegraphics[angle=0,width=8.500cm]{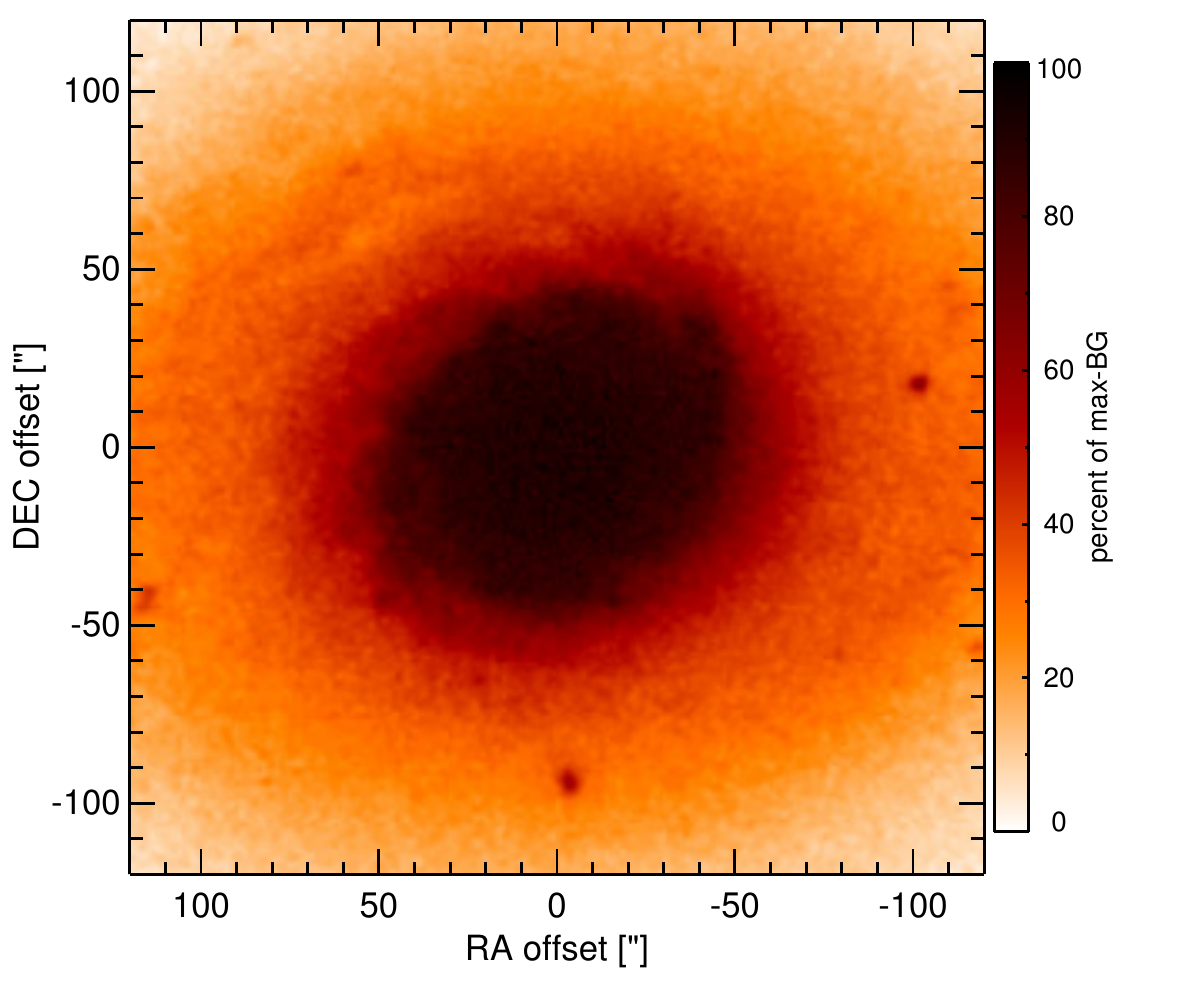}
    \caption{\label{fig:OPTim_NGC4736}
             Optical image (DSS, red filter) of NGC\,4736. Displayed are the central $4\arcmin$ with North up and East to the left. 
              The colour scaling is linear with white corresponding to the median background and black to the $0.01\%$ pixels with the highest intensity.  
           }
\end{figure}
\begin{figure}
   \centering
   \includegraphics[angle=0,height=3.11cm]{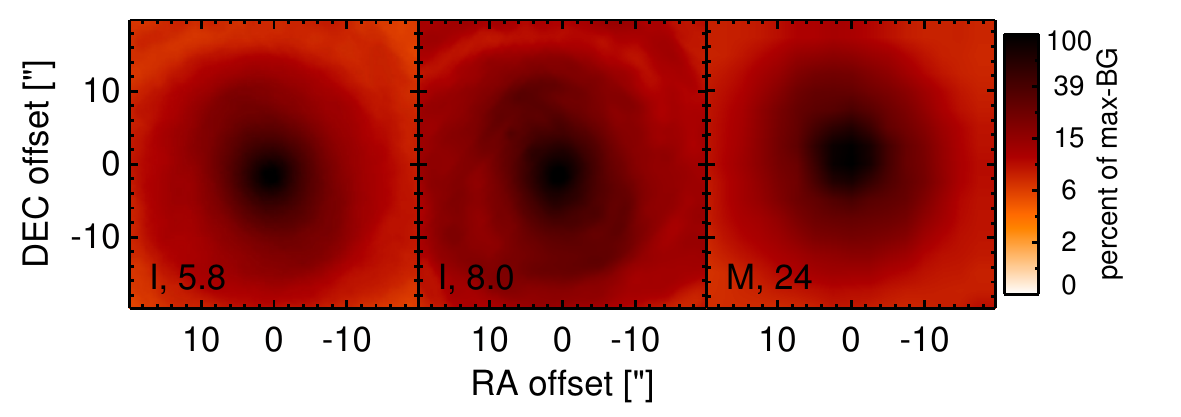}
    \caption{\label{fig:INTim_NGC4736}
             \spitzerr MIR images of NGC\,4736. Displayed are the inner $40\arcsec$ with North up and East to the left. The colour scaling is logarithmic with white corresponding to median background and black to the $0.1\%$ pixels with the highest intensity.
             The label in the bottom left states instrument and central wavelength of the filter in $\mu$m (I: IRAC, M: MIPS). 
           }
\end{figure}
\begin{figure}
   \centering
   \includegraphics[angle=0,height=3.11cm]{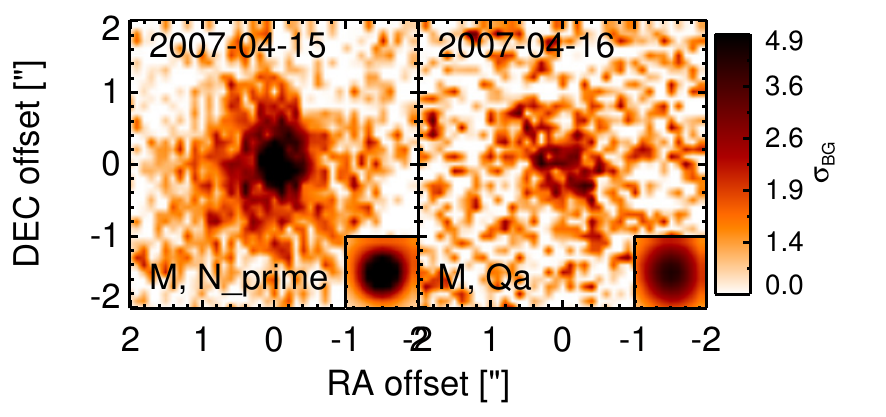}
    \caption{\label{fig:HARim_NGC4736}
             Subarcsecond-resolution MIR images of NGC\,4736 sorted by increasing filter wavelength. 
             Displayed are the inner $4\arcsec$ with North up and East to the left. 
             The colour scaling is logarithmic with white corresponding to median background and black to the $75\%$ of the highest intensity of all images in units of $\sigbg$.
             The inset image shows the central arcsecond of the PSF from the calibrator star, scaled to match the science target.
             The labels in the bottom left state instrument and filter names (C: COMICS, M: Michelle, T: T-ReCS, V: VISIR).
           }
\end{figure}
\begin{figure}
   \centering
   \includegraphics[angle=0,width=8.50cm]{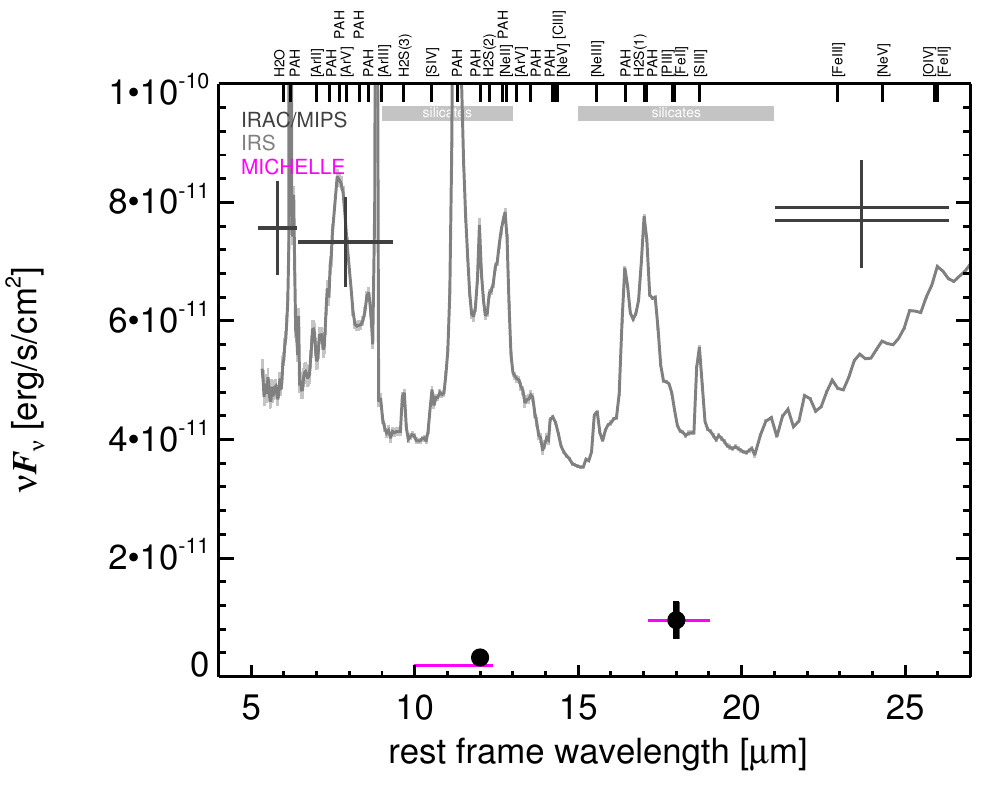}
   \caption{\label{fig:MISED_NGC4736}
      MIR SED of NGC\,4736. The description  of the symbols (if present) is the following.
      Grey crosses and  solid lines mark the \spitzer/IRAC, MIPS and IRS data. 
      The colour coding of the other symbols is: 
      green for COMICS, magenta for Michelle, blue for T-ReCS and red for VISIR data.
      Darker-coloured solid lines mark spectra of the corresponding instrument.
      The black filled circles mark the nuclear 12 and $18\,\mu$m  continuum emission estimate from the data.
      The ticks on the top axis mark positions of common MIR emission lines, while the light grey horizontal bars mark wavelength ranges affected by the silicate 10 and 18$\mu$m features.}
\end{figure}
\clearpage

\twocolumn[\begin{@twocolumnfalse}  
\subsection{NGC\,4746}\label{app:NGC4746}
NGC\,4746 is an  edge-on spiral galaxy at a distance of $D=$ $33.5 \pm 5.5\,$Mpc (NED redshift-independent) with a little-studied possibly active nucleus, which has been classified as a borderline LINER/H\,II \citep{decarli_census_2007}.
No pointed X -ray observations are published and no compact source was detected with subarcsecond-resolution radio observations \citep{irwin_high-resolution_2000}.
We  conservatively treat NGC\,4746 as an uncertain AGN.
No \spitzer observations are available for NGC\,4746.
The \wisee images show the extended flattened host emission without a clearly separable unresolved nuclear component.
The nuclear region of NGC\,4746 was observed with VISIR in the PAH2 filter in 2005, but nothing was detected \citep{siebenmorgen_nuclear_2008}.
This lack of data prohibits any conclusions about the presence of a weak AGN in NGC\,4746.
\newline\end{@twocolumnfalse}]

\begin{figure}
   \centering
   \includegraphics[angle=0,width=8.500cm]{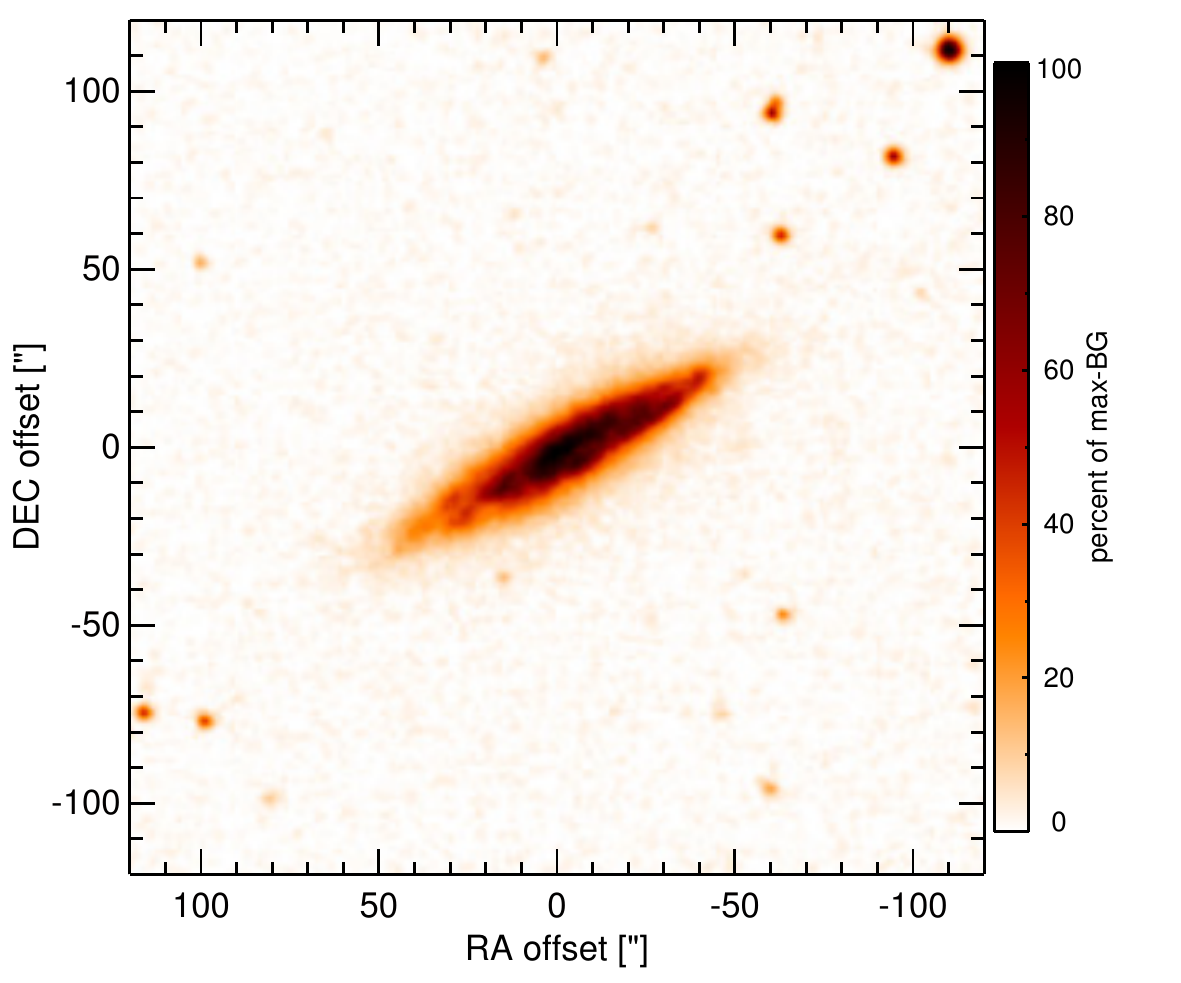}
    \caption{\label{fig:OPTim_NGC4746}
             Optical image (DSS, red filter) of NGC\,4746. Displayed are the central $4\arcmin$ with North up and East to the left. 
              The colour scaling is linear with white corresponding to the median background and black to the $0.01\%$ pixels with the highest intensity.  
           }
\end{figure}
\begin{figure}
   \centering
   \includegraphics[angle=0,width=8.50cm]{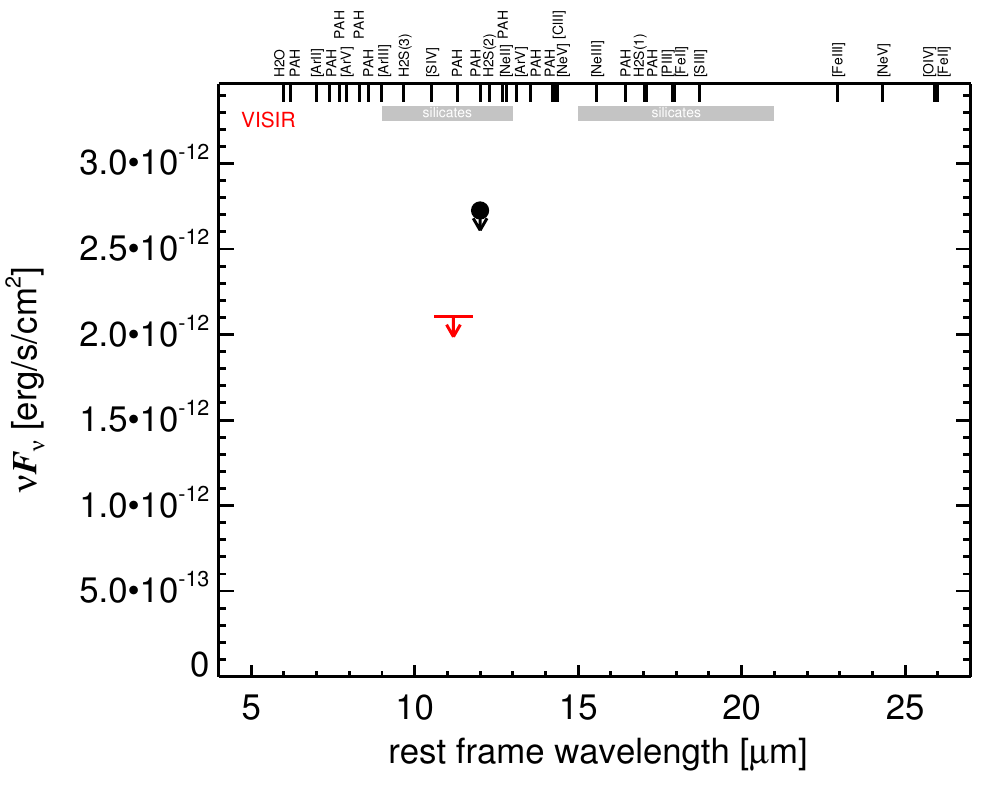}
   \caption{\label{fig:MISED_NGC4746}
      MIR SED of NGC\,4746. The description  of the symbols (if present) is the following.
      Grey crosses and  solid lines mark the \spitzer/IRAC, MIPS and IRS data. 
      The colour coding of the other symbols is: 
      green for COMICS, magenta for Michelle, blue for T-ReCS and red for VISIR data.
      Darker-coloured solid lines mark spectra of the corresponding instrument.
      The black filled circles mark the nuclear 12 and $18\,\mu$m  continuum emission estimate from the data.
      The ticks on the top axis mark positions of common MIR emission lines, while the light grey horizontal bars mark wavelength ranges affected by the silicate 10 and 18$\mu$m features.}
\end{figure}
\clearpage

\twocolumn[\begin{@twocolumnfalse}  
\subsection{NGC\,4785}\label{app:NGC4785}
NGC\,4785 is an inclined barred spiral galaxy at a redshift of $z=$ 0.0123 ($D\sim59\,$Mpc) with a very little-studied Sy\,2 nucleus \citep{veron-cetty_catalogue_2010}.
No high-angular resolution X-ray and radio observations have been published.
It was not detected with \textit{HEAO1} \citep{polletta_multiwavelength_1996}.
We conservatively treat NGC\,4785 as an uncertain AGN.
No \spitzer observations are available for NGC\,4785.
The \wisee images show the extended flattened host emission without a clearly separable unresolved nuclear component.
The nuclear region of NGC\,4785 was observed with VISIR in the PAH2 filter in 2005 but no nucleus was detected (unpublished, to our knowledge).
This lack of data prohibits any conclusions about the presence of a weak AGN in NGC\,4785.
 \newline\end{@twocolumnfalse}]

\begin{figure}
   \centering
   \includegraphics[angle=0,width=8.500cm]{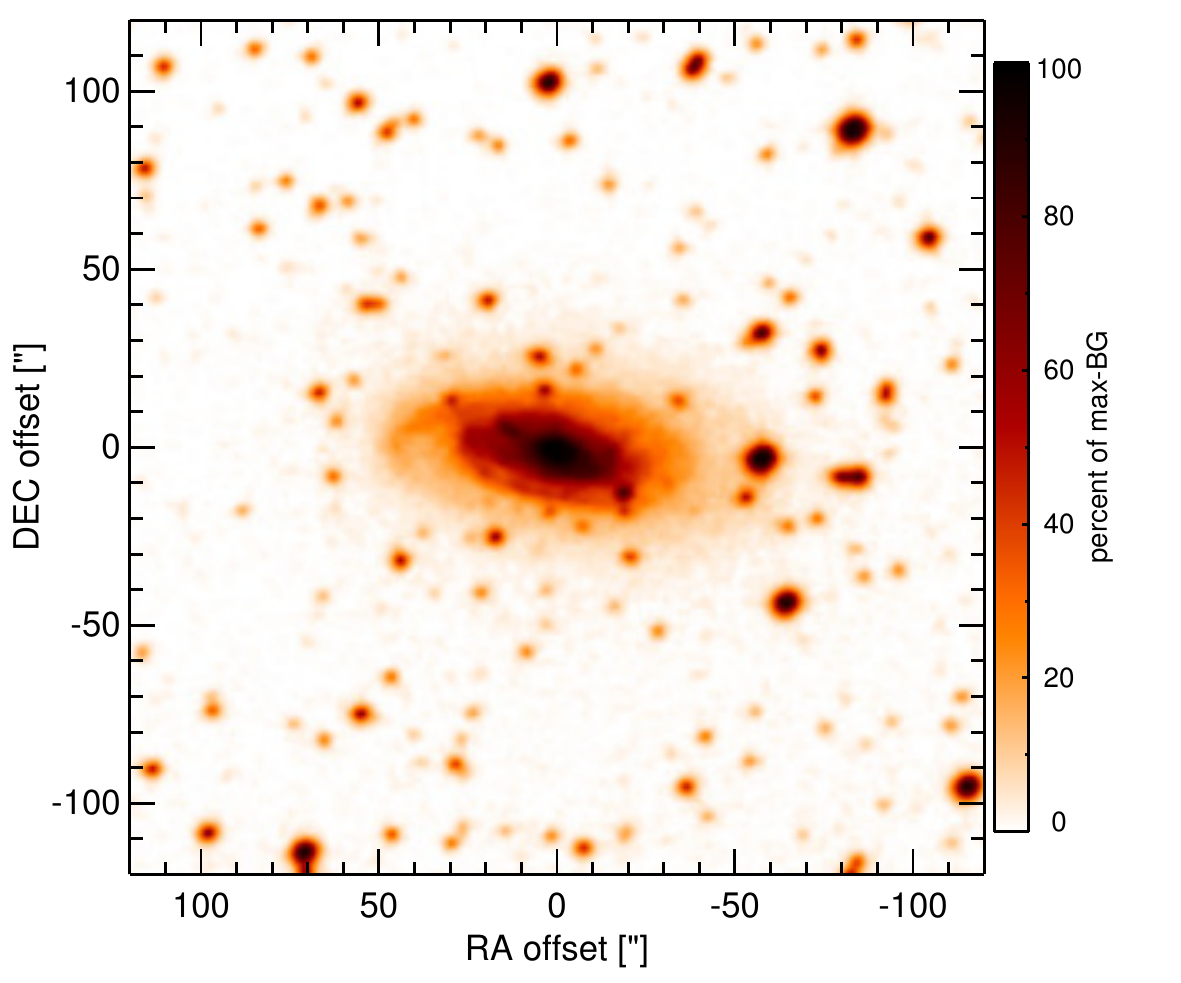}
    \caption{\label{fig:OPTim_NGC4785}
             Optical image (DSS, red filter) of NGC\,4785. Displayed are the central $4\arcmin$ with North up and East to the left. 
              The colour scaling is linear with white corresponding to the median background and black to the $0.01\%$ pixels with the highest intensity.  
           }
\end{figure}
\begin{figure}
   \centering
   \includegraphics[angle=0,width=8.50cm]{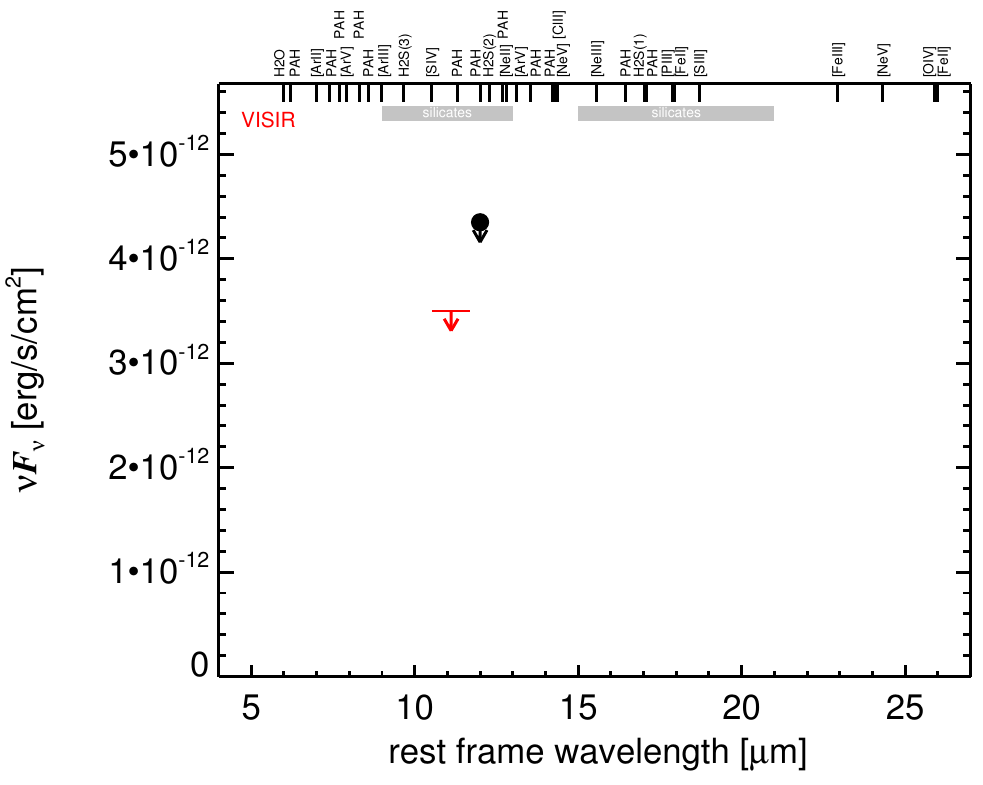}
   \caption{\label{fig:MISED_NGC4785}
      MIR SED of NGC\,4785. The description  of the symbols (if present) is the following.
      Grey crosses and  solid lines mark the \spitzer/IRAC, MIPS and IRS data. 
      The colour coding of the other symbols is: 
      green for COMICS, magenta for Michelle, blue for T-ReCS and red for VISIR data.
      Darker-coloured solid lines mark spectra of the corresponding instrument.
      The black filled circles mark the nuclear 12 and $18\,\mu$m  continuum emission estimate from the data.
      The ticks on the top axis mark positions of common MIR emission lines, while the light grey horizontal bars mark wavelength ranges affected by the silicate 10 and 18$\mu$m features.}
\end{figure}
\clearpage

\twocolumn[\begin{@twocolumnfalse}  
\subsection{NGC\,4941}\label{app:NGC4941}
NGC\,4941 is an inclined barred spiral galaxy at a distance of $D=$ $21.2 \pm 6.7\,$Mpc (NED redshift-independent median) with a Sy\,2.0 nucleus \citep{veron-cetty_catalogue_2010}.
At radio wavelengths, a compact double source was detected, which in total extends for $\sim0.5\arcsec\sim50\,$pc along a PA$\sim-25\degree$ \citep{schmitt_jet_2001}.
The \oiii emission is compact with a weak extended halo \citep{pogge_circumnuclear_1989}.
The first ground-based MIR observations were carried out with Palomar 5\,m/MIRLIN in 2000 \citep{gorjian_10_2004}, in which a compact nucleus was detected.
The \spitzer/IRAC and MIPS images show a compact nucleus surrounded by the lenticularly-formed host emission.
Our nuclear IRAC $5.8$ and $8.0\,\mu$m photometry is significantly lower than the values in \cite{gallimore_infrared_2010}.
The \spitzer/IRS LR staring-mode spectrum exhibits possible weak silicate 10 and $18\,\mu$m emission, very weak PAH features, prominent forbidden emission lines and a red spectral slope in $\nu F_\nu$-space (see also \citealt{shi_9.7_2006,wu_spitzer/irs_2009,tommasin_spitzer-irs_2010,gallimore_infrared_2010}).
This arcsecond-scale MIR SED, thus, indicates little star formation.
We observed the nuclear region of NGC\,4941 with VISIR in four narrow $N$ and one $Q$ band filter in 2006 \citep{horst_mid_2008,horst_mid-infrared_2009}, 2009 \citep{asmus_mid-infrared_2011} and 2011.
The corresponding images show an unresolved nucleus without any further host emission.
Our reanalysis of the published images provides unchanged nuclear fluxes, while the subarcsecond photometry in general is consistent with the \spitzerr spectrophotometry. 
Therefore, the MIR emission from the central $\sim0.4\,$kpc appears to be AGN-dominated.
\newline\end{@twocolumnfalse}]

\begin{figure}
   \centering
   \includegraphics[angle=0,width=8.500cm]{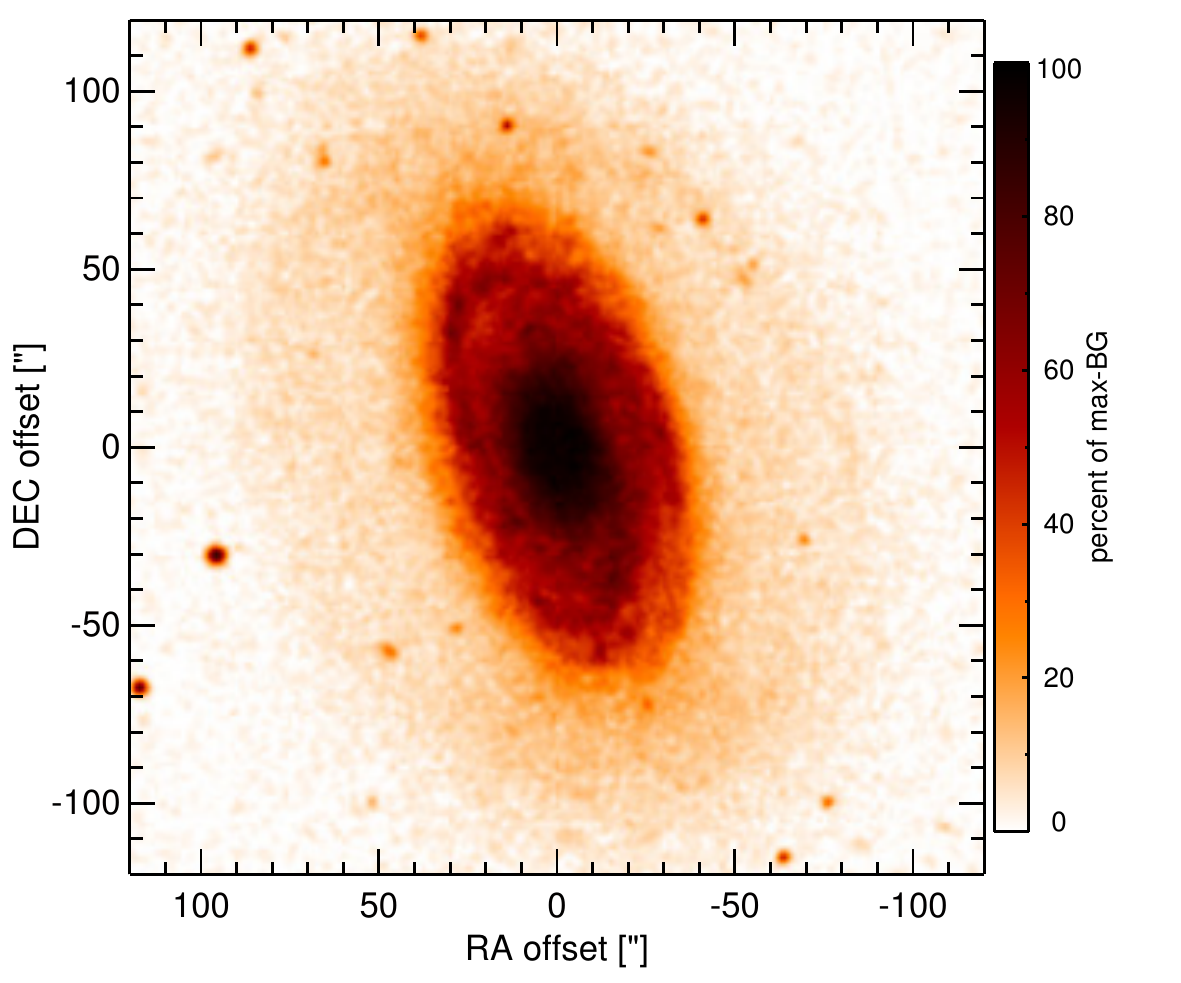}
    \caption{\label{fig:OPTim_NGC4941}
             Optical image (DSS, red filter) of NGC\,4941. Displayed are the central $4\arcmin$ with North up and East to the left. 
              The colour scaling is linear with white corresponding to the median background and black to the $0.01\%$ pixels with the highest intensity.  
           }
\end{figure}
\begin{figure}
   \centering
   \includegraphics[angle=0,height=3.11cm]{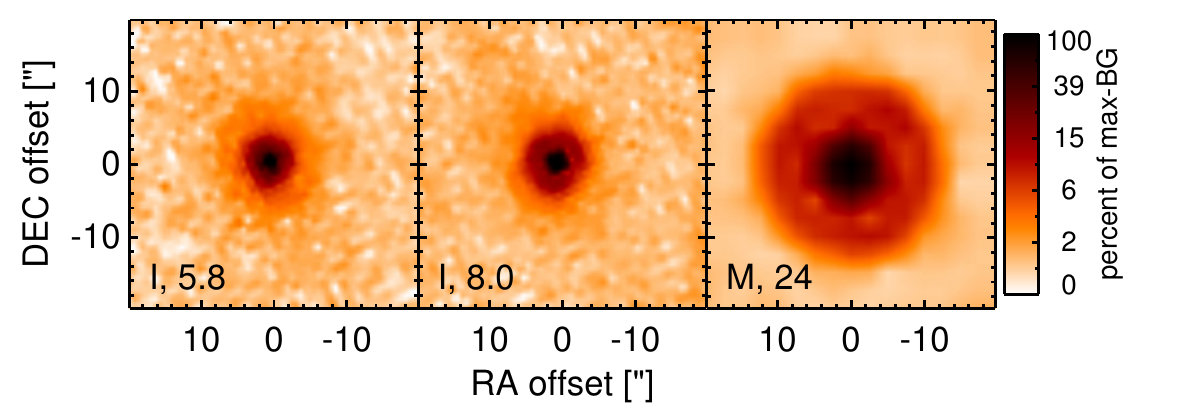}
    \caption{\label{fig:INTim_NGC4941}
             \spitzerr MIR images of NGC\,4941. Displayed are the inner $40\arcsec$ with North up and East to the left. The colour scaling is logarithmic with white corresponding to median background and black to the $0.1\%$ pixels with the highest intensity.
             The label in the bottom left states instrument and central wavelength of the filter in $\mu$m (I: IRAC, M: MIPS). 
           }
\end{figure}
\begin{figure}
   \centering
   \includegraphics[angle=0,width=8.500cm]{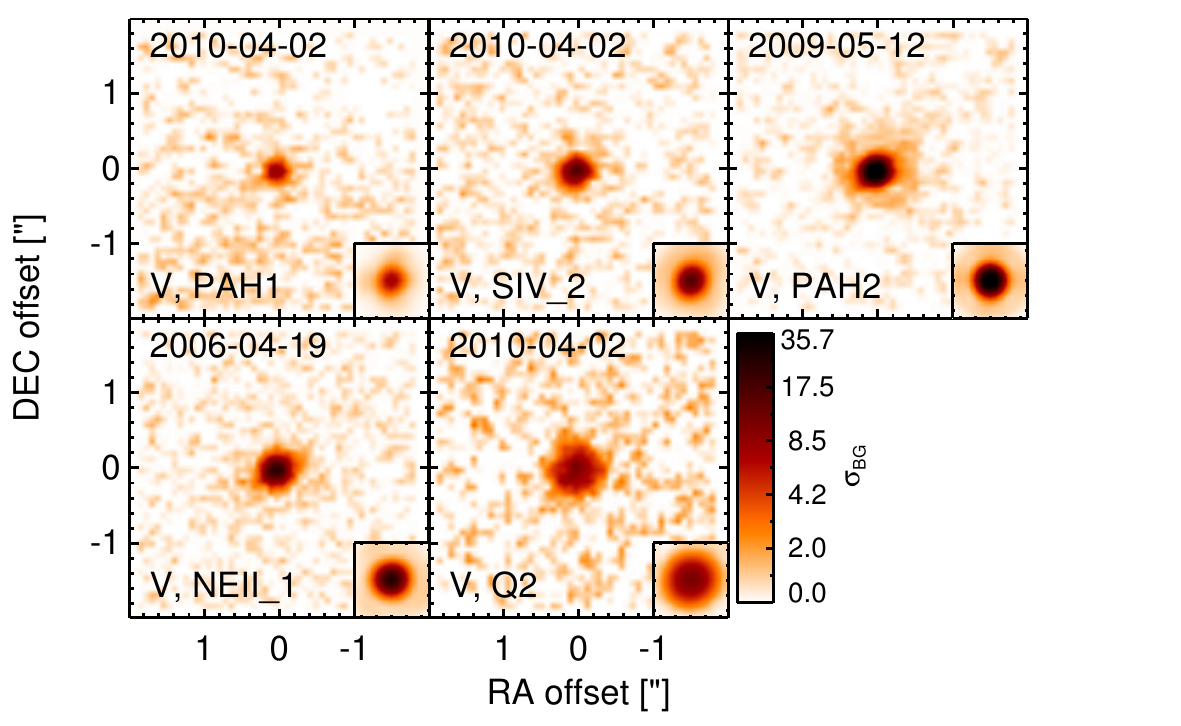}
    \caption{\label{fig:HARim_NGC4941}
             Subarcsecond-resolution MIR images of NGC\,4941 sorted by increasing filter wavelength. 
             Displayed are the inner $4\arcsec$ with North up and East to the left. 
             The colour scaling is logarithmic with white corresponding to median background and black to the $75\%$ of the highest intensity of all images in units of $\sigbg$.
             The inset image shows the central arcsecond of the PSF from the calibrator star, scaled to match the science target.
             The labels in the bottom left state instrument and filter names (C: COMICS, M: Michelle, T: T-ReCS, V: VISIR).
           }
\end{figure}
\begin{figure}
   \centering
   \includegraphics[angle=0,width=8.50cm]{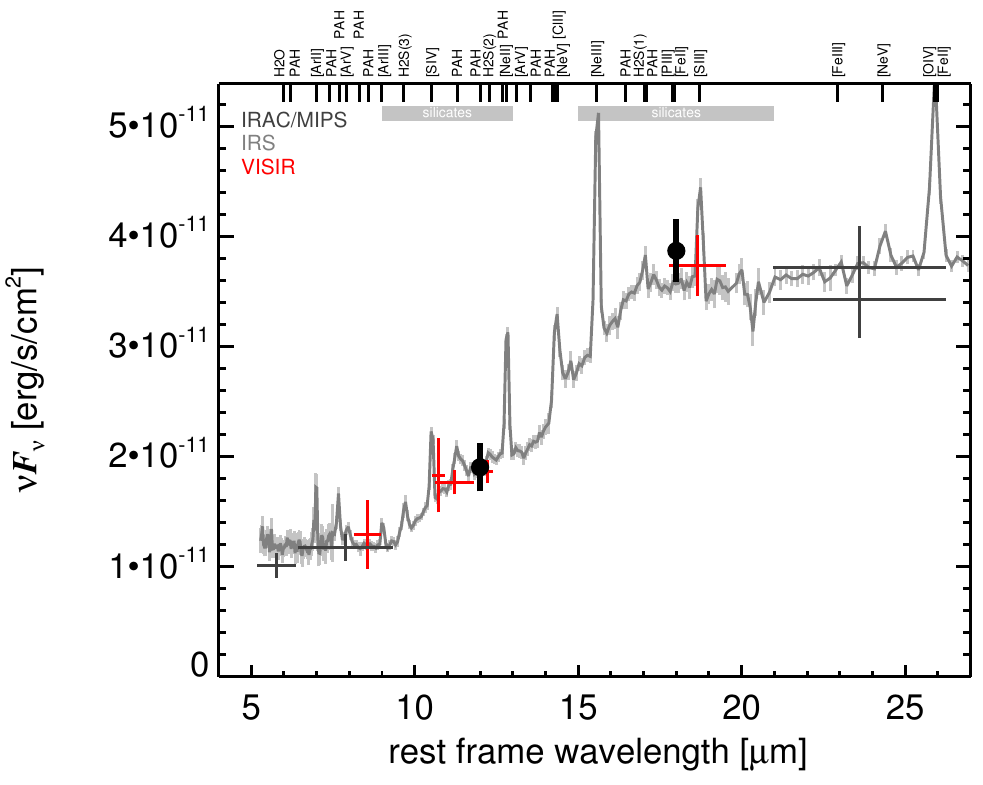}
   \caption{\label{fig:MISED_NGC4941}
      MIR SED of NGC\,4941. The description  of the symbols (if present) is the following.
      Grey crosses and  solid lines mark the \spitzer/IRAC, MIPS and IRS data. 
      The colour coding of the other symbols is: 
      green for COMICS, magenta for Michelle, blue for T-ReCS and red for VISIR data.
      Darker-coloured solid lines mark spectra of the corresponding instrument.
      The black filled circles mark the nuclear 12 and $18\,\mu$m  continuum emission estimate from the data.
      The ticks on the top axis mark positions of common MIR emission lines, while the light grey horizontal bars mark wavelength ranges affected by the silicate 10 and 18$\mu$m features.}
\end{figure}
\clearpage

\twocolumn[\begin{@twocolumnfalse}  
\subsection{NGC\,4945}\label{app:NGC4945}
NGC\,4945 is a nearby almost edge-on spiral galaxy (distance of $D=$ $3.7 \pm 0.8\,$Mpc; NED redshift-independent median) with a highly obscured nucleus containing both a starburst and an AGN \citep{moorwood_extended_1994,moorwood_starburst_1996}.
In the optical, the nucleus is almost completely obscured and appears as an H\,II nucleus \citep{moorwood_starburst_1996,goulding_towards_2009}. 
The \oiii emission seems to be completely obscured in the nuclear region \citep{moorwood_starburst_1996}.
On the other hand, the AGN is evident in hard X-rays, where it is one of the flux-brightest and most variable sources with very complex spectral properties (e.g.,, \citealt{yaqoob_nature_2012,marinucci_x-ray_2012}).
It appears as a compact radio source at subarcsecond scales \citep{elmouttie_radio_1997}, and an additional jet-like structure extending $\sim5\,$pc southwest has been detected recently as well \citep{lenc_sub-parsec_2009}. 
The nucleus also hosts water mega-maser emission on subparsec scales (PA$\sim45\degree$; \citealt{dos_santos_detection_1979,greenhill_distribution_1997}).
The nucleus is surrounded by a starburst disc with $\sim 11\arcsec\sim200\,$pc diameter orientated  in the same direction as the galactic disc \citep{marconi_elusive_2000}.
The first ground-based MIR observations were performed by \cite{moorwood_infrared_1984} and followed up with \isoo observations \citep{rigopoulou_large_1999, spoon_mid-infrared_2000,lutz_relation_2004}.
The corresponding MIR spectra show deep silicate 10$\,\mu$m absorption and indicate heavy absorption even in the MIR \citep{spoon_mid-infrared_2000,krabbe_n-band_2001}.
The first ground-based $N$-band images were obtained with ESO 3.6\,m/TIMMI in 1996 \citep{marconi_elusive_2000} and with ESO 2.2\,m/MANIAC in 1998 \citep{krabbe_n-band_2001}.
These show a compact nucleus embedded within elongated emission parallel to the galaxy major axis.
A similar galactic morphology is evident in the \spitzer/IRAC and MIPS images however with lower spatial resolution.
Moreover, the nucleus is saturated in the IRAC $5.8$ and $8.0\,\mu$m PBCD images and, thus, not analysed.
The \spitzer/IRS LR spectrum verifies the deep silicate 10$\,\mu$m absorption and in addition indicates silicate 18\,$\mu$m absorption, weak PAH emission and a in general shallow red spectral slope in $\nu F_\nu$-space (see also \citealt{brandl_mid-infrared_2006,bernard-salas_spitzer_2009,goulding_towards_2009,perez-beaupuits_deeply_2011}).
Thus, the arcsecond-scale MIR SED has AGN/starburst composite characteristics.
The nuclear region of NGC\,4945 was observed with T-ReCS in the Si2 and Qa filters in 2006 (partly published in \citealt{imanishi_subaru_2011}), and with VISIR in two narrow $N$ and one $Q$-band filter also in 2006 (unpublished, to our knowledge).
The $N$-band images show a weakly detected elongated structure (major axis diameter $\sim3\arcsec\sim54\,$pc; PA$\sim45\degree$) with a compact component located roughly in the middle.
We assume this to be the location of the AGN and measure the unresolved nuclear component with manual PSF-scaling. 
The resulting fluxes are on average $\sim 81\%$ lower than the \spitzerr spectrophotometry.
In the $Q$-band images, the MIR structure is barely visible, and we  treat the measured fluxes as upper limits (but see \citealt{imanishi_subaru_2011}).
Owing to the composite nature and the nucleus that cannot be clearly separated from the surrounding emission, our subarcsecond MIR fluxes have to be regarded as upper limits on any AGN-powered MIR emission in NGC\,4945. 
The nuclear MIR coverage is not sufficient to distinguish, whether at subarcsecond resolution resolution the silicate absorption is deeper or the continuum emission is lower.
We conclude that the observed MIR emission of the central $\sim70\,$pc in NGC\,4945 is star-formation dominated in agreement with previous results \citep{spoon_mid-infrared_2000}.
\newline\end{@twocolumnfalse}]

\begin{figure}
   \centering
   \includegraphics[angle=0,width=8.500cm]{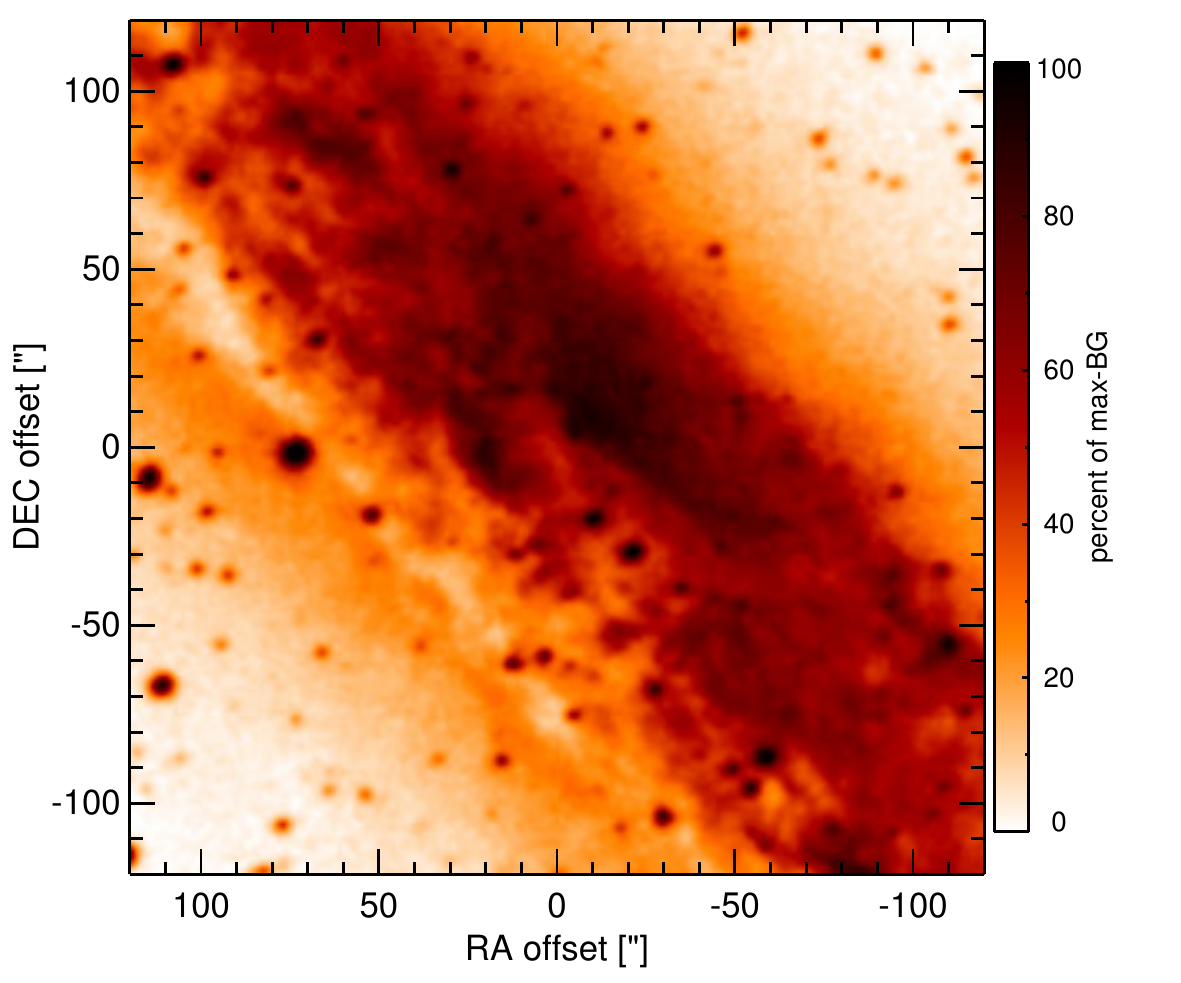}
    \caption{\label{fig:OPTim_NGC4945}
             Optical image (DSS, red filter) of NGC\,4945. Displayed are the central $4\arcmin$ with North up and East to the left. 
              The colour scaling is linear with white corresponding to the median background and black to the $0.01\%$ pixels with the highest intensity.  
           }
\end{figure}
\begin{figure}
   \centering
   \includegraphics[angle=0,height=3.11cm]{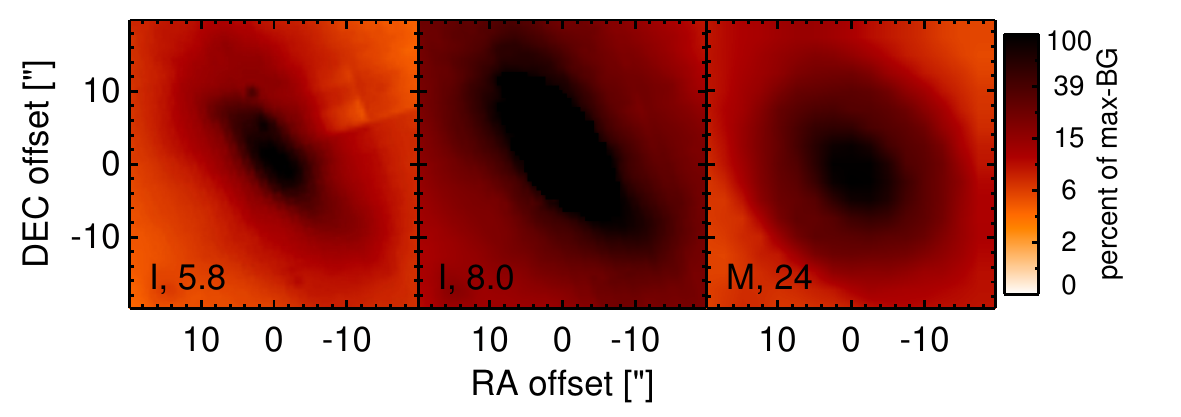}
    \caption{\label{fig:INTim_NGC4945}
             \spitzerr MIR images of NGC\,4945. Displayed are the inner $40\arcsec$ with North up and East to the left. The colour scaling is logarithmic with white corresponding to median background and black to the $0.1\%$ pixels with the highest intensity.
             The label in the bottom left states instrument and central wavelength of the filter in $\mu$m (I: IRAC, M: MIPS). 
             Note that the central region in the IRAC 5.8 and $8.0\,\mu$m images is completely saturated.
           }
\end{figure}
\begin{figure}
   \centering
   \includegraphics[angle=0,width=8.500cm]{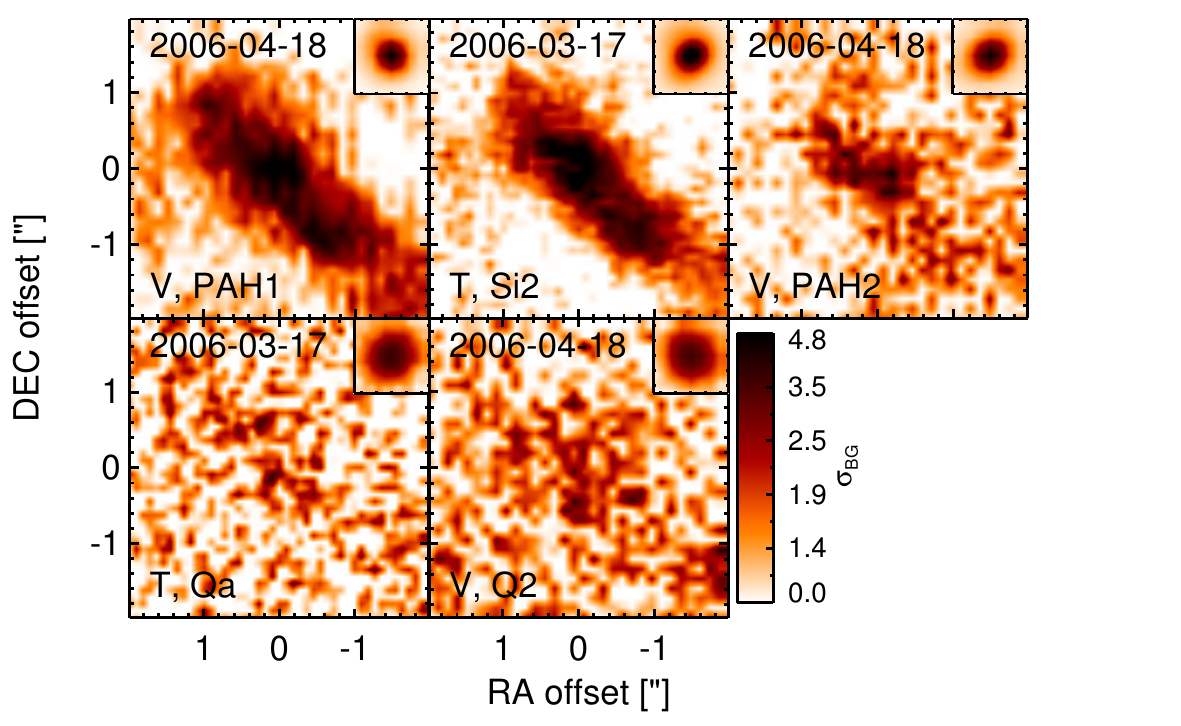}
    \caption{\label{fig:HARim_NGC4945}
             Subarcsecond-resolution MIR images of NGC\,4945 sorted by increasing filter wavelength. 
             Displayed are the inner $4\arcsec$ with North up and East to the left. 
             The colour scaling is logarithmic with white corresponding to median background and black to the $75\%$ of the highest intensity of all images in units of $\sigbg$.
             The inset image shows the central arcsecond of the PSF from the calibrator star, scaled to match the science target.
             The labels in the bottom left state instrument and filter names (C: COMICS, M: Michelle, T: T-ReCS, V: VISIR).
           }
\end{figure}
\begin{figure}
   \centering
   \includegraphics[angle=0,width=8.50cm]{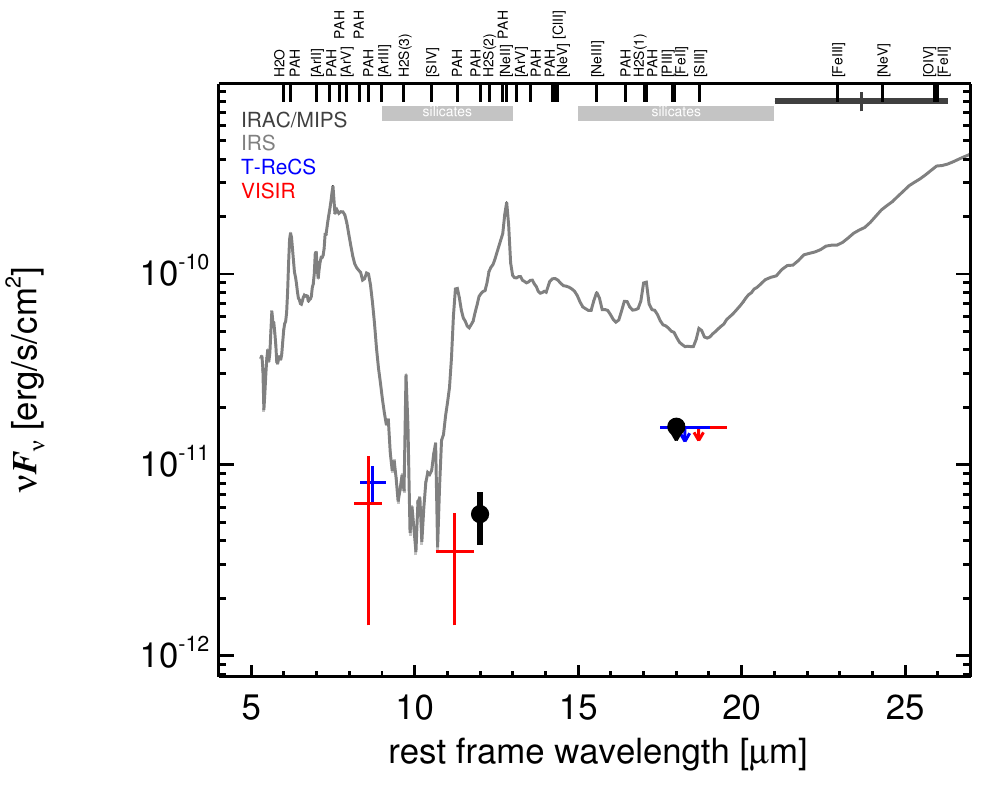}
   \caption{\label{fig:MISED_NGC4945}
      MIR SED of NGC\,4945. The description  of the symbols (if present) is the following.
      Grey crosses and  solid lines mark the \spitzer/IRAC, MIPS and IRS data. 
      The colour coding of the other symbols is: 
      green for COMICS, magenta for Michelle, blue for T-ReCS and red for VISIR data.
      Darker-coloured solid lines mark spectra of the corresponding instrument.
      The black filled circles mark the nuclear 12 and $18\,\mu$m  continuum emission estimate from the data.
      The ticks on the top axis mark positions of common MIR emission lines, while the light grey horizontal bars mark wavelength ranges affected by the silicate 10 and 18$\mu$m features.}
\end{figure}
\clearpage

\twocolumn[\begin{@twocolumnfalse}  
\subsection{NGC\,4992}\label{app:NGC4992}
NGC\,4992 is an inclined spiral galaxy at a redshift of $z=$ 0.0251 ($D\sim119\,$Mpc) with a little-studied AGN, which was discovered in hard X-rays with \textit{INTEGRAL} \citep{sazonov_identification_2005} and belongs to the nine-month BAT AGN sample.
It has contradicting optical classifications as inactive nucleus \citep{masetti_unveiling_2006}, LINER \citep{winter_optical_2010} and Sy\,1.0 nucleus \citep{veron-cetty_catalogue_2010}.
The X-ray properties strongly suggest an obscured AGN \citep{sazonov_identification_2005}.
NGC\,4992 was also detected with low-angular resolution radio observations \citep{white_catalog_1997}.
No detection was reported by \iras, but it was also observed with \spitzer/IRAC and IRS and appears as a compact nucleus surrounded by weak spiral-like host emission in the corresponding images.
The IRS LR staring-mode spectrum exhibits deep silicate 10 and 18\,$\mu$m absorption and a blue spectral slope in $\nu F_\nu$-space but  no PAH emission or forbidden emission lines.
This arcsecond-scale MIR SED, thus, also favours an obscured AGN.
We observed NGC\,4992 with VISIR in four narrow $N$-band filters in 2008 (partly published in \citealt{gandhi_resolving_2009}) and weakly detected a compact nucleus in all cases.
In the three images with sufficient S/N, the nucleus appears possibly resolved but with inconsistent elongations/orientations.
Therefore, we classify its subarcsecond MIR extension as uncertain.
Our nuclear photometry is consistent with the previously published values and the \spitzerr spectrophotometry.
Therefore, we use the latter to compute the 12\,$\mu$m continuum emission estimate.
\newline\end{@twocolumnfalse}]

\begin{figure}
   \centering
   \includegraphics[angle=0,width=8.500cm]{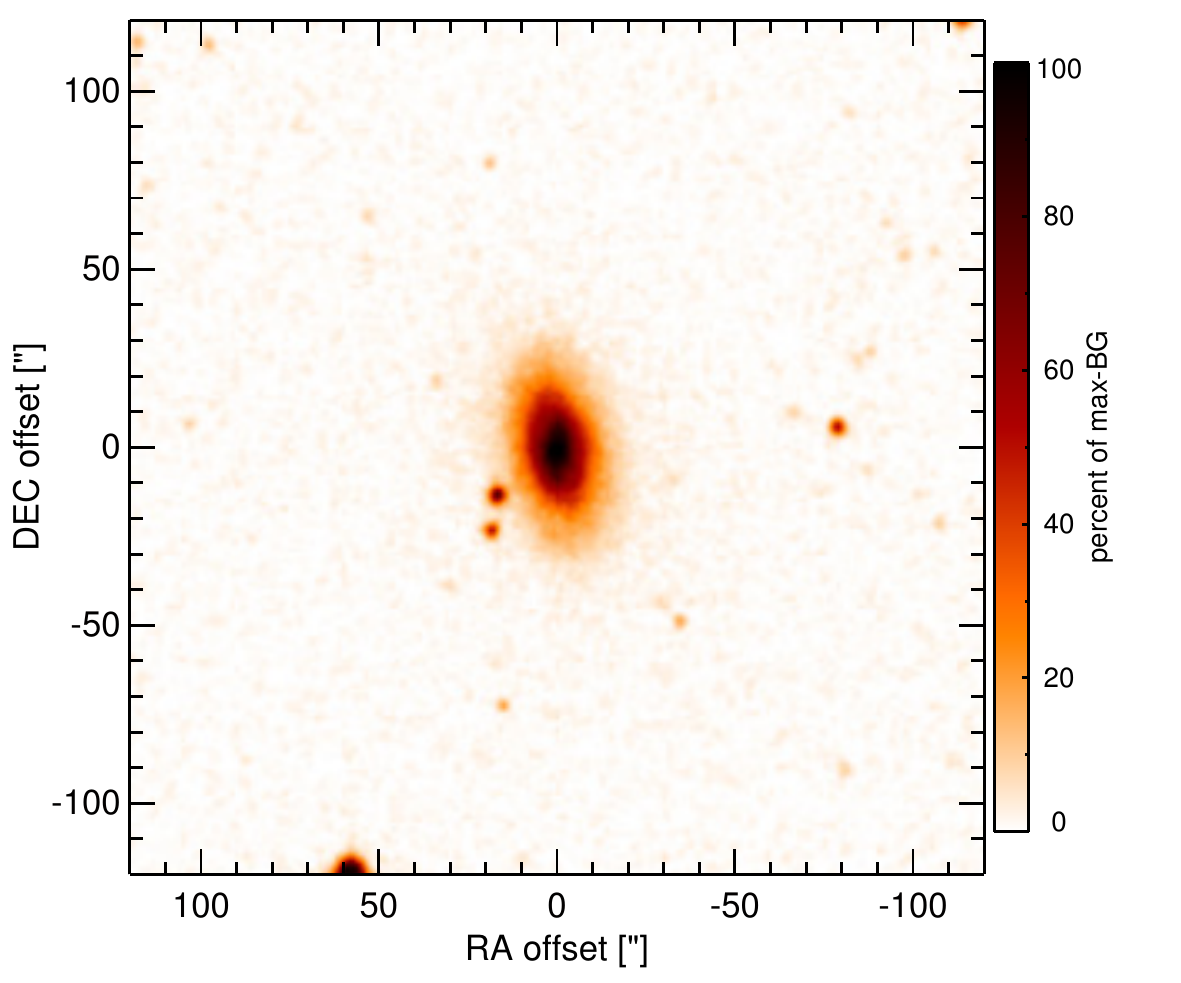}
    \caption{\label{fig:OPTim_NGC4992}
             Optical image (DSS, red filter) of NGC\,4992. Displayed are the central $4\arcmin$ with North up and East to the left. 
              The colour scaling is linear with white corresponding to the median background and black to the $0.01\%$ pixels with the highest intensity.  
           }
\end{figure}
\begin{figure}
   \centering
   \includegraphics[angle=0,height=3.11cm]{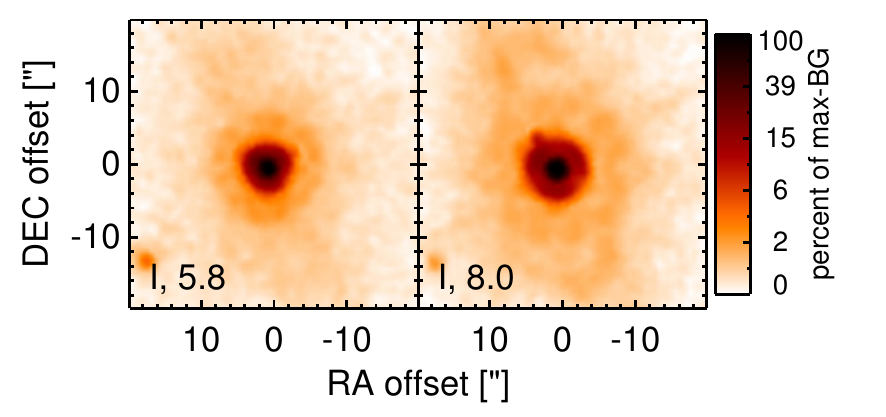}
    \caption{\label{fig:INTim_NGC4992}
             \spitzerr MIR images of NGC\,4992. Displayed are the inner $40\arcsec$ with North up and East to the left. The colour scaling is logarithmic with white corresponding to median background and black to the $0.1\%$ pixels with the highest intensity.
             The label in the bottom left states instrument and central wavelength of the filter in $\mu$m (I: IRAC, M: MIPS). 
           }
\end{figure}
\begin{figure}
   \centering
   \includegraphics[angle=0,width=8.500cm]{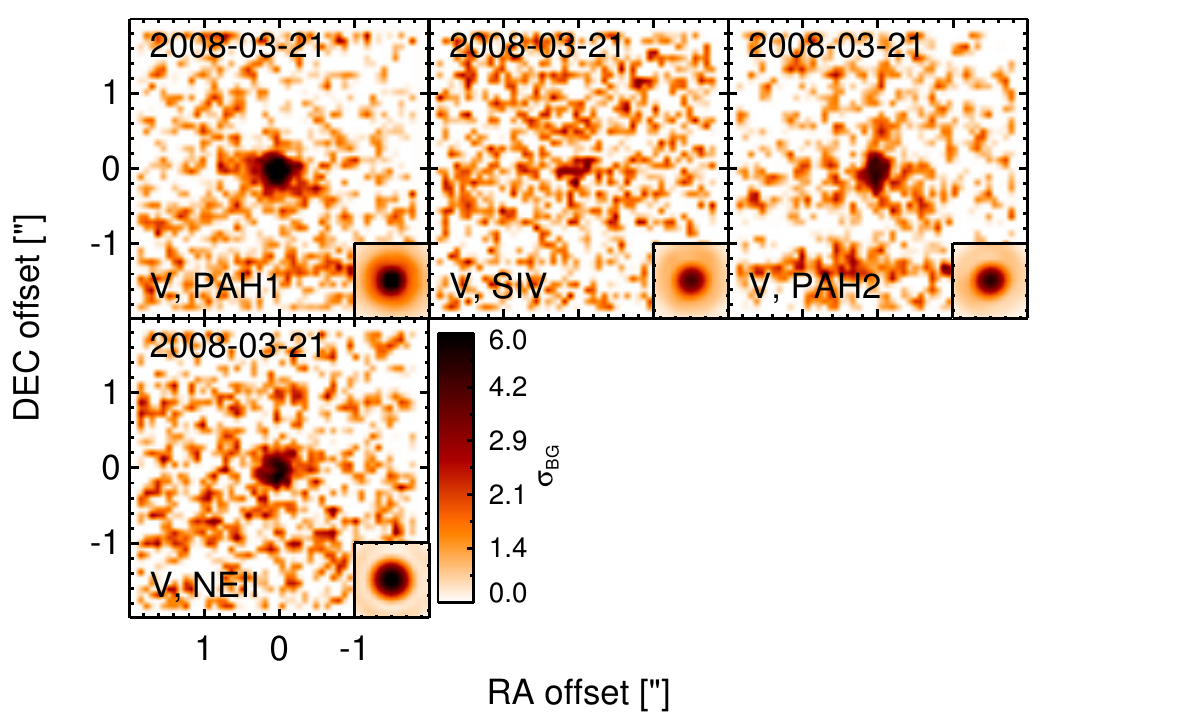}
    \caption{\label{fig:HARim_NGC4992}
             Subarcsecond-resolution MIR images of NGC\,4992 sorted by increasing filter wavelength. 
             Displayed are the inner $4\arcsec$ with North up and East to the left. 
             The colour scaling is logarithmic with white corresponding to median background and black to the $75\%$ of the highest intensity of all images in units of $\sigbg$.
             The inset image shows the central arcsecond of the PSF from the calibrator star, scaled to match the science target.
             The labels in the bottom left state instrument and filter names (C: COMICS, M: Michelle, T: T-ReCS, V: VISIR).
           }
\end{figure}
\begin{figure}
   \centering
   \includegraphics[angle=0,width=8.50cm]{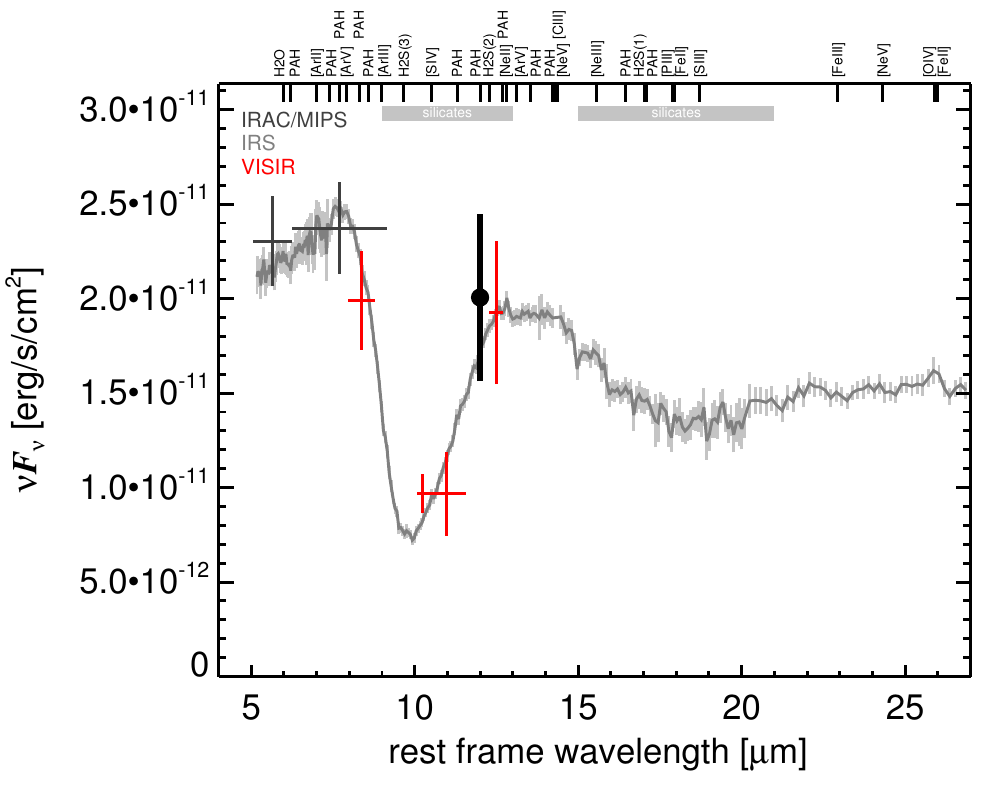}
   \caption{\label{fig:MISED_NGC4992}
      MIR SED of NGC\,4992. The description  of the symbols (if present) is the following.
      Grey crosses and  solid lines mark the \spitzer/IRAC, MIPS and IRS data. 
      The colour coding of the other symbols is: 
      green for COMICS, magenta for Michelle, blue for T-ReCS and red for VISIR data.
      Darker-coloured solid lines mark spectra of the corresponding instrument.
      The black filled circles mark the nuclear 12 and $18\,\mu$m  continuum emission estimate from the data.
      The ticks on the top axis mark positions of common MIR emission lines, while the light grey horizontal bars mark wavelength ranges affected by the silicate 10 and 18$\mu$m features.}
\end{figure}
\clearpage

\twocolumn[\begin{@twocolumnfalse}  
\subsection{NGC\,5005}\label{app:NGC5005}
NGC\,5005 is an inclined barred spiral galaxy at a distance of $D=$ $16.9 \pm 6.7\,$Mpc (NED redshift-independent median) with a broad-line LINER nucleus \citep{ho_search_1997-1}.
In X-rays, it appears as a compact source embedded within extended emission \citep{dudik_chandra_2005,gonzalez-martin_x-ray_2009}.
An elongated radio nucleus was detected in arcsecond-scale radio observations (PA$\sim135\degree$; \citealt{vila_compact_1990}; not detected at 2cm \citealt{nagar_radio_2000}).
Furthermore, NGC\,5005 features a wide NLR cone extending $\sim 3\arcsec\sim240\,$pc south-east (PA$\sim135\degree$; \citealt{pogge_narrow-line_2000}) and a circum-nuclear clumpy, dusty star-forming region parallel to the galaxy major axis \citep{barth_search_1998}.
The first ground-based MIR observations were performed by \cite{cizdziel_multiaperture_1985}, followed by \cite{devereux_spatial_1987} and \cite{maiolino_new_1995}.
A first subarcsecond-resolution $N$-band imaging attempt was reported by \cite{gorjian_10_2004} with Palomar 5\,m/MIRLIN, but the nucleus remained undetected.
The \spitzer/IRAC and MIPS images show a nucleus elongated parallel to the major-axis.
It is embedded within the spiral-like host emission. 
Our nuclear IRAC $5.8$ and $8.0\,\mu$m fluxes are significantly lower than the fluxes given in \cite{gallimore_infrared_2010} but roughly match the \spitzer/IRS flux levels.
The IRS LR staring-mode spectrum is dominated by PAH features with possibly weak silicate  10\,$\mu$m absorption and a red spectral slope in $\nu F_\nu$-space (see also \citealt{buchanan_spitzer_2006,shi_9.7_2006,wu_spitzer/irs_2009,tommasin_spitzer-irs_2010,gallimore_infrared_2010}).
Therefore, the arcsecond-scale MIR SED  appears to be completely star-formation dominated.
The nuclear region of NGC\,5005 was observed with Michelle in the N' filter in 2008 \citep{mason_nuclear_2012}, and a compact nucleus, embedded within $\sim6\arcsec\sim0.5\,$kpc elongated emission, is weakly detected (PA$\sim 62\degree$). 
Our manually-scaled PSF flux is consistent with \cite{mason_nuclear_2012} and $\sim 95\%$ lower than the \spitzerr spectrophotometry.
Therefore, we conclude that the central $\sim0.4$\,kpc MIR emission of NGC\,5005 is completely star-formation dominated and our nuclear flux has to be regarded as an upper limit to the AGN-powered emission.
\newline\end{@twocolumnfalse}]

\begin{figure}
   \centering
   \includegraphics[angle=0,width=8.500cm]{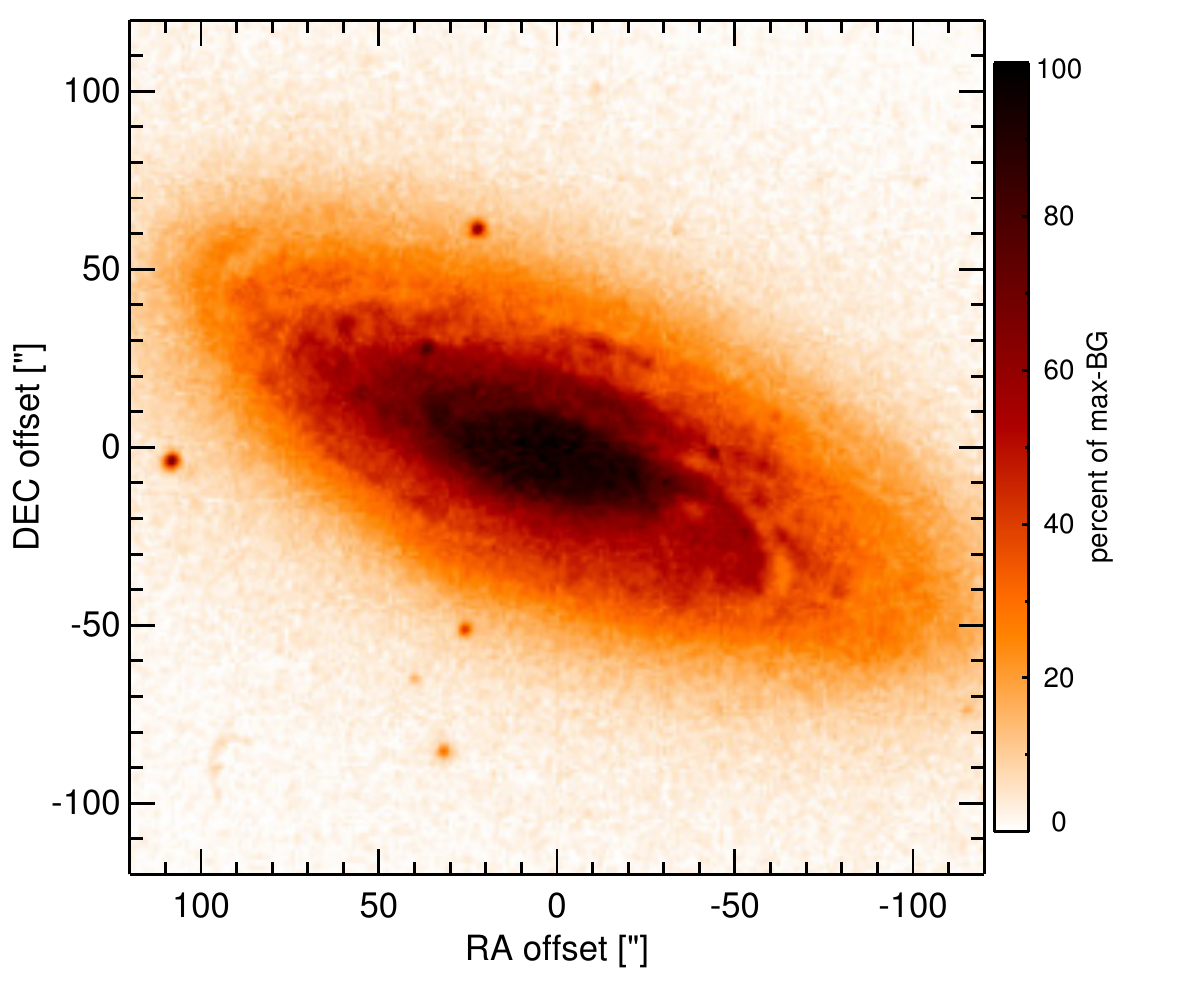}
    \caption{\label{fig:OPTim_NGC5005}
             Optical image (DSS, red filter) of NGC\,5005. Displayed are the central $4\arcmin$ with North up and East to the left. 
              The colour scaling is linear with white corresponding to the median background and black to the $0.01\%$ pixels with the highest intensity.  
           }
\end{figure}
\begin{figure}
   \centering
   \includegraphics[angle=0,height=3.11cm]{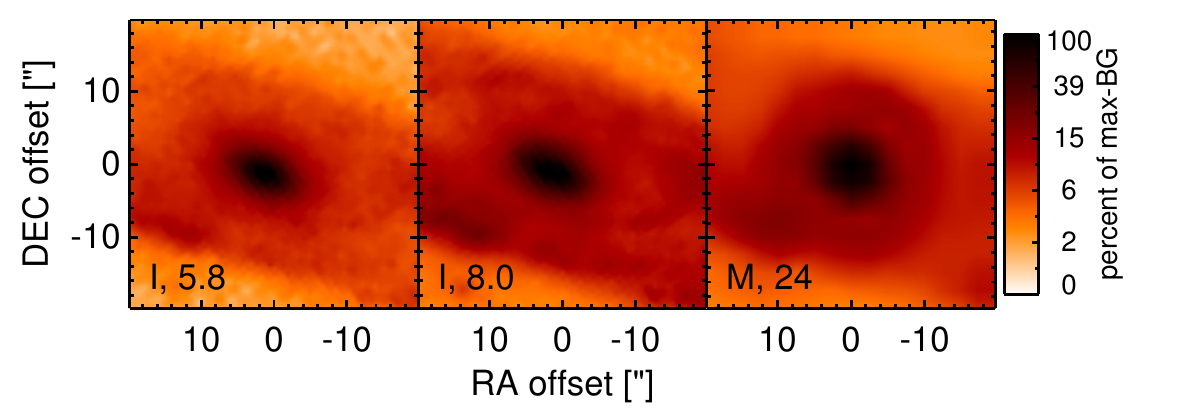}
    \caption{\label{fig:INTim_NGC5005}
             \spitzerr MIR images of NGC\,5005. Displayed are the inner $40\arcsec$ with North up and East to the left. The colour scaling is logarithmic with white corresponding to median background and black to the $0.1\%$ pixels with the highest intensity.
             The label in the bottom left states instrument and central wavelength of the filter in $\mu$m (I: IRAC, M: MIPS). 
           }
\end{figure}
\begin{figure}
   \centering
   \includegraphics[angle=0,height=3.11cm]{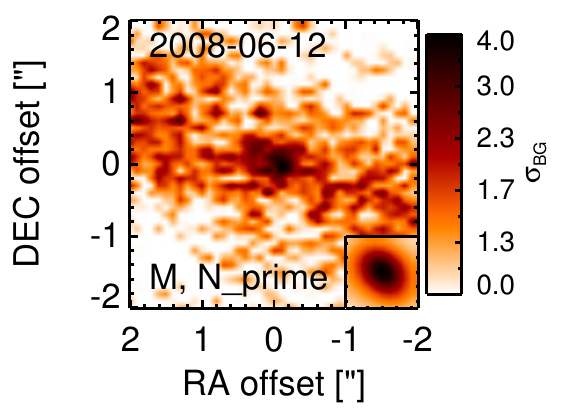}
    \caption{\label{fig:HARim_NGC5005}
             Subarcsecond-resolution MIR images of NGC\,5005 sorted by increasing filter wavelength. 
             Displayed are the inner $4\arcsec$ with North up and East to the left. 
             The colour scaling is logarithmic with white corresponding to median background and black to the $75\%$ of the highest intensity of all images in units of $\sigbg$.
             The inset image shows the central arcsecond of the PSF from the calibrator star, scaled to match the science target.
             The labels in the bottom left state instrument and filter names (C: COMICS, M: Michelle, T: T-ReCS, V: VISIR).
           }
\end{figure}
\begin{figure}
   \centering
   \includegraphics[angle=0,width=8.50cm]{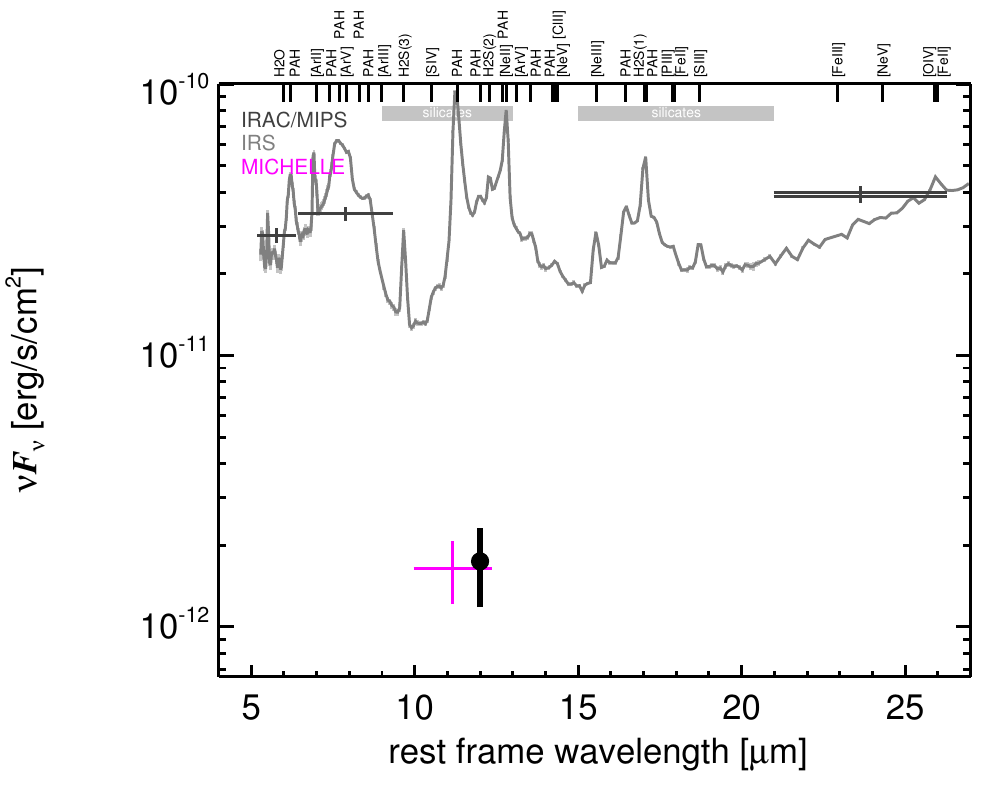}
   \caption{\label{fig:MISED_NGC5005}
      MIR SED of NGC\,5005. The description  of the symbols (if present) is the following.
      Grey crosses and  solid lines mark the \spitzer/IRAC, MIPS and IRS data. 
      The colour coding of the other symbols is: 
      green for COMICS, magenta for Michelle, blue for T-ReCS and red for VISIR data.
      Darker-coloured solid lines mark spectra of the corresponding instrument.
      The black filled circles mark the nuclear 12 and $18\,\mu$m  continuum emission estimate from the data.
      The ticks on the top axis mark positions of common MIR emission lines, while the light grey horizontal bars mark wavelength ranges affected by the silicate 10 and 18$\mu$m features.}
\end{figure}
\clearpage

\twocolumn[\begin{@twocolumnfalse}  
\subsection{NGC\,5033}\label{app:NGC5033}
NGC\,5033 is an inclined spiral galaxy at a distance of $D=$ $18.1 \pm 7.4\,$Mpc (NED redshift-independent median) hosting an AGN, which has optically been classified either as a Sy\,1.2, 1.5, 1.8 or 1.9 (see discussion in \citealt{trippe_multi-wavelength_2010}).
It features a compact radio core with jet-like structure extending $\sim0.5\arcsec\sim45\,$pc to the east (PA$\sim95\degree$; e.g.,  \citealt{ho_radio_2001}) and a one-sided very wide kiloparsec-scale \oiii cone towards PA$\sim80\degree$ \citep{mediavilla_asymmetrical_2005}.
The first MIR observations of NGC\,5033 were performed by \cite{rieke_10_1978}, but the nucleus remained undetected.
\cite{lawrence_observations_1985} report the first detection in $N-$ and $Q$-band, followed by \cite{devereux_spatial_1987}, \cite{edelson_broad-band_1987} and \cite{maiolino_new_1995}.
NGC\,5033 was also observed with \isoo \citep{clavel_2.5-11_2000,ramos_almeida_mid-infrared_2007}.
The first subarcsecond-resolution $N$-band image was obtained with Palomar 5\,m/MIRLIN in 2000 \citep{gorjian_10_2004}.
The \spitzer/IRAC and MIPS images show an extended nucleus embedded within the spiral-like host emission.
Our nuclear  IRAC $5.8$ and $8.0\,\mu$m fluxes are significantly lower than the values given in \cite{gallimore_infrared_2010} but in agreement with the \spitzer/IRS LR mapping PBCD spectrum.
The latter exhibits prominent PAH emission and a rather flat spectral slop in $\nu F_\nu$-space (see also \citealt{dale_spitzer_2009,wu_spitzer/irs_2009,goulding_towards_2009,gallimore_infrared_2010,marble_aromatic_2010}).
Thus, the arcsecond-scale MIR SED  is dominated by significant star formation.
The nuclear region of NGC\,5033 was observed with Michelle in the N' filter in 2008 \citep{mason_nuclear_2012} and in the Si-5 filter in 2010 (this work).
A compact nucleus was detected that is embedded within very weak emission extending in north-south direction. This extended emission coincides with the edges of the NLR cone and was detected in both images (diameter$\sim2\arcsec\sim 180\,$pc; PA$\sim-5\degree$).
In the sharper N' image, the nucleus is possibly marginally resolved (FWHM(major axis)$\sim 0.42\arcsec\sim37\,$pc; PA$\sim87\degree$). 
Additional deeper subarcsecond MIR imaging is required to confirm this morphology.
Our nuclear photometry is consistent with \cite{mason_nuclear_2012} and on average $\sim 87\%$ lower than the \spitzerr spectrophotometry.
The low spectral coverage does not permit characterisation of the nature of the nuclear MIR emission. 
We conclude that the MIR emission of the central $\sim\,350\,$pc in NGC\,5033 is dominated by extended emission from star formation and the NLR.
\newline\end{@twocolumnfalse}]

\begin{figure}
   \centering
   \includegraphics[angle=0,width=8.500cm]{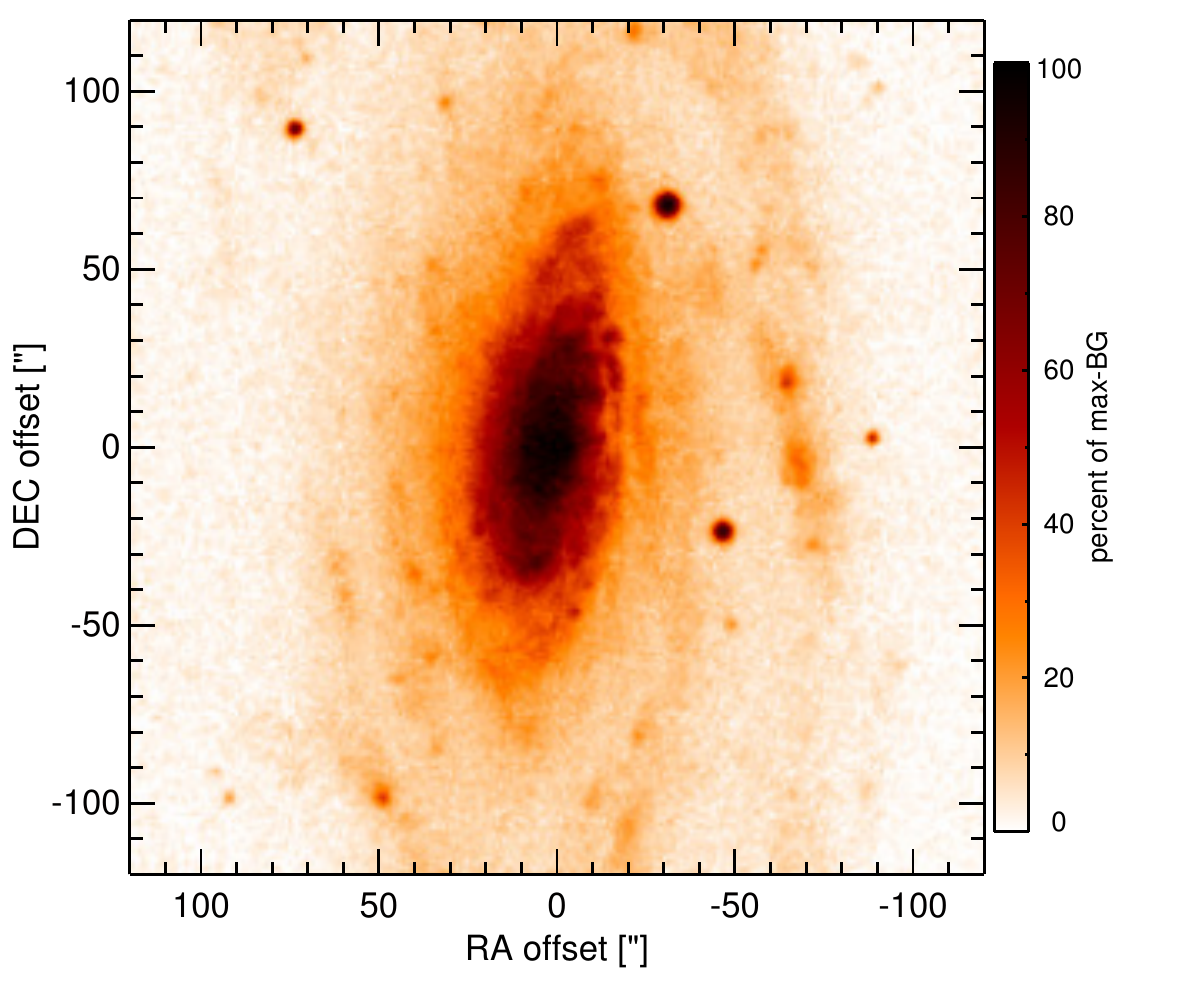}
    \caption{\label{fig:OPTim_NGC5033}
             Optical image (DSS, red filter) of NGC\,5033. Displayed are the central $4\arcmin$ with North up and East to the left. 
              The colour scaling is linear with white corresponding to the median background and black to the $0.01\%$ pixels with the highest intensity.  
           }
\end{figure}
\begin{figure}
   \centering
   \includegraphics[angle=0,height=3.11cm]{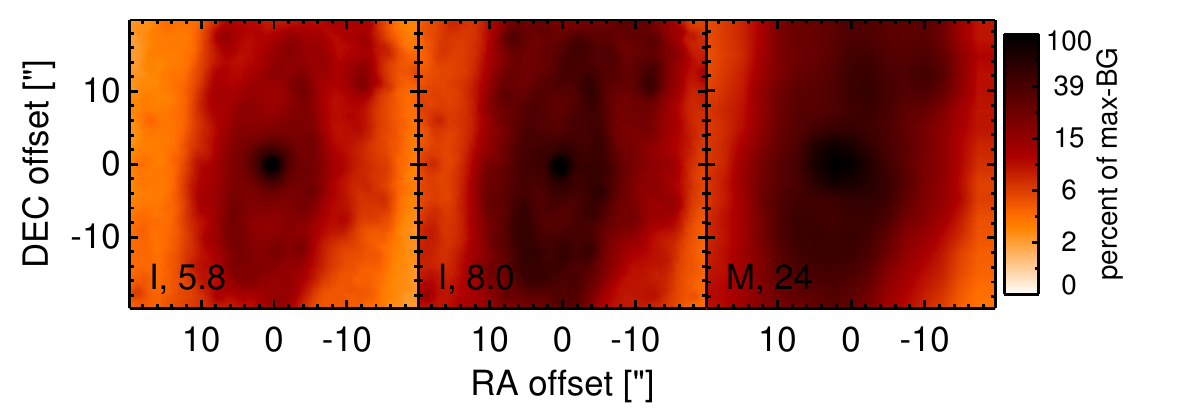}
    \caption{\label{fig:INTim_NGC5033}
             \spitzerr MIR images of NGC\,5033. Displayed are the inner $40\arcsec$ with North up and East to the left. The colour scaling is logarithmic with white corresponding to median background and black to the $0.1\%$ pixels with the highest intensity.
             The label in the bottom left states instrument and central wavelength of the filter in $\mu$m (I: IRAC, M: MIPS). 
           }
\end{figure}
\begin{figure}
   \centering
   \includegraphics[angle=0,height=3.11cm]{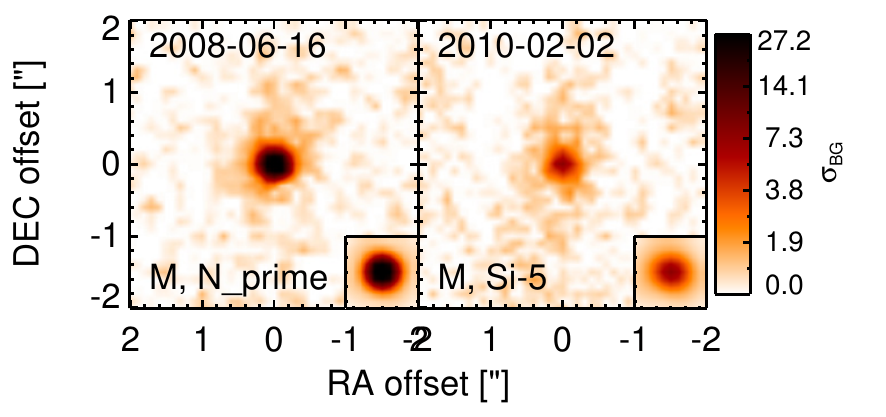}
    \caption{\label{fig:HARim_NGC5033}
             Subarcsecond-resolution MIR images of NGC\,5033 sorted by increasing filter wavelength. 
             Displayed are the inner $4\arcsec$ with North up and East to the left. 
             The colour scaling is logarithmic with white corresponding to median background and black to the $75\%$ of the highest intensity of all images in units of $\sigbg$.
             The inset image shows the central arcsecond of the PSF from the calibrator star, scaled to match the science target.
             The labels in the bottom left state instrument and filter names (C: COMICS, M: Michelle, T: T-ReCS, V: VISIR).
           }
\end{figure}
\begin{figure}
   \centering
   \includegraphics[angle=0,width=8.50cm]{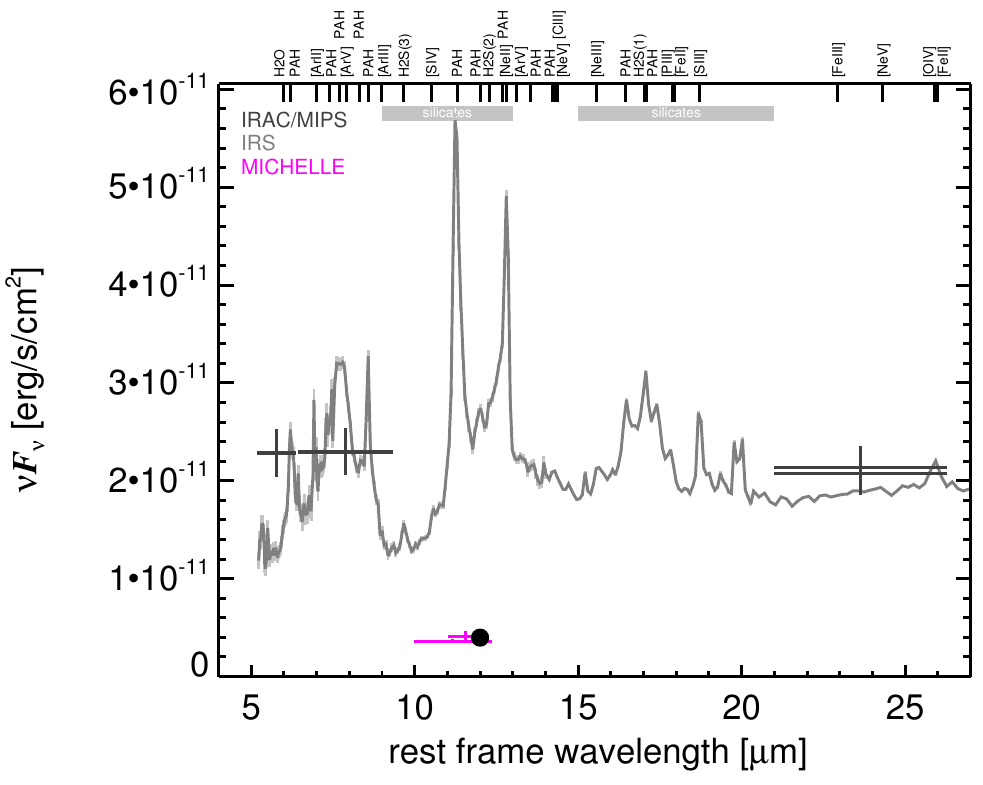}
   \caption{\label{fig:MISED_NGC5033}
      MIR SED of NGC\,5033. The description  of the symbols (if present) is the following.
      Grey crosses and  solid lines mark the \spitzer/IRAC, MIPS and IRS data. 
      The colour coding of the other symbols is: 
      green for COMICS, magenta for Michelle, blue for T-ReCS and red for VISIR data.
      Darker-coloured solid lines mark spectra of the corresponding instrument.
      The black filled circles mark the nuclear 12 and $18\,\mu$m  continuum emission estimate from the data.
      The ticks on the top axis mark positions of common MIR emission lines, while the light grey horizontal bars mark wavelength ranges affected by the silicate 10 and 18$\mu$m features.}
\end{figure}
\clearpage

\twocolumn[\begin{@twocolumnfalse}  
\subsection{NGC\,5135}\label{app:NGC5135}
NGC\,5135 is an infrared-luminous face-on barred spiral galaxy at a redshift of $z=$ 0.0137 ($D\sim$\,Mpc). It harbours a Sy\,2 nucleus \citep{veron-cetty_catalogue_2010} and a circum-nuclear starburst \citep{gonzalez_delgado_ultraviolet-optical_1998,bedregal_near-ir_2009}.  
The latter has a knotty morphology and extends over $\sim 3\arcsec \sim 0.9\,$kpc.
A compact radio source with faint extended emission up to $\sim9\arcsec\sim2.8\,$kpc to the north-east (PA$\sim30\degree$) was detected at lower resolution \citep{ulvestad_radio_1989} and is presumably associated with a compact star-forming region  $\sim 3\arcsec\sim0.3\,$kpc south of the nucleus and not with the AGN \citep{levenson_accretion_2004}.
This agrees with the non-detection in subarcsecond-resolution radio observations \citep{thean_high-resolution_2000}.
The  \oiii emission is extended in north-south direction on a scale of $\sim2\arcsec\sim $\,pc \citep{gonzalez_delgado_ultraviolet-optical_1998}.
The first ground-based MIR observations of NGC\,5135 were performed by \cite{wynn-williams_luminous_1993}, followed by \cite{maiolino_new_1995} and \cite{gorjian_10_2004}, the latter authors reporting the first subarcsecond-resolution $N$-band photometry.
The \spitzer/IRAC and MIPS images show an extended nucleus embedded within the spiral-like host emission. 
The IRAC $8.0\,\mu$m PBCD image is saturated in the nucleus and, thus, not analysed (but see \citealt{gallimore_infrared_2010}).
Our nuclear IRAC $5.8\,\mu$m flux is consistent with \cite{alonso-herrero_high_2006} and \cite{u_spectral_2012} but not \cite{gallimore_infrared_2010}.
The \spitzer/IRS LR staring-mode spectrum is dominated by PAH emission with prominent silicate $10\,\mu$m absorption and a red spectral slope in $\nu F_\nu$-space (see also \citealt{shi_9.7_2006,wu_spitzer/irs_2009,tommasin_spitzer-irs_2010,gallimore_infrared_2010,alonso-herrero_local_2012}).
Thus, the arcsecond-scale MIR SED  is star-formation dominated.
The nuclear region of NGC\,5135 was observed with T-ReCS in the broad N filter in 2006 \citep{alonso-herrero_high_2006}, and with VISIR in five different narrow $N$-band filters in 2006 and 2008 (partly published in \citealt{horst_mid_2008,horst_mid-infrared_2009}).
In addition, a T-ReCS LR $N$-band spectrum was obtained by \cite{diaz-santos_high_2010}.
In all cases, a compact nucleus partly surrounded by banana-shaped, clumpy extended emission was detected. 
The inner radius of this emission is located at $\sim1\arcsec\sim300\,$pc from the nucleus, and the outer radius is located at $\sim2.5\arcsec\sim750\,$pc from the nucleus, extending from PA$\sim65\degree$ to PA$\sim214\degree$.
The nucleus itself is unresolved in the sharpest images (NEII\_1 and NEII from 2006). 
Therefore, we assume that the apparent extensions found in several images are non-intrinsic and the nucleus of NGC\,5135 is generally unresolved at subarcsecond resolution in the MIR.
Our nuclear photometry is consistent with the values published by \cite{horst_mid_2008} and the T-ReCS spectrum as published by \cite{gonzalez-martin_dust_2013}, while being on average $\sim 72\%$ lower than the \spitzerr spectrophotometry.
The nuclear MIR SED exhibits deep silicate absorption but no PAH features, which indicates that the AGN-powered emission has been isolated.
However, star-formation dominates the MIR emission of the central $\sim 1$\,kpc of NGC\,5135.
\newline\end{@twocolumnfalse}]

\begin{figure}
   \centering
   \includegraphics[angle=0,width=8.500cm]{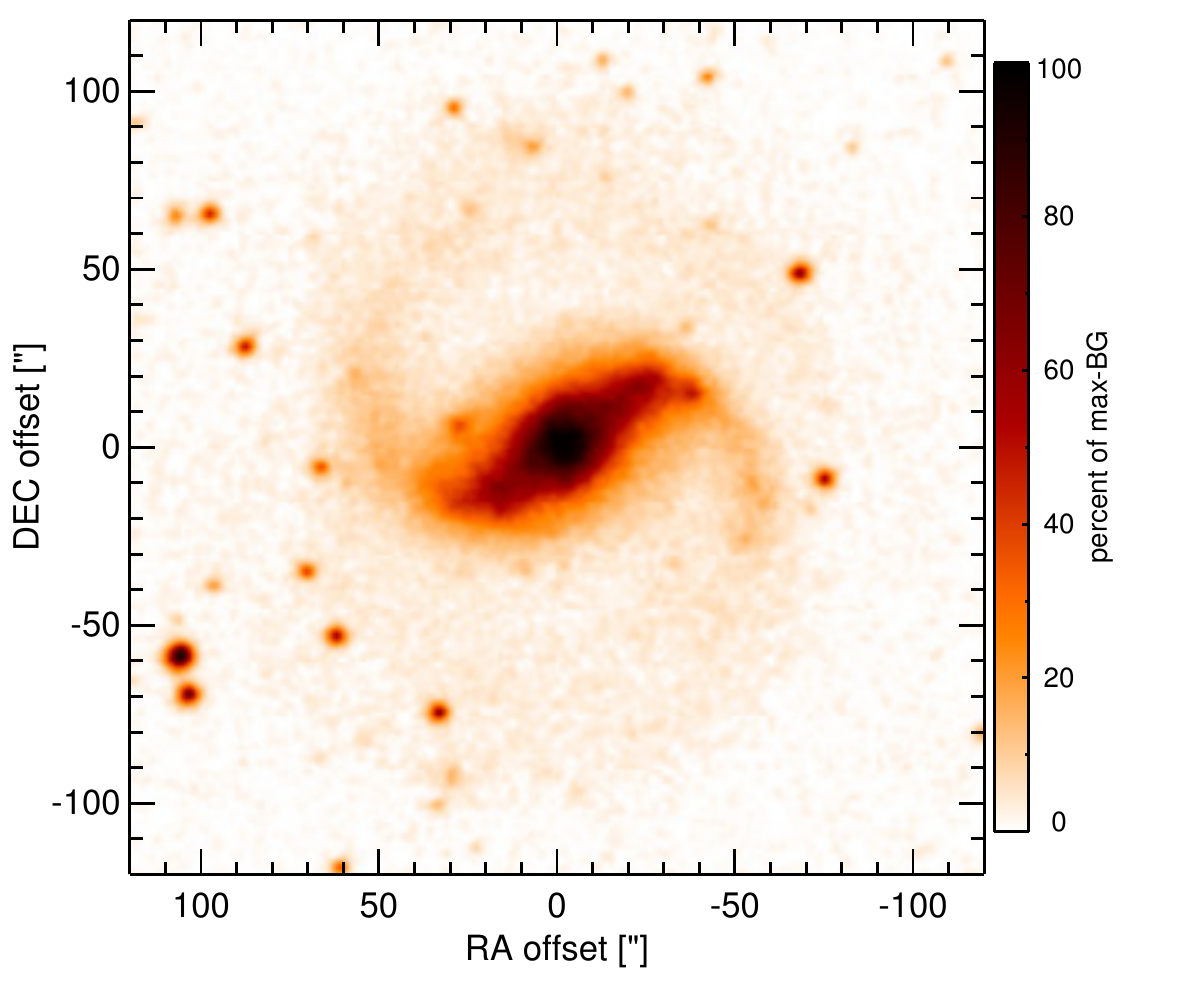}
    \caption{\label{fig:OPTim_NGC5135}
             Optical image (DSS, red filter) of NGC\,5135. Displayed are the central $4\arcmin$ with North up and East to the left. 
              The colour scaling is linear with white corresponding to the median background and black to the $0.01\%$ pixels with the highest intensity.  
           }
\end{figure}
\begin{figure}
   \centering
   \includegraphics[angle=0,height=3.11cm]{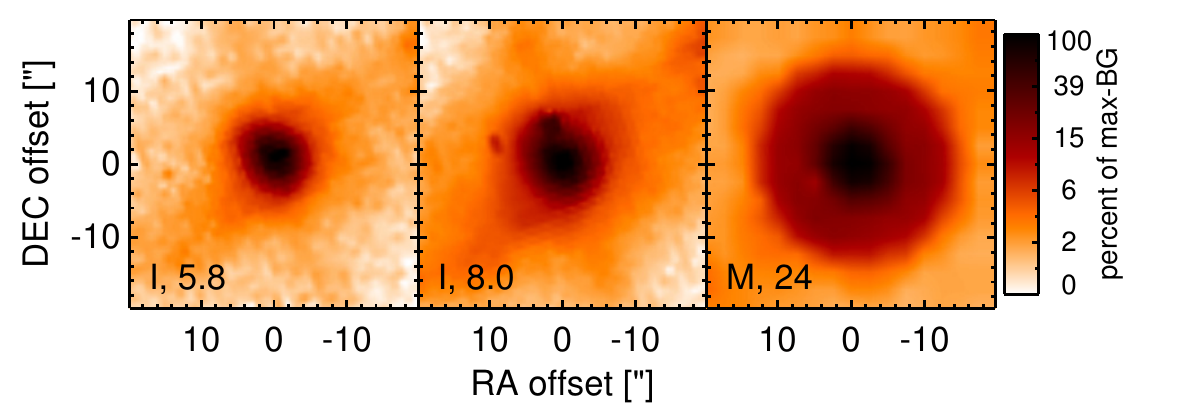}
    \caption{\label{fig:INTim_NGC5135}
             \spitzerr MIR images of NGC\,5135. Displayed are the inner $40\arcsec$ with North up and East to the left. The colour scaling is logarithmic with white corresponding to median background and black to the $0.1\%$ pixels with the highest intensity.
             The label in the bottom left states instrument and central wavelength of the filter in $\mu$m (I: IRAC, M: MIPS). 
             Note that the apparent off-nuclear compact source in the IRAC $8.0\,\mu$m image is an instrumental artefact.
           }
\end{figure}
\begin{figure}
   \centering
   \includegraphics[angle=0,width=8.500cm]{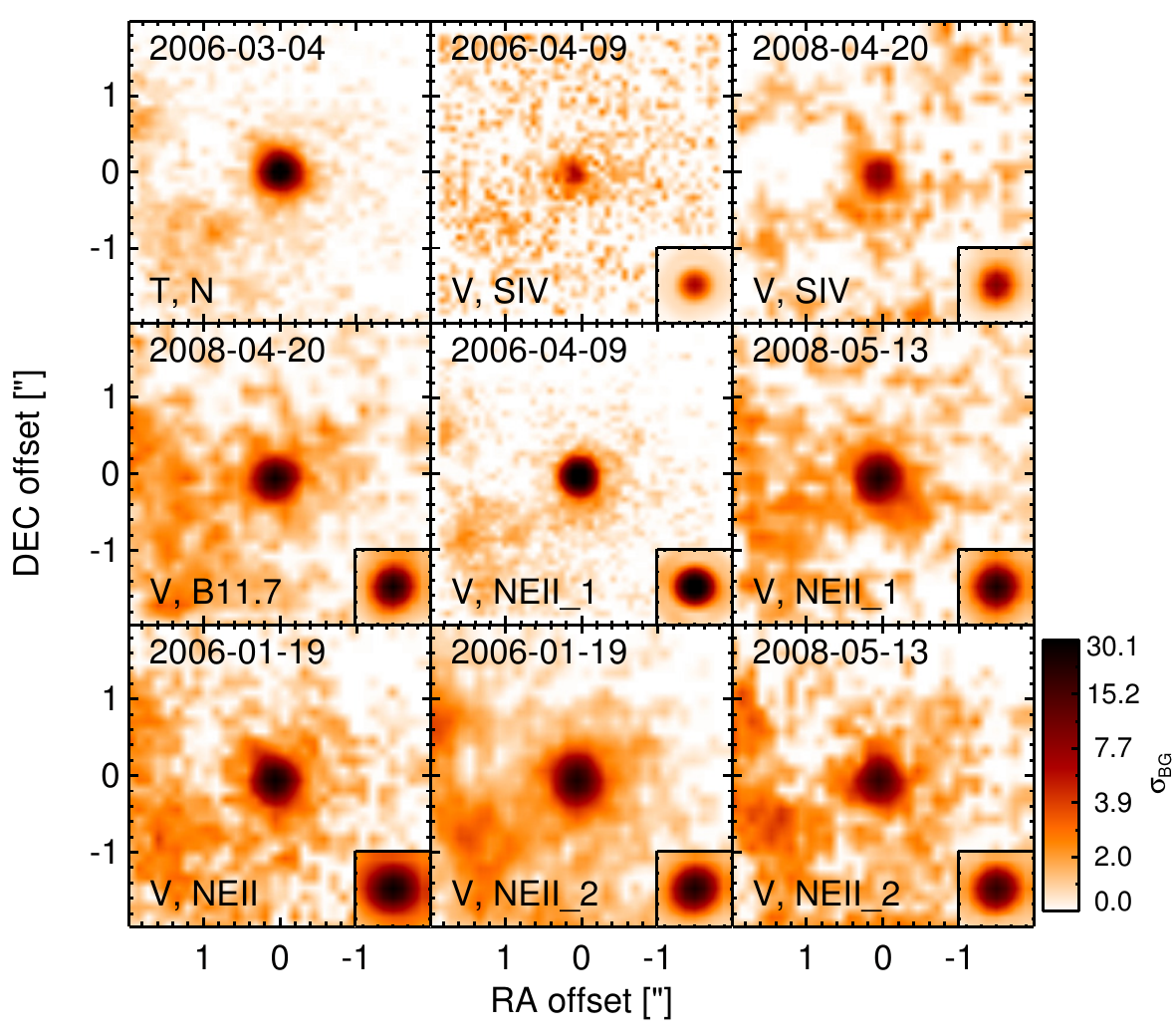}
    \caption{\label{fig:HARim_NGC5135}
             Subarcsecond-resolution MIR images of NGC\,5135 sorted by increasing filter wavelength. 
             Displayed are the inner $4\arcsec$ with North up and East to the left. 
             The colour scaling is logarithmic with white corresponding to median background and black to the $75\%$ of the highest intensity of all images in units of $\sigbg$.
             The inset image shows the central arcsecond of the PSF from the calibrator star, scaled to match the science target.
             The labels in the bottom left state instrument and filter names (C: COMICS, M: Michelle, T: T-ReCS, V: VISIR).
           }
\end{figure}
\begin{figure}
   \centering
   \includegraphics[angle=0,width=8.50cm]{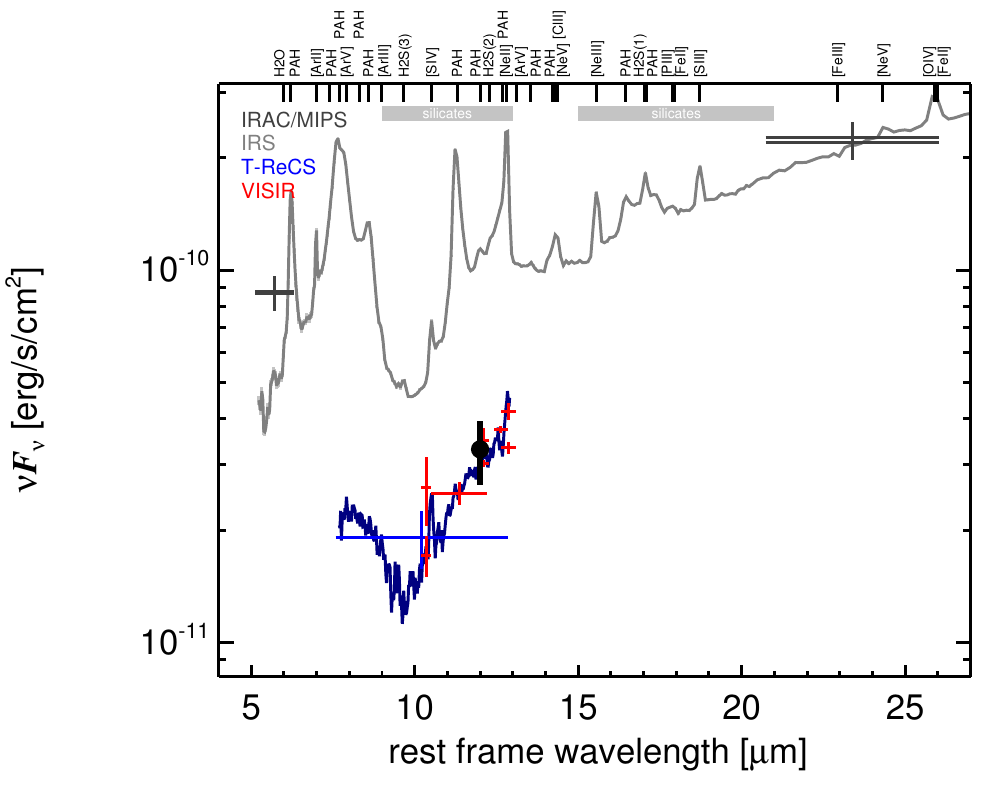}
   \caption{\label{fig:MISED_NGC5135}
      MIR SED of NGC\,5135. The description  of the symbols (if present) is the following.
      Grey crosses and  solid lines mark the \spitzer/IRAC, MIPS and IRS data. 
      The colour coding of the other symbols is: 
      green for COMICS, magenta for Michelle, blue for T-ReCS and red for VISIR data.
      Darker-coloured solid lines mark spectra of the corresponding instrument.
      The black filled circles mark the nuclear 12 and $18\,\mu$m  continuum emission estimate from the data.
      The ticks on the top axis mark positions of common MIR emission lines, while the light grey horizontal bars mark wavelength ranges affected by the silicate 10 and 18$\mu$m features.}
\end{figure}
\clearpage

\twocolumn[\begin{@twocolumnfalse}  
\subsection{NGC\,5252}\label{app:NGC5252}
NGC\,5252 is a lenticular galaxy at a redshift of $z=$ 0.0230 ($D\sim109\,$Mpc) with an AGN optically classified as either a Sy\,1.9 \citep{osterbrock_spectroscopic_1993,trippe_multi-wavelength_2010} or a Sy\,2.0 \citep{veron-cetty_catalogue_2010}. It is listed in the nine-month BAT AGN sample.
The galaxy features a compact radio core with jet-like extended emission (diameter $\sim 4\arcsec\sim 2\,$kpc, PA$\sim 175\degree$)  \citep{wilson_ionization_1994} and very prominent, kiloparsec-scale, shell-like ionization cones along a PA$\sim167\degree$ \citep{tadhunter_anisotropic_1989}.
NGC\,5252 was not detected by \iras, instead the first MIR detection is reported by \cite{maiolino_new_1995}, followed by \isoo observations by \cite{prieto_infrared_2003}.
In addition, \spitzer/IRS and MIPS observations are available, and the corresponding images show a compact nucleus embedded within very faint host emission.
The IRS LR staring-mode spectrum exhibits prominent silicate 10\,$\mu$m and 18\,$\mu$m emission and a blue spectral slope in $\nu F_\nu$-space. Howver, no PAH emission is detected.
The arcsecond-scale MIR SED is, therefore, free of star formation and resembles an unobscured AGN SED.
We observed NGC\,5252 with VISIR in three narrow $N$-band filters in 2010 and detected a compact MIR nucleus without any further host emission.
The nucleus is possibly marginally resolved in the NEII and SIV filter images, but no elongation is detected in the PAH2\_2 filter image. 
Because the SIV image possess the best resolution, we classify the MIR extension of NGC\,5252 as uncertain in the MIR at subarcsecond resolution. 
However, we note that the PAs measured for the elongated SIV and NEII sources would be consistent with the outflow orientation.
Our nuclear photometry is systematically high by $\sim18\%$ as compared to the \spitzerr spectrophotometry for unknown reasons.
However, it confirms the silicate emission in the central $\sim150$\,pc of NGC\,5252. Thus, we use the IRS spectrum to compute the 12\,$\mu$m continuum emission estimate corrected for the silicate feature.
\newline\end{@twocolumnfalse}]

\begin{figure}
   \centering
   \includegraphics[angle=0,width=8.500cm]{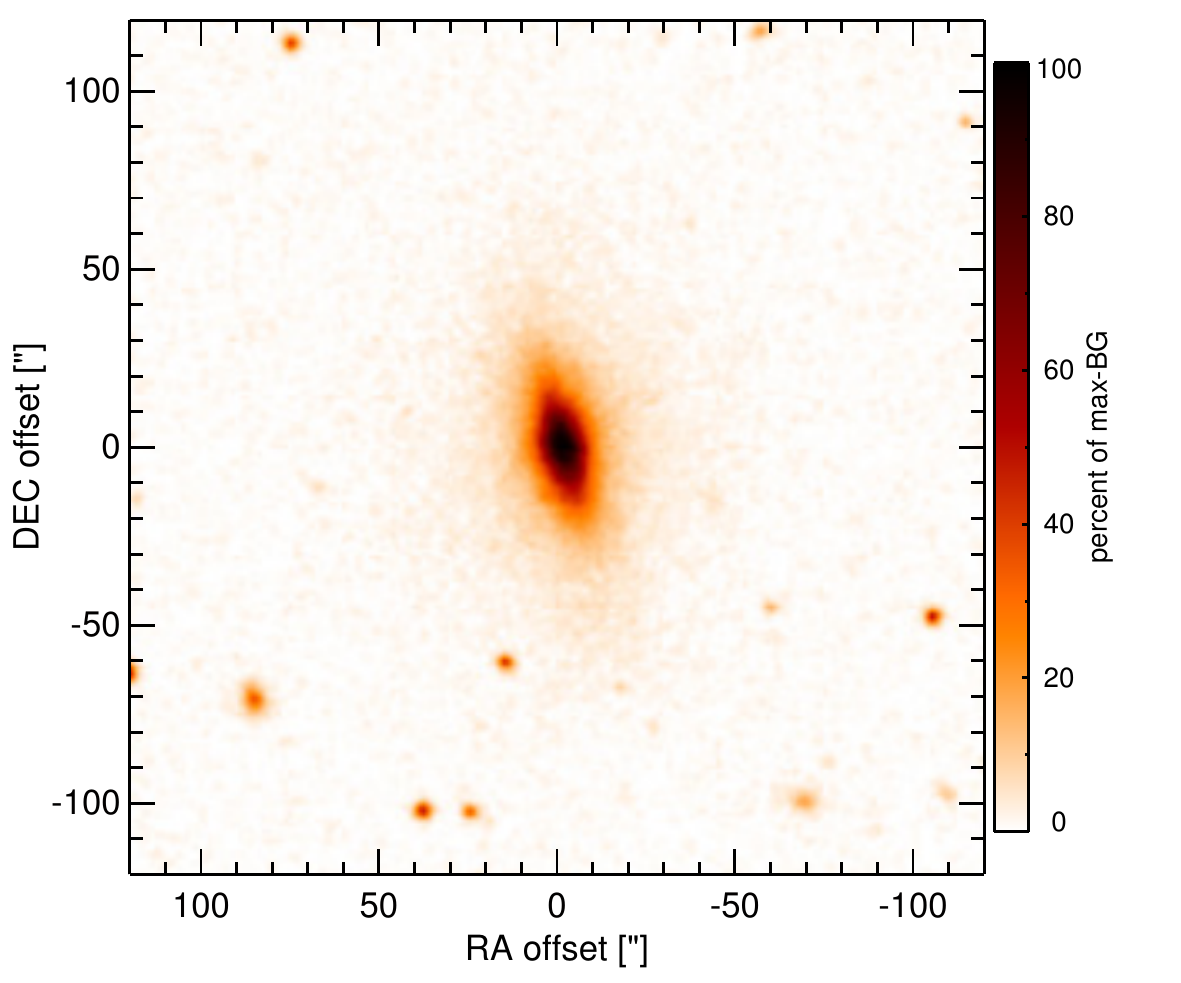}
    \caption{\label{fig:OPTim_NGC5252}
             Optical image (DSS, red filter) of NGC\,5252. Displayed are the central $4\arcmin$ with North up and East to the left. 
              The colour scaling is linear with white corresponding to the median background and black to the $0.01\%$ pixels with the highest intensity.  
           }
\end{figure}
\begin{figure}
   \centering
   \includegraphics[angle=0,height=3.11cm]{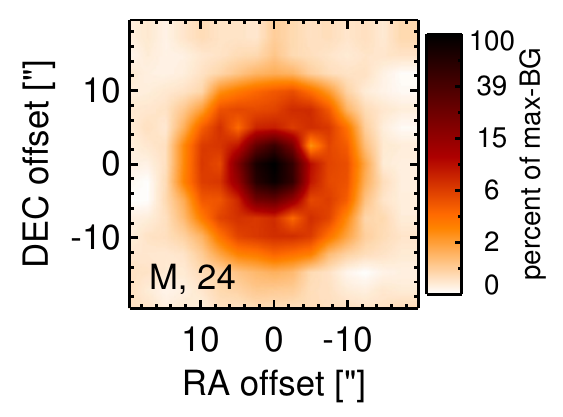}
    \caption{\label{fig:INTim_NGC5252}
             \spitzerr MIR images of NGC\,5252. Displayed are the inner $40\arcsec$ with North up and East to the left. The colour scaling is logarithmic with white corresponding to median background and black to the $0.1\%$ pixels with the highest intensity.
             The label in the bottom left states instrument and central wavelength of the filter in $\mu$m (I: IRAC, M: MIPS). 
           }
\end{figure}
\begin{figure}
   \centering
   \includegraphics[angle=0,height=3.11cm]{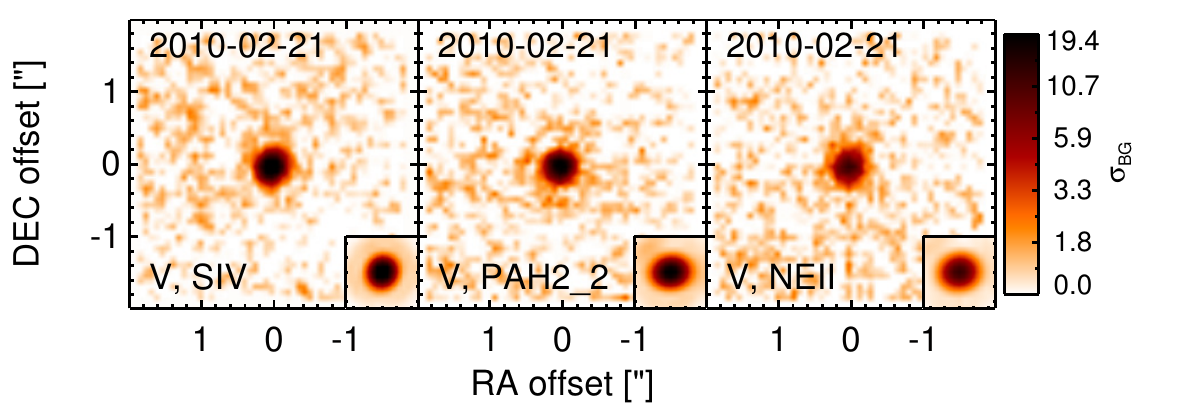}
    \caption{\label{fig:HARim_NGC5252}
             Subarcsecond-resolution MIR images of NGC\,5252 sorted by increasing filter wavelength. 
             Displayed are the inner $4\arcsec$ with North up and East to the left. 
             The colour scaling is logarithmic with white corresponding to median background and black to the $75\%$ of the highest intensity of all images in units of $\sigbg$.
             The inset image shows the central arcsecond of the PSF from the calibrator star, scaled to match the science target.
             The labels in the bottom left state instrument and filter names (C: COMICS, M: Michelle, T: T-ReCS, V: VISIR).
           }
\end{figure}
\begin{figure}
   \centering
   \includegraphics[angle=0,width=8.50cm]{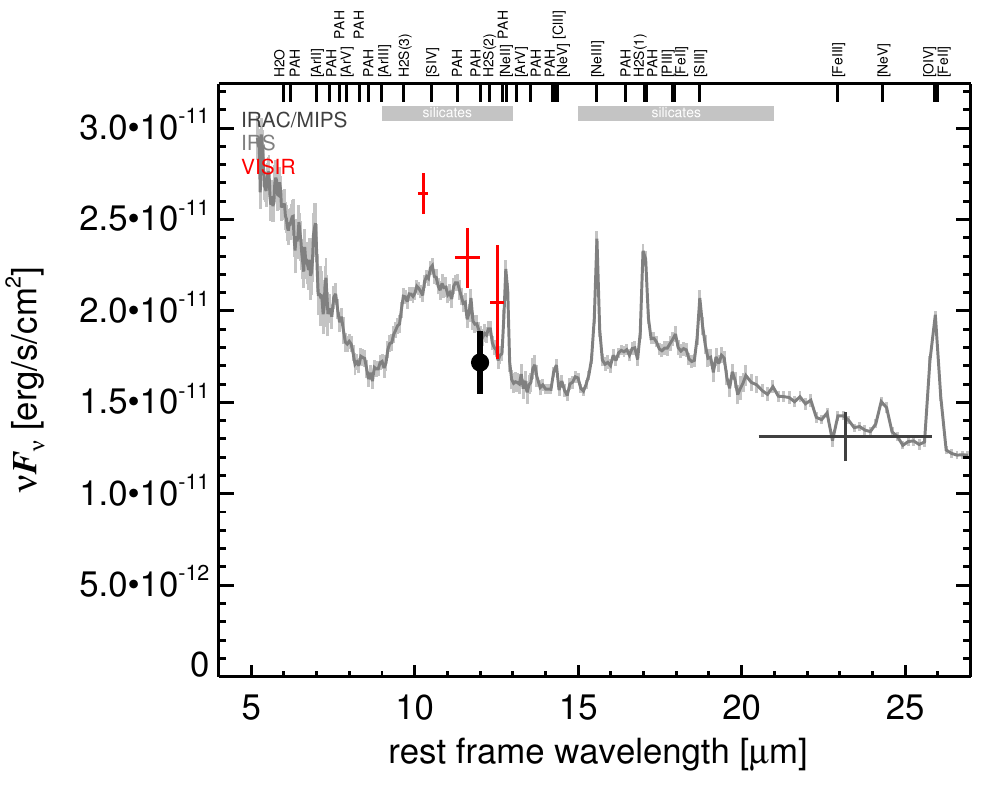}
   \caption{\label{fig:MISED_NGC5252}
      MIR SED of NGC\,5252. The description  of the symbols (if present) is the following.
      Grey crosses and  solid lines mark the \spitzer/IRAC, MIPS and IRS data. 
      The colour coding of the other symbols is: 
      green for COMICS, magenta for Michelle, blue for T-ReCS and red for VISIR data.
      Darker-coloured solid lines mark spectra of the corresponding instrument.
      The black filled circles mark the nuclear 12 and $18\,\mu$m  continuum emission estimate from the data.
      The ticks on the top axis mark positions of common MIR emission lines, while the light grey horizontal bars mark wavelength ranges affected by the silicate 10 and 18$\mu$m features.}
\end{figure}
\clearpage

\twocolumn[\begin{@twocolumnfalse}  
\subsection{NGC\,5258}\label{app:NGC5258}
NGC\,5258 is a disturbed spiral galaxy interacting with NGC\,5257 (1.3\arcmin\ towards the north-west) at a redshift of $z=$ 0.0225 ($D\sim173\,$Mpc).
It possibly harbours an AGN (according to NED) classified as a borderline LINER/H\,II nucleus (no reference could be retrieved), while \cite{yuan_role_2010} classify it as H\,II nucleus.
No X-ray detection or subarcsecond-resolution radio observation are reported in the literature.
After its detection in \iras, \cite{cutri_statistical_1985}, \cite{carico_iras_1988} and \cite{bushouse_distribution_1998} unsuccessfully attempted to detect the nucleus of NGC\,5258.
The \spitzer/IRAC and MIPS images show extended spiral-like host emission but no nucleus. 
Instead, the brightest emission region is the inner part of the western spiral arm (closest to NGC\,5257; see also \citealt{smith_spitzer_2007}). 
We extract the unresolved part of the emission at the estimated position of the nucleus from the IRAC and MIPS images. 
However, the MIPS 24\,$\mu$m measurement is affected by the bright off-nuclear region and, therefore, unreliable.
The \spitzer/IRS LR staring-mode spectrum is dominated by PAH emission with a possible weak silicate 10$\,\mu$m absorption feature and a flat spectral slope in $\nu F_\nu$-space (see also \citealt{stierwalt_mid-infrared_2013}).
Thus, the arcsecond-scale MIR SED is completely star-formation dominated.
The nuclear region of NGC\,5258 was observed with VISIR in the PAH2 filter in 2005 \citep{siebenmorgen_nuclear_2008} but no nuclear source was detected.
Our derived flux upper limit is more conservative than that given in \cite{siebenmorgen_nuclear_2008} but still $\sim67\%$ lower than the \spitzerr spectrophotometry.
This confirms that the central $\sim2\,$kpc region of NGC\,5258 is completely star-formation dominated.
\newline\end{@twocolumnfalse}]

\begin{figure}
   \centering
   \includegraphics[angle=0,width=8.500cm]{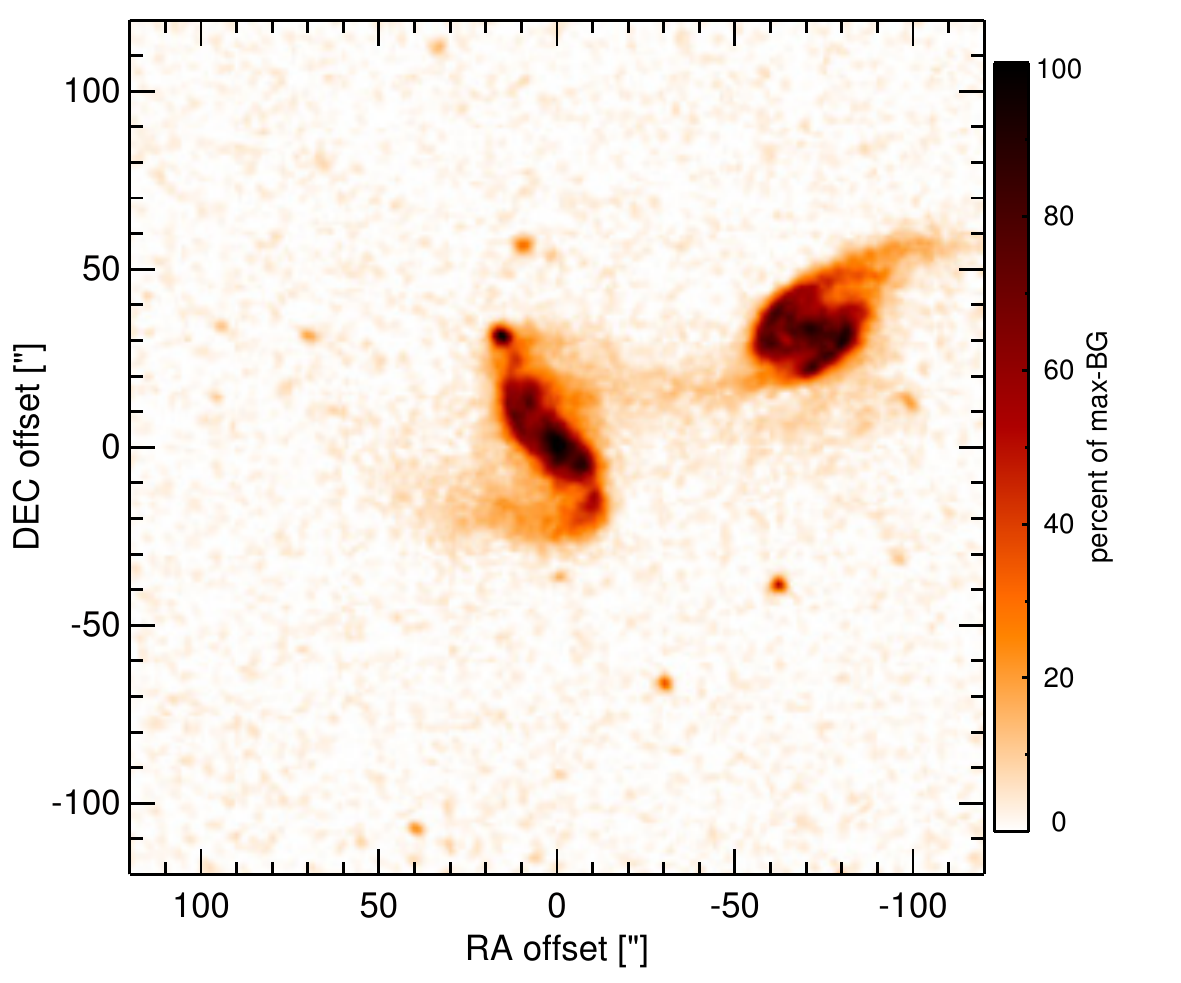}
    \caption{\label{fig:OPTim_NGC5258}
             Optical image (DSS, red filter) of NGC\,5258. Displayed are the central $4\arcmin$ with North up and East to the left. 
              The colour scaling is linear with white corresponding to the median background and black to the $0.01\%$ pixels with the highest intensity.  
           }
\end{figure}
\begin{figure}
   \centering
   \includegraphics[angle=0,height=3.11cm]{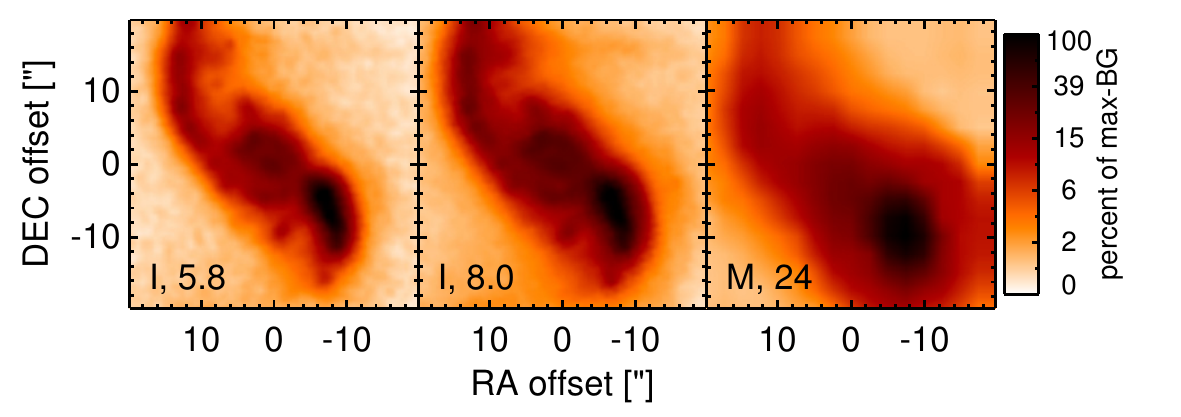}
    \caption{\label{fig:INTim_NGC5258}
             \spitzerr MIR images of NGC\,5258. Displayed are the inner $40\arcsec$ with North up and East to the left. The colour scaling is logarithmic with white corresponding to median background and black to the $0.1\%$ pixels with the highest intensity.
             The label in the bottom left states instrument and central wavelength of the filter in $\mu$m (I: IRAC, M: MIPS). 
           }
\end{figure}
\begin{figure}
   \centering
   \includegraphics[angle=0,width=8.50cm]{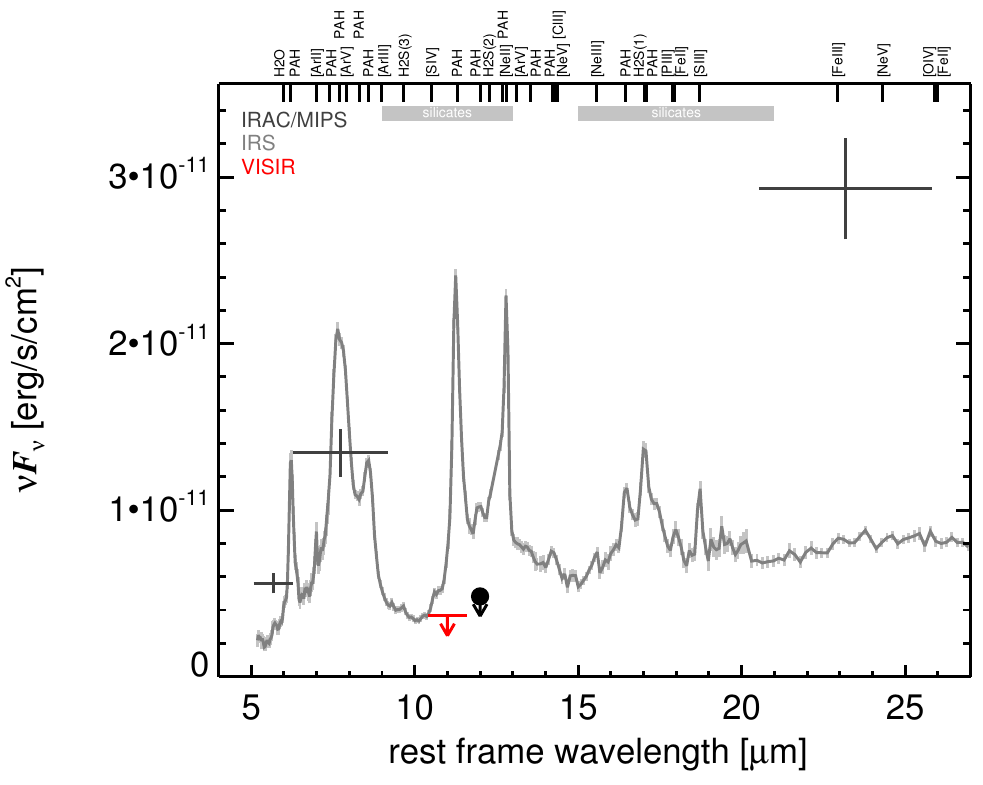}
   \caption{\label{fig:MISED_NGC5258}
      MIR SED of NGC\,5258. The description  of the symbols (if present) is the following.
      Grey crosses and  solid lines mark the \spitzer/IRAC, MIPS and IRS data. 
      The colour coding of the other symbols is: 
      green for COMICS, magenta for Michelle, blue for T-ReCS and red for VISIR data.
      Darker-coloured solid lines mark spectra of the corresponding instrument.
      The black filled circles mark the nuclear 12 and $18\,\mu$m  continuum emission estimate from the data.
      The ticks on the top axis mark positions of common MIR emission lines, while the light grey horizontal bars mark wavelength ranges affected by the silicate 10 and 18$\mu$m features.}
\end{figure}
\clearpage

\twocolumn[\begin{@twocolumnfalse}  
\subsection{NGC\,5273}\label{app:NGC5273}
NGC\,5273 is a lenticular galaxy at a distance of $D=$ $15.3 \pm 3.3\,$Mpc (NED redshift-independent median) hosting an AGN, that has been optically classified either as a Sy\,1.5 or a Sy\,1.9 (see discussion in \citealt{trippe_multi-wavelength_2010}).
It features a slightly elongated compact radio source (PA$\sim5\degree$; \citealt{ulvestad_radio_1984,nagar_radio_1999}) and a cospatial one-sided NLR cone $\sim1\arcsec\sim74\,$pc to the south (PA$\sim180\degree$; \citealt{ferruit_hubble_2000}).
The first ground-based MIR observations are reported in \cite{maiolino_new_1995}. 
In addition, NGC\,5273 was observed with \isoo \citep{clavel_2.5-11_2000,ramos_almeida_mid-infrared_2007} and \spitzer/IRAC, IRS and MIPS.
The corresponding IRAC and MIPS images show a compact nucleus embedded within weak diffuse host emission.
Our nuclear MIPS 24\,$\mu$m photometry provides a flux consistent with \cite{temi_spitzer_2009}.
The IRS LR staring-mode spectrum suffers from a low S/N but indicates significant PAH emission and a red spectral slope in $\nu F_\nu$-space.
The arcsecond-scale MIR SED is thus significantly affected by star formation.
Note that \cite{oi_comparison_2010} detect PAH 3.3\,$\mu$m emission in the central $\sim120$\,pc of NGC\,5273.
We observed the nuclear region of NGC\,5273 with Michelle in two $N$-band filters in 2010 and detected a compact MIR nucleus without further host emission.
The nucleus appears marginally resolved in both images (FWHM(major axis)$\sim0.53\arcsec\sim40\,$pc; PA$\sim90\degree$).
However, the Si-5 image is affected by strong vertical stripe-like sky residuals, which make this result uncertain. We classify the nucleus as possibly extended at subarcsecond resolution in the MIR.
Our nuclear Michelle photometry is on average $\sim 33\%$ lower than the \spitzerr spectrophotometry.
We can not exclude that the nuclear MIR SED ($\sim40\,$pc) is still significantly affected by star formation.
\newline\end{@twocolumnfalse}]

\begin{figure}
   \centering
   \includegraphics[angle=0,width=8.500cm]{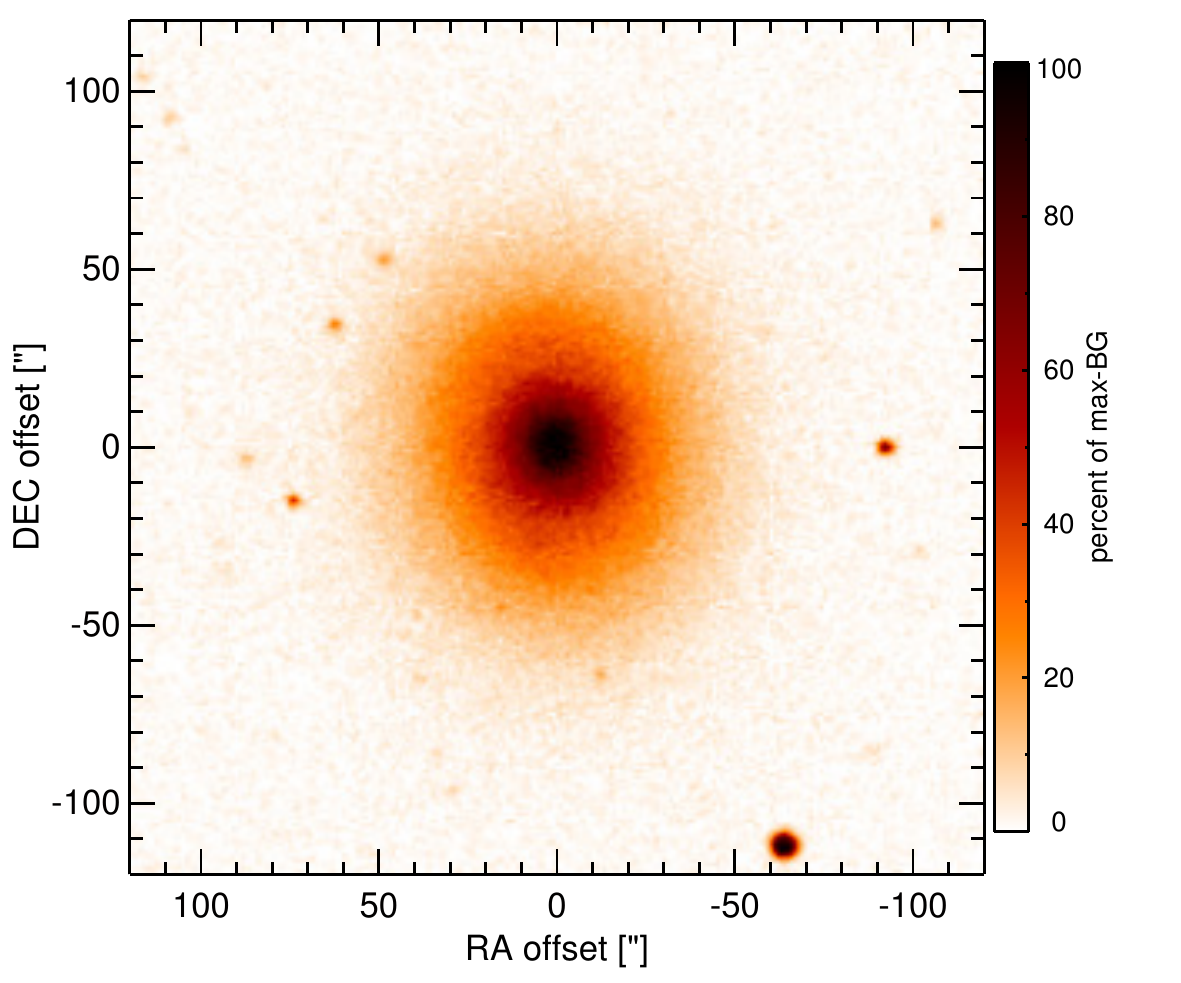}
    \caption{\label{fig:OPTim_NGC5273}
             Optical image (DSS, red filter) of NGC\,5273. Displayed are the central $4\arcmin$ with North up and East to the left. 
              The colour scaling is linear with white corresponding to the median background and black to the $0.01\%$ pixels with the highest intensity.  
           }
\end{figure}
\begin{figure}
   \centering
   \includegraphics[angle=0,height=3.11cm]{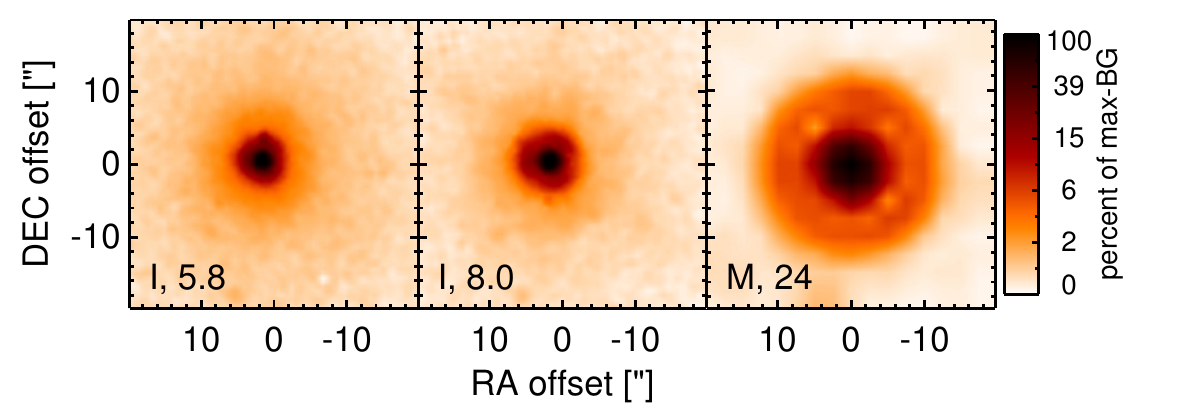}
    \caption{\label{fig:INTim_NGC5273}
             \spitzerr MIR images of NGC\,5273. Displayed are the inner $40\arcsec$ with North up and East to the left. The colour scaling is logarithmic with white corresponding to median background and black to the $0.1\%$ pixels with the highest intensity.
             The label in the bottom left states instrument and central wavelength of the filter in $\mu$m (I: IRAC, M: MIPS). 
           }
\end{figure}
\begin{figure}
   \centering
   \includegraphics[angle=0,height=3.11cm]{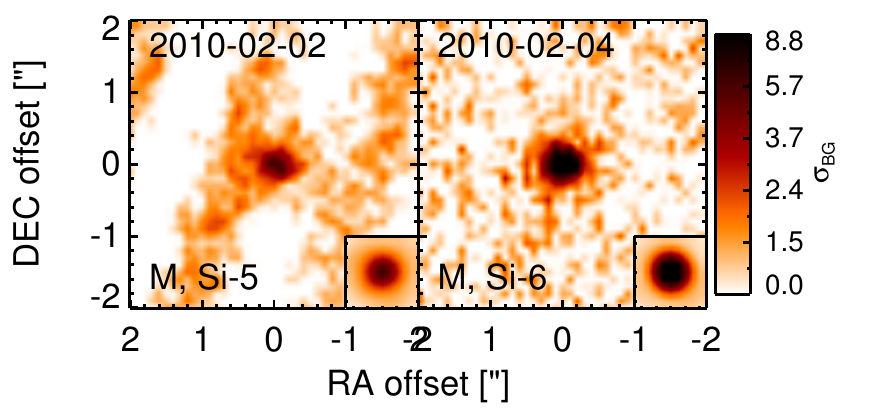}
    \caption{\label{fig:HARim_NGC5273}
             Subarcsecond-resolution MIR images of NGC\,5273 sorted by increasing filter wavelength. 
             Displayed are the inner $4\arcsec$ with North up and East to the left. 
             The colour scaling is logarithmic with white corresponding to median background and black to the $75\%$ of the highest intensity of all images in units of $\sigbg$.
             The inset image shows the central arcsecond of the PSF from the calibrator star, scaled to match the science target.
             The labels in the bottom left state instrument and filter names (C: COMICS, M: Michelle, T: T-ReCS, V: VISIR).
           }
\end{figure}
\begin{figure}
   \centering
   \includegraphics[angle=0,width=8.50cm]{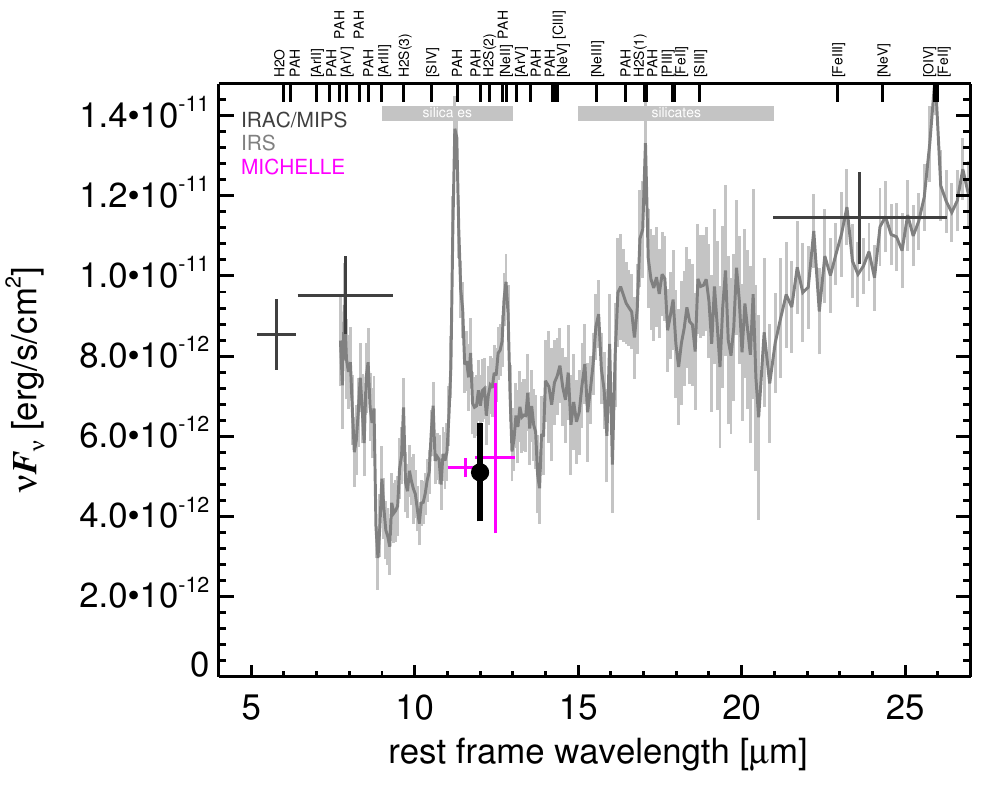}
   \caption{\label{fig:MISED_NGC5273}
      MIR SED of NGC\,5273. The description  of the symbols (if present) is the following.
      Grey crosses and  solid lines mark the \spitzer/IRAC, MIPS and IRS data. 
      The colour coding of the other symbols is: 
      green for COMICS, magenta for Michelle, blue for T-ReCS and red for VISIR data.
      Darker-coloured solid lines mark spectra of the corresponding instrument.
      The black filled circles mark the nuclear 12 and $18\,\mu$m  continuum emission estimate from the data.
      The ticks on the top axis mark positions of common MIR emission lines, while the light grey horizontal bars mark wavelength ranges affected by the silicate 10 and 18$\mu$m features.}
\end{figure}
\clearpage

\twocolumn[\begin{@twocolumnfalse}  
\subsection{NGC\,5347}\label{app:NGC5347}
NGC\,5347 is a low-inclination barred grand-design spiral galaxy at a redshift of $z=$ 0.0078 ($D\sim38.1\,$Mpc) harbouring a Sy\,2 nucleus \citep{veron-cetty_catalogue_2010} with polarized broad emission lines \citep{moran_composite_2001}.
It features an unresolved radio core \citep{schmitt_jet_2001}, a nuclear   water maser \citep{braatz_survey_2003}, and a one-sided NLR cone $\sim2\arcsec\sim360\,$pc to the north (PA$\sim30\degree$; \citealt{pogge_circumnuclear_1989,schmitt_hubble_2003}).
The first ground-based MIR observations were reported by \cite{maiolino_new_1995}, and the first subarcsecond-resolution $N$-band photometry was published by \cite{gorjian_10_2004} using Palomar 5\,m/MIRLIN.
In the \spitzer/IRAC and MIPS images, NGC\,5347 appears as a bright compact nucleus embedded within very faint host emission.
Our nuclear IRAC $5.8$ and $8.0\,\mu$m photometry is consistent with \cite{gallimore_infrared_2010}.
The \spitzer/IRS LR staring-mode spectrum exhibits a very smooth continuum with weak silicate $10\,\mu$m absorption, faint PAH features and a wide emission peak around $\sim20\,\mu$m  in $\nu F_\nu$-space (see also \citealt{buchanan_spitzer_2006,shi_9.7_2006,wu_spitzer/irs_2009,tommasin_spitzer-irs_2010,gallimore_infrared_2010}).
The nuclear region of NGC\,5347 was observed with Michelle in the N' and Qa filters in 2006 (unpublished, to our knowledge), and a compact nucleus without further host emission is detect in both images.
A matching standard star observation could be retrieved only for the N' image, and this standard star shows an elongated PSF that is not visible in the science image.
Therefore, we classify the MIR extension of NGC\,5347 as uncertain at subarcsecond resolution.
The N' flux is consistent with the \spitzerr spectrophotometry, while the Qa flux is significantly higher.
This is presumably caused by the use of the median conversion factor in the absence of a matching standard star observation.
Therefore, we use the IRS spectrum to compute the nuclear 12 and 18\,$\mu$m continuum emission estimates.
\newline\end{@twocolumnfalse}]

\begin{figure}
   \centering
   \includegraphics[angle=0,width=8.500cm]{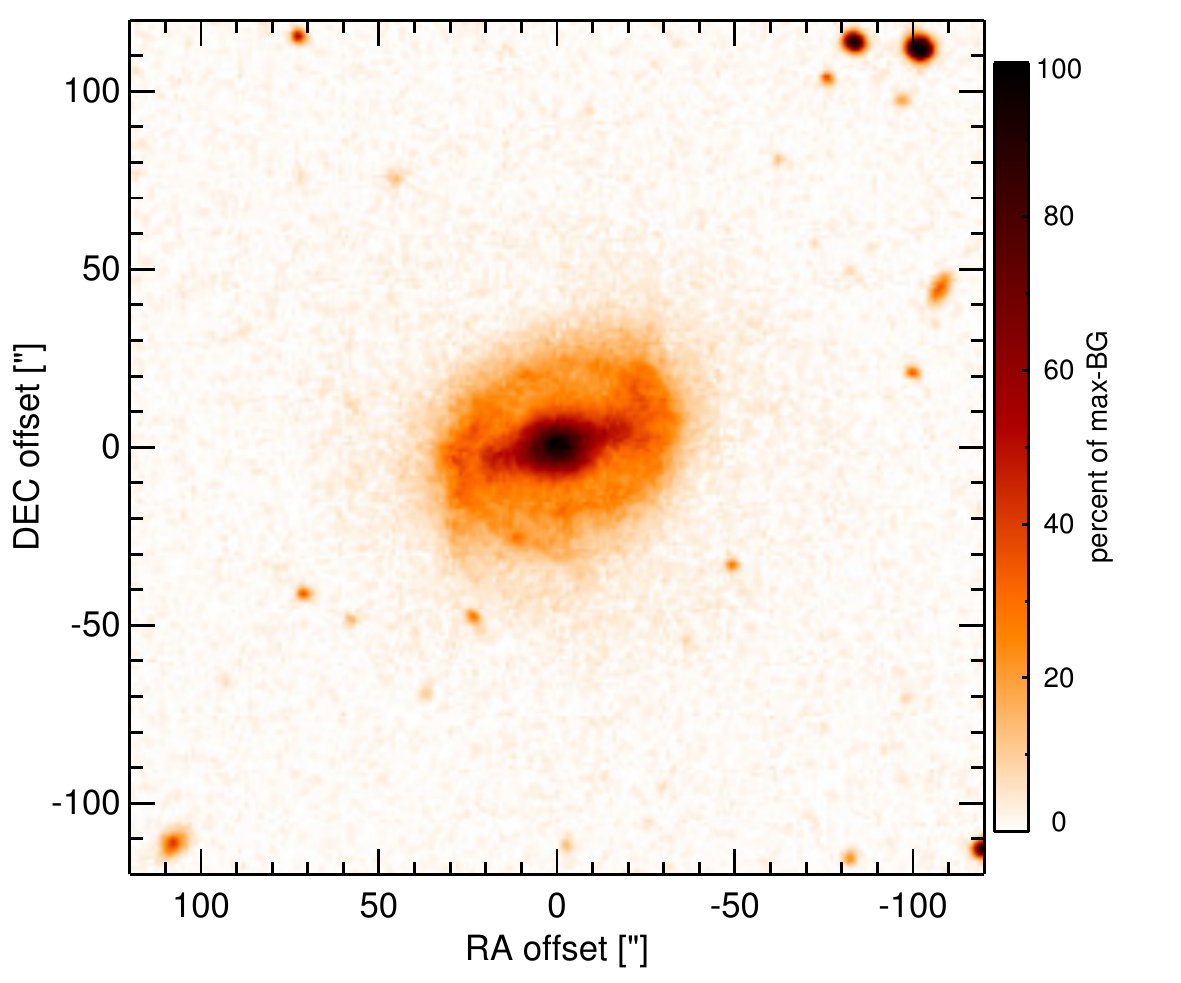}
    \caption{\label{fig:OPTim_NGC5347}
             Optical image (DSS, red filter) of NGC\,5347. Displayed are the central $4\arcmin$ with North up and East to the left. 
              The colour scaling is linear with white corresponding to the median background and black to the $0.01\%$ pixels with the highest intensity.  
           }
\end{figure}
\begin{figure}
   \centering
   \includegraphics[angle=0,height=3.11cm]{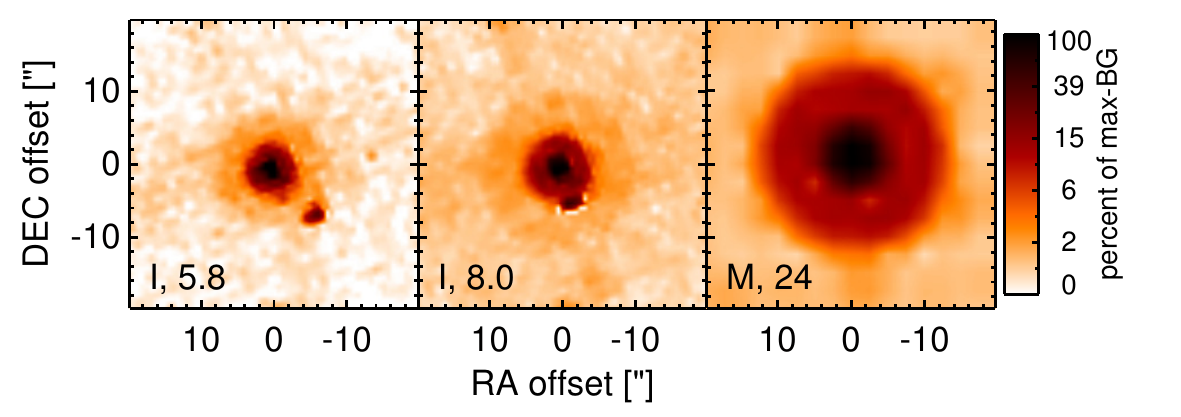}
    \caption{\label{fig:INTim_NGC5347}
             \spitzerr MIR images of NGC\,5347. Displayed are the inner $40\arcsec$ with North up and East to the left. The colour scaling is logarithmic with white corresponding to median background and black to the $0.1\%$ pixels with the highest intensity.
             The label in the bottom left states instrument and central wavelength of the filter in $\mu$m (I: IRAC, M: MIPS).
             Note that the apparent off-nuclear compact sources in the IRAC 5.8 and $8.0\,\mu$m images are instrumental artefacts.
           }
\end{figure}
\begin{figure}
   \centering
   \includegraphics[angle=0,height=3.11cm]{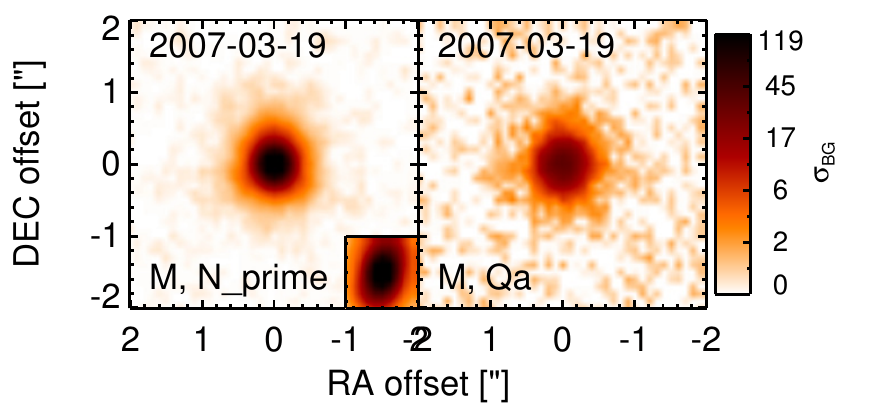}
    \caption{\label{fig:HARim_NGC5347}
             Subarcsecond-resolution MIR images of NGC\,5347 sorted by increasing filter wavelength. 
             Displayed are the inner $4\arcsec$ with North up and East to the left. 
             The colour scaling is logarithmic with white corresponding to median background and black to the $75\%$ of the highest intensity of all images in units of $\sigbg$.
             The inset image shows the central arcsecond of the PSF from the calibrator star, scaled to match the science target.
             The labels in the bottom left state instrument and filter names (C: COMICS, M: Michelle, T: T-ReCS, V: VISIR).
           }
\end{figure}
\begin{figure}
   \centering
   \includegraphics[angle=0,width=8.50cm]{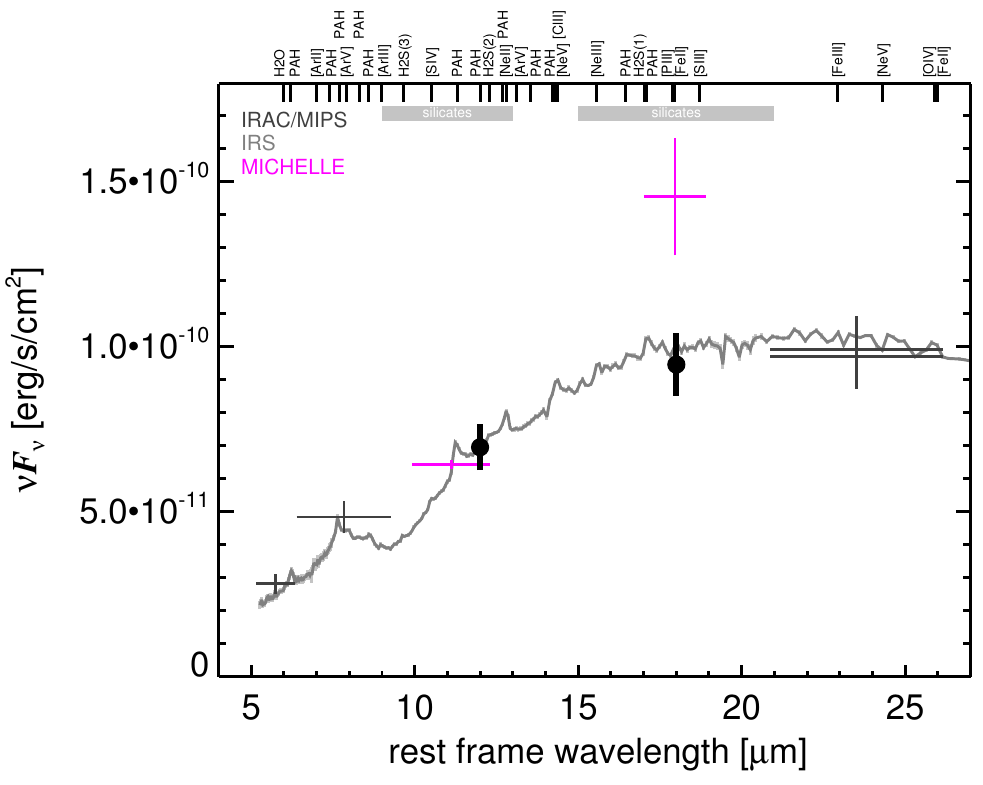}
   \caption{\label{fig:MISED_NGC5347}
      MIR SED of NGC\,5347. The description  of the symbols (if present) is the following.
      Grey crosses and  solid lines mark the \spitzer/IRAC, MIPS and IRS data. 
      The colour coding of the other symbols is: 
      green for COMICS, magenta for Michelle, blue for T-ReCS and red for VISIR data.
      Darker-coloured solid lines mark spectra of the corresponding instrument.
      The black filled circles mark the nuclear 12 and $18\,\mu$m  continuum emission estimate from the data.
      The ticks on the top axis mark positions of common MIR emission lines, while the light grey horizontal bars mark wavelength ranges affected by the silicate 10 and 18$\mu$m features.}
\end{figure}
\clearpage

\twocolumn[\begin{@twocolumnfalse}  
\subsection{NGC\,5363}\label{app:NGC5363}
NGC\,5363 is a dusty peculiar early-type galaxy at a redshift of $z=$ 0.0038 ($D\sim21.0\,$Mpc) with a LINER nucleus \citep{ho_search_1997-1}.
The presence of an AGN is supported by the X-ray properties \citep{gonzalez-martin_x-ray_2009}. Furthermore, a compact radio core with biconical elongated emission along a PA$\sim145\degree$ was detected on arcsecond-scales \citep{hummel_central_1984}, and jet-like elongation can be seen on milliarcsecond-scale along a PA$\sim-100\degree$ \citep{nagar_radio_2005}.
NGC\,5363 was observed with \isoo \citep{xilouris_dust_2004,temi_cold_2004} and \spitzer/IRAC, IRS and MIPS.
It appears as an extended nucleus (diameter$\sim15\arcsec\sim1.5\,$kpc; PA$\sim35\degree$) embedded within larger-scale host emission (see also \citealt{pahre_mid-infrared_2004,pahre_spatial_2004}).
In addition, a bright compact source $\sim6.4\arcsec\sim650\,$pc to the west of the nucleus is visible (PA$\sim-108\degree$).
The IRS LR mapping-mode PBCD spectrum consists of the shortest wavelength setting only.
It indicates the presence of the PAH 6.2\,$\mu$m feature as typical for star formation (see also \citealt{mason_nuclear_2012}).
The nuclear region of NGC\,5363 was observed with Michelle in the N' filter in 2008 \citep{mason_nuclear_2012}, and with VISIR in two $N$-band filters in 2009 \citep{asmus_mid-infrared_2011}.
A compact nucleus is only detected in the Michelle image with a low S/N.
The nucleus is embedded within very weak bar-like emission extending several arcsecond along a PA$\sim 45\degree$.
In addition, the compact source visible in the IRAC images is also weakly detected. 
To our knowledge, this compact source has not been reported so far at any wavelength apart from its detection in the 2MASS point-source catalogue.
Because it is also visible in the SDSS images and appears very blue, it is presumably a foreground star.
We perform manual PSF-scaling to measure the unresolved flux of the nucleus, which provides a value consistent with \cite{mason_nuclear_2012} and much lower than the \spitzerr spectrophotometry.
The two derived flux upper limits from the VISIR images are consistent with the N' flux.
\newline\end{@twocolumnfalse}]

\begin{figure}
   \centering
   \includegraphics[angle=0,width=8.500cm]{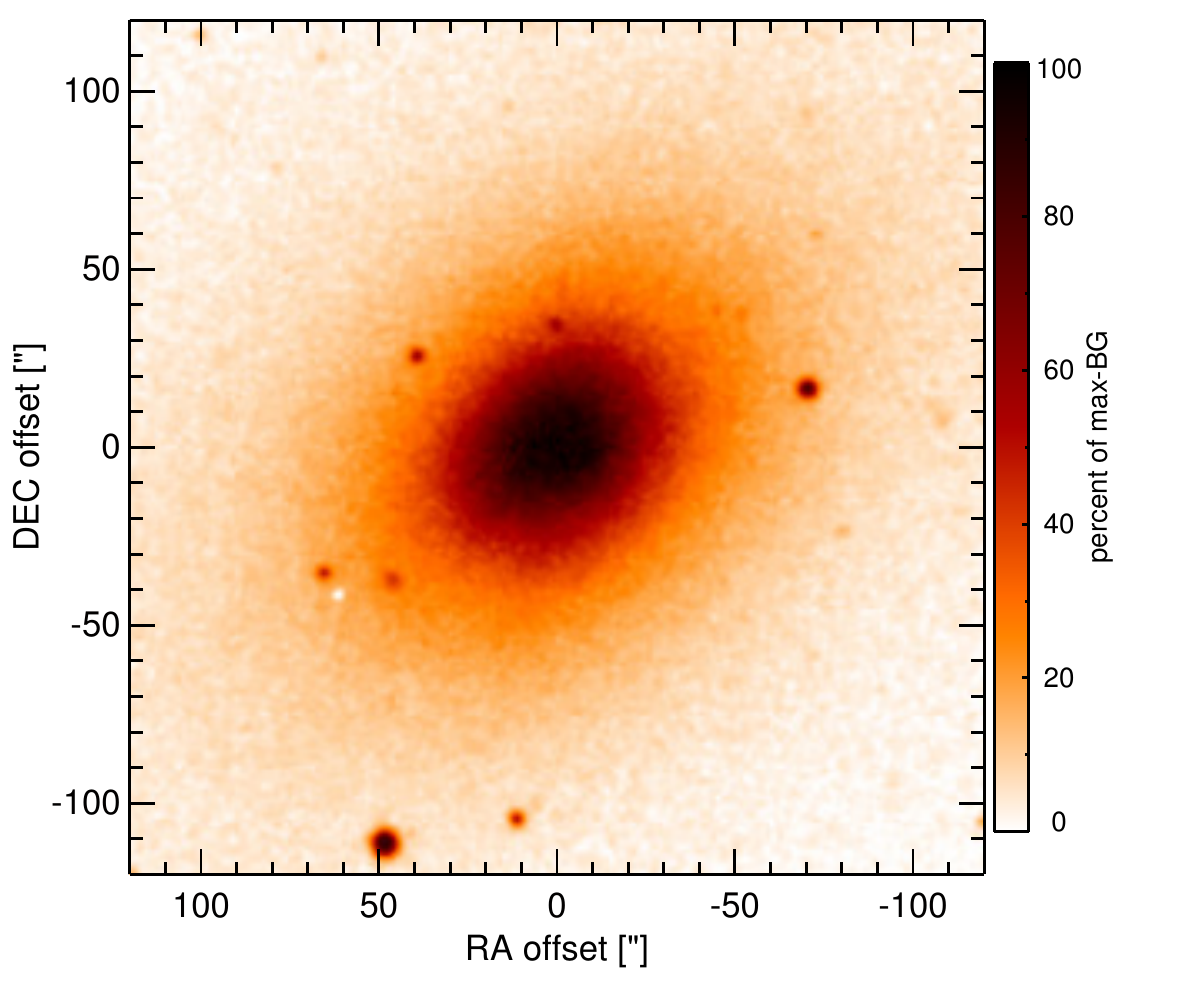}
    \caption{\label{fig:OPTim_NGC5363}
             Optical image (DSS, red filter) of NGC\,5363. Displayed are the central $4\arcmin$ with North up and East to the left. 
              The colour scaling is linear with white corresponding to the median background and black to the $0.01\%$ pixels with the highest intensity.  
           }
\end{figure}
\begin{figure}
   \centering
   \includegraphics[angle=0,height=3.11cm]{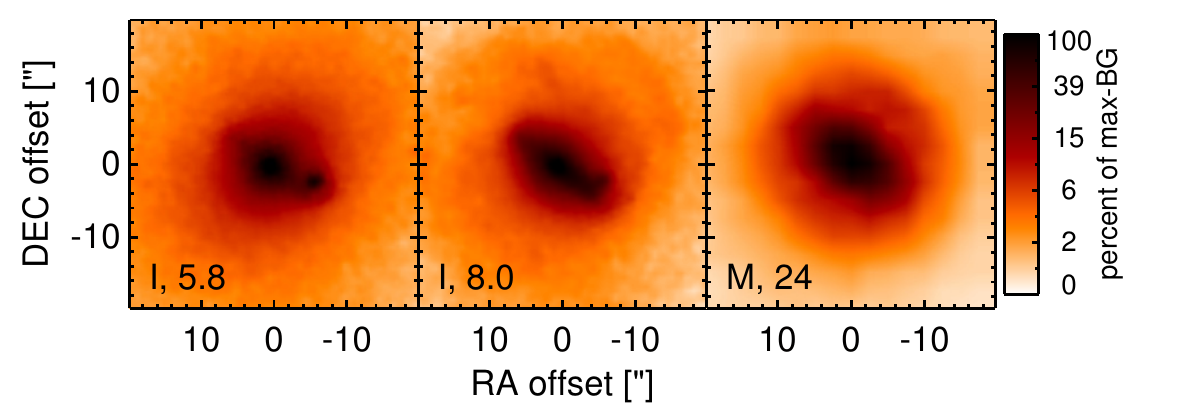}
    \caption{\label{fig:INTim_NGC5363}
             \spitzerr MIR images of NGC\,5363. Displayed are the inner $40\arcsec$ with North up and East to the left. The colour scaling is logarithmic with white corresponding to median background and black to the $0.1\%$ pixels with the highest intensity.
             The label in the bottom left states instrument and central wavelength of the filter in $\mu$m (I: IRAC, M: MIPS). 
           }
\end{figure}
\begin{figure}
   \centering
   \includegraphics[angle=0,height=3.11cm]{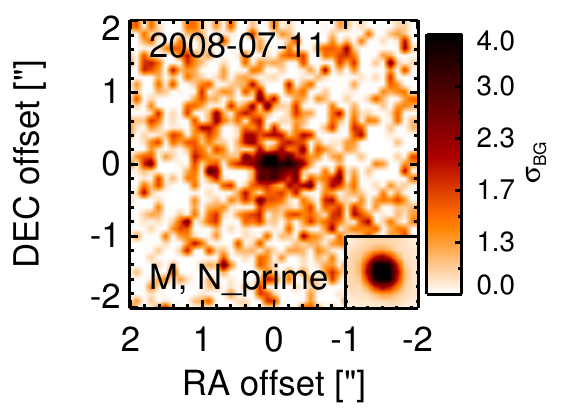}
    \caption{\label{fig:HARim_NGC5363}
             Subarcsecond-resolution MIR images of NGC\,5363 sorted by increasing filter wavelength. 
             Displayed are the inner $4\arcsec$ with North up and East to the left. 
             The colour scaling is logarithmic with white corresponding to median background and black to the $75\%$ of the highest intensity of all images in units of $\sigbg$.
             The inset image shows the central arcsecond of the PSF from the calibrator star, scaled to match the science target.
             The labels in the bottom left state instrument and filter names (C: COMICS, M: Michelle, T: T-ReCS, V: VISIR).
           }
\end{figure}
\begin{figure}
   \centering
   \includegraphics[angle=0,width=8.50cm]{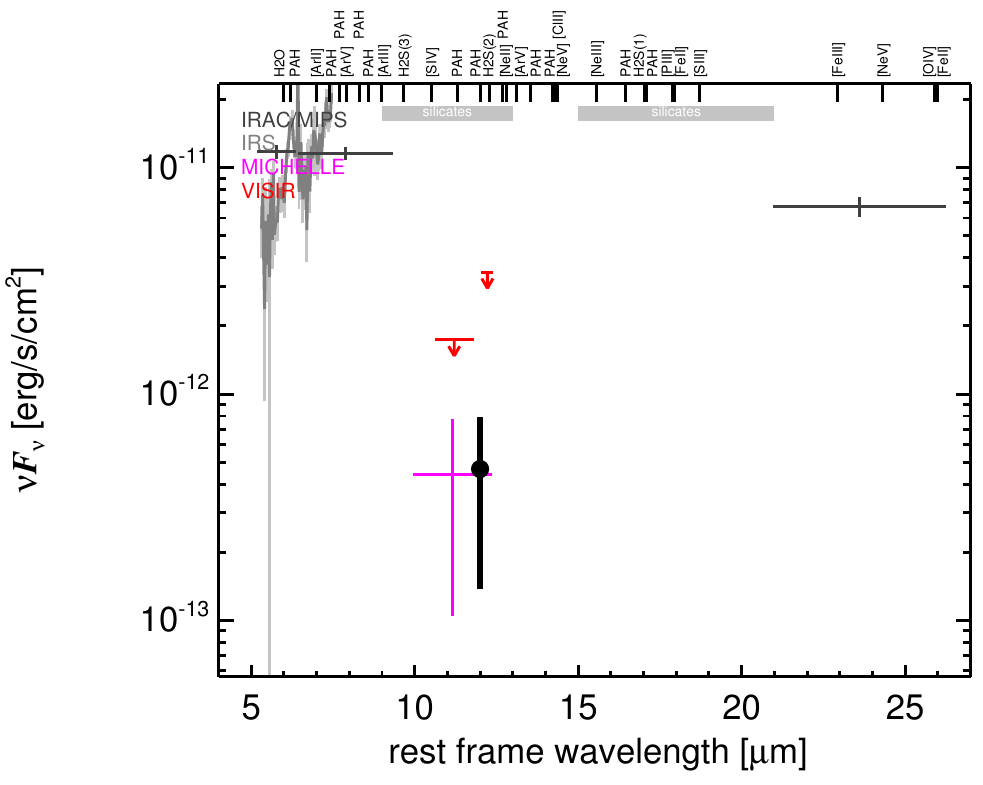}
   \caption{\label{fig:MISED_NGC5363}
      MIR SED of NGC\,5363. The description  of the symbols (if present) is the following.
      Grey crosses and  solid lines mark the \spitzer/IRAC, MIPS and IRS data. 
      The colour coding of the other symbols is: 
      green for COMICS, magenta for Michelle, blue for T-ReCS and red for VISIR data.
      Darker-coloured solid lines mark spectra of the corresponding instrument.
      The black filled circles mark the nuclear 12 and $18\,\mu$m  continuum emission estimate from the data.
      The ticks on the top axis mark positions of common MIR emission lines, while the light grey horizontal bars mark wavelength ranges affected by the silicate 10 and 18$\mu$m features.}
\end{figure}
\clearpage

\twocolumn[\begin{@twocolumnfalse}  
\subsection{NGC\,5427}\label{app:NGC5427}
NGC\,5427 is a face-on grand-design spiral at a distance of $D=$ $29.4\pm 13.5\,$Mpc \citep{terry_local_2002}, which is interacting with NGC\,5426 2.3\arcmin\, to the south.
It hosts a Sy\,2 nucleus \citep{veron-cetty_catalogue_2010}, which appears as a compact radio core \citep{morganti_radio_1999} and is embedded within a compact \oiii emission region, possibly slightly elongated to the north \citep{gonzalez_delgado_h_1997}.
The first ground-based MIR observations are reported in \cite{devereux_spatial_1987}.
NGC\,5427 was also observed with \spitzer/IRAC, IRS and MIPS, and the corresponding images show a compact nucleus embedded within bright spiral-like host emission that blends with the nucleus, in particular in the MIPS 24\,$\mu$m image (see also \citealt{smith_spitzer_2007}).
The IRS LR staring-mode spectrum exhibits strong PAH emission, prominent forbidden emission lines and a red spectral slope in $\nu F_\nu$-space (see also \citealt{pereira-santaella_mid-infrared_2010}).
Thus, the arcsecond-scale MIR SED is  significantly affected by star formation.
The nuclear region of NGC\,5427 was observed with VISIR in the PAH2 filter in 2005 \citep{haas_visir_2007} but no source was detected.
Our derived flux upper limit is much more conservative than the value given by \cite{haas_visir_2007} but still $\sim74\%$ lower than the \spitzerr spectrophotometry.
Therefore, the central $\sim 0.5$\,kpc of NGC\,5427 are dominated by star-formation in the MIR.
\newline\end{@twocolumnfalse}]

\begin{figure}
   \centering
   \includegraphics[angle=0,width=8.500cm]{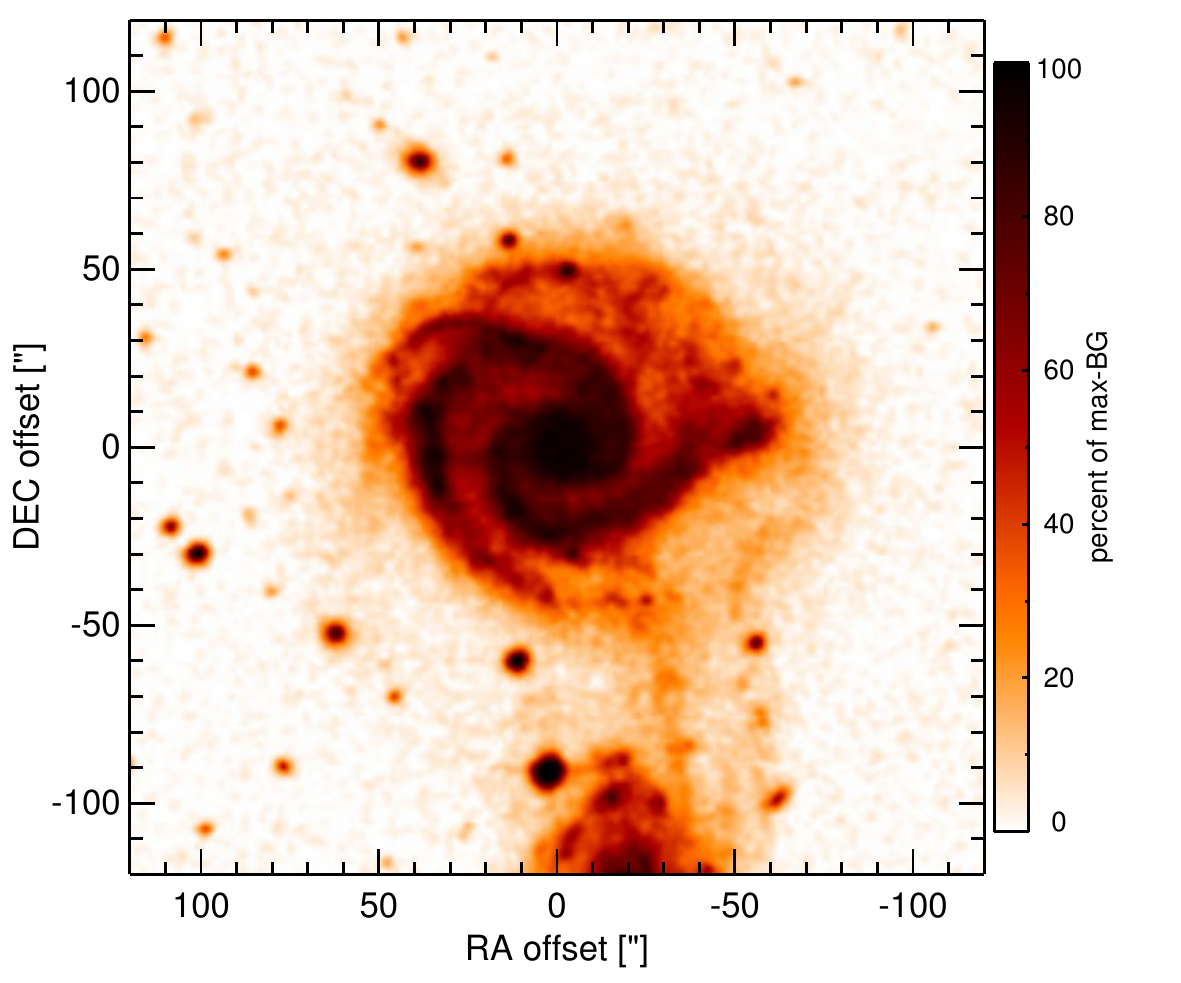}
    \caption{\label{fig:OPTim_NGC5427}
             Optical image (DSS, red filter) of NGC\,5427. Displayed are the central $4\arcmin$ with North up and East to the left. 
              The colour scaling is linear with white corresponding to the median background and black to the $0.01\%$ pixels with the highest intensity.  
           }
\end{figure}
\begin{figure}
   \centering
   \includegraphics[angle=0,height=3.11cm]{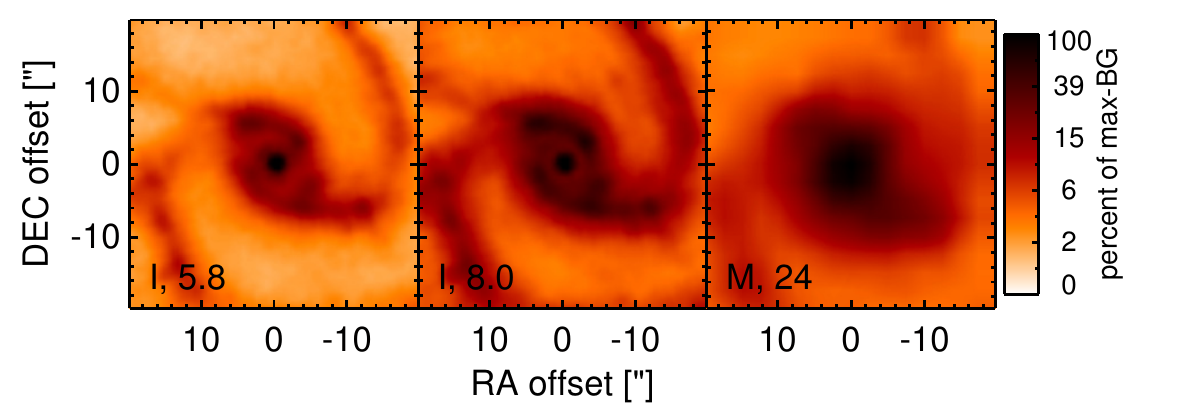}
    \caption{\label{fig:INTim_NGC5427}
             \spitzerr MIR images of NGC\,5427. Displayed are the inner $40\arcsec$ with North up and East to the left. The colour scaling is logarithmic with white corresponding to median background and black to the $0.1\%$ pixels with the highest intensity.
             The label in the bottom left states instrument and central wavelength of the filter in $\mu$m (I: IRAC, M: MIPS). 
           }
\end{figure}
\begin{figure}
   \centering
   \includegraphics[angle=0,width=8.50cm]{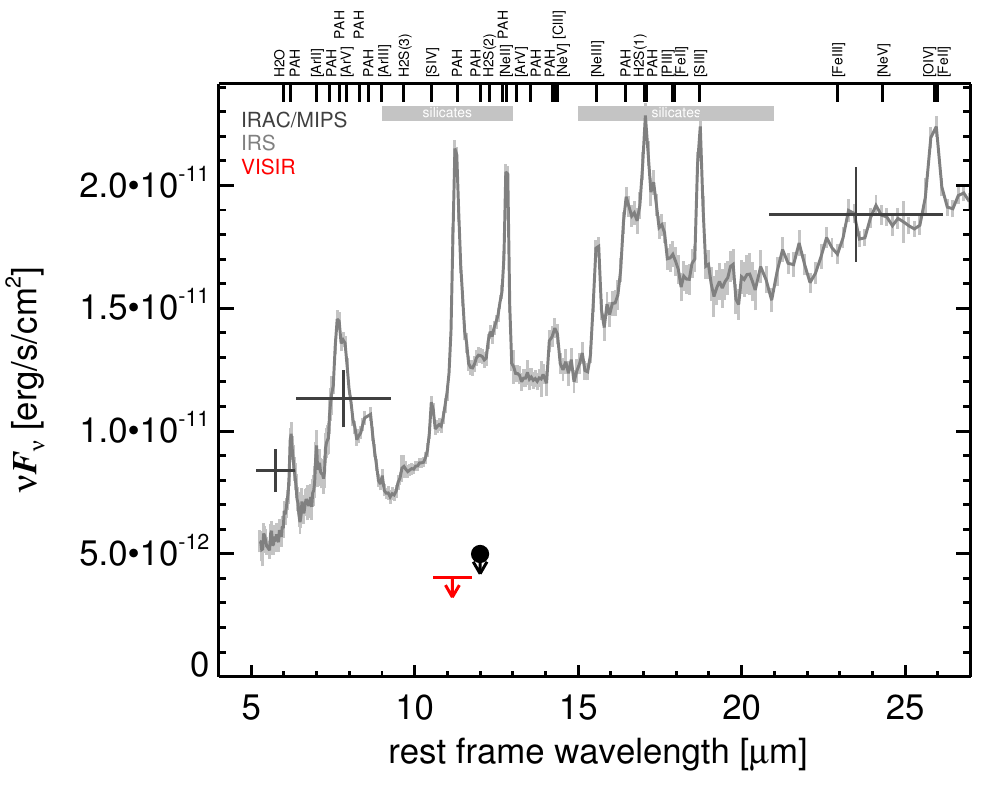}
   \caption{\label{fig:MISED_NGC5427}
      MIR SED of NGC\,5427. The description  of the symbols (if present) is the following.
      Grey crosses and  solid lines mark the \spitzer/IRAC, MIPS and IRS data. 
      The colour coding of the other symbols is: 
      green for COMICS, magenta for Michelle, blue for T-ReCS and red for VISIR data.
      Darker-coloured solid lines mark spectra of the corresponding instrument.
      The black filled circles mark the nuclear 12 and $18\,\mu$m  continuum emission estimate from the data.
      The ticks on the top axis mark positions of common MIR emission lines, while the light grey horizontal bars mark wavelength ranges affected by the silicate 10 and 18$\mu$m features.}
\end{figure}
\clearpage

\twocolumn[\begin{@twocolumnfalse}  
\subsection{NGC\,5506 -- Mrk\,1376}\label{app:NGC5506}
NGC\,5506 is a peculiar edge-on spiral galaxy at a redshift of $z=$ 0.0062 ($D\sim31.6\,$Mpc) with an AGN classified as a Sy\,1.9, a narrow-line Sy\,1 or a Sy\,2 nucleus (see discussion in \citealt{trippe_multi-wavelength_2010}).
Broad emission lines are detected in the near-infrared \citep{nagar_ngc_2002}.
NGC\,5506 is very bright in X-rays and is included in the nine-month BAT AGN sample.
At radio wavelengths, it features a compact nucleus embedded within a halo elongated towards the North with $\sim3\arcsec\sim450\,$pc extent  \citep{wehrle_unusual_1987}.
On milliarcsecond-scales, the nucleus is further resolved into a double source along a PA$\sim70\degree$ \citep{kinney_jet_2000}.
In addition, a nuclear water maser was detected \citep{braatz_discovery_1994}.
The \oiii emission extends over several kiloparsecs and is elongated in the north-south direction \citep{wilson_kinematics_1985}.
The first ground-based MIR observations of NGC\,5506 were performed by \cite{roche_8-13_1984} in 1981 and 1982, followed by \cite{ward_continuum_1987}, \cite{telesco_genesis_1993}, and \cite{maiolino_new_1995}.
Additional low-angular resolution spectrophotometry was obtained with \isoo \citep{rigopoulou_large_1999,tran_isocam-cvf_2001}.
The first subarcsecond $N$-band images were obtained with Palomar 5\,m/MIRLIN in 2000 \citep{gorjian_10_2004}, followed by $N$-band spectrophotometry with ESO 3.6\,m/TIMMI2 in 2001 and 2002 \citep{siebenmorgen_isocam_2004,raban_core_2008}. 
An unresolved nucleus was detected in all these images, but the 11.9\,$\mu$m image from \cite{raban_core_2008} indicates  extended emission towards the north-east.
The \spitzer/IRAC and MIPS images are dominated by a bright compact nucleus with  extended host emission visible in the IRAC images.
The nucleus is saturated in the PBCD IRAC images and, thus, not analysed here (but see \citealt{gallimore_infrared_2010}).
The \spitzer/IRS LR staring-mode spectrum exhibits deep silicate 10$\,\mu$m and weak silicate 18\,$\mu$m absorption, very weak PAH emission and a flat spectral slope in $\nu F_\nu$-space (see also \citealt{shi_9.7_2006,wu_spitzer/irs_2009,tommasin_spitzer-irs_2010,gallimore_infrared_2010,mullaney_defining_2011}).
Therefore, contribution of star formation to the arcsecond-scale MIR SED is probably minor.
The SED flux levels agree with the historical data within the uncertainties, which are dominated by the steep $N$-band slopes caused by the silicate absorption.
The nuclear region of NGC\,5506 was observed with T-ReCS in $N$-band imaging and spectroscopic modes in 2004 \citep{roche_silicate_2007}, with Michelle in the N' and Qa filters \citep{ramos_almeida_infrared_2009}, and with VISIR in four $N$ and one $Q$ band filters in 2005, 2006 and 2010 (partly published in \citealt{haas_visir_2007,reunanen_vlt_2010}).
In all cases, a compact nucleus without further host emission was detected.
The nucleus appears marginally resolved but is not intrinsically elongated in the sharpest images (FWHM$\sim 0.4\arcsec\sim 60\,$pc). The only exception might be the Si6 image, for which no matching standard star observation could be found.
Therefore, we treat the Si6 measurement as an upper limit for the unresolved nuclear emission.
In neither of the images did we clearly detect the extended emission seen in the TIMMI2 11.9\,$\mu$m, although some images may hint towards a slight extension in eastern direction.
Our nuclear photometry is generally consistent with \cite{ramos_almeida_infrared_2009} and \cite{reunanen_vlt_2010}, and on average $\sim 26\%$ lower than the \spitzerr spectrophotometry.
Note that the Michelle measurements are significantly higher than the other subarcsecond measurements because they were obtained in substandard seeing conditions.
The nuclear MIR SED indicates the same silicate 10\,$\mu$m absorption feature as seen in the arcsecond-scale MIR SED.
Note that the T-ReCS $N$-band LR spectrum  as published in \cite{gonzalez-martin_dust_2013} exhibits $\sim25\%$ higher flux levels than our photometry for unknown reasons and agrees with the IRS spectrum within the uncertainties.
The difference between the arcsecond-scale and nuclear MIR SEDs is presumably caused by weak circum-nuclear star formation.
The nuclear MIR emission of NGC\,5506 could be further resolved with MIDI interferometric observations and was modelled as two components, one extending over $\gtrsim10\,$pc, and a brighter one remaining unresolved ($< 1$\,pc; \citealt{burtscher_diversity_2013}).
\newline\end{@twocolumnfalse}]

\begin{figure}
   \centering
   \includegraphics[angle=0,width=8.500cm]{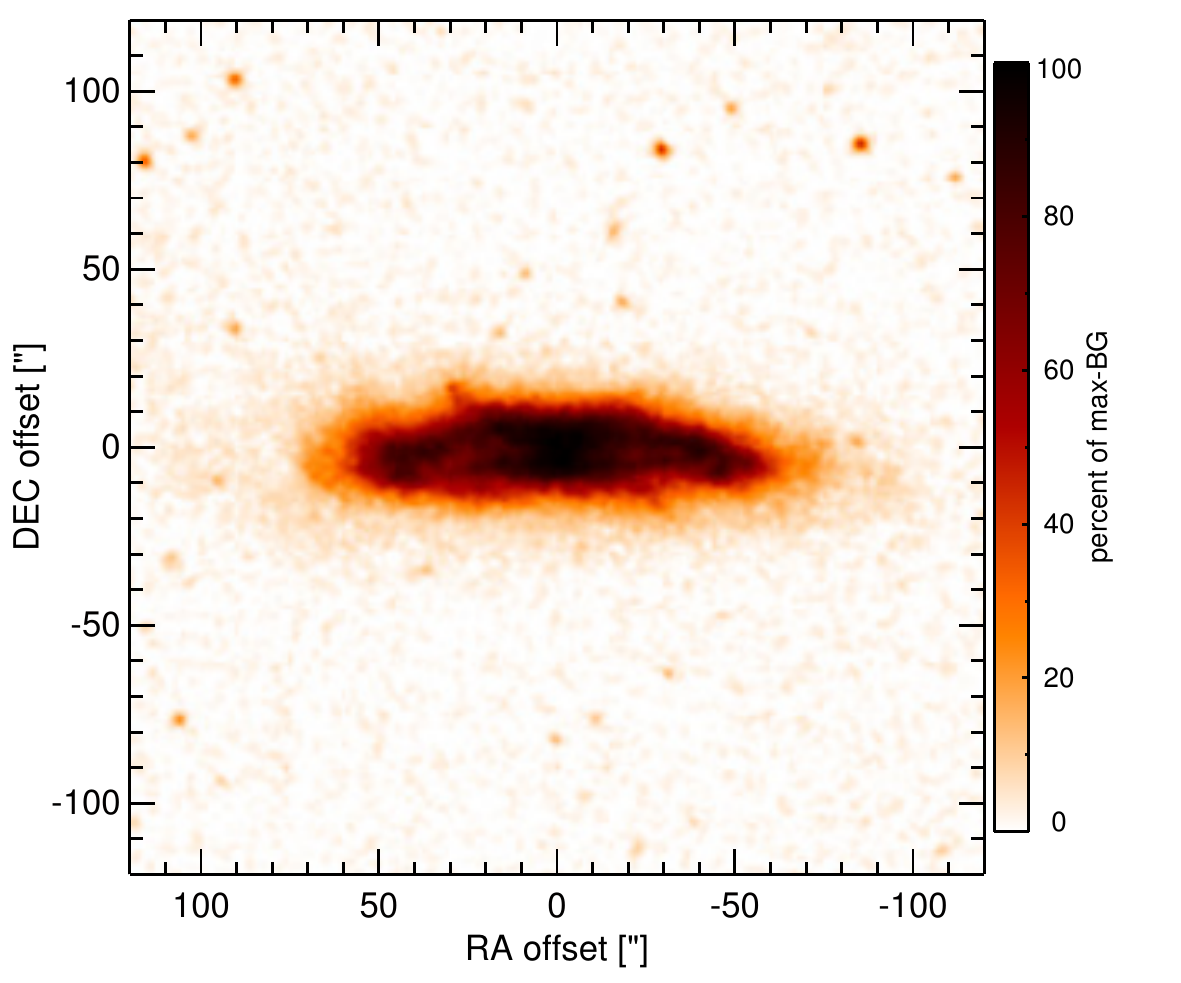}
    \caption{\label{fig:OPTim_NGC5506}
             Optical image (DSS, red filter) of NGC\,5506. Displayed are the central $4\arcmin$ with North up and East to the left. 
              The colour scaling is linear with white corresponding to the median background and black to the $0.01\%$ pixels with the highest intensity.  
           }
\end{figure}
\begin{figure}
   \centering
   \includegraphics[angle=0,height=3.11cm]{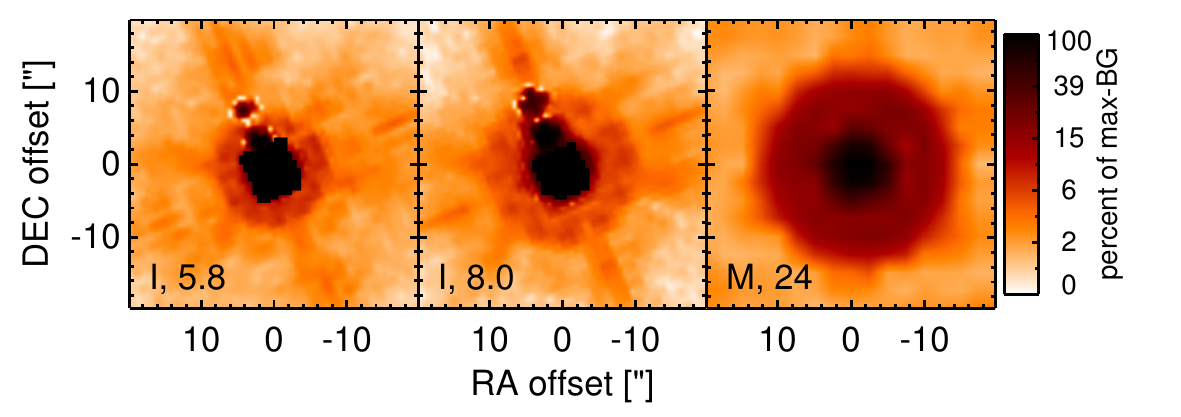}
    \caption{\label{fig:INTim_NGC5506}
             \spitzerr MIR images of NGC\,5506. Displayed are the inner $40\arcsec$ with North up and East to the left. The colour scaling is logarithmic with white corresponding to median background and black to the $0.1\%$ pixels with the highest intensity.
             The label in the bottom left states instrument and central wavelength of the filter in $\mu$m (I: IRAC, M: MIPS).
             Note that the apparent off-nuclear compact sources in the IRAC 5.8 and $8.0\,\mu$m images are instrumental artefacts.
           }
\end{figure}
\begin{figure}
   \centering
   \includegraphics[angle=0,width=8.500cm]{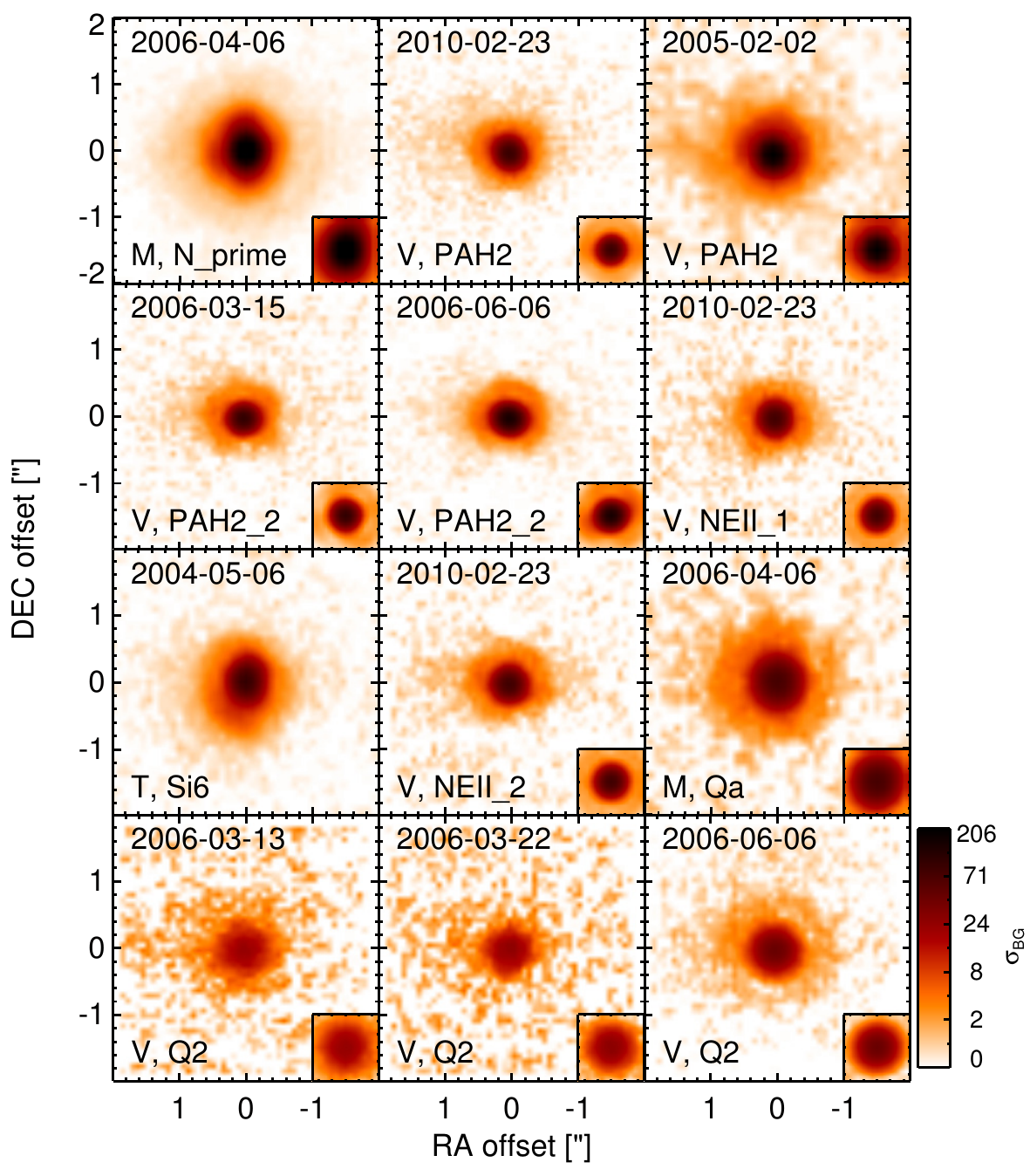}
    \caption{\label{fig:HARim_NGC5506}
             Subarcsecond-resolution MIR images of NGC\,5506 sorted by increasing filter wavelength. 
             Displayed are the inner $4\arcsec$ with North up and East to the left. 
             The colour scaling is logarithmic with white corresponding to median background and black to the $75\%$ of the highest intensity of all images in units of $\sigbg$.
             The inset image shows the central arcsecond of the PSF from the calibrator star, scaled to match the science target.
             The labels in the bottom left state instrument and filter names (C: COMICS, M: Michelle, T: T-ReCS, V: VISIR).
           }
\end{figure}
\begin{figure}
   \centering
   \includegraphics[angle=0,width=8.50cm]{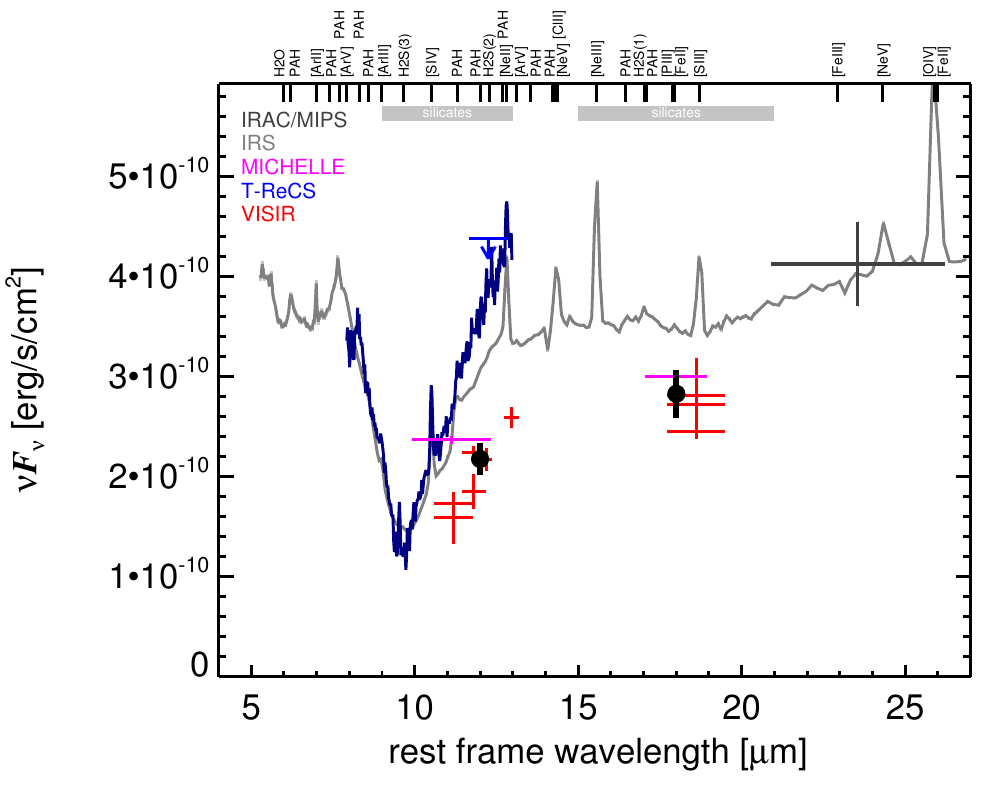}
   \caption{\label{fig:MISED_NGC5506}
      MIR SED of NGC\,5506. The description  of the symbols (if present) is the following.
      Grey crosses and  solid lines mark the \spitzer/IRAC, MIPS and IRS data. 
      The colour coding of the other symbols is: 
      green for COMICS, magenta for Michelle, blue for T-ReCS and red for VISIR data.
      Darker-coloured solid lines mark spectra of the corresponding instrument.
      The black filled circles mark the nuclear 12 and $18\,\mu$m  continuum emission estimate from the data.
      The ticks on the top axis mark positions of common MIR emission lines, while the light grey horizontal bars mark wavelength ranges affected by the silicate 10 and 18$\mu$m features.}
\end{figure}
\clearpage

\twocolumn[\begin{@twocolumnfalse}  
\subsection{NGC\,5548 -- Mrk\,1509}\label{app:NGC5548}
NGC\,5548 is a face-on spiral galaxy at a redshift of $z=$ 0.0172 ($D\sim80.7\,$Mpc) with a well-studied Sy\,1.5 nucleus \citep{veron-cetty_catalogue_2010}.
It is highly variable in optical and X-rays (e.g. \citealt{peterson_steps_2002}) and belongs to the nine-month BAT AGN sample.
A compact radio-core with biconical structure was detected along a PA$\sim160\degree$ and $\sim15\arcsec$ extent \citep{wilson_radio_1982-1,ho_radio_2001}.
The \oiii emission extends to kiloparsec scales in north-south direction, roughly aligned with the radio emission \citep{wilson_kinematics_1989,schmitt_hubble_2003}.
The first MIR observation of NGC\,5548 were carried out by \cite{kleinmann_infrared_1970}, followed by \cite{rieke_infrared_1972,rieke_infrared_1978}, and \citep{ward_continuum_1987}.
A first subarcsecond-resolution image was obtained with Palomar 5\,m/MIRLIN in 2000 \citep{gorjian_10_2004}.
NGC\,5548 was also observed with \isoo \citep{clavel_2.5-11_2000,ramos_almeida_mid-infrared_2007} and \spitzer/IRAC, IRS and MIPS.
The corresponding IRAC and MIPS images are dominated by a bright nuclear source, with the extended host emission being faintly visible in the 5.8 and $8.0\,\mu$m images.
Our nuclear IRAC 5.8 and $8.0\,\mu$m photometry is consistent with \cite{gallimore_infrared_2010}.
The IRS LR staring-mode spectrum shows silicate 10 and 18\,$\mu$m emission, a prominent PAH 11.3\,$\mu$m feature, and an emission peak at $\sim 17\,\mu$m in $\nu F_\nu$-space (see also \citealt{buchanan_spitzer_2006,shi_9.7_2006,wu_spitzer/irs_2009,tommasin_spitzer-irs_2010,gallimore_infrared_2010}).
Thus, the arcsecond-scale MIR SED  appears to be AGN-dominated.
The nuclear region of NGC\,5548 was observed with VISIR in two narrow $N$-band filters in 2010 \citep{kishimoto_mapping_2011}.
The nucleus appears to be elongated in north-south direction in both images. However, they were taken at a high airmass without matched standard star observations.
Therefore, at least another epoch of MIR subarcsecond imaging under good ambient conditions is required to verify this extension, so that we classify the nucleus as possibly extended.
Owing to this possible extension, the corresponding nuclear VISIR photometry is very uncertain and $\sim43\%$ lower than the \spitzerr spectrophotometry.
In addition, all nuclear $N$-band flux measurements over the last $\sim40$ years are consistent with no flux variations within the uncertainties. The only exception is the MIRLIN N flux, which is higher than the long-term average.
\newline\end{@twocolumnfalse}]

\begin{figure}
   \centering
   \includegraphics[angle=0,width=8.500cm]{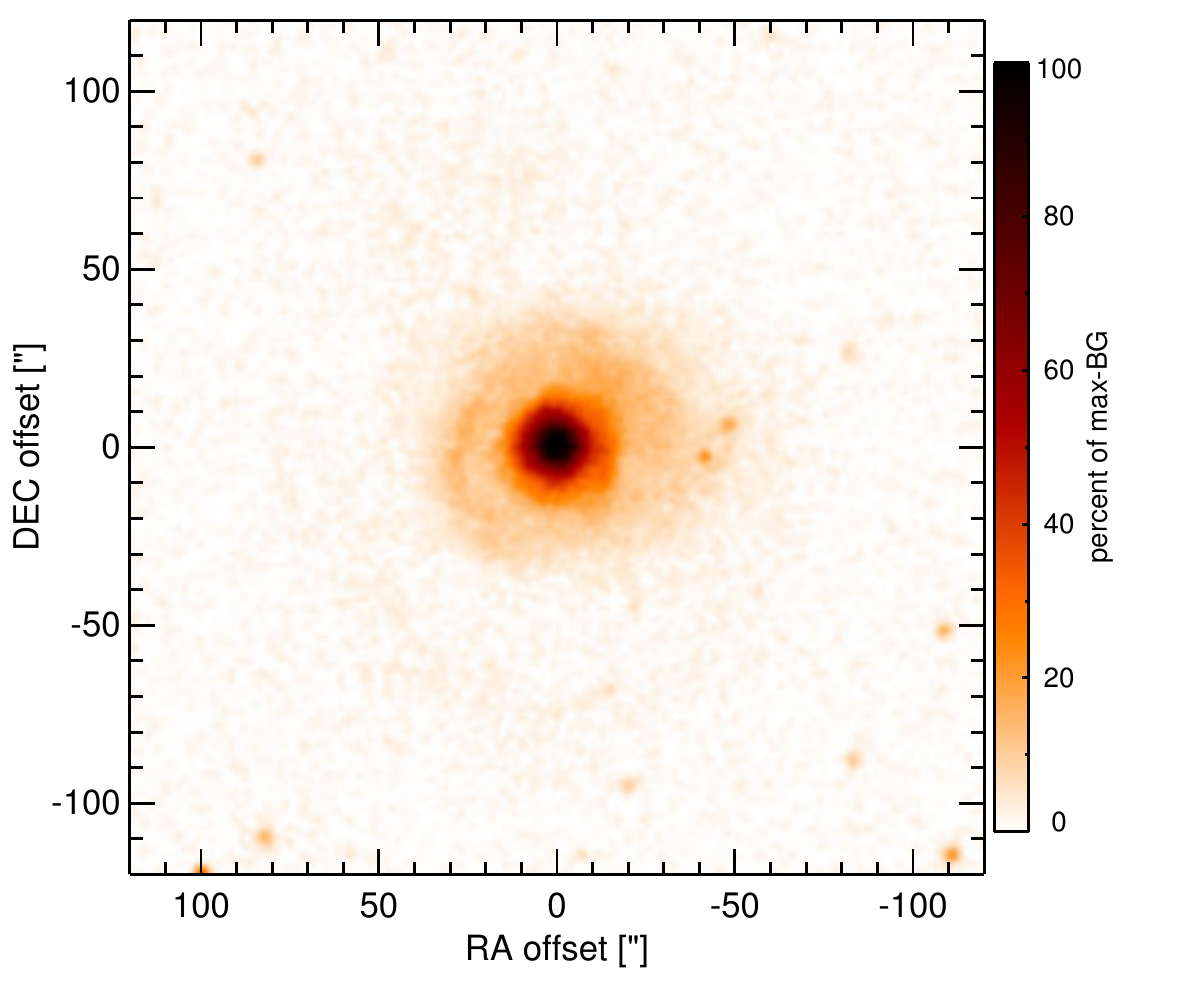}
    \caption{\label{fig:OPTim_NGC5548}
             Optical image (DSS, red filter) of NGC\,5548. Displayed are the central $4\arcmin$ with North up and East to the left. 
              The colour scaling is linear with white corresponding to the median background and black to the $0.01\%$ pixels with the highest intensity.  
           }
\end{figure}
\begin{figure}
   \centering
   \includegraphics[angle=0,height=3.11cm]{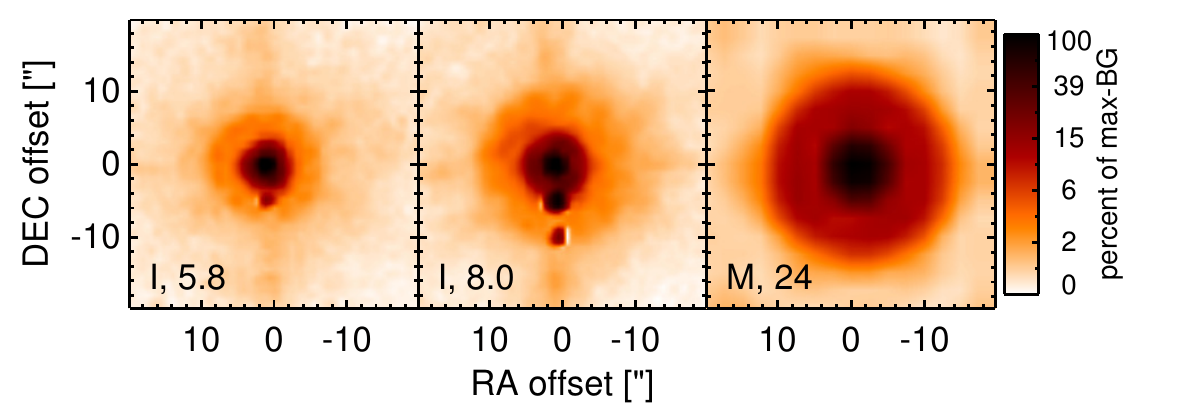}
    \caption{\label{fig:INTim_NGC5548}
             \spitzerr MIR images of NGC\,5548. Displayed are the inner $40\arcsec$ with North up and East to the left. The colour scaling is logarithmic with white corresponding to median background and black to the $0.1\%$ pixels with the highest intensity.
             The label in the bottom left states instrument and central wavelength of the filter in $\mu$m (I: IRAC, M: MIPS). 
             Note that the apparent off-nuclear compact sources in the IRAC 5.8 and $8.0\,\mu$m images are instrumental artefacts.
           }
\end{figure}
\begin{figure}
   \centering
   \includegraphics[angle=0,height=3.11cm]{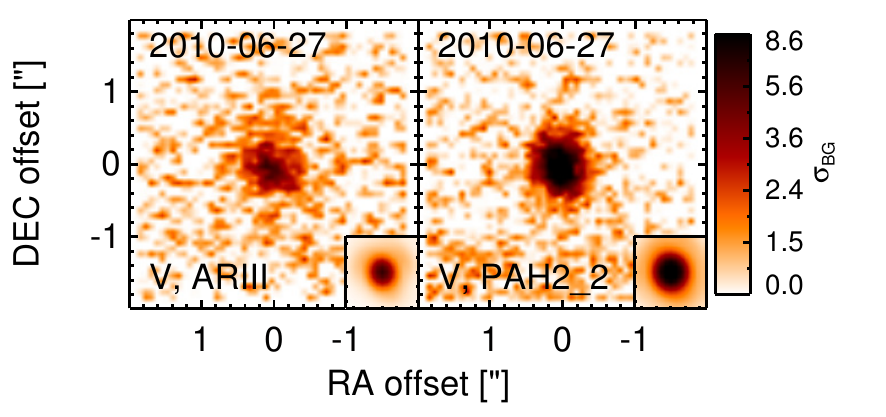}
    \caption{\label{fig:HARim_NGC5548}
             Subarcsecond-resolution MIR images of NGC\,5548 sorted by increasing filter wavelength. 
             Displayed are the inner $4\arcsec$ with North up and East to the left. 
             The colour scaling is logarithmic with white corresponding to median background and black to the $75\%$ of the highest intensity of all images in units of $\sigbg$.
             The inset image shows the central arcsecond of the PSF from the calibrator star, scaled to match the science target.
             The labels in the bottom left state instrument and filter names (C: COMICS, M: Michelle, T: T-ReCS, V: VISIR).
           }
\end{figure}
\begin{figure}
   \centering
   \includegraphics[angle=0,width=8.50cm]{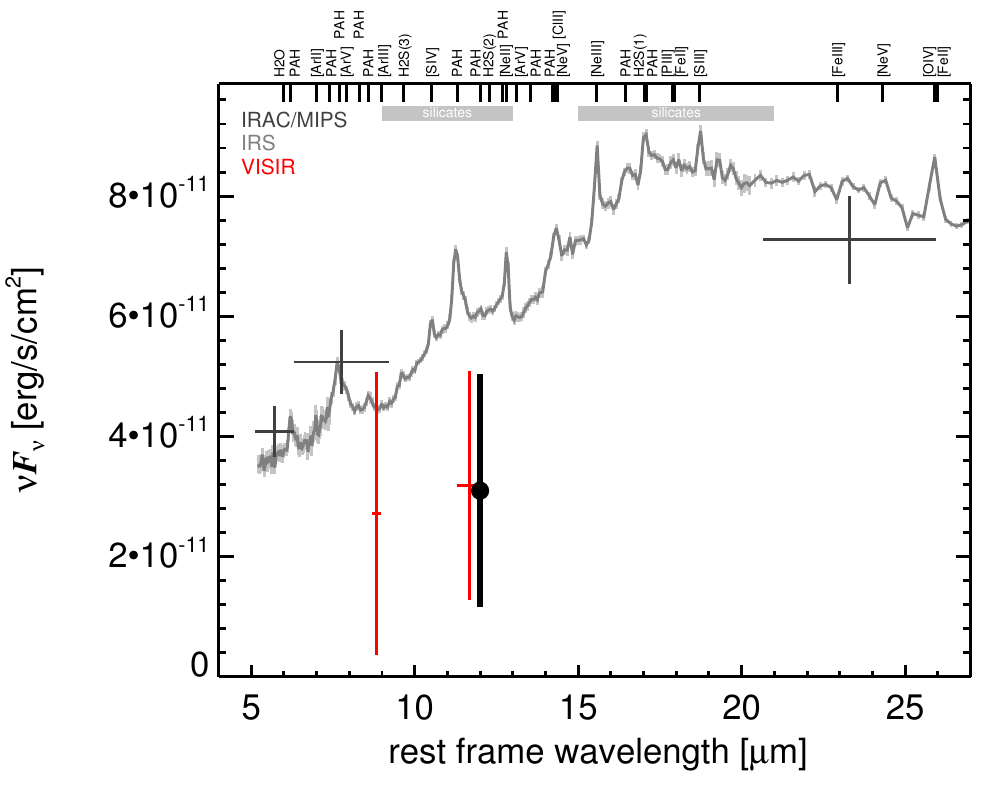}
   \caption{\label{fig:MISED_NGC5548}
      MIR SED of NGC\,5548. The description  of the symbols (if present) is the following.
      Grey crosses and  solid lines mark the \spitzer/IRAC, MIPS and IRS data. 
      The colour coding of the other symbols is: 
      green for COMICS, magenta for Michelle, blue for T-ReCS and red for VISIR data.
      Darker-coloured solid lines mark spectra of the corresponding instrument.
      The black filled circles mark the nuclear 12 and $18\,\mu$m  continuum emission estimate from the data.
      The ticks on the top axis mark positions of common MIR emission lines, while the light grey horizontal bars mark wavelength ranges affected by the silicate 10 and 18$\mu$m features.}
\end{figure}
\clearpage

\twocolumn[\begin{@twocolumnfalse}  
\subsection{NGC\,5643}\label{app:NGC5643}
NGC\,5643 is a face-on barred spiral galaxy at a distance of $D=$ $16.9 \pm 3.4\,$Mpc \citep{tully_nearby_1988} with a Sy\,2 nucleus \citep{veron-cetty_catalogue_2010}.
It features a compact radio core with two-sided, kiloparsec-scale lobes in east-west direction (PA$\sim87\degree$; \citealt{morris_velocity_1985, leipski_radio_2006}) and a cospatial one-sided, kiloparsec-scale NLR cone (PA$\sim84\degree$; e.g. \citealt{simpson_one-sided_1997}).
In addition, water maser emission was detected in NGC\,5643 \citep{greenhill_discovery_2003}.
The first ground-based MIR observations of NGC\,5643 were performed by \cite{roche_atlas_1991}, while first arcsecond-resolution $N$-band data obtained with ESO 3.6\,m/TIMMI2 were presented in \cite{siebenmorgen_mid-infrared_2004}.
NGC\,5643 was also observed with \isoo \citep{rigopoulou_large_1999,siebenmorgen_mid-infrared_2004} and \spitzer/IRAC, IRS and MIPS.
The corresponding IRAC and MIPS images show a compact MIR nucleus embedded within the spiral-like host emission. 
The IRS LR staring-mode spectrum exhibits silicate 10\,$\mu$m absorption, prominent PAH emission, and a steep red spectral slope in $\nu F_\nu$-space (see also \citealt{shi_9.7_2006,goulding_towards_2009}).
Thus, the arcsecond-scale MIR SED is significantly affected by star formation.
We observed the nuclear region of NGC\,5643 with VISIR in 2009 and obtained three narrow $N$-band images and a LR $N$-band spectrum \citep{honig_dusty_2010-1}.
In addition, a T-ReCS LR $N$-band spectrum is presented in \cite{gonzalez-martin_dust_2013}.
All three VISIR images show a compact nucleus as well as additional emission from the host.
The nucleus appears elongated with a PA$\sim50\degree$ (FWHM(major axis)$\sim 0.43\arcsec\sim43\,$pc).
However, at least a second epoch of subarcsecond MIR imaging is required to confirm this extension.
Our remeasured nuclear ARIII and PAH2\_2 fluxes are unchanged with respect to \cite{honig_dusty_2010-1}.
The nuclear photometry and spectroscopy is in general consistent, except for the shortest wavelengths where deviations occur.
The resulting nuclear MIR SED has on average $\sim 34\%$ lower flux levels than the \spitzerr spectrophotometry and exhibits much weaker PAH emission.
Therefore, we attribute the flux difference to circum-nuclear star formation as already demonstrated in \cite{honig_dusty_2010-1}.
The silicate 10\,$\mu$m absorption feature possesses the same depth at both spatial scales, which we intrepret as generic to the (projected) central $\sim35\,$pc of NGC\,5643.
\newline\end{@twocolumnfalse}]

\begin{figure}
   \centering
   \includegraphics[angle=0,width=8.500cm]{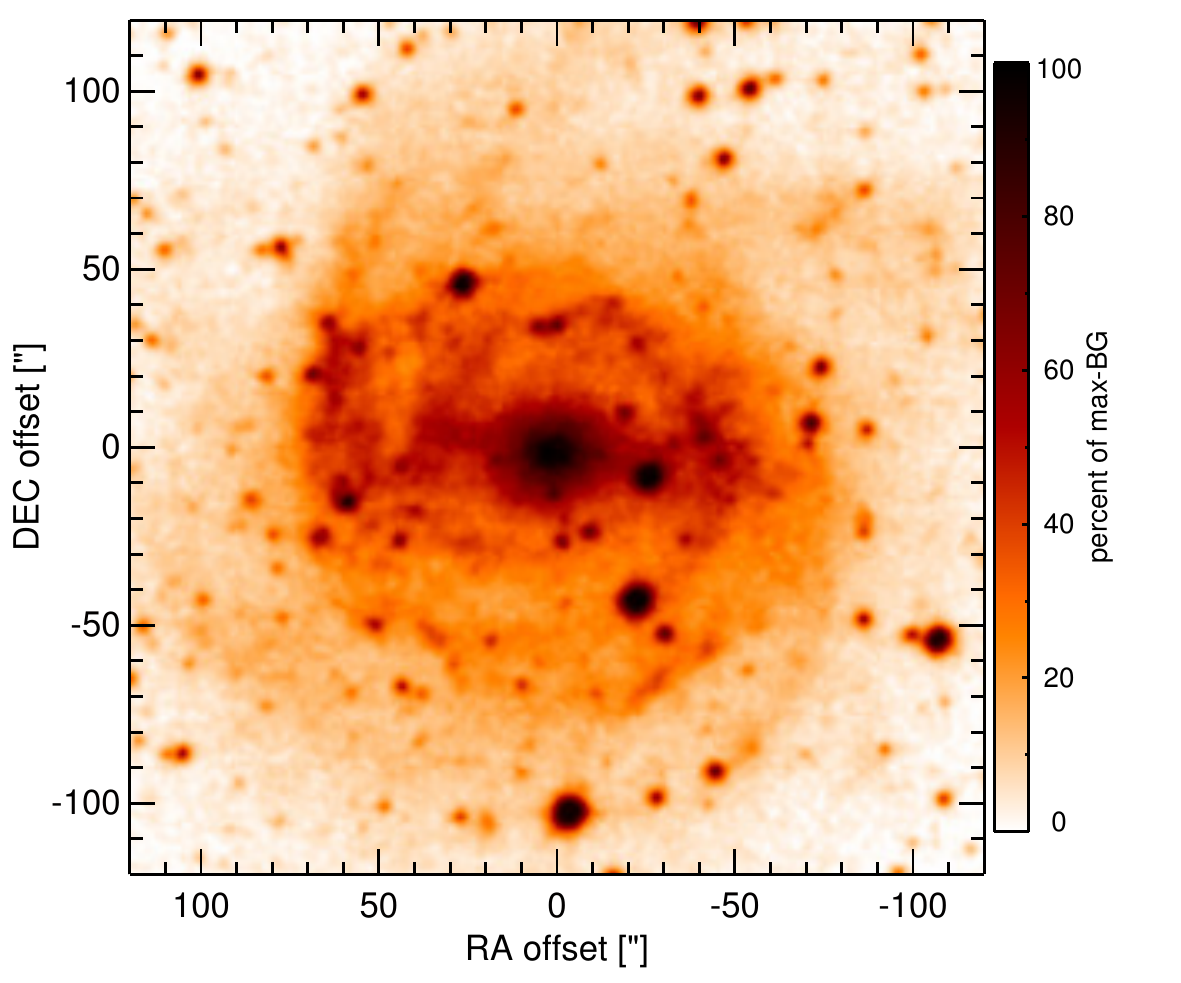}
    \caption{\label{fig:OPTim_NGC5643}
             Optical image (DSS, red filter) of NGC\,5643. Displayed are the central $4\arcmin$ with North up and East to the left. 
              The colour scaling is linear with white corresponding to the median background and black to the $0.01\%$ pixels with the highest intensity.  
           }
\end{figure}
\begin{figure}
   \centering
   \includegraphics[angle=0,height=3.11cm]{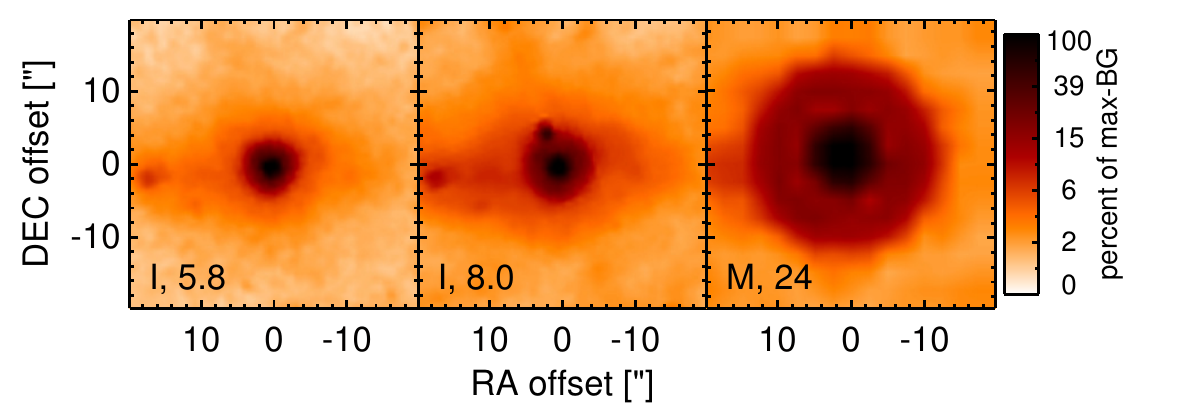}
    \caption{\label{fig:INTim_NGC5643}
             \spitzerr MIR images of NGC\,5643. Displayed are the inner $40\arcsec$ with North up and East to the left. The colour scaling is logarithmic with white corresponding to median background and black to the $0.1\%$ pixels with the highest intensity.
             The label in the bottom left states instrument and central wavelength of the filter in $\mu$m (I: IRAC, M: MIPS). 
             Note that the apparent off-nuclear compact source in the IRAC $8.0\,\mu$m image is an instrumental artefact.
           }
\end{figure}
\begin{figure}
   \centering
   \includegraphics[angle=0,height=3.11cm]{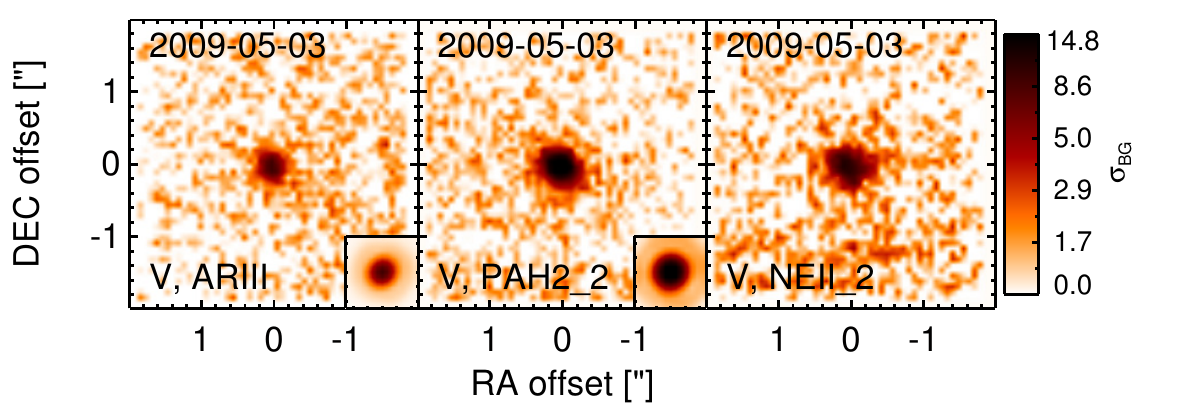}
    \caption{\label{fig:HARim_NGC5643}
             Subarcsecond-resolution MIR images of NGC\,5643 sorted by increasing filter wavelength. 
             Displayed are the inner $4\arcsec$ with North up and East to the left. 
             The colour scaling is logarithmic with white corresponding to median background and black to the $75\%$ of the highest intensity of all images in units of $\sigbg$.
             The inset image shows the central arcsecond of the PSF from the calibrator star, scaled to match the science target.
             The labels in the bottom left state instrument and filter names (C: COMICS, M: Michelle, T: T-ReCS, V: VISIR).
           }
\end{figure}
\begin{figure}
   \centering
   \includegraphics[angle=0,width=8.50cm]{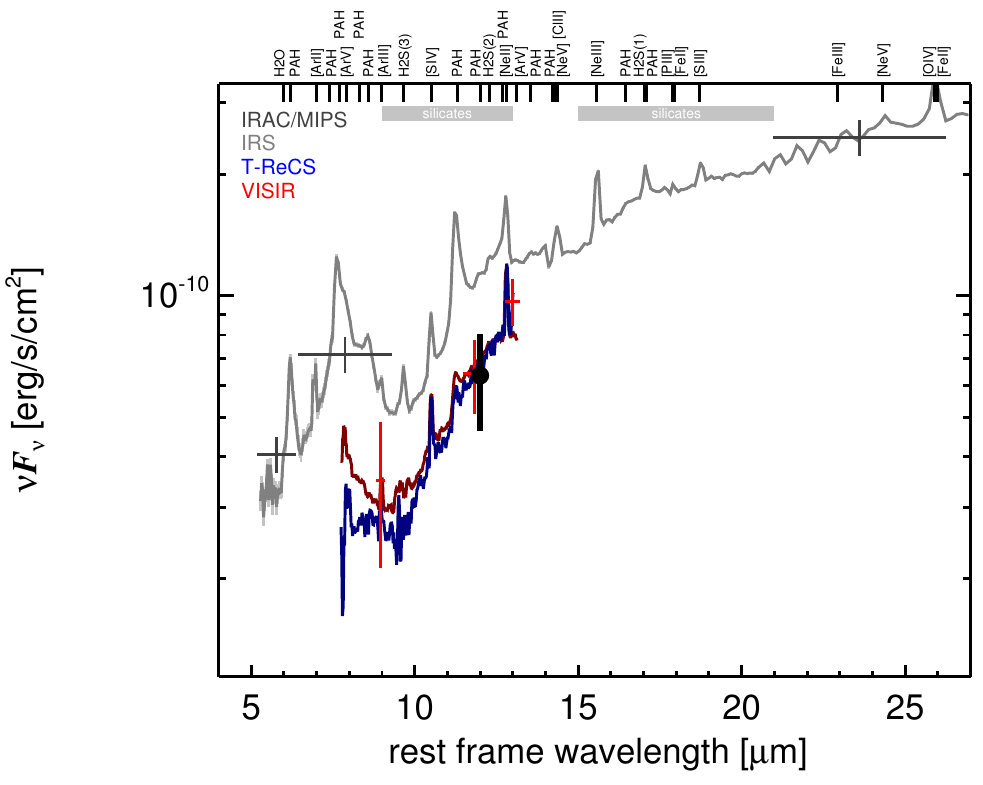}
   \caption{\label{fig:MISED_NGC5643}
      MIR SED of NGC\,5643. The description  of the symbols (if present) is the following.
      Grey crosses and  solid lines mark the \spitzer/IRAC, MIPS and IRS data. 
      The colour coding of the other symbols is: 
      green for COMICS, magenta for Michelle, blue for T-ReCS and red for VISIR data.
      Darker-coloured solid lines mark spectra of the corresponding instrument.
      The black filled circles mark the nuclear 12 and $18\,\mu$m  continuum emission estimate from the data.
      The ticks on the top axis mark positions of common MIR emission lines, while the light grey horizontal bars mark wavelength ranges affected by the silicate 10 and 18$\mu$m features.}
\end{figure}
\clearpage

\twocolumn[\begin{@twocolumnfalse}  
\subsection{NGC\,5728}\label{app:NGC5728}
NGC\,5728 is an inclined barred spiral galaxy at a redshift of $z=$ 0.0094 ($D\sim 45.4\,$Mpc) hosting a Sy\,1.9 nucleus \citep{pecontal_observation_1990}, which is commonly referred to as a Sy\,2 type in the literature (e.g. \citealt{tueller_swift_2008}).
It belongs to the nine-month BAT AGN sample.
The nucleus features a compact radio core with one-sided extension up to $\sim5\arcsec\sim1\,$kpc to the north-west (PA$\sim-53\degree$; \citealt{schommer_ionized_1988}) and a cospatial biconical kiloparsec-scale NLR  (PA$\sim130\degree$; \citealt{pogge_circumnuclear_1989,wilson_ionization_1993}).
In addition, disc-like water maser emission was detected \citep{braatz_green_2004}.
After being detected in the MIR with \irass for the first time, NGC\,5728 was followed up with \isoo \citep{clavel_2.5-11_2000,ramos_almeida_mid-infrared_2007} and \spitzer/IRAC, IRS and MIPS.
The corresponding IRAC and MIPS images show a compact nucleus embedded within bright disc-like emission with $\sim13\arcsec\sim2.8\,$kpc diameter and fainter spiral-like host emission on larger scales.
The IRS LR staring-mode spectrum exhibits deep silicate 10\,$\mu$m absorption, prominent PAH emission, forbidden emission lines, and a red spectral slope in $\nu F_\nu$-space (see also \citealt{weaver_mid-infrared_2010,pereira-santaella_mid-infrared_2010}).
Thus, the arcsecond-scale MIR SED  appears to be star-formation dominated.
The nuclear region of NGC\,5728 was observed with T-ReCS in the Si2 and Qa filters in 2005 \citep{ramos_almeida_infrared_2009}, and with VISIR in five narrow $N$-band filters in 2008 \citep{gandhi_resolving_2009} and 2010 (this work).
A compact nucleus without further host emission was detected in all images, except PAH1 and SIV.
The nucleus appears marginally resolved and roundish in the images, but the nuclear FWHM of the individual images do not match.
We measure only the unresolved nuclear fluxes, which are on average $\sim30\%$ lower than the nuclear fluxes published in \cite{ramos_almeida_infrared_2009} and \cite{gandhi_resolving_2009} and also $\sim 62\%$ lower than the \spitzerr spectrophotometry.
The nuclear MIR SED indicates a silicate 10\,$\mu$m absorption depth similar to the \spitzerr data, but the PAH features are much.
However, it might still be significantly affected by star-formation caused MIR emission.
We conclude that the silicate absorption is produced in the projected central $\sim75\,$pc of NGC\,5728.
\newline\end{@twocolumnfalse}]

\begin{figure}
   \centering
   \includegraphics[angle=0,width=8.500cm]{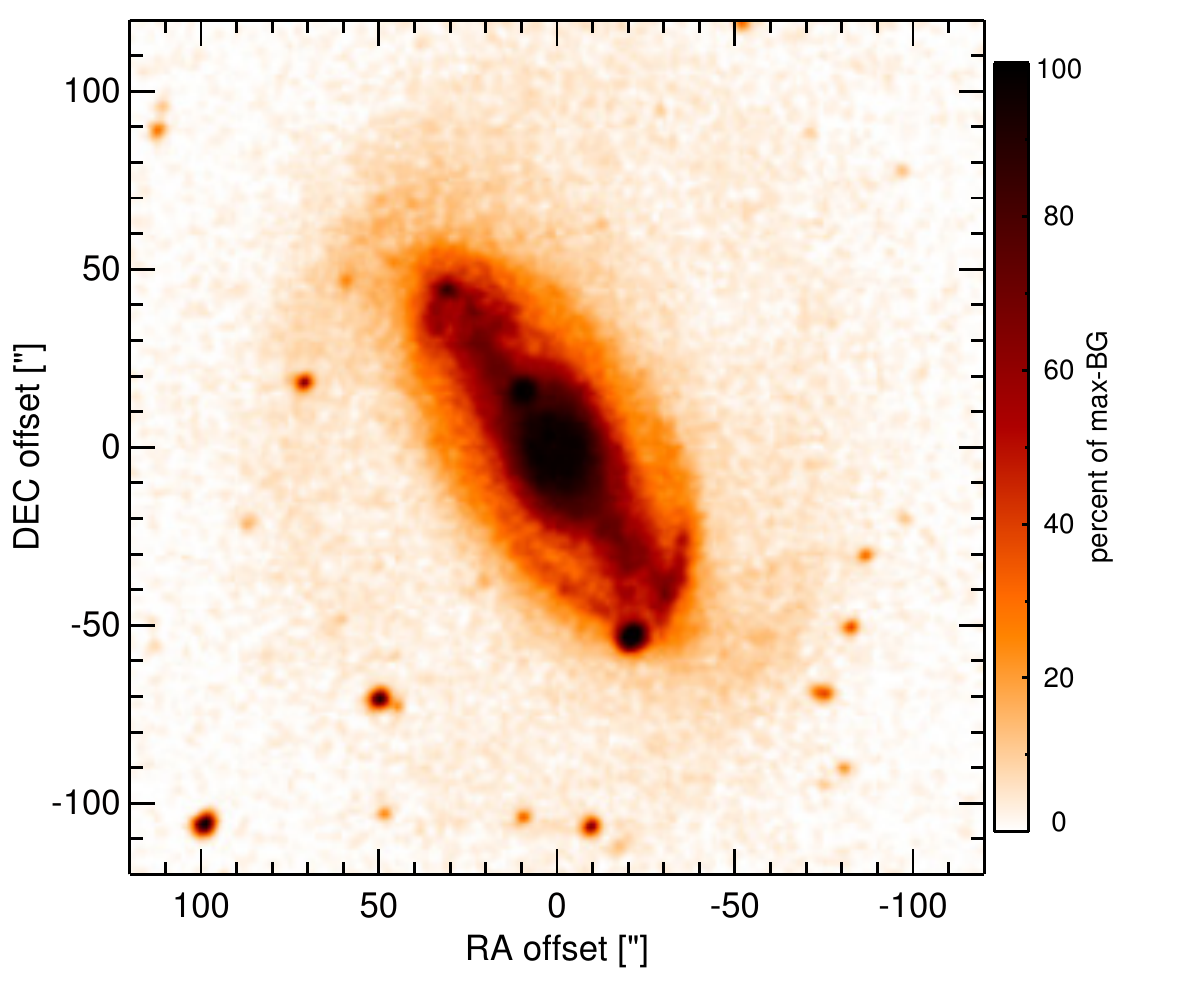}
    \caption{\label{fig:OPTim_NGC5728}
             Optical image (DSS, red filter) of NGC\,5728. Displayed are the central $4\arcmin$ with North up and East to the left. 
              The colour scaling is linear with white corresponding to the median background and black to the $0.01\%$ pixels with the highest intensity.  
           }
\end{figure}
\begin{figure}
   \centering
   \includegraphics[angle=0,height=3.11cm]{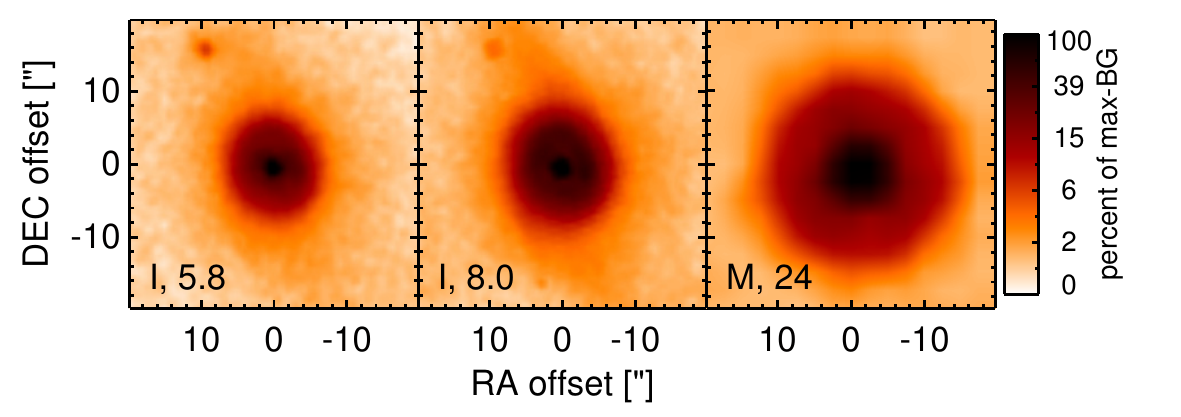}
    \caption{\label{fig:INTim_NGC5728}
             \spitzerr MIR images of NGC\,5728. Displayed are the inner $40\arcsec$ with North up and East to the left. The colour scaling is logarithmic with white corresponding to median background and black to the $0.1\%$ pixels with the highest intensity.
             The label in the bottom left states instrument and central wavelength of the filter in $\mu$m (I: IRAC, M: MIPS). 
           }
\end{figure}
\begin{figure}
   \centering
   \includegraphics[angle=0,width=8.500cm]{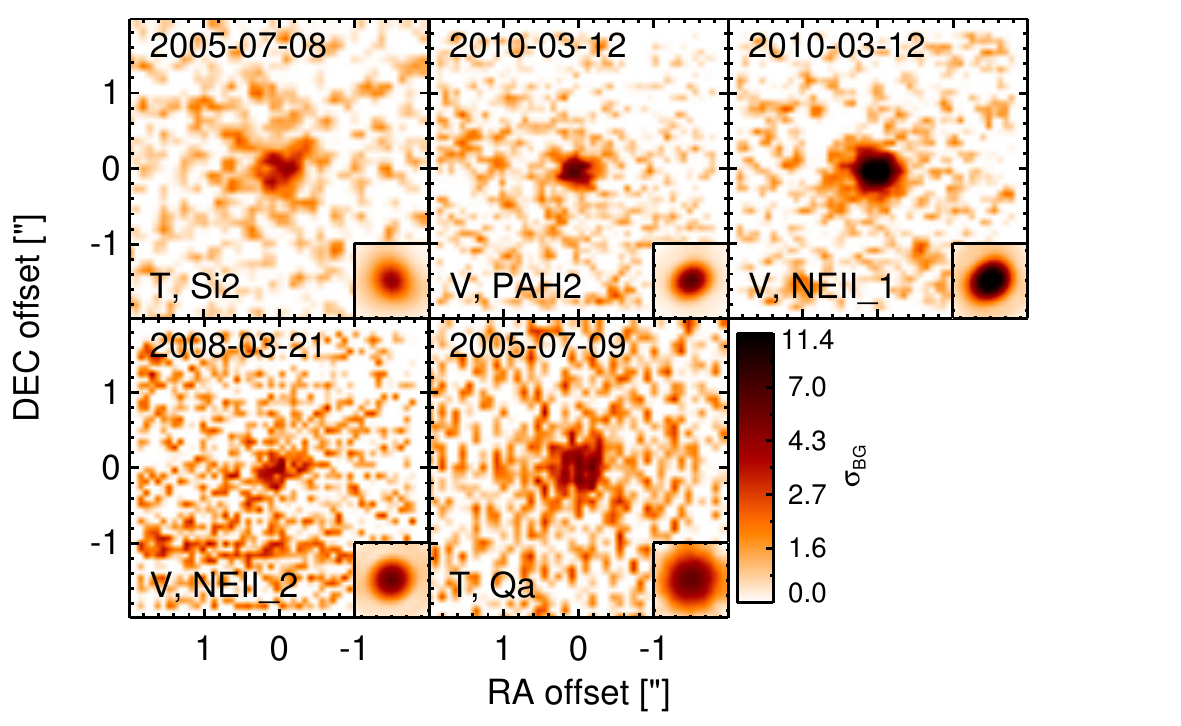}
    \caption{\label{fig:HARim_NGC5728}
             Subarcsecond-resolution MIR images of NGC\,5728 sorted by increasing filter wavelength. 
             Displayed are the inner $4\arcsec$ with North up and East to the left. 
             The colour scaling is logarithmic with white corresponding to median background and black to the $75\%$ of the highest intensity of all images in units of $\sigbg$.
             The inset image shows the central arcsecond of the PSF from the calibrator star, scaled to match the science target.
             The labels in the bottom left state instrument and filter names (C: COMICS, M: Michelle, T: T-ReCS, V: VISIR).
           }
\end{figure}
\begin{figure}
   \centering
   \includegraphics[angle=0,width=8.50cm]{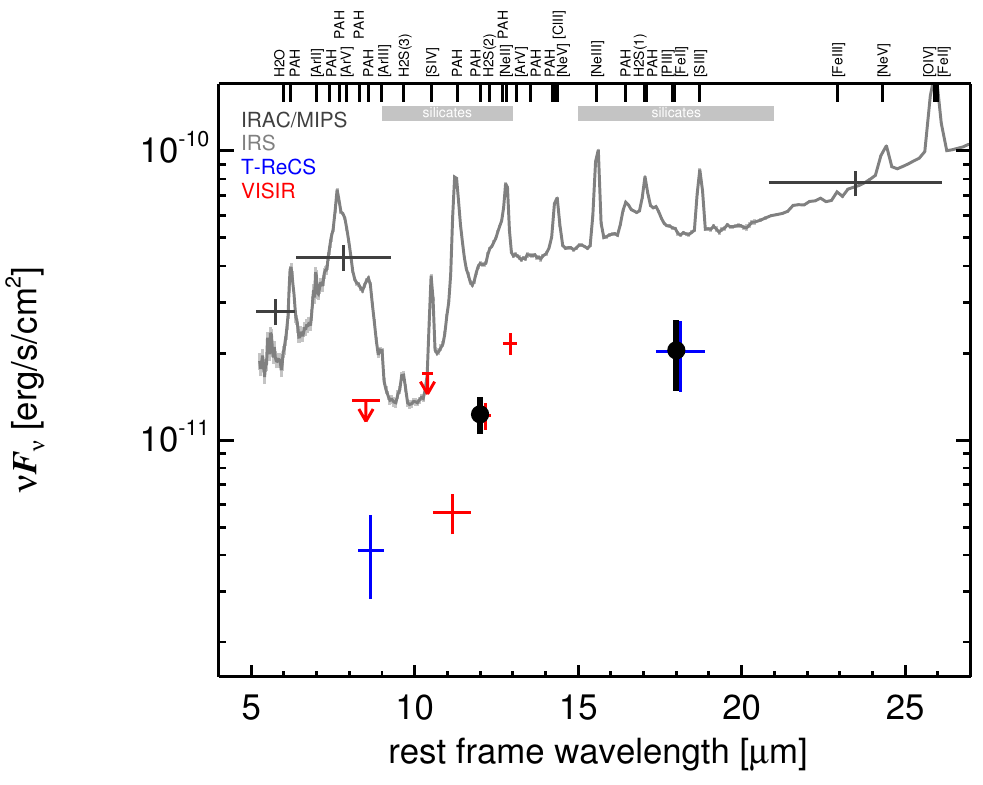}
   \caption{\label{fig:MISED_NGC5728}
      MIR SED of NGC\,5728. The description  of the symbols (if present) is the following.
      Grey crosses and  solid lines mark the \spitzer/IRAC, MIPS and IRS data. 
      The colour coding of the other symbols is: 
      green for COMICS, magenta for Michelle, blue for T-ReCS and red for VISIR data.
      Darker-coloured solid lines mark spectra of the corresponding instrument.
      The black filled circles mark the nuclear 12 and $18\,\mu$m  continuum emission estimate from the data.
      The ticks on the top axis mark positions of common MIR emission lines, while the light grey horizontal bars mark wavelength ranges affected by the silicate 10 and 18$\mu$m features.}
\end{figure}
\clearpage

\twocolumn[\begin{@twocolumnfalse}  
\subsection{NGC\,5813 -- UGC\,9655}\label{app:NGC5813}
NGC\,5813 is a passive elliptical galaxy at a distance of $D=$ $29.7 \pm 1.8\,$Mpc (NED redshift-independent median) with evidence for past AGN outbursts \citep{randall_shocks_2011}.
It possibly still contains a LINER nucleus \citep{ho_search_1997-1}.
However, indications for a nuclear X-ray compact source are inconclusive \citep{gonzalez-martin_x-ray_2009,gonzalez-martin_fitting_2009,randall_shocks_2011}.
On the other hand, a compact radio core was detected \citep{wrobel_radio-continuum_1991,nagar_radio_2005}.
We treat NGC\,5813 conservatively as an uncertain AGN.
The nucleus contains a dusty circum-nuclear disc with $\sim1.4\arcsec\sim200\,$pc diameter and PA$\sim150\degree$ \citep{tran_dusty_2001}.
The first attempt to detect the nucleus of NGC\,5813 in the MIR from the ground failed \citep{impey_infrared_1986}. 
Successful MIR observations were carried out with \isoo \citep{ferrari_survey_2002} and \spitzer/IRAC, IRS and MIPS.
The corresponding IRAC and MIPS images show centrally-peaked, yet diffuse host emission without a clearly separable nuclear component.
We measure the flux of the central four (seven for MIPS) arcsecond region, which provides IRAC $5.8$ and $8.0\,\mu$m and MIPS $24\,\mu$m fluxes significantly lower than the fluxes in the literature \citep{temi_mid-infrared_2005,temi_far-infrared_2007,temi_spitzer_2009}.
The IRS LR staring-mode spectrum exhibits a MIR SED typical of a passive old stellar population without any prominent spectral features and a blue spectral slope  in $\nu F_\nu$-space (see also \citealt{bregman_ages_2006}).
We observed the nuclear region of NGC\,5813 with VISIR in 2009 with two narrow $N$-band filters but failed to detect any MIR emission \citep{asmus_mid-infrared_2011}.
The corresponding flux upper limits are well above the \spitzerr spectrophotometry, and thus we use the latter as an upper limit for the 12\,$\mu$m continuum emission of any putative AGN component in NGC\,5813.
\newline\end{@twocolumnfalse}]

\begin{figure}
   \centering
   \includegraphics[angle=0,width=8.500cm]{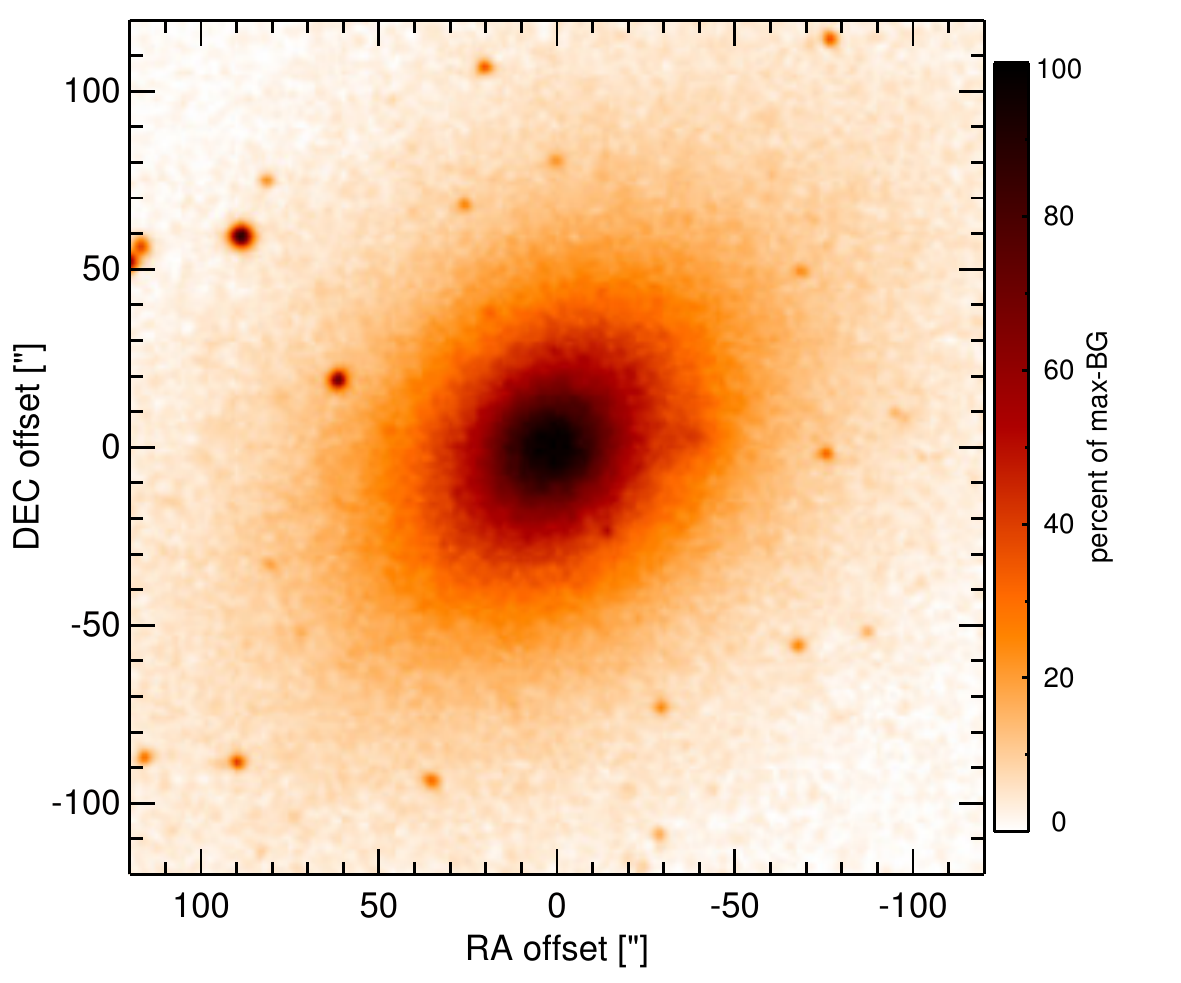}
    \caption{\label{fig:OPTim_NGC5813}
             Optical image (DSS, red filter) of NGC\,5813. Displayed are the central $4\arcmin$ with North up and East to the left. 
              The colour scaling is linear with white corresponding to the median background and black to the $0.01\%$ pixels with the highest intensity.  
           }
\end{figure}
\begin{figure}
   \centering
   \includegraphics[angle=0,height=3.11cm]{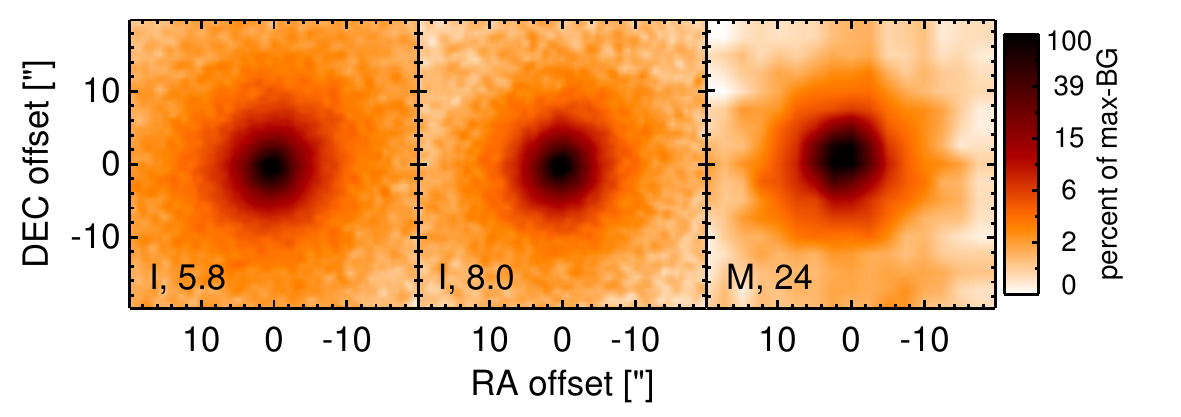}
    \caption{\label{fig:INTim_NGC5813}
             \spitzerr MIR images of NGC\,5813. Displayed are the inner $40\arcsec$ with North up and East to the left. The colour scaling is logarithmic with white corresponding to median background and black to the $0.1\%$ pixels with the highest intensity.
             The label in the bottom left states instrument and central wavelength of the filter in $\mu$m (I: IRAC, M: MIPS). 
           }
\end{figure}
\begin{figure}
   \centering
   \includegraphics[angle=0,width=8.50cm]{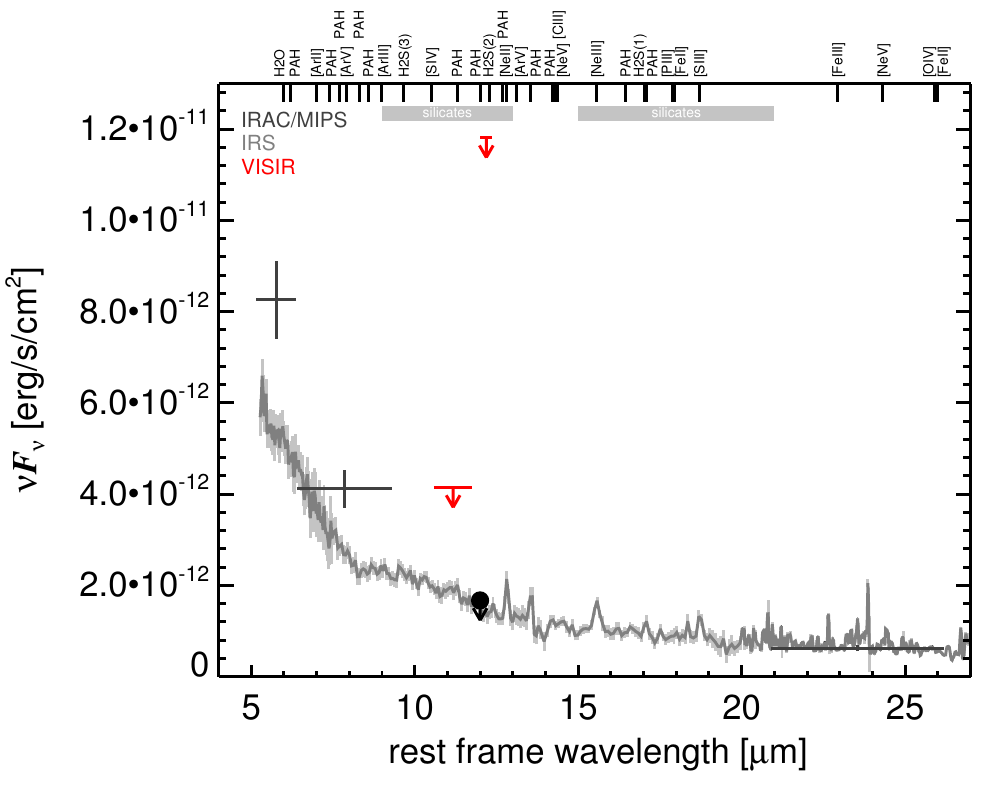}
   \caption{\label{fig:MISED_NGC5813}
      MIR SED of NGC\,5813. The description  of the symbols (if present) is the following.
      Grey crosses and  solid lines mark the \spitzer/IRAC, MIPS and IRS data. 
      The colour coding of the other symbols is: 
      green for COMICS, magenta for Michelle, blue for T-ReCS and red for VISIR data.
      Darker-coloured solid lines mark spectra of the corresponding instrument.
      The black filled circles mark the nuclear 12 and $18\,\mu$m  continuum emission estimate from the data.
      The ticks on the top axis mark positions of common MIR emission lines, while the light grey horizontal bars mark wavelength ranges affected by the silicate 10 and 18$\mu$m features.}
\end{figure}
\clearpage

\twocolumn[\begin{@twocolumnfalse}  
\subsection{NGC\,5866 -- UGC\,9723}\label{app:NGC5866}
NGC\,5866 is an edge-on lenticular galaxy at a distance of $D=$ $14.1 \pm 2.7\,$Mpc (NED redshift-independent median) with a borderline LINER/H\,II nucleus \citep{ho_search_1997-1}.
The central bulge and nucleus are crossed by a prominent narrow dust lane.
The X-ray evidence for an AGN remains inconclusive \citep{gonzalez-martin_x-ray_2006,gonzalez-martin_x-ray_2009,flohic_central_2006}, while a 
compact core was detected at radio wavelengths \citep{nagar_radio_2000,falcke_radio_2000}.
We conservatively treat NGC\,5866 as an uncertain AGN.
After first being detected in the MIR with \iras, NGC\,5866 was observed with \isoo \citep{dale_iso_2000,malhotra_probing_2000,lu_infrared_2003,xilouris_dust_2004} and \spitzer/IRAC, IRS and MIPS.
The corresponding IRAC and MIPS images are dominated by the prominent dust lane or edge-on disc with an extended nuclear bulge without a separable nuclear component.
Therefore, our extracted IRAC $5.8$ and $8.0\,\mu$m and MIPS $24\,\mu$m fluxes of the central four-arcsecond region (seven for MIPS) are much lower than the total fluxes in the literature \citep{temi_ages_2005,dale_infrared_2005,dale_ultraviolet--radio_2007,munoz-mateos_radial_2009,marble_aromatic_2010}.
The IRS LR mapping-mode spectrum suffers from low S/N but shows strong PAH emission and a possibly a weak silicate 10\,$\mu$m absorption feature from star-formation, but no red spectral slope in $\nu F_\nu$-space (see also \citealt{smith_mid-infrared_2007,marble_aromatic_2010}).
The nuclear region of NGC\,5866 was observed with Michelle in 2008 in the N' filter but no compact nucleus was detected.
Instead, the dust lane is weakly visible in the image.
Our derived flux upper limit of any unresolved nuclear emission is $\sim80\%$ lower than the \spitzerr spectrophotometry.
The projected central $\sim250$\,pc of NGC\,5866 are, thus, star-formation dominated in the MIR.
\newline\end{@twocolumnfalse}]

\begin{figure}
   \centering
   \includegraphics[angle=0,width=8.500cm]{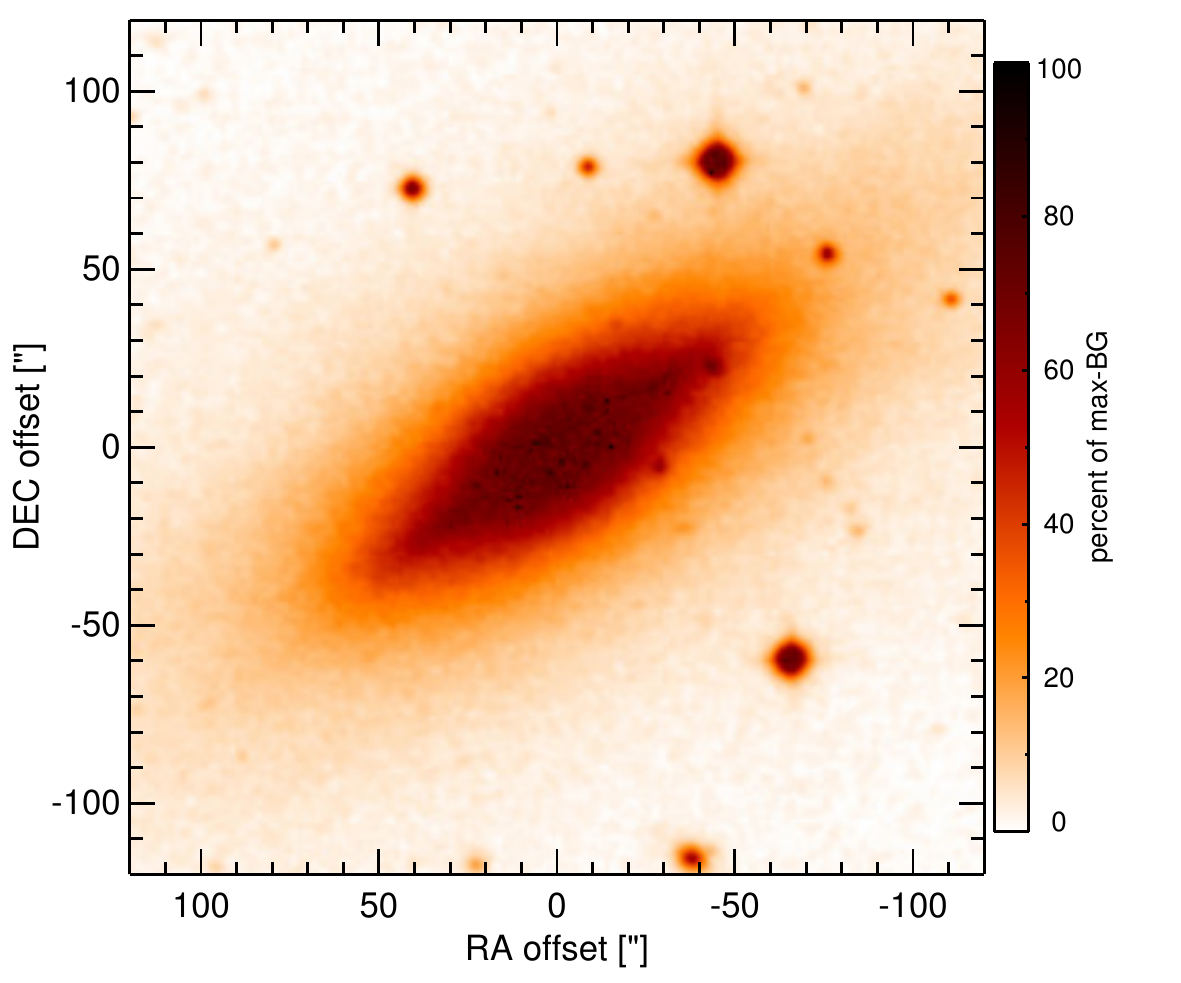}
    \caption{\label{fig:OPTim_NGC5866}
             Optical image (DSS, red filter) of NGC\,5866. Displayed are the central $4\arcmin$ with North up and East to the left. 
              The colour scaling is linear with white corresponding to the median background and black to the $0.01\%$ pixels with the highest intensity.  
           }
\end{figure}
\begin{figure}
   \centering
   \includegraphics[angle=0,height=3.11cm]{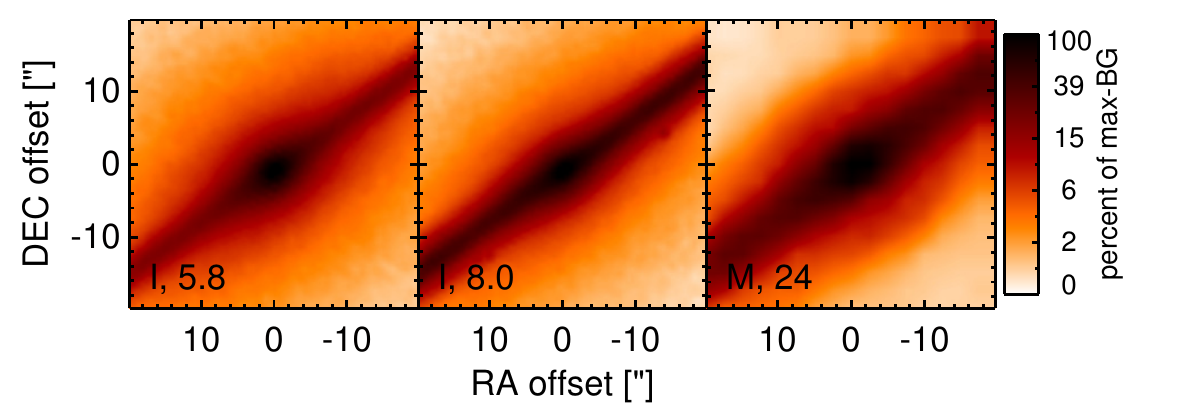}
    \caption{\label{fig:INTim_NGC5866}
             \spitzerr MIR images of NGC\,5866. Displayed are the inner $40\arcsec$ with North up and East to the left. The colour scaling is logarithmic with white corresponding to median background and black to the $0.1\%$ pixels with the highest intensity.
             The label in the bottom left states instrument and central wavelength of the filter in $\mu$m (I: IRAC, M: MIPS). 
           }
\end{figure}
\begin{figure}
   \centering
   \includegraphics[angle=0,width=8.50cm]{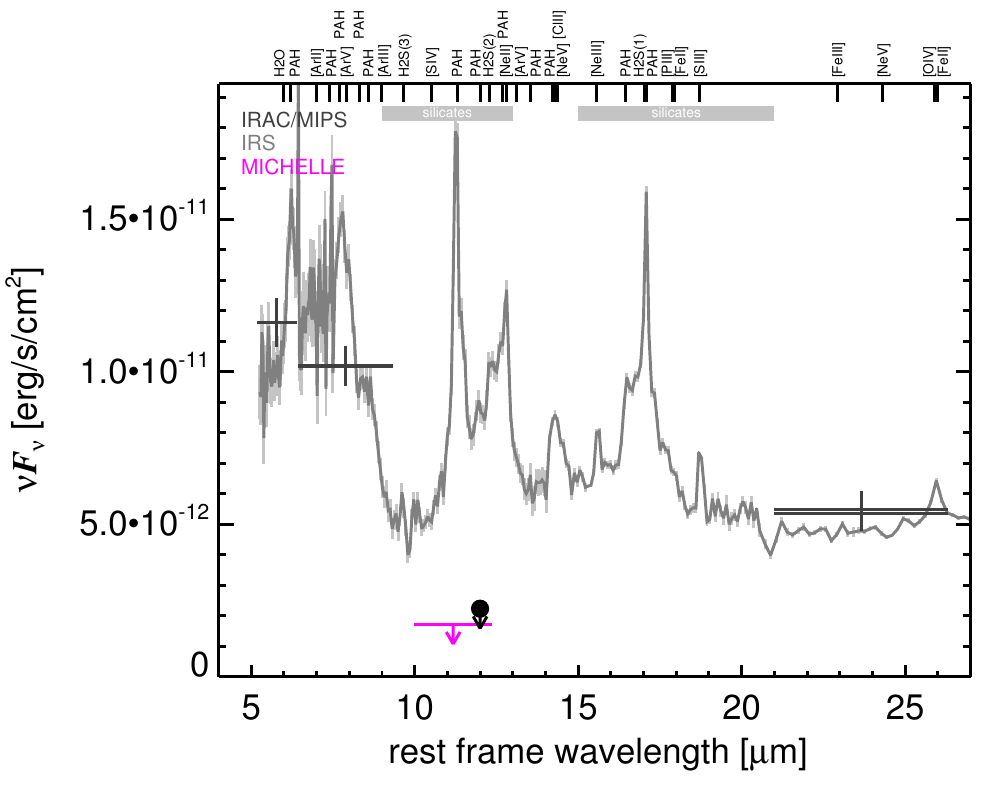}
   \caption{\label{fig:MISED_NGC5866}
      MIR SED of NGC\,5866. The description  of the symbols (if present) is the following.
      Grey crosses and  solid lines mark the \spitzer/IRAC, MIPS and IRS data. 
      The colour coding of the other symbols is: 
      green for COMICS, magenta for Michelle, blue for T-ReCS and red for VISIR data.
      Darker-coloured solid lines mark spectra of the corresponding instrument.
      The black filled circles mark the nuclear 12 and $18\,\mu$m  continuum emission estimate from the data.
      The ticks on the top axis mark positions of common MIR emission lines, while the light grey horizontal bars mark wavelength ranges affected by the silicate 10 and 18$\mu$m features.}
\end{figure}
\clearpage

\twocolumn[\begin{@twocolumnfalse}  
\subsection{NGC\,5953}\label{app:NGC5953}
NGC\,5953 is a lenticular galaxy at a redshift of $z=$ 0.0066 ($D\sim31.4\,$Mpc) which heavily interacts with NGC\,5954 to the north-east at a projected distance of $D=$ $\sim43\arcsec$ ($\sim6\,$kpc). BOth galaxies form the pair Arp\,91 (see \citealt{casasola_molecular_2010} for a recent detailed study). 
NGC\,5953 hosts an AGN optically classified either as a Sy\,2 \citep{gonzalez_delgado_circumnuclear_1996,goncalves_agns_1999}, LINER \citep{veilleux_optical_1995,kim_optical_1995}, or AGN/starburst composite \citep{veron_agns_1997}.
The AGN is embedded within a disc-like circum-nuclear starburst with a diameter of $\sim12\arcsec\sim1.8\,$kpc \citep{gonzalez_delgado_circumnuclear_1996,casasola_molecular_2010}
Nuclear X-ray emission is only detected in the soft band \citep{guainazzi_x-ray_2005,lamassa_uncovering_2011}
A compact radio core with jet-like extension of $\sim 1\arcsec$ in PA$\sim 20\degree$ \citep{jenkins_arp_1984,krips_nuclei_2007} and roughly cospatial \oiii emission are present. \citep{gonzalez_delgado_circumnuclear_1996}.
After its detection with \iras, NGC\,5953 was followed up with ground-based MIR photometry \citep{cutri_statistical_1985,wright_recent_1988,maiolino_new_1995} and space-based MIR spectrophotometry with \isoo \citep{clavel_2.5-11_2000,ramos_almeida_mid-infrared_2007}.
The nucleus remained undetected in the first subarcsecond MIR imaging observations \citep{gorjian_10_2004}.
A compact nucleus embedded within the star-formation disc with a diameter of $\sim12\arcsec$ is visible in the \spitzer/IRAC $5.8\,\mu$m image.
In the \spitzer/IRAC $8.0\,\mu$m and MIPS $24\,\mu$m image, this disc dominates the MIR emission with the brightest location being a compact region at the south-western tip of the disc.
Therefore, our corresponding photometry of the unresolved nuclear component is significantly lower than the fluxes published in the literature \citep{engelbracht_metallicity_2008,gallimore_infrared_2010}.
The \spitzer/IRS LR staring-mode spectrum is dominated by strong PAH emission and shows a red spectral slope in $\nu F_\nu$-space (see also \citealt{buchanan_spitzer_2006,wu_spitzer/irs_2009,gallimore_infrared_2010}.
Thus, the arcsecond-scale MIR SED is  completely star-formation dominated.
The nuclear region of NGC\,5953 was observed with T-ReCS in the broad N filter in 2004 \citep{videla_nuclear_2013}).
In the image, an extended nucleus is faintly visible (S/N$\sim1$), surrounded by a ring-like structure corresponding to the star-formation disc.
Therefore, we use a measurement of the unresolved central component to determine an upper limit on the nuclear flux, which makes up only $\sim14\%$ of the flux seen in the arcsecond-scale MIR SED as probed by \spitzer.
This nuclear value is presumably still affected by star formation.
On the other hand, the AGN-indicative \nev was detected in the IRS spectrum (e.g. \citealt{gallimore_infrared_2010}).
\newline\end{@twocolumnfalse}]

\begin{figure}
   \centering
   \includegraphics[angle=0,width=8.500cm]{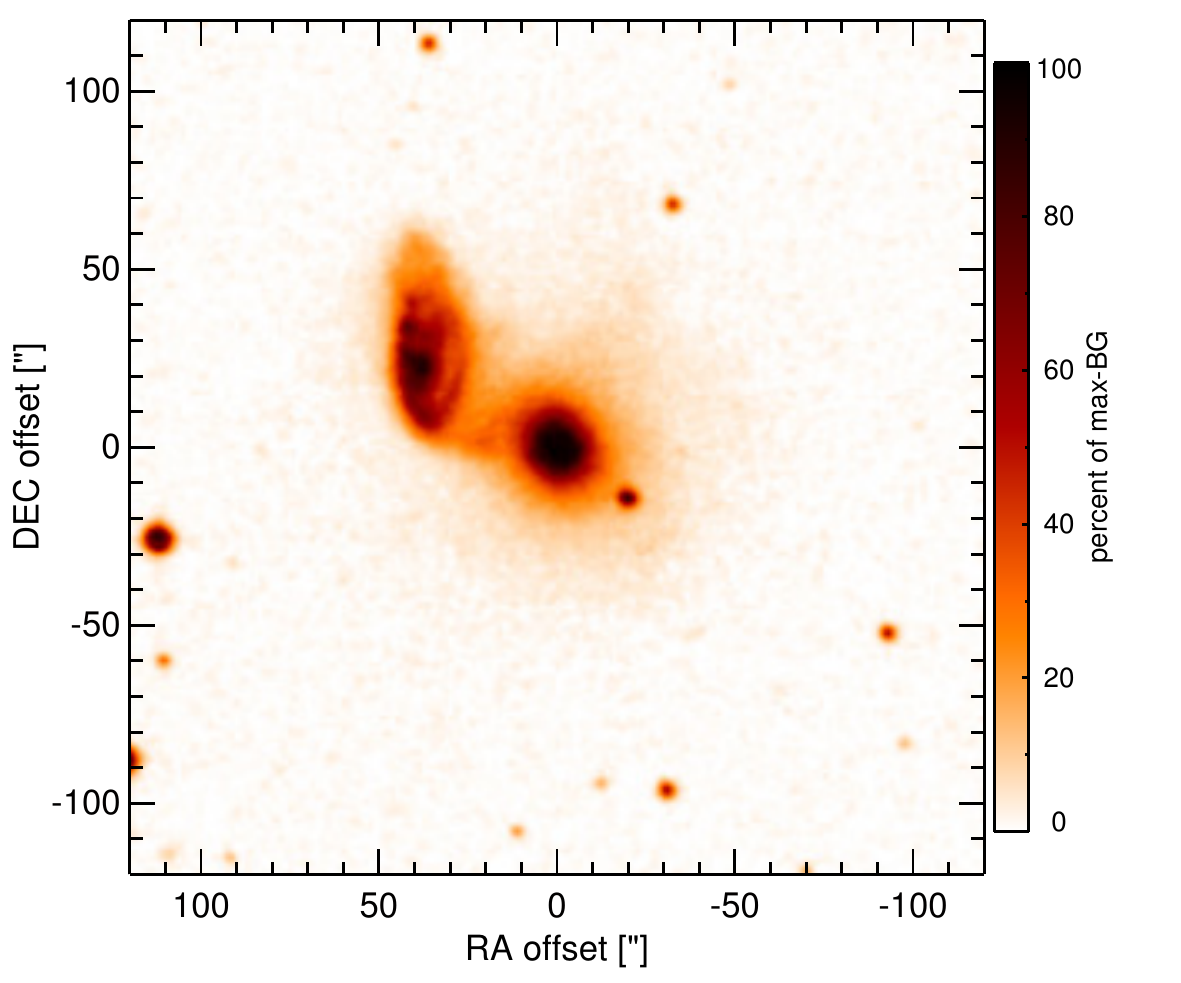}
    \caption{\label{fig:OPTim_NGC5953}
             Optical image (DSS, red filter) of NGC\,5953. Displayed are the central $4\arcmin$ with North up and East to the left. 
              The colour scaling is linear with white corresponding to the median background and black to the $0.01\%$ pixels with the highest intensity.  
           }
\end{figure}
\begin{figure}
   \centering
   \includegraphics[angle=0,height=3.11cm]{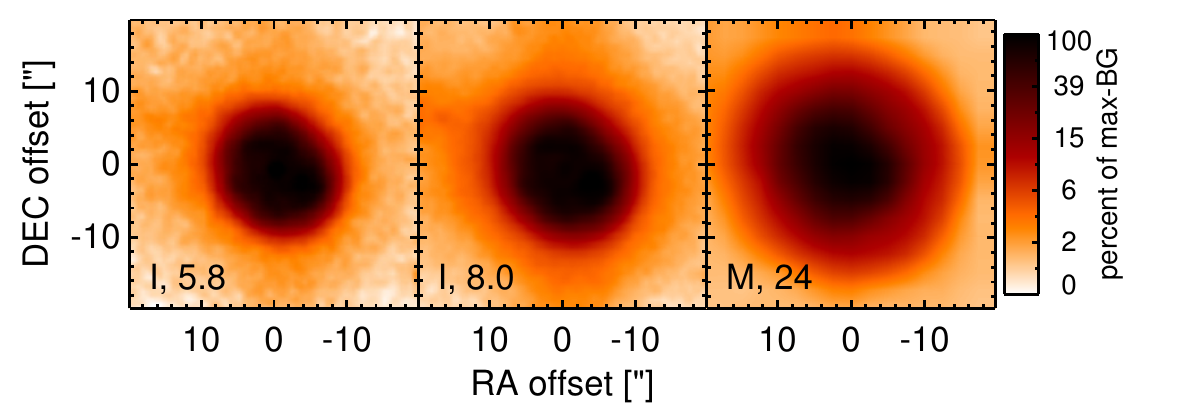}
    \caption{\label{fig:INTim_NGC5953}
             \spitzerr MIR images of NGC\,5953. Displayed are the inner $40\arcsec$ with North up and East to the left. The colour scaling is logarithmic with white corresponding to median background and black to the $0.1\%$ pixels with the highest intensity.
             The label in the bottom left states instrument and central wavelength of the filter in $\mu$m (I: IRAC, M: MIPS). 
           }
\end{figure}
\begin{figure}
   \centering
   \includegraphics[angle=0,height=3.11cm]{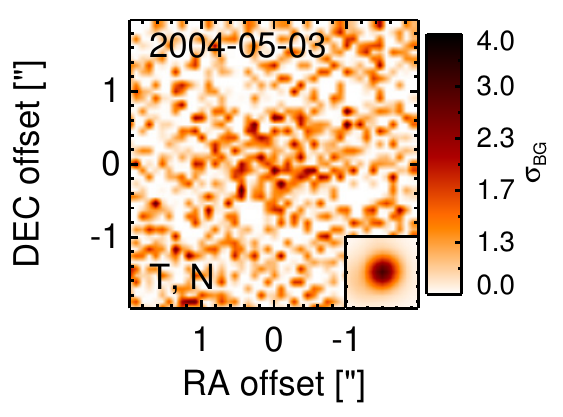}
    \caption{\label{fig:HARim_NGC5953}
             Subarcsecond-resolution MIR images of NGC\,5953 sorted by increasing filter wavelength. 
             Displayed are the inner $4\arcsec$ with North up and East to the left. 
             The colour scaling is logarithmic with white corresponding to median background and black to the $75\%$ of the highest intensity of all images in units of $\sigbg$.
             The inset image shows the central arcsecond of the PSF from the calibrator star, scaled to match the science target.
             The labels in the bottom left state instrument and filter names (C: COMICS, M: Michelle, T: T-ReCS, V: VISIR).
           }
\end{figure}
\begin{figure}
   \centering
   \includegraphics[angle=0,width=8.50cm]{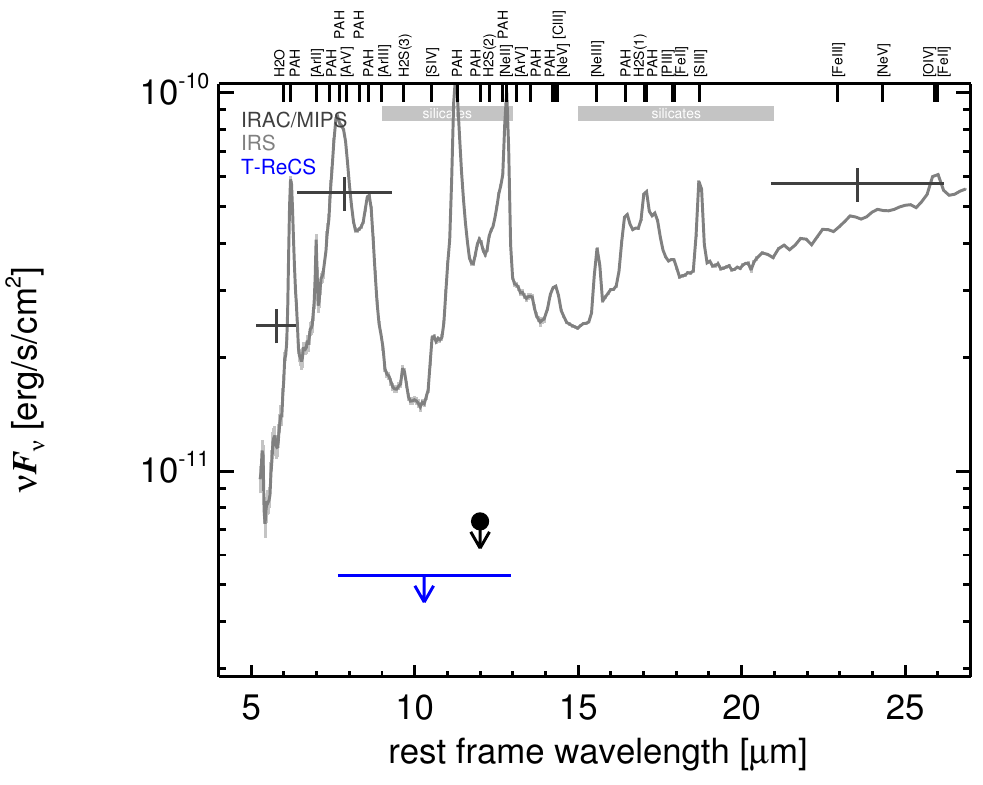}
   \caption{\label{fig:MISED_NGC5953}
      MIR SED of NGC\,5953. The description  of the symbols (if present) is the following.
      Grey crosses and  solid lines mark the \spitzer/IRAC, MIPS and IRS data. 
      The colour coding of the other symbols is: 
      green for COMICS, magenta for Michelle, blue for T-ReCS and red for VISIR data.
      Darker-coloured solid lines mark spectra of the corresponding instrument.
      The black filled circles mark the nuclear 12 and $18\,\mu$m  continuum emission estimate from the data.
      The ticks on the top axis mark positions of common MIR emission lines, while the light grey horizontal bars mark wavelength ranges affected by the silicate 10 and 18$\mu$m features.}
\end{figure}
\clearpage

\twocolumn[\begin{@twocolumnfalse}  
\subsection{NGC\,5995 -- MCG-2-40-4}\label{app:NGC5995}
NGC\,5995 is an inclined barred spiral galaxy at a redshift of $z=$ 0.0252 ($D\sim$117\,Mpc) with a Sy\,1.9 nucleus \citep{veron-cetty_catalogue_2010}, which was detected as a compact radio source \citep{thean_high-resolution_2000}.
After its detection in \iras, this object was followed up with Palomar 5\,m/MIRLIN \citep{gorjian_10_2004} and ESO 3.6\,m/TIMMI2 \citep{galliano_mid-infrared_2005}.
An unresolved nucleus at arcsecond resolution was detected in both data sets.
The \spitzer/IRAC and MIPS images are also dominated by a compact nucleus without significant host emission.
Our corresponding nuclear IRAC 5.8\,$\mu$m photometry agrees with \cite{gallimore_infrared_2010}, while our IRAC 8.0\,$\mu$m flux is $\sim30\%$ lower than published by these authors. In comparison, our fluxes are consistent with the \spitzer/IRS spectrum.
Not surprisingly, our MIPS 24$\,\mu$m flux is significantly higher than the star-formation subtracted nuclear flux from \cite{shi_unobscured_2010}.
The \spitzer/IRS LR mapping-mode spectrum exhibits silicate 10\,$\mu$m absorption,  PAH emission and a flat spectral slope in $\nu F_\nu$-space (see also \citealt{wu_spitzer/irs_2009,tommasin_spitzer-irs_2010,shi_unobscured_2010,gallimore_infrared_2010}).
Thus, the arcsecond-scale MIR SED  seems to be affected by star formation.
We observed the nuclear region of NGC\,5995 with VISIR in three narrow $N$-band filters in 2006 \citep{horst_mid_2008,horst_mid-infrared_2009} and another two in 2010 (unpublished, to our knowledge), and also obtained a LR $N$-band spectrum in 2009 \citep{honig_dusty_2010-1}. 
A compact MIR nucleus was detected in all images, and the diffraction-limited high-S/N images from 2010 show that the nucleus is unresolved, which has not been clear form the 2006 images.
Our nuclear photometry is consistent with the previously published values, yet systematically higher by $\sim10\%$ compared to the VISIR spectrum. However, this offset is well within the calibration uncertainties.
The corresponding nuclear MIR SED has on average $\sim 13\%$ lower flux levels than the arcsecond-scale MIR SED as traced by \spitzer, while exhibiting the same silicate 10\,$\mu$m absorption and much weaker PAH emission. 
Therefore, we conclude that the silicate feature is generic to the (projected) central $\sim 200\,$pc of NGC\,5995, and star-formation contribution towards the nuclear MIR SED is presumably very low. 
The nuclear MIR emission of NGC\,5995 could be resolved with MIDI interferometric observations and was modelled as two approximately equally bright components, one extending for $\sim20\,$pc, and the other one remaining unresolved ($< 5$\,pc; \citealt{burtscher_diversity_2013}).
\newline\end{@twocolumnfalse}]

\begin{figure}
   \centering
   \includegraphics[angle=0,width=8.500cm]{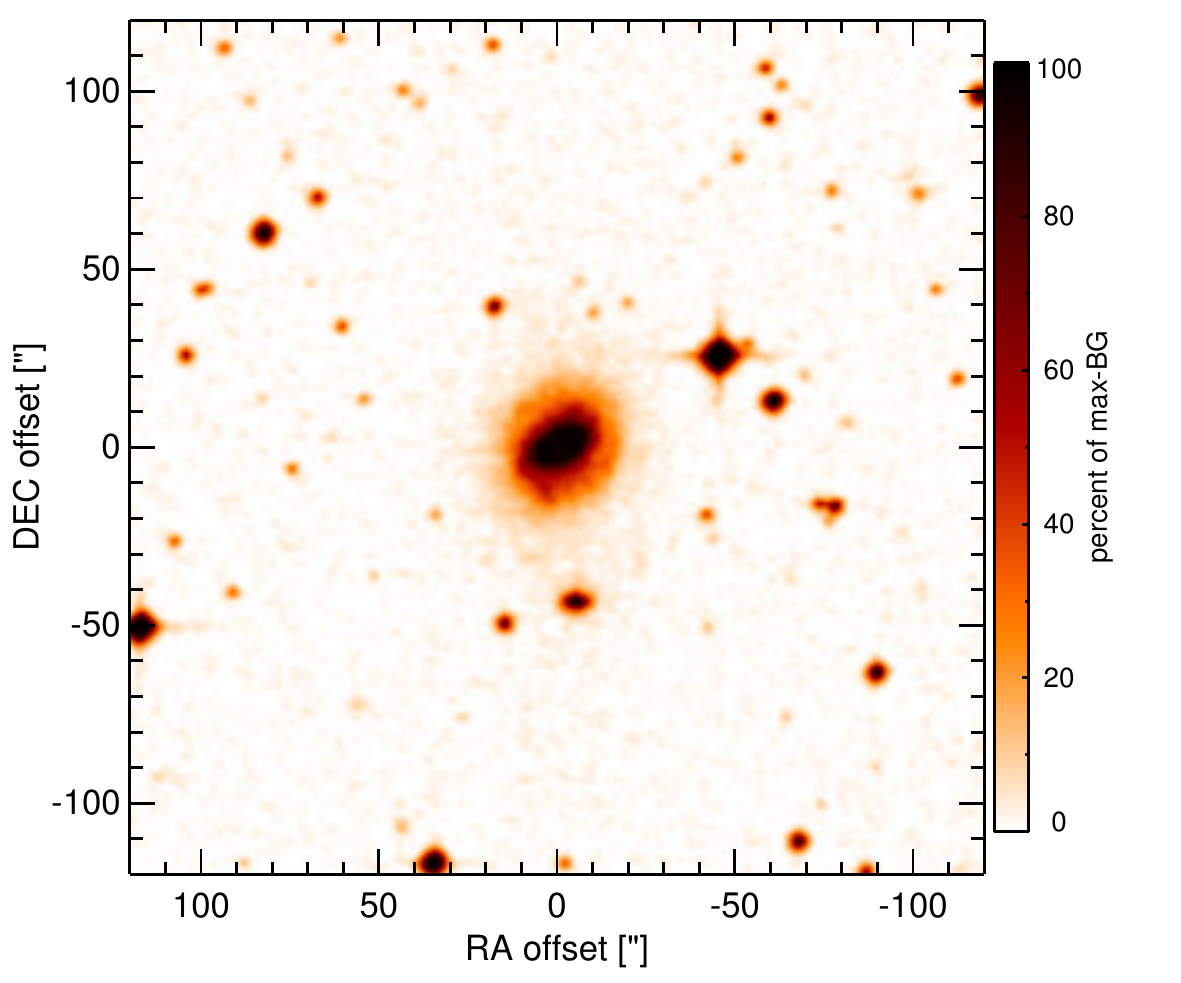}
    \caption{\label{fig:OPTim_NGC5995}
             Optical image (DSS, red filter) of NGC\,5995. Displayed are the central $4\arcmin$ with North up and East to the left. 
              The colour scaling is linear with white corresponding to the median background and black to the $0.01\%$ pixels with the highest intensity.  
           }
\end{figure}
\begin{figure}
   \centering
   \includegraphics[angle=0,height=3.11cm]{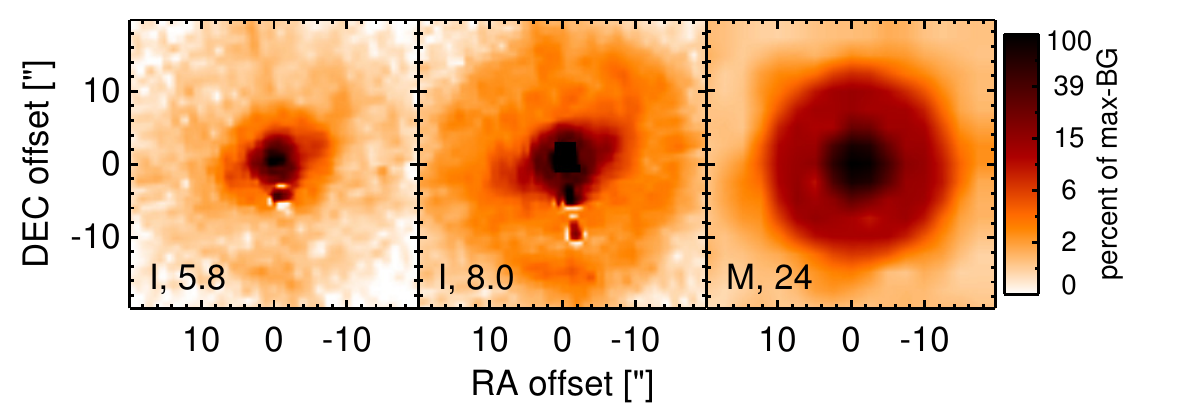}
    \caption{\label{fig:INTim_NGC5995}
             \spitzerr MIR images of NGC\,5995. Displayed are the inner $40\arcsec$ with North up and East to the left. The colour scaling is logarithmic with white corresponding to median background and black to the $0.1\%$ pixels with the highest intensity.
             The label in the bottom left states instrument and central wavelength of the filter in $\mu$m (I: IRAC, M: MIPS). 
             Note that the apparent off-nuclear compact sources in the IRAC 5.8 and $8.0\,\mu$m images are instrumental artefacts.
           }
\end{figure}
\begin{figure}
   \centering
   \includegraphics[angle=0,width=8.500cm]{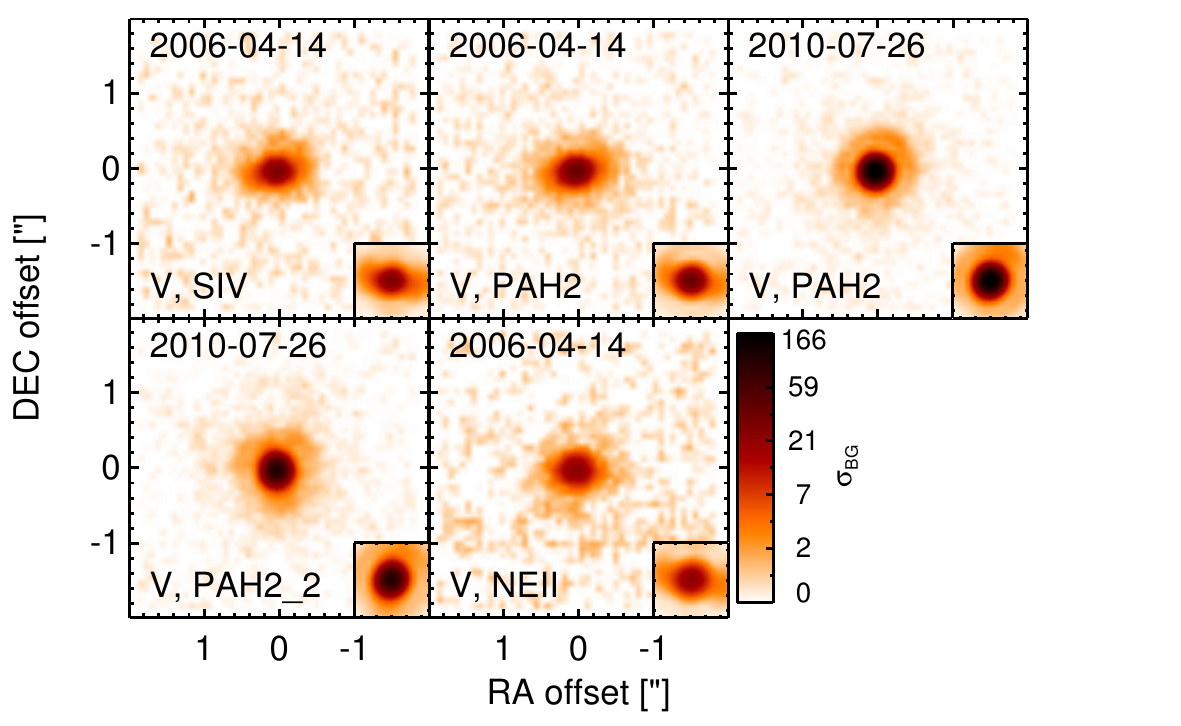}
    \caption{\label{fig:HARim_NGC5995}
             Subarcsecond-resolution MIR images of NGC\,5995 sorted by increasing filter wavelength. 
             Displayed are the inner $4\arcsec$ with North up and East to the left. 
             The colour scaling is logarithmic with white corresponding to median background and black to the $75\%$ of the highest intensity of all images in units of $\sigbg$.
             The inset image shows the central arcsecond of the PSF from the calibrator star, scaled to match the science target.
             The labels in the bottom left state instrument and filter names (C: COMICS, M: Michelle, T: T-ReCS, V: VISIR).
           }
\end{figure}
\begin{figure}
   \centering
   \includegraphics[angle=0,width=8.50cm]{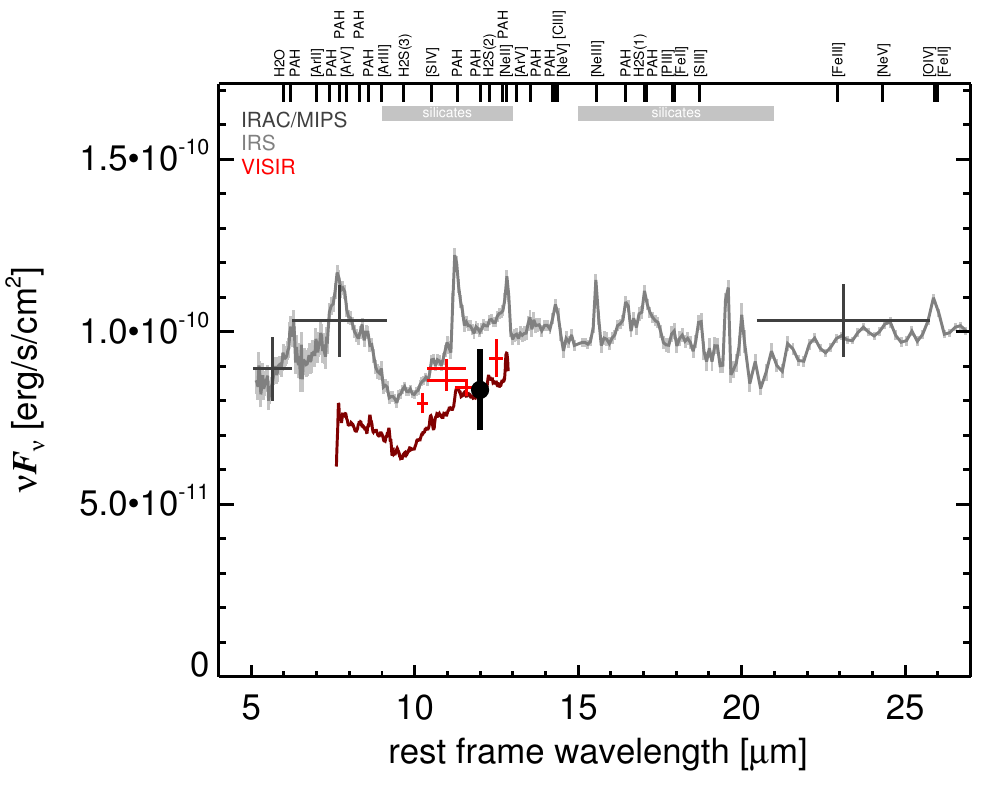}
   \caption{\label{fig:MISED_NGC5995}
      MIR SED of NGC\,5995. The description  of the symbols (if present) is the following.
      Grey crosses and  solid lines mark the \spitzer/IRAC, MIPS and IRS data. 
      The colour coding of the other symbols is: 
      green for COMICS, magenta for Michelle, blue for T-ReCS and red for VISIR data.
      Darker-coloured solid lines mark spectra of the corresponding instrument.
      The black filled circles mark the nuclear 12 and $18\,\mu$m  continuum emission estimate from the data.
      The ticks on the top axis mark positions of common MIR emission lines, while the light grey horizontal bars mark wavelength ranges affected by the silicate 10 and 18$\mu$m features.}
\end{figure}
\clearpage

\twocolumn[\begin{@twocolumnfalse}  
\subsection{NGC\,6221}\label{app:NGC6221}
NGC\,6221 is a barred late-type spiral galaxy at a low Galactic latitude at a distance of $D=$ $10.7 \pm 4.3\,$Mpc (NED redshift-independent median) with an active nucleus containing a clumpy starburst and a Sy\,2 nucleus (see \citealt{levenson_obscuring_2001} for a detailed study). This results in a AGN/starburst composite classification (e.g. \citealt{veron_agns_1997}).
In particular, the AGN is surrounded by several compact star clusters.
A compact radio core with linear extended emission of $\sim6\arcsec\sim300\,$pc along PA$\sim40\degree$ was detected at radio wavelengths \citep{morganti_radio_1999}.
We conservatively treat NGC\,6221 as an uncertain AGN.
The first ground-based MIR observations were performed by \cite{glass_mid-infrared_1982}, followed by \cite{roche_atlas_1991}.
In addition, NGC\,6221 was observed with \isoo \citep{bendo_infrared_2002} and \spitzer/IRAC, IRS and MIPS.
The corresponding IRAC and MIPS images show an elongated, bright nuclear emission region (diameter$\sim7\arcsec\sim$360\,pc; PA$\sim30\degree$) embedded within the spiral-like host emission.
The IRS LR staring-mode spectrum exhibits strong PAH emission, a possible silicate 10\,$\mu$m absorption feature, and a steep red spectral slope in $\nu F_\nu$-space (see also \citealt{pereira-santaella_mid-infrared_2010}).
Thus, the arcsecond-scale MIR SED  seems to be completely star-formation dominated.
The nuclear region of NGC\,6221 was observed with T-ReCS in the Si2 and Qa filters in 2009 \citep{ramos_almeida_testing_2011}.
In both images, an extended nucleus (FWHM(major axis)$\sim 0.6\arcsec\sim30\,$pc; PA$\sim 32\degree$) embedded within aligned arc-like extended emission and a second compact source $\sim 1.75\arcsec\sim90$\,pc to the south-west (PA$\sim305\degree$) are detected. 
The latter source is associated with a compact star-forming region \citep{ramos_almeida_testing_2011}.
We use manually-scaled PSF photometry to measure the nuclear flux. This methods provide values that are consistent with \cite{ramos_almeida_testing_2011} and on average $\sim 87\%$ lower than the \spitzerr spectrophotometry.
The resulting nuclear MIR SED might, however, still be affected by star-formation emission.
\newline\end{@twocolumnfalse}]

\begin{figure}
   \centering
   \includegraphics[angle=0,width=8.500cm]{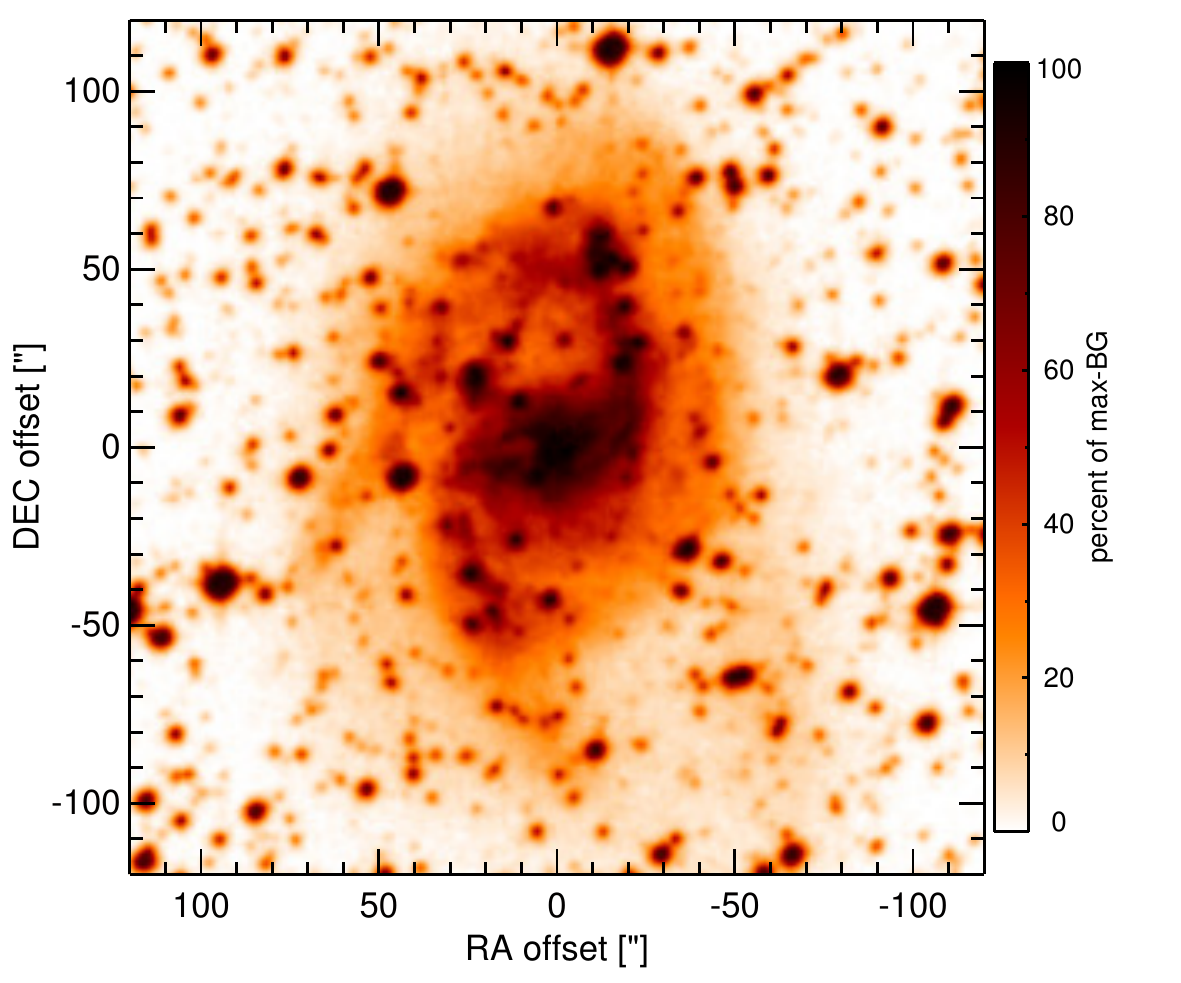}
    \caption{\label{fig:OPTim_NGC6221}
             Optical image (DSS, red filter) of NGC\,6221. Displayed are the central $4\arcmin$ with North up and East to the left. 
              The colour scaling is linear with white corresponding to the median background and black to the $0.01\%$ pixels with the highest intensity.  
           }
\end{figure}
\begin{figure}
   \centering
   \includegraphics[angle=0,height=3.11cm]{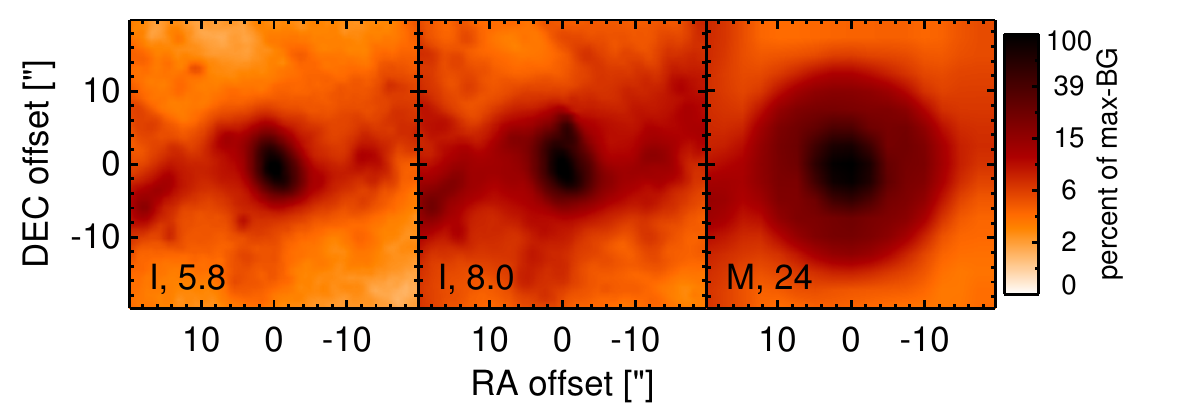}
    \caption{\label{fig:INTim_NGC6221}
             \spitzerr MIR images of NGC\,6221. Displayed are the inner $40\arcsec$ with North up and East to the left. The colour scaling is logarithmic with white corresponding to median background and black to the $0.1\%$ pixels with the highest intensity.
             The label in the bottom left states instrument and central wavelength of the filter in $\mu$m (I: IRAC, M: MIPS). 
             Note that the apparent off-nuclear compact source in the IRAC $8.0\,\mu$m image is an instrumental artefact.
           }
\end{figure}
\begin{figure}
   \centering
   \includegraphics[angle=0,height=3.11cm]{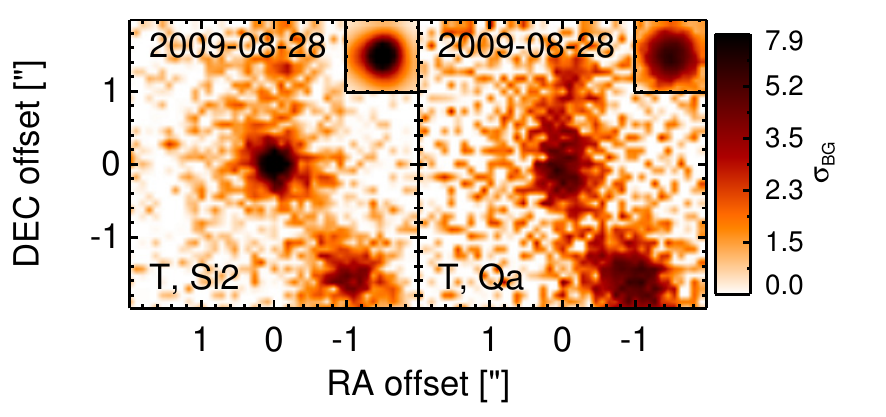}
    \caption{\label{fig:HARim_NGC6221}
             Subarcsecond-resolution MIR images of NGC\,6221 sorted by increasing filter wavelength. 
             Displayed are the inner $4\arcsec$ with North up and East to the left. 
             The colour scaling is logarithmic with white corresponding to median background and black to the $75\%$ of the highest intensity of all images in units of $\sigbg$.
             The inset image shows the central arcsecond of the PSF from the calibrator star, scaled to match the science target.
             The labels in the bottom left state instrument and filter names (C: COMICS, M: Michelle, T: T-ReCS, V: VISIR).
           }
\end{figure}
\begin{figure}
   \centering
   \includegraphics[angle=0,width=8.50cm]{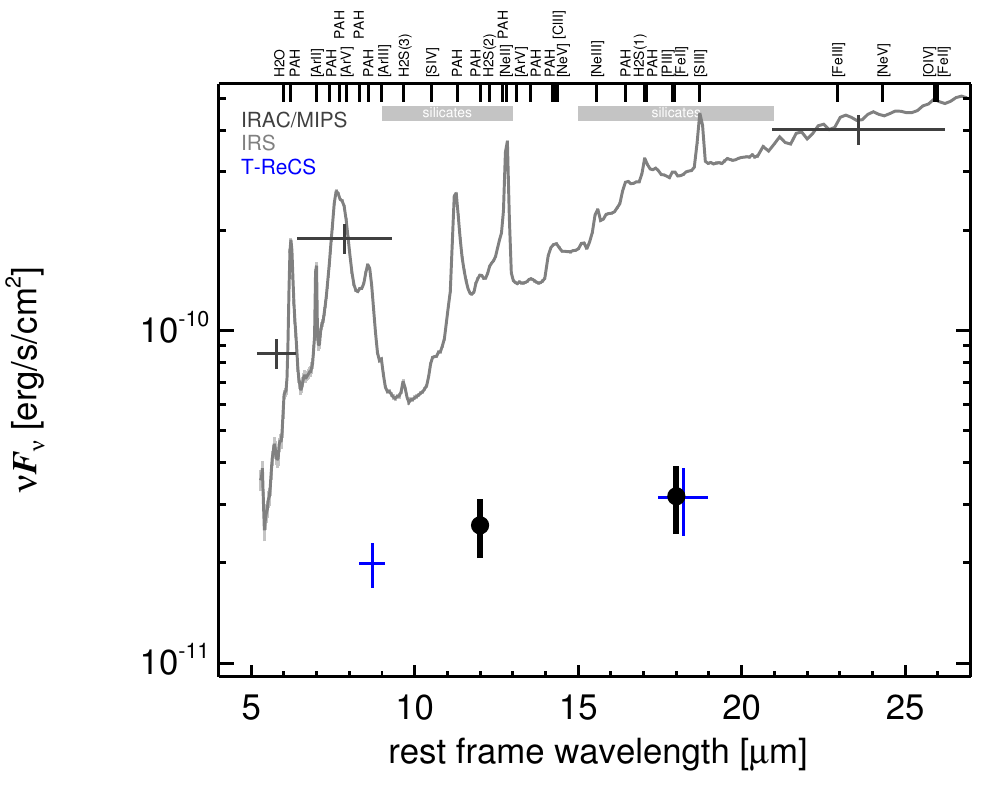}
   \caption{\label{fig:MISED_NGC6221}
      MIR SED of NGC\,6221. The description  of the symbols (if present) is the following.
      Grey crosses and  solid lines mark the \spitzer/IRAC, MIPS and IRS data. 
      The colour coding of the other symbols is: 
      green for COMICS, magenta for Michelle, blue for T-ReCS and red for VISIR data.
      Darker-coloured solid lines mark spectra of the corresponding instrument.
      The black filled circles mark the nuclear 12 and $18\,\mu$m  continuum emission estimate from the data.
      The ticks on the top axis mark positions of common MIR emission lines, while the light grey horizontal bars mark wavelength ranges affected by the silicate 10 and 18$\mu$m features.}
\end{figure}
\clearpage

\twocolumn[\begin{@twocolumnfalse}  
\subsection{NGC\,6240}\label{app:NGC6240N}\label{app:NGC6240S}
NGC\,6240 is a very infrared-luminous, late-stage merger system consisting of two gas-rich massive spirals at a redshift of $z=$ 0.025 ($D\sim114\,$Mpc) with a massive starburst and a binary AGN.
Owing to this unique nature (discovered by \cite{fosbury_unusual_1979,fried_ngc_1983}), NGC\,6240 has been extensively studied at all wavelengths over the last decades.
It belongs to the nine-month BAT AGN sample.
Both active nuclei have been optically classified as LINERs, while the southern one, NGC\,6240S, might also be classified as a Sy\,2 or H\,II nucleus \citep{rafanelli_subarcsec_1997}.
We treat them as AGN/starburst composites (see also discussion in \citealt{yuan_role_2010}).
The double AGN nature was discovered in X-rays \citep{komossa_discovery_2003} and verified at radio \citep{gallimore_parsec-scale_2004} and infrared wavelengths \citep{risaliti_double_2006}.
Both nuclei are separated by $\sim1.5\arcsec \sim 0.8\,$kpc in the north-south direction (PA$\sim10\degree$; \citealt{max_locating_2007}) and are embedded within the kiloparsec-sized starburst \citep{tecza_stellar_2000,engel_ngc_2010}.
At radio wavelengths, both nuclei appear point-like at arcsecond resolution \citep{colbert_radio_1994,beswick_merlin_2001}, but the northern nucleus, NGC\,6240N, shows jet-like east-west elongation at milliarcsecond resolution \citep{gallimore_parsec-scale_2004,hagiwara_very_2011}.
NGC\,6240S emits strong water maser emission \citep{hagiwara_search_2002,nakai_detection_2002,hagiwara_location_2003,hagiwara_two_2010}.
The first attempt to detect NGC\,6240 in the MIR failed \cite{allen_near-infrared_1976}.
Instead, it was first detected with \irass \citep{wright_ultraluminous_1984} and followed up with many ground-based MIR observations  \citep{rieke_10_1985,wright_recent_1988,smith_nature_1989,roche_atlas_1991,wynn-williams_luminous_1993,keto_infrared_1997,dudley_new_1999}.
NGC\,6240 has also been extensively studied \isoo  \citep{klaas_infrared_1997,genzel_what_1998,rigopoulou_large_1999,charmandaris_mid-ir_1999,thornley_massive_2000,tran_isocam-cvf_2001,lutz_iso_2003,forster_schreiber_warm_2004}.
The first arcsecond-resolution MIR images resolving both nuclei were obtained with ESO MPI 2.2\,m/MANIAC in 1997 \citep{krabbe_n-band_2001}, followed by ESO 3.6\,m/TIMMI2 images and spectroscopy \citep{siebenmorgen_mid-infrared_2004}.
\cite{egami_subarcsecond_2006} reported the first subarcsecond-resolution MIR images using Keck/MIRLIN and an LR $N$-band spectrum using Keck/LWS.
The images show both nuclei embedded within extended emission connecting them, while NGC\,6240S completely dominates the total MIR emission of the system.  
They claim that the spectrum of NGC\,6240S is consistent with pure star formation. 
No spectrum of NGC\,6240N is presented.
The nuclear structure is not resolved in the \spitzer/IRAC and MIPS images, which show a bright elongated nucleus embedded within diffuse host emission in the IRAC $5.8$ and $8.0\,\mu$m bands (see also \citealt{bush_structure_2008}), and only a point source in the MIPS 24\,$\mu$m band.
Our MIPS 24\,$\mu$m photometry agrees with the value published by \cite{marshall_decomposing_2007}.
The \spitzer/IRS LR staring-mode spectrum exhibits deep silicate 10$\,\mu$m and weak silicate 18\,$\mu$m absorption, strong PAH emission, and a red spectral slope in $\nu F_\nu$-space (see also \citealt{armus_detection_2006,armus_observations_2007,farrah_high-resolution_2007}).
The detection of AGN-associated \nev emission indicates a significant AGN contribution to the arcsecond-scale MIR SED \cite{armus_observations_2007}.
Because both nuclei are blended in the \spitzerr data, the resulting spectrophotometry is regarded as arcsecond-scale MIR SED for both.
The nuclear region of NGC\,6240 was observed with VISIR in two narrow $N$ and one $Q$-band filter in 2005 (unpublished, to our knowledge).
The VISIR images show a similar morphology as the MIRLIN images with both nuclei embedded within extended emission.
NGC\,6240S is elongated along PA$\sim 165\degree$.
We perform manual PSF-scaling to measure the unresolved nuclear fluxes in both nuclei.
The resulting nuclear MIR SED of NGC\,6240S is on average $\sim 70\%$ lower than the \spitzerr spectrophotometry ($\sim 93\%$ lower for NGC\,6240N).
Because these nuclear MIR SEDs might still be contaminated by star formation, this result is consistent with the $\sim20\%$ AGN contribution estimate of \cite{armus_observations_2007}.
\newline\end{@twocolumnfalse}]

\begin{figure}
   \centering
   \includegraphics[angle=0,width=8.500cm]{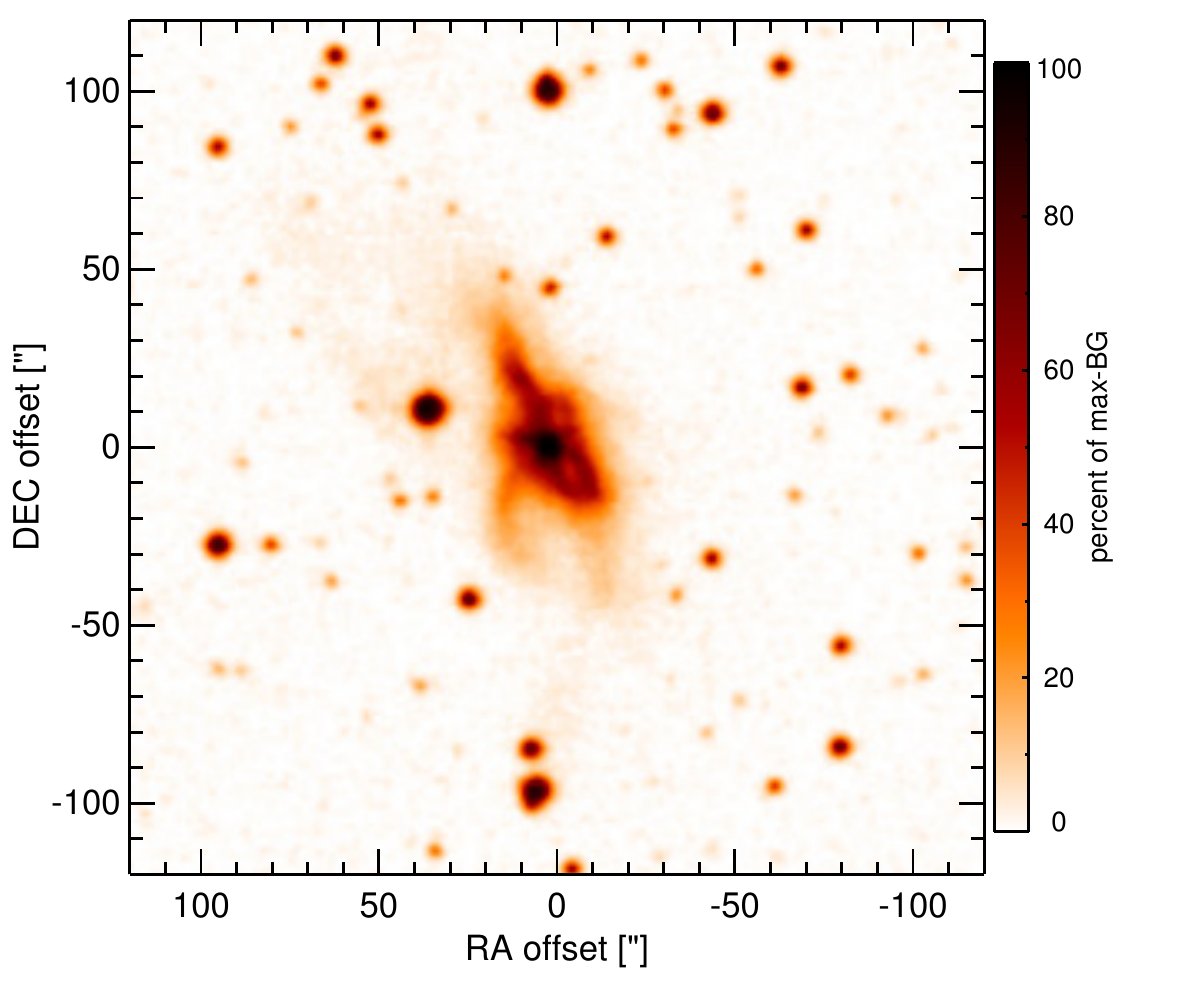}
    \caption{\label{fig:OPTim_NGC6240S}
             Optical image (DSS, red filter) of NGC\,6240, centred on the southern nucleus. Displayed are the central $4\arcmin$ with North up and East to the left. 
              The colour scaling is linear with white corresponding to the median background and black to the $0.01\%$ pixels with the highest intensity.  
           }
\end{figure}
\begin{figure}
   \centering
   \includegraphics[angle=0,height=3.11cm]{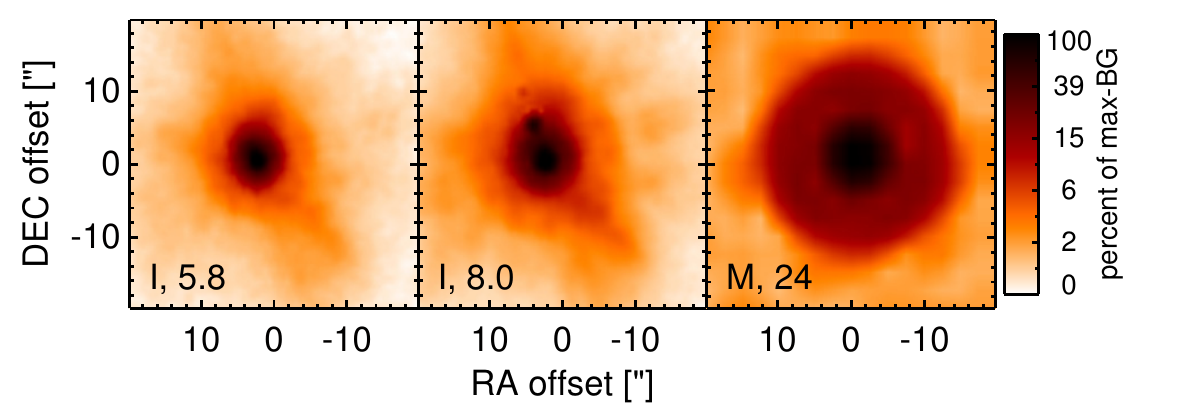}
    \caption{\label{fig:INTim_NGC6240S}
             \spitzerr MIR images of NGC\,6240, centred on the southern nucleus. Displayed are the inner $40\arcsec$ with North up and East to the left. The colour scaling is logarithmic with white corresponding to median background and black to the $0.1\%$ pixels with the highest intensity.
             The label in the bottom left states instrument and central wavelength of the filter in $\mu$m (I: IRAC, M: MIPS). 
             Note that the apparent off-nuclear compact source in the IRAC $8.0\,\mu$m image is an instrumental artefact.
           }
\end{figure}
\begin{figure}
   \centering
   \includegraphics[angle=0,height=3.11cm]{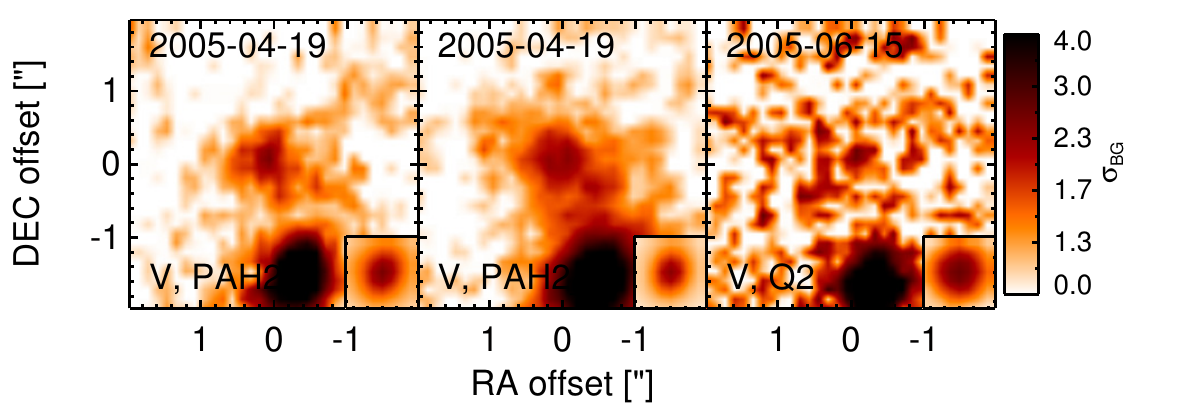}
    \caption{\label{fig:HARim_NGC6240N}
             Subarcsecond-resolution MIR images of NGC\,6240N sorted by increasing filter wavelength. 
             Displayed are the inner $4\arcsec$ with North up and East to the left. 
             The colour scaling is logarithmic with white corresponding to median background and black to the $75\%$ of the highest intensity of all images in units of $\sigbg$.
             The inset image shows the central arcsecond of the PSF from the calibrator star, scaled to match the science target.
             The labels in the bottom left state instrument and filter names (C: COMICS, M: Michelle, T: T-ReCS, V: VISIR).
           }
\end{figure}
\begin{figure}
   \centering
   \includegraphics[angle=0,width=8.50cm]{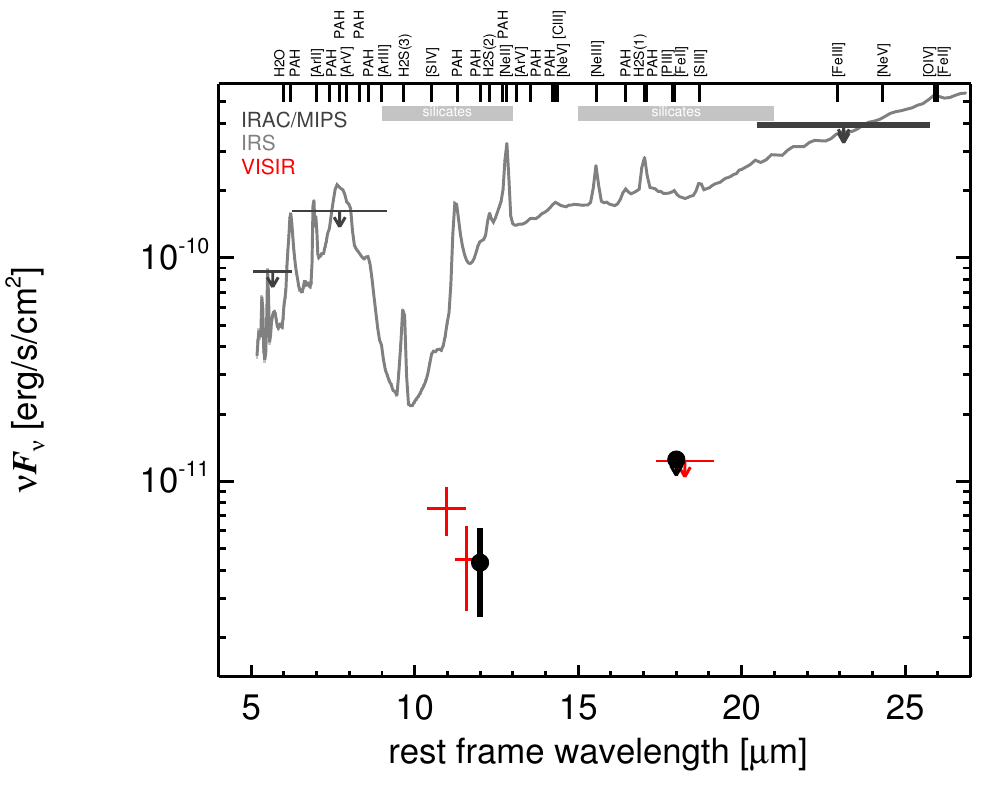}
   \caption{\label{fig:MISED_NGC6240N}
      MIR SED of NGC\,6240N. The description  of the symbols (if present) is the following.
      Grey crosses and  solid lines mark the \spitzer/IRAC, MIPS and IRS data. 
      The colour coding of the other symbols is: 
      green for COMICS, magenta for Michelle, blue for T-ReCS and red for VISIR data.
      Darker-coloured solid lines mark spectra of the corresponding instrument.
      The black filled circles mark the nuclear 12 and $18\,\mu$m  continuum emission estimate from the data.
      The ticks on the top axis mark positions of common MIR emission lines, while the light grey horizontal bars mark wavelength ranges affected by the silicate 10 and 18$\mu$m features.}
\end{figure}
\clearpage

\begin{figure}
   \centering
   \includegraphics[angle=0,height=3.11cm]{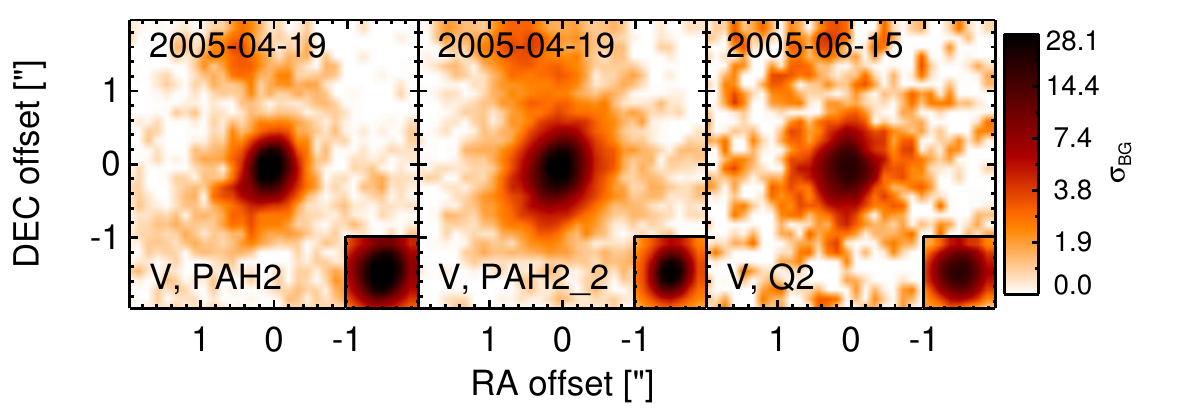}
    \caption{\label{fig:HARim_NGC6240S}
             Subarcsecond-resolution MIR images of NGC\,6240S sorted by increasing filter wavelength. 
             Displayed are the inner $4\arcsec$ with North up and East to the left. 
             The colour scaling is logarithmic with white corresponding to median background and black to the $75\%$ of the highest intensity of all images in units of $\sigbg$.
             The inset image shows the central arcsecond of the PSF from the calibrator star, scaled to match the science target.
             The labels in the bottom left state instrument and filter names (C: COMICS, M: Michelle, T: T-ReCS, V: VISIR).
           }
\end{figure}
\begin{figure}
   \centering
   \includegraphics[angle=0,width=8.50cm]{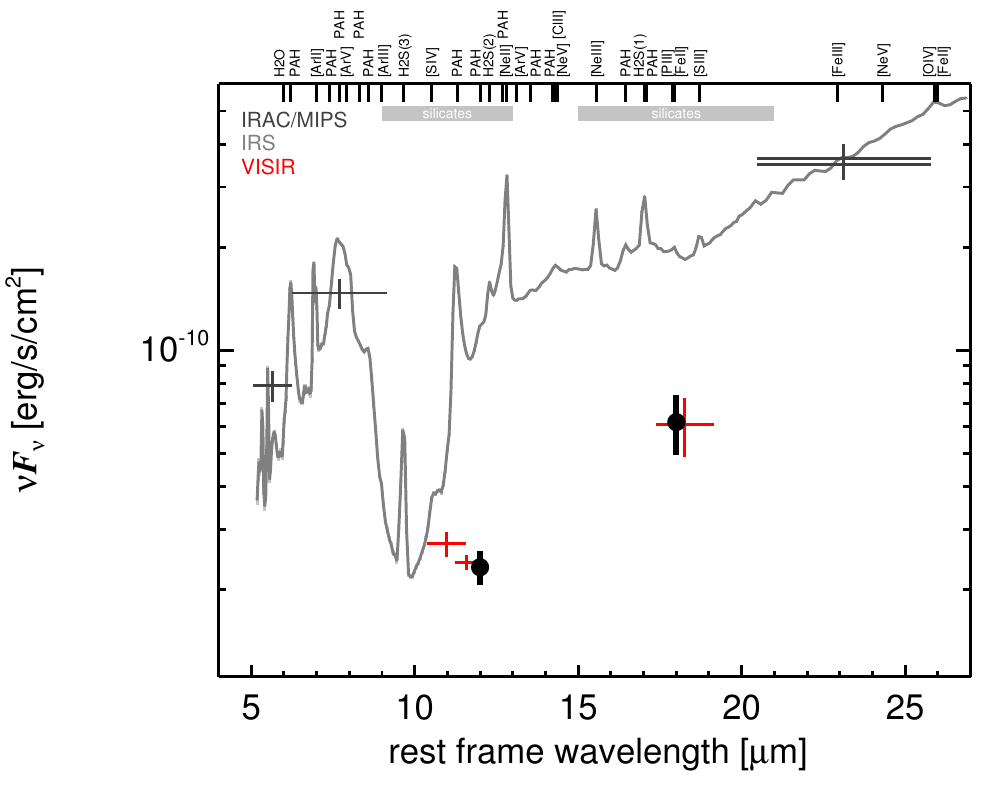}
   \caption{\label{fig:MISED_NGC6240S}
      MIR SED of NGC\,6240S. The description  of the symbols (if present) is the following.
      Grey crosses and  solid lines mark the \spitzer/IRAC, MIPS and IRS data. 
      The colour coding of the other symbols is: 
      green for COMICS, magenta for Michelle, blue for T-ReCS and red for VISIR data.
      Darker-coloured solid lines mark spectra of the corresponding instrument.
      The black filled circles mark the nuclear 12 and $18\,\mu$m  continuum emission estimate from the data.
      The ticks on the top axis mark positions of common MIR emission lines, while the light grey horizontal bars mark wavelength ranges affected by the silicate 10 and 18$\mu$m features.}
\end{figure}
\clearpage

\twocolumn[\begin{@twocolumnfalse}  
\subsection{NGC\,6251}\label{app:NGC6251}
NGC\,6251 is an elliptical galaxy at a redshift of $z=$ 0.0247 ($D\sim112\,$Mpc) hosting one of the largest radio sources in the sky \citep{waggett_ngc_1977} and classified as FR\,I/II. It consists of giant double lobes with an S-shaped geometry (see \citealt{migliori_implications_2011} for a recent SED study).
Its nucleus is active but with uncertain classification.
Initially, it was optically classified as a Sy\,2 \citep{shuder_empirical_1981}, while its optical emission line ratios are more LINER-like \citep{ferrarese_nuclear_1999}, and the detection of a broad H$\alpha$ component would lead to a Sy\,1 classification (see discussion in \citealt{shi_unobscured_2010}).
At radio wavelengths, the nucleus is unresolved with a one-sided narrow jet extending on kiloparsec scales toward the north-west (PA$\sim296\degree$; \citealt{perley_high-resolution_1984,jones_high_1986}).
A highly inclined dust disc with a major axis diameter of $\sim1.4\arcsec\sim0.7\,$kpc surrounds the nucleus (PA$\sim0\degree$; \citealt{ferrarese_nuclear_1999}).
NGC\,6251 was first observed in the MIR with \irass and later followed up with \spitzer/IRAC, IRS and MIPS.
The corresponding IRAC and MIPS images show a compact nucleus embedded within diffuse host emission with no signs of jet emission.
Our nuclear MIPS 24$\,\mu$m photometry is consistent with \cite{shi_unobscured_2010}.
The IRS LR staring-mode spectrum exhibits strong silicate 10 and 18\,$\mu$m emission, a very weak PAH 11.3\,$\mu$m feature and a shallow blue spectral slope in $\nu F_\nu$-space (see also \citealt{leipski_spitzer_2009,shi_unobscured_2010}).
Thus, the arcsecond-scale MIR SED indicates the presence of large amounts of warm dust and supports the unobscured LINER or type~I scenario.
The nuclear region of NGC\,6251 was imaged with COMICS in the N11.7 filter in 2009 (unpublished, to our knowledge).  
In the image, an elongated nucleus was detected without any further emission (FWHM(major axis)$\sim 0.57\arcsec$; PA$\sim143\degree$).
We consider it likely that this apparent extension is in fact an artefact, because only a single low S/N image taken at very high airmass is available, and the nuclear source morphology is suspicious while not matching any other wavelength.
Therefore, the nuclear extension of NGC\,6251 remains uncertain at subarcsecond scales in the MIR.
The nuclear N11.7 flux is $\sim62\%$ lower than the \spitzerr spectrophotometry.
From the current data, no inference about the properties of the nuclear MIR emission can be made.
\newline\end{@twocolumnfalse}]

\begin{figure}
   \centering
   \includegraphics[angle=0,width=8.500cm]{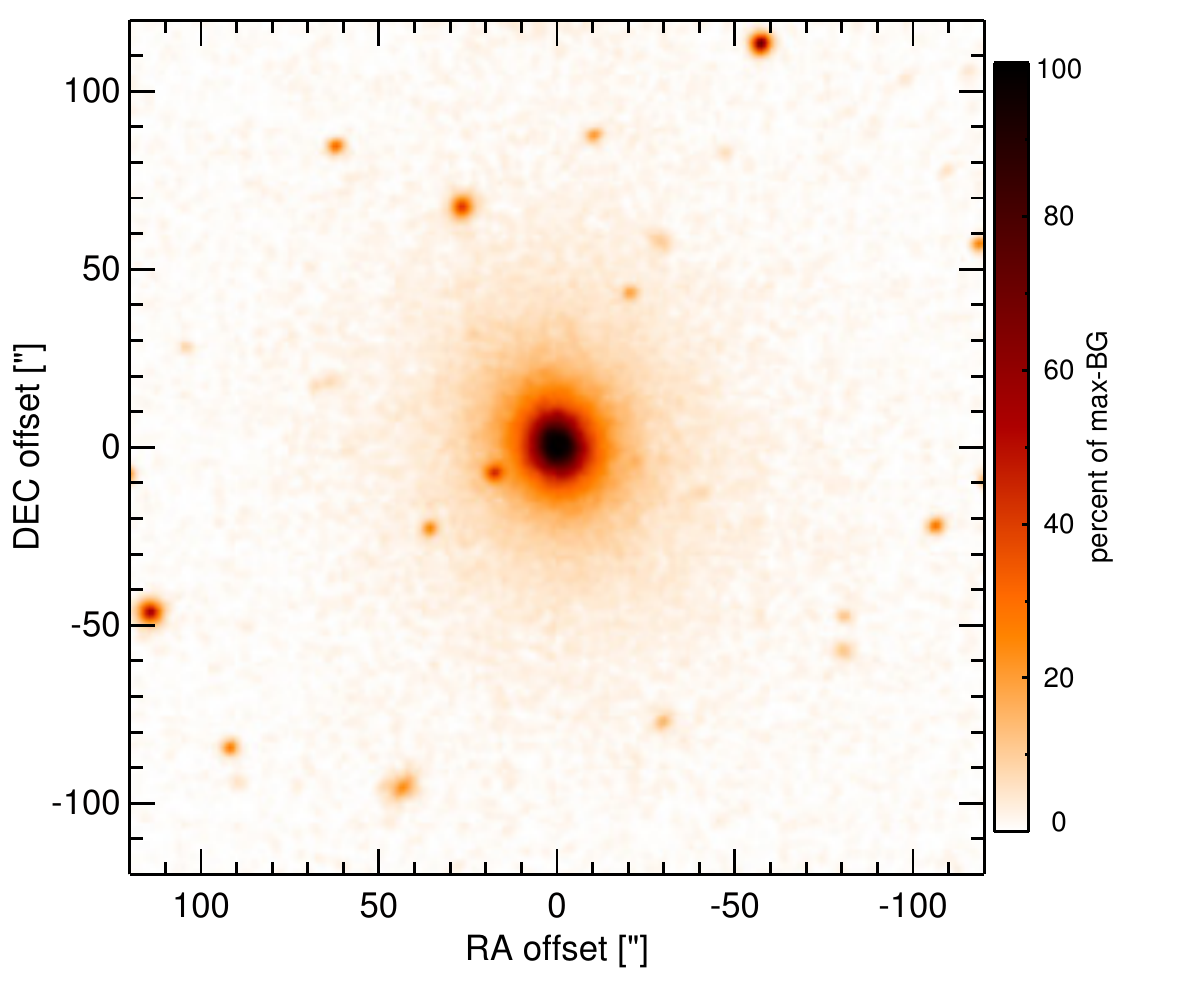}
    \caption{\label{fig:OPTim_NGC6251}
             Optical image (DSS, red filter) of NGC\,6251. Displayed are the central $4\arcmin$ with North up and East to the left. 
              The colour scaling is linear with white corresponding to the median background and black to the $0.01\%$ pixels with the highest intensity.  
           }
\end{figure}
\begin{figure}
   \centering
   \includegraphics[angle=0,height=3.11cm]{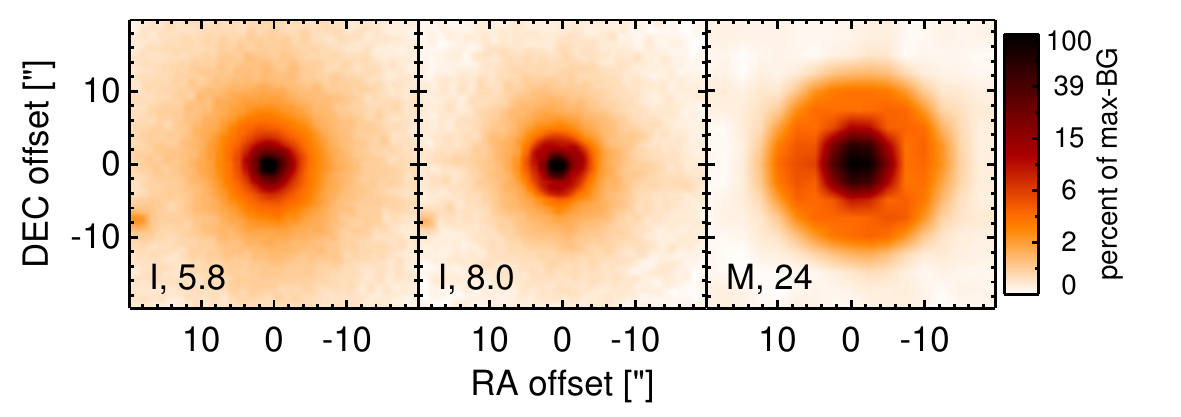}
    \caption{\label{fig:INTim_NGC6251}
             \spitzerr MIR images of NGC\,6251. Displayed are the inner $40\arcsec$ with North up and East to the left. The colour scaling is logarithmic with white corresponding to median background and black to the $0.1\%$ pixels with the highest intensity.
             The label in the bottom left states instrument and central wavelength of the filter in $\mu$m (I: IRAC, M: MIPS). 
           }
\end{figure}
\begin{figure}
   \centering
   \includegraphics[angle=0,height=3.11cm]{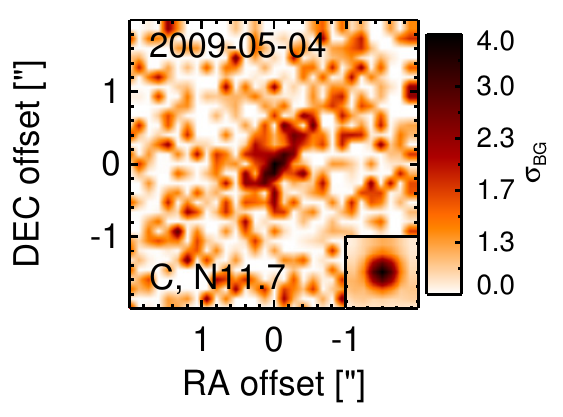}
    \caption{\label{fig:HARim_NGC6251}
             Subarcsecond-resolution MIR images of NGC\,6251 sorted by increasing filter wavelength. 
             Displayed are the inner $4\arcsec$ with North up and East to the left. 
             The colour scaling is logarithmic with white corresponding to median background and black to the $75\%$ of the highest intensity of all images in units of $\sigbg$.
             The inset image shows the central arcsecond of the PSF from the calibrator star, scaled to match the science target.
             The labels in the bottom left state instrument and filter names (C: COMICS, M: Michelle, T: T-ReCS, V: VISIR).
           }
\end{figure}
\begin{figure}
   \centering
   \includegraphics[angle=0,width=8.50cm]{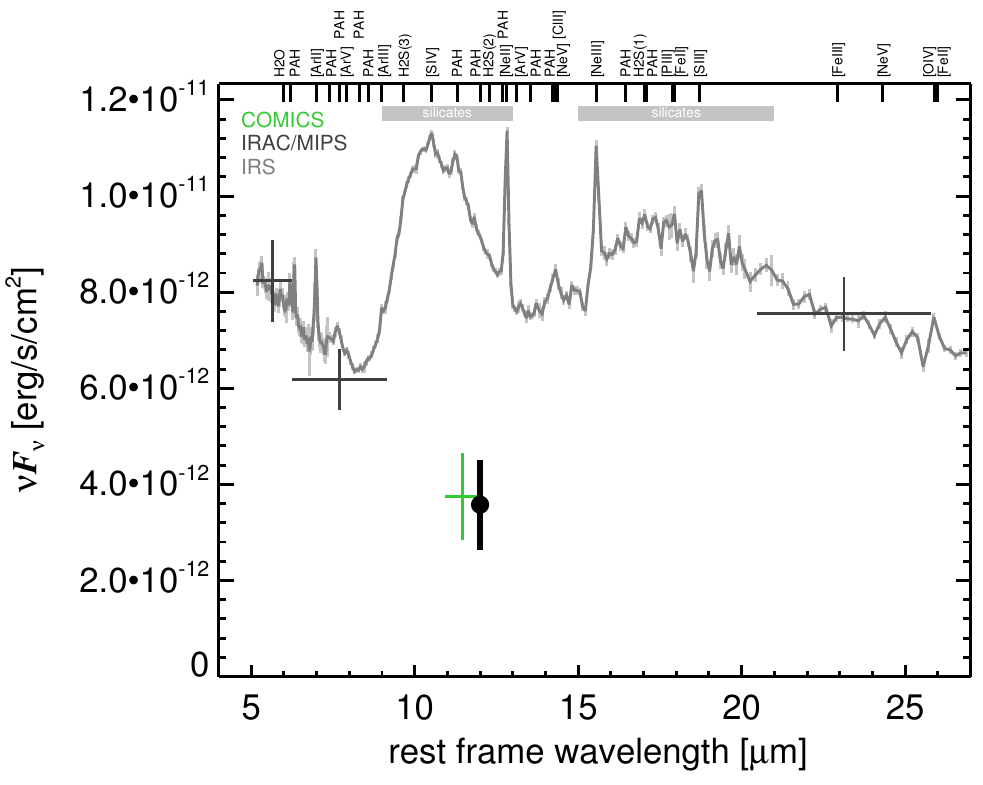}
   \caption{\label{fig:MISED_NGC6251}
      MIR SED of NGC\,6251. The description  of the symbols (if present) is the following.
      Grey crosses and  solid lines mark the \spitzer/IRAC, MIPS and IRS data. 
      The colour coding of the other symbols is: 
      green for COMICS, magenta for Michelle, blue for T-ReCS and red for VISIR data.
      Darker-coloured solid lines mark spectra of the corresponding instrument.
      The black filled circles mark the nuclear 12 and $18\,\mu$m  continuum emission estimate from the data.
      The ticks on the top axis mark positions of common MIR emission lines, while the light grey horizontal bars mark wavelength ranges affected by the silicate 10 and 18$\mu$m features.}
\end{figure}
\clearpage

\twocolumn[\begin{@twocolumnfalse}  
\subsection{NGC\,6300}\label{app:NGC6300}
NGC\,6300 is a low-inclined barred spiral galaxy at a low Galactic latitude and a distance of $D=$ $14.3 \pm 3.3\,$Mpc (NED redshift-independent median) hosting a Sy\,2 nucleus \citep{veron-cetty_catalogue_2010}.
It features a slightly resolved radio core at arcsecond resolution with extension in southern direction (PA$\sim190\degree$; \citealt{morganti_radio_1999}) and water maser emission \citep{greenhill_discovery_2003}.
After \iras, NGC\,6300 was observed with \isoo \citep{bendo_infrared_2002} and \spitzer/IRAC, IRS and MIPS.
The corresponding IRAC and MIPS images show a compact nucleus surrounded by spiral-like host emission and a few foreground stars, which are brightest at shortest wavelengths.
The IRS LR staring-mode spectrum exhibits deep silicate 10 and 18$\,\mu$m absorption, very weak PAH emission, and a steep red spectral slope in $\nu F_\nu$-space (see also \citealt{goulding_towards_2009,pereira-santaella_mid-infrared_2010}).
Thus, the arcsecond-scale MIR SED  does not appear to be affected by significant star formation.
We observed the nuclear region of NGC\,6300 with VISIR in four narrow $N$-band filters in 2008 (partly published in \citealt{gandhi_resolving_2009}), and detected an elongated nucleus (FWHM(major axis)$\sim 0.46\arcsec\sim 32\,$pc; PA$\sim111\degree$) without further host emission in all images.
However, at least a second epoch of subarcsecond MIR imaging is required to verify this morphology.
The nuclear photometry is on average $\sim 37\%$ lower than the \spitzerr spectrophotometry and exhibits the same silicate 10\,$\mu$m absorption depth.
Therefore, the silicate absorption originates in the projected central $\sim25\,$pc of NGC\,6300, and we use the IRS spectrum to compute the 12\,$\mu$m continuum emission estimate corrected for the silicate feature.
\newline\end{@twocolumnfalse}]

\begin{figure}
   \centering
   \includegraphics[angle=0,width=8.500cm]{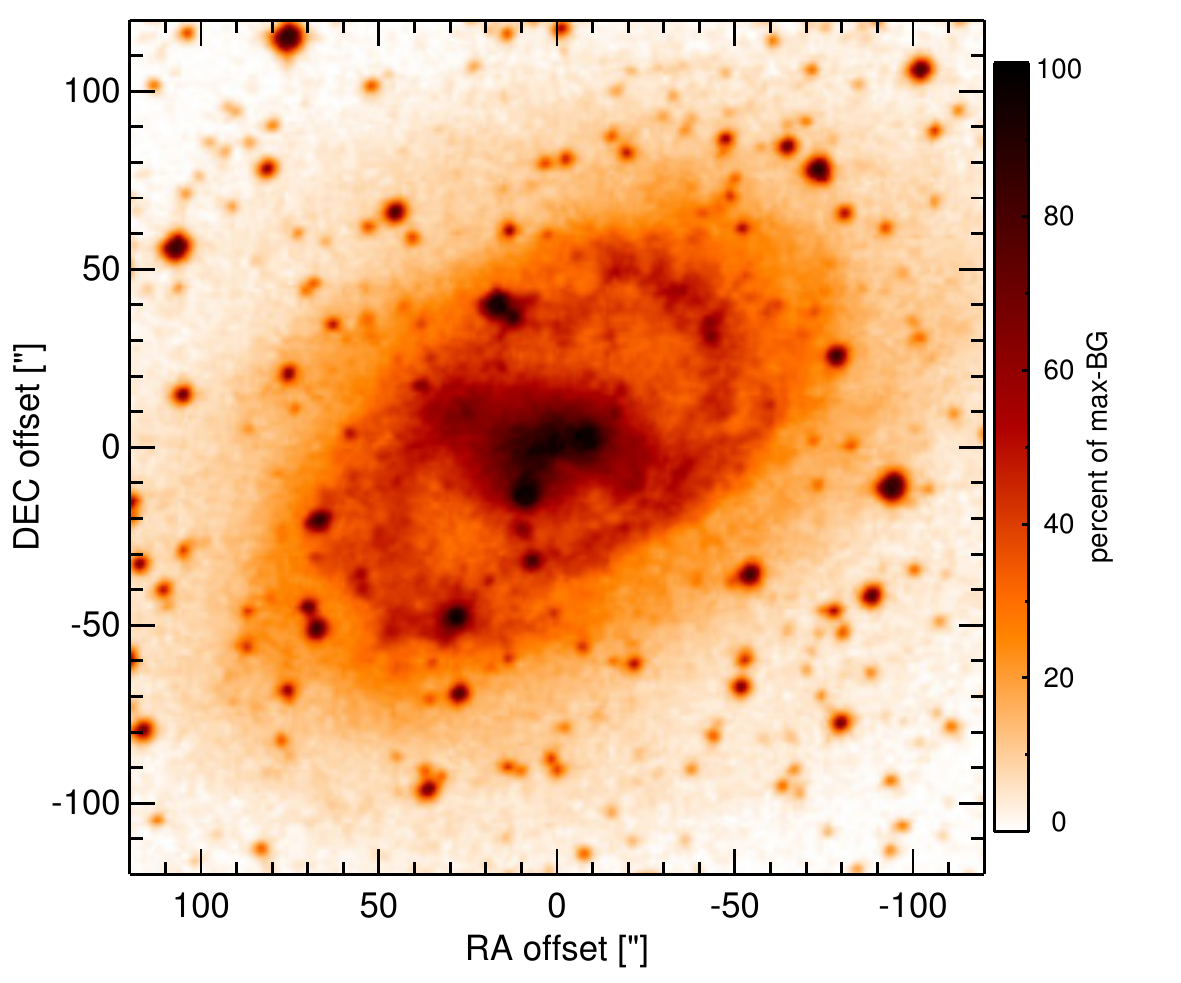}
    \caption{\label{fig:OPTim_NGC6300}
             Optical image (DSS, red filter) of NGC\,6300. Displayed are the central $4\arcmin$ with North up and East to the left. 
              The colour scaling is linear with white corresponding to the median background and black to the $0.01\%$ pixels with the highest intensity.  
           }
\end{figure}
\begin{figure}
   \centering
   \includegraphics[angle=0,height=3.11cm]{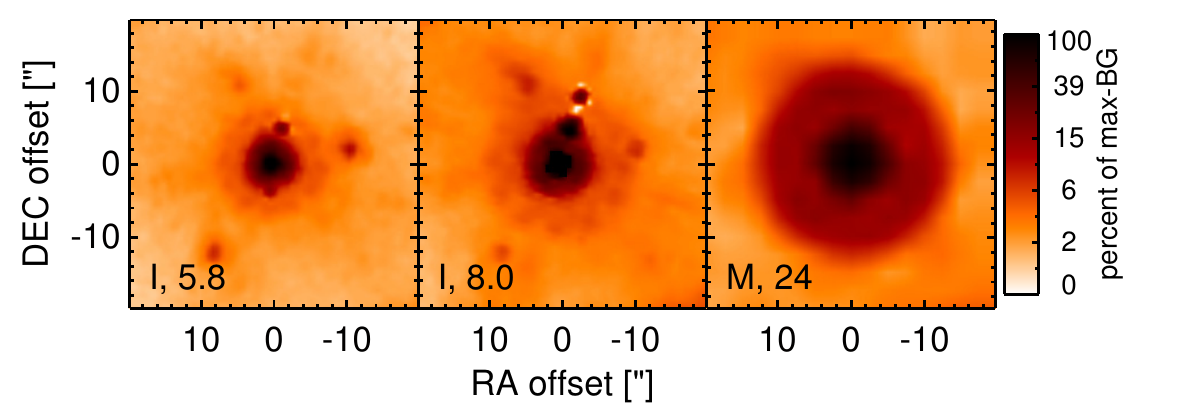}
    \caption{\label{fig:INTim_NGC6300}
             \spitzerr MIR images of NGC\,6300. Displayed are the inner $40\arcsec$ with North up and East to the left. The colour scaling is logarithmic with white corresponding to median background and black to the $0.1\%$ pixels with the highest intensity.
             The label in the bottom left states instrument and central wavelength of the filter in $\mu$m (I: IRAC, M: MIPS).
             Note that the apparent off-nuclear compact sources in the IRAC 5.8 and  $8.0\,\mu$m images are instrumental artefacts.
           }
\end{figure}
\begin{figure}
   \centering
   \includegraphics[angle=0,width=8.500cm]{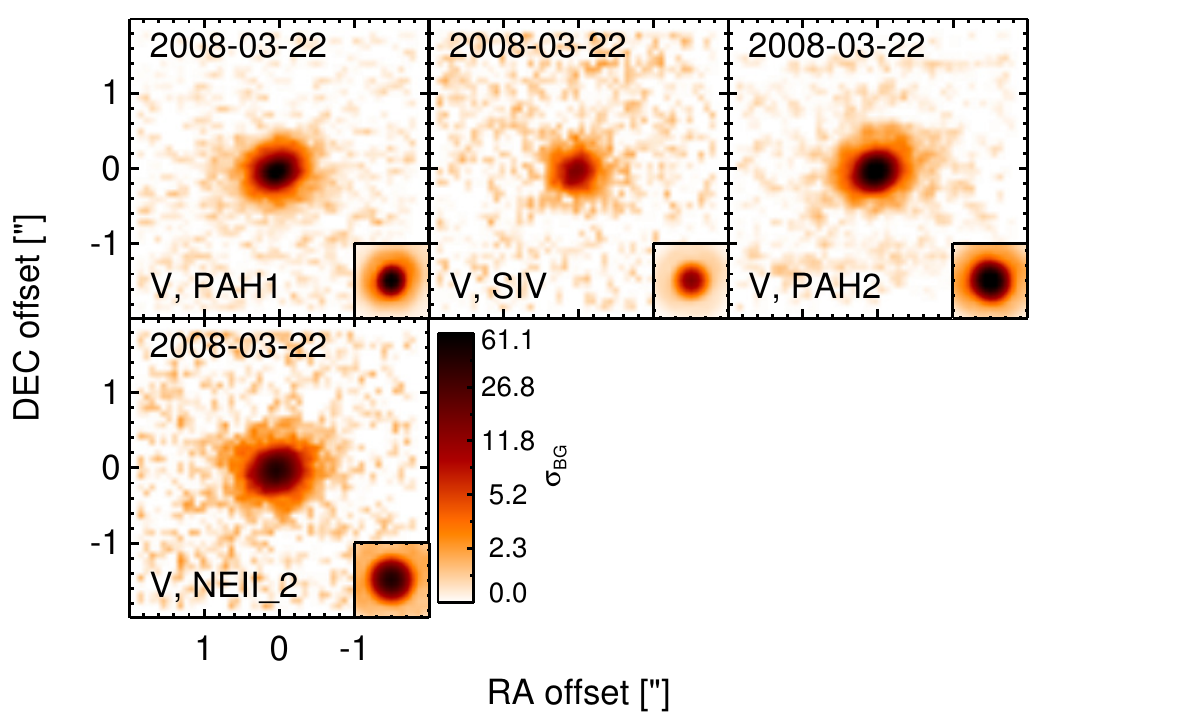}
    \caption{\label{fig:HARim_NGC6300}
             Subarcsecond-resolution MIR images of NGC\,6300 sorted by increasing filter wavelength. 
             Displayed are the inner $4\arcsec$ with North up and East to the left. 
             The colour scaling is logarithmic with white corresponding to median background and black to the $75\%$ of the highest intensity of all images in units of $\sigbg$.
             The inset image shows the central arcsecond of the PSF from the calibrator star, scaled to match the science target.
             The labels in the bottom left state instrument and filter names (C: COMICS, M: Michelle, T: T-ReCS, V: VISIR).
           }
\end{figure}
\begin{figure}
   \centering
   \includegraphics[angle=0,width=8.50cm]{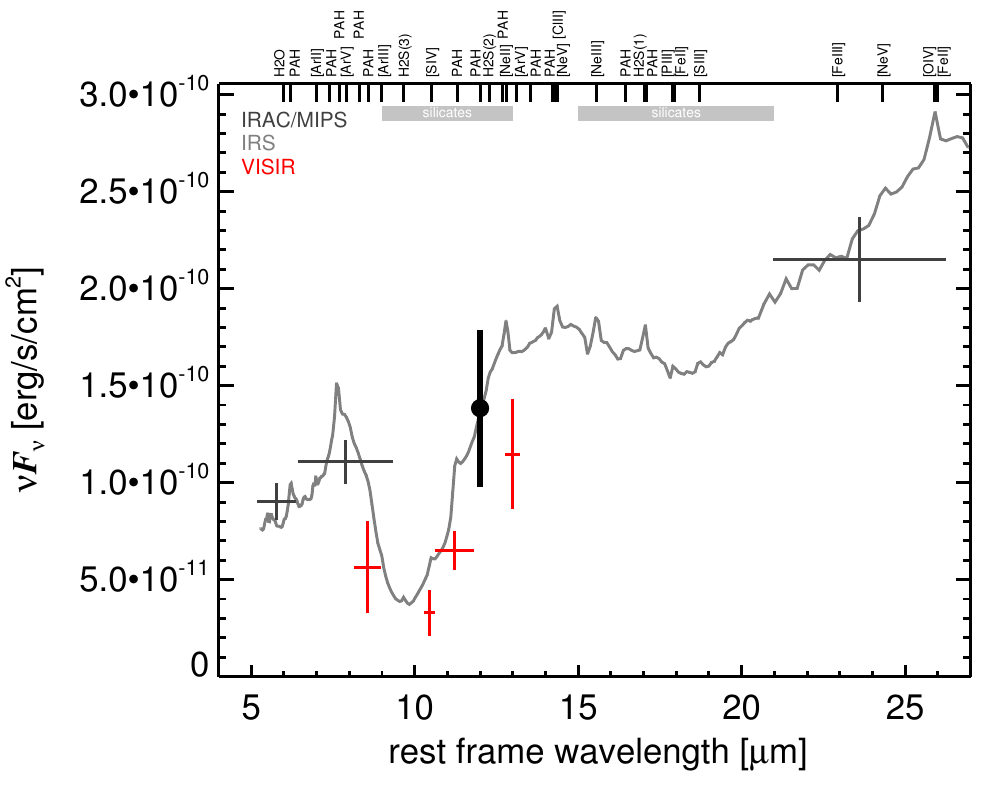}
   \caption{\label{fig:MISED_NGC6300}
      MIR SED of NGC\,6300. The description  of the symbols (if present) is the following.
      Grey crosses and  solid lines mark the \spitzer/IRAC, MIPS and IRS data. 
      The colour coding of the other symbols is: 
      green for COMICS, magenta for Michelle, blue for T-ReCS and red for VISIR data.
      Darker-coloured solid lines mark spectra of the corresponding instrument.
      The black filled circles mark the nuclear 12 and $18\,\mu$m  continuum emission estimate from the data.
      The ticks on the top axis mark positions of common MIR emission lines, while the light grey horizontal bars mark wavelength ranges affected by the silicate 10 and 18$\mu$m features.}
\end{figure}
\clearpage

\twocolumn[\begin{@twocolumnfalse}  
\subsection{NGC\,6810}\label{app:NGC6810}
NGC\,6810 is an edge-on spiral galaxy at a distance of $D=$ $28.6 \pm 5.2$\,Mpc (NED redshift-independent median) with a poorly studied active nucleus.
Initially, it was classified optically as a Sy\,2 \citep{kirhakos_x-ray_1990}, while it turned out later to be either an H\,II or H\,II/AGN transition or composite nucleus \citep{strickland_new_2007,brightman_nature_2008,brightman_xmm-newton_2011-1,yuan_role_2010}.
In particular, \cite{strickland_new_2007} find no evidence for an AGN in X-rays and based on other diagnostics.
No compact radio source was detected in the nucleus either \citep{kewley_compact_2000}.
Kiloparsec-scale H$\alpha$ outflow filaments along the minor galaxy axis have been detected \citep{hameed_h_1999}.
We conservatively treat NGC\,6810 as an uncertain AGN.
After being detected in the MIR with \irass for the first time, NGC\,6810 was followed up with \spitzer/IRAC, IRS and MIPS.
The corresponding IRAC images show a nucleus elongated in north-south direction and embedded within complex host emission. The nucleus remains mostly unresolved in the MIPS 24\,$\mu$m images.
Particularly bright knots of star-formation are seen $\sim10\arcsec\sim1.4\,$kpc to the south-east and $\sim11.5\arcsec\sim1.6\,$kpc to the north-west. The nucleus and both knots are aligned roughly along PA$\sim146\degree=326\degree$.
Our nuclear IRAC 5.8 and 8.0$\,\mu$m photometry is significantly lower than the values by \cite{gallimore_infrared_2010} but consistent with the IRS mapping-mode post-BCD spectrum.
The latter exhibits strong PAH emission, possibly silicate 10$\,\mu$m absorption and a steep red spectral slope in $\nu F_\nu$-space (see also \citealt{buchanan_spitzer_2006,wu_spitzer/irs_2009,tommasin_spitzer-irs_2010,gallimore_infrared_2010}).
Thus, the arcsecond-scale MIR SED  appears to be completely star-formation dominated.
Note that no AGN-indicative \nev emission was detected in the IRS spectrum (e.g. \citealt{tommasin_spitzer-irs_2010}).
The nuclear region of NGC\,6810 was imaged with T-ReCS in the broad N filter in 2004 \citep{videla_nuclear_2013}, and a nucleus extended in north-south direction was detected (FWHM(major axis)$\sim0.77\arcsec\sim107\,$pc; PA$\sim164\degree$).
The slightly S-shaped extended emission has a total diameter of at least $\sim4\arcsec\sim0.6\,$kpc along the major axis, and the south-eastern star-formation knots are faintly visible as well.
We fit only the unresolved nuclear component with manual PSF-scaling so that the residual emission becomes smooth.
The resulting flux is $\sim92\%$ lower than the \spitzerr spectrophotometry and also $\sim30\%$ lower than the value published by \cite{videla_nuclear_2013}.
Note, however, that the flux measurement method is very uncertain, and applying slightly different parameters results in flux uncertainties of $\sim$30\%.
The nuclear flux might still be affected or dominated by star-formation.
Therefore, there is no evidence for an AGN in NGC\,6810 from the MIR point of view.
\newline\end{@twocolumnfalse}]

\begin{figure}
   \centering
   \includegraphics[angle=0,width=8.500cm]{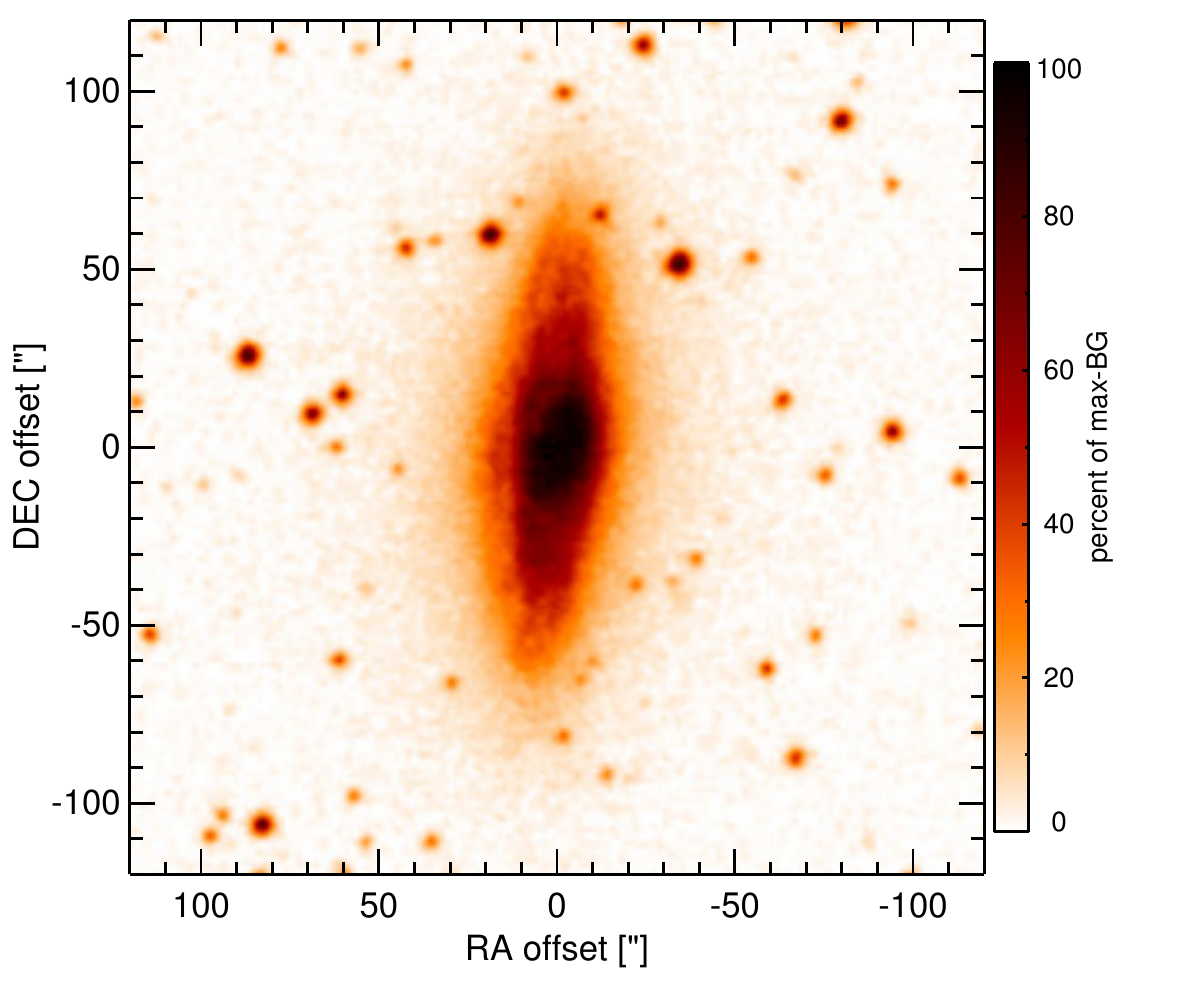}
    \caption{\label{fig:OPTim_NGC6810}
             Optical image (DSS, red filter) of NGC\,6810. Displayed are the central $4\arcmin$ with North up and East to the left. 
              The colour scaling is linear with white corresponding to the median background and black to the $0.01\%$ pixels with the highest intensity.  
           }
\end{figure}
\begin{figure}
   \centering
   \includegraphics[angle=0,height=3.11cm]{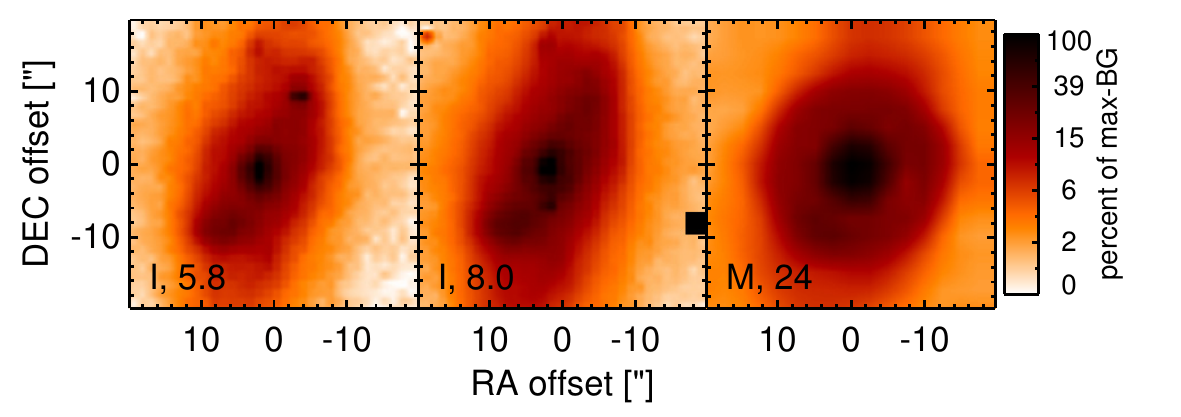}
    \caption{\label{fig:INTim_NGC6810}
             \spitzerr MIR images of NGC\,6810. Displayed are the inner $40\arcsec$ with North up and East to the left. The colour scaling is logarithmic with white corresponding to median background and black to the $0.1\%$ pixels with the highest intensity.
             The label in the bottom left states instrument and central wavelength of the filter in $\mu$m (I: IRAC, M: MIPS).
             Note that the apparent off-nuclear compact sources in the IRAC 5.8 and  $8.0\,\mu$m images are instrumental artefacts.
           }
\end{figure}
\begin{figure}
   \centering
   \includegraphics[angle=0,height=3.11cm]{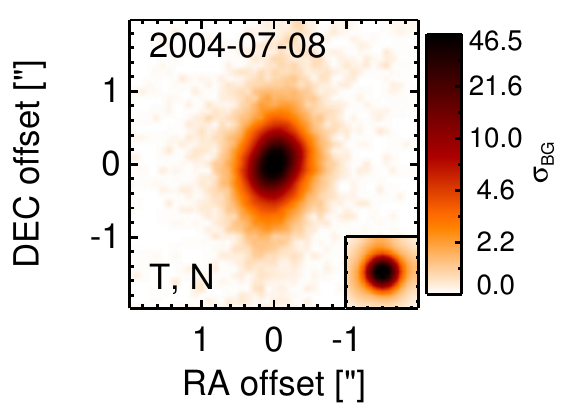}
    \caption{\label{fig:HARim_NGC6810}
             Subarcsecond-resolution MIR images of NGC\,6810 sorted by increasing filter wavelength. 
             Displayed are the inner $4\arcsec$ with North up and East to the left. 
             The colour scaling is logarithmic with white corresponding to median background and black to the $75\%$ of the highest intensity of all images in units of $\sigbg$.
             The inset image shows the central arcsecond of the PSF from the calibrator star, scaled to match the science target.
             The labels in the bottom left state instrument and filter names (C: COMICS, M: Michelle, T: T-ReCS, V: VISIR).
           }
\end{figure}
\begin{figure}
   \centering
   \includegraphics[angle=0,width=8.50cm]{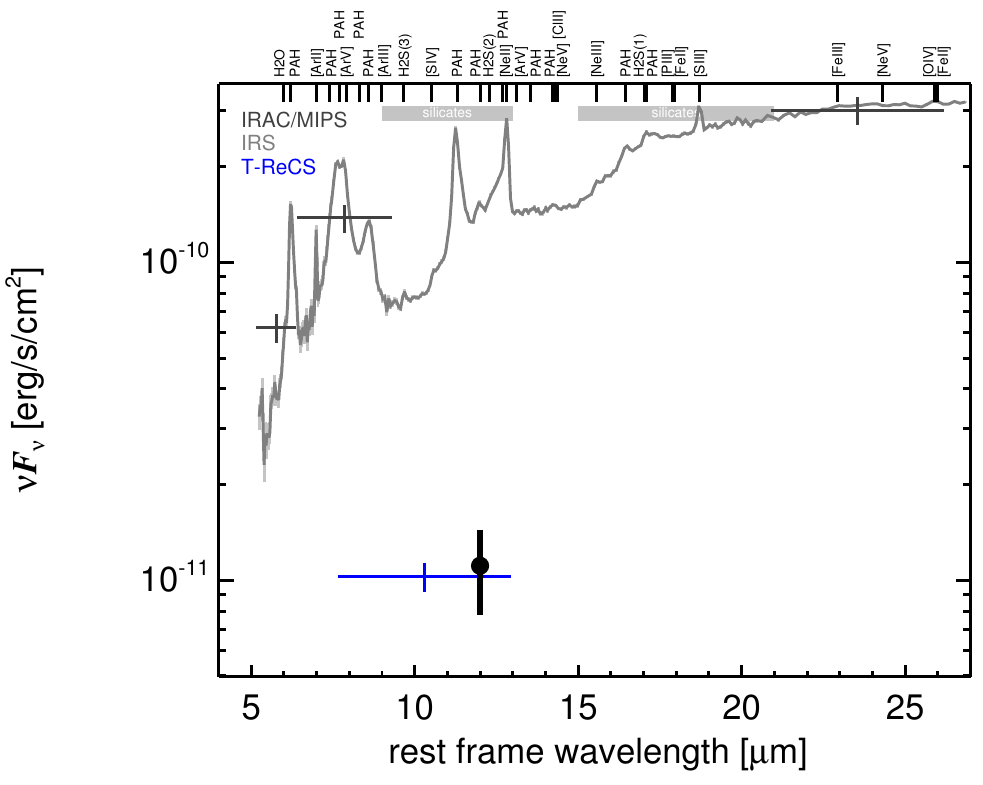}
   \caption{\label{fig:MISED_NGC6810}
      MIR SED of NGC\,6810. The description  of the symbols (if present) is the following.
      Grey crosses and  solid lines mark the \spitzer/IRAC, MIPS and IRS data. 
      The colour coding of the other symbols is: 
      green for COMICS, magenta for Michelle, blue for T-ReCS and red for VISIR data.
      Darker-coloured solid lines mark spectra of the corresponding instrument.
      The black filled circles mark the nuclear 12 and $18\,\mu$m  continuum emission estimate from the data.
      The ticks on the top axis mark positions of common MIR emission lines, while the light grey horizontal bars mark wavelength ranges affected by the silicate 10 and 18$\mu$m features.}
\end{figure}
\clearpage

\twocolumn[\begin{@twocolumnfalse}  
\subsection{NGC\,6814}\label{app:NGC6814}
NGC\,6814 is a face-on grand-design spiral galaxy at a redshift of $z=$ 0.0052 ($D\sim21.2\,$Mpc) with a Sy\,1.5 nucleus \citep{veron-cetty_catalogue_2010}. This nucleus is variable by one order of magnitude variable in X-rays \citep{koenig_seyfert_1997,mukai_x-ray_2003} and belongs to the nine-month BAT AGN sample.
In addition, optical type changes from Sy\,1.0 to Sy\,1.9 have been reported \citep{sekiguchi_spectral_1990}.
It features a slightly resolved radio core with north-south extension of $\sim 4\arcsec\sim0.4\,$kpc and a minor-axis extension of $\sim2\arcsec\sim$200\,pc to the west \citep{ulvestad_radio_1984}. Additionally, extended \oiii emission $\sim1\arcsec\sim100\,$pc along a PA$\sim150\degree$ has been reported\citep{schmitt_comparison_1996}.
Pioneering MIR observations of NGC\,6814 were performed by \cite{kleinmann_observations_1970}, followed by \cite{rieke_infrared_1972,rieke_infrared_1978} and \cite{glass_mid-infrared_1982}.
After \iras, NGC\,6814 was followed up in the MIR with \isoo \citep{laurent_mid-infrared_2000} and \spitzer/IRAC, IRS and MIPS.
The corresponding IRAC and MIPS images show a dominating compact nucleus embedded within spiral-like host emission.
The IRS LR staring-mode spectrum exhibits silicate 10 and 18$\mu$m emission, very weak PAH features, and a flat spectral slope in $\nu F_\nu$-space (see also \citealt{pereira-santaella_mid-infrared_2010}).
Thus, the arcsecond-scale MIR SED  appears to be AGN-dominated with at best minor star-formation contribution.
We observed the nuclear region of NGC\,6814 with VISIR in four narrow $N$-band filters in 2007 (partly published in \citealt{gandhi_resolving_2009}).
In addition, T-ReCS imaging in the Si2 and Qa filters was performed in 2009 \citep{ramos_almeida_testing_2011}.
In all cases, a compact nucleus without further host emission was detected.
The nucleus is unresolved in the T-ReCS images, while it appears elongated in all the single-epoch VISIR images (PA$\sim100\degree$).
Because the T-ReCS images were taken during better MIR seeing conditions, we classify the nucleus as unresolved.
Our nuclear photometry is consistent with the previously published values of the same data and also with the \spitzerr spectrophotometry.
Therefore, we use the latter to compute the 12 and 18$\,\mu$m continuum emission estimates corrected for the silicate features.
The combination of all available MIR data of NGC\,6814 is consistent with no $N$-band flux variations during the last $\sim40\,$years, except for the very first measurement by \cite{kleinmann_observations_1970} which might be caused by calibration systematics.
\newline\end{@twocolumnfalse}]

\begin{figure}
   \centering
   \includegraphics[angle=0,width=8.500cm]{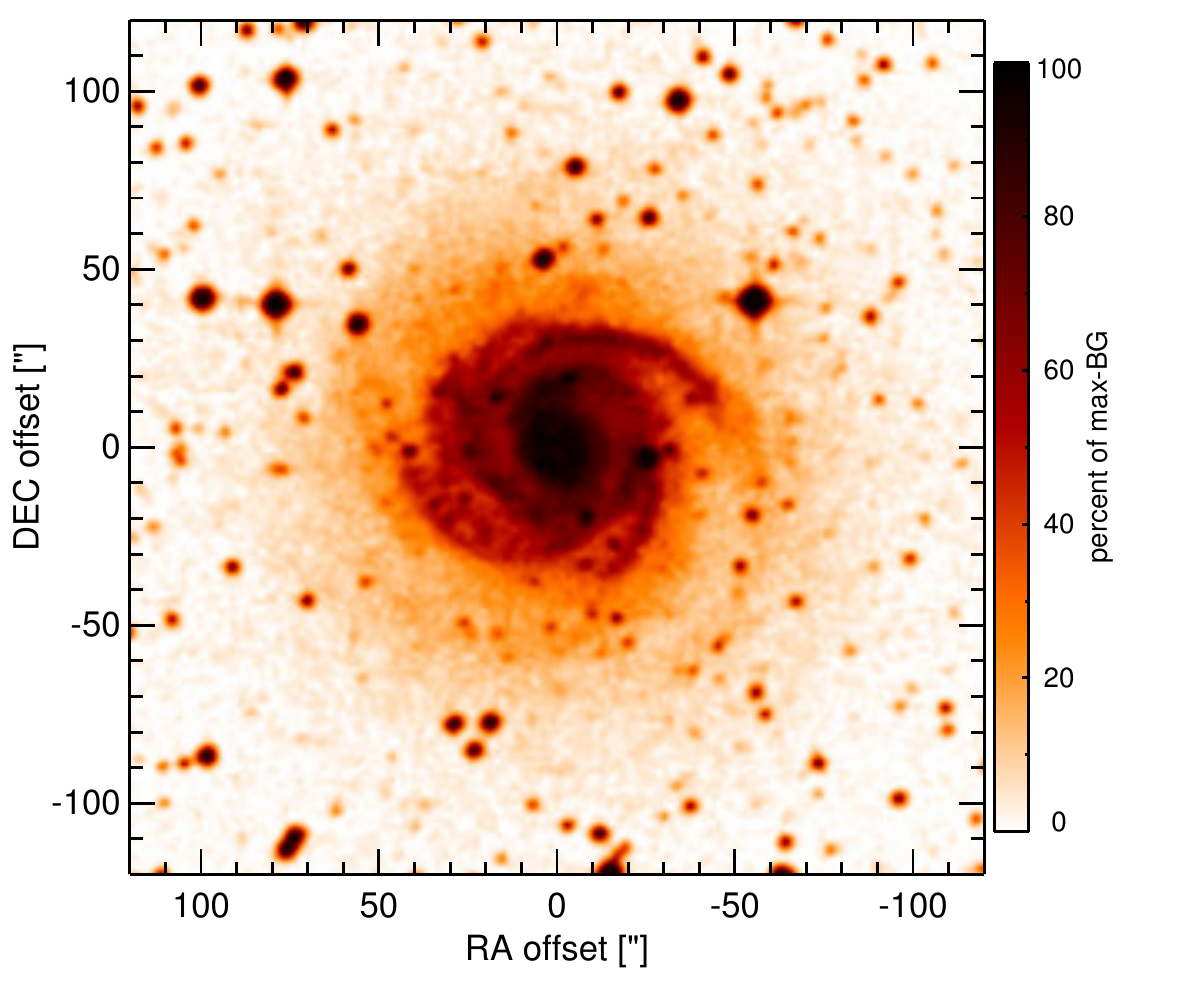}
    \caption{\label{fig:OPTim_NGC6814}
             Optical image (DSS, red filter) of NGC\,6814. Displayed are the central $4\arcmin$ with North up and East to the left. 
              The colour scaling is linear with white corresponding to the median background and black to the $0.01\%$ pixels with the highest intensity.  
           }
\end{figure}
\begin{figure}
   \centering
   \includegraphics[angle=0,height=3.11cm]{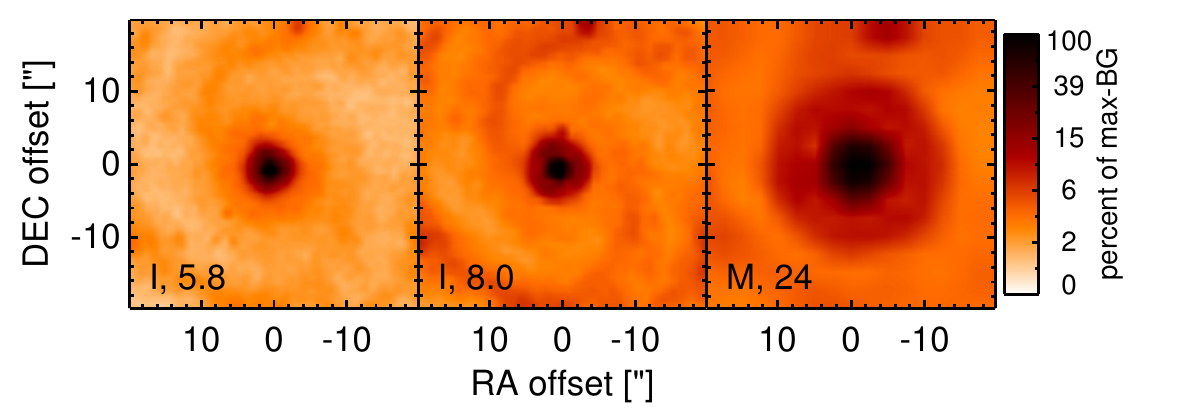}
    \caption{\label{fig:INTim_NGC6814}
             \spitzerr MIR images of NGC\,6814. Displayed are the inner $40\arcsec$ with North up and East to the left. The colour scaling is logarithmic with white corresponding to median background and black to the $0.1\%$ pixels with the highest intensity.
             The label in the bottom left states instrument and central wavelength of the filter in $\mu$m (I: IRAC, M: MIPS). 
           }
\end{figure}
\begin{figure}
   \centering
   \includegraphics[angle=0,width=8.500cm]{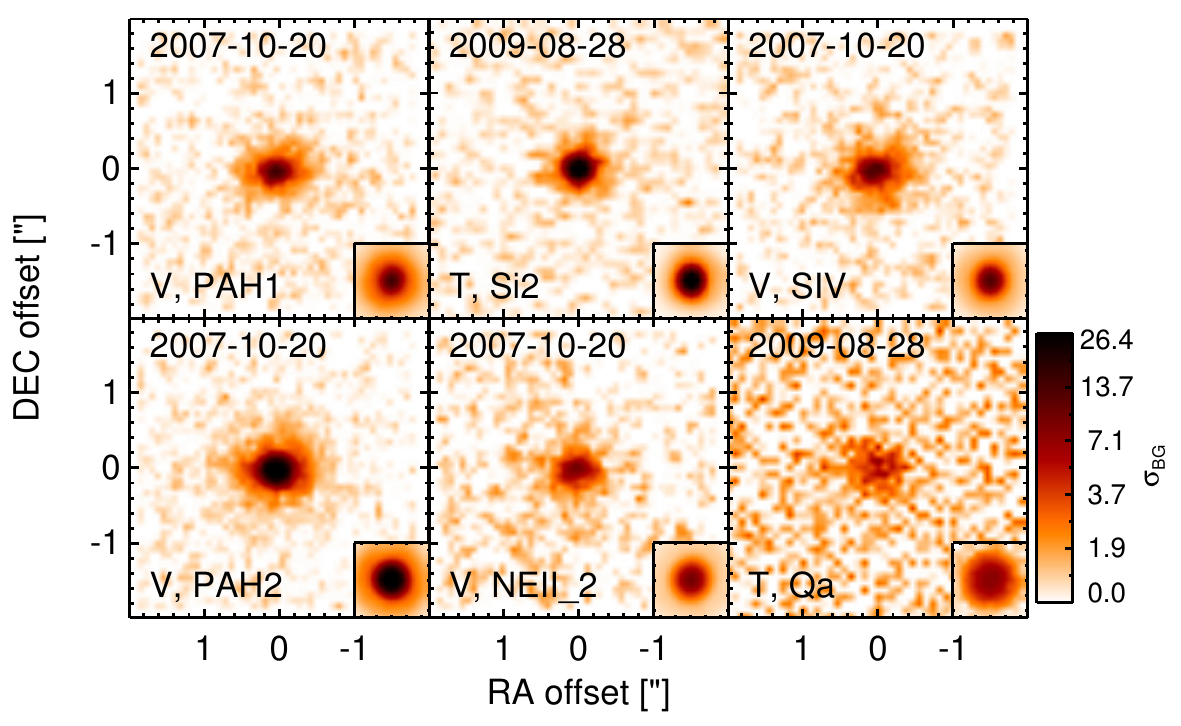}
    \caption{\label{fig:HARim_NGC6814}
             Subarcsecond-resolution MIR images of NGC\,6814 sorted by increasing filter wavelength. 
             Displayed are the inner $4\arcsec$ with North up and East to the left. 
             The colour scaling is logarithmic with white corresponding to median background and black to the $75\%$ of the highest intensity of all images in units of $\sigbg$.
             The inset image shows the central arcsecond of the PSF from the calibrator star, scaled to match the science target.
             The labels in the bottom left state instrument and filter names (C: COMICS, M: Michelle, T: T-ReCS, V: VISIR).
           }
\end{figure}
\begin{figure}
   \centering
   \includegraphics[angle=0,width=8.50cm]{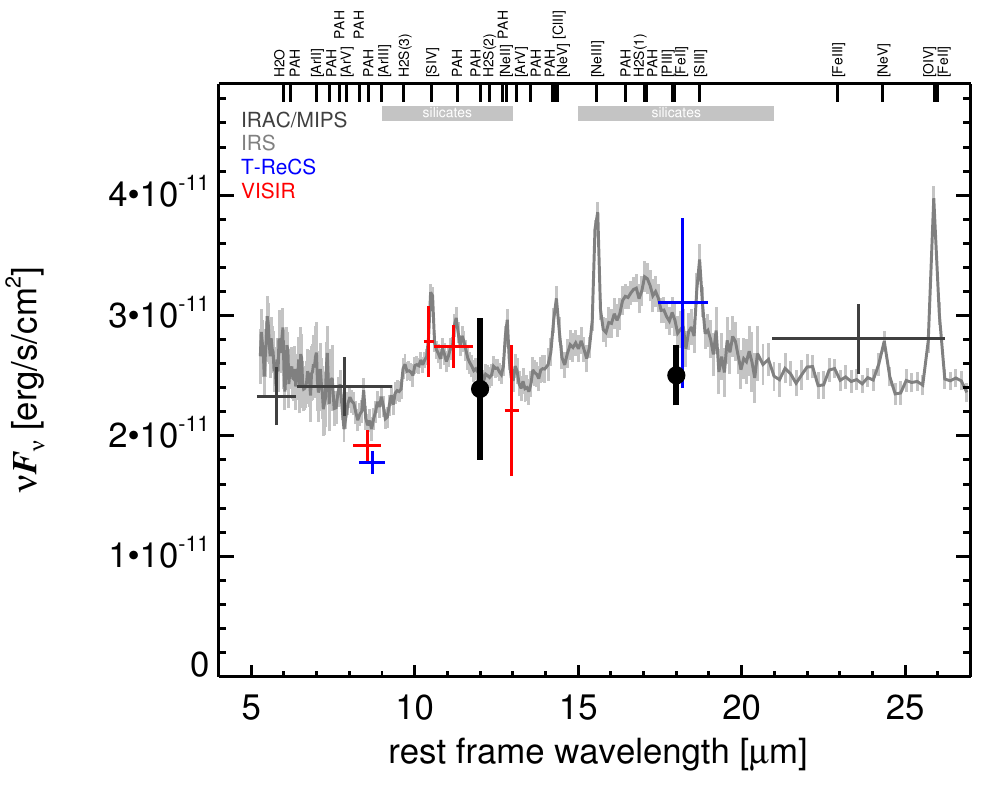}
   \caption{\label{fig:MISED_NGC6814}
      MIR SED of NGC\,6814. The description  of the symbols (if present) is the following.
      Grey crosses and  solid lines mark the \spitzer/IRAC, MIPS and IRS data. 
      The colour coding of the other symbols is: 
      green for COMICS, magenta for Michelle, blue for T-ReCS and red for VISIR data.
      Darker-coloured solid lines mark spectra of the corresponding instrument.
      The black filled circles mark the nuclear 12 and $18\,\mu$m  continuum emission estimate from the data.
      The ticks on the top axis mark positions of common MIR emission lines, while the light grey horizontal bars mark wavelength ranges affected by the silicate 10 and 18$\mu$m features.}
\end{figure}
\clearpage

\twocolumn[\begin{@twocolumnfalse}  
\subsection{NGC\,6860}\label{app:NGC6860}
NGC\,6860 is an inclined barred spiral galaxy at a redshift of $z=$ 0.0149 ($D\sim65.8\,$Mpc) with a Sy\,1.5 nucleus \citep{veron-cetty_catalogue_2010} and circum-nuclear star formation \citep{lipari_high-resolution_1993}.
It belongs to the nine-month BAT AGN sample and features nuclear east-west elongated \oiii emission ($\sim6\arcsec\sim2\,$kpc; PA$\sim90\degree$ \citealt{lipari_high-resolution_1993,schmitt_hubble_2003}).
No high-angular resolution radio observations are described in the literature.
NGC\,6860 was first observed in the MIR with \iras, which led to the discovery of its AGN.
Follow-up observations were performed with \spitzer/IRAC, IRS and MIPS.
The corresponding IRAC and MIPS images show a dominating compact nucleus embedded within weak host emission.
Our nuclear IRAC photometry is significantly lower than the values published in \cite{gallimore_infrared_2010} for unknown reasons.
The IRS LR mapping-mode spectrum suffers from low S/N but indicates weak silicate 10 and 18\,$\mu$m emission, weak PAH features, and a blue spectral slope in $\nu F_\nu$-space (see also \citealt{buchanan_spitzer_2006,wu_spitzer/irs_2009,tommasin_spitzer-irs_2010}).
Thus, the arcsecond-scale MIR SED appears to be marginally affected by star formation.
The nuclear region of NGC\,6860 was observed with T-ReCS in the Si2 and Qa filters in 2008 and 2007, respectively (unpublished, to our knowledge).
In addition, we observed it with VISIR in three narrow $N$-band filters in 2009.
A compact nucleus without further host emission was detected in all images, which is unresolved in all cases but Si2.
However, the latter observations suffer from sub-average MIR seeing and, thus, we classify the nucleus of NGC\,6860 as unresolved in the MIR.
The nuclear photometry is consistent with the \spitzerr spectrophotometry but has systematically $\sim 6\%$ higher flux levels.
This might indicate slight flux variations between 2004 and 2009.
\newline\end{@twocolumnfalse}]

\begin{figure}
   \centering
   \includegraphics[angle=0,width=8.500cm]{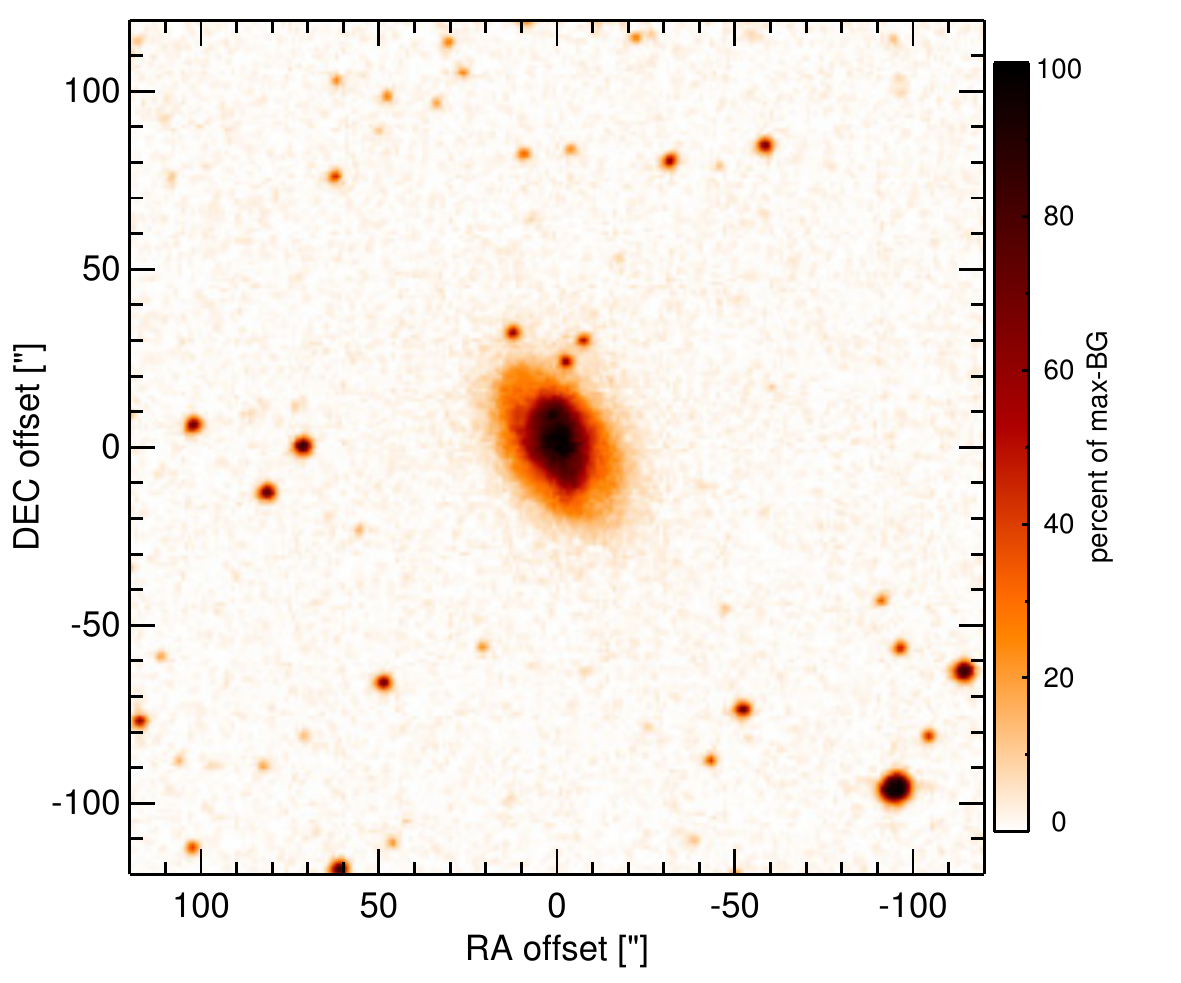}
    \caption{\label{fig:OPTim_NGC6860}
             Optical image (DSS, red filter) of NGC\,6860. Displayed are the central $4\arcmin$ with North up and East to the left. 
              The colour scaling is linear with white corresponding to the median background and black to the $0.01\%$ pixels with the highest intensity.  
           }
\end{figure}
\begin{figure}
   \centering
   \includegraphics[angle=0,height=3.11cm]{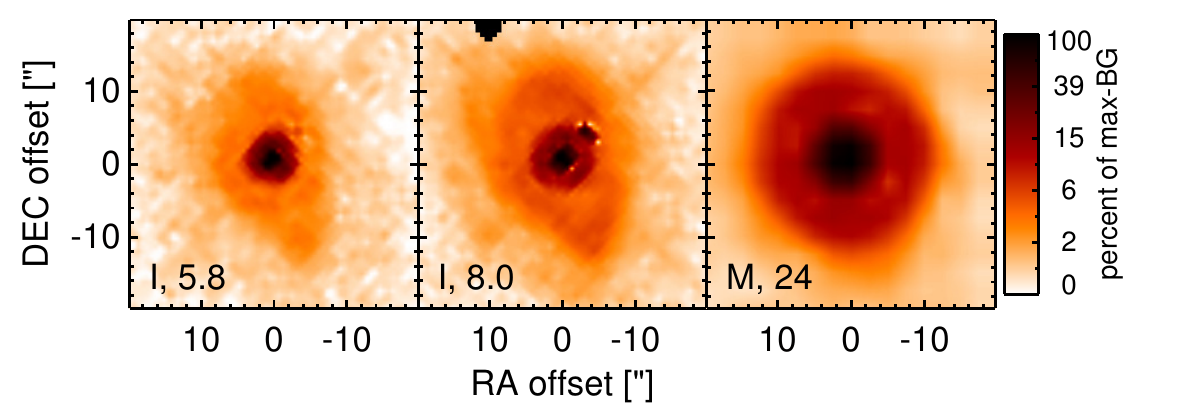}
    \caption{\label{fig:INTim_NGC6860}
             \spitzerr MIR images of NGC\,6860. Displayed are the inner $40\arcsec$ with North up and East to the left. The colour scaling is logarithmic with white corresponding to median background and black to the $0.1\%$ pixels with the highest intensity.
             The label in the bottom left states instrument and central wavelength of the filter in $\mu$m (I: IRAC, M: MIPS). 
             Note that the apparent off-nuclear compact source in the IRAC $8.0\,\mu$m image is an instrumental artefact.
           }
\end{figure}
\begin{figure}
   \centering
   \includegraphics[angle=0,width=8.500cm]{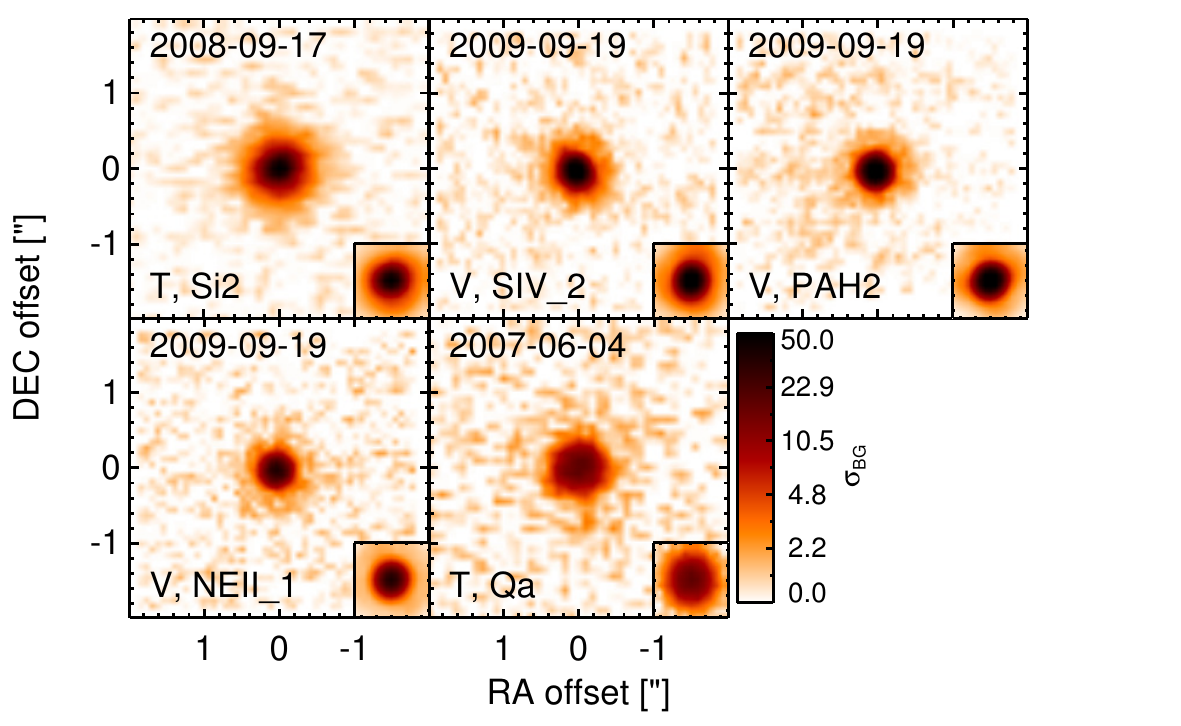}
    \caption{\label{fig:HARim_NGC6860}
             Subarcsecond-resolution MIR images of NGC\,6860 sorted by increasing filter wavelength. 
             Displayed are the inner $4\arcsec$ with North up and East to the left. 
             The colour scaling is logarithmic with white corresponding to median background and black to the $75\%$ of the highest intensity of all images in units of $\sigbg$.
             The inset image shows the central arcsecond of the PSF from the calibrator star, scaled to match the science target.
             The labels in the bottom left state instrument and filter names (C: COMICS, M: Michelle, T: T-ReCS, V: VISIR).
           }
\end{figure}
\begin{figure}
   \centering
   \includegraphics[angle=0,width=8.50cm]{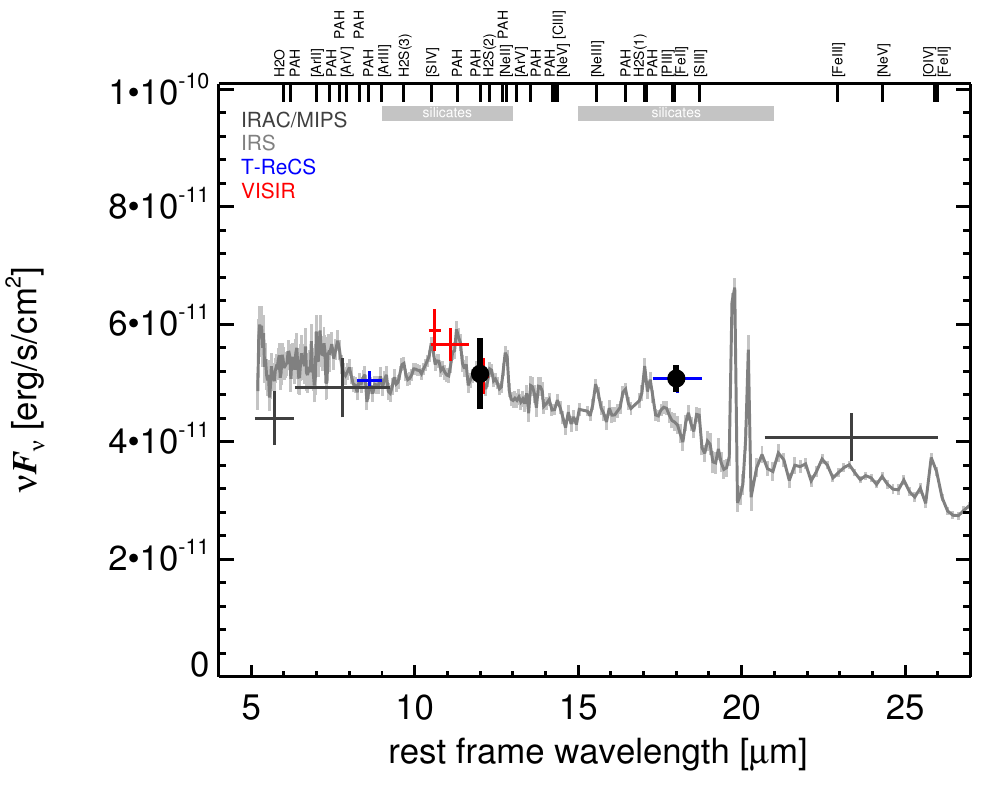}
   \caption{\label{fig:MISED_NGC6860}
      MIR SED of NGC\,6860. The description  of the symbols (if present) is the following.
      Grey crosses and  solid lines mark the \spitzer/IRAC, MIPS and IRS data. 
      The colour coding of the other symbols is: 
      green for COMICS, magenta for Michelle, blue for T-ReCS and red for VISIR data.
      Darker-coloured solid lines mark spectra of the corresponding instrument.
      The black filled circles mark the nuclear 12 and $18\,\mu$m  continuum emission estimate from the data.
      The ticks on the top axis mark positions of common MIR emission lines, while the light grey horizontal bars mark wavelength ranges affected by the silicate 10 and 18$\mu$m features.}
\end{figure}
\clearpage

\twocolumn[\begin{@twocolumnfalse}  
\subsection{NGC\,6890}\label{app:NGC6890}
NGC\,6890 is a low-inclination grand-design spiral galaxy at a redshift of $z=$ 0.0081($D\sim33.8\,$Mpc) with a Sy\,1.9/2 nucleus \citep{phillips_nearby_1983,storchi-bergmann_stellar_1990}.
The nucleus is detected but unresolved at arcsecond resolution at radio wavelengths \citep{ulvestad_radio_1989,morganti_radio_1999}.
After its detection with \iras, NGC\,6890 was followed up in the MIR with \spitzer/IRAC, IRS and MIPS.
The corresponding IRAC and MIPS images show a dominant compact nucleus embedded within faint spiral-like host emission.
Our nuclear IRAC 5.8 and 8.0$\mu$m photometry is significantly lower than the values published by \cite{gallimore_infrared_2010} but agree with the \spitzer/IRS mapping-mode post-BCD spectrum.
The latter suffers from low S/N but indicates weak silicate 10 and 18$\,\mu$m emission, weak PAH features, and a red spectral slope in $\nu F_\nu$-space (see also \citealt{buchanan_spitzer_2006,tommasin_spitzer_2008,tommasin_spitzer-irs_2010,wu_spitzer/irs_2009,gallimore_infrared_2010}).
Thus, the arcsecond-scale MIR SED indicates unobscured AGN emission with minor star-formation contribution.
The nuclear region of NGC\,6890 was observed with T-ReCS in the broad N filter in 2004 \citep{videla_nuclear_2013} and an apparently elongated nucleus without further host emission was detected (FWHM(major axis)$\sim0.42\arcsec\sim68\,$pc; PA$\sim 102\degree$). 
However, at least a second epoch of MIR subarcsecond imaging is required to confirm this elongation.
Our nuclear photometry is consistent with the value published by \cite{videla_nuclear_2013} and also marginally consistent with the \spitzerr spectrophotometry ($\sim18\%$ lower).
Therefore, we use the latter to compute the nuclear 12\,$\mu$m continuum emission estimate.
\newline\end{@twocolumnfalse}]

\begin{figure}
   \centering
   \includegraphics[angle=0,width=8.500cm]{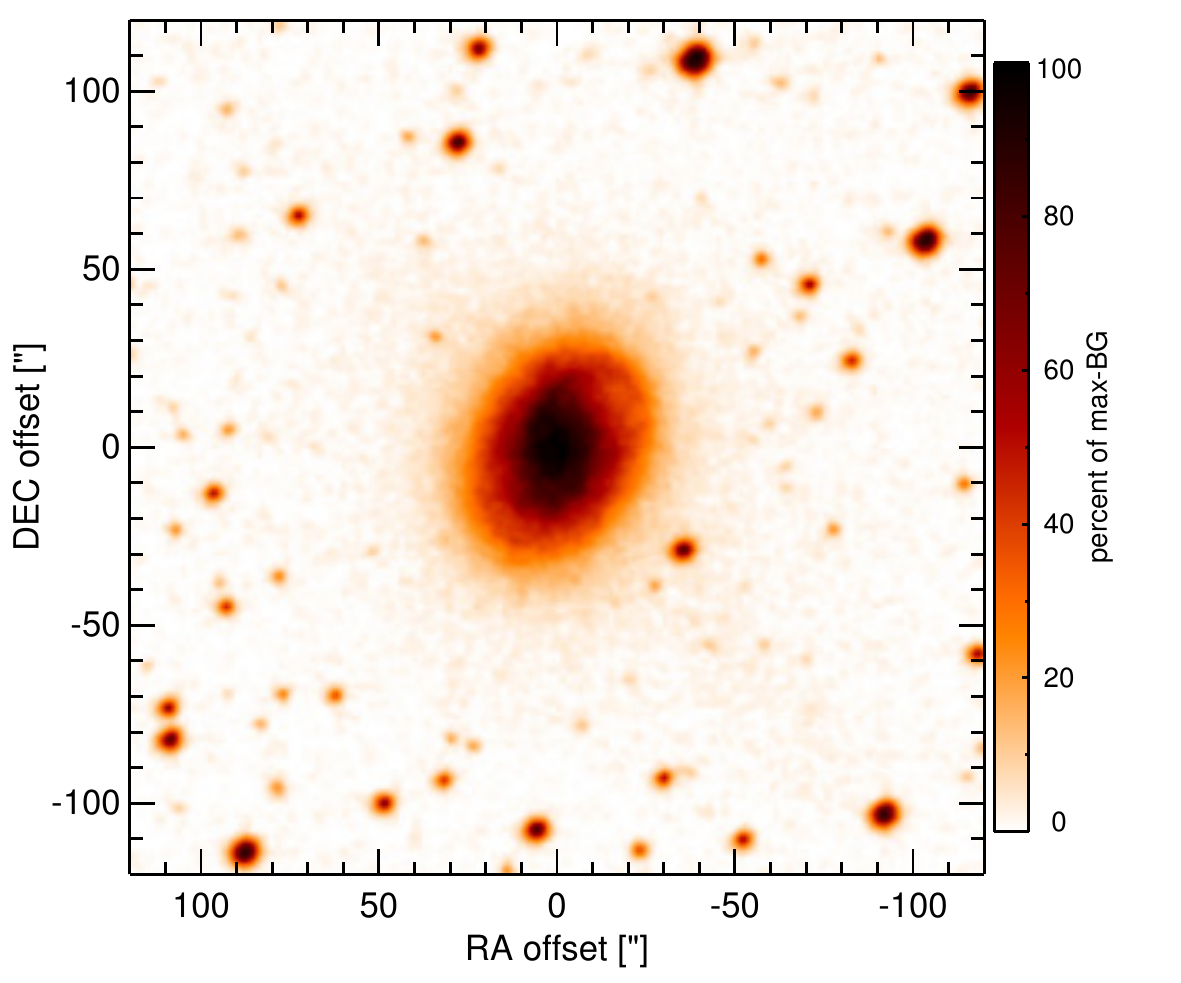}
    \caption{\label{fig:OPTim_NGC6890}
             Optical image (DSS, red filter) of NGC\,6890. Displayed are the central $4\arcmin$ with North up and East to the left. 
              The colour scaling is linear with white corresponding to the median background and black to the $0.01\%$ pixels with the highest intensity.  
           }
\end{figure}
\begin{figure}
   \centering
   \includegraphics[angle=0,height=3.11cm]{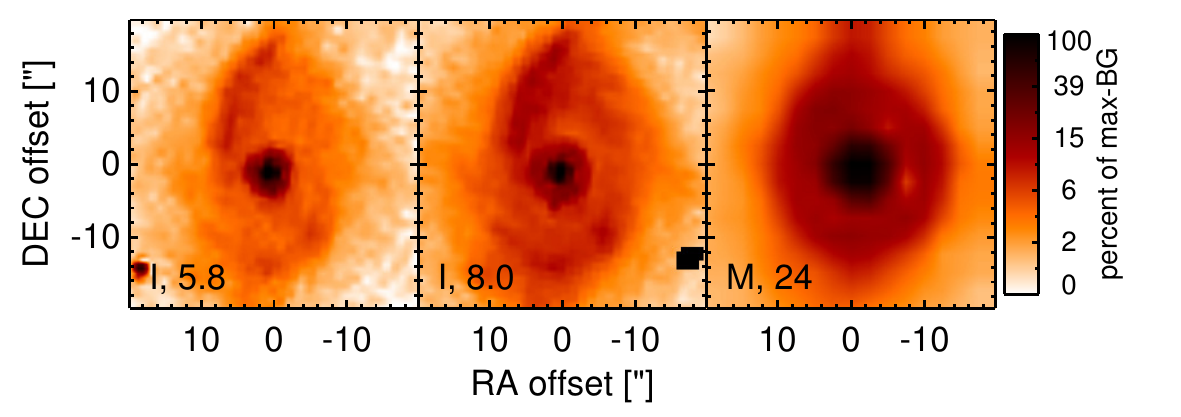}
    \caption{\label{fig:INTim_NGC6890}
             \spitzerr MIR images of NGC\,6890. Displayed are the inner $40\arcsec$ with North up and East to the left. The colour scaling is logarithmic with white corresponding to median background and black to the $0.1\%$ pixels with the highest intensity.
             The label in the bottom left states instrument and central wavelength of the filter in $\mu$m (I: IRAC, M: MIPS). 
           }
\end{figure}
\begin{figure}
   \centering
   \includegraphics[angle=0,height=3.11cm]{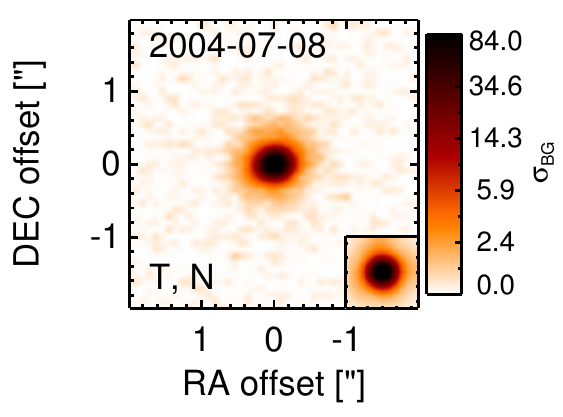}
    \caption{\label{fig:HARim_NGC6890}
             Subarcsecond-resolution MIR images of NGC\,6890 sorted by increasing filter wavelength. 
             Displayed are the inner $4\arcsec$ with North up and East to the left. 
             The colour scaling is logarithmic with white corresponding to median background and black to the $75\%$ of the highest intensity of all images in units of $\sigbg$.
             The inset image shows the central arcsecond of the PSF from the calibrator star, scaled to match the science target.
             The labels in the bottom left state instrument and filter names (C: COMICS, M: Michelle, T: T-ReCS, V: VISIR).
           }
\end{figure}
\begin{figure}
   \centering
   \includegraphics[angle=0,width=8.50cm]{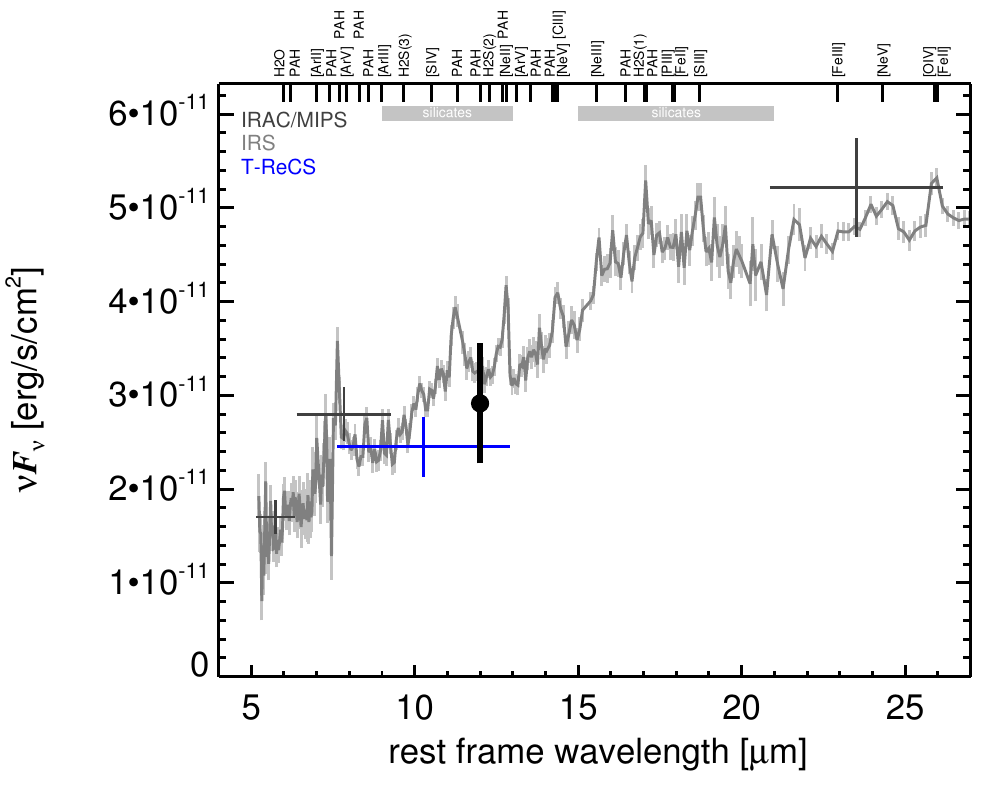}
   \caption{\label{fig:MISED_NGC6890}
      MIR SED of NGC\,6890. The description  of the symbols (if present) is the following.
      Grey crosses and  solid lines mark the \spitzer/IRAC, MIPS and IRS data. 
      The colour coding of the other symbols is: 
      green for COMICS, magenta for Michelle, blue for T-ReCS and red for VISIR data.
      Darker-coloured solid lines mark spectra of the corresponding instrument.
      The black filled circles mark the nuclear 12 and $18\,\mu$m  continuum emission estimate from the data.
      The ticks on the top axis mark positions of common MIR emission lines, while the light grey horizontal bars mark wavelength ranges affected by the silicate 10 and 18$\mu$m features.}
\end{figure}
\clearpage

\twocolumn[\begin{@twocolumnfalse}  
\subsection{NGC\,7130 -- IC\,5135 -- IRAS\,21453-3511}\label{app:NGC7130}
NGC\,7130 is a peculiar low-inclination spiral galaxy at a redshift of $z=$ 0.0162 ($D\sim68.9\,$Mpc) with an active nucleus containing both an AGN and a powerful compact starburst \citep{thuan_ultraviolet_1984,gonzalez_delgado_ultraviolet-optical_1998,levenson_deconstructing_2005}.
The nuclear starburst extends over the central $\sim1\arcsec\sim0.3\,$kpc.
The AGN is optically classified as a Sy\,1.9 nucleus \citep{veron-cetty_catalogue_2010},
Sy\,2 \citep{phillips_nearby_1983} and LINER \citep{kim_optical_1995,veilleux_optical_1995}, and we treat it as AGN/starburst composite.
NGC\,7130 features a compact radio core, which has a slightly elongated component with $\sim3\arcsec\sim$1\,kpc length along a PA$\sim0\degree$ \citep{norris_compact_1990,thean_high-resolution_2000}.
The \oiii emission is compact and unresolved on arcsecond scales \citep{shields_emission-line_1990}.
After being detected in the MIR with \irass for the first time, NGC\,7130 was followed up in the MIR with IRTF \citep{wynn-williams_luminous_1993} and \spitzer/IRAC, IRS and MIPS.
The corresponding IRAC and MIPS images show an extended, bright nucleus embedded within spiral-like host emission.
Our nuclear IRAC 5.8 and 8.0$\mu$m photometry is significantly lower than the values published by \cite{gallimore_infrared_2010} but roughly agree with the \spitzer/IRS staring-mode post-BCD spectrum.
The latter exhibits silicate 10$\,\mu$m absorption, strong PAH features and a steep red spectral slope in $\nu F_\nu$-space (see also \citealt{buchanan_spitzer_2006,shi_9.7_2006,wu_spitzer/irs_2009,tommasin_spitzer-irs_2010,gallimore_infrared_2010}).
Thus, the arcsecond-scale MIR SED indicates obscured AGN emission with a high star-formation contribution (see also \citealt{alonso-herrero_local_2012}).
The nuclear region of NGC\,7130 was imaged with T-ReCS in the broad $N$-band filter \citep{alonso-herrero_high_2006,diaz-santos_understanding_2008}, and a T-ReCS LR $N$-band spectrum was taken subsequently \citep{diaz-santos_high_2010,gonzalez-martin_dust_2013}.
In addition, it was imaged with VISIR in five different narrow $N$-band filters in 2009 (unpublished, to our knowledge).
A compact nucleus was detected in all images, while the largest-FOV T-ReCS image also shows  extended diffuse emission to the north as well as a compact source $\sim10\arcsec\sim3.2\,$kpc to the north (PA$\sim-12\degree$).
\cite{alonso-herrero_high_2006} identifies the off-nuclear emission with H\,II regions.
The nucleus appears marginally resolved (but not significantly elongated) in all images (median FWHM$\sim0.56\arcsec\sim180\,$pc) and, thus, is classified as extended.
Accordingly, we measure only the unresolved nuclear flux, which provides 
results generally consistent with the PSF-extracted T-ReCS spectrum from \cite{gonzalez-martin_dust_2013}, but is systematically lower by a few percent.
The exceptions to this are the PAH-probing PAH1 and B11.7 fluxes, which are significantly lower and thus possibly hint at lower PAH emission in the central $\sim100$\,pc.
The nuclear MIR SED has on average $\sim 59\%$ lower fluxes than the \spitzerr spectrophotometry while exhibiting a similar silicate 10\,$\mu$m absorption strength. 
Despite being weaker, PAH emission might still be present on subarcsecond resolution.
We conclude that the nuclear MIR SED is presumably still affected by significant star-formation emission and, thus, should be regarded as an upper limit to the MIR emission by the AGN (but see also \citealt{gonzalez-martin_dust_2013}).
\newline\end{@twocolumnfalse}]

\begin{figure}
   \centering
   \includegraphics[angle=0,width=8.500cm]{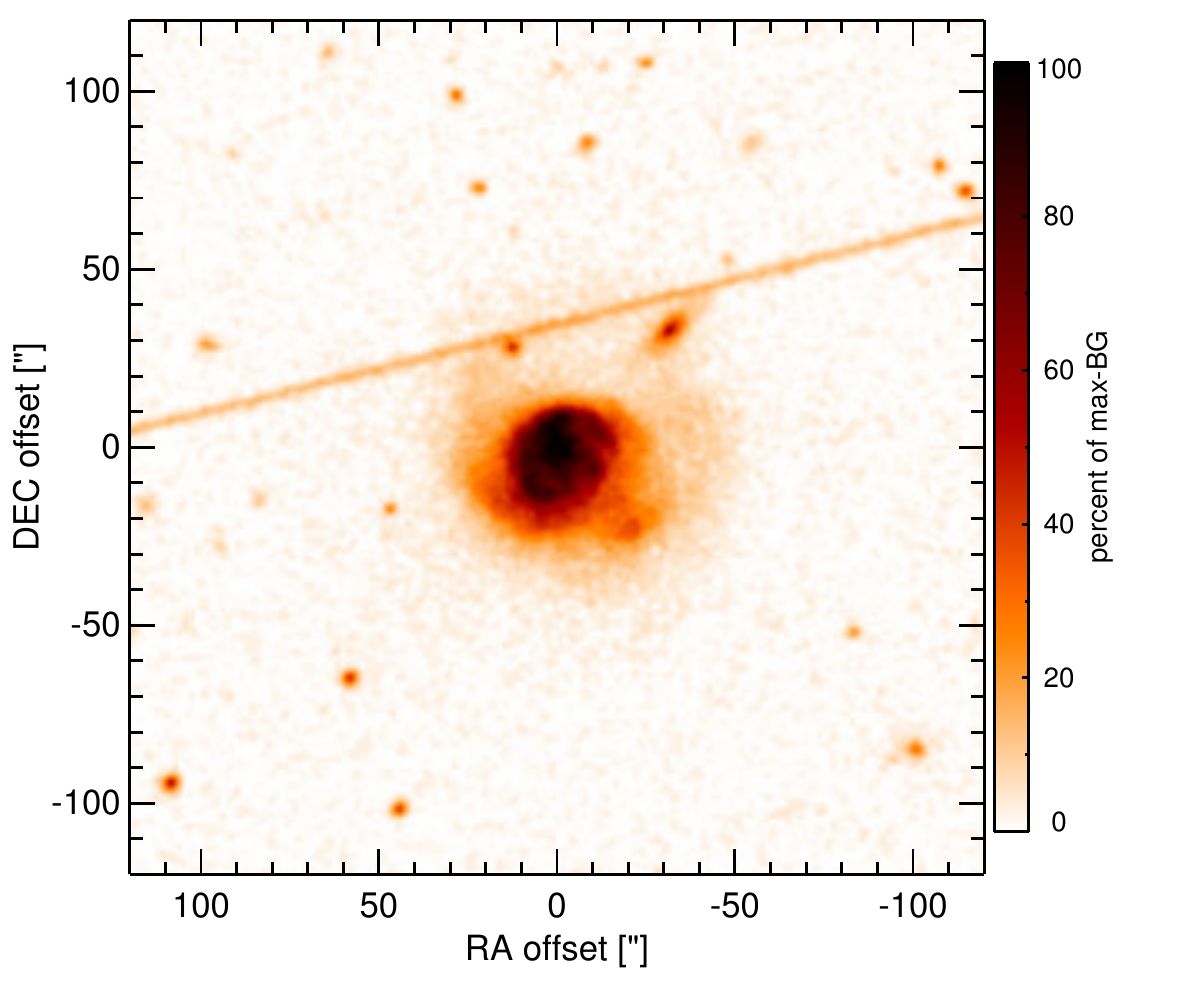}
    \caption{\label{fig:OPTim_NGC7130}
             Optical image (DSS, red filter) of NGC\,7130. Displayed are the central $4\arcmin$ with North up and East to the left. 
              The colour scaling is linear with white corresponding to the median background and black to the $0.01\%$ pixels with the highest intensity.  
           }
\end{figure}
\begin{figure}
   \centering
   \includegraphics[angle=0,height=3.11cm]{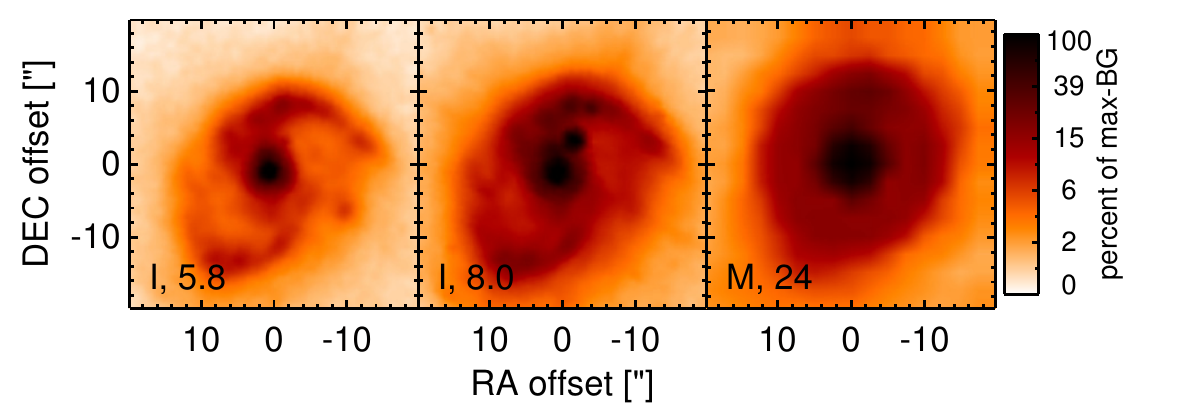}
    \caption{\label{fig:INTim_NGC7130}
             \spitzerr MIR images of NGC\,7130. Displayed are the inner $40\arcsec$ with North up and East to the left. The colour scaling is logarithmic with white corresponding to median background and black to the $0.1\%$ pixels with the highest intensity.
             The label in the bottom left states instrument and central wavelength of the filter in $\mu$m (I: IRAC, M: MIPS). 
             Note that the apparent off-nuclear compact source in the IRAC $8.0\,\mu$m image is an instrumental artefact.
           }
\end{figure}
\begin{figure}
   \centering
   \includegraphics[angle=0,width=8.500cm]{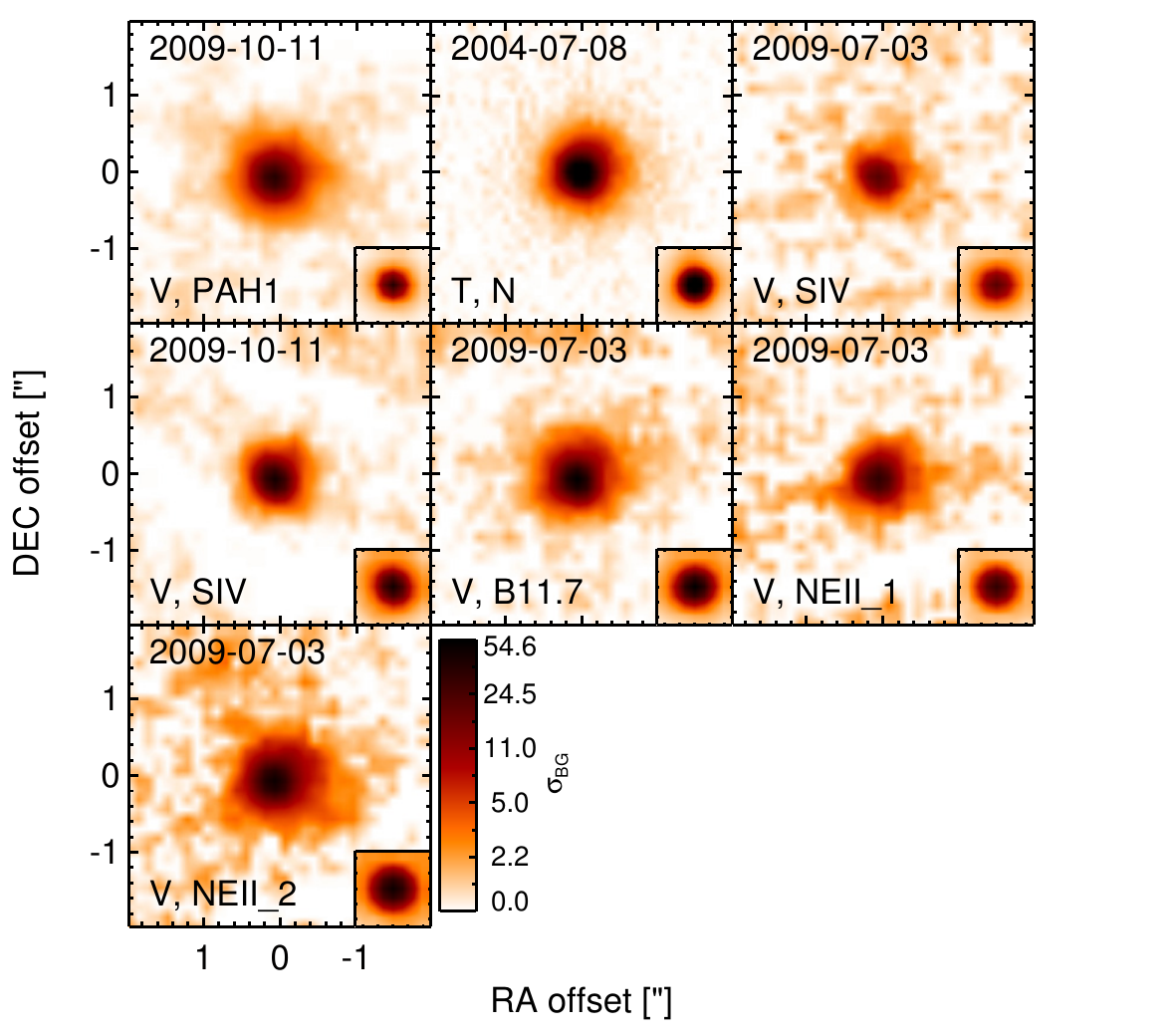}
    \caption{\label{fig:HARim_NGC7130}
             Subarcsecond-resolution MIR images of NGC\,7130 sorted by increasing filter wavelength. 
             Displayed are the inner $4\arcsec$ with North up and East to the left. 
             The colour scaling is logarithmic with white corresponding to median background and black to the $75\%$ of the highest intensity of all images in units of $\sigbg$.
             The inset image shows the central arcsecond of the PSF from the calibrator star, scaled to match the science target.
             The labels in the bottom left state instrument and filter names (C: COMICS, M: Michelle, T: T-ReCS, V: VISIR).
           }
\end{figure}
\begin{figure}
   \centering
   \includegraphics[angle=0,width=8.50cm]{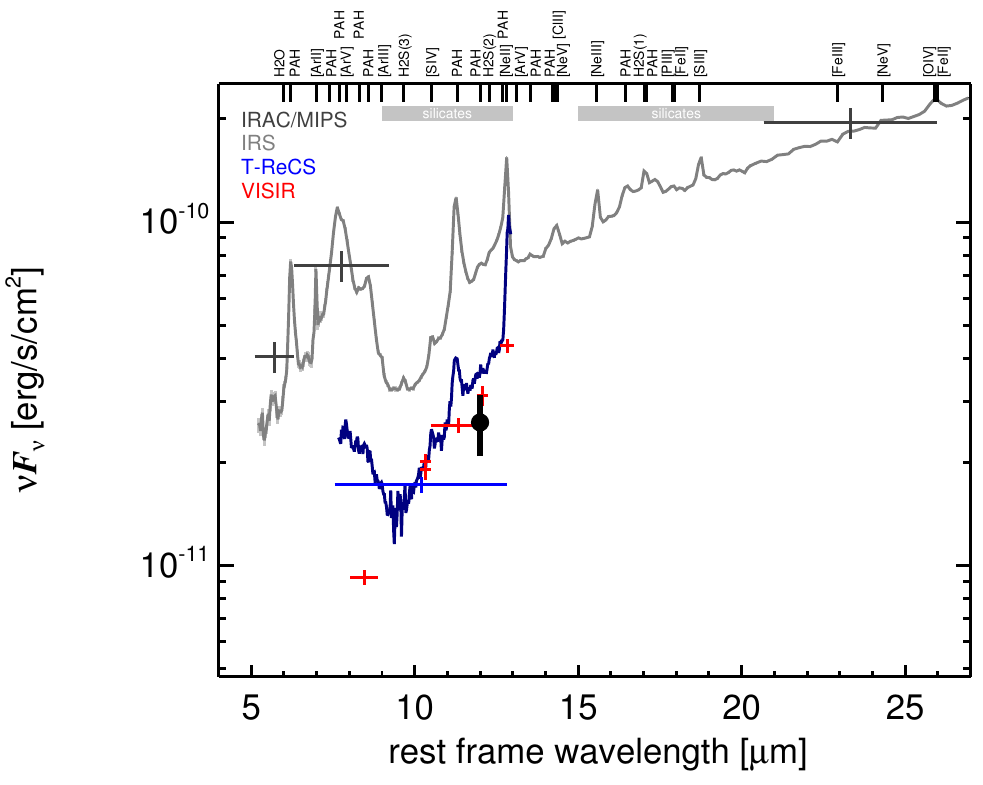}
   \caption{\label{fig:MISED_NGC7130}
      MIR SED of NGC\,7130. The description  of the symbols (if present) is the following.
      Grey crosses and  solid lines mark the \spitzer/IRAC, MIPS and IRS data. 
      The colour coding of the other symbols is: 
      green for COMICS, magenta for Michelle, blue for T-ReCS and red for VISIR data.
      Darker-coloured solid lines mark spectra of the corresponding instrument.
      The black filled circles mark the nuclear 12 and $18\,\mu$m  continuum emission estimate from the data.
      The ticks on the top axis mark positions of common MIR emission lines, while the light grey horizontal bars mark wavelength ranges affected by the silicate 10 and 18$\mu$m features.}
\end{figure}
\clearpage

\twocolumn[\begin{@twocolumnfalse}  
\subsection{NGC\,7172}\label{app:NGC7172}
NGC\,7172 is an edge-on lenticular galaxy at a redshift of $z=$ 0.0087 ($D\sim34.8\,$Mpc) with a Sy\,2 nucleus \citep{veron-cetty_catalogue_2010} and a prominent dust lane projected onto the nucleus (\citealt{sharples_ngc_1984}; see \citealt{smajic_unveiling_2012} for a recent detailed study).
It is variable by a factor of $\sim$ two in X-rays (e.g. \citealt{akylas_bepposax_2001}) and belongs to the nine-month BAT AGN sample. Furthermore, it is an X-ray ``buried" AGN candidate \citep{noguchi_new_2009}.
The nucleus is slightly resolved in arcsecond-resolution radio observations with extended emission of $\sim6\arcsec\sim1\,$kpc size along the host major axis (PA$\sim90\degree$; \citealt{unger_radio_1987,morganti_radio_1999}).
After first being detected in the MIR by \iras, NGC\,7172 was followed up with ground-based MIR observations \citep{ward_continuum_1987,roche_atlas_1991}.
The first subarcsecond $N$-band image was obtained with Palomar 5\,m/MIRLIN in 2000 \citep{gorjian_10_2004}.
The \spitzer/IRAC and MIPS images show a compact nucleus embedded within the edge-on host emission.
Our nuclear IRAC $5.8$ and $8.0\,\mu$m photometry provides values significantly lower than those values in \cite{johnson_infrared_2007,gallimore_infrared_2010} but significantly higher than the nuclear fluxes in \cite{gallagher_revealing_2008}, while roughly consistent with the \spitzer/IRS LR staring-mode post-BCD spectrum.
The latter exhibits extremely deep silicate 10$\,\mu$m and 18$\,\mu$m absorption, significant PAH emission, and a generally blue spectral shape in $\nu F_\nu$-space (see also \citealt{shi_9.7_2006,wu_spitzer/irs_2009,tommasin_spitzer-irs_2010,gallimore_infrared_2010}).
Thus, the arcsecond-scale MIR SED  is presumably affected by significant star formation.
The nuclear region of NGC\,7172 was imaged in the Si6 filter, and a LR $N$-band spectrum was taken with T-ReCS in 2004 \citep{roche_silicate_2007,gonzalez-martin_dust_2013}.
Note that \cite{ramos_almeida_infrared_2009} report additional T-ReCS data in the broad N filter, which could not be successfully reduced with the \texttt{midir} and thus is not used. 
In addition, we imaged the nucleus with VISIR in total with three narrow $N$-band filters in 2006 \citep{horst_mid_2008,horst_mid-infrared_2009} and in 2009.
A compact nucleus was detected in all images, while part of the dust lane  appears to be faintly visible in the Si6 image (diameter$\sim1\arcsec\sim170\,$pc; PA$\sim 90\degree$).
Note, however, that no matching standard star observation could be retrieved for the Si6 image, and, thus, it can only be used qualitatively for the extension analysis and be treated as upper limit on the unresolved nuclear emission.
Similar to T-ReCS, the nucleus appears elongated also in the VISIR images (FWHM(major axis)$\sim0.45\arcsec\sim75\,$pc; PA$\sim 90\degree$).
Therefore, we measure only the unresolved nuclear flux, which provides values generally consistent with the T-ReCS $N$-band spectrum from \cite{gonzalez-martin_dust_2013}, but is systematically lower by $\sim10\%$.
The resulting nuclear MIR SED is on average $\sim 56\%$ lower than the \spitzerr spectrophotometry, exhibits a similar silicate 10\,$\mu$m absorption feature but no PAH emission.
Therefore, we assume that the nuclear MIR SED is free of star-formation contamination and use the T-ReCS spectrum to calculate the 12\,$\mu$m continuum emission estimate corrected for the silicate feature.
\newline\end{@twocolumnfalse}]

\begin{figure}
   \centering
   \includegraphics[angle=0,width=8.500cm]{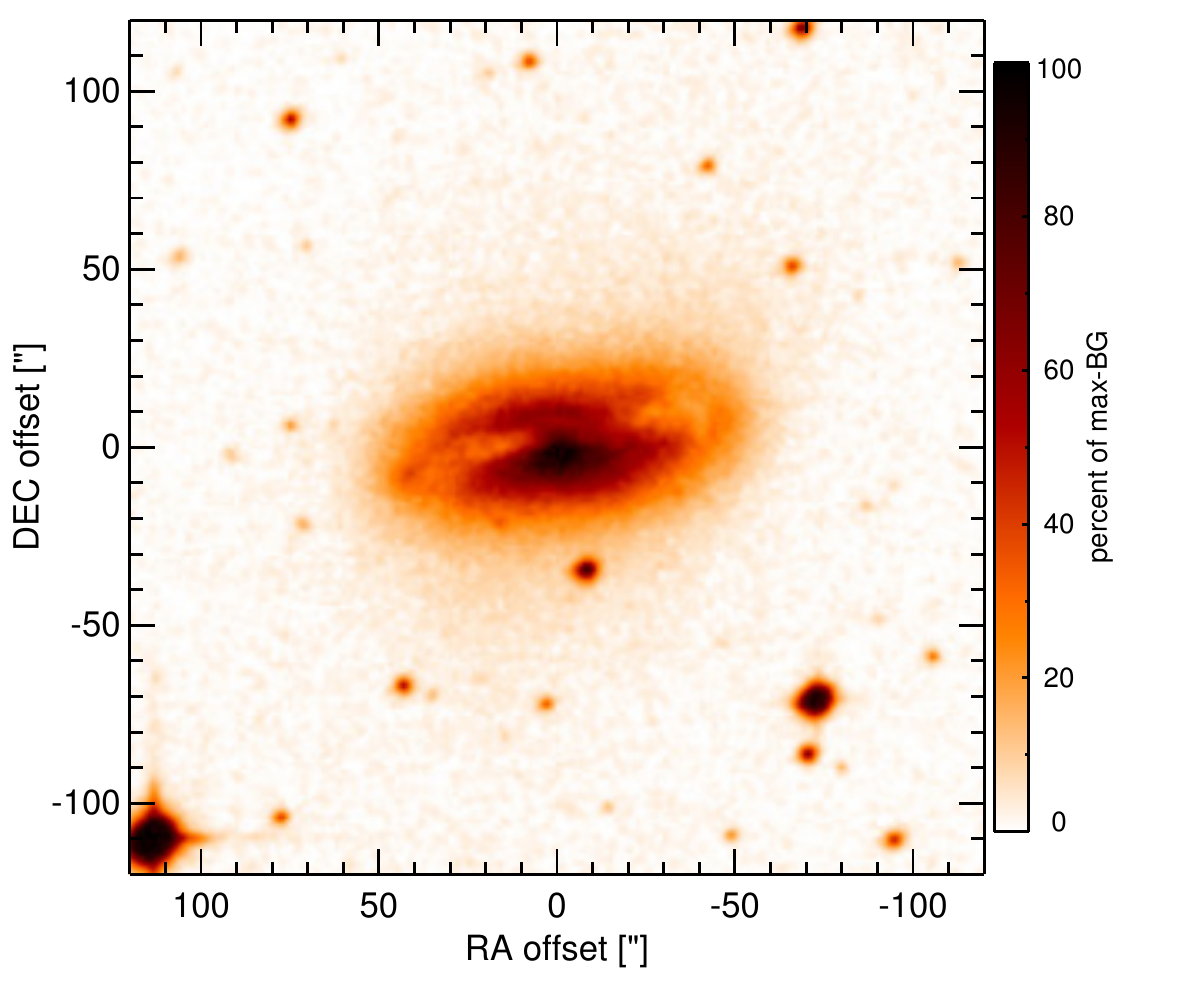}
    \caption{\label{fig:OPTim_NGC7172}
             Optical image (DSS, red filter) of NGC\,7172. Displayed are the central $4\arcmin$ with North up and East to the left. 
              The colour scaling is linear with white corresponding to the median background and black to the $0.01\%$ pixels with the highest intensity.  
           }
\end{figure}
\begin{figure}
   \centering
   \includegraphics[angle=0,height=3.11cm]{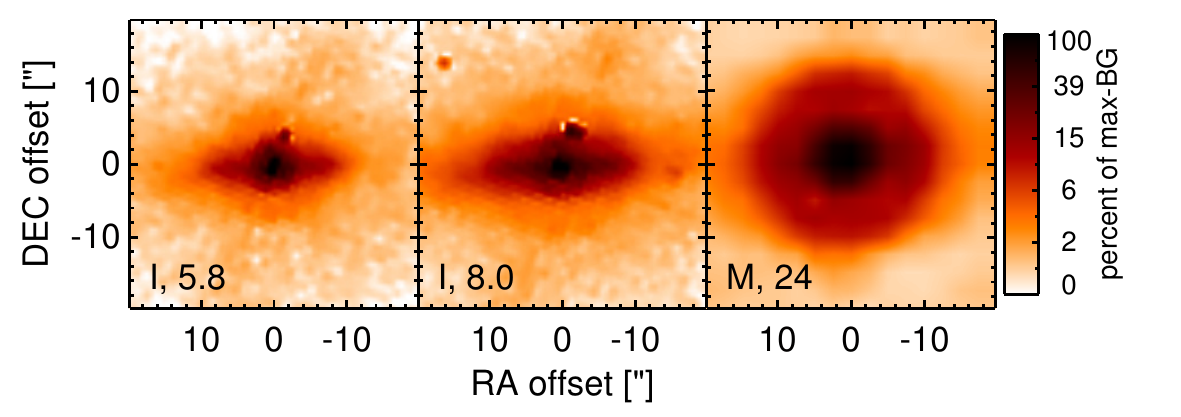}
    \caption{\label{fig:INTim_NGC7172}
             \spitzerr MIR images of NGC\,7172. Displayed are the inner $40\arcsec$ with North up and East to the left. The colour scaling is logarithmic with white corresponding to median background and black to the $0.1\%$ pixels with the highest intensity.
             The label in the bottom left states instrument and central wavelength of the filter in $\mu$m (I: IRAC, M: MIPS). 
             Note that the apparent off-nuclear compact sources in the IRAC 5.8 and  $8.0\,\mu$m images are instrumental artefacts.
           }
\end{figure}
\begin{figure}
   \centering
   \includegraphics[angle=0,width=8.500cm]{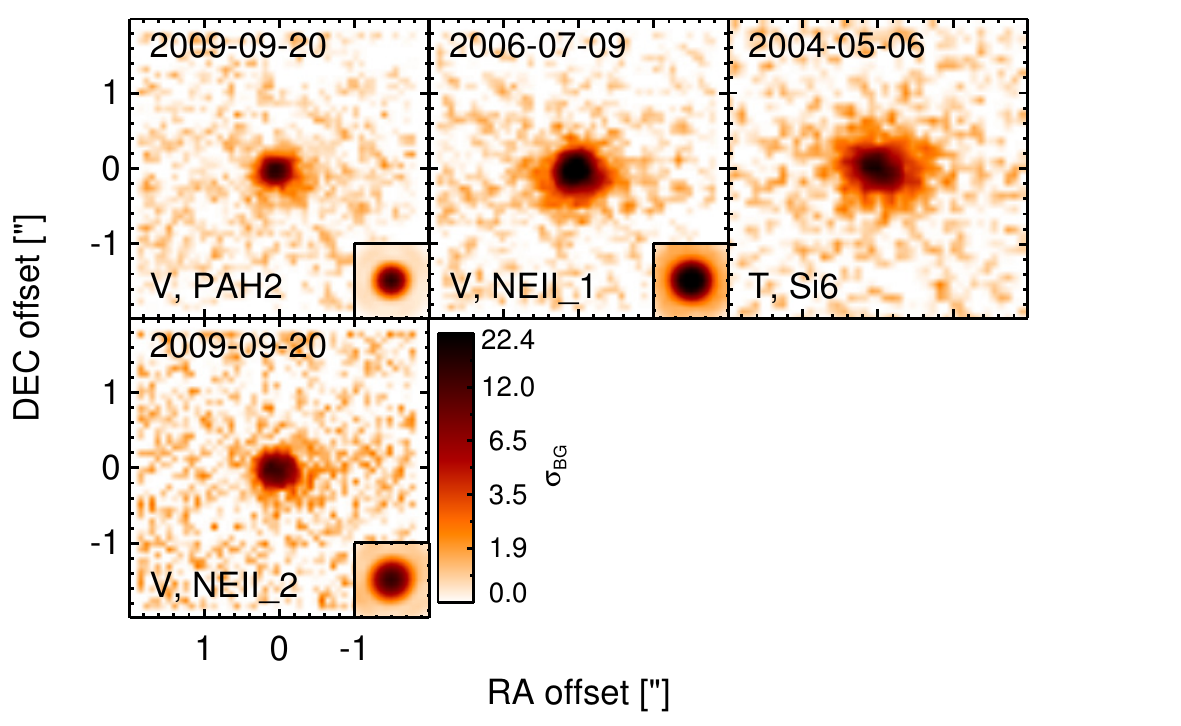}
    \caption{\label{fig:HARim_NGC7172}
             Subarcsecond-resolution MIR images of NGC\,7172 sorted by increasing filter wavelength. 
             Displayed are the inner $4\arcsec$ with North up and East to the left. 
             The colour scaling is logarithmic with white corresponding to median background and black to the $75\%$ of the highest intensity of all images in units of $\sigbg$.
             The inset image shows the central arcsecond of the PSF from the calibrator star, scaled to match the science target.
             The labels in the bottom left state instrument and filter names (C: COMICS, M: Michelle, T: T-ReCS, V: VISIR).
           }
\end{figure}
\begin{figure}
   \centering
   \includegraphics[angle=0,width=8.50cm]{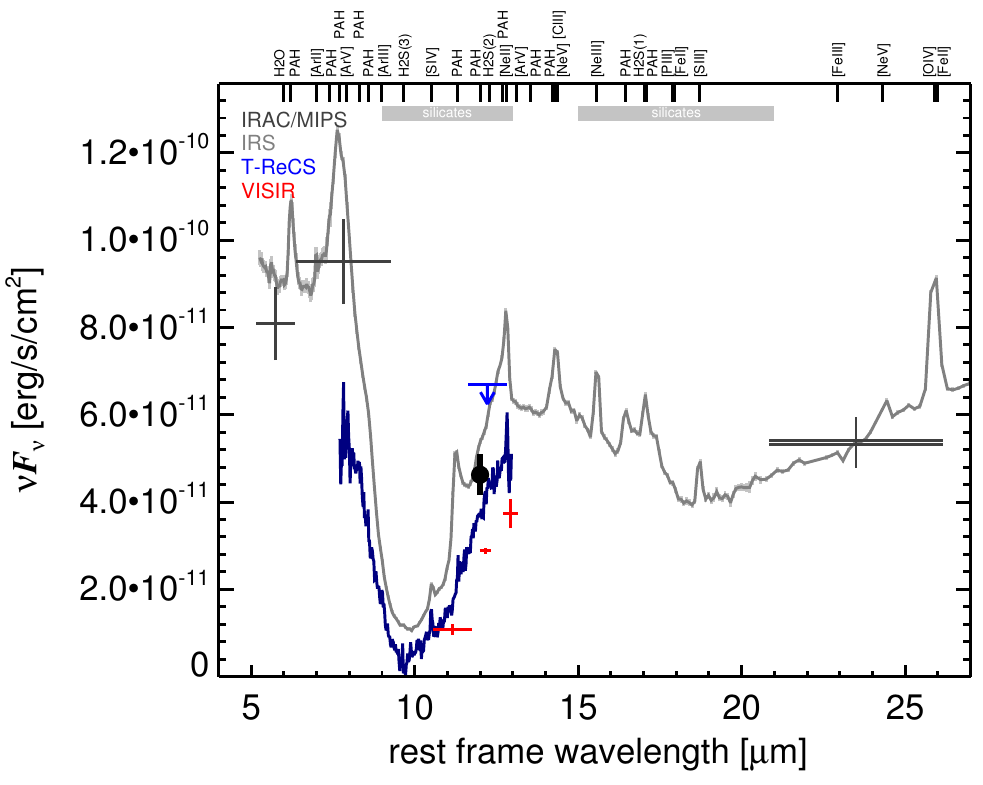}
   \caption{\label{fig:MISED_NGC7172}
      MIR SED of NGC\,7172. The description  of the symbols (if present) is the following.
      Grey crosses and  solid lines mark the \spitzer/IRAC, MIPS and IRS data. 
      The colour coding of the other symbols is: 
      green for COMICS, magenta for Michelle, blue for T-ReCS and red for VISIR data.
      Darker-coloured solid lines mark spectra of the corresponding instrument.
      The black filled circles mark the nuclear 12 and $18\,\mu$m  continuum emission estimate from the data.
      The ticks on the top axis mark positions of common MIR emission lines, while the light grey horizontal bars mark wavelength ranges affected by the silicate 10 and 18$\mu$m features.}
\end{figure}
\clearpage

\twocolumn[\begin{@twocolumnfalse}  
\subsection{NGC\,7212}\label{app:NGC7212}
NGC\,7212 is a system of three interacting galaxies with one containing an AGN \citep{wasilewski_new_1981}.
In the following we refer to the largest, central galaxy, NGC\,7212 NED02 as NGC\,7172. This galaxy is an inclined spiral at a redshift of $z=$ 0.0266 ($D\sim116\,$Mpc) containing a Sy\,2 nucleus \citep{de_robertis_analysis_1986} with broad emission lines in polarized light \citep{tran_nature_1995}.
A double radio source along PA$\sim170\degree$ with $\sim1\arcsec\sim0.5\,$kpc length was detected in its nucleus \citep{falcke_hst_1998} with an aligned extended kiloparsec-scale NLR ,(e.g. \citealt{schmitt_hubble_2003,cracco_origin_2011}).
After being detected in the MIR with \irass for the first time, NGC\,7212 was followed up with \spitzer/IRS and was also imaged with \wise.
All three galaxies are detected in the corresponding \wisee images but no substructure is visible.
The IRS LR staring-mode spectrum exhibits weak silicate 10\,$\mu$m absorption, weak PAH features, prominent forbidden emission lines, and a continuum emission peak at $\sim18\,\mu$m (see also \citealt{sargsyan_infrared_2011}).
Thus, the arcsecond-scale MIR SED appears to be AGN dominated.
The nuclear region of NGC\,7212 was observed with T-ReCS in the Qa filter in 2007 (unpublished, to our knowledge).
In the image, an elongated nucleus without further host emission was detected (FWHM(major axis)$\sim0.79\arcsec\sim0.4\,$kpc; PA$\sim 177\degree$).
At least a second epoch of subarcsecond MIR imaging is required to verify this extension.
However, it is parallel to the ionization cone direction and indicates dust in the ionization cone as suggested by \cite{cracco_origin_2011}.
The nuclear Qa flux is consistent with the IRS spectrum, and, thus, we use the latter to calculate the 12\,$\mu$m continuum emission estimate.
Note, however, that the nuclear flux would be significantly lower if the subarcsecond-extended emission can be verified.
For now, the resulting synthetic flux is scaled by half of the ratio between $\Fpsf$ and $\Fgau$ to account for the possibly nuclear extension.
\newline\end{@twocolumnfalse}]

\begin{figure}
   \centering
   \includegraphics[angle=0,width=8.500cm]{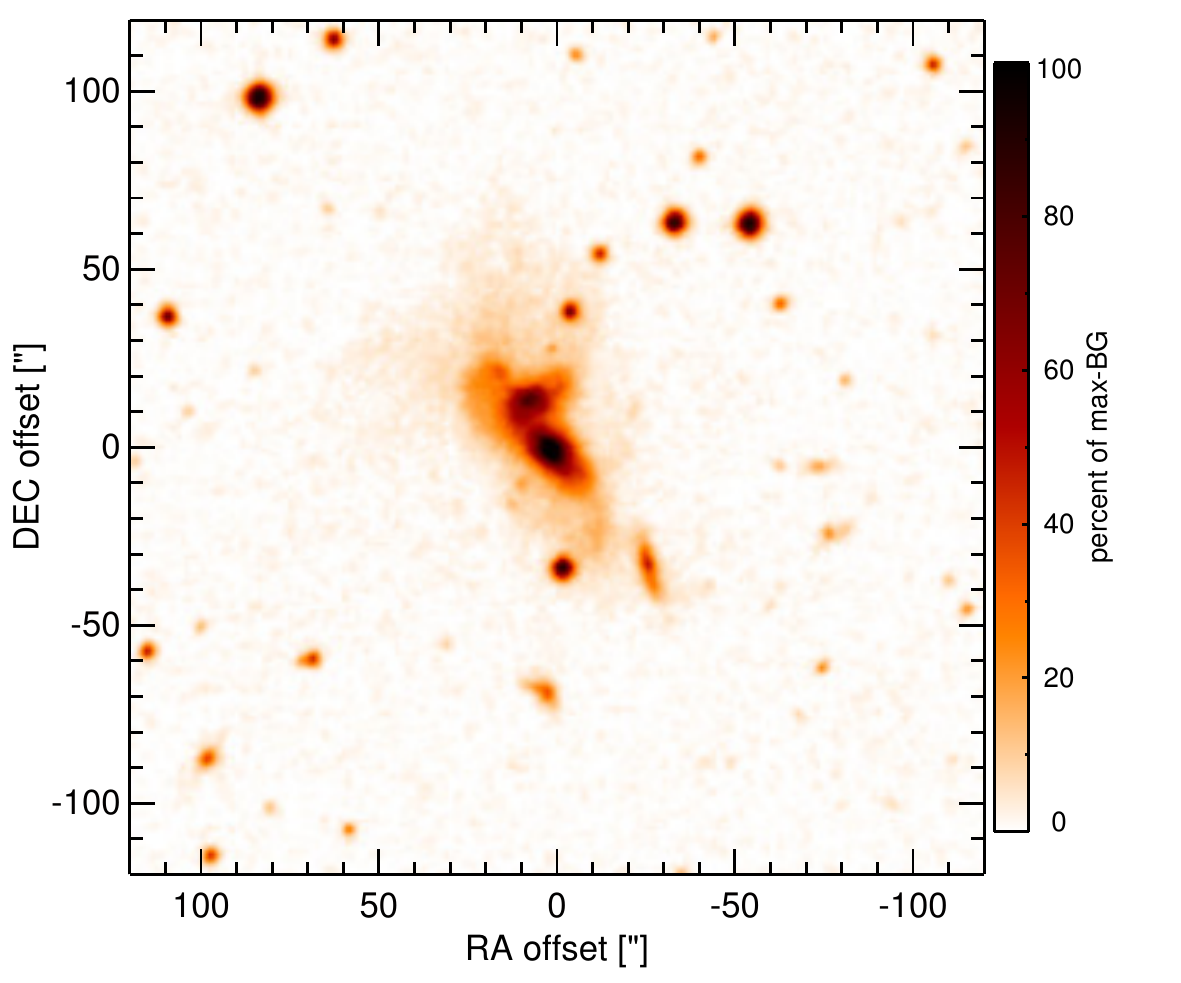}
    \caption{\label{fig:OPTim_NGC7212}
             Optical image (DSS, red filter) of NGC\,7212. Displayed are the central $4\arcmin$ with North up and East to the left. 
              The colour scaling is linear with white corresponding to the median background and black to the $0.01\%$ pixels with the highest intensity.  
           }
\end{figure}
\begin{figure}
   \centering
   \includegraphics[angle=0,height=3.11cm]{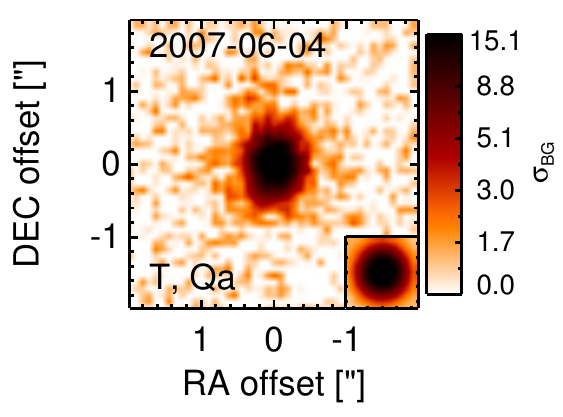}
    \caption{\label{fig:HARim_NGC7212}
             Subarcsecond-resolution MIR images of NGC\,7212 sorted by increasing filter wavelength. 
             Displayed are the inner $4\arcsec$ with North up and East to the left. 
             The colour scaling is logarithmic with white corresponding to median background and black to the $75\%$ of the highest intensity of all images in units of $\sigbg$.
             The inset image shows the central arcsecond of the PSF from the calibrator star, scaled to match the science target.
             The labels in the bottom left state instrument and filter names (C: COMICS, M: Michelle, T: T-ReCS, V: VISIR).
           }
\end{figure}
\begin{figure}
   \centering
   \includegraphics[angle=0,width=8.50cm]{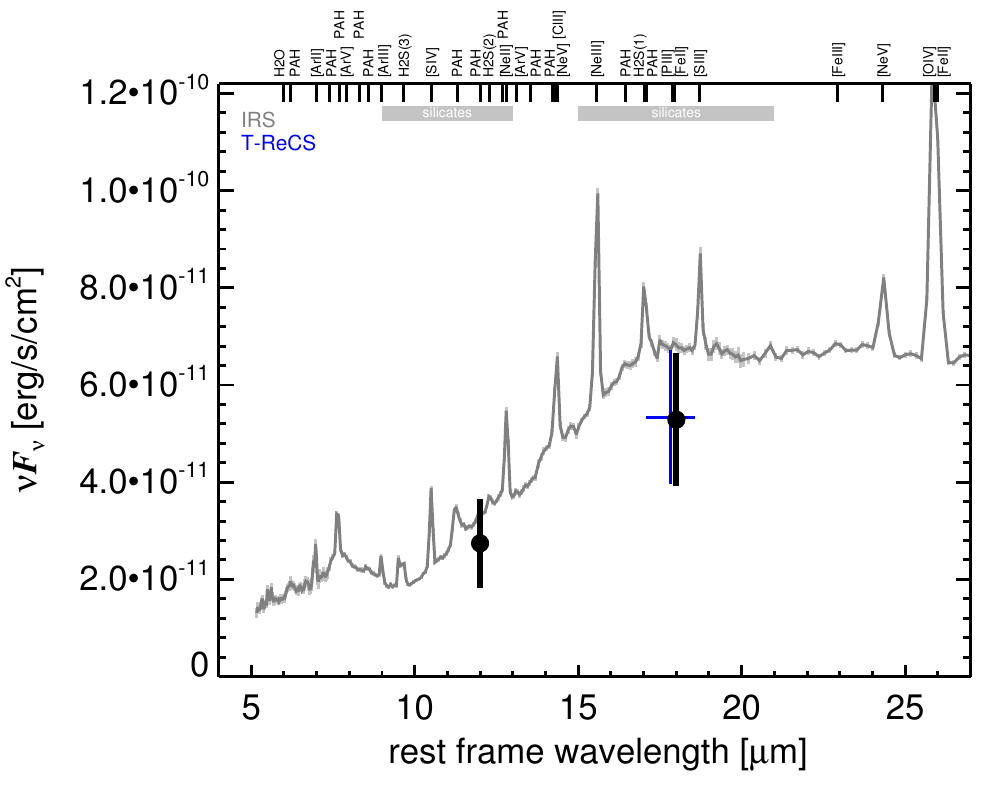}
   \caption{\label{fig:MISED_NGC7212}
      MIR SED of NGC\,7212. The description  of the symbols (if present) is the following.
      Grey crosses and  solid lines mark the \spitzer/IRAC, MIPS and IRS data. 
      The colour coding of the other symbols is: 
      green for COMICS, magenta for Michelle, blue for T-ReCS and red for VISIR data.
      Darker-coloured solid lines mark spectra of the corresponding instrument.
      The black filled circles mark the nuclear 12 and $18\,\mu$m  continuum emission estimate from the data.
      The ticks on the top axis mark positions of common MIR emission lines, while the light grey horizontal bars mark wavelength ranges affected by the silicate 10 and 18$\mu$m features.}
\end{figure}
\clearpage

\twocolumn[\begin{@twocolumnfalse}  
\subsection{NGC\,7213}\label{app:NGC7213}
NGC\,7213 is a face-on lenticular galaxy at a redshift of $z=$ 0.0058 ($D\sim23\,$Mpc) with a borderline Sy\,1.5/LINER nucleus \citep{phillips_optical_1979,filippenko_ngc_1984}.
It has been studied well in X-rays and also belongs to the nine-month BAT AGN sample (see \citealt{bell_x-ray_2011} and \citealt{emmanoulopoulos_harder_2012} for a recent detailed study).
The nucleus is unresolved at radio wavelengths down to milliarcsecond scales, has a flat radio spectrum, and has an intermediate radio-loudness \citep{bransford_radio-luminous_1998,schmitt_jet_2001,thean_high-resolution_2000, blank_8.4-ghz_2005}.
The \oiii morphology is halo-like with a diameter of $\sim1\arcsec\sim110\,$pc, slightly elongated into the east-west direction \citep{schmitt_hubble_2003}.
After first being detected in the MIR with \iras, NGC\,7213 was followed up from the ground \citep{glass_mid-infrared_1982} and from space with \spitzer/IRAC, IRS and MIPS.
The corresponding IRAC and MIPS images show a dominating compact nucleus embedded within weak spiral-like host emission.
Our nuclear IRAC 5.8 and 8.0\,$\mu$m photometry is significantly lower than the values published by \cite{gallimore_infrared_2010} but in better agreement with the IRS LR staring-mode post-BCD spectrum.
The latter is dominated by strong silicate 10 and 18\,$\mu$m emission, a weak PAH 11.3\,$\mu$m feature, and a flat spectral continuum slope in  $\nu F_\nu$-space (see also \citealt{shi_9.7_2006,wu_spitzer/irs_2009,deo_mid-infrared_2009,tommasin_spitzer-irs_2010,gallimore_infrared_2010}).
Thus, the arcsecond-scale MIR SED indicates large amounts of AGN-heated dust and no significant star formation.
We imaged the nuclear region of NGC\,7213 with VISIR in total in four narrow $N$-band and one $Q$-band filters in 2006 \citep{horst_mid_2008,horst_mid-infrared_2009} and 2009 (partly published in \citealt{asmus_mid-infrared_2011}), and obtained a VISIR LR $N$-band spectrum in 2008 \citep{honig_dusty_2010-1}.
In addition, it was imaged in the Si2 and Qa filters with T-ReCS in 2007, 2008 and 2010 (unpublished, to our knowledge).
In all images, a compact nucleus without further host emission was detected.
The nucleus appears essentially unresolved in the sharpest images (PAH2 from 2009, SIV from 2006 and NEII\_1 from May 2009). Therefore we classify is as generally unresolved at subarcsecond resolution in the MIR.
Note that the nucleus appears elongated in other PAH2 and NEII\_1 images, which conclusively is not object intrinsic.
The nuclear photometry indicates scatter of up to $\sim20\%$ between lowest and highest fluxes (PAH2 2006 versus 2009), while being in general consistent with the VISIR $N$-band spectrum.
The latter exhibits a $\sim20\%$ lower continuum flux level than the \spitzerr spectrophotometry with a similar silicate 10\,$\mu$m emission strength and without and PAH emission. 
We use the VISIR spectrum to compute the 12\,$\mu$m continuum emission estimate corrected for the silicate feature, which conclusively originates in the projected central $\sim 40$\,pc of NGC\,7213.
\newline\end{@twocolumnfalse}]

\begin{figure}
   \centering
   \includegraphics[angle=0,width=8.500cm]{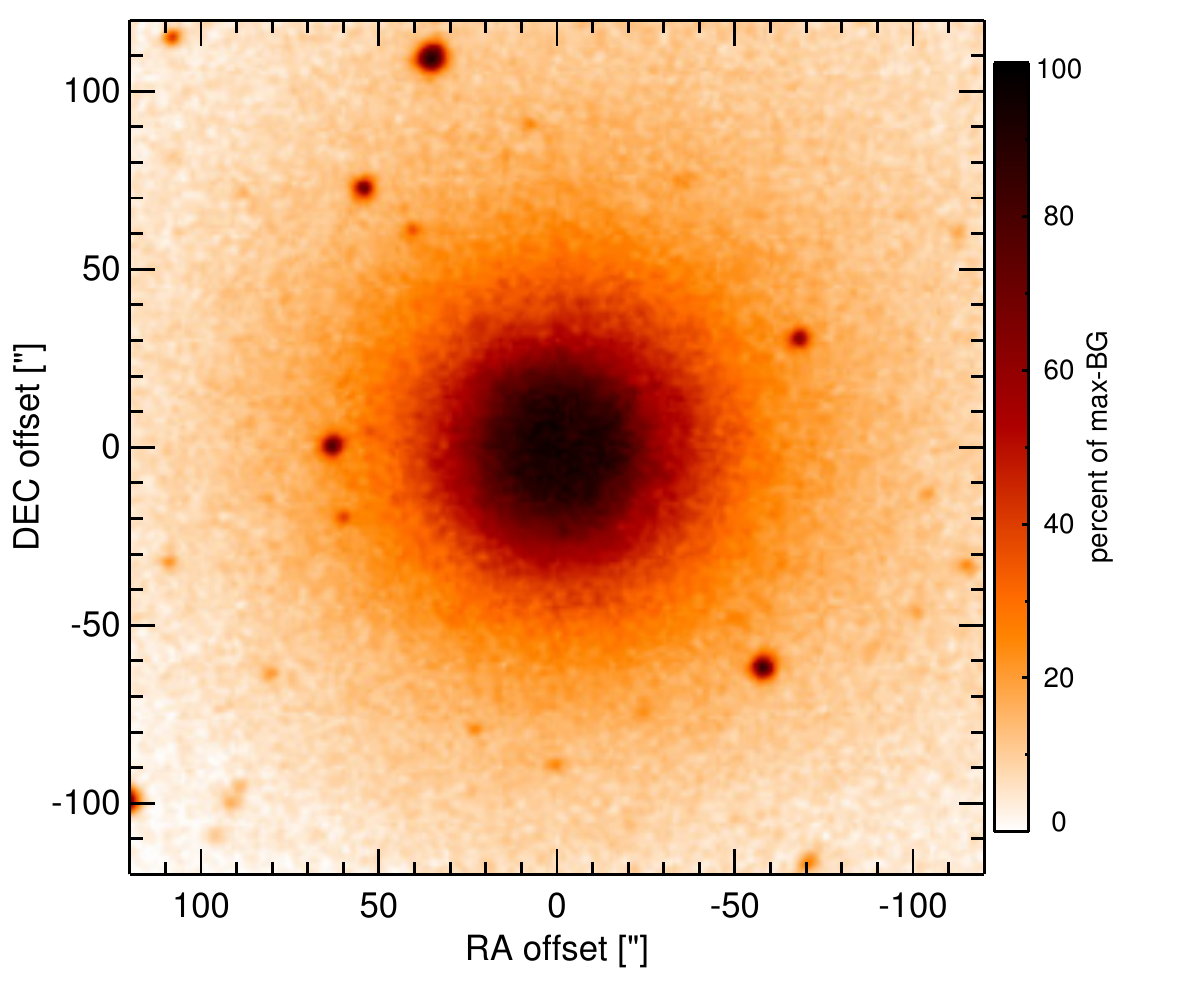}
    \caption{\label{fig:OPTim_NGC7213}
             Optical image (DSS, red filter) of NGC\,7213. Displayed are the central $4\arcmin$ with North up and East to the left. 
              The colour scaling is linear with white corresponding to the median background and black to the $0.01\%$ pixels with the highest intensity.  
           }
\end{figure}
\begin{figure}
   \centering
   \includegraphics[angle=0,height=3.11cm]{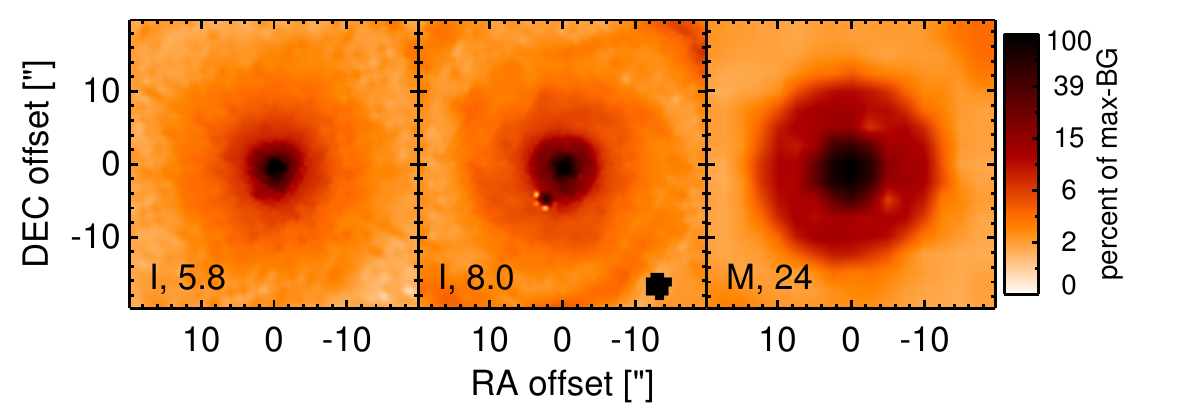}
    \caption{\label{fig:INTim_NGC7213}
             \spitzerr MIR images of NGC\,7213. Displayed are the inner $40\arcsec$ with North up and East to the left. The colour scaling is logarithmic with white corresponding to median background and black to the $0.1\%$ pixels with the highest intensity.
             The label in the bottom left states instrument and central wavelength of the filter in $\mu$m (I: IRAC, M: MIPS). 
             Note that the apparent off-nuclear compact source in the IRAC $8.0\,\mu$m image is an instrumental artefact.
           }
\end{figure}
\begin{figure}
   \centering
   \includegraphics[angle=0,width=8.500cm]{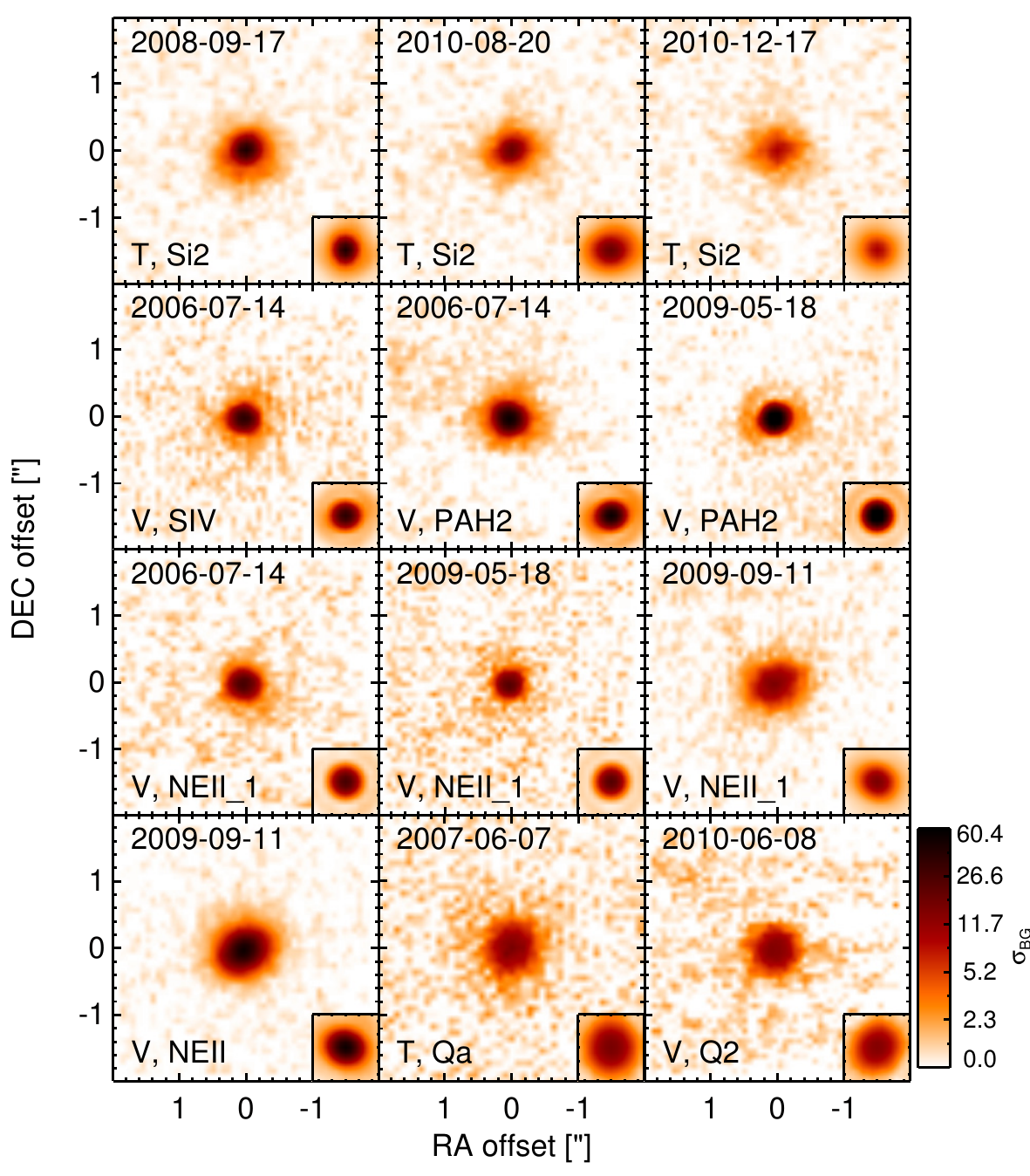}
    \caption{\label{fig:HARim_NGC7213}
             Subarcsecond-resolution MIR images of NGC\,7213 sorted by increasing filter wavelength. 
             Displayed are the inner $4\arcsec$ with North up and East to the left. 
             The colour scaling is logarithmic with white corresponding to median background and black to the $75\%$ of the highest intensity of all images in units of $\sigbg$.
             The inset image shows the central arcsecond of the PSF from the calibrator star, scaled to match the science target.
             The labels in the bottom left state instrument and filter names (C: COMICS, M: Michelle, T: T-ReCS, V: VISIR).
           }
\end{figure}
\begin{figure}
   \centering
   \includegraphics[angle=0,width=8.50cm]{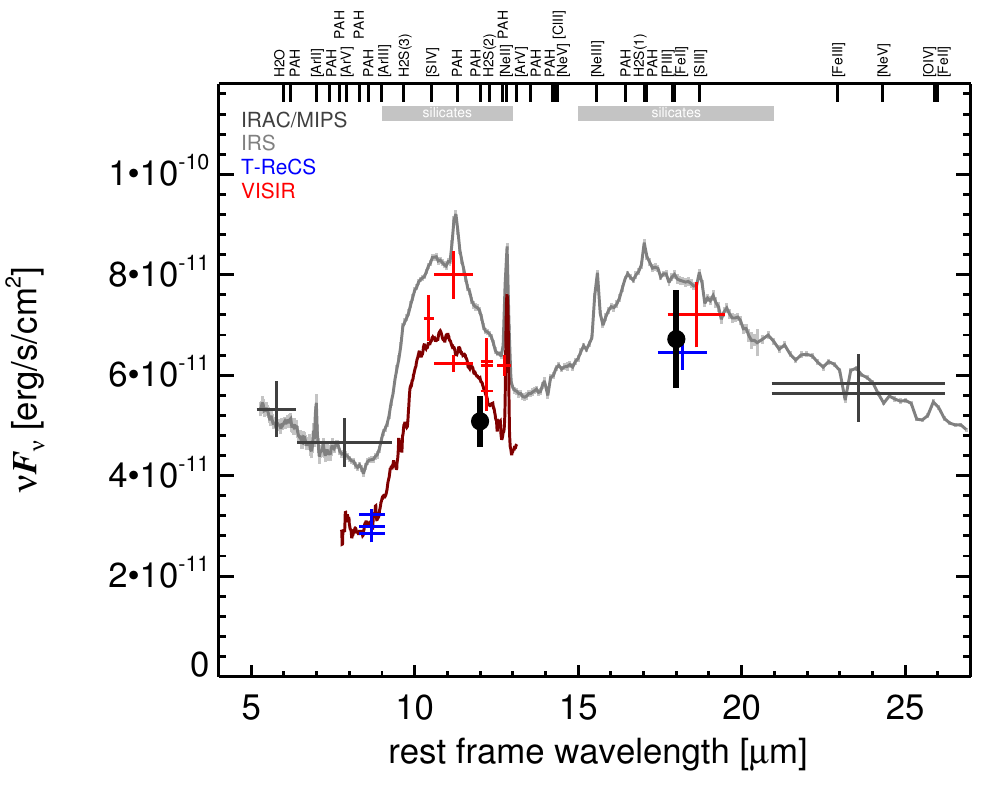}
   \caption{\label{fig:MISED_NGC7213}
      MIR SED of NGC\,7213. The description  of the symbols (if present) is the following.
      Grey crosses and  solid lines mark the \spitzer/IRAC, MIPS and IRS data. 
      The colour coding of the other symbols is: 
      green for COMICS, magenta for Michelle, blue for T-ReCS and red for VISIR data.
      Darker-coloured solid lines mark spectra of the corresponding instrument.
      The black filled circles mark the nuclear 12 and $18\,\mu$m  continuum emission estimate from the data.
      The ticks on the top axis mark positions of common MIR emission lines, while the light grey horizontal bars mark wavelength ranges affected by the silicate 10 and 18$\mu$m features.}
\end{figure}
\clearpage

\twocolumn[\begin{@twocolumnfalse}  
\subsection{NGC\,7314}\label{app:NGC7314}
NGC\,7314 is an inclined late-type barred spiral galaxy at a distance of $D=$ $ 18.3\pm2.1\,$Mpc (NED redshift-independent median) with a Sy\,1.9-2 nucleus with polarized broad emission lines \citep{hughes_atlas_2003,morris_spectrophotometry_1988,lumsden_spectropolarimetry_2004}.
It is highly variable in X-rays by up to a factor of four \citep{turner_dramatic_1987} and belongs to the nine-month BAT AGN sample.
\cite{morganti_radio_1999} report an unresolved radio core at arcsecond resolution, while \cite{thean_high-resolution_2000} report a double source with $\sim4\arcsec\sim360\,$pc separation along a PA$\sim170\degree$.
After first being detected in the MIR with \iras, NGC\,7314 was followed up with \isoo \citep{clavel_2.5-11_2000,ramos_almeida_mid-infrared_2007} and Palomar 5\,m/MIRLIN \citep{gorjian_10_2004}.
It was also observed with \spitzer/IRAC, IRS and MIPS, and the corresponding IRAC and MIPS images show a compact nucleus surrounded by spiral-like host emission.
Our nuclear IRAC 5.8 an 8.0\,$\mu$m photometry is significantly lower than the values reported in \cite{gallimore_infrared_2010} but in better agreement with the IRS LR staring-mode post-BCD spectrum.
The latter exhibits silicate 10\,$\mu$m absorption, very weak PAH features, prominent forbidden emission lines, and a red spectral slope in $\nu F_\nu$-space (see also \citealt{shi_9.7_2006,wu_spitzer/irs_2009,tommasin_spitzer-irs_2010,gallimore_infrared_2010,mullaney_defining_2011}).
Thus, the arcsecond-scale MIR SED appears to be AGN dominated with only weak star-formation.
We observed the nuclear region of NGC\,7314 with VISIR in two narrow $N$-band filters in 2005 (two epochs; \citealt{horst_small_2006,horst_mid-infrared_2009}).
In addition, T-ReCS imaging in the Si2 and Qa filters from 2010 is available (unpublished, to our knowledge).
In all images, a compact nucleus without further host emission was detected.
The nucleus appears marginally resolved in all images (FWHM(major axis)$\sim0.53\arcsec\sim\,47$pc; PA$\sim110\degree$).
The second epoch of VISIR images were taken under better MIR seeing conditions, leading to more reliable estimates for the corresponding unresolved nuclear fluxes.
The unresolved nuclear fluxes are on average $\sim 32\%$ lower than the \spitzerr spectrophotometry with a comparable silicate 10\,$\mu$m absorption depth, while the total fluxes are consistent with \cite{horst_small_2006}.
\newline\end{@twocolumnfalse}]

\begin{figure}
   \centering
   \includegraphics[angle=0,width=8.500cm]{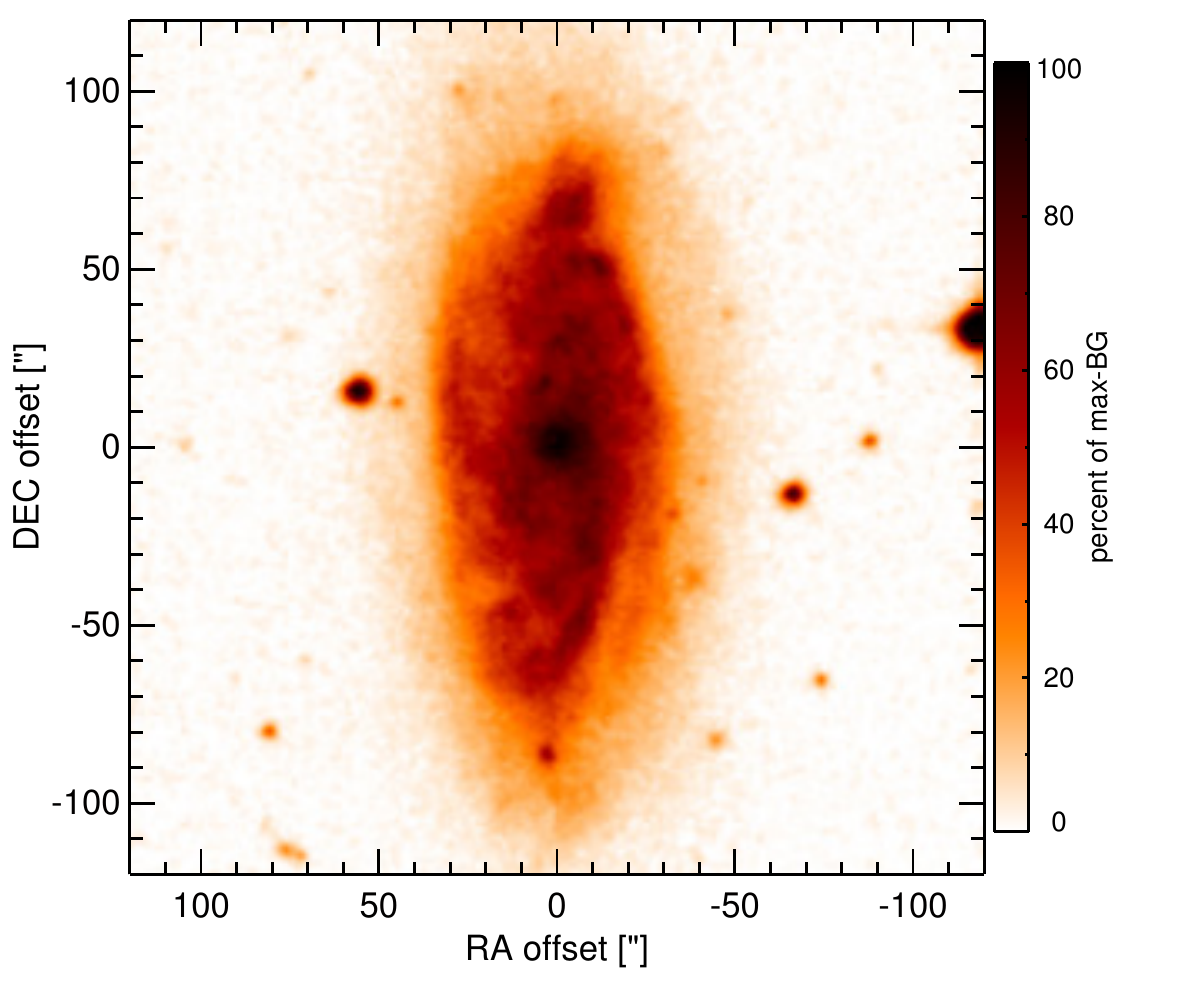}
    \caption{\label{fig:OPTim_NGC7314}
             Optical image (DSS, red filter) of NGC\,7314. Displayed are the central $4\arcmin$ with North up and East to the left. 
              The colour scaling is linear with white corresponding to the median background and black to the $0.01\%$ pixels with the highest intensity.  
           }
\end{figure}
\begin{figure}
   \centering
   \includegraphics[angle=0,height=3.11cm]{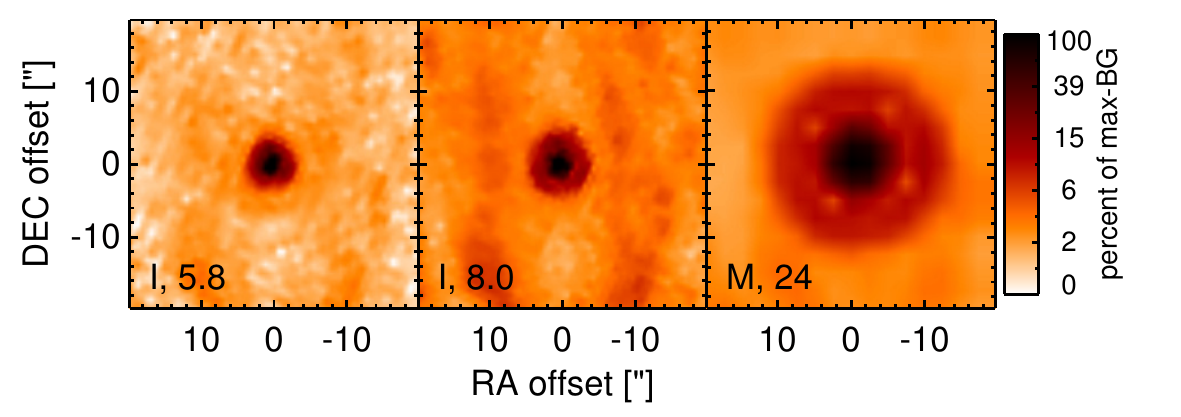}
    \caption{\label{fig:INTim_NGC7314}
             \spitzerr MIR images of NGC\,7314. Displayed are the inner $40\arcsec$ with North up and East to the left. The colour scaling is logarithmic with white corresponding to median background and black to the $0.1\%$ pixels with the highest intensity.
             The label in the bottom left states instrument and central wavelength of the filter in $\mu$m (I: IRAC, M: MIPS). 
           }
\end{figure}
\begin{figure}
   \centering
   \includegraphics[angle=0,width=8.500cm]{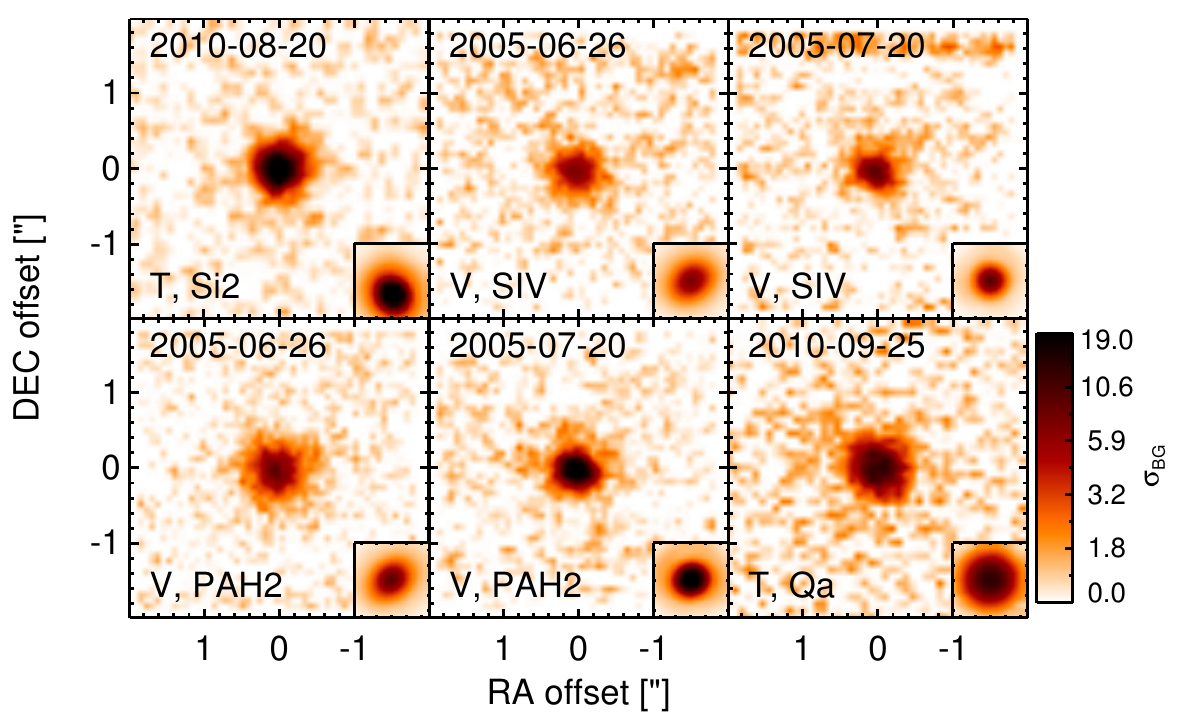}
    \caption{\label{fig:HARim_NGC7314}
             Subarcsecond-resolution MIR images of NGC\,7314 sorted by increasing filter wavelength. 
             Displayed are the inner $4\arcsec$ with North up and East to the left. 
             The colour scaling is logarithmic with white corresponding to median background and black to the $75\%$ of the highest intensity of all images in units of $\sigbg$.
             The inset image shows the central arcsecond of the PSF from the calibrator star, scaled to match the science target.
             The labels in the bottom left state instrument and filter names (C: COMICS, M: Michelle, T: T-ReCS, V: VISIR).
           }
\end{figure}
\begin{figure}
   \centering
   \includegraphics[angle=0,width=8.50cm]{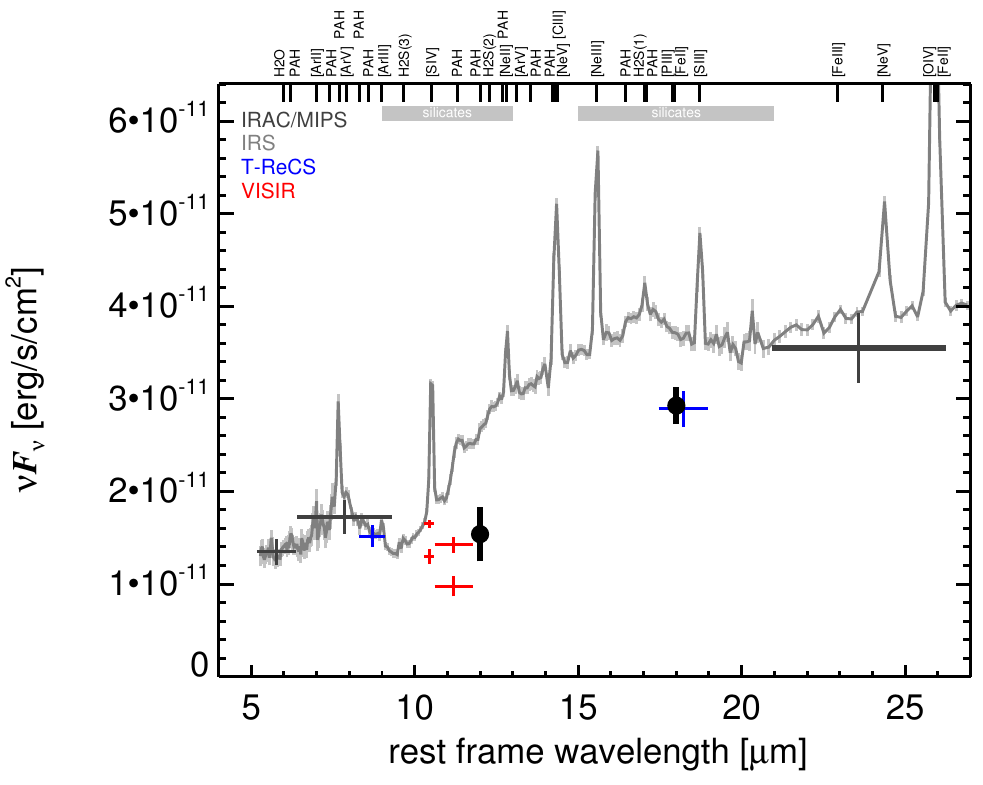}
   \caption{\label{fig:MISED_NGC7314}
      MIR SED of NGC\,7314. The description  of the symbols (if present) is the following.
      Grey crosses and  solid lines mark the \spitzer/IRAC, MIPS and IRS data. 
      The colour coding of the other symbols is: 
      green for COMICS, magenta for Michelle, blue for T-ReCS and red for VISIR data.
      Darker-coloured solid lines mark spectra of the corresponding instrument.
      The black filled circles mark the nuclear 12 and $18\,\mu$m  continuum emission estimate from the data.
      The ticks on the top axis mark positions of common MIR emission lines, while the light grey horizontal bars mark wavelength ranges affected by the silicate 10 and 18$\mu$m features.}
\end{figure}
\clearpage

\twocolumn[\begin{@twocolumnfalse}  
\subsection{NGC\,7469 -- IRAS\,23007+0836}\label{app:NGC7469}
NGC\,7469 is a low-inclination infrared-luminous barred spiral galaxy at a redshift of $z=$ 0.0163 ($D\sim67.9\,$Mpc) and one of the original six Seyfert galaxies \citep{seyfert_nuclear_1943}.
It is presumably interacting with the nearby irregular galaxy IC\,5283 $\sim22\,$kpc to the north, forming the pair Arp\,298.
The AGN is optically classified either as Sy\,1.0, Sy\,1.2 or Sy\,1.5  (see \citealt{osterbrock_spectroscopic_1993}) and surrounded by a powerful starburst with $\sim5\arcsec\sim1.6\,$kpc diameter (e.g. \citealt{wilson_starburst_1991,genzel_infrared_1995,diaz-santos_resolving_2007}).
NGC\,7469 also belongs to the nine-month BAT AGN sample.
It features a compact radio core at arcsecond resolution, which is slightly elongated in the east-west direction \citep{wilson_starburst_1991}. The radio core is resolved into several compact sources with a maximum separation of $\sim0.17\arcsec\sim54\,$pc along the same direction at milliarcsecond scales (PA$\sim85\degree$; \citealt{lonsdale_starburst-agn_1993,lonsdale_vlbi_2003}).
This objects has been extensively studied in the MIR since the pioneering observations by \citeauthor{kleinmann_observations_1970} (1970; \citealt{rieke_infrared_1972,rieke_infrared_1978,lebofsky_extinction_1979,aitken_question_1981,malkan_stellar_1983,ward_continuum_1987,carico_iras_1988,roche_atlas_1991,keto_subarcsecond_1992,wynn-williams_luminous_1993}).
The first subarcsecond-resolution MIR image obtained with Palomar 5\,m/SpectroCam-10 was presented by \cite{miles_high-resolution_1994,miles_high-resolution_1996}.
This image resolved for the first time the compact nucleus clearly from the patchy starburst ring with $\sim3\arcsec\sim1\,$kpc diameter in the MIR.
The same structure was also visible in following MIR images \citep{soifer_high_2003,gorjian_10_2004,galliano_mid-infrared_2005,raban_core_2008}.
In addition, NGC\,7469 was observed with the space-based \isoo \citep{rigopoulou_large_1999,thornley_massive_2000,tran_isocam-cvf_2001} and \spitzer/IRAC, IRS and MIPS.
The corresponding IRAC and MIPS images are completely dominated by a bright nucleus with the spiral-like host emission being only visible in the former (see also \citealt{smith_spitzer_2007}).
The post-BCD IRAC images are saturated and thus not analysed (but see \citealt{gallimore_infrared_2010}).
Our nuclear MIPS $24\,\mu$m photometry agrees with \cite{u_spectral_2012}.
The IRS LR staring-mode spectrum exhibits strong PAH emission and a red spectral slope in $\nu F_\nu$-space. Weak 10 and 18\,$\mu$m silicate emission features might be present but are diluted by the PAH features (see also \citealt{buchanan_spitzer_2006,wu_spitzer/irs_2009,gallimore_infrared_2010}).
Thus, the arcsecond-scale MIR SED is heavily affected and possibly even dominated by star formation (see also \citealt{alonso-herrero_local_2012}).
The complex morphology and strong PAH emission complicate a comparison with the historical MIR photometry, but the data indicates no significant flux variations in the last $\sim40$ years. 
The nuclear region of NGC\,7469 was imaged with VISIR in four different $N$-band and three $Q$-band filters between 2006 and 2008 (partly published in \citealt{horst_mid_2008,horst_mid-infrared_2009,reunanen_vlt_2010}). In addition, a VISIR LR $N$-band spectrum was obtained \citep{honig_dusty_2010-1}.
Further imaging data was taken by T-ReCS in the Si2 and Qa filters in 2009 \citep{ramos_almeida_testing_2011}.
In all cases, the compact nucleus was detected, while only the brightest knots of the starburst ring are visible in most images (see \citealt{ramos_almeida_testing_2011} for an analysis of the starburst knots).
The nucleus appears marginally resolved in the VISIR $N$-band images (FWHM(major axis)$\sim0.4\arcsec\sim130\,$pc; PA$\sim85\degree$), while it remains unresolved in the sharpest $Q$-band images.
Note that the elongation is roughly parallel to the radio morphology.
Our corresponding unresolved nuclear fluxes are generally consistent with the PSF-scaled flux measurements from \cite{reunanen_vlt_2010} and \cite{ramos_almeida_testing_2011}, while the total nuclear fluxes agree with the photometry from \cite{horst_mid_2008} and the VISIR spectrum from \cite{honig_dusty_2010-1} within the uncertainties.
Note that the VISIR NEII\_2 measurement is treated as an upper limit on the unresolved flux because it lacks a suitable standard star observation.
The resulting nuclear MIR SED is $\sim64\%$ lower than the \spitzerr spectrophotometry and does not exhibit significant PAH emission, which indicates that the AGN has been successfully isolated.
Furthermore, we can conclude that star formation indeed dominates the MIR emission of the central $\sim1\,$kpc in NGC\,7469.
Note that the MIR nucleus of NGC\,7469 was recently further resolved with MIR interferometric observations and was modelled as two approximately equally bright components, one extending by $\sim28\,$pc, and one remaining unresolved ($\sim 8$\,pc;  \citealt{tristram_parsec-scale_2009,burtscher_diversity_2013}).
\newline\end{@twocolumnfalse}]

\begin{figure}
   \centering
   \includegraphics[angle=0,width=8.500cm]{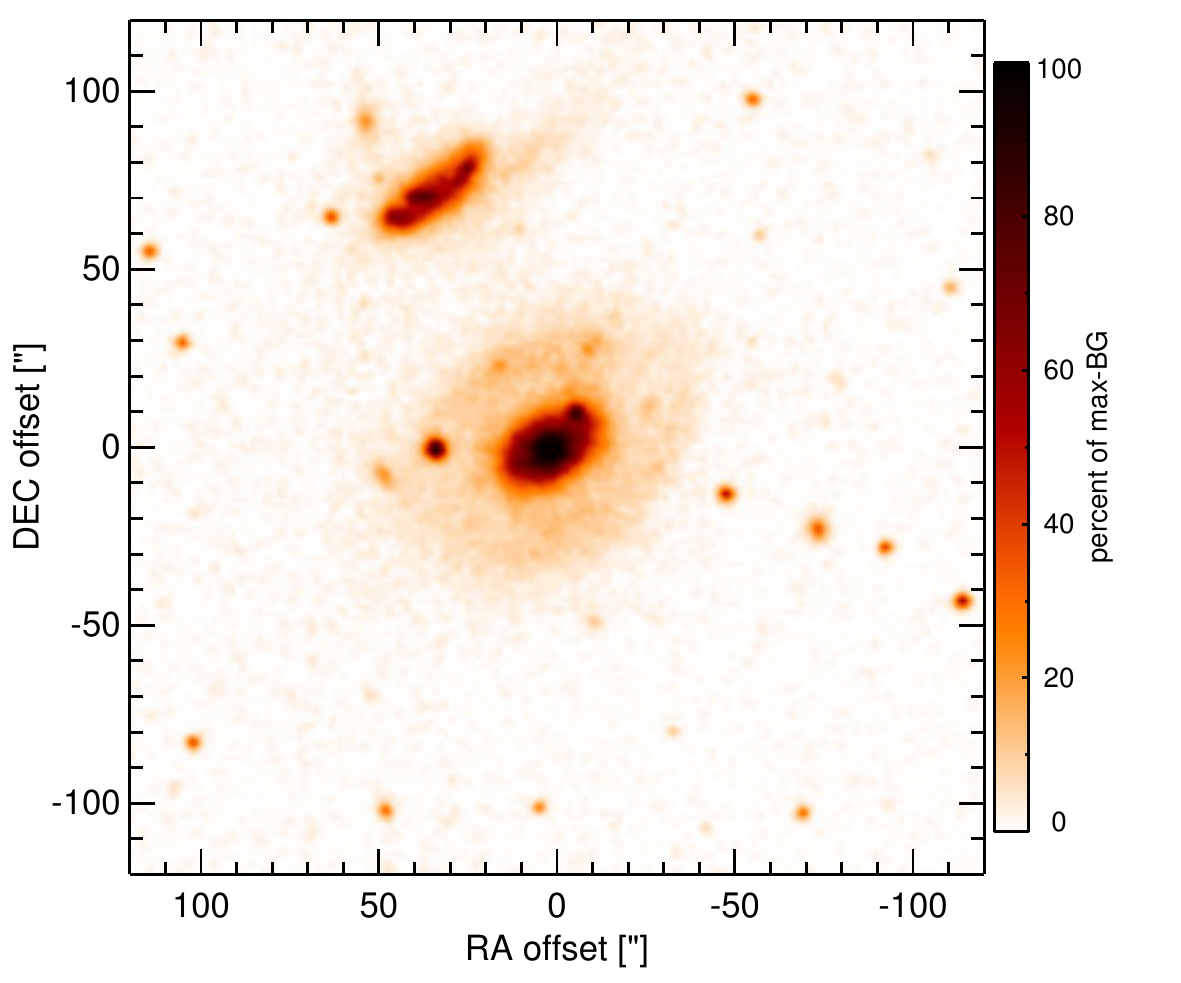}
    \caption{\label{fig:OPTim_NGC7469}
             Optical image (DSS, red filter) of NGC\,7469. Displayed are the central $4\arcmin$ with North up and East to the left. 
              The colour scaling is linear with white corresponding to the median background and black to the $0.01\%$ pixels with the highest intensity.  
           }
\end{figure}
\begin{figure}
   \centering
   \includegraphics[angle=0,height=3.11cm]{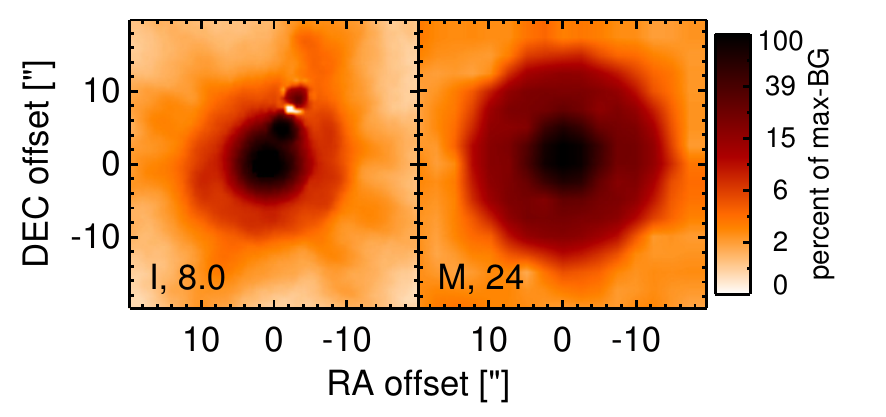}
    \caption{\label{fig:INTim_NGC7469}
             \spitzerr MIR images of NGC\,7469. Displayed are the inner $40\arcsec$ with North up and East to the left. The colour scaling is logarithmic with white corresponding to median background and black to the $0.1\%$ pixels with the highest intensity.
             The label in the bottom left states instrument and central wavelength of the filter in $\mu$m (I: IRAC, M: MIPS).
             Note that the apparent off-nuclear compact sources in the IRAC $8.0\,\mu$m image are instrumental artefacts.
           }
\end{figure}
\begin{figure}
   \centering
   \includegraphics[angle=0,width=8.500cm]{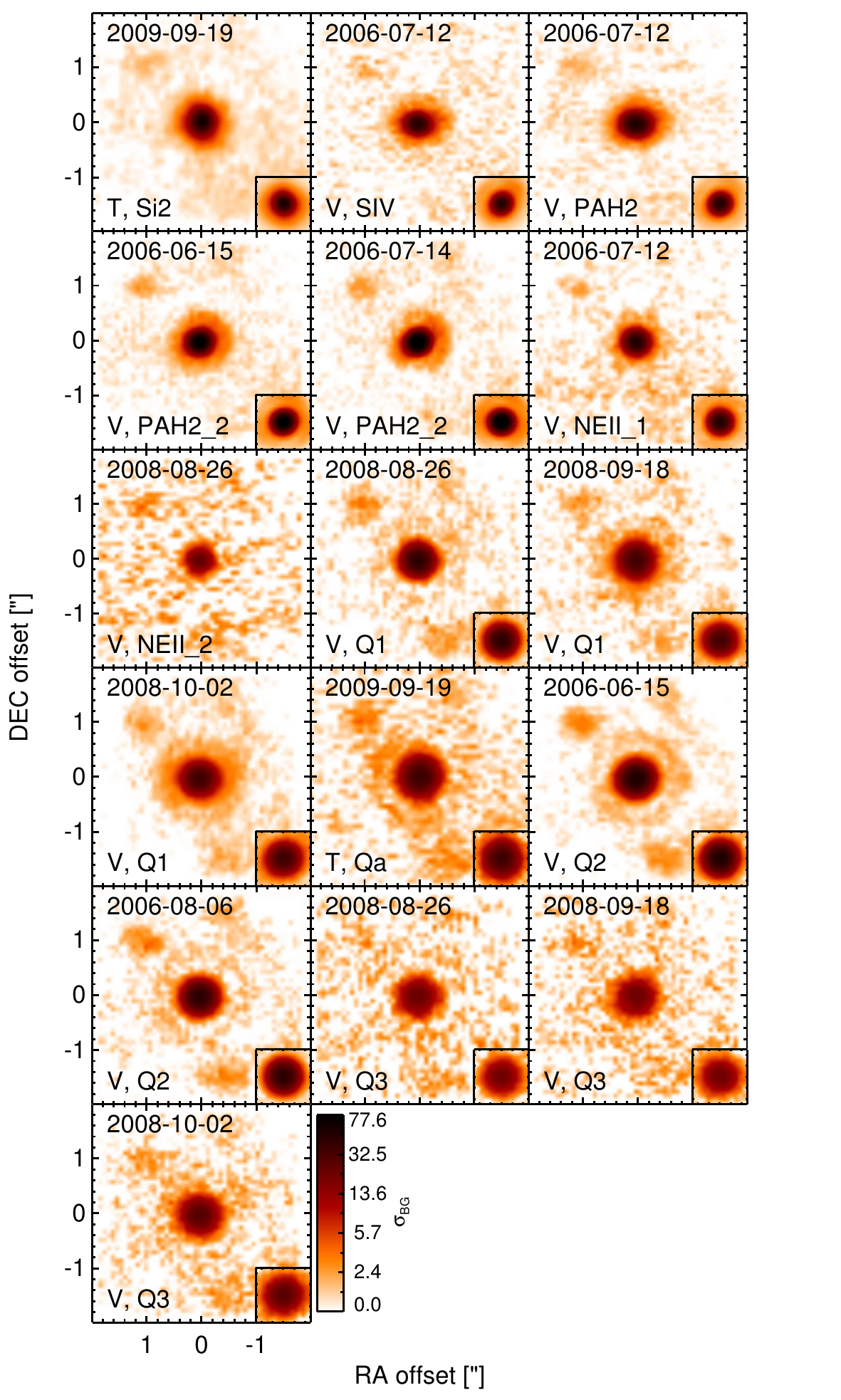}
    \caption{\label{fig:HARim_NGC7469}
             Subarcsecond-resolution MIR images of NGC\,7469 sorted by increasing filter wavelength. 
             Displayed are the inner $4\arcsec$ with North up and East to the left. 
             The colour scaling is logarithmic with white corresponding to median background and black to the $75\%$ of the highest intensity of all images in units of $\sigbg$.
             The inset image shows the central arcsecond of the PSF from the calibrator star, scaled to match the science target.
             The labels in the bottom left state instrument and filter names (C: COMICS, M: Michelle, T: T-ReCS, V: VISIR).
           }
\end{figure}
\begin{figure}
   \centering
   \includegraphics[angle=0,width=8.50cm]{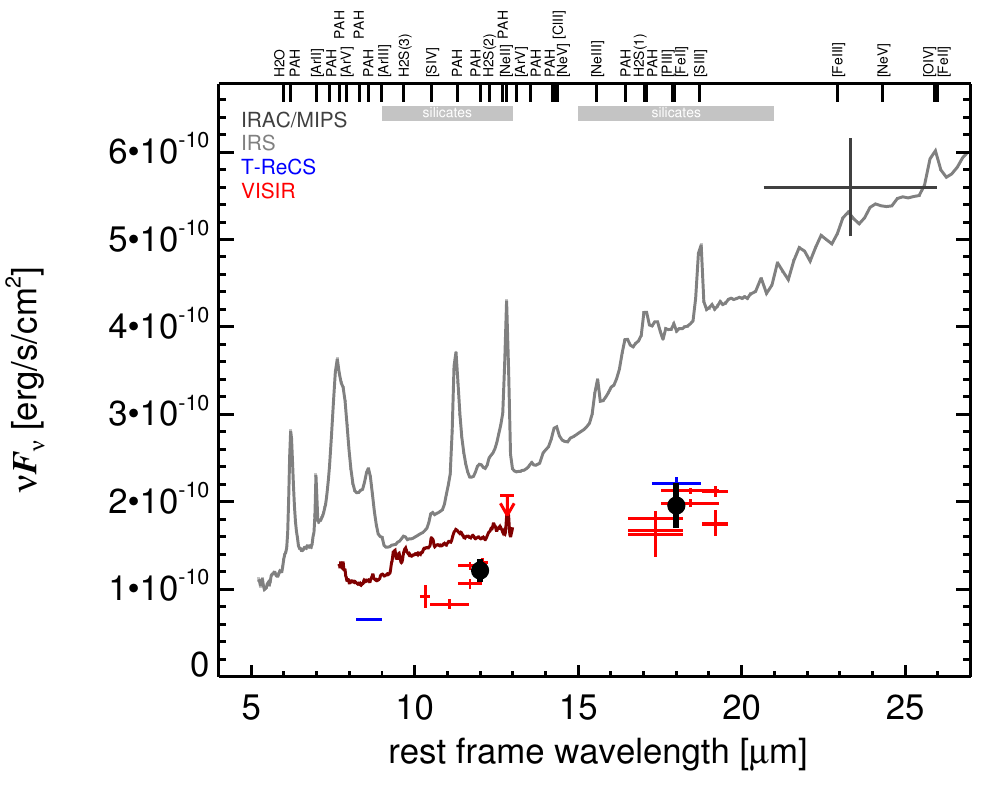}
   \caption{\label{fig:MISED_NGC7469}
      MIR SED of NGC\,7469. The description  of the symbols (if present) is the following.
      Grey crosses and  solid lines mark the \spitzer/IRAC, MIPS and IRS data. 
      The colour coding of the other symbols is: 
      green for COMICS, magenta for Michelle, blue for T-ReCS and red for VISIR data.
      Darker-coloured solid lines mark spectra of the corresponding instrument.
      The black filled circles mark the nuclear 12 and $18\,\mu$m  continuum emission estimate from the data.
      The ticks on the top axis mark positions of common MIR emission lines, while the light grey horizontal bars mark wavelength ranges affected by the silicate 10 and 18$\mu$m features.}
\end{figure}
\clearpage

\twocolumn[\begin{@twocolumnfalse}  
\subsection{NGC\,7479}\label{app:NGC7479}
NGC\,7479 is a low-inclination disturbed barred grand-design spiral galaxy at a redshift of $z=$ 0.0079 ($D\sim30\,$Mpc), which presumably underwent a recent minor merger \citep{quillen_estimate_1995,laine_minor-merger_1999}.
Its nucleus is active and was initially classified as a LINER, then as Sy\,1.9 and most recently as Sy\,2.0 (see discussion in \citealt{trippe_multi-wavelength_2010}).
In addition, star formation is reported to occur in the nucleus \citep{martin_star_1997}.
At radio wavelengths, the nucleus appears as a compact source with a possible kiloparsec-scale bended jet to the north \cite{ho_radio_2001,laine_radio_2008}.
Detected water maser emission is possibly originating in the nucleus \citep{braatz_discovery_2008}.
After first being detected in the MIR with \iras, NGC\,7479 was followed up with IRTF \citep{willner_infrared_1985,devereux_spatial_1987,telesco_genesis_1993}, AAT/UKIRT \citep{roche_atlas_1991}, and Keck/LWS \citep{soifer_high_2004}.
The latter observation resulted in the first subarcsecond MIR image, which showed a point-like nucleus.
The \spitzer/IRAC and MIPS images are also dominated by compact nuclear emission embedded within bar-like host emission (see also \citealt{zhou_star_2011}.
The \spitzer/IRS LR staring-mode spectrum exhibits very deep silicate 10 and 18\,$\mu$m absorption features, no significant PAH emission or atomic forbidden emission lines, and a steep red spectral slope in $\nu F_\nu$-space (see also \citealt{dartois_carbonaceous_2007,pereira-santaella_mid-infrared_2010}).
Therefore, a MIR-dominating, highly-obscured AGN might be present.
On the other hand, the absence of any strong MIR emission lines indicates that putative ionized gas is heavily extincted in NGC\,7479 (similar to NGC\,4945; \citealt{perez-beaupuits_deeply_2011}).
Its nuclear region was imaged with Michelle in the Si-2 filter in 2008 and an unresolved nucleus without further host emission was detected \citep{mason_nuclear_2012}.
In addition, a T-ReCS LR $N$-band spectrum was obtained \citep{gonzalez-martin_dust_2013}, which agrees well with the \spitzerr spectrophotometry, exhibiting the same continuum and silicate absorption.
The silicate absorption is, thus, produced in the projected central $\sim50\,$pc. 
Our nuclear Si-2 photometry is consistent with the value by \cite{mason_nuclear_2012} and the T-ReCS spectrum.
Therefore, we use the latter to compute the $12\,\mu$m continuum emission estimate corrected for the silicate feature.
\newline\end{@twocolumnfalse}]

\begin{figure}
   \centering
   \includegraphics[angle=0,width=8.500cm]{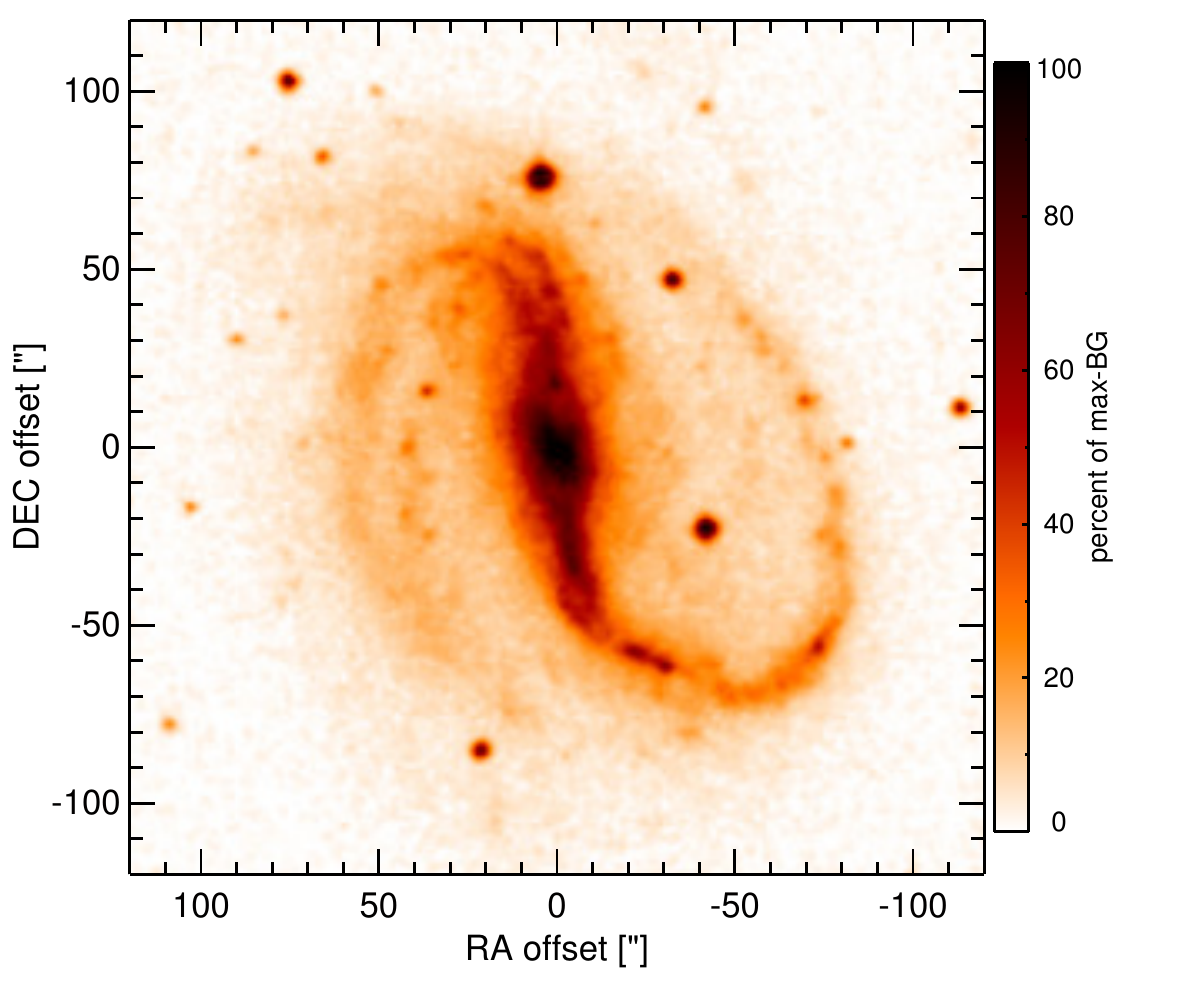}
    \caption{\label{fig:OPTim_NGC7479}
             Optical image (DSS, red filter) of NGC\,7479. Displayed are the central $4\arcmin$ with North up and East to the left. 
              The colour scaling is linear with white corresponding to the median background and black to the $0.01\%$ pixels with the highest intensity.  
           }
\end{figure}
\begin{figure}
   \centering
   \includegraphics[angle=0,height=3.11cm]{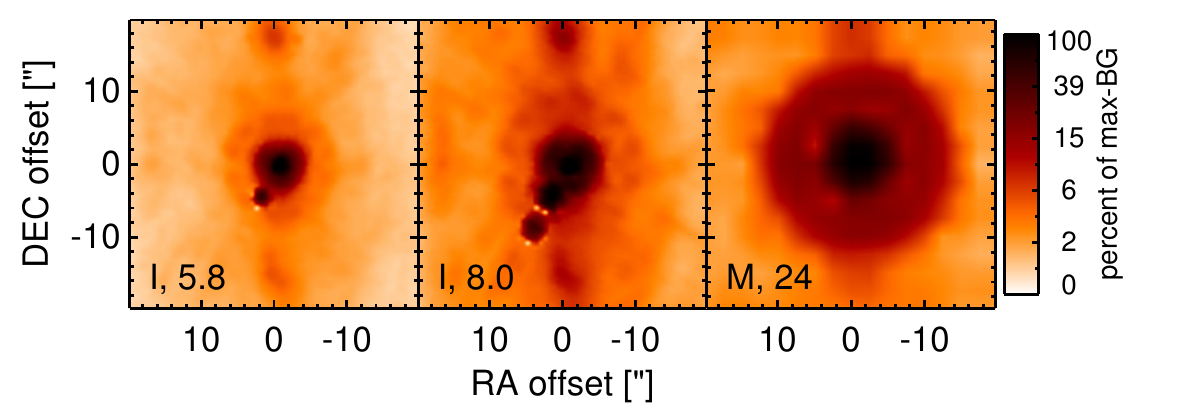}
    \caption{\label{fig:INTim_NGC7479}
             \spitzerr MIR images of NGC\,7479. Displayed are the inner $40\arcsec$ with North up and East to the left. The colour scaling is logarithmic with white corresponding to median background and black to the $0.1\%$ pixels with the highest intensity.
             The label in the bottom left states instrument and central wavelength of the filter in $\mu$m (I: IRAC, M: MIPS).
             Note that the apparent off-nuclear compact sources in the IRAC 5.8 and  $8.0\,\mu$m images are instrumental artefacts.
           }
\end{figure}
\begin{figure}
   \centering
   \includegraphics[angle=0,height=3.11cm]{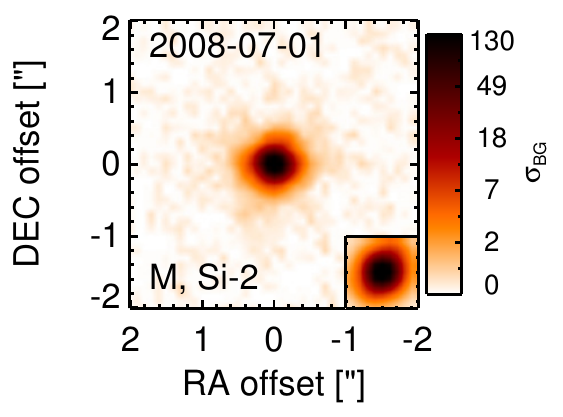}
    \caption{\label{fig:HARim_NGC7479}
             Subarcsecond-resolution MIR images of NGC\,7479 sorted by increasing filter wavelength. 
             Displayed are the inner $4\arcsec$ with North up and East to the left. 
             The colour scaling is logarithmic with white corresponding to median background and black to the $75\%$ of the highest intensity of all images in units of $\sigbg$.
             The inset image shows the central arcsecond of the PSF from the calibrator star, scaled to match the science target.
             The labels in the bottom left state instrument and filter names (C: COMICS, M: Michelle, T: T-ReCS, V: VISIR).
           }
\end{figure}
\begin{figure}
   \centering
   \includegraphics[angle=0,width=8.50cm]{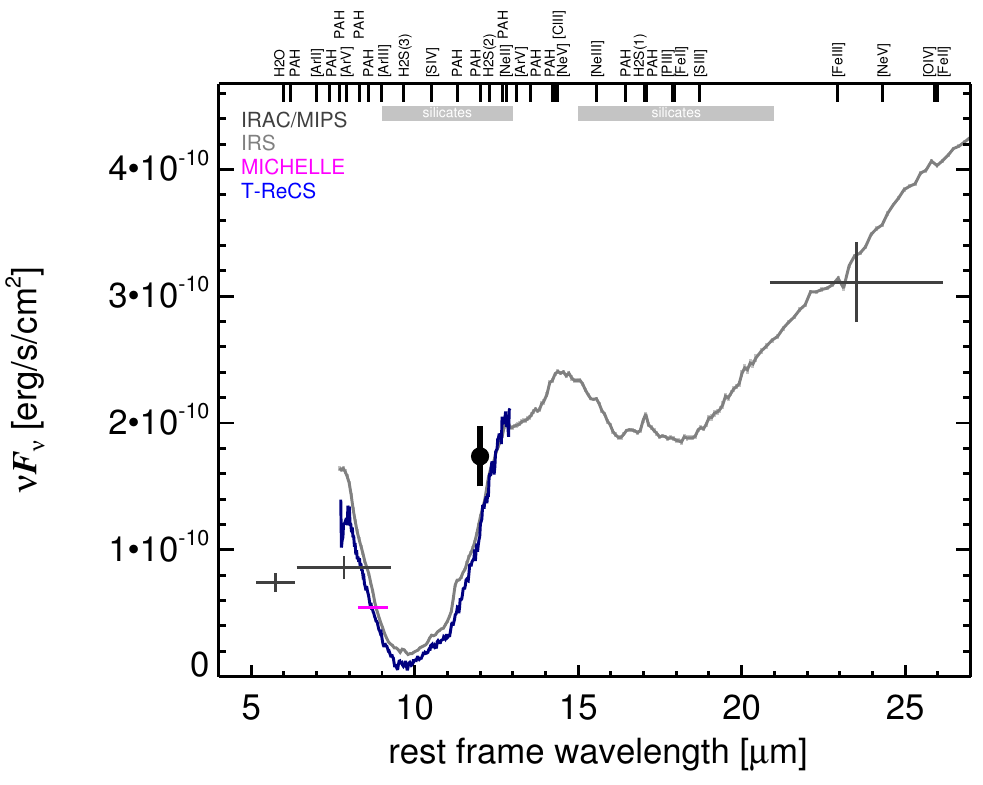}
   \caption{\label{fig:MISED_NGC7479}
      MIR SED of NGC\,7479. The description  of the symbols (if present) is the following.
      Grey crosses and  solid lines mark the \spitzer/IRAC, MIPS and IRS data. 
      The colour coding of the other symbols is: 
      green for COMICS, magenta for Michelle, blue for T-ReCS and red for VISIR data.
      Darker-coloured solid lines mark spectra of the corresponding instrument.
      The black filled circles mark the nuclear 12 and $18\,\mu$m  continuum emission estimate from the data.
      The ticks on the top axis mark positions of common MIR emission lines, while the light grey horizontal bars mark wavelength ranges affected by the silicate 10 and 18$\mu$m features.}
\end{figure}
\clearpage

\twocolumn[\begin{@twocolumnfalse}  
\subsection{NGC\,7496}\label{app:NGC7496}
NGC\,7496 is a low-inclination barred late-type spiral at a redshift of $z=$ 0.0055 ($D\sim2.1\,$Mpc) with an active nucleus containing both an AGN and a starburst \citep{schmitt_multiwavelength_2006,munoz_marin_atlas_2007}.
It has been optically classified as a Sy\,2 \citep{veron_how_1981} and H\,II nucleus \citep{yuan_role_2010}, and conclusively is an AGN/starburst composite \citep{veron_agns_1997}.
The nucleus is unresolved in arcsecond-resolution radio observations \citep{norris_compact_1990,morganti_radio_1999}.
After first being detected in the MIR with \iras, NGC\,7496 was followed up with \spitzer/IRAC, IRS and MIPS observations.
The corresponding IRAC and MIPS images show a dominating compact nucleus embedded within the much weak bar and spiral-like host emission. 
Our nuclear IRAC 5.8\,$\mu$m photometry agrees with \cite{gallimore_infrared_2010}. On the other hand, our IRAC 8.0\,$\mu$m value is significantly lower than the published value, yet in better agreement with the IRS LR mapping-mode post-BCD spectrum. 
The latter exhibits strong PAH emission, no clearly determinable silicate features, and a steep red spectral slope in $\nu F_\nu$-space (see also \citealt{buchanan_spitzer_2006,wu_spitzer/irs_2009,tommasin_spitzer_2008,tommasin_spitzer-irs_2010,gallimore_infrared_2010}).
Thus, the arcsecond-scale MIR SED is significantly affected or even dominated by star-formation emission.
The nuclear region of NGC\,7496 was observed with T-ReCS in the Si2 and Qa filters in 2010 (unpublished, to our knowledge).
In the images, a marginally resolved roundish nucleus was detected (FWHM$\sim23\%$ larger than standard star).
The unresolved nuclear component fluxes are $\sim 45\%$ lower than the \spitzerr spectrophotometry.
We conclude that star formation dominates over AGN-related MIR emission in NGC\,7496 and possible still affects the nuclear MIR SED.
\newline\end{@twocolumnfalse}]

\begin{figure}
   \centering
   \includegraphics[angle=0,width=8.500cm]{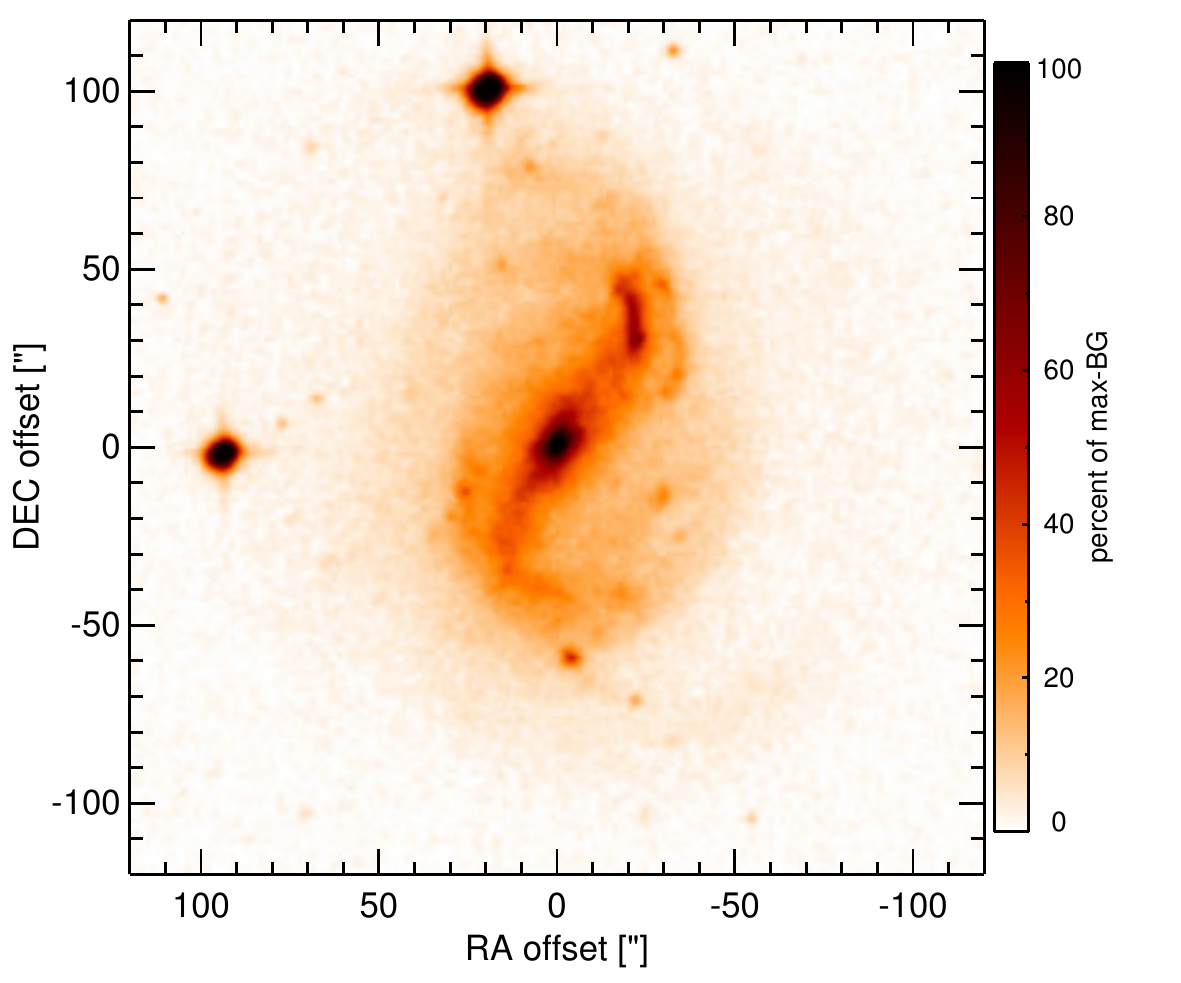}
    \caption{\label{fig:OPTim_NGC7496}
             Optical image (DSS, red filter) of NGC\,7496. Displayed are the central $4\arcmin$ with North up and East to the left. 
              The colour scaling is linear with white corresponding to the median background and black to the $0.01\%$ pixels with the highest intensity.  
           }
\end{figure}
\begin{figure}
   \centering
   \includegraphics[angle=0,height=3.11cm]{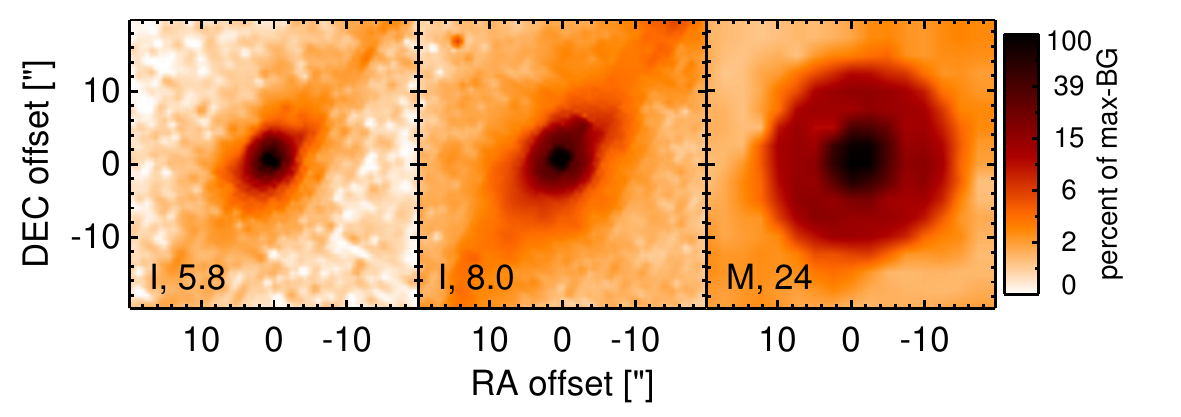}
    \caption{\label{fig:INTim_NGC7496}
             \spitzerr MIR images of NGC\,7496. Displayed are the inner $40\arcsec$ with North up and East to the left. The colour scaling is logarithmic with white corresponding to median background and black to the $0.1\%$ pixels with the highest intensity.
             The label in the bottom left states instrument and central wavelength of the filter in $\mu$m (I: IRAC, M: MIPS). 
             Note that the apparent off-nuclear compact source in the IRAC $8.0\,\mu$m image is an instrumental artefact.
           }
\end{figure}
\begin{figure}
   \centering
   \includegraphics[angle=0,height=3.11cm]{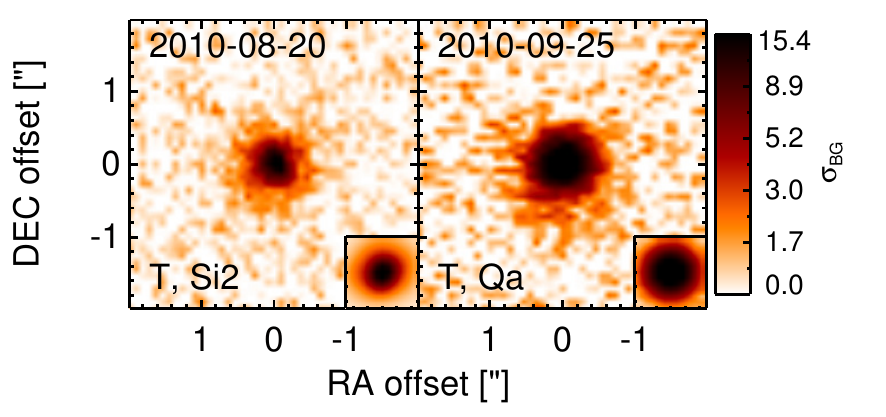}
    \caption{\label{fig:HARim_NGC7496}
             Subarcsecond-resolution MIR images of NGC\,7496 sorted by increasing filter wavelength. 
             Displayed are the inner $4\arcsec$ with North up and East to the left. 
             The colour scaling is logarithmic with white corresponding to median background and black to the $75\%$ of the highest intensity of all images in units of $\sigbg$.
             The inset image shows the central arcsecond of the PSF from the calibrator star, scaled to match the science target.
             The labels in the bottom left state instrument and filter names (C: COMICS, M: Michelle, T: T-ReCS, V: VISIR).
           }
\end{figure}
\begin{figure}
   \centering
   \includegraphics[angle=0,width=8.50cm]{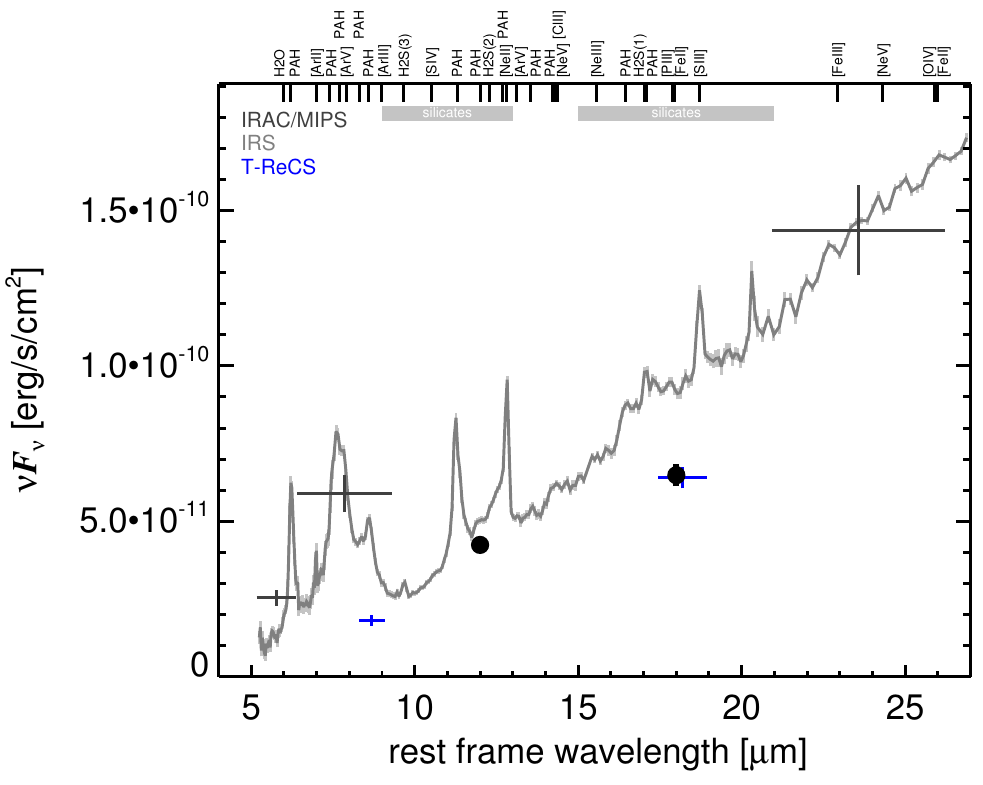}
   \caption{\label{fig:MISED_NGC7496}
      MIR SED of NGC\,7496. The description  of the symbols (if present) is the following.
      Grey crosses and  solid lines mark the \spitzer/IRAC, MIPS and IRS data. 
      The colour coding of the other symbols is: 
      green for COMICS, magenta for Michelle, blue for T-ReCS and red for VISIR data.
      Darker-coloured solid lines mark spectra of the corresponding instrument.
      The black filled circles mark the nuclear 12 and $18\,\mu$m  continuum emission estimate from the data.
      The ticks on the top axis mark positions of common MIR emission lines, while the light grey horizontal bars mark wavelength ranges affected by the silicate 10 and 18$\mu$m features.}
\end{figure}
\clearpage

\twocolumn[\begin{@twocolumnfalse}  
\subsection{NGC\,7552}\label{app:NGC7552}
NGC\,7552 is an infrared-luminous, nearly face-on, barred spiral galaxy at a redshift of $z=$ 0.0054 ($D\sim20.4\,$Mpc) with a possibe active nucleus surrounded by a powerful circum-nuclear starburst ring  (diameter$\sim5\arcsec\sim0.5\,$kpc; \citealt{forbes_nuclear_1994,forbes_ngc_1994}; see  \citealt{pan_formation_2013,brandl_high_2012} for recent detailed studies).
The nucleus has been optically classified as a LINER \citep{durret_imaging_1987} or H\,II \citep{yuan_role_2010}.
The presence of an AGN is supported by the detection of a compact nuclear source in hard X-rays \citep{grier_discovery_2011,brightman_xmm-newton_2011}.
However, we conservatively treat NGC\,7552 as an uncertain AGN.
Initial MIR observations of NGC\,7552 are reported in \cite{kleinmann_10-micron_1974}, followed by many ground-based observations \citep{frogel_8-13_1982,phillips_8-13_1984,roche_atlas_1991,schinnerer_circumnuclear_1997,siebenmorgen_mid-infrared_2004}.
The latter two works present the first subarcsecond-resolution MIR images, which resolve the starburst ring as a clumpy structure.
NGC\,7552 was also observed with the space-based \isoo \citep{rigopoulou_large_1999,roussel_atlas_2001,verma_mid-infrared_2003} and \spitzer/IRAC, IRS and MIPS.
The corresponding IRAC and MIPS images do not completely resolve the starburst ring, which appears as a dominating central source embedded within the spiral-like host galaxy emission (see also \citealt{brandl_high_2012}).
We measure the nuclear flux component in the non-saturated IRAC 5.8\,$\mu$m and MIPS 24\,$\mu$m post-BCD images, which conclusively results in significantly lower values than in the literature (e.g. \citealt{dale_infrared_2005,munoz-mateos_radial_2009}).
The IRS LR mapping-mode spectrum shows strong PAH emission and possibly weak silicate 10\,$\mu$m absorption with a red spectral slope in $\nu F_\nu$-space (see also \citealt{smith_mid-infrared_2007,marble_aromatic_2010}).
As expected, the arcsecond-scale MIR SED appears to be star-formation dominated.
The nuclear region of NGC\,7552 was imaged with VISIR in eleven $N$-band and one $Q$-band filters in 2005 (partly published in \citealt{brandl_high_2012}).
At least the brightest knots of the starburst ring are detected in all images, while the nucleus remains undetected (see \citealt{brandl_high_2012} for the analysis of the ring).
A compact source is faintly visible at the nuclear position ($<2\sigma$) only in the longest-wavelength $N$-band filters.
Therefore, we determine flux upper limits on the nuclear emission, which are on average $\sim94\%$ lower than the \spitzerr spectrophotometry.
We conclude that a very weak AGN might be present in NGC\,7552 but star formation is dominating its MIR emission completely. 
\newline\end{@twocolumnfalse}]

\begin{figure}
   \centering
   \includegraphics[angle=0,width=8.500cm]{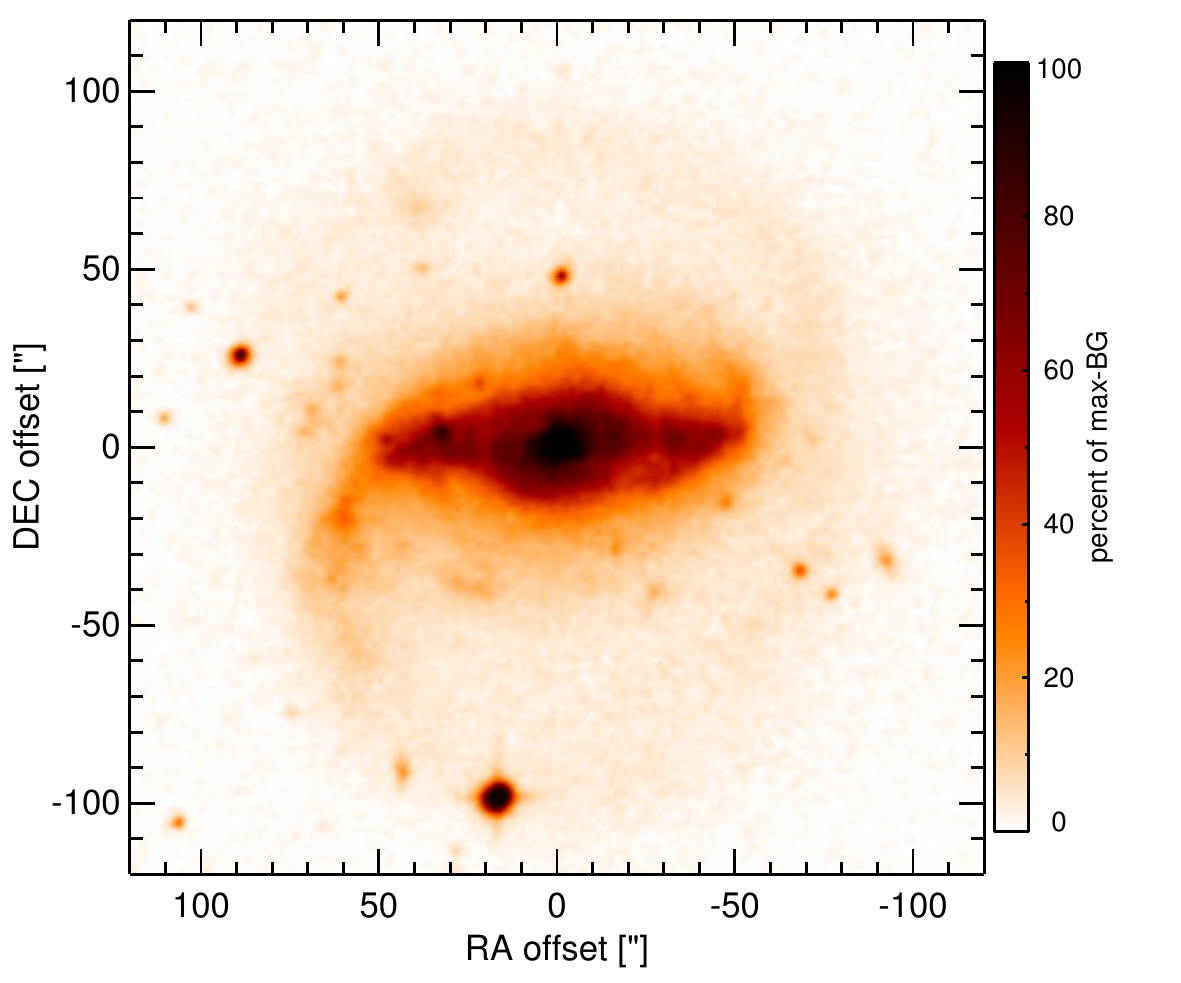}
    \caption{\label{fig:OPTim_NGC7552}
             Optical image (DSS, red filter) of NGC\,7552. Displayed are the central $4\arcmin$ with North up and East to the left. 
              The colour scaling is linear with white corresponding to the median background and black to the $0.01\%$ pixels with the highest intensity.  
           }
\end{figure}
\begin{figure}
   \centering
   \includegraphics[angle=0,height=3.11cm]{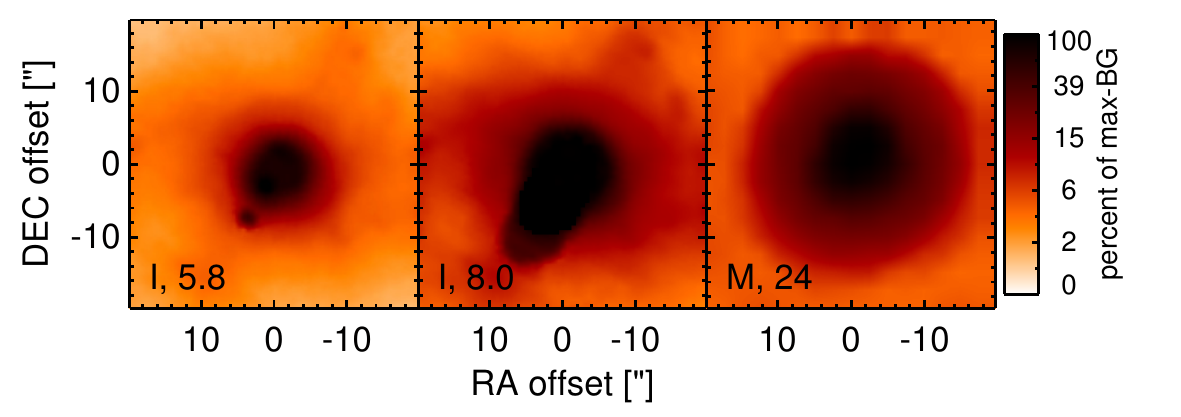}
    \caption{\label{fig:INTim_NGC7552}
             \spitzerr MIR images of NGC\,7552. Displayed are the inner $40\arcsec$ with North up and East to the left. The colour scaling is logarithmic with white corresponding to median background and black to the $0.1\%$ pixels with the highest intensity.
             The label in the bottom left states instrument and central wavelength of the filter in $\mu$m (I: IRAC, M: MIPS).
             Note that the apparent off-nuclear sources in the IRAC 5.8 and  $8.0\,\mu$m images are instrumental artefacts.
           }
\end{figure}
\begin{figure}
   \centering
   \includegraphics[angle=0,width=8.50cm]{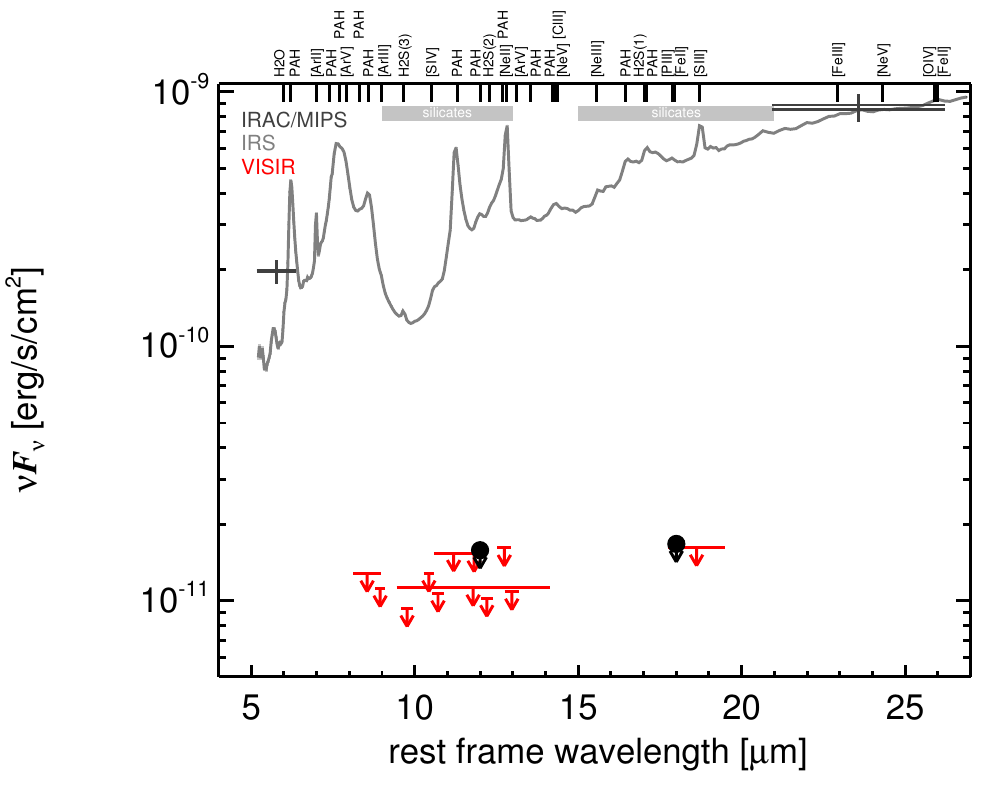}
   \caption{\label{fig:MISED_NGC7552}
      MIR SED of NGC\,7552. The description  of the symbols (if present) is the following.
      Grey crosses and  solid lines mark the \spitzer/IRAC, MIPS and IRS data. 
      The colour coding of the other symbols is: 
      green for COMICS, magenta for Michelle, blue for T-ReCS and red for VISIR data.
      Darker-coloured solid lines mark spectra of the corresponding instrument.
      The black filled circles mark the nuclear 12 and $18\,\mu$m  continuum emission estimate from the data.
      The ticks on the top axis mark positions of common MIR emission lines, while the light grey horizontal bars mark wavelength ranges affected by the silicate 10 and 18$\mu$m features.}
\end{figure}
\clearpage

\twocolumn[\begin{@twocolumnfalse}  
\subsection{NGC\,7582}\label{app:NGC7582}
NGC\,7582 is a highly inclined barred spiral galaxy at a distance of $D=$ $23 \pm 4.6$\,Mpc \citep{springob_erratum:_2009} with an obscured AGN surrounded by a powerful star formation disc (major axis diameter$\sim4\arcsec\sim0.4$\,kpc) and a dust lane crossing over the nucleus \citep{morris_velocity_1985,regan_using_1999,riffel_agn-starburst_2009}.
The AGN belongs to nine-month BAT AGN sample and has complex X-ray source with varying obscuration (e.g. \citealt{piconcelli_xmm-newton_2007,bianchi_how_2009}).
The optical nuclear spectrum is a superposition of a Sy\,2 and a starburst, hence we treat it as an AGN/starburst composite \citep{veron_agns_1997}.
Transitional broad emission lines were observed \citep{aretxaga_seyfert_1999}, in agreement with the varying obscuration scenario.
The nucleus appears as a weak compact source embedded within extended starburst-related emission at radio wavelengths with arcsecond resolution  \citep{ulvestad_radio_1984,morganti_radio_1999,thean_high-resolution_2000}.
It features a one-sided kiloparsec-scale \oiii ionization cone to the south-west (PA$\sim250\degree$; \citealt{morris_velocity_1985,storchi-bergmann_detection_1991}).
Pioneering ground-based MIR observations of NGC\,7582 were performed by \cite{frogel_8-13_1982} and \cite{glass_mid-infrared_1982}, followed by \cite{roche_8-13_1984,roche_atlas_1991}.
The first subarcsecond-resolution MIR images were taken with ESO 3.6\,m/TIMMI2 \citep{siebenmorgen_mid-infrared_2004,raban_core_2008} and CTIO 4\,m/OSCIR \citep{ramos_almeida_infrared_2009} and show a compact nucleus embedded within complex extended emission mainly along the north-south direction (diameter$\sim4\arcsec\sim0.4\,$kpc).
NGC\,7582 was also observed with the space-based \isoo \citep{genzel_what_1998,radovich_10-200_1999,rigopoulou_large_1999}
and \spitzer/IRAC, IRS and MIPS.
The corresponding IRAC and MIPS images are dominated by a very bright extended nucleus comprising AGN and starburst emission, which outshines the remaining spiral-like host emission.
The IRAC $5.8$ and $8.0\,\mu$m post-BCD images are saturated and, thus, not used (but see \citealt{gallimore_infrared_2010}).
The IRS LR starring-mode spectrum exhibits deep silicate 10\,$\mu$m absorption, bright PAH features, and a steep red spectral slope in  $\nu F_\nu$-space (see also \citealt{wu_spitzer/irs_2009,tommasin_spitzer-irs_2010,gallimore_infrared_2010}).
Thus, the arcsecond-scale MIR SED is consistent with the dual AGN/starburst nature of NGC\,7582.
Its nuclear region was observed with VISIR in four different $N$-band and one $Q$-band filter images, HR \neii spectra, and a LR $N$-band spectrum, all obtained between 2004 and 2008 \citep{wold_nuclear_2006-1,wold_nuclear_2006,honig_dusty_2010-1,reunanen_vlt_2010}.
In addition, a broad N filter image (unpublished, to our knowledge) and a LR $N$-band spectrum \citep{gonzalez-martin_dust_2013} were obtained with T-ReCS in 2005.
A compact nucleus, embedded within extended emission on $<3\sigma$ level, was detected in all images.
In the sharpest image (PAH2\_2 from 2006), the nucleus is unresolved, while it appears to be marginally resolved in several other cases (PA$\sim80\degree$).
Therefore, we classify the nucleus as generally unresolved in the MIR at subarcsecond resolution.
The extended emission is best visible in the longest-wavelength images. 
It extends $\sim1.5\sim170\,$pc to the south with two bright emission knots (PA$\sim175\degree$ and $\sim197\degree$, respectively), and $\sim 2\arcsec\sim220\,$pc to the north-west with several knotty features (PA$\sim-26\degree$; see \citealt{wold_nuclear_2006-1} for a detailed analysis).
Our nuclear photometry is consistent with the previously published values \citep{honig_dusty_2010-1,reunanen_vlt_2010}, and also with the VISIR and T-ReCS spectra.
The resulting nuclear MIR SED exhibits on average $\sim 65\%$ lower flux levels than the \spitzerr spectrophotometry, while maintaining a deep silicate 10$\,\mu$m absorption feature.
We use the T-ReCS spectrum to compute the nuclear, silicate feature-corrected 12\,$\mu$m continuum emission estimate.
In conclusion, the silicate absorption is originating in the projected central $\sim 40$\,pc of NGC\,7582, either in the nucleus or the foreground dust lane, and the starburst dominates its total MIR emission.
The nuclear MIR SED might still be affected by a minor star-formation contribution as indicated by the weak PAH\,11.3\,$\mu$m feature in the subarcsecond-resolution spectra.
Note that NGC\,7582 was unsuccessfully attempted interferometrically with MIDI \citep{tristram_parsec-scale_2009}.
\newline\end{@twocolumnfalse}]

\begin{figure}
   \centering
   \includegraphics[angle=0,width=8.500cm]{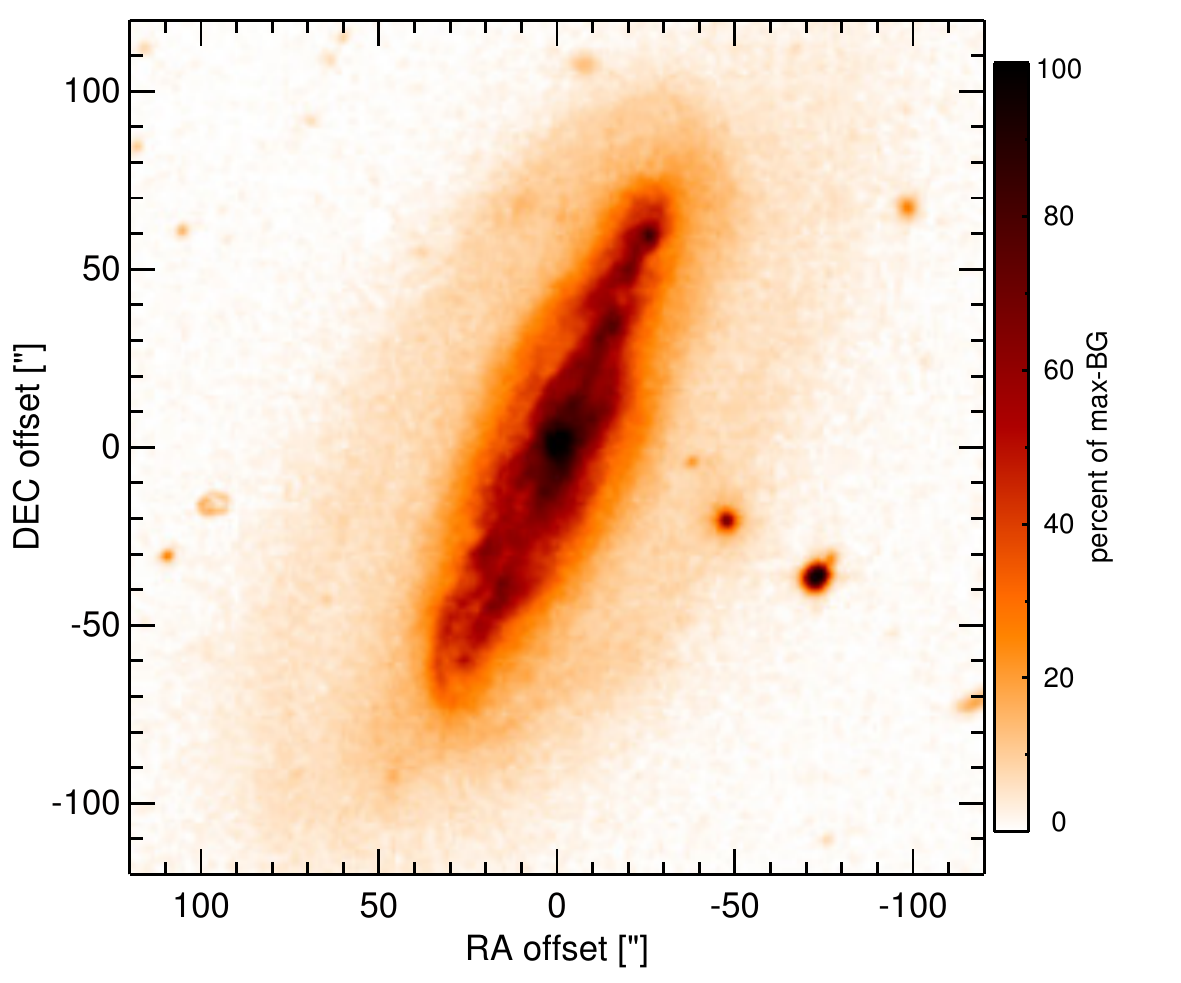}
    \caption{\label{fig:OPTim_NGC7582}
             Optical image (DSS, red filter) of NGC\,7582. Displayed are the central $4\arcmin$ with North up and East to the left. 
              The colour scaling is linear with white corresponding to the median background and black to the $0.01\%$ pixels with the highest intensity.  
           }
\end{figure}
\begin{figure}
   \centering
   \includegraphics[angle=0,height=3.11cm]{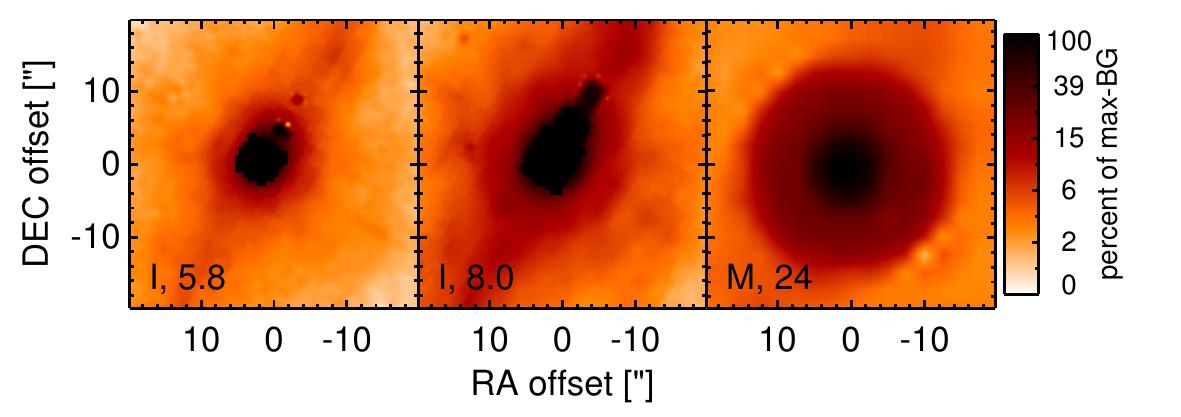}
    \caption{\label{fig:INTim_NGC7582}
             \spitzerr MIR images of NGC\,7582. Displayed are the inner $40\arcsec$ with North up and East to the left. The colour scaling is logarithmic with white corresponding to median background and black to the $0.1\%$ pixels with the highest intensity.
             The label in the bottom left states instrument and central wavelength of the filter in $\mu$m (I: IRAC, M: MIPS). 
             Note that the apparent off-nuclear compact sources in the IRAC 5.8 and  $8.0\,\mu$m images are instrumental artefacts.
           }
\end{figure}
\begin{figure}
   \centering
   \includegraphics[angle=0,width=8.500cm]{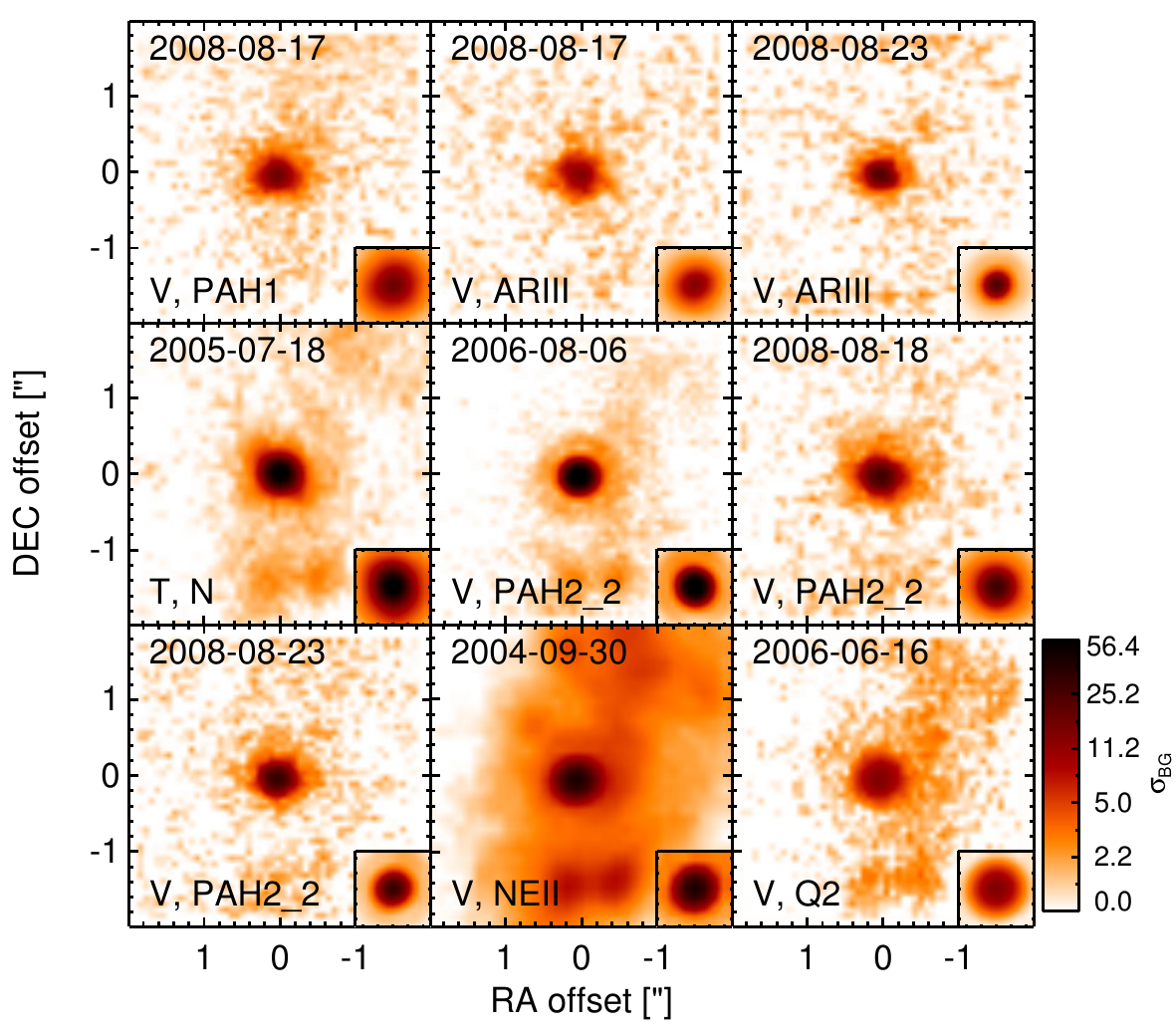}
    \caption{\label{fig:HARim_NGC7582}
             Subarcsecond-resolution MIR images of NGC\,7582 sorted by increasing filter wavelength. 
             Displayed are the inner $4\arcsec$ with North up and East to the left. 
             The colour scaling is logarithmic with white corresponding to median background and black to the $75\%$ of the highest intensity of all images in units of $\sigbg$.
             The inset image shows the central arcsecond of the PSF from the calibrator star, scaled to match the science target.
             The labels in the bottom left state instrument and filter names (C: COMICS, M: Michelle, T: T-ReCS, V: VISIR).
           }
\end{figure}
\begin{figure}
   \centering
   \includegraphics[angle=0,width=8.50cm]{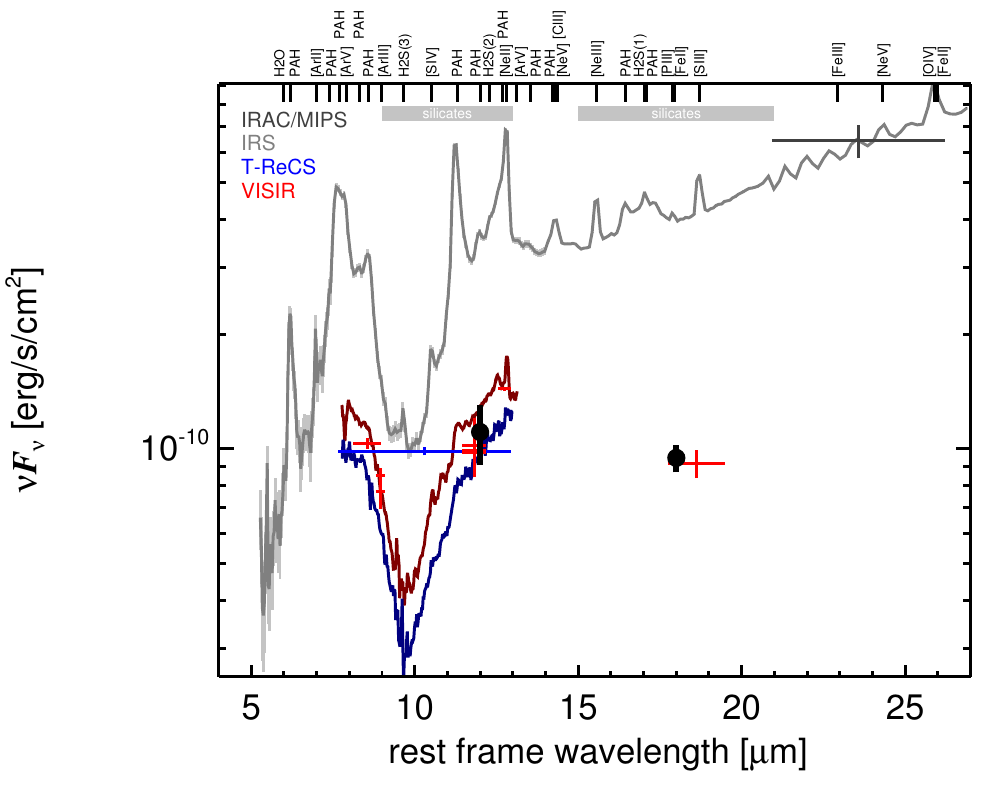}
   \caption{\label{fig:MISED_NGC7582}
      MIR SED of NGC\,7582. The description  of the symbols (if present) is the following.
      Grey crosses and  solid lines mark the \spitzer/IRAC, MIPS and IRS data. 
      The colour coding of the other symbols is: 
      green for COMICS, magenta for Michelle, blue for T-ReCS and red for VISIR data.
      Darker-coloured solid lines mark spectra of the corresponding instrument.
      The black filled circles mark the nuclear 12 and $18\,\mu$m  continuum emission estimate from the data.
      The ticks on the top axis mark positions of common MIR emission lines, while the light grey horizontal bars mark wavelength ranges affected by the silicate 10 and 18$\mu$m features.}
\end{figure}
\clearpage

\twocolumn[\begin{@twocolumnfalse}  
\subsection{NGC\,7590}\label{app:NGC7590}
NGC\,7590 is a highly-inclined late-type spiral galaxy at a distance of $D=$ $26.5\pm4.8\,$Mpc hosting a Sy\,2 nucleus and circum-nuclear star formation \citep{shi_unobscured_2010}.
\cite{tran_unified_2003} reclassified the nucleus as non-AGN, while we treat this object as an uncertain AGN/starburst composite.
In fact, the X-ray morphology is complex and the nature of the nuclear source is uncertain \citep{shu_xmm-newton_2010}.
Furthermore, the nucleus has not been detected with arcsecond-resolution radio observations so far \citep{morganti_radio_1999}.
After its first MIR detection, NGC\,7590 was followed up in the MIR with \spitzer/IRAC, IRS and MIPS observations.
The corresponding IRAC images show an extended nucleus, elongated along the host major axis, and embedded within comparably bright spiral-like host emission, which dominates towards longer wavelengths.
The nucleus is not visible in the MIPS 24\,$\mu$m image, and we determine an upper limit on the nuclear flux (see also \citealt{shi_unobscured_2010}).
Our nuclear IRAC $5.8$ and $8.0\,\mu$m photometry is significantly lower than the values published in \cite{gallimore_infrared_2010}.
The IRS LR mapping-mode spectrum suffers from low S/N but exhibits strong PAH emission features and a red spectral slope in $\nu F_\nu$-space (see also \citealt{tommasin_spitzer_2008,tommasin_spitzer-irs_2010,wu_spitzer/irs_2009,gallimore_infrared_2010}).
It is significantly contaminated or even dominated by emission from the circum-nuclear starburst ring in the slit and should be regarded as an upper limit to the arcsecond-sale MIR SED.
No AGN-indicative \nev emission was detected in the IRS spectrum.
The nuclear region of NGC\,7590 was observed with T-ReCS in the broad N filter in 2005 (unpublished, to our knowledge), and with VISIR in two narrow $N$-band filters in 2009 \citep{asmus_mid-infrared_2011}.
No emission structures are detected in the images, and our corresponding flux upper limits are comparable to the \spitzerr spectrophotometry.
We conclude that there is no MIR evidence for an AGN in NGC\,7590.
However, the flux upper limits are consistent with the MIR--X-ray correlation even for the Compton-thick AGN scenario proposed by \citeauthor{shu_xmm-newton_2010} (2010; \citealt{asmus_mid-infrared_2011}).
 \newline\end{@twocolumnfalse}]

\begin{figure}
   \centering
   \includegraphics[angle=0,width=8.500cm]{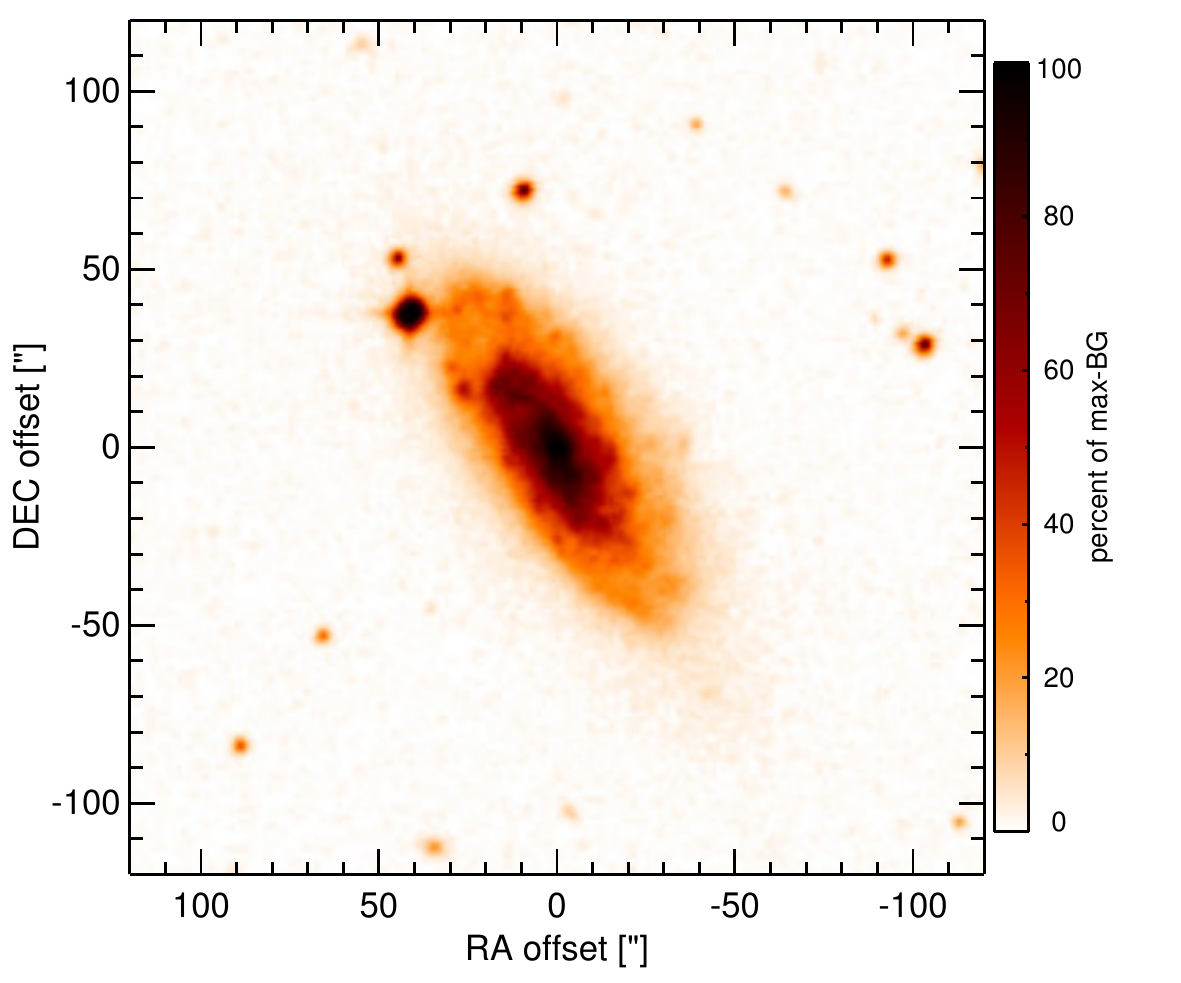}
    \caption{\label{fig:OPTim_NGC7590}
             Optical image (DSS, red filter) of NGC\,7590. Displayed are the central $4\arcmin$ with North up and East to the left. 
              The colour scaling is linear with white corresponding to the median background and black to the $0.01\%$ pixels with the highest intensity.  
           }
\end{figure}
\begin{figure}
   \centering
   \includegraphics[angle=0,height=3.11cm]{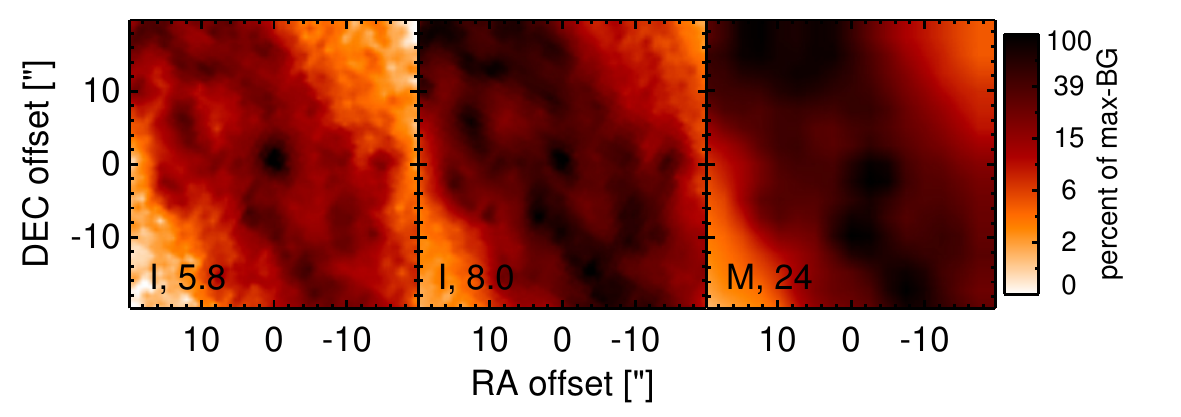}
    \caption{\label{fig:INTim_NGC7590}
             \spitzerr MIR images of NGC\,7590. Displayed are the inner $40\arcsec$ with North up and East to the left. The colour scaling is logarithmic with white corresponding to median background and black to the $0.1\%$ pixels with the highest intensity.
             The label in the bottom left states instrument and central wavelength of the filter in $\mu$m (I: IRAC, M: MIPS). 
           }
\end{figure}
\begin{figure}
   \centering
   \includegraphics[angle=0,width=8.50cm]{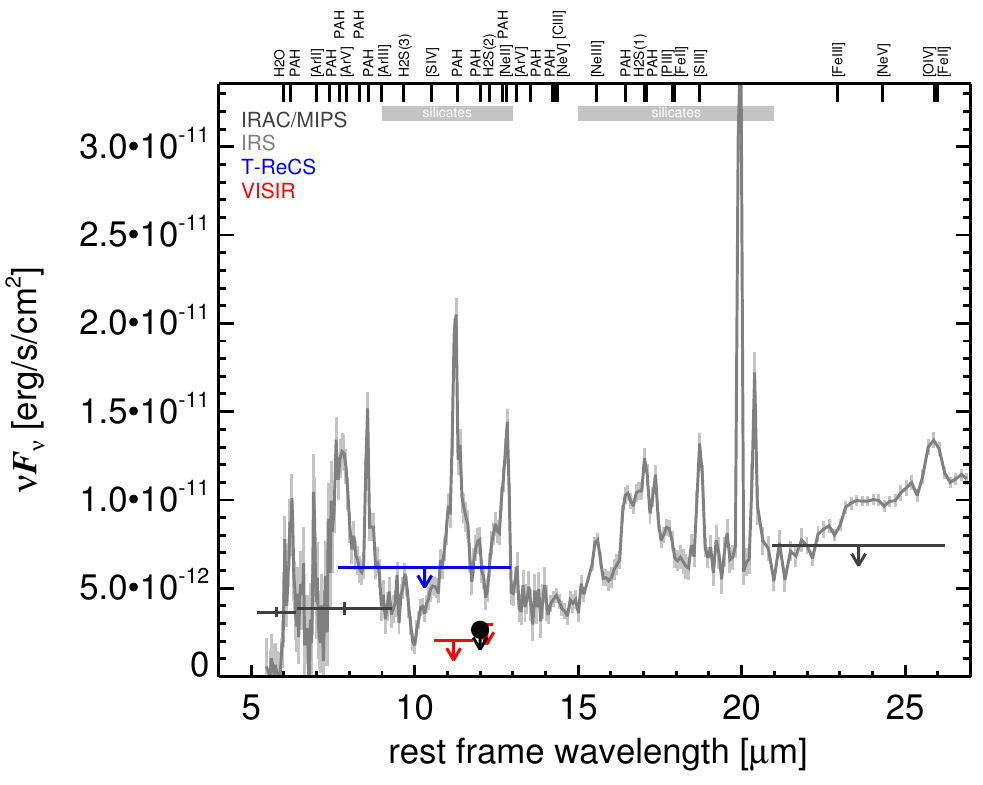}
   \caption{\label{fig:MISED_NGC7590}
      MIR SED of NGC\,7590. The description  of the symbols (if present) is the following.
      Grey crosses and  solid lines mark the \spitzer/IRAC, MIPS and IRS data. 
      The colour coding of the other symbols is: 
      green for COMICS, magenta for Michelle, blue for T-ReCS and red for VISIR data.
      Darker-coloured solid lines mark spectra of the corresponding instrument.
      The black filled circles mark the nuclear 12 and $18\,\mu$m  continuum emission estimate from the data.
      The ticks on the top axis mark positions of common MIR emission lines, while the light grey horizontal bars mark wavelength ranges affected by the silicate 10 and 18$\mu$m features.}
\end{figure}
\clearpage

\twocolumn[\begin{@twocolumnfalse}  
\subsection{NGC\,7592W -- NGC\,7592A -- Mrk\,928b}\label{app:NGC7592W}
NGC\,7592 is an infrared-luminous merger system of two, possibly three, galaxies at a redshift of $z=$ 0.0246 ($D\sim105\,$Mpc; \citealt{hattori_tridimensional_2002}).
The two main nuclei are separated by $\sim 12\arcsec\sim\,6$kpc in the east-west directions (PA$\sim100\degree$; \citealt{rafanelli_complex_1992,dopita_star_2002}).
The western, early-type component, NGC\,7592W, contains both an AGN and a circum-nuclear starburst \citep{hattori_tridimensional_2002} and is optically classified as a Sy\,2/starburst composite. The eastern, late-type, component, NGC\,7592E, is apparently a pure starburst \citep{dahari_statistical_1988,rafanelli_complex_1992,hattori_tridimensional_2002}.
The detection of a compact hard-X-ray and radio source verifies the presence of an AGN in NGC\,7592W \citep{wang_chandra_2010,lonsdale_vlbi_1992}. 
Furthermore, bipolar kiloparsec-scale ionization cones have been reported in NGC\,7592W (PA$\sim160\degree$; \citealt{hattori_tridimensional_2002}).
After first being detected in the MIR with \iras, NGC\,7592 was followed up 
with Palomar 5\,m bolometer MIR photometry of both main nuclei \citep{carico_iras_1988}, \isoo MIR \citep{hwang_mid-infrared_1999,clavel_2.5-11_2000,ramos_almeida_mid-infrared_2007} and \spitzer/IRAC, IRS and MIPS observations.
In the corresponding IRAC and MIPS images, NGC\,7592W appears as a compact source dominating the MIR emission of the system, while NGC\,7592E appears elongated without a clearly separable unresolved nuclear component.
Our nuclear MIPS 24\,$\mu$m photometry of the unresolved core of NGC\,7592W is affected by blending emission from NGC\,7592E and, thus, should be treated as an upper limit. 
The IRS LR mapping-mode spectrum of NGC\,7592W exhibits silicate 10$\,\mu$m absorption, prominent PAH emission, and a red spectral slope in $\nu F_\nu$-space (see also \citealt{stierwalt_mid-infrared_2013}).
Thus, the arcsecond-scale MIR SED indicates the composition of obscured AGN and starburst emission.
NGC\,7592 was imaged with T-ReCS in the Qa filter in 2010 \citep{imanishi_subaru_2011}.
Only NGC\,7592W was detected in the image and appears as a possibly resolved nucleus (FWHM(major axis)$\sim0.71\arcsec\sim0.4\,$kpc; PA$\sim85\degree$) without further host emission.
However, the current subarcsecond MIR data are insufficient to reach any robust conclusion about the nuclear extension at subarcsecond scales in the MIR.
Our nuclear Qa photometry is $\sim18\%$ lower than the value published in \cite{imanishi_subaru_2011} and also $\sim25\%$ lower than the \spitzerr spectrophotometry.
Therefore, we extrapolate from the Qa measurement towards shorter wavelengths in order to compute the nuclear $12\,\mu$m continuum emission estimate as described in Sect.~\ref{sec:cont}.
\newline\end{@twocolumnfalse}]

\begin{figure}
   \centering
   \includegraphics[angle=0,width=8.500cm]{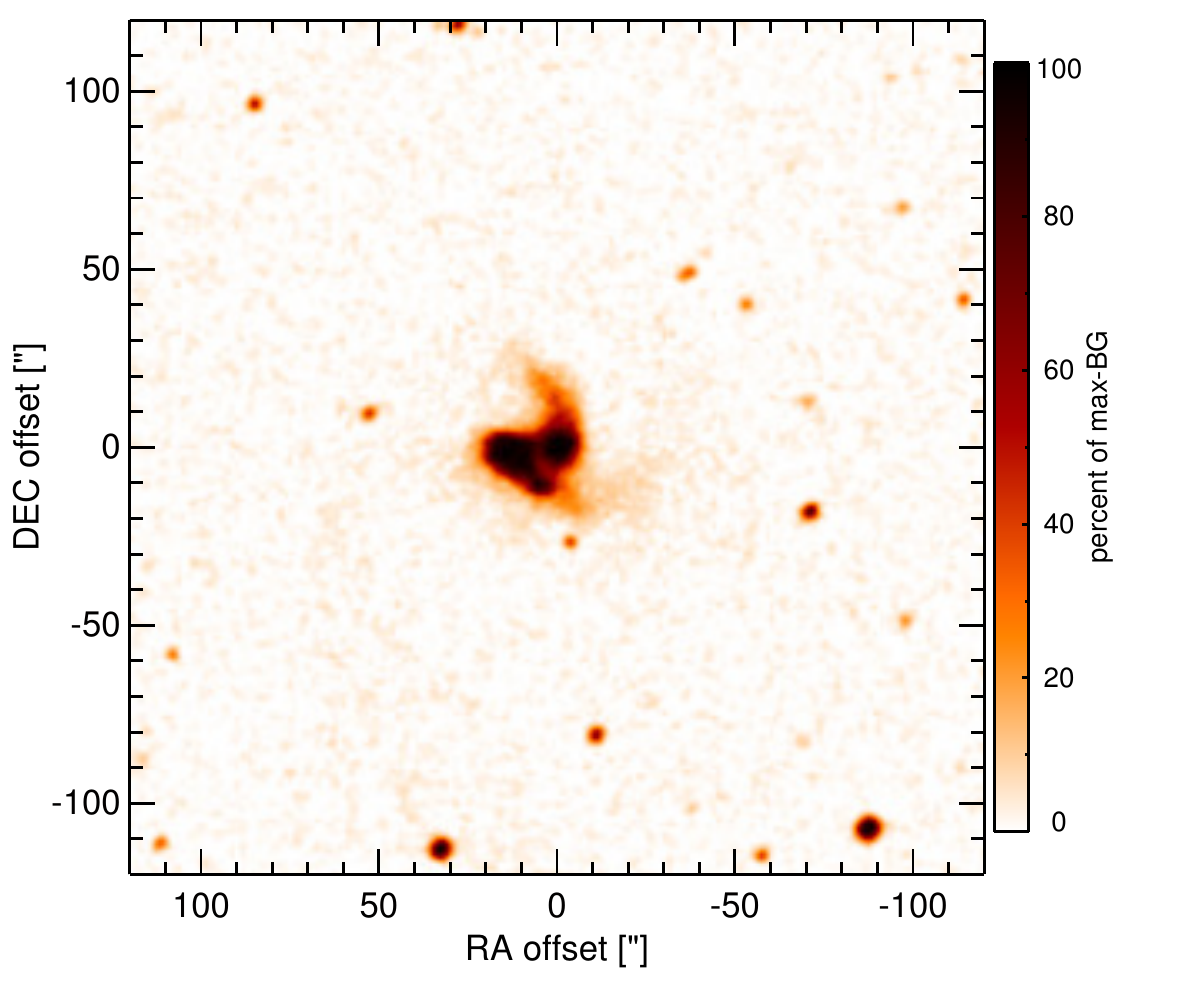}
    \caption{\label{fig:OPTim_NGC7592W}
             Optical image (DSS, red filter) of NGC\,7592W. Displayed are the central $4\arcmin$ with North up and East to the left. 
              The colour scaling is linear with white corresponding to the median background and black to the $0.01\%$ pixels with the highest intensity.  
           }
\end{figure}
\begin{figure}
   \centering
   \includegraphics[angle=0,height=3.11cm]{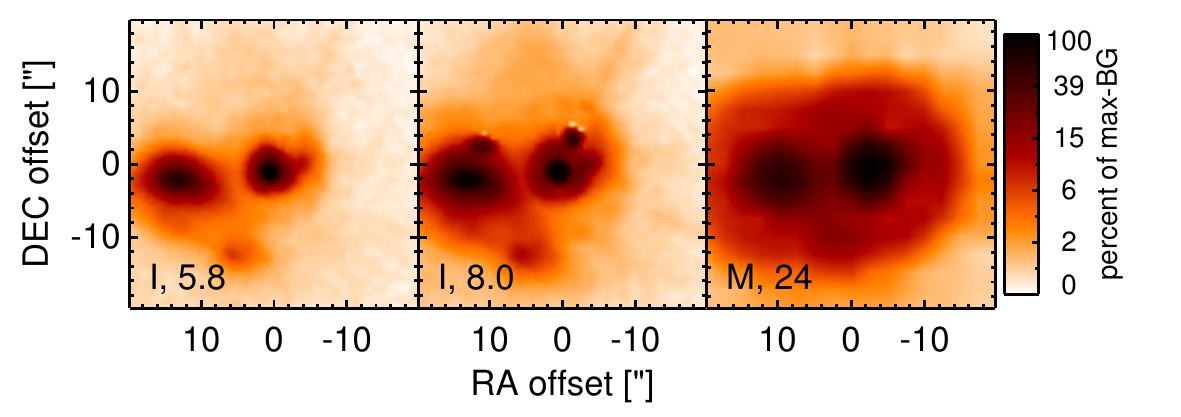}
    \caption{\label{fig:INTim_NGC7592W}
             \spitzerr MIR images of NGC\,7592W. Displayed are the inner $40\arcsec$ with North up and East to the left. The colour scaling is logarithmic with white corresponding to median background and black to the $0.1\%$ pixels with the highest intensity.
             The label in the bottom left states instrument and central wavelength of the filter in $\mu$m (I: IRAC, M: MIPS). 
             Note that the apparent off-nuclear compact source to the north-west in the IRAC 5.8 and  $8.0\,\mu$m images is an  instrumental artefact.
           }
\end{figure}
\begin{figure}
   \centering
   \includegraphics[angle=0,height=3.11cm]{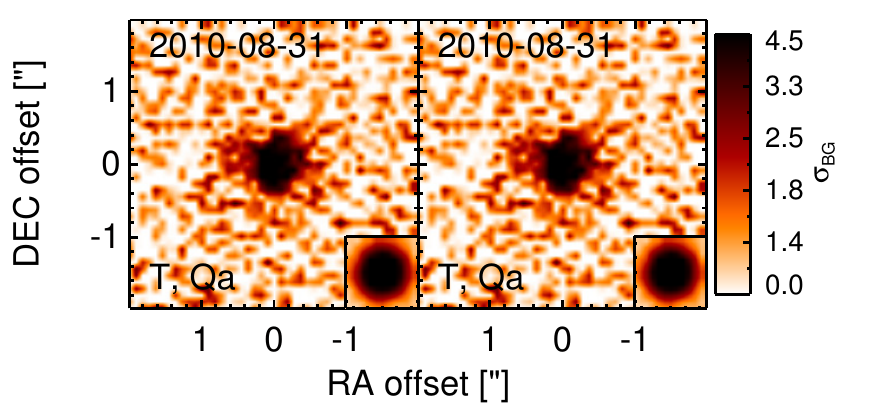}
    \caption{\label{fig:HARim_NGC7592W}
             Subarcsecond-resolution MIR images of NGC\,7592W sorted by increasing filter wavelength. 
             Displayed are the inner $4\arcsec$ with North up and East to the left. 
             The colour scaling is logarithmic with white corresponding to median background and black to the $75\%$ of the highest intensity of all images in units of $\sigbg$.
             The inset image shows the central arcsecond of the PSF from the calibrator star, scaled to match the science target.
             The labels in the bottom left state instrument and filter names (C: COMICS, M: Michelle, T: T-ReCS, V: VISIR).
           }
\end{figure}
\begin{figure}
   \centering
   \includegraphics[angle=0,width=8.50cm]{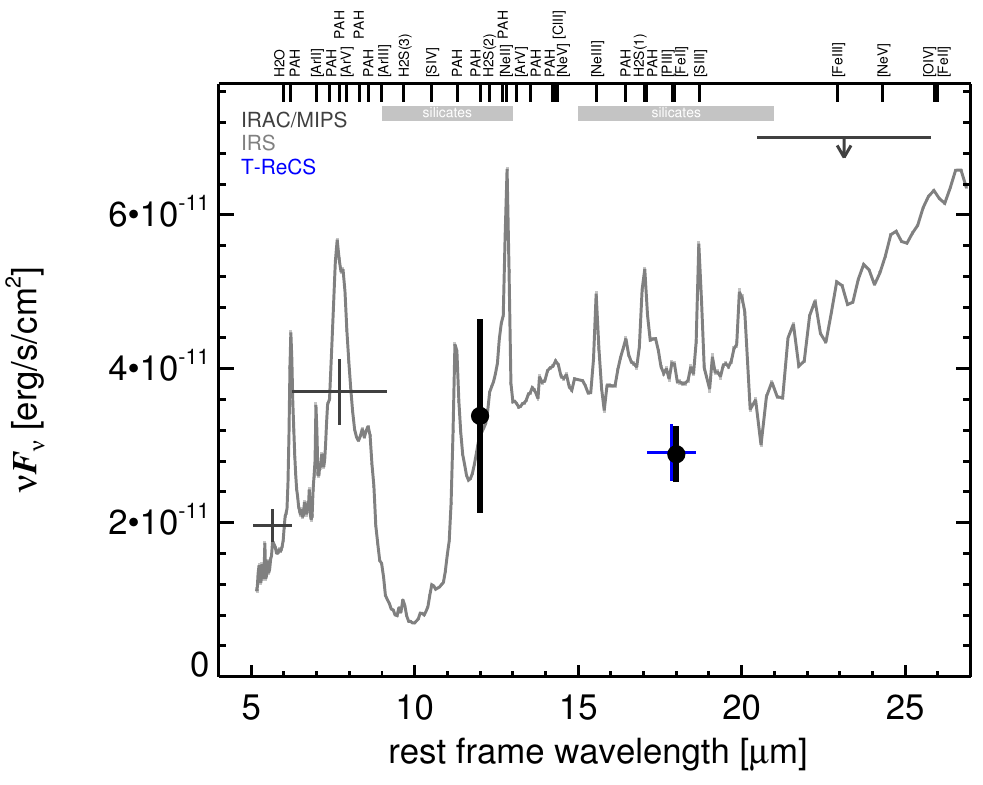}
   \caption{\label{fig:MISED_NGC7592W}
      MIR SED of NGC\,7592W. The description  of the symbols (if present) is the following.
      Grey crosses and  solid lines mark the \spitzer/IRAC, MIPS and IRS data. 
      The colour coding of the other symbols is: 
      green for COMICS, magenta for Michelle, blue for T-ReCS and red for VISIR data.
      Darker-coloured solid lines mark spectra of the corresponding instrument.
      The black filled circles mark the nuclear 12 and $18\,\mu$m  continuum emission estimate from the data.
      The ticks on the top axis mark positions of common MIR emission lines, while the light grey horizontal bars mark wavelength ranges affected by the silicate 10 and 18$\mu$m features.}
\end{figure}
\clearpage

\twocolumn[\begin{@twocolumnfalse}  
\subsection{NGC\,7626}\label{app:NGC7626}
NGC\,7626 is a peculiar giant elliptical galaxy at a redshift of $z=$ 0.0114 ($D\sim45.4\,$Mpc) with a FR\,I radio morphology and a possibly active LINER nucleus \citep{ho_search_1997-1}. 
Its detection in X-rays is reported by \cite{ho_radiatively_2009}. At radio wavelengths, a compact core with a prominent biconical kiloparsec-scale radio jet has been found (PA$\sim35\degree$; \citealt{birkinshaw_orientations_1985,nagar_radio_2005}.
These evidence heavily support the presence of an AGN in NGC\,7626.
Furthermore, a nuclear warped dust lane with $\sim1\arcsec\sim216\,$pc extent obscures the nucleus in the north-south direction (PA$\sim167\degree$;\citealt{forbes_ellipticals_1995,verdoes_kleijn_hubble_1999}).
NGC\,7626 remained undetected in the MIR with \irass and ground-based IRTF observations \citep{sparks_infrared_1986}.
The first MIR detection could be achieved with \isoo \citep{ferrari_survey_2002,temi_ages_2005}, followed by \spitzer/IRAC, IRS and MIPS observations.
The corresponding IRAC and MIPS images show extended elliptical emission without a clearly separable unresolved nuclear component.
In addition, two compact emission sources are visible in the IRAC 8.0$\,\mu$m images at distances $\sim7\arcsec\sim1.5\,$kpc north-east and $\sim5\arcsec\sim1\,$kpc south-west of the nucleus (PA$\sim16\degree$).
These might be related to the large-scale jets.
Our nuclear MIPS 24\,$\mu$m photometric flux is significantly lower than the total flux given in \cite{temi_spitzer_2009}.
The IRS HR staring-mode spectrum suffers from low S/N and does not indicate any spectral features apart from possibly PAH emission.
Note also that no background subtraction was performed for this spectrum.
\cite{dudik_spitzer_2009} provide upper limits on the \nev and \oiv high-ionization lines.
The IRAC and MIPS photometry is more trustworthy and resembles a blue arcsecond-scale MIR SED of a rather passive galaxy \citep{bressan_spitzer_2006}.
We observed NGC\,7626 with VISIR in two narrow $N$-band filters in 2009 \citep{asmus_mid-infrared_2011} but failed to detect any MIR emission.
Our new measurement routines provide more constraining upper limits, which, however, are still insufficient to reach any conclusions about the existence of a weak AGN in NGC\,7626 from the MIR point of view.
\newline\end{@twocolumnfalse}]

\begin{figure}
   \centering
   \includegraphics[angle=0,width=8.500cm]{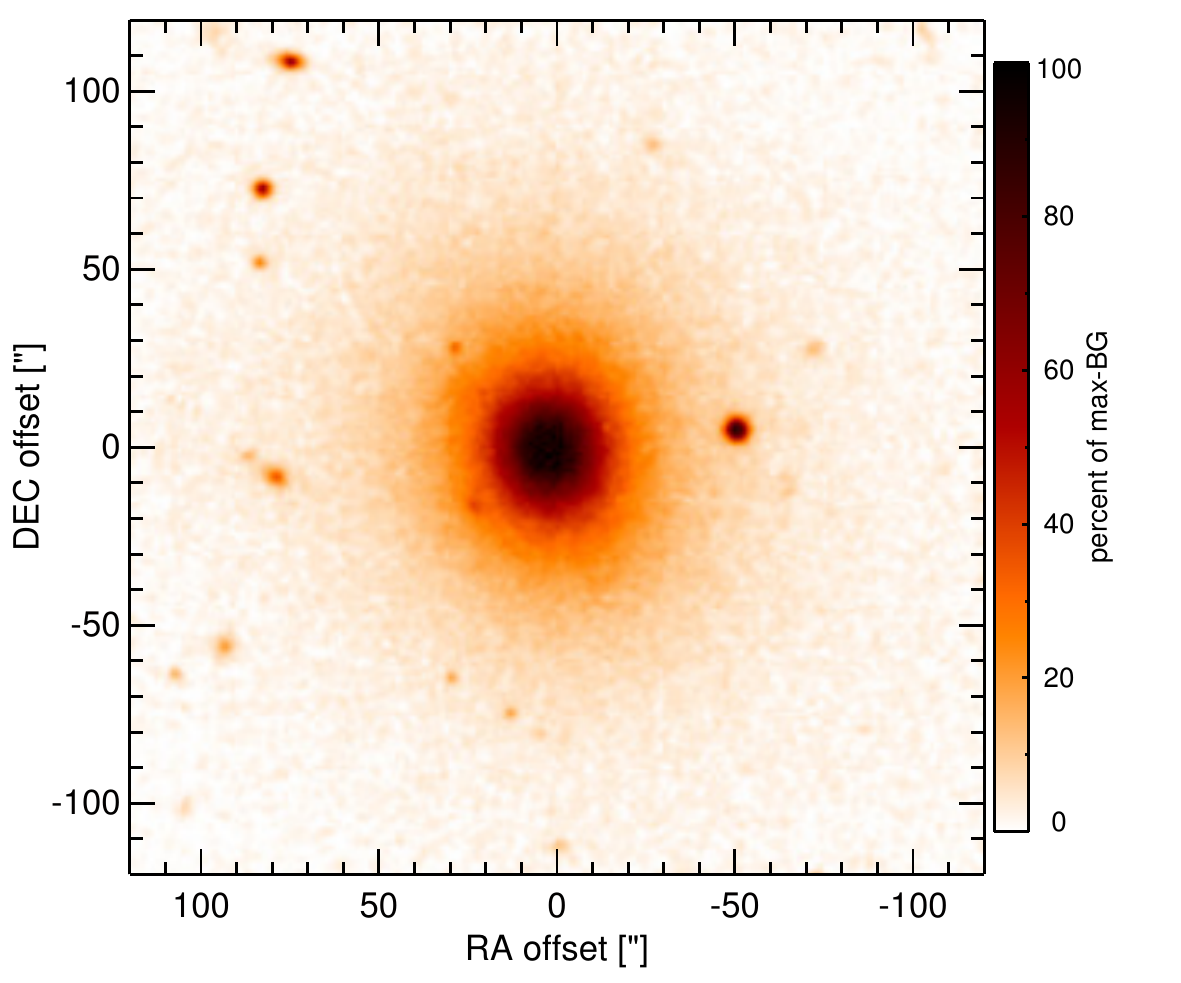}
    \caption{\label{fig:OPTim_NGC7626}
             Optical image (DSS, red filter) of NGC\,7626. Displayed are the central $4\arcmin$ with North up and East to the left. 
              The colour scaling is linear with white corresponding to the median background and black to the $0.01\%$ pixels with the highest intensity.  
           }
\end{figure}
\begin{figure}
   \centering
   \includegraphics[angle=0,height=3.11cm]{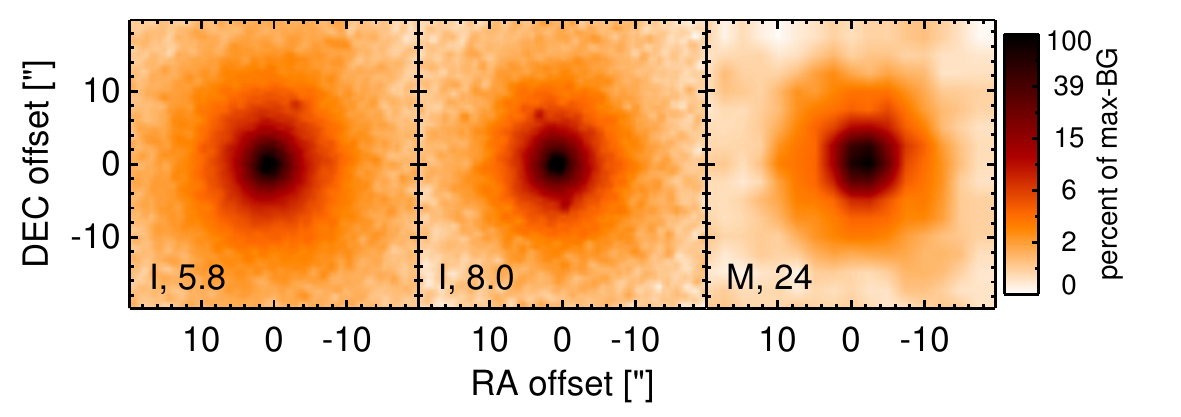}
    \caption{\label{fig:INTim_NGC7626}
             \spitzerr MIR images of NGC\,7626. Displayed are the inner $40\arcsec$ with North up and East to the left. The colour scaling is logarithmic with white corresponding to median background and black to the $0.1\%$ pixels with the highest intensity.
             The label in the bottom left states instrument and central wavelength of the filter in $\mu$m (I: IRAC, M: MIPS). 
           }
\end{figure}
\begin{figure}
   \centering
   \includegraphics[angle=0,width=8.50cm]{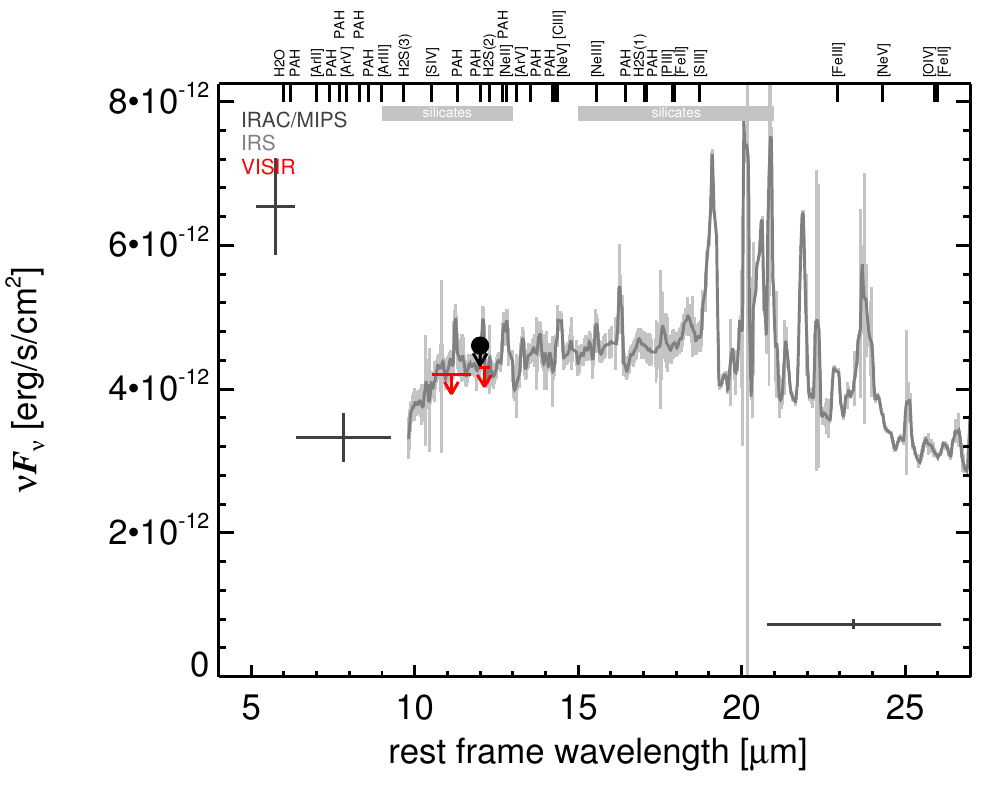}
   \caption{\label{fig:MISED_NGC7626}
      MIR SED of NGC\,7626. The description  of the symbols (if present) is the following.
      Grey crosses and  solid lines mark the \spitzer/IRAC, MIPS and IRS data. 
      The colour coding of the other symbols is: 
      green for COMICS, magenta for Michelle, blue for T-ReCS and red for VISIR data.
      Darker-coloured solid lines mark spectra of the corresponding instrument.
      The black filled circles mark the nuclear 12 and $18\,\mu$m  continuum emission estimate from the data.
      The ticks on the top axis mark positions of common MIR emission lines, while the light grey horizontal bars mark wavelength ranges affected by the silicate 10 and 18$\mu$m features.}
\end{figure}
\clearpage

\twocolumn[\begin{@twocolumnfalse}  
\subsection{NGC\,7674 -- Mrk\,533}\label{app:NGC7674}
NGC\,7674 is an infrared-luminous, nearly face-on late-type spiral galaxy at a redshift of $z=$ 0.0289 ($D\sim126\,$Mpc) hosting a Sy\,2 nucleus \citep{veron-cetty_catalogue_2010} with polarized broad emission lines \citep{miller_spectropolarimetry_1990}.
It features a compact radio core with kiloparsec-scale, bended jet-like emission to the east and west \citep{unger_radio_1988,momjian_sensitive_2003}, which is roughly aligned with the biconical $\sim4.4arcsec\sim2.5\,$kpc extended \oiii emission  (PA$\sim120\degree$; \citealt{schmitt_hubble_2003}).
After first being detected in the MIR with \iras, NGC\,7674 was followed up with ground-based MIR observations \citep{aitken_8-13_1985,edelson_broad-band_1987,carico_iras_1988,roche_atlas_1991,maiolino_new_1995,miles_high-resolution_1996}.
The latter presented the first subarcsecond MIR observations of NGC\,7674 and detected an unresolved nucleus, while \cite{siebenmorgen_mid-infrared_2004} report an east-west elongation of $\sim0.5\arcsec\sim240\,$pc in their ESO 3.6\,m/TIMMI2 image.
\cite{gorjian_10_2004} mention no extension in the comparable Palomar 5\,m/MIRLIN $N$-band image.
NGC\,7674 was followed up in space with \isoo \citep{clavel_2.5-11_2000,ramos_almeida_mid-infrared_2007} and \spitzer/IRAC, IRS and MIPS observations.
The corresponding IRAC and MIPS images are dominated by a bright compact nucleus embedded within the much fainter spiral-like host emission.
Our nuclear IRAC 5.8$\,\mu$m photometry agrees with \cite{gallimore_infrared_2010}. On the other hand, our IRAC $8.0\,\mu$m flux is significantly lower than the published value, yet in agreement with the IRS LR mapping-mode spectrum.
The latter exhibits silicate 10$\,\mu$m absorption, weak PAH features, and a broad emission peak at $\sim 18\,\mu$m in $\nu F_\nu$-space (see also \citealt{buchanan_spitzer_2006,wu_spitzer/irs_2009,tommasin_spitzer-irs_2010,gallimore_infrared_2010,stierwalt_mid-infrared_2013}).
Thus, the arcsecond-scale MIR SED appears to be AGN-dominated.
We observed NGC\,7674 with VISIR with three narrow $N$-band images and one LR $N$-band spectrum between 2006 and 2009 (partly published in \cite{horst_mid_2008,horst_mid-infrared_2009,honig_dusty_2010-1}).
In addition, a Qa filter image was obtained with T-ReCS in 2007 (unpublished, to our knowledge).
A compact nucleus without further host emission was detected in all cases.
The nucleus is marginally resolved in the sharpest image (ARIII) as well as all remaining images except for the PAH2\_2 filter. The nuclear elongation is consistently detected at PA$\sim 125\degree$ with a FWHM $\sim15\%$ larger than the corresponding standard stars.
Therefore, we classify the nucleus of NGC\,7674 as generally extended in the MIR at subarcsecond resolution.
Note that this extension aligns roughly with the ionization cone.
The resulting unresolved nuclear fluxes are $\sim 20\%$ lower than the \spitzerr spectrophotometry, the 0.75\arcsec-extracted VISIR spectrum \citep{honig_dusty_2010-1}, and the published total nuclear NEII flux measurement \citep{horst_mid_2008}.  
This flux difference may either be attributed to nuclear star formation as indicated by the weak PAH 11.3\,$\mu$m feature in the VISIR spectrum or to dust in the NLR as suggested by the elongation.
\newline\end{@twocolumnfalse}]

\begin{figure}
   \centering
   \includegraphics[angle=0,width=8.500cm]{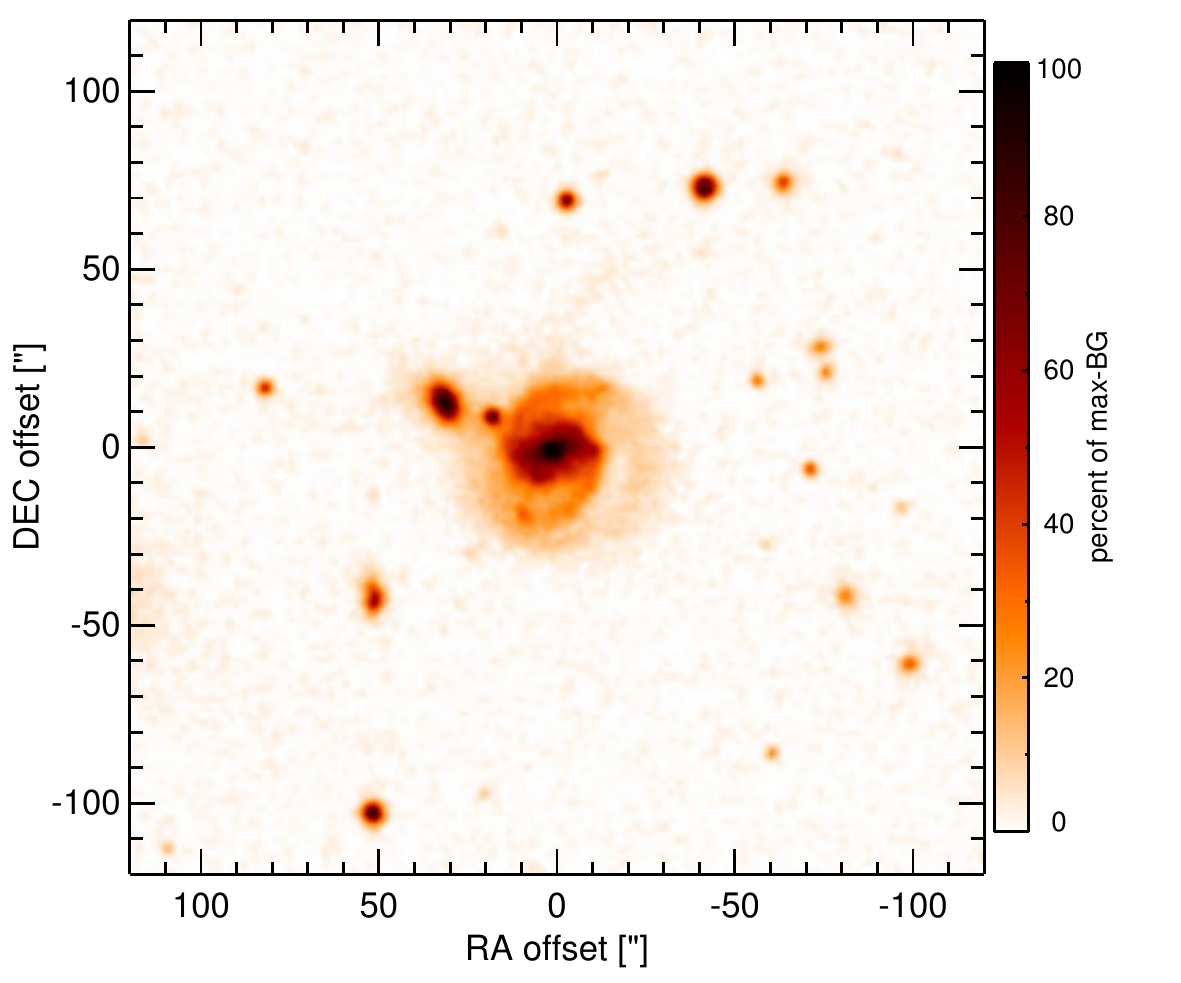}
    \caption{\label{fig:OPTim_NGC7674}
             Optical image (DSS, red filter) of NGC\,7674. Displayed are the central $4\arcmin$ with North up and East to the left. 
              The colour scaling is linear with white corresponding to the median background and black to the $0.01\%$ pixels with the highest intensity.  
           }
\end{figure}
\begin{figure}
   \centering
   \includegraphics[angle=0,height=3.11cm]{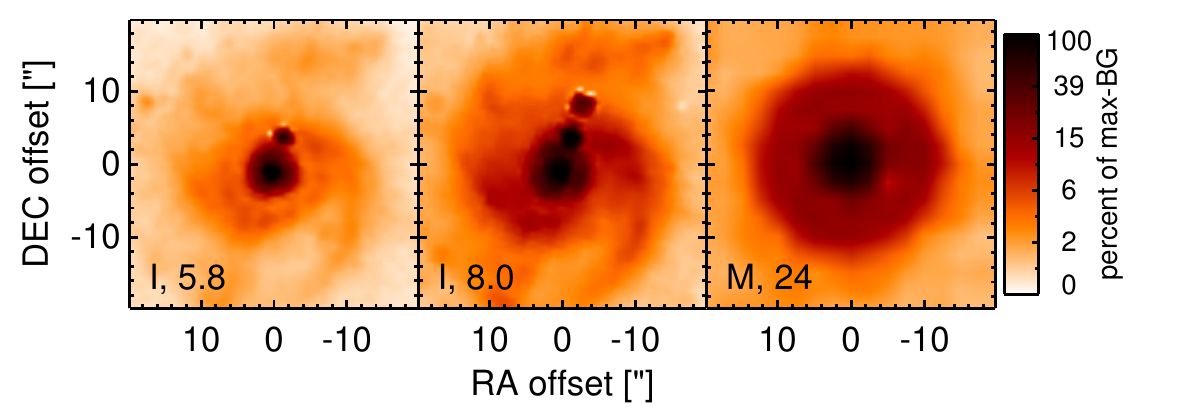}
    \caption{\label{fig:INTim_NGC7674}
             \spitzerr MIR images of NGC\,7674. Displayed are the inner $40\arcsec$ with North up and East to the left. The colour scaling is logarithmic with white corresponding to median background and black to the $0.1\%$ pixels with the highest intensity.
             The label in the bottom left states instrument and central wavelength of the filter in $\mu$m (I: IRAC, M: MIPS). 
             Note that the apparent off-nuclear compact sources in the IRAC 5.8 and  $8.0\,\mu$m images are instrumental artefacts.
           }
\end{figure}
\begin{figure}
   \centering
   \includegraphics[angle=0,width=8.500cm]{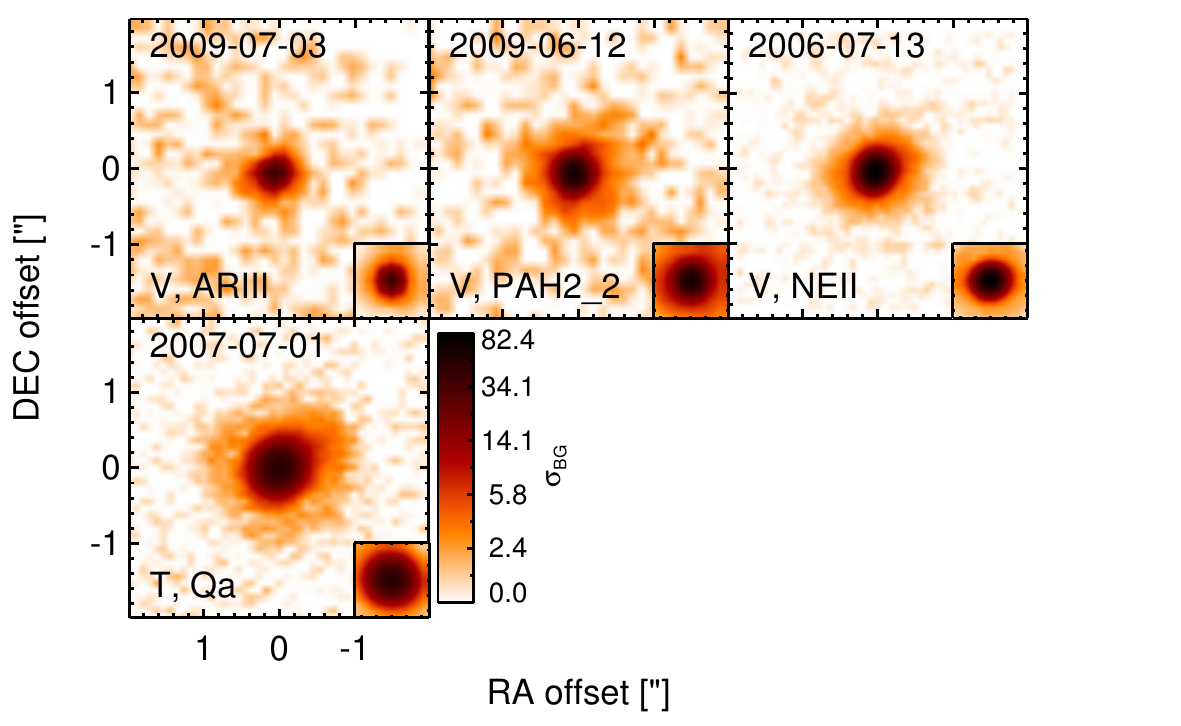}
    \caption{\label{fig:HARim_NGC7674}
             Subarcsecond-resolution MIR images of NGC\,7674 sorted by increasing filter wavelength. 
             Displayed are the inner $4\arcsec$ with North up and East to the left. 
             The colour scaling is logarithmic with white corresponding to median background and black to the $75\%$ of the highest intensity of all images in units of $\sigbg$.
             The inset image shows the central arcsecond of the PSF from the calibrator star, scaled to match the science target.
             The labels in the bottom left state instrument and filter names (C: COMICS, M: Michelle, T: T-ReCS, V: VISIR).
           }
\end{figure}
\begin{figure}
   \centering
   \includegraphics[angle=0,width=8.50cm]{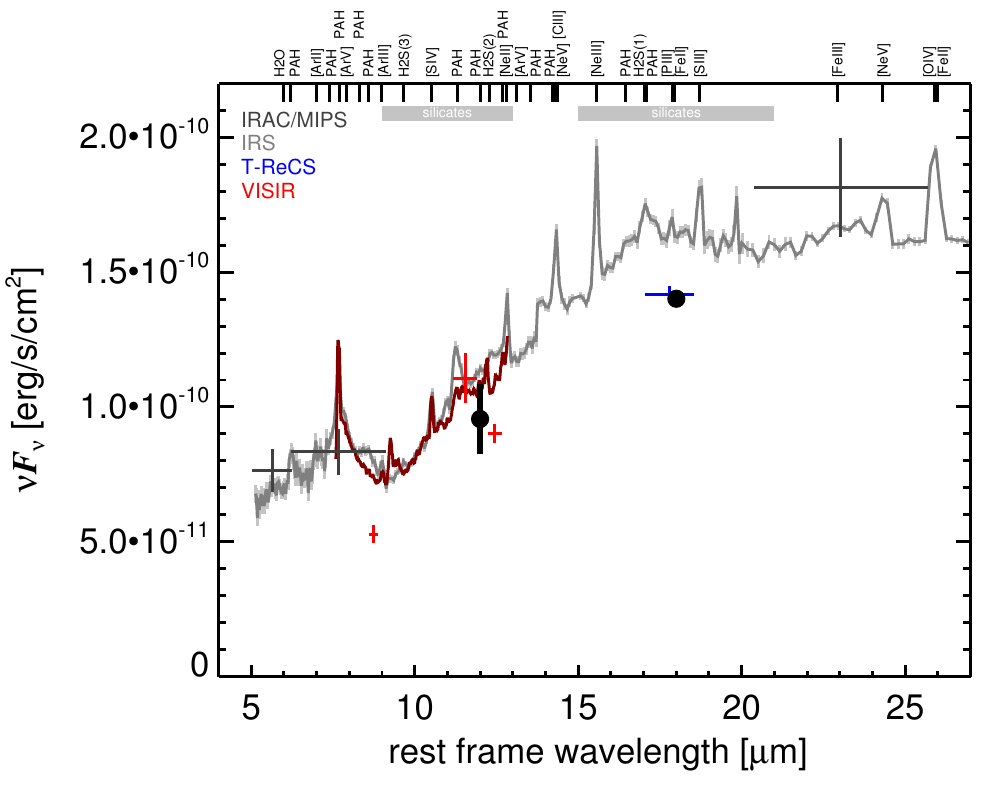}
   \caption{\label{fig:MISED_NGC7674}
      MIR SED of NGC\,7674. The description  of the symbols (if present) is the following.
      Grey crosses and  solid lines mark the \spitzer/IRAC, MIPS and IRS data. 
      The colour coding of the other symbols is: 
      green for COMICS, magenta for Michelle, blue for T-ReCS and red for VISIR data.
      Darker-coloured solid lines mark spectra of the corresponding instrument.
      The black filled circles mark the nuclear 12 and $18\,\mu$m  continuum emission estimate from the data.
      The ticks on the top axis mark positions of common MIR emission lines, while the light grey horizontal bars mark wavelength ranges affected by the silicate 10 and 18$\mu$m features.}
\end{figure}
\clearpage

\twocolumn[\begin{@twocolumnfalse}  
\subsection{NGC\,7679 -- Mrk\,534 -- IRAS\,23262+0314 -- UGC\,12618}\label{app:NGC7679}
NGC\,7679 is an infrared-luminous, low-inclination peculiar barred spiral galaxy at a redshift of $z=$ 0.0171 ($D\sim71.7\,$Mpc), forming a wide pair with NGC\,7682 $\sim4.5\arcmin$ to the east (Arp\,216).
It was initially classified as a LINER/H\,II transition object \citep{dahari_nuclear_1985} and contains both a circum-nuclear and a nuclear starburst, as well as an AGN \citep{pogge_star_1993,gu_nuclear_2001,buson_ngc_2006}. 
The AGN was later optically classified either as Sy\,1.9 or Sy\,2 nucleus \citep{veron-cetty_catalogue_2010,veilleux_optical_1995}.
Following \cite{kewley_optical_2001} and \cite{buson_ngc_2006}, we treat the nucleus as AGN/starburst composite.
Interestingly, the AGN appears unabsorbed in X-rays \citep{della_ceca_unveiling_2001}, which is very atypical for an AGN/starburst composite object.
A compact radio core embedded within extended clumpy emission \citep{parra_cola._2010} and kiloparsec-scale extended \oiii emission have been detected (PA$\sim80\degree$; \citealt{yankulova_luminous_2007}).
After its identification as an infrared-luminous galaxy with \iras, NGC\,7679 was followed up in the MIR with IRTF \citep{telesco_genesis_1993} and \spitzer/IRAC, IRS and MIPS.
The corresponding IRAC and MIPS images show an extended, elongated bright nucleus embedded within much fainter host emission (major axis diameter $\sim 8\arcsec\sim3\,$kpc; PA$\sim120\degree$). 
The IRS spectrum is dominated by strong PAH emission and has a steep red spectral slope in the $N$-band in $\nu F_\nu$-space, while the silicate features appear to be weak if present at all (see also \citealt{shi_unobscured_2010,stierwalt_mid-infrared_2013}).
The arcsecond-scale MIR SED appears to be completely star-formation dominated.
We imaged NGC\,7679 with VISIR in three narrow $N$-band filters in 2006 \citep{horst_mid_2008,horst_mid-infrared_2009} and weakly detected a compact nucleus without further extended emission.
The nucleus appears marginally resolved in the PAH2 image but not in the sharper SIV filter.
Therefore, we classify the general MIR extension of the nucleus at subarcsecond resolution as uncertain.
Deeper MIR imaging is required.
Our new photometry optimized for faint detections results in marginally consistent values with \cite{horst_mid_2008}.
The corresponding nuclear MIR SED is $75\%$ lower than the \spitzerr spectrophotometry, verifying that star formation is indeed dominating the MIR emission of NGC\,7679.
Furthermore, the nuclear MIR SED indicates PAH 11.3\,$\mu$m emission and is likely still containing star-formation emission.
 \newline\end{@twocolumnfalse}]

\begin{figure}
   \centering
   \includegraphics[angle=0,width=8.500cm]{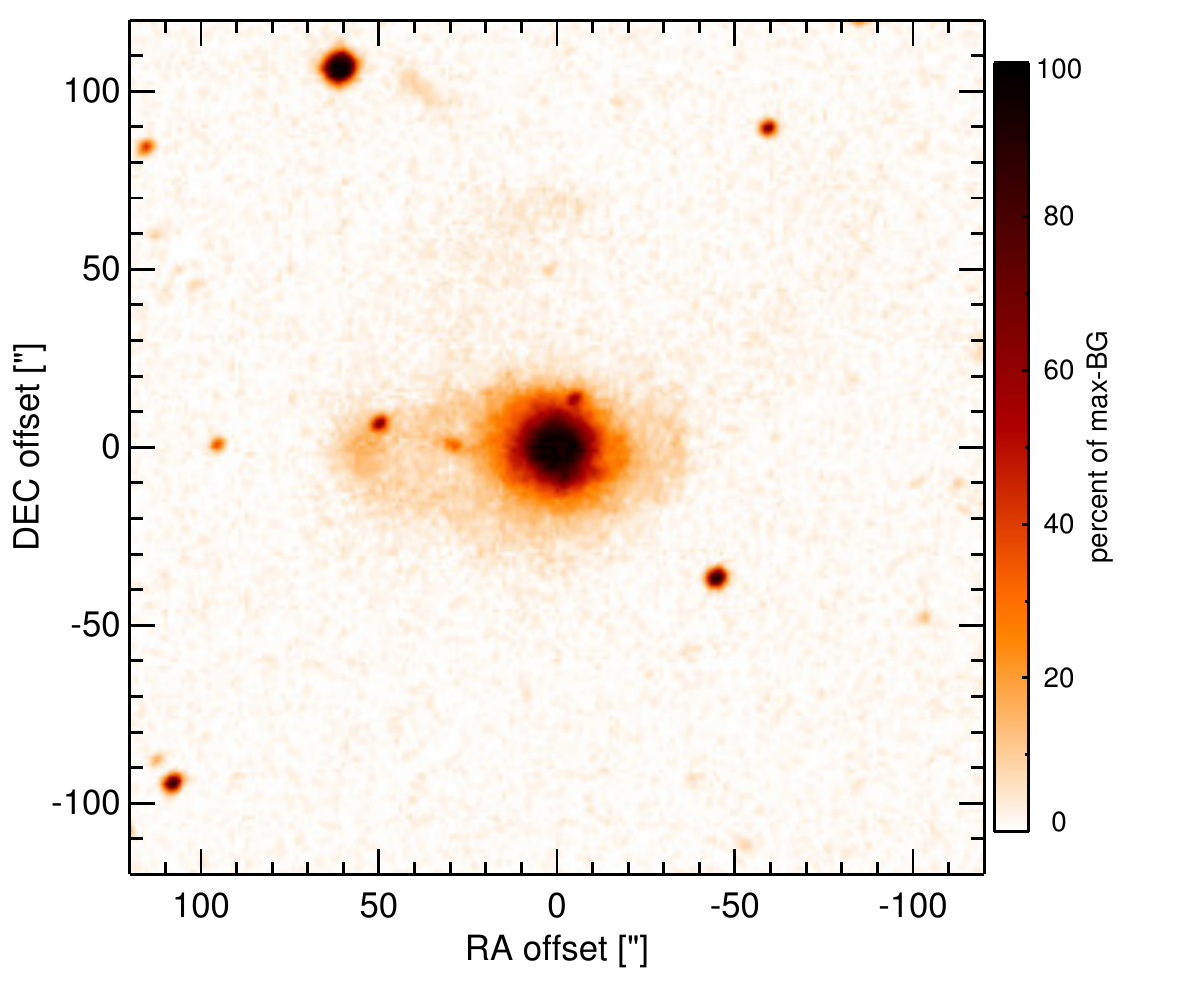}
    \caption{\label{fig:OPTim_NGC7679}
             Optical image (DSS, red filter) of NGC\,7679. Displayed are the central $4\arcmin$ with North up and East to the left. 
              The colour scaling is linear with white corresponding to the median background and black to the $0.01\%$ pixels with the highest intensity.  
           }
\end{figure}
\begin{figure}
   \centering
   \includegraphics[angle=0,height=3.11cm]{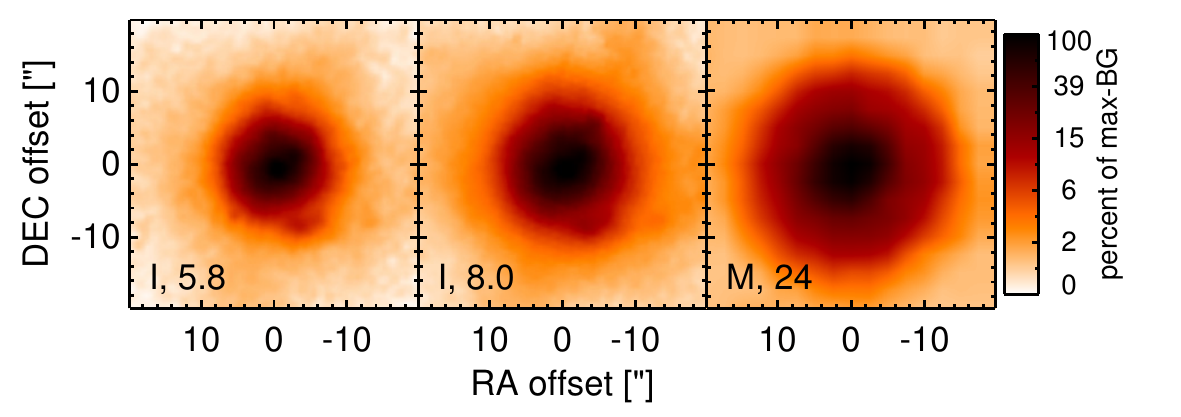}
    \caption{\label{fig:INTim_NGC7679}
             \spitzerr MIR images of NGC\,7679. Displayed are the inner $40\arcsec$ with North up and East to the left. The colour scaling is logarithmic with white corresponding to median background and black to the $0.1\%$ pixels with the highest intensity.
             The label in the bottom left states instrument and central wavelength of the filter in $\mu$m (I: IRAC, M: MIPS). 
           }
\end{figure}
\begin{figure}
   \centering
   \includegraphics[angle=0,height=3.11cm]{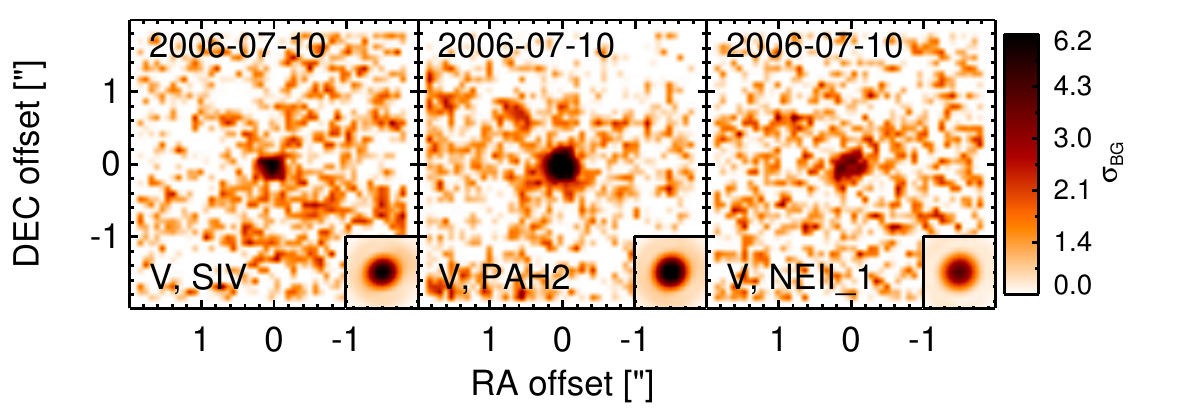}
    \caption{\label{fig:HARim_NGC7679}
             Subarcsecond-resolution MIR images of NGC\,7679 sorted by increasing filter wavelength. 
             Displayed are the inner $4\arcsec$ with North up and East to the left. 
             The colour scaling is logarithmic with white corresponding to median background and black to the $75\%$ of the highest intensity of all images in units of $\sigbg$.
             The inset image shows the central arcsecond of the PSF from the calibrator star, scaled to match the science target.
             The labels in the bottom left state instrument and filter names (C: COMICS, M: Michelle, T: T-ReCS, V: VISIR).
           }
\end{figure}
\begin{figure}
   \centering
   \includegraphics[angle=0,width=8.50cm]{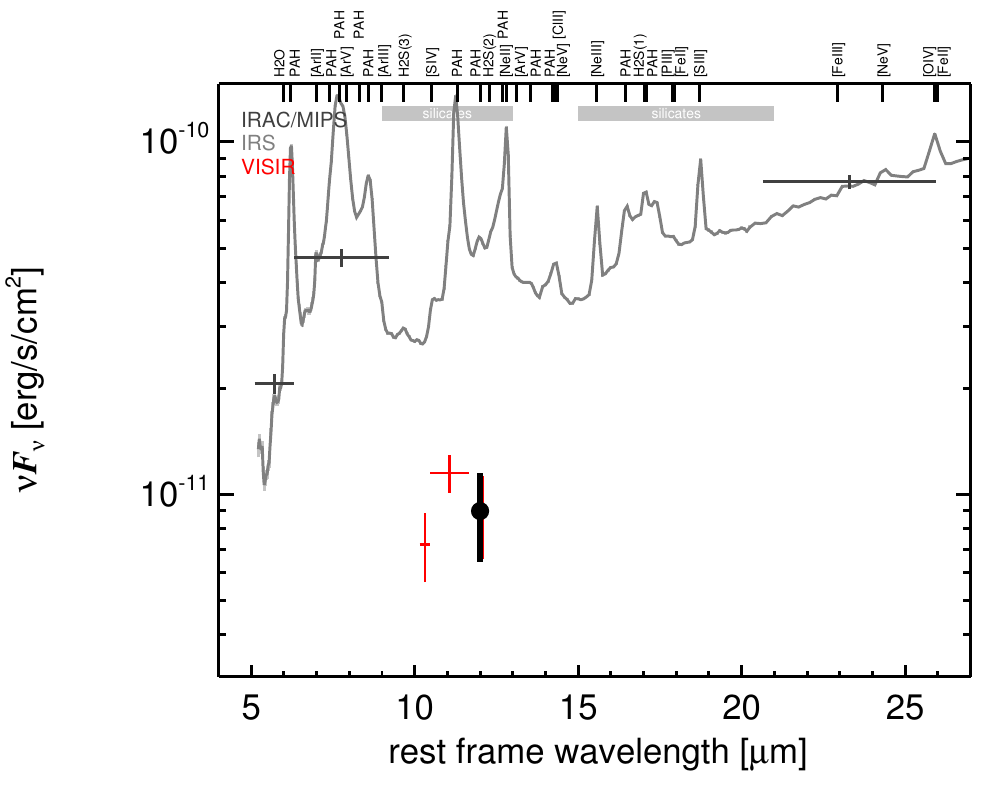}
   \caption{\label{fig:MISED_NGC7679}
      MIR SED of NGC\,7679. The description  of the symbols (if present) is the following.
      Grey crosses and  solid lines mark the \spitzer/IRAC, MIPS and IRS data. 
      The colour coding of the other symbols is: 
      green for COMICS, magenta for Michelle, blue for T-ReCS and red for VISIR data.
      Darker-coloured solid lines mark spectra of the corresponding instrument.
      The black filled circles mark the nuclear 12 and $18\,\mu$m  continuum emission estimate from the data.
      The ticks on the top axis mark positions of common MIR emission lines, while the light grey horizontal bars mark wavelength ranges affected by the silicate 10 and 18$\mu$m features.}
\end{figure}
\clearpage

\twocolumn[\begin{@twocolumnfalse}  
\subsection{NGC\,7743}\label{app:NGC7743}
NGC\,7743 is a low-inclination barred lenticular galaxy at a distance of $D=$ $19.2 \pm 1.6\,$Mpc \citep{jensen_measuring_2003} with a borderline Sy\,2/LINER nucleus \citep{ho_reevaluation_1993,ho_search_1997-1,alonso-herrero_nature_2000}.
At radio wavelengths, it appears as a compact source at arcsecond resolution \citep{nagar_radio_1999,ho_radio_2001}.
NGC\,7743 was weakly detected with \irass in the MIR and followed up with \spitzer/IRAC, IRS and MIPS observations.
The corresponding IRAC and MIPS images show an extended nucleus embedded within elliptical diffuse host emission.
The IRS LR staring-mode spectrum suffers from a very low S/N but indicates PAH emission and a red spectral slope in $\nu F_\nu$-space.
Thus, the arcsecond-scale MIR SED is significantly affected by star formation.
We observed the nuclear region of NGC\,7743 with VISIR in four narrow $N$-band filters in 2009 but failed to detect the nucleus (partly published in \citealt{asmus_mid-infrared_2011}).
Owing to our optimized flux measurement method, we are able to derive more stringent flux upper limits than in the previous work.
These are on average $\sim 63\%$ lower than the \spitzerr spectrophotometry and thus indicate that the MIR emission of NGC\,7743 is indeed dominated by star formation.
Note, however, that \cite{pereira-santaella_mid-infrared_2010} report the detection of AGN-indicative \nev emission in the IRS spectrum.
\newline\end{@twocolumnfalse}]

\begin{figure}
   \centering
   \includegraphics[angle=0,width=8.500cm]{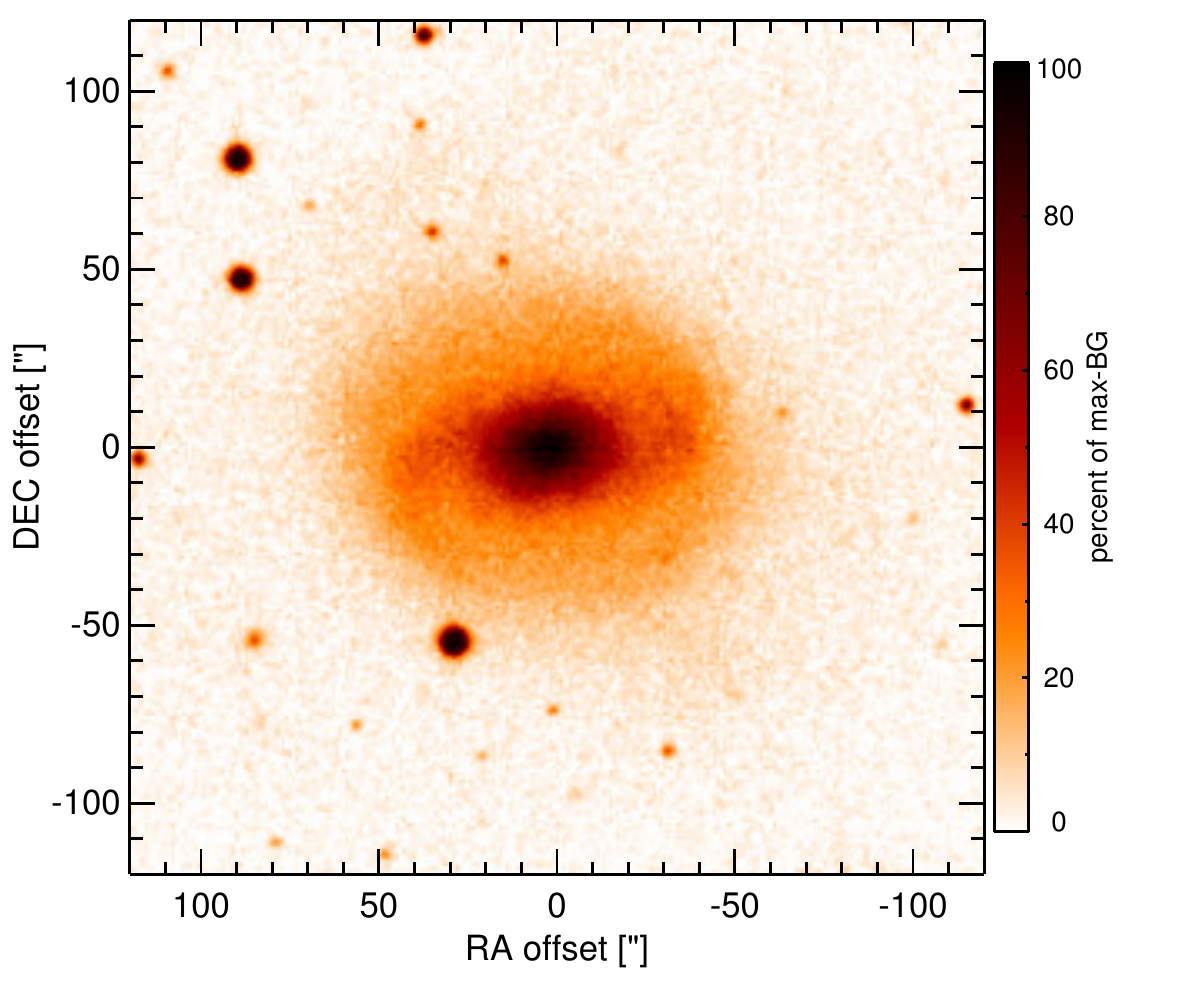}
    \caption{\label{fig:OPTim_NGC7743}
             Optical image (DSS, red filter) of NGC\,7743. Displayed are the central $4\arcmin$ with North up and East to the left. 
              The colour scaling is linear with white corresponding to the median background and black to the $0.01\%$ pixels with the highest intensity.  
           }
\end{figure}
\begin{figure}
   \centering
   \includegraphics[angle=0,height=3.11cm]{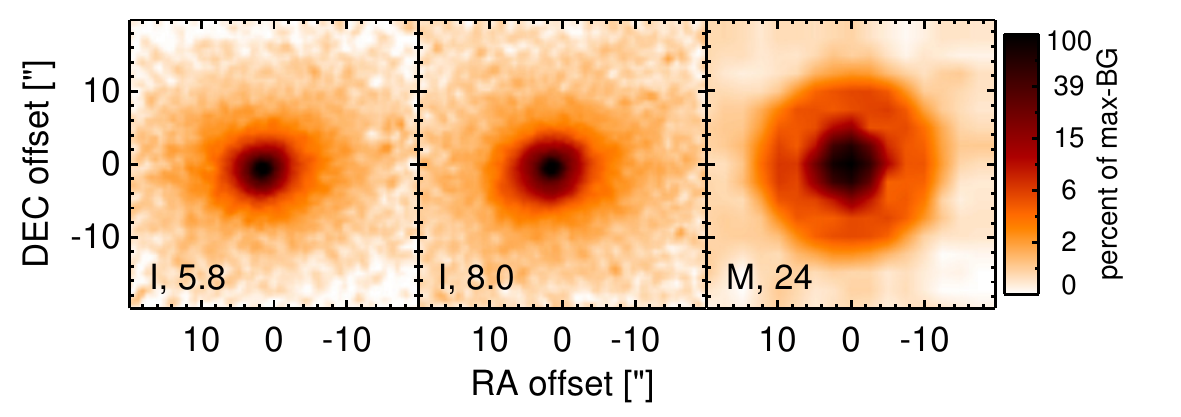}
    \caption{\label{fig:INTim_NGC7743}
             \spitzerr MIR images of NGC\,7743. Displayed are the inner $40\arcsec$ with North up and East to the left. The colour scaling is logarithmic with white corresponding to median background and black to the $0.1\%$ pixels with the highest intensity.
             The label in the bottom left states instrument and central wavelength of the filter in $\mu$m (I: IRAC, M: MIPS). 
           }
\end{figure}
\begin{figure}
   \centering
   \includegraphics[angle=0,width=8.50cm]{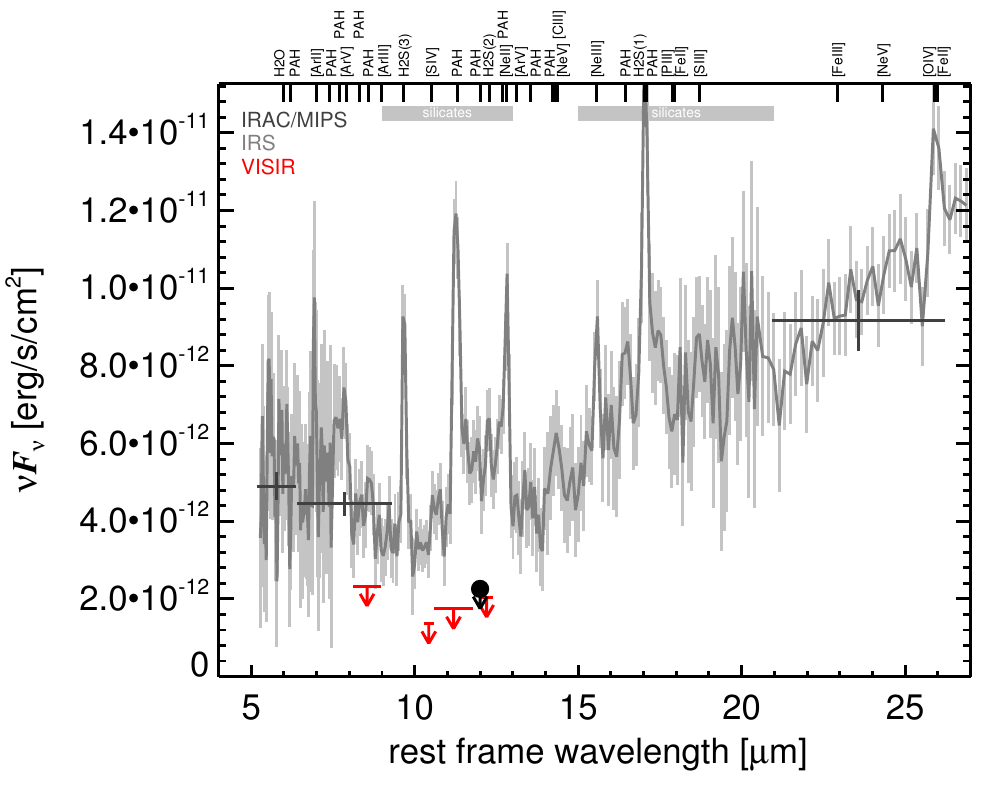}
   \caption{\label{fig:MISED_NGC7743}
      MIR SED of NGC\,7743. The description  of the symbols (if present) is the following.
      Grey crosses and  solid lines mark the \spitzer/IRAC, MIPS and IRS data. 
      The colour coding of the other symbols is: 
      green for COMICS, magenta for Michelle, blue for T-ReCS and red for VISIR data.
      Darker-coloured solid lines mark spectra of the corresponding instrument.
      The black filled circles mark the nuclear 12 and $18\,\mu$m  continuum emission estimate from the data.
      The ticks on the top axis mark positions of common MIR emission lines, while the light grey horizontal bars mark wavelength ranges affected by the silicate 10 and 18$\mu$m features.}
\end{figure}
\clearpage

\twocolumn[\begin{@twocolumnfalse}  
\subsection{PG\,0026+129}\label{app:PG0026+129}
PG\,0026+129 is a radio-quiet quasar at a redshift of $z=$ 0.142 (D$\sim691\,$Mpc) with an optical Sy\,1.2 classification \citep{veron-cetty_catalogue_2010}.
It was detected at radio wavelengths with a compact core and extended kiloparsec-scale biconical emission in the north-south direction while its \oiii emission is unresolved \citep{leipski_radio_2006}.
Pioneering MIR observations were performed by \cite{neugebauer_absolute_1979,neugebauer_continuum_1987}, followed by \cite{elvis_atlas_1994} and \cite{neugebauer_variability_1999}.
PG\,0026+129 was also observed in the MIR with \isoo \citep{haas_dust_2000} and \spitzer/IRS and MIPS, and appears as a point source in the corresponding MIPS 24\,$\mu$m image.
The IRS LR staring-mode spectrum exhibits prominent silicate 10 and 18$\,\mu$m emission and a blue spectral slope in $\nu F_\nu$-space, but no PAH features were detected (see also \citealt{schweitzer_spitzer_2006,schweitzer_extended_2008}).
PG\,0026+129 was imaged with COMICS in the N11.7 filter in 2006 (unpublished, to our knowledge), and a compact nucleus without further host emission has been weakly detected.
More and deeper subarcsecond MIR imaging is required to investigate the source extension.
Our nuclear N11.7 photometry is higher but still consistent with the \spitzerr spectrophotometry.
Therefore, we use the latter to compute the nuclear 12\,$\mu$m continuum emission estimate corrected for the silicate feature.
Comparison with the historical $N$-band photometry shows apparent flux variations on the order of $\sim17\%$ during the last $\sim30$ years.
This indicates intrinsic variability in the $N$-band emission of PG\,0026+129 despite uncertainties and systematics of the different instruments, filters and measurement methods.
\newline\end{@twocolumnfalse}]

\begin{figure}
   \centering
   \includegraphics[angle=0,width=8.500cm]{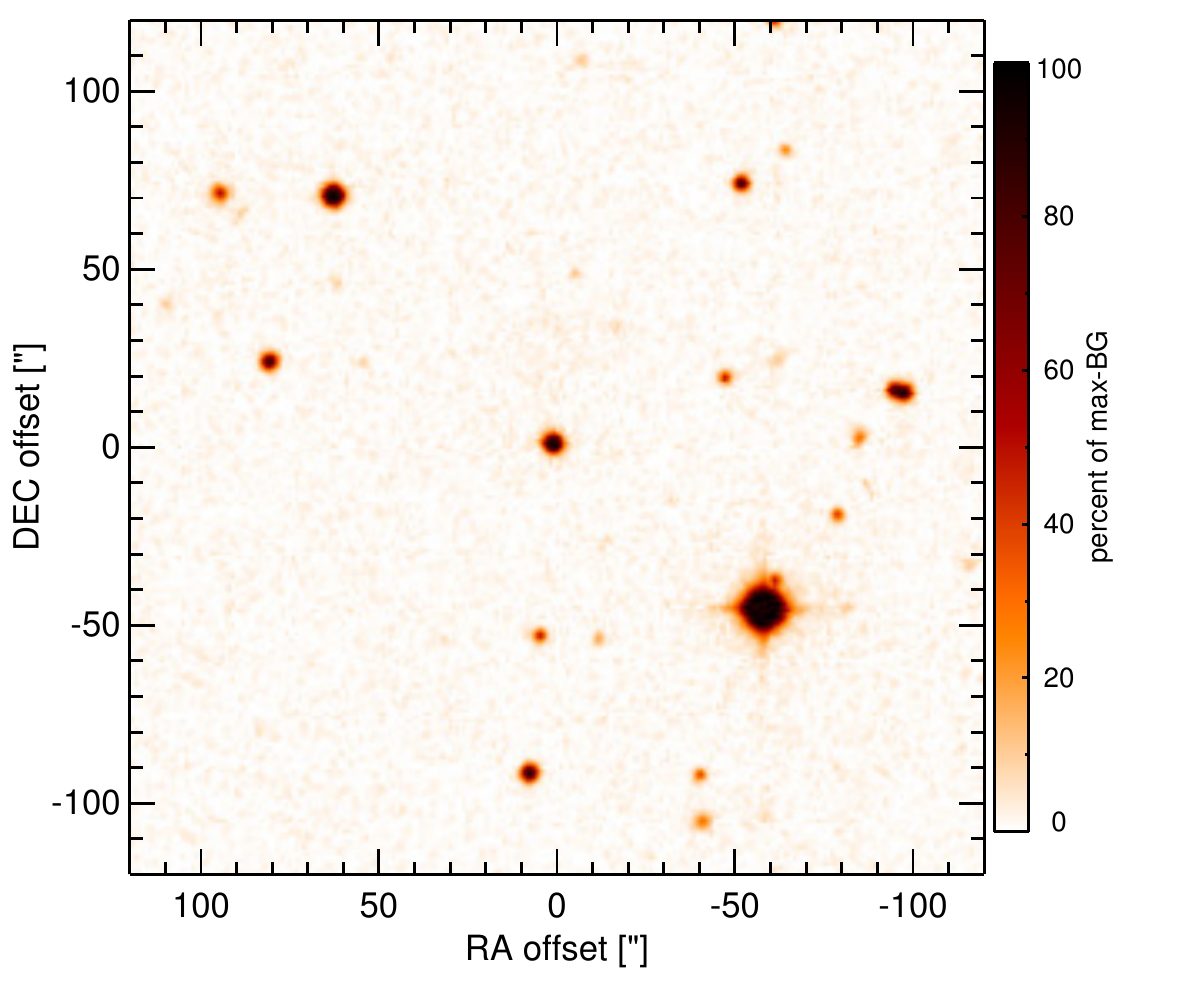}
    \caption{\label{fig:OPTim_PG0026+129}
             Optical image (DSS, red filter) of PG\,0026+129. Displayed are the central $4\arcmin$ with North up and East to the left. 
              The colour scaling is linear with white corresponding to the median background and black to the $0.01\%$ pixels with the highest intensity.  
           }
\end{figure}
\begin{figure}
   \centering
   \includegraphics[angle=0,height=3.11cm]{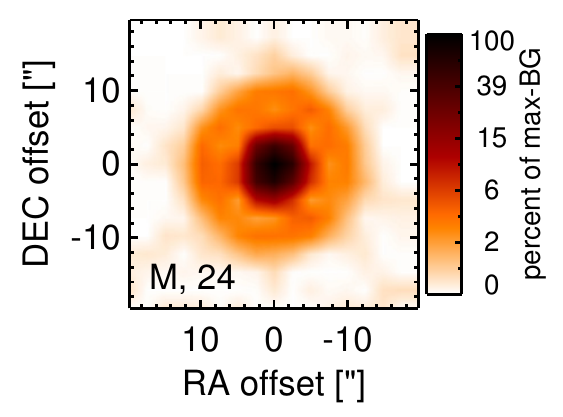}
    \caption{\label{fig:INTim_PG0026+129}
             \spitzerr MIR images of PG\,0026+129. Displayed are the inner $40\arcsec$ with North up and East to the left. The colour scaling is logarithmic with white corresponding to median background and black to the $0.1\%$ pixels with the highest intensity.
             The label in the bottom left states instrument and central wavelength of the filter in $\mu$m (I: IRAC, M: MIPS). 
           }
\end{figure}
\begin{figure}
   \centering
   \includegraphics[angle=0,height=3.11cm]{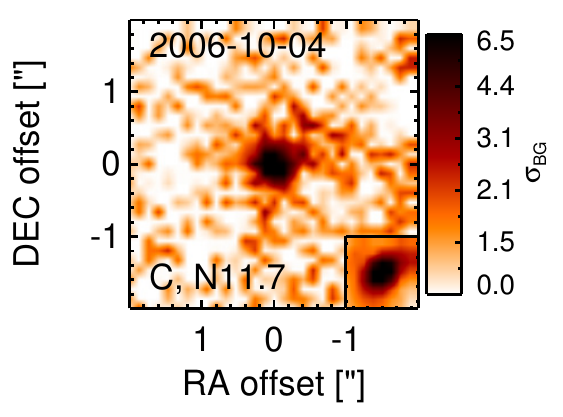}
    \caption{\label{fig:HARim_PG0026+129}
             Subarcsecond-resolution MIR images of PG\,0026+129 sorted by increasing filter wavelength. 
             Displayed are the inner $4\arcsec$ with North up and East to the left. 
             The colour scaling is logarithmic with white corresponding to median background and black to the $75\%$ of the highest intensity of all images in units of $\sigbg$.
             The inset image shows the central arcsecond of the PSF from the calibrator star, scaled to match the science target.
             The labels in the bottom left state instrument and filter names (C: COMICS, M: Michelle, T: T-ReCS, V: VISIR).
           }
\end{figure}
\begin{figure}
   \centering
   \includegraphics[angle=0,width=8.50cm]{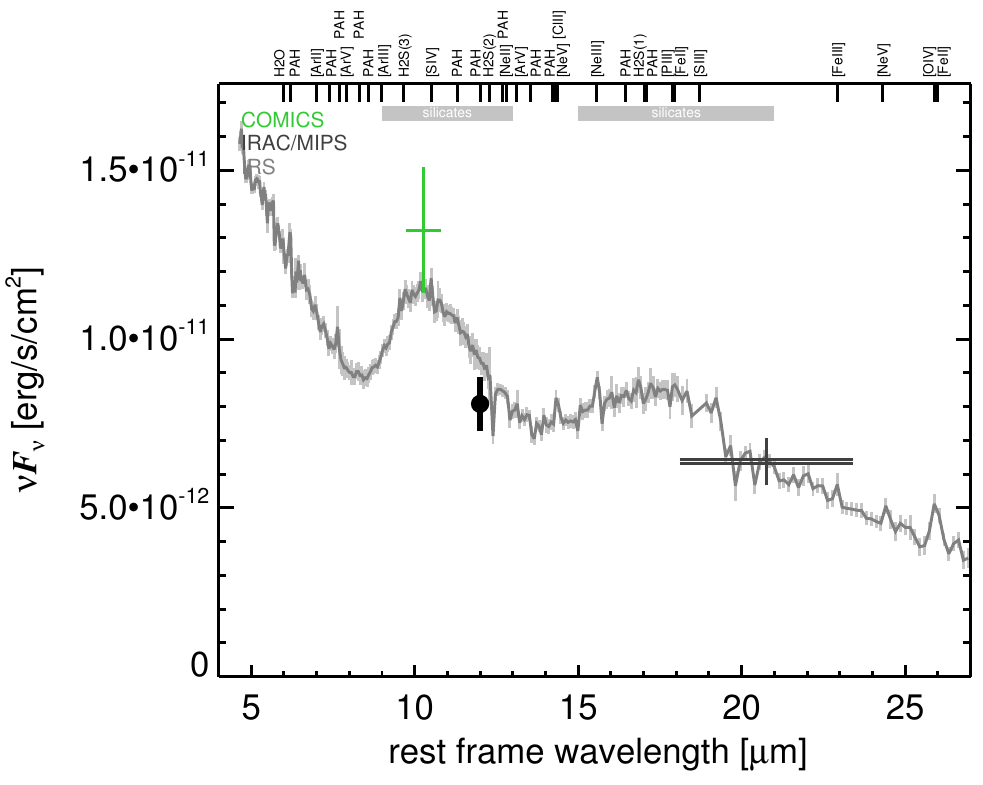}
   \caption{\label{fig:MISED_PG0026+129}
      MIR SED of PG\,0026+129. The description  of the symbols (if present) is the following.
      Grey crosses and  solid lines mark the \spitzer/IRAC, MIPS and IRS data. 
      The colour coding of the other symbols is: 
      green for COMICS, magenta for Michelle, blue for T-ReCS and red for VISIR data.
      Darker-coloured solid lines mark spectra of the corresponding instrument.
      The black filled circles mark the nuclear 12 and $18\,\mu$m  continuum emission estimate from the data.
      The ticks on the top axis mark positions of common MIR emission lines, while the light grey horizontal bars mark wavelength ranges affected by the silicate 10 and 18$\mu$m features.}
\end{figure}
\clearpage

\twocolumn[\begin{@twocolumnfalse}  
\subsection{PG\,0052+251}\label{app:PG0052+251}
PG\,0052+251 is a radio-quiet quasar at a redshift of $z=$ 0.154 (D$\sim759\,$Mpc) with an optical Sy\,1.2 classification \citep{veron-cetty_catalogue_2010} in a low-inclination spiral host galaxy \citep{bahcall_apparently_1996}.
It was detected at radio wavelengths with a slightly elongated core in the east-west direction, while its \oiii emission is extended in the north-south direction \citep{leipski_radio_2006,bennert_size_2002}.
PG\,0052+251 was first detected in the MIR by \cite{neugebauer_continuum_1987}, and then followed up with \isoo \citep{haas_dust_2000} and \spitzer/IRS and MIPS.
It appears as a point source in the corresponding MIPS 24\,$\mu$m image, while the IRS LR staring-mode spectrum exhibits prominent silicate 10 and 18$\,\mu$m emission, a very weak PAH 11.3\,$\mu$m feature, and a blue spectral slope in $\nu F_\nu$-space (see also \citealt{shi_9.7_2006,shi_aromatic_2007}).
PG\,0052+252 was imaged with COMICS in the N11.7 filter in 2006 (unpublished, to our knowledge), and a compact nucleus without further host emission has been weakly detected.
However, the current  MIR data are insufficient to reach any robust conclusion about the nuclear extension at subarcsecond scales in the MIR.
Our nuclear N11.7 photometry is $\sim29\%$ lower than the \spitzerr spectrophotometry, indicating that host emission contributes to the arcsecond-scale MIR SED.
\newline\end{@twocolumnfalse}]

\begin{figure}
   \centering
   \includegraphics[angle=0,width=8.500cm]{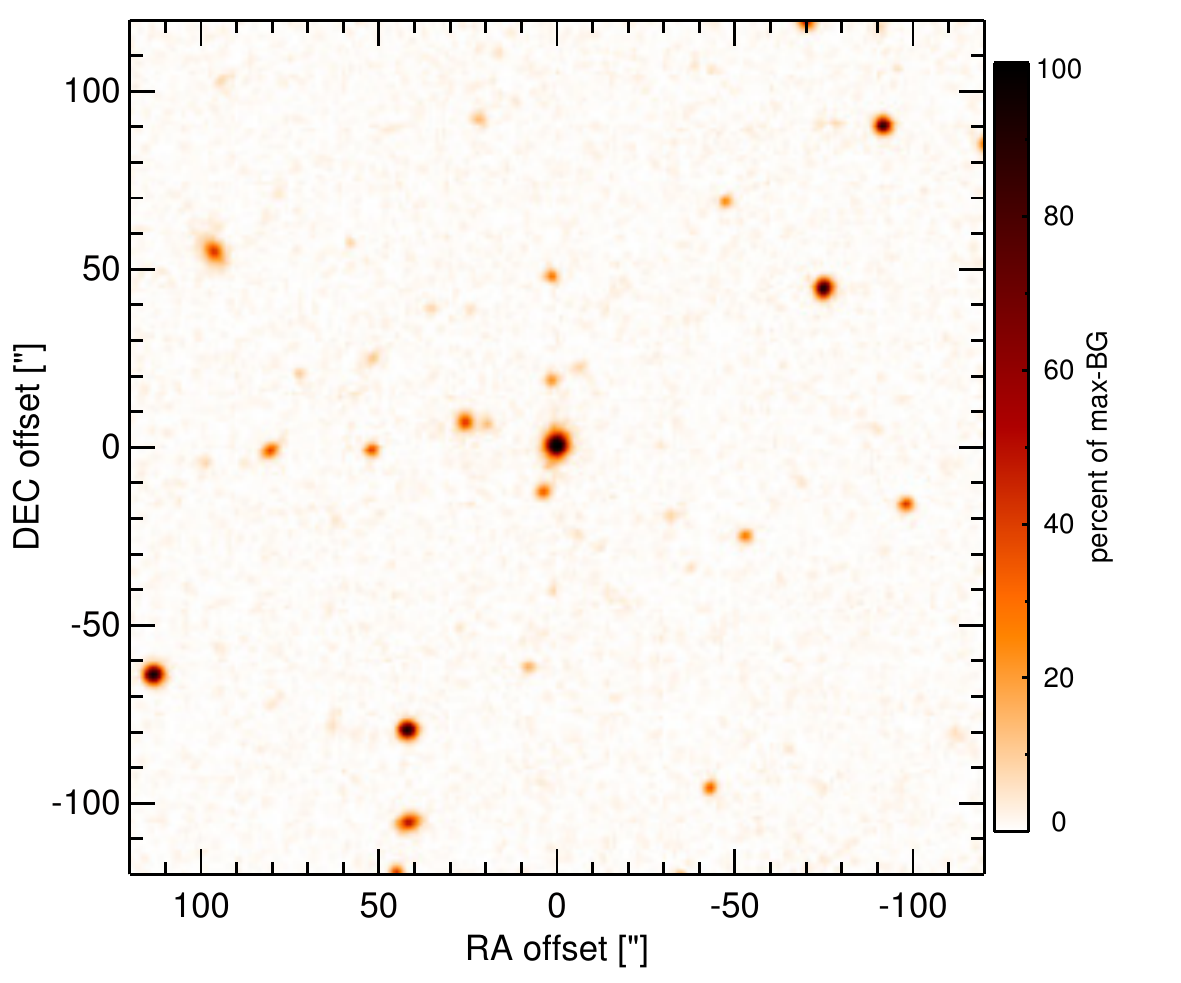}
    \caption{\label{fig:OPTim_PG0052+251}
             Optical image (DSS, red filter) of PG\,0052+251. Displayed are the central $4\arcmin$ with North up and East to the left. 
              The colour scaling is linear with white corresponding to the median background and black to the $0.01\%$ pixels with the highest intensity.  
           }
\end{figure}
\begin{figure}
   \centering
   \includegraphics[angle=0,height=3.11cm]{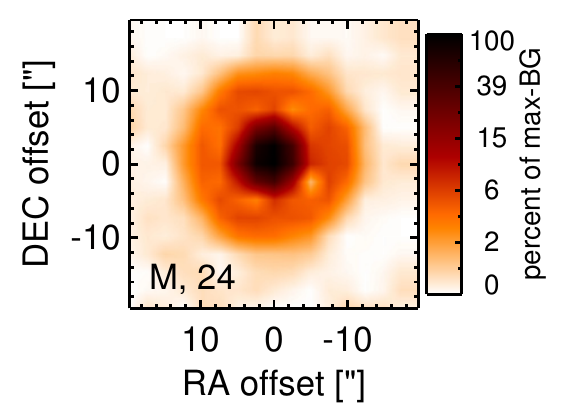}
    \caption{\label{fig:INTim_PG0052+251}
             \spitzerr MIR images of PG\,0052+251. Displayed are the inner $40\arcsec$ with North up and East to the left. The colour scaling is logarithmic with white corresponding to median background and black to the $0.1\%$ pixels with the highest intensity.
             The label in the bottom left states instrument and central wavelength of the filter in $\mu$m (I: IRAC, M: MIPS). 
           }
\end{figure}
\begin{figure}
   \centering
   \includegraphics[angle=0,height=3.11cm]{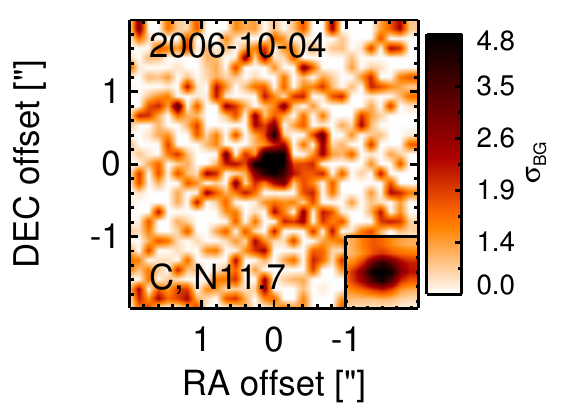}
    \caption{\label{fig:HARim_PG0052+251}
             Subarcsecond-resolution MIR images of PG\,0052+251 sorted by increasing filter wavelength. 
             Displayed are the inner $4\arcsec$ with North up and East to the left. 
             The colour scaling is logarithmic with white corresponding to median background and black to the $75\%$ of the highest intensity of all images in units of $\sigbg$.
             The inset image shows the central arcsecond of the PSF from the calibrator star, scaled to match the science target.
             The labels in the bottom left state instrument and filter names (C: COMICS, M: Michelle, T: T-ReCS, V: VISIR).
           }
\end{figure}
\begin{figure}
   \centering
   \includegraphics[angle=0,width=8.50cm]{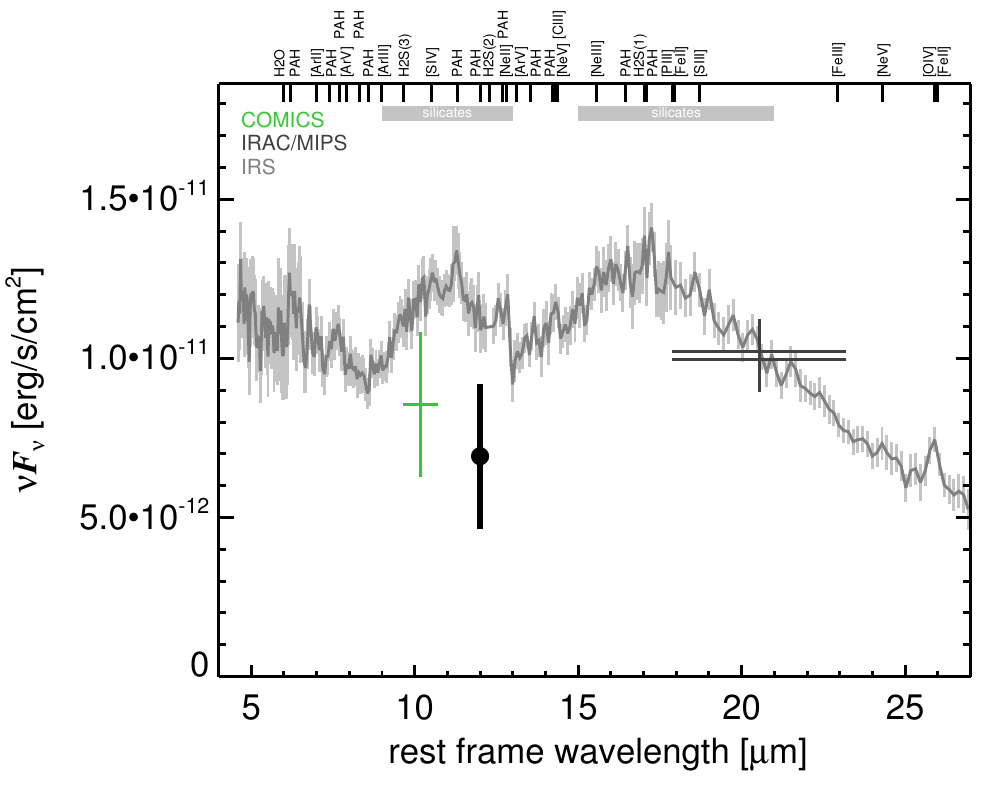}
   \caption{\label{fig:MISED_PG0052+251}
      MIR SED of PG\,0052+251. The description  of the symbols (if present) is the following.
      Grey crosses and  solid lines mark the \spitzer/IRAC, MIPS and IRS data. 
      The colour coding of the other symbols is: 
      green for COMICS, magenta for Michelle, blue for T-ReCS and red for VISIR data.
      Darker-coloured solid lines mark spectra of the corresponding instrument.
      The black filled circles mark the nuclear 12 and $18\,\mu$m  continuum emission estimate from the data.
      The ticks on the top axis mark positions of common MIR emission lines, while the light grey horizontal bars mark wavelength ranges affected by the silicate 10 and 18$\mu$m features.}
\end{figure}
\clearpage

\twocolumn[\begin{@twocolumnfalse}  
\subsection{PG\,0844+349 -- Ton\,951}\label{app:PG0844+349}
PG\,0844+349 is a radio-quiet quasar at a redshift of $z=$ 0.0640 (D$\sim302\,$Mpc) with an optical Sy\,1.0 classification \citep{veron-cetty_catalogue_2010} in an interacting barred spiral host galaxy \citep{hutchings_images_1990}.
Its properties are quite similar to a narrow-line Sy\,1 type AGN\citep{brinkmann_xmm-newton_2003}.
At radio wavelengths, the nucleus is unresolved \citep{kellermann_radio_1994}.
PG\,0844+349 was first detected in the MIR with \irass \citep{neugebauer_quasars_1986}, and followed up with the Palomar Observatory \citep{neugebauer_continuum_1987, neugebauer_variability_1999} and \spitzer/IRAC, IRS and MIPS.
The corresponding IRAC and MIPS images show a compact nucleus without significant host emission.
The IRS LR staring-mode spectrum exhibits prominent silicate 10 and 18$\,\mu$m emission, a very weak PAH 11.3\,$\mu$m feature, and a shallow blue spectral slope in $\nu F_\nu$-space (see also \citealt{shi_aromatic_2007}).
PG\,0844+349 was imaged with COMICS in the N11.7 filter in 2006 \citep{imanishi_subaru_2011}, and an elongated nucleus was detected (FWHM(major axis)$\sim0.75\arcsec\sim1\,$kpc; PA$\sim108\degree$).
However, at least a second epoch of subarcsecond MIR imaging is required to verify this extension.
Our nuclear N11.7 photometry is consistent with the value by \cite{imanishi_subaru_2011} and the \spitzerr spectrophotometry.
Therefore, we use the latter to compute the nuclear 12\,$\mu$m continuum emission estimate corrected for the silicate feature.
Note however, that the nuclear flux would be significantly lower if the subarcsecond-extended emission can be verified.
For now, the resulting synthetic flux is then scaled by half of the ratio between $\Fpsf$ and $\Fgau$ to account for the possibly nuclear extension.
\newline\end{@twocolumnfalse}]

\begin{figure}
   \centering
   \includegraphics[angle=0,width=8.500cm]{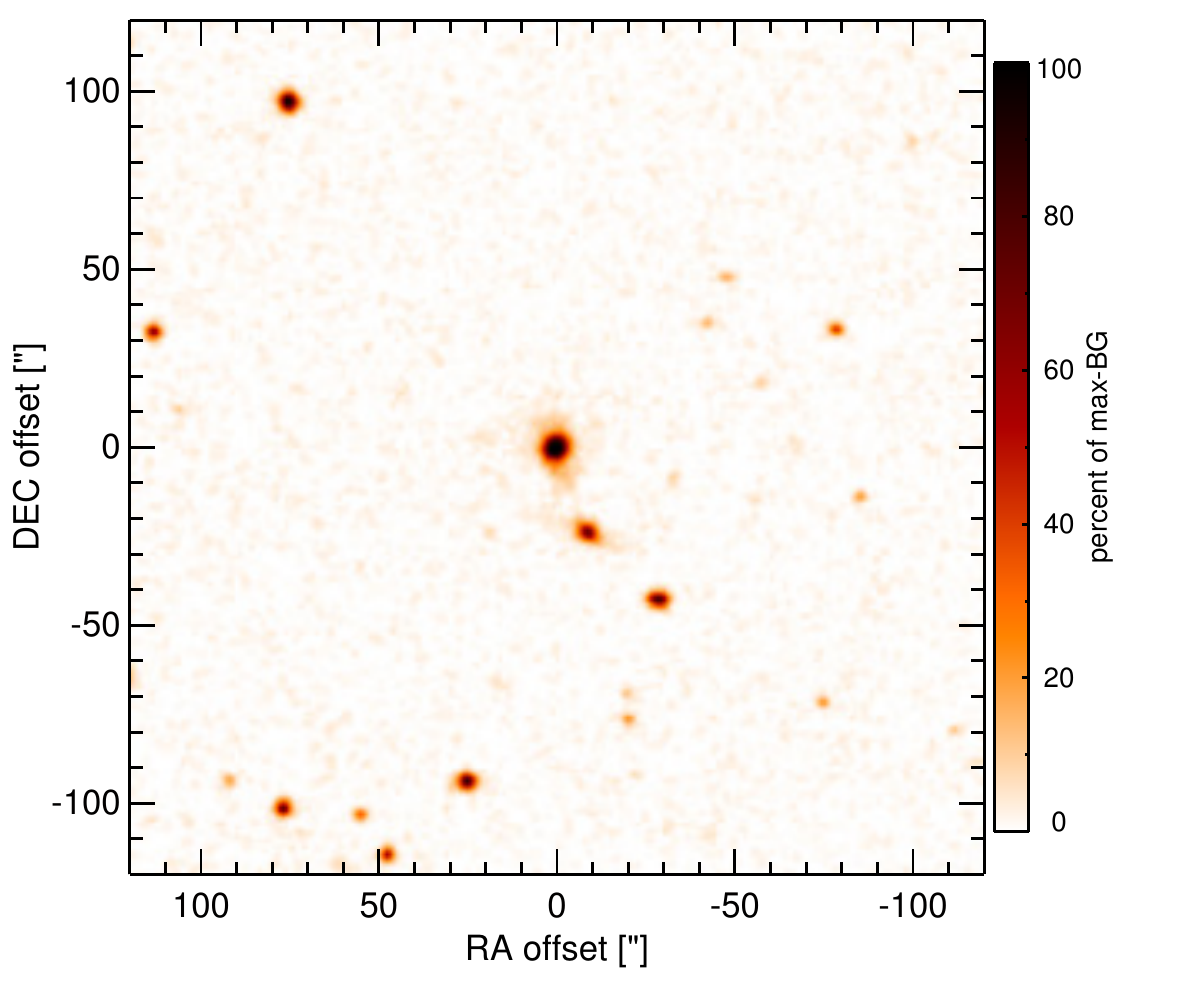}
    \caption{\label{fig:OPTim_PG0844+349}
             Optical image (DSS, red filter) of PG\,0844+349. Displayed are the central $4\arcmin$ with North up and East to the left. 
              The colour scaling is linear with white corresponding to the median background and black to the $0.01\%$ pixels with the highest intensity.  
           }
\end{figure}
\begin{figure}
   \centering
   \includegraphics[angle=0,height=3.11cm]{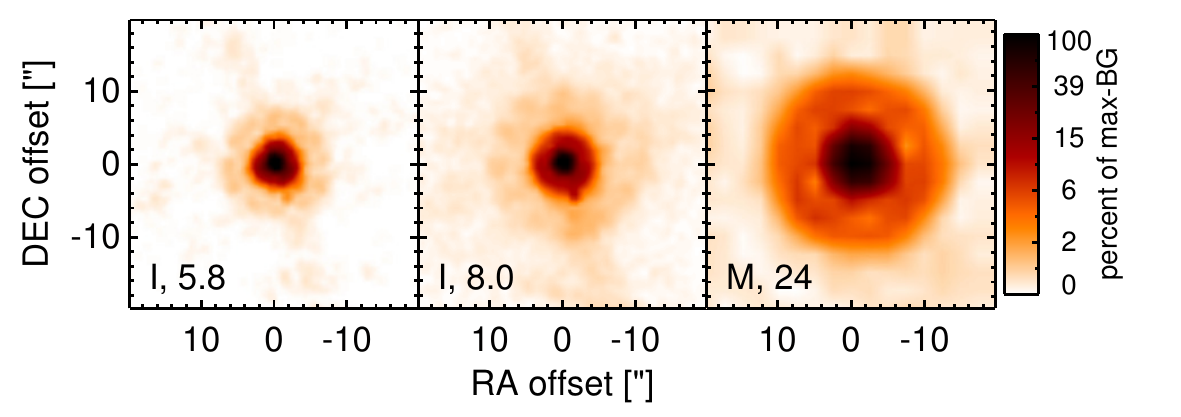}
    \caption{\label{fig:INTim_PG0844+349}
             \spitzerr MIR images of PG\,0844+349. Displayed are the inner $40\arcsec$ with North up and East to the left. The colour scaling is logarithmic with white corresponding to median background and black to the $0.1\%$ pixels with the highest intensity.
             The label in the bottom left states instrument and central wavelength of the filter in $\mu$m (I: IRAC, M: MIPS). 
           }
\end{figure}
\begin{figure}
   \centering
   \includegraphics[angle=0,height=3.11cm]{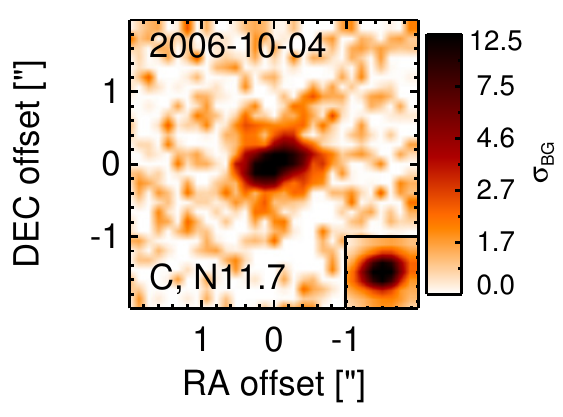}
    \caption{\label{fig:HARim_PG0844+349}
             Subarcsecond-resolution MIR images of PG\,0844+349 sorted by increasing filter wavelength. 
             Displayed are the inner $4\arcsec$ with North up and East to the left. 
             The colour scaling is logarithmic with white corresponding to median background and black to the $75\%$ of the highest intensity of all images in units of $\sigbg$.
             The inset image shows the central arcsecond of the PSF from the calibrator star, scaled to match the science target.
             The labels in the bottom left state instrument and filter names (C: COMICS, M: Michelle, T: T-ReCS, V: VISIR).
           }
\end{figure}
\begin{figure}
   \centering
   \includegraphics[angle=0,width=8.50cm]{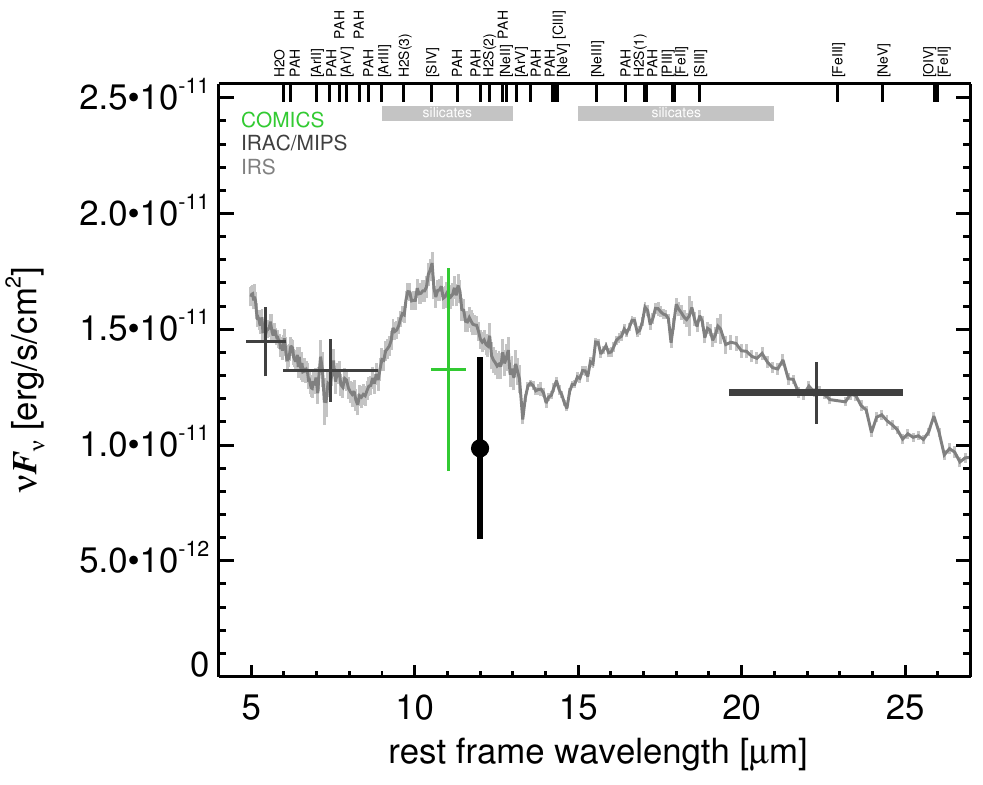}
   \caption{\label{fig:MISED_PG0844+349}
      MIR SED of PG\,0844+349. The description  of the symbols (if present) is the following.
      Grey crosses and  solid lines mark the \spitzer/IRAC, MIPS and IRS data. 
      The colour coding of the other symbols is: 
      green for COMICS, magenta for Michelle, blue for T-ReCS and red for VISIR data.
      Darker-coloured solid lines mark spectra of the corresponding instrument.
      The black filled circles mark the nuclear 12 and $18\,\mu$m  continuum emission estimate from the data.
      The ticks on the top axis mark positions of common MIR emission lines, while the light grey horizontal bars mark wavelength ranges affected by the silicate 10 and 18$\mu$m features.}
\end{figure}
\clearpage

\twocolumn[\begin{@twocolumnfalse}  
\subsection{PG\,2130+099 -- Mrk\,1513 -- II\,Zw\,136 -- UGC\,11763}\label{app:PG2130+099}
PG\,2130+099 is a borderline quasar/Seyfert object at a redshift of $z=$ 0.063 (D$\sim288\,$Mpc) with an optical classification as Sy\,1.5 \citep{veron-cetty_catalogue_2010} in a highly-inclined peculiar spiral host galaxy.
At radio wavelengths, it appears as a compact radio core with two off-nuclear compact sources at $\sim1\arcsec\sim\,1.2\,$ kpc to the south-east and $\sim1.5\arcsec\sim\,1.8\,$ kpc to the north-west (PA$\sim135\degree$;\citealt{kukula_radio_1998,leipski_radio_2006}).
Pioneering MIR observations of PG\,2130+099 were performed by \cite{rieke_infrared_1978}, followed by \cite{neugebauer_continuum_1987}, \citep{elvis_atlas_1994}, \cite{neugebauer_variability_1999}, and \cite{galliano_mid-infrared_2005}.
The latter authors report the first subarcsecond-resolution MIR observations, in which PG\,2130+099 appears unresolved.
This object was also observed with \isoo \citep{rigopoulou_large_1999,haas_iso_2003} and \spitzer/IRS and MIPS, and appears point-like in the corresponding MIPS 24\,$\mu$m image.
The IRS LR staring-mode spectrum exhibits weak silicate 10 and 18\,$\mu$m emission, very weak PAH features and a very shallow blue spectral slope in $\nu F_\nu$-space (see also \citealt{shi_aromatic_2007}).
We observed PG\,2130+099 with VISIR in three narrow $N$-band filters in 2005 \citep{horst_small_2006,horst_mid-infrared_2009}, and an additional COMICS image in the N11.7 filter was taken in 2006 \citep{imanishi_subaru_2011}.
A compact nucleus without further host emission was detected in all cases.
The nuclear morphology is inconsistent between the VISIR and COMICS images, and we classify its MIR extension at subarcsecond resolution as uncertain.
Our nuclear photometry is consistent with \cite{horst_small_2006}, 32\% higher than the value from \cite{imanishi_subaru_2011}, and consistent with, yet systematically higher than, the \spitzerr spectrophotometry.
Comparison with the historical $N$-band photometry shows apparent flux variations on the order of $\sim6\%$ during the last $\sim30$ years, well within the uncertainties and systematics due to different instruments, filters and measurement methods.
\newline\end{@twocolumnfalse}]

\begin{figure}
   \centering
   \includegraphics[angle=0,width=8.500cm]{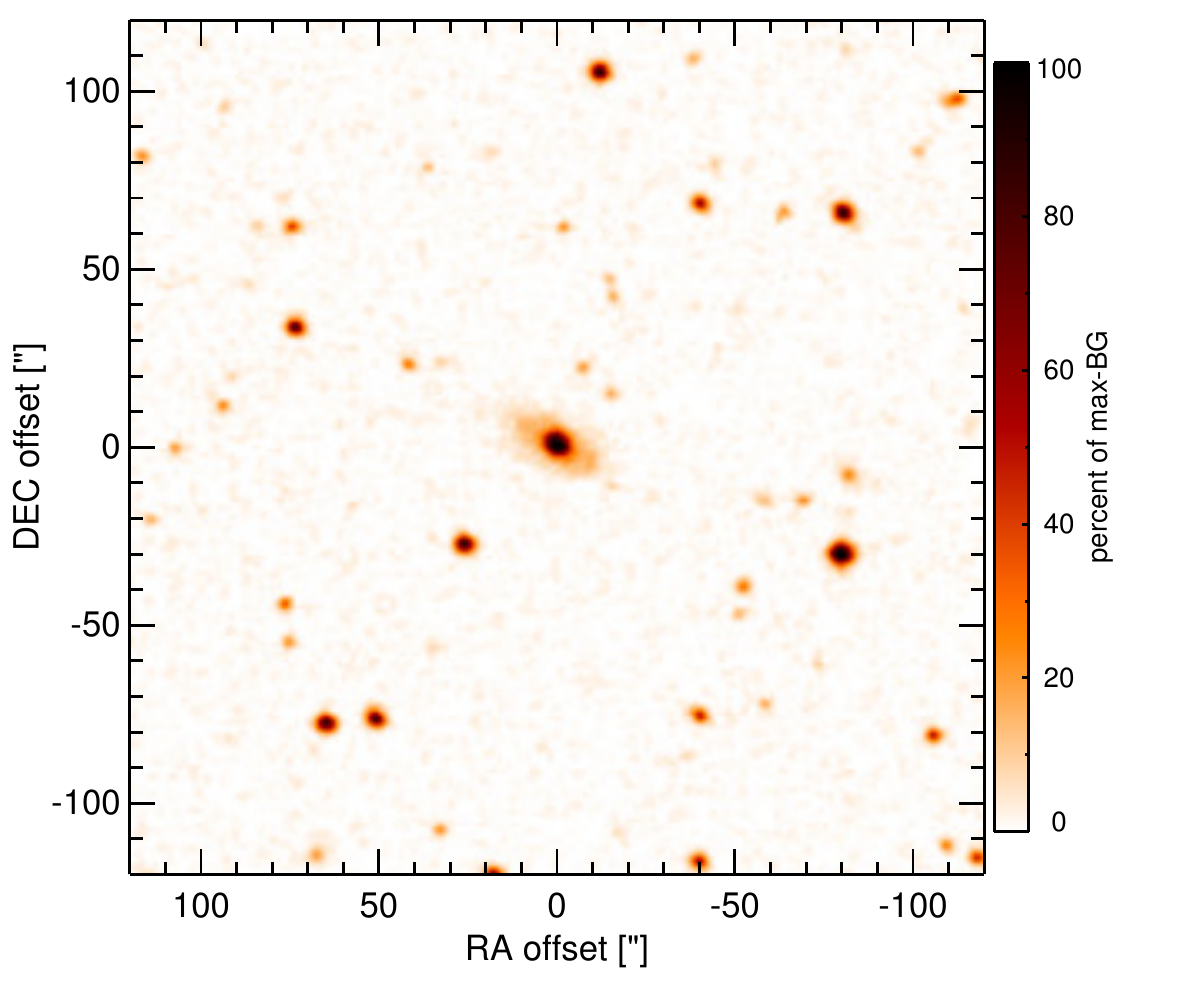}
    \caption{\label{fig:OPTim_PG2130+099}
             Optical image (DSS, red filter) of PG\,2130+099. Displayed are the central $4\arcmin$ with North up and East to the left. 
              The colour scaling is linear with white corresponding to the median background and black to the $0.01\%$ pixels with the highest intensity.  
           }
\end{figure}
\begin{figure}
   \centering
   \includegraphics[angle=0,height=3.11cm]{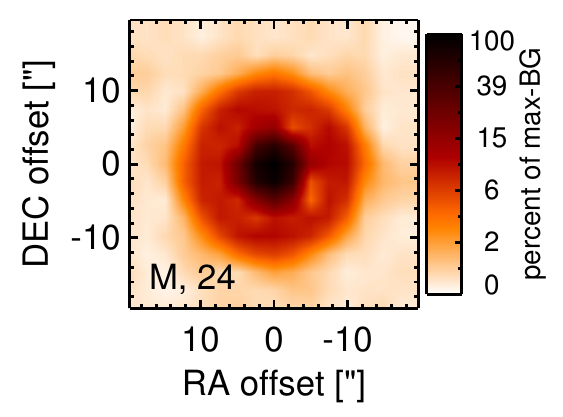}
    \caption{\label{fig:INTim_PG2130+099}
             \spitzerr MIR images of PG\,2130+099. Displayed are the inner $40\arcsec$ with North up and East to the left. The colour scaling is logarithmic with white corresponding to median background and black to the $0.1\%$ pixels with the highest intensity.
             The label in the bottom left states instrument and central wavelength of the filter in $\mu$m (I: IRAC, M: MIPS). 
           }
\end{figure}
\begin{figure}
   \centering
   \includegraphics[angle=0,width=8.500cm]{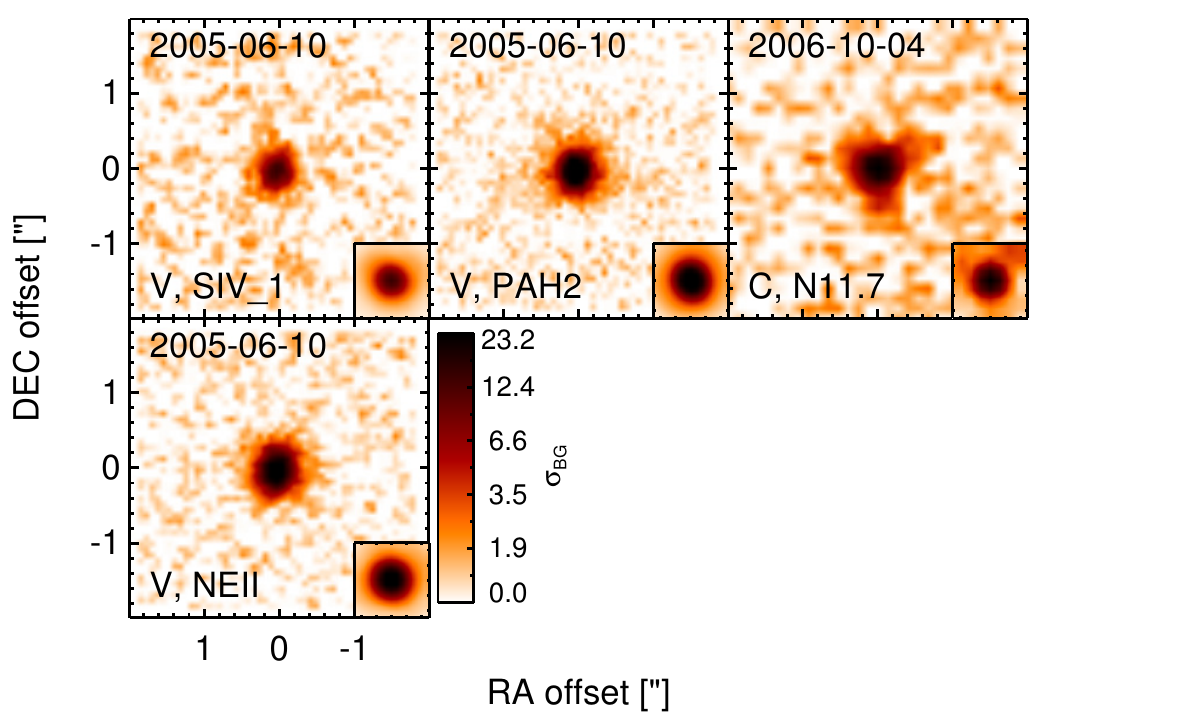}
    \caption{\label{fig:HARim_PG2130+099}
             Subarcsecond-resolution MIR images of PG\,2130+099 sorted by increasing filter wavelength. 
             Displayed are the inner $4\arcsec$ with North up and East to the left. 
             The colour scaling is logarithmic with white corresponding to median background and black to the $75\%$ of the highest intensity of all images in units of $\sigbg$.
             The inset image shows the central arcsecond of the PSF from the calibrator star, scaled to match the science target.
             The labels in the bottom left state instrument and filter names (C: COMICS, M: Michelle, T: T-ReCS, V: VISIR).
           }
\end{figure}
\begin{figure}
   \centering
   \includegraphics[angle=0,width=8.50cm]{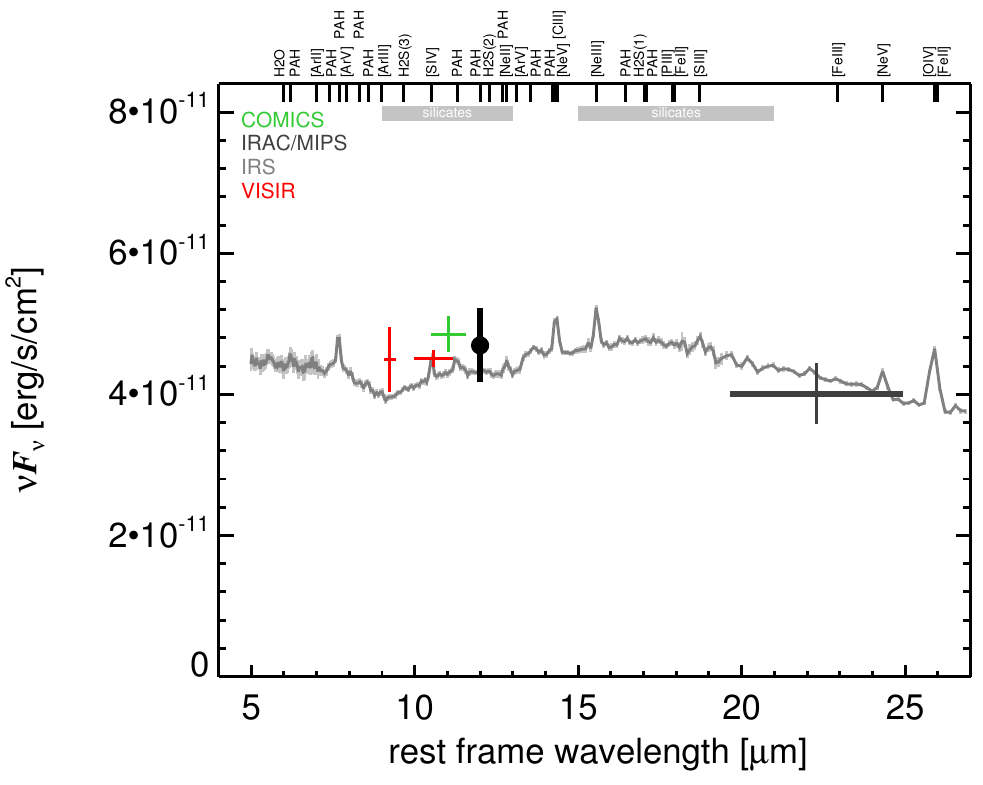}
   \caption{\label{fig:MISED_PG2130+099}
      MIR SED of PG\,2130+099. The description  of the symbols (if present) is the following.
      Grey crosses and  solid lines mark the \spitzer/IRAC, MIPS and IRS data. 
      The colour coding of the other symbols is: 
      green for COMICS, magenta for Michelle, blue for T-ReCS and red for VISIR data.
      Darker-coloured solid lines mark spectra of the corresponding instrument.
      The black filled circles mark the nuclear 12 and $18\,\mu$m  continuum emission estimate from the data.
      The ticks on the top axis mark positions of common MIR emission lines, while the light grey horizontal bars mark wavelength ranges affected by the silicate 10 and 18$\mu$m features.}
\end{figure}
\clearpage

\twocolumn[\begin{@twocolumnfalse}  
\subsection{Pictor\,A -- ESO\,252-18 -- PKS\,0518-45}\label{app:PictorA}
Pictor\,A is a radio-loud inclined lenticular galaxy at a redshift of $z=$ 0.0351 ($D\sim161\,$Mpc) with an FR\,II radio morphology.
It contains an AGN optically classified as a borderline Sy\,1.5/LINER \citep{veron-cetty_catalogue_1998,tueller_swift_2008} that belongs to the nine-month BAT AGN sample.
Pictor\,A features two well-studied giant radio lobes extending $\sim0.5\,$Mpc from the nucleus.
A narrow jet extends to the western lobe and is also visible in X-rays (PA$\sim281\degree$; e.g.\citealt{perley_radio_1997,wilson_chandra_2001}). 
The nucleus itself is a compact flat-spectrum radio source. 
The \oiii emission is extended along the north-west direction by $\sim0.3\arcsec\sim200\,$pc \citep{simkin_pictor_1999}.
After its first MIR detection in \iras, Pictor\,A was followed up with \spitzer/IRAC, IRS and MIPS.
The corresponding IRAC and MIPS images are dominated by a compact nucleus without any significant non-nuclear emission.
Our nuclear MIPS 24\,$\mu$m photometry is consistent with \cite{shi_far-infrared_2005}.
The IRS LR staring-mode spectrum exhibits strong silicate 10 and 18\,$\mu$m emission and a flat spectral slope in $\nu F_\nu$-space. No PAH features are detected (see also \citealt{shi_9.7_2006}).
Thus, the arcsecond-scale MIR SED is dominated by AGN-heated dust emission.
Pictor\,A was observed with VISIR in the broad SIC filter in 2006 \citep{van_der_wolk_dust_2010}, in three narrow $N$-band filters in 2009 (this work), and with T-ReCS in two $N$-band and one $Q$-band filters in 2007 and 2008 (unpublished, to our knowledge).
The nucleus appears unresolved in the sharpest images (SIC and Si2), while it appears extended in the longest wavelength filters, NEII and Q2 (FWHM(major axis) $\sim 1\arcsec \sim 0.7\,$kpc; PA$\sim125\degree$).
The elongation has a low significance owing to the low S/N of the NEII and Q2 images, but corresponds roughly to the \oiii morphology.
Deeper imaging is, therefore, required to confirm extension, and we classify the nuclear MIR extension at subarcsecond resolution as uncertain for now.
The nuclear photometry is 25\% higher than the value by \cite{van_der_wolk_dust_2010} for unknown reasons, and consistent with the \spitzerr spectrophotometry.
Therefore, the silicate emission originates in the projected central $\sim0.3$\,kpc of Pictor\,A, and we use the IRS spectrum to correct our nuclear 12\,$\mu$m continuum emission estimate for this feature.
\newline\end{@twocolumnfalse}]

\begin{figure}
   \centering
   \includegraphics[angle=0,width=8.500cm]{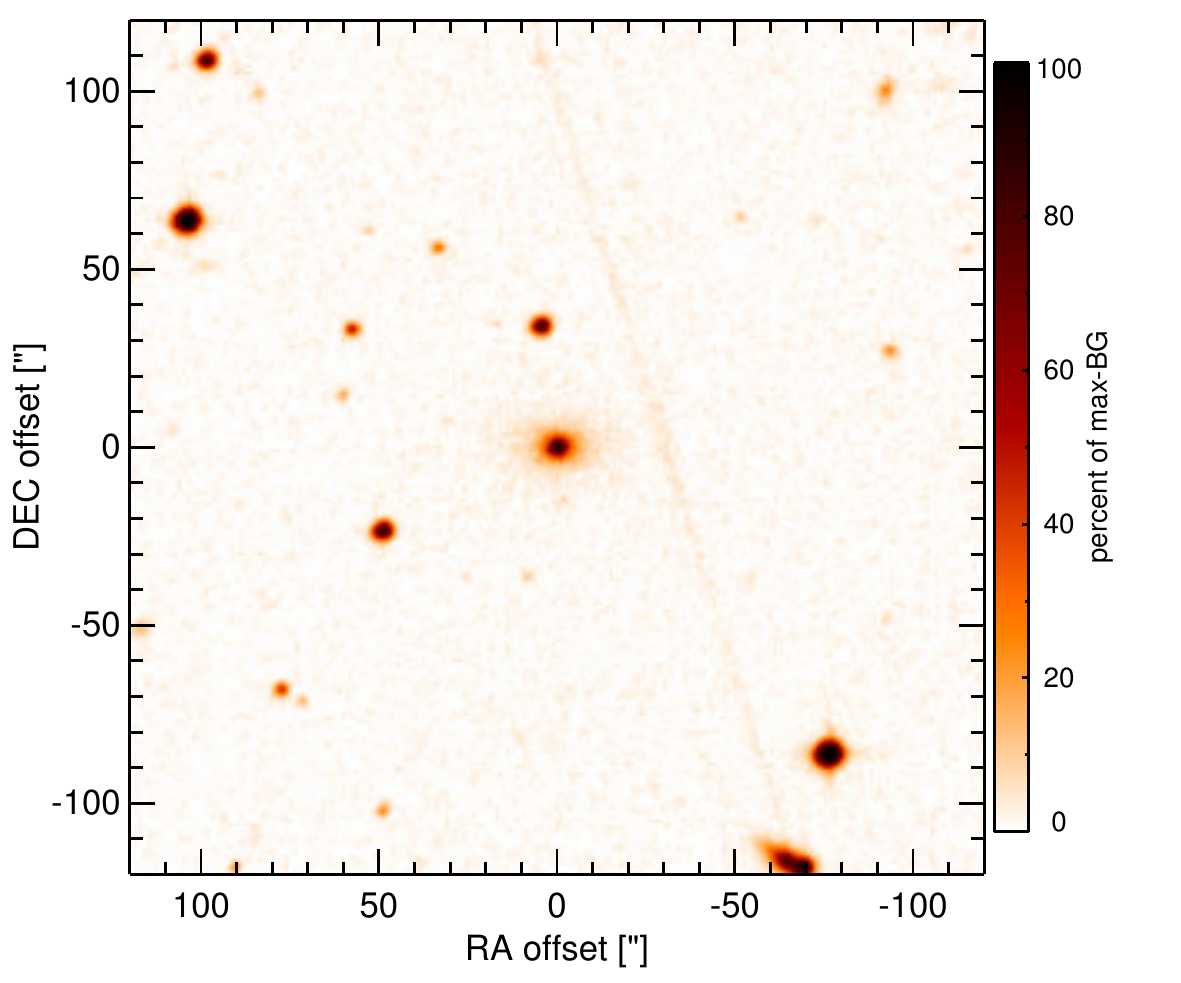}
    \caption{\label{fig:OPTim_PictorA}
             Optical image (DSS, red filter) of Pictor\,A. Displayed are the central $4\arcmin$ with North up and East to the left. 
              The colour scaling is linear with white corresponding to the median background and black to the $0.01\%$ pixels with the highest intensity.  
           }
\end{figure}
\begin{figure}
   \centering
   \includegraphics[angle=0,height=3.11cm]{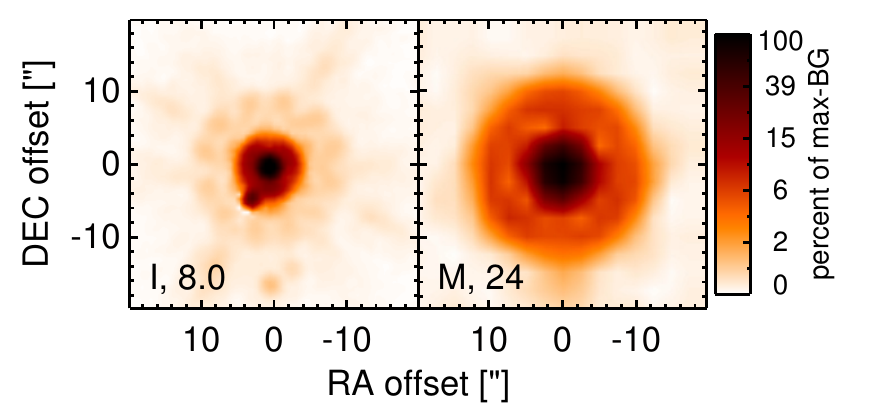}
    \caption{\label{fig:INTim_PictorA}
             \spitzerr MIR images of Pictor\,A. Displayed are the inner $40\arcsec$ with North up and East to the left. The colour scaling is logarithmic with white corresponding to median background and black to the $0.1\%$ pixels with the highest intensity.
             The label in the bottom left states instrument and central wavelength of the filter in $\mu$m (I: IRAC, M: MIPS).
             Note that the apparent off-nuclear compact source in the IRAC $8.0\,\mu$m image is an instrumental artefact.
           }
\end{figure}
\begin{figure}
   \centering
   \includegraphics[angle=0,width=8.500cm]{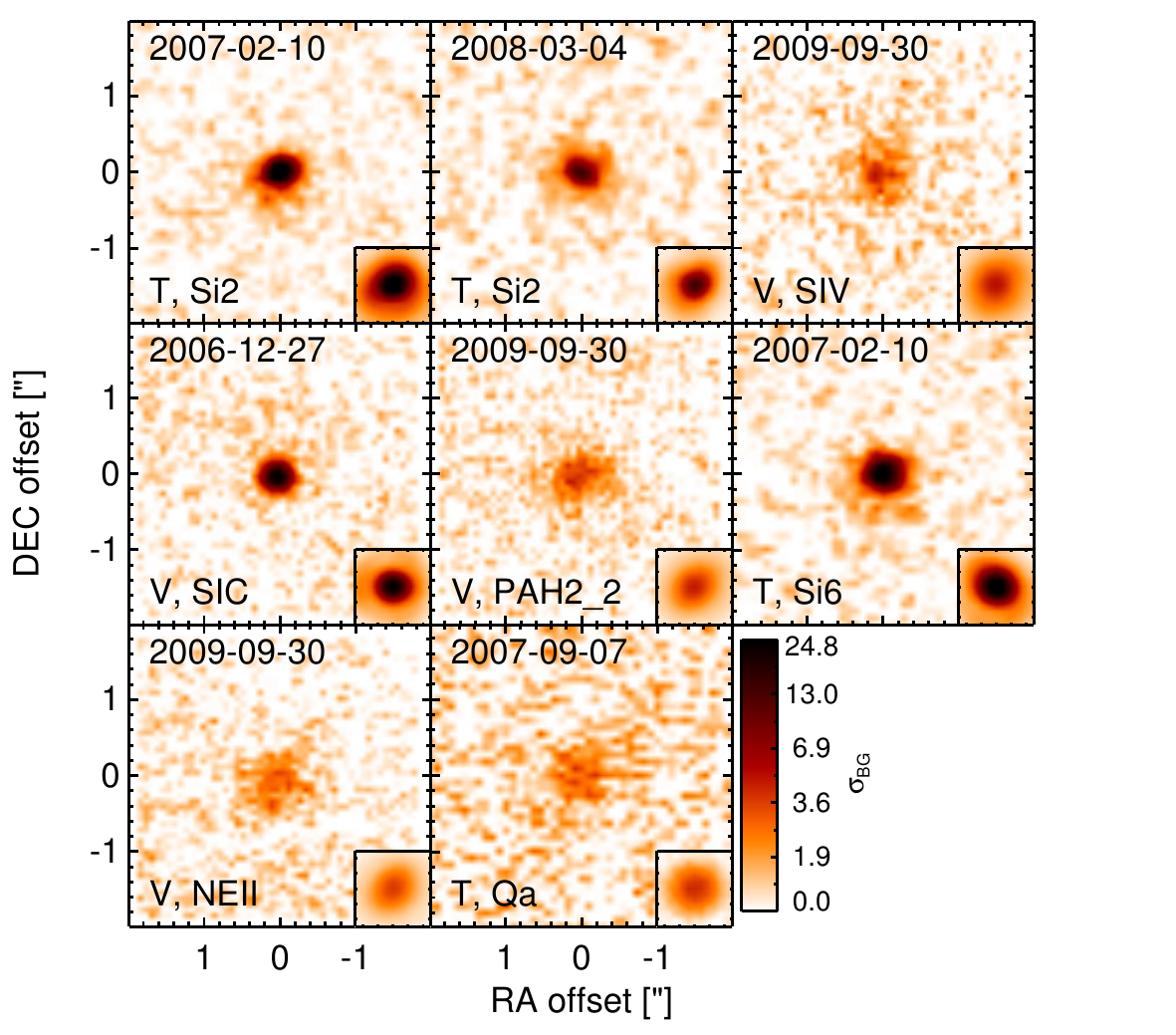}
    \caption{\label{fig:HARim_PictorA}
             Subarcsecond-resolution MIR images of Pictor\,A sorted by increasing filter wavelength. 
             Displayed are the inner $4\arcsec$ with North up and East to the left. 
             The colour scaling is logarithmic with white corresponding to median background and black to the $75\%$ of the highest intensity of all images in units of $\sigbg$.
             The inset image shows the central arcsecond of the PSF from the calibrator star, scaled to match the science target.
             The labels in the bottom left state instrument and filter names (C: COMICS, M: Michelle, T: T-ReCS, V: VISIR).
           }
\end{figure}
\begin{figure}
   \centering
   \includegraphics[angle=0,width=8.50cm]{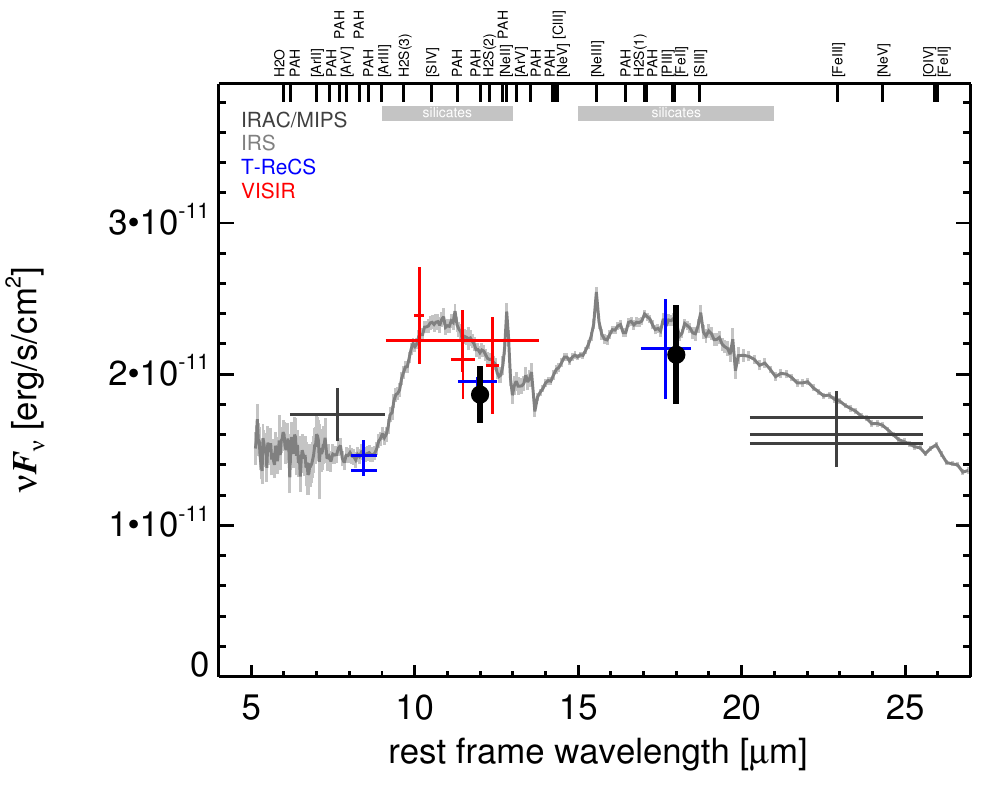}
   \caption{\label{fig:MISED_PictorA}
      MIR SED of Pictor\,A. The description  of the symbols (if present) is the following.
      Grey crosses and  solid lines mark the \spitzer/IRAC, MIPS and IRS data. 
      The colour coding of the other symbols is: 
      green for COMICS, magenta for Michelle, blue for T-ReCS and red for VISIR data.
      Darker-coloured solid lines mark spectra of the corresponding instrument.
      The black filled circles mark the nuclear 12 and $18\,\mu$m  continuum emission estimate from the data.
      The ticks on the top axis mark positions of common MIR emission lines, while the light grey horizontal bars mark wavelength ranges affected by the silicate 10 and 18$\mu$m features.}
\end{figure}
\clearpage

\twocolumn[\begin{@twocolumnfalse}  
\subsection{PKS\,1417-19 -- LEDA\,51213}\label{app:PKS1417-19}
PKS\,1417-19/LEDA\,51213 is a radio-loud galaxy at a redshift of $z=$ 0.1200 ($D\sim791\,$Mpc) with a Sy\,1.5 nucleus \citep{veron-cetty_catalogue_2010}.
It possesses a biconical large-scale radio morphology with a jet along the north-east direction (PA$\sim20\degree$; \citealt{antonucci_vla_1985}).
PKS\,1417-19 has not been detected with \iras, and no \spitzerr observations are available. 
It appears as a point source in the \wisee images.
\cite{van_der_wolk_dust_2010} observed this object with VISIR in the broad SIC filter in 2006 and weakly detected a compact nucleus.
The emission is possibly extended (FWHM(major axis)$\sim0.5\arcsec\sim1.2\,$kpc; PA$\sim97\degree$) perpendicular to the jet axis. 
However, at least a second epoch of deeper subarcsecond MIR imaging is required to verify this extension.
The nuclear MIR data alone is insufficient for any conclusion about its nature but \cite{van_der_wolk_dust_2010} attribute a large amount of the nuclear MIR emission to thermal radiation.
\newline\end{@twocolumnfalse}]

\begin{figure}
   \centering
   \includegraphics[angle=0,width=8.500cm]{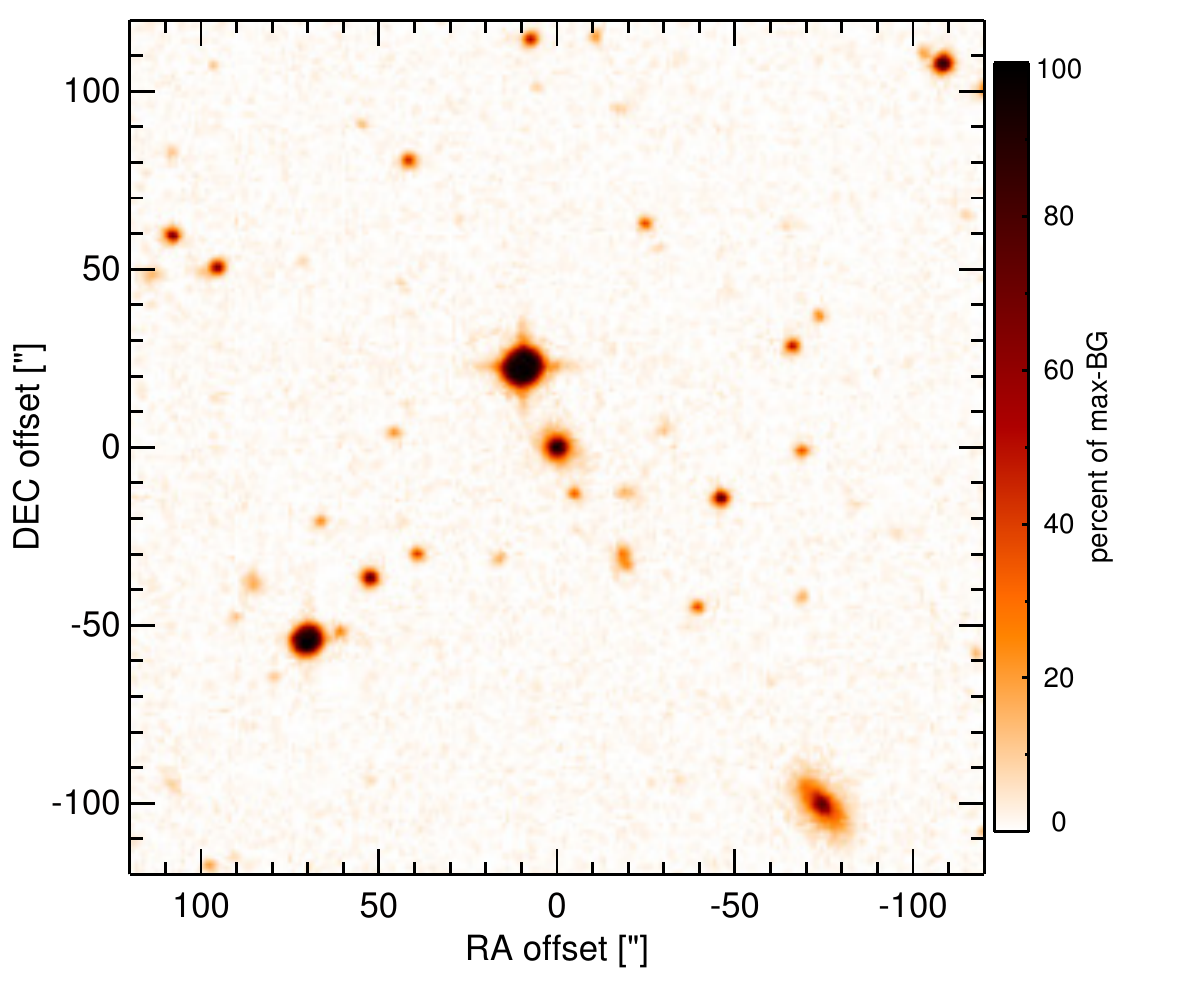}
    \caption{\label{fig:OPTim_PKS1417-19}
             Optical image (DSS, red filter) of PKS\,1417-19. Displayed are the central $4\arcmin$ with North up and East to the left. 
              The colour scaling is linear with white corresponding to the median background and black to the $0.01\%$ pixels with the highest intensity.  
           }
\end{figure}
\begin{figure}
   \centering
   \includegraphics[angle=0,height=3.11cm]{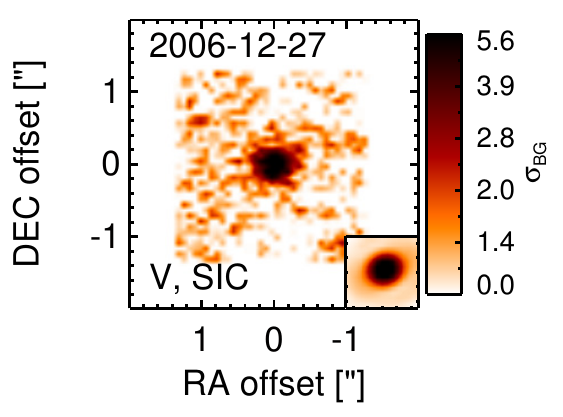}
    \caption{\label{fig:HARim_PKS1417-19}
             Subarcsecond-resolution MIR images of PKS\,1417-19 sorted by increasing filter wavelength. 
             Displayed are the inner $4\arcsec$ with North up and East to the left. 
             The colour scaling is logarithmic with white corresponding to median background and black to the $75\%$ of the highest intensity of all images in units of $\sigbg$.
             The inset image shows the central arcsecond of the PSF from the calibrator star, scaled to match the science target.
             The labels in the bottom left state instrument and filter names (C: COMICS, M: Michelle, T: T-ReCS, V: VISIR).
           }
\end{figure}
\begin{figure}
   \centering
   \includegraphics[angle=0,width=8.50cm]{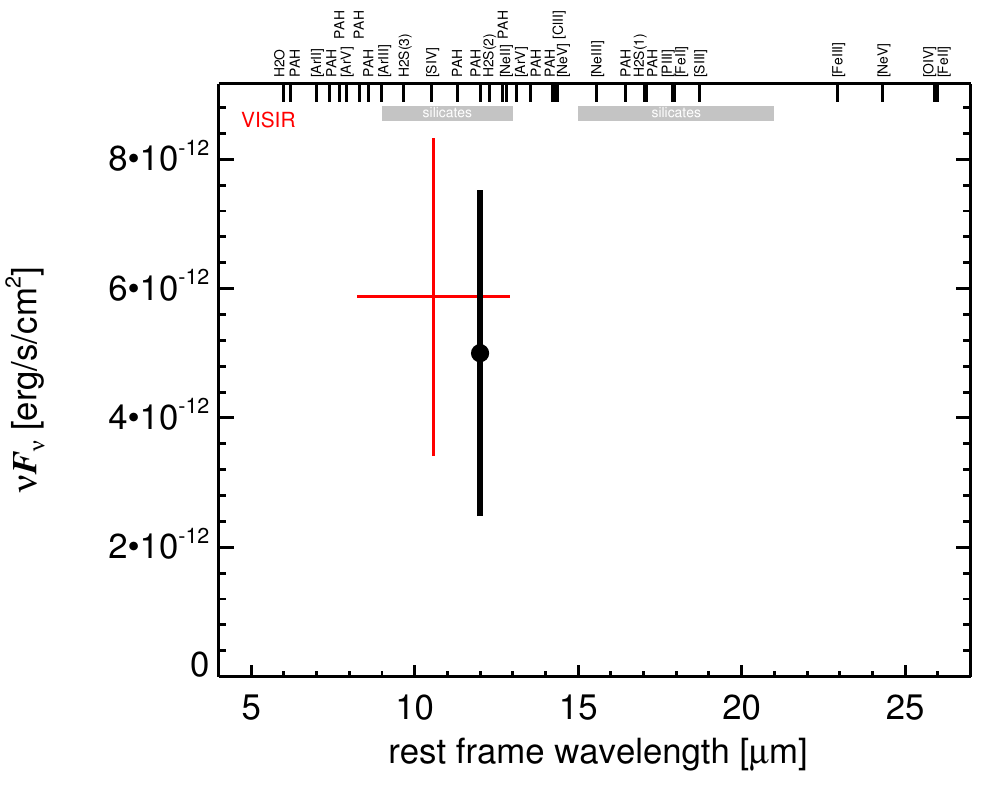}
   \caption{\label{fig:MISED_PKS1417-19}
      MIR SED of PKS\,1417-19. The description  of the symbols (if present) is the following.
      Grey crosses and  solid lines mark the \spitzer/IRAC, MIPS and IRS data. 
      The colour coding of the other symbols is: 
      green for COMICS, magenta for Michelle, blue for T-ReCS and red for VISIR data.
      Darker-coloured solid lines mark spectra of the corresponding instrument.
      The black filled circles mark the nuclear 12 and $18\,\mu$m  continuum emission estimate from the data.
      The ticks on the top axis mark positions of common MIR emission lines, while the light grey horizontal bars mark wavelength ranges affected by the silicate 10 and 18$\mu$m features.}
\end{figure}
\clearpage

\twocolumn[\begin{@twocolumnfalse}  
\subsection{PKS\,1814-63 -- LEDA\,329451}\label{app:PKS1814-63}
PKS\,1814-63/LEDA\,329451 is a radio-loud disturbed edge-on lenticular galaxy at a redshift of $z=$ 0.0627 ($D\sim302\,$Mpc) with a Sy\,2 nucleus \citep{veron-cetty_catalogue_2010} and a  dust lane crossing the galaxy's centre \citep{ramos_almeida_optical_2011}.
The nucleus is extended in the north-south direction in subarcsecond-resolution radio observations and has a steep radio spectrum (e.g. \citealt{tzioumis_sample_2002}).
The first MIR detection of this object was achieved with \spitzer/MIPS where it appears nearly unresolved \citep{dicken_origin_2008}.
Our nuclear MIPS 24\,$\mu$m photometry is consistent with the latter work.
The \spitzer/IRS LR staring-mode spectrum exhibits silicate absorption, weak PAH features, and a shallow red spectral slope in $\nu F_\nu$-space.
The arcsecond-scale MIR SED might, thus, be affected by star-formation emission.
PKS\,1814-63 was observed with VISIR in the broad SIC filter in 2010 (unpublished, to our knowledge), and a compact nucleus is weakly detected.
It appears to be elongated in the north-south direction (FWHM $\sim 0.56\arcsec \sim 0.7\,$kpc; PA$\sim26\degree$).
However, the current MIR data are insufficient to reach any robust conclusion about the nuclear extension at subarcsecond scales in the MIR.
The nuclear VISIR photometry is consistent with the \spitzerr spectrophotometry.
\newline\end{@twocolumnfalse}]

\begin{figure}
   \centering
   \includegraphics[angle=0,width=8.500cm]{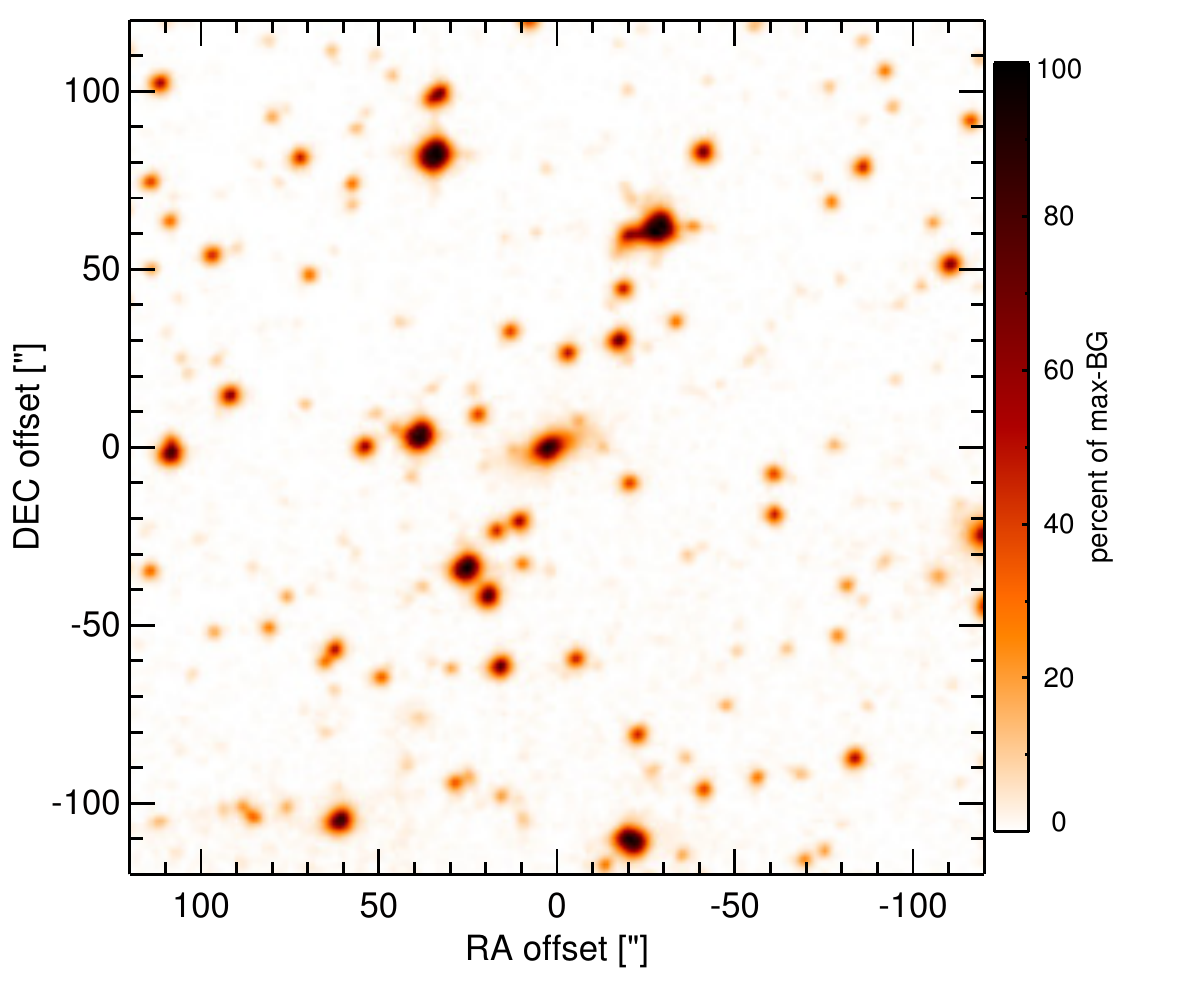}
    \caption{\label{fig:OPTim_PKS1814-63}
             Optical image (DSS, red filter) of PKS\,1814-63. Displayed are the central $4\arcmin$ with North up and East to the left. 
              The colour scaling is linear with white corresponding to the median background and black to the $0.01\%$ pixels with the highest intensity.  
           }
\end{figure}
\begin{figure}
   \centering
   \includegraphics[angle=0,height=3.11cm]{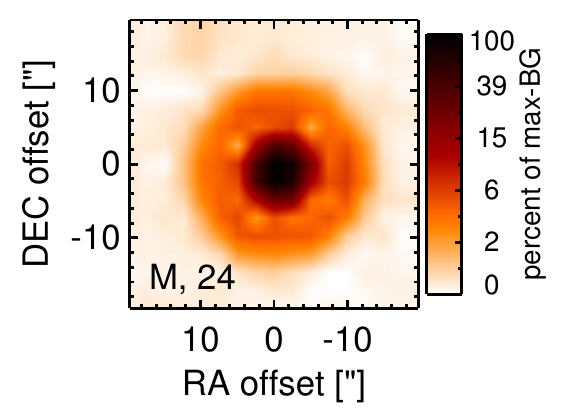}
    \caption{\label{fig:INTim_PKS1814-63}
             \spitzerr MIR images of PKS\,1814-63. Displayed are the inner $40\arcsec$ with North up and East to the left. The colour scaling is logarithmic with white corresponding to median background and black to the $0.1\%$ pixels with the highest intensity.
             The label in the bottom left states instrument and central wavelength of the filter in $\mu$m (I: IRAC, M: MIPS). 
           }
\end{figure}
\begin{figure}
   \centering
   \includegraphics[angle=0,height=3.11cm]{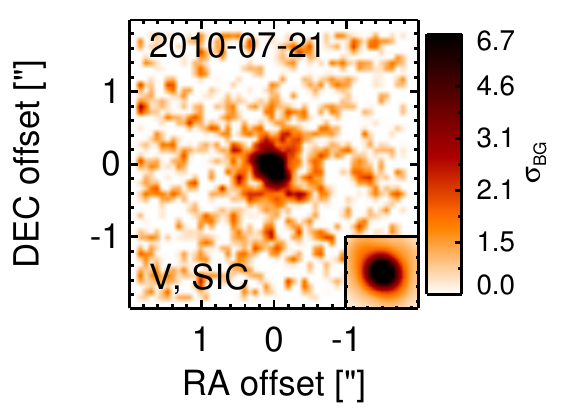}
    \caption{\label{fig:HARim_PKS1814-63}
             Subarcsecond-resolution MIR images of PKS\,1814-63 sorted by increasing filter wavelength. 
             Displayed are the inner $4\arcsec$ with North up and East to the left. 
             The colour scaling is logarithmic with white corresponding to median background and black to the $75\%$ of the highest intensity of all images in units of $\sigbg$.
             The inset image shows the central arcsecond of the PSF from the calibrator star, scaled to match the science target.
             The labels in the bottom left state instrument and filter names (C: COMICS, M: Michelle, T: T-ReCS, V: VISIR).
           }
\end{figure}
\begin{figure}
   \centering
   \includegraphics[angle=0,width=8.50cm]{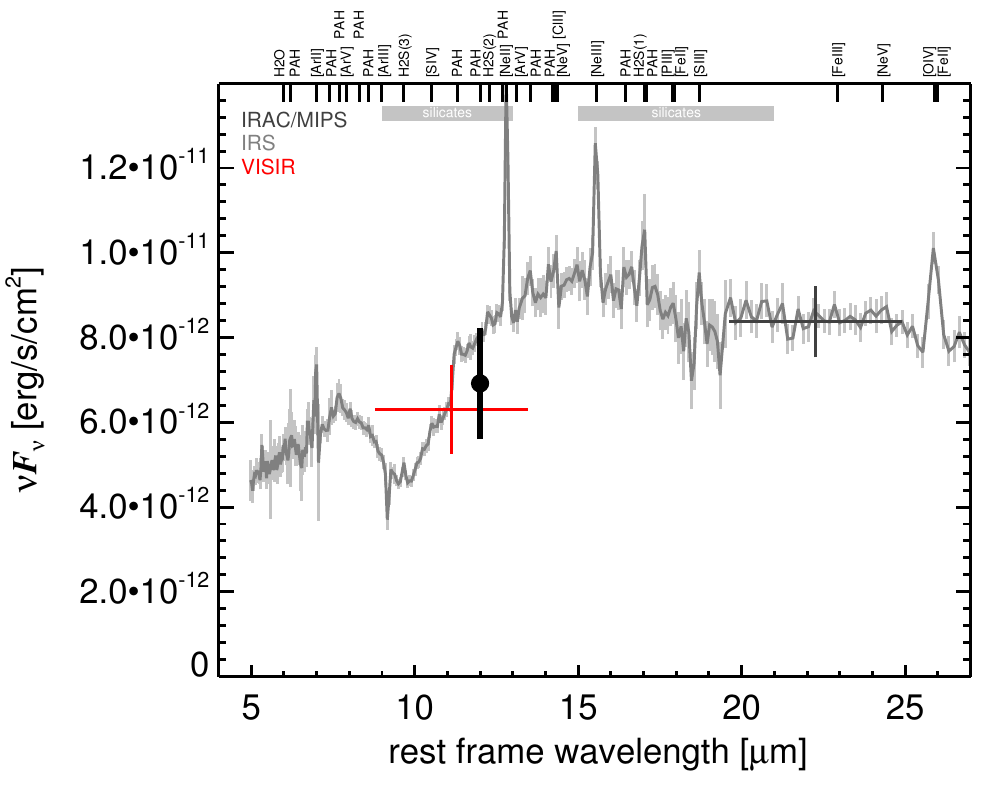}
   \caption{\label{fig:MISED_PKS1814-63}
      MIR SED of PKS\,1814-63. The description  of the symbols (if present) is the following.
      Grey crosses and  solid lines mark the \spitzer/IRAC, MIPS and IRS data. 
      The colour coding of the other symbols is: 
      green for COMICS, magenta for Michelle, blue for T-ReCS and red for VISIR data.
      Darker-coloured solid lines mark spectra of the corresponding instrument.
      The black filled circles mark the nuclear 12 and $18\,\mu$m  continuum emission estimate from the data.
      The ticks on the top axis mark positions of common MIR emission lines, while the light grey horizontal bars mark wavelength ranges affected by the silicate 10 and 18$\mu$m features.}
\end{figure}
\clearpage

\twocolumn[\begin{@twocolumnfalse}  
\subsection{PKS\,1932-46}\label{app:PKS1932-46}
PKS\,1932-46 is an FR\,II radio source in an elliptical galaxy at a redshift of $z=$ 0.2307 ($D\sim1191\,$Mpc) with a Sy\,1.9 nucleus \citep{veron-cetty_catalogue_2010}.
The galaxy has a disturbed optical morphology, presumably caused by a past merger or the close-by star-forming companion galaxy  \citep{villar-martin_giant_2005,inskip_pks1932-46:_2007}.
PKS\,1932-46 features a galactic-scale extended emission line region, which is just partly belonging to the AGN ionization cone, and extended double-lobe radio emission along a PA$\sim-72\degree$ \citep{villar-martin_pks_1998,villar-martin_giant_2005}.
The first MIR detection of this object was achieved with \spitzer/MIPS where it appears as a compact nucleus embedded within extended emission coinciding with the radio extension \citep{inskip_pks1932-46:_2007}.
Our nuclear MIPS 24\,$\mu$m photometry is consistent with the latter work.
The \spitzer/IRS LR staring-mode spectrum suffers from very low S/N but indicates PAH and possibly silicate emission and a flat spectral slope in $\nu F_\nu$-space.
The arcsecond-scale MIR SED might, thus, be affected by star-formation emission.
PKS\,1932-46 was observed with VISIR in the broad SIC filter in 2010 (unpublished, to our knowledge), but no emission source could be detected.
The  corresponding flux upper limit is much less stringent than the \spitzerr spectrophotometry, which is why we us the latter to derive an upper limit on any AGN-caused 12\,$\mu$m continuum emission.
\newline\end{@twocolumnfalse}]

\begin{figure}
   \centering
   \includegraphics[angle=0,width=8.500cm]{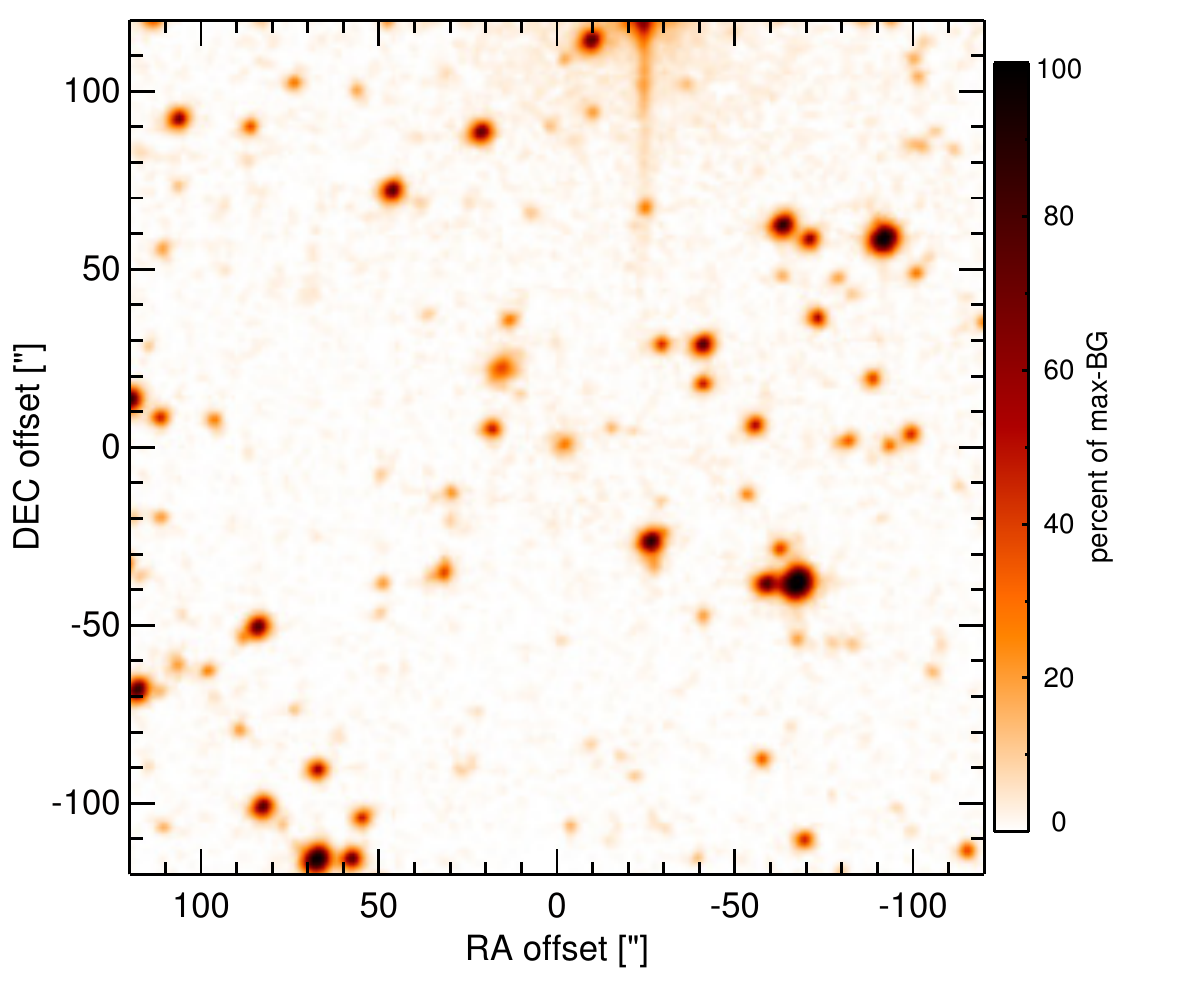}
    \caption{\label{fig:OPTim_PKS1932-46}
             Optical image (DSS, red filter) of PKS\,1932-46. Displayed are the central $4\arcmin$ with North up and East to the left. 
              The colour scaling is linear with white corresponding to the median background and black to the $0.01\%$ pixels with the highest intensity.  
           }
\end{figure}
\begin{figure}
   \centering
   \includegraphics[angle=0,height=3.11cm]{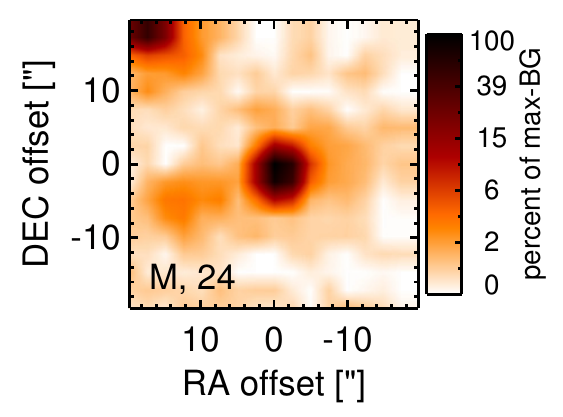}
    \caption{\label{fig:INTim_PKS1932-46}
             \spitzerr MIR images of PKS\,1932-46. Displayed are the inner $40\arcsec$ with North up and East to the left. The colour scaling is logarithmic with white corresponding to median background and black to the $0.1\%$ pixels with the highest intensity.
             The label in the bottom left states instrument and central wavelength of the filter in $\mu$m (I: IRAC, M: MIPS). 
           }
\end{figure}
\begin{figure}
   \centering
   \includegraphics[angle=0,width=8.50cm]{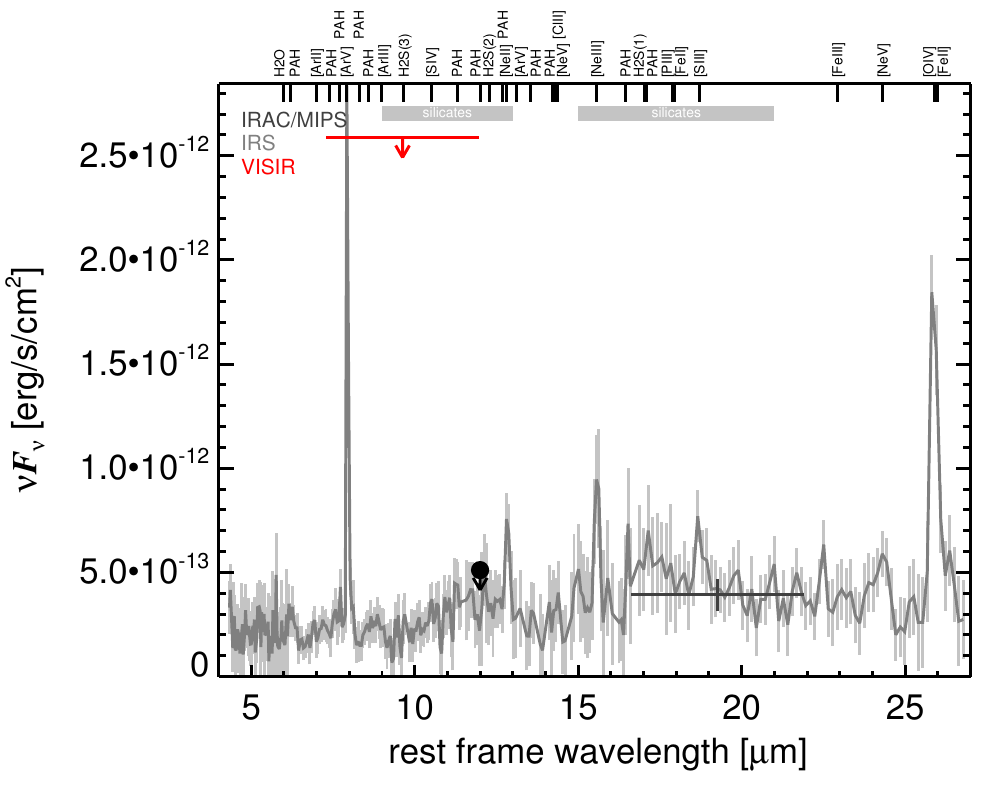}
   \caption{\label{fig:MISED_PKS1932-46}
      MIR SED of PKS\,1932-46. The description  of the symbols (if present) is the following.
      Grey crosses and  solid lines mark the \spitzer/IRAC, MIPS and IRS data. 
      The colour coding of the other symbols is: 
      green for COMICS, magenta for Michelle, blue for T-ReCS and red for VISIR data.
      Darker-coloured solid lines mark spectra of the corresponding instrument.
      The black filled circles mark the nuclear 12 and $18\,\mu$m  continuum emission estimate from the data.
      The ticks on the top axis mark positions of common MIR emission lines, while the light grey horizontal bars mark wavelength ranges affected by the silicate 10 and 18$\mu$m features.}
\end{figure}
\clearpage

\twocolumn[\begin{@twocolumnfalse}  
\subsection{PKS\,2158-380 -- MCG-6-48-13}\label{app:PKS2158-380}
PKS\,2158-380/MCG-6-48-13 is a radio-loud lenticular galaxy at a redshift of $z=$ 0.0334 ($D\sim$149\,Mpc) with a Sy\,2 nucleus \citep{veron-cetty_catalogue_2010} and was first studied in detail by \cite{fosbury_very_1982}.
HST observations revealed three compact but resolved sources in the nucleus instead of one central source (total extend $\sim 1\arcsec\sim0.7$\,kpc; PA$\sim 90\degree$; \citealt{boyce_faint_1996,zirbel_ultraviolet_1998}).
In addition, water maser emission was detected in this object \citep{kondratko_discovery_2006}.
No \spitzerr data are available for PKS\,2158-380, which was imaged with VISIR in the SIC filter in 2006 \citep{van_der_wolk_dust_2010}.
A compact MIR nucleus is weakly detected in the image.
The low S/N prevents a quantitative analyses of the source morphology but the latter seems different than that seen in HST, as only one source was detected. 
Our nuclear photometry is consistent with the value in \cite{van_der_wolk_dust_2010}.
\newline\end{@twocolumnfalse}]

\begin{figure}
   \centering
   \includegraphics[angle=0,width=8.500cm]{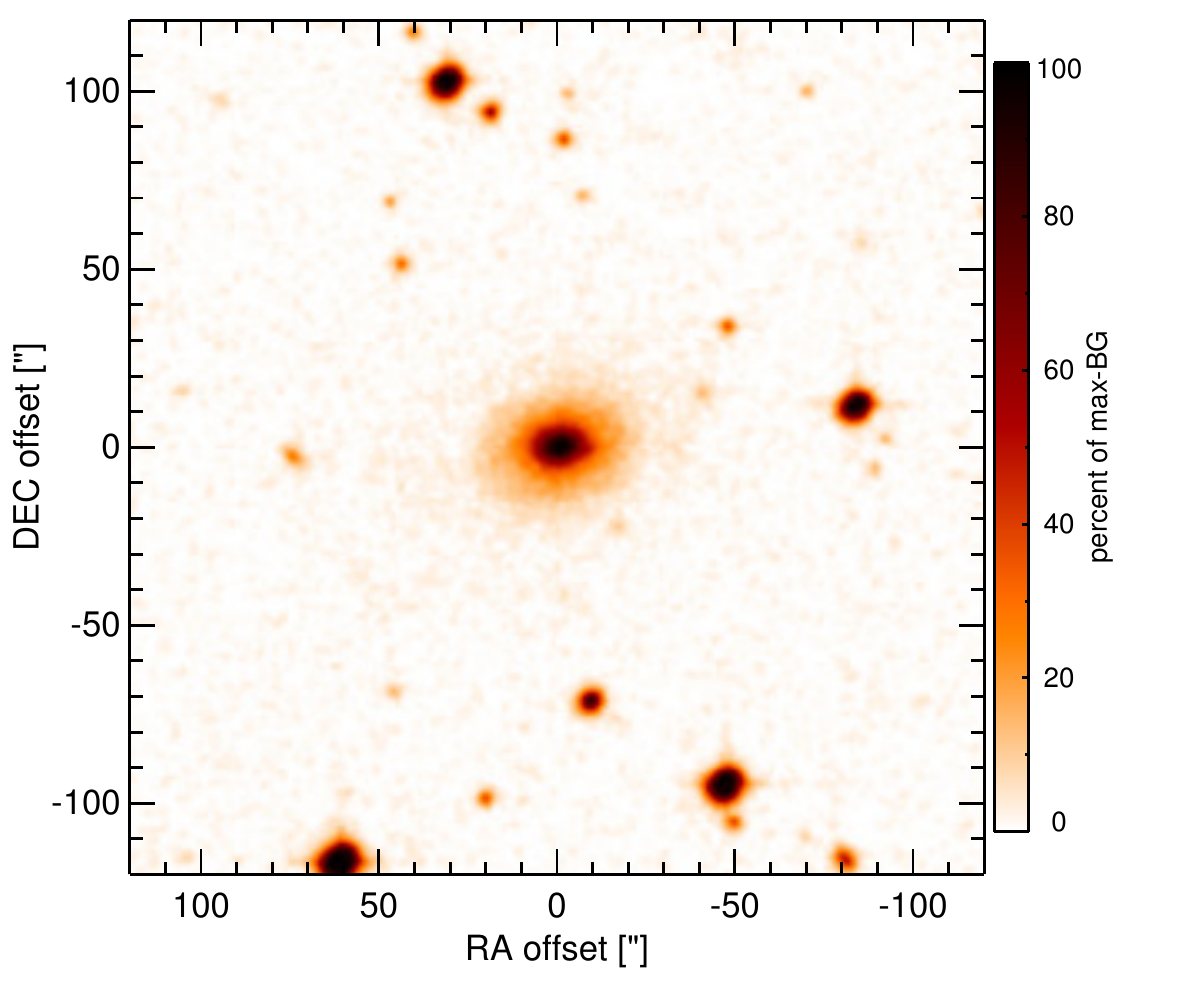}
    \caption{\label{fig:OPTim_PKS2158-380}
             Optical image (DSS, red filter) of PKS\,2158-380. Displayed are the central $4\arcmin$ with North up and East to the left. 
              The colour scaling is linear with white corresponding to the median background and black to the $0.01\%$ pixels with the highest intensity.  
           }
\end{figure}
\begin{figure}
   \centering
   \includegraphics[angle=0,height=3.11cm]{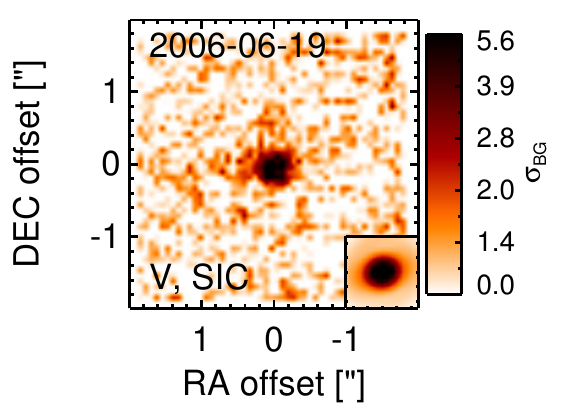}
    \caption{\label{fig:HARim_PKS2158-380}
             Subarcsecond-resolution MIR images of PKS\,2158-380 sorted by increasing filter wavelength. 
             Displayed are the inner $4\arcsec$ with North up and East to the left. 
             The colour scaling is logarithmic with white corresponding to median background and black to the $75\%$ of the highest intensity of all images in units of $\sigbg$.
             The inset image shows the central arcsecond of the PSF from the calibrator star, scaled to match the science target.
             The labels in the bottom left state instrument and filter names (C: COMICS, M: Michelle, T: T-ReCS, V: VISIR).
           }
\end{figure}
\begin{figure}
   \centering
   \includegraphics[angle=0,width=8.50cm]{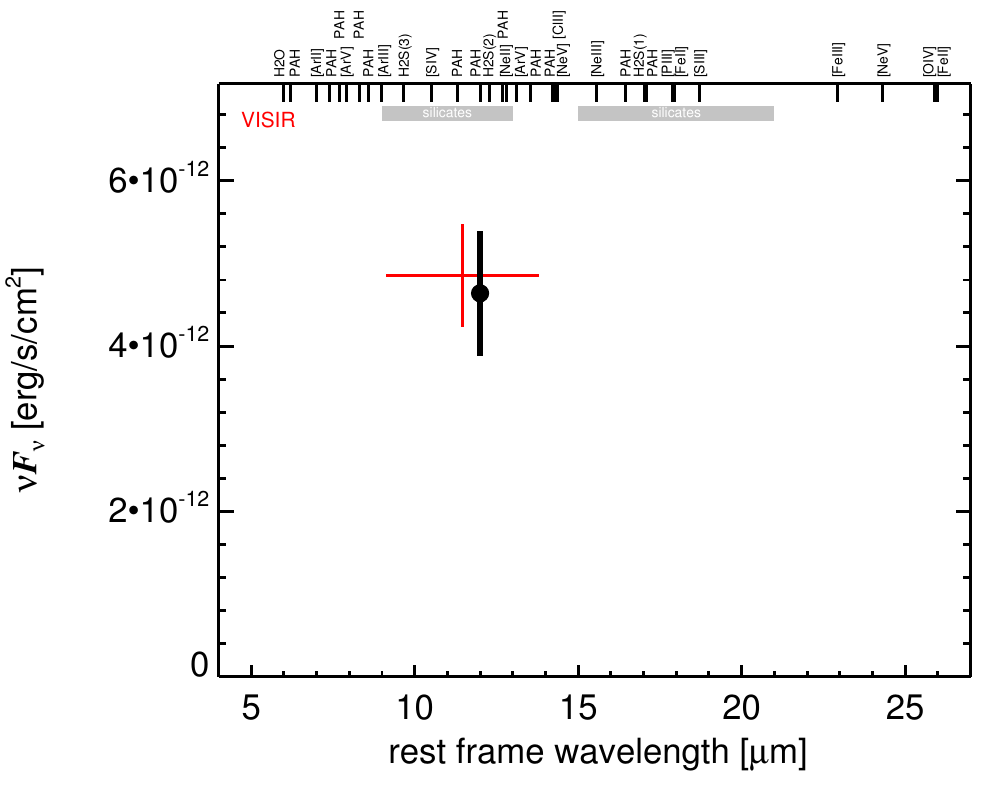}
   \caption{\label{fig:MISED_PKS2158-380}
      MIR SED of PKS\,2158-380. The description  of the symbols (if present) is the following.
      Grey crosses and  solid lines mark the \spitzer/IRAC, MIPS and IRS data. 
      The colour coding of the other symbols is: 
      green for COMICS, magenta for Michelle, blue for T-ReCS and red for VISIR data.
      Darker-coloured solid lines mark spectra of the corresponding instrument.
      The black filled circles mark the nuclear 12 and $18\,\mu$m  continuum emission estimate from the data.
      The ticks on the top axis mark positions of common MIR emission lines, while the light grey horizontal bars mark wavelength ranges affected by the silicate 10 and 18$\mu$m features.}
\end{figure}
\clearpage

\twocolumn[\begin{@twocolumnfalse}  
\subsection{PKS\,2354-35 -- Abell\,4059 -- ESO\,349-10}\label{app:PKS2354-35}
PKS\,2354-35 is an ultra-steep spectrum FR\,I radio source identified with the elliptical galaxy ESO\,349-10 at a redshift of $z=$ 0.0491 ($D\sim222$\,Mpc). It contains an optical LINER-type nucleus \citep{veron-cetty_catalogue_2010}. 
It was observed with \spitzer/IRAC and MIPS and appears as a compact nuclear source embedded within diffuse host emission in the IRAC $5.8$ and $8.0\,\mu$m images, while only a weak point source was detected in the MIPS $24\,\mu$m image.
The latter image is affected by artefacts, which appear as extended boxy emission south-west of the nucleus.
The \spitzerr images have also been analysed in \cite{quillen_infrared_2008} who find significantly higher fluxes than we do for the nuclear point source. 
The cause for this discrepancy is probably related to the intrinsic faintness of the object and the different flux measurement methods.
One narrow $N$-band filter image of PKS\,2354-35 was taken in 2006 with VISIR \citep{van_der_wolk_dust_2010}. 
As expected from the milli-Jansky fluxes found by \spitzer, the source has not been detected.
Therefore, we use the \spitzerr photometry to derive an upper limit for the $12\,\mu$m continuum flux, which is $\sim3$ times lower than our upper limit obtained from the VISIR photometry.
\newline\end{@twocolumnfalse}]

\begin{figure}
   \centering
   \includegraphics[angle=0,width=8.500cm]{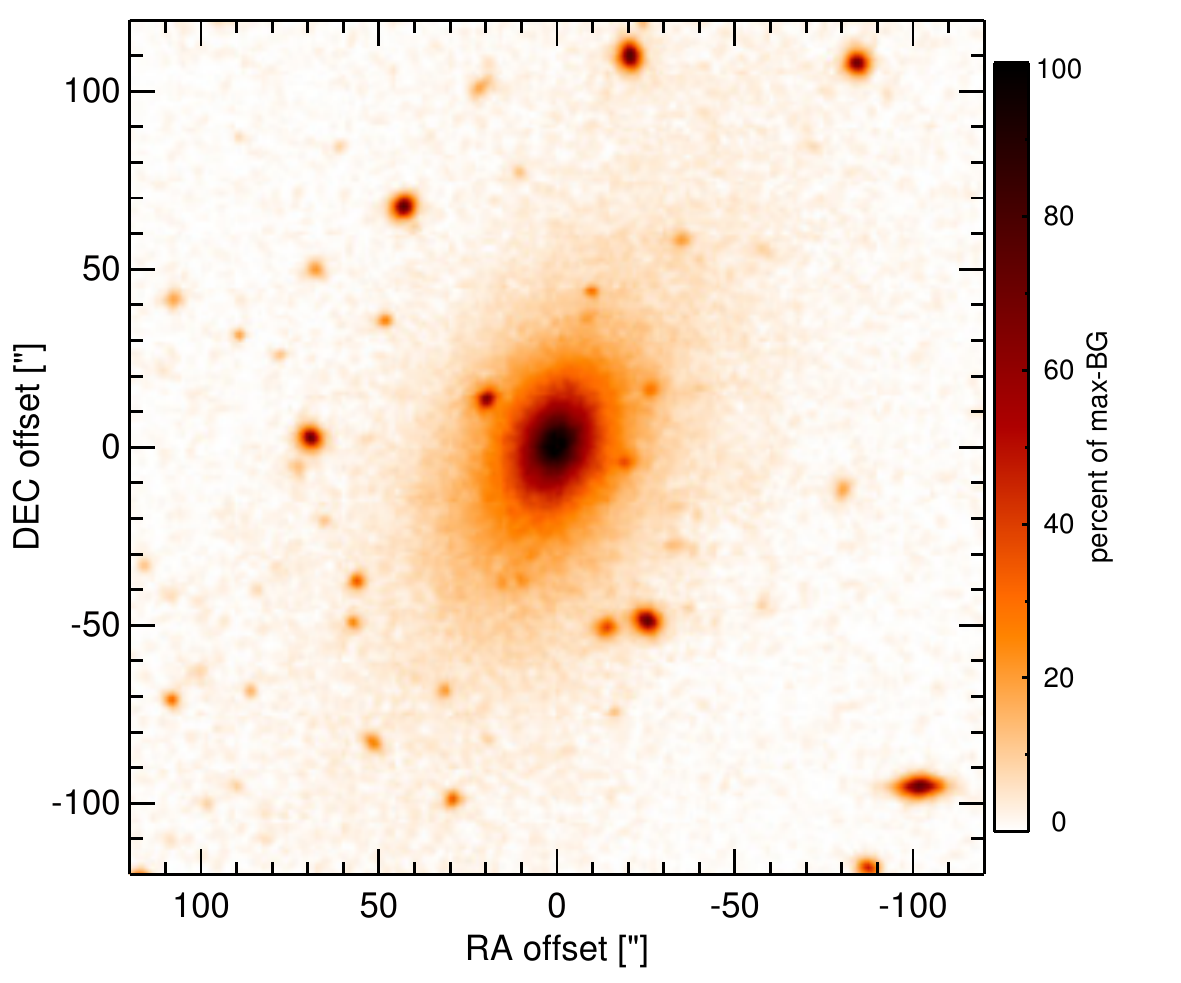}
    \caption{\label{fig:OPTim_PKS2354-35}
             Optical image (DSS, red filter) of PKS\,2354-35. Displayed are the central $4\arcmin$ with North up and East to the left. 
              The colour scaling is linear with white corresponding to the median background and black to the $0.01\%$ pixels with the highest intensity.  
           }
\end{figure}
\begin{figure}
   \centering
   \includegraphics[angle=0,height=3.11cm]{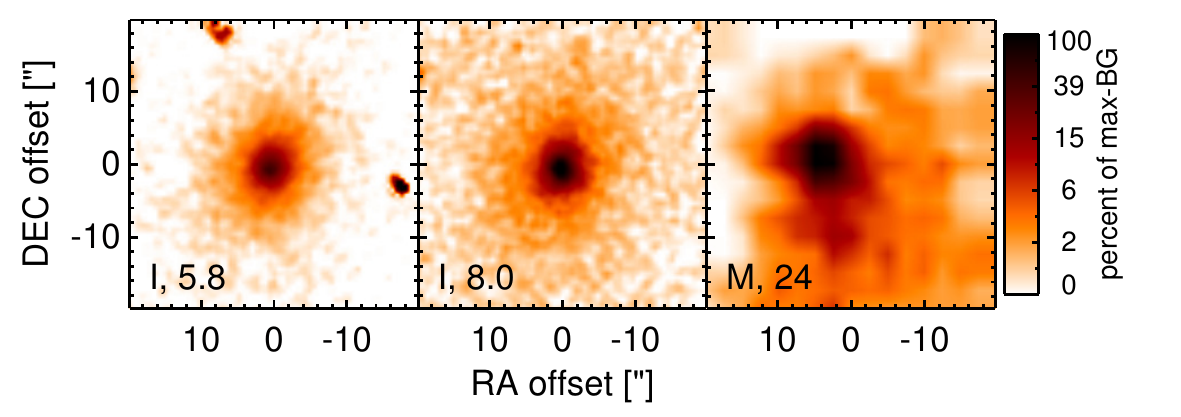}
    \caption{\label{fig:INTim_PKS2354-35}
             \spitzerr MIR images of PKS\,2354-35. Displayed are the inner $40\arcsec$ with North up and East to the left. The colour scaling is logarithmic with white corresponding to median background and black to the $0.1\%$ pixels with the highest intensity.
             The label in the bottom left states instrument and central wavelength of the filter in $\mu$m (I: IRAC, M: MIPS). 
           }
\end{figure}
\begin{figure}
   \centering
   \includegraphics[angle=0,width=8.50cm]{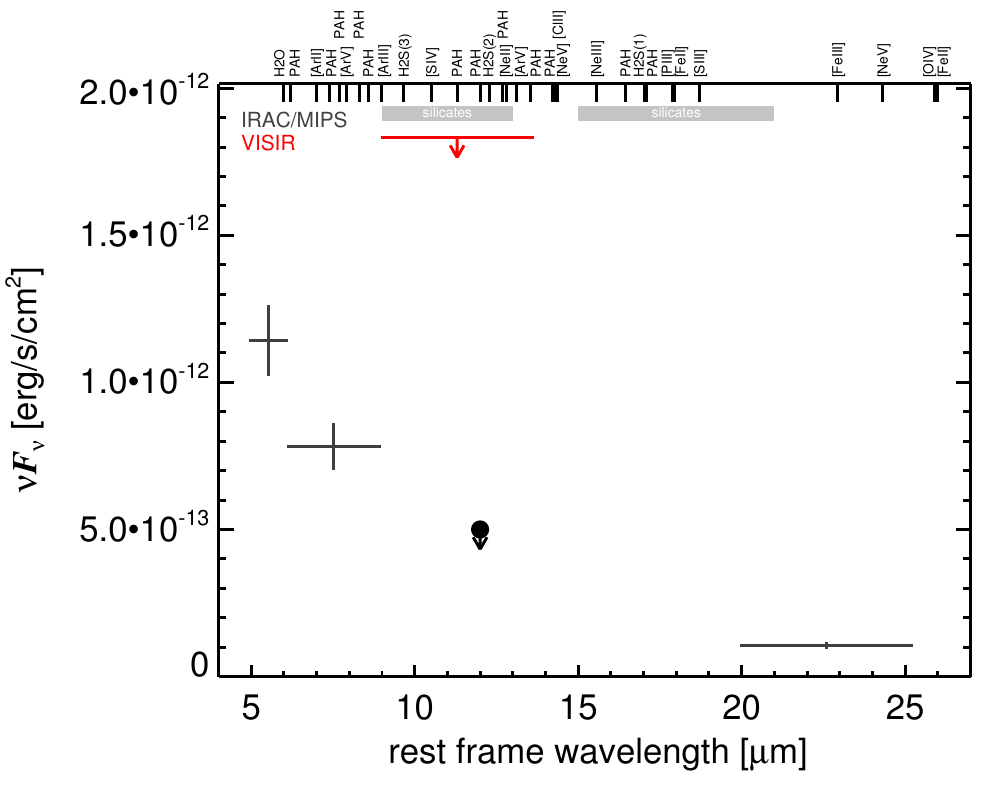}
   \caption{\label{fig:MISED_PKS2354-35}
      MIR SED of PKS\,2354-35. The description  of the symbols (if present) is the following.
      Grey crosses and  solid lines mark the \spitzer/IRAC, MIPS and IRS data. 
      The colour coding of the other symbols is: 
      green for COMICS, magenta for Michelle, blue for T-ReCS and red for VISIR data.
      Darker-coloured solid lines mark spectra of the corresponding instrument.
      The black filled circles mark the nuclear 12 and $18\,\mu$m  continuum emission estimate from the data.
      The ticks on the top axis mark positions of common MIR emission lines, while the light grey horizontal bars mark wavelength ranges affected by the silicate 10 and 18$\mu$m features.}
\end{figure}
\clearpage

\twocolumn[\begin{@twocolumnfalse}  
\subsection{Superantennae -- IRAS\,19254-7245 / AM\,1925-724}\label{app:SuperantennaeS}
The Superantennae is a ultra-luminous infrared system of a pair of merging galaxies at a redshift of $z=$ 0.0617 ($D\sim287\,$Mpc) with a nuclear separation of $\sim8\arcsec\sim10\,$kpc in the north-south direction (PA$\sim-10\degree$; \citealt{melnick_ntt_1990}).
The southern component, Superantennae\,S (AM\,1925-724 NED02), contains a Sy\,2 nucleus \citep{veron-cetty_catalogue_2010}, while the northern exhibits apparently pure star formation \citep{mirabel_superantennae_1991,colina_massive_1991,vanzi_multi-wavelength_2002,berta_spatially-resolved_2003}.
After first being detected in the MIR with \iras, the Superantennae were followed up with ground-based $N$-band photometry \citep{mirabel_superantennae_1991} and space-based MIR observations with \isoo \citep{rigopoulou_large_1999,lutz_mid-infrared_1999,tran_isocam-cvf_2001,klaas_infrared_2001,charmandaris_mid-infrared_2002} and \spitzer/IRAC, IRS and MIPS. 
The corresponding IRAC and MIPS images show a compact southern nucleus dominating the MIR emission completely, while the northern nucleus is only faintly detected as a compact source.
No significant extended emission is visible.
The IRS LR staring-mode spectrum of Superantennae\,S exhibits silicate 10\,$\mu$m absorption, significant PAH emission, and a steep red spectral slope in $\nu F_\nu$-space (see also \citealt{wu_spitzer/irs_2009}).
Thus, the arcsecond-scale MIR SED appears to be significantly affected by star formation.
The Superantennae were imaged with T-ReCS in the broad N filter in 2004 (unpublished, to our knowledge) and with the broad Qa filter in 2008 and 2009 \citep{imanishi_subaru_2011}.
In both cases, the southern nucleus was detected as a compact source, while the northern nucleus is faintly detected only in the N image as a compact source (flux: $6 \pm 1$\,mJy; FWHM(major axis)$\sim0,47\arcsec$; PA$\sim-18\degree$).
The southern nucleus appears marginally resolved in the sharpest image (N; FWHM$\sim0.46\arcsec\sim0.6\,$kpc) but not in the Qa images.
Therefore, at least a second epoch of subarcsecond-resolution $N$-band imaging is required to verify whether Superantennae\,S is truly extended in the $N$-band at subarcsecond resolution.
Our nuclear photometry is consistent with \cite{imanishi_subaru_2011} and the \spitzerr spectrophotometry.
Therefore, we use the latter to compute the nuclear 12\,$\mu$m continuum emission estimate, which however is presumably still star-formation contaminated.
\newline\end{@twocolumnfalse}]

\begin{figure}
   \centering
   \includegraphics[angle=0,width=8.500cm]{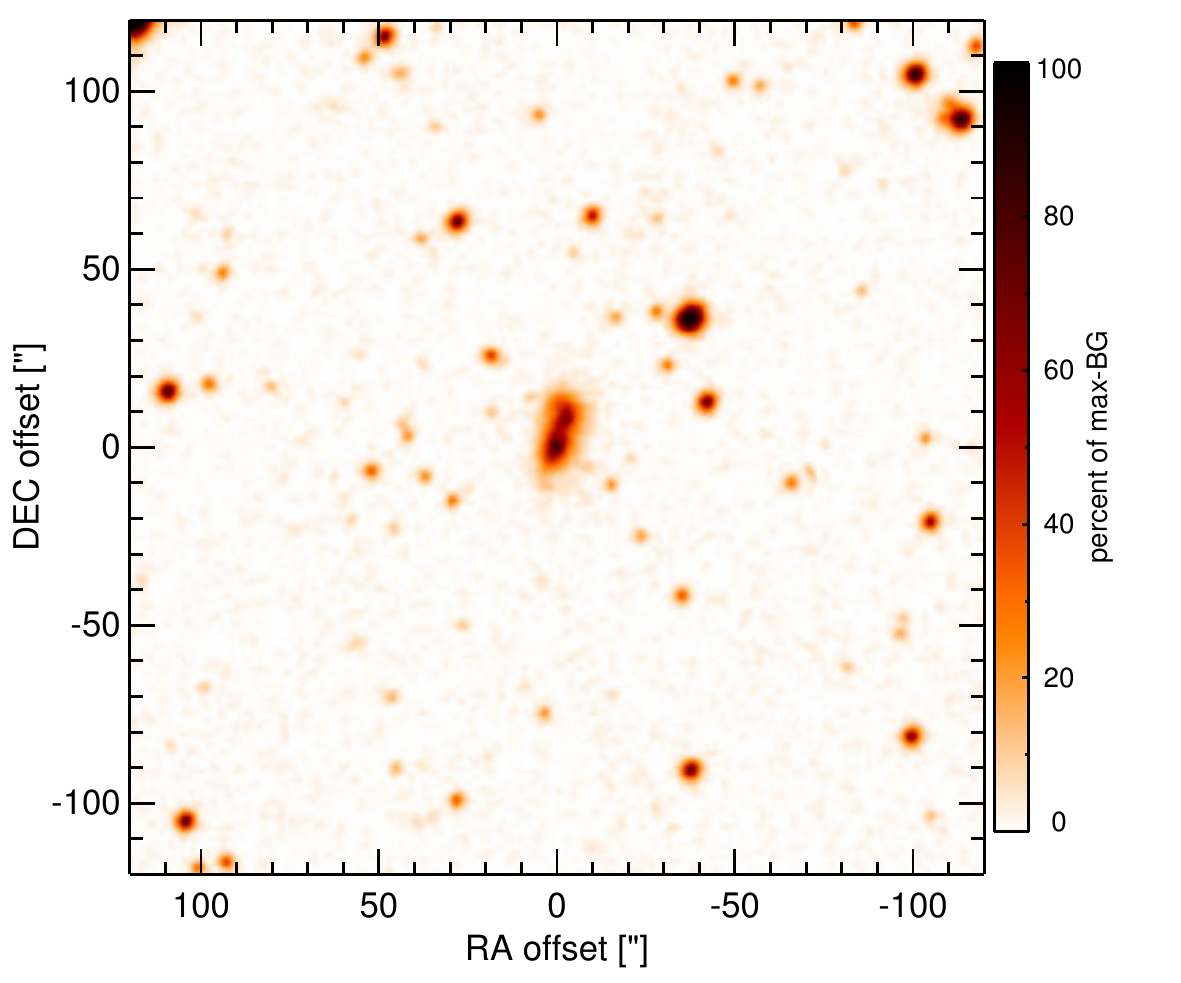}
    \caption{\label{fig:OPTim_SuperantennaeS}
             Optical image (DSS, red filter) of Superantennae\,S. Displayed are the central $4\arcmin$ with North up and East to the left. 
              The colour scaling is linear with white corresponding to the median background and black to the $0.01\%$ pixels with the highest intensity.  
           }
\end{figure}
\begin{figure}
   \centering
   \includegraphics[angle=0,height=3.11cm]{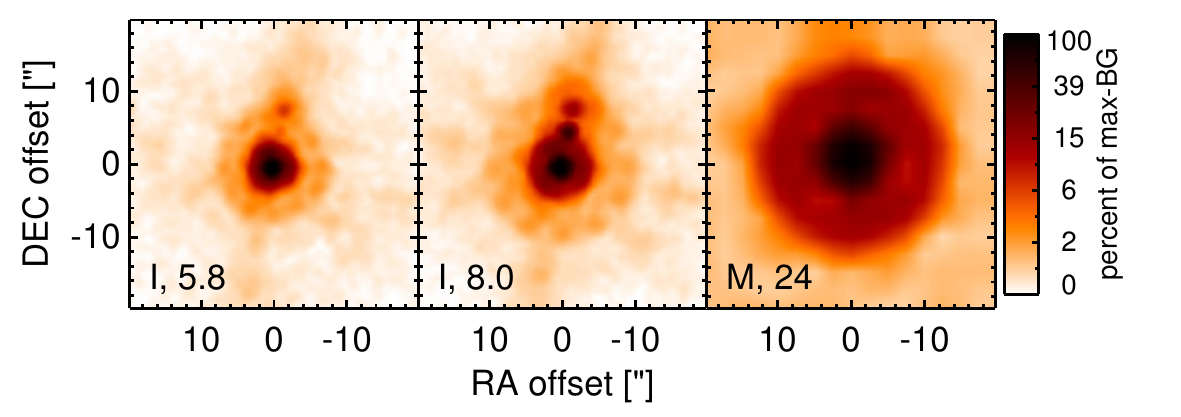}
    \caption{\label{fig:INTim_SuperantennaeS}
             \spitzerr MIR images of Superantennae\,S. Displayed are the inner $40\arcsec$ with North up and East to the left. The colour scaling is logarithmic with white corresponding to median background and black to the $0.1\%$ pixels with the highest intensity.
             The label in the bottom left states instrument and central wavelength of the filter in $\mu$m (I: IRAC, M: MIPS).
             Note that the apparent off-nuclear compact sources in the IRAC 5.8 and  $8.0\,\mu$m images are instrumental artefacts.
           }
\end{figure}
\begin{figure}
   \centering
   \includegraphics[angle=0,width=8.500cm]{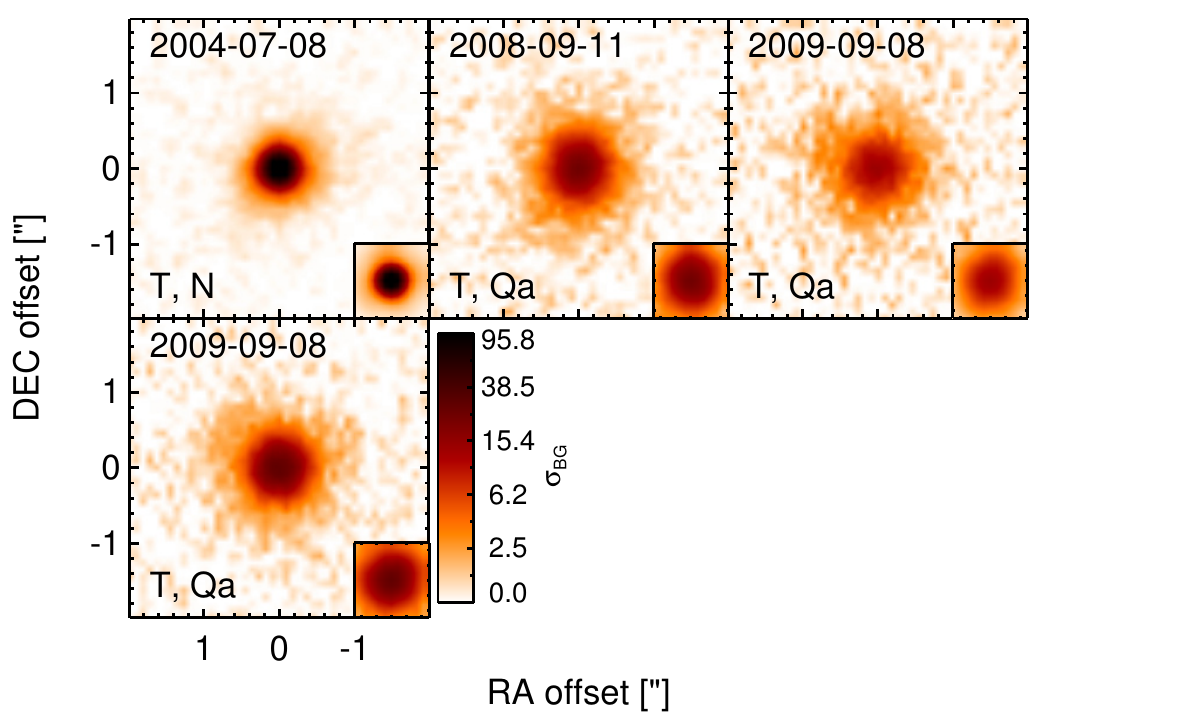}
    \caption{\label{fig:HARim_SuperantennaeS}
             Subarcsecond-resolution MIR images of Superantennae\,S sorted by increasing filter wavelength. 
             Displayed are the inner $4\arcsec$ with North up and East to the left. 
             The colour scaling is logarithmic with white corresponding to median background and black to the $75\%$ of the highest intensity of all images in units of $\sigbg$.
             The inset image shows the central arcsecond of the PSF from the calibrator star, scaled to match the science target.
             The labels in the bottom left state instrument and filter names (C: COMICS, M: Michelle, T: T-ReCS, V: VISIR).
           }
\end{figure}
\begin{figure}
   \centering
   \includegraphics[angle=0,width=8.50cm]{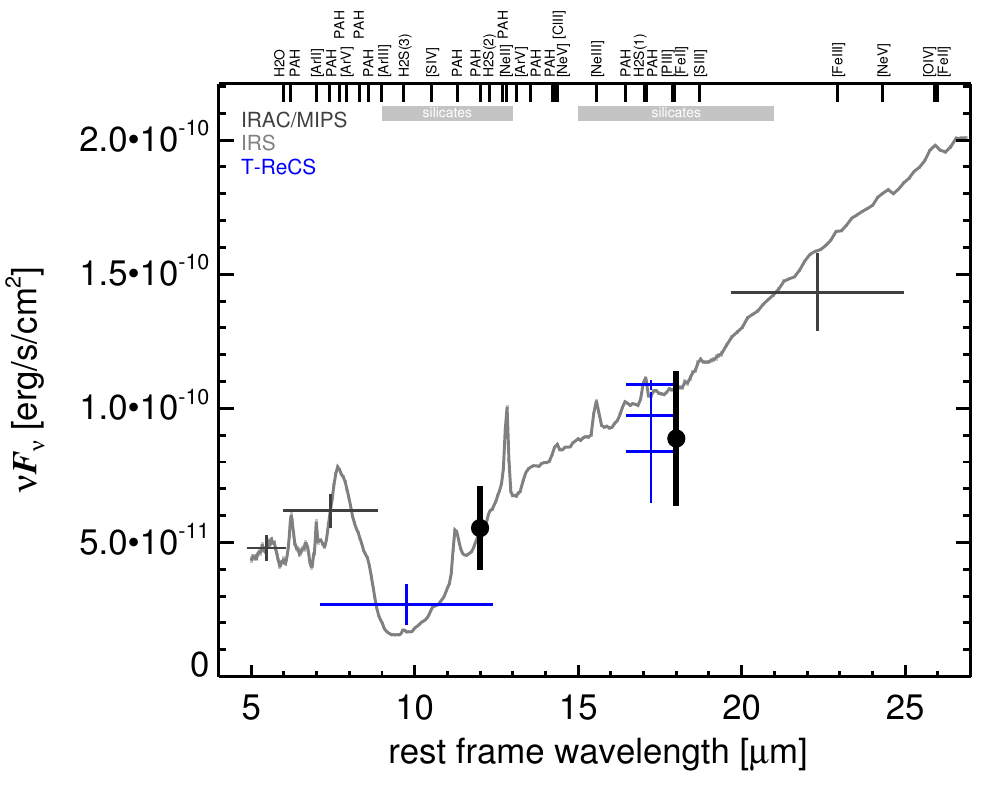}
   \caption{\label{fig:MISED_SuperantennaeS}
      MIR SED of Superantennae\,S. The description  of the symbols (if present) is the following.
      Grey crosses and  solid lines mark the \spitzer/IRAC, MIPS and IRS data. 
      The colour coding of the other symbols is: 
      green for COMICS, magenta for Michelle, blue for T-ReCS and red for VISIR data.
      Darker-coloured solid lines mark spectra of the corresponding instrument.
      The black filled circles mark the nuclear 12 and $18\,\mu$m  continuum emission estimate from the data.
      The ticks on the top axis mark positions of common MIR emission lines, while the light grey horizontal bars mark wavelength ranges affected by the silicate 10 and 18$\mu$m features.}
\end{figure}
\clearpage

\twocolumn[\begin{@twocolumnfalse}  
\subsection{UGC\,5101 -- IRAS\,09320+6134}\label{app:UGC05101}
UGC\,5101 is an infrared-ultra-luminous disturbed highly-inclined spiral galaxy \citep{sanders_ultraluminous_1988} at a redshift of $z=$ 0.0394 ($D\sim182\,$Mpc) with an active nucleus containing both a starburst and an AGN.
This lead to various optical classifications in the literature ranging from Sy\,1 to Sy\,2 or LINER, while we follow \cite{goncalves_agns_1999} and treat the nucleus as AGN/starburst composite.
The $L$-band and X-ray properties provide evidence for a highly obscured AGN in UGC\,5101 \citep{imanishi_strong_2001,imanishi_x-ray_2003}.
A compact core with elliptical extended emission $\sim2\arcsec\sim1.6\,$kpc in the east-west direction was detected in subarcsecond-resolution radio observations (PA$\sim90\degree$; \citealt{condon_compact_1991,crawford_vla_1996}).
Furthermore OH mega-maser emission was detected in UGC\,5101 \citep{martin_discovery_1989}.
After first being detected in the MIR with \iras, UGC\,5101 was followed up with ground-based MIR observations \citep{carico_iras_1988,sanders_ultraluminous_1988,maiolino_new_1995,gorjian_10_2004}.
The first subarcsecond-resolution MIR images were presented by
\cite{soifer_high_2000} obtained with Keck/LWS and MIRLIN in 1998, and show a compact nucleus embedded within elliptical $\sim2\arcsec\sim1.6\,$kpc extended emission along PA$\sim80\degree$.
This is cospatial with the radio extension.
UGC\,5101 was also observed in the MIR from space with \isoo \citep{rigopoulou_large_1999,lutz_mid-infrared_1999,klaas_infrared_2001,tran_isocam-cvf_2001,spoon_ice_2002} and \spitzer/IRAC, IRS and MIPS.
The corresponding IRAC 5.8 and 8.0\,$\mu$m images show an elongated nucleus in the same direction (FWHM(major axis)$\sim4\arcsec\sim3.3\,$kpc; PA$\sim80\degree$), while the nucleus is nearly unresolved in the MIPS 24\,$\mu$m image. 
The IRS LR staring-mode spectrum exhibits deep silicate 10 and 18\,$\mu$m absorption, strong PAH emission, and a steep red spectral slope in in $\nu F_\nu$-space (see also \citealt{armus_observations_2004,armus_observations_2007,dartois_carbonaceous_2007,farrah_high-resolution_2007,wu_spitzer/irs_2009}).
Thus, the arcsecond-scale MIR SED indicates both a buried AGN and intense star formation.
UGC\,5101 was observed with COMICS in the N8.8 and Q17.7 filters in 2009 (unpublished, to our knowledge), and a compact nucleus was weakly detected in both cases.
The nucleus appears to be similarly elongated in N8.8 image as in the earlier Keck images of \cite{soifer_high_2000}. 
Therefore, we classify the nucleus of UGC\,5101 as extended in the MIR at subarcsecond resolution.
The nuclear MIR photometry is on average $\sim 15\%$ lower than the \spitzerr spectrophotometry. 
However, it is likely that the former is still significantly affected by star formation.
\newline\end{@twocolumnfalse}]

\begin{figure}
   \centering
   \includegraphics[angle=0,width=8.500cm]{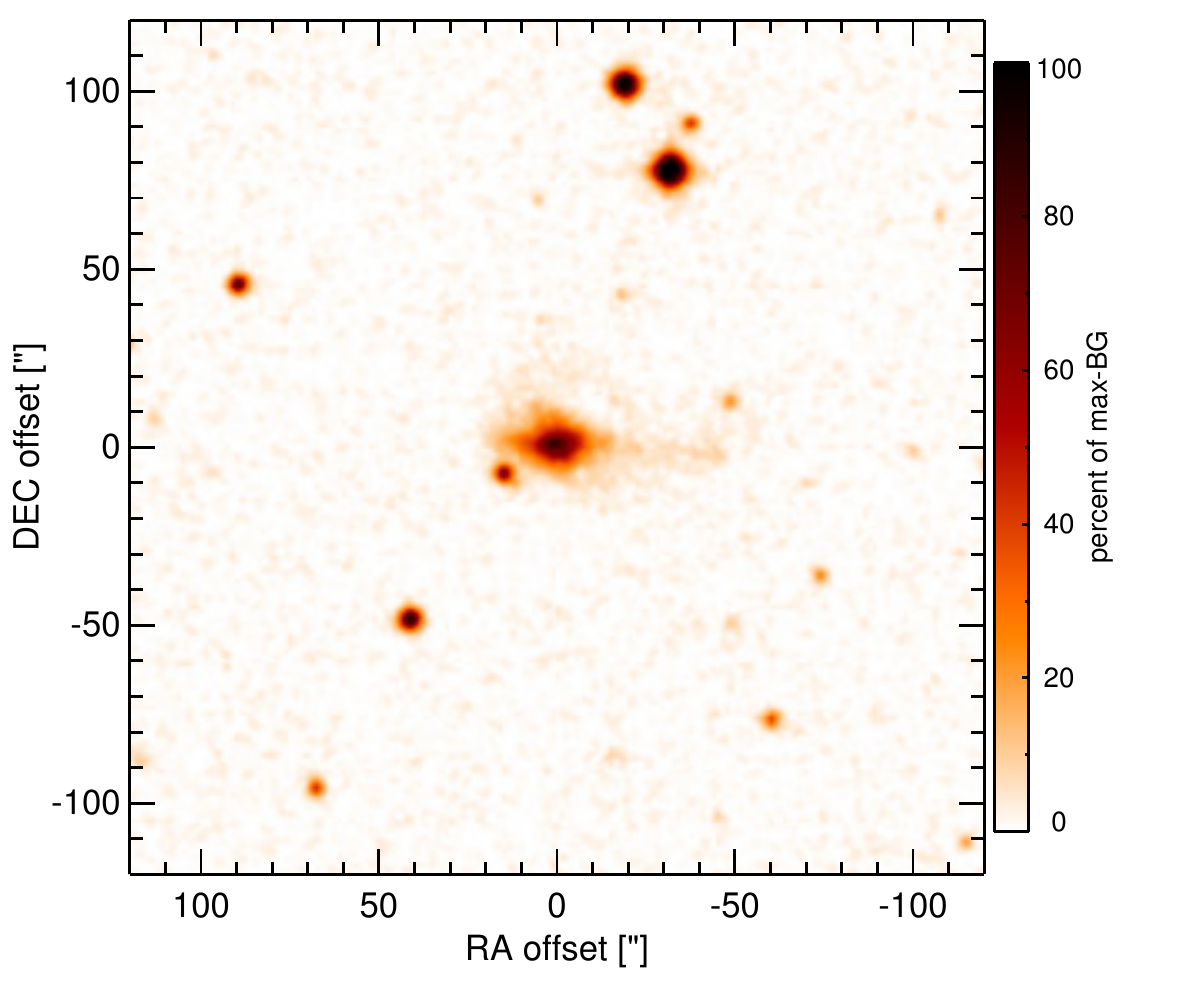}
    \caption{\label{fig:OPTim_UGC05101}
             Optical image (DSS, red filter) of UGC\,5101. Displayed are the central $4\arcmin$ with North up and East to the left. 
              The colour scaling is linear with white corresponding to the median background and black to the $0.01\%$ pixels with the highest intensity.  
           }
\end{figure}
\begin{figure}
   \centering
   \includegraphics[angle=0,height=3.11cm]{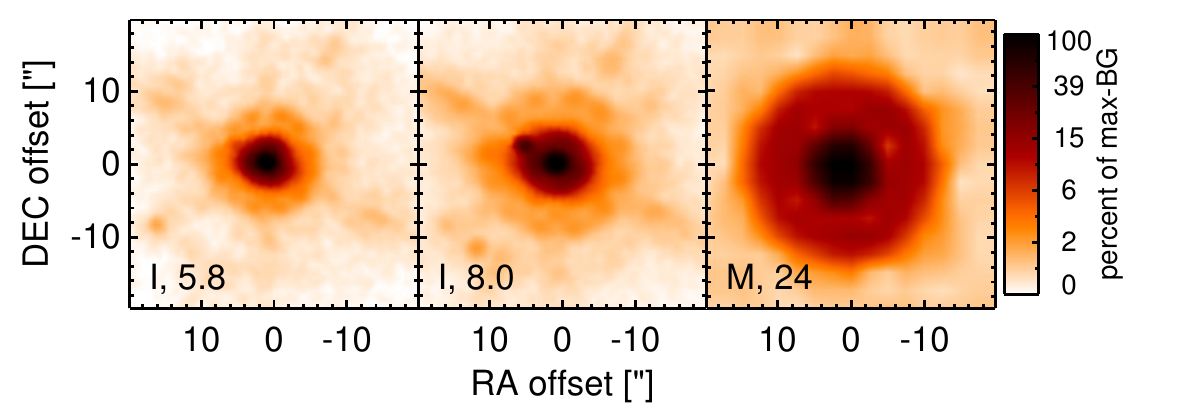}
    \caption{\label{fig:INTim_UGC05101}
             \spitzerr MIR images of UGC\,5101. Displayed are the inner $40\arcsec$ with North up and East to the left. The colour scaling is logarithmic with white corresponding to median background and black to the $0.1\%$ pixels with the highest intensity.
             The label in the bottom left states instrument and central wavelength of the filter in $\mu$m (I: IRAC, M: MIPS). 
             Note that the apparent off-nuclear compact source in the IRAC $8.0\,\mu$m image is an instrumental artefact.
           }
\end{figure}
\begin{figure}
   \centering
   \includegraphics[angle=0,height=3.11cm]{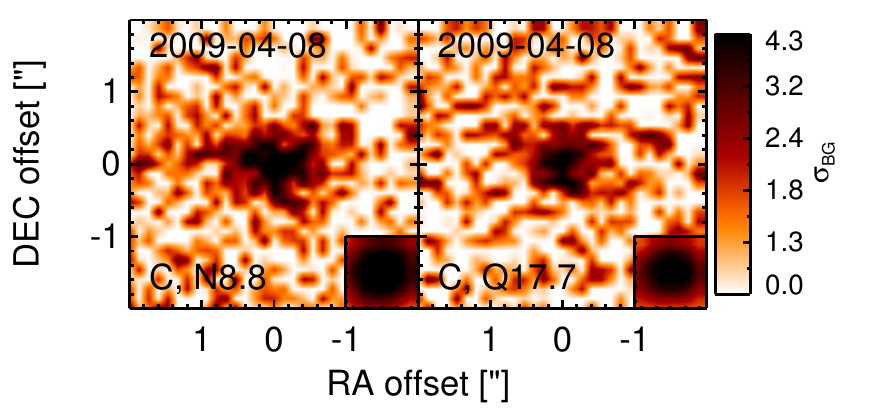}
    \caption{\label{fig:HARim_UGC05101}
             Subarcsecond-resolution MIR images of UGC\,5101 sorted by increasing filter wavelength. 
             Displayed are the inner $4\arcsec$ with North up and East to the left. 
             The colour scaling is logarithmic with white corresponding to median background and black to the $75\%$ of the highest intensity of all images in units of $\sigbg$.
             The inset image shows the central arcsecond of the PSF from the calibrator star, scaled to match the science target.
             The labels in the bottom left state instrument and filter names (C: COMICS, M: Michelle, T: T-ReCS, V: VISIR).
           }
\end{figure}
\begin{figure}
   \centering
   \includegraphics[angle=0,width=8.50cm]{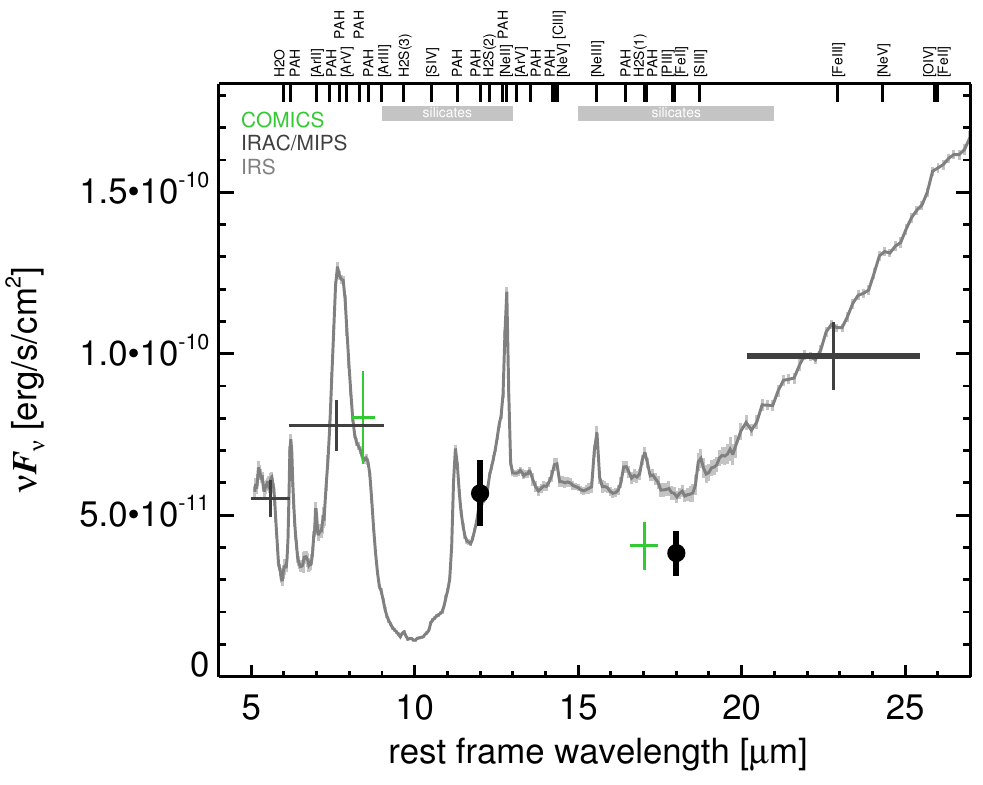}
   \caption{\label{fig:MISED_UGC05101}
      MIR SED of UGC\,5101. The description  of the symbols (if present) is the following.
      Grey crosses and  solid lines mark the \spitzer/IRAC, MIPS and IRS data. 
      The colour coding of the other symbols is: 
      green for COMICS, magenta for Michelle, blue for T-ReCS and red for VISIR data.
      Darker-coloured solid lines mark spectra of the corresponding instrument.
      The black filled circles mark the nuclear 12 and $18\,\mu$m  continuum emission estimate from the data.
      The ticks on the top axis mark positions of common MIR emission lines, while the light grey horizontal bars mark wavelength ranges affected by the silicate 10 and 18$\mu$m features.}
\end{figure}
\clearpage

\twocolumn[\begin{@twocolumnfalse}  
\subsection{UGC\,12348}\label{app:UGC12348}
UGC\,12348 is a highly-inclined spiral galaxy at a redshift of $z=$ 0.0255 ($D\sim110\,$Mpc) with a Sy\,2 nucleus \citep{veron-cetty_catalogue_2010}.
It features an unresolved radio nucleus \citep{schmitt_jet_2001} and east-west elongated, extended \oiii emission (diameter$\sim2.2\arcsec\sim1.1\,$kpc; PA$\sim100\degree$; \citealt{schmitt_hubble_2003}).
UGC\,12348 was detected with \irass but has not been followed up with \isoo or \spitzer.
The \wisee images show a bright nucleus embedded in disc-like host emission.
Furthermore, this object was imaged with T-ReCS in the Qa filter in 2007  (unpublished, to our knowledge), and a possibly east-west elongated nucleus without further host emission was detected (FWHM(major axis)$\sim0.61\arcsec\sim310\,$pc; PA$\sim70\degree$).
However, at least another epoch of subarcsecond MIR imaging is required to confirm this extension.
\newline\end{@twocolumnfalse}]

\begin{figure}
   \centering
   \includegraphics[angle=0,width=8.500cm]{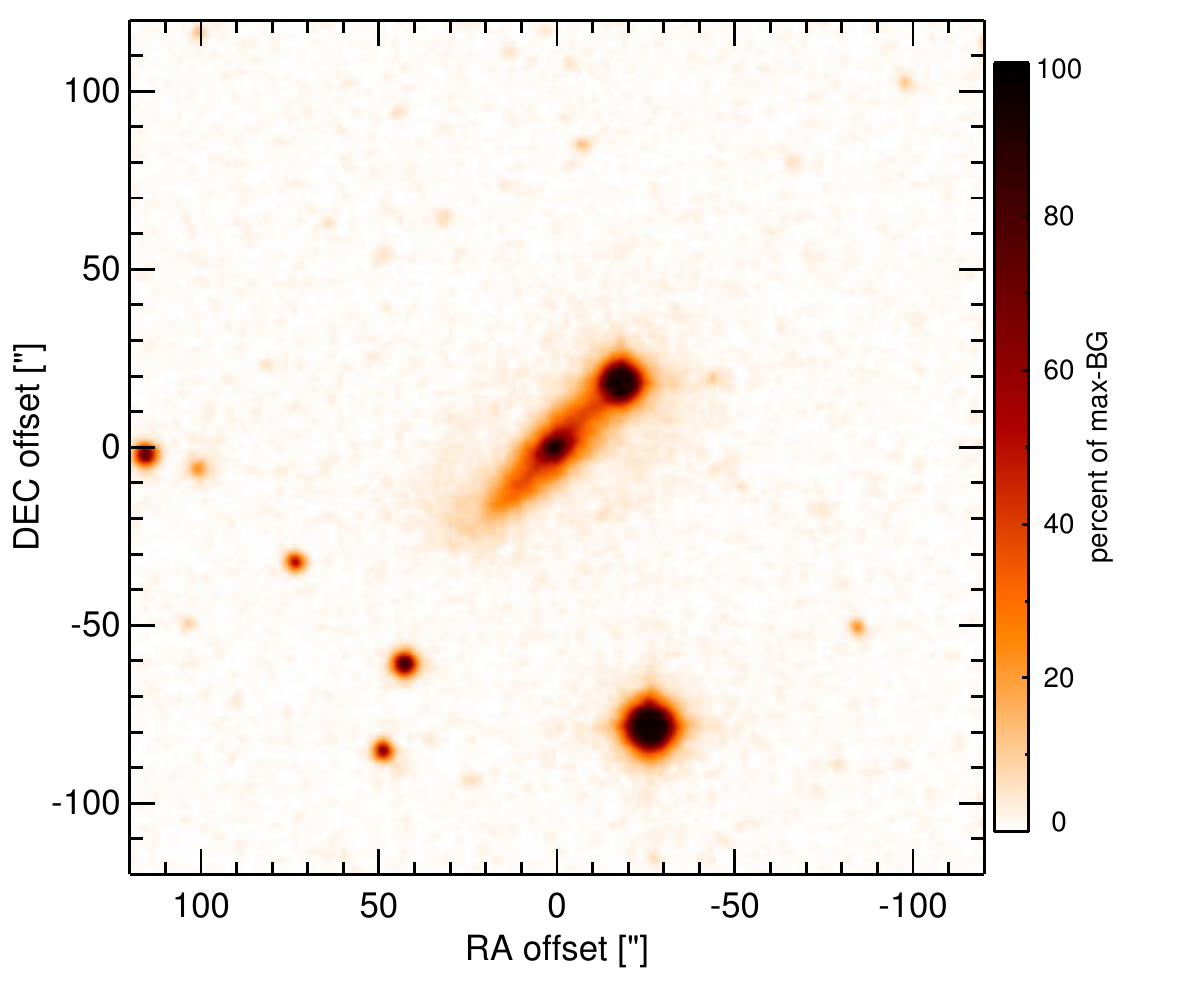}
    \caption{\label{fig:OPTim_UGC12348}
             Optical image (DSS, red filter) of UGC\,12348. Displayed are the central $4\arcmin$ with North up and East to the left. 
              The colour scaling is linear with white corresponding to the median background and black to the $0.01\%$ pixels with the highest intensity.  
           }
\end{figure}
\begin{figure}
   \centering
   \includegraphics[angle=0,height=3.11cm]{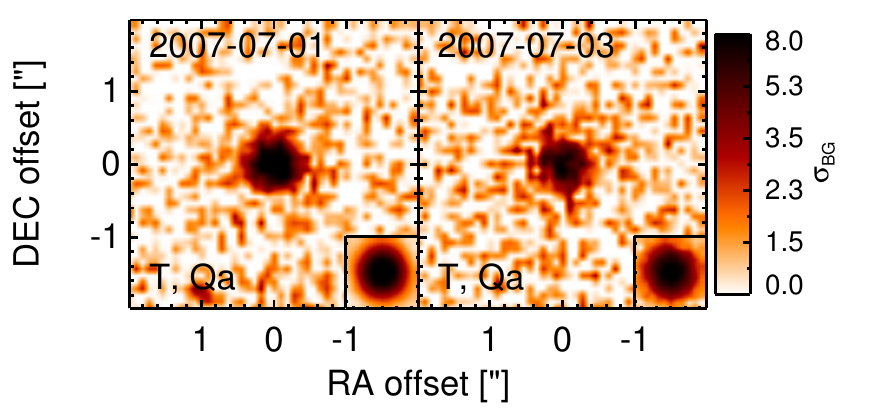}
    \caption{\label{fig:HARim_UGC12348}
             Subarcsecond-resolution MIR images of UGC\,12348 sorted by increasing filter wavelength. 
             Displayed are the inner $4\arcsec$ with North up and East to the left. 
             The colour scaling is logarithmic with white corresponding to median background and black to the $75\%$ of the highest intensity of all images in units of $\sigbg$.
             The inset image shows the central arcsecond of the PSF from the calibrator star, scaled to match the science target.
             The labels in the bottom left state instrument and filter names (C: COMICS, M: Michelle, T: T-ReCS, V: VISIR).
           }
\end{figure}
\begin{figure}
   \centering
   \includegraphics[angle=0,width=8.50cm]{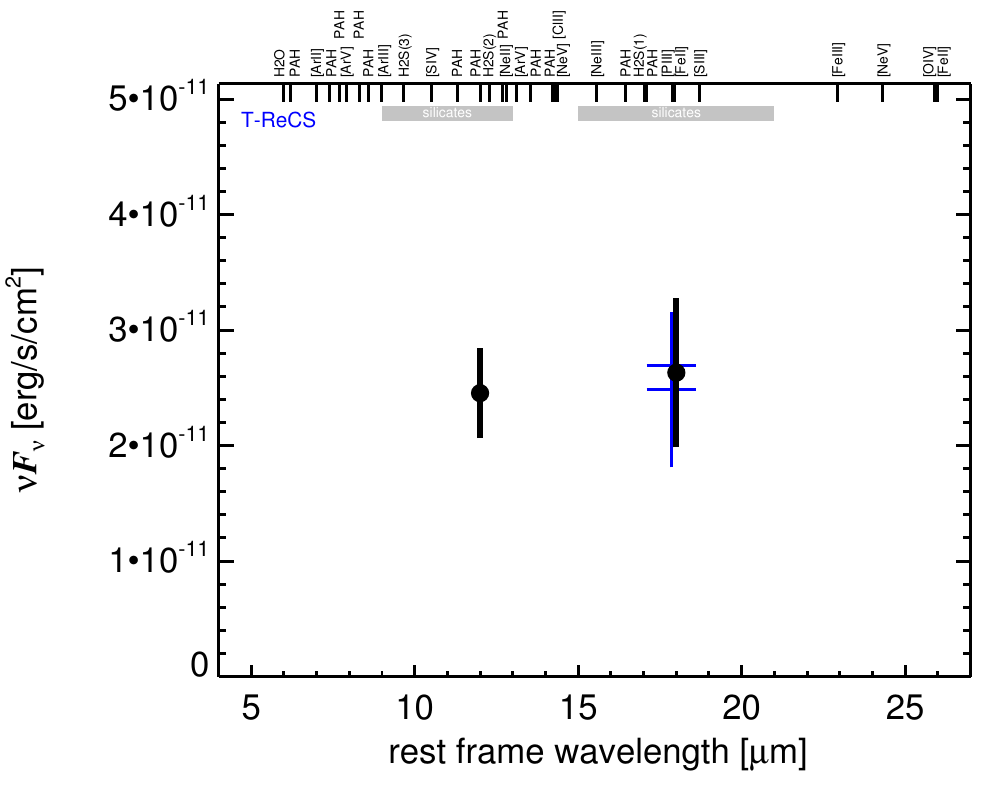}
   \caption{\label{fig:MISED_UGC12348}
      MIR SED of UGC\,12348. The description  of the symbols (if present) is the following.
      Grey crosses and  solid lines mark the \spitzer/IRAC, MIPS and IRS data. 
      The colour coding of the other symbols is: 
      green for COMICS, magenta for Michelle, blue for T-ReCS and red for VISIR data.
      Darker-coloured solid lines mark spectra of the corresponding instrument.
      The black filled circles mark the nuclear 12 and $18\,\mu$m  continuum emission estimate from the data.
      The ticks on the top axis mark positions of common MIR emission lines, while the light grey horizontal bars mark wavelength ranges affected by the silicate 10 and 18$\mu$m features.}
\end{figure}
\clearpage

\twocolumn[\begin{@twocolumnfalse}  
\subsection{Z\,41-20 -- CGCG\,41-20 -- SWIFT\,J1200.8+0650}\label{app:Z041-020}
Z\,41-20 is a galaxy at low-inclination at a redshift of $z=$ 0.0360 ($D\sim170\,$Mpc) with a little-studied Sy\,2 nucleus \citep{veron-cetty_catalogue_2010}, discovered with \textit{Swift}/BAT in X-rays \citep{landi_agn_2007}.
It belongs to the nine-month BAT AGN sample.
This object remained undetected in the MIR with \irass and appears as a compact source in the \wisee images.
A \spitzer/IRS LR staring-mode spectrum is available \citep{sargsyan_infrared_2011} and exhibits silicate 10\,$\mu$m absorption, strong PAH features, and a red spectral slope in $\nu F_\nu$-space.
Thus, the arcsecond-scale MIR SED indicates significant star formation in Z\,41-20. 
We imaged this object with VISIR in three narrow $N$-band filters in 2009 and weakly detected a compact nucleus.
The low S/N of the detection does not allow for a quantitative extension analysis.
The nuclear VISIR is on average $\sim50\%$ lower than the IRS spectrum.
Therefore, the total MIR emission of Z\,41-20 is indeed star formation dominated.
The nuclear MIR SED indicates much weaker PAH 11.3\,$\mu$m emission in the central $\sim0.3$\,kpc.
\newline\end{@twocolumnfalse}]

\begin{figure}
   \centering
   \includegraphics[angle=0,width=8.500cm]{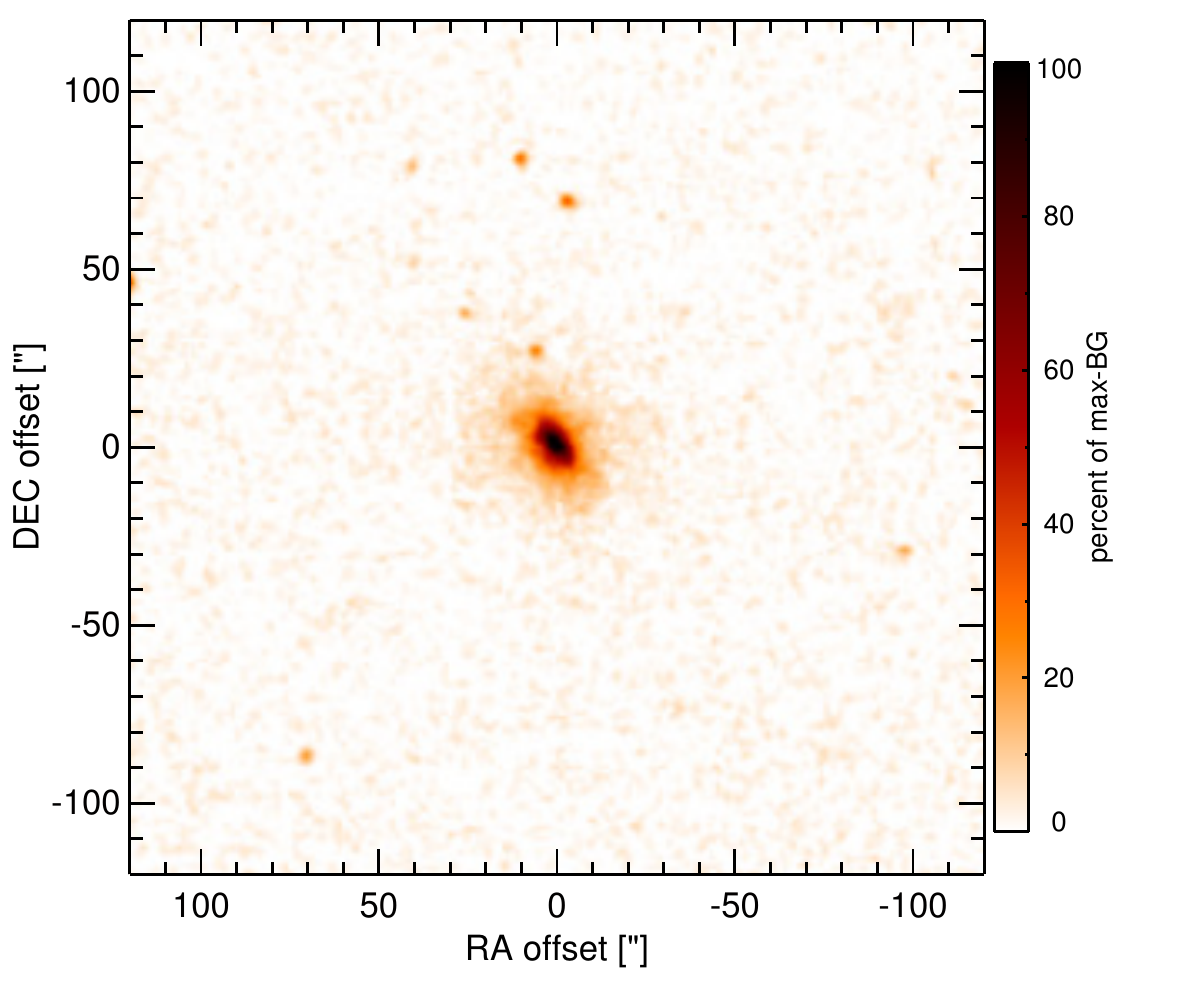}
    \caption{\label{fig:OPTim_Z041-020}
             Optical image (DSS, red filter) of Z\,41-20. Displayed are the central $4\arcmin$ with North up and East to the left. 
              The colour scaling is linear with white corresponding to the median background and black to the $0.01\%$ pixels with the highest intensity.  
           }
\end{figure}
\begin{figure}
   \centering
   \includegraphics[angle=0,height=3.11cm]{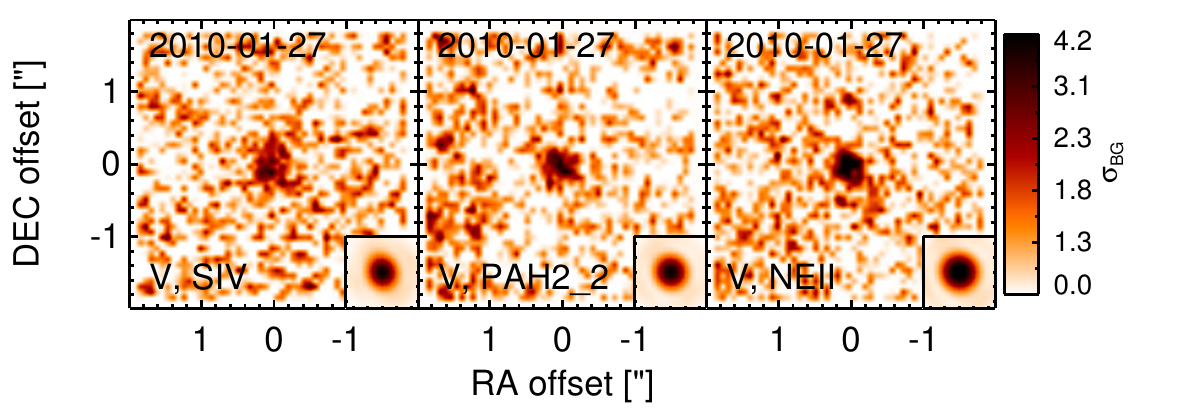}
    \caption{\label{fig:HARim_Z041-020}
             Subarcsecond-resolution MIR images of Z\,41-20 sorted by increasing filter wavelength. 
             Displayed are the inner $4\arcsec$ with North up and East to the left. 
             The colour scaling is logarithmic with white corresponding to median background and black to the $75\%$ of the highest intensity of all images in units of $\sigbg$.
             The inset image shows the central arcsecond of the PSF from the calibrator star, scaled to match the science target.
             The labels in the bottom left state instrument and filter names (C: COMICS, M: Michelle, T: T-ReCS, V: VISIR).
           }
\end{figure}
\begin{figure}
   \centering
   \includegraphics[angle=0,width=8.50cm]{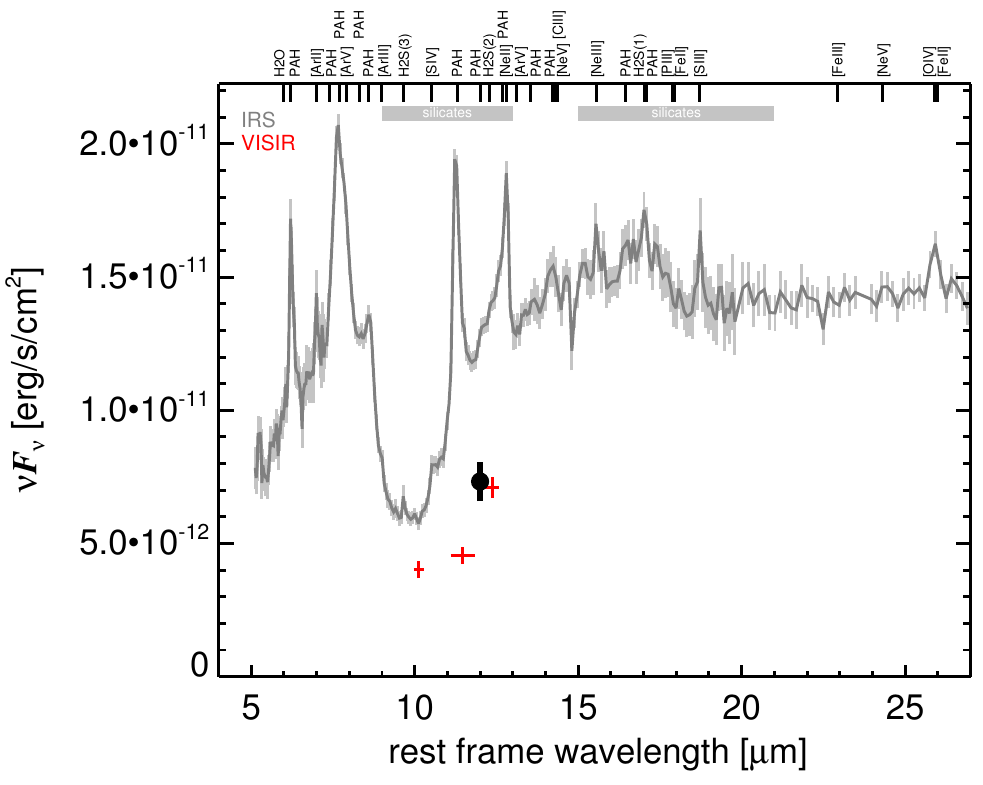}
   \caption{\label{fig:MISED_Z041-020}
      MIR SED of Z\,41-20. The description  of the symbols (if present) is the following.
      Grey crosses and  solid lines mark the \spitzer/IRAC, MIPS and IRS data. 
      The colour coding of the other symbols is: 
      green for COMICS, magenta for Michelle, blue for T-ReCS and red for VISIR data.
      Darker-coloured solid lines mark spectra of the corresponding instrument.
      The black filled circles mark the nuclear 12 and $18\,\mu$m  continuum emission estimate from the data.
      The ticks on the top axis mark positions of common MIR emission lines, while the light grey horizontal bars mark wavelength ranges affected by the silicate 10 and 18$\mu$m features.}
\end{figure}
\label{lastpage}

\end{document}